\documentclass[12pt]{iopart}

\usepackage{bm}
\usepackage{setstack}
\usepackage{amssymb}

\newcommand{\SSS}{\hat{S}}

\newcommand{\tfrac}[2]{{\textstyle{\frac{#1}{#2}}}}
\newcommand{\thf}{\tfrac{1}{2}}
\newcommand{\thq}{\tfrac{1}{4}}
\newcommand{\thh}{\tfrac{1}{8}}
\newcommand{\bfr}{\mathbf r}
\newcommand{\bfs}{\mathbf s}

\newcommand{\bfnab}{\boldsymbol\nabla}
\newcommand{\mathi}{{\mathrm i}}
\newcommand{\bfj}{\mathbf j}
\newcommand{\bfJ}{\mathbf J}
\newcommand{\bbJ}{\mathsf J}
\newcommand{\bfT}{\mathbf T}

\begin{document}

\title[Nonlocal low-energy nuclear EDF]{Nonlocal energy density functionals for low-energy nuclear structure}

\author{F Raimondi$^1$, K Bennaceur$^{2,3}$, J Dobaczewski$^{3,4}$}

\address{$^1$TRIUMF, 4004 Wesbrook Mall, Vancouver, British Columbia V6T 2A3, Canada}
\address{$^2$ Universit\'e de Lyon, F-69003 Lyon, France;
Institut de Physique Nucl\'eaire de Lyon, CNRS/IN2P3, Universit\'e Lyon 1,
F-69622 Villeurbanne Cedex, France}
\address{$^3$ Department of Physics, PO Box 35 (YFL), FI-40014
University of Jyv{\"a}skyl{\"a}, Finland}
\address{$^4$ Institute of Theoretical Physics, Faculty of Physics, University of Warsaw, ul. Ho{\.z}a 69, PL-00681 Warsaw, Poland}

\ead{fraimondi@triumf.ca}
\begin{abstract}
We introduce a finite-range pseudopotential built as an expansion in
derivatives up to next-to-next-to-next-to-leading order (N$^3$LO) and
we calculate the corresponding nonlocal energy density functional
(EDF). The coupling constants of the nonlocal EDF, for both finite
nuclei and infinite nuclear matter, are expressed through the
parameters of the pseudopotential. All central, spin-orbit, and
tensor terms of the pseudopotential are derived both in the
spherical-tensor and Cartesian representation. At next-to-leading
order (NLO), we also derive relations between the nonlocal EDF
expressed in the spherical-tensor and Cartesian formalism.
Finally, a simplified version of the finite-range pseudopotential is
considered, which generates the EDF identical to that generated by a
local potential. \end{abstract}

\maketitle

\section{Introduction}

Since the discovery of the neutron by Chadwick in
1932, the quest for nuclear interactions has been driving an intense research in
nuclear physics. Nuclei are quantum objects composed of correlated components,
the nucleons, whose mutual interactions are governed by the low-energy limit of quantum
chromodynamics. On the one hand, the contact between the field theory of quarks and
gluons and low-energy nuclear structure has been set in the
framework of chiral effective field theory~\cite{[Epe09],[Mac11]}. On
the other hand, the concept of an effective theory (ET) is powerful enough to
embrace a larger set of nuclear structure models, including
nuclear potentials employed in phenomenological self-consistent
approaches~\cite{[Dob12a]}.

One of the main features of nuclear forces in the vacuum is the fact that
nucleons interact over distances that are comparable with their sizes.
Therefore, even when contact or zero-range forces might be adopted for
computational reasons, to obtain accurate description of nuclear
observables the introduction of finite-range effects is mandatory.
In the domain of phenomenological models, different approaches resort
to some degree of approximation in describing how nucleons
effectively interact in nuclear medium. In mean-field calculations, the use of
finite-range interactions was pioneered by
Brink and Boecker~\cite{[Bri67w]} and extended by
Gogny~\cite{[Gog75],[Gog75c]}, who argued that long and intermediate
range of the nuclear in-medium effective force should be included explicitly.
Nonetheless, the zero-range approximation
was also widely and successfully employed in mean-field calculations, e.g., for determining
binding energies and radii, using the zero-range Skyrme
interactions~\cite{[Sky56],[Sky59]}. In this case, the finite-range
character of the interaction was mimicked through a dependence on
relative momenta of the interacting nucleons.

At present, an increased availability of powerful computational
resources makes the option of finite-range interaction more and more
viable. Furthermore, we tend to regard phenomenological interactions
as low-energy in-medium manifestations of the nuclear force, rather
than competing and mutually-exclusive approximations of an underlying
'true' potential.  This explains the intense ongoing work devoted to
improving the form of the phenomenological energy density functionals
and interactions~\cite{[Car08],[Gez10],[Rai11a],[Rai11b],[Sad13]},
along with the recent progress in elaborating new parametrizations of
existing ones~\cite{[Kor10b],[Kor12],[Was12]}. The zero-range
standard Skyrme interaction has been enriched with higher-order
derivative terms~\cite{[Rai11a]}, whereas
extensions to three-body component of the force have been envisaged
in both nonlocal~\cite{[Gez10]} and contact~\cite{[Sad13]} forces.

In a recent work~\cite{[Dob12a]}, we showed that it is possible to
apply in a consistent way the ET methodology to
low-energy nuclear physics. There we presented a proof-of-principle study
concerning the convergence, independence of regularization scale, and
naturalness of a new class of effective interactions, introduced as
regularized zero-range pseudopotentials. In another study~\cite{[Ben13]},
we introduced a regularized interaction that was a genuine force,
that is, it was density independent, and built as an expansion in relative momenta.
This latter interaction has proven to be able to capture the
relevant physics in the mechanism of saturation but the effective mass.
The absence of
density-dependent term makes the regularized pseudopotentials suitable
for beyond-mean-field calculations, without divergences at high momenta
and other pathological behaviors affecting the functionals
in multi-reference EDF calculations~\cite{[Dob07d],[Dug09]}, that is,
when standard techniques, like symmetry restoration or Generator Coordinate
Method in its most general form, are employed.

In this paper we introduce the general formalism describing the finite-range
regularized pseudopotential and nonlocal EDFs obtained by the
averaging procedure over the uncorrelated nuclear wavefunction. By
considering a finite-range interaction, we get a functional which is
bilinear in densities that are built by coupling
differential operator to the one-body density matrix. We consider
here the expansion in derivatives up the sixth order, corresponding
to the N$^3$LO functional. Such a functional is genuinely nonlocal,
as compared to the quasilocal form of the EDF stemming from the
zero-range pseudopotential. The derived EDF is
richer in numbers of terms and variety of tensorial couplings of
densities to bilinear scalars, as compared to its quasilocal version.
This implies, in principle, a functional that has the
potential of being more flexible and predictive, but it also implies a
larger parameter space to be optimized to nuclear data.
The derived regularized pseudopotentials,
along with the corresponding nonlocal EDFs, are ready for the optimization procedure,
which is the natural follow-up of this formal development,
and is now being carried out using the new implementations built
within the solver HFODD~\cite{[Sch12],[Sch14],[Dob07c]}.

The contents of the paper is as follows. In Sec.~\ref{NL_EDF}
we present the construction of the nonlocal EDF from
the finite-range pseudopotential. In particular, in Sec.~\ref{Pseudo}
we introduce the main features of the
finite-range pseudopotential and in Sec.~\ref{EDF} we present the
nonlocal EDF up to sixth order in derivatives. In Sec.~\ref{NuMatt}
we reduce our results to the case of infinite nuclear matter, whereas
in Sec.~\ref{Cartesian} we compare the NLO EDF derived in the spherical-tensor
formalism to that derived in the Cartesian representation. In
Sec.~\ref{Central} we consider a reduced version of the finite-range
pseudopotential, for which only the central part of the force is taken
into account and a simplified dependence on the relative momenta is
assumed. Section~\ref{SOtensor} presents construction of higher-order
spin-orbit (SO) and tensor terms in the Cartesian representation.
Finally, our conclusions are presented in Sec.~\ref{end}.

\section{General form of the nonlocal nuclear functional in
spherical-tensor formalism\label{NL_EDF}}

In this section, we first build the finite-range momentum-dependent
two-body pseudo\-potential, based on general symmetry conditions, and
next we derive a nonlocal nuclear EDF by the Hartree-Fock (HF) averaging of the
pseudopotential over an uncorrelated many-body wavefunction.
In this sense, the pseudopotential can be regarded as an EDF generator,
and not as a true in-medium interaction. The derivation from the
pseudopotential guarantees the functional to be free from the self-interaction problem. Indeed, after the EDF is
built, all results are obtained by minimizing the EDF with respect
to Kohn-Sham orbitals, whereupon the underlying pseudopotential
is never more explicitly invoked. Moreover, the proposed nonlocal EDF
is meant to be an approximation of the exact functional whose existence is
guaranteed by the variational principle. As such, the corresponding results
must be regarded as full solutions and not as pertaining to the
HF approximation, which is the first order of the many-body perturbation theory.

\subsection{Finite-range pseudopotential with higher-order momenta\label{Pseudo}}
The pseudopotential that we introduce here can be regarded as a
common extension of the most successful nuclear phenomenological
interactions: the Skyrme and Gogny forces. Indeed, the central part of our
pseudopotential is finite-range and momentum-dependent, whereas the one of the
Skyrme force is momentum-dependent but zero-range and the one of the Gogny
force is finite-range but momentum-independent. Similarly as for the Skyrme
force, we build the pseudopotential in terms of an expansion in
relative momenta, and similarly as for the Gogny force, we use
finite-range formfactors. We note here that the constructed
pseudopotential has only one finite-range parameter, allowing us to
interpret that parameter as a regularization scale, according to the
ET philosophy~\cite{[Dob12a]}.

Following closely the methodology of introducing higher-order
derivatives of zero-range pseudopotentials~\cite{[Rai11a]}, we obtain
the general form of the finite-range pseudopotential by coupling
derivative operators with spin operators in the spherical-tensor
formalism. The complete list of the building blocks of the tensors
can be found in Ref.~\cite{[Car08]}. We obtain the
finite-range version by simply substituting the zero-range Dirac
delta with a regularized (smeared) delta,
\begin{equation}
\label{reg_delta}
g_a(\bm{r})=g_a(r)=
\frac{e^{-\frac{r^2}{a^2}}}{\left(a\sqrt{\pi}\right)^3}  ,
\end{equation}
where $r=|\bm{r}|=|\bm{r}_2-\bm{r}_1|$ and $\bm{r}_2$  and $\bm{r}_1$ are positions of the two
interacting particles.

In such a formalism, the general expression for the antisymmetrized finite-range
pseudopotential reads,
\begin{equation}
 \hat{V}=\sum_{\tilde{n}' \tilde{L}', \tilde{t} \atop
 \tilde{n} \tilde{L}, v_{12} \tilde{S}}
  C_{\tilde{n} \tilde{L},v_{12} \tilde{S}}^{\tilde{n}' \tilde{L}',  \tilde{t} }
\hat{V}_{\tilde{n} \tilde{L},v_{12} \tilde{S}}^{\tilde{n}' \tilde{L}',  \tilde{t} }
\label{eq:0finiterange},
\end{equation}
where each term, which is built according to the symmetries discussed
in Ref.~\cite{[Rai11a]} and multiplied by the corresponding strength
parameter $C_{\tilde{n} \tilde{L},v_{12} \tilde{S}}^{\tilde{n}'
\tilde{L}',  \tilde{t} }$, is given by
\begin{eqnarray}
 \nonumber \hat{V}_{\tilde{n} \tilde{L},v_{12} \tilde{S}}^{\tilde{n}' \tilde{L}', \tilde{t} }=
&&\frac{1}{2}i^{v_{12}} \left( \left[ \left[K'_{\tilde{n}'\tilde{L}'}
                                            K_ {\tilde{n }\tilde{L}}\right]_{\tilde{S}}
\SSS_{v_{12} \tilde{S}}\right]_{0} \right. \nonumber \\
&& \nonumber  \left. + (-1)^{v_{12}+\tilde{S}} \left[ \left[K'_{\tilde{n} \tilde{L}}
                                        K_ {\tilde{n}'\tilde{L}'}\right]_{\tilde{S}}
\SSS_{v_{12} \tilde{S}}\right]_{0}   \right)\left(\hat{P}^{\tau}\right)^{ \tilde{t} } \\
&&\times \left(1-\hat{P}^{M}\hat{P}^{\sigma}\hat{P}^{\tau}\right)
\delta(\bm{r}'_1\!-\! \bm{r}_1)
\delta(\bm{r}'_2\!-\! \bm{r}_2) g_a(r)
\label{eq:1finiterange}.
\end{eqnarray}

For the sake of completeness, we recall here definitions of
tensors that appear in Eq.~(\ref{eq:1finiterange}). Each term of the
pseudopotential is a scalar with a composite internal structure and is
built of higher-rank spherical tensors in the space, spin, and isospin
coordinates. Higher-order tensor derivatives $K_ {\tilde{n }\tilde{L}}$
are built of
relative momenta $\bm{k}=(\bm{\nabla}_1-\bm{\nabla}_2)/2i$, with spherical
components
\begin{equation}
\label{eq:05a}
k_{1,\mu=\left\{-1,0,1\right\}} =
         -i       \left\{\tfrac{ 1}{\sqrt{2}}\left(k_x-ik_y\right),   k_z,
                         \tfrac{-1}{\sqrt{2}}\left(k_x+ik_y\right)\right\} ,
\end{equation}
that are coupled to rank $\tilde{L}$, where symbol $\tilde{n}$
denotes the order of the tensor, that is, the number of operators
$\bm{k}$. Operators acting on the primed coordinates, $K'_{\tilde{n}
\tilde{L}}$, are built in the same way of relative momenta
$\bm{k}'=(\bm{\nabla}'_1-\bm{\nabla}'_2)/2i$.

In the spin space, for particles numbered by $i=$1 or 2, we have as building
blocks the spin unity matrix of rank 0, denoted as
$\sigma^{(i)}_{00}$, and
the Pauli matrices $\sigma^{(i)}_{ x,y,z}$, which
define the spin spherical tensor of rank 1,
\begin{equation}
\label{eq:08a}
\sigma^{(i)}_{ 1,\mu=\left\{-1,0,1\right\}} =
     -i \left\{\tfrac{ 1}{\sqrt{2}}\left(\sigma^{(i)}_{ x}
                                       -i\sigma^{(i)}_{ y}\right),    \sigma^{(i)}_{ z},
               \tfrac{-1}{\sqrt{2}}\left(\sigma^{(i)}_{ x}
                                       +i\sigma^{(i)}_{ y}\right)\right\}.
\end{equation}
From these definitions, we can construct tensors of rank up to 2,
according to the following symmetrized expression
\begin{eqnarray}
\label{eq:spin}
\SSS_{v_{12} \tilde{S}} =\left(1-\tfrac{1}{2}\delta_{v_1,v_2}\right)\left(
[\sigma^{(1)}_{v_1}\sigma^{(2)}_{v_2}]_{\tilde{S}} +
[\sigma^{(1)}_{v_2}\sigma^{(2)}_{v_1}]_{\tilde{S}} \right),
\end{eqnarray}
where $v_{12}=v_1+v_2$ and $\sigma^{(i)}_{v}$ are the spherical-tensor
rank-$v$ Pauli matrices.

The locality delta functions $\delta(\bm{r}'_1\!-\!
\bm{r}_1)\delta(\bm{r}'_2\!-\! \bm{r}_2)$ together with the
regularized delta $g_a(r)$ must be understood as distributions, whereupon the
derivatives in $K_ {\tilde{n }\tilde{L}}$ and $K'_ {\tilde{n
}\tilde{L}}$ act. The locality deltas ensure the locality of the
energy density only in the zero-range case, whereas it
is in principle nonlocal due to the presence of the
so called Majorana operator $\hat{P}^{M}$ in the exchange part of the
pseudopotential.

In Eq.~(\ref{eq:1finiterange}), index $\tilde{t}$, taking either
value 0 or 1, defines the dependence of the pseudopotential on the isospin.
This dependence is in addition to that stemming from the
exchange term, and is a new element as compared to
the case of the zero-range pseudopotential~\cite{[Rai11a]}.

In principle, one could write down an equivalent version of the
antisymmetrized
finite-range pseudopotential, where the new dependence on the isospin
exchange operator would have been substituted by the Majorana
operator exchanging space coordinates, $\hat{P}^{M}$. This would be
possible in virtue of the relation
$\hat{P}^{\tau}\equiv-\hat{P}^{\sigma}\hat{P}^{M}$, valid when acting on a
fermionic wave function. However, the presence
of the Majorana operator in the pseudopotential would break the
correspondence between the exchange part of the antisymmetrized
pseudopotential and the nonlocal sector of the EDF, according to the common
understanding of the nonlocal part of the EDF as resulting from the exchange
part of the antisymmetrized interaction. While in a zero-range pseudopotential
this operator reduces to a phase depending on the order of the relative-momenta
tensors~\cite{[Per04],[Rai11a]}, in a finite-range interaction this
operator acts explicitly, and switches the space coordinates of the
two interacting nucleons. Therefore the finite-range character of the
interaction, complemented with the action of the Majorana operator,
would have given rise to the nonlocal form of the functional stemming
from the direct interaction. Although such a construction would be
perfectly correct and exactly equivalent to that presented below, it
would also be counterintuitive, and thus in our study it is not further
pursued.

The constructed pseudopotential possesses all symmetries of the
nuclear interaction. Referring the reader to Ref.~\cite{[Rai11a]} for
a comprehensive discussion, here we only list these symmetries, which are
rotational symmetry, time-reversal, parity, and Galilean invariance,
supplemented with the hermiticity of the pseudopotential operator and
indistinguishability principle operating as invariance under exchange
of particles 1 and 2. The pseudopotential~(\ref{eq:0finiterange}) is
indeed invariant under all the transformations corresponding to the
symmetries listed above. This guarantees that the resulting EDF is
also invariant with respect to the same symmetries.

All terms of the pseudopotential, derived at zero, second, fourth,
and sixth order, $\tilde{n}+\tilde{n}'=0$, 2, 4, and 6, respectively,
are collected in the supplemental material. The supplemental
material contains all detailed results of the present study. They
were derived by means of symbolic programing, and are presented
in readable format as well as in the form directly usable in
computer programming.
In \Tref{tab_pseudo} we
show numbers of different terms of pseudopotential
(\ref{eq:0finiterange}), where we also distinguish between central
($\tilde{S}=0$), SO ($\tilde{S}=1$), and tensor
($\tilde{S}=2$) terms, corresponding to different ranks in the
coupling of the relative momenta tensors $K_ {\tilde{n }\tilde{L}}$
with the spin operator $\SSS_{v_{12} \tilde{S}}$. At each order, the
overall numbers of terms equal 4, 14, 30, and 52, giving the total
number of 100 terms up to N$^3$LO. At each order, these numbers are
twice larger than the corresponding numbers for the zero-range
pseudopotential, which reflects the addition of the new quantum
number $\tilde{t}$.

We note, that here we classified tensor terms as those corresponding
to $\tilde{S}=2$; however, in the pseudopotential
(\ref{eq:1finiterange}) there also appear terms that have purely
space-tensor character, e.g., those for $\tilde{S}=0$ and
$\tilde{L}=\tilde{L}'=2$. However, we know from the generalized
Cayley-Hamilton (GCH) theorem~\cite{[Roh10]}, that any scalar function of
two vectors $\bm{k}$ and $\bm{k}'$ must have form of a function of three elementary scalars
$\bm{k}^2$, $\bm{k}'^2$, and $\bm{k}'\cdot\bm{k}$. Therefore, the
central term ($\tilde{S}=0$) can always be recoupled to the form
where terms with $\tilde{L},\tilde{L}'>1$ do not appear. An explicit
construction of such a form is presented in the Cartesian coordinates
in Sec.~\ref{Cartesian}.

\begin{table}
\caption{\label{tab_pseudo}Numbers of terms of pseudopotential
(\ref{eq:0finiterange}) at different orders up to N$^3$LO. In the second,
third, and fourth column, numbers of central ($\tilde{S}=0$),
SO ($\tilde{S}=1$), and tensor ($\tilde{S}=2$) terms,
respectively, are displayed.}
\begin{indented}
\lineup
\item[]\begin{tabular}{@{}*{5}{l}}
\br
Order&$\tilde{S}=0$ &$\tilde{S}=1$ & $\tilde{S}=2$ & Total\cr
\mr
0 & 4 & 0 & 0 & 4\cr
2& 8 & 2 &  4 & 14\cr
4 & 16  & 4  & 10  &  30 \cr
6 & 24 & 8 & 20  & 52\cr
N$^3$LO &  52  &  14 &  34 & 100 \cr
\br
\end{tabular}
\end{indented}
\end{table}

\subsection{Nonlocal nuclear EDF\label{EDF}}

The interaction energy, which is the potential part of the EDF, is
derived by averaging the pseudopotential $\hat{V}$~(\ref{eq:0finiterange}) over the
uncorrelated nuclear wave function expressed as Slater determinant.
The functional obtained in this way is in general nonlocal, meaning
that it contains terms depending on one-body densities
non-diagonal with respect to the spatial coordinates.

If we leave out the exchange operator from the
pseudopotential~(\ref{eq:0finiterange}), we can define it as $\hat{V}=\mathcal{V}
\left(1-\hat{P}^{M}\hat{P}^{\sigma}\hat{P}^{\tau}\right)$. This allows us to
express the EDF in the following way,
\begin{eqnarray}\hspace*{-2cm}
{\cal E}&=&\frac{1}{2} \int  \rmd{\bm{r}'_1}\rmd{\bm{r}'_2}\rmd{\bm{r}_1}\rmd{\bm{r}_2}\,
\sum_{{\sigma_1}{\sigma_2} \atop {\sigma'_1}{\sigma'_2}}
\sum_{{\tau_1}{\tau_2} \atop {\tau'_1}{\tau'_2}}
\mathcal{V}(\bm{r}'_{1} \sigma'_1 \tau'_1 \bm{r}'_{2} \sigma'_2 \tau'_2, \bm{r}_{1} \sigma_1 \tau_1 \bm{r}_{2} \sigma_2 \tau_2) \nonumber \\\hspace*{-2cm}
&&(\rho(\bm{r}_{1} \sigma_1 \tau_1,\bm{r}'_{1} \sigma'_1 \tau'_1)\rho(\bm{r}_{2} \sigma_2 \tau_2,\bm{r}'_{2} \sigma'_2 \tau'_2)
  -\rho(\bm{r}_{2} \sigma_2 \tau_2,\bm{r}'_{1} \sigma'_1 \tau'_1)\rho(\bm{r}_{1} \sigma_1 \tau_1,\bm{r}'_{2} \sigma'_2 \tau'_2)),
  \label{e202bis}
\end{eqnarray}
where the two-body spin-isospin (non-antisymmetrized) matrix elements of $\mathcal{V}$ are defined as
\begin{eqnarray}
\label{eq:16a}
\mathcal{V} (\bm{r}'_{1} s'_1 t'_1 \bm{r}'_{2} s'_2 t'_2, \bm{r}_{1} s_1 t_1 \bm{r}_{2} s_2 t_2)
=\langle s'_1 t'_1, s'_2 t'_2 |\mathcal{V} (\bm{r}'_{1},\bm{r}'_{2},\bm{r}_{1}, \bm{r}_{2}) |s_1 t_1 , s_2 t_2\rangle  ,
\end{eqnarray}
and
$\rho(\bm{r}_{1} s_1 t_1,\bm{r}'_{1} s'_1 t'_1)$ and $\rho(\bm{r}_{2} s_2 t_2,\bm{r}'_{2} s'_2 t'_2)$,
are the one-body density matrices in spin-isospin channels.
In Eq.~(\ref{e202bis}), the two terms that are bilinear in densities
lead to the standard direct and exchange terms, which are
respectively, local and nonlocal in space coordinates.

We performed derivations of average energies (\ref{e202bis})
separately for all terms of the pseudopotential~(\ref{eq:0finiterange}).
The final result of this derivation is given by linear
combinations of terms of the EDF appearing on the rhs of the
following expression,
\begin{equation}
\langle  C_{\tilde{n} \tilde{L},v_{12} \tilde{S}}^{\tilde{n}' \tilde{L}',\tilde{t}}
\hat{V}_{\tilde{n} \tilde{L},v_{12} \tilde{S}}^{\tilde{n}' \tilde{L}',\tilde{t}} \rangle =
 \sum C_{a,\alpha,Q}^{a',\alpha',t,\mathcal{L}} T_{a, \alpha,Q}^{a',\alpha', t,\mathcal{L}}.
 \label{eq:17}
\end{equation}
In this expression, $C_{a,\alpha,Q}^{a',\alpha',t,\mathcal{L}} $ and
$T_{a, \alpha,Q}^{a',\alpha', t,\mathcal{L}} $ denote, respectively,
the coupling constants and terms of the EDF according to the compact
notation introduced in  Ref.~\cite{[Car10d]}, where the Greek indices
$\alpha=\left\{ n_{\alpha}S_{\alpha}v_{\alpha}J_{\alpha}\right\}$ and
Roman indices $a=\left\{m_a I_a\right\}$ combine all the quantum
numbers of the local densities $\rho_{\alpha}(\bm{r})$ and derivative
operators $D_{a}$, as defined below. The spherical-tensor formalism
for the higher-order EDF has been developed in Ref.~\cite{[Car08]}.
Because here we treat the isospin degree of freedom explicitly, and
because we deal with a nonlocal functional, we enriched the notation
by adding superscripts $t$, which denote the isoscalar ($t=0$) or
isovector ($t=1$) channels, and by adding labels $\mathcal{L}$ that
distinguish between local ($\mathcal{L}=L$) and nonlocal
($\mathcal{L}=N$) terms of the functional. Index $Q$ corresponds to
the total rank of densities, which are coupled to a scalar.

The formalism developed in Ref.~\cite{[Car08]}, along with the
straightforward extension to the isospace introduced in
Ref.~\cite{[Rai11b]}, which originally pertained to quasilocal
higher-order EDF and zero-range pseudopotential, can easily be
accommodated to express the EDF discussed in this paper. Then, direct (local) terms
of the functional, bilinear in local densities, read
\begin{eqnarray}
\label{eq:termEDF2_non_local}
T^{a',\alpha', t,L}_{a,\alpha,Q}=\int \rmd{\bm{r}_1}\rmd{\bm{r}_2}\,
g_a(\bm{r}) \left[ \left[\left[D_{a'} \rho_{\alpha'}^t(\bm{r}_1)\right]_Q [D_{a}\rho_{\alpha}^t(\bm{r}_2)]_{Q} \right]^0 \right]_0.
\end{eqnarray}
They have been obtained using the integration by parts to transfer
all derivatives onto the density matrices, and then employing the
locality deltas to perform integrations over two out of four space
coordinates. In Eq.~(\ref{eq:termEDF2_non_local}),
subscripts and superscripts denote the standard
coupling of the angular momentum and isospin, respectively, and
the higher-order derivative operators $D_{a}$ of order $m_a$ and rank $I_a$,
are built from the order-one, rank-one derivative operators,
\begin{equation}
\label{D11LOC}
D_{11}= 2\bfnab_{\!1} \quad\mbox{or}\quad
D_{11}= 2\bfnab_{\!2},
\end{equation}
depending on whether they act on variables $\bm{r}_1$ or $\bm{r}_2$.

In an analogous way, exchange (nonlocal) terms of the functional read
\begin{eqnarray}
\label{eq:termEDF2_non_local_bili_nonloc}
T^{a',\alpha', t,N}_{a,\alpha,Q}&=&\!\!\int\!\! \rmd{\bm{r}_1}\rmd{\bm{r}_2}\,
g_a(\bm{r})  \left[ \left[ \left[D_{a'}\rho_{\alpha'}^t(\bm{r}_{1},\bm{r}_{2})\right]_Q [D_{a}\rho_{\alpha}^t(\bm{r}_{2},\bm{r}_{1})]_{Q} \right]^0 \right]_0,
\nonumber \\
\end{eqnarray}
where the nonlocal densities are defined as,
\begin{equation}
\label{eq:nLvJ_nonloc}
\rho_{\alpha}^{t}(\bm{r}_{1},\bm{r}_{2})=[K_{nS}\rho_v^t(\bm{r}_{1},\bm{r}_{2})]_J ,
\end{equation}
with the order-$n$ and rank-$S$ relative derivative operators
$K_{nS}$ acting on nonlocal densities and built from the relative-momentum operators,
\begin{equation}
\label{K11}
K_{11}= \frac{1}{2 \mathi } (\bfnab_{\!1}-\bfnab_{\!2}).
\end{equation}
Derivative operators $D_{a}$ of order $m_a$ and rank $I_a$,
are built from the building blocks,
\begin{equation}
\label{D11}
D_{11}= \bfnab_{\!1}+\bfnab_{\!2}.
\end{equation}
We do not use different notations for operators (\ref{D11LOC}) and (\ref{D11}) --
which one of them is used is clear from the context.

The coupling constants of the functional that we consider in the
present study do not depend on densities. Therefore, in principle, in
expressions~(\ref{eq:termEDF2_non_local}) and
(\ref{eq:termEDF2_non_local_bili_nonloc}) one could perform
integrations by parts and thus, in an attempt to achieve the same
form as that for the quasilocal EDF considered in
Ref.~\cite{[Car08]}, transfer the derivative operators $D_{a'}$ onto
the second density. However, this transformation would have created a
series of extra terms produced by the action of the derivative
operators onto the regularized delta, in such a way that the
functional would have had a mixed form composed of derivatives of
densities and an expansion in the parameter $a$ of~(\ref{reg_delta}).
For this reason, we keep the form of
Eqs.~(\ref{eq:termEDF2_non_local})
and~(\ref{eq:termEDF2_non_local_bili_nonloc}) with the derivatives
operators acting on both densities.

Explicit
calculations of linear combinations in Eq.~(\ref{eq:17})
are formally identical to those performed for the
zero-range pseudopotential (see Sec.~III of Ref.~\cite{[Rai11a]} for
details). In this Section, we present general results for the
finite-range pseudopotential (\ref{eq:0finiterange}), which we also supplement
by those pertaining to the symmetric spin-saturated nuclear matter.
\Tref{tableEDF} lists the numbers of independent
terms~(\ref{eq:termEDF2_non_local})
and~(\ref{eq:termEDF2_non_local_bili_nonloc})  of the functional
obtained from the finite-range pseudopotential. Each allowed
combination of indexes ($a',\alpha', t,a,\alpha,Q$) gives a pair of
EDF terms, one local and another one nonlocal; therefore, the numbers shown
are twice the numbers of such allowed combinations. The fact that the
isospin does not couple with operators belonging to spin and
position-coordinate space, along with the requirement that the EDF is
isoscalar, implies that isoscalar and isovector densities,
respectively $t$=0 and $t$=1,
give rise to two isospin channels in the
functional having the same structure. Therefore only one isospin
channel is accounted for in~\tref{tableEDF}. For the symmetric nuclear
matter, the isovector terms
do not contribute.

\Tref{tableEDF} also displays numbers of EDF
independent terms obtained separately from the central
($\tilde{S}$=0), SO ($\tilde{S}$=1), and tensor ($\tilde{S}$=2) terms of the
finite-range pseudopotential. Strictly speaking, the EDF cannot be
divided into central and tensor contributions, even though the SO
part of the EDF is decoupled from the other two. This is at
the origin of the functional terms that mix scalar and vector densities in
spin spaces. However there are terms of the EDF that can be produced by
both central and tensor terms of the pseudopotential. This explains
why the sums of terms in the second, third, and fourth columns do not
equal to the corresponding values in the fifth column.

\begin{table}
\caption{\label{tableEDF}Numbers of terms defined in
Eqs.~(\ref{eq:termEDF2_non_local})
and~(\ref{eq:termEDF2_non_local_bili_nonloc}) of different orders in
the EDF up to N$^3$LO, given for one isospin channel. In the second,
third, and fourth columns, numbers terms stemming from central
($\tilde{S}$=0), SO ($\tilde{S}$=1), and tensor ($\tilde{S}$=2)
finite-range pseudopotential are given, respectively. The last column
gives the numbers of terms of the EDF when it is applied to the
symmetric spin-saturated nuclear matter.}
\begin{indented}
\lineup
\item[]\begin{tabular}{@{}*{6}{l}}
\br
Order&$\tilde{S}$=0&$\tilde{S}$=1& $\tilde{S}$=2 &Total &Nuclear matter\cr
\mr
0 &4&0& 0 &4 & 2\cr
2&24&8& 16  &36&3\cr
4 & 144   & 64   &  114  &222   & 8  \cr
6 &  640  &  336 &   564  &1010  &12 \cr
N$^3$LO &  776  &  408 &  694  &1272  & 25 \cr
\br
\end{tabular}
\end{indented}
\end{table}

\subsubsection{Zero-order (LO) EDF.\label{ZeroEDF}}

In order to illustrate how the nonlocal EDF derives from a
finite-range pseudopotential, we show explicitly the functional
obtained from the averaging of the zero-order finite-range
pseudopotential $\hat{V}^{(0)}$ over the nuclear Slater
determinant. Below, symbols $\hat{V}^{(n)}$ and ${\cal E}^{(n)}$
denote, respectively, the order-$n$ regularized pseudopotential and
functional.

The four terms of the pseudopotential at zero order are,
  \begin{equation}
 \hat{V}^{(0)}=C_{00,00}^{00,0}
\hat{V}_{00,00}^{00,0}+C_{00,00}^{00,1}
\hat{V}_{00,00}^{00,1}+C_{00,20}^{00,0}
\hat{V}_{00,20}^{00,0}+C_{00,20}^{00,1}
\hat{V}_{00,20}^{00,1}
\label{eq:Zero_order_pseudo}.
\end{equation}

The complete expression for the zero-order nonlocal EDF obtained
from~(\ref{eq:Zero_order_pseudo}) reads
\begin{eqnarray}
\label{eq:zero_order_EDF_final}
{\cal E}^{(0)}&=&
\left(\frac{C_{00,00}^{00,0}}{2}+\frac{C_{00,00}^{00,1}}{4} \right)  T^{00,0000, 0,L}_{00,000,0} \nonumber \\
\nonumber &&+\left( -\frac{C_{00,20}^{00,0}}{2} -\frac{C_{00,20}^{00,1}}{4}\right)  T^{00,0011, 0,L}_{00,0011,1}\\
\nonumber &&+\left(\frac{\sqrt{3}}{4}C_{00,00}^{00,1} \right) T^{00,0000, 1,L}_{00,0000,0}\\
\nonumber &&+\left( - \frac{\sqrt{3}}{4}C_{00,20}^{00,1}\right)  T^{00,0011, 1,L}_{00,0011,1}\\
\nonumber &&+
\left(-\frac{C_{00,00}^{00,0}}{8} - \frac{C_{00,00}^{00,1}}{4} +\frac{\sqrt{3}}{8}C_{00,20}^{00,0}+\frac{\sqrt{3}}{4}C_{00,20}^{00,1}\right) T^{00,0000, 0,N}_{00,000,0}\\
\nonumber &&+\left(-\frac{\sqrt{3}}{8}C_{00,00}^{00,0}-\frac{\sqrt{3}}{4}C_{00,00}^{00,1} -\frac{C_{00,20}^{00,0}}{8} -\frac{C_{00,20}^{00,1}}{4}\right)T^{00,0011, 0,N}_{00,0011,1}\\
 &&+\left(-\frac{\sqrt{3}}{8}C_{00,00}^{00,0}+\frac{3}{8}C_{00,20}^{00,0}  \right) T^{00,0000, 1,N}_{00,0000,0} \nonumber \\
 &&+\left(-\frac{3}{8}C_{00,00}^{00,0}-\frac{\sqrt{3}}{8}C_{00,20}^{00,0}  \right)  T^{00,0011, 1,N}_{00,0011,1}.
\end{eqnarray}
In Eq.~(\ref{eq:zero_order_EDF_final}), we have eight zero-order, 4 local
and 4 nonlocal, or 4 isoscalar and 4 isovector. coupling constants expressed explicitly as linear
combinations of the four zero-order pseudopotential parameters given
in Eq.~(\ref{eq:Zero_order_pseudo}).

\subsubsection{Beyond the LO EDF.\label{beyond_EDF}}

The 72 second-order isoscalar and isovector coupling constants of the
nonlocal EDF, expressed by the 14 second-order finite-range
pseudo\-potential parameters, are collected in the supplemental
material according to the following formula using the grouped index
defined as $(a,\alpha,Q,a', \alpha',t,\mathcal{L})\equiv \mathcal{A}$,
\begin{eqnarray}
\label{eq:sec_order_rel}
C_{a,\alpha,Q}^{a', \alpha',t,\mathcal{L}} &=&  \sum_{\tilde{t}=0}^1  a_{\mathcal{A},\tilde{t}} C_{00,00}^{20,\tilde{t}}
+ b_{\mathcal{A},\tilde{t}} C_{00,20}^{20,\tilde{t}}+  c_{\mathcal{A},\tilde{t}} C_{00,22}^{22,\tilde{t}}+
 d_{\mathcal{A},\tilde{t}} C_{11,00}^{11,\tilde{t}}+e_{\mathcal{A},\tilde{t}} C_{11,11}^{11,\tilde{t}} \nonumber  \\
&&  +f_{\mathcal{A},\tilde{t}} C_{11,20}^{11,\tilde{t}}+g_{\mathcal{A},\tilde{t}} C_{11,22}^{11,\tilde{t}}.
\end{eqnarray}
In Eq.~(\ref{eq:sec_order_rel}), and
in Eqs.~(\ref{eq:fourth_order_rel}) and (\ref{eq:sixth_order_rel}) below,
numerical coefficients
$a_{\mathcal{A},\tilde{t}}$,$b_{\mathcal{A},\tilde{t}}$,\ldots,
express the EDF coupling constants as linear combinations of the
finite-range pseudopotential parameters.

Similar expressions for the fourth-order (sixth-order)  444 (2020)
coupling constants that are related to 30 (52) parameters of the finite-range pseudopotential
can also be found in the supplemental material,
according to the formulas
\begin{eqnarray}
\label{eq:fourth_order_rel}
C_{a,\alpha,Q}^{a',\alpha',t,\mathcal{L}}&=& \sum_{\tilde{t}=0}^1  a_{\mathcal{A},\tilde{t}} C_{00,00}^{40,\tilde{t}}+b_{\mathcal{A},\tilde{t}} C_{00,20}^{40,\tilde{t}}+c_{\mathcal{A},\tilde{t}} C_{00,22}^{42,\tilde{t}}+ d_{\mathcal{A},\tilde{t}} C_{11,00}^{31,\tilde{t}}+e_{\mathcal{A},\tilde{t}} C_{11,11}^{31,\tilde{t}}  \nonumber \\ && + f_{\mathcal{A},\tilde{t}} C_{11,20}^{31,\tilde{t}}+g_{\mathcal{A},\tilde{t}} C_{11,22}^{31,\tilde{t}}+h_{\mathcal{A},\tilde{t}} C_{11,22}^{33,\tilde{t}}+i_{\mathcal{A},\tilde{t}} C_{20,00}^{20,\tilde{t}}+j_{\mathcal{A},\tilde{t}} C_{20,20}^{20,\tilde{t}}\nonumber \\
&& +k_{\mathcal{A},\tilde{t}} C_{20,22}^{22,\tilde{t}} +l_{\mathcal{A},\tilde{t}} C_{22,00}^{22,\tilde{t}}+  m_{\mathcal{A},\tilde{t}} C_{22,11}^{22,\tilde{t}}+n_{\mathcal{A},\tilde{t}} C_{22,20}^{22,\tilde{t}}+o_{\mathcal{A},\tilde{t}} C_{22,22}^{22,\tilde{t}},
\end{eqnarray}
and
 \begin{eqnarray}
\label{eq:sixth_order_rel}
C_{a,\alpha,Q}^{a', \alpha',t,\mathcal{L}}&=&  \sum_{\tilde{t}=0}^1 a_{\mathcal{A},\tilde{t}} C_{00,00}^{60,\tilde{t}}+b_{\mathcal{A},\tilde{t}} C_{00,20}^{60,\tilde{t}}+c_{\mathcal{A},\tilde{t}} C_{00,22}^{62,\tilde{t}}+d_{\mathcal{A},\tilde{t}} C_{11,00}^{51,\tilde{t}}+e_{\mathcal{A},\tilde{t}} C_{11,11}^{51,\tilde{t}} \nonumber \\ && + f_{\mathcal{A},\tilde{t}} C_{11,20}^{31,\tilde{t}}+g_{\mathcal{A},\tilde{t}} C_{11,22}^{51,\tilde{t}}+h_{\mathcal{A},\tilde{t}} C_{11,22}^{53,\tilde{t}}+i_{\mathcal{A},\tilde{t}} C_{20,00}^{40,\tilde{t}}+j_{\mathcal{A},\tilde{t}} C_{20,20}^{40,\tilde{t}} \nonumber \\
&& + k_{\mathcal{A},\tilde{t}} C_{20,22}^{42,\tilde{t}}+l_{\mathcal{A},\tilde{t}} C_{22,00}^{42,\tilde{t}}+m_{\mathcal{A},\tilde{t}} C_{22,11}^{42,\tilde{t}}+n_{\mathcal{A},\tilde{t}} C_{22,20}^{42,\tilde{t}}+o_{\mathcal{A},\tilde{t}} C_{22,22}^{40,\tilde{t}}\nonumber \\
&& +p_{\mathcal{A},\tilde{t}} C_{22,22}^{42,\tilde{t}}+ q_{\mathcal{A},\tilde{t}} C_{22,22}^{44,\tilde{t}}+r_{\mathcal{A},\tilde{t}} C_{31,00}^{31,\tilde{t}}+ s_{\mathcal{A},\tilde{t}} C_{31,11}^{31,\tilde{t}}+ u_{\mathcal{A},\tilde{t}} C_{31,20}^{31,\tilde{t}} \nonumber \\
&& + v_{\mathcal{A},\tilde{t}} C_{31,22}^{31,\tilde{t}}+w_{\mathcal{A},\tilde{t}} C_{31,22}^{33,\tilde{t}}+x_{\mathcal{A},\tilde{t}} C_{33,00}^{33,\tilde{t}}+y_{\mathcal{A},\tilde{t}} C_{33,11}^{33,\tilde{t}}+z_{\mathcal{A},\tilde{t}} C_{33,20}^{33,\tilde{t}} \nonumber \\
&& + \beta_{\mathcal{A},\tilde{t}} C_{33,22}^{33,\tilde{t}},
\end{eqnarray}
respectively.

Besides the coupling constants expressed through the linear combinations
in Eqs.~(\ref{eq:fourth_order_rel}) and (\ref{eq:sixth_order_rel}) there are
6 (38) more coupling constants for each isospin
channel at fourth-order (sixth-order),
corresponding to EDF terms that are allowed by the general symmetries of the functional, but that are forced to be equal to zero once the functional is derived from the pseudopotential. Therefore, the number of terms of the fourth-order (sixth-order) EDF when not derived from the pseudopotential is 228 (1048) in each isospin channel. The complete lists of
these fourth-order and sixth-order vanishing coupling constants are included
in the supplemental material. Indeed as we verified for the quasilocal EDF obtained from the zero-range
pseudopotential, also for the finite-range one, the explicit Galilean
invariance of the pseudopotential~(\ref{eq:1finiterange}) yields the
Galilean invariance of the derived nonlocal functional. The coupling constants
of the terms which are not Galilean invariant turn out to be equal to zero
when the derivation of the functional from the pseudopotential is performed.  Moreover using
expressions~(\ref{eq:sec_order_rel})--(\ref{eq:sixth_order_rel}), one
can, in principle, determine constraints between the nonvanishing coupling constants
of the EDF that enforce Galilean symmetry.
For example, at NLO this task can be realized by inverting two subsets of
14 linear combinations~(\ref{eq:sec_order_rel}), chosen arbitrarily
within each local and nonlocal part separately. The only requirement
is in selecting linear combinations that lead to a nonsingular matrix. In this way,
one can express the remaining coupling constants through the 14 selected
as independent ones. The two sets of linear
combinations, one for the local component and one for the nonlocal
component of the EDF, form together the constraints on the functional
obtained by imposing the Galilean invariance.

These linear combinations form subsets of equivalent relations in
each isospin space. This is owing to the specularity of the EDF with respect to the
isospin. In the local and nonlocal sectors of the functionals, this is owing
to the fact that linear combinations between EDF terms can be
established only for terms with the same kind of densities, local or
nonlocal.

Then, considering at NLO both isospin channels ($t=0, 1$) and both
local and nonlocal sectors of the functional ($\mathcal{L}=L, N$), 44
dependent coupling constants are equal to specific linear
combinations of the 28 ones, that is,
\begin{eqnarray}
C_{00,1110,0}^{00,1110,t,\mathcal{L}}&=& -\frac{C_{00,2011,1}^{00,0011,t,\mathcal{L}}}{3}-\frac{\sqrt{5} C_{00,2211,1}^{00,0011,t,\mathcal{L}}}{3} ,  \label{subeq: 1of1gal}  \\
C_{00,1111,1}^{00,1111,t,\mathcal{L}}  &=&   -\frac{C_{00,2011,1}^{00,0011,t,\mathcal{L}}}{\sqrt{3}}+\frac{1}{2} \sqrt{\frac{5}{3}} C_{00,2211,1}^{00,0011,t,\mathcal{L}} ,  \label{subeq: 1of2gal}  \\
C_{00,1112,2}^{00,1112,t,\mathcal{L}}   &=&  -\frac{\sqrt{5} C_{00,2011,1}^{00,0011,t,\mathcal{L}}}{3}-\frac{C_{00,2211,1}^{00,0011,t,\mathcal{L}}}{6} ,  \label{subeq: 1of3gal}  \\
C_{00,2000,0}^{00,0000,t,\mathcal{L}} &=&  -C_{00,1101,1}^{00,1101,t,\mathcal{L}} ,  \label{subeq: 1of4gal}  \\
C_{11,0000,1}^{00,1111,t,\mathcal{L}}  &=&  -C_{11,1101,1}^{00,0011,t,\mathcal{L}} ,  \label{subeq: 1of5gal}  \\
C_{11,0011,1}^{00,1101,t,\mathcal{L}}   &=&  C_{11,1101,1}^{00,0011,t,\mathcal{L}} ,  \label{subeq: 1of6gal}  \\
C_{11,0011,1}^{11,0011,t,\mathcal{L}}  &=&   -\frac{2 C_{11,0011,0}^{11,0011,t,\mathcal{L}}}{\sqrt{3}}+\sqrt{\frac{5}{3}} C_{11,0011,2}^{11,0011,t,\mathcal{L}}  ,  \label{subeq: 1of7gal}  \\
C_{11,1111,0}^{00,0000,t,\mathcal{L}}   &=&   C_{11,1101,1}^{00,0011,t,\mathcal{L}}  ,  \label{subeq: 1of8gal}  \\
C_{20,0000,0}^{00,0000,t,\mathcal{L}}   &=&  -C_{11,0000,1}^{11,0000,t,\mathcal{L}}  ,  \label{subeq: 1of9gal}  \\
C_{20,0011,1}^{00,0011,t,\mathcal{L}}    &=&  \frac{C_{11,0011,0}^{11,0011,t,\mathcal{L}}}{3}-\frac{2 \sqrt{5} C_{11,0011,2}^{11,0011,t,\mathcal{L}}}{3}  ,  \label{subeq: 1of10gal}  \\
C_{22,0011,1}^{00,0011,t,\mathcal{L}}  &=&   -\frac{2 \sqrt{5} C_{11,0011,0}^{11,0011,t,\mathcal{L}}}{3}+\frac{2 C_{11,0011,2}^{11,0011,t,\mathcal{L}}}{3}.  \label{subeq: 1of11gal}
\end{eqnarray}

Equations~(\ref{subeq: 1of1gal})--(\ref{subeq: 1of5gal}) are formally
equivalent to the constraints found for the same symmetry in the case
of quasilocal EDF~\cite{[Car08]}. This is so, because the
corresponding terms have exactly
the same tensor forms as those found in the functional stemming
from the zero-range pseudopotential. The remaining linear
combinations~(\ref{subeq: 1of6gal})--(\ref{subeq: 1of11gal}) have no
analogs in the quasilocal EDF, because they involve terms that are
absent in the quasilocal version of the functional. We also note that
there are no terms corresponding to unrestricted coupling constants; indeed,
the Galilean symmetry forces all the nonvanishing coupling constants to enter
in specific linear combinations.

\section{Nonlocal EDF for the symmetric spin-saturated nuclear matter\label{NuMatt}}

For the symmetric spin-saturated homogeneous nuclear matter, the EDF coupling
constants expressed in terms of the pseudopotential parameters can
easily be obtained from those derived for finite nuclei. Indeed,
in this former system, isovector densities are zero because of the
equality of proton and neutron densities, gradients of local densities
are zero because of its homogeneity, and all tensors of rank greater
than zero vanish because of its isotropy.

At zero-order, expressions relating the EDF non-vanishing coupling constants
to pseudopotential parameters can be read from the first and fifth lines
of Eq.~(\ref{eq:zero_order_EDF_final}). They explicitly read,
\begin{eqnarray}
\label{eq:zero_order_results}
C_{00,0000,0}^{00,0000,0,L}&=&\frac{C_{00,00}^{00,0}}{2}+\frac{C_{00,00}^{00,1}}{4} ,\\
C_{00,0000,0}^{00,0000,0,N}&=&-\frac{C_{00,00}^{00,0}}{8} - \frac{C_{00,00}^{00,1}}{4} +\frac{\sqrt{3}}{8}C_{00,20}^{00,0}+\frac{\sqrt{3}}{4}C_{00,20}^{00,1}.
\end{eqnarray}

For higher-order terms of the EDF,
we note that terms depending on gradients $D_{m 0}$ of nonlocal densities
do contribute provided that they are coupled to a scalar. There are 1, 4, and 8
such a terms at second, fourth, and sixth order, respectively;
below they are
shown in bold face. For these
densities, the local limit does not give rise to the cancellation of
the corresponding term of the functional.

At second-order, we obtain three nonvanishing coupling constants,
which are given by,
\begin{eqnarray}
\label{infinite_IV}
C_{00,2000,0}^{00,0000,0,L}&=& \frac{1}{4} C_{00,00}^{20,0}+\frac{1}{8}
C_{00,00}^{20,1}+
\frac{1}{4} C_{11,00}^{11,0}+\frac{1}{8} C_{11,00}^{11,1}\\
C_{00,2000,0}^{00,0000,0,N}&=& -\frac{1}{16} C_{00,00}^{20,0}-\frac{1}{8} C_{00,00}^{20,1}+\frac{1}{16} \sqrt{3} C_{00,20}^{20,0} +\frac{1}{8} \sqrt{3} C_{00,20}^{20,1}
\nonumber \\ && +\frac{1}{16} C_{11,00}^{11,0}+\frac{1}{8} C_{11,00}^{11,1} -\frac{1}{16} \sqrt{3} C_{11,20}^{11,0}-\frac{1}{8} \sqrt{3} C_{11,20}^{11,1}\\
\label{infinite_II} \bm{C_{20,0000,0}^{00,0000,0,\bf{N}}}&=&
\frac{1}{64}  C_{00,00}^{20,0}+\frac{1}{32} C_{00,00}^{20,1}
-\frac{1}{64} \sqrt{3} C_{00,20}^{20,0}-\frac{1}{32} \sqrt{3} C_{00,20}^{20,1}
\nonumber \\ &&+\frac{1}{64} C_{11,00}^{11,0}+\frac{1}{32} C_{11,00}^{11,1} -\frac{1}{64} \sqrt{3} C_{11,20}^{11,0}  -\frac{1}{32} \sqrt{3} C_{11,20}^{11,1}.
\end{eqnarray}

At fourth (sixth) order, there are 8 (12) non-vanishing
coupling constants, which are linear
combinations of the 16 (24) parameters corresponding to the central terms of the pseudopotential. These lengthy
expressions are collected in the supplemental material. Here we only list
these non-vanishing coupling constants, which at fourth and sixth order are
$C_ {00, 2000, 0}^{00, 2000, 0, L}$,
$C_ {00, 2000, 0}^{00, 2000, 0, N}$,
$C_ {00, 4000, 0}^{00, 0000, 0, L}$,
$C_ {00, 4000, 0}^{00, 0000, 0, N}$,
$\bm{C_ {20, 0000, 0}^{00, 2000,  0, \bf {N}}}$,
$\bm{C_ {20, 0000, 0}^{20, 0000,  0, \bf {N}}}$,
$\bm{C_ {20, 2000, 0}^{00, 0000,  0, \bf {N}}}$,
$\bm{C_ {40, 0000, 0}^{00, 0000,  0, \bf {N}}}$
and
$C_ {00, 4000, 0}^{00, 2000, 0, L}$,
$C_ {00, 4000, 0}^{00, 2000, 0, N}$,
$C_ {00, 6000, 0}^{00, 0000, 0, L}$,
$C_ {00, 6000, 0}^{00, 0000, 0, N}$,
$\bm{C_ {20, 0000, 0}^{00, 4000, 0, \bf {N}}}$,
$\bm{C_ {20, 2000, 0}^{00, 2000, 0, \bf {N}}}$,
$\bm{C_ {20, 2000, 0}^{20, 0000, 0, \bf {N}}}$,
$\bm{C_ {20, 4000, 0}^{00, 0000, 0, \bf {N}}}$,
$\bm{C_ {40, 0000, 0}^{00, 2000, 0, \bf {N}}}$,
$\bm{C_ {40, 0000, 0}^{20, 0000, 0, \bf {N}}}$,
$\bm{C_ {40, 2000, 0}^{00, 0000, 0, \bf {N}}}$,
$\bm{C_ {60, 0000, 0}^{00, 0000, 0, \bf {N}}}$, respectively.

\section{Cartesian representation of the central pseudopotential\label{Cartesian}}

In this section, we present derivation in the Cartesian representation
of the nonlocal EDF that stems from the central
($\tilde{S}$=0) component of the regularized
pseudopotential. This derivation is useful, because it establishes a
clear connection of results presented in Sec.~\ref{NL_EDF} with the
Skyrme functional, which, in fact, constitutes the local limit of the
nonlocal EDF. We used the explicit expressions for the Cartesian NLO
functional, derived from the central finite-range pseudopotential, to
benchmark their spherical counterparts. We checked explicitly that
the two representations are equivalent when the relations between
Cartesian and spherical local densities~\cite{[Car08]} and parameters
of the interactions~\cite{[Rai11a]} were applied.

Following Refs.~\cite{[Dob12a],[Ben13],[Dav13]}, we define the Cartesian form
of the (non-anti\-symmetrized) central
pseudopotential~(\ref{eq:0finiterange})--(\ref{eq:1finiterange})
in two equivalent representations as
\begin{eqnarray}
\label{eq:1}
 \mathcal{V}_C &=& \sum_{nj}
   \left(W^{(n)}_j \hat{1}_\sigma\hat{1}_\tau+B^{(n)}_j \hat{1}_\tau  \hat{P}^\sigma
        -H^{(n)}_j \hat{1}_\sigma\hat{P}^\tau-M^{(n)}_j \hat{P}^\sigma\hat{P}^\tau\right)
\nonumber \\
&&~~~~~~~~~~~~\times \hat{O}^{(n)}_j(\bm{k}',\bm{k})\delta(\bm{r}'_1-\bm{r}_1)\delta(\bm{r}'_2-\bm{r}_2)
g_a(\bm{r}_1-\bm{r}_2)
\end{eqnarray}
and
\begin{eqnarray}
\label{eq:1a}
 \mathcal{V}_C &=& \sum_{nj} t^{(n)}_j
   \left(          \hat{1}_\sigma\hat{1}_\tau+x^{(n)}_j \hat{1}_\tau  \hat{P}^\sigma
        -y^{(n)}_j \hat{1}_\sigma\hat{P}^\tau-z^{(n)}_j \hat{P}^\sigma\hat{P}^\tau\right)
\nonumber \\
&&~~~~~~~~\times \hat{O}^{(n)}_j(\bm{k}',\bm{k})\delta(\bm{r}'_1-\bm{r}_1)\delta(\bm{r}'_2-\bm{r}_2)
g_a(\bm{r}_1-\bm{r}_2) ,
\end{eqnarray}
which contain the standard identity ($\hat{1}_{\sigma,\tau}$) and
exchange ($\hat{P}^{\sigma,\tau}$) operators in spin and isospin
spaces. In these expressions, indexes $n$ denote the orders of
differential operators $\hat{O}^{(n)}_j(\bm{k}',\bm{k})$ and indexes $j$
number different operators of the same order. The first form,
Eq.~(\ref{eq:1}), generalizes the Gogny interaction by adding terms of
higher orders $n>0$, whereas the second form, Eq.~(\ref{eq:1a}),
generalizes the Skyrme interaction by adding two additional exchange
terms. Obviously, the two forms are simply related one to another by
the following relations between their
strength parameters:
$W^{(n)}_j=t^{(n)}_j$,
$B^{(n)}_j=t^{(n)}_j x^{(n)}_j$,
$H^{(n)}_j=t^{(n)}_j y^{(n)}_j$, and
$M^{(n)}_j=t^{(n)}_j z^{(n)}_j$.

Differential operators $\hat{O}^{(n)}_j(\bm{k}',\bm{k})$ are scalar
polynomial functions of two vectors, so owing to the GCH
theorem~\cite{[Roh10]}, they must be polynomials of three elementary
scalars: $\bm{k}^2$, $\bm{k}'^2$, and $\bm{k}'\cdot\bm{k}$. Hermiticity of
the operators $\hat{O}^{(n)}_j(\bm{k}',\bm{k})$ can be enforced by
using expressions symmetric with respect to exchanging $\bm{k}'^*$ and $\bm{k}$; therefore,
it is convenient to build them from the following three scalars,
\begin{eqnarray}
\label{eq:2}
 \hat{T}_1 &=& \tfrac{1}{2}(\bm{k}'^*{}^2+  \bm{k}^2) , \\
 \hat{T}_2 &=&              \bm{k}'^*  \cdot\bm{k}      , \\
 \hat{T}_3 &=& \tfrac{1}{2}(\bm{k}'^*{}^2-  \bm{k}^2) ,
\label{eq:2a}
\end{eqnarray}
with the condition that only even powers of $\hat{T}_3$ can appear.
In terms of $\hat{T}_1$,
$\hat{T}_2$, and  $\hat{T}_3$, we now can define the following
differential operators:
\begin{eqnarray}
\label{eq:3a}   \hat{O}^{(0)}_1(\bm{k}',\bm{k}) &=&\hat{1}                             , \\
\label{eq:3b}   \hat{O}^{(2)}_1(\bm{k}',\bm{k}) &=&\hat{T}_1                           , \\
\label{eq:3c}   \hat{O}^{(2)}_2(\bm{k}',\bm{k}) &=&\hat{T}_2                           , \\
                \hat{O}^{(4)}_1(\bm{k}',\bm{k}) &=&\hat{T}_1^2+ \hat{T}_2^2            , \\
                \hat{O}^{(4)}_2(\bm{k}',\bm{k}) &=&2\hat{T}_1   \hat{T}_2              , \\
                \hat{O}^{(4)}_3(\bm{k}',\bm{k}) &=&\hat{T}_1^2- \hat{T}_2^2            , \\
                \hat{O}^{(4)}_4(\bm{k}',\bm{k}) &=&\hat{T}_3^2                         , \\
                \hat{O}^{(6)}_1(\bm{k}',\bm{k}) &=&\hat{T}_1^3+3\hat{T}_1  \hat{T}_2^2 , \\
                \hat{O}^{(6)}_2(\bm{k}',\bm{k}) &=&\hat{T}_2^3+3\hat{T}_1^2\hat{T}_2   , \\
                \hat{O}^{(6)}_3(\bm{k}',\bm{k}) &=&\hat{T}_1^3- \hat{T}_1  \hat{T}_2^2 , \\
                \hat{O}^{(6)}_4(\bm{k}',\bm{k}) &=&\hat{T}_2^3- \hat{T}_1^2\hat{T}_2   , \\
                \hat{O}^{(6)}_5(\bm{k}',\bm{k}) &=&\hat{T}_3^2\hat{T}_1                , \\
\label{eq:3j}   \hat{O}^{(6)}_6(\bm{k}',\bm{k}) &=&\hat{T}_3^2\hat{T}_2                .
\end{eqnarray}
Because for every operator $\hat{O}^{(n)}_j(\bm{k}',\bm{k})$ there
appear in Eqs.~(\ref{eq:1}) and~(\ref{eq:1a}) four different
spin-isospin terms, we recover here the numbers of central terms
shown in \Tref{tab_pseudo}.

Of course, at any given order, the choice of polynomials of
$\hat{T}_1$, $\hat{T}_2$, and $\hat{T}_3$ is quite arbitrary -- with
only requirement that these polynomials be linearly independent.
Definitions (\ref{eq:3a})--(\ref{eq:3j}) were chosen so as to
naturally link them to the standard Skyrme interaction, for which we
have
\begin{eqnarray}
\label{eq:4}
t_0 &=& t^{(0)}_1,  \quad  x_0 = x^{(0)}_1  , \\
t_1 &=& t^{(2)}_1,  \quad  x_1 = x^{(2)}_1  , \\
t_2 &=& t^{(2)}_2,  \quad  x_2 = x^{(2)}_2  ,
\end{eqnarray}
and to encompass definitions of fourth-order parameters
$t^{(4)}_1$, $t^{(4)}_2$, $x^{(4)}_1$, and $x^{(4)}_2$ introduced
in Ref.~\cite{[Dav13]}.
For a lighter notation used below, we also define the analogous parameters:
\begin{eqnarray}
y_0 &=& y^{(0)}_1,  \quad  z_0 = z^{(0)}_1  , \\
y_1 &=& y^{(2)}_1,  \quad  z_1 = z^{(2)}_1  , \\
y_2 &=& y^{(2)}_2,  \quad  z_2 = z^{(2)}_2  .
\label{eq:4a}
\end{eqnarray}

At higher orders, we picked the $j=1$ and 2 terms so as to
have at any order $n>0$,
\begin{eqnarray}
\label{eq:5}\hspace*{-1cm}
\hat{O}^{(n)}_1(\bm{k}',\bm{k}) - \hat{O}^{(n)}_2(\bm{k}',\bm{k}) &=& \left(\hat{T}_1-\hat{T}_2\right)^{n/2}
       = \frac{1}{2^n}(\bm{k}'^*-\bm{k})^n  \equiv \frac{1}{2^n}(\bm{k}'+\bm{k})^n .
\end{eqnarray}
These particular polynomials of relative momenta are special~\cite{[Dob12a]},
because, as it turns out, the sum of relative-momentum operators
$\bm{k}'+\bm{k}$ commutes
with the locality deltas, see Sec.~\ref{Central}.

In the following we give separate expressions for the functional
derived from three lowest-order terms~(\ref{eq:3a})--(\ref{eq:3c}) of
the pseudopotential, denoting them by $\langle V_i \rangle$ for $i=0$,
1, and 2. We also separate the local and non local terms, denoting them,
respectively, by $\langle V_i^L\rangle$ and $\langle V_i^{N}\rangle$.
To have more compact expressions, we also introduced the following
combinations of parameters of the regularized interaction
(\ref{eq:4})--(\ref{eq:4a}):
\begin{eqnarray}
\label{AB_to_tx}
 A^{\rho_0}_i&=& \phantom{-}\thf\, t_i\left(1+\thf\,x_i-\thf\,y_i-\thq\,z_i\right), \\
 A^{\rho_1}_i&=&         - \thf\, t_i\left(\thf\,y_i+\thq\,z_i\right)             , \\
 A^{\bm{s}_0}_i&=& \phantom{-}\thf\, t_i\left(\thf\,x_i-\thq\,z_i\right)          , \\
 A^{\bm{s}_1}_i&=&      - \thh\, t_iz_i                                           , \\
 B^{\rho_0}_i&=& -\thf\, t_i\left(\thq+\thf\,x_i-\thf\,y_i-z_i\right)             , \\
 B^{\rho_1}_i&=& -\thf\, t_i\left(\thq+\thf\,x_i\right)                           , \\
 B^{\bm{s}_0}_i&=& -\thf\, t_i\left(\thq-\thf\,y_i\right)                         , \\
 B^{\bm{s}_1}_i&=& -\thh\, t_i \label{AB_to_tx_fin}                               .
\end{eqnarray}

\subsection{Zero-order functional\label{zero_cartesian}}

The leading order functional, which
depends on four parameters $t_0$, $x_0$, $y_0$, and $z_0$ of the
regularized interaction (\ref{eq:1a}), has the form
\begin{equation}
\label{V0}
\langle V_0\rangle=\langle V_0^L \rangle + \langle V_0^{N} \rangle  ,
\end{equation}
with
\begin{eqnarray}
\label{eq:8}
\langle V_0^L \rangle&=&\int\!
\mathrm d\bfr_1\,\mathrm d\bfr_2\,g_a(\bfr_1 - \bfr_2)
\Bigl[
 A^{\rho_0}_0 \rho_0(\bfr_1)\rho_0(\bfr_2)
+ A^{\rho_1}_0 \rho_1(\bfr_1)\rho_1(\bfr_2) \nonumber \\
&&+ A^{\bfs_0}_0 \bfs_0(\bfr_1)\cdot\bfs_0(\bfr_2)
+ A^{\bfs_1}_0 \bfs_1(\bfr_1)\cdot\bfs_1(\bfr_2)
\Bigr]
\end{eqnarray}
and
\begin{eqnarray}
\label{eq:9}
\langle V_0^{N} \rangle&=&\int\!
\mathrm d\bfr_1\,\mathrm d\bfr_2\, g_a(\bfr_1-\bfr_2)
\Bigl[
   B^{\rho_0}_0 \rho_0(\bfr_2,\bfr_1)\rho_0(\bfr_1,\bfr_2) \nonumber \\
&&+B^{\rho_1}_0 \rho_1(\bfr_2,\bfr_1)\rho_1(\bfr_1,\bfr_2) \nonumber \\
&&+B^{\bfs_0}_0 \bfs_0(\bfr_2,\bfr_1)\cdot\bfs_0(\bfr_1,\bfr_2)
+ B^{\bfs_1}_0\bfs_1(\bfr_2,\bfr_1)\cdot\bfs_1(\bfr_1,\bfr_2)
\Bigr].
\end{eqnarray}
Densities $\rho_0(\bm{r})$, $\rho_1(\bm{r})$, $\bfs_0(\bm{r})$, and
$\bfs_1(\bm{r})$, together with their nonlocal counterparts,
are the standard scalar and vector densities in spin and
isospin spaces, as defined in Ref.~\cite{[Per04]}.
By taking the zero-range limit, which amounts to bringing in
Eqs.~(\ref{eq:8}) and~(\ref{eq:9}) the regularized delta
function $g_a(\bm{r})$ to its Dirac delta limit,
and using relations (\ref{AB_to_tx}),
one recovers the standard local form of the zero-order functional,
\begin{eqnarray}
\label{eq:10}
\langle V_0\rangle=&\thf\,t_0\int\!
\mathrm d\bfr\,
\Bigl\{\tfrac{3}{4}\left(1+z_0\right)\rho_0^2
-\left[\thq\left(1+z_0\right)+\thf\left(x_0+y_0\right)\right]\rho_1^2 \nonumber \\
&-\left[\thq\left(1+z_0\right)-\thf\left(x_0+y_0\right)\right]\bfs_0^2
-\thq\left(1+ z_0\right)\bfs_1^2
\Bigr\}.
\end{eqnarray}

We explicitly verified that the Cartesian zero-order nonlocal
EDF~(\ref{V0}) is exactly equivalent to that in the spherical-tensor
representation. To show this equivalence, we made use of the
relations of conversions between spherical and Cartesian local
densities~\cite{[Car08]}, which are also valid for nonlocal
densities. In the same way, one finds the relations of conversions
between the parameters of the zero-order regularized pseudopotential (\ref{eq:1finiterange})
and those of its Cartesian form~(\ref{eq:1a}), which read
\begin{eqnarray}
\label{eq:t0_conv}
C_{00,00}^{00,0}&=&  t_0+ \frac{1}{2}t_0x_0,\\
C_{00,00}^{00,1}&=&  -\frac{1}{2}t_0z_0-t_0y_0,\\
C_{00,20}^{00,0}&=& -\frac{\sqrt{3}}{2} t_0x_0 ,\\
C_{00,20}^{00,1}&=& \frac{\sqrt{3}}{2} t_0z_0.
\end{eqnarray}

\subsection{Second-order functional\label{Second_cartesian}}

At NLO, one obtains the EDF corresponding to the $i=1$ term of the
regularized pseudo\-potential~(\ref{eq:3b}) by applying on
all possible bilinear densities the relative-momentum operator:
\begin{eqnarray}
\bm{k}'^*{}^2+  \bm{k}^2
&=&-\thq\left(\bm{\nabla}'_1{}^2-2\bm{\nabla}'_1\cdot\bm{\nabla}'_2+\bm{\nabla}'_2{}^2
             +\bm{\nabla} _1  ^2-2\bm{\nabla} _1\cdot\bm{\nabla} _2+\bm{\nabla} _2^2\right).
\end{eqnarray}
Using once again for densities the same notations as in
Ref.~\cite{[Per04]}, we get the quasilocal term,
\begin{eqnarray}
\label{V_1_L}\hspace*{-2cm}
\langle V_1^L \rangle&=&-\thf\int\!
\mathrm d\bfr_1\,\mathrm d\bfr_2\,g_a(\bfr_1-\bfr_2)
\nonumber \\
\hspace*{-2cm}&\times& \biggl\{
A_1^{\rho_0}
\Bigl[
\left[\thf\,\Delta\rho_0(\bfr_1)-\tau_0(\bfr_1)\right]\rho_0(\bfr_2)
-\thq\,\bfnab\rho_0(\bfr_1)\cdot\bfnab\rho_0(\bfr_2)
+\bfj_0(\bfr_1)\cdot\bfj_0(\bfr_2)\Bigr]  \nonumber \\
\hspace*{-2cm}&&+ A_1^{\rho_1}
\Bigl[
\left[\thf\,\Delta\rho_1(\bfr_1)-\tau_1(\bfr_1)\right]\rho_1(\bfr_2)
 -\thq\,\bfnab\rho_1(\bfr_1)\cdot\bfnab\rho_1(\bfr_2)
 +\bfj_1(\bfr_1)\cdot\bfj_1(\bfr_2)\Bigr] \nonumber \\
\hspace*{-2cm}&&+ A_1^{\bfs_0}
\Bigl[
\left[\thf\,\Delta\bfs_0(\bfr_1)-\bfT_0(\bfr_1)\right]\cdot\bfs_0(\bfr_2)
\nonumber \\ \hspace*{-2cm}&&~~~~~~~~
 -\thq\,\bfnab\otimes\bfs_0(\bfr_1)\cdot\bfnab\otimes\bfs_0(\bfr_2)
+\bbJ_0(\bfr_1)\cdot\bbJ_0(\bfr_2)
\Bigr] \nonumber \\
\hspace*{-2cm}&&+ A_1^{\bfs_1}
\Bigl[
\left[\thf\,\Delta\bfs_1(\bfr_1)-\bfT_1(\bfr_1)\right]\cdot\bfs_1(\bfr_2)
\nonumber \\ \hspace*{-2cm}&&~~~~~~~~
-\thq\,\bfnab\otimes\bfs_1(\bfr_1)\cdot\bfnab\otimes\bfs_1(\bfr_2)
+\bbJ_1(\bfr_1)\cdot\bbJ_1(\bfr_2)
\Bigr] \biggr\}
\end{eqnarray}
and the nonlocal term,
\begin{eqnarray}
\label{V_1_NL}\hspace*{-2cm}
\langle V_1^{N}\rangle&=&-\thq\int\!
\mathrm d\bfr_1\,\mathrm d\bfr_2\,g_a(\bfr_1-\bfr_2) \nonumber \\
\hspace*{-2cm}&\times& \biggl\{
   B_1^{\rho_0}
\left[
\rho_0(\bfr_2,\bfr_1)\Delta\rho_0(\bfr_1,\bfr_2)
-2\rho_0(\bfr_2,\bfr_1)\tau_0(\bfr_1,\bfr_2) \right. \nonumber \\
 \hspace*{-2cm}&& \left. ~~~~~~-\thf \bfnab\rho_0(\bfr_2,\bfr_1)\cdot\bfnab\rho_0(\bfr_1,\bfr_2)
\right]  \nonumber \\
\hspace*{-2cm}&& + B_1^{\rho_1}
\left[
\rho_1(\bfr_2,\bfr_1)\Delta\rho_1(\bfr_1,\bfr_2)
-2\rho_1(\bfr_2,\bfr_1)\tau_1(\bfr_1,\bfr_2)  \right. \nonumber \\
\hspace*{-2cm}&&\left. ~~~~~~-\thf \bfnab\rho_1(\bfr_2,\bfr_1)\cdot\bfnab\rho_1(\bfr_1,\bfr_2)
  \right]
 \nonumber \\
\hspace*{-2cm}&& + B_1^{\bfs_0}
\left[
 \bfs_0(\bfr_2,\bfr_1)\cdot\Delta \bfs_0(\bfr_1,\bfr_2)
 -2  \bfs_0(\bfr_2,\bfr_1)\cdot \bfT_0(\bfr_1,\bfr_2) \right.  \nonumber \\
  \hspace*{-2cm}&& \left.~~~~~~ -\thf  \bfnab \otimes\bfs_0(\bfr_2,\bfr_1) \cdot \bfnab \otimes\bfs_0(\bfr_1,\bfr_2)
 +2 \bbJ_0(\bfr_2,\bfr_1)\cdot \bbJ_0(\bfr_1,\bfr_2)
  \right]
 \nonumber \\
 \hspace*{-2cm}&&+ B_1^{\bfs_1}
\left[
 \bfs_1(\bfr_2,\bfr_1)\cdot\Delta \bfs_1(\bfr_1,\bfr_2)
 -2  \bfs_1(\bfr_2,\bfr_1)\cdot \bfT_1(\bfr_1,\bfr_2)  \right. \nonumber \\
 \hspace*{-2cm}&&\left.~~~~~~ -\thf  \bfnab \otimes\bfs_1(\bfr_2,\bfr_1) \cdot \bfnab \otimes\bfs_1(\bfr_1,\bfr_2)
 +2 \bbJ_1(\bfr_2,\bfr_1)\cdot\bbJ_1(\bfr_1,\bfr_2)
 \right]\biggr\}.
\end{eqnarray}
In Eqs.~(\ref{V_1_L}) and (\ref{V_1_NL}), similarly as in Sec.~\ref{NL_EDF},
depending on the context, the nabla operator $\bfnab$ acts on the local densities
as $\bfnab_1$ or $\bfnab_2$ and on the nonlocal densities
as $\bfnab_1+\bfnab_2$.

By taking the local limit of
functionals~(\ref{V_1_L}) and (\ref{V_1_NL}), we again recover the
quasilocal Cartesian EDF, which reads
\begin{eqnarray}
\langle V_1\rangle=
\thq\,t_1\int\!
\mathrm d\bfr\,&
\tfrac{3}{4}\left(1+z_1\right)
\left[\tau_0\rho_0-\bfj_0^2-\tfrac{3}{4}\rho_0\Delta\rho_0\right] \nonumber \\
&-\left[\thq\left(1+z_1\right)+\thf\left(x_1+y_1\right)\right]
\left[ \tau_1\rho_1-\bfj_1^2-\tfrac{3}{4}\rho_1\Delta\rho_1\right] \nonumber \\
&-\left[\thq\left(1+z_1\right)-\thf\left(x_1+y_1\right)\right]
\left[\bfT_0\cdot\bfs_0-\bbJ_0^2-\tfrac{3}{4}\,\bfs_0\cdot\Delta\bfs_0 \right] \nonumber \\
&-\thq\left(1+z_1\right)
\left[\bfT_1\cdot\bfs_1-\bbJ_1^2-\tfrac{3}{4}\,\bfs_1\cdot\Delta\bfs_1 \right].
\end{eqnarray}

Finally, one obtains the EDF corresponding to the $i=2$ term of the
regularized pseudopotential~(\ref{eq:3c}) using the relative-momentum operator:
\begin{eqnarray}
\bm{k}'^*\cdot\bm{k}
&=&\thq\left(\bm{\nabla}'_1\cdot\bm{\nabla}_1-\bm{\nabla}'_2\cdot\bm{\nabla}_1
            +\bm{\nabla}'_2\cdot\bm{\nabla}_2-\bm{\nabla}'_1\cdot\bm{\nabla}_2\right),
\end{eqnarray}
with the result
\begin{eqnarray}
\label{V_2_L}\hspace*{-2cm}
\langle V_2^L \rangle&=&
\thf\int\!\mathrm d\bfr_1\,\mathrm d\bfr_2\, g_a(\bfr_1-\bfr_2)  \nonumber \\
\hspace*{-2cm}&\times&\biggl\{
    A_2^{\rho_0}
\left[ \rho_0(\bfr_1)\tau_0(\bfr_2)
-\thq\,\bfnab\rho_0(\bfr_1)\cdot\bfnab\rho_0(\bfr_2)
-\bfj_0(\bfr_1)\cdot\bfj_0(\bfr_2)
\right] \nonumber \\
\hspace*{-2cm}&&+  A_2^{\rho_1}
\left[ \rho_1(\bfr_1)\tau_1(\bfr_2)
-\thq\,\bfnab\rho_1(\bfr_1)\cdot\bfnab\rho_1(\bfr_2)
-\bfj_1(\bfr_1)\cdot\bfj_1(\bfr_2) \right] \nonumber \\
\hspace*{-2cm}&&+  A_2^{\bfs_0}
 \left[ \bfs_0(\bfr_1)\cdot\bfT_0(\bfr_2)
-\thq\,\bfnab\otimes\bfs_0(\bfr_1)\cdot\bfnab\otimes\bfs_0(\bfr_2)
-\bbJ_0(\bfr_1)\cdot\bbJ_0(\bfr_2) \right] \nonumber \\
\hspace*{-2cm}&& +  A_2^{\bfs_1}
 \left[ \bfs_1(\bfr_1)\cdot\bfT_1(\bfr_2)
-\thq\,\bfnab\otimes\bfs_1(\bfr_1)\cdot\bfnab\otimes\bfs_1(\bfr_2)
-\bbJ_1(\bfr_1)\cdot\bbJ_1(\bfr_2) \right]
\biggr\}
\end{eqnarray}
and
\begin{eqnarray}
\label{V_2_NL} \hspace*{-2cm}
\langle V_2^{N}\rangle&=&\thf
\int\!\mathrm d\bfr_1\,\mathrm d\bfr_2\,  g_a(\bfr_1-\bfr_2) \nonumber \\
\hspace*{-2cm}&\times&\!\!\biggl\{
B_2^{\rho_0}
\left[\thq
\bfnab\rho_0(\bfr_2,\bfr_1)\cdot\bfnab\rho_0(\bfr_1,\bfr_2)
+ \bfj_0(\bfr_2,\bfr_1)\cdot\bfj_0(\bfr_1,\bfr_2)
-\rho_0(\bfr_2,\bfr_1)\tau_0(\bfr_1,\bfr_2) \right] \nonumber \\
\hspace*{-2cm}&&\!\!\!\!\!+
B_2^{\rho_1}
\left[\thq
\bfnab\rho_1(\bfr_2,\bfr_1)\cdot\bfnab\rho_1(\bfr_1,\bfr_2)
+ \bfj_1(\bfr_2,\bfr_1)\cdot\bfj_1(\bfr_1,\bfr_2)
-\rho_1(\bfr_2,\bfr_1)\tau_1(\bfr_1,\bfr_2)
 \right] \nonumber \\
\hspace*{-2cm}&&+
B_2^{\bfs_0}
 \left[ \thq  \bfnab \otimes\bfs_0(\bfr_2,\bfr_1) \cdot \bfnab \otimes\bfs_0(\bfr_1,\bfr_2)
 + \bbJ_0(\bfr_2,\bfr_1)\cdot \bbJ_0(\bfr_1,\bfr_2)  \right. \nonumber \\
\hspace*{-2cm}&& \left.~~~~~~ - \bfs_0(\bfr_2,\bfr_1)\cdot \bfT_0(\bfr_1,\bfr_2)\right]  \nonumber \\
\hspace*{-2cm}&& +
B_2^{\bfs_1}
 \left[
 \thq  \bfnab \otimes\bfs_1(\bfr_2,\bfr_1) \cdot \bfnab \otimes\bfs_1(\bfr_1,\bfr_2)
 + \bbJ_1(\bfr_2,\bfr_1)\cdot \bbJ_1(\bfr_1,\bfr_2) \right.  \nonumber \\
\hspace*{-2cm}&& \left.~~~~~~ - \bfs_1(\bfr_2,\bfr_1)\cdot \bfT_1(\bfr_1,\bfr_2)
 \right]
\biggr\},
\end{eqnarray}
whereas the corresponding quasilocal Cartesian EDF reads,
\begin{eqnarray}
\langle V_2\rangle=\thq\,t_2
\int\!\mathrm d\bfr
\biggl\{&
\left[\tfrac{5}{4}(1-z_2)+x_2-y_2\right]
\left[ \rho_0\tau_0+\thq\,\rho_0\Delta\rho_0-\bfj_0^2 \right]\nonumber \\
&+
\left[\thq(1-z_2)+\thf(x_2-y_2)\right]
\left[ \rho_1\tau_1+\thq\,\rho_1\Delta\rho_1-\bfj_1^2 \right] \nonumber \\
&+
\left[\thq(1-z_2)+\thf(x_2-y_2)\right]
 \left[ \bfs_0\cdot\bfT_0+\thq\,\bfs_0\cdot\Delta\bfs_0-\bfJ_0^2 \right]\nonumber \\
&+
\thq(1-z_2)
 \left[ \bfs_1\cdot\bfT_1+\thq\,\bfs_1\cdot\Delta\bfs_1-\bfJ_1^2 \right]
\biggr\}.
\end{eqnarray}

The second-order Cartesian EDF~(\ref{V_1_L}-\ref{V_1_NL})
and~(\ref{V_2_L}-\ref{V_2_NL}) is exactly equivalent to that in the
spherical-tensor representation. The corresponding relations of
conversions between the parameters of the second-order regularized
pseudopotential (\ref{eq:1finiterange}) and those of its Cartesian
form~(\ref{eq:1a}) read
\begin{eqnarray}
\label{eq:t1t2_conv}
C_{00,00}^{20,0}&=&  \sqrt{3}t_1+ \frac{\sqrt{3}}{2}t_1x_1,\\
C_{00,00}^{20,1}&=&  -\sqrt{3}t_1y_1- \frac{\sqrt{3}}{2}t_1z_1 ,\\
C_{00,20}^{20,0}&=& -\frac{3}{2} t_1x_1 ,\\
C_{00,20}^{20,1}&=& \frac{3}{2} t_1z_1, \\
C_{11,00}^{11,0}&=&  \sqrt{3}t_2+ \frac{\sqrt{3}}{2}t_2x_2 , \\
C_{11,00}^{11,1}&=&   -\sqrt{3}t_2y_2- \frac{\sqrt{3}}{2}t_2z_2, \\
C_{11,20}^{11,0}&=&  -\frac{3}{2} t_2x_2 , \\
C_{11,20}^{11,1}&=&  \frac{3}{2} t_2z_2.
\end{eqnarray}

\section{Central pseudopotential depending on the sum of relative momenta\label{Central}}

In a recent paper~\cite{[Dob12a]}, we studied the regularized
finite-range interaction as an application of the ET
principles to low-energy nuclear phenomena.
To give an illustrative version of such effective theory, we
considered a restricted form of the finite-range pseudopotential,
neglecting the SO and tensor parts, and treating the
remaining central part as depending only on the sum of relative
momenta. In the spherical-tensor pseudopotential, this amounts to
considering only terms of the form
\begin{eqnarray}
 \nonumber \hat{V}_{v_{12} 0}^{N,\bar{t}}=
&& \frac{1}{2}i^{v_{12}}  \left[ \left[\left(K'_{11} + K_{11} \right)^N \right]_0
\SSS_{v_{12} 0}\right]_{0}  \left(\hat{P}^{\tau}\right)^{\bar{t}}  \nonumber \\
&&\times \left(1-\hat{P}^{M}\hat{P}^{\sigma}\hat{P}^{\tau}\right)
\delta(\bm{r}'_1-\bm{r}_1)\delta(\bm{r}'_2-\bm{r}_2)g_a(r)
\label{eq:1finiterange_central},
\end{eqnarray}
with $N$=0, 2, 4, and 6.

Dependence on the sum of relative momenta only, that is, on
$\left[\left(K'_{11} + K_{11} \right)^N \right]_0$\,, is a crucial
feature to obtain a functional in the form of an expansion in the length scale $a$
of the regularized delta, which we used for the study of the
pseudopotentials in the ET framework. Indeed, the series in $a$ is
obtained by acting with the relative-momentum operators directly and
only on $g_a(r)$, which is possible, because the sums of
relative momenta do commute with the
locality deltas. This fact can be explicitly demonstrated by calculating
the action of $K'_{11}+K_{11}$ on the
locality deltas, that is,
\begin{eqnarray}
\label{eq:6a}
&&\Big[K'_{11}+K_{11}\Big]  \delta(\bm{r}'_1-\bm{r}_1)\delta(\bm{r}'_2-\bm{r}_2)
 =\Big[\bm{k}'+\bm{k}\Big]  \delta(\bm{r}'_1-\bm{r}_1)\delta(\bm{r}'_2-\bm{r}_2) \nonumber \\
&&~~ = \frac{1}{2i}\Big[ \bm{\nabla}'_1-\bm{\nabla}'_2+\bm{\nabla}_1-\bm{\nabla}_2\Big] \delta(\bm{r}'_1-\bm{r}_1)\delta(\bm{r}'_2-\bm{r}_2)  \nonumber \\
&&~~ = \frac{1}{2i}\Big[  \delta\,'(\bm{r}'_1-\bm{r}_1)\delta   (\bm{r}'_2-\bm{r}_2)
                         -\delta   (\bm{r}'_1-\bm{r}_1)\delta\,'(\bm{r}'_2-\bm{r}_2)         \nonumber \\
&&~~~~~~~                -\delta\,'(\bm{r}'_1-\bm{r}_1)\delta   (\bm{r}'_2-\bm{r}_2)
                         +\delta   (\bm{r}'_1-\bm{r}_1)\delta\,'(\bm{r}'_2-\bm{r}_2)\Big] \equiv 0.
\end{eqnarray}

As a consequence, pseudopotentials defined by~(\ref{eq:1finiterange_central})
are strictly equivalent to ordinary local potentials given by a
series of powers of Laplacians acting on the regularized delta
$g_a(r)$~\cite{[Dob12a]}.

This particular restriction of the relative-momentum operators
gives rise to a set of constraints
on parameters of the general second-, fourth-, and sixth-order
pseudopotential~(\ref{eq:0finiterange}), expressed in spherical
or Cartesian forms, which we list below.

At second order, corresponding to the conditions
\begin{eqnarray}
\label{eq:7}
t^{(2)}_2  &=& - t^{(2)}_1,  \quad
x^{(2)}_2   =    x^{(2)}_1,  \quad
y^{(2)}_2   =    y^{(2)}_1,  \quad
x^{(2)}_2   =    x^{(2)}_1,
\end{eqnarray}
valid for combination of the Cartesian relative momenta~\cite{[Dob12a]},
we get the constraints,
\begin{eqnarray}
\label{eq:second_order_results_central}
C_{11,00}^{11,\bar{t}}&=& C_{00,00}^{20,\bar{t}},\\
C_{11,20}^{11,\bar{t}}&=& C_{00,20}^{20,\bar{t}}.
\end{eqnarray}

The analogous relations at fourth and sixth order are found by
applying the binomial expansion of the term  $\left[\left(K'_{11} +
K_{11} \right)^N \right]_0$ for $N$ = 4, 6 respectively. At fourth
order for $v_{12}$ = 0, 2 and $t^{(4)}_j=0$ for $j>2$,  we find
\begin{eqnarray}
\label{eq:fourth_order_results_central}
C_{11,v_{12}0}^{31,\bar{t}}&=& 4  \, C_{00,v_{12}0}^{40,\bar{t}},\\
C_{20,v_{12}0}^{20,\bar{t}}&=& 3  \, C_{00,v_{12}0}^{40,\bar{t}},\\
C_{22,v_{12}0}^{22,\bar{t}}&=& 3  \, C_{00,v_{12}0}^{40,\bar{t}},
\end{eqnarray}
and at sixth order we have,
\begin{eqnarray}
\label{eq:sixth_order_results_central}
C_{11,v_{12}0}^{51,\bar{t}}&=& 6  \, C_{00,v_{12}0}^{60,\bar{t}},\\
C_{20,v_{12}0}^{40,\bar{t}}&=& 15 \, C_{00,v_{12}0}^{60,\bar{t}},\\
C_{22,v_{12}0}^{42,\bar{t}}&=& 15 \, C_{00,v_{12}0}^{60,\bar{t}},\\
C_{31,v_{12}0}^{31,\bar{t}}&=& 10 \, C_{00,v_{12}0}^{60,\bar{t}},\\
C_{33,v_{12}0}^{33,\bar{t}}&=& 10 \, C_{00,v_{12}0}^{60,\bar{t}},
\end{eqnarray}
whereas all the other parameters $  C_{\tilde{n} \tilde{L},v_{12}
S}^{\tilde{n}' \tilde{L}', \bar{t}}$ with $S\neq$0 are set to zero.

\section{Cartesian representation of the SO and tensor pseudopotentials\label{SOtensor}}

In this Section, we present construction of the SO
($\tilde{S}$=1) and tensor ($\tilde{S}$=2) components of the regularized
pseudopotential. Following the same methodology as that introduced for central terms in Sec.~\ref{Central},
we define the Cartesian forms
of the (non-anti\-symmetrized) SO and tensor
pseudopotential, respectively, as
\begin{eqnarray}
\label{eq:101a}
 \mathcal{V}_{SO} &=& \sum_{nj} p^{(n)}_j
   \left(          \hat{1}_\sigma\hat{1}_\tau
        -w^{(n)}_j \hat{1}_\sigma\hat{P}^\tau\right)
\nonumber \\
&&~~~~~~~~\times \hat{P}^{(n)}_j(\bm{k}',\bm{k})\delta(\bm{r}'_1-\bm{r}_1)\delta(\bm{r}'_2-\bm{r}_2)
g_a(\bm{r}_1-\bm{r}_2)
\end{eqnarray}
and
\begin{eqnarray}
\label{eq:101b}
 \mathcal{V}_T &=& \sum_{nj} q^{(n)}_j
   \left(          \hat{1}_\sigma\hat{1}_\tau
        -v^{(n)}_j \hat{1}_\sigma\hat{P}^\tau\right)
\nonumber \\
&&~~~~~~~~\times \hat{Q}^{(n)}_j(\bm{k}',\bm{k})\delta(\bm{r}'_1-\bm{r}_1)\delta(\bm{r}'_2-\bm{r}_2)
g_a(\bm{r}_1-\bm{r}_2) .
\end{eqnarray}

The spin-dependent differential operators $\hat{P}^{(n)}_j(\bm{k}',\bm{k})$ and  $\hat{Q}^{(n)}_j(\bm{k}',\bm{k})$
are built in the following way. First of all, they must be scalar, hermitian, and time-even operators
that are obtained by coupling the space part to standard spin-vector and spin-tensor operators,
\begin{eqnarray}
\label{eq:102}
 \hat{S}_m              &=& \sigma^{(1)}_m + \sigma^{(2)}_m  ,\\
 \hat{{\mathsf S}}_{mn} &=& \tfrac{3}{2}\left(\sigma^{(1)}_m\sigma^{(2)}_n + \sigma^{(1)}_n\sigma^{(2)}_m\right)
                             -\delta_{mn} \bm{\sigma}^{(1)}\cdot\bm{\sigma}^{(2)} ,
\end{eqnarray}
respectively,
where indices $m$ and $n$ denote Cartesian components of the Pauli spin matrices.
At this point we note that $\hat{\bm{S}}\hat{P}^\sigma\equiv\hat{\bm{S}}$
and $\hat{{\mathsf S}}\hat{P}^\sigma\equiv\hat{{\mathsf S}}$; therefore, the spin-exchange
operators $\hat{P}^\sigma$ do not give independent terms and thus do not appear in Eqs.~(\ref{eq:101a}) and (\ref{eq:101b}).

Because spin-vector operator $\hat{S}_m$ is even-parity and time-odd, we must build from relative-momentum operators
an elementary even-parity and time-odd vector, which is only one,
\begin{eqnarray}
\label{eq:103}
 i(\bm{k}'^*\times\bm{k})_m &=&  i\sum_{nl}\varepsilon_{mnl} k_n'^* k_l.
\end{eqnarray}
Similarly, because spin-tensor operator $\hat{{\mathsf S}}_{mn}$ is even-parity
and time-even, we must built from relative-momentum operators all
elementary even-parity and time-even traceless tensors, which are
three,
\begin{eqnarray}
\label{eq:104}
 (\bm{k}   \otimes\bm{k}   )_{mn} &=& k   _m k   _n - \tfrac{1}{3} \delta_{mn} \bm{k}   \cdot\bm{k}    , \\
 (\bm{k}'^*\otimes\bm{k}   )_{mn} &=& k'^*_m k   _n - \tfrac{1}{3} \delta_{mn} \bm{k}'^*\cdot\bm{k}    , \\
 (\bm{k}'^*\otimes\bm{k}'^*)_{mn} &=& k'^*_m k'^*_n - \tfrac{1}{3} \delta_{mn} \bm{k}'^*\cdot\bm{k}'^* .
\end{eqnarray}

Contracting spin-vector and space-vector, as well as spin-tensor and space-tensor operators,
we now obtain all suitable elementary scalar operators. Selecting convenient combinations
of tensor terms, cf.~Eqs.~(\ref{eq:2})--(\ref{eq:2a}), we define them as,
\begin{eqnarray}
\label{eq:105a}
 \hat{T}_s  &=&  i(\bm{k}'^*\times\bm{k})\cdot \left(\bm{\sigma}^{(1)} + \bm{\sigma}^{(2)}\right) , \\
\label{eq:105b}
 \hat{T}_e  &=&  \tfrac{1}{2}(\bm{k}   \cdot\hat{{\mathsf S}}\cdot\bm{k}
                             +\bm{k}'^*\cdot\hat{{\mathsf S}}\cdot\bm{k}'^*), \\
\label{eq:105c}
 \hat{T}_o  &=&               \bm{k}'^*\cdot\hat{{\mathsf S}}\cdot\bm{k}    , \\
\label{eq:105d}
 \hat{T}_a  &=&  \tfrac{1}{2}(\bm{k}   \cdot\hat{{\mathsf S}}\cdot\bm{k}
                             -\bm{k}'^*\cdot\hat{{\mathsf S}}\cdot\bm{k}'^*).
\end{eqnarray}
Of course, at second order we have the standard SO (\ref{eq:105a}),
tensor-even  (\ref{eq:105b}), and tensor-odd (\ref{eq:105c})
elementary operators, see, e.g., Ref.~\cite{[Per04]}, whereas the
tensor-antihermitian operator (\ref{eq:105d}) does not appear. Owing
to the GCH theorem~\cite{[Roh10]}, we can now build all higher-order
operators by multiplying elementary operators
(\ref{eq:105a})--(\ref{eq:105d}) and scalars
(\ref{eq:2})--(\ref{eq:2a}). The hermiticity is then enforced by always
using even numbers of antihermitian factors.

Finally, up to N$^3$LO, we obtain all possible SO terms,
\begin{eqnarray}
\label{eq:106a} \hat{P}^{(2)}_1(\bm{k}',\bm{k}) &=&                                    \hat{T}_s , \\
                \hat{P}^{(4)}_1(\bm{k}',\bm{k}) &=& \hat{T}_1                          \hat{T}_s , \\
                \hat{P}^{(4)}_2(\bm{k}',\bm{k}) &=& \hat{T}_2                          \hat{T}_s , \\
                \hat{P}^{(6)}_1(\bm{k}',\bm{k}) &=&(\hat{T}_1^2+ \hat{T}_2^2)          \hat{T}_s , \\
                \hat{P}^{(6)}_2(\bm{k}',\bm{k}) &=& 2\hat{T}_1   \hat{T}_2             \hat{T}_s , \\
                \hat{P}^{(6)}_3(\bm{k}',\bm{k}) &=&(\hat{T}_1^2- \hat{T}_2^2)          \hat{T}_s , \\
                \hat{P}^{(6)}_4(\bm{k}',\bm{k}) &=& \hat{T}_3^2                        \hat{T}_s ,
\end{eqnarray}
and all possible tensor terms,
\begin{eqnarray}
\label{eq:107a} \hat{Q}^{(2)}_1   (\bm{k}',\bm{k}) &=&                                        \hat{T}_e  , \\
                \hat{Q}^{(2)}_2   (\bm{k}',\bm{k}) &=&                                        \hat{T}_o  , \\
                \hat{Q}^{(4)}_1   (\bm{k}',\bm{k}) &=& \hat{T}_1                 (\hat{T}_e - \hat{T}_o) , \\
                \hat{Q}^{(4)}_2   (\bm{k}',\bm{k}) &=& \hat{T}_2                 (\hat{T}_e - \hat{T}_o) , \\
                \hat{Q}^{(4)}_3   (\bm{k}',\bm{k}) &=& \hat{T}_1                 (\hat{T}_e + \hat{T}_o) , \\
                \hat{Q}^{(4)}_4   (\bm{k}',\bm{k}) &=& \hat{T}_2                 (\hat{T}_e + \hat{T}_o) , \\
                \hat{Q}^{(4)}_5   (\bm{k}',\bm{k}) &=& \hat{T}_3                              \hat{T}_a  , \\
                \hat{Q}^{(6)}_1   (\bm{k}',\bm{k}) &=&(\hat{T}_1^2+ \hat{T}_2^2) (\hat{T}_e - \hat{T}_o) , \\
                \hat{Q}^{(6)}_2   (\bm{k}',\bm{k}) &=& 2\hat{T}_1   \hat{T}_2    (\hat{T}_e - \hat{T}_o) , \\
                \hat{Q}^{(6)}_3   (\bm{k}',\bm{k}) &=&(\hat{T}_1^2- \hat{T}_2^2) (\hat{T}_e - \hat{T}_o) , \\
                \hat{Q}^{(6)}_4   (\bm{k}',\bm{k}) &=& \hat{T}_3^2               (\hat{T}_e - \hat{T}_o) , \\
                \hat{Q}^{(6)}_5   (\bm{k}',\bm{k}) &=&(\hat{T}_1^2+ \hat{T}_2^2) (\hat{T}_e + \hat{T}_o) , \\
                \hat{Q}^{(6)}_6   (\bm{k}',\bm{k}) &=& 2\hat{T}_1   \hat{T}_2    (\hat{T}_e + \hat{T}_o) , \\
                \hat{Q}^{(6)}_7   (\bm{k}',\bm{k}) &=&(\hat{T}_1^2- \hat{T}_2^2) (\hat{T}_e + \hat{T}_o) , \\
                \hat{Q}^{(6)}_8   (\bm{k}',\bm{k}) &=& \hat{T}_3^2               (\hat{T}_e + \hat{T}_o) , \\
                \hat{Q}^{(6)}_9   (\bm{k}',\bm{k}) &=& \hat{T}_1    \hat{T}_3                 \hat{T}_a  , \\
                \hat{Q}^{(6)}_{10}(\bm{k}',\bm{k}) &=& \hat{T}_2    \hat{T}_3                 \hat{T}_a  .
\end{eqnarray}
We note that we here recover the numbers of SO and tensor terms given in \Tref{tab_pseudo}.
We also note that the difference of the tensor-even and tensor-odd elementary
operators depends only on the sum of relative momenta, that is,
\begin{eqnarray}
\label{eq:108a}
\hat{T}_e - \hat{T}_o &=& \tfrac{1}{2}(\bm{k}'+\bm{k})   \cdot\hat{{\mathsf S}}\cdot(\bm{k}'+\bm{k}) .
\end{eqnarray}
This has motivated our choice of the $j=1$ and $j=2$ tensor terms, for which at $n>0$ we have,
\begin{eqnarray}
\label{eq:108}\hspace*{-1cm}
\hat{Q}^{(n)}_1(\bm{k}',\bm{k}) - \hat{Q}^{(n)}_2(\bm{k}',\bm{k}) &\equiv&
   \frac{1}{2^n}(\bm{k}'+\bm{k})^{n-2}\left[(\bm{k}'+\bm{k})   \cdot\hat{{\mathsf S}}\cdot(\bm{k}'+\bm{k})\right] .
\end{eqnarray}
Therefore, these particular combinations of terms commute with the locality deltas, see Sec.~\ref{Central}, and thus
are equivalent to local tensor potentials. On the other hand, none of the SO terms has such a property.

The tensor interaction presented here may be compared with the one
discussed in a recent work by Davesne {\em et al.}~\cite{[Dav14]},
which extends to N$^3$LO the zero-range Cartesian pseudopotential.
First we note that the $\hat T_e$ and $\hat T_o$ operators defined in
the aforementioned article differ from those in Eqs.~(\ref{eq:105b})
and~(\ref{eq:105c}) by factors of 2. When $a\rightarrow 0$, one has
the following correspondence between the coupling constants
$q_j^{(n)}$ appearing in Eq.~(\ref{eq:101b}) and those used in
Ref.~\cite{[Dav14]}, denoted by $t_e^{(n)}$ and $t_o^{(n)}$. (The
coupling constants $w_j^{(n)}$ can be disregarded, because at the
zero-range limit, the action of operator $\hat P^\tau$ reduces to a
phase.) At second order, one recovers the pseudopotential
from Ref.~\cite{[Dav14]} with
\begin{eqnarray}
q_1^{(2)}&=&t_e^{(2)}\,, \\
q_2^{(2)}&=&t_o^{(2)}\,,
\end{eqnarray}
at fourth order with
\begin{eqnarray}
q_1^{(4)}&=&2\left(t_e^{(4)}-t_o^{(4)}\right)\,, \\
q_2^{(4)}&=&2\left(t_o^{(4)}-t_e^{(4)}\right)\,, \\
q_3^{(4)}&=&2\left(t_e^{(4)}+t_o^{(4)}\right)\,, \\
q_4^{(4)}&=&2\left(t_e^{(4)}+t_o^{(4)}\right)\,, \\
q_5^{(4)}&=&0\,,
\end{eqnarray}
and at sixth order with
\begin{eqnarray}
q_1^{(6)}&=&t_e^{(6)}-t_o^{(6)}\,, \\
q_2^{(6)}&=&t_o^{(6)}-t_e^{(6)}\,, \\
q_5^{(6)}&=&t_e^{(6)}+t_o^{(6)}\,, \\
q_6^{(6)}&=&t_e^{(6)}+t_o^{(6)}
\end{eqnarray}
and $q_3^{(6)}=q_4^{(6)}=q_7^{(6)}=q_8^{(6)}=q_9^{(6)}=q_{10}^{(6)}=0$.

\section{Conclusions\label{end}}

In summary, in this work, for the first time we constructed the
finite-range, higher-order in derivatives pseudopotential, for which
we built all terms up to sixth order N$^3$LO, and then we derived the
corresponding nonlocal nuclear N$^3$LO EDF by calculating the HF
average energy over a nuclear Slater determinant. The proposed
pseudopotential can thus be regarded as a generator of the EDF, which
has several advantages for calculations beyond the simple HF level.
First, the use of a regulator introduces a natural cut-off at high
momenta, preventing divergences of pairing energies or correlation
energies in the (Q)RPA limit. Then, for multi-reference EDF
calculations, the link with a pseudopotential guaranties the absence of
poles in energy kernels. Moreover, the fact that the pseudopotential
does not depend on density guaranties that average energies are not
polluted by self-interaction or self-pairing.

The ability of this pseudopotential to generate attractive pairing
has not been checked yet -- this is of course of crucial importance
for future developments. However, the fact that at NLO it has all the
features of the Gogny interaction gives us hope that this will indeed
be the case. Let us also mention that all the central
terms of the pseudopotential and
the corresponding mean fields have already been implemented in the
solver HFODD~\cite{[Sch12],[Sch14],[Dob07c]} and numerical studies of finite
nuclei are under way. Moreover, a preliminary study has
demonstrated that the NLO functional can provide binding energy of
doubly-magic nuclei with a reasonable accuracy~\cite{[Ben13]}.

\bigskip
This work has been supported in part by the Academy of Finland and
University of Jyv\"askyl\"a within the FIDIPRO programme and by the
Polish National Science Center under Contract No.\
2012/07/B/ST2/03907. The research of FR is supported by the 
Natural Science and Engineering Research Council (NSERC).

\bigskip
\bibliographystyle{unsrt}

\end{document}


\noindent\ \textbf{Explicit expressions of the finite-range pseudopotential at 0,2,4 and 6 order.}

\vspace{1.0cm}
\noindent\ \textbf{Zero order}

\vspace{0.3cm}
\noindent\(C_{00,00}^{00,0} V_{00,00}^{00,0}+C_{00,00}^{00,1} V_{00,00}^{00,1}+C_{00,20}^{00,0} V_{00,20}^{00,0}+C_{00,20}^{00,1} V_{00,20}^{00,1}\),

\vspace{1.0cm}

\noindent\ \textbf{Second order}

\vspace{0.3cm}
\noindent\(C_{00,00}^{20,0} V_{00,00}^{20,0}+C_{00,00}^{20,1} V_{00,00}^{20,1}+C_{00,20}^{20,0} V_{00,20}^{20,0}+C_{00,20}^{20,1} V_{00,20}^{20,1}+C_{00,22}^{22,0} V_{00,22}^{22,0}+C_{00,22}^{22,1} V_{00,22}^{22,1}+C_{11,00}^{11,0} V_{11,00}^{11,0}+C_{11,00}^{11,1} V_{11,00}^{11,1}+C_{11,11}^{11,0} V_{11,11}^{11,0}+C_{11,11}^{11,1} V_{11,11}^{11,1}+C_{11,20}^{11,0} V_{11,20}^{11,0}+C_{11,20}^{11,1} V_{11,20}^{11,1}+C_{11,22}^{11,0} V_{11,22}^{11,0}+C_{11,22}^{11,1} V_{11,22}^{11,1}\),

\vspace{1.0cm}
\noindent\ \textbf{Fourth order}

\vspace{0.3cm}
\noindent\(C_{00,00}^{40,0} V_{00,00}^{40,0}+C_{00,00}^{40,1} V_{00,00}^{40,1}+C_{00,20}^{40,0} V_{00,20}^{40,0}+C_{00,20}^{40,1} V_{00,20}^{40,1}+C_{00,22}^{42,0} V_{00,22}^{42,0}+C_{00,22}^{42,1} V_{00,22}^{42,1}+C_{11,00}^{31,0} V_{11,00}^{31,0}+C_{11,00}^{31,1} V_{11,00}^{31,1}+C_{11,11}^{31,0} V_{11,11}^{31,0}+C_{11,11}^{31,1} V_{11,11}^{31,1}+C_{11,20}^{31,0} V_{11,20}^{31,0}+C_{11,20}^{31,1} V_{11,20}^{31,1}+C_{11,22}^{31,0} V_{11,22}^{31,0}+C_{11,22}^{31,1} V_{11,22}^{31,1}+C_{11,22}^{33,0} V_{11,22}^{33,0}+C_{11,22}^{33,1} V_{11,22}^{33,1}+C_{20,00}^{20,0} V_{20,00}^{20,0}+C_{20,00}^{20,1} V_{20,00}^{20,1}+C_{20,20}^{20,0} V_{20,20}^{20,0}+C_{20,20}^{20,1} V_{20,20}^{20,1}+C_{20,22}^{22,0} V_{20,22}^{22,0}+C_{20,22}^{22,1} V_{20,22}^{22,1}+C_{22,00}^{22,0} V_{22,00}^{22,0}+C_{22,00}^{22,1} V_{22,00}^{22,1}+C_{22,11}^{22,0} V_{22,11}^{22,0}+C_{22,11}^{22,1} V_{22,11}^{22,1}+C_{22,20}^{22,0} V_{22,20}^{22,0}+C_{22,20}^{22,1} V_{22,20}^{22,1}+C_{22,22}^{22,0} V_{22,22}^{22,0}+C_{22,22}^{22,1} V_{22,22}^{22,1}\),

\vspace{1.0cm}
\noindent\ \textbf{Sixth order}

\vspace{0.3cm}
\noindent\(C_{00,00}^{60,0} V_{00,00}^{60,0}+C_{00,00}^{60,1} V_{00,00}^{60,1}+C_{00,20}^{60,0} V_{00,20}^{60,0}+C_{00,20}^{60,1} V_{00,20}^{60,1}+C_{00,22}^{62,0} V_{00,22}^{62,0}+C_{00,22}^{62,1} V_{00,22}^{62,1}+C_{11,00}^{51,0} V_{11,00}^{51,0}+C_{11,00}^{51,1} V_{11,00}^{51,1}+C_{11,11}^{51,0} V_{11,11}^{51,0}+C_{11,11}^{51,1} V_{11,11}^{51,1}+C_{11,20}^{31,0} V_{11,20}^{31,0}+C_{11,20}^{31,1} V_{11,20}^{31,1}+C_{11,22}^{51,0} V_{11,22}^{51,0}+C_{11,22}^{51,1} V_{11,22}^{51,1}+C_{11,22}^{53,0} V_{11,22}^{53,0}+C_{11,22}^{53,1} V_{11,22}^{53,1}+C_{20,00}^{40,0} V_{20,00}^{40,0}+C_{20,00}^{40,1} V_{20,00}^{40,1}+C_{20,20}^{40,0} V_{20,20}^{40,0}+C_{20,20}^{40,1} V_{20,20}^{40,1}+C_{20,22}^{42,0} V_{20,22}^{42,0}+C_{20,22}^{42,1} V_{20,22}^{42,1}+C_{22,00}^{42,0} V_{22,00}^{42,0}+C_{22,00}^{42,1} V_{22,00}^{42,1}+C_{22,11}^{42,0} V_{22,11}^{42,0}+C_{22,11}^{42,1} V_{22,11}^{42,1}+C_{22,20}^{42,0} V_{22,20}^{42,0}+C_{22,20}^{42,1} V_{22,20}^{42,1}+C_{22,22}^{40,0} V_{22,22}^{40,0}+C_{22,22}^{40,1} V_{22,22}^{40,1}+C_{22,22}^{42,0} V_{22,22}^{42,0}+C_{22,22}^{42,1} V_{22,22}^{42,1}+C_{22,22}^{44,0} V_{22,22}^{44,0}+C_{22,22}^{44,1} V_{22,22}^{44,1}+C_{31,00}^{31,0} V_{31,00}^{31,0}+C_{31,00}^{31,1} V_{31,00}^{31,1}+C_{31,11}^{31,0} V_{31,11}^{31,0}+C_{31,11}^{31,1} V_{31,11}^{31,1}+C_{31,20}^{31,0} V_{31,20}^{31,0}+C_{31,20}^{31,1} V_{31,20}^{31,1}+C_{31,22}^{31,0} V_{31,22}^{31,0}+C_{31,22}^{31,1} V_{31,22}^{31,1}+C_{31,22}^{33,0} V_{31,22}^{33,0}+C_{31,22}^{33,1} V_{31,22}^{33,1}+C_{33,00}^{33,0} V_{33,00}^{33,0}+C_{33,00}^{33,1} V_{33,00}^{33,1}+C_{33,11}^{33,0} V_{33,11}^{33,0}+C_{33,11}^{33,1} V_{33,11}^{33,1}+C_{33,20}^{33,0} V_{33,20}^{33,0}+C_{33,20}^{33,1} V_{33,20}^{33,1}+C_{33,22}^{33,0} V_{33,22}^{33,0}+C_{33,22}^{33,1} V_{33,22}^{33,1}\).


\noindent\ \textbf{444 fourth-order isoscalar and isovector coupling constants of the nonlocal EDF expressed by 30 fourth-order finite-range pseudopotential parameters.}

\vspace{1.0cm}

\noindent\(C_{00,2000,0}^{00,2000,0,\text{Loc}}= \frac{5 C_{00,00}^{40,0}}{48}+\frac{5 C_{00,00}^{40,1}}{96}+\frac{5 C_{11,00}^{31,0}}{48}+\frac{5
C_{11,00}^{31,1}}{96}+\frac{5 C_{20,00}^{20,0}}{48}+\frac{5 C_{20,00}^{20,1}}{96}+\frac{\sqrt{5} C_{22,00}^{22,0}}{24}+\frac{\sqrt{5} C_{22,00}^{22,1}}{48}\)

\noindent\(C_{00,2000,0}^{00,2000,0,\text{NonLoc}}= -\frac{5 C_{00,00}^{40,0}}{192}-\frac{5 C_{00,00}^{40,1}}{96}+\frac{5 C_{00,20}^{40,0}}{64
\sqrt{3}}+\frac{5 C_{00,20}^{40,1}}{32 \sqrt{3}}+\frac{5 C_{11,00}^{31,0}}{192}+\frac{5 C_{11,00}^{31,1}}{96}-\frac{5 C_{11,20}^{31,0}}{64 \sqrt{3}}-\frac{5
C_{11,20}^{31,1}}{32 \sqrt{3}}-\frac{5 C_{20,00}^{20,0}}{192}-\frac{5 C_{20,00}^{20,1}}{96}+\frac{5 C_{20,20}^{20,0}}{64 \sqrt{3}}+\frac{5 C_{20,20}^{20,1}}{32
\sqrt{3}}-\frac{\sqrt{5} C_{22,00}^{22,0}}{96}-\frac{\sqrt{5} C_{22,00}^{22,1}}{48}+\frac{1}{32} \sqrt{\frac{5}{3}} C_{22,20}^{22,0}+\frac{1}{16}
\sqrt{\frac{5}{3}} C_{22,20}^{22,1}\)

\noindent\(C_{00,2000,0}^{00,2000,1,\text{Loc}}= \frac{5 C_{00,00}^{40,1}}{32 \sqrt{3}}+\frac{5 C_{11,00}^{31,1}}{32 \sqrt{3}}+\frac{5
C_{20,00}^{20,1}}{32 \sqrt{3}}+\frac{1}{16} \sqrt{\frac{5}{3}} C_{22,00}^{22,1}\)

\noindent\(C_{00,2000,0}^{00,2000,1,\text{NonLoc}}= -\frac{5 C_{00,00}^{40,0}}{64 \sqrt{3}}+\frac{5 C_{00,20}^{40,0}}{64}+\frac{5 C_{11,00}^{31,0}}{64
\sqrt{3}}-\frac{5 C_{11,20}^{31,0}}{64}-\frac{5 C_{20,00}^{20,0}}{64 \sqrt{3}}+\frac{5 C_{20,20}^{20,0}}{64}-\frac{1}{32} \sqrt{\frac{5}{3}} C_{22,00}^{22,0}+\frac{\sqrt{5}
C_{22,20}^{22,0}}{32}\)

\noindent\(C_{00,2011,1}^{00,2011,0,\text{Loc}}= -\frac{5 C_{00,20}^{40,0}}{48}-\frac{5 C_{00,20}^{40,1}}{96}-\frac{5 C_{11,20}^{31,0}}{48}-\frac{5
C_{11,20}^{31,1}}{96}-\frac{5 C_{20,20}^{20,0}}{48}-\frac{5 C_{20,20}^{20,1}}{96}-\frac{\sqrt{5} C_{22,20}^{22,0}}{24}-\frac{\sqrt{5} C_{22,20}^{22,1}}{48}\)

\noindent\(C_{00,2011,1}^{00,2011,0,\text{NonLoc}}= -\frac{5 C_{00,00}^{40,0}}{64 \sqrt{3}}-\frac{5 C_{00,00}^{40,1}}{32 \sqrt{3}}-\frac{5
C_{00,20}^{40,0}}{192}-\frac{5 C_{00,20}^{40,1}}{96}+\frac{5 C_{11,00}^{31,0}}{64 \sqrt{3}}+\frac{5 C_{11,00}^{31,1}}{32 \sqrt{3}}+\frac{5 C_{11,20}^{31,0}}{192}+\frac{5
C_{11,20}^{31,1}}{96}-\frac{5 C_{20,00}^{20,0}}{64 \sqrt{3}}-\frac{5 C_{20,00}^{20,1}}{32 \sqrt{3}}-\frac{5 C_{20,20}^{20,0}}{192}-\frac{5 C_{20,20}^{20,1}}{96}-\frac{1}{32}
\sqrt{\frac{5}{3}} C_{22,00}^{22,0}-\frac{1}{16} \sqrt{\frac{5}{3}} C_{22,00}^{22,1}-\frac{\sqrt{5} C_{22,20}^{22,0}}{96}-\frac{\sqrt{5} C_{22,20}^{22,1}}{48}\)

\noindent\(C_{00,2011,1}^{00,2011,1,\text{Loc}}= -\frac{5 C_{00,20}^{40,1}}{32 \sqrt{3}}-\frac{5 C_{11,20}^{31,1}}{32 \sqrt{3}}-\frac{5
C_{20,20}^{20,1}}{32 \sqrt{3}}-\frac{1}{16} \sqrt{\frac{5}{3}} C_{22,20}^{22,1}\)

\noindent\(C_{00,2011,1}^{00,2011,1,\text{NonLoc}}= -\frac{5 C_{00,00}^{40,0}}{64}-\frac{5 C_{00,20}^{40,0}}{64 \sqrt{3}}+\frac{5 C_{11,00}^{31,0}}{64}+\frac{5
C_{11,20}^{31,0}}{64 \sqrt{3}}-\frac{5 C_{20,00}^{20,0}}{64}-\frac{5 C_{20,20}^{20,0}}{64 \sqrt{3}}-\frac{\sqrt{5} C_{22,00}^{22,0}}{32}-\frac{1}{32}
\sqrt{\frac{5}{3}} C_{22,20}^{22,0}\)

\noindent\(C_{00,2202,2}^{00,2202,0,\text{Loc}}= \frac{\sqrt{5} C_{00,00}^{40,0}}{24}+\frac{\sqrt{5} C_{00,00}^{40,1}}{48}+\frac{\sqrt{5}
C_{11,00}^{31,0}}{24}+\frac{\sqrt{5} C_{11,00}^{31,1}}{48}+\frac{\sqrt{5} C_{20,00}^{20,0}}{24}+\frac{\sqrt{5} C_{20,00}^{20,1}}{48}+\frac{C_{22,00}^{22,0}}{12}+\frac{C_{22,00}^{22,1}}{24}\)

\noindent\(C_{00,2202,2}^{00,2202,0,\text{NonLoc}}= -\frac{\sqrt{5} C_{00,00}^{40,0}}{96}-\frac{\sqrt{5} C_{00,00}^{40,1}}{48}+\frac{1}{32}
\sqrt{\frac{5}{3}} C_{00,20}^{40,0}+\frac{1}{16} \sqrt{\frac{5}{3}} C_{00,20}^{40,1}+\frac{\sqrt{5} C_{11,00}^{31,0}}{96}+\frac{\sqrt{5} C_{11,00}^{31,1}}{48}-\frac{1}{32}
\sqrt{\frac{5}{3}} C_{11,20}^{31,0}-\frac{1}{16} \sqrt{\frac{5}{3}} C_{11,20}^{31,1}-\frac{\sqrt{5} C_{20,00}^{20,0}}{96}-\frac{\sqrt{5} C_{20,00}^{20,1}}{48}+\frac{1}{32}
\sqrt{\frac{5}{3}} C_{20,20}^{20,0}+\frac{1}{16} \sqrt{\frac{5}{3}} C_{20,20}^{20,1}-\frac{C_{22,00}^{22,0}}{48}-\frac{C_{22,00}^{22,1}}{24}+\frac{C_{22,20}^{22,0}}{16
\sqrt{3}}+\frac{C_{22,20}^{22,1}}{8 \sqrt{3}}\)

\noindent\(C_{00,2202,2}^{00,2202,1,\text{Loc}}= \frac{1}{16} \sqrt{\frac{5}{3}} C_{00,00}^{40,1}+\frac{1}{16} \sqrt{\frac{5}{3}} C_{11,00}^{31,1}+\frac{1}{16}
\sqrt{\frac{5}{3}} C_{20,00}^{20,1}+\frac{C_{22,00}^{22,1}}{8 \sqrt{3}}\)

\noindent\(C_{00,2202,2}^{00,2202,1,\text{NonLoc}}= -\frac{1}{32} \sqrt{\frac{5}{3}} C_{00,00}^{40,0}+\frac{\sqrt{5} C_{00,20}^{40,0}}{32}+\frac{1}{32}
\sqrt{\frac{5}{3}} C_{11,00}^{31,0}-\frac{\sqrt{5} C_{11,20}^{31,0}}{32}-\frac{1}{32} \sqrt{\frac{5}{3}} C_{20,00}^{20,0}+\frac{\sqrt{5} C_{20,20}^{20,0}}{32}-\frac{C_{22,00}^{22,0}}{16
\sqrt{3}}+\frac{C_{22,20}^{22,0}}{16}\)

\noindent\(C_{00,2211,1}^{00,2011,0,\text{Loc}}= -\frac{7 C_{00,22}^{42,0}}{48}-\frac{7 C_{00,22}^{42,1}}{96}-\frac{7 C_{11,22}^{31,0}}{48}-\frac{7
C_{11,22}^{31,1}}{96}-\frac{1}{16} \sqrt{\frac{21}{5}} C_{11,22}^{33,0}-\frac{1}{32} \sqrt{\frac{21}{5}} C_{11,22}^{33,1}-\frac{7 C_{20,22}^{22,0}}{48}-\frac{7
C_{20,22}^{22,1}}{96}-\frac{\sqrt{7} C_{22,22}^{22,0}}{24}-\frac{\sqrt{7} C_{22,22}^{22,1}}{48}\)

\noindent\(C_{00,2211,1}^{00,2011,0,\text{NonLoc}}= \frac{7 C_{00,22}^{42,0}}{96}+\frac{7 C_{00,22}^{42,1}}{48}-\frac{7 C_{11,22}^{31,0}}{96}-\frac{7
C_{11,22}^{31,1}}{48}-\frac{1}{32} \sqrt{\frac{21}{5}} C_{11,22}^{33,0}-\frac{1}{16} \sqrt{\frac{21}{5}} C_{11,22}^{33,1}+\frac{7 C_{20,22}^{22,0}}{96}+\frac{7
C_{20,22}^{22,1}}{48}+\frac{\sqrt{7} C_{22,22}^{22,0}}{48}+\frac{\sqrt{7} C_{22,22}^{22,1}}{24}\)

\noindent\(C_{00,2211,1}^{00,2011,1,\text{Loc}}= -\frac{7 C_{00,22}^{42,1}}{32 \sqrt{3}}-\frac{7 C_{11,22}^{31,1}}{32 \sqrt{3}}-\frac{3}{32}
\sqrt{\frac{7}{5}} C_{11,22}^{33,1}-\frac{7 C_{20,22}^{22,1}}{32 \sqrt{3}}-\frac{1}{16} \sqrt{\frac{7}{3}} C_{22,22}^{22,1}\)

\noindent\(C_{00,2211,1}^{00,2011,1,\text{NonLoc}}= \frac{7 C_{00,22}^{42,0}}{32 \sqrt{3}}-\frac{7 C_{11,22}^{31,0}}{32 \sqrt{3}}-\frac{3}{32}
\sqrt{\frac{7}{5}} C_{11,22}^{33,0}+\frac{7 C_{20,22}^{22,0}}{32 \sqrt{3}}+\frac{1}{16} \sqrt{\frac{7}{3}} C_{22,22}^{22,0}\)

\noindent\(C_{00,2211,1}^{00,2211,0,\text{Loc}}= -\frac{C_{00,20}^{40,0}}{24}-\frac{C_{00,20}^{40,1}}{48}-\frac{7 C_{00,22}^{42,0}}{96
\sqrt{5}}-\frac{7 C_{00,22}^{42,1}}{192 \sqrt{5}}-\frac{C_{11,20}^{31,0}}{24}-\frac{C_{11,20}^{31,1}}{48}-\frac{7 C_{11,22}^{31,0}}{96 \sqrt{5}}-\frac{7
C_{11,22}^{31,1}}{192 \sqrt{5}}-\frac{\sqrt{21} C_{11,22}^{33,0}}{160}-\frac{\sqrt{21} C_{11,22}^{33,1}}{320}-\frac{C_{20,20}^{20,0}}{24}-\frac{C_{20,20}^{20,1}}{48}-\frac{7
C_{20,22}^{22,0}}{96 \sqrt{5}}-\frac{7 C_{20,22}^{22,1}}{192 \sqrt{5}}-\frac{C_{22,20}^{22,0}}{12 \sqrt{5}}-\frac{C_{22,20}^{22,1}}{24 \sqrt{5}}-\frac{1}{48}
\sqrt{\frac{7}{5}} C_{22,22}^{22,0}-\frac{1}{96} \sqrt{\frac{7}{5}} C_{22,22}^{22,1}\)

\noindent\(C_{00,2211,1}^{00,2211,0,\text{NonLoc}}= -\frac{C_{00,00}^{40,0}}{32 \sqrt{3}}-\frac{C_{00,00}^{40,1}}{16 \sqrt{3}}-\frac{C_{00,20}^{40,0}}{96}-\frac{C_{00,20}^{40,1}}{48}+\frac{7
C_{00,22}^{42,0}}{192 \sqrt{5}}+\frac{7 C_{00,22}^{42,1}}{96 \sqrt{5}}+\frac{C_{11,00}^{31,0}}{32 \sqrt{3}}+\frac{C_{11,00}^{31,1}}{16 \sqrt{3}}+\frac{C_{11,20}^{31,0}}{96}+\frac{C_{11,20}^{31,1}}{48}-\frac{7
C_{11,22}^{31,0}}{192 \sqrt{5}}-\frac{7 C_{11,22}^{31,1}}{96 \sqrt{5}}-\frac{\sqrt{21} C_{11,22}^{33,0}}{320}-\frac{\sqrt{21} C_{11,22}^{33,1}}{160}-\frac{C_{20,00}^{20,0}}{32
\sqrt{3}}-\frac{C_{20,00}^{20,1}}{16 \sqrt{3}}-\frac{C_{20,20}^{20,0}}{96}-\frac{C_{20,20}^{20,1}}{48}+\frac{7 C_{20,22}^{22,0}}{192 \sqrt{5}}+\frac{7
C_{20,22}^{22,1}}{96 \sqrt{5}}-\frac{C_{22,00}^{22,0}}{16 \sqrt{15}}-\frac{C_{22,00}^{22,1}}{8 \sqrt{15}}-\frac{C_{22,20}^{22,0}}{48 \sqrt{5}}-\frac{C_{22,20}^{22,1}}{24
\sqrt{5}}+\frac{1}{96} \sqrt{\frac{7}{5}} C_{22,22}^{22,0}+\frac{1}{48} \sqrt{\frac{7}{5}} C_{22,22}^{22,1}\)

\noindent\(C_{00,2211,1}^{00,2211,1,\text{Loc}}= -\frac{C_{00,20}^{40,1}}{16 \sqrt{3}}-\frac{7 C_{00,22}^{42,1}}{64 \sqrt{15}}-\frac{C_{11,20}^{31,1}}{16
\sqrt{3}}-\frac{7 C_{11,22}^{31,1}}{64 \sqrt{15}}-\frac{3 \sqrt{7} C_{11,22}^{33,1}}{320}-\frac{C_{20,20}^{20,1}}{16 \sqrt{3}}-\frac{7 C_{20,22}^{22,1}}{64
\sqrt{15}}-\frac{C_{22,20}^{22,1}}{8 \sqrt{15}}-\frac{1}{32} \sqrt{\frac{7}{15}} C_{22,22}^{22,1}\)

\noindent\(C_{00,2211,1}^{00,2211,1,\text{NonLoc}}= -\frac{C_{00,00}^{40,0}}{32}-\frac{C_{00,20}^{40,0}}{32 \sqrt{3}}+\frac{7 C_{00,22}^{42,0}}{64
\sqrt{15}}+\frac{C_{11,00}^{31,0}}{32}+\frac{C_{11,20}^{31,0}}{32 \sqrt{3}}-\frac{7 C_{11,22}^{31,0}}{64 \sqrt{15}}-\frac{3 \sqrt{7} C_{11,22}^{33,0}}{320}-\frac{C_{20,00}^{20,0}}{32}-\frac{C_{20,20}^{20,0}}{32
\sqrt{3}}+\frac{7 C_{20,22}^{22,0}}{64 \sqrt{15}}-\frac{C_{22,00}^{22,0}}{16 \sqrt{5}}-\frac{C_{22,20}^{22,0}}{16 \sqrt{15}}+\frac{1}{32} \sqrt{\frac{7}{15}}
C_{22,22}^{22,0}\)

\noindent\(C_{00,2212,2}^{00,2212,0,\text{Loc}}= -\frac{1}{24} \sqrt{\frac{5}{3}} C_{00,20}^{40,0}-\frac{1}{48} \sqrt{\frac{5}{3}}
C_{00,20}^{40,1}+\frac{7 C_{00,22}^{42,0}}{96 \sqrt{3}}+\frac{7 C_{00,22}^{42,1}}{192 \sqrt{3}}-\frac{1}{24} \sqrt{\frac{5}{3}} C_{11,20}^{31,0}-\frac{1}{48}
\sqrt{\frac{5}{3}} C_{11,20}^{31,1}+\frac{7 C_{11,22}^{31,0}}{96 \sqrt{3}}+\frac{7 C_{11,22}^{31,1}}{192 \sqrt{3}}+\frac{1}{32} \sqrt{\frac{7}{5}}
C_{11,22}^{33,0}+\frac{1}{64} \sqrt{\frac{7}{5}} C_{11,22}^{33,1}-\frac{1}{24} \sqrt{\frac{5}{3}} C_{20,20}^{20,0}-\frac{1}{48} \sqrt{\frac{5}{3}}
C_{20,20}^{20,1}+\frac{7 C_{20,22}^{22,0}}{96 \sqrt{3}}+\frac{7 C_{20,22}^{22,1}}{192 \sqrt{3}}-\frac{C_{22,20}^{22,0}}{12 \sqrt{3}}-\frac{C_{22,20}^{22,1}}{24
\sqrt{3}}+\frac{1}{48} \sqrt{\frac{7}{3}} C_{22,22}^{22,0}+\frac{1}{96} \sqrt{\frac{7}{3}} C_{22,22}^{22,1}\)

\noindent\(C_{00,2212,2}^{00,2212,0,\text{NonLoc}}= -\frac{\sqrt{5} C_{00,00}^{40,0}}{96}-\frac{\sqrt{5} C_{00,00}^{40,1}}{48}-\frac{1}{96}
\sqrt{\frac{5}{3}} C_{00,20}^{40,0}-\frac{1}{48} \sqrt{\frac{5}{3}} C_{00,20}^{40,1}-\frac{7 C_{00,22}^{42,0}}{192 \sqrt{3}}-\frac{7 C_{00,22}^{42,1}}{96
\sqrt{3}}+\frac{\sqrt{5} C_{11,00}^{31,0}}{96}+\frac{\sqrt{5} C_{11,00}^{31,1}}{48}+\frac{1}{96} \sqrt{\frac{5}{3}} C_{11,20}^{31,0}+\frac{1}{48}
\sqrt{\frac{5}{3}} C_{11,20}^{31,1}+\frac{7 C_{11,22}^{31,0}}{192 \sqrt{3}}+\frac{7 C_{11,22}^{31,1}}{96 \sqrt{3}}+\frac{1}{64} \sqrt{\frac{7}{5}}
C_{11,22}^{33,0}+\frac{1}{32} \sqrt{\frac{7}{5}} C_{11,22}^{33,1}-\frac{\sqrt{5} C_{20,00}^{20,0}}{96}-\frac{\sqrt{5} C_{20,00}^{20,1}}{48}-\frac{1}{96}
\sqrt{\frac{5}{3}} C_{20,20}^{20,0}-\frac{1}{48} \sqrt{\frac{5}{3}} C_{20,20}^{20,1}-\frac{7 C_{20,22}^{22,0}}{192 \sqrt{3}}-\frac{7 C_{20,22}^{22,1}}{96
\sqrt{3}}-\frac{C_{22,00}^{22,0}}{48}-\frac{C_{22,00}^{22,1}}{24}-\frac{C_{22,20}^{22,0}}{48 \sqrt{3}}-\frac{C_{22,20}^{22,1}}{24 \sqrt{3}}-\frac{1}{96}
\sqrt{\frac{7}{3}} C_{22,22}^{22,0}-\frac{1}{48} \sqrt{\frac{7}{3}} C_{22,22}^{22,1}\)

\noindent\(C_{00,2212,2}^{00,2212,1,\text{Loc}}= -\frac{\sqrt{5} C_{00,20}^{40,1}}{48}+\frac{7 C_{00,22}^{42,1}}{192}-\frac{\sqrt{5}
C_{11,20}^{31,1}}{48}+\frac{7 C_{11,22}^{31,1}}{192}+\frac{1}{64} \sqrt{\frac{21}{5}} C_{11,22}^{33,1}-\frac{\sqrt{5} C_{20,20}^{20,1}}{48}+\frac{7
C_{20,22}^{22,1}}{192}-\frac{C_{22,20}^{22,1}}{24}+\frac{\sqrt{7} C_{22,22}^{22,1}}{96}\)

\noindent\(C_{00,2212,2}^{00,2212,1,\text{NonLoc}}= -\frac{1}{32} \sqrt{\frac{5}{3}} C_{00,00}^{40,0}-\frac{\sqrt{5} C_{00,20}^{40,0}}{96}-\frac{7
C_{00,22}^{42,0}}{192}+\frac{1}{32} \sqrt{\frac{5}{3}} C_{11,00}^{31,0}+\frac{\sqrt{5} C_{11,20}^{31,0}}{96}+\frac{7 C_{11,22}^{31,0}}{192}+\frac{1}{64}
\sqrt{\frac{21}{5}} C_{11,22}^{33,0}-\frac{1}{32} \sqrt{\frac{5}{3}} C_{20,00}^{20,0}-\frac{\sqrt{5} C_{20,20}^{20,0}}{96}-\frac{7 C_{20,22}^{22,0}}{192}-\frac{C_{22,00}^{22,0}}{16
\sqrt{3}}-\frac{C_{22,20}^{22,0}}{48}-\frac{\sqrt{7} C_{22,22}^{22,0}}{96}\)

\noindent\(C_{00,2213,3}^{00,2213,0,\text{Loc}}= -\frac{1}{24} \sqrt{\frac{7}{3}} C_{00,20}^{40,0}-\frac{1}{48} \sqrt{\frac{7}{3}}
C_{00,20}^{40,1}-\frac{1}{48} \sqrt{\frac{7}{15}} C_{00,22}^{42,0}-\frac{1}{96} \sqrt{\frac{7}{15}} C_{00,22}^{42,1}-\frac{1}{24} \sqrt{\frac{7}{3}}
C_{11,20}^{31,0}-\frac{1}{48} \sqrt{\frac{7}{3}} C_{11,20}^{31,1}-\frac{1}{48} \sqrt{\frac{7}{15}} C_{11,22}^{31,0}-\frac{1}{96} \sqrt{\frac{7}{15}}
C_{11,22}^{31,1}-\frac{C_{11,22}^{33,0}}{80}-\frac{C_{11,22}^{33,1}}{160}-\frac{1}{24} \sqrt{\frac{7}{3}} C_{20,20}^{20,0}-\frac{1}{48} \sqrt{\frac{7}{3}}
C_{20,20}^{20,1}-\frac{1}{48} \sqrt{\frac{7}{15}} C_{20,22}^{22,0}-\frac{1}{96} \sqrt{\frac{7}{15}} C_{20,22}^{22,1}-\frac{1}{12} \sqrt{\frac{7}{15}}
C_{22,20}^{22,0}-\frac{1}{24} \sqrt{\frac{7}{15}} C_{22,20}^{22,1}-\frac{C_{22,22}^{22,0}}{24 \sqrt{15}}-\frac{C_{22,22}^{22,1}}{48 \sqrt{15}}\)

\noindent\(C_{00,2213,3}^{00,2213,0,\text{NonLoc}}= -\frac{\sqrt{7} C_{00,00}^{40,0}}{96}-\frac{\sqrt{7} C_{00,00}^{40,1}}{48}-\frac{1}{96}
\sqrt{\frac{7}{3}} C_{00,20}^{40,0}-\frac{1}{48} \sqrt{\frac{7}{3}} C_{00,20}^{40,1}+\frac{1}{96} \sqrt{\frac{7}{15}} C_{00,22}^{42,0}+\frac{1}{48}
\sqrt{\frac{7}{15}} C_{00,22}^{42,1}+\frac{\sqrt{7} C_{11,00}^{31,0}}{96}+\frac{\sqrt{7} C_{11,00}^{31,1}}{48}+\frac{1}{96} \sqrt{\frac{7}{3}} C_{11,20}^{31,0}+\frac{1}{48}
\sqrt{\frac{7}{3}} C_{11,20}^{31,1}-\frac{1}{96} \sqrt{\frac{7}{15}} C_{11,22}^{31,0}-\frac{1}{48} \sqrt{\frac{7}{15}} C_{11,22}^{31,1}-\frac{C_{11,22}^{33,0}}{160}-\frac{C_{11,22}^{33,1}}{80}-\frac{\sqrt{7}
C_{20,00}^{20,0}}{96}-\frac{\sqrt{7} C_{20,00}^{20,1}}{48}-\frac{1}{96} \sqrt{\frac{7}{3}} C_{20,20}^{20,0}-\frac{1}{48} \sqrt{\frac{7}{3}} C_{20,20}^{20,1}+\frac{1}{96}
\sqrt{\frac{7}{15}} C_{20,22}^{22,0}+\frac{1}{48} \sqrt{\frac{7}{15}} C_{20,22}^{22,1}-\frac{1}{48} \sqrt{\frac{7}{5}} C_{22,00}^{22,0}-\frac{1}{24}
\sqrt{\frac{7}{5}} C_{22,00}^{22,1}-\frac{1}{48} \sqrt{\frac{7}{15}} C_{22,20}^{22,0}-\frac{1}{24} \sqrt{\frac{7}{15}} C_{22,20}^{22,1}+\frac{C_{22,22}^{22,0}}{48
\sqrt{15}}+\frac{C_{22,22}^{22,1}}{24 \sqrt{15}}\)

\noindent\(C_{00,2213,3}^{00,2213,1,\text{Loc}}= -\frac{\sqrt{7} C_{00,20}^{40,1}}{48}-\frac{1}{96} \sqrt{\frac{7}{5}} C_{00,22}^{42,1}-\frac{\sqrt{7}
C_{11,20}^{31,1}}{48}-\frac{1}{96} \sqrt{\frac{7}{5}} C_{11,22}^{31,1}-\frac{\sqrt{3} C_{11,22}^{33,1}}{160}-\frac{\sqrt{7} C_{20,20}^{20,1}}{48}-\frac{1}{96}
\sqrt{\frac{7}{5}} C_{20,22}^{22,1}-\frac{1}{24} \sqrt{\frac{7}{5}} C_{22,20}^{22,1}-\frac{C_{22,22}^{22,1}}{48 \sqrt{5}}\)

\noindent\(C_{00,2213,3}^{00,2213,1,\text{NonLoc}}= -\frac{1}{32} \sqrt{\frac{7}{3}} C_{00,00}^{40,0}-\frac{\sqrt{7} C_{00,20}^{40,0}}{96}+\frac{1}{96}
\sqrt{\frac{7}{5}} C_{00,22}^{42,0}+\frac{1}{32} \sqrt{\frac{7}{3}} C_{11,00}^{31,0}+\frac{\sqrt{7} C_{11,20}^{31,0}}{96}-\frac{1}{96} \sqrt{\frac{7}{5}}
C_{11,22}^{31,0}-\frac{\sqrt{3} C_{11,22}^{33,0}}{160}-\frac{1}{32} \sqrt{\frac{7}{3}} C_{20,00}^{20,0}-\frac{\sqrt{7} C_{20,20}^{20,0}}{96}+\frac{1}{96}
\sqrt{\frac{7}{5}} C_{20,22}^{22,0}-\frac{1}{16} \sqrt{\frac{7}{15}} C_{22,00}^{22,0}-\frac{1}{48} \sqrt{\frac{7}{5}} C_{22,20}^{22,0}+\frac{C_{22,22}^{22,0}}{48
\sqrt{5}}\)

\noindent\(C_{00,3101,1}^{00,1101,0,\text{Loc}}= -\frac{C_{00,00}^{40,0}}{4}-\frac{C_{00,00}^{40,1}}{8}-\frac{C_{11,00}^{31,0}}{4}-\frac{C_{11,00}^{31,1}}{8}-\frac{C_{20,00}^{20,0}}{4}-\frac{C_{20,00}^{20,1}}{8}-\frac{C_{22,00}^{22,0}}{2
\sqrt{5}}-\frac{C_{22,00}^{22,1}}{4 \sqrt{5}}\)

\noindent\(C_{00,3101,1}^{00,1101,0,\text{NonLoc}}= \frac{C_{00,00}^{40,0}}{16}+\frac{C_{00,00}^{40,1}}{8}-\frac{\sqrt{3} C_{00,20}^{40,0}}{16}-\frac{\sqrt{3}
C_{00,20}^{40,1}}{8}-\frac{C_{11,00}^{31,0}}{16}-\frac{C_{11,00}^{31,1}}{8}+\frac{\sqrt{3} C_{11,20}^{31,0}}{16}+\frac{\sqrt{3} C_{11,20}^{31,1}}{8}+\frac{C_{20,00}^{20,0}}{16}+\frac{C_{20,00}^{20,1}}{8}-\frac{\sqrt{3}
C_{20,20}^{20,0}}{16}-\frac{\sqrt{3} C_{20,20}^{20,1}}{8}+\frac{C_{22,00}^{22,0}}{8 \sqrt{5}}+\frac{C_{22,00}^{22,1}}{4 \sqrt{5}}-\frac{1}{8} \sqrt{\frac{3}{5}}
C_{22,20}^{22,0}-\frac{1}{4} \sqrt{\frac{3}{5}} C_{22,20}^{22,1}\)

\noindent\(C_{00,3101,1}^{00,1101,1,\text{Loc}}= -\frac{\sqrt{3} C_{00,00}^{40,1}}{8}-\frac{\sqrt{3} C_{11,00}^{31,1}}{8}-\frac{\sqrt{3}
C_{20,00}^{20,1}}{8}-\frac{1}{4} \sqrt{\frac{3}{5}} C_{22,00}^{22,1}\)

\noindent\(C_{00,3101,1}^{00,1101,1,\text{NonLoc}}= \frac{\sqrt{3} C_{00,00}^{40,0}}{16}-\frac{3 C_{00,20}^{40,0}}{16}-\frac{\sqrt{3}
C_{11,00}^{31,0}}{16}+\frac{3 C_{11,20}^{31,0}}{16}+\frac{\sqrt{3} C_{20,00}^{20,0}}{16}-\frac{3 C_{20,20}^{20,0}}{16}+\frac{1}{8} \sqrt{\frac{3}{5}}
C_{22,00}^{22,0}-\frac{3 C_{22,20}^{22,0}}{8 \sqrt{5}}\)

\noindent\(C_{00,3110,0}^{00,1110,0,\text{Loc}}= \frac{C_{00,20}^{40,0}}{12}+\frac{C_{00,20}^{40,1}}{24}+\frac{7 C_{00,22}^{42,0}}{24
\sqrt{5}}+\frac{7 C_{00,22}^{42,1}}{48 \sqrt{5}}+\frac{C_{11,20}^{31,0}}{12}+\frac{C_{11,20}^{31,1}}{24}+\frac{7 C_{11,22}^{31,0}}{24 \sqrt{5}}+\frac{7
C_{11,22}^{31,1}}{48 \sqrt{5}}+\frac{\sqrt{21} C_{11,22}^{33,0}}{40}+\frac{\sqrt{21} C_{11,22}^{33,1}}{80}+\frac{C_{20,20}^{20,0}}{12}+\frac{C_{20,20}^{20,1}}{24}+\frac{7
C_{20,22}^{22,0}}{24 \sqrt{5}}+\frac{7 C_{20,22}^{22,1}}{48 \sqrt{5}}+\frac{C_{22,20}^{22,0}}{6 \sqrt{5}}+\frac{C_{22,20}^{22,1}}{12 \sqrt{5}}+\frac{1}{12}
\sqrt{\frac{7}{5}} C_{22,22}^{22,0}+\frac{1}{24} \sqrt{\frac{7}{5}} C_{22,22}^{22,1}\)

\noindent\(C_{00,3110,0}^{00,1110,0,\text{NonLoc}}= \frac{C_{00,00}^{40,0}}{16 \sqrt{3}}+\frac{C_{00,00}^{40,1}}{8 \sqrt{3}}+\frac{C_{00,20}^{40,0}}{48}+\frac{C_{00,20}^{40,1}}{24}-\frac{7
C_{00,22}^{42,0}}{48 \sqrt{5}}-\frac{7 C_{00,22}^{42,1}}{24 \sqrt{5}}-\frac{C_{11,00}^{31,0}}{16 \sqrt{3}}-\frac{C_{11,00}^{31,1}}{8 \sqrt{3}}-\frac{C_{11,20}^{31,0}}{48}-\frac{C_{11,20}^{31,1}}{24}+\frac{7
C_{11,22}^{31,0}}{48 \sqrt{5}}+\frac{7 C_{11,22}^{31,1}}{24 \sqrt{5}}+\frac{\sqrt{21} C_{11,22}^{33,0}}{80}+\frac{\sqrt{21} C_{11,22}^{33,1}}{40}+\frac{C_{20,00}^{20,0}}{16
\sqrt{3}}+\frac{C_{20,00}^{20,1}}{8 \sqrt{3}}+\frac{C_{20,20}^{20,0}}{48}+\frac{C_{20,20}^{20,1}}{24}-\frac{7 C_{20,22}^{22,0}}{48 \sqrt{5}}-\frac{7
C_{20,22}^{22,1}}{24 \sqrt{5}}+\frac{C_{22,00}^{22,0}}{8 \sqrt{15}}+\frac{C_{22,00}^{22,1}}{4 \sqrt{15}}+\frac{C_{22,20}^{22,0}}{24 \sqrt{5}}+\frac{C_{22,20}^{22,1}}{12
\sqrt{5}}-\frac{1}{24} \sqrt{\frac{7}{5}} C_{22,22}^{22,0}-\frac{1}{12} \sqrt{\frac{7}{5}} C_{22,22}^{22,1}\)

\noindent\(C_{00,3110,0}^{00,1110,1,\text{Loc}}= \frac{C_{00,20}^{40,1}}{8 \sqrt{3}}+\frac{7 C_{00,22}^{42,1}}{16 \sqrt{15}}+\frac{C_{11,20}^{31,1}}{8
\sqrt{3}}+\frac{7 C_{11,22}^{31,1}}{16 \sqrt{15}}+\frac{3 \sqrt{7} C_{11,22}^{33,1}}{80}+\frac{C_{20,20}^{20,1}}{8 \sqrt{3}}+\frac{7 C_{20,22}^{22,1}}{16
\sqrt{15}}+\frac{C_{22,20}^{22,1}}{4 \sqrt{15}}+\frac{1}{8} \sqrt{\frac{7}{15}} C_{22,22}^{22,1}\)

\noindent\(C_{00,3110,0}^{00,1110,1,\text{NonLoc}}= \frac{C_{00,00}^{40,0}}{16}+\frac{C_{00,20}^{40,0}}{16 \sqrt{3}}-\frac{7 C_{00,22}^{42,0}}{16
\sqrt{15}}-\frac{C_{11,00}^{31,0}}{16}-\frac{C_{11,20}^{31,0}}{16 \sqrt{3}}+\frac{7 C_{11,22}^{31,0}}{16 \sqrt{15}}+\frac{3 \sqrt{7} C_{11,22}^{33,0}}{80}+\frac{C_{20,00}^{20,0}}{16}+\frac{C_{20,20}^{20,0}}{16
\sqrt{3}}-\frac{7 C_{20,22}^{22,0}}{16 \sqrt{15}}+\frac{C_{22,00}^{22,0}}{8 \sqrt{5}}+\frac{C_{22,20}^{22,0}}{8 \sqrt{15}}-\frac{1}{8} \sqrt{\frac{7}{15}}
C_{22,22}^{22,0}\)

\noindent\(C_{00,3111,1}^{00,1111,0,\text{Loc}}= \frac{C_{00,20}^{40,0}}{4 \sqrt{3}}+\frac{C_{00,20}^{40,1}}{8 \sqrt{3}}-\frac{7 C_{00,22}^{42,0}}{16
\sqrt{15}}-\frac{7 C_{00,22}^{42,1}}{32 \sqrt{15}}+\frac{C_{11,20}^{31,0}}{4 \sqrt{3}}+\frac{C_{11,20}^{31,1}}{8 \sqrt{3}}-\frac{7 C_{11,22}^{31,0}}{16
\sqrt{15}}-\frac{7 C_{11,22}^{31,1}}{32 \sqrt{15}}-\frac{3 \sqrt{7} C_{11,22}^{33,0}}{80}-\frac{3 \sqrt{7} C_{11,22}^{33,1}}{160}+\frac{C_{20,20}^{20,0}}{4
\sqrt{3}}+\frac{C_{20,20}^{20,1}}{8 \sqrt{3}}-\frac{7 C_{20,22}^{22,0}}{16 \sqrt{15}}-\frac{7 C_{20,22}^{22,1}}{32 \sqrt{15}}+\frac{C_{22,20}^{22,0}}{2
\sqrt{15}}+\frac{C_{22,20}^{22,1}}{4 \sqrt{15}}-\frac{1}{8} \sqrt{\frac{7}{15}} C_{22,22}^{22,0}-\frac{1}{16} \sqrt{\frac{7}{15}} C_{22,22}^{22,1}\)

\noindent\(C_{00,3111,1}^{00,1111,0,\text{NonLoc}}= \frac{C_{00,00}^{40,0}}{16}+\frac{C_{00,00}^{40,1}}{8}+\frac{C_{00,20}^{40,0}}{16
\sqrt{3}}+\frac{C_{00,20}^{40,1}}{8 \sqrt{3}}+\frac{7 C_{00,22}^{42,0}}{32 \sqrt{15}}+\frac{7 C_{00,22}^{42,1}}{16 \sqrt{15}}-\frac{C_{11,00}^{31,0}}{16}-\frac{C_{11,00}^{31,1}}{8}-\frac{C_{11,20}^{31,0}}{16
\sqrt{3}}-\frac{C_{11,20}^{31,1}}{8 \sqrt{3}}-\frac{7 C_{11,22}^{31,0}}{32 \sqrt{15}}-\frac{7 C_{11,22}^{31,1}}{16 \sqrt{15}}-\frac{3 \sqrt{7} C_{11,22}^{33,0}}{160}-\frac{3
\sqrt{7} C_{11,22}^{33,1}}{80}+\frac{C_{20,00}^{20,0}}{16}+\frac{C_{20,00}^{20,1}}{8}+\frac{C_{20,20}^{20,0}}{16 \sqrt{3}}+\frac{C_{20,20}^{20,1}}{8
\sqrt{3}}+\frac{7 C_{20,22}^{22,0}}{32 \sqrt{15}}+\frac{7 C_{20,22}^{22,1}}{16 \sqrt{15}}+\frac{C_{22,00}^{22,0}}{8 \sqrt{5}}+\frac{C_{22,00}^{22,1}}{4
\sqrt{5}}+\frac{C_{22,20}^{22,0}}{8 \sqrt{15}}+\frac{C_{22,20}^{22,1}}{4 \sqrt{15}}+\frac{1}{16} \sqrt{\frac{7}{15}} C_{22,22}^{22,0}+\frac{1}{8}
\sqrt{\frac{7}{15}} C_{22,22}^{22,1}\)

\noindent\(C_{00,3111,1}^{00,1111,1,\text{Loc}}= \frac{C_{00,20}^{40,1}}{8}-\frac{7 C_{00,22}^{42,1}}{32 \sqrt{5}}+\frac{C_{11,20}^{31,1}}{8}-\frac{7
C_{11,22}^{31,1}}{32 \sqrt{5}}-\frac{3 \sqrt{21} C_{11,22}^{33,1}}{160}+\frac{C_{20,20}^{20,1}}{8}-\frac{7 C_{20,22}^{22,1}}{32 \sqrt{5}}+\frac{C_{22,20}^{22,1}}{4
\sqrt{5}}-\frac{1}{16} \sqrt{\frac{7}{5}} C_{22,22}^{22,1}\)

\noindent\(C_{00,3111,1}^{00,1111,1,\text{NonLoc}}= \frac{\sqrt{3} C_{00,00}^{40,0}}{16}+\frac{C_{00,20}^{40,0}}{16}+\frac{7 C_{00,22}^{42,0}}{32
\sqrt{5}}-\frac{\sqrt{3} C_{11,00}^{31,0}}{16}-\frac{C_{11,20}^{31,0}}{16}-\frac{7 C_{11,22}^{31,0}}{32 \sqrt{5}}-\frac{3 \sqrt{21} C_{11,22}^{33,0}}{160}+\frac{\sqrt{3}
C_{20,00}^{20,0}}{16}+\frac{C_{20,20}^{20,0}}{16}+\frac{7 C_{20,22}^{22,0}}{32 \sqrt{5}}+\frac{1}{8} \sqrt{\frac{3}{5}} C_{22,00}^{22,0}+\frac{C_{22,20}^{22,0}}{8
\sqrt{5}}+\frac{1}{16} \sqrt{\frac{7}{5}} C_{22,22}^{22,0}\)

\noindent\(C_{00,3112,2}^{00,1112,0,\text{Loc}}= \frac{\sqrt{5} C_{00,20}^{40,0}}{12}+\frac{\sqrt{5} C_{00,20}^{40,1}}{24}+\frac{7
C_{00,22}^{42,0}}{240}+\frac{7 C_{00,22}^{42,1}}{480}+\frac{\sqrt{5} C_{11,20}^{31,0}}{12}+\frac{\sqrt{5} C_{11,20}^{31,1}}{24}+\frac{7 C_{11,22}^{31,0}}{240}+\frac{7
C_{11,22}^{31,1}}{480}+\frac{1}{80} \sqrt{\frac{21}{5}} C_{11,22}^{33,0}+\frac{1}{160} \sqrt{\frac{21}{5}} C_{11,22}^{33,1}+\frac{\sqrt{5} C_{20,20}^{20,0}}{12}+\frac{\sqrt{5}
C_{20,20}^{20,1}}{24}+\frac{7 C_{20,22}^{22,0}}{240}+\frac{7 C_{20,22}^{22,1}}{480}+\frac{C_{22,20}^{22,0}}{6}+\frac{C_{22,20}^{22,1}}{12}+\frac{\sqrt{7}
C_{22,22}^{22,0}}{120}+\frac{\sqrt{7} C_{22,22}^{22,1}}{240}\)

\noindent\(C_{00,3112,2}^{00,1112,0,\text{NonLoc}}= \frac{1}{16} \sqrt{\frac{5}{3}} C_{00,00}^{40,0}+\frac{1}{8} \sqrt{\frac{5}{3}}
C_{00,00}^{40,1}+\frac{\sqrt{5} C_{00,20}^{40,0}}{48}+\frac{\sqrt{5} C_{00,20}^{40,1}}{24}-\frac{7 C_{00,22}^{42,0}}{480}-\frac{7 C_{00,22}^{42,1}}{240}-\frac{1}{16}
\sqrt{\frac{5}{3}} C_{11,00}^{31,0}-\frac{1}{8} \sqrt{\frac{5}{3}} C_{11,00}^{31,1}-\frac{\sqrt{5} C_{11,20}^{31,0}}{48}-\frac{\sqrt{5} C_{11,20}^{31,1}}{24}+\frac{7
C_{11,22}^{31,0}}{480}+\frac{7 C_{11,22}^{31,1}}{240}+\frac{1}{160} \sqrt{\frac{21}{5}} C_{11,22}^{33,0}+\frac{1}{80} \sqrt{\frac{21}{5}} C_{11,22}^{33,1}+\frac{1}{16}
\sqrt{\frac{5}{3}} C_{20,00}^{20,0}+\frac{1}{8} \sqrt{\frac{5}{3}} C_{20,00}^{20,1}+\frac{\sqrt{5} C_{20,20}^{20,0}}{48}+\frac{\sqrt{5} C_{20,20}^{20,1}}{24}-\frac{7
C_{20,22}^{22,0}}{480}-\frac{7 C_{20,22}^{22,1}}{240}+\frac{C_{22,00}^{22,0}}{8 \sqrt{3}}+\frac{C_{22,00}^{22,1}}{4 \sqrt{3}}+\frac{C_{22,20}^{22,0}}{24}+\frac{C_{22,20}^{22,1}}{12}-\frac{\sqrt{7}
C_{22,22}^{22,0}}{240}-\frac{\sqrt{7} C_{22,22}^{22,1}}{120}\)

\noindent\(C_{00,3112,2}^{00,1112,1,\text{Loc}}= \frac{1}{8} \sqrt{\frac{5}{3}} C_{00,20}^{40,1}+\frac{7 C_{00,22}^{42,1}}{160 \sqrt{3}}+\frac{1}{8}
\sqrt{\frac{5}{3}} C_{11,20}^{31,1}+\frac{7 C_{11,22}^{31,1}}{160 \sqrt{3}}+\frac{3}{160} \sqrt{\frac{7}{5}} C_{11,22}^{33,1}+\frac{1}{8} \sqrt{\frac{5}{3}}
C_{20,20}^{20,1}+\frac{7 C_{20,22}^{22,1}}{160 \sqrt{3}}+\frac{C_{22,20}^{22,1}}{4 \sqrt{3}}+\frac{1}{80} \sqrt{\frac{7}{3}} C_{22,22}^{22,1}\)

\noindent\(C_{00,3112,2}^{00,1112,1,\text{NonLoc}}= \frac{\sqrt{5} C_{00,00}^{40,0}}{16}+\frac{1}{16} \sqrt{\frac{5}{3}} C_{00,20}^{40,0}-\frac{7
C_{00,22}^{42,0}}{160 \sqrt{3}}-\frac{\sqrt{5} C_{11,00}^{31,0}}{16}-\frac{1}{16} \sqrt{\frac{5}{3}} C_{11,20}^{31,0}+\frac{7 C_{11,22}^{31,0}}{160
\sqrt{3}}+\frac{3}{160} \sqrt{\frac{7}{5}} C_{11,22}^{33,0}+\frac{\sqrt{5} C_{20,00}^{20,0}}{16}+\frac{1}{16} \sqrt{\frac{5}{3}} C_{20,20}^{20,0}-\frac{7
C_{20,22}^{22,0}}{160 \sqrt{3}}+\frac{C_{22,00}^{22,0}}{8}+\frac{C_{22,20}^{22,0}}{8 \sqrt{3}}-\frac{1}{80} \sqrt{\frac{7}{3}} C_{22,22}^{22,0}\)

\noindent\(C_{00,3312,2}^{00,1112,0,\text{Loc}}= \frac{1}{8} \sqrt{\frac{7}{15}} C_{00,22}^{42,0}+\frac{1}{16} \sqrt{\frac{7}{15}}
C_{00,22}^{42,1}+\frac{1}{8} \sqrt{\frac{7}{15}} C_{11,22}^{31,0}+\frac{1}{16} \sqrt{\frac{7}{15}} C_{11,22}^{31,1}+\frac{3 C_{11,22}^{33,0}}{40}+\frac{3
C_{11,22}^{33,1}}{80}+\frac{1}{8} \sqrt{\frac{7}{15}} C_{20,22}^{22,0}+\frac{1}{16} \sqrt{\frac{7}{15}} C_{20,22}^{22,1}+\frac{C_{22,22}^{22,0}}{4
\sqrt{15}}+\frac{C_{22,22}^{22,1}}{8 \sqrt{15}}\)

\noindent\(C_{00,3312,2}^{00,1112,0,\text{NonLoc}}= -\frac{1}{16} \sqrt{\frac{7}{15}} C_{00,22}^{42,0}-\frac{1}{8} \sqrt{\frac{7}{15}}
C_{00,22}^{42,1}+\frac{1}{16} \sqrt{\frac{7}{15}} C_{11,22}^{31,0}+\frac{1}{8} \sqrt{\frac{7}{15}} C_{11,22}^{31,1}+\frac{3 C_{11,22}^{33,0}}{80}+\frac{3
C_{11,22}^{33,1}}{40}-\frac{1}{16} \sqrt{\frac{7}{15}} C_{20,22}^{22,0}-\frac{1}{8} \sqrt{\frac{7}{15}} C_{20,22}^{22,1}-\frac{C_{22,22}^{22,0}}{8
\sqrt{15}}-\frac{C_{22,22}^{22,1}}{4 \sqrt{15}}\)

\noindent\(C_{00,3312,2}^{00,1112,1,\text{Loc}}= \frac{1}{16} \sqrt{\frac{7}{5}} C_{00,22}^{42,1}+\frac{1}{16} \sqrt{\frac{7}{5}} C_{11,22}^{31,1}+\frac{3
\sqrt{3} C_{11,22}^{33,1}}{80}+\frac{1}{16} \sqrt{\frac{7}{5}} C_{20,22}^{22,1}+\frac{C_{22,22}^{22,1}}{8 \sqrt{5}}\)

\noindent\(C_{00,3312,2}^{00,1112,1,\text{NonLoc}}= -\frac{1}{16} \sqrt{\frac{7}{5}} C_{00,22}^{42,0}+\frac{1}{16} \sqrt{\frac{7}{5}}
C_{11,22}^{31,0}+\frac{3 \sqrt{3} C_{11,22}^{33,0}}{80}-\frac{1}{16} \sqrt{\frac{7}{5}} C_{20,22}^{22,0}-\frac{C_{22,22}^{22,0}}{8 \sqrt{5}}\)

\noindent\(C_{00,4000,0}^{00,0000,0,\text{Loc}}= \frac{C_{00,00}^{40,0}}{16}+\frac{C_{00,00}^{40,1}}{32}+\frac{C_{11,00}^{31,0}}{16}+\frac{C_{11,00}^{31,1}}{32}+\frac{C_{20,00}^{20,0}}{16}+\frac{C_{20,00}^{20,1}}{32}+\frac{C_{22,00}^{22,0}}{8
\sqrt{5}}+\frac{C_{22,00}^{22,1}}{16 \sqrt{5}}\)

\noindent\(C_{00,4000,0}^{00,0000,0,\text{NonLoc}}= -\frac{C_{00,00}^{40,0}}{64}-\frac{C_{00,00}^{40,1}}{32}+\frac{\sqrt{3} C_{00,20}^{40,0}}{64}+\frac{\sqrt{3}
C_{00,20}^{40,1}}{32}+\frac{C_{11,00}^{31,0}}{64}+\frac{C_{11,00}^{31,1}}{32}-\frac{\sqrt{3} C_{11,20}^{31,0}}{64}-\frac{\sqrt{3} C_{11,20}^{31,1}}{32}-\frac{C_{20,00}^{20,0}}{64}-\frac{C_{20,00}^{20,1}}{32}+\frac{\sqrt{3}
C_{20,20}^{20,0}}{64}+\frac{\sqrt{3} C_{20,20}^{20,1}}{32}-\frac{C_{22,00}^{22,0}}{32 \sqrt{5}}-\frac{C_{22,00}^{22,1}}{16 \sqrt{5}}+\frac{1}{32}
\sqrt{\frac{3}{5}} C_{22,20}^{22,0}+\frac{1}{16} \sqrt{\frac{3}{5}} C_{22,20}^{22,1}\)

\noindent\(C_{00,4000,0}^{00,0000,1,\text{Loc}}= \frac{\sqrt{3} C_{00,00}^{40,1}}{32}+\frac{\sqrt{3} C_{11,00}^{31,1}}{32}+\frac{\sqrt{3}
C_{20,00}^{20,1}}{32}+\frac{1}{16} \sqrt{\frac{3}{5}} C_{22,00}^{22,1}\)

\noindent\(C_{00,4000,0}^{00,0000,1,\text{NonLoc}}= -\frac{\sqrt{3} C_{00,00}^{40,0}}{64}+\frac{3 C_{00,20}^{40,0}}{64}+\frac{\sqrt{3}
C_{11,00}^{31,0}}{64}-\frac{3 C_{11,20}^{31,0}}{64}-\frac{\sqrt{3} C_{20,00}^{20,0}}{64}+\frac{3 C_{20,20}^{20,0}}{64}-\frac{1}{32} \sqrt{\frac{3}{5}}
C_{22,00}^{22,0}+\frac{3 C_{22,20}^{22,0}}{32 \sqrt{5}}\)

\noindent\(C_{00,4011,1}^{00,0011,0,\text{Loc}}= -\frac{C_{00,20}^{40,0}}{16}-\frac{C_{00,20}^{40,1}}{32}-\frac{C_{11,20}^{31,0}}{16}-\frac{C_{11,20}^{31,1}}{32}-\frac{C_{20,20}^{20,0}}{16}-\frac{C_{20,20}^{20,1}}{32}-\frac{C_{22,20}^{22,0}}{8
\sqrt{5}}-\frac{C_{22,20}^{22,1}}{16 \sqrt{5}}\)

\noindent\(C_{00,4011,1}^{00,0011,0,\text{NonLoc}}= -\frac{\sqrt{3} C_{00,00}^{40,0}}{64}-\frac{\sqrt{3} C_{00,00}^{40,1}}{32}-\frac{C_{00,20}^{40,0}}{64}-\frac{C_{00,20}^{40,1}}{32}+\frac{\sqrt{3}
C_{11,00}^{31,0}}{64}+\frac{\sqrt{3} C_{11,00}^{31,1}}{32}+\frac{C_{11,20}^{31,0}}{64}+\frac{C_{11,20}^{31,1}}{32}-\frac{\sqrt{3} C_{20,00}^{20,0}}{64}-\frac{\sqrt{3}
C_{20,00}^{20,1}}{32}-\frac{C_{20,20}^{20,0}}{64}-\frac{C_{20,20}^{20,1}}{32}-\frac{1}{32} \sqrt{\frac{3}{5}} C_{22,00}^{22,0}-\frac{1}{16} \sqrt{\frac{3}{5}}
C_{22,00}^{22,1}-\frac{C_{22,20}^{22,0}}{32 \sqrt{5}}-\frac{C_{22,20}^{22,1}}{16 \sqrt{5}}\)

\noindent\(C_{00,4011,1}^{00,0011,1,\text{Loc}}= -\frac{\sqrt{3} C_{00,20}^{40,1}}{32}-\frac{\sqrt{3} C_{11,20}^{31,1}}{32}-\frac{\sqrt{3}
C_{20,20}^{20,1}}{32}-\frac{1}{16} \sqrt{\frac{3}{5}} C_{22,20}^{22,1}\)

\noindent\(C_{00,4011,1}^{00,0011,1,\text{NonLoc}}= -\frac{3 C_{00,00}^{40,0}}{64}-\frac{\sqrt{3} C_{00,20}^{40,0}}{64}+\frac{3 C_{11,00}^{31,0}}{64}+\frac{\sqrt{3}
C_{11,20}^{31,0}}{64}-\frac{3 C_{20,00}^{20,0}}{64}-\frac{\sqrt{3} C_{20,20}^{20,0}}{64}-\frac{3 C_{22,00}^{22,0}}{32 \sqrt{5}}-\frac{1}{32} \sqrt{\frac{3}{5}}
C_{22,20}^{22,0}\)

\noindent\(C_{00,4211,1}^{00,0011,0,\text{Loc}}= -\frac{C_{00,22}^{42,0}}{16}-\frac{C_{00,22}^{42,1}}{32}-\frac{C_{11,22}^{31,0}}{16}-\frac{C_{11,22}^{31,1}}{32}-\frac{3}{16}
\sqrt{\frac{3}{35}} C_{11,22}^{33,0}-\frac{3}{32} \sqrt{\frac{3}{35}} C_{11,22}^{33,1}-\frac{C_{20,22}^{22,0}}{16}-\frac{C_{20,22}^{22,1}}{32}-\frac{C_{22,22}^{22,0}}{8
\sqrt{7}}-\frac{C_{22,22}^{22,1}}{16 \sqrt{7}}\)

\noindent\(C_{00,4211,1}^{00,0011,0,\text{NonLoc}}= \frac{C_{00,22}^{42,0}}{32}+\frac{C_{00,22}^{42,1}}{16}-\frac{C_{11,22}^{31,0}}{32}-\frac{C_{11,22}^{31,1}}{16}-\frac{3}{32}
\sqrt{\frac{3}{35}} C_{11,22}^{33,0}-\frac{3}{16} \sqrt{\frac{3}{35}} C_{11,22}^{33,1}+\frac{C_{20,22}^{22,0}}{32}+\frac{C_{20,22}^{22,1}}{16}+\frac{C_{22,22}^{22,0}}{16
\sqrt{7}}+\frac{C_{22,22}^{22,1}}{8 \sqrt{7}}\)

\noindent\(C_{00,4211,1}^{00,0011,1,\text{Loc}}= -\frac{\sqrt{3} C_{00,22}^{42,1}}{32}-\frac{\sqrt{3} C_{11,22}^{31,1}}{32}-\frac{9
C_{11,22}^{33,1}}{32 \sqrt{35}}-\frac{\sqrt{3} C_{20,22}^{22,1}}{32}-\frac{1}{16} \sqrt{\frac{3}{7}} C_{22,22}^{22,1}\)

\noindent\(C_{00,4211,1}^{00,0011,1,\text{NonLoc}}= \frac{\sqrt{3} C_{00,22}^{42,0}}{32}-\frac{\sqrt{3} C_{11,22}^{31,0}}{32}-\frac{9
C_{11,22}^{33,0}}{32 \sqrt{35}}+\frac{\sqrt{3} C_{20,22}^{22,0}}{32}+\frac{1}{16} \sqrt{\frac{3}{7}} C_{22,22}^{22,0}\)

\noindent\(C_{11,0000,1}^{00,3111,0,\text{Loc}}= \frac{C_{11,11}^{31,0}}{16}+\frac{C_{11,11}^{31,1}}{32}+\frac{1}{8} \sqrt{\frac{3}{5}}
C_{22,11}^{22,0}+\frac{1}{16} \sqrt{\frac{3}{5}} C_{22,11}^{22,1}\)

\noindent\(C_{11,0000,1}^{00,3111,0,\text{NonLoc}}= \frac{C_{11,11}^{31,0}}{32}+\frac{C_{11,11}^{31,1}}{16}-\frac{1}{16} \sqrt{\frac{3}{5}}
C_{22,11}^{22,0}-\frac{1}{8} \sqrt{\frac{3}{5}} C_{22,11}^{22,1}\)

\noindent\(C_{11,0000,1}^{00,3111,1,\text{Loc}}= \frac{\sqrt{3} C_{11,11}^{31,1}}{32}+\frac{3 C_{22,11}^{22,1}}{16 \sqrt{5}}\)

\noindent\(C_{11,0000,1}^{00,3111,1,\text{NonLoc}}= \frac{\sqrt{3} C_{11,11}^{31,0}}{32}-\frac{3 C_{22,11}^{22,0}}{16 \sqrt{5}}\)

\noindent\(C_{11,0011,1}^{00,3101,0,\text{Loc}}= -\frac{C_{11,11}^{31,0}}{16}-\frac{C_{11,11}^{31,1}}{32}-\frac{1}{8} \sqrt{\frac{3}{5}}
C_{22,11}^{22,0}-\frac{1}{16} \sqrt{\frac{3}{5}} C_{22,11}^{22,1}\)

\noindent\(C_{11,0011,1}^{00,3101,0,\text{NonLoc}}= -\frac{C_{11,11}^{31,0}}{32}-\frac{C_{11,11}^{31,1}}{16}+\frac{1}{16} \sqrt{\frac{3}{5}}
C_{22,11}^{22,0}+\frac{1}{8} \sqrt{\frac{3}{5}} C_{22,11}^{22,1}\)

\noindent\(C_{11,0011,1}^{00,3101,1,\text{Loc}}= -\frac{\sqrt{3} C_{11,11}^{31,1}}{32}-\frac{3 C_{22,11}^{22,1}}{16 \sqrt{5}}\)

\noindent\(C_{11,0011,1}^{00,3101,1,\text{NonLoc}}= -\frac{\sqrt{3} C_{11,11}^{31,0}}{32}+\frac{3 C_{22,11}^{22,0}}{16 \sqrt{5}}\)

\noindent\(C_{11,1101,0}^{11,1101,0,\text{Loc}}= -\frac{5 C_{00,00}^{40,0}}{48}-\frac{5 C_{00,00}^{40,1}}{96}+\frac{C_{20,00}^{20,0}}{16}+\frac{C_{20,00}^{20,1}}{32}\)

\noindent\(C_{11,1101,0}^{11,1101,0,\text{NonLoc}}= \frac{5 C_{00,00}^{40,0}}{192}+\frac{5 C_{00,00}^{40,1}}{96}-\frac{5 C_{00,20}^{40,0}}{64
\sqrt{3}}-\frac{5 C_{00,20}^{40,1}}{32 \sqrt{3}}-\frac{C_{20,00}^{20,0}}{64}-\frac{C_{20,00}^{20,1}}{32}+\frac{\sqrt{3} C_{20,20}^{20,0}}{64}+\frac{\sqrt{3}
C_{20,20}^{20,1}}{32}\)

\noindent\(C_{11,1101,0}^{11,1101,1,\text{Loc}}= -\frac{5 C_{00,00}^{40,1}}{32 \sqrt{3}}+\frac{\sqrt{3} C_{20,00}^{20,1}}{32}\)

\noindent\(C_{11,1101,0}^{11,1101,1,\text{NonLoc}}= \frac{5 C_{00,00}^{40,0}}{64 \sqrt{3}}-\frac{5 C_{00,20}^{40,0}}{64}-\frac{\sqrt{3}
C_{20,00}^{20,0}}{64}+\frac{3 C_{20,20}^{20,0}}{64}\)

\noindent\(C_{11,1101,1}^{00,2011,0,\text{Loc}}= -\frac{5 C_{11,11}^{31,0}}{48}-\frac{5 C_{11,11}^{31,1}}{96}-\frac{1}{8} \sqrt{\frac{5}{3}}
C_{22,11}^{22,0}-\frac{1}{16} \sqrt{\frac{5}{3}} C_{22,11}^{22,1}\)

\noindent\(C_{11,1101,1}^{00,2011,0,\text{NonLoc}}= -\frac{5 C_{11,11}^{31,0}}{96}-\frac{5 C_{11,11}^{31,1}}{48}+\frac{1}{16} \sqrt{\frac{5}{3}}
C_{22,11}^{22,0}+\frac{1}{8} \sqrt{\frac{5}{3}} C_{22,11}^{22,1}\)

\noindent\(C_{11,1101,1}^{00,2011,1,\text{Loc}}= -\frac{5 C_{11,11}^{31,1}}{32 \sqrt{3}}-\frac{\sqrt{5} C_{22,11}^{22,1}}{16}\)

\noindent\(C_{11,1101,1}^{00,2011,1,\text{NonLoc}}= -\frac{5 C_{11,11}^{31,0}}{32 \sqrt{3}}+\frac{\sqrt{5} C_{22,11}^{22,0}}{16}\)

\noindent\(C_{11,1101,1}^{00,2211,0,\text{Loc}}= \frac{\sqrt{5} C_{11,11}^{31,0}}{48}+\frac{\sqrt{5} C_{11,11}^{31,1}}{96}+\frac{C_{22,11}^{22,0}}{8
\sqrt{3}}+\frac{C_{22,11}^{22,1}}{16 \sqrt{3}}\)

\noindent\(C_{11,1101,1}^{00,2211,0,\text{NonLoc}}= \frac{\sqrt{5} C_{11,11}^{31,0}}{96}+\frac{\sqrt{5} C_{11,11}^{31,1}}{48}-\frac{C_{22,11}^{22,0}}{16
\sqrt{3}}-\frac{C_{22,11}^{22,1}}{8 \sqrt{3}}\)

\noindent\(C_{11,1101,1}^{00,2211,1,\text{Loc}}= \frac{1}{32} \sqrt{\frac{5}{3}} C_{11,11}^{31,1}+\frac{C_{22,11}^{22,1}}{16}\)

\noindent\(C_{11,1101,1}^{00,2211,1,\text{NonLoc}}= \frac{1}{32} \sqrt{\frac{5}{3}} C_{11,11}^{31,0}-\frac{C_{22,11}^{22,0}}{16}\)

\noindent\(C_{11,1101,1}^{11,1101,0,\text{Loc}}= -\frac{C_{20,00}^{20,0}}{8 \sqrt{3}}-\frac{C_{20,00}^{20,1}}{16 \sqrt{3}}+\frac{1}{16}
\sqrt{\frac{5}{3}} C_{22,00}^{22,0}+\frac{1}{32} \sqrt{\frac{5}{3}} C_{22,00}^{22,1}\)

\noindent\(C_{11,1101,1}^{11,1101,0,\text{NonLoc}}= \frac{C_{20,00}^{20,0}}{32 \sqrt{3}}+\frac{C_{20,00}^{20,1}}{16 \sqrt{3}}-\frac{C_{20,20}^{20,0}}{32}-\frac{C_{20,20}^{20,1}}{16}-\frac{1}{64}
\sqrt{\frac{5}{3}} C_{22,00}^{22,0}-\frac{1}{32} \sqrt{\frac{5}{3}} C_{22,00}^{22,1}+\frac{\sqrt{5} C_{22,20}^{22,0}}{64}+\frac{\sqrt{5} C_{22,20}^{22,1}}{32}\)

\noindent\(C_{11,1101,1}^{11,1101,1,\text{Loc}}= -\frac{C_{20,00}^{20,1}}{16}+\frac{\sqrt{5} C_{22,00}^{22,1}}{32}\)

\noindent\(C_{11,1101,1}^{11,1101,1,\text{NonLoc}}= \frac{C_{20,00}^{20,0}}{32}-\frac{\sqrt{3} C_{20,20}^{20,0}}{32}-\frac{\sqrt{5}
C_{22,00}^{22,0}}{64}+\frac{\sqrt{15} C_{22,20}^{22,0}}{64}\)

\noindent\(C_{11,1101,2}^{00,2212,0,\text{Loc}}= -\frac{1}{16} \sqrt{\frac{5}{3}} C_{11,11}^{31,0}-\frac{1}{32} \sqrt{\frac{5}{3}}
C_{11,11}^{31,1}-\frac{C_{22,11}^{22,0}}{8}-\frac{C_{22,11}^{22,1}}{16}\)

\noindent\(C_{11,1101,2}^{00,2212,0,\text{NonLoc}}= -\frac{1}{32} \sqrt{\frac{5}{3}} C_{11,11}^{31,0}-\frac{1}{16} \sqrt{\frac{5}{3}}
C_{11,11}^{31,1}+\frac{C_{22,11}^{22,0}}{16}+\frac{C_{22,11}^{22,1}}{8}\)

\noindent\(C_{11,1101,2}^{00,2212,1,\text{Loc}}= -\frac{\sqrt{5} C_{11,11}^{31,1}}{32}-\frac{\sqrt{3} C_{22,11}^{22,1}}{16}\)

\noindent\(C_{11,1101,2}^{00,2212,1,\text{NonLoc}}= -\frac{\sqrt{5} C_{11,11}^{31,0}}{32}+\frac{\sqrt{3} C_{22,11}^{22,0}}{16}\)

\noindent\(C_{11,1101,2}^{11,1101,0,\text{Loc}}= -\frac{\sqrt{5} C_{00,00}^{40,0}}{24}-\frac{\sqrt{5} C_{00,00}^{40,1}}{48}+\frac{C_{22,00}^{22,0}}{16}+\frac{C_{22,00}^{22,1}}{32}\)

\noindent\(C_{11,1101,2}^{11,1101,0,\text{NonLoc}}= \frac{\sqrt{5} C_{00,00}^{40,0}}{96}+\frac{\sqrt{5} C_{00,00}^{40,1}}{48}-\frac{1}{32}
\sqrt{\frac{5}{3}} C_{00,20}^{40,0}-\frac{1}{16} \sqrt{\frac{5}{3}} C_{00,20}^{40,1}-\frac{C_{22,00}^{22,0}}{64}-\frac{C_{22,00}^{22,1}}{32}+\frac{\sqrt{3}
C_{22,20}^{22,0}}{64}+\frac{\sqrt{3} C_{22,20}^{22,1}}{32}\)

\noindent\(C_{11,1101,2}^{11,1101,1,\text{Loc}}= -\frac{1}{16} \sqrt{\frac{5}{3}} C_{00,00}^{40,1}+\frac{\sqrt{3} C_{22,00}^{22,1}}{32}\)

\noindent\(C_{11,1101,2}^{11,1101,1,\text{NonLoc}}= \frac{1}{32} \sqrt{\frac{5}{3}} C_{00,00}^{40,0}-\frac{\sqrt{5} C_{00,20}^{40,0}}{32}-\frac{\sqrt{3}
C_{22,00}^{22,0}}{64}+\frac{3 C_{22,20}^{22,0}}{64}\)

\noindent\(C_{11,1110,1}^{11,1110,0,\text{Loc}}= \frac{5 C_{00,20}^{40,0}}{144}+\frac{5 C_{00,20}^{40,1}}{288}+\frac{7 \sqrt{5} C_{00,22}^{42,0}}{288}+\frac{7
\sqrt{5} C_{00,22}^{42,1}}{576}+\frac{\sqrt{5} C_{11,22}^{31,0}}{96}+\frac{\sqrt{5} C_{11,22}^{31,1}}{192}-\frac{1}{32} \sqrt{\frac{7}{3}} C_{11,22}^{33,0}-\frac{1}{64}
\sqrt{\frac{7}{3}} C_{11,22}^{33,1}+\frac{C_{20,20}^{20,0}}{144}+\frac{C_{20,20}^{20,1}}{288}-\frac{\sqrt{5} C_{20,22}^{22,0}}{288}-\frac{\sqrt{5}
C_{20,22}^{22,1}}{576}-\frac{\sqrt{5} C_{22,20}^{22,0}}{72}-\frac{\sqrt{5} C_{22,20}^{22,1}}{144}-\frac{\sqrt{35} C_{22,22}^{22,0}}{144}-\frac{\sqrt{35}
C_{22,22}^{22,1}}{288}\)

\noindent\(C_{11,1110,1}^{11,1110,0,\text{NonLoc}}= \frac{5 C_{00,00}^{40,0}}{192 \sqrt{3}}+\frac{5 C_{00,00}^{40,1}}{96 \sqrt{3}}+\frac{5
C_{00,20}^{40,0}}{576}+\frac{5 C_{00,20}^{40,1}}{288}-\frac{7 \sqrt{5} C_{00,22}^{42,0}}{576}-\frac{7 \sqrt{5} C_{00,22}^{42,1}}{288}+\frac{\sqrt{5}
C_{11,22}^{31,0}}{192}+\frac{\sqrt{5} C_{11,22}^{31,1}}{96}-\frac{1}{64} \sqrt{\frac{7}{3}} C_{11,22}^{33,0}-\frac{1}{32} \sqrt{\frac{7}{3}} C_{11,22}^{33,1}+\frac{C_{20,00}^{20,0}}{192
\sqrt{3}}+\frac{C_{20,00}^{20,1}}{96 \sqrt{3}}+\frac{C_{20,20}^{20,0}}{576}+\frac{C_{20,20}^{20,1}}{288}+\frac{\sqrt{5} C_{20,22}^{22,0}}{576}+\frac{\sqrt{5}
C_{20,22}^{22,1}}{288}-\frac{1}{96} \sqrt{\frac{5}{3}} C_{22,00}^{22,0}-\frac{1}{48} \sqrt{\frac{5}{3}} C_{22,00}^{22,1}-\frac{\sqrt{5} C_{22,20}^{22,0}}{288}-\frac{\sqrt{5}
C_{22,20}^{22,1}}{144}+\frac{\sqrt{35} C_{22,22}^{22,0}}{288}+\frac{\sqrt{35} C_{22,22}^{22,1}}{144}\)

\noindent\(C_{11,1110,1}^{11,1110,1,\text{Loc}}= \frac{5 C_{00,20}^{40,1}}{96 \sqrt{3}}+\frac{7}{192} \sqrt{\frac{5}{3}} C_{00,22}^{42,1}+\frac{1}{64}
\sqrt{\frac{5}{3}} C_{11,22}^{31,1}-\frac{\sqrt{7} C_{11,22}^{33,1}}{64}+\frac{C_{20,20}^{20,1}}{96 \sqrt{3}}-\frac{1}{192} \sqrt{\frac{5}{3}} C_{20,22}^{22,1}-\frac{1}{48}
\sqrt{\frac{5}{3}} C_{22,20}^{22,1}-\frac{1}{96} \sqrt{\frac{35}{3}} C_{22,22}^{22,1}\)

\noindent\(C_{11,1110,1}^{11,1110,1,\text{NonLoc}}= \frac{5 C_{00,00}^{40,0}}{192}+\frac{5 C_{00,20}^{40,0}}{192 \sqrt{3}}-\frac{7}{192}
\sqrt{\frac{5}{3}} C_{00,22}^{42,0}+\frac{1}{64} \sqrt{\frac{5}{3}} C_{11,22}^{31,0}-\frac{\sqrt{7} C_{11,22}^{33,0}}{64}+\frac{C_{20,00}^{20,0}}{192}+\frac{C_{20,20}^{20,0}}{192
\sqrt{3}}+\frac{1}{192} \sqrt{\frac{5}{3}} C_{20,22}^{22,0}-\frac{\sqrt{5} C_{22,00}^{22,0}}{96}-\frac{1}{96} \sqrt{\frac{5}{3}} C_{22,20}^{22,0}+\frac{1}{96}
\sqrt{\frac{35}{3}} C_{22,22}^{22,0}\)

\noindent\(C_{11,1111,0}^{00,2000,0,\text{Loc}}= -\frac{5 C_{11,11}^{31,0}}{48}-\frac{5 C_{11,11}^{31,1}}{96}-\frac{1}{8} \sqrt{\frac{5}{3}}
C_{22,11}^{22,0}-\frac{1}{16} \sqrt{\frac{5}{3}} C_{22,11}^{22,1}\)

\noindent\(C_{11,1111,0}^{00,2000,0,\text{NonLoc}}= -\frac{5 C_{11,11}^{31,0}}{96}-\frac{5 C_{11,11}^{31,1}}{48}+\frac{1}{16} \sqrt{\frac{5}{3}}
C_{22,11}^{22,0}+\frac{1}{8} \sqrt{\frac{5}{3}} C_{22,11}^{22,1}\)

\noindent\(C_{11,1111,0}^{00,2000,1,\text{Loc}}= -\frac{5 C_{11,11}^{31,1}}{32 \sqrt{3}}-\frac{\sqrt{5} C_{22,11}^{22,1}}{16}\)

\noindent\(C_{11,1111,0}^{00,2000,1,\text{NonLoc}}= -\frac{5 C_{11,11}^{31,0}}{32 \sqrt{3}}+\frac{\sqrt{5} C_{22,11}^{22,0}}{16}\)

\noindent\(C_{11,1111,0}^{11,1111,0,\text{Loc}}= \frac{C_{20,20}^{20,0}}{24 \sqrt{3}}+\frac{C_{20,20}^{20,1}}{48 \sqrt{3}}-\frac{1}{48}
\sqrt{\frac{5}{3}} C_{20,22}^{22,0}-\frac{1}{96} \sqrt{\frac{5}{3}} C_{20,22}^{22,1}-\frac{1}{48} \sqrt{\frac{5}{3}} C_{22,20}^{22,0}-\frac{1}{96}
\sqrt{\frac{5}{3}} C_{22,20}^{22,1}+\frac{1}{48} \sqrt{\frac{35}{3}} C_{22,22}^{22,0}+\frac{1}{96} \sqrt{\frac{35}{3}} C_{22,22}^{22,1}\)

\noindent\(C_{11,1111,0}^{11,1111,0,\text{NonLoc}}= \frac{C_{20,00}^{20,0}}{96}+\frac{C_{20,00}^{20,1}}{48}+\frac{C_{20,20}^{20,0}}{96
\sqrt{3}}+\frac{C_{20,20}^{20,1}}{48 \sqrt{3}}+\frac{1}{96} \sqrt{\frac{5}{3}} C_{20,22}^{22,0}+\frac{1}{48} \sqrt{\frac{5}{3}} C_{20,22}^{22,1}-\frac{\sqrt{5}
C_{22,00}^{22,0}}{192}-\frac{\sqrt{5} C_{22,00}^{22,1}}{96}-\frac{1}{192} \sqrt{\frac{5}{3}} C_{22,20}^{22,0}-\frac{1}{96} \sqrt{\frac{5}{3}} C_{22,20}^{22,1}-\frac{1}{96}
\sqrt{\frac{35}{3}} C_{22,22}^{22,0}-\frac{1}{48} \sqrt{\frac{35}{3}} C_{22,22}^{22,1}\)

\noindent\(C_{11,1111,0}^{11,1111,1,\text{Loc}}= \frac{C_{20,20}^{20,1}}{48}-\frac{\sqrt{5} C_{20,22}^{22,1}}{96}-\frac{\sqrt{5} C_{22,20}^{22,1}}{96}+\frac{\sqrt{35}
C_{22,22}^{22,1}}{96}\)

\noindent\(C_{11,1111,0}^{11,1111,1,\text{NonLoc}}= \frac{C_{20,00}^{20,0}}{32 \sqrt{3}}+\frac{C_{20,20}^{20,0}}{96}+\frac{\sqrt{5}
C_{20,22}^{22,0}}{96}-\frac{1}{64} \sqrt{\frac{5}{3}} C_{22,00}^{22,0}-\frac{\sqrt{5} C_{22,20}^{22,0}}{192}-\frac{\sqrt{35} C_{22,22}^{22,0}}{96}\)

\noindent\(C_{11,1111,1}^{11,1111,0,\text{Loc}}= \frac{5 C_{00,20}^{40,0}}{96}+\frac{5 C_{00,20}^{40,1}}{192}-\frac{7 \sqrt{5} C_{00,22}^{42,0}}{384}-\frac{7
\sqrt{5} C_{00,22}^{42,1}}{768}-\frac{\sqrt{5} C_{11,22}^{31,0}}{128}-\frac{\sqrt{5} C_{11,22}^{31,1}}{256}+\frac{\sqrt{21} C_{11,22}^{33,0}}{128}+\frac{\sqrt{21}
C_{11,22}^{33,1}}{256}-\frac{C_{20,20}^{20,0}}{96}-\frac{C_{20,20}^{20,1}}{192}+\frac{5 \sqrt{5} C_{20,22}^{22,0}}{384}+\frac{5 \sqrt{5} C_{20,22}^{22,1}}{768}-\frac{\sqrt{5}
C_{22,20}^{22,0}}{96}-\frac{\sqrt{5} C_{22,20}^{22,1}}{192}-\frac{\sqrt{35} C_{22,22}^{22,0}}{192}-\frac{\sqrt{35} C_{22,22}^{22,1}}{384}\)

\noindent\(C_{11,1111,1}^{11,1111,0,\text{NonLoc}}= \frac{5 C_{00,00}^{40,0}}{128 \sqrt{3}}+\frac{5 C_{00,00}^{40,1}}{64 \sqrt{3}}+\frac{5
C_{00,20}^{40,0}}{384}+\frac{5 C_{00,20}^{40,1}}{192}+\frac{7 \sqrt{5} C_{00,22}^{42,0}}{768}+\frac{7 \sqrt{5} C_{00,22}^{42,1}}{384}-\frac{\sqrt{5}
C_{11,22}^{31,0}}{256}-\frac{\sqrt{5} C_{11,22}^{31,1}}{128}+\frac{\sqrt{21} C_{11,22}^{33,0}}{256}+\frac{\sqrt{21} C_{11,22}^{33,1}}{128}-\frac{C_{20,00}^{20,0}}{128
\sqrt{3}}-\frac{C_{20,00}^{20,1}}{64 \sqrt{3}}-\frac{C_{20,20}^{20,0}}{384}-\frac{C_{20,20}^{20,1}}{192}-\frac{5 \sqrt{5} C_{20,22}^{22,0}}{768}-\frac{5
\sqrt{5} C_{20,22}^{22,1}}{384}-\frac{1}{128} \sqrt{\frac{5}{3}} C_{22,00}^{22,0}-\frac{1}{64} \sqrt{\frac{5}{3}} C_{22,00}^{22,1}-\frac{\sqrt{5}
C_{22,20}^{22,0}}{384}-\frac{\sqrt{5} C_{22,20}^{22,1}}{192}+\frac{\sqrt{35} C_{22,22}^{22,0}}{384}+\frac{\sqrt{35} C_{22,22}^{22,1}}{192}\)

\noindent\(C_{11,1111,1}^{11,1111,1,\text{Loc}}= \frac{5 C_{00,20}^{40,1}}{64 \sqrt{3}}-\frac{7}{256} \sqrt{\frac{5}{3}} C_{00,22}^{42,1}-\frac{\sqrt{15}
C_{11,22}^{31,1}}{256}+\frac{3 \sqrt{7} C_{11,22}^{33,1}}{256}-\frac{C_{20,20}^{20,1}}{64 \sqrt{3}}+\frac{5}{256} \sqrt{\frac{5}{3}} C_{20,22}^{22,1}-\frac{1}{64}
\sqrt{\frac{5}{3}} C_{22,20}^{22,1}-\frac{1}{128} \sqrt{\frac{35}{3}} C_{22,22}^{22,1}\)

\noindent\(C_{11,1111,1}^{11,1111,1,\text{NonLoc}}= \frac{5 C_{00,00}^{40,0}}{128}+\frac{5 C_{00,20}^{40,0}}{128 \sqrt{3}}+\frac{7}{256}
\sqrt{\frac{5}{3}} C_{00,22}^{42,0}-\frac{\sqrt{15} C_{11,22}^{31,0}}{256}+\frac{3 \sqrt{7} C_{11,22}^{33,0}}{256}-\frac{C_{20,00}^{20,0}}{128}-\frac{C_{20,20}^{20,0}}{128
\sqrt{3}}-\frac{5}{256} \sqrt{\frac{5}{3}} C_{20,22}^{22,0}-\frac{\sqrt{5} C_{22,00}^{22,0}}{128}-\frac{1}{128} \sqrt{\frac{5}{3}} C_{22,20}^{22,0}+\frac{1}{128}
\sqrt{\frac{35}{3}} C_{22,22}^{22,0}\)

\noindent\(C_{11,1111,2}^{00,2202,0,\text{Loc}}= \frac{\sqrt{5} C_{11,11}^{31,0}}{48}+\frac{\sqrt{5} C_{11,11}^{31,1}}{96}+\frac{C_{22,11}^{22,0}}{8
\sqrt{3}}+\frac{C_{22,11}^{22,1}}{16 \sqrt{3}}\)

\noindent\(C_{11,1111,2}^{00,2202,0,\text{NonLoc}}= \frac{\sqrt{5} C_{11,11}^{31,0}}{96}+\frac{\sqrt{5} C_{11,11}^{31,1}}{48}-\frac{C_{22,11}^{22,0}}{16
\sqrt{3}}-\frac{C_{22,11}^{22,1}}{8 \sqrt{3}}\)

\noindent\(C_{11,1111,2}^{00,2202,1,\text{Loc}}= \frac{1}{32} \sqrt{\frac{5}{3}} C_{11,11}^{31,1}+\frac{C_{22,11}^{22,1}}{16}\)

\noindent\(C_{11,1111,2}^{00,2202,1,\text{NonLoc}}= \frac{1}{32} \sqrt{\frac{5}{3}} C_{11,11}^{31,0}-\frac{C_{22,11}^{22,0}}{16}\)

\noindent\(C_{11,1111,2}^{11,1111,0,\text{Loc}}= \frac{1}{32} \sqrt{\frac{5}{3}} C_{00,20}^{40,0}+\frac{1}{64} \sqrt{\frac{5}{3}} C_{00,20}^{40,1}-\frac{7
C_{00,22}^{42,0}}{128 \sqrt{3}}-\frac{7 C_{00,22}^{42,1}}{256 \sqrt{3}}-\frac{\sqrt{3} C_{11,22}^{31,0}}{128}-\frac{\sqrt{3} C_{11,22}^{31,1}}{256}+\frac{3}{128}
\sqrt{\frac{7}{5}} C_{11,22}^{33,0}+\frac{3}{256} \sqrt{\frac{7}{5}} C_{11,22}^{33,1}+\frac{1}{96} \sqrt{\frac{5}{3}} C_{20,20}^{20,0}+\frac{1}{192}
\sqrt{\frac{5}{3}} C_{20,20}^{20,1}-\frac{C_{20,22}^{22,0}}{384 \sqrt{3}}-\frac{C_{20,22}^{22,1}}{768 \sqrt{3}}-\frac{7 C_{22,20}^{22,0}}{96 \sqrt{3}}-\frac{7
C_{22,20}^{22,1}}{192 \sqrt{3}}+\frac{5}{192} \sqrt{\frac{7}{3}} C_{22,22}^{22,0}+\frac{5}{384} \sqrt{\frac{7}{3}} C_{22,22}^{22,1}\)

\noindent\(C_{11,1111,2}^{11,1111,0,\text{NonLoc}}= \frac{\sqrt{5} C_{00,00}^{40,0}}{128}+\frac{\sqrt{5} C_{00,00}^{40,1}}{64}+\frac{1}{128}
\sqrt{\frac{5}{3}} C_{00,20}^{40,0}+\frac{1}{64} \sqrt{\frac{5}{3}} C_{00,20}^{40,1}+\frac{7 C_{00,22}^{42,0}}{256 \sqrt{3}}+\frac{7 C_{00,22}^{42,1}}{128
\sqrt{3}}-\frac{\sqrt{3} C_{11,22}^{31,0}}{256}-\frac{\sqrt{3} C_{11,22}^{31,1}}{128}+\frac{3}{256} \sqrt{\frac{7}{5}} C_{11,22}^{33,0}+\frac{3}{128}
\sqrt{\frac{7}{5}} C_{11,22}^{33,1}+\frac{\sqrt{5} C_{20,00}^{20,0}}{384}+\frac{\sqrt{5} C_{20,00}^{20,1}}{192}+\frac{1}{384} \sqrt{\frac{5}{3}}
C_{20,20}^{20,0}+\frac{1}{192} \sqrt{\frac{5}{3}} C_{20,20}^{20,1}+\frac{C_{20,22}^{22,0}}{768 \sqrt{3}}+\frac{C_{20,22}^{22,1}}{384 \sqrt{3}}-\frac{7
C_{22,00}^{22,0}}{384}-\frac{7 C_{22,00}^{22,1}}{192}-\frac{7 C_{22,20}^{22,0}}{384 \sqrt{3}}-\frac{7 C_{22,20}^{22,1}}{192 \sqrt{3}}-\frac{5}{384}
\sqrt{\frac{7}{3}} C_{22,22}^{22,0}-\frac{5}{192} \sqrt{\frac{7}{3}} C_{22,22}^{22,1}\)

\noindent\(C_{11,1111,2}^{11,1111,1,\text{Loc}}= \frac{\sqrt{5} C_{00,20}^{40,1}}{64}-\frac{7 C_{00,22}^{42,1}}{256}-\frac{3 C_{11,22}^{31,1}}{256}+\frac{3}{256}
\sqrt{\frac{21}{5}} C_{11,22}^{33,1}+\frac{\sqrt{5} C_{20,20}^{20,1}}{192}-\frac{C_{20,22}^{22,1}}{768}-\frac{7 C_{22,20}^{22,1}}{192}+\frac{5 \sqrt{7}
C_{22,22}^{22,1}}{384}\)

\noindent\(C_{11,1111,2}^{11,1111,1,\text{NonLoc}}= \frac{\sqrt{15} C_{00,00}^{40,0}}{128}+\frac{\sqrt{5} C_{00,20}^{40,0}}{128}+\frac{7
C_{00,22}^{42,0}}{256}-\frac{3 C_{11,22}^{31,0}}{256}+\frac{3}{256} \sqrt{\frac{21}{5}} C_{11,22}^{33,0}+\frac{1}{128} \sqrt{\frac{5}{3}} C_{20,00}^{20,0}+\frac{\sqrt{5}
C_{20,20}^{20,0}}{384}+\frac{C_{20,22}^{22,0}}{768}-\frac{7 C_{22,00}^{22,0}}{128 \sqrt{3}}-\frac{7 C_{22,20}^{22,0}}{384}-\frac{5 \sqrt{7} C_{22,22}^{22,0}}{384}\)

\noindent\(C_{11,1112,1}^{11,1110,0,\text{Loc}}= \frac{\sqrt{5} C_{00,20}^{40,0}}{36}+\frac{\sqrt{5} C_{00,20}^{40,1}}{72}+\frac{7
C_{00,22}^{42,0}}{72}+\frac{7 C_{00,22}^{42,1}}{144}-\frac{C_{11,22}^{31,0}}{48}-\frac{C_{11,22}^{31,1}}{96}+\frac{1}{16} \sqrt{\frac{7}{15}} C_{11,22}^{33,0}+\frac{1}{32}
\sqrt{\frac{7}{15}} C_{11,22}^{33,1}-\frac{\sqrt{5} C_{20,20}^{20,0}}{36}-\frac{\sqrt{5} C_{20,20}^{20,1}}{72}-\frac{C_{20,22}^{22,0}}{18}-\frac{C_{20,22}^{22,1}}{36}+\frac{C_{22,20}^{22,0}}{36}+\frac{C_{22,20}^{22,1}}{72}+\frac{\sqrt{7}
C_{22,22}^{22,0}}{72}+\frac{\sqrt{7} C_{22,22}^{22,1}}{144}\)

\noindent\(C_{11,1112,1}^{11,1110,0,\text{NonLoc}}= \frac{1}{48} \sqrt{\frac{5}{3}} C_{00,00}^{40,0}+\frac{1}{24} \sqrt{\frac{5}{3}}
C_{00,00}^{40,1}+\frac{\sqrt{5} C_{00,20}^{40,0}}{144}+\frac{\sqrt{5} C_{00,20}^{40,1}}{72}-\frac{7 C_{00,22}^{42,0}}{144}-\frac{7 C_{00,22}^{42,1}}{72}-\frac{C_{11,22}^{31,0}}{96}-\frac{C_{11,22}^{31,1}}{48}+\frac{1}{32}
\sqrt{\frac{7}{15}} C_{11,22}^{33,0}+\frac{1}{16} \sqrt{\frac{7}{15}} C_{11,22}^{33,1}-\frac{1}{48} \sqrt{\frac{5}{3}} C_{20,00}^{20,0}-\frac{1}{24}
\sqrt{\frac{5}{3}} C_{20,00}^{20,1}-\frac{\sqrt{5} C_{20,20}^{20,0}}{144}-\frac{\sqrt{5} C_{20,20}^{20,1}}{72}+\frac{C_{20,22}^{22,0}}{36}+\frac{C_{20,22}^{22,1}}{18}+\frac{C_{22,00}^{22,0}}{48
\sqrt{3}}+\frac{C_{22,00}^{22,1}}{24 \sqrt{3}}+\frac{C_{22,20}^{22,0}}{144}+\frac{C_{22,20}^{22,1}}{72}-\frac{\sqrt{7} C_{22,22}^{22,0}}{144}-\frac{\sqrt{7}
C_{22,22}^{22,1}}{72}\)

\noindent\(C_{11,1112,1}^{11,1110,1,\text{Loc}}= \frac{1}{24} \sqrt{\frac{5}{3}} C_{00,20}^{40,1}+\frac{7 C_{00,22}^{42,1}}{48 \sqrt{3}}-\frac{C_{11,22}^{31,1}}{32
\sqrt{3}}+\frac{1}{32} \sqrt{\frac{7}{5}} C_{11,22}^{33,1}-\frac{1}{24} \sqrt{\frac{5}{3}} C_{20,20}^{20,1}-\frac{C_{20,22}^{22,1}}{12 \sqrt{3}}+\frac{C_{22,20}^{22,1}}{24
\sqrt{3}}+\frac{1}{48} \sqrt{\frac{7}{3}} C_{22,22}^{22,1}\)

\noindent\(C_{11,1112,1}^{11,1110,1,\text{NonLoc}}= \frac{\sqrt{5} C_{00,00}^{40,0}}{48}+\frac{1}{48} \sqrt{\frac{5}{3}} C_{00,20}^{40,0}-\frac{7
C_{00,22}^{42,0}}{48 \sqrt{3}}-\frac{C_{11,22}^{31,0}}{32 \sqrt{3}}+\frac{1}{32} \sqrt{\frac{7}{5}} C_{11,22}^{33,0}-\frac{\sqrt{5} C_{20,00}^{20,0}}{48}-\frac{1}{48}
\sqrt{\frac{5}{3}} C_{20,20}^{20,0}+\frac{C_{20,22}^{22,0}}{12 \sqrt{3}}+\frac{C_{22,00}^{22,0}}{48}+\frac{C_{22,20}^{22,0}}{48 \sqrt{3}}-\frac{1}{48}
\sqrt{\frac{7}{3}} C_{22,22}^{22,0}\)

\noindent\(C_{11,1112,1}^{11,1111,0,\text{Loc}}= -\frac{1}{16} \sqrt{\frac{5}{3}} C_{00,20}^{40,0}-\frac{1}{32} \sqrt{\frac{5}{3}}
C_{00,20}^{40,1}+\frac{7 C_{00,22}^{42,0}}{64 \sqrt{3}}+\frac{7 C_{00,22}^{42,1}}{128 \sqrt{3}}-\frac{\sqrt{3} C_{11,22}^{31,0}}{64}-\frac{\sqrt{3}
C_{11,22}^{31,1}}{128}+\frac{3}{64} \sqrt{\frac{7}{5}} C_{11,22}^{33,0}+\frac{3}{128} \sqrt{\frac{7}{5}} C_{11,22}^{33,1}+\frac{1}{16} \sqrt{\frac{5}{3}}
C_{20,20}^{20,0}+\frac{1}{32} \sqrt{\frac{5}{3}} C_{20,20}^{20,1}-\frac{C_{20,22}^{22,0}}{64 \sqrt{3}}-\frac{C_{20,22}^{22,1}}{128 \sqrt{3}}-\frac{C_{22,20}^{22,0}}{16
\sqrt{3}}-\frac{C_{22,20}^{22,1}}{32 \sqrt{3}}-\frac{1}{32} \sqrt{\frac{7}{3}} C_{22,22}^{22,0}-\frac{1}{64} \sqrt{\frac{7}{3}} C_{22,22}^{22,1}\)

\noindent\(C_{11,1112,1}^{11,1111,0,\text{NonLoc}}= -\frac{\sqrt{5} C_{00,00}^{40,0}}{64}-\frac{\sqrt{5} C_{00,00}^{40,1}}{32}-\frac{1}{64}
\sqrt{\frac{5}{3}} C_{00,20}^{40,0}-\frac{1}{32} \sqrt{\frac{5}{3}} C_{00,20}^{40,1}-\frac{7 C_{00,22}^{42,0}}{128 \sqrt{3}}-\frac{7 C_{00,22}^{42,1}}{64
\sqrt{3}}-\frac{\sqrt{3} C_{11,22}^{31,0}}{128}-\frac{\sqrt{3} C_{11,22}^{31,1}}{64}+\frac{3}{128} \sqrt{\frac{7}{5}} C_{11,22}^{33,0}+\frac{3}{64}
\sqrt{\frac{7}{5}} C_{11,22}^{33,1}+\frac{\sqrt{5} C_{20,00}^{20,0}}{64}+\frac{\sqrt{5} C_{20,00}^{20,1}}{32}+\frac{1}{64} \sqrt{\frac{5}{3}} C_{20,20}^{20,0}+\frac{1}{32}
\sqrt{\frac{5}{3}} C_{20,20}^{20,1}+\frac{C_{20,22}^{22,0}}{128 \sqrt{3}}+\frac{C_{20,22}^{22,1}}{64 \sqrt{3}}-\frac{C_{22,00}^{22,0}}{64}-\frac{C_{22,00}^{22,1}}{32}-\frac{C_{22,20}^{22,0}}{64
\sqrt{3}}-\frac{C_{22,20}^{22,1}}{32 \sqrt{3}}+\frac{1}{64} \sqrt{\frac{7}{3}} C_{22,22}^{22,0}+\frac{1}{32} \sqrt{\frac{7}{3}} C_{22,22}^{22,1}\)

\noindent\(C_{11,1112,1}^{11,1111,1,\text{Loc}}= -\frac{\sqrt{5} C_{00,20}^{40,1}}{32}+\frac{7 C_{00,22}^{42,1}}{128}-\frac{3 C_{11,22}^{31,1}}{128}+\frac{3}{128}
\sqrt{\frac{21}{5}} C_{11,22}^{33,1}+\frac{\sqrt{5} C_{20,20}^{20,1}}{32}-\frac{C_{20,22}^{22,1}}{128}-\frac{C_{22,20}^{22,1}}{32}-\frac{\sqrt{7}
C_{22,22}^{22,1}}{64}\)

\noindent\(C_{11,1112,1}^{11,1111,1,\text{NonLoc}}= -\frac{\sqrt{15} C_{00,00}^{40,0}}{64}-\frac{\sqrt{5} C_{00,20}^{40,0}}{64}-\frac{7
C_{00,22}^{42,0}}{128}-\frac{3 C_{11,22}^{31,0}}{128}+\frac{3}{128} \sqrt{\frac{21}{5}} C_{11,22}^{33,0}+\frac{\sqrt{15} C_{20,00}^{20,0}}{64}+\frac{\sqrt{5}
C_{20,20}^{20,0}}{64}+\frac{C_{20,22}^{22,0}}{128}-\frac{\sqrt{3} C_{22,00}^{22,0}}{64}-\frac{C_{22,20}^{22,0}}{64}+\frac{\sqrt{7} C_{22,22}^{22,0}}{64}\)

\noindent\(C_{11,1112,1}^{11,1112,0,\text{Loc}}= \frac{17 C_{00,20}^{40,0}}{288}+\frac{17 C_{00,20}^{40,1}}{576}+\frac{49 C_{00,22}^{42,0}}{1152
\sqrt{5}}+\frac{49 C_{00,22}^{42,1}}{2304 \sqrt{5}}-\frac{\sqrt{5} C_{11,22}^{31,0}}{384}-\frac{\sqrt{5} C_{11,22}^{31,1}}{768}+\frac{1}{128} \sqrt{\frac{7}{3}}
C_{11,22}^{33,0}+\frac{1}{256} \sqrt{\frac{7}{3}} C_{11,22}^{33,1}-\frac{5 C_{20,20}^{20,0}}{288}-\frac{5 C_{20,20}^{20,1}}{576}+\frac{\sqrt{5} C_{20,22}^{22,0}}{1152}+\frac{\sqrt{5}
C_{20,22}^{22,1}}{2304}-\frac{13 C_{22,20}^{22,0}}{288 \sqrt{5}}-\frac{13 C_{22,20}^{22,1}}{576 \sqrt{5}}-\frac{13}{576} \sqrt{\frac{7}{5}} C_{22,22}^{22,0}-\frac{13
\sqrt{\frac{7}{5}} C_{22,22}^{22,1}}{1152}\)

\noindent\(C_{11,1112,1}^{11,1112,0,\text{NonLoc}}= \frac{17 C_{00,00}^{40,0}}{384 \sqrt{3}}+\frac{17 C_{00,00}^{40,1}}{192 \sqrt{3}}+\frac{17
C_{00,20}^{40,0}}{1152}+\frac{17 C_{00,20}^{40,1}}{576}-\frac{49 C_{00,22}^{42,0}}{2304 \sqrt{5}}-\frac{49 C_{00,22}^{42,1}}{1152 \sqrt{5}}-\frac{\sqrt{5}
C_{11,22}^{31,0}}{768}-\frac{\sqrt{5} C_{11,22}^{31,1}}{384}+\frac{1}{256} \sqrt{\frac{7}{3}} C_{11,22}^{33,0}+\frac{1}{128} \sqrt{\frac{7}{3}} C_{11,22}^{33,1}-\frac{5
C_{20,00}^{20,0}}{384 \sqrt{3}}-\frac{5 C_{20,00}^{20,1}}{192 \sqrt{3}}-\frac{5 C_{20,20}^{20,0}}{1152}-\frac{5 C_{20,20}^{20,1}}{576}-\frac{\sqrt{5}
C_{20,22}^{22,0}}{2304}-\frac{\sqrt{5} C_{20,22}^{22,1}}{1152}-\frac{13 C_{22,00}^{22,0}}{384 \sqrt{15}}-\frac{13 C_{22,00}^{22,1}}{192 \sqrt{15}}-\frac{13
C_{22,20}^{22,0}}{1152 \sqrt{5}}-\frac{13 C_{22,20}^{22,1}}{576 \sqrt{5}}+\frac{13 \sqrt{\frac{7}{5}} C_{22,22}^{22,0}}{1152}+\frac{13}{576} \sqrt{\frac{7}{5}}
C_{22,22}^{22,1}\)

\noindent\(C_{11,1112,1}^{11,1112,1,\text{Loc}}= \frac{17 C_{00,20}^{40,1}}{192 \sqrt{3}}+\frac{49 C_{00,22}^{42,1}}{768 \sqrt{15}}-\frac{1}{256}
\sqrt{\frac{5}{3}} C_{11,22}^{31,1}+\frac{\sqrt{7} C_{11,22}^{33,1}}{256}-\frac{5 C_{20,20}^{20,1}}{192 \sqrt{3}}+\frac{1}{768} \sqrt{\frac{5}{3}}
C_{20,22}^{22,1}-\frac{13 C_{22,20}^{22,1}}{192 \sqrt{15}}-\frac{13}{384} \sqrt{\frac{7}{15}} C_{22,22}^{22,1}\)

\noindent\(C_{11,1112,1}^{11,1112,1,\text{NonLoc}}= \frac{17 C_{00,00}^{40,0}}{384}+\frac{17 C_{00,20}^{40,0}}{384 \sqrt{3}}-\frac{49
C_{00,22}^{42,0}}{768 \sqrt{15}}-\frac{1}{256} \sqrt{\frac{5}{3}} C_{11,22}^{31,0}+\frac{\sqrt{7} C_{11,22}^{33,0}}{256}-\frac{5 C_{20,00}^{20,0}}{384}-\frac{5
C_{20,20}^{20,0}}{384 \sqrt{3}}-\frac{1}{768} \sqrt{\frac{5}{3}} C_{20,22}^{22,0}-\frac{13 C_{22,00}^{22,0}}{384 \sqrt{5}}-\frac{13 C_{22,20}^{22,0}}{384
\sqrt{15}}+\frac{13}{384} \sqrt{\frac{7}{15}} C_{22,22}^{22,0}\)

\noindent\(C_{11,1112,2}^{00,2202,0,\text{Loc}}= \frac{1}{16} \sqrt{\frac{5}{3}} C_{11,11}^{31,0}+\frac{1}{32} \sqrt{\frac{5}{3}} C_{11,11}^{31,1}+\frac{C_{22,11}^{22,0}}{8}+\frac{C_{22,11}^{22,1}}{16}\)

\noindent\(C_{11,1112,2}^{00,2202,0,\text{NonLoc}}= \frac{1}{32} \sqrt{\frac{5}{3}} C_{11,11}^{31,0}+\frac{1}{16} \sqrt{\frac{5}{3}}
C_{11,11}^{31,1}-\frac{C_{22,11}^{22,0}}{16}-\frac{C_{22,11}^{22,1}}{8}\)

\noindent\(C_{11,1112,2}^{00,2202,1,\text{Loc}}= \frac{\sqrt{5} C_{11,11}^{31,1}}{32}+\frac{\sqrt{3} C_{22,11}^{22,1}}{16}\)

\noindent\(C_{11,1112,2}^{00,2202,1,\text{NonLoc}}= \frac{\sqrt{5} C_{11,11}^{31,0}}{32}-\frac{\sqrt{3} C_{22,11}^{22,0}}{16}\)

\noindent\(C_{11,1112,2}^{11,1111,0,\text{Loc}}= \frac{\sqrt{5} C_{00,20}^{40,0}}{48}+\frac{\sqrt{5} C_{00,20}^{40,1}}{96}-\frac{7
C_{00,22}^{42,0}}{192}-\frac{7 C_{00,22}^{42,1}}{384}+\frac{C_{11,22}^{31,0}}{64}+\frac{C_{11,22}^{31,1}}{128}-\frac{1}{64} \sqrt{\frac{21}{5}} C_{11,22}^{33,0}-\frac{1}{128}
\sqrt{\frac{21}{5}} C_{11,22}^{33,1}-\frac{\sqrt{5} C_{20,20}^{20,0}}{48}-\frac{\sqrt{5} C_{20,20}^{20,1}}{96}+\frac{C_{20,22}^{22,0}}{192}+\frac{C_{20,22}^{22,1}}{384}+\frac{C_{22,20}^{22,0}}{48}+\frac{C_{22,20}^{22,1}}{96}+\frac{\sqrt{7}
C_{22,22}^{22,0}}{96}+\frac{\sqrt{7} C_{22,22}^{22,1}}{192}\)

\noindent\(C_{11,1112,2}^{11,1111,0,\text{NonLoc}}= \frac{1}{64} \sqrt{\frac{5}{3}} C_{00,00}^{40,0}+\frac{1}{32} \sqrt{\frac{5}{3}}
C_{00,00}^{40,1}+\frac{\sqrt{5} C_{00,20}^{40,0}}{192}+\frac{\sqrt{5} C_{00,20}^{40,1}}{96}+\frac{7 C_{00,22}^{42,0}}{384}+\frac{7 C_{00,22}^{42,1}}{192}+\frac{C_{11,22}^{31,0}}{128}+\frac{C_{11,22}^{31,1}}{64}-\frac{1}{128}
\sqrt{\frac{21}{5}} C_{11,22}^{33,0}-\frac{1}{64} \sqrt{\frac{21}{5}} C_{11,22}^{33,1}-\frac{1}{64} \sqrt{\frac{5}{3}} C_{20,00}^{20,0}-\frac{1}{32}
\sqrt{\frac{5}{3}} C_{20,00}^{20,1}-\frac{\sqrt{5} C_{20,20}^{20,0}}{192}-\frac{\sqrt{5} C_{20,20}^{20,1}}{96}-\frac{C_{20,22}^{22,0}}{384}-\frac{C_{20,22}^{22,1}}{192}+\frac{C_{22,00}^{22,0}}{64
\sqrt{3}}+\frac{C_{22,00}^{22,1}}{32 \sqrt{3}}+\frac{C_{22,20}^{22,0}}{192}+\frac{C_{22,20}^{22,1}}{96}-\frac{\sqrt{7} C_{22,22}^{22,0}}{192}-\frac{\sqrt{7}
C_{22,22}^{22,1}}{96}\)

\noindent\(C_{11,1112,2}^{11,1111,1,\text{Loc}}= \frac{1}{32} \sqrt{\frac{5}{3}} C_{00,20}^{40,1}-\frac{7 C_{00,22}^{42,1}}{128 \sqrt{3}}+\frac{\sqrt{3}
C_{11,22}^{31,1}}{128}-\frac{3}{128} \sqrt{\frac{7}{5}} C_{11,22}^{33,1}-\frac{1}{32} \sqrt{\frac{5}{3}} C_{20,20}^{20,1}+\frac{C_{20,22}^{22,1}}{128
\sqrt{3}}+\frac{C_{22,20}^{22,1}}{32 \sqrt{3}}+\frac{1}{64} \sqrt{\frac{7}{3}} C_{22,22}^{22,1}\)

\noindent\(C_{11,1112,2}^{11,1111,1,\text{NonLoc}}= \frac{\sqrt{5} C_{00,00}^{40,0}}{64}+\frac{1}{64} \sqrt{\frac{5}{3}} C_{00,20}^{40,0}+\frac{7
C_{00,22}^{42,0}}{128 \sqrt{3}}+\frac{\sqrt{3} C_{11,22}^{31,0}}{128}-\frac{3}{128} \sqrt{\frac{7}{5}} C_{11,22}^{33,0}-\frac{\sqrt{5} C_{20,00}^{20,0}}{64}-\frac{1}{64}
\sqrt{\frac{5}{3}} C_{20,20}^{20,0}-\frac{C_{20,22}^{22,0}}{128 \sqrt{3}}+\frac{C_{22,00}^{22,0}}{64}+\frac{C_{22,20}^{22,0}}{64 \sqrt{3}}-\frac{1}{64}
\sqrt{\frac{7}{3}} C_{22,22}^{22,0}\)

\noindent\(C_{11,1112,2}^{11,1112,0,\text{Loc}}= \frac{1}{96} \sqrt{\frac{5}{3}} C_{00,20}^{40,0}+\frac{1}{192} \sqrt{\frac{5}{3}}
C_{00,20}^{40,1}-\frac{7 C_{00,22}^{42,0}}{384 \sqrt{3}}-\frac{7 C_{00,22}^{42,1}}{768 \sqrt{3}}+\frac{\sqrt{3} C_{11,22}^{31,0}}{128}+\frac{\sqrt{3}
C_{11,22}^{31,1}}{256}-\frac{3}{128} \sqrt{\frac{7}{5}} C_{11,22}^{33,0}-\frac{3}{256} \sqrt{\frac{7}{5}} C_{11,22}^{33,1}+\frac{1}{32} \sqrt{\frac{5}{3}}
C_{20,20}^{20,0}+\frac{1}{64} \sqrt{\frac{5}{3}} C_{20,20}^{20,1}-\frac{C_{20,22}^{22,0}}{128 \sqrt{3}}-\frac{C_{20,22}^{22,1}}{256 \sqrt{3}}-\frac{\sqrt{3}
C_{22,20}^{22,0}}{32}-\frac{\sqrt{3} C_{22,20}^{22,1}}{64}+\frac{1}{64} \sqrt{\frac{7}{3}} C_{22,22}^{22,0}+\frac{1}{128} \sqrt{\frac{7}{3}} C_{22,22}^{22,1}\)

\noindent\(C_{11,1112,2}^{11,1112,0,\text{NonLoc}}= \frac{\sqrt{5} C_{00,00}^{40,0}}{384}+\frac{\sqrt{5} C_{00,00}^{40,1}}{192}+\frac{1}{384}
\sqrt{\frac{5}{3}} C_{00,20}^{40,0}+\frac{1}{192} \sqrt{\frac{5}{3}} C_{00,20}^{40,1}+\frac{7 C_{00,22}^{42,0}}{768 \sqrt{3}}+\frac{7 C_{00,22}^{42,1}}{384
\sqrt{3}}+\frac{\sqrt{3} C_{11,22}^{31,0}}{256}+\frac{\sqrt{3} C_{11,22}^{31,1}}{128}-\frac{3}{256} \sqrt{\frac{7}{5}} C_{11,22}^{33,0}-\frac{3}{128}
\sqrt{\frac{7}{5}} C_{11,22}^{33,1}+\frac{\sqrt{5} C_{20,00}^{20,0}}{128}+\frac{\sqrt{5} C_{20,00}^{20,1}}{64}+\frac{1}{128} \sqrt{\frac{5}{3}} C_{20,20}^{20,0}+\frac{1}{64}
\sqrt{\frac{5}{3}} C_{20,20}^{20,1}+\frac{C_{20,22}^{22,0}}{256 \sqrt{3}}+\frac{C_{20,22}^{22,1}}{128 \sqrt{3}}-\frac{3 C_{22,00}^{22,0}}{128}-\frac{3
C_{22,00}^{22,1}}{64}-\frac{\sqrt{3} C_{22,20}^{22,0}}{128}-\frac{\sqrt{3} C_{22,20}^{22,1}}{64}-\frac{1}{128} \sqrt{\frac{7}{3}} C_{22,22}^{22,0}-\frac{1}{64}
\sqrt{\frac{7}{3}} C_{22,22}^{22,1}\)

\noindent\(C_{11,1112,2}^{11,1112,1,\text{Loc}}= \frac{\sqrt{5} C_{00,20}^{40,1}}{192}-\frac{7 C_{00,22}^{42,1}}{768}+\frac{3 C_{11,22}^{31,1}}{256}-\frac{3}{256}
\sqrt{\frac{21}{5}} C_{11,22}^{33,1}+\frac{\sqrt{5} C_{20,20}^{20,1}}{64}-\frac{C_{20,22}^{22,1}}{256}-\frac{3 C_{22,20}^{22,1}}{64}+\frac{\sqrt{7}
C_{22,22}^{22,1}}{128}\)

\noindent\(C_{11,1112,2}^{11,1112,1,\text{NonLoc}}= \frac{1}{128} \sqrt{\frac{5}{3}} C_{00,00}^{40,0}+\frac{\sqrt{5} C_{00,20}^{40,0}}{384}+\frac{7
C_{00,22}^{42,0}}{768}+\frac{3 C_{11,22}^{31,0}}{256}-\frac{3}{256} \sqrt{\frac{21}{5}} C_{11,22}^{33,0}+\frac{\sqrt{15} C_{20,00}^{20,0}}{128}+\frac{\sqrt{5}
C_{20,20}^{20,0}}{128}+\frac{C_{20,22}^{22,0}}{256}-\frac{3 \sqrt{3} C_{22,00}^{22,0}}{128}-\frac{3 C_{22,20}^{22,0}}{128}-\frac{\sqrt{7} C_{22,22}^{22,0}}{128}\)

\noindent\(C_{11,1112,3}^{11,1112,0,\text{Loc}}= \frac{1}{24} \sqrt{\frac{7}{3}} C_{00,20}^{40,0}+\frac{1}{48} \sqrt{\frac{7}{3}} C_{00,20}^{40,1}+\frac{1}{48}
\sqrt{\frac{7}{15}} C_{00,22}^{42,0}+\frac{1}{96} \sqrt{\frac{7}{15}} C_{00,22}^{42,1}-\frac{1}{16} \sqrt{\frac{7}{15}} C_{22,20}^{22,0}-\frac{1}{32}
\sqrt{\frac{7}{15}} C_{22,20}^{22,1}-\frac{C_{22,22}^{22,0}}{16 \sqrt{15}}-\frac{C_{22,22}^{22,1}}{32 \sqrt{15}}\)

\noindent\(C_{11,1112,3}^{11,1112,0,\text{NonLoc}}= \frac{\sqrt{7} C_{00,00}^{40,0}}{96}+\frac{\sqrt{7} C_{00,00}^{40,1}}{48}+\frac{1}{96}
\sqrt{\frac{7}{3}} C_{00,20}^{40,0}+\frac{1}{48} \sqrt{\frac{7}{3}} C_{00,20}^{40,1}-\frac{1}{96} \sqrt{\frac{7}{15}} C_{00,22}^{42,0}-\frac{1}{48}
\sqrt{\frac{7}{15}} C_{00,22}^{42,1}-\frac{1}{64} \sqrt{\frac{7}{5}} C_{22,00}^{22,0}-\frac{1}{32} \sqrt{\frac{7}{5}} C_{22,00}^{22,1}-\frac{1}{64}
\sqrt{\frac{7}{15}} C_{22,20}^{22,0}-\frac{1}{32} \sqrt{\frac{7}{15}} C_{22,20}^{22,1}+\frac{C_{22,22}^{22,0}}{32 \sqrt{15}}+\frac{C_{22,22}^{22,1}}{16
\sqrt{15}}\)

\noindent\(C_{11,1112,3}^{11,1112,1,\text{Loc}}= \frac{\sqrt{7} C_{00,20}^{40,1}}{48}+\frac{1}{96} \sqrt{\frac{7}{5}} C_{00,22}^{42,1}-\frac{1}{32}
\sqrt{\frac{7}{5}} C_{22,20}^{22,1}-\frac{C_{22,22}^{22,1}}{32 \sqrt{5}}\)

\noindent\(C_{11,1112,3}^{11,1112,1,\text{NonLoc}}= \frac{1}{32} \sqrt{\frac{7}{3}} C_{00,00}^{40,0}+\frac{\sqrt{7} C_{00,20}^{40,0}}{96}-\frac{1}{96}
\sqrt{\frac{7}{5}} C_{00,22}^{42,0}-\frac{1}{64} \sqrt{\frac{21}{5}} C_{22,00}^{22,0}-\frac{1}{64} \sqrt{\frac{7}{5}} C_{22,20}^{22,0}+\frac{C_{22,22}^{22,0}}{32
\sqrt{5}}\)

\noindent\(C_{11,2000,1}^{00,1111,0,\text{Loc}}= \frac{5 C_{11,11}^{31,0}}{48}+\frac{5 C_{11,11}^{31,1}}{96}+\frac{1}{8} \sqrt{\frac{5}{3}}
C_{22,11}^{22,0}+\frac{1}{16} \sqrt{\frac{5}{3}} C_{22,11}^{22,1}\)

\noindent\(C_{11,2000,1}^{00,1111,0,\text{NonLoc}}= \frac{5 C_{11,11}^{31,0}}{96}+\frac{5 C_{11,11}^{31,1}}{48}-\frac{1}{16} \sqrt{\frac{5}{3}}
C_{22,11}^{22,0}-\frac{1}{8} \sqrt{\frac{5}{3}} C_{22,11}^{22,1}\)

\noindent\(C_{11,2000,1}^{00,1111,1,\text{Loc}}= \frac{5 C_{11,11}^{31,1}}{32 \sqrt{3}}+\frac{\sqrt{5} C_{22,11}^{22,1}}{16}\)

\noindent\(C_{11,2000,1}^{00,1111,1,\text{NonLoc}}= \frac{5 C_{11,11}^{31,0}}{32 \sqrt{3}}-\frac{\sqrt{5} C_{22,11}^{22,0}}{16}\)

\noindent\(C_{11,2000,1}^{11,0000,0,\text{Loc}}= \frac{5 C_{00,00}^{40,0}}{48}+\frac{5 C_{00,00}^{40,1}}{96}+\frac{C_{20,00}^{20,0}}{48}+\frac{C_{20,00}^{20,1}}{96}-\frac{\sqrt{5}
C_{22,00}^{22,0}}{24}-\frac{\sqrt{5} C_{22,00}^{22,1}}{48}\)

\noindent\(C_{11,2000,1}^{11,0000,0,\text{NonLoc}}= -\frac{5 C_{00,00}^{40,0}}{192}-\frac{5 C_{00,00}^{40,1}}{96}+\frac{5 C_{00,20}^{40,0}}{64
\sqrt{3}}+\frac{5 C_{00,20}^{40,1}}{32 \sqrt{3}}-\frac{C_{20,00}^{20,0}}{192}-\frac{C_{20,00}^{20,1}}{96}+\frac{C_{20,20}^{20,0}}{64 \sqrt{3}}+\frac{C_{20,20}^{20,1}}{32
\sqrt{3}}+\frac{\sqrt{5} C_{22,00}^{22,0}}{96}+\frac{\sqrt{5} C_{22,00}^{22,1}}{48}-\frac{1}{32} \sqrt{\frac{5}{3}} C_{22,20}^{22,0}-\frac{1}{16}
\sqrt{\frac{5}{3}} C_{22,20}^{22,1}\)

\noindent\(C_{11,2000,1}^{11,0000,1,\text{Loc}}= \frac{5 C_{00,00}^{40,1}}{32 \sqrt{3}}+\frac{C_{20,00}^{20,1}}{32 \sqrt{3}}-\frac{1}{16}
\sqrt{\frac{5}{3}} C_{22,00}^{22,1}\)

\noindent\(C_{11,2000,1}^{11,0000,1,\text{NonLoc}}= -\frac{5 C_{00,00}^{40,0}}{64 \sqrt{3}}+\frac{5 C_{00,20}^{40,0}}{64}-\frac{C_{20,00}^{20,0}}{64
\sqrt{3}}+\frac{C_{20,20}^{20,0}}{64}+\frac{1}{32} \sqrt{\frac{5}{3}} C_{22,00}^{22,0}-\frac{\sqrt{5} C_{22,20}^{22,0}}{32}\)

\noindent\(C_{11,2011,0}^{11,0011,0,\text{Loc}}= -\frac{5 C_{00,20}^{40,0}}{144}-\frac{5 C_{00,20}^{40,1}}{288}-\frac{7 \sqrt{5} C_{00,22}^{42,0}}{288}-\frac{7
\sqrt{5} C_{00,22}^{42,1}}{576}+\frac{\sqrt{5} C_{11,22}^{31,0}}{96}+\frac{\sqrt{5} C_{11,22}^{31,1}}{192}-\frac{1}{32} \sqrt{\frac{7}{3}} C_{11,22}^{33,0}-\frac{1}{64}
\sqrt{\frac{7}{3}} C_{11,22}^{33,1}-\frac{C_{20,20}^{20,0}}{144}-\frac{C_{20,20}^{20,1}}{288}+\frac{\sqrt{5} C_{20,22}^{22,0}}{288}+\frac{\sqrt{5}
C_{20,22}^{22,1}}{576}+\frac{\sqrt{5} C_{22,20}^{22,0}}{72}+\frac{\sqrt{5} C_{22,20}^{22,1}}{144}+\frac{\sqrt{35} C_{22,22}^{22,0}}{144}+\frac{\sqrt{35}
C_{22,22}^{22,1}}{288}\)

\noindent\(C_{11,2011,0}^{11,0011,0,\text{NonLoc}}= -\frac{5 C_{00,00}^{40,0}}{192 \sqrt{3}}-\frac{5 C_{00,00}^{40,1}}{96 \sqrt{3}}-\frac{5
C_{00,20}^{40,0}}{576}-\frac{5 C_{00,20}^{40,1}}{288}+\frac{7 \sqrt{5} C_{00,22}^{42,0}}{576}+\frac{7 \sqrt{5} C_{00,22}^{42,1}}{288}+\frac{\sqrt{5}
C_{11,22}^{31,0}}{192}+\frac{\sqrt{5} C_{11,22}^{31,1}}{96}-\frac{1}{64} \sqrt{\frac{7}{3}} C_{11,22}^{33,0}-\frac{1}{32} \sqrt{\frac{7}{3}} C_{11,22}^{33,1}-\frac{C_{20,00}^{20,0}}{192
\sqrt{3}}-\frac{C_{20,00}^{20,1}}{96 \sqrt{3}}-\frac{C_{20,20}^{20,0}}{576}-\frac{C_{20,20}^{20,1}}{288}-\frac{\sqrt{5} C_{20,22}^{22,0}}{576}-\frac{\sqrt{5}
C_{20,22}^{22,1}}{288}+\frac{1}{96} \sqrt{\frac{5}{3}} C_{22,00}^{22,0}+\frac{1}{48} \sqrt{\frac{5}{3}} C_{22,00}^{22,1}+\frac{\sqrt{5} C_{22,20}^{22,0}}{288}+\frac{\sqrt{5}
C_{22,20}^{22,1}}{144}-\frac{\sqrt{35} C_{22,22}^{22,0}}{288}-\frac{\sqrt{35} C_{22,22}^{22,1}}{144}\)

\noindent\(C_{11,2011,0}^{11,0011,1,\text{Loc}}= -\frac{5 C_{00,20}^{40,1}}{96 \sqrt{3}}-\frac{7}{192} \sqrt{\frac{5}{3}} C_{00,22}^{42,1}+\frac{1}{64}
\sqrt{\frac{5}{3}} C_{11,22}^{31,1}-\frac{\sqrt{7} C_{11,22}^{33,1}}{64}-\frac{C_{20,20}^{20,1}}{96 \sqrt{3}}+\frac{1}{192} \sqrt{\frac{5}{3}} C_{20,22}^{22,1}+\frac{1}{48}
\sqrt{\frac{5}{3}} C_{22,20}^{22,1}+\frac{1}{96} \sqrt{\frac{35}{3}} C_{22,22}^{22,1}\)

\noindent\(C_{11,2011,0}^{11,0011,1,\text{NonLoc}}= -\frac{5 C_{00,00}^{40,0}}{192}-\frac{5 C_{00,20}^{40,0}}{192 \sqrt{3}}+\frac{7}{192}
\sqrt{\frac{5}{3}} C_{00,22}^{42,0}+\frac{1}{64} \sqrt{\frac{5}{3}} C_{11,22}^{31,0}-\frac{\sqrt{7} C_{11,22}^{33,0}}{64}-\frac{C_{20,00}^{20,0}}{192}-\frac{C_{20,20}^{20,0}}{192
\sqrt{3}}-\frac{1}{192} \sqrt{\frac{5}{3}} C_{20,22}^{22,0}+\frac{\sqrt{5} C_{22,00}^{22,0}}{96}+\frac{1}{96} \sqrt{\frac{5}{3}} C_{22,20}^{22,0}-\frac{1}{96}
\sqrt{\frac{35}{3}} C_{22,22}^{22,0}\)

\noindent\(C_{11,2011,1}^{00,1101,0,\text{Loc}}= -\frac{5 C_{11,11}^{31,0}}{48}-\frac{5 C_{11,11}^{31,1}}{96}-\frac{1}{8} \sqrt{\frac{5}{3}}
C_{22,11}^{22,0}-\frac{1}{16} \sqrt{\frac{5}{3}} C_{22,11}^{22,1}\)

\noindent\(C_{11,2011,1}^{00,1101,0,\text{NonLoc}}= -\frac{5 C_{11,11}^{31,0}}{96}-\frac{5 C_{11,11}^{31,1}}{48}+\frac{1}{16} \sqrt{\frac{5}{3}}
C_{22,11}^{22,0}+\frac{1}{8} \sqrt{\frac{5}{3}} C_{22,11}^{22,1}\)

\noindent\(C_{11,2011,1}^{00,1101,1,\text{Loc}}= -\frac{5 C_{11,11}^{31,1}}{32 \sqrt{3}}-\frac{\sqrt{5} C_{22,11}^{22,1}}{16}\)

\noindent\(C_{11,2011,1}^{00,1101,1,\text{NonLoc}}= -\frac{5 C_{11,11}^{31,0}}{32 \sqrt{3}}+\frac{\sqrt{5} C_{22,11}^{22,0}}{16}\)

\noindent\(C_{11,2011,1}^{11,0011,0,\text{Loc}}= -\frac{5 C_{00,20}^{40,0}}{48 \sqrt{3}}-\frac{5 C_{00,20}^{40,1}}{96 \sqrt{3}}+\frac{7}{192}
\sqrt{\frac{5}{3}} C_{00,22}^{42,0}+\frac{7}{384} \sqrt{\frac{5}{3}} C_{00,22}^{42,1}-\frac{1}{64} \sqrt{\frac{5}{3}} C_{11,22}^{31,0}-\frac{1}{128}
\sqrt{\frac{5}{3}} C_{11,22}^{31,1}+\frac{\sqrt{7} C_{11,22}^{33,0}}{64}+\frac{\sqrt{7} C_{11,22}^{33,1}}{128}-\frac{C_{20,20}^{20,0}}{48 \sqrt{3}}-\frac{C_{20,20}^{20,1}}{96
\sqrt{3}}-\frac{1}{192} \sqrt{\frac{5}{3}} C_{20,22}^{22,0}-\frac{1}{384} \sqrt{\frac{5}{3}} C_{20,22}^{22,1}+\frac{1}{24} \sqrt{\frac{5}{3}} C_{22,20}^{22,0}+\frac{1}{48}
\sqrt{\frac{5}{3}} C_{22,20}^{22,1}-\frac{1}{96} \sqrt{\frac{35}{3}} C_{22,22}^{22,0}-\frac{1}{192} \sqrt{\frac{35}{3}} C_{22,22}^{22,1}\)

\noindent\(C_{11,2011,1}^{11,0011,0,\text{NonLoc}}= -\frac{5 C_{00,00}^{40,0}}{192}-\frac{5 C_{00,00}^{40,1}}{96}-\frac{5 C_{00,20}^{40,0}}{192
\sqrt{3}}-\frac{5 C_{00,20}^{40,1}}{96 \sqrt{3}}-\frac{7}{384} \sqrt{\frac{5}{3}} C_{00,22}^{42,0}-\frac{7}{192} \sqrt{\frac{5}{3}} C_{00,22}^{42,1}-\frac{1}{128}
\sqrt{\frac{5}{3}} C_{11,22}^{31,0}-\frac{1}{64} \sqrt{\frac{5}{3}} C_{11,22}^{31,1}+\frac{\sqrt{7} C_{11,22}^{33,0}}{128}+\frac{\sqrt{7} C_{11,22}^{33,1}}{64}-\frac{C_{20,00}^{20,0}}{192}-\frac{C_{20,00}^{20,1}}{96}-\frac{C_{20,20}^{20,0}}{192
\sqrt{3}}-\frac{C_{20,20}^{20,1}}{96 \sqrt{3}}+\frac{1}{384} \sqrt{\frac{5}{3}} C_{20,22}^{22,0}+\frac{1}{192} \sqrt{\frac{5}{3}} C_{20,22}^{22,1}+\frac{\sqrt{5}
C_{22,00}^{22,0}}{96}+\frac{\sqrt{5} C_{22,00}^{22,1}}{48}+\frac{1}{96} \sqrt{\frac{5}{3}} C_{22,20}^{22,0}+\frac{1}{48} \sqrt{\frac{5}{3}} C_{22,20}^{22,1}+\frac{1}{192}
\sqrt{\frac{35}{3}} C_{22,22}^{22,0}+\frac{1}{96} \sqrt{\frac{35}{3}} C_{22,22}^{22,1}\)

\noindent\(C_{11,2011,1}^{11,0011,1,\text{Loc}}= -\frac{5 C_{00,20}^{40,1}}{96}+\frac{7 \sqrt{5} C_{00,22}^{42,1}}{384}-\frac{\sqrt{5}
C_{11,22}^{31,1}}{128}+\frac{\sqrt{21} C_{11,22}^{33,1}}{128}-\frac{C_{20,20}^{20,1}}{96}-\frac{\sqrt{5} C_{20,22}^{22,1}}{384}+\frac{\sqrt{5} C_{22,20}^{22,1}}{48}-\frac{\sqrt{35}
C_{22,22}^{22,1}}{192}\)

\noindent\(C_{11,2011,1}^{11,0011,1,\text{NonLoc}}= -\frac{5 C_{00,00}^{40,0}}{64 \sqrt{3}}-\frac{5 C_{00,20}^{40,0}}{192}-\frac{7
\sqrt{5} C_{00,22}^{42,0}}{384}-\frac{\sqrt{5} C_{11,22}^{31,0}}{128}+\frac{\sqrt{21} C_{11,22}^{33,0}}{128}-\frac{C_{20,00}^{20,0}}{64 \sqrt{3}}-\frac{C_{20,20}^{20,0}}{192}+\frac{\sqrt{5}
C_{20,22}^{22,0}}{384}+\frac{1}{32} \sqrt{\frac{5}{3}} C_{22,00}^{22,0}+\frac{\sqrt{5} C_{22,20}^{22,0}}{96}+\frac{\sqrt{35} C_{22,22}^{22,0}}{192}\)

\noindent\(C_{11,2011,2}^{11,0011,0,\text{Loc}}= -\frac{5 \sqrt{5} C_{00,20}^{40,0}}{144}-\frac{5 \sqrt{5} C_{00,20}^{40,1}}{288}-\frac{7
C_{00,22}^{42,0}}{576}-\frac{7 C_{00,22}^{42,1}}{1152}+\frac{C_{11,22}^{31,0}}{192}+\frac{C_{11,22}^{31,1}}{384}-\frac{1}{64} \sqrt{\frac{7}{15}}
C_{11,22}^{33,0}-\frac{1}{128} \sqrt{\frac{7}{15}} C_{11,22}^{33,1}-\frac{\sqrt{5} C_{20,20}^{20,0}}{144}-\frac{\sqrt{5} C_{20,20}^{20,1}}{288}+\frac{C_{20,22}^{22,0}}{576}+\frac{C_{20,22}^{22,1}}{1152}+\frac{5
C_{22,20}^{22,0}}{72}+\frac{5 C_{22,20}^{22,1}}{144}+\frac{\sqrt{7} C_{22,22}^{22,0}}{288}+\frac{\sqrt{7} C_{22,22}^{22,1}}{576}\)

\noindent\(C_{11,2011,2}^{11,0011,0,\text{NonLoc}}= -\frac{5}{192} \sqrt{\frac{5}{3}} C_{00,00}^{40,0}-\frac{5}{96} \sqrt{\frac{5}{3}}
C_{00,00}^{40,1}-\frac{5 \sqrt{5} C_{00,20}^{40,0}}{576}-\frac{5 \sqrt{5} C_{00,20}^{40,1}}{288}+\frac{7 C_{00,22}^{42,0}}{1152}+\frac{7 C_{00,22}^{42,1}}{576}+\frac{C_{11,22}^{31,0}}{384}+\frac{C_{11,22}^{31,1}}{192}-\frac{1}{128}
\sqrt{\frac{7}{15}} C_{11,22}^{33,0}-\frac{1}{64} \sqrt{\frac{7}{15}} C_{11,22}^{33,1}-\frac{1}{192} \sqrt{\frac{5}{3}} C_{20,00}^{20,0}-\frac{1}{96}
\sqrt{\frac{5}{3}} C_{20,00}^{20,1}-\frac{\sqrt{5} C_{20,20}^{20,0}}{576}-\frac{\sqrt{5} C_{20,20}^{20,1}}{288}-\frac{C_{20,22}^{22,0}}{1152}-\frac{C_{20,22}^{22,1}}{576}+\frac{5
C_{22,00}^{22,0}}{96 \sqrt{3}}+\frac{5 C_{22,00}^{22,1}}{48 \sqrt{3}}+\frac{5 C_{22,20}^{22,0}}{288}+\frac{5 C_{22,20}^{22,1}}{144}-\frac{\sqrt{7}
C_{22,22}^{22,0}}{576}-\frac{\sqrt{7} C_{22,22}^{22,1}}{288}\)

\noindent\(C_{11,2011,2}^{11,0011,1,\text{Loc}}= -\frac{5}{96} \sqrt{\frac{5}{3}} C_{00,20}^{40,1}-\frac{7 C_{00,22}^{42,1}}{384 \sqrt{3}}+\frac{C_{11,22}^{31,1}}{128
\sqrt{3}}-\frac{1}{128} \sqrt{\frac{7}{5}} C_{11,22}^{33,1}-\frac{1}{96} \sqrt{\frac{5}{3}} C_{20,20}^{20,1}+\frac{C_{20,22}^{22,1}}{384 \sqrt{3}}+\frac{5
C_{22,20}^{22,1}}{48 \sqrt{3}}+\frac{1}{192} \sqrt{\frac{7}{3}} C_{22,22}^{22,1}\)

\noindent\(C_{11,2011,2}^{11,0011,1,\text{NonLoc}}= -\frac{5 \sqrt{5} C_{00,00}^{40,0}}{192}-\frac{5}{192} \sqrt{\frac{5}{3}} C_{00,20}^{40,0}+\frac{7
C_{00,22}^{42,0}}{384 \sqrt{3}}+\frac{C_{11,22}^{31,0}}{128 \sqrt{3}}-\frac{1}{128} \sqrt{\frac{7}{5}} C_{11,22}^{33,0}-\frac{\sqrt{5} C_{20,00}^{20,0}}{192}-\frac{1}{192}
\sqrt{\frac{5}{3}} C_{20,20}^{20,0}-\frac{C_{20,22}^{22,0}}{384 \sqrt{3}}+\frac{5 C_{22,00}^{22,0}}{96}+\frac{5 C_{22,20}^{22,0}}{96 \sqrt{3}}-\frac{1}{192}
\sqrt{\frac{7}{3}} C_{22,22}^{22,0}\)

\noindent\(C_{11,2202,1}^{00,1111,0,\text{Loc}}= -\frac{\sqrt{5} C_{11,11}^{31,0}}{48}-\frac{\sqrt{5} C_{11,11}^{31,1}}{96}-\frac{C_{22,11}^{22,0}}{8
\sqrt{3}}-\frac{C_{22,11}^{22,1}}{16 \sqrt{3}}\)

\noindent\(C_{11,2202,1}^{00,1111,0,\text{NonLoc}}= -\frac{\sqrt{5} C_{11,11}^{31,0}}{96}-\frac{\sqrt{5} C_{11,11}^{31,1}}{48}+\frac{C_{22,11}^{22,0}}{16
\sqrt{3}}+\frac{C_{22,11}^{22,1}}{8 \sqrt{3}}\)

\noindent\(C_{11,2202,1}^{00,1111,1,\text{Loc}}= -\frac{1}{32} \sqrt{\frac{5}{3}} C_{11,11}^{31,1}-\frac{C_{22,11}^{22,1}}{16}\)

\noindent\(C_{11,2202,1}^{00,1111,1,\text{NonLoc}}= -\frac{1}{32} \sqrt{\frac{5}{3}} C_{11,11}^{31,0}+\frac{C_{22,11}^{22,0}}{16}\)

\noindent\(C_{11,2202,1}^{11,0000,0,\text{Loc}}= \frac{\sqrt{5} C_{00,00}^{40,0}}{24}+\frac{\sqrt{5} C_{00,00}^{40,1}}{48}-\frac{\sqrt{5}
C_{20,00}^{20,0}}{24}-\frac{\sqrt{5} C_{20,00}^{20,1}}{48}+\frac{C_{22,00}^{22,0}}{24}+\frac{C_{22,00}^{22,1}}{48}\)

\noindent\(C_{11,2202,1}^{11,0000,0,\text{NonLoc}}= -\frac{\sqrt{5} C_{00,00}^{40,0}}{96}-\frac{\sqrt{5} C_{00,00}^{40,1}}{48}+\frac{1}{32}
\sqrt{\frac{5}{3}} C_{00,20}^{40,0}+\frac{1}{16} \sqrt{\frac{5}{3}} C_{00,20}^{40,1}+\frac{\sqrt{5} C_{20,00}^{20,0}}{96}+\frac{\sqrt{5} C_{20,00}^{20,1}}{48}-\frac{1}{32}
\sqrt{\frac{5}{3}} C_{20,20}^{20,0}-\frac{1}{16} \sqrt{\frac{5}{3}} C_{20,20}^{20,1}-\frac{C_{22,00}^{22,0}}{96}-\frac{C_{22,00}^{22,1}}{48}+\frac{C_{22,20}^{22,0}}{32
\sqrt{3}}+\frac{C_{22,20}^{22,1}}{16 \sqrt{3}}\)

\noindent\(C_{11,2202,1}^{11,0000,1,\text{Loc}}= \frac{1}{16} \sqrt{\frac{5}{3}} C_{00,00}^{40,1}-\frac{1}{16} \sqrt{\frac{5}{3}} C_{20,00}^{20,1}+\frac{C_{22,00}^{22,1}}{16
\sqrt{3}}\)

\noindent\(C_{11,2202,1}^{11,0000,1,\text{NonLoc}}= -\frac{1}{32} \sqrt{\frac{5}{3}} C_{00,00}^{40,0}+\frac{\sqrt{5} C_{00,20}^{40,0}}{32}+\frac{1}{32}
\sqrt{\frac{5}{3}} C_{20,00}^{20,0}-\frac{\sqrt{5} C_{20,20}^{20,0}}{32}-\frac{C_{22,00}^{22,0}}{32 \sqrt{3}}+\frac{C_{22,20}^{22,0}}{32}\)

\noindent\(C_{11,2202,2}^{00,1112,0,\text{Loc}}= \frac{1}{16} \sqrt{\frac{5}{3}} C_{11,11}^{31,0}+\frac{1}{32} \sqrt{\frac{5}{3}} C_{11,11}^{31,1}+\frac{C_{22,11}^{22,0}}{8}+\frac{C_{22,11}^{22,1}}{16}\)

\noindent\(C_{11,2202,2}^{00,1112,0,\text{NonLoc}}= \frac{1}{32} \sqrt{\frac{5}{3}} C_{11,11}^{31,0}+\frac{1}{16} \sqrt{\frac{5}{3}}
C_{11,11}^{31,1}-\frac{C_{22,11}^{22,0}}{16}-\frac{C_{22,11}^{22,1}}{8}\)

\noindent\(C_{11,2202,2}^{00,1112,1,\text{Loc}}= \frac{\sqrt{5} C_{11,11}^{31,1}}{32}+\frac{\sqrt{3} C_{22,11}^{22,1}}{16}\)

\noindent\(C_{11,2202,2}^{00,1112,1,\text{NonLoc}}= \frac{\sqrt{5} C_{11,11}^{31,0}}{32}-\frac{\sqrt{3} C_{22,11}^{22,0}}{16}\)

\noindent\(C_{11,2211,0}^{11,0011,0,\text{Loc}}= -\frac{\sqrt{5} C_{00,20}^{40,0}}{72}-\frac{\sqrt{5} C_{00,20}^{40,1}}{144}-\frac{7
C_{00,22}^{42,0}}{144}-\frac{7 C_{00,22}^{42,1}}{288}-\frac{C_{11,22}^{31,0}}{96}-\frac{C_{11,22}^{31,1}}{192}+\frac{1}{32} \sqrt{\frac{7}{15}} C_{11,22}^{33,0}+\frac{1}{64}
\sqrt{\frac{7}{15}} C_{11,22}^{33,1}+\frac{\sqrt{5} C_{20,20}^{20,0}}{72}+\frac{\sqrt{5} C_{20,20}^{20,1}}{144}+\frac{C_{20,22}^{22,0}}{36}+\frac{C_{20,22}^{22,1}}{72}-\frac{C_{22,20}^{22,0}}{72}-\frac{C_{22,20}^{22,1}}{144}-\frac{\sqrt{7}
C_{22,22}^{22,0}}{144}-\frac{\sqrt{7} C_{22,22}^{22,1}}{288}\)

\noindent\(C_{11,2211,0}^{11,0011,0,\text{NonLoc}}= -\frac{1}{96} \sqrt{\frac{5}{3}} C_{00,00}^{40,0}-\frac{1}{48} \sqrt{\frac{5}{3}}
C_{00,00}^{40,1}-\frac{\sqrt{5} C_{00,20}^{40,0}}{288}-\frac{\sqrt{5} C_{00,20}^{40,1}}{144}+\frac{7 C_{00,22}^{42,0}}{288}+\frac{7 C_{00,22}^{42,1}}{144}-\frac{C_{11,22}^{31,0}}{192}-\frac{C_{11,22}^{31,1}}{96}+\frac{1}{64}
\sqrt{\frac{7}{15}} C_{11,22}^{33,0}+\frac{1}{32} \sqrt{\frac{7}{15}} C_{11,22}^{33,1}+\frac{1}{96} \sqrt{\frac{5}{3}} C_{20,00}^{20,0}+\frac{1}{48}
\sqrt{\frac{5}{3}} C_{20,00}^{20,1}+\frac{\sqrt{5} C_{20,20}^{20,0}}{288}+\frac{\sqrt{5} C_{20,20}^{20,1}}{144}-\frac{C_{20,22}^{22,0}}{72}-\frac{C_{20,22}^{22,1}}{36}-\frac{C_{22,00}^{22,0}}{96
\sqrt{3}}-\frac{C_{22,00}^{22,1}}{48 \sqrt{3}}-\frac{C_{22,20}^{22,0}}{288}-\frac{C_{22,20}^{22,1}}{144}+\frac{\sqrt{7} C_{22,22}^{22,0}}{288}+\frac{\sqrt{7}
C_{22,22}^{22,1}}{144}\)

\noindent\(C_{11,2211,0}^{11,0011,1,\text{Loc}}= -\frac{1}{48} \sqrt{\frac{5}{3}} C_{00,20}^{40,1}-\frac{7 C_{00,22}^{42,1}}{96 \sqrt{3}}-\frac{C_{11,22}^{31,1}}{64
\sqrt{3}}+\frac{1}{64} \sqrt{\frac{7}{5}} C_{11,22}^{33,1}+\frac{1}{48} \sqrt{\frac{5}{3}} C_{20,20}^{20,1}+\frac{C_{20,22}^{22,1}}{24 \sqrt{3}}-\frac{C_{22,20}^{22,1}}{48
\sqrt{3}}-\frac{1}{96} \sqrt{\frac{7}{3}} C_{22,22}^{22,1}\)

\noindent\(C_{11,2211,0}^{11,0011,1,\text{NonLoc}}= -\frac{\sqrt{5} C_{00,00}^{40,0}}{96}-\frac{1}{96} \sqrt{\frac{5}{3}} C_{00,20}^{40,0}+\frac{7
C_{00,22}^{42,0}}{96 \sqrt{3}}-\frac{C_{11,22}^{31,0}}{64 \sqrt{3}}+\frac{1}{64} \sqrt{\frac{7}{5}} C_{11,22}^{33,0}+\frac{\sqrt{5} C_{20,00}^{20,0}}{96}+\frac{1}{96}
\sqrt{\frac{5}{3}} C_{20,20}^{20,0}-\frac{C_{20,22}^{22,0}}{24 \sqrt{3}}-\frac{C_{22,00}^{22,0}}{96}-\frac{C_{22,20}^{22,0}}{96 \sqrt{3}}+\frac{1}{96}
\sqrt{\frac{7}{3}} C_{22,22}^{22,0}\)

\noindent\(C_{11,2211,1}^{00,1101,0,\text{Loc}}= \frac{\sqrt{5} C_{11,11}^{31,0}}{48}+\frac{\sqrt{5} C_{11,11}^{31,1}}{96}+\frac{C_{22,11}^{22,0}}{8
\sqrt{3}}+\frac{C_{22,11}^{22,1}}{16 \sqrt{3}}\)

\noindent\(C_{11,2211,1}^{00,1101,0,\text{NonLoc}}= \frac{\sqrt{5} C_{11,11}^{31,0}}{96}+\frac{\sqrt{5} C_{11,11}^{31,1}}{48}-\frac{C_{22,11}^{22,0}}{16
\sqrt{3}}-\frac{C_{22,11}^{22,1}}{8 \sqrt{3}}\)

\noindent\(C_{11,2211,1}^{00,1101,1,\text{Loc}}= \frac{1}{32} \sqrt{\frac{5}{3}} C_{11,11}^{31,1}+\frac{C_{22,11}^{22,1}}{16}\)

\noindent\(C_{11,2211,1}^{00,1101,1,\text{NonLoc}}= \frac{1}{32} \sqrt{\frac{5}{3}} C_{11,11}^{31,0}-\frac{C_{22,11}^{22,0}}{16}\)

\noindent\(C_{11,2211,1}^{11,0011,0,\text{Loc}}= \frac{1}{48} \sqrt{\frac{5}{3}} C_{00,20}^{40,0}+\frac{1}{96} \sqrt{\frac{5}{3}} C_{00,20}^{40,1}-\frac{7
C_{00,22}^{42,0}}{192 \sqrt{3}}-\frac{7 C_{00,22}^{42,1}}{384 \sqrt{3}}-\frac{C_{11,22}^{31,0}}{32 \sqrt{3}}-\frac{C_{11,22}^{31,1}}{64 \sqrt{3}}+\frac{1}{32}
\sqrt{\frac{7}{5}} C_{11,22}^{33,0}+\frac{1}{64} \sqrt{\frac{7}{5}} C_{11,22}^{33,1}-\frac{1}{48} \sqrt{\frac{5}{3}} C_{20,20}^{20,0}-\frac{1}{96}
\sqrt{\frac{5}{3}} C_{20,20}^{20,1}-\frac{5 C_{20,22}^{22,0}}{192 \sqrt{3}}-\frac{5 C_{20,22}^{22,1}}{384 \sqrt{3}}+\frac{C_{22,20}^{22,0}}{48 \sqrt{3}}+\frac{C_{22,20}^{22,1}}{96
\sqrt{3}}+\frac{1}{24} \sqrt{\frac{7}{3}} C_{22,22}^{22,0}+\frac{1}{48} \sqrt{\frac{7}{3}} C_{22,22}^{22,1}\)

\noindent\(C_{11,2211,1}^{11,0011,0,\text{NonLoc}}= \frac{\sqrt{5} C_{00,00}^{40,0}}{192}+\frac{\sqrt{5} C_{00,00}^{40,1}}{96}+\frac{1}{192}
\sqrt{\frac{5}{3}} C_{00,20}^{40,0}+\frac{1}{96} \sqrt{\frac{5}{3}} C_{00,20}^{40,1}+\frac{7 C_{00,22}^{42,0}}{384 \sqrt{3}}+\frac{7 C_{00,22}^{42,1}}{192
\sqrt{3}}-\frac{C_{11,22}^{31,0}}{64 \sqrt{3}}-\frac{C_{11,22}^{31,1}}{32 \sqrt{3}}+\frac{1}{64} \sqrt{\frac{7}{5}} C_{11,22}^{33,0}+\frac{1}{32}
\sqrt{\frac{7}{5}} C_{11,22}^{33,1}-\frac{\sqrt{5} C_{20,00}^{20,0}}{192}-\frac{\sqrt{5} C_{20,00}^{20,1}}{96}-\frac{1}{192} \sqrt{\frac{5}{3}} C_{20,20}^{20,0}-\frac{1}{96}
\sqrt{\frac{5}{3}} C_{20,20}^{20,1}+\frac{5 C_{20,22}^{22,0}}{384 \sqrt{3}}+\frac{5 C_{20,22}^{22,1}}{192 \sqrt{3}}+\frac{C_{22,00}^{22,0}}{192}+\frac{C_{22,00}^{22,1}}{96}+\frac{C_{22,20}^{22,0}}{192
\sqrt{3}}+\frac{C_{22,20}^{22,1}}{96 \sqrt{3}}-\frac{1}{48} \sqrt{\frac{7}{3}} C_{22,22}^{22,0}-\frac{1}{24} \sqrt{\frac{7}{3}} C_{22,22}^{22,1}\)

\noindent\(C_{11,2211,1}^{11,0011,1,\text{Loc}}= \frac{\sqrt{5} C_{00,20}^{40,1}}{96}-\frac{7 C_{00,22}^{42,1}}{384}-\frac{C_{11,22}^{31,1}}{64}+\frac{1}{64}
\sqrt{\frac{21}{5}} C_{11,22}^{33,1}-\frac{\sqrt{5} C_{20,20}^{20,1}}{96}-\frac{5 C_{20,22}^{22,1}}{384}+\frac{C_{22,20}^{22,1}}{96}+\frac{\sqrt{7}
C_{22,22}^{22,1}}{48}\)

\noindent\(C_{11,2211,1}^{11,0011,1,\text{NonLoc}}= \frac{1}{64} \sqrt{\frac{5}{3}} C_{00,00}^{40,0}+\frac{\sqrt{5} C_{00,20}^{40,0}}{192}+\frac{7
C_{00,22}^{42,0}}{384}-\frac{C_{11,22}^{31,0}}{64}+\frac{1}{64} \sqrt{\frac{21}{5}} C_{11,22}^{33,0}-\frac{1}{64} \sqrt{\frac{5}{3}} C_{20,00}^{20,0}-\frac{\sqrt{5}
C_{20,20}^{20,0}}{192}+\frac{5 C_{20,22}^{22,0}}{384}+\frac{C_{22,00}^{22,0}}{64 \sqrt{3}}+\frac{C_{22,20}^{22,0}}{192}-\frac{\sqrt{7} C_{22,22}^{22,0}}{48}\)

\noindent\(C_{11,2211,2}^{11,0011,0,\text{Loc}}= -\frac{C_{00,20}^{40,0}}{144}-\frac{C_{00,20}^{40,1}}{288}-\frac{77 C_{00,22}^{42,0}}{576
\sqrt{5}}-\frac{77 C_{00,22}^{42,1}}{1152 \sqrt{5}}-\frac{\sqrt{5} C_{11,22}^{31,0}}{96}-\frac{\sqrt{5} C_{11,22}^{31,1}}{192}+\frac{1}{32} \sqrt{\frac{7}{3}}
C_{11,22}^{33,0}+\frac{1}{64} \sqrt{\frac{7}{3}} C_{11,22}^{33,1}+\frac{C_{20,20}^{20,0}}{144}+\frac{C_{20,20}^{20,1}}{288}+\frac{17 C_{20,22}^{22,0}}{576
\sqrt{5}}+\frac{17 C_{20,22}^{22,1}}{1152 \sqrt{5}}-\frac{C_{22,20}^{22,0}}{144 \sqrt{5}}-\frac{C_{22,20}^{22,1}}{288 \sqrt{5}}+\frac{1}{36} \sqrt{\frac{7}{5}}
C_{22,22}^{22,0}+\frac{1}{72} \sqrt{\frac{7}{5}} C_{22,22}^{22,1}\)

\noindent\(C_{11,2211,2}^{11,0011,0,\text{NonLoc}}= -\frac{C_{00,00}^{40,0}}{192 \sqrt{3}}-\frac{C_{00,00}^{40,1}}{96 \sqrt{3}}-\frac{C_{00,20}^{40,0}}{576}-\frac{C_{00,20}^{40,1}}{288}+\frac{77
C_{00,22}^{42,0}}{1152 \sqrt{5}}+\frac{77 C_{00,22}^{42,1}}{576 \sqrt{5}}-\frac{\sqrt{5} C_{11,22}^{31,0}}{192}-\frac{\sqrt{5} C_{11,22}^{31,1}}{96}+\frac{1}{64}
\sqrt{\frac{7}{3}} C_{11,22}^{33,0}+\frac{1}{32} \sqrt{\frac{7}{3}} C_{11,22}^{33,1}+\frac{C_{20,00}^{20,0}}{192 \sqrt{3}}+\frac{C_{20,00}^{20,1}}{96
\sqrt{3}}+\frac{C_{20,20}^{20,0}}{576}+\frac{C_{20,20}^{20,1}}{288}-\frac{17 C_{20,22}^{22,0}}{1152 \sqrt{5}}-\frac{17 C_{20,22}^{22,1}}{576 \sqrt{5}}-\frac{C_{22,00}^{22,0}}{192
\sqrt{15}}-\frac{C_{22,00}^{22,1}}{96 \sqrt{15}}-\frac{C_{22,20}^{22,0}}{576 \sqrt{5}}-\frac{C_{22,20}^{22,1}}{288 \sqrt{5}}-\frac{1}{72} \sqrt{\frac{7}{5}}
C_{22,22}^{22,0}-\frac{1}{36} \sqrt{\frac{7}{5}} C_{22,22}^{22,1}\)

\noindent\(C_{11,2211,2}^{11,0011,1,\text{Loc}}= -\frac{C_{00,20}^{40,1}}{96 \sqrt{3}}-\frac{77 C_{00,22}^{42,1}}{384 \sqrt{15}}-\frac{1}{64}
\sqrt{\frac{5}{3}} C_{11,22}^{31,1}+\frac{\sqrt{7} C_{11,22}^{33,1}}{64}+\frac{C_{20,20}^{20,1}}{96 \sqrt{3}}+\frac{17 C_{20,22}^{22,1}}{384 \sqrt{15}}-\frac{C_{22,20}^{22,1}}{96
\sqrt{15}}+\frac{1}{24} \sqrt{\frac{7}{15}} C_{22,22}^{22,1}\)

\noindent\(C_{11,2211,2}^{11,0011,1,\text{NonLoc}}= -\frac{C_{00,00}^{40,0}}{192}-\frac{C_{00,20}^{40,0}}{192 \sqrt{3}}+\frac{77 C_{00,22}^{42,0}}{384
\sqrt{15}}-\frac{1}{64} \sqrt{\frac{5}{3}} C_{11,22}^{31,0}+\frac{\sqrt{7} C_{11,22}^{33,0}}{64}+\frac{C_{20,00}^{20,0}}{192}+\frac{C_{20,20}^{20,0}}{192
\sqrt{3}}-\frac{17 C_{20,22}^{22,0}}{384 \sqrt{15}}-\frac{C_{22,00}^{22,0}}{192 \sqrt{5}}-\frac{C_{22,20}^{22,0}}{192 \sqrt{15}}-\frac{1}{24} \sqrt{\frac{7}{15}}
C_{22,22}^{22,0}\)

\noindent\(C_{11,2212,1}^{00,1101,0,\text{Loc}}= \frac{1}{16} \sqrt{\frac{5}{3}} C_{11,11}^{31,0}+\frac{1}{32} \sqrt{\frac{5}{3}} C_{11,11}^{31,1}+\frac{C_{22,11}^{22,0}}{8}+\frac{C_{22,11}^{22,1}}{16}\)

\noindent\(C_{11,2212,1}^{00,1101,0,\text{NonLoc}}= \frac{1}{32} \sqrt{\frac{5}{3}} C_{11,11}^{31,0}+\frac{1}{16} \sqrt{\frac{5}{3}}
C_{11,11}^{31,1}-\frac{C_{22,11}^{22,0}}{16}-\frac{C_{22,11}^{22,1}}{8}\)

\noindent\(C_{11,2212,1}^{00,1101,1,\text{Loc}}= \frac{\sqrt{5} C_{11,11}^{31,1}}{32}+\frac{\sqrt{3} C_{22,11}^{22,1}}{16}\)

\noindent\(C_{11,2212,1}^{00,1101,1,\text{NonLoc}}= \frac{\sqrt{5} C_{11,11}^{31,0}}{32}-\frac{\sqrt{3} C_{22,11}^{22,0}}{16}\)

\noindent\(C_{11,2212,1}^{11,0011,0,\text{Loc}}= -\frac{\sqrt{5} C_{00,20}^{40,0}}{48}-\frac{\sqrt{5} C_{00,20}^{40,1}}{96}+\frac{7
C_{00,22}^{42,0}}{192}+\frac{7 C_{00,22}^{42,1}}{384}+\frac{\sqrt{5} C_{20,20}^{20,0}}{48}+\frac{\sqrt{5} C_{20,20}^{20,1}}{96}-\frac{7 C_{20,22}^{22,0}}{192}-\frac{7
C_{20,22}^{22,1}}{384}-\frac{C_{22,20}^{22,0}}{48}-\frac{C_{22,20}^{22,1}}{96}+\frac{\sqrt{7} C_{22,22}^{22,0}}{48}+\frac{\sqrt{7} C_{22,22}^{22,1}}{96}\)

\noindent\(C_{11,2212,1}^{11,0011,0,\text{NonLoc}}= -\frac{1}{64} \sqrt{\frac{5}{3}} C_{00,00}^{40,0}-\frac{1}{32} \sqrt{\frac{5}{3}}
C_{00,00}^{40,1}-\frac{\sqrt{5} C_{00,20}^{40,0}}{192}-\frac{\sqrt{5} C_{00,20}^{40,1}}{96}-\frac{7 C_{00,22}^{42,0}}{384}-\frac{7 C_{00,22}^{42,1}}{192}+\frac{1}{64}
\sqrt{\frac{5}{3}} C_{20,00}^{20,0}+\frac{1}{32} \sqrt{\frac{5}{3}} C_{20,00}^{20,1}+\frac{\sqrt{5} C_{20,20}^{20,0}}{192}+\frac{\sqrt{5} C_{20,20}^{20,1}}{96}+\frac{7
C_{20,22}^{22,0}}{384}+\frac{7 C_{20,22}^{22,1}}{192}-\frac{C_{22,00}^{22,0}}{64 \sqrt{3}}-\frac{C_{22,00}^{22,1}}{32 \sqrt{3}}-\frac{C_{22,20}^{22,0}}{192}-\frac{C_{22,20}^{22,1}}{96}-\frac{\sqrt{7}
C_{22,22}^{22,0}}{96}-\frac{\sqrt{7} C_{22,22}^{22,1}}{48}\)

\noindent\(C_{11,2212,1}^{11,0011,1,\text{Loc}}= -\frac{1}{32} \sqrt{\frac{5}{3}} C_{00,20}^{40,1}+\frac{7 C_{00,22}^{42,1}}{128 \sqrt{3}}+\frac{1}{32}
\sqrt{\frac{5}{3}} C_{20,20}^{20,1}-\frac{7 C_{20,22}^{22,1}}{128 \sqrt{3}}-\frac{C_{22,20}^{22,1}}{32 \sqrt{3}}+\frac{1}{32} \sqrt{\frac{7}{3}}
C_{22,22}^{22,1}\)

\noindent\(C_{11,2212,1}^{11,0011,1,\text{NonLoc}}= -\frac{\sqrt{5} C_{00,00}^{40,0}}{64}-\frac{1}{64} \sqrt{\frac{5}{3}} C_{00,20}^{40,0}-\frac{7
C_{00,22}^{42,0}}{128 \sqrt{3}}+\frac{\sqrt{5} C_{20,00}^{20,0}}{64}+\frac{1}{64} \sqrt{\frac{5}{3}} C_{20,20}^{20,0}+\frac{7 C_{20,22}^{22,0}}{128
\sqrt{3}}-\frac{C_{22,00}^{22,0}}{64}-\frac{C_{22,20}^{22,0}}{64 \sqrt{3}}-\frac{1}{32} \sqrt{\frac{7}{3}} C_{22,22}^{22,0}\)

\noindent\(C_{11,2212,2}^{11,0011,0,\text{Loc}}= \frac{1}{48} \sqrt{\frac{5}{3}} C_{00,20}^{40,0}+\frac{1}{96} \sqrt{\frac{5}{3}} C_{00,20}^{40,1}-\frac{7
C_{00,22}^{42,0}}{192 \sqrt{3}}-\frac{7 C_{00,22}^{42,1}}{384 \sqrt{3}}-\frac{1}{48} \sqrt{\frac{5}{3}} C_{20,20}^{20,0}-\frac{1}{96} \sqrt{\frac{5}{3}}
C_{20,20}^{20,1}+\frac{7 C_{20,22}^{22,0}}{192 \sqrt{3}}+\frac{7 C_{20,22}^{22,1}}{384 \sqrt{3}}+\frac{C_{22,20}^{22,0}}{48 \sqrt{3}}+\frac{C_{22,20}^{22,1}}{96
\sqrt{3}}-\frac{1}{48} \sqrt{\frac{7}{3}} C_{22,22}^{22,0}-\frac{1}{96} \sqrt{\frac{7}{3}} C_{22,22}^{22,1}\)

\noindent\(C_{11,2212,2}^{11,0011,0,\text{NonLoc}}= \frac{\sqrt{5} C_{00,00}^{40,0}}{192}+\frac{\sqrt{5} C_{00,00}^{40,1}}{96}+\frac{1}{192}
\sqrt{\frac{5}{3}} C_{00,20}^{40,0}+\frac{1}{96} \sqrt{\frac{5}{3}} C_{00,20}^{40,1}+\frac{7 C_{00,22}^{42,0}}{384 \sqrt{3}}+\frac{7 C_{00,22}^{42,1}}{192
\sqrt{3}}-\frac{\sqrt{5} C_{20,00}^{20,0}}{192}-\frac{\sqrt{5} C_{20,00}^{20,1}}{96}-\frac{1}{192} \sqrt{\frac{5}{3}} C_{20,20}^{20,0}-\frac{1}{96}
\sqrt{\frac{5}{3}} C_{20,20}^{20,1}-\frac{7 C_{20,22}^{22,0}}{384 \sqrt{3}}-\frac{7 C_{20,22}^{22,1}}{192 \sqrt{3}}+\frac{C_{22,00}^{22,0}}{192}+\frac{C_{22,00}^{22,1}}{96}+\frac{C_{22,20}^{22,0}}{192
\sqrt{3}}+\frac{C_{22,20}^{22,1}}{96 \sqrt{3}}+\frac{1}{96} \sqrt{\frac{7}{3}} C_{22,22}^{22,0}+\frac{1}{48} \sqrt{\frac{7}{3}} C_{22,22}^{22,1}\)

\noindent\(C_{11,2212,2}^{11,0011,1,\text{Loc}}= \frac{\sqrt{5} C_{00,20}^{40,1}}{96}-\frac{7 C_{00,22}^{42,1}}{384}-\frac{\sqrt{5}
C_{20,20}^{20,1}}{96}+\frac{7 C_{20,22}^{22,1}}{384}+\frac{C_{22,20}^{22,1}}{96}-\frac{\sqrt{7} C_{22,22}^{22,1}}{96}\)

\noindent\(C_{11,2212,2}^{11,0011,1,\text{NonLoc}}= \frac{1}{64} \sqrt{\frac{5}{3}} C_{00,00}^{40,0}+\frac{\sqrt{5} C_{00,20}^{40,0}}{192}+\frac{7
C_{00,22}^{42,0}}{384}-\frac{1}{64} \sqrt{\frac{5}{3}} C_{20,00}^{20,0}-\frac{\sqrt{5} C_{20,20}^{20,0}}{192}-\frac{7 C_{20,22}^{22,0}}{384}+\frac{C_{22,00}^{22,0}}{64
\sqrt{3}}+\frac{C_{22,20}^{22,0}}{192}+\frac{\sqrt{7} C_{22,22}^{22,0}}{96}\)

\noindent\(C_{11,2213,2}^{11,0011,0,\text{Loc}}= -\frac{1}{24} \sqrt{\frac{7}{3}} C_{00,20}^{40,0}-\frac{1}{48} \sqrt{\frac{7}{3}}
C_{00,20}^{40,1}-\frac{1}{48} \sqrt{\frac{7}{15}} C_{00,22}^{42,0}-\frac{1}{96} \sqrt{\frac{7}{15}} C_{00,22}^{42,1}+\frac{1}{24} \sqrt{\frac{7}{3}}
C_{20,20}^{20,0}+\frac{1}{48} \sqrt{\frac{7}{3}} C_{20,20}^{20,1}+\frac{1}{48} \sqrt{\frac{7}{15}} C_{20,22}^{22,0}+\frac{1}{96} \sqrt{\frac{7}{15}}
C_{20,22}^{22,1}-\frac{1}{24} \sqrt{\frac{7}{15}} C_{22,20}^{22,0}-\frac{1}{48} \sqrt{\frac{7}{15}} C_{22,20}^{22,1}-\frac{C_{22,22}^{22,0}}{12 \sqrt{15}}-\frac{C_{22,22}^{22,1}}{24
\sqrt{15}}\)

\noindent\(C_{11,2213,2}^{11,0011,0,\text{NonLoc}}= -\frac{\sqrt{7} C_{00,00}^{40,0}}{96}-\frac{\sqrt{7} C_{00,00}^{40,1}}{48}-\frac{1}{96}
\sqrt{\frac{7}{3}} C_{00,20}^{40,0}-\frac{1}{48} \sqrt{\frac{7}{3}} C_{00,20}^{40,1}+\frac{1}{96} \sqrt{\frac{7}{15}} C_{00,22}^{42,0}+\frac{1}{48}
\sqrt{\frac{7}{15}} C_{00,22}^{42,1}+\frac{\sqrt{7} C_{20,00}^{20,0}}{96}+\frac{\sqrt{7} C_{20,00}^{20,1}}{48}+\frac{1}{96} \sqrt{\frac{7}{3}} C_{20,20}^{20,0}+\frac{1}{48}
\sqrt{\frac{7}{3}} C_{20,20}^{20,1}-\frac{1}{96} \sqrt{\frac{7}{15}} C_{20,22}^{22,0}-\frac{1}{48} \sqrt{\frac{7}{15}} C_{20,22}^{22,1}-\frac{1}{96}
\sqrt{\frac{7}{5}} C_{22,00}^{22,0}-\frac{1}{48} \sqrt{\frac{7}{5}} C_{22,00}^{22,1}-\frac{1}{96} \sqrt{\frac{7}{15}} C_{22,20}^{22,0}-\frac{1}{48}
\sqrt{\frac{7}{15}} C_{22,20}^{22,1}+\frac{C_{22,22}^{22,0}}{24 \sqrt{15}}+\frac{C_{22,22}^{22,1}}{12 \sqrt{15}}\)

\noindent\(C_{11,2213,2}^{11,0011,1,\text{Loc}}= -\frac{\sqrt{7} C_{00,20}^{40,1}}{48}-\frac{1}{96} \sqrt{\frac{7}{5}} C_{00,22}^{42,1}+\frac{\sqrt{7}
C_{20,20}^{20,1}}{48}+\frac{1}{96} \sqrt{\frac{7}{5}} C_{20,22}^{22,1}-\frac{1}{48} \sqrt{\frac{7}{5}} C_{22,20}^{22,1}-\frac{C_{22,22}^{22,1}}{24
\sqrt{5}}\)

\noindent\(C_{11,2213,2}^{11,0011,1,\text{NonLoc}}= -\frac{1}{32} \sqrt{\frac{7}{3}} C_{00,00}^{40,0}-\frac{\sqrt{7} C_{00,20}^{40,0}}{96}+\frac{1}{96}
\sqrt{\frac{7}{5}} C_{00,22}^{42,0}+\frac{1}{32} \sqrt{\frac{7}{3}} C_{20,00}^{20,0}+\frac{\sqrt{7} C_{20,20}^{20,0}}{96}-\frac{1}{96} \sqrt{\frac{7}{5}}
C_{20,22}^{22,0}-\frac{1}{32} \sqrt{\frac{7}{15}} C_{22,00}^{22,0}-\frac{1}{96} \sqrt{\frac{7}{5}} C_{22,20}^{22,0}+\frac{C_{22,22}^{22,0}}{24 \sqrt{5}}\)

\noindent\(C_{11,3101,1}^{00,0011,0,\text{Loc}}= -\frac{C_{11,11}^{31,0}}{16}-\frac{C_{11,11}^{31,1}}{32}-\frac{1}{8} \sqrt{\frac{3}{5}}
C_{22,11}^{22,0}-\frac{1}{16} \sqrt{\frac{3}{5}} C_{22,11}^{22,1}\)

\noindent\(C_{11,3101,1}^{00,0011,0,\text{NonLoc}}= -\frac{C_{11,11}^{31,0}}{32}-\frac{C_{11,11}^{31,1}}{16}+\frac{1}{16} \sqrt{\frac{3}{5}}
C_{22,11}^{22,0}+\frac{1}{8} \sqrt{\frac{3}{5}} C_{22,11}^{22,1}\)

\noindent\(C_{11,3101,1}^{00,0011,1,\text{Loc}}= -\frac{\sqrt{3} C_{11,11}^{31,1}}{32}-\frac{3 C_{22,11}^{22,1}}{16 \sqrt{5}}\)

\noindent\(C_{11,3101,1}^{00,0011,1,\text{NonLoc}}= -\frac{\sqrt{3} C_{11,11}^{31,0}}{32}+\frac{3 C_{22,11}^{22,0}}{16 \sqrt{5}}\)

\noindent\(C_{11,3111,0}^{00,0000,0,\text{Loc}}= -\frac{C_{11,11}^{31,0}}{16}-\frac{C_{11,11}^{31,1}}{32}-\frac{1}{8} \sqrt{\frac{3}{5}}
C_{22,11}^{22,0}-\frac{1}{16} \sqrt{\frac{3}{5}} C_{22,11}^{22,1}\)

\noindent\(C_{11,3111,0}^{00,0000,0,\text{NonLoc}}= -\frac{C_{11,11}^{31,0}}{32}-\frac{C_{11,11}^{31,1}}{16}+\frac{1}{16} \sqrt{\frac{3}{5}}
C_{22,11}^{22,0}+\frac{1}{8} \sqrt{\frac{3}{5}} C_{22,11}^{22,1}\)

\noindent\(C_{11,3111,0}^{00,0000,1,\text{Loc}}= -\frac{\sqrt{3} C_{11,11}^{31,1}}{32}-\frac{3 C_{22,11}^{22,1}}{16 \sqrt{5}}\)

\noindent\(C_{11,3111,0}^{00,0000,1,\text{NonLoc}}= -\frac{\sqrt{3} C_{11,11}^{31,0}}{32}+\frac{3 C_{22,11}^{22,0}}{16 \sqrt{5}}\)

\noindent\(C_{20,0000,0}^{00,2000,0,\text{Loc}}= -\frac{5 C_{00,00}^{40,0}}{96}-\frac{5 C_{00,00}^{40,1}}{192}-\frac{C_{20,00}^{20,0}}{96}-\frac{C_{20,00}^{20,1}}{192}+\frac{\sqrt{5}
C_{22,00}^{22,0}}{48}+\frac{\sqrt{5} C_{22,00}^{22,1}}{96}\)

\noindent\(C_{20,0000,0}^{00,2000,0,\text{NonLoc}}= \frac{5 C_{00,00}^{40,0}}{384}+\frac{5 C_{00,00}^{40,1}}{192}-\frac{5 C_{00,20}^{40,0}}{128
\sqrt{3}}-\frac{5 C_{00,20}^{40,1}}{64 \sqrt{3}}+\frac{C_{20,00}^{20,0}}{384}+\frac{C_{20,00}^{20,1}}{192}-\frac{C_{20,20}^{20,0}}{128 \sqrt{3}}-\frac{C_{20,20}^{20,1}}{64
\sqrt{3}}-\frac{\sqrt{5} C_{22,00}^{22,0}}{192}-\frac{\sqrt{5} C_{22,00}^{22,1}}{96}+\frac{1}{64} \sqrt{\frac{5}{3}} C_{22,20}^{22,0}+\frac{1}{32}
\sqrt{\frac{5}{3}} C_{22,20}^{22,1}\)

\noindent\(C_{20,0000,0}^{00,2000,1,\text{Loc}}= -\frac{5 C_{00,00}^{40,1}}{64 \sqrt{3}}-\frac{C_{20,00}^{20,1}}{64 \sqrt{3}}+\frac{1}{32}
\sqrt{\frac{5}{3}} C_{22,00}^{22,1}\)

\noindent\(C_{20,0000,0}^{00,2000,1,\text{NonLoc}}= \frac{5 C_{00,00}^{40,0}}{128 \sqrt{3}}-\frac{5 C_{00,20}^{40,0}}{128}+\frac{C_{20,00}^{20,0}}{128
\sqrt{3}}-\frac{C_{20,20}^{20,0}}{128}-\frac{1}{64} \sqrt{\frac{5}{3}} C_{22,00}^{22,0}+\frac{\sqrt{5} C_{22,20}^{22,0}}{64}\)

\noindent\(C_{20,0000,0}^{11,1111,0,\text{Loc}}= \frac{5 C_{11,11}^{31,0}}{192}+\frac{5 C_{11,11}^{31,1}}{384}-\frac{1}{32} \sqrt{\frac{5}{3}}
C_{22,11}^{22,0}-\frac{1}{64} \sqrt{\frac{5}{3}} C_{22,11}^{22,1}\)

\noindent\(C_{20,0000,0}^{11,1111,0,\text{NonLoc}}= \frac{5 C_{11,11}^{31,0}}{384}+\frac{5 C_{11,11}^{31,1}}{192}+\frac{1}{64} \sqrt{\frac{5}{3}}
C_{22,11}^{22,0}+\frac{1}{32} \sqrt{\frac{5}{3}} C_{22,11}^{22,1}\)

\noindent\(C_{20,0000,0}^{11,1111,1,\text{Loc}}= \frac{5 C_{11,11}^{31,1}}{128 \sqrt{3}}-\frac{\sqrt{5} C_{22,11}^{22,1}}{64}\)

\noindent\(C_{20,0000,0}^{11,1111,1,\text{NonLoc}}= \frac{5 C_{11,11}^{31,0}}{128 \sqrt{3}}+\frac{\sqrt{5} C_{22,11}^{22,0}}{64}\)

\noindent\(C_{20,0000,0}^{20,0000,0,\text{Loc}}= \frac{5 C_{00,00}^{40,0}}{768}+\frac{5 C_{00,00}^{40,1}}{1536}-\frac{5 C_{11,00}^{31,0}}{768}-\frac{5
C_{11,00}^{31,1}}{1536}+\frac{5 C_{20,00}^{20,0}}{768}+\frac{5 C_{20,00}^{20,1}}{1536}+\frac{\sqrt{5} C_{22,00}^{22,0}}{384}+\frac{\sqrt{5} C_{22,00}^{22,1}}{768}\)

\noindent\(C_{20,0000,0}^{20,0000,0,\text{NonLoc}}= -\frac{5 C_{00,00}^{40,0}}{3072}-\frac{5 C_{00,00}^{40,1}}{1536}+\frac{5 C_{00,20}^{40,0}}{1024
\sqrt{3}}+\frac{5 C_{00,20}^{40,1}}{512 \sqrt{3}}-\frac{5 C_{11,00}^{31,0}}{3072}-\frac{5 C_{11,00}^{31,1}}{1536}+\frac{5 C_{11,20}^{31,0}}{1024
\sqrt{3}}+\frac{5 C_{11,20}^{31,1}}{512 \sqrt{3}}-\frac{5 C_{20,00}^{20,0}}{3072}-\frac{5 C_{20,00}^{20,1}}{1536}+\frac{5 C_{20,20}^{20,0}}{1024
\sqrt{3}}+\frac{5 C_{20,20}^{20,1}}{512 \sqrt{3}}-\frac{\sqrt{5} C_{22,00}^{22,0}}{1536}-\frac{\sqrt{5} C_{22,00}^{22,1}}{768}+\frac{1}{512} \sqrt{\frac{5}{3}}
C_{22,20}^{22,0}+\frac{1}{256} \sqrt{\frac{5}{3}} C_{22,20}^{22,1}\)

\noindent\(C_{20,0000,0}^{20,0000,1,\text{Loc}}= \frac{5 C_{00,00}^{40,1}}{512 \sqrt{3}}-\frac{5 C_{11,00}^{31,1}}{512 \sqrt{3}}+\frac{5
C_{20,00}^{20,1}}{512 \sqrt{3}}+\frac{1}{256} \sqrt{\frac{5}{3}} C_{22,00}^{22,1}\)

\noindent\(C_{20,0000,0}^{20,0000,1,\text{NonLoc}}= -\frac{5 C_{00,00}^{40,0}}{1024 \sqrt{3}}+\frac{5 C_{00,20}^{40,0}}{1024}-\frac{5
C_{11,00}^{31,0}}{1024 \sqrt{3}}+\frac{5 C_{11,20}^{31,0}}{1024}-\frac{5 C_{20,00}^{20,0}}{1024 \sqrt{3}}+\frac{5 C_{20,20}^{20,0}}{1024}-\frac{1}{512}
\sqrt{\frac{5}{3}} C_{22,00}^{22,0}+\frac{\sqrt{5} C_{22,20}^{22,0}}{512}\)

\noindent\(C_{20,0011,1}^{00,2011,0,\text{Loc}}= \frac{5 C_{00,20}^{40,0}}{96}+\frac{5 C_{00,20}^{40,1}}{192}+\frac{C_{20,20}^{20,0}}{96}+\frac{C_{20,20}^{20,1}}{192}-\frac{\sqrt{5}
C_{22,20}^{22,0}}{48}-\frac{\sqrt{5} C_{22,20}^{22,1}}{96}\)

\noindent\(C_{20,0011,1}^{00,2011,0,\text{NonLoc}}= \frac{5 C_{00,00}^{40,0}}{128 \sqrt{3}}+\frac{5 C_{00,00}^{40,1}}{64 \sqrt{3}}+\frac{5
C_{00,20}^{40,0}}{384}+\frac{5 C_{00,20}^{40,1}}{192}+\frac{C_{20,00}^{20,0}}{128 \sqrt{3}}+\frac{C_{20,00}^{20,1}}{64 \sqrt{3}}+\frac{C_{20,20}^{20,0}}{384}+\frac{C_{20,20}^{20,1}}{192}-\frac{1}{64}
\sqrt{\frac{5}{3}} C_{22,00}^{22,0}-\frac{1}{32} \sqrt{\frac{5}{3}} C_{22,00}^{22,1}-\frac{\sqrt{5} C_{22,20}^{22,0}}{192}-\frac{\sqrt{5} C_{22,20}^{22,1}}{96}\)

\noindent\(C_{20,0011,1}^{00,2011,1,\text{Loc}}= \frac{5 C_{00,20}^{40,1}}{64 \sqrt{3}}+\frac{C_{20,20}^{20,1}}{64 \sqrt{3}}-\frac{1}{32}
\sqrt{\frac{5}{3}} C_{22,20}^{22,1}\)

\noindent\(C_{20,0011,1}^{00,2011,1,\text{NonLoc}}= \frac{5 C_{00,00}^{40,0}}{128}+\frac{5 C_{00,20}^{40,0}}{128 \sqrt{3}}+\frac{C_{20,00}^{20,0}}{128}+\frac{C_{20,20}^{20,0}}{128
\sqrt{3}}-\frac{\sqrt{5} C_{22,00}^{22,0}}{64}-\frac{1}{64} \sqrt{\frac{5}{3}} C_{22,20}^{22,0}\)

\noindent\(C_{20,0011,1}^{00,2211,0,\text{Loc}}= \frac{7 C_{00,22}^{42,0}}{192}+\frac{7 C_{00,22}^{42,1}}{384}+\frac{C_{11,22}^{31,0}}{64}+\frac{C_{11,22}^{31,1}}{128}-\frac{1}{64}
\sqrt{\frac{21}{5}} C_{11,22}^{33,0}-\frac{1}{128} \sqrt{\frac{21}{5}} C_{11,22}^{33,1}-\frac{C_{20,22}^{22,0}}{192}-\frac{C_{20,22}^{22,1}}{384}-\frac{\sqrt{7}
C_{22,22}^{22,0}}{96}-\frac{\sqrt{7} C_{22,22}^{22,1}}{192}\)

\noindent\(C_{20,0011,1}^{00,2211,0,\text{NonLoc}}= -\frac{7 C_{00,22}^{42,0}}{384}-\frac{7 C_{00,22}^{42,1}}{192}+\frac{C_{11,22}^{31,0}}{128}+\frac{C_{11,22}^{31,1}}{64}-\frac{1}{128}
\sqrt{\frac{21}{5}} C_{11,22}^{33,0}-\frac{1}{64} \sqrt{\frac{21}{5}} C_{11,22}^{33,1}+\frac{C_{20,22}^{22,0}}{384}+\frac{C_{20,22}^{22,1}}{192}+\frac{\sqrt{7}
C_{22,22}^{22,0}}{192}+\frac{\sqrt{7} C_{22,22}^{22,1}}{96}\)

\noindent\(C_{20,0011,1}^{00,2211,1,\text{Loc}}= \frac{7 C_{00,22}^{42,1}}{128 \sqrt{3}}+\frac{\sqrt{3} C_{11,22}^{31,1}}{128}-\frac{3}{128}
\sqrt{\frac{7}{5}} C_{11,22}^{33,1}-\frac{C_{20,22}^{22,1}}{128 \sqrt{3}}-\frac{1}{64} \sqrt{\frac{7}{3}} C_{22,22}^{22,1}\)

\noindent\(C_{20,0011,1}^{00,2211,1,\text{NonLoc}}= -\frac{7 C_{00,22}^{42,0}}{128 \sqrt{3}}+\frac{\sqrt{3} C_{11,22}^{31,0}}{128}-\frac{3}{128}
\sqrt{\frac{7}{5}} C_{11,22}^{33,0}+\frac{C_{20,22}^{22,0}}{128 \sqrt{3}}+\frac{1}{64} \sqrt{\frac{7}{3}} C_{22,22}^{22,0}\)

\noindent\(C_{20,0011,1}^{11,1101,0,\text{Loc}}= \frac{5 C_{11,11}^{31,0}}{192}+\frac{5 C_{11,11}^{31,1}}{384}-\frac{1}{32} \sqrt{\frac{5}{3}}
C_{22,11}^{22,0}-\frac{1}{64} \sqrt{\frac{5}{3}} C_{22,11}^{22,1}\)

\noindent\(C_{20,0011,1}^{11,1101,0,\text{NonLoc}}= \frac{5 C_{11,11}^{31,0}}{384}+\frac{5 C_{11,11}^{31,1}}{192}+\frac{1}{64} \sqrt{\frac{5}{3}}
C_{22,11}^{22,0}+\frac{1}{32} \sqrt{\frac{5}{3}} C_{22,11}^{22,1}\)

\noindent\(C_{20,0011,1}^{11,1101,1,\text{Loc}}= \frac{5 C_{11,11}^{31,1}}{128 \sqrt{3}}-\frac{\sqrt{5} C_{22,11}^{22,1}}{64}\)

\noindent\(C_{20,0011,1}^{11,1101,1,\text{NonLoc}}= \frac{5 C_{11,11}^{31,0}}{128 \sqrt{3}}+\frac{\sqrt{5} C_{22,11}^{22,0}}{64}\)

\noindent\(C_{20,0011,1}^{20,0011,0,\text{Loc}}= -\frac{5 C_{00,20}^{40,0}}{768}-\frac{5 C_{00,20}^{40,1}}{1536}+\frac{5 C_{11,20}^{31,0}}{768}+\frac{5
C_{11,20}^{31,1}}{1536}-\frac{5 C_{20,20}^{20,0}}{768}-\frac{5 C_{20,20}^{20,1}}{1536}-\frac{\sqrt{5} C_{22,20}^{22,0}}{384}-\frac{\sqrt{5} C_{22,20}^{22,1}}{768}\)

\noindent\(C_{20,0011,1}^{20,0011,0,\text{NonLoc}}= -\frac{5 C_{00,00}^{40,0}}{1024 \sqrt{3}}-\frac{5 C_{00,00}^{40,1}}{512 \sqrt{3}}-\frac{5
C_{00,20}^{40,0}}{3072}-\frac{5 C_{00,20}^{40,1}}{1536}-\frac{5 C_{11,00}^{31,0}}{1024 \sqrt{3}}-\frac{5 C_{11,00}^{31,1}}{512 \sqrt{3}}-\frac{5
C_{11,20}^{31,0}}{3072}-\frac{5 C_{11,20}^{31,1}}{1536}-\frac{5 C_{20,00}^{20,0}}{1024 \sqrt{3}}-\frac{5 C_{20,00}^{20,1}}{512 \sqrt{3}}-\frac{5
C_{20,20}^{20,0}}{3072}-\frac{5 C_{20,20}^{20,1}}{1536}-\frac{1}{512} \sqrt{\frac{5}{3}} C_{22,00}^{22,0}-\frac{1}{256} \sqrt{\frac{5}{3}} C_{22,00}^{22,1}-\frac{\sqrt{5}
C_{22,20}^{22,0}}{1536}-\frac{\sqrt{5} C_{22,20}^{22,1}}{768}\)

\noindent\(C_{20,0011,1}^{20,0011,1,\text{Loc}}= -\frac{5 C_{00,20}^{40,1}}{512 \sqrt{3}}+\frac{5 C_{11,20}^{31,1}}{512 \sqrt{3}}-\frac{5
C_{20,20}^{20,1}}{512 \sqrt{3}}-\frac{1}{256} \sqrt{\frac{5}{3}} C_{22,20}^{22,1}\)

\noindent\(C_{20,0011,1}^{20,0011,1,\text{NonLoc}}= -\frac{5 C_{00,00}^{40,0}}{1024}-\frac{5 C_{00,20}^{40,0}}{1024 \sqrt{3}}-\frac{5
C_{11,00}^{31,0}}{1024}-\frac{5 C_{11,20}^{31,0}}{1024 \sqrt{3}}-\frac{5 C_{20,00}^{20,0}}{1024}-\frac{5 C_{20,20}^{20,0}}{1024 \sqrt{3}}-\frac{\sqrt{5}
C_{22,00}^{22,0}}{512}-\frac{1}{512} \sqrt{\frac{5}{3}} C_{22,20}^{22,0}\)

\noindent\(C_{20,1101,1}^{00,1101,0,\text{Loc}}= \frac{5 C_{00,00}^{40,0}}{48}+\frac{5 C_{00,00}^{40,1}}{96}+\frac{C_{20,00}^{20,0}}{48}+\frac{C_{20,00}^{20,1}}{96}-\frac{\sqrt{5}
C_{22,00}^{22,0}}{24}-\frac{\sqrt{5} C_{22,00}^{22,1}}{48}\)

\noindent\(C_{20,1101,1}^{00,1101,0,\text{NonLoc}}= -\frac{5 C_{00,00}^{40,0}}{192}-\frac{5 C_{00,00}^{40,1}}{96}+\frac{5 C_{00,20}^{40,0}}{64
\sqrt{3}}+\frac{5 C_{00,20}^{40,1}}{32 \sqrt{3}}-\frac{C_{20,00}^{20,0}}{192}-\frac{C_{20,00}^{20,1}}{96}+\frac{C_{20,20}^{20,0}}{64 \sqrt{3}}+\frac{C_{20,20}^{20,1}}{32
\sqrt{3}}+\frac{\sqrt{5} C_{22,00}^{22,0}}{96}+\frac{\sqrt{5} C_{22,00}^{22,1}}{48}-\frac{1}{32} \sqrt{\frac{5}{3}} C_{22,20}^{22,0}-\frac{1}{16}
\sqrt{\frac{5}{3}} C_{22,20}^{22,1}\)

\noindent\(C_{20,1101,1}^{00,1101,1,\text{Loc}}= \frac{5 C_{00,00}^{40,1}}{32 \sqrt{3}}+\frac{C_{20,00}^{20,1}}{32 \sqrt{3}}-\frac{1}{16}
\sqrt{\frac{5}{3}} C_{22,00}^{22,1}\)

\noindent\(C_{20,1101,1}^{00,1101,1,\text{NonLoc}}= -\frac{5 C_{00,00}^{40,0}}{64 \sqrt{3}}+\frac{5 C_{00,20}^{40,0}}{64}-\frac{C_{20,00}^{20,0}}{64
\sqrt{3}}+\frac{C_{20,20}^{20,0}}{64}+\frac{1}{32} \sqrt{\frac{5}{3}} C_{22,00}^{22,0}-\frac{\sqrt{5} C_{22,20}^{22,0}}{32}\)

\noindent\(C_{20,1101,1}^{11,0011,0,\text{Loc}}= \frac{5 C_{11,11}^{31,0}}{192}+\frac{5 C_{11,11}^{31,1}}{384}-\frac{1}{32} \sqrt{\frac{5}{3}}
C_{22,11}^{22,0}-\frac{1}{64} \sqrt{\frac{5}{3}} C_{22,11}^{22,1}\)

\noindent\(C_{20,1101,1}^{11,0011,0,\text{NonLoc}}= \frac{5 C_{11,11}^{31,0}}{384}+\frac{5 C_{11,11}^{31,1}}{192}+\frac{1}{64} \sqrt{\frac{5}{3}}
C_{22,11}^{22,0}+\frac{1}{32} \sqrt{\frac{5}{3}} C_{22,11}^{22,1}\)

\noindent\(C_{20,1101,1}^{11,0011,1,\text{Loc}}= \frac{5 C_{11,11}^{31,1}}{128 \sqrt{3}}-\frac{\sqrt{5} C_{22,11}^{22,1}}{64}\)

\noindent\(C_{20,1101,1}^{11,0011,1,\text{NonLoc}}= \frac{5 C_{11,11}^{31,0}}{128 \sqrt{3}}+\frac{\sqrt{5} C_{22,11}^{22,0}}{64}\)

\noindent\(C_{20,1110,0}^{00,1110,0,\text{Loc}}= -\frac{5 C_{00,20}^{40,0}}{144}-\frac{5 C_{00,20}^{40,1}}{288}-\frac{7 \sqrt{5} C_{00,22}^{42,0}}{288}-\frac{7
\sqrt{5} C_{00,22}^{42,1}}{576}-\frac{\sqrt{5} C_{11,22}^{31,0}}{96}-\frac{\sqrt{5} C_{11,22}^{31,1}}{192}+\frac{1}{32} \sqrt{\frac{7}{3}} C_{11,22}^{33,0}+\frac{1}{64}
\sqrt{\frac{7}{3}} C_{11,22}^{33,1}-\frac{C_{20,20}^{20,0}}{144}-\frac{C_{20,20}^{20,1}}{288}+\frac{\sqrt{5} C_{20,22}^{22,0}}{288}+\frac{\sqrt{5}
C_{20,22}^{22,1}}{576}+\frac{\sqrt{5} C_{22,20}^{22,0}}{72}+\frac{\sqrt{5} C_{22,20}^{22,1}}{144}+\frac{\sqrt{35} C_{22,22}^{22,0}}{144}+\frac{\sqrt{35}
C_{22,22}^{22,1}}{288}\)

\noindent\(C_{20,1110,0}^{00,1110,0,\text{NonLoc}}= -\frac{5 C_{00,00}^{40,0}}{192 \sqrt{3}}-\frac{5 C_{00,00}^{40,1}}{96 \sqrt{3}}-\frac{5
C_{00,20}^{40,0}}{576}-\frac{5 C_{00,20}^{40,1}}{288}+\frac{7 \sqrt{5} C_{00,22}^{42,0}}{576}+\frac{7 \sqrt{5} C_{00,22}^{42,1}}{288}-\frac{\sqrt{5}
C_{11,22}^{31,0}}{192}-\frac{\sqrt{5} C_{11,22}^{31,1}}{96}+\frac{1}{64} \sqrt{\frac{7}{3}} C_{11,22}^{33,0}+\frac{1}{32} \sqrt{\frac{7}{3}} C_{11,22}^{33,1}-\frac{C_{20,00}^{20,0}}{192
\sqrt{3}}-\frac{C_{20,00}^{20,1}}{96 \sqrt{3}}-\frac{C_{20,20}^{20,0}}{576}-\frac{C_{20,20}^{20,1}}{288}-\frac{\sqrt{5} C_{20,22}^{22,0}}{576}-\frac{\sqrt{5}
C_{20,22}^{22,1}}{288}+\frac{1}{96} \sqrt{\frac{5}{3}} C_{22,00}^{22,0}+\frac{1}{48} \sqrt{\frac{5}{3}} C_{22,00}^{22,1}+\frac{\sqrt{5} C_{22,20}^{22,0}}{288}+\frac{\sqrt{5}
C_{22,20}^{22,1}}{144}-\frac{\sqrt{35} C_{22,22}^{22,0}}{288}-\frac{\sqrt{35} C_{22,22}^{22,1}}{144}\)

\noindent\(C_{20,1110,0}^{00,1110,1,\text{Loc}}= -\frac{5 C_{00,20}^{40,1}}{96 \sqrt{3}}-\frac{7}{192} \sqrt{\frac{5}{3}} C_{00,22}^{42,1}-\frac{1}{64}
\sqrt{\frac{5}{3}} C_{11,22}^{31,1}+\frac{\sqrt{7} C_{11,22}^{33,1}}{64}-\frac{C_{20,20}^{20,1}}{96 \sqrt{3}}+\frac{1}{192} \sqrt{\frac{5}{3}} C_{20,22}^{22,1}+\frac{1}{48}
\sqrt{\frac{5}{3}} C_{22,20}^{22,1}+\frac{1}{96} \sqrt{\frac{35}{3}} C_{22,22}^{22,1}\)

\noindent\(C_{20,1110,0}^{00,1110,1,\text{NonLoc}}= -\frac{5 C_{00,00}^{40,0}}{192}-\frac{5 C_{00,20}^{40,0}}{192 \sqrt{3}}+\frac{7}{192}
\sqrt{\frac{5}{3}} C_{00,22}^{42,0}-\frac{1}{64} \sqrt{\frac{5}{3}} C_{11,22}^{31,0}+\frac{\sqrt{7} C_{11,22}^{33,0}}{64}-\frac{C_{20,00}^{20,0}}{192}-\frac{C_{20,20}^{20,0}}{192
\sqrt{3}}-\frac{1}{192} \sqrt{\frac{5}{3}} C_{20,22}^{22,0}+\frac{\sqrt{5} C_{22,00}^{22,0}}{96}+\frac{1}{96} \sqrt{\frac{5}{3}} C_{22,20}^{22,0}-\frac{1}{96}
\sqrt{\frac{35}{3}} C_{22,22}^{22,0}\)

\noindent\(C_{20,1111,1}^{00,1111,0,\text{Loc}}= -\frac{5 C_{00,20}^{40,0}}{48 \sqrt{3}}-\frac{5 C_{00,20}^{40,1}}{96 \sqrt{3}}+\frac{7}{192}
\sqrt{\frac{5}{3}} C_{00,22}^{42,0}+\frac{7}{384} \sqrt{\frac{5}{3}} C_{00,22}^{42,1}+\frac{1}{64} \sqrt{\frac{5}{3}} C_{11,22}^{31,0}+\frac{1}{128}
\sqrt{\frac{5}{3}} C_{11,22}^{31,1}-\frac{\sqrt{7} C_{11,22}^{33,0}}{64}-\frac{\sqrt{7} C_{11,22}^{33,1}}{128}-\frac{C_{20,20}^{20,0}}{48 \sqrt{3}}-\frac{C_{20,20}^{20,1}}{96
\sqrt{3}}-\frac{1}{192} \sqrt{\frac{5}{3}} C_{20,22}^{22,0}-\frac{1}{384} \sqrt{\frac{5}{3}} C_{20,22}^{22,1}+\frac{1}{24} \sqrt{\frac{5}{3}} C_{22,20}^{22,0}+\frac{1}{48}
\sqrt{\frac{5}{3}} C_{22,20}^{22,1}-\frac{1}{96} \sqrt{\frac{35}{3}} C_{22,22}^{22,0}-\frac{1}{192} \sqrt{\frac{35}{3}} C_{22,22}^{22,1}\)

\noindent\(C_{20,1111,1}^{00,1111,0,\text{NonLoc}}= -\frac{5 C_{00,00}^{40,0}}{192}-\frac{5 C_{00,00}^{40,1}}{96}-\frac{5 C_{00,20}^{40,0}}{192
\sqrt{3}}-\frac{5 C_{00,20}^{40,1}}{96 \sqrt{3}}-\frac{7}{384} \sqrt{\frac{5}{3}} C_{00,22}^{42,0}-\frac{7}{192} \sqrt{\frac{5}{3}} C_{00,22}^{42,1}+\frac{1}{128}
\sqrt{\frac{5}{3}} C_{11,22}^{31,0}+\frac{1}{64} \sqrt{\frac{5}{3}} C_{11,22}^{31,1}-\frac{\sqrt{7} C_{11,22}^{33,0}}{128}-\frac{\sqrt{7} C_{11,22}^{33,1}}{64}-\frac{C_{20,00}^{20,0}}{192}-\frac{C_{20,00}^{20,1}}{96}-\frac{C_{20,20}^{20,0}}{192
\sqrt{3}}-\frac{C_{20,20}^{20,1}}{96 \sqrt{3}}+\frac{1}{384} \sqrt{\frac{5}{3}} C_{20,22}^{22,0}+\frac{1}{192} \sqrt{\frac{5}{3}} C_{20,22}^{22,1}+\frac{\sqrt{5}
C_{22,00}^{22,0}}{96}+\frac{\sqrt{5} C_{22,00}^{22,1}}{48}+\frac{1}{96} \sqrt{\frac{5}{3}} C_{22,20}^{22,0}+\frac{1}{48} \sqrt{\frac{5}{3}} C_{22,20}^{22,1}+\frac{1}{192}
\sqrt{\frac{35}{3}} C_{22,22}^{22,0}+\frac{1}{96} \sqrt{\frac{35}{3}} C_{22,22}^{22,1}\)

\noindent\(C_{20,1111,1}^{00,1111,1,\text{Loc}}= -\frac{5 C_{00,20}^{40,1}}{96}+\frac{7 \sqrt{5} C_{00,22}^{42,1}}{384}+\frac{\sqrt{5}
C_{11,22}^{31,1}}{128}-\frac{\sqrt{21} C_{11,22}^{33,1}}{128}-\frac{C_{20,20}^{20,1}}{96}-\frac{\sqrt{5} C_{20,22}^{22,1}}{384}+\frac{\sqrt{5} C_{22,20}^{22,1}}{48}-\frac{\sqrt{35}
C_{22,22}^{22,1}}{192}\)

\noindent\(C_{20,1111,1}^{00,1111,1,\text{NonLoc}}= -\frac{5 C_{00,00}^{40,0}}{64 \sqrt{3}}-\frac{5 C_{00,20}^{40,0}}{192}-\frac{7
\sqrt{5} C_{00,22}^{42,0}}{384}+\frac{\sqrt{5} C_{11,22}^{31,0}}{128}-\frac{\sqrt{21} C_{11,22}^{33,0}}{128}-\frac{C_{20,00}^{20,0}}{64 \sqrt{3}}-\frac{C_{20,20}^{20,0}}{192}+\frac{\sqrt{5}
C_{20,22}^{22,0}}{384}+\frac{1}{32} \sqrt{\frac{5}{3}} C_{22,00}^{22,0}+\frac{\sqrt{5} C_{22,20}^{22,0}}{96}+\frac{\sqrt{35} C_{22,22}^{22,0}}{192}\)

\noindent\(C_{20,1111,1}^{11,0000,0,\text{Loc}}= -\frac{5 C_{11,11}^{31,0}}{192}-\frac{5 C_{11,11}^{31,1}}{384}+\frac{1}{32} \sqrt{\frac{5}{3}}
C_{22,11}^{22,0}+\frac{1}{64} \sqrt{\frac{5}{3}} C_{22,11}^{22,1}\)

\noindent\(C_{20,1111,1}^{11,0000,0,\text{NonLoc}}= -\frac{5 C_{11,11}^{31,0}}{384}-\frac{5 C_{11,11}^{31,1}}{192}-\frac{1}{64} \sqrt{\frac{5}{3}}
C_{22,11}^{22,0}-\frac{1}{32} \sqrt{\frac{5}{3}} C_{22,11}^{22,1}\)

\noindent\(C_{20,1111,1}^{11,0000,1,\text{Loc}}= -\frac{5 C_{11,11}^{31,1}}{128 \sqrt{3}}+\frac{\sqrt{5} C_{22,11}^{22,1}}{64}\)

\noindent\(C_{20,1111,1}^{11,0000,1,\text{NonLoc}}= -\frac{5 C_{11,11}^{31,0}}{128 \sqrt{3}}-\frac{\sqrt{5} C_{22,11}^{22,0}}{64}\)

\noindent\(C_{20,1112,2}^{00,1112,0,\text{Loc}}= -\frac{5 \sqrt{5} C_{00,20}^{40,0}}{144}-\frac{5 \sqrt{5} C_{00,20}^{40,1}}{288}-\frac{7
C_{00,22}^{42,0}}{576}-\frac{7 C_{00,22}^{42,1}}{1152}-\frac{C_{11,22}^{31,0}}{192}-\frac{C_{11,22}^{31,1}}{384}+\frac{1}{64} \sqrt{\frac{7}{15}}
C_{11,22}^{33,0}+\frac{1}{128} \sqrt{\frac{7}{15}} C_{11,22}^{33,1}-\frac{\sqrt{5} C_{20,20}^{20,0}}{144}-\frac{\sqrt{5} C_{20,20}^{20,1}}{288}+\frac{C_{20,22}^{22,0}}{576}+\frac{C_{20,22}^{22,1}}{1152}+\frac{5
C_{22,20}^{22,0}}{72}+\frac{5 C_{22,20}^{22,1}}{144}+\frac{\sqrt{7} C_{22,22}^{22,0}}{288}+\frac{\sqrt{7} C_{22,22}^{22,1}}{576}\)

\noindent\(C_{20,1112,2}^{00,1112,0,\text{NonLoc}}= -\frac{5}{192} \sqrt{\frac{5}{3}} C_{00,00}^{40,0}-\frac{5}{96} \sqrt{\frac{5}{3}}
C_{00,00}^{40,1}-\frac{5 \sqrt{5} C_{00,20}^{40,0}}{576}-\frac{5 \sqrt{5} C_{00,20}^{40,1}}{288}+\frac{7 C_{00,22}^{42,0}}{1152}+\frac{7 C_{00,22}^{42,1}}{576}-\frac{C_{11,22}^{31,0}}{384}-\frac{C_{11,22}^{31,1}}{192}+\frac{1}{128}
\sqrt{\frac{7}{15}} C_{11,22}^{33,0}+\frac{1}{64} \sqrt{\frac{7}{15}} C_{11,22}^{33,1}-\frac{1}{192} \sqrt{\frac{5}{3}} C_{20,00}^{20,0}-\frac{1}{96}
\sqrt{\frac{5}{3}} C_{20,00}^{20,1}-\frac{\sqrt{5} C_{20,20}^{20,0}}{576}-\frac{\sqrt{5} C_{20,20}^{20,1}}{288}-\frac{C_{20,22}^{22,0}}{1152}-\frac{C_{20,22}^{22,1}}{576}+\frac{5
C_{22,00}^{22,0}}{96 \sqrt{3}}+\frac{5 C_{22,00}^{22,1}}{48 \sqrt{3}}+\frac{5 C_{22,20}^{22,0}}{288}+\frac{5 C_{22,20}^{22,1}}{144}-\frac{\sqrt{7}
C_{22,22}^{22,0}}{576}-\frac{\sqrt{7} C_{22,22}^{22,1}}{288}\)

\noindent\(C_{20,1112,2}^{00,1112,1,\text{Loc}}= -\frac{5}{96} \sqrt{\frac{5}{3}} C_{00,20}^{40,1}-\frac{7 C_{00,22}^{42,1}}{384 \sqrt{3}}-\frac{C_{11,22}^{31,1}}{128
\sqrt{3}}+\frac{1}{128} \sqrt{\frac{7}{5}} C_{11,22}^{33,1}-\frac{1}{96} \sqrt{\frac{5}{3}} C_{20,20}^{20,1}+\frac{C_{20,22}^{22,1}}{384 \sqrt{3}}+\frac{5
C_{22,20}^{22,1}}{48 \sqrt{3}}+\frac{1}{192} \sqrt{\frac{7}{3}} C_{22,22}^{22,1}\)

\noindent\(C_{20,1112,2}^{00,1112,1,\text{NonLoc}}= -\frac{5 \sqrt{5} C_{00,00}^{40,0}}{192}-\frac{5}{192} \sqrt{\frac{5}{3}} C_{00,20}^{40,0}+\frac{7
C_{00,22}^{42,0}}{384 \sqrt{3}}-\frac{C_{11,22}^{31,0}}{128 \sqrt{3}}+\frac{1}{128} \sqrt{\frac{7}{5}} C_{11,22}^{33,0}-\frac{\sqrt{5} C_{20,00}^{20,0}}{192}-\frac{1}{192}
\sqrt{\frac{5}{3}} C_{20,20}^{20,0}-\frac{C_{20,22}^{22,0}}{384 \sqrt{3}}+\frac{5 C_{22,00}^{22,0}}{96}+\frac{5 C_{22,20}^{22,0}}{96 \sqrt{3}}-\frac{1}{192}
\sqrt{\frac{7}{3}} C_{22,22}^{22,0}\)

\noindent\(C_{20,2000,0}^{00,0000,0,\text{Loc}}= -\frac{5 C_{00,00}^{40,0}}{96}-\frac{5 C_{00,00}^{40,1}}{192}-\frac{C_{20,00}^{20,0}}{96}-\frac{C_{20,00}^{20,1}}{192}+\frac{\sqrt{5}
C_{22,00}^{22,0}}{48}+\frac{\sqrt{5} C_{22,00}^{22,1}}{96}\)

\noindent\(C_{20,2000,0}^{00,0000,0,\text{NonLoc}}= \frac{5 C_{00,00}^{40,0}}{384}+\frac{5 C_{00,00}^{40,1}}{192}-\frac{5 C_{00,20}^{40,0}}{128
\sqrt{3}}-\frac{5 C_{00,20}^{40,1}}{64 \sqrt{3}}+\frac{C_{20,00}^{20,0}}{384}+\frac{C_{20,00}^{20,1}}{192}-\frac{C_{20,20}^{20,0}}{128 \sqrt{3}}-\frac{C_{20,20}^{20,1}}{64
\sqrt{3}}-\frac{\sqrt{5} C_{22,00}^{22,0}}{192}-\frac{\sqrt{5} C_{22,00}^{22,1}}{96}+\frac{1}{64} \sqrt{\frac{5}{3}} C_{22,20}^{22,0}+\frac{1}{32}
\sqrt{\frac{5}{3}} C_{22,20}^{22,1}\)

\noindent\(C_{20,2000,0}^{00,0000,1,\text{Loc}}= -\frac{5 C_{00,00}^{40,1}}{64 \sqrt{3}}-\frac{C_{20,00}^{20,1}}{64 \sqrt{3}}+\frac{1}{32}
\sqrt{\frac{5}{3}} C_{22,00}^{22,1}\)

\noindent\(C_{20,2000,0}^{00,0000,1,\text{NonLoc}}= \frac{5 C_{00,00}^{40,0}}{128 \sqrt{3}}-\frac{5 C_{00,20}^{40,0}}{128}+\frac{C_{20,00}^{20,0}}{128
\sqrt{3}}-\frac{C_{20,20}^{20,0}}{128}-\frac{1}{64} \sqrt{\frac{5}{3}} C_{22,00}^{22,0}+\frac{\sqrt{5} C_{22,20}^{22,0}}{64}\)

\noindent\(C_{20,2011,1}^{00,0011,0,\text{Loc}}= \frac{5 C_{00,20}^{40,0}}{96}+\frac{5 C_{00,20}^{40,1}}{192}+\frac{C_{20,20}^{20,0}}{96}+\frac{C_{20,20}^{20,1}}{192}-\frac{\sqrt{5}
C_{22,20}^{22,0}}{48}-\frac{\sqrt{5} C_{22,20}^{22,1}}{96}\)

\noindent\(C_{20,2011,1}^{00,0011,0,\text{NonLoc}}= \frac{5 C_{00,00}^{40,0}}{128 \sqrt{3}}+\frac{5 C_{00,00}^{40,1}}{64 \sqrt{3}}+\frac{5
C_{00,20}^{40,0}}{384}+\frac{5 C_{00,20}^{40,1}}{192}+\frac{C_{20,00}^{20,0}}{128 \sqrt{3}}+\frac{C_{20,00}^{20,1}}{64 \sqrt{3}}+\frac{C_{20,20}^{20,0}}{384}+\frac{C_{20,20}^{20,1}}{192}-\frac{1}{64}
\sqrt{\frac{5}{3}} C_{22,00}^{22,0}-\frac{1}{32} \sqrt{\frac{5}{3}} C_{22,00}^{22,1}-\frac{\sqrt{5} C_{22,20}^{22,0}}{192}-\frac{\sqrt{5} C_{22,20}^{22,1}}{96}\)

\noindent\(C_{20,2011,1}^{00,0011,1,\text{Loc}}= \frac{5 C_{00,20}^{40,1}}{64 \sqrt{3}}+\frac{C_{20,20}^{20,1}}{64 \sqrt{3}}-\frac{1}{32}
\sqrt{\frac{5}{3}} C_{22,20}^{22,1}\)

\noindent\(C_{20,2011,1}^{00,0011,1,\text{NonLoc}}= \frac{5 C_{00,00}^{40,0}}{128}+\frac{5 C_{00,20}^{40,0}}{128 \sqrt{3}}+\frac{C_{20,00}^{20,0}}{128}+\frac{C_{20,20}^{20,0}}{128
\sqrt{3}}-\frac{\sqrt{5} C_{22,00}^{22,0}}{64}-\frac{1}{64} \sqrt{\frac{5}{3}} C_{22,20}^{22,0}\)

\noindent\(C_{20,2211,1}^{00,0011,0,\text{Loc}}= \frac{7 C_{00,22}^{42,0}}{192}+\frac{7 C_{00,22}^{42,1}}{384}+\frac{C_{11,22}^{31,0}}{64}+\frac{C_{11,22}^{31,1}}{128}-\frac{1}{64}
\sqrt{\frac{21}{5}} C_{11,22}^{33,0}-\frac{1}{128} \sqrt{\frac{21}{5}} C_{11,22}^{33,1}-\frac{C_{20,22}^{22,0}}{192}-\frac{C_{20,22}^{22,1}}{384}-\frac{\sqrt{7}
C_{22,22}^{22,0}}{96}-\frac{\sqrt{7} C_{22,22}^{22,1}}{192}\)

\noindent\(C_{20,2211,1}^{00,0011,0,\text{NonLoc}}= -\frac{7 C_{00,22}^{42,0}}{384}-\frac{7 C_{00,22}^{42,1}}{192}+\frac{C_{11,22}^{31,0}}{128}+\frac{C_{11,22}^{31,1}}{64}-\frac{1}{128}
\sqrt{\frac{21}{5}} C_{11,22}^{33,0}-\frac{1}{64} \sqrt{\frac{21}{5}} C_{11,22}^{33,1}+\frac{C_{20,22}^{22,0}}{384}+\frac{C_{20,22}^{22,1}}{192}+\frac{\sqrt{7}
C_{22,22}^{22,0}}{192}+\frac{\sqrt{7} C_{22,22}^{22,1}}{96}\)

\noindent\(C_{20,2211,1}^{00,0011,1,\text{Loc}}= \frac{7 C_{00,22}^{42,1}}{128 \sqrt{3}}+\frac{\sqrt{3} C_{11,22}^{31,1}}{128}-\frac{3}{128}
\sqrt{\frac{7}{5}} C_{11,22}^{33,1}-\frac{C_{20,22}^{22,1}}{128 \sqrt{3}}-\frac{1}{64} \sqrt{\frac{7}{3}} C_{22,22}^{22,1}\)

\noindent\(C_{20,2211,1}^{00,0011,1,\text{NonLoc}}= -\frac{7 C_{00,22}^{42,0}}{128 \sqrt{3}}+\frac{\sqrt{3} C_{11,22}^{31,0}}{128}-\frac{3}{128}
\sqrt{\frac{7}{5}} C_{11,22}^{33,0}+\frac{C_{20,22}^{22,0}}{128 \sqrt{3}}+\frac{1}{64} \sqrt{\frac{7}{3}} C_{22,22}^{22,0}\)

\noindent\(C_{22,0000,2}^{00,2202,0,\text{Loc}}= -\frac{\sqrt{5} C_{00,00}^{40,0}}{48}-\frac{\sqrt{5} C_{00,00}^{40,1}}{96}+\frac{\sqrt{5}
C_{20,00}^{20,0}}{48}+\frac{\sqrt{5} C_{20,00}^{20,1}}{96}-\frac{C_{22,00}^{22,0}}{48}-\frac{C_{22,00}^{22,1}}{96}\)

\noindent\(C_{22,0000,2}^{00,2202,0,\text{NonLoc}}= \frac{\sqrt{5} C_{00,00}^{40,0}}{192}+\frac{\sqrt{5} C_{00,00}^{40,1}}{96}-\frac{1}{64}
\sqrt{\frac{5}{3}} C_{00,20}^{40,0}-\frac{1}{32} \sqrt{\frac{5}{3}} C_{00,20}^{40,1}-\frac{\sqrt{5} C_{20,00}^{20,0}}{192}-\frac{\sqrt{5} C_{20,00}^{20,1}}{96}+\frac{1}{64}
\sqrt{\frac{5}{3}} C_{20,20}^{20,0}+\frac{1}{32} \sqrt{\frac{5}{3}} C_{20,20}^{20,1}+\frac{C_{22,00}^{22,0}}{192}+\frac{C_{22,00}^{22,1}}{96}-\frac{C_{22,20}^{22,0}}{64
\sqrt{3}}-\frac{C_{22,20}^{22,1}}{32 \sqrt{3}}\)

\noindent\(C_{22,0000,2}^{00,2202,1,\text{Loc}}= -\frac{1}{32} \sqrt{\frac{5}{3}} C_{00,00}^{40,1}+\frac{1}{32} \sqrt{\frac{5}{3}}
C_{20,00}^{20,1}-\frac{C_{22,00}^{22,1}}{32 \sqrt{3}}\)

\noindent\(C_{22,0000,2}^{00,2202,1,\text{NonLoc}}= \frac{1}{64} \sqrt{\frac{5}{3}} C_{00,00}^{40,0}-\frac{\sqrt{5} C_{00,20}^{40,0}}{64}-\frac{1}{64}
\sqrt{\frac{5}{3}} C_{20,00}^{20,0}+\frac{\sqrt{5} C_{20,20}^{20,0}}{64}+\frac{C_{22,00}^{22,0}}{64 \sqrt{3}}-\frac{C_{22,20}^{22,0}}{64}\)

\noindent\(C_{22,0000,2}^{11,1111,0,\text{Loc}}= \frac{\sqrt{5} C_{11,11}^{31,0}}{96}+\frac{\sqrt{5} C_{11,11}^{31,1}}{192}-\frac{C_{22,11}^{22,0}}{16
\sqrt{3}}-\frac{C_{22,11}^{22,1}}{32 \sqrt{3}}\)

\noindent\(C_{22,0000,2}^{11,1111,0,\text{NonLoc}}= \frac{\sqrt{5} C_{11,11}^{31,0}}{192}+\frac{\sqrt{5} C_{11,11}^{31,1}}{96}+\frac{C_{22,11}^{22,0}}{32
\sqrt{3}}+\frac{C_{22,11}^{22,1}}{16 \sqrt{3}}\)

\noindent\(C_{22,0000,2}^{11,1111,1,\text{Loc}}= \frac{1}{64} \sqrt{\frac{5}{3}} C_{11,11}^{31,1}-\frac{C_{22,11}^{22,1}}{32}\)

\noindent\(C_{22,0000,2}^{11,1111,1,\text{NonLoc}}= \frac{1}{64} \sqrt{\frac{5}{3}} C_{11,11}^{31,0}+\frac{C_{22,11}^{22,0}}{32}\)

\noindent\(C_{22,0000,2}^{22,0000,0,\text{Loc}}= \frac{\sqrt{5} C_{00,00}^{40,0}}{384}+\frac{\sqrt{5} C_{00,00}^{40,1}}{768}-\frac{\sqrt{5}
C_{11,00}^{31,0}}{384}-\frac{\sqrt{5} C_{11,00}^{31,1}}{768}+\frac{\sqrt{5} C_{20,00}^{20,0}}{384}+\frac{\sqrt{5} C_{20,00}^{20,1}}{768}+\frac{C_{22,00}^{22,0}}{192}+\frac{C_{22,00}^{22,1}}{384}\)

\noindent\(C_{22,0000,2}^{22,0000,0,\text{NonLoc}}= -\frac{\sqrt{5} C_{00,00}^{40,0}}{1536}-\frac{\sqrt{5} C_{00,00}^{40,1}}{768}+\frac{1}{512}
\sqrt{\frac{5}{3}} C_{00,20}^{40,0}+\frac{1}{256} \sqrt{\frac{5}{3}} C_{00,20}^{40,1}-\frac{\sqrt{5} C_{11,00}^{31,0}}{1536}-\frac{\sqrt{5} C_{11,00}^{31,1}}{768}+\frac{1}{512}
\sqrt{\frac{5}{3}} C_{11,20}^{31,0}+\frac{1}{256} \sqrt{\frac{5}{3}} C_{11,20}^{31,1}-\frac{\sqrt{5} C_{20,00}^{20,0}}{1536}-\frac{\sqrt{5} C_{20,00}^{20,1}}{768}+\frac{1}{512}
\sqrt{\frac{5}{3}} C_{20,20}^{20,0}+\frac{1}{256} \sqrt{\frac{5}{3}} C_{20,20}^{20,1}-\frac{C_{22,00}^{22,0}}{768}-\frac{C_{22,00}^{22,1}}{384}+\frac{C_{22,20}^{22,0}}{256
\sqrt{3}}+\frac{C_{22,20}^{22,1}}{128 \sqrt{3}}\)

\noindent\(C_{22,0000,2}^{22,0000,1,\text{Loc}}= \frac{1}{256} \sqrt{\frac{5}{3}} C_{00,00}^{40,1}-\frac{1}{256} \sqrt{\frac{5}{3}}
C_{11,00}^{31,1}+\frac{1}{256} \sqrt{\frac{5}{3}} C_{20,00}^{20,1}+\frac{C_{22,00}^{22,1}}{128 \sqrt{3}}\)

\noindent\(C_{22,0000,2}^{22,0000,1,\text{NonLoc}}= -\frac{1}{512} \sqrt{\frac{5}{3}} C_{00,00}^{40,0}+\frac{\sqrt{5} C_{00,20}^{40,0}}{512}-\frac{1}{512}
\sqrt{\frac{5}{3}} C_{11,00}^{31,0}+\frac{\sqrt{5} C_{11,20}^{31,0}}{512}-\frac{1}{512} \sqrt{\frac{5}{3}} C_{20,00}^{20,0}+\frac{\sqrt{5} C_{20,20}^{20,0}}{512}-\frac{C_{22,00}^{22,0}}{256
\sqrt{3}}+\frac{C_{22,20}^{22,0}}{256}\)

\noindent\(C_{22,0011,1}^{00,2011,0,\text{Loc}}= \frac{7 C_{00,22}^{42,0}}{192}+\frac{7 C_{00,22}^{42,1}}{384}-\frac{C_{11,22}^{31,0}}{64}-\frac{C_{11,22}^{31,1}}{128}+\frac{1}{64}
\sqrt{\frac{21}{5}} C_{11,22}^{33,0}+\frac{1}{128} \sqrt{\frac{21}{5}} C_{11,22}^{33,1}-\frac{C_{20,22}^{22,0}}{192}-\frac{C_{20,22}^{22,1}}{384}-\frac{\sqrt{7}
C_{22,22}^{22,0}}{96}-\frac{\sqrt{7} C_{22,22}^{22,1}}{192}\)

\noindent\(C_{22,0011,1}^{00,2011,0,\text{NonLoc}}= -\frac{7 C_{00,22}^{42,0}}{384}-\frac{7 C_{00,22}^{42,1}}{192}-\frac{C_{11,22}^{31,0}}{128}-\frac{C_{11,22}^{31,1}}{64}+\frac{1}{128}
\sqrt{\frac{21}{5}} C_{11,22}^{33,0}+\frac{1}{64} \sqrt{\frac{21}{5}} C_{11,22}^{33,1}+\frac{C_{20,22}^{22,0}}{384}+\frac{C_{20,22}^{22,1}}{192}+\frac{\sqrt{7}
C_{22,22}^{22,0}}{192}+\frac{\sqrt{7} C_{22,22}^{22,1}}{96}\)

\noindent\(C_{22,0011,1}^{00,2011,1,\text{Loc}}= \frac{7 C_{00,22}^{42,1}}{128 \sqrt{3}}-\frac{\sqrt{3} C_{11,22}^{31,1}}{128}+\frac{3}{128}
\sqrt{\frac{7}{5}} C_{11,22}^{33,1}-\frac{C_{20,22}^{22,1}}{128 \sqrt{3}}-\frac{1}{64} \sqrt{\frac{7}{3}} C_{22,22}^{22,1}\)

\noindent\(C_{22,0011,1}^{00,2011,1,\text{NonLoc}}= -\frac{7 C_{00,22}^{42,0}}{128 \sqrt{3}}-\frac{\sqrt{3} C_{11,22}^{31,0}}{128}+\frac{3}{128}
\sqrt{\frac{7}{5}} C_{11,22}^{33,0}+\frac{C_{20,22}^{22,0}}{128 \sqrt{3}}+\frac{1}{64} \sqrt{\frac{7}{3}} C_{22,22}^{22,0}\)

\noindent\(C_{22,0011,1}^{00,2211,0,\text{Loc}}= \frac{C_{00,20}^{40,0}}{48}+\frac{C_{00,20}^{40,1}}{96}+\frac{7 C_{00,22}^{42,0}}{192
\sqrt{5}}+\frac{7 C_{00,22}^{42,1}}{384 \sqrt{5}}-\frac{C_{20,20}^{20,0}}{48}-\frac{C_{20,20}^{20,1}}{96}-\frac{7 C_{20,22}^{22,0}}{192 \sqrt{5}}-\frac{7
C_{20,22}^{22,1}}{384 \sqrt{5}}+\frac{C_{22,20}^{22,0}}{48 \sqrt{5}}+\frac{C_{22,20}^{22,1}}{96 \sqrt{5}}+\frac{1}{48} \sqrt{\frac{7}{5}} C_{22,22}^{22,0}+\frac{1}{96}
\sqrt{\frac{7}{5}} C_{22,22}^{22,1}\)

\noindent\(C_{22,0011,1}^{00,2211,0,\text{NonLoc}}= \frac{C_{00,00}^{40,0}}{64 \sqrt{3}}+\frac{C_{00,00}^{40,1}}{32 \sqrt{3}}+\frac{C_{00,20}^{40,0}}{192}+\frac{C_{00,20}^{40,1}}{96}-\frac{7
C_{00,22}^{42,0}}{384 \sqrt{5}}-\frac{7 C_{00,22}^{42,1}}{192 \sqrt{5}}-\frac{C_{20,00}^{20,0}}{64 \sqrt{3}}-\frac{C_{20,00}^{20,1}}{32 \sqrt{3}}-\frac{C_{20,20}^{20,0}}{192}-\frac{C_{20,20}^{20,1}}{96}+\frac{7
C_{20,22}^{22,0}}{384 \sqrt{5}}+\frac{7 C_{20,22}^{22,1}}{192 \sqrt{5}}+\frac{C_{22,00}^{22,0}}{64 \sqrt{15}}+\frac{C_{22,00}^{22,1}}{32 \sqrt{15}}+\frac{C_{22,20}^{22,0}}{192
\sqrt{5}}+\frac{C_{22,20}^{22,1}}{96 \sqrt{5}}-\frac{1}{96} \sqrt{\frac{7}{5}} C_{22,22}^{22,0}-\frac{1}{48} \sqrt{\frac{7}{5}} C_{22,22}^{22,1}\)

\noindent\(C_{22,0011,1}^{00,2211,1,\text{Loc}}= \frac{C_{00,20}^{40,1}}{32 \sqrt{3}}+\frac{7 C_{00,22}^{42,1}}{128 \sqrt{15}}-\frac{C_{20,20}^{20,1}}{32
\sqrt{3}}-\frac{7 C_{20,22}^{22,1}}{128 \sqrt{15}}+\frac{C_{22,20}^{22,1}}{32 \sqrt{15}}+\frac{1}{32} \sqrt{\frac{7}{15}} C_{22,22}^{22,1}\)

\noindent\(C_{22,0011,1}^{00,2211,1,\text{NonLoc}}= \frac{C_{00,00}^{40,0}}{64}+\frac{C_{00,20}^{40,0}}{64 \sqrt{3}}-\frac{7 C_{00,22}^{42,0}}{128
\sqrt{15}}-\frac{C_{20,00}^{20,0}}{64}-\frac{C_{20,20}^{20,0}}{64 \sqrt{3}}+\frac{7 C_{20,22}^{22,0}}{128 \sqrt{15}}+\frac{C_{22,00}^{22,0}}{64 \sqrt{5}}+\frac{C_{22,20}^{22,0}}{64
\sqrt{15}}-\frac{1}{32} \sqrt{\frac{7}{15}} C_{22,22}^{22,0}\)

\noindent\(C_{22,0011,1}^{11,1101,0,\text{Loc}}= -\frac{\sqrt{5} C_{11,11}^{31,0}}{192}-\frac{\sqrt{5} C_{11,11}^{31,1}}{384}+\frac{C_{22,11}^{22,0}}{32
\sqrt{3}}+\frac{C_{22,11}^{22,1}}{64 \sqrt{3}}\)

\noindent\(C_{22,0011,1}^{11,1101,0,\text{NonLoc}}= -\frac{\sqrt{5} C_{11,11}^{31,0}}{384}-\frac{\sqrt{5} C_{11,11}^{31,1}}{192}-\frac{C_{22,11}^{22,0}}{64
\sqrt{3}}-\frac{C_{22,11}^{22,1}}{32 \sqrt{3}}\)

\noindent\(C_{22,0011,1}^{11,1101,1,\text{Loc}}= -\frac{1}{128} \sqrt{\frac{5}{3}} C_{11,11}^{31,1}+\frac{C_{22,11}^{22,1}}{64}\)

\noindent\(C_{22,0011,1}^{11,1101,1,\text{NonLoc}}= -\frac{1}{128} \sqrt{\frac{5}{3}} C_{11,11}^{31,0}-\frac{C_{22,11}^{22,0}}{64}\)

\noindent\(C_{22,0011,1}^{20,0011,0,\text{Loc}}= -\frac{7 C_{00,22}^{42,0}}{768}-\frac{7 C_{00,22}^{42,1}}{1536}+\frac{7 C_{11,22}^{31,0}}{768}+\frac{7
C_{11,22}^{31,1}}{1536}+\frac{1}{256} \sqrt{\frac{21}{5}} C_{11,22}^{33,0}+\frac{1}{512} \sqrt{\frac{21}{5}} C_{11,22}^{33,1}-\frac{7 C_{20,22}^{22,0}}{768}-\frac{7
C_{20,22}^{22,1}}{1536}-\frac{\sqrt{7} C_{22,22}^{22,0}}{384}-\frac{\sqrt{7} C_{22,22}^{22,1}}{768}\)

\noindent\(C_{22,0011,1}^{20,0011,0,\text{NonLoc}}= \frac{7 C_{00,22}^{42,0}}{1536}+\frac{7 C_{00,22}^{42,1}}{768}+\frac{7 C_{11,22}^{31,0}}{1536}+\frac{7
C_{11,22}^{31,1}}{768}+\frac{1}{512} \sqrt{\frac{21}{5}} C_{11,22}^{33,0}+\frac{1}{256} \sqrt{\frac{21}{5}} C_{11,22}^{33,1}+\frac{7 C_{20,22}^{22,0}}{1536}+\frac{7
C_{20,22}^{22,1}}{768}+\frac{\sqrt{7} C_{22,22}^{22,0}}{768}+\frac{\sqrt{7} C_{22,22}^{22,1}}{384}\)

\noindent\(C_{22,0011,1}^{20,0011,1,\text{Loc}}= -\frac{7 C_{00,22}^{42,1}}{512 \sqrt{3}}+\frac{7 C_{11,22}^{31,1}}{512 \sqrt{3}}+\frac{3}{512}
\sqrt{\frac{7}{5}} C_{11,22}^{33,1}-\frac{7 C_{20,22}^{22,1}}{512 \sqrt{3}}-\frac{1}{256} \sqrt{\frac{7}{3}} C_{22,22}^{22,1}\)

\noindent\(C_{22,0011,1}^{20,0011,1,\text{NonLoc}}= \frac{7 C_{00,22}^{42,0}}{512 \sqrt{3}}+\frac{7 C_{11,22}^{31,0}}{512 \sqrt{3}}+\frac{3}{512}
\sqrt{\frac{7}{5}} C_{11,22}^{33,0}+\frac{7 C_{20,22}^{22,0}}{512 \sqrt{3}}+\frac{1}{256} \sqrt{\frac{7}{3}} C_{22,22}^{22,0}\)

\noindent\(C_{22,0011,1}^{22,0011,0,\text{Loc}}= -\frac{C_{00,20}^{40,0}}{384}-\frac{C_{00,20}^{40,1}}{768}-\frac{7 C_{00,22}^{42,0}}{1536
\sqrt{5}}-\frac{7 C_{00,22}^{42,1}}{3072 \sqrt{5}}+\frac{C_{11,20}^{31,0}}{384}+\frac{C_{11,20}^{31,1}}{768}+\frac{7 C_{11,22}^{31,0}}{1536 \sqrt{5}}+\frac{7
C_{11,22}^{31,1}}{3072 \sqrt{5}}+\frac{\sqrt{21} C_{11,22}^{33,0}}{2560}+\frac{\sqrt{21} C_{11,22}^{33,1}}{5120}-\frac{C_{20,20}^{20,0}}{384}-\frac{C_{20,20}^{20,1}}{768}-\frac{7
C_{20,22}^{22,0}}{1536 \sqrt{5}}-\frac{7 C_{20,22}^{22,1}}{3072 \sqrt{5}}-\frac{C_{22,20}^{22,0}}{192 \sqrt{5}}-\frac{C_{22,20}^{22,1}}{384 \sqrt{5}}-\frac{1}{768}
\sqrt{\frac{7}{5}} C_{22,22}^{22,0}-\frac{\sqrt{\frac{7}{5}} C_{22,22}^{22,1}}{1536}\)

\noindent\(C_{22,0011,1}^{22,0011,0,\text{NonLoc}}= -\frac{C_{00,00}^{40,0}}{512 \sqrt{3}}-\frac{C_{00,00}^{40,1}}{256 \sqrt{3}}-\frac{C_{00,20}^{40,0}}{1536}-\frac{C_{00,20}^{40,1}}{768}+\frac{7
C_{00,22}^{42,0}}{3072 \sqrt{5}}+\frac{7 C_{00,22}^{42,1}}{1536 \sqrt{5}}-\frac{C_{11,00}^{31,0}}{512 \sqrt{3}}-\frac{C_{11,00}^{31,1}}{256 \sqrt{3}}-\frac{C_{11,20}^{31,0}}{1536}-\frac{C_{11,20}^{31,1}}{768}+\frac{7
C_{11,22}^{31,0}}{3072 \sqrt{5}}+\frac{7 C_{11,22}^{31,1}}{1536 \sqrt{5}}+\frac{\sqrt{21} C_{11,22}^{33,0}}{5120}+\frac{\sqrt{21} C_{11,22}^{33,1}}{2560}-\frac{C_{20,00}^{20,0}}{512
\sqrt{3}}-\frac{C_{20,00}^{20,1}}{256 \sqrt{3}}-\frac{C_{20,20}^{20,0}}{1536}-\frac{C_{20,20}^{20,1}}{768}+\frac{7 C_{20,22}^{22,0}}{3072 \sqrt{5}}+\frac{7
C_{20,22}^{22,1}}{1536 \sqrt{5}}-\frac{C_{22,00}^{22,0}}{256 \sqrt{15}}-\frac{C_{22,00}^{22,1}}{128 \sqrt{15}}-\frac{C_{22,20}^{22,0}}{768 \sqrt{5}}-\frac{C_{22,20}^{22,1}}{384
\sqrt{5}}+\frac{\sqrt{\frac{7}{5}} C_{22,22}^{22,0}}{1536}+\frac{1}{768} \sqrt{\frac{7}{5}} C_{22,22}^{22,1}\)

\noindent\(C_{22,0011,1}^{22,0011,1,\text{Loc}}= -\frac{C_{00,20}^{40,1}}{256 \sqrt{3}}-\frac{7 C_{00,22}^{42,1}}{1024 \sqrt{15}}+\frac{C_{11,20}^{31,1}}{256
\sqrt{3}}+\frac{7 C_{11,22}^{31,1}}{1024 \sqrt{15}}+\frac{3 \sqrt{7} C_{11,22}^{33,1}}{5120}-\frac{C_{20,20}^{20,1}}{256 \sqrt{3}}-\frac{7 C_{20,22}^{22,1}}{1024
\sqrt{15}}-\frac{C_{22,20}^{22,1}}{128 \sqrt{15}}-\frac{1}{512} \sqrt{\frac{7}{15}} C_{22,22}^{22,1}\)

\noindent\(C_{22,0011,1}^{22,0011,1,\text{NonLoc}}= -\frac{C_{00,00}^{40,0}}{512}-\frac{C_{00,20}^{40,0}}{512 \sqrt{3}}+\frac{7 C_{00,22}^{42,0}}{1024
\sqrt{15}}-\frac{C_{11,00}^{31,0}}{512}-\frac{C_{11,20}^{31,0}}{512 \sqrt{3}}+\frac{7 C_{11,22}^{31,0}}{1024 \sqrt{15}}+\frac{3 \sqrt{7} C_{11,22}^{33,0}}{5120}-\frac{C_{20,00}^{20,0}}{512}-\frac{C_{20,20}^{20,0}}{512
\sqrt{3}}+\frac{7 C_{20,22}^{22,0}}{1024 \sqrt{15}}-\frac{C_{22,00}^{22,0}}{256 \sqrt{5}}-\frac{C_{22,20}^{22,0}}{256 \sqrt{15}}+\frac{1}{512} \sqrt{\frac{7}{15}}
C_{22,22}^{22,0}\)

\noindent\(C_{22,0011,2}^{00,2212,0,\text{Loc}}= \frac{1}{48} \sqrt{\frac{5}{3}} C_{00,20}^{40,0}+\frac{1}{96} \sqrt{\frac{5}{3}} C_{00,20}^{40,1}-\frac{7
C_{00,22}^{42,0}}{192 \sqrt{3}}-\frac{7 C_{00,22}^{42,1}}{384 \sqrt{3}}-\frac{1}{48} \sqrt{\frac{5}{3}} C_{20,20}^{20,0}-\frac{1}{96} \sqrt{\frac{5}{3}}
C_{20,20}^{20,1}+\frac{7 C_{20,22}^{22,0}}{192 \sqrt{3}}+\frac{7 C_{20,22}^{22,1}}{384 \sqrt{3}}+\frac{C_{22,20}^{22,0}}{48 \sqrt{3}}+\frac{C_{22,20}^{22,1}}{96
\sqrt{3}}-\frac{1}{48} \sqrt{\frac{7}{3}} C_{22,22}^{22,0}-\frac{1}{96} \sqrt{\frac{7}{3}} C_{22,22}^{22,1}\)

\noindent\(C_{22,0011,2}^{00,2212,0,\text{NonLoc}}= \frac{\sqrt{5} C_{00,00}^{40,0}}{192}+\frac{\sqrt{5} C_{00,00}^{40,1}}{96}+\frac{1}{192}
\sqrt{\frac{5}{3}} C_{00,20}^{40,0}+\frac{1}{96} \sqrt{\frac{5}{3}} C_{00,20}^{40,1}+\frac{7 C_{00,22}^{42,0}}{384 \sqrt{3}}+\frac{7 C_{00,22}^{42,1}}{192
\sqrt{3}}-\frac{\sqrt{5} C_{20,00}^{20,0}}{192}-\frac{\sqrt{5} C_{20,00}^{20,1}}{96}-\frac{1}{192} \sqrt{\frac{5}{3}} C_{20,20}^{20,0}-\frac{1}{96}
\sqrt{\frac{5}{3}} C_{20,20}^{20,1}-\frac{7 C_{20,22}^{22,0}}{384 \sqrt{3}}-\frac{7 C_{20,22}^{22,1}}{192 \sqrt{3}}+\frac{C_{22,00}^{22,0}}{192}+\frac{C_{22,00}^{22,1}}{96}+\frac{C_{22,20}^{22,0}}{192
\sqrt{3}}+\frac{C_{22,20}^{22,1}}{96 \sqrt{3}}+\frac{1}{96} \sqrt{\frac{7}{3}} C_{22,22}^{22,0}+\frac{1}{48} \sqrt{\frac{7}{3}} C_{22,22}^{22,1}\)

\noindent\(C_{22,0011,2}^{00,2212,1,\text{Loc}}= \frac{\sqrt{5} C_{00,20}^{40,1}}{96}-\frac{7 C_{00,22}^{42,1}}{384}-\frac{\sqrt{5}
C_{20,20}^{20,1}}{96}+\frac{7 C_{20,22}^{22,1}}{384}+\frac{C_{22,20}^{22,1}}{96}-\frac{\sqrt{7} C_{22,22}^{22,1}}{96}\)

\noindent\(C_{22,0011,2}^{00,2212,1,\text{NonLoc}}= \frac{1}{64} \sqrt{\frac{5}{3}} C_{00,00}^{40,0}+\frac{\sqrt{5} C_{00,20}^{40,0}}{192}+\frac{7
C_{00,22}^{42,0}}{384}-\frac{1}{64} \sqrt{\frac{5}{3}} C_{20,00}^{20,0}-\frac{\sqrt{5} C_{20,20}^{20,0}}{192}-\frac{7 C_{20,22}^{22,0}}{384}+\frac{C_{22,00}^{22,0}}{64
\sqrt{3}}+\frac{C_{22,20}^{22,0}}{192}+\frac{\sqrt{7} C_{22,22}^{22,0}}{96}\)

\noindent\(C_{22,0011,2}^{11,1101,0,\text{Loc}}= -\frac{1}{64} \sqrt{\frac{5}{3}} C_{11,11}^{31,0}-\frac{1}{128} \sqrt{\frac{5}{3}}
C_{11,11}^{31,1}+\frac{C_{22,11}^{22,0}}{32}+\frac{C_{22,11}^{22,1}}{64}\)

\noindent\(C_{22,0011,2}^{11,1101,0,\text{NonLoc}}= -\frac{1}{128} \sqrt{\frac{5}{3}} C_{11,11}^{31,0}-\frac{1}{64} \sqrt{\frac{5}{3}}
C_{11,11}^{31,1}-\frac{C_{22,11}^{22,0}}{64}-\frac{C_{22,11}^{22,1}}{32}\)

\noindent\(C_{22,0011,2}^{11,1101,1,\text{Loc}}= -\frac{\sqrt{5} C_{11,11}^{31,1}}{128}+\frac{\sqrt{3} C_{22,11}^{22,1}}{64}\)

\noindent\(C_{22,0011,2}^{11,1101,1,\text{NonLoc}}= -\frac{\sqrt{5} C_{11,11}^{31,0}}{128}-\frac{\sqrt{3} C_{22,11}^{22,0}}{64}\)

\noindent\(C_{22,0011,2}^{22,0011,0,\text{Loc}}= -\frac{1}{384} \sqrt{\frac{5}{3}} C_{00,20}^{40,0}-\frac{1}{768} \sqrt{\frac{5}{3}}
C_{00,20}^{40,1}+\frac{7 C_{00,22}^{42,0}}{1536 \sqrt{3}}+\frac{7 C_{00,22}^{42,1}}{3072 \sqrt{3}}+\frac{1}{384} \sqrt{\frac{5}{3}} C_{11,20}^{31,0}+\frac{1}{768}
\sqrt{\frac{5}{3}} C_{11,20}^{31,1}-\frac{7 C_{11,22}^{31,0}}{1536 \sqrt{3}}-\frac{7 C_{11,22}^{31,1}}{3072 \sqrt{3}}-\frac{1}{512} \sqrt{\frac{7}{5}}
C_{11,22}^{33,0}-\frac{\sqrt{\frac{7}{5}} C_{11,22}^{33,1}}{1024}-\frac{1}{384} \sqrt{\frac{5}{3}} C_{20,20}^{20,0}-\frac{1}{768} \sqrt{\frac{5}{3}}
C_{20,20}^{20,1}+\frac{7 C_{20,22}^{22,0}}{1536 \sqrt{3}}+\frac{7 C_{20,22}^{22,1}}{3072 \sqrt{3}}-\frac{C_{22,20}^{22,0}}{192 \sqrt{3}}-\frac{C_{22,20}^{22,1}}{384
\sqrt{3}}+\frac{1}{768} \sqrt{\frac{7}{3}} C_{22,22}^{22,0}+\frac{\sqrt{\frac{7}{3}} C_{22,22}^{22,1}}{1536}\)

\noindent\(C_{22,0011,2}^{22,0011,0,\text{NonLoc}}= -\frac{\sqrt{5} C_{00,00}^{40,0}}{1536}-\frac{\sqrt{5} C_{00,00}^{40,1}}{768}-\frac{\sqrt{\frac{5}{3}}
C_{00,20}^{40,0}}{1536}-\frac{1}{768} \sqrt{\frac{5}{3}} C_{00,20}^{40,1}-\frac{7 C_{00,22}^{42,0}}{3072 \sqrt{3}}-\frac{7 C_{00,22}^{42,1}}{1536
\sqrt{3}}-\frac{\sqrt{5} C_{11,00}^{31,0}}{1536}-\frac{\sqrt{5} C_{11,00}^{31,1}}{768}-\frac{\sqrt{\frac{5}{3}} C_{11,20}^{31,0}}{1536}-\frac{1}{768}
\sqrt{\frac{5}{3}} C_{11,20}^{31,1}-\frac{7 C_{11,22}^{31,0}}{3072 \sqrt{3}}-\frac{7 C_{11,22}^{31,1}}{1536 \sqrt{3}}-\frac{\sqrt{\frac{7}{5}} C_{11,22}^{33,0}}{1024}-\frac{1}{512}
\sqrt{\frac{7}{5}} C_{11,22}^{33,1}-\frac{\sqrt{5} C_{20,00}^{20,0}}{1536}-\frac{\sqrt{5} C_{20,00}^{20,1}}{768}-\frac{\sqrt{\frac{5}{3}} C_{20,20}^{20,0}}{1536}-\frac{1}{768}
\sqrt{\frac{5}{3}} C_{20,20}^{20,1}-\frac{7 C_{20,22}^{22,0}}{3072 \sqrt{3}}-\frac{7 C_{20,22}^{22,1}}{1536 \sqrt{3}}-\frac{C_{22,00}^{22,0}}{768}-\frac{C_{22,00}^{22,1}}{384}-\frac{C_{22,20}^{22,0}}{768
\sqrt{3}}-\frac{C_{22,20}^{22,1}}{384 \sqrt{3}}-\frac{\sqrt{\frac{7}{3}} C_{22,22}^{22,0}}{1536}-\frac{1}{768} \sqrt{\frac{7}{3}} C_{22,22}^{22,1}\)

\noindent\(C_{22,0011,2}^{22,0011,1,\text{Loc}}= -\frac{\sqrt{5} C_{00,20}^{40,1}}{768}+\frac{7 C_{00,22}^{42,1}}{3072}+\frac{\sqrt{5}
C_{11,20}^{31,1}}{768}-\frac{7 C_{11,22}^{31,1}}{3072}-\frac{\sqrt{\frac{21}{5}} C_{11,22}^{33,1}}{1024}-\frac{\sqrt{5} C_{20,20}^{20,1}}{768}+\frac{7
C_{20,22}^{22,1}}{3072}-\frac{C_{22,20}^{22,1}}{384}+\frac{\sqrt{7} C_{22,22}^{22,1}}{1536}\)

\noindent\(C_{22,0011,2}^{22,0011,1,\text{NonLoc}}= -\frac{1}{512} \sqrt{\frac{5}{3}} C_{00,00}^{40,0}-\frac{\sqrt{5} C_{00,20}^{40,0}}{1536}-\frac{7
C_{00,22}^{42,0}}{3072}-\frac{1}{512} \sqrt{\frac{5}{3}} C_{11,00}^{31,0}-\frac{\sqrt{5} C_{11,20}^{31,0}}{1536}-\frac{7 C_{11,22}^{31,0}}{3072}-\frac{\sqrt{\frac{21}{5}}
C_{11,22}^{33,0}}{1024}-\frac{1}{512} \sqrt{\frac{5}{3}} C_{20,00}^{20,0}-\frac{\sqrt{5} C_{20,20}^{20,0}}{1536}-\frac{7 C_{20,22}^{22,0}}{3072}-\frac{C_{22,00}^{22,0}}{256
\sqrt{3}}-\frac{C_{22,20}^{22,0}}{768}-\frac{\sqrt{7} C_{22,22}^{22,0}}{1536}\)

\noindent\(C_{22,0011,3}^{00,2213,0,\text{Loc}}= \frac{1}{48} \sqrt{\frac{7}{3}} C_{00,20}^{40,0}+\frac{1}{96} \sqrt{\frac{7}{3}} C_{00,20}^{40,1}+\frac{1}{96}
\sqrt{\frac{7}{15}} C_{00,22}^{42,0}+\frac{1}{192} \sqrt{\frac{7}{15}} C_{00,22}^{42,1}-\frac{1}{48} \sqrt{\frac{7}{3}} C_{20,20}^{20,0}-\frac{1}{96}
\sqrt{\frac{7}{3}} C_{20,20}^{20,1}-\frac{1}{96} \sqrt{\frac{7}{15}} C_{20,22}^{22,0}-\frac{1}{192} \sqrt{\frac{7}{15}} C_{20,22}^{22,1}+\frac{1}{48}
\sqrt{\frac{7}{15}} C_{22,20}^{22,0}+\frac{1}{96} \sqrt{\frac{7}{15}} C_{22,20}^{22,1}+\frac{C_{22,22}^{22,0}}{24 \sqrt{15}}+\frac{C_{22,22}^{22,1}}{48
\sqrt{15}}\)

\noindent\(C_{22,0011,3}^{00,2213,0,\text{NonLoc}}= \frac{\sqrt{7} C_{00,00}^{40,0}}{192}+\frac{\sqrt{7} C_{00,00}^{40,1}}{96}+\frac{1}{192}
\sqrt{\frac{7}{3}} C_{00,20}^{40,0}+\frac{1}{96} \sqrt{\frac{7}{3}} C_{00,20}^{40,1}-\frac{1}{192} \sqrt{\frac{7}{15}} C_{00,22}^{42,0}-\frac{1}{96}
\sqrt{\frac{7}{15}} C_{00,22}^{42,1}-\frac{\sqrt{7} C_{20,00}^{20,0}}{192}-\frac{\sqrt{7} C_{20,00}^{20,1}}{96}-\frac{1}{192} \sqrt{\frac{7}{3}}
C_{20,20}^{20,0}-\frac{1}{96} \sqrt{\frac{7}{3}} C_{20,20}^{20,1}+\frac{1}{192} \sqrt{\frac{7}{15}} C_{20,22}^{22,0}+\frac{1}{96} \sqrt{\frac{7}{15}}
C_{20,22}^{22,1}+\frac{1}{192} \sqrt{\frac{7}{5}} C_{22,00}^{22,0}+\frac{1}{96} \sqrt{\frac{7}{5}} C_{22,00}^{22,1}+\frac{1}{192} \sqrt{\frac{7}{15}}
C_{22,20}^{22,0}+\frac{1}{96} \sqrt{\frac{7}{15}} C_{22,20}^{22,1}-\frac{C_{22,22}^{22,0}}{48 \sqrt{15}}-\frac{C_{22,22}^{22,1}}{24 \sqrt{15}}\)

\noindent\(C_{22,0011,3}^{00,2213,1,\text{Loc}}= \frac{\sqrt{7} C_{00,20}^{40,1}}{96}+\frac{1}{192} \sqrt{\frac{7}{5}} C_{00,22}^{42,1}-\frac{\sqrt{7}
C_{20,20}^{20,1}}{96}-\frac{1}{192} \sqrt{\frac{7}{5}} C_{20,22}^{22,1}+\frac{1}{96} \sqrt{\frac{7}{5}} C_{22,20}^{22,1}+\frac{C_{22,22}^{22,1}}{48
\sqrt{5}}\)

\noindent\(C_{22,0011,3}^{00,2213,1,\text{NonLoc}}= \frac{1}{64} \sqrt{\frac{7}{3}} C_{00,00}^{40,0}+\frac{\sqrt{7} C_{00,20}^{40,0}}{192}-\frac{1}{192}
\sqrt{\frac{7}{5}} C_{00,22}^{42,0}-\frac{1}{64} \sqrt{\frac{7}{3}} C_{20,00}^{20,0}-\frac{\sqrt{7} C_{20,20}^{20,0}}{192}+\frac{1}{192} \sqrt{\frac{7}{5}}
C_{20,22}^{22,0}+\frac{1}{64} \sqrt{\frac{7}{15}} C_{22,00}^{22,0}+\frac{1}{192} \sqrt{\frac{7}{5}} C_{22,20}^{22,0}-\frac{C_{22,22}^{22,0}}{48 \sqrt{5}}\)

\noindent\(C_{22,0011,3}^{22,0011,0,\text{Loc}}= -\frac{1}{384} \sqrt{\frac{7}{3}} C_{00,20}^{40,0}-\frac{1}{768} \sqrt{\frac{7}{3}}
C_{00,20}^{40,1}-\frac{1}{768} \sqrt{\frac{7}{15}} C_{00,22}^{42,0}-\frac{\sqrt{\frac{7}{15}} C_{00,22}^{42,1}}{1536}+\frac{1}{384} \sqrt{\frac{7}{3}}
C_{11,20}^{31,0}+\frac{1}{768} \sqrt{\frac{7}{3}} C_{11,20}^{31,1}+\frac{1}{768} \sqrt{\frac{7}{15}} C_{11,22}^{31,0}+\frac{\sqrt{\frac{7}{15}} C_{11,22}^{31,1}}{1536}+\frac{C_{11,22}^{33,0}}{1280}+\frac{C_{11,22}^{33,1}}{2560}-\frac{1}{384}
\sqrt{\frac{7}{3}} C_{20,20}^{20,0}-\frac{1}{768} \sqrt{\frac{7}{3}} C_{20,20}^{20,1}-\frac{1}{768} \sqrt{\frac{7}{15}} C_{20,22}^{22,0}-\frac{\sqrt{\frac{7}{15}}
C_{20,22}^{22,1}}{1536}-\frac{1}{192} \sqrt{\frac{7}{15}} C_{22,20}^{22,0}-\frac{1}{384} \sqrt{\frac{7}{15}} C_{22,20}^{22,1}-\frac{C_{22,22}^{22,0}}{384
\sqrt{15}}-\frac{C_{22,22}^{22,1}}{768 \sqrt{15}}\)

\noindent\(C_{22,0011,3}^{22,0011,0,\text{NonLoc}}= -\frac{\sqrt{7} C_{00,00}^{40,0}}{1536}-\frac{\sqrt{7} C_{00,00}^{40,1}}{768}-\frac{\sqrt{\frac{7}{3}}
C_{00,20}^{40,0}}{1536}-\frac{1}{768} \sqrt{\frac{7}{3}} C_{00,20}^{40,1}+\frac{\sqrt{\frac{7}{15}} C_{00,22}^{42,0}}{1536}+\frac{1}{768} \sqrt{\frac{7}{15}}
C_{00,22}^{42,1}-\frac{\sqrt{7} C_{11,00}^{31,0}}{1536}-\frac{\sqrt{7} C_{11,00}^{31,1}}{768}-\frac{\sqrt{\frac{7}{3}} C_{11,20}^{31,0}}{1536}-\frac{1}{768}
\sqrt{\frac{7}{3}} C_{11,20}^{31,1}+\frac{\sqrt{\frac{7}{15}} C_{11,22}^{31,0}}{1536}+\frac{1}{768} \sqrt{\frac{7}{15}} C_{11,22}^{31,1}+\frac{C_{11,22}^{33,0}}{2560}+\frac{C_{11,22}^{33,1}}{1280}-\frac{\sqrt{7}
C_{20,00}^{20,0}}{1536}-\frac{\sqrt{7} C_{20,00}^{20,1}}{768}-\frac{\sqrt{\frac{7}{3}} C_{20,20}^{20,0}}{1536}-\frac{1}{768} \sqrt{\frac{7}{3}} C_{20,20}^{20,1}+\frac{\sqrt{\frac{7}{15}}
C_{20,22}^{22,0}}{1536}+\frac{1}{768} \sqrt{\frac{7}{15}} C_{20,22}^{22,1}-\frac{1}{768} \sqrt{\frac{7}{5}} C_{22,00}^{22,0}-\frac{1}{384} \sqrt{\frac{7}{5}}
C_{22,00}^{22,1}-\frac{1}{768} \sqrt{\frac{7}{15}} C_{22,20}^{22,0}-\frac{1}{384} \sqrt{\frac{7}{15}} C_{22,20}^{22,1}+\frac{C_{22,22}^{22,0}}{768
\sqrt{15}}+\frac{C_{22,22}^{22,1}}{384 \sqrt{15}}\)

\noindent\(C_{22,0011,3}^{22,0011,1,\text{Loc}}= -\frac{\sqrt{7} C_{00,20}^{40,1}}{768}-\frac{\sqrt{\frac{7}{5}} C_{00,22}^{42,1}}{1536}+\frac{\sqrt{7}
C_{11,20}^{31,1}}{768}+\frac{\sqrt{\frac{7}{5}} C_{11,22}^{31,1}}{1536}+\frac{\sqrt{3} C_{11,22}^{33,1}}{2560}-\frac{\sqrt{7} C_{20,20}^{20,1}}{768}-\frac{\sqrt{\frac{7}{5}}
C_{20,22}^{22,1}}{1536}-\frac{1}{384} \sqrt{\frac{7}{5}} C_{22,20}^{22,1}-\frac{C_{22,22}^{22,1}}{768 \sqrt{5}}\)

\noindent\(C_{22,0011,3}^{22,0011,1,\text{NonLoc}}= -\frac{1}{512} \sqrt{\frac{7}{3}} C_{00,00}^{40,0}-\frac{\sqrt{7} C_{00,20}^{40,0}}{1536}+\frac{\sqrt{\frac{7}{5}}
C_{00,22}^{42,0}}{1536}-\frac{1}{512} \sqrt{\frac{7}{3}} C_{11,00}^{31,0}-\frac{\sqrt{7} C_{11,20}^{31,0}}{1536}+\frac{\sqrt{\frac{7}{5}} C_{11,22}^{31,0}}{1536}+\frac{\sqrt{3}
C_{11,22}^{33,0}}{2560}-\frac{1}{512} \sqrt{\frac{7}{3}} C_{20,00}^{20,0}-\frac{\sqrt{7} C_{20,20}^{20,0}}{1536}+\frac{\sqrt{\frac{7}{5}} C_{20,22}^{22,0}}{1536}-\frac{1}{256}
\sqrt{\frac{7}{15}} C_{22,00}^{22,0}-\frac{1}{768} \sqrt{\frac{7}{5}} C_{22,20}^{22,0}+\frac{C_{22,22}^{22,0}}{768 \sqrt{5}}\)

\noindent\(C_{22,1101,1}^{00,1101,0,\text{Loc}}= \frac{\sqrt{5} C_{00,00}^{40,0}}{24}+\frac{\sqrt{5} C_{00,00}^{40,1}}{48}-\frac{\sqrt{5}
C_{20,00}^{20,0}}{24}-\frac{\sqrt{5} C_{20,00}^{20,1}}{48}+\frac{C_{22,00}^{22,0}}{24}+\frac{C_{22,00}^{22,1}}{48}\)

\noindent\(C_{22,1101,1}^{00,1101,0,\text{NonLoc}}= -\frac{\sqrt{5} C_{00,00}^{40,0}}{96}-\frac{\sqrt{5} C_{00,00}^{40,1}}{48}+\frac{1}{32}
\sqrt{\frac{5}{3}} C_{00,20}^{40,0}+\frac{1}{16} \sqrt{\frac{5}{3}} C_{00,20}^{40,1}+\frac{\sqrt{5} C_{20,00}^{20,0}}{96}+\frac{\sqrt{5} C_{20,00}^{20,1}}{48}-\frac{1}{32}
\sqrt{\frac{5}{3}} C_{20,20}^{20,0}-\frac{1}{16} \sqrt{\frac{5}{3}} C_{20,20}^{20,1}-\frac{C_{22,00}^{22,0}}{96}-\frac{C_{22,00}^{22,1}}{48}+\frac{C_{22,20}^{22,0}}{32
\sqrt{3}}+\frac{C_{22,20}^{22,1}}{16 \sqrt{3}}\)

\noindent\(C_{22,1101,1}^{00,1101,1,\text{Loc}}= \frac{1}{16} \sqrt{\frac{5}{3}} C_{00,00}^{40,1}-\frac{1}{16} \sqrt{\frac{5}{3}} C_{20,00}^{20,1}+\frac{C_{22,00}^{22,1}}{16
\sqrt{3}}\)

\noindent\(C_{22,1101,1}^{00,1101,1,\text{NonLoc}}= -\frac{1}{32} \sqrt{\frac{5}{3}} C_{00,00}^{40,0}+\frac{\sqrt{5} C_{00,20}^{40,0}}{32}+\frac{1}{32}
\sqrt{\frac{5}{3}} C_{20,00}^{20,0}-\frac{\sqrt{5} C_{20,20}^{20,0}}{32}-\frac{C_{22,00}^{22,0}}{32 \sqrt{3}}+\frac{C_{22,20}^{22,0}}{32}\)

\noindent\(C_{22,1101,1}^{11,0011,0,\text{Loc}}= -\frac{\sqrt{5} C_{11,11}^{31,0}}{192}-\frac{\sqrt{5} C_{11,11}^{31,1}}{384}+\frac{C_{22,11}^{22,0}}{32
\sqrt{3}}+\frac{C_{22,11}^{22,1}}{64 \sqrt{3}}\)

\noindent\(C_{22,1101,1}^{11,0011,0,\text{NonLoc}}= -\frac{\sqrt{5} C_{11,11}^{31,0}}{384}-\frac{\sqrt{5} C_{11,11}^{31,1}}{192}-\frac{C_{22,11}^{22,0}}{64
\sqrt{3}}-\frac{C_{22,11}^{22,1}}{32 \sqrt{3}}\)

\noindent\(C_{22,1101,1}^{11,0011,1,\text{Loc}}= -\frac{1}{128} \sqrt{\frac{5}{3}} C_{11,11}^{31,1}+\frac{C_{22,11}^{22,1}}{64}\)

\noindent\(C_{22,1101,1}^{11,0011,1,\text{NonLoc}}= -\frac{1}{128} \sqrt{\frac{5}{3}} C_{11,11}^{31,0}-\frac{C_{22,11}^{22,0}}{64}\)

\noindent\(C_{22,1101,2}^{11,0011,0,\text{Loc}}= -\frac{1}{64} \sqrt{\frac{5}{3}} C_{11,11}^{31,0}-\frac{1}{128} \sqrt{\frac{5}{3}}
C_{11,11}^{31,1}+\frac{C_{22,11}^{22,0}}{32}+\frac{C_{22,11}^{22,1}}{64}\)

\noindent\(C_{22,1101,2}^{11,0011,0,\text{NonLoc}}= -\frac{1}{128} \sqrt{\frac{5}{3}} C_{11,11}^{31,0}-\frac{1}{64} \sqrt{\frac{5}{3}}
C_{11,11}^{31,1}-\frac{C_{22,11}^{22,0}}{64}-\frac{C_{22,11}^{22,1}}{32}\)

\noindent\(C_{22,1101,2}^{11,0011,1,\text{Loc}}= -\frac{\sqrt{5} C_{11,11}^{31,1}}{128}+\frac{\sqrt{3} C_{22,11}^{22,1}}{64}\)

\noindent\(C_{22,1101,2}^{11,0011,1,\text{NonLoc}}= -\frac{\sqrt{5} C_{11,11}^{31,0}}{128}-\frac{\sqrt{3} C_{22,11}^{22,0}}{64}\)

\noindent\(C_{22,1110,2}^{00,1112,0,\text{Loc}}= -\frac{\sqrt{5} C_{00,20}^{40,0}}{72}-\frac{\sqrt{5} C_{00,20}^{40,1}}{144}-\frac{7
C_{00,22}^{42,0}}{144}-\frac{7 C_{00,22}^{42,1}}{288}+\frac{C_{11,22}^{31,0}}{96}+\frac{C_{11,22}^{31,1}}{192}-\frac{1}{32} \sqrt{\frac{7}{15}} C_{11,22}^{33,0}-\frac{1}{64}
\sqrt{\frac{7}{15}} C_{11,22}^{33,1}+\frac{\sqrt{5} C_{20,20}^{20,0}}{72}+\frac{\sqrt{5} C_{20,20}^{20,1}}{144}+\frac{C_{20,22}^{22,0}}{36}+\frac{C_{20,22}^{22,1}}{72}-\frac{C_{22,20}^{22,0}}{72}-\frac{C_{22,20}^{22,1}}{144}-\frac{\sqrt{7}
C_{22,22}^{22,0}}{144}-\frac{\sqrt{7} C_{22,22}^{22,1}}{288}\)

\noindent\(C_{22,1110,2}^{00,1112,0,\text{NonLoc}}= -\frac{1}{96} \sqrt{\frac{5}{3}} C_{00,00}^{40,0}-\frac{1}{48} \sqrt{\frac{5}{3}}
C_{00,00}^{40,1}-\frac{\sqrt{5} C_{00,20}^{40,0}}{288}-\frac{\sqrt{5} C_{00,20}^{40,1}}{144}+\frac{7 C_{00,22}^{42,0}}{288}+\frac{7 C_{00,22}^{42,1}}{144}+\frac{C_{11,22}^{31,0}}{192}+\frac{C_{11,22}^{31,1}}{96}-\frac{1}{64}
\sqrt{\frac{7}{15}} C_{11,22}^{33,0}-\frac{1}{32} \sqrt{\frac{7}{15}} C_{11,22}^{33,1}+\frac{1}{96} \sqrt{\frac{5}{3}} C_{20,00}^{20,0}+\frac{1}{48}
\sqrt{\frac{5}{3}} C_{20,00}^{20,1}+\frac{\sqrt{5} C_{20,20}^{20,0}}{288}+\frac{\sqrt{5} C_{20,20}^{20,1}}{144}-\frac{C_{20,22}^{22,0}}{72}-\frac{C_{20,22}^{22,1}}{36}-\frac{C_{22,00}^{22,0}}{96
\sqrt{3}}-\frac{C_{22,00}^{22,1}}{48 \sqrt{3}}-\frac{C_{22,20}^{22,0}}{288}-\frac{C_{22,20}^{22,1}}{144}+\frac{\sqrt{7} C_{22,22}^{22,0}}{288}+\frac{\sqrt{7}
C_{22,22}^{22,1}}{144}\)

\noindent\(C_{22,1110,2}^{00,1112,1,\text{Loc}}= -\frac{1}{48} \sqrt{\frac{5}{3}} C_{00,20}^{40,1}-\frac{7 C_{00,22}^{42,1}}{96 \sqrt{3}}+\frac{C_{11,22}^{31,1}}{64
\sqrt{3}}-\frac{1}{64} \sqrt{\frac{7}{5}} C_{11,22}^{33,1}+\frac{1}{48} \sqrt{\frac{5}{3}} C_{20,20}^{20,1}+\frac{C_{20,22}^{22,1}}{24 \sqrt{3}}-\frac{C_{22,20}^{22,1}}{48
\sqrt{3}}-\frac{1}{96} \sqrt{\frac{7}{3}} C_{22,22}^{22,1}\)

\noindent\(C_{22,1110,2}^{00,1112,1,\text{NonLoc}}= -\frac{\sqrt{5} C_{00,00}^{40,0}}{96}-\frac{1}{96} \sqrt{\frac{5}{3}} C_{00,20}^{40,0}+\frac{7
C_{00,22}^{42,0}}{96 \sqrt{3}}+\frac{C_{11,22}^{31,0}}{64 \sqrt{3}}-\frac{1}{64} \sqrt{\frac{7}{5}} C_{11,22}^{33,0}+\frac{\sqrt{5} C_{20,00}^{20,0}}{96}+\frac{1}{96}
\sqrt{\frac{5}{3}} C_{20,20}^{20,0}-\frac{C_{20,22}^{22,0}}{24 \sqrt{3}}-\frac{C_{22,00}^{22,0}}{96}-\frac{C_{22,20}^{22,0}}{96 \sqrt{3}}+\frac{1}{96}
\sqrt{\frac{7}{3}} C_{22,22}^{22,0}\)

\noindent\(C_{22,1111,1}^{00,1111,0,\text{Loc}}= \frac{1}{48} \sqrt{\frac{5}{3}} C_{00,20}^{40,0}+\frac{1}{96} \sqrt{\frac{5}{3}} C_{00,20}^{40,1}-\frac{7
C_{00,22}^{42,0}}{192 \sqrt{3}}-\frac{7 C_{00,22}^{42,1}}{384 \sqrt{3}}-\frac{C_{11,22}^{31,0}}{64 \sqrt{3}}-\frac{C_{11,22}^{31,1}}{128 \sqrt{3}}+\frac{1}{64}
\sqrt{\frac{7}{5}} C_{11,22}^{33,0}+\frac{1}{128} \sqrt{\frac{7}{5}} C_{11,22}^{33,1}-\frac{1}{48} \sqrt{\frac{5}{3}} C_{20,20}^{20,0}-\frac{1}{96}
\sqrt{\frac{5}{3}} C_{20,20}^{20,1}+\frac{13 C_{20,22}^{22,0}}{192 \sqrt{3}}+\frac{13 C_{20,22}^{22,1}}{384 \sqrt{3}}+\frac{C_{22,20}^{22,0}}{48
\sqrt{3}}+\frac{C_{22,20}^{22,1}}{96 \sqrt{3}}-\frac{5}{96} \sqrt{\frac{7}{3}} C_{22,22}^{22,0}-\frac{5}{192} \sqrt{\frac{7}{3}} C_{22,22}^{22,1}\)

\noindent\(C_{22,1111,1}^{00,1111,0,\text{NonLoc}}= \frac{\sqrt{5} C_{00,00}^{40,0}}{192}+\frac{\sqrt{5} C_{00,00}^{40,1}}{96}+\frac{1}{192}
\sqrt{\frac{5}{3}} C_{00,20}^{40,0}+\frac{1}{96} \sqrt{\frac{5}{3}} C_{00,20}^{40,1}+\frac{7 C_{00,22}^{42,0}}{384 \sqrt{3}}+\frac{7 C_{00,22}^{42,1}}{192
\sqrt{3}}-\frac{C_{11,22}^{31,0}}{128 \sqrt{3}}-\frac{C_{11,22}^{31,1}}{64 \sqrt{3}}+\frac{1}{128} \sqrt{\frac{7}{5}} C_{11,22}^{33,0}+\frac{1}{64}
\sqrt{\frac{7}{5}} C_{11,22}^{33,1}-\frac{\sqrt{5} C_{20,00}^{20,0}}{192}-\frac{\sqrt{5} C_{20,00}^{20,1}}{96}-\frac{1}{192} \sqrt{\frac{5}{3}} C_{20,20}^{20,0}-\frac{1}{96}
\sqrt{\frac{5}{3}} C_{20,20}^{20,1}-\frac{13 C_{20,22}^{22,0}}{384 \sqrt{3}}-\frac{13 C_{20,22}^{22,1}}{192 \sqrt{3}}+\frac{C_{22,00}^{22,0}}{192}+\frac{C_{22,00}^{22,1}}{96}+\frac{C_{22,20}^{22,0}}{192
\sqrt{3}}+\frac{C_{22,20}^{22,1}}{96 \sqrt{3}}+\frac{5}{192} \sqrt{\frac{7}{3}} C_{22,22}^{22,0}+\frac{5}{96} \sqrt{\frac{7}{3}} C_{22,22}^{22,1}\)

\noindent\(C_{22,1111,1}^{00,1111,1,\text{Loc}}= \frac{\sqrt{5} C_{00,20}^{40,1}}{96}-\frac{7 C_{00,22}^{42,1}}{384}-\frac{C_{11,22}^{31,1}}{128}+\frac{1}{128}
\sqrt{\frac{21}{5}} C_{11,22}^{33,1}-\frac{\sqrt{5} C_{20,20}^{20,1}}{96}+\frac{13 C_{20,22}^{22,1}}{384}+\frac{C_{22,20}^{22,1}}{96}-\frac{5 \sqrt{7}
C_{22,22}^{22,1}}{192}\)

\noindent\(C_{22,1111,1}^{00,1111,1,\text{NonLoc}}= \frac{1}{64} \sqrt{\frac{5}{3}} C_{00,00}^{40,0}+\frac{\sqrt{5} C_{00,20}^{40,0}}{192}+\frac{7
C_{00,22}^{42,0}}{384}-\frac{C_{11,22}^{31,0}}{128}+\frac{1}{128} \sqrt{\frac{21}{5}} C_{11,22}^{33,0}-\frac{1}{64} \sqrt{\frac{5}{3}} C_{20,00}^{20,0}-\frac{\sqrt{5}
C_{20,20}^{20,0}}{192}-\frac{13 C_{20,22}^{22,0}}{384}+\frac{C_{22,00}^{22,0}}{64 \sqrt{3}}+\frac{C_{22,20}^{22,0}}{192}+\frac{5 \sqrt{7} C_{22,22}^{22,0}}{192}\)

\noindent\(C_{22,1111,1}^{11,0000,0,\text{Loc}}= -\frac{\sqrt{5} C_{11,11}^{31,0}}{96}-\frac{\sqrt{5} C_{11,11}^{31,1}}{192}+\frac{C_{22,11}^{22,0}}{16
\sqrt{3}}+\frac{C_{22,11}^{22,1}}{32 \sqrt{3}}\)

\noindent\(C_{22,1111,1}^{11,0000,0,\text{NonLoc}}= -\frac{\sqrt{5} C_{11,11}^{31,0}}{192}-\frac{\sqrt{5} C_{11,11}^{31,1}}{96}-\frac{C_{22,11}^{22,0}}{32
\sqrt{3}}-\frac{C_{22,11}^{22,1}}{16 \sqrt{3}}\)

\noindent\(C_{22,1111,1}^{11,0000,1,\text{Loc}}= -\frac{1}{64} \sqrt{\frac{5}{3}} C_{11,11}^{31,1}+\frac{C_{22,11}^{22,1}}{32}\)

\noindent\(C_{22,1111,1}^{11,0000,1,\text{NonLoc}}= -\frac{1}{64} \sqrt{\frac{5}{3}} C_{11,11}^{31,0}-\frac{C_{22,11}^{22,0}}{32}\)

\noindent\(C_{22,1111,2}^{00,1112,0,\text{Loc}}= \frac{\sqrt{5} C_{00,20}^{40,0}}{48}+\frac{\sqrt{5} C_{00,20}^{40,1}}{96}-\frac{7
C_{00,22}^{42,0}}{192}-\frac{7 C_{00,22}^{42,1}}{384}+\frac{C_{11,22}^{31,0}}{64}+\frac{C_{11,22}^{31,1}}{128}-\frac{1}{64} \sqrt{\frac{21}{5}} C_{11,22}^{33,0}-\frac{1}{128}
\sqrt{\frac{21}{5}} C_{11,22}^{33,1}-\frac{\sqrt{5} C_{20,20}^{20,0}}{48}-\frac{\sqrt{5} C_{20,20}^{20,1}}{96}+\frac{C_{20,22}^{22,0}}{192}+\frac{C_{20,22}^{22,1}}{384}+\frac{C_{22,20}^{22,0}}{48}+\frac{C_{22,20}^{22,1}}{96}+\frac{\sqrt{7}
C_{22,22}^{22,0}}{96}+\frac{\sqrt{7} C_{22,22}^{22,1}}{192}\)

\noindent\(C_{22,1111,2}^{00,1112,0,\text{NonLoc}}= \frac{1}{64} \sqrt{\frac{5}{3}} C_{00,00}^{40,0}+\frac{1}{32} \sqrt{\frac{5}{3}}
C_{00,00}^{40,1}+\frac{\sqrt{5} C_{00,20}^{40,0}}{192}+\frac{\sqrt{5} C_{00,20}^{40,1}}{96}+\frac{7 C_{00,22}^{42,0}}{384}+\frac{7 C_{00,22}^{42,1}}{192}+\frac{C_{11,22}^{31,0}}{128}+\frac{C_{11,22}^{31,1}}{64}-\frac{1}{128}
\sqrt{\frac{21}{5}} C_{11,22}^{33,0}-\frac{1}{64} \sqrt{\frac{21}{5}} C_{11,22}^{33,1}-\frac{1}{64} \sqrt{\frac{5}{3}} C_{20,00}^{20,0}-\frac{1}{32}
\sqrt{\frac{5}{3}} C_{20,00}^{20,1}-\frac{\sqrt{5} C_{20,20}^{20,0}}{192}-\frac{\sqrt{5} C_{20,20}^{20,1}}{96}-\frac{C_{20,22}^{22,0}}{384}-\frac{C_{20,22}^{22,1}}{192}+\frac{C_{22,00}^{22,0}}{64
\sqrt{3}}+\frac{C_{22,00}^{22,1}}{32 \sqrt{3}}+\frac{C_{22,20}^{22,0}}{192}+\frac{C_{22,20}^{22,1}}{96}-\frac{\sqrt{7} C_{22,22}^{22,0}}{192}-\frac{\sqrt{7}
C_{22,22}^{22,1}}{96}\)

\noindent\(C_{22,1111,2}^{00,1112,1,\text{Loc}}= \frac{1}{32} \sqrt{\frac{5}{3}} C_{00,20}^{40,1}-\frac{7 C_{00,22}^{42,1}}{128 \sqrt{3}}+\frac{\sqrt{3}
C_{11,22}^{31,1}}{128}-\frac{3}{128} \sqrt{\frac{7}{5}} C_{11,22}^{33,1}-\frac{1}{32} \sqrt{\frac{5}{3}} C_{20,20}^{20,1}+\frac{C_{20,22}^{22,1}}{128
\sqrt{3}}+\frac{C_{22,20}^{22,1}}{32 \sqrt{3}}+\frac{1}{64} \sqrt{\frac{7}{3}} C_{22,22}^{22,1}\)

\noindent\(C_{22,1111,2}^{00,1112,1,\text{NonLoc}}= \frac{\sqrt{5} C_{00,00}^{40,0}}{64}+\frac{1}{64} \sqrt{\frac{5}{3}} C_{00,20}^{40,0}+\frac{7
C_{00,22}^{42,0}}{128 \sqrt{3}}+\frac{\sqrt{3} C_{11,22}^{31,0}}{128}-\frac{3}{128} \sqrt{\frac{7}{5}} C_{11,22}^{33,0}-\frac{\sqrt{5} C_{20,00}^{20,0}}{64}-\frac{1}{64}
\sqrt{\frac{5}{3}} C_{20,20}^{20,0}-\frac{C_{20,22}^{22,0}}{128 \sqrt{3}}+\frac{C_{22,00}^{22,0}}{64}+\frac{C_{22,20}^{22,0}}{64 \sqrt{3}}-\frac{1}{64}
\sqrt{\frac{7}{3}} C_{22,22}^{22,0}\)

\noindent\(C_{22,1112,0}^{00,1110,0,\text{Loc}}= -\frac{\sqrt{5} C_{00,20}^{40,0}}{72}-\frac{\sqrt{5} C_{00,20}^{40,1}}{144}-\frac{7
C_{00,22}^{42,0}}{144}-\frac{7 C_{00,22}^{42,1}}{288}+\frac{C_{11,22}^{31,0}}{96}+\frac{C_{11,22}^{31,1}}{192}-\frac{1}{32} \sqrt{\frac{7}{15}} C_{11,22}^{33,0}-\frac{1}{64}
\sqrt{\frac{7}{15}} C_{11,22}^{33,1}+\frac{\sqrt{5} C_{20,20}^{20,0}}{72}+\frac{\sqrt{5} C_{20,20}^{20,1}}{144}+\frac{C_{20,22}^{22,0}}{36}+\frac{C_{20,22}^{22,1}}{72}-\frac{C_{22,20}^{22,0}}{72}-\frac{C_{22,20}^{22,1}}{144}-\frac{\sqrt{7}
C_{22,22}^{22,0}}{144}-\frac{\sqrt{7} C_{22,22}^{22,1}}{288}\)

\noindent\(C_{22,1112,0}^{00,1110,0,\text{NonLoc}}= -\frac{1}{96} \sqrt{\frac{5}{3}} C_{00,00}^{40,0}-\frac{1}{48} \sqrt{\frac{5}{3}}
C_{00,00}^{40,1}-\frac{\sqrt{5} C_{00,20}^{40,0}}{288}-\frac{\sqrt{5} C_{00,20}^{40,1}}{144}+\frac{7 C_{00,22}^{42,0}}{288}+\frac{7 C_{00,22}^{42,1}}{144}+\frac{C_{11,22}^{31,0}}{192}+\frac{C_{11,22}^{31,1}}{96}-\frac{1}{64}
\sqrt{\frac{7}{15}} C_{11,22}^{33,0}-\frac{1}{32} \sqrt{\frac{7}{15}} C_{11,22}^{33,1}+\frac{1}{96} \sqrt{\frac{5}{3}} C_{20,00}^{20,0}+\frac{1}{48}
\sqrt{\frac{5}{3}} C_{20,00}^{20,1}+\frac{\sqrt{5} C_{20,20}^{20,0}}{288}+\frac{\sqrt{5} C_{20,20}^{20,1}}{144}-\frac{C_{20,22}^{22,0}}{72}-\frac{C_{20,22}^{22,1}}{36}-\frac{C_{22,00}^{22,0}}{96
\sqrt{3}}-\frac{C_{22,00}^{22,1}}{48 \sqrt{3}}-\frac{C_{22,20}^{22,0}}{288}-\frac{C_{22,20}^{22,1}}{144}+\frac{\sqrt{7} C_{22,22}^{22,0}}{288}+\frac{\sqrt{7}
C_{22,22}^{22,1}}{144}\)

\noindent\(C_{22,1112,0}^{00,1110,1,\text{Loc}}= -\frac{1}{48} \sqrt{\frac{5}{3}} C_{00,20}^{40,1}-\frac{7 C_{00,22}^{42,1}}{96 \sqrt{3}}+\frac{C_{11,22}^{31,1}}{64
\sqrt{3}}-\frac{1}{64} \sqrt{\frac{7}{5}} C_{11,22}^{33,1}+\frac{1}{48} \sqrt{\frac{5}{3}} C_{20,20}^{20,1}+\frac{C_{20,22}^{22,1}}{24 \sqrt{3}}-\frac{C_{22,20}^{22,1}}{48
\sqrt{3}}-\frac{1}{96} \sqrt{\frac{7}{3}} C_{22,22}^{22,1}\)

\noindent\(C_{22,1112,0}^{00,1110,1,\text{NonLoc}}= -\frac{\sqrt{5} C_{00,00}^{40,0}}{96}-\frac{1}{96} \sqrt{\frac{5}{3}} C_{00,20}^{40,0}+\frac{7
C_{00,22}^{42,0}}{96 \sqrt{3}}+\frac{C_{11,22}^{31,0}}{64 \sqrt{3}}-\frac{1}{64} \sqrt{\frac{7}{5}} C_{11,22}^{33,0}+\frac{\sqrt{5} C_{20,00}^{20,0}}{96}+\frac{1}{96}
\sqrt{\frac{5}{3}} C_{20,20}^{20,0}-\frac{C_{20,22}^{22,0}}{24 \sqrt{3}}-\frac{C_{22,00}^{22,0}}{96}-\frac{C_{22,20}^{22,0}}{96 \sqrt{3}}+\frac{1}{96}
\sqrt{\frac{7}{3}} C_{22,22}^{22,0}\)

\noindent\(C_{22,1112,1}^{00,1111,0,\text{Loc}}= -\frac{\sqrt{5} C_{00,20}^{40,0}}{48}-\frac{\sqrt{5} C_{00,20}^{40,1}}{96}+\frac{7
C_{00,22}^{42,0}}{192}+\frac{7 C_{00,22}^{42,1}}{384}-\frac{C_{11,22}^{31,0}}{64}-\frac{C_{11,22}^{31,1}}{128}+\frac{1}{64} \sqrt{\frac{21}{5}} C_{11,22}^{33,0}+\frac{1}{128}
\sqrt{\frac{21}{5}} C_{11,22}^{33,1}+\frac{\sqrt{5} C_{20,20}^{20,0}}{48}+\frac{\sqrt{5} C_{20,20}^{20,1}}{96}-\frac{C_{20,22}^{22,0}}{192}-\frac{C_{20,22}^{22,1}}{384}-\frac{C_{22,20}^{22,0}}{48}-\frac{C_{22,20}^{22,1}}{96}-\frac{\sqrt{7}
C_{22,22}^{22,0}}{96}-\frac{\sqrt{7} C_{22,22}^{22,1}}{192}\)

\noindent\(C_{22,1112,1}^{00,1111,0,\text{NonLoc}}= -\frac{1}{64} \sqrt{\frac{5}{3}} C_{00,00}^{40,0}-\frac{1}{32} \sqrt{\frac{5}{3}}
C_{00,00}^{40,1}-\frac{\sqrt{5} C_{00,20}^{40,0}}{192}-\frac{\sqrt{5} C_{00,20}^{40,1}}{96}-\frac{7 C_{00,22}^{42,0}}{384}-\frac{7 C_{00,22}^{42,1}}{192}-\frac{C_{11,22}^{31,0}}{128}-\frac{C_{11,22}^{31,1}}{64}+\frac{1}{128}
\sqrt{\frac{21}{5}} C_{11,22}^{33,0}+\frac{1}{64} \sqrt{\frac{21}{5}} C_{11,22}^{33,1}+\frac{1}{64} \sqrt{\frac{5}{3}} C_{20,00}^{20,0}+\frac{1}{32}
\sqrt{\frac{5}{3}} C_{20,00}^{20,1}+\frac{\sqrt{5} C_{20,20}^{20,0}}{192}+\frac{\sqrt{5} C_{20,20}^{20,1}}{96}+\frac{C_{20,22}^{22,0}}{384}+\frac{C_{20,22}^{22,1}}{192}-\frac{C_{22,00}^{22,0}}{64
\sqrt{3}}-\frac{C_{22,00}^{22,1}}{32 \sqrt{3}}-\frac{C_{22,20}^{22,0}}{192}-\frac{C_{22,20}^{22,1}}{96}+\frac{\sqrt{7} C_{22,22}^{22,0}}{192}+\frac{\sqrt{7}
C_{22,22}^{22,1}}{96}\)

\noindent\(C_{22,1112,1}^{00,1111,1,\text{Loc}}= -\frac{1}{32} \sqrt{\frac{5}{3}} C_{00,20}^{40,1}+\frac{7 C_{00,22}^{42,1}}{128 \sqrt{3}}-\frac{\sqrt{3}
C_{11,22}^{31,1}}{128}+\frac{3}{128} \sqrt{\frac{7}{5}} C_{11,22}^{33,1}+\frac{1}{32} \sqrt{\frac{5}{3}} C_{20,20}^{20,1}-\frac{C_{20,22}^{22,1}}{128
\sqrt{3}}-\frac{C_{22,20}^{22,1}}{32 \sqrt{3}}-\frac{1}{64} \sqrt{\frac{7}{3}} C_{22,22}^{22,1}\)

\noindent\(C_{22,1112,1}^{00,1111,1,\text{NonLoc}}= -\frac{\sqrt{5} C_{00,00}^{40,0}}{64}-\frac{1}{64} \sqrt{\frac{5}{3}} C_{00,20}^{40,0}-\frac{7
C_{00,22}^{42,0}}{128 \sqrt{3}}-\frac{\sqrt{3} C_{11,22}^{31,0}}{128}+\frac{3}{128} \sqrt{\frac{7}{5}} C_{11,22}^{33,0}+\frac{\sqrt{5} C_{20,00}^{20,0}}{64}+\frac{1}{64}
\sqrt{\frac{5}{3}} C_{20,20}^{20,0}+\frac{C_{20,22}^{22,0}}{128 \sqrt{3}}-\frac{C_{22,00}^{22,0}}{64}-\frac{C_{22,20}^{22,0}}{64 \sqrt{3}}+\frac{1}{64}
\sqrt{\frac{7}{3}} C_{22,22}^{22,0}\)

\noindent\(C_{22,1112,2}^{00,1112,0,\text{Loc}}= -\frac{\sqrt{35} C_{00,20}^{40,0}}{144}-\frac{\sqrt{35} C_{00,20}^{40,1}}{288}-\frac{5
\sqrt{7} C_{00,22}^{42,0}}{576}-\frac{5 \sqrt{7} C_{00,22}^{42,1}}{1152}+\frac{\sqrt{7} C_{11,22}^{31,0}}{192}+\frac{\sqrt{7} C_{11,22}^{31,1}}{384}-\frac{7
C_{11,22}^{33,0}}{64 \sqrt{15}}-\frac{7 C_{11,22}^{33,1}}{128 \sqrt{15}}+\frac{\sqrt{35} C_{20,20}^{20,0}}{144}+\frac{\sqrt{35} C_{20,20}^{20,1}}{288}-\frac{\sqrt{7}
C_{20,22}^{22,0}}{576}-\frac{\sqrt{7} C_{20,22}^{22,1}}{1152}-\frac{\sqrt{7} C_{22,20}^{22,0}}{144}-\frac{\sqrt{7} C_{22,20}^{22,1}}{288}+\frac{11
C_{22,22}^{22,0}}{288}+\frac{11 C_{22,22}^{22,1}}{576}\)

\noindent\(C_{22,1112,2}^{00,1112,0,\text{NonLoc}}= -\frac{1}{192} \sqrt{\frac{35}{3}} C_{00,00}^{40,0}-\frac{1}{96} \sqrt{\frac{35}{3}}
C_{00,00}^{40,1}-\frac{\sqrt{35} C_{00,20}^{40,0}}{576}-\frac{\sqrt{35} C_{00,20}^{40,1}}{288}+\frac{5 \sqrt{7} C_{00,22}^{42,0}}{1152}+\frac{5 \sqrt{7}
C_{00,22}^{42,1}}{576}+\frac{\sqrt{7} C_{11,22}^{31,0}}{384}+\frac{\sqrt{7} C_{11,22}^{31,1}}{192}-\frac{7 C_{11,22}^{33,0}}{128 \sqrt{15}}-\frac{7
C_{11,22}^{33,1}}{64 \sqrt{15}}+\frac{1}{192} \sqrt{\frac{35}{3}} C_{20,00}^{20,0}+\frac{1}{96} \sqrt{\frac{35}{3}} C_{20,00}^{20,1}+\frac{\sqrt{35}
C_{20,20}^{20,0}}{576}+\frac{\sqrt{35} C_{20,20}^{20,1}}{288}+\frac{\sqrt{7} C_{20,22}^{22,0}}{1152}+\frac{\sqrt{7} C_{20,22}^{22,1}}{576}-\frac{1}{192}
\sqrt{\frac{7}{3}} C_{22,00}^{22,0}-\frac{1}{96} \sqrt{\frac{7}{3}} C_{22,00}^{22,1}-\frac{\sqrt{7} C_{22,20}^{22,0}}{576}-\frac{\sqrt{7} C_{22,20}^{22,1}}{288}-\frac{11
C_{22,22}^{22,0}}{576}-\frac{11 C_{22,22}^{22,1}}{288}\)

\noindent\(C_{22,1112,2}^{00,1112,1,\text{Loc}}= -\frac{1}{96} \sqrt{\frac{35}{3}} C_{00,20}^{40,1}-\frac{5}{384} \sqrt{\frac{7}{3}}
C_{00,22}^{42,1}+\frac{1}{128} \sqrt{\frac{7}{3}} C_{11,22}^{31,1}-\frac{7 C_{11,22}^{33,1}}{128 \sqrt{5}}+\frac{1}{96} \sqrt{\frac{35}{3}} C_{20,20}^{20,1}-\frac{1}{384}
\sqrt{\frac{7}{3}} C_{20,22}^{22,1}-\frac{1}{96} \sqrt{\frac{7}{3}} C_{22,20}^{22,1}+\frac{11 C_{22,22}^{22,1}}{192 \sqrt{3}}\)

\noindent\(C_{22,1112,2}^{00,1112,1,\text{NonLoc}}= -\frac{\sqrt{35} C_{00,00}^{40,0}}{192}-\frac{1}{192} \sqrt{\frac{35}{3}} C_{00,20}^{40,0}+\frac{5}{384}
\sqrt{\frac{7}{3}} C_{00,22}^{42,0}+\frac{1}{128} \sqrt{\frac{7}{3}} C_{11,22}^{31,0}-\frac{7 C_{11,22}^{33,0}}{128 \sqrt{5}}+\frac{\sqrt{35} C_{20,00}^{20,0}}{192}+\frac{1}{192}
\sqrt{\frac{35}{3}} C_{20,20}^{20,0}+\frac{1}{384} \sqrt{\frac{7}{3}} C_{20,22}^{22,0}-\frac{\sqrt{7} C_{22,00}^{22,0}}{192}-\frac{1}{192} \sqrt{\frac{7}{3}}
C_{22,20}^{22,0}-\frac{11 C_{22,22}^{22,0}}{192 \sqrt{3}}\)

\noindent\(C_{22,2011,1}^{00,0011,0,\text{Loc}}= \frac{7 C_{00,22}^{42,0}}{192}+\frac{7 C_{00,22}^{42,1}}{384}-\frac{C_{11,22}^{31,0}}{64}-\frac{C_{11,22}^{31,1}}{128}+\frac{1}{64}
\sqrt{\frac{21}{5}} C_{11,22}^{33,0}+\frac{1}{128} \sqrt{\frac{21}{5}} C_{11,22}^{33,1}-\frac{C_{20,22}^{22,0}}{192}-\frac{C_{20,22}^{22,1}}{384}-\frac{\sqrt{7}
C_{22,22}^{22,0}}{96}-\frac{\sqrt{7} C_{22,22}^{22,1}}{192}\)

\noindent\(C_{22,2011,1}^{00,0011,0,\text{NonLoc}}= -\frac{7 C_{00,22}^{42,0}}{384}-\frac{7 C_{00,22}^{42,1}}{192}-\frac{C_{11,22}^{31,0}}{128}-\frac{C_{11,22}^{31,1}}{64}+\frac{1}{128}
\sqrt{\frac{21}{5}} C_{11,22}^{33,0}+\frac{1}{64} \sqrt{\frac{21}{5}} C_{11,22}^{33,1}+\frac{C_{20,22}^{22,0}}{384}+\frac{C_{20,22}^{22,1}}{192}+\frac{\sqrt{7}
C_{22,22}^{22,0}}{192}+\frac{\sqrt{7} C_{22,22}^{22,1}}{96}\)

\noindent\(C_{22,2011,1}^{00,0011,1,\text{Loc}}= \frac{7 C_{00,22}^{42,1}}{128 \sqrt{3}}-\frac{\sqrt{3} C_{11,22}^{31,1}}{128}+\frac{3}{128}
\sqrt{\frac{7}{5}} C_{11,22}^{33,1}-\frac{C_{20,22}^{22,1}}{128 \sqrt{3}}-\frac{1}{64} \sqrt{\frac{7}{3}} C_{22,22}^{22,1}\)

\noindent\(C_{22,2011,1}^{00,0011,1,\text{NonLoc}}= -\frac{7 C_{00,22}^{42,0}}{128 \sqrt{3}}-\frac{\sqrt{3} C_{11,22}^{31,0}}{128}+\frac{3}{128}
\sqrt{\frac{7}{5}} C_{11,22}^{33,0}+\frac{C_{20,22}^{22,0}}{128 \sqrt{3}}+\frac{1}{64} \sqrt{\frac{7}{3}} C_{22,22}^{22,0}\)

\noindent\(C_{22,2202,0}^{00,0000,0,\text{Loc}}= -\frac{\sqrt{5} C_{00,00}^{40,0}}{48}-\frac{\sqrt{5} C_{00,00}^{40,1}}{96}+\frac{\sqrt{5}
C_{20,00}^{20,0}}{48}+\frac{\sqrt{5} C_{20,00}^{20,1}}{96}-\frac{C_{22,00}^{22,0}}{48}-\frac{C_{22,00}^{22,1}}{96}\)

\noindent\(C_{22,2202,0}^{00,0000,0,\text{NonLoc}}= \frac{\sqrt{5} C_{00,00}^{40,0}}{192}+\frac{\sqrt{5} C_{00,00}^{40,1}}{96}-\frac{1}{64}
\sqrt{\frac{5}{3}} C_{00,20}^{40,0}-\frac{1}{32} \sqrt{\frac{5}{3}} C_{00,20}^{40,1}-\frac{\sqrt{5} C_{20,00}^{20,0}}{192}-\frac{\sqrt{5} C_{20,00}^{20,1}}{96}+\frac{1}{64}
\sqrt{\frac{5}{3}} C_{20,20}^{20,0}+\frac{1}{32} \sqrt{\frac{5}{3}} C_{20,20}^{20,1}+\frac{C_{22,00}^{22,0}}{192}+\frac{C_{22,00}^{22,1}}{96}-\frac{C_{22,20}^{22,0}}{64
\sqrt{3}}-\frac{C_{22,20}^{22,1}}{32 \sqrt{3}}\)

\noindent\(C_{22,2202,0}^{00,0000,1,\text{Loc}}= -\frac{1}{32} \sqrt{\frac{5}{3}} C_{00,00}^{40,1}+\frac{1}{32} \sqrt{\frac{5}{3}}
C_{20,00}^{20,1}-\frac{C_{22,00}^{22,1}}{32 \sqrt{3}}\)

\noindent\(C_{22,2202,0}^{00,0000,1,\text{NonLoc}}= \frac{1}{64} \sqrt{\frac{5}{3}} C_{00,00}^{40,0}-\frac{\sqrt{5} C_{00,20}^{40,0}}{64}-\frac{1}{64}
\sqrt{\frac{5}{3}} C_{20,00}^{20,0}+\frac{\sqrt{5} C_{20,20}^{20,0}}{64}+\frac{C_{22,00}^{22,0}}{64 \sqrt{3}}-\frac{C_{22,20}^{22,0}}{64}\)

\noindent\(C_{22,2211,1}^{00,0011,0,\text{Loc}}= \frac{C_{00,20}^{40,0}}{48}+\frac{C_{00,20}^{40,1}}{96}+\frac{7 C_{00,22}^{42,0}}{192
\sqrt{5}}+\frac{7 C_{00,22}^{42,1}}{384 \sqrt{5}}-\frac{C_{20,20}^{20,0}}{48}-\frac{C_{20,20}^{20,1}}{96}-\frac{7 C_{20,22}^{22,0}}{192 \sqrt{5}}-\frac{7
C_{20,22}^{22,1}}{384 \sqrt{5}}+\frac{C_{22,20}^{22,0}}{48 \sqrt{5}}+\frac{C_{22,20}^{22,1}}{96 \sqrt{5}}+\frac{1}{48} \sqrt{\frac{7}{5}} C_{22,22}^{22,0}+\frac{1}{96}
\sqrt{\frac{7}{5}} C_{22,22}^{22,1}\)

\noindent\(C_{22,2211,1}^{00,0011,0,\text{NonLoc}}= \frac{C_{00,00}^{40,0}}{64 \sqrt{3}}+\frac{C_{00,00}^{40,1}}{32 \sqrt{3}}+\frac{C_{00,20}^{40,0}}{192}+\frac{C_{00,20}^{40,1}}{96}-\frac{7
C_{00,22}^{42,0}}{384 \sqrt{5}}-\frac{7 C_{00,22}^{42,1}}{192 \sqrt{5}}-\frac{C_{20,00}^{20,0}}{64 \sqrt{3}}-\frac{C_{20,00}^{20,1}}{32 \sqrt{3}}-\frac{C_{20,20}^{20,0}}{192}-\frac{C_{20,20}^{20,1}}{96}+\frac{7
C_{20,22}^{22,0}}{384 \sqrt{5}}+\frac{7 C_{20,22}^{22,1}}{192 \sqrt{5}}+\frac{C_{22,00}^{22,0}}{64 \sqrt{15}}+\frac{C_{22,00}^{22,1}}{32 \sqrt{15}}+\frac{C_{22,20}^{22,0}}{192
\sqrt{5}}+\frac{C_{22,20}^{22,1}}{96 \sqrt{5}}-\frac{1}{96} \sqrt{\frac{7}{5}} C_{22,22}^{22,0}-\frac{1}{48} \sqrt{\frac{7}{5}} C_{22,22}^{22,1}\)

\noindent\(C_{22,2211,1}^{00,0011,1,\text{Loc}}= \frac{C_{00,20}^{40,1}}{32 \sqrt{3}}+\frac{7 C_{00,22}^{42,1}}{128 \sqrt{15}}-\frac{C_{20,20}^{20,1}}{32
\sqrt{3}}-\frac{7 C_{20,22}^{22,1}}{128 \sqrt{15}}+\frac{C_{22,20}^{22,1}}{32 \sqrt{15}}+\frac{1}{32} \sqrt{\frac{7}{15}} C_{22,22}^{22,1}\)

\noindent\(C_{22,2211,1}^{00,0011,1,\text{NonLoc}}= \frac{C_{00,00}^{40,0}}{64}+\frac{C_{00,20}^{40,0}}{64 \sqrt{3}}-\frac{7 C_{00,22}^{42,0}}{128
\sqrt{15}}-\frac{C_{20,00}^{20,0}}{64}-\frac{C_{20,20}^{20,0}}{64 \sqrt{3}}+\frac{7 C_{20,22}^{22,0}}{128 \sqrt{15}}+\frac{C_{22,00}^{22,0}}{64 \sqrt{5}}+\frac{C_{22,20}^{22,0}}{64
\sqrt{15}}-\frac{1}{32} \sqrt{\frac{7}{15}} C_{22,22}^{22,0}\)

\noindent\(C_{22,2212,1}^{00,0011,0,\text{Loc}}= -\frac{1}{48} \sqrt{\frac{5}{3}} C_{00,20}^{40,0}-\frac{1}{96} \sqrt{\frac{5}{3}}
C_{00,20}^{40,1}+\frac{7 C_{00,22}^{42,0}}{192 \sqrt{3}}+\frac{7 C_{00,22}^{42,1}}{384 \sqrt{3}}+\frac{1}{48} \sqrt{\frac{5}{3}} C_{20,20}^{20,0}+\frac{1}{96}
\sqrt{\frac{5}{3}} C_{20,20}^{20,1}-\frac{7 C_{20,22}^{22,0}}{192 \sqrt{3}}-\frac{7 C_{20,22}^{22,1}}{384 \sqrt{3}}-\frac{C_{22,20}^{22,0}}{48 \sqrt{3}}-\frac{C_{22,20}^{22,1}}{96
\sqrt{3}}+\frac{1}{48} \sqrt{\frac{7}{3}} C_{22,22}^{22,0}+\frac{1}{96} \sqrt{\frac{7}{3}} C_{22,22}^{22,1}\)

\noindent\(C_{22,2212,1}^{00,0011,0,\text{NonLoc}}= -\frac{\sqrt{5} C_{00,00}^{40,0}}{192}-\frac{\sqrt{5} C_{00,00}^{40,1}}{96}-\frac{1}{192}
\sqrt{\frac{5}{3}} C_{00,20}^{40,0}-\frac{1}{96} \sqrt{\frac{5}{3}} C_{00,20}^{40,1}-\frac{7 C_{00,22}^{42,0}}{384 \sqrt{3}}-\frac{7 C_{00,22}^{42,1}}{192
\sqrt{3}}+\frac{\sqrt{5} C_{20,00}^{20,0}}{192}+\frac{\sqrt{5} C_{20,00}^{20,1}}{96}+\frac{1}{192} \sqrt{\frac{5}{3}} C_{20,20}^{20,0}+\frac{1}{96}
\sqrt{\frac{5}{3}} C_{20,20}^{20,1}+\frac{7 C_{20,22}^{22,0}}{384 \sqrt{3}}+\frac{7 C_{20,22}^{22,1}}{192 \sqrt{3}}-\frac{C_{22,00}^{22,0}}{192}-\frac{C_{22,00}^{22,1}}{96}-\frac{C_{22,20}^{22,0}}{192
\sqrt{3}}-\frac{C_{22,20}^{22,1}}{96 \sqrt{3}}-\frac{1}{96} \sqrt{\frac{7}{3}} C_{22,22}^{22,0}-\frac{1}{48} \sqrt{\frac{7}{3}} C_{22,22}^{22,1}\)

\noindent\(C_{22,2212,1}^{00,0011,1,\text{Loc}}= -\frac{\sqrt{5} C_{00,20}^{40,1}}{96}+\frac{7 C_{00,22}^{42,1}}{384}+\frac{\sqrt{5}
C_{20,20}^{20,1}}{96}-\frac{7 C_{20,22}^{22,1}}{384}-\frac{C_{22,20}^{22,1}}{96}+\frac{\sqrt{7} C_{22,22}^{22,1}}{96}\)

\noindent\(C_{22,2212,1}^{00,0011,1,\text{NonLoc}}= -\frac{1}{64} \sqrt{\frac{5}{3}} C_{00,00}^{40,0}-\frac{\sqrt{5} C_{00,20}^{40,0}}{192}-\frac{7
C_{00,22}^{42,0}}{384}+\frac{1}{64} \sqrt{\frac{5}{3}} C_{20,00}^{20,0}+\frac{\sqrt{5} C_{20,20}^{20,0}}{192}+\frac{7 C_{20,22}^{22,0}}{384}-\frac{C_{22,00}^{22,0}}{64
\sqrt{3}}-\frac{C_{22,20}^{22,0}}{192}-\frac{\sqrt{7} C_{22,22}^{22,0}}{96}\)

\noindent\(C_{22,2213,1}^{00,0011,0,\text{Loc}}= \frac{1}{48} \sqrt{\frac{7}{3}} C_{00,20}^{40,0}+\frac{1}{96} \sqrt{\frac{7}{3}} C_{00,20}^{40,1}+\frac{1}{96}
\sqrt{\frac{7}{15}} C_{00,22}^{42,0}+\frac{1}{192} \sqrt{\frac{7}{15}} C_{00,22}^{42,1}-\frac{1}{48} \sqrt{\frac{7}{3}} C_{20,20}^{20,0}-\frac{1}{96}
\sqrt{\frac{7}{3}} C_{20,20}^{20,1}-\frac{1}{96} \sqrt{\frac{7}{15}} C_{20,22}^{22,0}-\frac{1}{192} \sqrt{\frac{7}{15}} C_{20,22}^{22,1}+\frac{1}{48}
\sqrt{\frac{7}{15}} C_{22,20}^{22,0}+\frac{1}{96} \sqrt{\frac{7}{15}} C_{22,20}^{22,1}+\frac{C_{22,22}^{22,0}}{24 \sqrt{15}}+\frac{C_{22,22}^{22,1}}{48
\sqrt{15}}\)

\noindent\(C_{22,2213,1}^{00,0011,0,\text{NonLoc}}= \frac{\sqrt{7} C_{00,00}^{40,0}}{192}+\frac{\sqrt{7} C_{00,00}^{40,1}}{96}+\frac{1}{192}
\sqrt{\frac{7}{3}} C_{00,20}^{40,0}+\frac{1}{96} \sqrt{\frac{7}{3}} C_{00,20}^{40,1}-\frac{1}{192} \sqrt{\frac{7}{15}} C_{00,22}^{42,0}-\frac{1}{96}
\sqrt{\frac{7}{15}} C_{00,22}^{42,1}-\frac{\sqrt{7} C_{20,00}^{20,0}}{192}-\frac{\sqrt{7} C_{20,00}^{20,1}}{96}-\frac{1}{192} \sqrt{\frac{7}{3}}
C_{20,20}^{20,0}-\frac{1}{96} \sqrt{\frac{7}{3}} C_{20,20}^{20,1}+\frac{1}{192} \sqrt{\frac{7}{15}} C_{20,22}^{22,0}+\frac{1}{96} \sqrt{\frac{7}{15}}
C_{20,22}^{22,1}+\frac{1}{192} \sqrt{\frac{7}{5}} C_{22,00}^{22,0}+\frac{1}{96} \sqrt{\frac{7}{5}} C_{22,00}^{22,1}+\frac{1}{192} \sqrt{\frac{7}{15}}
C_{22,20}^{22,0}+\frac{1}{96} \sqrt{\frac{7}{15}} C_{22,20}^{22,1}-\frac{C_{22,22}^{22,0}}{48 \sqrt{15}}-\frac{C_{22,22}^{22,1}}{24 \sqrt{15}}\)

\noindent\(C_{22,2213,1}^{00,0011,1,\text{Loc}}= \frac{\sqrt{7} C_{00,20}^{40,1}}{96}+\frac{1}{192} \sqrt{\frac{7}{5}} C_{00,22}^{42,1}-\frac{\sqrt{7}
C_{20,20}^{20,1}}{96}-\frac{1}{192} \sqrt{\frac{7}{5}} C_{20,22}^{22,1}+\frac{1}{96} \sqrt{\frac{7}{5}} C_{22,20}^{22,1}+\frac{C_{22,22}^{22,1}}{48
\sqrt{5}}\)

\noindent\(C_{22,2213,1}^{00,0011,1,\text{NonLoc}}= \frac{1}{64} \sqrt{\frac{7}{3}} C_{00,00}^{40,0}+\frac{\sqrt{7} C_{00,20}^{40,0}}{192}-\frac{1}{192}
\sqrt{\frac{7}{5}} C_{00,22}^{42,0}-\frac{1}{64} \sqrt{\frac{7}{3}} C_{20,00}^{20,0}-\frac{\sqrt{7} C_{20,20}^{20,0}}{192}+\frac{1}{192} \sqrt{\frac{7}{5}}
C_{20,22}^{22,0}+\frac{1}{64} \sqrt{\frac{7}{15}} C_{22,00}^{22,0}+\frac{1}{192} \sqrt{\frac{7}{5}} C_{22,20}^{22,0}-\frac{C_{22,22}^{22,0}}{48 \sqrt{5}}\)

\noindent\(C_{31,0000,1}^{00,1111,0,\text{Loc}}= -\frac{C_{11,11}^{31,0}}{64}-\frac{C_{11,11}^{31,1}}{128}+\frac{1}{32} \sqrt{\frac{3}{5}}
C_{22,11}^{22,0}+\frac{1}{64} \sqrt{\frac{3}{5}} C_{22,11}^{22,1}\)

\noindent\(C_{31,0000,1}^{00,1111,0,\text{NonLoc}}= -\frac{C_{11,11}^{31,0}}{128}-\frac{C_{11,11}^{31,1}}{64}-\frac{1}{64} \sqrt{\frac{3}{5}}
C_{22,11}^{22,0}-\frac{1}{32} \sqrt{\frac{3}{5}} C_{22,11}^{22,1}\)

\noindent\(C_{31,0000,1}^{00,1111,1,\text{Loc}}= -\frac{\sqrt{3} C_{11,11}^{31,1}}{128}+\frac{3 C_{22,11}^{22,1}}{64 \sqrt{5}}\)

\noindent\(C_{31,0000,1}^{00,1111,1,\text{NonLoc}}= -\frac{\sqrt{3} C_{11,11}^{31,0}}{128}-\frac{3 C_{22,11}^{22,0}}{64 \sqrt{5}}\)

\noindent\(C_{31,0000,1}^{11,0000,0,\text{Loc}}= -\frac{C_{00,00}^{40,0}}{64}-\frac{C_{00,00}^{40,1}}{128}+\frac{C_{11,00}^{31,0}}{64}+\frac{C_{11,00}^{31,1}}{128}-\frac{C_{20,00}^{20,0}}{64}-\frac{C_{20,00}^{20,1}}{128}-\frac{C_{22,00}^{22,0}}{32
\sqrt{5}}-\frac{C_{22,00}^{22,1}}{64 \sqrt{5}}\)

\noindent\(C_{31,0000,1}^{11,0000,0,\text{NonLoc}}= \frac{C_{00,00}^{40,0}}{256}+\frac{C_{00,00}^{40,1}}{128}-\frac{\sqrt{3} C_{00,20}^{40,0}}{256}-\frac{\sqrt{3}
C_{00,20}^{40,1}}{128}+\frac{C_{11,00}^{31,0}}{256}+\frac{C_{11,00}^{31,1}}{128}-\frac{\sqrt{3} C_{11,20}^{31,0}}{256}-\frac{\sqrt{3} C_{11,20}^{31,1}}{128}+\frac{C_{20,00}^{20,0}}{256}+\frac{C_{20,00}^{20,1}}{128}-\frac{\sqrt{3}
C_{20,20}^{20,0}}{256}-\frac{\sqrt{3} C_{20,20}^{20,1}}{128}+\frac{C_{22,00}^{22,0}}{128 \sqrt{5}}+\frac{C_{22,00}^{22,1}}{64 \sqrt{5}}-\frac{1}{128}
\sqrt{\frac{3}{5}} C_{22,20}^{22,0}-\frac{1}{64} \sqrt{\frac{3}{5}} C_{22,20}^{22,1}\)

\noindent\(C_{31,0000,1}^{11,0000,1,\text{Loc}}= -\frac{\sqrt{3} C_{00,00}^{40,1}}{128}+\frac{\sqrt{3} C_{11,00}^{31,1}}{128}-\frac{\sqrt{3}
C_{20,00}^{20,1}}{128}-\frac{1}{64} \sqrt{\frac{3}{5}} C_{22,00}^{22,1}\)

\noindent\(C_{31,0000,1}^{11,0000,1,\text{NonLoc}}= \frac{\sqrt{3} C_{00,00}^{40,0}}{256}-\frac{3 C_{00,20}^{40,0}}{256}+\frac{\sqrt{3}
C_{11,00}^{31,0}}{256}-\frac{3 C_{11,20}^{31,0}}{256}+\frac{\sqrt{3} C_{20,00}^{20,0}}{256}-\frac{3 C_{20,20}^{20,0}}{256}+\frac{1}{128} \sqrt{\frac{3}{5}}
C_{22,00}^{22,0}-\frac{3 C_{22,20}^{22,0}}{128 \sqrt{5}}\)

\noindent\(C_{31,0011,0}^{11,0011,0,\text{Loc}}= \frac{C_{00,20}^{40,0}}{192}+\frac{C_{00,20}^{40,1}}{384}+\frac{7 C_{00,22}^{42,0}}{384
\sqrt{5}}+\frac{7 C_{00,22}^{42,1}}{768 \sqrt{5}}-\frac{C_{11,20}^{31,0}}{192}-\frac{C_{11,20}^{31,1}}{384}-\frac{7 C_{11,22}^{31,0}}{384 \sqrt{5}}-\frac{7
C_{11,22}^{31,1}}{768 \sqrt{5}}-\frac{\sqrt{21} C_{11,22}^{33,0}}{640}-\frac{\sqrt{21} C_{11,22}^{33,1}}{1280}+\frac{C_{20,20}^{20,0}}{192}+\frac{C_{20,20}^{20,1}}{384}+\frac{7
C_{20,22}^{22,0}}{384 \sqrt{5}}+\frac{7 C_{20,22}^{22,1}}{768 \sqrt{5}}+\frac{C_{22,20}^{22,0}}{96 \sqrt{5}}+\frac{C_{22,20}^{22,1}}{192 \sqrt{5}}+\frac{1}{192}
\sqrt{\frac{7}{5}} C_{22,22}^{22,0}+\frac{1}{384} \sqrt{\frac{7}{5}} C_{22,22}^{22,1}\)

\noindent\(C_{31,0011,0}^{11,0011,0,\text{NonLoc}}= \frac{C_{00,00}^{40,0}}{256 \sqrt{3}}+\frac{C_{00,00}^{40,1}}{128 \sqrt{3}}+\frac{C_{00,20}^{40,0}}{768}+\frac{C_{00,20}^{40,1}}{384}-\frac{7
C_{00,22}^{42,0}}{768 \sqrt{5}}-\frac{7 C_{00,22}^{42,1}}{384 \sqrt{5}}+\frac{C_{11,00}^{31,0}}{256 \sqrt{3}}+\frac{C_{11,00}^{31,1}}{128 \sqrt{3}}+\frac{C_{11,20}^{31,0}}{768}+\frac{C_{11,20}^{31,1}}{384}-\frac{7
C_{11,22}^{31,0}}{768 \sqrt{5}}-\frac{7 C_{11,22}^{31,1}}{384 \sqrt{5}}-\frac{\sqrt{21} C_{11,22}^{33,0}}{1280}-\frac{\sqrt{21} C_{11,22}^{33,1}}{640}+\frac{C_{20,00}^{20,0}}{256
\sqrt{3}}+\frac{C_{20,00}^{20,1}}{128 \sqrt{3}}+\frac{C_{20,20}^{20,0}}{768}+\frac{C_{20,20}^{20,1}}{384}-\frac{7 C_{20,22}^{22,0}}{768 \sqrt{5}}-\frac{7
C_{20,22}^{22,1}}{384 \sqrt{5}}+\frac{C_{22,00}^{22,0}}{128 \sqrt{15}}+\frac{C_{22,00}^{22,1}}{64 \sqrt{15}}+\frac{C_{22,20}^{22,0}}{384 \sqrt{5}}+\frac{C_{22,20}^{22,1}}{192
\sqrt{5}}-\frac{1}{384} \sqrt{\frac{7}{5}} C_{22,22}^{22,0}-\frac{1}{192} \sqrt{\frac{7}{5}} C_{22,22}^{22,1}\)

\noindent\(C_{31,0011,0}^{11,0011,1,\text{Loc}}= \frac{C_{00,20}^{40,1}}{128 \sqrt{3}}+\frac{7 C_{00,22}^{42,1}}{256 \sqrt{15}}-\frac{C_{11,20}^{31,1}}{128
\sqrt{3}}-\frac{7 C_{11,22}^{31,1}}{256 \sqrt{15}}-\frac{3 \sqrt{7} C_{11,22}^{33,1}}{1280}+\frac{C_{20,20}^{20,1}}{128 \sqrt{3}}+\frac{7 C_{20,22}^{22,1}}{256
\sqrt{15}}+\frac{C_{22,20}^{22,1}}{64 \sqrt{15}}+\frac{1}{128} \sqrt{\frac{7}{15}} C_{22,22}^{22,1}\)

\noindent\(C_{31,0011,0}^{11,0011,1,\text{NonLoc}}= \frac{C_{00,00}^{40,0}}{256}+\frac{C_{00,20}^{40,0}}{256 \sqrt{3}}-\frac{7 C_{00,22}^{42,0}}{256
\sqrt{15}}+\frac{C_{11,00}^{31,0}}{256}+\frac{C_{11,20}^{31,0}}{256 \sqrt{3}}-\frac{7 C_{11,22}^{31,0}}{256 \sqrt{15}}-\frac{3 \sqrt{7} C_{11,22}^{33,0}}{1280}+\frac{C_{20,00}^{20,0}}{256}+\frac{C_{20,20}^{20,0}}{256
\sqrt{3}}-\frac{7 C_{20,22}^{22,0}}{256 \sqrt{15}}+\frac{C_{22,00}^{22,0}}{128 \sqrt{5}}+\frac{C_{22,20}^{22,0}}{128 \sqrt{15}}-\frac{1}{128} \sqrt{\frac{7}{15}}
C_{22,22}^{22,0}\)

\noindent\(C_{31,0011,1}^{00,1101,0,\text{Loc}}= \frac{C_{11,11}^{31,0}}{64}+\frac{C_{11,11}^{31,1}}{128}-\frac{1}{32} \sqrt{\frac{3}{5}}
C_{22,11}^{22,0}-\frac{1}{64} \sqrt{\frac{3}{5}} C_{22,11}^{22,1}\)

\noindent\(C_{31,0011,1}^{00,1101,0,\text{NonLoc}}= \frac{C_{11,11}^{31,0}}{128}+\frac{C_{11,11}^{31,1}}{64}+\frac{1}{64} \sqrt{\frac{3}{5}}
C_{22,11}^{22,0}+\frac{1}{32} \sqrt{\frac{3}{5}} C_{22,11}^{22,1}\)

\noindent\(C_{31,0011,1}^{00,1101,1,\text{Loc}}= \frac{\sqrt{3} C_{11,11}^{31,1}}{128}-\frac{3 C_{22,11}^{22,1}}{64 \sqrt{5}}\)

\noindent\(C_{31,0011,1}^{00,1101,1,\text{NonLoc}}= \frac{\sqrt{3} C_{11,11}^{31,0}}{128}+\frac{3 C_{22,11}^{22,0}}{64 \sqrt{5}}\)

\noindent\(C_{31,0011,1}^{11,0011,0,\text{Loc}}= \frac{C_{00,20}^{40,0}}{64 \sqrt{3}}+\frac{C_{00,20}^{40,1}}{128 \sqrt{3}}-\frac{7
C_{00,22}^{42,0}}{256 \sqrt{15}}-\frac{7 C_{00,22}^{42,1}}{512 \sqrt{15}}-\frac{C_{11,20}^{31,0}}{64 \sqrt{3}}-\frac{C_{11,20}^{31,1}}{128 \sqrt{3}}+\frac{7
C_{11,22}^{31,0}}{256 \sqrt{15}}+\frac{7 C_{11,22}^{31,1}}{512 \sqrt{15}}+\frac{3 \sqrt{7} C_{11,22}^{33,0}}{1280}+\frac{3 \sqrt{7} C_{11,22}^{33,1}}{2560}+\frac{C_{20,20}^{20,0}}{64
\sqrt{3}}+\frac{C_{20,20}^{20,1}}{128 \sqrt{3}}-\frac{7 C_{20,22}^{22,0}}{256 \sqrt{15}}-\frac{7 C_{20,22}^{22,1}}{512 \sqrt{15}}+\frac{C_{22,20}^{22,0}}{32
\sqrt{15}}+\frac{C_{22,20}^{22,1}}{64 \sqrt{15}}-\frac{1}{128} \sqrt{\frac{7}{15}} C_{22,22}^{22,0}-\frac{1}{256} \sqrt{\frac{7}{15}} C_{22,22}^{22,1}\)

\noindent\(C_{31,0011,1}^{11,0011,0,\text{NonLoc}}= \frac{C_{00,00}^{40,0}}{256}+\frac{C_{00,00}^{40,1}}{128}+\frac{C_{00,20}^{40,0}}{256
\sqrt{3}}+\frac{C_{00,20}^{40,1}}{128 \sqrt{3}}+\frac{7 C_{00,22}^{42,0}}{512 \sqrt{15}}+\frac{7 C_{00,22}^{42,1}}{256 \sqrt{15}}+\frac{C_{11,00}^{31,0}}{256}+\frac{C_{11,00}^{31,1}}{128}+\frac{C_{11,20}^{31,0}}{256
\sqrt{3}}+\frac{C_{11,20}^{31,1}}{128 \sqrt{3}}+\frac{7 C_{11,22}^{31,0}}{512 \sqrt{15}}+\frac{7 C_{11,22}^{31,1}}{256 \sqrt{15}}+\frac{3 \sqrt{7}
C_{11,22}^{33,0}}{2560}+\frac{3 \sqrt{7} C_{11,22}^{33,1}}{1280}+\frac{C_{20,00}^{20,0}}{256}+\frac{C_{20,00}^{20,1}}{128}+\frac{C_{20,20}^{20,0}}{256
\sqrt{3}}+\frac{C_{20,20}^{20,1}}{128 \sqrt{3}}+\frac{7 C_{20,22}^{22,0}}{512 \sqrt{15}}+\frac{7 C_{20,22}^{22,1}}{256 \sqrt{15}}+\frac{C_{22,00}^{22,0}}{128
\sqrt{5}}+\frac{C_{22,00}^{22,1}}{64 \sqrt{5}}+\frac{C_{22,20}^{22,0}}{128 \sqrt{15}}+\frac{C_{22,20}^{22,1}}{64 \sqrt{15}}+\frac{1}{256} \sqrt{\frac{7}{15}}
C_{22,22}^{22,0}+\frac{1}{128} \sqrt{\frac{7}{15}} C_{22,22}^{22,1}\)

\noindent\(C_{31,0011,1}^{11,0011,1,\text{Loc}}= \frac{C_{00,20}^{40,1}}{128}-\frac{7 C_{00,22}^{42,1}}{512 \sqrt{5}}-\frac{C_{11,20}^{31,1}}{128}+\frac{7
C_{11,22}^{31,1}}{512 \sqrt{5}}+\frac{3 \sqrt{21} C_{11,22}^{33,1}}{2560}+\frac{C_{20,20}^{20,1}}{128}-\frac{7 C_{20,22}^{22,1}}{512 \sqrt{5}}+\frac{C_{22,20}^{22,1}}{64
\sqrt{5}}-\frac{1}{256} \sqrt{\frac{7}{5}} C_{22,22}^{22,1}\)

\noindent\(C_{31,0011,1}^{11,0011,1,\text{NonLoc}}= \frac{\sqrt{3} C_{00,00}^{40,0}}{256}+\frac{C_{00,20}^{40,0}}{256}+\frac{7 C_{00,22}^{42,0}}{512
\sqrt{5}}+\frac{\sqrt{3} C_{11,00}^{31,0}}{256}+\frac{C_{11,20}^{31,0}}{256}+\frac{7 C_{11,22}^{31,0}}{512 \sqrt{5}}+\frac{3 \sqrt{21} C_{11,22}^{33,0}}{2560}+\frac{\sqrt{3}
C_{20,00}^{20,0}}{256}+\frac{C_{20,20}^{20,0}}{256}+\frac{7 C_{20,22}^{22,0}}{512 \sqrt{5}}+\frac{1}{128} \sqrt{\frac{3}{5}} C_{22,00}^{22,0}+\frac{C_{22,20}^{22,0}}{128
\sqrt{5}}+\frac{1}{256} \sqrt{\frac{7}{5}} C_{22,22}^{22,0}\)

\noindent\(C_{31,0011,2}^{11,0011,0,\text{Loc}}= \frac{\sqrt{5} C_{00,20}^{40,0}}{192}+\frac{\sqrt{5} C_{00,20}^{40,1}}{384}+\frac{7
C_{00,22}^{42,0}}{3840}+\frac{7 C_{00,22}^{42,1}}{7680}-\frac{\sqrt{5} C_{11,20}^{31,0}}{192}-\frac{\sqrt{5} C_{11,20}^{31,1}}{384}-\frac{7 C_{11,22}^{31,0}}{3840}-\frac{7
C_{11,22}^{31,1}}{7680}-\frac{\sqrt{\frac{21}{5}} C_{11,22}^{33,0}}{1280}-\frac{\sqrt{\frac{21}{5}} C_{11,22}^{33,1}}{2560}+\frac{\sqrt{5} C_{20,20}^{20,0}}{192}+\frac{\sqrt{5}
C_{20,20}^{20,1}}{384}+\frac{7 C_{20,22}^{22,0}}{3840}+\frac{7 C_{20,22}^{22,1}}{7680}+\frac{C_{22,20}^{22,0}}{96}+\frac{C_{22,20}^{22,1}}{192}+\frac{\sqrt{7}
C_{22,22}^{22,0}}{1920}+\frac{\sqrt{7} C_{22,22}^{22,1}}{3840}\)

\noindent\(C_{31,0011,2}^{11,0011,0,\text{NonLoc}}= \frac{1}{256} \sqrt{\frac{5}{3}} C_{00,00}^{40,0}+\frac{1}{128} \sqrt{\frac{5}{3}}
C_{00,00}^{40,1}+\frac{\sqrt{5} C_{00,20}^{40,0}}{768}+\frac{\sqrt{5} C_{00,20}^{40,1}}{384}-\frac{7 C_{00,22}^{42,0}}{7680}-\frac{7 C_{00,22}^{42,1}}{3840}+\frac{1}{256}
\sqrt{\frac{5}{3}} C_{11,00}^{31,0}+\frac{1}{128} \sqrt{\frac{5}{3}} C_{11,00}^{31,1}+\frac{\sqrt{5} C_{11,20}^{31,0}}{768}+\frac{\sqrt{5} C_{11,20}^{31,1}}{384}-\frac{7
C_{11,22}^{31,0}}{7680}-\frac{7 C_{11,22}^{31,1}}{3840}-\frac{\sqrt{\frac{21}{5}} C_{11,22}^{33,0}}{2560}-\frac{\sqrt{\frac{21}{5}} C_{11,22}^{33,1}}{1280}+\frac{1}{256}
\sqrt{\frac{5}{3}} C_{20,00}^{20,0}+\frac{1}{128} \sqrt{\frac{5}{3}} C_{20,00}^{20,1}+\frac{\sqrt{5} C_{20,20}^{20,0}}{768}+\frac{\sqrt{5} C_{20,20}^{20,1}}{384}-\frac{7
C_{20,22}^{22,0}}{7680}-\frac{7 C_{20,22}^{22,1}}{3840}+\frac{C_{22,00}^{22,0}}{128 \sqrt{3}}+\frac{C_{22,00}^{22,1}}{64 \sqrt{3}}+\frac{C_{22,20}^{22,0}}{384}+\frac{C_{22,20}^{22,1}}{192}-\frac{\sqrt{7}
C_{22,22}^{22,0}}{3840}-\frac{\sqrt{7} C_{22,22}^{22,1}}{1920}\)

\noindent\(C_{31,0011,2}^{11,0011,1,\text{Loc}}= \frac{1}{128} \sqrt{\frac{5}{3}} C_{00,20}^{40,1}+\frac{7 C_{00,22}^{42,1}}{2560 \sqrt{3}}-\frac{1}{128}
\sqrt{\frac{5}{3}} C_{11,20}^{31,1}-\frac{7 C_{11,22}^{31,1}}{2560 \sqrt{3}}-\frac{3 \sqrt{\frac{7}{5}} C_{11,22}^{33,1}}{2560}+\frac{1}{128} \sqrt{\frac{5}{3}}
C_{20,20}^{20,1}+\frac{7 C_{20,22}^{22,1}}{2560 \sqrt{3}}+\frac{C_{22,20}^{22,1}}{64 \sqrt{3}}+\frac{\sqrt{\frac{7}{3}} C_{22,22}^{22,1}}{1280}\)

\noindent\(C_{31,0011,2}^{11,0011,1,\text{NonLoc}}= \frac{\sqrt{5} C_{00,00}^{40,0}}{256}+\frac{1}{256} \sqrt{\frac{5}{3}} C_{00,20}^{40,0}-\frac{7
C_{00,22}^{42,0}}{2560 \sqrt{3}}+\frac{\sqrt{5} C_{11,00}^{31,0}}{256}+\frac{1}{256} \sqrt{\frac{5}{3}} C_{11,20}^{31,0}-\frac{7 C_{11,22}^{31,0}}{2560
\sqrt{3}}-\frac{3 \sqrt{\frac{7}{5}} C_{11,22}^{33,0}}{2560}+\frac{\sqrt{5} C_{20,00}^{20,0}}{256}+\frac{1}{256} \sqrt{\frac{5}{3}} C_{20,20}^{20,0}-\frac{7
C_{20,22}^{22,0}}{2560 \sqrt{3}}+\frac{C_{22,00}^{22,0}}{128}+\frac{C_{22,20}^{22,0}}{128 \sqrt{3}}-\frac{\sqrt{\frac{7}{3}} C_{22,22}^{22,0}}{1280}\)

\noindent\(C_{31,1101,1}^{00,0011,0,\text{Loc}}= \frac{C_{11,11}^{31,0}}{64}+\frac{C_{11,11}^{31,1}}{128}-\frac{1}{32} \sqrt{\frac{3}{5}}
C_{22,11}^{22,0}-\frac{1}{64} \sqrt{\frac{3}{5}} C_{22,11}^{22,1}\)

\noindent\(C_{31,1101,1}^{00,0011,0,\text{NonLoc}}= \frac{C_{11,11}^{31,0}}{128}+\frac{C_{11,11}^{31,1}}{64}+\frac{1}{64} \sqrt{\frac{3}{5}}
C_{22,11}^{22,0}+\frac{1}{32} \sqrt{\frac{3}{5}} C_{22,11}^{22,1}\)

\noindent\(C_{31,1101,1}^{00,0011,1,\text{Loc}}= \frac{\sqrt{3} C_{11,11}^{31,1}}{128}-\frac{3 C_{22,11}^{22,1}}{64 \sqrt{5}}\)

\noindent\(C_{31,1101,1}^{00,0011,1,\text{NonLoc}}= \frac{\sqrt{3} C_{11,11}^{31,0}}{128}+\frac{3 C_{22,11}^{22,0}}{64 \sqrt{5}}\)

\noindent\(C_{31,1111,0}^{00,0000,0,\text{Loc}}= \frac{C_{11,11}^{31,0}}{64}+\frac{C_{11,11}^{31,1}}{128}-\frac{1}{32} \sqrt{\frac{3}{5}}
C_{22,11}^{22,0}-\frac{1}{64} \sqrt{\frac{3}{5}} C_{22,11}^{22,1}\)

\noindent\(C_{31,1111,0}^{00,0000,0,\text{NonLoc}}= \frac{C_{11,11}^{31,0}}{128}+\frac{C_{11,11}^{31,1}}{64}+\frac{1}{64} \sqrt{\frac{3}{5}}
C_{22,11}^{22,0}+\frac{1}{32} \sqrt{\frac{3}{5}} C_{22,11}^{22,1}\)

\noindent\(C_{31,1111,0}^{00,0000,1,\text{Loc}}= \frac{\sqrt{3} C_{11,11}^{31,1}}{128}-\frac{3 C_{22,11}^{22,1}}{64 \sqrt{5}}\)

\noindent\(C_{31,1111,0}^{00,0000,1,\text{NonLoc}}= \frac{\sqrt{3} C_{11,11}^{31,0}}{128}+\frac{3 C_{22,11}^{22,0}}{64 \sqrt{5}}\)

\noindent\(C_{33,0011,2}^{11,0011,0,\text{Loc}}= \frac{1}{128} \sqrt{\frac{7}{15}} C_{00,22}^{42,0}+\frac{1}{256} \sqrt{\frac{7}{15}}
C_{00,22}^{42,1}-\frac{1}{128} \sqrt{\frac{7}{15}} C_{11,22}^{31,0}-\frac{1}{256} \sqrt{\frac{7}{15}} C_{11,22}^{31,1}-\frac{3 C_{11,22}^{33,0}}{640}-\frac{3
C_{11,22}^{33,1}}{1280}+\frac{1}{128} \sqrt{\frac{7}{15}} C_{20,22}^{22,0}+\frac{1}{256} \sqrt{\frac{7}{15}} C_{20,22}^{22,1}+\frac{C_{22,22}^{22,0}}{64
\sqrt{15}}+\frac{C_{22,22}^{22,1}}{128 \sqrt{15}}\)

\noindent\(C_{33,0011,2}^{11,0011,0,\text{NonLoc}}= -\frac{1}{256} \sqrt{\frac{7}{15}} C_{00,22}^{42,0}-\frac{1}{128} \sqrt{\frac{7}{15}}
C_{00,22}^{42,1}-\frac{1}{256} \sqrt{\frac{7}{15}} C_{11,22}^{31,0}-\frac{1}{128} \sqrt{\frac{7}{15}} C_{11,22}^{31,1}-\frac{3 C_{11,22}^{33,0}}{1280}-\frac{3
C_{11,22}^{33,1}}{640}-\frac{1}{256} \sqrt{\frac{7}{15}} C_{20,22}^{22,0}-\frac{1}{128} \sqrt{\frac{7}{15}} C_{20,22}^{22,1}-\frac{C_{22,22}^{22,0}}{128
\sqrt{15}}-\frac{C_{22,22}^{22,1}}{64 \sqrt{15}}\)

\noindent\(C_{33,0011,2}^{11,0011,1,\text{Loc}}= \frac{1}{256} \sqrt{\frac{7}{5}} C_{00,22}^{42,1}-\frac{1}{256} \sqrt{\frac{7}{5}}
C_{11,22}^{31,1}-\frac{3 \sqrt{3} C_{11,22}^{33,1}}{1280}+\frac{1}{256} \sqrt{\frac{7}{5}} C_{20,22}^{22,1}+\frac{C_{22,22}^{22,1}}{128 \sqrt{5}}\)

\noindent\(C_{33,0011,2}^{11,0011,1,\text{NonLoc}}= -\frac{1}{256} \sqrt{\frac{7}{5}} C_{00,22}^{42,0}-\frac{1}{256} \sqrt{\frac{7}{5}}
C_{11,22}^{31,0}-\frac{3 \sqrt{3} C_{11,22}^{33,0}}{1280}-\frac{1}{256} \sqrt{\frac{7}{5}} C_{20,22}^{22,0}-\frac{C_{22,22}^{22,0}}{128 \sqrt{5}}\)

\noindent\(C_{40,0000,0}^{00,0000,0,\text{Loc}}= \frac{C_{00,00}^{40,0}}{256}+\frac{C_{00,00}^{40,1}}{512}-\frac{C_{11,00}^{31,0}}{256}-\frac{C_{11,00}^{31,1}}{512}+\frac{C_{20,00}^{20,0}}{256}+\frac{C_{20,00}^{20,1}}{512}+\frac{C_{22,00}^{22,0}}{128
\sqrt{5}}+\frac{C_{22,00}^{22,1}}{256 \sqrt{5}}\)

\noindent\(C_{40,0000,0}^{00,0000,0,\text{NonLoc}}= -\frac{C_{00,00}^{40,0}}{1024}-\frac{C_{00,00}^{40,1}}{512}+\frac{\sqrt{3} C_{00,20}^{40,0}}{1024}+\frac{\sqrt{3}
C_{00,20}^{40,1}}{512}-\frac{C_{11,00}^{31,0}}{1024}-\frac{C_{11,00}^{31,1}}{512}+\frac{\sqrt{3} C_{11,20}^{31,0}}{1024}+\frac{\sqrt{3} C_{11,20}^{31,1}}{512}-\frac{C_{20,00}^{20,0}}{1024}-\frac{C_{20,00}^{20,1}}{512}+\frac{\sqrt{3}
C_{20,20}^{20,0}}{1024}+\frac{\sqrt{3} C_{20,20}^{20,1}}{512}-\frac{C_{22,00}^{22,0}}{512 \sqrt{5}}-\frac{C_{22,00}^{22,1}}{256 \sqrt{5}}+\frac{1}{512}
\sqrt{\frac{3}{5}} C_{22,20}^{22,0}+\frac{1}{256} \sqrt{\frac{3}{5}} C_{22,20}^{22,1}\)

\noindent\(C_{40,0000,0}^{00,0000,1,\text{Loc}}= \frac{\sqrt{3} C_{00,00}^{40,1}}{512}-\frac{\sqrt{3} C_{11,00}^{31,1}}{512}+\frac{\sqrt{3}
C_{20,00}^{20,1}}{512}+\frac{1}{256} \sqrt{\frac{3}{5}} C_{22,00}^{22,1}\)

\noindent\(C_{40,0000,0}^{00,0000,1,\text{NonLoc}}= -\frac{\sqrt{3} C_{00,00}^{40,0}}{1024}+\frac{3 C_{00,20}^{40,0}}{1024}-\frac{\sqrt{3}
C_{11,00}^{31,0}}{1024}+\frac{3 C_{11,20}^{31,0}}{1024}-\frac{\sqrt{3} C_{20,00}^{20,0}}{1024}+\frac{3 C_{20,20}^{20,0}}{1024}-\frac{1}{512} \sqrt{\frac{3}{5}}
C_{22,00}^{22,0}+\frac{3 C_{22,20}^{22,0}}{512 \sqrt{5}}\)

\noindent\(C_{40,0011,1}^{00,0011,0,\text{Loc}}= -\frac{C_{00,20}^{40,0}}{256}-\frac{C_{00,20}^{40,1}}{512}+\frac{C_{11,20}^{31,0}}{256}+\frac{C_{11,20}^{31,1}}{512}-\frac{C_{20,20}^{20,0}}{256}-\frac{C_{20,20}^{20,1}}{512}-\frac{C_{22,20}^{22,0}}{128
\sqrt{5}}-\frac{C_{22,20}^{22,1}}{256 \sqrt{5}}\)

\noindent\(C_{40,0011,1}^{00,0011,0,\text{NonLoc}}= -\frac{\sqrt{3} C_{00,00}^{40,0}}{1024}-\frac{\sqrt{3} C_{00,00}^{40,1}}{512}-\frac{C_{00,20}^{40,0}}{1024}-\frac{C_{00,20}^{40,1}}{512}-\frac{\sqrt{3}
C_{11,00}^{31,0}}{1024}-\frac{\sqrt{3} C_{11,00}^{31,1}}{512}-\frac{C_{11,20}^{31,0}}{1024}-\frac{C_{11,20}^{31,1}}{512}-\frac{\sqrt{3} C_{20,00}^{20,0}}{1024}-\frac{\sqrt{3}
C_{20,00}^{20,1}}{512}-\frac{C_{20,20}^{20,0}}{1024}-\frac{C_{20,20}^{20,1}}{512}-\frac{1}{512} \sqrt{\frac{3}{5}} C_{22,00}^{22,0}-\frac{1}{256}
\sqrt{\frac{3}{5}} C_{22,00}^{22,1}-\frac{C_{22,20}^{22,0}}{512 \sqrt{5}}-\frac{C_{22,20}^{22,1}}{256 \sqrt{5}}\)

\noindent\(C_{40,0011,1}^{00,0011,1,\text{Loc}}= -\frac{\sqrt{3} C_{00,20}^{40,1}}{512}+\frac{\sqrt{3} C_{11,20}^{31,1}}{512}-\frac{\sqrt{3}
C_{20,20}^{20,1}}{512}-\frac{1}{256} \sqrt{\frac{3}{5}} C_{22,20}^{22,1}\)

\noindent\(C_{40,0011,1}^{00,0011,1,\text{NonLoc}}= -\frac{3 C_{00,00}^{40,0}}{1024}-\frac{\sqrt{3} C_{00,20}^{40,0}}{1024}-\frac{3
C_{11,00}^{31,0}}{1024}-\frac{\sqrt{3} C_{11,20}^{31,0}}{1024}-\frac{3 C_{20,00}^{20,0}}{1024}-\frac{\sqrt{3} C_{20,20}^{20,0}}{1024}-\frac{3 C_{22,00}^{22,0}}{512
\sqrt{5}}-\frac{1}{512} \sqrt{\frac{3}{5}} C_{22,20}^{22,0}\)

\noindent\(C_{42,0011,1}^{00,0011,0,\text{Loc}}= -\frac{C_{00,22}^{42,0}}{256}-\frac{C_{00,22}^{42,1}}{512}+\frac{C_{11,22}^{31,0}}{256}+\frac{C_{11,22}^{31,1}}{512}+\frac{3}{256}
\sqrt{\frac{3}{35}} C_{11,22}^{33,0}+\frac{3}{512} \sqrt{\frac{3}{35}} C_{11,22}^{33,1}-\frac{C_{20,22}^{22,0}}{256}-\frac{C_{20,22}^{22,1}}{512}-\frac{C_{22,22}^{22,0}}{128
\sqrt{7}}-\frac{C_{22,22}^{22,1}}{256 \sqrt{7}}\)

\noindent\(C_{42,0011,1}^{00,0011,0,\text{NonLoc}}= \frac{C_{00,22}^{42,0}}{512}+\frac{C_{00,22}^{42,1}}{256}+\frac{C_{11,22}^{31,0}}{512}+\frac{C_{11,22}^{31,1}}{256}+\frac{3}{512}
\sqrt{\frac{3}{35}} C_{11,22}^{33,0}+\frac{3}{256} \sqrt{\frac{3}{35}} C_{11,22}^{33,1}+\frac{C_{20,22}^{22,0}}{512}+\frac{C_{20,22}^{22,1}}{256}+\frac{C_{22,22}^{22,0}}{256
\sqrt{7}}+\frac{C_{22,22}^{22,1}}{128 \sqrt{7}}\)

\noindent\(C_{42,0011,1}^{00,0011,1,\text{Loc}}= -\frac{\sqrt{3} C_{00,22}^{42,1}}{512}+\frac{\sqrt{3} C_{11,22}^{31,1}}{512}+\frac{9
C_{11,22}^{33,1}}{512 \sqrt{35}}-\frac{\sqrt{3} C_{20,22}^{22,1}}{512}-\frac{1}{256} \sqrt{\frac{3}{7}} C_{22,22}^{22,1}\)

\noindent\(C_{42,0011,1}^{00,0011,1,\text{NonLoc}}= \frac{\sqrt{3} C_{00,22}^{42,0}}{512}+\frac{\sqrt{3} C_{11,22}^{31,0}}{512}+\frac{9
C_{11,22}^{33,0}}{512 \sqrt{35}}+\frac{\sqrt{3} C_{20,22}^{22,0}}{512}+\frac{1}{256} \sqrt{\frac{3}{7}} C_{22,22}^{22,0}\)

\vspace{1.0 cm}
\noindent\ \textbf{12 vanishing fourth-order coupling constants of the nonlocal EDF.}
\vspace{1.0 cm}

C_{11,1111,1}^{11,1110,0,\text{Loc}}

C_{11,1111,1}^{11,1110,0,\text{NonLoc}}

C_{11,1111,1}^{11,1110,1,\text{Loc}}

C_{11,1111,1}^{11,1110,1,\text{NonLoc}}

C_{22,0000,2}^{11,1112,0,\text{Loc}}

C_{22,0000,2}^{11,1112,0,\text{NonLoc}}

C_{22,0000,2}^{11,1112,1,\text{Loc}}

C_{22,0000,2}^{11,1112,1,\text{NonLoc}}

C_{22,1112,1}^{11,0000,0,\text{Loc}}

C_{22,1112,1}^{11,0000,0,\text{NonLoc}}

C_{22,1112,1}^{11,0000,1,\text{Loc}}

C_{22,1112,1}^{11,0000,1,\text{NonLoc}}

\vspace{1.0cm}


\noindent\ \textbf{2020 sixth-order isoscalar and isovector coupling constants of the nonlocal EDF expressed by 52 sixth-order finite-range pseudopotential parameters.}

\vspace{1.0cm}

\noindent\(C_{00,3101,1}^{00,3101,0,\text{Loc}}= -\frac{21 C_{00,00}^{60,0}}{160}-\frac{21 C_{00,00}^{60,1}}{320}-\frac{21 C_{11,00}^{51,0}}{160}-\frac{21
C_{11,00}^{51,1}}{320}-\frac{21 C_{20,00}^{40,0}}{160}-\frac{21 C_{20,00}^{40,1}}{320}-\frac{21 C_{22,00}^{42,0}}{80 \sqrt{5}}-\frac{21 C_{22,00}^{42,1}}{160
\sqrt{5}}-\frac{21 C_{31,00}^{31,0}}{160}-\frac{21 C_{31,00}^{31,1}}{320}-\frac{9 \sqrt{21} C_{33,00}^{33,0}}{400}-\frac{9 \sqrt{21} C_{33,00}^{33,1}}{800}\)

\noindent\(C_{00,3101,1}^{00,3101,0,\text{NonLoc}}= \frac{21 C_{00,00}^{60,0}}{640}+\frac{21 C_{00,00}^{60,1}}{320}-\frac{21 \sqrt{3}
C_{00,20}^{60,0}}{640}-\frac{21 \sqrt{3} C_{00,20}^{60,1}}{320}-\frac{21 C_{11,00}^{51,0}}{640}-\frac{21 C_{11,00}^{51,1}}{320}+\frac{21 C_{20,00}^{40,0}}{640}+\frac{21
C_{20,00}^{40,1}}{320}-\frac{21 \sqrt{3} C_{20,20}^{40,0}}{640}-\frac{21 \sqrt{3} C_{20,20}^{40,1}}{320}+\frac{21 C_{22,00}^{42,0}}{320 \sqrt{5}}+\frac{21
C_{22,00}^{42,1}}{160 \sqrt{5}}-\frac{21}{320} \sqrt{\frac{3}{5}} C_{22,20}^{42,0}-\frac{21}{160} \sqrt{\frac{3}{5}} C_{22,20}^{42,1}-\frac{21 C_{31,00}^{31,0}}{640}-\frac{21
C_{31,00}^{31,1}}{320}+\frac{21 \sqrt{3} C_{31,20}^{31,0}}{640}+\frac{21 \sqrt{3} C_{31,20}^{31,1}}{320}-\frac{9 \sqrt{21} C_{33,00}^{33,0}}{1600}-\frac{9
\sqrt{21} C_{33,00}^{33,1}}{800}+\frac{27 \sqrt{7} C_{33,20}^{33,0}}{1600}+\frac{27 \sqrt{7} C_{33,20}^{33,1}}{800}\)

\noindent\(C_{00,3101,1}^{00,3101,1,\text{Loc}}= -\frac{21 \sqrt{3} C_{00,00}^{60,1}}{320}-\frac{21 \sqrt{3} C_{11,00}^{51,1}}{320}-\frac{21
\sqrt{3} C_{20,00}^{40,1}}{320}-\frac{21}{160} \sqrt{\frac{3}{5}} C_{22,00}^{42,1}-\frac{21 \sqrt{3} C_{31,00}^{31,1}}{320}-\frac{27 \sqrt{7} C_{33,00}^{33,1}}{800}\)

\noindent\(C_{00,3101,1}^{00,3101,1,\text{NonLoc}}= \frac{21 \sqrt{3} C_{00,00}^{60,0}}{640}-\frac{63 C_{00,20}^{60,0}}{640}-\frac{21
\sqrt{3} C_{11,00}^{51,0}}{640}+\frac{21 \sqrt{3} C_{20,00}^{40,0}}{640}-\frac{63 C_{20,20}^{40,0}}{640}+\frac{21}{320} \sqrt{\frac{3}{5}} C_{22,00}^{42,0}-\frac{63
C_{22,20}^{42,0}}{320 \sqrt{5}}-\frac{21 \sqrt{3} C_{31,00}^{31,0}}{640}+\frac{63 C_{31,20}^{31,0}}{640}-\frac{27 \sqrt{7} C_{33,00}^{33,0}}{1600}+\frac{27
\sqrt{21} C_{33,20}^{33,0}}{1600}\)

\noindent\(C_{00,3110,0}^{00,3110,0,\text{Loc}}= \frac{7 C_{00,20}^{60,0}}{160}+\frac{7 C_{00,20}^{60,1}}{320}+\frac{21 C_{00,22}^{62,0}}{160
\sqrt{5}}+\frac{21 C_{00,22}^{62,1}}{320 \sqrt{5}}+\frac{21 C_{11,22}^{51,0}}{160 \sqrt{5}}+\frac{21 C_{11,22}^{51,1}}{320 \sqrt{5}}+\frac{9 \sqrt{21}
C_{11,22}^{53,0}}{800}+\frac{9 \sqrt{21} C_{11,22}^{53,1}}{1600}+\frac{7 C_{20,20}^{40,0}}{160}+\frac{7 C_{20,20}^{40,1}}{320}+\frac{21 C_{20,22}^{42,0}}{160
\sqrt{5}}+\frac{21 C_{20,22}^{42,1}}{320 \sqrt{5}}+\frac{7 C_{22,20}^{42,0}}{80 \sqrt{5}}+\frac{7 C_{22,20}^{42,1}}{160 \sqrt{5}}+\frac{21 C_{22,22}^{40,0}}{160
\sqrt{5}}+\frac{21 C_{22,22}^{40,1}}{320 \sqrt{5}}+\frac{3}{80} \sqrt{\frac{7}{5}} C_{22,22}^{42,0}+\frac{3}{160} \sqrt{\frac{7}{5}} C_{22,22}^{42,1}+\frac{9
C_{22,22}^{44,0}}{200}+\frac{9 C_{22,22}^{44,1}}{400}+\frac{7 C_{31,20}^{31,0}}{160}+\frac{7 C_{31,20}^{31,1}}{320}+\frac{21 C_{31,22}^{31,0}}{160
\sqrt{5}}+\frac{21 C_{31,22}^{31,1}}{320 \sqrt{5}}+\frac{9 \sqrt{21} C_{31,22}^{33,0}}{800}+\frac{9 \sqrt{21} C_{31,22}^{33,1}}{1600}+\frac{3 \sqrt{21}
C_{33,20}^{33,0}}{400}+\frac{3 \sqrt{21} C_{33,20}^{33,1}}{800}+\frac{9}{200} \sqrt{\frac{7}{10}} C_{33,22}^{33,0}+\frac{9}{400} \sqrt{\frac{7}{10}}
C_{33,22}^{33,1}\)

\noindent\(C_{00,3110,0}^{00,3110,0,\text{NonLoc}}= \frac{7 \sqrt{3} C_{00,00}^{60,0}}{640}+\frac{7 \sqrt{3} C_{00,00}^{60,1}}{320}+\frac{7
C_{00,20}^{60,0}}{640}+\frac{7 C_{00,20}^{60,1}}{320}-\frac{21 C_{00,22}^{62,0}}{320 \sqrt{5}}-\frac{21 C_{00,22}^{62,1}}{160 \sqrt{5}}-\frac{7 \sqrt{3}
C_{11,00}^{51,0}}{640}-\frac{7 \sqrt{3} C_{11,00}^{51,1}}{320}+\frac{21 C_{11,22}^{51,0}}{320 \sqrt{5}}+\frac{21 C_{11,22}^{51,1}}{160 \sqrt{5}}+\frac{9
\sqrt{21} C_{11,22}^{53,0}}{1600}+\frac{9 \sqrt{21} C_{11,22}^{53,1}}{800}+\frac{7 \sqrt{3} C_{20,00}^{40,0}}{640}+\frac{7 \sqrt{3} C_{20,00}^{40,1}}{320}+\frac{7
C_{20,20}^{40,0}}{640}+\frac{7 C_{20,20}^{40,1}}{320}-\frac{21 C_{20,22}^{42,0}}{320 \sqrt{5}}-\frac{21 C_{20,22}^{42,1}}{160 \sqrt{5}}+\frac{7}{320}
\sqrt{\frac{3}{5}} C_{22,00}^{42,0}+\frac{7}{160} \sqrt{\frac{3}{5}} C_{22,00}^{42,1}+\frac{7 C_{22,20}^{42,0}}{320 \sqrt{5}}+\frac{7 C_{22,20}^{42,1}}{160
\sqrt{5}}-\frac{21 C_{22,22}^{40,0}}{320 \sqrt{5}}-\frac{21 C_{22,22}^{40,1}}{160 \sqrt{5}}-\frac{3}{160} \sqrt{\frac{7}{5}} C_{22,22}^{42,0}-\frac{3}{80}
\sqrt{\frac{7}{5}} C_{22,22}^{42,1}-\frac{9 C_{22,22}^{44,0}}{400}-\frac{9 C_{22,22}^{44,1}}{200}-\frac{7 \sqrt{3} C_{31,00}^{31,0}}{640}-\frac{7
\sqrt{3} C_{31,00}^{31,1}}{320}-\frac{7 C_{31,20}^{31,0}}{640}-\frac{7 C_{31,20}^{31,1}}{320}+\frac{21 C_{31,22}^{31,0}}{320 \sqrt{5}}+\frac{21 C_{31,22}^{31,1}}{160
\sqrt{5}}+\frac{9 \sqrt{21} C_{31,22}^{33,0}}{1600}+\frac{9 \sqrt{21} C_{31,22}^{33,1}}{800}-\frac{9 \sqrt{7} C_{33,00}^{33,0}}{1600}-\frac{9 \sqrt{7}
C_{33,00}^{33,1}}{800}-\frac{3 \sqrt{21} C_{33,20}^{33,0}}{1600}-\frac{3 \sqrt{21} C_{33,20}^{33,1}}{800}+\frac{9}{400} \sqrt{\frac{7}{10}} C_{33,22}^{33,0}+\frac{9}{200}
\sqrt{\frac{7}{10}} C_{33,22}^{33,1}\)

\noindent\(C_{00,3110,0}^{00,3110,1,\text{Loc}}= \frac{7 \sqrt{3} C_{00,20}^{60,1}}{320}+\frac{21}{320} \sqrt{\frac{3}{5}} C_{00,22}^{62,1}+\frac{21}{320}
\sqrt{\frac{3}{5}} C_{11,22}^{51,1}+\frac{27 \sqrt{7} C_{11,22}^{53,1}}{1600}+\frac{7 \sqrt{3} C_{20,20}^{40,1}}{320}+\frac{21}{320} \sqrt{\frac{3}{5}}
C_{20,22}^{42,1}+\frac{7}{160} \sqrt{\frac{3}{5}} C_{22,20}^{42,1}+\frac{21}{320} \sqrt{\frac{3}{5}} C_{22,22}^{40,1}+\frac{3}{160} \sqrt{\frac{21}{5}}
C_{22,22}^{42,1}+\frac{9 \sqrt{3} C_{22,22}^{44,1}}{400}+\frac{7 \sqrt{3} C_{31,20}^{31,1}}{320}+\frac{21}{320} \sqrt{\frac{3}{5}} C_{31,22}^{31,1}+\frac{27
\sqrt{7} C_{31,22}^{33,1}}{1600}+\frac{9 \sqrt{7} C_{33,20}^{33,1}}{800}+\frac{9}{400} \sqrt{\frac{21}{10}} C_{33,22}^{33,1}\)

\noindent\(C_{00,3110,0}^{00,3110,1,\text{NonLoc}}= \frac{21 C_{00,00}^{60,0}}{640}+\frac{7 \sqrt{3} C_{00,20}^{60,0}}{640}-\frac{21}{320}
\sqrt{\frac{3}{5}} C_{00,22}^{62,0}-\frac{21 C_{11,00}^{51,0}}{640}+\frac{21}{320} \sqrt{\frac{3}{5}} C_{11,22}^{51,0}+\frac{27 \sqrt{7} C_{11,22}^{53,0}}{1600}+\frac{21
C_{20,00}^{40,0}}{640}+\frac{7 \sqrt{3} C_{20,20}^{40,0}}{640}-\frac{21}{320} \sqrt{\frac{3}{5}} C_{20,22}^{42,0}+\frac{21 C_{22,00}^{42,0}}{320
\sqrt{5}}+\frac{7}{320} \sqrt{\frac{3}{5}} C_{22,20}^{42,0}-\frac{21}{320} \sqrt{\frac{3}{5}} C_{22,22}^{40,0}-\frac{3}{160} \sqrt{\frac{21}{5}}
C_{22,22}^{42,0}-\frac{9 \sqrt{3} C_{22,22}^{44,0}}{400}-\frac{21 C_{31,00}^{31,0}}{640}-\frac{7 \sqrt{3} C_{31,20}^{31,0}}{640}+\frac{21}{320} \sqrt{\frac{3}{5}}
C_{31,22}^{31,0}+\frac{27 \sqrt{7} C_{31,22}^{33,0}}{1600}-\frac{9 \sqrt{21} C_{33,00}^{33,0}}{1600}-\frac{9 \sqrt{7} C_{33,20}^{33,0}}{1600}+\frac{9}{400}
\sqrt{\frac{21}{10}} C_{33,22}^{33,0}\)

\noindent\(C_{00,3111,1}^{00,3111,0,\text{Loc}}= \frac{7 \sqrt{3} C_{00,20}^{60,0}}{160}+\frac{7 \sqrt{3} C_{00,20}^{60,1}}{320}-\frac{21}{320}
\sqrt{\frac{3}{5}} C_{00,22}^{62,0}-\frac{21}{640} \sqrt{\frac{3}{5}} C_{00,22}^{62,1}-\frac{21}{320} \sqrt{\frac{3}{5}} C_{11,22}^{51,0}-\frac{21}{640}
\sqrt{\frac{3}{5}} C_{11,22}^{51,1}-\frac{27 \sqrt{7} C_{11,22}^{53,0}}{1600}-\frac{27 \sqrt{7} C_{11,22}^{53,1}}{3200}+\frac{7 \sqrt{3} C_{20,20}^{40,0}}{160}+\frac{7
\sqrt{3} C_{20,20}^{40,1}}{320}-\frac{21}{320} \sqrt{\frac{3}{5}} C_{20,22}^{42,0}-\frac{21}{640} \sqrt{\frac{3}{5}} C_{20,22}^{42,1}+\frac{7}{80}
\sqrt{\frac{3}{5}} C_{22,20}^{42,0}+\frac{7}{160} \sqrt{\frac{3}{5}} C_{22,20}^{42,1}-\frac{21}{320} \sqrt{\frac{3}{5}} C_{22,22}^{40,0}-\frac{21}{640}
\sqrt{\frac{3}{5}} C_{22,22}^{40,1}-\frac{3}{160} \sqrt{\frac{21}{5}} C_{22,22}^{42,0}-\frac{3}{320} \sqrt{\frac{21}{5}} C_{22,22}^{42,1}-\frac{9
\sqrt{3} C_{22,22}^{44,0}}{400}-\frac{9 \sqrt{3} C_{22,22}^{44,1}}{800}+\frac{7 \sqrt{3} C_{31,20}^{31,0}}{160}+\frac{7 \sqrt{3} C_{31,20}^{31,1}}{320}-\frac{21}{320}
\sqrt{\frac{3}{5}} C_{31,22}^{31,0}-\frac{21}{640} \sqrt{\frac{3}{5}} C_{31,22}^{31,1}-\frac{27 \sqrt{7} C_{31,22}^{33,0}}{1600}-\frac{27 \sqrt{7}
C_{31,22}^{33,1}}{3200}+\frac{9 \sqrt{7} C_{33,20}^{33,0}}{400}+\frac{9 \sqrt{7} C_{33,20}^{33,1}}{800}-\frac{9}{400} \sqrt{\frac{21}{10}} C_{33,22}^{33,0}-\frac{9}{800}
\sqrt{\frac{21}{10}} C_{33,22}^{33,1}\)

\noindent\(C_{00,3111,1}^{00,3111,0,\text{NonLoc}}= \frac{21 C_{00,00}^{60,0}}{640}+\frac{21 C_{00,00}^{60,1}}{320}+\frac{7 \sqrt{3}
C_{00,20}^{60,0}}{640}+\frac{7 \sqrt{3} C_{00,20}^{60,1}}{320}+\frac{21}{640} \sqrt{\frac{3}{5}} C_{00,22}^{62,0}+\frac{21}{320} \sqrt{\frac{3}{5}}
C_{00,22}^{62,1}-\frac{21 C_{11,00}^{51,0}}{640}-\frac{21 C_{11,00}^{51,1}}{320}-\frac{21}{640} \sqrt{\frac{3}{5}} C_{11,22}^{51,0}-\frac{21}{320}
\sqrt{\frac{3}{5}} C_{11,22}^{51,1}-\frac{27 \sqrt{7} C_{11,22}^{53,0}}{3200}-\frac{27 \sqrt{7} C_{11,22}^{53,1}}{1600}+\frac{21 C_{20,00}^{40,0}}{640}+\frac{21
C_{20,00}^{40,1}}{320}+\frac{7 \sqrt{3} C_{20,20}^{40,0}}{640}+\frac{7 \sqrt{3} C_{20,20}^{40,1}}{320}+\frac{21}{640} \sqrt{\frac{3}{5}} C_{20,22}^{42,0}+\frac{21}{320}
\sqrt{\frac{3}{5}} C_{20,22}^{42,1}+\frac{21 C_{22,00}^{42,0}}{320 \sqrt{5}}+\frac{21 C_{22,00}^{42,1}}{160 \sqrt{5}}+\frac{7}{320} \sqrt{\frac{3}{5}}
C_{22,20}^{42,0}+\frac{7}{160} \sqrt{\frac{3}{5}} C_{22,20}^{42,1}+\frac{21}{640} \sqrt{\frac{3}{5}} C_{22,22}^{40,0}+\frac{21}{320} \sqrt{\frac{3}{5}}
C_{22,22}^{40,1}+\frac{3}{320} \sqrt{\frac{21}{5}} C_{22,22}^{42,0}+\frac{3}{160} \sqrt{\frac{21}{5}} C_{22,22}^{42,1}+\frac{9 \sqrt{3} C_{22,22}^{44,0}}{800}+\frac{9
\sqrt{3} C_{22,22}^{44,1}}{400}-\frac{21 C_{31,00}^{31,0}}{640}-\frac{21 C_{31,00}^{31,1}}{320}-\frac{7 \sqrt{3} C_{31,20}^{31,0}}{640}-\frac{7 \sqrt{3}
C_{31,20}^{31,1}}{320}-\frac{21}{640} \sqrt{\frac{3}{5}} C_{31,22}^{31,0}-\frac{21}{320} \sqrt{\frac{3}{5}} C_{31,22}^{31,1}-\frac{27 \sqrt{7} C_{31,22}^{33,0}}{3200}-\frac{27
\sqrt{7} C_{31,22}^{33,1}}{1600}-\frac{9 \sqrt{21} C_{33,00}^{33,0}}{1600}-\frac{9 \sqrt{21} C_{33,00}^{33,1}}{800}-\frac{9 \sqrt{7} C_{33,20}^{33,0}}{1600}-\frac{9
\sqrt{7} C_{33,20}^{33,1}}{800}-\frac{9}{800} \sqrt{\frac{21}{10}} C_{33,22}^{33,0}-\frac{9}{400} \sqrt{\frac{21}{10}} C_{33,22}^{33,1}\)

\noindent\(C_{00,3111,1}^{00,3111,1,\text{Loc}}= \frac{21 C_{00,20}^{60,1}}{320}-\frac{63 C_{00,22}^{62,1}}{640 \sqrt{5}}-\frac{63
C_{11,22}^{51,1}}{640 \sqrt{5}}-\frac{27 \sqrt{21} C_{11,22}^{53,1}}{3200}+\frac{21 C_{20,20}^{40,1}}{320}-\frac{63 C_{20,22}^{42,1}}{640 \sqrt{5}}+\frac{21
C_{22,20}^{42,1}}{160 \sqrt{5}}-\frac{63 C_{22,22}^{40,1}}{640 \sqrt{5}}-\frac{9}{320} \sqrt{\frac{7}{5}} C_{22,22}^{42,1}-\frac{27 C_{22,22}^{44,1}}{800}+\frac{21
C_{31,20}^{31,1}}{320}-\frac{63 C_{31,22}^{31,1}}{640 \sqrt{5}}-\frac{27 \sqrt{21} C_{31,22}^{33,1}}{3200}+\frac{9 \sqrt{21} C_{33,20}^{33,1}}{800}-\frac{27}{800}
\sqrt{\frac{7}{10}} C_{33,22}^{33,1}\)

\noindent\(C_{00,3111,1}^{00,3111,1,\text{NonLoc}}= \frac{21 \sqrt{3} C_{00,00}^{60,0}}{640}+\frac{21 C_{00,20}^{60,0}}{640}+\frac{63
C_{00,22}^{62,0}}{640 \sqrt{5}}-\frac{21 \sqrt{3} C_{11,00}^{51,0}}{640}-\frac{63 C_{11,22}^{51,0}}{640 \sqrt{5}}-\frac{27 \sqrt{21} C_{11,22}^{53,0}}{3200}+\frac{21
\sqrt{3} C_{20,00}^{40,0}}{640}+\frac{21 C_{20,20}^{40,0}}{640}+\frac{63 C_{20,22}^{42,0}}{640 \sqrt{5}}+\frac{21}{320} \sqrt{\frac{3}{5}} C_{22,00}^{42,0}+\frac{21
C_{22,20}^{42,0}}{320 \sqrt{5}}+\frac{63 C_{22,22}^{40,0}}{640 \sqrt{5}}+\frac{9}{320} \sqrt{\frac{7}{5}} C_{22,22}^{42,0}+\frac{27 C_{22,22}^{44,0}}{800}-\frac{21
\sqrt{3} C_{31,00}^{31,0}}{640}-\frac{21 C_{31,20}^{31,0}}{640}-\frac{63 C_{31,22}^{31,0}}{640 \sqrt{5}}-\frac{27 \sqrt{21} C_{31,22}^{33,0}}{3200}-\frac{27
\sqrt{7} C_{33,00}^{33,0}}{1600}-\frac{9 \sqrt{21} C_{33,20}^{33,0}}{1600}-\frac{27}{800} \sqrt{\frac{7}{10}} C_{33,22}^{33,0}\)

\noindent\(C_{00,3112,2}^{00,3112,0,\text{Loc}}= \frac{7 C_{00,20}^{60,0}}{32 \sqrt{5}}+\frac{7 C_{00,20}^{60,1}}{64 \sqrt{5}}+\frac{21
C_{00,22}^{62,0}}{1600}+\frac{21 C_{00,22}^{62,1}}{3200}+\frac{21 C_{11,22}^{51,0}}{1600}+\frac{21 C_{11,22}^{51,1}}{3200}+\frac{9 \sqrt{\frac{21}{5}}
C_{11,22}^{53,0}}{1600}+\frac{9 \sqrt{\frac{21}{5}} C_{11,22}^{53,1}}{3200}+\frac{7 C_{20,20}^{40,0}}{32 \sqrt{5}}+\frac{7 C_{20,20}^{40,1}}{64 \sqrt{5}}+\frac{21
C_{20,22}^{42,0}}{1600}+\frac{21 C_{20,22}^{42,1}}{3200}+\frac{7 C_{22,20}^{42,0}}{80}+\frac{7 C_{22,20}^{42,1}}{160}+\frac{21 C_{22,22}^{40,0}}{1600}+\frac{21
C_{22,22}^{40,1}}{3200}+\frac{3 \sqrt{7} C_{22,22}^{42,0}}{800}+\frac{3 \sqrt{7} C_{22,22}^{42,1}}{1600}+\frac{9 C_{22,22}^{44,0}}{400 \sqrt{5}}+\frac{9
C_{22,22}^{44,1}}{800 \sqrt{5}}+\frac{7 C_{31,20}^{31,0}}{32 \sqrt{5}}+\frac{7 C_{31,20}^{31,1}}{64 \sqrt{5}}+\frac{21 C_{31,22}^{31,0}}{1600}+\frac{21
C_{31,22}^{31,1}}{3200}+\frac{9 \sqrt{\frac{21}{5}} C_{31,22}^{33,0}}{1600}+\frac{9 \sqrt{\frac{21}{5}} C_{31,22}^{33,1}}{3200}+\frac{3}{80} \sqrt{\frac{21}{5}}
C_{33,20}^{33,0}+\frac{3}{160} \sqrt{\frac{21}{5}} C_{33,20}^{33,1}+\frac{9 \sqrt{\frac{7}{2}} C_{33,22}^{33,0}}{2000}+\frac{9 \sqrt{\frac{7}{2}}
C_{33,22}^{33,1}}{4000}\)

\noindent\(C_{00,3112,2}^{00,3112,0,\text{NonLoc}}= \frac{7}{128} \sqrt{\frac{3}{5}} C_{00,00}^{60,0}+\frac{7}{64} \sqrt{\frac{3}{5}}
C_{00,00}^{60,1}+\frac{7 C_{00,20}^{60,0}}{128 \sqrt{5}}+\frac{7 C_{00,20}^{60,1}}{64 \sqrt{5}}-\frac{21 C_{00,22}^{62,0}}{3200}-\frac{21 C_{00,22}^{62,1}}{1600}-\frac{7}{128}
\sqrt{\frac{3}{5}} C_{11,00}^{51,0}-\frac{7}{64} \sqrt{\frac{3}{5}} C_{11,00}^{51,1}+\frac{21 C_{11,22}^{51,0}}{3200}+\frac{21 C_{11,22}^{51,1}}{1600}+\frac{9
\sqrt{\frac{21}{5}} C_{11,22}^{53,0}}{3200}+\frac{9 \sqrt{\frac{21}{5}} C_{11,22}^{53,1}}{1600}+\frac{7}{128} \sqrt{\frac{3}{5}} C_{20,00}^{40,0}+\frac{7}{64}
\sqrt{\frac{3}{5}} C_{20,00}^{40,1}+\frac{7 C_{20,20}^{40,0}}{128 \sqrt{5}}+\frac{7 C_{20,20}^{40,1}}{64 \sqrt{5}}-\frac{21 C_{20,22}^{42,0}}{3200}-\frac{21
C_{20,22}^{42,1}}{1600}+\frac{7 \sqrt{3} C_{22,00}^{42,0}}{320}+\frac{7 \sqrt{3} C_{22,00}^{42,1}}{160}+\frac{7 C_{22,20}^{42,0}}{320}+\frac{7 C_{22,20}^{42,1}}{160}-\frac{21
C_{22,22}^{40,0}}{3200}-\frac{21 C_{22,22}^{40,1}}{1600}-\frac{3 \sqrt{7} C_{22,22}^{42,0}}{1600}-\frac{3 \sqrt{7} C_{22,22}^{42,1}}{800}-\frac{9
C_{22,22}^{44,0}}{800 \sqrt{5}}-\frac{9 C_{22,22}^{44,1}}{400 \sqrt{5}}-\frac{7}{128} \sqrt{\frac{3}{5}} C_{31,00}^{31,0}-\frac{7}{64} \sqrt{\frac{3}{5}}
C_{31,00}^{31,1}-\frac{7 C_{31,20}^{31,0}}{128 \sqrt{5}}-\frac{7 C_{31,20}^{31,1}}{64 \sqrt{5}}+\frac{21 C_{31,22}^{31,0}}{3200}+\frac{21 C_{31,22}^{31,1}}{1600}+\frac{9
\sqrt{\frac{21}{5}} C_{31,22}^{33,0}}{3200}+\frac{9 \sqrt{\frac{21}{5}} C_{31,22}^{33,1}}{1600}-\frac{9}{320} \sqrt{\frac{7}{5}} C_{33,00}^{33,0}-\frac{9}{160}
\sqrt{\frac{7}{5}} C_{33,00}^{33,1}-\frac{3}{320} \sqrt{\frac{21}{5}} C_{33,20}^{33,0}-\frac{3}{160} \sqrt{\frac{21}{5}} C_{33,20}^{33,1}+\frac{9
\sqrt{\frac{7}{2}} C_{33,22}^{33,0}}{4000}+\frac{9 \sqrt{\frac{7}{2}} C_{33,22}^{33,1}}{2000}\)

\noindent\(C_{00,3112,2}^{00,3112,1,\text{Loc}}= \frac{7}{64} \sqrt{\frac{3}{5}} C_{00,20}^{60,1}+\frac{21 \sqrt{3} C_{00,22}^{62,1}}{3200}+\frac{21
\sqrt{3} C_{11,22}^{51,1}}{3200}+\frac{27 \sqrt{\frac{7}{5}} C_{11,22}^{53,1}}{3200}+\frac{7}{64} \sqrt{\frac{3}{5}} C_{20,20}^{40,1}+\frac{21 \sqrt{3}
C_{20,22}^{42,1}}{3200}+\frac{7 \sqrt{3} C_{22,20}^{42,1}}{160}+\frac{21 \sqrt{3} C_{22,22}^{40,1}}{3200}+\frac{3 \sqrt{21} C_{22,22}^{42,1}}{1600}+\frac{9}{800}
\sqrt{\frac{3}{5}} C_{22,22}^{44,1}+\frac{7}{64} \sqrt{\frac{3}{5}} C_{31,20}^{31,1}+\frac{21 \sqrt{3} C_{31,22}^{31,1}}{3200}+\frac{27 \sqrt{\frac{7}{5}}
C_{31,22}^{33,1}}{3200}+\frac{9}{160} \sqrt{\frac{7}{5}} C_{33,20}^{33,1}+\frac{9 \sqrt{\frac{21}{2}} C_{33,22}^{33,1}}{4000}\)

\noindent\(C_{00,3112,2}^{00,3112,1,\text{NonLoc}}= \frac{21 C_{00,00}^{60,0}}{128 \sqrt{5}}+\frac{7}{128} \sqrt{\frac{3}{5}} C_{00,20}^{60,0}-\frac{21
\sqrt{3} C_{00,22}^{62,0}}{3200}-\frac{21 C_{11,00}^{51,0}}{128 \sqrt{5}}+\frac{21 \sqrt{3} C_{11,22}^{51,0}}{3200}+\frac{27 \sqrt{\frac{7}{5}} C_{11,22}^{53,0}}{3200}+\frac{21
C_{20,00}^{40,0}}{128 \sqrt{5}}+\frac{7}{128} \sqrt{\frac{3}{5}} C_{20,20}^{40,0}-\frac{21 \sqrt{3} C_{20,22}^{42,0}}{3200}+\frac{21 C_{22,00}^{42,0}}{320}+\frac{7
\sqrt{3} C_{22,20}^{42,0}}{320}-\frac{21 \sqrt{3} C_{22,22}^{40,0}}{3200}-\frac{3 \sqrt{21} C_{22,22}^{42,0}}{1600}-\frac{9}{800} \sqrt{\frac{3}{5}}
C_{22,22}^{44,0}-\frac{21 C_{31,00}^{31,0}}{128 \sqrt{5}}-\frac{7}{128} \sqrt{\frac{3}{5}} C_{31,20}^{31,0}+\frac{21 \sqrt{3} C_{31,22}^{31,0}}{3200}+\frac{27
\sqrt{\frac{7}{5}} C_{31,22}^{33,0}}{3200}-\frac{9}{320} \sqrt{\frac{21}{5}} C_{33,00}^{33,0}-\frac{9}{320} \sqrt{\frac{7}{5}} C_{33,20}^{33,0}+\frac{9
\sqrt{\frac{21}{2}} C_{33,22}^{33,0}}{4000}\)

\noindent\(C_{00,3303,3}^{00,3303,0,\text{Loc}}= -\frac{1}{48} \sqrt{\frac{7}{3}} C_{00,00}^{60,0}-\frac{1}{96} \sqrt{\frac{7}{3}}
C_{00,00}^{60,1}-\frac{1}{48} \sqrt{\frac{7}{3}} C_{11,00}^{51,0}-\frac{1}{96} \sqrt{\frac{7}{3}} C_{11,00}^{51,1}-\frac{1}{48} \sqrt{\frac{7}{3}}
C_{20,00}^{40,0}-\frac{1}{96} \sqrt{\frac{7}{3}} C_{20,00}^{40,1}-\frac{1}{24} \sqrt{\frac{7}{15}} C_{22,00}^{42,0}-\frac{1}{48} \sqrt{\frac{7}{15}}
C_{22,00}^{42,1}-\frac{1}{48} \sqrt{\frac{7}{3}} C_{31,00}^{31,0}-\frac{1}{96} \sqrt{\frac{7}{3}} C_{31,00}^{31,1}-\frac{C_{33,00}^{33,0}}{40}-\frac{C_{33,00}^{33,1}}{80}\)

\noindent\(C_{00,3303,3}^{00,3303,0,\text{NonLoc}}= \frac{1}{192} \sqrt{\frac{7}{3}} C_{00,00}^{60,0}+\frac{1}{96} \sqrt{\frac{7}{3}}
C_{00,00}^{60,1}-\frac{\sqrt{7} C_{00,20}^{60,0}}{192}-\frac{\sqrt{7} C_{00,20}^{60,1}}{96}-\frac{1}{192} \sqrt{\frac{7}{3}} C_{11,00}^{51,0}-\frac{1}{96}
\sqrt{\frac{7}{3}} C_{11,00}^{51,1}+\frac{1}{192} \sqrt{\frac{7}{3}} C_{20,00}^{40,0}+\frac{1}{96} \sqrt{\frac{7}{3}} C_{20,00}^{40,1}-\frac{\sqrt{7}
C_{20,20}^{40,0}}{192}-\frac{\sqrt{7} C_{20,20}^{40,1}}{96}+\frac{1}{96} \sqrt{\frac{7}{15}} C_{22,00}^{42,0}+\frac{1}{48} \sqrt{\frac{7}{15}} C_{22,00}^{42,1}-\frac{1}{96}
\sqrt{\frac{7}{5}} C_{22,20}^{42,0}-\frac{1}{48} \sqrt{\frac{7}{5}} C_{22,20}^{42,1}-\frac{1}{192} \sqrt{\frac{7}{3}} C_{31,00}^{31,0}-\frac{1}{96}
\sqrt{\frac{7}{3}} C_{31,00}^{31,1}+\frac{\sqrt{7} C_{31,20}^{31,0}}{192}+\frac{\sqrt{7} C_{31,20}^{31,1}}{96}-\frac{C_{33,00}^{33,0}}{160}-\frac{C_{33,00}^{33,1}}{80}+\frac{\sqrt{3}
C_{33,20}^{33,0}}{160}+\frac{\sqrt{3} C_{33,20}^{33,1}}{80}\)

\noindent\(C_{00,3303,3}^{00,3303,1,\text{Loc}}= -\frac{\sqrt{7} C_{00,00}^{60,1}}{96}-\frac{\sqrt{7} C_{11,00}^{51,1}}{96}-\frac{\sqrt{7}
C_{20,00}^{40,1}}{96}-\frac{1}{48} \sqrt{\frac{7}{5}} C_{22,00}^{42,1}-\frac{\sqrt{7} C_{31,00}^{31,1}}{96}-\frac{\sqrt{3} C_{33,00}^{33,1}}{80}\)

\noindent\(C_{00,3303,3}^{00,3303,1,\text{NonLoc}}= \frac{\sqrt{7} C_{00,00}^{60,0}}{192}-\frac{1}{64} \sqrt{\frac{7}{3}} C_{00,20}^{60,0}-\frac{\sqrt{7}
C_{11,00}^{51,0}}{192}+\frac{\sqrt{7} C_{20,00}^{40,0}}{192}-\frac{1}{64} \sqrt{\frac{7}{3}} C_{20,20}^{40,0}+\frac{1}{96} \sqrt{\frac{7}{5}} C_{22,00}^{42,0}-\frac{1}{32}
\sqrt{\frac{7}{15}} C_{22,20}^{42,0}-\frac{\sqrt{7} C_{31,00}^{31,0}}{192}+\frac{1}{64} \sqrt{\frac{7}{3}} C_{31,20}^{31,0}-\frac{\sqrt{3} C_{33,00}^{33,0}}{160}+\frac{3
C_{33,20}^{33,0}}{160}\)

\noindent\(C_{00,3312,2}^{00,3112,0,\text{Loc}}= \frac{3}{80} \sqrt{\frac{21}{5}} C_{00,22}^{62,0}+\frac{3}{160} \sqrt{\frac{21}{5}}
C_{00,22}^{62,1}+\frac{3}{80} \sqrt{\frac{21}{5}} C_{11,22}^{51,0}+\frac{3}{160} \sqrt{\frac{21}{5}} C_{11,22}^{51,1}+\frac{27 C_{11,22}^{53,0}}{400}+\frac{27
C_{11,22}^{53,1}}{800}+\frac{3}{80} \sqrt{\frac{21}{5}} C_{20,22}^{42,0}+\frac{3}{160} \sqrt{\frac{21}{5}} C_{20,22}^{42,1}+\frac{3}{80} \sqrt{\frac{21}{5}}
C_{22,22}^{40,0}+\frac{3}{160} \sqrt{\frac{21}{5}} C_{22,22}^{40,1}+\frac{3}{40} \sqrt{\frac{3}{5}} C_{22,22}^{42,0}+\frac{3}{80} \sqrt{\frac{3}{5}}
C_{22,22}^{42,1}+\frac{9}{100} \sqrt{\frac{3}{7}} C_{22,22}^{44,0}+\frac{9}{200} \sqrt{\frac{3}{7}} C_{22,22}^{44,1}+\frac{3}{80} \sqrt{\frac{21}{5}}
C_{31,22}^{31,0}+\frac{3}{160} \sqrt{\frac{21}{5}} C_{31,22}^{31,1}+\frac{27 C_{31,22}^{33,0}}{400}+\frac{27 C_{31,22}^{33,1}}{800}+\frac{9}{100}
\sqrt{\frac{3}{10}} C_{33,22}^{33,0}+\frac{9}{200} \sqrt{\frac{3}{10}} C_{33,22}^{33,1}\)

\noindent\(C_{00,3312,2}^{00,3112,0,\text{NonLoc}}= -\frac{3}{160} \sqrt{\frac{21}{5}} C_{00,22}^{62,0}-\frac{3}{80} \sqrt{\frac{21}{5}}
C_{00,22}^{62,1}+\frac{3}{160} \sqrt{\frac{21}{5}} C_{11,22}^{51,0}+\frac{3}{80} \sqrt{\frac{21}{5}} C_{11,22}^{51,1}+\frac{27 C_{11,22}^{53,0}}{800}+\frac{27
C_{11,22}^{53,1}}{400}-\frac{3}{160} \sqrt{\frac{21}{5}} C_{20,22}^{42,0}-\frac{3}{80} \sqrt{\frac{21}{5}} C_{20,22}^{42,1}-\frac{3}{160} \sqrt{\frac{21}{5}}
C_{22,22}^{40,0}-\frac{3}{80} \sqrt{\frac{21}{5}} C_{22,22}^{40,1}-\frac{3}{80} \sqrt{\frac{3}{5}} C_{22,22}^{42,0}-\frac{3}{40} \sqrt{\frac{3}{5}}
C_{22,22}^{42,1}-\frac{9}{200} \sqrt{\frac{3}{7}} C_{22,22}^{44,0}-\frac{9}{100} \sqrt{\frac{3}{7}} C_{22,22}^{44,1}+\frac{3}{160} \sqrt{\frac{21}{5}}
C_{31,22}^{31,0}+\frac{3}{80} \sqrt{\frac{21}{5}} C_{31,22}^{31,1}+\frac{27 C_{31,22}^{33,0}}{800}+\frac{27 C_{31,22}^{33,1}}{400}+\frac{9}{200}
\sqrt{\frac{3}{10}} C_{33,22}^{33,0}+\frac{9}{100} \sqrt{\frac{3}{10}} C_{33,22}^{33,1}\)

\noindent\(C_{00,3312,2}^{00,3112,1,\text{Loc}}= \frac{9}{160} \sqrt{\frac{7}{5}} C_{00,22}^{62,1}+\frac{9}{160} \sqrt{\frac{7}{5}}
C_{11,22}^{51,1}+\frac{27 \sqrt{3} C_{11,22}^{53,1}}{800}+\frac{9}{160} \sqrt{\frac{7}{5}} C_{20,22}^{42,1}+\frac{9}{160} \sqrt{\frac{7}{5}} C_{22,22}^{40,1}+\frac{9
C_{22,22}^{42,1}}{80 \sqrt{5}}+\frac{27 C_{22,22}^{44,1}}{200 \sqrt{7}}+\frac{9}{160} \sqrt{\frac{7}{5}} C_{31,22}^{31,1}+\frac{27 \sqrt{3} C_{31,22}^{33,1}}{800}+\frac{27
C_{33,22}^{33,1}}{200 \sqrt{10}}\)

\noindent\(C_{00,3312,2}^{00,3112,1,\text{NonLoc}}= -\frac{9}{160} \sqrt{\frac{7}{5}} C_{00,22}^{62,0}+\frac{9}{160} \sqrt{\frac{7}{5}}
C_{11,22}^{51,0}+\frac{27 \sqrt{3} C_{11,22}^{53,0}}{800}-\frac{9}{160} \sqrt{\frac{7}{5}} C_{20,22}^{42,0}-\frac{9}{160} \sqrt{\frac{7}{5}} C_{22,22}^{40,0}-\frac{9
C_{22,22}^{42,0}}{80 \sqrt{5}}-\frac{27 C_{22,22}^{44,0}}{200 \sqrt{7}}+\frac{9}{160} \sqrt{\frac{7}{5}} C_{31,22}^{31,0}+\frac{27 \sqrt{3} C_{31,22}^{33,0}}{800}+\frac{27
C_{33,22}^{33,0}}{200 \sqrt{10}}\)

\noindent\(C_{00,3312,2}^{00,3312,0,\text{Loc}}= \frac{\sqrt{5} C_{00,20}^{60,0}}{144}+\frac{\sqrt{5} C_{00,20}^{60,1}}{288}+\frac{C_{00,22}^{62,0}}{120}+\frac{C_{00,22}^{62,1}}{240}+\frac{C_{11,22}^{51,0}}{120}+\frac{C_{11,22}^{51,1}}{240}+\frac{1}{40}
\sqrt{\frac{3}{35}} C_{11,22}^{53,0}+\frac{1}{80} \sqrt{\frac{3}{35}} C_{11,22}^{53,1}+\frac{\sqrt{5} C_{20,20}^{40,0}}{144}+\frac{\sqrt{5} C_{20,20}^{40,1}}{288}+\frac{C_{20,22}^{42,0}}{120}+\frac{C_{20,22}^{42,1}}{240}+\frac{C_{22,20}^{42,0}}{72}+\frac{C_{22,20}^{42,1}}{144}+\frac{C_{22,22}^{40,0}}{120}+\frac{C_{22,22}^{40,1}}{240}+\frac{C_{22,22}^{42,0}}{60
\sqrt{7}}+\frac{C_{22,22}^{42,1}}{120 \sqrt{7}}+\frac{C_{22,22}^{44,0}}{70 \sqrt{5}}+\frac{C_{22,22}^{44,1}}{140 \sqrt{5}}+\frac{\sqrt{5} C_{31,20}^{31,0}}{144}+\frac{\sqrt{5}
C_{31,20}^{31,1}}{288}+\frac{C_{31,22}^{31,0}}{120}+\frac{C_{31,22}^{31,1}}{240}+\frac{1}{40} \sqrt{\frac{3}{35}} C_{31,22}^{33,0}+\frac{1}{80} \sqrt{\frac{3}{35}}
C_{31,22}^{33,1}+\frac{C_{33,20}^{33,0}}{8 \sqrt{105}}+\frac{C_{33,20}^{33,1}}{16 \sqrt{105}}+\frac{C_{33,22}^{33,0}}{50 \sqrt{14}}+\frac{C_{33,22}^{33,1}}{100
\sqrt{14}}\)

\noindent\(C_{00,3312,2}^{00,3312,0,\text{NonLoc}}= \frac{1}{192} \sqrt{\frac{5}{3}} C_{00,00}^{60,0}+\frac{1}{96} \sqrt{\frac{5}{3}}
C_{00,00}^{60,1}+\frac{\sqrt{5} C_{00,20}^{60,0}}{576}+\frac{\sqrt{5} C_{00,20}^{60,1}}{288}-\frac{C_{00,22}^{62,0}}{240}-\frac{C_{00,22}^{62,1}}{120}-\frac{1}{192}
\sqrt{\frac{5}{3}} C_{11,00}^{51,0}-\frac{1}{96} \sqrt{\frac{5}{3}} C_{11,00}^{51,1}+\frac{C_{11,22}^{51,0}}{240}+\frac{C_{11,22}^{51,1}}{120}+\frac{1}{80}
\sqrt{\frac{3}{35}} C_{11,22}^{53,0}+\frac{1}{40} \sqrt{\frac{3}{35}} C_{11,22}^{53,1}+\frac{1}{192} \sqrt{\frac{5}{3}} C_{20,00}^{40,0}+\frac{1}{96}
\sqrt{\frac{5}{3}} C_{20,00}^{40,1}+\frac{\sqrt{5} C_{20,20}^{40,0}}{576}+\frac{\sqrt{5} C_{20,20}^{40,1}}{288}-\frac{C_{20,22}^{42,0}}{240}-\frac{C_{20,22}^{42,1}}{120}+\frac{C_{22,00}^{42,0}}{96
\sqrt{3}}+\frac{C_{22,00}^{42,1}}{48 \sqrt{3}}+\frac{C_{22,20}^{42,0}}{288}+\frac{C_{22,20}^{42,1}}{144}-\frac{C_{22,22}^{40,0}}{240}-\frac{C_{22,22}^{40,1}}{120}-\frac{C_{22,22}^{42,0}}{120
\sqrt{7}}-\frac{C_{22,22}^{42,1}}{60 \sqrt{7}}-\frac{C_{22,22}^{44,0}}{140 \sqrt{5}}-\frac{C_{22,22}^{44,1}}{70 \sqrt{5}}-\frac{1}{192} \sqrt{\frac{5}{3}}
C_{31,00}^{31,0}-\frac{1}{96} \sqrt{\frac{5}{3}} C_{31,00}^{31,1}-\frac{\sqrt{5} C_{31,20}^{31,0}}{576}-\frac{\sqrt{5} C_{31,20}^{31,1}}{288}+\frac{C_{31,22}^{31,0}}{240}+\frac{C_{31,22}^{31,1}}{120}+\frac{1}{80}
\sqrt{\frac{3}{35}} C_{31,22}^{33,0}+\frac{1}{40} \sqrt{\frac{3}{35}} C_{31,22}^{33,1}-\frac{C_{33,00}^{33,0}}{32 \sqrt{35}}-\frac{C_{33,00}^{33,1}}{16
\sqrt{35}}-\frac{C_{33,20}^{33,0}}{32 \sqrt{105}}-\frac{C_{33,20}^{33,1}}{16 \sqrt{105}}+\frac{C_{33,22}^{33,0}}{100 \sqrt{14}}+\frac{C_{33,22}^{33,1}}{50
\sqrt{14}}\)

\noindent\(C_{00,3312,2}^{00,3312,1,\text{Loc}}= \frac{1}{96} \sqrt{\frac{5}{3}} C_{00,20}^{60,1}+\frac{C_{00,22}^{62,1}}{80 \sqrt{3}}+\frac{C_{11,22}^{51,1}}{80
\sqrt{3}}+\frac{3 C_{11,22}^{53,1}}{80 \sqrt{35}}+\frac{1}{96} \sqrt{\frac{5}{3}} C_{20,20}^{40,1}+\frac{C_{20,22}^{42,1}}{80 \sqrt{3}}+\frac{C_{22,20}^{42,1}}{48
\sqrt{3}}+\frac{C_{22,22}^{40,1}}{80 \sqrt{3}}+\frac{C_{22,22}^{42,1}}{40 \sqrt{21}}+\frac{1}{140} \sqrt{\frac{3}{5}} C_{22,22}^{44,1}+\frac{1}{96}
\sqrt{\frac{5}{3}} C_{31,20}^{31,1}+\frac{C_{31,22}^{31,1}}{80 \sqrt{3}}+\frac{3 C_{31,22}^{33,1}}{80 \sqrt{35}}+\frac{C_{33,20}^{33,1}}{16 \sqrt{35}}+\frac{1}{100}
\sqrt{\frac{3}{14}} C_{33,22}^{33,1}\)

\noindent\(C_{00,3312,2}^{00,3312,1,\text{NonLoc}}= \frac{\sqrt{5} C_{00,00}^{60,0}}{192}+\frac{1}{192} \sqrt{\frac{5}{3}} C_{00,20}^{60,0}-\frac{C_{00,22}^{62,0}}{80
\sqrt{3}}-\frac{\sqrt{5} C_{11,00}^{51,0}}{192}+\frac{C_{11,22}^{51,0}}{80 \sqrt{3}}+\frac{3 C_{11,22}^{53,0}}{80 \sqrt{35}}+\frac{\sqrt{5} C_{20,00}^{40,0}}{192}+\frac{1}{192}
\sqrt{\frac{5}{3}} C_{20,20}^{40,0}-\frac{C_{20,22}^{42,0}}{80 \sqrt{3}}+\frac{C_{22,00}^{42,0}}{96}+\frac{C_{22,20}^{42,0}}{96 \sqrt{3}}-\frac{C_{22,22}^{40,0}}{80
\sqrt{3}}-\frac{C_{22,22}^{42,0}}{40 \sqrt{21}}-\frac{1}{140} \sqrt{\frac{3}{5}} C_{22,22}^{44,0}-\frac{\sqrt{5} C_{31,00}^{31,0}}{192}-\frac{1}{192}
\sqrt{\frac{5}{3}} C_{31,20}^{31,0}+\frac{C_{31,22}^{31,0}}{80 \sqrt{3}}+\frac{3 C_{31,22}^{33,0}}{80 \sqrt{35}}-\frac{1}{32} \sqrt{\frac{3}{35}}
C_{33,00}^{33,0}-\frac{C_{33,20}^{33,0}}{32 \sqrt{35}}+\frac{1}{100} \sqrt{\frac{3}{14}} C_{33,22}^{33,0}\)

\noindent\(C_{00,3313,3}^{00,3313,0,\text{Loc}}= \frac{\sqrt{7} C_{00,20}^{60,0}}{144}+\frac{\sqrt{7} C_{00,20}^{60,1}}{288}-\frac{1}{96}
\sqrt{\frac{7}{5}} C_{00,22}^{62,0}-\frac{1}{192} \sqrt{\frac{7}{5}} C_{00,22}^{62,1}-\frac{1}{96} \sqrt{\frac{7}{5}} C_{11,22}^{51,0}-\frac{1}{192}
\sqrt{\frac{7}{5}} C_{11,22}^{51,1}-\frac{\sqrt{3} C_{11,22}^{53,0}}{160}-\frac{\sqrt{3} C_{11,22}^{53,1}}{320}+\frac{\sqrt{7} C_{20,20}^{40,0}}{144}+\frac{\sqrt{7}
C_{20,20}^{40,1}}{288}-\frac{1}{96} \sqrt{\frac{7}{5}} C_{20,22}^{42,0}-\frac{1}{192} \sqrt{\frac{7}{5}} C_{20,22}^{42,1}+\frac{1}{72} \sqrt{\frac{7}{5}}
C_{22,20}^{42,0}+\frac{1}{144} \sqrt{\frac{7}{5}} C_{22,20}^{42,1}-\frac{1}{96} \sqrt{\frac{7}{5}} C_{22,22}^{40,0}-\frac{1}{192} \sqrt{\frac{7}{5}}
C_{22,22}^{40,1}-\frac{C_{22,22}^{42,0}}{48 \sqrt{5}}-\frac{C_{22,22}^{42,1}}{96 \sqrt{5}}-\frac{C_{22,22}^{44,0}}{40 \sqrt{7}}-\frac{C_{22,22}^{44,1}}{80
\sqrt{7}}+\frac{\sqrt{7} C_{31,20}^{31,0}}{144}+\frac{\sqrt{7} C_{31,20}^{31,1}}{288}-\frac{1}{96} \sqrt{\frac{7}{5}} C_{31,22}^{31,0}-\frac{1}{192}
\sqrt{\frac{7}{5}} C_{31,22}^{31,1}-\frac{\sqrt{3} C_{31,22}^{33,0}}{160}-\frac{\sqrt{3} C_{31,22}^{33,1}}{320}+\frac{C_{33,20}^{33,0}}{40 \sqrt{3}}+\frac{C_{33,20}^{33,1}}{80
\sqrt{3}}-\frac{C_{33,22}^{33,0}}{40 \sqrt{10}}-\frac{C_{33,22}^{33,1}}{80 \sqrt{10}}\)

\noindent\(C_{00,3313,3}^{00,3313,0,\text{NonLoc}}= \frac{1}{192} \sqrt{\frac{7}{3}} C_{00,00}^{60,0}+\frac{1}{96} \sqrt{\frac{7}{3}}
C_{00,00}^{60,1}+\frac{\sqrt{7} C_{00,20}^{60,0}}{576}+\frac{\sqrt{7} C_{00,20}^{60,1}}{288}+\frac{1}{192} \sqrt{\frac{7}{5}} C_{00,22}^{62,0}+\frac{1}{96}
\sqrt{\frac{7}{5}} C_{00,22}^{62,1}-\frac{1}{192} \sqrt{\frac{7}{3}} C_{11,00}^{51,0}-\frac{1}{96} \sqrt{\frac{7}{3}} C_{11,00}^{51,1}-\frac{1}{192}
\sqrt{\frac{7}{5}} C_{11,22}^{51,0}-\frac{1}{96} \sqrt{\frac{7}{5}} C_{11,22}^{51,1}-\frac{\sqrt{3} C_{11,22}^{53,0}}{320}-\frac{\sqrt{3} C_{11,22}^{53,1}}{160}+\frac{1}{192}
\sqrt{\frac{7}{3}} C_{20,00}^{40,0}+\frac{1}{96} \sqrt{\frac{7}{3}} C_{20,00}^{40,1}+\frac{\sqrt{7} C_{20,20}^{40,0}}{576}+\frac{\sqrt{7} C_{20,20}^{40,1}}{288}+\frac{1}{192}
\sqrt{\frac{7}{5}} C_{20,22}^{42,0}+\frac{1}{96} \sqrt{\frac{7}{5}} C_{20,22}^{42,1}+\frac{1}{96} \sqrt{\frac{7}{15}} C_{22,00}^{42,0}+\frac{1}{48}
\sqrt{\frac{7}{15}} C_{22,00}^{42,1}+\frac{1}{288} \sqrt{\frac{7}{5}} C_{22,20}^{42,0}+\frac{1}{144} \sqrt{\frac{7}{5}} C_{22,20}^{42,1}+\frac{1}{192}
\sqrt{\frac{7}{5}} C_{22,22}^{40,0}+\frac{1}{96} \sqrt{\frac{7}{5}} C_{22,22}^{40,1}+\frac{C_{22,22}^{42,0}}{96 \sqrt{5}}+\frac{C_{22,22}^{42,1}}{48
\sqrt{5}}+\frac{C_{22,22}^{44,0}}{80 \sqrt{7}}+\frac{C_{22,22}^{44,1}}{40 \sqrt{7}}-\frac{1}{192} \sqrt{\frac{7}{3}} C_{31,00}^{31,0}-\frac{1}{96}
\sqrt{\frac{7}{3}} C_{31,00}^{31,1}-\frac{\sqrt{7} C_{31,20}^{31,0}}{576}-\frac{\sqrt{7} C_{31,20}^{31,1}}{288}-\frac{1}{192} \sqrt{\frac{7}{5}}
C_{31,22}^{31,0}-\frac{1}{96} \sqrt{\frac{7}{5}} C_{31,22}^{31,1}-\frac{\sqrt{3} C_{31,22}^{33,0}}{320}-\frac{\sqrt{3} C_{31,22}^{33,1}}{160}-\frac{C_{33,00}^{33,0}}{160}-\frac{C_{33,00}^{33,1}}{80}-\frac{C_{33,20}^{33,0}}{160
\sqrt{3}}-\frac{C_{33,20}^{33,1}}{80 \sqrt{3}}-\frac{C_{33,22}^{33,0}}{80 \sqrt{10}}-\frac{C_{33,22}^{33,1}}{40 \sqrt{10}}\)

\noindent\(C_{00,3313,3}^{00,3313,1,\text{Loc}}= \frac{1}{96} \sqrt{\frac{7}{3}} C_{00,20}^{60,1}-\frac{1}{64} \sqrt{\frac{7}{15}}
C_{00,22}^{62,1}-\frac{1}{64} \sqrt{\frac{7}{15}} C_{11,22}^{51,1}-\frac{3 C_{11,22}^{53,1}}{320}+\frac{1}{96} \sqrt{\frac{7}{3}} C_{20,20}^{40,1}-\frac{1}{64}
\sqrt{\frac{7}{15}} C_{20,22}^{42,1}+\frac{1}{48} \sqrt{\frac{7}{15}} C_{22,20}^{42,1}-\frac{1}{64} \sqrt{\frac{7}{15}} C_{22,22}^{40,1}-\frac{C_{22,22}^{42,1}}{32
\sqrt{15}}-\frac{1}{80} \sqrt{\frac{3}{7}} C_{22,22}^{44,1}+\frac{1}{96} \sqrt{\frac{7}{3}} C_{31,20}^{31,1}-\frac{1}{64} \sqrt{\frac{7}{15}} C_{31,22}^{31,1}-\frac{3
C_{31,22}^{33,1}}{320}+\frac{C_{33,20}^{33,1}}{80}-\frac{1}{80} \sqrt{\frac{3}{10}} C_{33,22}^{33,1}\)

\noindent\(C_{00,3313,3}^{00,3313,1,\text{NonLoc}}= \frac{\sqrt{7} C_{00,00}^{60,0}}{192}+\frac{1}{192} \sqrt{\frac{7}{3}} C_{00,20}^{60,0}+\frac{1}{64}
\sqrt{\frac{7}{15}} C_{00,22}^{62,0}-\frac{\sqrt{7} C_{11,00}^{51,0}}{192}-\frac{1}{64} \sqrt{\frac{7}{15}} C_{11,22}^{51,0}-\frac{3 C_{11,22}^{53,0}}{320}+\frac{\sqrt{7}
C_{20,00}^{40,0}}{192}+\frac{1}{192} \sqrt{\frac{7}{3}} C_{20,20}^{40,0}+\frac{1}{64} \sqrt{\frac{7}{15}} C_{20,22}^{42,0}+\frac{1}{96} \sqrt{\frac{7}{5}}
C_{22,00}^{42,0}+\frac{1}{96} \sqrt{\frac{7}{15}} C_{22,20}^{42,0}+\frac{1}{64} \sqrt{\frac{7}{15}} C_{22,22}^{40,0}+\frac{C_{22,22}^{42,0}}{32 \sqrt{15}}+\frac{1}{80}
\sqrt{\frac{3}{7}} C_{22,22}^{44,0}-\frac{\sqrt{7} C_{31,00}^{31,0}}{192}-\frac{1}{192} \sqrt{\frac{7}{3}} C_{31,20}^{31,0}-\frac{1}{64} \sqrt{\frac{7}{15}}
C_{31,22}^{31,0}-\frac{3 C_{31,22}^{33,0}}{320}-\frac{\sqrt{3} C_{33,00}^{33,0}}{160}-\frac{C_{33,20}^{33,0}}{160}-\frac{1}{80} \sqrt{\frac{3}{10}}
C_{33,22}^{33,0}\)

\noindent\(C_{00,3314,4}^{00,3314,0,\text{Loc}}= \frac{C_{00,20}^{60,0}}{48}+\frac{C_{00,20}^{60,1}}{96}+\frac{C_{00,22}^{62,0}}{96
\sqrt{5}}+\frac{C_{00,22}^{62,1}}{192 \sqrt{5}}+\frac{C_{11,22}^{51,0}}{96 \sqrt{5}}+\frac{C_{11,22}^{51,1}}{192 \sqrt{5}}+\frac{1}{160} \sqrt{\frac{3}{7}}
C_{11,22}^{53,0}+\frac{1}{320} \sqrt{\frac{3}{7}} C_{11,22}^{53,1}+\frac{C_{20,20}^{40,0}}{48}+\frac{C_{20,20}^{40,1}}{96}+\frac{C_{20,22}^{42,0}}{96
\sqrt{5}}+\frac{C_{20,22}^{42,1}}{192 \sqrt{5}}+\frac{C_{22,20}^{42,0}}{24 \sqrt{5}}+\frac{C_{22,20}^{42,1}}{48 \sqrt{5}}+\frac{C_{22,22}^{40,0}}{96
\sqrt{5}}+\frac{C_{22,22}^{40,1}}{192 \sqrt{5}}+\frac{C_{22,22}^{42,0}}{48 \sqrt{35}}+\frac{C_{22,22}^{42,1}}{96 \sqrt{35}}+\frac{C_{22,22}^{44,0}}{280}+\frac{C_{22,22}^{44,1}}{560}+\frac{C_{31,20}^{31,0}}{48}+\frac{C_{31,20}^{31,1}}{96}+\frac{C_{31,22}^{31,0}}{96
\sqrt{5}}+\frac{C_{31,22}^{31,1}}{192 \sqrt{5}}+\frac{1}{160} \sqrt{\frac{3}{7}} C_{31,22}^{33,0}+\frac{1}{320} \sqrt{\frac{3}{7}} C_{31,22}^{33,1}+\frac{1}{40}
\sqrt{\frac{3}{7}} C_{33,20}^{33,0}+\frac{1}{80} \sqrt{\frac{3}{7}} C_{33,20}^{33,1}+\frac{C_{33,22}^{33,0}}{40 \sqrt{70}}+\frac{C_{33,22}^{33,1}}{80
\sqrt{70}}\)

\noindent\(C_{00,3314,4}^{00,3314,0,\text{NonLoc}}= \frac{C_{00,00}^{60,0}}{64 \sqrt{3}}+\frac{C_{00,00}^{60,1}}{32 \sqrt{3}}+\frac{C_{00,20}^{60,0}}{192}+\frac{C_{00,20}^{60,1}}{96}-\frac{C_{00,22}^{62,0}}{192
\sqrt{5}}-\frac{C_{00,22}^{62,1}}{96 \sqrt{5}}-\frac{C_{11,00}^{51,0}}{64 \sqrt{3}}-\frac{C_{11,00}^{51,1}}{32 \sqrt{3}}+\frac{C_{11,22}^{51,0}}{192
\sqrt{5}}+\frac{C_{11,22}^{51,1}}{96 \sqrt{5}}+\frac{1}{320} \sqrt{\frac{3}{7}} C_{11,22}^{53,0}+\frac{1}{160} \sqrt{\frac{3}{7}} C_{11,22}^{53,1}+\frac{C_{20,00}^{40,0}}{64
\sqrt{3}}+\frac{C_{20,00}^{40,1}}{32 \sqrt{3}}+\frac{C_{20,20}^{40,0}}{192}+\frac{C_{20,20}^{40,1}}{96}-\frac{C_{20,22}^{42,0}}{192 \sqrt{5}}-\frac{C_{20,22}^{42,1}}{96
\sqrt{5}}+\frac{C_{22,00}^{42,0}}{32 \sqrt{15}}+\frac{C_{22,00}^{42,1}}{16 \sqrt{15}}+\frac{C_{22,20}^{42,0}}{96 \sqrt{5}}+\frac{C_{22,20}^{42,1}}{48
\sqrt{5}}-\frac{C_{22,22}^{40,0}}{192 \sqrt{5}}-\frac{C_{22,22}^{40,1}}{96 \sqrt{5}}-\frac{C_{22,22}^{42,0}}{96 \sqrt{35}}-\frac{C_{22,22}^{42,1}}{48
\sqrt{35}}-\frac{C_{22,22}^{44,0}}{560}-\frac{C_{22,22}^{44,1}}{280}-\frac{C_{31,00}^{31,0}}{64 \sqrt{3}}-\frac{C_{31,00}^{31,1}}{32 \sqrt{3}}-\frac{C_{31,20}^{31,0}}{192}-\frac{C_{31,20}^{31,1}}{96}+\frac{C_{31,22}^{31,0}}{192
\sqrt{5}}+\frac{C_{31,22}^{31,1}}{96 \sqrt{5}}+\frac{1}{320} \sqrt{\frac{3}{7}} C_{31,22}^{33,0}+\frac{1}{160} \sqrt{\frac{3}{7}} C_{31,22}^{33,1}-\frac{3
C_{33,00}^{33,0}}{160 \sqrt{7}}-\frac{3 C_{33,00}^{33,1}}{80 \sqrt{7}}-\frac{1}{160} \sqrt{\frac{3}{7}} C_{33,20}^{33,0}-\frac{1}{80} \sqrt{\frac{3}{7}}
C_{33,20}^{33,1}+\frac{C_{33,22}^{33,0}}{80 \sqrt{70}}+\frac{C_{33,22}^{33,1}}{40 \sqrt{70}}\)

\noindent\(C_{00,3314,4}^{00,3314,1,\text{Loc}}= \frac{C_{00,20}^{60,1}}{32 \sqrt{3}}+\frac{C_{00,22}^{62,1}}{64 \sqrt{15}}+\frac{C_{11,22}^{51,1}}{64
\sqrt{15}}+\frac{3 C_{11,22}^{53,1}}{320 \sqrt{7}}+\frac{C_{20,20}^{40,1}}{32 \sqrt{3}}+\frac{C_{20,22}^{42,1}}{64 \sqrt{15}}+\frac{C_{22,20}^{42,1}}{16
\sqrt{15}}+\frac{C_{22,22}^{40,1}}{64 \sqrt{15}}+\frac{C_{22,22}^{42,1}}{32 \sqrt{105}}+\frac{\sqrt{3} C_{22,22}^{44,1}}{560}+\frac{C_{31,20}^{31,1}}{32
\sqrt{3}}+\frac{C_{31,22}^{31,1}}{64 \sqrt{15}}+\frac{3 C_{31,22}^{33,1}}{320 \sqrt{7}}+\frac{3 C_{33,20}^{33,1}}{80 \sqrt{7}}+\frac{1}{80} \sqrt{\frac{3}{70}}
C_{33,22}^{33,1}\)

\noindent\(C_{00,3314,4}^{00,3314,1,\text{NonLoc}}= \frac{C_{00,00}^{60,0}}{64}+\frac{C_{00,20}^{60,0}}{64 \sqrt{3}}-\frac{C_{00,22}^{62,0}}{64
\sqrt{15}}-\frac{C_{11,00}^{51,0}}{64}+\frac{C_{11,22}^{51,0}}{64 \sqrt{15}}+\frac{3 C_{11,22}^{53,0}}{320 \sqrt{7}}+\frac{C_{20,00}^{40,0}}{64}+\frac{C_{20,20}^{40,0}}{64
\sqrt{3}}-\frac{C_{20,22}^{42,0}}{64 \sqrt{15}}+\frac{C_{22,00}^{42,0}}{32 \sqrt{5}}+\frac{C_{22,20}^{42,0}}{32 \sqrt{15}}-\frac{C_{22,22}^{40,0}}{64
\sqrt{15}}-\frac{C_{22,22}^{42,0}}{32 \sqrt{105}}-\frac{\sqrt{3} C_{22,22}^{44,0}}{560}-\frac{C_{31,00}^{31,0}}{64}-\frac{C_{31,20}^{31,0}}{64 \sqrt{3}}+\frac{C_{31,22}^{31,0}}{64
\sqrt{15}}+\frac{3 C_{31,22}^{33,0}}{320 \sqrt{7}}-\frac{3}{160} \sqrt{\frac{3}{7}} C_{33,00}^{33,0}-\frac{3 C_{33,20}^{33,0}}{160 \sqrt{7}}+\frac{1}{80}
\sqrt{\frac{3}{70}} C_{33,22}^{33,0}\)

\noindent\(C_{00,4000,0}^{00,2000,0,\text{Loc}}= \frac{7 C_{00,00}^{60,0}}{64}+\frac{7 C_{00,00}^{60,1}}{128}+\frac{7 C_{11,00}^{51,0}}{64}+\frac{7
C_{11,00}^{51,1}}{128}+\frac{7 C_{20,00}^{40,0}}{64}+\frac{7 C_{20,00}^{40,1}}{128}+\frac{7 C_{22,00}^{42,0}}{32 \sqrt{5}}+\frac{7 C_{22,00}^{42,1}}{64
\sqrt{5}}+\frac{7 C_{31,00}^{31,0}}{64}+\frac{7 C_{31,00}^{31,1}}{128}+\frac{3 \sqrt{21} C_{33,00}^{33,0}}{160}+\frac{3 \sqrt{21} C_{33,00}^{33,1}}{320}\)

\noindent\(C_{00,4000,0}^{00,2000,0,\text{NonLoc}}= -\frac{7 C_{00,00}^{60,0}}{256}-\frac{7 C_{00,00}^{60,1}}{128}+\frac{7 \sqrt{3}
C_{00,20}^{60,0}}{256}+\frac{7 \sqrt{3} C_{00,20}^{60,1}}{128}+\frac{7 C_{11,00}^{51,0}}{256}+\frac{7 C_{11,00}^{51,1}}{128}-\frac{7 \sqrt{3} C_{11,20}^{31,0}}{512}-\frac{7
\sqrt{3} C_{11,20}^{31,1}}{256}-\frac{7 C_{20,00}^{40,0}}{256}-\frac{7 C_{20,00}^{40,1}}{128}+\frac{7 \sqrt{3} C_{20,20}^{40,0}}{256}+\frac{7 \sqrt{3}
C_{20,20}^{40,1}}{128}-\frac{7 C_{22,00}^{42,0}}{128 \sqrt{5}}-\frac{7 C_{22,00}^{42,1}}{64 \sqrt{5}}+\frac{7}{128} \sqrt{\frac{3}{5}} C_{22,20}^{42,0}+\frac{7}{64}
\sqrt{\frac{3}{5}} C_{22,20}^{42,1}+\frac{7 C_{31,00}^{31,0}}{256}+\frac{7 C_{31,00}^{31,1}}{128}-\frac{7 \sqrt{3} C_{31,20}^{31,0}}{256}-\frac{7
\sqrt{3} C_{31,20}^{31,1}}{128}+\frac{3 \sqrt{21} C_{33,00}^{33,0}}{640}+\frac{3 \sqrt{21} C_{33,00}^{33,1}}{320}-\frac{9 \sqrt{7} C_{33,20}^{33,0}}{640}-\frac{9
\sqrt{7} C_{33,20}^{33,1}}{320}\)

\noindent\(C_{00,4000,0}^{00,2000,1,\text{Loc}}= \frac{7 \sqrt{3} C_{00,00}^{60,1}}{128}+\frac{7 \sqrt{3} C_{11,00}^{51,1}}{128}+\frac{7
\sqrt{3} C_{20,00}^{40,1}}{128}+\frac{7}{64} \sqrt{\frac{3}{5}} C_{22,00}^{42,1}+\frac{7 \sqrt{3} C_{31,00}^{31,1}}{128}+\frac{9 \sqrt{7} C_{33,00}^{33,1}}{320}\)

\noindent\(C_{00,4000,0}^{00,2000,1,\text{NonLoc}}= -\frac{7 \sqrt{3} C_{00,00}^{60,0}}{256}+\frac{21 C_{00,20}^{60,0}}{256}+\frac{7
\sqrt{3} C_{11,00}^{51,0}}{256}-\frac{21 C_{11,20}^{31,0}}{512}-\frac{7 \sqrt{3} C_{20,00}^{40,0}}{256}+\frac{21 C_{20,20}^{40,0}}{256}-\frac{7}{128}
\sqrt{\frac{3}{5}} C_{22,00}^{42,0}+\frac{21 C_{22,20}^{42,0}}{128 \sqrt{5}}+\frac{7 \sqrt{3} C_{31,00}^{31,0}}{256}-\frac{21 C_{31,20}^{31,0}}{256}+\frac{9
\sqrt{7} C_{33,00}^{33,0}}{640}-\frac{9 \sqrt{21} C_{33,20}^{33,0}}{640}\)

\noindent\(C_{00,4011,1}^{00,2011,0,\text{Loc}}= -\frac{7 C_{00,20}^{60,0}}{64}-\frac{7 C_{00,20}^{60,1}}{128}-\frac{7 C_{11,20}^{31,0}}{128}-\frac{7
C_{11,20}^{31,1}}{256}-\frac{7 C_{20,20}^{40,0}}{64}-\frac{7 C_{20,20}^{40,1}}{128}-\frac{7 C_{22,20}^{42,0}}{32 \sqrt{5}}-\frac{7 C_{22,20}^{42,1}}{64
\sqrt{5}}-\frac{7 C_{31,20}^{31,0}}{64}-\frac{7 C_{31,20}^{31,1}}{128}-\frac{3 \sqrt{21} C_{33,20}^{33,0}}{160}-\frac{3 \sqrt{21} C_{33,20}^{33,1}}{320}\)

\noindent\(C_{00,4011,1}^{00,2011,0,\text{NonLoc}}= -\frac{7 \sqrt{3} C_{00,00}^{60,0}}{256}-\frac{7 \sqrt{3} C_{00,00}^{60,1}}{128}-\frac{7
C_{00,20}^{60,0}}{256}-\frac{7 C_{00,20}^{60,1}}{128}+\frac{7 \sqrt{3} C_{11,00}^{51,0}}{256}+\frac{7 \sqrt{3} C_{11,00}^{51,1}}{128}+\frac{7 C_{11,20}^{31,0}}{512}+\frac{7
C_{11,20}^{31,1}}{256}-\frac{7 \sqrt{3} C_{20,00}^{40,0}}{256}-\frac{7 \sqrt{3} C_{20,00}^{40,1}}{128}-\frac{7 C_{20,20}^{40,0}}{256}-\frac{7 C_{20,20}^{40,1}}{128}-\frac{7}{128}
\sqrt{\frac{3}{5}} C_{22,00}^{42,0}-\frac{7}{64} \sqrt{\frac{3}{5}} C_{22,00}^{42,1}-\frac{7 C_{22,20}^{42,0}}{128 \sqrt{5}}-\frac{7 C_{22,20}^{42,1}}{64
\sqrt{5}}+\frac{7 \sqrt{3} C_{31,00}^{31,0}}{256}+\frac{7 \sqrt{3} C_{31,00}^{31,1}}{128}+\frac{7 C_{31,20}^{31,0}}{256}+\frac{7 C_{31,20}^{31,1}}{128}+\frac{9
\sqrt{7} C_{33,00}^{33,0}}{640}+\frac{9 \sqrt{7} C_{33,00}^{33,1}}{320}+\frac{3 \sqrt{21} C_{33,20}^{33,0}}{640}+\frac{3 \sqrt{21} C_{33,20}^{33,1}}{320}\)

\noindent\(C_{00,4011,1}^{00,2011,1,\text{Loc}}= -\frac{7 \sqrt{3} C_{00,20}^{60,1}}{128}-\frac{7 \sqrt{3} C_{11,20}^{31,1}}{256}-\frac{7
\sqrt{3} C_{20,20}^{40,1}}{128}-\frac{7}{64} \sqrt{\frac{3}{5}} C_{22,20}^{42,1}-\frac{7 \sqrt{3} C_{31,20}^{31,1}}{128}-\frac{9 \sqrt{7} C_{33,20}^{33,1}}{320}\)

\noindent\(C_{00,4011,1}^{00,2011,1,\text{NonLoc}}= -\frac{21 C_{00,00}^{60,0}}{256}-\frac{7 \sqrt{3} C_{00,20}^{60,0}}{256}+\frac{21
C_{11,00}^{51,0}}{256}+\frac{7 \sqrt{3} C_{11,20}^{31,0}}{512}-\frac{21 C_{20,00}^{40,0}}{256}-\frac{7 \sqrt{3} C_{20,20}^{40,0}}{256}-\frac{21 C_{22,00}^{42,0}}{128
\sqrt{5}}-\frac{7}{128} \sqrt{\frac{3}{5}} C_{22,20}^{42,0}+\frac{21 C_{31,00}^{31,0}}{256}+\frac{7 \sqrt{3} C_{31,20}^{31,0}}{256}+\frac{9 \sqrt{21}
C_{33,00}^{33,0}}{640}+\frac{9 \sqrt{7} C_{33,20}^{33,0}}{640}\)

\noindent\(C_{00,4011,1}^{00,2211,0,\text{Loc}}= -\frac{21 C_{00,22}^{62,0}}{320}-\frac{21 C_{00,22}^{62,1}}{640}-\frac{21 C_{11,22}^{51,0}}{320}-\frac{21
C_{11,22}^{51,1}}{640}-\frac{9}{320} \sqrt{\frac{21}{5}} C_{11,22}^{53,0}-\frac{9}{640} \sqrt{\frac{21}{5}} C_{11,22}^{53,1}-\frac{21 C_{20,22}^{42,0}}{320}-\frac{21
C_{20,22}^{42,1}}{640}-\frac{21 C_{22,22}^{40,0}}{320}-\frac{21 C_{22,22}^{40,1}}{640}-\frac{3 \sqrt{7} C_{22,22}^{42,0}}{160}-\frac{3 \sqrt{7} C_{22,22}^{42,1}}{320}-\frac{9
C_{22,22}^{44,0}}{80 \sqrt{5}}-\frac{9 C_{22,22}^{44,1}}{160 \sqrt{5}}-\frac{21 C_{31,22}^{31,0}}{320}-\frac{21 C_{31,22}^{31,1}}{640}-\frac{9}{320}
\sqrt{\frac{21}{5}} C_{31,22}^{33,0}-\frac{9}{640} \sqrt{\frac{21}{5}} C_{31,22}^{33,1}-\frac{9}{400} \sqrt{\frac{7}{2}} C_{33,22}^{33,0}-\frac{9}{800}
\sqrt{\frac{7}{2}} C_{33,22}^{33,1}\)

\noindent\(C_{00,4011,1}^{00,2211,0,\text{NonLoc}}= \frac{21 C_{00,22}^{62,0}}{640}+\frac{21 C_{00,22}^{62,1}}{320}-\frac{21 C_{11,22}^{51,0}}{640}-\frac{21
C_{11,22}^{51,1}}{320}-\frac{9}{640} \sqrt{\frac{21}{5}} C_{11,22}^{53,0}-\frac{9}{320} \sqrt{\frac{21}{5}} C_{11,22}^{53,1}+\frac{21 C_{20,22}^{42,0}}{640}+\frac{21
C_{20,22}^{42,1}}{320}+\frac{21 C_{22,22}^{40,0}}{640}+\frac{21 C_{22,22}^{40,1}}{320}+\frac{3 \sqrt{7} C_{22,22}^{42,0}}{320}+\frac{3 \sqrt{7} C_{22,22}^{42,1}}{160}+\frac{9
C_{22,22}^{44,0}}{160 \sqrt{5}}+\frac{9 C_{22,22}^{44,1}}{80 \sqrt{5}}-\frac{21 C_{31,22}^{31,0}}{640}-\frac{21 C_{31,22}^{31,1}}{320}-\frac{9}{640}
\sqrt{\frac{21}{5}} C_{31,22}^{33,0}-\frac{9}{320} \sqrt{\frac{21}{5}} C_{31,22}^{33,1}-\frac{9}{800} \sqrt{\frac{7}{2}} C_{33,22}^{33,0}-\frac{9}{400}
\sqrt{\frac{7}{2}} C_{33,22}^{33,1}\)

\noindent\(C_{00,4011,1}^{00,2211,1,\text{Loc}}= -\frac{21 \sqrt{3} C_{00,22}^{62,1}}{640}-\frac{21 \sqrt{3} C_{11,22}^{51,1}}{640}-\frac{27}{640}
\sqrt{\frac{7}{5}} C_{11,22}^{53,1}-\frac{21 \sqrt{3} C_{20,22}^{42,1}}{640}-\frac{21 \sqrt{3} C_{22,22}^{40,1}}{640}-\frac{3 \sqrt{21} C_{22,22}^{42,1}}{320}-\frac{9}{160}
\sqrt{\frac{3}{5}} C_{22,22}^{44,1}-\frac{21 \sqrt{3} C_{31,22}^{31,1}}{640}-\frac{27}{640} \sqrt{\frac{7}{5}} C_{31,22}^{33,1}-\frac{9}{800} \sqrt{\frac{21}{2}}
C_{33,22}^{33,1}\)

\noindent\(C_{00,4011,1}^{00,2211,1,\text{NonLoc}}= \frac{21 \sqrt{3} C_{00,22}^{62,0}}{640}-\frac{21 \sqrt{3} C_{11,22}^{51,0}}{640}-\frac{27}{640}
\sqrt{\frac{7}{5}} C_{11,22}^{53,0}+\frac{21 \sqrt{3} C_{20,22}^{42,0}}{640}+\frac{21 \sqrt{3} C_{22,22}^{40,0}}{640}+\frac{3 \sqrt{21} C_{22,22}^{42,0}}{320}+\frac{9}{160}
\sqrt{\frac{3}{5}} C_{22,22}^{44,0}-\frac{21 \sqrt{3} C_{31,22}^{31,0}}{640}-\frac{27}{640} \sqrt{\frac{7}{5}} C_{31,22}^{33,0}-\frac{9}{800} \sqrt{\frac{21}{2}}
C_{33,22}^{33,0}\)

\noindent\(C_{00,4202,2}^{00,2202,0,\text{Loc}}= \frac{\sqrt{5} C_{00,00}^{60,0}}{16}+\frac{\sqrt{5} C_{00,00}^{60,1}}{32}+\frac{\sqrt{5}
C_{11,00}^{51,0}}{16}+\frac{\sqrt{5} C_{11,00}^{51,1}}{32}+\frac{\sqrt{5} C_{20,00}^{40,0}}{16}+\frac{\sqrt{5} C_{20,00}^{40,1}}{32}+\frac{C_{22,00}^{42,0}}{8}+\frac{C_{22,00}^{42,1}}{16}+\frac{\sqrt{5}
C_{31,00}^{31,0}}{16}+\frac{\sqrt{5} C_{31,00}^{31,1}}{32}+\frac{3}{8} \sqrt{\frac{3}{35}} C_{33,00}^{33,0}+\frac{3}{16} \sqrt{\frac{3}{35}} C_{33,00}^{33,1}\)

\noindent\(C_{00,4202,2}^{00,2202,0,\text{NonLoc}}= -\frac{\sqrt{5} C_{00,00}^{60,0}}{64}-\frac{\sqrt{5} C_{00,00}^{60,1}}{32}+\frac{\sqrt{15}
C_{00,20}^{60,0}}{64}+\frac{\sqrt{15} C_{00,20}^{60,1}}{32}+\frac{\sqrt{5} C_{11,00}^{51,0}}{64}+\frac{\sqrt{5} C_{11,00}^{51,1}}{32}-\frac{\sqrt{5}
C_{20,00}^{40,0}}{64}-\frac{\sqrt{5} C_{20,00}^{40,1}}{32}+\frac{\sqrt{15} C_{20,20}^{40,0}}{64}+\frac{\sqrt{15} C_{20,20}^{40,1}}{32}-\frac{C_{22,00}^{42,0}}{32}-\frac{C_{22,00}^{42,1}}{16}+\frac{\sqrt{3}
C_{22,20}^{42,0}}{32}+\frac{\sqrt{3} C_{22,20}^{42,1}}{16}+\frac{\sqrt{5} C_{31,00}^{31,0}}{64}+\frac{\sqrt{5} C_{31,00}^{31,1}}{32}-\frac{\sqrt{15}
C_{31,20}^{31,0}}{64}-\frac{\sqrt{15} C_{31,20}^{31,1}}{32}+\frac{3}{32} \sqrt{\frac{3}{35}} C_{33,00}^{33,0}+\frac{3}{16} \sqrt{\frac{3}{35}} C_{33,00}^{33,1}-\frac{9
C_{33,20}^{33,0}}{32 \sqrt{35}}-\frac{9 C_{33,20}^{33,1}}{16 \sqrt{35}}\)

\noindent\(C_{00,4202,2}^{00,2202,1,\text{Loc}}= \frac{\sqrt{15} C_{00,00}^{60,1}}{32}+\frac{\sqrt{15} C_{11,00}^{51,1}}{32}+\frac{\sqrt{15}
C_{20,00}^{40,1}}{32}+\frac{\sqrt{3} C_{22,00}^{42,1}}{16}+\frac{\sqrt{15} C_{31,00}^{31,1}}{32}+\frac{9 C_{33,00}^{33,1}}{16 \sqrt{35}}\)

\noindent\(C_{00,4202,2}^{00,2202,1,\text{NonLoc}}= -\frac{\sqrt{15} C_{00,00}^{60,0}}{64}+\frac{3 \sqrt{5} C_{00,20}^{60,0}}{64}+\frac{\sqrt{15}
C_{11,00}^{51,0}}{64}-\frac{\sqrt{15} C_{20,00}^{40,0}}{64}+\frac{3 \sqrt{5} C_{20,20}^{40,0}}{64}-\frac{\sqrt{3} C_{22,00}^{42,0}}{32}+\frac{3 C_{22,20}^{42,0}}{32}+\frac{\sqrt{15}
C_{31,00}^{31,0}}{64}-\frac{3 \sqrt{5} C_{31,20}^{31,0}}{64}+\frac{9 C_{33,00}^{33,0}}{32 \sqrt{35}}-\frac{9}{32} \sqrt{\frac{3}{35}} C_{33,20}^{33,0}\)

\noindent\(C_{00,4211,1}^{00,2011,0,\text{Loc}}= -\frac{3 C_{00,22}^{62,0}}{32}-\frac{3 C_{00,22}^{62,1}}{64}-\frac{3 C_{11,22}^{51,0}}{32}-\frac{3
C_{11,22}^{51,1}}{64}-\frac{9}{32} \sqrt{\frac{3}{35}} C_{11,22}^{53,0}-\frac{9}{64} \sqrt{\frac{3}{35}} C_{11,22}^{53,1}-\frac{3 C_{20,22}^{42,0}}{32}-\frac{3
C_{20,22}^{42,1}}{64}-\frac{3 C_{22,22}^{40,0}}{32}-\frac{3 C_{22,22}^{40,1}}{64}-\frac{3 C_{22,22}^{42,0}}{16 \sqrt{7}}-\frac{3 C_{22,22}^{42,1}}{32
\sqrt{7}}-\frac{9 C_{22,22}^{44,0}}{56 \sqrt{5}}-\frac{9 C_{22,22}^{44,1}}{112 \sqrt{5}}-\frac{3 C_{31,22}^{31,0}}{32}-\frac{3 C_{31,22}^{31,1}}{64}-\frac{9}{32}
\sqrt{\frac{3}{35}} C_{31,22}^{33,0}-\frac{9}{64} \sqrt{\frac{3}{35}} C_{31,22}^{33,1}-\frac{9 C_{33,22}^{33,0}}{40 \sqrt{14}}-\frac{9 C_{33,22}^{33,1}}{80
\sqrt{14}}\)

\noindent\(C_{00,4211,1}^{00,2011,0,\text{NonLoc}}= \frac{3 C_{00,22}^{62,0}}{64}+\frac{3 C_{00,22}^{62,1}}{32}-\frac{3 C_{11,22}^{51,0}}{64}-\frac{3
C_{11,22}^{51,1}}{32}-\frac{9}{64} \sqrt{\frac{3}{35}} C_{11,22}^{53,0}-\frac{9}{32} \sqrt{\frac{3}{35}} C_{11,22}^{53,1}+\frac{3 C_{20,22}^{42,0}}{64}+\frac{3
C_{20,22}^{42,1}}{32}+\frac{3 C_{22,22}^{40,0}}{64}+\frac{3 C_{22,22}^{40,1}}{32}+\frac{3 C_{22,22}^{42,0}}{32 \sqrt{7}}+\frac{3 C_{22,22}^{42,1}}{16
\sqrt{7}}+\frac{9 C_{22,22}^{44,0}}{112 \sqrt{5}}+\frac{9 C_{22,22}^{44,1}}{56 \sqrt{5}}-\frac{3 C_{31,22}^{31,0}}{64}-\frac{3 C_{31,22}^{31,1}}{32}-\frac{9}{64}
\sqrt{\frac{3}{35}} C_{31,22}^{33,0}-\frac{9}{32} \sqrt{\frac{3}{35}} C_{31,22}^{33,1}-\frac{9 C_{33,22}^{33,0}}{80 \sqrt{14}}-\frac{9 C_{33,22}^{33,1}}{40
\sqrt{14}}\)

\noindent\(C_{00,4211,1}^{00,2011,1,\text{Loc}}= -\frac{3 \sqrt{3} C_{00,22}^{62,1}}{64}-\frac{3 \sqrt{3} C_{11,22}^{51,1}}{64}-\frac{27
C_{11,22}^{53,1}}{64 \sqrt{35}}-\frac{3 \sqrt{3} C_{20,22}^{42,1}}{64}-\frac{3 \sqrt{3} C_{22,22}^{40,1}}{64}-\frac{3}{32} \sqrt{\frac{3}{7}} C_{22,22}^{42,1}-\frac{9}{112}
\sqrt{\frac{3}{5}} C_{22,22}^{44,1}-\frac{3 \sqrt{3} C_{31,22}^{31,1}}{64}-\frac{27 C_{31,22}^{33,1}}{64 \sqrt{35}}-\frac{9}{80} \sqrt{\frac{3}{14}}
C_{33,22}^{33,1}\)

\noindent\(C_{00,4211,1}^{00,2011,1,\text{NonLoc}}= \frac{3 \sqrt{3} C_{00,22}^{62,0}}{64}-\frac{3 \sqrt{3} C_{11,22}^{51,0}}{64}-\frac{27
C_{11,22}^{53,0}}{64 \sqrt{35}}+\frac{3 \sqrt{3} C_{20,22}^{42,0}}{64}+\frac{3 \sqrt{3} C_{22,22}^{40,0}}{64}+\frac{3}{32} \sqrt{\frac{3}{7}} C_{22,22}^{42,0}+\frac{9}{112}
\sqrt{\frac{3}{5}} C_{22,22}^{44,0}-\frac{3 \sqrt{3} C_{31,22}^{31,0}}{64}-\frac{27 C_{31,22}^{33,0}}{64 \sqrt{35}}-\frac{9}{80} \sqrt{\frac{3}{14}}
C_{33,22}^{33,0}\)

\noindent\(C_{00,4211,1}^{00,2211,0,\text{Loc}}= -\frac{C_{00,20}^{60,0}}{16}-\frac{C_{00,20}^{60,1}}{32}-\frac{3 C_{00,22}^{62,0}}{32
\sqrt{5}}-\frac{3 C_{00,22}^{62,1}}{64 \sqrt{5}}-\frac{3 C_{11,22}^{51,0}}{32 \sqrt{5}}-\frac{3 C_{11,22}^{51,1}}{64 \sqrt{5}}-\frac{9}{160} \sqrt{\frac{3}{7}}
C_{11,22}^{53,0}-\frac{9}{320} \sqrt{\frac{3}{7}} C_{11,22}^{53,1}-\frac{C_{20,20}^{40,0}}{16}-\frac{C_{20,20}^{40,1}}{32}-\frac{3 C_{20,22}^{42,0}}{32
\sqrt{5}}-\frac{3 C_{20,22}^{42,1}}{64 \sqrt{5}}-\frac{C_{22,20}^{42,0}}{8 \sqrt{5}}-\frac{C_{22,20}^{42,1}}{16 \sqrt{5}}-\frac{3 C_{22,22}^{40,0}}{32
\sqrt{5}}-\frac{3 C_{22,22}^{40,1}}{64 \sqrt{5}}-\frac{3 C_{22,22}^{42,0}}{16 \sqrt{35}}-\frac{3 C_{22,22}^{42,1}}{32 \sqrt{35}}-\frac{9 C_{22,22}^{44,0}}{280}-\frac{9
C_{22,22}^{44,1}}{560}-\frac{C_{31,20}^{31,0}}{16}-\frac{C_{31,20}^{31,1}}{32}-\frac{3 C_{31,22}^{31,0}}{32 \sqrt{5}}-\frac{3 C_{31,22}^{31,1}}{64
\sqrt{5}}-\frac{9}{160} \sqrt{\frac{3}{7}} C_{31,22}^{33,0}-\frac{9}{320} \sqrt{\frac{3}{7}} C_{31,22}^{33,1}-\frac{3}{40} \sqrt{\frac{3}{7}} C_{33,20}^{33,0}-\frac{3}{80}
\sqrt{\frac{3}{7}} C_{33,20}^{33,1}-\frac{9 C_{33,22}^{33,0}}{40 \sqrt{70}}-\frac{9 C_{33,22}^{33,1}}{80 \sqrt{70}}\)

\noindent\(C_{00,4211,1}^{00,2211,0,\text{NonLoc}}= -\frac{\sqrt{3} C_{00,00}^{60,0}}{64}-\frac{\sqrt{3} C_{00,00}^{60,1}}{32}-\frac{C_{00,20}^{60,0}}{64}-\frac{C_{00,20}^{60,1}}{32}+\frac{3
C_{00,22}^{62,0}}{64 \sqrt{5}}+\frac{3 C_{00,22}^{62,1}}{32 \sqrt{5}}+\frac{\sqrt{3} C_{11,00}^{51,0}}{64}+\frac{\sqrt{3} C_{11,00}^{51,1}}{32}-\frac{3
C_{11,22}^{51,0}}{64 \sqrt{5}}-\frac{3 C_{11,22}^{51,1}}{32 \sqrt{5}}-\frac{9}{320} \sqrt{\frac{3}{7}} C_{11,22}^{53,0}-\frac{9}{160} \sqrt{\frac{3}{7}}
C_{11,22}^{53,1}-\frac{\sqrt{3} C_{20,00}^{40,0}}{64}-\frac{\sqrt{3} C_{20,00}^{40,1}}{32}-\frac{C_{20,20}^{40,0}}{64}-\frac{C_{20,20}^{40,1}}{32}+\frac{3
C_{20,22}^{42,0}}{64 \sqrt{5}}+\frac{3 C_{20,22}^{42,1}}{32 \sqrt{5}}-\frac{1}{32} \sqrt{\frac{3}{5}} C_{22,00}^{42,0}-\frac{1}{16} \sqrt{\frac{3}{5}}
C_{22,00}^{42,1}-\frac{C_{22,20}^{42,0}}{32 \sqrt{5}}-\frac{C_{22,20}^{42,1}}{16 \sqrt{5}}+\frac{3 C_{22,22}^{40,0}}{64 \sqrt{5}}+\frac{3 C_{22,22}^{40,1}}{32
\sqrt{5}}+\frac{3 C_{22,22}^{42,0}}{32 \sqrt{35}}+\frac{3 C_{22,22}^{42,1}}{16 \sqrt{35}}+\frac{9 C_{22,22}^{44,0}}{560}+\frac{9 C_{22,22}^{44,1}}{280}+\frac{\sqrt{3}
C_{31,00}^{31,0}}{64}+\frac{\sqrt{3} C_{31,00}^{31,1}}{32}+\frac{C_{31,20}^{31,0}}{64}+\frac{C_{31,20}^{31,1}}{32}-\frac{3 C_{31,22}^{31,0}}{64 \sqrt{5}}-\frac{3
C_{31,22}^{31,1}}{32 \sqrt{5}}-\frac{9}{320} \sqrt{\frac{3}{7}} C_{31,22}^{33,0}-\frac{9}{160} \sqrt{\frac{3}{7}} C_{31,22}^{33,1}+\frac{9 C_{33,00}^{33,0}}{160
\sqrt{7}}+\frac{9 C_{33,00}^{33,1}}{80 \sqrt{7}}+\frac{3}{160} \sqrt{\frac{3}{7}} C_{33,20}^{33,0}+\frac{3}{80} \sqrt{\frac{3}{7}} C_{33,20}^{33,1}-\frac{9
C_{33,22}^{33,0}}{80 \sqrt{70}}-\frac{9 C_{33,22}^{33,1}}{40 \sqrt{70}}\)

\noindent\(C_{00,4211,1}^{00,2211,1,\text{Loc}}= -\frac{\sqrt{3} C_{00,20}^{60,1}}{32}-\frac{3}{64} \sqrt{\frac{3}{5}} C_{00,22}^{62,1}-\frac{3}{64}
\sqrt{\frac{3}{5}} C_{11,22}^{51,1}-\frac{27 C_{11,22}^{53,1}}{320 \sqrt{7}}-\frac{\sqrt{3} C_{20,20}^{40,1}}{32}-\frac{3}{64} \sqrt{\frac{3}{5}}
C_{20,22}^{42,1}-\frac{1}{16} \sqrt{\frac{3}{5}} C_{22,20}^{42,1}-\frac{3}{64} \sqrt{\frac{3}{5}} C_{22,22}^{40,1}-\frac{3}{32} \sqrt{\frac{3}{35}}
C_{22,22}^{42,1}-\frac{9 \sqrt{3} C_{22,22}^{44,1}}{560}-\frac{\sqrt{3} C_{31,20}^{31,1}}{32}-\frac{3}{64} \sqrt{\frac{3}{5}} C_{31,22}^{31,1}-\frac{27
C_{31,22}^{33,1}}{320 \sqrt{7}}-\frac{9 C_{33,20}^{33,1}}{80 \sqrt{7}}-\frac{9}{80} \sqrt{\frac{3}{70}} C_{33,22}^{33,1}\)

\noindent\(C_{00,4211,1}^{00,2211,1,\text{NonLoc}}= -\frac{3 C_{00,00}^{60,0}}{64}-\frac{\sqrt{3} C_{00,20}^{60,0}}{64}+\frac{3}{64}
\sqrt{\frac{3}{5}} C_{00,22}^{62,0}+\frac{3 C_{11,00}^{51,0}}{64}-\frac{3}{64} \sqrt{\frac{3}{5}} C_{11,22}^{51,0}-\frac{27 C_{11,22}^{53,0}}{320
\sqrt{7}}-\frac{3 C_{20,00}^{40,0}}{64}-\frac{\sqrt{3} C_{20,20}^{40,0}}{64}+\frac{3}{64} \sqrt{\frac{3}{5}} C_{20,22}^{42,0}-\frac{3 C_{22,00}^{42,0}}{32
\sqrt{5}}-\frac{1}{32} \sqrt{\frac{3}{5}} C_{22,20}^{42,0}+\frac{3}{64} \sqrt{\frac{3}{5}} C_{22,22}^{40,0}+\frac{3}{32} \sqrt{\frac{3}{35}} C_{22,22}^{42,0}+\frac{9
\sqrt{3} C_{22,22}^{44,0}}{560}+\frac{3 C_{31,00}^{31,0}}{64}+\frac{\sqrt{3} C_{31,20}^{31,0}}{64}-\frac{3}{64} \sqrt{\frac{3}{5}} C_{31,22}^{31,0}-\frac{27
C_{31,22}^{33,0}}{320 \sqrt{7}}+\frac{9}{160} \sqrt{\frac{3}{7}} C_{33,00}^{33,0}+\frac{9 C_{33,20}^{33,0}}{160 \sqrt{7}}-\frac{9}{80} \sqrt{\frac{3}{70}}
C_{33,22}^{33,0}\)

\noindent\(C_{00,4212,2}^{00,2212,0,\text{Loc}}= -\frac{1}{16} \sqrt{\frac{5}{3}} C_{00,20}^{60,0}-\frac{1}{32} \sqrt{\frac{5}{3}}
C_{00,20}^{60,1}+\frac{\sqrt{3} C_{00,22}^{62,0}}{32}+\frac{\sqrt{3} C_{00,22}^{62,1}}{64}+\frac{\sqrt{3} C_{11,22}^{51,0}}{32}+\frac{\sqrt{3} C_{11,22}^{51,1}}{64}+\frac{9
C_{11,22}^{53,0}}{32 \sqrt{35}}+\frac{9 C_{11,22}^{53,1}}{64 \sqrt{35}}-\frac{1}{16} \sqrt{\frac{5}{3}} C_{20,20}^{40,0}-\frac{1}{32} \sqrt{\frac{5}{3}}
C_{20,20}^{40,1}+\frac{\sqrt{3} C_{20,22}^{42,0}}{32}+\frac{\sqrt{3} C_{20,22}^{42,1}}{64}-\frac{C_{22,20}^{42,0}}{8 \sqrt{3}}-\frac{C_{22,20}^{42,1}}{16
\sqrt{3}}+\frac{\sqrt{3} C_{22,22}^{40,0}}{32}+\frac{\sqrt{3} C_{22,22}^{40,1}}{64}+\frac{1}{16} \sqrt{\frac{3}{7}} C_{22,22}^{42,0}+\frac{1}{32}
\sqrt{\frac{3}{7}} C_{22,22}^{42,1}+\frac{3}{56} \sqrt{\frac{3}{5}} C_{22,22}^{44,0}+\frac{3}{112} \sqrt{\frac{3}{5}} C_{22,22}^{44,1}-\frac{1}{16}
\sqrt{\frac{5}{3}} C_{31,20}^{31,0}-\frac{1}{32} \sqrt{\frac{5}{3}} C_{31,20}^{31,1}+\frac{\sqrt{3} C_{31,22}^{31,0}}{32}+\frac{\sqrt{3} C_{31,22}^{31,1}}{64}+\frac{9
C_{31,22}^{33,0}}{32 \sqrt{35}}+\frac{9 C_{31,22}^{33,1}}{64 \sqrt{35}}-\frac{3 C_{33,20}^{33,0}}{8 \sqrt{35}}-\frac{3 C_{33,20}^{33,1}}{16 \sqrt{35}}+\frac{3}{40}
\sqrt{\frac{3}{14}} C_{33,22}^{33,0}+\frac{3}{80} \sqrt{\frac{3}{14}} C_{33,22}^{33,1}\)

\noindent\(C_{00,4212,2}^{00,2212,0,\text{NonLoc}}= -\frac{\sqrt{5} C_{00,00}^{60,0}}{64}-\frac{\sqrt{5} C_{00,00}^{60,1}}{32}-\frac{1}{64}
\sqrt{\frac{5}{3}} C_{00,20}^{60,0}-\frac{1}{32} \sqrt{\frac{5}{3}} C_{00,20}^{60,1}-\frac{\sqrt{3} C_{00,22}^{62,0}}{64}-\frac{\sqrt{3} C_{00,22}^{62,1}}{32}+\frac{\sqrt{5}
C_{11,00}^{51,0}}{64}+\frac{\sqrt{5} C_{11,00}^{51,1}}{32}+\frac{\sqrt{3} C_{11,22}^{51,0}}{64}+\frac{\sqrt{3} C_{11,22}^{51,1}}{32}+\frac{9 C_{11,22}^{53,0}}{64
\sqrt{35}}+\frac{9 C_{11,22}^{53,1}}{32 \sqrt{35}}-\frac{\sqrt{5} C_{20,00}^{40,0}}{64}-\frac{\sqrt{5} C_{20,00}^{40,1}}{32}-\frac{1}{64} \sqrt{\frac{5}{3}}
C_{20,20}^{40,0}-\frac{1}{32} \sqrt{\frac{5}{3}} C_{20,20}^{40,1}-\frac{\sqrt{3} C_{20,22}^{42,0}}{64}-\frac{\sqrt{3} C_{20,22}^{42,1}}{32}-\frac{C_{22,00}^{42,0}}{32}-\frac{C_{22,00}^{42,1}}{16}-\frac{C_{22,20}^{42,0}}{32
\sqrt{3}}-\frac{C_{22,20}^{42,1}}{16 \sqrt{3}}-\frac{\sqrt{3} C_{22,22}^{40,0}}{64}-\frac{\sqrt{3} C_{22,22}^{40,1}}{32}-\frac{1}{32} \sqrt{\frac{3}{7}}
C_{22,22}^{42,0}-\frac{1}{16} \sqrt{\frac{3}{7}} C_{22,22}^{42,1}-\frac{3}{112} \sqrt{\frac{3}{5}} C_{22,22}^{44,0}-\frac{3}{56} \sqrt{\frac{3}{5}}
C_{22,22}^{44,1}+\frac{\sqrt{5} C_{31,00}^{31,0}}{64}+\frac{\sqrt{5} C_{31,00}^{31,1}}{32}+\frac{1}{64} \sqrt{\frac{5}{3}} C_{31,20}^{31,0}+\frac{1}{32}
\sqrt{\frac{5}{3}} C_{31,20}^{31,1}+\frac{\sqrt{3} C_{31,22}^{31,0}}{64}+\frac{\sqrt{3} C_{31,22}^{31,1}}{32}+\frac{9 C_{31,22}^{33,0}}{64 \sqrt{35}}+\frac{9
C_{31,22}^{33,1}}{32 \sqrt{35}}+\frac{3}{32} \sqrt{\frac{3}{35}} C_{33,00}^{33,0}+\frac{3}{16} \sqrt{\frac{3}{35}} C_{33,00}^{33,1}+\frac{3 C_{33,20}^{33,0}}{32
\sqrt{35}}+\frac{3 C_{33,20}^{33,1}}{16 \sqrt{35}}+\frac{3}{80} \sqrt{\frac{3}{14}} C_{33,22}^{33,0}+\frac{3}{40} \sqrt{\frac{3}{14}} C_{33,22}^{33,1}\)

\noindent\(C_{00,4212,2}^{00,2212,1,\text{Loc}}= -\frac{\sqrt{5} C_{00,20}^{60,1}}{32}+\frac{3 C_{00,22}^{62,1}}{64}+\frac{3 C_{11,22}^{51,1}}{64}+\frac{9}{64}
\sqrt{\frac{3}{35}} C_{11,22}^{53,1}-\frac{\sqrt{5} C_{20,20}^{40,1}}{32}+\frac{3 C_{20,22}^{42,1}}{64}-\frac{C_{22,20}^{42,1}}{16}+\frac{3 C_{22,22}^{40,1}}{64}+\frac{3
C_{22,22}^{42,1}}{32 \sqrt{7}}+\frac{9 C_{22,22}^{44,1}}{112 \sqrt{5}}-\frac{\sqrt{5} C_{31,20}^{31,1}}{32}+\frac{3 C_{31,22}^{31,1}}{64}+\frac{9}{64}
\sqrt{\frac{3}{35}} C_{31,22}^{33,1}-\frac{3}{16} \sqrt{\frac{3}{35}} C_{33,20}^{33,1}+\frac{9 C_{33,22}^{33,1}}{80 \sqrt{14}}\)

\noindent\(C_{00,4212,2}^{00,2212,1,\text{NonLoc}}= -\frac{\sqrt{15} C_{00,00}^{60,0}}{64}-\frac{\sqrt{5} C_{00,20}^{60,0}}{64}-\frac{3
C_{00,22}^{62,0}}{64}+\frac{\sqrt{15} C_{11,00}^{51,0}}{64}+\frac{3 C_{11,22}^{51,0}}{64}+\frac{9}{64} \sqrt{\frac{3}{35}} C_{11,22}^{53,0}-\frac{\sqrt{15}
C_{20,00}^{40,0}}{64}-\frac{\sqrt{5} C_{20,20}^{40,0}}{64}-\frac{3 C_{20,22}^{42,0}}{64}-\frac{\sqrt{3} C_{22,00}^{42,0}}{32}-\frac{C_{22,20}^{42,0}}{32}-\frac{3
C_{22,22}^{40,0}}{64}-\frac{3 C_{22,22}^{42,0}}{32 \sqrt{7}}-\frac{9 C_{22,22}^{44,0}}{112 \sqrt{5}}+\frac{\sqrt{15} C_{31,00}^{31,0}}{64}+\frac{\sqrt{5}
C_{31,20}^{31,0}}{64}+\frac{3 C_{31,22}^{31,0}}{64}+\frac{9}{64} \sqrt{\frac{3}{35}} C_{31,22}^{33,0}+\frac{9 C_{33,00}^{33,0}}{32 \sqrt{35}}+\frac{3}{32}
\sqrt{\frac{3}{35}} C_{33,20}^{33,0}+\frac{9 C_{33,22}^{33,0}}{80 \sqrt{14}}\)

\noindent\(C_{00,4213,3}^{00,2213,0,\text{Loc}}= -\frac{1}{16} \sqrt{\frac{7}{3}} C_{00,20}^{60,0}-\frac{1}{32} \sqrt{\frac{7}{3}}
C_{00,20}^{60,1}-\frac{1}{16} \sqrt{\frac{3}{35}} C_{00,22}^{62,0}-\frac{1}{32} \sqrt{\frac{3}{35}} C_{00,22}^{62,1}-\frac{1}{16} \sqrt{\frac{3}{35}}
C_{11,22}^{51,0}-\frac{1}{32} \sqrt{\frac{3}{35}} C_{11,22}^{51,1}-\frac{9 C_{11,22}^{53,0}}{560}-\frac{9 C_{11,22}^{53,1}}{1120}-\frac{1}{16} \sqrt{\frac{7}{3}}
C_{20,20}^{40,0}-\frac{1}{32} \sqrt{\frac{7}{3}} C_{20,20}^{40,1}-\frac{1}{16} \sqrt{\frac{3}{35}} C_{20,22}^{42,0}-\frac{1}{32} \sqrt{\frac{3}{35}}
C_{20,22}^{42,1}-\frac{1}{8} \sqrt{\frac{7}{15}} C_{22,20}^{42,0}-\frac{1}{16} \sqrt{\frac{7}{15}} C_{22,20}^{42,1}-\frac{1}{16} \sqrt{\frac{3}{35}}
C_{22,22}^{40,0}-\frac{1}{32} \sqrt{\frac{3}{35}} C_{22,22}^{40,1}-\frac{1}{56} \sqrt{\frac{3}{5}} C_{22,22}^{42,0}-\frac{1}{112} \sqrt{\frac{3}{5}}
C_{22,22}^{42,1}-\frac{3}{140} \sqrt{\frac{3}{7}} C_{22,22}^{44,0}-\frac{3}{280} \sqrt{\frac{3}{7}} C_{22,22}^{44,1}-\frac{1}{16} \sqrt{\frac{7}{3}}
C_{31,20}^{31,0}-\frac{1}{32} \sqrt{\frac{7}{3}} C_{31,20}^{31,1}-\frac{1}{16} \sqrt{\frac{3}{35}} C_{31,22}^{31,0}-\frac{1}{32} \sqrt{\frac{3}{35}}
C_{31,22}^{31,1}-\frac{9 C_{31,22}^{33,0}}{560}-\frac{9 C_{31,22}^{33,1}}{1120}-\frac{3 C_{33,20}^{33,0}}{40}-\frac{3 C_{33,20}^{33,1}}{80}-\frac{3}{140}
\sqrt{\frac{3}{10}} C_{33,22}^{33,0}-\frac{3}{280} \sqrt{\frac{3}{10}} C_{33,22}^{33,1}\)

\noindent\(C_{00,4213,3}^{00,2213,0,\text{NonLoc}}= -\frac{\sqrt{7} C_{00,00}^{60,0}}{64}-\frac{\sqrt{7} C_{00,00}^{60,1}}{32}-\frac{1}{64}
\sqrt{\frac{7}{3}} C_{00,20}^{60,0}-\frac{1}{32} \sqrt{\frac{7}{3}} C_{00,20}^{60,1}+\frac{1}{32} \sqrt{\frac{3}{35}} C_{00,22}^{62,0}+\frac{1}{16}
\sqrt{\frac{3}{35}} C_{00,22}^{62,1}+\frac{\sqrt{7} C_{11,00}^{51,0}}{64}+\frac{\sqrt{7} C_{11,00}^{51,1}}{32}-\frac{1}{32} \sqrt{\frac{3}{35}} C_{11,22}^{51,0}-\frac{1}{16}
\sqrt{\frac{3}{35}} C_{11,22}^{51,1}-\frac{9 C_{11,22}^{53,0}}{1120}-\frac{9 C_{11,22}^{53,1}}{560}-\frac{\sqrt{7} C_{20,00}^{40,0}}{64}-\frac{\sqrt{7}
C_{20,00}^{40,1}}{32}-\frac{1}{64} \sqrt{\frac{7}{3}} C_{20,20}^{40,0}-\frac{1}{32} \sqrt{\frac{7}{3}} C_{20,20}^{40,1}+\frac{1}{32} \sqrt{\frac{3}{35}}
C_{20,22}^{42,0}+\frac{1}{16} \sqrt{\frac{3}{35}} C_{20,22}^{42,1}-\frac{1}{32} \sqrt{\frac{7}{5}} C_{22,00}^{42,0}-\frac{1}{16} \sqrt{\frac{7}{5}}
C_{22,00}^{42,1}-\frac{1}{32} \sqrt{\frac{7}{15}} C_{22,20}^{42,0}-\frac{1}{16} \sqrt{\frac{7}{15}} C_{22,20}^{42,1}+\frac{1}{32} \sqrt{\frac{3}{35}}
C_{22,22}^{40,0}+\frac{1}{16} \sqrt{\frac{3}{35}} C_{22,22}^{40,1}+\frac{1}{112} \sqrt{\frac{3}{5}} C_{22,22}^{42,0}+\frac{1}{56} \sqrt{\frac{3}{5}}
C_{22,22}^{42,1}+\frac{3}{280} \sqrt{\frac{3}{7}} C_{22,22}^{44,0}+\frac{3}{140} \sqrt{\frac{3}{7}} C_{22,22}^{44,1}+\frac{\sqrt{7} C_{31,00}^{31,0}}{64}+\frac{\sqrt{7}
C_{31,00}^{31,1}}{32}+\frac{1}{64} \sqrt{\frac{7}{3}} C_{31,20}^{31,0}+\frac{1}{32} \sqrt{\frac{7}{3}} C_{31,20}^{31,1}-\frac{1}{32} \sqrt{\frac{3}{35}}
C_{31,22}^{31,0}-\frac{1}{16} \sqrt{\frac{3}{35}} C_{31,22}^{31,1}-\frac{9 C_{31,22}^{33,0}}{1120}-\frac{9 C_{31,22}^{33,1}}{560}+\frac{3 \sqrt{3}
C_{33,00}^{33,0}}{160}+\frac{3 \sqrt{3} C_{33,00}^{33,1}}{80}+\frac{3 C_{33,20}^{33,0}}{160}+\frac{3 C_{33,20}^{33,1}}{80}-\frac{3}{280} \sqrt{\frac{3}{10}}
C_{33,22}^{33,0}-\frac{3}{140} \sqrt{\frac{3}{10}} C_{33,22}^{33,1}\)

\noindent\(C_{00,4213,3}^{00,2213,1,\text{Loc}}= -\frac{\sqrt{7} C_{00,20}^{60,1}}{32}-\frac{3 C_{00,22}^{62,1}}{32 \sqrt{35}}-\frac{3
C_{11,22}^{51,1}}{32 \sqrt{35}}-\frac{9 \sqrt{3} C_{11,22}^{53,1}}{1120}-\frac{\sqrt{7} C_{20,20}^{40,1}}{32}-\frac{3 C_{20,22}^{42,1}}{32 \sqrt{35}}-\frac{1}{16}
\sqrt{\frac{7}{5}} C_{22,20}^{42,1}-\frac{3 C_{22,22}^{40,1}}{32 \sqrt{35}}-\frac{3 C_{22,22}^{42,1}}{112 \sqrt{5}}-\frac{9 C_{22,22}^{44,1}}{280
\sqrt{7}}-\frac{\sqrt{7} C_{31,20}^{31,1}}{32}-\frac{3 C_{31,22}^{31,1}}{32 \sqrt{35}}-\frac{9 \sqrt{3} C_{31,22}^{33,1}}{1120}-\frac{3 \sqrt{3}
C_{33,20}^{33,1}}{80}-\frac{9 C_{33,22}^{33,1}}{280 \sqrt{10}}\)

\noindent\(C_{00,4213,3}^{00,2213,1,\text{NonLoc}}= -\frac{\sqrt{21} C_{00,00}^{60,0}}{64}-\frac{\sqrt{7} C_{00,20}^{60,0}}{64}+\frac{3
C_{00,22}^{62,0}}{32 \sqrt{35}}+\frac{\sqrt{21} C_{11,00}^{51,0}}{64}-\frac{3 C_{11,22}^{51,0}}{32 \sqrt{35}}-\frac{9 \sqrt{3} C_{11,22}^{53,0}}{1120}-\frac{\sqrt{21}
C_{20,00}^{40,0}}{64}-\frac{\sqrt{7} C_{20,20}^{40,0}}{64}+\frac{3 C_{20,22}^{42,0}}{32 \sqrt{35}}-\frac{1}{32} \sqrt{\frac{21}{5}} C_{22,00}^{42,0}-\frac{1}{32}
\sqrt{\frac{7}{5}} C_{22,20}^{42,0}+\frac{3 C_{22,22}^{40,0}}{32 \sqrt{35}}+\frac{3 C_{22,22}^{42,0}}{112 \sqrt{5}}+\frac{9 C_{22,22}^{44,0}}{280
\sqrt{7}}+\frac{\sqrt{21} C_{31,00}^{31,0}}{64}+\frac{\sqrt{7} C_{31,20}^{31,0}}{64}-\frac{3 C_{31,22}^{31,0}}{32 \sqrt{35}}-\frac{9 \sqrt{3} C_{31,22}^{33,0}}{1120}+\frac{9
C_{33,00}^{33,0}}{160}+\frac{3 \sqrt{3} C_{33,20}^{33,0}}{160}-\frac{9 C_{33,22}^{33,0}}{280 \sqrt{10}}\)

\noindent\(C_{00,4413,3}^{00,2213,0,\text{Loc}}= -\frac{C_{00,22}^{62,0}}{16 \sqrt{5}}-\frac{C_{00,22}^{62,1}}{32 \sqrt{5}}-\frac{C_{11,22}^{51,0}}{16
\sqrt{5}}-\frac{C_{11,22}^{51,1}}{32 \sqrt{5}}-\frac{3}{80} \sqrt{\frac{3}{7}} C_{11,22}^{53,0}-\frac{3}{160} \sqrt{\frac{3}{7}} C_{11,22}^{53,1}-\frac{C_{20,22}^{42,0}}{16
\sqrt{5}}-\frac{C_{20,22}^{42,1}}{32 \sqrt{5}}-\frac{C_{22,22}^{40,0}}{16 \sqrt{5}}-\frac{C_{22,22}^{40,1}}{32 \sqrt{5}}-\frac{C_{22,22}^{42,0}}{8
\sqrt{35}}-\frac{C_{22,22}^{42,1}}{16 \sqrt{35}}-\frac{3 C_{22,22}^{44,0}}{140}-\frac{3 C_{22,22}^{44,1}}{280}-\frac{C_{31,22}^{31,0}}{16 \sqrt{5}}-\frac{C_{31,22}^{31,1}}{32
\sqrt{5}}-\frac{3}{80} \sqrt{\frac{3}{7}} C_{31,22}^{33,0}-\frac{3}{160} \sqrt{\frac{3}{7}} C_{31,22}^{33,1}-\frac{3 C_{33,22}^{33,0}}{20 \sqrt{70}}-\frac{3
C_{33,22}^{33,1}}{40 \sqrt{70}}\)

\noindent\(C_{00,4413,3}^{00,2213,0,\text{NonLoc}}= \frac{C_{00,22}^{62,0}}{32 \sqrt{5}}+\frac{C_{00,22}^{62,1}}{16 \sqrt{5}}-\frac{C_{11,22}^{51,0}}{32
\sqrt{5}}-\frac{C_{11,22}^{51,1}}{16 \sqrt{5}}-\frac{3}{160} \sqrt{\frac{3}{7}} C_{11,22}^{53,0}-\frac{3}{80} \sqrt{\frac{3}{7}} C_{11,22}^{53,1}+\frac{C_{20,22}^{42,0}}{32
\sqrt{5}}+\frac{C_{20,22}^{42,1}}{16 \sqrt{5}}+\frac{C_{22,22}^{40,0}}{32 \sqrt{5}}+\frac{C_{22,22}^{40,1}}{16 \sqrt{5}}+\frac{C_{22,22}^{42,0}}{16
\sqrt{35}}+\frac{C_{22,22}^{42,1}}{8 \sqrt{35}}+\frac{3 C_{22,22}^{44,0}}{280}+\frac{3 C_{22,22}^{44,1}}{140}-\frac{C_{31,22}^{31,0}}{32 \sqrt{5}}-\frac{C_{31,22}^{31,1}}{16
\sqrt{5}}-\frac{3}{160} \sqrt{\frac{3}{7}} C_{31,22}^{33,0}-\frac{3}{80} \sqrt{\frac{3}{7}} C_{31,22}^{33,1}-\frac{3 C_{33,22}^{33,0}}{40 \sqrt{70}}-\frac{3
C_{33,22}^{33,1}}{20 \sqrt{70}}\)

\noindent\(C_{00,4413,3}^{00,2213,1,\text{Loc}}= -\frac{1}{32} \sqrt{\frac{3}{5}} C_{00,22}^{62,1}-\frac{1}{32} \sqrt{\frac{3}{5}}
C_{11,22}^{51,1}-\frac{9 C_{11,22}^{53,1}}{160 \sqrt{7}}-\frac{1}{32} \sqrt{\frac{3}{5}} C_{20,22}^{42,1}-\frac{1}{32} \sqrt{\frac{3}{5}} C_{22,22}^{40,1}-\frac{1}{16}
\sqrt{\frac{3}{35}} C_{22,22}^{42,1}-\frac{3 \sqrt{3} C_{22,22}^{44,1}}{280}-\frac{1}{32} \sqrt{\frac{3}{5}} C_{31,22}^{31,1}-\frac{9 C_{31,22}^{33,1}}{160
\sqrt{7}}-\frac{3}{40} \sqrt{\frac{3}{70}} C_{33,22}^{33,1}\)

\noindent\(C_{00,4413,3}^{00,2213,1,\text{NonLoc}}= \frac{1}{32} \sqrt{\frac{3}{5}} C_{00,22}^{62,0}-\frac{1}{32} \sqrt{\frac{3}{5}}
C_{11,22}^{51,0}-\frac{9 C_{11,22}^{53,0}}{160 \sqrt{7}}+\frac{1}{32} \sqrt{\frac{3}{5}} C_{20,22}^{42,0}+\frac{1}{32} \sqrt{\frac{3}{5}} C_{22,22}^{40,0}+\frac{1}{16}
\sqrt{\frac{3}{35}} C_{22,22}^{42,0}+\frac{3 \sqrt{3} C_{22,22}^{44,0}}{280}-\frac{1}{32} \sqrt{\frac{3}{5}} C_{31,22}^{31,0}-\frac{9 C_{31,22}^{33,0}}{160
\sqrt{7}}-\frac{3}{40} \sqrt{\frac{3}{70}} C_{33,22}^{33,0}\)

\noindent\(C_{00,5101,1}^{00,1101,0,\text{Loc}}= -\frac{3 C_{00,00}^{60,0}}{32}-\frac{3 C_{00,00}^{60,1}}{64}-\frac{3 C_{11,00}^{51,0}}{32}-\frac{3
C_{11,00}^{51,1}}{64}-\frac{3 C_{20,00}^{40,0}}{32}-\frac{3 C_{20,00}^{40,1}}{64}-\frac{3 C_{22,00}^{42,0}}{16 \sqrt{5}}-\frac{3 C_{22,00}^{42,1}}{32
\sqrt{5}}-\frac{3 C_{31,00}^{31,0}}{32}-\frac{3 C_{31,00}^{31,1}}{64}-\frac{9}{80} \sqrt{\frac{3}{7}} C_{33,00}^{33,0}-\frac{9}{160} \sqrt{\frac{3}{7}}
C_{33,00}^{33,1}\)

\noindent\(C_{00,5101,1}^{00,1101,0,\text{NonLoc}}= \frac{3 C_{00,00}^{60,0}}{128}+\frac{3 C_{00,00}^{60,1}}{64}-\frac{3 \sqrt{3} C_{00,20}^{60,0}}{128}-\frac{3
\sqrt{3} C_{00,20}^{60,1}}{64}-\frac{3 C_{11,00}^{51,0}}{128}-\frac{3 C_{11,00}^{51,1}}{64}+\frac{3 \sqrt{3} C_{11,20}^{31,0}}{128}+\frac{3 \sqrt{3}
C_{11,20}^{31,1}}{64}+\frac{3 C_{20,00}^{40,0}}{128}+\frac{3 C_{20,00}^{40,1}}{64}-\frac{3 \sqrt{3} C_{20,20}^{40,0}}{128}-\frac{3 \sqrt{3} C_{20,20}^{40,1}}{64}+\frac{3
C_{22,00}^{42,0}}{64 \sqrt{5}}+\frac{3 C_{22,00}^{42,1}}{32 \sqrt{5}}-\frac{3}{64} \sqrt{\frac{3}{5}} C_{22,20}^{42,0}-\frac{3}{32} \sqrt{\frac{3}{5}}
C_{22,20}^{42,1}-\frac{3 C_{31,00}^{31,0}}{128}-\frac{3 C_{31,00}^{31,1}}{64}+\frac{3 \sqrt{3} C_{31,20}^{31,0}}{128}+\frac{3 \sqrt{3} C_{31,20}^{31,1}}{64}-\frac{9}{320}
\sqrt{\frac{3}{7}} C_{33,00}^{33,0}-\frac{9}{160} \sqrt{\frac{3}{7}} C_{33,00}^{33,1}+\frac{27 C_{33,20}^{33,0}}{320 \sqrt{7}}+\frac{27 C_{33,20}^{33,1}}{160
\sqrt{7}}\)

\noindent\(C_{00,5101,1}^{00,1101,1,\text{Loc}}= -\frac{3 \sqrt{3} C_{00,00}^{60,1}}{64}-\frac{3 \sqrt{3} C_{11,00}^{51,1}}{64}-\frac{3
\sqrt{3} C_{20,00}^{40,1}}{64}-\frac{3}{32} \sqrt{\frac{3}{5}} C_{22,00}^{42,1}-\frac{3 \sqrt{3} C_{31,00}^{31,1}}{64}-\frac{27 C_{33,00}^{33,1}}{160
\sqrt{7}}\)

\noindent\(C_{00,5101,1}^{00,1101,1,\text{NonLoc}}= \frac{3 \sqrt{3} C_{00,00}^{60,0}}{128}-\frac{9 C_{00,20}^{60,0}}{128}-\frac{3
\sqrt{3} C_{11,00}^{51,0}}{128}+\frac{9 C_{11,20}^{31,0}}{128}+\frac{3 \sqrt{3} C_{20,00}^{40,0}}{128}-\frac{9 C_{20,20}^{40,0}}{128}+\frac{3}{64}
\sqrt{\frac{3}{5}} C_{22,00}^{42,0}-\frac{9 C_{22,20}^{42,0}}{64 \sqrt{5}}-\frac{3 \sqrt{3} C_{31,00}^{31,0}}{128}+\frac{9 C_{31,20}^{31,0}}{128}-\frac{27
C_{33,00}^{33,0}}{320 \sqrt{7}}+\frac{27}{320} \sqrt{\frac{3}{7}} C_{33,20}^{33,0}\)

\noindent\(C_{00,5110,0}^{00,1110,0,\text{Loc}}= \frac{C_{00,20}^{60,0}}{32}+\frac{C_{00,20}^{60,1}}{64}+\frac{3 C_{00,22}^{62,0}}{32
\sqrt{5}}+\frac{3 C_{00,22}^{62,1}}{64 \sqrt{5}}+\frac{C_{11,20}^{31,0}}{32}+\frac{C_{11,20}^{31,1}}{64}+\frac{3 C_{11,22}^{51,0}}{32 \sqrt{5}}+\frac{3
C_{11,22}^{51,1}}{64 \sqrt{5}}+\frac{9}{160} \sqrt{\frac{3}{7}} C_{11,22}^{53,0}+\frac{9}{320} \sqrt{\frac{3}{7}} C_{11,22}^{53,1}+\frac{C_{20,20}^{40,0}}{32}+\frac{C_{20,20}^{40,1}}{64}+\frac{3
C_{20,22}^{42,0}}{32 \sqrt{5}}+\frac{3 C_{20,22}^{42,1}}{64 \sqrt{5}}+\frac{C_{22,20}^{42,0}}{16 \sqrt{5}}+\frac{C_{22,20}^{42,1}}{32 \sqrt{5}}+\frac{3
C_{22,22}^{40,0}}{32 \sqrt{5}}+\frac{3 C_{22,22}^{40,1}}{64 \sqrt{5}}+\frac{3 C_{22,22}^{42,0}}{16 \sqrt{35}}+\frac{3 C_{22,22}^{42,1}}{32 \sqrt{35}}+\frac{9
C_{22,22}^{44,0}}{280}+\frac{9 C_{22,22}^{44,1}}{560}+\frac{C_{31,20}^{31,0}}{32}+\frac{C_{31,20}^{31,1}}{64}+\frac{3 C_{31,22}^{31,0}}{32 \sqrt{5}}+\frac{3
C_{31,22}^{31,1}}{64 \sqrt{5}}+\frac{9}{160} \sqrt{\frac{3}{7}} C_{31,22}^{33,0}+\frac{9}{320} \sqrt{\frac{3}{7}} C_{31,22}^{33,1}+\frac{3}{80} \sqrt{\frac{3}{7}}
C_{33,20}^{33,0}+\frac{3}{160} \sqrt{\frac{3}{7}} C_{33,20}^{33,1}+\frac{9 C_{33,22}^{33,0}}{40 \sqrt{70}}+\frac{9 C_{33,22}^{33,1}}{80 \sqrt{70}}\)

\noindent\(C_{00,5110,0}^{00,1110,0,\text{NonLoc}}= \frac{\sqrt{3} C_{00,00}^{60,0}}{128}+\frac{\sqrt{3} C_{00,00}^{60,1}}{64}+\frac{C_{00,20}^{60,0}}{128}+\frac{C_{00,20}^{60,1}}{64}-\frac{3
C_{00,22}^{62,0}}{64 \sqrt{5}}-\frac{3 C_{00,22}^{62,1}}{32 \sqrt{5}}-\frac{\sqrt{3} C_{11,00}^{51,0}}{128}-\frac{\sqrt{3} C_{11,00}^{51,1}}{64}-\frac{C_{11,20}^{31,0}}{128}-\frac{C_{11,20}^{31,1}}{64}+\frac{3
C_{11,22}^{51,0}}{64 \sqrt{5}}+\frac{3 C_{11,22}^{51,1}}{32 \sqrt{5}}+\frac{9}{320} \sqrt{\frac{3}{7}} C_{11,22}^{53,0}+\frac{9}{160} \sqrt{\frac{3}{7}}
C_{11,22}^{53,1}+\frac{\sqrt{3} C_{20,00}^{40,0}}{128}+\frac{\sqrt{3} C_{20,00}^{40,1}}{64}+\frac{C_{20,20}^{40,0}}{128}+\frac{C_{20,20}^{40,1}}{64}-\frac{3
C_{20,22}^{42,0}}{64 \sqrt{5}}-\frac{3 C_{20,22}^{42,1}}{32 \sqrt{5}}+\frac{1}{64} \sqrt{\frac{3}{5}} C_{22,00}^{42,0}+\frac{1}{32} \sqrt{\frac{3}{5}}
C_{22,00}^{42,1}+\frac{C_{22,20}^{42,0}}{64 \sqrt{5}}+\frac{C_{22,20}^{42,1}}{32 \sqrt{5}}-\frac{3 C_{22,22}^{40,0}}{64 \sqrt{5}}-\frac{3 C_{22,22}^{40,1}}{32
\sqrt{5}}-\frac{3 C_{22,22}^{42,0}}{32 \sqrt{35}}-\frac{3 C_{22,22}^{42,1}}{16 \sqrt{35}}-\frac{9 C_{22,22}^{44,0}}{560}-\frac{9 C_{22,22}^{44,1}}{280}-\frac{\sqrt{3}
C_{31,00}^{31,0}}{128}-\frac{\sqrt{3} C_{31,00}^{31,1}}{64}-\frac{C_{31,20}^{31,0}}{128}-\frac{C_{31,20}^{31,1}}{64}+\frac{3 C_{31,22}^{31,0}}{64
\sqrt{5}}+\frac{3 C_{31,22}^{31,1}}{32 \sqrt{5}}+\frac{9}{320} \sqrt{\frac{3}{7}} C_{31,22}^{33,0}+\frac{9}{160} \sqrt{\frac{3}{7}} C_{31,22}^{33,1}-\frac{9
C_{33,00}^{33,0}}{320 \sqrt{7}}-\frac{9 C_{33,00}^{33,1}}{160 \sqrt{7}}-\frac{3}{320} \sqrt{\frac{3}{7}} C_{33,20}^{33,0}-\frac{3}{160} \sqrt{\frac{3}{7}}
C_{33,20}^{33,1}+\frac{9 C_{33,22}^{33,0}}{80 \sqrt{70}}+\frac{9 C_{33,22}^{33,1}}{40 \sqrt{70}}\)

\noindent\(C_{00,5110,0}^{00,1110,1,\text{Loc}}= \frac{\sqrt{3} C_{00,20}^{60,1}}{64}+\frac{3}{64} \sqrt{\frac{3}{5}} C_{00,22}^{62,1}+\frac{\sqrt{3}
C_{11,20}^{31,1}}{64}+\frac{3}{64} \sqrt{\frac{3}{5}} C_{11,22}^{51,1}+\frac{27 C_{11,22}^{53,1}}{320 \sqrt{7}}+\frac{\sqrt{3} C_{20,20}^{40,1}}{64}+\frac{3}{64}
\sqrt{\frac{3}{5}} C_{20,22}^{42,1}+\frac{1}{32} \sqrt{\frac{3}{5}} C_{22,20}^{42,1}+\frac{3}{64} \sqrt{\frac{3}{5}} C_{22,22}^{40,1}+\frac{3}{32}
\sqrt{\frac{3}{35}} C_{22,22}^{42,1}+\frac{9 \sqrt{3} C_{22,22}^{44,1}}{560}+\frac{\sqrt{3} C_{31,20}^{31,1}}{64}+\frac{3}{64} \sqrt{\frac{3}{5}}
C_{31,22}^{31,1}+\frac{27 C_{31,22}^{33,1}}{320 \sqrt{7}}+\frac{9 C_{33,20}^{33,1}}{160 \sqrt{7}}+\frac{9}{80} \sqrt{\frac{3}{70}} C_{33,22}^{33,1}\)

\noindent\(C_{00,5110,0}^{00,1110,1,\text{NonLoc}}= \frac{3 C_{00,00}^{60,0}}{128}+\frac{\sqrt{3} C_{00,20}^{60,0}}{128}-\frac{3}{64}
\sqrt{\frac{3}{5}} C_{00,22}^{62,0}-\frac{3 C_{11,00}^{51,0}}{128}-\frac{\sqrt{3} C_{11,20}^{31,0}}{128}+\frac{3}{64} \sqrt{\frac{3}{5}} C_{11,22}^{51,0}+\frac{27
C_{11,22}^{53,0}}{320 \sqrt{7}}+\frac{3 C_{20,00}^{40,0}}{128}+\frac{\sqrt{3} C_{20,20}^{40,0}}{128}-\frac{3}{64} \sqrt{\frac{3}{5}} C_{20,22}^{42,0}+\frac{3
C_{22,00}^{42,0}}{64 \sqrt{5}}+\frac{1}{64} \sqrt{\frac{3}{5}} C_{22,20}^{42,0}-\frac{3}{64} \sqrt{\frac{3}{5}} C_{22,22}^{40,0}-\frac{3}{32} \sqrt{\frac{3}{35}}
C_{22,22}^{42,0}-\frac{9 \sqrt{3} C_{22,22}^{44,0}}{560}-\frac{3 C_{31,00}^{31,0}}{128}-\frac{\sqrt{3} C_{31,20}^{31,0}}{128}+\frac{3}{64} \sqrt{\frac{3}{5}}
C_{31,22}^{31,0}+\frac{27 C_{31,22}^{33,0}}{320 \sqrt{7}}-\frac{9}{320} \sqrt{\frac{3}{7}} C_{33,00}^{33,0}-\frac{9 C_{33,20}^{33,0}}{320 \sqrt{7}}+\frac{9}{80}
\sqrt{\frac{3}{70}} C_{33,22}^{33,0}\)

\noindent\(C_{00,5111,1}^{00,1111,0,\text{Loc}}= \frac{\sqrt{3} C_{00,20}^{60,0}}{32}+\frac{\sqrt{3} C_{00,20}^{60,1}}{64}-\frac{3}{64}
\sqrt{\frac{3}{5}} C_{00,22}^{62,0}-\frac{3}{128} \sqrt{\frac{3}{5}} C_{00,22}^{62,1}+\frac{\sqrt{3} C_{11,20}^{31,0}}{32}+\frac{\sqrt{3} C_{11,20}^{31,1}}{64}-\frac{3}{64}
\sqrt{\frac{3}{5}} C_{11,22}^{51,0}-\frac{3}{128} \sqrt{\frac{3}{5}} C_{11,22}^{51,1}-\frac{27 C_{11,22}^{53,0}}{320 \sqrt{7}}-\frac{27 C_{11,22}^{53,1}}{640
\sqrt{7}}+\frac{\sqrt{3} C_{20,20}^{40,0}}{32}+\frac{\sqrt{3} C_{20,20}^{40,1}}{64}-\frac{3}{64} \sqrt{\frac{3}{5}} C_{20,22}^{42,0}-\frac{3}{128}
\sqrt{\frac{3}{5}} C_{20,22}^{42,1}+\frac{1}{16} \sqrt{\frac{3}{5}} C_{22,20}^{42,0}+\frac{1}{32} \sqrt{\frac{3}{5}} C_{22,20}^{42,1}-\frac{3}{64}
\sqrt{\frac{3}{5}} C_{22,22}^{40,0}-\frac{3}{128} \sqrt{\frac{3}{5}} C_{22,22}^{40,1}-\frac{3}{32} \sqrt{\frac{3}{35}} C_{22,22}^{42,0}-\frac{3}{64}
\sqrt{\frac{3}{35}} C_{22,22}^{42,1}-\frac{9 \sqrt{3} C_{22,22}^{44,0}}{560}-\frac{9 \sqrt{3} C_{22,22}^{44,1}}{1120}+\frac{\sqrt{3} C_{31,20}^{31,0}}{32}+\frac{\sqrt{3}
C_{31,20}^{31,1}}{64}-\frac{3}{64} \sqrt{\frac{3}{5}} C_{31,22}^{31,0}-\frac{3}{128} \sqrt{\frac{3}{5}} C_{31,22}^{31,1}-\frac{27 C_{31,22}^{33,0}}{320
\sqrt{7}}-\frac{27 C_{31,22}^{33,1}}{640 \sqrt{7}}+\frac{9 C_{33,20}^{33,0}}{80 \sqrt{7}}+\frac{9 C_{33,20}^{33,1}}{160 \sqrt{7}}-\frac{9}{80} \sqrt{\frac{3}{70}}
C_{33,22}^{33,0}-\frac{9}{160} \sqrt{\frac{3}{70}} C_{33,22}^{33,1}\)

\noindent\(C_{00,5111,1}^{00,1111,0,\text{NonLoc}}= \frac{3 C_{00,00}^{60,0}}{128}+\frac{3 C_{00,00}^{60,1}}{64}+\frac{\sqrt{3} C_{00,20}^{60,0}}{128}+\frac{\sqrt{3}
C_{00,20}^{60,1}}{64}+\frac{3}{128} \sqrt{\frac{3}{5}} C_{00,22}^{62,0}+\frac{3}{64} \sqrt{\frac{3}{5}} C_{00,22}^{62,1}-\frac{3 C_{11,00}^{51,0}}{128}-\frac{3
C_{11,00}^{51,1}}{64}-\frac{\sqrt{3} C_{11,20}^{31,0}}{128}-\frac{\sqrt{3} C_{11,20}^{31,1}}{64}-\frac{3}{128} \sqrt{\frac{3}{5}} C_{11,22}^{51,0}-\frac{3}{64}
\sqrt{\frac{3}{5}} C_{11,22}^{51,1}-\frac{27 C_{11,22}^{53,0}}{640 \sqrt{7}}-\frac{27 C_{11,22}^{53,1}}{320 \sqrt{7}}+\frac{3 C_{20,00}^{40,0}}{128}+\frac{3
C_{20,00}^{40,1}}{64}+\frac{\sqrt{3} C_{20,20}^{40,0}}{128}+\frac{\sqrt{3} C_{20,20}^{40,1}}{64}+\frac{3}{128} \sqrt{\frac{3}{5}} C_{20,22}^{42,0}+\frac{3}{64}
\sqrt{\frac{3}{5}} C_{20,22}^{42,1}+\frac{3 C_{22,00}^{42,0}}{64 \sqrt{5}}+\frac{3 C_{22,00}^{42,1}}{32 \sqrt{5}}+\frac{1}{64} \sqrt{\frac{3}{5}}
C_{22,20}^{42,0}+\frac{1}{32} \sqrt{\frac{3}{5}} C_{22,20}^{42,1}+\frac{3}{128} \sqrt{\frac{3}{5}} C_{22,22}^{40,0}+\frac{3}{64} \sqrt{\frac{3}{5}}
C_{22,22}^{40,1}+\frac{3}{64} \sqrt{\frac{3}{35}} C_{22,22}^{42,0}+\frac{3}{32} \sqrt{\frac{3}{35}} C_{22,22}^{42,1}+\frac{9 \sqrt{3} C_{22,22}^{44,0}}{1120}+\frac{9
\sqrt{3} C_{22,22}^{44,1}}{560}-\frac{3 C_{31,00}^{31,0}}{128}-\frac{3 C_{31,00}^{31,1}}{64}-\frac{\sqrt{3} C_{31,20}^{31,0}}{128}-\frac{\sqrt{3}
C_{31,20}^{31,1}}{64}-\frac{3}{128} \sqrt{\frac{3}{5}} C_{31,22}^{31,0}-\frac{3}{64} \sqrt{\frac{3}{5}} C_{31,22}^{31,1}-\frac{27 C_{31,22}^{33,0}}{640
\sqrt{7}}-\frac{27 C_{31,22}^{33,1}}{320 \sqrt{7}}-\frac{9}{320} \sqrt{\frac{3}{7}} C_{33,00}^{33,0}-\frac{9}{160} \sqrt{\frac{3}{7}} C_{33,00}^{33,1}-\frac{9
C_{33,20}^{33,0}}{320 \sqrt{7}}-\frac{9 C_{33,20}^{33,1}}{160 \sqrt{7}}-\frac{9}{160} \sqrt{\frac{3}{70}} C_{33,22}^{33,0}-\frac{9}{80} \sqrt{\frac{3}{70}}
C_{33,22}^{33,1}\)

\noindent\(C_{00,5111,1}^{00,1111,1,\text{Loc}}= \frac{3 C_{00,20}^{60,1}}{64}-\frac{9 C_{00,22}^{62,1}}{128 \sqrt{5}}+\frac{3 C_{11,20}^{31,1}}{64}-\frac{9
C_{11,22}^{51,1}}{128 \sqrt{5}}-\frac{27}{640} \sqrt{\frac{3}{7}} C_{11,22}^{53,1}+\frac{3 C_{20,20}^{40,1}}{64}-\frac{9 C_{20,22}^{42,1}}{128 \sqrt{5}}+\frac{3
C_{22,20}^{42,1}}{32 \sqrt{5}}-\frac{9 C_{22,22}^{40,1}}{128 \sqrt{5}}-\frac{9 C_{22,22}^{42,1}}{64 \sqrt{35}}-\frac{27 C_{22,22}^{44,1}}{1120}+\frac{3
C_{31,20}^{31,1}}{64}-\frac{9 C_{31,22}^{31,1}}{128 \sqrt{5}}-\frac{27}{640} \sqrt{\frac{3}{7}} C_{31,22}^{33,1}+\frac{9}{160} \sqrt{\frac{3}{7}}
C_{33,20}^{33,1}-\frac{27 C_{33,22}^{33,1}}{160 \sqrt{70}}\)

\noindent\(C_{00,5111,1}^{00,1111,1,\text{NonLoc}}= \frac{3 \sqrt{3} C_{00,00}^{60,0}}{128}+\frac{3 C_{00,20}^{60,0}}{128}+\frac{9
C_{00,22}^{62,0}}{128 \sqrt{5}}-\frac{3 \sqrt{3} C_{11,00}^{51,0}}{128}-\frac{3 C_{11,20}^{31,0}}{128}-\frac{9 C_{11,22}^{51,0}}{128 \sqrt{5}}-\frac{27}{640}
\sqrt{\frac{3}{7}} C_{11,22}^{53,0}+\frac{3 \sqrt{3} C_{20,00}^{40,0}}{128}+\frac{3 C_{20,20}^{40,0}}{128}+\frac{9 C_{20,22}^{42,0}}{128 \sqrt{5}}+\frac{3}{64}
\sqrt{\frac{3}{5}} C_{22,00}^{42,0}+\frac{3 C_{22,20}^{42,0}}{64 \sqrt{5}}+\frac{9 C_{22,22}^{40,0}}{128 \sqrt{5}}+\frac{9 C_{22,22}^{42,0}}{64 \sqrt{35}}+\frac{27
C_{22,22}^{44,0}}{1120}-\frac{3 \sqrt{3} C_{31,00}^{31,0}}{128}-\frac{3 C_{31,20}^{31,0}}{128}-\frac{9 C_{31,22}^{31,0}}{128 \sqrt{5}}-\frac{27}{640}
\sqrt{\frac{3}{7}} C_{31,22}^{33,0}-\frac{27 C_{33,00}^{33,0}}{320 \sqrt{7}}-\frac{9}{320} \sqrt{\frac{3}{7}} C_{33,20}^{33,0}-\frac{27 C_{33,22}^{33,0}}{160
\sqrt{70}}\)

\noindent\(C_{00,5112,2}^{00,1112,0,\text{Loc}}= \frac{\sqrt{5} C_{00,20}^{60,0}}{32}+\frac{\sqrt{5} C_{00,20}^{60,1}}{64}+\frac{3
C_{00,22}^{62,0}}{320}+\frac{3 C_{00,22}^{62,1}}{640}+\frac{\sqrt{5} C_{11,20}^{31,0}}{32}+\frac{\sqrt{5} C_{11,20}^{31,1}}{64}+\frac{3 C_{11,22}^{51,0}}{320}+\frac{3
C_{11,22}^{51,1}}{640}+\frac{9}{320} \sqrt{\frac{3}{35}} C_{11,22}^{53,0}+\frac{9}{640} \sqrt{\frac{3}{35}} C_{11,22}^{53,1}+\frac{\sqrt{5} C_{20,20}^{40,0}}{32}+\frac{\sqrt{5}
C_{20,20}^{40,1}}{64}+\frac{3 C_{20,22}^{42,0}}{320}+\frac{3 C_{20,22}^{42,1}}{640}+\frac{C_{22,20}^{42,0}}{16}+\frac{C_{22,20}^{42,1}}{32}+\frac{3
C_{22,22}^{40,0}}{320}+\frac{3 C_{22,22}^{40,1}}{640}+\frac{3 C_{22,22}^{42,0}}{160 \sqrt{7}}+\frac{3 C_{22,22}^{42,1}}{320 \sqrt{7}}+\frac{9 C_{22,22}^{44,0}}{560
\sqrt{5}}+\frac{9 C_{22,22}^{44,1}}{1120 \sqrt{5}}+\frac{\sqrt{5} C_{31,20}^{31,0}}{32}+\frac{\sqrt{5} C_{31,20}^{31,1}}{64}+\frac{3 C_{31,22}^{31,0}}{320}+\frac{3
C_{31,22}^{31,1}}{640}+\frac{9}{320} \sqrt{\frac{3}{35}} C_{31,22}^{33,0}+\frac{9}{640} \sqrt{\frac{3}{35}} C_{31,22}^{33,1}+\frac{3}{16} \sqrt{\frac{3}{35}}
C_{33,20}^{33,0}+\frac{3}{32} \sqrt{\frac{3}{35}} C_{33,20}^{33,1}+\frac{9 C_{33,22}^{33,0}}{400 \sqrt{14}}+\frac{9 C_{33,22}^{33,1}}{800 \sqrt{14}}\)

\noindent\(C_{00,5112,2}^{00,1112,0,\text{NonLoc}}= \frac{\sqrt{15} C_{00,00}^{60,0}}{128}+\frac{\sqrt{15} C_{00,00}^{60,1}}{64}+\frac{\sqrt{5}
C_{00,20}^{60,0}}{128}+\frac{\sqrt{5} C_{00,20}^{60,1}}{64}-\frac{3 C_{00,22}^{62,0}}{640}-\frac{3 C_{00,22}^{62,1}}{320}-\frac{\sqrt{15} C_{11,00}^{51,0}}{128}-\frac{\sqrt{15}
C_{11,00}^{51,1}}{64}-\frac{\sqrt{5} C_{11,20}^{31,0}}{128}-\frac{\sqrt{5} C_{11,20}^{31,1}}{64}+\frac{3 C_{11,22}^{51,0}}{640}+\frac{3 C_{11,22}^{51,1}}{320}+\frac{9}{640}
\sqrt{\frac{3}{35}} C_{11,22}^{53,0}+\frac{9}{320} \sqrt{\frac{3}{35}} C_{11,22}^{53,1}+\frac{\sqrt{15} C_{20,00}^{40,0}}{128}+\frac{\sqrt{15} C_{20,00}^{40,1}}{64}+\frac{\sqrt{5}
C_{20,20}^{40,0}}{128}+\frac{\sqrt{5} C_{20,20}^{40,1}}{64}-\frac{3 C_{20,22}^{42,0}}{640}-\frac{3 C_{20,22}^{42,1}}{320}+\frac{\sqrt{3} C_{22,00}^{42,0}}{64}+\frac{\sqrt{3}
C_{22,00}^{42,1}}{32}+\frac{C_{22,20}^{42,0}}{64}+\frac{C_{22,20}^{42,1}}{32}-\frac{3 C_{22,22}^{40,0}}{640}-\frac{3 C_{22,22}^{40,1}}{320}-\frac{3
C_{22,22}^{42,0}}{320 \sqrt{7}}-\frac{3 C_{22,22}^{42,1}}{160 \sqrt{7}}-\frac{9 C_{22,22}^{44,0}}{1120 \sqrt{5}}-\frac{9 C_{22,22}^{44,1}}{560 \sqrt{5}}-\frac{\sqrt{15}
C_{31,00}^{31,0}}{128}-\frac{\sqrt{15} C_{31,00}^{31,1}}{64}-\frac{\sqrt{5} C_{31,20}^{31,0}}{128}-\frac{\sqrt{5} C_{31,20}^{31,1}}{64}+\frac{3 C_{31,22}^{31,0}}{640}+\frac{3
C_{31,22}^{31,1}}{320}+\frac{9}{640} \sqrt{\frac{3}{35}} C_{31,22}^{33,0}+\frac{9}{320} \sqrt{\frac{3}{35}} C_{31,22}^{33,1}-\frac{9 C_{33,00}^{33,0}}{64
\sqrt{35}}-\frac{9 C_{33,00}^{33,1}}{32 \sqrt{35}}-\frac{3}{64} \sqrt{\frac{3}{35}} C_{33,20}^{33,0}-\frac{3}{32} \sqrt{\frac{3}{35}} C_{33,20}^{33,1}+\frac{9
C_{33,22}^{33,0}}{800 \sqrt{14}}+\frac{9 C_{33,22}^{33,1}}{400 \sqrt{14}}\)

\noindent\(C_{00,5112,2}^{00,1112,1,\text{Loc}}= \frac{\sqrt{15} C_{00,20}^{60,1}}{64}+\frac{3 \sqrt{3} C_{00,22}^{62,1}}{640}+\frac{\sqrt{15}
C_{11,20}^{31,1}}{64}+\frac{3 \sqrt{3} C_{11,22}^{51,1}}{640}+\frac{27 C_{11,22}^{53,1}}{640 \sqrt{35}}+\frac{\sqrt{15} C_{20,20}^{40,1}}{64}+\frac{3
\sqrt{3} C_{20,22}^{42,1}}{640}+\frac{\sqrt{3} C_{22,20}^{42,1}}{32}+\frac{3 \sqrt{3} C_{22,22}^{40,1}}{640}+\frac{3}{320} \sqrt{\frac{3}{7}} C_{22,22}^{42,1}+\frac{9
\sqrt{\frac{3}{5}} C_{22,22}^{44,1}}{1120}+\frac{\sqrt{15} C_{31,20}^{31,1}}{64}+\frac{3 \sqrt{3} C_{31,22}^{31,1}}{640}+\frac{27 C_{31,22}^{33,1}}{640
\sqrt{35}}+\frac{9 C_{33,20}^{33,1}}{32 \sqrt{35}}+\frac{9}{800} \sqrt{\frac{3}{14}} C_{33,22}^{33,1}\)

\noindent\(C_{00,5112,2}^{00,1112,1,\text{NonLoc}}= \frac{3 \sqrt{5} C_{00,00}^{60,0}}{128}+\frac{\sqrt{15} C_{00,20}^{60,0}}{128}-\frac{3
\sqrt{3} C_{00,22}^{62,0}}{640}-\frac{3 \sqrt{5} C_{11,00}^{51,0}}{128}-\frac{\sqrt{15} C_{11,20}^{31,0}}{128}+\frac{3 \sqrt{3} C_{11,22}^{51,0}}{640}+\frac{27
C_{11,22}^{53,0}}{640 \sqrt{35}}+\frac{3 \sqrt{5} C_{20,00}^{40,0}}{128}+\frac{\sqrt{15} C_{20,20}^{40,0}}{128}-\frac{3 \sqrt{3} C_{20,22}^{42,0}}{640}+\frac{3
C_{22,00}^{42,0}}{64}+\frac{\sqrt{3} C_{22,20}^{42,0}}{64}-\frac{3 \sqrt{3} C_{22,22}^{40,0}}{640}-\frac{3}{320} \sqrt{\frac{3}{7}} C_{22,22}^{42,0}-\frac{9
\sqrt{\frac{3}{5}} C_{22,22}^{44,0}}{1120}-\frac{3 \sqrt{5} C_{31,00}^{31,0}}{128}-\frac{\sqrt{15} C_{31,20}^{31,0}}{128}+\frac{3 \sqrt{3} C_{31,22}^{31,0}}{640}+\frac{27
C_{31,22}^{33,0}}{640 \sqrt{35}}-\frac{9}{64} \sqrt{\frac{3}{35}} C_{33,00}^{33,0}-\frac{9 C_{33,20}^{33,0}}{64 \sqrt{35}}+\frac{9}{800} \sqrt{\frac{3}{14}}
C_{33,22}^{33,0}\)

\noindent\(C_{00,5312,2}^{00,1112,0,\text{Loc}}= \frac{1}{16} \sqrt{\frac{7}{15}} C_{00,22}^{62,0}+\frac{1}{32} \sqrt{\frac{7}{15}}
C_{00,22}^{62,1}+\frac{1}{16} \sqrt{\frac{7}{15}} C_{11,22}^{51,0}+\frac{1}{32} \sqrt{\frac{7}{15}} C_{11,22}^{51,1}+\frac{3 C_{11,22}^{53,0}}{80}+\frac{3
C_{11,22}^{53,1}}{160}+\frac{1}{16} \sqrt{\frac{7}{15}} C_{20,22}^{42,0}+\frac{1}{32} \sqrt{\frac{7}{15}} C_{20,22}^{42,1}+\frac{1}{16} \sqrt{\frac{7}{15}}
C_{22,22}^{40,0}+\frac{1}{32} \sqrt{\frac{7}{15}} C_{22,22}^{40,1}+\frac{C_{22,22}^{42,0}}{8 \sqrt{15}}+\frac{C_{22,22}^{42,1}}{16 \sqrt{15}}+\frac{1}{20}
\sqrt{\frac{3}{7}} C_{22,22}^{44,0}+\frac{1}{40} \sqrt{\frac{3}{7}} C_{22,22}^{44,1}+\frac{1}{16} \sqrt{\frac{7}{15}} C_{31,22}^{31,0}+\frac{1}{32}
\sqrt{\frac{7}{15}} C_{31,22}^{31,1}+\frac{3 C_{31,22}^{33,0}}{80}+\frac{3 C_{31,22}^{33,1}}{160}+\frac{1}{20} \sqrt{\frac{3}{10}} C_{33,22}^{33,0}+\frac{1}{40}
\sqrt{\frac{3}{10}} C_{33,22}^{33,1}\)

\noindent\(C_{00,5312,2}^{00,1112,0,\text{NonLoc}}= -\frac{1}{32} \sqrt{\frac{7}{15}} C_{00,22}^{62,0}-\frac{1}{16} \sqrt{\frac{7}{15}}
C_{00,22}^{62,1}+\frac{1}{32} \sqrt{\frac{7}{15}} C_{11,22}^{51,0}+\frac{1}{16} \sqrt{\frac{7}{15}} C_{11,22}^{51,1}+\frac{3 C_{11,22}^{53,0}}{160}+\frac{3
C_{11,22}^{53,1}}{80}-\frac{1}{32} \sqrt{\frac{7}{15}} C_{20,22}^{42,0}-\frac{1}{16} \sqrt{\frac{7}{15}} C_{20,22}^{42,1}-\frac{1}{32} \sqrt{\frac{7}{15}}
C_{22,22}^{40,0}-\frac{1}{16} \sqrt{\frac{7}{15}} C_{22,22}^{40,1}-\frac{C_{22,22}^{42,0}}{16 \sqrt{15}}-\frac{C_{22,22}^{42,1}}{8 \sqrt{15}}-\frac{1}{40}
\sqrt{\frac{3}{7}} C_{22,22}^{44,0}-\frac{1}{20} \sqrt{\frac{3}{7}} C_{22,22}^{44,1}+\frac{1}{32} \sqrt{\frac{7}{15}} C_{31,22}^{31,0}+\frac{1}{16}
\sqrt{\frac{7}{15}} C_{31,22}^{31,1}+\frac{3 C_{31,22}^{33,0}}{160}+\frac{3 C_{31,22}^{33,1}}{80}+\frac{1}{40} \sqrt{\frac{3}{10}} C_{33,22}^{33,0}+\frac{1}{20}
\sqrt{\frac{3}{10}} C_{33,22}^{33,1}\)

\noindent\(C_{00,5312,2}^{00,1112,1,\text{Loc}}= \frac{1}{32} \sqrt{\frac{7}{5}} C_{00,22}^{62,1}+\frac{1}{32} \sqrt{\frac{7}{5}} C_{11,22}^{51,1}+\frac{3
\sqrt{3} C_{11,22}^{53,1}}{160}+\frac{1}{32} \sqrt{\frac{7}{5}} C_{20,22}^{42,1}+\frac{1}{32} \sqrt{\frac{7}{5}} C_{22,22}^{40,1}+\frac{C_{22,22}^{42,1}}{16
\sqrt{5}}+\frac{3 C_{22,22}^{44,1}}{40 \sqrt{7}}+\frac{1}{32} \sqrt{\frac{7}{5}} C_{31,22}^{31,1}+\frac{3 \sqrt{3} C_{31,22}^{33,1}}{160}+\frac{3
C_{33,22}^{33,1}}{40 \sqrt{10}}\)

\noindent\(C_{00,5312,2}^{00,1112,1,\text{NonLoc}}= -\frac{1}{32} \sqrt{\frac{7}{5}} C_{00,22}^{62,0}+\frac{1}{32} \sqrt{\frac{7}{5}}
C_{11,22}^{51,0}+\frac{3 \sqrt{3} C_{11,22}^{53,0}}{160}-\frac{1}{32} \sqrt{\frac{7}{5}} C_{20,22}^{42,0}-\frac{1}{32} \sqrt{\frac{7}{5}} C_{22,22}^{40,0}-\frac{C_{22,22}^{42,0}}{16
\sqrt{5}}-\frac{3 C_{22,22}^{44,0}}{40 \sqrt{7}}+\frac{1}{32} \sqrt{\frac{7}{5}} C_{31,22}^{31,0}+\frac{3 \sqrt{3} C_{31,22}^{33,0}}{160}+\frac{3
C_{33,22}^{33,0}}{40 \sqrt{10}}\)

\noindent\(C_{00,6000,0}^{00,0000,0,\text{Loc}}= \frac{C_{00,00}^{60,0}}{64}+\frac{C_{00,00}^{60,1}}{128}+\frac{C_{11,00}^{51,0}}{64}+\frac{C_{11,00}^{51,1}}{128}+\frac{C_{20,00}^{40,0}}{64}+\frac{C_{20,00}^{40,1}}{128}+\frac{C_{22,00}^{42,0}}{32
\sqrt{5}}+\frac{C_{22,00}^{42,1}}{64 \sqrt{5}}+\frac{C_{31,00}^{31,0}}{64}+\frac{C_{31,00}^{31,1}}{128}+\frac{3}{160} \sqrt{\frac{3}{7}} C_{33,00}^{33,0}+\frac{3}{320}
\sqrt{\frac{3}{7}} C_{33,00}^{33,1}\)

\noindent\(C_{00,6000,0}^{00,0000,0,\text{NonLoc}}= -\frac{C_{00,00}^{60,0}}{256}-\frac{C_{00,00}^{60,1}}{128}+\frac{\sqrt{3} C_{00,20}^{60,0}}{256}+\frac{\sqrt{3}
C_{00,20}^{60,1}}{128}+\frac{C_{11,00}^{51,0}}{256}+\frac{C_{11,00}^{51,1}}{128}-\frac{\sqrt{3} C_{11,20}^{31,0}}{256}-\frac{\sqrt{3} C_{11,20}^{31,1}}{128}-\frac{C_{20,00}^{40,0}}{256}-\frac{C_{20,00}^{40,1}}{128}+\frac{\sqrt{3}
C_{20,20}^{40,0}}{256}+\frac{\sqrt{3} C_{20,20}^{40,1}}{128}-\frac{C_{22,00}^{42,0}}{128 \sqrt{5}}-\frac{C_{22,00}^{42,1}}{64 \sqrt{5}}+\frac{1}{128}
\sqrt{\frac{3}{5}} C_{22,20}^{42,0}+\frac{1}{64} \sqrt{\frac{3}{5}} C_{22,20}^{42,1}+\frac{C_{31,00}^{31,0}}{256}+\frac{C_{31,00}^{31,1}}{128}-\frac{\sqrt{3}
C_{31,20}^{31,0}}{256}-\frac{\sqrt{3} C_{31,20}^{31,1}}{128}+\frac{3}{640} \sqrt{\frac{3}{7}} C_{33,00}^{33,0}+\frac{3}{320} \sqrt{\frac{3}{7}} C_{33,00}^{33,1}-\frac{9
C_{33,20}^{33,0}}{640 \sqrt{7}}-\frac{9 C_{33,20}^{33,1}}{320 \sqrt{7}}\)

\noindent\(C_{00,6000,0}^{00,0000,1,\text{Loc}}= \frac{\sqrt{3} C_{00,00}^{60,1}}{128}+\frac{\sqrt{3} C_{11,00}^{51,1}}{128}+\frac{\sqrt{3}
C_{20,00}^{40,1}}{128}+\frac{1}{64} \sqrt{\frac{3}{5}} C_{22,00}^{42,1}+\frac{\sqrt{3} C_{31,00}^{31,1}}{128}+\frac{9 C_{33,00}^{33,1}}{320 \sqrt{7}}\)

\noindent\(C_{00,6000,0}^{00,0000,1,\text{NonLoc}}= -\frac{\sqrt{3} C_{00,00}^{60,0}}{256}+\frac{3 C_{00,20}^{60,0}}{256}+\frac{\sqrt{3}
C_{11,00}^{51,0}}{256}-\frac{3 C_{11,20}^{31,0}}{256}-\frac{\sqrt{3} C_{20,00}^{40,0}}{256}+\frac{3 C_{20,20}^{40,0}}{256}-\frac{1}{128} \sqrt{\frac{3}{5}}
C_{22,00}^{42,0}+\frac{3 C_{22,20}^{42,0}}{128 \sqrt{5}}+\frac{\sqrt{3} C_{31,00}^{31,0}}{256}-\frac{3 C_{31,20}^{31,0}}{256}+\frac{9 C_{33,00}^{33,0}}{640
\sqrt{7}}-\frac{9}{640} \sqrt{\frac{3}{7}} C_{33,20}^{33,0}\)

\noindent\(C_{00,6011,1}^{00,0011,0,\text{Loc}}= -\frac{C_{00,20}^{60,0}}{64}-\frac{C_{00,20}^{60,1}}{128}-\frac{C_{11,20}^{31,0}}{64}-\frac{C_{11,20}^{31,1}}{128}-\frac{C_{20,20}^{40,0}}{64}-\frac{C_{20,20}^{40,1}}{128}-\frac{C_{22,20}^{42,0}}{32
\sqrt{5}}-\frac{C_{22,20}^{42,1}}{64 \sqrt{5}}-\frac{C_{31,20}^{31,0}}{64}-\frac{C_{31,20}^{31,1}}{128}-\frac{3}{160} \sqrt{\frac{3}{7}} C_{33,20}^{33,0}-\frac{3}{320}
\sqrt{\frac{3}{7}} C_{33,20}^{33,1}\)

\noindent\(C_{00,6011,1}^{00,0011,0,\text{NonLoc}}= -\frac{\sqrt{3} C_{00,00}^{60,0}}{256}-\frac{\sqrt{3} C_{00,00}^{60,1}}{128}-\frac{C_{00,20}^{60,0}}{256}-\frac{C_{00,20}^{60,1}}{128}+\frac{\sqrt{3}
C_{11,00}^{51,0}}{256}+\frac{\sqrt{3} C_{11,00}^{51,1}}{128}+\frac{C_{11,20}^{31,0}}{256}+\frac{C_{11,20}^{31,1}}{128}-\frac{\sqrt{3} C_{20,00}^{40,0}}{256}-\frac{\sqrt{3}
C_{20,00}^{40,1}}{128}-\frac{C_{20,20}^{40,0}}{256}-\frac{C_{20,20}^{40,1}}{128}-\frac{1}{128} \sqrt{\frac{3}{5}} C_{22,00}^{42,0}-\frac{1}{64} \sqrt{\frac{3}{5}}
C_{22,00}^{42,1}-\frac{C_{22,20}^{42,0}}{128 \sqrt{5}}-\frac{C_{22,20}^{42,1}}{64 \sqrt{5}}+\frac{\sqrt{3} C_{31,00}^{31,0}}{256}+\frac{\sqrt{3}
C_{31,00}^{31,1}}{128}+\frac{C_{31,20}^{31,0}}{256}+\frac{C_{31,20}^{31,1}}{128}+\frac{9 C_{33,00}^{33,0}}{640 \sqrt{7}}+\frac{9 C_{33,00}^{33,1}}{320
\sqrt{7}}+\frac{3}{640} \sqrt{\frac{3}{7}} C_{33,20}^{33,0}+\frac{3}{320} \sqrt{\frac{3}{7}} C_{33,20}^{33,1}\)

\noindent\(C_{00,6011,1}^{00,0011,1,\text{Loc}}= -\frac{\sqrt{3} C_{00,20}^{60,1}}{128}-\frac{\sqrt{3} C_{11,20}^{31,1}}{128}-\frac{\sqrt{3}
C_{20,20}^{40,1}}{128}-\frac{1}{64} \sqrt{\frac{3}{5}} C_{22,20}^{42,1}-\frac{\sqrt{3} C_{31,20}^{31,1}}{128}-\frac{9 C_{33,20}^{33,1}}{320 \sqrt{7}}\)

\noindent\(C_{00,6011,1}^{00,0011,1,\text{NonLoc}}= -\frac{3 C_{00,00}^{60,0}}{256}-\frac{\sqrt{3} C_{00,20}^{60,0}}{256}+\frac{3 C_{11,00}^{51,0}}{256}+\frac{\sqrt{3}
C_{11,20}^{31,0}}{256}-\frac{3 C_{20,00}^{40,0}}{256}-\frac{\sqrt{3} C_{20,20}^{40,0}}{256}-\frac{3 C_{22,00}^{42,0}}{128 \sqrt{5}}-\frac{1}{128}
\sqrt{\frac{3}{5}} C_{22,20}^{42,0}+\frac{3 C_{31,00}^{31,0}}{256}+\frac{\sqrt{3} C_{31,20}^{31,0}}{256}+\frac{9}{640} \sqrt{\frac{3}{7}} C_{33,00}^{33,0}+\frac{9
C_{33,20}^{33,0}}{640 \sqrt{7}}\)

\noindent\(C_{00,6211,1}^{00,0011,0,\text{Loc}}= -\frac{C_{00,22}^{62,0}}{64}-\frac{C_{00,22}^{62,1}}{128}-\frac{C_{11,22}^{51,0}}{64}-\frac{C_{11,22}^{51,1}}{128}-\frac{3}{64}
\sqrt{\frac{3}{35}} C_{11,22}^{53,0}-\frac{3}{128} \sqrt{\frac{3}{35}} C_{11,22}^{53,1}-\frac{C_{20,22}^{42,0}}{64}-\frac{C_{20,22}^{42,1}}{128}-\frac{C_{22,22}^{40,0}}{64}-\frac{C_{22,22}^{40,1}}{128}-\frac{C_{22,22}^{42,0}}{32
\sqrt{7}}-\frac{C_{22,22}^{42,1}}{64 \sqrt{7}}-\frac{3 C_{22,22}^{44,0}}{112 \sqrt{5}}-\frac{3 C_{22,22}^{44,1}}{224 \sqrt{5}}-\frac{C_{31,22}^{31,0}}{64}-\frac{C_{31,22}^{31,1}}{128}-\frac{3}{64}
\sqrt{\frac{3}{35}} C_{31,22}^{33,0}-\frac{3}{128} \sqrt{\frac{3}{35}} C_{31,22}^{33,1}-\frac{3 C_{33,22}^{33,0}}{80 \sqrt{14}}-\frac{3 C_{33,22}^{33,1}}{160
\sqrt{14}}\)

\noindent\(C_{00,6211,1}^{00,0011,0,\text{NonLoc}}= \frac{C_{00,22}^{62,0}}{128}+\frac{C_{00,22}^{62,1}}{64}-\frac{C_{11,22}^{51,0}}{128}-\frac{C_{11,22}^{51,1}}{64}-\frac{3}{128}
\sqrt{\frac{3}{35}} C_{11,22}^{53,0}-\frac{3}{64} \sqrt{\frac{3}{35}} C_{11,22}^{53,1}+\frac{C_{20,22}^{42,0}}{128}+\frac{C_{20,22}^{42,1}}{64}+\frac{C_{22,22}^{40,0}}{128}+\frac{C_{22,22}^{40,1}}{64}+\frac{C_{22,22}^{42,0}}{64
\sqrt{7}}+\frac{C_{22,22}^{42,1}}{32 \sqrt{7}}+\frac{3 C_{22,22}^{44,0}}{224 \sqrt{5}}+\frac{3 C_{22,22}^{44,1}}{112 \sqrt{5}}-\frac{C_{31,22}^{31,0}}{128}-\frac{C_{31,22}^{31,1}}{64}-\frac{3}{128}
\sqrt{\frac{3}{35}} C_{31,22}^{33,0}-\frac{3}{64} \sqrt{\frac{3}{35}} C_{31,22}^{33,1}-\frac{3 C_{33,22}^{33,0}}{160 \sqrt{14}}-\frac{3 C_{33,22}^{33,1}}{80
\sqrt{14}}\)

\noindent\(C_{00,6211,1}^{00,0011,1,\text{Loc}}= -\frac{\sqrt{3} C_{00,22}^{62,1}}{128}-\frac{\sqrt{3} C_{11,22}^{51,1}}{128}-\frac{9
C_{11,22}^{53,1}}{128 \sqrt{35}}-\frac{\sqrt{3} C_{20,22}^{42,1}}{128}-\frac{\sqrt{3} C_{22,22}^{40,1}}{128}-\frac{1}{64} \sqrt{\frac{3}{7}} C_{22,22}^{42,1}-\frac{3}{224}
\sqrt{\frac{3}{5}} C_{22,22}^{44,1}-\frac{\sqrt{3} C_{31,22}^{31,1}}{128}-\frac{9 C_{31,22}^{33,1}}{128 \sqrt{35}}-\frac{3}{160} \sqrt{\frac{3}{14}}
C_{33,22}^{33,1}\)

\noindent\(C_{00,6211,1}^{00,0011,1,\text{NonLoc}}= \frac{\sqrt{3} C_{00,22}^{62,0}}{128}-\frac{\sqrt{3} C_{11,22}^{51,0}}{128}-\frac{9
C_{11,22}^{53,0}}{128 \sqrt{35}}+\frac{\sqrt{3} C_{20,22}^{42,0}}{128}+\frac{\sqrt{3} C_{22,22}^{40,0}}{128}+\frac{1}{64} \sqrt{\frac{3}{7}} C_{22,22}^{42,0}+\frac{3}{224}
\sqrt{\frac{3}{5}} C_{22,22}^{44,0}-\frac{\sqrt{3} C_{31,22}^{31,0}}{128}-\frac{9 C_{31,22}^{33,0}}{128 \sqrt{35}}-\frac{3}{160} \sqrt{\frac{3}{14}}
C_{33,22}^{33,0}\)

\noindent\(C_{11,0000,1}^{00,5111,0,\text{Loc}}= \frac{C_{11,11}^{51,0}}{64}+\frac{C_{11,11}^{51,1}}{128}+\frac{1}{32} \sqrt{\frac{3}{5}}
C_{22,11}^{42,0}+\frac{1}{64} \sqrt{\frac{3}{5}} C_{22,11}^{42,1}+\frac{C_{31,11}^{31,0}}{64}+\frac{C_{31,11}^{31,1}}{128}+\frac{9 C_{33,11}^{33,0}}{80
\sqrt{14}}+\frac{9 C_{33,11}^{33,1}}{160 \sqrt{14}}\)

\noindent\(C_{11,0000,1}^{00,5111,0,\text{NonLoc}}= \frac{C_{11,11}^{51,0}}{128}+\frac{C_{11,11}^{51,1}}{64}-\frac{1}{64} \sqrt{\frac{3}{5}}
C_{22,11}^{42,0}-\frac{1}{32} \sqrt{\frac{3}{5}} C_{22,11}^{42,1}+\frac{C_{31,11}^{31,0}}{128}+\frac{C_{31,11}^{31,1}}{64}+\frac{9 C_{33,11}^{33,0}}{160
\sqrt{14}}+\frac{9 C_{33,11}^{33,1}}{80 \sqrt{14}}\)

\noindent\(C_{11,0000,1}^{00,5111,1,\text{Loc}}= \frac{\sqrt{3} C_{11,11}^{51,1}}{128}+\frac{3 C_{22,11}^{42,1}}{64 \sqrt{5}}+\frac{\sqrt{3}
C_{31,11}^{31,1}}{128}+\frac{9}{160} \sqrt{\frac{3}{14}} C_{33,11}^{33,1}\)

\noindent\(C_{11,0000,1}^{00,5111,1,\text{NonLoc}}= \frac{\sqrt{3} C_{11,11}^{51,0}}{128}-\frac{3 C_{22,11}^{42,0}}{64 \sqrt{5}}+\frac{\sqrt{3}
C_{31,11}^{31,0}}{128}+\frac{9}{160} \sqrt{\frac{3}{14}} C_{33,11}^{33,0}\)

\noindent\(C_{11,0011,1}^{00,5101,0,\text{Loc}}= -\frac{C_{11,11}^{51,0}}{64}-\frac{C_{11,11}^{51,1}}{128}-\frac{1}{32} \sqrt{\frac{3}{5}}
C_{22,11}^{42,0}-\frac{1}{64} \sqrt{\frac{3}{5}} C_{22,11}^{42,1}-\frac{C_{31,11}^{31,0}}{64}-\frac{C_{31,11}^{31,1}}{128}-\frac{9 C_{33,11}^{33,0}}{80
\sqrt{14}}-\frac{9 C_{33,11}^{33,1}}{160 \sqrt{14}}\)

\noindent\(C_{11,0011,1}^{00,5101,0,\text{NonLoc}}= -\frac{C_{11,11}^{51,0}}{128}-\frac{C_{11,11}^{51,1}}{64}+\frac{1}{64} \sqrt{\frac{3}{5}}
C_{22,11}^{42,0}+\frac{1}{32} \sqrt{\frac{3}{5}} C_{22,11}^{42,1}-\frac{C_{31,11}^{31,0}}{128}-\frac{C_{31,11}^{31,1}}{64}-\frac{9 C_{33,11}^{33,0}}{160
\sqrt{14}}-\frac{9 C_{33,11}^{33,1}}{80 \sqrt{14}}\)

\noindent\(C_{11,0011,1}^{00,5101,1,\text{Loc}}= -\frac{\sqrt{3} C_{11,11}^{51,1}}{128}-\frac{3 C_{22,11}^{42,1}}{64 \sqrt{5}}-\frac{\sqrt{3}
C_{31,11}^{31,1}}{128}-\frac{9}{160} \sqrt{\frac{3}{14}} C_{33,11}^{33,1}\)

\noindent\(C_{11,0011,1}^{00,5101,1,\text{NonLoc}}= -\frac{\sqrt{3} C_{11,11}^{51,0}}{128}+\frac{3 C_{22,11}^{42,0}}{64 \sqrt{5}}-\frac{\sqrt{3}
C_{31,11}^{31,0}}{128}-\frac{9}{160} \sqrt{\frac{3}{14}} C_{33,11}^{33,0}\)

\noindent\(C_{11,1101,1}^{00,4011,0,\text{Loc}}= -\frac{7 C_{11,11}^{51,0}}{192}-\frac{7 C_{11,11}^{51,1}}{384}-\frac{7 C_{22,11}^{42,0}}{32
\sqrt{15}}-\frac{7 C_{22,11}^{42,1}}{64 \sqrt{15}}-\frac{7 C_{31,11}^{31,0}}{192}-\frac{7 C_{31,11}^{31,1}}{384}-\frac{3}{80} \sqrt{\frac{7}{2}}
C_{33,11}^{33,0}-\frac{3}{160} \sqrt{\frac{7}{2}} C_{33,11}^{33,1}\)

\noindent\(C_{11,1101,1}^{00,4011,0,\text{NonLoc}}= -\frac{7 C_{11,11}^{51,0}}{384}-\frac{7 C_{11,11}^{51,1}}{192}+\frac{7 C_{22,11}^{42,0}}{64
\sqrt{15}}+\frac{7 C_{22,11}^{42,1}}{32 \sqrt{15}}-\frac{7 C_{31,11}^{31,0}}{384}-\frac{7 C_{31,11}^{31,1}}{192}-\frac{3}{160} \sqrt{\frac{7}{2}}
C_{33,11}^{33,0}-\frac{3}{80} \sqrt{\frac{7}{2}} C_{33,11}^{33,1}\)

\noindent\(C_{11,1101,1}^{00,4011,1,\text{Loc}}= -\frac{7 C_{11,11}^{51,1}}{128 \sqrt{3}}-\frac{7 C_{22,11}^{42,1}}{64 \sqrt{5}}-\frac{7
C_{31,11}^{31,1}}{128 \sqrt{3}}-\frac{3}{160} \sqrt{\frac{21}{2}} C_{33,11}^{33,1}\)

\noindent\(C_{11,1101,1}^{00,4011,1,\text{NonLoc}}= -\frac{7 C_{11,11}^{51,0}}{128 \sqrt{3}}+\frac{7 C_{22,11}^{42,0}}{64 \sqrt{5}}-\frac{7
C_{31,11}^{31,0}}{128 \sqrt{3}}-\frac{3}{160} \sqrt{\frac{21}{2}} C_{33,11}^{33,0}\)

\noindent\(C_{11,1101,1}^{00,4211,0,\text{Loc}}= \frac{\sqrt{5} C_{11,11}^{51,0}}{96}+\frac{\sqrt{5} C_{11,11}^{51,1}}{192}+\frac{C_{22,11}^{42,0}}{16
\sqrt{3}}+\frac{C_{22,11}^{42,1}}{32 \sqrt{3}}+\frac{\sqrt{5} C_{31,11}^{31,0}}{96}+\frac{\sqrt{5} C_{31,11}^{31,1}}{192}+\frac{3 C_{33,11}^{33,0}}{8
\sqrt{70}}+\frac{3 C_{33,11}^{33,1}}{16 \sqrt{70}}\)

\noindent\(C_{11,1101,1}^{00,4211,0,\text{NonLoc}}= \frac{\sqrt{5} C_{11,11}^{51,0}}{192}+\frac{\sqrt{5} C_{11,11}^{51,1}}{96}-\frac{C_{22,11}^{42,0}}{32
\sqrt{3}}-\frac{C_{22,11}^{42,1}}{16 \sqrt{3}}+\frac{\sqrt{5} C_{31,11}^{31,0}}{192}+\frac{\sqrt{5} C_{31,11}^{31,1}}{96}+\frac{3 C_{33,11}^{33,0}}{16
\sqrt{70}}+\frac{3 C_{33,11}^{33,1}}{8 \sqrt{70}}\)

\noindent\(C_{11,1101,1}^{00,4211,1,\text{Loc}}= \frac{1}{64} \sqrt{\frac{5}{3}} C_{11,11}^{51,1}+\frac{C_{22,11}^{42,1}}{32}+\frac{1}{64}
\sqrt{\frac{5}{3}} C_{31,11}^{31,1}+\frac{3}{16} \sqrt{\frac{3}{70}} C_{33,11}^{33,1}\)

\noindent\(C_{11,1101,1}^{00,4211,1,\text{NonLoc}}= \frac{1}{64} \sqrt{\frac{5}{3}} C_{11,11}^{51,0}-\frac{C_{22,11}^{42,0}}{32}+\frac{1}{64}
\sqrt{\frac{5}{3}} C_{31,11}^{31,0}+\frac{3}{16} \sqrt{\frac{3}{70}} C_{33,11}^{33,0}\)

\noindent\(C_{11,1101,2}^{00,4212,0,\text{Loc}}= -\frac{1}{32} \sqrt{\frac{5}{3}} C_{11,11}^{51,0}-\frac{1}{64} \sqrt{\frac{5}{3}}
C_{11,11}^{51,1}-\frac{C_{22,11}^{42,0}}{16}-\frac{C_{22,11}^{42,1}}{32}-\frac{1}{32} \sqrt{\frac{5}{3}} C_{31,11}^{31,0}-\frac{1}{64} \sqrt{\frac{5}{3}}
C_{31,11}^{31,1}-\frac{3}{8} \sqrt{\frac{3}{70}} C_{33,11}^{33,0}-\frac{3}{16} \sqrt{\frac{3}{70}} C_{33,11}^{33,1}\)

\noindent\(C_{11,1101,2}^{00,4212,0,\text{NonLoc}}= -\frac{1}{64} \sqrt{\frac{5}{3}} C_{11,11}^{51,0}-\frac{1}{32} \sqrt{\frac{5}{3}}
C_{11,11}^{51,1}+\frac{C_{22,11}^{42,0}}{32}+\frac{C_{22,11}^{42,1}}{16}-\frac{1}{64} \sqrt{\frac{5}{3}} C_{31,11}^{31,0}-\frac{1}{32} \sqrt{\frac{5}{3}}
C_{31,11}^{31,1}-\frac{3}{16} \sqrt{\frac{3}{70}} C_{33,11}^{33,0}-\frac{3}{8} \sqrt{\frac{3}{70}} C_{33,11}^{33,1}\)

\noindent\(C_{11,1101,2}^{00,4212,1,\text{Loc}}= -\frac{\sqrt{5} C_{11,11}^{51,1}}{64}-\frac{\sqrt{3} C_{22,11}^{42,1}}{32}-\frac{\sqrt{5}
C_{31,11}^{31,1}}{64}-\frac{9 C_{33,11}^{33,1}}{16 \sqrt{70}}\)

\noindent\(C_{11,1101,2}^{00,4212,1,\text{NonLoc}}= -\frac{\sqrt{5} C_{11,11}^{51,0}}{64}+\frac{\sqrt{3} C_{22,11}^{42,0}}{32}-\frac{\sqrt{5}
C_{31,11}^{31,0}}{64}-\frac{9 C_{33,11}^{33,0}}{16 \sqrt{70}}\)

\noindent\(C_{11,1111,0}^{00,4000,0,\text{Loc}}= -\frac{7 C_{11,11}^{51,0}}{192}-\frac{7 C_{11,11}^{51,1}}{384}-\frac{7 C_{22,11}^{42,0}}{32
\sqrt{15}}-\frac{7 C_{22,11}^{42,1}}{64 \sqrt{15}}-\frac{7 C_{31,11}^{31,0}}{192}-\frac{7 C_{31,11}^{31,1}}{384}-\frac{3}{80} \sqrt{\frac{7}{2}}
C_{33,11}^{33,0}-\frac{3}{160} \sqrt{\frac{7}{2}} C_{33,11}^{33,1}\)

\noindent\(C_{11,1111,0}^{00,4000,0,\text{NonLoc}}= -\frac{7 C_{11,11}^{51,0}}{384}-\frac{7 C_{11,11}^{51,1}}{192}+\frac{7 C_{22,11}^{42,0}}{64
\sqrt{15}}+\frac{7 C_{22,11}^{42,1}}{32 \sqrt{15}}-\frac{7 C_{31,11}^{31,0}}{384}-\frac{7 C_{31,11}^{31,1}}{192}-\frac{3}{160} \sqrt{\frac{7}{2}}
C_{33,11}^{33,0}-\frac{3}{80} \sqrt{\frac{7}{2}} C_{33,11}^{33,1}\)

\noindent\(C_{11,1111,0}^{00,4000,1,\text{Loc}}= -\frac{7 C_{11,11}^{51,1}}{128 \sqrt{3}}-\frac{7 C_{22,11}^{42,1}}{64 \sqrt{5}}-\frac{7
C_{31,11}^{31,1}}{128 \sqrt{3}}-\frac{3}{160} \sqrt{\frac{21}{2}} C_{33,11}^{33,1}\)

\noindent\(C_{11,1111,0}^{00,4000,1,\text{NonLoc}}= -\frac{7 C_{11,11}^{51,0}}{128 \sqrt{3}}+\frac{7 C_{22,11}^{42,0}}{64 \sqrt{5}}-\frac{7
C_{31,11}^{31,0}}{128 \sqrt{3}}-\frac{3}{160} \sqrt{\frac{21}{2}} C_{33,11}^{33,0}\)

\noindent\(C_{11,1111,2}^{00,4202,0,\text{Loc}}= \frac{\sqrt{5} C_{11,11}^{51,0}}{96}+\frac{\sqrt{5} C_{11,11}^{51,1}}{192}+\frac{C_{22,11}^{42,0}}{16
\sqrt{3}}+\frac{C_{22,11}^{42,1}}{32 \sqrt{3}}+\frac{\sqrt{5} C_{31,11}^{31,0}}{96}+\frac{\sqrt{5} C_{31,11}^{31,1}}{192}+\frac{3 C_{33,11}^{33,0}}{8
\sqrt{70}}+\frac{3 C_{33,11}^{33,1}}{16 \sqrt{70}}\)

\noindent\(C_{11,1111,2}^{00,4202,0,\text{NonLoc}}= \frac{\sqrt{5} C_{11,11}^{51,0}}{192}+\frac{\sqrt{5} C_{11,11}^{51,1}}{96}-\frac{C_{22,11}^{42,0}}{32
\sqrt{3}}-\frac{C_{22,11}^{42,1}}{16 \sqrt{3}}+\frac{\sqrt{5} C_{31,11}^{31,0}}{192}+\frac{\sqrt{5} C_{31,11}^{31,1}}{96}+\frac{3 C_{33,11}^{33,0}}{16
\sqrt{70}}+\frac{3 C_{33,11}^{33,1}}{8 \sqrt{70}}\)

\noindent\(C_{11,1111,2}^{00,4202,1,\text{Loc}}= \frac{1}{64} \sqrt{\frac{5}{3}} C_{11,11}^{51,1}+\frac{C_{22,11}^{42,1}}{32}+\frac{1}{64}
\sqrt{\frac{5}{3}} C_{31,11}^{31,1}+\frac{3}{16} \sqrt{\frac{3}{70}} C_{33,11}^{33,1}\)

\noindent\(C_{11,1111,2}^{00,4202,1,\text{NonLoc}}= \frac{1}{64} \sqrt{\frac{5}{3}} C_{11,11}^{51,0}-\frac{C_{22,11}^{42,0}}{32}+\frac{1}{64}
\sqrt{\frac{5}{3}} C_{31,11}^{31,0}+\frac{3}{16} \sqrt{\frac{3}{70}} C_{33,11}^{33,0}\)

\noindent\(C_{11,1112,2}^{00,4202,0,\text{Loc}}= \frac{1}{32} \sqrt{\frac{5}{3}} C_{11,11}^{51,0}+\frac{1}{64} \sqrt{\frac{5}{3}} C_{11,11}^{51,1}+\frac{C_{22,11}^{42,0}}{16}+\frac{C_{22,11}^{42,1}}{32}+\frac{1}{32}
\sqrt{\frac{5}{3}} C_{31,11}^{31,0}+\frac{1}{64} \sqrt{\frac{5}{3}} C_{31,11}^{31,1}+\frac{3}{8} \sqrt{\frac{3}{70}} C_{33,11}^{33,0}+\frac{3}{16}
\sqrt{\frac{3}{70}} C_{33,11}^{33,1}\)

\noindent\(C_{11,1112,2}^{00,4202,0,\text{NonLoc}}= \frac{1}{64} \sqrt{\frac{5}{3}} C_{11,11}^{51,0}+\frac{1}{32} \sqrt{\frac{5}{3}}
C_{11,11}^{51,1}-\frac{C_{22,11}^{42,0}}{32}-\frac{C_{22,11}^{42,1}}{16}+\frac{1}{64} \sqrt{\frac{5}{3}} C_{31,11}^{31,0}+\frac{1}{32} \sqrt{\frac{5}{3}}
C_{31,11}^{31,1}+\frac{3}{16} \sqrt{\frac{3}{70}} C_{33,11}^{33,0}+\frac{3}{8} \sqrt{\frac{3}{70}} C_{33,11}^{33,1}\)

\noindent\(C_{11,1112,2}^{00,4202,1,\text{Loc}}= \frac{\sqrt{5} C_{11,11}^{51,1}}{64}+\frac{\sqrt{3} C_{22,11}^{42,1}}{32}+\frac{\sqrt{5}
C_{31,11}^{31,1}}{64}+\frac{9 C_{33,11}^{33,1}}{16 \sqrt{70}}\)

\noindent\(C_{11,1112,2}^{00,4202,1,\text{NonLoc}}= \frac{\sqrt{5} C_{11,11}^{51,0}}{64}-\frac{\sqrt{3} C_{22,11}^{42,0}}{32}+\frac{\sqrt{5}
C_{31,11}^{31,0}}{64}+\frac{9 C_{33,11}^{33,0}}{16 \sqrt{70}}\)

\noindent\(C_{11,2000,1}^{00,3111,0,\text{Loc}}= \frac{7 C_{11,11}^{51,0}}{96}+\frac{7 C_{11,11}^{51,1}}{192}+\frac{7 C_{22,11}^{42,0}}{16
\sqrt{15}}+\frac{7 C_{22,11}^{42,1}}{32 \sqrt{15}}+\frac{7 C_{31,11}^{31,0}}{96}+\frac{7 C_{31,11}^{31,1}}{192}+\frac{3}{40} \sqrt{\frac{7}{2}} C_{33,11}^{33,0}+\frac{3}{80}
\sqrt{\frac{7}{2}} C_{33,11}^{33,1}\)

\noindent\(C_{11,2000,1}^{00,3111,0,\text{NonLoc}}= \frac{7 C_{11,11}^{51,0}}{192}+\frac{7 C_{11,11}^{51,1}}{96}-\frac{7 C_{22,11}^{42,0}}{32
\sqrt{15}}-\frac{7 C_{22,11}^{42,1}}{16 \sqrt{15}}+\frac{7 C_{31,11}^{31,0}}{192}+\frac{7 C_{31,11}^{31,1}}{96}+\frac{3}{80} \sqrt{\frac{7}{2}} C_{33,11}^{33,0}+\frac{3}{40}
\sqrt{\frac{7}{2}} C_{33,11}^{33,1}\)

\noindent\(C_{11,2000,1}^{00,3111,1,\text{Loc}}= \frac{7 C_{11,11}^{51,1}}{64 \sqrt{3}}+\frac{7 C_{22,11}^{42,1}}{32 \sqrt{5}}+\frac{7
C_{31,11}^{31,1}}{64 \sqrt{3}}+\frac{3}{80} \sqrt{\frac{21}{2}} C_{33,11}^{33,1}\)

\noindent\(C_{11,2000,1}^{00,3111,1,\text{NonLoc}}= \frac{7 C_{11,11}^{51,0}}{64 \sqrt{3}}-\frac{7 C_{22,11}^{42,0}}{32 \sqrt{5}}+\frac{7
C_{31,11}^{31,0}}{64 \sqrt{3}}+\frac{3}{80} \sqrt{\frac{21}{2}} C_{33,11}^{33,0}\)

\noindent\(C_{11,2000,1}^{11,2000,0,\text{Loc}}= \frac{35 C_{00,00}^{60,0}}{384}+\frac{35 C_{00,00}^{60,1}}{768}+\frac{35 C_{11,00}^{51,0}}{1152}+\frac{35
C_{11,00}^{51,1}}{2304}+\frac{25 C_{20,00}^{40,0}}{1152}+\frac{25 C_{20,00}^{40,1}}{2304}-\frac{7 \sqrt{5} C_{22,00}^{42,0}}{576}-\frac{7 \sqrt{5}
C_{22,00}^{42,1}}{1152}-\frac{5 C_{31,00}^{31,0}}{1152}-\frac{5 C_{31,00}^{31,1}}{2304}-\frac{\sqrt{21} C_{33,00}^{33,0}}{64}-\frac{\sqrt{21} C_{33,00}^{33,1}}{128}\)

\noindent\(C_{11,2000,1}^{11,2000,0,\text{NonLoc}}= -\frac{35 C_{00,00}^{60,0}}{1536}-\frac{35 C_{00,00}^{60,1}}{768}+\frac{35 C_{00,20}^{60,0}}{512
\sqrt{3}}+\frac{35 C_{00,20}^{60,1}}{256 \sqrt{3}}+\frac{35 C_{11,00}^{51,0}}{4608}+\frac{35 C_{11,00}^{51,1}}{2304}-\frac{25 C_{20,00}^{40,0}}{4608}-\frac{25
C_{20,00}^{40,1}}{2304}+\frac{25 C_{20,20}^{40,0}}{1536 \sqrt{3}}+\frac{25 C_{20,20}^{40,1}}{768 \sqrt{3}}+\frac{7 \sqrt{5} C_{22,00}^{42,0}}{2304}+\frac{7
\sqrt{5} C_{22,00}^{42,1}}{1152}-\frac{7}{768} \sqrt{\frac{5}{3}} C_{22,20}^{42,0}-\frac{7}{384} \sqrt{\frac{5}{3}} C_{22,20}^{42,1}-\frac{5 C_{31,00}^{31,0}}{4608}-\frac{5
C_{31,00}^{31,1}}{2304}+\frac{5 C_{31,20}^{31,0}}{1536 \sqrt{3}}+\frac{5 C_{31,20}^{31,1}}{768 \sqrt{3}}-\frac{\sqrt{21} C_{33,00}^{33,0}}{256}-\frac{\sqrt{21}
C_{33,00}^{33,1}}{128}+\frac{3 \sqrt{7} C_{33,20}^{33,0}}{256}+\frac{3 \sqrt{7} C_{33,20}^{33,1}}{128}\)

\noindent\(C_{11,2000,1}^{11,2000,1,\text{Loc}}= \frac{35 C_{00,00}^{60,1}}{256 \sqrt{3}}+\frac{35 C_{11,00}^{51,1}}{768 \sqrt{3}}+\frac{25
C_{20,00}^{40,1}}{768 \sqrt{3}}-\frac{7}{384} \sqrt{\frac{5}{3}} C_{22,00}^{42,1}-\frac{5 C_{31,00}^{31,1}}{768 \sqrt{3}}-\frac{3 \sqrt{7} C_{33,00}^{33,1}}{128}\)

\noindent\(C_{11,2000,1}^{11,2000,1,\text{NonLoc}}= -\frac{35 C_{00,00}^{60,0}}{512 \sqrt{3}}+\frac{35 C_{00,20}^{60,0}}{512}+\frac{35
C_{11,00}^{51,0}}{1536 \sqrt{3}}-\frac{25 C_{20,00}^{40,0}}{1536 \sqrt{3}}+\frac{25 C_{20,20}^{40,0}}{1536}+\frac{7}{768} \sqrt{\frac{5}{3}} C_{22,00}^{42,0}-\frac{7
\sqrt{5} C_{22,20}^{42,0}}{768}-\frac{5 C_{31,00}^{31,0}}{1536 \sqrt{3}}+\frac{5 C_{31,20}^{31,0}}{1536}-\frac{3 \sqrt{7} C_{33,00}^{33,0}}{256}+\frac{3
\sqrt{21} C_{33,20}^{33,0}}{256}\)

\noindent\(C_{11,2011,0}^{11,2011,0,\text{Loc}}= -\frac{35 C_{00,20}^{60,0}}{1152}-\frac{35 C_{00,20}^{60,1}}{2304}-\frac{7 \sqrt{5}
C_{00,22}^{62,0}}{384}-\frac{7 \sqrt{5} C_{00,22}^{62,1}}{768}+\frac{7 \sqrt{5} C_{11,22}^{51,0}}{3456}+\frac{7 \sqrt{5} C_{11,22}^{51,1}}{6912}-\frac{\sqrt{21}
C_{11,22}^{53,0}}{128}-\frac{\sqrt{21} C_{11,22}^{53,1}}{256}-\frac{25 C_{20,20}^{40,0}}{3456}-\frac{25 C_{20,20}^{40,1}}{6912}-\frac{7 \sqrt{5}
C_{20,22}^{42,0}}{3456}-\frac{7 \sqrt{5} C_{20,22}^{42,1}}{6912}+\frac{7 \sqrt{5} C_{22,20}^{42,0}}{1728}+\frac{7 \sqrt{5} C_{22,20}^{42,1}}{3456}+\frac{17
\sqrt{5} C_{22,22}^{40,0}}{3456}+\frac{17 \sqrt{5} C_{22,22}^{40,1}}{6912}+\frac{5 \sqrt{35} C_{22,22}^{42,0}}{1728}+\frac{5 \sqrt{35} C_{22,22}^{42,1}}{3456}-\frac{C_{22,22}^{44,0}}{32}-\frac{C_{22,22}^{44,1}}{64}+\frac{5
C_{31,20}^{31,0}}{3456}+\frac{5 C_{31,20}^{31,1}}{6912}+\frac{23 \sqrt{5} C_{31,22}^{31,0}}{3456}+\frac{23 \sqrt{5} C_{31,22}^{31,1}}{6912}-\frac{1}{384}
\sqrt{\frac{7}{3}} C_{31,22}^{33,0}-\frac{1}{768} \sqrt{\frac{7}{3}} C_{31,22}^{33,1}+\frac{1}{64} \sqrt{\frac{7}{3}} C_{33,20}^{33,0}+\frac{1}{128}
\sqrt{\frac{7}{3}} C_{33,20}^{33,1}+\frac{1}{32} \sqrt{\frac{7}{10}} C_{33,22}^{33,0}+\frac{1}{64} \sqrt{\frac{7}{10}} C_{33,22}^{33,1}\)

\noindent\(C_{11,2011,0}^{11,2011,0,\text{NonLoc}}= -\frac{35 C_{00,00}^{60,0}}{1536 \sqrt{3}}-\frac{35 C_{00,00}^{60,1}}{768 \sqrt{3}}-\frac{35
C_{00,20}^{60,0}}{4608}-\frac{35 C_{00,20}^{60,1}}{2304}+\frac{7 \sqrt{5} C_{00,22}^{62,0}}{768}+\frac{7 \sqrt{5} C_{00,22}^{62,1}}{384}+\frac{35
C_{11,00}^{51,0}}{4608 \sqrt{3}}+\frac{35 C_{11,00}^{51,1}}{2304 \sqrt{3}}+\frac{7 \sqrt{5} C_{11,22}^{51,0}}{6912}+\frac{7 \sqrt{5} C_{11,22}^{51,1}}{3456}-\frac{\sqrt{21}
C_{11,22}^{53,0}}{256}-\frac{\sqrt{21} C_{11,22}^{53,1}}{128}-\frac{25 C_{20,00}^{40,0}}{4608 \sqrt{3}}-\frac{25 C_{20,00}^{40,1}}{2304 \sqrt{3}}-\frac{25
C_{20,20}^{40,0}}{13824}-\frac{25 C_{20,20}^{40,1}}{6912}+\frac{7 \sqrt{5} C_{20,22}^{42,0}}{6912}+\frac{7 \sqrt{5} C_{20,22}^{42,1}}{3456}+\frac{7
\sqrt{\frac{5}{3}} C_{22,00}^{42,0}}{2304}+\frac{7 \sqrt{\frac{5}{3}} C_{22,00}^{42,1}}{1152}+\frac{7 \sqrt{5} C_{22,20}^{42,0}}{6912}+\frac{7 \sqrt{5}
C_{22,20}^{42,1}}{3456}-\frac{17 \sqrt{5} C_{22,22}^{40,0}}{6912}-\frac{17 \sqrt{5} C_{22,22}^{40,1}}{3456}-\frac{5 \sqrt{35} C_{22,22}^{42,0}}{3456}-\frac{5
\sqrt{35} C_{22,22}^{42,1}}{1728}+\frac{C_{22,22}^{44,0}}{64}+\frac{C_{22,22}^{44,1}}{32}-\frac{5 C_{31,00}^{31,0}}{4608 \sqrt{3}}-\frac{5 C_{31,00}^{31,1}}{2304
\sqrt{3}}-\frac{5 C_{31,20}^{31,0}}{13824}-\frac{5 C_{31,20}^{31,1}}{6912}+\frac{23 \sqrt{5} C_{31,22}^{31,0}}{6912}+\frac{23 \sqrt{5} C_{31,22}^{31,1}}{3456}-\frac{1}{768}
\sqrt{\frac{7}{3}} C_{31,22}^{33,0}-\frac{1}{384} \sqrt{\frac{7}{3}} C_{31,22}^{33,1}-\frac{\sqrt{7} C_{33,00}^{33,0}}{256}-\frac{\sqrt{7} C_{33,00}^{33,1}}{128}-\frac{1}{256}
\sqrt{\frac{7}{3}} C_{33,20}^{33,0}-\frac{1}{128} \sqrt{\frac{7}{3}} C_{33,20}^{33,1}+\frac{1}{64} \sqrt{\frac{7}{10}} C_{33,22}^{33,0}+\frac{1}{32}
\sqrt{\frac{7}{10}} C_{33,22}^{33,1}\)

\noindent\(C_{11,2011,0}^{11,2011,1,\text{Loc}}= -\frac{35 C_{00,20}^{60,1}}{768 \sqrt{3}}-\frac{7}{256} \sqrt{\frac{5}{3}} C_{00,22}^{62,1}+\frac{7
\sqrt{\frac{5}{3}} C_{11,22}^{51,1}}{2304}-\frac{3 \sqrt{7} C_{11,22}^{53,1}}{256}-\frac{25 C_{20,20}^{40,1}}{2304 \sqrt{3}}-\frac{7 \sqrt{\frac{5}{3}}
C_{20,22}^{42,1}}{2304}+\frac{7 \sqrt{\frac{5}{3}} C_{22,20}^{42,1}}{1152}+\frac{17 \sqrt{\frac{5}{3}} C_{22,22}^{40,1}}{2304}+\frac{5 \sqrt{\frac{35}{3}}
C_{22,22}^{42,1}}{1152}-\frac{\sqrt{3} C_{22,22}^{44,1}}{64}+\frac{5 C_{31,20}^{31,1}}{2304 \sqrt{3}}+\frac{23 \sqrt{\frac{5}{3}} C_{31,22}^{31,1}}{2304}-\frac{\sqrt{7}
C_{31,22}^{33,1}}{768}+\frac{\sqrt{7} C_{33,20}^{33,1}}{128}+\frac{1}{64} \sqrt{\frac{21}{10}} C_{33,22}^{33,1}\)

\noindent\(C_{11,2011,0}^{11,2011,1,\text{NonLoc}}= -\frac{35 C_{00,00}^{60,0}}{1536}-\frac{35 C_{00,20}^{60,0}}{1536 \sqrt{3}}+\frac{7}{256}
\sqrt{\frac{5}{3}} C_{00,22}^{62,0}+\frac{35 C_{11,00}^{51,0}}{4608}+\frac{7 \sqrt{\frac{5}{3}} C_{11,22}^{51,0}}{2304}-\frac{3 \sqrt{7} C_{11,22}^{53,0}}{256}-\frac{25
C_{20,00}^{40,0}}{4608}-\frac{25 C_{20,20}^{40,0}}{4608 \sqrt{3}}+\frac{7 \sqrt{\frac{5}{3}} C_{20,22}^{42,0}}{2304}+\frac{7 \sqrt{5} C_{22,00}^{42,0}}{2304}+\frac{7
\sqrt{\frac{5}{3}} C_{22,20}^{42,0}}{2304}-\frac{17 \sqrt{\frac{5}{3}} C_{22,22}^{40,0}}{2304}-\frac{5 \sqrt{\frac{35}{3}} C_{22,22}^{42,0}}{1152}+\frac{\sqrt{3}
C_{22,22}^{44,0}}{64}-\frac{5 C_{31,00}^{31,0}}{4608}-\frac{5 C_{31,20}^{31,0}}{4608 \sqrt{3}}+\frac{23 \sqrt{\frac{5}{3}} C_{31,22}^{31,0}}{2304}-\frac{\sqrt{7}
C_{31,22}^{33,0}}{768}-\frac{\sqrt{21} C_{33,00}^{33,0}}{256}-\frac{\sqrt{7} C_{33,20}^{33,0}}{256}+\frac{1}{64} \sqrt{\frac{21}{10}} C_{33,22}^{33,0}\)

\noindent\(C_{11,2011,1}^{00,3101,0,\text{Loc}}= -\frac{7 C_{11,11}^{51,0}}{96}-\frac{7 C_{11,11}^{51,1}}{192}-\frac{7 C_{22,11}^{42,0}}{16
\sqrt{15}}-\frac{7 C_{22,11}^{42,1}}{32 \sqrt{15}}-\frac{7 C_{31,11}^{31,0}}{96}-\frac{7 C_{31,11}^{31,1}}{192}-\frac{3}{40} \sqrt{\frac{7}{2}} C_{33,11}^{33,0}-\frac{3}{80}
\sqrt{\frac{7}{2}} C_{33,11}^{33,1}\)

\noindent\(C_{11,2011,1}^{00,3101,0,\text{NonLoc}}= -\frac{7 C_{11,11}^{51,0}}{192}-\frac{7 C_{11,11}^{51,1}}{96}+\frac{7 C_{22,11}^{42,0}}{32
\sqrt{15}}+\frac{7 C_{22,11}^{42,1}}{16 \sqrt{15}}-\frac{7 C_{31,11}^{31,0}}{192}-\frac{7 C_{31,11}^{31,1}}{96}-\frac{3}{80} \sqrt{\frac{7}{2}} C_{33,11}^{33,0}-\frac{3}{40}
\sqrt{\frac{7}{2}} C_{33,11}^{33,1}\)

\noindent\(C_{11,2011,1}^{00,3101,1,\text{Loc}}= -\frac{7 C_{11,11}^{51,1}}{64 \sqrt{3}}-\frac{7 C_{22,11}^{42,1}}{32 \sqrt{5}}-\frac{7
C_{31,11}^{31,1}}{64 \sqrt{3}}-\frac{3}{80} \sqrt{\frac{21}{2}} C_{33,11}^{33,1}\)

\noindent\(C_{11,2011,1}^{00,3101,1,\text{NonLoc}}= -\frac{7 C_{11,11}^{51,0}}{64 \sqrt{3}}+\frac{7 C_{22,11}^{42,0}}{32 \sqrt{5}}-\frac{7
C_{31,11}^{31,0}}{64 \sqrt{3}}-\frac{3}{80} \sqrt{\frac{21}{2}} C_{33,11}^{33,0}\)

\noindent\(C_{11,2011,1}^{11,2011,0,\text{Loc}}= -\frac{35 C_{00,20}^{60,0}}{384 \sqrt{3}}-\frac{35 C_{00,20}^{60,1}}{768 \sqrt{3}}+\frac{7}{256}
\sqrt{\frac{5}{3}} C_{00,22}^{62,0}+\frac{7}{512} \sqrt{\frac{5}{3}} C_{00,22}^{62,1}-\frac{7 \sqrt{\frac{5}{3}} C_{11,22}^{51,0}}{2304}-\frac{7
\sqrt{\frac{5}{3}} C_{11,22}^{51,1}}{4608}+\frac{3 \sqrt{7} C_{11,22}^{53,0}}{256}+\frac{3 \sqrt{7} C_{11,22}^{53,1}}{512}-\frac{25 C_{20,20}^{40,0}}{1152
\sqrt{3}}-\frac{25 C_{20,20}^{40,1}}{2304 \sqrt{3}}+\frac{7 \sqrt{\frac{5}{3}} C_{20,22}^{42,0}}{2304}+\frac{7 \sqrt{\frac{5}{3}} C_{20,22}^{42,1}}{4608}+\frac{7}{576}
\sqrt{\frac{5}{3}} C_{22,20}^{42,0}+\frac{7 \sqrt{\frac{5}{3}} C_{22,20}^{42,1}}{1152}-\frac{17 \sqrt{\frac{5}{3}} C_{22,22}^{40,0}}{2304}-\frac{17
\sqrt{\frac{5}{3}} C_{22,22}^{40,1}}{4608}-\frac{5 \sqrt{\frac{35}{3}} C_{22,22}^{42,0}}{1152}-\frac{5 \sqrt{\frac{35}{3}} C_{22,22}^{42,1}}{2304}+\frac{\sqrt{3}
C_{22,22}^{44,0}}{64}+\frac{\sqrt{3} C_{22,22}^{44,1}}{128}+\frac{5 C_{31,20}^{31,0}}{1152 \sqrt{3}}+\frac{5 C_{31,20}^{31,1}}{2304 \sqrt{3}}-\frac{23
\sqrt{\frac{5}{3}} C_{31,22}^{31,0}}{2304}-\frac{23 \sqrt{\frac{5}{3}} C_{31,22}^{31,1}}{4608}+\frac{\sqrt{7} C_{31,22}^{33,0}}{768}+\frac{\sqrt{7}
C_{31,22}^{33,1}}{1536}+\frac{\sqrt{7} C_{33,20}^{33,0}}{64}+\frac{\sqrt{7} C_{33,20}^{33,1}}{128}-\frac{1}{64} \sqrt{\frac{21}{10}} C_{33,22}^{33,0}-\frac{1}{128}
\sqrt{\frac{21}{10}} C_{33,22}^{33,1}\)

\noindent\(C_{11,2011,1}^{11,2011,0,\text{NonLoc}}= -\frac{35 C_{00,00}^{60,0}}{1536}-\frac{35 C_{00,00}^{60,1}}{768}-\frac{35 C_{00,20}^{60,0}}{1536
\sqrt{3}}-\frac{35 C_{00,20}^{60,1}}{768 \sqrt{3}}-\frac{7}{512} \sqrt{\frac{5}{3}} C_{00,22}^{62,0}-\frac{7}{256} \sqrt{\frac{5}{3}} C_{00,22}^{62,1}+\frac{35
C_{11,00}^{51,0}}{4608}+\frac{35 C_{11,00}^{51,1}}{2304}-\frac{7 \sqrt{\frac{5}{3}} C_{11,22}^{51,0}}{4608}-\frac{7 \sqrt{\frac{5}{3}} C_{11,22}^{51,1}}{2304}+\frac{3
\sqrt{7} C_{11,22}^{53,0}}{512}+\frac{3 \sqrt{7} C_{11,22}^{53,1}}{256}-\frac{25 C_{20,00}^{40,0}}{4608}-\frac{25 C_{20,00}^{40,1}}{2304}-\frac{25
C_{20,20}^{40,0}}{4608 \sqrt{3}}-\frac{25 C_{20,20}^{40,1}}{2304 \sqrt{3}}-\frac{7 \sqrt{\frac{5}{3}} C_{20,22}^{42,0}}{4608}-\frac{7 \sqrt{\frac{5}{3}}
C_{20,22}^{42,1}}{2304}+\frac{7 \sqrt{5} C_{22,00}^{42,0}}{2304}+\frac{7 \sqrt{5} C_{22,00}^{42,1}}{1152}+\frac{7 \sqrt{\frac{5}{3}} C_{22,20}^{42,0}}{2304}+\frac{7
\sqrt{\frac{5}{3}} C_{22,20}^{42,1}}{1152}+\frac{17 \sqrt{\frac{5}{3}} C_{22,22}^{40,0}}{4608}+\frac{17 \sqrt{\frac{5}{3}} C_{22,22}^{40,1}}{2304}+\frac{5
\sqrt{\frac{35}{3}} C_{22,22}^{42,0}}{2304}+\frac{5 \sqrt{\frac{35}{3}} C_{22,22}^{42,1}}{1152}-\frac{\sqrt{3} C_{22,22}^{44,0}}{128}-\frac{\sqrt{3}
C_{22,22}^{44,1}}{64}-\frac{5 C_{31,00}^{31,0}}{4608}-\frac{5 C_{31,00}^{31,1}}{2304}-\frac{5 C_{31,20}^{31,0}}{4608 \sqrt{3}}-\frac{5 C_{31,20}^{31,1}}{2304
\sqrt{3}}-\frac{23 \sqrt{\frac{5}{3}} C_{31,22}^{31,0}}{4608}-\frac{23 \sqrt{\frac{5}{3}} C_{31,22}^{31,1}}{2304}+\frac{\sqrt{7} C_{31,22}^{33,0}}{1536}+\frac{\sqrt{7}
C_{31,22}^{33,1}}{768}-\frac{\sqrt{21} C_{33,00}^{33,0}}{256}-\frac{\sqrt{21} C_{33,00}^{33,1}}{128}-\frac{\sqrt{7} C_{33,20}^{33,0}}{256}-\frac{\sqrt{7}
C_{33,20}^{33,1}}{128}-\frac{1}{128} \sqrt{\frac{21}{10}} C_{33,22}^{33,0}-\frac{1}{64} \sqrt{\frac{21}{10}} C_{33,22}^{33,1}\)

\noindent\(C_{11,2011,1}^{11,2011,1,\text{Loc}}= -\frac{35 C_{00,20}^{60,1}}{768}+\frac{7 \sqrt{5} C_{00,22}^{62,1}}{512}-\frac{7 \sqrt{5}
C_{11,22}^{51,1}}{4608}+\frac{3 \sqrt{21} C_{11,22}^{53,1}}{512}-\frac{25 C_{20,20}^{40,1}}{2304}+\frac{7 \sqrt{5} C_{20,22}^{42,1}}{4608}+\frac{7
\sqrt{5} C_{22,20}^{42,1}}{1152}-\frac{17 \sqrt{5} C_{22,22}^{40,1}}{4608}-\frac{5 \sqrt{35} C_{22,22}^{42,1}}{2304}+\frac{3 C_{22,22}^{44,1}}{128}+\frac{5
C_{31,20}^{31,1}}{2304}-\frac{23 \sqrt{5} C_{31,22}^{31,1}}{4608}+\frac{1}{512} \sqrt{\frac{7}{3}} C_{31,22}^{33,1}+\frac{\sqrt{21} C_{33,20}^{33,1}}{128}-\frac{3}{128}
\sqrt{\frac{7}{10}} C_{33,22}^{33,1}\)

\noindent\(C_{11,2011,1}^{11,2011,1,\text{NonLoc}}= -\frac{35 C_{00,00}^{60,0}}{512 \sqrt{3}}-\frac{35 C_{00,20}^{60,0}}{1536}-\frac{7
\sqrt{5} C_{00,22}^{62,0}}{512}+\frac{35 C_{11,00}^{51,0}}{1536 \sqrt{3}}-\frac{7 \sqrt{5} C_{11,22}^{51,0}}{4608}+\frac{3 \sqrt{21} C_{11,22}^{53,0}}{512}-\frac{25
C_{20,00}^{40,0}}{1536 \sqrt{3}}-\frac{25 C_{20,20}^{40,0}}{4608}-\frac{7 \sqrt{5} C_{20,22}^{42,0}}{4608}+\frac{7}{768} \sqrt{\frac{5}{3}} C_{22,00}^{42,0}+\frac{7
\sqrt{5} C_{22,20}^{42,0}}{2304}+\frac{17 \sqrt{5} C_{22,22}^{40,0}}{4608}+\frac{5 \sqrt{35} C_{22,22}^{42,0}}{2304}-\frac{3 C_{22,22}^{44,0}}{128}-\frac{5
C_{31,00}^{31,0}}{1536 \sqrt{3}}-\frac{5 C_{31,20}^{31,0}}{4608}-\frac{23 \sqrt{5} C_{31,22}^{31,0}}{4608}+\frac{1}{512} \sqrt{\frac{7}{3}} C_{31,22}^{33,0}-\frac{3
\sqrt{7} C_{33,00}^{33,0}}{256}-\frac{\sqrt{21} C_{33,20}^{33,0}}{256}-\frac{3}{128} \sqrt{\frac{7}{10}} C_{33,22}^{33,0}\)

\noindent\(C_{11,2011,2}^{11,2011,0,\text{Loc}}= -\frac{35 \sqrt{5} C_{00,20}^{60,0}}{1152}-\frac{35 \sqrt{5} C_{00,20}^{60,1}}{2304}-\frac{7
C_{00,22}^{62,0}}{768}-\frac{7 C_{00,22}^{62,1}}{1536}+\frac{7 C_{11,22}^{51,0}}{6912}+\frac{7 C_{11,22}^{51,1}}{13824}-\frac{1}{256} \sqrt{\frac{21}{5}}
C_{11,22}^{53,0}-\frac{1}{512} \sqrt{\frac{21}{5}} C_{11,22}^{53,1}-\frac{25 \sqrt{5} C_{20,20}^{40,0}}{3456}-\frac{25 \sqrt{5} C_{20,20}^{40,1}}{6912}-\frac{7
C_{20,22}^{42,0}}{6912}-\frac{7 C_{20,22}^{42,1}}{13824}+\frac{35 C_{22,20}^{42,0}}{1728}+\frac{35 C_{22,20}^{42,1}}{3456}+\frac{17 C_{22,22}^{40,0}}{6912}+\frac{17
C_{22,22}^{40,1}}{13824}+\frac{5 \sqrt{7} C_{22,22}^{42,0}}{3456}+\frac{5 \sqrt{7} C_{22,22}^{42,1}}{6912}-\frac{C_{22,22}^{44,0}}{64 \sqrt{5}}-\frac{C_{22,22}^{44,1}}{128
\sqrt{5}}+\frac{5 \sqrt{5} C_{31,20}^{31,0}}{3456}+\frac{5 \sqrt{5} C_{31,20}^{31,1}}{6912}+\frac{23 C_{31,22}^{31,0}}{6912}+\frac{23 C_{31,22}^{31,1}}{13824}-\frac{1}{768}
\sqrt{\frac{7}{15}} C_{31,22}^{33,0}-\frac{\sqrt{\frac{7}{15}} C_{31,22}^{33,1}}{1536}+\frac{1}{64} \sqrt{\frac{35}{3}} C_{33,20}^{33,0}+\frac{1}{128}
\sqrt{\frac{35}{3}} C_{33,20}^{33,1}+\frac{1}{320} \sqrt{\frac{7}{2}} C_{33,22}^{33,0}+\frac{1}{640} \sqrt{\frac{7}{2}} C_{33,22}^{33,1}\)

\noindent\(C_{11,2011,2}^{11,2011,0,\text{NonLoc}}= -\frac{35 \sqrt{\frac{5}{3}} C_{00,00}^{60,0}}{1536}-\frac{35}{768} \sqrt{\frac{5}{3}}
C_{00,00}^{60,1}-\frac{35 \sqrt{5} C_{00,20}^{60,0}}{4608}-\frac{35 \sqrt{5} C_{00,20}^{60,1}}{2304}+\frac{7 C_{00,22}^{62,0}}{1536}+\frac{7 C_{00,22}^{62,1}}{768}+\frac{35
\sqrt{\frac{5}{3}} C_{11,00}^{51,0}}{4608}+\frac{35 \sqrt{\frac{5}{3}} C_{11,00}^{51,1}}{2304}+\frac{7 C_{11,22}^{51,0}}{13824}+\frac{7 C_{11,22}^{51,1}}{6912}-\frac{1}{512}
\sqrt{\frac{21}{5}} C_{11,22}^{53,0}-\frac{1}{256} \sqrt{\frac{21}{5}} C_{11,22}^{53,1}-\frac{25 \sqrt{\frac{5}{3}} C_{20,00}^{40,0}}{4608}-\frac{25
\sqrt{\frac{5}{3}} C_{20,00}^{40,1}}{2304}-\frac{25 \sqrt{5} C_{20,20}^{40,0}}{13824}-\frac{25 \sqrt{5} C_{20,20}^{40,1}}{6912}+\frac{7 C_{20,22}^{42,0}}{13824}+\frac{7
C_{20,22}^{42,1}}{6912}+\frac{35 C_{22,00}^{42,0}}{2304 \sqrt{3}}+\frac{35 C_{22,00}^{42,1}}{1152 \sqrt{3}}+\frac{35 C_{22,20}^{42,0}}{6912}+\frac{35
C_{22,20}^{42,1}}{3456}-\frac{17 C_{22,22}^{40,0}}{13824}-\frac{17 C_{22,22}^{40,1}}{6912}-\frac{5 \sqrt{7} C_{22,22}^{42,0}}{6912}-\frac{5 \sqrt{7}
C_{22,22}^{42,1}}{3456}+\frac{C_{22,22}^{44,0}}{128 \sqrt{5}}+\frac{C_{22,22}^{44,1}}{64 \sqrt{5}}-\frac{5 \sqrt{\frac{5}{3}} C_{31,00}^{31,0}}{4608}-\frac{5
\sqrt{\frac{5}{3}} C_{31,00}^{31,1}}{2304}-\frac{5 \sqrt{5} C_{31,20}^{31,0}}{13824}-\frac{5 \sqrt{5} C_{31,20}^{31,1}}{6912}+\frac{23 C_{31,22}^{31,0}}{13824}+\frac{23
C_{31,22}^{31,1}}{6912}-\frac{\sqrt{\frac{7}{15}} C_{31,22}^{33,0}}{1536}-\frac{1}{768} \sqrt{\frac{7}{15}} C_{31,22}^{33,1}-\frac{\sqrt{35} C_{33,00}^{33,0}}{256}-\frac{\sqrt{35}
C_{33,00}^{33,1}}{128}-\frac{1}{256} \sqrt{\frac{35}{3}} C_{33,20}^{33,0}-\frac{1}{128} \sqrt{\frac{35}{3}} C_{33,20}^{33,1}+\frac{1}{640} \sqrt{\frac{7}{2}}
C_{33,22}^{33,0}+\frac{1}{320} \sqrt{\frac{7}{2}} C_{33,22}^{33,1}\)

\noindent\(C_{11,2011,2}^{11,2011,1,\text{Loc}}= -\frac{35}{768} \sqrt{\frac{5}{3}} C_{00,20}^{60,1}-\frac{7 C_{00,22}^{62,1}}{512
\sqrt{3}}+\frac{7 C_{11,22}^{51,1}}{4608 \sqrt{3}}-\frac{3}{512} \sqrt{\frac{7}{5}} C_{11,22}^{53,1}-\frac{25 \sqrt{\frac{5}{3}} C_{20,20}^{40,1}}{2304}-\frac{7
C_{20,22}^{42,1}}{4608 \sqrt{3}}+\frac{35 C_{22,20}^{42,1}}{1152 \sqrt{3}}+\frac{17 C_{22,22}^{40,1}}{4608 \sqrt{3}}+\frac{5 \sqrt{\frac{7}{3}} C_{22,22}^{42,1}}{2304}-\frac{1}{128}
\sqrt{\frac{3}{5}} C_{22,22}^{44,1}+\frac{5 \sqrt{\frac{5}{3}} C_{31,20}^{31,1}}{2304}+\frac{23 C_{31,22}^{31,1}}{4608 \sqrt{3}}-\frac{\sqrt{\frac{7}{5}}
C_{31,22}^{33,1}}{1536}+\frac{\sqrt{35} C_{33,20}^{33,1}}{128}+\frac{1}{640} \sqrt{\frac{21}{2}} C_{33,22}^{33,1}\)

\noindent\(C_{11,2011,2}^{11,2011,1,\text{NonLoc}}= -\frac{35 \sqrt{5} C_{00,00}^{60,0}}{1536}-\frac{35 \sqrt{\frac{5}{3}} C_{00,20}^{60,0}}{1536}+\frac{7
C_{00,22}^{62,0}}{512 \sqrt{3}}+\frac{35 \sqrt{5} C_{11,00}^{51,0}}{4608}+\frac{7 C_{11,22}^{51,0}}{4608 \sqrt{3}}-\frac{3}{512} \sqrt{\frac{7}{5}}
C_{11,22}^{53,0}-\frac{25 \sqrt{5} C_{20,00}^{40,0}}{4608}-\frac{25 \sqrt{\frac{5}{3}} C_{20,20}^{40,0}}{4608}+\frac{7 C_{20,22}^{42,0}}{4608 \sqrt{3}}+\frac{35
C_{22,00}^{42,0}}{2304}+\frac{35 C_{22,20}^{42,0}}{2304 \sqrt{3}}-\frac{17 C_{22,22}^{40,0}}{4608 \sqrt{3}}-\frac{5 \sqrt{\frac{7}{3}} C_{22,22}^{42,0}}{2304}+\frac{1}{128}
\sqrt{\frac{3}{5}} C_{22,22}^{44,0}-\frac{5 \sqrt{5} C_{31,00}^{31,0}}{4608}-\frac{5 \sqrt{\frac{5}{3}} C_{31,20}^{31,0}}{4608}+\frac{23 C_{31,22}^{31,0}}{4608
\sqrt{3}}-\frac{\sqrt{\frac{7}{5}} C_{31,22}^{33,0}}{1536}-\frac{\sqrt{105} C_{33,00}^{33,0}}{256}-\frac{\sqrt{35} C_{33,20}^{33,0}}{256}+\frac{1}{640}
\sqrt{\frac{21}{2}} C_{33,22}^{33,0}\)

\noindent\(C_{11,2202,1}^{00,3111,0,\text{Loc}}= -\frac{7 C_{11,11}^{51,0}}{96 \sqrt{5}}-\frac{7 C_{11,11}^{51,1}}{192 \sqrt{5}}-\frac{7
C_{22,11}^{42,0}}{80 \sqrt{3}}-\frac{7 C_{22,11}^{42,1}}{160 \sqrt{3}}-\frac{7 C_{31,11}^{31,0}}{96 \sqrt{5}}-\frac{7 C_{31,11}^{31,1}}{192 \sqrt{5}}-\frac{3}{40}
\sqrt{\frac{7}{10}} C_{33,11}^{33,0}-\frac{3}{80} \sqrt{\frac{7}{10}} C_{33,11}^{33,1}\)

\noindent\(C_{11,2202,1}^{00,3111,0,\text{NonLoc}}= -\frac{7 C_{11,11}^{51,0}}{192 \sqrt{5}}-\frac{7 C_{11,11}^{51,1}}{96 \sqrt{5}}+\frac{7
C_{22,11}^{42,0}}{160 \sqrt{3}}+\frac{7 C_{22,11}^{42,1}}{80 \sqrt{3}}-\frac{7 C_{31,11}^{31,0}}{192 \sqrt{5}}-\frac{7 C_{31,11}^{31,1}}{96 \sqrt{5}}-\frac{3}{80}
\sqrt{\frac{7}{10}} C_{33,11}^{33,0}-\frac{3}{40} \sqrt{\frac{7}{10}} C_{33,11}^{33,1}\)

\noindent\(C_{11,2202,1}^{00,3111,1,\text{Loc}}= -\frac{7 C_{11,11}^{51,1}}{64 \sqrt{15}}-\frac{7 C_{22,11}^{42,1}}{160}-\frac{7 C_{31,11}^{31,1}}{64
\sqrt{15}}-\frac{3}{80} \sqrt{\frac{21}{10}} C_{33,11}^{33,1}\)

\noindent\(C_{11,2202,1}^{00,3111,1,\text{NonLoc}}= -\frac{7 C_{11,11}^{51,0}}{64 \sqrt{15}}+\frac{7 C_{22,11}^{42,0}}{160}-\frac{7
C_{31,11}^{31,0}}{64 \sqrt{15}}-\frac{3}{80} \sqrt{\frac{21}{10}} C_{33,11}^{33,0}\)

\noindent\(C_{11,2202,1}^{11,2000,0,\text{Loc}}= \frac{7 \sqrt{5} C_{00,00}^{60,0}}{96}+\frac{7 \sqrt{5} C_{00,00}^{60,1}}{192}+\frac{7
\sqrt{5} C_{11,00}^{51,0}}{288}+\frac{7 \sqrt{5} C_{11,00}^{51,1}}{576}-\frac{7 \sqrt{5} C_{20,00}^{40,0}}{288}-\frac{7 \sqrt{5} C_{20,00}^{40,1}}{576}+\frac{7
C_{22,00}^{42,0}}{288}+\frac{7 C_{22,00}^{42,1}}{576}-\frac{7 \sqrt{5} C_{31,00}^{31,0}}{288}-\frac{7 \sqrt{5} C_{31,00}^{31,1}}{576}+\frac{1}{32}
\sqrt{\frac{21}{5}} C_{33,00}^{33,0}+\frac{1}{64} \sqrt{\frac{21}{5}} C_{33,00}^{33,1}\)

\noindent\(C_{11,2202,1}^{11,2000,0,\text{NonLoc}}= -\frac{7 \sqrt{5} C_{00,00}^{60,0}}{384}-\frac{7 \sqrt{5} C_{00,00}^{60,1}}{192}+\frac{7}{128}
\sqrt{\frac{5}{3}} C_{00,20}^{60,0}+\frac{7}{64} \sqrt{\frac{5}{3}} C_{00,20}^{60,1}+\frac{7 \sqrt{5} C_{11,00}^{51,0}}{1152}+\frac{7 \sqrt{5} C_{11,00}^{51,1}}{576}+\frac{7
\sqrt{5} C_{20,00}^{40,0}}{1152}+\frac{7 \sqrt{5} C_{20,00}^{40,1}}{576}-\frac{7}{384} \sqrt{\frac{5}{3}} C_{20,20}^{40,0}-\frac{7}{192} \sqrt{\frac{5}{3}}
C_{20,20}^{40,1}-\frac{7 C_{22,00}^{42,0}}{1152}-\frac{7 C_{22,00}^{42,1}}{576}+\frac{7 C_{22,20}^{42,0}}{384 \sqrt{3}}+\frac{7 C_{22,20}^{42,1}}{192
\sqrt{3}}-\frac{7 \sqrt{5} C_{31,00}^{31,0}}{1152}-\frac{7 \sqrt{5} C_{31,00}^{31,1}}{576}+\frac{7}{384} \sqrt{\frac{5}{3}} C_{31,20}^{31,0}+\frac{7}{192}
\sqrt{\frac{5}{3}} C_{31,20}^{31,1}+\frac{1}{128} \sqrt{\frac{21}{5}} C_{33,00}^{33,0}+\frac{1}{64} \sqrt{\frac{21}{5}} C_{33,00}^{33,1}-\frac{3}{128}
\sqrt{\frac{7}{5}} C_{33,20}^{33,0}-\frac{3}{64} \sqrt{\frac{7}{5}} C_{33,20}^{33,1}\)

\noindent\(C_{11,2202,1}^{11,2000,1,\text{Loc}}= \frac{7}{64} \sqrt{\frac{5}{3}} C_{00,00}^{60,1}+\frac{7}{192} \sqrt{\frac{5}{3}}
C_{11,00}^{51,1}-\frac{7}{192} \sqrt{\frac{5}{3}} C_{20,00}^{40,1}+\frac{7 C_{22,00}^{42,1}}{192 \sqrt{3}}-\frac{7}{192} \sqrt{\frac{5}{3}} C_{31,00}^{31,1}+\frac{3}{64}
\sqrt{\frac{7}{5}} C_{33,00}^{33,1}\)

\noindent\(C_{11,2202,1}^{11,2000,1,\text{NonLoc}}= -\frac{7}{128} \sqrt{\frac{5}{3}} C_{00,00}^{60,0}+\frac{7 \sqrt{5} C_{00,20}^{60,0}}{128}+\frac{7}{384}
\sqrt{\frac{5}{3}} C_{11,00}^{51,0}+\frac{7}{384} \sqrt{\frac{5}{3}} C_{20,00}^{40,0}-\frac{7 \sqrt{5} C_{20,20}^{40,0}}{384}-\frac{7 C_{22,00}^{42,0}}{384
\sqrt{3}}+\frac{7 C_{22,20}^{42,0}}{384}-\frac{7}{384} \sqrt{\frac{5}{3}} C_{31,00}^{31,0}+\frac{7 \sqrt{5} C_{31,20}^{31,0}}{384}+\frac{3}{128}
\sqrt{\frac{7}{5}} C_{33,00}^{33,0}-\frac{3}{128} \sqrt{\frac{21}{5}} C_{33,20}^{33,0}\)

\noindent\(C_{11,2202,1}^{11,2202,0,\text{Loc}}= \frac{7 C_{00,00}^{60,0}}{96}+\frac{7 C_{00,00}^{60,1}}{192}+\frac{7 C_{11,00}^{51,0}}{288}+\frac{7
C_{11,00}^{51,1}}{576}-\frac{C_{20,00}^{40,0}}{288}-\frac{C_{20,00}^{40,1}}{576}-\frac{7 C_{22,00}^{42,0}}{576 \sqrt{5}}-\frac{7 C_{22,00}^{42,1}}{1152
\sqrt{5}}-\frac{C_{31,00}^{31,0}}{72}-\frac{C_{31,00}^{31,1}}{144}-\frac{\sqrt{21} C_{33,00}^{33,0}}{320}-\frac{\sqrt{21} C_{33,00}^{33,1}}{640}\)

\noindent\(C_{11,2202,1}^{11,2202,0,\text{NonLoc}}= -\frac{7 C_{00,00}^{60,0}}{384}-\frac{7 C_{00,00}^{60,1}}{192}+\frac{7 C_{00,20}^{60,0}}{128
\sqrt{3}}+\frac{7 C_{00,20}^{60,1}}{64 \sqrt{3}}+\frac{7 C_{11,00}^{51,0}}{1152}+\frac{7 C_{11,00}^{51,1}}{576}+\frac{C_{20,00}^{40,0}}{1152}+\frac{C_{20,00}^{40,1}}{576}-\frac{C_{20,20}^{40,0}}{384
\sqrt{3}}-\frac{C_{20,20}^{40,1}}{192 \sqrt{3}}+\frac{7 C_{22,00}^{42,0}}{2304 \sqrt{5}}+\frac{7 C_{22,00}^{42,1}}{1152 \sqrt{5}}-\frac{7 C_{22,20}^{42,0}}{768
\sqrt{15}}-\frac{7 C_{22,20}^{42,1}}{384 \sqrt{15}}-\frac{C_{31,00}^{31,0}}{288}-\frac{C_{31,00}^{31,1}}{144}+\frac{C_{31,20}^{31,0}}{96 \sqrt{3}}+\frac{C_{31,20}^{31,1}}{48
\sqrt{3}}-\frac{\sqrt{21} C_{33,00}^{33,0}}{1280}-\frac{\sqrt{21} C_{33,00}^{33,1}}{640}+\frac{3 \sqrt{7} C_{33,20}^{33,0}}{1280}+\frac{3 \sqrt{7}
C_{33,20}^{33,1}}{640}\)

\noindent\(C_{11,2202,1}^{11,2202,1,\text{Loc}}= \frac{7 C_{00,00}^{60,1}}{64 \sqrt{3}}+\frac{7 C_{11,00}^{51,1}}{192 \sqrt{3}}-\frac{C_{20,00}^{40,1}}{192
\sqrt{3}}-\frac{7 C_{22,00}^{42,1}}{384 \sqrt{15}}-\frac{C_{31,00}^{31,1}}{48 \sqrt{3}}-\frac{3 \sqrt{7} C_{33,00}^{33,1}}{640}\)

\noindent\(C_{11,2202,1}^{11,2202,1,\text{NonLoc}}= -\frac{7 C_{00,00}^{60,0}}{128 \sqrt{3}}+\frac{7 C_{00,20}^{60,0}}{128}+\frac{7
C_{11,00}^{51,0}}{384 \sqrt{3}}+\frac{C_{20,00}^{40,0}}{384 \sqrt{3}}-\frac{C_{20,20}^{40,0}}{384}+\frac{7 C_{22,00}^{42,0}}{768 \sqrt{15}}-\frac{7
C_{22,20}^{42,0}}{768 \sqrt{5}}-\frac{C_{31,00}^{31,0}}{96 \sqrt{3}}+\frac{C_{31,20}^{31,0}}{96}-\frac{3 \sqrt{7} C_{33,00}^{33,0}}{1280}+\frac{3
\sqrt{21} C_{33,20}^{33,0}}{1280}\)

\noindent\(C_{11,2202,2}^{00,3112,0,\text{Loc}}= \frac{7 C_{11,11}^{51,0}}{32 \sqrt{15}}+\frac{7 C_{11,11}^{51,1}}{64 \sqrt{15}}+\frac{7
C_{22,11}^{42,0}}{80}+\frac{7 C_{22,11}^{42,1}}{160}+\frac{7 C_{31,11}^{31,0}}{32 \sqrt{15}}+\frac{7 C_{31,11}^{31,1}}{64 \sqrt{15}}+\frac{3}{40}
\sqrt{\frac{21}{10}} C_{33,11}^{33,0}+\frac{3}{80} \sqrt{\frac{21}{10}} C_{33,11}^{33,1}\)

\noindent\(C_{11,2202,2}^{00,3112,0,\text{NonLoc}}= \frac{7 C_{11,11}^{51,0}}{64 \sqrt{15}}+\frac{7 C_{11,11}^{51,1}}{32 \sqrt{15}}-\frac{7
C_{22,11}^{42,0}}{160}-\frac{7 C_{22,11}^{42,1}}{80}+\frac{7 C_{31,11}^{31,0}}{64 \sqrt{15}}+\frac{7 C_{31,11}^{31,1}}{32 \sqrt{15}}+\frac{3}{80}
\sqrt{\frac{21}{10}} C_{33,11}^{33,0}+\frac{3}{40} \sqrt{\frac{21}{10}} C_{33,11}^{33,1}\)

\noindent\(C_{11,2202,2}^{00,3112,1,\text{Loc}}= \frac{7 C_{11,11}^{51,1}}{64 \sqrt{5}}+\frac{7 \sqrt{3} C_{22,11}^{42,1}}{160}+\frac{7
C_{31,11}^{31,1}}{64 \sqrt{5}}+\frac{9}{80} \sqrt{\frac{7}{10}} C_{33,11}^{33,1}\)

\noindent\(C_{11,2202,2}^{00,3112,1,\text{NonLoc}}= \frac{7 C_{11,11}^{51,0}}{64 \sqrt{5}}-\frac{7 \sqrt{3} C_{22,11}^{42,0}}{160}+\frac{7
C_{31,11}^{31,0}}{64 \sqrt{5}}+\frac{9}{80} \sqrt{\frac{7}{10}} C_{33,11}^{33,0}\)

\noindent\(C_{11,2202,2}^{00,3312,0,\text{Loc}}= -\frac{\sqrt{7} C_{11,11}^{51,0}}{144}-\frac{\sqrt{7} C_{11,11}^{51,1}}{288}-\frac{1}{24}
\sqrt{\frac{7}{15}} C_{22,11}^{42,0}-\frac{1}{48} \sqrt{\frac{7}{15}} C_{22,11}^{42,1}-\frac{\sqrt{7} C_{31,11}^{31,0}}{144}-\frac{\sqrt{7} C_{31,11}^{31,1}}{288}-\frac{C_{33,11}^{33,0}}{20
\sqrt{2}}-\frac{C_{33,11}^{33,1}}{40 \sqrt{2}}\)

\noindent\(C_{11,2202,2}^{00,3312,0,\text{NonLoc}}= -\frac{\sqrt{7} C_{11,11}^{51,0}}{288}-\frac{\sqrt{7} C_{11,11}^{51,1}}{144}+\frac{1}{48}
\sqrt{\frac{7}{15}} C_{22,11}^{42,0}+\frac{1}{24} \sqrt{\frac{7}{15}} C_{22,11}^{42,1}-\frac{\sqrt{7} C_{31,11}^{31,0}}{288}-\frac{\sqrt{7} C_{31,11}^{31,1}}{144}-\frac{C_{33,11}^{33,0}}{40
\sqrt{2}}-\frac{C_{33,11}^{33,1}}{20 \sqrt{2}}\)

\noindent\(C_{11,2202,2}^{00,3312,1,\text{Loc}}= -\frac{1}{96} \sqrt{\frac{7}{3}} C_{11,11}^{51,1}-\frac{1}{48} \sqrt{\frac{7}{5}}
C_{22,11}^{42,1}-\frac{1}{96} \sqrt{\frac{7}{3}} C_{31,11}^{31,1}-\frac{1}{40} \sqrt{\frac{3}{2}} C_{33,11}^{33,1}\)

\noindent\(C_{11,2202,2}^{00,3312,1,\text{NonLoc}}= -\frac{1}{96} \sqrt{\frac{7}{3}} C_{11,11}^{51,0}+\frac{1}{48} \sqrt{\frac{7}{5}}
C_{22,11}^{42,0}-\frac{1}{96} \sqrt{\frac{7}{3}} C_{31,11}^{31,0}-\frac{1}{40} \sqrt{\frac{3}{2}} C_{33,11}^{33,0}\)

\noindent\(C_{11,2202,2}^{11,2202,0,\text{Loc}}= \frac{1}{48} \sqrt{\frac{5}{3}} C_{20,00}^{40,0}+\frac{1}{96} \sqrt{\frac{5}{3}} C_{20,00}^{40,1}-\frac{7
C_{22,00}^{42,0}}{192 \sqrt{3}}-\frac{7 C_{22,00}^{42,1}}{384 \sqrt{3}}+\frac{1}{96} \sqrt{\frac{5}{3}} C_{31,00}^{31,0}+\frac{1}{192} \sqrt{\frac{5}{3}}
C_{31,00}^{31,1}-\frac{3}{64} \sqrt{\frac{7}{5}} C_{33,00}^{33,0}-\frac{3}{128} \sqrt{\frac{7}{5}} C_{33,00}^{33,1}\)

\noindent\(C_{11,2202,2}^{11,2202,0,\text{NonLoc}}= -\frac{1}{192} \sqrt{\frac{5}{3}} C_{20,00}^{40,0}-\frac{1}{96} \sqrt{\frac{5}{3}}
C_{20,00}^{40,1}+\frac{\sqrt{5} C_{20,20}^{40,0}}{192}+\frac{\sqrt{5} C_{20,20}^{40,1}}{96}+\frac{7 C_{22,00}^{42,0}}{768 \sqrt{3}}+\frac{7 C_{22,00}^{42,1}}{384
\sqrt{3}}-\frac{7 C_{22,20}^{42,0}}{768}-\frac{7 C_{22,20}^{42,1}}{384}+\frac{1}{384} \sqrt{\frac{5}{3}} C_{31,00}^{31,0}+\frac{1}{192} \sqrt{\frac{5}{3}}
C_{31,00}^{31,1}-\frac{\sqrt{5} C_{31,20}^{31,0}}{384}-\frac{\sqrt{5} C_{31,20}^{31,1}}{192}-\frac{3}{256} \sqrt{\frac{7}{5}} C_{33,00}^{33,0}-\frac{3}{128}
\sqrt{\frac{7}{5}} C_{33,00}^{33,1}+\frac{3}{256} \sqrt{\frac{21}{5}} C_{33,20}^{33,0}+\frac{3}{128} \sqrt{\frac{21}{5}} C_{33,20}^{33,1}\)

\noindent\(C_{11,2202,2}^{11,2202,1,\text{Loc}}= \frac{\sqrt{5} C_{20,00}^{40,1}}{96}-\frac{7 C_{22,00}^{42,1}}{384}+\frac{\sqrt{5}
C_{31,00}^{31,1}}{192}-\frac{3}{128} \sqrt{\frac{21}{5}} C_{33,00}^{33,1}\)

\noindent\(C_{11,2202,2}^{11,2202,1,\text{NonLoc}}= -\frac{\sqrt{5} C_{20,00}^{40,0}}{192}+\frac{1}{64} \sqrt{\frac{5}{3}} C_{20,20}^{40,0}+\frac{7
C_{22,00}^{42,0}}{768}-\frac{7 C_{22,20}^{42,0}}{256 \sqrt{3}}+\frac{\sqrt{5} C_{31,00}^{31,0}}{384}-\frac{1}{128} \sqrt{\frac{5}{3}} C_{31,20}^{31,0}-\frac{3}{256}
\sqrt{\frac{21}{5}} C_{33,00}^{33,0}+\frac{9}{256} \sqrt{\frac{7}{5}} C_{33,20}^{33,0}\)

\noindent\(C_{11,2202,3}^{00,3313,0,\text{Loc}}= \frac{1}{72} \sqrt{\frac{7}{2}} C_{11,11}^{51,0}+\frac{1}{144} \sqrt{\frac{7}{2}}
C_{11,11}^{51,1}+\frac{1}{12} \sqrt{\frac{7}{30}} C_{22,11}^{42,0}+\frac{1}{24} \sqrt{\frac{7}{30}} C_{22,11}^{42,1}+\frac{1}{72} \sqrt{\frac{7}{2}}
C_{31,11}^{31,0}+\frac{1}{144} \sqrt{\frac{7}{2}} C_{31,11}^{31,1}+\frac{C_{33,11}^{33,0}}{20}+\frac{C_{33,11}^{33,1}}{40}\)

\noindent\(C_{11,2202,3}^{00,3313,0,\text{NonLoc}}= \frac{1}{144} \sqrt{\frac{7}{2}} C_{11,11}^{51,0}+\frac{1}{72} \sqrt{\frac{7}{2}}
C_{11,11}^{51,1}-\frac{1}{24} \sqrt{\frac{7}{30}} C_{22,11}^{42,0}-\frac{1}{12} \sqrt{\frac{7}{30}} C_{22,11}^{42,1}+\frac{1}{144} \sqrt{\frac{7}{2}}
C_{31,11}^{31,0}+\frac{1}{72} \sqrt{\frac{7}{2}} C_{31,11}^{31,1}+\frac{C_{33,11}^{33,0}}{40}+\frac{C_{33,11}^{33,1}}{20}\)

\noindent\(C_{11,2202,3}^{00,3313,1,\text{Loc}}= \frac{1}{48} \sqrt{\frac{7}{6}} C_{11,11}^{51,1}+\frac{1}{24} \sqrt{\frac{7}{10}}
C_{22,11}^{42,1}+\frac{1}{48} \sqrt{\frac{7}{6}} C_{31,11}^{31,1}+\frac{\sqrt{3} C_{33,11}^{33,1}}{40}\)

\noindent\(C_{11,2202,3}^{00,3313,1,\text{NonLoc}}= \frac{1}{48} \sqrt{\frac{7}{6}} C_{11,11}^{51,0}-\frac{1}{24} \sqrt{\frac{7}{10}}
C_{22,11}^{42,0}+\frac{1}{48} \sqrt{\frac{7}{6}} C_{31,11}^{31,0}+\frac{\sqrt{3} C_{33,11}^{33,0}}{40}\)

\noindent\(C_{11,2202,3}^{11,2202,0,\text{Loc}}= \frac{\sqrt{21} C_{00,00}^{60,0}}{64}+\frac{\sqrt{21} C_{00,00}^{60,1}}{128}+\frac{1}{64}
\sqrt{\frac{7}{3}} C_{11,00}^{51,0}+\frac{1}{128} \sqrt{\frac{7}{3}} C_{11,00}^{51,1}+\frac{1}{192} \sqrt{\frac{7}{3}} C_{20,00}^{40,0}+\frac{1}{384}
\sqrt{\frac{7}{3}} C_{20,00}^{40,1}-\frac{1}{48} \sqrt{\frac{7}{15}} C_{22,00}^{42,0}-\frac{1}{96} \sqrt{\frac{7}{15}} C_{22,00}^{42,1}-\frac{1}{192}
\sqrt{\frac{7}{3}} C_{31,00}^{31,0}-\frac{1}{384} \sqrt{\frac{7}{3}} C_{31,00}^{31,1}-\frac{3 C_{33,00}^{33,0}}{80}-\frac{3 C_{33,00}^{33,1}}{160}\)

\noindent\(C_{11,2202,3}^{11,2202,0,\text{NonLoc}}= -\frac{\sqrt{21} C_{00,00}^{60,0}}{256}-\frac{\sqrt{21} C_{00,00}^{60,1}}{128}+\frac{3
\sqrt{7} C_{00,20}^{60,0}}{256}+\frac{3 \sqrt{7} C_{00,20}^{60,1}}{128}+\frac{1}{256} \sqrt{\frac{7}{3}} C_{11,00}^{51,0}+\frac{1}{128} \sqrt{\frac{7}{3}}
C_{11,00}^{51,1}-\frac{1}{768} \sqrt{\frac{7}{3}} C_{20,00}^{40,0}-\frac{1}{384} \sqrt{\frac{7}{3}} C_{20,00}^{40,1}+\frac{\sqrt{7} C_{20,20}^{40,0}}{768}+\frac{\sqrt{7}
C_{20,20}^{40,1}}{384}+\frac{1}{192} \sqrt{\frac{7}{15}} C_{22,00}^{42,0}+\frac{1}{96} \sqrt{\frac{7}{15}} C_{22,00}^{42,1}-\frac{1}{192} \sqrt{\frac{7}{5}}
C_{22,20}^{42,0}-\frac{1}{96} \sqrt{\frac{7}{5}} C_{22,20}^{42,1}-\frac{1}{768} \sqrt{\frac{7}{3}} C_{31,00}^{31,0}-\frac{1}{384} \sqrt{\frac{7}{3}}
C_{31,00}^{31,1}+\frac{\sqrt{7} C_{31,20}^{31,0}}{768}+\frac{\sqrt{7} C_{31,20}^{31,1}}{384}-\frac{3 C_{33,00}^{33,0}}{320}-\frac{3 C_{33,00}^{33,1}}{160}+\frac{3
\sqrt{3} C_{33,20}^{33,0}}{320}+\frac{3 \sqrt{3} C_{33,20}^{33,1}}{160}\)

\noindent\(C_{11,2202,3}^{11,2202,1,\text{Loc}}= \frac{3 \sqrt{7} C_{00,00}^{60,1}}{128}+\frac{\sqrt{7} C_{11,00}^{51,1}}{128}+\frac{\sqrt{7}
C_{20,00}^{40,1}}{384}-\frac{1}{96} \sqrt{\frac{7}{5}} C_{22,00}^{42,1}-\frac{\sqrt{7} C_{31,00}^{31,1}}{384}-\frac{3 \sqrt{3} C_{33,00}^{33,1}}{160}\)

\noindent\(C_{11,2202,3}^{11,2202,1,\text{NonLoc}}= -\frac{3 \sqrt{7} C_{00,00}^{60,0}}{256}+\frac{3 \sqrt{21} C_{00,20}^{60,0}}{256}+\frac{\sqrt{7}
C_{11,00}^{51,0}}{256}-\frac{\sqrt{7} C_{20,00}^{40,0}}{768}+\frac{1}{256} \sqrt{\frac{7}{3}} C_{20,20}^{40,0}+\frac{1}{192} \sqrt{\frac{7}{5}} C_{22,00}^{42,0}-\frac{1}{64}
\sqrt{\frac{7}{15}} C_{22,20}^{42,0}-\frac{\sqrt{7} C_{31,00}^{31,0}}{768}+\frac{1}{256} \sqrt{\frac{7}{3}} C_{31,20}^{31,0}-\frac{3 \sqrt{3} C_{33,00}^{33,0}}{320}+\frac{9
C_{33,20}^{33,0}}{320}\)

\noindent\(C_{11,2211,0}^{11,2011,0,\text{Loc}}= -\frac{7 \sqrt{5} C_{00,20}^{60,0}}{288}-\frac{7 \sqrt{5} C_{00,20}^{60,1}}{576}-\frac{7
C_{00,22}^{62,0}}{96}-\frac{7 C_{00,22}^{62,1}}{192}-\frac{49 C_{11,22}^{51,0}}{1728}-\frac{49 C_{11,22}^{51,1}}{3456}-\frac{1}{128} \sqrt{\frac{21}{5}}
C_{11,22}^{53,0}-\frac{1}{256} \sqrt{\frac{21}{5}} C_{11,22}^{53,1}+\frac{7 \sqrt{5} C_{20,20}^{40,0}}{864}+\frac{7 \sqrt{5} C_{20,20}^{40,1}}{1728}+\frac{7
C_{20,22}^{42,0}}{1728}+\frac{7 C_{20,22}^{42,1}}{3456}-\frac{7 C_{22,20}^{42,0}}{864}-\frac{7 C_{22,20}^{42,1}}{1728}+\frac{7 C_{22,22}^{40,0}}{432}+\frac{7
C_{22,22}^{40,1}}{864}-\frac{5 \sqrt{7} C_{22,22}^{42,0}}{864}-\frac{5 \sqrt{7} C_{22,22}^{42,1}}{1728}+\frac{C_{22,22}^{44,0}}{16 \sqrt{5}}+\frac{C_{22,22}^{44,1}}{32
\sqrt{5}}+\frac{7 \sqrt{5} C_{31,20}^{31,0}}{864}+\frac{7 \sqrt{5} C_{31,20}^{31,1}}{1728}+\frac{7 C_{31,22}^{31,0}}{1728}+\frac{7 C_{31,22}^{31,1}}{3456}+\frac{17}{384}
\sqrt{\frac{7}{15}} C_{31,22}^{33,0}+\frac{17}{768} \sqrt{\frac{7}{15}} C_{31,22}^{33,1}-\frac{1}{32} \sqrt{\frac{7}{15}} C_{33,20}^{33,0}-\frac{1}{64}
\sqrt{\frac{7}{15}} C_{33,20}^{33,1}-\frac{1}{80} \sqrt{\frac{7}{2}} C_{33,22}^{33,0}-\frac{1}{160} \sqrt{\frac{7}{2}} C_{33,22}^{33,1}\)

\noindent\(C_{11,2211,0}^{11,2011,0,\text{NonLoc}}= -\frac{7}{384} \sqrt{\frac{5}{3}} C_{00,00}^{60,0}-\frac{7}{192} \sqrt{\frac{5}{3}}
C_{00,00}^{60,1}-\frac{7 \sqrt{5} C_{00,20}^{60,0}}{1152}-\frac{7 \sqrt{5} C_{00,20}^{60,1}}{576}+\frac{7 C_{00,22}^{62,0}}{192}+\frac{7 C_{00,22}^{62,1}}{96}+\frac{7
\sqrt{\frac{5}{3}} C_{11,00}^{51,0}}{1152}+\frac{7}{576} \sqrt{\frac{5}{3}} C_{11,00}^{51,1}-\frac{49 C_{11,22}^{51,0}}{3456}-\frac{49 C_{11,22}^{51,1}}{1728}-\frac{1}{256}
\sqrt{\frac{21}{5}} C_{11,22}^{53,0}-\frac{1}{128} \sqrt{\frac{21}{5}} C_{11,22}^{53,1}+\frac{7 \sqrt{\frac{5}{3}} C_{20,00}^{40,0}}{1152}+\frac{7}{576}
\sqrt{\frac{5}{3}} C_{20,00}^{40,1}+\frac{7 \sqrt{5} C_{20,20}^{40,0}}{3456}+\frac{7 \sqrt{5} C_{20,20}^{40,1}}{1728}-\frac{7 C_{20,22}^{42,0}}{3456}-\frac{7
C_{20,22}^{42,1}}{1728}-\frac{7 C_{22,00}^{42,0}}{1152 \sqrt{3}}-\frac{7 C_{22,00}^{42,1}}{576 \sqrt{3}}-\frac{7 C_{22,20}^{42,0}}{3456}-\frac{7
C_{22,20}^{42,1}}{1728}-\frac{7 C_{22,22}^{40,0}}{864}-\frac{7 C_{22,22}^{40,1}}{432}+\frac{5 \sqrt{7} C_{22,22}^{42,0}}{1728}+\frac{5 \sqrt{7} C_{22,22}^{42,1}}{864}-\frac{C_{22,22}^{44,0}}{32
\sqrt{5}}-\frac{C_{22,22}^{44,1}}{16 \sqrt{5}}-\frac{7 \sqrt{\frac{5}{3}} C_{31,00}^{31,0}}{1152}-\frac{7}{576} \sqrt{\frac{5}{3}} C_{31,00}^{31,1}-\frac{7
\sqrt{5} C_{31,20}^{31,0}}{3456}-\frac{7 \sqrt{5} C_{31,20}^{31,1}}{1728}+\frac{7 C_{31,22}^{31,0}}{3456}+\frac{7 C_{31,22}^{31,1}}{1728}+\frac{17}{768}
\sqrt{\frac{7}{15}} C_{31,22}^{33,0}+\frac{17}{384} \sqrt{\frac{7}{15}} C_{31,22}^{33,1}+\frac{1}{128} \sqrt{\frac{7}{5}} C_{33,00}^{33,0}+\frac{1}{64}
\sqrt{\frac{7}{5}} C_{33,00}^{33,1}+\frac{1}{128} \sqrt{\frac{7}{15}} C_{33,20}^{33,0}+\frac{1}{64} \sqrt{\frac{7}{15}} C_{33,20}^{33,1}-\frac{1}{160}
\sqrt{\frac{7}{2}} C_{33,22}^{33,0}-\frac{1}{80} \sqrt{\frac{7}{2}} C_{33,22}^{33,1}\)

\noindent\(C_{11,2211,0}^{11,2011,1,\text{Loc}}= -\frac{7}{192} \sqrt{\frac{5}{3}} C_{00,20}^{60,1}-\frac{7 C_{00,22}^{62,1}}{64 \sqrt{3}}-\frac{49
C_{11,22}^{51,1}}{1152 \sqrt{3}}-\frac{3}{256} \sqrt{\frac{7}{5}} C_{11,22}^{53,1}+\frac{7}{576} \sqrt{\frac{5}{3}} C_{20,20}^{40,1}+\frac{7 C_{20,22}^{42,1}}{1152
\sqrt{3}}-\frac{7 C_{22,20}^{42,1}}{576 \sqrt{3}}+\frac{7 C_{22,22}^{40,1}}{288 \sqrt{3}}-\frac{5}{576} \sqrt{\frac{7}{3}} C_{22,22}^{42,1}+\frac{1}{32}
\sqrt{\frac{3}{5}} C_{22,22}^{44,1}+\frac{7}{576} \sqrt{\frac{5}{3}} C_{31,20}^{31,1}+\frac{7 C_{31,22}^{31,1}}{1152 \sqrt{3}}+\frac{17}{768} \sqrt{\frac{7}{5}}
C_{31,22}^{33,1}-\frac{1}{64} \sqrt{\frac{7}{5}} C_{33,20}^{33,1}-\frac{1}{160} \sqrt{\frac{21}{2}} C_{33,22}^{33,1}\)

\noindent\(C_{11,2211,0}^{11,2011,1,\text{NonLoc}}= -\frac{7 \sqrt{5} C_{00,00}^{60,0}}{384}-\frac{7}{384} \sqrt{\frac{5}{3}} C_{00,20}^{60,0}+\frac{7
C_{00,22}^{62,0}}{64 \sqrt{3}}+\frac{7 \sqrt{5} C_{11,00}^{51,0}}{1152}-\frac{49 C_{11,22}^{51,0}}{1152 \sqrt{3}}-\frac{3}{256} \sqrt{\frac{7}{5}}
C_{11,22}^{53,0}+\frac{7 \sqrt{5} C_{20,00}^{40,0}}{1152}+\frac{7 \sqrt{\frac{5}{3}} C_{20,20}^{40,0}}{1152}-\frac{7 C_{20,22}^{42,0}}{1152 \sqrt{3}}-\frac{7
C_{22,00}^{42,0}}{1152}-\frac{7 C_{22,20}^{42,0}}{1152 \sqrt{3}}-\frac{7 C_{22,22}^{40,0}}{288 \sqrt{3}}+\frac{5}{576} \sqrt{\frac{7}{3}} C_{22,22}^{42,0}-\frac{1}{32}
\sqrt{\frac{3}{5}} C_{22,22}^{44,0}-\frac{7 \sqrt{5} C_{31,00}^{31,0}}{1152}-\frac{7 \sqrt{\frac{5}{3}} C_{31,20}^{31,0}}{1152}+\frac{7 C_{31,22}^{31,0}}{1152
\sqrt{3}}+\frac{17}{768} \sqrt{\frac{7}{5}} C_{31,22}^{33,0}+\frac{1}{128} \sqrt{\frac{21}{5}} C_{33,00}^{33,0}+\frac{1}{128} \sqrt{\frac{7}{5}}
C_{33,20}^{33,0}-\frac{1}{160} \sqrt{\frac{21}{2}} C_{33,22}^{33,0}\)

\noindent\(C_{11,2211,0}^{11,2211,0,\text{Loc}}= -\frac{7 C_{00,20}^{60,0}}{288}-\frac{7 C_{00,20}^{60,1}}{576}-\frac{7 C_{00,22}^{62,0}}{96
\sqrt{5}}-\frac{7 C_{00,22}^{62,1}}{192 \sqrt{5}}-\frac{7 C_{11,22}^{51,0}}{108 \sqrt{5}}-\frac{7 C_{11,22}^{51,1}}{216 \sqrt{5}}+\frac{\sqrt{21}
C_{11,22}^{53,0}}{320}+\frac{\sqrt{21} C_{11,22}^{53,1}}{640}+\frac{C_{20,20}^{40,0}}{864}+\frac{C_{20,20}^{40,1}}{1728}+\frac{91 C_{20,22}^{42,0}}{1728
\sqrt{5}}+\frac{91 C_{20,22}^{42,1}}{3456 \sqrt{5}}+\frac{7 C_{22,20}^{42,0}}{1728 \sqrt{5}}+\frac{7 C_{22,20}^{42,1}}{3456 \sqrt{5}}-\frac{13 C_{22,22}^{40,0}}{216
\sqrt{5}}-\frac{13 C_{22,22}^{40,1}}{432 \sqrt{5}}+\frac{\sqrt{35} C_{22,22}^{42,0}}{1728}+\frac{\sqrt{35} C_{22,22}^{42,1}}{3456}-\frac{C_{22,22}^{44,0}}{160}-\frac{C_{22,22}^{44,1}}{320}+\frac{C_{31,20}^{31,0}}{216}+\frac{C_{31,20}^{31,1}}{432}+\frac{47
C_{31,22}^{31,0}}{864 \sqrt{5}}+\frac{47 C_{31,22}^{31,1}}{1728 \sqrt{5}}-\frac{1}{120} \sqrt{\frac{7}{3}} C_{31,22}^{33,0}-\frac{1}{240} \sqrt{\frac{7}{3}}
C_{31,22}^{33,1}+\frac{1}{320} \sqrt{\frac{7}{3}} C_{33,20}^{33,0}+\frac{1}{640} \sqrt{\frac{7}{3}} C_{33,20}^{33,1}+\frac{1}{160} \sqrt{\frac{7}{10}}
C_{33,22}^{33,0}+\frac{1}{320} \sqrt{\frac{7}{10}} C_{33,22}^{33,1}\)

\noindent\(C_{11,2211,0}^{11,2211,0,\text{NonLoc}}= -\frac{7 C_{00,00}^{60,0}}{384 \sqrt{3}}-\frac{7 C_{00,00}^{60,1}}{192 \sqrt{3}}-\frac{7
C_{00,20}^{60,0}}{1152}-\frac{7 C_{00,20}^{60,1}}{576}+\frac{7 C_{00,22}^{62,0}}{192 \sqrt{5}}+\frac{7 C_{00,22}^{62,1}}{96 \sqrt{5}}+\frac{7 C_{11,00}^{51,0}}{1152
\sqrt{3}}+\frac{7 C_{11,00}^{51,1}}{576 \sqrt{3}}-\frac{7 C_{11,22}^{51,0}}{216 \sqrt{5}}-\frac{7 C_{11,22}^{51,1}}{108 \sqrt{5}}+\frac{\sqrt{21}
C_{11,22}^{53,0}}{640}+\frac{\sqrt{21} C_{11,22}^{53,1}}{320}+\frac{C_{20,00}^{40,0}}{1152 \sqrt{3}}+\frac{C_{20,00}^{40,1}}{576 \sqrt{3}}+\frac{C_{20,20}^{40,0}}{3456}+\frac{C_{20,20}^{40,1}}{1728}-\frac{91
C_{20,22}^{42,0}}{3456 \sqrt{5}}-\frac{91 C_{20,22}^{42,1}}{1728 \sqrt{5}}+\frac{7 C_{22,00}^{42,0}}{2304 \sqrt{15}}+\frac{7 C_{22,00}^{42,1}}{1152
\sqrt{15}}+\frac{7 C_{22,20}^{42,0}}{6912 \sqrt{5}}+\frac{7 C_{22,20}^{42,1}}{3456 \sqrt{5}}+\frac{13 C_{22,22}^{40,0}}{432 \sqrt{5}}+\frac{13 C_{22,22}^{40,1}}{216
\sqrt{5}}-\frac{\sqrt{35} C_{22,22}^{42,0}}{3456}-\frac{\sqrt{35} C_{22,22}^{42,1}}{1728}+\frac{C_{22,22}^{44,0}}{320}+\frac{C_{22,22}^{44,1}}{160}-\frac{C_{31,00}^{31,0}}{288
\sqrt{3}}-\frac{C_{31,00}^{31,1}}{144 \sqrt{3}}-\frac{C_{31,20}^{31,0}}{864}-\frac{C_{31,20}^{31,1}}{432}+\frac{47 C_{31,22}^{31,0}}{1728 \sqrt{5}}+\frac{47
C_{31,22}^{31,1}}{864 \sqrt{5}}-\frac{1}{240} \sqrt{\frac{7}{3}} C_{31,22}^{33,0}-\frac{1}{120} \sqrt{\frac{7}{3}} C_{31,22}^{33,1}-\frac{\sqrt{7}
C_{33,00}^{33,0}}{1280}-\frac{\sqrt{7} C_{33,00}^{33,1}}{640}-\frac{\sqrt{\frac{7}{3}} C_{33,20}^{33,0}}{1280}-\frac{1}{640} \sqrt{\frac{7}{3}} C_{33,20}^{33,1}+\frac{1}{320}
\sqrt{\frac{7}{10}} C_{33,22}^{33,0}+\frac{1}{160} \sqrt{\frac{7}{10}} C_{33,22}^{33,1}\)

\noindent\(C_{11,2211,0}^{11,2211,1,\text{Loc}}= -\frac{7 C_{00,20}^{60,1}}{192 \sqrt{3}}-\frac{7 C_{00,22}^{62,1}}{64 \sqrt{15}}-\frac{7
C_{11,22}^{51,1}}{72 \sqrt{15}}+\frac{3 \sqrt{7} C_{11,22}^{53,1}}{640}+\frac{C_{20,20}^{40,1}}{576 \sqrt{3}}+\frac{91 C_{20,22}^{42,1}}{1152 \sqrt{15}}+\frac{7
C_{22,20}^{42,1}}{1152 \sqrt{15}}-\frac{13 C_{22,22}^{40,1}}{144 \sqrt{15}}+\frac{\sqrt{\frac{35}{3}} C_{22,22}^{42,1}}{1152}-\frac{\sqrt{3} C_{22,22}^{44,1}}{320}+\frac{C_{31,20}^{31,1}}{144
\sqrt{3}}+\frac{47 C_{31,22}^{31,1}}{576 \sqrt{15}}-\frac{\sqrt{7} C_{31,22}^{33,1}}{240}+\frac{\sqrt{7} C_{33,20}^{33,1}}{640}+\frac{1}{320} \sqrt{\frac{21}{10}}
C_{33,22}^{33,1}\)

\noindent\(C_{11,2211,0}^{11,2211,1,\text{NonLoc}}= -\frac{7 C_{00,00}^{60,0}}{384}-\frac{7 C_{00,20}^{60,0}}{384 \sqrt{3}}+\frac{7
C_{00,22}^{62,0}}{64 \sqrt{15}}+\frac{7 C_{11,00}^{51,0}}{1152}-\frac{7 C_{11,22}^{51,0}}{72 \sqrt{15}}+\frac{3 \sqrt{7} C_{11,22}^{53,0}}{640}+\frac{C_{20,00}^{40,0}}{1152}+\frac{C_{20,20}^{40,0}}{1152
\sqrt{3}}-\frac{91 C_{20,22}^{42,0}}{1152 \sqrt{15}}+\frac{7 C_{22,00}^{42,0}}{2304 \sqrt{5}}+\frac{7 C_{22,20}^{42,0}}{2304 \sqrt{15}}+\frac{13
C_{22,22}^{40,0}}{144 \sqrt{15}}-\frac{\sqrt{\frac{35}{3}} C_{22,22}^{42,0}}{1152}+\frac{\sqrt{3} C_{22,22}^{44,0}}{320}-\frac{C_{31,00}^{31,0}}{288}-\frac{C_{31,20}^{31,0}}{288
\sqrt{3}}+\frac{47 C_{31,22}^{31,0}}{576 \sqrt{15}}-\frac{\sqrt{7} C_{31,22}^{33,0}}{240}-\frac{\sqrt{21} C_{33,00}^{33,0}}{1280}-\frac{\sqrt{7}
C_{33,20}^{33,0}}{1280}+\frac{1}{320} \sqrt{\frac{21}{10}} C_{33,22}^{33,0}\)

\noindent\(C_{11,2211,1}^{00,3101,0,\text{Loc}}= \frac{7 C_{11,11}^{51,0}}{96 \sqrt{5}}+\frac{7 C_{11,11}^{51,1}}{192 \sqrt{5}}+\frac{7
C_{22,11}^{42,0}}{80 \sqrt{3}}+\frac{7 C_{22,11}^{42,1}}{160 \sqrt{3}}+\frac{7 C_{31,11}^{31,0}}{96 \sqrt{5}}+\frac{7 C_{31,11}^{31,1}}{192 \sqrt{5}}+\frac{3}{40}
\sqrt{\frac{7}{10}} C_{33,11}^{33,0}+\frac{3}{80} \sqrt{\frac{7}{10}} C_{33,11}^{33,1}\)

\noindent\(C_{11,2211,1}^{00,3101,0,\text{NonLoc}}= \frac{7 C_{11,11}^{51,0}}{192 \sqrt{5}}+\frac{7 C_{11,11}^{51,1}}{96 \sqrt{5}}-\frac{7
C_{22,11}^{42,0}}{160 \sqrt{3}}-\frac{7 C_{22,11}^{42,1}}{80 \sqrt{3}}+\frac{7 C_{31,11}^{31,0}}{192 \sqrt{5}}+\frac{7 C_{31,11}^{31,1}}{96 \sqrt{5}}+\frac{3}{80}
\sqrt{\frac{7}{10}} C_{33,11}^{33,0}+\frac{3}{40} \sqrt{\frac{7}{10}} C_{33,11}^{33,1}\)

\noindent\(C_{11,2211,1}^{00,3101,1,\text{Loc}}= \frac{7 C_{11,11}^{51,1}}{64 \sqrt{15}}+\frac{7 C_{22,11}^{42,1}}{160}+\frac{7 C_{31,11}^{31,1}}{64
\sqrt{15}}+\frac{3}{80} \sqrt{\frac{21}{10}} C_{33,11}^{33,1}\)

\noindent\(C_{11,2211,1}^{00,3101,1,\text{NonLoc}}= \frac{7 C_{11,11}^{51,0}}{64 \sqrt{15}}-\frac{7 C_{22,11}^{42,0}}{160}+\frac{7
C_{31,11}^{31,0}}{64 \sqrt{15}}+\frac{3}{80} \sqrt{\frac{21}{10}} C_{33,11}^{33,0}\)

\noindent\(C_{11,2211,1}^{11,2011,0,\text{Loc}}= \frac{7}{192} \sqrt{\frac{5}{3}} C_{00,20}^{60,0}+\frac{7}{384} \sqrt{\frac{5}{3}}
C_{00,20}^{60,1}-\frac{7 C_{00,22}^{62,0}}{128 \sqrt{3}}-\frac{7 C_{00,22}^{62,1}}{256 \sqrt{3}}-\frac{7 C_{11,22}^{51,0}}{144 \sqrt{3}}-\frac{7
C_{11,22}^{51,1}}{288 \sqrt{3}}+\frac{3}{256} \sqrt{\frac{7}{5}} C_{11,22}^{53,0}+\frac{3}{512} \sqrt{\frac{7}{5}} C_{11,22}^{53,1}-\frac{7}{576}
\sqrt{\frac{5}{3}} C_{20,20}^{40,0}-\frac{7 \sqrt{\frac{5}{3}} C_{20,20}^{40,1}}{1152}-\frac{7 C_{20,22}^{42,0}}{288 \sqrt{3}}-\frac{7 C_{20,22}^{42,1}}{576
\sqrt{3}}+\frac{7 C_{22,20}^{42,0}}{576 \sqrt{3}}+\frac{7 C_{22,20}^{42,1}}{1152 \sqrt{3}}-\frac{49 C_{22,22}^{40,0}}{1152 \sqrt{3}}-\frac{49 C_{22,22}^{40,1}}{2304
\sqrt{3}}+\frac{11}{576} \sqrt{\frac{7}{3}} C_{22,22}^{42,0}+\frac{11 \sqrt{\frac{7}{3}} C_{22,22}^{42,1}}{1152}+\frac{1}{16} \sqrt{\frac{3}{5}}
C_{22,22}^{44,0}+\frac{1}{32} \sqrt{\frac{3}{5}} C_{22,22}^{44,1}-\frac{7}{576} \sqrt{\frac{5}{3}} C_{31,20}^{31,0}-\frac{7 \sqrt{\frac{5}{3}} C_{31,20}^{31,1}}{1152}-\frac{7
C_{31,22}^{31,0}}{288 \sqrt{3}}-\frac{7 C_{31,22}^{31,1}}{576 \sqrt{3}}+\frac{7}{768} \sqrt{\frac{7}{5}} C_{31,22}^{33,0}+\frac{7 \sqrt{\frac{7}{5}}
C_{31,22}^{33,1}}{1536}+\frac{1}{64} \sqrt{\frac{7}{5}} C_{33,20}^{33,0}+\frac{1}{128} \sqrt{\frac{7}{5}} C_{33,20}^{33,1}+\frac{1}{40} \sqrt{\frac{21}{2}}
C_{33,22}^{33,0}+\frac{1}{80} \sqrt{\frac{21}{2}} C_{33,22}^{33,1}\)

\noindent\(C_{11,2211,1}^{11,2011,0,\text{NonLoc}}= \frac{7 \sqrt{5} C_{00,00}^{60,0}}{768}+\frac{7 \sqrt{5} C_{00,00}^{60,1}}{384}+\frac{7}{768}
\sqrt{\frac{5}{3}} C_{00,20}^{60,0}+\frac{7}{384} \sqrt{\frac{5}{3}} C_{00,20}^{60,1}+\frac{7 C_{00,22}^{62,0}}{256 \sqrt{3}}+\frac{7 C_{00,22}^{62,1}}{128
\sqrt{3}}-\frac{7 \sqrt{5} C_{11,00}^{51,0}}{2304}-\frac{7 \sqrt{5} C_{11,00}^{51,1}}{1152}-\frac{7 C_{11,22}^{51,0}}{288 \sqrt{3}}-\frac{7 C_{11,22}^{51,1}}{144
\sqrt{3}}+\frac{3}{512} \sqrt{\frac{7}{5}} C_{11,22}^{53,0}+\frac{3}{256} \sqrt{\frac{7}{5}} C_{11,22}^{53,1}-\frac{7 \sqrt{5} C_{20,00}^{40,0}}{2304}-\frac{7
\sqrt{5} C_{20,00}^{40,1}}{1152}-\frac{7 \sqrt{\frac{5}{3}} C_{20,20}^{40,0}}{2304}-\frac{7 \sqrt{\frac{5}{3}} C_{20,20}^{40,1}}{1152}+\frac{7 C_{20,22}^{42,0}}{576
\sqrt{3}}+\frac{7 C_{20,22}^{42,1}}{288 \sqrt{3}}+\frac{7 C_{22,00}^{42,0}}{2304}+\frac{7 C_{22,00}^{42,1}}{1152}+\frac{7 C_{22,20}^{42,0}}{2304
\sqrt{3}}+\frac{7 C_{22,20}^{42,1}}{1152 \sqrt{3}}+\frac{49 C_{22,22}^{40,0}}{2304 \sqrt{3}}+\frac{49 C_{22,22}^{40,1}}{1152 \sqrt{3}}-\frac{11 \sqrt{\frac{7}{3}}
C_{22,22}^{42,0}}{1152}-\frac{11}{576} \sqrt{\frac{7}{3}} C_{22,22}^{42,1}-\frac{1}{32} \sqrt{\frac{3}{5}} C_{22,22}^{44,0}-\frac{1}{16} \sqrt{\frac{3}{5}}
C_{22,22}^{44,1}+\frac{7 \sqrt{5} C_{31,00}^{31,0}}{2304}+\frac{7 \sqrt{5} C_{31,00}^{31,1}}{1152}+\frac{7 \sqrt{\frac{5}{3}} C_{31,20}^{31,0}}{2304}+\frac{7
\sqrt{\frac{5}{3}} C_{31,20}^{31,1}}{1152}-\frac{7 C_{31,22}^{31,0}}{576 \sqrt{3}}-\frac{7 C_{31,22}^{31,1}}{288 \sqrt{3}}+\frac{7 \sqrt{\frac{7}{5}}
C_{31,22}^{33,0}}{1536}+\frac{7}{768} \sqrt{\frac{7}{5}} C_{31,22}^{33,1}-\frac{1}{256} \sqrt{\frac{21}{5}} C_{33,00}^{33,0}-\frac{1}{128} \sqrt{\frac{21}{5}}
C_{33,00}^{33,1}-\frac{1}{256} \sqrt{\frac{7}{5}} C_{33,20}^{33,0}-\frac{1}{128} \sqrt{\frac{7}{5}} C_{33,20}^{33,1}+\frac{1}{80} \sqrt{\frac{21}{2}}
C_{33,22}^{33,0}+\frac{1}{40} \sqrt{\frac{21}{2}} C_{33,22}^{33,1}\)

\noindent\(C_{11,2211,1}^{11,2011,1,\text{Loc}}= \frac{7 \sqrt{5} C_{00,20}^{60,1}}{384}-\frac{7 C_{00,22}^{62,1}}{256}-\frac{7 C_{11,22}^{51,1}}{288}+\frac{3}{512}
\sqrt{\frac{21}{5}} C_{11,22}^{53,1}-\frac{7 \sqrt{5} C_{20,20}^{40,1}}{1152}-\frac{7 C_{20,22}^{42,1}}{576}+\frac{7 C_{22,20}^{42,1}}{1152}-\frac{49
C_{22,22}^{40,1}}{2304}+\frac{11 \sqrt{7} C_{22,22}^{42,1}}{1152}+\frac{3 C_{22,22}^{44,1}}{32 \sqrt{5}}-\frac{7 \sqrt{5} C_{31,20}^{31,1}}{1152}-\frac{7
C_{31,22}^{31,1}}{576}+\frac{7}{512} \sqrt{\frac{7}{15}} C_{31,22}^{33,1}+\frac{1}{128} \sqrt{\frac{21}{5}} C_{33,20}^{33,1}+\frac{3}{80} \sqrt{\frac{7}{2}}
C_{33,22}^{33,1}\)

\noindent\(C_{11,2211,1}^{11,2011,1,\text{NonLoc}}= \frac{7}{256} \sqrt{\frac{5}{3}} C_{00,00}^{60,0}+\frac{7 \sqrt{5} C_{00,20}^{60,0}}{768}+\frac{7
C_{00,22}^{62,0}}{256}-\frac{7}{768} \sqrt{\frac{5}{3}} C_{11,00}^{51,0}-\frac{7 C_{11,22}^{51,0}}{288}+\frac{3}{512} \sqrt{\frac{21}{5}} C_{11,22}^{53,0}-\frac{7}{768}
\sqrt{\frac{5}{3}} C_{20,00}^{40,0}-\frac{7 \sqrt{5} C_{20,20}^{40,0}}{2304}+\frac{7 C_{20,22}^{42,0}}{576}+\frac{7 C_{22,00}^{42,0}}{768 \sqrt{3}}+\frac{7
C_{22,20}^{42,0}}{2304}+\frac{49 C_{22,22}^{40,0}}{2304}-\frac{11 \sqrt{7} C_{22,22}^{42,0}}{1152}-\frac{3 C_{22,22}^{44,0}}{32 \sqrt{5}}+\frac{7}{768}
\sqrt{\frac{5}{3}} C_{31,00}^{31,0}+\frac{7 \sqrt{5} C_{31,20}^{31,0}}{2304}-\frac{7 C_{31,22}^{31,0}}{576}+\frac{7}{512} \sqrt{\frac{7}{15}} C_{31,22}^{33,0}-\frac{3}{256}
\sqrt{\frac{7}{5}} C_{33,00}^{33,0}-\frac{1}{256} \sqrt{\frac{21}{5}} C_{33,20}^{33,0}+\frac{3}{80} \sqrt{\frac{7}{2}} C_{33,22}^{33,0}\)

\noindent\(C_{11,2211,1}^{11,2211,0,\text{Loc}}= -\frac{7 C_{00,20}^{60,0}}{384 \sqrt{3}}-\frac{7 C_{00,20}^{60,1}}{768 \sqrt{3}}+\frac{7
C_{00,22}^{62,0}}{256 \sqrt{15}}+\frac{7 C_{00,22}^{62,1}}{512 \sqrt{15}}+\frac{119 C_{11,22}^{51,0}}{2304 \sqrt{15}}+\frac{119 C_{11,22}^{51,1}}{4608
\sqrt{15}}-\frac{3 \sqrt{7} C_{11,22}^{53,0}}{640}-\frac{3 \sqrt{7} C_{11,22}^{53,1}}{1280}-\frac{17 C_{20,20}^{40,0}}{1152 \sqrt{3}}-\frac{17 C_{20,20}^{40,1}}{2304
\sqrt{3}}-\frac{203 C_{20,22}^{42,0}}{2304 \sqrt{15}}-\frac{203 C_{20,22}^{42,1}}{4608 \sqrt{15}}+\frac{7 \sqrt{\frac{5}{3}} C_{22,20}^{42,0}}{1152}+\frac{7
\sqrt{\frac{5}{3}} C_{22,20}^{42,1}}{2304}+\frac{187 C_{22,22}^{40,0}}{2304 \sqrt{15}}+\frac{187 C_{22,22}^{40,1}}{4608 \sqrt{15}}+\frac{\sqrt{\frac{7}{15}}
C_{22,22}^{42,0}}{1152}+\frac{\sqrt{\frac{7}{15}} C_{22,22}^{42,1}}{2304}+\frac{\sqrt{3} C_{22,22}^{44,0}}{80}+\frac{\sqrt{3} C_{22,22}^{44,1}}{160}-\frac{5
C_{31,20}^{31,0}}{1152 \sqrt{3}}-\frac{5 C_{31,20}^{31,1}}{2304 \sqrt{3}}-\frac{209 C_{31,22}^{31,0}}{2304 \sqrt{15}}-\frac{209 C_{31,22}^{31,1}}{4608
\sqrt{15}}+\frac{7 \sqrt{7} C_{31,22}^{33,0}}{960}+\frac{7 \sqrt{7} C_{31,22}^{33,1}}{1920}+\frac{\sqrt{7} C_{33,20}^{33,0}}{128}+\frac{\sqrt{7}
C_{33,20}^{33,1}}{256}+\frac{1}{160} \sqrt{\frac{21}{10}} C_{33,22}^{33,0}+\frac{1}{320} \sqrt{\frac{21}{10}} C_{33,22}^{33,1}\)

\noindent\(C_{11,2211,1}^{11,2211,0,\text{NonLoc}}= -\frac{7 C_{00,00}^{60,0}}{1536}-\frac{7 C_{00,00}^{60,1}}{768}-\frac{7 C_{00,20}^{60,0}}{1536
\sqrt{3}}-\frac{7 C_{00,20}^{60,1}}{768 \sqrt{3}}-\frac{7 C_{00,22}^{62,0}}{512 \sqrt{15}}-\frac{7 C_{00,22}^{62,1}}{256 \sqrt{15}}+\frac{7 C_{11,00}^{51,0}}{4608}+\frac{7
C_{11,00}^{51,1}}{2304}+\frac{119 C_{11,22}^{51,0}}{4608 \sqrt{15}}+\frac{119 C_{11,22}^{51,1}}{2304 \sqrt{15}}-\frac{3 \sqrt{7} C_{11,22}^{53,0}}{1280}-\frac{3
\sqrt{7} C_{11,22}^{53,1}}{640}-\frac{17 C_{20,00}^{40,0}}{4608}-\frac{17 C_{20,00}^{40,1}}{2304}-\frac{17 C_{20,20}^{40,0}}{4608 \sqrt{3}}-\frac{17
C_{20,20}^{40,1}}{2304 \sqrt{3}}+\frac{203 C_{20,22}^{42,0}}{4608 \sqrt{15}}+\frac{203 C_{20,22}^{42,1}}{2304 \sqrt{15}}+\frac{7 \sqrt{5} C_{22,00}^{42,0}}{4608}+\frac{7
\sqrt{5} C_{22,00}^{42,1}}{2304}+\frac{7 \sqrt{\frac{5}{3}} C_{22,20}^{42,0}}{4608}+\frac{7 \sqrt{\frac{5}{3}} C_{22,20}^{42,1}}{2304}-\frac{187
C_{22,22}^{40,0}}{4608 \sqrt{15}}-\frac{187 C_{22,22}^{40,1}}{2304 \sqrt{15}}-\frac{\sqrt{\frac{7}{15}} C_{22,22}^{42,0}}{2304}-\frac{\sqrt{\frac{7}{15}}
C_{22,22}^{42,1}}{1152}-\frac{\sqrt{3} C_{22,22}^{44,0}}{160}-\frac{\sqrt{3} C_{22,22}^{44,1}}{80}+\frac{5 C_{31,00}^{31,0}}{4608}+\frac{5 C_{31,00}^{31,1}}{2304}+\frac{5
C_{31,20}^{31,0}}{4608 \sqrt{3}}+\frac{5 C_{31,20}^{31,1}}{2304 \sqrt{3}}-\frac{209 C_{31,22}^{31,0}}{4608 \sqrt{15}}-\frac{209 C_{31,22}^{31,1}}{2304
\sqrt{15}}+\frac{7 \sqrt{7} C_{31,22}^{33,0}}{1920}+\frac{7 \sqrt{7} C_{31,22}^{33,1}}{960}-\frac{\sqrt{21} C_{33,00}^{33,0}}{512}-\frac{\sqrt{21}
C_{33,00}^{33,1}}{256}-\frac{\sqrt{7} C_{33,20}^{33,0}}{512}-\frac{\sqrt{7} C_{33,20}^{33,1}}{256}+\frac{1}{320} \sqrt{\frac{21}{10}} C_{33,22}^{33,0}+\frac{1}{160}
\sqrt{\frac{21}{10}} C_{33,22}^{33,1}\)

\noindent\(C_{11,2211,1}^{11,2211,1,\text{Loc}}= -\frac{7 C_{00,20}^{60,1}}{768}+\frac{7 C_{00,22}^{62,1}}{512 \sqrt{5}}+\frac{119
C_{11,22}^{51,1}}{4608 \sqrt{5}}-\frac{3 \sqrt{21} C_{11,22}^{53,1}}{1280}-\frac{17 C_{20,20}^{40,1}}{2304}-\frac{203 C_{20,22}^{42,1}}{4608 \sqrt{5}}+\frac{7
\sqrt{5} C_{22,20}^{42,1}}{2304}+\frac{187 C_{22,22}^{40,1}}{4608 \sqrt{5}}+\frac{\sqrt{\frac{7}{5}} C_{22,22}^{42,1}}{2304}+\frac{3 C_{22,22}^{44,1}}{160}-\frac{5
C_{31,20}^{31,1}}{2304}-\frac{209 C_{31,22}^{31,1}}{4608 \sqrt{5}}+\frac{7}{640} \sqrt{\frac{7}{3}} C_{31,22}^{33,1}+\frac{\sqrt{21} C_{33,20}^{33,1}}{256}+\frac{3}{320}
\sqrt{\frac{7}{10}} C_{33,22}^{33,1}\)

\noindent\(C_{11,2211,1}^{11,2211,1,\text{NonLoc}}= -\frac{7 C_{00,00}^{60,0}}{512 \sqrt{3}}-\frac{7 C_{00,20}^{60,0}}{1536}-\frac{7
C_{00,22}^{62,0}}{512 \sqrt{5}}+\frac{7 C_{11,00}^{51,0}}{1536 \sqrt{3}}+\frac{119 C_{11,22}^{51,0}}{4608 \sqrt{5}}-\frac{3 \sqrt{21} C_{11,22}^{53,0}}{1280}-\frac{17
C_{20,00}^{40,0}}{1536 \sqrt{3}}-\frac{17 C_{20,20}^{40,0}}{4608}+\frac{203 C_{20,22}^{42,0}}{4608 \sqrt{5}}+\frac{7 \sqrt{\frac{5}{3}} C_{22,00}^{42,0}}{1536}+\frac{7
\sqrt{5} C_{22,20}^{42,0}}{4608}-\frac{187 C_{22,22}^{40,0}}{4608 \sqrt{5}}-\frac{\sqrt{\frac{7}{5}} C_{22,22}^{42,0}}{2304}-\frac{3 C_{22,22}^{44,0}}{160}+\frac{5
C_{31,00}^{31,0}}{1536 \sqrt{3}}+\frac{5 C_{31,20}^{31,0}}{4608}-\frac{209 C_{31,22}^{31,0}}{4608 \sqrt{5}}+\frac{7}{640} \sqrt{\frac{7}{3}} C_{31,22}^{33,0}-\frac{3
\sqrt{7} C_{33,00}^{33,0}}{512}-\frac{\sqrt{21} C_{33,20}^{33,0}}{512}+\frac{3}{320} \sqrt{\frac{7}{10}} C_{33,22}^{33,0}\)

\noindent\(C_{11,2211,2}^{11,2011,0,\text{Loc}}= -\frac{7 C_{00,20}^{60,0}}{576}-\frac{7 C_{00,20}^{60,1}}{1152}-\frac{77 C_{00,22}^{62,0}}{384
\sqrt{5}}-\frac{77 C_{00,22}^{62,1}}{768 \sqrt{5}}-\frac{91 C_{11,22}^{51,0}}{864 \sqrt{5}}-\frac{91 C_{11,22}^{51,1}}{1728 \sqrt{5}}-\frac{\sqrt{21}
C_{11,22}^{53,0}}{1280}-\frac{\sqrt{21} C_{11,22}^{53,1}}{2560}+\frac{7 C_{20,20}^{40,0}}{1728}+\frac{7 C_{20,20}^{40,1}}{3456}-\frac{7 C_{20,22}^{42,0}}{432
\sqrt{5}}-\frac{7 C_{20,22}^{42,1}}{864 \sqrt{5}}-\frac{7 C_{22,20}^{42,0}}{1728 \sqrt{5}}-\frac{7 C_{22,20}^{42,1}}{3456 \sqrt{5}}-\frac{7 \sqrt{5}
C_{22,22}^{40,0}}{3456}-\frac{7 \sqrt{5} C_{22,22}^{40,1}}{6912}+\frac{13 \sqrt{\frac{7}{5}} C_{22,22}^{42,0}}{1728}+\frac{13 \sqrt{\frac{7}{5}}
C_{22,22}^{42,1}}{3456}+\frac{C_{22,22}^{44,0}}{16}+\frac{C_{22,22}^{44,1}}{32}+\frac{7 C_{31,20}^{31,0}}{1728}+\frac{7 C_{31,20}^{31,1}}{3456}-\frac{7
C_{31,22}^{31,0}}{432 \sqrt{5}}-\frac{7 C_{31,22}^{31,1}}{864 \sqrt{5}}+\frac{89 \sqrt{\frac{7}{3}} C_{31,22}^{33,0}}{3840}+\frac{89 \sqrt{\frac{7}{3}}
C_{31,22}^{33,1}}{7680}-\frac{1}{320} \sqrt{\frac{7}{3}} C_{33,20}^{33,0}-\frac{1}{640} \sqrt{\frac{7}{3}} C_{33,20}^{33,1}+\frac{1}{20} \sqrt{\frac{7}{10}}
C_{33,22}^{33,0}+\frac{1}{40} \sqrt{\frac{7}{10}} C_{33,22}^{33,1}\)

\noindent\(C_{11,2211,2}^{11,2011,0,\text{NonLoc}}= -\frac{7 C_{00,00}^{60,0}}{768 \sqrt{3}}-\frac{7 C_{00,00}^{60,1}}{384 \sqrt{3}}-\frac{7
C_{00,20}^{60,0}}{2304}-\frac{7 C_{00,20}^{60,1}}{1152}+\frac{77 C_{00,22}^{62,0}}{768 \sqrt{5}}+\frac{77 C_{00,22}^{62,1}}{384 \sqrt{5}}+\frac{7
C_{11,00}^{51,0}}{2304 \sqrt{3}}+\frac{7 C_{11,00}^{51,1}}{1152 \sqrt{3}}-\frac{91 C_{11,22}^{51,0}}{1728 \sqrt{5}}-\frac{91 C_{11,22}^{51,1}}{864
\sqrt{5}}-\frac{\sqrt{21} C_{11,22}^{53,0}}{2560}-\frac{\sqrt{21} C_{11,22}^{53,1}}{1280}+\frac{7 C_{20,00}^{40,0}}{2304 \sqrt{3}}+\frac{7 C_{20,00}^{40,1}}{1152
\sqrt{3}}+\frac{7 C_{20,20}^{40,0}}{6912}+\frac{7 C_{20,20}^{40,1}}{3456}+\frac{7 C_{20,22}^{42,0}}{864 \sqrt{5}}+\frac{7 C_{20,22}^{42,1}}{432 \sqrt{5}}-\frac{7
C_{22,00}^{42,0}}{2304 \sqrt{15}}-\frac{7 C_{22,00}^{42,1}}{1152 \sqrt{15}}-\frac{7 C_{22,20}^{42,0}}{6912 \sqrt{5}}-\frac{7 C_{22,20}^{42,1}}{3456
\sqrt{5}}+\frac{7 \sqrt{5} C_{22,22}^{40,0}}{6912}+\frac{7 \sqrt{5} C_{22,22}^{40,1}}{3456}-\frac{13 \sqrt{\frac{7}{5}} C_{22,22}^{42,0}}{3456}-\frac{13
\sqrt{\frac{7}{5}} C_{22,22}^{42,1}}{1728}-\frac{C_{22,22}^{44,0}}{32}-\frac{C_{22,22}^{44,1}}{16}-\frac{7 C_{31,00}^{31,0}}{2304 \sqrt{3}}-\frac{7
C_{31,00}^{31,1}}{1152 \sqrt{3}}-\frac{7 C_{31,20}^{31,0}}{6912}-\frac{7 C_{31,20}^{31,1}}{3456}-\frac{7 C_{31,22}^{31,0}}{864 \sqrt{5}}-\frac{7
C_{31,22}^{31,1}}{432 \sqrt{5}}+\frac{89 \sqrt{\frac{7}{3}} C_{31,22}^{33,0}}{7680}+\frac{89 \sqrt{\frac{7}{3}} C_{31,22}^{33,1}}{3840}+\frac{\sqrt{7}
C_{33,00}^{33,0}}{1280}+\frac{\sqrt{7} C_{33,00}^{33,1}}{640}+\frac{\sqrt{\frac{7}{3}} C_{33,20}^{33,0}}{1280}+\frac{1}{640} \sqrt{\frac{7}{3}} C_{33,20}^{33,1}+\frac{1}{40}
\sqrt{\frac{7}{10}} C_{33,22}^{33,0}+\frac{1}{20} \sqrt{\frac{7}{10}} C_{33,22}^{33,1}\)

\noindent\(C_{11,2211,2}^{11,2011,1,\text{Loc}}= -\frac{7 C_{00,20}^{60,1}}{384 \sqrt{3}}-\frac{77 C_{00,22}^{62,1}}{256 \sqrt{15}}-\frac{91
C_{11,22}^{51,1}}{576 \sqrt{15}}-\frac{3 \sqrt{7} C_{11,22}^{53,1}}{2560}+\frac{7 C_{20,20}^{40,1}}{1152 \sqrt{3}}-\frac{7 C_{20,22}^{42,1}}{288
\sqrt{15}}-\frac{7 C_{22,20}^{42,1}}{1152 \sqrt{15}}-\frac{7 \sqrt{\frac{5}{3}} C_{22,22}^{40,1}}{2304}+\frac{13 \sqrt{\frac{7}{15}} C_{22,22}^{42,1}}{1152}+\frac{\sqrt{3}
C_{22,22}^{44,1}}{32}+\frac{7 C_{31,20}^{31,1}}{1152 \sqrt{3}}-\frac{7 C_{31,22}^{31,1}}{288 \sqrt{15}}+\frac{89 \sqrt{7} C_{31,22}^{33,1}}{7680}-\frac{\sqrt{7}
C_{33,20}^{33,1}}{640}+\frac{1}{40} \sqrt{\frac{21}{10}} C_{33,22}^{33,1}\)

\noindent\(C_{11,2211,2}^{11,2011,1,\text{NonLoc}}= -\frac{7 C_{00,00}^{60,0}}{768}-\frac{7 C_{00,20}^{60,0}}{768 \sqrt{3}}+\frac{77
C_{00,22}^{62,0}}{256 \sqrt{15}}+\frac{7 C_{11,00}^{51,0}}{2304}-\frac{91 C_{11,22}^{51,0}}{576 \sqrt{15}}-\frac{3 \sqrt{7} C_{11,22}^{53,0}}{2560}+\frac{7
C_{20,00}^{40,0}}{2304}+\frac{7 C_{20,20}^{40,0}}{2304 \sqrt{3}}+\frac{7 C_{20,22}^{42,0}}{288 \sqrt{15}}-\frac{7 C_{22,00}^{42,0}}{2304 \sqrt{5}}-\frac{7
C_{22,20}^{42,0}}{2304 \sqrt{15}}+\frac{7 \sqrt{\frac{5}{3}} C_{22,22}^{40,0}}{2304}-\frac{13 \sqrt{\frac{7}{15}} C_{22,22}^{42,0}}{1152}-\frac{\sqrt{3}
C_{22,22}^{44,0}}{32}-\frac{7 C_{31,00}^{31,0}}{2304}-\frac{7 C_{31,20}^{31,0}}{2304 \sqrt{3}}-\frac{7 C_{31,22}^{31,0}}{288 \sqrt{15}}+\frac{89
\sqrt{7} C_{31,22}^{33,0}}{7680}+\frac{\sqrt{21} C_{33,00}^{33,0}}{1280}+\frac{\sqrt{7} C_{33,20}^{33,0}}{1280}+\frac{1}{40} \sqrt{\frac{21}{10}}
C_{33,22}^{33,0}\)

\noindent\(C_{11,2211,2}^{11,2211,0,\text{Loc}}= -\frac{77 C_{00,20}^{60,0}}{1152 \sqrt{5}}-\frac{77 C_{00,20}^{60,1}}{2304 \sqrt{5}}-\frac{91
C_{00,22}^{62,0}}{3840}-\frac{91 C_{00,22}^{62,1}}{7680}-\frac{539 C_{11,22}^{51,0}}{34560}-\frac{539 C_{11,22}^{51,1}}{69120}+\frac{1}{640} \sqrt{\frac{21}{5}}
C_{11,22}^{53,0}+\frac{\sqrt{\frac{21}{5}} C_{11,22}^{53,1}}{1280}-\frac{43 C_{20,20}^{40,0}}{3456 \sqrt{5}}-\frac{43 C_{20,20}^{40,1}}{6912 \sqrt{5}}+\frac{119
C_{20,22}^{42,0}}{34560}+\frac{119 C_{20,22}^{42,1}}{69120}+\frac{133 C_{22,20}^{42,0}}{17280}+\frac{133 C_{22,20}^{42,1}}{34560}-\frac{271 C_{22,22}^{40,0}}{34560}-\frac{271
C_{22,22}^{40,1}}{69120}+\frac{23 \sqrt{7} C_{22,22}^{42,0}}{17280}+\frac{23 \sqrt{7} C_{22,22}^{42,1}}{34560}+\frac{C_{22,22}^{44,0}}{40 \sqrt{5}}+\frac{C_{22,22}^{44,1}}{80
\sqrt{5}}+\frac{17 C_{31,20}^{31,0}}{3456 \sqrt{5}}+\frac{17 C_{31,20}^{31,1}}{6912 \sqrt{5}}+\frac{25 C_{31,22}^{31,0}}{6912}+\frac{25 C_{31,22}^{31,1}}{13824}+\frac{1}{192}
\sqrt{\frac{7}{15}} C_{31,22}^{33,0}+\frac{1}{384} \sqrt{\frac{7}{15}} C_{31,22}^{33,1}+\frac{19}{640} \sqrt{\frac{7}{15}} C_{33,20}^{33,0}+\frac{19
\sqrt{\frac{7}{15}} C_{33,20}^{33,1}}{1280}+\frac{1}{160} \sqrt{\frac{7}{2}} C_{33,22}^{33,0}+\frac{1}{320} \sqrt{\frac{7}{2}} C_{33,22}^{33,1}\)

\noindent\(C_{11,2211,2}^{11,2211,0,\text{NonLoc}}= -\frac{77 C_{00,00}^{60,0}}{1536 \sqrt{15}}-\frac{77 C_{00,00}^{60,1}}{768 \sqrt{15}}-\frac{77
C_{00,20}^{60,0}}{4608 \sqrt{5}}-\frac{77 C_{00,20}^{60,1}}{2304 \sqrt{5}}+\frac{91 C_{00,22}^{62,0}}{7680}+\frac{91 C_{00,22}^{62,1}}{3840}+\frac{77
C_{11,00}^{51,0}}{4608 \sqrt{15}}+\frac{77 C_{11,00}^{51,1}}{2304 \sqrt{15}}-\frac{539 C_{11,22}^{51,0}}{69120}-\frac{539 C_{11,22}^{51,1}}{34560}+\frac{\sqrt{\frac{21}{5}}
C_{11,22}^{53,0}}{1280}+\frac{1}{640} \sqrt{\frac{21}{5}} C_{11,22}^{53,1}-\frac{43 C_{20,00}^{40,0}}{4608 \sqrt{15}}-\frac{43 C_{20,00}^{40,1}}{2304
\sqrt{15}}-\frac{43 C_{20,20}^{40,0}}{13824 \sqrt{5}}-\frac{43 C_{20,20}^{40,1}}{6912 \sqrt{5}}-\frac{119 C_{20,22}^{42,0}}{69120}-\frac{119 C_{20,22}^{42,1}}{34560}+\frac{133
C_{22,00}^{42,0}}{23040 \sqrt{3}}+\frac{133 C_{22,00}^{42,1}}{11520 \sqrt{3}}+\frac{133 C_{22,20}^{42,0}}{69120}+\frac{133 C_{22,20}^{42,1}}{34560}+\frac{271
C_{22,22}^{40,0}}{69120}+\frac{271 C_{22,22}^{40,1}}{34560}-\frac{23 \sqrt{7} C_{22,22}^{42,0}}{34560}-\frac{23 \sqrt{7} C_{22,22}^{42,1}}{17280}-\frac{C_{22,22}^{44,0}}{80
\sqrt{5}}-\frac{C_{22,22}^{44,1}}{40 \sqrt{5}}-\frac{17 C_{31,00}^{31,0}}{4608 \sqrt{15}}-\frac{17 C_{31,00}^{31,1}}{2304 \sqrt{15}}-\frac{17 C_{31,20}^{31,0}}{13824
\sqrt{5}}-\frac{17 C_{31,20}^{31,1}}{6912 \sqrt{5}}+\frac{25 C_{31,22}^{31,0}}{13824}+\frac{25 C_{31,22}^{31,1}}{6912}+\frac{1}{384} \sqrt{\frac{7}{15}}
C_{31,22}^{33,0}+\frac{1}{192} \sqrt{\frac{7}{15}} C_{31,22}^{33,1}-\frac{19 \sqrt{\frac{7}{5}} C_{33,00}^{33,0}}{2560}-\frac{19 \sqrt{\frac{7}{5}}
C_{33,00}^{33,1}}{1280}-\frac{19 \sqrt{\frac{7}{15}} C_{33,20}^{33,0}}{2560}-\frac{19 \sqrt{\frac{7}{15}} C_{33,20}^{33,1}}{1280}+\frac{1}{320} \sqrt{\frac{7}{2}}
C_{33,22}^{33,0}+\frac{1}{160} \sqrt{\frac{7}{2}} C_{33,22}^{33,1}\)

\noindent\(C_{11,2211,2}^{11,2211,1,\text{Loc}}= -\frac{77 C_{00,20}^{60,1}}{768 \sqrt{15}}-\frac{91 C_{00,22}^{62,1}}{2560 \sqrt{3}}-\frac{539
C_{11,22}^{51,1}}{23040 \sqrt{3}}+\frac{3 \sqrt{\frac{7}{5}} C_{11,22}^{53,1}}{1280}-\frac{43 C_{20,20}^{40,1}}{2304 \sqrt{15}}+\frac{119 C_{20,22}^{42,1}}{23040
\sqrt{3}}+\frac{133 C_{22,20}^{42,1}}{11520 \sqrt{3}}-\frac{271 C_{22,22}^{40,1}}{23040 \sqrt{3}}+\frac{23 \sqrt{\frac{7}{3}} C_{22,22}^{42,1}}{11520}+\frac{1}{80}
\sqrt{\frac{3}{5}} C_{22,22}^{44,1}+\frac{17 C_{31,20}^{31,1}}{2304 \sqrt{15}}+\frac{25 C_{31,22}^{31,1}}{4608 \sqrt{3}}+\frac{1}{384} \sqrt{\frac{7}{5}}
C_{31,22}^{33,1}+\frac{19 \sqrt{\frac{7}{5}} C_{33,20}^{33,1}}{1280}+\frac{1}{320} \sqrt{\frac{21}{2}} C_{33,22}^{33,1}\)

\noindent\(C_{11,2211,2}^{11,2211,1,\text{NonLoc}}= -\frac{77 C_{00,00}^{60,0}}{1536 \sqrt{5}}-\frac{77 C_{00,20}^{60,0}}{1536 \sqrt{15}}+\frac{91
C_{00,22}^{62,0}}{2560 \sqrt{3}}+\frac{77 C_{11,00}^{51,0}}{4608 \sqrt{5}}-\frac{539 C_{11,22}^{51,0}}{23040 \sqrt{3}}+\frac{3 \sqrt{\frac{7}{5}}
C_{11,22}^{53,0}}{1280}-\frac{43 C_{20,00}^{40,0}}{4608 \sqrt{5}}-\frac{43 C_{20,20}^{40,0}}{4608 \sqrt{15}}-\frac{119 C_{20,22}^{42,0}}{23040 \sqrt{3}}+\frac{133
C_{22,00}^{42,0}}{23040}+\frac{133 C_{22,20}^{42,0}}{23040 \sqrt{3}}+\frac{271 C_{22,22}^{40,0}}{23040 \sqrt{3}}-\frac{23 \sqrt{\frac{7}{3}} C_{22,22}^{42,0}}{11520}-\frac{1}{80}
\sqrt{\frac{3}{5}} C_{22,22}^{44,0}-\frac{17 C_{31,00}^{31,0}}{4608 \sqrt{5}}-\frac{17 C_{31,20}^{31,0}}{4608 \sqrt{15}}+\frac{25 C_{31,22}^{31,0}}{4608
\sqrt{3}}+\frac{1}{384} \sqrt{\frac{7}{5}} C_{31,22}^{33,0}-\frac{19 \sqrt{\frac{21}{5}} C_{33,00}^{33,0}}{2560}-\frac{19 \sqrt{\frac{7}{5}} C_{33,20}^{33,0}}{2560}+\frac{1}{320}
\sqrt{\frac{21}{2}} C_{33,22}^{33,0}\)

\noindent\(C_{11,2212,1}^{00,3101,0,\text{Loc}}= \frac{7 C_{11,11}^{51,0}}{32 \sqrt{15}}+\frac{7 C_{11,11}^{51,1}}{64 \sqrt{15}}+\frac{7
C_{22,11}^{42,0}}{80}+\frac{7 C_{22,11}^{42,1}}{160}+\frac{7 C_{31,11}^{31,0}}{32 \sqrt{15}}+\frac{7 C_{31,11}^{31,1}}{64 \sqrt{15}}+\frac{3}{40}
\sqrt{\frac{21}{10}} C_{33,11}^{33,0}+\frac{3}{80} \sqrt{\frac{21}{10}} C_{33,11}^{33,1}\)

\noindent\(C_{11,2212,1}^{00,3101,0,\text{NonLoc}}= \frac{7 C_{11,11}^{51,0}}{64 \sqrt{15}}+\frac{7 C_{11,11}^{51,1}}{32 \sqrt{15}}-\frac{7
C_{22,11}^{42,0}}{160}-\frac{7 C_{22,11}^{42,1}}{80}+\frac{7 C_{31,11}^{31,0}}{64 \sqrt{15}}+\frac{7 C_{31,11}^{31,1}}{32 \sqrt{15}}+\frac{3}{80}
\sqrt{\frac{21}{10}} C_{33,11}^{33,0}+\frac{3}{40} \sqrt{\frac{21}{10}} C_{33,11}^{33,1}\)

\noindent\(C_{11,2212,1}^{00,3101,1,\text{Loc}}= \frac{7 C_{11,11}^{51,1}}{64 \sqrt{5}}+\frac{7 \sqrt{3} C_{22,11}^{42,1}}{160}+\frac{7
C_{31,11}^{31,1}}{64 \sqrt{5}}+\frac{9}{80} \sqrt{\frac{7}{10}} C_{33,11}^{33,1}\)

\noindent\(C_{11,2212,1}^{00,3101,1,\text{NonLoc}}= \frac{7 C_{11,11}^{51,0}}{64 \sqrt{5}}-\frac{7 \sqrt{3} C_{22,11}^{42,0}}{160}+\frac{7
C_{31,11}^{31,0}}{64 \sqrt{5}}+\frac{9}{80} \sqrt{\frac{7}{10}} C_{33,11}^{33,0}\)

\noindent\(C_{11,2212,1}^{11,2011,0,\text{Loc}}= -\frac{7 \sqrt{5} C_{00,20}^{60,0}}{192}-\frac{7 \sqrt{5} C_{00,20}^{60,1}}{384}+\frac{7
C_{00,22}^{62,0}}{128}+\frac{7 C_{00,22}^{62,1}}{256}+\frac{7 C_{11,22}^{51,0}}{576}+\frac{7 C_{11,22}^{51,1}}{1152}+\frac{3}{256} \sqrt{\frac{21}{5}}
C_{11,22}^{53,0}+\frac{3}{512} \sqrt{\frac{21}{5}} C_{11,22}^{53,1}+\frac{7 \sqrt{5} C_{20,20}^{40,0}}{576}+\frac{7 \sqrt{5} C_{20,20}^{40,1}}{1152}-\frac{7
C_{20,22}^{42,0}}{576}-\frac{7 C_{20,22}^{42,1}}{1152}-\frac{7 C_{22,20}^{42,0}}{576}-\frac{7 C_{22,20}^{42,1}}{1152}-\frac{35 C_{22,22}^{40,0}}{1152}-\frac{35
C_{22,22}^{40,1}}{2304}+\frac{7 \sqrt{7} C_{22,22}^{42,0}}{576}+\frac{7 \sqrt{7} C_{22,22}^{42,1}}{1152}+\frac{7 \sqrt{5} C_{31,20}^{31,0}}{576}+\frac{7
\sqrt{5} C_{31,20}^{31,1}}{1152}-\frac{7 C_{31,22}^{31,0}}{576}-\frac{7 C_{31,22}^{31,1}}{1152}-\frac{3}{256} \sqrt{\frac{21}{5}} C_{31,22}^{33,0}-\frac{3}{512}
\sqrt{\frac{21}{5}} C_{31,22}^{33,1}-\frac{1}{64} \sqrt{\frac{21}{5}} C_{33,20}^{33,0}-\frac{1}{128} \sqrt{\frac{21}{5}} C_{33,20}^{33,1}+\frac{3}{80}
\sqrt{\frac{7}{2}} C_{33,22}^{33,0}+\frac{3}{160} \sqrt{\frac{7}{2}} C_{33,22}^{33,1}\)

\noindent\(C_{11,2212,1}^{11,2011,0,\text{NonLoc}}= -\frac{7}{256} \sqrt{\frac{5}{3}} C_{00,00}^{60,0}-\frac{7}{128} \sqrt{\frac{5}{3}}
C_{00,00}^{60,1}-\frac{7 \sqrt{5} C_{00,20}^{60,0}}{768}-\frac{7 \sqrt{5} C_{00,20}^{60,1}}{384}-\frac{7 C_{00,22}^{62,0}}{256}-\frac{7 C_{00,22}^{62,1}}{128}+\frac{7}{768}
\sqrt{\frac{5}{3}} C_{11,00}^{51,0}+\frac{7}{384} \sqrt{\frac{5}{3}} C_{11,00}^{51,1}+\frac{7 C_{11,22}^{51,0}}{1152}+\frac{7 C_{11,22}^{51,1}}{576}+\frac{3}{512}
\sqrt{\frac{21}{5}} C_{11,22}^{53,0}+\frac{3}{256} \sqrt{\frac{21}{5}} C_{11,22}^{53,1}+\frac{7}{768} \sqrt{\frac{5}{3}} C_{20,00}^{40,0}+\frac{7}{384}
\sqrt{\frac{5}{3}} C_{20,00}^{40,1}+\frac{7 \sqrt{5} C_{20,20}^{40,0}}{2304}+\frac{7 \sqrt{5} C_{20,20}^{40,1}}{1152}+\frac{7 C_{20,22}^{42,0}}{1152}+\frac{7
C_{20,22}^{42,1}}{576}-\frac{7 C_{22,00}^{42,0}}{768 \sqrt{3}}-\frac{7 C_{22,00}^{42,1}}{384 \sqrt{3}}-\frac{7 C_{22,20}^{42,0}}{2304}-\frac{7 C_{22,20}^{42,1}}{1152}+\frac{35
C_{22,22}^{40,0}}{2304}+\frac{35 C_{22,22}^{40,1}}{1152}-\frac{7 \sqrt{7} C_{22,22}^{42,0}}{1152}-\frac{7 \sqrt{7} C_{22,22}^{42,1}}{576}-\frac{7}{768}
\sqrt{\frac{5}{3}} C_{31,00}^{31,0}-\frac{7}{384} \sqrt{\frac{5}{3}} C_{31,00}^{31,1}-\frac{7 \sqrt{5} C_{31,20}^{31,0}}{2304}-\frac{7 \sqrt{5} C_{31,20}^{31,1}}{1152}-\frac{7
C_{31,22}^{31,0}}{1152}-\frac{7 C_{31,22}^{31,1}}{576}-\frac{3}{512} \sqrt{\frac{21}{5}} C_{31,22}^{33,0}-\frac{3}{256} \sqrt{\frac{21}{5}} C_{31,22}^{33,1}+\frac{3}{256}
\sqrt{\frac{7}{5}} C_{33,00}^{33,0}+\frac{3}{128} \sqrt{\frac{7}{5}} C_{33,00}^{33,1}+\frac{1}{256} \sqrt{\frac{21}{5}} C_{33,20}^{33,0}+\frac{1}{128}
\sqrt{\frac{21}{5}} C_{33,20}^{33,1}+\frac{3}{160} \sqrt{\frac{7}{2}} C_{33,22}^{33,0}+\frac{3}{80} \sqrt{\frac{7}{2}} C_{33,22}^{33,1}\)

\noindent\(C_{11,2212,1}^{11,2011,1,\text{Loc}}= -\frac{7}{128} \sqrt{\frac{5}{3}} C_{00,20}^{60,1}+\frac{7 \sqrt{3} C_{00,22}^{62,1}}{256}+\frac{7
C_{11,22}^{51,1}}{384 \sqrt{3}}+\frac{9}{512} \sqrt{\frac{7}{5}} C_{11,22}^{53,1}+\frac{7}{384} \sqrt{\frac{5}{3}} C_{20,20}^{40,1}-\frac{7 C_{20,22}^{42,1}}{384
\sqrt{3}}-\frac{7 C_{22,20}^{42,1}}{384 \sqrt{3}}-\frac{35 C_{22,22}^{40,1}}{768 \sqrt{3}}+\frac{7}{384} \sqrt{\frac{7}{3}} C_{22,22}^{42,1}+\frac{7}{384}
\sqrt{\frac{5}{3}} C_{31,20}^{31,1}-\frac{7 C_{31,22}^{31,1}}{384 \sqrt{3}}-\frac{9}{512} \sqrt{\frac{7}{5}} C_{31,22}^{33,1}-\frac{3}{128} \sqrt{\frac{7}{5}}
C_{33,20}^{33,1}+\frac{3}{160} \sqrt{\frac{21}{2}} C_{33,22}^{33,1}\)

\noindent\(C_{11,2212,1}^{11,2011,1,\text{NonLoc}}= -\frac{7 \sqrt{5} C_{00,00}^{60,0}}{256}-\frac{7}{256} \sqrt{\frac{5}{3}} C_{00,20}^{60,0}-\frac{7
\sqrt{3} C_{00,22}^{62,0}}{256}+\frac{7 \sqrt{5} C_{11,00}^{51,0}}{768}+\frac{7 C_{11,22}^{51,0}}{384 \sqrt{3}}+\frac{9}{512} \sqrt{\frac{7}{5}}
C_{11,22}^{53,0}+\frac{7 \sqrt{5} C_{20,00}^{40,0}}{768}+\frac{7}{768} \sqrt{\frac{5}{3}} C_{20,20}^{40,0}+\frac{7 C_{20,22}^{42,0}}{384 \sqrt{3}}-\frac{7
C_{22,00}^{42,0}}{768}-\frac{7 C_{22,20}^{42,0}}{768 \sqrt{3}}+\frac{35 C_{22,22}^{40,0}}{768 \sqrt{3}}-\frac{7}{384} \sqrt{\frac{7}{3}} C_{22,22}^{42,0}-\frac{7
\sqrt{5} C_{31,00}^{31,0}}{768}-\frac{7}{768} \sqrt{\frac{5}{3}} C_{31,20}^{31,0}-\frac{7 C_{31,22}^{31,0}}{384 \sqrt{3}}-\frac{9}{512} \sqrt{\frac{7}{5}}
C_{31,22}^{33,0}+\frac{3}{256} \sqrt{\frac{21}{5}} C_{33,00}^{33,0}+\frac{3}{256} \sqrt{\frac{7}{5}} C_{33,20}^{33,0}+\frac{3}{160} \sqrt{\frac{21}{2}}
C_{33,22}^{33,0}\)

\noindent\(C_{11,2212,1}^{11,2211,0,\text{Loc}}= \frac{7 C_{00,20}^{60,0}}{192}+\frac{7 C_{00,20}^{60,1}}{384}-\frac{7 C_{00,22}^{62,0}}{128
\sqrt{5}}-\frac{7 C_{00,22}^{62,1}}{256 \sqrt{5}}-\frac{77 C_{11,22}^{51,0}}{1152 \sqrt{5}}-\frac{77 C_{11,22}^{51,1}}{2304 \sqrt{5}}+\frac{3 \sqrt{21}
C_{11,22}^{53,0}}{640}+\frac{3 \sqrt{21} C_{11,22}^{53,1}}{1280}-\frac{7 C_{20,20}^{40,0}}{576}-\frac{7 C_{20,20}^{40,1}}{1152}+\frac{77 C_{20,22}^{42,0}}{1152
\sqrt{5}}+\frac{77 C_{20,22}^{42,1}}{2304 \sqrt{5}}+\frac{7 C_{22,20}^{42,0}}{576 \sqrt{5}}+\frac{7 C_{22,20}^{42,1}}{1152 \sqrt{5}}-\frac{11 \sqrt{5}
C_{22,22}^{40,0}}{1152}-\frac{11 \sqrt{5} C_{22,22}^{40,1}}{2304}-\frac{7}{576} \sqrt{\frac{7}{5}} C_{22,22}^{42,0}-\frac{7 \sqrt{\frac{7}{5}} C_{22,22}^{42,1}}{1152}-\frac{7
C_{31,20}^{31,0}}{576}-\frac{7 C_{31,20}^{31,1}}{1152}+\frac{59 C_{31,22}^{31,0}}{1152 \sqrt{5}}+\frac{59 C_{31,22}^{31,1}}{2304 \sqrt{5}}-\frac{\sqrt{21}
C_{31,22}^{33,0}}{640}-\frac{\sqrt{21} C_{31,22}^{33,1}}{1280}+\frac{\sqrt{21} C_{33,20}^{33,0}}{320}+\frac{\sqrt{21} C_{33,20}^{33,1}}{640}-\frac{3}{80}
\sqrt{\frac{7}{10}} C_{33,22}^{33,0}-\frac{3}{160} \sqrt{\frac{7}{10}} C_{33,22}^{33,1}\)

\noindent\(C_{11,2212,1}^{11,2211,0,\text{NonLoc}}= \frac{7 C_{00,00}^{60,0}}{256 \sqrt{3}}+\frac{7 C_{00,00}^{60,1}}{128 \sqrt{3}}+\frac{7
C_{00,20}^{60,0}}{768}+\frac{7 C_{00,20}^{60,1}}{384}+\frac{7 C_{00,22}^{62,0}}{256 \sqrt{5}}+\frac{7 C_{00,22}^{62,1}}{128 \sqrt{5}}-\frac{7 C_{11,00}^{51,0}}{768
\sqrt{3}}-\frac{7 C_{11,00}^{51,1}}{384 \sqrt{3}}-\frac{77 C_{11,22}^{51,0}}{2304 \sqrt{5}}-\frac{77 C_{11,22}^{51,1}}{1152 \sqrt{5}}+\frac{3 \sqrt{21}
C_{11,22}^{53,0}}{1280}+\frac{3 \sqrt{21} C_{11,22}^{53,1}}{640}-\frac{7 C_{20,00}^{40,0}}{768 \sqrt{3}}-\frac{7 C_{20,00}^{40,1}}{384 \sqrt{3}}-\frac{7
C_{20,20}^{40,0}}{2304}-\frac{7 C_{20,20}^{40,1}}{1152}-\frac{77 C_{20,22}^{42,0}}{2304 \sqrt{5}}-\frac{77 C_{20,22}^{42,1}}{1152 \sqrt{5}}+\frac{7
C_{22,00}^{42,0}}{768 \sqrt{15}}+\frac{7 C_{22,00}^{42,1}}{384 \sqrt{15}}+\frac{7 C_{22,20}^{42,0}}{2304 \sqrt{5}}+\frac{7 C_{22,20}^{42,1}}{1152
\sqrt{5}}+\frac{11 \sqrt{5} C_{22,22}^{40,0}}{2304}+\frac{11 \sqrt{5} C_{22,22}^{40,1}}{1152}+\frac{7 \sqrt{\frac{7}{5}} C_{22,22}^{42,0}}{1152}+\frac{7}{576}
\sqrt{\frac{7}{5}} C_{22,22}^{42,1}+\frac{7 C_{31,00}^{31,0}}{768 \sqrt{3}}+\frac{7 C_{31,00}^{31,1}}{384 \sqrt{3}}+\frac{7 C_{31,20}^{31,0}}{2304}+\frac{7
C_{31,20}^{31,1}}{1152}+\frac{59 C_{31,22}^{31,0}}{2304 \sqrt{5}}+\frac{59 C_{31,22}^{31,1}}{1152 \sqrt{5}}-\frac{\sqrt{21} C_{31,22}^{33,0}}{1280}-\frac{\sqrt{21}
C_{31,22}^{33,1}}{640}-\frac{3 \sqrt{7} C_{33,00}^{33,0}}{1280}-\frac{3 \sqrt{7} C_{33,00}^{33,1}}{640}-\frac{\sqrt{21} C_{33,20}^{33,0}}{1280}-\frac{\sqrt{21}
C_{33,20}^{33,1}}{640}-\frac{3}{160} \sqrt{\frac{7}{10}} C_{33,22}^{33,0}-\frac{3}{80} \sqrt{\frac{7}{10}} C_{33,22}^{33,1}\)

\noindent\(C_{11,2212,1}^{11,2211,1,\text{Loc}}= \frac{7 C_{00,20}^{60,1}}{128 \sqrt{3}}-\frac{7}{256} \sqrt{\frac{3}{5}} C_{00,22}^{62,1}-\frac{77
C_{11,22}^{51,1}}{768 \sqrt{15}}+\frac{9 \sqrt{7} C_{11,22}^{53,1}}{1280}-\frac{7 C_{20,20}^{40,1}}{384 \sqrt{3}}+\frac{77 C_{20,22}^{42,1}}{768
\sqrt{15}}+\frac{7 C_{22,20}^{42,1}}{384 \sqrt{15}}-\frac{11}{768} \sqrt{\frac{5}{3}} C_{22,22}^{40,1}-\frac{7}{384} \sqrt{\frac{7}{15}} C_{22,22}^{42,1}-\frac{7
C_{31,20}^{31,1}}{384 \sqrt{3}}+\frac{59 C_{31,22}^{31,1}}{768 \sqrt{15}}-\frac{3 \sqrt{7} C_{31,22}^{33,1}}{1280}+\frac{3 \sqrt{7} C_{33,20}^{33,1}}{640}-\frac{3}{160}
\sqrt{\frac{21}{10}} C_{33,22}^{33,1}\)

\noindent\(C_{11,2212,1}^{11,2211,1,\text{NonLoc}}= \frac{7 C_{00,00}^{60,0}}{256}+\frac{7 C_{00,20}^{60,0}}{256 \sqrt{3}}+\frac{7}{256}
\sqrt{\frac{3}{5}} C_{00,22}^{62,0}-\frac{7 C_{11,00}^{51,0}}{768}-\frac{77 C_{11,22}^{51,0}}{768 \sqrt{15}}+\frac{9 \sqrt{7} C_{11,22}^{53,0}}{1280}-\frac{7
C_{20,00}^{40,0}}{768}-\frac{7 C_{20,20}^{40,0}}{768 \sqrt{3}}-\frac{77 C_{20,22}^{42,0}}{768 \sqrt{15}}+\frac{7 C_{22,00}^{42,0}}{768 \sqrt{5}}+\frac{7
C_{22,20}^{42,0}}{768 \sqrt{15}}+\frac{11}{768} \sqrt{\frac{5}{3}} C_{22,22}^{40,0}+\frac{7}{384} \sqrt{\frac{7}{15}} C_{22,22}^{42,0}+\frac{7 C_{31,00}^{31,0}}{768}+\frac{7
C_{31,20}^{31,0}}{768 \sqrt{3}}+\frac{59 C_{31,22}^{31,0}}{768 \sqrt{15}}-\frac{3 \sqrt{7} C_{31,22}^{33,0}}{1280}-\frac{3 \sqrt{21} C_{33,00}^{33,0}}{1280}-\frac{3
\sqrt{7} C_{33,20}^{33,0}}{1280}-\frac{3}{160} \sqrt{\frac{21}{10}} C_{33,22}^{33,0}\)

\noindent\(C_{11,2212,1}^{11,2212,0,\text{Loc}}= -\frac{7 C_{00,20}^{60,0}}{128 \sqrt{3}}-\frac{7 C_{00,20}^{60,1}}{256 \sqrt{3}}+\frac{7}{256}
\sqrt{\frac{3}{5}} C_{00,22}^{62,0}+\frac{7}{512} \sqrt{\frac{3}{5}} C_{00,22}^{62,1}+\frac{7}{768} \sqrt{\frac{5}{3}} C_{11,22}^{51,0}+\frac{7 \sqrt{\frac{5}{3}}
C_{11,22}^{51,1}}{1536}-\frac{C_{20,20}^{40,0}}{384 \sqrt{3}}-\frac{C_{20,20}^{40,1}}{768 \sqrt{3}}-\frac{7 C_{20,22}^{42,0}}{768 \sqrt{15}}-\frac{7
C_{20,22}^{42,1}}{1536 \sqrt{15}}+\frac{7 C_{22,20}^{42,0}}{384 \sqrt{15}}+\frac{7 C_{22,20}^{42,1}}{768 \sqrt{15}}+\frac{7}{768} \sqrt{\frac{5}{3}}
C_{22,22}^{40,0}+\frac{7 \sqrt{\frac{5}{3}} C_{22,22}^{40,1}}{1536}-\frac{7}{384} \sqrt{\frac{7}{15}} C_{22,22}^{42,0}-\frac{7}{768} \sqrt{\frac{7}{15}}
C_{22,22}^{42,1}+\frac{C_{31,20}^{31,0}}{128 \sqrt{3}}+\frac{C_{31,20}^{31,1}}{256 \sqrt{3}}-\frac{7 C_{31,22}^{31,0}}{256 \sqrt{15}}-\frac{7 C_{31,22}^{31,1}}{512
\sqrt{15}}+\frac{\sqrt{7} C_{31,22}^{33,0}}{640}+\frac{\sqrt{7} C_{31,22}^{33,1}}{1280}+\frac{3 \sqrt{7} C_{33,20}^{33,0}}{640}+\frac{3 \sqrt{7}
C_{33,20}^{33,1}}{1280}-\frac{3}{160} \sqrt{\frac{21}{10}} C_{33,22}^{33,0}-\frac{3}{320} \sqrt{\frac{21}{10}} C_{33,22}^{33,1}\)

\noindent\(C_{11,2212,1}^{11,2212,0,\text{NonLoc}}= -\frac{7 C_{00,00}^{60,0}}{512}-\frac{7 C_{00,00}^{60,1}}{256}-\frac{7 C_{00,20}^{60,0}}{512
\sqrt{3}}-\frac{7 C_{00,20}^{60,1}}{256 \sqrt{3}}-\frac{7}{512} \sqrt{\frac{3}{5}} C_{00,22}^{62,0}-\frac{7}{256} \sqrt{\frac{3}{5}} C_{00,22}^{62,1}+\frac{7
C_{11,00}^{51,0}}{1536}+\frac{7 C_{11,00}^{51,1}}{768}+\frac{7 \sqrt{\frac{5}{3}} C_{11,22}^{51,0}}{1536}+\frac{7}{768} \sqrt{\frac{5}{3}} C_{11,22}^{51,1}-\frac{C_{20,00}^{40,0}}{1536}-\frac{C_{20,00}^{40,1}}{768}-\frac{C_{20,20}^{40,0}}{1536
\sqrt{3}}-\frac{C_{20,20}^{40,1}}{768 \sqrt{3}}+\frac{7 C_{20,22}^{42,0}}{1536 \sqrt{15}}+\frac{7 C_{20,22}^{42,1}}{768 \sqrt{15}}+\frac{7 C_{22,00}^{42,0}}{1536
\sqrt{5}}+\frac{7 C_{22,00}^{42,1}}{768 \sqrt{5}}+\frac{7 C_{22,20}^{42,0}}{1536 \sqrt{15}}+\frac{7 C_{22,20}^{42,1}}{768 \sqrt{15}}-\frac{7 \sqrt{\frac{5}{3}}
C_{22,22}^{40,0}}{1536}-\frac{7}{768} \sqrt{\frac{5}{3}} C_{22,22}^{40,1}+\frac{7}{768} \sqrt{\frac{7}{15}} C_{22,22}^{42,0}+\frac{7}{384} \sqrt{\frac{7}{15}}
C_{22,22}^{42,1}-\frac{C_{31,00}^{31,0}}{512}-\frac{C_{31,00}^{31,1}}{256}-\frac{C_{31,20}^{31,0}}{512 \sqrt{3}}-\frac{C_{31,20}^{31,1}}{256 \sqrt{3}}-\frac{7
C_{31,22}^{31,0}}{512 \sqrt{15}}-\frac{7 C_{31,22}^{31,1}}{256 \sqrt{15}}+\frac{\sqrt{7} C_{31,22}^{33,0}}{1280}+\frac{\sqrt{7} C_{31,22}^{33,1}}{640}-\frac{3
\sqrt{21} C_{33,00}^{33,0}}{2560}-\frac{3 \sqrt{21} C_{33,00}^{33,1}}{1280}-\frac{3 \sqrt{7} C_{33,20}^{33,0}}{2560}-\frac{3 \sqrt{7} C_{33,20}^{33,1}}{1280}-\frac{3}{320}
\sqrt{\frac{21}{10}} C_{33,22}^{33,0}-\frac{3}{160} \sqrt{\frac{21}{10}} C_{33,22}^{33,1}\)

\noindent\(C_{11,2212,1}^{11,2212,1,\text{Loc}}= -\frac{7 C_{00,20}^{60,1}}{256}+\frac{21 C_{00,22}^{62,1}}{512 \sqrt{5}}+\frac{7 \sqrt{5}
C_{11,22}^{51,1}}{1536}-\frac{C_{20,20}^{40,1}}{768}-\frac{7 C_{20,22}^{42,1}}{1536 \sqrt{5}}+\frac{7 C_{22,20}^{42,1}}{768 \sqrt{5}}+\frac{7 \sqrt{5}
C_{22,22}^{40,1}}{1536}-\frac{7}{768} \sqrt{\frac{7}{5}} C_{22,22}^{42,1}+\frac{C_{31,20}^{31,1}}{256}-\frac{7 C_{31,22}^{31,1}}{512 \sqrt{5}}+\frac{\sqrt{21}
C_{31,22}^{33,1}}{1280}+\frac{3 \sqrt{21} C_{33,20}^{33,1}}{1280}-\frac{9}{320} \sqrt{\frac{7}{10}} C_{33,22}^{33,1}\)

\noindent\(C_{11,2212,1}^{11,2212,1,\text{NonLoc}}= -\frac{7 \sqrt{3} C_{00,00}^{60,0}}{512}-\frac{7 C_{00,20}^{60,0}}{512}-\frac{21
C_{00,22}^{62,0}}{512 \sqrt{5}}+\frac{7 C_{11,00}^{51,0}}{512 \sqrt{3}}+\frac{7 \sqrt{5} C_{11,22}^{51,0}}{1536}-\frac{C_{20,00}^{40,0}}{512 \sqrt{3}}-\frac{C_{20,20}^{40,0}}{1536}+\frac{7
C_{20,22}^{42,0}}{1536 \sqrt{5}}+\frac{7 C_{22,00}^{42,0}}{512 \sqrt{15}}+\frac{7 C_{22,20}^{42,0}}{1536 \sqrt{5}}-\frac{7 \sqrt{5} C_{22,22}^{40,0}}{1536}+\frac{7}{768}
\sqrt{\frac{7}{5}} C_{22,22}^{42,0}-\frac{\sqrt{3} C_{31,00}^{31,0}}{512}-\frac{C_{31,20}^{31,0}}{512}-\frac{7 C_{31,22}^{31,0}}{512 \sqrt{5}}+\frac{\sqrt{21}
C_{31,22}^{33,0}}{1280}-\frac{9 \sqrt{7} C_{33,00}^{33,0}}{2560}-\frac{3 \sqrt{21} C_{33,20}^{33,0}}{2560}-\frac{9}{320} \sqrt{\frac{7}{10}} C_{33,22}^{33,0}\)

\noindent\(C_{11,2212,2}^{11,2011,0,\text{Loc}}= \frac{7}{192} \sqrt{\frac{5}{3}} C_{00,20}^{60,0}+\frac{7}{384} \sqrt{\frac{5}{3}}
C_{00,20}^{60,1}-\frac{7 C_{00,22}^{62,0}}{128 \sqrt{3}}-\frac{7 C_{00,22}^{62,1}}{256 \sqrt{3}}-\frac{7 C_{11,22}^{51,0}}{576 \sqrt{3}}-\frac{7
C_{11,22}^{51,1}}{1152 \sqrt{3}}-\frac{3}{256} \sqrt{\frac{7}{5}} C_{11,22}^{53,0}-\frac{3}{512} \sqrt{\frac{7}{5}} C_{11,22}^{53,1}-\frac{7}{576}
\sqrt{\frac{5}{3}} C_{20,20}^{40,0}-\frac{7 \sqrt{\frac{5}{3}} C_{20,20}^{40,1}}{1152}+\frac{7 C_{20,22}^{42,0}}{576 \sqrt{3}}+\frac{7 C_{20,22}^{42,1}}{1152
\sqrt{3}}+\frac{7 C_{22,20}^{42,0}}{576 \sqrt{3}}+\frac{7 C_{22,20}^{42,1}}{1152 \sqrt{3}}+\frac{35 C_{22,22}^{40,0}}{1152 \sqrt{3}}+\frac{35 C_{22,22}^{40,1}}{2304
\sqrt{3}}-\frac{7}{576} \sqrt{\frac{7}{3}} C_{22,22}^{42,0}-\frac{7 \sqrt{\frac{7}{3}} C_{22,22}^{42,1}}{1152}-\frac{7}{576} \sqrt{\frac{5}{3}} C_{31,20}^{31,0}-\frac{7
\sqrt{\frac{5}{3}} C_{31,20}^{31,1}}{1152}+\frac{7 C_{31,22}^{31,0}}{576 \sqrt{3}}+\frac{7 C_{31,22}^{31,1}}{1152 \sqrt{3}}+\frac{3}{256} \sqrt{\frac{7}{5}}
C_{31,22}^{33,0}+\frac{3}{512} \sqrt{\frac{7}{5}} C_{31,22}^{33,1}+\frac{1}{64} \sqrt{\frac{7}{5}} C_{33,20}^{33,0}+\frac{1}{128} \sqrt{\frac{7}{5}}
C_{33,20}^{33,1}-\frac{1}{80} \sqrt{\frac{21}{2}} C_{33,22}^{33,0}-\frac{1}{160} \sqrt{\frac{21}{2}} C_{33,22}^{33,1}\)

\noindent\(C_{11,2212,2}^{11,2011,0,\text{NonLoc}}= \frac{7 \sqrt{5} C_{00,00}^{60,0}}{768}+\frac{7 \sqrt{5} C_{00,00}^{60,1}}{384}+\frac{7}{768}
\sqrt{\frac{5}{3}} C_{00,20}^{60,0}+\frac{7}{384} \sqrt{\frac{5}{3}} C_{00,20}^{60,1}+\frac{7 C_{00,22}^{62,0}}{256 \sqrt{3}}+\frac{7 C_{00,22}^{62,1}}{128
\sqrt{3}}-\frac{7 \sqrt{5} C_{11,00}^{51,0}}{2304}-\frac{7 \sqrt{5} C_{11,00}^{51,1}}{1152}-\frac{7 C_{11,22}^{51,0}}{1152 \sqrt{3}}-\frac{7 C_{11,22}^{51,1}}{576
\sqrt{3}}-\frac{3}{512} \sqrt{\frac{7}{5}} C_{11,22}^{53,0}-\frac{3}{256} \sqrt{\frac{7}{5}} C_{11,22}^{53,1}-\frac{7 \sqrt{5} C_{20,00}^{40,0}}{2304}-\frac{7
\sqrt{5} C_{20,00}^{40,1}}{1152}-\frac{7 \sqrt{\frac{5}{3}} C_{20,20}^{40,0}}{2304}-\frac{7 \sqrt{\frac{5}{3}} C_{20,20}^{40,1}}{1152}-\frac{7 C_{20,22}^{42,0}}{1152
\sqrt{3}}-\frac{7 C_{20,22}^{42,1}}{576 \sqrt{3}}+\frac{7 C_{22,00}^{42,0}}{2304}+\frac{7 C_{22,00}^{42,1}}{1152}+\frac{7 C_{22,20}^{42,0}}{2304
\sqrt{3}}+\frac{7 C_{22,20}^{42,1}}{1152 \sqrt{3}}-\frac{35 C_{22,22}^{40,0}}{2304 \sqrt{3}}-\frac{35 C_{22,22}^{40,1}}{1152 \sqrt{3}}+\frac{7 \sqrt{\frac{7}{3}}
C_{22,22}^{42,0}}{1152}+\frac{7}{576} \sqrt{\frac{7}{3}} C_{22,22}^{42,1}+\frac{7 \sqrt{5} C_{31,00}^{31,0}}{2304}+\frac{7 \sqrt{5} C_{31,00}^{31,1}}{1152}+\frac{7
\sqrt{\frac{5}{3}} C_{31,20}^{31,0}}{2304}+\frac{7 \sqrt{\frac{5}{3}} C_{31,20}^{31,1}}{1152}+\frac{7 C_{31,22}^{31,0}}{1152 \sqrt{3}}+\frac{7 C_{31,22}^{31,1}}{576
\sqrt{3}}+\frac{3}{512} \sqrt{\frac{7}{5}} C_{31,22}^{33,0}+\frac{3}{256} \sqrt{\frac{7}{5}} C_{31,22}^{33,1}-\frac{1}{256} \sqrt{\frac{21}{5}} C_{33,00}^{33,0}-\frac{1}{128}
\sqrt{\frac{21}{5}} C_{33,00}^{33,1}-\frac{1}{256} \sqrt{\frac{7}{5}} C_{33,20}^{33,0}-\frac{1}{128} \sqrt{\frac{7}{5}} C_{33,20}^{33,1}-\frac{1}{160}
\sqrt{\frac{21}{2}} C_{33,22}^{33,0}-\frac{1}{80} \sqrt{\frac{21}{2}} C_{33,22}^{33,1}\)

\noindent\(C_{11,2212,2}^{11,2011,1,\text{Loc}}= \frac{7 \sqrt{5} C_{00,20}^{60,1}}{384}-\frac{7 C_{00,22}^{62,1}}{256}-\frac{7 C_{11,22}^{51,1}}{1152}-\frac{3}{512}
\sqrt{\frac{21}{5}} C_{11,22}^{53,1}-\frac{7 \sqrt{5} C_{20,20}^{40,1}}{1152}+\frac{7 C_{20,22}^{42,1}}{1152}+\frac{7 C_{22,20}^{42,1}}{1152}+\frac{35
C_{22,22}^{40,1}}{2304}-\frac{7 \sqrt{7} C_{22,22}^{42,1}}{1152}-\frac{7 \sqrt{5} C_{31,20}^{31,1}}{1152}+\frac{7 C_{31,22}^{31,1}}{1152}+\frac{3}{512}
\sqrt{\frac{21}{5}} C_{31,22}^{33,1}+\frac{1}{128} \sqrt{\frac{21}{5}} C_{33,20}^{33,1}-\frac{3}{160} \sqrt{\frac{7}{2}} C_{33,22}^{33,1}\)

\noindent\(C_{11,2212,2}^{11,2011,1,\text{NonLoc}}= \frac{7}{256} \sqrt{\frac{5}{3}} C_{00,00}^{60,0}+\frac{7 \sqrt{5} C_{00,20}^{60,0}}{768}+\frac{7
C_{00,22}^{62,0}}{256}-\frac{7}{768} \sqrt{\frac{5}{3}} C_{11,00}^{51,0}-\frac{7 C_{11,22}^{51,0}}{1152}-\frac{3}{512} \sqrt{\frac{21}{5}} C_{11,22}^{53,0}-\frac{7}{768}
\sqrt{\frac{5}{3}} C_{20,00}^{40,0}-\frac{7 \sqrt{5} C_{20,20}^{40,0}}{2304}-\frac{7 C_{20,22}^{42,0}}{1152}+\frac{7 C_{22,00}^{42,0}}{768 \sqrt{3}}+\frac{7
C_{22,20}^{42,0}}{2304}-\frac{35 C_{22,22}^{40,0}}{2304}+\frac{7 \sqrt{7} C_{22,22}^{42,0}}{1152}+\frac{7}{768} \sqrt{\frac{5}{3}} C_{31,00}^{31,0}+\frac{7
\sqrt{5} C_{31,20}^{31,0}}{2304}+\frac{7 C_{31,22}^{31,0}}{1152}+\frac{3}{512} \sqrt{\frac{21}{5}} C_{31,22}^{33,0}-\frac{3}{256} \sqrt{\frac{7}{5}}
C_{33,00}^{33,0}-\frac{1}{256} \sqrt{\frac{21}{5}} C_{33,20}^{33,0}-\frac{3}{160} \sqrt{\frac{7}{2}} C_{33,22}^{33,0}\)

\noindent\(C_{11,2212,2}^{11,2211,0,\text{Loc}}= -\frac{7 C_{00,20}^{60,0}}{192 \sqrt{3}}-\frac{7 C_{00,20}^{60,1}}{384 \sqrt{3}}+\frac{7
C_{00,22}^{62,0}}{128 \sqrt{15}}+\frac{7 C_{00,22}^{62,1}}{256 \sqrt{15}}+\frac{77 C_{11,22}^{51,0}}{1152 \sqrt{15}}+\frac{77 C_{11,22}^{51,1}}{2304
\sqrt{15}}-\frac{3 \sqrt{7} C_{11,22}^{53,0}}{640}-\frac{3 \sqrt{7} C_{11,22}^{53,1}}{1280}+\frac{7 C_{20,20}^{40,0}}{576 \sqrt{3}}+\frac{7 C_{20,20}^{40,1}}{1152
\sqrt{3}}-\frac{77 C_{20,22}^{42,0}}{1152 \sqrt{15}}-\frac{77 C_{20,22}^{42,1}}{2304 \sqrt{15}}-\frac{7 C_{22,20}^{42,0}}{576 \sqrt{15}}-\frac{7
C_{22,20}^{42,1}}{1152 \sqrt{15}}+\frac{11 \sqrt{\frac{5}{3}} C_{22,22}^{40,0}}{1152}+\frac{11 \sqrt{\frac{5}{3}} C_{22,22}^{40,1}}{2304}+\frac{7}{576}
\sqrt{\frac{7}{15}} C_{22,22}^{42,0}+\frac{7 \sqrt{\frac{7}{15}} C_{22,22}^{42,1}}{1152}+\frac{7 C_{31,20}^{31,0}}{576 \sqrt{3}}+\frac{7 C_{31,20}^{31,1}}{1152
\sqrt{3}}-\frac{59 C_{31,22}^{31,0}}{1152 \sqrt{15}}-\frac{59 C_{31,22}^{31,1}}{2304 \sqrt{15}}+\frac{\sqrt{7} C_{31,22}^{33,0}}{640}+\frac{\sqrt{7}
C_{31,22}^{33,1}}{1280}-\frac{\sqrt{7} C_{33,20}^{33,0}}{320}-\frac{\sqrt{7} C_{33,20}^{33,1}}{640}+\frac{1}{80} \sqrt{\frac{21}{10}} C_{33,22}^{33,0}+\frac{1}{160}
\sqrt{\frac{21}{10}} C_{33,22}^{33,1}\)

\noindent\(C_{11,2212,2}^{11,2211,0,\text{NonLoc}}= -\frac{7 C_{00,00}^{60,0}}{768}-\frac{7 C_{00,00}^{60,1}}{384}-\frac{7 C_{00,20}^{60,0}}{768
\sqrt{3}}-\frac{7 C_{00,20}^{60,1}}{384 \sqrt{3}}-\frac{7 C_{00,22}^{62,0}}{256 \sqrt{15}}-\frac{7 C_{00,22}^{62,1}}{128 \sqrt{15}}+\frac{7 C_{11,00}^{51,0}}{2304}+\frac{7
C_{11,00}^{51,1}}{1152}+\frac{77 C_{11,22}^{51,0}}{2304 \sqrt{15}}+\frac{77 C_{11,22}^{51,1}}{1152 \sqrt{15}}-\frac{3 \sqrt{7} C_{11,22}^{53,0}}{1280}-\frac{3
\sqrt{7} C_{11,22}^{53,1}}{640}+\frac{7 C_{20,00}^{40,0}}{2304}+\frac{7 C_{20,00}^{40,1}}{1152}+\frac{7 C_{20,20}^{40,0}}{2304 \sqrt{3}}+\frac{7
C_{20,20}^{40,1}}{1152 \sqrt{3}}+\frac{77 C_{20,22}^{42,0}}{2304 \sqrt{15}}+\frac{77 C_{20,22}^{42,1}}{1152 \sqrt{15}}-\frac{7 C_{22,00}^{42,0}}{2304
\sqrt{5}}-\frac{7 C_{22,00}^{42,1}}{1152 \sqrt{5}}-\frac{7 C_{22,20}^{42,0}}{2304 \sqrt{15}}-\frac{7 C_{22,20}^{42,1}}{1152 \sqrt{15}}-\frac{11 \sqrt{\frac{5}{3}}
C_{22,22}^{40,0}}{2304}-\frac{11 \sqrt{\frac{5}{3}} C_{22,22}^{40,1}}{1152}-\frac{7 \sqrt{\frac{7}{15}} C_{22,22}^{42,0}}{1152}-\frac{7}{576} \sqrt{\frac{7}{15}}
C_{22,22}^{42,1}-\frac{7 C_{31,00}^{31,0}}{2304}-\frac{7 C_{31,00}^{31,1}}{1152}-\frac{7 C_{31,20}^{31,0}}{2304 \sqrt{3}}-\frac{7 C_{31,20}^{31,1}}{1152
\sqrt{3}}-\frac{59 C_{31,22}^{31,0}}{2304 \sqrt{15}}-\frac{59 C_{31,22}^{31,1}}{1152 \sqrt{15}}+\frac{\sqrt{7} C_{31,22}^{33,0}}{1280}+\frac{\sqrt{7}
C_{31,22}^{33,1}}{640}+\frac{\sqrt{21} C_{33,00}^{33,0}}{1280}+\frac{\sqrt{21} C_{33,00}^{33,1}}{640}+\frac{\sqrt{7} C_{33,20}^{33,0}}{1280}+\frac{\sqrt{7}
C_{33,20}^{33,1}}{640}+\frac{1}{160} \sqrt{\frac{21}{10}} C_{33,22}^{33,0}+\frac{1}{80} \sqrt{\frac{21}{10}} C_{33,22}^{33,1}\)

\noindent\(C_{11,2212,2}^{11,2211,1,\text{Loc}}= -\frac{7 C_{00,20}^{60,1}}{384}+\frac{7 C_{00,22}^{62,1}}{256 \sqrt{5}}+\frac{77 C_{11,22}^{51,1}}{2304
\sqrt{5}}-\frac{3 \sqrt{21} C_{11,22}^{53,1}}{1280}+\frac{7 C_{20,20}^{40,1}}{1152}-\frac{77 C_{20,22}^{42,1}}{2304 \sqrt{5}}-\frac{7 C_{22,20}^{42,1}}{1152
\sqrt{5}}+\frac{11 \sqrt{5} C_{22,22}^{40,1}}{2304}+\frac{7 \sqrt{\frac{7}{5}} C_{22,22}^{42,1}}{1152}+\frac{7 C_{31,20}^{31,1}}{1152}-\frac{59 C_{31,22}^{31,1}}{2304
\sqrt{5}}+\frac{\sqrt{21} C_{31,22}^{33,1}}{1280}-\frac{\sqrt{21} C_{33,20}^{33,1}}{640}+\frac{3}{160} \sqrt{\frac{7}{10}} C_{33,22}^{33,1}\)

\noindent\(C_{11,2212,2}^{11,2211,1,\text{NonLoc}}= -\frac{7 C_{00,00}^{60,0}}{256 \sqrt{3}}-\frac{7 C_{00,20}^{60,0}}{768}-\frac{7
C_{00,22}^{62,0}}{256 \sqrt{5}}+\frac{7 C_{11,00}^{51,0}}{768 \sqrt{3}}+\frac{77 C_{11,22}^{51,0}}{2304 \sqrt{5}}-\frac{3 \sqrt{21} C_{11,22}^{53,0}}{1280}+\frac{7
C_{20,00}^{40,0}}{768 \sqrt{3}}+\frac{7 C_{20,20}^{40,0}}{2304}+\frac{77 C_{20,22}^{42,0}}{2304 \sqrt{5}}-\frac{7 C_{22,00}^{42,0}}{768 \sqrt{15}}-\frac{7
C_{22,20}^{42,0}}{2304 \sqrt{5}}-\frac{11 \sqrt{5} C_{22,22}^{40,0}}{2304}-\frac{7 \sqrt{\frac{7}{5}} C_{22,22}^{42,0}}{1152}-\frac{7 C_{31,00}^{31,0}}{768
\sqrt{3}}-\frac{7 C_{31,20}^{31,0}}{2304}-\frac{59 C_{31,22}^{31,0}}{2304 \sqrt{5}}+\frac{\sqrt{21} C_{31,22}^{33,0}}{1280}+\frac{3 \sqrt{7} C_{33,00}^{33,0}}{1280}+\frac{\sqrt{21}
C_{33,20}^{33,0}}{1280}+\frac{3}{160} \sqrt{\frac{7}{10}} C_{33,22}^{33,0}\)

\noindent\(C_{11,2212,2}^{11,2212,0,\text{Loc}}= -\frac{7 \sqrt{5} C_{00,20}^{60,0}}{1152}-\frac{7 \sqrt{5} C_{00,20}^{60,1}}{2304}+\frac{7
C_{00,22}^{62,0}}{768}+\frac{7 C_{00,22}^{62,1}}{1536}+\frac{35 C_{11,22}^{51,0}}{6912}+\frac{35 C_{11,22}^{51,1}}{13824}-\frac{17 \sqrt{5} C_{20,20}^{40,0}}{3456}-\frac{17
\sqrt{5} C_{20,20}^{40,1}}{6912}+\frac{49 C_{20,22}^{42,0}}{6912}+\frac{49 C_{20,22}^{42,1}}{13824}+\frac{35 C_{22,20}^{42,0}}{3456}+\frac{35 C_{22,20}^{42,1}}{6912}-\frac{77
C_{22,22}^{40,0}}{6912}-\frac{77 C_{22,22}^{40,1}}{13824}+\frac{13 \sqrt{7} C_{22,22}^{42,0}}{3456}+\frac{13 \sqrt{7} C_{22,22}^{42,1}}{6912}-\frac{C_{22,22}^{44,0}}{16
\sqrt{5}}-\frac{C_{22,22}^{44,1}}{32 \sqrt{5}}-\frac{5 \sqrt{5} C_{31,20}^{31,0}}{3456}-\frac{5 \sqrt{5} C_{31,20}^{31,1}}{6912}+\frac{91 C_{31,22}^{31,0}}{6912}+\frac{91
C_{31,22}^{31,1}}{13824}-\frac{11}{384} \sqrt{\frac{7}{15}} C_{31,22}^{33,0}-\frac{11}{768} \sqrt{\frac{7}{15}} C_{31,22}^{33,1}+\frac{1}{128} \sqrt{\frac{35}{3}}
C_{33,20}^{33,0}+\frac{1}{256} \sqrt{\frac{35}{3}} C_{33,20}^{33,1}+\frac{1}{160} \sqrt{\frac{7}{2}} C_{33,22}^{33,0}+\frac{1}{320} \sqrt{\frac{7}{2}}
C_{33,22}^{33,1}\)

\noindent\(C_{11,2212,2}^{11,2212,0,\text{NonLoc}}= -\frac{7 \sqrt{\frac{5}{3}} C_{00,00}^{60,0}}{1536}-\frac{7}{768} \sqrt{\frac{5}{3}}
C_{00,00}^{60,1}-\frac{7 \sqrt{5} C_{00,20}^{60,0}}{4608}-\frac{7 \sqrt{5} C_{00,20}^{60,1}}{2304}-\frac{7 C_{00,22}^{62,0}}{1536}-\frac{7 C_{00,22}^{62,1}}{768}+\frac{7
\sqrt{\frac{5}{3}} C_{11,00}^{51,0}}{4608}+\frac{7 \sqrt{\frac{5}{3}} C_{11,00}^{51,1}}{2304}+\frac{35 C_{11,22}^{51,0}}{13824}+\frac{35 C_{11,22}^{51,1}}{6912}-\frac{17
\sqrt{\frac{5}{3}} C_{20,00}^{40,0}}{4608}-\frac{17 \sqrt{\frac{5}{3}} C_{20,00}^{40,1}}{2304}-\frac{17 \sqrt{5} C_{20,20}^{40,0}}{13824}-\frac{17
\sqrt{5} C_{20,20}^{40,1}}{6912}-\frac{49 C_{20,22}^{42,0}}{13824}-\frac{49 C_{20,22}^{42,1}}{6912}+\frac{35 C_{22,00}^{42,0}}{4608 \sqrt{3}}+\frac{35
C_{22,00}^{42,1}}{2304 \sqrt{3}}+\frac{35 C_{22,20}^{42,0}}{13824}+\frac{35 C_{22,20}^{42,1}}{6912}+\frac{77 C_{22,22}^{40,0}}{13824}+\frac{77 C_{22,22}^{40,1}}{6912}-\frac{13
\sqrt{7} C_{22,22}^{42,0}}{6912}-\frac{13 \sqrt{7} C_{22,22}^{42,1}}{3456}+\frac{C_{22,22}^{44,0}}{32 \sqrt{5}}+\frac{C_{22,22}^{44,1}}{16 \sqrt{5}}+\frac{5
\sqrt{\frac{5}{3}} C_{31,00}^{31,0}}{4608}+\frac{5 \sqrt{\frac{5}{3}} C_{31,00}^{31,1}}{2304}+\frac{5 \sqrt{5} C_{31,20}^{31,0}}{13824}+\frac{5 \sqrt{5}
C_{31,20}^{31,1}}{6912}+\frac{91 C_{31,22}^{31,0}}{13824}+\frac{91 C_{31,22}^{31,1}}{6912}-\frac{11}{768} \sqrt{\frac{7}{15}} C_{31,22}^{33,0}-\frac{11}{384}
\sqrt{\frac{7}{15}} C_{31,22}^{33,1}-\frac{\sqrt{35} C_{33,00}^{33,0}}{512}-\frac{\sqrt{35} C_{33,00}^{33,1}}{256}-\frac{1}{512} \sqrt{\frac{35}{3}}
C_{33,20}^{33,0}-\frac{1}{256} \sqrt{\frac{35}{3}} C_{33,20}^{33,1}+\frac{1}{320} \sqrt{\frac{7}{2}} C_{33,22}^{33,0}+\frac{1}{160} \sqrt{\frac{7}{2}}
C_{33,22}^{33,1}\)

\noindent\(C_{11,2212,2}^{11,2212,1,\text{Loc}}= -\frac{7}{768} \sqrt{\frac{5}{3}} C_{00,20}^{60,1}+\frac{7 C_{00,22}^{62,1}}{512 \sqrt{3}}+\frac{35
C_{11,22}^{51,1}}{4608 \sqrt{3}}-\frac{17 \sqrt{\frac{5}{3}} C_{20,20}^{40,1}}{2304}+\frac{49 C_{20,22}^{42,1}}{4608 \sqrt{3}}+\frac{35 C_{22,20}^{42,1}}{2304
\sqrt{3}}-\frac{77 C_{22,22}^{40,1}}{4608 \sqrt{3}}+\frac{13 \sqrt{\frac{7}{3}} C_{22,22}^{42,1}}{2304}-\frac{1}{32} \sqrt{\frac{3}{5}} C_{22,22}^{44,1}-\frac{5
\sqrt{\frac{5}{3}} C_{31,20}^{31,1}}{2304}+\frac{91 C_{31,22}^{31,1}}{4608 \sqrt{3}}-\frac{11}{768} \sqrt{\frac{7}{5}} C_{31,22}^{33,1}+\frac{\sqrt{35}
C_{33,20}^{33,1}}{256}+\frac{1}{320} \sqrt{\frac{21}{2}} C_{33,22}^{33,1}\)

\noindent\(C_{11,2212,2}^{11,2212,1,\text{NonLoc}}= -\frac{7 \sqrt{5} C_{00,00}^{60,0}}{1536}-\frac{7 \sqrt{\frac{5}{3}} C_{00,20}^{60,0}}{1536}-\frac{7
C_{00,22}^{62,0}}{512 \sqrt{3}}+\frac{7 \sqrt{5} C_{11,00}^{51,0}}{4608}+\frac{35 C_{11,22}^{51,0}}{4608 \sqrt{3}}-\frac{17 \sqrt{5} C_{20,00}^{40,0}}{4608}-\frac{17
\sqrt{\frac{5}{3}} C_{20,20}^{40,0}}{4608}-\frac{49 C_{20,22}^{42,0}}{4608 \sqrt{3}}+\frac{35 C_{22,00}^{42,0}}{4608}+\frac{35 C_{22,20}^{42,0}}{4608
\sqrt{3}}+\frac{77 C_{22,22}^{40,0}}{4608 \sqrt{3}}-\frac{13 \sqrt{\frac{7}{3}} C_{22,22}^{42,0}}{2304}+\frac{1}{32} \sqrt{\frac{3}{5}} C_{22,22}^{44,0}+\frac{5
\sqrt{5} C_{31,00}^{31,0}}{4608}+\frac{5 \sqrt{\frac{5}{3}} C_{31,20}^{31,0}}{4608}+\frac{91 C_{31,22}^{31,0}}{4608 \sqrt{3}}-\frac{11}{768} \sqrt{\frac{7}{5}}
C_{31,22}^{33,0}-\frac{\sqrt{105} C_{33,00}^{33,0}}{512}-\frac{\sqrt{35} C_{33,20}^{33,0}}{512}+\frac{1}{320} \sqrt{\frac{21}{2}} C_{33,22}^{33,0}\)

\noindent\(C_{11,2212,3}^{00,3303,0,\text{Loc}}= -\frac{\sqrt{7} C_{11,11}^{51,0}}{144}-\frac{\sqrt{7} C_{11,11}^{51,1}}{288}-\frac{1}{24}
\sqrt{\frac{7}{15}} C_{22,11}^{42,0}-\frac{1}{48} \sqrt{\frac{7}{15}} C_{22,11}^{42,1}-\frac{\sqrt{7} C_{31,11}^{31,0}}{144}-\frac{\sqrt{7} C_{31,11}^{31,1}}{288}-\frac{C_{33,11}^{33,0}}{20
\sqrt{2}}-\frac{C_{33,11}^{33,1}}{40 \sqrt{2}}\)

\noindent\(C_{11,2212,3}^{00,3303,0,\text{NonLoc}}= -\frac{\sqrt{7} C_{11,11}^{51,0}}{288}-\frac{\sqrt{7} C_{11,11}^{51,1}}{144}+\frac{1}{48}
\sqrt{\frac{7}{15}} C_{22,11}^{42,0}+\frac{1}{24} \sqrt{\frac{7}{15}} C_{22,11}^{42,1}-\frac{\sqrt{7} C_{31,11}^{31,0}}{288}-\frac{\sqrt{7} C_{31,11}^{31,1}}{144}-\frac{C_{33,11}^{33,0}}{40
\sqrt{2}}-\frac{C_{33,11}^{33,1}}{20 \sqrt{2}}\)

\noindent\(C_{11,2212,3}^{00,3303,1,\text{Loc}}= -\frac{1}{96} \sqrt{\frac{7}{3}} C_{11,11}^{51,1}-\frac{1}{48} \sqrt{\frac{7}{5}}
C_{22,11}^{42,1}-\frac{1}{96} \sqrt{\frac{7}{3}} C_{31,11}^{31,1}-\frac{1}{40} \sqrt{\frac{3}{2}} C_{33,11}^{33,1}\)

\noindent\(C_{11,2212,3}^{00,3303,1,\text{NonLoc}}= -\frac{1}{96} \sqrt{\frac{7}{3}} C_{11,11}^{51,0}+\frac{1}{48} \sqrt{\frac{7}{5}}
C_{22,11}^{42,0}-\frac{1}{96} \sqrt{\frac{7}{3}} C_{31,11}^{31,0}-\frac{1}{40} \sqrt{\frac{3}{2}} C_{33,11}^{33,0}\)

\noindent\(C_{11,2212,3}^{11,2212,0,\text{Loc}}= -\frac{\sqrt{7} C_{00,20}^{60,0}}{72}-\frac{\sqrt{7} C_{00,20}^{60,1}}{144}+\frac{1}{48}
\sqrt{\frac{7}{5}} C_{00,22}^{62,0}+\frac{1}{96} \sqrt{\frac{7}{5}} C_{00,22}^{62,1}+\frac{\sqrt{35} C_{11,22}^{51,0}}{432}+\frac{\sqrt{35} C_{11,22}^{51,1}}{864}-\frac{\sqrt{7}
C_{20,20}^{40,0}}{432}-\frac{\sqrt{7} C_{20,20}^{40,1}}{864}+\frac{\sqrt{\frac{7}{5}} C_{20,22}^{42,0}}{1728}+\frac{\sqrt{\frac{7}{5}} C_{20,22}^{42,1}}{3456}+\frac{13
\sqrt{\frac{7}{5}} C_{22,20}^{42,0}}{1728}+\frac{13 \sqrt{\frac{7}{5}} C_{22,20}^{42,1}}{3456}+\frac{\sqrt{35} C_{22,22}^{40,0}}{864}+\frac{\sqrt{35}
C_{22,22}^{40,1}}{1728}-\frac{31 C_{22,22}^{42,0}}{1728 \sqrt{5}}-\frac{31 C_{22,22}^{42,1}}{3456 \sqrt{5}}-\frac{C_{22,22}^{44,0}}{32 \sqrt{7}}-\frac{C_{22,22}^{44,1}}{64
\sqrt{7}}+\frac{\sqrt{7} C_{31,20}^{31,0}}{864}+\frac{\sqrt{7} C_{31,20}^{31,1}}{1728}-\frac{1}{864} \sqrt{\frac{7}{5}} C_{31,22}^{31,0}-\frac{\sqrt{\frac{7}{5}}
C_{31,22}^{31,1}}{1728}-\frac{7 C_{31,22}^{33,0}}{960 \sqrt{3}}-\frac{7 C_{31,22}^{33,1}}{1920 \sqrt{3}}+\frac{13 C_{33,20}^{33,0}}{320 \sqrt{3}}+\frac{13
C_{33,20}^{33,1}}{640 \sqrt{3}}-\frac{11 C_{33,22}^{33,0}}{160 \sqrt{10}}-\frac{11 C_{33,22}^{33,1}}{320 \sqrt{10}}\)

\noindent\(C_{11,2212,3}^{11,2212,0,\text{NonLoc}}= -\frac{1}{96} \sqrt{\frac{7}{3}} C_{00,00}^{60,0}-\frac{1}{48} \sqrt{\frac{7}{3}}
C_{00,00}^{60,1}-\frac{\sqrt{7} C_{00,20}^{60,0}}{288}-\frac{\sqrt{7} C_{00,20}^{60,1}}{144}-\frac{1}{96} \sqrt{\frac{7}{5}} C_{00,22}^{62,0}-\frac{1}{48}
\sqrt{\frac{7}{5}} C_{00,22}^{62,1}+\frac{1}{288} \sqrt{\frac{7}{3}} C_{11,00}^{51,0}+\frac{1}{144} \sqrt{\frac{7}{3}} C_{11,00}^{51,1}+\frac{\sqrt{35}
C_{11,22}^{51,0}}{864}+\frac{\sqrt{35} C_{11,22}^{51,1}}{432}-\frac{1}{576} \sqrt{\frac{7}{3}} C_{20,00}^{40,0}-\frac{1}{288} \sqrt{\frac{7}{3}}
C_{20,00}^{40,1}-\frac{\sqrt{7} C_{20,20}^{40,0}}{1728}-\frac{\sqrt{7} C_{20,20}^{40,1}}{864}-\frac{\sqrt{\frac{7}{5}} C_{20,22}^{42,0}}{3456}-\frac{\sqrt{\frac{7}{5}}
C_{20,22}^{42,1}}{1728}+\frac{13 \sqrt{\frac{7}{15}} C_{22,00}^{42,0}}{2304}+\frac{13 \sqrt{\frac{7}{15}} C_{22,00}^{42,1}}{1152}+\frac{13 \sqrt{\frac{7}{5}}
C_{22,20}^{42,0}}{6912}+\frac{13 \sqrt{\frac{7}{5}} C_{22,20}^{42,1}}{3456}-\frac{\sqrt{35} C_{22,22}^{40,0}}{1728}-\frac{\sqrt{35} C_{22,22}^{40,1}}{864}+\frac{31
C_{22,22}^{42,0}}{3456 \sqrt{5}}+\frac{31 C_{22,22}^{42,1}}{1728 \sqrt{5}}+\frac{C_{22,22}^{44,0}}{64 \sqrt{7}}+\frac{C_{22,22}^{44,1}}{32 \sqrt{7}}-\frac{\sqrt{\frac{7}{3}}
C_{31,00}^{31,0}}{1152}-\frac{1}{576} \sqrt{\frac{7}{3}} C_{31,00}^{31,1}-\frac{\sqrt{7} C_{31,20}^{31,0}}{3456}-\frac{\sqrt{7} C_{31,20}^{31,1}}{1728}-\frac{\sqrt{\frac{7}{5}}
C_{31,22}^{31,0}}{1728}-\frac{1}{864} \sqrt{\frac{7}{5}} C_{31,22}^{31,1}-\frac{7 C_{31,22}^{33,0}}{1920 \sqrt{3}}-\frac{7 C_{31,22}^{33,1}}{960
\sqrt{3}}-\frac{13 C_{33,00}^{33,0}}{1280}-\frac{13 C_{33,00}^{33,1}}{640}-\frac{13 C_{33,20}^{33,0}}{1280 \sqrt{3}}-\frac{13 C_{33,20}^{33,1}}{640
\sqrt{3}}-\frac{11 C_{33,22}^{33,0}}{320 \sqrt{10}}-\frac{11 C_{33,22}^{33,1}}{160 \sqrt{10}}\)

\noindent\(C_{11,2212,3}^{11,2212,1,\text{Loc}}= -\frac{1}{48} \sqrt{\frac{7}{3}} C_{00,20}^{60,1}+\frac{1}{32} \sqrt{\frac{7}{15}}
C_{00,22}^{62,1}+\frac{1}{288} \sqrt{\frac{35}{3}} C_{11,22}^{51,1}-\frac{1}{288} \sqrt{\frac{7}{3}} C_{20,20}^{40,1}+\frac{\sqrt{\frac{7}{15}} C_{20,22}^{42,1}}{1152}+\frac{13
\sqrt{\frac{7}{15}} C_{22,20}^{42,1}}{1152}+\frac{1}{576} \sqrt{\frac{35}{3}} C_{22,22}^{40,1}-\frac{31 C_{22,22}^{42,1}}{1152 \sqrt{15}}-\frac{1}{64}
\sqrt{\frac{3}{7}} C_{22,22}^{44,1}+\frac{1}{576} \sqrt{\frac{7}{3}} C_{31,20}^{31,1}-\frac{1}{576} \sqrt{\frac{7}{15}} C_{31,22}^{31,1}-\frac{7
C_{31,22}^{33,1}}{1920}+\frac{13 C_{33,20}^{33,1}}{640}-\frac{11}{320} \sqrt{\frac{3}{10}} C_{33,22}^{33,1}\)

\noindent\(C_{11,2212,3}^{11,2212,1,\text{NonLoc}}= -\frac{\sqrt{7} C_{00,00}^{60,0}}{96}-\frac{1}{96} \sqrt{\frac{7}{3}} C_{00,20}^{60,0}-\frac{1}{32}
\sqrt{\frac{7}{15}} C_{00,22}^{62,0}+\frac{\sqrt{7} C_{11,00}^{51,0}}{288}+\frac{1}{288} \sqrt{\frac{35}{3}} C_{11,22}^{51,0}-\frac{\sqrt{7} C_{20,00}^{40,0}}{576}-\frac{1}{576}
\sqrt{\frac{7}{3}} C_{20,20}^{40,0}-\frac{\sqrt{\frac{7}{15}} C_{20,22}^{42,0}}{1152}+\frac{13 \sqrt{\frac{7}{5}} C_{22,00}^{42,0}}{2304}+\frac{13
\sqrt{\frac{7}{15}} C_{22,20}^{42,0}}{2304}-\frac{1}{576} \sqrt{\frac{35}{3}} C_{22,22}^{40,0}+\frac{31 C_{22,22}^{42,0}}{1152 \sqrt{15}}+\frac{1}{64}
\sqrt{\frac{3}{7}} C_{22,22}^{44,0}-\frac{\sqrt{7} C_{31,00}^{31,0}}{1152}-\frac{\sqrt{\frac{7}{3}} C_{31,20}^{31,0}}{1152}-\frac{1}{576} \sqrt{\frac{7}{15}}
C_{31,22}^{31,0}-\frac{7 C_{31,22}^{33,0}}{1920}-\frac{13 \sqrt{3} C_{33,00}^{33,0}}{1280}-\frac{13 C_{33,20}^{33,0}}{1280}-\frac{11}{320} \sqrt{\frac{3}{10}}
C_{33,22}^{33,0}\)

\noindent\(C_{11,2213,2}^{11,2011,0,\text{Loc}}= -\frac{7}{96} \sqrt{\frac{7}{3}} C_{00,20}^{60,0}-\frac{7}{192} \sqrt{\frac{7}{3}}
C_{00,20}^{60,1}-\frac{1}{32} \sqrt{\frac{7}{15}} C_{00,22}^{62,0}-\frac{1}{64} \sqrt{\frac{7}{15}} C_{00,22}^{62,1}-\frac{1}{144} \sqrt{\frac{7}{15}}
C_{11,22}^{51,0}-\frac{1}{288} \sqrt{\frac{7}{15}} C_{11,22}^{51,1}-\frac{3 C_{11,22}^{53,0}}{320}-\frac{3 C_{11,22}^{53,1}}{640}+\frac{7}{288} \sqrt{\frac{7}{3}}
C_{20,20}^{40,0}+\frac{7}{576} \sqrt{\frac{7}{3}} C_{20,20}^{40,1}+\frac{1}{144} \sqrt{\frac{7}{15}} C_{20,22}^{42,0}+\frac{1}{288} \sqrt{\frac{7}{15}}
C_{20,22}^{42,1}-\frac{7}{288} \sqrt{\frac{7}{15}} C_{22,20}^{42,0}-\frac{7}{576} \sqrt{\frac{7}{15}} C_{22,20}^{42,1}+\frac{1}{288} \sqrt{\frac{35}{3}}
C_{22,22}^{40,0}+\frac{1}{576} \sqrt{\frac{35}{3}} C_{22,22}^{40,1}-\frac{7 C_{22,22}^{42,0}}{144 \sqrt{15}}-\frac{7 C_{22,22}^{42,1}}{288 \sqrt{15}}+\frac{7}{288}
\sqrt{\frac{7}{3}} C_{31,20}^{31,0}+\frac{7}{576} \sqrt{\frac{7}{3}} C_{31,20}^{31,1}+\frac{1}{144} \sqrt{\frac{7}{15}} C_{31,22}^{31,0}+\frac{1}{288}
\sqrt{\frac{7}{15}} C_{31,22}^{31,1}+\frac{3 C_{31,22}^{33,0}}{320}+\frac{3 C_{31,22}^{33,1}}{640}-\frac{7 C_{33,20}^{33,0}}{160}-\frac{7 C_{33,20}^{33,1}}{320}-\frac{1}{20}
\sqrt{\frac{3}{10}} C_{33,22}^{33,0}-\frac{1}{40} \sqrt{\frac{3}{10}} C_{33,22}^{33,1}\)

\noindent\(C_{11,2213,2}^{11,2011,0,\text{NonLoc}}= -\frac{7 \sqrt{7} C_{00,00}^{60,0}}{384}-\frac{7 \sqrt{7} C_{00,00}^{60,1}}{192}-\frac{7}{384}
\sqrt{\frac{7}{3}} C_{00,20}^{60,0}-\frac{7}{192} \sqrt{\frac{7}{3}} C_{00,20}^{60,1}+\frac{1}{64} \sqrt{\frac{7}{15}} C_{00,22}^{62,0}+\frac{1}{32}
\sqrt{\frac{7}{15}} C_{00,22}^{62,1}+\frac{7 \sqrt{7} C_{11,00}^{51,0}}{1152}+\frac{7 \sqrt{7} C_{11,00}^{51,1}}{576}-\frac{1}{288} \sqrt{\frac{7}{15}}
C_{11,22}^{51,0}-\frac{1}{144} \sqrt{\frac{7}{15}} C_{11,22}^{51,1}-\frac{3 C_{11,22}^{53,0}}{640}-\frac{3 C_{11,22}^{53,1}}{320}+\frac{7 \sqrt{7}
C_{20,00}^{40,0}}{1152}+\frac{7 \sqrt{7} C_{20,00}^{40,1}}{576}+\frac{7 \sqrt{\frac{7}{3}} C_{20,20}^{40,0}}{1152}+\frac{7}{576} \sqrt{\frac{7}{3}}
C_{20,20}^{40,1}-\frac{1}{288} \sqrt{\frac{7}{15}} C_{20,22}^{42,0}-\frac{1}{144} \sqrt{\frac{7}{15}} C_{20,22}^{42,1}-\frac{7 \sqrt{\frac{7}{5}}
C_{22,00}^{42,0}}{1152}-\frac{7}{576} \sqrt{\frac{7}{5}} C_{22,00}^{42,1}-\frac{7 \sqrt{\frac{7}{15}} C_{22,20}^{42,0}}{1152}-\frac{7}{576} \sqrt{\frac{7}{15}}
C_{22,20}^{42,1}-\frac{1}{576} \sqrt{\frac{35}{3}} C_{22,22}^{40,0}-\frac{1}{288} \sqrt{\frac{35}{3}} C_{22,22}^{40,1}+\frac{7 C_{22,22}^{42,0}}{288
\sqrt{15}}+\frac{7 C_{22,22}^{42,1}}{144 \sqrt{15}}-\frac{7 \sqrt{7} C_{31,00}^{31,0}}{1152}-\frac{7 \sqrt{7} C_{31,00}^{31,1}}{576}-\frac{7 \sqrt{\frac{7}{3}}
C_{31,20}^{31,0}}{1152}-\frac{7}{576} \sqrt{\frac{7}{3}} C_{31,20}^{31,1}+\frac{1}{288} \sqrt{\frac{7}{15}} C_{31,22}^{31,0}+\frac{1}{144} \sqrt{\frac{7}{15}}
C_{31,22}^{31,1}+\frac{3 C_{31,22}^{33,0}}{640}+\frac{3 C_{31,22}^{33,1}}{320}+\frac{7 \sqrt{3} C_{33,00}^{33,0}}{640}+\frac{7 \sqrt{3} C_{33,00}^{33,1}}{320}+\frac{7
C_{33,20}^{33,0}}{640}+\frac{7 C_{33,20}^{33,1}}{320}-\frac{1}{40} \sqrt{\frac{3}{10}} C_{33,22}^{33,0}-\frac{1}{20} \sqrt{\frac{3}{10}} C_{33,22}^{33,1}\)

\noindent\(C_{11,2213,2}^{11,2011,1,\text{Loc}}= -\frac{7 \sqrt{7} C_{00,20}^{60,1}}{192}-\frac{1}{64} \sqrt{\frac{7}{5}} C_{00,22}^{62,1}-\frac{1}{288}
\sqrt{\frac{7}{5}} C_{11,22}^{51,1}-\frac{3 \sqrt{3} C_{11,22}^{53,1}}{640}+\frac{7 \sqrt{7} C_{20,20}^{40,1}}{576}+\frac{1}{288} \sqrt{\frac{7}{5}}
C_{20,22}^{42,1}-\frac{7}{576} \sqrt{\frac{7}{5}} C_{22,20}^{42,1}+\frac{\sqrt{35} C_{22,22}^{40,1}}{576}-\frac{7 C_{22,22}^{42,1}}{288 \sqrt{5}}+\frac{7
\sqrt{7} C_{31,20}^{31,1}}{576}+\frac{1}{288} \sqrt{\frac{7}{5}} C_{31,22}^{31,1}+\frac{3 \sqrt{3} C_{31,22}^{33,1}}{640}-\frac{7 \sqrt{3} C_{33,20}^{33,1}}{320}-\frac{3
C_{33,22}^{33,1}}{40 \sqrt{10}}\)

\noindent\(C_{11,2213,2}^{11,2011,1,\text{NonLoc}}= -\frac{7}{128} \sqrt{\frac{7}{3}} C_{00,00}^{60,0}-\frac{7 \sqrt{7} C_{00,20}^{60,0}}{384}+\frac{1}{64}
\sqrt{\frac{7}{5}} C_{00,22}^{62,0}+\frac{7}{384} \sqrt{\frac{7}{3}} C_{11,00}^{51,0}-\frac{1}{288} \sqrt{\frac{7}{5}} C_{11,22}^{51,0}-\frac{3 \sqrt{3}
C_{11,22}^{53,0}}{640}+\frac{7}{384} \sqrt{\frac{7}{3}} C_{20,00}^{40,0}+\frac{7 \sqrt{7} C_{20,20}^{40,0}}{1152}-\frac{1}{288} \sqrt{\frac{7}{5}}
C_{20,22}^{42,0}-\frac{7}{384} \sqrt{\frac{7}{15}} C_{22,00}^{42,0}-\frac{7 \sqrt{\frac{7}{5}} C_{22,20}^{42,0}}{1152}-\frac{\sqrt{35} C_{22,22}^{40,0}}{576}+\frac{7
C_{22,22}^{42,0}}{288 \sqrt{5}}-\frac{7}{384} \sqrt{\frac{7}{3}} C_{31,00}^{31,0}-\frac{7 \sqrt{7} C_{31,20}^{31,0}}{1152}+\frac{1}{288} \sqrt{\frac{7}{5}}
C_{31,22}^{31,0}+\frac{3 \sqrt{3} C_{31,22}^{33,0}}{640}+\frac{21 C_{33,00}^{33,0}}{640}+\frac{7 \sqrt{3} C_{33,20}^{33,0}}{640}-\frac{3 C_{33,22}^{33,0}}{40
\sqrt{10}}\)

\noindent\(C_{11,2213,2}^{11,2211,0,\text{Loc}}= -\frac{1}{48} \sqrt{\frac{7}{15}} C_{00,20}^{60,0}-\frac{1}{96} \sqrt{\frac{7}{15}}
C_{00,20}^{60,1}-\frac{1}{20} \sqrt{\frac{7}{3}} C_{00,22}^{62,0}-\frac{1}{40} \sqrt{\frac{7}{3}} C_{00,22}^{62,1}-\frac{7 \sqrt{\frac{7}{3}} C_{11,22}^{51,0}}{1440}-\frac{7
\sqrt{\frac{7}{3}} C_{11,22}^{51,1}}{2880}-\frac{33 C_{11,22}^{53,0}}{320 \sqrt{5}}-\frac{33 C_{11,22}^{53,1}}{640 \sqrt{5}}+\frac{1}{144} \sqrt{\frac{7}{15}}
C_{20,20}^{40,0}+\frac{1}{288} \sqrt{\frac{7}{15}} C_{20,20}^{40,1}+\frac{7 \sqrt{\frac{7}{3}} C_{20,22}^{42,0}}{1440}+\frac{7 \sqrt{\frac{7}{3}}
C_{20,22}^{42,1}}{2880}-\frac{1}{720} \sqrt{\frac{7}{3}} C_{22,20}^{42,0}-\frac{\sqrt{\frac{7}{3}} C_{22,20}^{42,1}}{1440}+\frac{11}{720} \sqrt{\frac{7}{3}}
C_{22,22}^{40,0}+\frac{11 \sqrt{\frac{7}{3}} C_{22,22}^{40,1}}{1440}-\frac{C_{22,22}^{42,0}}{360 \sqrt{3}}-\frac{C_{22,22}^{42,1}}{720 \sqrt{3}}-\frac{9}{80}
\sqrt{\frac{3}{35}} C_{22,22}^{44,0}-\frac{9}{160} \sqrt{\frac{3}{35}} C_{22,22}^{44,1}+\frac{1}{144} \sqrt{\frac{7}{15}} C_{31,20}^{31,0}+\frac{1}{288}
\sqrt{\frac{7}{15}} C_{31,20}^{31,1}+\frac{5}{288} \sqrt{\frac{7}{3}} C_{31,22}^{31,0}+\frac{5}{576} \sqrt{\frac{7}{3}} C_{31,22}^{31,1}+\frac{C_{31,22}^{33,0}}{64
\sqrt{5}}+\frac{C_{31,22}^{33,1}}{128 \sqrt{5}}-\frac{C_{33,20}^{33,0}}{80 \sqrt{5}}-\frac{C_{33,20}^{33,1}}{160 \sqrt{5}}-\frac{1}{80} \sqrt{\frac{3}{2}}
C_{33,22}^{33,0}-\frac{1}{160} \sqrt{\frac{3}{2}} C_{33,22}^{33,1}\)

\noindent\(C_{11,2213,2}^{11,2211,0,\text{NonLoc}}= -\frac{1}{192} \sqrt{\frac{7}{5}} C_{00,00}^{60,0}-\frac{1}{96} \sqrt{\frac{7}{5}}
C_{00,00}^{60,1}-\frac{1}{192} \sqrt{\frac{7}{15}} C_{00,20}^{60,0}-\frac{1}{96} \sqrt{\frac{7}{15}} C_{00,20}^{60,1}+\frac{1}{40} \sqrt{\frac{7}{3}}
C_{00,22}^{62,0}+\frac{1}{20} \sqrt{\frac{7}{3}} C_{00,22}^{62,1}+\frac{1}{576} \sqrt{\frac{7}{5}} C_{11,00}^{51,0}+\frac{1}{288} \sqrt{\frac{7}{5}}
C_{11,00}^{51,1}-\frac{7 \sqrt{\frac{7}{3}} C_{11,22}^{51,0}}{2880}-\frac{7 \sqrt{\frac{7}{3}} C_{11,22}^{51,1}}{1440}-\frac{33 C_{11,22}^{53,0}}{640
\sqrt{5}}-\frac{33 C_{11,22}^{53,1}}{320 \sqrt{5}}+\frac{1}{576} \sqrt{\frac{7}{5}} C_{20,00}^{40,0}+\frac{1}{288} \sqrt{\frac{7}{5}} C_{20,00}^{40,1}+\frac{1}{576}
\sqrt{\frac{7}{15}} C_{20,20}^{40,0}+\frac{1}{288} \sqrt{\frac{7}{15}} C_{20,20}^{40,1}-\frac{7 \sqrt{\frac{7}{3}} C_{20,22}^{42,0}}{2880}-\frac{7
\sqrt{\frac{7}{3}} C_{20,22}^{42,1}}{1440}-\frac{\sqrt{7} C_{22,00}^{42,0}}{2880}-\frac{\sqrt{7} C_{22,00}^{42,1}}{1440}-\frac{\sqrt{\frac{7}{3}}
C_{22,20}^{42,0}}{2880}-\frac{\sqrt{\frac{7}{3}} C_{22,20}^{42,1}}{1440}-\frac{11 \sqrt{\frac{7}{3}} C_{22,22}^{40,0}}{1440}-\frac{11}{720} \sqrt{\frac{7}{3}}
C_{22,22}^{40,1}+\frac{C_{22,22}^{42,0}}{720 \sqrt{3}}+\frac{C_{22,22}^{42,1}}{360 \sqrt{3}}+\frac{9}{160} \sqrt{\frac{3}{35}} C_{22,22}^{44,0}+\frac{9}{80}
\sqrt{\frac{3}{35}} C_{22,22}^{44,1}-\frac{1}{576} \sqrt{\frac{7}{5}} C_{31,00}^{31,0}-\frac{1}{288} \sqrt{\frac{7}{5}} C_{31,00}^{31,1}-\frac{1}{576}
\sqrt{\frac{7}{15}} C_{31,20}^{31,0}-\frac{1}{288} \sqrt{\frac{7}{15}} C_{31,20}^{31,1}+\frac{5}{576} \sqrt{\frac{7}{3}} C_{31,22}^{31,0}+\frac{5}{288}
\sqrt{\frac{7}{3}} C_{31,22}^{31,1}+\frac{C_{31,22}^{33,0}}{128 \sqrt{5}}+\frac{C_{31,22}^{33,1}}{64 \sqrt{5}}+\frac{1}{320} \sqrt{\frac{3}{5}} C_{33,00}^{33,0}+\frac{1}{160}
\sqrt{\frac{3}{5}} C_{33,00}^{33,1}+\frac{C_{33,20}^{33,0}}{320 \sqrt{5}}+\frac{C_{33,20}^{33,1}}{160 \sqrt{5}}-\frac{1}{160} \sqrt{\frac{3}{2}}
C_{33,22}^{33,0}-\frac{1}{80} \sqrt{\frac{3}{2}} C_{33,22}^{33,1}\)

\noindent\(C_{11,2213,2}^{11,2211,1,\text{Loc}}= -\frac{1}{96} \sqrt{\frac{7}{5}} C_{00,20}^{60,1}-\frac{\sqrt{7} C_{00,22}^{62,1}}{40}-\frac{7
\sqrt{7} C_{11,22}^{51,1}}{2880}-\frac{33}{640} \sqrt{\frac{3}{5}} C_{11,22}^{53,1}+\frac{1}{288} \sqrt{\frac{7}{5}} C_{20,20}^{40,1}+\frac{7 \sqrt{7}
C_{20,22}^{42,1}}{2880}-\frac{\sqrt{7} C_{22,20}^{42,1}}{1440}+\frac{11 \sqrt{7} C_{22,22}^{40,1}}{1440}-\frac{C_{22,22}^{42,1}}{720}-\frac{27 C_{22,22}^{44,1}}{160
\sqrt{35}}+\frac{1}{288} \sqrt{\frac{7}{5}} C_{31,20}^{31,1}+\frac{5 \sqrt{7} C_{31,22}^{31,1}}{576}+\frac{1}{128} \sqrt{\frac{3}{5}} C_{31,22}^{33,1}-\frac{1}{160}
\sqrt{\frac{3}{5}} C_{33,20}^{33,1}-\frac{3 C_{33,22}^{33,1}}{160 \sqrt{2}}\)

\noindent\(C_{11,2213,2}^{11,2211,1,\text{NonLoc}}= -\frac{1}{64} \sqrt{\frac{7}{15}} C_{00,00}^{60,0}-\frac{1}{192} \sqrt{\frac{7}{5}}
C_{00,20}^{60,0}+\frac{\sqrt{7} C_{00,22}^{62,0}}{40}+\frac{1}{192} \sqrt{\frac{7}{15}} C_{11,00}^{51,0}-\frac{7 \sqrt{7} C_{11,22}^{51,0}}{2880}-\frac{33}{640}
\sqrt{\frac{3}{5}} C_{11,22}^{53,0}+\frac{1}{192} \sqrt{\frac{7}{15}} C_{20,00}^{40,0}+\frac{1}{576} \sqrt{\frac{7}{5}} C_{20,20}^{40,0}-\frac{7
\sqrt{7} C_{20,22}^{42,0}}{2880}-\frac{1}{960} \sqrt{\frac{7}{3}} C_{22,00}^{42,0}-\frac{\sqrt{7} C_{22,20}^{42,0}}{2880}-\frac{11 \sqrt{7} C_{22,22}^{40,0}}{1440}+\frac{C_{22,22}^{42,0}}{720}+\frac{27
C_{22,22}^{44,0}}{160 \sqrt{35}}-\frac{1}{192} \sqrt{\frac{7}{15}} C_{31,00}^{31,0}-\frac{1}{576} \sqrt{\frac{7}{5}} C_{31,20}^{31,0}+\frac{5 \sqrt{7}
C_{31,22}^{31,0}}{576}+\frac{1}{128} \sqrt{\frac{3}{5}} C_{31,22}^{33,0}+\frac{3 C_{33,00}^{33,0}}{320 \sqrt{5}}+\frac{1}{320} \sqrt{\frac{3}{5}}
C_{33,20}^{33,0}-\frac{3 C_{33,22}^{33,0}}{160 \sqrt{2}}\)

\noindent\(C_{11,2213,2}^{11,2212,0,\text{Loc}}= \frac{\sqrt{7} C_{00,20}^{60,0}}{72}+\frac{\sqrt{7} C_{00,20}^{60,1}}{144}-\frac{1}{48}
\sqrt{\frac{7}{5}} C_{00,22}^{62,0}-\frac{1}{96} \sqrt{\frac{7}{5}} C_{00,22}^{62,1}+\frac{1}{108} \sqrt{\frac{7}{5}} C_{11,22}^{51,0}+\frac{1}{216}
\sqrt{\frac{7}{5}} C_{11,22}^{51,1}-\frac{3 \sqrt{3} C_{11,22}^{53,0}}{160}-\frac{3 \sqrt{3} C_{11,22}^{53,1}}{320}-\frac{\sqrt{7} C_{20,20}^{40,0}}{216}-\frac{\sqrt{7}
C_{20,20}^{40,1}}{432}-\frac{1}{108} \sqrt{\frac{7}{5}} C_{20,22}^{42,0}-\frac{1}{216} \sqrt{\frac{7}{5}} C_{20,22}^{42,1}+\frac{1}{216} \sqrt{\frac{7}{5}}
C_{22,20}^{42,0}+\frac{1}{432} \sqrt{\frac{7}{5}} C_{22,20}^{42,1}+\frac{\sqrt{35} C_{22,22}^{40,0}}{432}+\frac{\sqrt{35} C_{22,22}^{40,1}}{864}+\frac{C_{22,22}^{42,0}}{27
\sqrt{5}}+\frac{C_{22,22}^{42,1}}{54 \sqrt{5}}-\frac{C_{22,22}^{44,0}}{16 \sqrt{7}}-\frac{C_{22,22}^{44,1}}{32 \sqrt{7}}-\frac{\sqrt{7} C_{31,20}^{31,0}}{216}-\frac{\sqrt{7}
C_{31,20}^{31,1}}{432}+\frac{1}{216} \sqrt{\frac{7}{5}} C_{31,22}^{31,0}+\frac{1}{432} \sqrt{\frac{7}{5}} C_{31,22}^{31,1}-\frac{C_{31,22}^{33,0}}{480
\sqrt{3}}-\frac{C_{31,22}^{33,1}}{960 \sqrt{3}}+\frac{C_{33,20}^{33,0}}{40 \sqrt{3}}+\frac{C_{33,20}^{33,1}}{80 \sqrt{3}}+\frac{7 C_{33,22}^{33,0}}{80
\sqrt{10}}+\frac{7 C_{33,22}^{33,1}}{160 \sqrt{10}}\)

\noindent\(C_{11,2213,2}^{11,2212,0,\text{NonLoc}}= \frac{1}{96} \sqrt{\frac{7}{3}} C_{00,00}^{60,0}+\frac{1}{48} \sqrt{\frac{7}{3}}
C_{00,00}^{60,1}+\frac{\sqrt{7} C_{00,20}^{60,0}}{288}+\frac{\sqrt{7} C_{00,20}^{60,1}}{144}+\frac{1}{96} \sqrt{\frac{7}{5}} C_{00,22}^{62,0}+\frac{1}{48}
\sqrt{\frac{7}{5}} C_{00,22}^{62,1}-\frac{1}{288} \sqrt{\frac{7}{3}} C_{11,00}^{51,0}-\frac{1}{144} \sqrt{\frac{7}{3}} C_{11,00}^{51,1}+\frac{1}{216}
\sqrt{\frac{7}{5}} C_{11,22}^{51,0}+\frac{1}{108} \sqrt{\frac{7}{5}} C_{11,22}^{51,1}-\frac{3 \sqrt{3} C_{11,22}^{53,0}}{320}-\frac{3 \sqrt{3} C_{11,22}^{53,1}}{160}-\frac{1}{288}
\sqrt{\frac{7}{3}} C_{20,00}^{40,0}-\frac{1}{144} \sqrt{\frac{7}{3}} C_{20,00}^{40,1}-\frac{\sqrt{7} C_{20,20}^{40,0}}{864}-\frac{\sqrt{7} C_{20,20}^{40,1}}{432}+\frac{1}{216}
\sqrt{\frac{7}{5}} C_{20,22}^{42,0}+\frac{1}{108} \sqrt{\frac{7}{5}} C_{20,22}^{42,1}+\frac{1}{288} \sqrt{\frac{7}{15}} C_{22,00}^{42,0}+\frac{1}{144}
\sqrt{\frac{7}{15}} C_{22,00}^{42,1}+\frac{1}{864} \sqrt{\frac{7}{5}} C_{22,20}^{42,0}+\frac{1}{432} \sqrt{\frac{7}{5}} C_{22,20}^{42,1}-\frac{\sqrt{35}
C_{22,22}^{40,0}}{864}-\frac{\sqrt{35} C_{22,22}^{40,1}}{432}-\frac{C_{22,22}^{42,0}}{54 \sqrt{5}}-\frac{C_{22,22}^{42,1}}{27 \sqrt{5}}+\frac{C_{22,22}^{44,0}}{32
\sqrt{7}}+\frac{C_{22,22}^{44,1}}{16 \sqrt{7}}+\frac{1}{288} \sqrt{\frac{7}{3}} C_{31,00}^{31,0}+\frac{1}{144} \sqrt{\frac{7}{3}} C_{31,00}^{31,1}+\frac{\sqrt{7}
C_{31,20}^{31,0}}{864}+\frac{\sqrt{7} C_{31,20}^{31,1}}{432}+\frac{1}{432} \sqrt{\frac{7}{5}} C_{31,22}^{31,0}+\frac{1}{216} \sqrt{\frac{7}{5}} C_{31,22}^{31,1}-\frac{C_{31,22}^{33,0}}{960
\sqrt{3}}-\frac{C_{31,22}^{33,1}}{480 \sqrt{3}}-\frac{C_{33,00}^{33,0}}{160}-\frac{C_{33,00}^{33,1}}{80}-\frac{C_{33,20}^{33,0}}{160 \sqrt{3}}-\frac{C_{33,20}^{33,1}}{80
\sqrt{3}}+\frac{7 C_{33,22}^{33,0}}{160 \sqrt{10}}+\frac{7 C_{33,22}^{33,1}}{80 \sqrt{10}}\)

\noindent\(C_{11,2213,2}^{11,2212,1,\text{Loc}}= \frac{1}{48} \sqrt{\frac{7}{3}} C_{00,20}^{60,1}-\frac{1}{32} \sqrt{\frac{7}{15}}
C_{00,22}^{62,1}+\frac{1}{72} \sqrt{\frac{7}{15}} C_{11,22}^{51,1}-\frac{9 C_{11,22}^{53,1}}{320}-\frac{1}{144} \sqrt{\frac{7}{3}} C_{20,20}^{40,1}-\frac{1}{72}
\sqrt{\frac{7}{15}} C_{20,22}^{42,1}+\frac{1}{144} \sqrt{\frac{7}{15}} C_{22,20}^{42,1}+\frac{1}{288} \sqrt{\frac{35}{3}} C_{22,22}^{40,1}+\frac{C_{22,22}^{42,1}}{18
\sqrt{15}}-\frac{1}{32} \sqrt{\frac{3}{7}} C_{22,22}^{44,1}-\frac{1}{144} \sqrt{\frac{7}{3}} C_{31,20}^{31,1}+\frac{1}{144} \sqrt{\frac{7}{15}} C_{31,22}^{31,1}-\frac{C_{31,22}^{33,1}}{960}+\frac{C_{33,20}^{33,1}}{80}+\frac{7}{160}
\sqrt{\frac{3}{10}} C_{33,22}^{33,1}\)

\noindent\(C_{11,2213,2}^{11,2212,1,\text{NonLoc}}= \frac{\sqrt{7} C_{00,00}^{60,0}}{96}+\frac{1}{96} \sqrt{\frac{7}{3}} C_{00,20}^{60,0}+\frac{1}{32}
\sqrt{\frac{7}{15}} C_{00,22}^{62,0}-\frac{\sqrt{7} C_{11,00}^{51,0}}{288}+\frac{1}{72} \sqrt{\frac{7}{15}} C_{11,22}^{51,0}-\frac{9 C_{11,22}^{53,0}}{320}-\frac{\sqrt{7}
C_{20,00}^{40,0}}{288}-\frac{1}{288} \sqrt{\frac{7}{3}} C_{20,20}^{40,0}+\frac{1}{72} \sqrt{\frac{7}{15}} C_{20,22}^{42,0}+\frac{1}{288} \sqrt{\frac{7}{5}}
C_{22,00}^{42,0}+\frac{1}{288} \sqrt{\frac{7}{15}} C_{22,20}^{42,0}-\frac{1}{288} \sqrt{\frac{35}{3}} C_{22,22}^{40,0}-\frac{C_{22,22}^{42,0}}{18
\sqrt{15}}+\frac{1}{32} \sqrt{\frac{3}{7}} C_{22,22}^{44,0}+\frac{\sqrt{7} C_{31,00}^{31,0}}{288}+\frac{1}{288} \sqrt{\frac{7}{3}} C_{31,20}^{31,0}+\frac{1}{144}
\sqrt{\frac{7}{15}} C_{31,22}^{31,0}-\frac{C_{31,22}^{33,0}}{960}-\frac{\sqrt{3} C_{33,00}^{33,0}}{160}-\frac{C_{33,20}^{33,0}}{160}+\frac{7}{160}
\sqrt{\frac{3}{10}} C_{33,22}^{33,0}\)

\noindent\(C_{11,2213,2}^{11,2213,0,\text{Loc}}= -\frac{59 C_{00,20}^{60,0}}{576 \sqrt{5}}-\frac{59 C_{00,20}^{60,1}}{1152 \sqrt{5}}-\frac{11
C_{00,22}^{62,0}}{960}-\frac{11 C_{00,22}^{62,1}}{1920}-\frac{19 C_{11,22}^{51,0}}{8640}-\frac{19 C_{11,22}^{51,1}}{17280}-\frac{3}{160} \sqrt{\frac{3}{35}}
C_{11,22}^{53,0}-\frac{3}{320} \sqrt{\frac{3}{35}} C_{11,22}^{53,1}-\frac{C_{20,20}^{40,0}}{1728 \sqrt{5}}-\frac{C_{20,20}^{40,1}}{3456 \sqrt{5}}-\frac{11
C_{20,22}^{42,0}}{8640}-\frac{11 C_{20,22}^{42,1}}{17280}+\frac{23 C_{22,20}^{42,0}}{4320}+\frac{23 C_{22,20}^{42,1}}{8640}-\frac{11 C_{22,22}^{40,0}}{8640}-\frac{11
C_{22,22}^{40,1}}{17280}+\frac{19 C_{22,22}^{42,0}}{1080 \sqrt{7}}+\frac{19 C_{22,22}^{42,1}}{2160 \sqrt{7}}-\frac{C_{22,22}^{44,0}}{140 \sqrt{5}}-\frac{C_{22,22}^{44,1}}{280
\sqrt{5}}+\frac{29 C_{31,20}^{31,0}}{1728 \sqrt{5}}+\frac{29 C_{31,20}^{31,1}}{3456 \sqrt{5}}+\frac{5 C_{31,22}^{31,0}}{1728}+\frac{5 C_{31,22}^{31,1}}{3456}-\frac{C_{31,22}^{33,0}}{96
\sqrt{105}}-\frac{C_{31,22}^{33,1}}{192 \sqrt{105}}+\frac{23 C_{33,20}^{33,0}}{160 \sqrt{105}}+\frac{23 C_{33,20}^{33,1}}{320 \sqrt{105}}+\frac{C_{33,22}^{33,0}}{20
\sqrt{14}}+\frac{C_{33,22}^{33,1}}{40 \sqrt{14}}\)

\noindent\(C_{11,2213,2}^{11,2213,0,\text{NonLoc}}= -\frac{59 C_{00,00}^{60,0}}{768 \sqrt{15}}-\frac{59 C_{00,00}^{60,1}}{384 \sqrt{15}}-\frac{59
C_{00,20}^{60,0}}{2304 \sqrt{5}}-\frac{59 C_{00,20}^{60,1}}{1152 \sqrt{5}}+\frac{11 C_{00,22}^{62,0}}{1920}+\frac{11 C_{00,22}^{62,1}}{960}+\frac{59
C_{11,00}^{51,0}}{2304 \sqrt{15}}+\frac{59 C_{11,00}^{51,1}}{1152 \sqrt{15}}-\frac{19 C_{11,22}^{51,0}}{17280}-\frac{19 C_{11,22}^{51,1}}{8640}-\frac{3}{320}
\sqrt{\frac{3}{35}} C_{11,22}^{53,0}-\frac{3}{160} \sqrt{\frac{3}{35}} C_{11,22}^{53,1}-\frac{C_{20,00}^{40,0}}{2304 \sqrt{15}}-\frac{C_{20,00}^{40,1}}{1152
\sqrt{15}}-\frac{C_{20,20}^{40,0}}{6912 \sqrt{5}}-\frac{C_{20,20}^{40,1}}{3456 \sqrt{5}}+\frac{11 C_{20,22}^{42,0}}{17280}+\frac{11 C_{20,22}^{42,1}}{8640}+\frac{23
C_{22,00}^{42,0}}{5760 \sqrt{3}}+\frac{23 C_{22,00}^{42,1}}{2880 \sqrt{3}}+\frac{23 C_{22,20}^{42,0}}{17280}+\frac{23 C_{22,20}^{42,1}}{8640}+\frac{11
C_{22,22}^{40,0}}{17280}+\frac{11 C_{22,22}^{40,1}}{8640}-\frac{19 C_{22,22}^{42,0}}{2160 \sqrt{7}}-\frac{19 C_{22,22}^{42,1}}{1080 \sqrt{7}}+\frac{C_{22,22}^{44,0}}{280
\sqrt{5}}+\frac{C_{22,22}^{44,1}}{140 \sqrt{5}}-\frac{29 C_{31,00}^{31,0}}{2304 \sqrt{15}}-\frac{29 C_{31,00}^{31,1}}{1152 \sqrt{15}}-\frac{29 C_{31,20}^{31,0}}{6912
\sqrt{5}}-\frac{29 C_{31,20}^{31,1}}{3456 \sqrt{5}}+\frac{5 C_{31,22}^{31,0}}{3456}+\frac{5 C_{31,22}^{31,1}}{1728}-\frac{C_{31,22}^{33,0}}{192 \sqrt{105}}-\frac{C_{31,22}^{33,1}}{96
\sqrt{105}}-\frac{23 C_{33,00}^{33,0}}{640 \sqrt{35}}-\frac{23 C_{33,00}^{33,1}}{320 \sqrt{35}}-\frac{23 C_{33,20}^{33,0}}{640 \sqrt{105}}-\frac{23
C_{33,20}^{33,1}}{320 \sqrt{105}}+\frac{C_{33,22}^{33,0}}{40 \sqrt{14}}+\frac{C_{33,22}^{33,1}}{20 \sqrt{14}}\)

\noindent\(C_{11,2213,2}^{11,2213,1,\text{Loc}}= -\frac{59 C_{00,20}^{60,1}}{384 \sqrt{15}}-\frac{11 C_{00,22}^{62,1}}{640 \sqrt{3}}-\frac{19
C_{11,22}^{51,1}}{5760 \sqrt{3}}-\frac{9 C_{11,22}^{53,1}}{320 \sqrt{35}}-\frac{C_{20,20}^{40,1}}{1152 \sqrt{15}}-\frac{11 C_{20,22}^{42,1}}{5760
\sqrt{3}}+\frac{23 C_{22,20}^{42,1}}{2880 \sqrt{3}}-\frac{11 C_{22,22}^{40,1}}{5760 \sqrt{3}}+\frac{19 C_{22,22}^{42,1}}{720 \sqrt{21}}-\frac{1}{280}
\sqrt{\frac{3}{5}} C_{22,22}^{44,1}+\frac{29 C_{31,20}^{31,1}}{1152 \sqrt{15}}+\frac{5 C_{31,22}^{31,1}}{1152 \sqrt{3}}-\frac{C_{31,22}^{33,1}}{192
\sqrt{35}}+\frac{23 C_{33,20}^{33,1}}{320 \sqrt{35}}+\frac{1}{40} \sqrt{\frac{3}{14}} C_{33,22}^{33,1}\)

\noindent\(C_{11,2213,2}^{11,2213,1,\text{NonLoc}}= -\frac{59 C_{00,00}^{60,0}}{768 \sqrt{5}}-\frac{59 C_{00,20}^{60,0}}{768 \sqrt{15}}+\frac{11
C_{00,22}^{62,0}}{640 \sqrt{3}}+\frac{59 C_{11,00}^{51,0}}{2304 \sqrt{5}}-\frac{19 C_{11,22}^{51,0}}{5760 \sqrt{3}}-\frac{9 C_{11,22}^{53,0}}{320
\sqrt{35}}-\frac{C_{20,00}^{40,0}}{2304 \sqrt{5}}-\frac{C_{20,20}^{40,0}}{2304 \sqrt{15}}+\frac{11 C_{20,22}^{42,0}}{5760 \sqrt{3}}+\frac{23 C_{22,00}^{42,0}}{5760}+\frac{23
C_{22,20}^{42,0}}{5760 \sqrt{3}}+\frac{11 C_{22,22}^{40,0}}{5760 \sqrt{3}}-\frac{19 C_{22,22}^{42,0}}{720 \sqrt{21}}+\frac{1}{280} \sqrt{\frac{3}{5}}
C_{22,22}^{44,0}-\frac{29 C_{31,00}^{31,0}}{2304 \sqrt{5}}-\frac{29 C_{31,20}^{31,0}}{2304 \sqrt{15}}+\frac{5 C_{31,22}^{31,0}}{1152 \sqrt{3}}-\frac{C_{31,22}^{33,0}}{192
\sqrt{35}}-\frac{23}{640} \sqrt{\frac{3}{35}} C_{33,00}^{33,0}-\frac{23 C_{33,20}^{33,0}}{640 \sqrt{35}}+\frac{1}{40} \sqrt{\frac{3}{14}} C_{33,22}^{33,0}\)

\noindent\(C_{11,2213,3}^{00,3303,0,\text{Loc}}= -\frac{1}{72} \sqrt{\frac{7}{2}} C_{11,11}^{51,0}-\frac{1}{144} \sqrt{\frac{7}{2}}
C_{11,11}^{51,1}-\frac{1}{12} \sqrt{\frac{7}{30}} C_{22,11}^{42,0}-\frac{1}{24} \sqrt{\frac{7}{30}} C_{22,11}^{42,1}-\frac{1}{72} \sqrt{\frac{7}{2}}
C_{31,11}^{31,0}-\frac{1}{144} \sqrt{\frac{7}{2}} C_{31,11}^{31,1}-\frac{C_{33,11}^{33,0}}{20}-\frac{C_{33,11}^{33,1}}{40}\)

\noindent\(C_{11,2213,3}^{00,3303,0,\text{NonLoc}}= -\frac{1}{144} \sqrt{\frac{7}{2}} C_{11,11}^{51,0}-\frac{1}{72} \sqrt{\frac{7}{2}}
C_{11,11}^{51,1}+\frac{1}{24} \sqrt{\frac{7}{30}} C_{22,11}^{42,0}+\frac{1}{12} \sqrt{\frac{7}{30}} C_{22,11}^{42,1}-\frac{1}{144} \sqrt{\frac{7}{2}}
C_{31,11}^{31,0}-\frac{1}{72} \sqrt{\frac{7}{2}} C_{31,11}^{31,1}-\frac{C_{33,11}^{33,0}}{40}-\frac{C_{33,11}^{33,1}}{20}\)

\noindent\(C_{11,2213,3}^{00,3303,1,\text{Loc}}= -\frac{1}{48} \sqrt{\frac{7}{6}} C_{11,11}^{51,1}-\frac{1}{24} \sqrt{\frac{7}{10}}
C_{22,11}^{42,1}-\frac{1}{48} \sqrt{\frac{7}{6}} C_{31,11}^{31,1}-\frac{\sqrt{3} C_{33,11}^{33,1}}{40}\)

\noindent\(C_{11,2213,3}^{00,3303,1,\text{NonLoc}}= -\frac{1}{48} \sqrt{\frac{7}{6}} C_{11,11}^{51,0}+\frac{1}{24} \sqrt{\frac{7}{10}}
C_{22,11}^{42,0}-\frac{1}{48} \sqrt{\frac{7}{6}} C_{31,11}^{31,0}-\frac{\sqrt{3} C_{33,11}^{33,0}}{40}\)

\noindent\(C_{11,2213,3}^{11,2212,0,\text{Loc}}= -\frac{1}{72} \sqrt{\frac{7}{2}} C_{00,20}^{60,0}-\frac{1}{144} \sqrt{\frac{7}{2}}
C_{00,20}^{60,1}+\frac{1}{48} \sqrt{\frac{7}{10}} C_{00,22}^{62,0}+\frac{1}{96} \sqrt{\frac{7}{10}} C_{00,22}^{62,1}-\frac{1}{108} \sqrt{\frac{7}{10}}
C_{11,22}^{51,0}-\frac{1}{216} \sqrt{\frac{7}{10}} C_{11,22}^{51,1}+\frac{3}{160} \sqrt{\frac{3}{2}} C_{11,22}^{53,0}+\frac{3}{320} \sqrt{\frac{3}{2}}
C_{11,22}^{53,1}+\frac{1}{216} \sqrt{\frac{7}{2}} C_{20,20}^{40,0}+\frac{1}{432} \sqrt{\frac{7}{2}} C_{20,20}^{40,1}+\frac{1}{108} \sqrt{\frac{7}{10}}
C_{20,22}^{42,0}+\frac{1}{216} \sqrt{\frac{7}{10}} C_{20,22}^{42,1}-\frac{1}{216} \sqrt{\frac{7}{10}} C_{22,20}^{42,0}-\frac{1}{432} \sqrt{\frac{7}{10}}
C_{22,20}^{42,1}-\frac{1}{432} \sqrt{\frac{35}{2}} C_{22,22}^{40,0}-\frac{1}{864} \sqrt{\frac{35}{2}} C_{22,22}^{40,1}-\frac{C_{22,22}^{42,0}}{27
\sqrt{10}}-\frac{C_{22,22}^{42,1}}{54 \sqrt{10}}+\frac{C_{22,22}^{44,0}}{16 \sqrt{14}}+\frac{C_{22,22}^{44,1}}{32 \sqrt{14}}+\frac{1}{216} \sqrt{\frac{7}{2}}
C_{31,20}^{31,0}+\frac{1}{432} \sqrt{\frac{7}{2}} C_{31,20}^{31,1}-\frac{1}{216} \sqrt{\frac{7}{10}} C_{31,22}^{31,0}-\frac{1}{432} \sqrt{\frac{7}{10}}
C_{31,22}^{31,1}+\frac{C_{31,22}^{33,0}}{480 \sqrt{6}}+\frac{C_{31,22}^{33,1}}{960 \sqrt{6}}-\frac{C_{33,20}^{33,0}}{40 \sqrt{6}}-\frac{C_{33,20}^{33,1}}{80
\sqrt{6}}-\frac{7 C_{33,22}^{33,0}}{160 \sqrt{5}}-\frac{7 C_{33,22}^{33,1}}{320 \sqrt{5}}\)

\noindent\(C_{11,2213,3}^{11,2212,0,\text{NonLoc}}= -\frac{1}{96} \sqrt{\frac{7}{6}} C_{00,00}^{60,0}-\frac{1}{48} \sqrt{\frac{7}{6}}
C_{00,00}^{60,1}-\frac{1}{288} \sqrt{\frac{7}{2}} C_{00,20}^{60,0}-\frac{1}{144} \sqrt{\frac{7}{2}} C_{00,20}^{60,1}-\frac{1}{96} \sqrt{\frac{7}{10}}
C_{00,22}^{62,0}-\frac{1}{48} \sqrt{\frac{7}{10}} C_{00,22}^{62,1}+\frac{1}{288} \sqrt{\frac{7}{6}} C_{11,00}^{51,0}+\frac{1}{144} \sqrt{\frac{7}{6}}
C_{11,00}^{51,1}-\frac{1}{216} \sqrt{\frac{7}{10}} C_{11,22}^{51,0}-\frac{1}{108} \sqrt{\frac{7}{10}} C_{11,22}^{51,1}+\frac{3}{320} \sqrt{\frac{3}{2}}
C_{11,22}^{53,0}+\frac{3}{160} \sqrt{\frac{3}{2}} C_{11,22}^{53,1}+\frac{1}{288} \sqrt{\frac{7}{6}} C_{20,00}^{40,0}+\frac{1}{144} \sqrt{\frac{7}{6}}
C_{20,00}^{40,1}+\frac{1}{864} \sqrt{\frac{7}{2}} C_{20,20}^{40,0}+\frac{1}{432} \sqrt{\frac{7}{2}} C_{20,20}^{40,1}-\frac{1}{216} \sqrt{\frac{7}{10}}
C_{20,22}^{42,0}-\frac{1}{108} \sqrt{\frac{7}{10}} C_{20,22}^{42,1}-\frac{1}{288} \sqrt{\frac{7}{30}} C_{22,00}^{42,0}-\frac{1}{144} \sqrt{\frac{7}{30}}
C_{22,00}^{42,1}-\frac{1}{864} \sqrt{\frac{7}{10}} C_{22,20}^{42,0}-\frac{1}{432} \sqrt{\frac{7}{10}} C_{22,20}^{42,1}+\frac{1}{864} \sqrt{\frac{35}{2}}
C_{22,22}^{40,0}+\frac{1}{432} \sqrt{\frac{35}{2}} C_{22,22}^{40,1}+\frac{C_{22,22}^{42,0}}{54 \sqrt{10}}+\frac{C_{22,22}^{42,1}}{27 \sqrt{10}}-\frac{C_{22,22}^{44,0}}{32
\sqrt{14}}-\frac{C_{22,22}^{44,1}}{16 \sqrt{14}}-\frac{1}{288} \sqrt{\frac{7}{6}} C_{31,00}^{31,0}-\frac{1}{144} \sqrt{\frac{7}{6}} C_{31,00}^{31,1}-\frac{1}{864}
\sqrt{\frac{7}{2}} C_{31,20}^{31,0}-\frac{1}{432} \sqrt{\frac{7}{2}} C_{31,20}^{31,1}-\frac{1}{432} \sqrt{\frac{7}{10}} C_{31,22}^{31,0}-\frac{1}{216}
\sqrt{\frac{7}{10}} C_{31,22}^{31,1}+\frac{C_{31,22}^{33,0}}{960 \sqrt{6}}+\frac{C_{31,22}^{33,1}}{480 \sqrt{6}}+\frac{C_{33,00}^{33,0}}{160 \sqrt{2}}+\frac{C_{33,00}^{33,1}}{80
\sqrt{2}}+\frac{C_{33,20}^{33,0}}{160 \sqrt{6}}+\frac{C_{33,20}^{33,1}}{80 \sqrt{6}}-\frac{7 C_{33,22}^{33,0}}{320 \sqrt{5}}-\frac{7 C_{33,22}^{33,1}}{160
\sqrt{5}}\)

\noindent\(C_{11,2213,3}^{11,2212,1,\text{Loc}}= -\frac{1}{48} \sqrt{\frac{7}{6}} C_{00,20}^{60,1}+\frac{1}{32} \sqrt{\frac{7}{30}}
C_{00,22}^{62,1}-\frac{1}{72} \sqrt{\frac{7}{30}} C_{11,22}^{51,1}+\frac{9 C_{11,22}^{53,1}}{320 \sqrt{2}}+\frac{1}{144} \sqrt{\frac{7}{6}} C_{20,20}^{40,1}+\frac{1}{72}
\sqrt{\frac{7}{30}} C_{20,22}^{42,1}-\frac{1}{144} \sqrt{\frac{7}{30}} C_{22,20}^{42,1}-\frac{1}{288} \sqrt{\frac{35}{6}} C_{22,22}^{40,1}-\frac{C_{22,22}^{42,1}}{18
\sqrt{30}}+\frac{1}{32} \sqrt{\frac{3}{14}} C_{22,22}^{44,1}+\frac{1}{144} \sqrt{\frac{7}{6}} C_{31,20}^{31,1}-\frac{1}{144} \sqrt{\frac{7}{30}}
C_{31,22}^{31,1}+\frac{C_{31,22}^{33,1}}{960 \sqrt{2}}-\frac{C_{33,20}^{33,1}}{80 \sqrt{2}}-\frac{7}{320} \sqrt{\frac{3}{5}} C_{33,22}^{33,1}\)

\noindent\(C_{11,2213,3}^{11,2212,1,\text{NonLoc}}= -\frac{1}{96} \sqrt{\frac{7}{2}} C_{00,00}^{60,0}-\frac{1}{96} \sqrt{\frac{7}{6}}
C_{00,20}^{60,0}-\frac{1}{32} \sqrt{\frac{7}{30}} C_{00,22}^{62,0}+\frac{1}{288} \sqrt{\frac{7}{2}} C_{11,00}^{51,0}-\frac{1}{72} \sqrt{\frac{7}{30}}
C_{11,22}^{51,0}+\frac{9 C_{11,22}^{53,0}}{320 \sqrt{2}}+\frac{1}{288} \sqrt{\frac{7}{2}} C_{20,00}^{40,0}+\frac{1}{288} \sqrt{\frac{7}{6}} C_{20,20}^{40,0}-\frac{1}{72}
\sqrt{\frac{7}{30}} C_{20,22}^{42,0}-\frac{1}{288} \sqrt{\frac{7}{10}} C_{22,00}^{42,0}-\frac{1}{288} \sqrt{\frac{7}{30}} C_{22,20}^{42,0}+\frac{1}{288}
\sqrt{\frac{35}{6}} C_{22,22}^{40,0}+\frac{C_{22,22}^{42,0}}{18 \sqrt{30}}-\frac{1}{32} \sqrt{\frac{3}{14}} C_{22,22}^{44,0}-\frac{1}{288} \sqrt{\frac{7}{2}}
C_{31,00}^{31,0}-\frac{1}{288} \sqrt{\frac{7}{6}} C_{31,20}^{31,0}-\frac{1}{144} \sqrt{\frac{7}{30}} C_{31,22}^{31,0}+\frac{C_{31,22}^{33,0}}{960
\sqrt{2}}+\frac{1}{160} \sqrt{\frac{3}{2}} C_{33,00}^{33,0}+\frac{C_{33,20}^{33,0}}{160 \sqrt{2}}-\frac{7}{320} \sqrt{\frac{3}{5}} C_{33,22}^{33,0}\)

\noindent\(C_{11,2213,3}^{11,2213,0,\text{Loc}}= -\frac{\sqrt{7} C_{00,20}^{60,0}}{576}-\frac{\sqrt{7} C_{00,20}^{60,1}}{1152}+\frac{1}{384}
\sqrt{\frac{7}{5}} C_{00,22}^{62,0}+\frac{1}{768} \sqrt{\frac{7}{5}} C_{00,22}^{62,1}-\frac{13 \sqrt{\frac{7}{5}} C_{11,22}^{51,0}}{3456}-\frac{13
\sqrt{\frac{7}{5}} C_{11,22}^{51,1}}{6912}+\frac{3 \sqrt{3} C_{11,22}^{53,0}}{640}+\frac{3 \sqrt{3} C_{11,22}^{53,1}}{1280}-\frac{11 \sqrt{7} C_{20,20}^{40,0}}{1728}-\frac{11
\sqrt{7} C_{20,20}^{40,1}}{3456}+\frac{\sqrt{\frac{7}{5}} C_{20,22}^{42,0}}{3456}+\frac{\sqrt{\frac{7}{5}} C_{20,22}^{42,1}}{6912}+\frac{\sqrt{35}
C_{22,20}^{42,0}}{432}+\frac{\sqrt{35} C_{22,20}^{42,1}}{864}+\frac{\sqrt{\frac{7}{5}} C_{22,22}^{40,0}}{3456}+\frac{\sqrt{\frac{7}{5}} C_{22,22}^{40,1}}{6912}-\frac{29
C_{22,22}^{42,0}}{1728 \sqrt{5}}-\frac{29 C_{22,22}^{42,1}}{3456 \sqrt{5}}+\frac{13 C_{22,22}^{44,0}}{320 \sqrt{7}}+\frac{13 C_{22,22}^{44,1}}{640
\sqrt{7}}-\frac{5 \sqrt{7} C_{31,20}^{31,0}}{1728}-\frac{5 \sqrt{7} C_{31,20}^{31,1}}{3456}-\frac{17 \sqrt{\frac{7}{5}} C_{31,22}^{31,0}}{3456}-\frac{17
\sqrt{\frac{7}{5}} C_{31,22}^{31,1}}{6912}+\frac{23 C_{31,22}^{33,0}}{1920 \sqrt{3}}+\frac{23 C_{31,22}^{33,1}}{3840 \sqrt{3}}+\frac{C_{33,20}^{33,0}}{16
\sqrt{3}}+\frac{C_{33,20}^{33,1}}{32 \sqrt{3}}-\frac{11 C_{33,22}^{33,0}}{320 \sqrt{10}}-\frac{11 C_{33,22}^{33,1}}{640 \sqrt{10}}\)

\noindent\(C_{11,2213,3}^{11,2213,0,\text{NonLoc}}= -\frac{1}{768} \sqrt{\frac{7}{3}} C_{00,00}^{60,0}-\frac{1}{384} \sqrt{\frac{7}{3}}
C_{00,00}^{60,1}-\frac{\sqrt{7} C_{00,20}^{60,0}}{2304}-\frac{\sqrt{7} C_{00,20}^{60,1}}{1152}-\frac{1}{768} \sqrt{\frac{7}{5}} C_{00,22}^{62,0}-\frac{1}{384}
\sqrt{\frac{7}{5}} C_{00,22}^{62,1}+\frac{\sqrt{\frac{7}{3}} C_{11,00}^{51,0}}{2304}+\frac{\sqrt{\frac{7}{3}} C_{11,00}^{51,1}}{1152}-\frac{13 \sqrt{\frac{7}{5}}
C_{11,22}^{51,0}}{6912}-\frac{13 \sqrt{\frac{7}{5}} C_{11,22}^{51,1}}{3456}+\frac{3 \sqrt{3} C_{11,22}^{53,0}}{1280}+\frac{3 \sqrt{3} C_{11,22}^{53,1}}{640}-\frac{11
\sqrt{\frac{7}{3}} C_{20,00}^{40,0}}{2304}-\frac{11 \sqrt{\frac{7}{3}} C_{20,00}^{40,1}}{1152}-\frac{11 \sqrt{7} C_{20,20}^{40,0}}{6912}-\frac{11
\sqrt{7} C_{20,20}^{40,1}}{3456}-\frac{\sqrt{\frac{7}{5}} C_{20,22}^{42,0}}{6912}-\frac{\sqrt{\frac{7}{5}} C_{20,22}^{42,1}}{3456}+\frac{1}{576}
\sqrt{\frac{35}{3}} C_{22,00}^{42,0}+\frac{1}{288} \sqrt{\frac{35}{3}} C_{22,00}^{42,1}+\frac{\sqrt{35} C_{22,20}^{42,0}}{1728}+\frac{\sqrt{35} C_{22,20}^{42,1}}{864}-\frac{\sqrt{\frac{7}{5}}
C_{22,22}^{40,0}}{6912}-\frac{\sqrt{\frac{7}{5}} C_{22,22}^{40,1}}{3456}+\frac{29 C_{22,22}^{42,0}}{3456 \sqrt{5}}+\frac{29 C_{22,22}^{42,1}}{1728
\sqrt{5}}-\frac{13 C_{22,22}^{44,0}}{640 \sqrt{7}}-\frac{13 C_{22,22}^{44,1}}{320 \sqrt{7}}+\frac{5 \sqrt{\frac{7}{3}} C_{31,00}^{31,0}}{2304}+\frac{5
\sqrt{\frac{7}{3}} C_{31,00}^{31,1}}{1152}+\frac{5 \sqrt{7} C_{31,20}^{31,0}}{6912}+\frac{5 \sqrt{7} C_{31,20}^{31,1}}{3456}-\frac{17 \sqrt{\frac{7}{5}}
C_{31,22}^{31,0}}{6912}-\frac{17 \sqrt{\frac{7}{5}} C_{31,22}^{31,1}}{3456}+\frac{23 C_{31,22}^{33,0}}{3840 \sqrt{3}}+\frac{23 C_{31,22}^{33,1}}{1920
\sqrt{3}}-\frac{C_{33,00}^{33,0}}{64}-\frac{C_{33,00}^{33,1}}{32}-\frac{C_{33,20}^{33,0}}{64 \sqrt{3}}-\frac{C_{33,20}^{33,1}}{32 \sqrt{3}}-\frac{11
C_{33,22}^{33,0}}{640 \sqrt{10}}-\frac{11 C_{33,22}^{33,1}}{320 \sqrt{10}}\)

\noindent\(C_{11,2213,3}^{11,2213,1,\text{Loc}}= -\frac{1}{384} \sqrt{\frac{7}{3}} C_{00,20}^{60,1}+\frac{1}{256} \sqrt{\frac{7}{15}}
C_{00,22}^{62,1}-\frac{13 \sqrt{\frac{7}{15}} C_{11,22}^{51,1}}{2304}+\frac{9 C_{11,22}^{53,1}}{1280}-\frac{11 \sqrt{\frac{7}{3}} C_{20,20}^{40,1}}{1152}+\frac{\sqrt{\frac{7}{15}}
C_{20,22}^{42,1}}{2304}+\frac{1}{288} \sqrt{\frac{35}{3}} C_{22,20}^{42,1}+\frac{\sqrt{\frac{7}{15}} C_{22,22}^{40,1}}{2304}-\frac{29 C_{22,22}^{42,1}}{1152
\sqrt{15}}+\frac{13}{640} \sqrt{\frac{3}{7}} C_{22,22}^{44,1}-\frac{5 \sqrt{\frac{7}{3}} C_{31,20}^{31,1}}{1152}-\frac{17 \sqrt{\frac{7}{15}} C_{31,22}^{31,1}}{2304}+\frac{23
C_{31,22}^{33,1}}{3840}+\frac{C_{33,20}^{33,1}}{32}-\frac{11}{640} \sqrt{\frac{3}{10}} C_{33,22}^{33,1}\)

\noindent\(C_{11,2213,3}^{11,2213,1,\text{NonLoc}}= -\frac{\sqrt{7} C_{00,00}^{60,0}}{768}-\frac{1}{768} \sqrt{\frac{7}{3}} C_{00,20}^{60,0}-\frac{1}{256}
\sqrt{\frac{7}{15}} C_{00,22}^{62,0}+\frac{\sqrt{7} C_{11,00}^{51,0}}{2304}-\frac{13 \sqrt{\frac{7}{15}} C_{11,22}^{51,0}}{2304}+\frac{9 C_{11,22}^{53,0}}{1280}-\frac{11
\sqrt{7} C_{20,00}^{40,0}}{2304}-\frac{11 \sqrt{\frac{7}{3}} C_{20,20}^{40,0}}{2304}-\frac{\sqrt{\frac{7}{15}} C_{20,22}^{42,0}}{2304}+\frac{\sqrt{35}
C_{22,00}^{42,0}}{576}+\frac{1}{576} \sqrt{\frac{35}{3}} C_{22,20}^{42,0}-\frac{\sqrt{\frac{7}{15}} C_{22,22}^{40,0}}{2304}+\frac{29 C_{22,22}^{42,0}}{1152
\sqrt{15}}-\frac{13}{640} \sqrt{\frac{3}{7}} C_{22,22}^{44,0}+\frac{5 \sqrt{7} C_{31,00}^{31,0}}{2304}+\frac{5 \sqrt{\frac{7}{3}} C_{31,20}^{31,0}}{2304}-\frac{17
\sqrt{\frac{7}{15}} C_{31,22}^{31,0}}{2304}+\frac{23 C_{31,22}^{33,0}}{3840}-\frac{\sqrt{3} C_{33,00}^{33,0}}{64}-\frac{C_{33,20}^{33,0}}{64}-\frac{11}{640}
\sqrt{\frac{3}{10}} C_{33,22}^{33,0}\)

\noindent\(C_{11,2213,4}^{11,2213,0,\text{Loc}}= -\frac{3 C_{00,20}^{60,0}}{64}-\frac{3 C_{00,20}^{60,1}}{128}-\frac{3 C_{00,22}^{62,0}}{128
\sqrt{5}}-\frac{3 C_{00,22}^{62,1}}{256 \sqrt{5}}-\frac{C_{11,22}^{51,0}}{128 \sqrt{5}}-\frac{C_{11,22}^{51,1}}{256 \sqrt{5}}-\frac{3}{640} \sqrt{\frac{3}{7}}
C_{11,22}^{53,0}-\frac{3 \sqrt{\frac{3}{7}} C_{11,22}^{53,1}}{1280}-\frac{C_{20,20}^{40,0}}{192}-\frac{C_{20,20}^{40,1}}{384}-\frac{C_{20,22}^{42,0}}{384
\sqrt{5}}-\frac{C_{20,22}^{42,1}}{768 \sqrt{5}}+\frac{C_{22,20}^{42,0}}{48 \sqrt{5}}+\frac{C_{22,20}^{42,1}}{96 \sqrt{5}}-\frac{C_{22,22}^{40,0}}{384
\sqrt{5}}-\frac{C_{22,22}^{40,1}}{768 \sqrt{5}}+\frac{1}{192} \sqrt{\frac{5}{7}} C_{22,22}^{42,0}+\frac{1}{384} \sqrt{\frac{5}{7}} C_{22,22}^{42,1}+\frac{3
C_{22,22}^{44,0}}{2240}+\frac{3 C_{22,22}^{44,1}}{4480}+\frac{C_{31,20}^{31,0}}{192}+\frac{C_{31,20}^{31,1}}{384}+\frac{C_{31,22}^{31,0}}{384 \sqrt{5}}+\frac{C_{31,22}^{31,1}}{768
\sqrt{5}}+\frac{1}{640} \sqrt{\frac{3}{7}} C_{31,22}^{33,0}+\frac{\sqrt{\frac{3}{7}} C_{31,22}^{33,1}}{1280}+\frac{3}{80} \sqrt{\frac{3}{7}} C_{33,20}^{33,0}+\frac{3}{160}
\sqrt{\frac{3}{7}} C_{33,20}^{33,1}+\frac{27 C_{33,22}^{33,0}}{320 \sqrt{70}}+\frac{27 C_{33,22}^{33,1}}{640 \sqrt{70}}\)

\noindent\(C_{11,2213,4}^{11,2213,0,\text{NonLoc}}= -\frac{3 \sqrt{3} C_{00,00}^{60,0}}{256}-\frac{3 \sqrt{3} C_{00,00}^{60,1}}{128}-\frac{3
C_{00,20}^{60,0}}{256}-\frac{3 C_{00,20}^{60,1}}{128}+\frac{3 C_{00,22}^{62,0}}{256 \sqrt{5}}+\frac{3 C_{00,22}^{62,1}}{128 \sqrt{5}}+\frac{\sqrt{3}
C_{11,00}^{51,0}}{256}+\frac{\sqrt{3} C_{11,00}^{51,1}}{128}-\frac{C_{11,22}^{51,0}}{256 \sqrt{5}}-\frac{C_{11,22}^{51,1}}{128 \sqrt{5}}-\frac{3
\sqrt{\frac{3}{7}} C_{11,22}^{53,0}}{1280}-\frac{3}{640} \sqrt{\frac{3}{7}} C_{11,22}^{53,1}-\frac{C_{20,00}^{40,0}}{256 \sqrt{3}}-\frac{C_{20,00}^{40,1}}{128
\sqrt{3}}-\frac{C_{20,20}^{40,0}}{768}-\frac{C_{20,20}^{40,1}}{384}+\frac{C_{20,22}^{42,0}}{768 \sqrt{5}}+\frac{C_{20,22}^{42,1}}{384 \sqrt{5}}+\frac{C_{22,00}^{42,0}}{64
\sqrt{15}}+\frac{C_{22,00}^{42,1}}{32 \sqrt{15}}+\frac{C_{22,20}^{42,0}}{192 \sqrt{5}}+\frac{C_{22,20}^{42,1}}{96 \sqrt{5}}+\frac{C_{22,22}^{40,0}}{768
\sqrt{5}}+\frac{C_{22,22}^{40,1}}{384 \sqrt{5}}-\frac{1}{384} \sqrt{\frac{5}{7}} C_{22,22}^{42,0}-\frac{1}{192} \sqrt{\frac{5}{7}} C_{22,22}^{42,1}-\frac{3
C_{22,22}^{44,0}}{4480}-\frac{3 C_{22,22}^{44,1}}{2240}-\frac{C_{31,00}^{31,0}}{256 \sqrt{3}}-\frac{C_{31,00}^{31,1}}{128 \sqrt{3}}-\frac{C_{31,20}^{31,0}}{768}-\frac{C_{31,20}^{31,1}}{384}+\frac{C_{31,22}^{31,0}}{768
\sqrt{5}}+\frac{C_{31,22}^{31,1}}{384 \sqrt{5}}+\frac{\sqrt{\frac{3}{7}} C_{31,22}^{33,0}}{1280}+\frac{1}{640} \sqrt{\frac{3}{7}} C_{31,22}^{33,1}-\frac{9
C_{33,00}^{33,0}}{320 \sqrt{7}}-\frac{9 C_{33,00}^{33,1}}{160 \sqrt{7}}-\frac{3}{320} \sqrt{\frac{3}{7}} C_{33,20}^{33,0}-\frac{3}{160} \sqrt{\frac{3}{7}}
C_{33,20}^{33,1}+\frac{27 C_{33,22}^{33,0}}{640 \sqrt{70}}+\frac{27 C_{33,22}^{33,1}}{320 \sqrt{70}}\)

\noindent\(C_{11,2213,4}^{11,2213,1,\text{Loc}}= -\frac{3 \sqrt{3} C_{00,20}^{60,1}}{128}-\frac{3}{256} \sqrt{\frac{3}{5}} C_{00,22}^{62,1}-\frac{1}{256}
\sqrt{\frac{3}{5}} C_{11,22}^{51,1}-\frac{9 C_{11,22}^{53,1}}{1280 \sqrt{7}}-\frac{C_{20,20}^{40,1}}{128 \sqrt{3}}-\frac{C_{20,22}^{42,1}}{256 \sqrt{15}}+\frac{C_{22,20}^{42,1}}{32
\sqrt{15}}-\frac{C_{22,22}^{40,1}}{256 \sqrt{15}}+\frac{1}{128} \sqrt{\frac{5}{21}} C_{22,22}^{42,1}+\frac{3 \sqrt{3} C_{22,22}^{44,1}}{4480}+\frac{C_{31,20}^{31,1}}{128
\sqrt{3}}+\frac{C_{31,22}^{31,1}}{256 \sqrt{15}}+\frac{3 C_{31,22}^{33,1}}{1280 \sqrt{7}}+\frac{9 C_{33,20}^{33,1}}{160 \sqrt{7}}+\frac{27}{640}
\sqrt{\frac{3}{70}} C_{33,22}^{33,1}\)

\noindent\(C_{11,2213,4}^{11,2213,1,\text{NonLoc}}= -\frac{9 C_{00,00}^{60,0}}{256}-\frac{3 \sqrt{3} C_{00,20}^{60,0}}{256}+\frac{3}{256}
\sqrt{\frac{3}{5}} C_{00,22}^{62,0}+\frac{3 C_{11,00}^{51,0}}{256}-\frac{1}{256} \sqrt{\frac{3}{5}} C_{11,22}^{51,0}-\frac{9 C_{11,22}^{53,0}}{1280
\sqrt{7}}-\frac{C_{20,00}^{40,0}}{256}-\frac{C_{20,20}^{40,0}}{256 \sqrt{3}}+\frac{C_{20,22}^{42,0}}{256 \sqrt{15}}+\frac{C_{22,00}^{42,0}}{64 \sqrt{5}}+\frac{C_{22,20}^{42,0}}{64
\sqrt{15}}+\frac{C_{22,22}^{40,0}}{256 \sqrt{15}}-\frac{1}{128} \sqrt{\frac{5}{21}} C_{22,22}^{42,0}-\frac{3 \sqrt{3} C_{22,22}^{44,0}}{4480}-\frac{C_{31,00}^{31,0}}{256}-\frac{C_{31,20}^{31,0}}{256
\sqrt{3}}+\frac{C_{31,22}^{31,0}}{256 \sqrt{15}}+\frac{3 C_{31,22}^{33,0}}{1280 \sqrt{7}}-\frac{9}{320} \sqrt{\frac{3}{7}} C_{33,00}^{33,0}-\frac{9
C_{33,20}^{33,0}}{320 \sqrt{7}}+\frac{27}{640} \sqrt{\frac{3}{70}} C_{33,22}^{33,0}\)

\noindent\(C_{11,3101,0}^{11,1101,0,\text{Loc}}= -\frac{7 C_{00,00}^{60,0}}{32}-\frac{7 C_{00,00}^{60,1}}{64}-\frac{7 C_{11,00}^{51,0}}{96}-\frac{7
C_{11,00}^{51,1}}{192}+\frac{C_{20,00}^{40,0}}{32}+\frac{C_{20,00}^{40,1}}{64}+\frac{5 C_{31,00}^{31,0}}{96}+\frac{5 C_{31,00}^{31,1}}{192}\)

\noindent\(C_{11,3101,0}^{11,1101,0,\text{NonLoc}}= \frac{7 C_{00,00}^{60,0}}{128}+\frac{7 C_{00,00}^{60,1}}{64}-\frac{7 \sqrt{3} C_{00,20}^{60,0}}{128}-\frac{7
\sqrt{3} C_{00,20}^{60,1}}{64}-\frac{7 C_{11,00}^{51,0}}{384}-\frac{7 C_{11,00}^{51,1}}{192}+\frac{7 C_{11,20}^{31,0}}{128 \sqrt{3}}+\frac{7 C_{11,20}^{31,1}}{64
\sqrt{3}}-\frac{C_{20,00}^{40,0}}{128}-\frac{C_{20,00}^{40,1}}{64}+\frac{\sqrt{3} C_{20,20}^{40,0}}{128}+\frac{\sqrt{3} C_{20,20}^{40,1}}{64}+\frac{5
C_{31,00}^{31,0}}{384}+\frac{5 C_{31,00}^{31,1}}{192}-\frac{5 C_{31,20}^{31,0}}{128 \sqrt{3}}-\frac{5 C_{31,20}^{31,1}}{64 \sqrt{3}}\)

\noindent\(C_{11,3101,0}^{11,1101,1,\text{Loc}}= -\frac{7 \sqrt{3} C_{00,00}^{60,1}}{64}-\frac{7 C_{11,00}^{51,1}}{64 \sqrt{3}}+\frac{\sqrt{3}
C_{20,00}^{40,1}}{64}+\frac{5 C_{31,00}^{31,1}}{64 \sqrt{3}}\)

\noindent\(C_{11,3101,0}^{11,1101,1,\text{NonLoc}}= \frac{7 \sqrt{3} C_{00,00}^{60,0}}{128}-\frac{21 C_{00,20}^{60,0}}{128}-\frac{7
C_{11,00}^{51,0}}{128 \sqrt{3}}+\frac{7 C_{11,20}^{31,0}}{128}-\frac{\sqrt{3} C_{20,00}^{40,0}}{128}+\frac{3 C_{20,20}^{40,0}}{128}+\frac{5 C_{31,00}^{31,0}}{128
\sqrt{3}}-\frac{5 C_{31,20}^{31,0}}{128}\)

\noindent\(C_{11,3101,1}^{00,2011,0,\text{Loc}}= -\frac{7 C_{11,11}^{51,0}}{96}-\frac{7 C_{11,11}^{51,1}}{192}-\frac{7 C_{22,11}^{42,0}}{16
\sqrt{15}}-\frac{7 C_{22,11}^{42,1}}{32 \sqrt{15}}-\frac{7 C_{31,11}^{31,0}}{96}-\frac{7 C_{31,11}^{31,1}}{192}-\frac{3}{40} \sqrt{\frac{7}{2}} C_{33,11}^{33,0}-\frac{3}{80}
\sqrt{\frac{7}{2}} C_{33,11}^{33,1}\)

\noindent\(C_{11,3101,1}^{00,2011,0,\text{NonLoc}}= -\frac{7 C_{11,11}^{51,0}}{192}-\frac{7 C_{11,11}^{51,1}}{96}+\frac{7 C_{22,11}^{42,0}}{32
\sqrt{15}}+\frac{7 C_{22,11}^{42,1}}{16 \sqrt{15}}-\frac{7 C_{31,11}^{31,0}}{192}-\frac{7 C_{31,11}^{31,1}}{96}-\frac{3}{80} \sqrt{\frac{7}{2}} C_{33,11}^{33,0}-\frac{3}{40}
\sqrt{\frac{7}{2}} C_{33,11}^{33,1}\)

\noindent\(C_{11,3101,1}^{00,2011,1,\text{Loc}}= -\frac{7 C_{11,11}^{51,1}}{64 \sqrt{3}}-\frac{7 C_{22,11}^{42,1}}{32 \sqrt{5}}-\frac{7
C_{31,11}^{31,1}}{64 \sqrt{3}}-\frac{3}{80} \sqrt{\frac{21}{2}} C_{33,11}^{33,1}\)

\noindent\(C_{11,3101,1}^{00,2011,1,\text{NonLoc}}= -\frac{7 C_{11,11}^{51,0}}{64 \sqrt{3}}+\frac{7 C_{22,11}^{42,0}}{32 \sqrt{5}}-\frac{7
C_{31,11}^{31,0}}{64 \sqrt{3}}-\frac{3}{80} \sqrt{\frac{21}{2}} C_{33,11}^{33,0}\)

\noindent\(C_{11,3101,1}^{00,2211,0,\text{Loc}}= \frac{7 C_{11,11}^{51,0}}{96 \sqrt{5}}+\frac{7 C_{11,11}^{51,1}}{192 \sqrt{5}}+\frac{7
C_{22,11}^{42,0}}{80 \sqrt{3}}+\frac{7 C_{22,11}^{42,1}}{160 \sqrt{3}}+\frac{7 C_{31,11}^{31,0}}{96 \sqrt{5}}+\frac{7 C_{31,11}^{31,1}}{192 \sqrt{5}}+\frac{3}{40}
\sqrt{\frac{7}{10}} C_{33,11}^{33,0}+\frac{3}{80} \sqrt{\frac{7}{10}} C_{33,11}^{33,1}\)

\noindent\(C_{11,3101,1}^{00,2211,0,\text{NonLoc}}= \frac{7 C_{11,11}^{51,0}}{192 \sqrt{5}}+\frac{7 C_{11,11}^{51,1}}{96 \sqrt{5}}-\frac{7
C_{22,11}^{42,0}}{160 \sqrt{3}}-\frac{7 C_{22,11}^{42,1}}{80 \sqrt{3}}+\frac{7 C_{31,11}^{31,0}}{192 \sqrt{5}}+\frac{7 C_{31,11}^{31,1}}{96 \sqrt{5}}+\frac{3}{80}
\sqrt{\frac{7}{10}} C_{33,11}^{33,0}+\frac{3}{40} \sqrt{\frac{7}{10}} C_{33,11}^{33,1}\)

\noindent\(C_{11,3101,1}^{00,2211,1,\text{Loc}}= \frac{7 C_{11,11}^{51,1}}{64 \sqrt{15}}+\frac{7 C_{22,11}^{42,1}}{160}+\frac{7 C_{31,11}^{31,1}}{64
\sqrt{15}}+\frac{3}{80} \sqrt{\frac{21}{10}} C_{33,11}^{33,1}\)

\noindent\(C_{11,3101,1}^{00,2211,1,\text{NonLoc}}= \frac{7 C_{11,11}^{51,0}}{64 \sqrt{15}}-\frac{7 C_{22,11}^{42,0}}{160}+\frac{7
C_{31,11}^{31,0}}{64 \sqrt{15}}+\frac{3}{80} \sqrt{\frac{21}{10}} C_{33,11}^{33,0}\)

\noindent\(C_{11,3101,1}^{11,1101,0,\text{Loc}}= -\frac{C_{20,00}^{40,0}}{8 \sqrt{3}}-\frac{C_{20,00}^{40,1}}{16 \sqrt{3}}+\frac{7
C_{22,00}^{42,0}}{32 \sqrt{15}}+\frac{7 C_{22,00}^{42,1}}{64 \sqrt{15}}-\frac{C_{31,00}^{31,0}}{16 \sqrt{3}}-\frac{C_{31,00}^{31,1}}{32 \sqrt{3}}+\frac{9
\sqrt{7} C_{33,00}^{33,0}}{160}+\frac{9 \sqrt{7} C_{33,00}^{33,1}}{320}\)

\noindent\(C_{11,3101,1}^{11,1101,0,\text{NonLoc}}= \frac{C_{20,00}^{40,0}}{32 \sqrt{3}}+\frac{C_{20,00}^{40,1}}{16 \sqrt{3}}-\frac{C_{20,20}^{40,0}}{32}-\frac{C_{20,20}^{40,1}}{16}-\frac{7
C_{22,00}^{42,0}}{128 \sqrt{15}}-\frac{7 C_{22,00}^{42,1}}{64 \sqrt{15}}+\frac{7 C_{22,20}^{42,0}}{128 \sqrt{5}}+\frac{7 C_{22,20}^{42,1}}{64 \sqrt{5}}-\frac{C_{31,00}^{31,0}}{64
\sqrt{3}}-\frac{C_{31,00}^{31,1}}{32 \sqrt{3}}+\frac{C_{31,20}^{31,0}}{64}+\frac{C_{31,20}^{31,1}}{32}+\frac{9 \sqrt{7} C_{33,00}^{33,0}}{640}+\frac{9
\sqrt{7} C_{33,00}^{33,1}}{320}-\frac{9 \sqrt{21} C_{33,20}^{33,0}}{640}-\frac{9 \sqrt{21} C_{33,20}^{33,1}}{320}\)

\noindent\(C_{11,3101,1}^{11,1101,1,\text{Loc}}= -\frac{C_{20,00}^{40,1}}{16}+\frac{7 C_{22,00}^{42,1}}{64 \sqrt{5}}-\frac{C_{31,00}^{31,1}}{32}+\frac{9
\sqrt{21} C_{33,00}^{33,1}}{320}\)

\noindent\(C_{11,3101,1}^{11,1101,1,\text{NonLoc}}= \frac{C_{20,00}^{40,0}}{32}-\frac{\sqrt{3} C_{20,20}^{40,0}}{32}-\frac{7 C_{22,00}^{42,0}}{128
\sqrt{5}}+\frac{7}{128} \sqrt{\frac{3}{5}} C_{22,20}^{42,0}-\frac{C_{31,00}^{31,0}}{64}+\frac{\sqrt{3} C_{31,20}^{31,0}}{64}+\frac{9 \sqrt{21} C_{33,00}^{33,0}}{640}-\frac{27
\sqrt{7} C_{33,20}^{33,0}}{640}\)

\noindent\(C_{11,3101,2}^{00,2212,0,\text{Loc}}= -\frac{7 C_{11,11}^{51,0}}{32 \sqrt{15}}-\frac{7 C_{11,11}^{51,1}}{64 \sqrt{15}}-\frac{7
C_{22,11}^{42,0}}{80}-\frac{7 C_{22,11}^{42,1}}{160}-\frac{7 C_{31,11}^{31,0}}{32 \sqrt{15}}-\frac{7 C_{31,11}^{31,1}}{64 \sqrt{15}}-\frac{3}{40}
\sqrt{\frac{21}{10}} C_{33,11}^{33,0}-\frac{3}{80} \sqrt{\frac{21}{10}} C_{33,11}^{33,1}\)

\noindent\(C_{11,3101,2}^{00,2212,0,\text{NonLoc}}= -\frac{7 C_{11,11}^{51,0}}{64 \sqrt{15}}-\frac{7 C_{11,11}^{51,1}}{32 \sqrt{15}}+\frac{7
C_{22,11}^{42,0}}{160}+\frac{7 C_{22,11}^{42,1}}{80}-\frac{7 C_{31,11}^{31,0}}{64 \sqrt{15}}-\frac{7 C_{31,11}^{31,1}}{32 \sqrt{15}}-\frac{3}{80}
\sqrt{\frac{21}{10}} C_{33,11}^{33,0}-\frac{3}{40} \sqrt{\frac{21}{10}} C_{33,11}^{33,1}\)

\noindent\(C_{11,3101,2}^{00,2212,1,\text{Loc}}= -\frac{7 C_{11,11}^{51,1}}{64 \sqrt{5}}-\frac{7 \sqrt{3} C_{22,11}^{42,1}}{160}-\frac{7
C_{31,11}^{31,1}}{64 \sqrt{5}}-\frac{9}{80} \sqrt{\frac{7}{10}} C_{33,11}^{33,1}\)

\noindent\(C_{11,3101,2}^{00,2212,1,\text{NonLoc}}= -\frac{7 C_{11,11}^{51,0}}{64 \sqrt{5}}+\frac{7 \sqrt{3} C_{22,11}^{42,0}}{160}-\frac{7
C_{31,11}^{31,0}}{64 \sqrt{5}}-\frac{9}{80} \sqrt{\frac{7}{10}} C_{33,11}^{33,0}\)

\noindent\(C_{11,3101,2}^{11,1101,0,\text{Loc}}= -\frac{7 C_{00,00}^{60,0}}{16 \sqrt{5}}-\frac{7 C_{00,00}^{60,1}}{32 \sqrt{5}}-\frac{7
C_{11,00}^{51,0}}{48 \sqrt{5}}-\frac{7 C_{11,00}^{51,1}}{96 \sqrt{5}}-\frac{C_{20,00}^{40,0}}{16 \sqrt{5}}-\frac{C_{20,00}^{40,1}}{32 \sqrt{5}}+\frac{7
C_{22,00}^{42,0}}{160}+\frac{7 C_{22,00}^{42,1}}{320}+\frac{C_{31,00}^{31,0}}{24 \sqrt{5}}+\frac{C_{31,00}^{31,1}}{48 \sqrt{5}}+\frac{9}{160} \sqrt{\frac{21}{5}}
C_{33,00}^{33,0}+\frac{9}{320} \sqrt{\frac{21}{5}} C_{33,00}^{33,1}\)

\noindent\(C_{11,3101,2}^{11,1101,0,\text{NonLoc}}= \frac{7 C_{00,00}^{60,0}}{64 \sqrt{5}}+\frac{7 C_{00,00}^{60,1}}{32 \sqrt{5}}-\frac{7}{64}
\sqrt{\frac{3}{5}} C_{00,20}^{60,0}-\frac{7}{32} \sqrt{\frac{3}{5}} C_{00,20}^{60,1}-\frac{7 C_{11,00}^{51,0}}{192 \sqrt{5}}-\frac{7 C_{11,00}^{51,1}}{96
\sqrt{5}}+\frac{7 C_{11,20}^{31,0}}{64 \sqrt{15}}+\frac{7 C_{11,20}^{31,1}}{32 \sqrt{15}}+\frac{C_{20,00}^{40,0}}{64 \sqrt{5}}+\frac{C_{20,00}^{40,1}}{32
\sqrt{5}}-\frac{1}{64} \sqrt{\frac{3}{5}} C_{20,20}^{40,0}-\frac{1}{32} \sqrt{\frac{3}{5}} C_{20,20}^{40,1}-\frac{7 C_{22,00}^{42,0}}{640}-\frac{7
C_{22,00}^{42,1}}{320}+\frac{7 \sqrt{3} C_{22,20}^{42,0}}{640}+\frac{7 \sqrt{3} C_{22,20}^{42,1}}{320}+\frac{C_{31,00}^{31,0}}{96 \sqrt{5}}+\frac{C_{31,00}^{31,1}}{48
\sqrt{5}}-\frac{C_{31,20}^{31,0}}{32 \sqrt{15}}-\frac{C_{31,20}^{31,1}}{16 \sqrt{15}}+\frac{9}{640} \sqrt{\frac{21}{5}} C_{33,00}^{33,0}+\frac{9}{320}
\sqrt{\frac{21}{5}} C_{33,00}^{33,1}-\frac{27}{640} \sqrt{\frac{7}{5}} C_{33,20}^{33,0}-\frac{27}{320} \sqrt{\frac{7}{5}} C_{33,20}^{33,1}\)

\noindent\(C_{11,3101,2}^{11,1101,1,\text{Loc}}= -\frac{7}{32} \sqrt{\frac{3}{5}} C_{00,00}^{60,1}-\frac{7 C_{11,00}^{51,1}}{32 \sqrt{15}}-\frac{1}{32}
\sqrt{\frac{3}{5}} C_{20,00}^{40,1}+\frac{7 \sqrt{3} C_{22,00}^{42,1}}{320}+\frac{C_{31,00}^{31,1}}{16 \sqrt{15}}+\frac{27}{320} \sqrt{\frac{7}{5}}
C_{33,00}^{33,1}\)

\noindent\(C_{11,3101,2}^{11,1101,1,\text{NonLoc}}= \frac{7}{64} \sqrt{\frac{3}{5}} C_{00,00}^{60,0}-\frac{21 C_{00,20}^{60,0}}{64
\sqrt{5}}-\frac{7 C_{11,00}^{51,0}}{64 \sqrt{15}}+\frac{7 C_{11,20}^{31,0}}{64 \sqrt{5}}+\frac{1}{64} \sqrt{\frac{3}{5}} C_{20,00}^{40,0}-\frac{3
C_{20,20}^{40,0}}{64 \sqrt{5}}-\frac{7 \sqrt{3} C_{22,00}^{42,0}}{640}+\frac{21 C_{22,20}^{42,0}}{640}+\frac{C_{31,00}^{31,0}}{32 \sqrt{15}}-\frac{C_{31,20}^{31,0}}{32
\sqrt{5}}+\frac{27}{640} \sqrt{\frac{7}{5}} C_{33,00}^{33,0}-\frac{27}{640} \sqrt{\frac{21}{5}} C_{33,20}^{33,0}\)

\noindent\(C_{11,3110,1}^{11,1110,0,\text{Loc}}= \frac{7 C_{00,20}^{60,0}}{96}+\frac{7 C_{00,20}^{60,1}}{192}+\frac{7 C_{00,22}^{62,0}}{32
\sqrt{5}}+\frac{7 C_{00,22}^{62,1}}{64 \sqrt{5}}+\frac{7 C_{11,20}^{31,0}}{288}+\frac{7 C_{11,20}^{31,1}}{576}+\frac{7 \sqrt{5} C_{11,22}^{51,0}}{288}+\frac{7
\sqrt{5} C_{11,22}^{51,1}}{576}+\frac{5 C_{20,20}^{40,0}}{288}+\frac{5 C_{20,20}^{40,1}}{576}+\frac{7 C_{20,22}^{42,0}}{288 \sqrt{5}}+\frac{7 C_{20,22}^{42,1}}{576
\sqrt{5}}-\frac{7 C_{22,20}^{42,0}}{144 \sqrt{5}}-\frac{7 C_{22,20}^{42,1}}{288 \sqrt{5}}+\frac{7 C_{22,22}^{40,0}}{288 \sqrt{5}}+\frac{7 C_{22,22}^{40,1}}{576
\sqrt{5}}-\frac{1}{72} \sqrt{\frac{7}{5}} C_{22,22}^{42,0}-\frac{1}{144} \sqrt{\frac{7}{5}} C_{22,22}^{42,1}-\frac{3 C_{22,22}^{44,0}}{40}-\frac{3
C_{22,22}^{44,1}}{80}-\frac{C_{31,20}^{31,0}}{288}-\frac{C_{31,20}^{31,1}}{576}+\frac{7 C_{31,22}^{31,0}}{288 \sqrt{5}}+\frac{7 C_{31,22}^{31,1}}{576
\sqrt{5}}-\frac{1}{40} \sqrt{\frac{7}{3}} C_{31,22}^{33,0}-\frac{1}{80} \sqrt{\frac{7}{3}} C_{31,22}^{33,1}-\frac{\sqrt{21} C_{33,20}^{33,0}}{80}-\frac{\sqrt{21}
C_{33,20}^{33,1}}{160}-\frac{3}{40} \sqrt{\frac{7}{10}} C_{33,22}^{33,0}-\frac{3}{80} \sqrt{\frac{7}{10}} C_{33,22}^{33,1}\)

\noindent\(C_{11,3110,1}^{11,1110,0,\text{NonLoc}}= \frac{7 C_{00,00}^{60,0}}{128 \sqrt{3}}+\frac{7 C_{00,00}^{60,1}}{64 \sqrt{3}}+\frac{7
C_{00,20}^{60,0}}{384}+\frac{7 C_{00,20}^{60,1}}{192}-\frac{7 C_{00,22}^{62,0}}{64 \sqrt{5}}-\frac{7 C_{00,22}^{62,1}}{32 \sqrt{5}}-\frac{7 C_{11,00}^{51,0}}{384
\sqrt{3}}-\frac{7 C_{11,00}^{51,1}}{192 \sqrt{3}}-\frac{7 C_{11,20}^{31,0}}{1152}-\frac{7 C_{11,20}^{31,1}}{576}+\frac{7 \sqrt{5} C_{11,22}^{51,0}}{576}+\frac{7
\sqrt{5} C_{11,22}^{51,1}}{288}+\frac{5 C_{20,00}^{40,0}}{384 \sqrt{3}}+\frac{5 C_{20,00}^{40,1}}{192 \sqrt{3}}+\frac{5 C_{20,20}^{40,0}}{1152}+\frac{5
C_{20,20}^{40,1}}{576}-\frac{7 C_{20,22}^{42,0}}{576 \sqrt{5}}-\frac{7 C_{20,22}^{42,1}}{288 \sqrt{5}}-\frac{7 C_{22,00}^{42,0}}{192 \sqrt{15}}-\frac{7
C_{22,00}^{42,1}}{96 \sqrt{15}}-\frac{7 C_{22,20}^{42,0}}{576 \sqrt{5}}-\frac{7 C_{22,20}^{42,1}}{288 \sqrt{5}}-\frac{7 C_{22,22}^{40,0}}{576 \sqrt{5}}-\frac{7
C_{22,22}^{40,1}}{288 \sqrt{5}}+\frac{1}{144} \sqrt{\frac{7}{5}} C_{22,22}^{42,0}+\frac{1}{72} \sqrt{\frac{7}{5}} C_{22,22}^{42,1}+\frac{3 C_{22,22}^{44,0}}{80}+\frac{3
C_{22,22}^{44,1}}{40}+\frac{C_{31,00}^{31,0}}{384 \sqrt{3}}+\frac{C_{31,00}^{31,1}}{192 \sqrt{3}}+\frac{C_{31,20}^{31,0}}{1152}+\frac{C_{31,20}^{31,1}}{576}+\frac{7
C_{31,22}^{31,0}}{576 \sqrt{5}}+\frac{7 C_{31,22}^{31,1}}{288 \sqrt{5}}-\frac{1}{80} \sqrt{\frac{7}{3}} C_{31,22}^{33,0}-\frac{1}{40} \sqrt{\frac{7}{3}}
C_{31,22}^{33,1}+\frac{3 \sqrt{7} C_{33,00}^{33,0}}{320}+\frac{3 \sqrt{7} C_{33,00}^{33,1}}{160}+\frac{\sqrt{21} C_{33,20}^{33,0}}{320}+\frac{\sqrt{21}
C_{33,20}^{33,1}}{160}-\frac{3}{80} \sqrt{\frac{7}{10}} C_{33,22}^{33,0}-\frac{3}{40} \sqrt{\frac{7}{10}} C_{33,22}^{33,1}\)

\noindent\(C_{11,3110,1}^{11,1110,1,\text{Loc}}= \frac{7 C_{00,20}^{60,1}}{64 \sqrt{3}}+\frac{7}{64} \sqrt{\frac{3}{5}} C_{00,22}^{62,1}+\frac{7
C_{11,20}^{31,1}}{192 \sqrt{3}}+\frac{7}{192} \sqrt{\frac{5}{3}} C_{11,22}^{51,1}+\frac{5 C_{20,20}^{40,1}}{192 \sqrt{3}}+\frac{7 C_{20,22}^{42,1}}{192
\sqrt{15}}-\frac{7 C_{22,20}^{42,1}}{96 \sqrt{15}}+\frac{7 C_{22,22}^{40,1}}{192 \sqrt{15}}-\frac{1}{48} \sqrt{\frac{7}{15}} C_{22,22}^{42,1}-\frac{3
\sqrt{3} C_{22,22}^{44,1}}{80}-\frac{C_{31,20}^{31,1}}{192 \sqrt{3}}+\frac{7 C_{31,22}^{31,1}}{192 \sqrt{15}}-\frac{\sqrt{7} C_{31,22}^{33,1}}{80}-\frac{3
\sqrt{7} C_{33,20}^{33,1}}{160}-\frac{3}{80} \sqrt{\frac{21}{10}} C_{33,22}^{33,1}\)

\noindent\(C_{11,3110,1}^{11,1110,1,\text{NonLoc}}= \frac{7 C_{00,00}^{60,0}}{128}+\frac{7 C_{00,20}^{60,0}}{128 \sqrt{3}}-\frac{7}{64}
\sqrt{\frac{3}{5}} C_{00,22}^{62,0}-\frac{7 C_{11,00}^{51,0}}{384}-\frac{7 C_{11,20}^{31,0}}{384 \sqrt{3}}+\frac{7}{192} \sqrt{\frac{5}{3}} C_{11,22}^{51,0}+\frac{5
C_{20,00}^{40,0}}{384}+\frac{5 C_{20,20}^{40,0}}{384 \sqrt{3}}-\frac{7 C_{20,22}^{42,0}}{192 \sqrt{15}}-\frac{7 C_{22,00}^{42,0}}{192 \sqrt{5}}-\frac{7
C_{22,20}^{42,0}}{192 \sqrt{15}}-\frac{7 C_{22,22}^{40,0}}{192 \sqrt{15}}+\frac{1}{48} \sqrt{\frac{7}{15}} C_{22,22}^{42,0}+\frac{3 \sqrt{3} C_{22,22}^{44,0}}{80}+\frac{C_{31,00}^{31,0}}{384}+\frac{C_{31,20}^{31,0}}{384
\sqrt{3}}+\frac{7 C_{31,22}^{31,0}}{192 \sqrt{15}}-\frac{\sqrt{7} C_{31,22}^{33,0}}{80}+\frac{3 \sqrt{21} C_{33,00}^{33,0}}{320}+\frac{3 \sqrt{7}
C_{33,20}^{33,0}}{320}-\frac{3}{80} \sqrt{\frac{21}{10}} C_{33,22}^{33,0}\)

\noindent\(C_{11,3110,1}^{11,1112,0,\text{Loc}}= \frac{7 C_{00,20}^{60,0}}{48 \sqrt{5}}+\frac{7 C_{00,20}^{60,1}}{96 \sqrt{5}}+\frac{7
C_{00,22}^{62,0}}{80}+\frac{7 C_{00,22}^{62,1}}{160}+\frac{7 C_{11,20}^{31,0}}{144 \sqrt{5}}+\frac{7 C_{11,20}^{31,1}}{288 \sqrt{5}}+\frac{7 C_{11,22}^{51,0}}{1440}+\frac{7
C_{11,22}^{51,1}}{2880}+\frac{9}{320} \sqrt{\frac{21}{5}} C_{11,22}^{53,0}+\frac{9}{640} \sqrt{\frac{21}{5}} C_{11,22}^{53,1}-\frac{7 C_{20,20}^{40,0}}{144
\sqrt{5}}-\frac{7 C_{20,20}^{40,1}}{288 \sqrt{5}}-\frac{7 C_{20,22}^{42,0}}{1440}-\frac{7 C_{20,22}^{42,1}}{2880}+\frac{7 C_{22,20}^{42,0}}{720}+\frac{7
C_{22,20}^{42,1}}{1440}-\frac{13 C_{22,22}^{40,0}}{360}-\frac{13 C_{22,22}^{40,1}}{720}+\frac{\sqrt{7} C_{22,22}^{42,0}}{360}+\frac{\sqrt{7} C_{22,22}^{42,1}}{720}+\frac{3
C_{22,22}^{44,0}}{40 \sqrt{5}}+\frac{3 C_{22,22}^{44,1}}{80 \sqrt{5}}-\frac{7 C_{31,20}^{31,0}}{144 \sqrt{5}}-\frac{7 C_{31,20}^{31,1}}{288 \sqrt{5}}-\frac{37
C_{31,22}^{31,0}}{1440}-\frac{37 C_{31,22}^{31,1}}{2880}-\frac{7}{320} \sqrt{\frac{7}{15}} C_{31,22}^{33,0}-\frac{7}{640} \sqrt{\frac{7}{15}} C_{31,22}^{33,1}+\frac{1}{80}
\sqrt{\frac{21}{5}} C_{33,20}^{33,0}+\frac{1}{160} \sqrt{\frac{21}{5}} C_{33,20}^{33,1}+\frac{3}{200} \sqrt{\frac{7}{2}} C_{33,22}^{33,0}+\frac{3}{400}
\sqrt{\frac{7}{2}} C_{33,22}^{33,1}\)

\noindent\(C_{11,3110,1}^{11,1112,0,\text{NonLoc}}= \frac{7 C_{00,00}^{60,0}}{64 \sqrt{15}}+\frac{7 C_{00,00}^{60,1}}{32 \sqrt{15}}+\frac{7
C_{00,20}^{60,0}}{192 \sqrt{5}}+\frac{7 C_{00,20}^{60,1}}{96 \sqrt{5}}-\frac{7 C_{00,22}^{62,0}}{160}-\frac{7 C_{00,22}^{62,1}}{80}-\frac{7 C_{11,00}^{51,0}}{192
\sqrt{15}}-\frac{7 C_{11,00}^{51,1}}{96 \sqrt{15}}-\frac{7 C_{11,20}^{31,0}}{576 \sqrt{5}}-\frac{7 C_{11,20}^{31,1}}{288 \sqrt{5}}+\frac{7 C_{11,22}^{51,0}}{2880}+\frac{7
C_{11,22}^{51,1}}{1440}+\frac{9}{640} \sqrt{\frac{21}{5}} C_{11,22}^{53,0}+\frac{9}{320} \sqrt{\frac{21}{5}} C_{11,22}^{53,1}-\frac{7 C_{20,00}^{40,0}}{192
\sqrt{15}}-\frac{7 C_{20,00}^{40,1}}{96 \sqrt{15}}-\frac{7 C_{20,20}^{40,0}}{576 \sqrt{5}}-\frac{7 C_{20,20}^{40,1}}{288 \sqrt{5}}+\frac{7 C_{20,22}^{42,0}}{2880}+\frac{7
C_{20,22}^{42,1}}{1440}+\frac{7 C_{22,00}^{42,0}}{960 \sqrt{3}}+\frac{7 C_{22,00}^{42,1}}{480 \sqrt{3}}+\frac{7 C_{22,20}^{42,0}}{2880}+\frac{7 C_{22,20}^{42,1}}{1440}+\frac{13
C_{22,22}^{40,0}}{720}+\frac{13 C_{22,22}^{40,1}}{360}-\frac{\sqrt{7} C_{22,22}^{42,0}}{720}-\frac{\sqrt{7} C_{22,22}^{42,1}}{360}-\frac{3 C_{22,22}^{44,0}}{80
\sqrt{5}}-\frac{3 C_{22,22}^{44,1}}{40 \sqrt{5}}+\frac{7 C_{31,00}^{31,0}}{192 \sqrt{15}}+\frac{7 C_{31,00}^{31,1}}{96 \sqrt{15}}+\frac{7 C_{31,20}^{31,0}}{576
\sqrt{5}}+\frac{7 C_{31,20}^{31,1}}{288 \sqrt{5}}-\frac{37 C_{31,22}^{31,0}}{2880}-\frac{37 C_{31,22}^{31,1}}{1440}-\frac{7}{640} \sqrt{\frac{7}{15}}
C_{31,22}^{33,0}-\frac{7}{320} \sqrt{\frac{7}{15}} C_{31,22}^{33,1}-\frac{3}{320} \sqrt{\frac{7}{5}} C_{33,00}^{33,0}-\frac{3}{160} \sqrt{\frac{7}{5}}
C_{33,00}^{33,1}-\frac{1}{320} \sqrt{\frac{21}{5}} C_{33,20}^{33,0}-\frac{1}{160} \sqrt{\frac{21}{5}} C_{33,20}^{33,1}+\frac{3}{400} \sqrt{\frac{7}{2}}
C_{33,22}^{33,0}+\frac{3}{200} \sqrt{\frac{7}{2}} C_{33,22}^{33,1}\)

\noindent\(C_{11,3110,1}^{11,1112,1,\text{Loc}}= \frac{7 C_{00,20}^{60,1}}{32 \sqrt{15}}+\frac{7 \sqrt{3} C_{00,22}^{62,1}}{160}+\frac{7
C_{11,20}^{31,1}}{96 \sqrt{15}}+\frac{7 C_{11,22}^{51,1}}{960 \sqrt{3}}+\frac{27}{640} \sqrt{\frac{7}{5}} C_{11,22}^{53,1}-\frac{7 C_{20,20}^{40,1}}{96
\sqrt{15}}-\frac{7 C_{20,22}^{42,1}}{960 \sqrt{3}}+\frac{7 C_{22,20}^{42,1}}{480 \sqrt{3}}-\frac{13 C_{22,22}^{40,1}}{240 \sqrt{3}}+\frac{1}{240}
\sqrt{\frac{7}{3}} C_{22,22}^{42,1}+\frac{3}{80} \sqrt{\frac{3}{5}} C_{22,22}^{44,1}-\frac{7 C_{31,20}^{31,1}}{96 \sqrt{15}}-\frac{37 C_{31,22}^{31,1}}{960
\sqrt{3}}-\frac{7}{640} \sqrt{\frac{7}{5}} C_{31,22}^{33,1}+\frac{3}{160} \sqrt{\frac{7}{5}} C_{33,20}^{33,1}+\frac{3}{400} \sqrt{\frac{21}{2}} C_{33,22}^{33,1}\)

\noindent\(C_{11,3110,1}^{11,1112,1,\text{NonLoc}}= \frac{7 C_{00,00}^{60,0}}{64 \sqrt{5}}+\frac{7 C_{00,20}^{60,0}}{64 \sqrt{15}}-\frac{7
\sqrt{3} C_{00,22}^{62,0}}{160}-\frac{7 C_{11,00}^{51,0}}{192 \sqrt{5}}-\frac{7 C_{11,20}^{31,0}}{192 \sqrt{15}}+\frac{7 C_{11,22}^{51,0}}{960 \sqrt{3}}+\frac{27}{640}
\sqrt{\frac{7}{5}} C_{11,22}^{53,0}-\frac{7 C_{20,00}^{40,0}}{192 \sqrt{5}}-\frac{7 C_{20,20}^{40,0}}{192 \sqrt{15}}+\frac{7 C_{20,22}^{42,0}}{960
\sqrt{3}}+\frac{7 C_{22,00}^{42,0}}{960}+\frac{7 C_{22,20}^{42,0}}{960 \sqrt{3}}+\frac{13 C_{22,22}^{40,0}}{240 \sqrt{3}}-\frac{1}{240} \sqrt{\frac{7}{3}}
C_{22,22}^{42,0}-\frac{3}{80} \sqrt{\frac{3}{5}} C_{22,22}^{44,0}+\frac{7 C_{31,00}^{31,0}}{192 \sqrt{5}}+\frac{7 C_{31,20}^{31,0}}{192 \sqrt{15}}-\frac{37
C_{31,22}^{31,0}}{960 \sqrt{3}}-\frac{7}{640} \sqrt{\frac{7}{5}} C_{31,22}^{33,0}-\frac{3}{320} \sqrt{\frac{21}{5}} C_{33,00}^{33,0}-\frac{3}{320}
\sqrt{\frac{7}{5}} C_{33,20}^{33,0}+\frac{3}{400} \sqrt{\frac{21}{2}} C_{33,22}^{33,0}\)

\noindent\(C_{11,3111,0}^{00,2000,0,\text{Loc}}= -\frac{7 C_{11,11}^{51,0}}{96}-\frac{7 C_{11,11}^{51,1}}{192}-\frac{7 C_{22,11}^{42,0}}{16
\sqrt{15}}-\frac{7 C_{22,11}^{42,1}}{32 \sqrt{15}}-\frac{7 C_{31,11}^{31,0}}{96}-\frac{7 C_{31,11}^{31,1}}{192}-\frac{3}{40} \sqrt{\frac{7}{2}} C_{33,11}^{33,0}-\frac{3}{80}
\sqrt{\frac{7}{2}} C_{33,11}^{33,1}\)

\noindent\(C_{11,3111,0}^{00,2000,0,\text{NonLoc}}= -\frac{7 C_{11,11}^{51,0}}{192}-\frac{7 C_{11,11}^{51,1}}{96}+\frac{7 C_{22,11}^{42,0}}{32
\sqrt{15}}+\frac{7 C_{22,11}^{42,1}}{16 \sqrt{15}}-\frac{7 C_{31,11}^{31,0}}{192}-\frac{7 C_{31,11}^{31,1}}{96}-\frac{3}{80} \sqrt{\frac{7}{2}} C_{33,11}^{33,0}-\frac{3}{40}
\sqrt{\frac{7}{2}} C_{33,11}^{33,1}\)

\noindent\(C_{11,3111,0}^{00,2000,1,\text{Loc}}= -\frac{7 C_{11,11}^{51,1}}{64 \sqrt{3}}-\frac{7 C_{22,11}^{42,1}}{32 \sqrt{5}}-\frac{7
C_{31,11}^{31,1}}{64 \sqrt{3}}-\frac{3}{80} \sqrt{\frac{21}{2}} C_{33,11}^{33,1}\)

\noindent\(C_{11,3111,0}^{00,2000,1,\text{NonLoc}}= -\frac{7 C_{11,11}^{51,0}}{64 \sqrt{3}}+\frac{7 C_{22,11}^{42,0}}{32 \sqrt{5}}-\frac{7
C_{31,11}^{31,0}}{64 \sqrt{3}}-\frac{3}{80} \sqrt{\frac{21}{2}} C_{33,11}^{33,0}\)

\noindent\(C_{11,3111,0}^{11,1111,0,\text{Loc}}= \frac{C_{20,20}^{40,0}}{24 \sqrt{3}}+\frac{C_{20,20}^{40,1}}{48 \sqrt{3}}-\frac{7
C_{20,22}^{42,0}}{96 \sqrt{15}}-\frac{7 C_{20,22}^{42,1}}{192 \sqrt{15}}-\frac{7 C_{22,20}^{42,0}}{96 \sqrt{15}}-\frac{7 C_{22,20}^{42,1}}{192 \sqrt{15}}-\frac{1}{48}
\sqrt{\frac{5}{3}} C_{22,22}^{40,0}-\frac{1}{96} \sqrt{\frac{5}{3}} C_{22,22}^{40,1}+\frac{7}{96} \sqrt{\frac{7}{15}} C_{22,22}^{42,0}+\frac{7}{192}
\sqrt{\frac{7}{15}} C_{22,22}^{42,1}+\frac{C_{31,20}^{31,0}}{48 \sqrt{3}}+\frac{C_{31,20}^{31,1}}{96 \sqrt{3}}-\frac{C_{31,22}^{31,0}}{48 \sqrt{15}}-\frac{C_{31,22}^{31,1}}{96
\sqrt{15}}-\frac{\sqrt{7} C_{31,22}^{33,0}}{160}-\frac{\sqrt{7} C_{31,22}^{33,1}}{320}-\frac{3 \sqrt{7} C_{33,20}^{33,0}}{160}-\frac{3 \sqrt{7} C_{33,20}^{33,1}}{320}+\frac{3}{40}
\sqrt{\frac{21}{10}} C_{33,22}^{33,0}+\frac{3}{80} \sqrt{\frac{21}{10}} C_{33,22}^{33,1}\)

\noindent\(C_{11,3111,0}^{11,1111,0,\text{NonLoc}}= \frac{C_{20,00}^{40,0}}{96}+\frac{C_{20,00}^{40,1}}{48}+\frac{C_{20,20}^{40,0}}{96
\sqrt{3}}+\frac{C_{20,20}^{40,1}}{48 \sqrt{3}}+\frac{7 C_{20,22}^{42,0}}{192 \sqrt{15}}+\frac{7 C_{20,22}^{42,1}}{96 \sqrt{15}}-\frac{7 C_{22,00}^{42,0}}{384
\sqrt{5}}-\frac{7 C_{22,00}^{42,1}}{192 \sqrt{5}}-\frac{7 C_{22,20}^{42,0}}{384 \sqrt{15}}-\frac{7 C_{22,20}^{42,1}}{192 \sqrt{15}}+\frac{1}{96}
\sqrt{\frac{5}{3}} C_{22,22}^{40,0}+\frac{1}{48} \sqrt{\frac{5}{3}} C_{22,22}^{40,1}-\frac{7}{192} \sqrt{\frac{7}{15}} C_{22,22}^{42,0}-\frac{7}{96}
\sqrt{\frac{7}{15}} C_{22,22}^{42,1}-\frac{C_{31,00}^{31,0}}{192}-\frac{C_{31,00}^{31,1}}{96}-\frac{C_{31,20}^{31,0}}{192 \sqrt{3}}-\frac{C_{31,20}^{31,1}}{96
\sqrt{3}}-\frac{C_{31,22}^{31,0}}{96 \sqrt{15}}-\frac{C_{31,22}^{31,1}}{48 \sqrt{15}}-\frac{\sqrt{7} C_{31,22}^{33,0}}{320}-\frac{\sqrt{7} C_{31,22}^{33,1}}{160}+\frac{3
\sqrt{21} C_{33,00}^{33,0}}{640}+\frac{3 \sqrt{21} C_{33,00}^{33,1}}{320}+\frac{3 \sqrt{7} C_{33,20}^{33,0}}{640}+\frac{3 \sqrt{7} C_{33,20}^{33,1}}{320}+\frac{3}{80}
\sqrt{\frac{21}{10}} C_{33,22}^{33,0}+\frac{3}{40} \sqrt{\frac{21}{10}} C_{33,22}^{33,1}\)

\noindent\(C_{11,3111,0}^{11,1111,1,\text{Loc}}= \frac{C_{20,20}^{40,1}}{48}-\frac{7 C_{20,22}^{42,1}}{192 \sqrt{5}}-\frac{7 C_{22,20}^{42,1}}{192
\sqrt{5}}-\frac{\sqrt{5} C_{22,22}^{40,1}}{96}+\frac{7}{192} \sqrt{\frac{7}{5}} C_{22,22}^{42,1}+\frac{C_{31,20}^{31,1}}{96}-\frac{C_{31,22}^{31,1}}{96
\sqrt{5}}-\frac{\sqrt{21} C_{31,22}^{33,1}}{320}-\frac{3 \sqrt{21} C_{33,20}^{33,1}}{320}+\frac{9}{80} \sqrt{\frac{7}{10}} C_{33,22}^{33,1}\)

\noindent\(C_{11,3111,0}^{11,1111,1,\text{NonLoc}}= \frac{C_{20,00}^{40,0}}{32 \sqrt{3}}+\frac{C_{20,20}^{40,0}}{96}+\frac{7 C_{20,22}^{42,0}}{192
\sqrt{5}}-\frac{7 C_{22,00}^{42,0}}{128 \sqrt{15}}-\frac{7 C_{22,20}^{42,0}}{384 \sqrt{5}}+\frac{\sqrt{5} C_{22,22}^{40,0}}{96}-\frac{7}{192} \sqrt{\frac{7}{5}}
C_{22,22}^{42,0}-\frac{C_{31,00}^{31,0}}{64 \sqrt{3}}-\frac{C_{31,20}^{31,0}}{192}-\frac{C_{31,22}^{31,0}}{96 \sqrt{5}}-\frac{\sqrt{21} C_{31,22}^{33,0}}{320}+\frac{9
\sqrt{7} C_{33,00}^{33,0}}{640}+\frac{3 \sqrt{21} C_{33,20}^{33,0}}{640}+\frac{9}{80} \sqrt{\frac{7}{10}} C_{33,22}^{33,0}\)

\noindent\(C_{11,3111,1}^{11,1111,0,\text{Loc}}= \frac{7 C_{00,20}^{60,0}}{64}+\frac{7 C_{00,20}^{60,1}}{128}-\frac{21 C_{00,22}^{62,0}}{128
\sqrt{5}}-\frac{21 C_{00,22}^{62,1}}{256 \sqrt{5}}+\frac{7 C_{11,20}^{31,0}}{192}+\frac{7 C_{11,20}^{31,1}}{384}-\frac{7 \sqrt{5} C_{11,22}^{51,0}}{384}-\frac{7
\sqrt{5} C_{11,22}^{51,1}}{768}+\frac{C_{20,20}^{40,0}}{192}+\frac{C_{20,20}^{40,1}}{384}+\frac{7 C_{20,22}^{42,0}}{384 \sqrt{5}}+\frac{7 C_{20,22}^{42,1}}{768
\sqrt{5}}-\frac{7 C_{22,20}^{42,0}}{192 \sqrt{5}}-\frac{7 C_{22,20}^{42,1}}{384 \sqrt{5}}+\frac{13 C_{22,22}^{40,0}}{384 \sqrt{5}}+\frac{13 C_{22,22}^{40,1}}{768
\sqrt{5}}-\frac{\sqrt{35} C_{22,22}^{42,0}}{192}-\frac{\sqrt{35} C_{22,22}^{42,1}}{384}+\frac{9 C_{22,22}^{44,0}}{160}+\frac{9 C_{22,22}^{44,1}}{320}-\frac{C_{31,20}^{31,0}}{64}-\frac{C_{31,20}^{31,1}}{128}-\frac{C_{31,22}^{31,0}}{128
\sqrt{5}}-\frac{C_{31,22}^{31,1}}{256 \sqrt{5}}+\frac{3 \sqrt{21} C_{31,22}^{33,0}}{320}+\frac{3 \sqrt{21} C_{31,22}^{33,1}}{640}-\frac{3 \sqrt{21}
C_{33,20}^{33,0}}{320}-\frac{3 \sqrt{21} C_{33,20}^{33,1}}{640}-\frac{9}{160} \sqrt{\frac{7}{10}} C_{33,22}^{33,0}-\frac{9}{320} \sqrt{\frac{7}{10}}
C_{33,22}^{33,1}\)

\noindent\(C_{11,3111,1}^{11,1111,0,\text{NonLoc}}= \frac{7 \sqrt{3} C_{00,00}^{60,0}}{256}+\frac{7 \sqrt{3} C_{00,00}^{60,1}}{128}+\frac{7
C_{00,20}^{60,0}}{256}+\frac{7 C_{00,20}^{60,1}}{128}+\frac{21 C_{00,22}^{62,0}}{256 \sqrt{5}}+\frac{21 C_{00,22}^{62,1}}{128 \sqrt{5}}-\frac{7 C_{11,00}^{51,0}}{256
\sqrt{3}}-\frac{7 C_{11,00}^{51,1}}{128 \sqrt{3}}-\frac{7 C_{11,20}^{31,0}}{768}-\frac{7 C_{11,20}^{31,1}}{384}-\frac{7 \sqrt{5} C_{11,22}^{51,0}}{768}-\frac{7
\sqrt{5} C_{11,22}^{51,1}}{384}+\frac{C_{20,00}^{40,0}}{256 \sqrt{3}}+\frac{C_{20,00}^{40,1}}{128 \sqrt{3}}+\frac{C_{20,20}^{40,0}}{768}+\frac{C_{20,20}^{40,1}}{384}-\frac{7
C_{20,22}^{42,0}}{768 \sqrt{5}}-\frac{7 C_{20,22}^{42,1}}{384 \sqrt{5}}-\frac{7 C_{22,00}^{42,0}}{256 \sqrt{15}}-\frac{7 C_{22,00}^{42,1}}{128 \sqrt{15}}-\frac{7
C_{22,20}^{42,0}}{768 \sqrt{5}}-\frac{7 C_{22,20}^{42,1}}{384 \sqrt{5}}-\frac{13 C_{22,22}^{40,0}}{768 \sqrt{5}}-\frac{13 C_{22,22}^{40,1}}{384 \sqrt{5}}+\frac{\sqrt{35}
C_{22,22}^{42,0}}{384}+\frac{\sqrt{35} C_{22,22}^{42,1}}{192}-\frac{9 C_{22,22}^{44,0}}{320}-\frac{9 C_{22,22}^{44,1}}{160}+\frac{\sqrt{3} C_{31,00}^{31,0}}{256}+\frac{\sqrt{3}
C_{31,00}^{31,1}}{128}+\frac{C_{31,20}^{31,0}}{256}+\frac{C_{31,20}^{31,1}}{128}-\frac{C_{31,22}^{31,0}}{256 \sqrt{5}}-\frac{C_{31,22}^{31,1}}{128
\sqrt{5}}+\frac{3 \sqrt{21} C_{31,22}^{33,0}}{640}+\frac{3 \sqrt{21} C_{31,22}^{33,1}}{320}+\frac{9 \sqrt{7} C_{33,00}^{33,0}}{1280}+\frac{9 \sqrt{7}
C_{33,00}^{33,1}}{640}+\frac{3 \sqrt{21} C_{33,20}^{33,0}}{1280}+\frac{3 \sqrt{21} C_{33,20}^{33,1}}{640}-\frac{9}{320} \sqrt{\frac{7}{10}} C_{33,22}^{33,0}-\frac{9}{160}
\sqrt{\frac{7}{10}} C_{33,22}^{33,1}\)

\noindent\(C_{11,3111,1}^{11,1111,1,\text{Loc}}= \frac{7 \sqrt{3} C_{00,20}^{60,1}}{128}-\frac{21}{256} \sqrt{\frac{3}{5}} C_{00,22}^{62,1}+\frac{7
C_{11,20}^{31,1}}{128 \sqrt{3}}-\frac{7}{256} \sqrt{\frac{5}{3}} C_{11,22}^{51,1}+\frac{C_{20,20}^{40,1}}{128 \sqrt{3}}+\frac{7 C_{20,22}^{42,1}}{256
\sqrt{15}}-\frac{7 C_{22,20}^{42,1}}{128 \sqrt{15}}+\frac{13 C_{22,22}^{40,1}}{256 \sqrt{15}}-\frac{1}{128} \sqrt{\frac{35}{3}} C_{22,22}^{42,1}+\frac{9
\sqrt{3} C_{22,22}^{44,1}}{320}-\frac{\sqrt{3} C_{31,20}^{31,1}}{128}-\frac{1}{256} \sqrt{\frac{3}{5}} C_{31,22}^{31,1}+\frac{9 \sqrt{7} C_{31,22}^{33,1}}{640}-\frac{9
\sqrt{7} C_{33,20}^{33,1}}{640}-\frac{9}{320} \sqrt{\frac{21}{10}} C_{33,22}^{33,1}\)

\noindent\(C_{11,3111,1}^{11,1111,1,\text{NonLoc}}= \frac{21 C_{00,00}^{60,0}}{256}+\frac{7 \sqrt{3} C_{00,20}^{60,0}}{256}+\frac{21}{256}
\sqrt{\frac{3}{5}} C_{00,22}^{62,0}-\frac{7 C_{11,00}^{51,0}}{256}-\frac{7 C_{11,20}^{31,0}}{256 \sqrt{3}}-\frac{7}{256} \sqrt{\frac{5}{3}} C_{11,22}^{51,0}+\frac{C_{20,00}^{40,0}}{256}+\frac{C_{20,20}^{40,0}}{256
\sqrt{3}}-\frac{7 C_{20,22}^{42,0}}{256 \sqrt{15}}-\frac{7 C_{22,00}^{42,0}}{256 \sqrt{5}}-\frac{7 C_{22,20}^{42,0}}{256 \sqrt{15}}-\frac{13 C_{22,22}^{40,0}}{256
\sqrt{15}}+\frac{1}{128} \sqrt{\frac{35}{3}} C_{22,22}^{42,0}-\frac{9 \sqrt{3} C_{22,22}^{44,0}}{320}+\frac{3 C_{31,00}^{31,0}}{256}+\frac{\sqrt{3}
C_{31,20}^{31,0}}{256}-\frac{1}{256} \sqrt{\frac{3}{5}} C_{31,22}^{31,0}+\frac{9 \sqrt{7} C_{31,22}^{33,0}}{640}+\frac{9 \sqrt{21} C_{33,00}^{33,0}}{1280}+\frac{9
\sqrt{7} C_{33,20}^{33,0}}{1280}-\frac{9}{320} \sqrt{\frac{21}{10}} C_{33,22}^{33,0}\)

\noindent\(C_{11,3111,1}^{11,1112,0,\text{Loc}}= -\frac{7}{64} \sqrt{\frac{3}{5}} C_{00,20}^{60,0}-\frac{7}{128} \sqrt{\frac{3}{5}}
C_{00,20}^{60,1}+\frac{21 \sqrt{3} C_{00,22}^{62,0}}{640}+\frac{21 \sqrt{3} C_{00,22}^{62,1}}{1280}-\frac{7 C_{11,20}^{31,0}}{64 \sqrt{15}}-\frac{7
C_{11,20}^{31,1}}{128 \sqrt{15}}-\frac{7 C_{11,22}^{51,0}}{640 \sqrt{3}}-\frac{7 C_{11,22}^{51,1}}{1280 \sqrt{3}}+\frac{27}{640} \sqrt{\frac{7}{5}}
C_{11,22}^{53,0}+\frac{27 \sqrt{\frac{7}{5}} C_{11,22}^{53,1}}{1280}+\frac{7 C_{20,20}^{40,0}}{64 \sqrt{15}}+\frac{7 C_{20,20}^{40,1}}{128 \sqrt{15}}+\frac{7
C_{20,22}^{42,0}}{640 \sqrt{3}}+\frac{7 C_{20,22}^{42,1}}{1280 \sqrt{3}}-\frac{7 C_{22,20}^{42,0}}{320 \sqrt{3}}-\frac{7 C_{22,20}^{42,1}}{640 \sqrt{3}}-\frac{17
C_{22,22}^{40,0}}{640 \sqrt{3}}-\frac{17 C_{22,22}^{40,1}}{1280 \sqrt{3}}-\frac{1}{64} \sqrt{\frac{7}{3}} C_{22,22}^{42,0}-\frac{1}{128} \sqrt{\frac{7}{3}}
C_{22,22}^{42,1}+\frac{9}{160} \sqrt{\frac{3}{5}} C_{22,22}^{44,0}+\frac{9}{320} \sqrt{\frac{3}{5}} C_{22,22}^{44,1}+\frac{7 C_{31,20}^{31,0}}{64
\sqrt{15}}+\frac{7 C_{31,20}^{31,1}}{128 \sqrt{15}}-\frac{23 C_{31,22}^{31,0}}{640 \sqrt{3}}-\frac{23 C_{31,22}^{31,1}}{1280 \sqrt{3}}+\frac{3}{640}
\sqrt{\frac{7}{5}} C_{31,22}^{33,0}+\frac{3 \sqrt{\frac{7}{5}} C_{31,22}^{33,1}}{1280}-\frac{9}{320} \sqrt{\frac{7}{5}} C_{33,20}^{33,0}-\frac{9}{640}
\sqrt{\frac{7}{5}} C_{33,20}^{33,1}-\frac{9}{800} \sqrt{\frac{21}{2}} C_{33,22}^{33,0}-\frac{9 \sqrt{\frac{21}{2}} C_{33,22}^{33,1}}{1600}\)

\noindent\(C_{11,3111,1}^{11,1112,0,\text{NonLoc}}= -\frac{21 C_{00,00}^{60,0}}{256 \sqrt{5}}-\frac{21 C_{00,00}^{60,1}}{128 \sqrt{5}}-\frac{7}{256}
\sqrt{\frac{3}{5}} C_{00,20}^{60,0}-\frac{7}{128} \sqrt{\frac{3}{5}} C_{00,20}^{60,1}-\frac{21 \sqrt{3} C_{00,22}^{62,0}}{1280}-\frac{21 \sqrt{3}
C_{00,22}^{62,1}}{640}+\frac{7 C_{11,00}^{51,0}}{256 \sqrt{5}}+\frac{7 C_{11,00}^{51,1}}{128 \sqrt{5}}+\frac{7 C_{11,20}^{31,0}}{256 \sqrt{15}}+\frac{7
C_{11,20}^{31,1}}{128 \sqrt{15}}-\frac{7 C_{11,22}^{51,0}}{1280 \sqrt{3}}-\frac{7 C_{11,22}^{51,1}}{640 \sqrt{3}}+\frac{27 \sqrt{\frac{7}{5}} C_{11,22}^{53,0}}{1280}+\frac{27}{640}
\sqrt{\frac{7}{5}} C_{11,22}^{53,1}+\frac{7 C_{20,00}^{40,0}}{256 \sqrt{5}}+\frac{7 C_{20,00}^{40,1}}{128 \sqrt{5}}+\frac{7 C_{20,20}^{40,0}}{256
\sqrt{15}}+\frac{7 C_{20,20}^{40,1}}{128 \sqrt{15}}-\frac{7 C_{20,22}^{42,0}}{1280 \sqrt{3}}-\frac{7 C_{20,22}^{42,1}}{640 \sqrt{3}}-\frac{7 C_{22,00}^{42,0}}{1280}-\frac{7
C_{22,00}^{42,1}}{640}-\frac{7 C_{22,20}^{42,0}}{1280 \sqrt{3}}-\frac{7 C_{22,20}^{42,1}}{640 \sqrt{3}}+\frac{17 C_{22,22}^{40,0}}{1280 \sqrt{3}}+\frac{17
C_{22,22}^{40,1}}{640 \sqrt{3}}+\frac{1}{128} \sqrt{\frac{7}{3}} C_{22,22}^{42,0}+\frac{1}{64} \sqrt{\frac{7}{3}} C_{22,22}^{42,1}-\frac{9}{320}
\sqrt{\frac{3}{5}} C_{22,22}^{44,0}-\frac{9}{160} \sqrt{\frac{3}{5}} C_{22,22}^{44,1}-\frac{7 C_{31,00}^{31,0}}{256 \sqrt{5}}-\frac{7 C_{31,00}^{31,1}}{128
\sqrt{5}}-\frac{7 C_{31,20}^{31,0}}{256 \sqrt{15}}-\frac{7 C_{31,20}^{31,1}}{128 \sqrt{15}}-\frac{23 C_{31,22}^{31,0}}{1280 \sqrt{3}}-\frac{23 C_{31,22}^{31,1}}{640
\sqrt{3}}+\frac{3 \sqrt{\frac{7}{5}} C_{31,22}^{33,0}}{1280}+\frac{3}{640} \sqrt{\frac{7}{5}} C_{31,22}^{33,1}+\frac{9 \sqrt{\frac{21}{5}} C_{33,00}^{33,0}}{1280}+\frac{9}{640}
\sqrt{\frac{21}{5}} C_{33,00}^{33,1}+\frac{9 \sqrt{\frac{7}{5}} C_{33,20}^{33,0}}{1280}+\frac{9}{640} \sqrt{\frac{7}{5}} C_{33,20}^{33,1}-\frac{9
\sqrt{\frac{21}{2}} C_{33,22}^{33,0}}{1600}-\frac{9}{800} \sqrt{\frac{21}{2}} C_{33,22}^{33,1}\)

\noindent\(C_{11,3111,1}^{11,1112,1,\text{Loc}}= -\frac{21 C_{00,20}^{60,1}}{128 \sqrt{5}}+\frac{63 C_{00,22}^{62,1}}{1280}-\frac{7
C_{11,20}^{31,1}}{128 \sqrt{5}}-\frac{7 C_{11,22}^{51,1}}{1280}+\frac{27 \sqrt{\frac{21}{5}} C_{11,22}^{53,1}}{1280}+\frac{7 C_{20,20}^{40,1}}{128
\sqrt{5}}+\frac{7 C_{20,22}^{42,1}}{1280}-\frac{7 C_{22,20}^{42,1}}{640}-\frac{17 C_{22,22}^{40,1}}{1280}-\frac{\sqrt{7} C_{22,22}^{42,1}}{128}+\frac{27
C_{22,22}^{44,1}}{320 \sqrt{5}}+\frac{7 C_{31,20}^{31,1}}{128 \sqrt{5}}-\frac{23 C_{31,22}^{31,1}}{1280}+\frac{3 \sqrt{\frac{21}{5}} C_{31,22}^{33,1}}{1280}-\frac{9}{640}
\sqrt{\frac{21}{5}} C_{33,20}^{33,1}-\frac{27 \sqrt{\frac{7}{2}} C_{33,22}^{33,1}}{1600}\)

\noindent\(C_{11,3111,1}^{11,1112,1,\text{NonLoc}}= -\frac{21}{256} \sqrt{\frac{3}{5}} C_{00,00}^{60,0}-\frac{21 C_{00,20}^{60,0}}{256
\sqrt{5}}-\frac{63 C_{00,22}^{62,0}}{1280}+\frac{7}{256} \sqrt{\frac{3}{5}} C_{11,00}^{51,0}+\frac{7 C_{11,20}^{31,0}}{256 \sqrt{5}}-\frac{7 C_{11,22}^{51,0}}{1280}+\frac{27
\sqrt{\frac{21}{5}} C_{11,22}^{53,0}}{1280}+\frac{7}{256} \sqrt{\frac{3}{5}} C_{20,00}^{40,0}+\frac{7 C_{20,20}^{40,0}}{256 \sqrt{5}}-\frac{7 C_{20,22}^{42,0}}{1280}-\frac{7
\sqrt{3} C_{22,00}^{42,0}}{1280}-\frac{7 C_{22,20}^{42,0}}{1280}+\frac{17 C_{22,22}^{40,0}}{1280}+\frac{\sqrt{7} C_{22,22}^{42,0}}{128}-\frac{27
C_{22,22}^{44,0}}{320 \sqrt{5}}-\frac{7}{256} \sqrt{\frac{3}{5}} C_{31,00}^{31,0}-\frac{7 C_{31,20}^{31,0}}{256 \sqrt{5}}-\frac{23 C_{31,22}^{31,0}}{1280}+\frac{3
\sqrt{\frac{21}{5}} C_{31,22}^{33,0}}{1280}+\frac{27 \sqrt{\frac{7}{5}} C_{33,00}^{33,0}}{1280}+\frac{9 \sqrt{\frac{21}{5}} C_{33,20}^{33,0}}{1280}-\frac{27
\sqrt{\frac{7}{2}} C_{33,22}^{33,0}}{1600}\)

\noindent\(C_{11,3111,2}^{00,2202,0,\text{Loc}}= \frac{7 C_{11,11}^{51,0}}{96 \sqrt{5}}+\frac{7 C_{11,11}^{51,1}}{192 \sqrt{5}}+\frac{7
C_{22,11}^{42,0}}{80 \sqrt{3}}+\frac{7 C_{22,11}^{42,1}}{160 \sqrt{3}}+\frac{7 C_{31,11}^{31,0}}{96 \sqrt{5}}+\frac{7 C_{31,11}^{31,1}}{192 \sqrt{5}}+\frac{3}{40}
\sqrt{\frac{7}{10}} C_{33,11}^{33,0}+\frac{3}{80} \sqrt{\frac{7}{10}} C_{33,11}^{33,1}\)

\noindent\(C_{11,3111,2}^{00,2202,0,\text{NonLoc}}= \frac{7 C_{11,11}^{51,0}}{192 \sqrt{5}}+\frac{7 C_{11,11}^{51,1}}{96 \sqrt{5}}-\frac{7
C_{22,11}^{42,0}}{160 \sqrt{3}}-\frac{7 C_{22,11}^{42,1}}{80 \sqrt{3}}+\frac{7 C_{31,11}^{31,0}}{192 \sqrt{5}}+\frac{7 C_{31,11}^{31,1}}{96 \sqrt{5}}+\frac{3}{80}
\sqrt{\frac{7}{10}} C_{33,11}^{33,0}+\frac{3}{40} \sqrt{\frac{7}{10}} C_{33,11}^{33,1}\)

\noindent\(C_{11,3111,2}^{00,2202,1,\text{Loc}}= \frac{7 C_{11,11}^{51,1}}{64 \sqrt{15}}+\frac{7 C_{22,11}^{42,1}}{160}+\frac{7 C_{31,11}^{31,1}}{64
\sqrt{15}}+\frac{3}{80} \sqrt{\frac{21}{10}} C_{33,11}^{33,1}\)

\noindent\(C_{11,3111,2}^{00,2202,1,\text{NonLoc}}= \frac{7 C_{11,11}^{51,0}}{64 \sqrt{15}}-\frac{7 C_{22,11}^{42,0}}{160}+\frac{7
C_{31,11}^{31,0}}{64 \sqrt{15}}+\frac{3}{80} \sqrt{\frac{21}{10}} C_{33,11}^{33,0}\)

\noindent\(C_{11,3111,2}^{11,1111,0,\text{Loc}}= \frac{7}{64} \sqrt{\frac{3}{5}} C_{00,20}^{60,0}+\frac{7}{128} \sqrt{\frac{3}{5}}
C_{00,20}^{60,1}-\frac{21 \sqrt{3} C_{00,22}^{62,0}}{640}-\frac{21 \sqrt{3} C_{00,22}^{62,1}}{1280}+\frac{7 C_{11,20}^{31,0}}{64 \sqrt{15}}+\frac{7
C_{11,20}^{31,1}}{128 \sqrt{15}}-\frac{7 C_{11,22}^{51,0}}{128 \sqrt{3}}-\frac{7 C_{11,22}^{51,1}}{256 \sqrt{3}}+\frac{19 C_{20,20}^{40,0}}{192 \sqrt{15}}+\frac{19
C_{20,20}^{40,1}}{384 \sqrt{15}}-\frac{7 C_{20,22}^{42,0}}{384 \sqrt{3}}-\frac{7 C_{20,22}^{42,1}}{768 \sqrt{3}}-\frac{49 C_{22,20}^{42,0}}{960 \sqrt{3}}-\frac{49
C_{22,20}^{42,1}}{1920 \sqrt{3}}-\frac{41 C_{22,22}^{40,0}}{1920 \sqrt{3}}-\frac{41 C_{22,22}^{40,1}}{3840 \sqrt{3}}+\frac{13}{960} \sqrt{\frac{7}{3}}
C_{22,22}^{42,0}+\frac{13 \sqrt{\frac{7}{3}} C_{22,22}^{42,1}}{1920}+\frac{9}{160} \sqrt{\frac{3}{5}} C_{22,22}^{44,0}+\frac{9}{320} \sqrt{\frac{3}{5}}
C_{22,22}^{44,1}-\frac{C_{31,20}^{31,0}}{192 \sqrt{15}}-\frac{C_{31,20}^{31,1}}{384 \sqrt{15}}-\frac{5 C_{31,22}^{31,0}}{384 \sqrt{3}}-\frac{5 C_{31,22}^{31,1}}{768
\sqrt{3}}+\frac{1}{64} \sqrt{\frac{7}{5}} C_{31,22}^{33,0}+\frac{1}{128} \sqrt{\frac{7}{5}} C_{31,22}^{33,1}-\frac{21}{320} \sqrt{\frac{7}{5}} C_{33,20}^{33,0}-\frac{21}{640}
\sqrt{\frac{7}{5}} C_{33,20}^{33,1}+\frac{3}{160} \sqrt{\frac{21}{2}} C_{33,22}^{33,0}+\frac{3}{320} \sqrt{\frac{21}{2}} C_{33,22}^{33,1}\)

\noindent\(C_{11,3111,2}^{11,1111,0,\text{NonLoc}}= \frac{21 C_{00,00}^{60,0}}{256 \sqrt{5}}+\frac{21 C_{00,00}^{60,1}}{128 \sqrt{5}}+\frac{7}{256}
\sqrt{\frac{3}{5}} C_{00,20}^{60,0}+\frac{7}{128} \sqrt{\frac{3}{5}} C_{00,20}^{60,1}+\frac{21 \sqrt{3} C_{00,22}^{62,0}}{1280}+\frac{21 \sqrt{3}
C_{00,22}^{62,1}}{640}-\frac{7 C_{11,00}^{51,0}}{256 \sqrt{5}}-\frac{7 C_{11,00}^{51,1}}{128 \sqrt{5}}-\frac{7 C_{11,20}^{31,0}}{256 \sqrt{15}}-\frac{7
C_{11,20}^{31,1}}{128 \sqrt{15}}-\frac{7 C_{11,22}^{51,0}}{256 \sqrt{3}}-\frac{7 C_{11,22}^{51,1}}{128 \sqrt{3}}+\frac{19 C_{20,00}^{40,0}}{768 \sqrt{5}}+\frac{19
C_{20,00}^{40,1}}{384 \sqrt{5}}+\frac{19 C_{20,20}^{40,0}}{768 \sqrt{15}}+\frac{19 C_{20,20}^{40,1}}{384 \sqrt{15}}+\frac{7 C_{20,22}^{42,0}}{768
\sqrt{3}}+\frac{7 C_{20,22}^{42,1}}{384 \sqrt{3}}-\frac{49 C_{22,00}^{42,0}}{3840}-\frac{49 C_{22,00}^{42,1}}{1920}-\frac{49 C_{22,20}^{42,0}}{3840
\sqrt{3}}-\frac{49 C_{22,20}^{42,1}}{1920 \sqrt{3}}+\frac{41 C_{22,22}^{40,0}}{3840 \sqrt{3}}+\frac{41 C_{22,22}^{40,1}}{1920 \sqrt{3}}-\frac{13
\sqrt{\frac{7}{3}} C_{22,22}^{42,0}}{1920}-\frac{13}{960} \sqrt{\frac{7}{3}} C_{22,22}^{42,1}-\frac{9}{320} \sqrt{\frac{3}{5}} C_{22,22}^{44,0}-\frac{9}{160}
\sqrt{\frac{3}{5}} C_{22,22}^{44,1}+\frac{C_{31,00}^{31,0}}{768 \sqrt{5}}+\frac{C_{31,00}^{31,1}}{384 \sqrt{5}}+\frac{C_{31,20}^{31,0}}{768 \sqrt{15}}+\frac{C_{31,20}^{31,1}}{384
\sqrt{15}}-\frac{5 C_{31,22}^{31,0}}{768 \sqrt{3}}-\frac{5 C_{31,22}^{31,1}}{384 \sqrt{3}}+\frac{1}{128} \sqrt{\frac{7}{5}} C_{31,22}^{33,0}+\frac{1}{64}
\sqrt{\frac{7}{5}} C_{31,22}^{33,1}+\frac{21 \sqrt{\frac{21}{5}} C_{33,00}^{33,0}}{1280}+\frac{21}{640} \sqrt{\frac{21}{5}} C_{33,00}^{33,1}+\frac{21
\sqrt{\frac{7}{5}} C_{33,20}^{33,0}}{1280}+\frac{21}{640} \sqrt{\frac{7}{5}} C_{33,20}^{33,1}+\frac{3}{320} \sqrt{\frac{21}{2}} C_{33,22}^{33,0}+\frac{3}{160}
\sqrt{\frac{21}{2}} C_{33,22}^{33,1}\)

\noindent\(C_{11,3111,2}^{11,1111,1,\text{Loc}}= \frac{21 C_{00,20}^{60,1}}{128 \sqrt{5}}-\frac{63 C_{00,22}^{62,1}}{1280}+\frac{7
C_{11,20}^{31,1}}{128 \sqrt{5}}-\frac{7 C_{11,22}^{51,1}}{256}+\frac{19 C_{20,20}^{40,1}}{384 \sqrt{5}}-\frac{7 C_{20,22}^{42,1}}{768}-\frac{49 C_{22,20}^{42,1}}{1920}-\frac{41
C_{22,22}^{40,1}}{3840}+\frac{13 \sqrt{7} C_{22,22}^{42,1}}{1920}+\frac{27 C_{22,22}^{44,1}}{320 \sqrt{5}}-\frac{C_{31,20}^{31,1}}{384 \sqrt{5}}-\frac{5
C_{31,22}^{31,1}}{768}+\frac{1}{128} \sqrt{\frac{21}{5}} C_{31,22}^{33,1}-\frac{21}{640} \sqrt{\frac{21}{5}} C_{33,20}^{33,1}+\frac{9}{320} \sqrt{\frac{7}{2}}
C_{33,22}^{33,1}\)

\noindent\(C_{11,3111,2}^{11,1111,1,\text{NonLoc}}= \frac{21}{256} \sqrt{\frac{3}{5}} C_{00,00}^{60,0}+\frac{21 C_{00,20}^{60,0}}{256
\sqrt{5}}+\frac{63 C_{00,22}^{62,0}}{1280}-\frac{7}{256} \sqrt{\frac{3}{5}} C_{11,00}^{51,0}-\frac{7 C_{11,20}^{31,0}}{256 \sqrt{5}}-\frac{7 C_{11,22}^{51,0}}{256}+\frac{19
C_{20,00}^{40,0}}{256 \sqrt{15}}+\frac{19 C_{20,20}^{40,0}}{768 \sqrt{5}}+\frac{7 C_{20,22}^{42,0}}{768}-\frac{49 C_{22,00}^{42,0}}{1280 \sqrt{3}}-\frac{49
C_{22,20}^{42,0}}{3840}+\frac{41 C_{22,22}^{40,0}}{3840}-\frac{13 \sqrt{7} C_{22,22}^{42,0}}{1920}-\frac{27 C_{22,22}^{44,0}}{320 \sqrt{5}}+\frac{C_{31,00}^{31,0}}{256
\sqrt{15}}+\frac{C_{31,20}^{31,0}}{768 \sqrt{5}}-\frac{5 C_{31,22}^{31,0}}{768}+\frac{1}{128} \sqrt{\frac{21}{5}} C_{31,22}^{33,0}+\frac{63 \sqrt{\frac{7}{5}}
C_{33,00}^{33,0}}{1280}+\frac{21 \sqrt{\frac{21}{5}} C_{33,20}^{33,0}}{1280}+\frac{9}{320} \sqrt{\frac{7}{2}} C_{33,22}^{33,0}\)

\noindent\(C_{11,3111,2}^{11,1112,0,\text{Loc}}= \frac{7 C_{00,20}^{60,0}}{64 \sqrt{5}}+\frac{7 C_{00,20}^{60,1}}{128 \sqrt{5}}-\frac{21
C_{00,22}^{62,0}}{640}-\frac{21 C_{00,22}^{62,1}}{1280}+\frac{7 C_{11,20}^{31,0}}{192 \sqrt{5}}+\frac{7 C_{11,20}^{31,1}}{384 \sqrt{5}}+\frac{7 C_{11,22}^{51,0}}{1920}+\frac{7
C_{11,22}^{51,1}}{3840}-\frac{9}{640} \sqrt{\frac{21}{5}} C_{11,22}^{53,0}-\frac{9 \sqrt{\frac{21}{5}} C_{11,22}^{53,1}}{1280}-\frac{7 C_{20,20}^{40,0}}{192
\sqrt{5}}-\frac{7 C_{20,20}^{40,1}}{384 \sqrt{5}}-\frac{7 C_{20,22}^{42,0}}{1920}-\frac{7 C_{20,22}^{42,1}}{3840}+\frac{7 C_{22,20}^{42,0}}{960}+\frac{7
C_{22,20}^{42,1}}{1920}+\frac{17 C_{22,22}^{40,0}}{1920}+\frac{17 C_{22,22}^{40,1}}{3840}+\frac{\sqrt{7} C_{22,22}^{42,0}}{192}+\frac{\sqrt{7} C_{22,22}^{42,1}}{384}-\frac{9
C_{22,22}^{44,0}}{160 \sqrt{5}}-\frac{9 C_{22,22}^{44,1}}{320 \sqrt{5}}-\frac{7 C_{31,20}^{31,0}}{192 \sqrt{5}}-\frac{7 C_{31,20}^{31,1}}{384 \sqrt{5}}+\frac{23
C_{31,22}^{31,0}}{1920}+\frac{23 C_{31,22}^{31,1}}{3840}-\frac{1}{640} \sqrt{\frac{21}{5}} C_{31,22}^{33,0}-\frac{\sqrt{\frac{21}{5}} C_{31,22}^{33,1}}{1280}+\frac{3}{320}
\sqrt{\frac{21}{5}} C_{33,20}^{33,0}+\frac{3}{640} \sqrt{\frac{21}{5}} C_{33,20}^{33,1}+\frac{9}{800} \sqrt{\frac{7}{2}} C_{33,22}^{33,0}+\frac{9
\sqrt{\frac{7}{2}} C_{33,22}^{33,1}}{1600}\)

\noindent\(C_{11,3111,2}^{11,1112,0,\text{NonLoc}}= \frac{7}{256} \sqrt{\frac{3}{5}} C_{00,00}^{60,0}+\frac{7}{128} \sqrt{\frac{3}{5}}
C_{00,00}^{60,1}+\frac{7 C_{00,20}^{60,0}}{256 \sqrt{5}}+\frac{7 C_{00,20}^{60,1}}{128 \sqrt{5}}+\frac{21 C_{00,22}^{62,0}}{1280}+\frac{21 C_{00,22}^{62,1}}{640}-\frac{7
C_{11,00}^{51,0}}{256 \sqrt{15}}-\frac{7 C_{11,00}^{51,1}}{128 \sqrt{15}}-\frac{7 C_{11,20}^{31,0}}{768 \sqrt{5}}-\frac{7 C_{11,20}^{31,1}}{384 \sqrt{5}}+\frac{7
C_{11,22}^{51,0}}{3840}+\frac{7 C_{11,22}^{51,1}}{1920}-\frac{9 \sqrt{\frac{21}{5}} C_{11,22}^{53,0}}{1280}-\frac{9}{640} \sqrt{\frac{21}{5}} C_{11,22}^{53,1}-\frac{7
C_{20,00}^{40,0}}{256 \sqrt{15}}-\frac{7 C_{20,00}^{40,1}}{128 \sqrt{15}}-\frac{7 C_{20,20}^{40,0}}{768 \sqrt{5}}-\frac{7 C_{20,20}^{40,1}}{384 \sqrt{5}}+\frac{7
C_{20,22}^{42,0}}{3840}+\frac{7 C_{20,22}^{42,1}}{1920}+\frac{7 C_{22,00}^{42,0}}{1280 \sqrt{3}}+\frac{7 C_{22,00}^{42,1}}{640 \sqrt{3}}+\frac{7
C_{22,20}^{42,0}}{3840}+\frac{7 C_{22,20}^{42,1}}{1920}-\frac{17 C_{22,22}^{40,0}}{3840}-\frac{17 C_{22,22}^{40,1}}{1920}-\frac{\sqrt{7} C_{22,22}^{42,0}}{384}-\frac{\sqrt{7}
C_{22,22}^{42,1}}{192}+\frac{9 C_{22,22}^{44,0}}{320 \sqrt{5}}+\frac{9 C_{22,22}^{44,1}}{160 \sqrt{5}}+\frac{7 C_{31,00}^{31,0}}{256 \sqrt{15}}+\frac{7
C_{31,00}^{31,1}}{128 \sqrt{15}}+\frac{7 C_{31,20}^{31,0}}{768 \sqrt{5}}+\frac{7 C_{31,20}^{31,1}}{384 \sqrt{5}}+\frac{23 C_{31,22}^{31,0}}{3840}+\frac{23
C_{31,22}^{31,1}}{1920}-\frac{\sqrt{\frac{21}{5}} C_{31,22}^{33,0}}{1280}-\frac{1}{640} \sqrt{\frac{21}{5}} C_{31,22}^{33,1}-\frac{9 \sqrt{\frac{7}{5}}
C_{33,00}^{33,0}}{1280}-\frac{9}{640} \sqrt{\frac{7}{5}} C_{33,00}^{33,1}-\frac{3 \sqrt{\frac{21}{5}} C_{33,20}^{33,0}}{1280}-\frac{3}{640} \sqrt{\frac{21}{5}}
C_{33,20}^{33,1}+\frac{9 \sqrt{\frac{7}{2}} C_{33,22}^{33,0}}{1600}+\frac{9}{800} \sqrt{\frac{7}{2}} C_{33,22}^{33,1}\)

\noindent\(C_{11,3111,2}^{11,1112,1,\text{Loc}}= \frac{7}{128} \sqrt{\frac{3}{5}} C_{00,20}^{60,1}-\frac{21 \sqrt{3} C_{00,22}^{62,1}}{1280}+\frac{7
C_{11,20}^{31,1}}{128 \sqrt{15}}+\frac{7 C_{11,22}^{51,1}}{1280 \sqrt{3}}-\frac{27 \sqrt{\frac{7}{5}} C_{11,22}^{53,1}}{1280}-\frac{7 C_{20,20}^{40,1}}{128
\sqrt{15}}-\frac{7 C_{20,22}^{42,1}}{1280 \sqrt{3}}+\frac{7 C_{22,20}^{42,1}}{640 \sqrt{3}}+\frac{17 C_{22,22}^{40,1}}{1280 \sqrt{3}}+\frac{1}{128}
\sqrt{\frac{7}{3}} C_{22,22}^{42,1}-\frac{9}{320} \sqrt{\frac{3}{5}} C_{22,22}^{44,1}-\frac{7 C_{31,20}^{31,1}}{128 \sqrt{15}}+\frac{23 C_{31,22}^{31,1}}{1280
\sqrt{3}}-\frac{3 \sqrt{\frac{7}{5}} C_{31,22}^{33,1}}{1280}+\frac{9}{640} \sqrt{\frac{7}{5}} C_{33,20}^{33,1}+\frac{9 \sqrt{\frac{21}{2}} C_{33,22}^{33,1}}{1600}\)

\noindent\(C_{11,3111,2}^{11,1112,1,\text{NonLoc}}= \frac{21 C_{00,00}^{60,0}}{256 \sqrt{5}}+\frac{7}{256} \sqrt{\frac{3}{5}} C_{00,20}^{60,0}+\frac{21
\sqrt{3} C_{00,22}^{62,0}}{1280}-\frac{7 C_{11,00}^{51,0}}{256 \sqrt{5}}-\frac{7 C_{11,20}^{31,0}}{256 \sqrt{15}}+\frac{7 C_{11,22}^{51,0}}{1280
\sqrt{3}}-\frac{27 \sqrt{\frac{7}{5}} C_{11,22}^{53,0}}{1280}-\frac{7 C_{20,00}^{40,0}}{256 \sqrt{5}}-\frac{7 C_{20,20}^{40,0}}{256 \sqrt{15}}+\frac{7
C_{20,22}^{42,0}}{1280 \sqrt{3}}+\frac{7 C_{22,00}^{42,0}}{1280}+\frac{7 C_{22,20}^{42,0}}{1280 \sqrt{3}}-\frac{17 C_{22,22}^{40,0}}{1280 \sqrt{3}}-\frac{1}{128}
\sqrt{\frac{7}{3}} C_{22,22}^{42,0}+\frac{9}{320} \sqrt{\frac{3}{5}} C_{22,22}^{44,0}+\frac{7 C_{31,00}^{31,0}}{256 \sqrt{5}}+\frac{7 C_{31,20}^{31,0}}{256
\sqrt{15}}+\frac{23 C_{31,22}^{31,0}}{1280 \sqrt{3}}-\frac{3 \sqrt{\frac{7}{5}} C_{31,22}^{33,0}}{1280}-\frac{9 \sqrt{\frac{21}{5}} C_{33,00}^{33,0}}{1280}-\frac{9
\sqrt{\frac{7}{5}} C_{33,20}^{33,0}}{1280}+\frac{9 \sqrt{\frac{21}{2}} C_{33,22}^{33,0}}{1600}\)

\noindent\(C_{11,3112,1}^{11,1110,0,\text{Loc}}= \frac{7 C_{00,20}^{60,0}}{48 \sqrt{5}}+\frac{7 C_{00,20}^{60,1}}{96 \sqrt{5}}+\frac{7
C_{00,22}^{62,0}}{80}+\frac{7 C_{00,22}^{62,1}}{160}+\frac{7 C_{11,20}^{31,0}}{144 \sqrt{5}}+\frac{7 C_{11,20}^{31,1}}{288 \sqrt{5}}+\frac{7 C_{11,22}^{51,0}}{1440}+\frac{7
C_{11,22}^{51,1}}{2880}+\frac{9}{320} \sqrt{\frac{21}{5}} C_{11,22}^{53,0}+\frac{9}{640} \sqrt{\frac{21}{5}} C_{11,22}^{53,1}-\frac{7 C_{20,20}^{40,0}}{144
\sqrt{5}}-\frac{7 C_{20,20}^{40,1}}{288 \sqrt{5}}-\frac{7 C_{20,22}^{42,0}}{1440}-\frac{7 C_{20,22}^{42,1}}{2880}+\frac{7 C_{22,20}^{42,0}}{720}+\frac{7
C_{22,20}^{42,1}}{1440}-\frac{13 C_{22,22}^{40,0}}{360}-\frac{13 C_{22,22}^{40,1}}{720}+\frac{\sqrt{7} C_{22,22}^{42,0}}{360}+\frac{\sqrt{7} C_{22,22}^{42,1}}{720}+\frac{3
C_{22,22}^{44,0}}{40 \sqrt{5}}+\frac{3 C_{22,22}^{44,1}}{80 \sqrt{5}}-\frac{7 C_{31,20}^{31,0}}{144 \sqrt{5}}-\frac{7 C_{31,20}^{31,1}}{288 \sqrt{5}}-\frac{37
C_{31,22}^{31,0}}{1440}-\frac{37 C_{31,22}^{31,1}}{2880}-\frac{7}{320} \sqrt{\frac{7}{15}} C_{31,22}^{33,0}-\frac{7}{640} \sqrt{\frac{7}{15}} C_{31,22}^{33,1}+\frac{1}{80}
\sqrt{\frac{21}{5}} C_{33,20}^{33,0}+\frac{1}{160} \sqrt{\frac{21}{5}} C_{33,20}^{33,1}+\frac{3}{200} \sqrt{\frac{7}{2}} C_{33,22}^{33,0}+\frac{3}{400}
\sqrt{\frac{7}{2}} C_{33,22}^{33,1}\)

\noindent\(C_{11,3112,1}^{11,1110,0,\text{NonLoc}}= \frac{7 C_{00,00}^{60,0}}{64 \sqrt{15}}+\frac{7 C_{00,00}^{60,1}}{32 \sqrt{15}}+\frac{7
C_{00,20}^{60,0}}{192 \sqrt{5}}+\frac{7 C_{00,20}^{60,1}}{96 \sqrt{5}}-\frac{7 C_{00,22}^{62,0}}{160}-\frac{7 C_{00,22}^{62,1}}{80}-\frac{7 C_{11,00}^{51,0}}{192
\sqrt{15}}-\frac{7 C_{11,00}^{51,1}}{96 \sqrt{15}}-\frac{7 C_{11,20}^{31,0}}{576 \sqrt{5}}-\frac{7 C_{11,20}^{31,1}}{288 \sqrt{5}}+\frac{7 C_{11,22}^{51,0}}{2880}+\frac{7
C_{11,22}^{51,1}}{1440}+\frac{9}{640} \sqrt{\frac{21}{5}} C_{11,22}^{53,0}+\frac{9}{320} \sqrt{\frac{21}{5}} C_{11,22}^{53,1}-\frac{7 C_{20,00}^{40,0}}{192
\sqrt{15}}-\frac{7 C_{20,00}^{40,1}}{96 \sqrt{15}}-\frac{7 C_{20,20}^{40,0}}{576 \sqrt{5}}-\frac{7 C_{20,20}^{40,1}}{288 \sqrt{5}}+\frac{7 C_{20,22}^{42,0}}{2880}+\frac{7
C_{20,22}^{42,1}}{1440}+\frac{7 C_{22,00}^{42,0}}{960 \sqrt{3}}+\frac{7 C_{22,00}^{42,1}}{480 \sqrt{3}}+\frac{7 C_{22,20}^{42,0}}{2880}+\frac{7 C_{22,20}^{42,1}}{1440}+\frac{13
C_{22,22}^{40,0}}{720}+\frac{13 C_{22,22}^{40,1}}{360}-\frac{\sqrt{7} C_{22,22}^{42,0}}{720}-\frac{\sqrt{7} C_{22,22}^{42,1}}{360}-\frac{3 C_{22,22}^{44,0}}{80
\sqrt{5}}-\frac{3 C_{22,22}^{44,1}}{40 \sqrt{5}}+\frac{7 C_{31,00}^{31,0}}{192 \sqrt{15}}+\frac{7 C_{31,00}^{31,1}}{96 \sqrt{15}}+\frac{7 C_{31,20}^{31,0}}{576
\sqrt{5}}+\frac{7 C_{31,20}^{31,1}}{288 \sqrt{5}}-\frac{37 C_{31,22}^{31,0}}{2880}-\frac{37 C_{31,22}^{31,1}}{1440}-\frac{7}{640} \sqrt{\frac{7}{15}}
C_{31,22}^{33,0}-\frac{7}{320} \sqrt{\frac{7}{15}} C_{31,22}^{33,1}-\frac{3}{320} \sqrt{\frac{7}{5}} C_{33,00}^{33,0}-\frac{3}{160} \sqrt{\frac{7}{5}}
C_{33,00}^{33,1}-\frac{1}{320} \sqrt{\frac{21}{5}} C_{33,20}^{33,0}-\frac{1}{160} \sqrt{\frac{21}{5}} C_{33,20}^{33,1}+\frac{3}{400} \sqrt{\frac{7}{2}}
C_{33,22}^{33,0}+\frac{3}{200} \sqrt{\frac{7}{2}} C_{33,22}^{33,1}\)

\noindent\(C_{11,3112,1}^{11,1110,1,\text{Loc}}= \frac{7 C_{00,20}^{60,1}}{32 \sqrt{15}}+\frac{7 \sqrt{3} C_{00,22}^{62,1}}{160}+\frac{7
C_{11,20}^{31,1}}{96 \sqrt{15}}+\frac{7 C_{11,22}^{51,1}}{960 \sqrt{3}}+\frac{27}{640} \sqrt{\frac{7}{5}} C_{11,22}^{53,1}-\frac{7 C_{20,20}^{40,1}}{96
\sqrt{15}}-\frac{7 C_{20,22}^{42,1}}{960 \sqrt{3}}+\frac{7 C_{22,20}^{42,1}}{480 \sqrt{3}}-\frac{13 C_{22,22}^{40,1}}{240 \sqrt{3}}+\frac{1}{240}
\sqrt{\frac{7}{3}} C_{22,22}^{42,1}+\frac{3}{80} \sqrt{\frac{3}{5}} C_{22,22}^{44,1}-\frac{7 C_{31,20}^{31,1}}{96 \sqrt{15}}-\frac{37 C_{31,22}^{31,1}}{960
\sqrt{3}}-\frac{7}{640} \sqrt{\frac{7}{5}} C_{31,22}^{33,1}+\frac{3}{160} \sqrt{\frac{7}{5}} C_{33,20}^{33,1}+\frac{3}{400} \sqrt{\frac{21}{2}} C_{33,22}^{33,1}\)

\noindent\(C_{11,3112,1}^{11,1110,1,\text{NonLoc}}= \frac{7 C_{00,00}^{60,0}}{64 \sqrt{5}}+\frac{7 C_{00,20}^{60,0}}{64 \sqrt{15}}-\frac{7
\sqrt{3} C_{00,22}^{62,0}}{160}-\frac{7 C_{11,00}^{51,0}}{192 \sqrt{5}}-\frac{7 C_{11,20}^{31,0}}{192 \sqrt{15}}+\frac{7 C_{11,22}^{51,0}}{960 \sqrt{3}}+\frac{27}{640}
\sqrt{\frac{7}{5}} C_{11,22}^{53,0}-\frac{7 C_{20,00}^{40,0}}{192 \sqrt{5}}-\frac{7 C_{20,20}^{40,0}}{192 \sqrt{15}}+\frac{7 C_{20,22}^{42,0}}{960
\sqrt{3}}+\frac{7 C_{22,00}^{42,0}}{960}+\frac{7 C_{22,20}^{42,0}}{960 \sqrt{3}}+\frac{13 C_{22,22}^{40,0}}{240 \sqrt{3}}-\frac{1}{240} \sqrt{\frac{7}{3}}
C_{22,22}^{42,0}-\frac{3}{80} \sqrt{\frac{3}{5}} C_{22,22}^{44,0}+\frac{7 C_{31,00}^{31,0}}{192 \sqrt{5}}+\frac{7 C_{31,20}^{31,0}}{192 \sqrt{15}}-\frac{37
C_{31,22}^{31,0}}{960 \sqrt{3}}-\frac{7}{640} \sqrt{\frac{7}{5}} C_{31,22}^{33,0}-\frac{3}{320} \sqrt{\frac{21}{5}} C_{33,00}^{33,0}-\frac{3}{320}
\sqrt{\frac{7}{5}} C_{33,20}^{33,0}+\frac{3}{400} \sqrt{\frac{21}{2}} C_{33,22}^{33,0}\)

\noindent\(C_{11,3112,1}^{11,1111,0,\text{Loc}}= -\frac{7}{64} \sqrt{\frac{3}{5}} C_{00,20}^{60,0}-\frac{7}{128} \sqrt{\frac{3}{5}}
C_{00,20}^{60,1}+\frac{21 \sqrt{3} C_{00,22}^{62,0}}{640}+\frac{21 \sqrt{3} C_{00,22}^{62,1}}{1280}-\frac{7 C_{11,20}^{31,0}}{64 \sqrt{15}}-\frac{7
C_{11,20}^{31,1}}{128 \sqrt{15}}-\frac{7 C_{11,22}^{51,0}}{640 \sqrt{3}}-\frac{7 C_{11,22}^{51,1}}{1280 \sqrt{3}}+\frac{27}{640} \sqrt{\frac{7}{5}}
C_{11,22}^{53,0}+\frac{27 \sqrt{\frac{7}{5}} C_{11,22}^{53,1}}{1280}+\frac{7 C_{20,20}^{40,0}}{64 \sqrt{15}}+\frac{7 C_{20,20}^{40,1}}{128 \sqrt{15}}+\frac{7
C_{20,22}^{42,0}}{640 \sqrt{3}}+\frac{7 C_{20,22}^{42,1}}{1280 \sqrt{3}}-\frac{7 C_{22,20}^{42,0}}{320 \sqrt{3}}-\frac{7 C_{22,20}^{42,1}}{640 \sqrt{3}}-\frac{17
C_{22,22}^{40,0}}{640 \sqrt{3}}-\frac{17 C_{22,22}^{40,1}}{1280 \sqrt{3}}-\frac{1}{64} \sqrt{\frac{7}{3}} C_{22,22}^{42,0}-\frac{1}{128} \sqrt{\frac{7}{3}}
C_{22,22}^{42,1}+\frac{9}{160} \sqrt{\frac{3}{5}} C_{22,22}^{44,0}+\frac{9}{320} \sqrt{\frac{3}{5}} C_{22,22}^{44,1}+\frac{7 C_{31,20}^{31,0}}{64
\sqrt{15}}+\frac{7 C_{31,20}^{31,1}}{128 \sqrt{15}}-\frac{23 C_{31,22}^{31,0}}{640 \sqrt{3}}-\frac{23 C_{31,22}^{31,1}}{1280 \sqrt{3}}+\frac{3}{640}
\sqrt{\frac{7}{5}} C_{31,22}^{33,0}+\frac{3 \sqrt{\frac{7}{5}} C_{31,22}^{33,1}}{1280}-\frac{9}{320} \sqrt{\frac{7}{5}} C_{33,20}^{33,0}-\frac{9}{640}
\sqrt{\frac{7}{5}} C_{33,20}^{33,1}-\frac{9}{800} \sqrt{\frac{21}{2}} C_{33,22}^{33,0}-\frac{9 \sqrt{\frac{21}{2}} C_{33,22}^{33,1}}{1600}\)

\noindent\(C_{11,3112,1}^{11,1111,0,\text{NonLoc}}= -\frac{21 C_{00,00}^{60,0}}{256 \sqrt{5}}-\frac{21 C_{00,00}^{60,1}}{128 \sqrt{5}}-\frac{7}{256}
\sqrt{\frac{3}{5}} C_{00,20}^{60,0}-\frac{7}{128} \sqrt{\frac{3}{5}} C_{00,20}^{60,1}-\frac{21 \sqrt{3} C_{00,22}^{62,0}}{1280}-\frac{21 \sqrt{3}
C_{00,22}^{62,1}}{640}+\frac{7 C_{11,00}^{51,0}}{256 \sqrt{5}}+\frac{7 C_{11,00}^{51,1}}{128 \sqrt{5}}+\frac{7 C_{11,20}^{31,0}}{256 \sqrt{15}}+\frac{7
C_{11,20}^{31,1}}{128 \sqrt{15}}-\frac{7 C_{11,22}^{51,0}}{1280 \sqrt{3}}-\frac{7 C_{11,22}^{51,1}}{640 \sqrt{3}}+\frac{27 \sqrt{\frac{7}{5}} C_{11,22}^{53,0}}{1280}+\frac{27}{640}
\sqrt{\frac{7}{5}} C_{11,22}^{53,1}+\frac{7 C_{20,00}^{40,0}}{256 \sqrt{5}}+\frac{7 C_{20,00}^{40,1}}{128 \sqrt{5}}+\frac{7 C_{20,20}^{40,0}}{256
\sqrt{15}}+\frac{7 C_{20,20}^{40,1}}{128 \sqrt{15}}-\frac{7 C_{20,22}^{42,0}}{1280 \sqrt{3}}-\frac{7 C_{20,22}^{42,1}}{640 \sqrt{3}}-\frac{7 C_{22,00}^{42,0}}{1280}-\frac{7
C_{22,00}^{42,1}}{640}-\frac{7 C_{22,20}^{42,0}}{1280 \sqrt{3}}-\frac{7 C_{22,20}^{42,1}}{640 \sqrt{3}}+\frac{17 C_{22,22}^{40,0}}{1280 \sqrt{3}}+\frac{17
C_{22,22}^{40,1}}{640 \sqrt{3}}+\frac{1}{128} \sqrt{\frac{7}{3}} C_{22,22}^{42,0}+\frac{1}{64} \sqrt{\frac{7}{3}} C_{22,22}^{42,1}-\frac{9}{320}
\sqrt{\frac{3}{5}} C_{22,22}^{44,0}-\frac{9}{160} \sqrt{\frac{3}{5}} C_{22,22}^{44,1}-\frac{7 C_{31,00}^{31,0}}{256 \sqrt{5}}-\frac{7 C_{31,00}^{31,1}}{128
\sqrt{5}}-\frac{7 C_{31,20}^{31,0}}{256 \sqrt{15}}-\frac{7 C_{31,20}^{31,1}}{128 \sqrt{15}}-\frac{23 C_{31,22}^{31,0}}{1280 \sqrt{3}}-\frac{23 C_{31,22}^{31,1}}{640
\sqrt{3}}+\frac{3 \sqrt{\frac{7}{5}} C_{31,22}^{33,0}}{1280}+\frac{3}{640} \sqrt{\frac{7}{5}} C_{31,22}^{33,1}+\frac{9 \sqrt{\frac{21}{5}} C_{33,00}^{33,0}}{1280}+\frac{9}{640}
\sqrt{\frac{21}{5}} C_{33,00}^{33,1}+\frac{9 \sqrt{\frac{7}{5}} C_{33,20}^{33,0}}{1280}+\frac{9}{640} \sqrt{\frac{7}{5}} C_{33,20}^{33,1}-\frac{9
\sqrt{\frac{21}{2}} C_{33,22}^{33,0}}{1600}-\frac{9}{800} \sqrt{\frac{21}{2}} C_{33,22}^{33,1}\)

\noindent\(C_{11,3112,1}^{11,1111,1,\text{Loc}}= -\frac{21 C_{00,20}^{60,1}}{128 \sqrt{5}}+\frac{63 C_{00,22}^{62,1}}{1280}-\frac{7
C_{11,20}^{31,1}}{128 \sqrt{5}}-\frac{7 C_{11,22}^{51,1}}{1280}+\frac{27 \sqrt{\frac{21}{5}} C_{11,22}^{53,1}}{1280}+\frac{7 C_{20,20}^{40,1}}{128
\sqrt{5}}+\frac{7 C_{20,22}^{42,1}}{1280}-\frac{7 C_{22,20}^{42,1}}{640}-\frac{17 C_{22,22}^{40,1}}{1280}-\frac{\sqrt{7} C_{22,22}^{42,1}}{128}+\frac{27
C_{22,22}^{44,1}}{320 \sqrt{5}}+\frac{7 C_{31,20}^{31,1}}{128 \sqrt{5}}-\frac{23 C_{31,22}^{31,1}}{1280}+\frac{3 \sqrt{\frac{21}{5}} C_{31,22}^{33,1}}{1280}-\frac{9}{640}
\sqrt{\frac{21}{5}} C_{33,20}^{33,1}-\frac{27 \sqrt{\frac{7}{2}} C_{33,22}^{33,1}}{1600}\)

\noindent\(C_{11,3112,1}^{11,1111,1,\text{NonLoc}}= -\frac{21}{256} \sqrt{\frac{3}{5}} C_{00,00}^{60,0}-\frac{21 C_{00,20}^{60,0}}{256
\sqrt{5}}-\frac{63 C_{00,22}^{62,0}}{1280}+\frac{7}{256} \sqrt{\frac{3}{5}} C_{11,00}^{51,0}+\frac{7 C_{11,20}^{31,0}}{256 \sqrt{5}}-\frac{7 C_{11,22}^{51,0}}{1280}+\frac{27
\sqrt{\frac{21}{5}} C_{11,22}^{53,0}}{1280}+\frac{7}{256} \sqrt{\frac{3}{5}} C_{20,00}^{40,0}+\frac{7 C_{20,20}^{40,0}}{256 \sqrt{5}}-\frac{7 C_{20,22}^{42,0}}{1280}-\frac{7
\sqrt{3} C_{22,00}^{42,0}}{1280}-\frac{7 C_{22,20}^{42,0}}{1280}+\frac{17 C_{22,22}^{40,0}}{1280}+\frac{\sqrt{7} C_{22,22}^{42,0}}{128}-\frac{27
C_{22,22}^{44,0}}{320 \sqrt{5}}-\frac{7}{256} \sqrt{\frac{3}{5}} C_{31,00}^{31,0}-\frac{7 C_{31,20}^{31,0}}{256 \sqrt{5}}-\frac{23 C_{31,22}^{31,0}}{1280}+\frac{3
\sqrt{\frac{21}{5}} C_{31,22}^{33,0}}{1280}+\frac{27 \sqrt{\frac{7}{5}} C_{33,00}^{33,0}}{1280}+\frac{9 \sqrt{\frac{21}{5}} C_{33,20}^{33,0}}{1280}-\frac{27
\sqrt{\frac{7}{2}} C_{33,22}^{33,0}}{1600}\)

\noindent\(C_{11,3112,1}^{11,1112,0,\text{Loc}}= \frac{119 C_{00,20}^{60,0}}{960}+\frac{119 C_{00,20}^{60,1}}{1920}+\frac{49 C_{00,22}^{62,0}}{640
\sqrt{5}}+\frac{49 C_{00,22}^{62,1}}{1280 \sqrt{5}}+\frac{119 C_{11,20}^{31,0}}{2880}+\frac{119 C_{11,20}^{31,1}}{5760}-\frac{7 C_{11,22}^{51,0}}{5760
\sqrt{5}}-\frac{7 C_{11,22}^{51,1}}{11520 \sqrt{5}}+\frac{9 \sqrt{21} C_{11,22}^{53,0}}{1600}+\frac{9 \sqrt{21} C_{11,22}^{53,1}}{3200}+\frac{C_{20,20}^{40,0}}{2880}+\frac{C_{20,20}^{40,1}}{5760}+\frac{91
C_{20,22}^{42,0}}{5760 \sqrt{5}}+\frac{91 C_{20,22}^{42,1}}{11520 \sqrt{5}}-\frac{91 C_{22,20}^{42,0}}{2880 \sqrt{5}}-\frac{91 C_{22,20}^{42,1}}{5760
\sqrt{5}}-\frac{7 C_{22,22}^{40,0}}{1152 \sqrt{5}}-\frac{7 C_{22,22}^{40,1}}{2304 \sqrt{5}}-\frac{53 \sqrt{\frac{7}{5}} C_{22,22}^{42,0}}{2880}-\frac{53
\sqrt{\frac{7}{5}} C_{22,22}^{42,1}}{5760}+\frac{3 C_{22,22}^{44,0}}{160}+\frac{3 C_{22,22}^{44,1}}{320}-\frac{59 C_{31,20}^{31,0}}{2880}-\frac{59
C_{31,20}^{31,1}}{5760}-\frac{119 C_{31,22}^{31,0}}{5760 \sqrt{5}}-\frac{119 C_{31,22}^{31,1}}{11520 \sqrt{5}}+\frac{1}{400} \sqrt{\frac{7}{3}} C_{31,22}^{33,0}+\frac{1}{800}
\sqrt{\frac{7}{3}} C_{31,22}^{33,1}-\frac{13 \sqrt{21} C_{33,20}^{33,0}}{1600}-\frac{13 \sqrt{21} C_{33,20}^{33,1}}{3200}-\frac{39}{800} \sqrt{\frac{7}{10}}
C_{33,22}^{33,0}-\frac{39 \sqrt{\frac{7}{10}} C_{33,22}^{33,1}}{1600}\)

\noindent\(C_{11,3112,1}^{11,1112,0,\text{NonLoc}}= \frac{119 C_{00,00}^{60,0}}{1280 \sqrt{3}}+\frac{119 C_{00,00}^{60,1}}{640 \sqrt{3}}+\frac{119
C_{00,20}^{60,0}}{3840}+\frac{119 C_{00,20}^{60,1}}{1920}-\frac{49 C_{00,22}^{62,0}}{1280 \sqrt{5}}-\frac{49 C_{00,22}^{62,1}}{640 \sqrt{5}}-\frac{119
C_{11,00}^{51,0}}{3840 \sqrt{3}}-\frac{119 C_{11,00}^{51,1}}{1920 \sqrt{3}}-\frac{119 C_{11,20}^{31,0}}{11520}-\frac{119 C_{11,20}^{31,1}}{5760}-\frac{7
C_{11,22}^{51,0}}{11520 \sqrt{5}}-\frac{7 C_{11,22}^{51,1}}{5760 \sqrt{5}}+\frac{9 \sqrt{21} C_{11,22}^{53,0}}{3200}+\frac{9 \sqrt{21} C_{11,22}^{53,1}}{1600}+\frac{C_{20,00}^{40,0}}{3840
\sqrt{3}}+\frac{C_{20,00}^{40,1}}{1920 \sqrt{3}}+\frac{C_{20,20}^{40,0}}{11520}+\frac{C_{20,20}^{40,1}}{5760}-\frac{91 C_{20,22}^{42,0}}{11520 \sqrt{5}}-\frac{91
C_{20,22}^{42,1}}{5760 \sqrt{5}}-\frac{91 C_{22,00}^{42,0}}{3840 \sqrt{15}}-\frac{91 C_{22,00}^{42,1}}{1920 \sqrt{15}}-\frac{91 C_{22,20}^{42,0}}{11520
\sqrt{5}}-\frac{91 C_{22,20}^{42,1}}{5760 \sqrt{5}}+\frac{7 C_{22,22}^{40,0}}{2304 \sqrt{5}}+\frac{7 C_{22,22}^{40,1}}{1152 \sqrt{5}}+\frac{53 \sqrt{\frac{7}{5}}
C_{22,22}^{42,0}}{5760}+\frac{53 \sqrt{\frac{7}{5}} C_{22,22}^{42,1}}{2880}-\frac{3 C_{22,22}^{44,0}}{320}-\frac{3 C_{22,22}^{44,1}}{160}+\frac{59
C_{31,00}^{31,0}}{3840 \sqrt{3}}+\frac{59 C_{31,00}^{31,1}}{1920 \sqrt{3}}+\frac{59 C_{31,20}^{31,0}}{11520}+\frac{59 C_{31,20}^{31,1}}{5760}-\frac{119
C_{31,22}^{31,0}}{11520 \sqrt{5}}-\frac{119 C_{31,22}^{31,1}}{5760 \sqrt{5}}+\frac{1}{800} \sqrt{\frac{7}{3}} C_{31,22}^{33,0}+\frac{1}{400} \sqrt{\frac{7}{3}}
C_{31,22}^{33,1}+\frac{39 \sqrt{7} C_{33,00}^{33,0}}{6400}+\frac{39 \sqrt{7} C_{33,00}^{33,1}}{3200}+\frac{13 \sqrt{21} C_{33,20}^{33,0}}{6400}+\frac{13
\sqrt{21} C_{33,20}^{33,1}}{3200}-\frac{39 \sqrt{\frac{7}{10}} C_{33,22}^{33,0}}{1600}-\frac{39}{800} \sqrt{\frac{7}{10}} C_{33,22}^{33,1}\)

\noindent\(C_{11,3112,1}^{11,1112,1,\text{Loc}}= \frac{119 C_{00,20}^{60,1}}{640 \sqrt{3}}+\frac{49 \sqrt{\frac{3}{5}} C_{00,22}^{62,1}}{1280}+\frac{119
C_{11,20}^{31,1}}{1920 \sqrt{3}}-\frac{7 C_{11,22}^{51,1}}{3840 \sqrt{15}}+\frac{27 \sqrt{7} C_{11,22}^{53,1}}{3200}+\frac{C_{20,20}^{40,1}}{1920
\sqrt{3}}+\frac{91 C_{20,22}^{42,1}}{3840 \sqrt{15}}-\frac{91 C_{22,20}^{42,1}}{1920 \sqrt{15}}-\frac{7 C_{22,22}^{40,1}}{768 \sqrt{15}}-\frac{53
\sqrt{\frac{7}{15}} C_{22,22}^{42,1}}{1920}+\frac{3 \sqrt{3} C_{22,22}^{44,1}}{320}-\frac{59 C_{31,20}^{31,1}}{1920 \sqrt{3}}-\frac{119 C_{31,22}^{31,1}}{3840
\sqrt{15}}+\frac{\sqrt{7} C_{31,22}^{33,1}}{800}-\frac{39 \sqrt{7} C_{33,20}^{33,1}}{3200}-\frac{39 \sqrt{\frac{21}{10}} C_{33,22}^{33,1}}{1600}\)

\noindent\(C_{11,3112,1}^{11,1112,1,\text{NonLoc}}= \frac{119 C_{00,00}^{60,0}}{1280}+\frac{119 C_{00,20}^{60,0}}{1280 \sqrt{3}}-\frac{49
\sqrt{\frac{3}{5}} C_{00,22}^{62,0}}{1280}-\frac{119 C_{11,00}^{51,0}}{3840}-\frac{119 C_{11,20}^{31,0}}{3840 \sqrt{3}}-\frac{7 C_{11,22}^{51,0}}{3840
\sqrt{15}}+\frac{27 \sqrt{7} C_{11,22}^{53,0}}{3200}+\frac{C_{20,00}^{40,0}}{3840}+\frac{C_{20,20}^{40,0}}{3840 \sqrt{3}}-\frac{91 C_{20,22}^{42,0}}{3840
\sqrt{15}}-\frac{91 C_{22,00}^{42,0}}{3840 \sqrt{5}}-\frac{91 C_{22,20}^{42,0}}{3840 \sqrt{15}}+\frac{7 C_{22,22}^{40,0}}{768 \sqrt{15}}+\frac{53
\sqrt{\frac{7}{15}} C_{22,22}^{42,0}}{1920}-\frac{3 \sqrt{3} C_{22,22}^{44,0}}{320}+\frac{59 C_{31,00}^{31,0}}{3840}+\frac{59 C_{31,20}^{31,0}}{3840
\sqrt{3}}-\frac{119 C_{31,22}^{31,0}}{3840 \sqrt{15}}+\frac{\sqrt{7} C_{31,22}^{33,0}}{800}+\frac{39 \sqrt{21} C_{33,00}^{33,0}}{6400}+\frac{39 \sqrt{7}
C_{33,20}^{33,0}}{6400}-\frac{39 \sqrt{\frac{21}{10}} C_{33,22}^{33,0}}{1600}\)

\noindent\(C_{11,3112,2}^{00,2202,0,\text{Loc}}= \frac{7 C_{11,11}^{51,0}}{32 \sqrt{15}}+\frac{7 C_{11,11}^{51,1}}{64 \sqrt{15}}+\frac{7
C_{22,11}^{42,0}}{80}+\frac{7 C_{22,11}^{42,1}}{160}+\frac{7 C_{31,11}^{31,0}}{32 \sqrt{15}}+\frac{7 C_{31,11}^{31,1}}{64 \sqrt{15}}+\frac{3}{40}
\sqrt{\frac{21}{10}} C_{33,11}^{33,0}+\frac{3}{80} \sqrt{\frac{21}{10}} C_{33,11}^{33,1}\)

\noindent\(C_{11,3112,2}^{00,2202,0,\text{NonLoc}}= \frac{7 C_{11,11}^{51,0}}{64 \sqrt{15}}+\frac{7 C_{11,11}^{51,1}}{32 \sqrt{15}}-\frac{7
C_{22,11}^{42,0}}{160}-\frac{7 C_{22,11}^{42,1}}{80}+\frac{7 C_{31,11}^{31,0}}{64 \sqrt{15}}+\frac{7 C_{31,11}^{31,1}}{32 \sqrt{15}}+\frac{3}{80}
\sqrt{\frac{21}{10}} C_{33,11}^{33,0}+\frac{3}{40} \sqrt{\frac{21}{10}} C_{33,11}^{33,1}\)

\noindent\(C_{11,3112,2}^{00,2202,1,\text{Loc}}= \frac{7 C_{11,11}^{51,1}}{64 \sqrt{5}}+\frac{7 \sqrt{3} C_{22,11}^{42,1}}{160}+\frac{7
C_{31,11}^{31,1}}{64 \sqrt{5}}+\frac{9}{80} \sqrt{\frac{7}{10}} C_{33,11}^{33,1}\)

\noindent\(C_{11,3112,2}^{00,2202,1,\text{NonLoc}}= \frac{7 C_{11,11}^{51,0}}{64 \sqrt{5}}-\frac{7 \sqrt{3} C_{22,11}^{42,0}}{160}+\frac{7
C_{31,11}^{31,0}}{64 \sqrt{5}}+\frac{9}{80} \sqrt{\frac{7}{10}} C_{33,11}^{33,0}\)

\noindent\(C_{11,3112,2}^{11,1111,0,\text{Loc}}= \frac{7 C_{00,20}^{60,0}}{64 \sqrt{5}}+\frac{7 C_{00,20}^{60,1}}{128 \sqrt{5}}-\frac{21
C_{00,22}^{62,0}}{640}-\frac{21 C_{00,22}^{62,1}}{1280}+\frac{7 C_{11,20}^{31,0}}{192 \sqrt{5}}+\frac{7 C_{11,20}^{31,1}}{384 \sqrt{5}}+\frac{7 C_{11,22}^{51,0}}{1920}+\frac{7
C_{11,22}^{51,1}}{3840}-\frac{9}{640} \sqrt{\frac{21}{5}} C_{11,22}^{53,0}-\frac{9 \sqrt{\frac{21}{5}} C_{11,22}^{53,1}}{1280}-\frac{7 C_{20,20}^{40,0}}{192
\sqrt{5}}-\frac{7 C_{20,20}^{40,1}}{384 \sqrt{5}}-\frac{7 C_{20,22}^{42,0}}{1920}-\frac{7 C_{20,22}^{42,1}}{3840}+\frac{7 C_{22,20}^{42,0}}{960}+\frac{7
C_{22,20}^{42,1}}{1920}+\frac{17 C_{22,22}^{40,0}}{1920}+\frac{17 C_{22,22}^{40,1}}{3840}+\frac{\sqrt{7} C_{22,22}^{42,0}}{192}+\frac{\sqrt{7} C_{22,22}^{42,1}}{384}-\frac{9
C_{22,22}^{44,0}}{160 \sqrt{5}}-\frac{9 C_{22,22}^{44,1}}{320 \sqrt{5}}-\frac{7 C_{31,20}^{31,0}}{192 \sqrt{5}}-\frac{7 C_{31,20}^{31,1}}{384 \sqrt{5}}+\frac{23
C_{31,22}^{31,0}}{1920}+\frac{23 C_{31,22}^{31,1}}{3840}-\frac{1}{640} \sqrt{\frac{21}{5}} C_{31,22}^{33,0}-\frac{\sqrt{\frac{21}{5}} C_{31,22}^{33,1}}{1280}+\frac{3}{320}
\sqrt{\frac{21}{5}} C_{33,20}^{33,0}+\frac{3}{640} \sqrt{\frac{21}{5}} C_{33,20}^{33,1}+\frac{9}{800} \sqrt{\frac{7}{2}} C_{33,22}^{33,0}+\frac{9
\sqrt{\frac{7}{2}} C_{33,22}^{33,1}}{1600}\)

\noindent\(C_{11,3112,2}^{11,1111,0,\text{NonLoc}}= \frac{7}{256} \sqrt{\frac{3}{5}} C_{00,00}^{60,0}+\frac{7}{128} \sqrt{\frac{3}{5}}
C_{00,00}^{60,1}+\frac{7 C_{00,20}^{60,0}}{256 \sqrt{5}}+\frac{7 C_{00,20}^{60,1}}{128 \sqrt{5}}+\frac{21 C_{00,22}^{62,0}}{1280}+\frac{21 C_{00,22}^{62,1}}{640}-\frac{7
C_{11,00}^{51,0}}{256 \sqrt{15}}-\frac{7 C_{11,00}^{51,1}}{128 \sqrt{15}}-\frac{7 C_{11,20}^{31,0}}{768 \sqrt{5}}-\frac{7 C_{11,20}^{31,1}}{384 \sqrt{5}}+\frac{7
C_{11,22}^{51,0}}{3840}+\frac{7 C_{11,22}^{51,1}}{1920}-\frac{9 \sqrt{\frac{21}{5}} C_{11,22}^{53,0}}{1280}-\frac{9}{640} \sqrt{\frac{21}{5}} C_{11,22}^{53,1}-\frac{7
C_{20,00}^{40,0}}{256 \sqrt{15}}-\frac{7 C_{20,00}^{40,1}}{128 \sqrt{15}}-\frac{7 C_{20,20}^{40,0}}{768 \sqrt{5}}-\frac{7 C_{20,20}^{40,1}}{384 \sqrt{5}}+\frac{7
C_{20,22}^{42,0}}{3840}+\frac{7 C_{20,22}^{42,1}}{1920}+\frac{7 C_{22,00}^{42,0}}{1280 \sqrt{3}}+\frac{7 C_{22,00}^{42,1}}{640 \sqrt{3}}+\frac{7
C_{22,20}^{42,0}}{3840}+\frac{7 C_{22,20}^{42,1}}{1920}-\frac{17 C_{22,22}^{40,0}}{3840}-\frac{17 C_{22,22}^{40,1}}{1920}-\frac{\sqrt{7} C_{22,22}^{42,0}}{384}-\frac{\sqrt{7}
C_{22,22}^{42,1}}{192}+\frac{9 C_{22,22}^{44,0}}{320 \sqrt{5}}+\frac{9 C_{22,22}^{44,1}}{160 \sqrt{5}}+\frac{7 C_{31,00}^{31,0}}{256 \sqrt{15}}+\frac{7
C_{31,00}^{31,1}}{128 \sqrt{15}}+\frac{7 C_{31,20}^{31,0}}{768 \sqrt{5}}+\frac{7 C_{31,20}^{31,1}}{384 \sqrt{5}}+\frac{23 C_{31,22}^{31,0}}{3840}+\frac{23
C_{31,22}^{31,1}}{1920}-\frac{\sqrt{\frac{21}{5}} C_{31,22}^{33,0}}{1280}-\frac{1}{640} \sqrt{\frac{21}{5}} C_{31,22}^{33,1}-\frac{9 \sqrt{\frac{7}{5}}
C_{33,00}^{33,0}}{1280}-\frac{9}{640} \sqrt{\frac{7}{5}} C_{33,00}^{33,1}-\frac{3 \sqrt{\frac{21}{5}} C_{33,20}^{33,0}}{1280}-\frac{3}{640} \sqrt{\frac{21}{5}}
C_{33,20}^{33,1}+\frac{9 \sqrt{\frac{7}{2}} C_{33,22}^{33,0}}{1600}+\frac{9}{800} \sqrt{\frac{7}{2}} C_{33,22}^{33,1}\)

\noindent\(C_{11,3112,2}^{11,1111,1,\text{Loc}}= \frac{7}{128} \sqrt{\frac{3}{5}} C_{00,20}^{60,1}-\frac{21 \sqrt{3} C_{00,22}^{62,1}}{1280}+\frac{7
C_{11,20}^{31,1}}{128 \sqrt{15}}+\frac{7 C_{11,22}^{51,1}}{1280 \sqrt{3}}-\frac{27 \sqrt{\frac{7}{5}} C_{11,22}^{53,1}}{1280}-\frac{7 C_{20,20}^{40,1}}{128
\sqrt{15}}-\frac{7 C_{20,22}^{42,1}}{1280 \sqrt{3}}+\frac{7 C_{22,20}^{42,1}}{640 \sqrt{3}}+\frac{17 C_{22,22}^{40,1}}{1280 \sqrt{3}}+\frac{1}{128}
\sqrt{\frac{7}{3}} C_{22,22}^{42,1}-\frac{9}{320} \sqrt{\frac{3}{5}} C_{22,22}^{44,1}-\frac{7 C_{31,20}^{31,1}}{128 \sqrt{15}}+\frac{23 C_{31,22}^{31,1}}{1280
\sqrt{3}}-\frac{3 \sqrt{\frac{7}{5}} C_{31,22}^{33,1}}{1280}+\frac{9}{640} \sqrt{\frac{7}{5}} C_{33,20}^{33,1}+\frac{9 \sqrt{\frac{21}{2}} C_{33,22}^{33,1}}{1600}\)

\noindent\(C_{11,3112,2}^{11,1111,1,\text{NonLoc}}= \frac{21 C_{00,00}^{60,0}}{256 \sqrt{5}}+\frac{7}{256} \sqrt{\frac{3}{5}} C_{00,20}^{60,0}+\frac{21
\sqrt{3} C_{00,22}^{62,0}}{1280}-\frac{7 C_{11,00}^{51,0}}{256 \sqrt{5}}-\frac{7 C_{11,20}^{31,0}}{256 \sqrt{15}}+\frac{7 C_{11,22}^{51,0}}{1280
\sqrt{3}}-\frac{27 \sqrt{\frac{7}{5}} C_{11,22}^{53,0}}{1280}-\frac{7 C_{20,00}^{40,0}}{256 \sqrt{5}}-\frac{7 C_{20,20}^{40,0}}{256 \sqrt{15}}+\frac{7
C_{20,22}^{42,0}}{1280 \sqrt{3}}+\frac{7 C_{22,00}^{42,0}}{1280}+\frac{7 C_{22,20}^{42,0}}{1280 \sqrt{3}}-\frac{17 C_{22,22}^{40,0}}{1280 \sqrt{3}}-\frac{1}{128}
\sqrt{\frac{7}{3}} C_{22,22}^{42,0}+\frac{9}{320} \sqrt{\frac{3}{5}} C_{22,22}^{44,0}+\frac{7 C_{31,00}^{31,0}}{256 \sqrt{5}}+\frac{7 C_{31,20}^{31,0}}{256
\sqrt{15}}+\frac{23 C_{31,22}^{31,0}}{1280 \sqrt{3}}-\frac{3 \sqrt{\frac{7}{5}} C_{31,22}^{33,0}}{1280}-\frac{9 \sqrt{\frac{21}{5}} C_{33,00}^{33,0}}{1280}-\frac{9
\sqrt{\frac{7}{5}} C_{33,20}^{33,0}}{1280}+\frac{9 \sqrt{\frac{21}{2}} C_{33,22}^{33,0}}{1600}\)

\noindent\(C_{11,3112,2}^{11,1112,0,\text{Loc}}= \frac{7 C_{00,20}^{60,0}}{64 \sqrt{15}}+\frac{7 C_{00,20}^{60,1}}{128 \sqrt{15}}-\frac{7
\sqrt{3} C_{00,22}^{62,0}}{640}-\frac{7 \sqrt{3} C_{00,22}^{62,1}}{1280}+\frac{7 C_{11,20}^{31,0}}{192 \sqrt{15}}+\frac{7 C_{11,20}^{31,1}}{384 \sqrt{15}}+\frac{49
C_{11,22}^{51,0}}{1920 \sqrt{3}}+\frac{49 C_{11,22}^{51,1}}{3840 \sqrt{3}}-\frac{9}{320} \sqrt{\frac{7}{5}} C_{11,22}^{53,0}-\frac{9}{640} \sqrt{\frac{7}{5}}
C_{11,22}^{53,1}+\frac{11 C_{20,20}^{40,0}}{64 \sqrt{15}}+\frac{11 C_{20,20}^{40,1}}{128 \sqrt{15}}-\frac{7 C_{20,22}^{42,0}}{640 \sqrt{3}}-\frac{7
C_{20,22}^{42,1}}{1280 \sqrt{3}}-\frac{7 \sqrt{3} C_{22,20}^{42,0}}{320}-\frac{7 \sqrt{3} C_{22,20}^{42,1}}{640}+\frac{7 C_{22,22}^{40,0}}{640 \sqrt{3}}+\frac{7
C_{22,22}^{40,1}}{1280 \sqrt{3}}+\frac{1}{64} \sqrt{\frac{7}{3}} C_{22,22}^{42,0}+\frac{1}{128} \sqrt{\frac{7}{3}} C_{22,22}^{42,1}-\frac{9}{160}
\sqrt{\frac{3}{5}} C_{22,22}^{44,0}-\frac{9}{320} \sqrt{\frac{3}{5}} C_{22,22}^{44,1}+\frac{13 C_{31,20}^{31,0}}{192 \sqrt{15}}+\frac{13 C_{31,20}^{31,1}}{384
\sqrt{15}}+\frac{49 C_{31,22}^{31,0}}{1920 \sqrt{3}}+\frac{49 C_{31,22}^{31,1}}{3840 \sqrt{3}}-\frac{1}{80} \sqrt{\frac{7}{5}} C_{31,22}^{33,0}-\frac{1}{160}
\sqrt{\frac{7}{5}} C_{31,22}^{33,1}-\frac{27}{320} \sqrt{\frac{7}{5}} C_{33,20}^{33,0}-\frac{27}{640} \sqrt{\frac{7}{5}} C_{33,20}^{33,1}+\frac{9}{800}
\sqrt{\frac{21}{2}} C_{33,22}^{33,0}+\frac{9 \sqrt{\frac{21}{2}} C_{33,22}^{33,1}}{1600}\)

\noindent\(C_{11,3112,2}^{11,1112,0,\text{NonLoc}}= \frac{7 C_{00,00}^{60,0}}{256 \sqrt{5}}+\frac{7 C_{00,00}^{60,1}}{128 \sqrt{5}}+\frac{7
C_{00,20}^{60,0}}{256 \sqrt{15}}+\frac{7 C_{00,20}^{60,1}}{128 \sqrt{15}}+\frac{7 \sqrt{3} C_{00,22}^{62,0}}{1280}+\frac{7 \sqrt{3} C_{00,22}^{62,1}}{640}-\frac{7
C_{11,00}^{51,0}}{768 \sqrt{5}}-\frac{7 C_{11,00}^{51,1}}{384 \sqrt{5}}-\frac{7 C_{11,20}^{31,0}}{768 \sqrt{15}}-\frac{7 C_{11,20}^{31,1}}{384 \sqrt{15}}+\frac{49
C_{11,22}^{51,0}}{3840 \sqrt{3}}+\frac{49 C_{11,22}^{51,1}}{1920 \sqrt{3}}-\frac{9}{640} \sqrt{\frac{7}{5}} C_{11,22}^{53,0}-\frac{9}{320} \sqrt{\frac{7}{5}}
C_{11,22}^{53,1}+\frac{11 C_{20,00}^{40,0}}{256 \sqrt{5}}+\frac{11 C_{20,00}^{40,1}}{128 \sqrt{5}}+\frac{11 C_{20,20}^{40,0}}{256 \sqrt{15}}+\frac{11
C_{20,20}^{40,1}}{128 \sqrt{15}}+\frac{7 C_{20,22}^{42,0}}{1280 \sqrt{3}}+\frac{7 C_{20,22}^{42,1}}{640 \sqrt{3}}-\frac{21 C_{22,00}^{42,0}}{1280}-\frac{21
C_{22,00}^{42,1}}{640}-\frac{7 \sqrt{3} C_{22,20}^{42,0}}{1280}-\frac{7 \sqrt{3} C_{22,20}^{42,1}}{640}-\frac{7 C_{22,22}^{40,0}}{1280 \sqrt{3}}-\frac{7
C_{22,22}^{40,1}}{640 \sqrt{3}}-\frac{1}{128} \sqrt{\frac{7}{3}} C_{22,22}^{42,0}-\frac{1}{64} \sqrt{\frac{7}{3}} C_{22,22}^{42,1}+\frac{9}{320}
\sqrt{\frac{3}{5}} C_{22,22}^{44,0}+\frac{9}{160} \sqrt{\frac{3}{5}} C_{22,22}^{44,1}-\frac{13 C_{31,00}^{31,0}}{768 \sqrt{5}}-\frac{13 C_{31,00}^{31,1}}{384
\sqrt{5}}-\frac{13 C_{31,20}^{31,0}}{768 \sqrt{15}}-\frac{13 C_{31,20}^{31,1}}{384 \sqrt{15}}+\frac{49 C_{31,22}^{31,0}}{3840 \sqrt{3}}+\frac{49
C_{31,22}^{31,1}}{1920 \sqrt{3}}-\frac{1}{160} \sqrt{\frac{7}{5}} C_{31,22}^{33,0}-\frac{1}{80} \sqrt{\frac{7}{5}} C_{31,22}^{33,1}+\frac{27 \sqrt{\frac{21}{5}}
C_{33,00}^{33,0}}{1280}+\frac{27}{640} \sqrt{\frac{21}{5}} C_{33,00}^{33,1}+\frac{27 \sqrt{\frac{7}{5}} C_{33,20}^{33,0}}{1280}+\frac{27}{640} \sqrt{\frac{7}{5}}
C_{33,20}^{33,1}+\frac{9 \sqrt{\frac{21}{2}} C_{33,22}^{33,0}}{1600}+\frac{9}{800} \sqrt{\frac{21}{2}} C_{33,22}^{33,1}\)

\noindent\(C_{11,3112,2}^{11,1112,1,\text{Loc}}= \frac{7 C_{00,20}^{60,1}}{128 \sqrt{5}}-\frac{21 C_{00,22}^{62,1}}{1280}+\frac{7 C_{11,20}^{31,1}}{384
\sqrt{5}}+\frac{49 C_{11,22}^{51,1}}{3840}-\frac{9}{640} \sqrt{\frac{21}{5}} C_{11,22}^{53,1}+\frac{11 C_{20,20}^{40,1}}{128 \sqrt{5}}-\frac{7 C_{20,22}^{42,1}}{1280}-\frac{21
C_{22,20}^{42,1}}{640}+\frac{7 C_{22,22}^{40,1}}{1280}+\frac{\sqrt{7} C_{22,22}^{42,1}}{128}-\frac{27 C_{22,22}^{44,1}}{320 \sqrt{5}}+\frac{13 C_{31,20}^{31,1}}{384
\sqrt{5}}+\frac{49 C_{31,22}^{31,1}}{3840}-\frac{1}{160} \sqrt{\frac{21}{5}} C_{31,22}^{33,1}-\frac{27}{640} \sqrt{\frac{21}{5}} C_{33,20}^{33,1}+\frac{27
\sqrt{\frac{7}{2}} C_{33,22}^{33,1}}{1600}\)

\noindent\(C_{11,3112,2}^{11,1112,1,\text{NonLoc}}= \frac{7}{256} \sqrt{\frac{3}{5}} C_{00,00}^{60,0}+\frac{7 C_{00,20}^{60,0}}{256
\sqrt{5}}+\frac{21 C_{00,22}^{62,0}}{1280}-\frac{7 C_{11,00}^{51,0}}{256 \sqrt{15}}-\frac{7 C_{11,20}^{31,0}}{768 \sqrt{5}}+\frac{49 C_{11,22}^{51,0}}{3840}-\frac{9}{640}
\sqrt{\frac{21}{5}} C_{11,22}^{53,0}+\frac{11}{256} \sqrt{\frac{3}{5}} C_{20,00}^{40,0}+\frac{11 C_{20,20}^{40,0}}{256 \sqrt{5}}+\frac{7 C_{20,22}^{42,0}}{1280}-\frac{21
\sqrt{3} C_{22,00}^{42,0}}{1280}-\frac{21 C_{22,20}^{42,0}}{1280}-\frac{7 C_{22,22}^{40,0}}{1280}-\frac{\sqrt{7} C_{22,22}^{42,0}}{128}+\frac{27
C_{22,22}^{44,0}}{320 \sqrt{5}}-\frac{13 C_{31,00}^{31,0}}{256 \sqrt{15}}-\frac{13 C_{31,20}^{31,0}}{768 \sqrt{5}}+\frac{49 C_{31,22}^{31,0}}{3840}-\frac{1}{160}
\sqrt{\frac{21}{5}} C_{31,22}^{33,0}+\frac{81 \sqrt{\frac{7}{5}} C_{33,00}^{33,0}}{1280}+\frac{27 \sqrt{\frac{21}{5}} C_{33,20}^{33,0}}{1280}+\frac{27
\sqrt{\frac{7}{2}} C_{33,22}^{33,0}}{1600}\)

\noindent\(C_{11,3112,3}^{11,1112,0,\text{Loc}}= \frac{7}{80} \sqrt{\frac{7}{3}} C_{00,20}^{60,0}+\frac{7}{160} \sqrt{\frac{7}{3}}
C_{00,20}^{60,1}+\frac{1}{80} \sqrt{\frac{21}{5}} C_{00,22}^{62,0}+\frac{1}{160} \sqrt{\frac{21}{5}} C_{00,22}^{62,1}+\frac{7}{240} \sqrt{\frac{7}{3}}
C_{11,20}^{31,0}+\frac{7}{480} \sqrt{\frac{7}{3}} C_{11,20}^{31,1}+\frac{1}{120} \sqrt{\frac{7}{15}} C_{11,22}^{51,0}+\frac{1}{240} \sqrt{\frac{7}{15}}
C_{11,22}^{51,1}+\frac{9 C_{11,22}^{53,0}}{800}+\frac{9 C_{11,22}^{53,1}}{1600}+\frac{1}{80} \sqrt{\frac{7}{3}} C_{20,20}^{40,0}+\frac{1}{160} \sqrt{\frac{7}{3}}
C_{20,20}^{40,1}+\frac{1}{160} \sqrt{\frac{7}{15}} C_{20,22}^{42,0}+\frac{1}{320} \sqrt{\frac{7}{15}} C_{20,22}^{42,1}-\frac{7}{160} \sqrt{\frac{7}{15}}
C_{22,20}^{42,0}-\frac{7}{320} \sqrt{\frac{7}{15}} C_{22,20}^{42,1}-\frac{7 C_{22,22}^{42,0}}{160 \sqrt{15}}-\frac{7 C_{22,22}^{42,1}}{320 \sqrt{15}}-\frac{1}{120}
\sqrt{\frac{7}{3}} C_{31,20}^{31,0}-\frac{1}{240} \sqrt{\frac{7}{3}} C_{31,20}^{31,1}-\frac{1}{240} \sqrt{\frac{7}{15}} C_{31,22}^{31,0}-\frac{1}{480}
\sqrt{\frac{7}{15}} C_{31,22}^{31,1}-\frac{C_{31,22}^{33,0}}{400}-\frac{C_{31,22}^{33,1}}{800}-\frac{63 C_{33,20}^{33,0}}{800}-\frac{63 C_{33,20}^{33,1}}{1600}-\frac{9}{200}
\sqrt{\frac{3}{10}} C_{33,22}^{33,0}-\frac{9}{400} \sqrt{\frac{3}{10}} C_{33,22}^{33,1}\)

\noindent\(C_{11,3112,3}^{11,1112,0,\text{NonLoc}}= \frac{7 \sqrt{7} C_{00,00}^{60,0}}{320}+\frac{7 \sqrt{7} C_{00,00}^{60,1}}{160}+\frac{7}{320}
\sqrt{\frac{7}{3}} C_{00,20}^{60,0}+\frac{7}{160} \sqrt{\frac{7}{3}} C_{00,20}^{60,1}-\frac{1}{160} \sqrt{\frac{21}{5}} C_{00,22}^{62,0}-\frac{1}{80}
\sqrt{\frac{21}{5}} C_{00,22}^{62,1}-\frac{7 \sqrt{7} C_{11,00}^{51,0}}{960}-\frac{7 \sqrt{7} C_{11,00}^{51,1}}{480}-\frac{7}{960} \sqrt{\frac{7}{3}}
C_{11,20}^{31,0}-\frac{7}{480} \sqrt{\frac{7}{3}} C_{11,20}^{31,1}+\frac{1}{240} \sqrt{\frac{7}{15}} C_{11,22}^{51,0}+\frac{1}{120} \sqrt{\frac{7}{15}}
C_{11,22}^{51,1}+\frac{9 C_{11,22}^{53,0}}{1600}+\frac{9 C_{11,22}^{53,1}}{800}+\frac{\sqrt{7} C_{20,00}^{40,0}}{320}+\frac{\sqrt{7} C_{20,00}^{40,1}}{160}+\frac{1}{320}
\sqrt{\frac{7}{3}} C_{20,20}^{40,0}+\frac{1}{160} \sqrt{\frac{7}{3}} C_{20,20}^{40,1}-\frac{1}{320} \sqrt{\frac{7}{15}} C_{20,22}^{42,0}-\frac{1}{160}
\sqrt{\frac{7}{15}} C_{20,22}^{42,1}-\frac{7}{640} \sqrt{\frac{7}{5}} C_{22,00}^{42,0}-\frac{7}{320} \sqrt{\frac{7}{5}} C_{22,00}^{42,1}-\frac{7}{640}
\sqrt{\frac{7}{15}} C_{22,20}^{42,0}-\frac{7}{320} \sqrt{\frac{7}{15}} C_{22,20}^{42,1}+\frac{7 C_{22,22}^{42,0}}{320 \sqrt{15}}+\frac{7 C_{22,22}^{42,1}}{160
\sqrt{15}}+\frac{\sqrt{7} C_{31,00}^{31,0}}{480}+\frac{\sqrt{7} C_{31,00}^{31,1}}{240}+\frac{1}{480} \sqrt{\frac{7}{3}} C_{31,20}^{31,0}+\frac{1}{240}
\sqrt{\frac{7}{3}} C_{31,20}^{31,1}-\frac{1}{480} \sqrt{\frac{7}{15}} C_{31,22}^{31,0}-\frac{1}{240} \sqrt{\frac{7}{15}} C_{31,22}^{31,1}-\frac{C_{31,22}^{33,0}}{800}-\frac{C_{31,22}^{33,1}}{400}+\frac{63
\sqrt{3} C_{33,00}^{33,0}}{3200}+\frac{63 \sqrt{3} C_{33,00}^{33,1}}{1600}+\frac{63 C_{33,20}^{33,0}}{3200}+\frac{63 C_{33,20}^{33,1}}{1600}-\frac{9}{400}
\sqrt{\frac{3}{10}} C_{33,22}^{33,0}-\frac{9}{200} \sqrt{\frac{3}{10}} C_{33,22}^{33,1}\)

\noindent\(C_{11,3112,3}^{11,1112,1,\text{Loc}}= \frac{7 \sqrt{7} C_{00,20}^{60,1}}{160}+\frac{3}{160} \sqrt{\frac{7}{5}} C_{00,22}^{62,1}+\frac{7
\sqrt{7} C_{11,20}^{31,1}}{480}+\frac{1}{240} \sqrt{\frac{7}{5}} C_{11,22}^{51,1}+\frac{9 \sqrt{3} C_{11,22}^{53,1}}{1600}+\frac{\sqrt{7} C_{20,20}^{40,1}}{160}+\frac{1}{320}
\sqrt{\frac{7}{5}} C_{20,22}^{42,1}-\frac{7}{320} \sqrt{\frac{7}{5}} C_{22,20}^{42,1}-\frac{7 C_{22,22}^{42,1}}{320 \sqrt{5}}-\frac{\sqrt{7} C_{31,20}^{31,1}}{240}-\frac{1}{480}
\sqrt{\frac{7}{5}} C_{31,22}^{31,1}-\frac{\sqrt{3} C_{31,22}^{33,1}}{800}-\frac{63 \sqrt{3} C_{33,20}^{33,1}}{1600}-\frac{27 C_{33,22}^{33,1}}{400
\sqrt{10}}\)

\noindent\(C_{11,3112,3}^{11,1112,1,\text{NonLoc}}= \frac{7 \sqrt{21} C_{00,00}^{60,0}}{320}+\frac{7 \sqrt{7} C_{00,20}^{60,0}}{320}-\frac{3}{160}
\sqrt{\frac{7}{5}} C_{00,22}^{62,0}-\frac{7}{320} \sqrt{\frac{7}{3}} C_{11,00}^{51,0}-\frac{7 \sqrt{7} C_{11,20}^{31,0}}{960}+\frac{1}{240} \sqrt{\frac{7}{5}}
C_{11,22}^{51,0}+\frac{9 \sqrt{3} C_{11,22}^{53,0}}{1600}+\frac{\sqrt{21} C_{20,00}^{40,0}}{320}+\frac{\sqrt{7} C_{20,20}^{40,0}}{320}-\frac{1}{320}
\sqrt{\frac{7}{5}} C_{20,22}^{42,0}-\frac{7}{640} \sqrt{\frac{21}{5}} C_{22,00}^{42,0}-\frac{7}{640} \sqrt{\frac{7}{5}} C_{22,20}^{42,0}+\frac{7
C_{22,22}^{42,0}}{320 \sqrt{5}}+\frac{1}{160} \sqrt{\frac{7}{3}} C_{31,00}^{31,0}+\frac{\sqrt{7} C_{31,20}^{31,0}}{480}-\frac{1}{480} \sqrt{\frac{7}{5}}
C_{31,22}^{31,0}-\frac{\sqrt{3} C_{31,22}^{33,0}}{800}+\frac{189 C_{33,00}^{33,0}}{3200}+\frac{63 \sqrt{3} C_{33,20}^{33,0}}{3200}-\frac{27 C_{33,22}^{33,0}}{400
\sqrt{10}}\)

\noindent\(C_{11,3303,2}^{00,2212,0,\text{Loc}}= \frac{\sqrt{7} C_{11,11}^{51,0}}{144}+\frac{\sqrt{7} C_{11,11}^{51,1}}{288}+\frac{1}{24}
\sqrt{\frac{7}{15}} C_{22,11}^{42,0}+\frac{1}{48} \sqrt{\frac{7}{15}} C_{22,11}^{42,1}+\frac{\sqrt{7} C_{31,11}^{31,0}}{144}+\frac{\sqrt{7} C_{31,11}^{31,1}}{288}+\frac{C_{33,11}^{33,0}}{20
\sqrt{2}}+\frac{C_{33,11}^{33,1}}{40 \sqrt{2}}\)

\noindent\(C_{11,3303,2}^{00,2212,0,\text{NonLoc}}= \frac{\sqrt{7} C_{11,11}^{51,0}}{288}+\frac{\sqrt{7} C_{11,11}^{51,1}}{144}-\frac{1}{48}
\sqrt{\frac{7}{15}} C_{22,11}^{42,0}-\frac{1}{24} \sqrt{\frac{7}{15}} C_{22,11}^{42,1}+\frac{\sqrt{7} C_{31,11}^{31,0}}{288}+\frac{\sqrt{7} C_{31,11}^{31,1}}{144}+\frac{C_{33,11}^{33,0}}{40
\sqrt{2}}+\frac{C_{33,11}^{33,1}}{20 \sqrt{2}}\)

\noindent\(C_{11,3303,2}^{00,2212,1,\text{Loc}}= \frac{1}{96} \sqrt{\frac{7}{3}} C_{11,11}^{51,1}+\frac{1}{48} \sqrt{\frac{7}{5}} C_{22,11}^{42,1}+\frac{1}{96}
\sqrt{\frac{7}{3}} C_{31,11}^{31,1}+\frac{1}{40} \sqrt{\frac{3}{2}} C_{33,11}^{33,1}\)

\noindent\(C_{11,3303,2}^{00,2212,1,\text{NonLoc}}= \frac{1}{96} \sqrt{\frac{7}{3}} C_{11,11}^{51,0}-\frac{1}{48} \sqrt{\frac{7}{5}}
C_{22,11}^{42,0}+\frac{1}{96} \sqrt{\frac{7}{3}} C_{31,11}^{31,0}+\frac{1}{40} \sqrt{\frac{3}{2}} C_{33,11}^{33,0}\)

\noindent\(C_{11,3303,2}^{11,1101,0,\text{Loc}}= -\frac{1}{16} \sqrt{\frac{7}{3}} C_{00,00}^{60,0}-\frac{1}{32} \sqrt{\frac{7}{3}}
C_{00,00}^{60,1}-\frac{1}{48} \sqrt{\frac{7}{3}} C_{11,00}^{51,0}-\frac{1}{96} \sqrt{\frac{7}{3}} C_{11,00}^{51,1}+\frac{1}{48} \sqrt{\frac{7}{3}}
C_{20,00}^{40,0}+\frac{1}{96} \sqrt{\frac{7}{3}} C_{20,00}^{40,1}-\frac{1}{48} \sqrt{\frac{7}{15}} C_{22,00}^{42,0}-\frac{1}{96} \sqrt{\frac{7}{15}}
C_{22,00}^{42,1}+\frac{1}{48} \sqrt{\frac{7}{3}} C_{31,00}^{31,0}+\frac{1}{96} \sqrt{\frac{7}{3}} C_{31,00}^{31,1}-\frac{3 C_{33,00}^{33,0}}{80}-\frac{3
C_{33,00}^{33,1}}{160}\)

\noindent\(C_{11,3303,2}^{11,1101,0,\text{NonLoc}}= \frac{1}{64} \sqrt{\frac{7}{3}} C_{00,00}^{60,0}+\frac{1}{32} \sqrt{\frac{7}{3}}
C_{00,00}^{60,1}-\frac{\sqrt{7} C_{00,20}^{60,0}}{64}-\frac{\sqrt{7} C_{00,20}^{60,1}}{32}-\frac{1}{192} \sqrt{\frac{7}{3}} C_{11,00}^{51,0}-\frac{1}{96}
\sqrt{\frac{7}{3}} C_{11,00}^{51,1}-\frac{1}{192} \sqrt{\frac{7}{3}} C_{20,00}^{40,0}-\frac{1}{96} \sqrt{\frac{7}{3}} C_{20,00}^{40,1}+\frac{\sqrt{7}
C_{20,20}^{40,0}}{192}+\frac{\sqrt{7} C_{20,20}^{40,1}}{96}+\frac{1}{192} \sqrt{\frac{7}{15}} C_{22,00}^{42,0}+\frac{1}{96} \sqrt{\frac{7}{15}} C_{22,00}^{42,1}-\frac{1}{192}
\sqrt{\frac{7}{5}} C_{22,20}^{42,0}-\frac{1}{96} \sqrt{\frac{7}{5}} C_{22,20}^{42,1}+\frac{1}{192} \sqrt{\frac{7}{3}} C_{31,00}^{31,0}+\frac{1}{96}
\sqrt{\frac{7}{3}} C_{31,00}^{31,1}-\frac{\sqrt{7} C_{31,20}^{31,0}}{192}-\frac{\sqrt{7} C_{31,20}^{31,1}}{96}-\frac{3 C_{33,00}^{33,0}}{320}-\frac{3
C_{33,00}^{33,1}}{160}+\frac{3 \sqrt{3} C_{33,20}^{33,0}}{320}+\frac{3 \sqrt{3} C_{33,20}^{33,1}}{160}\)

\noindent\(C_{11,3303,2}^{11,1101,1,\text{Loc}}= -\frac{\sqrt{7} C_{00,00}^{60,1}}{32}-\frac{\sqrt{7} C_{11,00}^{51,1}}{96}+\frac{\sqrt{7}
C_{20,00}^{40,1}}{96}-\frac{1}{96} \sqrt{\frac{7}{5}} C_{22,00}^{42,1}+\frac{\sqrt{7} C_{31,00}^{31,1}}{96}-\frac{3 \sqrt{3} C_{33,00}^{33,1}}{160}\)

\noindent\(C_{11,3303,2}^{11,1101,1,\text{NonLoc}}= \frac{\sqrt{7} C_{00,00}^{60,0}}{64}-\frac{\sqrt{21} C_{00,20}^{60,0}}{64}-\frac{\sqrt{7}
C_{11,00}^{51,0}}{192}-\frac{\sqrt{7} C_{20,00}^{40,0}}{192}+\frac{1}{64} \sqrt{\frac{7}{3}} C_{20,20}^{40,0}+\frac{1}{192} \sqrt{\frac{7}{5}} C_{22,00}^{42,0}-\frac{1}{64}
\sqrt{\frac{7}{15}} C_{22,20}^{42,0}+\frac{\sqrt{7} C_{31,00}^{31,0}}{192}-\frac{1}{64} \sqrt{\frac{7}{3}} C_{31,20}^{31,0}-\frac{3 \sqrt{3} C_{33,00}^{33,0}}{320}+\frac{9
C_{33,20}^{33,0}}{320}\)

\noindent\(C_{11,3303,3}^{00,2213,0,\text{Loc}}= -\frac{1}{72} \sqrt{\frac{7}{2}} C_{11,11}^{51,0}-\frac{1}{144} \sqrt{\frac{7}{2}}
C_{11,11}^{51,1}-\frac{1}{12} \sqrt{\frac{7}{30}} C_{22,11}^{42,0}-\frac{1}{24} \sqrt{\frac{7}{30}} C_{22,11}^{42,1}-\frac{1}{72} \sqrt{\frac{7}{2}}
C_{31,11}^{31,0}-\frac{1}{144} \sqrt{\frac{7}{2}} C_{31,11}^{31,1}-\frac{C_{33,11}^{33,0}}{20}-\frac{C_{33,11}^{33,1}}{40}\)

\noindent\(C_{11,3303,3}^{00,2213,0,\text{NonLoc}}= -\frac{1}{144} \sqrt{\frac{7}{2}} C_{11,11}^{51,0}-\frac{1}{72} \sqrt{\frac{7}{2}}
C_{11,11}^{51,1}+\frac{1}{24} \sqrt{\frac{7}{30}} C_{22,11}^{42,0}+\frac{1}{12} \sqrt{\frac{7}{30}} C_{22,11}^{42,1}-\frac{1}{144} \sqrt{\frac{7}{2}}
C_{31,11}^{31,0}-\frac{1}{72} \sqrt{\frac{7}{2}} C_{31,11}^{31,1}-\frac{C_{33,11}^{33,0}}{40}-\frac{C_{33,11}^{33,1}}{20}\)

\noindent\(C_{11,3303,3}^{00,2213,1,\text{Loc}}= -\frac{1}{48} \sqrt{\frac{7}{6}} C_{11,11}^{51,1}-\frac{1}{24} \sqrt{\frac{7}{10}}
C_{22,11}^{42,1}-\frac{1}{48} \sqrt{\frac{7}{6}} C_{31,11}^{31,1}-\frac{\sqrt{3} C_{33,11}^{33,1}}{40}\)

\noindent\(C_{11,3303,3}^{00,2213,1,\text{NonLoc}}= -\frac{1}{48} \sqrt{\frac{7}{6}} C_{11,11}^{51,0}+\frac{1}{24} \sqrt{\frac{7}{10}}
C_{22,11}^{42,0}-\frac{1}{48} \sqrt{\frac{7}{6}} C_{31,11}^{31,0}-\frac{\sqrt{3} C_{33,11}^{33,0}}{40}\)

\noindent\(C_{11,3312,1}^{11,1110,0,\text{Loc}}= \frac{1}{48} \sqrt{\frac{7}{3}} C_{00,20}^{60,0}+\frac{1}{96} \sqrt{\frac{7}{3}} C_{00,20}^{60,1}+\frac{1}{16}
\sqrt{\frac{7}{15}} C_{00,22}^{62,0}+\frac{1}{32} \sqrt{\frac{7}{15}} C_{00,22}^{62,1}+\frac{1}{18} \sqrt{\frac{7}{15}} C_{11,22}^{51,0}+\frac{1}{36}
\sqrt{\frac{7}{15}} C_{11,22}^{51,1}-\frac{3 C_{11,22}^{53,0}}{160}-\frac{3 C_{11,22}^{53,1}}{320}-\frac{1}{144} \sqrt{\frac{7}{3}} C_{20,20}^{40,0}-\frac{1}{288}
\sqrt{\frac{7}{3}} C_{20,20}^{40,1}-\frac{1}{18} \sqrt{\frac{7}{15}} C_{20,22}^{42,0}-\frac{1}{36} \sqrt{\frac{7}{15}} C_{20,22}^{42,1}+\frac{1}{144}
\sqrt{\frac{7}{15}} C_{22,20}^{42,0}+\frac{1}{288} \sqrt{\frac{7}{15}} C_{22,20}^{42,1}+\frac{7}{144} \sqrt{\frac{7}{15}} C_{22,22}^{40,0}+\frac{7}{288}
\sqrt{\frac{7}{15}} C_{22,22}^{40,1}+\frac{C_{22,22}^{42,0}}{72 \sqrt{15}}+\frac{C_{22,22}^{42,1}}{144 \sqrt{15}}+\frac{1}{40} \sqrt{\frac{3}{7}}
C_{22,22}^{44,0}+\frac{1}{80} \sqrt{\frac{3}{7}} C_{22,22}^{44,1}-\frac{1}{144} \sqrt{\frac{7}{3}} C_{31,20}^{31,0}-\frac{1}{288} \sqrt{\frac{7}{3}}
C_{31,20}^{31,1}-\frac{1}{18} \sqrt{\frac{7}{15}} C_{31,22}^{31,0}-\frac{1}{36} \sqrt{\frac{7}{15}} C_{31,22}^{31,1}+\frac{3 C_{31,22}^{33,0}}{160}+\frac{3
C_{31,22}^{33,1}}{320}+\frac{C_{33,20}^{33,0}}{80}+\frac{C_{33,20}^{33,1}}{160}+\frac{1}{40} \sqrt{\frac{3}{10}} C_{33,22}^{33,0}+\frac{1}{80} \sqrt{\frac{3}{10}}
C_{33,22}^{33,1}\)

\noindent\(C_{11,3312,1}^{11,1110,0,\text{NonLoc}}= \frac{\sqrt{7} C_{00,00}^{60,0}}{192}+\frac{\sqrt{7} C_{00,00}^{60,1}}{96}+\frac{1}{192}
\sqrt{\frac{7}{3}} C_{00,20}^{60,0}+\frac{1}{96} \sqrt{\frac{7}{3}} C_{00,20}^{60,1}-\frac{1}{32} \sqrt{\frac{7}{15}} C_{00,22}^{62,0}-\frac{1}{16}
\sqrt{\frac{7}{15}} C_{00,22}^{62,1}-\frac{\sqrt{7} C_{11,00}^{51,0}}{576}-\frac{\sqrt{7} C_{11,00}^{51,1}}{288}+\frac{1}{36} \sqrt{\frac{7}{15}}
C_{11,22}^{51,0}+\frac{1}{18} \sqrt{\frac{7}{15}} C_{11,22}^{51,1}-\frac{3 C_{11,22}^{53,0}}{320}-\frac{3 C_{11,22}^{53,1}}{160}-\frac{\sqrt{7} C_{20,00}^{40,0}}{576}-\frac{\sqrt{7}
C_{20,00}^{40,1}}{288}-\frac{1}{576} \sqrt{\frac{7}{3}} C_{20,20}^{40,0}-\frac{1}{288} \sqrt{\frac{7}{3}} C_{20,20}^{40,1}+\frac{1}{36} \sqrt{\frac{7}{15}}
C_{20,22}^{42,0}+\frac{1}{18} \sqrt{\frac{7}{15}} C_{20,22}^{42,1}+\frac{1}{576} \sqrt{\frac{7}{5}} C_{22,00}^{42,0}+\frac{1}{288} \sqrt{\frac{7}{5}}
C_{22,00}^{42,1}+\frac{1}{576} \sqrt{\frac{7}{15}} C_{22,20}^{42,0}+\frac{1}{288} \sqrt{\frac{7}{15}} C_{22,20}^{42,1}-\frac{7}{288} \sqrt{\frac{7}{15}}
C_{22,22}^{40,0}-\frac{7}{144} \sqrt{\frac{7}{15}} C_{22,22}^{40,1}-\frac{C_{22,22}^{42,0}}{144 \sqrt{15}}-\frac{C_{22,22}^{42,1}}{72 \sqrt{15}}-\frac{1}{80}
\sqrt{\frac{3}{7}} C_{22,22}^{44,0}-\frac{1}{40} \sqrt{\frac{3}{7}} C_{22,22}^{44,1}+\frac{\sqrt{7} C_{31,00}^{31,0}}{576}+\frac{\sqrt{7} C_{31,00}^{31,1}}{288}+\frac{1}{576}
\sqrt{\frac{7}{3}} C_{31,20}^{31,0}+\frac{1}{288} \sqrt{\frac{7}{3}} C_{31,20}^{31,1}-\frac{1}{36} \sqrt{\frac{7}{15}} C_{31,22}^{31,0}-\frac{1}{18}
\sqrt{\frac{7}{15}} C_{31,22}^{31,1}+\frac{3 C_{31,22}^{33,0}}{320}+\frac{3 C_{31,22}^{33,1}}{160}-\frac{\sqrt{3} C_{33,00}^{33,0}}{320}-\frac{\sqrt{3}
C_{33,00}^{33,1}}{160}-\frac{C_{33,20}^{33,0}}{320}-\frac{C_{33,20}^{33,1}}{160}+\frac{1}{80} \sqrt{\frac{3}{10}} C_{33,22}^{33,0}+\frac{1}{40} \sqrt{\frac{3}{10}}
C_{33,22}^{33,1}\)

\noindent\(C_{11,3312,1}^{11,1110,1,\text{Loc}}= \frac{\sqrt{7} C_{00,20}^{60,1}}{96}+\frac{1}{32} \sqrt{\frac{7}{5}} C_{00,22}^{62,1}+\frac{1}{36}
\sqrt{\frac{7}{5}} C_{11,22}^{51,1}-\frac{3 \sqrt{3} C_{11,22}^{53,1}}{320}-\frac{\sqrt{7} C_{20,20}^{40,1}}{288}-\frac{1}{36} \sqrt{\frac{7}{5}}
C_{20,22}^{42,1}+\frac{1}{288} \sqrt{\frac{7}{5}} C_{22,20}^{42,1}+\frac{7}{288} \sqrt{\frac{7}{5}} C_{22,22}^{40,1}+\frac{C_{22,22}^{42,1}}{144
\sqrt{5}}+\frac{3 C_{22,22}^{44,1}}{80 \sqrt{7}}-\frac{\sqrt{7} C_{31,20}^{31,1}}{288}-\frac{1}{36} \sqrt{\frac{7}{5}} C_{31,22}^{31,1}+\frac{3 \sqrt{3}
C_{31,22}^{33,1}}{320}+\frac{\sqrt{3} C_{33,20}^{33,1}}{160}+\frac{3 C_{33,22}^{33,1}}{80 \sqrt{10}}\)

\noindent\(C_{11,3312,1}^{11,1110,1,\text{NonLoc}}= \frac{1}{64} \sqrt{\frac{7}{3}} C_{00,00}^{60,0}+\frac{\sqrt{7} C_{00,20}^{60,0}}{192}-\frac{1}{32}
\sqrt{\frac{7}{5}} C_{00,22}^{62,0}-\frac{1}{192} \sqrt{\frac{7}{3}} C_{11,00}^{51,0}+\frac{1}{36} \sqrt{\frac{7}{5}} C_{11,22}^{51,0}-\frac{3 \sqrt{3}
C_{11,22}^{53,0}}{320}-\frac{1}{192} \sqrt{\frac{7}{3}} C_{20,00}^{40,0}-\frac{\sqrt{7} C_{20,20}^{40,0}}{576}+\frac{1}{36} \sqrt{\frac{7}{5}} C_{20,22}^{42,0}+\frac{1}{192}
\sqrt{\frac{7}{15}} C_{22,00}^{42,0}+\frac{1}{576} \sqrt{\frac{7}{5}} C_{22,20}^{42,0}-\frac{7}{288} \sqrt{\frac{7}{5}} C_{22,22}^{40,0}-\frac{C_{22,22}^{42,0}}{144
\sqrt{5}}-\frac{3 C_{22,22}^{44,0}}{80 \sqrt{7}}+\frac{1}{192} \sqrt{\frac{7}{3}} C_{31,00}^{31,0}+\frac{\sqrt{7} C_{31,20}^{31,0}}{576}-\frac{1}{36}
\sqrt{\frac{7}{5}} C_{31,22}^{31,0}+\frac{3 \sqrt{3} C_{31,22}^{33,0}}{320}-\frac{3 C_{33,00}^{33,0}}{320}-\frac{\sqrt{3} C_{33,20}^{33,0}}{320}+\frac{3
C_{33,22}^{33,0}}{80 \sqrt{10}}\)

\noindent\(C_{11,3312,1}^{11,1111,0,\text{Loc}}= \frac{\sqrt{7} C_{00,20}^{60,0}}{96}+\frac{\sqrt{7} C_{00,20}^{60,1}}{192}-\frac{1}{64}
\sqrt{\frac{7}{5}} C_{00,22}^{62,0}-\frac{1}{128} \sqrt{\frac{7}{5}} C_{00,22}^{62,1}-\frac{7}{288} \sqrt{\frac{7}{5}} C_{11,22}^{51,0}-\frac{7}{576}
\sqrt{\frac{7}{5}} C_{11,22}^{51,1}+\frac{9 \sqrt{3} C_{11,22}^{53,0}}{640}+\frac{9 \sqrt{3} C_{11,22}^{53,1}}{1280}-\frac{\sqrt{7} C_{20,20}^{40,0}}{288}-\frac{\sqrt{7}
C_{20,20}^{40,1}}{576}+\frac{7}{288} \sqrt{\frac{7}{5}} C_{20,22}^{42,0}+\frac{7}{576} \sqrt{\frac{7}{5}} C_{20,22}^{42,1}+\frac{1}{288} \sqrt{\frac{7}{5}}
C_{22,20}^{42,0}+\frac{1}{576} \sqrt{\frac{7}{5}} C_{22,20}^{42,1}-\frac{19}{576} \sqrt{\frac{7}{5}} C_{22,22}^{40,0}-\frac{19 \sqrt{\frac{7}{5}}
C_{22,22}^{40,1}}{1152}+\frac{\sqrt{5} C_{22,22}^{42,0}}{288}+\frac{\sqrt{5} C_{22,22}^{42,1}}{576}-\frac{3 C_{22,22}^{44,0}}{80 \sqrt{7}}-\frac{3
C_{22,22}^{44,1}}{160 \sqrt{7}}-\frac{\sqrt{7} C_{31,20}^{31,0}}{288}-\frac{\sqrt{7} C_{31,20}^{31,1}}{576}+\frac{7}{288} \sqrt{\frac{7}{5}} C_{31,22}^{31,0}+\frac{7}{576}
\sqrt{\frac{7}{5}} C_{31,22}^{31,1}-\frac{9 \sqrt{3} C_{31,22}^{33,0}}{640}-\frac{9 \sqrt{3} C_{31,22}^{33,1}}{1280}+\frac{\sqrt{3} C_{33,20}^{33,0}}{160}+\frac{\sqrt{3}
C_{33,20}^{33,1}}{320}+\frac{3 C_{33,22}^{33,0}}{80 \sqrt{10}}+\frac{3 C_{33,22}^{33,1}}{160 \sqrt{10}}\)

\noindent\(C_{11,3312,1}^{11,1111,0,\text{NonLoc}}= \frac{1}{128} \sqrt{\frac{7}{3}} C_{00,00}^{60,0}+\frac{1}{64} \sqrt{\frac{7}{3}}
C_{00,00}^{60,1}+\frac{\sqrt{7} C_{00,20}^{60,0}}{384}+\frac{\sqrt{7} C_{00,20}^{60,1}}{192}+\frac{1}{128} \sqrt{\frac{7}{5}} C_{00,22}^{62,0}+\frac{1}{64}
\sqrt{\frac{7}{5}} C_{00,22}^{62,1}-\frac{1}{384} \sqrt{\frac{7}{3}} C_{11,00}^{51,0}-\frac{1}{192} \sqrt{\frac{7}{3}} C_{11,00}^{51,1}-\frac{7}{576}
\sqrt{\frac{7}{5}} C_{11,22}^{51,0}-\frac{7}{288} \sqrt{\frac{7}{5}} C_{11,22}^{51,1}+\frac{9 \sqrt{3} C_{11,22}^{53,0}}{1280}+\frac{9 \sqrt{3} C_{11,22}^{53,1}}{640}-\frac{1}{384}
\sqrt{\frac{7}{3}} C_{20,00}^{40,0}-\frac{1}{192} \sqrt{\frac{7}{3}} C_{20,00}^{40,1}-\frac{\sqrt{7} C_{20,20}^{40,0}}{1152}-\frac{\sqrt{7} C_{20,20}^{40,1}}{576}-\frac{7}{576}
\sqrt{\frac{7}{5}} C_{20,22}^{42,0}-\frac{7}{288} \sqrt{\frac{7}{5}} C_{20,22}^{42,1}+\frac{1}{384} \sqrt{\frac{7}{15}} C_{22,00}^{42,0}+\frac{1}{192}
\sqrt{\frac{7}{15}} C_{22,00}^{42,1}+\frac{\sqrt{\frac{7}{5}} C_{22,20}^{42,0}}{1152}+\frac{1}{576} \sqrt{\frac{7}{5}} C_{22,20}^{42,1}+\frac{19
\sqrt{\frac{7}{5}} C_{22,22}^{40,0}}{1152}+\frac{19}{576} \sqrt{\frac{7}{5}} C_{22,22}^{40,1}-\frac{\sqrt{5} C_{22,22}^{42,0}}{576}-\frac{\sqrt{5}
C_{22,22}^{42,1}}{288}+\frac{3 C_{22,22}^{44,0}}{160 \sqrt{7}}+\frac{3 C_{22,22}^{44,1}}{80 \sqrt{7}}+\frac{1}{384} \sqrt{\frac{7}{3}} C_{31,00}^{31,0}+\frac{1}{192}
\sqrt{\frac{7}{3}} C_{31,00}^{31,1}+\frac{\sqrt{7} C_{31,20}^{31,0}}{1152}+\frac{\sqrt{7} C_{31,20}^{31,1}}{576}+\frac{7}{576} \sqrt{\frac{7}{5}}
C_{31,22}^{31,0}+\frac{7}{288} \sqrt{\frac{7}{5}} C_{31,22}^{31,1}-\frac{9 \sqrt{3} C_{31,22}^{33,0}}{1280}-\frac{9 \sqrt{3} C_{31,22}^{33,1}}{640}-\frac{3
C_{33,00}^{33,0}}{640}-\frac{3 C_{33,00}^{33,1}}{320}-\frac{\sqrt{3} C_{33,20}^{33,0}}{640}-\frac{\sqrt{3} C_{33,20}^{33,1}}{320}+\frac{3 C_{33,22}^{33,0}}{160
\sqrt{10}}+\frac{3 C_{33,22}^{33,1}}{80 \sqrt{10}}\)

\noindent\(C_{11,3312,1}^{11,1111,1,\text{Loc}}= \frac{1}{64} \sqrt{\frac{7}{3}} C_{00,20}^{60,1}-\frac{1}{128} \sqrt{\frac{21}{5}}
C_{00,22}^{62,1}-\frac{7}{192} \sqrt{\frac{7}{15}} C_{11,22}^{51,1}+\frac{27 C_{11,22}^{53,1}}{1280}-\frac{1}{192} \sqrt{\frac{7}{3}} C_{20,20}^{40,1}+\frac{7}{192}
\sqrt{\frac{7}{15}} C_{20,22}^{42,1}+\frac{1}{192} \sqrt{\frac{7}{15}} C_{22,20}^{42,1}-\frac{19}{384} \sqrt{\frac{7}{15}} C_{22,22}^{40,1}+\frac{1}{192}
\sqrt{\frac{5}{3}} C_{22,22}^{42,1}-\frac{3}{160} \sqrt{\frac{3}{7}} C_{22,22}^{44,1}-\frac{1}{192} \sqrt{\frac{7}{3}} C_{31,20}^{31,1}+\frac{7}{192}
\sqrt{\frac{7}{15}} C_{31,22}^{31,1}-\frac{27 C_{31,22}^{33,1}}{1280}+\frac{3 C_{33,20}^{33,1}}{320}+\frac{3}{160} \sqrt{\frac{3}{10}} C_{33,22}^{33,1}\)

\noindent\(C_{11,3312,1}^{11,1111,1,\text{NonLoc}}= \frac{\sqrt{7} C_{00,00}^{60,0}}{128}+\frac{1}{128} \sqrt{\frac{7}{3}} C_{00,20}^{60,0}+\frac{1}{128}
\sqrt{\frac{21}{5}} C_{00,22}^{62,0}-\frac{\sqrt{7} C_{11,00}^{51,0}}{384}-\frac{7}{192} \sqrt{\frac{7}{15}} C_{11,22}^{51,0}+\frac{27 C_{11,22}^{53,0}}{1280}-\frac{\sqrt{7}
C_{20,00}^{40,0}}{384}-\frac{1}{384} \sqrt{\frac{7}{3}} C_{20,20}^{40,0}-\frac{7}{192} \sqrt{\frac{7}{15}} C_{20,22}^{42,0}+\frac{1}{384} \sqrt{\frac{7}{5}}
C_{22,00}^{42,0}+\frac{1}{384} \sqrt{\frac{7}{15}} C_{22,20}^{42,0}+\frac{19}{384} \sqrt{\frac{7}{15}} C_{22,22}^{40,0}-\frac{1}{192} \sqrt{\frac{5}{3}}
C_{22,22}^{42,0}+\frac{3}{160} \sqrt{\frac{3}{7}} C_{22,22}^{44,0}+\frac{\sqrt{7} C_{31,00}^{31,0}}{384}+\frac{1}{384} \sqrt{\frac{7}{3}} C_{31,20}^{31,0}+\frac{7}{192}
\sqrt{\frac{7}{15}} C_{31,22}^{31,0}-\frac{27 C_{31,22}^{33,0}}{1280}-\frac{3 \sqrt{3} C_{33,00}^{33,0}}{640}-\frac{3 C_{33,20}^{33,0}}{640}+\frac{3}{160}
\sqrt{\frac{3}{10}} C_{33,22}^{33,0}\)

\noindent\(C_{11,3312,1}^{11,1112,0,\text{Loc}}= \frac{1}{96} \sqrt{\frac{7}{15}} C_{00,20}^{60,0}+\frac{1}{192} \sqrt{\frac{7}{15}}
C_{00,20}^{60,1}+\frac{11}{320} \sqrt{\frac{7}{3}} C_{00,22}^{62,0}+\frac{11}{640} \sqrt{\frac{7}{3}} C_{00,22}^{62,1}+\frac{13}{720} \sqrt{\frac{7}{3}}
C_{11,22}^{51,0}+\frac{13 \sqrt{\frac{7}{3}} C_{11,22}^{51,1}}{1440}+\frac{3 C_{11,22}^{53,0}}{640 \sqrt{5}}+\frac{3 C_{11,22}^{53,1}}{1280 \sqrt{5}}-\frac{1}{288}
\sqrt{\frac{7}{15}} C_{20,20}^{40,0}-\frac{1}{576} \sqrt{\frac{7}{15}} C_{20,20}^{40,1}-\frac{1}{180} \sqrt{\frac{7}{3}} C_{20,22}^{42,0}-\frac{1}{360}
\sqrt{\frac{7}{3}} C_{20,22}^{42,1}+\frac{\sqrt{\frac{7}{3}} C_{22,20}^{42,0}}{1440}+\frac{\sqrt{\frac{7}{3}} C_{22,20}^{42,1}}{2880}+\frac{1}{576}
\sqrt{\frac{7}{3}} C_{22,22}^{40,0}+\frac{\sqrt{\frac{7}{3}} C_{22,22}^{40,1}}{1152}+\frac{11 C_{22,22}^{42,0}}{1440 \sqrt{3}}+\frac{11 C_{22,22}^{42,1}}{2880
\sqrt{3}}-\frac{1}{16} \sqrt{\frac{3}{35}} C_{22,22}^{44,0}-\frac{1}{32} \sqrt{\frac{3}{35}} C_{22,22}^{44,1}-\frac{1}{288} \sqrt{\frac{7}{15}} C_{31,20}^{31,0}-\frac{1}{576}
\sqrt{\frac{7}{15}} C_{31,20}^{31,1}-\frac{1}{180} \sqrt{\frac{7}{3}} C_{31,22}^{31,0}-\frac{1}{360} \sqrt{\frac{7}{3}} C_{31,22}^{31,1}-\frac{19
C_{31,22}^{33,0}}{640 \sqrt{5}}-\frac{19 C_{31,22}^{33,1}}{1280 \sqrt{5}}+\frac{C_{33,20}^{33,0}}{160 \sqrt{5}}+\frac{C_{33,20}^{33,1}}{320 \sqrt{5}}+\frac{1}{400}
\sqrt{\frac{3}{2}} C_{33,22}^{33,0}+\frac{1}{800} \sqrt{\frac{3}{2}} C_{33,22}^{33,1}\)

\noindent\(C_{11,3312,1}^{11,1112,0,\text{NonLoc}}= \frac{1}{384} \sqrt{\frac{7}{5}} C_{00,00}^{60,0}+\frac{1}{192} \sqrt{\frac{7}{5}}
C_{00,00}^{60,1}+\frac{1}{384} \sqrt{\frac{7}{15}} C_{00,20}^{60,0}+\frac{1}{192} \sqrt{\frac{7}{15}} C_{00,20}^{60,1}-\frac{11}{640} \sqrt{\frac{7}{3}}
C_{00,22}^{62,0}-\frac{11}{320} \sqrt{\frac{7}{3}} C_{00,22}^{62,1}-\frac{\sqrt{\frac{7}{5}} C_{11,00}^{51,0}}{1152}-\frac{1}{576} \sqrt{\frac{7}{5}}
C_{11,00}^{51,1}+\frac{13 \sqrt{\frac{7}{3}} C_{11,22}^{51,0}}{1440}+\frac{13}{720} \sqrt{\frac{7}{3}} C_{11,22}^{51,1}+\frac{3 C_{11,22}^{53,0}}{1280
\sqrt{5}}+\frac{3 C_{11,22}^{53,1}}{640 \sqrt{5}}-\frac{\sqrt{\frac{7}{5}} C_{20,00}^{40,0}}{1152}-\frac{1}{576} \sqrt{\frac{7}{5}} C_{20,00}^{40,1}-\frac{\sqrt{\frac{7}{15}}
C_{20,20}^{40,0}}{1152}-\frac{1}{576} \sqrt{\frac{7}{15}} C_{20,20}^{40,1}+\frac{1}{360} \sqrt{\frac{7}{3}} C_{20,22}^{42,0}+\frac{1}{180} \sqrt{\frac{7}{3}}
C_{20,22}^{42,1}+\frac{\sqrt{7} C_{22,00}^{42,0}}{5760}+\frac{\sqrt{7} C_{22,00}^{42,1}}{2880}+\frac{\sqrt{\frac{7}{3}} C_{22,20}^{42,0}}{5760}+\frac{\sqrt{\frac{7}{3}}
C_{22,20}^{42,1}}{2880}-\frac{\sqrt{\frac{7}{3}} C_{22,22}^{40,0}}{1152}-\frac{1}{576} \sqrt{\frac{7}{3}} C_{22,22}^{40,1}-\frac{11 C_{22,22}^{42,0}}{2880
\sqrt{3}}-\frac{11 C_{22,22}^{42,1}}{1440 \sqrt{3}}+\frac{1}{32} \sqrt{\frac{3}{35}} C_{22,22}^{44,0}+\frac{1}{16} \sqrt{\frac{3}{35}} C_{22,22}^{44,1}+\frac{\sqrt{\frac{7}{5}}
C_{31,00}^{31,0}}{1152}+\frac{1}{576} \sqrt{\frac{7}{5}} C_{31,00}^{31,1}+\frac{\sqrt{\frac{7}{15}} C_{31,20}^{31,0}}{1152}+\frac{1}{576} \sqrt{\frac{7}{15}}
C_{31,20}^{31,1}-\frac{1}{360} \sqrt{\frac{7}{3}} C_{31,22}^{31,0}-\frac{1}{180} \sqrt{\frac{7}{3}} C_{31,22}^{31,1}-\frac{19 C_{31,22}^{33,0}}{1280
\sqrt{5}}-\frac{19 C_{31,22}^{33,1}}{640 \sqrt{5}}-\frac{1}{640} \sqrt{\frac{3}{5}} C_{33,00}^{33,0}-\frac{1}{320} \sqrt{\frac{3}{5}} C_{33,00}^{33,1}-\frac{C_{33,20}^{33,0}}{640
\sqrt{5}}-\frac{C_{33,20}^{33,1}}{320 \sqrt{5}}+\frac{1}{800} \sqrt{\frac{3}{2}} C_{33,22}^{33,0}+\frac{1}{400} \sqrt{\frac{3}{2}} C_{33,22}^{33,1}\)

\noindent\(C_{11,3312,1}^{11,1112,1,\text{Loc}}= \frac{1}{192} \sqrt{\frac{7}{5}} C_{00,20}^{60,1}+\frac{11 \sqrt{7} C_{00,22}^{62,1}}{640}+\frac{13
\sqrt{7} C_{11,22}^{51,1}}{1440}+\frac{3 \sqrt{\frac{3}{5}} C_{11,22}^{53,1}}{1280}-\frac{1}{576} \sqrt{\frac{7}{5}} C_{20,20}^{40,1}-\frac{\sqrt{7}
C_{20,22}^{42,1}}{360}+\frac{\sqrt{7} C_{22,20}^{42,1}}{2880}+\frac{\sqrt{7} C_{22,22}^{40,1}}{1152}+\frac{11 C_{22,22}^{42,1}}{2880}-\frac{3 C_{22,22}^{44,1}}{32
\sqrt{35}}-\frac{1}{576} \sqrt{\frac{7}{5}} C_{31,20}^{31,1}-\frac{\sqrt{7} C_{31,22}^{31,1}}{360}-\frac{19 \sqrt{\frac{3}{5}} C_{31,22}^{33,1}}{1280}+\frac{1}{320}
\sqrt{\frac{3}{5}} C_{33,20}^{33,1}+\frac{3 C_{33,22}^{33,1}}{800 \sqrt{2}}\)

\noindent\(C_{11,3312,1}^{11,1112,1,\text{NonLoc}}= \frac{1}{128} \sqrt{\frac{7}{15}} C_{00,00}^{60,0}+\frac{1}{384} \sqrt{\frac{7}{5}}
C_{00,20}^{60,0}-\frac{11 \sqrt{7} C_{00,22}^{62,0}}{640}-\frac{1}{384} \sqrt{\frac{7}{15}} C_{11,00}^{51,0}+\frac{13 \sqrt{7} C_{11,22}^{51,0}}{1440}+\frac{3
\sqrt{\frac{3}{5}} C_{11,22}^{53,0}}{1280}-\frac{1}{384} \sqrt{\frac{7}{15}} C_{20,00}^{40,0}-\frac{\sqrt{\frac{7}{5}} C_{20,20}^{40,0}}{1152}+\frac{\sqrt{7}
C_{20,22}^{42,0}}{360}+\frac{\sqrt{\frac{7}{3}} C_{22,00}^{42,0}}{1920}+\frac{\sqrt{7} C_{22,20}^{42,0}}{5760}-\frac{\sqrt{7} C_{22,22}^{40,0}}{1152}-\frac{11
C_{22,22}^{42,0}}{2880}+\frac{3 C_{22,22}^{44,0}}{32 \sqrt{35}}+\frac{1}{384} \sqrt{\frac{7}{15}} C_{31,00}^{31,0}+\frac{\sqrt{\frac{7}{5}} C_{31,20}^{31,0}}{1152}-\frac{\sqrt{7}
C_{31,22}^{31,0}}{360}-\frac{19 \sqrt{\frac{3}{5}} C_{31,22}^{33,0}}{1280}-\frac{3 C_{33,00}^{33,0}}{640 \sqrt{5}}-\frac{1}{640} \sqrt{\frac{3}{5}}
C_{33,20}^{33,0}+\frac{3 C_{33,22}^{33,0}}{800 \sqrt{2}}\)

\noindent\(C_{11,3312,2}^{00,2202,0,\text{Loc}}= -\frac{\sqrt{7} C_{11,11}^{51,0}}{144}-\frac{\sqrt{7} C_{11,11}^{51,1}}{288}-\frac{1}{24}
\sqrt{\frac{7}{15}} C_{22,11}^{42,0}-\frac{1}{48} \sqrt{\frac{7}{15}} C_{22,11}^{42,1}-\frac{\sqrt{7} C_{31,11}^{31,0}}{144}-\frac{\sqrt{7} C_{31,11}^{31,1}}{288}-\frac{C_{33,11}^{33,0}}{20
\sqrt{2}}-\frac{C_{33,11}^{33,1}}{40 \sqrt{2}}\)

\noindent\(C_{11,3312,2}^{00,2202,0,\text{NonLoc}}= -\frac{\sqrt{7} C_{11,11}^{51,0}}{288}-\frac{\sqrt{7} C_{11,11}^{51,1}}{144}+\frac{1}{48}
\sqrt{\frac{7}{15}} C_{22,11}^{42,0}+\frac{1}{24} \sqrt{\frac{7}{15}} C_{22,11}^{42,1}-\frac{\sqrt{7} C_{31,11}^{31,0}}{288}-\frac{\sqrt{7} C_{31,11}^{31,1}}{144}-\frac{C_{33,11}^{33,0}}{40
\sqrt{2}}-\frac{C_{33,11}^{33,1}}{20 \sqrt{2}}\)

\noindent\(C_{11,3312,2}^{00,2202,1,\text{Loc}}= -\frac{1}{96} \sqrt{\frac{7}{3}} C_{11,11}^{51,1}-\frac{1}{48} \sqrt{\frac{7}{5}}
C_{22,11}^{42,1}-\frac{1}{96} \sqrt{\frac{7}{3}} C_{31,11}^{31,1}-\frac{1}{40} \sqrt{\frac{3}{2}} C_{33,11}^{33,1}\)

\noindent\(C_{11,3312,2}^{00,2202,1,\text{NonLoc}}= -\frac{1}{96} \sqrt{\frac{7}{3}} C_{11,11}^{51,0}+\frac{1}{48} \sqrt{\frac{7}{5}}
C_{22,11}^{42,0}-\frac{1}{96} \sqrt{\frac{7}{3}} C_{31,11}^{31,0}-\frac{1}{40} \sqrt{\frac{3}{2}} C_{33,11}^{33,0}\)

\noindent\(C_{11,3312,2}^{11,1111,0,\text{Loc}}= -\frac{1}{96} \sqrt{\frac{7}{3}} C_{00,20}^{60,0}-\frac{1}{192} \sqrt{\frac{7}{3}}
C_{00,20}^{60,1}+\frac{1}{64} \sqrt{\frac{7}{15}} C_{00,22}^{62,0}+\frac{1}{128} \sqrt{\frac{7}{15}} C_{00,22}^{62,1}+\frac{7}{288} \sqrt{\frac{7}{15}}
C_{11,22}^{51,0}+\frac{7}{576} \sqrt{\frac{7}{15}} C_{11,22}^{51,1}-\frac{9 C_{11,22}^{53,0}}{640}-\frac{9 C_{11,22}^{53,1}}{1280}+\frac{1}{288}
\sqrt{\frac{7}{3}} C_{20,20}^{40,0}+\frac{1}{576} \sqrt{\frac{7}{3}} C_{20,20}^{40,1}-\frac{7}{288} \sqrt{\frac{7}{15}} C_{20,22}^{42,0}-\frac{7}{576}
\sqrt{\frac{7}{15}} C_{20,22}^{42,1}-\frac{1}{288} \sqrt{\frac{7}{15}} C_{22,20}^{42,0}-\frac{1}{576} \sqrt{\frac{7}{15}} C_{22,20}^{42,1}+\frac{19}{576}
\sqrt{\frac{7}{15}} C_{22,22}^{40,0}+\frac{19 \sqrt{\frac{7}{15}} C_{22,22}^{40,1}}{1152}-\frac{1}{288} \sqrt{\frac{5}{3}} C_{22,22}^{42,0}-\frac{1}{576}
\sqrt{\frac{5}{3}} C_{22,22}^{42,1}+\frac{1}{80} \sqrt{\frac{3}{7}} C_{22,22}^{44,0}+\frac{1}{160} \sqrt{\frac{3}{7}} C_{22,22}^{44,1}+\frac{1}{288}
\sqrt{\frac{7}{3}} C_{31,20}^{31,0}+\frac{1}{576} \sqrt{\frac{7}{3}} C_{31,20}^{31,1}-\frac{7}{288} \sqrt{\frac{7}{15}} C_{31,22}^{31,0}-\frac{7}{576}
\sqrt{\frac{7}{15}} C_{31,22}^{31,1}+\frac{9 C_{31,22}^{33,0}}{640}+\frac{9 C_{31,22}^{33,1}}{1280}-\frac{C_{33,20}^{33,0}}{160}-\frac{C_{33,20}^{33,1}}{320}-\frac{1}{80}
\sqrt{\frac{3}{10}} C_{33,22}^{33,0}-\frac{1}{160} \sqrt{\frac{3}{10}} C_{33,22}^{33,1}\)

\noindent\(C_{11,3312,2}^{11,1111,0,\text{NonLoc}}= -\frac{\sqrt{7} C_{00,00}^{60,0}}{384}-\frac{\sqrt{7} C_{00,00}^{60,1}}{192}-\frac{1}{384}
\sqrt{\frac{7}{3}} C_{00,20}^{60,0}-\frac{1}{192} \sqrt{\frac{7}{3}} C_{00,20}^{60,1}-\frac{1}{128} \sqrt{\frac{7}{15}} C_{00,22}^{62,0}-\frac{1}{64}
\sqrt{\frac{7}{15}} C_{00,22}^{62,1}+\frac{\sqrt{7} C_{11,00}^{51,0}}{1152}+\frac{\sqrt{7} C_{11,00}^{51,1}}{576}+\frac{7}{576} \sqrt{\frac{7}{15}}
C_{11,22}^{51,0}+\frac{7}{288} \sqrt{\frac{7}{15}} C_{11,22}^{51,1}-\frac{9 C_{11,22}^{53,0}}{1280}-\frac{9 C_{11,22}^{53,1}}{640}+\frac{\sqrt{7}
C_{20,00}^{40,0}}{1152}+\frac{\sqrt{7} C_{20,00}^{40,1}}{576}+\frac{\sqrt{\frac{7}{3}} C_{20,20}^{40,0}}{1152}+\frac{1}{576} \sqrt{\frac{7}{3}} C_{20,20}^{40,1}+\frac{7}{576}
\sqrt{\frac{7}{15}} C_{20,22}^{42,0}+\frac{7}{288} \sqrt{\frac{7}{15}} C_{20,22}^{42,1}-\frac{\sqrt{\frac{7}{5}} C_{22,00}^{42,0}}{1152}-\frac{1}{576}
\sqrt{\frac{7}{5}} C_{22,00}^{42,1}-\frac{\sqrt{\frac{7}{15}} C_{22,20}^{42,0}}{1152}-\frac{1}{576} \sqrt{\frac{7}{15}} C_{22,20}^{42,1}-\frac{19
\sqrt{\frac{7}{15}} C_{22,22}^{40,0}}{1152}-\frac{19}{576} \sqrt{\frac{7}{15}} C_{22,22}^{40,1}+\frac{1}{576} \sqrt{\frac{5}{3}} C_{22,22}^{42,0}+\frac{1}{288}
\sqrt{\frac{5}{3}} C_{22,22}^{42,1}-\frac{1}{160} \sqrt{\frac{3}{7}} C_{22,22}^{44,0}-\frac{1}{80} \sqrt{\frac{3}{7}} C_{22,22}^{44,1}-\frac{\sqrt{7}
C_{31,00}^{31,0}}{1152}-\frac{\sqrt{7} C_{31,00}^{31,1}}{576}-\frac{\sqrt{\frac{7}{3}} C_{31,20}^{31,0}}{1152}-\frac{1}{576} \sqrt{\frac{7}{3}} C_{31,20}^{31,1}-\frac{7}{576}
\sqrt{\frac{7}{15}} C_{31,22}^{31,0}-\frac{7}{288} \sqrt{\frac{7}{15}} C_{31,22}^{31,1}+\frac{9 C_{31,22}^{33,0}}{1280}+\frac{9 C_{31,22}^{33,1}}{640}+\frac{\sqrt{3}
C_{33,00}^{33,0}}{640}+\frac{\sqrt{3} C_{33,00}^{33,1}}{320}+\frac{C_{33,20}^{33,0}}{640}+\frac{C_{33,20}^{33,1}}{320}-\frac{1}{160} \sqrt{\frac{3}{10}}
C_{33,22}^{33,0}-\frac{1}{80} \sqrt{\frac{3}{10}} C_{33,22}^{33,1}\)

\noindent\(C_{11,3312,2}^{11,1111,1,\text{Loc}}= -\frac{\sqrt{7} C_{00,20}^{60,1}}{192}+\frac{1}{128} \sqrt{\frac{7}{5}} C_{00,22}^{62,1}+\frac{7}{576}
\sqrt{\frac{7}{5}} C_{11,22}^{51,1}-\frac{9 \sqrt{3} C_{11,22}^{53,1}}{1280}+\frac{\sqrt{7} C_{20,20}^{40,1}}{576}-\frac{7}{576} \sqrt{\frac{7}{5}}
C_{20,22}^{42,1}-\frac{1}{576} \sqrt{\frac{7}{5}} C_{22,20}^{42,1}+\frac{19 \sqrt{\frac{7}{5}} C_{22,22}^{40,1}}{1152}-\frac{\sqrt{5} C_{22,22}^{42,1}}{576}+\frac{3
C_{22,22}^{44,1}}{160 \sqrt{7}}+\frac{\sqrt{7} C_{31,20}^{31,1}}{576}-\frac{7}{576} \sqrt{\frac{7}{5}} C_{31,22}^{31,1}+\frac{9 \sqrt{3} C_{31,22}^{33,1}}{1280}-\frac{\sqrt{3}
C_{33,20}^{33,1}}{320}-\frac{3 C_{33,22}^{33,1}}{160 \sqrt{10}}\)

\noindent\(C_{11,3312,2}^{11,1111,1,\text{NonLoc}}= -\frac{1}{128} \sqrt{\frac{7}{3}} C_{00,00}^{60,0}-\frac{\sqrt{7} C_{00,20}^{60,0}}{384}-\frac{1}{128}
\sqrt{\frac{7}{5}} C_{00,22}^{62,0}+\frac{1}{384} \sqrt{\frac{7}{3}} C_{11,00}^{51,0}+\frac{7}{576} \sqrt{\frac{7}{5}} C_{11,22}^{51,0}-\frac{9 \sqrt{3}
C_{11,22}^{53,0}}{1280}+\frac{1}{384} \sqrt{\frac{7}{3}} C_{20,00}^{40,0}+\frac{\sqrt{7} C_{20,20}^{40,0}}{1152}+\frac{7}{576} \sqrt{\frac{7}{5}}
C_{20,22}^{42,0}-\frac{1}{384} \sqrt{\frac{7}{15}} C_{22,00}^{42,0}-\frac{\sqrt{\frac{7}{5}} C_{22,20}^{42,0}}{1152}-\frac{19 \sqrt{\frac{7}{5}}
C_{22,22}^{40,0}}{1152}+\frac{\sqrt{5} C_{22,22}^{42,0}}{576}-\frac{3 C_{22,22}^{44,0}}{160 \sqrt{7}}-\frac{1}{384} \sqrt{\frac{7}{3}} C_{31,00}^{31,0}-\frac{\sqrt{7}
C_{31,20}^{31,0}}{1152}-\frac{7}{576} \sqrt{\frac{7}{5}} C_{31,22}^{31,0}+\frac{9 \sqrt{3} C_{31,22}^{33,0}}{1280}+\frac{3 C_{33,00}^{33,0}}{640}+\frac{\sqrt{3}
C_{33,20}^{33,0}}{640}-\frac{3 C_{33,22}^{33,0}}{160 \sqrt{10}}\)

\noindent\(C_{11,3312,2}^{11,1112,0,\text{Loc}}= -\frac{\sqrt{7} C_{00,20}^{60,0}}{288}-\frac{\sqrt{7} C_{00,20}^{60,1}}{576}+\frac{1}{192}
\sqrt{\frac{7}{5}} C_{00,22}^{62,0}+\frac{1}{384} \sqrt{\frac{7}{5}} C_{00,22}^{62,1}+\frac{1}{216} \sqrt{\frac{7}{5}} C_{11,22}^{51,0}+\frac{1}{432}
\sqrt{\frac{7}{5}} C_{11,22}^{51,1}-\frac{\sqrt{3} C_{11,22}^{53,0}}{640}-\frac{\sqrt{3} C_{11,22}^{53,1}}{1280}+\frac{\sqrt{7} C_{20,20}^{40,0}}{864}+\frac{\sqrt{7}
C_{20,20}^{40,1}}{1728}+\frac{7}{432} \sqrt{\frac{7}{5}} C_{20,22}^{42,0}+\frac{7}{864} \sqrt{\frac{7}{5}} C_{20,22}^{42,1}-\frac{1}{864} \sqrt{\frac{7}{5}}
C_{22,20}^{42,0}-\frac{\sqrt{\frac{7}{5}} C_{22,20}^{42,1}}{1728}+\frac{7 \sqrt{\frac{7}{5}} C_{22,22}^{40,0}}{1728}+\frac{7 \sqrt{\frac{7}{5}} C_{22,22}^{40,1}}{3456}-\frac{7
\sqrt{5} C_{22,22}^{42,0}}{864}-\frac{7 \sqrt{5} C_{22,22}^{42,1}}{1728}-\frac{\sqrt{7} C_{22,22}^{44,0}}{80}-\frac{\sqrt{7} C_{22,22}^{44,1}}{160}+\frac{\sqrt{7}
C_{31,20}^{31,0}}{864}+\frac{\sqrt{7} C_{31,20}^{31,1}}{1728}+\frac{7}{432} \sqrt{\frac{7}{5}} C_{31,22}^{31,0}+\frac{7}{864} \sqrt{\frac{7}{5}}
C_{31,22}^{31,1}-\frac{13 C_{31,22}^{33,0}}{640 \sqrt{3}}-\frac{13 C_{31,22}^{33,1}}{1280 \sqrt{3}}-\frac{C_{33,20}^{33,0}}{160 \sqrt{3}}-\frac{C_{33,20}^{33,1}}{320
\sqrt{3}}-\frac{13 C_{33,22}^{33,0}}{80 \sqrt{10}}-\frac{13 C_{33,22}^{33,1}}{160 \sqrt{10}}\)

\noindent\(C_{11,3312,2}^{11,1112,0,\text{NonLoc}}= -\frac{1}{384} \sqrt{\frac{7}{3}} C_{00,00}^{60,0}-\frac{1}{192} \sqrt{\frac{7}{3}}
C_{00,00}^{60,1}-\frac{\sqrt{7} C_{00,20}^{60,0}}{1152}-\frac{\sqrt{7} C_{00,20}^{60,1}}{576}-\frac{1}{384} \sqrt{\frac{7}{5}} C_{00,22}^{62,0}-\frac{1}{192}
\sqrt{\frac{7}{5}} C_{00,22}^{62,1}+\frac{\sqrt{\frac{7}{3}} C_{11,00}^{51,0}}{1152}+\frac{1}{576} \sqrt{\frac{7}{3}} C_{11,00}^{51,1}+\frac{1}{432}
\sqrt{\frac{7}{5}} C_{11,22}^{51,0}+\frac{1}{216} \sqrt{\frac{7}{5}} C_{11,22}^{51,1}-\frac{\sqrt{3} C_{11,22}^{53,0}}{1280}-\frac{\sqrt{3} C_{11,22}^{53,1}}{640}+\frac{\sqrt{\frac{7}{3}}
C_{20,00}^{40,0}}{1152}+\frac{1}{576} \sqrt{\frac{7}{3}} C_{20,00}^{40,1}+\frac{\sqrt{7} C_{20,20}^{40,0}}{3456}+\frac{\sqrt{7} C_{20,20}^{40,1}}{1728}-\frac{7}{864}
\sqrt{\frac{7}{5}} C_{20,22}^{42,0}-\frac{7}{432} \sqrt{\frac{7}{5}} C_{20,22}^{42,1}-\frac{\sqrt{\frac{7}{15}} C_{22,00}^{42,0}}{1152}-\frac{1}{576}
\sqrt{\frac{7}{15}} C_{22,00}^{42,1}-\frac{\sqrt{\frac{7}{5}} C_{22,20}^{42,0}}{3456}-\frac{\sqrt{\frac{7}{5}} C_{22,20}^{42,1}}{1728}-\frac{7 \sqrt{\frac{7}{5}}
C_{22,22}^{40,0}}{3456}-\frac{7 \sqrt{\frac{7}{5}} C_{22,22}^{40,1}}{1728}+\frac{7 \sqrt{5} C_{22,22}^{42,0}}{1728}+\frac{7 \sqrt{5} C_{22,22}^{42,1}}{864}+\frac{\sqrt{7}
C_{22,22}^{44,0}}{160}+\frac{\sqrt{7} C_{22,22}^{44,1}}{80}-\frac{\sqrt{\frac{7}{3}} C_{31,00}^{31,0}}{1152}-\frac{1}{576} \sqrt{\frac{7}{3}} C_{31,00}^{31,1}-\frac{\sqrt{7}
C_{31,20}^{31,0}}{3456}-\frac{\sqrt{7} C_{31,20}^{31,1}}{1728}+\frac{7}{864} \sqrt{\frac{7}{5}} C_{31,22}^{31,0}+\frac{7}{432} \sqrt{\frac{7}{5}}
C_{31,22}^{31,1}-\frac{13 C_{31,22}^{33,0}}{1280 \sqrt{3}}-\frac{13 C_{31,22}^{33,1}}{640 \sqrt{3}}+\frac{C_{33,00}^{33,0}}{640}+\frac{C_{33,00}^{33,1}}{320}+\frac{C_{33,20}^{33,0}}{640
\sqrt{3}}+\frac{C_{33,20}^{33,1}}{320 \sqrt{3}}-\frac{13 C_{33,22}^{33,0}}{160 \sqrt{10}}-\frac{13 C_{33,22}^{33,1}}{80 \sqrt{10}}\)

\noindent\(C_{11,3312,2}^{11,1112,1,\text{Loc}}= -\frac{1}{192} \sqrt{\frac{7}{3}} C_{00,20}^{60,1}+\frac{1}{128} \sqrt{\frac{7}{15}}
C_{00,22}^{62,1}+\frac{1}{144} \sqrt{\frac{7}{15}} C_{11,22}^{51,1}-\frac{3 C_{11,22}^{53,1}}{1280}+\frac{1}{576} \sqrt{\frac{7}{3}} C_{20,20}^{40,1}+\frac{7}{288}
\sqrt{\frac{7}{15}} C_{20,22}^{42,1}-\frac{1}{576} \sqrt{\frac{7}{15}} C_{22,20}^{42,1}+\frac{7 \sqrt{\frac{7}{15}} C_{22,22}^{40,1}}{1152}-\frac{7}{576}
\sqrt{\frac{5}{3}} C_{22,22}^{42,1}-\frac{\sqrt{21} C_{22,22}^{44,1}}{160}+\frac{1}{576} \sqrt{\frac{7}{3}} C_{31,20}^{31,1}+\frac{7}{288} \sqrt{\frac{7}{15}}
C_{31,22}^{31,1}-\frac{13 C_{31,22}^{33,1}}{1280}-\frac{C_{33,20}^{33,1}}{320}-\frac{13}{160} \sqrt{\frac{3}{10}} C_{33,22}^{33,1}\)

\noindent\(C_{11,3312,2}^{11,1112,1,\text{NonLoc}}= -\frac{\sqrt{7} C_{00,00}^{60,0}}{384}-\frac{1}{384} \sqrt{\frac{7}{3}} C_{00,20}^{60,0}-\frac{1}{128}
\sqrt{\frac{7}{15}} C_{00,22}^{62,0}+\frac{\sqrt{7} C_{11,00}^{51,0}}{1152}+\frac{1}{144} \sqrt{\frac{7}{15}} C_{11,22}^{51,0}-\frac{3 C_{11,22}^{53,0}}{1280}+\frac{\sqrt{7}
C_{20,00}^{40,0}}{1152}+\frac{\sqrt{\frac{7}{3}} C_{20,20}^{40,0}}{1152}-\frac{7}{288} \sqrt{\frac{7}{15}} C_{20,22}^{42,0}-\frac{\sqrt{\frac{7}{5}}
C_{22,00}^{42,0}}{1152}-\frac{\sqrt{\frac{7}{15}} C_{22,20}^{42,0}}{1152}-\frac{7 \sqrt{\frac{7}{15}} C_{22,22}^{40,0}}{1152}+\frac{7}{576} \sqrt{\frac{5}{3}}
C_{22,22}^{42,0}+\frac{\sqrt{21} C_{22,22}^{44,0}}{160}-\frac{\sqrt{7} C_{31,00}^{31,0}}{1152}-\frac{\sqrt{\frac{7}{3}} C_{31,20}^{31,0}}{1152}+\frac{7}{288}
\sqrt{\frac{7}{15}} C_{31,22}^{31,0}-\frac{13 C_{31,22}^{33,0}}{1280}+\frac{\sqrt{3} C_{33,00}^{33,0}}{640}+\frac{C_{33,20}^{33,0}}{640}-\frac{13}{160}
\sqrt{\frac{3}{10}} C_{33,22}^{33,0}\)

\noindent\(C_{11,3312,3}^{11,1112,0,\text{Loc}}= \frac{C_{00,20}^{60,0}}{144 \sqrt{5}}+\frac{C_{00,20}^{60,1}}{288 \sqrt{5}}+\frac{13
C_{00,22}^{62,0}}{240}+\frac{13 C_{00,22}^{62,1}}{480}+\frac{127 C_{11,22}^{51,0}}{4320}+\frac{127 C_{11,22}^{51,1}}{8640}+\frac{1}{320} \sqrt{\frac{3}{35}}
C_{11,22}^{53,0}+\frac{1}{640} \sqrt{\frac{3}{35}} C_{11,22}^{53,1}-\frac{C_{20,20}^{40,0}}{432 \sqrt{5}}-\frac{C_{20,20}^{40,1}}{864 \sqrt{5}}-\frac{C_{20,22}^{42,0}}{4320}-\frac{C_{20,22}^{42,1}}{8640}+\frac{C_{22,20}^{42,0}}{2160}+\frac{C_{22,20}^{42,1}}{4320}+\frac{C_{22,22}^{40,0}}{216}+\frac{C_{22,22}^{40,1}}{432}-\frac{19
C_{22,22}^{42,0}}{2160 \sqrt{7}}-\frac{19 C_{22,22}^{42,1}}{4320 \sqrt{7}}-\frac{C_{22,22}^{44,0}}{14 \sqrt{5}}-\frac{C_{22,22}^{44,1}}{28 \sqrt{5}}-\frac{C_{31,20}^{31,0}}{432
\sqrt{5}}-\frac{C_{31,20}^{31,1}}{864 \sqrt{5}}-\frac{C_{31,22}^{31,0}}{4320}-\frac{C_{31,22}^{31,1}}{8640}-\frac{59 C_{31,22}^{33,0}}{320 \sqrt{105}}-\frac{59
C_{31,22}^{33,1}}{640 \sqrt{105}}+\frac{C_{33,20}^{33,0}}{80 \sqrt{105}}+\frac{C_{33,20}^{33,1}}{160 \sqrt{105}}-\frac{1}{100} \sqrt{\frac{7}{2}}
C_{33,22}^{33,0}-\frac{1}{200} \sqrt{\frac{7}{2}} C_{33,22}^{33,1}\)

\noindent\(C_{11,3312,3}^{11,1112,0,\text{NonLoc}}= \frac{C_{00,00}^{60,0}}{192 \sqrt{15}}+\frac{C_{00,00}^{60,1}}{96 \sqrt{15}}+\frac{C_{00,20}^{60,0}}{576
\sqrt{5}}+\frac{C_{00,20}^{60,1}}{288 \sqrt{5}}-\frac{13 C_{00,22}^{62,0}}{480}-\frac{13 C_{00,22}^{62,1}}{240}-\frac{C_{11,00}^{51,0}}{576 \sqrt{15}}-\frac{C_{11,00}^{51,1}}{288
\sqrt{15}}+\frac{127 C_{11,22}^{51,0}}{8640}+\frac{127 C_{11,22}^{51,1}}{4320}+\frac{1}{640} \sqrt{\frac{3}{35}} C_{11,22}^{53,0}+\frac{1}{320} \sqrt{\frac{3}{35}}
C_{11,22}^{53,1}-\frac{C_{20,00}^{40,0}}{576 \sqrt{15}}-\frac{C_{20,00}^{40,1}}{288 \sqrt{15}}-\frac{C_{20,20}^{40,0}}{1728 \sqrt{5}}-\frac{C_{20,20}^{40,1}}{864
\sqrt{5}}+\frac{C_{20,22}^{42,0}}{8640}+\frac{C_{20,22}^{42,1}}{4320}+\frac{C_{22,00}^{42,0}}{2880 \sqrt{3}}+\frac{C_{22,00}^{42,1}}{1440 \sqrt{3}}+\frac{C_{22,20}^{42,0}}{8640}+\frac{C_{22,20}^{42,1}}{4320}-\frac{C_{22,22}^{40,0}}{432}-\frac{C_{22,22}^{40,1}}{216}+\frac{19
C_{22,22}^{42,0}}{4320 \sqrt{7}}+\frac{19 C_{22,22}^{42,1}}{2160 \sqrt{7}}+\frac{C_{22,22}^{44,0}}{28 \sqrt{5}}+\frac{C_{22,22}^{44,1}}{14 \sqrt{5}}+\frac{C_{31,00}^{31,0}}{576
\sqrt{15}}+\frac{C_{31,00}^{31,1}}{288 \sqrt{15}}+\frac{C_{31,20}^{31,0}}{1728 \sqrt{5}}+\frac{C_{31,20}^{31,1}}{864 \sqrt{5}}-\frac{C_{31,22}^{31,0}}{8640}-\frac{C_{31,22}^{31,1}}{4320}-\frac{59
C_{31,22}^{33,0}}{640 \sqrt{105}}-\frac{59 C_{31,22}^{33,1}}{320 \sqrt{105}}-\frac{C_{33,00}^{33,0}}{320 \sqrt{35}}-\frac{C_{33,00}^{33,1}}{160 \sqrt{35}}-\frac{C_{33,20}^{33,0}}{320
\sqrt{105}}-\frac{C_{33,20}^{33,1}}{160 \sqrt{105}}-\frac{1}{200} \sqrt{\frac{7}{2}} C_{33,22}^{33,0}-\frac{1}{100} \sqrt{\frac{7}{2}} C_{33,22}^{33,1}\)

\noindent\(C_{11,3312,3}^{11,1112,1,\text{Loc}}= \frac{C_{00,20}^{60,1}}{96 \sqrt{15}}+\frac{13 C_{00,22}^{62,1}}{160 \sqrt{3}}+\frac{127
C_{11,22}^{51,1}}{2880 \sqrt{3}}+\frac{3 C_{11,22}^{53,1}}{640 \sqrt{35}}-\frac{C_{20,20}^{40,1}}{288 \sqrt{15}}-\frac{C_{20,22}^{42,1}}{2880 \sqrt{3}}+\frac{C_{22,20}^{42,1}}{1440
\sqrt{3}}+\frac{C_{22,22}^{40,1}}{144 \sqrt{3}}-\frac{19 C_{22,22}^{42,1}}{1440 \sqrt{21}}-\frac{1}{28} \sqrt{\frac{3}{5}} C_{22,22}^{44,1}-\frac{C_{31,20}^{31,1}}{288
\sqrt{15}}-\frac{C_{31,22}^{31,1}}{2880 \sqrt{3}}-\frac{59 C_{31,22}^{33,1}}{640 \sqrt{35}}+\frac{C_{33,20}^{33,1}}{160 \sqrt{35}}-\frac{1}{200}
\sqrt{\frac{21}{2}} C_{33,22}^{33,1}\)

\noindent\(C_{11,3312,3}^{11,1112,1,\text{NonLoc}}= \frac{C_{00,00}^{60,0}}{192 \sqrt{5}}+\frac{C_{00,20}^{60,0}}{192 \sqrt{15}}-\frac{13
C_{00,22}^{62,0}}{160 \sqrt{3}}-\frac{C_{11,00}^{51,0}}{576 \sqrt{5}}+\frac{127 C_{11,22}^{51,0}}{2880 \sqrt{3}}+\frac{3 C_{11,22}^{53,0}}{640 \sqrt{35}}-\frac{C_{20,00}^{40,0}}{576
\sqrt{5}}-\frac{C_{20,20}^{40,0}}{576 \sqrt{15}}+\frac{C_{20,22}^{42,0}}{2880 \sqrt{3}}+\frac{C_{22,00}^{42,0}}{2880}+\frac{C_{22,20}^{42,0}}{2880
\sqrt{3}}-\frac{C_{22,22}^{40,0}}{144 \sqrt{3}}+\frac{19 C_{22,22}^{42,0}}{1440 \sqrt{21}}+\frac{1}{28} \sqrt{\frac{3}{5}} C_{22,22}^{44,0}+\frac{C_{31,00}^{31,0}}{576
\sqrt{5}}+\frac{C_{31,20}^{31,0}}{576 \sqrt{15}}-\frac{C_{31,22}^{31,0}}{2880 \sqrt{3}}-\frac{59 C_{31,22}^{33,0}}{640 \sqrt{35}}-\frac{1}{320} \sqrt{\frac{3}{35}}
C_{33,00}^{33,0}-\frac{C_{33,20}^{33,0}}{320 \sqrt{35}}-\frac{1}{200} \sqrt{\frac{21}{2}} C_{33,22}^{33,0}\)

\noindent\(C_{11,3313,2}^{00,2202,0,\text{Loc}}= -\frac{1}{72} \sqrt{\frac{7}{2}} C_{11,11}^{51,0}-\frac{1}{144} \sqrt{\frac{7}{2}}
C_{11,11}^{51,1}-\frac{1}{12} \sqrt{\frac{7}{30}} C_{22,11}^{42,0}-\frac{1}{24} \sqrt{\frac{7}{30}} C_{22,11}^{42,1}-\frac{1}{72} \sqrt{\frac{7}{2}}
C_{31,11}^{31,0}-\frac{1}{144} \sqrt{\frac{7}{2}} C_{31,11}^{31,1}-\frac{C_{33,11}^{33,0}}{20}-\frac{C_{33,11}^{33,1}}{40}\)

\noindent\(C_{11,3313,2}^{00,2202,0,\text{NonLoc}}= -\frac{1}{144} \sqrt{\frac{7}{2}} C_{11,11}^{51,0}-\frac{1}{72} \sqrt{\frac{7}{2}}
C_{11,11}^{51,1}+\frac{1}{24} \sqrt{\frac{7}{30}} C_{22,11}^{42,0}+\frac{1}{12} \sqrt{\frac{7}{30}} C_{22,11}^{42,1}-\frac{1}{144} \sqrt{\frac{7}{2}}
C_{31,11}^{31,0}-\frac{1}{72} \sqrt{\frac{7}{2}} C_{31,11}^{31,1}-\frac{C_{33,11}^{33,0}}{40}-\frac{C_{33,11}^{33,1}}{20}\)

\noindent\(C_{11,3313,2}^{00,2202,1,\text{Loc}}= -\frac{1}{48} \sqrt{\frac{7}{6}} C_{11,11}^{51,1}-\frac{1}{24} \sqrt{\frac{7}{10}}
C_{22,11}^{42,1}-\frac{1}{48} \sqrt{\frac{7}{6}} C_{31,11}^{31,1}-\frac{\sqrt{3} C_{33,11}^{33,1}}{40}\)

\noindent\(C_{11,3313,2}^{00,2202,1,\text{NonLoc}}= -\frac{1}{48} \sqrt{\frac{7}{6}} C_{11,11}^{51,0}+\frac{1}{24} \sqrt{\frac{7}{10}}
C_{22,11}^{42,0}-\frac{1}{48} \sqrt{\frac{7}{6}} C_{31,11}^{31,0}-\frac{\sqrt{3} C_{33,11}^{33,0}}{40}\)

\noindent\(C_{11,3313,2}^{11,1111,0,\text{Loc}}= \frac{1}{24} \sqrt{\frac{7}{6}} C_{00,20}^{60,0}+\frac{1}{48} \sqrt{\frac{7}{6}} C_{00,20}^{60,1}-\frac{1}{16}
\sqrt{\frac{7}{30}} C_{00,22}^{62,0}-\frac{1}{32} \sqrt{\frac{7}{30}} C_{00,22}^{62,1}-\frac{1}{144} \sqrt{\frac{35}{6}} C_{11,22}^{51,0}-\frac{1}{288}
\sqrt{\frac{35}{6}} C_{11,22}^{51,1}-\frac{1}{72} \sqrt{\frac{7}{6}} C_{20,20}^{40,0}-\frac{1}{144} \sqrt{\frac{7}{6}} C_{20,20}^{40,1}+\frac{1}{144}
\sqrt{\frac{35}{6}} C_{20,22}^{42,0}+\frac{1}{288} \sqrt{\frac{35}{6}} C_{20,22}^{42,1}+\frac{1}{72} \sqrt{\frac{7}{30}} C_{22,20}^{42,0}+\frac{1}{144}
\sqrt{\frac{7}{30}} C_{22,20}^{42,1}-\frac{1}{144} \sqrt{\frac{7}{30}} C_{22,22}^{40,0}-\frac{1}{288} \sqrt{\frac{7}{30}} C_{22,22}^{40,1}-\frac{C_{22,22}^{42,0}}{18
\sqrt{30}}-\frac{C_{22,22}^{42,1}}{36 \sqrt{30}}-\frac{1}{80} \sqrt{\frac{3}{14}} C_{22,22}^{44,0}-\frac{1}{160} \sqrt{\frac{3}{14}} C_{22,22}^{44,1}-\frac{1}{72}
\sqrt{\frac{7}{6}} C_{31,20}^{31,0}-\frac{1}{144} \sqrt{\frac{7}{6}} C_{31,20}^{31,1}+\frac{1}{144} \sqrt{\frac{35}{6}} C_{31,22}^{31,0}+\frac{1}{288}
\sqrt{\frac{35}{6}} C_{31,22}^{31,1}+\frac{C_{33,20}^{33,0}}{40 \sqrt{2}}+\frac{C_{33,20}^{33,1}}{80 \sqrt{2}}-\frac{1}{32} \sqrt{\frac{3}{5}} C_{33,22}^{33,0}-\frac{1}{64}
\sqrt{\frac{3}{5}} C_{33,22}^{33,1}\)

\noindent\(C_{11,3313,2}^{11,1111,0,\text{NonLoc}}= \frac{1}{96} \sqrt{\frac{7}{2}} C_{00,00}^{60,0}+\frac{1}{48} \sqrt{\frac{7}{2}}
C_{00,00}^{60,1}+\frac{1}{96} \sqrt{\frac{7}{6}} C_{00,20}^{60,0}+\frac{1}{48} \sqrt{\frac{7}{6}} C_{00,20}^{60,1}+\frac{1}{32} \sqrt{\frac{7}{30}}
C_{00,22}^{62,0}+\frac{1}{16} \sqrt{\frac{7}{30}} C_{00,22}^{62,1}-\frac{1}{288} \sqrt{\frac{7}{2}} C_{11,00}^{51,0}-\frac{1}{144} \sqrt{\frac{7}{2}}
C_{11,00}^{51,1}-\frac{1}{288} \sqrt{\frac{35}{6}} C_{11,22}^{51,0}-\frac{1}{144} \sqrt{\frac{35}{6}} C_{11,22}^{51,1}-\frac{1}{288} \sqrt{\frac{7}{2}}
C_{20,00}^{40,0}-\frac{1}{144} \sqrt{\frac{7}{2}} C_{20,00}^{40,1}-\frac{1}{288} \sqrt{\frac{7}{6}} C_{20,20}^{40,0}-\frac{1}{144} \sqrt{\frac{7}{6}}
C_{20,20}^{40,1}-\frac{1}{288} \sqrt{\frac{35}{6}} C_{20,22}^{42,0}-\frac{1}{144} \sqrt{\frac{35}{6}} C_{20,22}^{42,1}+\frac{1}{288} \sqrt{\frac{7}{10}}
C_{22,00}^{42,0}+\frac{1}{144} \sqrt{\frac{7}{10}} C_{22,00}^{42,1}+\frac{1}{288} \sqrt{\frac{7}{30}} C_{22,20}^{42,0}+\frac{1}{144} \sqrt{\frac{7}{30}}
C_{22,20}^{42,1}+\frac{1}{288} \sqrt{\frac{7}{30}} C_{22,22}^{40,0}+\frac{1}{144} \sqrt{\frac{7}{30}} C_{22,22}^{40,1}+\frac{C_{22,22}^{42,0}}{36
\sqrt{30}}+\frac{C_{22,22}^{42,1}}{18 \sqrt{30}}+\frac{1}{160} \sqrt{\frac{3}{14}} C_{22,22}^{44,0}+\frac{1}{80} \sqrt{\frac{3}{14}} C_{22,22}^{44,1}+\frac{1}{288}
\sqrt{\frac{7}{2}} C_{31,00}^{31,0}+\frac{1}{144} \sqrt{\frac{7}{2}} C_{31,00}^{31,1}+\frac{1}{288} \sqrt{\frac{7}{6}} C_{31,20}^{31,0}+\frac{1}{144}
\sqrt{\frac{7}{6}} C_{31,20}^{31,1}+\frac{1}{288} \sqrt{\frac{35}{6}} C_{31,22}^{31,0}+\frac{1}{144} \sqrt{\frac{35}{6}} C_{31,22}^{31,1}-\frac{1}{160}
\sqrt{\frac{3}{2}} C_{33,00}^{33,0}-\frac{1}{80} \sqrt{\frac{3}{2}} C_{33,00}^{33,1}-\frac{C_{33,20}^{33,0}}{160 \sqrt{2}}-\frac{C_{33,20}^{33,1}}{80
\sqrt{2}}-\frac{1}{64} \sqrt{\frac{3}{5}} C_{33,22}^{33,0}-\frac{1}{32} \sqrt{\frac{3}{5}} C_{33,22}^{33,1}\)

\noindent\(C_{11,3313,2}^{11,1111,1,\text{Loc}}= \frac{1}{48} \sqrt{\frac{7}{2}} C_{00,20}^{60,1}-\frac{1}{32} \sqrt{\frac{7}{10}}
C_{00,22}^{62,1}-\frac{1}{288} \sqrt{\frac{35}{2}} C_{11,22}^{51,1}-\frac{1}{144} \sqrt{\frac{7}{2}} C_{20,20}^{40,1}+\frac{1}{288} \sqrt{\frac{35}{2}}
C_{20,22}^{42,1}+\frac{1}{144} \sqrt{\frac{7}{10}} C_{22,20}^{42,1}-\frac{1}{288} \sqrt{\frac{7}{10}} C_{22,22}^{40,1}-\frac{C_{22,22}^{42,1}}{36
\sqrt{10}}-\frac{3 C_{22,22}^{44,1}}{160 \sqrt{14}}-\frac{1}{144} \sqrt{\frac{7}{2}} C_{31,20}^{31,1}+\frac{1}{288} \sqrt{\frac{35}{2}} C_{31,22}^{31,1}+\frac{1}{80}
\sqrt{\frac{3}{2}} C_{33,20}^{33,1}-\frac{3 C_{33,22}^{33,1}}{64 \sqrt{5}}\)

\noindent\(C_{11,3313,2}^{11,1111,1,\text{NonLoc}}= \frac{1}{32} \sqrt{\frac{7}{6}} C_{00,00}^{60,0}+\frac{1}{96} \sqrt{\frac{7}{2}}
C_{00,20}^{60,0}+\frac{1}{32} \sqrt{\frac{7}{10}} C_{00,22}^{62,0}-\frac{1}{96} \sqrt{\frac{7}{6}} C_{11,00}^{51,0}-\frac{1}{288} \sqrt{\frac{35}{2}}
C_{11,22}^{51,0}-\frac{1}{96} \sqrt{\frac{7}{6}} C_{20,00}^{40,0}-\frac{1}{288} \sqrt{\frac{7}{2}} C_{20,20}^{40,0}-\frac{1}{288} \sqrt{\frac{35}{2}}
C_{20,22}^{42,0}+\frac{1}{96} \sqrt{\frac{7}{30}} C_{22,00}^{42,0}+\frac{1}{288} \sqrt{\frac{7}{10}} C_{22,20}^{42,0}+\frac{1}{288} \sqrt{\frac{7}{10}}
C_{22,22}^{40,0}+\frac{C_{22,22}^{42,0}}{36 \sqrt{10}}+\frac{3 C_{22,22}^{44,0}}{160 \sqrt{14}}+\frac{1}{96} \sqrt{\frac{7}{6}} C_{31,00}^{31,0}+\frac{1}{288}
\sqrt{\frac{7}{2}} C_{31,20}^{31,0}+\frac{1}{288} \sqrt{\frac{35}{2}} C_{31,22}^{31,0}-\frac{3 C_{33,00}^{33,0}}{160 \sqrt{2}}-\frac{1}{160} \sqrt{\frac{3}{2}}
C_{33,20}^{33,0}-\frac{3 C_{33,22}^{33,0}}{64 \sqrt{5}}\)

\noindent\(C_{11,3313,2}^{11,1112,0,\text{Loc}}= \frac{1}{72} \sqrt{\frac{7}{2}} C_{00,20}^{60,0}+\frac{1}{144} \sqrt{\frac{7}{2}}
C_{00,20}^{60,1}-\frac{1}{48} \sqrt{\frac{7}{10}} C_{00,22}^{62,0}-\frac{1}{96} \sqrt{\frac{7}{10}} C_{00,22}^{62,1}+\frac{1}{432} \sqrt{\frac{7}{10}}
C_{11,22}^{51,0}+\frac{1}{864} \sqrt{\frac{7}{10}} C_{11,22}^{51,1}-\frac{1}{80} \sqrt{\frac{3}{2}} C_{11,22}^{53,0}-\frac{1}{160} \sqrt{\frac{3}{2}}
C_{11,22}^{53,1}-\frac{1}{216} \sqrt{\frac{7}{2}} C_{20,20}^{40,0}-\frac{1}{432} \sqrt{\frac{7}{2}} C_{20,20}^{40,1}-\frac{1}{432} \sqrt{\frac{7}{10}}
C_{20,22}^{42,0}-\frac{1}{864} \sqrt{\frac{7}{10}} C_{20,22}^{42,1}+\frac{1}{216} \sqrt{\frac{7}{10}} C_{22,20}^{42,0}+\frac{1}{432} \sqrt{\frac{7}{10}}
C_{22,20}^{42,1}+\frac{11}{432} \sqrt{\frac{7}{10}} C_{22,22}^{40,0}+\frac{11}{864} \sqrt{\frac{7}{10}} C_{22,22}^{40,1}-\frac{1}{108} \sqrt{\frac{5}{2}}
C_{22,22}^{42,0}-\frac{1}{216} \sqrt{\frac{5}{2}} C_{22,22}^{42,1}+\frac{C_{22,22}^{44,0}}{80 \sqrt{14}}+\frac{C_{22,22}^{44,1}}{160 \sqrt{14}}-\frac{1}{216}
\sqrt{\frac{7}{2}} C_{31,20}^{31,0}-\frac{1}{432} \sqrt{\frac{7}{2}} C_{31,20}^{31,1}-\frac{1}{432} \sqrt{\frac{7}{10}} C_{31,22}^{31,0}-\frac{1}{864}
\sqrt{\frac{7}{10}} C_{31,22}^{31,1}+\frac{1}{80} \sqrt{\frac{3}{2}} C_{31,22}^{33,0}+\frac{1}{160} \sqrt{\frac{3}{2}} C_{31,22}^{33,1}+\frac{C_{33,20}^{33,0}}{40
\sqrt{6}}+\frac{C_{33,20}^{33,1}}{80 \sqrt{6}}-\frac{11 C_{33,22}^{33,0}}{160 \sqrt{5}}-\frac{11 C_{33,22}^{33,1}}{320 \sqrt{5}}\)

\noindent\(C_{11,3313,2}^{11,1112,0,\text{NonLoc}}= \frac{1}{96} \sqrt{\frac{7}{6}} C_{00,00}^{60,0}+\frac{1}{48} \sqrt{\frac{7}{6}}
C_{00,00}^{60,1}+\frac{1}{288} \sqrt{\frac{7}{2}} C_{00,20}^{60,0}+\frac{1}{144} \sqrt{\frac{7}{2}} C_{00,20}^{60,1}+\frac{1}{96} \sqrt{\frac{7}{10}}
C_{00,22}^{62,0}+\frac{1}{48} \sqrt{\frac{7}{10}} C_{00,22}^{62,1}-\frac{1}{288} \sqrt{\frac{7}{6}} C_{11,00}^{51,0}-\frac{1}{144} \sqrt{\frac{7}{6}}
C_{11,00}^{51,1}+\frac{1}{864} \sqrt{\frac{7}{10}} C_{11,22}^{51,0}+\frac{1}{432} \sqrt{\frac{7}{10}} C_{11,22}^{51,1}-\frac{1}{160} \sqrt{\frac{3}{2}}
C_{11,22}^{53,0}-\frac{1}{80} \sqrt{\frac{3}{2}} C_{11,22}^{53,1}-\frac{1}{288} \sqrt{\frac{7}{6}} C_{20,00}^{40,0}-\frac{1}{144} \sqrt{\frac{7}{6}}
C_{20,00}^{40,1}-\frac{1}{864} \sqrt{\frac{7}{2}} C_{20,20}^{40,0}-\frac{1}{432} \sqrt{\frac{7}{2}} C_{20,20}^{40,1}+\frac{1}{864} \sqrt{\frac{7}{10}}
C_{20,22}^{42,0}+\frac{1}{432} \sqrt{\frac{7}{10}} C_{20,22}^{42,1}+\frac{1}{288} \sqrt{\frac{7}{30}} C_{22,00}^{42,0}+\frac{1}{144} \sqrt{\frac{7}{30}}
C_{22,00}^{42,1}+\frac{1}{864} \sqrt{\frac{7}{10}} C_{22,20}^{42,0}+\frac{1}{432} \sqrt{\frac{7}{10}} C_{22,20}^{42,1}-\frac{11}{864} \sqrt{\frac{7}{10}}
C_{22,22}^{40,0}-\frac{11}{432} \sqrt{\frac{7}{10}} C_{22,22}^{40,1}+\frac{1}{216} \sqrt{\frac{5}{2}} C_{22,22}^{42,0}+\frac{1}{108} \sqrt{\frac{5}{2}}
C_{22,22}^{42,1}-\frac{C_{22,22}^{44,0}}{160 \sqrt{14}}-\frac{C_{22,22}^{44,1}}{80 \sqrt{14}}+\frac{1}{288} \sqrt{\frac{7}{6}} C_{31,00}^{31,0}+\frac{1}{144}
\sqrt{\frac{7}{6}} C_{31,00}^{31,1}+\frac{1}{864} \sqrt{\frac{7}{2}} C_{31,20}^{31,0}+\frac{1}{432} \sqrt{\frac{7}{2}} C_{31,20}^{31,1}-\frac{1}{864}
\sqrt{\frac{7}{10}} C_{31,22}^{31,0}-\frac{1}{432} \sqrt{\frac{7}{10}} C_{31,22}^{31,1}+\frac{1}{160} \sqrt{\frac{3}{2}} C_{31,22}^{33,0}+\frac{1}{80}
\sqrt{\frac{3}{2}} C_{31,22}^{33,1}-\frac{C_{33,00}^{33,0}}{160 \sqrt{2}}-\frac{C_{33,00}^{33,1}}{80 \sqrt{2}}-\frac{C_{33,20}^{33,0}}{160 \sqrt{6}}-\frac{C_{33,20}^{33,1}}{80
\sqrt{6}}-\frac{11 C_{33,22}^{33,0}}{320 \sqrt{5}}-\frac{11 C_{33,22}^{33,1}}{160 \sqrt{5}}\)

\noindent\(C_{11,3313,2}^{11,1112,1,\text{Loc}}= \frac{1}{48} \sqrt{\frac{7}{6}} C_{00,20}^{60,1}-\frac{1}{32} \sqrt{\frac{7}{30}}
C_{00,22}^{62,1}+\frac{1}{288} \sqrt{\frac{7}{30}} C_{11,22}^{51,1}-\frac{3 C_{11,22}^{53,1}}{160 \sqrt{2}}-\frac{1}{144} \sqrt{\frac{7}{6}} C_{20,20}^{40,1}-\frac{1}{288}
\sqrt{\frac{7}{30}} C_{20,22}^{42,1}+\frac{1}{144} \sqrt{\frac{7}{30}} C_{22,20}^{42,1}+\frac{11}{288} \sqrt{\frac{7}{30}} C_{22,22}^{40,1}-\frac{1}{72}
\sqrt{\frac{5}{6}} C_{22,22}^{42,1}+\frac{1}{160} \sqrt{\frac{3}{14}} C_{22,22}^{44,1}-\frac{1}{144} \sqrt{\frac{7}{6}} C_{31,20}^{31,1}-\frac{1}{288}
\sqrt{\frac{7}{30}} C_{31,22}^{31,1}+\frac{3 C_{31,22}^{33,1}}{160 \sqrt{2}}+\frac{C_{33,20}^{33,1}}{80 \sqrt{2}}-\frac{11}{320} \sqrt{\frac{3}{5}}
C_{33,22}^{33,1}\)

\noindent\(C_{11,3313,2}^{11,1112,1,\text{NonLoc}}= \frac{1}{96} \sqrt{\frac{7}{2}} C_{00,00}^{60,0}+\frac{1}{96} \sqrt{\frac{7}{6}}
C_{00,20}^{60,0}+\frac{1}{32} \sqrt{\frac{7}{30}} C_{00,22}^{62,0}-\frac{1}{288} \sqrt{\frac{7}{2}} C_{11,00}^{51,0}+\frac{1}{288} \sqrt{\frac{7}{30}}
C_{11,22}^{51,0}-\frac{3 C_{11,22}^{53,0}}{160 \sqrt{2}}-\frac{1}{288} \sqrt{\frac{7}{2}} C_{20,00}^{40,0}-\frac{1}{288} \sqrt{\frac{7}{6}} C_{20,20}^{40,0}+\frac{1}{288}
\sqrt{\frac{7}{30}} C_{20,22}^{42,0}+\frac{1}{288} \sqrt{\frac{7}{10}} C_{22,00}^{42,0}+\frac{1}{288} \sqrt{\frac{7}{30}} C_{22,20}^{42,0}-\frac{11}{288}
\sqrt{\frac{7}{30}} C_{22,22}^{40,0}+\frac{1}{72} \sqrt{\frac{5}{6}} C_{22,22}^{42,0}-\frac{1}{160} \sqrt{\frac{3}{14}} C_{22,22}^{44,0}+\frac{1}{288}
\sqrt{\frac{7}{2}} C_{31,00}^{31,0}+\frac{1}{288} \sqrt{\frac{7}{6}} C_{31,20}^{31,0}-\frac{1}{288} \sqrt{\frac{7}{30}} C_{31,22}^{31,0}+\frac{3
C_{31,22}^{33,0}}{160 \sqrt{2}}-\frac{1}{160} \sqrt{\frac{3}{2}} C_{33,00}^{33,0}-\frac{C_{33,20}^{33,0}}{160 \sqrt{2}}-\frac{11}{320} \sqrt{\frac{3}{5}}
C_{33,22}^{33,0}\)

\noindent\(C_{11,3313,3}^{11,1112,0,\text{Loc}}= -\frac{\sqrt{7} C_{00,20}^{60,0}}{144}-\frac{\sqrt{7} C_{00,20}^{60,1}}{288}+\frac{1}{96}
\sqrt{\frac{7}{5}} C_{00,22}^{62,0}+\frac{1}{192} \sqrt{\frac{7}{5}} C_{00,22}^{62,1}-\frac{1}{864} \sqrt{\frac{7}{5}} C_{11,22}^{51,0}-\frac{\sqrt{\frac{7}{5}}
C_{11,22}^{51,1}}{1728}+\frac{\sqrt{3} C_{11,22}^{53,0}}{160}+\frac{\sqrt{3} C_{11,22}^{53,1}}{320}+\frac{\sqrt{7} C_{20,20}^{40,0}}{432}+\frac{\sqrt{7}
C_{20,20}^{40,1}}{864}+\frac{1}{864} \sqrt{\frac{7}{5}} C_{20,22}^{42,0}+\frac{\sqrt{\frac{7}{5}} C_{20,22}^{42,1}}{1728}-\frac{1}{432} \sqrt{\frac{7}{5}}
C_{22,20}^{42,0}-\frac{1}{864} \sqrt{\frac{7}{5}} C_{22,20}^{42,1}-\frac{11}{864} \sqrt{\frac{7}{5}} C_{22,22}^{40,0}-\frac{11 \sqrt{\frac{7}{5}}
C_{22,22}^{40,1}}{1728}+\frac{\sqrt{5} C_{22,22}^{42,0}}{216}+\frac{\sqrt{5} C_{22,22}^{42,1}}{432}-\frac{C_{22,22}^{44,0}}{160 \sqrt{7}}-\frac{C_{22,22}^{44,1}}{320
\sqrt{7}}+\frac{\sqrt{7} C_{31,20}^{31,0}}{432}+\frac{\sqrt{7} C_{31,20}^{31,1}}{864}+\frac{1}{864} \sqrt{\frac{7}{5}} C_{31,22}^{31,0}+\frac{\sqrt{\frac{7}{5}}
C_{31,22}^{31,1}}{1728}-\frac{\sqrt{3} C_{31,22}^{33,0}}{160}-\frac{\sqrt{3} C_{31,22}^{33,1}}{320}-\frac{C_{33,20}^{33,0}}{80 \sqrt{3}}-\frac{C_{33,20}^{33,1}}{160
\sqrt{3}}+\frac{11 C_{33,22}^{33,0}}{160 \sqrt{10}}+\frac{11 C_{33,22}^{33,1}}{320 \sqrt{10}}\)

\noindent\(C_{11,3313,3}^{11,1112,0,\text{NonLoc}}= -\frac{1}{192} \sqrt{\frac{7}{3}} C_{00,00}^{60,0}-\frac{1}{96} \sqrt{\frac{7}{3}}
C_{00,00}^{60,1}-\frac{\sqrt{7} C_{00,20}^{60,0}}{576}-\frac{\sqrt{7} C_{00,20}^{60,1}}{288}-\frac{1}{192} \sqrt{\frac{7}{5}} C_{00,22}^{62,0}-\frac{1}{96}
\sqrt{\frac{7}{5}} C_{00,22}^{62,1}+\frac{1}{576} \sqrt{\frac{7}{3}} C_{11,00}^{51,0}+\frac{1}{288} \sqrt{\frac{7}{3}} C_{11,00}^{51,1}-\frac{\sqrt{\frac{7}{5}}
C_{11,22}^{51,0}}{1728}-\frac{1}{864} \sqrt{\frac{7}{5}} C_{11,22}^{51,1}+\frac{\sqrt{3} C_{11,22}^{53,0}}{320}+\frac{\sqrt{3} C_{11,22}^{53,1}}{160}+\frac{1}{576}
\sqrt{\frac{7}{3}} C_{20,00}^{40,0}+\frac{1}{288} \sqrt{\frac{7}{3}} C_{20,00}^{40,1}+\frac{\sqrt{7} C_{20,20}^{40,0}}{1728}+\frac{\sqrt{7} C_{20,20}^{40,1}}{864}-\frac{\sqrt{\frac{7}{5}}
C_{20,22}^{42,0}}{1728}-\frac{1}{864} \sqrt{\frac{7}{5}} C_{20,22}^{42,1}-\frac{1}{576} \sqrt{\frac{7}{15}} C_{22,00}^{42,0}-\frac{1}{288} \sqrt{\frac{7}{15}}
C_{22,00}^{42,1}-\frac{\sqrt{\frac{7}{5}} C_{22,20}^{42,0}}{1728}-\frac{1}{864} \sqrt{\frac{7}{5}} C_{22,20}^{42,1}+\frac{11 \sqrt{\frac{7}{5}} C_{22,22}^{40,0}}{1728}+\frac{11}{864}
\sqrt{\frac{7}{5}} C_{22,22}^{40,1}-\frac{\sqrt{5} C_{22,22}^{42,0}}{432}-\frac{\sqrt{5} C_{22,22}^{42,1}}{216}+\frac{C_{22,22}^{44,0}}{320 \sqrt{7}}+\frac{C_{22,22}^{44,1}}{160
\sqrt{7}}-\frac{1}{576} \sqrt{\frac{7}{3}} C_{31,00}^{31,0}-\frac{1}{288} \sqrt{\frac{7}{3}} C_{31,00}^{31,1}-\frac{\sqrt{7} C_{31,20}^{31,0}}{1728}-\frac{\sqrt{7}
C_{31,20}^{31,1}}{864}+\frac{\sqrt{\frac{7}{5}} C_{31,22}^{31,0}}{1728}+\frac{1}{864} \sqrt{\frac{7}{5}} C_{31,22}^{31,1}-\frac{\sqrt{3} C_{31,22}^{33,0}}{320}-\frac{\sqrt{3}
C_{31,22}^{33,1}}{160}+\frac{C_{33,00}^{33,0}}{320}+\frac{C_{33,00}^{33,1}}{160}+\frac{C_{33,20}^{33,0}}{320 \sqrt{3}}+\frac{C_{33,20}^{33,1}}{160
\sqrt{3}}+\frac{11 C_{33,22}^{33,0}}{320 \sqrt{10}}+\frac{11 C_{33,22}^{33,1}}{160 \sqrt{10}}\)

\noindent\(C_{11,3313,3}^{11,1112,1,\text{Loc}}= -\frac{1}{96} \sqrt{\frac{7}{3}} C_{00,20}^{60,1}+\frac{1}{64} \sqrt{\frac{7}{15}}
C_{00,22}^{62,1}-\frac{1}{576} \sqrt{\frac{7}{15}} C_{11,22}^{51,1}+\frac{3 C_{11,22}^{53,1}}{320}+\frac{1}{288} \sqrt{\frac{7}{3}} C_{20,20}^{40,1}+\frac{1}{576}
\sqrt{\frac{7}{15}} C_{20,22}^{42,1}-\frac{1}{288} \sqrt{\frac{7}{15}} C_{22,20}^{42,1}-\frac{11}{576} \sqrt{\frac{7}{15}} C_{22,22}^{40,1}+\frac{1}{144}
\sqrt{\frac{5}{3}} C_{22,22}^{42,1}-\frac{1}{320} \sqrt{\frac{3}{7}} C_{22,22}^{44,1}+\frac{1}{288} \sqrt{\frac{7}{3}} C_{31,20}^{31,1}+\frac{1}{576}
\sqrt{\frac{7}{15}} C_{31,22}^{31,1}-\frac{3 C_{31,22}^{33,1}}{320}-\frac{C_{33,20}^{33,1}}{160}+\frac{11}{320} \sqrt{\frac{3}{10}} C_{33,22}^{33,1}\)

\noindent\(C_{11,3313,3}^{11,1112,1,\text{NonLoc}}= -\frac{\sqrt{7} C_{00,00}^{60,0}}{192}-\frac{1}{192} \sqrt{\frac{7}{3}} C_{00,20}^{60,0}-\frac{1}{64}
\sqrt{\frac{7}{15}} C_{00,22}^{62,0}+\frac{\sqrt{7} C_{11,00}^{51,0}}{576}-\frac{1}{576} \sqrt{\frac{7}{15}} C_{11,22}^{51,0}+\frac{3 C_{11,22}^{53,0}}{320}+\frac{\sqrt{7}
C_{20,00}^{40,0}}{576}+\frac{1}{576} \sqrt{\frac{7}{3}} C_{20,20}^{40,0}-\frac{1}{576} \sqrt{\frac{7}{15}} C_{20,22}^{42,0}-\frac{1}{576} \sqrt{\frac{7}{5}}
C_{22,00}^{42,0}-\frac{1}{576} \sqrt{\frac{7}{15}} C_{22,20}^{42,0}+\frac{11}{576} \sqrt{\frac{7}{15}} C_{22,22}^{40,0}-\frac{1}{144} \sqrt{\frac{5}{3}}
C_{22,22}^{42,0}+\frac{1}{320} \sqrt{\frac{3}{7}} C_{22,22}^{44,0}-\frac{\sqrt{7} C_{31,00}^{31,0}}{576}-\frac{1}{576} \sqrt{\frac{7}{3}} C_{31,20}^{31,0}+\frac{1}{576}
\sqrt{\frac{7}{15}} C_{31,22}^{31,0}-\frac{3 C_{31,22}^{33,0}}{320}+\frac{\sqrt{3} C_{33,00}^{33,0}}{320}+\frac{C_{33,20}^{33,0}}{320}+\frac{11}{320}
\sqrt{\frac{3}{10}} C_{33,22}^{33,0}\)

\noindent\(C_{11,3314,3}^{11,1112,0,\text{Loc}}= \frac{C_{00,20}^{60,0}}{16}+\frac{C_{00,20}^{60,1}}{32}+\frac{C_{00,22}^{62,0}}{32
\sqrt{5}}+\frac{C_{00,22}^{62,1}}{64 \sqrt{5}}+\frac{C_{11,22}^{51,0}}{96 \sqrt{5}}+\frac{C_{11,22}^{51,1}}{192 \sqrt{5}}+\frac{1}{160} \sqrt{\frac{3}{7}}
C_{11,22}^{53,0}+\frac{1}{320} \sqrt{\frac{3}{7}} C_{11,22}^{53,1}-\frac{C_{20,20}^{40,0}}{48}-\frac{C_{20,20}^{40,1}}{96}-\frac{C_{20,22}^{42,0}}{96
\sqrt{5}}-\frac{C_{20,22}^{42,1}}{192 \sqrt{5}}+\frac{C_{22,20}^{42,0}}{48 \sqrt{5}}+\frac{C_{22,20}^{42,1}}{96 \sqrt{5}}-\frac{C_{22,22}^{40,0}}{96
\sqrt{5}}-\frac{C_{22,22}^{40,1}}{192 \sqrt{5}}+\frac{C_{22,22}^{42,0}}{24 \sqrt{35}}+\frac{C_{22,22}^{42,1}}{48 \sqrt{35}}+\frac{C_{22,22}^{44,0}}{1120}+\frac{C_{22,22}^{44,1}}{2240}-\frac{C_{31,20}^{31,0}}{48}-\frac{C_{31,20}^{31,1}}{96}-\frac{C_{31,22}^{31,0}}{96
\sqrt{5}}-\frac{C_{31,22}^{31,1}}{192 \sqrt{5}}-\frac{1}{160} \sqrt{\frac{3}{7}} C_{31,22}^{33,0}-\frac{1}{320} \sqrt{\frac{3}{7}} C_{31,22}^{33,1}+\frac{3}{80}
\sqrt{\frac{3}{7}} C_{33,20}^{33,0}+\frac{3}{160} \sqrt{\frac{3}{7}} C_{33,20}^{33,1}+\frac{3}{160} \sqrt{\frac{7}{10}} C_{33,22}^{33,0}+\frac{3}{320}
\sqrt{\frac{7}{10}} C_{33,22}^{33,1}\)

\noindent\(C_{11,3314,3}^{11,1112,0,\text{NonLoc}}= \frac{\sqrt{3} C_{00,00}^{60,0}}{64}+\frac{\sqrt{3} C_{00,00}^{60,1}}{32}+\frac{C_{00,20}^{60,0}}{64}+\frac{C_{00,20}^{60,1}}{32}-\frac{C_{00,22}^{62,0}}{64
\sqrt{5}}-\frac{C_{00,22}^{62,1}}{32 \sqrt{5}}-\frac{C_{11,00}^{51,0}}{64 \sqrt{3}}-\frac{C_{11,00}^{51,1}}{32 \sqrt{3}}+\frac{C_{11,22}^{51,0}}{192
\sqrt{5}}+\frac{C_{11,22}^{51,1}}{96 \sqrt{5}}+\frac{1}{320} \sqrt{\frac{3}{7}} C_{11,22}^{53,0}+\frac{1}{160} \sqrt{\frac{3}{7}} C_{11,22}^{53,1}-\frac{C_{20,00}^{40,0}}{64
\sqrt{3}}-\frac{C_{20,00}^{40,1}}{32 \sqrt{3}}-\frac{C_{20,20}^{40,0}}{192}-\frac{C_{20,20}^{40,1}}{96}+\frac{C_{20,22}^{42,0}}{192 \sqrt{5}}+\frac{C_{20,22}^{42,1}}{96
\sqrt{5}}+\frac{C_{22,00}^{42,0}}{64 \sqrt{15}}+\frac{C_{22,00}^{42,1}}{32 \sqrt{15}}+\frac{C_{22,20}^{42,0}}{192 \sqrt{5}}+\frac{C_{22,20}^{42,1}}{96
\sqrt{5}}+\frac{C_{22,22}^{40,0}}{192 \sqrt{5}}+\frac{C_{22,22}^{40,1}}{96 \sqrt{5}}-\frac{C_{22,22}^{42,0}}{48 \sqrt{35}}-\frac{C_{22,22}^{42,1}}{24
\sqrt{35}}-\frac{C_{22,22}^{44,0}}{2240}-\frac{C_{22,22}^{44,1}}{1120}+\frac{C_{31,00}^{31,0}}{64 \sqrt{3}}+\frac{C_{31,00}^{31,1}}{32 \sqrt{3}}+\frac{C_{31,20}^{31,0}}{192}+\frac{C_{31,20}^{31,1}}{96}-\frac{C_{31,22}^{31,0}}{192
\sqrt{5}}-\frac{C_{31,22}^{31,1}}{96 \sqrt{5}}-\frac{1}{320} \sqrt{\frac{3}{7}} C_{31,22}^{33,0}-\frac{1}{160} \sqrt{\frac{3}{7}} C_{31,22}^{33,1}-\frac{9
C_{33,00}^{33,0}}{320 \sqrt{7}}-\frac{9 C_{33,00}^{33,1}}{160 \sqrt{7}}-\frac{3}{320} \sqrt{\frac{3}{7}} C_{33,20}^{33,0}-\frac{3}{160} \sqrt{\frac{3}{7}}
C_{33,20}^{33,1}+\frac{3}{320} \sqrt{\frac{7}{10}} C_{33,22}^{33,0}+\frac{3}{160} \sqrt{\frac{7}{10}} C_{33,22}^{33,1}\)

\noindent\(C_{11,3314,3}^{11,1112,1,\text{Loc}}= \frac{\sqrt{3} C_{00,20}^{60,1}}{32}+\frac{1}{64} \sqrt{\frac{3}{5}} C_{00,22}^{62,1}+\frac{C_{11,22}^{51,1}}{64
\sqrt{15}}+\frac{3 C_{11,22}^{53,1}}{320 \sqrt{7}}-\frac{C_{20,20}^{40,1}}{32 \sqrt{3}}-\frac{C_{20,22}^{42,1}}{64 \sqrt{15}}+\frac{C_{22,20}^{42,1}}{32
\sqrt{15}}-\frac{C_{22,22}^{40,1}}{64 \sqrt{15}}+\frac{C_{22,22}^{42,1}}{16 \sqrt{105}}+\frac{\sqrt{3} C_{22,22}^{44,1}}{2240}-\frac{C_{31,20}^{31,1}}{32
\sqrt{3}}-\frac{C_{31,22}^{31,1}}{64 \sqrt{15}}-\frac{3 C_{31,22}^{33,1}}{320 \sqrt{7}}+\frac{9 C_{33,20}^{33,1}}{160 \sqrt{7}}+\frac{3}{320} \sqrt{\frac{21}{10}}
C_{33,22}^{33,1}\)

\noindent\(C_{11,3314,3}^{11,1112,1,\text{NonLoc}}= \frac{3 C_{00,00}^{60,0}}{64}+\frac{\sqrt{3} C_{00,20}^{60,0}}{64}-\frac{1}{64}
\sqrt{\frac{3}{5}} C_{00,22}^{62,0}-\frac{C_{11,00}^{51,0}}{64}+\frac{C_{11,22}^{51,0}}{64 \sqrt{15}}+\frac{3 C_{11,22}^{53,0}}{320 \sqrt{7}}-\frac{C_{20,00}^{40,0}}{64}-\frac{C_{20,20}^{40,0}}{64
\sqrt{3}}+\frac{C_{20,22}^{42,0}}{64 \sqrt{15}}+\frac{C_{22,00}^{42,0}}{64 \sqrt{5}}+\frac{C_{22,20}^{42,0}}{64 \sqrt{15}}+\frac{C_{22,22}^{40,0}}{64
\sqrt{15}}-\frac{C_{22,22}^{42,0}}{16 \sqrt{105}}-\frac{\sqrt{3} C_{22,22}^{44,0}}{2240}+\frac{C_{31,00}^{31,0}}{64}+\frac{C_{31,20}^{31,0}}{64 \sqrt{3}}-\frac{C_{31,22}^{31,0}}{64
\sqrt{15}}-\frac{3 C_{31,22}^{33,0}}{320 \sqrt{7}}-\frac{9}{320} \sqrt{\frac{3}{7}} C_{33,00}^{33,0}-\frac{9 C_{33,20}^{33,0}}{320 \sqrt{7}}+\frac{3}{320}
\sqrt{\frac{21}{10}} C_{33,22}^{33,0}\)

\noindent\(C_{11,4000,1}^{00,1111,0,\text{Loc}}= \frac{7 C_{11,11}^{51,0}}{192}+\frac{7 C_{11,11}^{51,1}}{384}+\frac{7 C_{22,11}^{42,0}}{32
\sqrt{15}}+\frac{7 C_{22,11}^{42,1}}{64 \sqrt{15}}+\frac{7 C_{31,11}^{31,0}}{192}+\frac{7 C_{31,11}^{31,1}}{384}+\frac{3}{80} \sqrt{\frac{7}{2}}
C_{33,11}^{33,0}+\frac{3}{160} \sqrt{\frac{7}{2}} C_{33,11}^{33,1}\)

\noindent\(C_{11,4000,1}^{00,1111,0,\text{NonLoc}}= \frac{7 C_{11,11}^{51,0}}{384}+\frac{7 C_{11,11}^{51,1}}{192}-\frac{7 C_{22,11}^{42,0}}{64
\sqrt{15}}-\frac{7 C_{22,11}^{42,1}}{32 \sqrt{15}}+\frac{7 C_{31,11}^{31,0}}{384}+\frac{7 C_{31,11}^{31,1}}{192}+\frac{3}{160} \sqrt{\frac{7}{2}}
C_{33,11}^{33,0}+\frac{3}{80} \sqrt{\frac{7}{2}} C_{33,11}^{33,1}\)

\noindent\(C_{11,4000,1}^{00,1111,1,\text{Loc}}= \frac{7 C_{11,11}^{51,1}}{128 \sqrt{3}}+\frac{7 C_{22,11}^{42,1}}{64 \sqrt{5}}+\frac{7
C_{31,11}^{31,1}}{128 \sqrt{3}}+\frac{3}{160} \sqrt{\frac{21}{2}} C_{33,11}^{33,1}\)

\noindent\(C_{11,4000,1}^{00,1111,1,\text{NonLoc}}= \frac{7 C_{11,11}^{51,0}}{128 \sqrt{3}}-\frac{7 C_{22,11}^{42,0}}{64 \sqrt{5}}+\frac{7
C_{31,11}^{31,0}}{128 \sqrt{3}}+\frac{3}{160} \sqrt{\frac{21}{2}} C_{33,11}^{33,0}\)

\noindent\(C_{11,4000,1}^{11,0000,0,\text{Loc}}= \frac{7 C_{00,00}^{60,0}}{128}+\frac{7 C_{00,00}^{60,1}}{256}+\frac{7 C_{11,00}^{51,0}}{384}+\frac{7
C_{11,00}^{51,1}}{768}+\frac{5 C_{20,00}^{40,0}}{384}+\frac{5 C_{20,00}^{40,1}}{768}-\frac{7 C_{22,00}^{42,0}}{192 \sqrt{5}}-\frac{7 C_{22,00}^{42,1}}{384
\sqrt{5}}-\frac{C_{31,00}^{31,0}}{384}-\frac{C_{31,00}^{31,1}}{768}-\frac{3 \sqrt{21} C_{33,00}^{33,0}}{320}-\frac{3 \sqrt{21} C_{33,00}^{33,1}}{640}\)

\noindent\(C_{11,4000,1}^{11,0000,0,\text{NonLoc}}= -\frac{7 C_{00,00}^{60,0}}{512}-\frac{7 C_{00,00}^{60,1}}{256}+\frac{7 \sqrt{3}
C_{00,20}^{60,0}}{512}+\frac{7 \sqrt{3} C_{00,20}^{60,1}}{256}+\frac{7 C_{11,00}^{51,0}}{1536}+\frac{7 C_{11,00}^{51,1}}{768}-\frac{7 C_{11,20}^{31,0}}{512
\sqrt{3}}-\frac{7 C_{11,20}^{31,1}}{256 \sqrt{3}}-\frac{5 C_{20,00}^{40,0}}{1536}-\frac{5 C_{20,00}^{40,1}}{768}+\frac{5 C_{20,20}^{40,0}}{512 \sqrt{3}}+\frac{5
C_{20,20}^{40,1}}{256 \sqrt{3}}+\frac{7 C_{22,00}^{42,0}}{768 \sqrt{5}}+\frac{7 C_{22,00}^{42,1}}{384 \sqrt{5}}-\frac{7 C_{22,20}^{42,0}}{256 \sqrt{15}}-\frac{7
C_{22,20}^{42,1}}{128 \sqrt{15}}-\frac{C_{31,00}^{31,0}}{1536}-\frac{C_{31,00}^{31,1}}{768}+\frac{C_{31,20}^{31,0}}{512 \sqrt{3}}+\frac{C_{31,20}^{31,1}}{256
\sqrt{3}}-\frac{3 \sqrt{21} C_{33,00}^{33,0}}{1280}-\frac{3 \sqrt{21} C_{33,00}^{33,1}}{640}+\frac{9 \sqrt{7} C_{33,20}^{33,0}}{1280}+\frac{9 \sqrt{7}
C_{33,20}^{33,1}}{640}\)

\noindent\(C_{11,4000,1}^{11,0000,1,\text{Loc}}= \frac{7 \sqrt{3} C_{00,00}^{60,1}}{256}+\frac{7 C_{11,00}^{51,1}}{256 \sqrt{3}}+\frac{5
C_{20,00}^{40,1}}{256 \sqrt{3}}-\frac{7 C_{22,00}^{42,1}}{128 \sqrt{15}}-\frac{C_{31,00}^{31,1}}{256 \sqrt{3}}-\frac{9 \sqrt{7} C_{33,00}^{33,1}}{640}\)

\noindent\(C_{11,4000,1}^{11,0000,1,\text{NonLoc}}= -\frac{7 \sqrt{3} C_{00,00}^{60,0}}{512}+\frac{21 C_{00,20}^{60,0}}{512}+\frac{7
C_{11,00}^{51,0}}{512 \sqrt{3}}-\frac{7 C_{11,20}^{31,0}}{512}-\frac{5 C_{20,00}^{40,0}}{512 \sqrt{3}}+\frac{5 C_{20,20}^{40,0}}{512}+\frac{7 C_{22,00}^{42,0}}{256
\sqrt{15}}-\frac{7 C_{22,20}^{42,0}}{256 \sqrt{5}}-\frac{C_{31,00}^{31,0}}{512 \sqrt{3}}+\frac{C_{31,20}^{31,0}}{512}-\frac{9 \sqrt{7} C_{33,00}^{33,0}}{1280}+\frac{9
\sqrt{21} C_{33,20}^{33,0}}{1280}\)

\noindent\(C_{11,4011,0}^{11,0011,0,\text{Loc}}= -\frac{7 C_{00,20}^{60,0}}{384}-\frac{7 C_{00,20}^{60,1}}{768}-\frac{7 C_{00,22}^{62,0}}{128
\sqrt{5}}-\frac{7 C_{00,22}^{62,1}}{256 \sqrt{5}}-\frac{7 C_{11,20}^{31,0}}{1152}-\frac{7 C_{11,20}^{31,1}}{2304}+\frac{7 C_{11,22}^{51,0}}{1152
\sqrt{5}}+\frac{7 C_{11,22}^{51,1}}{2304 \sqrt{5}}-\frac{3 \sqrt{21} C_{11,22}^{53,0}}{640}-\frac{3 \sqrt{21} C_{11,22}^{53,1}}{1280}-\frac{5 C_{20,20}^{40,0}}{1152}-\frac{5
C_{20,20}^{40,1}}{2304}-\frac{7 C_{20,22}^{42,0}}{1152 \sqrt{5}}-\frac{7 C_{20,22}^{42,1}}{2304 \sqrt{5}}+\frac{7 C_{22,20}^{42,0}}{576 \sqrt{5}}+\frac{7
C_{22,20}^{42,1}}{1152 \sqrt{5}}+\frac{17 C_{22,22}^{40,0}}{1152 \sqrt{5}}+\frac{17 C_{22,22}^{40,1}}{2304 \sqrt{5}}+\frac{\sqrt{35} C_{22,22}^{42,0}}{576}+\frac{\sqrt{35}
C_{22,22}^{42,1}}{1152}-\frac{3 C_{22,22}^{44,0}}{160}-\frac{3 C_{22,22}^{44,1}}{320}+\frac{C_{31,20}^{31,0}}{1152}+\frac{C_{31,20}^{31,1}}{2304}+\frac{23
C_{31,22}^{31,0}}{1152 \sqrt{5}}+\frac{23 C_{31,22}^{31,1}}{2304 \sqrt{5}}-\frac{1}{640} \sqrt{\frac{7}{3}} C_{31,22}^{33,0}-\frac{\sqrt{\frac{7}{3}}
C_{31,22}^{33,1}}{1280}+\frac{\sqrt{21} C_{33,20}^{33,0}}{320}+\frac{\sqrt{21} C_{33,20}^{33,1}}{640}+\frac{3}{160} \sqrt{\frac{7}{10}} C_{33,22}^{33,0}+\frac{3}{320}
\sqrt{\frac{7}{10}} C_{33,22}^{33,1}\)

\noindent\(C_{11,4011,0}^{11,0011,0,\text{NonLoc}}= -\frac{7 C_{00,00}^{60,0}}{512 \sqrt{3}}-\frac{7 C_{00,00}^{60,1}}{256 \sqrt{3}}-\frac{7
C_{00,20}^{60,0}}{1536}-\frac{7 C_{00,20}^{60,1}}{768}+\frac{7 C_{00,22}^{62,0}}{256 \sqrt{5}}+\frac{7 C_{00,22}^{62,1}}{128 \sqrt{5}}+\frac{7 C_{11,00}^{51,0}}{1536
\sqrt{3}}+\frac{7 C_{11,00}^{51,1}}{768 \sqrt{3}}+\frac{7 C_{11,20}^{31,0}}{4608}+\frac{7 C_{11,20}^{31,1}}{2304}+\frac{7 C_{11,22}^{51,0}}{2304
\sqrt{5}}+\frac{7 C_{11,22}^{51,1}}{1152 \sqrt{5}}-\frac{3 \sqrt{21} C_{11,22}^{53,0}}{1280}-\frac{3 \sqrt{21} C_{11,22}^{53,1}}{640}-\frac{5 C_{20,00}^{40,0}}{1536
\sqrt{3}}-\frac{5 C_{20,00}^{40,1}}{768 \sqrt{3}}-\frac{5 C_{20,20}^{40,0}}{4608}-\frac{5 C_{20,20}^{40,1}}{2304}+\frac{7 C_{20,22}^{42,0}}{2304
\sqrt{5}}+\frac{7 C_{20,22}^{42,1}}{1152 \sqrt{5}}+\frac{7 C_{22,00}^{42,0}}{768 \sqrt{15}}+\frac{7 C_{22,00}^{42,1}}{384 \sqrt{15}}+\frac{7 C_{22,20}^{42,0}}{2304
\sqrt{5}}+\frac{7 C_{22,20}^{42,1}}{1152 \sqrt{5}}-\frac{17 C_{22,22}^{40,0}}{2304 \sqrt{5}}-\frac{17 C_{22,22}^{40,1}}{1152 \sqrt{5}}-\frac{\sqrt{35}
C_{22,22}^{42,0}}{1152}-\frac{\sqrt{35} C_{22,22}^{42,1}}{576}+\frac{3 C_{22,22}^{44,0}}{320}+\frac{3 C_{22,22}^{44,1}}{160}-\frac{C_{31,00}^{31,0}}{1536
\sqrt{3}}-\frac{C_{31,00}^{31,1}}{768 \sqrt{3}}-\frac{C_{31,20}^{31,0}}{4608}-\frac{C_{31,20}^{31,1}}{2304}+\frac{23 C_{31,22}^{31,0}}{2304 \sqrt{5}}+\frac{23
C_{31,22}^{31,1}}{1152 \sqrt{5}}-\frac{\sqrt{\frac{7}{3}} C_{31,22}^{33,0}}{1280}-\frac{1}{640} \sqrt{\frac{7}{3}} C_{31,22}^{33,1}-\frac{3 \sqrt{7}
C_{33,00}^{33,0}}{1280}-\frac{3 \sqrt{7} C_{33,00}^{33,1}}{640}-\frac{\sqrt{21} C_{33,20}^{33,0}}{1280}-\frac{\sqrt{21} C_{33,20}^{33,1}}{640}+\frac{3}{320}
\sqrt{\frac{7}{10}} C_{33,22}^{33,0}+\frac{3}{160} \sqrt{\frac{7}{10}} C_{33,22}^{33,1}\)

\noindent\(C_{11,4011,0}^{11,0011,1,\text{Loc}}= -\frac{7 C_{00,20}^{60,1}}{256 \sqrt{3}}-\frac{7}{256} \sqrt{\frac{3}{5}} C_{00,22}^{62,1}-\frac{7
C_{11,20}^{31,1}}{768 \sqrt{3}}+\frac{7 C_{11,22}^{51,1}}{768 \sqrt{15}}-\frac{9 \sqrt{7} C_{11,22}^{53,1}}{1280}-\frac{5 C_{20,20}^{40,1}}{768 \sqrt{3}}-\frac{7
C_{20,22}^{42,1}}{768 \sqrt{15}}+\frac{7 C_{22,20}^{42,1}}{384 \sqrt{15}}+\frac{17 C_{22,22}^{40,1}}{768 \sqrt{15}}+\frac{1}{384} \sqrt{\frac{35}{3}}
C_{22,22}^{42,1}-\frac{3 \sqrt{3} C_{22,22}^{44,1}}{320}+\frac{C_{31,20}^{31,1}}{768 \sqrt{3}}+\frac{23 C_{31,22}^{31,1}}{768 \sqrt{15}}-\frac{\sqrt{7}
C_{31,22}^{33,1}}{1280}+\frac{3 \sqrt{7} C_{33,20}^{33,1}}{640}+\frac{3}{320} \sqrt{\frac{21}{10}} C_{33,22}^{33,1}\)

\noindent\(C_{11,4011,0}^{11,0011,1,\text{NonLoc}}= -\frac{7 C_{00,00}^{60,0}}{512}-\frac{7 C_{00,20}^{60,0}}{512 \sqrt{3}}+\frac{7}{256}
\sqrt{\frac{3}{5}} C_{00,22}^{62,0}+\frac{7 C_{11,00}^{51,0}}{1536}+\frac{7 C_{11,20}^{31,0}}{1536 \sqrt{3}}+\frac{7 C_{11,22}^{51,0}}{768 \sqrt{15}}-\frac{9
\sqrt{7} C_{11,22}^{53,0}}{1280}-\frac{5 C_{20,00}^{40,0}}{1536}-\frac{5 C_{20,20}^{40,0}}{1536 \sqrt{3}}+\frac{7 C_{20,22}^{42,0}}{768 \sqrt{15}}+\frac{7
C_{22,00}^{42,0}}{768 \sqrt{5}}+\frac{7 C_{22,20}^{42,0}}{768 \sqrt{15}}-\frac{17 C_{22,22}^{40,0}}{768 \sqrt{15}}-\frac{1}{384} \sqrt{\frac{35}{3}}
C_{22,22}^{42,0}+\frac{3 \sqrt{3} C_{22,22}^{44,0}}{320}-\frac{C_{31,00}^{31,0}}{1536}-\frac{C_{31,20}^{31,0}}{1536 \sqrt{3}}+\frac{23 C_{31,22}^{31,0}}{768
\sqrt{15}}-\frac{\sqrt{7} C_{31,22}^{33,0}}{1280}-\frac{3 \sqrt{21} C_{33,00}^{33,0}}{1280}-\frac{3 \sqrt{7} C_{33,20}^{33,0}}{1280}+\frac{3}{320}
\sqrt{\frac{21}{10}} C_{33,22}^{33,0}\)

\noindent\(C_{11,4011,1}^{00,1101,0,\text{Loc}}= -\frac{7 C_{11,11}^{51,0}}{192}-\frac{7 C_{11,11}^{51,1}}{384}-\frac{7 C_{22,11}^{42,0}}{32
\sqrt{15}}-\frac{7 C_{22,11}^{42,1}}{64 \sqrt{15}}-\frac{7 C_{31,11}^{31,0}}{192}-\frac{7 C_{31,11}^{31,1}}{384}-\frac{3}{80} \sqrt{\frac{7}{2}}
C_{33,11}^{33,0}-\frac{3}{160} \sqrt{\frac{7}{2}} C_{33,11}^{33,1}\)

\noindent\(C_{11,4011,1}^{00,1101,0,\text{NonLoc}}= -\frac{7 C_{11,11}^{51,0}}{384}-\frac{7 C_{11,11}^{51,1}}{192}+\frac{7 C_{22,11}^{42,0}}{64
\sqrt{15}}+\frac{7 C_{22,11}^{42,1}}{32 \sqrt{15}}-\frac{7 C_{31,11}^{31,0}}{384}-\frac{7 C_{31,11}^{31,1}}{192}-\frac{3}{160} \sqrt{\frac{7}{2}}
C_{33,11}^{33,0}-\frac{3}{80} \sqrt{\frac{7}{2}} C_{33,11}^{33,1}\)

\noindent\(C_{11,4011,1}^{00,1101,1,\text{Loc}}= -\frac{7 C_{11,11}^{51,1}}{128 \sqrt{3}}-\frac{7 C_{22,11}^{42,1}}{64 \sqrt{5}}-\frac{7
C_{31,11}^{31,1}}{128 \sqrt{3}}-\frac{3}{160} \sqrt{\frac{21}{2}} C_{33,11}^{33,1}\)

\noindent\(C_{11,4011,1}^{00,1101,1,\text{NonLoc}}= -\frac{7 C_{11,11}^{51,0}}{128 \sqrt{3}}+\frac{7 C_{22,11}^{42,0}}{64 \sqrt{5}}-\frac{7
C_{31,11}^{31,0}}{128 \sqrt{3}}-\frac{3}{160} \sqrt{\frac{21}{2}} C_{33,11}^{33,0}\)

\noindent\(C_{11,4011,1}^{11,0011,0,\text{Loc}}= -\frac{7 C_{00,20}^{60,0}}{128 \sqrt{3}}-\frac{7 C_{00,20}^{60,1}}{256 \sqrt{3}}+\frac{7}{256}
\sqrt{\frac{3}{5}} C_{00,22}^{62,0}+\frac{7}{512} \sqrt{\frac{3}{5}} C_{00,22}^{62,1}-\frac{7 C_{11,20}^{31,0}}{384 \sqrt{3}}-\frac{7 C_{11,20}^{31,1}}{768
\sqrt{3}}-\frac{7 C_{11,22}^{51,0}}{768 \sqrt{15}}-\frac{7 C_{11,22}^{51,1}}{1536 \sqrt{15}}+\frac{9 \sqrt{7} C_{11,22}^{53,0}}{1280}+\frac{9 \sqrt{7}
C_{11,22}^{53,1}}{2560}-\frac{5 C_{20,20}^{40,0}}{384 \sqrt{3}}-\frac{5 C_{20,20}^{40,1}}{768 \sqrt{3}}+\frac{7 C_{20,22}^{42,0}}{768 \sqrt{15}}+\frac{7
C_{20,22}^{42,1}}{1536 \sqrt{15}}+\frac{7 C_{22,20}^{42,0}}{192 \sqrt{15}}+\frac{7 C_{22,20}^{42,1}}{384 \sqrt{15}}-\frac{17 C_{22,22}^{40,0}}{768
\sqrt{15}}-\frac{17 C_{22,22}^{40,1}}{1536 \sqrt{15}}-\frac{1}{384} \sqrt{\frac{35}{3}} C_{22,22}^{42,0}-\frac{1}{768} \sqrt{\frac{35}{3}} C_{22,22}^{42,1}+\frac{3
\sqrt{3} C_{22,22}^{44,0}}{320}+\frac{3 \sqrt{3} C_{22,22}^{44,1}}{640}+\frac{C_{31,20}^{31,0}}{384 \sqrt{3}}+\frac{C_{31,20}^{31,1}}{768 \sqrt{3}}-\frac{23
C_{31,22}^{31,0}}{768 \sqrt{15}}-\frac{23 C_{31,22}^{31,1}}{1536 \sqrt{15}}+\frac{\sqrt{7} C_{31,22}^{33,0}}{1280}+\frac{\sqrt{7} C_{31,22}^{33,1}}{2560}+\frac{3
\sqrt{7} C_{33,20}^{33,0}}{320}+\frac{3 \sqrt{7} C_{33,20}^{33,1}}{640}-\frac{3}{320} \sqrt{\frac{21}{10}} C_{33,22}^{33,0}-\frac{3}{640} \sqrt{\frac{21}{10}}
C_{33,22}^{33,1}\)

\noindent\(C_{11,4011,1}^{11,0011,0,\text{NonLoc}}= -\frac{7 C_{00,00}^{60,0}}{512}-\frac{7 C_{00,00}^{60,1}}{256}-\frac{7 C_{00,20}^{60,0}}{512
\sqrt{3}}-\frac{7 C_{00,20}^{60,1}}{256 \sqrt{3}}-\frac{7}{512} \sqrt{\frac{3}{5}} C_{00,22}^{62,0}-\frac{7}{256} \sqrt{\frac{3}{5}} C_{00,22}^{62,1}+\frac{7
C_{11,00}^{51,0}}{1536}+\frac{7 C_{11,00}^{51,1}}{768}+\frac{7 C_{11,20}^{31,0}}{1536 \sqrt{3}}+\frac{7 C_{11,20}^{31,1}}{768 \sqrt{3}}-\frac{7 C_{11,22}^{51,0}}{1536
\sqrt{15}}-\frac{7 C_{11,22}^{51,1}}{768 \sqrt{15}}+\frac{9 \sqrt{7} C_{11,22}^{53,0}}{2560}+\frac{9 \sqrt{7} C_{11,22}^{53,1}}{1280}-\frac{5 C_{20,00}^{40,0}}{1536}-\frac{5
C_{20,00}^{40,1}}{768}-\frac{5 C_{20,20}^{40,0}}{1536 \sqrt{3}}-\frac{5 C_{20,20}^{40,1}}{768 \sqrt{3}}-\frac{7 C_{20,22}^{42,0}}{1536 \sqrt{15}}-\frac{7
C_{20,22}^{42,1}}{768 \sqrt{15}}+\frac{7 C_{22,00}^{42,0}}{768 \sqrt{5}}+\frac{7 C_{22,00}^{42,1}}{384 \sqrt{5}}+\frac{7 C_{22,20}^{42,0}}{768 \sqrt{15}}+\frac{7
C_{22,20}^{42,1}}{384 \sqrt{15}}+\frac{17 C_{22,22}^{40,0}}{1536 \sqrt{15}}+\frac{17 C_{22,22}^{40,1}}{768 \sqrt{15}}+\frac{1}{768} \sqrt{\frac{35}{3}}
C_{22,22}^{42,0}+\frac{1}{384} \sqrt{\frac{35}{3}} C_{22,22}^{42,1}-\frac{3 \sqrt{3} C_{22,22}^{44,0}}{640}-\frac{3 \sqrt{3} C_{22,22}^{44,1}}{320}-\frac{C_{31,00}^{31,0}}{1536}-\frac{C_{31,00}^{31,1}}{768}-\frac{C_{31,20}^{31,0}}{1536
\sqrt{3}}-\frac{C_{31,20}^{31,1}}{768 \sqrt{3}}-\frac{23 C_{31,22}^{31,0}}{1536 \sqrt{15}}-\frac{23 C_{31,22}^{31,1}}{768 \sqrt{15}}+\frac{\sqrt{7}
C_{31,22}^{33,0}}{2560}+\frac{\sqrt{7} C_{31,22}^{33,1}}{1280}-\frac{3 \sqrt{21} C_{33,00}^{33,0}}{1280}-\frac{3 \sqrt{21} C_{33,00}^{33,1}}{640}-\frac{3
\sqrt{7} C_{33,20}^{33,0}}{1280}-\frac{3 \sqrt{7} C_{33,20}^{33,1}}{640}-\frac{3}{640} \sqrt{\frac{21}{10}} C_{33,22}^{33,0}-\frac{3}{320} \sqrt{\frac{21}{10}}
C_{33,22}^{33,1}\)

\noindent\(C_{11,4011,1}^{11,0011,1,\text{Loc}}= -\frac{7 C_{00,20}^{60,1}}{256}+\frac{21 C_{00,22}^{62,1}}{512 \sqrt{5}}-\frac{7 C_{11,20}^{31,1}}{768}-\frac{7
C_{11,22}^{51,1}}{1536 \sqrt{5}}+\frac{9 \sqrt{21} C_{11,22}^{53,1}}{2560}-\frac{5 C_{20,20}^{40,1}}{768}+\frac{7 C_{20,22}^{42,1}}{1536 \sqrt{5}}+\frac{7
C_{22,20}^{42,1}}{384 \sqrt{5}}-\frac{17 C_{22,22}^{40,1}}{1536 \sqrt{5}}-\frac{\sqrt{35} C_{22,22}^{42,1}}{768}+\frac{9 C_{22,22}^{44,1}}{640}+\frac{C_{31,20}^{31,1}}{768}-\frac{23
C_{31,22}^{31,1}}{1536 \sqrt{5}}+\frac{\sqrt{21} C_{31,22}^{33,1}}{2560}+\frac{3 \sqrt{21} C_{33,20}^{33,1}}{640}-\frac{9}{640} \sqrt{\frac{7}{10}}
C_{33,22}^{33,1}\)

\noindent\(C_{11,4011,1}^{11,0011,1,\text{NonLoc}}= -\frac{7 \sqrt{3} C_{00,00}^{60,0}}{512}-\frac{7 C_{00,20}^{60,0}}{512}-\frac{21
C_{00,22}^{62,0}}{512 \sqrt{5}}+\frac{7 C_{11,00}^{51,0}}{512 \sqrt{3}}+\frac{7 C_{11,20}^{31,0}}{1536}-\frac{7 C_{11,22}^{51,0}}{1536 \sqrt{5}}+\frac{9
\sqrt{21} C_{11,22}^{53,0}}{2560}-\frac{5 C_{20,00}^{40,0}}{512 \sqrt{3}}-\frac{5 C_{20,20}^{40,0}}{1536}-\frac{7 C_{20,22}^{42,0}}{1536 \sqrt{5}}+\frac{7
C_{22,00}^{42,0}}{256 \sqrt{15}}+\frac{7 C_{22,20}^{42,0}}{768 \sqrt{5}}+\frac{17 C_{22,22}^{40,0}}{1536 \sqrt{5}}+\frac{\sqrt{35} C_{22,22}^{42,0}}{768}-\frac{9
C_{22,22}^{44,0}}{640}-\frac{C_{31,00}^{31,0}}{512 \sqrt{3}}-\frac{C_{31,20}^{31,0}}{1536}-\frac{23 C_{31,22}^{31,0}}{1536 \sqrt{5}}+\frac{\sqrt{21}
C_{31,22}^{33,0}}{2560}-\frac{9 \sqrt{7} C_{33,00}^{33,0}}{1280}-\frac{3 \sqrt{21} C_{33,20}^{33,0}}{1280}-\frac{9}{640} \sqrt{\frac{7}{10}} C_{33,22}^{33,0}\)

\noindent\(C_{11,4011,2}^{11,0011,0,\text{Loc}}= -\frac{7 \sqrt{5} C_{00,20}^{60,0}}{384}-\frac{7 \sqrt{5} C_{00,20}^{60,1}}{768}-\frac{7
C_{00,22}^{62,0}}{1280}-\frac{7 C_{00,22}^{62,1}}{2560}-\frac{7 \sqrt{5} C_{11,20}^{31,0}}{1152}-\frac{7 \sqrt{5} C_{11,20}^{31,1}}{2304}+\frac{7
C_{11,22}^{51,0}}{11520}+\frac{7 C_{11,22}^{51,1}}{23040}-\frac{3 \sqrt{\frac{21}{5}} C_{11,22}^{53,0}}{1280}-\frac{3 \sqrt{\frac{21}{5}} C_{11,22}^{53,1}}{2560}-\frac{5
\sqrt{5} C_{20,20}^{40,0}}{1152}-\frac{5 \sqrt{5} C_{20,20}^{40,1}}{2304}-\frac{7 C_{20,22}^{42,0}}{11520}-\frac{7 C_{20,22}^{42,1}}{23040}+\frac{7
C_{22,20}^{42,0}}{576}+\frac{7 C_{22,20}^{42,1}}{1152}+\frac{17 C_{22,22}^{40,0}}{11520}+\frac{17 C_{22,22}^{40,1}}{23040}+\frac{\sqrt{7} C_{22,22}^{42,0}}{1152}+\frac{\sqrt{7}
C_{22,22}^{42,1}}{2304}-\frac{3 C_{22,22}^{44,0}}{320 \sqrt{5}}-\frac{3 C_{22,22}^{44,1}}{640 \sqrt{5}}+\frac{\sqrt{5} C_{31,20}^{31,0}}{1152}+\frac{\sqrt{5}
C_{31,20}^{31,1}}{2304}+\frac{23 C_{31,22}^{31,0}}{11520}+\frac{23 C_{31,22}^{31,1}}{23040}-\frac{\sqrt{\frac{7}{15}} C_{31,22}^{33,0}}{1280}-\frac{\sqrt{\frac{7}{15}}
C_{31,22}^{33,1}}{2560}+\frac{1}{64} \sqrt{\frac{21}{5}} C_{33,20}^{33,0}+\frac{1}{128} \sqrt{\frac{21}{5}} C_{33,20}^{33,1}+\frac{3 \sqrt{\frac{7}{2}}
C_{33,22}^{33,0}}{1600}+\frac{3 \sqrt{\frac{7}{2}} C_{33,22}^{33,1}}{3200}\)

\noindent\(C_{11,4011,2}^{11,0011,0,\text{NonLoc}}= -\frac{7}{512} \sqrt{\frac{5}{3}} C_{00,00}^{60,0}-\frac{7}{256} \sqrt{\frac{5}{3}}
C_{00,00}^{60,1}-\frac{7 \sqrt{5} C_{00,20}^{60,0}}{1536}-\frac{7 \sqrt{5} C_{00,20}^{60,1}}{768}+\frac{7 C_{00,22}^{62,0}}{2560}+\frac{7 C_{00,22}^{62,1}}{1280}+\frac{7
\sqrt{\frac{5}{3}} C_{11,00}^{51,0}}{1536}+\frac{7}{768} \sqrt{\frac{5}{3}} C_{11,00}^{51,1}+\frac{7 \sqrt{5} C_{11,20}^{31,0}}{4608}+\frac{7 \sqrt{5}
C_{11,20}^{31,1}}{2304}+\frac{7 C_{11,22}^{51,0}}{23040}+\frac{7 C_{11,22}^{51,1}}{11520}-\frac{3 \sqrt{\frac{21}{5}} C_{11,22}^{53,0}}{2560}-\frac{3
\sqrt{\frac{21}{5}} C_{11,22}^{53,1}}{1280}-\frac{5 \sqrt{\frac{5}{3}} C_{20,00}^{40,0}}{1536}-\frac{5}{768} \sqrt{\frac{5}{3}} C_{20,00}^{40,1}-\frac{5
\sqrt{5} C_{20,20}^{40,0}}{4608}-\frac{5 \sqrt{5} C_{20,20}^{40,1}}{2304}+\frac{7 C_{20,22}^{42,0}}{23040}+\frac{7 C_{20,22}^{42,1}}{11520}+\frac{7
C_{22,00}^{42,0}}{768 \sqrt{3}}+\frac{7 C_{22,00}^{42,1}}{384 \sqrt{3}}+\frac{7 C_{22,20}^{42,0}}{2304}+\frac{7 C_{22,20}^{42,1}}{1152}-\frac{17
C_{22,22}^{40,0}}{23040}-\frac{17 C_{22,22}^{40,1}}{11520}-\frac{\sqrt{7} C_{22,22}^{42,0}}{2304}-\frac{\sqrt{7} C_{22,22}^{42,1}}{1152}+\frac{3
C_{22,22}^{44,0}}{640 \sqrt{5}}+\frac{3 C_{22,22}^{44,1}}{320 \sqrt{5}}-\frac{\sqrt{\frac{5}{3}} C_{31,00}^{31,0}}{1536}-\frac{1}{768} \sqrt{\frac{5}{3}}
C_{31,00}^{31,1}-\frac{\sqrt{5} C_{31,20}^{31,0}}{4608}-\frac{\sqrt{5} C_{31,20}^{31,1}}{2304}+\frac{23 C_{31,22}^{31,0}}{23040}+\frac{23 C_{31,22}^{31,1}}{11520}-\frac{\sqrt{\frac{7}{15}}
C_{31,22}^{33,0}}{2560}-\frac{\sqrt{\frac{7}{15}} C_{31,22}^{33,1}}{1280}-\frac{3}{256} \sqrt{\frac{7}{5}} C_{33,00}^{33,0}-\frac{3}{128} \sqrt{\frac{7}{5}}
C_{33,00}^{33,1}-\frac{1}{256} \sqrt{\frac{21}{5}} C_{33,20}^{33,0}-\frac{1}{128} \sqrt{\frac{21}{5}} C_{33,20}^{33,1}+\frac{3 \sqrt{\frac{7}{2}}
C_{33,22}^{33,0}}{3200}+\frac{3 \sqrt{\frac{7}{2}} C_{33,22}^{33,1}}{1600}\)

\noindent\(C_{11,4011,2}^{11,0011,1,\text{Loc}}= -\frac{7}{256} \sqrt{\frac{5}{3}} C_{00,20}^{60,1}-\frac{7 \sqrt{3} C_{00,22}^{62,1}}{2560}-\frac{7}{768}
\sqrt{\frac{5}{3}} C_{11,20}^{31,1}+\frac{7 C_{11,22}^{51,1}}{7680 \sqrt{3}}-\frac{9 \sqrt{\frac{7}{5}} C_{11,22}^{53,1}}{2560}-\frac{5}{768} \sqrt{\frac{5}{3}}
C_{20,20}^{40,1}-\frac{7 C_{20,22}^{42,1}}{7680 \sqrt{3}}+\frac{7 C_{22,20}^{42,1}}{384 \sqrt{3}}+\frac{17 C_{22,22}^{40,1}}{7680 \sqrt{3}}+\frac{1}{768}
\sqrt{\frac{7}{3}} C_{22,22}^{42,1}-\frac{3}{640} \sqrt{\frac{3}{5}} C_{22,22}^{44,1}+\frac{1}{768} \sqrt{\frac{5}{3}} C_{31,20}^{31,1}+\frac{23
C_{31,22}^{31,1}}{7680 \sqrt{3}}-\frac{\sqrt{\frac{7}{5}} C_{31,22}^{33,1}}{2560}+\frac{3}{128} \sqrt{\frac{7}{5}} C_{33,20}^{33,1}+\frac{3 \sqrt{\frac{21}{2}}
C_{33,22}^{33,1}}{3200}\)

\noindent\(C_{11,4011,2}^{11,0011,1,\text{NonLoc}}= -\frac{7 \sqrt{5} C_{00,00}^{60,0}}{512}-\frac{7}{512} \sqrt{\frac{5}{3}} C_{00,20}^{60,0}+\frac{7
\sqrt{3} C_{00,22}^{62,0}}{2560}+\frac{7 \sqrt{5} C_{11,00}^{51,0}}{1536}+\frac{7 \sqrt{\frac{5}{3}} C_{11,20}^{31,0}}{1536}+\frac{7 C_{11,22}^{51,0}}{7680
\sqrt{3}}-\frac{9 \sqrt{\frac{7}{5}} C_{11,22}^{53,0}}{2560}-\frac{5 \sqrt{5} C_{20,00}^{40,0}}{1536}-\frac{5 \sqrt{\frac{5}{3}} C_{20,20}^{40,0}}{1536}+\frac{7
C_{20,22}^{42,0}}{7680 \sqrt{3}}+\frac{7 C_{22,00}^{42,0}}{768}+\frac{7 C_{22,20}^{42,0}}{768 \sqrt{3}}-\frac{17 C_{22,22}^{40,0}}{7680 \sqrt{3}}-\frac{1}{768}
\sqrt{\frac{7}{3}} C_{22,22}^{42,0}+\frac{3}{640} \sqrt{\frac{3}{5}} C_{22,22}^{44,0}-\frac{\sqrt{5} C_{31,00}^{31,0}}{1536}-\frac{\sqrt{\frac{5}{3}}
C_{31,20}^{31,0}}{1536}+\frac{23 C_{31,22}^{31,0}}{7680 \sqrt{3}}-\frac{\sqrt{\frac{7}{5}} C_{31,22}^{33,0}}{2560}-\frac{3}{256} \sqrt{\frac{21}{5}}
C_{33,00}^{33,0}-\frac{3}{256} \sqrt{\frac{7}{5}} C_{33,20}^{33,0}+\frac{3 \sqrt{\frac{21}{2}} C_{33,22}^{33,0}}{3200}\)

\noindent\(C_{11,4202,1}^{00,1111,0,\text{Loc}}= -\frac{\sqrt{5} C_{11,11}^{51,0}}{96}-\frac{\sqrt{5} C_{11,11}^{51,1}}{192}-\frac{C_{22,11}^{42,0}}{16
\sqrt{3}}-\frac{C_{22,11}^{42,1}}{32 \sqrt{3}}-\frac{\sqrt{5} C_{31,11}^{31,0}}{96}-\frac{\sqrt{5} C_{31,11}^{31,1}}{192}-\frac{3 C_{33,11}^{33,0}}{8
\sqrt{70}}-\frac{3 C_{33,11}^{33,1}}{16 \sqrt{70}}\)

\noindent\(C_{11,4202,1}^{00,1111,0,\text{NonLoc}}= -\frac{\sqrt{5} C_{11,11}^{51,0}}{192}-\frac{\sqrt{5} C_{11,11}^{51,1}}{96}+\frac{C_{22,11}^{42,0}}{32
\sqrt{3}}+\frac{C_{22,11}^{42,1}}{16 \sqrt{3}}-\frac{\sqrt{5} C_{31,11}^{31,0}}{192}-\frac{\sqrt{5} C_{31,11}^{31,1}}{96}-\frac{3 C_{33,11}^{33,0}}{16
\sqrt{70}}-\frac{3 C_{33,11}^{33,1}}{8 \sqrt{70}}\)

\noindent\(C_{11,4202,1}^{00,1111,1,\text{Loc}}= -\frac{1}{64} \sqrt{\frac{5}{3}} C_{11,11}^{51,1}-\frac{C_{22,11}^{42,1}}{32}-\frac{1}{64}
\sqrt{\frac{5}{3}} C_{31,11}^{31,1}-\frac{3}{16} \sqrt{\frac{3}{70}} C_{33,11}^{33,1}\)

\noindent\(C_{11,4202,1}^{00,1111,1,\text{NonLoc}}= -\frac{1}{64} \sqrt{\frac{5}{3}} C_{11,11}^{51,0}+\frac{C_{22,11}^{42,0}}{32}-\frac{1}{64}
\sqrt{\frac{5}{3}} C_{31,11}^{31,0}-\frac{3}{16} \sqrt{\frac{3}{70}} C_{33,11}^{33,0}\)

\noindent\(C_{11,4202,1}^{11,0000,0,\text{Loc}}= \frac{\sqrt{5} C_{00,00}^{60,0}}{32}+\frac{\sqrt{5} C_{00,00}^{60,1}}{64}+\frac{\sqrt{5}
C_{11,00}^{51,0}}{96}+\frac{\sqrt{5} C_{11,00}^{51,1}}{192}-\frac{\sqrt{5} C_{20,00}^{40,0}}{96}-\frac{\sqrt{5} C_{20,00}^{40,1}}{192}+\frac{C_{22,00}^{42,0}}{96}+\frac{C_{22,00}^{42,1}}{192}-\frac{\sqrt{5}
C_{31,00}^{31,0}}{96}-\frac{\sqrt{5} C_{31,00}^{31,1}}{192}+\frac{3}{32} \sqrt{\frac{3}{35}} C_{33,00}^{33,0}+\frac{3}{64} \sqrt{\frac{3}{35}} C_{33,00}^{33,1}\)

\noindent\(C_{11,4202,1}^{11,0000,0,\text{NonLoc}}= -\frac{\sqrt{5} C_{00,00}^{60,0}}{128}-\frac{\sqrt{5} C_{00,00}^{60,1}}{64}+\frac{\sqrt{15}
C_{00,20}^{60,0}}{128}+\frac{\sqrt{15} C_{00,20}^{60,1}}{64}+\frac{\sqrt{5} C_{11,00}^{51,0}}{384}+\frac{\sqrt{5} C_{11,00}^{51,1}}{192}-\frac{1}{128}
\sqrt{\frac{5}{3}} C_{11,20}^{31,0}-\frac{1}{64} \sqrt{\frac{5}{3}} C_{11,20}^{31,1}+\frac{\sqrt{5} C_{20,00}^{40,0}}{384}+\frac{\sqrt{5} C_{20,00}^{40,1}}{192}-\frac{1}{128}
\sqrt{\frac{5}{3}} C_{20,20}^{40,0}-\frac{1}{64} \sqrt{\frac{5}{3}} C_{20,20}^{40,1}-\frac{C_{22,00}^{42,0}}{384}-\frac{C_{22,00}^{42,1}}{192}+\frac{C_{22,20}^{42,0}}{128
\sqrt{3}}+\frac{C_{22,20}^{42,1}}{64 \sqrt{3}}-\frac{\sqrt{5} C_{31,00}^{31,0}}{384}-\frac{\sqrt{5} C_{31,00}^{31,1}}{192}+\frac{1}{128} \sqrt{\frac{5}{3}}
C_{31,20}^{31,0}+\frac{1}{64} \sqrt{\frac{5}{3}} C_{31,20}^{31,1}+\frac{3}{128} \sqrt{\frac{3}{35}} C_{33,00}^{33,0}+\frac{3}{64} \sqrt{\frac{3}{35}}
C_{33,00}^{33,1}-\frac{9 C_{33,20}^{33,0}}{128 \sqrt{35}}-\frac{9 C_{33,20}^{33,1}}{64 \sqrt{35}}\)

\noindent\(C_{11,4202,1}^{11,0000,1,\text{Loc}}= \frac{\sqrt{15} C_{00,00}^{60,1}}{64}+\frac{1}{64} \sqrt{\frac{5}{3}} C_{11,00}^{51,1}-\frac{1}{64}
\sqrt{\frac{5}{3}} C_{20,00}^{40,1}+\frac{C_{22,00}^{42,1}}{64 \sqrt{3}}-\frac{1}{64} \sqrt{\frac{5}{3}} C_{31,00}^{31,1}+\frac{9 C_{33,00}^{33,1}}{64
\sqrt{35}}\)

\noindent\(C_{11,4202,1}^{11,0000,1,\text{NonLoc}}= -\frac{\sqrt{15} C_{00,00}^{60,0}}{128}+\frac{3 \sqrt{5} C_{00,20}^{60,0}}{128}+\frac{1}{128}
\sqrt{\frac{5}{3}} C_{11,00}^{51,0}-\frac{\sqrt{5} C_{11,20}^{31,0}}{128}+\frac{1}{128} \sqrt{\frac{5}{3}} C_{20,00}^{40,0}-\frac{\sqrt{5} C_{20,20}^{40,0}}{128}-\frac{C_{22,00}^{42,0}}{128
\sqrt{3}}+\frac{C_{22,20}^{42,0}}{128}-\frac{1}{128} \sqrt{\frac{5}{3}} C_{31,00}^{31,0}+\frac{\sqrt{5} C_{31,20}^{31,0}}{128}+\frac{9 C_{33,00}^{33,0}}{128
\sqrt{35}}-\frac{9}{128} \sqrt{\frac{3}{35}} C_{33,20}^{33,0}\)

\noindent\(C_{11,4202,2}^{00,1112,0,\text{Loc}}= \frac{1}{32} \sqrt{\frac{5}{3}} C_{11,11}^{51,0}+\frac{1}{64} \sqrt{\frac{5}{3}} C_{11,11}^{51,1}+\frac{C_{22,11}^{42,0}}{16}+\frac{C_{22,11}^{42,1}}{32}+\frac{1}{32}
\sqrt{\frac{5}{3}} C_{31,11}^{31,0}+\frac{1}{64} \sqrt{\frac{5}{3}} C_{31,11}^{31,1}+\frac{3}{8} \sqrt{\frac{3}{70}} C_{33,11}^{33,0}+\frac{3}{16}
\sqrt{\frac{3}{70}} C_{33,11}^{33,1}\)

\noindent\(C_{11,4202,2}^{00,1112,0,\text{NonLoc}}= \frac{1}{64} \sqrt{\frac{5}{3}} C_{11,11}^{51,0}+\frac{1}{32} \sqrt{\frac{5}{3}}
C_{11,11}^{51,1}-\frac{C_{22,11}^{42,0}}{32}-\frac{C_{22,11}^{42,1}}{16}+\frac{1}{64} \sqrt{\frac{5}{3}} C_{31,11}^{31,0}+\frac{1}{32} \sqrt{\frac{5}{3}}
C_{31,11}^{31,1}+\frac{3}{16} \sqrt{\frac{3}{70}} C_{33,11}^{33,0}+\frac{3}{8} \sqrt{\frac{3}{70}} C_{33,11}^{33,1}\)

\noindent\(C_{11,4202,2}^{00,1112,1,\text{Loc}}= \frac{\sqrt{5} C_{11,11}^{51,1}}{64}+\frac{\sqrt{3} C_{22,11}^{42,1}}{32}+\frac{\sqrt{5}
C_{31,11}^{31,1}}{64}+\frac{9 C_{33,11}^{33,1}}{16 \sqrt{70}}\)

\noindent\(C_{11,4202,2}^{00,1112,1,\text{NonLoc}}= \frac{\sqrt{5} C_{11,11}^{51,0}}{64}-\frac{\sqrt{3} C_{22,11}^{42,0}}{32}+\frac{\sqrt{5}
C_{31,11}^{31,0}}{64}+\frac{9 C_{33,11}^{33,0}}{16 \sqrt{70}}\)

\noindent\(C_{11,4211,0}^{11,0011,0,\text{Loc}}= -\frac{\sqrt{5} C_{00,20}^{60,0}}{96}-\frac{\sqrt{5} C_{00,20}^{60,1}}{192}-\frac{C_{00,22}^{62,0}}{32}-\frac{C_{00,22}^{62,1}}{64}-\frac{\sqrt{5}
C_{11,20}^{31,0}}{288}-\frac{\sqrt{5} C_{11,20}^{31,1}}{576}-\frac{7 C_{11,22}^{51,0}}{576}-\frac{7 C_{11,22}^{51,1}}{1152}-\frac{3}{128} \sqrt{\frac{3}{35}}
C_{11,22}^{53,0}-\frac{3}{256} \sqrt{\frac{3}{35}} C_{11,22}^{53,1}+\frac{\sqrt{5} C_{20,20}^{40,0}}{288}+\frac{\sqrt{5} C_{20,20}^{40,1}}{576}+\frac{C_{20,22}^{42,0}}{576}+\frac{C_{20,22}^{42,1}}{1152}-\frac{C_{22,20}^{42,0}}{288}-\frac{C_{22,20}^{42,1}}{576}+\frac{C_{22,22}^{40,0}}{144}+\frac{C_{22,22}^{40,1}}{288}-\frac{5
C_{22,22}^{42,0}}{288 \sqrt{7}}-\frac{5 C_{22,22}^{42,1}}{576 \sqrt{7}}+\frac{3 C_{22,22}^{44,0}}{112 \sqrt{5}}+\frac{3 C_{22,22}^{44,1}}{224 \sqrt{5}}+\frac{\sqrt{5}
C_{31,20}^{31,0}}{288}+\frac{\sqrt{5} C_{31,20}^{31,1}}{576}+\frac{C_{31,22}^{31,0}}{576}+\frac{C_{31,22}^{31,1}}{1152}+\frac{17 C_{31,22}^{33,0}}{128
\sqrt{105}}+\frac{17 C_{31,22}^{33,1}}{256 \sqrt{105}}-\frac{1}{32} \sqrt{\frac{3}{35}} C_{33,20}^{33,0}-\frac{1}{64} \sqrt{\frac{3}{35}} C_{33,20}^{33,1}-\frac{3
C_{33,22}^{33,0}}{80 \sqrt{14}}-\frac{3 C_{33,22}^{33,1}}{160 \sqrt{14}}\)

\noindent\(C_{11,4211,0}^{11,0011,0,\text{NonLoc}}= -\frac{1}{128} \sqrt{\frac{5}{3}} C_{00,00}^{60,0}-\frac{1}{64} \sqrt{\frac{5}{3}}
C_{00,00}^{60,1}-\frac{\sqrt{5} C_{00,20}^{60,0}}{384}-\frac{\sqrt{5} C_{00,20}^{60,1}}{192}+\frac{C_{00,22}^{62,0}}{64}+\frac{C_{00,22}^{62,1}}{32}+\frac{1}{384}
\sqrt{\frac{5}{3}} C_{11,00}^{51,0}+\frac{1}{192} \sqrt{\frac{5}{3}} C_{11,00}^{51,1}+\frac{\sqrt{5} C_{11,20}^{31,0}}{1152}+\frac{\sqrt{5} C_{11,20}^{31,1}}{576}-\frac{7
C_{11,22}^{51,0}}{1152}-\frac{7 C_{11,22}^{51,1}}{576}-\frac{3}{256} \sqrt{\frac{3}{35}} C_{11,22}^{53,0}-\frac{3}{128} \sqrt{\frac{3}{35}} C_{11,22}^{53,1}+\frac{1}{384}
\sqrt{\frac{5}{3}} C_{20,00}^{40,0}+\frac{1}{192} \sqrt{\frac{5}{3}} C_{20,00}^{40,1}+\frac{\sqrt{5} C_{20,20}^{40,0}}{1152}+\frac{\sqrt{5} C_{20,20}^{40,1}}{576}-\frac{C_{20,22}^{42,0}}{1152}-\frac{C_{20,22}^{42,1}}{576}-\frac{C_{22,00}^{42,0}}{384
\sqrt{3}}-\frac{C_{22,00}^{42,1}}{192 \sqrt{3}}-\frac{C_{22,20}^{42,0}}{1152}-\frac{C_{22,20}^{42,1}}{576}-\frac{C_{22,22}^{40,0}}{288}-\frac{C_{22,22}^{40,1}}{144}+\frac{5
C_{22,22}^{42,0}}{576 \sqrt{7}}+\frac{5 C_{22,22}^{42,1}}{288 \sqrt{7}}-\frac{3 C_{22,22}^{44,0}}{224 \sqrt{5}}-\frac{3 C_{22,22}^{44,1}}{112 \sqrt{5}}-\frac{1}{384}
\sqrt{\frac{5}{3}} C_{31,00}^{31,0}-\frac{1}{192} \sqrt{\frac{5}{3}} C_{31,00}^{31,1}-\frac{\sqrt{5} C_{31,20}^{31,0}}{1152}-\frac{\sqrt{5} C_{31,20}^{31,1}}{576}+\frac{C_{31,22}^{31,0}}{1152}+\frac{C_{31,22}^{31,1}}{576}+\frac{17
C_{31,22}^{33,0}}{256 \sqrt{105}}+\frac{17 C_{31,22}^{33,1}}{128 \sqrt{105}}+\frac{3 C_{33,00}^{33,0}}{128 \sqrt{35}}+\frac{3 C_{33,00}^{33,1}}{64
\sqrt{35}}+\frac{1}{128} \sqrt{\frac{3}{35}} C_{33,20}^{33,0}+\frac{1}{64} \sqrt{\frac{3}{35}} C_{33,20}^{33,1}-\frac{3 C_{33,22}^{33,0}}{160 \sqrt{14}}-\frac{3
C_{33,22}^{33,1}}{80 \sqrt{14}}\)

\noindent\(C_{11,4211,0}^{11,0011,1,\text{Loc}}= -\frac{1}{64} \sqrt{\frac{5}{3}} C_{00,20}^{60,1}-\frac{\sqrt{3} C_{00,22}^{62,1}}{64}-\frac{1}{192}
\sqrt{\frac{5}{3}} C_{11,20}^{31,1}-\frac{7 C_{11,22}^{51,1}}{384 \sqrt{3}}-\frac{9 C_{11,22}^{53,1}}{256 \sqrt{35}}+\frac{1}{192} \sqrt{\frac{5}{3}}
C_{20,20}^{40,1}+\frac{C_{20,22}^{42,1}}{384 \sqrt{3}}-\frac{C_{22,20}^{42,1}}{192 \sqrt{3}}+\frac{C_{22,22}^{40,1}}{96 \sqrt{3}}-\frac{5 C_{22,22}^{42,1}}{192
\sqrt{21}}+\frac{3}{224} \sqrt{\frac{3}{5}} C_{22,22}^{44,1}+\frac{1}{192} \sqrt{\frac{5}{3}} C_{31,20}^{31,1}+\frac{C_{31,22}^{31,1}}{384 \sqrt{3}}+\frac{17
C_{31,22}^{33,1}}{256 \sqrt{35}}-\frac{3 C_{33,20}^{33,1}}{64 \sqrt{35}}-\frac{3}{160} \sqrt{\frac{3}{14}} C_{33,22}^{33,1}\)

\noindent\(C_{11,4211,0}^{11,0011,1,\text{NonLoc}}= -\frac{\sqrt{5} C_{00,00}^{60,0}}{128}-\frac{1}{128} \sqrt{\frac{5}{3}} C_{00,20}^{60,0}+\frac{\sqrt{3}
C_{00,22}^{62,0}}{64}+\frac{\sqrt{5} C_{11,00}^{51,0}}{384}+\frac{1}{384} \sqrt{\frac{5}{3}} C_{11,20}^{31,0}-\frac{7 C_{11,22}^{51,0}}{384 \sqrt{3}}-\frac{9
C_{11,22}^{53,0}}{256 \sqrt{35}}+\frac{\sqrt{5} C_{20,00}^{40,0}}{384}+\frac{1}{384} \sqrt{\frac{5}{3}} C_{20,20}^{40,0}-\frac{C_{20,22}^{42,0}}{384
\sqrt{3}}-\frac{C_{22,00}^{42,0}}{384}-\frac{C_{22,20}^{42,0}}{384 \sqrt{3}}-\frac{C_{22,22}^{40,0}}{96 \sqrt{3}}+\frac{5 C_{22,22}^{42,0}}{192 \sqrt{21}}-\frac{3}{224}
\sqrt{\frac{3}{5}} C_{22,22}^{44,0}-\frac{\sqrt{5} C_{31,00}^{31,0}}{384}-\frac{1}{384} \sqrt{\frac{5}{3}} C_{31,20}^{31,0}+\frac{C_{31,22}^{31,0}}{384
\sqrt{3}}+\frac{17 C_{31,22}^{33,0}}{256 \sqrt{35}}+\frac{3}{128} \sqrt{\frac{3}{35}} C_{33,00}^{33,0}+\frac{3 C_{33,20}^{33,0}}{128 \sqrt{35}}-\frac{3}{160}
\sqrt{\frac{3}{14}} C_{33,22}^{33,0}\)

\noindent\(C_{11,4211,1}^{00,1101,0,\text{Loc}}= \frac{\sqrt{5} C_{11,11}^{51,0}}{96}+\frac{\sqrt{5} C_{11,11}^{51,1}}{192}+\frac{C_{22,11}^{42,0}}{16
\sqrt{3}}+\frac{C_{22,11}^{42,1}}{32 \sqrt{3}}+\frac{\sqrt{5} C_{31,11}^{31,0}}{96}+\frac{\sqrt{5} C_{31,11}^{31,1}}{192}+\frac{3 C_{33,11}^{33,0}}{8
\sqrt{70}}+\frac{3 C_{33,11}^{33,1}}{16 \sqrt{70}}\)

\noindent\(C_{11,4211,1}^{00,1101,0,\text{NonLoc}}= \frac{\sqrt{5} C_{11,11}^{51,0}}{192}+\frac{\sqrt{5} C_{11,11}^{51,1}}{96}-\frac{C_{22,11}^{42,0}}{32
\sqrt{3}}-\frac{C_{22,11}^{42,1}}{16 \sqrt{3}}+\frac{\sqrt{5} C_{31,11}^{31,0}}{192}+\frac{\sqrt{5} C_{31,11}^{31,1}}{96}+\frac{3 C_{33,11}^{33,0}}{16
\sqrt{70}}+\frac{3 C_{33,11}^{33,1}}{8 \sqrt{70}}\)

\noindent\(C_{11,4211,1}^{00,1101,1,\text{Loc}}= \frac{1}{64} \sqrt{\frac{5}{3}} C_{11,11}^{51,1}+\frac{C_{22,11}^{42,1}}{32}+\frac{1}{64}
\sqrt{\frac{5}{3}} C_{31,11}^{31,1}+\frac{3}{16} \sqrt{\frac{3}{70}} C_{33,11}^{33,1}\)

\noindent\(C_{11,4211,1}^{00,1101,1,\text{NonLoc}}= \frac{1}{64} \sqrt{\frac{5}{3}} C_{11,11}^{51,0}-\frac{C_{22,11}^{42,0}}{32}+\frac{1}{64}
\sqrt{\frac{5}{3}} C_{31,11}^{31,0}+\frac{3}{16} \sqrt{\frac{3}{70}} C_{33,11}^{33,0}\)

\noindent\(C_{11,4211,1}^{11,0011,0,\text{Loc}}= \frac{1}{64} \sqrt{\frac{5}{3}} C_{00,20}^{60,0}+\frac{1}{128} \sqrt{\frac{5}{3}}
C_{00,20}^{60,1}-\frac{\sqrt{3} C_{00,22}^{62,0}}{128}-\frac{\sqrt{3} C_{00,22}^{62,1}}{256}+\frac{1}{192} \sqrt{\frac{5}{3}} C_{11,20}^{31,0}+\frac{1}{384}
\sqrt{\frac{5}{3}} C_{11,20}^{31,1}-\frac{C_{11,22}^{51,0}}{48 \sqrt{3}}-\frac{C_{11,22}^{51,1}}{96 \sqrt{3}}+\frac{9 C_{11,22}^{53,0}}{256 \sqrt{35}}+\frac{9
C_{11,22}^{53,1}}{512 \sqrt{35}}-\frac{1}{192} \sqrt{\frac{5}{3}} C_{20,20}^{40,0}-\frac{1}{384} \sqrt{\frac{5}{3}} C_{20,20}^{40,1}-\frac{C_{20,22}^{42,0}}{96
\sqrt{3}}-\frac{C_{20,22}^{42,1}}{192 \sqrt{3}}+\frac{C_{22,20}^{42,0}}{192 \sqrt{3}}+\frac{C_{22,20}^{42,1}}{384 \sqrt{3}}-\frac{7 C_{22,22}^{40,0}}{384
\sqrt{3}}-\frac{7 C_{22,22}^{40,1}}{768 \sqrt{3}}+\frac{11 C_{22,22}^{42,0}}{192 \sqrt{21}}+\frac{11 C_{22,22}^{42,1}}{384 \sqrt{21}}+\frac{3}{112}
\sqrt{\frac{3}{5}} C_{22,22}^{44,0}+\frac{3}{224} \sqrt{\frac{3}{5}} C_{22,22}^{44,1}-\frac{1}{192} \sqrt{\frac{5}{3}} C_{31,20}^{31,0}-\frac{1}{384}
\sqrt{\frac{5}{3}} C_{31,20}^{31,1}-\frac{C_{31,22}^{31,0}}{96 \sqrt{3}}-\frac{C_{31,22}^{31,1}}{192 \sqrt{3}}+\frac{1}{256} \sqrt{\frac{7}{5}} C_{31,22}^{33,0}+\frac{1}{512}
\sqrt{\frac{7}{5}} C_{31,22}^{33,1}+\frac{3 C_{33,20}^{33,0}}{64 \sqrt{35}}+\frac{3 C_{33,20}^{33,1}}{128 \sqrt{35}}+\frac{3}{40} \sqrt{\frac{3}{14}}
C_{33,22}^{33,0}+\frac{3}{80} \sqrt{\frac{3}{14}} C_{33,22}^{33,1}\)

\noindent\(C_{11,4211,1}^{11,0011,0,\text{NonLoc}}= \frac{\sqrt{5} C_{00,00}^{60,0}}{256}+\frac{\sqrt{5} C_{00,00}^{60,1}}{128}+\frac{1}{256}
\sqrt{\frac{5}{3}} C_{00,20}^{60,0}+\frac{1}{128} \sqrt{\frac{5}{3}} C_{00,20}^{60,1}+\frac{\sqrt{3} C_{00,22}^{62,0}}{256}+\frac{\sqrt{3} C_{00,22}^{62,1}}{128}-\frac{\sqrt{5}
C_{11,00}^{51,0}}{768}-\frac{\sqrt{5} C_{11,00}^{51,1}}{384}-\frac{1}{768} \sqrt{\frac{5}{3}} C_{11,20}^{31,0}-\frac{1}{384} \sqrt{\frac{5}{3}} C_{11,20}^{31,1}-\frac{C_{11,22}^{51,0}}{96
\sqrt{3}}-\frac{C_{11,22}^{51,1}}{48 \sqrt{3}}+\frac{9 C_{11,22}^{53,0}}{512 \sqrt{35}}+\frac{9 C_{11,22}^{53,1}}{256 \sqrt{35}}-\frac{\sqrt{5} C_{20,00}^{40,0}}{768}-\frac{\sqrt{5}
C_{20,00}^{40,1}}{384}-\frac{1}{768} \sqrt{\frac{5}{3}} C_{20,20}^{40,0}-\frac{1}{384} \sqrt{\frac{5}{3}} C_{20,20}^{40,1}+\frac{C_{20,22}^{42,0}}{192
\sqrt{3}}+\frac{C_{20,22}^{42,1}}{96 \sqrt{3}}+\frac{C_{22,00}^{42,0}}{768}+\frac{C_{22,00}^{42,1}}{384}+\frac{C_{22,20}^{42,0}}{768 \sqrt{3}}+\frac{C_{22,20}^{42,1}}{384
\sqrt{3}}+\frac{7 C_{22,22}^{40,0}}{768 \sqrt{3}}+\frac{7 C_{22,22}^{40,1}}{384 \sqrt{3}}-\frac{11 C_{22,22}^{42,0}}{384 \sqrt{21}}-\frac{11 C_{22,22}^{42,1}}{192
\sqrt{21}}-\frac{3}{224} \sqrt{\frac{3}{5}} C_{22,22}^{44,0}-\frac{3}{112} \sqrt{\frac{3}{5}} C_{22,22}^{44,1}+\frac{\sqrt{5} C_{31,00}^{31,0}}{768}+\frac{\sqrt{5}
C_{31,00}^{31,1}}{384}+\frac{1}{768} \sqrt{\frac{5}{3}} C_{31,20}^{31,0}+\frac{1}{384} \sqrt{\frac{5}{3}} C_{31,20}^{31,1}-\frac{C_{31,22}^{31,0}}{192
\sqrt{3}}-\frac{C_{31,22}^{31,1}}{96 \sqrt{3}}+\frac{1}{512} \sqrt{\frac{7}{5}} C_{31,22}^{33,0}+\frac{1}{256} \sqrt{\frac{7}{5}} C_{31,22}^{33,1}-\frac{3}{256}
\sqrt{\frac{3}{35}} C_{33,00}^{33,0}-\frac{3}{128} \sqrt{\frac{3}{35}} C_{33,00}^{33,1}-\frac{3 C_{33,20}^{33,0}}{256 \sqrt{35}}-\frac{3 C_{33,20}^{33,1}}{128
\sqrt{35}}+\frac{3}{80} \sqrt{\frac{3}{14}} C_{33,22}^{33,0}+\frac{3}{40} \sqrt{\frac{3}{14}} C_{33,22}^{33,1}\)

\noindent\(C_{11,4211,1}^{11,0011,1,\text{Loc}}= \frac{\sqrt{5} C_{00,20}^{60,1}}{128}-\frac{3 C_{00,22}^{62,1}}{256}+\frac{\sqrt{5}
C_{11,20}^{31,1}}{384}-\frac{C_{11,22}^{51,1}}{96}+\frac{9}{512} \sqrt{\frac{3}{35}} C_{11,22}^{53,1}-\frac{\sqrt{5} C_{20,20}^{40,1}}{384}-\frac{C_{20,22}^{42,1}}{192}+\frac{C_{22,20}^{42,1}}{384}-\frac{7
C_{22,22}^{40,1}}{768}+\frac{11 C_{22,22}^{42,1}}{384 \sqrt{7}}+\frac{9 C_{22,22}^{44,1}}{224 \sqrt{5}}-\frac{\sqrt{5} C_{31,20}^{31,1}}{384}-\frac{C_{31,22}^{31,1}}{192}+\frac{1}{512}
\sqrt{\frac{21}{5}} C_{31,22}^{33,1}+\frac{3}{128} \sqrt{\frac{3}{35}} C_{33,20}^{33,1}+\frac{9 C_{33,22}^{33,1}}{80 \sqrt{14}}\)

\noindent\(C_{11,4211,1}^{11,0011,1,\text{NonLoc}}= \frac{\sqrt{15} C_{00,00}^{60,0}}{256}+\frac{\sqrt{5} C_{00,20}^{60,0}}{256}+\frac{3
C_{00,22}^{62,0}}{256}-\frac{1}{256} \sqrt{\frac{5}{3}} C_{11,00}^{51,0}-\frac{\sqrt{5} C_{11,20}^{31,0}}{768}-\frac{C_{11,22}^{51,0}}{96}+\frac{9}{512}
\sqrt{\frac{3}{35}} C_{11,22}^{53,0}-\frac{1}{256} \sqrt{\frac{5}{3}} C_{20,00}^{40,0}-\frac{\sqrt{5} C_{20,20}^{40,0}}{768}+\frac{C_{20,22}^{42,0}}{192}+\frac{C_{22,00}^{42,0}}{256
\sqrt{3}}+\frac{C_{22,20}^{42,0}}{768}+\frac{7 C_{22,22}^{40,0}}{768}-\frac{11 C_{22,22}^{42,0}}{384 \sqrt{7}}-\frac{9 C_{22,22}^{44,0}}{224 \sqrt{5}}+\frac{1}{256}
\sqrt{\frac{5}{3}} C_{31,00}^{31,0}+\frac{\sqrt{5} C_{31,20}^{31,0}}{768}-\frac{C_{31,22}^{31,0}}{192}+\frac{1}{512} \sqrt{\frac{21}{5}} C_{31,22}^{33,0}-\frac{9
C_{33,00}^{33,0}}{256 \sqrt{35}}-\frac{3}{256} \sqrt{\frac{3}{35}} C_{33,20}^{33,0}+\frac{9 C_{33,22}^{33,0}}{80 \sqrt{14}}\)

\noindent\(C_{11,4211,2}^{11,0011,0,\text{Loc}}= -\frac{C_{00,20}^{60,0}}{192}-\frac{C_{00,20}^{60,1}}{384}-\frac{11 C_{00,22}^{62,0}}{128
\sqrt{5}}-\frac{11 C_{00,22}^{62,1}}{256 \sqrt{5}}-\frac{C_{11,20}^{31,0}}{576}-\frac{C_{11,20}^{31,1}}{1152}-\frac{13 C_{11,22}^{51,0}}{288 \sqrt{5}}-\frac{13
C_{11,22}^{51,1}}{576 \sqrt{5}}-\frac{3 \sqrt{\frac{3}{7}} C_{11,22}^{53,0}}{1280}-\frac{3 \sqrt{\frac{3}{7}} C_{11,22}^{53,1}}{2560}+\frac{C_{20,20}^{40,0}}{576}+\frac{C_{20,20}^{40,1}}{1152}-\frac{C_{20,22}^{42,0}}{144
\sqrt{5}}-\frac{C_{20,22}^{42,1}}{288 \sqrt{5}}-\frac{C_{22,20}^{42,0}}{576 \sqrt{5}}-\frac{C_{22,20}^{42,1}}{1152 \sqrt{5}}-\frac{\sqrt{5} C_{22,22}^{40,0}}{1152}-\frac{\sqrt{5}
C_{22,22}^{40,1}}{2304}+\frac{13 C_{22,22}^{42,0}}{576 \sqrt{35}}+\frac{13 C_{22,22}^{42,1}}{1152 \sqrt{35}}+\frac{3 C_{22,22}^{44,0}}{112}+\frac{3
C_{22,22}^{44,1}}{224}+\frac{C_{31,20}^{31,0}}{576}+\frac{C_{31,20}^{31,1}}{1152}-\frac{C_{31,22}^{31,0}}{144 \sqrt{5}}-\frac{C_{31,22}^{31,1}}{288
\sqrt{5}}+\frac{89 C_{31,22}^{33,0}}{1280 \sqrt{21}}+\frac{89 C_{31,22}^{33,1}}{2560 \sqrt{21}}-\frac{1}{320} \sqrt{\frac{3}{7}} C_{33,20}^{33,0}-\frac{1}{640}
\sqrt{\frac{3}{7}} C_{33,20}^{33,1}+\frac{3 C_{33,22}^{33,0}}{20 \sqrt{70}}+\frac{3 C_{33,22}^{33,1}}{40 \sqrt{70}}\)

\noindent\(C_{11,4211,2}^{11,0011,0,\text{NonLoc}}= -\frac{C_{00,00}^{60,0}}{256 \sqrt{3}}-\frac{C_{00,00}^{60,1}}{128 \sqrt{3}}-\frac{C_{00,20}^{60,0}}{768}-\frac{C_{00,20}^{60,1}}{384}+\frac{11
C_{00,22}^{62,0}}{256 \sqrt{5}}+\frac{11 C_{00,22}^{62,1}}{128 \sqrt{5}}+\frac{C_{11,00}^{51,0}}{768 \sqrt{3}}+\frac{C_{11,00}^{51,1}}{384 \sqrt{3}}+\frac{C_{11,20}^{31,0}}{2304}+\frac{C_{11,20}^{31,1}}{1152}-\frac{13
C_{11,22}^{51,0}}{576 \sqrt{5}}-\frac{13 C_{11,22}^{51,1}}{288 \sqrt{5}}-\frac{3 \sqrt{\frac{3}{7}} C_{11,22}^{53,0}}{2560}-\frac{3 \sqrt{\frac{3}{7}}
C_{11,22}^{53,1}}{1280}+\frac{C_{20,00}^{40,0}}{768 \sqrt{3}}+\frac{C_{20,00}^{40,1}}{384 \sqrt{3}}+\frac{C_{20,20}^{40,0}}{2304}+\frac{C_{20,20}^{40,1}}{1152}+\frac{C_{20,22}^{42,0}}{288
\sqrt{5}}+\frac{C_{20,22}^{42,1}}{144 \sqrt{5}}-\frac{C_{22,00}^{42,0}}{768 \sqrt{15}}-\frac{C_{22,00}^{42,1}}{384 \sqrt{15}}-\frac{C_{22,20}^{42,0}}{2304
\sqrt{5}}-\frac{C_{22,20}^{42,1}}{1152 \sqrt{5}}+\frac{\sqrt{5} C_{22,22}^{40,0}}{2304}+\frac{\sqrt{5} C_{22,22}^{40,1}}{1152}-\frac{13 C_{22,22}^{42,0}}{1152
\sqrt{35}}-\frac{13 C_{22,22}^{42,1}}{576 \sqrt{35}}-\frac{3 C_{22,22}^{44,0}}{224}-\frac{3 C_{22,22}^{44,1}}{112}-\frac{C_{31,00}^{31,0}}{768 \sqrt{3}}-\frac{C_{31,00}^{31,1}}{384
\sqrt{3}}-\frac{C_{31,20}^{31,0}}{2304}-\frac{C_{31,20}^{31,1}}{1152}-\frac{C_{31,22}^{31,0}}{288 \sqrt{5}}-\frac{C_{31,22}^{31,1}}{144 \sqrt{5}}+\frac{89
C_{31,22}^{33,0}}{2560 \sqrt{21}}+\frac{89 C_{31,22}^{33,1}}{1280 \sqrt{21}}+\frac{3 C_{33,00}^{33,0}}{1280 \sqrt{7}}+\frac{3 C_{33,00}^{33,1}}{640
\sqrt{7}}+\frac{\sqrt{\frac{3}{7}} C_{33,20}^{33,0}}{1280}+\frac{1}{640} \sqrt{\frac{3}{7}} C_{33,20}^{33,1}+\frac{3 C_{33,22}^{33,0}}{40 \sqrt{70}}+\frac{3
C_{33,22}^{33,1}}{20 \sqrt{70}}\)

\noindent\(C_{11,4211,2}^{11,0011,1,\text{Loc}}= -\frac{C_{00,20}^{60,1}}{128 \sqrt{3}}-\frac{11}{256} \sqrt{\frac{3}{5}} C_{00,22}^{62,1}-\frac{C_{11,20}^{31,1}}{384
\sqrt{3}}-\frac{13 C_{11,22}^{51,1}}{192 \sqrt{15}}-\frac{9 C_{11,22}^{53,1}}{2560 \sqrt{7}}+\frac{C_{20,20}^{40,1}}{384 \sqrt{3}}-\frac{C_{20,22}^{42,1}}{96
\sqrt{15}}-\frac{C_{22,20}^{42,1}}{384 \sqrt{15}}-\frac{1}{768} \sqrt{\frac{5}{3}} C_{22,22}^{40,1}+\frac{13 C_{22,22}^{42,1}}{384 \sqrt{105}}+\frac{3
\sqrt{3} C_{22,22}^{44,1}}{224}+\frac{C_{31,20}^{31,1}}{384 \sqrt{3}}-\frac{C_{31,22}^{31,1}}{96 \sqrt{15}}+\frac{89 C_{31,22}^{33,1}}{2560 \sqrt{7}}-\frac{3
C_{33,20}^{33,1}}{640 \sqrt{7}}+\frac{3}{40} \sqrt{\frac{3}{70}} C_{33,22}^{33,1}\)

\noindent\(C_{11,4211,2}^{11,0011,1,\text{NonLoc}}= -\frac{C_{00,00}^{60,0}}{256}-\frac{C_{00,20}^{60,0}}{256 \sqrt{3}}+\frac{11}{256}
\sqrt{\frac{3}{5}} C_{00,22}^{62,0}+\frac{C_{11,00}^{51,0}}{768}+\frac{C_{11,20}^{31,0}}{768 \sqrt{3}}-\frac{13 C_{11,22}^{51,0}}{192 \sqrt{15}}-\frac{9
C_{11,22}^{53,0}}{2560 \sqrt{7}}+\frac{C_{20,00}^{40,0}}{768}+\frac{C_{20,20}^{40,0}}{768 \sqrt{3}}+\frac{C_{20,22}^{42,0}}{96 \sqrt{15}}-\frac{C_{22,00}^{42,0}}{768
\sqrt{5}}-\frac{C_{22,20}^{42,0}}{768 \sqrt{15}}+\frac{1}{768} \sqrt{\frac{5}{3}} C_{22,22}^{40,0}-\frac{13 C_{22,22}^{42,0}}{384 \sqrt{105}}-\frac{3
\sqrt{3} C_{22,22}^{44,0}}{224}-\frac{C_{31,00}^{31,0}}{768}-\frac{C_{31,20}^{31,0}}{768 \sqrt{3}}-\frac{C_{31,22}^{31,0}}{96 \sqrt{15}}+\frac{89
C_{31,22}^{33,0}}{2560 \sqrt{7}}+\frac{3 \sqrt{\frac{3}{7}} C_{33,00}^{33,0}}{1280}+\frac{3 C_{33,20}^{33,0}}{1280 \sqrt{7}}+\frac{3}{40} \sqrt{\frac{3}{70}}
C_{33,22}^{33,0}\)

\noindent\(C_{11,4212,1}^{00,1101,0,\text{Loc}}= \frac{1}{32} \sqrt{\frac{5}{3}} C_{11,11}^{51,0}+\frac{1}{64} \sqrt{\frac{5}{3}} C_{11,11}^{51,1}+\frac{C_{22,11}^{42,0}}{16}+\frac{C_{22,11}^{42,1}}{32}+\frac{1}{32}
\sqrt{\frac{5}{3}} C_{31,11}^{31,0}+\frac{1}{64} \sqrt{\frac{5}{3}} C_{31,11}^{31,1}+\frac{3}{8} \sqrt{\frac{3}{70}} C_{33,11}^{33,0}+\frac{3}{16}
\sqrt{\frac{3}{70}} C_{33,11}^{33,1}\)

\noindent\(C_{11,4212,1}^{00,1101,0,\text{NonLoc}}= \frac{1}{64} \sqrt{\frac{5}{3}} C_{11,11}^{51,0}+\frac{1}{32} \sqrt{\frac{5}{3}}
C_{11,11}^{51,1}-\frac{C_{22,11}^{42,0}}{32}-\frac{C_{22,11}^{42,1}}{16}+\frac{1}{64} \sqrt{\frac{5}{3}} C_{31,11}^{31,0}+\frac{1}{32} \sqrt{\frac{5}{3}}
C_{31,11}^{31,1}+\frac{3}{16} \sqrt{\frac{3}{70}} C_{33,11}^{33,0}+\frac{3}{8} \sqrt{\frac{3}{70}} C_{33,11}^{33,1}\)

\noindent\(C_{11,4212,1}^{00,1101,1,\text{Loc}}= \frac{\sqrt{5} C_{11,11}^{51,1}}{64}+\frac{\sqrt{3} C_{22,11}^{42,1}}{32}+\frac{\sqrt{5}
C_{31,11}^{31,1}}{64}+\frac{9 C_{33,11}^{33,1}}{16 \sqrt{70}}\)

\noindent\(C_{11,4212,1}^{00,1101,1,\text{NonLoc}}= \frac{\sqrt{5} C_{11,11}^{51,0}}{64}-\frac{\sqrt{3} C_{22,11}^{42,0}}{32}+\frac{\sqrt{5}
C_{31,11}^{31,0}}{64}+\frac{9 C_{33,11}^{33,0}}{16 \sqrt{70}}\)

\noindent\(C_{11,4212,1}^{11,0011,0,\text{Loc}}= -\frac{\sqrt{5} C_{00,20}^{60,0}}{64}-\frac{\sqrt{5} C_{00,20}^{60,1}}{128}+\frac{3
C_{00,22}^{62,0}}{128}+\frac{3 C_{00,22}^{62,1}}{256}-\frac{\sqrt{5} C_{11,20}^{31,0}}{192}-\frac{\sqrt{5} C_{11,20}^{31,1}}{384}+\frac{C_{11,22}^{51,0}}{192}+\frac{C_{11,22}^{51,1}}{384}+\frac{9}{256}
\sqrt{\frac{3}{35}} C_{11,22}^{53,0}+\frac{9}{512} \sqrt{\frac{3}{35}} C_{11,22}^{53,1}+\frac{\sqrt{5} C_{20,20}^{40,0}}{192}+\frac{\sqrt{5} C_{20,20}^{40,1}}{384}-\frac{C_{20,22}^{42,0}}{192}-\frac{C_{20,22}^{42,1}}{384}-\frac{C_{22,20}^{42,0}}{192}-\frac{C_{22,20}^{42,1}}{384}-\frac{5
C_{22,22}^{40,0}}{384}-\frac{5 C_{22,22}^{40,1}}{768}+\frac{\sqrt{7} C_{22,22}^{42,0}}{192}+\frac{\sqrt{7} C_{22,22}^{42,1}}{384}+\frac{\sqrt{5}
C_{31,20}^{31,0}}{192}+\frac{\sqrt{5} C_{31,20}^{31,1}}{384}-\frac{C_{31,22}^{31,0}}{192}-\frac{C_{31,22}^{31,1}}{384}-\frac{9}{256} \sqrt{\frac{3}{35}}
C_{31,22}^{33,0}-\frac{9}{512} \sqrt{\frac{3}{35}} C_{31,22}^{33,1}-\frac{3}{64} \sqrt{\frac{3}{35}} C_{33,20}^{33,0}-\frac{3}{128} \sqrt{\frac{3}{35}}
C_{33,20}^{33,1}+\frac{9 C_{33,22}^{33,0}}{80 \sqrt{14}}+\frac{9 C_{33,22}^{33,1}}{160 \sqrt{14}}\)

\noindent\(C_{11,4212,1}^{11,0011,0,\text{NonLoc}}= -\frac{\sqrt{15} C_{00,00}^{60,0}}{256}-\frac{\sqrt{15} C_{00,00}^{60,1}}{128}-\frac{\sqrt{5}
C_{00,20}^{60,0}}{256}-\frac{\sqrt{5} C_{00,20}^{60,1}}{128}-\frac{3 C_{00,22}^{62,0}}{256}-\frac{3 C_{00,22}^{62,1}}{128}+\frac{1}{256} \sqrt{\frac{5}{3}}
C_{11,00}^{51,0}+\frac{1}{128} \sqrt{\frac{5}{3}} C_{11,00}^{51,1}+\frac{\sqrt{5} C_{11,20}^{31,0}}{768}+\frac{\sqrt{5} C_{11,20}^{31,1}}{384}+\frac{C_{11,22}^{51,0}}{384}+\frac{C_{11,22}^{51,1}}{192}+\frac{9}{512}
\sqrt{\frac{3}{35}} C_{11,22}^{53,0}+\frac{9}{256} \sqrt{\frac{3}{35}} C_{11,22}^{53,1}+\frac{1}{256} \sqrt{\frac{5}{3}} C_{20,00}^{40,0}+\frac{1}{128}
\sqrt{\frac{5}{3}} C_{20,00}^{40,1}+\frac{\sqrt{5} C_{20,20}^{40,0}}{768}+\frac{\sqrt{5} C_{20,20}^{40,1}}{384}+\frac{C_{20,22}^{42,0}}{384}+\frac{C_{20,22}^{42,1}}{192}-\frac{C_{22,00}^{42,0}}{256
\sqrt{3}}-\frac{C_{22,00}^{42,1}}{128 \sqrt{3}}-\frac{C_{22,20}^{42,0}}{768}-\frac{C_{22,20}^{42,1}}{384}+\frac{5 C_{22,22}^{40,0}}{768}+\frac{5
C_{22,22}^{40,1}}{384}-\frac{\sqrt{7} C_{22,22}^{42,0}}{384}-\frac{\sqrt{7} C_{22,22}^{42,1}}{192}-\frac{1}{256} \sqrt{\frac{5}{3}} C_{31,00}^{31,0}-\frac{1}{128}
\sqrt{\frac{5}{3}} C_{31,00}^{31,1}-\frac{\sqrt{5} C_{31,20}^{31,0}}{768}-\frac{\sqrt{5} C_{31,20}^{31,1}}{384}-\frac{C_{31,22}^{31,0}}{384}-\frac{C_{31,22}^{31,1}}{192}-\frac{9}{512}
\sqrt{\frac{3}{35}} C_{31,22}^{33,0}-\frac{9}{256} \sqrt{\frac{3}{35}} C_{31,22}^{33,1}+\frac{9 C_{33,00}^{33,0}}{256 \sqrt{35}}+\frac{9 C_{33,00}^{33,1}}{128
\sqrt{35}}+\frac{3}{256} \sqrt{\frac{3}{35}} C_{33,20}^{33,0}+\frac{3}{128} \sqrt{\frac{3}{35}} C_{33,20}^{33,1}+\frac{9 C_{33,22}^{33,0}}{160 \sqrt{14}}+\frac{9
C_{33,22}^{33,1}}{80 \sqrt{14}}\)

\noindent\(C_{11,4212,1}^{11,0011,1,\text{Loc}}= -\frac{\sqrt{15} C_{00,20}^{60,1}}{128}+\frac{3 \sqrt{3} C_{00,22}^{62,1}}{256}-\frac{1}{128}
\sqrt{\frac{5}{3}} C_{11,20}^{31,1}+\frac{C_{11,22}^{51,1}}{128 \sqrt{3}}+\frac{27 C_{11,22}^{53,1}}{512 \sqrt{35}}+\frac{1}{128} \sqrt{\frac{5}{3}}
C_{20,20}^{40,1}-\frac{C_{20,22}^{42,1}}{128 \sqrt{3}}-\frac{C_{22,20}^{42,1}}{128 \sqrt{3}}-\frac{5 C_{22,22}^{40,1}}{256 \sqrt{3}}+\frac{1}{128}
\sqrt{\frac{7}{3}} C_{22,22}^{42,1}+\frac{1}{128} \sqrt{\frac{5}{3}} C_{31,20}^{31,1}-\frac{C_{31,22}^{31,1}}{128 \sqrt{3}}-\frac{27 C_{31,22}^{33,1}}{512
\sqrt{35}}-\frac{9 C_{33,20}^{33,1}}{128 \sqrt{35}}+\frac{9}{160} \sqrt{\frac{3}{14}} C_{33,22}^{33,1}\)

\noindent\(C_{11,4212,1}^{11,0011,1,\text{NonLoc}}= -\frac{3 \sqrt{5} C_{00,00}^{60,0}}{256}-\frac{\sqrt{15} C_{00,20}^{60,0}}{256}-\frac{3
\sqrt{3} C_{00,22}^{62,0}}{256}+\frac{\sqrt{5} C_{11,00}^{51,0}}{256}+\frac{1}{256} \sqrt{\frac{5}{3}} C_{11,20}^{31,0}+\frac{C_{11,22}^{51,0}}{128
\sqrt{3}}+\frac{27 C_{11,22}^{53,0}}{512 \sqrt{35}}+\frac{\sqrt{5} C_{20,00}^{40,0}}{256}+\frac{1}{256} \sqrt{\frac{5}{3}} C_{20,20}^{40,0}+\frac{C_{20,22}^{42,0}}{128
\sqrt{3}}-\frac{C_{22,00}^{42,0}}{256}-\frac{C_{22,20}^{42,0}}{256 \sqrt{3}}+\frac{5 C_{22,22}^{40,0}}{256 \sqrt{3}}-\frac{1}{128} \sqrt{\frac{7}{3}}
C_{22,22}^{42,0}-\frac{\sqrt{5} C_{31,00}^{31,0}}{256}-\frac{1}{256} \sqrt{\frac{5}{3}} C_{31,20}^{31,0}-\frac{C_{31,22}^{31,0}}{128 \sqrt{3}}-\frac{27
C_{31,22}^{33,0}}{512 \sqrt{35}}+\frac{9}{256} \sqrt{\frac{3}{35}} C_{33,00}^{33,0}+\frac{9 C_{33,20}^{33,0}}{256 \sqrt{35}}+\frac{9}{160} \sqrt{\frac{3}{14}}
C_{33,22}^{33,0}\)

\noindent\(C_{11,4212,2}^{11,0011,0,\text{Loc}}= \frac{1}{64} \sqrt{\frac{5}{3}} C_{00,20}^{60,0}+\frac{1}{128} \sqrt{\frac{5}{3}}
C_{00,20}^{60,1}-\frac{\sqrt{3} C_{00,22}^{62,0}}{128}-\frac{\sqrt{3} C_{00,22}^{62,1}}{256}+\frac{1}{192} \sqrt{\frac{5}{3}} C_{11,20}^{31,0}+\frac{1}{384}
\sqrt{\frac{5}{3}} C_{11,20}^{31,1}-\frac{C_{11,22}^{51,0}}{192 \sqrt{3}}-\frac{C_{11,22}^{51,1}}{384 \sqrt{3}}-\frac{9 C_{11,22}^{53,0}}{256 \sqrt{35}}-\frac{9
C_{11,22}^{53,1}}{512 \sqrt{35}}-\frac{1}{192} \sqrt{\frac{5}{3}} C_{20,20}^{40,0}-\frac{1}{384} \sqrt{\frac{5}{3}} C_{20,20}^{40,1}+\frac{C_{20,22}^{42,0}}{192
\sqrt{3}}+\frac{C_{20,22}^{42,1}}{384 \sqrt{3}}+\frac{C_{22,20}^{42,0}}{192 \sqrt{3}}+\frac{C_{22,20}^{42,1}}{384 \sqrt{3}}+\frac{5 C_{22,22}^{40,0}}{384
\sqrt{3}}+\frac{5 C_{22,22}^{40,1}}{768 \sqrt{3}}-\frac{1}{192} \sqrt{\frac{7}{3}} C_{22,22}^{42,0}-\frac{1}{384} \sqrt{\frac{7}{3}} C_{22,22}^{42,1}-\frac{1}{192}
\sqrt{\frac{5}{3}} C_{31,20}^{31,0}-\frac{1}{384} \sqrt{\frac{5}{3}} C_{31,20}^{31,1}+\frac{C_{31,22}^{31,0}}{192 \sqrt{3}}+\frac{C_{31,22}^{31,1}}{384
\sqrt{3}}+\frac{9 C_{31,22}^{33,0}}{256 \sqrt{35}}+\frac{9 C_{31,22}^{33,1}}{512 \sqrt{35}}+\frac{3 C_{33,20}^{33,0}}{64 \sqrt{35}}+\frac{3 C_{33,20}^{33,1}}{128
\sqrt{35}}-\frac{3}{80} \sqrt{\frac{3}{14}} C_{33,22}^{33,0}-\frac{3}{160} \sqrt{\frac{3}{14}} C_{33,22}^{33,1}\)

\noindent\(C_{11,4212,2}^{11,0011,0,\text{NonLoc}}= \frac{\sqrt{5} C_{00,00}^{60,0}}{256}+\frac{\sqrt{5} C_{00,00}^{60,1}}{128}+\frac{1}{256}
\sqrt{\frac{5}{3}} C_{00,20}^{60,0}+\frac{1}{128} \sqrt{\frac{5}{3}} C_{00,20}^{60,1}+\frac{\sqrt{3} C_{00,22}^{62,0}}{256}+\frac{\sqrt{3} C_{00,22}^{62,1}}{128}-\frac{\sqrt{5}
C_{11,00}^{51,0}}{768}-\frac{\sqrt{5} C_{11,00}^{51,1}}{384}-\frac{1}{768} \sqrt{\frac{5}{3}} C_{11,20}^{31,0}-\frac{1}{384} \sqrt{\frac{5}{3}} C_{11,20}^{31,1}-\frac{C_{11,22}^{51,0}}{384
\sqrt{3}}-\frac{C_{11,22}^{51,1}}{192 \sqrt{3}}-\frac{9 C_{11,22}^{53,0}}{512 \sqrt{35}}-\frac{9 C_{11,22}^{53,1}}{256 \sqrt{35}}-\frac{\sqrt{5}
C_{20,00}^{40,0}}{768}-\frac{\sqrt{5} C_{20,00}^{40,1}}{384}-\frac{1}{768} \sqrt{\frac{5}{3}} C_{20,20}^{40,0}-\frac{1}{384} \sqrt{\frac{5}{3}} C_{20,20}^{40,1}-\frac{C_{20,22}^{42,0}}{384
\sqrt{3}}-\frac{C_{20,22}^{42,1}}{192 \sqrt{3}}+\frac{C_{22,00}^{42,0}}{768}+\frac{C_{22,00}^{42,1}}{384}+\frac{C_{22,20}^{42,0}}{768 \sqrt{3}}+\frac{C_{22,20}^{42,1}}{384
\sqrt{3}}-\frac{5 C_{22,22}^{40,0}}{768 \sqrt{3}}-\frac{5 C_{22,22}^{40,1}}{384 \sqrt{3}}+\frac{1}{384} \sqrt{\frac{7}{3}} C_{22,22}^{42,0}+\frac{1}{192}
\sqrt{\frac{7}{3}} C_{22,22}^{42,1}+\frac{\sqrt{5} C_{31,00}^{31,0}}{768}+\frac{\sqrt{5} C_{31,00}^{31,1}}{384}+\frac{1}{768} \sqrt{\frac{5}{3}}
C_{31,20}^{31,0}+\frac{1}{384} \sqrt{\frac{5}{3}} C_{31,20}^{31,1}+\frac{C_{31,22}^{31,0}}{384 \sqrt{3}}+\frac{C_{31,22}^{31,1}}{192 \sqrt{3}}+\frac{9
C_{31,22}^{33,0}}{512 \sqrt{35}}+\frac{9 C_{31,22}^{33,1}}{256 \sqrt{35}}-\frac{3}{256} \sqrt{\frac{3}{35}} C_{33,00}^{33,0}-\frac{3}{128} \sqrt{\frac{3}{35}}
C_{33,00}^{33,1}-\frac{3 C_{33,20}^{33,0}}{256 \sqrt{35}}-\frac{3 C_{33,20}^{33,1}}{128 \sqrt{35}}-\frac{3}{160} \sqrt{\frac{3}{14}} C_{33,22}^{33,0}-\frac{3}{80}
\sqrt{\frac{3}{14}} C_{33,22}^{33,1}\)

\noindent\(C_{11,4212,2}^{11,0011,1,\text{Loc}}= \frac{\sqrt{5} C_{00,20}^{60,1}}{128}-\frac{3 C_{00,22}^{62,1}}{256}+\frac{\sqrt{5}
C_{11,20}^{31,1}}{384}-\frac{C_{11,22}^{51,1}}{384}-\frac{9}{512} \sqrt{\frac{3}{35}} C_{11,22}^{53,1}-\frac{\sqrt{5} C_{20,20}^{40,1}}{384}+\frac{C_{20,22}^{42,1}}{384}+\frac{C_{22,20}^{42,1}}{384}+\frac{5
C_{22,22}^{40,1}}{768}-\frac{\sqrt{7} C_{22,22}^{42,1}}{384}-\frac{\sqrt{5} C_{31,20}^{31,1}}{384}+\frac{C_{31,22}^{31,1}}{384}+\frac{9}{512} \sqrt{\frac{3}{35}}
C_{31,22}^{33,1}+\frac{3}{128} \sqrt{\frac{3}{35}} C_{33,20}^{33,1}-\frac{9 C_{33,22}^{33,1}}{160 \sqrt{14}}\)

\noindent\(C_{11,4212,2}^{11,0011,1,\text{NonLoc}}= \frac{\sqrt{15} C_{00,00}^{60,0}}{256}+\frac{\sqrt{5} C_{00,20}^{60,0}}{256}+\frac{3
C_{00,22}^{62,0}}{256}-\frac{1}{256} \sqrt{\frac{5}{3}} C_{11,00}^{51,0}-\frac{\sqrt{5} C_{11,20}^{31,0}}{768}-\frac{C_{11,22}^{51,0}}{384}-\frac{9}{512}
\sqrt{\frac{3}{35}} C_{11,22}^{53,0}-\frac{1}{256} \sqrt{\frac{5}{3}} C_{20,00}^{40,0}-\frac{\sqrt{5} C_{20,20}^{40,0}}{768}-\frac{C_{20,22}^{42,0}}{384}+\frac{C_{22,00}^{42,0}}{256
\sqrt{3}}+\frac{C_{22,20}^{42,0}}{768}-\frac{5 C_{22,22}^{40,0}}{768}+\frac{\sqrt{7} C_{22,22}^{42,0}}{384}+\frac{1}{256} \sqrt{\frac{5}{3}} C_{31,00}^{31,0}+\frac{\sqrt{5}
C_{31,20}^{31,0}}{768}+\frac{C_{31,22}^{31,0}}{384}+\frac{9}{512} \sqrt{\frac{3}{35}} C_{31,22}^{33,0}-\frac{9 C_{33,00}^{33,0}}{256 \sqrt{35}}-\frac{3}{256}
\sqrt{\frac{3}{35}} C_{33,20}^{33,0}-\frac{9 C_{33,22}^{33,0}}{160 \sqrt{14}}\)

\noindent\(C_{11,4213,2}^{11,0011,0,\text{Loc}}= -\frac{1}{32} \sqrt{\frac{7}{3}} C_{00,20}^{60,0}-\frac{1}{64} \sqrt{\frac{7}{3}}
C_{00,20}^{60,1}-\frac{1}{32} \sqrt{\frac{3}{35}} C_{00,22}^{62,0}-\frac{1}{64} \sqrt{\frac{3}{35}} C_{00,22}^{62,1}-\frac{1}{96} \sqrt{\frac{7}{3}}
C_{11,20}^{31,0}-\frac{1}{192} \sqrt{\frac{7}{3}} C_{11,20}^{31,1}-\frac{C_{11,22}^{51,0}}{48 \sqrt{105}}-\frac{C_{11,22}^{51,1}}{96 \sqrt{105}}-\frac{9
C_{11,22}^{53,0}}{2240}-\frac{9 C_{11,22}^{53,1}}{4480}+\frac{1}{96} \sqrt{\frac{7}{3}} C_{20,20}^{40,0}+\frac{1}{192} \sqrt{\frac{7}{3}} C_{20,20}^{40,1}+\frac{C_{20,22}^{42,0}}{48
\sqrt{105}}+\frac{C_{20,22}^{42,1}}{96 \sqrt{105}}-\frac{1}{96} \sqrt{\frac{7}{15}} C_{22,20}^{42,0}-\frac{1}{192} \sqrt{\frac{7}{15}} C_{22,20}^{42,1}+\frac{1}{96}
\sqrt{\frac{5}{21}} C_{22,22}^{40,0}+\frac{1}{192} \sqrt{\frac{5}{21}} C_{22,22}^{40,1}-\frac{C_{22,22}^{42,0}}{48 \sqrt{15}}-\frac{C_{22,22}^{42,1}}{96
\sqrt{15}}+\frac{1}{96} \sqrt{\frac{7}{3}} C_{31,20}^{31,0}+\frac{1}{192} \sqrt{\frac{7}{3}} C_{31,20}^{31,1}+\frac{C_{31,22}^{31,0}}{48 \sqrt{105}}+\frac{C_{31,22}^{31,1}}{96
\sqrt{105}}+\frac{9 C_{31,22}^{33,0}}{2240}+\frac{9 C_{31,22}^{33,1}}{4480}-\frac{3 C_{33,20}^{33,0}}{160}-\frac{3 C_{33,20}^{33,1}}{320}-\frac{3}{140}
\sqrt{\frac{3}{10}} C_{33,22}^{33,0}-\frac{3}{280} \sqrt{\frac{3}{10}} C_{33,22}^{33,1}\)

\noindent\(C_{11,4213,2}^{11,0011,0,\text{NonLoc}}= -\frac{\sqrt{7} C_{00,00}^{60,0}}{128}-\frac{\sqrt{7} C_{00,00}^{60,1}}{64}-\frac{1}{128}
\sqrt{\frac{7}{3}} C_{00,20}^{60,0}-\frac{1}{64} \sqrt{\frac{7}{3}} C_{00,20}^{60,1}+\frac{1}{64} \sqrt{\frac{3}{35}} C_{00,22}^{62,0}+\frac{1}{32}
\sqrt{\frac{3}{35}} C_{00,22}^{62,1}+\frac{\sqrt{7} C_{11,00}^{51,0}}{384}+\frac{\sqrt{7} C_{11,00}^{51,1}}{192}+\frac{1}{384} \sqrt{\frac{7}{3}}
C_{11,20}^{31,0}+\frac{1}{192} \sqrt{\frac{7}{3}} C_{11,20}^{31,1}-\frac{C_{11,22}^{51,0}}{96 \sqrt{105}}-\frac{C_{11,22}^{51,1}}{48 \sqrt{105}}-\frac{9
C_{11,22}^{53,0}}{4480}-\frac{9 C_{11,22}^{53,1}}{2240}+\frac{\sqrt{7} C_{20,00}^{40,0}}{384}+\frac{\sqrt{7} C_{20,00}^{40,1}}{192}+\frac{1}{384}
\sqrt{\frac{7}{3}} C_{20,20}^{40,0}+\frac{1}{192} \sqrt{\frac{7}{3}} C_{20,20}^{40,1}-\frac{C_{20,22}^{42,0}}{96 \sqrt{105}}-\frac{C_{20,22}^{42,1}}{48
\sqrt{105}}-\frac{1}{384} \sqrt{\frac{7}{5}} C_{22,00}^{42,0}-\frac{1}{192} \sqrt{\frac{7}{5}} C_{22,00}^{42,1}-\frac{1}{384} \sqrt{\frac{7}{15}}
C_{22,20}^{42,0}-\frac{1}{192} \sqrt{\frac{7}{15}} C_{22,20}^{42,1}-\frac{1}{192} \sqrt{\frac{5}{21}} C_{22,22}^{40,0}-\frac{1}{96} \sqrt{\frac{5}{21}}
C_{22,22}^{40,1}+\frac{C_{22,22}^{42,0}}{96 \sqrt{15}}+\frac{C_{22,22}^{42,1}}{48 \sqrt{15}}-\frac{\sqrt{7} C_{31,00}^{31,0}}{384}-\frac{\sqrt{7}
C_{31,00}^{31,1}}{192}-\frac{1}{384} \sqrt{\frac{7}{3}} C_{31,20}^{31,0}-\frac{1}{192} \sqrt{\frac{7}{3}} C_{31,20}^{31,1}+\frac{C_{31,22}^{31,0}}{96
\sqrt{105}}+\frac{C_{31,22}^{31,1}}{48 \sqrt{105}}+\frac{9 C_{31,22}^{33,0}}{4480}+\frac{9 C_{31,22}^{33,1}}{2240}+\frac{3 \sqrt{3} C_{33,00}^{33,0}}{640}+\frac{3
\sqrt{3} C_{33,00}^{33,1}}{320}+\frac{3 C_{33,20}^{33,0}}{640}+\frac{3 C_{33,20}^{33,1}}{320}-\frac{3}{280} \sqrt{\frac{3}{10}} C_{33,22}^{33,0}-\frac{3}{140}
\sqrt{\frac{3}{10}} C_{33,22}^{33,1}\)

\noindent\(C_{11,4213,2}^{11,0011,1,\text{Loc}}= -\frac{\sqrt{7} C_{00,20}^{60,1}}{64}-\frac{3 C_{00,22}^{62,1}}{64 \sqrt{35}}-\frac{\sqrt{7}
C_{11,20}^{31,1}}{192}-\frac{C_{11,22}^{51,1}}{96 \sqrt{35}}-\frac{9 \sqrt{3} C_{11,22}^{53,1}}{4480}+\frac{\sqrt{7} C_{20,20}^{40,1}}{192}+\frac{C_{20,22}^{42,1}}{96
\sqrt{35}}-\frac{1}{192} \sqrt{\frac{7}{5}} C_{22,20}^{42,1}+\frac{1}{192} \sqrt{\frac{5}{7}} C_{22,22}^{40,1}-\frac{C_{22,22}^{42,1}}{96 \sqrt{5}}+\frac{\sqrt{7}
C_{31,20}^{31,1}}{192}+\frac{C_{31,22}^{31,1}}{96 \sqrt{35}}+\frac{9 \sqrt{3} C_{31,22}^{33,1}}{4480}-\frac{3 \sqrt{3} C_{33,20}^{33,1}}{320}-\frac{9
C_{33,22}^{33,1}}{280 \sqrt{10}}\)

\noindent\(C_{11,4213,2}^{11,0011,1,\text{NonLoc}}= -\frac{\sqrt{21} C_{00,00}^{60,0}}{128}-\frac{\sqrt{7} C_{00,20}^{60,0}}{128}+\frac{3
C_{00,22}^{62,0}}{64 \sqrt{35}}+\frac{1}{128} \sqrt{\frac{7}{3}} C_{11,00}^{51,0}+\frac{\sqrt{7} C_{11,20}^{31,0}}{384}-\frac{C_{11,22}^{51,0}}{96
\sqrt{35}}-\frac{9 \sqrt{3} C_{11,22}^{53,0}}{4480}+\frac{1}{128} \sqrt{\frac{7}{3}} C_{20,00}^{40,0}+\frac{\sqrt{7} C_{20,20}^{40,0}}{384}-\frac{C_{20,22}^{42,0}}{96
\sqrt{35}}-\frac{1}{128} \sqrt{\frac{7}{15}} C_{22,00}^{42,0}-\frac{1}{384} \sqrt{\frac{7}{5}} C_{22,20}^{42,0}-\frac{1}{192} \sqrt{\frac{5}{7}}
C_{22,22}^{40,0}+\frac{C_{22,22}^{42,0}}{96 \sqrt{5}}-\frac{1}{128} \sqrt{\frac{7}{3}} C_{31,00}^{31,0}-\frac{\sqrt{7} C_{31,20}^{31,0}}{384}+\frac{C_{31,22}^{31,0}}{96
\sqrt{35}}+\frac{9 \sqrt{3} C_{31,22}^{33,0}}{4480}+\frac{9 C_{33,00}^{33,0}}{640}+\frac{3 \sqrt{3} C_{33,20}^{33,0}}{640}-\frac{9 C_{33,22}^{33,0}}{280
\sqrt{10}}\)

\noindent\(C_{11,4413,2}^{11,0011,0,\text{Loc}}= -\frac{C_{00,22}^{62,0}}{32 \sqrt{5}}-\frac{C_{00,22}^{62,1}}{64 \sqrt{5}}-\frac{C_{11,22}^{51,0}}{32
\sqrt{5}}-\frac{C_{11,22}^{51,1}}{64 \sqrt{5}}+\frac{1}{80} \sqrt{\frac{3}{7}} C_{11,22}^{53,0}+\frac{1}{160} \sqrt{\frac{3}{7}} C_{11,22}^{53,1}+\frac{C_{20,22}^{42,0}}{32
\sqrt{5}}+\frac{C_{20,22}^{42,1}}{64 \sqrt{5}}-\frac{C_{22,22}^{40,0}}{32 \sqrt{5}}-\frac{C_{22,22}^{40,1}}{64 \sqrt{5}}-\frac{C_{22,22}^{44,0}}{160}-\frac{C_{22,22}^{44,1}}{320}+\frac{C_{31,22}^{31,0}}{32
\sqrt{5}}+\frac{C_{31,22}^{31,1}}{64 \sqrt{5}}-\frac{1}{80} \sqrt{\frac{3}{7}} C_{31,22}^{33,0}-\frac{1}{160} \sqrt{\frac{3}{7}} C_{31,22}^{33,1}-\frac{3
C_{33,22}^{33,0}}{160 \sqrt{70}}-\frac{3 C_{33,22}^{33,1}}{320 \sqrt{70}}\)

\noindent\(C_{11,4413,2}^{11,0011,0,\text{NonLoc}}= \frac{C_{00,22}^{62,0}}{64 \sqrt{5}}+\frac{C_{00,22}^{62,1}}{32 \sqrt{5}}-\frac{C_{11,22}^{51,0}}{64
\sqrt{5}}-\frac{C_{11,22}^{51,1}}{32 \sqrt{5}}+\frac{1}{160} \sqrt{\frac{3}{7}} C_{11,22}^{53,0}+\frac{1}{80} \sqrt{\frac{3}{7}} C_{11,22}^{53,1}-\frac{C_{20,22}^{42,0}}{64
\sqrt{5}}-\frac{C_{20,22}^{42,1}}{32 \sqrt{5}}+\frac{C_{22,22}^{40,0}}{64 \sqrt{5}}+\frac{C_{22,22}^{40,1}}{32 \sqrt{5}}+\frac{C_{22,22}^{44,0}}{320}+\frac{C_{22,22}^{44,1}}{160}+\frac{C_{31,22}^{31,0}}{64
\sqrt{5}}+\frac{C_{31,22}^{31,1}}{32 \sqrt{5}}-\frac{1}{160} \sqrt{\frac{3}{7}} C_{31,22}^{33,0}-\frac{1}{80} \sqrt{\frac{3}{7}} C_{31,22}^{33,1}-\frac{3
C_{33,22}^{33,0}}{320 \sqrt{70}}-\frac{3 C_{33,22}^{33,1}}{160 \sqrt{70}}\)

\noindent\(C_{11,4413,2}^{11,0011,1,\text{Loc}}= -\frac{1}{64} \sqrt{\frac{3}{5}} C_{00,22}^{62,1}-\frac{1}{64} \sqrt{\frac{3}{5}}
C_{11,22}^{51,1}+\frac{3 C_{11,22}^{53,1}}{160 \sqrt{7}}+\frac{1}{64} \sqrt{\frac{3}{5}} C_{20,22}^{42,1}-\frac{1}{64} \sqrt{\frac{3}{5}} C_{22,22}^{40,1}-\frac{\sqrt{3}
C_{22,22}^{44,1}}{320}+\frac{1}{64} \sqrt{\frac{3}{5}} C_{31,22}^{31,1}-\frac{3 C_{31,22}^{33,1}}{160 \sqrt{7}}-\frac{3}{320} \sqrt{\frac{3}{70}}
C_{33,22}^{33,1}\)

\noindent\(C_{11,4413,2}^{11,0011,1,\text{NonLoc}}= \frac{1}{64} \sqrt{\frac{3}{5}} C_{00,22}^{62,0}-\frac{1}{64} \sqrt{\frac{3}{5}}
C_{11,22}^{51,0}+\frac{3 C_{11,22}^{53,0}}{160 \sqrt{7}}-\frac{1}{64} \sqrt{\frac{3}{5}} C_{20,22}^{42,0}+\frac{1}{64} \sqrt{\frac{3}{5}} C_{22,22}^{40,0}+\frac{\sqrt{3}
C_{22,22}^{44,0}}{320}+\frac{1}{64} \sqrt{\frac{3}{5}} C_{31,22}^{31,0}-\frac{3 C_{31,22}^{33,0}}{160 \sqrt{7}}-\frac{3}{320} \sqrt{\frac{3}{70}}
C_{33,22}^{33,0}\)

\noindent\(C_{11,5101,1}^{00,0011,0,\text{Loc}}= -\frac{C_{11,11}^{51,0}}{64}-\frac{C_{11,11}^{51,1}}{128}-\frac{1}{32} \sqrt{\frac{3}{5}}
C_{22,11}^{42,0}-\frac{1}{64} \sqrt{\frac{3}{5}} C_{22,11}^{42,1}-\frac{C_{31,11}^{31,0}}{64}-\frac{C_{31,11}^{31,1}}{128}-\frac{9 C_{33,11}^{33,0}}{80
\sqrt{14}}-\frac{9 C_{33,11}^{33,1}}{160 \sqrt{14}}\)

\noindent\(C_{11,5101,1}^{00,0011,0,\text{NonLoc}}= -\frac{C_{11,11}^{51,0}}{128}-\frac{C_{11,11}^{51,1}}{64}+\frac{1}{64} \sqrt{\frac{3}{5}}
C_{22,11}^{42,0}+\frac{1}{32} \sqrt{\frac{3}{5}} C_{22,11}^{42,1}-\frac{C_{31,11}^{31,0}}{128}-\frac{C_{31,11}^{31,1}}{64}-\frac{9 C_{33,11}^{33,0}}{160
\sqrt{14}}-\frac{9 C_{33,11}^{33,1}}{80 \sqrt{14}}\)

\noindent\(C_{11,5101,1}^{00,0011,1,\text{Loc}}= -\frac{\sqrt{3} C_{11,11}^{51,1}}{128}-\frac{3 C_{22,11}^{42,1}}{64 \sqrt{5}}-\frac{\sqrt{3}
C_{31,11}^{31,1}}{128}-\frac{9}{160} \sqrt{\frac{3}{14}} C_{33,11}^{33,1}\)

\noindent\(C_{11,5101,1}^{00,0011,1,\text{NonLoc}}= -\frac{\sqrt{3} C_{11,11}^{51,0}}{128}+\frac{3 C_{22,11}^{42,0}}{64 \sqrt{5}}-\frac{\sqrt{3}
C_{31,11}^{31,0}}{128}-\frac{9}{160} \sqrt{\frac{3}{14}} C_{33,11}^{33,0}\)

\noindent\(C_{11,5111,0}^{00,0000,0,\text{Loc}}= -\frac{C_{11,11}^{51,0}}{64}-\frac{C_{11,11}^{51,1}}{128}-\frac{1}{32} \sqrt{\frac{3}{5}}
C_{22,11}^{42,0}-\frac{1}{64} \sqrt{\frac{3}{5}} C_{22,11}^{42,1}-\frac{C_{31,11}^{31,0}}{64}-\frac{C_{31,11}^{31,1}}{128}-\frac{9 C_{33,11}^{33,0}}{80
\sqrt{14}}-\frac{9 C_{33,11}^{33,1}}{160 \sqrt{14}}\)

\noindent\(C_{11,5111,0}^{00,0000,0,\text{NonLoc}}= -\frac{C_{11,11}^{51,0}}{128}-\frac{C_{11,11}^{51,1}}{64}+\frac{1}{64} \sqrt{\frac{3}{5}}
C_{22,11}^{42,0}+\frac{1}{32} \sqrt{\frac{3}{5}} C_{22,11}^{42,1}-\frac{C_{31,11}^{31,0}}{128}-\frac{C_{31,11}^{31,1}}{64}-\frac{9 C_{33,11}^{33,0}}{160
\sqrt{14}}-\frac{9 C_{33,11}^{33,1}}{80 \sqrt{14}}\)

\noindent\(C_{11,5111,0}^{00,0000,1,\text{Loc}}= -\frac{\sqrt{3} C_{11,11}^{51,1}}{128}-\frac{3 C_{22,11}^{42,1}}{64 \sqrt{5}}-\frac{\sqrt{3}
C_{31,11}^{31,1}}{128}-\frac{9}{160} \sqrt{\frac{3}{14}} C_{33,11}^{33,1}\)

\noindent\(C_{11,5111,0}^{00,0000,1,\text{NonLoc}}= -\frac{\sqrt{3} C_{11,11}^{51,0}}{128}+\frac{3 C_{22,11}^{42,0}}{64 \sqrt{5}}-\frac{\sqrt{3}
C_{31,11}^{31,0}}{128}-\frac{9}{160} \sqrt{\frac{3}{14}} C_{33,11}^{33,0}\)

\noindent\(C_{20,0000,0}^{00,4000,0,\text{Loc}}= -\frac{7 C_{00,00}^{60,0}}{256}-\frac{7 C_{00,00}^{60,1}}{512}-\frac{7 C_{11,00}^{51,0}}{768}-\frac{7
C_{11,00}^{51,1}}{1536}-\frac{5 C_{20,00}^{40,0}}{768}-\frac{5 C_{20,00}^{40,1}}{1536}+\frac{7 C_{22,00}^{42,0}}{384 \sqrt{5}}+\frac{7 C_{22,00}^{42,1}}{768
\sqrt{5}}+\frac{C_{31,00}^{31,0}}{768}+\frac{C_{31,00}^{31,1}}{1536}+\frac{3 \sqrt{21} C_{33,00}^{33,0}}{640}+\frac{3 \sqrt{21} C_{33,00}^{33,1}}{1280}\)

\noindent\(C_{20,0000,0}^{00,4000,0,\text{NonLoc}}= \frac{7 C_{00,00}^{60,0}}{1024}+\frac{7 C_{00,00}^{60,1}}{512}-\frac{7 \sqrt{3}
C_{00,20}^{60,0}}{1024}-\frac{7 \sqrt{3} C_{00,20}^{60,1}}{512}-\frac{7 C_{11,00}^{51,0}}{3072}-\frac{7 C_{11,00}^{51,1}}{1536}+\frac{7 C_{11,20}^{31,0}}{2048
\sqrt{3}}+\frac{7 C_{11,20}^{31,1}}{1024 \sqrt{3}}+\frac{5 C_{20,00}^{40,0}}{3072}+\frac{5 C_{20,00}^{40,1}}{1536}-\frac{5 C_{20,20}^{40,0}}{1024
\sqrt{3}}-\frac{5 C_{20,20}^{40,1}}{512 \sqrt{3}}-\frac{7 C_{22,00}^{42,0}}{1536 \sqrt{5}}-\frac{7 C_{22,00}^{42,1}}{768 \sqrt{5}}+\frac{7 C_{22,20}^{42,0}}{512
\sqrt{15}}+\frac{7 C_{22,20}^{42,1}}{256 \sqrt{15}}+\frac{C_{31,00}^{31,0}}{3072}+\frac{C_{31,00}^{31,1}}{1536}-\frac{C_{31,20}^{31,0}}{1024 \sqrt{3}}-\frac{C_{31,20}^{31,1}}{512
\sqrt{3}}+\frac{3 \sqrt{21} C_{33,00}^{33,0}}{2560}+\frac{3 \sqrt{21} C_{33,00}^{33,1}}{1280}-\frac{9 \sqrt{7} C_{33,20}^{33,0}}{2560}-\frac{9 \sqrt{7}
C_{33,20}^{33,1}}{1280}\)

\noindent\(C_{20,0000,0}^{00,4000,1,\text{Loc}}= -\frac{7 \sqrt{3} C_{00,00}^{60,1}}{512}-\frac{7 C_{11,00}^{51,1}}{512 \sqrt{3}}-\frac{5
C_{20,00}^{40,1}}{512 \sqrt{3}}+\frac{7 C_{22,00}^{42,1}}{256 \sqrt{15}}+\frac{C_{31,00}^{31,1}}{512 \sqrt{3}}+\frac{9 \sqrt{7} C_{33,00}^{33,1}}{1280}\)

\noindent\(C_{20,0000,0}^{00,4000,1,\text{NonLoc}}= \frac{7 \sqrt{3} C_{00,00}^{60,0}}{1024}-\frac{21 C_{00,20}^{60,0}}{1024}-\frac{7
C_{11,00}^{51,0}}{1024 \sqrt{3}}+\frac{7 C_{11,20}^{31,0}}{2048}+\frac{5 C_{20,00}^{40,0}}{1024 \sqrt{3}}-\frac{5 C_{20,20}^{40,0}}{1024}-\frac{7
C_{22,00}^{42,0}}{512 \sqrt{15}}+\frac{7 C_{22,20}^{42,0}}{512 \sqrt{5}}+\frac{C_{31,00}^{31,0}}{1024 \sqrt{3}}-\frac{C_{31,20}^{31,0}}{1024}+\frac{9
\sqrt{7} C_{33,00}^{33,0}}{2560}-\frac{9 \sqrt{21} C_{33,20}^{33,0}}{2560}\)

\noindent\(C_{20,0000,0}^{11,3111,0,\text{Loc}}= \frac{7 C_{11,11}^{51,0}}{384}+\frac{7 C_{11,11}^{51,1}}{768}+\frac{C_{31,11}^{31,0}}{128}+\frac{C_{31,11}^{31,1}}{256}-\frac{3}{160}
\sqrt{\frac{7}{2}} C_{33,11}^{33,0}-\frac{3}{320} \sqrt{\frac{7}{2}} C_{33,11}^{33,1}\)

\noindent\(C_{20,0000,0}^{11,3111,0,\text{NonLoc}}= \frac{7 C_{11,11}^{51,0}}{768}+\frac{7 C_{11,11}^{51,1}}{384}+\frac{C_{31,11}^{31,0}}{256}+\frac{C_{31,11}^{31,1}}{128}-\frac{3}{320}
\sqrt{\frac{7}{2}} C_{33,11}^{33,0}-\frac{3}{160} \sqrt{\frac{7}{2}} C_{33,11}^{33,1}\)

\noindent\(C_{20,0000,0}^{11,3111,1,\text{Loc}}= \frac{7 C_{11,11}^{51,1}}{256 \sqrt{3}}+\frac{\sqrt{3} C_{31,11}^{31,1}}{256}-\frac{3}{320}
\sqrt{\frac{21}{2}} C_{33,11}^{33,1}\)

\noindent\(C_{20,0000,0}^{11,3111,1,\text{NonLoc}}= \frac{7 C_{11,11}^{51,0}}{256 \sqrt{3}}+\frac{\sqrt{3} C_{31,11}^{31,0}}{256}-\frac{3}{320}
\sqrt{\frac{21}{2}} C_{33,11}^{33,0}\)

\noindent\(C_{20,0011,1}^{00,4011,0,\text{Loc}}= \frac{7 C_{00,20}^{60,0}}{256}+\frac{7 C_{00,20}^{60,1}}{512}+\frac{7 C_{11,20}^{31,0}}{1536}+\frac{7
C_{11,20}^{31,1}}{3072}+\frac{5 C_{20,20}^{40,0}}{768}+\frac{5 C_{20,20}^{40,1}}{1536}-\frac{7 C_{22,20}^{42,0}}{384 \sqrt{5}}-\frac{7 C_{22,20}^{42,1}}{768
\sqrt{5}}-\frac{C_{31,20}^{31,0}}{768}-\frac{C_{31,20}^{31,1}}{1536}-\frac{3 \sqrt{21} C_{33,20}^{33,0}}{640}-\frac{3 \sqrt{21} C_{33,20}^{33,1}}{1280}\)

\noindent\(C_{20,0011,1}^{00,4011,0,\text{NonLoc}}= \frac{7 \sqrt{3} C_{00,00}^{60,0}}{1024}+\frac{7 \sqrt{3} C_{00,00}^{60,1}}{512}+\frac{7
C_{00,20}^{60,0}}{1024}+\frac{7 C_{00,20}^{60,1}}{512}-\frac{7 C_{11,00}^{51,0}}{1024 \sqrt{3}}-\frac{7 C_{11,00}^{51,1}}{512 \sqrt{3}}-\frac{7 C_{11,20}^{31,0}}{6144}-\frac{7
C_{11,20}^{31,1}}{3072}+\frac{5 C_{20,00}^{40,0}}{1024 \sqrt{3}}+\frac{5 C_{20,00}^{40,1}}{512 \sqrt{3}}+\frac{5 C_{20,20}^{40,0}}{3072}+\frac{5
C_{20,20}^{40,1}}{1536}-\frac{7 C_{22,00}^{42,0}}{512 \sqrt{15}}-\frac{7 C_{22,00}^{42,1}}{256 \sqrt{15}}-\frac{7 C_{22,20}^{42,0}}{1536 \sqrt{5}}-\frac{7
C_{22,20}^{42,1}}{768 \sqrt{5}}+\frac{C_{31,00}^{31,0}}{1024 \sqrt{3}}+\frac{C_{31,00}^{31,1}}{512 \sqrt{3}}+\frac{C_{31,20}^{31,0}}{3072}+\frac{C_{31,20}^{31,1}}{1536}+\frac{9
\sqrt{7} C_{33,00}^{33,0}}{2560}+\frac{9 \sqrt{7} C_{33,00}^{33,1}}{1280}+\frac{3 \sqrt{21} C_{33,20}^{33,0}}{2560}+\frac{3 \sqrt{21} C_{33,20}^{33,1}}{1280}\)

\noindent\(C_{20,0011,1}^{00,4011,1,\text{Loc}}= \frac{7 \sqrt{3} C_{00,20}^{60,1}}{512}+\frac{7 C_{11,20}^{31,1}}{1024 \sqrt{3}}+\frac{5
C_{20,20}^{40,1}}{512 \sqrt{3}}-\frac{7 C_{22,20}^{42,1}}{256 \sqrt{15}}-\frac{C_{31,20}^{31,1}}{512 \sqrt{3}}-\frac{9 \sqrt{7} C_{33,20}^{33,1}}{1280}\)

\noindent\(C_{20,0011,1}^{00,4011,1,\text{NonLoc}}= \frac{21 C_{00,00}^{60,0}}{1024}+\frac{7 \sqrt{3} C_{00,20}^{60,0}}{1024}-\frac{7
C_{11,00}^{51,0}}{1024}-\frac{7 C_{11,20}^{31,0}}{2048 \sqrt{3}}+\frac{5 C_{20,00}^{40,0}}{1024}+\frac{5 C_{20,20}^{40,0}}{1024 \sqrt{3}}-\frac{7
C_{22,00}^{42,0}}{512 \sqrt{5}}-\frac{7 C_{22,20}^{42,0}}{512 \sqrt{15}}+\frac{C_{31,00}^{31,0}}{1024}+\frac{C_{31,20}^{31,0}}{1024 \sqrt{3}}+\frac{9
\sqrt{21} C_{33,00}^{33,0}}{2560}+\frac{9 \sqrt{7} C_{33,20}^{33,0}}{2560}\)

\noindent\(C_{20,0011,1}^{00,4211,0,\text{Loc}}= \frac{3 C_{00,22}^{62,0}}{128}+\frac{3 C_{00,22}^{62,1}}{256}+\frac{5 C_{11,22}^{51,0}}{384}+\frac{5
C_{11,22}^{51,1}}{768}+\frac{C_{20,22}^{42,0}}{384}+\frac{C_{20,22}^{42,1}}{768}+\frac{C_{22,22}^{40,0}}{384}+\frac{C_{22,22}^{40,1}}{768}-\frac{C_{22,22}^{42,0}}{96
\sqrt{7}}-\frac{C_{22,22}^{42,1}}{192 \sqrt{7}}-\frac{9 C_{22,22}^{44,0}}{224 \sqrt{5}}-\frac{9 C_{22,22}^{44,1}}{448 \sqrt{5}}+\frac{C_{31,22}^{31,0}}{384}+\frac{C_{31,22}^{31,1}}{768}-\frac{1}{32}
\sqrt{\frac{3}{35}} C_{31,22}^{33,0}-\frac{1}{64} \sqrt{\frac{3}{35}} C_{31,22}^{33,1}-\frac{9 C_{33,22}^{33,0}}{160 \sqrt{14}}-\frac{9 C_{33,22}^{33,1}}{320
\sqrt{14}}\)

\noindent\(C_{20,0011,1}^{00,4211,0,\text{NonLoc}}= -\frac{3 C_{00,22}^{62,0}}{256}-\frac{3 C_{00,22}^{62,1}}{128}+\frac{5 C_{11,22}^{51,0}}{768}+\frac{5
C_{11,22}^{51,1}}{384}-\frac{C_{20,22}^{42,0}}{768}-\frac{C_{20,22}^{42,1}}{384}-\frac{C_{22,22}^{40,0}}{768}-\frac{C_{22,22}^{40,1}}{384}+\frac{C_{22,22}^{42,0}}{192
\sqrt{7}}+\frac{C_{22,22}^{42,1}}{96 \sqrt{7}}+\frac{9 C_{22,22}^{44,0}}{448 \sqrt{5}}+\frac{9 C_{22,22}^{44,1}}{224 \sqrt{5}}+\frac{C_{31,22}^{31,0}}{768}+\frac{C_{31,22}^{31,1}}{384}-\frac{1}{64}
\sqrt{\frac{3}{35}} C_{31,22}^{33,0}-\frac{1}{32} \sqrt{\frac{3}{35}} C_{31,22}^{33,1}-\frac{9 C_{33,22}^{33,0}}{320 \sqrt{14}}-\frac{9 C_{33,22}^{33,1}}{160
\sqrt{14}}\)

\noindent\(C_{20,0011,1}^{00,4211,1,\text{Loc}}= \frac{3 \sqrt{3} C_{00,22}^{62,1}}{256}+\frac{5 C_{11,22}^{51,1}}{256 \sqrt{3}}+\frac{C_{20,22}^{42,1}}{256
\sqrt{3}}+\frac{C_{22,22}^{40,1}}{256 \sqrt{3}}-\frac{C_{22,22}^{42,1}}{64 \sqrt{21}}-\frac{9}{448} \sqrt{\frac{3}{5}} C_{22,22}^{44,1}+\frac{C_{31,22}^{31,1}}{256
\sqrt{3}}-\frac{3 C_{31,22}^{33,1}}{64 \sqrt{35}}-\frac{9}{320} \sqrt{\frac{3}{14}} C_{33,22}^{33,1}\)

\noindent\(C_{20,0011,1}^{00,4211,1,\text{NonLoc}}= -\frac{3 \sqrt{3} C_{00,22}^{62,0}}{256}+\frac{5 C_{11,22}^{51,0}}{256 \sqrt{3}}-\frac{C_{20,22}^{42,0}}{256
\sqrt{3}}-\frac{C_{22,22}^{40,0}}{256 \sqrt{3}}+\frac{C_{22,22}^{42,0}}{64 \sqrt{21}}+\frac{9}{448} \sqrt{\frac{3}{5}} C_{22,22}^{44,0}+\frac{C_{31,22}^{31,0}}{256
\sqrt{3}}-\frac{3 C_{31,22}^{33,0}}{64 \sqrt{35}}-\frac{9}{320} \sqrt{\frac{3}{14}} C_{33,22}^{33,0}\)

\noindent\(C_{20,0011,1}^{11,3101,0,\text{Loc}}= \frac{7 C_{11,11}^{51,0}}{384}+\frac{7 C_{11,11}^{51,1}}{768}+\frac{C_{31,11}^{31,0}}{128}+\frac{C_{31,11}^{31,1}}{256}-\frac{3}{160}
\sqrt{\frac{7}{2}} C_{33,11}^{33,0}-\frac{3}{320} \sqrt{\frac{7}{2}} C_{33,11}^{33,1}\)

\noindent\(C_{20,0011,1}^{11,3101,0,\text{NonLoc}}= \frac{7 C_{11,11}^{51,0}}{768}+\frac{7 C_{11,11}^{51,1}}{384}+\frac{C_{31,11}^{31,0}}{256}+\frac{C_{31,11}^{31,1}}{128}-\frac{3}{320}
\sqrt{\frac{7}{2}} C_{33,11}^{33,0}-\frac{3}{160} \sqrt{\frac{7}{2}} C_{33,11}^{33,1}\)

\noindent\(C_{20,0011,1}^{11,3101,1,\text{Loc}}= \frac{7 C_{11,11}^{51,1}}{256 \sqrt{3}}+\frac{\sqrt{3} C_{31,11}^{31,1}}{256}-\frac{3}{320}
\sqrt{\frac{21}{2}} C_{33,11}^{33,1}\)

\noindent\(C_{20,0011,1}^{11,3101,1,\text{NonLoc}}= \frac{7 C_{11,11}^{51,0}}{256 \sqrt{3}}+\frac{\sqrt{3} C_{31,11}^{31,0}}{256}-\frac{3}{320}
\sqrt{\frac{21}{2}} C_{33,11}^{33,0}\)

\noindent\(C_{20,1101,1}^{00,3101,0,\text{Loc}}= \frac{7 C_{00,00}^{60,0}}{64}+\frac{7 C_{00,00}^{60,1}}{128}+\frac{7 C_{11,00}^{51,0}}{192}+\frac{7
C_{11,00}^{51,1}}{384}+\frac{5 C_{20,00}^{40,0}}{192}+\frac{5 C_{20,00}^{40,1}}{384}-\frac{7 C_{22,00}^{42,0}}{96 \sqrt{5}}-\frac{7 C_{22,00}^{42,1}}{192
\sqrt{5}}-\frac{C_{31,00}^{31,0}}{192}-\frac{C_{31,00}^{31,1}}{384}-\frac{3 \sqrt{21} C_{33,00}^{33,0}}{160}-\frac{3 \sqrt{21} C_{33,00}^{33,1}}{320}\)

\noindent\(C_{20,1101,1}^{00,3101,0,\text{NonLoc}}= -\frac{7 C_{00,00}^{60,0}}{256}-\frac{7 C_{00,00}^{60,1}}{128}+\frac{7 \sqrt{3}
C_{00,20}^{60,0}}{256}+\frac{7 \sqrt{3} C_{00,20}^{60,1}}{128}+\frac{7 C_{11,00}^{51,0}}{768}+\frac{7 C_{11,00}^{51,1}}{384}-\frac{5 C_{20,00}^{40,0}}{768}-\frac{5
C_{20,00}^{40,1}}{384}+\frac{5 C_{20,20}^{40,0}}{256 \sqrt{3}}+\frac{5 C_{20,20}^{40,1}}{128 \sqrt{3}}+\frac{7 C_{22,00}^{42,0}}{384 \sqrt{5}}+\frac{7
C_{22,00}^{42,1}}{192 \sqrt{5}}-\frac{7 C_{22,20}^{42,0}}{128 \sqrt{15}}-\frac{7 C_{22,20}^{42,1}}{64 \sqrt{15}}-\frac{C_{31,00}^{31,0}}{768}-\frac{C_{31,00}^{31,1}}{384}+\frac{C_{31,20}^{31,0}}{256
\sqrt{3}}+\frac{C_{31,20}^{31,1}}{128 \sqrt{3}}-\frac{3 \sqrt{21} C_{33,00}^{33,0}}{640}-\frac{3 \sqrt{21} C_{33,00}^{33,1}}{320}+\frac{9 \sqrt{7}
C_{33,20}^{33,0}}{640}+\frac{9 \sqrt{7} C_{33,20}^{33,1}}{320}\)

\noindent\(C_{20,1101,1}^{00,3101,1,\text{Loc}}= \frac{7 \sqrt{3} C_{00,00}^{60,1}}{128}+\frac{7 C_{11,00}^{51,1}}{128 \sqrt{3}}+\frac{5
C_{20,00}^{40,1}}{128 \sqrt{3}}-\frac{7 C_{22,00}^{42,1}}{64 \sqrt{15}}-\frac{C_{31,00}^{31,1}}{128 \sqrt{3}}-\frac{9 \sqrt{7} C_{33,00}^{33,1}}{320}\)

\noindent\(C_{20,1101,1}^{00,3101,1,\text{NonLoc}}= -\frac{7 \sqrt{3} C_{00,00}^{60,0}}{256}+\frac{21 C_{00,20}^{60,0}}{256}+\frac{7
C_{11,00}^{51,0}}{256 \sqrt{3}}-\frac{5 C_{20,00}^{40,0}}{256 \sqrt{3}}+\frac{5 C_{20,20}^{40,0}}{256}+\frac{7 C_{22,00}^{42,0}}{128 \sqrt{15}}-\frac{7
C_{22,20}^{42,0}}{128 \sqrt{5}}-\frac{C_{31,00}^{31,0}}{256 \sqrt{3}}+\frac{C_{31,20}^{31,0}}{256}-\frac{9 \sqrt{7} C_{33,00}^{33,0}}{640}+\frac{9
\sqrt{21} C_{33,20}^{33,0}}{640}\)

\noindent\(C_{20,1101,1}^{11,2011,0,\text{Loc}}= \frac{35 C_{11,11}^{51,0}}{1152}+\frac{35 C_{11,11}^{51,1}}{2304}+\frac{5 C_{31,11}^{31,0}}{384}+\frac{5
C_{31,11}^{31,1}}{768}-\frac{1}{32} \sqrt{\frac{7}{2}} C_{33,11}^{33,0}-\frac{1}{64} \sqrt{\frac{7}{2}} C_{33,11}^{33,1}\)

\noindent\(C_{20,1101,1}^{11,2011,0,\text{NonLoc}}= \frac{35 C_{11,11}^{51,0}}{2304}+\frac{35 C_{11,11}^{51,1}}{1152}+\frac{5 C_{31,11}^{31,0}}{768}+\frac{5
C_{31,11}^{31,1}}{384}-\frac{1}{64} \sqrt{\frac{7}{2}} C_{33,11}^{33,0}-\frac{1}{32} \sqrt{\frac{7}{2}} C_{33,11}^{33,1}\)

\noindent\(C_{20,1101,1}^{11,2011,1,\text{Loc}}= \frac{35 C_{11,11}^{51,1}}{768 \sqrt{3}}+\frac{5 C_{31,11}^{31,1}}{256 \sqrt{3}}-\frac{1}{64}
\sqrt{\frac{21}{2}} C_{33,11}^{33,1}\)

\noindent\(C_{20,1101,1}^{11,2011,1,\text{NonLoc}}= \frac{35 C_{11,11}^{51,0}}{768 \sqrt{3}}+\frac{5 C_{31,11}^{31,0}}{256 \sqrt{3}}-\frac{1}{64}
\sqrt{\frac{21}{2}} C_{33,11}^{33,0}\)

\noindent\(C_{20,1101,1}^{11,2211,0,\text{Loc}}= -\frac{7 \sqrt{5} C_{11,11}^{51,0}}{1152}-\frac{7 \sqrt{5} C_{11,11}^{51,1}}{2304}-\frac{\sqrt{5}
C_{31,11}^{31,0}}{384}-\frac{\sqrt{5} C_{31,11}^{31,1}}{768}+\frac{1}{32} \sqrt{\frac{7}{10}} C_{33,11}^{33,0}+\frac{1}{64} \sqrt{\frac{7}{10}} C_{33,11}^{33,1}\)

\noindent\(C_{20,1101,1}^{11,2211,0,\text{NonLoc}}= -\frac{7 \sqrt{5} C_{11,11}^{51,0}}{2304}-\frac{7 \sqrt{5} C_{11,11}^{51,1}}{1152}-\frac{\sqrt{5}
C_{31,11}^{31,0}}{768}-\frac{\sqrt{5} C_{31,11}^{31,1}}{384}+\frac{1}{64} \sqrt{\frac{7}{10}} C_{33,11}^{33,0}+\frac{1}{32} \sqrt{\frac{7}{10}} C_{33,11}^{33,1}\)

\noindent\(C_{20,1101,1}^{11,2211,1,\text{Loc}}= -\frac{7}{768} \sqrt{\frac{5}{3}} C_{11,11}^{51,1}-\frac{1}{256} \sqrt{\frac{5}{3}}
C_{31,11}^{31,1}+\frac{1}{64} \sqrt{\frac{21}{10}} C_{33,11}^{33,1}\)

\noindent\(C_{20,1101,1}^{11,2211,1,\text{NonLoc}}= -\frac{7}{768} \sqrt{\frac{5}{3}} C_{11,11}^{51,0}-\frac{1}{256} \sqrt{\frac{5}{3}}
C_{31,11}^{31,0}+\frac{1}{64} \sqrt{\frac{21}{10}} C_{33,11}^{33,0}\)

\noindent\(C_{20,1101,1}^{11,2212,0,\text{Loc}}= -\frac{7}{384} \sqrt{\frac{5}{3}} C_{11,11}^{51,0}-\frac{7}{768} \sqrt{\frac{5}{3}}
C_{11,11}^{51,1}-\frac{1}{128} \sqrt{\frac{5}{3}} C_{31,11}^{31,0}-\frac{1}{256} \sqrt{\frac{5}{3}} C_{31,11}^{31,1}+\frac{1}{32} \sqrt{\frac{21}{10}}
C_{33,11}^{33,0}+\frac{1}{64} \sqrt{\frac{21}{10}} C_{33,11}^{33,1}\)

\noindent\(C_{20,1101,1}^{11,2212,0,\text{NonLoc}}= -\frac{7}{768} \sqrt{\frac{5}{3}} C_{11,11}^{51,0}-\frac{7}{384} \sqrt{\frac{5}{3}}
C_{11,11}^{51,1}-\frac{1}{256} \sqrt{\frac{5}{3}} C_{31,11}^{31,0}-\frac{1}{128} \sqrt{\frac{5}{3}} C_{31,11}^{31,1}+\frac{1}{64} \sqrt{\frac{21}{10}}
C_{33,11}^{33,0}+\frac{1}{32} \sqrt{\frac{21}{10}} C_{33,11}^{33,1}\)

\noindent\(C_{20,1101,1}^{11,2212,1,\text{Loc}}= -\frac{7 \sqrt{5} C_{11,11}^{51,1}}{768}-\frac{\sqrt{5} C_{31,11}^{31,1}}{256}+\frac{3}{64}
\sqrt{\frac{7}{10}} C_{33,11}^{33,1}\)

\noindent\(C_{20,1101,1}^{11,2212,1,\text{NonLoc}}= -\frac{7 \sqrt{5} C_{11,11}^{51,0}}{768}-\frac{\sqrt{5} C_{31,11}^{31,0}}{256}+\frac{3}{64}
\sqrt{\frac{7}{10}} C_{33,11}^{33,0}\)

\noindent\(C_{20,1101,1}^{20,1101,0,\text{Loc}}= -\frac{35 C_{00,00}^{60,0}}{1536}-\frac{35 C_{00,00}^{60,1}}{3072}+\frac{35 C_{11,00}^{51,0}}{4608}+\frac{35
C_{11,00}^{51,1}}{9216}-\frac{25 C_{20,00}^{40,0}}{4608}-\frac{25 C_{20,00}^{40,1}}{9216}+\frac{7 \sqrt{5} C_{22,00}^{42,0}}{2304}+\frac{7 \sqrt{5}
C_{22,00}^{42,1}}{4608}-\frac{5 C_{31,00}^{31,0}}{4608}-\frac{5 C_{31,00}^{31,1}}{9216}-\frac{\sqrt{21} C_{33,00}^{33,0}}{256}-\frac{\sqrt{21} C_{33,00}^{33,1}}{512}\)

\noindent\(C_{20,1101,1}^{20,1101,0,\text{NonLoc}}= \frac{35 C_{00,00}^{60,0}}{6144}+\frac{35 C_{00,00}^{60,1}}{3072}-\frac{35 C_{00,20}^{60,0}}{2048
\sqrt{3}}-\frac{35 C_{00,20}^{60,1}}{1024 \sqrt{3}}+\frac{35 C_{11,00}^{51,0}}{18432}+\frac{35 C_{11,00}^{51,1}}{9216}+\frac{25 C_{20,00}^{40,0}}{18432}+\frac{25
C_{20,00}^{40,1}}{9216}-\frac{25 C_{20,20}^{40,0}}{6144 \sqrt{3}}-\frac{25 C_{20,20}^{40,1}}{3072 \sqrt{3}}-\frac{7 \sqrt{5} C_{22,00}^{42,0}}{9216}-\frac{7
\sqrt{5} C_{22,00}^{42,1}}{4608}+\frac{7 \sqrt{\frac{5}{3}} C_{22,20}^{42,0}}{3072}+\frac{7 \sqrt{\frac{5}{3}} C_{22,20}^{42,1}}{1536}-\frac{5 C_{31,00}^{31,0}}{18432}-\frac{5
C_{31,00}^{31,1}}{9216}+\frac{5 C_{31,20}^{31,0}}{6144 \sqrt{3}}+\frac{5 C_{31,20}^{31,1}}{3072 \sqrt{3}}-\frac{\sqrt{21} C_{33,00}^{33,0}}{1024}-\frac{\sqrt{21}
C_{33,00}^{33,1}}{512}+\frac{3 \sqrt{7} C_{33,20}^{33,0}}{1024}+\frac{3 \sqrt{7} C_{33,20}^{33,1}}{512}\)

\noindent\(C_{20,1101,1}^{20,1101,1,\text{Loc}}= -\frac{35 C_{00,00}^{60,1}}{1024 \sqrt{3}}+\frac{35 C_{11,00}^{51,1}}{3072 \sqrt{3}}-\frac{25
C_{20,00}^{40,1}}{3072 \sqrt{3}}+\frac{7 \sqrt{\frac{5}{3}} C_{22,00}^{42,1}}{1536}-\frac{5 C_{31,00}^{31,1}}{3072 \sqrt{3}}-\frac{3 \sqrt{7} C_{33,00}^{33,1}}{512}\)

\noindent\(C_{20,1101,1}^{20,1101,1,\text{NonLoc}}= \frac{35 C_{00,00}^{60,0}}{2048 \sqrt{3}}-\frac{35 C_{00,20}^{60,0}}{2048}+\frac{35
C_{11,00}^{51,0}}{6144 \sqrt{3}}+\frac{25 C_{20,00}^{40,0}}{6144 \sqrt{3}}-\frac{25 C_{20,20}^{40,0}}{6144}-\frac{7 \sqrt{\frac{5}{3}} C_{22,00}^{42,0}}{3072}+\frac{7
\sqrt{5} C_{22,20}^{42,0}}{3072}-\frac{5 C_{31,00}^{31,0}}{6144 \sqrt{3}}+\frac{5 C_{31,20}^{31,0}}{6144}-\frac{3 \sqrt{7} C_{33,00}^{33,0}}{1024}+\frac{3
\sqrt{21} C_{33,20}^{33,0}}{1024}\)

\noindent\(C_{20,1110,0}^{00,3110,0,\text{Loc}}= -\frac{7 C_{00,20}^{60,0}}{192}-\frac{7 C_{00,20}^{60,1}}{384}-\frac{7 C_{00,22}^{62,0}}{64
\sqrt{5}}-\frac{7 C_{00,22}^{62,1}}{128 \sqrt{5}}-\frac{7 \sqrt{5} C_{11,22}^{51,0}}{576}-\frac{7 \sqrt{5} C_{11,22}^{51,1}}{1152}-\frac{5 C_{20,20}^{40,0}}{576}-\frac{5
C_{20,20}^{40,1}}{1152}-\frac{7 C_{20,22}^{42,0}}{576 \sqrt{5}}-\frac{7 C_{20,22}^{42,1}}{1152 \sqrt{5}}+\frac{7 C_{22,20}^{42,0}}{288 \sqrt{5}}+\frac{7
C_{22,20}^{42,1}}{576 \sqrt{5}}-\frac{7 C_{22,22}^{40,0}}{576 \sqrt{5}}-\frac{7 C_{22,22}^{40,1}}{1152 \sqrt{5}}+\frac{1}{144} \sqrt{\frac{7}{5}}
C_{22,22}^{42,0}+\frac{1}{288} \sqrt{\frac{7}{5}} C_{22,22}^{42,1}+\frac{3 C_{22,22}^{44,0}}{80}+\frac{3 C_{22,22}^{44,1}}{160}+\frac{C_{31,20}^{31,0}}{576}+\frac{C_{31,20}^{31,1}}{1152}-\frac{7
C_{31,22}^{31,0}}{576 \sqrt{5}}-\frac{7 C_{31,22}^{31,1}}{1152 \sqrt{5}}+\frac{1}{80} \sqrt{\frac{7}{3}} C_{31,22}^{33,0}+\frac{1}{160} \sqrt{\frac{7}{3}}
C_{31,22}^{33,1}+\frac{\sqrt{21} C_{33,20}^{33,0}}{160}+\frac{\sqrt{21} C_{33,20}^{33,1}}{320}+\frac{3}{80} \sqrt{\frac{7}{10}} C_{33,22}^{33,0}+\frac{3}{160}
\sqrt{\frac{7}{10}} C_{33,22}^{33,1}\)

\noindent\(C_{20,1110,0}^{00,3110,0,\text{NonLoc}}= -\frac{7 C_{00,00}^{60,0}}{256 \sqrt{3}}-\frac{7 C_{00,00}^{60,1}}{128 \sqrt{3}}-\frac{7
C_{00,20}^{60,0}}{768}-\frac{7 C_{00,20}^{60,1}}{384}+\frac{7 C_{00,22}^{62,0}}{128 \sqrt{5}}+\frac{7 C_{00,22}^{62,1}}{64 \sqrt{5}}+\frac{7 C_{11,00}^{51,0}}{768
\sqrt{3}}+\frac{7 C_{11,00}^{51,1}}{384 \sqrt{3}}-\frac{7 \sqrt{5} C_{11,22}^{51,0}}{1152}-\frac{7 \sqrt{5} C_{11,22}^{51,1}}{576}-\frac{5 C_{20,00}^{40,0}}{768
\sqrt{3}}-\frac{5 C_{20,00}^{40,1}}{384 \sqrt{3}}-\frac{5 C_{20,20}^{40,0}}{2304}-\frac{5 C_{20,20}^{40,1}}{1152}+\frac{7 C_{20,22}^{42,0}}{1152
\sqrt{5}}+\frac{7 C_{20,22}^{42,1}}{576 \sqrt{5}}+\frac{7 C_{22,00}^{42,0}}{384 \sqrt{15}}+\frac{7 C_{22,00}^{42,1}}{192 \sqrt{15}}+\frac{7 C_{22,20}^{42,0}}{1152
\sqrt{5}}+\frac{7 C_{22,20}^{42,1}}{576 \sqrt{5}}+\frac{7 C_{22,22}^{40,0}}{1152 \sqrt{5}}+\frac{7 C_{22,22}^{40,1}}{576 \sqrt{5}}-\frac{1}{288}
\sqrt{\frac{7}{5}} C_{22,22}^{42,0}-\frac{1}{144} \sqrt{\frac{7}{5}} C_{22,22}^{42,1}-\frac{3 C_{22,22}^{44,0}}{160}-\frac{3 C_{22,22}^{44,1}}{80}-\frac{C_{31,00}^{31,0}}{768
\sqrt{3}}-\frac{C_{31,00}^{31,1}}{384 \sqrt{3}}-\frac{C_{31,20}^{31,0}}{2304}-\frac{C_{31,20}^{31,1}}{1152}-\frac{7 C_{31,22}^{31,0}}{1152 \sqrt{5}}-\frac{7
C_{31,22}^{31,1}}{576 \sqrt{5}}+\frac{1}{160} \sqrt{\frac{7}{3}} C_{31,22}^{33,0}+\frac{1}{80} \sqrt{\frac{7}{3}} C_{31,22}^{33,1}-\frac{3 \sqrt{7}
C_{33,00}^{33,0}}{640}-\frac{3 \sqrt{7} C_{33,00}^{33,1}}{320}-\frac{\sqrt{21} C_{33,20}^{33,0}}{640}-\frac{\sqrt{21} C_{33,20}^{33,1}}{320}+\frac{3}{160}
\sqrt{\frac{7}{10}} C_{33,22}^{33,0}+\frac{3}{80} \sqrt{\frac{7}{10}} C_{33,22}^{33,1}\)

\noindent\(C_{20,1110,0}^{00,3110,1,\text{Loc}}= -\frac{7 C_{00,20}^{60,1}}{128 \sqrt{3}}-\frac{7}{128} \sqrt{\frac{3}{5}} C_{00,22}^{62,1}-\frac{7}{384}
\sqrt{\frac{5}{3}} C_{11,22}^{51,1}-\frac{5 C_{20,20}^{40,1}}{384 \sqrt{3}}-\frac{7 C_{20,22}^{42,1}}{384 \sqrt{15}}+\frac{7 C_{22,20}^{42,1}}{192
\sqrt{15}}-\frac{7 C_{22,22}^{40,1}}{384 \sqrt{15}}+\frac{1}{96} \sqrt{\frac{7}{15}} C_{22,22}^{42,1}+\frac{3 \sqrt{3} C_{22,22}^{44,1}}{160}+\frac{C_{31,20}^{31,1}}{384
\sqrt{3}}-\frac{7 C_{31,22}^{31,1}}{384 \sqrt{15}}+\frac{\sqrt{7} C_{31,22}^{33,1}}{160}+\frac{3 \sqrt{7} C_{33,20}^{33,1}}{320}+\frac{3}{160} \sqrt{\frac{21}{10}}
C_{33,22}^{33,1}\)

\noindent\(C_{20,1110,0}^{00,3110,1,\text{NonLoc}}= -\frac{7 C_{00,00}^{60,0}}{256}-\frac{7 C_{00,20}^{60,0}}{256 \sqrt{3}}+\frac{7}{128}
\sqrt{\frac{3}{5}} C_{00,22}^{62,0}+\frac{7 C_{11,00}^{51,0}}{768}-\frac{7}{384} \sqrt{\frac{5}{3}} C_{11,22}^{51,0}-\frac{5 C_{20,00}^{40,0}}{768}-\frac{5
C_{20,20}^{40,0}}{768 \sqrt{3}}+\frac{7 C_{20,22}^{42,0}}{384 \sqrt{15}}+\frac{7 C_{22,00}^{42,0}}{384 \sqrt{5}}+\frac{7 C_{22,20}^{42,0}}{384 \sqrt{15}}+\frac{7
C_{22,22}^{40,0}}{384 \sqrt{15}}-\frac{1}{96} \sqrt{\frac{7}{15}} C_{22,22}^{42,0}-\frac{3 \sqrt{3} C_{22,22}^{44,0}}{160}-\frac{C_{31,00}^{31,0}}{768}-\frac{C_{31,20}^{31,0}}{768
\sqrt{3}}-\frac{7 C_{31,22}^{31,0}}{384 \sqrt{15}}+\frac{\sqrt{7} C_{31,22}^{33,0}}{160}-\frac{3 \sqrt{21} C_{33,00}^{33,0}}{640}-\frac{3 \sqrt{7}
C_{33,20}^{33,0}}{640}+\frac{3}{160} \sqrt{\frac{21}{10}} C_{33,22}^{33,0}\)

\noindent\(C_{20,1110,0}^{20,1110,0,\text{Loc}}= \frac{35 C_{00,20}^{60,0}}{4608}+\frac{35 C_{00,20}^{60,1}}{9216}+\frac{7 \sqrt{5}
C_{00,22}^{62,0}}{1536}+\frac{7 \sqrt{5} C_{00,22}^{62,1}}{3072}+\frac{7 \sqrt{5} C_{11,22}^{51,0}}{13824}+\frac{7 \sqrt{5} C_{11,22}^{51,1}}{27648}-\frac{\sqrt{21}
C_{11,22}^{53,0}}{512}-\frac{\sqrt{21} C_{11,22}^{53,1}}{1024}+\frac{25 C_{20,20}^{40,0}}{13824}+\frac{25 C_{20,20}^{40,1}}{27648}+\frac{7 \sqrt{5}
C_{20,22}^{42,0}}{13824}+\frac{7 \sqrt{5} C_{20,22}^{42,1}}{27648}-\frac{7 \sqrt{5} C_{22,20}^{42,0}}{6912}-\frac{7 \sqrt{5} C_{22,20}^{42,1}}{13824}-\frac{17
\sqrt{5} C_{22,22}^{40,0}}{13824}-\frac{17 \sqrt{5} C_{22,22}^{40,1}}{27648}-\frac{5 \sqrt{35} C_{22,22}^{42,0}}{6912}-\frac{5 \sqrt{35} C_{22,22}^{42,1}}{13824}+\frac{C_{22,22}^{44,0}}{128}+\frac{C_{22,22}^{44,1}}{256}+\frac{5
C_{31,20}^{31,0}}{13824}+\frac{5 C_{31,20}^{31,1}}{27648}+\frac{23 \sqrt{5} C_{31,22}^{31,0}}{13824}+\frac{23 \sqrt{5} C_{31,22}^{31,1}}{27648}-\frac{\sqrt{\frac{7}{3}}
C_{31,22}^{33,0}}{1536}-\frac{\sqrt{\frac{7}{3}} C_{31,22}^{33,1}}{3072}+\frac{1}{256} \sqrt{\frac{7}{3}} C_{33,20}^{33,0}+\frac{1}{512} \sqrt{\frac{7}{3}}
C_{33,20}^{33,1}+\frac{1}{128} \sqrt{\frac{7}{10}} C_{33,22}^{33,0}+\frac{1}{256} \sqrt{\frac{7}{10}} C_{33,22}^{33,1}\)

\noindent\(C_{20,1110,0}^{20,1110,0,\text{NonLoc}}= \frac{35 C_{00,00}^{60,0}}{6144 \sqrt{3}}+\frac{35 C_{00,00}^{60,1}}{3072 \sqrt{3}}+\frac{35
C_{00,20}^{60,0}}{18432}+\frac{35 C_{00,20}^{60,1}}{9216}-\frac{7 \sqrt{5} C_{00,22}^{62,0}}{3072}-\frac{7 \sqrt{5} C_{00,22}^{62,1}}{1536}+\frac{35
C_{11,00}^{51,0}}{18432 \sqrt{3}}+\frac{35 C_{11,00}^{51,1}}{9216 \sqrt{3}}+\frac{7 \sqrt{5} C_{11,22}^{51,0}}{27648}+\frac{7 \sqrt{5} C_{11,22}^{51,1}}{13824}-\frac{\sqrt{21}
C_{11,22}^{53,0}}{1024}-\frac{\sqrt{21} C_{11,22}^{53,1}}{512}+\frac{25 C_{20,00}^{40,0}}{18432 \sqrt{3}}+\frac{25 C_{20,00}^{40,1}}{9216 \sqrt{3}}+\frac{25
C_{20,20}^{40,0}}{55296}+\frac{25 C_{20,20}^{40,1}}{27648}-\frac{7 \sqrt{5} C_{20,22}^{42,0}}{27648}-\frac{7 \sqrt{5} C_{20,22}^{42,1}}{13824}-\frac{7
\sqrt{\frac{5}{3}} C_{22,00}^{42,0}}{9216}-\frac{7 \sqrt{\frac{5}{3}} C_{22,00}^{42,1}}{4608}-\frac{7 \sqrt{5} C_{22,20}^{42,0}}{27648}-\frac{7 \sqrt{5}
C_{22,20}^{42,1}}{13824}+\frac{17 \sqrt{5} C_{22,22}^{40,0}}{27648}+\frac{17 \sqrt{5} C_{22,22}^{40,1}}{13824}+\frac{5 \sqrt{35} C_{22,22}^{42,0}}{13824}+\frac{5
\sqrt{35} C_{22,22}^{42,1}}{6912}-\frac{C_{22,22}^{44,0}}{256}-\frac{C_{22,22}^{44,1}}{128}-\frac{5 C_{31,00}^{31,0}}{18432 \sqrt{3}}-\frac{5 C_{31,00}^{31,1}}{9216
\sqrt{3}}-\frac{5 C_{31,20}^{31,0}}{55296}-\frac{5 C_{31,20}^{31,1}}{27648}+\frac{23 \sqrt{5} C_{31,22}^{31,0}}{27648}+\frac{23 \sqrt{5} C_{31,22}^{31,1}}{13824}-\frac{\sqrt{\frac{7}{3}}
C_{31,22}^{33,0}}{3072}-\frac{\sqrt{\frac{7}{3}} C_{31,22}^{33,1}}{1536}-\frac{\sqrt{7} C_{33,00}^{33,0}}{1024}-\frac{\sqrt{7} C_{33,00}^{33,1}}{512}-\frac{\sqrt{\frac{7}{3}}
C_{33,20}^{33,0}}{1024}-\frac{1}{512} \sqrt{\frac{7}{3}} C_{33,20}^{33,1}+\frac{1}{256} \sqrt{\frac{7}{10}} C_{33,22}^{33,0}+\frac{1}{128} \sqrt{\frac{7}{10}}
C_{33,22}^{33,1}\)

\noindent\(C_{20,1110,0}^{20,1110,1,\text{Loc}}= \frac{35 C_{00,20}^{60,1}}{3072 \sqrt{3}}+\frac{7 \sqrt{\frac{5}{3}} C_{00,22}^{62,1}}{1024}+\frac{7
\sqrt{\frac{5}{3}} C_{11,22}^{51,1}}{9216}-\frac{3 \sqrt{7} C_{11,22}^{53,1}}{1024}+\frac{25 C_{20,20}^{40,1}}{9216 \sqrt{3}}+\frac{7 \sqrt{\frac{5}{3}}
C_{20,22}^{42,1}}{9216}-\frac{7 \sqrt{\frac{5}{3}} C_{22,20}^{42,1}}{4608}-\frac{17 \sqrt{\frac{5}{3}} C_{22,22}^{40,1}}{9216}-\frac{5 \sqrt{\frac{35}{3}}
C_{22,22}^{42,1}}{4608}+\frac{\sqrt{3} C_{22,22}^{44,1}}{256}+\frac{5 C_{31,20}^{31,1}}{9216 \sqrt{3}}+\frac{23 \sqrt{\frac{5}{3}} C_{31,22}^{31,1}}{9216}-\frac{\sqrt{7}
C_{31,22}^{33,1}}{3072}+\frac{\sqrt{7} C_{33,20}^{33,1}}{512}+\frac{1}{256} \sqrt{\frac{21}{10}} C_{33,22}^{33,1}\)

\noindent\(C_{20,1110,0}^{20,1110,1,\text{NonLoc}}= \frac{35 C_{00,00}^{60,0}}{6144}+\frac{35 C_{00,20}^{60,0}}{6144 \sqrt{3}}-\frac{7
\sqrt{\frac{5}{3}} C_{00,22}^{62,0}}{1024}+\frac{35 C_{11,00}^{51,0}}{18432}+\frac{7 \sqrt{\frac{5}{3}} C_{11,22}^{51,0}}{9216}-\frac{3 \sqrt{7}
C_{11,22}^{53,0}}{1024}+\frac{25 C_{20,00}^{40,0}}{18432}+\frac{25 C_{20,20}^{40,0}}{18432 \sqrt{3}}-\frac{7 \sqrt{\frac{5}{3}} C_{20,22}^{42,0}}{9216}-\frac{7
\sqrt{5} C_{22,00}^{42,0}}{9216}-\frac{7 \sqrt{\frac{5}{3}} C_{22,20}^{42,0}}{9216}+\frac{17 \sqrt{\frac{5}{3}} C_{22,22}^{40,0}}{9216}+\frac{5 \sqrt{\frac{35}{3}}
C_{22,22}^{42,0}}{4608}-\frac{\sqrt{3} C_{22,22}^{44,0}}{256}-\frac{5 C_{31,00}^{31,0}}{18432}-\frac{5 C_{31,20}^{31,0}}{18432 \sqrt{3}}+\frac{23
\sqrt{\frac{5}{3}} C_{31,22}^{31,0}}{9216}-\frac{\sqrt{7} C_{31,22}^{33,0}}{3072}-\frac{\sqrt{21} C_{33,00}^{33,0}}{1024}-\frac{\sqrt{7} C_{33,20}^{33,0}}{1024}+\frac{1}{256}
\sqrt{\frac{21}{10}} C_{33,22}^{33,0}\)

\noindent\(C_{20,1111,1}^{00,3111,0,\text{Loc}}= -\frac{7 C_{00,20}^{60,0}}{64 \sqrt{3}}-\frac{7 C_{00,20}^{60,1}}{128 \sqrt{3}}+\frac{7}{128}
\sqrt{\frac{3}{5}} C_{00,22}^{62,0}+\frac{7}{256} \sqrt{\frac{3}{5}} C_{00,22}^{62,1}+\frac{7}{384} \sqrt{\frac{5}{3}} C_{11,22}^{51,0}+\frac{7}{768}
\sqrt{\frac{5}{3}} C_{11,22}^{51,1}-\frac{5 C_{20,20}^{40,0}}{192 \sqrt{3}}-\frac{5 C_{20,20}^{40,1}}{384 \sqrt{3}}+\frac{7 C_{20,22}^{42,0}}{384
\sqrt{15}}+\frac{7 C_{20,22}^{42,1}}{768 \sqrt{15}}+\frac{7 C_{22,20}^{42,0}}{96 \sqrt{15}}+\frac{7 C_{22,20}^{42,1}}{192 \sqrt{15}}+\frac{7 C_{22,22}^{40,0}}{384
\sqrt{15}}+\frac{7 C_{22,22}^{40,1}}{768 \sqrt{15}}-\frac{1}{96} \sqrt{\frac{7}{15}} C_{22,22}^{42,0}-\frac{1}{192} \sqrt{\frac{7}{15}} C_{22,22}^{42,1}-\frac{3
\sqrt{3} C_{22,22}^{44,0}}{160}-\frac{3 \sqrt{3} C_{22,22}^{44,1}}{320}+\frac{C_{31,20}^{31,0}}{192 \sqrt{3}}+\frac{C_{31,20}^{31,1}}{384 \sqrt{3}}+\frac{7
C_{31,22}^{31,0}}{384 \sqrt{15}}+\frac{7 C_{31,22}^{31,1}}{768 \sqrt{15}}-\frac{\sqrt{7} C_{31,22}^{33,0}}{160}-\frac{\sqrt{7} C_{31,22}^{33,1}}{320}+\frac{3
\sqrt{7} C_{33,20}^{33,0}}{160}+\frac{3 \sqrt{7} C_{33,20}^{33,1}}{320}-\frac{3}{160} \sqrt{\frac{21}{10}} C_{33,22}^{33,0}-\frac{3}{320} \sqrt{\frac{21}{10}}
C_{33,22}^{33,1}\)

\noindent\(C_{20,1111,1}^{00,3111,0,\text{NonLoc}}= -\frac{7 C_{00,00}^{60,0}}{256}-\frac{7 C_{00,00}^{60,1}}{128}-\frac{7 C_{00,20}^{60,0}}{256
\sqrt{3}}-\frac{7 C_{00,20}^{60,1}}{128 \sqrt{3}}-\frac{7}{256} \sqrt{\frac{3}{5}} C_{00,22}^{62,0}-\frac{7}{128} \sqrt{\frac{3}{5}} C_{00,22}^{62,1}+\frac{7
C_{11,00}^{51,0}}{768}+\frac{7 C_{11,00}^{51,1}}{384}+\frac{7}{768} \sqrt{\frac{5}{3}} C_{11,22}^{51,0}+\frac{7}{384} \sqrt{\frac{5}{3}} C_{11,22}^{51,1}-\frac{5
C_{20,00}^{40,0}}{768}-\frac{5 C_{20,00}^{40,1}}{384}-\frac{5 C_{20,20}^{40,0}}{768 \sqrt{3}}-\frac{5 C_{20,20}^{40,1}}{384 \sqrt{3}}-\frac{7 C_{20,22}^{42,0}}{768
\sqrt{15}}-\frac{7 C_{20,22}^{42,1}}{384 \sqrt{15}}+\frac{7 C_{22,00}^{42,0}}{384 \sqrt{5}}+\frac{7 C_{22,00}^{42,1}}{192 \sqrt{5}}+\frac{7 C_{22,20}^{42,0}}{384
\sqrt{15}}+\frac{7 C_{22,20}^{42,1}}{192 \sqrt{15}}-\frac{7 C_{22,22}^{40,0}}{768 \sqrt{15}}-\frac{7 C_{22,22}^{40,1}}{384 \sqrt{15}}+\frac{1}{192}
\sqrt{\frac{7}{15}} C_{22,22}^{42,0}+\frac{1}{96} \sqrt{\frac{7}{15}} C_{22,22}^{42,1}+\frac{3 \sqrt{3} C_{22,22}^{44,0}}{320}+\frac{3 \sqrt{3} C_{22,22}^{44,1}}{160}-\frac{C_{31,00}^{31,0}}{768}-\frac{C_{31,00}^{31,1}}{384}-\frac{C_{31,20}^{31,0}}{768
\sqrt{3}}-\frac{C_{31,20}^{31,1}}{384 \sqrt{3}}+\frac{7 C_{31,22}^{31,0}}{768 \sqrt{15}}+\frac{7 C_{31,22}^{31,1}}{384 \sqrt{15}}-\frac{\sqrt{7}
C_{31,22}^{33,0}}{320}-\frac{\sqrt{7} C_{31,22}^{33,1}}{160}-\frac{3 \sqrt{21} C_{33,00}^{33,0}}{640}-\frac{3 \sqrt{21} C_{33,00}^{33,1}}{320}-\frac{3
\sqrt{7} C_{33,20}^{33,0}}{640}-\frac{3 \sqrt{7} C_{33,20}^{33,1}}{320}-\frac{3}{320} \sqrt{\frac{21}{10}} C_{33,22}^{33,0}-\frac{3}{160} \sqrt{\frac{21}{10}}
C_{33,22}^{33,1}\)

\noindent\(C_{20,1111,1}^{00,3111,1,\text{Loc}}= -\frac{7 C_{00,20}^{60,1}}{128}+\frac{21 C_{00,22}^{62,1}}{256 \sqrt{5}}+\frac{7 \sqrt{5}
C_{11,22}^{51,1}}{768}-\frac{5 C_{20,20}^{40,1}}{384}+\frac{7 C_{20,22}^{42,1}}{768 \sqrt{5}}+\frac{7 C_{22,20}^{42,1}}{192 \sqrt{5}}+\frac{7 C_{22,22}^{40,1}}{768
\sqrt{5}}-\frac{1}{192} \sqrt{\frac{7}{5}} C_{22,22}^{42,1}-\frac{9 C_{22,22}^{44,1}}{320}+\frac{C_{31,20}^{31,1}}{384}+\frac{7 C_{31,22}^{31,1}}{768
\sqrt{5}}-\frac{\sqrt{21} C_{31,22}^{33,1}}{320}+\frac{3 \sqrt{21} C_{33,20}^{33,1}}{320}-\frac{9}{320} \sqrt{\frac{7}{10}} C_{33,22}^{33,1}\)

\noindent\(C_{20,1111,1}^{00,3111,1,\text{NonLoc}}= -\frac{7 \sqrt{3} C_{00,00}^{60,0}}{256}-\frac{7 C_{00,20}^{60,0}}{256}-\frac{21
C_{00,22}^{62,0}}{256 \sqrt{5}}+\frac{7 C_{11,00}^{51,0}}{256 \sqrt{3}}+\frac{7 \sqrt{5} C_{11,22}^{51,0}}{768}-\frac{5 C_{20,00}^{40,0}}{256 \sqrt{3}}-\frac{5
C_{20,20}^{40,0}}{768}-\frac{7 C_{20,22}^{42,0}}{768 \sqrt{5}}+\frac{7 C_{22,00}^{42,0}}{128 \sqrt{15}}+\frac{7 C_{22,20}^{42,0}}{384 \sqrt{5}}-\frac{7
C_{22,22}^{40,0}}{768 \sqrt{5}}+\frac{1}{192} \sqrt{\frac{7}{5}} C_{22,22}^{42,0}+\frac{9 C_{22,22}^{44,0}}{320}-\frac{C_{31,00}^{31,0}}{256 \sqrt{3}}-\frac{C_{31,20}^{31,0}}{768}+\frac{7
C_{31,22}^{31,0}}{768 \sqrt{5}}-\frac{\sqrt{21} C_{31,22}^{33,0}}{320}-\frac{9 \sqrt{7} C_{33,00}^{33,0}}{640}-\frac{3 \sqrt{21} C_{33,20}^{33,0}}{640}-\frac{9}{320}
\sqrt{\frac{7}{10}} C_{33,22}^{33,0}\)

\noindent\(C_{20,1111,1}^{11,2000,0,\text{Loc}}= -\frac{35 C_{11,11}^{51,0}}{1152}-\frac{35 C_{11,11}^{51,1}}{2304}-\frac{5 C_{31,11}^{31,0}}{384}-\frac{5
C_{31,11}^{31,1}}{768}+\frac{1}{32} \sqrt{\frac{7}{2}} C_{33,11}^{33,0}+\frac{1}{64} \sqrt{\frac{7}{2}} C_{33,11}^{33,1}\)

\noindent\(C_{20,1111,1}^{11,2000,0,\text{NonLoc}}= -\frac{35 C_{11,11}^{51,0}}{2304}-\frac{35 C_{11,11}^{51,1}}{1152}-\frac{5 C_{31,11}^{31,0}}{768}-\frac{5
C_{31,11}^{31,1}}{384}+\frac{1}{64} \sqrt{\frac{7}{2}} C_{33,11}^{33,0}+\frac{1}{32} \sqrt{\frac{7}{2}} C_{33,11}^{33,1}\)

\noindent\(C_{20,1111,1}^{11,2000,1,\text{Loc}}= -\frac{35 C_{11,11}^{51,1}}{768 \sqrt{3}}-\frac{5 C_{31,11}^{31,1}}{256 \sqrt{3}}+\frac{1}{64}
\sqrt{\frac{21}{2}} C_{33,11}^{33,1}\)

\noindent\(C_{20,1111,1}^{11,2000,1,\text{NonLoc}}= -\frac{35 C_{11,11}^{51,0}}{768 \sqrt{3}}-\frac{5 C_{31,11}^{31,0}}{256 \sqrt{3}}+\frac{1}{64}
\sqrt{\frac{21}{2}} C_{33,11}^{33,0}\)

\noindent\(C_{20,1111,1}^{11,2202,0,\text{Loc}}= \frac{7 \sqrt{5} C_{11,11}^{51,0}}{1152}+\frac{7 \sqrt{5} C_{11,11}^{51,1}}{2304}+\frac{\sqrt{5}
C_{31,11}^{31,0}}{384}+\frac{\sqrt{5} C_{31,11}^{31,1}}{768}-\frac{1}{32} \sqrt{\frac{7}{10}} C_{33,11}^{33,0}-\frac{1}{64} \sqrt{\frac{7}{10}} C_{33,11}^{33,1}\)

\noindent\(C_{20,1111,1}^{11,2202,0,\text{NonLoc}}= \frac{7 \sqrt{5} C_{11,11}^{51,0}}{2304}+\frac{7 \sqrt{5} C_{11,11}^{51,1}}{1152}+\frac{\sqrt{5}
C_{31,11}^{31,0}}{768}+\frac{\sqrt{5} C_{31,11}^{31,1}}{384}-\frac{1}{64} \sqrt{\frac{7}{10}} C_{33,11}^{33,0}-\frac{1}{32} \sqrt{\frac{7}{10}} C_{33,11}^{33,1}\)

\noindent\(C_{20,1111,1}^{11,2202,1,\text{Loc}}= \frac{7}{768} \sqrt{\frac{5}{3}} C_{11,11}^{51,1}+\frac{1}{256} \sqrt{\frac{5}{3}}
C_{31,11}^{31,1}-\frac{1}{64} \sqrt{\frac{21}{10}} C_{33,11}^{33,1}\)

\noindent\(C_{20,1111,1}^{11,2202,1,\text{NonLoc}}= \frac{7}{768} \sqrt{\frac{5}{3}} C_{11,11}^{51,0}+\frac{1}{256} \sqrt{\frac{5}{3}}
C_{31,11}^{31,0}-\frac{1}{64} \sqrt{\frac{21}{10}} C_{33,11}^{33,0}\)

\noindent\(C_{20,1111,1}^{20,1111,0,\text{Loc}}= \frac{35 C_{00,20}^{60,0}}{1536 \sqrt{3}}+\frac{35 C_{00,20}^{60,1}}{3072 \sqrt{3}}-\frac{7
\sqrt{\frac{5}{3}} C_{00,22}^{62,0}}{1024}-\frac{7 \sqrt{\frac{5}{3}} C_{00,22}^{62,1}}{2048}-\frac{7 \sqrt{\frac{5}{3}} C_{11,22}^{51,0}}{9216}-\frac{7
\sqrt{\frac{5}{3}} C_{11,22}^{51,1}}{18432}+\frac{3 \sqrt{7} C_{11,22}^{53,0}}{1024}+\frac{3 \sqrt{7} C_{11,22}^{53,1}}{2048}+\frac{25 C_{20,20}^{40,0}}{4608
\sqrt{3}}+\frac{25 C_{20,20}^{40,1}}{9216 \sqrt{3}}-\frac{7 \sqrt{\frac{5}{3}} C_{20,22}^{42,0}}{9216}-\frac{7 \sqrt{\frac{5}{3}} C_{20,22}^{42,1}}{18432}-\frac{7
\sqrt{\frac{5}{3}} C_{22,20}^{42,0}}{2304}-\frac{7 \sqrt{\frac{5}{3}} C_{22,20}^{42,1}}{4608}+\frac{17 \sqrt{\frac{5}{3}} C_{22,22}^{40,0}}{9216}+\frac{17
\sqrt{\frac{5}{3}} C_{22,22}^{40,1}}{18432}+\frac{5 \sqrt{\frac{35}{3}} C_{22,22}^{42,0}}{4608}+\frac{5 \sqrt{\frac{35}{3}} C_{22,22}^{42,1}}{9216}-\frac{\sqrt{3}
C_{22,22}^{44,0}}{256}-\frac{\sqrt{3} C_{22,22}^{44,1}}{512}+\frac{5 C_{31,20}^{31,0}}{4608 \sqrt{3}}+\frac{5 C_{31,20}^{31,1}}{9216 \sqrt{3}}-\frac{23
\sqrt{\frac{5}{3}} C_{31,22}^{31,0}}{9216}-\frac{23 \sqrt{\frac{5}{3}} C_{31,22}^{31,1}}{18432}+\frac{\sqrt{7} C_{31,22}^{33,0}}{3072}+\frac{\sqrt{7}
C_{31,22}^{33,1}}{6144}+\frac{\sqrt{7} C_{33,20}^{33,0}}{256}+\frac{\sqrt{7} C_{33,20}^{33,1}}{512}-\frac{1}{256} \sqrt{\frac{21}{10}} C_{33,22}^{33,0}-\frac{1}{512}
\sqrt{\frac{21}{10}} C_{33,22}^{33,1}\)

\noindent\(C_{20,1111,1}^{20,1111,0,\text{NonLoc}}= \frac{35 C_{00,00}^{60,0}}{6144}+\frac{35 C_{00,00}^{60,1}}{3072}+\frac{35 C_{00,20}^{60,0}}{6144
\sqrt{3}}+\frac{35 C_{00,20}^{60,1}}{3072 \sqrt{3}}+\frac{7 \sqrt{\frac{5}{3}} C_{00,22}^{62,0}}{2048}+\frac{7 \sqrt{\frac{5}{3}} C_{00,22}^{62,1}}{1024}+\frac{35
C_{11,00}^{51,0}}{18432}+\frac{35 C_{11,00}^{51,1}}{9216}-\frac{7 \sqrt{\frac{5}{3}} C_{11,22}^{51,0}}{18432}-\frac{7 \sqrt{\frac{5}{3}} C_{11,22}^{51,1}}{9216}+\frac{3
\sqrt{7} C_{11,22}^{53,0}}{2048}+\frac{3 \sqrt{7} C_{11,22}^{53,1}}{1024}+\frac{25 C_{20,00}^{40,0}}{18432}+\frac{25 C_{20,00}^{40,1}}{9216}+\frac{25
C_{20,20}^{40,0}}{18432 \sqrt{3}}+\frac{25 C_{20,20}^{40,1}}{9216 \sqrt{3}}+\frac{7 \sqrt{\frac{5}{3}} C_{20,22}^{42,0}}{18432}+\frac{7 \sqrt{\frac{5}{3}}
C_{20,22}^{42,1}}{9216}-\frac{7 \sqrt{5} C_{22,00}^{42,0}}{9216}-\frac{7 \sqrt{5} C_{22,00}^{42,1}}{4608}-\frac{7 \sqrt{\frac{5}{3}} C_{22,20}^{42,0}}{9216}-\frac{7
\sqrt{\frac{5}{3}} C_{22,20}^{42,1}}{4608}-\frac{17 \sqrt{\frac{5}{3}} C_{22,22}^{40,0}}{18432}-\frac{17 \sqrt{\frac{5}{3}} C_{22,22}^{40,1}}{9216}-\frac{5
\sqrt{\frac{35}{3}} C_{22,22}^{42,0}}{9216}-\frac{5 \sqrt{\frac{35}{3}} C_{22,22}^{42,1}}{4608}+\frac{\sqrt{3} C_{22,22}^{44,0}}{512}+\frac{\sqrt{3}
C_{22,22}^{44,1}}{256}-\frac{5 C_{31,00}^{31,0}}{18432}-\frac{5 C_{31,00}^{31,1}}{9216}-\frac{5 C_{31,20}^{31,0}}{18432 \sqrt{3}}-\frac{5 C_{31,20}^{31,1}}{9216
\sqrt{3}}-\frac{23 \sqrt{\frac{5}{3}} C_{31,22}^{31,0}}{18432}-\frac{23 \sqrt{\frac{5}{3}} C_{31,22}^{31,1}}{9216}+\frac{\sqrt{7} C_{31,22}^{33,0}}{6144}+\frac{\sqrt{7}
C_{31,22}^{33,1}}{3072}-\frac{\sqrt{21} C_{33,00}^{33,0}}{1024}-\frac{\sqrt{21} C_{33,00}^{33,1}}{512}-\frac{\sqrt{7} C_{33,20}^{33,0}}{1024}-\frac{\sqrt{7}
C_{33,20}^{33,1}}{512}-\frac{1}{512} \sqrt{\frac{21}{10}} C_{33,22}^{33,0}-\frac{1}{256} \sqrt{\frac{21}{10}} C_{33,22}^{33,1}\)

\noindent\(C_{20,1111,1}^{20,1111,1,\text{Loc}}= \frac{35 C_{00,20}^{60,1}}{3072}-\frac{7 \sqrt{5} C_{00,22}^{62,1}}{2048}-\frac{7
\sqrt{5} C_{11,22}^{51,1}}{18432}+\frac{3 \sqrt{21} C_{11,22}^{53,1}}{2048}+\frac{25 C_{20,20}^{40,1}}{9216}-\frac{7 \sqrt{5} C_{20,22}^{42,1}}{18432}-\frac{7
\sqrt{5} C_{22,20}^{42,1}}{4608}+\frac{17 \sqrt{5} C_{22,22}^{40,1}}{18432}+\frac{5 \sqrt{35} C_{22,22}^{42,1}}{9216}-\frac{3 C_{22,22}^{44,1}}{512}+\frac{5
C_{31,20}^{31,1}}{9216}-\frac{23 \sqrt{5} C_{31,22}^{31,1}}{18432}+\frac{\sqrt{\frac{7}{3}} C_{31,22}^{33,1}}{2048}+\frac{\sqrt{21} C_{33,20}^{33,1}}{512}-\frac{3}{512}
\sqrt{\frac{7}{10}} C_{33,22}^{33,1}\)

\noindent\(C_{20,1111,1}^{20,1111,1,\text{NonLoc}}= \frac{35 C_{00,00}^{60,0}}{2048 \sqrt{3}}+\frac{35 C_{00,20}^{60,0}}{6144}+\frac{7
\sqrt{5} C_{00,22}^{62,0}}{2048}+\frac{35 C_{11,00}^{51,0}}{6144 \sqrt{3}}-\frac{7 \sqrt{5} C_{11,22}^{51,0}}{18432}+\frac{3 \sqrt{21} C_{11,22}^{53,0}}{2048}+\frac{25
C_{20,00}^{40,0}}{6144 \sqrt{3}}+\frac{25 C_{20,20}^{40,0}}{18432}+\frac{7 \sqrt{5} C_{20,22}^{42,0}}{18432}-\frac{7 \sqrt{\frac{5}{3}} C_{22,00}^{42,0}}{3072}-\frac{7
\sqrt{5} C_{22,20}^{42,0}}{9216}-\frac{17 \sqrt{5} C_{22,22}^{40,0}}{18432}-\frac{5 \sqrt{35} C_{22,22}^{42,0}}{9216}+\frac{3 C_{22,22}^{44,0}}{512}-\frac{5
C_{31,00}^{31,0}}{6144 \sqrt{3}}-\frac{5 C_{31,20}^{31,0}}{18432}-\frac{23 \sqrt{5} C_{31,22}^{31,0}}{18432}+\frac{\sqrt{\frac{7}{3}} C_{31,22}^{33,0}}{2048}-\frac{3
\sqrt{7} C_{33,00}^{33,0}}{1024}-\frac{\sqrt{21} C_{33,20}^{33,0}}{1024}-\frac{3}{512} \sqrt{\frac{7}{10}} C_{33,22}^{33,0}\)

\noindent\(C_{20,1112,2}^{00,3112,0,\text{Loc}}= -\frac{7 \sqrt{5} C_{00,20}^{60,0}}{192}-\frac{7 \sqrt{5} C_{00,20}^{60,1}}{384}-\frac{7
C_{00,22}^{62,0}}{640}-\frac{7 C_{00,22}^{62,1}}{1280}-\frac{7 C_{11,22}^{51,0}}{1152}-\frac{7 C_{11,22}^{51,1}}{2304}-\frac{5 \sqrt{5} C_{20,20}^{40,0}}{576}-\frac{5
\sqrt{5} C_{20,20}^{40,1}}{1152}-\frac{7 C_{20,22}^{42,0}}{5760}-\frac{7 C_{20,22}^{42,1}}{11520}+\frac{7 C_{22,20}^{42,0}}{288}+\frac{7 C_{22,20}^{42,1}}{576}-\frac{7
C_{22,22}^{40,0}}{5760}-\frac{7 C_{22,22}^{40,1}}{11520}+\frac{\sqrt{7} C_{22,22}^{42,0}}{1440}+\frac{\sqrt{7} C_{22,22}^{42,1}}{2880}+\frac{3 C_{22,22}^{44,0}}{160
\sqrt{5}}+\frac{3 C_{22,22}^{44,1}}{320 \sqrt{5}}+\frac{\sqrt{5} C_{31,20}^{31,0}}{576}+\frac{\sqrt{5} C_{31,20}^{31,1}}{1152}-\frac{7 C_{31,22}^{31,0}}{5760}-\frac{7
C_{31,22}^{31,1}}{11520}+\frac{1}{160} \sqrt{\frac{7}{15}} C_{31,22}^{33,0}+\frac{1}{320} \sqrt{\frac{7}{15}} C_{31,22}^{33,1}+\frac{1}{32} \sqrt{\frac{21}{5}}
C_{33,20}^{33,0}+\frac{1}{64} \sqrt{\frac{21}{5}} C_{33,20}^{33,1}+\frac{3}{800} \sqrt{\frac{7}{2}} C_{33,22}^{33,0}+\frac{3 \sqrt{\frac{7}{2}} C_{33,22}^{33,1}}{1600}\)

\noindent\(C_{20,1112,2}^{00,3112,0,\text{NonLoc}}= -\frac{7}{256} \sqrt{\frac{5}{3}} C_{00,00}^{60,0}-\frac{7}{128} \sqrt{\frac{5}{3}}
C_{00,00}^{60,1}-\frac{7 \sqrt{5} C_{00,20}^{60,0}}{768}-\frac{7 \sqrt{5} C_{00,20}^{60,1}}{384}+\frac{7 C_{00,22}^{62,0}}{1280}+\frac{7 C_{00,22}^{62,1}}{640}+\frac{7}{768}
\sqrt{\frac{5}{3}} C_{11,00}^{51,0}+\frac{7}{384} \sqrt{\frac{5}{3}} C_{11,00}^{51,1}-\frac{7 C_{11,22}^{51,0}}{2304}-\frac{7 C_{11,22}^{51,1}}{1152}-\frac{5}{768}
\sqrt{\frac{5}{3}} C_{20,00}^{40,0}-\frac{5}{384} \sqrt{\frac{5}{3}} C_{20,00}^{40,1}-\frac{5 \sqrt{5} C_{20,20}^{40,0}}{2304}-\frac{5 \sqrt{5} C_{20,20}^{40,1}}{1152}+\frac{7
C_{20,22}^{42,0}}{11520}+\frac{7 C_{20,22}^{42,1}}{5760}+\frac{7 C_{22,00}^{42,0}}{384 \sqrt{3}}+\frac{7 C_{22,00}^{42,1}}{192 \sqrt{3}}+\frac{7
C_{22,20}^{42,0}}{1152}+\frac{7 C_{22,20}^{42,1}}{576}+\frac{7 C_{22,22}^{40,0}}{11520}+\frac{7 C_{22,22}^{40,1}}{5760}-\frac{\sqrt{7} C_{22,22}^{42,0}}{2880}-\frac{\sqrt{7}
C_{22,22}^{42,1}}{1440}-\frac{3 C_{22,22}^{44,0}}{320 \sqrt{5}}-\frac{3 C_{22,22}^{44,1}}{160 \sqrt{5}}-\frac{1}{768} \sqrt{\frac{5}{3}} C_{31,00}^{31,0}-\frac{1}{384}
\sqrt{\frac{5}{3}} C_{31,00}^{31,1}-\frac{\sqrt{5} C_{31,20}^{31,0}}{2304}-\frac{\sqrt{5} C_{31,20}^{31,1}}{1152}-\frac{7 C_{31,22}^{31,0}}{11520}-\frac{7
C_{31,22}^{31,1}}{5760}+\frac{1}{320} \sqrt{\frac{7}{15}} C_{31,22}^{33,0}+\frac{1}{160} \sqrt{\frac{7}{15}} C_{31,22}^{33,1}-\frac{3}{128} \sqrt{\frac{7}{5}}
C_{33,00}^{33,0}-\frac{3}{64} \sqrt{\frac{7}{5}} C_{33,00}^{33,1}-\frac{1}{128} \sqrt{\frac{21}{5}} C_{33,20}^{33,0}-\frac{1}{64} \sqrt{\frac{21}{5}}
C_{33,20}^{33,1}+\frac{3 \sqrt{\frac{7}{2}} C_{33,22}^{33,0}}{1600}+\frac{3}{800} \sqrt{\frac{7}{2}} C_{33,22}^{33,1}\)

\noindent\(C_{20,1112,2}^{00,3112,1,\text{Loc}}= -\frac{7}{128} \sqrt{\frac{5}{3}} C_{00,20}^{60,1}-\frac{7 \sqrt{3} C_{00,22}^{62,1}}{1280}-\frac{7
C_{11,22}^{51,1}}{768 \sqrt{3}}-\frac{5}{384} \sqrt{\frac{5}{3}} C_{20,20}^{40,1}-\frac{7 C_{20,22}^{42,1}}{3840 \sqrt{3}}+\frac{7 C_{22,20}^{42,1}}{192
\sqrt{3}}-\frac{7 C_{22,22}^{40,1}}{3840 \sqrt{3}}+\frac{1}{960} \sqrt{\frac{7}{3}} C_{22,22}^{42,1}+\frac{3}{320} \sqrt{\frac{3}{5}} C_{22,22}^{44,1}+\frac{1}{384}
\sqrt{\frac{5}{3}} C_{31,20}^{31,1}-\frac{7 C_{31,22}^{31,1}}{3840 \sqrt{3}}+\frac{1}{320} \sqrt{\frac{7}{5}} C_{31,22}^{33,1}+\frac{3}{64} \sqrt{\frac{7}{5}}
C_{33,20}^{33,1}+\frac{3 \sqrt{\frac{21}{2}} C_{33,22}^{33,1}}{1600}\)

\noindent\(C_{20,1112,2}^{00,3112,1,\text{NonLoc}}= -\frac{7 \sqrt{5} C_{00,00}^{60,0}}{256}-\frac{7}{256} \sqrt{\frac{5}{3}} C_{00,20}^{60,0}+\frac{7
\sqrt{3} C_{00,22}^{62,0}}{1280}+\frac{7 \sqrt{5} C_{11,00}^{51,0}}{768}-\frac{7 C_{11,22}^{51,0}}{768 \sqrt{3}}-\frac{5 \sqrt{5} C_{20,00}^{40,0}}{768}-\frac{5}{768}
\sqrt{\frac{5}{3}} C_{20,20}^{40,0}+\frac{7 C_{20,22}^{42,0}}{3840 \sqrt{3}}+\frac{7 C_{22,00}^{42,0}}{384}+\frac{7 C_{22,20}^{42,0}}{384 \sqrt{3}}+\frac{7
C_{22,22}^{40,0}}{3840 \sqrt{3}}-\frac{1}{960} \sqrt{\frac{7}{3}} C_{22,22}^{42,0}-\frac{3}{320} \sqrt{\frac{3}{5}} C_{22,22}^{44,0}-\frac{\sqrt{5}
C_{31,00}^{31,0}}{768}-\frac{1}{768} \sqrt{\frac{5}{3}} C_{31,20}^{31,0}-\frac{7 C_{31,22}^{31,0}}{3840 \sqrt{3}}+\frac{1}{320} \sqrt{\frac{7}{5}}
C_{31,22}^{33,0}-\frac{3}{128} \sqrt{\frac{21}{5}} C_{33,00}^{33,0}-\frac{3}{128} \sqrt{\frac{7}{5}} C_{33,20}^{33,0}+\frac{3 \sqrt{\frac{21}{2}}
C_{33,22}^{33,0}}{1600}\)

\noindent\(C_{20,1112,2}^{00,3312,0,\text{Loc}}= -\frac{1}{64} \sqrt{\frac{21}{5}} C_{00,22}^{62,0}-\frac{1}{128} \sqrt{\frac{21}{5}}
C_{00,22}^{62,1}-\frac{1}{192} \sqrt{\frac{35}{3}} C_{11,22}^{51,0}-\frac{1}{384} \sqrt{\frac{35}{3}} C_{11,22}^{51,1}-\frac{1}{192} \sqrt{\frac{7}{15}}
C_{20,22}^{42,0}-\frac{1}{384} \sqrt{\frac{7}{15}} C_{20,22}^{42,1}-\frac{1}{192} \sqrt{\frac{7}{15}} C_{22,22}^{40,0}-\frac{1}{384} \sqrt{\frac{7}{15}}
C_{22,22}^{40,1}+\frac{C_{22,22}^{42,0}}{48 \sqrt{15}}+\frac{C_{22,22}^{42,1}}{96 \sqrt{15}}+\frac{3}{80} \sqrt{\frac{3}{7}} C_{22,22}^{44,0}+\frac{3}{160}
\sqrt{\frac{3}{7}} C_{22,22}^{44,1}-\frac{1}{192} \sqrt{\frac{7}{15}} C_{31,22}^{31,0}-\frac{1}{384} \sqrt{\frac{7}{15}} C_{31,22}^{31,1}+\frac{C_{31,22}^{33,0}}{80}+\frac{C_{31,22}^{33,1}}{160}+\frac{3}{80}
\sqrt{\frac{3}{10}} C_{33,22}^{33,0}+\frac{3}{160} \sqrt{\frac{3}{10}} C_{33,22}^{33,1}\)

\noindent\(C_{20,1112,2}^{00,3312,0,\text{NonLoc}}= \frac{1}{128} \sqrt{\frac{21}{5}} C_{00,22}^{62,0}+\frac{1}{64} \sqrt{\frac{21}{5}}
C_{00,22}^{62,1}-\frac{1}{384} \sqrt{\frac{35}{3}} C_{11,22}^{51,0}-\frac{1}{192} \sqrt{\frac{35}{3}} C_{11,22}^{51,1}+\frac{1}{384} \sqrt{\frac{7}{15}}
C_{20,22}^{42,0}+\frac{1}{192} \sqrt{\frac{7}{15}} C_{20,22}^{42,1}+\frac{1}{384} \sqrt{\frac{7}{15}} C_{22,22}^{40,0}+\frac{1}{192} \sqrt{\frac{7}{15}}
C_{22,22}^{40,1}-\frac{C_{22,22}^{42,0}}{96 \sqrt{15}}-\frac{C_{22,22}^{42,1}}{48 \sqrt{15}}-\frac{3}{160} \sqrt{\frac{3}{7}} C_{22,22}^{44,0}-\frac{3}{80}
\sqrt{\frac{3}{7}} C_{22,22}^{44,1}-\frac{1}{384} \sqrt{\frac{7}{15}} C_{31,22}^{31,0}-\frac{1}{192} \sqrt{\frac{7}{15}} C_{31,22}^{31,1}+\frac{C_{31,22}^{33,0}}{160}+\frac{C_{31,22}^{33,1}}{80}+\frac{3}{160}
\sqrt{\frac{3}{10}} C_{33,22}^{33,0}+\frac{3}{80} \sqrt{\frac{3}{10}} C_{33,22}^{33,1}\)

\noindent\(C_{20,1112,2}^{00,3312,1,\text{Loc}}= -\frac{3}{128} \sqrt{\frac{7}{5}} C_{00,22}^{62,1}-\frac{\sqrt{35} C_{11,22}^{51,1}}{384}-\frac{1}{384}
\sqrt{\frac{7}{5}} C_{20,22}^{42,1}-\frac{1}{384} \sqrt{\frac{7}{5}} C_{22,22}^{40,1}+\frac{C_{22,22}^{42,1}}{96 \sqrt{5}}+\frac{9 C_{22,22}^{44,1}}{160
\sqrt{7}}-\frac{1}{384} \sqrt{\frac{7}{5}} C_{31,22}^{31,1}+\frac{\sqrt{3} C_{31,22}^{33,1}}{160}+\frac{9 C_{33,22}^{33,1}}{160 \sqrt{10}}\)

\noindent\(C_{20,1112,2}^{00,3312,1,\text{NonLoc}}= \frac{3}{128} \sqrt{\frac{7}{5}} C_{00,22}^{62,0}-\frac{\sqrt{35} C_{11,22}^{51,0}}{384}+\frac{1}{384}
\sqrt{\frac{7}{5}} C_{20,22}^{42,0}+\frac{1}{384} \sqrt{\frac{7}{5}} C_{22,22}^{40,0}-\frac{C_{22,22}^{42,0}}{96 \sqrt{5}}-\frac{9 C_{22,22}^{44,0}}{160
\sqrt{7}}-\frac{1}{384} \sqrt{\frac{7}{5}} C_{31,22}^{31,0}+\frac{\sqrt{3} C_{31,22}^{33,0}}{160}+\frac{9 C_{33,22}^{33,0}}{160 \sqrt{10}}\)

\noindent\(C_{20,1112,2}^{11,2202,0,\text{Loc}}= -\frac{7}{384} \sqrt{\frac{5}{3}} C_{11,11}^{51,0}-\frac{7}{768} \sqrt{\frac{5}{3}}
C_{11,11}^{51,1}-\frac{1}{128} \sqrt{\frac{5}{3}} C_{31,11}^{31,0}-\frac{1}{256} \sqrt{\frac{5}{3}} C_{31,11}^{31,1}+\frac{1}{32} \sqrt{\frac{21}{10}}
C_{33,11}^{33,0}+\frac{1}{64} \sqrt{\frac{21}{10}} C_{33,11}^{33,1}\)

\noindent\(C_{20,1112,2}^{11,2202,0,\text{NonLoc}}= -\frac{7}{768} \sqrt{\frac{5}{3}} C_{11,11}^{51,0}-\frac{7}{384} \sqrt{\frac{5}{3}}
C_{11,11}^{51,1}-\frac{1}{256} \sqrt{\frac{5}{3}} C_{31,11}^{31,0}-\frac{1}{128} \sqrt{\frac{5}{3}} C_{31,11}^{31,1}+\frac{1}{64} \sqrt{\frac{21}{10}}
C_{33,11}^{33,0}+\frac{1}{32} \sqrt{\frac{21}{10}} C_{33,11}^{33,1}\)

\noindent\(C_{20,1112,2}^{11,2202,1,\text{Loc}}= -\frac{7 \sqrt{5} C_{11,11}^{51,1}}{768}-\frac{\sqrt{5} C_{31,11}^{31,1}}{256}+\frac{3}{64}
\sqrt{\frac{7}{10}} C_{33,11}^{33,1}\)

\noindent\(C_{20,1112,2}^{11,2202,1,\text{NonLoc}}= -\frac{7 \sqrt{5} C_{11,11}^{51,0}}{768}-\frac{\sqrt{5} C_{31,11}^{31,0}}{256}+\frac{3}{64}
\sqrt{\frac{7}{10}} C_{33,11}^{33,0}\)

\noindent\(C_{20,1112,2}^{20,1112,0,\text{Loc}}= \frac{35 \sqrt{5} C_{00,20}^{60,0}}{4608}+\frac{35 \sqrt{5} C_{00,20}^{60,1}}{9216}+\frac{7
C_{00,22}^{62,0}}{3072}+\frac{7 C_{00,22}^{62,1}}{6144}+\frac{7 C_{11,22}^{51,0}}{27648}+\frac{7 C_{11,22}^{51,1}}{55296}-\frac{\sqrt{\frac{21}{5}}
C_{11,22}^{53,0}}{1024}-\frac{\sqrt{\frac{21}{5}} C_{11,22}^{53,1}}{2048}+\frac{25 \sqrt{5} C_{20,20}^{40,0}}{13824}+\frac{25 \sqrt{5} C_{20,20}^{40,1}}{27648}+\frac{7
C_{20,22}^{42,0}}{27648}+\frac{7 C_{20,22}^{42,1}}{55296}-\frac{35 C_{22,20}^{42,0}}{6912}-\frac{35 C_{22,20}^{42,1}}{13824}-\frac{17 C_{22,22}^{40,0}}{27648}-\frac{17
C_{22,22}^{40,1}}{55296}-\frac{5 \sqrt{7} C_{22,22}^{42,0}}{13824}-\frac{5 \sqrt{7} C_{22,22}^{42,1}}{27648}+\frac{C_{22,22}^{44,0}}{256 \sqrt{5}}+\frac{C_{22,22}^{44,1}}{512
\sqrt{5}}+\frac{5 \sqrt{5} C_{31,20}^{31,0}}{13824}+\frac{5 \sqrt{5} C_{31,20}^{31,1}}{27648}+\frac{23 C_{31,22}^{31,0}}{27648}+\frac{23 C_{31,22}^{31,1}}{55296}-\frac{\sqrt{\frac{7}{15}}
C_{31,22}^{33,0}}{3072}-\frac{\sqrt{\frac{7}{15}} C_{31,22}^{33,1}}{6144}+\frac{1}{256} \sqrt{\frac{35}{3}} C_{33,20}^{33,0}+\frac{1}{512} \sqrt{\frac{35}{3}}
C_{33,20}^{33,1}+\frac{\sqrt{\frac{7}{2}} C_{33,22}^{33,0}}{1280}+\frac{\sqrt{\frac{7}{2}} C_{33,22}^{33,1}}{2560}\)

\noindent\(C_{20,1112,2}^{20,1112,0,\text{NonLoc}}= \frac{35 \sqrt{\frac{5}{3}} C_{00,00}^{60,0}}{6144}+\frac{35 \sqrt{\frac{5}{3}}
C_{00,00}^{60,1}}{3072}+\frac{35 \sqrt{5} C_{00,20}^{60,0}}{18432}+\frac{35 \sqrt{5} C_{00,20}^{60,1}}{9216}-\frac{7 C_{00,22}^{62,0}}{6144}-\frac{7
C_{00,22}^{62,1}}{3072}+\frac{35 \sqrt{\frac{5}{3}} C_{11,00}^{51,0}}{18432}+\frac{35 \sqrt{\frac{5}{3}} C_{11,00}^{51,1}}{9216}+\frac{7 C_{11,22}^{51,0}}{55296}+\frac{7
C_{11,22}^{51,1}}{27648}-\frac{\sqrt{\frac{21}{5}} C_{11,22}^{53,0}}{2048}-\frac{\sqrt{\frac{21}{5}} C_{11,22}^{53,1}}{1024}+\frac{25 \sqrt{\frac{5}{3}}
C_{20,00}^{40,0}}{18432}+\frac{25 \sqrt{\frac{5}{3}} C_{20,00}^{40,1}}{9216}+\frac{25 \sqrt{5} C_{20,20}^{40,0}}{55296}+\frac{25 \sqrt{5} C_{20,20}^{40,1}}{27648}-\frac{7
C_{20,22}^{42,0}}{55296}-\frac{7 C_{20,22}^{42,1}}{27648}-\frac{35 C_{22,00}^{42,0}}{9216 \sqrt{3}}-\frac{35 C_{22,00}^{42,1}}{4608 \sqrt{3}}-\frac{35
C_{22,20}^{42,0}}{27648}-\frac{35 C_{22,20}^{42,1}}{13824}+\frac{17 C_{22,22}^{40,0}}{55296}+\frac{17 C_{22,22}^{40,1}}{27648}+\frac{5 \sqrt{7} C_{22,22}^{42,0}}{27648}+\frac{5
\sqrt{7} C_{22,22}^{42,1}}{13824}-\frac{C_{22,22}^{44,0}}{512 \sqrt{5}}-\frac{C_{22,22}^{44,1}}{256 \sqrt{5}}-\frac{5 \sqrt{\frac{5}{3}} C_{31,00}^{31,0}}{18432}-\frac{5
\sqrt{\frac{5}{3}} C_{31,00}^{31,1}}{9216}-\frac{5 \sqrt{5} C_{31,20}^{31,0}}{55296}-\frac{5 \sqrt{5} C_{31,20}^{31,1}}{27648}+\frac{23 C_{31,22}^{31,0}}{55296}+\frac{23
C_{31,22}^{31,1}}{27648}-\frac{\sqrt{\frac{7}{15}} C_{31,22}^{33,0}}{6144}-\frac{\sqrt{\frac{7}{15}} C_{31,22}^{33,1}}{3072}-\frac{\sqrt{35} C_{33,00}^{33,0}}{1024}-\frac{\sqrt{35}
C_{33,00}^{33,1}}{512}-\frac{\sqrt{\frac{35}{3}} C_{33,20}^{33,0}}{1024}-\frac{1}{512} \sqrt{\frac{35}{3}} C_{33,20}^{33,1}+\frac{\sqrt{\frac{7}{2}}
C_{33,22}^{33,0}}{2560}+\frac{\sqrt{\frac{7}{2}} C_{33,22}^{33,1}}{1280}\)

\noindent\(C_{20,1112,2}^{20,1112,1,\text{Loc}}= \frac{35 \sqrt{\frac{5}{3}} C_{00,20}^{60,1}}{3072}+\frac{7 C_{00,22}^{62,1}}{2048
\sqrt{3}}+\frac{7 C_{11,22}^{51,1}}{18432 \sqrt{3}}-\frac{3 \sqrt{\frac{7}{5}} C_{11,22}^{53,1}}{2048}+\frac{25 \sqrt{\frac{5}{3}} C_{20,20}^{40,1}}{9216}+\frac{7
C_{20,22}^{42,1}}{18432 \sqrt{3}}-\frac{35 C_{22,20}^{42,1}}{4608 \sqrt{3}}-\frac{17 C_{22,22}^{40,1}}{18432 \sqrt{3}}-\frac{5 \sqrt{\frac{7}{3}}
C_{22,22}^{42,1}}{9216}+\frac{1}{512} \sqrt{\frac{3}{5}} C_{22,22}^{44,1}+\frac{5 \sqrt{\frac{5}{3}} C_{31,20}^{31,1}}{9216}+\frac{23 C_{31,22}^{31,1}}{18432
\sqrt{3}}-\frac{\sqrt{\frac{7}{5}} C_{31,22}^{33,1}}{6144}+\frac{\sqrt{35} C_{33,20}^{33,1}}{512}+\frac{\sqrt{\frac{21}{2}} C_{33,22}^{33,1}}{2560}\)

\noindent\(C_{20,1112,2}^{20,1112,1,\text{NonLoc}}= \frac{35 \sqrt{5} C_{00,00}^{60,0}}{6144}+\frac{35 \sqrt{\frac{5}{3}} C_{00,20}^{60,0}}{6144}-\frac{7
C_{00,22}^{62,0}}{2048 \sqrt{3}}+\frac{35 \sqrt{5} C_{11,00}^{51,0}}{18432}+\frac{7 C_{11,22}^{51,0}}{18432 \sqrt{3}}-\frac{3 \sqrt{\frac{7}{5}}
C_{11,22}^{53,0}}{2048}+\frac{25 \sqrt{5} C_{20,00}^{40,0}}{18432}+\frac{25 \sqrt{\frac{5}{3}} C_{20,20}^{40,0}}{18432}-\frac{7 C_{20,22}^{42,0}}{18432
\sqrt{3}}-\frac{35 C_{22,00}^{42,0}}{9216}-\frac{35 C_{22,20}^{42,0}}{9216 \sqrt{3}}+\frac{17 C_{22,22}^{40,0}}{18432 \sqrt{3}}+\frac{5 \sqrt{\frac{7}{3}}
C_{22,22}^{42,0}}{9216}-\frac{1}{512} \sqrt{\frac{3}{5}} C_{22,22}^{44,0}-\frac{5 \sqrt{5} C_{31,00}^{31,0}}{18432}-\frac{5 \sqrt{\frac{5}{3}} C_{31,20}^{31,0}}{18432}+\frac{23
C_{31,22}^{31,0}}{18432 \sqrt{3}}-\frac{\sqrt{\frac{7}{5}} C_{31,22}^{33,0}}{6144}-\frac{\sqrt{105} C_{33,00}^{33,0}}{1024}-\frac{\sqrt{35} C_{33,20}^{33,0}}{1024}+\frac{\sqrt{\frac{21}{2}}
C_{33,22}^{33,0}}{2560}\)

\noindent\(C_{20,2000,0}^{00,2000,0,\text{Loc}}= -\frac{35 C_{00,00}^{60,0}}{384}-\frac{35 C_{00,00}^{60,1}}{768}-\frac{35 C_{11,00}^{51,0}}{1152}-\frac{35
C_{11,00}^{51,1}}{2304}-\frac{25 C_{20,00}^{40,0}}{1152}-\frac{25 C_{20,00}^{40,1}}{2304}+\frac{7 \sqrt{5} C_{22,00}^{42,0}}{576}+\frac{7 \sqrt{5}
C_{22,00}^{42,1}}{1152}+\frac{5 C_{31,00}^{31,0}}{1152}+\frac{5 C_{31,00}^{31,1}}{2304}+\frac{\sqrt{21} C_{33,00}^{33,0}}{64}+\frac{\sqrt{21} C_{33,00}^{33,1}}{128}\)

\noindent\(C_{20,2000,0}^{00,2000,0,\text{NonLoc}}= \frac{35 C_{00,00}^{60,0}}{1536}+\frac{35 C_{00,00}^{60,1}}{768}-\frac{35 C_{00,20}^{60,0}}{512
\sqrt{3}}-\frac{35 C_{00,20}^{60,1}}{256 \sqrt{3}}-\frac{35 C_{11,00}^{51,0}}{4608}-\frac{35 C_{11,00}^{51,1}}{2304}+\frac{35 C_{11,20}^{31,0}}{1536
\sqrt{3}}+\frac{35 C_{11,20}^{31,1}}{768 \sqrt{3}}+\frac{25 C_{20,00}^{40,0}}{4608}+\frac{25 C_{20,00}^{40,1}}{2304}-\frac{25 C_{20,20}^{40,0}}{1536
\sqrt{3}}-\frac{25 C_{20,20}^{40,1}}{768 \sqrt{3}}-\frac{7 \sqrt{5} C_{22,00}^{42,0}}{2304}-\frac{7 \sqrt{5} C_{22,00}^{42,1}}{1152}+\frac{7}{768}
\sqrt{\frac{5}{3}} C_{22,20}^{42,0}+\frac{7}{384} \sqrt{\frac{5}{3}} C_{22,20}^{42,1}+\frac{5 C_{31,00}^{31,0}}{4608}+\frac{5 C_{31,00}^{31,1}}{2304}-\frac{5
C_{31,20}^{31,0}}{1536 \sqrt{3}}-\frac{5 C_{31,20}^{31,1}}{768 \sqrt{3}}+\frac{\sqrt{21} C_{33,00}^{33,0}}{256}+\frac{\sqrt{21} C_{33,00}^{33,1}}{128}-\frac{3
\sqrt{7} C_{33,20}^{33,0}}{256}-\frac{3 \sqrt{7} C_{33,20}^{33,1}}{128}\)

\noindent\(C_{20,2000,0}^{00,2000,1,\text{Loc}}= -\frac{35 C_{00,00}^{60,1}}{256 \sqrt{3}}-\frac{35 C_{11,00}^{51,1}}{768 \sqrt{3}}-\frac{25
C_{20,00}^{40,1}}{768 \sqrt{3}}+\frac{7}{384} \sqrt{\frac{5}{3}} C_{22,00}^{42,1}+\frac{5 C_{31,00}^{31,1}}{768 \sqrt{3}}+\frac{3 \sqrt{7} C_{33,00}^{33,1}}{128}\)

\noindent\(C_{20,2000,0}^{00,2000,1,\text{NonLoc}}= \frac{35 C_{00,00}^{60,0}}{512 \sqrt{3}}-\frac{35 C_{00,20}^{60,0}}{512}-\frac{35
C_{11,00}^{51,0}}{1536 \sqrt{3}}+\frac{35 C_{11,20}^{31,0}}{1536}+\frac{25 C_{20,00}^{40,0}}{1536 \sqrt{3}}-\frac{25 C_{20,20}^{40,0}}{1536}-\frac{7}{768}
\sqrt{\frac{5}{3}} C_{22,00}^{42,0}+\frac{7 \sqrt{5} C_{22,20}^{42,0}}{768}+\frac{5 C_{31,00}^{31,0}}{1536 \sqrt{3}}-\frac{5 C_{31,20}^{31,0}}{1536}+\frac{3
\sqrt{7} C_{33,00}^{33,0}}{256}-\frac{3 \sqrt{21} C_{33,20}^{33,0}}{256}\)

\noindent\(C_{20,2000,0}^{11,1111,0,\text{Loc}}= \frac{35 C_{11,11}^{51,0}}{1152}+\frac{35 C_{11,11}^{51,1}}{2304}+\frac{5 C_{31,11}^{31,0}}{384}+\frac{5
C_{31,11}^{31,1}}{768}-\frac{1}{32} \sqrt{\frac{7}{2}} C_{33,11}^{33,0}-\frac{1}{64} \sqrt{\frac{7}{2}} C_{33,11}^{33,1}\)

\noindent\(C_{20,2000,0}^{11,1111,0,\text{NonLoc}}= \frac{35 C_{11,11}^{51,0}}{2304}+\frac{35 C_{11,11}^{51,1}}{1152}+\frac{5 C_{31,11}^{31,0}}{768}+\frac{5
C_{31,11}^{31,1}}{384}-\frac{1}{64} \sqrt{\frac{7}{2}} C_{33,11}^{33,0}-\frac{1}{32} \sqrt{\frac{7}{2}} C_{33,11}^{33,1}\)

\noindent\(C_{20,2000,0}^{11,1111,1,\text{Loc}}= \frac{35 C_{11,11}^{51,1}}{768 \sqrt{3}}+\frac{5 C_{31,11}^{31,1}}{256 \sqrt{3}}-\frac{1}{64}
\sqrt{\frac{21}{2}} C_{33,11}^{33,1}\)

\noindent\(C_{20,2000,0}^{11,1111,1,\text{NonLoc}}= \frac{35 C_{11,11}^{51,0}}{768 \sqrt{3}}+\frac{5 C_{31,11}^{31,0}}{256 \sqrt{3}}-\frac{1}{64}
\sqrt{\frac{21}{2}} C_{33,11}^{33,0}\)

\noindent\(C_{20,2000,0}^{20,0000,0,\text{Loc}}= \frac{35 C_{00,00}^{60,0}}{1536}+\frac{35 C_{00,00}^{60,1}}{3072}-\frac{35 C_{11,00}^{51,0}}{4608}-\frac{35
C_{11,00}^{51,1}}{9216}+\frac{25 C_{20,00}^{40,0}}{4608}+\frac{25 C_{20,00}^{40,1}}{9216}-\frac{7 \sqrt{5} C_{22,00}^{42,0}}{2304}-\frac{7 \sqrt{5}
C_{22,00}^{42,1}}{4608}+\frac{5 C_{31,00}^{31,0}}{4608}+\frac{5 C_{31,00}^{31,1}}{9216}+\frac{\sqrt{21} C_{33,00}^{33,0}}{256}+\frac{\sqrt{21} C_{33,00}^{33,1}}{512}\)

\noindent\(C_{20,2000,0}^{20,0000,0,\text{NonLoc}}= -\frac{35 C_{00,00}^{60,0}}{6144}-\frac{35 C_{00,00}^{60,1}}{3072}+\frac{35 C_{00,20}^{60,0}}{2048
\sqrt{3}}+\frac{35 C_{00,20}^{60,1}}{1024 \sqrt{3}}-\frac{35 C_{11,00}^{51,0}}{18432}-\frac{35 C_{11,00}^{51,1}}{9216}+\frac{35 C_{11,20}^{31,0}}{6144
\sqrt{3}}+\frac{35 C_{11,20}^{31,1}}{3072 \sqrt{3}}-\frac{25 C_{20,00}^{40,0}}{18432}-\frac{25 C_{20,00}^{40,1}}{9216}+\frac{25 C_{20,20}^{40,0}}{6144
\sqrt{3}}+\frac{25 C_{20,20}^{40,1}}{3072 \sqrt{3}}+\frac{7 \sqrt{5} C_{22,00}^{42,0}}{9216}+\frac{7 \sqrt{5} C_{22,00}^{42,1}}{4608}-\frac{7 \sqrt{\frac{5}{3}}
C_{22,20}^{42,0}}{3072}-\frac{7 \sqrt{\frac{5}{3}} C_{22,20}^{42,1}}{1536}+\frac{5 C_{31,00}^{31,0}}{18432}+\frac{5 C_{31,00}^{31,1}}{9216}-\frac{5
C_{31,20}^{31,0}}{6144 \sqrt{3}}-\frac{5 C_{31,20}^{31,1}}{3072 \sqrt{3}}+\frac{\sqrt{21} C_{33,00}^{33,0}}{1024}+\frac{\sqrt{21} C_{33,00}^{33,1}}{512}-\frac{3
\sqrt{7} C_{33,20}^{33,0}}{1024}-\frac{3 \sqrt{7} C_{33,20}^{33,1}}{512}\)

\noindent\(C_{20,2000,0}^{20,0000,1,\text{Loc}}= \frac{35 C_{00,00}^{60,1}}{1024 \sqrt{3}}-\frac{35 C_{11,00}^{51,1}}{3072 \sqrt{3}}+\frac{25
C_{20,00}^{40,1}}{3072 \sqrt{3}}-\frac{7 \sqrt{\frac{5}{3}} C_{22,00}^{42,1}}{1536}+\frac{5 C_{31,00}^{31,1}}{3072 \sqrt{3}}+\frac{3 \sqrt{7} C_{33,00}^{33,1}}{512}\)

\noindent\(C_{20,2000,0}^{20,0000,1,\text{NonLoc}}= -\frac{35 C_{00,00}^{60,0}}{2048 \sqrt{3}}+\frac{35 C_{00,20}^{60,0}}{2048}-\frac{35
C_{11,00}^{51,0}}{6144 \sqrt{3}}+\frac{35 C_{11,20}^{31,0}}{6144}-\frac{25 C_{20,00}^{40,0}}{6144 \sqrt{3}}+\frac{25 C_{20,20}^{40,0}}{6144}+\frac{7
\sqrt{\frac{5}{3}} C_{22,00}^{42,0}}{3072}-\frac{7 \sqrt{5} C_{22,20}^{42,0}}{3072}+\frac{5 C_{31,00}^{31,0}}{6144 \sqrt{3}}-\frac{5 C_{31,20}^{31,0}}{6144}+\frac{3
\sqrt{7} C_{33,00}^{33,0}}{1024}-\frac{3 \sqrt{21} C_{33,20}^{33,0}}{1024}\)

\noindent\(C_{20,2011,1}^{00,2011,0,\text{Loc}}= \frac{35 C_{00,20}^{60,0}}{384}+\frac{35 C_{00,20}^{60,1}}{768}+\frac{35 C_{11,20}^{31,0}}{1152}+\frac{35
C_{11,20}^{31,1}}{2304}+\frac{25 C_{20,20}^{40,0}}{1152}+\frac{25 C_{20,20}^{40,1}}{2304}-\frac{7 \sqrt{5} C_{22,20}^{42,0}}{576}-\frac{7 \sqrt{5}
C_{22,20}^{42,1}}{1152}-\frac{5 C_{31,20}^{31,0}}{1152}-\frac{5 C_{31,20}^{31,1}}{2304}-\frac{\sqrt{21} C_{33,20}^{33,0}}{64}-\frac{\sqrt{21} C_{33,20}^{33,1}}{128}\)

\noindent\(C_{20,2011,1}^{00,2011,0,\text{NonLoc}}= \frac{35 C_{00,00}^{60,0}}{512 \sqrt{3}}+\frac{35 C_{00,00}^{60,1}}{256 \sqrt{3}}+\frac{35
C_{00,20}^{60,0}}{1536}+\frac{35 C_{00,20}^{60,1}}{768}-\frac{35 C_{11,00}^{51,0}}{1536 \sqrt{3}}-\frac{35 C_{11,00}^{51,1}}{768 \sqrt{3}}-\frac{35
C_{11,20}^{31,0}}{4608}-\frac{35 C_{11,20}^{31,1}}{2304}+\frac{25 C_{20,00}^{40,0}}{1536 \sqrt{3}}+\frac{25 C_{20,00}^{40,1}}{768 \sqrt{3}}+\frac{25
C_{20,20}^{40,0}}{4608}+\frac{25 C_{20,20}^{40,1}}{2304}-\frac{7}{768} \sqrt{\frac{5}{3}} C_{22,00}^{42,0}-\frac{7}{384} \sqrt{\frac{5}{3}} C_{22,00}^{42,1}-\frac{7
\sqrt{5} C_{22,20}^{42,0}}{2304}-\frac{7 \sqrt{5} C_{22,20}^{42,1}}{1152}+\frac{5 C_{31,00}^{31,0}}{1536 \sqrt{3}}+\frac{5 C_{31,00}^{31,1}}{768
\sqrt{3}}+\frac{5 C_{31,20}^{31,0}}{4608}+\frac{5 C_{31,20}^{31,1}}{2304}+\frac{3 \sqrt{7} C_{33,00}^{33,0}}{256}+\frac{3 \sqrt{7} C_{33,00}^{33,1}}{128}+\frac{\sqrt{21}
C_{33,20}^{33,0}}{256}+\frac{\sqrt{21} C_{33,20}^{33,1}}{128}\)

\noindent\(C_{20,2011,1}^{00,2011,1,\text{Loc}}= \frac{35 C_{00,20}^{60,1}}{256 \sqrt{3}}+\frac{35 C_{11,20}^{31,1}}{768 \sqrt{3}}+\frac{25
C_{20,20}^{40,1}}{768 \sqrt{3}}-\frac{7}{384} \sqrt{\frac{5}{3}} C_{22,20}^{42,1}-\frac{5 C_{31,20}^{31,1}}{768 \sqrt{3}}-\frac{3 \sqrt{7} C_{33,20}^{33,1}}{128}\)

\noindent\(C_{20,2011,1}^{00,2011,1,\text{NonLoc}}= \frac{35 C_{00,00}^{60,0}}{512}+\frac{35 C_{00,20}^{60,0}}{512 \sqrt{3}}-\frac{35
C_{11,00}^{51,0}}{1536}-\frac{35 C_{11,20}^{31,0}}{1536 \sqrt{3}}+\frac{25 C_{20,00}^{40,0}}{1536}+\frac{25 C_{20,20}^{40,0}}{1536 \sqrt{3}}-\frac{7
\sqrt{5} C_{22,00}^{42,0}}{768}-\frac{7}{768} \sqrt{\frac{5}{3}} C_{22,20}^{42,0}+\frac{5 C_{31,00}^{31,0}}{1536}+\frac{5 C_{31,20}^{31,0}}{1536
\sqrt{3}}+\frac{3 \sqrt{21} C_{33,00}^{33,0}}{256}+\frac{3 \sqrt{7} C_{33,20}^{33,0}}{256}\)

\noindent\(C_{20,2011,1}^{00,2211,0,\text{Loc}}= \frac{7 C_{00,22}^{62,0}}{128}+\frac{7 C_{00,22}^{62,1}}{256}+\frac{35 C_{11,22}^{51,0}}{1152}+\frac{35
C_{11,22}^{51,1}}{2304}+\frac{7 C_{20,22}^{42,0}}{1152}+\frac{7 C_{20,22}^{42,1}}{2304}+\frac{7 C_{22,22}^{40,0}}{1152}+\frac{7 C_{22,22}^{40,1}}{2304}-\frac{\sqrt{7}
C_{22,22}^{42,0}}{288}-\frac{\sqrt{7} C_{22,22}^{42,1}}{576}-\frac{3 C_{22,22}^{44,0}}{32 \sqrt{5}}-\frac{3 C_{22,22}^{44,1}}{64 \sqrt{5}}+\frac{7
C_{31,22}^{31,0}}{1152}+\frac{7 C_{31,22}^{31,1}}{2304}-\frac{1}{32} \sqrt{\frac{7}{15}} C_{31,22}^{33,0}-\frac{1}{64} \sqrt{\frac{7}{15}} C_{31,22}^{33,1}-\frac{3}{160}
\sqrt{\frac{7}{2}} C_{33,22}^{33,0}-\frac{3}{320} \sqrt{\frac{7}{2}} C_{33,22}^{33,1}\)

\noindent\(C_{20,2011,1}^{00,2211,0,\text{NonLoc}}= -\frac{7 C_{00,22}^{62,0}}{256}-\frac{7 C_{00,22}^{62,1}}{128}+\frac{35 C_{11,22}^{51,0}}{2304}+\frac{35
C_{11,22}^{51,1}}{1152}-\frac{7 C_{20,22}^{42,0}}{2304}-\frac{7 C_{20,22}^{42,1}}{1152}-\frac{7 C_{22,22}^{40,0}}{2304}-\frac{7 C_{22,22}^{40,1}}{1152}+\frac{\sqrt{7}
C_{22,22}^{42,0}}{576}+\frac{\sqrt{7} C_{22,22}^{42,1}}{288}+\frac{3 C_{22,22}^{44,0}}{64 \sqrt{5}}+\frac{3 C_{22,22}^{44,1}}{32 \sqrt{5}}+\frac{7
C_{31,22}^{31,0}}{2304}+\frac{7 C_{31,22}^{31,1}}{1152}-\frac{1}{64} \sqrt{\frac{7}{15}} C_{31,22}^{33,0}-\frac{1}{32} \sqrt{\frac{7}{15}} C_{31,22}^{33,1}-\frac{3}{320}
\sqrt{\frac{7}{2}} C_{33,22}^{33,0}-\frac{3}{160} \sqrt{\frac{7}{2}} C_{33,22}^{33,1}\)

\noindent\(C_{20,2011,1}^{00,2211,1,\text{Loc}}= \frac{7 \sqrt{3} C_{00,22}^{62,1}}{256}+\frac{35 C_{11,22}^{51,1}}{768 \sqrt{3}}+\frac{7
C_{20,22}^{42,1}}{768 \sqrt{3}}+\frac{7 C_{22,22}^{40,1}}{768 \sqrt{3}}-\frac{1}{192} \sqrt{\frac{7}{3}} C_{22,22}^{42,1}-\frac{3}{64} \sqrt{\frac{3}{5}}
C_{22,22}^{44,1}+\frac{7 C_{31,22}^{31,1}}{768 \sqrt{3}}-\frac{1}{64} \sqrt{\frac{7}{5}} C_{31,22}^{33,1}-\frac{3}{320} \sqrt{\frac{21}{2}} C_{33,22}^{33,1}\)

\noindent\(C_{20,2011,1}^{00,2211,1,\text{NonLoc}}= -\frac{7 \sqrt{3} C_{00,22}^{62,0}}{256}+\frac{35 C_{11,22}^{51,0}}{768 \sqrt{3}}-\frac{7
C_{20,22}^{42,0}}{768 \sqrt{3}}-\frac{7 C_{22,22}^{40,0}}{768 \sqrt{3}}+\frac{1}{192} \sqrt{\frac{7}{3}} C_{22,22}^{42,0}+\frac{3}{64} \sqrt{\frac{3}{5}}
C_{22,22}^{44,0}+\frac{7 C_{31,22}^{31,0}}{768 \sqrt{3}}-\frac{1}{64} \sqrt{\frac{7}{5}} C_{31,22}^{33,0}-\frac{3}{320} \sqrt{\frac{21}{2}} C_{33,22}^{33,0}\)

\noindent\(C_{20,2011,1}^{11,1101,0,\text{Loc}}= \frac{35 C_{11,11}^{51,0}}{1152}+\frac{35 C_{11,11}^{51,1}}{2304}+\frac{5 C_{31,11}^{31,0}}{384}+\frac{5
C_{31,11}^{31,1}}{768}-\frac{1}{32} \sqrt{\frac{7}{2}} C_{33,11}^{33,0}-\frac{1}{64} \sqrt{\frac{7}{2}} C_{33,11}^{33,1}\)

\noindent\(C_{20,2011,1}^{11,1101,0,\text{NonLoc}}= \frac{35 C_{11,11}^{51,0}}{2304}+\frac{35 C_{11,11}^{51,1}}{1152}+\frac{5 C_{31,11}^{31,0}}{768}+\frac{5
C_{31,11}^{31,1}}{384}-\frac{1}{64} \sqrt{\frac{7}{2}} C_{33,11}^{33,0}-\frac{1}{32} \sqrt{\frac{7}{2}} C_{33,11}^{33,1}\)

\noindent\(C_{20,2011,1}^{11,1101,1,\text{Loc}}= \frac{35 C_{11,11}^{51,1}}{768 \sqrt{3}}+\frac{5 C_{31,11}^{31,1}}{256 \sqrt{3}}-\frac{1}{64}
\sqrt{\frac{21}{2}} C_{33,11}^{33,1}\)

\noindent\(C_{20,2011,1}^{11,1101,1,\text{NonLoc}}= \frac{35 C_{11,11}^{51,0}}{768 \sqrt{3}}+\frac{5 C_{31,11}^{31,0}}{256 \sqrt{3}}-\frac{1}{64}
\sqrt{\frac{21}{2}} C_{33,11}^{33,0}\)

\noindent\(C_{20,2011,1}^{20,0011,0,\text{Loc}}= -\frac{35 C_{00,20}^{60,0}}{1536}-\frac{35 C_{00,20}^{60,1}}{3072}+\frac{35 C_{11,20}^{31,0}}{4608}+\frac{35
C_{11,20}^{31,1}}{9216}-\frac{25 C_{20,20}^{40,0}}{4608}-\frac{25 C_{20,20}^{40,1}}{9216}+\frac{7 \sqrt{5} C_{22,20}^{42,0}}{2304}+\frac{7 \sqrt{5}
C_{22,20}^{42,1}}{4608}-\frac{5 C_{31,20}^{31,0}}{4608}-\frac{5 C_{31,20}^{31,1}}{9216}-\frac{\sqrt{21} C_{33,20}^{33,0}}{256}-\frac{\sqrt{21} C_{33,20}^{33,1}}{512}\)

\noindent\(C_{20,2011,1}^{20,0011,0,\text{NonLoc}}= -\frac{35 C_{00,00}^{60,0}}{2048 \sqrt{3}}-\frac{35 C_{00,00}^{60,1}}{1024 \sqrt{3}}-\frac{35
C_{00,20}^{60,0}}{6144}-\frac{35 C_{00,20}^{60,1}}{3072}-\frac{35 C_{11,00}^{51,0}}{6144 \sqrt{3}}-\frac{35 C_{11,00}^{51,1}}{3072 \sqrt{3}}-\frac{35
C_{11,20}^{31,0}}{18432}-\frac{35 C_{11,20}^{31,1}}{9216}-\frac{25 C_{20,00}^{40,0}}{6144 \sqrt{3}}-\frac{25 C_{20,00}^{40,1}}{3072 \sqrt{3}}-\frac{25
C_{20,20}^{40,0}}{18432}-\frac{25 C_{20,20}^{40,1}}{9216}+\frac{7 \sqrt{\frac{5}{3}} C_{22,00}^{42,0}}{3072}+\frac{7 \sqrt{\frac{5}{3}} C_{22,00}^{42,1}}{1536}+\frac{7
\sqrt{5} C_{22,20}^{42,0}}{9216}+\frac{7 \sqrt{5} C_{22,20}^{42,1}}{4608}+\frac{5 C_{31,00}^{31,0}}{6144 \sqrt{3}}+\frac{5 C_{31,00}^{31,1}}{3072
\sqrt{3}}+\frac{5 C_{31,20}^{31,0}}{18432}+\frac{5 C_{31,20}^{31,1}}{9216}+\frac{3 \sqrt{7} C_{33,00}^{33,0}}{1024}+\frac{3 \sqrt{7} C_{33,00}^{33,1}}{512}+\frac{\sqrt{21}
C_{33,20}^{33,0}}{1024}+\frac{\sqrt{21} C_{33,20}^{33,1}}{512}\)

\noindent\(C_{20,2011,1}^{20,0011,1,\text{Loc}}= -\frac{35 C_{00,20}^{60,1}}{1024 \sqrt{3}}+\frac{35 C_{11,20}^{31,1}}{3072 \sqrt{3}}-\frac{25
C_{20,20}^{40,1}}{3072 \sqrt{3}}+\frac{7 \sqrt{\frac{5}{3}} C_{22,20}^{42,1}}{1536}-\frac{5 C_{31,20}^{31,1}}{3072 \sqrt{3}}-\frac{3 \sqrt{7} C_{33,20}^{33,1}}{512}\)

\noindent\(C_{20,2011,1}^{20,0011,1,\text{NonLoc}}= -\frac{35 C_{00,00}^{60,0}}{2048}-\frac{35 C_{00,20}^{60,0}}{2048 \sqrt{3}}-\frac{35
C_{11,00}^{51,0}}{6144}-\frac{35 C_{11,20}^{31,0}}{6144 \sqrt{3}}-\frac{25 C_{20,00}^{40,0}}{6144}-\frac{25 C_{20,20}^{40,0}}{6144 \sqrt{3}}+\frac{7
\sqrt{5} C_{22,00}^{42,0}}{3072}+\frac{7 \sqrt{\frac{5}{3}} C_{22,20}^{42,0}}{3072}+\frac{5 C_{31,00}^{31,0}}{6144}+\frac{5 C_{31,20}^{31,0}}{6144
\sqrt{3}}+\frac{3 \sqrt{21} C_{33,00}^{33,0}}{1024}+\frac{3 \sqrt{7} C_{33,20}^{33,0}}{1024}\)

\noindent\(C_{20,2202,2}^{00,2202,0,\text{Loc}}= -\frac{7 \sqrt{5} C_{00,00}^{60,0}}{192}-\frac{7 \sqrt{5} C_{00,00}^{60,1}}{384}-\frac{7
\sqrt{5} C_{11,00}^{51,0}}{576}-\frac{7 \sqrt{5} C_{11,00}^{51,1}}{1152}-\frac{5 \sqrt{5} C_{20,00}^{40,0}}{576}-\frac{5 \sqrt{5} C_{20,00}^{40,1}}{1152}+\frac{7
C_{22,00}^{42,0}}{288}+\frac{7 C_{22,00}^{42,1}}{576}+\frac{\sqrt{5} C_{31,00}^{31,0}}{576}+\frac{\sqrt{5} C_{31,00}^{31,1}}{1152}+\frac{1}{32} \sqrt{\frac{21}{5}}
C_{33,00}^{33,0}+\frac{1}{64} \sqrt{\frac{21}{5}} C_{33,00}^{33,1}\)

\noindent\(C_{20,2202,2}^{00,2202,0,\text{NonLoc}}= \frac{7 \sqrt{5} C_{00,00}^{60,0}}{768}+\frac{7 \sqrt{5} C_{00,00}^{60,1}}{384}-\frac{7}{256}
\sqrt{\frac{5}{3}} C_{00,20}^{60,0}-\frac{7}{128} \sqrt{\frac{5}{3}} C_{00,20}^{60,1}-\frac{7 \sqrt{5} C_{11,00}^{51,0}}{2304}-\frac{7 \sqrt{5} C_{11,00}^{51,1}}{1152}+\frac{5
\sqrt{5} C_{20,00}^{40,0}}{2304}+\frac{5 \sqrt{5} C_{20,00}^{40,1}}{1152}-\frac{5}{768} \sqrt{\frac{5}{3}} C_{20,20}^{40,0}-\frac{5}{384} \sqrt{\frac{5}{3}}
C_{20,20}^{40,1}-\frac{7 C_{22,00}^{42,0}}{1152}-\frac{7 C_{22,00}^{42,1}}{576}+\frac{7 C_{22,20}^{42,0}}{384 \sqrt{3}}+\frac{7 C_{22,20}^{42,1}}{192
\sqrt{3}}+\frac{\sqrt{5} C_{31,00}^{31,0}}{2304}+\frac{\sqrt{5} C_{31,00}^{31,1}}{1152}-\frac{1}{768} \sqrt{\frac{5}{3}} C_{31,20}^{31,0}-\frac{1}{384}
\sqrt{\frac{5}{3}} C_{31,20}^{31,1}+\frac{1}{128} \sqrt{\frac{21}{5}} C_{33,00}^{33,0}+\frac{1}{64} \sqrt{\frac{21}{5}} C_{33,00}^{33,1}-\frac{3}{128}
\sqrt{\frac{7}{5}} C_{33,20}^{33,0}-\frac{3}{64} \sqrt{\frac{7}{5}} C_{33,20}^{33,1}\)

\noindent\(C_{20,2202,2}^{00,2202,1,\text{Loc}}= -\frac{7}{128} \sqrt{\frac{5}{3}} C_{00,00}^{60,1}-\frac{7}{384} \sqrt{\frac{5}{3}}
C_{11,00}^{51,1}-\frac{5}{384} \sqrt{\frac{5}{3}} C_{20,00}^{40,1}+\frac{7 C_{22,00}^{42,1}}{192 \sqrt{3}}+\frac{1}{384} \sqrt{\frac{5}{3}} C_{31,00}^{31,1}+\frac{3}{64}
\sqrt{\frac{7}{5}} C_{33,00}^{33,1}\)

\noindent\(C_{20,2202,2}^{00,2202,1,\text{NonLoc}}= \frac{7}{256} \sqrt{\frac{5}{3}} C_{00,00}^{60,0}-\frac{7 \sqrt{5} C_{00,20}^{60,0}}{256}-\frac{7}{768}
\sqrt{\frac{5}{3}} C_{11,00}^{51,0}+\frac{5}{768} \sqrt{\frac{5}{3}} C_{20,00}^{40,0}-\frac{5 \sqrt{5} C_{20,20}^{40,0}}{768}-\frac{7 C_{22,00}^{42,0}}{384
\sqrt{3}}+\frac{7 C_{22,20}^{42,0}}{384}+\frac{1}{768} \sqrt{\frac{5}{3}} C_{31,00}^{31,0}-\frac{\sqrt{5} C_{31,20}^{31,0}}{768}+\frac{3}{128} \sqrt{\frac{7}{5}}
C_{33,00}^{33,0}-\frac{3}{128} \sqrt{\frac{21}{5}} C_{33,20}^{33,0}\)

\noindent\(C_{20,2202,2}^{11,1111,0,\text{Loc}}= -\frac{7 \sqrt{5} C_{11,11}^{51,0}}{1152}-\frac{7 \sqrt{5} C_{11,11}^{51,1}}{2304}-\frac{\sqrt{5}
C_{31,11}^{31,0}}{384}-\frac{\sqrt{5} C_{31,11}^{31,1}}{768}+\frac{1}{32} \sqrt{\frac{7}{10}} C_{33,11}^{33,0}+\frac{1}{64} \sqrt{\frac{7}{10}} C_{33,11}^{33,1}\)

\noindent\(C_{20,2202,2}^{11,1111,0,\text{NonLoc}}= -\frac{7 \sqrt{5} C_{11,11}^{51,0}}{2304}-\frac{7 \sqrt{5} C_{11,11}^{51,1}}{1152}-\frac{\sqrt{5}
C_{31,11}^{31,0}}{768}-\frac{\sqrt{5} C_{31,11}^{31,1}}{384}+\frac{1}{64} \sqrt{\frac{7}{10}} C_{33,11}^{33,0}+\frac{1}{32} \sqrt{\frac{7}{10}} C_{33,11}^{33,1}\)

\noindent\(C_{20,2202,2}^{11,1111,1,\text{Loc}}= -\frac{7}{768} \sqrt{\frac{5}{3}} C_{11,11}^{51,1}-\frac{1}{256} \sqrt{\frac{5}{3}}
C_{31,11}^{31,1}+\frac{1}{64} \sqrt{\frac{21}{10}} C_{33,11}^{33,1}\)

\noindent\(C_{20,2202,2}^{11,1111,1,\text{NonLoc}}= -\frac{7}{768} \sqrt{\frac{5}{3}} C_{11,11}^{51,0}-\frac{1}{256} \sqrt{\frac{5}{3}}
C_{31,11}^{31,0}+\frac{1}{64} \sqrt{\frac{21}{10}} C_{33,11}^{33,0}\)

\noindent\(C_{20,2202,2}^{11,1112,0,\text{Loc}}= -\frac{7}{384} \sqrt{\frac{5}{3}} C_{11,11}^{51,0}-\frac{7}{768} \sqrt{\frac{5}{3}}
C_{11,11}^{51,1}-\frac{1}{128} \sqrt{\frac{5}{3}} C_{31,11}^{31,0}-\frac{1}{256} \sqrt{\frac{5}{3}} C_{31,11}^{31,1}+\frac{1}{32} \sqrt{\frac{21}{10}}
C_{33,11}^{33,0}+\frac{1}{64} \sqrt{\frac{21}{10}} C_{33,11}^{33,1}\)

\noindent\(C_{20,2202,2}^{11,1112,0,\text{NonLoc}}= -\frac{7}{768} \sqrt{\frac{5}{3}} C_{11,11}^{51,0}-\frac{7}{384} \sqrt{\frac{5}{3}}
C_{11,11}^{51,1}-\frac{1}{256} \sqrt{\frac{5}{3}} C_{31,11}^{31,0}-\frac{1}{128} \sqrt{\frac{5}{3}} C_{31,11}^{31,1}+\frac{1}{64} \sqrt{\frac{21}{10}}
C_{33,11}^{33,0}+\frac{1}{32} \sqrt{\frac{21}{10}} C_{33,11}^{33,1}\)

\noindent\(C_{20,2202,2}^{11,1112,1,\text{Loc}}= -\frac{7 \sqrt{5} C_{11,11}^{51,1}}{768}-\frac{\sqrt{5} C_{31,11}^{31,1}}{256}+\frac{3}{64}
\sqrt{\frac{7}{10}} C_{33,11}^{33,1}\)

\noindent\(C_{20,2202,2}^{11,1112,1,\text{NonLoc}}= -\frac{7 \sqrt{5} C_{11,11}^{51,0}}{768}-\frac{\sqrt{5} C_{31,11}^{31,0}}{256}+\frac{3}{64}
\sqrt{\frac{7}{10}} C_{33,11}^{33,0}\)

\noindent\(C_{20,2211,1}^{00,2011,0,\text{Loc}}= \frac{7 C_{00,22}^{62,0}}{128}+\frac{7 C_{00,22}^{62,1}}{256}+\frac{35 C_{11,22}^{51,0}}{1152}+\frac{35
C_{11,22}^{51,1}}{2304}+\frac{7 C_{20,22}^{42,0}}{1152}+\frac{7 C_{20,22}^{42,1}}{2304}+\frac{7 C_{22,22}^{40,0}}{1152}+\frac{7 C_{22,22}^{40,1}}{2304}-\frac{\sqrt{7}
C_{22,22}^{42,0}}{288}-\frac{\sqrt{7} C_{22,22}^{42,1}}{576}-\frac{3 C_{22,22}^{44,0}}{32 \sqrt{5}}-\frac{3 C_{22,22}^{44,1}}{64 \sqrt{5}}+\frac{7
C_{31,22}^{31,0}}{1152}+\frac{7 C_{31,22}^{31,1}}{2304}-\frac{1}{32} \sqrt{\frac{7}{15}} C_{31,22}^{33,0}-\frac{1}{64} \sqrt{\frac{7}{15}} C_{31,22}^{33,1}-\frac{3}{160}
\sqrt{\frac{7}{2}} C_{33,22}^{33,0}-\frac{3}{320} \sqrt{\frac{7}{2}} C_{33,22}^{33,1}\)

\noindent\(C_{20,2211,1}^{00,2011,0,\text{NonLoc}}= -\frac{7 C_{00,22}^{62,0}}{256}-\frac{7 C_{00,22}^{62,1}}{128}+\frac{35 C_{11,22}^{51,0}}{2304}+\frac{35
C_{11,22}^{51,1}}{1152}-\frac{7 C_{20,22}^{42,0}}{2304}-\frac{7 C_{20,22}^{42,1}}{1152}-\frac{7 C_{22,22}^{40,0}}{2304}-\frac{7 C_{22,22}^{40,1}}{1152}+\frac{\sqrt{7}
C_{22,22}^{42,0}}{576}+\frac{\sqrt{7} C_{22,22}^{42,1}}{288}+\frac{3 C_{22,22}^{44,0}}{64 \sqrt{5}}+\frac{3 C_{22,22}^{44,1}}{32 \sqrt{5}}+\frac{7
C_{31,22}^{31,0}}{2304}+\frac{7 C_{31,22}^{31,1}}{1152}-\frac{1}{64} \sqrt{\frac{7}{15}} C_{31,22}^{33,0}-\frac{1}{32} \sqrt{\frac{7}{15}} C_{31,22}^{33,1}-\frac{3}{320}
\sqrt{\frac{7}{2}} C_{33,22}^{33,0}-\frac{3}{160} \sqrt{\frac{7}{2}} C_{33,22}^{33,1}\)

\noindent\(C_{20,2211,1}^{00,2011,1,\text{Loc}}= \frac{7 \sqrt{3} C_{00,22}^{62,1}}{256}+\frac{35 C_{11,22}^{51,1}}{768 \sqrt{3}}+\frac{7
C_{20,22}^{42,1}}{768 \sqrt{3}}+\frac{7 C_{22,22}^{40,1}}{768 \sqrt{3}}-\frac{1}{192} \sqrt{\frac{7}{3}} C_{22,22}^{42,1}-\frac{3}{64} \sqrt{\frac{3}{5}}
C_{22,22}^{44,1}+\frac{7 C_{31,22}^{31,1}}{768 \sqrt{3}}-\frac{1}{64} \sqrt{\frac{7}{5}} C_{31,22}^{33,1}-\frac{3}{320} \sqrt{\frac{21}{2}} C_{33,22}^{33,1}\)

\noindent\(C_{20,2211,1}^{00,2011,1,\text{NonLoc}}= -\frac{7 \sqrt{3} C_{00,22}^{62,0}}{256}+\frac{35 C_{11,22}^{51,0}}{768 \sqrt{3}}-\frac{7
C_{20,22}^{42,0}}{768 \sqrt{3}}-\frac{7 C_{22,22}^{40,0}}{768 \sqrt{3}}+\frac{1}{192} \sqrt{\frac{7}{3}} C_{22,22}^{42,0}+\frac{3}{64} \sqrt{\frac{3}{5}}
C_{22,22}^{44,0}+\frac{7 C_{31,22}^{31,0}}{768 \sqrt{3}}-\frac{1}{64} \sqrt{\frac{7}{5}} C_{31,22}^{33,0}-\frac{3}{320} \sqrt{\frac{21}{2}} C_{33,22}^{33,0}\)

\noindent\(C_{20,2211,1}^{00,2211,0,\text{Loc}}= \frac{7 C_{00,20}^{60,0}}{192}+\frac{7 C_{00,20}^{60,1}}{384}+\frac{7 C_{00,22}^{62,0}}{128
\sqrt{5}}+\frac{7 C_{00,22}^{62,1}}{256 \sqrt{5}}+\frac{7 \sqrt{5} C_{11,22}^{51,0}}{1152}+\frac{7 \sqrt{5} C_{11,22}^{51,1}}{2304}+\frac{5 C_{20,20}^{40,0}}{576}+\frac{5
C_{20,20}^{40,1}}{1152}+\frac{7 C_{20,22}^{42,0}}{1152 \sqrt{5}}+\frac{7 C_{20,22}^{42,1}}{2304 \sqrt{5}}-\frac{7 C_{22,20}^{42,0}}{288 \sqrt{5}}-\frac{7
C_{22,20}^{42,1}}{576 \sqrt{5}}+\frac{7 C_{22,22}^{40,0}}{1152 \sqrt{5}}+\frac{7 C_{22,22}^{40,1}}{2304 \sqrt{5}}-\frac{1}{288} \sqrt{\frac{7}{5}}
C_{22,22}^{42,0}-\frac{1}{576} \sqrt{\frac{7}{5}} C_{22,22}^{42,1}-\frac{3 C_{22,22}^{44,0}}{160}-\frac{3 C_{22,22}^{44,1}}{320}-\frac{C_{31,20}^{31,0}}{576}-\frac{C_{31,20}^{31,1}}{1152}+\frac{7
C_{31,22}^{31,0}}{1152 \sqrt{5}}+\frac{7 C_{31,22}^{31,1}}{2304 \sqrt{5}}-\frac{1}{160} \sqrt{\frac{7}{3}} C_{31,22}^{33,0}-\frac{1}{320} \sqrt{\frac{7}{3}}
C_{31,22}^{33,1}-\frac{\sqrt{21} C_{33,20}^{33,0}}{160}-\frac{\sqrt{21} C_{33,20}^{33,1}}{320}-\frac{3}{160} \sqrt{\frac{7}{10}} C_{33,22}^{33,0}-\frac{3}{320}
\sqrt{\frac{7}{10}} C_{33,22}^{33,1}\)

\noindent\(C_{20,2211,1}^{00,2211,0,\text{NonLoc}}= \frac{7 C_{00,00}^{60,0}}{256 \sqrt{3}}+\frac{7 C_{00,00}^{60,1}}{128 \sqrt{3}}+\frac{7
C_{00,20}^{60,0}}{768}+\frac{7 C_{00,20}^{60,1}}{384}-\frac{7 C_{00,22}^{62,0}}{256 \sqrt{5}}-\frac{7 C_{00,22}^{62,1}}{128 \sqrt{5}}-\frac{7 C_{11,00}^{51,0}}{768
\sqrt{3}}-\frac{7 C_{11,00}^{51,1}}{384 \sqrt{3}}+\frac{7 \sqrt{5} C_{11,22}^{51,0}}{2304}+\frac{7 \sqrt{5} C_{11,22}^{51,1}}{1152}+\frac{5 C_{20,00}^{40,0}}{768
\sqrt{3}}+\frac{5 C_{20,00}^{40,1}}{384 \sqrt{3}}+\frac{5 C_{20,20}^{40,0}}{2304}+\frac{5 C_{20,20}^{40,1}}{1152}-\frac{7 C_{20,22}^{42,0}}{2304
\sqrt{5}}-\frac{7 C_{20,22}^{42,1}}{1152 \sqrt{5}}-\frac{7 C_{22,00}^{42,0}}{384 \sqrt{15}}-\frac{7 C_{22,00}^{42,1}}{192 \sqrt{15}}-\frac{7 C_{22,20}^{42,0}}{1152
\sqrt{5}}-\frac{7 C_{22,20}^{42,1}}{576 \sqrt{5}}-\frac{7 C_{22,22}^{40,0}}{2304 \sqrt{5}}-\frac{7 C_{22,22}^{40,1}}{1152 \sqrt{5}}+\frac{1}{576}
\sqrt{\frac{7}{5}} C_{22,22}^{42,0}+\frac{1}{288} \sqrt{\frac{7}{5}} C_{22,22}^{42,1}+\frac{3 C_{22,22}^{44,0}}{320}+\frac{3 C_{22,22}^{44,1}}{160}+\frac{C_{31,00}^{31,0}}{768
\sqrt{3}}+\frac{C_{31,00}^{31,1}}{384 \sqrt{3}}+\frac{C_{31,20}^{31,0}}{2304}+\frac{C_{31,20}^{31,1}}{1152}+\frac{7 C_{31,22}^{31,0}}{2304 \sqrt{5}}+\frac{7
C_{31,22}^{31,1}}{1152 \sqrt{5}}-\frac{1}{320} \sqrt{\frac{7}{3}} C_{31,22}^{33,0}-\frac{1}{160} \sqrt{\frac{7}{3}} C_{31,22}^{33,1}+\frac{3 \sqrt{7}
C_{33,00}^{33,0}}{640}+\frac{3 \sqrt{7} C_{33,00}^{33,1}}{320}+\frac{\sqrt{21} C_{33,20}^{33,0}}{640}+\frac{\sqrt{21} C_{33,20}^{33,1}}{320}-\frac{3}{320}
\sqrt{\frac{7}{10}} C_{33,22}^{33,0}-\frac{3}{160} \sqrt{\frac{7}{10}} C_{33,22}^{33,1}\)

\noindent\(C_{20,2211,1}^{00,2211,1,\text{Loc}}= \frac{7 C_{00,20}^{60,1}}{128 \sqrt{3}}+\frac{7}{256} \sqrt{\frac{3}{5}} C_{00,22}^{62,1}+\frac{7}{768}
\sqrt{\frac{5}{3}} C_{11,22}^{51,1}+\frac{5 C_{20,20}^{40,1}}{384 \sqrt{3}}+\frac{7 C_{20,22}^{42,1}}{768 \sqrt{15}}-\frac{7 C_{22,20}^{42,1}}{192
\sqrt{15}}+\frac{7 C_{22,22}^{40,1}}{768 \sqrt{15}}-\frac{1}{192} \sqrt{\frac{7}{15}} C_{22,22}^{42,1}-\frac{3 \sqrt{3} C_{22,22}^{44,1}}{320}-\frac{C_{31,20}^{31,1}}{384
\sqrt{3}}+\frac{7 C_{31,22}^{31,1}}{768 \sqrt{15}}-\frac{\sqrt{7} C_{31,22}^{33,1}}{320}-\frac{3 \sqrt{7} C_{33,20}^{33,1}}{320}-\frac{3}{320} \sqrt{\frac{21}{10}}
C_{33,22}^{33,1}\)

\noindent\(C_{20,2211,1}^{00,2211,1,\text{NonLoc}}= \frac{7 C_{00,00}^{60,0}}{256}+\frac{7 C_{00,20}^{60,0}}{256 \sqrt{3}}-\frac{7}{256}
\sqrt{\frac{3}{5}} C_{00,22}^{62,0}-\frac{7 C_{11,00}^{51,0}}{768}+\frac{7}{768} \sqrt{\frac{5}{3}} C_{11,22}^{51,0}+\frac{5 C_{20,00}^{40,0}}{768}+\frac{5
C_{20,20}^{40,0}}{768 \sqrt{3}}-\frac{7 C_{20,22}^{42,0}}{768 \sqrt{15}}-\frac{7 C_{22,00}^{42,0}}{384 \sqrt{5}}-\frac{7 C_{22,20}^{42,0}}{384 \sqrt{15}}-\frac{7
C_{22,22}^{40,0}}{768 \sqrt{15}}+\frac{1}{192} \sqrt{\frac{7}{15}} C_{22,22}^{42,0}+\frac{3 \sqrt{3} C_{22,22}^{44,0}}{320}+\frac{C_{31,00}^{31,0}}{768}+\frac{C_{31,20}^{31,0}}{768
\sqrt{3}}+\frac{7 C_{31,22}^{31,0}}{768 \sqrt{15}}-\frac{\sqrt{7} C_{31,22}^{33,0}}{320}+\frac{3 \sqrt{21} C_{33,00}^{33,0}}{640}+\frac{3 \sqrt{7}
C_{33,20}^{33,0}}{640}-\frac{3}{320} \sqrt{\frac{21}{10}} C_{33,22}^{33,0}\)

\noindent\(C_{20,2211,1}^{11,1101,0,\text{Loc}}= -\frac{7 \sqrt{5} C_{11,11}^{51,0}}{1152}-\frac{7 \sqrt{5} C_{11,11}^{51,1}}{2304}-\frac{\sqrt{5}
C_{31,11}^{31,0}}{384}-\frac{\sqrt{5} C_{31,11}^{31,1}}{768}+\frac{1}{32} \sqrt{\frac{7}{10}} C_{33,11}^{33,0}+\frac{1}{64} \sqrt{\frac{7}{10}} C_{33,11}^{33,1}\)

\noindent\(C_{20,2211,1}^{11,1101,0,\text{NonLoc}}= -\frac{7 \sqrt{5} C_{11,11}^{51,0}}{2304}-\frac{7 \sqrt{5} C_{11,11}^{51,1}}{1152}-\frac{\sqrt{5}
C_{31,11}^{31,0}}{768}-\frac{\sqrt{5} C_{31,11}^{31,1}}{384}+\frac{1}{64} \sqrt{\frac{7}{10}} C_{33,11}^{33,0}+\frac{1}{32} \sqrt{\frac{7}{10}} C_{33,11}^{33,1}\)

\noindent\(C_{20,2211,1}^{11,1101,1,\text{Loc}}= -\frac{7}{768} \sqrt{\frac{5}{3}} C_{11,11}^{51,1}-\frac{1}{256} \sqrt{\frac{5}{3}}
C_{31,11}^{31,1}+\frac{1}{64} \sqrt{\frac{21}{10}} C_{33,11}^{33,1}\)

\noindent\(C_{20,2211,1}^{11,1101,1,\text{NonLoc}}= -\frac{7}{768} \sqrt{\frac{5}{3}} C_{11,11}^{51,0}-\frac{1}{256} \sqrt{\frac{5}{3}}
C_{31,11}^{31,0}+\frac{1}{64} \sqrt{\frac{21}{10}} C_{33,11}^{33,0}\)

\noindent\(C_{20,2211,1}^{20,0011,0,\text{Loc}}= -\frac{7 C_{00,22}^{62,0}}{512}-\frac{7 C_{00,22}^{62,1}}{1024}-\frac{7 C_{11,22}^{51,0}}{4608}-\frac{7
C_{11,22}^{51,1}}{9216}+\frac{3}{512} \sqrt{\frac{21}{5}} C_{11,22}^{53,0}+\frac{3 \sqrt{\frac{21}{5}} C_{11,22}^{53,1}}{1024}-\frac{7 C_{20,22}^{42,0}}{4608}-\frac{7
C_{20,22}^{42,1}}{9216}+\frac{17 C_{22,22}^{40,0}}{4608}+\frac{17 C_{22,22}^{40,1}}{9216}+\frac{5 \sqrt{7} C_{22,22}^{42,0}}{2304}+\frac{5 \sqrt{7}
C_{22,22}^{42,1}}{4608}-\frac{3 C_{22,22}^{44,0}}{128 \sqrt{5}}-\frac{3 C_{22,22}^{44,1}}{256 \sqrt{5}}-\frac{23 C_{31,22}^{31,0}}{4608}-\frac{23
C_{31,22}^{31,1}}{9216}+\frac{1}{512} \sqrt{\frac{7}{15}} C_{31,22}^{33,0}+\frac{\sqrt{\frac{7}{15}} C_{31,22}^{33,1}}{1024}-\frac{3}{640} \sqrt{\frac{7}{2}}
C_{33,22}^{33,0}-\frac{3 \sqrt{\frac{7}{2}} C_{33,22}^{33,1}}{1280}\)

\noindent\(C_{20,2211,1}^{20,0011,0,\text{NonLoc}}= \frac{7 C_{00,22}^{62,0}}{1024}+\frac{7 C_{00,22}^{62,1}}{512}-\frac{7 C_{11,22}^{51,0}}{9216}-\frac{7
C_{11,22}^{51,1}}{4608}+\frac{3 \sqrt{\frac{21}{5}} C_{11,22}^{53,0}}{1024}+\frac{3}{512} \sqrt{\frac{21}{5}} C_{11,22}^{53,1}+\frac{7 C_{20,22}^{42,0}}{9216}+\frac{7
C_{20,22}^{42,1}}{4608}-\frac{17 C_{22,22}^{40,0}}{9216}-\frac{17 C_{22,22}^{40,1}}{4608}-\frac{5 \sqrt{7} C_{22,22}^{42,0}}{4608}-\frac{5 \sqrt{7}
C_{22,22}^{42,1}}{2304}+\frac{3 C_{22,22}^{44,0}}{256 \sqrt{5}}+\frac{3 C_{22,22}^{44,1}}{128 \sqrt{5}}-\frac{23 C_{31,22}^{31,0}}{9216}-\frac{23
C_{31,22}^{31,1}}{4608}+\frac{\sqrt{\frac{7}{15}} C_{31,22}^{33,0}}{1024}+\frac{1}{512} \sqrt{\frac{7}{15}} C_{31,22}^{33,1}-\frac{3 \sqrt{\frac{7}{2}}
C_{33,22}^{33,0}}{1280}-\frac{3}{640} \sqrt{\frac{7}{2}} C_{33,22}^{33,1}\)

\noindent\(C_{20,2211,1}^{20,0011,1,\text{Loc}}= -\frac{7 \sqrt{3} C_{00,22}^{62,1}}{1024}-\frac{7 C_{11,22}^{51,1}}{3072 \sqrt{3}}+\frac{9
\sqrt{\frac{7}{5}} C_{11,22}^{53,1}}{1024}-\frac{7 C_{20,22}^{42,1}}{3072 \sqrt{3}}+\frac{17 C_{22,22}^{40,1}}{3072 \sqrt{3}}+\frac{5 \sqrt{\frac{7}{3}}
C_{22,22}^{42,1}}{1536}-\frac{3}{256} \sqrt{\frac{3}{5}} C_{22,22}^{44,1}-\frac{23 C_{31,22}^{31,1}}{3072 \sqrt{3}}+\frac{\sqrt{\frac{7}{5}} C_{31,22}^{33,1}}{1024}-\frac{3
\sqrt{\frac{21}{2}} C_{33,22}^{33,1}}{1280}\)

\noindent\(C_{20,2211,1}^{20,0011,1,\text{NonLoc}}= \frac{7 \sqrt{3} C_{00,22}^{62,0}}{1024}-\frac{7 C_{11,22}^{51,0}}{3072 \sqrt{3}}+\frac{9
\sqrt{\frac{7}{5}} C_{11,22}^{53,0}}{1024}+\frac{7 C_{20,22}^{42,0}}{3072 \sqrt{3}}-\frac{17 C_{22,22}^{40,0}}{3072 \sqrt{3}}-\frac{5 \sqrt{\frac{7}{3}}
C_{22,22}^{42,0}}{1536}+\frac{3}{256} \sqrt{\frac{3}{5}} C_{22,22}^{44,0}-\frac{23 C_{31,22}^{31,0}}{3072 \sqrt{3}}+\frac{\sqrt{\frac{7}{5}} C_{31,22}^{33,0}}{1024}-\frac{3
\sqrt{\frac{21}{2}} C_{33,22}^{33,0}}{1280}\)

\noindent\(C_{20,2212,2}^{00,2212,0,\text{Loc}}= \frac{7}{192} \sqrt{\frac{5}{3}} C_{00,20}^{60,0}+\frac{7}{384} \sqrt{\frac{5}{3}}
C_{00,20}^{60,1}-\frac{7 C_{00,22}^{62,0}}{128 \sqrt{3}}-\frac{7 C_{00,22}^{62,1}}{256 \sqrt{3}}-\frac{35 C_{11,22}^{51,0}}{1152 \sqrt{3}}-\frac{35
C_{11,22}^{51,1}}{2304 \sqrt{3}}+\frac{5}{576} \sqrt{\frac{5}{3}} C_{20,20}^{40,0}+\frac{5 \sqrt{\frac{5}{3}} C_{20,20}^{40,1}}{1152}-\frac{7 C_{20,22}^{42,0}}{1152
\sqrt{3}}-\frac{7 C_{20,22}^{42,1}}{2304 \sqrt{3}}-\frac{7 C_{22,20}^{42,0}}{288 \sqrt{3}}-\frac{7 C_{22,20}^{42,1}}{576 \sqrt{3}}-\frac{7 C_{22,22}^{40,0}}{1152
\sqrt{3}}-\frac{7 C_{22,22}^{40,1}}{2304 \sqrt{3}}+\frac{1}{288} \sqrt{\frac{7}{3}} C_{22,22}^{42,0}+\frac{1}{576} \sqrt{\frac{7}{3}} C_{22,22}^{42,1}+\frac{1}{32}
\sqrt{\frac{3}{5}} C_{22,22}^{44,0}+\frac{1}{64} \sqrt{\frac{3}{5}} C_{22,22}^{44,1}-\frac{1}{576} \sqrt{\frac{5}{3}} C_{31,20}^{31,0}-\frac{\sqrt{\frac{5}{3}}
C_{31,20}^{31,1}}{1152}-\frac{7 C_{31,22}^{31,0}}{1152 \sqrt{3}}-\frac{7 C_{31,22}^{31,1}}{2304 \sqrt{3}}+\frac{1}{96} \sqrt{\frac{7}{5}} C_{31,22}^{33,0}+\frac{1}{192}
\sqrt{\frac{7}{5}} C_{31,22}^{33,1}-\frac{1}{32} \sqrt{\frac{7}{5}} C_{33,20}^{33,0}-\frac{1}{64} \sqrt{\frac{7}{5}} C_{33,20}^{33,1}+\frac{1}{160}
\sqrt{\frac{21}{2}} C_{33,22}^{33,0}+\frac{1}{320} \sqrt{\frac{21}{2}} C_{33,22}^{33,1}\)

\noindent\(C_{20,2212,2}^{00,2212,0,\text{NonLoc}}= \frac{7 \sqrt{5} C_{00,00}^{60,0}}{768}+\frac{7 \sqrt{5} C_{00,00}^{60,1}}{384}+\frac{7}{768}
\sqrt{\frac{5}{3}} C_{00,20}^{60,0}+\frac{7}{384} \sqrt{\frac{5}{3}} C_{00,20}^{60,1}+\frac{7 C_{00,22}^{62,0}}{256 \sqrt{3}}+\frac{7 C_{00,22}^{62,1}}{128
\sqrt{3}}-\frac{7 \sqrt{5} C_{11,00}^{51,0}}{2304}-\frac{7 \sqrt{5} C_{11,00}^{51,1}}{1152}-\frac{35 C_{11,22}^{51,0}}{2304 \sqrt{3}}-\frac{35 C_{11,22}^{51,1}}{1152
\sqrt{3}}+\frac{5 \sqrt{5} C_{20,00}^{40,0}}{2304}+\frac{5 \sqrt{5} C_{20,00}^{40,1}}{1152}+\frac{5 \sqrt{\frac{5}{3}} C_{20,20}^{40,0}}{2304}+\frac{5
\sqrt{\frac{5}{3}} C_{20,20}^{40,1}}{1152}+\frac{7 C_{20,22}^{42,0}}{2304 \sqrt{3}}+\frac{7 C_{20,22}^{42,1}}{1152 \sqrt{3}}-\frac{7 C_{22,00}^{42,0}}{1152}-\frac{7
C_{22,00}^{42,1}}{576}-\frac{7 C_{22,20}^{42,0}}{1152 \sqrt{3}}-\frac{7 C_{22,20}^{42,1}}{576 \sqrt{3}}+\frac{7 C_{22,22}^{40,0}}{2304 \sqrt{3}}+\frac{7
C_{22,22}^{40,1}}{1152 \sqrt{3}}-\frac{1}{576} \sqrt{\frac{7}{3}} C_{22,22}^{42,0}-\frac{1}{288} \sqrt{\frac{7}{3}} C_{22,22}^{42,1}-\frac{1}{64}
\sqrt{\frac{3}{5}} C_{22,22}^{44,0}-\frac{1}{32} \sqrt{\frac{3}{5}} C_{22,22}^{44,1}+\frac{\sqrt{5} C_{31,00}^{31,0}}{2304}+\frac{\sqrt{5} C_{31,00}^{31,1}}{1152}+\frac{\sqrt{\frac{5}{3}}
C_{31,20}^{31,0}}{2304}+\frac{\sqrt{\frac{5}{3}} C_{31,20}^{31,1}}{1152}-\frac{7 C_{31,22}^{31,0}}{2304 \sqrt{3}}-\frac{7 C_{31,22}^{31,1}}{1152
\sqrt{3}}+\frac{1}{192} \sqrt{\frac{7}{5}} C_{31,22}^{33,0}+\frac{1}{96} \sqrt{\frac{7}{5}} C_{31,22}^{33,1}+\frac{1}{128} \sqrt{\frac{21}{5}} C_{33,00}^{33,0}+\frac{1}{64}
\sqrt{\frac{21}{5}} C_{33,00}^{33,1}+\frac{1}{128} \sqrt{\frac{7}{5}} C_{33,20}^{33,0}+\frac{1}{64} \sqrt{\frac{7}{5}} C_{33,20}^{33,1}+\frac{1}{320}
\sqrt{\frac{21}{2}} C_{33,22}^{33,0}+\frac{1}{160} \sqrt{\frac{21}{2}} C_{33,22}^{33,1}\)

\noindent\(C_{20,2212,2}^{00,2212,1,\text{Loc}}= \frac{7 \sqrt{5} C_{00,20}^{60,1}}{384}-\frac{7 C_{00,22}^{62,1}}{256}-\frac{35 C_{11,22}^{51,1}}{2304}+\frac{5
\sqrt{5} C_{20,20}^{40,1}}{1152}-\frac{7 C_{20,22}^{42,1}}{2304}-\frac{7 C_{22,20}^{42,1}}{576}-\frac{7 C_{22,22}^{40,1}}{2304}+\frac{\sqrt{7} C_{22,22}^{42,1}}{576}+\frac{3
C_{22,22}^{44,1}}{64 \sqrt{5}}-\frac{\sqrt{5} C_{31,20}^{31,1}}{1152}-\frac{7 C_{31,22}^{31,1}}{2304}+\frac{1}{64} \sqrt{\frac{7}{15}} C_{31,22}^{33,1}-\frac{1}{64}
\sqrt{\frac{21}{5}} C_{33,20}^{33,1}+\frac{3}{320} \sqrt{\frac{7}{2}} C_{33,22}^{33,1}\)

\noindent\(C_{20,2212,2}^{00,2212,1,\text{NonLoc}}= \frac{7}{256} \sqrt{\frac{5}{3}} C_{00,00}^{60,0}+\frac{7 \sqrt{5} C_{00,20}^{60,0}}{768}+\frac{7
C_{00,22}^{62,0}}{256}-\frac{7}{768} \sqrt{\frac{5}{3}} C_{11,00}^{51,0}-\frac{35 C_{11,22}^{51,0}}{2304}+\frac{5}{768} \sqrt{\frac{5}{3}} C_{20,00}^{40,0}+\frac{5
\sqrt{5} C_{20,20}^{40,0}}{2304}+\frac{7 C_{20,22}^{42,0}}{2304}-\frac{7 C_{22,00}^{42,0}}{384 \sqrt{3}}-\frac{7 C_{22,20}^{42,0}}{1152}+\frac{7
C_{22,22}^{40,0}}{2304}-\frac{\sqrt{7} C_{22,22}^{42,0}}{576}-\frac{3 C_{22,22}^{44,0}}{64 \sqrt{5}}+\frac{1}{768} \sqrt{\frac{5}{3}} C_{31,00}^{31,0}+\frac{\sqrt{5}
C_{31,20}^{31,0}}{2304}-\frac{7 C_{31,22}^{31,0}}{2304}+\frac{1}{64} \sqrt{\frac{7}{15}} C_{31,22}^{33,0}+\frac{3}{128} \sqrt{\frac{7}{5}} C_{33,00}^{33,0}+\frac{1}{128}
\sqrt{\frac{21}{5}} C_{33,20}^{33,0}+\frac{3}{320} \sqrt{\frac{7}{2}} C_{33,22}^{33,0}\)

\noindent\(C_{20,2212,2}^{11,1101,0,\text{Loc}}= \frac{7}{384} \sqrt{\frac{5}{3}} C_{11,11}^{51,0}+\frac{7}{768} \sqrt{\frac{5}{3}}
C_{11,11}^{51,1}+\frac{1}{128} \sqrt{\frac{5}{3}} C_{31,11}^{31,0}+\frac{1}{256} \sqrt{\frac{5}{3}} C_{31,11}^{31,1}-\frac{1}{32} \sqrt{\frac{21}{10}}
C_{33,11}^{33,0}-\frac{1}{64} \sqrt{\frac{21}{10}} C_{33,11}^{33,1}\)

\noindent\(C_{20,2212,2}^{11,1101,0,\text{NonLoc}}= \frac{7}{768} \sqrt{\frac{5}{3}} C_{11,11}^{51,0}+\frac{7}{384} \sqrt{\frac{5}{3}}
C_{11,11}^{51,1}+\frac{1}{256} \sqrt{\frac{5}{3}} C_{31,11}^{31,0}+\frac{1}{128} \sqrt{\frac{5}{3}} C_{31,11}^{31,1}-\frac{1}{64} \sqrt{\frac{21}{10}}
C_{33,11}^{33,0}-\frac{1}{32} \sqrt{\frac{21}{10}} C_{33,11}^{33,1}\)

\noindent\(C_{20,2212,2}^{11,1101,1,\text{Loc}}= \frac{7 \sqrt{5} C_{11,11}^{51,1}}{768}+\frac{\sqrt{5} C_{31,11}^{31,1}}{256}-\frac{3}{64}
\sqrt{\frac{7}{10}} C_{33,11}^{33,1}\)

\noindent\(C_{20,2212,2}^{11,1101,1,\text{NonLoc}}= \frac{7 \sqrt{5} C_{11,11}^{51,0}}{768}+\frac{\sqrt{5} C_{31,11}^{31,0}}{256}-\frac{3}{64}
\sqrt{\frac{7}{10}} C_{33,11}^{33,0}\)

\noindent\(C_{20,2213,3}^{00,2213,0,\text{Loc}}= \frac{7}{192} \sqrt{\frac{7}{3}} C_{00,20}^{60,0}+\frac{7}{384} \sqrt{\frac{7}{3}}
C_{00,20}^{60,1}+\frac{1}{64} \sqrt{\frac{7}{15}} C_{00,22}^{62,0}+\frac{1}{128} \sqrt{\frac{7}{15}} C_{00,22}^{62,1}+\frac{1}{576} \sqrt{\frac{35}{3}}
C_{11,22}^{51,0}+\frac{\sqrt{\frac{35}{3}} C_{11,22}^{51,1}}{1152}+\frac{5}{576} \sqrt{\frac{7}{3}} C_{20,20}^{40,0}+\frac{5 \sqrt{\frac{7}{3}} C_{20,20}^{40,1}}{1152}+\frac{1}{576}
\sqrt{\frac{7}{15}} C_{20,22}^{42,0}+\frac{\sqrt{\frac{7}{15}} C_{20,22}^{42,1}}{1152}-\frac{7}{288} \sqrt{\frac{7}{15}} C_{22,20}^{42,0}-\frac{7}{576}
\sqrt{\frac{7}{15}} C_{22,20}^{42,1}+\frac{1}{576} \sqrt{\frac{7}{15}} C_{22,22}^{40,0}+\frac{\sqrt{\frac{7}{15}} C_{22,22}^{40,1}}{1152}-\frac{C_{22,22}^{42,0}}{144
\sqrt{15}}-\frac{C_{22,22}^{42,1}}{288 \sqrt{15}}-\frac{1}{80} \sqrt{\frac{3}{7}} C_{22,22}^{44,0}-\frac{1}{160} \sqrt{\frac{3}{7}} C_{22,22}^{44,1}-\frac{1}{576}
\sqrt{\frac{7}{3}} C_{31,20}^{31,0}-\frac{\sqrt{\frac{7}{3}} C_{31,20}^{31,1}}{1152}+\frac{1}{576} \sqrt{\frac{7}{15}} C_{31,22}^{31,0}+\frac{\sqrt{\frac{7}{15}}
C_{31,22}^{31,1}}{1152}-\frac{C_{31,22}^{33,0}}{240}-\frac{C_{31,22}^{33,1}}{480}-\frac{7 C_{33,20}^{33,0}}{160}-\frac{7 C_{33,20}^{33,1}}{320}-\frac{1}{80}
\sqrt{\frac{3}{10}} C_{33,22}^{33,0}-\frac{1}{160} \sqrt{\frac{3}{10}} C_{33,22}^{33,1}\)

\noindent\(C_{20,2213,3}^{00,2213,0,\text{NonLoc}}= \frac{7 \sqrt{7} C_{00,00}^{60,0}}{768}+\frac{7 \sqrt{7} C_{00,00}^{60,1}}{384}+\frac{7}{768}
\sqrt{\frac{7}{3}} C_{00,20}^{60,0}+\frac{7}{384} \sqrt{\frac{7}{3}} C_{00,20}^{60,1}-\frac{1}{128} \sqrt{\frac{7}{15}} C_{00,22}^{62,0}-\frac{1}{64}
\sqrt{\frac{7}{15}} C_{00,22}^{62,1}-\frac{7 \sqrt{7} C_{11,00}^{51,0}}{2304}-\frac{7 \sqrt{7} C_{11,00}^{51,1}}{1152}+\frac{\sqrt{\frac{35}{3}}
C_{11,22}^{51,0}}{1152}+\frac{1}{576} \sqrt{\frac{35}{3}} C_{11,22}^{51,1}+\frac{5 \sqrt{7} C_{20,00}^{40,0}}{2304}+\frac{5 \sqrt{7} C_{20,00}^{40,1}}{1152}+\frac{5
\sqrt{\frac{7}{3}} C_{20,20}^{40,0}}{2304}+\frac{5 \sqrt{\frac{7}{3}} C_{20,20}^{40,1}}{1152}-\frac{\sqrt{\frac{7}{15}} C_{20,22}^{42,0}}{1152}-\frac{1}{576}
\sqrt{\frac{7}{15}} C_{20,22}^{42,1}-\frac{7 \sqrt{\frac{7}{5}} C_{22,00}^{42,0}}{1152}-\frac{7}{576} \sqrt{\frac{7}{5}} C_{22,00}^{42,1}-\frac{7
\sqrt{\frac{7}{15}} C_{22,20}^{42,0}}{1152}-\frac{7}{576} \sqrt{\frac{7}{15}} C_{22,20}^{42,1}-\frac{\sqrt{\frac{7}{15}} C_{22,22}^{40,0}}{1152}-\frac{1}{576}
\sqrt{\frac{7}{15}} C_{22,22}^{40,1}+\frac{C_{22,22}^{42,0}}{288 \sqrt{15}}+\frac{C_{22,22}^{42,1}}{144 \sqrt{15}}+\frac{1}{160} \sqrt{\frac{3}{7}}
C_{22,22}^{44,0}+\frac{1}{80} \sqrt{\frac{3}{7}} C_{22,22}^{44,1}+\frac{\sqrt{7} C_{31,00}^{31,0}}{2304}+\frac{\sqrt{7} C_{31,00}^{31,1}}{1152}+\frac{\sqrt{\frac{7}{3}}
C_{31,20}^{31,0}}{2304}+\frac{\sqrt{\frac{7}{3}} C_{31,20}^{31,1}}{1152}+\frac{\sqrt{\frac{7}{15}} C_{31,22}^{31,0}}{1152}+\frac{1}{576} \sqrt{\frac{7}{15}}
C_{31,22}^{31,1}-\frac{C_{31,22}^{33,0}}{480}-\frac{C_{31,22}^{33,1}}{240}+\frac{7 \sqrt{3} C_{33,00}^{33,0}}{640}+\frac{7 \sqrt{3} C_{33,00}^{33,1}}{320}+\frac{7
C_{33,20}^{33,0}}{640}+\frac{7 C_{33,20}^{33,1}}{320}-\frac{1}{160} \sqrt{\frac{3}{10}} C_{33,22}^{33,0}-\frac{1}{80} \sqrt{\frac{3}{10}} C_{33,22}^{33,1}\)

\noindent\(C_{20,2213,3}^{00,2213,1,\text{Loc}}= \frac{7 \sqrt{7} C_{00,20}^{60,1}}{384}+\frac{1}{128} \sqrt{\frac{7}{5}} C_{00,22}^{62,1}+\frac{\sqrt{35}
C_{11,22}^{51,1}}{1152}+\frac{5 \sqrt{7} C_{20,20}^{40,1}}{1152}+\frac{\sqrt{\frac{7}{5}} C_{20,22}^{42,1}}{1152}-\frac{7}{576} \sqrt{\frac{7}{5}}
C_{22,20}^{42,1}+\frac{\sqrt{\frac{7}{5}} C_{22,22}^{40,1}}{1152}-\frac{C_{22,22}^{42,1}}{288 \sqrt{5}}-\frac{3 C_{22,22}^{44,1}}{160 \sqrt{7}}-\frac{\sqrt{7}
C_{31,20}^{31,1}}{1152}+\frac{\sqrt{\frac{7}{5}} C_{31,22}^{31,1}}{1152}-\frac{C_{31,22}^{33,1}}{160 \sqrt{3}}-\frac{7 \sqrt{3} C_{33,20}^{33,1}}{320}-\frac{3
C_{33,22}^{33,1}}{160 \sqrt{10}}\)

\noindent\(C_{20,2213,3}^{00,2213,1,\text{NonLoc}}= \frac{7}{256} \sqrt{\frac{7}{3}} C_{00,00}^{60,0}+\frac{7 \sqrt{7} C_{00,20}^{60,0}}{768}-\frac{1}{128}
\sqrt{\frac{7}{5}} C_{00,22}^{62,0}-\frac{7}{768} \sqrt{\frac{7}{3}} C_{11,00}^{51,0}+\frac{\sqrt{35} C_{11,22}^{51,0}}{1152}+\frac{5}{768} \sqrt{\frac{7}{3}}
C_{20,00}^{40,0}+\frac{5 \sqrt{7} C_{20,20}^{40,0}}{2304}-\frac{\sqrt{\frac{7}{5}} C_{20,22}^{42,0}}{1152}-\frac{7}{384} \sqrt{\frac{7}{15}} C_{22,00}^{42,0}-\frac{7
\sqrt{\frac{7}{5}} C_{22,20}^{42,0}}{1152}-\frac{\sqrt{\frac{7}{5}} C_{22,22}^{40,0}}{1152}+\frac{C_{22,22}^{42,0}}{288 \sqrt{5}}+\frac{3 C_{22,22}^{44,0}}{160
\sqrt{7}}+\frac{1}{768} \sqrt{\frac{7}{3}} C_{31,00}^{31,0}+\frac{\sqrt{7} C_{31,20}^{31,0}}{2304}+\frac{\sqrt{\frac{7}{5}} C_{31,22}^{31,0}}{1152}-\frac{C_{31,22}^{33,0}}{160
\sqrt{3}}+\frac{21 C_{33,00}^{33,0}}{640}+\frac{7 \sqrt{3} C_{33,20}^{33,0}}{640}-\frac{3 C_{33,22}^{33,0}}{160 \sqrt{10}}\)

\noindent\(C_{20,3101,1}^{00,1101,0,\text{Loc}}= \frac{7 C_{00,00}^{60,0}}{64}+\frac{7 C_{00,00}^{60,1}}{128}+\frac{7 C_{11,00}^{51,0}}{192}+\frac{7
C_{11,00}^{51,1}}{384}+\frac{5 C_{20,00}^{40,0}}{192}+\frac{5 C_{20,00}^{40,1}}{384}-\frac{7 C_{22,00}^{42,0}}{96 \sqrt{5}}-\frac{7 C_{22,00}^{42,1}}{192
\sqrt{5}}-\frac{C_{31,00}^{31,0}}{192}-\frac{C_{31,00}^{31,1}}{384}-\frac{3 \sqrt{21} C_{33,00}^{33,0}}{160}-\frac{3 \sqrt{21} C_{33,00}^{33,1}}{320}\)

\noindent\(C_{20,3101,1}^{00,1101,0,\text{NonLoc}}= -\frac{7 C_{00,00}^{60,0}}{256}-\frac{7 C_{00,00}^{60,1}}{128}+\frac{7 \sqrt{3}
C_{00,20}^{60,0}}{256}+\frac{7 \sqrt{3} C_{00,20}^{60,1}}{128}+\frac{7 C_{11,00}^{51,0}}{768}+\frac{7 C_{11,00}^{51,1}}{384}-\frac{7 C_{11,20}^{31,0}}{256
\sqrt{3}}-\frac{7 C_{11,20}^{31,1}}{128 \sqrt{3}}-\frac{5 C_{20,00}^{40,0}}{768}-\frac{5 C_{20,00}^{40,1}}{384}+\frac{5 C_{20,20}^{40,0}}{256 \sqrt{3}}+\frac{5
C_{20,20}^{40,1}}{128 \sqrt{3}}+\frac{7 C_{22,00}^{42,0}}{384 \sqrt{5}}+\frac{7 C_{22,00}^{42,1}}{192 \sqrt{5}}-\frac{7 C_{22,20}^{42,0}}{128 \sqrt{15}}-\frac{7
C_{22,20}^{42,1}}{64 \sqrt{15}}-\frac{C_{31,00}^{31,0}}{768}-\frac{C_{31,00}^{31,1}}{384}+\frac{C_{31,20}^{31,0}}{256 \sqrt{3}}+\frac{C_{31,20}^{31,1}}{128
\sqrt{3}}-\frac{3 \sqrt{21} C_{33,00}^{33,0}}{640}-\frac{3 \sqrt{21} C_{33,00}^{33,1}}{320}+\frac{9 \sqrt{7} C_{33,20}^{33,0}}{640}+\frac{9 \sqrt{7}
C_{33,20}^{33,1}}{320}\)

\noindent\(C_{20,3101,1}^{00,1101,1,\text{Loc}}= \frac{7 \sqrt{3} C_{00,00}^{60,1}}{128}+\frac{7 C_{11,00}^{51,1}}{128 \sqrt{3}}+\frac{5
C_{20,00}^{40,1}}{128 \sqrt{3}}-\frac{7 C_{22,00}^{42,1}}{64 \sqrt{15}}-\frac{C_{31,00}^{31,1}}{128 \sqrt{3}}-\frac{9 \sqrt{7} C_{33,00}^{33,1}}{320}\)

\noindent\(C_{20,3101,1}^{00,1101,1,\text{NonLoc}}= -\frac{7 \sqrt{3} C_{00,00}^{60,0}}{256}+\frac{21 C_{00,20}^{60,0}}{256}+\frac{7
C_{11,00}^{51,0}}{256 \sqrt{3}}-\frac{7 C_{11,20}^{31,0}}{256}-\frac{5 C_{20,00}^{40,0}}{256 \sqrt{3}}+\frac{5 C_{20,20}^{40,0}}{256}+\frac{7 C_{22,00}^{42,0}}{128
\sqrt{15}}-\frac{7 C_{22,20}^{42,0}}{128 \sqrt{5}}-\frac{C_{31,00}^{31,0}}{256 \sqrt{3}}+\frac{C_{31,20}^{31,0}}{256}-\frac{9 \sqrt{7} C_{33,00}^{33,0}}{640}+\frac{9
\sqrt{21} C_{33,20}^{33,0}}{640}\)

\noindent\(C_{20,3101,1}^{11,0011,0,\text{Loc}}= \frac{7 C_{11,11}^{51,0}}{384}+\frac{7 C_{11,11}^{51,1}}{768}+\frac{C_{31,11}^{31,0}}{128}+\frac{C_{31,11}^{31,1}}{256}-\frac{3}{160}
\sqrt{\frac{7}{2}} C_{33,11}^{33,0}-\frac{3}{320} \sqrt{\frac{7}{2}} C_{33,11}^{33,1}\)

\noindent\(C_{20,3101,1}^{11,0011,0,\text{NonLoc}}= \frac{7 C_{11,11}^{51,0}}{768}+\frac{7 C_{11,11}^{51,1}}{384}+\frac{C_{31,11}^{31,0}}{256}+\frac{C_{31,11}^{31,1}}{128}-\frac{3}{320}
\sqrt{\frac{7}{2}} C_{33,11}^{33,0}-\frac{3}{160} \sqrt{\frac{7}{2}} C_{33,11}^{33,1}\)

\noindent\(C_{20,3101,1}^{11,0011,1,\text{Loc}}= \frac{7 C_{11,11}^{51,1}}{256 \sqrt{3}}+\frac{\sqrt{3} C_{31,11}^{31,1}}{256}-\frac{3}{320}
\sqrt{\frac{21}{2}} C_{33,11}^{33,1}\)

\noindent\(C_{20,3101,1}^{11,0011,1,\text{NonLoc}}= \frac{7 C_{11,11}^{51,0}}{256 \sqrt{3}}+\frac{\sqrt{3} C_{31,11}^{31,0}}{256}-\frac{3}{320}
\sqrt{\frac{21}{2}} C_{33,11}^{33,0}\)

\noindent\(C_{20,3110,0}^{00,1110,0,\text{Loc}}= -\frac{7 C_{00,20}^{60,0}}{192}-\frac{7 C_{00,20}^{60,1}}{384}-\frac{7 C_{00,22}^{62,0}}{64
\sqrt{5}}-\frac{7 C_{00,22}^{62,1}}{128 \sqrt{5}}-\frac{7 C_{11,20}^{31,0}}{576}-\frac{7 C_{11,20}^{31,1}}{1152}-\frac{7 \sqrt{5} C_{11,22}^{51,0}}{576}-\frac{7
\sqrt{5} C_{11,22}^{51,1}}{1152}-\frac{5 C_{20,20}^{40,0}}{576}-\frac{5 C_{20,20}^{40,1}}{1152}-\frac{7 C_{20,22}^{42,0}}{576 \sqrt{5}}-\frac{7 C_{20,22}^{42,1}}{1152
\sqrt{5}}+\frac{7 C_{22,20}^{42,0}}{288 \sqrt{5}}+\frac{7 C_{22,20}^{42,1}}{576 \sqrt{5}}-\frac{7 C_{22,22}^{40,0}}{576 \sqrt{5}}-\frac{7 C_{22,22}^{40,1}}{1152
\sqrt{5}}+\frac{1}{144} \sqrt{\frac{7}{5}} C_{22,22}^{42,0}+\frac{1}{288} \sqrt{\frac{7}{5}} C_{22,22}^{42,1}+\frac{3 C_{22,22}^{44,0}}{80}+\frac{3
C_{22,22}^{44,1}}{160}+\frac{C_{31,20}^{31,0}}{576}+\frac{C_{31,20}^{31,1}}{1152}-\frac{7 C_{31,22}^{31,0}}{576 \sqrt{5}}-\frac{7 C_{31,22}^{31,1}}{1152
\sqrt{5}}+\frac{1}{80} \sqrt{\frac{7}{3}} C_{31,22}^{33,0}+\frac{1}{160} \sqrt{\frac{7}{3}} C_{31,22}^{33,1}+\frac{\sqrt{21} C_{33,20}^{33,0}}{160}+\frac{\sqrt{21}
C_{33,20}^{33,1}}{320}+\frac{3}{80} \sqrt{\frac{7}{10}} C_{33,22}^{33,0}+\frac{3}{160} \sqrt{\frac{7}{10}} C_{33,22}^{33,1}\)

\noindent\(C_{20,3110,0}^{00,1110,0,\text{NonLoc}}= -\frac{7 C_{00,00}^{60,0}}{256 \sqrt{3}}-\frac{7 C_{00,00}^{60,1}}{128 \sqrt{3}}-\frac{7
C_{00,20}^{60,0}}{768}-\frac{7 C_{00,20}^{60,1}}{384}+\frac{7 C_{00,22}^{62,0}}{128 \sqrt{5}}+\frac{7 C_{00,22}^{62,1}}{64 \sqrt{5}}+\frac{7 C_{11,00}^{51,0}}{768
\sqrt{3}}+\frac{7 C_{11,00}^{51,1}}{384 \sqrt{3}}+\frac{7 C_{11,20}^{31,0}}{2304}+\frac{7 C_{11,20}^{31,1}}{1152}-\frac{7 \sqrt{5} C_{11,22}^{51,0}}{1152}-\frac{7
\sqrt{5} C_{11,22}^{51,1}}{576}-\frac{5 C_{20,00}^{40,0}}{768 \sqrt{3}}-\frac{5 C_{20,00}^{40,1}}{384 \sqrt{3}}-\frac{5 C_{20,20}^{40,0}}{2304}-\frac{5
C_{20,20}^{40,1}}{1152}+\frac{7 C_{20,22}^{42,0}}{1152 \sqrt{5}}+\frac{7 C_{20,22}^{42,1}}{576 \sqrt{5}}+\frac{7 C_{22,00}^{42,0}}{384 \sqrt{15}}+\frac{7
C_{22,00}^{42,1}}{192 \sqrt{15}}+\frac{7 C_{22,20}^{42,0}}{1152 \sqrt{5}}+\frac{7 C_{22,20}^{42,1}}{576 \sqrt{5}}+\frac{7 C_{22,22}^{40,0}}{1152
\sqrt{5}}+\frac{7 C_{22,22}^{40,1}}{576 \sqrt{5}}-\frac{1}{288} \sqrt{\frac{7}{5}} C_{22,22}^{42,0}-\frac{1}{144} \sqrt{\frac{7}{5}} C_{22,22}^{42,1}-\frac{3
C_{22,22}^{44,0}}{160}-\frac{3 C_{22,22}^{44,1}}{80}-\frac{C_{31,00}^{31,0}}{768 \sqrt{3}}-\frac{C_{31,00}^{31,1}}{384 \sqrt{3}}-\frac{C_{31,20}^{31,0}}{2304}-\frac{C_{31,20}^{31,1}}{1152}-\frac{7
C_{31,22}^{31,0}}{1152 \sqrt{5}}-\frac{7 C_{31,22}^{31,1}}{576 \sqrt{5}}+\frac{1}{160} \sqrt{\frac{7}{3}} C_{31,22}^{33,0}+\frac{1}{80} \sqrt{\frac{7}{3}}
C_{31,22}^{33,1}-\frac{3 \sqrt{7} C_{33,00}^{33,0}}{640}-\frac{3 \sqrt{7} C_{33,00}^{33,1}}{320}-\frac{\sqrt{21} C_{33,20}^{33,0}}{640}-\frac{\sqrt{21}
C_{33,20}^{33,1}}{320}+\frac{3}{160} \sqrt{\frac{7}{10}} C_{33,22}^{33,0}+\frac{3}{80} \sqrt{\frac{7}{10}} C_{33,22}^{33,1}\)

\noindent\(C_{20,3110,0}^{00,1110,1,\text{Loc}}= -\frac{7 C_{00,20}^{60,1}}{128 \sqrt{3}}-\frac{7}{128} \sqrt{\frac{3}{5}} C_{00,22}^{62,1}-\frac{7
C_{11,20}^{31,1}}{384 \sqrt{3}}-\frac{7}{384} \sqrt{\frac{5}{3}} C_{11,22}^{51,1}-\frac{5 C_{20,20}^{40,1}}{384 \sqrt{3}}-\frac{7 C_{20,22}^{42,1}}{384
\sqrt{15}}+\frac{7 C_{22,20}^{42,1}}{192 \sqrt{15}}-\frac{7 C_{22,22}^{40,1}}{384 \sqrt{15}}+\frac{1}{96} \sqrt{\frac{7}{15}} C_{22,22}^{42,1}+\frac{3
\sqrt{3} C_{22,22}^{44,1}}{160}+\frac{C_{31,20}^{31,1}}{384 \sqrt{3}}-\frac{7 C_{31,22}^{31,1}}{384 \sqrt{15}}+\frac{\sqrt{7} C_{31,22}^{33,1}}{160}+\frac{3
\sqrt{7} C_{33,20}^{33,1}}{320}+\frac{3}{160} \sqrt{\frac{21}{10}} C_{33,22}^{33,1}\)

\noindent\(C_{20,3110,0}^{00,1110,1,\text{NonLoc}}= -\frac{7 C_{00,00}^{60,0}}{256}-\frac{7 C_{00,20}^{60,0}}{256 \sqrt{3}}+\frac{7}{128}
\sqrt{\frac{3}{5}} C_{00,22}^{62,0}+\frac{7 C_{11,00}^{51,0}}{768}+\frac{7 C_{11,20}^{31,0}}{768 \sqrt{3}}-\frac{7}{384} \sqrt{\frac{5}{3}} C_{11,22}^{51,0}-\frac{5
C_{20,00}^{40,0}}{768}-\frac{5 C_{20,20}^{40,0}}{768 \sqrt{3}}+\frac{7 C_{20,22}^{42,0}}{384 \sqrt{15}}+\frac{7 C_{22,00}^{42,0}}{384 \sqrt{5}}+\frac{7
C_{22,20}^{42,0}}{384 \sqrt{15}}+\frac{7 C_{22,22}^{40,0}}{384 \sqrt{15}}-\frac{1}{96} \sqrt{\frac{7}{15}} C_{22,22}^{42,0}-\frac{3 \sqrt{3} C_{22,22}^{44,0}}{160}-\frac{C_{31,00}^{31,0}}{768}-\frac{C_{31,20}^{31,0}}{768
\sqrt{3}}-\frac{7 C_{31,22}^{31,0}}{384 \sqrt{15}}+\frac{\sqrt{7} C_{31,22}^{33,0}}{160}-\frac{3 \sqrt{21} C_{33,00}^{33,0}}{640}-\frac{3 \sqrt{7}
C_{33,20}^{33,0}}{640}+\frac{3}{160} \sqrt{\frac{21}{10}} C_{33,22}^{33,0}\)

\noindent\(C_{20,3111,1}^{00,1111,0,\text{Loc}}= -\frac{7 C_{00,20}^{60,0}}{64 \sqrt{3}}-\frac{7 C_{00,20}^{60,1}}{128 \sqrt{3}}+\frac{7}{128}
\sqrt{\frac{3}{5}} C_{00,22}^{62,0}+\frac{7}{256} \sqrt{\frac{3}{5}} C_{00,22}^{62,1}-\frac{7 C_{11,20}^{31,0}}{192 \sqrt{3}}-\frac{7 C_{11,20}^{31,1}}{384
\sqrt{3}}+\frac{7}{384} \sqrt{\frac{5}{3}} C_{11,22}^{51,0}+\frac{7}{768} \sqrt{\frac{5}{3}} C_{11,22}^{51,1}-\frac{5 C_{20,20}^{40,0}}{192 \sqrt{3}}-\frac{5
C_{20,20}^{40,1}}{384 \sqrt{3}}+\frac{7 C_{20,22}^{42,0}}{384 \sqrt{15}}+\frac{7 C_{20,22}^{42,1}}{768 \sqrt{15}}+\frac{7 C_{22,20}^{42,0}}{96 \sqrt{15}}+\frac{7
C_{22,20}^{42,1}}{192 \sqrt{15}}+\frac{7 C_{22,22}^{40,0}}{384 \sqrt{15}}+\frac{7 C_{22,22}^{40,1}}{768 \sqrt{15}}-\frac{1}{96} \sqrt{\frac{7}{15}}
C_{22,22}^{42,0}-\frac{1}{192} \sqrt{\frac{7}{15}} C_{22,22}^{42,1}-\frac{3 \sqrt{3} C_{22,22}^{44,0}}{160}-\frac{3 \sqrt{3} C_{22,22}^{44,1}}{320}+\frac{C_{31,20}^{31,0}}{192
\sqrt{3}}+\frac{C_{31,20}^{31,1}}{384 \sqrt{3}}+\frac{7 C_{31,22}^{31,0}}{384 \sqrt{15}}+\frac{7 C_{31,22}^{31,1}}{768 \sqrt{15}}-\frac{\sqrt{7}
C_{31,22}^{33,0}}{160}-\frac{\sqrt{7} C_{31,22}^{33,1}}{320}+\frac{3 \sqrt{7} C_{33,20}^{33,0}}{160}+\frac{3 \sqrt{7} C_{33,20}^{33,1}}{320}-\frac{3}{160}
\sqrt{\frac{21}{10}} C_{33,22}^{33,0}-\frac{3}{320} \sqrt{\frac{21}{10}} C_{33,22}^{33,1}\)

\noindent\(C_{20,3111,1}^{00,1111,0,\text{NonLoc}}= -\frac{7 C_{00,00}^{60,0}}{256}-\frac{7 C_{00,00}^{60,1}}{128}-\frac{7 C_{00,20}^{60,0}}{256
\sqrt{3}}-\frac{7 C_{00,20}^{60,1}}{128 \sqrt{3}}-\frac{7}{256} \sqrt{\frac{3}{5}} C_{00,22}^{62,0}-\frac{7}{128} \sqrt{\frac{3}{5}} C_{00,22}^{62,1}+\frac{7
C_{11,00}^{51,0}}{768}+\frac{7 C_{11,00}^{51,1}}{384}+\frac{7 C_{11,20}^{31,0}}{768 \sqrt{3}}+\frac{7 C_{11,20}^{31,1}}{384 \sqrt{3}}+\frac{7}{768}
\sqrt{\frac{5}{3}} C_{11,22}^{51,0}+\frac{7}{384} \sqrt{\frac{5}{3}} C_{11,22}^{51,1}-\frac{5 C_{20,00}^{40,0}}{768}-\frac{5 C_{20,00}^{40,1}}{384}-\frac{5
C_{20,20}^{40,0}}{768 \sqrt{3}}-\frac{5 C_{20,20}^{40,1}}{384 \sqrt{3}}-\frac{7 C_{20,22}^{42,0}}{768 \sqrt{15}}-\frac{7 C_{20,22}^{42,1}}{384 \sqrt{15}}+\frac{7
C_{22,00}^{42,0}}{384 \sqrt{5}}+\frac{7 C_{22,00}^{42,1}}{192 \sqrt{5}}+\frac{7 C_{22,20}^{42,0}}{384 \sqrt{15}}+\frac{7 C_{22,20}^{42,1}}{192 \sqrt{15}}-\frac{7
C_{22,22}^{40,0}}{768 \sqrt{15}}-\frac{7 C_{22,22}^{40,1}}{384 \sqrt{15}}+\frac{1}{192} \sqrt{\frac{7}{15}} C_{22,22}^{42,0}+\frac{1}{96} \sqrt{\frac{7}{15}}
C_{22,22}^{42,1}+\frac{3 \sqrt{3} C_{22,22}^{44,0}}{320}+\frac{3 \sqrt{3} C_{22,22}^{44,1}}{160}-\frac{C_{31,00}^{31,0}}{768}-\frac{C_{31,00}^{31,1}}{384}-\frac{C_{31,20}^{31,0}}{768
\sqrt{3}}-\frac{C_{31,20}^{31,1}}{384 \sqrt{3}}+\frac{7 C_{31,22}^{31,0}}{768 \sqrt{15}}+\frac{7 C_{31,22}^{31,1}}{384 \sqrt{15}}-\frac{\sqrt{7}
C_{31,22}^{33,0}}{320}-\frac{\sqrt{7} C_{31,22}^{33,1}}{160}-\frac{3 \sqrt{21} C_{33,00}^{33,0}}{640}-\frac{3 \sqrt{21} C_{33,00}^{33,1}}{320}-\frac{3
\sqrt{7} C_{33,20}^{33,0}}{640}-\frac{3 \sqrt{7} C_{33,20}^{33,1}}{320}-\frac{3}{320} \sqrt{\frac{21}{10}} C_{33,22}^{33,0}-\frac{3}{160} \sqrt{\frac{21}{10}}
C_{33,22}^{33,1}\)

\noindent\(C_{20,3111,1}^{00,1111,1,\text{Loc}}= -\frac{7 C_{00,20}^{60,1}}{128}+\frac{21 C_{00,22}^{62,1}}{256 \sqrt{5}}-\frac{7 C_{11,20}^{31,1}}{384}+\frac{7
\sqrt{5} C_{11,22}^{51,1}}{768}-\frac{5 C_{20,20}^{40,1}}{384}+\frac{7 C_{20,22}^{42,1}}{768 \sqrt{5}}+\frac{7 C_{22,20}^{42,1}}{192 \sqrt{5}}+\frac{7
C_{22,22}^{40,1}}{768 \sqrt{5}}-\frac{1}{192} \sqrt{\frac{7}{5}} C_{22,22}^{42,1}-\frac{9 C_{22,22}^{44,1}}{320}+\frac{C_{31,20}^{31,1}}{384}+\frac{7
C_{31,22}^{31,1}}{768 \sqrt{5}}-\frac{\sqrt{21} C_{31,22}^{33,1}}{320}+\frac{3 \sqrt{21} C_{33,20}^{33,1}}{320}-\frac{9}{320} \sqrt{\frac{7}{10}}
C_{33,22}^{33,1}\)

\noindent\(C_{20,3111,1}^{00,1111,1,\text{NonLoc}}= -\frac{7 \sqrt{3} C_{00,00}^{60,0}}{256}-\frac{7 C_{00,20}^{60,0}}{256}-\frac{21
C_{00,22}^{62,0}}{256 \sqrt{5}}+\frac{7 C_{11,00}^{51,0}}{256 \sqrt{3}}+\frac{7 C_{11,20}^{31,0}}{768}+\frac{7 \sqrt{5} C_{11,22}^{51,0}}{768}-\frac{5
C_{20,00}^{40,0}}{256 \sqrt{3}}-\frac{5 C_{20,20}^{40,0}}{768}-\frac{7 C_{20,22}^{42,0}}{768 \sqrt{5}}+\frac{7 C_{22,00}^{42,0}}{128 \sqrt{15}}+\frac{7
C_{22,20}^{42,0}}{384 \sqrt{5}}-\frac{7 C_{22,22}^{40,0}}{768 \sqrt{5}}+\frac{1}{192} \sqrt{\frac{7}{5}} C_{22,22}^{42,0}+\frac{9 C_{22,22}^{44,0}}{320}-\frac{C_{31,00}^{31,0}}{256
\sqrt{3}}-\frac{C_{31,20}^{31,0}}{768}+\frac{7 C_{31,22}^{31,0}}{768 \sqrt{5}}-\frac{\sqrt{21} C_{31,22}^{33,0}}{320}-\frac{9 \sqrt{7} C_{33,00}^{33,0}}{640}-\frac{3
\sqrt{21} C_{33,20}^{33,0}}{640}-\frac{9}{320} \sqrt{\frac{7}{10}} C_{33,22}^{33,0}\)

\noindent\(C_{20,3111,1}^{11,0000,0,\text{Loc}}= -\frac{7 C_{11,11}^{51,0}}{384}-\frac{7 C_{11,11}^{51,1}}{768}-\frac{C_{31,11}^{31,0}}{128}-\frac{C_{31,11}^{31,1}}{256}+\frac{3}{160}
\sqrt{\frac{7}{2}} C_{33,11}^{33,0}+\frac{3}{320} \sqrt{\frac{7}{2}} C_{33,11}^{33,1}\)

\noindent\(C_{20,3111,1}^{11,0000,0,\text{NonLoc}}= -\frac{7 C_{11,11}^{51,0}}{768}-\frac{7 C_{11,11}^{51,1}}{384}-\frac{C_{31,11}^{31,0}}{256}-\frac{C_{31,11}^{31,1}}{128}+\frac{3}{320}
\sqrt{\frac{7}{2}} C_{33,11}^{33,0}+\frac{3}{160} \sqrt{\frac{7}{2}} C_{33,11}^{33,1}\)

\noindent\(C_{20,3111,1}^{11,0000,1,\text{Loc}}= -\frac{7 C_{11,11}^{51,1}}{256 \sqrt{3}}-\frac{\sqrt{3} C_{31,11}^{31,1}}{256}+\frac{3}{320}
\sqrt{\frac{21}{2}} C_{33,11}^{33,1}\)

\noindent\(C_{20,3111,1}^{11,0000,1,\text{NonLoc}}= -\frac{7 C_{11,11}^{51,0}}{256 \sqrt{3}}-\frac{\sqrt{3} C_{31,11}^{31,0}}{256}+\frac{3}{320}
\sqrt{\frac{21}{2}} C_{33,11}^{33,0}\)

\noindent\(C_{20,3112,2}^{00,1112,0,\text{Loc}}= -\frac{7 \sqrt{5} C_{00,20}^{60,0}}{192}-\frac{7 \sqrt{5} C_{00,20}^{60,1}}{384}-\frac{7
C_{00,22}^{62,0}}{640}-\frac{7 C_{00,22}^{62,1}}{1280}-\frac{7 \sqrt{5} C_{11,20}^{31,0}}{576}-\frac{7 \sqrt{5} C_{11,20}^{31,1}}{1152}-\frac{7 C_{11,22}^{51,0}}{1152}-\frac{7
C_{11,22}^{51,1}}{2304}-\frac{5 \sqrt{5} C_{20,20}^{40,0}}{576}-\frac{5 \sqrt{5} C_{20,20}^{40,1}}{1152}-\frac{7 C_{20,22}^{42,0}}{5760}-\frac{7
C_{20,22}^{42,1}}{11520}+\frac{7 C_{22,20}^{42,0}}{288}+\frac{7 C_{22,20}^{42,1}}{576}-\frac{7 C_{22,22}^{40,0}}{5760}-\frac{7 C_{22,22}^{40,1}}{11520}+\frac{\sqrt{7}
C_{22,22}^{42,0}}{1440}+\frac{\sqrt{7} C_{22,22}^{42,1}}{2880}+\frac{3 C_{22,22}^{44,0}}{160 \sqrt{5}}+\frac{3 C_{22,22}^{44,1}}{320 \sqrt{5}}+\frac{\sqrt{5}
C_{31,20}^{31,0}}{576}+\frac{\sqrt{5} C_{31,20}^{31,1}}{1152}-\frac{7 C_{31,22}^{31,0}}{5760}-\frac{7 C_{31,22}^{31,1}}{11520}+\frac{1}{160} \sqrt{\frac{7}{15}}
C_{31,22}^{33,0}+\frac{1}{320} \sqrt{\frac{7}{15}} C_{31,22}^{33,1}+\frac{1}{32} \sqrt{\frac{21}{5}} C_{33,20}^{33,0}+\frac{1}{64} \sqrt{\frac{21}{5}}
C_{33,20}^{33,1}+\frac{3}{800} \sqrt{\frac{7}{2}} C_{33,22}^{33,0}+\frac{3 \sqrt{\frac{7}{2}} C_{33,22}^{33,1}}{1600}\)

\noindent\(C_{20,3112,2}^{00,1112,0,\text{NonLoc}}= -\frac{7}{256} \sqrt{\frac{5}{3}} C_{00,00}^{60,0}-\frac{7}{128} \sqrt{\frac{5}{3}}
C_{00,00}^{60,1}-\frac{7 \sqrt{5} C_{00,20}^{60,0}}{768}-\frac{7 \sqrt{5} C_{00,20}^{60,1}}{384}+\frac{7 C_{00,22}^{62,0}}{1280}+\frac{7 C_{00,22}^{62,1}}{640}+\frac{7}{768}
\sqrt{\frac{5}{3}} C_{11,00}^{51,0}+\frac{7}{384} \sqrt{\frac{5}{3}} C_{11,00}^{51,1}+\frac{7 \sqrt{5} C_{11,20}^{31,0}}{2304}+\frac{7 \sqrt{5} C_{11,20}^{31,1}}{1152}-\frac{7
C_{11,22}^{51,0}}{2304}-\frac{7 C_{11,22}^{51,1}}{1152}-\frac{5}{768} \sqrt{\frac{5}{3}} C_{20,00}^{40,0}-\frac{5}{384} \sqrt{\frac{5}{3}} C_{20,00}^{40,1}-\frac{5
\sqrt{5} C_{20,20}^{40,0}}{2304}-\frac{5 \sqrt{5} C_{20,20}^{40,1}}{1152}+\frac{7 C_{20,22}^{42,0}}{11520}+\frac{7 C_{20,22}^{42,1}}{5760}+\frac{7
C_{22,00}^{42,0}}{384 \sqrt{3}}+\frac{7 C_{22,00}^{42,1}}{192 \sqrt{3}}+\frac{7 C_{22,20}^{42,0}}{1152}+\frac{7 C_{22,20}^{42,1}}{576}+\frac{7 C_{22,22}^{40,0}}{11520}+\frac{7
C_{22,22}^{40,1}}{5760}-\frac{\sqrt{7} C_{22,22}^{42,0}}{2880}-\frac{\sqrt{7} C_{22,22}^{42,1}}{1440}-\frac{3 C_{22,22}^{44,0}}{320 \sqrt{5}}-\frac{3
C_{22,22}^{44,1}}{160 \sqrt{5}}-\frac{1}{768} \sqrt{\frac{5}{3}} C_{31,00}^{31,0}-\frac{1}{384} \sqrt{\frac{5}{3}} C_{31,00}^{31,1}-\frac{\sqrt{5}
C_{31,20}^{31,0}}{2304}-\frac{\sqrt{5} C_{31,20}^{31,1}}{1152}-\frac{7 C_{31,22}^{31,0}}{11520}-\frac{7 C_{31,22}^{31,1}}{5760}+\frac{1}{320} \sqrt{\frac{7}{15}}
C_{31,22}^{33,0}+\frac{1}{160} \sqrt{\frac{7}{15}} C_{31,22}^{33,1}-\frac{3}{128} \sqrt{\frac{7}{5}} C_{33,00}^{33,0}-\frac{3}{64} \sqrt{\frac{7}{5}}
C_{33,00}^{33,1}-\frac{1}{128} \sqrt{\frac{21}{5}} C_{33,20}^{33,0}-\frac{1}{64} \sqrt{\frac{21}{5}} C_{33,20}^{33,1}+\frac{3 \sqrt{\frac{7}{2}}
C_{33,22}^{33,0}}{1600}+\frac{3}{800} \sqrt{\frac{7}{2}} C_{33,22}^{33,1}\)

\noindent\(C_{20,3112,2}^{00,1112,1,\text{Loc}}= -\frac{7}{128} \sqrt{\frac{5}{3}} C_{00,20}^{60,1}-\frac{7 \sqrt{3} C_{00,22}^{62,1}}{1280}-\frac{7}{384}
\sqrt{\frac{5}{3}} C_{11,20}^{31,1}-\frac{7 C_{11,22}^{51,1}}{768 \sqrt{3}}-\frac{5}{384} \sqrt{\frac{5}{3}} C_{20,20}^{40,1}-\frac{7 C_{20,22}^{42,1}}{3840
\sqrt{3}}+\frac{7 C_{22,20}^{42,1}}{192 \sqrt{3}}-\frac{7 C_{22,22}^{40,1}}{3840 \sqrt{3}}+\frac{1}{960} \sqrt{\frac{7}{3}} C_{22,22}^{42,1}+\frac{3}{320}
\sqrt{\frac{3}{5}} C_{22,22}^{44,1}+\frac{1}{384} \sqrt{\frac{5}{3}} C_{31,20}^{31,1}-\frac{7 C_{31,22}^{31,1}}{3840 \sqrt{3}}+\frac{1}{320} \sqrt{\frac{7}{5}}
C_{31,22}^{33,1}+\frac{3}{64} \sqrt{\frac{7}{5}} C_{33,20}^{33,1}+\frac{3 \sqrt{\frac{21}{2}} C_{33,22}^{33,1}}{1600}\)

\noindent\(C_{20,3112,2}^{00,1112,1,\text{NonLoc}}= -\frac{7 \sqrt{5} C_{00,00}^{60,0}}{256}-\frac{7}{256} \sqrt{\frac{5}{3}} C_{00,20}^{60,0}+\frac{7
\sqrt{3} C_{00,22}^{62,0}}{1280}+\frac{7 \sqrt{5} C_{11,00}^{51,0}}{768}+\frac{7}{768} \sqrt{\frac{5}{3}} C_{11,20}^{31,0}-\frac{7 C_{11,22}^{51,0}}{768
\sqrt{3}}-\frac{5 \sqrt{5} C_{20,00}^{40,0}}{768}-\frac{5}{768} \sqrt{\frac{5}{3}} C_{20,20}^{40,0}+\frac{7 C_{20,22}^{42,0}}{3840 \sqrt{3}}+\frac{7
C_{22,00}^{42,0}}{384}+\frac{7 C_{22,20}^{42,0}}{384 \sqrt{3}}+\frac{7 C_{22,22}^{40,0}}{3840 \sqrt{3}}-\frac{1}{960} \sqrt{\frac{7}{3}} C_{22,22}^{42,0}-\frac{3}{320}
\sqrt{\frac{3}{5}} C_{22,22}^{44,0}-\frac{\sqrt{5} C_{31,00}^{31,0}}{768}-\frac{1}{768} \sqrt{\frac{5}{3}} C_{31,20}^{31,0}-\frac{7 C_{31,22}^{31,0}}{3840
\sqrt{3}}+\frac{1}{320} \sqrt{\frac{7}{5}} C_{31,22}^{33,0}-\frac{3}{128} \sqrt{\frac{21}{5}} C_{33,00}^{33,0}-\frac{3}{128} \sqrt{\frac{7}{5}} C_{33,20}^{33,0}+\frac{3
\sqrt{\frac{21}{2}} C_{33,22}^{33,0}}{1600}\)

\noindent\(C_{20,3312,2}^{00,1112,0,\text{Loc}}= -\frac{1}{64} \sqrt{\frac{21}{5}} C_{00,22}^{62,0}-\frac{1}{128} \sqrt{\frac{21}{5}}
C_{00,22}^{62,1}-\frac{1}{192} \sqrt{\frac{35}{3}} C_{11,22}^{51,0}-\frac{1}{384} \sqrt{\frac{35}{3}} C_{11,22}^{51,1}-\frac{1}{192} \sqrt{\frac{7}{15}}
C_{20,22}^{42,0}-\frac{1}{384} \sqrt{\frac{7}{15}} C_{20,22}^{42,1}-\frac{1}{192} \sqrt{\frac{7}{15}} C_{22,22}^{40,0}-\frac{1}{384} \sqrt{\frac{7}{15}}
C_{22,22}^{40,1}+\frac{C_{22,22}^{42,0}}{48 \sqrt{15}}+\frac{C_{22,22}^{42,1}}{96 \sqrt{15}}+\frac{3}{80} \sqrt{\frac{3}{7}} C_{22,22}^{44,0}+\frac{3}{160}
\sqrt{\frac{3}{7}} C_{22,22}^{44,1}-\frac{1}{192} \sqrt{\frac{7}{15}} C_{31,22}^{31,0}-\frac{1}{384} \sqrt{\frac{7}{15}} C_{31,22}^{31,1}+\frac{C_{31,22}^{33,0}}{80}+\frac{C_{31,22}^{33,1}}{160}+\frac{3}{80}
\sqrt{\frac{3}{10}} C_{33,22}^{33,0}+\frac{3}{160} \sqrt{\frac{3}{10}} C_{33,22}^{33,1}\)

\noindent\(C_{20,3312,2}^{00,1112,0,\text{NonLoc}}= \frac{1}{128} \sqrt{\frac{21}{5}} C_{00,22}^{62,0}+\frac{1}{64} \sqrt{\frac{21}{5}}
C_{00,22}^{62,1}-\frac{1}{384} \sqrt{\frac{35}{3}} C_{11,22}^{51,0}-\frac{1}{192} \sqrt{\frac{35}{3}} C_{11,22}^{51,1}+\frac{1}{384} \sqrt{\frac{7}{15}}
C_{20,22}^{42,0}+\frac{1}{192} \sqrt{\frac{7}{15}} C_{20,22}^{42,1}+\frac{1}{384} \sqrt{\frac{7}{15}} C_{22,22}^{40,0}+\frac{1}{192} \sqrt{\frac{7}{15}}
C_{22,22}^{40,1}-\frac{C_{22,22}^{42,0}}{96 \sqrt{15}}-\frac{C_{22,22}^{42,1}}{48 \sqrt{15}}-\frac{3}{160} \sqrt{\frac{3}{7}} C_{22,22}^{44,0}-\frac{3}{80}
\sqrt{\frac{3}{7}} C_{22,22}^{44,1}-\frac{1}{384} \sqrt{\frac{7}{15}} C_{31,22}^{31,0}-\frac{1}{192} \sqrt{\frac{7}{15}} C_{31,22}^{31,1}+\frac{C_{31,22}^{33,0}}{160}+\frac{C_{31,22}^{33,1}}{80}+\frac{3}{160}
\sqrt{\frac{3}{10}} C_{33,22}^{33,0}+\frac{3}{80} \sqrt{\frac{3}{10}} C_{33,22}^{33,1}\)

\noindent\(C_{20,3312,2}^{00,1112,1,\text{Loc}}= -\frac{3}{128} \sqrt{\frac{7}{5}} C_{00,22}^{62,1}-\frac{\sqrt{35} C_{11,22}^{51,1}}{384}-\frac{1}{384}
\sqrt{\frac{7}{5}} C_{20,22}^{42,1}-\frac{1}{384} \sqrt{\frac{7}{5}} C_{22,22}^{40,1}+\frac{C_{22,22}^{42,1}}{96 \sqrt{5}}+\frac{9 C_{22,22}^{44,1}}{160
\sqrt{7}}-\frac{1}{384} \sqrt{\frac{7}{5}} C_{31,22}^{31,1}+\frac{\sqrt{3} C_{31,22}^{33,1}}{160}+\frac{9 C_{33,22}^{33,1}}{160 \sqrt{10}}\)

\noindent\(C_{20,3312,2}^{00,1112,1,\text{NonLoc}}= \frac{3}{128} \sqrt{\frac{7}{5}} C_{00,22}^{62,0}-\frac{\sqrt{35} C_{11,22}^{51,0}}{384}+\frac{1}{384}
\sqrt{\frac{7}{5}} C_{20,22}^{42,0}+\frac{1}{384} \sqrt{\frac{7}{5}} C_{22,22}^{40,0}-\frac{C_{22,22}^{42,0}}{96 \sqrt{5}}-\frac{9 C_{22,22}^{44,0}}{160
\sqrt{7}}-\frac{1}{384} \sqrt{\frac{7}{5}} C_{31,22}^{31,0}+\frac{\sqrt{3} C_{31,22}^{33,0}}{160}+\frac{9 C_{33,22}^{33,0}}{160 \sqrt{10}}\)

\noindent\(C_{20,4000,0}^{00,0000,0,\text{Loc}}= -\frac{7 C_{00,00}^{60,0}}{256}-\frac{7 C_{00,00}^{60,1}}{512}-\frac{7 C_{11,00}^{51,0}}{768}-\frac{7
C_{11,00}^{51,1}}{1536}-\frac{5 C_{20,00}^{40,0}}{768}-\frac{5 C_{20,00}^{40,1}}{1536}+\frac{7 C_{22,00}^{42,0}}{384 \sqrt{5}}+\frac{7 C_{22,00}^{42,1}}{768
\sqrt{5}}+\frac{C_{31,00}^{31,0}}{768}+\frac{C_{31,00}^{31,1}}{1536}+\frac{3 \sqrt{21} C_{33,00}^{33,0}}{640}+\frac{3 \sqrt{21} C_{33,00}^{33,1}}{1280}\)

\noindent\(C_{20,4000,0}^{00,0000,0,\text{NonLoc}}= \frac{7 C_{00,00}^{60,0}}{1024}+\frac{7 C_{00,00}^{60,1}}{512}-\frac{7 \sqrt{3}
C_{00,20}^{60,0}}{1024}-\frac{7 \sqrt{3} C_{00,20}^{60,1}}{512}-\frac{7 C_{11,00}^{51,0}}{3072}-\frac{7 C_{11,00}^{51,1}}{1536}+\frac{7 C_{11,20}^{31,0}}{1024
\sqrt{3}}+\frac{7 C_{11,20}^{31,1}}{512 \sqrt{3}}+\frac{5 C_{20,00}^{40,0}}{3072}+\frac{5 C_{20,00}^{40,1}}{1536}-\frac{5 C_{20,20}^{40,0}}{1024
\sqrt{3}}-\frac{5 C_{20,20}^{40,1}}{512 \sqrt{3}}-\frac{7 C_{22,00}^{42,0}}{1536 \sqrt{5}}-\frac{7 C_{22,00}^{42,1}}{768 \sqrt{5}}+\frac{7 C_{22,20}^{42,0}}{512
\sqrt{15}}+\frac{7 C_{22,20}^{42,1}}{256 \sqrt{15}}+\frac{C_{31,00}^{31,0}}{3072}+\frac{C_{31,00}^{31,1}}{1536}-\frac{C_{31,20}^{31,0}}{1024 \sqrt{3}}-\frac{C_{31,20}^{31,1}}{512
\sqrt{3}}+\frac{3 \sqrt{21} C_{33,00}^{33,0}}{2560}+\frac{3 \sqrt{21} C_{33,00}^{33,1}}{1280}-\frac{9 \sqrt{7} C_{33,20}^{33,0}}{2560}-\frac{9 \sqrt{7}
C_{33,20}^{33,1}}{1280}\)

\noindent\(C_{20,4000,0}^{00,0000,1,\text{Loc}}= -\frac{7 \sqrt{3} C_{00,00}^{60,1}}{512}-\frac{7 C_{11,00}^{51,1}}{512 \sqrt{3}}-\frac{5
C_{20,00}^{40,1}}{512 \sqrt{3}}+\frac{7 C_{22,00}^{42,1}}{256 \sqrt{15}}+\frac{C_{31,00}^{31,1}}{512 \sqrt{3}}+\frac{9 \sqrt{7} C_{33,00}^{33,1}}{1280}\)

\noindent\(C_{20,4000,0}^{00,0000,1,\text{NonLoc}}= \frac{7 \sqrt{3} C_{00,00}^{60,0}}{1024}-\frac{21 C_{00,20}^{60,0}}{1024}-\frac{7
C_{11,00}^{51,0}}{1024 \sqrt{3}}+\frac{7 C_{11,20}^{31,0}}{1024}+\frac{5 C_{20,00}^{40,0}}{1024 \sqrt{3}}-\frac{5 C_{20,20}^{40,0}}{1024}-\frac{7
C_{22,00}^{42,0}}{512 \sqrt{15}}+\frac{7 C_{22,20}^{42,0}}{512 \sqrt{5}}+\frac{C_{31,00}^{31,0}}{1024 \sqrt{3}}-\frac{C_{31,20}^{31,0}}{1024}+\frac{9
\sqrt{7} C_{33,00}^{33,0}}{2560}-\frac{9 \sqrt{21} C_{33,20}^{33,0}}{2560}\)

\noindent\(C_{20,4011,1}^{00,0011,0,\text{Loc}}= \frac{7 C_{00,20}^{60,0}}{256}+\frac{7 C_{00,20}^{60,1}}{512}+\frac{7 C_{11,20}^{31,0}}{768}+\frac{7
C_{11,20}^{31,1}}{1536}+\frac{5 C_{20,20}^{40,0}}{768}+\frac{5 C_{20,20}^{40,1}}{1536}-\frac{7 C_{22,20}^{42,0}}{384 \sqrt{5}}-\frac{7 C_{22,20}^{42,1}}{768
\sqrt{5}}-\frac{C_{31,20}^{31,0}}{768}-\frac{C_{31,20}^{31,1}}{1536}-\frac{3 \sqrt{21} C_{33,20}^{33,0}}{640}-\frac{3 \sqrt{21} C_{33,20}^{33,1}}{1280}\)

\noindent\(C_{20,4011,1}^{00,0011,0,\text{NonLoc}}= \frac{7 \sqrt{3} C_{00,00}^{60,0}}{1024}+\frac{7 \sqrt{3} C_{00,00}^{60,1}}{512}+\frac{7
C_{00,20}^{60,0}}{1024}+\frac{7 C_{00,20}^{60,1}}{512}-\frac{7 C_{11,00}^{51,0}}{1024 \sqrt{3}}-\frac{7 C_{11,00}^{51,1}}{512 \sqrt{3}}-\frac{7 C_{11,20}^{31,0}}{3072}-\frac{7
C_{11,20}^{31,1}}{1536}+\frac{5 C_{20,00}^{40,0}}{1024 \sqrt{3}}+\frac{5 C_{20,00}^{40,1}}{512 \sqrt{3}}+\frac{5 C_{20,20}^{40,0}}{3072}+\frac{5
C_{20,20}^{40,1}}{1536}-\frac{7 C_{22,00}^{42,0}}{512 \sqrt{15}}-\frac{7 C_{22,00}^{42,1}}{256 \sqrt{15}}-\frac{7 C_{22,20}^{42,0}}{1536 \sqrt{5}}-\frac{7
C_{22,20}^{42,1}}{768 \sqrt{5}}+\frac{C_{31,00}^{31,0}}{1024 \sqrt{3}}+\frac{C_{31,00}^{31,1}}{512 \sqrt{3}}+\frac{C_{31,20}^{31,0}}{3072}+\frac{C_{31,20}^{31,1}}{1536}+\frac{9
\sqrt{7} C_{33,00}^{33,0}}{2560}+\frac{9 \sqrt{7} C_{33,00}^{33,1}}{1280}+\frac{3 \sqrt{21} C_{33,20}^{33,0}}{2560}+\frac{3 \sqrt{21} C_{33,20}^{33,1}}{1280}\)

\noindent\(C_{20,4011,1}^{00,0011,1,\text{Loc}}= \frac{7 \sqrt{3} C_{00,20}^{60,1}}{512}+\frac{7 C_{11,20}^{31,1}}{512 \sqrt{3}}+\frac{5
C_{20,20}^{40,1}}{512 \sqrt{3}}-\frac{7 C_{22,20}^{42,1}}{256 \sqrt{15}}-\frac{C_{31,20}^{31,1}}{512 \sqrt{3}}-\frac{9 \sqrt{7} C_{33,20}^{33,1}}{1280}\)

\noindent\(C_{20,4011,1}^{00,0011,1,\text{NonLoc}}= \frac{21 C_{00,00}^{60,0}}{1024}+\frac{7 \sqrt{3} C_{00,20}^{60,0}}{1024}-\frac{7
C_{11,00}^{51,0}}{1024}-\frac{7 C_{11,20}^{31,0}}{1024 \sqrt{3}}+\frac{5 C_{20,00}^{40,0}}{1024}+\frac{5 C_{20,20}^{40,0}}{1024 \sqrt{3}}-\frac{7
C_{22,00}^{42,0}}{512 \sqrt{5}}-\frac{7 C_{22,20}^{42,0}}{512 \sqrt{15}}+\frac{C_{31,00}^{31,0}}{1024}+\frac{C_{31,20}^{31,0}}{1024 \sqrt{3}}+\frac{9
\sqrt{21} C_{33,00}^{33,0}}{2560}+\frac{9 \sqrt{7} C_{33,20}^{33,0}}{2560}\)

\noindent\(C_{20,4211,1}^{00,0011,0,\text{Loc}}= \frac{3 C_{00,22}^{62,0}}{128}+\frac{3 C_{00,22}^{62,1}}{256}+\frac{5 C_{11,22}^{51,0}}{384}+\frac{5
C_{11,22}^{51,1}}{768}+\frac{C_{20,22}^{42,0}}{384}+\frac{C_{20,22}^{42,1}}{768}+\frac{C_{22,22}^{40,0}}{384}+\frac{C_{22,22}^{40,1}}{768}-\frac{C_{22,22}^{42,0}}{96
\sqrt{7}}-\frac{C_{22,22}^{42,1}}{192 \sqrt{7}}-\frac{9 C_{22,22}^{44,0}}{224 \sqrt{5}}-\frac{9 C_{22,22}^{44,1}}{448 \sqrt{5}}+\frac{C_{31,22}^{31,0}}{384}+\frac{C_{31,22}^{31,1}}{768}-\frac{1}{32}
\sqrt{\frac{3}{35}} C_{31,22}^{33,0}-\frac{1}{64} \sqrt{\frac{3}{35}} C_{31,22}^{33,1}-\frac{9 C_{33,22}^{33,0}}{160 \sqrt{14}}-\frac{9 C_{33,22}^{33,1}}{320
\sqrt{14}}\)

\noindent\(C_{20,4211,1}^{00,0011,0,\text{NonLoc}}= -\frac{3 C_{00,22}^{62,0}}{256}-\frac{3 C_{00,22}^{62,1}}{128}+\frac{5 C_{11,22}^{51,0}}{768}+\frac{5
C_{11,22}^{51,1}}{384}-\frac{C_{20,22}^{42,0}}{768}-\frac{C_{20,22}^{42,1}}{384}-\frac{C_{22,22}^{40,0}}{768}-\frac{C_{22,22}^{40,1}}{384}+\frac{C_{22,22}^{42,0}}{192
\sqrt{7}}+\frac{C_{22,22}^{42,1}}{96 \sqrt{7}}+\frac{9 C_{22,22}^{44,0}}{448 \sqrt{5}}+\frac{9 C_{22,22}^{44,1}}{224 \sqrt{5}}+\frac{C_{31,22}^{31,0}}{768}+\frac{C_{31,22}^{31,1}}{384}-\frac{1}{64}
\sqrt{\frac{3}{35}} C_{31,22}^{33,0}-\frac{1}{32} \sqrt{\frac{3}{35}} C_{31,22}^{33,1}-\frac{9 C_{33,22}^{33,0}}{320 \sqrt{14}}-\frac{9 C_{33,22}^{33,1}}{160
\sqrt{14}}\)

\noindent\(C_{20,4211,1}^{00,0011,1,\text{Loc}}= \frac{3 \sqrt{3} C_{00,22}^{62,1}}{256}+\frac{5 C_{11,22}^{51,1}}{256 \sqrt{3}}+\frac{C_{20,22}^{42,1}}{256
\sqrt{3}}+\frac{C_{22,22}^{40,1}}{256 \sqrt{3}}-\frac{C_{22,22}^{42,1}}{64 \sqrt{21}}-\frac{9}{448} \sqrt{\frac{3}{5}} C_{22,22}^{44,1}+\frac{C_{31,22}^{31,1}}{256
\sqrt{3}}-\frac{3 C_{31,22}^{33,1}}{64 \sqrt{35}}-\frac{9}{320} \sqrt{\frac{3}{14}} C_{33,22}^{33,1}\)

\noindent\(C_{20,4211,1}^{00,0011,1,\text{NonLoc}}= -\frac{3 \sqrt{3} C_{00,22}^{62,0}}{256}+\frac{5 C_{11,22}^{51,0}}{256 \sqrt{3}}-\frac{C_{20,22}^{42,0}}{256
\sqrt{3}}-\frac{C_{22,22}^{40,0}}{256 \sqrt{3}}+\frac{C_{22,22}^{42,0}}{64 \sqrt{21}}+\frac{9}{448} \sqrt{\frac{3}{5}} C_{22,22}^{44,0}+\frac{C_{31,22}^{31,0}}{256
\sqrt{3}}-\frac{3 C_{31,22}^{33,0}}{64 \sqrt{35}}-\frac{9}{320} \sqrt{\frac{3}{14}} C_{33,22}^{33,0}\)

\noindent\(C_{22,0000,2}^{00,4202,0,\text{Loc}}= -\frac{\sqrt{5} C_{00,00}^{60,0}}{64}-\frac{\sqrt{5} C_{00,00}^{60,1}}{128}-\frac{\sqrt{5}
C_{11,00}^{51,0}}{192}-\frac{\sqrt{5} C_{11,00}^{51,1}}{384}+\frac{\sqrt{5} C_{20,00}^{40,0}}{192}+\frac{\sqrt{5} C_{20,00}^{40,1}}{384}-\frac{C_{22,00}^{42,0}}{192}-\frac{C_{22,00}^{42,1}}{384}+\frac{\sqrt{5}
C_{31,00}^{31,0}}{192}+\frac{\sqrt{5} C_{31,00}^{31,1}}{384}-\frac{3}{64} \sqrt{\frac{3}{35}} C_{33,00}^{33,0}-\frac{3}{128} \sqrt{\frac{3}{35}}
C_{33,00}^{33,1}\)

\noindent\(C_{22,0000,2}^{00,4202,0,\text{NonLoc}}= \frac{\sqrt{5} C_{00,00}^{60,0}}{256}+\frac{\sqrt{5} C_{00,00}^{60,1}}{128}-\frac{\sqrt{15}
C_{00,20}^{60,0}}{256}-\frac{\sqrt{15} C_{00,20}^{60,1}}{128}-\frac{\sqrt{5} C_{11,00}^{51,0}}{768}-\frac{\sqrt{5} C_{11,00}^{51,1}}{384}-\frac{\sqrt{5}
C_{20,00}^{40,0}}{768}-\frac{\sqrt{5} C_{20,00}^{40,1}}{384}+\frac{1}{256} \sqrt{\frac{5}{3}} C_{20,20}^{40,0}+\frac{1}{128} \sqrt{\frac{5}{3}} C_{20,20}^{40,1}+\frac{C_{22,00}^{42,0}}{768}+\frac{C_{22,00}^{42,1}}{384}-\frac{C_{22,20}^{42,0}}{256
\sqrt{3}}-\frac{C_{22,20}^{42,1}}{128 \sqrt{3}}+\frac{\sqrt{5} C_{31,00}^{31,0}}{768}+\frac{\sqrt{5} C_{31,00}^{31,1}}{384}-\frac{1}{256} \sqrt{\frac{5}{3}}
C_{31,20}^{31,0}-\frac{1}{128} \sqrt{\frac{5}{3}} C_{31,20}^{31,1}-\frac{3}{256} \sqrt{\frac{3}{35}} C_{33,00}^{33,0}-\frac{3}{128} \sqrt{\frac{3}{35}}
C_{33,00}^{33,1}+\frac{9 C_{33,20}^{33,0}}{256 \sqrt{35}}+\frac{9 C_{33,20}^{33,1}}{128 \sqrt{35}}\)

\noindent\(C_{22,0000,2}^{00,4202,1,\text{Loc}}= -\frac{\sqrt{15} C_{00,00}^{60,1}}{128}-\frac{1}{128} \sqrt{\frac{5}{3}} C_{11,00}^{51,1}+\frac{1}{128}
\sqrt{\frac{5}{3}} C_{20,00}^{40,1}-\frac{C_{22,00}^{42,1}}{128 \sqrt{3}}+\frac{1}{128} \sqrt{\frac{5}{3}} C_{31,00}^{31,1}-\frac{9 C_{33,00}^{33,1}}{128
\sqrt{35}}\)

\noindent\(C_{22,0000,2}^{00,4202,1,\text{NonLoc}}= \frac{\sqrt{15} C_{00,00}^{60,0}}{256}-\frac{3 \sqrt{5} C_{00,20}^{60,0}}{256}-\frac{1}{256}
\sqrt{\frac{5}{3}} C_{11,00}^{51,0}-\frac{1}{256} \sqrt{\frac{5}{3}} C_{20,00}^{40,0}+\frac{\sqrt{5} C_{20,20}^{40,0}}{256}+\frac{C_{22,00}^{42,0}}{256
\sqrt{3}}-\frac{C_{22,20}^{42,0}}{256}+\frac{1}{256} \sqrt{\frac{5}{3}} C_{31,00}^{31,0}-\frac{\sqrt{5} C_{31,20}^{31,0}}{256}-\frac{9 C_{33,00}^{33,0}}{256
\sqrt{35}}+\frac{9}{256} \sqrt{\frac{3}{35}} C_{33,20}^{33,0}\)

\noindent\(C_{22,0000,2}^{11,3111,0,\text{Loc}}= \frac{7 C_{11,11}^{51,0}}{192 \sqrt{5}}+\frac{7 C_{11,11}^{51,1}}{384 \sqrt{5}}+\frac{C_{31,11}^{31,0}}{64
\sqrt{5}}+\frac{C_{31,11}^{31,1}}{128 \sqrt{5}}-\frac{3}{80} \sqrt{\frac{7}{10}} C_{33,11}^{33,0}-\frac{3}{160} \sqrt{\frac{7}{10}} C_{33,11}^{33,1}\)

\noindent\(C_{22,0000,2}^{11,3111,0,\text{NonLoc}}= \frac{7 C_{11,11}^{51,0}}{384 \sqrt{5}}+\frac{7 C_{11,11}^{51,1}}{192 \sqrt{5}}+\frac{C_{31,11}^{31,0}}{128
\sqrt{5}}+\frac{C_{31,11}^{31,1}}{64 \sqrt{5}}-\frac{3}{160} \sqrt{\frac{7}{10}} C_{33,11}^{33,0}-\frac{3}{80} \sqrt{\frac{7}{10}} C_{33,11}^{33,1}\)

\noindent\(C_{22,0000,2}^{11,3111,1,\text{Loc}}= \frac{7 C_{11,11}^{51,1}}{128 \sqrt{15}}+\frac{1}{128} \sqrt{\frac{3}{5}} C_{31,11}^{31,1}-\frac{3}{160}
\sqrt{\frac{21}{10}} C_{33,11}^{33,1}\)

\noindent\(C_{22,0000,2}^{11,3111,1,\text{NonLoc}}= \frac{7 C_{11,11}^{51,0}}{128 \sqrt{15}}+\frac{1}{128} \sqrt{\frac{3}{5}} C_{31,11}^{31,0}-\frac{3}{160}
\sqrt{\frac{21}{10}} C_{33,11}^{33,0}\)

\noindent\(C_{22,0000,2}^{11,3313,0,\text{Loc}}= \frac{1}{96} \sqrt{\frac{7}{2}} C_{11,11}^{51,0}+\frac{1}{192} \sqrt{\frac{7}{2}}
C_{11,11}^{51,1}-\frac{1}{96} \sqrt{\frac{7}{2}} C_{31,11}^{31,0}-\frac{1}{192} \sqrt{\frac{7}{2}} C_{31,11}^{31,1}+\frac{3 C_{33,11}^{33,0}}{320}+\frac{3
C_{33,11}^{33,1}}{640}\)

\noindent\(C_{22,0000,2}^{11,3313,0,\text{NonLoc}}= \frac{1}{192} \sqrt{\frac{7}{2}} C_{11,11}^{51,0}+\frac{1}{96} \sqrt{\frac{7}{2}}
C_{11,11}^{51,1}-\frac{1}{192} \sqrt{\frac{7}{2}} C_{31,11}^{31,0}-\frac{1}{96} \sqrt{\frac{7}{2}} C_{31,11}^{31,1}+\frac{3 C_{33,11}^{33,0}}{640}+\frac{3
C_{33,11}^{33,1}}{320}\)

\noindent\(C_{22,0000,2}^{11,3313,1,\text{Loc}}= \frac{1}{64} \sqrt{\frac{7}{6}} C_{11,11}^{51,1}-\frac{1}{64} \sqrt{\frac{7}{6}} C_{31,11}^{31,1}+\frac{3
\sqrt{3} C_{33,11}^{33,1}}{640}\)

\noindent\(C_{22,0000,2}^{11,3313,1,\text{NonLoc}}= \frac{1}{64} \sqrt{\frac{7}{6}} C_{11,11}^{51,0}-\frac{1}{64} \sqrt{\frac{7}{6}}
C_{31,11}^{31,0}+\frac{3 \sqrt{3} C_{33,11}^{33,0}}{640}\)

\noindent\(C_{22,0000,2}^{20,2202,0,\text{Loc}}= \frac{7 \sqrt{5} C_{00,00}^{60,0}}{768}+\frac{7 \sqrt{5} C_{00,00}^{60,1}}{1536}-\frac{7
\sqrt{5} C_{11,00}^{51,0}}{2304}-\frac{7 \sqrt{5} C_{11,00}^{51,1}}{4608}-\frac{7 \sqrt{5} C_{20,00}^{40,0}}{2304}-\frac{7 \sqrt{5} C_{20,00}^{40,1}}{4608}+\frac{7
C_{22,00}^{42,0}}{2304}+\frac{7 C_{22,00}^{42,1}}{4608}+\frac{7 \sqrt{5} C_{31,00}^{31,0}}{2304}+\frac{7 \sqrt{5} C_{31,00}^{31,1}}{4608}-\frac{1}{256}
\sqrt{\frac{21}{5}} C_{33,00}^{33,0}-\frac{1}{512} \sqrt{\frac{21}{5}} C_{33,00}^{33,1}\)

\noindent\(C_{22,0000,2}^{20,2202,0,\text{NonLoc}}= -\frac{7 \sqrt{5} C_{00,00}^{60,0}}{3072}-\frac{7 \sqrt{5} C_{00,00}^{60,1}}{1536}+\frac{7
\sqrt{\frac{5}{3}} C_{00,20}^{60,0}}{1024}+\frac{7}{512} \sqrt{\frac{5}{3}} C_{00,20}^{60,1}-\frac{7 \sqrt{5} C_{11,00}^{51,0}}{9216}-\frac{7 \sqrt{5}
C_{11,00}^{51,1}}{4608}+\frac{7 \sqrt{5} C_{20,00}^{40,0}}{9216}+\frac{7 \sqrt{5} C_{20,00}^{40,1}}{4608}-\frac{7 \sqrt{\frac{5}{3}} C_{20,20}^{40,0}}{3072}-\frac{7
\sqrt{\frac{5}{3}} C_{20,20}^{40,1}}{1536}-\frac{7 C_{22,00}^{42,0}}{9216}-\frac{7 C_{22,00}^{42,1}}{4608}+\frac{7 C_{22,20}^{42,0}}{3072 \sqrt{3}}+\frac{7
C_{22,20}^{42,1}}{1536 \sqrt{3}}+\frac{7 \sqrt{5} C_{31,00}^{31,0}}{9216}+\frac{7 \sqrt{5} C_{31,00}^{31,1}}{4608}-\frac{7 \sqrt{\frac{5}{3}} C_{31,20}^{31,0}}{3072}-\frac{7
\sqrt{\frac{5}{3}} C_{31,20}^{31,1}}{1536}-\frac{\sqrt{\frac{21}{5}} C_{33,00}^{33,0}}{1024}-\frac{1}{512} \sqrt{\frac{21}{5}} C_{33,00}^{33,1}+\frac{3
\sqrt{\frac{7}{5}} C_{33,20}^{33,0}}{1024}+\frac{3}{512} \sqrt{\frac{7}{5}} C_{33,20}^{33,1}\)

\noindent\(C_{22,0000,2}^{20,2202,1,\text{Loc}}= \frac{7}{512} \sqrt{\frac{5}{3}} C_{00,00}^{60,1}-\frac{7 \sqrt{\frac{5}{3}} C_{11,00}^{51,1}}{1536}-\frac{7
\sqrt{\frac{5}{3}} C_{20,00}^{40,1}}{1536}+\frac{7 C_{22,00}^{42,1}}{1536 \sqrt{3}}+\frac{7 \sqrt{\frac{5}{3}} C_{31,00}^{31,1}}{1536}-\frac{3}{512}
\sqrt{\frac{7}{5}} C_{33,00}^{33,1}\)

\noindent\(C_{22,0000,2}^{20,2202,1,\text{NonLoc}}= -\frac{7 \sqrt{\frac{5}{3}} C_{00,00}^{60,0}}{1024}+\frac{7 \sqrt{5} C_{00,20}^{60,0}}{1024}-\frac{7
\sqrt{\frac{5}{3}} C_{11,00}^{51,0}}{3072}+\frac{7 \sqrt{\frac{5}{3}} C_{20,00}^{40,0}}{3072}-\frac{7 \sqrt{5} C_{20,20}^{40,0}}{3072}-\frac{7 C_{22,00}^{42,0}}{3072
\sqrt{3}}+\frac{7 C_{22,20}^{42,0}}{3072}+\frac{7 \sqrt{\frac{5}{3}} C_{31,00}^{31,0}}{3072}-\frac{7 \sqrt{5} C_{31,20}^{31,0}}{3072}-\frac{3 \sqrt{\frac{7}{5}}
C_{33,00}^{33,0}}{1024}+\frac{3 \sqrt{\frac{21}{5}} C_{33,20}^{33,0}}{1024}\)

\noindent\(C_{22,0011,1}^{00,4011,0,\text{Loc}}= \frac{21 C_{00,22}^{62,0}}{1280}+\frac{21 C_{00,22}^{62,1}}{2560}-\frac{7 C_{11,22}^{51,0}}{3840}-\frac{7
C_{11,22}^{51,1}}{7680}+\frac{9 \sqrt{\frac{21}{5}} C_{11,22}^{53,0}}{1280}+\frac{9 \sqrt{\frac{21}{5}} C_{11,22}^{53,1}}{2560}+\frac{7 C_{20,22}^{42,0}}{3840}+\frac{7
C_{20,22}^{42,1}}{7680}-\frac{17 C_{22,22}^{40,0}}{3840}-\frac{17 C_{22,22}^{40,1}}{7680}-\frac{\sqrt{7} C_{22,22}^{42,0}}{384}-\frac{\sqrt{7} C_{22,22}^{42,1}}{768}+\frac{9
C_{22,22}^{44,0}}{320 \sqrt{5}}+\frac{9 C_{22,22}^{44,1}}{640 \sqrt{5}}-\frac{23 C_{31,22}^{31,0}}{3840}-\frac{23 C_{31,22}^{31,1}}{7680}+\frac{\sqrt{\frac{21}{5}}
C_{31,22}^{33,0}}{1280}+\frac{\sqrt{\frac{21}{5}} C_{31,22}^{33,1}}{2560}-\frac{9 \sqrt{\frac{7}{2}} C_{33,22}^{33,0}}{1600}-\frac{9 \sqrt{\frac{7}{2}}
C_{33,22}^{33,1}}{3200}\)

\noindent\(C_{22,0011,1}^{00,4011,0,\text{NonLoc}}= -\frac{21 C_{00,22}^{62,0}}{2560}-\frac{21 C_{00,22}^{62,1}}{1280}-\frac{7 C_{11,22}^{51,0}}{7680}-\frac{7
C_{11,22}^{51,1}}{3840}+\frac{9 \sqrt{\frac{21}{5}} C_{11,22}^{53,0}}{2560}+\frac{9 \sqrt{\frac{21}{5}} C_{11,22}^{53,1}}{1280}-\frac{7 C_{20,22}^{42,0}}{7680}-\frac{7
C_{20,22}^{42,1}}{3840}+\frac{17 C_{22,22}^{40,0}}{7680}+\frac{17 C_{22,22}^{40,1}}{3840}+\frac{\sqrt{7} C_{22,22}^{42,0}}{768}+\frac{\sqrt{7} C_{22,22}^{42,1}}{384}-\frac{9
C_{22,22}^{44,0}}{640 \sqrt{5}}-\frac{9 C_{22,22}^{44,1}}{320 \sqrt{5}}-\frac{23 C_{31,22}^{31,0}}{7680}-\frac{23 C_{31,22}^{31,1}}{3840}+\frac{\sqrt{\frac{21}{5}}
C_{31,22}^{33,0}}{2560}+\frac{\sqrt{\frac{21}{5}} C_{31,22}^{33,1}}{1280}-\frac{9 \sqrt{\frac{7}{2}} C_{33,22}^{33,0}}{3200}-\frac{9 \sqrt{\frac{7}{2}}
C_{33,22}^{33,1}}{1600}\)

\noindent\(C_{22,0011,1}^{00,4011,1,\text{Loc}}= \frac{21 \sqrt{3} C_{00,22}^{62,1}}{2560}-\frac{7 C_{11,22}^{51,1}}{2560 \sqrt{3}}+\frac{27
\sqrt{\frac{7}{5}} C_{11,22}^{53,1}}{2560}+\frac{7 C_{20,22}^{42,1}}{2560 \sqrt{3}}-\frac{17 C_{22,22}^{40,1}}{2560 \sqrt{3}}-\frac{1}{256} \sqrt{\frac{7}{3}}
C_{22,22}^{42,1}+\frac{9}{640} \sqrt{\frac{3}{5}} C_{22,22}^{44,1}-\frac{23 C_{31,22}^{31,1}}{2560 \sqrt{3}}+\frac{3 \sqrt{\frac{7}{5}} C_{31,22}^{33,1}}{2560}-\frac{9
\sqrt{\frac{21}{2}} C_{33,22}^{33,1}}{3200}\)

\noindent\(C_{22,0011,1}^{00,4011,1,\text{NonLoc}}= -\frac{21 \sqrt{3} C_{00,22}^{62,0}}{2560}-\frac{7 C_{11,22}^{51,0}}{2560 \sqrt{3}}+\frac{27
\sqrt{\frac{7}{5}} C_{11,22}^{53,0}}{2560}-\frac{7 C_{20,22}^{42,0}}{2560 \sqrt{3}}+\frac{17 C_{22,22}^{40,0}}{2560 \sqrt{3}}+\frac{1}{256} \sqrt{\frac{7}{3}}
C_{22,22}^{42,0}-\frac{9}{640} \sqrt{\frac{3}{5}} C_{22,22}^{44,0}-\frac{23 C_{31,22}^{31,0}}{2560 \sqrt{3}}+\frac{3 \sqrt{\frac{7}{5}} C_{31,22}^{33,0}}{2560}-\frac{9
\sqrt{\frac{21}{2}} C_{33,22}^{33,0}}{3200}\)

\noindent\(C_{22,0011,1}^{00,4211,0,\text{Loc}}= \frac{C_{00,20}^{60,0}}{64}+\frac{C_{00,20}^{60,1}}{128}+\frac{3 C_{00,22}^{62,0}}{128
\sqrt{5}}+\frac{3 C_{00,22}^{62,1}}{256 \sqrt{5}}+\frac{C_{11,22}^{51,0}}{192 \sqrt{5}}+\frac{C_{11,22}^{51,1}}{384 \sqrt{5}}+\frac{9 \sqrt{\frac{3}{7}}
C_{11,22}^{53,0}}{1280}+\frac{9 \sqrt{\frac{3}{7}} C_{11,22}^{53,1}}{2560}-\frac{C_{20,20}^{40,0}}{192}-\frac{C_{20,20}^{40,1}}{384}-\frac{C_{20,22}^{42,0}}{192
\sqrt{5}}-\frac{C_{20,22}^{42,1}}{384 \sqrt{5}}+\frac{C_{22,20}^{42,0}}{192 \sqrt{5}}+\frac{C_{22,20}^{42,1}}{384 \sqrt{5}}-\frac{\sqrt{5} C_{22,22}^{40,0}}{384}-\frac{\sqrt{5}
C_{22,22}^{40,1}}{768}+\frac{1}{192} \sqrt{\frac{7}{5}} C_{22,22}^{42,0}+\frac{1}{384} \sqrt{\frac{7}{5}} C_{22,22}^{42,1}-\frac{C_{31,20}^{31,0}}{192}-\frac{C_{31,20}^{31,1}}{384}-\frac{C_{31,22}^{31,0}}{192
\sqrt{5}}-\frac{C_{31,22}^{31,1}}{384 \sqrt{5}}-\frac{9 \sqrt{\frac{3}{7}} C_{31,22}^{33,0}}{1280}-\frac{9 \sqrt{\frac{3}{7}} C_{31,22}^{33,1}}{2560}+\frac{3}{320}
\sqrt{\frac{3}{7}} C_{33,20}^{33,0}+\frac{3}{640} \sqrt{\frac{3}{7}} C_{33,20}^{33,1}+\frac{9 C_{33,22}^{33,0}}{80 \sqrt{70}}+\frac{9 C_{33,22}^{33,1}}{160
\sqrt{70}}\)

\noindent\(C_{22,0011,1}^{00,4211,0,\text{NonLoc}}= \frac{\sqrt{3} C_{00,00}^{60,0}}{256}+\frac{\sqrt{3} C_{00,00}^{60,1}}{128}+\frac{C_{00,20}^{60,0}}{256}+\frac{C_{00,20}^{60,1}}{128}-\frac{3
C_{00,22}^{62,0}}{256 \sqrt{5}}-\frac{3 C_{00,22}^{62,1}}{128 \sqrt{5}}-\frac{C_{11,00}^{51,0}}{256 \sqrt{3}}-\frac{C_{11,00}^{51,1}}{128 \sqrt{3}}+\frac{C_{11,22}^{51,0}}{384
\sqrt{5}}+\frac{C_{11,22}^{51,1}}{192 \sqrt{5}}+\frac{9 \sqrt{\frac{3}{7}} C_{11,22}^{53,0}}{2560}+\frac{9 \sqrt{\frac{3}{7}} C_{11,22}^{53,1}}{1280}-\frac{C_{20,00}^{40,0}}{256
\sqrt{3}}-\frac{C_{20,00}^{40,1}}{128 \sqrt{3}}-\frac{C_{20,20}^{40,0}}{768}-\frac{C_{20,20}^{40,1}}{384}+\frac{C_{20,22}^{42,0}}{384 \sqrt{5}}+\frac{C_{20,22}^{42,1}}{192
\sqrt{5}}+\frac{C_{22,00}^{42,0}}{256 \sqrt{15}}+\frac{C_{22,00}^{42,1}}{128 \sqrt{15}}+\frac{C_{22,20}^{42,0}}{768 \sqrt{5}}+\frac{C_{22,20}^{42,1}}{384
\sqrt{5}}+\frac{\sqrt{5} C_{22,22}^{40,0}}{768}+\frac{\sqrt{5} C_{22,22}^{40,1}}{384}-\frac{1}{384} \sqrt{\frac{7}{5}} C_{22,22}^{42,0}-\frac{1}{192}
\sqrt{\frac{7}{5}} C_{22,22}^{42,1}+\frac{C_{31,00}^{31,0}}{256 \sqrt{3}}+\frac{C_{31,00}^{31,1}}{128 \sqrt{3}}+\frac{C_{31,20}^{31,0}}{768}+\frac{C_{31,20}^{31,1}}{384}-\frac{C_{31,22}^{31,0}}{384
\sqrt{5}}-\frac{C_{31,22}^{31,1}}{192 \sqrt{5}}-\frac{9 \sqrt{\frac{3}{7}} C_{31,22}^{33,0}}{2560}-\frac{9 \sqrt{\frac{3}{7}} C_{31,22}^{33,1}}{1280}-\frac{9
C_{33,00}^{33,0}}{1280 \sqrt{7}}-\frac{9 C_{33,00}^{33,1}}{640 \sqrt{7}}-\frac{3 \sqrt{\frac{3}{7}} C_{33,20}^{33,0}}{1280}-\frac{3}{640} \sqrt{\frac{3}{7}}
C_{33,20}^{33,1}+\frac{9 C_{33,22}^{33,0}}{160 \sqrt{70}}+\frac{9 C_{33,22}^{33,1}}{80 \sqrt{70}}\)

\noindent\(C_{22,0011,1}^{00,4211,1,\text{Loc}}= \frac{\sqrt{3} C_{00,20}^{60,1}}{128}+\frac{3}{256} \sqrt{\frac{3}{5}} C_{00,22}^{62,1}+\frac{C_{11,22}^{51,1}}{128
\sqrt{15}}+\frac{27 C_{11,22}^{53,1}}{2560 \sqrt{7}}-\frac{C_{20,20}^{40,1}}{128 \sqrt{3}}-\frac{C_{20,22}^{42,1}}{128 \sqrt{15}}+\frac{C_{22,20}^{42,1}}{128
\sqrt{15}}-\frac{1}{256} \sqrt{\frac{5}{3}} C_{22,22}^{40,1}+\frac{1}{128} \sqrt{\frac{7}{15}} C_{22,22}^{42,1}-\frac{C_{31,20}^{31,1}}{128 \sqrt{3}}-\frac{C_{31,22}^{31,1}}{128
\sqrt{15}}-\frac{27 C_{31,22}^{33,1}}{2560 \sqrt{7}}+\frac{9 C_{33,20}^{33,1}}{640 \sqrt{7}}+\frac{9}{160} \sqrt{\frac{3}{70}} C_{33,22}^{33,1}\)

\noindent\(C_{22,0011,1}^{00,4211,1,\text{NonLoc}}= \frac{3 C_{00,00}^{60,0}}{256}+\frac{\sqrt{3} C_{00,20}^{60,0}}{256}-\frac{3}{256}
\sqrt{\frac{3}{5}} C_{00,22}^{62,0}-\frac{C_{11,00}^{51,0}}{256}+\frac{C_{11,22}^{51,0}}{128 \sqrt{15}}+\frac{27 C_{11,22}^{53,0}}{2560 \sqrt{7}}-\frac{C_{20,00}^{40,0}}{256}-\frac{C_{20,20}^{40,0}}{256
\sqrt{3}}+\frac{C_{20,22}^{42,0}}{128 \sqrt{15}}+\frac{C_{22,00}^{42,0}}{256 \sqrt{5}}+\frac{C_{22,20}^{42,0}}{256 \sqrt{15}}+\frac{1}{256} \sqrt{\frac{5}{3}}
C_{22,22}^{40,0}-\frac{1}{128} \sqrt{\frac{7}{15}} C_{22,22}^{42,0}+\frac{C_{31,00}^{31,0}}{256}+\frac{C_{31,20}^{31,0}}{256 \sqrt{3}}-\frac{C_{31,22}^{31,0}}{128
\sqrt{15}}-\frac{27 C_{31,22}^{33,0}}{2560 \sqrt{7}}-\frac{9 \sqrt{\frac{3}{7}} C_{33,00}^{33,0}}{1280}-\frac{9 C_{33,20}^{33,0}}{1280 \sqrt{7}}+\frac{9}{160}
\sqrt{\frac{3}{70}} C_{33,22}^{33,0}\)

\noindent\(C_{22,0011,1}^{11,3101,0,\text{Loc}}= -\frac{7 C_{11,11}^{51,0}}{384 \sqrt{5}}-\frac{7 C_{11,11}^{51,1}}{768 \sqrt{5}}-\frac{C_{31,11}^{31,0}}{128
\sqrt{5}}-\frac{C_{31,11}^{31,1}}{256 \sqrt{5}}+\frac{3}{160} \sqrt{\frac{7}{10}} C_{33,11}^{33,0}+\frac{3}{320} \sqrt{\frac{7}{10}} C_{33,11}^{33,1}\)

\noindent\(C_{22,0011,1}^{11,3101,0,\text{NonLoc}}= -\frac{7 C_{11,11}^{51,0}}{768 \sqrt{5}}-\frac{7 C_{11,11}^{51,1}}{384 \sqrt{5}}-\frac{C_{31,11}^{31,0}}{256
\sqrt{5}}-\frac{C_{31,11}^{31,1}}{128 \sqrt{5}}+\frac{3}{320} \sqrt{\frac{7}{10}} C_{33,11}^{33,0}+\frac{3}{160} \sqrt{\frac{7}{10}} C_{33,11}^{33,1}\)

\noindent\(C_{22,0011,1}^{11,3101,1,\text{Loc}}= -\frac{7 C_{11,11}^{51,1}}{256 \sqrt{15}}-\frac{1}{256} \sqrt{\frac{3}{5}} C_{31,11}^{31,1}+\frac{3}{320}
\sqrt{\frac{21}{10}} C_{33,11}^{33,1}\)

\noindent\(C_{22,0011,1}^{11,3101,1,\text{NonLoc}}= -\frac{7 C_{11,11}^{51,0}}{256 \sqrt{15}}-\frac{1}{256} \sqrt{\frac{3}{5}} C_{31,11}^{31,0}+\frac{3}{320}
\sqrt{\frac{21}{10}} C_{33,11}^{33,0}\)

\noindent\(C_{22,0011,1}^{20,2011,0,\text{Loc}}= -\frac{7 C_{00,22}^{62,0}}{512}-\frac{7 C_{00,22}^{62,1}}{1024}+\frac{35 C_{11,22}^{51,0}}{4608}+\frac{35
C_{11,22}^{51,1}}{9216}-\frac{7 C_{20,22}^{42,0}}{4608}-\frac{7 C_{20,22}^{42,1}}{9216}-\frac{7 C_{22,22}^{40,0}}{4608}-\frac{7 C_{22,22}^{40,1}}{9216}+\frac{\sqrt{7}
C_{22,22}^{42,0}}{1152}+\frac{\sqrt{7} C_{22,22}^{42,1}}{2304}+\frac{3 C_{22,22}^{44,0}}{128 \sqrt{5}}+\frac{3 C_{22,22}^{44,1}}{256 \sqrt{5}}+\frac{7
C_{31,22}^{31,0}}{4608}+\frac{7 C_{31,22}^{31,1}}{9216}-\frac{1}{128} \sqrt{\frac{7}{15}} C_{31,22}^{33,0}-\frac{1}{256} \sqrt{\frac{7}{15}} C_{31,22}^{33,1}-\frac{3}{640}
\sqrt{\frac{7}{2}} C_{33,22}^{33,0}-\frac{3 \sqrt{\frac{7}{2}} C_{33,22}^{33,1}}{1280}\)

\noindent\(C_{22,0011,1}^{20,2011,0,\text{NonLoc}}= \frac{7 C_{00,22}^{62,0}}{1024}+\frac{7 C_{00,22}^{62,1}}{512}+\frac{35 C_{11,22}^{51,0}}{9216}+\frac{35
C_{11,22}^{51,1}}{4608}+\frac{7 C_{20,22}^{42,0}}{9216}+\frac{7 C_{20,22}^{42,1}}{4608}+\frac{7 C_{22,22}^{40,0}}{9216}+\frac{7 C_{22,22}^{40,1}}{4608}-\frac{\sqrt{7}
C_{22,22}^{42,0}}{2304}-\frac{\sqrt{7} C_{22,22}^{42,1}}{1152}-\frac{3 C_{22,22}^{44,0}}{256 \sqrt{5}}-\frac{3 C_{22,22}^{44,1}}{128 \sqrt{5}}+\frac{7
C_{31,22}^{31,0}}{9216}+\frac{7 C_{31,22}^{31,1}}{4608}-\frac{1}{256} \sqrt{\frac{7}{15}} C_{31,22}^{33,0}-\frac{1}{128} \sqrt{\frac{7}{15}} C_{31,22}^{33,1}-\frac{3
\sqrt{\frac{7}{2}} C_{33,22}^{33,0}}{1280}-\frac{3}{640} \sqrt{\frac{7}{2}} C_{33,22}^{33,1}\)

\noindent\(C_{22,0011,1}^{20,2011,1,\text{Loc}}= -\frac{7 \sqrt{3} C_{00,22}^{62,1}}{1024}+\frac{35 C_{11,22}^{51,1}}{3072 \sqrt{3}}-\frac{7
C_{20,22}^{42,1}}{3072 \sqrt{3}}-\frac{7 C_{22,22}^{40,1}}{3072 \sqrt{3}}+\frac{1}{768} \sqrt{\frac{7}{3}} C_{22,22}^{42,1}+\frac{3}{256} \sqrt{\frac{3}{5}}
C_{22,22}^{44,1}+\frac{7 C_{31,22}^{31,1}}{3072 \sqrt{3}}-\frac{1}{256} \sqrt{\frac{7}{5}} C_{31,22}^{33,1}-\frac{3 \sqrt{\frac{21}{2}} C_{33,22}^{33,1}}{1280}\)

\noindent\(C_{22,0011,1}^{20,2011,1,\text{NonLoc}}= \frac{7 \sqrt{3} C_{00,22}^{62,0}}{1024}+\frac{35 C_{11,22}^{51,0}}{3072 \sqrt{3}}+\frac{7
C_{20,22}^{42,0}}{3072 \sqrt{3}}+\frac{7 C_{22,22}^{40,0}}{3072 \sqrt{3}}-\frac{1}{768} \sqrt{\frac{7}{3}} C_{22,22}^{42,0}-\frac{3}{256} \sqrt{\frac{3}{5}}
C_{22,22}^{44,0}+\frac{7 C_{31,22}^{31,0}}{3072 \sqrt{3}}-\frac{1}{256} \sqrt{\frac{7}{5}} C_{31,22}^{33,0}-\frac{3 \sqrt{\frac{21}{2}} C_{33,22}^{33,0}}{1280}\)

\noindent\(C_{22,0011,1}^{20,2211,0,\text{Loc}}= -\frac{7 C_{00,20}^{60,0}}{768}-\frac{7 C_{00,20}^{60,1}}{1536}-\frac{7 C_{00,22}^{62,0}}{512
\sqrt{5}}-\frac{7 C_{00,22}^{62,1}}{1024 \sqrt{5}}+\frac{7 C_{11,22}^{51,0}}{2304 \sqrt{5}}+\frac{7 C_{11,22}^{51,1}}{4608 \sqrt{5}}+\frac{3 \sqrt{21}
C_{11,22}^{53,0}}{5120}+\frac{3 \sqrt{21} C_{11,22}^{53,1}}{10240}+\frac{7 C_{20,20}^{40,0}}{2304}+\frac{7 C_{20,20}^{40,1}}{4608}+\frac{7 C_{20,22}^{42,0}}{2304
\sqrt{5}}+\frac{7 C_{20,22}^{42,1}}{4608 \sqrt{5}}-\frac{7 C_{22,20}^{42,0}}{2304 \sqrt{5}}-\frac{7 C_{22,20}^{42,1}}{4608 \sqrt{5}}+\frac{7 \sqrt{5}
C_{22,22}^{40,0}}{4608}+\frac{7 \sqrt{5} C_{22,22}^{40,1}}{9216}-\frac{7 \sqrt{\frac{7}{5}} C_{22,22}^{42,0}}{2304}-\frac{7 \sqrt{\frac{7}{5}} C_{22,22}^{42,1}}{4608}-\frac{7
C_{31,20}^{31,0}}{2304}-\frac{7 C_{31,20}^{31,1}}{4608}-\frac{7 C_{31,22}^{31,0}}{2304 \sqrt{5}}-\frac{7 C_{31,22}^{31,1}}{4608 \sqrt{5}}-\frac{3
\sqrt{21} C_{31,22}^{33,0}}{5120}-\frac{3 \sqrt{21} C_{31,22}^{33,1}}{10240}+\frac{\sqrt{21} C_{33,20}^{33,0}}{1280}+\frac{\sqrt{21} C_{33,20}^{33,1}}{2560}+\frac{3}{320}
\sqrt{\frac{7}{10}} C_{33,22}^{33,0}+\frac{3}{640} \sqrt{\frac{7}{10}} C_{33,22}^{33,1}\)

\noindent\(C_{22,0011,1}^{20,2211,0,\text{NonLoc}}= -\frac{7 C_{00,00}^{60,0}}{1024 \sqrt{3}}-\frac{7 C_{00,00}^{60,1}}{512 \sqrt{3}}-\frac{7
C_{00,20}^{60,0}}{3072}-\frac{7 C_{00,20}^{60,1}}{1536}+\frac{7 C_{00,22}^{62,0}}{1024 \sqrt{5}}+\frac{7 C_{00,22}^{62,1}}{512 \sqrt{5}}-\frac{7
C_{11,00}^{51,0}}{3072 \sqrt{3}}-\frac{7 C_{11,00}^{51,1}}{1536 \sqrt{3}}+\frac{7 C_{11,22}^{51,0}}{4608 \sqrt{5}}+\frac{7 C_{11,22}^{51,1}}{2304
\sqrt{5}}+\frac{3 \sqrt{21} C_{11,22}^{53,0}}{10240}+\frac{3 \sqrt{21} C_{11,22}^{53,1}}{5120}+\frac{7 C_{20,00}^{40,0}}{3072 \sqrt{3}}+\frac{7 C_{20,00}^{40,1}}{1536
\sqrt{3}}+\frac{7 C_{20,20}^{40,0}}{9216}+\frac{7 C_{20,20}^{40,1}}{4608}-\frac{7 C_{20,22}^{42,0}}{4608 \sqrt{5}}-\frac{7 C_{20,22}^{42,1}}{2304
\sqrt{5}}-\frac{7 C_{22,00}^{42,0}}{3072 \sqrt{15}}-\frac{7 C_{22,00}^{42,1}}{1536 \sqrt{15}}-\frac{7 C_{22,20}^{42,0}}{9216 \sqrt{5}}-\frac{7 C_{22,20}^{42,1}}{4608
\sqrt{5}}-\frac{7 \sqrt{5} C_{22,22}^{40,0}}{9216}-\frac{7 \sqrt{5} C_{22,22}^{40,1}}{4608}+\frac{7 \sqrt{\frac{7}{5}} C_{22,22}^{42,0}}{4608}+\frac{7
\sqrt{\frac{7}{5}} C_{22,22}^{42,1}}{2304}+\frac{7 C_{31,00}^{31,0}}{3072 \sqrt{3}}+\frac{7 C_{31,00}^{31,1}}{1536 \sqrt{3}}+\frac{7 C_{31,20}^{31,0}}{9216}+\frac{7
C_{31,20}^{31,1}}{4608}-\frac{7 C_{31,22}^{31,0}}{4608 \sqrt{5}}-\frac{7 C_{31,22}^{31,1}}{2304 \sqrt{5}}-\frac{3 \sqrt{21} C_{31,22}^{33,0}}{10240}-\frac{3
\sqrt{21} C_{31,22}^{33,1}}{5120}-\frac{3 \sqrt{7} C_{33,00}^{33,0}}{5120}-\frac{3 \sqrt{7} C_{33,00}^{33,1}}{2560}-\frac{\sqrt{21} C_{33,20}^{33,0}}{5120}-\frac{\sqrt{21}
C_{33,20}^{33,1}}{2560}+\frac{3}{640} \sqrt{\frac{7}{10}} C_{33,22}^{33,0}+\frac{3}{320} \sqrt{\frac{7}{10}} C_{33,22}^{33,1}\)

\noindent\(C_{22,0011,1}^{20,2211,1,\text{Loc}}= -\frac{7 C_{00,20}^{60,1}}{512 \sqrt{3}}-\frac{7 \sqrt{\frac{3}{5}} C_{00,22}^{62,1}}{1024}+\frac{7
C_{11,22}^{51,1}}{1536 \sqrt{15}}+\frac{9 \sqrt{7} C_{11,22}^{53,1}}{10240}+\frac{7 C_{20,20}^{40,1}}{1536 \sqrt{3}}+\frac{7 C_{20,22}^{42,1}}{1536
\sqrt{15}}-\frac{7 C_{22,20}^{42,1}}{1536 \sqrt{15}}+\frac{7 \sqrt{\frac{5}{3}} C_{22,22}^{40,1}}{3072}-\frac{7 \sqrt{\frac{7}{15}} C_{22,22}^{42,1}}{1536}-\frac{7
C_{31,20}^{31,1}}{1536 \sqrt{3}}-\frac{7 C_{31,22}^{31,1}}{1536 \sqrt{15}}-\frac{9 \sqrt{7} C_{31,22}^{33,1}}{10240}+\frac{3 \sqrt{7} C_{33,20}^{33,1}}{2560}+\frac{3}{640}
\sqrt{\frac{21}{10}} C_{33,22}^{33,1}\)

\noindent\(C_{22,0011,1}^{20,2211,1,\text{NonLoc}}= -\frac{7 C_{00,00}^{60,0}}{1024}-\frac{7 C_{00,20}^{60,0}}{1024 \sqrt{3}}+\frac{7
\sqrt{\frac{3}{5}} C_{00,22}^{62,0}}{1024}-\frac{7 C_{11,00}^{51,0}}{3072}+\frac{7 C_{11,22}^{51,0}}{1536 \sqrt{15}}+\frac{9 \sqrt{7} C_{11,22}^{53,0}}{10240}+\frac{7
C_{20,00}^{40,0}}{3072}+\frac{7 C_{20,20}^{40,0}}{3072 \sqrt{3}}-\frac{7 C_{20,22}^{42,0}}{1536 \sqrt{15}}-\frac{7 C_{22,00}^{42,0}}{3072 \sqrt{5}}-\frac{7
C_{22,20}^{42,0}}{3072 \sqrt{15}}-\frac{7 \sqrt{\frac{5}{3}} C_{22,22}^{40,0}}{3072}+\frac{7 \sqrt{\frac{7}{15}} C_{22,22}^{42,0}}{1536}+\frac{7
C_{31,00}^{31,0}}{3072}+\frac{7 C_{31,20}^{31,0}}{3072 \sqrt{3}}-\frac{7 C_{31,22}^{31,0}}{1536 \sqrt{15}}-\frac{9 \sqrt{7} C_{31,22}^{33,0}}{10240}-\frac{3
\sqrt{21} C_{33,00}^{33,0}}{5120}-\frac{3 \sqrt{7} C_{33,20}^{33,0}}{5120}+\frac{3}{640} \sqrt{\frac{21}{10}} C_{33,22}^{33,0}\)

\noindent\(C_{22,0011,2}^{00,4212,0,\text{Loc}}= \frac{1}{64} \sqrt{\frac{5}{3}} C_{00,20}^{60,0}+\frac{1}{128} \sqrt{\frac{5}{3}}
C_{00,20}^{60,1}-\frac{\sqrt{3} C_{00,22}^{62,0}}{128}-\frac{\sqrt{3} C_{00,22}^{62,1}}{256}-\frac{C_{11,22}^{51,0}}{192 \sqrt{3}}-\frac{C_{11,22}^{51,1}}{384
\sqrt{3}}-\frac{9 C_{11,22}^{53,0}}{256 \sqrt{35}}-\frac{9 C_{11,22}^{53,1}}{512 \sqrt{35}}-\frac{1}{192} \sqrt{\frac{5}{3}} C_{20,20}^{40,0}-\frac{1}{384}
\sqrt{\frac{5}{3}} C_{20,20}^{40,1}+\frac{C_{20,22}^{42,0}}{192 \sqrt{3}}+\frac{C_{20,22}^{42,1}}{384 \sqrt{3}}+\frac{C_{22,20}^{42,0}}{192 \sqrt{3}}+\frac{C_{22,20}^{42,1}}{384
\sqrt{3}}+\frac{5 C_{22,22}^{40,0}}{384 \sqrt{3}}+\frac{5 C_{22,22}^{40,1}}{768 \sqrt{3}}-\frac{1}{192} \sqrt{\frac{7}{3}} C_{22,22}^{42,0}-\frac{1}{384}
\sqrt{\frac{7}{3}} C_{22,22}^{42,1}-\frac{1}{192} \sqrt{\frac{5}{3}} C_{31,20}^{31,0}-\frac{1}{384} \sqrt{\frac{5}{3}} C_{31,20}^{31,1}+\frac{C_{31,22}^{31,0}}{192
\sqrt{3}}+\frac{C_{31,22}^{31,1}}{384 \sqrt{3}}+\frac{9 C_{31,22}^{33,0}}{256 \sqrt{35}}+\frac{9 C_{31,22}^{33,1}}{512 \sqrt{35}}+\frac{3 C_{33,20}^{33,0}}{64
\sqrt{35}}+\frac{3 C_{33,20}^{33,1}}{128 \sqrt{35}}-\frac{3}{80} \sqrt{\frac{3}{14}} C_{33,22}^{33,0}-\frac{3}{160} \sqrt{\frac{3}{14}} C_{33,22}^{33,1}\)

\noindent\(C_{22,0011,2}^{00,4212,0,\text{NonLoc}}= \frac{\sqrt{5} C_{00,00}^{60,0}}{256}+\frac{\sqrt{5} C_{00,00}^{60,1}}{128}+\frac{1}{256}
\sqrt{\frac{5}{3}} C_{00,20}^{60,0}+\frac{1}{128} \sqrt{\frac{5}{3}} C_{00,20}^{60,1}+\frac{\sqrt{3} C_{00,22}^{62,0}}{256}+\frac{\sqrt{3} C_{00,22}^{62,1}}{128}-\frac{\sqrt{5}
C_{11,00}^{51,0}}{768}-\frac{\sqrt{5} C_{11,00}^{51,1}}{384}-\frac{C_{11,22}^{51,0}}{384 \sqrt{3}}-\frac{C_{11,22}^{51,1}}{192 \sqrt{3}}-\frac{9
C_{11,22}^{53,0}}{512 \sqrt{35}}-\frac{9 C_{11,22}^{53,1}}{256 \sqrt{35}}-\frac{\sqrt{5} C_{20,00}^{40,0}}{768}-\frac{\sqrt{5} C_{20,00}^{40,1}}{384}-\frac{1}{768}
\sqrt{\frac{5}{3}} C_{20,20}^{40,0}-\frac{1}{384} \sqrt{\frac{5}{3}} C_{20,20}^{40,1}-\frac{C_{20,22}^{42,0}}{384 \sqrt{3}}-\frac{C_{20,22}^{42,1}}{192
\sqrt{3}}+\frac{C_{22,00}^{42,0}}{768}+\frac{C_{22,00}^{42,1}}{384}+\frac{C_{22,20}^{42,0}}{768 \sqrt{3}}+\frac{C_{22,20}^{42,1}}{384 \sqrt{3}}-\frac{5
C_{22,22}^{40,0}}{768 \sqrt{3}}-\frac{5 C_{22,22}^{40,1}}{384 \sqrt{3}}+\frac{1}{384} \sqrt{\frac{7}{3}} C_{22,22}^{42,0}+\frac{1}{192} \sqrt{\frac{7}{3}}
C_{22,22}^{42,1}+\frac{\sqrt{5} C_{31,00}^{31,0}}{768}+\frac{\sqrt{5} C_{31,00}^{31,1}}{384}+\frac{1}{768} \sqrt{\frac{5}{3}} C_{31,20}^{31,0}+\frac{1}{384}
\sqrt{\frac{5}{3}} C_{31,20}^{31,1}+\frac{C_{31,22}^{31,0}}{384 \sqrt{3}}+\frac{C_{31,22}^{31,1}}{192 \sqrt{3}}+\frac{9 C_{31,22}^{33,0}}{512 \sqrt{35}}+\frac{9
C_{31,22}^{33,1}}{256 \sqrt{35}}-\frac{3}{256} \sqrt{\frac{3}{35}} C_{33,00}^{33,0}-\frac{3}{128} \sqrt{\frac{3}{35}} C_{33,00}^{33,1}-\frac{3 C_{33,20}^{33,0}}{256
\sqrt{35}}-\frac{3 C_{33,20}^{33,1}}{128 \sqrt{35}}-\frac{3}{160} \sqrt{\frac{3}{14}} C_{33,22}^{33,0}-\frac{3}{80} \sqrt{\frac{3}{14}} C_{33,22}^{33,1}\)

\noindent\(C_{22,0011,2}^{00,4212,1,\text{Loc}}= \frac{\sqrt{5} C_{00,20}^{60,1}}{128}-\frac{3 C_{00,22}^{62,1}}{256}-\frac{C_{11,22}^{51,1}}{384}-\frac{9}{512}
\sqrt{\frac{3}{35}} C_{11,22}^{53,1}-\frac{\sqrt{5} C_{20,20}^{40,1}}{384}+\frac{C_{20,22}^{42,1}}{384}+\frac{C_{22,20}^{42,1}}{384}+\frac{5 C_{22,22}^{40,1}}{768}-\frac{\sqrt{7}
C_{22,22}^{42,1}}{384}-\frac{\sqrt{5} C_{31,20}^{31,1}}{384}+\frac{C_{31,22}^{31,1}}{384}+\frac{9}{512} \sqrt{\frac{3}{35}} C_{31,22}^{33,1}+\frac{3}{128}
\sqrt{\frac{3}{35}} C_{33,20}^{33,1}-\frac{9 C_{33,22}^{33,1}}{160 \sqrt{14}}\)

\noindent\(C_{22,0011,2}^{00,4212,1,\text{NonLoc}}= \frac{\sqrt{15} C_{00,00}^{60,0}}{256}+\frac{\sqrt{5} C_{00,20}^{60,0}}{256}+\frac{3
C_{00,22}^{62,0}}{256}-\frac{1}{256} \sqrt{\frac{5}{3}} C_{11,00}^{51,0}-\frac{C_{11,22}^{51,0}}{384}-\frac{9}{512} \sqrt{\frac{3}{35}} C_{11,22}^{53,0}-\frac{1}{256}
\sqrt{\frac{5}{3}} C_{20,00}^{40,0}-\frac{\sqrt{5} C_{20,20}^{40,0}}{768}-\frac{C_{20,22}^{42,0}}{384}+\frac{C_{22,00}^{42,0}}{256 \sqrt{3}}+\frac{C_{22,20}^{42,0}}{768}-\frac{5
C_{22,22}^{40,0}}{768}+\frac{\sqrt{7} C_{22,22}^{42,0}}{384}+\frac{1}{256} \sqrt{\frac{5}{3}} C_{31,00}^{31,0}+\frac{\sqrt{5} C_{31,20}^{31,0}}{768}+\frac{C_{31,22}^{31,0}}{384}+\frac{9}{512}
\sqrt{\frac{3}{35}} C_{31,22}^{33,0}-\frac{9 C_{33,00}^{33,0}}{256 \sqrt{35}}-\frac{3}{256} \sqrt{\frac{3}{35}} C_{33,20}^{33,0}-\frac{9 C_{33,22}^{33,0}}{160
\sqrt{14}}\)

\noindent\(C_{22,0011,2}^{11,3101,0,\text{Loc}}= -\frac{7 C_{11,11}^{51,0}}{128 \sqrt{15}}-\frac{7 C_{11,11}^{51,1}}{256 \sqrt{15}}-\frac{1}{128}
\sqrt{\frac{3}{5}} C_{31,11}^{31,0}-\frac{1}{256} \sqrt{\frac{3}{5}} C_{31,11}^{31,1}+\frac{3}{160} \sqrt{\frac{21}{10}} C_{33,11}^{33,0}+\frac{3}{320}
\sqrt{\frac{21}{10}} C_{33,11}^{33,1}\)

\noindent\(C_{22,0011,2}^{11,3101,0,\text{NonLoc}}= -\frac{7 C_{11,11}^{51,0}}{256 \sqrt{15}}-\frac{7 C_{11,11}^{51,1}}{128 \sqrt{15}}-\frac{1}{256}
\sqrt{\frac{3}{5}} C_{31,11}^{31,0}-\frac{1}{128} \sqrt{\frac{3}{5}} C_{31,11}^{31,1}+\frac{3}{320} \sqrt{\frac{21}{10}} C_{33,11}^{33,0}+\frac{3}{160}
\sqrt{\frac{21}{10}} C_{33,11}^{33,1}\)

\noindent\(C_{22,0011,2}^{11,3101,1,\text{Loc}}= -\frac{7 C_{11,11}^{51,1}}{256 \sqrt{5}}-\frac{3 C_{31,11}^{31,1}}{256 \sqrt{5}}+\frac{9}{320}
\sqrt{\frac{7}{10}} C_{33,11}^{33,1}\)

\noindent\(C_{22,0011,2}^{11,3101,1,\text{NonLoc}}= -\frac{7 C_{11,11}^{51,0}}{256 \sqrt{5}}-\frac{3 C_{31,11}^{31,0}}{256 \sqrt{5}}+\frac{9}{320}
\sqrt{\frac{7}{10}} C_{33,11}^{33,0}\)

\noindent\(C_{22,0011,2}^{11,3303,0,\text{Loc}}= -\frac{\sqrt{7} C_{11,11}^{51,0}}{144}-\frac{\sqrt{7} C_{11,11}^{51,1}}{288}+\frac{\sqrt{7}
C_{31,11}^{31,0}}{144}+\frac{\sqrt{7} C_{31,11}^{31,1}}{288}-\frac{C_{33,11}^{33,0}}{80 \sqrt{2}}-\frac{C_{33,11}^{33,1}}{160 \sqrt{2}}\)

\noindent\(C_{22,0011,2}^{11,3303,0,\text{NonLoc}}= -\frac{\sqrt{7} C_{11,11}^{51,0}}{288}-\frac{\sqrt{7} C_{11,11}^{51,1}}{144}+\frac{\sqrt{7}
C_{31,11}^{31,0}}{288}+\frac{\sqrt{7} C_{31,11}^{31,1}}{144}-\frac{C_{33,11}^{33,0}}{160 \sqrt{2}}-\frac{C_{33,11}^{33,1}}{80 \sqrt{2}}\)

\noindent\(C_{22,0011,2}^{11,3303,1,\text{Loc}}= -\frac{1}{96} \sqrt{\frac{7}{3}} C_{11,11}^{51,1}+\frac{1}{96} \sqrt{\frac{7}{3}}
C_{31,11}^{31,1}-\frac{1}{160} \sqrt{\frac{3}{2}} C_{33,11}^{33,1}\)

\noindent\(C_{22,0011,2}^{11,3303,1,\text{NonLoc}}= -\frac{1}{96} \sqrt{\frac{7}{3}} C_{11,11}^{51,0}+\frac{1}{96} \sqrt{\frac{7}{3}}
C_{31,11}^{31,0}-\frac{1}{160} \sqrt{\frac{3}{2}} C_{33,11}^{33,0}\)

\noindent\(C_{22,0011,2}^{20,2212,0,\text{Loc}}= -\frac{7}{768} \sqrt{\frac{5}{3}} C_{00,20}^{60,0}-\frac{7 \sqrt{\frac{5}{3}} C_{00,20}^{60,1}}{1536}+\frac{7
C_{00,22}^{62,0}}{512 \sqrt{3}}+\frac{7 C_{00,22}^{62,1}}{1024 \sqrt{3}}-\frac{7 C_{11,22}^{51,0}}{2304 \sqrt{3}}-\frac{7 C_{11,22}^{51,1}}{4608
\sqrt{3}}-\frac{3 \sqrt{\frac{7}{5}} C_{11,22}^{53,0}}{1024}-\frac{3 \sqrt{\frac{7}{5}} C_{11,22}^{53,1}}{2048}+\frac{7 \sqrt{\frac{5}{3}} C_{20,20}^{40,0}}{2304}+\frac{7
\sqrt{\frac{5}{3}} C_{20,20}^{40,1}}{4608}-\frac{7 C_{20,22}^{42,0}}{2304 \sqrt{3}}-\frac{7 C_{20,22}^{42,1}}{4608 \sqrt{3}}-\frac{7 C_{22,20}^{42,0}}{2304
\sqrt{3}}-\frac{7 C_{22,20}^{42,1}}{4608 \sqrt{3}}-\frac{35 C_{22,22}^{40,0}}{4608 \sqrt{3}}-\frac{35 C_{22,22}^{40,1}}{9216 \sqrt{3}}+\frac{7 \sqrt{\frac{7}{3}}
C_{22,22}^{42,0}}{2304}+\frac{7 \sqrt{\frac{7}{3}} C_{22,22}^{42,1}}{4608}-\frac{7 \sqrt{\frac{5}{3}} C_{31,20}^{31,0}}{2304}-\frac{7 \sqrt{\frac{5}{3}}
C_{31,20}^{31,1}}{4608}+\frac{7 C_{31,22}^{31,0}}{2304 \sqrt{3}}+\frac{7 C_{31,22}^{31,1}}{4608 \sqrt{3}}+\frac{3 \sqrt{\frac{7}{5}} C_{31,22}^{33,0}}{1024}+\frac{3
\sqrt{\frac{7}{5}} C_{31,22}^{33,1}}{2048}+\frac{1}{256} \sqrt{\frac{7}{5}} C_{33,20}^{33,0}+\frac{1}{512} \sqrt{\frac{7}{5}} C_{33,20}^{33,1}-\frac{1}{320}
\sqrt{\frac{21}{2}} C_{33,22}^{33,0}-\frac{1}{640} \sqrt{\frac{21}{2}} C_{33,22}^{33,1}\)

\noindent\(C_{22,0011,2}^{20,2212,0,\text{NonLoc}}= -\frac{7 \sqrt{5} C_{00,00}^{60,0}}{3072}-\frac{7 \sqrt{5} C_{00,00}^{60,1}}{1536}-\frac{7
\sqrt{\frac{5}{3}} C_{00,20}^{60,0}}{3072}-\frac{7 \sqrt{\frac{5}{3}} C_{00,20}^{60,1}}{1536}-\frac{7 C_{00,22}^{62,0}}{1024 \sqrt{3}}-\frac{7 C_{00,22}^{62,1}}{512
\sqrt{3}}-\frac{7 \sqrt{5} C_{11,00}^{51,0}}{9216}-\frac{7 \sqrt{5} C_{11,00}^{51,1}}{4608}-\frac{7 C_{11,22}^{51,0}}{4608 \sqrt{3}}-\frac{7 C_{11,22}^{51,1}}{2304
\sqrt{3}}-\frac{3 \sqrt{\frac{7}{5}} C_{11,22}^{53,0}}{2048}-\frac{3 \sqrt{\frac{7}{5}} C_{11,22}^{53,1}}{1024}+\frac{7 \sqrt{5} C_{20,00}^{40,0}}{9216}+\frac{7
\sqrt{5} C_{20,00}^{40,1}}{4608}+\frac{7 \sqrt{\frac{5}{3}} C_{20,20}^{40,0}}{9216}+\frac{7 \sqrt{\frac{5}{3}} C_{20,20}^{40,1}}{4608}+\frac{7 C_{20,22}^{42,0}}{4608
\sqrt{3}}+\frac{7 C_{20,22}^{42,1}}{2304 \sqrt{3}}-\frac{7 C_{22,00}^{42,0}}{9216}-\frac{7 C_{22,00}^{42,1}}{4608}-\frac{7 C_{22,20}^{42,0}}{9216
\sqrt{3}}-\frac{7 C_{22,20}^{42,1}}{4608 \sqrt{3}}+\frac{35 C_{22,22}^{40,0}}{9216 \sqrt{3}}+\frac{35 C_{22,22}^{40,1}}{4608 \sqrt{3}}-\frac{7 \sqrt{\frac{7}{3}}
C_{22,22}^{42,0}}{4608}-\frac{7 \sqrt{\frac{7}{3}} C_{22,22}^{42,1}}{2304}+\frac{7 \sqrt{5} C_{31,00}^{31,0}}{9216}+\frac{7 \sqrt{5} C_{31,00}^{31,1}}{4608}+\frac{7
\sqrt{\frac{5}{3}} C_{31,20}^{31,0}}{9216}+\frac{7 \sqrt{\frac{5}{3}} C_{31,20}^{31,1}}{4608}+\frac{7 C_{31,22}^{31,0}}{4608 \sqrt{3}}+\frac{7 C_{31,22}^{31,1}}{2304
\sqrt{3}}+\frac{3 \sqrt{\frac{7}{5}} C_{31,22}^{33,0}}{2048}+\frac{3 \sqrt{\frac{7}{5}} C_{31,22}^{33,1}}{1024}-\frac{\sqrt{\frac{21}{5}} C_{33,00}^{33,0}}{1024}-\frac{1}{512}
\sqrt{\frac{21}{5}} C_{33,00}^{33,1}-\frac{\sqrt{\frac{7}{5}} C_{33,20}^{33,0}}{1024}-\frac{1}{512} \sqrt{\frac{7}{5}} C_{33,20}^{33,1}-\frac{1}{640}
\sqrt{\frac{21}{2}} C_{33,22}^{33,0}-\frac{1}{320} \sqrt{\frac{21}{2}} C_{33,22}^{33,1}\)

\noindent\(C_{22,0011,2}^{20,2212,1,\text{Loc}}= -\frac{7 \sqrt{5} C_{00,20}^{60,1}}{1536}+\frac{7 C_{00,22}^{62,1}}{1024}-\frac{7
C_{11,22}^{51,1}}{4608}-\frac{3 \sqrt{\frac{21}{5}} C_{11,22}^{53,1}}{2048}+\frac{7 \sqrt{5} C_{20,20}^{40,1}}{4608}-\frac{7 C_{20,22}^{42,1}}{4608}-\frac{7
C_{22,20}^{42,1}}{4608}-\frac{35 C_{22,22}^{40,1}}{9216}+\frac{7 \sqrt{7} C_{22,22}^{42,1}}{4608}-\frac{7 \sqrt{5} C_{31,20}^{31,1}}{4608}+\frac{7
C_{31,22}^{31,1}}{4608}+\frac{3 \sqrt{\frac{21}{5}} C_{31,22}^{33,1}}{2048}+\frac{1}{512} \sqrt{\frac{21}{5}} C_{33,20}^{33,1}-\frac{3}{640} \sqrt{\frac{7}{2}}
C_{33,22}^{33,1}\)

\noindent\(C_{22,0011,2}^{20,2212,1,\text{NonLoc}}= -\frac{7 \sqrt{\frac{5}{3}} C_{00,00}^{60,0}}{1024}-\frac{7 \sqrt{5} C_{00,20}^{60,0}}{3072}-\frac{7
C_{00,22}^{62,0}}{1024}-\frac{7 \sqrt{\frac{5}{3}} C_{11,00}^{51,0}}{3072}-\frac{7 C_{11,22}^{51,0}}{4608}-\frac{3 \sqrt{\frac{21}{5}} C_{11,22}^{53,0}}{2048}+\frac{7
\sqrt{\frac{5}{3}} C_{20,00}^{40,0}}{3072}+\frac{7 \sqrt{5} C_{20,20}^{40,0}}{9216}+\frac{7 C_{20,22}^{42,0}}{4608}-\frac{7 C_{22,00}^{42,0}}{3072
\sqrt{3}}-\frac{7 C_{22,20}^{42,0}}{9216}+\frac{35 C_{22,22}^{40,0}}{9216}-\frac{7 \sqrt{7} C_{22,22}^{42,0}}{4608}+\frac{7 \sqrt{\frac{5}{3}} C_{31,00}^{31,0}}{3072}+\frac{7
\sqrt{5} C_{31,20}^{31,0}}{9216}+\frac{7 C_{31,22}^{31,0}}{4608}+\frac{3 \sqrt{\frac{21}{5}} C_{31,22}^{33,0}}{2048}-\frac{3 \sqrt{\frac{7}{5}} C_{33,00}^{33,0}}{1024}-\frac{\sqrt{\frac{21}{5}}
C_{33,20}^{33,0}}{1024}-\frac{3}{640} \sqrt{\frac{7}{2}} C_{33,22}^{33,0}\)

\noindent\(C_{22,0011,3}^{00,4213,0,\text{Loc}}= \frac{1}{64} \sqrt{\frac{7}{3}} C_{00,20}^{60,0}+\frac{1}{128} \sqrt{\frac{7}{3}}
C_{00,20}^{60,1}+\frac{1}{64} \sqrt{\frac{3}{35}} C_{00,22}^{62,0}+\frac{1}{128} \sqrt{\frac{3}{35}} C_{00,22}^{62,1}+\frac{C_{11,22}^{51,0}}{96
\sqrt{105}}+\frac{C_{11,22}^{51,1}}{192 \sqrt{105}}+\frac{9 C_{11,22}^{53,0}}{4480}+\frac{9 C_{11,22}^{53,1}}{8960}-\frac{1}{192} \sqrt{\frac{7}{3}}
C_{20,20}^{40,0}-\frac{1}{384} \sqrt{\frac{7}{3}} C_{20,20}^{40,1}-\frac{C_{20,22}^{42,0}}{96 \sqrt{105}}-\frac{C_{20,22}^{42,1}}{192 \sqrt{105}}+\frac{1}{192}
\sqrt{\frac{7}{15}} C_{22,20}^{42,0}+\frac{1}{384} \sqrt{\frac{7}{15}} C_{22,20}^{42,1}-\frac{1}{192} \sqrt{\frac{5}{21}} C_{22,22}^{40,0}-\frac{1}{384}
\sqrt{\frac{5}{21}} C_{22,22}^{40,1}+\frac{C_{22,22}^{42,0}}{96 \sqrt{15}}+\frac{C_{22,22}^{42,1}}{192 \sqrt{15}}-\frac{1}{192} \sqrt{\frac{7}{3}}
C_{31,20}^{31,0}-\frac{1}{384} \sqrt{\frac{7}{3}} C_{31,20}^{31,1}-\frac{C_{31,22}^{31,0}}{96 \sqrt{105}}-\frac{C_{31,22}^{31,1}}{192 \sqrt{105}}-\frac{9
C_{31,22}^{33,0}}{4480}-\frac{9 C_{31,22}^{33,1}}{8960}+\frac{3 C_{33,20}^{33,0}}{320}+\frac{3 C_{33,20}^{33,1}}{640}+\frac{3}{280} \sqrt{\frac{3}{10}}
C_{33,22}^{33,0}+\frac{3}{560} \sqrt{\frac{3}{10}} C_{33,22}^{33,1}\)

\noindent\(C_{22,0011,3}^{00,4213,0,\text{NonLoc}}= \frac{\sqrt{7} C_{00,00}^{60,0}}{256}+\frac{\sqrt{7} C_{00,00}^{60,1}}{128}+\frac{1}{256}
\sqrt{\frac{7}{3}} C_{00,20}^{60,0}+\frac{1}{128} \sqrt{\frac{7}{3}} C_{00,20}^{60,1}-\frac{1}{128} \sqrt{\frac{3}{35}} C_{00,22}^{62,0}-\frac{1}{64}
\sqrt{\frac{3}{35}} C_{00,22}^{62,1}-\frac{\sqrt{7} C_{11,00}^{51,0}}{768}-\frac{\sqrt{7} C_{11,00}^{51,1}}{384}+\frac{C_{11,22}^{51,0}}{192 \sqrt{105}}+\frac{C_{11,22}^{51,1}}{96
\sqrt{105}}+\frac{9 C_{11,22}^{53,0}}{8960}+\frac{9 C_{11,22}^{53,1}}{4480}-\frac{\sqrt{7} C_{20,00}^{40,0}}{768}-\frac{\sqrt{7} C_{20,00}^{40,1}}{384}-\frac{1}{768}
\sqrt{\frac{7}{3}} C_{20,20}^{40,0}-\frac{1}{384} \sqrt{\frac{7}{3}} C_{20,20}^{40,1}+\frac{C_{20,22}^{42,0}}{192 \sqrt{105}}+\frac{C_{20,22}^{42,1}}{96
\sqrt{105}}+\frac{1}{768} \sqrt{\frac{7}{5}} C_{22,00}^{42,0}+\frac{1}{384} \sqrt{\frac{7}{5}} C_{22,00}^{42,1}+\frac{1}{768} \sqrt{\frac{7}{15}}
C_{22,20}^{42,0}+\frac{1}{384} \sqrt{\frac{7}{15}} C_{22,20}^{42,1}+\frac{1}{384} \sqrt{\frac{5}{21}} C_{22,22}^{40,0}+\frac{1}{192} \sqrt{\frac{5}{21}}
C_{22,22}^{40,1}-\frac{C_{22,22}^{42,0}}{192 \sqrt{15}}-\frac{C_{22,22}^{42,1}}{96 \sqrt{15}}+\frac{\sqrt{7} C_{31,00}^{31,0}}{768}+\frac{\sqrt{7}
C_{31,00}^{31,1}}{384}+\frac{1}{768} \sqrt{\frac{7}{3}} C_{31,20}^{31,0}+\frac{1}{384} \sqrt{\frac{7}{3}} C_{31,20}^{31,1}-\frac{C_{31,22}^{31,0}}{192
\sqrt{105}}-\frac{C_{31,22}^{31,1}}{96 \sqrt{105}}-\frac{9 C_{31,22}^{33,0}}{8960}-\frac{9 C_{31,22}^{33,1}}{4480}-\frac{3 \sqrt{3} C_{33,00}^{33,0}}{1280}-\frac{3
\sqrt{3} C_{33,00}^{33,1}}{640}-\frac{3 C_{33,20}^{33,0}}{1280}-\frac{3 C_{33,20}^{33,1}}{640}+\frac{3}{560} \sqrt{\frac{3}{10}} C_{33,22}^{33,0}+\frac{3}{280}
\sqrt{\frac{3}{10}} C_{33,22}^{33,1}\)

\noindent\(C_{22,0011,3}^{00,4213,1,\text{Loc}}= \frac{\sqrt{7} C_{00,20}^{60,1}}{128}+\frac{3 C_{00,22}^{62,1}}{128 \sqrt{35}}+\frac{C_{11,22}^{51,1}}{192
\sqrt{35}}+\frac{9 \sqrt{3} C_{11,22}^{53,1}}{8960}-\frac{\sqrt{7} C_{20,20}^{40,1}}{384}-\frac{C_{20,22}^{42,1}}{192 \sqrt{35}}+\frac{1}{384} \sqrt{\frac{7}{5}}
C_{22,20}^{42,1}-\frac{1}{384} \sqrt{\frac{5}{7}} C_{22,22}^{40,1}+\frac{C_{22,22}^{42,1}}{192 \sqrt{5}}-\frac{\sqrt{7} C_{31,20}^{31,1}}{384}-\frac{C_{31,22}^{31,1}}{192
\sqrt{35}}-\frac{9 \sqrt{3} C_{31,22}^{33,1}}{8960}+\frac{3 \sqrt{3} C_{33,20}^{33,1}}{640}+\frac{9 C_{33,22}^{33,1}}{560 \sqrt{10}}\)

\noindent\(C_{22,0011,3}^{00,4213,1,\text{NonLoc}}= \frac{\sqrt{21} C_{00,00}^{60,0}}{256}+\frac{\sqrt{7} C_{00,20}^{60,0}}{256}-\frac{3
C_{00,22}^{62,0}}{128 \sqrt{35}}-\frac{1}{256} \sqrt{\frac{7}{3}} C_{11,00}^{51,0}+\frac{C_{11,22}^{51,0}}{192 \sqrt{35}}+\frac{9 \sqrt{3} C_{11,22}^{53,0}}{8960}-\frac{1}{256}
\sqrt{\frac{7}{3}} C_{20,00}^{40,0}-\frac{\sqrt{7} C_{20,20}^{40,0}}{768}+\frac{C_{20,22}^{42,0}}{192 \sqrt{35}}+\frac{1}{256} \sqrt{\frac{7}{15}}
C_{22,00}^{42,0}+\frac{1}{768} \sqrt{\frac{7}{5}} C_{22,20}^{42,0}+\frac{1}{384} \sqrt{\frac{5}{7}} C_{22,22}^{40,0}-\frac{C_{22,22}^{42,0}}{192
\sqrt{5}}+\frac{1}{256} \sqrt{\frac{7}{3}} C_{31,00}^{31,0}+\frac{\sqrt{7} C_{31,20}^{31,0}}{768}-\frac{C_{31,22}^{31,0}}{192 \sqrt{35}}-\frac{9
\sqrt{3} C_{31,22}^{33,0}}{8960}-\frac{9 C_{33,00}^{33,0}}{1280}-\frac{3 \sqrt{3} C_{33,20}^{33,0}}{1280}+\frac{9 C_{33,22}^{33,0}}{560 \sqrt{10}}\)

\noindent\(C_{22,0011,3}^{00,4413,0,\text{Loc}}= \frac{C_{00,22}^{62,0}}{64 \sqrt{5}}+\frac{C_{00,22}^{62,1}}{128 \sqrt{5}}+\frac{C_{11,22}^{51,0}}{64
\sqrt{5}}+\frac{C_{11,22}^{51,1}}{128 \sqrt{5}}-\frac{1}{160} \sqrt{\frac{3}{7}} C_{11,22}^{53,0}-\frac{1}{320} \sqrt{\frac{3}{7}} C_{11,22}^{53,1}-\frac{C_{20,22}^{42,0}}{64
\sqrt{5}}-\frac{C_{20,22}^{42,1}}{128 \sqrt{5}}+\frac{C_{22,22}^{40,0}}{64 \sqrt{5}}+\frac{C_{22,22}^{40,1}}{128 \sqrt{5}}+\frac{C_{22,22}^{44,0}}{320}+\frac{C_{22,22}^{44,1}}{640}-\frac{C_{31,22}^{31,0}}{64
\sqrt{5}}-\frac{C_{31,22}^{31,1}}{128 \sqrt{5}}+\frac{1}{160} \sqrt{\frac{3}{7}} C_{31,22}^{33,0}+\frac{1}{320} \sqrt{\frac{3}{7}} C_{31,22}^{33,1}+\frac{3
C_{33,22}^{33,0}}{320 \sqrt{70}}+\frac{3 C_{33,22}^{33,1}}{640 \sqrt{70}}\)

\noindent\(C_{22,0011,3}^{00,4413,0,\text{NonLoc}}= -\frac{C_{00,22}^{62,0}}{128 \sqrt{5}}-\frac{C_{00,22}^{62,1}}{64 \sqrt{5}}+\frac{C_{11,22}^{51,0}}{128
\sqrt{5}}+\frac{C_{11,22}^{51,1}}{64 \sqrt{5}}-\frac{1}{320} \sqrt{\frac{3}{7}} C_{11,22}^{53,0}-\frac{1}{160} \sqrt{\frac{3}{7}} C_{11,22}^{53,1}+\frac{C_{20,22}^{42,0}}{128
\sqrt{5}}+\frac{C_{20,22}^{42,1}}{64 \sqrt{5}}-\frac{C_{22,22}^{40,0}}{128 \sqrt{5}}-\frac{C_{22,22}^{40,1}}{64 \sqrt{5}}-\frac{C_{22,22}^{44,0}}{640}-\frac{C_{22,22}^{44,1}}{320}-\frac{C_{31,22}^{31,0}}{128
\sqrt{5}}-\frac{C_{31,22}^{31,1}}{64 \sqrt{5}}+\frac{1}{320} \sqrt{\frac{3}{7}} C_{31,22}^{33,0}+\frac{1}{160} \sqrt{\frac{3}{7}} C_{31,22}^{33,1}+\frac{3
C_{33,22}^{33,0}}{640 \sqrt{70}}+\frac{3 C_{33,22}^{33,1}}{320 \sqrt{70}}\)

\noindent\(C_{22,0011,3}^{00,4413,1,\text{Loc}}= \frac{1}{128} \sqrt{\frac{3}{5}} C_{00,22}^{62,1}+\frac{1}{128} \sqrt{\frac{3}{5}}
C_{11,22}^{51,1}-\frac{3 C_{11,22}^{53,1}}{320 \sqrt{7}}-\frac{1}{128} \sqrt{\frac{3}{5}} C_{20,22}^{42,1}+\frac{1}{128} \sqrt{\frac{3}{5}} C_{22,22}^{40,1}+\frac{\sqrt{3}
C_{22,22}^{44,1}}{640}-\frac{1}{128} \sqrt{\frac{3}{5}} C_{31,22}^{31,1}+\frac{3 C_{31,22}^{33,1}}{320 \sqrt{7}}+\frac{3}{640} \sqrt{\frac{3}{70}}
C_{33,22}^{33,1}\)

\noindent\(C_{22,0011,3}^{00,4413,1,\text{NonLoc}}= -\frac{1}{128} \sqrt{\frac{3}{5}} C_{00,22}^{62,0}+\frac{1}{128} \sqrt{\frac{3}{5}}
C_{11,22}^{51,0}-\frac{3 C_{11,22}^{53,0}}{320 \sqrt{7}}+\frac{1}{128} \sqrt{\frac{3}{5}} C_{20,22}^{42,0}-\frac{1}{128} \sqrt{\frac{3}{5}} C_{22,22}^{40,0}-\frac{\sqrt{3}
C_{22,22}^{44,0}}{640}-\frac{1}{128} \sqrt{\frac{3}{5}} C_{31,22}^{31,0}+\frac{3 C_{31,22}^{33,0}}{320 \sqrt{7}}+\frac{3}{640} \sqrt{\frac{3}{70}}
C_{33,22}^{33,0}\)

\noindent\(C_{22,0011,3}^{11,3303,0,\text{Loc}}= \frac{1}{288} \sqrt{\frac{7}{2}} C_{11,11}^{51,0}+\frac{1}{576} \sqrt{\frac{7}{2}}
C_{11,11}^{51,1}-\frac{1}{288} \sqrt{\frac{7}{2}} C_{31,11}^{31,0}-\frac{1}{576} \sqrt{\frac{7}{2}} C_{31,11}^{31,1}+\frac{C_{33,11}^{33,0}}{320}+\frac{C_{33,11}^{33,1}}{640}\)

\noindent\(C_{22,0011,3}^{11,3303,0,\text{NonLoc}}= \frac{1}{576} \sqrt{\frac{7}{2}} C_{11,11}^{51,0}+\frac{1}{288} \sqrt{\frac{7}{2}}
C_{11,11}^{51,1}-\frac{1}{576} \sqrt{\frac{7}{2}} C_{31,11}^{31,0}-\frac{1}{288} \sqrt{\frac{7}{2}} C_{31,11}^{31,1}+\frac{C_{33,11}^{33,0}}{640}+\frac{C_{33,11}^{33,1}}{320}\)

\noindent\(C_{22,0011,3}^{11,3303,1,\text{Loc}}= \frac{1}{192} \sqrt{\frac{7}{6}} C_{11,11}^{51,1}-\frac{1}{192} \sqrt{\frac{7}{6}}
C_{31,11}^{31,1}+\frac{\sqrt{3} C_{33,11}^{33,1}}{640}\)

\noindent\(C_{22,0011,3}^{11,3303,1,\text{NonLoc}}= \frac{1}{192} \sqrt{\frac{7}{6}} C_{11,11}^{51,0}-\frac{1}{192} \sqrt{\frac{7}{6}}
C_{31,11}^{31,0}+\frac{\sqrt{3} C_{33,11}^{33,0}}{640}\)

\noindent\(C_{22,0011,3}^{20,2213,0,\text{Loc}}= -\frac{7}{768} \sqrt{\frac{7}{3}} C_{00,20}^{60,0}-\frac{7 \sqrt{\frac{7}{3}} C_{00,20}^{60,1}}{1536}-\frac{1}{256}
\sqrt{\frac{7}{15}} C_{00,22}^{62,0}-\frac{1}{512} \sqrt{\frac{7}{15}} C_{00,22}^{62,1}+\frac{\sqrt{\frac{7}{15}} C_{11,22}^{51,0}}{1152}+\frac{\sqrt{\frac{7}{15}}
C_{11,22}^{51,1}}{2304}+\frac{3 C_{11,22}^{53,0}}{2560}+\frac{3 C_{11,22}^{53,1}}{5120}+\frac{7 \sqrt{\frac{7}{3}} C_{20,20}^{40,0}}{2304}+\frac{7
\sqrt{\frac{7}{3}} C_{20,20}^{40,1}}{4608}+\frac{\sqrt{\frac{7}{15}} C_{20,22}^{42,0}}{1152}+\frac{\sqrt{\frac{7}{15}} C_{20,22}^{42,1}}{2304}-\frac{7
\sqrt{\frac{7}{15}} C_{22,20}^{42,0}}{2304}-\frac{7 \sqrt{\frac{7}{15}} C_{22,20}^{42,1}}{4608}+\frac{\sqrt{\frac{35}{3}} C_{22,22}^{40,0}}{2304}+\frac{\sqrt{\frac{35}{3}}
C_{22,22}^{40,1}}{4608}-\frac{7 C_{22,22}^{42,0}}{1152 \sqrt{15}}-\frac{7 C_{22,22}^{42,1}}{2304 \sqrt{15}}-\frac{7 \sqrt{\frac{7}{3}} C_{31,20}^{31,0}}{2304}-\frac{7
\sqrt{\frac{7}{3}} C_{31,20}^{31,1}}{4608}-\frac{\sqrt{\frac{7}{15}} C_{31,22}^{31,0}}{1152}-\frac{\sqrt{\frac{7}{15}} C_{31,22}^{31,1}}{2304}-\frac{3
C_{31,22}^{33,0}}{2560}-\frac{3 C_{31,22}^{33,1}}{5120}+\frac{7 C_{33,20}^{33,0}}{1280}+\frac{7 C_{33,20}^{33,1}}{2560}+\frac{1}{160} \sqrt{\frac{3}{10}}
C_{33,22}^{33,0}+\frac{1}{320} \sqrt{\frac{3}{10}} C_{33,22}^{33,1}\)

\noindent\(C_{22,0011,3}^{20,2213,0,\text{NonLoc}}= -\frac{7 \sqrt{7} C_{00,00}^{60,0}}{3072}-\frac{7 \sqrt{7} C_{00,00}^{60,1}}{1536}-\frac{7
\sqrt{\frac{7}{3}} C_{00,20}^{60,0}}{3072}-\frac{7 \sqrt{\frac{7}{3}} C_{00,20}^{60,1}}{1536}+\frac{1}{512} \sqrt{\frac{7}{15}} C_{00,22}^{62,0}+\frac{1}{256}
\sqrt{\frac{7}{15}} C_{00,22}^{62,1}-\frac{7 \sqrt{7} C_{11,00}^{51,0}}{9216}-\frac{7 \sqrt{7} C_{11,00}^{51,1}}{4608}+\frac{\sqrt{\frac{7}{15}}
C_{11,22}^{51,0}}{2304}+\frac{\sqrt{\frac{7}{15}} C_{11,22}^{51,1}}{1152}+\frac{3 C_{11,22}^{53,0}}{5120}+\frac{3 C_{11,22}^{53,1}}{2560}+\frac{7
\sqrt{7} C_{20,00}^{40,0}}{9216}+\frac{7 \sqrt{7} C_{20,00}^{40,1}}{4608}+\frac{7 \sqrt{\frac{7}{3}} C_{20,20}^{40,0}}{9216}+\frac{7 \sqrt{\frac{7}{3}}
C_{20,20}^{40,1}}{4608}-\frac{\sqrt{\frac{7}{15}} C_{20,22}^{42,0}}{2304}-\frac{\sqrt{\frac{7}{15}} C_{20,22}^{42,1}}{1152}-\frac{7 \sqrt{\frac{7}{5}}
C_{22,00}^{42,0}}{9216}-\frac{7 \sqrt{\frac{7}{5}} C_{22,00}^{42,1}}{4608}-\frac{7 \sqrt{\frac{7}{15}} C_{22,20}^{42,0}}{9216}-\frac{7 \sqrt{\frac{7}{15}}
C_{22,20}^{42,1}}{4608}-\frac{\sqrt{\frac{35}{3}} C_{22,22}^{40,0}}{4608}-\frac{\sqrt{\frac{35}{3}} C_{22,22}^{40,1}}{2304}+\frac{7 C_{22,22}^{42,0}}{2304
\sqrt{15}}+\frac{7 C_{22,22}^{42,1}}{1152 \sqrt{15}}+\frac{7 \sqrt{7} C_{31,00}^{31,0}}{9216}+\frac{7 \sqrt{7} C_{31,00}^{31,1}}{4608}+\frac{7 \sqrt{\frac{7}{3}}
C_{31,20}^{31,0}}{9216}+\frac{7 \sqrt{\frac{7}{3}} C_{31,20}^{31,1}}{4608}-\frac{\sqrt{\frac{7}{15}} C_{31,22}^{31,0}}{2304}-\frac{\sqrt{\frac{7}{15}}
C_{31,22}^{31,1}}{1152}-\frac{3 C_{31,22}^{33,0}}{5120}-\frac{3 C_{31,22}^{33,1}}{2560}-\frac{7 \sqrt{3} C_{33,00}^{33,0}}{5120}-\frac{7 \sqrt{3}
C_{33,00}^{33,1}}{2560}-\frac{7 C_{33,20}^{33,0}}{5120}-\frac{7 C_{33,20}^{33,1}}{2560}+\frac{1}{320} \sqrt{\frac{3}{10}} C_{33,22}^{33,0}+\frac{1}{160}
\sqrt{\frac{3}{10}} C_{33,22}^{33,1}\)

\noindent\(C_{22,0011,3}^{20,2213,1,\text{Loc}}= -\frac{7 \sqrt{7} C_{00,20}^{60,1}}{1536}-\frac{1}{512} \sqrt{\frac{7}{5}} C_{00,22}^{62,1}+\frac{\sqrt{\frac{7}{5}}
C_{11,22}^{51,1}}{2304}+\frac{3 \sqrt{3} C_{11,22}^{53,1}}{5120}+\frac{7 \sqrt{7} C_{20,20}^{40,1}}{4608}+\frac{\sqrt{\frac{7}{5}} C_{20,22}^{42,1}}{2304}-\frac{7
\sqrt{\frac{7}{5}} C_{22,20}^{42,1}}{4608}+\frac{\sqrt{35} C_{22,22}^{40,1}}{4608}-\frac{7 C_{22,22}^{42,1}}{2304 \sqrt{5}}-\frac{7 \sqrt{7} C_{31,20}^{31,1}}{4608}-\frac{\sqrt{\frac{7}{5}}
C_{31,22}^{31,1}}{2304}-\frac{3 \sqrt{3} C_{31,22}^{33,1}}{5120}+\frac{7 \sqrt{3} C_{33,20}^{33,1}}{2560}+\frac{3 C_{33,22}^{33,1}}{320 \sqrt{10}}\)

\noindent\(C_{22,0011,3}^{20,2213,1,\text{NonLoc}}= -\frac{7 \sqrt{\frac{7}{3}} C_{00,00}^{60,0}}{1024}-\frac{7 \sqrt{7} C_{00,20}^{60,0}}{3072}+\frac{1}{512}
\sqrt{\frac{7}{5}} C_{00,22}^{62,0}-\frac{7 \sqrt{\frac{7}{3}} C_{11,00}^{51,0}}{3072}+\frac{\sqrt{\frac{7}{5}} C_{11,22}^{51,0}}{2304}+\frac{3 \sqrt{3}
C_{11,22}^{53,0}}{5120}+\frac{7 \sqrt{\frac{7}{3}} C_{20,00}^{40,0}}{3072}+\frac{7 \sqrt{7} C_{20,20}^{40,0}}{9216}-\frac{\sqrt{\frac{7}{5}} C_{20,22}^{42,0}}{2304}-\frac{7
\sqrt{\frac{7}{15}} C_{22,00}^{42,0}}{3072}-\frac{7 \sqrt{\frac{7}{5}} C_{22,20}^{42,0}}{9216}-\frac{\sqrt{35} C_{22,22}^{40,0}}{4608}+\frac{7 C_{22,22}^{42,0}}{2304
\sqrt{5}}+\frac{7 \sqrt{\frac{7}{3}} C_{31,00}^{31,0}}{3072}+\frac{7 \sqrt{7} C_{31,20}^{31,0}}{9216}-\frac{\sqrt{\frac{7}{5}} C_{31,22}^{31,0}}{2304}-\frac{3
\sqrt{3} C_{31,22}^{33,0}}{5120}-\frac{21 C_{33,00}^{33,0}}{5120}-\frac{7 \sqrt{3} C_{33,20}^{33,0}}{5120}+\frac{3 C_{33,22}^{33,0}}{320 \sqrt{10}}\)

\noindent\(C_{22,1101,1}^{00,3101,0,\text{Loc}}= \frac{7 C_{00,00}^{60,0}}{32 \sqrt{5}}+\frac{7 C_{00,00}^{60,1}}{64 \sqrt{5}}+\frac{7
C_{11,00}^{51,0}}{96 \sqrt{5}}+\frac{7 C_{11,00}^{51,1}}{192 \sqrt{5}}-\frac{7 C_{20,00}^{40,0}}{96 \sqrt{5}}-\frac{7 C_{20,00}^{40,1}}{192 \sqrt{5}}+\frac{7
C_{22,00}^{42,0}}{480}+\frac{7 C_{22,00}^{42,1}}{960}-\frac{7 C_{31,00}^{31,0}}{96 \sqrt{5}}-\frac{7 C_{31,00}^{31,1}}{192 \sqrt{5}}+\frac{3}{160}
\sqrt{\frac{21}{5}} C_{33,00}^{33,0}+\frac{3}{320} \sqrt{\frac{21}{5}} C_{33,00}^{33,1}\)

\noindent\(C_{22,1101,1}^{00,3101,0,\text{NonLoc}}= -\frac{7 C_{00,00}^{60,0}}{128 \sqrt{5}}-\frac{7 C_{00,00}^{60,1}}{64 \sqrt{5}}+\frac{7}{128}
\sqrt{\frac{3}{5}} C_{00,20}^{60,0}+\frac{7}{64} \sqrt{\frac{3}{5}} C_{00,20}^{60,1}+\frac{7 C_{11,00}^{51,0}}{384 \sqrt{5}}+\frac{7 C_{11,00}^{51,1}}{192
\sqrt{5}}+\frac{7 C_{20,00}^{40,0}}{384 \sqrt{5}}+\frac{7 C_{20,00}^{40,1}}{192 \sqrt{5}}-\frac{7 C_{20,20}^{40,0}}{128 \sqrt{15}}-\frac{7 C_{20,20}^{40,1}}{64
\sqrt{15}}-\frac{7 C_{22,00}^{42,0}}{1920}-\frac{7 C_{22,00}^{42,1}}{960}+\frac{7 C_{22,20}^{42,0}}{640 \sqrt{3}}+\frac{7 C_{22,20}^{42,1}}{320 \sqrt{3}}-\frac{7
C_{31,00}^{31,0}}{384 \sqrt{5}}-\frac{7 C_{31,00}^{31,1}}{192 \sqrt{5}}+\frac{7 C_{31,20}^{31,0}}{128 \sqrt{15}}+\frac{7 C_{31,20}^{31,1}}{64 \sqrt{15}}+\frac{3}{640}
\sqrt{\frac{21}{5}} C_{33,00}^{33,0}+\frac{3}{320} \sqrt{\frac{21}{5}} C_{33,00}^{33,1}-\frac{9}{640} \sqrt{\frac{7}{5}} C_{33,20}^{33,0}-\frac{9}{320}
\sqrt{\frac{7}{5}} C_{33,20}^{33,1}\)

\noindent\(C_{22,1101,1}^{00,3101,1,\text{Loc}}= \frac{7}{64} \sqrt{\frac{3}{5}} C_{00,00}^{60,1}+\frac{7 C_{11,00}^{51,1}}{64 \sqrt{15}}-\frac{7
C_{20,00}^{40,1}}{64 \sqrt{15}}+\frac{7 C_{22,00}^{42,1}}{320 \sqrt{3}}-\frac{7 C_{31,00}^{31,1}}{64 \sqrt{15}}+\frac{9}{320} \sqrt{\frac{7}{5}}
C_{33,00}^{33,1}\)

\noindent\(C_{22,1101,1}^{00,3101,1,\text{NonLoc}}= -\frac{7}{128} \sqrt{\frac{3}{5}} C_{00,00}^{60,0}+\frac{21 C_{00,20}^{60,0}}{128
\sqrt{5}}+\frac{7 C_{11,00}^{51,0}}{128 \sqrt{15}}+\frac{7 C_{20,00}^{40,0}}{128 \sqrt{15}}-\frac{7 C_{20,20}^{40,0}}{128 \sqrt{5}}-\frac{7 C_{22,00}^{42,0}}{640
\sqrt{3}}+\frac{7 C_{22,20}^{42,0}}{640}-\frac{7 C_{31,00}^{31,0}}{128 \sqrt{15}}+\frac{7 C_{31,20}^{31,0}}{128 \sqrt{5}}+\frac{9}{640} \sqrt{\frac{7}{5}}
C_{33,00}^{33,0}-\frac{9}{640} \sqrt{\frac{21}{5}} C_{33,20}^{33,0}\)

\noindent\(C_{22,1101,1}^{11,2011,0,\text{Loc}}= -\frac{7 \sqrt{5} C_{11,11}^{51,0}}{1152}-\frac{7 \sqrt{5} C_{11,11}^{51,1}}{2304}-\frac{\sqrt{5}
C_{31,11}^{31,0}}{384}-\frac{\sqrt{5} C_{31,11}^{31,1}}{768}+\frac{1}{32} \sqrt{\frac{7}{10}} C_{33,11}^{33,0}+\frac{1}{64} \sqrt{\frac{7}{10}} C_{33,11}^{33,1}\)

\noindent\(C_{22,1101,1}^{11,2011,0,\text{NonLoc}}= -\frac{7 \sqrt{5} C_{11,11}^{51,0}}{2304}-\frac{7 \sqrt{5} C_{11,11}^{51,1}}{1152}-\frac{\sqrt{5}
C_{31,11}^{31,0}}{768}-\frac{\sqrt{5} C_{31,11}^{31,1}}{384}+\frac{1}{64} \sqrt{\frac{7}{10}} C_{33,11}^{33,0}+\frac{1}{32} \sqrt{\frac{7}{10}} C_{33,11}^{33,1}\)

\noindent\(C_{22,1101,1}^{11,2011,1,\text{Loc}}= -\frac{7}{768} \sqrt{\frac{5}{3}} C_{11,11}^{51,1}-\frac{1}{256} \sqrt{\frac{5}{3}}
C_{31,11}^{31,1}+\frac{1}{64} \sqrt{\frac{21}{10}} C_{33,11}^{33,1}\)

\noindent\(C_{22,1101,1}^{11,2011,1,\text{NonLoc}}= -\frac{7}{768} \sqrt{\frac{5}{3}} C_{11,11}^{51,0}-\frac{1}{256} \sqrt{\frac{5}{3}}
C_{31,11}^{31,0}+\frac{1}{64} \sqrt{\frac{21}{10}} C_{33,11}^{33,0}\)

\noindent\(C_{22,1101,1}^{11,2211,0,\text{Loc}}= \frac{7 C_{11,11}^{51,0}}{1152}+\frac{7 C_{11,11}^{51,1}}{2304}+\frac{C_{31,11}^{31,0}}{384}+\frac{C_{31,11}^{31,1}}{768}-\frac{1}{160}
\sqrt{\frac{7}{2}} C_{33,11}^{33,0}-\frac{1}{320} \sqrt{\frac{7}{2}} C_{33,11}^{33,1}\)

\noindent\(C_{22,1101,1}^{11,2211,0,\text{NonLoc}}= \frac{7 C_{11,11}^{51,0}}{2304}+\frac{7 C_{11,11}^{51,1}}{1152}+\frac{C_{31,11}^{31,0}}{768}+\frac{C_{31,11}^{31,1}}{384}-\frac{1}{320}
\sqrt{\frac{7}{2}} C_{33,11}^{33,0}-\frac{1}{160} \sqrt{\frac{7}{2}} C_{33,11}^{33,1}\)

\noindent\(C_{22,1101,1}^{11,2211,1,\text{Loc}}= \frac{7 C_{11,11}^{51,1}}{768 \sqrt{3}}+\frac{C_{31,11}^{31,1}}{256 \sqrt{3}}-\frac{1}{320}
\sqrt{\frac{21}{2}} C_{33,11}^{33,1}\)

\noindent\(C_{22,1101,1}^{11,2211,1,\text{NonLoc}}= \frac{7 C_{11,11}^{51,0}}{768 \sqrt{3}}+\frac{C_{31,11}^{31,0}}{256 \sqrt{3}}-\frac{1}{320}
\sqrt{\frac{21}{2}} C_{33,11}^{33,0}\)

\noindent\(C_{22,1101,1}^{11,2212,0,\text{Loc}}= -\frac{35 C_{11,11}^{51,0}}{384 \sqrt{3}}-\frac{35 C_{11,11}^{51,1}}{768 \sqrt{3}}+\frac{11
C_{31,11}^{31,0}}{128 \sqrt{3}}+\frac{11 C_{31,11}^{31,1}}{256 \sqrt{3}}-\frac{1}{160} \sqrt{\frac{21}{2}} C_{33,11}^{33,0}-\frac{1}{320} \sqrt{\frac{21}{2}}
C_{33,11}^{33,1}\)

\noindent\(C_{22,1101,1}^{11,2212,0,\text{NonLoc}}= -\frac{35 C_{11,11}^{51,0}}{768 \sqrt{3}}-\frac{35 C_{11,11}^{51,1}}{384 \sqrt{3}}+\frac{11
C_{31,11}^{31,0}}{256 \sqrt{3}}+\frac{11 C_{31,11}^{31,1}}{128 \sqrt{3}}-\frac{1}{320} \sqrt{\frac{21}{2}} C_{33,11}^{33,0}-\frac{1}{160} \sqrt{\frac{21}{2}}
C_{33,11}^{33,1}\)

\noindent\(C_{22,1101,1}^{11,2212,1,\text{Loc}}= -\frac{35 C_{11,11}^{51,1}}{768}+\frac{11 C_{31,11}^{31,1}}{256}-\frac{3}{320} \sqrt{\frac{7}{2}}
C_{33,11}^{33,1}\)

\noindent\(C_{22,1101,1}^{11,2212,1,\text{NonLoc}}= -\frac{35 C_{11,11}^{51,0}}{768}+\frac{11 C_{31,11}^{31,0}}{256}-\frac{3}{320}
\sqrt{\frac{7}{2}} C_{33,11}^{33,0}\)

\noindent\(C_{22,1101,1}^{20,1101,0,\text{Loc}}= -\frac{7 \sqrt{5} C_{00,00}^{60,0}}{384}-\frac{7 \sqrt{5} C_{00,00}^{60,1}}{768}+\frac{7
\sqrt{5} C_{11,00}^{51,0}}{1152}+\frac{7 \sqrt{5} C_{11,00}^{51,1}}{2304}+\frac{7 \sqrt{5} C_{20,00}^{40,0}}{1152}+\frac{7 \sqrt{5} C_{20,00}^{40,1}}{2304}-\frac{7
C_{22,00}^{42,0}}{1152}-\frac{7 C_{22,00}^{42,1}}{2304}-\frac{7 \sqrt{5} C_{31,00}^{31,0}}{1152}-\frac{7 \sqrt{5} C_{31,00}^{31,1}}{2304}+\frac{1}{128}
\sqrt{\frac{21}{5}} C_{33,00}^{33,0}+\frac{1}{256} \sqrt{\frac{21}{5}} C_{33,00}^{33,1}\)

\noindent\(C_{22,1101,1}^{20,1101,0,\text{NonLoc}}= \frac{7 \sqrt{5} C_{00,00}^{60,0}}{1536}+\frac{7 \sqrt{5} C_{00,00}^{60,1}}{768}-\frac{7}{512}
\sqrt{\frac{5}{3}} C_{00,20}^{60,0}-\frac{7}{256} \sqrt{\frac{5}{3}} C_{00,20}^{60,1}+\frac{7 \sqrt{5} C_{11,00}^{51,0}}{4608}+\frac{7 \sqrt{5} C_{11,00}^{51,1}}{2304}-\frac{7
\sqrt{5} C_{20,00}^{40,0}}{4608}-\frac{7 \sqrt{5} C_{20,00}^{40,1}}{2304}+\frac{7 \sqrt{\frac{5}{3}} C_{20,20}^{40,0}}{1536}+\frac{7}{768} \sqrt{\frac{5}{3}}
C_{20,20}^{40,1}+\frac{7 C_{22,00}^{42,0}}{4608}+\frac{7 C_{22,00}^{42,1}}{2304}-\frac{7 C_{22,20}^{42,0}}{1536 \sqrt{3}}-\frac{7 C_{22,20}^{42,1}}{768
\sqrt{3}}-\frac{7 \sqrt{5} C_{31,00}^{31,0}}{4608}-\frac{7 \sqrt{5} C_{31,00}^{31,1}}{2304}+\frac{7 \sqrt{\frac{5}{3}} C_{31,20}^{31,0}}{1536}+\frac{7}{768}
\sqrt{\frac{5}{3}} C_{31,20}^{31,1}+\frac{1}{512} \sqrt{\frac{21}{5}} C_{33,00}^{33,0}+\frac{1}{256} \sqrt{\frac{21}{5}} C_{33,00}^{33,1}-\frac{3}{512}
\sqrt{\frac{7}{5}} C_{33,20}^{33,0}-\frac{3}{256} \sqrt{\frac{7}{5}} C_{33,20}^{33,1}\)

\noindent\(C_{22,1101,1}^{20,1101,1,\text{Loc}}= -\frac{7}{256} \sqrt{\frac{5}{3}} C_{00,00}^{60,1}+\frac{7}{768} \sqrt{\frac{5}{3}}
C_{11,00}^{51,1}+\frac{7}{768} \sqrt{\frac{5}{3}} C_{20,00}^{40,1}-\frac{7 C_{22,00}^{42,1}}{768 \sqrt{3}}-\frac{7}{768} \sqrt{\frac{5}{3}} C_{31,00}^{31,1}+\frac{3}{256}
\sqrt{\frac{7}{5}} C_{33,00}^{33,1}\)

\noindent\(C_{22,1101,1}^{20,1101,1,\text{NonLoc}}= \frac{7}{512} \sqrt{\frac{5}{3}} C_{00,00}^{60,0}-\frac{7 \sqrt{5} C_{00,20}^{60,0}}{512}+\frac{7
\sqrt{\frac{5}{3}} C_{11,00}^{51,0}}{1536}-\frac{7 \sqrt{\frac{5}{3}} C_{20,00}^{40,0}}{1536}+\frac{7 \sqrt{5} C_{20,20}^{40,0}}{1536}+\frac{7 C_{22,00}^{42,0}}{1536
\sqrt{3}}-\frac{7 C_{22,20}^{42,0}}{1536}-\frac{7 \sqrt{\frac{5}{3}} C_{31,00}^{31,0}}{1536}+\frac{7 \sqrt{5} C_{31,20}^{31,0}}{1536}+\frac{3}{512}
\sqrt{\frac{7}{5}} C_{33,00}^{33,0}-\frac{3}{512} \sqrt{\frac{21}{5}} C_{33,20}^{33,0}\)

\noindent\(C_{22,1101,1}^{22,1101,0,\text{Loc}}= -\frac{7 C_{00,00}^{60,0}}{384}-\frac{7 C_{00,00}^{60,1}}{768}+\frac{7 C_{11,00}^{51,0}}{1152}+\frac{7
C_{11,00}^{51,1}}{2304}+\frac{C_{20,00}^{40,0}}{1152}+\frac{C_{20,00}^{40,1}}{2304}+\frac{7 C_{22,00}^{42,0}}{2304 \sqrt{5}}+\frac{7 C_{22,00}^{42,1}}{4608
\sqrt{5}}-\frac{C_{31,00}^{31,0}}{288}-\frac{C_{31,00}^{31,1}}{576}-\frac{\sqrt{21} C_{33,00}^{33,0}}{1280}-\frac{\sqrt{21} C_{33,00}^{33,1}}{2560}\)

\noindent\(C_{22,1101,1}^{22,1101,0,\text{NonLoc}}= \frac{7 C_{00,00}^{60,0}}{1536}+\frac{7 C_{00,00}^{60,1}}{768}-\frac{7 C_{00,20}^{60,0}}{512
\sqrt{3}}-\frac{7 C_{00,20}^{60,1}}{256 \sqrt{3}}+\frac{7 C_{11,00}^{51,0}}{4608}+\frac{7 C_{11,00}^{51,1}}{2304}-\frac{C_{20,00}^{40,0}}{4608}-\frac{C_{20,00}^{40,1}}{2304}+\frac{C_{20,20}^{40,0}}{1536
\sqrt{3}}+\frac{C_{20,20}^{40,1}}{768 \sqrt{3}}-\frac{7 C_{22,00}^{42,0}}{9216 \sqrt{5}}-\frac{7 C_{22,00}^{42,1}}{4608 \sqrt{5}}+\frac{7 C_{22,20}^{42,0}}{3072
\sqrt{15}}+\frac{7 C_{22,20}^{42,1}}{1536 \sqrt{15}}-\frac{C_{31,00}^{31,0}}{1152}-\frac{C_{31,00}^{31,1}}{576}+\frac{C_{31,20}^{31,0}}{384 \sqrt{3}}+\frac{C_{31,20}^{31,1}}{192
\sqrt{3}}-\frac{\sqrt{21} C_{33,00}^{33,0}}{5120}-\frac{\sqrt{21} C_{33,00}^{33,1}}{2560}+\frac{3 \sqrt{7} C_{33,20}^{33,0}}{5120}+\frac{3 \sqrt{7}
C_{33,20}^{33,1}}{2560}\)

\noindent\(C_{22,1101,1}^{22,1101,1,\text{Loc}}= -\frac{7 C_{00,00}^{60,1}}{256 \sqrt{3}}+\frac{7 C_{11,00}^{51,1}}{768 \sqrt{3}}+\frac{C_{20,00}^{40,1}}{768
\sqrt{3}}+\frac{7 C_{22,00}^{42,1}}{1536 \sqrt{15}}-\frac{C_{31,00}^{31,1}}{192 \sqrt{3}}-\frac{3 \sqrt{7} C_{33,00}^{33,1}}{2560}\)

\noindent\(C_{22,1101,1}^{22,1101,1,\text{NonLoc}}= \frac{7 C_{00,00}^{60,0}}{512 \sqrt{3}}-\frac{7 C_{00,20}^{60,0}}{512}+\frac{7
C_{11,00}^{51,0}}{1536 \sqrt{3}}-\frac{C_{20,00}^{40,0}}{1536 \sqrt{3}}+\frac{C_{20,20}^{40,0}}{1536}-\frac{7 C_{22,00}^{42,0}}{3072 \sqrt{15}}+\frac{7
C_{22,20}^{42,0}}{3072 \sqrt{5}}-\frac{C_{31,00}^{31,0}}{384 \sqrt{3}}+\frac{C_{31,20}^{31,0}}{384}-\frac{3 \sqrt{7} C_{33,00}^{33,0}}{5120}+\frac{3
\sqrt{21} C_{33,20}^{33,0}}{5120}\)

\noindent\(C_{22,1101,2}^{11,2011,0,\text{Loc}}= -\frac{7}{384} \sqrt{\frac{5}{3}} C_{11,11}^{51,0}-\frac{7}{768} \sqrt{\frac{5}{3}}
C_{11,11}^{51,1}-\frac{1}{128} \sqrt{\frac{5}{3}} C_{31,11}^{31,0}-\frac{1}{256} \sqrt{\frac{5}{3}} C_{31,11}^{31,1}+\frac{1}{32} \sqrt{\frac{21}{10}}
C_{33,11}^{33,0}+\frac{1}{64} \sqrt{\frac{21}{10}} C_{33,11}^{33,1}\)

\noindent\(C_{22,1101,2}^{11,2011,0,\text{NonLoc}}= -\frac{7}{768} \sqrt{\frac{5}{3}} C_{11,11}^{51,0}-\frac{7}{384} \sqrt{\frac{5}{3}}
C_{11,11}^{51,1}-\frac{1}{256} \sqrt{\frac{5}{3}} C_{31,11}^{31,0}-\frac{1}{128} \sqrt{\frac{5}{3}} C_{31,11}^{31,1}+\frac{1}{64} \sqrt{\frac{21}{10}}
C_{33,11}^{33,0}+\frac{1}{32} \sqrt{\frac{21}{10}} C_{33,11}^{33,1}\)

\noindent\(C_{22,1101,2}^{11,2011,1,\text{Loc}}= -\frac{7 \sqrt{5} C_{11,11}^{51,1}}{768}-\frac{\sqrt{5} C_{31,11}^{31,1}}{256}+\frac{3}{64}
\sqrt{\frac{7}{10}} C_{33,11}^{33,1}\)

\noindent\(C_{22,1101,2}^{11,2011,1,\text{NonLoc}}= -\frac{7 \sqrt{5} C_{11,11}^{51,0}}{768}-\frac{\sqrt{5} C_{31,11}^{31,0}}{256}+\frac{3}{64}
\sqrt{\frac{7}{10}} C_{33,11}^{33,0}\)

\noindent\(C_{22,1101,2}^{11,2211,0,\text{Loc}}= \frac{7 C_{11,11}^{51,0}}{384 \sqrt{3}}+\frac{7 C_{11,11}^{51,1}}{768 \sqrt{3}}+\frac{C_{31,11}^{31,0}}{128
\sqrt{3}}+\frac{C_{31,11}^{31,1}}{256 \sqrt{3}}-\frac{1}{160} \sqrt{\frac{21}{2}} C_{33,11}^{33,0}-\frac{1}{320} \sqrt{\frac{21}{2}} C_{33,11}^{33,1}\)

\noindent\(C_{22,1101,2}^{11,2211,0,\text{NonLoc}}= \frac{7 C_{11,11}^{51,0}}{768 \sqrt{3}}+\frac{7 C_{11,11}^{51,1}}{384 \sqrt{3}}+\frac{C_{31,11}^{31,0}}{256
\sqrt{3}}+\frac{C_{31,11}^{31,1}}{128 \sqrt{3}}-\frac{1}{320} \sqrt{\frac{21}{2}} C_{33,11}^{33,0}-\frac{1}{160} \sqrt{\frac{21}{2}} C_{33,11}^{33,1}\)

\noindent\(C_{22,1101,2}^{11,2211,1,\text{Loc}}= \frac{7 C_{11,11}^{51,1}}{768}+\frac{C_{31,11}^{31,1}}{256}-\frac{3}{320} \sqrt{\frac{7}{2}}
C_{33,11}^{33,1}\)

\noindent\(C_{22,1101,2}^{11,2211,1,\text{NonLoc}}= \frac{7 C_{11,11}^{51,0}}{768}+\frac{C_{31,11}^{31,0}}{256}-\frac{3}{320} \sqrt{\frac{7}{2}}
C_{33,11}^{33,0}\)

\noindent\(C_{22,1101,2}^{11,2212,0,\text{Loc}}= \frac{7 \sqrt{5} C_{11,11}^{51,0}}{1152}+\frac{7 \sqrt{5} C_{11,11}^{51,1}}{2304}-\frac{13
\sqrt{5} C_{31,11}^{31,0}}{1152}-\frac{13 \sqrt{5} C_{31,11}^{31,1}}{2304}+\frac{1}{32} \sqrt{\frac{7}{10}} C_{33,11}^{33,0}+\frac{1}{64} \sqrt{\frac{7}{10}}
C_{33,11}^{33,1}\)

\noindent\(C_{22,1101,2}^{11,2212,0,\text{NonLoc}}= \frac{7 \sqrt{5} C_{11,11}^{51,0}}{2304}+\frac{7 \sqrt{5} C_{11,11}^{51,1}}{1152}-\frac{13
\sqrt{5} C_{31,11}^{31,0}}{2304}-\frac{13 \sqrt{5} C_{31,11}^{31,1}}{1152}+\frac{1}{64} \sqrt{\frac{7}{10}} C_{33,11}^{33,0}+\frac{1}{32} \sqrt{\frac{7}{10}}
C_{33,11}^{33,1}\)

\noindent\(C_{22,1101,2}^{11,2212,1,\text{Loc}}= \frac{7}{768} \sqrt{\frac{5}{3}} C_{11,11}^{51,1}-\frac{13}{768} \sqrt{\frac{5}{3}}
C_{31,11}^{31,1}+\frac{1}{64} \sqrt{\frac{21}{10}} C_{33,11}^{33,1}\)

\noindent\(C_{22,1101,2}^{11,2212,1,\text{NonLoc}}= \frac{7}{768} \sqrt{\frac{5}{3}} C_{11,11}^{51,0}-\frac{13}{768} \sqrt{\frac{5}{3}}
C_{31,11}^{31,0}+\frac{1}{64} \sqrt{\frac{21}{10}} C_{33,11}^{33,0}\)

\noindent\(C_{22,1101,2}^{11,2213,0,\text{Loc}}= -\frac{\sqrt{7} C_{11,11}^{51,0}}{144}-\frac{\sqrt{7} C_{11,11}^{51,1}}{288}+\frac{\sqrt{7}
C_{31,11}^{31,0}}{144}+\frac{\sqrt{7} C_{31,11}^{31,1}}{288}-\frac{C_{33,11}^{33,0}}{80 \sqrt{2}}-\frac{C_{33,11}^{33,1}}{160 \sqrt{2}}\)

\noindent\(C_{22,1101,2}^{11,2213,0,\text{NonLoc}}= -\frac{\sqrt{7} C_{11,11}^{51,0}}{288}-\frac{\sqrt{7} C_{11,11}^{51,1}}{144}+\frac{\sqrt{7}
C_{31,11}^{31,0}}{288}+\frac{\sqrt{7} C_{31,11}^{31,1}}{144}-\frac{C_{33,11}^{33,0}}{160 \sqrt{2}}-\frac{C_{33,11}^{33,1}}{80 \sqrt{2}}\)

\noindent\(C_{22,1101,2}^{11,2213,1,\text{Loc}}= -\frac{1}{96} \sqrt{\frac{7}{3}} C_{11,11}^{51,1}+\frac{1}{96} \sqrt{\frac{7}{3}}
C_{31,11}^{31,1}-\frac{1}{160} \sqrt{\frac{3}{2}} C_{33,11}^{33,1}\)

\noindent\(C_{22,1101,2}^{11,2213,1,\text{NonLoc}}= -\frac{1}{96} \sqrt{\frac{7}{3}} C_{11,11}^{51,0}+\frac{1}{96} \sqrt{\frac{7}{3}}
C_{31,11}^{31,0}-\frac{1}{160} \sqrt{\frac{3}{2}} C_{33,11}^{33,0}\)

\noindent\(C_{22,1101,2}^{22,1101,0,\text{Loc}}= -\frac{1}{192} \sqrt{\frac{5}{3}} C_{20,00}^{40,0}-\frac{1}{384} \sqrt{\frac{5}{3}}
C_{20,00}^{40,1}+\frac{7 C_{22,00}^{42,0}}{768 \sqrt{3}}+\frac{7 C_{22,00}^{42,1}}{1536 \sqrt{3}}+\frac{1}{384} \sqrt{\frac{5}{3}} C_{31,00}^{31,0}+\frac{1}{768}
\sqrt{\frac{5}{3}} C_{31,00}^{31,1}-\frac{3}{256} \sqrt{\frac{7}{5}} C_{33,00}^{33,0}-\frac{3}{512} \sqrt{\frac{7}{5}} C_{33,00}^{33,1}\)

\noindent\(C_{22,1101,2}^{22,1101,0,\text{NonLoc}}= \frac{1}{768} \sqrt{\frac{5}{3}} C_{20,00}^{40,0}+\frac{1}{384} \sqrt{\frac{5}{3}}
C_{20,00}^{40,1}-\frac{\sqrt{5} C_{20,20}^{40,0}}{768}-\frac{\sqrt{5} C_{20,20}^{40,1}}{384}-\frac{7 C_{22,00}^{42,0}}{3072 \sqrt{3}}-\frac{7 C_{22,00}^{42,1}}{1536
\sqrt{3}}+\frac{7 C_{22,20}^{42,0}}{3072}+\frac{7 C_{22,20}^{42,1}}{1536}+\frac{\sqrt{\frac{5}{3}} C_{31,00}^{31,0}}{1536}+\frac{1}{768} \sqrt{\frac{5}{3}}
C_{31,00}^{31,1}-\frac{\sqrt{5} C_{31,20}^{31,0}}{1536}-\frac{\sqrt{5} C_{31,20}^{31,1}}{768}-\frac{3 \sqrt{\frac{7}{5}} C_{33,00}^{33,0}}{1024}-\frac{3}{512}
\sqrt{\frac{7}{5}} C_{33,00}^{33,1}+\frac{3 \sqrt{\frac{21}{5}} C_{33,20}^{33,0}}{1024}+\frac{3}{512} \sqrt{\frac{21}{5}} C_{33,20}^{33,1}\)

\noindent\(C_{22,1101,2}^{22,1101,1,\text{Loc}}= -\frac{\sqrt{5} C_{20,00}^{40,1}}{384}+\frac{7 C_{22,00}^{42,1}}{1536}+\frac{\sqrt{5}
C_{31,00}^{31,1}}{768}-\frac{3}{512} \sqrt{\frac{21}{5}} C_{33,00}^{33,1}\)

\noindent\(C_{22,1101,2}^{22,1101,1,\text{NonLoc}}= \frac{\sqrt{5} C_{20,00}^{40,0}}{768}-\frac{1}{256} \sqrt{\frac{5}{3}} C_{20,20}^{40,0}-\frac{7
C_{22,00}^{42,0}}{3072}+\frac{7 C_{22,20}^{42,0}}{1024 \sqrt{3}}+\frac{\sqrt{5} C_{31,00}^{31,0}}{1536}-\frac{1}{512} \sqrt{\frac{5}{3}} C_{31,20}^{31,0}-\frac{3
\sqrt{\frac{21}{5}} C_{33,00}^{33,0}}{1024}+\frac{9 \sqrt{\frac{7}{5}} C_{33,20}^{33,0}}{1024}\)

\noindent\(C_{22,1101,3}^{00,3303,0,\text{Loc}}= \frac{1}{32} \sqrt{\frac{7}{3}} C_{00,00}^{60,0}+\frac{1}{64} \sqrt{\frac{7}{3}} C_{00,00}^{60,1}+\frac{1}{96}
\sqrt{\frac{7}{3}} C_{11,00}^{51,0}+\frac{1}{192} \sqrt{\frac{7}{3}} C_{11,00}^{51,1}-\frac{1}{96} \sqrt{\frac{7}{3}} C_{20,00}^{40,0}-\frac{1}{192}
\sqrt{\frac{7}{3}} C_{20,00}^{40,1}+\frac{1}{96} \sqrt{\frac{7}{15}} C_{22,00}^{42,0}+\frac{1}{192} \sqrt{\frac{7}{15}} C_{22,00}^{42,1}-\frac{1}{96}
\sqrt{\frac{7}{3}} C_{31,00}^{31,0}-\frac{1}{192} \sqrt{\frac{7}{3}} C_{31,00}^{31,1}+\frac{3 C_{33,00}^{33,0}}{160}+\frac{3 C_{33,00}^{33,1}}{320}\)

\noindent\(C_{22,1101,3}^{00,3303,0,\text{NonLoc}}= -\frac{1}{128} \sqrt{\frac{7}{3}} C_{00,00}^{60,0}-\frac{1}{64} \sqrt{\frac{7}{3}}
C_{00,00}^{60,1}+\frac{\sqrt{7} C_{00,20}^{60,0}}{128}+\frac{\sqrt{7} C_{00,20}^{60,1}}{64}+\frac{1}{384} \sqrt{\frac{7}{3}} C_{11,00}^{51,0}+\frac{1}{192}
\sqrt{\frac{7}{3}} C_{11,00}^{51,1}+\frac{1}{384} \sqrt{\frac{7}{3}} C_{20,00}^{40,0}+\frac{1}{192} \sqrt{\frac{7}{3}} C_{20,00}^{40,1}-\frac{\sqrt{7}
C_{20,20}^{40,0}}{384}-\frac{\sqrt{7} C_{20,20}^{40,1}}{192}-\frac{1}{384} \sqrt{\frac{7}{15}} C_{22,00}^{42,0}-\frac{1}{192} \sqrt{\frac{7}{15}}
C_{22,00}^{42,1}+\frac{1}{384} \sqrt{\frac{7}{5}} C_{22,20}^{42,0}+\frac{1}{192} \sqrt{\frac{7}{5}} C_{22,20}^{42,1}-\frac{1}{384} \sqrt{\frac{7}{3}}
C_{31,00}^{31,0}-\frac{1}{192} \sqrt{\frac{7}{3}} C_{31,00}^{31,1}+\frac{\sqrt{7} C_{31,20}^{31,0}}{384}+\frac{\sqrt{7} C_{31,20}^{31,1}}{192}+\frac{3
C_{33,00}^{33,0}}{640}+\frac{3 C_{33,00}^{33,1}}{320}-\frac{3 \sqrt{3} C_{33,20}^{33,0}}{640}-\frac{3 \sqrt{3} C_{33,20}^{33,1}}{320}\)

\noindent\(C_{22,1101,3}^{00,3303,1,\text{Loc}}= \frac{\sqrt{7} C_{00,00}^{60,1}}{64}+\frac{\sqrt{7} C_{11,00}^{51,1}}{192}-\frac{\sqrt{7}
C_{20,00}^{40,1}}{192}+\frac{1}{192} \sqrt{\frac{7}{5}} C_{22,00}^{42,1}-\frac{\sqrt{7} C_{31,00}^{31,1}}{192}+\frac{3 \sqrt{3} C_{33,00}^{33,1}}{320}\)

\noindent\(C_{22,1101,3}^{00,3303,1,\text{NonLoc}}= -\frac{\sqrt{7} C_{00,00}^{60,0}}{128}+\frac{\sqrt{21} C_{00,20}^{60,0}}{128}+\frac{\sqrt{7}
C_{11,00}^{51,0}}{384}+\frac{\sqrt{7} C_{20,00}^{40,0}}{384}-\frac{1}{128} \sqrt{\frac{7}{3}} C_{20,20}^{40,0}-\frac{1}{384} \sqrt{\frac{7}{5}} C_{22,00}^{42,0}+\frac{1}{128}
\sqrt{\frac{7}{15}} C_{22,20}^{42,0}-\frac{\sqrt{7} C_{31,00}^{31,0}}{384}+\frac{1}{128} \sqrt{\frac{7}{3}} C_{31,20}^{31,0}+\frac{3 \sqrt{3} C_{33,00}^{33,0}}{640}-\frac{9
C_{33,20}^{33,0}}{640}\)

\noindent\(C_{22,1101,3}^{11,2212,0,\text{Loc}}= -\frac{5 \sqrt{7} C_{11,11}^{51,0}}{576}-\frac{5 \sqrt{7} C_{11,11}^{51,1}}{1152}-\frac{\sqrt{7}
C_{31,11}^{31,0}}{576}-\frac{\sqrt{7} C_{31,11}^{31,1}}{1152}+\frac{C_{33,11}^{33,0}}{20 \sqrt{2}}+\frac{C_{33,11}^{33,1}}{40 \sqrt{2}}\)

\noindent\(C_{22,1101,3}^{11,2212,0,\text{NonLoc}}= -\frac{5 \sqrt{7} C_{11,11}^{51,0}}{1152}-\frac{5 \sqrt{7} C_{11,11}^{51,1}}{576}-\frac{\sqrt{7}
C_{31,11}^{31,0}}{1152}-\frac{\sqrt{7} C_{31,11}^{31,1}}{576}+\frac{C_{33,11}^{33,0}}{40 \sqrt{2}}+\frac{C_{33,11}^{33,1}}{20 \sqrt{2}}\)

\noindent\(C_{22,1101,3}^{11,2212,1,\text{Loc}}= -\frac{5}{384} \sqrt{\frac{7}{3}} C_{11,11}^{51,1}-\frac{1}{384} \sqrt{\frac{7}{3}}
C_{31,11}^{31,1}+\frac{1}{40} \sqrt{\frac{3}{2}} C_{33,11}^{33,1}\)

\noindent\(C_{22,1101,3}^{11,2212,1,\text{NonLoc}}= -\frac{5}{384} \sqrt{\frac{7}{3}} C_{11,11}^{51,0}-\frac{1}{384} \sqrt{\frac{7}{3}}
C_{31,11}^{31,0}+\frac{1}{40} \sqrt{\frac{3}{2}} C_{33,11}^{33,0}\)

\noindent\(C_{22,1101,3}^{11,2213,0,\text{Loc}}= \frac{1}{288} \sqrt{\frac{7}{2}} C_{11,11}^{51,0}+\frac{1}{576} \sqrt{\frac{7}{2}}
C_{11,11}^{51,1}-\frac{1}{288} \sqrt{\frac{7}{2}} C_{31,11}^{31,0}-\frac{1}{576} \sqrt{\frac{7}{2}} C_{31,11}^{31,1}+\frac{C_{33,11}^{33,0}}{320}+\frac{C_{33,11}^{33,1}}{640}\)

\noindent\(C_{22,1101,3}^{11,2213,0,\text{NonLoc}}= \frac{1}{576} \sqrt{\frac{7}{2}} C_{11,11}^{51,0}+\frac{1}{288} \sqrt{\frac{7}{2}}
C_{11,11}^{51,1}-\frac{1}{576} \sqrt{\frac{7}{2}} C_{31,11}^{31,0}-\frac{1}{288} \sqrt{\frac{7}{2}} C_{31,11}^{31,1}+\frac{C_{33,11}^{33,0}}{640}+\frac{C_{33,11}^{33,1}}{320}\)

\noindent\(C_{22,1101,3}^{11,2213,1,\text{Loc}}= \frac{1}{192} \sqrt{\frac{7}{6}} C_{11,11}^{51,1}-\frac{1}{192} \sqrt{\frac{7}{6}}
C_{31,11}^{31,1}+\frac{\sqrt{3} C_{33,11}^{33,1}}{640}\)

\noindent\(C_{22,1101,3}^{11,2213,1,\text{NonLoc}}= \frac{1}{192} \sqrt{\frac{7}{6}} C_{11,11}^{51,0}-\frac{1}{192} \sqrt{\frac{7}{6}}
C_{31,11}^{31,0}+\frac{\sqrt{3} C_{33,11}^{33,0}}{640}\)

\noindent\(C_{22,1101,3}^{22,1101,0,\text{Loc}}= -\frac{\sqrt{21} C_{00,00}^{60,0}}{256}-\frac{\sqrt{21} C_{00,00}^{60,1}}{512}+\frac{1}{256}
\sqrt{\frac{7}{3}} C_{11,00}^{51,0}+\frac{1}{512} \sqrt{\frac{7}{3}} C_{11,00}^{51,1}-\frac{1}{768} \sqrt{\frac{7}{3}} C_{20,00}^{40,0}-\frac{\sqrt{\frac{7}{3}}
C_{20,00}^{40,1}}{1536}+\frac{1}{192} \sqrt{\frac{7}{15}} C_{22,00}^{42,0}+\frac{1}{384} \sqrt{\frac{7}{15}} C_{22,00}^{42,1}-\frac{1}{768} \sqrt{\frac{7}{3}}
C_{31,00}^{31,0}-\frac{\sqrt{\frac{7}{3}} C_{31,00}^{31,1}}{1536}-\frac{3 C_{33,00}^{33,0}}{320}-\frac{3 C_{33,00}^{33,1}}{640}\)

\noindent\(C_{22,1101,3}^{22,1101,0,\text{NonLoc}}= \frac{\sqrt{21} C_{00,00}^{60,0}}{1024}+\frac{\sqrt{21} C_{00,00}^{60,1}}{512}-\frac{3
\sqrt{7} C_{00,20}^{60,0}}{1024}-\frac{3 \sqrt{7} C_{00,20}^{60,1}}{512}+\frac{\sqrt{\frac{7}{3}} C_{11,00}^{51,0}}{1024}+\frac{1}{512} \sqrt{\frac{7}{3}}
C_{11,00}^{51,1}+\frac{\sqrt{\frac{7}{3}} C_{20,00}^{40,0}}{3072}+\frac{\sqrt{\frac{7}{3}} C_{20,00}^{40,1}}{1536}-\frac{\sqrt{7} C_{20,20}^{40,0}}{3072}-\frac{\sqrt{7}
C_{20,20}^{40,1}}{1536}-\frac{1}{768} \sqrt{\frac{7}{15}} C_{22,00}^{42,0}-\frac{1}{384} \sqrt{\frac{7}{15}} C_{22,00}^{42,1}+\frac{1}{768} \sqrt{\frac{7}{5}}
C_{22,20}^{42,0}+\frac{1}{384} \sqrt{\frac{7}{5}} C_{22,20}^{42,1}-\frac{\sqrt{\frac{7}{3}} C_{31,00}^{31,0}}{3072}-\frac{\sqrt{\frac{7}{3}} C_{31,00}^{31,1}}{1536}+\frac{\sqrt{7}
C_{31,20}^{31,0}}{3072}+\frac{\sqrt{7} C_{31,20}^{31,1}}{1536}-\frac{3 C_{33,00}^{33,0}}{1280}-\frac{3 C_{33,00}^{33,1}}{640}+\frac{3 \sqrt{3} C_{33,20}^{33,0}}{1280}+\frac{3
\sqrt{3} C_{33,20}^{33,1}}{640}\)

\noindent\(C_{22,1101,3}^{22,1101,1,\text{Loc}}= -\frac{3 \sqrt{7} C_{00,00}^{60,1}}{512}+\frac{\sqrt{7} C_{11,00}^{51,1}}{512}-\frac{\sqrt{7}
C_{20,00}^{40,1}}{1536}+\frac{1}{384} \sqrt{\frac{7}{5}} C_{22,00}^{42,1}-\frac{\sqrt{7} C_{31,00}^{31,1}}{1536}-\frac{3 \sqrt{3} C_{33,00}^{33,1}}{640}\)

\noindent\(C_{22,1101,3}^{22,1101,1,\text{NonLoc}}= \frac{3 \sqrt{7} C_{00,00}^{60,0}}{1024}-\frac{3 \sqrt{21} C_{00,20}^{60,0}}{1024}+\frac{\sqrt{7}
C_{11,00}^{51,0}}{1024}+\frac{\sqrt{7} C_{20,00}^{40,0}}{3072}-\frac{\sqrt{\frac{7}{3}} C_{20,20}^{40,0}}{1024}-\frac{1}{768} \sqrt{\frac{7}{5}}
C_{22,00}^{42,0}+\frac{1}{256} \sqrt{\frac{7}{15}} C_{22,20}^{42,0}-\frac{\sqrt{7} C_{31,00}^{31,0}}{3072}+\frac{\sqrt{\frac{7}{3}} C_{31,20}^{31,0}}{1024}-\frac{3
\sqrt{3} C_{33,00}^{33,0}}{1280}+\frac{9 C_{33,20}^{33,0}}{1280}\)

\noindent\(C_{22,1110,2}^{00,3112,0,\text{Loc}}= -\frac{7 C_{00,20}^{60,0}}{96 \sqrt{5}}-\frac{7 C_{00,20}^{60,1}}{192 \sqrt{5}}-\frac{7
C_{00,22}^{62,0}}{160}-\frac{7 C_{00,22}^{62,1}}{320}-\frac{7 C_{11,22}^{51,0}}{2880}-\frac{7 C_{11,22}^{51,1}}{5760}-\frac{9}{640} \sqrt{\frac{21}{5}}
C_{11,22}^{53,0}-\frac{9 \sqrt{\frac{21}{5}} C_{11,22}^{53,1}}{1280}+\frac{7 C_{20,20}^{40,0}}{288 \sqrt{5}}+\frac{7 C_{20,20}^{40,1}}{576 \sqrt{5}}+\frac{7
C_{20,22}^{42,0}}{2880}+\frac{7 C_{20,22}^{42,1}}{5760}-\frac{7 C_{22,20}^{42,0}}{1440}-\frac{7 C_{22,20}^{42,1}}{2880}+\frac{13 C_{22,22}^{40,0}}{720}+\frac{13
C_{22,22}^{40,1}}{1440}-\frac{\sqrt{7} C_{22,22}^{42,0}}{720}-\frac{\sqrt{7} C_{22,22}^{42,1}}{1440}-\frac{3 C_{22,22}^{44,0}}{80 \sqrt{5}}-\frac{3
C_{22,22}^{44,1}}{160 \sqrt{5}}+\frac{7 C_{31,20}^{31,0}}{288 \sqrt{5}}+\frac{7 C_{31,20}^{31,1}}{576 \sqrt{5}}+\frac{37 C_{31,22}^{31,0}}{2880}+\frac{37
C_{31,22}^{31,1}}{5760}+\frac{7}{640} \sqrt{\frac{7}{15}} C_{31,22}^{33,0}+\frac{7 \sqrt{\frac{7}{15}} C_{31,22}^{33,1}}{1280}-\frac{1}{160} \sqrt{\frac{21}{5}}
C_{33,20}^{33,0}-\frac{1}{320} \sqrt{\frac{21}{5}} C_{33,20}^{33,1}-\frac{3}{400} \sqrt{\frac{7}{2}} C_{33,22}^{33,0}-\frac{3}{800} \sqrt{\frac{7}{2}}
C_{33,22}^{33,1}\)

\noindent\(C_{22,1110,2}^{00,3112,0,\text{NonLoc}}= -\frac{7 C_{00,00}^{60,0}}{128 \sqrt{15}}-\frac{7 C_{00,00}^{60,1}}{64 \sqrt{15}}-\frac{7
C_{00,20}^{60,0}}{384 \sqrt{5}}-\frac{7 C_{00,20}^{60,1}}{192 \sqrt{5}}+\frac{7 C_{00,22}^{62,0}}{320}+\frac{7 C_{00,22}^{62,1}}{160}+\frac{7 C_{11,00}^{51,0}}{384
\sqrt{15}}+\frac{7 C_{11,00}^{51,1}}{192 \sqrt{15}}-\frac{7 C_{11,22}^{51,0}}{5760}-\frac{7 C_{11,22}^{51,1}}{2880}-\frac{9 \sqrt{\frac{21}{5}} C_{11,22}^{53,0}}{1280}-\frac{9}{640}
\sqrt{\frac{21}{5}} C_{11,22}^{53,1}+\frac{7 C_{20,00}^{40,0}}{384 \sqrt{15}}+\frac{7 C_{20,00}^{40,1}}{192 \sqrt{15}}+\frac{7 C_{20,20}^{40,0}}{1152
\sqrt{5}}+\frac{7 C_{20,20}^{40,1}}{576 \sqrt{5}}-\frac{7 C_{20,22}^{42,0}}{5760}-\frac{7 C_{20,22}^{42,1}}{2880}-\frac{7 C_{22,00}^{42,0}}{1920
\sqrt{3}}-\frac{7 C_{22,00}^{42,1}}{960 \sqrt{3}}-\frac{7 C_{22,20}^{42,0}}{5760}-\frac{7 C_{22,20}^{42,1}}{2880}-\frac{13 C_{22,22}^{40,0}}{1440}-\frac{13
C_{22,22}^{40,1}}{720}+\frac{\sqrt{7} C_{22,22}^{42,0}}{1440}+\frac{\sqrt{7} C_{22,22}^{42,1}}{720}+\frac{3 C_{22,22}^{44,0}}{160 \sqrt{5}}+\frac{3
C_{22,22}^{44,1}}{80 \sqrt{5}}-\frac{7 C_{31,00}^{31,0}}{384 \sqrt{15}}-\frac{7 C_{31,00}^{31,1}}{192 \sqrt{15}}-\frac{7 C_{31,20}^{31,0}}{1152 \sqrt{5}}-\frac{7
C_{31,20}^{31,1}}{576 \sqrt{5}}+\frac{37 C_{31,22}^{31,0}}{5760}+\frac{37 C_{31,22}^{31,1}}{2880}+\frac{7 \sqrt{\frac{7}{15}} C_{31,22}^{33,0}}{1280}+\frac{7}{640}
\sqrt{\frac{7}{15}} C_{31,22}^{33,1}+\frac{3}{640} \sqrt{\frac{7}{5}} C_{33,00}^{33,0}+\frac{3}{320} \sqrt{\frac{7}{5}} C_{33,00}^{33,1}+\frac{1}{640}
\sqrt{\frac{21}{5}} C_{33,20}^{33,0}+\frac{1}{320} \sqrt{\frac{21}{5}} C_{33,20}^{33,1}-\frac{3}{800} \sqrt{\frac{7}{2}} C_{33,22}^{33,0}-\frac{3}{400}
\sqrt{\frac{7}{2}} C_{33,22}^{33,1}\)

\noindent\(C_{22,1110,2}^{00,3112,1,\text{Loc}}= -\frac{7 C_{00,20}^{60,1}}{64 \sqrt{15}}-\frac{7 \sqrt{3} C_{00,22}^{62,1}}{320}-\frac{7
C_{11,22}^{51,1}}{1920 \sqrt{3}}-\frac{27 \sqrt{\frac{7}{5}} C_{11,22}^{53,1}}{1280}+\frac{7 C_{20,20}^{40,1}}{192 \sqrt{15}}+\frac{7 C_{20,22}^{42,1}}{1920
\sqrt{3}}-\frac{7 C_{22,20}^{42,1}}{960 \sqrt{3}}+\frac{13 C_{22,22}^{40,1}}{480 \sqrt{3}}-\frac{1}{480} \sqrt{\frac{7}{3}} C_{22,22}^{42,1}-\frac{3}{160}
\sqrt{\frac{3}{5}} C_{22,22}^{44,1}+\frac{7 C_{31,20}^{31,1}}{192 \sqrt{15}}+\frac{37 C_{31,22}^{31,1}}{1920 \sqrt{3}}+\frac{7 \sqrt{\frac{7}{5}}
C_{31,22}^{33,1}}{1280}-\frac{3}{320} \sqrt{\frac{7}{5}} C_{33,20}^{33,1}-\frac{3}{800} \sqrt{\frac{21}{2}} C_{33,22}^{33,1}\)

\noindent\(C_{22,1110,2}^{00,3112,1,\text{NonLoc}}= -\frac{7 C_{00,00}^{60,0}}{128 \sqrt{5}}-\frac{7 C_{00,20}^{60,0}}{128 \sqrt{15}}+\frac{7
\sqrt{3} C_{00,22}^{62,0}}{320}+\frac{7 C_{11,00}^{51,0}}{384 \sqrt{5}}-\frac{7 C_{11,22}^{51,0}}{1920 \sqrt{3}}-\frac{27 \sqrt{\frac{7}{5}} C_{11,22}^{53,0}}{1280}+\frac{7
C_{20,00}^{40,0}}{384 \sqrt{5}}+\frac{7 C_{20,20}^{40,0}}{384 \sqrt{15}}-\frac{7 C_{20,22}^{42,0}}{1920 \sqrt{3}}-\frac{7 C_{22,00}^{42,0}}{1920}-\frac{7
C_{22,20}^{42,0}}{1920 \sqrt{3}}-\frac{13 C_{22,22}^{40,0}}{480 \sqrt{3}}+\frac{1}{480} \sqrt{\frac{7}{3}} C_{22,22}^{42,0}+\frac{3}{160} \sqrt{\frac{3}{5}}
C_{22,22}^{44,0}-\frac{7 C_{31,00}^{31,0}}{384 \sqrt{5}}-\frac{7 C_{31,20}^{31,0}}{384 \sqrt{15}}+\frac{37 C_{31,22}^{31,0}}{1920 \sqrt{3}}+\frac{7
\sqrt{\frac{7}{5}} C_{31,22}^{33,0}}{1280}+\frac{3}{640} \sqrt{\frac{21}{5}} C_{33,00}^{33,0}+\frac{3}{640} \sqrt{\frac{7}{5}} C_{33,20}^{33,0}-\frac{3}{800}
\sqrt{\frac{21}{2}} C_{33,22}^{33,0}\)

\noindent\(C_{22,1110,2}^{00,3312,0,\text{Loc}}= -\frac{1}{96} \sqrt{\frac{7}{3}} C_{00,20}^{60,0}-\frac{1}{192} \sqrt{\frac{7}{3}}
C_{00,20}^{60,1}-\frac{1}{32} \sqrt{\frac{7}{15}} C_{00,22}^{62,0}-\frac{1}{64} \sqrt{\frac{7}{15}} C_{00,22}^{62,1}-\frac{1}{36} \sqrt{\frac{7}{15}}
C_{11,22}^{51,0}-\frac{1}{72} \sqrt{\frac{7}{15}} C_{11,22}^{51,1}+\frac{3 C_{11,22}^{53,0}}{320}+\frac{3 C_{11,22}^{53,1}}{640}+\frac{1}{288} \sqrt{\frac{7}{3}}
C_{20,20}^{40,0}+\frac{1}{576} \sqrt{\frac{7}{3}} C_{20,20}^{40,1}+\frac{1}{36} \sqrt{\frac{7}{15}} C_{20,22}^{42,0}+\frac{1}{72} \sqrt{\frac{7}{15}}
C_{20,22}^{42,1}-\frac{1}{288} \sqrt{\frac{7}{15}} C_{22,20}^{42,0}-\frac{1}{576} \sqrt{\frac{7}{15}} C_{22,20}^{42,1}-\frac{7}{288} \sqrt{\frac{7}{15}}
C_{22,22}^{40,0}-\frac{7}{576} \sqrt{\frac{7}{15}} C_{22,22}^{40,1}-\frac{C_{22,22}^{42,0}}{144 \sqrt{15}}-\frac{C_{22,22}^{42,1}}{288 \sqrt{15}}-\frac{1}{80}
\sqrt{\frac{3}{7}} C_{22,22}^{44,0}-\frac{1}{160} \sqrt{\frac{3}{7}} C_{22,22}^{44,1}+\frac{1}{288} \sqrt{\frac{7}{3}} C_{31,20}^{31,0}+\frac{1}{576}
\sqrt{\frac{7}{3}} C_{31,20}^{31,1}+\frac{1}{36} \sqrt{\frac{7}{15}} C_{31,22}^{31,0}+\frac{1}{72} \sqrt{\frac{7}{15}} C_{31,22}^{31,1}-\frac{3 C_{31,22}^{33,0}}{320}-\frac{3
C_{31,22}^{33,1}}{640}-\frac{C_{33,20}^{33,0}}{160}-\frac{C_{33,20}^{33,1}}{320}-\frac{1}{80} \sqrt{\frac{3}{10}} C_{33,22}^{33,0}-\frac{1}{160}
\sqrt{\frac{3}{10}} C_{33,22}^{33,1}\)

\noindent\(C_{22,1110,2}^{00,3312,0,\text{NonLoc}}= -\frac{\sqrt{7} C_{00,00}^{60,0}}{384}-\frac{\sqrt{7} C_{00,00}^{60,1}}{192}-\frac{1}{384}
\sqrt{\frac{7}{3}} C_{00,20}^{60,0}-\frac{1}{192} \sqrt{\frac{7}{3}} C_{00,20}^{60,1}+\frac{1}{64} \sqrt{\frac{7}{15}} C_{00,22}^{62,0}+\frac{1}{32}
\sqrt{\frac{7}{15}} C_{00,22}^{62,1}+\frac{\sqrt{7} C_{11,00}^{51,0}}{1152}+\frac{\sqrt{7} C_{11,00}^{51,1}}{576}-\frac{1}{72} \sqrt{\frac{7}{15}}
C_{11,22}^{51,0}-\frac{1}{36} \sqrt{\frac{7}{15}} C_{11,22}^{51,1}+\frac{3 C_{11,22}^{53,0}}{640}+\frac{3 C_{11,22}^{53,1}}{320}+\frac{\sqrt{7} C_{20,00}^{40,0}}{1152}+\frac{\sqrt{7}
C_{20,00}^{40,1}}{576}+\frac{\sqrt{\frac{7}{3}} C_{20,20}^{40,0}}{1152}+\frac{1}{576} \sqrt{\frac{7}{3}} C_{20,20}^{40,1}-\frac{1}{72} \sqrt{\frac{7}{15}}
C_{20,22}^{42,0}-\frac{1}{36} \sqrt{\frac{7}{15}} C_{20,22}^{42,1}-\frac{\sqrt{\frac{7}{5}} C_{22,00}^{42,0}}{1152}-\frac{1}{576} \sqrt{\frac{7}{5}}
C_{22,00}^{42,1}-\frac{\sqrt{\frac{7}{15}} C_{22,20}^{42,0}}{1152}-\frac{1}{576} \sqrt{\frac{7}{15}} C_{22,20}^{42,1}+\frac{7}{576} \sqrt{\frac{7}{15}}
C_{22,22}^{40,0}+\frac{7}{288} \sqrt{\frac{7}{15}} C_{22,22}^{40,1}+\frac{C_{22,22}^{42,0}}{288 \sqrt{15}}+\frac{C_{22,22}^{42,1}}{144 \sqrt{15}}+\frac{1}{160}
\sqrt{\frac{3}{7}} C_{22,22}^{44,0}+\frac{1}{80} \sqrt{\frac{3}{7}} C_{22,22}^{44,1}-\frac{\sqrt{7} C_{31,00}^{31,0}}{1152}-\frac{\sqrt{7} C_{31,00}^{31,1}}{576}-\frac{\sqrt{\frac{7}{3}}
C_{31,20}^{31,0}}{1152}-\frac{1}{576} \sqrt{\frac{7}{3}} C_{31,20}^{31,1}+\frac{1}{72} \sqrt{\frac{7}{15}} C_{31,22}^{31,0}+\frac{1}{36} \sqrt{\frac{7}{15}}
C_{31,22}^{31,1}-\frac{3 C_{31,22}^{33,0}}{640}-\frac{3 C_{31,22}^{33,1}}{320}+\frac{\sqrt{3} C_{33,00}^{33,0}}{640}+\frac{\sqrt{3} C_{33,00}^{33,1}}{320}+\frac{C_{33,20}^{33,0}}{640}+\frac{C_{33,20}^{33,1}}{320}-\frac{1}{160}
\sqrt{\frac{3}{10}} C_{33,22}^{33,0}-\frac{1}{80} \sqrt{\frac{3}{10}} C_{33,22}^{33,1}\)

\noindent\(C_{22,1110,2}^{00,3312,1,\text{Loc}}= -\frac{\sqrt{7} C_{00,20}^{60,1}}{192}-\frac{1}{64} \sqrt{\frac{7}{5}} C_{00,22}^{62,1}-\frac{1}{72}
\sqrt{\frac{7}{5}} C_{11,22}^{51,1}+\frac{3 \sqrt{3} C_{11,22}^{53,1}}{640}+\frac{\sqrt{7} C_{20,20}^{40,1}}{576}+\frac{1}{72} \sqrt{\frac{7}{5}}
C_{20,22}^{42,1}-\frac{1}{576} \sqrt{\frac{7}{5}} C_{22,20}^{42,1}-\frac{7}{576} \sqrt{\frac{7}{5}} C_{22,22}^{40,1}-\frac{C_{22,22}^{42,1}}{288
\sqrt{5}}-\frac{3 C_{22,22}^{44,1}}{160 \sqrt{7}}+\frac{\sqrt{7} C_{31,20}^{31,1}}{576}+\frac{1}{72} \sqrt{\frac{7}{5}} C_{31,22}^{31,1}-\frac{3
\sqrt{3} C_{31,22}^{33,1}}{640}-\frac{\sqrt{3} C_{33,20}^{33,1}}{320}-\frac{3 C_{33,22}^{33,1}}{160 \sqrt{10}}\)

\noindent\(C_{22,1110,2}^{00,3312,1,\text{NonLoc}}= -\frac{1}{128} \sqrt{\frac{7}{3}} C_{00,00}^{60,0}-\frac{\sqrt{7} C_{00,20}^{60,0}}{384}+\frac{1}{64}
\sqrt{\frac{7}{5}} C_{00,22}^{62,0}+\frac{1}{384} \sqrt{\frac{7}{3}} C_{11,00}^{51,0}-\frac{1}{72} \sqrt{\frac{7}{5}} C_{11,22}^{51,0}+\frac{3 \sqrt{3}
C_{11,22}^{53,0}}{640}+\frac{1}{384} \sqrt{\frac{7}{3}} C_{20,00}^{40,0}+\frac{\sqrt{7} C_{20,20}^{40,0}}{1152}-\frac{1}{72} \sqrt{\frac{7}{5}} C_{20,22}^{42,0}-\frac{1}{384}
\sqrt{\frac{7}{15}} C_{22,00}^{42,0}-\frac{\sqrt{\frac{7}{5}} C_{22,20}^{42,0}}{1152}+\frac{7}{576} \sqrt{\frac{7}{5}} C_{22,22}^{40,0}+\frac{C_{22,22}^{42,0}}{288
\sqrt{5}}+\frac{3 C_{22,22}^{44,0}}{160 \sqrt{7}}-\frac{1}{384} \sqrt{\frac{7}{3}} C_{31,00}^{31,0}-\frac{\sqrt{7} C_{31,20}^{31,0}}{1152}+\frac{1}{72}
\sqrt{\frac{7}{5}} C_{31,22}^{31,0}-\frac{3 \sqrt{3} C_{31,22}^{33,0}}{640}+\frac{3 C_{33,00}^{33,0}}{640}+\frac{\sqrt{3} C_{33,20}^{33,0}}{640}-\frac{3
C_{33,22}^{33,0}}{160 \sqrt{10}}\)

\noindent\(C_{22,1110,2}^{20,1112,0,\text{Loc}}= \frac{7 \sqrt{5} C_{00,20}^{60,0}}{1152}+\frac{7 \sqrt{5} C_{00,20}^{60,1}}{2304}+\frac{7
C_{00,22}^{62,0}}{384}+\frac{7 C_{00,22}^{62,1}}{768}-\frac{49 C_{11,22}^{51,0}}{6912}-\frac{49 C_{11,22}^{51,1}}{13824}-\frac{1}{512} \sqrt{\frac{21}{5}}
C_{11,22}^{53,0}-\frac{\sqrt{\frac{21}{5}} C_{11,22}^{53,1}}{1024}-\frac{7 \sqrt{5} C_{20,20}^{40,0}}{3456}-\frac{7 \sqrt{5} C_{20,20}^{40,1}}{6912}-\frac{7
C_{20,22}^{42,0}}{6912}-\frac{7 C_{20,22}^{42,1}}{13824}+\frac{7 C_{22,20}^{42,0}}{3456}+\frac{7 C_{22,20}^{42,1}}{6912}-\frac{7 C_{22,22}^{40,0}}{1728}-\frac{7
C_{22,22}^{40,1}}{3456}+\frac{5 \sqrt{7} C_{22,22}^{42,0}}{3456}+\frac{5 \sqrt{7} C_{22,22}^{42,1}}{6912}-\frac{C_{22,22}^{44,0}}{64 \sqrt{5}}-\frac{C_{22,22}^{44,1}}{128
\sqrt{5}}+\frac{7 \sqrt{5} C_{31,20}^{31,0}}{3456}+\frac{7 \sqrt{5} C_{31,20}^{31,1}}{6912}+\frac{7 C_{31,22}^{31,0}}{6912}+\frac{7 C_{31,22}^{31,1}}{13824}+\frac{17
\sqrt{\frac{7}{15}} C_{31,22}^{33,0}}{1536}+\frac{17 \sqrt{\frac{7}{15}} C_{31,22}^{33,1}}{3072}-\frac{1}{128} \sqrt{\frac{7}{15}} C_{33,20}^{33,0}-\frac{1}{256}
\sqrt{\frac{7}{15}} C_{33,20}^{33,1}-\frac{1}{320} \sqrt{\frac{7}{2}} C_{33,22}^{33,0}-\frac{1}{640} \sqrt{\frac{7}{2}} C_{33,22}^{33,1}\)

\noindent\(C_{22,1110,2}^{20,1112,0,\text{NonLoc}}= \frac{7 \sqrt{\frac{5}{3}} C_{00,00}^{60,0}}{1536}+\frac{7}{768} \sqrt{\frac{5}{3}}
C_{00,00}^{60,1}+\frac{7 \sqrt{5} C_{00,20}^{60,0}}{4608}+\frac{7 \sqrt{5} C_{00,20}^{60,1}}{2304}-\frac{7 C_{00,22}^{62,0}}{768}-\frac{7 C_{00,22}^{62,1}}{384}+\frac{7
\sqrt{\frac{5}{3}} C_{11,00}^{51,0}}{4608}+\frac{7 \sqrt{\frac{5}{3}} C_{11,00}^{51,1}}{2304}-\frac{49 C_{11,22}^{51,0}}{13824}-\frac{49 C_{11,22}^{51,1}}{6912}-\frac{\sqrt{\frac{21}{5}}
C_{11,22}^{53,0}}{1024}-\frac{1}{512} \sqrt{\frac{21}{5}} C_{11,22}^{53,1}-\frac{7 \sqrt{\frac{5}{3}} C_{20,00}^{40,0}}{4608}-\frac{7 \sqrt{\frac{5}{3}}
C_{20,00}^{40,1}}{2304}-\frac{7 \sqrt{5} C_{20,20}^{40,0}}{13824}-\frac{7 \sqrt{5} C_{20,20}^{40,1}}{6912}+\frac{7 C_{20,22}^{42,0}}{13824}+\frac{7
C_{20,22}^{42,1}}{6912}+\frac{7 C_{22,00}^{42,0}}{4608 \sqrt{3}}+\frac{7 C_{22,00}^{42,1}}{2304 \sqrt{3}}+\frac{7 C_{22,20}^{42,0}}{13824}+\frac{7
C_{22,20}^{42,1}}{6912}+\frac{7 C_{22,22}^{40,0}}{3456}+\frac{7 C_{22,22}^{40,1}}{1728}-\frac{5 \sqrt{7} C_{22,22}^{42,0}}{6912}-\frac{5 \sqrt{7}
C_{22,22}^{42,1}}{3456}+\frac{C_{22,22}^{44,0}}{128 \sqrt{5}}+\frac{C_{22,22}^{44,1}}{64 \sqrt{5}}-\frac{7 \sqrt{\frac{5}{3}} C_{31,00}^{31,0}}{4608}-\frac{7
\sqrt{\frac{5}{3}} C_{31,00}^{31,1}}{2304}-\frac{7 \sqrt{5} C_{31,20}^{31,0}}{13824}-\frac{7 \sqrt{5} C_{31,20}^{31,1}}{6912}+\frac{7 C_{31,22}^{31,0}}{13824}+\frac{7
C_{31,22}^{31,1}}{6912}+\frac{17 \sqrt{\frac{7}{15}} C_{31,22}^{33,0}}{3072}+\frac{17 \sqrt{\frac{7}{15}} C_{31,22}^{33,1}}{1536}+\frac{1}{512} \sqrt{\frac{7}{5}}
C_{33,00}^{33,0}+\frac{1}{256} \sqrt{\frac{7}{5}} C_{33,00}^{33,1}+\frac{1}{512} \sqrt{\frac{7}{15}} C_{33,20}^{33,0}+\frac{1}{256} \sqrt{\frac{7}{15}}
C_{33,20}^{33,1}-\frac{1}{640} \sqrt{\frac{7}{2}} C_{33,22}^{33,0}-\frac{1}{320} \sqrt{\frac{7}{2}} C_{33,22}^{33,1}\)

\noindent\(C_{22,1110,2}^{20,1112,1,\text{Loc}}= \frac{7}{768} \sqrt{\frac{5}{3}} C_{00,20}^{60,1}+\frac{7 C_{00,22}^{62,1}}{256 \sqrt{3}}-\frac{49
C_{11,22}^{51,1}}{4608 \sqrt{3}}-\frac{3 \sqrt{\frac{7}{5}} C_{11,22}^{53,1}}{1024}-\frac{7 \sqrt{\frac{5}{3}} C_{20,20}^{40,1}}{2304}-\frac{7 C_{20,22}^{42,1}}{4608
\sqrt{3}}+\frac{7 C_{22,20}^{42,1}}{2304 \sqrt{3}}-\frac{7 C_{22,22}^{40,1}}{1152 \sqrt{3}}+\frac{5 \sqrt{\frac{7}{3}} C_{22,22}^{42,1}}{2304}-\frac{1}{128}
\sqrt{\frac{3}{5}} C_{22,22}^{44,1}+\frac{7 \sqrt{\frac{5}{3}} C_{31,20}^{31,1}}{2304}+\frac{7 C_{31,22}^{31,1}}{4608 \sqrt{3}}+\frac{17 \sqrt{\frac{7}{5}}
C_{31,22}^{33,1}}{3072}-\frac{1}{256} \sqrt{\frac{7}{5}} C_{33,20}^{33,1}-\frac{1}{640} \sqrt{\frac{21}{2}} C_{33,22}^{33,1}\)

\noindent\(C_{22,1110,2}^{20,1112,1,\text{NonLoc}}= \frac{7 \sqrt{5} C_{00,00}^{60,0}}{1536}+\frac{7 \sqrt{\frac{5}{3}} C_{00,20}^{60,0}}{1536}-\frac{7
C_{00,22}^{62,0}}{256 \sqrt{3}}+\frac{7 \sqrt{5} C_{11,00}^{51,0}}{4608}-\frac{49 C_{11,22}^{51,0}}{4608 \sqrt{3}}-\frac{3 \sqrt{\frac{7}{5}} C_{11,22}^{53,0}}{1024}-\frac{7
\sqrt{5} C_{20,00}^{40,0}}{4608}-\frac{7 \sqrt{\frac{5}{3}} C_{20,20}^{40,0}}{4608}+\frac{7 C_{20,22}^{42,0}}{4608 \sqrt{3}}+\frac{7 C_{22,00}^{42,0}}{4608}+\frac{7
C_{22,20}^{42,0}}{4608 \sqrt{3}}+\frac{7 C_{22,22}^{40,0}}{1152 \sqrt{3}}-\frac{5 \sqrt{\frac{7}{3}} C_{22,22}^{42,0}}{2304}+\frac{1}{128} \sqrt{\frac{3}{5}}
C_{22,22}^{44,0}-\frac{7 \sqrt{5} C_{31,00}^{31,0}}{4608}-\frac{7 \sqrt{\frac{5}{3}} C_{31,20}^{31,0}}{4608}+\frac{7 C_{31,22}^{31,0}}{4608 \sqrt{3}}+\frac{17
\sqrt{\frac{7}{5}} C_{31,22}^{33,0}}{3072}+\frac{1}{512} \sqrt{\frac{21}{5}} C_{33,00}^{33,0}+\frac{1}{512} \sqrt{\frac{7}{5}} C_{33,20}^{33,0}-\frac{1}{640}
\sqrt{\frac{21}{2}} C_{33,22}^{33,0}\)

\noindent\(C_{22,1110,2}^{22,1110,0,\text{Loc}}= \frac{7 \sqrt{5} C_{00,20}^{60,0}}{2304}+\frac{7 \sqrt{5} C_{00,20}^{60,1}}{4608}+\frac{7
C_{00,22}^{62,0}}{768}+\frac{7 C_{00,22}^{62,1}}{1536}+\frac{7 C_{11,22}^{51,0}}{6912}+\frac{7 C_{11,22}^{51,1}}{13824}-\frac{1}{256} \sqrt{\frac{21}{5}}
C_{11,22}^{53,0}-\frac{1}{512} \sqrt{\frac{21}{5}} C_{11,22}^{53,1}+\frac{5 \sqrt{5} C_{20,20}^{40,0}}{6912}+\frac{5 \sqrt{5} C_{20,20}^{40,1}}{13824}+\frac{7
C_{20,22}^{42,0}}{6912}+\frac{7 C_{20,22}^{42,1}}{13824}-\frac{7 C_{22,20}^{42,0}}{3456}-\frac{7 C_{22,20}^{42,1}}{6912}-\frac{17 C_{22,22}^{40,0}}{6912}-\frac{17
C_{22,22}^{40,1}}{13824}-\frac{5 \sqrt{7} C_{22,22}^{42,0}}{3456}-\frac{5 \sqrt{7} C_{22,22}^{42,1}}{6912}+\frac{C_{22,22}^{44,0}}{64 \sqrt{5}}+\frac{C_{22,22}^{44,1}}{128
\sqrt{5}}+\frac{\sqrt{5} C_{31,20}^{31,0}}{6912}+\frac{\sqrt{5} C_{31,20}^{31,1}}{13824}+\frac{23 C_{31,22}^{31,0}}{6912}+\frac{23 C_{31,22}^{31,1}}{13824}-\frac{1}{768}
\sqrt{\frac{7}{15}} C_{31,22}^{33,0}-\frac{\sqrt{\frac{7}{15}} C_{31,22}^{33,1}}{1536}+\frac{1}{128} \sqrt{\frac{7}{15}} C_{33,20}^{33,0}+\frac{1}{256}
\sqrt{\frac{7}{15}} C_{33,20}^{33,1}+\frac{1}{320} \sqrt{\frac{7}{2}} C_{33,22}^{33,0}+\frac{1}{640} \sqrt{\frac{7}{2}} C_{33,22}^{33,1}\)

\noindent\(C_{22,1110,2}^{22,1110,0,\text{NonLoc}}= \frac{7 \sqrt{\frac{5}{3}} C_{00,00}^{60,0}}{3072}+\frac{7 \sqrt{\frac{5}{3}} C_{00,00}^{60,1}}{1536}+\frac{7
\sqrt{5} C_{00,20}^{60,0}}{9216}+\frac{7 \sqrt{5} C_{00,20}^{60,1}}{4608}-\frac{7 C_{00,22}^{62,0}}{1536}-\frac{7 C_{00,22}^{62,1}}{768}+\frac{7
\sqrt{\frac{5}{3}} C_{11,00}^{51,0}}{9216}+\frac{7 \sqrt{\frac{5}{3}} C_{11,00}^{51,1}}{4608}+\frac{7 C_{11,22}^{51,0}}{13824}+\frac{7 C_{11,22}^{51,1}}{6912}-\frac{1}{512}
\sqrt{\frac{21}{5}} C_{11,22}^{53,0}-\frac{1}{256} \sqrt{\frac{21}{5}} C_{11,22}^{53,1}+\frac{5 \sqrt{\frac{5}{3}} C_{20,00}^{40,0}}{9216}+\frac{5
\sqrt{\frac{5}{3}} C_{20,00}^{40,1}}{4608}+\frac{5 \sqrt{5} C_{20,20}^{40,0}}{27648}+\frac{5 \sqrt{5} C_{20,20}^{40,1}}{13824}-\frac{7 C_{20,22}^{42,0}}{13824}-\frac{7
C_{20,22}^{42,1}}{6912}-\frac{7 C_{22,00}^{42,0}}{4608 \sqrt{3}}-\frac{7 C_{22,00}^{42,1}}{2304 \sqrt{3}}-\frac{7 C_{22,20}^{42,0}}{13824}-\frac{7
C_{22,20}^{42,1}}{6912}+\frac{17 C_{22,22}^{40,0}}{13824}+\frac{17 C_{22,22}^{40,1}}{6912}+\frac{5 \sqrt{7} C_{22,22}^{42,0}}{6912}+\frac{5 \sqrt{7}
C_{22,22}^{42,1}}{3456}-\frac{C_{22,22}^{44,0}}{128 \sqrt{5}}-\frac{C_{22,22}^{44,1}}{64 \sqrt{5}}-\frac{\sqrt{\frac{5}{3}} C_{31,00}^{31,0}}{9216}-\frac{\sqrt{\frac{5}{3}}
C_{31,00}^{31,1}}{4608}-\frac{\sqrt{5} C_{31,20}^{31,0}}{27648}-\frac{\sqrt{5} C_{31,20}^{31,1}}{13824}+\frac{23 C_{31,22}^{31,0}}{13824}+\frac{23
C_{31,22}^{31,1}}{6912}-\frac{\sqrt{\frac{7}{15}} C_{31,22}^{33,0}}{1536}-\frac{1}{768} \sqrt{\frac{7}{15}} C_{31,22}^{33,1}-\frac{1}{512} \sqrt{\frac{7}{5}}
C_{33,00}^{33,0}-\frac{1}{256} \sqrt{\frac{7}{5}} C_{33,00}^{33,1}-\frac{1}{512} \sqrt{\frac{7}{15}} C_{33,20}^{33,0}-\frac{1}{256} \sqrt{\frac{7}{15}}
C_{33,20}^{33,1}+\frac{1}{640} \sqrt{\frac{7}{2}} C_{33,22}^{33,0}+\frac{1}{320} \sqrt{\frac{7}{2}} C_{33,22}^{33,1}\)

\noindent\(C_{22,1110,2}^{22,1110,1,\text{Loc}}= \frac{7 \sqrt{\frac{5}{3}} C_{00,20}^{60,1}}{1536}+\frac{7 C_{00,22}^{62,1}}{512 \sqrt{3}}+\frac{7
C_{11,22}^{51,1}}{4608 \sqrt{3}}-\frac{3}{512} \sqrt{\frac{7}{5}} C_{11,22}^{53,1}+\frac{5 \sqrt{\frac{5}{3}} C_{20,20}^{40,1}}{4608}+\frac{7 C_{20,22}^{42,1}}{4608
\sqrt{3}}-\frac{7 C_{22,20}^{42,1}}{2304 \sqrt{3}}-\frac{17 C_{22,22}^{40,1}}{4608 \sqrt{3}}-\frac{5 \sqrt{\frac{7}{3}} C_{22,22}^{42,1}}{2304}+\frac{1}{128}
\sqrt{\frac{3}{5}} C_{22,22}^{44,1}+\frac{\sqrt{\frac{5}{3}} C_{31,20}^{31,1}}{4608}+\frac{23 C_{31,22}^{31,1}}{4608 \sqrt{3}}-\frac{\sqrt{\frac{7}{5}}
C_{31,22}^{33,1}}{1536}+\frac{1}{256} \sqrt{\frac{7}{5}} C_{33,20}^{33,1}+\frac{1}{640} \sqrt{\frac{21}{2}} C_{33,22}^{33,1}\)

\noindent\(C_{22,1110,2}^{22,1110,1,\text{NonLoc}}= \frac{7 \sqrt{5} C_{00,00}^{60,0}}{3072}+\frac{7 \sqrt{\frac{5}{3}} C_{00,20}^{60,0}}{3072}-\frac{7
C_{00,22}^{62,0}}{512 \sqrt{3}}+\frac{7 \sqrt{5} C_{11,00}^{51,0}}{9216}+\frac{7 C_{11,22}^{51,0}}{4608 \sqrt{3}}-\frac{3}{512} \sqrt{\frac{7}{5}}
C_{11,22}^{53,0}+\frac{5 \sqrt{5} C_{20,00}^{40,0}}{9216}+\frac{5 \sqrt{\frac{5}{3}} C_{20,20}^{40,0}}{9216}-\frac{7 C_{20,22}^{42,0}}{4608 \sqrt{3}}-\frac{7
C_{22,00}^{42,0}}{4608}-\frac{7 C_{22,20}^{42,0}}{4608 \sqrt{3}}+\frac{17 C_{22,22}^{40,0}}{4608 \sqrt{3}}+\frac{5 \sqrt{\frac{7}{3}} C_{22,22}^{42,0}}{2304}-\frac{1}{128}
\sqrt{\frac{3}{5}} C_{22,22}^{44,0}-\frac{\sqrt{5} C_{31,00}^{31,0}}{9216}-\frac{\sqrt{\frac{5}{3}} C_{31,20}^{31,0}}{9216}+\frac{23 C_{31,22}^{31,0}}{4608
\sqrt{3}}-\frac{\sqrt{\frac{7}{5}} C_{31,22}^{33,0}}{1536}-\frac{1}{512} \sqrt{\frac{21}{5}} C_{33,00}^{33,0}-\frac{1}{512} \sqrt{\frac{7}{5}} C_{33,20}^{33,0}+\frac{1}{640}
\sqrt{\frac{21}{2}} C_{33,22}^{33,0}\)

\noindent\(C_{22,1111,1}^{00,3111,0,\text{Loc}}= \frac{7 C_{00,20}^{60,0}}{64 \sqrt{15}}+\frac{7 C_{00,20}^{60,1}}{128 \sqrt{15}}-\frac{7
\sqrt{3} C_{00,22}^{62,0}}{640}-\frac{7 \sqrt{3} C_{00,22}^{62,1}}{1280}-\frac{7 C_{11,22}^{51,0}}{384 \sqrt{3}}-\frac{7 C_{11,22}^{51,1}}{768 \sqrt{3}}-\frac{7
C_{20,20}^{40,0}}{192 \sqrt{15}}-\frac{7 C_{20,20}^{40,1}}{384 \sqrt{15}}+\frac{7 C_{20,22}^{42,0}}{384 \sqrt{3}}+\frac{7 C_{20,22}^{42,1}}{768 \sqrt{3}}+\frac{7
C_{22,20}^{42,0}}{960 \sqrt{3}}+\frac{7 C_{22,20}^{42,1}}{1920 \sqrt{3}}+\frac{53 C_{22,22}^{40,0}}{1920 \sqrt{3}}+\frac{53 C_{22,22}^{40,1}}{3840
\sqrt{3}}-\frac{19}{960} \sqrt{\frac{7}{3}} C_{22,22}^{42,0}-\frac{19 \sqrt{\frac{7}{3}} C_{22,22}^{42,1}}{1920}+\frac{3}{160} \sqrt{\frac{3}{5}}
C_{22,22}^{44,0}+\frac{3}{320} \sqrt{\frac{3}{5}} C_{22,22}^{44,1}-\frac{7 C_{31,20}^{31,0}}{192 \sqrt{15}}-\frac{7 C_{31,20}^{31,1}}{384 \sqrt{15}}+\frac{C_{31,22}^{31,0}}{384
\sqrt{3}}+\frac{C_{31,22}^{31,1}}{768 \sqrt{3}}+\frac{1}{64} \sqrt{\frac{7}{5}} C_{31,22}^{33,0}+\frac{1}{128} \sqrt{\frac{7}{5}} C_{31,22}^{33,1}+\frac{3}{320}
\sqrt{\frac{7}{5}} C_{33,20}^{33,0}+\frac{3}{640} \sqrt{\frac{7}{5}} C_{33,20}^{33,1}-\frac{3}{160} \sqrt{\frac{21}{2}} C_{33,22}^{33,0}-\frac{3}{320}
\sqrt{\frac{21}{2}} C_{33,22}^{33,1}\)

\noindent\(C_{22,1111,1}^{00,3111,0,\text{NonLoc}}= \frac{7 C_{00,00}^{60,0}}{256 \sqrt{5}}+\frac{7 C_{00,00}^{60,1}}{128 \sqrt{5}}+\frac{7
C_{00,20}^{60,0}}{256 \sqrt{15}}+\frac{7 C_{00,20}^{60,1}}{128 \sqrt{15}}+\frac{7 \sqrt{3} C_{00,22}^{62,0}}{1280}+\frac{7 \sqrt{3} C_{00,22}^{62,1}}{640}-\frac{7
C_{11,00}^{51,0}}{768 \sqrt{5}}-\frac{7 C_{11,00}^{51,1}}{384 \sqrt{5}}-\frac{7 C_{11,22}^{51,0}}{768 \sqrt{3}}-\frac{7 C_{11,22}^{51,1}}{384 \sqrt{3}}-\frac{7
C_{20,00}^{40,0}}{768 \sqrt{5}}-\frac{7 C_{20,00}^{40,1}}{384 \sqrt{5}}-\frac{7 C_{20,20}^{40,0}}{768 \sqrt{15}}-\frac{7 C_{20,20}^{40,1}}{384 \sqrt{15}}-\frac{7
C_{20,22}^{42,0}}{768 \sqrt{3}}-\frac{7 C_{20,22}^{42,1}}{384 \sqrt{3}}+\frac{7 C_{22,00}^{42,0}}{3840}+\frac{7 C_{22,00}^{42,1}}{1920}+\frac{7 C_{22,20}^{42,0}}{3840
\sqrt{3}}+\frac{7 C_{22,20}^{42,1}}{1920 \sqrt{3}}-\frac{53 C_{22,22}^{40,0}}{3840 \sqrt{3}}-\frac{53 C_{22,22}^{40,1}}{1920 \sqrt{3}}+\frac{19 \sqrt{\frac{7}{3}}
C_{22,22}^{42,0}}{1920}+\frac{19}{960} \sqrt{\frac{7}{3}} C_{22,22}^{42,1}-\frac{3}{320} \sqrt{\frac{3}{5}} C_{22,22}^{44,0}-\frac{3}{160} \sqrt{\frac{3}{5}}
C_{22,22}^{44,1}+\frac{7 C_{31,00}^{31,0}}{768 \sqrt{5}}+\frac{7 C_{31,00}^{31,1}}{384 \sqrt{5}}+\frac{7 C_{31,20}^{31,0}}{768 \sqrt{15}}+\frac{7
C_{31,20}^{31,1}}{384 \sqrt{15}}+\frac{C_{31,22}^{31,0}}{768 \sqrt{3}}+\frac{C_{31,22}^{31,1}}{384 \sqrt{3}}+\frac{1}{128} \sqrt{\frac{7}{5}} C_{31,22}^{33,0}+\frac{1}{64}
\sqrt{\frac{7}{5}} C_{31,22}^{33,1}-\frac{3 \sqrt{\frac{21}{5}} C_{33,00}^{33,0}}{1280}-\frac{3}{640} \sqrt{\frac{21}{5}} C_{33,00}^{33,1}-\frac{3
\sqrt{\frac{7}{5}} C_{33,20}^{33,0}}{1280}-\frac{3}{640} \sqrt{\frac{7}{5}} C_{33,20}^{33,1}-\frac{3}{320} \sqrt{\frac{21}{2}} C_{33,22}^{33,0}-\frac{3}{160}
\sqrt{\frac{21}{2}} C_{33,22}^{33,1}\)

\noindent\(C_{22,1111,1}^{00,3111,1,\text{Loc}}= \frac{7 C_{00,20}^{60,1}}{128 \sqrt{5}}-\frac{21 C_{00,22}^{62,1}}{1280}-\frac{7 C_{11,22}^{51,1}}{768}-\frac{7
C_{20,20}^{40,1}}{384 \sqrt{5}}+\frac{7 C_{20,22}^{42,1}}{768}+\frac{7 C_{22,20}^{42,1}}{1920}+\frac{53 C_{22,22}^{40,1}}{3840}-\frac{19 \sqrt{7}
C_{22,22}^{42,1}}{1920}+\frac{9 C_{22,22}^{44,1}}{320 \sqrt{5}}-\frac{7 C_{31,20}^{31,1}}{384 \sqrt{5}}+\frac{C_{31,22}^{31,1}}{768}+\frac{1}{128}
\sqrt{\frac{21}{5}} C_{31,22}^{33,1}+\frac{3}{640} \sqrt{\frac{21}{5}} C_{33,20}^{33,1}-\frac{9}{320} \sqrt{\frac{7}{2}} C_{33,22}^{33,1}\)

\noindent\(C_{22,1111,1}^{00,3111,1,\text{NonLoc}}= \frac{7}{256} \sqrt{\frac{3}{5}} C_{00,00}^{60,0}+\frac{7 C_{00,20}^{60,0}}{256
\sqrt{5}}+\frac{21 C_{00,22}^{62,0}}{1280}-\frac{7 C_{11,00}^{51,0}}{256 \sqrt{15}}-\frac{7 C_{11,22}^{51,0}}{768}-\frac{7 C_{20,00}^{40,0}}{256
\sqrt{15}}-\frac{7 C_{20,20}^{40,0}}{768 \sqrt{5}}-\frac{7 C_{20,22}^{42,0}}{768}+\frac{7 C_{22,00}^{42,0}}{1280 \sqrt{3}}+\frac{7 C_{22,20}^{42,0}}{3840}-\frac{53
C_{22,22}^{40,0}}{3840}+\frac{19 \sqrt{7} C_{22,22}^{42,0}}{1920}-\frac{9 C_{22,22}^{44,0}}{320 \sqrt{5}}+\frac{7 C_{31,00}^{31,0}}{256 \sqrt{15}}+\frac{7
C_{31,20}^{31,0}}{768 \sqrt{5}}+\frac{C_{31,22}^{31,0}}{768}+\frac{1}{128} \sqrt{\frac{21}{5}} C_{31,22}^{33,0}-\frac{9 \sqrt{\frac{7}{5}} C_{33,00}^{33,0}}{1280}-\frac{3
\sqrt{\frac{21}{5}} C_{33,20}^{33,0}}{1280}-\frac{9}{320} \sqrt{\frac{7}{2}} C_{33,22}^{33,0}\)

\noindent\(C_{22,1111,1}^{11,2000,0,\text{Loc}}= -\frac{7 \sqrt{5} C_{11,11}^{51,0}}{576}-\frac{7 \sqrt{5} C_{11,11}^{51,1}}{1152}-\frac{\sqrt{5}
C_{31,11}^{31,0}}{192}-\frac{\sqrt{5} C_{31,11}^{31,1}}{384}+\frac{1}{16} \sqrt{\frac{7}{10}} C_{33,11}^{33,0}+\frac{1}{32} \sqrt{\frac{7}{10}} C_{33,11}^{33,1}\)

\noindent\(C_{22,1111,1}^{11,2000,0,\text{NonLoc}}= -\frac{7 \sqrt{5} C_{11,11}^{51,0}}{1152}-\frac{7 \sqrt{5} C_{11,11}^{51,1}}{576}-\frac{\sqrt{5}
C_{31,11}^{31,0}}{384}-\frac{\sqrt{5} C_{31,11}^{31,1}}{192}+\frac{1}{32} \sqrt{\frac{7}{10}} C_{33,11}^{33,0}+\frac{1}{16} \sqrt{\frac{7}{10}} C_{33,11}^{33,1}\)

\noindent\(C_{22,1111,1}^{11,2000,1,\text{Loc}}= -\frac{7}{384} \sqrt{\frac{5}{3}} C_{11,11}^{51,1}-\frac{1}{128} \sqrt{\frac{5}{3}}
C_{31,11}^{31,1}+\frac{1}{32} \sqrt{\frac{21}{10}} C_{33,11}^{33,1}\)

\noindent\(C_{22,1111,1}^{11,2000,1,\text{NonLoc}}= -\frac{7}{384} \sqrt{\frac{5}{3}} C_{11,11}^{51,0}-\frac{1}{128} \sqrt{\frac{5}{3}}
C_{31,11}^{31,0}+\frac{1}{32} \sqrt{\frac{21}{10}} C_{33,11}^{33,0}\)

\noindent\(C_{22,1111,1}^{11,2202,0,\text{Loc}}= -\frac{49 C_{11,11}^{51,0}}{1152}-\frac{49 C_{11,11}^{51,1}}{2304}+\frac{17 C_{31,11}^{31,0}}{384}+\frac{17
C_{31,11}^{31,1}}{768}-\frac{1}{80} \sqrt{\frac{7}{2}} C_{33,11}^{33,0}-\frac{1}{160} \sqrt{\frac{7}{2}} C_{33,11}^{33,1}\)

\noindent\(C_{22,1111,1}^{11,2202,0,\text{NonLoc}}= -\frac{49 C_{11,11}^{51,0}}{2304}-\frac{49 C_{11,11}^{51,1}}{1152}+\frac{17 C_{31,11}^{31,0}}{768}+\frac{17
C_{31,11}^{31,1}}{384}-\frac{1}{160} \sqrt{\frac{7}{2}} C_{33,11}^{33,0}-\frac{1}{80} \sqrt{\frac{7}{2}} C_{33,11}^{33,1}\)

\noindent\(C_{22,1111,1}^{11,2202,1,\text{Loc}}= -\frac{49 C_{11,11}^{51,1}}{768 \sqrt{3}}+\frac{17 C_{31,11}^{31,1}}{256 \sqrt{3}}-\frac{1}{160}
\sqrt{\frac{21}{2}} C_{33,11}^{33,1}\)

\noindent\(C_{22,1111,1}^{11,2202,1,\text{NonLoc}}= -\frac{49 C_{11,11}^{51,0}}{768 \sqrt{3}}+\frac{17 C_{31,11}^{31,0}}{256 \sqrt{3}}-\frac{1}{160}
\sqrt{\frac{21}{2}} C_{33,11}^{33,0}\)

\noindent\(C_{22,1111,1}^{20,1111,0,\text{Loc}}= -\frac{7}{768} \sqrt{\frac{5}{3}} C_{00,20}^{60,0}-\frac{7 \sqrt{\frac{5}{3}} C_{00,20}^{60,1}}{1536}+\frac{7
C_{00,22}^{62,0}}{512 \sqrt{3}}+\frac{7 C_{00,22}^{62,1}}{1024 \sqrt{3}}+\frac{7 C_{11,22}^{51,0}}{4608 \sqrt{3}}+\frac{7 C_{11,22}^{51,1}}{9216
\sqrt{3}}-\frac{3}{512} \sqrt{\frac{7}{5}} C_{11,22}^{53,0}-\frac{3 \sqrt{\frac{7}{5}} C_{11,22}^{53,1}}{1024}+\frac{7 \sqrt{\frac{5}{3}} C_{20,20}^{40,0}}{2304}+\frac{7
\sqrt{\frac{5}{3}} C_{20,20}^{40,1}}{4608}-\frac{35 C_{20,22}^{42,0}}{4608 \sqrt{3}}-\frac{35 C_{20,22}^{42,1}}{9216 \sqrt{3}}-\frac{7 C_{22,20}^{42,0}}{2304
\sqrt{3}}-\frac{7 C_{22,20}^{42,1}}{4608 \sqrt{3}}-\frac{77 C_{22,22}^{40,0}}{4608 \sqrt{3}}-\frac{77 C_{22,22}^{40,1}}{9216 \sqrt{3}}+\frac{1}{144}
\sqrt{\frac{7}{3}} C_{22,22}^{42,0}+\frac{1}{288} \sqrt{\frac{7}{3}} C_{22,22}^{42,1}+\frac{1}{128} \sqrt{\frac{3}{5}} C_{22,22}^{44,0}+\frac{1}{256}
\sqrt{\frac{3}{5}} C_{22,22}^{44,1}-\frac{7 \sqrt{\frac{5}{3}} C_{31,20}^{31,0}}{2304}-\frac{7 \sqrt{\frac{5}{3}} C_{31,20}^{31,1}}{4608}+\frac{35
C_{31,22}^{31,0}}{4608 \sqrt{3}}+\frac{35 C_{31,22}^{31,1}}{9216 \sqrt{3}}+\frac{\sqrt{35} C_{31,22}^{33,0}}{1536}+\frac{\sqrt{35} C_{31,22}^{33,1}}{3072}+\frac{1}{256}
\sqrt{\frac{7}{5}} C_{33,20}^{33,0}+\frac{1}{512} \sqrt{\frac{7}{5}} C_{33,20}^{33,1}-\frac{1}{128} \sqrt{\frac{21}{2}} C_{33,22}^{33,0}-\frac{1}{256}
\sqrt{\frac{21}{2}} C_{33,22}^{33,1}\)

\noindent\(C_{22,1111,1}^{20,1111,0,\text{NonLoc}}= -\frac{7 \sqrt{5} C_{00,00}^{60,0}}{3072}-\frac{7 \sqrt{5} C_{00,00}^{60,1}}{1536}-\frac{7
\sqrt{\frac{5}{3}} C_{00,20}^{60,0}}{3072}-\frac{7 \sqrt{\frac{5}{3}} C_{00,20}^{60,1}}{1536}-\frac{7 C_{00,22}^{62,0}}{1024 \sqrt{3}}-\frac{7 C_{00,22}^{62,1}}{512
\sqrt{3}}-\frac{7 \sqrt{5} C_{11,00}^{51,0}}{9216}-\frac{7 \sqrt{5} C_{11,00}^{51,1}}{4608}+\frac{7 C_{11,22}^{51,0}}{9216 \sqrt{3}}+\frac{7 C_{11,22}^{51,1}}{4608
\sqrt{3}}-\frac{3 \sqrt{\frac{7}{5}} C_{11,22}^{53,0}}{1024}-\frac{3}{512} \sqrt{\frac{7}{5}} C_{11,22}^{53,1}+\frac{7 \sqrt{5} C_{20,00}^{40,0}}{9216}+\frac{7
\sqrt{5} C_{20,00}^{40,1}}{4608}+\frac{7 \sqrt{\frac{5}{3}} C_{20,20}^{40,0}}{9216}+\frac{7 \sqrt{\frac{5}{3}} C_{20,20}^{40,1}}{4608}+\frac{35 C_{20,22}^{42,0}}{9216
\sqrt{3}}+\frac{35 C_{20,22}^{42,1}}{4608 \sqrt{3}}-\frac{7 C_{22,00}^{42,0}}{9216}-\frac{7 C_{22,00}^{42,1}}{4608}-\frac{7 C_{22,20}^{42,0}}{9216
\sqrt{3}}-\frac{7 C_{22,20}^{42,1}}{4608 \sqrt{3}}+\frac{77 C_{22,22}^{40,0}}{9216 \sqrt{3}}+\frac{77 C_{22,22}^{40,1}}{4608 \sqrt{3}}-\frac{1}{288}
\sqrt{\frac{7}{3}} C_{22,22}^{42,0}-\frac{1}{144} \sqrt{\frac{7}{3}} C_{22,22}^{42,1}-\frac{1}{256} \sqrt{\frac{3}{5}} C_{22,22}^{44,0}-\frac{1}{128}
\sqrt{\frac{3}{5}} C_{22,22}^{44,1}+\frac{7 \sqrt{5} C_{31,00}^{31,0}}{9216}+\frac{7 \sqrt{5} C_{31,00}^{31,1}}{4608}+\frac{7 \sqrt{\frac{5}{3}}
C_{31,20}^{31,0}}{9216}+\frac{7 \sqrt{\frac{5}{3}} C_{31,20}^{31,1}}{4608}+\frac{35 C_{31,22}^{31,0}}{9216 \sqrt{3}}+\frac{35 C_{31,22}^{31,1}}{4608
\sqrt{3}}+\frac{\sqrt{35} C_{31,22}^{33,0}}{3072}+\frac{\sqrt{35} C_{31,22}^{33,1}}{1536}-\frac{\sqrt{\frac{21}{5}} C_{33,00}^{33,0}}{1024}-\frac{1}{512}
\sqrt{\frac{21}{5}} C_{33,00}^{33,1}-\frac{\sqrt{\frac{7}{5}} C_{33,20}^{33,0}}{1024}-\frac{1}{512} \sqrt{\frac{7}{5}} C_{33,20}^{33,1}-\frac{1}{256}
\sqrt{\frac{21}{2}} C_{33,22}^{33,0}-\frac{1}{128} \sqrt{\frac{21}{2}} C_{33,22}^{33,1}\)

\noindent\(C_{22,1111,1}^{20,1111,1,\text{Loc}}= -\frac{7 \sqrt{5} C_{00,20}^{60,1}}{1536}+\frac{7 C_{00,22}^{62,1}}{1024}+\frac{7
C_{11,22}^{51,1}}{9216}-\frac{3 \sqrt{\frac{21}{5}} C_{11,22}^{53,1}}{1024}+\frac{7 \sqrt{5} C_{20,20}^{40,1}}{4608}-\frac{35 C_{20,22}^{42,1}}{9216}-\frac{7
C_{22,20}^{42,1}}{4608}-\frac{77 C_{22,22}^{40,1}}{9216}+\frac{\sqrt{7} C_{22,22}^{42,1}}{288}+\frac{3 C_{22,22}^{44,1}}{256 \sqrt{5}}-\frac{7 \sqrt{5}
C_{31,20}^{31,1}}{4608}+\frac{35 C_{31,22}^{31,1}}{9216}+\frac{\sqrt{\frac{35}{3}} C_{31,22}^{33,1}}{1024}+\frac{1}{512} \sqrt{\frac{21}{5}} C_{33,20}^{33,1}-\frac{3}{256}
\sqrt{\frac{7}{2}} C_{33,22}^{33,1}\)

\noindent\(C_{22,1111,1}^{20,1111,1,\text{NonLoc}}= -\frac{7 \sqrt{\frac{5}{3}} C_{00,00}^{60,0}}{1024}-\frac{7 \sqrt{5} C_{00,20}^{60,0}}{3072}-\frac{7
C_{00,22}^{62,0}}{1024}-\frac{7 \sqrt{\frac{5}{3}} C_{11,00}^{51,0}}{3072}+\frac{7 C_{11,22}^{51,0}}{9216}-\frac{3 \sqrt{\frac{21}{5}} C_{11,22}^{53,0}}{1024}+\frac{7
\sqrt{\frac{5}{3}} C_{20,00}^{40,0}}{3072}+\frac{7 \sqrt{5} C_{20,20}^{40,0}}{9216}+\frac{35 C_{20,22}^{42,0}}{9216}-\frac{7 C_{22,00}^{42,0}}{3072
\sqrt{3}}-\frac{7 C_{22,20}^{42,0}}{9216}+\frac{77 C_{22,22}^{40,0}}{9216}-\frac{\sqrt{7} C_{22,22}^{42,0}}{288}-\frac{3 C_{22,22}^{44,0}}{256 \sqrt{5}}+\frac{7
\sqrt{\frac{5}{3}} C_{31,00}^{31,0}}{3072}+\frac{7 \sqrt{5} C_{31,20}^{31,0}}{9216}+\frac{35 C_{31,22}^{31,0}}{9216}+\frac{\sqrt{\frac{35}{3}} C_{31,22}^{33,0}}{1024}-\frac{3
\sqrt{\frac{7}{5}} C_{33,00}^{33,0}}{1024}-\frac{\sqrt{\frac{21}{5}} C_{33,20}^{33,0}}{1024}-\frac{3}{256} \sqrt{\frac{7}{2}} C_{33,22}^{33,0}\)

\noindent\(C_{22,1111,1}^{22,1111,0,\text{Loc}}= \frac{7 C_{00,20}^{60,0}}{1536 \sqrt{3}}+\frac{7 C_{00,20}^{60,1}}{3072 \sqrt{3}}-\frac{7
C_{00,22}^{62,0}}{1024 \sqrt{15}}-\frac{7 C_{00,22}^{62,1}}{2048 \sqrt{15}}-\frac{7 C_{11,22}^{51,0}}{9216 \sqrt{15}}-\frac{7 C_{11,22}^{51,1}}{18432
\sqrt{15}}+\frac{3 \sqrt{7} C_{11,22}^{53,0}}{5120}+\frac{3 \sqrt{7} C_{11,22}^{53,1}}{10240}+\frac{17 C_{20,20}^{40,0}}{4608 \sqrt{3}}+\frac{17
C_{20,20}^{40,1}}{9216 \sqrt{3}}-\frac{49 C_{20,22}^{42,0}}{9216 \sqrt{15}}-\frac{49 C_{20,22}^{42,1}}{18432 \sqrt{15}}-\frac{7 \sqrt{\frac{5}{3}}
C_{22,20}^{42,0}}{4608}-\frac{7 \sqrt{\frac{5}{3}} C_{22,20}^{42,1}}{9216}-\frac{43 C_{22,22}^{40,0}}{9216 \sqrt{15}}-\frac{43 C_{22,22}^{40,1}}{18432
\sqrt{15}}+\frac{13 \sqrt{\frac{7}{15}} C_{22,22}^{42,0}}{2304}+\frac{13 \sqrt{\frac{7}{15}} C_{22,22}^{42,1}}{4608}-\frac{\sqrt{3} C_{22,22}^{44,0}}{1280}-\frac{\sqrt{3}
C_{22,22}^{44,1}}{2560}-\frac{5 C_{31,20}^{31,0}}{4608 \sqrt{3}}-\frac{5 C_{31,20}^{31,1}}{9216 \sqrt{3}}-\frac{11 C_{31,22}^{31,0}}{9216 \sqrt{15}}-\frac{11
C_{31,22}^{31,1}}{18432 \sqrt{15}}+\frac{7 \sqrt{7} C_{31,22}^{33,0}}{15360}+\frac{7 \sqrt{7} C_{31,22}^{33,1}}{30720}+\frac{\sqrt{7} C_{33,20}^{33,0}}{512}+\frac{\sqrt{7}
C_{33,20}^{33,1}}{1024}-\frac{7 \sqrt{\frac{21}{10}} C_{33,22}^{33,0}}{1280}-\frac{7 \sqrt{\frac{21}{10}} C_{33,22}^{33,1}}{2560}\)

\noindent\(C_{22,1111,1}^{22,1111,0,\text{NonLoc}}= \frac{7 C_{00,00}^{60,0}}{6144}+\frac{7 C_{00,00}^{60,1}}{3072}+\frac{7 C_{00,20}^{60,0}}{6144
\sqrt{3}}+\frac{7 C_{00,20}^{60,1}}{3072 \sqrt{3}}+\frac{7 C_{00,22}^{62,0}}{2048 \sqrt{15}}+\frac{7 C_{00,22}^{62,1}}{1024 \sqrt{15}}+\frac{7 C_{11,00}^{51,0}}{18432}+\frac{7
C_{11,00}^{51,1}}{9216}-\frac{7 C_{11,22}^{51,0}}{18432 \sqrt{15}}-\frac{7 C_{11,22}^{51,1}}{9216 \sqrt{15}}+\frac{3 \sqrt{7} C_{11,22}^{53,0}}{10240}+\frac{3
\sqrt{7} C_{11,22}^{53,1}}{5120}+\frac{17 C_{20,00}^{40,0}}{18432}+\frac{17 C_{20,00}^{40,1}}{9216}+\frac{17 C_{20,20}^{40,0}}{18432 \sqrt{3}}+\frac{17
C_{20,20}^{40,1}}{9216 \sqrt{3}}+\frac{49 C_{20,22}^{42,0}}{18432 \sqrt{15}}+\frac{49 C_{20,22}^{42,1}}{9216 \sqrt{15}}-\frac{7 \sqrt{5} C_{22,00}^{42,0}}{18432}-\frac{7
\sqrt{5} C_{22,00}^{42,1}}{9216}-\frac{7 \sqrt{\frac{5}{3}} C_{22,20}^{42,0}}{18432}-\frac{7 \sqrt{\frac{5}{3}} C_{22,20}^{42,1}}{9216}+\frac{43
C_{22,22}^{40,0}}{18432 \sqrt{15}}+\frac{43 C_{22,22}^{40,1}}{9216 \sqrt{15}}-\frac{13 \sqrt{\frac{7}{15}} C_{22,22}^{42,0}}{4608}-\frac{13 \sqrt{\frac{7}{15}}
C_{22,22}^{42,1}}{2304}+\frac{\sqrt{3} C_{22,22}^{44,0}}{2560}+\frac{\sqrt{3} C_{22,22}^{44,1}}{1280}+\frac{5 C_{31,00}^{31,0}}{18432}+\frac{5 C_{31,00}^{31,1}}{9216}+\frac{5
C_{31,20}^{31,0}}{18432 \sqrt{3}}+\frac{5 C_{31,20}^{31,1}}{9216 \sqrt{3}}-\frac{11 C_{31,22}^{31,0}}{18432 \sqrt{15}}-\frac{11 C_{31,22}^{31,1}}{9216
\sqrt{15}}+\frac{7 \sqrt{7} C_{31,22}^{33,0}}{30720}+\frac{7 \sqrt{7} C_{31,22}^{33,1}}{15360}-\frac{\sqrt{21} C_{33,00}^{33,0}}{2048}-\frac{\sqrt{21}
C_{33,00}^{33,1}}{1024}-\frac{\sqrt{7} C_{33,20}^{33,0}}{2048}-\frac{\sqrt{7} C_{33,20}^{33,1}}{1024}-\frac{7 \sqrt{\frac{21}{10}} C_{33,22}^{33,0}}{2560}-\frac{7
\sqrt{\frac{21}{10}} C_{33,22}^{33,1}}{1280}\)

\noindent\(C_{22,1111,1}^{22,1111,1,\text{Loc}}= \frac{7 C_{00,20}^{60,1}}{3072}-\frac{7 C_{00,22}^{62,1}}{2048 \sqrt{5}}-\frac{7 C_{11,22}^{51,1}}{18432
\sqrt{5}}+\frac{3 \sqrt{21} C_{11,22}^{53,1}}{10240}+\frac{17 C_{20,20}^{40,1}}{9216}-\frac{49 C_{20,22}^{42,1}}{18432 \sqrt{5}}-\frac{7 \sqrt{5}
C_{22,20}^{42,1}}{9216}-\frac{43 C_{22,22}^{40,1}}{18432 \sqrt{5}}+\frac{13 \sqrt{\frac{7}{5}} C_{22,22}^{42,1}}{4608}-\frac{3 C_{22,22}^{44,1}}{2560}-\frac{5
C_{31,20}^{31,1}}{9216}-\frac{11 C_{31,22}^{31,1}}{18432 \sqrt{5}}+\frac{7 \sqrt{\frac{7}{3}} C_{31,22}^{33,1}}{10240}+\frac{\sqrt{21} C_{33,20}^{33,1}}{1024}-\frac{21
\sqrt{\frac{7}{10}} C_{33,22}^{33,1}}{2560}\)

\noindent\(C_{22,1111,1}^{22,1111,1,\text{NonLoc}}= \frac{7 C_{00,00}^{60,0}}{2048 \sqrt{3}}+\frac{7 C_{00,20}^{60,0}}{6144}+\frac{7
C_{00,22}^{62,0}}{2048 \sqrt{5}}+\frac{7 C_{11,00}^{51,0}}{6144 \sqrt{3}}-\frac{7 C_{11,22}^{51,0}}{18432 \sqrt{5}}+\frac{3 \sqrt{21} C_{11,22}^{53,0}}{10240}+\frac{17
C_{20,00}^{40,0}}{6144 \sqrt{3}}+\frac{17 C_{20,20}^{40,0}}{18432}+\frac{49 C_{20,22}^{42,0}}{18432 \sqrt{5}}-\frac{7 \sqrt{\frac{5}{3}} C_{22,00}^{42,0}}{6144}-\frac{7
\sqrt{5} C_{22,20}^{42,0}}{18432}+\frac{43 C_{22,22}^{40,0}}{18432 \sqrt{5}}-\frac{13 \sqrt{\frac{7}{5}} C_{22,22}^{42,0}}{4608}+\frac{3 C_{22,22}^{44,0}}{2560}+\frac{5
C_{31,00}^{31,0}}{6144 \sqrt{3}}+\frac{5 C_{31,20}^{31,0}}{18432}-\frac{11 C_{31,22}^{31,0}}{18432 \sqrt{5}}+\frac{7 \sqrt{\frac{7}{3}} C_{31,22}^{33,0}}{10240}-\frac{3
\sqrt{7} C_{33,00}^{33,0}}{2048}-\frac{\sqrt{21} C_{33,20}^{33,0}}{2048}-\frac{21 \sqrt{\frac{7}{10}} C_{33,22}^{33,0}}{2560}\)

\noindent\(C_{22,1111,2}^{00,3112,0,\text{Loc}}= \frac{7 C_{00,20}^{60,0}}{64 \sqrt{5}}+\frac{7 C_{00,20}^{60,1}}{128 \sqrt{5}}-\frac{21
C_{00,22}^{62,0}}{640}-\frac{21 C_{00,22}^{62,1}}{1280}+\frac{7 C_{11,22}^{51,0}}{1920}+\frac{7 C_{11,22}^{51,1}}{3840}-\frac{9}{640} \sqrt{\frac{21}{5}}
C_{11,22}^{53,0}-\frac{9 \sqrt{\frac{21}{5}} C_{11,22}^{53,1}}{1280}-\frac{7 C_{20,20}^{40,0}}{192 \sqrt{5}}-\frac{7 C_{20,20}^{40,1}}{384 \sqrt{5}}-\frac{7
C_{20,22}^{42,0}}{1920}-\frac{7 C_{20,22}^{42,1}}{3840}+\frac{7 C_{22,20}^{42,0}}{960}+\frac{7 C_{22,20}^{42,1}}{1920}+\frac{17 C_{22,22}^{40,0}}{1920}+\frac{17
C_{22,22}^{40,1}}{3840}+\frac{\sqrt{7} C_{22,22}^{42,0}}{192}+\frac{\sqrt{7} C_{22,22}^{42,1}}{384}-\frac{9 C_{22,22}^{44,0}}{160 \sqrt{5}}-\frac{9
C_{22,22}^{44,1}}{320 \sqrt{5}}-\frac{7 C_{31,20}^{31,0}}{192 \sqrt{5}}-\frac{7 C_{31,20}^{31,1}}{384 \sqrt{5}}+\frac{23 C_{31,22}^{31,0}}{1920}+\frac{23
C_{31,22}^{31,1}}{3840}-\frac{1}{640} \sqrt{\frac{21}{5}} C_{31,22}^{33,0}-\frac{\sqrt{\frac{21}{5}} C_{31,22}^{33,1}}{1280}+\frac{3}{320} \sqrt{\frac{21}{5}}
C_{33,20}^{33,0}+\frac{3}{640} \sqrt{\frac{21}{5}} C_{33,20}^{33,1}+\frac{9}{800} \sqrt{\frac{7}{2}} C_{33,22}^{33,0}+\frac{9 \sqrt{\frac{7}{2}}
C_{33,22}^{33,1}}{1600}\)

\noindent\(C_{22,1111,2}^{00,3112,0,\text{NonLoc}}= \frac{7}{256} \sqrt{\frac{3}{5}} C_{00,00}^{60,0}+\frac{7}{128} \sqrt{\frac{3}{5}}
C_{00,00}^{60,1}+\frac{7 C_{00,20}^{60,0}}{256 \sqrt{5}}+\frac{7 C_{00,20}^{60,1}}{128 \sqrt{5}}+\frac{21 C_{00,22}^{62,0}}{1280}+\frac{21 C_{00,22}^{62,1}}{640}-\frac{7
C_{11,00}^{51,0}}{256 \sqrt{15}}-\frac{7 C_{11,00}^{51,1}}{128 \sqrt{15}}+\frac{7 C_{11,22}^{51,0}}{3840}+\frac{7 C_{11,22}^{51,1}}{1920}-\frac{9
\sqrt{\frac{21}{5}} C_{11,22}^{53,0}}{1280}-\frac{9}{640} \sqrt{\frac{21}{5}} C_{11,22}^{53,1}-\frac{7 C_{20,00}^{40,0}}{256 \sqrt{15}}-\frac{7 C_{20,00}^{40,1}}{128
\sqrt{15}}-\frac{7 C_{20,20}^{40,0}}{768 \sqrt{5}}-\frac{7 C_{20,20}^{40,1}}{384 \sqrt{5}}+\frac{7 C_{20,22}^{42,0}}{3840}+\frac{7 C_{20,22}^{42,1}}{1920}+\frac{7
C_{22,00}^{42,0}}{1280 \sqrt{3}}+\frac{7 C_{22,00}^{42,1}}{640 \sqrt{3}}+\frac{7 C_{22,20}^{42,0}}{3840}+\frac{7 C_{22,20}^{42,1}}{1920}-\frac{17
C_{22,22}^{40,0}}{3840}-\frac{17 C_{22,22}^{40,1}}{1920}-\frac{\sqrt{7} C_{22,22}^{42,0}}{384}-\frac{\sqrt{7} C_{22,22}^{42,1}}{192}+\frac{9 C_{22,22}^{44,0}}{320
\sqrt{5}}+\frac{9 C_{22,22}^{44,1}}{160 \sqrt{5}}+\frac{7 C_{31,00}^{31,0}}{256 \sqrt{15}}+\frac{7 C_{31,00}^{31,1}}{128 \sqrt{15}}+\frac{7 C_{31,20}^{31,0}}{768
\sqrt{5}}+\frac{7 C_{31,20}^{31,1}}{384 \sqrt{5}}+\frac{23 C_{31,22}^{31,0}}{3840}+\frac{23 C_{31,22}^{31,1}}{1920}-\frac{\sqrt{\frac{21}{5}} C_{31,22}^{33,0}}{1280}-\frac{1}{640}
\sqrt{\frac{21}{5}} C_{31,22}^{33,1}-\frac{9 \sqrt{\frac{7}{5}} C_{33,00}^{33,0}}{1280}-\frac{9}{640} \sqrt{\frac{7}{5}} C_{33,00}^{33,1}-\frac{3
\sqrt{\frac{21}{5}} C_{33,20}^{33,0}}{1280}-\frac{3}{640} \sqrt{\frac{21}{5}} C_{33,20}^{33,1}+\frac{9 \sqrt{\frac{7}{2}} C_{33,22}^{33,0}}{1600}+\frac{9}{800}
\sqrt{\frac{7}{2}} C_{33,22}^{33,1}\)

\noindent\(C_{22,1111,2}^{00,3112,1,\text{Loc}}= \frac{7}{128} \sqrt{\frac{3}{5}} C_{00,20}^{60,1}-\frac{21 \sqrt{3} C_{00,22}^{62,1}}{1280}+\frac{7
C_{11,22}^{51,1}}{1280 \sqrt{3}}-\frac{27 \sqrt{\frac{7}{5}} C_{11,22}^{53,1}}{1280}-\frac{7 C_{20,20}^{40,1}}{128 \sqrt{15}}-\frac{7 C_{20,22}^{42,1}}{1280
\sqrt{3}}+\frac{7 C_{22,20}^{42,1}}{640 \sqrt{3}}+\frac{17 C_{22,22}^{40,1}}{1280 \sqrt{3}}+\frac{1}{128} \sqrt{\frac{7}{3}} C_{22,22}^{42,1}-\frac{9}{320}
\sqrt{\frac{3}{5}} C_{22,22}^{44,1}-\frac{7 C_{31,20}^{31,1}}{128 \sqrt{15}}+\frac{23 C_{31,22}^{31,1}}{1280 \sqrt{3}}-\frac{3 \sqrt{\frac{7}{5}}
C_{31,22}^{33,1}}{1280}+\frac{9}{640} \sqrt{\frac{7}{5}} C_{33,20}^{33,1}+\frac{9 \sqrt{\frac{21}{2}} C_{33,22}^{33,1}}{1600}\)

\noindent\(C_{22,1111,2}^{00,3112,1,\text{NonLoc}}= \frac{21 C_{00,00}^{60,0}}{256 \sqrt{5}}+\frac{7}{256} \sqrt{\frac{3}{5}} C_{00,20}^{60,0}+\frac{21
\sqrt{3} C_{00,22}^{62,0}}{1280}-\frac{7 C_{11,00}^{51,0}}{256 \sqrt{5}}+\frac{7 C_{11,22}^{51,0}}{1280 \sqrt{3}}-\frac{27 \sqrt{\frac{7}{5}} C_{11,22}^{53,0}}{1280}-\frac{7
C_{20,00}^{40,0}}{256 \sqrt{5}}-\frac{7 C_{20,20}^{40,0}}{256 \sqrt{15}}+\frac{7 C_{20,22}^{42,0}}{1280 \sqrt{3}}+\frac{7 C_{22,00}^{42,0}}{1280}+\frac{7
C_{22,20}^{42,0}}{1280 \sqrt{3}}-\frac{17 C_{22,22}^{40,0}}{1280 \sqrt{3}}-\frac{1}{128} \sqrt{\frac{7}{3}} C_{22,22}^{42,0}+\frac{9}{320} \sqrt{\frac{3}{5}}
C_{22,22}^{44,0}+\frac{7 C_{31,00}^{31,0}}{256 \sqrt{5}}+\frac{7 C_{31,20}^{31,0}}{256 \sqrt{15}}+\frac{23 C_{31,22}^{31,0}}{1280 \sqrt{3}}-\frac{3
\sqrt{\frac{7}{5}} C_{31,22}^{33,0}}{1280}-\frac{9 \sqrt{\frac{21}{5}} C_{33,00}^{33,0}}{1280}-\frac{9 \sqrt{\frac{7}{5}} C_{33,20}^{33,0}}{1280}+\frac{9
\sqrt{\frac{21}{2}} C_{33,22}^{33,0}}{1600}\)

\noindent\(C_{22,1111,2}^{00,3312,0,\text{Loc}}= -\frac{1}{96} \sqrt{\frac{7}{3}} C_{00,20}^{60,0}-\frac{1}{192} \sqrt{\frac{7}{3}}
C_{00,20}^{60,1}+\frac{1}{64} \sqrt{\frac{7}{15}} C_{00,22}^{62,0}+\frac{1}{128} \sqrt{\frac{7}{15}} C_{00,22}^{62,1}+\frac{7}{288} \sqrt{\frac{7}{15}}
C_{11,22}^{51,0}+\frac{7}{576} \sqrt{\frac{7}{15}} C_{11,22}^{51,1}-\frac{9 C_{11,22}^{53,0}}{640}-\frac{9 C_{11,22}^{53,1}}{1280}+\frac{1}{288}
\sqrt{\frac{7}{3}} C_{20,20}^{40,0}+\frac{1}{576} \sqrt{\frac{7}{3}} C_{20,20}^{40,1}-\frac{7}{288} \sqrt{\frac{7}{15}} C_{20,22}^{42,0}-\frac{7}{576}
\sqrt{\frac{7}{15}} C_{20,22}^{42,1}-\frac{1}{288} \sqrt{\frac{7}{15}} C_{22,20}^{42,0}-\frac{1}{576} \sqrt{\frac{7}{15}} C_{22,20}^{42,1}+\frac{19}{576}
\sqrt{\frac{7}{15}} C_{22,22}^{40,0}+\frac{19 \sqrt{\frac{7}{15}} C_{22,22}^{40,1}}{1152}-\frac{1}{288} \sqrt{\frac{5}{3}} C_{22,22}^{42,0}-\frac{1}{576}
\sqrt{\frac{5}{3}} C_{22,22}^{42,1}+\frac{1}{80} \sqrt{\frac{3}{7}} C_{22,22}^{44,0}+\frac{1}{160} \sqrt{\frac{3}{7}} C_{22,22}^{44,1}+\frac{1}{288}
\sqrt{\frac{7}{3}} C_{31,20}^{31,0}+\frac{1}{576} \sqrt{\frac{7}{3}} C_{31,20}^{31,1}-\frac{7}{288} \sqrt{\frac{7}{15}} C_{31,22}^{31,0}-\frac{7}{576}
\sqrt{\frac{7}{15}} C_{31,22}^{31,1}+\frac{9 C_{31,22}^{33,0}}{640}+\frac{9 C_{31,22}^{33,1}}{1280}-\frac{C_{33,20}^{33,0}}{160}-\frac{C_{33,20}^{33,1}}{320}-\frac{1}{80}
\sqrt{\frac{3}{10}} C_{33,22}^{33,0}-\frac{1}{160} \sqrt{\frac{3}{10}} C_{33,22}^{33,1}\)

\noindent\(C_{22,1111,2}^{00,3312,0,\text{NonLoc}}= -\frac{\sqrt{7} C_{00,00}^{60,0}}{384}-\frac{\sqrt{7} C_{00,00}^{60,1}}{192}-\frac{1}{384}
\sqrt{\frac{7}{3}} C_{00,20}^{60,0}-\frac{1}{192} \sqrt{\frac{7}{3}} C_{00,20}^{60,1}-\frac{1}{128} \sqrt{\frac{7}{15}} C_{00,22}^{62,0}-\frac{1}{64}
\sqrt{\frac{7}{15}} C_{00,22}^{62,1}+\frac{\sqrt{7} C_{11,00}^{51,0}}{1152}+\frac{\sqrt{7} C_{11,00}^{51,1}}{576}+\frac{7}{576} \sqrt{\frac{7}{15}}
C_{11,22}^{51,0}+\frac{7}{288} \sqrt{\frac{7}{15}} C_{11,22}^{51,1}-\frac{9 C_{11,22}^{53,0}}{1280}-\frac{9 C_{11,22}^{53,1}}{640}+\frac{\sqrt{7}
C_{20,00}^{40,0}}{1152}+\frac{\sqrt{7} C_{20,00}^{40,1}}{576}+\frac{\sqrt{\frac{7}{3}} C_{20,20}^{40,0}}{1152}+\frac{1}{576} \sqrt{\frac{7}{3}} C_{20,20}^{40,1}+\frac{7}{576}
\sqrt{\frac{7}{15}} C_{20,22}^{42,0}+\frac{7}{288} \sqrt{\frac{7}{15}} C_{20,22}^{42,1}-\frac{\sqrt{\frac{7}{5}} C_{22,00}^{42,0}}{1152}-\frac{1}{576}
\sqrt{\frac{7}{5}} C_{22,00}^{42,1}-\frac{\sqrt{\frac{7}{15}} C_{22,20}^{42,0}}{1152}-\frac{1}{576} \sqrt{\frac{7}{15}} C_{22,20}^{42,1}-\frac{19
\sqrt{\frac{7}{15}} C_{22,22}^{40,0}}{1152}-\frac{19}{576} \sqrt{\frac{7}{15}} C_{22,22}^{40,1}+\frac{1}{576} \sqrt{\frac{5}{3}} C_{22,22}^{42,0}+\frac{1}{288}
\sqrt{\frac{5}{3}} C_{22,22}^{42,1}-\frac{1}{160} \sqrt{\frac{3}{7}} C_{22,22}^{44,0}-\frac{1}{80} \sqrt{\frac{3}{7}} C_{22,22}^{44,1}-\frac{\sqrt{7}
C_{31,00}^{31,0}}{1152}-\frac{\sqrt{7} C_{31,00}^{31,1}}{576}-\frac{\sqrt{\frac{7}{3}} C_{31,20}^{31,0}}{1152}-\frac{1}{576} \sqrt{\frac{7}{3}} C_{31,20}^{31,1}-\frac{7}{576}
\sqrt{\frac{7}{15}} C_{31,22}^{31,0}-\frac{7}{288} \sqrt{\frac{7}{15}} C_{31,22}^{31,1}+\frac{9 C_{31,22}^{33,0}}{1280}+\frac{9 C_{31,22}^{33,1}}{640}+\frac{\sqrt{3}
C_{33,00}^{33,0}}{640}+\frac{\sqrt{3} C_{33,00}^{33,1}}{320}+\frac{C_{33,20}^{33,0}}{640}+\frac{C_{33,20}^{33,1}}{320}-\frac{1}{160} \sqrt{\frac{3}{10}}
C_{33,22}^{33,0}-\frac{1}{80} \sqrt{\frac{3}{10}} C_{33,22}^{33,1}\)

\noindent\(C_{22,1111,2}^{00,3312,1,\text{Loc}}= -\frac{\sqrt{7} C_{00,20}^{60,1}}{192}+\frac{1}{128} \sqrt{\frac{7}{5}} C_{00,22}^{62,1}+\frac{7}{576}
\sqrt{\frac{7}{5}} C_{11,22}^{51,1}-\frac{9 \sqrt{3} C_{11,22}^{53,1}}{1280}+\frac{\sqrt{7} C_{20,20}^{40,1}}{576}-\frac{7}{576} \sqrt{\frac{7}{5}}
C_{20,22}^{42,1}-\frac{1}{576} \sqrt{\frac{7}{5}} C_{22,20}^{42,1}+\frac{19 \sqrt{\frac{7}{5}} C_{22,22}^{40,1}}{1152}-\frac{\sqrt{5} C_{22,22}^{42,1}}{576}+\frac{3
C_{22,22}^{44,1}}{160 \sqrt{7}}+\frac{\sqrt{7} C_{31,20}^{31,1}}{576}-\frac{7}{576} \sqrt{\frac{7}{5}} C_{31,22}^{31,1}+\frac{9 \sqrt{3} C_{31,22}^{33,1}}{1280}-\frac{\sqrt{3}
C_{33,20}^{33,1}}{320}-\frac{3 C_{33,22}^{33,1}}{160 \sqrt{10}}\)

\noindent\(C_{22,1111,2}^{00,3312,1,\text{NonLoc}}= -\frac{1}{128} \sqrt{\frac{7}{3}} C_{00,00}^{60,0}-\frac{\sqrt{7} C_{00,20}^{60,0}}{384}-\frac{1}{128}
\sqrt{\frac{7}{5}} C_{00,22}^{62,0}+\frac{1}{384} \sqrt{\frac{7}{3}} C_{11,00}^{51,0}+\frac{7}{576} \sqrt{\frac{7}{5}} C_{11,22}^{51,0}-\frac{9 \sqrt{3}
C_{11,22}^{53,0}}{1280}+\frac{1}{384} \sqrt{\frac{7}{3}} C_{20,00}^{40,0}+\frac{\sqrt{7} C_{20,20}^{40,0}}{1152}+\frac{7}{576} \sqrt{\frac{7}{5}}
C_{20,22}^{42,0}-\frac{1}{384} \sqrt{\frac{7}{15}} C_{22,00}^{42,0}-\frac{\sqrt{\frac{7}{5}} C_{22,20}^{42,0}}{1152}-\frac{19 \sqrt{\frac{7}{5}}
C_{22,22}^{40,0}}{1152}+\frac{\sqrt{5} C_{22,22}^{42,0}}{576}-\frac{3 C_{22,22}^{44,0}}{160 \sqrt{7}}-\frac{1}{384} \sqrt{\frac{7}{3}} C_{31,00}^{31,0}-\frac{\sqrt{7}
C_{31,20}^{31,0}}{1152}-\frac{7}{576} \sqrt{\frac{7}{5}} C_{31,22}^{31,0}+\frac{9 \sqrt{3} C_{31,22}^{33,0}}{1280}+\frac{3 C_{33,00}^{33,0}}{640}+\frac{\sqrt{3}
C_{33,20}^{33,0}}{640}-\frac{3 C_{33,22}^{33,0}}{160 \sqrt{10}}\)

\noindent\(C_{22,1111,2}^{11,2202,0,\text{Loc}}= \frac{7}{384} \sqrt{\frac{5}{3}} C_{11,11}^{51,0}+\frac{7}{768} \sqrt{\frac{5}{3}}
C_{11,11}^{51,1}-\frac{5}{384} \sqrt{\frac{5}{3}} C_{31,11}^{31,0}-\frac{5}{768} \sqrt{\frac{5}{3}} C_{31,11}^{31,1}\)

\noindent\(C_{22,1111,2}^{11,2202,0,\text{NonLoc}}= \frac{7}{768} \sqrt{\frac{5}{3}} C_{11,11}^{51,0}+\frac{7}{384} \sqrt{\frac{5}{3}}
C_{11,11}^{51,1}-\frac{5}{768} \sqrt{\frac{5}{3}} C_{31,11}^{31,0}-\frac{5}{384} \sqrt{\frac{5}{3}} C_{31,11}^{31,1}\)

\noindent\(C_{22,1111,2}^{11,2202,1,\text{Loc}}= \frac{7 \sqrt{5} C_{11,11}^{51,1}}{768}-\frac{5 \sqrt{5} C_{31,11}^{31,1}}{768}\)

\noindent\(C_{22,1111,2}^{11,2202,1,\text{NonLoc}}= \frac{7 \sqrt{5} C_{11,11}^{51,0}}{768}-\frac{5 \sqrt{5} C_{31,11}^{31,0}}{768}\)

\noindent\(C_{22,1111,2}^{20,1112,0,\text{Loc}}= -\frac{7 \sqrt{5} C_{00,20}^{60,0}}{768}-\frac{7 \sqrt{5} C_{00,20}^{60,1}}{1536}+\frac{7
C_{00,22}^{62,0}}{512}+\frac{7 C_{00,22}^{62,1}}{1024}-\frac{35 C_{11,22}^{51,0}}{4608}-\frac{35 C_{11,22}^{51,1}}{9216}+\frac{7 \sqrt{5} C_{20,20}^{40,0}}{2304}+\frac{7
\sqrt{5} C_{20,20}^{40,1}}{4608}+\frac{7 C_{20,22}^{42,0}}{4608}+\frac{7 C_{20,22}^{42,1}}{9216}-\frac{7 C_{22,20}^{42,0}}{2304}-\frac{7 C_{22,20}^{42,1}}{4608}+\frac{7
C_{22,22}^{40,0}}{4608}+\frac{7 C_{22,22}^{40,1}}{9216}-\frac{\sqrt{7} C_{22,22}^{42,0}}{1152}-\frac{\sqrt{7} C_{22,22}^{42,1}}{2304}-\frac{3 C_{22,22}^{44,0}}{128
\sqrt{5}}-\frac{3 C_{22,22}^{44,1}}{256 \sqrt{5}}-\frac{7 \sqrt{5} C_{31,20}^{31,0}}{2304}-\frac{7 \sqrt{5} C_{31,20}^{31,1}}{4608}-\frac{7 C_{31,22}^{31,0}}{4608}-\frac{7
C_{31,22}^{31,1}}{9216}+\frac{1}{128} \sqrt{\frac{7}{15}} C_{31,22}^{33,0}+\frac{1}{256} \sqrt{\frac{7}{15}} C_{31,22}^{33,1}+\frac{1}{256} \sqrt{\frac{21}{5}}
C_{33,20}^{33,0}+\frac{1}{512} \sqrt{\frac{21}{5}} C_{33,20}^{33,1}+\frac{3}{640} \sqrt{\frac{7}{2}} C_{33,22}^{33,0}+\frac{3 \sqrt{\frac{7}{2}}
C_{33,22}^{33,1}}{1280}\)

\noindent\(C_{22,1111,2}^{20,1112,0,\text{NonLoc}}= -\frac{7 \sqrt{\frac{5}{3}} C_{00,00}^{60,0}}{1024}-\frac{7}{512} \sqrt{\frac{5}{3}}
C_{00,00}^{60,1}-\frac{7 \sqrt{5} C_{00,20}^{60,0}}{3072}-\frac{7 \sqrt{5} C_{00,20}^{60,1}}{1536}-\frac{7 C_{00,22}^{62,0}}{1024}-\frac{7 C_{00,22}^{62,1}}{512}-\frac{7
\sqrt{\frac{5}{3}} C_{11,00}^{51,0}}{3072}-\frac{7 \sqrt{\frac{5}{3}} C_{11,00}^{51,1}}{1536}-\frac{35 C_{11,22}^{51,0}}{9216}-\frac{35 C_{11,22}^{51,1}}{4608}+\frac{7
\sqrt{\frac{5}{3}} C_{20,00}^{40,0}}{3072}+\frac{7 \sqrt{\frac{5}{3}} C_{20,00}^{40,1}}{1536}+\frac{7 \sqrt{5} C_{20,20}^{40,0}}{9216}+\frac{7 \sqrt{5}
C_{20,20}^{40,1}}{4608}-\frac{7 C_{20,22}^{42,0}}{9216}-\frac{7 C_{20,22}^{42,1}}{4608}-\frac{7 C_{22,00}^{42,0}}{3072 \sqrt{3}}-\frac{7 C_{22,00}^{42,1}}{1536
\sqrt{3}}-\frac{7 C_{22,20}^{42,0}}{9216}-\frac{7 C_{22,20}^{42,1}}{4608}-\frac{7 C_{22,22}^{40,0}}{9216}-\frac{7 C_{22,22}^{40,1}}{4608}+\frac{\sqrt{7}
C_{22,22}^{42,0}}{2304}+\frac{\sqrt{7} C_{22,22}^{42,1}}{1152}+\frac{3 C_{22,22}^{44,0}}{256 \sqrt{5}}+\frac{3 C_{22,22}^{44,1}}{128 \sqrt{5}}+\frac{7
\sqrt{\frac{5}{3}} C_{31,00}^{31,0}}{3072}+\frac{7 \sqrt{\frac{5}{3}} C_{31,00}^{31,1}}{1536}+\frac{7 \sqrt{5} C_{31,20}^{31,0}}{9216}+\frac{7 \sqrt{5}
C_{31,20}^{31,1}}{4608}-\frac{7 C_{31,22}^{31,0}}{9216}-\frac{7 C_{31,22}^{31,1}}{4608}+\frac{1}{256} \sqrt{\frac{7}{15}} C_{31,22}^{33,0}+\frac{1}{128}
\sqrt{\frac{7}{15}} C_{31,22}^{33,1}-\frac{3 \sqrt{\frac{7}{5}} C_{33,00}^{33,0}}{1024}-\frac{3}{512} \sqrt{\frac{7}{5}} C_{33,00}^{33,1}-\frac{\sqrt{\frac{21}{5}}
C_{33,20}^{33,0}}{1024}-\frac{1}{512} \sqrt{\frac{21}{5}} C_{33,20}^{33,1}+\frac{3 \sqrt{\frac{7}{2}} C_{33,22}^{33,0}}{1280}+\frac{3}{640} \sqrt{\frac{7}{2}}
C_{33,22}^{33,1}\)

\noindent\(C_{22,1111,2}^{20,1112,1,\text{Loc}}= -\frac{7}{512} \sqrt{\frac{5}{3}} C_{00,20}^{60,1}+\frac{7 \sqrt{3} C_{00,22}^{62,1}}{1024}-\frac{35
C_{11,22}^{51,1}}{3072 \sqrt{3}}+\frac{7 \sqrt{\frac{5}{3}} C_{20,20}^{40,1}}{1536}+\frac{7 C_{20,22}^{42,1}}{3072 \sqrt{3}}-\frac{7 C_{22,20}^{42,1}}{1536
\sqrt{3}}+\frac{7 C_{22,22}^{40,1}}{3072 \sqrt{3}}-\frac{1}{768} \sqrt{\frac{7}{3}} C_{22,22}^{42,1}-\frac{3}{256} \sqrt{\frac{3}{5}} C_{22,22}^{44,1}-\frac{7
\sqrt{\frac{5}{3}} C_{31,20}^{31,1}}{1536}-\frac{7 C_{31,22}^{31,1}}{3072 \sqrt{3}}+\frac{1}{256} \sqrt{\frac{7}{5}} C_{31,22}^{33,1}+\frac{3}{512}
\sqrt{\frac{7}{5}} C_{33,20}^{33,1}+\frac{3 \sqrt{\frac{21}{2}} C_{33,22}^{33,1}}{1280}\)

\noindent\(C_{22,1111,2}^{20,1112,1,\text{NonLoc}}= -\frac{7 \sqrt{5} C_{00,00}^{60,0}}{1024}-\frac{7 \sqrt{\frac{5}{3}} C_{00,20}^{60,0}}{1024}-\frac{7
\sqrt{3} C_{00,22}^{62,0}}{1024}-\frac{7 \sqrt{5} C_{11,00}^{51,0}}{3072}-\frac{35 C_{11,22}^{51,0}}{3072 \sqrt{3}}+\frac{7 \sqrt{5} C_{20,00}^{40,0}}{3072}+\frac{7
\sqrt{\frac{5}{3}} C_{20,20}^{40,0}}{3072}-\frac{7 C_{20,22}^{42,0}}{3072 \sqrt{3}}-\frac{7 C_{22,00}^{42,0}}{3072}-\frac{7 C_{22,20}^{42,0}}{3072
\sqrt{3}}-\frac{7 C_{22,22}^{40,0}}{3072 \sqrt{3}}+\frac{1}{768} \sqrt{\frac{7}{3}} C_{22,22}^{42,0}+\frac{3}{256} \sqrt{\frac{3}{5}} C_{22,22}^{44,0}+\frac{7
\sqrt{5} C_{31,00}^{31,0}}{3072}+\frac{7 \sqrt{\frac{5}{3}} C_{31,20}^{31,0}}{3072}-\frac{7 C_{31,22}^{31,0}}{3072 \sqrt{3}}+\frac{1}{256} \sqrt{\frac{7}{5}}
C_{31,22}^{33,0}-\frac{3 \sqrt{\frac{21}{5}} C_{33,00}^{33,0}}{1024}-\frac{3 \sqrt{\frac{7}{5}} C_{33,20}^{33,0}}{1024}+\frac{3 \sqrt{\frac{21}{2}}
C_{33,22}^{33,0}}{1280}\)

\noindent\(C_{22,1111,2}^{22,1111,0,\text{Loc}}= \frac{7 \sqrt{5} C_{00,20}^{60,0}}{1536}+\frac{7 \sqrt{5} C_{00,20}^{60,1}}{3072}-\frac{7
C_{00,22}^{62,0}}{1024}-\frac{7 C_{00,22}^{62,1}}{2048}-\frac{7 C_{11,22}^{51,0}}{9216}-\frac{7 C_{11,22}^{51,1}}{18432}+\frac{3 \sqrt{\frac{21}{5}}
C_{11,22}^{53,0}}{1024}+\frac{3 \sqrt{\frac{21}{5}} C_{11,22}^{53,1}}{2048}+\frac{\sqrt{5} C_{20,20}^{40,0}}{4608}+\frac{\sqrt{5} C_{20,20}^{40,1}}{9216}+\frac{7
C_{20,22}^{42,0}}{9216}+\frac{7 C_{20,22}^{42,1}}{18432}-\frac{7 C_{22,20}^{42,0}}{4608}-\frac{7 C_{22,20}^{42,1}}{9216}+\frac{37 C_{22,22}^{40,0}}{9216}+\frac{37
C_{22,22}^{40,1}}{18432}-\frac{\sqrt{7} C_{22,22}^{42,0}}{2304}-\frac{\sqrt{7} C_{22,22}^{42,1}}{4608}-\frac{3 C_{22,22}^{44,0}}{256 \sqrt{5}}-\frac{3
C_{22,22}^{44,1}}{512 \sqrt{5}}+\frac{\sqrt{5} C_{31,20}^{31,0}}{1536}+\frac{\sqrt{5} C_{31,20}^{31,1}}{3072}-\frac{3 C_{31,22}^{31,0}}{1024}-\frac{3
C_{31,22}^{31,1}}{2048}-\frac{\sqrt{\frac{7}{15}} C_{31,22}^{33,0}}{1024}-\frac{\sqrt{\frac{7}{15}} C_{31,22}^{33,1}}{2048}+\frac{1}{512} \sqrt{\frac{21}{5}}
C_{33,20}^{33,0}+\frac{\sqrt{\frac{21}{5}} C_{33,20}^{33,1}}{1024}+\frac{3 \sqrt{\frac{7}{2}} C_{33,22}^{33,0}}{1280}+\frac{3 \sqrt{\frac{7}{2}}
C_{33,22}^{33,1}}{2560}\)

\noindent\(C_{22,1111,2}^{22,1111,0,\text{NonLoc}}= \frac{7 \sqrt{\frac{5}{3}} C_{00,00}^{60,0}}{2048}+\frac{7 \sqrt{\frac{5}{3}} C_{00,00}^{60,1}}{1024}+\frac{7
\sqrt{5} C_{00,20}^{60,0}}{6144}+\frac{7 \sqrt{5} C_{00,20}^{60,1}}{3072}+\frac{7 C_{00,22}^{62,0}}{2048}+\frac{7 C_{00,22}^{62,1}}{1024}+\frac{7
\sqrt{\frac{5}{3}} C_{11,00}^{51,0}}{6144}+\frac{7 \sqrt{\frac{5}{3}} C_{11,00}^{51,1}}{3072}-\frac{7 C_{11,22}^{51,0}}{18432}-\frac{7 C_{11,22}^{51,1}}{9216}+\frac{3
\sqrt{\frac{21}{5}} C_{11,22}^{53,0}}{2048}+\frac{3 \sqrt{\frac{21}{5}} C_{11,22}^{53,1}}{1024}+\frac{\sqrt{\frac{5}{3}} C_{20,00}^{40,0}}{6144}+\frac{\sqrt{\frac{5}{3}}
C_{20,00}^{40,1}}{3072}+\frac{\sqrt{5} C_{20,20}^{40,0}}{18432}+\frac{\sqrt{5} C_{20,20}^{40,1}}{9216}-\frac{7 C_{20,22}^{42,0}}{18432}-\frac{7 C_{20,22}^{42,1}}{9216}-\frac{7
C_{22,00}^{42,0}}{6144 \sqrt{3}}-\frac{7 C_{22,00}^{42,1}}{3072 \sqrt{3}}-\frac{7 C_{22,20}^{42,0}}{18432}-\frac{7 C_{22,20}^{42,1}}{9216}-\frac{37
C_{22,22}^{40,0}}{18432}-\frac{37 C_{22,22}^{40,1}}{9216}+\frac{\sqrt{7} C_{22,22}^{42,0}}{4608}+\frac{\sqrt{7} C_{22,22}^{42,1}}{2304}+\frac{3 C_{22,22}^{44,0}}{512
\sqrt{5}}+\frac{3 C_{22,22}^{44,1}}{256 \sqrt{5}}-\frac{\sqrt{\frac{5}{3}} C_{31,00}^{31,0}}{2048}-\frac{\sqrt{\frac{5}{3}} C_{31,00}^{31,1}}{1024}-\frac{\sqrt{5}
C_{31,20}^{31,0}}{6144}-\frac{\sqrt{5} C_{31,20}^{31,1}}{3072}-\frac{3 C_{31,22}^{31,0}}{2048}-\frac{3 C_{31,22}^{31,1}}{1024}-\frac{\sqrt{\frac{7}{15}}
C_{31,22}^{33,0}}{2048}-\frac{\sqrt{\frac{7}{15}} C_{31,22}^{33,1}}{1024}-\frac{3 \sqrt{\frac{7}{5}} C_{33,00}^{33,0}}{2048}-\frac{3 \sqrt{\frac{7}{5}}
C_{33,00}^{33,1}}{1024}-\frac{\sqrt{\frac{21}{5}} C_{33,20}^{33,0}}{2048}-\frac{\sqrt{\frac{21}{5}} C_{33,20}^{33,1}}{1024}+\frac{3 \sqrt{\frac{7}{2}}
C_{33,22}^{33,0}}{2560}+\frac{3 \sqrt{\frac{7}{2}} C_{33,22}^{33,1}}{1280}\)

\noindent\(C_{22,1111,2}^{22,1111,1,\text{Loc}}= \frac{7 \sqrt{\frac{5}{3}} C_{00,20}^{60,1}}{1024}-\frac{7 \sqrt{3} C_{00,22}^{62,1}}{2048}-\frac{7
C_{11,22}^{51,1}}{6144 \sqrt{3}}+\frac{9 \sqrt{\frac{7}{5}} C_{11,22}^{53,1}}{2048}+\frac{\sqrt{\frac{5}{3}} C_{20,20}^{40,1}}{3072}+\frac{7 C_{20,22}^{42,1}}{6144
\sqrt{3}}-\frac{7 C_{22,20}^{42,1}}{3072 \sqrt{3}}+\frac{37 C_{22,22}^{40,1}}{6144 \sqrt{3}}-\frac{\sqrt{\frac{7}{3}} C_{22,22}^{42,1}}{1536}-\frac{3}{512}
\sqrt{\frac{3}{5}} C_{22,22}^{44,1}+\frac{\sqrt{\frac{5}{3}} C_{31,20}^{31,1}}{1024}-\frac{3 \sqrt{3} C_{31,22}^{31,1}}{2048}-\frac{\sqrt{\frac{7}{5}}
C_{31,22}^{33,1}}{2048}+\frac{3 \sqrt{\frac{7}{5}} C_{33,20}^{33,1}}{1024}+\frac{3 \sqrt{\frac{21}{2}} C_{33,22}^{33,1}}{2560}\)

\noindent\(C_{22,1111,2}^{22,1111,1,\text{NonLoc}}= \frac{7 \sqrt{5} C_{00,00}^{60,0}}{2048}+\frac{7 \sqrt{\frac{5}{3}} C_{00,20}^{60,0}}{2048}+\frac{7
\sqrt{3} C_{00,22}^{62,0}}{2048}+\frac{7 \sqrt{5} C_{11,00}^{51,0}}{6144}-\frac{7 C_{11,22}^{51,0}}{6144 \sqrt{3}}+\frac{9 \sqrt{\frac{7}{5}} C_{11,22}^{53,0}}{2048}+\frac{\sqrt{5}
C_{20,00}^{40,0}}{6144}+\frac{\sqrt{\frac{5}{3}} C_{20,20}^{40,0}}{6144}-\frac{7 C_{20,22}^{42,0}}{6144 \sqrt{3}}-\frac{7 C_{22,00}^{42,0}}{6144}-\frac{7
C_{22,20}^{42,0}}{6144 \sqrt{3}}-\frac{37 C_{22,22}^{40,0}}{6144 \sqrt{3}}+\frac{\sqrt{\frac{7}{3}} C_{22,22}^{42,0}}{1536}+\frac{3}{512} \sqrt{\frac{3}{5}}
C_{22,22}^{44,0}-\frac{\sqrt{5} C_{31,00}^{31,0}}{2048}-\frac{\sqrt{\frac{5}{3}} C_{31,20}^{31,0}}{2048}-\frac{3 \sqrt{3} C_{31,22}^{31,0}}{2048}-\frac{\sqrt{\frac{7}{5}}
C_{31,22}^{33,0}}{2048}-\frac{3 \sqrt{\frac{21}{5}} C_{33,00}^{33,0}}{2048}-\frac{3 \sqrt{\frac{7}{5}} C_{33,20}^{33,0}}{2048}+\frac{3 \sqrt{\frac{21}{2}}
C_{33,22}^{33,0}}{2560}\)

\noindent\(C_{22,1111,3}^{00,3313,0,\text{Loc}}= -\frac{1}{48} \sqrt{\frac{7}{6}} C_{00,20}^{60,0}-\frac{1}{96} \sqrt{\frac{7}{6}}
C_{00,20}^{60,1}+\frac{1}{32} \sqrt{\frac{7}{30}} C_{00,22}^{62,0}+\frac{1}{64} \sqrt{\frac{7}{30}} C_{00,22}^{62,1}+\frac{1}{288} \sqrt{\frac{35}{6}}
C_{11,22}^{51,0}+\frac{1}{576} \sqrt{\frac{35}{6}} C_{11,22}^{51,1}+\frac{1}{144} \sqrt{\frac{7}{6}} C_{20,20}^{40,0}+\frac{1}{288} \sqrt{\frac{7}{6}}
C_{20,20}^{40,1}-\frac{1}{288} \sqrt{\frac{35}{6}} C_{20,22}^{42,0}-\frac{1}{576} \sqrt{\frac{35}{6}} C_{20,22}^{42,1}-\frac{1}{144} \sqrt{\frac{7}{30}}
C_{22,20}^{42,0}-\frac{1}{288} \sqrt{\frac{7}{30}} C_{22,20}^{42,1}+\frac{1}{288} \sqrt{\frac{7}{30}} C_{22,22}^{40,0}+\frac{1}{576} \sqrt{\frac{7}{30}}
C_{22,22}^{40,1}+\frac{C_{22,22}^{42,0}}{36 \sqrt{30}}+\frac{C_{22,22}^{42,1}}{72 \sqrt{30}}+\frac{1}{160} \sqrt{\frac{3}{14}} C_{22,22}^{44,0}+\frac{1}{320}
\sqrt{\frac{3}{14}} C_{22,22}^{44,1}+\frac{1}{144} \sqrt{\frac{7}{6}} C_{31,20}^{31,0}+\frac{1}{288} \sqrt{\frac{7}{6}} C_{31,20}^{31,1}-\frac{1}{288}
\sqrt{\frac{35}{6}} C_{31,22}^{31,0}-\frac{1}{576} \sqrt{\frac{35}{6}} C_{31,22}^{31,1}-\frac{C_{33,20}^{33,0}}{80 \sqrt{2}}-\frac{C_{33,20}^{33,1}}{160
\sqrt{2}}+\frac{1}{64} \sqrt{\frac{3}{5}} C_{33,22}^{33,0}+\frac{1}{128} \sqrt{\frac{3}{5}} C_{33,22}^{33,1}\)

\noindent\(C_{22,1111,3}^{00,3313,0,\text{NonLoc}}= -\frac{1}{192} \sqrt{\frac{7}{2}} C_{00,00}^{60,0}-\frac{1}{96} \sqrt{\frac{7}{2}}
C_{00,00}^{60,1}-\frac{1}{192} \sqrt{\frac{7}{6}} C_{00,20}^{60,0}-\frac{1}{96} \sqrt{\frac{7}{6}} C_{00,20}^{60,1}-\frac{1}{64} \sqrt{\frac{7}{30}}
C_{00,22}^{62,0}-\frac{1}{32} \sqrt{\frac{7}{30}} C_{00,22}^{62,1}+\frac{1}{576} \sqrt{\frac{7}{2}} C_{11,00}^{51,0}+\frac{1}{288} \sqrt{\frac{7}{2}}
C_{11,00}^{51,1}+\frac{1}{576} \sqrt{\frac{35}{6}} C_{11,22}^{51,0}+\frac{1}{288} \sqrt{\frac{35}{6}} C_{11,22}^{51,1}+\frac{1}{576} \sqrt{\frac{7}{2}}
C_{20,00}^{40,0}+\frac{1}{288} \sqrt{\frac{7}{2}} C_{20,00}^{40,1}+\frac{1}{576} \sqrt{\frac{7}{6}} C_{20,20}^{40,0}+\frac{1}{288} \sqrt{\frac{7}{6}}
C_{20,20}^{40,1}+\frac{1}{576} \sqrt{\frac{35}{6}} C_{20,22}^{42,0}+\frac{1}{288} \sqrt{\frac{35}{6}} C_{20,22}^{42,1}-\frac{1}{576} \sqrt{\frac{7}{10}}
C_{22,00}^{42,0}-\frac{1}{288} \sqrt{\frac{7}{10}} C_{22,00}^{42,1}-\frac{1}{576} \sqrt{\frac{7}{30}} C_{22,20}^{42,0}-\frac{1}{288} \sqrt{\frac{7}{30}}
C_{22,20}^{42,1}-\frac{1}{576} \sqrt{\frac{7}{30}} C_{22,22}^{40,0}-\frac{1}{288} \sqrt{\frac{7}{30}} C_{22,22}^{40,1}-\frac{C_{22,22}^{42,0}}{72
\sqrt{30}}-\frac{C_{22,22}^{42,1}}{36 \sqrt{30}}-\frac{1}{320} \sqrt{\frac{3}{14}} C_{22,22}^{44,0}-\frac{1}{160} \sqrt{\frac{3}{14}} C_{22,22}^{44,1}-\frac{1}{576}
\sqrt{\frac{7}{2}} C_{31,00}^{31,0}-\frac{1}{288} \sqrt{\frac{7}{2}} C_{31,00}^{31,1}-\frac{1}{576} \sqrt{\frac{7}{6}} C_{31,20}^{31,0}-\frac{1}{288}
\sqrt{\frac{7}{6}} C_{31,20}^{31,1}-\frac{1}{576} \sqrt{\frac{35}{6}} C_{31,22}^{31,0}-\frac{1}{288} \sqrt{\frac{35}{6}} C_{31,22}^{31,1}+\frac{1}{320}
\sqrt{\frac{3}{2}} C_{33,00}^{33,0}+\frac{1}{160} \sqrt{\frac{3}{2}} C_{33,00}^{33,1}+\frac{C_{33,20}^{33,0}}{320 \sqrt{2}}+\frac{C_{33,20}^{33,1}}{160
\sqrt{2}}+\frac{1}{128} \sqrt{\frac{3}{5}} C_{33,22}^{33,0}+\frac{1}{64} \sqrt{\frac{3}{5}} C_{33,22}^{33,1}\)

\noindent\(C_{22,1111,3}^{00,3313,1,\text{Loc}}= -\frac{1}{96} \sqrt{\frac{7}{2}} C_{00,20}^{60,1}+\frac{1}{64} \sqrt{\frac{7}{10}}
C_{00,22}^{62,1}+\frac{1}{576} \sqrt{\frac{35}{2}} C_{11,22}^{51,1}+\frac{1}{288} \sqrt{\frac{7}{2}} C_{20,20}^{40,1}-\frac{1}{576} \sqrt{\frac{35}{2}}
C_{20,22}^{42,1}-\frac{1}{288} \sqrt{\frac{7}{10}} C_{22,20}^{42,1}+\frac{1}{576} \sqrt{\frac{7}{10}} C_{22,22}^{40,1}+\frac{C_{22,22}^{42,1}}{72
\sqrt{10}}+\frac{3 C_{22,22}^{44,1}}{320 \sqrt{14}}+\frac{1}{288} \sqrt{\frac{7}{2}} C_{31,20}^{31,1}-\frac{1}{576} \sqrt{\frac{35}{2}} C_{31,22}^{31,1}-\frac{1}{160}
\sqrt{\frac{3}{2}} C_{33,20}^{33,1}+\frac{3 C_{33,22}^{33,1}}{128 \sqrt{5}}\)

\noindent\(C_{22,1111,3}^{00,3313,1,\text{NonLoc}}= -\frac{1}{64} \sqrt{\frac{7}{6}} C_{00,00}^{60,0}-\frac{1}{192} \sqrt{\frac{7}{2}}
C_{00,20}^{60,0}-\frac{1}{64} \sqrt{\frac{7}{10}} C_{00,22}^{62,0}+\frac{1}{192} \sqrt{\frac{7}{6}} C_{11,00}^{51,0}+\frac{1}{576} \sqrt{\frac{35}{2}}
C_{11,22}^{51,0}+\frac{1}{192} \sqrt{\frac{7}{6}} C_{20,00}^{40,0}+\frac{1}{576} \sqrt{\frac{7}{2}} C_{20,20}^{40,0}+\frac{1}{576} \sqrt{\frac{35}{2}}
C_{20,22}^{42,0}-\frac{1}{192} \sqrt{\frac{7}{30}} C_{22,00}^{42,0}-\frac{1}{576} \sqrt{\frac{7}{10}} C_{22,20}^{42,0}-\frac{1}{576} \sqrt{\frac{7}{10}}
C_{22,22}^{40,0}-\frac{C_{22,22}^{42,0}}{72 \sqrt{10}}-\frac{3 C_{22,22}^{44,0}}{320 \sqrt{14}}-\frac{1}{192} \sqrt{\frac{7}{6}} C_{31,00}^{31,0}-\frac{1}{576}
\sqrt{\frac{7}{2}} C_{31,20}^{31,0}-\frac{1}{576} \sqrt{\frac{35}{2}} C_{31,22}^{31,0}+\frac{3 C_{33,00}^{33,0}}{320 \sqrt{2}}+\frac{1}{320} \sqrt{\frac{3}{2}}
C_{33,20}^{33,0}+\frac{3 C_{33,22}^{33,0}}{128 \sqrt{5}}\)

\noindent\(C_{22,1111,3}^{11,2202,0,\text{Loc}}= \frac{1}{192} \sqrt{\frac{7}{3}} C_{11,11}^{51,0}+\frac{1}{384} \sqrt{\frac{7}{3}}
C_{11,11}^{51,1}+\frac{1}{192} \sqrt{\frac{7}{3}} C_{31,11}^{31,0}+\frac{1}{384} \sqrt{\frac{7}{3}} C_{31,11}^{31,1}-\frac{3}{160} \sqrt{\frac{3}{2}}
C_{33,11}^{33,0}-\frac{3}{320} \sqrt{\frac{3}{2}} C_{33,11}^{33,1}\)

\noindent\(C_{22,1111,3}^{11,2202,0,\text{NonLoc}}= \frac{1}{384} \sqrt{\frac{7}{3}} C_{11,11}^{51,0}+\frac{1}{192} \sqrt{\frac{7}{3}}
C_{11,11}^{51,1}+\frac{1}{384} \sqrt{\frac{7}{3}} C_{31,11}^{31,0}+\frac{1}{192} \sqrt{\frac{7}{3}} C_{31,11}^{31,1}-\frac{3}{320} \sqrt{\frac{3}{2}}
C_{33,11}^{33,0}-\frac{3}{160} \sqrt{\frac{3}{2}} C_{33,11}^{33,1}\)

\noindent\(C_{22,1111,3}^{11,2202,1,\text{Loc}}= \frac{\sqrt{7} C_{11,11}^{51,1}}{384}+\frac{\sqrt{7} C_{31,11}^{31,1}}{384}-\frac{9
C_{33,11}^{33,1}}{320 \sqrt{2}}\)

\noindent\(C_{22,1111,3}^{11,2202,1,\text{NonLoc}}= \frac{\sqrt{7} C_{11,11}^{51,0}}{384}+\frac{\sqrt{7} C_{31,11}^{31,0}}{384}-\frac{9
C_{33,11}^{33,0}}{320 \sqrt{2}}\)

\noindent\(C_{22,1111,3}^{22,1111,0,\text{Loc}}= \frac{\sqrt{7} C_{00,20}^{60,0}}{384}+\frac{\sqrt{7} C_{00,20}^{60,1}}{768}-\frac{1}{256}
\sqrt{\frac{7}{5}} C_{00,22}^{62,0}-\frac{1}{512} \sqrt{\frac{7}{5}} C_{00,22}^{62,1}-\frac{\sqrt{\frac{7}{5}} C_{11,22}^{51,0}}{2304}-\frac{\sqrt{\frac{7}{5}}
C_{11,22}^{51,1}}{4608}+\frac{3 \sqrt{3} C_{11,22}^{53,0}}{1280}+\frac{3 \sqrt{3} C_{11,22}^{53,1}}{2560}+\frac{\sqrt{7} C_{20,20}^{40,0}}{1152}+\frac{\sqrt{7}
C_{20,20}^{40,1}}{2304}-\frac{\sqrt{\frac{7}{5}} C_{20,22}^{42,0}}{1152}-\frac{\sqrt{\frac{7}{5}} C_{20,22}^{42,1}}{2304}-\frac{\sqrt{35} C_{22,20}^{42,0}}{2304}-\frac{\sqrt{35}
C_{22,20}^{42,1}}{4608}+\frac{\sqrt{\frac{7}{5}} C_{22,22}^{40,0}}{2304}+\frac{\sqrt{\frac{7}{5}} C_{22,22}^{40,1}}{4608}+\frac{17 C_{22,22}^{42,0}}{2304
\sqrt{5}}+\frac{17 C_{22,22}^{42,1}}{4608 \sqrt{5}}-\frac{3 C_{22,22}^{44,0}}{320 \sqrt{7}}-\frac{3 C_{22,22}^{44,1}}{640 \sqrt{7}}-\frac{1}{768}
\sqrt{\frac{7}{5}} C_{31,22}^{31,0}-\frac{\sqrt{\frac{7}{5}} C_{31,22}^{31,1}}{1536}+\frac{C_{31,22}^{33,0}}{640 \sqrt{3}}+\frac{C_{31,22}^{33,1}}{1280
\sqrt{3}}+\frac{\sqrt{3} C_{33,20}^{33,0}}{256}+\frac{\sqrt{3} C_{33,20}^{33,1}}{512}-\frac{3 C_{33,22}^{33,0}}{160 \sqrt{10}}-\frac{3 C_{33,22}^{33,1}}{320
\sqrt{10}}\)

\noindent\(C_{22,1111,3}^{22,1111,0,\text{NonLoc}}= \frac{1}{512} \sqrt{\frac{7}{3}} C_{00,00}^{60,0}+\frac{1}{256} \sqrt{\frac{7}{3}}
C_{00,00}^{60,1}+\frac{\sqrt{7} C_{00,20}^{60,0}}{1536}+\frac{\sqrt{7} C_{00,20}^{60,1}}{768}+\frac{1}{512} \sqrt{\frac{7}{5}} C_{00,22}^{62,0}+\frac{1}{256}
\sqrt{\frac{7}{5}} C_{00,22}^{62,1}+\frac{\sqrt{\frac{7}{3}} C_{11,00}^{51,0}}{1536}+\frac{1}{768} \sqrt{\frac{7}{3}} C_{11,00}^{51,1}-\frac{\sqrt{\frac{7}{5}}
C_{11,22}^{51,0}}{4608}-\frac{\sqrt{\frac{7}{5}} C_{11,22}^{51,1}}{2304}+\frac{3 \sqrt{3} C_{11,22}^{53,0}}{2560}+\frac{3 \sqrt{3} C_{11,22}^{53,1}}{1280}+\frac{\sqrt{\frac{7}{3}}
C_{20,00}^{40,0}}{1536}+\frac{1}{768} \sqrt{\frac{7}{3}} C_{20,00}^{40,1}+\frac{\sqrt{7} C_{20,20}^{40,0}}{4608}+\frac{\sqrt{7} C_{20,20}^{40,1}}{2304}+\frac{\sqrt{\frac{7}{5}}
C_{20,22}^{42,0}}{2304}+\frac{\sqrt{\frac{7}{5}} C_{20,22}^{42,1}}{1152}-\frac{\sqrt{\frac{35}{3}} C_{22,00}^{42,0}}{3072}-\frac{\sqrt{\frac{35}{3}}
C_{22,00}^{42,1}}{1536}-\frac{\sqrt{35} C_{22,20}^{42,0}}{9216}-\frac{\sqrt{35} C_{22,20}^{42,1}}{4608}-\frac{\sqrt{\frac{7}{5}} C_{22,22}^{40,0}}{4608}-\frac{\sqrt{\frac{7}{5}}
C_{22,22}^{40,1}}{2304}-\frac{17 C_{22,22}^{42,0}}{4608 \sqrt{5}}-\frac{17 C_{22,22}^{42,1}}{2304 \sqrt{5}}+\frac{3 C_{22,22}^{44,0}}{640 \sqrt{7}}+\frac{3
C_{22,22}^{44,1}}{320 \sqrt{7}}-\frac{\sqrt{\frac{7}{5}} C_{31,22}^{31,0}}{1536}-\frac{1}{768} \sqrt{\frac{7}{5}} C_{31,22}^{31,1}+\frac{C_{31,22}^{33,0}}{1280
\sqrt{3}}+\frac{C_{31,22}^{33,1}}{640 \sqrt{3}}-\frac{3 C_{33,00}^{33,0}}{1024}-\frac{3 C_{33,00}^{33,1}}{512}-\frac{\sqrt{3} C_{33,20}^{33,0}}{1024}-\frac{\sqrt{3}
C_{33,20}^{33,1}}{512}-\frac{3 C_{33,22}^{33,0}}{320 \sqrt{10}}-\frac{3 C_{33,22}^{33,1}}{160 \sqrt{10}}\)

\noindent\(C_{22,1111,3}^{22,1111,1,\text{Loc}}= \frac{1}{256} \sqrt{\frac{7}{3}} C_{00,20}^{60,1}-\frac{1}{512} \sqrt{\frac{21}{5}}
C_{00,22}^{62,1}-\frac{\sqrt{\frac{7}{15}} C_{11,22}^{51,1}}{1536}+\frac{9 C_{11,22}^{53,1}}{2560}+\frac{1}{768} \sqrt{\frac{7}{3}} C_{20,20}^{40,1}-\frac{1}{768}
\sqrt{\frac{7}{15}} C_{20,22}^{42,1}-\frac{\sqrt{\frac{35}{3}} C_{22,20}^{42,1}}{1536}+\frac{\sqrt{\frac{7}{15}} C_{22,22}^{40,1}}{1536}+\frac{17
C_{22,22}^{42,1}}{1536 \sqrt{15}}-\frac{3}{640} \sqrt{\frac{3}{7}} C_{22,22}^{44,1}-\frac{1}{512} \sqrt{\frac{7}{15}} C_{31,22}^{31,1}+\frac{C_{31,22}^{33,1}}{1280}+\frac{3
C_{33,20}^{33,1}}{512}-\frac{3}{320} \sqrt{\frac{3}{10}} C_{33,22}^{33,1}\)

\noindent\(C_{22,1111,3}^{22,1111,1,\text{NonLoc}}= \frac{\sqrt{7} C_{00,00}^{60,0}}{512}+\frac{1}{512} \sqrt{\frac{7}{3}} C_{00,20}^{60,0}+\frac{1}{512}
\sqrt{\frac{21}{5}} C_{00,22}^{62,0}+\frac{\sqrt{7} C_{11,00}^{51,0}}{1536}-\frac{\sqrt{\frac{7}{15}} C_{11,22}^{51,0}}{1536}+\frac{9 C_{11,22}^{53,0}}{2560}+\frac{\sqrt{7}
C_{20,00}^{40,0}}{1536}+\frac{\sqrt{\frac{7}{3}} C_{20,20}^{40,0}}{1536}+\frac{1}{768} \sqrt{\frac{7}{15}} C_{20,22}^{42,0}-\frac{\sqrt{35} C_{22,00}^{42,0}}{3072}-\frac{\sqrt{\frac{35}{3}}
C_{22,20}^{42,0}}{3072}-\frac{\sqrt{\frac{7}{15}} C_{22,22}^{40,0}}{1536}-\frac{17 C_{22,22}^{42,0}}{1536 \sqrt{15}}+\frac{3}{640} \sqrt{\frac{3}{7}}
C_{22,22}^{44,0}-\frac{1}{512} \sqrt{\frac{7}{15}} C_{31,22}^{31,0}+\frac{C_{31,22}^{33,0}}{1280}-\frac{3 \sqrt{3} C_{33,00}^{33,0}}{1024}-\frac{3
C_{33,20}^{33,0}}{1024}-\frac{3}{320} \sqrt{\frac{3}{10}} C_{33,22}^{33,0}\)

\noindent\(C_{22,1112,0}^{00,3110,0,\text{Loc}}= -\frac{7 C_{00,20}^{60,0}}{96 \sqrt{5}}-\frac{7 C_{00,20}^{60,1}}{192 \sqrt{5}}-\frac{7
C_{00,22}^{62,0}}{160}-\frac{7 C_{00,22}^{62,1}}{320}-\frac{7 C_{11,22}^{51,0}}{2880}-\frac{7 C_{11,22}^{51,1}}{5760}-\frac{9}{640} \sqrt{\frac{21}{5}}
C_{11,22}^{53,0}-\frac{9 \sqrt{\frac{21}{5}} C_{11,22}^{53,1}}{1280}+\frac{7 C_{20,20}^{40,0}}{288 \sqrt{5}}+\frac{7 C_{20,20}^{40,1}}{576 \sqrt{5}}+\frac{7
C_{20,22}^{42,0}}{2880}+\frac{7 C_{20,22}^{42,1}}{5760}-\frac{7 C_{22,20}^{42,0}}{1440}-\frac{7 C_{22,20}^{42,1}}{2880}+\frac{13 C_{22,22}^{40,0}}{720}+\frac{13
C_{22,22}^{40,1}}{1440}-\frac{\sqrt{7} C_{22,22}^{42,0}}{720}-\frac{\sqrt{7} C_{22,22}^{42,1}}{1440}-\frac{3 C_{22,22}^{44,0}}{80 \sqrt{5}}-\frac{3
C_{22,22}^{44,1}}{160 \sqrt{5}}+\frac{7 C_{31,20}^{31,0}}{288 \sqrt{5}}+\frac{7 C_{31,20}^{31,1}}{576 \sqrt{5}}+\frac{37 C_{31,22}^{31,0}}{2880}+\frac{37
C_{31,22}^{31,1}}{5760}+\frac{7}{640} \sqrt{\frac{7}{15}} C_{31,22}^{33,0}+\frac{7 \sqrt{\frac{7}{15}} C_{31,22}^{33,1}}{1280}-\frac{1}{160} \sqrt{\frac{21}{5}}
C_{33,20}^{33,0}-\frac{1}{320} \sqrt{\frac{21}{5}} C_{33,20}^{33,1}-\frac{3}{400} \sqrt{\frac{7}{2}} C_{33,22}^{33,0}-\frac{3}{800} \sqrt{\frac{7}{2}}
C_{33,22}^{33,1}\)

\noindent\(C_{22,1112,0}^{00,3110,0,\text{NonLoc}}= -\frac{7 C_{00,00}^{60,0}}{128 \sqrt{15}}-\frac{7 C_{00,00}^{60,1}}{64 \sqrt{15}}-\frac{7
C_{00,20}^{60,0}}{384 \sqrt{5}}-\frac{7 C_{00,20}^{60,1}}{192 \sqrt{5}}+\frac{7 C_{00,22}^{62,0}}{320}+\frac{7 C_{00,22}^{62,1}}{160}+\frac{7 C_{11,00}^{51,0}}{384
\sqrt{15}}+\frac{7 C_{11,00}^{51,1}}{192 \sqrt{15}}-\frac{7 C_{11,22}^{51,0}}{5760}-\frac{7 C_{11,22}^{51,1}}{2880}-\frac{9 \sqrt{\frac{21}{5}} C_{11,22}^{53,0}}{1280}-\frac{9}{640}
\sqrt{\frac{21}{5}} C_{11,22}^{53,1}+\frac{7 C_{20,00}^{40,0}}{384 \sqrt{15}}+\frac{7 C_{20,00}^{40,1}}{192 \sqrt{15}}+\frac{7 C_{20,20}^{40,0}}{1152
\sqrt{5}}+\frac{7 C_{20,20}^{40,1}}{576 \sqrt{5}}-\frac{7 C_{20,22}^{42,0}}{5760}-\frac{7 C_{20,22}^{42,1}}{2880}-\frac{7 C_{22,00}^{42,0}}{1920
\sqrt{3}}-\frac{7 C_{22,00}^{42,1}}{960 \sqrt{3}}-\frac{7 C_{22,20}^{42,0}}{5760}-\frac{7 C_{22,20}^{42,1}}{2880}-\frac{13 C_{22,22}^{40,0}}{1440}-\frac{13
C_{22,22}^{40,1}}{720}+\frac{\sqrt{7} C_{22,22}^{42,0}}{1440}+\frac{\sqrt{7} C_{22,22}^{42,1}}{720}+\frac{3 C_{22,22}^{44,0}}{160 \sqrt{5}}+\frac{3
C_{22,22}^{44,1}}{80 \sqrt{5}}-\frac{7 C_{31,00}^{31,0}}{384 \sqrt{15}}-\frac{7 C_{31,00}^{31,1}}{192 \sqrt{15}}-\frac{7 C_{31,20}^{31,0}}{1152 \sqrt{5}}-\frac{7
C_{31,20}^{31,1}}{576 \sqrt{5}}+\frac{37 C_{31,22}^{31,0}}{5760}+\frac{37 C_{31,22}^{31,1}}{2880}+\frac{7 \sqrt{\frac{7}{15}} C_{31,22}^{33,0}}{1280}+\frac{7}{640}
\sqrt{\frac{7}{15}} C_{31,22}^{33,1}+\frac{3}{640} \sqrt{\frac{7}{5}} C_{33,00}^{33,0}+\frac{3}{320} \sqrt{\frac{7}{5}} C_{33,00}^{33,1}+\frac{1}{640}
\sqrt{\frac{21}{5}} C_{33,20}^{33,0}+\frac{1}{320} \sqrt{\frac{21}{5}} C_{33,20}^{33,1}-\frac{3}{800} \sqrt{\frac{7}{2}} C_{33,22}^{33,0}-\frac{3}{400}
\sqrt{\frac{7}{2}} C_{33,22}^{33,1}\)

\noindent\(C_{22,1112,0}^{00,3110,1,\text{Loc}}= -\frac{7 C_{00,20}^{60,1}}{64 \sqrt{15}}-\frac{7 \sqrt{3} C_{00,22}^{62,1}}{320}-\frac{7
C_{11,22}^{51,1}}{1920 \sqrt{3}}-\frac{27 \sqrt{\frac{7}{5}} C_{11,22}^{53,1}}{1280}+\frac{7 C_{20,20}^{40,1}}{192 \sqrt{15}}+\frac{7 C_{20,22}^{42,1}}{1920
\sqrt{3}}-\frac{7 C_{22,20}^{42,1}}{960 \sqrt{3}}+\frac{13 C_{22,22}^{40,1}}{480 \sqrt{3}}-\frac{1}{480} \sqrt{\frac{7}{3}} C_{22,22}^{42,1}-\frac{3}{160}
\sqrt{\frac{3}{5}} C_{22,22}^{44,1}+\frac{7 C_{31,20}^{31,1}}{192 \sqrt{15}}+\frac{37 C_{31,22}^{31,1}}{1920 \sqrt{3}}+\frac{7 \sqrt{\frac{7}{5}}
C_{31,22}^{33,1}}{1280}-\frac{3}{320} \sqrt{\frac{7}{5}} C_{33,20}^{33,1}-\frac{3}{800} \sqrt{\frac{21}{2}} C_{33,22}^{33,1}\)

\noindent\(C_{22,1112,0}^{00,3110,1,\text{NonLoc}}= -\frac{7 C_{00,00}^{60,0}}{128 \sqrt{5}}-\frac{7 C_{00,20}^{60,0}}{128 \sqrt{15}}+\frac{7
\sqrt{3} C_{00,22}^{62,0}}{320}+\frac{7 C_{11,00}^{51,0}}{384 \sqrt{5}}-\frac{7 C_{11,22}^{51,0}}{1920 \sqrt{3}}-\frac{27 \sqrt{\frac{7}{5}} C_{11,22}^{53,0}}{1280}+\frac{7
C_{20,00}^{40,0}}{384 \sqrt{5}}+\frac{7 C_{20,20}^{40,0}}{384 \sqrt{15}}-\frac{7 C_{20,22}^{42,0}}{1920 \sqrt{3}}-\frac{7 C_{22,00}^{42,0}}{1920}-\frac{7
C_{22,20}^{42,0}}{1920 \sqrt{3}}-\frac{13 C_{22,22}^{40,0}}{480 \sqrt{3}}+\frac{1}{480} \sqrt{\frac{7}{3}} C_{22,22}^{42,0}+\frac{3}{160} \sqrt{\frac{3}{5}}
C_{22,22}^{44,0}-\frac{7 C_{31,00}^{31,0}}{384 \sqrt{5}}-\frac{7 C_{31,20}^{31,0}}{384 \sqrt{15}}+\frac{37 C_{31,22}^{31,0}}{1920 \sqrt{3}}+\frac{7
\sqrt{\frac{7}{5}} C_{31,22}^{33,0}}{1280}+\frac{3}{640} \sqrt{\frac{21}{5}} C_{33,00}^{33,0}+\frac{3}{640} \sqrt{\frac{7}{5}} C_{33,20}^{33,0}-\frac{3}{800}
\sqrt{\frac{21}{2}} C_{33,22}^{33,0}\)

\noindent\(C_{22,1112,0}^{20,1110,0,\text{Loc}}= \frac{7 \sqrt{5} C_{00,20}^{60,0}}{1152}+\frac{7 \sqrt{5} C_{00,20}^{60,1}}{2304}+\frac{7
C_{00,22}^{62,0}}{384}+\frac{7 C_{00,22}^{62,1}}{768}-\frac{49 C_{11,22}^{51,0}}{6912}-\frac{49 C_{11,22}^{51,1}}{13824}-\frac{1}{512} \sqrt{\frac{21}{5}}
C_{11,22}^{53,0}-\frac{\sqrt{\frac{21}{5}} C_{11,22}^{53,1}}{1024}-\frac{7 \sqrt{5} C_{20,20}^{40,0}}{3456}-\frac{7 \sqrt{5} C_{20,20}^{40,1}}{6912}-\frac{7
C_{20,22}^{42,0}}{6912}-\frac{7 C_{20,22}^{42,1}}{13824}+\frac{7 C_{22,20}^{42,0}}{3456}+\frac{7 C_{22,20}^{42,1}}{6912}-\frac{7 C_{22,22}^{40,0}}{1728}-\frac{7
C_{22,22}^{40,1}}{3456}+\frac{5 \sqrt{7} C_{22,22}^{42,0}}{3456}+\frac{5 \sqrt{7} C_{22,22}^{42,1}}{6912}-\frac{C_{22,22}^{44,0}}{64 \sqrt{5}}-\frac{C_{22,22}^{44,1}}{128
\sqrt{5}}+\frac{7 \sqrt{5} C_{31,20}^{31,0}}{3456}+\frac{7 \sqrt{5} C_{31,20}^{31,1}}{6912}+\frac{7 C_{31,22}^{31,0}}{6912}+\frac{7 C_{31,22}^{31,1}}{13824}+\frac{17
\sqrt{\frac{7}{15}} C_{31,22}^{33,0}}{1536}+\frac{17 \sqrt{\frac{7}{15}} C_{31,22}^{33,1}}{3072}-\frac{1}{128} \sqrt{\frac{7}{15}} C_{33,20}^{33,0}-\frac{1}{256}
\sqrt{\frac{7}{15}} C_{33,20}^{33,1}-\frac{1}{320} \sqrt{\frac{7}{2}} C_{33,22}^{33,0}-\frac{1}{640} \sqrt{\frac{7}{2}} C_{33,22}^{33,1}\)

\noindent\(C_{22,1112,0}^{20,1110,0,\text{NonLoc}}= \frac{7 \sqrt{\frac{5}{3}} C_{00,00}^{60,0}}{1536}+\frac{7}{768} \sqrt{\frac{5}{3}}
C_{00,00}^{60,1}+\frac{7 \sqrt{5} C_{00,20}^{60,0}}{4608}+\frac{7 \sqrt{5} C_{00,20}^{60,1}}{2304}-\frac{7 C_{00,22}^{62,0}}{768}-\frac{7 C_{00,22}^{62,1}}{384}+\frac{7
\sqrt{\frac{5}{3}} C_{11,00}^{51,0}}{4608}+\frac{7 \sqrt{\frac{5}{3}} C_{11,00}^{51,1}}{2304}-\frac{49 C_{11,22}^{51,0}}{13824}-\frac{49 C_{11,22}^{51,1}}{6912}-\frac{\sqrt{\frac{21}{5}}
C_{11,22}^{53,0}}{1024}-\frac{1}{512} \sqrt{\frac{21}{5}} C_{11,22}^{53,1}-\frac{7 \sqrt{\frac{5}{3}} C_{20,00}^{40,0}}{4608}-\frac{7 \sqrt{\frac{5}{3}}
C_{20,00}^{40,1}}{2304}-\frac{7 \sqrt{5} C_{20,20}^{40,0}}{13824}-\frac{7 \sqrt{5} C_{20,20}^{40,1}}{6912}+\frac{7 C_{20,22}^{42,0}}{13824}+\frac{7
C_{20,22}^{42,1}}{6912}+\frac{7 C_{22,00}^{42,0}}{4608 \sqrt{3}}+\frac{7 C_{22,00}^{42,1}}{2304 \sqrt{3}}+\frac{7 C_{22,20}^{42,0}}{13824}+\frac{7
C_{22,20}^{42,1}}{6912}+\frac{7 C_{22,22}^{40,0}}{3456}+\frac{7 C_{22,22}^{40,1}}{1728}-\frac{5 \sqrt{7} C_{22,22}^{42,0}}{6912}-\frac{5 \sqrt{7}
C_{22,22}^{42,1}}{3456}+\frac{C_{22,22}^{44,0}}{128 \sqrt{5}}+\frac{C_{22,22}^{44,1}}{64 \sqrt{5}}-\frac{7 \sqrt{\frac{5}{3}} C_{31,00}^{31,0}}{4608}-\frac{7
\sqrt{\frac{5}{3}} C_{31,00}^{31,1}}{2304}-\frac{7 \sqrt{5} C_{31,20}^{31,0}}{13824}-\frac{7 \sqrt{5} C_{31,20}^{31,1}}{6912}+\frac{7 C_{31,22}^{31,0}}{13824}+\frac{7
C_{31,22}^{31,1}}{6912}+\frac{17 \sqrt{\frac{7}{15}} C_{31,22}^{33,0}}{3072}+\frac{17 \sqrt{\frac{7}{15}} C_{31,22}^{33,1}}{1536}+\frac{1}{512} \sqrt{\frac{7}{5}}
C_{33,00}^{33,0}+\frac{1}{256} \sqrt{\frac{7}{5}} C_{33,00}^{33,1}+\frac{1}{512} \sqrt{\frac{7}{15}} C_{33,20}^{33,0}+\frac{1}{256} \sqrt{\frac{7}{15}}
C_{33,20}^{33,1}-\frac{1}{640} \sqrt{\frac{7}{2}} C_{33,22}^{33,0}-\frac{1}{320} \sqrt{\frac{7}{2}} C_{33,22}^{33,1}\)

\noindent\(C_{22,1112,0}^{20,1110,1,\text{Loc}}= \frac{7}{768} \sqrt{\frac{5}{3}} C_{00,20}^{60,1}+\frac{7 C_{00,22}^{62,1}}{256 \sqrt{3}}-\frac{49
C_{11,22}^{51,1}}{4608 \sqrt{3}}-\frac{3 \sqrt{\frac{7}{5}} C_{11,22}^{53,1}}{1024}-\frac{7 \sqrt{\frac{5}{3}} C_{20,20}^{40,1}}{2304}-\frac{7 C_{20,22}^{42,1}}{4608
\sqrt{3}}+\frac{7 C_{22,20}^{42,1}}{2304 \sqrt{3}}-\frac{7 C_{22,22}^{40,1}}{1152 \sqrt{3}}+\frac{5 \sqrt{\frac{7}{3}} C_{22,22}^{42,1}}{2304}-\frac{1}{128}
\sqrt{\frac{3}{5}} C_{22,22}^{44,1}+\frac{7 \sqrt{\frac{5}{3}} C_{31,20}^{31,1}}{2304}+\frac{7 C_{31,22}^{31,1}}{4608 \sqrt{3}}+\frac{17 \sqrt{\frac{7}{5}}
C_{31,22}^{33,1}}{3072}-\frac{1}{256} \sqrt{\frac{7}{5}} C_{33,20}^{33,1}-\frac{1}{640} \sqrt{\frac{21}{2}} C_{33,22}^{33,1}\)

\noindent\(C_{22,1112,0}^{20,1110,1,\text{NonLoc}}= \frac{7 \sqrt{5} C_{00,00}^{60,0}}{1536}+\frac{7 \sqrt{\frac{5}{3}} C_{00,20}^{60,0}}{1536}-\frac{7
C_{00,22}^{62,0}}{256 \sqrt{3}}+\frac{7 \sqrt{5} C_{11,00}^{51,0}}{4608}-\frac{49 C_{11,22}^{51,0}}{4608 \sqrt{3}}-\frac{3 \sqrt{\frac{7}{5}} C_{11,22}^{53,0}}{1024}-\frac{7
\sqrt{5} C_{20,00}^{40,0}}{4608}-\frac{7 \sqrt{\frac{5}{3}} C_{20,20}^{40,0}}{4608}+\frac{7 C_{20,22}^{42,0}}{4608 \sqrt{3}}+\frac{7 C_{22,00}^{42,0}}{4608}+\frac{7
C_{22,20}^{42,0}}{4608 \sqrt{3}}+\frac{7 C_{22,22}^{40,0}}{1152 \sqrt{3}}-\frac{5 \sqrt{\frac{7}{3}} C_{22,22}^{42,0}}{2304}+\frac{1}{128} \sqrt{\frac{3}{5}}
C_{22,22}^{44,0}-\frac{7 \sqrt{5} C_{31,00}^{31,0}}{4608}-\frac{7 \sqrt{\frac{5}{3}} C_{31,20}^{31,0}}{4608}+\frac{7 C_{31,22}^{31,0}}{4608 \sqrt{3}}+\frac{17
\sqrt{\frac{7}{5}} C_{31,22}^{33,0}}{3072}+\frac{1}{512} \sqrt{\frac{21}{5}} C_{33,00}^{33,0}+\frac{1}{512} \sqrt{\frac{7}{5}} C_{33,20}^{33,0}-\frac{1}{640}
\sqrt{\frac{21}{2}} C_{33,22}^{33,0}\)

\noindent\(C_{22,1112,0}^{22,1112,0,\text{Loc}}= \frac{7 C_{00,20}^{60,0}}{1152}+\frac{7 C_{00,20}^{60,1}}{2304}+\frac{7 C_{00,22}^{62,0}}{384
\sqrt{5}}+\frac{7 C_{00,22}^{62,1}}{768 \sqrt{5}}-\frac{7 C_{11,22}^{51,0}}{432 \sqrt{5}}-\frac{7 C_{11,22}^{51,1}}{864 \sqrt{5}}+\frac{\sqrt{21}
C_{11,22}^{53,0}}{1280}+\frac{\sqrt{21} C_{11,22}^{53,1}}{2560}-\frac{C_{20,20}^{40,0}}{3456}-\frac{C_{20,20}^{40,1}}{6912}-\frac{91 C_{20,22}^{42,0}}{6912
\sqrt{5}}-\frac{91 C_{20,22}^{42,1}}{13824 \sqrt{5}}-\frac{7 C_{22,20}^{42,0}}{6912 \sqrt{5}}-\frac{7 C_{22,20}^{42,1}}{13824 \sqrt{5}}+\frac{13
C_{22,22}^{40,0}}{864 \sqrt{5}}+\frac{13 C_{22,22}^{40,1}}{1728 \sqrt{5}}-\frac{\sqrt{35} C_{22,22}^{42,0}}{6912}-\frac{\sqrt{35} C_{22,22}^{42,1}}{13824}+\frac{C_{22,22}^{44,0}}{640}+\frac{C_{22,22}^{44,1}}{1280}+\frac{C_{31,20}^{31,0}}{864}+\frac{C_{31,20}^{31,1}}{1728}+\frac{47
C_{31,22}^{31,0}}{3456 \sqrt{5}}+\frac{47 C_{31,22}^{31,1}}{6912 \sqrt{5}}-\frac{1}{480} \sqrt{\frac{7}{3}} C_{31,22}^{33,0}-\frac{1}{960} \sqrt{\frac{7}{3}}
C_{31,22}^{33,1}+\frac{\sqrt{\frac{7}{3}} C_{33,20}^{33,0}}{1280}+\frac{\sqrt{\frac{7}{3}} C_{33,20}^{33,1}}{2560}+\frac{1}{640} \sqrt{\frac{7}{10}}
C_{33,22}^{33,0}+\frac{\sqrt{\frac{7}{10}} C_{33,22}^{33,1}}{1280}\)

\noindent\(C_{22,1112,0}^{22,1112,0,\text{NonLoc}}= \frac{7 C_{00,00}^{60,0}}{1536 \sqrt{3}}+\frac{7 C_{00,00}^{60,1}}{768 \sqrt{3}}+\frac{7
C_{00,20}^{60,0}}{4608}+\frac{7 C_{00,20}^{60,1}}{2304}-\frac{7 C_{00,22}^{62,0}}{768 \sqrt{5}}-\frac{7 C_{00,22}^{62,1}}{384 \sqrt{5}}+\frac{7 C_{11,00}^{51,0}}{4608
\sqrt{3}}+\frac{7 C_{11,00}^{51,1}}{2304 \sqrt{3}}-\frac{7 C_{11,22}^{51,0}}{864 \sqrt{5}}-\frac{7 C_{11,22}^{51,1}}{432 \sqrt{5}}+\frac{\sqrt{21}
C_{11,22}^{53,0}}{2560}+\frac{\sqrt{21} C_{11,22}^{53,1}}{1280}-\frac{C_{20,00}^{40,0}}{4608 \sqrt{3}}-\frac{C_{20,00}^{40,1}}{2304 \sqrt{3}}-\frac{C_{20,20}^{40,0}}{13824}-\frac{C_{20,20}^{40,1}}{6912}+\frac{91
C_{20,22}^{42,0}}{13824 \sqrt{5}}+\frac{91 C_{20,22}^{42,1}}{6912 \sqrt{5}}-\frac{7 C_{22,00}^{42,0}}{9216 \sqrt{15}}-\frac{7 C_{22,00}^{42,1}}{4608
\sqrt{15}}-\frac{7 C_{22,20}^{42,0}}{27648 \sqrt{5}}-\frac{7 C_{22,20}^{42,1}}{13824 \sqrt{5}}-\frac{13 C_{22,22}^{40,0}}{1728 \sqrt{5}}-\frac{13
C_{22,22}^{40,1}}{864 \sqrt{5}}+\frac{\sqrt{35} C_{22,22}^{42,0}}{13824}+\frac{\sqrt{35} C_{22,22}^{42,1}}{6912}-\frac{C_{22,22}^{44,0}}{1280}-\frac{C_{22,22}^{44,1}}{640}-\frac{C_{31,00}^{31,0}}{1152
\sqrt{3}}-\frac{C_{31,00}^{31,1}}{576 \sqrt{3}}-\frac{C_{31,20}^{31,0}}{3456}-\frac{C_{31,20}^{31,1}}{1728}+\frac{47 C_{31,22}^{31,0}}{6912 \sqrt{5}}+\frac{47
C_{31,22}^{31,1}}{3456 \sqrt{5}}-\frac{1}{960} \sqrt{\frac{7}{3}} C_{31,22}^{33,0}-\frac{1}{480} \sqrt{\frac{7}{3}} C_{31,22}^{33,1}-\frac{\sqrt{7}
C_{33,00}^{33,0}}{5120}-\frac{\sqrt{7} C_{33,00}^{33,1}}{2560}-\frac{\sqrt{\frac{7}{3}} C_{33,20}^{33,0}}{5120}-\frac{\sqrt{\frac{7}{3}} C_{33,20}^{33,1}}{2560}+\frac{\sqrt{\frac{7}{10}}
C_{33,22}^{33,0}}{1280}+\frac{1}{640} \sqrt{\frac{7}{10}} C_{33,22}^{33,1}\)

\noindent\(C_{22,1112,0}^{22,1112,1,\text{Loc}}= \frac{7 C_{00,20}^{60,1}}{768 \sqrt{3}}+\frac{7 C_{00,22}^{62,1}}{256 \sqrt{15}}-\frac{7
C_{11,22}^{51,1}}{288 \sqrt{15}}+\frac{3 \sqrt{7} C_{11,22}^{53,1}}{2560}-\frac{C_{20,20}^{40,1}}{2304 \sqrt{3}}-\frac{91 C_{20,22}^{42,1}}{4608
\sqrt{15}}-\frac{7 C_{22,20}^{42,1}}{4608 \sqrt{15}}+\frac{13 C_{22,22}^{40,1}}{576 \sqrt{15}}-\frac{\sqrt{\frac{35}{3}} C_{22,22}^{42,1}}{4608}+\frac{\sqrt{3}
C_{22,22}^{44,1}}{1280}+\frac{C_{31,20}^{31,1}}{576 \sqrt{3}}+\frac{47 C_{31,22}^{31,1}}{2304 \sqrt{15}}-\frac{\sqrt{7} C_{31,22}^{33,1}}{960}+\frac{\sqrt{7}
C_{33,20}^{33,1}}{2560}+\frac{\sqrt{\frac{21}{10}} C_{33,22}^{33,1}}{1280}\)

\noindent\(C_{22,1112,0}^{22,1112,1,\text{NonLoc}}= \frac{7 C_{00,00}^{60,0}}{1536}+\frac{7 C_{00,20}^{60,0}}{1536 \sqrt{3}}-\frac{7
C_{00,22}^{62,0}}{256 \sqrt{15}}+\frac{7 C_{11,00}^{51,0}}{4608}-\frac{7 C_{11,22}^{51,0}}{288 \sqrt{15}}+\frac{3 \sqrt{7} C_{11,22}^{53,0}}{2560}-\frac{C_{20,00}^{40,0}}{4608}-\frac{C_{20,20}^{40,0}}{4608
\sqrt{3}}+\frac{91 C_{20,22}^{42,0}}{4608 \sqrt{15}}-\frac{7 C_{22,00}^{42,0}}{9216 \sqrt{5}}-\frac{7 C_{22,20}^{42,0}}{9216 \sqrt{15}}-\frac{13
C_{22,22}^{40,0}}{576 \sqrt{15}}+\frac{\sqrt{\frac{35}{3}} C_{22,22}^{42,0}}{4608}-\frac{\sqrt{3} C_{22,22}^{44,0}}{1280}-\frac{C_{31,00}^{31,0}}{1152}-\frac{C_{31,20}^{31,0}}{1152
\sqrt{3}}+\frac{47 C_{31,22}^{31,0}}{2304 \sqrt{15}}-\frac{\sqrt{7} C_{31,22}^{33,0}}{960}-\frac{\sqrt{21} C_{33,00}^{33,0}}{5120}-\frac{\sqrt{7}
C_{33,20}^{33,0}}{5120}+\frac{\sqrt{\frac{21}{10}} C_{33,22}^{33,0}}{1280}\)

\noindent\(C_{22,1112,1}^{00,3111,0,\text{Loc}}= -\frac{7 C_{00,20}^{60,0}}{64 \sqrt{5}}-\frac{7 C_{00,20}^{60,1}}{128 \sqrt{5}}+\frac{21
C_{00,22}^{62,0}}{640}+\frac{21 C_{00,22}^{62,1}}{1280}-\frac{7 C_{11,22}^{51,0}}{1920}-\frac{7 C_{11,22}^{51,1}}{3840}+\frac{9}{640} \sqrt{\frac{21}{5}}
C_{11,22}^{53,0}+\frac{9 \sqrt{\frac{21}{5}} C_{11,22}^{53,1}}{1280}+\frac{7 C_{20,20}^{40,0}}{192 \sqrt{5}}+\frac{7 C_{20,20}^{40,1}}{384 \sqrt{5}}+\frac{7
C_{20,22}^{42,0}}{1920}+\frac{7 C_{20,22}^{42,1}}{3840}-\frac{7 C_{22,20}^{42,0}}{960}-\frac{7 C_{22,20}^{42,1}}{1920}-\frac{17 C_{22,22}^{40,0}}{1920}-\frac{17
C_{22,22}^{40,1}}{3840}-\frac{\sqrt{7} C_{22,22}^{42,0}}{192}-\frac{\sqrt{7} C_{22,22}^{42,1}}{384}+\frac{9 C_{22,22}^{44,0}}{160 \sqrt{5}}+\frac{9
C_{22,22}^{44,1}}{320 \sqrt{5}}+\frac{7 C_{31,20}^{31,0}}{192 \sqrt{5}}+\frac{7 C_{31,20}^{31,1}}{384 \sqrt{5}}-\frac{23 C_{31,22}^{31,0}}{1920}-\frac{23
C_{31,22}^{31,1}}{3840}+\frac{1}{640} \sqrt{\frac{21}{5}} C_{31,22}^{33,0}+\frac{\sqrt{\frac{21}{5}} C_{31,22}^{33,1}}{1280}-\frac{3}{320} \sqrt{\frac{21}{5}}
C_{33,20}^{33,0}-\frac{3}{640} \sqrt{\frac{21}{5}} C_{33,20}^{33,1}-\frac{9}{800} \sqrt{\frac{7}{2}} C_{33,22}^{33,0}-\frac{9 \sqrt{\frac{7}{2}}
C_{33,22}^{33,1}}{1600}\)

\noindent\(C_{22,1112,1}^{00,3111,0,\text{NonLoc}}= -\frac{7}{256} \sqrt{\frac{3}{5}} C_{00,00}^{60,0}-\frac{7}{128} \sqrt{\frac{3}{5}}
C_{00,00}^{60,1}-\frac{7 C_{00,20}^{60,0}}{256 \sqrt{5}}-\frac{7 C_{00,20}^{60,1}}{128 \sqrt{5}}-\frac{21 C_{00,22}^{62,0}}{1280}-\frac{21 C_{00,22}^{62,1}}{640}+\frac{7
C_{11,00}^{51,0}}{256 \sqrt{15}}+\frac{7 C_{11,00}^{51,1}}{128 \sqrt{15}}-\frac{7 C_{11,22}^{51,0}}{3840}-\frac{7 C_{11,22}^{51,1}}{1920}+\frac{9
\sqrt{\frac{21}{5}} C_{11,22}^{53,0}}{1280}+\frac{9}{640} \sqrt{\frac{21}{5}} C_{11,22}^{53,1}+\frac{7 C_{20,00}^{40,0}}{256 \sqrt{15}}+\frac{7 C_{20,00}^{40,1}}{128
\sqrt{15}}+\frac{7 C_{20,20}^{40,0}}{768 \sqrt{5}}+\frac{7 C_{20,20}^{40,1}}{384 \sqrt{5}}-\frac{7 C_{20,22}^{42,0}}{3840}-\frac{7 C_{20,22}^{42,1}}{1920}-\frac{7
C_{22,00}^{42,0}}{1280 \sqrt{3}}-\frac{7 C_{22,00}^{42,1}}{640 \sqrt{3}}-\frac{7 C_{22,20}^{42,0}}{3840}-\frac{7 C_{22,20}^{42,1}}{1920}+\frac{17
C_{22,22}^{40,0}}{3840}+\frac{17 C_{22,22}^{40,1}}{1920}+\frac{\sqrt{7} C_{22,22}^{42,0}}{384}+\frac{\sqrt{7} C_{22,22}^{42,1}}{192}-\frac{9 C_{22,22}^{44,0}}{320
\sqrt{5}}-\frac{9 C_{22,22}^{44,1}}{160 \sqrt{5}}-\frac{7 C_{31,00}^{31,0}}{256 \sqrt{15}}-\frac{7 C_{31,00}^{31,1}}{128 \sqrt{15}}-\frac{7 C_{31,20}^{31,0}}{768
\sqrt{5}}-\frac{7 C_{31,20}^{31,1}}{384 \sqrt{5}}-\frac{23 C_{31,22}^{31,0}}{3840}-\frac{23 C_{31,22}^{31,1}}{1920}+\frac{\sqrt{\frac{21}{5}} C_{31,22}^{33,0}}{1280}+\frac{1}{640}
\sqrt{\frac{21}{5}} C_{31,22}^{33,1}+\frac{9 \sqrt{\frac{7}{5}} C_{33,00}^{33,0}}{1280}+\frac{9}{640} \sqrt{\frac{7}{5}} C_{33,00}^{33,1}+\frac{3
\sqrt{\frac{21}{5}} C_{33,20}^{33,0}}{1280}+\frac{3}{640} \sqrt{\frac{21}{5}} C_{33,20}^{33,1}-\frac{9 \sqrt{\frac{7}{2}} C_{33,22}^{33,0}}{1600}-\frac{9}{800}
\sqrt{\frac{7}{2}} C_{33,22}^{33,1}\)

\noindent\(C_{22,1112,1}^{00,3111,1,\text{Loc}}= -\frac{7}{128} \sqrt{\frac{3}{5}} C_{00,20}^{60,1}+\frac{21 \sqrt{3} C_{00,22}^{62,1}}{1280}-\frac{7
C_{11,22}^{51,1}}{1280 \sqrt{3}}+\frac{27 \sqrt{\frac{7}{5}} C_{11,22}^{53,1}}{1280}+\frac{7 C_{20,20}^{40,1}}{128 \sqrt{15}}+\frac{7 C_{20,22}^{42,1}}{1280
\sqrt{3}}-\frac{7 C_{22,20}^{42,1}}{640 \sqrt{3}}-\frac{17 C_{22,22}^{40,1}}{1280 \sqrt{3}}-\frac{1}{128} \sqrt{\frac{7}{3}} C_{22,22}^{42,1}+\frac{9}{320}
\sqrt{\frac{3}{5}} C_{22,22}^{44,1}+\frac{7 C_{31,20}^{31,1}}{128 \sqrt{15}}-\frac{23 C_{31,22}^{31,1}}{1280 \sqrt{3}}+\frac{3 \sqrt{\frac{7}{5}}
C_{31,22}^{33,1}}{1280}-\frac{9}{640} \sqrt{\frac{7}{5}} C_{33,20}^{33,1}-\frac{9 \sqrt{\frac{21}{2}} C_{33,22}^{33,1}}{1600}\)

\noindent\(C_{22,1112,1}^{00,3111,1,\text{NonLoc}}= -\frac{21 C_{00,00}^{60,0}}{256 \sqrt{5}}-\frac{7}{256} \sqrt{\frac{3}{5}} C_{00,20}^{60,0}-\frac{21
\sqrt{3} C_{00,22}^{62,0}}{1280}+\frac{7 C_{11,00}^{51,0}}{256 \sqrt{5}}-\frac{7 C_{11,22}^{51,0}}{1280 \sqrt{3}}+\frac{27 \sqrt{\frac{7}{5}} C_{11,22}^{53,0}}{1280}+\frac{7
C_{20,00}^{40,0}}{256 \sqrt{5}}+\frac{7 C_{20,20}^{40,0}}{256 \sqrt{15}}-\frac{7 C_{20,22}^{42,0}}{1280 \sqrt{3}}-\frac{7 C_{22,00}^{42,0}}{1280}-\frac{7
C_{22,20}^{42,0}}{1280 \sqrt{3}}+\frac{17 C_{22,22}^{40,0}}{1280 \sqrt{3}}+\frac{1}{128} \sqrt{\frac{7}{3}} C_{22,22}^{42,0}-\frac{9}{320} \sqrt{\frac{3}{5}}
C_{22,22}^{44,0}-\frac{7 C_{31,00}^{31,0}}{256 \sqrt{5}}-\frac{7 C_{31,20}^{31,0}}{256 \sqrt{15}}-\frac{23 C_{31,22}^{31,0}}{1280 \sqrt{3}}+\frac{3
\sqrt{\frac{7}{5}} C_{31,22}^{33,0}}{1280}+\frac{9 \sqrt{\frac{21}{5}} C_{33,00}^{33,0}}{1280}+\frac{9 \sqrt{\frac{7}{5}} C_{33,20}^{33,0}}{1280}-\frac{9
\sqrt{\frac{21}{2}} C_{33,22}^{33,0}}{1600}\)

\noindent\(C_{22,1112,1}^{11,2202,0,\text{Loc}}= \frac{7 C_{11,11}^{51,0}}{128 \sqrt{3}}+\frac{7 C_{11,11}^{51,1}}{256 \sqrt{3}}-\frac{5
C_{31,11}^{31,0}}{128 \sqrt{3}}-\frac{5 C_{31,11}^{31,1}}{256 \sqrt{3}}\)

\noindent\(C_{22,1112,1}^{11,2202,0,\text{NonLoc}}= \frac{7 C_{11,11}^{51,0}}{256 \sqrt{3}}+\frac{7 C_{11,11}^{51,1}}{128 \sqrt{3}}-\frac{5
C_{31,11}^{31,0}}{256 \sqrt{3}}-\frac{5 C_{31,11}^{31,1}}{128 \sqrt{3}}\)

\noindent\(C_{22,1112,1}^{11,2202,1,\text{Loc}}= \frac{7 C_{11,11}^{51,1}}{256}-\frac{5 C_{31,11}^{31,1}}{256}\)

\noindent\(C_{22,1112,1}^{11,2202,1,\text{NonLoc}}= \frac{7 C_{11,11}^{51,0}}{256}-\frac{5 C_{31,11}^{31,0}}{256}\)

\noindent\(C_{22,1112,1}^{20,1111,0,\text{Loc}}= \frac{7 \sqrt{5} C_{00,20}^{60,0}}{768}+\frac{7 \sqrt{5} C_{00,20}^{60,1}}{1536}-\frac{7
C_{00,22}^{62,0}}{512}-\frac{7 C_{00,22}^{62,1}}{1024}+\frac{35 C_{11,22}^{51,0}}{4608}+\frac{35 C_{11,22}^{51,1}}{9216}-\frac{7 \sqrt{5} C_{20,20}^{40,0}}{2304}-\frac{7
\sqrt{5} C_{20,20}^{40,1}}{4608}-\frac{7 C_{20,22}^{42,0}}{4608}-\frac{7 C_{20,22}^{42,1}}{9216}+\frac{7 C_{22,20}^{42,0}}{2304}+\frac{7 C_{22,20}^{42,1}}{4608}-\frac{7
C_{22,22}^{40,0}}{4608}-\frac{7 C_{22,22}^{40,1}}{9216}+\frac{\sqrt{7} C_{22,22}^{42,0}}{1152}+\frac{\sqrt{7} C_{22,22}^{42,1}}{2304}+\frac{3 C_{22,22}^{44,0}}{128
\sqrt{5}}+\frac{3 C_{22,22}^{44,1}}{256 \sqrt{5}}+\frac{7 \sqrt{5} C_{31,20}^{31,0}}{2304}+\frac{7 \sqrt{5} C_{31,20}^{31,1}}{4608}+\frac{7 C_{31,22}^{31,0}}{4608}+\frac{7
C_{31,22}^{31,1}}{9216}-\frac{1}{128} \sqrt{\frac{7}{15}} C_{31,22}^{33,0}-\frac{1}{256} \sqrt{\frac{7}{15}} C_{31,22}^{33,1}-\frac{1}{256} \sqrt{\frac{21}{5}}
C_{33,20}^{33,0}-\frac{1}{512} \sqrt{\frac{21}{5}} C_{33,20}^{33,1}-\frac{3}{640} \sqrt{\frac{7}{2}} C_{33,22}^{33,0}-\frac{3 \sqrt{\frac{7}{2}}
C_{33,22}^{33,1}}{1280}\)

\noindent\(C_{22,1112,1}^{20,1111,0,\text{NonLoc}}= \frac{7 \sqrt{\frac{5}{3}} C_{00,00}^{60,0}}{1024}+\frac{7}{512} \sqrt{\frac{5}{3}}
C_{00,00}^{60,1}+\frac{7 \sqrt{5} C_{00,20}^{60,0}}{3072}+\frac{7 \sqrt{5} C_{00,20}^{60,1}}{1536}+\frac{7 C_{00,22}^{62,0}}{1024}+\frac{7 C_{00,22}^{62,1}}{512}+\frac{7
\sqrt{\frac{5}{3}} C_{11,00}^{51,0}}{3072}+\frac{7 \sqrt{\frac{5}{3}} C_{11,00}^{51,1}}{1536}+\frac{35 C_{11,22}^{51,0}}{9216}+\frac{35 C_{11,22}^{51,1}}{4608}-\frac{7
\sqrt{\frac{5}{3}} C_{20,00}^{40,0}}{3072}-\frac{7 \sqrt{\frac{5}{3}} C_{20,00}^{40,1}}{1536}-\frac{7 \sqrt{5} C_{20,20}^{40,0}}{9216}-\frac{7 \sqrt{5}
C_{20,20}^{40,1}}{4608}+\frac{7 C_{20,22}^{42,0}}{9216}+\frac{7 C_{20,22}^{42,1}}{4608}+\frac{7 C_{22,00}^{42,0}}{3072 \sqrt{3}}+\frac{7 C_{22,00}^{42,1}}{1536
\sqrt{3}}+\frac{7 C_{22,20}^{42,0}}{9216}+\frac{7 C_{22,20}^{42,1}}{4608}+\frac{7 C_{22,22}^{40,0}}{9216}+\frac{7 C_{22,22}^{40,1}}{4608}-\frac{\sqrt{7}
C_{22,22}^{42,0}}{2304}-\frac{\sqrt{7} C_{22,22}^{42,1}}{1152}-\frac{3 C_{22,22}^{44,0}}{256 \sqrt{5}}-\frac{3 C_{22,22}^{44,1}}{128 \sqrt{5}}-\frac{7
\sqrt{\frac{5}{3}} C_{31,00}^{31,0}}{3072}-\frac{7 \sqrt{\frac{5}{3}} C_{31,00}^{31,1}}{1536}-\frac{7 \sqrt{5} C_{31,20}^{31,0}}{9216}-\frac{7 \sqrt{5}
C_{31,20}^{31,1}}{4608}+\frac{7 C_{31,22}^{31,0}}{9216}+\frac{7 C_{31,22}^{31,1}}{4608}-\frac{1}{256} \sqrt{\frac{7}{15}} C_{31,22}^{33,0}-\frac{1}{128}
\sqrt{\frac{7}{15}} C_{31,22}^{33,1}+\frac{3 \sqrt{\frac{7}{5}} C_{33,00}^{33,0}}{1024}+\frac{3}{512} \sqrt{\frac{7}{5}} C_{33,00}^{33,1}+\frac{\sqrt{\frac{21}{5}}
C_{33,20}^{33,0}}{1024}+\frac{1}{512} \sqrt{\frac{21}{5}} C_{33,20}^{33,1}-\frac{3 \sqrt{\frac{7}{2}} C_{33,22}^{33,0}}{1280}-\frac{3}{640} \sqrt{\frac{7}{2}}
C_{33,22}^{33,1}\)

\noindent\(C_{22,1112,1}^{20,1111,1,\text{Loc}}= \frac{7}{512} \sqrt{\frac{5}{3}} C_{00,20}^{60,1}-\frac{7 \sqrt{3} C_{00,22}^{62,1}}{1024}+\frac{35
C_{11,22}^{51,1}}{3072 \sqrt{3}}-\frac{7 \sqrt{\frac{5}{3}} C_{20,20}^{40,1}}{1536}-\frac{7 C_{20,22}^{42,1}}{3072 \sqrt{3}}+\frac{7 C_{22,20}^{42,1}}{1536
\sqrt{3}}-\frac{7 C_{22,22}^{40,1}}{3072 \sqrt{3}}+\frac{1}{768} \sqrt{\frac{7}{3}} C_{22,22}^{42,1}+\frac{3}{256} \sqrt{\frac{3}{5}} C_{22,22}^{44,1}+\frac{7
\sqrt{\frac{5}{3}} C_{31,20}^{31,1}}{1536}+\frac{7 C_{31,22}^{31,1}}{3072 \sqrt{3}}-\frac{1}{256} \sqrt{\frac{7}{5}} C_{31,22}^{33,1}-\frac{3}{512}
\sqrt{\frac{7}{5}} C_{33,20}^{33,1}-\frac{3 \sqrt{\frac{21}{2}} C_{33,22}^{33,1}}{1280}\)

\noindent\(C_{22,1112,1}^{20,1111,1,\text{NonLoc}}= \frac{7 \sqrt{5} C_{00,00}^{60,0}}{1024}+\frac{7 \sqrt{\frac{5}{3}} C_{00,20}^{60,0}}{1024}+\frac{7
\sqrt{3} C_{00,22}^{62,0}}{1024}+\frac{7 \sqrt{5} C_{11,00}^{51,0}}{3072}+\frac{35 C_{11,22}^{51,0}}{3072 \sqrt{3}}-\frac{7 \sqrt{5} C_{20,00}^{40,0}}{3072}-\frac{7
\sqrt{\frac{5}{3}} C_{20,20}^{40,0}}{3072}+\frac{7 C_{20,22}^{42,0}}{3072 \sqrt{3}}+\frac{7 C_{22,00}^{42,0}}{3072}+\frac{7 C_{22,20}^{42,0}}{3072
\sqrt{3}}+\frac{7 C_{22,22}^{40,0}}{3072 \sqrt{3}}-\frac{1}{768} \sqrt{\frac{7}{3}} C_{22,22}^{42,0}-\frac{3}{256} \sqrt{\frac{3}{5}} C_{22,22}^{44,0}-\frac{7
\sqrt{5} C_{31,00}^{31,0}}{3072}-\frac{7 \sqrt{\frac{5}{3}} C_{31,20}^{31,0}}{3072}+\frac{7 C_{31,22}^{31,0}}{3072 \sqrt{3}}-\frac{1}{256} \sqrt{\frac{7}{5}}
C_{31,22}^{33,0}+\frac{3 \sqrt{\frac{21}{5}} C_{33,00}^{33,0}}{1024}+\frac{3 \sqrt{\frac{7}{5}} C_{33,20}^{33,0}}{1024}-\frac{3 \sqrt{\frac{21}{2}}
C_{33,22}^{33,0}}{1280}\)

\noindent\(C_{22,1112,1}^{22,1111,0,\text{Loc}}= -\frac{7 C_{00,20}^{60,0}}{768}-\frac{7 C_{00,20}^{60,1}}{1536}+\frac{7 C_{00,22}^{62,0}}{512
\sqrt{5}}+\frac{7 C_{00,22}^{62,1}}{1024 \sqrt{5}}-\frac{7 \sqrt{5} C_{11,22}^{51,0}}{4608}-\frac{7 \sqrt{5} C_{11,22}^{51,1}}{9216}+\frac{7 C_{20,20}^{40,0}}{2304}+\frac{7
C_{20,20}^{40,1}}{4608}+\frac{7 C_{20,22}^{42,0}}{4608 \sqrt{5}}+\frac{7 C_{20,22}^{42,1}}{9216 \sqrt{5}}-\frac{7 C_{22,20}^{42,0}}{2304 \sqrt{5}}-\frac{7
C_{22,20}^{42,1}}{4608 \sqrt{5}}+\frac{7 C_{22,22}^{40,0}}{4608 \sqrt{5}}+\frac{7 C_{22,22}^{40,1}}{9216 \sqrt{5}}-\frac{\sqrt{\frac{7}{5}} C_{22,22}^{42,0}}{1152}-\frac{\sqrt{\frac{7}{5}}
C_{22,22}^{42,1}}{2304}-\frac{3 C_{22,22}^{44,0}}{640}-\frac{3 C_{22,22}^{44,1}}{1280}-\frac{7 C_{31,20}^{31,0}}{2304}-\frac{7 C_{31,20}^{31,1}}{4608}-\frac{7
C_{31,22}^{31,0}}{4608 \sqrt{5}}-\frac{7 C_{31,22}^{31,1}}{9216 \sqrt{5}}+\frac{1}{640} \sqrt{\frac{7}{3}} C_{31,22}^{33,0}+\frac{\sqrt{\frac{7}{3}}
C_{31,22}^{33,1}}{1280}+\frac{\sqrt{21} C_{33,20}^{33,0}}{1280}+\frac{\sqrt{21} C_{33,20}^{33,1}}{2560}+\frac{3}{640} \sqrt{\frac{7}{10}} C_{33,22}^{33,0}+\frac{3
\sqrt{\frac{7}{10}} C_{33,22}^{33,1}}{1280}\)

\noindent\(C_{22,1112,1}^{22,1111,0,\text{NonLoc}}= -\frac{7 C_{00,00}^{60,0}}{1024 \sqrt{3}}-\frac{7 C_{00,00}^{60,1}}{512 \sqrt{3}}-\frac{7
C_{00,20}^{60,0}}{3072}-\frac{7 C_{00,20}^{60,1}}{1536}-\frac{7 C_{00,22}^{62,0}}{1024 \sqrt{5}}-\frac{7 C_{00,22}^{62,1}}{512 \sqrt{5}}-\frac{7
C_{11,00}^{51,0}}{3072 \sqrt{3}}-\frac{7 C_{11,00}^{51,1}}{1536 \sqrt{3}}-\frac{7 \sqrt{5} C_{11,22}^{51,0}}{9216}-\frac{7 \sqrt{5} C_{11,22}^{51,1}}{4608}+\frac{7
C_{20,00}^{40,0}}{3072 \sqrt{3}}+\frac{7 C_{20,00}^{40,1}}{1536 \sqrt{3}}+\frac{7 C_{20,20}^{40,0}}{9216}+\frac{7 C_{20,20}^{40,1}}{4608}-\frac{7
C_{20,22}^{42,0}}{9216 \sqrt{5}}-\frac{7 C_{20,22}^{42,1}}{4608 \sqrt{5}}-\frac{7 C_{22,00}^{42,0}}{3072 \sqrt{15}}-\frac{7 C_{22,00}^{42,1}}{1536
\sqrt{15}}-\frac{7 C_{22,20}^{42,0}}{9216 \sqrt{5}}-\frac{7 C_{22,20}^{42,1}}{4608 \sqrt{5}}-\frac{7 C_{22,22}^{40,0}}{9216 \sqrt{5}}-\frac{7 C_{22,22}^{40,1}}{4608
\sqrt{5}}+\frac{\sqrt{\frac{7}{5}} C_{22,22}^{42,0}}{2304}+\frac{\sqrt{\frac{7}{5}} C_{22,22}^{42,1}}{1152}+\frac{3 C_{22,22}^{44,0}}{1280}+\frac{3
C_{22,22}^{44,1}}{640}+\frac{7 C_{31,00}^{31,0}}{3072 \sqrt{3}}+\frac{7 C_{31,00}^{31,1}}{1536 \sqrt{3}}+\frac{7 C_{31,20}^{31,0}}{9216}+\frac{7
C_{31,20}^{31,1}}{4608}-\frac{7 C_{31,22}^{31,0}}{9216 \sqrt{5}}-\frac{7 C_{31,22}^{31,1}}{4608 \sqrt{5}}+\frac{\sqrt{\frac{7}{3}} C_{31,22}^{33,0}}{1280}+\frac{1}{640}
\sqrt{\frac{7}{3}} C_{31,22}^{33,1}-\frac{3 \sqrt{7} C_{33,00}^{33,0}}{5120}-\frac{3 \sqrt{7} C_{33,00}^{33,1}}{2560}-\frac{\sqrt{21} C_{33,20}^{33,0}}{5120}-\frac{\sqrt{21}
C_{33,20}^{33,1}}{2560}+\frac{3 \sqrt{\frac{7}{10}} C_{33,22}^{33,0}}{1280}+\frac{3}{640} \sqrt{\frac{7}{10}} C_{33,22}^{33,1}\)

\noindent\(C_{22,1112,1}^{22,1111,1,\text{Loc}}= -\frac{7 C_{00,20}^{60,1}}{512 \sqrt{3}}+\frac{7 \sqrt{\frac{3}{5}} C_{00,22}^{62,1}}{1024}-\frac{7
\sqrt{\frac{5}{3}} C_{11,22}^{51,1}}{3072}+\frac{7 C_{20,20}^{40,1}}{1536 \sqrt{3}}+\frac{7 C_{20,22}^{42,1}}{3072 \sqrt{15}}-\frac{7 C_{22,20}^{42,1}}{1536
\sqrt{15}}+\frac{7 C_{22,22}^{40,1}}{3072 \sqrt{15}}-\frac{1}{768} \sqrt{\frac{7}{15}} C_{22,22}^{42,1}-\frac{3 \sqrt{3} C_{22,22}^{44,1}}{1280}-\frac{7
C_{31,20}^{31,1}}{1536 \sqrt{3}}-\frac{7 C_{31,22}^{31,1}}{3072 \sqrt{15}}+\frac{\sqrt{7} C_{31,22}^{33,1}}{1280}+\frac{3 \sqrt{7} C_{33,20}^{33,1}}{2560}+\frac{3
\sqrt{\frac{21}{10}} C_{33,22}^{33,1}}{1280}\)

\noindent\(C_{22,1112,1}^{22,1111,1,\text{NonLoc}}= -\frac{7 C_{00,00}^{60,0}}{1024}-\frac{7 C_{00,20}^{60,0}}{1024 \sqrt{3}}-\frac{7
\sqrt{\frac{3}{5}} C_{00,22}^{62,0}}{1024}-\frac{7 C_{11,00}^{51,0}}{3072}-\frac{7 \sqrt{\frac{5}{3}} C_{11,22}^{51,0}}{3072}+\frac{7 C_{20,00}^{40,0}}{3072}+\frac{7
C_{20,20}^{40,0}}{3072 \sqrt{3}}-\frac{7 C_{20,22}^{42,0}}{3072 \sqrt{15}}-\frac{7 C_{22,00}^{42,0}}{3072 \sqrt{5}}-\frac{7 C_{22,20}^{42,0}}{3072
\sqrt{15}}-\frac{7 C_{22,22}^{40,0}}{3072 \sqrt{15}}+\frac{1}{768} \sqrt{\frac{7}{15}} C_{22,22}^{42,0}+\frac{3 \sqrt{3} C_{22,22}^{44,0}}{1280}+\frac{7
C_{31,00}^{31,0}}{3072}+\frac{7 C_{31,20}^{31,0}}{3072 \sqrt{3}}-\frac{7 C_{31,22}^{31,0}}{3072 \sqrt{15}}+\frac{\sqrt{7} C_{31,22}^{33,0}}{1280}-\frac{3
\sqrt{21} C_{33,00}^{33,0}}{5120}-\frac{3 \sqrt{7} C_{33,20}^{33,0}}{5120}+\frac{3 \sqrt{\frac{21}{10}} C_{33,22}^{33,0}}{1280}\)

\noindent\(C_{22,1112,1}^{22,1112,0,\text{Loc}}= \frac{7 C_{00,20}^{60,0}}{512 \sqrt{3}}+\frac{7 C_{00,20}^{60,1}}{1024 \sqrt{3}}-\frac{7
\sqrt{\frac{3}{5}} C_{00,22}^{62,0}}{1024}-\frac{7 \sqrt{\frac{3}{5}} C_{00,22}^{62,1}}{2048}+\frac{77 C_{11,22}^{51,0}}{3072 \sqrt{15}}+\frac{77
C_{11,22}^{51,1}}{6144 \sqrt{15}}-\frac{9 \sqrt{7} C_{11,22}^{53,0}}{5120}-\frac{9 \sqrt{7} C_{11,22}^{53,1}}{10240}+\frac{C_{20,20}^{40,0}}{1536
\sqrt{3}}+\frac{C_{20,20}^{40,1}}{3072 \sqrt{3}}+\frac{91 C_{20,22}^{42,0}}{3072 \sqrt{15}}+\frac{91 C_{20,22}^{42,1}}{6144 \sqrt{15}}-\frac{7 C_{22,20}^{42,0}}{1536
\sqrt{15}}-\frac{7 C_{22,20}^{42,1}}{3072 \sqrt{15}}-\frac{83 C_{22,22}^{40,0}}{3072 \sqrt{15}}-\frac{83 C_{22,22}^{40,1}}{6144 \sqrt{15}}-\frac{1}{768}
\sqrt{\frac{7}{15}} C_{22,22}^{42,0}-\frac{\sqrt{\frac{7}{15}} C_{22,22}^{42,1}}{1536}-\frac{3 \sqrt{3} C_{22,22}^{44,0}}{1280}-\frac{3 \sqrt{3}
C_{22,22}^{44,1}}{2560}+\frac{C_{31,20}^{31,0}}{512 \sqrt{3}}+\frac{C_{31,20}^{31,1}}{1024 \sqrt{3}}-\frac{29 C_{31,22}^{31,0}}{1024 \sqrt{15}}-\frac{29
C_{31,22}^{31,1}}{2048 \sqrt{15}}+\frac{9 \sqrt{7} C_{31,22}^{33,0}}{5120}+\frac{9 \sqrt{7} C_{31,22}^{33,1}}{10240}+\frac{3 \sqrt{7} C_{33,20}^{33,0}}{2560}+\frac{3
\sqrt{7} C_{33,20}^{33,1}}{5120}+\frac{3 \sqrt{\frac{21}{10}} C_{33,22}^{33,0}}{1280}+\frac{3 \sqrt{\frac{21}{10}} C_{33,22}^{33,1}}{2560}\)

\noindent\(C_{22,1112,1}^{22,1112,0,\text{NonLoc}}= \frac{7 C_{00,00}^{60,0}}{2048}+\frac{7 C_{00,00}^{60,1}}{1024}+\frac{7 C_{00,20}^{60,0}}{2048
\sqrt{3}}+\frac{7 C_{00,20}^{60,1}}{1024 \sqrt{3}}+\frac{7 \sqrt{\frac{3}{5}} C_{00,22}^{62,0}}{2048}+\frac{7 \sqrt{\frac{3}{5}} C_{00,22}^{62,1}}{1024}+\frac{7
C_{11,00}^{51,0}}{6144}+\frac{7 C_{11,00}^{51,1}}{3072}+\frac{77 C_{11,22}^{51,0}}{6144 \sqrt{15}}+\frac{77 C_{11,22}^{51,1}}{3072 \sqrt{15}}-\frac{9
\sqrt{7} C_{11,22}^{53,0}}{10240}-\frac{9 \sqrt{7} C_{11,22}^{53,1}}{5120}+\frac{C_{20,00}^{40,0}}{6144}+\frac{C_{20,00}^{40,1}}{3072}+\frac{C_{20,20}^{40,0}}{6144
\sqrt{3}}+\frac{C_{20,20}^{40,1}}{3072 \sqrt{3}}-\frac{91 C_{20,22}^{42,0}}{6144 \sqrt{15}}-\frac{91 C_{20,22}^{42,1}}{3072 \sqrt{15}}-\frac{7 C_{22,00}^{42,0}}{6144
\sqrt{5}}-\frac{7 C_{22,00}^{42,1}}{3072 \sqrt{5}}-\frac{7 C_{22,20}^{42,0}}{6144 \sqrt{15}}-\frac{7 C_{22,20}^{42,1}}{3072 \sqrt{15}}+\frac{83 C_{22,22}^{40,0}}{6144
\sqrt{15}}+\frac{83 C_{22,22}^{40,1}}{3072 \sqrt{15}}+\frac{\sqrt{\frac{7}{15}} C_{22,22}^{42,0}}{1536}+\frac{1}{768} \sqrt{\frac{7}{15}} C_{22,22}^{42,1}+\frac{3
\sqrt{3} C_{22,22}^{44,0}}{2560}+\frac{3 \sqrt{3} C_{22,22}^{44,1}}{1280}-\frac{C_{31,00}^{31,0}}{2048}-\frac{C_{31,00}^{31,1}}{1024}-\frac{C_{31,20}^{31,0}}{2048
\sqrt{3}}-\frac{C_{31,20}^{31,1}}{1024 \sqrt{3}}-\frac{29 C_{31,22}^{31,0}}{2048 \sqrt{15}}-\frac{29 C_{31,22}^{31,1}}{1024 \sqrt{15}}+\frac{9 \sqrt{7}
C_{31,22}^{33,0}}{10240}+\frac{9 \sqrt{7} C_{31,22}^{33,1}}{5120}-\frac{3 \sqrt{21} C_{33,00}^{33,0}}{10240}-\frac{3 \sqrt{21} C_{33,00}^{33,1}}{5120}-\frac{3
\sqrt{7} C_{33,20}^{33,0}}{10240}-\frac{3 \sqrt{7} C_{33,20}^{33,1}}{5120}+\frac{3 \sqrt{\frac{21}{10}} C_{33,22}^{33,0}}{2560}+\frac{3 \sqrt{\frac{21}{10}}
C_{33,22}^{33,1}}{1280}\)

\noindent\(C_{22,1112,1}^{22,1112,1,\text{Loc}}= \frac{7 C_{00,20}^{60,1}}{1024}-\frac{21 C_{00,22}^{62,1}}{2048 \sqrt{5}}+\frac{77
C_{11,22}^{51,1}}{6144 \sqrt{5}}-\frac{9 \sqrt{21} C_{11,22}^{53,1}}{10240}+\frac{C_{20,20}^{40,1}}{3072}+\frac{91 C_{20,22}^{42,1}}{6144 \sqrt{5}}-\frac{7
C_{22,20}^{42,1}}{3072 \sqrt{5}}-\frac{83 C_{22,22}^{40,1}}{6144 \sqrt{5}}-\frac{\sqrt{\frac{7}{5}} C_{22,22}^{42,1}}{1536}-\frac{9 C_{22,22}^{44,1}}{2560}+\frac{C_{31,20}^{31,1}}{1024}-\frac{29
C_{31,22}^{31,1}}{2048 \sqrt{5}}+\frac{9 \sqrt{21} C_{31,22}^{33,1}}{10240}+\frac{3 \sqrt{21} C_{33,20}^{33,1}}{5120}+\frac{9 \sqrt{\frac{7}{10}}
C_{33,22}^{33,1}}{2560}\)

\noindent\(C_{22,1112,1}^{22,1112,1,\text{NonLoc}}= \frac{7 \sqrt{3} C_{00,00}^{60,0}}{2048}+\frac{7 C_{00,20}^{60,0}}{2048}+\frac{21
C_{00,22}^{62,0}}{2048 \sqrt{5}}+\frac{7 C_{11,00}^{51,0}}{2048 \sqrt{3}}+\frac{77 C_{11,22}^{51,0}}{6144 \sqrt{5}}-\frac{9 \sqrt{21} C_{11,22}^{53,0}}{10240}+\frac{C_{20,00}^{40,0}}{2048
\sqrt{3}}+\frac{C_{20,20}^{40,0}}{6144}-\frac{91 C_{20,22}^{42,0}}{6144 \sqrt{5}}-\frac{7 C_{22,00}^{42,0}}{2048 \sqrt{15}}-\frac{7 C_{22,20}^{42,0}}{6144
\sqrt{5}}+\frac{83 C_{22,22}^{40,0}}{6144 \sqrt{5}}+\frac{\sqrt{\frac{7}{5}} C_{22,22}^{42,0}}{1536}+\frac{9 C_{22,22}^{44,0}}{2560}-\frac{\sqrt{3}
C_{31,00}^{31,0}}{2048}-\frac{C_{31,20}^{31,0}}{2048}-\frac{29 C_{31,22}^{31,0}}{2048 \sqrt{5}}+\frac{9 \sqrt{21} C_{31,22}^{33,0}}{10240}-\frac{9
\sqrt{7} C_{33,00}^{33,0}}{10240}-\frac{3 \sqrt{21} C_{33,20}^{33,0}}{10240}+\frac{9 \sqrt{\frac{7}{10}} C_{33,22}^{33,0}}{2560}\)

\noindent\(C_{22,1112,2}^{00,3112,0,\text{Loc}}= -\frac{7}{192} \sqrt{\frac{7}{5}} C_{00,20}^{60,0}-\frac{7}{384} \sqrt{\frac{7}{5}}
C_{00,20}^{60,1}-\frac{\sqrt{7} C_{00,22}^{62,0}}{128}-\frac{\sqrt{7} C_{00,22}^{62,1}}{256}+\frac{11 \sqrt{7} C_{11,22}^{51,0}}{5760}+\frac{11 \sqrt{7}
C_{11,22}^{51,1}}{11520}-\frac{9}{320} \sqrt{\frac{3}{5}} C_{11,22}^{53,0}-\frac{9}{640} \sqrt{\frac{3}{5}} C_{11,22}^{53,1}+\frac{7}{576} \sqrt{\frac{7}{5}}
C_{20,20}^{40,0}+\frac{7 \sqrt{\frac{7}{5}} C_{20,20}^{40,1}}{1152}-\frac{11 \sqrt{7} C_{20,22}^{42,0}}{5760}-\frac{11 \sqrt{7} C_{20,22}^{42,1}}{11520}-\frac{7
\sqrt{7} C_{22,20}^{42,0}}{2880}-\frac{7 \sqrt{7} C_{22,20}^{42,1}}{5760}+\frac{7 \sqrt{7} C_{22,22}^{40,0}}{5760}+\frac{7 \sqrt{7} C_{22,22}^{40,1}}{11520}+\frac{49
C_{22,22}^{42,0}}{2880}+\frac{49 C_{22,22}^{42,1}}{5760}-\frac{3}{160} \sqrt{\frac{7}{5}} C_{22,22}^{44,0}-\frac{3}{320} \sqrt{\frac{7}{5}} C_{22,22}^{44,1}+\frac{7}{576}
\sqrt{\frac{7}{5}} C_{31,20}^{31,0}+\frac{7 \sqrt{\frac{7}{5}} C_{31,20}^{31,1}}{1152}+\frac{19 \sqrt{7} C_{31,22}^{31,0}}{5760}+\frac{19 \sqrt{7}
C_{31,22}^{31,1}}{11520}-\frac{C_{31,22}^{33,0}}{40 \sqrt{15}}-\frac{C_{31,22}^{33,1}}{80 \sqrt{15}}-\frac{7}{320} \sqrt{\frac{3}{5}} C_{33,20}^{33,0}-\frac{7}{640}
\sqrt{\frac{3}{5}} C_{33,20}^{33,1}+\frac{33 C_{33,22}^{33,0}}{800 \sqrt{2}}+\frac{33 C_{33,22}^{33,1}}{1600 \sqrt{2}}\)

\noindent\(C_{22,1112,2}^{00,3112,0,\text{NonLoc}}= -\frac{7}{256} \sqrt{\frac{7}{15}} C_{00,00}^{60,0}-\frac{7}{128} \sqrt{\frac{7}{15}}
C_{00,00}^{60,1}-\frac{7}{768} \sqrt{\frac{7}{5}} C_{00,20}^{60,0}-\frac{7}{384} \sqrt{\frac{7}{5}} C_{00,20}^{60,1}+\frac{\sqrt{7} C_{00,22}^{62,0}}{256}+\frac{\sqrt{7}
C_{00,22}^{62,1}}{128}+\frac{7}{768} \sqrt{\frac{7}{15}} C_{11,00}^{51,0}+\frac{7}{384} \sqrt{\frac{7}{15}} C_{11,00}^{51,1}+\frac{11 \sqrt{7} C_{11,22}^{51,0}}{11520}+\frac{11
\sqrt{7} C_{11,22}^{51,1}}{5760}-\frac{9}{640} \sqrt{\frac{3}{5}} C_{11,22}^{53,0}-\frac{9}{320} \sqrt{\frac{3}{5}} C_{11,22}^{53,1}+\frac{7}{768}
\sqrt{\frac{7}{15}} C_{20,00}^{40,0}+\frac{7}{384} \sqrt{\frac{7}{15}} C_{20,00}^{40,1}+\frac{7 \sqrt{\frac{7}{5}} C_{20,20}^{40,0}}{2304}+\frac{7
\sqrt{\frac{7}{5}} C_{20,20}^{40,1}}{1152}+\frac{11 \sqrt{7} C_{20,22}^{42,0}}{11520}+\frac{11 \sqrt{7} C_{20,22}^{42,1}}{5760}-\frac{7 \sqrt{\frac{7}{3}}
C_{22,00}^{42,0}}{3840}-\frac{7 \sqrt{\frac{7}{3}} C_{22,00}^{42,1}}{1920}-\frac{7 \sqrt{7} C_{22,20}^{42,0}}{11520}-\frac{7 \sqrt{7} C_{22,20}^{42,1}}{5760}-\frac{7
\sqrt{7} C_{22,22}^{40,0}}{11520}-\frac{7 \sqrt{7} C_{22,22}^{40,1}}{5760}-\frac{49 C_{22,22}^{42,0}}{5760}-\frac{49 C_{22,22}^{42,1}}{2880}+\frac{3}{320}
\sqrt{\frac{7}{5}} C_{22,22}^{44,0}+\frac{3}{160} \sqrt{\frac{7}{5}} C_{22,22}^{44,1}-\frac{7}{768} \sqrt{\frac{7}{15}} C_{31,00}^{31,0}-\frac{7}{384}
\sqrt{\frac{7}{15}} C_{31,00}^{31,1}-\frac{7 \sqrt{\frac{7}{5}} C_{31,20}^{31,0}}{2304}-\frac{7 \sqrt{\frac{7}{5}} C_{31,20}^{31,1}}{1152}+\frac{19
\sqrt{7} C_{31,22}^{31,0}}{11520}+\frac{19 \sqrt{7} C_{31,22}^{31,1}}{5760}-\frac{C_{31,22}^{33,0}}{80 \sqrt{15}}-\frac{C_{31,22}^{33,1}}{40 \sqrt{15}}+\frac{21
C_{33,00}^{33,0}}{1280 \sqrt{5}}+\frac{21 C_{33,00}^{33,1}}{640 \sqrt{5}}+\frac{7 \sqrt{\frac{3}{5}} C_{33,20}^{33,0}}{1280}+\frac{7}{640} \sqrt{\frac{3}{5}}
C_{33,20}^{33,1}+\frac{33 C_{33,22}^{33,0}}{1600 \sqrt{2}}+\frac{33 C_{33,22}^{33,1}}{800 \sqrt{2}}\)

\noindent\(C_{22,1112,2}^{00,3112,1,\text{Loc}}= -\frac{7}{128} \sqrt{\frac{7}{15}} C_{00,20}^{60,1}-\frac{\sqrt{21} C_{00,22}^{62,1}}{256}+\frac{11
\sqrt{\frac{7}{3}} C_{11,22}^{51,1}}{3840}-\frac{27 C_{11,22}^{53,1}}{640 \sqrt{5}}+\frac{7}{384} \sqrt{\frac{7}{15}} C_{20,20}^{40,1}-\frac{11 \sqrt{\frac{7}{3}}
C_{20,22}^{42,1}}{3840}-\frac{7 \sqrt{\frac{7}{3}} C_{22,20}^{42,1}}{1920}+\frac{7 \sqrt{\frac{7}{3}} C_{22,22}^{40,1}}{3840}+\frac{49 C_{22,22}^{42,1}}{1920
\sqrt{3}}-\frac{3}{320} \sqrt{\frac{21}{5}} C_{22,22}^{44,1}+\frac{7}{384} \sqrt{\frac{7}{15}} C_{31,20}^{31,1}+\frac{19 \sqrt{\frac{7}{3}} C_{31,22}^{31,1}}{3840}-\frac{C_{31,22}^{33,1}}{80
\sqrt{5}}-\frac{21 C_{33,20}^{33,1}}{640 \sqrt{5}}+\frac{33 \sqrt{\frac{3}{2}} C_{33,22}^{33,1}}{1600}\)

\noindent\(C_{22,1112,2}^{00,3112,1,\text{NonLoc}}= -\frac{7}{256} \sqrt{\frac{7}{5}} C_{00,00}^{60,0}-\frac{7}{256} \sqrt{\frac{7}{15}}
C_{00,20}^{60,0}+\frac{\sqrt{21} C_{00,22}^{62,0}}{256}+\frac{7}{768} \sqrt{\frac{7}{5}} C_{11,00}^{51,0}+\frac{11 \sqrt{\frac{7}{3}} C_{11,22}^{51,0}}{3840}-\frac{27
C_{11,22}^{53,0}}{640 \sqrt{5}}+\frac{7}{768} \sqrt{\frac{7}{5}} C_{20,00}^{40,0}+\frac{7}{768} \sqrt{\frac{7}{15}} C_{20,20}^{40,0}+\frac{11 \sqrt{\frac{7}{3}}
C_{20,22}^{42,0}}{3840}-\frac{7 \sqrt{7} C_{22,00}^{42,0}}{3840}-\frac{7 \sqrt{\frac{7}{3}} C_{22,20}^{42,0}}{3840}-\frac{7 \sqrt{\frac{7}{3}} C_{22,22}^{40,0}}{3840}-\frac{49
C_{22,22}^{42,0}}{1920 \sqrt{3}}+\frac{3}{320} \sqrt{\frac{21}{5}} C_{22,22}^{44,0}-\frac{7}{768} \sqrt{\frac{7}{5}} C_{31,00}^{31,0}-\frac{7}{768}
\sqrt{\frac{7}{15}} C_{31,20}^{31,0}+\frac{19 \sqrt{\frac{7}{3}} C_{31,22}^{31,0}}{3840}-\frac{C_{31,22}^{33,0}}{80 \sqrt{5}}+\frac{21 \sqrt{\frac{3}{5}}
C_{33,00}^{33,0}}{1280}+\frac{21 C_{33,20}^{33,0}}{1280 \sqrt{5}}+\frac{33 \sqrt{\frac{3}{2}} C_{33,22}^{33,0}}{1600}\)

\noindent\(C_{22,1112,2}^{00,3312,0,\text{Loc}}= -\frac{C_{00,20}^{60,0}}{96 \sqrt{3}}-\frac{C_{00,20}^{60,1}}{192 \sqrt{3}}-\frac{1}{64}
\sqrt{\frac{5}{3}} C_{00,22}^{62,0}-\frac{1}{128} \sqrt{\frac{5}{3}} C_{00,22}^{62,1}-\frac{11 C_{11,22}^{51,0}}{288 \sqrt{15}}-\frac{11 C_{11,22}^{51,1}}{576
\sqrt{15}}-\frac{3 C_{11,22}^{53,0}}{640 \sqrt{7}}-\frac{3 C_{11,22}^{53,1}}{1280 \sqrt{7}}+\frac{C_{20,20}^{40,0}}{288 \sqrt{3}}+\frac{C_{20,20}^{40,1}}{576
\sqrt{3}}+\frac{11 C_{20,22}^{42,0}}{288 \sqrt{15}}+\frac{11 C_{20,22}^{42,1}}{576 \sqrt{15}}-\frac{C_{22,20}^{42,0}}{288 \sqrt{15}}-\frac{C_{22,20}^{42,1}}{576
\sqrt{15}}+\frac{C_{22,22}^{40,0}}{576 \sqrt{15}}+\frac{C_{22,22}^{40,1}}{1152 \sqrt{15}}-\frac{23 C_{22,22}^{42,0}}{288 \sqrt{105}}-\frac{23 C_{22,22}^{42,1}}{576
\sqrt{105}}-\frac{\sqrt{3} C_{22,22}^{44,0}}{560}-\frac{\sqrt{3} C_{22,22}^{44,1}}{1120}+\frac{C_{31,20}^{31,0}}{288 \sqrt{3}}+\frac{C_{31,20}^{31,1}}{576
\sqrt{3}}+\frac{11 C_{31,22}^{31,0}}{288 \sqrt{15}}+\frac{11 C_{31,22}^{31,1}}{576 \sqrt{15}}+\frac{3 C_{31,22}^{33,0}}{640 \sqrt{7}}+\frac{3 C_{31,22}^{33,1}}{1280
\sqrt{7}}-\frac{C_{33,20}^{33,0}}{160 \sqrt{7}}-\frac{C_{33,20}^{33,1}}{320 \sqrt{7}}-\frac{1}{80} \sqrt{\frac{21}{10}} C_{33,22}^{33,0}-\frac{1}{160}
\sqrt{\frac{21}{10}} C_{33,22}^{33,1}\)

\noindent\(C_{22,1112,2}^{00,3312,0,\text{NonLoc}}= -\frac{C_{00,00}^{60,0}}{384}-\frac{C_{00,00}^{60,1}}{192}-\frac{C_{00,20}^{60,0}}{384
\sqrt{3}}-\frac{C_{00,20}^{60,1}}{192 \sqrt{3}}+\frac{1}{128} \sqrt{\frac{5}{3}} C_{00,22}^{62,0}+\frac{1}{64} \sqrt{\frac{5}{3}} C_{00,22}^{62,1}+\frac{C_{11,00}^{51,0}}{1152}+\frac{C_{11,00}^{51,1}}{576}-\frac{11
C_{11,22}^{51,0}}{576 \sqrt{15}}-\frac{11 C_{11,22}^{51,1}}{288 \sqrt{15}}-\frac{3 C_{11,22}^{53,0}}{1280 \sqrt{7}}-\frac{3 C_{11,22}^{53,1}}{640
\sqrt{7}}+\frac{C_{20,00}^{40,0}}{1152}+\frac{C_{20,00}^{40,1}}{576}+\frac{C_{20,20}^{40,0}}{1152 \sqrt{3}}+\frac{C_{20,20}^{40,1}}{576 \sqrt{3}}-\frac{11
C_{20,22}^{42,0}}{576 \sqrt{15}}-\frac{11 C_{20,22}^{42,1}}{288 \sqrt{15}}-\frac{C_{22,00}^{42,0}}{1152 \sqrt{5}}-\frac{C_{22,00}^{42,1}}{576 \sqrt{5}}-\frac{C_{22,20}^{42,0}}{1152
\sqrt{15}}-\frac{C_{22,20}^{42,1}}{576 \sqrt{15}}-\frac{C_{22,22}^{40,0}}{1152 \sqrt{15}}-\frac{C_{22,22}^{40,1}}{576 \sqrt{15}}+\frac{23 C_{22,22}^{42,0}}{576
\sqrt{105}}+\frac{23 C_{22,22}^{42,1}}{288 \sqrt{105}}+\frac{\sqrt{3} C_{22,22}^{44,0}}{1120}+\frac{\sqrt{3} C_{22,22}^{44,1}}{560}-\frac{C_{31,00}^{31,0}}{1152}-\frac{C_{31,00}^{31,1}}{576}-\frac{C_{31,20}^{31,0}}{1152
\sqrt{3}}-\frac{C_{31,20}^{31,1}}{576 \sqrt{3}}+\frac{11 C_{31,22}^{31,0}}{576 \sqrt{15}}+\frac{11 C_{31,22}^{31,1}}{288 \sqrt{15}}+\frac{3 C_{31,22}^{33,0}}{1280
\sqrt{7}}+\frac{3 C_{31,22}^{33,1}}{640 \sqrt{7}}+\frac{1}{640} \sqrt{\frac{3}{7}} C_{33,00}^{33,0}+\frac{1}{320} \sqrt{\frac{3}{7}} C_{33,00}^{33,1}+\frac{C_{33,20}^{33,0}}{640
\sqrt{7}}+\frac{C_{33,20}^{33,1}}{320 \sqrt{7}}-\frac{1}{160} \sqrt{\frac{21}{10}} C_{33,22}^{33,0}-\frac{1}{80} \sqrt{\frac{21}{10}} C_{33,22}^{33,1}\)

\noindent\(C_{22,1112,2}^{00,3312,1,\text{Loc}}= -\frac{C_{00,20}^{60,1}}{192}-\frac{\sqrt{5} C_{00,22}^{62,1}}{128}-\frac{11 C_{11,22}^{51,1}}{576
\sqrt{5}}-\frac{3 \sqrt{\frac{3}{7}} C_{11,22}^{53,1}}{1280}+\frac{C_{20,20}^{40,1}}{576}+\frac{11 C_{20,22}^{42,1}}{576 \sqrt{5}}-\frac{C_{22,20}^{42,1}}{576
\sqrt{5}}+\frac{C_{22,22}^{40,1}}{1152 \sqrt{5}}-\frac{23 C_{22,22}^{42,1}}{576 \sqrt{35}}-\frac{3 C_{22,22}^{44,1}}{1120}+\frac{C_{31,20}^{31,1}}{576}+\frac{11
C_{31,22}^{31,1}}{576 \sqrt{5}}+\frac{3 \sqrt{\frac{3}{7}} C_{31,22}^{33,1}}{1280}-\frac{1}{320} \sqrt{\frac{3}{7}} C_{33,20}^{33,1}-\frac{3}{160}
\sqrt{\frac{7}{10}} C_{33,22}^{33,1}\)

\noindent\(C_{22,1112,2}^{00,3312,1,\text{NonLoc}}= -\frac{C_{00,00}^{60,0}}{128 \sqrt{3}}-\frac{C_{00,20}^{60,0}}{384}+\frac{\sqrt{5}
C_{00,22}^{62,0}}{128}+\frac{C_{11,00}^{51,0}}{384 \sqrt{3}}-\frac{11 C_{11,22}^{51,0}}{576 \sqrt{5}}-\frac{3 \sqrt{\frac{3}{7}} C_{11,22}^{53,0}}{1280}+\frac{C_{20,00}^{40,0}}{384
\sqrt{3}}+\frac{C_{20,20}^{40,0}}{1152}-\frac{11 C_{20,22}^{42,0}}{576 \sqrt{5}}-\frac{C_{22,00}^{42,0}}{384 \sqrt{15}}-\frac{C_{22,20}^{42,0}}{1152
\sqrt{5}}-\frac{C_{22,22}^{40,0}}{1152 \sqrt{5}}+\frac{23 C_{22,22}^{42,0}}{576 \sqrt{35}}+\frac{3 C_{22,22}^{44,0}}{1120}-\frac{C_{31,00}^{31,0}}{384
\sqrt{3}}-\frac{C_{31,20}^{31,0}}{1152}+\frac{11 C_{31,22}^{31,0}}{576 \sqrt{5}}+\frac{3 \sqrt{\frac{3}{7}} C_{31,22}^{33,0}}{1280}+\frac{3 C_{33,00}^{33,0}}{640
\sqrt{7}}+\frac{1}{640} \sqrt{\frac{3}{7}} C_{33,20}^{33,0}-\frac{3}{160} \sqrt{\frac{7}{10}} C_{33,22}^{33,0}\)

\noindent\(C_{22,1112,2}^{11,2202,0,\text{Loc}}= -\frac{1}{384} \sqrt{\frac{35}{3}} C_{11,11}^{51,0}-\frac{1}{768} \sqrt{\frac{35}{3}}
C_{11,11}^{51,1}+\frac{1}{128} \sqrt{\frac{35}{3}} C_{31,11}^{31,0}+\frac{1}{256} \sqrt{\frac{35}{3}} C_{31,11}^{31,1}-\frac{1}{16} \sqrt{\frac{3}{10}}
C_{33,11}^{33,0}-\frac{1}{32} \sqrt{\frac{3}{10}} C_{33,11}^{33,1}\)

\noindent\(C_{22,1112,2}^{11,2202,0,\text{NonLoc}}= -\frac{1}{768} \sqrt{\frac{35}{3}} C_{11,11}^{51,0}-\frac{1}{384} \sqrt{\frac{35}{3}}
C_{11,11}^{51,1}+\frac{1}{256} \sqrt{\frac{35}{3}} C_{31,11}^{31,0}+\frac{1}{128} \sqrt{\frac{35}{3}} C_{31,11}^{31,1}-\frac{1}{32} \sqrt{\frac{3}{10}}
C_{33,11}^{33,0}-\frac{1}{16} \sqrt{\frac{3}{10}} C_{33,11}^{33,1}\)

\noindent\(C_{22,1112,2}^{11,2202,1,\text{Loc}}= -\frac{\sqrt{35} C_{11,11}^{51,1}}{768}+\frac{\sqrt{35} C_{31,11}^{31,1}}{256}-\frac{3
C_{33,11}^{33,1}}{32 \sqrt{10}}\)

\noindent\(C_{22,1112,2}^{11,2202,1,\text{NonLoc}}= -\frac{\sqrt{35} C_{11,11}^{51,0}}{768}+\frac{\sqrt{35} C_{31,11}^{31,0}}{256}-\frac{3
C_{33,11}^{33,0}}{32 \sqrt{10}}\)

\noindent\(C_{22,1112,2}^{20,1112,0,\text{Loc}}= \frac{7 \sqrt{35} C_{00,20}^{60,0}}{2304}+\frac{7 \sqrt{35} C_{00,20}^{60,1}}{4608}+\frac{5
\sqrt{7} C_{00,22}^{62,0}}{1536}+\frac{5 \sqrt{7} C_{00,22}^{62,1}}{3072}-\frac{31 \sqrt{7} C_{11,22}^{51,0}}{13824}-\frac{31 \sqrt{7} C_{11,22}^{51,1}}{27648}+\frac{1}{512}
\sqrt{\frac{3}{5}} C_{11,22}^{53,0}+\frac{\sqrt{\frac{3}{5}} C_{11,22}^{53,1}}{1024}-\frac{7 \sqrt{35} C_{20,20}^{40,0}}{6912}-\frac{7 \sqrt{35}
C_{20,20}^{40,1}}{13824}+\frac{11 \sqrt{7} C_{20,22}^{42,0}}{13824}+\frac{11 \sqrt{7} C_{20,22}^{42,1}}{27648}+\frac{7 \sqrt{7} C_{22,20}^{42,0}}{6912}+\frac{7
\sqrt{7} C_{22,20}^{42,1}}{13824}+\frac{17 \sqrt{7} C_{22,22}^{40,0}}{13824}+\frac{17 \sqrt{7} C_{22,22}^{40,1}}{27648}-\frac{7 C_{22,22}^{42,0}}{1728}-\frac{7
C_{22,22}^{42,1}}{3456}-\frac{1}{128} \sqrt{\frac{7}{5}} C_{22,22}^{44,0}-\frac{1}{256} \sqrt{\frac{7}{5}} C_{22,22}^{44,1}+\frac{7 \sqrt{35} C_{31,20}^{31,0}}{6912}+\frac{7
\sqrt{35} C_{31,20}^{31,1}}{13824}-\frac{11 \sqrt{7} C_{31,22}^{31,0}}{13824}-\frac{11 \sqrt{7} C_{31,22}^{31,1}}{27648}+\frac{19 C_{31,22}^{33,0}}{1536
\sqrt{15}}+\frac{19 C_{31,22}^{33,1}}{3072 \sqrt{15}}-\frac{7 C_{33,20}^{33,0}}{256 \sqrt{15}}-\frac{7 C_{33,20}^{33,1}}{512 \sqrt{15}}+\frac{11
C_{33,22}^{33,0}}{640 \sqrt{2}}+\frac{11 C_{33,22}^{33,1}}{1280 \sqrt{2}}\)

\noindent\(C_{22,1112,2}^{20,1112,0,\text{NonLoc}}= \frac{7 \sqrt{\frac{35}{3}} C_{00,00}^{60,0}}{3072}+\frac{7 \sqrt{\frac{35}{3}}
C_{00,00}^{60,1}}{1536}+\frac{7 \sqrt{35} C_{00,20}^{60,0}}{9216}+\frac{7 \sqrt{35} C_{00,20}^{60,1}}{4608}-\frac{5 \sqrt{7} C_{00,22}^{62,0}}{3072}-\frac{5
\sqrt{7} C_{00,22}^{62,1}}{1536}+\frac{7 \sqrt{\frac{35}{3}} C_{11,00}^{51,0}}{9216}+\frac{7 \sqrt{\frac{35}{3}} C_{11,00}^{51,1}}{4608}-\frac{31
\sqrt{7} C_{11,22}^{51,0}}{27648}-\frac{31 \sqrt{7} C_{11,22}^{51,1}}{13824}+\frac{\sqrt{\frac{3}{5}} C_{11,22}^{53,0}}{1024}+\frac{1}{512} \sqrt{\frac{3}{5}}
C_{11,22}^{53,1}-\frac{7 \sqrt{\frac{35}{3}} C_{20,00}^{40,0}}{9216}-\frac{7 \sqrt{\frac{35}{3}} C_{20,00}^{40,1}}{4608}-\frac{7 \sqrt{35} C_{20,20}^{40,0}}{27648}-\frac{7
\sqrt{35} C_{20,20}^{40,1}}{13824}-\frac{11 \sqrt{7} C_{20,22}^{42,0}}{27648}-\frac{11 \sqrt{7} C_{20,22}^{42,1}}{13824}+\frac{7 \sqrt{\frac{7}{3}}
C_{22,00}^{42,0}}{9216}+\frac{7 \sqrt{\frac{7}{3}} C_{22,00}^{42,1}}{4608}+\frac{7 \sqrt{7} C_{22,20}^{42,0}}{27648}+\frac{7 \sqrt{7} C_{22,20}^{42,1}}{13824}-\frac{17
\sqrt{7} C_{22,22}^{40,0}}{27648}-\frac{17 \sqrt{7} C_{22,22}^{40,1}}{13824}+\frac{7 C_{22,22}^{42,0}}{3456}+\frac{7 C_{22,22}^{42,1}}{1728}+\frac{1}{256}
\sqrt{\frac{7}{5}} C_{22,22}^{44,0}+\frac{1}{128} \sqrt{\frac{7}{5}} C_{22,22}^{44,1}-\frac{7 \sqrt{\frac{35}{3}} C_{31,00}^{31,0}}{9216}-\frac{7
\sqrt{\frac{35}{3}} C_{31,00}^{31,1}}{4608}-\frac{7 \sqrt{35} C_{31,20}^{31,0}}{27648}-\frac{7 \sqrt{35} C_{31,20}^{31,1}}{13824}-\frac{11 \sqrt{7}
C_{31,22}^{31,0}}{27648}-\frac{11 \sqrt{7} C_{31,22}^{31,1}}{13824}+\frac{19 C_{31,22}^{33,0}}{3072 \sqrt{15}}+\frac{19 C_{31,22}^{33,1}}{1536 \sqrt{15}}+\frac{7
C_{33,00}^{33,0}}{1024 \sqrt{5}}+\frac{7 C_{33,00}^{33,1}}{512 \sqrt{5}}+\frac{7 C_{33,20}^{33,0}}{1024 \sqrt{15}}+\frac{7 C_{33,20}^{33,1}}{512
\sqrt{15}}+\frac{11 C_{33,22}^{33,0}}{1280 \sqrt{2}}+\frac{11 C_{33,22}^{33,1}}{640 \sqrt{2}}\)

\noindent\(C_{22,1112,2}^{20,1112,1,\text{Loc}}= \frac{7 \sqrt{\frac{35}{3}} C_{00,20}^{60,1}}{1536}+\frac{5 \sqrt{\frac{7}{3}} C_{00,22}^{62,1}}{1024}-\frac{31
\sqrt{\frac{7}{3}} C_{11,22}^{51,1}}{9216}+\frac{3 C_{11,22}^{53,1}}{1024 \sqrt{5}}-\frac{7 \sqrt{\frac{35}{3}} C_{20,20}^{40,1}}{4608}+\frac{11
\sqrt{\frac{7}{3}} C_{20,22}^{42,1}}{9216}+\frac{7 \sqrt{\frac{7}{3}} C_{22,20}^{42,1}}{4608}+\frac{17 \sqrt{\frac{7}{3}} C_{22,22}^{40,1}}{9216}-\frac{7
C_{22,22}^{42,1}}{1152 \sqrt{3}}-\frac{1}{256} \sqrt{\frac{21}{5}} C_{22,22}^{44,1}+\frac{7 \sqrt{\frac{35}{3}} C_{31,20}^{31,1}}{4608}-\frac{11
\sqrt{\frac{7}{3}} C_{31,22}^{31,1}}{9216}+\frac{19 C_{31,22}^{33,1}}{3072 \sqrt{5}}-\frac{7 C_{33,20}^{33,1}}{512 \sqrt{5}}+\frac{11 \sqrt{\frac{3}{2}}
C_{33,22}^{33,1}}{1280}\)

\noindent\(C_{22,1112,2}^{20,1112,1,\text{NonLoc}}= \frac{7 \sqrt{35} C_{00,00}^{60,0}}{3072}+\frac{7 \sqrt{\frac{35}{3}} C_{00,20}^{60,0}}{3072}-\frac{5
\sqrt{\frac{7}{3}} C_{00,22}^{62,0}}{1024}+\frac{7 \sqrt{35} C_{11,00}^{51,0}}{9216}-\frac{31 \sqrt{\frac{7}{3}} C_{11,22}^{51,0}}{9216}+\frac{3
C_{11,22}^{53,0}}{1024 \sqrt{5}}-\frac{7 \sqrt{35} C_{20,00}^{40,0}}{9216}-\frac{7 \sqrt{\frac{35}{3}} C_{20,20}^{40,0}}{9216}-\frac{11 \sqrt{\frac{7}{3}}
C_{20,22}^{42,0}}{9216}+\frac{7 \sqrt{7} C_{22,00}^{42,0}}{9216}+\frac{7 \sqrt{\frac{7}{3}} C_{22,20}^{42,0}}{9216}-\frac{17 \sqrt{\frac{7}{3}} C_{22,22}^{40,0}}{9216}+\frac{7
C_{22,22}^{42,0}}{1152 \sqrt{3}}+\frac{1}{256} \sqrt{\frac{21}{5}} C_{22,22}^{44,0}-\frac{7 \sqrt{35} C_{31,00}^{31,0}}{9216}-\frac{7 \sqrt{\frac{35}{3}}
C_{31,20}^{31,0}}{9216}-\frac{11 \sqrt{\frac{7}{3}} C_{31,22}^{31,0}}{9216}+\frac{19 C_{31,22}^{33,0}}{3072 \sqrt{5}}+\frac{7 \sqrt{\frac{3}{5}}
C_{33,00}^{33,0}}{1024}+\frac{7 C_{33,20}^{33,0}}{1024 \sqrt{5}}+\frac{11 \sqrt{\frac{3}{2}} C_{33,22}^{33,0}}{1280}\)

\noindent\(C_{22,1112,2}^{22,1110,0,\text{Loc}}= \frac{\sqrt{35} C_{00,20}^{60,0}}{576}+\frac{\sqrt{35} C_{00,20}^{60,1}}{1152}+\frac{\sqrt{7}
C_{00,22}^{62,0}}{192}+\frac{\sqrt{7} C_{00,22}^{62,1}}{384}-\frac{7 \sqrt{7} C_{11,22}^{51,0}}{3456}-\frac{7 \sqrt{7} C_{11,22}^{51,1}}{6912}-\frac{1}{256}
\sqrt{\frac{3}{5}} C_{11,22}^{53,0}-\frac{1}{512} \sqrt{\frac{3}{5}} C_{11,22}^{53,1}-\frac{\sqrt{35} C_{20,20}^{40,0}}{1728}-\frac{\sqrt{35} C_{20,20}^{40,1}}{3456}-\frac{\sqrt{7}
C_{20,22}^{42,0}}{3456}-\frac{\sqrt{7} C_{20,22}^{42,1}}{6912}+\frac{\sqrt{7} C_{22,20}^{42,0}}{1728}+\frac{\sqrt{7} C_{22,20}^{42,1}}{3456}-\frac{\sqrt{7}
C_{22,22}^{40,0}}{864}-\frac{\sqrt{7} C_{22,22}^{40,1}}{1728}+\frac{5 C_{22,22}^{42,0}}{1728}+\frac{5 C_{22,22}^{42,1}}{3456}-\frac{C_{22,22}^{44,0}}{32
\sqrt{35}}-\frac{C_{22,22}^{44,1}}{64 \sqrt{35}}+\frac{\sqrt{35} C_{31,20}^{31,0}}{1728}+\frac{\sqrt{35} C_{31,20}^{31,1}}{3456}+\frac{\sqrt{7} C_{31,22}^{31,0}}{3456}+\frac{\sqrt{7}
C_{31,22}^{31,1}}{6912}+\frac{17 C_{31,22}^{33,0}}{768 \sqrt{15}}+\frac{17 C_{31,22}^{33,1}}{1536 \sqrt{15}}-\frac{C_{33,20}^{33,0}}{64 \sqrt{15}}-\frac{C_{33,20}^{33,1}}{128
\sqrt{15}}-\frac{C_{33,22}^{33,0}}{160 \sqrt{2}}-\frac{C_{33,22}^{33,1}}{320 \sqrt{2}}\)

\noindent\(C_{22,1112,2}^{22,1110,0,\text{NonLoc}}= \frac{1}{768} \sqrt{\frac{35}{3}} C_{00,00}^{60,0}+\frac{1}{384} \sqrt{\frac{35}{3}}
C_{00,00}^{60,1}+\frac{\sqrt{35} C_{00,20}^{60,0}}{2304}+\frac{\sqrt{35} C_{00,20}^{60,1}}{1152}-\frac{\sqrt{7} C_{00,22}^{62,0}}{384}-\frac{\sqrt{7}
C_{00,22}^{62,1}}{192}+\frac{\sqrt{\frac{35}{3}} C_{11,00}^{51,0}}{2304}+\frac{\sqrt{\frac{35}{3}} C_{11,00}^{51,1}}{1152}-\frac{7 \sqrt{7} C_{11,22}^{51,0}}{6912}-\frac{7
\sqrt{7} C_{11,22}^{51,1}}{3456}-\frac{1}{512} \sqrt{\frac{3}{5}} C_{11,22}^{53,0}-\frac{1}{256} \sqrt{\frac{3}{5}} C_{11,22}^{53,1}-\frac{\sqrt{\frac{35}{3}}
C_{20,00}^{40,0}}{2304}-\frac{\sqrt{\frac{35}{3}} C_{20,00}^{40,1}}{1152}-\frac{\sqrt{35} C_{20,20}^{40,0}}{6912}-\frac{\sqrt{35} C_{20,20}^{40,1}}{3456}+\frac{\sqrt{7}
C_{20,22}^{42,0}}{6912}+\frac{\sqrt{7} C_{20,22}^{42,1}}{3456}+\frac{\sqrt{\frac{7}{3}} C_{22,00}^{42,0}}{2304}+\frac{\sqrt{\frac{7}{3}} C_{22,00}^{42,1}}{1152}+\frac{\sqrt{7}
C_{22,20}^{42,0}}{6912}+\frac{\sqrt{7} C_{22,20}^{42,1}}{3456}+\frac{\sqrt{7} C_{22,22}^{40,0}}{1728}+\frac{\sqrt{7} C_{22,22}^{40,1}}{864}-\frac{5
C_{22,22}^{42,0}}{3456}-\frac{5 C_{22,22}^{42,1}}{1728}+\frac{C_{22,22}^{44,0}}{64 \sqrt{35}}+\frac{C_{22,22}^{44,1}}{32 \sqrt{35}}-\frac{\sqrt{\frac{35}{3}}
C_{31,00}^{31,0}}{2304}-\frac{\sqrt{\frac{35}{3}} C_{31,00}^{31,1}}{1152}-\frac{\sqrt{35} C_{31,20}^{31,0}}{6912}-\frac{\sqrt{35} C_{31,20}^{31,1}}{3456}+\frac{\sqrt{7}
C_{31,22}^{31,0}}{6912}+\frac{\sqrt{7} C_{31,22}^{31,1}}{3456}+\frac{17 C_{31,22}^{33,0}}{1536 \sqrt{15}}+\frac{17 C_{31,22}^{33,1}}{768 \sqrt{15}}+\frac{C_{33,00}^{33,0}}{256
\sqrt{5}}+\frac{C_{33,00}^{33,1}}{128 \sqrt{5}}+\frac{C_{33,20}^{33,0}}{256 \sqrt{15}}+\frac{C_{33,20}^{33,1}}{128 \sqrt{15}}-\frac{C_{33,22}^{33,0}}{320
\sqrt{2}}-\frac{C_{33,22}^{33,1}}{160 \sqrt{2}}\)

\noindent\(C_{22,1112,2}^{22,1110,1,\text{Loc}}= \frac{1}{384} \sqrt{\frac{35}{3}} C_{00,20}^{60,1}+\frac{1}{128} \sqrt{\frac{7}{3}}
C_{00,22}^{62,1}-\frac{7 \sqrt{\frac{7}{3}} C_{11,22}^{51,1}}{2304}-\frac{3 C_{11,22}^{53,1}}{512 \sqrt{5}}-\frac{\sqrt{\frac{35}{3}} C_{20,20}^{40,1}}{1152}-\frac{\sqrt{\frac{7}{3}}
C_{20,22}^{42,1}}{2304}+\frac{\sqrt{\frac{7}{3}} C_{22,20}^{42,1}}{1152}-\frac{1}{576} \sqrt{\frac{7}{3}} C_{22,22}^{40,1}+\frac{5 C_{22,22}^{42,1}}{1152
\sqrt{3}}-\frac{1}{64} \sqrt{\frac{3}{35}} C_{22,22}^{44,1}+\frac{\sqrt{\frac{35}{3}} C_{31,20}^{31,1}}{1152}+\frac{\sqrt{\frac{7}{3}} C_{31,22}^{31,1}}{2304}+\frac{17
C_{31,22}^{33,1}}{1536 \sqrt{5}}-\frac{C_{33,20}^{33,1}}{128 \sqrt{5}}-\frac{1}{320} \sqrt{\frac{3}{2}} C_{33,22}^{33,1}\)

\noindent\(C_{22,1112,2}^{22,1110,1,\text{NonLoc}}= \frac{\sqrt{35} C_{00,00}^{60,0}}{768}+\frac{1}{768} \sqrt{\frac{35}{3}} C_{00,20}^{60,0}-\frac{1}{128}
\sqrt{\frac{7}{3}} C_{00,22}^{62,0}+\frac{\sqrt{35} C_{11,00}^{51,0}}{2304}-\frac{7 \sqrt{\frac{7}{3}} C_{11,22}^{51,0}}{2304}-\frac{3 C_{11,22}^{53,0}}{512
\sqrt{5}}-\frac{\sqrt{35} C_{20,00}^{40,0}}{2304}-\frac{\sqrt{\frac{35}{3}} C_{20,20}^{40,0}}{2304}+\frac{\sqrt{\frac{7}{3}} C_{20,22}^{42,0}}{2304}+\frac{\sqrt{7}
C_{22,00}^{42,0}}{2304}+\frac{\sqrt{\frac{7}{3}} C_{22,20}^{42,0}}{2304}+\frac{1}{576} \sqrt{\frac{7}{3}} C_{22,22}^{40,0}-\frac{5 C_{22,22}^{42,0}}{1152
\sqrt{3}}+\frac{1}{64} \sqrt{\frac{3}{35}} C_{22,22}^{44,0}-\frac{\sqrt{35} C_{31,00}^{31,0}}{2304}-\frac{\sqrt{\frac{35}{3}} C_{31,20}^{31,0}}{2304}+\frac{\sqrt{\frac{7}{3}}
C_{31,22}^{31,0}}{2304}+\frac{17 C_{31,22}^{33,0}}{1536 \sqrt{5}}+\frac{1}{256} \sqrt{\frac{3}{5}} C_{33,00}^{33,0}+\frac{C_{33,20}^{33,0}}{256 \sqrt{5}}-\frac{1}{320}
\sqrt{\frac{3}{2}} C_{33,22}^{33,0}\)

\noindent\(C_{22,1112,2}^{22,1111,0,\text{Loc}}= -\frac{\sqrt{35} C_{00,20}^{60,0}}{768}-\frac{\sqrt{35} C_{00,20}^{60,1}}{1536}+\frac{\sqrt{7}
C_{00,22}^{62,0}}{512}+\frac{\sqrt{7} C_{00,22}^{62,1}}{1024}-\frac{5 \sqrt{7} C_{11,22}^{51,0}}{4608}-\frac{5 \sqrt{7} C_{11,22}^{51,1}}{9216}+\frac{\sqrt{35}
C_{20,20}^{40,0}}{2304}+\frac{\sqrt{35} C_{20,20}^{40,1}}{4608}+\frac{\sqrt{7} C_{20,22}^{42,0}}{4608}+\frac{\sqrt{7} C_{20,22}^{42,1}}{9216}-\frac{\sqrt{7}
C_{22,20}^{42,0}}{2304}-\frac{\sqrt{7} C_{22,20}^{42,1}}{4608}+\frac{\sqrt{7} C_{22,22}^{40,0}}{4608}+\frac{\sqrt{7} C_{22,22}^{40,1}}{9216}-\frac{C_{22,22}^{42,0}}{1152}-\frac{C_{22,22}^{42,1}}{2304}-\frac{3
C_{22,22}^{44,0}}{128 \sqrt{35}}-\frac{3 C_{22,22}^{44,1}}{256 \sqrt{35}}-\frac{\sqrt{35} C_{31,20}^{31,0}}{2304}-\frac{\sqrt{35} C_{31,20}^{31,1}}{4608}-\frac{\sqrt{7}
C_{31,22}^{31,0}}{4608}-\frac{\sqrt{7} C_{31,22}^{31,1}}{9216}+\frac{C_{31,22}^{33,0}}{128 \sqrt{15}}+\frac{C_{31,22}^{33,1}}{256 \sqrt{15}}+\frac{1}{256}
\sqrt{\frac{3}{5}} C_{33,20}^{33,0}+\frac{1}{512} \sqrt{\frac{3}{5}} C_{33,20}^{33,1}+\frac{3 C_{33,22}^{33,0}}{640 \sqrt{2}}+\frac{3 C_{33,22}^{33,1}}{1280
\sqrt{2}}\)

\noindent\(C_{22,1112,2}^{22,1111,0,\text{NonLoc}}= -\frac{\sqrt{\frac{35}{3}} C_{00,00}^{60,0}}{1024}-\frac{1}{512} \sqrt{\frac{35}{3}}
C_{00,00}^{60,1}-\frac{\sqrt{35} C_{00,20}^{60,0}}{3072}-\frac{\sqrt{35} C_{00,20}^{60,1}}{1536}-\frac{\sqrt{7} C_{00,22}^{62,0}}{1024}-\frac{\sqrt{7}
C_{00,22}^{62,1}}{512}-\frac{\sqrt{\frac{35}{3}} C_{11,00}^{51,0}}{3072}-\frac{\sqrt{\frac{35}{3}} C_{11,00}^{51,1}}{1536}-\frac{5 \sqrt{7} C_{11,22}^{51,0}}{9216}-\frac{5
\sqrt{7} C_{11,22}^{51,1}}{4608}+\frac{\sqrt{\frac{35}{3}} C_{20,00}^{40,0}}{3072}+\frac{\sqrt{\frac{35}{3}} C_{20,00}^{40,1}}{1536}+\frac{\sqrt{35}
C_{20,20}^{40,0}}{9216}+\frac{\sqrt{35} C_{20,20}^{40,1}}{4608}-\frac{\sqrt{7} C_{20,22}^{42,0}}{9216}-\frac{\sqrt{7} C_{20,22}^{42,1}}{4608}-\frac{\sqrt{\frac{7}{3}}
C_{22,00}^{42,0}}{3072}-\frac{\sqrt{\frac{7}{3}} C_{22,00}^{42,1}}{1536}-\frac{\sqrt{7} C_{22,20}^{42,0}}{9216}-\frac{\sqrt{7} C_{22,20}^{42,1}}{4608}-\frac{\sqrt{7}
C_{22,22}^{40,0}}{9216}-\frac{\sqrt{7} C_{22,22}^{40,1}}{4608}+\frac{C_{22,22}^{42,0}}{2304}+\frac{C_{22,22}^{42,1}}{1152}+\frac{3 C_{22,22}^{44,0}}{256
\sqrt{35}}+\frac{3 C_{22,22}^{44,1}}{128 \sqrt{35}}+\frac{\sqrt{\frac{35}{3}} C_{31,00}^{31,0}}{3072}+\frac{\sqrt{\frac{35}{3}} C_{31,00}^{31,1}}{1536}+\frac{\sqrt{35}
C_{31,20}^{31,0}}{9216}+\frac{\sqrt{35} C_{31,20}^{31,1}}{4608}-\frac{\sqrt{7} C_{31,22}^{31,0}}{9216}-\frac{\sqrt{7} C_{31,22}^{31,1}}{4608}+\frac{C_{31,22}^{33,0}}{256
\sqrt{15}}+\frac{C_{31,22}^{33,1}}{128 \sqrt{15}}-\frac{3 C_{33,00}^{33,0}}{1024 \sqrt{5}}-\frac{3 C_{33,00}^{33,1}}{512 \sqrt{5}}-\frac{\sqrt{\frac{3}{5}}
C_{33,20}^{33,0}}{1024}-\frac{1}{512} \sqrt{\frac{3}{5}} C_{33,20}^{33,1}+\frac{3 C_{33,22}^{33,0}}{1280 \sqrt{2}}+\frac{3 C_{33,22}^{33,1}}{640
\sqrt{2}}\)

\noindent\(C_{22,1112,2}^{22,1111,1,\text{Loc}}= -\frac{1}{512} \sqrt{\frac{35}{3}} C_{00,20}^{60,1}+\frac{\sqrt{21} C_{00,22}^{62,1}}{1024}-\frac{5
\sqrt{\frac{7}{3}} C_{11,22}^{51,1}}{3072}+\frac{\sqrt{\frac{35}{3}} C_{20,20}^{40,1}}{1536}+\frac{\sqrt{\frac{7}{3}} C_{20,22}^{42,1}}{3072}-\frac{\sqrt{\frac{7}{3}}
C_{22,20}^{42,1}}{1536}+\frac{\sqrt{\frac{7}{3}} C_{22,22}^{40,1}}{3072}-\frac{C_{22,22}^{42,1}}{768 \sqrt{3}}-\frac{3}{256} \sqrt{\frac{3}{35}}
C_{22,22}^{44,1}-\frac{\sqrt{\frac{35}{3}} C_{31,20}^{31,1}}{1536}-\frac{\sqrt{\frac{7}{3}} C_{31,22}^{31,1}}{3072}+\frac{C_{31,22}^{33,1}}{256 \sqrt{5}}+\frac{3
C_{33,20}^{33,1}}{512 \sqrt{5}}+\frac{3 \sqrt{\frac{3}{2}} C_{33,22}^{33,1}}{1280}\)

\noindent\(C_{22,1112,2}^{22,1111,1,\text{NonLoc}}= -\frac{\sqrt{35} C_{00,00}^{60,0}}{1024}-\frac{\sqrt{\frac{35}{3}} C_{00,20}^{60,0}}{1024}-\frac{\sqrt{21}
C_{00,22}^{62,0}}{1024}-\frac{\sqrt{35} C_{11,00}^{51,0}}{3072}-\frac{5 \sqrt{\frac{7}{3}} C_{11,22}^{51,0}}{3072}+\frac{\sqrt{35} C_{20,00}^{40,0}}{3072}+\frac{\sqrt{\frac{35}{3}}
C_{20,20}^{40,0}}{3072}-\frac{\sqrt{\frac{7}{3}} C_{20,22}^{42,0}}{3072}-\frac{\sqrt{7} C_{22,00}^{42,0}}{3072}-\frac{\sqrt{\frac{7}{3}} C_{22,20}^{42,0}}{3072}-\frac{\sqrt{\frac{7}{3}}
C_{22,22}^{40,0}}{3072}+\frac{C_{22,22}^{42,0}}{768 \sqrt{3}}+\frac{3}{256} \sqrt{\frac{3}{35}} C_{22,22}^{44,0}+\frac{\sqrt{35} C_{31,00}^{31,0}}{3072}+\frac{\sqrt{\frac{35}{3}}
C_{31,20}^{31,0}}{3072}-\frac{\sqrt{\frac{7}{3}} C_{31,22}^{31,0}}{3072}+\frac{C_{31,22}^{33,0}}{256 \sqrt{5}}-\frac{3 \sqrt{\frac{3}{5}} C_{33,00}^{33,0}}{1024}-\frac{3
C_{33,20}^{33,0}}{1024 \sqrt{5}}+\frac{3 \sqrt{\frac{3}{2}} C_{33,22}^{33,0}}{1280}\)

\noindent\(C_{22,1112,2}^{22,1112,0,\text{Loc}}= \frac{11 \sqrt{5} C_{00,20}^{60,0}}{4608}+\frac{11 \sqrt{5} C_{00,20}^{60,1}}{9216}+\frac{13
C_{00,22}^{62,0}}{3072}+\frac{13 C_{00,22}^{62,1}}{6144}-\frac{95 C_{11,22}^{51,0}}{27648}-\frac{95 C_{11,22}^{51,1}}{55296}+\frac{\sqrt{\frac{15}{7}}
C_{11,22}^{53,0}}{1024}+\frac{\sqrt{\frac{15}{7}} C_{11,22}^{53,1}}{2048}+\frac{13 \sqrt{5} C_{20,20}^{40,0}}{13824}+\frac{13 \sqrt{5} C_{20,20}^{40,1}}{27648}-\frac{113
C_{20,22}^{42,0}}{27648}-\frac{113 C_{20,22}^{42,1}}{55296}-\frac{31 C_{22,20}^{42,0}}{13824}-\frac{31 C_{22,20}^{42,1}}{27648}+\frac{97 C_{22,22}^{40,0}}{27648}+\frac{97
C_{22,22}^{40,1}}{55296}-\frac{C_{22,22}^{42,0}}{6912 \sqrt{7}}-\frac{C_{22,22}^{42,1}}{13824 \sqrt{7}}+\frac{13 C_{22,22}^{44,0}}{1792 \sqrt{5}}+\frac{13
C_{22,22}^{44,1}}{3584 \sqrt{5}}-\frac{\sqrt{5} C_{31,20}^{31,0}}{13824}-\frac{\sqrt{5} C_{31,20}^{31,1}}{27648}+\frac{125 C_{31,22}^{31,0}}{27648}+\frac{125
C_{31,22}^{31,1}}{55296}-\frac{17 \sqrt{\frac{5}{21}} C_{31,22}^{33,0}}{3072}-\frac{17 \sqrt{\frac{5}{21}} C_{31,22}^{33,1}}{6144}+\frac{31 C_{33,20}^{33,0}}{512
\sqrt{105}}+\frac{31 C_{33,20}^{33,1}}{1024 \sqrt{105}}-\frac{C_{33,22}^{33,0}}{256 \sqrt{14}}-\frac{C_{33,22}^{33,1}}{512 \sqrt{14}}\)

\noindent\(C_{22,1112,2}^{22,1112,0,\text{NonLoc}}= \frac{11 \sqrt{\frac{5}{3}} C_{00,00}^{60,0}}{6144}+\frac{11 \sqrt{\frac{5}{3}}
C_{00,00}^{60,1}}{3072}+\frac{11 \sqrt{5} C_{00,20}^{60,0}}{18432}+\frac{11 \sqrt{5} C_{00,20}^{60,1}}{9216}-\frac{13 C_{00,22}^{62,0}}{6144}-\frac{13
C_{00,22}^{62,1}}{3072}+\frac{11 \sqrt{\frac{5}{3}} C_{11,00}^{51,0}}{18432}+\frac{11 \sqrt{\frac{5}{3}} C_{11,00}^{51,1}}{9216}-\frac{95 C_{11,22}^{51,0}}{55296}-\frac{95
C_{11,22}^{51,1}}{27648}+\frac{\sqrt{\frac{15}{7}} C_{11,22}^{53,0}}{2048}+\frac{\sqrt{\frac{15}{7}} C_{11,22}^{53,1}}{1024}+\frac{13 \sqrt{\frac{5}{3}}
C_{20,00}^{40,0}}{18432}+\frac{13 \sqrt{\frac{5}{3}} C_{20,00}^{40,1}}{9216}+\frac{13 \sqrt{5} C_{20,20}^{40,0}}{55296}+\frac{13 \sqrt{5} C_{20,20}^{40,1}}{27648}+\frac{113
C_{20,22}^{42,0}}{55296}+\frac{113 C_{20,22}^{42,1}}{27648}-\frac{31 C_{22,00}^{42,0}}{18432 \sqrt{3}}-\frac{31 C_{22,00}^{42,1}}{9216 \sqrt{3}}-\frac{31
C_{22,20}^{42,0}}{55296}-\frac{31 C_{22,20}^{42,1}}{27648}-\frac{97 C_{22,22}^{40,0}}{55296}-\frac{97 C_{22,22}^{40,1}}{27648}+\frac{C_{22,22}^{42,0}}{13824
\sqrt{7}}+\frac{C_{22,22}^{42,1}}{6912 \sqrt{7}}-\frac{13 C_{22,22}^{44,0}}{3584 \sqrt{5}}-\frac{13 C_{22,22}^{44,1}}{1792 \sqrt{5}}+\frac{\sqrt{\frac{5}{3}}
C_{31,00}^{31,0}}{18432}+\frac{\sqrt{\frac{5}{3}} C_{31,00}^{31,1}}{9216}+\frac{\sqrt{5} C_{31,20}^{31,0}}{55296}+\frac{\sqrt{5} C_{31,20}^{31,1}}{27648}+\frac{125
C_{31,22}^{31,0}}{55296}+\frac{125 C_{31,22}^{31,1}}{27648}-\frac{17 \sqrt{\frac{5}{21}} C_{31,22}^{33,0}}{6144}-\frac{17 \sqrt{\frac{5}{21}} C_{31,22}^{33,1}}{3072}-\frac{31
C_{33,00}^{33,0}}{2048 \sqrt{35}}-\frac{31 C_{33,00}^{33,1}}{1024 \sqrt{35}}-\frac{31 C_{33,20}^{33,0}}{2048 \sqrt{105}}-\frac{31 C_{33,20}^{33,1}}{1024
\sqrt{105}}-\frac{C_{33,22}^{33,0}}{512 \sqrt{14}}-\frac{C_{33,22}^{33,1}}{256 \sqrt{14}}\)

\noindent\(C_{22,1112,2}^{22,1112,1,\text{Loc}}= \frac{11 \sqrt{\frac{5}{3}} C_{00,20}^{60,1}}{3072}+\frac{13 C_{00,22}^{62,1}}{2048
\sqrt{3}}-\frac{95 C_{11,22}^{51,1}}{18432 \sqrt{3}}+\frac{3 \sqrt{\frac{5}{7}} C_{11,22}^{53,1}}{2048}+\frac{13 \sqrt{\frac{5}{3}} C_{20,20}^{40,1}}{9216}-\frac{113
C_{20,22}^{42,1}}{18432 \sqrt{3}}-\frac{31 C_{22,20}^{42,1}}{9216 \sqrt{3}}+\frac{97 C_{22,22}^{40,1}}{18432 \sqrt{3}}-\frac{C_{22,22}^{42,1}}{4608
\sqrt{21}}+\frac{13 \sqrt{\frac{3}{5}} C_{22,22}^{44,1}}{3584}-\frac{\sqrt{\frac{5}{3}} C_{31,20}^{31,1}}{9216}+\frac{125 C_{31,22}^{31,1}}{18432
\sqrt{3}}-\frac{17 \sqrt{\frac{5}{7}} C_{31,22}^{33,1}}{6144}+\frac{31 C_{33,20}^{33,1}}{1024 \sqrt{35}}-\frac{1}{512} \sqrt{\frac{3}{14}} C_{33,22}^{33,1}\)

\noindent\(C_{22,1112,2}^{22,1112,1,\text{NonLoc}}= \frac{11 \sqrt{5} C_{00,00}^{60,0}}{6144}+\frac{11 \sqrt{\frac{5}{3}} C_{00,20}^{60,0}}{6144}-\frac{13
C_{00,22}^{62,0}}{2048 \sqrt{3}}+\frac{11 \sqrt{5} C_{11,00}^{51,0}}{18432}-\frac{95 C_{11,22}^{51,0}}{18432 \sqrt{3}}+\frac{3 \sqrt{\frac{5}{7}}
C_{11,22}^{53,0}}{2048}+\frac{13 \sqrt{5} C_{20,00}^{40,0}}{18432}+\frac{13 \sqrt{\frac{5}{3}} C_{20,20}^{40,0}}{18432}+\frac{113 C_{20,22}^{42,0}}{18432
\sqrt{3}}-\frac{31 C_{22,00}^{42,0}}{18432}-\frac{31 C_{22,20}^{42,0}}{18432 \sqrt{3}}-\frac{97 C_{22,22}^{40,0}}{18432 \sqrt{3}}+\frac{C_{22,22}^{42,0}}{4608
\sqrt{21}}-\frac{13 \sqrt{\frac{3}{5}} C_{22,22}^{44,0}}{3584}+\frac{\sqrt{5} C_{31,00}^{31,0}}{18432}+\frac{\sqrt{\frac{5}{3}} C_{31,20}^{31,0}}{18432}+\frac{125
C_{31,22}^{31,0}}{18432 \sqrt{3}}-\frac{17 \sqrt{\frac{5}{7}} C_{31,22}^{33,0}}{6144}-\frac{31 \sqrt{\frac{3}{35}} C_{33,00}^{33,0}}{2048}-\frac{31
C_{33,20}^{33,0}}{2048 \sqrt{35}}-\frac{1}{512} \sqrt{\frac{3}{14}} C_{33,22}^{33,0}\)

\noindent\(C_{22,1112,3}^{00,3313,0,\text{Loc}}= -\frac{1}{96} \sqrt{\frac{7}{3}} C_{00,20}^{60,0}-\frac{1}{192} \sqrt{\frac{7}{3}}
C_{00,20}^{60,1}+\frac{1}{64} \sqrt{\frac{7}{15}} C_{00,22}^{62,0}+\frac{1}{128} \sqrt{\frac{7}{15}} C_{00,22}^{62,1}-\frac{1}{576} \sqrt{\frac{7}{15}}
C_{11,22}^{51,0}-\frac{\sqrt{\frac{7}{15}} C_{11,22}^{51,1}}{1152}+\frac{3 C_{11,22}^{53,0}}{320}+\frac{3 C_{11,22}^{53,1}}{640}+\frac{1}{288} \sqrt{\frac{7}{3}}
C_{20,20}^{40,0}+\frac{1}{576} \sqrt{\frac{7}{3}} C_{20,20}^{40,1}+\frac{1}{576} \sqrt{\frac{7}{15}} C_{20,22}^{42,0}+\frac{\sqrt{\frac{7}{15}} C_{20,22}^{42,1}}{1152}-\frac{1}{288}
\sqrt{\frac{7}{15}} C_{22,20}^{42,0}-\frac{1}{576} \sqrt{\frac{7}{15}} C_{22,20}^{42,1}-\frac{11}{576} \sqrt{\frac{7}{15}} C_{22,22}^{40,0}-\frac{11
\sqrt{\frac{7}{15}} C_{22,22}^{40,1}}{1152}+\frac{1}{144} \sqrt{\frac{5}{3}} C_{22,22}^{42,0}+\frac{1}{288} \sqrt{\frac{5}{3}} C_{22,22}^{42,1}-\frac{1}{320}
\sqrt{\frac{3}{7}} C_{22,22}^{44,0}-\frac{1}{640} \sqrt{\frac{3}{7}} C_{22,22}^{44,1}+\frac{1}{288} \sqrt{\frac{7}{3}} C_{31,20}^{31,0}+\frac{1}{576}
\sqrt{\frac{7}{3}} C_{31,20}^{31,1}+\frac{1}{576} \sqrt{\frac{7}{15}} C_{31,22}^{31,0}+\frac{\sqrt{\frac{7}{15}} C_{31,22}^{31,1}}{1152}-\frac{3
C_{31,22}^{33,0}}{320}-\frac{3 C_{31,22}^{33,1}}{640}-\frac{C_{33,20}^{33,0}}{160}-\frac{C_{33,20}^{33,1}}{320}+\frac{11}{320} \sqrt{\frac{3}{10}}
C_{33,22}^{33,0}+\frac{11}{640} \sqrt{\frac{3}{10}} C_{33,22}^{33,1}\)

\noindent\(C_{22,1112,3}^{00,3313,0,\text{NonLoc}}= -\frac{\sqrt{7} C_{00,00}^{60,0}}{384}-\frac{\sqrt{7} C_{00,00}^{60,1}}{192}-\frac{1}{384}
\sqrt{\frac{7}{3}} C_{00,20}^{60,0}-\frac{1}{192} \sqrt{\frac{7}{3}} C_{00,20}^{60,1}-\frac{1}{128} \sqrt{\frac{7}{15}} C_{00,22}^{62,0}-\frac{1}{64}
\sqrt{\frac{7}{15}} C_{00,22}^{62,1}+\frac{\sqrt{7} C_{11,00}^{51,0}}{1152}+\frac{\sqrt{7} C_{11,00}^{51,1}}{576}-\frac{\sqrt{\frac{7}{15}} C_{11,22}^{51,0}}{1152}-\frac{1}{576}
\sqrt{\frac{7}{15}} C_{11,22}^{51,1}+\frac{3 C_{11,22}^{53,0}}{640}+\frac{3 C_{11,22}^{53,1}}{320}+\frac{\sqrt{7} C_{20,00}^{40,0}}{1152}+\frac{\sqrt{7}
C_{20,00}^{40,1}}{576}+\frac{\sqrt{\frac{7}{3}} C_{20,20}^{40,0}}{1152}+\frac{1}{576} \sqrt{\frac{7}{3}} C_{20,20}^{40,1}-\frac{\sqrt{\frac{7}{15}}
C_{20,22}^{42,0}}{1152}-\frac{1}{576} \sqrt{\frac{7}{15}} C_{20,22}^{42,1}-\frac{\sqrt{\frac{7}{5}} C_{22,00}^{42,0}}{1152}-\frac{1}{576} \sqrt{\frac{7}{5}}
C_{22,00}^{42,1}-\frac{\sqrt{\frac{7}{15}} C_{22,20}^{42,0}}{1152}-\frac{1}{576} \sqrt{\frac{7}{15}} C_{22,20}^{42,1}+\frac{11 \sqrt{\frac{7}{15}}
C_{22,22}^{40,0}}{1152}+\frac{11}{576} \sqrt{\frac{7}{15}} C_{22,22}^{40,1}-\frac{1}{288} \sqrt{\frac{5}{3}} C_{22,22}^{42,0}-\frac{1}{144} \sqrt{\frac{5}{3}}
C_{22,22}^{42,1}+\frac{1}{640} \sqrt{\frac{3}{7}} C_{22,22}^{44,0}+\frac{1}{320} \sqrt{\frac{3}{7}} C_{22,22}^{44,1}-\frac{\sqrt{7} C_{31,00}^{31,0}}{1152}-\frac{\sqrt{7}
C_{31,00}^{31,1}}{576}-\frac{\sqrt{\frac{7}{3}} C_{31,20}^{31,0}}{1152}-\frac{1}{576} \sqrt{\frac{7}{3}} C_{31,20}^{31,1}+\frac{\sqrt{\frac{7}{15}}
C_{31,22}^{31,0}}{1152}+\frac{1}{576} \sqrt{\frac{7}{15}} C_{31,22}^{31,1}-\frac{3 C_{31,22}^{33,0}}{640}-\frac{3 C_{31,22}^{33,1}}{320}+\frac{\sqrt{3}
C_{33,00}^{33,0}}{640}+\frac{\sqrt{3} C_{33,00}^{33,1}}{320}+\frac{C_{33,20}^{33,0}}{640}+\frac{C_{33,20}^{33,1}}{320}+\frac{11}{640} \sqrt{\frac{3}{10}}
C_{33,22}^{33,0}+\frac{11}{320} \sqrt{\frac{3}{10}} C_{33,22}^{33,1}\)

\noindent\(C_{22,1112,3}^{00,3313,1,\text{Loc}}= -\frac{\sqrt{7} C_{00,20}^{60,1}}{192}+\frac{1}{128} \sqrt{\frac{7}{5}} C_{00,22}^{62,1}-\frac{\sqrt{\frac{7}{5}}
C_{11,22}^{51,1}}{1152}+\frac{3 \sqrt{3} C_{11,22}^{53,1}}{640}+\frac{\sqrt{7} C_{20,20}^{40,1}}{576}+\frac{\sqrt{\frac{7}{5}} C_{20,22}^{42,1}}{1152}-\frac{1}{576}
\sqrt{\frac{7}{5}} C_{22,20}^{42,1}-\frac{11 \sqrt{\frac{7}{5}} C_{22,22}^{40,1}}{1152}+\frac{\sqrt{5} C_{22,22}^{42,1}}{288}-\frac{3 C_{22,22}^{44,1}}{640
\sqrt{7}}+\frac{\sqrt{7} C_{31,20}^{31,1}}{576}+\frac{\sqrt{\frac{7}{5}} C_{31,22}^{31,1}}{1152}-\frac{3 \sqrt{3} C_{31,22}^{33,1}}{640}-\frac{\sqrt{3}
C_{33,20}^{33,1}}{320}+\frac{33 C_{33,22}^{33,1}}{640 \sqrt{10}}\)

\noindent\(C_{22,1112,3}^{00,3313,1,\text{NonLoc}}= -\frac{1}{128} \sqrt{\frac{7}{3}} C_{00,00}^{60,0}-\frac{\sqrt{7} C_{00,20}^{60,0}}{384}-\frac{1}{128}
\sqrt{\frac{7}{5}} C_{00,22}^{62,0}+\frac{1}{384} \sqrt{\frac{7}{3}} C_{11,00}^{51,0}-\frac{\sqrt{\frac{7}{5}} C_{11,22}^{51,0}}{1152}+\frac{3 \sqrt{3}
C_{11,22}^{53,0}}{640}+\frac{1}{384} \sqrt{\frac{7}{3}} C_{20,00}^{40,0}+\frac{\sqrt{7} C_{20,20}^{40,0}}{1152}-\frac{\sqrt{\frac{7}{5}} C_{20,22}^{42,0}}{1152}-\frac{1}{384}
\sqrt{\frac{7}{15}} C_{22,00}^{42,0}-\frac{\sqrt{\frac{7}{5}} C_{22,20}^{42,0}}{1152}+\frac{11 \sqrt{\frac{7}{5}} C_{22,22}^{40,0}}{1152}-\frac{\sqrt{5}
C_{22,22}^{42,0}}{288}+\frac{3 C_{22,22}^{44,0}}{640 \sqrt{7}}-\frac{1}{384} \sqrt{\frac{7}{3}} C_{31,00}^{31,0}-\frac{\sqrt{7} C_{31,20}^{31,0}}{1152}+\frac{\sqrt{\frac{7}{5}}
C_{31,22}^{31,0}}{1152}-\frac{3 \sqrt{3} C_{31,22}^{33,0}}{640}+\frac{3 C_{33,00}^{33,0}}{640}+\frac{\sqrt{3} C_{33,20}^{33,0}}{640}+\frac{33 C_{33,22}^{33,0}}{640
\sqrt{10}}\)

\noindent\(C_{22,1112,3}^{11,2202,0,\text{Loc}}= \frac{1}{48} \sqrt{\frac{7}{6}} C_{11,11}^{51,0}+\frac{1}{96} \sqrt{\frac{7}{6}} C_{11,11}^{51,1}-\frac{\sqrt{3}
C_{33,11}^{33,0}}{64}-\frac{\sqrt{3} C_{33,11}^{33,1}}{128}\)

\noindent\(C_{22,1112,3}^{11,2202,0,\text{NonLoc}}= \frac{1}{96} \sqrt{\frac{7}{6}} C_{11,11}^{51,0}+\frac{1}{48} \sqrt{\frac{7}{6}}
C_{11,11}^{51,1}-\frac{\sqrt{3} C_{33,11}^{33,0}}{128}-\frac{\sqrt{3} C_{33,11}^{33,1}}{64}\)

\noindent\(C_{22,1112,3}^{11,2202,1,\text{Loc}}= \frac{1}{96} \sqrt{\frac{7}{2}} C_{11,11}^{51,1}-\frac{3 C_{33,11}^{33,1}}{128}\)

\noindent\(C_{22,1112,3}^{11,2202,1,\text{NonLoc}}= \frac{1}{96} \sqrt{\frac{7}{2}} C_{11,11}^{51,0}-\frac{3 C_{33,11}^{33,0}}{128}\)

\noindent\(C_{22,1112,3}^{22,1111,0,\text{Loc}}= \frac{1}{192} \sqrt{\frac{7}{2}} C_{00,20}^{60,0}+\frac{1}{384} \sqrt{\frac{7}{2}}
C_{00,20}^{60,1}-\frac{1}{128} \sqrt{\frac{7}{10}} C_{00,22}^{62,0}-\frac{1}{256} \sqrt{\frac{7}{10}} C_{00,22}^{62,1}+\frac{\sqrt{\frac{35}{2}}
C_{11,22}^{51,0}}{1152}+\frac{\sqrt{\frac{35}{2}} C_{11,22}^{51,1}}{2304}-\frac{1}{576} \sqrt{\frac{7}{2}} C_{20,20}^{40,0}-\frac{\sqrt{\frac{7}{2}}
C_{20,20}^{40,1}}{1152}-\frac{\sqrt{\frac{7}{10}} C_{20,22}^{42,0}}{1152}-\frac{\sqrt{\frac{7}{10}} C_{20,22}^{42,1}}{2304}+\frac{1}{576} \sqrt{\frac{7}{10}}
C_{22,20}^{42,0}+\frac{\sqrt{\frac{7}{10}} C_{22,20}^{42,1}}{1152}-\frac{\sqrt{\frac{7}{10}} C_{22,22}^{40,0}}{1152}-\frac{\sqrt{\frac{7}{10}} C_{22,22}^{40,1}}{2304}+\frac{C_{22,22}^{42,0}}{288
\sqrt{10}}+\frac{C_{22,22}^{42,1}}{576 \sqrt{10}}+\frac{3 C_{22,22}^{44,0}}{160 \sqrt{14}}+\frac{3 C_{22,22}^{44,1}}{320 \sqrt{14}}+\frac{1}{576}
\sqrt{\frac{7}{2}} C_{31,20}^{31,0}+\frac{\sqrt{\frac{7}{2}} C_{31,20}^{31,1}}{1152}+\frac{\sqrt{\frac{7}{10}} C_{31,22}^{31,0}}{1152}+\frac{\sqrt{\frac{7}{10}}
C_{31,22}^{31,1}}{2304}-\frac{C_{31,22}^{33,0}}{160 \sqrt{6}}-\frac{C_{31,22}^{33,1}}{320 \sqrt{6}}-\frac{1}{320} \sqrt{\frac{3}{2}} C_{33,20}^{33,0}-\frac{1}{640}
\sqrt{\frac{3}{2}} C_{33,20}^{33,1}-\frac{3 C_{33,22}^{33,0}}{320 \sqrt{5}}-\frac{3 C_{33,22}^{33,1}}{640 \sqrt{5}}\)

\noindent\(C_{22,1112,3}^{22,1111,0,\text{NonLoc}}= \frac{1}{256} \sqrt{\frac{7}{6}} C_{00,00}^{60,0}+\frac{1}{128} \sqrt{\frac{7}{6}}
C_{00,00}^{60,1}+\frac{1}{768} \sqrt{\frac{7}{2}} C_{00,20}^{60,0}+\frac{1}{384} \sqrt{\frac{7}{2}} C_{00,20}^{60,1}+\frac{1}{256} \sqrt{\frac{7}{10}}
C_{00,22}^{62,0}+\frac{1}{128} \sqrt{\frac{7}{10}} C_{00,22}^{62,1}+\frac{1}{768} \sqrt{\frac{7}{6}} C_{11,00}^{51,0}+\frac{1}{384} \sqrt{\frac{7}{6}}
C_{11,00}^{51,1}+\frac{\sqrt{\frac{35}{2}} C_{11,22}^{51,0}}{2304}+\frac{\sqrt{\frac{35}{2}} C_{11,22}^{51,1}}{1152}-\frac{1}{768} \sqrt{\frac{7}{6}}
C_{20,00}^{40,0}-\frac{1}{384} \sqrt{\frac{7}{6}} C_{20,00}^{40,1}-\frac{\sqrt{\frac{7}{2}} C_{20,20}^{40,0}}{2304}-\frac{\sqrt{\frac{7}{2}} C_{20,20}^{40,1}}{1152}+\frac{\sqrt{\frac{7}{10}}
C_{20,22}^{42,0}}{2304}+\frac{\sqrt{\frac{7}{10}} C_{20,22}^{42,1}}{1152}+\frac{1}{768} \sqrt{\frac{7}{30}} C_{22,00}^{42,0}+\frac{1}{384} \sqrt{\frac{7}{30}}
C_{22,00}^{42,1}+\frac{\sqrt{\frac{7}{10}} C_{22,20}^{42,0}}{2304}+\frac{\sqrt{\frac{7}{10}} C_{22,20}^{42,1}}{1152}+\frac{\sqrt{\frac{7}{10}} C_{22,22}^{40,0}}{2304}+\frac{\sqrt{\frac{7}{10}}
C_{22,22}^{40,1}}{1152}-\frac{C_{22,22}^{42,0}}{576 \sqrt{10}}-\frac{C_{22,22}^{42,1}}{288 \sqrt{10}}-\frac{3 C_{22,22}^{44,0}}{320 \sqrt{14}}-\frac{3
C_{22,22}^{44,1}}{160 \sqrt{14}}-\frac{1}{768} \sqrt{\frac{7}{6}} C_{31,00}^{31,0}-\frac{1}{384} \sqrt{\frac{7}{6}} C_{31,00}^{31,1}-\frac{\sqrt{\frac{7}{2}}
C_{31,20}^{31,0}}{2304}-\frac{\sqrt{\frac{7}{2}} C_{31,20}^{31,1}}{1152}+\frac{\sqrt{\frac{7}{10}} C_{31,22}^{31,0}}{2304}+\frac{\sqrt{\frac{7}{10}}
C_{31,22}^{31,1}}{1152}-\frac{C_{31,22}^{33,0}}{320 \sqrt{6}}-\frac{C_{31,22}^{33,1}}{160 \sqrt{6}}+\frac{3 C_{33,00}^{33,0}}{1280 \sqrt{2}}+\frac{3
C_{33,00}^{33,1}}{640 \sqrt{2}}+\frac{\sqrt{\frac{3}{2}} C_{33,20}^{33,0}}{1280}+\frac{1}{640} \sqrt{\frac{3}{2}} C_{33,20}^{33,1}-\frac{3 C_{33,22}^{33,0}}{640
\sqrt{5}}-\frac{3 C_{33,22}^{33,1}}{320 \sqrt{5}}\)

\noindent\(C_{22,1112,3}^{22,1111,1,\text{Loc}}= \frac{1}{128} \sqrt{\frac{7}{6}} C_{00,20}^{60,1}-\frac{1}{256} \sqrt{\frac{21}{10}}
C_{00,22}^{62,1}+\frac{1}{768} \sqrt{\frac{35}{6}} C_{11,22}^{51,1}-\frac{1}{384} \sqrt{\frac{7}{6}} C_{20,20}^{40,1}-\frac{1}{768} \sqrt{\frac{7}{30}}
C_{20,22}^{42,1}+\frac{1}{384} \sqrt{\frac{7}{30}} C_{22,20}^{42,1}-\frac{1}{768} \sqrt{\frac{7}{30}} C_{22,22}^{40,1}+\frac{C_{22,22}^{42,1}}{192
\sqrt{30}}+\frac{3}{320} \sqrt{\frac{3}{14}} C_{22,22}^{44,1}+\frac{1}{384} \sqrt{\frac{7}{6}} C_{31,20}^{31,1}+\frac{1}{768} \sqrt{\frac{7}{30}}
C_{31,22}^{31,1}-\frac{C_{31,22}^{33,1}}{320 \sqrt{2}}-\frac{3 C_{33,20}^{33,1}}{640 \sqrt{2}}-\frac{3}{640} \sqrt{\frac{3}{5}} C_{33,22}^{33,1}\)

\noindent\(C_{22,1112,3}^{22,1111,1,\text{NonLoc}}= \frac{1}{256} \sqrt{\frac{7}{2}} C_{00,00}^{60,0}+\frac{1}{256} \sqrt{\frac{7}{6}}
C_{00,20}^{60,0}+\frac{1}{256} \sqrt{\frac{21}{10}} C_{00,22}^{62,0}+\frac{1}{768} \sqrt{\frac{7}{2}} C_{11,00}^{51,0}+\frac{1}{768} \sqrt{\frac{35}{6}}
C_{11,22}^{51,0}-\frac{1}{768} \sqrt{\frac{7}{2}} C_{20,00}^{40,0}-\frac{1}{768} \sqrt{\frac{7}{6}} C_{20,20}^{40,0}+\frac{1}{768} \sqrt{\frac{7}{30}}
C_{20,22}^{42,0}+\frac{1}{768} \sqrt{\frac{7}{10}} C_{22,00}^{42,0}+\frac{1}{768} \sqrt{\frac{7}{30}} C_{22,20}^{42,0}+\frac{1}{768} \sqrt{\frac{7}{30}}
C_{22,22}^{40,0}-\frac{C_{22,22}^{42,0}}{192 \sqrt{30}}-\frac{3}{320} \sqrt{\frac{3}{14}} C_{22,22}^{44,0}-\frac{1}{768} \sqrt{\frac{7}{2}} C_{31,00}^{31,0}-\frac{1}{768}
\sqrt{\frac{7}{6}} C_{31,20}^{31,0}+\frac{1}{768} \sqrt{\frac{7}{30}} C_{31,22}^{31,0}-\frac{C_{31,22}^{33,0}}{320 \sqrt{2}}+\frac{3 \sqrt{\frac{3}{2}}
C_{33,00}^{33,0}}{1280}+\frac{3 C_{33,20}^{33,0}}{1280 \sqrt{2}}-\frac{3}{640} \sqrt{\frac{3}{5}} C_{33,22}^{33,0}\)

\noindent\(C_{22,1112,3}^{22,1112,0,\text{Loc}}= \frac{\sqrt{7} C_{00,20}^{60,0}}{768}+\frac{\sqrt{7} C_{00,20}^{60,1}}{1536}-\frac{1}{512}
\sqrt{\frac{7}{5}} C_{00,22}^{62,0}-\frac{\sqrt{\frac{7}{5}} C_{00,22}^{62,1}}{1024}+\frac{11 \sqrt{\frac{7}{5}} C_{11,22}^{51,0}}{4608}+\frac{11
\sqrt{\frac{7}{5}} C_{11,22}^{51,1}}{9216}-\frac{3 \sqrt{3} C_{11,22}^{53,0}}{2560}-\frac{3 \sqrt{3} C_{11,22}^{53,1}}{5120}+\frac{\sqrt{7} C_{20,20}^{40,0}}{768}+\frac{\sqrt{7}
C_{20,20}^{40,1}}{1536}+\frac{\sqrt{\frac{7}{5}} C_{20,22}^{42,0}}{1536}+\frac{\sqrt{\frac{7}{5}} C_{20,22}^{42,1}}{3072}-\frac{1}{384} \sqrt{\frac{7}{5}}
C_{22,20}^{42,0}-\frac{1}{768} \sqrt{\frac{7}{5}} C_{22,20}^{42,1}-\frac{1}{512} \sqrt{\frac{7}{5}} C_{22,22}^{40,0}-\frac{\sqrt{\frac{7}{5}} C_{22,22}^{40,1}}{1024}+\frac{C_{22,22}^{42,0}}{768
\sqrt{5}}+\frac{C_{22,22}^{42,1}}{1536 \sqrt{5}}+\frac{9 C_{22,22}^{44,0}}{1280 \sqrt{7}}+\frac{9 C_{22,22}^{44,1}}{2560 \sqrt{7}}-\frac{\sqrt{7}
C_{31,20}^{31,0}}{2304}-\frac{\sqrt{7} C_{31,20}^{31,1}}{4608}-\frac{\sqrt{\frac{7}{5}} C_{31,22}^{31,0}}{4608}-\frac{\sqrt{\frac{7}{5}} C_{31,22}^{31,1}}{9216}-\frac{C_{31,22}^{33,0}}{2560
\sqrt{3}}-\frac{C_{31,22}^{33,1}}{5120 \sqrt{3}}+\frac{3 \sqrt{3} C_{33,20}^{33,0}}{640}+\frac{3 \sqrt{3} C_{33,20}^{33,1}}{1280}-\frac{9 C_{33,22}^{33,0}}{1280
\sqrt{10}}-\frac{9 C_{33,22}^{33,1}}{2560 \sqrt{10}}\)

\noindent\(C_{22,1112,3}^{22,1112,0,\text{NonLoc}}= \frac{\sqrt{\frac{7}{3}} C_{00,00}^{60,0}}{1024}+\frac{1}{512} \sqrt{\frac{7}{3}}
C_{00,00}^{60,1}+\frac{\sqrt{7} C_{00,20}^{60,0}}{3072}+\frac{\sqrt{7} C_{00,20}^{60,1}}{1536}+\frac{\sqrt{\frac{7}{5}} C_{00,22}^{62,0}}{1024}+\frac{1}{512}
\sqrt{\frac{7}{5}} C_{00,22}^{62,1}+\frac{\sqrt{\frac{7}{3}} C_{11,00}^{51,0}}{3072}+\frac{\sqrt{\frac{7}{3}} C_{11,00}^{51,1}}{1536}+\frac{11 \sqrt{\frac{7}{5}}
C_{11,22}^{51,0}}{9216}+\frac{11 \sqrt{\frac{7}{5}} C_{11,22}^{51,1}}{4608}-\frac{3 \sqrt{3} C_{11,22}^{53,0}}{5120}-\frac{3 \sqrt{3} C_{11,22}^{53,1}}{2560}+\frac{\sqrt{\frac{7}{3}}
C_{20,00}^{40,0}}{1024}+\frac{1}{512} \sqrt{\frac{7}{3}} C_{20,00}^{40,1}+\frac{\sqrt{7} C_{20,20}^{40,0}}{3072}+\frac{\sqrt{7} C_{20,20}^{40,1}}{1536}-\frac{\sqrt{\frac{7}{5}}
C_{20,22}^{42,0}}{3072}-\frac{\sqrt{\frac{7}{5}} C_{20,22}^{42,1}}{1536}-\frac{1}{512} \sqrt{\frac{7}{15}} C_{22,00}^{42,0}-\frac{1}{256} \sqrt{\frac{7}{15}}
C_{22,00}^{42,1}-\frac{\sqrt{\frac{7}{5}} C_{22,20}^{42,0}}{1536}-\frac{1}{768} \sqrt{\frac{7}{5}} C_{22,20}^{42,1}+\frac{\sqrt{\frac{7}{5}} C_{22,22}^{40,0}}{1024}+\frac{1}{512}
\sqrt{\frac{7}{5}} C_{22,22}^{40,1}-\frac{C_{22,22}^{42,0}}{1536 \sqrt{5}}-\frac{C_{22,22}^{42,1}}{768 \sqrt{5}}-\frac{9 C_{22,22}^{44,0}}{2560 \sqrt{7}}-\frac{9
C_{22,22}^{44,1}}{1280 \sqrt{7}}+\frac{\sqrt{\frac{7}{3}} C_{31,00}^{31,0}}{3072}+\frac{\sqrt{\frac{7}{3}} C_{31,00}^{31,1}}{1536}+\frac{\sqrt{7}
C_{31,20}^{31,0}}{9216}+\frac{\sqrt{7} C_{31,20}^{31,1}}{4608}-\frac{\sqrt{\frac{7}{5}} C_{31,22}^{31,0}}{9216}-\frac{\sqrt{\frac{7}{5}} C_{31,22}^{31,1}}{4608}-\frac{C_{31,22}^{33,0}}{5120
\sqrt{3}}-\frac{C_{31,22}^{33,1}}{2560 \sqrt{3}}-\frac{9 C_{33,00}^{33,0}}{2560}-\frac{9 C_{33,00}^{33,1}}{1280}-\frac{3 \sqrt{3} C_{33,20}^{33,0}}{2560}-\frac{3
\sqrt{3} C_{33,20}^{33,1}}{1280}-\frac{9 C_{33,22}^{33,0}}{2560 \sqrt{10}}-\frac{9 C_{33,22}^{33,1}}{1280 \sqrt{10}}\)

\noindent\(C_{22,1112,3}^{22,1112,1,\text{Loc}}= \frac{1}{512} \sqrt{\frac{7}{3}} C_{00,20}^{60,1}-\frac{\sqrt{\frac{21}{5}} C_{00,22}^{62,1}}{1024}+\frac{11
\sqrt{\frac{7}{15}} C_{11,22}^{51,1}}{3072}-\frac{9 C_{11,22}^{53,1}}{5120}+\frac{1}{512} \sqrt{\frac{7}{3}} C_{20,20}^{40,1}+\frac{\sqrt{\frac{7}{15}}
C_{20,22}^{42,1}}{1024}-\frac{1}{256} \sqrt{\frac{7}{15}} C_{22,20}^{42,1}-\frac{\sqrt{\frac{21}{5}} C_{22,22}^{40,1}}{1024}+\frac{C_{22,22}^{42,1}}{512
\sqrt{15}}+\frac{9 \sqrt{\frac{3}{7}} C_{22,22}^{44,1}}{2560}-\frac{\sqrt{\frac{7}{3}} C_{31,20}^{31,1}}{1536}-\frac{\sqrt{\frac{7}{15}} C_{31,22}^{31,1}}{3072}-\frac{C_{31,22}^{33,1}}{5120}+\frac{9
C_{33,20}^{33,1}}{1280}-\frac{9 \sqrt{\frac{3}{10}} C_{33,22}^{33,1}}{2560}\)

\noindent\(C_{22,1112,3}^{22,1112,1,\text{NonLoc}}= \frac{\sqrt{7} C_{00,00}^{60,0}}{1024}+\frac{\sqrt{\frac{7}{3}} C_{00,20}^{60,0}}{1024}+\frac{\sqrt{\frac{21}{5}}
C_{00,22}^{62,0}}{1024}+\frac{\sqrt{7} C_{11,00}^{51,0}}{3072}+\frac{11 \sqrt{\frac{7}{15}} C_{11,22}^{51,0}}{3072}-\frac{9 C_{11,22}^{53,0}}{5120}+\frac{\sqrt{7}
C_{20,00}^{40,0}}{1024}+\frac{\sqrt{\frac{7}{3}} C_{20,20}^{40,0}}{1024}-\frac{\sqrt{\frac{7}{15}} C_{20,22}^{42,0}}{1024}-\frac{1}{512} \sqrt{\frac{7}{5}}
C_{22,00}^{42,0}-\frac{1}{512} \sqrt{\frac{7}{15}} C_{22,20}^{42,0}+\frac{\sqrt{\frac{21}{5}} C_{22,22}^{40,0}}{1024}-\frac{C_{22,22}^{42,0}}{512
\sqrt{15}}-\frac{9 \sqrt{\frac{3}{7}} C_{22,22}^{44,0}}{2560}+\frac{\sqrt{7} C_{31,00}^{31,0}}{3072}+\frac{\sqrt{\frac{7}{3}} C_{31,20}^{31,0}}{3072}-\frac{\sqrt{\frac{7}{15}}
C_{31,22}^{31,0}}{3072}-\frac{C_{31,22}^{33,0}}{5120}-\frac{9 \sqrt{3} C_{33,00}^{33,0}}{2560}-\frac{9 C_{33,20}^{33,0}}{2560}-\frac{9 \sqrt{\frac{3}{10}}
C_{33,22}^{33,0}}{2560}\)

\noindent\(C_{22,1112,4}^{00,3314,0,\text{Loc}}= -\frac{C_{00,20}^{60,0}}{32}-\frac{C_{00,20}^{60,1}}{64}-\frac{C_{00,22}^{62,0}}{64
\sqrt{5}}-\frac{C_{00,22}^{62,1}}{128 \sqrt{5}}-\frac{C_{11,22}^{51,0}}{192 \sqrt{5}}-\frac{C_{11,22}^{51,1}}{384 \sqrt{5}}-\frac{1}{320} \sqrt{\frac{3}{7}}
C_{11,22}^{53,0}-\frac{1}{640} \sqrt{\frac{3}{7}} C_{11,22}^{53,1}+\frac{C_{20,20}^{40,0}}{96}+\frac{C_{20,20}^{40,1}}{192}+\frac{C_{20,22}^{42,0}}{192
\sqrt{5}}+\frac{C_{20,22}^{42,1}}{384 \sqrt{5}}-\frac{C_{22,20}^{42,0}}{96 \sqrt{5}}-\frac{C_{22,20}^{42,1}}{192 \sqrt{5}}+\frac{C_{22,22}^{40,0}}{192
\sqrt{5}}+\frac{C_{22,22}^{40,1}}{384 \sqrt{5}}-\frac{C_{22,22}^{42,0}}{48 \sqrt{35}}-\frac{C_{22,22}^{42,1}}{96 \sqrt{35}}-\frac{C_{22,22}^{44,0}}{2240}-\frac{C_{22,22}^{44,1}}{4480}+\frac{C_{31,20}^{31,0}}{96}+\frac{C_{31,20}^{31,1}}{192}+\frac{C_{31,22}^{31,0}}{192
\sqrt{5}}+\frac{C_{31,22}^{31,1}}{384 \sqrt{5}}+\frac{1}{320} \sqrt{\frac{3}{7}} C_{31,22}^{33,0}+\frac{1}{640} \sqrt{\frac{3}{7}} C_{31,22}^{33,1}-\frac{3}{160}
\sqrt{\frac{3}{7}} C_{33,20}^{33,0}-\frac{3}{320} \sqrt{\frac{3}{7}} C_{33,20}^{33,1}-\frac{3}{320} \sqrt{\frac{7}{10}} C_{33,22}^{33,0}-\frac{3}{640}
\sqrt{\frac{7}{10}} C_{33,22}^{33,1}\)

\noindent\(C_{22,1112,4}^{00,3314,0,\text{NonLoc}}= -\frac{\sqrt{3} C_{00,00}^{60,0}}{128}-\frac{\sqrt{3} C_{00,00}^{60,1}}{64}-\frac{C_{00,20}^{60,0}}{128}-\frac{C_{00,20}^{60,1}}{64}+\frac{C_{00,22}^{62,0}}{128
\sqrt{5}}+\frac{C_{00,22}^{62,1}}{64 \sqrt{5}}+\frac{C_{11,00}^{51,0}}{128 \sqrt{3}}+\frac{C_{11,00}^{51,1}}{64 \sqrt{3}}-\frac{C_{11,22}^{51,0}}{384
\sqrt{5}}-\frac{C_{11,22}^{51,1}}{192 \sqrt{5}}-\frac{1}{640} \sqrt{\frac{3}{7}} C_{11,22}^{53,0}-\frac{1}{320} \sqrt{\frac{3}{7}} C_{11,22}^{53,1}+\frac{C_{20,00}^{40,0}}{128
\sqrt{3}}+\frac{C_{20,00}^{40,1}}{64 \sqrt{3}}+\frac{C_{20,20}^{40,0}}{384}+\frac{C_{20,20}^{40,1}}{192}-\frac{C_{20,22}^{42,0}}{384 \sqrt{5}}-\frac{C_{20,22}^{42,1}}{192
\sqrt{5}}-\frac{C_{22,00}^{42,0}}{128 \sqrt{15}}-\frac{C_{22,00}^{42,1}}{64 \sqrt{15}}-\frac{C_{22,20}^{42,0}}{384 \sqrt{5}}-\frac{C_{22,20}^{42,1}}{192
\sqrt{5}}-\frac{C_{22,22}^{40,0}}{384 \sqrt{5}}-\frac{C_{22,22}^{40,1}}{192 \sqrt{5}}+\frac{C_{22,22}^{42,0}}{96 \sqrt{35}}+\frac{C_{22,22}^{42,1}}{48
\sqrt{35}}+\frac{C_{22,22}^{44,0}}{4480}+\frac{C_{22,22}^{44,1}}{2240}-\frac{C_{31,00}^{31,0}}{128 \sqrt{3}}-\frac{C_{31,00}^{31,1}}{64 \sqrt{3}}-\frac{C_{31,20}^{31,0}}{384}-\frac{C_{31,20}^{31,1}}{192}+\frac{C_{31,22}^{31,0}}{384
\sqrt{5}}+\frac{C_{31,22}^{31,1}}{192 \sqrt{5}}+\frac{1}{640} \sqrt{\frac{3}{7}} C_{31,22}^{33,0}+\frac{1}{320} \sqrt{\frac{3}{7}} C_{31,22}^{33,1}+\frac{9
C_{33,00}^{33,0}}{640 \sqrt{7}}+\frac{9 C_{33,00}^{33,1}}{320 \sqrt{7}}+\frac{3}{640} \sqrt{\frac{3}{7}} C_{33,20}^{33,0}+\frac{3}{320} \sqrt{\frac{3}{7}}
C_{33,20}^{33,1}-\frac{3}{640} \sqrt{\frac{7}{10}} C_{33,22}^{33,0}-\frac{3}{320} \sqrt{\frac{7}{10}} C_{33,22}^{33,1}\)

\noindent\(C_{22,1112,4}^{00,3314,1,\text{Loc}}= -\frac{\sqrt{3} C_{00,20}^{60,1}}{64}-\frac{1}{128} \sqrt{\frac{3}{5}} C_{00,22}^{62,1}-\frac{C_{11,22}^{51,1}}{128
\sqrt{15}}-\frac{3 C_{11,22}^{53,1}}{640 \sqrt{7}}+\frac{C_{20,20}^{40,1}}{64 \sqrt{3}}+\frac{C_{20,22}^{42,1}}{128 \sqrt{15}}-\frac{C_{22,20}^{42,1}}{64
\sqrt{15}}+\frac{C_{22,22}^{40,1}}{128 \sqrt{15}}-\frac{C_{22,22}^{42,1}}{32 \sqrt{105}}-\frac{\sqrt{3} C_{22,22}^{44,1}}{4480}+\frac{C_{31,20}^{31,1}}{64
\sqrt{3}}+\frac{C_{31,22}^{31,1}}{128 \sqrt{15}}+\frac{3 C_{31,22}^{33,1}}{640 \sqrt{7}}-\frac{9 C_{33,20}^{33,1}}{320 \sqrt{7}}-\frac{3}{640} \sqrt{\frac{21}{10}}
C_{33,22}^{33,1}\)

\noindent\(C_{22,1112,4}^{00,3314,1,\text{NonLoc}}= -\frac{3 C_{00,00}^{60,0}}{128}-\frac{\sqrt{3} C_{00,20}^{60,0}}{128}+\frac{1}{128}
\sqrt{\frac{3}{5}} C_{00,22}^{62,0}+\frac{C_{11,00}^{51,0}}{128}-\frac{C_{11,22}^{51,0}}{128 \sqrt{15}}-\frac{3 C_{11,22}^{53,0}}{640 \sqrt{7}}+\frac{C_{20,00}^{40,0}}{128}+\frac{C_{20,20}^{40,0}}{128
\sqrt{3}}-\frac{C_{20,22}^{42,0}}{128 \sqrt{15}}-\frac{C_{22,00}^{42,0}}{128 \sqrt{5}}-\frac{C_{22,20}^{42,0}}{128 \sqrt{15}}-\frac{C_{22,22}^{40,0}}{128
\sqrt{15}}+\frac{C_{22,22}^{42,0}}{32 \sqrt{105}}+\frac{\sqrt{3} C_{22,22}^{44,0}}{4480}-\frac{C_{31,00}^{31,0}}{128}-\frac{C_{31,20}^{31,0}}{128
\sqrt{3}}+\frac{C_{31,22}^{31,0}}{128 \sqrt{15}}+\frac{3 C_{31,22}^{33,0}}{640 \sqrt{7}}+\frac{9}{640} \sqrt{\frac{3}{7}} C_{33,00}^{33,0}+\frac{9
C_{33,20}^{33,0}}{640 \sqrt{7}}-\frac{3}{640} \sqrt{\frac{21}{10}} C_{33,22}^{33,0}\)

\noindent\(C_{22,1112,4}^{22,1112,0,\text{Loc}}= \frac{3 C_{00,20}^{60,0}}{256}+\frac{3 C_{00,20}^{60,1}}{512}+\frac{3 C_{00,22}^{62,0}}{512
\sqrt{5}}+\frac{3 C_{00,22}^{62,1}}{1024 \sqrt{5}}-\frac{C_{11,22}^{51,0}}{512 \sqrt{5}}-\frac{C_{11,22}^{51,1}}{1024 \sqrt{5}}-\frac{3 \sqrt{\frac{3}{7}}
C_{11,22}^{53,0}}{2560}-\frac{3 \sqrt{\frac{3}{7}} C_{11,22}^{53,1}}{5120}+\frac{C_{20,20}^{40,0}}{768}+\frac{C_{20,20}^{40,1}}{1536}+\frac{C_{20,22}^{42,0}}{1536
\sqrt{5}}+\frac{C_{20,22}^{42,1}}{3072 \sqrt{5}}-\frac{C_{22,20}^{42,0}}{192 \sqrt{5}}-\frac{C_{22,20}^{42,1}}{384 \sqrt{5}}+\frac{C_{22,22}^{40,0}}{1536
\sqrt{5}}+\frac{C_{22,22}^{40,1}}{3072 \sqrt{5}}-\frac{1}{768} \sqrt{\frac{5}{7}} C_{22,22}^{42,0}-\frac{\sqrt{\frac{5}{7}} C_{22,22}^{42,1}}{1536}-\frac{3
C_{22,22}^{44,0}}{8960}-\frac{3 C_{22,22}^{44,1}}{17920}+\frac{C_{31,20}^{31,0}}{768}+\frac{C_{31,20}^{31,1}}{1536}+\frac{C_{31,22}^{31,0}}{1536
\sqrt{5}}+\frac{C_{31,22}^{31,1}}{3072 \sqrt{5}}+\frac{\sqrt{\frac{3}{7}} C_{31,22}^{33,0}}{2560}+\frac{\sqrt{\frac{3}{7}} C_{31,22}^{33,1}}{5120}+\frac{3}{320}
\sqrt{\frac{3}{7}} C_{33,20}^{33,0}+\frac{3}{640} \sqrt{\frac{3}{7}} C_{33,20}^{33,1}+\frac{27 C_{33,22}^{33,0}}{1280 \sqrt{70}}+\frac{27 C_{33,22}^{33,1}}{2560
\sqrt{70}}\)

\noindent\(C_{22,1112,4}^{22,1112,0,\text{NonLoc}}= \frac{3 \sqrt{3} C_{00,00}^{60,0}}{1024}+\frac{3 \sqrt{3} C_{00,00}^{60,1}}{512}+\frac{3
C_{00,20}^{60,0}}{1024}+\frac{3 C_{00,20}^{60,1}}{512}-\frac{3 C_{00,22}^{62,0}}{1024 \sqrt{5}}-\frac{3 C_{00,22}^{62,1}}{512 \sqrt{5}}+\frac{\sqrt{3}
C_{11,00}^{51,0}}{1024}+\frac{\sqrt{3} C_{11,00}^{51,1}}{512}-\frac{C_{11,22}^{51,0}}{1024 \sqrt{5}}-\frac{C_{11,22}^{51,1}}{512 \sqrt{5}}-\frac{3
\sqrt{\frac{3}{7}} C_{11,22}^{53,0}}{5120}-\frac{3 \sqrt{\frac{3}{7}} C_{11,22}^{53,1}}{2560}+\frac{C_{20,00}^{40,0}}{1024 \sqrt{3}}+\frac{C_{20,00}^{40,1}}{512
\sqrt{3}}+\frac{C_{20,20}^{40,0}}{3072}+\frac{C_{20,20}^{40,1}}{1536}-\frac{C_{20,22}^{42,0}}{3072 \sqrt{5}}-\frac{C_{20,22}^{42,1}}{1536 \sqrt{5}}-\frac{C_{22,00}^{42,0}}{256
\sqrt{15}}-\frac{C_{22,00}^{42,1}}{128 \sqrt{15}}-\frac{C_{22,20}^{42,0}}{768 \sqrt{5}}-\frac{C_{22,20}^{42,1}}{384 \sqrt{5}}-\frac{C_{22,22}^{40,0}}{3072
\sqrt{5}}-\frac{C_{22,22}^{40,1}}{1536 \sqrt{5}}+\frac{\sqrt{\frac{5}{7}} C_{22,22}^{42,0}}{1536}+\frac{1}{768} \sqrt{\frac{5}{7}} C_{22,22}^{42,1}+\frac{3
C_{22,22}^{44,0}}{17920}+\frac{3 C_{22,22}^{44,1}}{8960}-\frac{C_{31,00}^{31,0}}{1024 \sqrt{3}}-\frac{C_{31,00}^{31,1}}{512 \sqrt{3}}-\frac{C_{31,20}^{31,0}}{3072}-\frac{C_{31,20}^{31,1}}{1536}+\frac{C_{31,22}^{31,0}}{3072
\sqrt{5}}+\frac{C_{31,22}^{31,1}}{1536 \sqrt{5}}+\frac{\sqrt{\frac{3}{7}} C_{31,22}^{33,0}}{5120}+\frac{\sqrt{\frac{3}{7}} C_{31,22}^{33,1}}{2560}-\frac{9
C_{33,00}^{33,0}}{1280 \sqrt{7}}-\frac{9 C_{33,00}^{33,1}}{640 \sqrt{7}}-\frac{3 \sqrt{\frac{3}{7}} C_{33,20}^{33,0}}{1280}-\frac{3}{640} \sqrt{\frac{3}{7}}
C_{33,20}^{33,1}+\frac{27 C_{33,22}^{33,0}}{2560 \sqrt{70}}+\frac{27 C_{33,22}^{33,1}}{1280 \sqrt{70}}\)

\noindent\(C_{22,1112,4}^{22,1112,1,\text{Loc}}= \frac{3 \sqrt{3} C_{00,20}^{60,1}}{512}+\frac{3 \sqrt{\frac{3}{5}} C_{00,22}^{62,1}}{1024}-\frac{\sqrt{\frac{3}{5}}
C_{11,22}^{51,1}}{1024}-\frac{9 C_{11,22}^{53,1}}{5120 \sqrt{7}}+\frac{C_{20,20}^{40,1}}{512 \sqrt{3}}+\frac{C_{20,22}^{42,1}}{1024 \sqrt{15}}-\frac{C_{22,20}^{42,1}}{128
\sqrt{15}}+\frac{C_{22,22}^{40,1}}{1024 \sqrt{15}}-\frac{1}{512} \sqrt{\frac{5}{21}} C_{22,22}^{42,1}-\frac{3 \sqrt{3} C_{22,22}^{44,1}}{17920}+\frac{C_{31,20}^{31,1}}{512
\sqrt{3}}+\frac{C_{31,22}^{31,1}}{1024 \sqrt{15}}+\frac{3 C_{31,22}^{33,1}}{5120 \sqrt{7}}+\frac{9 C_{33,20}^{33,1}}{640 \sqrt{7}}+\frac{27 \sqrt{\frac{3}{70}}
C_{33,22}^{33,1}}{2560}\)

\noindent\(C_{22,1112,4}^{22,1112,1,\text{NonLoc}}= \frac{9 C_{00,00}^{60,0}}{1024}+\frac{3 \sqrt{3} C_{00,20}^{60,0}}{1024}-\frac{3
\sqrt{\frac{3}{5}} C_{00,22}^{62,0}}{1024}+\frac{3 C_{11,00}^{51,0}}{1024}-\frac{\sqrt{\frac{3}{5}} C_{11,22}^{51,0}}{1024}-\frac{9 C_{11,22}^{53,0}}{5120
\sqrt{7}}+\frac{C_{20,00}^{40,0}}{1024}+\frac{C_{20,20}^{40,0}}{1024 \sqrt{3}}-\frac{C_{20,22}^{42,0}}{1024 \sqrt{15}}-\frac{C_{22,00}^{42,0}}{256
\sqrt{5}}-\frac{C_{22,20}^{42,0}}{256 \sqrt{15}}-\frac{C_{22,22}^{40,0}}{1024 \sqrt{15}}+\frac{1}{512} \sqrt{\frac{5}{21}} C_{22,22}^{42,0}+\frac{3
\sqrt{3} C_{22,22}^{44,0}}{17920}-\frac{C_{31,00}^{31,0}}{1024}-\frac{C_{31,20}^{31,0}}{1024 \sqrt{3}}+\frac{C_{31,22}^{31,0}}{1024 \sqrt{15}}+\frac{3
C_{31,22}^{33,0}}{5120 \sqrt{7}}-\frac{9 \sqrt{\frac{3}{7}} C_{33,00}^{33,0}}{1280}-\frac{9 C_{33,20}^{33,0}}{1280 \sqrt{7}}+\frac{27 \sqrt{\frac{3}{70}}
C_{33,22}^{33,0}}{2560}\)

\noindent\(C_{22,2000,2}^{00,2202,0,\text{Loc}}= -\frac{7 \sqrt{5} C_{00,00}^{60,0}}{192}-\frac{7 \sqrt{5} C_{00,00}^{60,1}}{384}-\frac{7
\sqrt{5} C_{11,00}^{51,0}}{576}-\frac{7 \sqrt{5} C_{11,00}^{51,1}}{1152}+\frac{7 \sqrt{5} C_{20,00}^{40,0}}{576}+\frac{7 \sqrt{5} C_{20,00}^{40,1}}{1152}-\frac{7
C_{22,00}^{42,0}}{576}-\frac{7 C_{22,00}^{42,1}}{1152}+\frac{7 \sqrt{5} C_{31,00}^{31,0}}{576}+\frac{7 \sqrt{5} C_{31,00}^{31,1}}{1152}-\frac{1}{64}
\sqrt{\frac{21}{5}} C_{33,00}^{33,0}-\frac{1}{128} \sqrt{\frac{21}{5}} C_{33,00}^{33,1}\)

\noindent\(C_{22,2000,2}^{00,2202,0,\text{NonLoc}}= \frac{7 \sqrt{5} C_{00,00}^{60,0}}{768}+\frac{7 \sqrt{5} C_{00,00}^{60,1}}{384}-\frac{7}{256}
\sqrt{\frac{5}{3}} C_{00,20}^{60,0}-\frac{7}{128} \sqrt{\frac{5}{3}} C_{00,20}^{60,1}-\frac{7 \sqrt{5} C_{11,00}^{51,0}}{2304}-\frac{7 \sqrt{5} C_{11,00}^{51,1}}{1152}-\frac{7
\sqrt{5} C_{20,00}^{40,0}}{2304}-\frac{7 \sqrt{5} C_{20,00}^{40,1}}{1152}+\frac{7}{768} \sqrt{\frac{5}{3}} C_{20,20}^{40,0}+\frac{7}{384} \sqrt{\frac{5}{3}}
C_{20,20}^{40,1}+\frac{7 C_{22,00}^{42,0}}{2304}+\frac{7 C_{22,00}^{42,1}}{1152}-\frac{7 C_{22,20}^{42,0}}{768 \sqrt{3}}-\frac{7 C_{22,20}^{42,1}}{384
\sqrt{3}}+\frac{7 \sqrt{5} C_{31,00}^{31,0}}{2304}+\frac{7 \sqrt{5} C_{31,00}^{31,1}}{1152}-\frac{7}{768} \sqrt{\frac{5}{3}} C_{31,20}^{31,0}-\frac{7}{384}
\sqrt{\frac{5}{3}} C_{31,20}^{31,1}-\frac{1}{256} \sqrt{\frac{21}{5}} C_{33,00}^{33,0}-\frac{1}{128} \sqrt{\frac{21}{5}} C_{33,00}^{33,1}+\frac{3}{256}
\sqrt{\frac{7}{5}} C_{33,20}^{33,0}+\frac{3}{128} \sqrt{\frac{7}{5}} C_{33,20}^{33,1}\)

\noindent\(C_{22,2000,2}^{00,2202,1,\text{Loc}}= -\frac{7}{128} \sqrt{\frac{5}{3}} C_{00,00}^{60,1}-\frac{7}{384} \sqrt{\frac{5}{3}}
C_{11,00}^{51,1}+\frac{7}{384} \sqrt{\frac{5}{3}} C_{20,00}^{40,1}-\frac{7 C_{22,00}^{42,1}}{384 \sqrt{3}}+\frac{7}{384} \sqrt{\frac{5}{3}} C_{31,00}^{31,1}-\frac{3}{128}
\sqrt{\frac{7}{5}} C_{33,00}^{33,1}\)

\noindent\(C_{22,2000,2}^{00,2202,1,\text{NonLoc}}= \frac{7}{256} \sqrt{\frac{5}{3}} C_{00,00}^{60,0}-\frac{7 \sqrt{5} C_{00,20}^{60,0}}{256}-\frac{7}{768}
\sqrt{\frac{5}{3}} C_{11,00}^{51,0}-\frac{7}{768} \sqrt{\frac{5}{3}} C_{20,00}^{40,0}+\frac{7 \sqrt{5} C_{20,20}^{40,0}}{768}+\frac{7 C_{22,00}^{42,0}}{768
\sqrt{3}}-\frac{7 C_{22,20}^{42,0}}{768}+\frac{7}{768} \sqrt{\frac{5}{3}} C_{31,00}^{31,0}-\frac{7 \sqrt{5} C_{31,20}^{31,0}}{768}-\frac{3}{256}
\sqrt{\frac{7}{5}} C_{33,00}^{33,0}+\frac{3}{256} \sqrt{\frac{21}{5}} C_{33,20}^{33,0}\)

\noindent\(C_{22,2000,2}^{11,1111,0,\text{Loc}}= \frac{7 \sqrt{5} C_{11,11}^{51,0}}{576}+\frac{7 \sqrt{5} C_{11,11}^{51,1}}{1152}+\frac{\sqrt{5}
C_{31,11}^{31,0}}{192}+\frac{\sqrt{5} C_{31,11}^{31,1}}{384}-\frac{1}{16} \sqrt{\frac{7}{10}} C_{33,11}^{33,0}-\frac{1}{32} \sqrt{\frac{7}{10}} C_{33,11}^{33,1}\)

\noindent\(C_{22,2000,2}^{11,1111,0,\text{NonLoc}}= \frac{7 \sqrt{5} C_{11,11}^{51,0}}{1152}+\frac{7 \sqrt{5} C_{11,11}^{51,1}}{576}+\frac{\sqrt{5}
C_{31,11}^{31,0}}{384}+\frac{\sqrt{5} C_{31,11}^{31,1}}{192}-\frac{1}{32} \sqrt{\frac{7}{10}} C_{33,11}^{33,0}-\frac{1}{16} \sqrt{\frac{7}{10}} C_{33,11}^{33,1}\)

\noindent\(C_{22,2000,2}^{11,1111,1,\text{Loc}}= \frac{7}{384} \sqrt{\frac{5}{3}} C_{11,11}^{51,1}+\frac{1}{128} \sqrt{\frac{5}{3}}
C_{31,11}^{31,1}-\frac{1}{32} \sqrt{\frac{21}{10}} C_{33,11}^{33,1}\)

\noindent\(C_{22,2000,2}^{11,1111,1,\text{NonLoc}}= \frac{7}{384} \sqrt{\frac{5}{3}} C_{11,11}^{51,0}+\frac{1}{128} \sqrt{\frac{5}{3}}
C_{31,11}^{31,0}-\frac{1}{32} \sqrt{\frac{21}{10}} C_{33,11}^{33,0}\)

\noindent\(C_{22,2000,2}^{22,0000,0,\text{Loc}}= \frac{7 \sqrt{5} C_{00,00}^{60,0}}{768}+\frac{7 \sqrt{5} C_{00,00}^{60,1}}{1536}-\frac{7
\sqrt{5} C_{11,00}^{51,0}}{2304}-\frac{7 \sqrt{5} C_{11,00}^{51,1}}{4608}+\frac{5 \sqrt{5} C_{20,00}^{40,0}}{2304}+\frac{5 \sqrt{5} C_{20,00}^{40,1}}{4608}-\frac{7
C_{22,00}^{42,0}}{1152}-\frac{7 C_{22,00}^{42,1}}{2304}+\frac{\sqrt{5} C_{31,00}^{31,0}}{2304}+\frac{\sqrt{5} C_{31,00}^{31,1}}{4608}+\frac{1}{128}
\sqrt{\frac{21}{5}} C_{33,00}^{33,0}+\frac{1}{256} \sqrt{\frac{21}{5}} C_{33,00}^{33,1}\)

\noindent\(C_{22,2000,2}^{22,0000,0,\text{NonLoc}}= -\frac{7 \sqrt{5} C_{00,00}^{60,0}}{3072}-\frac{7 \sqrt{5} C_{00,00}^{60,1}}{1536}+\frac{7
\sqrt{\frac{5}{3}} C_{00,20}^{60,0}}{1024}+\frac{7}{512} \sqrt{\frac{5}{3}} C_{00,20}^{60,1}-\frac{7 \sqrt{5} C_{11,00}^{51,0}}{9216}-\frac{7 \sqrt{5}
C_{11,00}^{51,1}}{4608}-\frac{5 \sqrt{5} C_{20,00}^{40,0}}{9216}-\frac{5 \sqrt{5} C_{20,00}^{40,1}}{4608}+\frac{5 \sqrt{\frac{5}{3}} C_{20,20}^{40,0}}{3072}+\frac{5
\sqrt{\frac{5}{3}} C_{20,20}^{40,1}}{1536}+\frac{7 C_{22,00}^{42,0}}{4608}+\frac{7 C_{22,00}^{42,1}}{2304}-\frac{7 C_{22,20}^{42,0}}{1536 \sqrt{3}}-\frac{7
C_{22,20}^{42,1}}{768 \sqrt{3}}+\frac{\sqrt{5} C_{31,00}^{31,0}}{9216}+\frac{\sqrt{5} C_{31,00}^{31,1}}{4608}-\frac{\sqrt{\frac{5}{3}} C_{31,20}^{31,0}}{3072}-\frac{\sqrt{\frac{5}{3}}
C_{31,20}^{31,1}}{1536}+\frac{1}{512} \sqrt{\frac{21}{5}} C_{33,00}^{33,0}+\frac{1}{256} \sqrt{\frac{21}{5}} C_{33,00}^{33,1}-\frac{3}{512} \sqrt{\frac{7}{5}}
C_{33,20}^{33,0}-\frac{3}{256} \sqrt{\frac{7}{5}} C_{33,20}^{33,1}\)

\noindent\(C_{22,2000,2}^{22,0000,1,\text{Loc}}= \frac{7}{512} \sqrt{\frac{5}{3}} C_{00,00}^{60,1}-\frac{7 \sqrt{\frac{5}{3}} C_{11,00}^{51,1}}{1536}+\frac{5
\sqrt{\frac{5}{3}} C_{20,00}^{40,1}}{1536}-\frac{7 C_{22,00}^{42,1}}{768 \sqrt{3}}+\frac{\sqrt{\frac{5}{3}} C_{31,00}^{31,1}}{1536}+\frac{3}{256}
\sqrt{\frac{7}{5}} C_{33,00}^{33,1}\)

\noindent\(C_{22,2000,2}^{22,0000,1,\text{NonLoc}}= -\frac{7 \sqrt{\frac{5}{3}} C_{00,00}^{60,0}}{1024}+\frac{7 \sqrt{5} C_{00,20}^{60,0}}{1024}-\frac{7
\sqrt{\frac{5}{3}} C_{11,00}^{51,0}}{3072}-\frac{5 \sqrt{\frac{5}{3}} C_{20,00}^{40,0}}{3072}+\frac{5 \sqrt{5} C_{20,20}^{40,0}}{3072}+\frac{7 C_{22,00}^{42,0}}{1536
\sqrt{3}}-\frac{7 C_{22,20}^{42,0}}{1536}+\frac{\sqrt{\frac{5}{3}} C_{31,00}^{31,0}}{3072}-\frac{\sqrt{5} C_{31,20}^{31,0}}{3072}+\frac{3}{512} \sqrt{\frac{7}{5}}
C_{33,00}^{33,0}-\frac{3}{512} \sqrt{\frac{21}{5}} C_{33,20}^{33,0}\)

\noindent\(C_{22,2011,1}^{00,2011,0,\text{Loc}}= \frac{7 C_{00,22}^{62,0}}{128}+\frac{7 C_{00,22}^{62,1}}{256}-\frac{7 C_{11,22}^{51,0}}{1152}-\frac{7
C_{11,22}^{51,1}}{2304}+\frac{3}{128} \sqrt{\frac{21}{5}} C_{11,22}^{53,0}+\frac{3}{256} \sqrt{\frac{21}{5}} C_{11,22}^{53,1}+\frac{7 C_{20,22}^{42,0}}{1152}+\frac{7
C_{20,22}^{42,1}}{2304}-\frac{17 C_{22,22}^{40,0}}{1152}-\frac{17 C_{22,22}^{40,1}}{2304}-\frac{5 \sqrt{7} C_{22,22}^{42,0}}{576}-\frac{5 \sqrt{7}
C_{22,22}^{42,1}}{1152}+\frac{3 C_{22,22}^{44,0}}{32 \sqrt{5}}+\frac{3 C_{22,22}^{44,1}}{64 \sqrt{5}}-\frac{23 C_{31,22}^{31,0}}{1152}-\frac{23 C_{31,22}^{31,1}}{2304}+\frac{1}{128}
\sqrt{\frac{7}{15}} C_{31,22}^{33,0}+\frac{1}{256} \sqrt{\frac{7}{15}} C_{31,22}^{33,1}-\frac{3}{160} \sqrt{\frac{7}{2}} C_{33,22}^{33,0}-\frac{3}{320}
\sqrt{\frac{7}{2}} C_{33,22}^{33,1}\)

\noindent\(C_{22,2011,1}^{00,2011,0,\text{NonLoc}}= -\frac{7 C_{00,22}^{62,0}}{256}-\frac{7 C_{00,22}^{62,1}}{128}-\frac{7 C_{11,22}^{51,0}}{2304}-\frac{7
C_{11,22}^{51,1}}{1152}+\frac{3}{256} \sqrt{\frac{21}{5}} C_{11,22}^{53,0}+\frac{3}{128} \sqrt{\frac{21}{5}} C_{11,22}^{53,1}-\frac{7 C_{20,22}^{42,0}}{2304}-\frac{7
C_{20,22}^{42,1}}{1152}+\frac{17 C_{22,22}^{40,0}}{2304}+\frac{17 C_{22,22}^{40,1}}{1152}+\frac{5 \sqrt{7} C_{22,22}^{42,0}}{1152}+\frac{5 \sqrt{7}
C_{22,22}^{42,1}}{576}-\frac{3 C_{22,22}^{44,0}}{64 \sqrt{5}}-\frac{3 C_{22,22}^{44,1}}{32 \sqrt{5}}-\frac{23 C_{31,22}^{31,0}}{2304}-\frac{23 C_{31,22}^{31,1}}{1152}+\frac{1}{256}
\sqrt{\frac{7}{15}} C_{31,22}^{33,0}+\frac{1}{128} \sqrt{\frac{7}{15}} C_{31,22}^{33,1}-\frac{3}{320} \sqrt{\frac{7}{2}} C_{33,22}^{33,0}-\frac{3}{160}
\sqrt{\frac{7}{2}} C_{33,22}^{33,1}\)

\noindent\(C_{22,2011,1}^{00,2011,1,\text{Loc}}= \frac{7 \sqrt{3} C_{00,22}^{62,1}}{256}-\frac{7 C_{11,22}^{51,1}}{768 \sqrt{3}}+\frac{9}{256}
\sqrt{\frac{7}{5}} C_{11,22}^{53,1}+\frac{7 C_{20,22}^{42,1}}{768 \sqrt{3}}-\frac{17 C_{22,22}^{40,1}}{768 \sqrt{3}}-\frac{5}{384} \sqrt{\frac{7}{3}}
C_{22,22}^{42,1}+\frac{3}{64} \sqrt{\frac{3}{5}} C_{22,22}^{44,1}-\frac{23 C_{31,22}^{31,1}}{768 \sqrt{3}}+\frac{1}{256} \sqrt{\frac{7}{5}} C_{31,22}^{33,1}-\frac{3}{320}
\sqrt{\frac{21}{2}} C_{33,22}^{33,1}\)

\noindent\(C_{22,2011,1}^{00,2011,1,\text{NonLoc}}= -\frac{7 \sqrt{3} C_{00,22}^{62,0}}{256}-\frac{7 C_{11,22}^{51,0}}{768 \sqrt{3}}+\frac{9}{256}
\sqrt{\frac{7}{5}} C_{11,22}^{53,0}-\frac{7 C_{20,22}^{42,0}}{768 \sqrt{3}}+\frac{17 C_{22,22}^{40,0}}{768 \sqrt{3}}+\frac{5}{384} \sqrt{\frac{7}{3}}
C_{22,22}^{42,0}-\frac{3}{64} \sqrt{\frac{3}{5}} C_{22,22}^{44,0}-\frac{23 C_{31,22}^{31,0}}{768 \sqrt{3}}+\frac{1}{256} \sqrt{\frac{7}{5}} C_{31,22}^{33,0}-\frac{3}{320}
\sqrt{\frac{21}{2}} C_{33,22}^{33,0}\)

\noindent\(C_{22,2011,1}^{00,2211,0,\text{Loc}}= \frac{7 C_{00,20}^{60,0}}{192}+\frac{7 C_{00,20}^{60,1}}{384}+\frac{7 C_{00,22}^{62,0}}{128
\sqrt{5}}+\frac{7 C_{00,22}^{62,1}}{256 \sqrt{5}}+\frac{7 C_{11,22}^{51,0}}{576 \sqrt{5}}+\frac{7 C_{11,22}^{51,1}}{1152 \sqrt{5}}+\frac{3 \sqrt{21}
C_{11,22}^{53,0}}{1280}+\frac{3 \sqrt{21} C_{11,22}^{53,1}}{2560}-\frac{7 C_{20,20}^{40,0}}{576}-\frac{7 C_{20,20}^{40,1}}{1152}-\frac{7 C_{20,22}^{42,0}}{576
\sqrt{5}}-\frac{7 C_{20,22}^{42,1}}{1152 \sqrt{5}}+\frac{7 C_{22,20}^{42,0}}{576 \sqrt{5}}+\frac{7 C_{22,20}^{42,1}}{1152 \sqrt{5}}-\frac{7 \sqrt{5}
C_{22,22}^{40,0}}{1152}-\frac{7 \sqrt{5} C_{22,22}^{40,1}}{2304}+\frac{7}{576} \sqrt{\frac{7}{5}} C_{22,22}^{42,0}+\frac{7 \sqrt{\frac{7}{5}} C_{22,22}^{42,1}}{1152}-\frac{7
C_{31,20}^{31,0}}{576}-\frac{7 C_{31,20}^{31,1}}{1152}-\frac{7 C_{31,22}^{31,0}}{576 \sqrt{5}}-\frac{7 C_{31,22}^{31,1}}{1152 \sqrt{5}}-\frac{3 \sqrt{21}
C_{31,22}^{33,0}}{1280}-\frac{3 \sqrt{21} C_{31,22}^{33,1}}{2560}+\frac{\sqrt{21} C_{33,20}^{33,0}}{320}+\frac{\sqrt{21} C_{33,20}^{33,1}}{640}+\frac{3}{80}
\sqrt{\frac{7}{10}} C_{33,22}^{33,0}+\frac{3}{160} \sqrt{\frac{7}{10}} C_{33,22}^{33,1}\)

\noindent\(C_{22,2011,1}^{00,2211,0,\text{NonLoc}}= \frac{7 C_{00,00}^{60,0}}{256 \sqrt{3}}+\frac{7 C_{00,00}^{60,1}}{128 \sqrt{3}}+\frac{7
C_{00,20}^{60,0}}{768}+\frac{7 C_{00,20}^{60,1}}{384}-\frac{7 C_{00,22}^{62,0}}{256 \sqrt{5}}-\frac{7 C_{00,22}^{62,1}}{128 \sqrt{5}}-\frac{7 C_{11,00}^{51,0}}{768
\sqrt{3}}-\frac{7 C_{11,00}^{51,1}}{384 \sqrt{3}}+\frac{7 C_{11,22}^{51,0}}{1152 \sqrt{5}}+\frac{7 C_{11,22}^{51,1}}{576 \sqrt{5}}+\frac{3 \sqrt{21}
C_{11,22}^{53,0}}{2560}+\frac{3 \sqrt{21} C_{11,22}^{53,1}}{1280}-\frac{7 C_{20,00}^{40,0}}{768 \sqrt{3}}-\frac{7 C_{20,00}^{40,1}}{384 \sqrt{3}}-\frac{7
C_{20,20}^{40,0}}{2304}-\frac{7 C_{20,20}^{40,1}}{1152}+\frac{7 C_{20,22}^{42,0}}{1152 \sqrt{5}}+\frac{7 C_{20,22}^{42,1}}{576 \sqrt{5}}+\frac{7
C_{22,00}^{42,0}}{768 \sqrt{15}}+\frac{7 C_{22,00}^{42,1}}{384 \sqrt{15}}+\frac{7 C_{22,20}^{42,0}}{2304 \sqrt{5}}+\frac{7 C_{22,20}^{42,1}}{1152
\sqrt{5}}+\frac{7 \sqrt{5} C_{22,22}^{40,0}}{2304}+\frac{7 \sqrt{5} C_{22,22}^{40,1}}{1152}-\frac{7 \sqrt{\frac{7}{5}} C_{22,22}^{42,0}}{1152}-\frac{7}{576}
\sqrt{\frac{7}{5}} C_{22,22}^{42,1}+\frac{7 C_{31,00}^{31,0}}{768 \sqrt{3}}+\frac{7 C_{31,00}^{31,1}}{384 \sqrt{3}}+\frac{7 C_{31,20}^{31,0}}{2304}+\frac{7
C_{31,20}^{31,1}}{1152}-\frac{7 C_{31,22}^{31,0}}{1152 \sqrt{5}}-\frac{7 C_{31,22}^{31,1}}{576 \sqrt{5}}-\frac{3 \sqrt{21} C_{31,22}^{33,0}}{2560}-\frac{3
\sqrt{21} C_{31,22}^{33,1}}{1280}-\frac{3 \sqrt{7} C_{33,00}^{33,0}}{1280}-\frac{3 \sqrt{7} C_{33,00}^{33,1}}{640}-\frac{\sqrt{21} C_{33,20}^{33,0}}{1280}-\frac{\sqrt{21}
C_{33,20}^{33,1}}{640}+\frac{3}{160} \sqrt{\frac{7}{10}} C_{33,22}^{33,0}+\frac{3}{80} \sqrt{\frac{7}{10}} C_{33,22}^{33,1}\)

\noindent\(C_{22,2011,1}^{00,2211,1,\text{Loc}}= \frac{7 C_{00,20}^{60,1}}{128 \sqrt{3}}+\frac{7}{256} \sqrt{\frac{3}{5}} C_{00,22}^{62,1}+\frac{7
C_{11,22}^{51,1}}{384 \sqrt{15}}+\frac{9 \sqrt{7} C_{11,22}^{53,1}}{2560}-\frac{7 C_{20,20}^{40,1}}{384 \sqrt{3}}-\frac{7 C_{20,22}^{42,1}}{384 \sqrt{15}}+\frac{7
C_{22,20}^{42,1}}{384 \sqrt{15}}-\frac{7}{768} \sqrt{\frac{5}{3}} C_{22,22}^{40,1}+\frac{7}{384} \sqrt{\frac{7}{15}} C_{22,22}^{42,1}-\frac{7 C_{31,20}^{31,1}}{384
\sqrt{3}}-\frac{7 C_{31,22}^{31,1}}{384 \sqrt{15}}-\frac{9 \sqrt{7} C_{31,22}^{33,1}}{2560}+\frac{3 \sqrt{7} C_{33,20}^{33,1}}{640}+\frac{3}{160}
\sqrt{\frac{21}{10}} C_{33,22}^{33,1}\)

\noindent\(C_{22,2011,1}^{00,2211,1,\text{NonLoc}}= \frac{7 C_{00,00}^{60,0}}{256}+\frac{7 C_{00,20}^{60,0}}{256 \sqrt{3}}-\frac{7}{256}
\sqrt{\frac{3}{5}} C_{00,22}^{62,0}-\frac{7 C_{11,00}^{51,0}}{768}+\frac{7 C_{11,22}^{51,0}}{384 \sqrt{15}}+\frac{9 \sqrt{7} C_{11,22}^{53,0}}{2560}-\frac{7
C_{20,00}^{40,0}}{768}-\frac{7 C_{20,20}^{40,0}}{768 \sqrt{3}}+\frac{7 C_{20,22}^{42,0}}{384 \sqrt{15}}+\frac{7 C_{22,00}^{42,0}}{768 \sqrt{5}}+\frac{7
C_{22,20}^{42,0}}{768 \sqrt{15}}+\frac{7}{768} \sqrt{\frac{5}{3}} C_{22,22}^{40,0}-\frac{7}{384} \sqrt{\frac{7}{15}} C_{22,22}^{42,0}+\frac{7 C_{31,00}^{31,0}}{768}+\frac{7
C_{31,20}^{31,0}}{768 \sqrt{3}}-\frac{7 C_{31,22}^{31,0}}{384 \sqrt{15}}-\frac{9 \sqrt{7} C_{31,22}^{33,0}}{2560}-\frac{3 \sqrt{21} C_{33,00}^{33,0}}{1280}-\frac{3
\sqrt{7} C_{33,20}^{33,0}}{1280}+\frac{3}{160} \sqrt{\frac{21}{10}} C_{33,22}^{33,0}\)

\noindent\(C_{22,2011,1}^{11,1101,0,\text{Loc}}= -\frac{7 \sqrt{5} C_{11,11}^{51,0}}{1152}-\frac{7 \sqrt{5} C_{11,11}^{51,1}}{2304}-\frac{\sqrt{5}
C_{31,11}^{31,0}}{384}-\frac{\sqrt{5} C_{31,11}^{31,1}}{768}+\frac{1}{32} \sqrt{\frac{7}{10}} C_{33,11}^{33,0}+\frac{1}{64} \sqrt{\frac{7}{10}} C_{33,11}^{33,1}\)

\noindent\(C_{22,2011,1}^{11,1101,0,\text{NonLoc}}= -\frac{7 \sqrt{5} C_{11,11}^{51,0}}{2304}-\frac{7 \sqrt{5} C_{11,11}^{51,1}}{1152}-\frac{\sqrt{5}
C_{31,11}^{31,0}}{768}-\frac{\sqrt{5} C_{31,11}^{31,1}}{384}+\frac{1}{64} \sqrt{\frac{7}{10}} C_{33,11}^{33,0}+\frac{1}{32} \sqrt{\frac{7}{10}} C_{33,11}^{33,1}\)

\noindent\(C_{22,2011,1}^{11,1101,1,\text{Loc}}= -\frac{7}{768} \sqrt{\frac{5}{3}} C_{11,11}^{51,1}-\frac{1}{256} \sqrt{\frac{5}{3}}
C_{31,11}^{31,1}+\frac{1}{64} \sqrt{\frac{21}{10}} C_{33,11}^{33,1}\)

\noindent\(C_{22,2011,1}^{11,1101,1,\text{NonLoc}}= -\frac{7}{768} \sqrt{\frac{5}{3}} C_{11,11}^{51,0}-\frac{1}{256} \sqrt{\frac{5}{3}}
C_{31,11}^{31,0}+\frac{1}{64} \sqrt{\frac{21}{10}} C_{33,11}^{33,0}\)

\noindent\(C_{22,2011,1}^{20,0011,0,\text{Loc}}= -\frac{7 C_{00,22}^{62,0}}{512}-\frac{7 C_{00,22}^{62,1}}{1024}+\frac{35 C_{11,22}^{51,0}}{4608}+\frac{35
C_{11,22}^{51,1}}{9216}-\frac{7 C_{20,22}^{42,0}}{4608}-\frac{7 C_{20,22}^{42,1}}{9216}-\frac{7 C_{22,22}^{40,0}}{4608}-\frac{7 C_{22,22}^{40,1}}{9216}+\frac{\sqrt{7}
C_{22,22}^{42,0}}{1152}+\frac{\sqrt{7} C_{22,22}^{42,1}}{2304}+\frac{3 C_{22,22}^{44,0}}{128 \sqrt{5}}+\frac{3 C_{22,22}^{44,1}}{256 \sqrt{5}}+\frac{7
C_{31,22}^{31,0}}{4608}+\frac{7 C_{31,22}^{31,1}}{9216}-\frac{1}{128} \sqrt{\frac{7}{15}} C_{31,22}^{33,0}-\frac{1}{256} \sqrt{\frac{7}{15}} C_{31,22}^{33,1}-\frac{3}{640}
\sqrt{\frac{7}{2}} C_{33,22}^{33,0}-\frac{3 \sqrt{\frac{7}{2}} C_{33,22}^{33,1}}{1280}\)

\noindent\(C_{22,2011,1}^{20,0011,0,\text{NonLoc}}= \frac{7 C_{00,22}^{62,0}}{1024}+\frac{7 C_{00,22}^{62,1}}{512}+\frac{35 C_{11,22}^{51,0}}{9216}+\frac{35
C_{11,22}^{51,1}}{4608}+\frac{7 C_{20,22}^{42,0}}{9216}+\frac{7 C_{20,22}^{42,1}}{4608}+\frac{7 C_{22,22}^{40,0}}{9216}+\frac{7 C_{22,22}^{40,1}}{4608}-\frac{\sqrt{7}
C_{22,22}^{42,0}}{2304}-\frac{\sqrt{7} C_{22,22}^{42,1}}{1152}-\frac{3 C_{22,22}^{44,0}}{256 \sqrt{5}}-\frac{3 C_{22,22}^{44,1}}{128 \sqrt{5}}+\frac{7
C_{31,22}^{31,0}}{9216}+\frac{7 C_{31,22}^{31,1}}{4608}-\frac{1}{256} \sqrt{\frac{7}{15}} C_{31,22}^{33,0}-\frac{1}{128} \sqrt{\frac{7}{15}} C_{31,22}^{33,1}-\frac{3
\sqrt{\frac{7}{2}} C_{33,22}^{33,0}}{1280}-\frac{3}{640} \sqrt{\frac{7}{2}} C_{33,22}^{33,1}\)

\noindent\(C_{22,2011,1}^{20,0011,1,\text{Loc}}= -\frac{7 \sqrt{3} C_{00,22}^{62,1}}{1024}+\frac{35 C_{11,22}^{51,1}}{3072 \sqrt{3}}-\frac{7
C_{20,22}^{42,1}}{3072 \sqrt{3}}-\frac{7 C_{22,22}^{40,1}}{3072 \sqrt{3}}+\frac{1}{768} \sqrt{\frac{7}{3}} C_{22,22}^{42,1}+\frac{3}{256} \sqrt{\frac{3}{5}}
C_{22,22}^{44,1}+\frac{7 C_{31,22}^{31,1}}{3072 \sqrt{3}}-\frac{1}{256} \sqrt{\frac{7}{5}} C_{31,22}^{33,1}-\frac{3 \sqrt{\frac{21}{2}} C_{33,22}^{33,1}}{1280}\)

\noindent\(C_{22,2011,1}^{20,0011,1,\text{NonLoc}}= \frac{7 \sqrt{3} C_{00,22}^{62,0}}{1024}+\frac{35 C_{11,22}^{51,0}}{3072 \sqrt{3}}+\frac{7
C_{20,22}^{42,0}}{3072 \sqrt{3}}+\frac{7 C_{22,22}^{40,0}}{3072 \sqrt{3}}-\frac{1}{768} \sqrt{\frac{7}{3}} C_{22,22}^{42,0}-\frac{3}{256} \sqrt{\frac{3}{5}}
C_{22,22}^{44,0}+\frac{7 C_{31,22}^{31,0}}{3072 \sqrt{3}}-\frac{1}{256} \sqrt{\frac{7}{5}} C_{31,22}^{33,0}-\frac{3 \sqrt{\frac{21}{2}} C_{33,22}^{33,0}}{1280}\)

\noindent\(C_{22,2011,1}^{22,0011,0,\text{Loc}}= -\frac{7 C_{00,20}^{60,0}}{768}-\frac{7 C_{00,20}^{60,1}}{1536}-\frac{7 C_{00,22}^{62,0}}{512
\sqrt{5}}-\frac{7 C_{00,22}^{62,1}}{1024 \sqrt{5}}+\frac{7 \sqrt{5} C_{11,22}^{51,0}}{4608}+\frac{7 \sqrt{5} C_{11,22}^{51,1}}{9216}-\frac{5 C_{20,20}^{40,0}}{2304}-\frac{5
C_{20,20}^{40,1}}{4608}-\frac{7 C_{20,22}^{42,0}}{4608 \sqrt{5}}-\frac{7 C_{20,22}^{42,1}}{9216 \sqrt{5}}+\frac{7 C_{22,20}^{42,0}}{1152 \sqrt{5}}+\frac{7
C_{22,20}^{42,1}}{2304 \sqrt{5}}-\frac{7 C_{22,22}^{40,0}}{4608 \sqrt{5}}-\frac{7 C_{22,22}^{40,1}}{9216 \sqrt{5}}+\frac{\sqrt{\frac{7}{5}} C_{22,22}^{42,0}}{1152}+\frac{\sqrt{\frac{7}{5}}
C_{22,22}^{42,1}}{2304}+\frac{3 C_{22,22}^{44,0}}{640}+\frac{3 C_{22,22}^{44,1}}{1280}-\frac{C_{31,20}^{31,0}}{2304}-\frac{C_{31,20}^{31,1}}{4608}+\frac{7
C_{31,22}^{31,0}}{4608 \sqrt{5}}+\frac{7 C_{31,22}^{31,1}}{9216 \sqrt{5}}-\frac{1}{640} \sqrt{\frac{7}{3}} C_{31,22}^{33,0}-\frac{\sqrt{\frac{7}{3}}
C_{31,22}^{33,1}}{1280}-\frac{\sqrt{21} C_{33,20}^{33,0}}{640}-\frac{\sqrt{21} C_{33,20}^{33,1}}{1280}-\frac{3}{640} \sqrt{\frac{7}{10}} C_{33,22}^{33,0}-\frac{3
\sqrt{\frac{7}{10}} C_{33,22}^{33,1}}{1280}\)

\noindent\(C_{22,2011,1}^{22,0011,0,\text{NonLoc}}= -\frac{7 C_{00,00}^{60,0}}{1024 \sqrt{3}}-\frac{7 C_{00,00}^{60,1}}{512 \sqrt{3}}-\frac{7
C_{00,20}^{60,0}}{3072}-\frac{7 C_{00,20}^{60,1}}{1536}+\frac{7 C_{00,22}^{62,0}}{1024 \sqrt{5}}+\frac{7 C_{00,22}^{62,1}}{512 \sqrt{5}}-\frac{7
C_{11,00}^{51,0}}{3072 \sqrt{3}}-\frac{7 C_{11,00}^{51,1}}{1536 \sqrt{3}}+\frac{7 \sqrt{5} C_{11,22}^{51,0}}{9216}+\frac{7 \sqrt{5} C_{11,22}^{51,1}}{4608}-\frac{5
C_{20,00}^{40,0}}{3072 \sqrt{3}}-\frac{5 C_{20,00}^{40,1}}{1536 \sqrt{3}}-\frac{5 C_{20,20}^{40,0}}{9216}-\frac{5 C_{20,20}^{40,1}}{4608}+\frac{7
C_{20,22}^{42,0}}{9216 \sqrt{5}}+\frac{7 C_{20,22}^{42,1}}{4608 \sqrt{5}}+\frac{7 C_{22,00}^{42,0}}{1536 \sqrt{15}}+\frac{7 C_{22,00}^{42,1}}{768
\sqrt{15}}+\frac{7 C_{22,20}^{42,0}}{4608 \sqrt{5}}+\frac{7 C_{22,20}^{42,1}}{2304 \sqrt{5}}+\frac{7 C_{22,22}^{40,0}}{9216 \sqrt{5}}+\frac{7 C_{22,22}^{40,1}}{4608
\sqrt{5}}-\frac{\sqrt{\frac{7}{5}} C_{22,22}^{42,0}}{2304}-\frac{\sqrt{\frac{7}{5}} C_{22,22}^{42,1}}{1152}-\frac{3 C_{22,22}^{44,0}}{1280}-\frac{3
C_{22,22}^{44,1}}{640}+\frac{C_{31,00}^{31,0}}{3072 \sqrt{3}}+\frac{C_{31,00}^{31,1}}{1536 \sqrt{3}}+\frac{C_{31,20}^{31,0}}{9216}+\frac{C_{31,20}^{31,1}}{4608}+\frac{7
C_{31,22}^{31,0}}{9216 \sqrt{5}}+\frac{7 C_{31,22}^{31,1}}{4608 \sqrt{5}}-\frac{\sqrt{\frac{7}{3}} C_{31,22}^{33,0}}{1280}-\frac{1}{640} \sqrt{\frac{7}{3}}
C_{31,22}^{33,1}+\frac{3 \sqrt{7} C_{33,00}^{33,0}}{2560}+\frac{3 \sqrt{7} C_{33,00}^{33,1}}{1280}+\frac{\sqrt{21} C_{33,20}^{33,0}}{2560}+\frac{\sqrt{21}
C_{33,20}^{33,1}}{1280}-\frac{3 \sqrt{\frac{7}{10}} C_{33,22}^{33,0}}{1280}-\frac{3}{640} \sqrt{\frac{7}{10}} C_{33,22}^{33,1}\)

\noindent\(C_{22,2011,1}^{22,0011,1,\text{Loc}}= -\frac{7 C_{00,20}^{60,1}}{512 \sqrt{3}}-\frac{7 \sqrt{\frac{3}{5}} C_{00,22}^{62,1}}{1024}+\frac{7
\sqrt{\frac{5}{3}} C_{11,22}^{51,1}}{3072}-\frac{5 C_{20,20}^{40,1}}{1536 \sqrt{3}}-\frac{7 C_{20,22}^{42,1}}{3072 \sqrt{15}}+\frac{7 C_{22,20}^{42,1}}{768
\sqrt{15}}-\frac{7 C_{22,22}^{40,1}}{3072 \sqrt{15}}+\frac{1}{768} \sqrt{\frac{7}{15}} C_{22,22}^{42,1}+\frac{3 \sqrt{3} C_{22,22}^{44,1}}{1280}-\frac{C_{31,20}^{31,1}}{1536
\sqrt{3}}+\frac{7 C_{31,22}^{31,1}}{3072 \sqrt{15}}-\frac{\sqrt{7} C_{31,22}^{33,1}}{1280}-\frac{3 \sqrt{7} C_{33,20}^{33,1}}{1280}-\frac{3 \sqrt{\frac{21}{10}}
C_{33,22}^{33,1}}{1280}\)

\noindent\(C_{22,2011,1}^{22,0011,1,\text{NonLoc}}= -\frac{7 C_{00,00}^{60,0}}{1024}-\frac{7 C_{00,20}^{60,0}}{1024 \sqrt{3}}+\frac{7
\sqrt{\frac{3}{5}} C_{00,22}^{62,0}}{1024}-\frac{7 C_{11,00}^{51,0}}{3072}+\frac{7 \sqrt{\frac{5}{3}} C_{11,22}^{51,0}}{3072}-\frac{5 C_{20,00}^{40,0}}{3072}-\frac{5
C_{20,20}^{40,0}}{3072 \sqrt{3}}+\frac{7 C_{20,22}^{42,0}}{3072 \sqrt{15}}+\frac{7 C_{22,00}^{42,0}}{1536 \sqrt{5}}+\frac{7 C_{22,20}^{42,0}}{1536
\sqrt{15}}+\frac{7 C_{22,22}^{40,0}}{3072 \sqrt{15}}-\frac{1}{768} \sqrt{\frac{7}{15}} C_{22,22}^{42,0}-\frac{3 \sqrt{3} C_{22,22}^{44,0}}{1280}+\frac{C_{31,00}^{31,0}}{3072}+\frac{C_{31,20}^{31,0}}{3072
\sqrt{3}}+\frac{7 C_{31,22}^{31,0}}{3072 \sqrt{15}}-\frac{\sqrt{7} C_{31,22}^{33,0}}{1280}+\frac{3 \sqrt{21} C_{33,00}^{33,0}}{2560}+\frac{3 \sqrt{7}
C_{33,20}^{33,0}}{2560}-\frac{3 \sqrt{\frac{21}{10}} C_{33,22}^{33,0}}{1280}\)

\noindent\(C_{22,2011,2}^{00,2212,0,\text{Loc}}= \frac{7}{192} \sqrt{\frac{5}{3}} C_{00,20}^{60,0}+\frac{7}{384} \sqrt{\frac{5}{3}}
C_{00,20}^{60,1}-\frac{7 C_{00,22}^{62,0}}{128 \sqrt{3}}-\frac{7 C_{00,22}^{62,1}}{256 \sqrt{3}}-\frac{7 C_{11,22}^{51,0}}{576 \sqrt{3}}-\frac{7
C_{11,22}^{51,1}}{1152 \sqrt{3}}-\frac{3}{256} \sqrt{\frac{7}{5}} C_{11,22}^{53,0}-\frac{3}{512} \sqrt{\frac{7}{5}} C_{11,22}^{53,1}-\frac{7}{576}
\sqrt{\frac{5}{3}} C_{20,20}^{40,0}-\frac{7 \sqrt{\frac{5}{3}} C_{20,20}^{40,1}}{1152}+\frac{7 C_{20,22}^{42,0}}{576 \sqrt{3}}+\frac{7 C_{20,22}^{42,1}}{1152
\sqrt{3}}+\frac{7 C_{22,20}^{42,0}}{576 \sqrt{3}}+\frac{7 C_{22,20}^{42,1}}{1152 \sqrt{3}}+\frac{35 C_{22,22}^{40,0}}{1152 \sqrt{3}}+\frac{35 C_{22,22}^{40,1}}{2304
\sqrt{3}}-\frac{7}{576} \sqrt{\frac{7}{3}} C_{22,22}^{42,0}-\frac{7 \sqrt{\frac{7}{3}} C_{22,22}^{42,1}}{1152}-\frac{7}{576} \sqrt{\frac{5}{3}} C_{31,20}^{31,0}-\frac{7
\sqrt{\frac{5}{3}} C_{31,20}^{31,1}}{1152}+\frac{7 C_{31,22}^{31,0}}{576 \sqrt{3}}+\frac{7 C_{31,22}^{31,1}}{1152 \sqrt{3}}+\frac{3}{256} \sqrt{\frac{7}{5}}
C_{31,22}^{33,0}+\frac{3}{512} \sqrt{\frac{7}{5}} C_{31,22}^{33,1}+\frac{1}{64} \sqrt{\frac{7}{5}} C_{33,20}^{33,0}+\frac{1}{128} \sqrt{\frac{7}{5}}
C_{33,20}^{33,1}-\frac{1}{80} \sqrt{\frac{21}{2}} C_{33,22}^{33,0}-\frac{1}{160} \sqrt{\frac{21}{2}} C_{33,22}^{33,1}\)

\noindent\(C_{22,2011,2}^{00,2212,0,\text{NonLoc}}= \frac{7 \sqrt{5} C_{00,00}^{60,0}}{768}+\frac{7 \sqrt{5} C_{00,00}^{60,1}}{384}+\frac{7}{768}
\sqrt{\frac{5}{3}} C_{00,20}^{60,0}+\frac{7}{384} \sqrt{\frac{5}{3}} C_{00,20}^{60,1}+\frac{7 C_{00,22}^{62,0}}{256 \sqrt{3}}+\frac{7 C_{00,22}^{62,1}}{128
\sqrt{3}}-\frac{7 \sqrt{5} C_{11,00}^{51,0}}{2304}-\frac{7 \sqrt{5} C_{11,00}^{51,1}}{1152}-\frac{7 C_{11,22}^{51,0}}{1152 \sqrt{3}}-\frac{7 C_{11,22}^{51,1}}{576
\sqrt{3}}-\frac{3}{512} \sqrt{\frac{7}{5}} C_{11,22}^{53,0}-\frac{3}{256} \sqrt{\frac{7}{5}} C_{11,22}^{53,1}-\frac{7 \sqrt{5} C_{20,00}^{40,0}}{2304}-\frac{7
\sqrt{5} C_{20,00}^{40,1}}{1152}-\frac{7 \sqrt{\frac{5}{3}} C_{20,20}^{40,0}}{2304}-\frac{7 \sqrt{\frac{5}{3}} C_{20,20}^{40,1}}{1152}-\frac{7 C_{20,22}^{42,0}}{1152
\sqrt{3}}-\frac{7 C_{20,22}^{42,1}}{576 \sqrt{3}}+\frac{7 C_{22,00}^{42,0}}{2304}+\frac{7 C_{22,00}^{42,1}}{1152}+\frac{7 C_{22,20}^{42,0}}{2304
\sqrt{3}}+\frac{7 C_{22,20}^{42,1}}{1152 \sqrt{3}}-\frac{35 C_{22,22}^{40,0}}{2304 \sqrt{3}}-\frac{35 C_{22,22}^{40,1}}{1152 \sqrt{3}}+\frac{7 \sqrt{\frac{7}{3}}
C_{22,22}^{42,0}}{1152}+\frac{7}{576} \sqrt{\frac{7}{3}} C_{22,22}^{42,1}+\frac{7 \sqrt{5} C_{31,00}^{31,0}}{2304}+\frac{7 \sqrt{5} C_{31,00}^{31,1}}{1152}+\frac{7
\sqrt{\frac{5}{3}} C_{31,20}^{31,0}}{2304}+\frac{7 \sqrt{\frac{5}{3}} C_{31,20}^{31,1}}{1152}+\frac{7 C_{31,22}^{31,0}}{1152 \sqrt{3}}+\frac{7 C_{31,22}^{31,1}}{576
\sqrt{3}}+\frac{3}{512} \sqrt{\frac{7}{5}} C_{31,22}^{33,0}+\frac{3}{256} \sqrt{\frac{7}{5}} C_{31,22}^{33,1}-\frac{1}{256} \sqrt{\frac{21}{5}} C_{33,00}^{33,0}-\frac{1}{128}
\sqrt{\frac{21}{5}} C_{33,00}^{33,1}-\frac{1}{256} \sqrt{\frac{7}{5}} C_{33,20}^{33,0}-\frac{1}{128} \sqrt{\frac{7}{5}} C_{33,20}^{33,1}-\frac{1}{160}
\sqrt{\frac{21}{2}} C_{33,22}^{33,0}-\frac{1}{80} \sqrt{\frac{21}{2}} C_{33,22}^{33,1}\)

\noindent\(C_{22,2011,2}^{00,2212,1,\text{Loc}}= \frac{7 \sqrt{5} C_{00,20}^{60,1}}{384}-\frac{7 C_{00,22}^{62,1}}{256}-\frac{7 C_{11,22}^{51,1}}{1152}-\frac{3}{512}
\sqrt{\frac{21}{5}} C_{11,22}^{53,1}-\frac{7 \sqrt{5} C_{20,20}^{40,1}}{1152}+\frac{7 C_{20,22}^{42,1}}{1152}+\frac{7 C_{22,20}^{42,1}}{1152}+\frac{35
C_{22,22}^{40,1}}{2304}-\frac{7 \sqrt{7} C_{22,22}^{42,1}}{1152}-\frac{7 \sqrt{5} C_{31,20}^{31,1}}{1152}+\frac{7 C_{31,22}^{31,1}}{1152}+\frac{3}{512}
\sqrt{\frac{21}{5}} C_{31,22}^{33,1}+\frac{1}{128} \sqrt{\frac{21}{5}} C_{33,20}^{33,1}-\frac{3}{160} \sqrt{\frac{7}{2}} C_{33,22}^{33,1}\)

\noindent\(C_{22,2011,2}^{00,2212,1,\text{NonLoc}}= \frac{7}{256} \sqrt{\frac{5}{3}} C_{00,00}^{60,0}+\frac{7 \sqrt{5} C_{00,20}^{60,0}}{768}+\frac{7
C_{00,22}^{62,0}}{256}-\frac{7}{768} \sqrt{\frac{5}{3}} C_{11,00}^{51,0}-\frac{7 C_{11,22}^{51,0}}{1152}-\frac{3}{512} \sqrt{\frac{21}{5}} C_{11,22}^{53,0}-\frac{7}{768}
\sqrt{\frac{5}{3}} C_{20,00}^{40,0}-\frac{7 \sqrt{5} C_{20,20}^{40,0}}{2304}-\frac{7 C_{20,22}^{42,0}}{1152}+\frac{7 C_{22,00}^{42,0}}{768 \sqrt{3}}+\frac{7
C_{22,20}^{42,0}}{2304}-\frac{35 C_{22,22}^{40,0}}{2304}+\frac{7 \sqrt{7} C_{22,22}^{42,0}}{1152}+\frac{7}{768} \sqrt{\frac{5}{3}} C_{31,00}^{31,0}+\frac{7
\sqrt{5} C_{31,20}^{31,0}}{2304}+\frac{7 C_{31,22}^{31,0}}{1152}+\frac{3}{512} \sqrt{\frac{21}{5}} C_{31,22}^{33,0}-\frac{3}{256} \sqrt{\frac{7}{5}}
C_{33,00}^{33,0}-\frac{1}{256} \sqrt{\frac{21}{5}} C_{33,20}^{33,0}-\frac{3}{160} \sqrt{\frac{7}{2}} C_{33,22}^{33,0}\)

\noindent\(C_{22,2011,2}^{11,1101,0,\text{Loc}}= -\frac{7}{384} \sqrt{\frac{5}{3}} C_{11,11}^{51,0}-\frac{7}{768} \sqrt{\frac{5}{3}}
C_{11,11}^{51,1}-\frac{1}{128} \sqrt{\frac{5}{3}} C_{31,11}^{31,0}-\frac{1}{256} \sqrt{\frac{5}{3}} C_{31,11}^{31,1}+\frac{1}{32} \sqrt{\frac{21}{10}}
C_{33,11}^{33,0}+\frac{1}{64} \sqrt{\frac{21}{10}} C_{33,11}^{33,1}\)

\noindent\(C_{22,2011,2}^{11,1101,0,\text{NonLoc}}= -\frac{7}{768} \sqrt{\frac{5}{3}} C_{11,11}^{51,0}-\frac{7}{384} \sqrt{\frac{5}{3}}
C_{11,11}^{51,1}-\frac{1}{256} \sqrt{\frac{5}{3}} C_{31,11}^{31,0}-\frac{1}{128} \sqrt{\frac{5}{3}} C_{31,11}^{31,1}+\frac{1}{64} \sqrt{\frac{21}{10}}
C_{33,11}^{33,0}+\frac{1}{32} \sqrt{\frac{21}{10}} C_{33,11}^{33,1}\)

\noindent\(C_{22,2011,2}^{11,1101,1,\text{Loc}}= -\frac{7 \sqrt{5} C_{11,11}^{51,1}}{768}-\frac{\sqrt{5} C_{31,11}^{31,1}}{256}+\frac{3}{64}
\sqrt{\frac{7}{10}} C_{33,11}^{33,1}\)

\noindent\(C_{22,2011,2}^{11,1101,1,\text{NonLoc}}= -\frac{7 \sqrt{5} C_{11,11}^{51,0}}{768}-\frac{\sqrt{5} C_{31,11}^{31,0}}{256}+\frac{3}{64}
\sqrt{\frac{7}{10}} C_{33,11}^{33,0}\)

\noindent\(C_{22,2011,2}^{22,0011,0,\text{Loc}}= -\frac{7}{768} \sqrt{\frac{5}{3}} C_{00,20}^{60,0}-\frac{7 \sqrt{\frac{5}{3}} C_{00,20}^{60,1}}{1536}+\frac{7
C_{00,22}^{62,0}}{512 \sqrt{3}}+\frac{7 C_{00,22}^{62,1}}{1024 \sqrt{3}}-\frac{35 C_{11,22}^{51,0}}{4608 \sqrt{3}}-\frac{35 C_{11,22}^{51,1}}{9216
\sqrt{3}}-\frac{5 \sqrt{\frac{5}{3}} C_{20,20}^{40,0}}{2304}-\frac{5 \sqrt{\frac{5}{3}} C_{20,20}^{40,1}}{4608}+\frac{7 C_{20,22}^{42,0}}{4608 \sqrt{3}}+\frac{7
C_{20,22}^{42,1}}{9216 \sqrt{3}}+\frac{7 C_{22,20}^{42,0}}{1152 \sqrt{3}}+\frac{7 C_{22,20}^{42,1}}{2304 \sqrt{3}}+\frac{7 C_{22,22}^{40,0}}{4608
\sqrt{3}}+\frac{7 C_{22,22}^{40,1}}{9216 \sqrt{3}}-\frac{\sqrt{\frac{7}{3}} C_{22,22}^{42,0}}{1152}-\frac{\sqrt{\frac{7}{3}} C_{22,22}^{42,1}}{2304}-\frac{1}{128}
\sqrt{\frac{3}{5}} C_{22,22}^{44,0}-\frac{1}{256} \sqrt{\frac{3}{5}} C_{22,22}^{44,1}-\frac{\sqrt{\frac{5}{3}} C_{31,20}^{31,0}}{2304}-\frac{\sqrt{\frac{5}{3}}
C_{31,20}^{31,1}}{4608}-\frac{7 C_{31,22}^{31,0}}{4608 \sqrt{3}}-\frac{7 C_{31,22}^{31,1}}{9216 \sqrt{3}}+\frac{1}{384} \sqrt{\frac{7}{5}} C_{31,22}^{33,0}+\frac{1}{768}
\sqrt{\frac{7}{5}} C_{31,22}^{33,1}-\frac{1}{128} \sqrt{\frac{7}{5}} C_{33,20}^{33,0}-\frac{1}{256} \sqrt{\frac{7}{5}} C_{33,20}^{33,1}+\frac{1}{640}
\sqrt{\frac{21}{2}} C_{33,22}^{33,0}+\frac{\sqrt{\frac{21}{2}} C_{33,22}^{33,1}}{1280}\)

\noindent\(C_{22,2011,2}^{22,0011,0,\text{NonLoc}}= -\frac{7 \sqrt{5} C_{00,00}^{60,0}}{3072}-\frac{7 \sqrt{5} C_{00,00}^{60,1}}{1536}-\frac{7
\sqrt{\frac{5}{3}} C_{00,20}^{60,0}}{3072}-\frac{7 \sqrt{\frac{5}{3}} C_{00,20}^{60,1}}{1536}-\frac{7 C_{00,22}^{62,0}}{1024 \sqrt{3}}-\frac{7 C_{00,22}^{62,1}}{512
\sqrt{3}}-\frac{7 \sqrt{5} C_{11,00}^{51,0}}{9216}-\frac{7 \sqrt{5} C_{11,00}^{51,1}}{4608}-\frac{35 C_{11,22}^{51,0}}{9216 \sqrt{3}}-\frac{35 C_{11,22}^{51,1}}{4608
\sqrt{3}}-\frac{5 \sqrt{5} C_{20,00}^{40,0}}{9216}-\frac{5 \sqrt{5} C_{20,00}^{40,1}}{4608}-\frac{5 \sqrt{\frac{5}{3}} C_{20,20}^{40,0}}{9216}-\frac{5
\sqrt{\frac{5}{3}} C_{20,20}^{40,1}}{4608}-\frac{7 C_{20,22}^{42,0}}{9216 \sqrt{3}}-\frac{7 C_{20,22}^{42,1}}{4608 \sqrt{3}}+\frac{7 C_{22,00}^{42,0}}{4608}+\frac{7
C_{22,00}^{42,1}}{2304}+\frac{7 C_{22,20}^{42,0}}{4608 \sqrt{3}}+\frac{7 C_{22,20}^{42,1}}{2304 \sqrt{3}}-\frac{7 C_{22,22}^{40,0}}{9216 \sqrt{3}}-\frac{7
C_{22,22}^{40,1}}{4608 \sqrt{3}}+\frac{\sqrt{\frac{7}{3}} C_{22,22}^{42,0}}{2304}+\frac{\sqrt{\frac{7}{3}} C_{22,22}^{42,1}}{1152}+\frac{1}{256}
\sqrt{\frac{3}{5}} C_{22,22}^{44,0}+\frac{1}{128} \sqrt{\frac{3}{5}} C_{22,22}^{44,1}+\frac{\sqrt{5} C_{31,00}^{31,0}}{9216}+\frac{\sqrt{5} C_{31,00}^{31,1}}{4608}+\frac{\sqrt{\frac{5}{3}}
C_{31,20}^{31,0}}{9216}+\frac{\sqrt{\frac{5}{3}} C_{31,20}^{31,1}}{4608}-\frac{7 C_{31,22}^{31,0}}{9216 \sqrt{3}}-\frac{7 C_{31,22}^{31,1}}{4608
\sqrt{3}}+\frac{1}{768} \sqrt{\frac{7}{5}} C_{31,22}^{33,0}+\frac{1}{384} \sqrt{\frac{7}{5}} C_{31,22}^{33,1}+\frac{1}{512} \sqrt{\frac{21}{5}} C_{33,00}^{33,0}+\frac{1}{256}
\sqrt{\frac{21}{5}} C_{33,00}^{33,1}+\frac{1}{512} \sqrt{\frac{7}{5}} C_{33,20}^{33,0}+\frac{1}{256} \sqrt{\frac{7}{5}} C_{33,20}^{33,1}+\frac{\sqrt{\frac{21}{2}}
C_{33,22}^{33,0}}{1280}+\frac{1}{640} \sqrt{\frac{21}{2}} C_{33,22}^{33,1}\)

\noindent\(C_{22,2011,2}^{22,0011,1,\text{Loc}}= -\frac{7 \sqrt{5} C_{00,20}^{60,1}}{1536}+\frac{7 C_{00,22}^{62,1}}{1024}-\frac{35
C_{11,22}^{51,1}}{9216}-\frac{5 \sqrt{5} C_{20,20}^{40,1}}{4608}+\frac{7 C_{20,22}^{42,1}}{9216}+\frac{7 C_{22,20}^{42,1}}{2304}+\frac{7 C_{22,22}^{40,1}}{9216}-\frac{\sqrt{7}
C_{22,22}^{42,1}}{2304}-\frac{3 C_{22,22}^{44,1}}{256 \sqrt{5}}-\frac{\sqrt{5} C_{31,20}^{31,1}}{4608}-\frac{7 C_{31,22}^{31,1}}{9216}+\frac{1}{256}
\sqrt{\frac{7}{15}} C_{31,22}^{33,1}-\frac{1}{256} \sqrt{\frac{21}{5}} C_{33,20}^{33,1}+\frac{3 \sqrt{\frac{7}{2}} C_{33,22}^{33,1}}{1280}\)

\noindent\(C_{22,2011,2}^{22,0011,1,\text{NonLoc}}= -\frac{7 \sqrt{\frac{5}{3}} C_{00,00}^{60,0}}{1024}-\frac{7 \sqrt{5} C_{00,20}^{60,0}}{3072}-\frac{7
C_{00,22}^{62,0}}{1024}-\frac{7 \sqrt{\frac{5}{3}} C_{11,00}^{51,0}}{3072}-\frac{35 C_{11,22}^{51,0}}{9216}-\frac{5 \sqrt{\frac{5}{3}} C_{20,00}^{40,0}}{3072}-\frac{5
\sqrt{5} C_{20,20}^{40,0}}{9216}-\frac{7 C_{20,22}^{42,0}}{9216}+\frac{7 C_{22,00}^{42,0}}{1536 \sqrt{3}}+\frac{7 C_{22,20}^{42,0}}{4608}-\frac{7
C_{22,22}^{40,0}}{9216}+\frac{\sqrt{7} C_{22,22}^{42,0}}{2304}+\frac{3 C_{22,22}^{44,0}}{256 \sqrt{5}}+\frac{\sqrt{\frac{5}{3}} C_{31,00}^{31,0}}{3072}+\frac{\sqrt{5}
C_{31,20}^{31,0}}{9216}-\frac{7 C_{31,22}^{31,0}}{9216}+\frac{1}{256} \sqrt{\frac{7}{15}} C_{31,22}^{33,0}+\frac{3}{512} \sqrt{\frac{7}{5}} C_{33,00}^{33,0}+\frac{1}{512}
\sqrt{\frac{21}{5}} C_{33,20}^{33,0}+\frac{3 \sqrt{\frac{7}{2}} C_{33,22}^{33,0}}{1280}\)

\noindent\(C_{22,2011,3}^{00,2213,0,\text{Loc}}= \frac{7}{192} \sqrt{\frac{7}{3}} C_{00,20}^{60,0}+\frac{7}{384} \sqrt{\frac{7}{3}}
C_{00,20}^{60,1}+\frac{1}{64} \sqrt{\frac{7}{15}} C_{00,22}^{62,0}+\frac{1}{128} \sqrt{\frac{7}{15}} C_{00,22}^{62,1}+\frac{1}{288} \sqrt{\frac{7}{15}}
C_{11,22}^{51,0}+\frac{1}{576} \sqrt{\frac{7}{15}} C_{11,22}^{51,1}+\frac{3 C_{11,22}^{53,0}}{640}+\frac{3 C_{11,22}^{53,1}}{1280}-\frac{7}{576}
\sqrt{\frac{7}{3}} C_{20,20}^{40,0}-\frac{7 \sqrt{\frac{7}{3}} C_{20,20}^{40,1}}{1152}-\frac{1}{288} \sqrt{\frac{7}{15}} C_{20,22}^{42,0}-\frac{1}{576}
\sqrt{\frac{7}{15}} C_{20,22}^{42,1}+\frac{7}{576} \sqrt{\frac{7}{15}} C_{22,20}^{42,0}+\frac{7 \sqrt{\frac{7}{15}} C_{22,20}^{42,1}}{1152}-\frac{1}{576}
\sqrt{\frac{35}{3}} C_{22,22}^{40,0}-\frac{\sqrt{\frac{35}{3}} C_{22,22}^{40,1}}{1152}+\frac{7 C_{22,22}^{42,0}}{288 \sqrt{15}}+\frac{7 C_{22,22}^{42,1}}{576
\sqrt{15}}-\frac{7}{576} \sqrt{\frac{7}{3}} C_{31,20}^{31,0}-\frac{7 \sqrt{\frac{7}{3}} C_{31,20}^{31,1}}{1152}-\frac{1}{288} \sqrt{\frac{7}{15}}
C_{31,22}^{31,0}-\frac{1}{576} \sqrt{\frac{7}{15}} C_{31,22}^{31,1}-\frac{3 C_{31,22}^{33,0}}{640}-\frac{3 C_{31,22}^{33,1}}{1280}+\frac{7 C_{33,20}^{33,0}}{320}+\frac{7
C_{33,20}^{33,1}}{640}+\frac{1}{40} \sqrt{\frac{3}{10}} C_{33,22}^{33,0}+\frac{1}{80} \sqrt{\frac{3}{10}} C_{33,22}^{33,1}\)

\noindent\(C_{22,2011,3}^{00,2213,0,\text{NonLoc}}= \frac{7 \sqrt{7} C_{00,00}^{60,0}}{768}+\frac{7 \sqrt{7} C_{00,00}^{60,1}}{384}+\frac{7}{768}
\sqrt{\frac{7}{3}} C_{00,20}^{60,0}+\frac{7}{384} \sqrt{\frac{7}{3}} C_{00,20}^{60,1}-\frac{1}{128} \sqrt{\frac{7}{15}} C_{00,22}^{62,0}-\frac{1}{64}
\sqrt{\frac{7}{15}} C_{00,22}^{62,1}-\frac{7 \sqrt{7} C_{11,00}^{51,0}}{2304}-\frac{7 \sqrt{7} C_{11,00}^{51,1}}{1152}+\frac{1}{576} \sqrt{\frac{7}{15}}
C_{11,22}^{51,0}+\frac{1}{288} \sqrt{\frac{7}{15}} C_{11,22}^{51,1}+\frac{3 C_{11,22}^{53,0}}{1280}+\frac{3 C_{11,22}^{53,1}}{640}-\frac{7 \sqrt{7}
C_{20,00}^{40,0}}{2304}-\frac{7 \sqrt{7} C_{20,00}^{40,1}}{1152}-\frac{7 \sqrt{\frac{7}{3}} C_{20,20}^{40,0}}{2304}-\frac{7 \sqrt{\frac{7}{3}} C_{20,20}^{40,1}}{1152}+\frac{1}{576}
\sqrt{\frac{7}{15}} C_{20,22}^{42,0}+\frac{1}{288} \sqrt{\frac{7}{15}} C_{20,22}^{42,1}+\frac{7 \sqrt{\frac{7}{5}} C_{22,00}^{42,0}}{2304}+\frac{7
\sqrt{\frac{7}{5}} C_{22,00}^{42,1}}{1152}+\frac{7 \sqrt{\frac{7}{15}} C_{22,20}^{42,0}}{2304}+\frac{7 \sqrt{\frac{7}{15}} C_{22,20}^{42,1}}{1152}+\frac{\sqrt{\frac{35}{3}}
C_{22,22}^{40,0}}{1152}+\frac{1}{576} \sqrt{\frac{35}{3}} C_{22,22}^{40,1}-\frac{7 C_{22,22}^{42,0}}{576 \sqrt{15}}-\frac{7 C_{22,22}^{42,1}}{288
\sqrt{15}}+\frac{7 \sqrt{7} C_{31,00}^{31,0}}{2304}+\frac{7 \sqrt{7} C_{31,00}^{31,1}}{1152}+\frac{7 \sqrt{\frac{7}{3}} C_{31,20}^{31,0}}{2304}+\frac{7
\sqrt{\frac{7}{3}} C_{31,20}^{31,1}}{1152}-\frac{1}{576} \sqrt{\frac{7}{15}} C_{31,22}^{31,0}-\frac{1}{288} \sqrt{\frac{7}{15}} C_{31,22}^{31,1}-\frac{3
C_{31,22}^{33,0}}{1280}-\frac{3 C_{31,22}^{33,1}}{640}-\frac{7 \sqrt{3} C_{33,00}^{33,0}}{1280}-\frac{7 \sqrt{3} C_{33,00}^{33,1}}{640}-\frac{7 C_{33,20}^{33,0}}{1280}-\frac{7
C_{33,20}^{33,1}}{640}+\frac{1}{80} \sqrt{\frac{3}{10}} C_{33,22}^{33,0}+\frac{1}{40} \sqrt{\frac{3}{10}} C_{33,22}^{33,1}\)

\noindent\(C_{22,2011,3}^{00,2213,1,\text{Loc}}= \frac{7 \sqrt{7} C_{00,20}^{60,1}}{384}+\frac{1}{128} \sqrt{\frac{7}{5}} C_{00,22}^{62,1}+\frac{1}{576}
\sqrt{\frac{7}{5}} C_{11,22}^{51,1}+\frac{3 \sqrt{3} C_{11,22}^{53,1}}{1280}-\frac{7 \sqrt{7} C_{20,20}^{40,1}}{1152}-\frac{1}{576} \sqrt{\frac{7}{5}}
C_{20,22}^{42,1}+\frac{7 \sqrt{\frac{7}{5}} C_{22,20}^{42,1}}{1152}-\frac{\sqrt{35} C_{22,22}^{40,1}}{1152}+\frac{7 C_{22,22}^{42,1}}{576 \sqrt{5}}-\frac{7
\sqrt{7} C_{31,20}^{31,1}}{1152}-\frac{1}{576} \sqrt{\frac{7}{5}} C_{31,22}^{31,1}-\frac{3 \sqrt{3} C_{31,22}^{33,1}}{1280}+\frac{7 \sqrt{3} C_{33,20}^{33,1}}{640}+\frac{3
C_{33,22}^{33,1}}{80 \sqrt{10}}\)

\noindent\(C_{22,2011,3}^{00,2213,1,\text{NonLoc}}= \frac{7}{256} \sqrt{\frac{7}{3}} C_{00,00}^{60,0}+\frac{7 \sqrt{7} C_{00,20}^{60,0}}{768}-\frac{1}{128}
\sqrt{\frac{7}{5}} C_{00,22}^{62,0}-\frac{7}{768} \sqrt{\frac{7}{3}} C_{11,00}^{51,0}+\frac{1}{576} \sqrt{\frac{7}{5}} C_{11,22}^{51,0}+\frac{3 \sqrt{3}
C_{11,22}^{53,0}}{1280}-\frac{7}{768} \sqrt{\frac{7}{3}} C_{20,00}^{40,0}-\frac{7 \sqrt{7} C_{20,20}^{40,0}}{2304}+\frac{1}{576} \sqrt{\frac{7}{5}}
C_{20,22}^{42,0}+\frac{7}{768} \sqrt{\frac{7}{15}} C_{22,00}^{42,0}+\frac{7 \sqrt{\frac{7}{5}} C_{22,20}^{42,0}}{2304}+\frac{\sqrt{35} C_{22,22}^{40,0}}{1152}-\frac{7
C_{22,22}^{42,0}}{576 \sqrt{5}}+\frac{7}{768} \sqrt{\frac{7}{3}} C_{31,00}^{31,0}+\frac{7 \sqrt{7} C_{31,20}^{31,0}}{2304}-\frac{1}{576} \sqrt{\frac{7}{5}}
C_{31,22}^{31,0}-\frac{3 \sqrt{3} C_{31,22}^{33,0}}{1280}-\frac{21 C_{33,00}^{33,0}}{1280}-\frac{7 \sqrt{3} C_{33,20}^{33,0}}{1280}+\frac{3 C_{33,22}^{33,0}}{80
\sqrt{10}}\)

\noindent\(C_{22,2011,3}^{22,0011,0,\text{Loc}}= -\frac{7}{768} \sqrt{\frac{7}{3}} C_{00,20}^{60,0}-\frac{7 \sqrt{\frac{7}{3}} C_{00,20}^{60,1}}{1536}-\frac{1}{256}
\sqrt{\frac{7}{15}} C_{00,22}^{62,0}-\frac{1}{512} \sqrt{\frac{7}{15}} C_{00,22}^{62,1}+\frac{\sqrt{\frac{35}{3}} C_{11,22}^{51,0}}{2304}+\frac{\sqrt{\frac{35}{3}}
C_{11,22}^{51,1}}{4608}-\frac{5 \sqrt{\frac{7}{3}} C_{20,20}^{40,0}}{2304}-\frac{5 \sqrt{\frac{7}{3}} C_{20,20}^{40,1}}{4608}-\frac{\sqrt{\frac{7}{15}}
C_{20,22}^{42,0}}{2304}-\frac{\sqrt{\frac{7}{15}} C_{20,22}^{42,1}}{4608}+\frac{7 \sqrt{\frac{7}{15}} C_{22,20}^{42,0}}{1152}+\frac{7 \sqrt{\frac{7}{15}}
C_{22,20}^{42,1}}{2304}-\frac{\sqrt{\frac{7}{15}} C_{22,22}^{40,0}}{2304}-\frac{\sqrt{\frac{7}{15}} C_{22,22}^{40,1}}{4608}+\frac{C_{22,22}^{42,0}}{576
\sqrt{15}}+\frac{C_{22,22}^{42,1}}{1152 \sqrt{15}}+\frac{1}{320} \sqrt{\frac{3}{7}} C_{22,22}^{44,0}+\frac{1}{640} \sqrt{\frac{3}{7}} C_{22,22}^{44,1}-\frac{\sqrt{\frac{7}{3}}
C_{31,20}^{31,0}}{2304}-\frac{\sqrt{\frac{7}{3}} C_{31,20}^{31,1}}{4608}+\frac{\sqrt{\frac{7}{15}} C_{31,22}^{31,0}}{2304}+\frac{\sqrt{\frac{7}{15}}
C_{31,22}^{31,1}}{4608}-\frac{C_{31,22}^{33,0}}{960}-\frac{C_{31,22}^{33,1}}{1920}-\frac{7 C_{33,20}^{33,0}}{640}-\frac{7 C_{33,20}^{33,1}}{1280}-\frac{1}{320}
\sqrt{\frac{3}{10}} C_{33,22}^{33,0}-\frac{1}{640} \sqrt{\frac{3}{10}} C_{33,22}^{33,1}\)

\noindent\(C_{22,2011,3}^{22,0011,0,\text{NonLoc}}= -\frac{7 \sqrt{7} C_{00,00}^{60,0}}{3072}-\frac{7 \sqrt{7} C_{00,00}^{60,1}}{1536}-\frac{7
\sqrt{\frac{7}{3}} C_{00,20}^{60,0}}{3072}-\frac{7 \sqrt{\frac{7}{3}} C_{00,20}^{60,1}}{1536}+\frac{1}{512} \sqrt{\frac{7}{15}} C_{00,22}^{62,0}+\frac{1}{256}
\sqrt{\frac{7}{15}} C_{00,22}^{62,1}-\frac{7 \sqrt{7} C_{11,00}^{51,0}}{9216}-\frac{7 \sqrt{7} C_{11,00}^{51,1}}{4608}+\frac{\sqrt{\frac{35}{3}}
C_{11,22}^{51,0}}{4608}+\frac{\sqrt{\frac{35}{3}} C_{11,22}^{51,1}}{2304}-\frac{5 \sqrt{7} C_{20,00}^{40,0}}{9216}-\frac{5 \sqrt{7} C_{20,00}^{40,1}}{4608}-\frac{5
\sqrt{\frac{7}{3}} C_{20,20}^{40,0}}{9216}-\frac{5 \sqrt{\frac{7}{3}} C_{20,20}^{40,1}}{4608}+\frac{\sqrt{\frac{7}{15}} C_{20,22}^{42,0}}{4608}+\frac{\sqrt{\frac{7}{15}}
C_{20,22}^{42,1}}{2304}+\frac{7 \sqrt{\frac{7}{5}} C_{22,00}^{42,0}}{4608}+\frac{7 \sqrt{\frac{7}{5}} C_{22,00}^{42,1}}{2304}+\frac{7 \sqrt{\frac{7}{15}}
C_{22,20}^{42,0}}{4608}+\frac{7 \sqrt{\frac{7}{15}} C_{22,20}^{42,1}}{2304}+\frac{\sqrt{\frac{7}{15}} C_{22,22}^{40,0}}{4608}+\frac{\sqrt{\frac{7}{15}}
C_{22,22}^{40,1}}{2304}-\frac{C_{22,22}^{42,0}}{1152 \sqrt{15}}-\frac{C_{22,22}^{42,1}}{576 \sqrt{15}}-\frac{1}{640} \sqrt{\frac{3}{7}} C_{22,22}^{44,0}-\frac{1}{320}
\sqrt{\frac{3}{7}} C_{22,22}^{44,1}+\frac{\sqrt{7} C_{31,00}^{31,0}}{9216}+\frac{\sqrt{7} C_{31,00}^{31,1}}{4608}+\frac{\sqrt{\frac{7}{3}} C_{31,20}^{31,0}}{9216}+\frac{\sqrt{\frac{7}{3}}
C_{31,20}^{31,1}}{4608}+\frac{\sqrt{\frac{7}{15}} C_{31,22}^{31,0}}{4608}+\frac{\sqrt{\frac{7}{15}} C_{31,22}^{31,1}}{2304}-\frac{C_{31,22}^{33,0}}{1920}-\frac{C_{31,22}^{33,1}}{960}+\frac{7
\sqrt{3} C_{33,00}^{33,0}}{2560}+\frac{7 \sqrt{3} C_{33,00}^{33,1}}{1280}+\frac{7 C_{33,20}^{33,0}}{2560}+\frac{7 C_{33,20}^{33,1}}{1280}-\frac{1}{640}
\sqrt{\frac{3}{10}} C_{33,22}^{33,0}-\frac{1}{320} \sqrt{\frac{3}{10}} C_{33,22}^{33,1}\)

\noindent\(C_{22,2011,3}^{22,0011,1,\text{Loc}}= -\frac{7 \sqrt{7} C_{00,20}^{60,1}}{1536}-\frac{1}{512} \sqrt{\frac{7}{5}} C_{00,22}^{62,1}+\frac{\sqrt{35}
C_{11,22}^{51,1}}{4608}-\frac{5 \sqrt{7} C_{20,20}^{40,1}}{4608}-\frac{\sqrt{\frac{7}{5}} C_{20,22}^{42,1}}{4608}+\frac{7 \sqrt{\frac{7}{5}} C_{22,20}^{42,1}}{2304}-\frac{\sqrt{\frac{7}{5}}
C_{22,22}^{40,1}}{4608}+\frac{C_{22,22}^{42,1}}{1152 \sqrt{5}}+\frac{3 C_{22,22}^{44,1}}{640 \sqrt{7}}-\frac{\sqrt{7} C_{31,20}^{31,1}}{4608}+\frac{\sqrt{\frac{7}{5}}
C_{31,22}^{31,1}}{4608}-\frac{C_{31,22}^{33,1}}{640 \sqrt{3}}-\frac{7 \sqrt{3} C_{33,20}^{33,1}}{1280}-\frac{3 C_{33,22}^{33,1}}{640 \sqrt{10}}\)

\noindent\(C_{22,2011,3}^{22,0011,1,\text{NonLoc}}= -\frac{7 \sqrt{\frac{7}{3}} C_{00,00}^{60,0}}{1024}-\frac{7 \sqrt{7} C_{00,20}^{60,0}}{3072}+\frac{1}{512}
\sqrt{\frac{7}{5}} C_{00,22}^{62,0}-\frac{7 \sqrt{\frac{7}{3}} C_{11,00}^{51,0}}{3072}+\frac{\sqrt{35} C_{11,22}^{51,0}}{4608}-\frac{5 \sqrt{\frac{7}{3}}
C_{20,00}^{40,0}}{3072}-\frac{5 \sqrt{7} C_{20,20}^{40,0}}{9216}+\frac{\sqrt{\frac{7}{5}} C_{20,22}^{42,0}}{4608}+\frac{7 \sqrt{\frac{7}{15}} C_{22,00}^{42,0}}{1536}+\frac{7
\sqrt{\frac{7}{5}} C_{22,20}^{42,0}}{4608}+\frac{\sqrt{\frac{7}{5}} C_{22,22}^{40,0}}{4608}-\frac{C_{22,22}^{42,0}}{1152 \sqrt{5}}-\frac{3 C_{22,22}^{44,0}}{640
\sqrt{7}}+\frac{\sqrt{\frac{7}{3}} C_{31,00}^{31,0}}{3072}+\frac{\sqrt{7} C_{31,20}^{31,0}}{9216}+\frac{\sqrt{\frac{7}{5}} C_{31,22}^{31,0}}{4608}-\frac{C_{31,22}^{33,0}}{640
\sqrt{3}}+\frac{21 C_{33,00}^{33,0}}{2560}+\frac{7 \sqrt{3} C_{33,20}^{33,0}}{2560}-\frac{3 C_{33,22}^{33,0}}{640 \sqrt{10}}\)

\noindent\(C_{22,2202,0}^{00,2000,0,\text{Loc}}= -\frac{7 \sqrt{5} C_{00,00}^{60,0}}{192}-\frac{7 \sqrt{5} C_{00,00}^{60,1}}{384}-\frac{7
\sqrt{5} C_{11,00}^{51,0}}{576}-\frac{7 \sqrt{5} C_{11,00}^{51,1}}{1152}+\frac{7 \sqrt{5} C_{20,00}^{40,0}}{576}+\frac{7 \sqrt{5} C_{20,00}^{40,1}}{1152}-\frac{7
C_{22,00}^{42,0}}{576}-\frac{7 C_{22,00}^{42,1}}{1152}+\frac{7 \sqrt{5} C_{31,00}^{31,0}}{576}+\frac{7 \sqrt{5} C_{31,00}^{31,1}}{1152}-\frac{1}{64}
\sqrt{\frac{21}{5}} C_{33,00}^{33,0}-\frac{1}{128} \sqrt{\frac{21}{5}} C_{33,00}^{33,1}\)

\noindent\(C_{22,2202,0}^{00,2000,0,\text{NonLoc}}= \frac{7 \sqrt{5} C_{00,00}^{60,0}}{768}+\frac{7 \sqrt{5} C_{00,00}^{60,1}}{384}-\frac{7}{256}
\sqrt{\frac{5}{3}} C_{00,20}^{60,0}-\frac{7}{128} \sqrt{\frac{5}{3}} C_{00,20}^{60,1}-\frac{7 \sqrt{5} C_{11,00}^{51,0}}{2304}-\frac{7 \sqrt{5} C_{11,00}^{51,1}}{1152}+\frac{7}{768}
\sqrt{\frac{5}{3}} C_{11,20}^{31,0}+\frac{7}{384} \sqrt{\frac{5}{3}} C_{11,20}^{31,1}-\frac{7 \sqrt{5} C_{20,00}^{40,0}}{2304}-\frac{7 \sqrt{5} C_{20,00}^{40,1}}{1152}+\frac{7}{768}
\sqrt{\frac{5}{3}} C_{20,20}^{40,0}+\frac{7}{384} \sqrt{\frac{5}{3}} C_{20,20}^{40,1}+\frac{7 C_{22,00}^{42,0}}{2304}+\frac{7 C_{22,00}^{42,1}}{1152}-\frac{7
C_{22,20}^{42,0}}{768 \sqrt{3}}-\frac{7 C_{22,20}^{42,1}}{384 \sqrt{3}}+\frac{7 \sqrt{5} C_{31,00}^{31,0}}{2304}+\frac{7 \sqrt{5} C_{31,00}^{31,1}}{1152}-\frac{7}{768}
\sqrt{\frac{5}{3}} C_{31,20}^{31,0}-\frac{7}{384} \sqrt{\frac{5}{3}} C_{31,20}^{31,1}-\frac{1}{256} \sqrt{\frac{21}{5}} C_{33,00}^{33,0}-\frac{1}{128}
\sqrt{\frac{21}{5}} C_{33,00}^{33,1}+\frac{3}{256} \sqrt{\frac{7}{5}} C_{33,20}^{33,0}+\frac{3}{128} \sqrt{\frac{7}{5}} C_{33,20}^{33,1}\)

\noindent\(C_{22,2202,0}^{00,2000,1,\text{Loc}}= -\frac{7}{128} \sqrt{\frac{5}{3}} C_{00,00}^{60,1}-\frac{7}{384} \sqrt{\frac{5}{3}}
C_{11,00}^{51,1}+\frac{7}{384} \sqrt{\frac{5}{3}} C_{20,00}^{40,1}-\frac{7 C_{22,00}^{42,1}}{384 \sqrt{3}}+\frac{7}{384} \sqrt{\frac{5}{3}} C_{31,00}^{31,1}-\frac{3}{128}
\sqrt{\frac{7}{5}} C_{33,00}^{33,1}\)

\noindent\(C_{22,2202,0}^{00,2000,1,\text{NonLoc}}= \frac{7}{256} \sqrt{\frac{5}{3}} C_{00,00}^{60,0}-\frac{7 \sqrt{5} C_{00,20}^{60,0}}{256}-\frac{7}{768}
\sqrt{\frac{5}{3}} C_{11,00}^{51,0}+\frac{7 \sqrt{5} C_{11,20}^{31,0}}{768}-\frac{7}{768} \sqrt{\frac{5}{3}} C_{20,00}^{40,0}+\frac{7 \sqrt{5} C_{20,20}^{40,0}}{768}+\frac{7
C_{22,00}^{42,0}}{768 \sqrt{3}}-\frac{7 C_{22,20}^{42,0}}{768}+\frac{7}{768} \sqrt{\frac{5}{3}} C_{31,00}^{31,0}-\frac{7 \sqrt{5} C_{31,20}^{31,0}}{768}-\frac{3}{256}
\sqrt{\frac{7}{5}} C_{33,00}^{33,0}+\frac{3}{256} \sqrt{\frac{21}{5}} C_{33,20}^{33,0}\)

\noindent\(C_{22,2202,0}^{11,1111,0,\text{Loc}}= \frac{7 \sqrt{5} C_{11,11}^{51,0}}{576}+\frac{7 \sqrt{5} C_{11,11}^{51,1}}{1152}-\frac{\sqrt{5}
C_{31,11}^{31,0}}{64}-\frac{\sqrt{5} C_{31,11}^{31,1}}{128}+\frac{1}{32} \sqrt{\frac{7}{10}} C_{33,11}^{33,0}+\frac{1}{64} \sqrt{\frac{7}{10}} C_{33,11}^{33,1}\)

\noindent\(C_{22,2202,0}^{11,1111,0,\text{NonLoc}}= \frac{7 \sqrt{5} C_{11,11}^{51,0}}{1152}+\frac{7 \sqrt{5} C_{11,11}^{51,1}}{576}-\frac{\sqrt{5}
C_{31,11}^{31,0}}{128}-\frac{\sqrt{5} C_{31,11}^{31,1}}{64}+\frac{1}{64} \sqrt{\frac{7}{10}} C_{33,11}^{33,0}+\frac{1}{32} \sqrt{\frac{7}{10}} C_{33,11}^{33,1}\)

\noindent\(C_{22,2202,0}^{11,1111,1,\text{Loc}}= \frac{7}{384} \sqrt{\frac{5}{3}} C_{11,11}^{51,1}-\frac{\sqrt{15} C_{31,11}^{31,1}}{128}+\frac{1}{64}
\sqrt{\frac{21}{10}} C_{33,11}^{33,1}\)

\noindent\(C_{22,2202,0}^{11,1111,1,\text{NonLoc}}= \frac{7}{384} \sqrt{\frac{5}{3}} C_{11,11}^{51,0}-\frac{\sqrt{15} C_{31,11}^{31,0}}{128}+\frac{1}{64}
\sqrt{\frac{21}{10}} C_{33,11}^{33,0}\)

\noindent\(C_{22,2202,0}^{20,0000,0,\text{Loc}}= \frac{7 \sqrt{5} C_{00,00}^{60,0}}{768}+\frac{7 \sqrt{5} C_{00,00}^{60,1}}{1536}-\frac{7
\sqrt{5} C_{11,00}^{51,0}}{2304}-\frac{7 \sqrt{5} C_{11,00}^{51,1}}{4608}-\frac{7 \sqrt{5} C_{20,00}^{40,0}}{2304}-\frac{7 \sqrt{5} C_{20,00}^{40,1}}{4608}+\frac{7
C_{22,00}^{42,0}}{2304}+\frac{7 C_{22,00}^{42,1}}{4608}+\frac{7 \sqrt{5} C_{31,00}^{31,0}}{2304}+\frac{7 \sqrt{5} C_{31,00}^{31,1}}{4608}-\frac{1}{256}
\sqrt{\frac{21}{5}} C_{33,00}^{33,0}-\frac{1}{512} \sqrt{\frac{21}{5}} C_{33,00}^{33,1}\)

\noindent\(C_{22,2202,0}^{20,0000,0,\text{NonLoc}}= -\frac{7 \sqrt{5} C_{00,00}^{60,0}}{3072}-\frac{7 \sqrt{5} C_{00,00}^{60,1}}{1536}+\frac{7
\sqrt{\frac{5}{3}} C_{00,20}^{60,0}}{1024}+\frac{7}{512} \sqrt{\frac{5}{3}} C_{00,20}^{60,1}-\frac{7 \sqrt{5} C_{11,00}^{51,0}}{9216}-\frac{7 \sqrt{5}
C_{11,00}^{51,1}}{4608}+\frac{7 \sqrt{\frac{5}{3}} C_{11,20}^{31,0}}{3072}+\frac{7 \sqrt{\frac{5}{3}} C_{11,20}^{31,1}}{1536}+\frac{7 \sqrt{5} C_{20,00}^{40,0}}{9216}+\frac{7
\sqrt{5} C_{20,00}^{40,1}}{4608}-\frac{7 \sqrt{\frac{5}{3}} C_{20,20}^{40,0}}{3072}-\frac{7 \sqrt{\frac{5}{3}} C_{20,20}^{40,1}}{1536}-\frac{7 C_{22,00}^{42,0}}{9216}-\frac{7
C_{22,00}^{42,1}}{4608}+\frac{7 C_{22,20}^{42,0}}{3072 \sqrt{3}}+\frac{7 C_{22,20}^{42,1}}{1536 \sqrt{3}}+\frac{7 \sqrt{5} C_{31,00}^{31,0}}{9216}+\frac{7
\sqrt{5} C_{31,00}^{31,1}}{4608}-\frac{7 \sqrt{\frac{5}{3}} C_{31,20}^{31,0}}{3072}-\frac{7 \sqrt{\frac{5}{3}} C_{31,20}^{31,1}}{1536}-\frac{\sqrt{\frac{21}{5}}
C_{33,00}^{33,0}}{1024}-\frac{1}{512} \sqrt{\frac{21}{5}} C_{33,00}^{33,1}+\frac{3 \sqrt{\frac{7}{5}} C_{33,20}^{33,0}}{1024}+\frac{3}{512} \sqrt{\frac{7}{5}}
C_{33,20}^{33,1}\)

\noindent\(C_{22,2202,0}^{20,0000,1,\text{Loc}}= \frac{7}{512} \sqrt{\frac{5}{3}} C_{00,00}^{60,1}-\frac{7 \sqrt{\frac{5}{3}} C_{11,00}^{51,1}}{1536}-\frac{7
\sqrt{\frac{5}{3}} C_{20,00}^{40,1}}{1536}+\frac{7 C_{22,00}^{42,1}}{1536 \sqrt{3}}+\frac{7 \sqrt{\frac{5}{3}} C_{31,00}^{31,1}}{1536}-\frac{3}{512}
\sqrt{\frac{7}{5}} C_{33,00}^{33,1}\)

\noindent\(C_{22,2202,0}^{20,0000,1,\text{NonLoc}}= -\frac{7 \sqrt{\frac{5}{3}} C_{00,00}^{60,0}}{1024}+\frac{7 \sqrt{5} C_{00,20}^{60,0}}{1024}-\frac{7
\sqrt{\frac{5}{3}} C_{11,00}^{51,0}}{3072}+\frac{7 \sqrt{5} C_{11,20}^{31,0}}{3072}+\frac{7 \sqrt{\frac{5}{3}} C_{20,00}^{40,0}}{3072}-\frac{7 \sqrt{5}
C_{20,20}^{40,0}}{3072}-\frac{7 C_{22,00}^{42,0}}{3072 \sqrt{3}}+\frac{7 C_{22,20}^{42,0}}{3072}+\frac{7 \sqrt{\frac{5}{3}} C_{31,00}^{31,0}}{3072}-\frac{7
\sqrt{5} C_{31,20}^{31,0}}{3072}-\frac{3 \sqrt{\frac{7}{5}} C_{33,00}^{33,0}}{1024}+\frac{3 \sqrt{\frac{21}{5}} C_{33,20}^{33,0}}{1024}\)

\noindent\(C_{22,2202,1}^{11,1111,0,\text{Loc}}= -\frac{7 \sqrt{5} C_{11,11}^{51,0}}{384}-\frac{7 \sqrt{5} C_{11,11}^{51,1}}{768}+\frac{5
\sqrt{5} C_{31,11}^{31,0}}{384}+\frac{5 \sqrt{5} C_{31,11}^{31,1}}{768}\)

\noindent\(C_{22,2202,1}^{11,1111,0,\text{NonLoc}}= -\frac{7 \sqrt{5} C_{11,11}^{51,0}}{768}-\frac{7 \sqrt{5} C_{11,11}^{51,1}}{384}+\frac{5
\sqrt{5} C_{31,11}^{31,0}}{768}+\frac{5 \sqrt{5} C_{31,11}^{31,1}}{384}\)

\noindent\(C_{22,2202,1}^{11,1111,1,\text{Loc}}= -\frac{7}{256} \sqrt{\frac{5}{3}} C_{11,11}^{51,1}+\frac{5}{256} \sqrt{\frac{5}{3}}
C_{31,11}^{31,1}\)

\noindent\(C_{22,2202,1}^{11,1111,1,\text{NonLoc}}= -\frac{7}{256} \sqrt{\frac{5}{3}} C_{11,11}^{51,0}+\frac{5}{256} \sqrt{\frac{5}{3}}
C_{31,11}^{31,0}\)

\noindent\(C_{22,2202,1}^{11,1112,0,\text{Loc}}= \frac{7 C_{11,11}^{51,0}}{128 \sqrt{3}}+\frac{7 C_{11,11}^{51,1}}{256 \sqrt{3}}-\frac{5
C_{31,11}^{31,0}}{128 \sqrt{3}}-\frac{5 C_{31,11}^{31,1}}{256 \sqrt{3}}\)

\noindent\(C_{22,2202,1}^{11,1112,0,\text{NonLoc}}= \frac{7 C_{11,11}^{51,0}}{256 \sqrt{3}}+\frac{7 C_{11,11}^{51,1}}{128 \sqrt{3}}-\frac{5
C_{31,11}^{31,0}}{256 \sqrt{3}}-\frac{5 C_{31,11}^{31,1}}{128 \sqrt{3}}\)

\noindent\(C_{22,2202,1}^{11,1112,1,\text{Loc}}= \frac{7 C_{11,11}^{51,1}}{256}-\frac{5 C_{31,11}^{31,1}}{256}\)

\noindent\(C_{22,2202,1}^{11,1112,1,\text{NonLoc}}= \frac{7 C_{11,11}^{51,0}}{256}-\frac{5 C_{31,11}^{31,0}}{256}\)

\noindent\(C_{22,2202,2}^{00,2202,0,\text{Loc}}= -\frac{\sqrt{35} C_{00,00}^{60,0}}{96}-\frac{\sqrt{35} C_{00,00}^{60,1}}{192}-\frac{\sqrt{35}
C_{11,00}^{51,0}}{288}-\frac{\sqrt{35} C_{11,00}^{51,1}}{576}+\frac{\sqrt{35} C_{20,00}^{40,0}}{288}+\frac{\sqrt{35} C_{20,00}^{40,1}}{576}-\frac{\sqrt{7}
C_{22,00}^{42,0}}{288}-\frac{\sqrt{7} C_{22,00}^{42,1}}{576}+\frac{\sqrt{35} C_{31,00}^{31,0}}{288}+\frac{\sqrt{35} C_{31,00}^{31,1}}{576}-\frac{1}{32}
\sqrt{\frac{3}{5}} C_{33,00}^{33,0}-\frac{1}{64} \sqrt{\frac{3}{5}} C_{33,00}^{33,1}\)

\noindent\(C_{22,2202,2}^{00,2202,0,\text{NonLoc}}= \frac{\sqrt{35} C_{00,00}^{60,0}}{384}+\frac{\sqrt{35} C_{00,00}^{60,1}}{192}-\frac{1}{128}
\sqrt{\frac{35}{3}} C_{00,20}^{60,0}-\frac{1}{64} \sqrt{\frac{35}{3}} C_{00,20}^{60,1}-\frac{\sqrt{35} C_{11,00}^{51,0}}{1152}-\frac{\sqrt{35} C_{11,00}^{51,1}}{576}-\frac{\sqrt{35}
C_{20,00}^{40,0}}{1152}-\frac{\sqrt{35} C_{20,00}^{40,1}}{576}+\frac{1}{384} \sqrt{\frac{35}{3}} C_{20,20}^{40,0}+\frac{1}{192} \sqrt{\frac{35}{3}}
C_{20,20}^{40,1}+\frac{\sqrt{7} C_{22,00}^{42,0}}{1152}+\frac{\sqrt{7} C_{22,00}^{42,1}}{576}-\frac{1}{384} \sqrt{\frac{7}{3}} C_{22,20}^{42,0}-\frac{1}{192}
\sqrt{\frac{7}{3}} C_{22,20}^{42,1}+\frac{\sqrt{35} C_{31,00}^{31,0}}{1152}+\frac{\sqrt{35} C_{31,00}^{31,1}}{576}-\frac{1}{384} \sqrt{\frac{35}{3}}
C_{31,20}^{31,0}-\frac{1}{192} \sqrt{\frac{35}{3}} C_{31,20}^{31,1}-\frac{1}{128} \sqrt{\frac{3}{5}} C_{33,00}^{33,0}-\frac{1}{64} \sqrt{\frac{3}{5}}
C_{33,00}^{33,1}+\frac{3 C_{33,20}^{33,0}}{128 \sqrt{5}}+\frac{3 C_{33,20}^{33,1}}{64 \sqrt{5}}\)

\noindent\(C_{22,2202,2}^{00,2202,1,\text{Loc}}= -\frac{1}{64} \sqrt{\frac{35}{3}} C_{00,00}^{60,1}-\frac{1}{192} \sqrt{\frac{35}{3}}
C_{11,00}^{51,1}+\frac{1}{192} \sqrt{\frac{35}{3}} C_{20,00}^{40,1}-\frac{1}{192} \sqrt{\frac{7}{3}} C_{22,00}^{42,1}+\frac{1}{192} \sqrt{\frac{35}{3}}
C_{31,00}^{31,1}-\frac{3 C_{33,00}^{33,1}}{64 \sqrt{5}}\)

\noindent\(C_{22,2202,2}^{00,2202,1,\text{NonLoc}}= \frac{1}{128} \sqrt{\frac{35}{3}} C_{00,00}^{60,0}-\frac{\sqrt{35} C_{00,20}^{60,0}}{128}-\frac{1}{384}
\sqrt{\frac{35}{3}} C_{11,00}^{51,0}-\frac{1}{384} \sqrt{\frac{35}{3}} C_{20,00}^{40,0}+\frac{\sqrt{35} C_{20,20}^{40,0}}{384}+\frac{1}{384} \sqrt{\frac{7}{3}}
C_{22,00}^{42,0}-\frac{\sqrt{7} C_{22,20}^{42,0}}{384}+\frac{1}{384} \sqrt{\frac{35}{3}} C_{31,00}^{31,0}-\frac{\sqrt{35} C_{31,20}^{31,0}}{384}-\frac{3
C_{33,00}^{33,0}}{128 \sqrt{5}}+\frac{3}{128} \sqrt{\frac{3}{5}} C_{33,20}^{33,0}\)

\noindent\(C_{22,2202,2}^{11,1111,0,\text{Loc}}= \frac{\sqrt{35} C_{11,11}^{51,0}}{1152}+\frac{\sqrt{35} C_{11,11}^{51,1}}{2304}-\frac{\sqrt{35}
C_{31,11}^{31,0}}{384}-\frac{\sqrt{35} C_{31,11}^{31,1}}{768}+\frac{C_{33,11}^{33,0}}{16 \sqrt{10}}+\frac{C_{33,11}^{33,1}}{32 \sqrt{10}}\)

\noindent\(C_{22,2202,2}^{11,1111,0,\text{NonLoc}}= \frac{\sqrt{35} C_{11,11}^{51,0}}{2304}+\frac{\sqrt{35} C_{11,11}^{51,1}}{1152}-\frac{\sqrt{35}
C_{31,11}^{31,0}}{768}-\frac{\sqrt{35} C_{31,11}^{31,1}}{384}+\frac{C_{33,11}^{33,0}}{32 \sqrt{10}}+\frac{C_{33,11}^{33,1}}{16 \sqrt{10}}\)

\noindent\(C_{22,2202,2}^{11,1111,1,\text{Loc}}= \frac{1}{768} \sqrt{\frac{35}{3}} C_{11,11}^{51,1}-\frac{1}{256} \sqrt{\frac{35}{3}}
C_{31,11}^{31,1}+\frac{1}{32} \sqrt{\frac{3}{10}} C_{33,11}^{33,1}\)

\noindent\(C_{22,2202,2}^{11,1111,1,\text{NonLoc}}= \frac{1}{768} \sqrt{\frac{35}{3}} C_{11,11}^{51,0}-\frac{1}{256} \sqrt{\frac{35}{3}}
C_{31,11}^{31,0}+\frac{1}{32} \sqrt{\frac{3}{10}} C_{33,11}^{33,0}\)

\noindent\(C_{22,2202,2}^{11,1112,0,\text{Loc}}= -\frac{1}{384} \sqrt{\frac{35}{3}} C_{11,11}^{51,0}-\frac{1}{768} \sqrt{\frac{35}{3}}
C_{11,11}^{51,1}+\frac{1}{128} \sqrt{\frac{35}{3}} C_{31,11}^{31,0}+\frac{1}{256} \sqrt{\frac{35}{3}} C_{31,11}^{31,1}-\frac{1}{16} \sqrt{\frac{3}{10}}
C_{33,11}^{33,0}-\frac{1}{32} \sqrt{\frac{3}{10}} C_{33,11}^{33,1}\)

\noindent\(C_{22,2202,2}^{11,1112,0,\text{NonLoc}}= -\frac{1}{768} \sqrt{\frac{35}{3}} C_{11,11}^{51,0}-\frac{1}{384} \sqrt{\frac{35}{3}}
C_{11,11}^{51,1}+\frac{1}{256} \sqrt{\frac{35}{3}} C_{31,11}^{31,0}+\frac{1}{128} \sqrt{\frac{35}{3}} C_{31,11}^{31,1}-\frac{1}{32} \sqrt{\frac{3}{10}}
C_{33,11}^{33,0}-\frac{1}{16} \sqrt{\frac{3}{10}} C_{33,11}^{33,1}\)

\noindent\(C_{22,2202,2}^{11,1112,1,\text{Loc}}= -\frac{\sqrt{35} C_{11,11}^{51,1}}{768}+\frac{\sqrt{35} C_{31,11}^{31,1}}{256}-\frac{3
C_{33,11}^{33,1}}{32 \sqrt{10}}\)

\noindent\(C_{22,2202,2}^{11,1112,1,\text{NonLoc}}= -\frac{\sqrt{35} C_{11,11}^{51,0}}{768}+\frac{\sqrt{35} C_{31,11}^{31,0}}{256}-\frac{3
C_{33,11}^{33,0}}{32 \sqrt{10}}\)

\noindent\(C_{22,2202,2}^{22,0000,0,\text{Loc}}= \frac{\sqrt{35} C_{00,00}^{60,0}}{384}+\frac{\sqrt{35} C_{00,00}^{60,1}}{768}-\frac{\sqrt{35}
C_{11,00}^{51,0}}{1152}-\frac{\sqrt{35} C_{11,00}^{51,1}}{2304}-\frac{\sqrt{35} C_{20,00}^{40,0}}{1152}-\frac{\sqrt{35} C_{20,00}^{40,1}}{2304}+\frac{\sqrt{7}
C_{22,00}^{42,0}}{1152}+\frac{\sqrt{7} C_{22,00}^{42,1}}{2304}+\frac{\sqrt{35} C_{31,00}^{31,0}}{1152}+\frac{\sqrt{35} C_{31,00}^{31,1}}{2304}-\frac{1}{128}
\sqrt{\frac{3}{5}} C_{33,00}^{33,0}-\frac{1}{256} \sqrt{\frac{3}{5}} C_{33,00}^{33,1}\)

\noindent\(C_{22,2202,2}^{22,0000,0,\text{NonLoc}}= -\frac{\sqrt{35} C_{00,00}^{60,0}}{1536}-\frac{\sqrt{35} C_{00,00}^{60,1}}{768}+\frac{1}{512}
\sqrt{\frac{35}{3}} C_{00,20}^{60,0}+\frac{1}{256} \sqrt{\frac{35}{3}} C_{00,20}^{60,1}-\frac{\sqrt{35} C_{11,00}^{51,0}}{4608}-\frac{\sqrt{35} C_{11,00}^{51,1}}{2304}+\frac{\sqrt{35}
C_{20,00}^{40,0}}{4608}+\frac{\sqrt{35} C_{20,00}^{40,1}}{2304}-\frac{\sqrt{\frac{35}{3}} C_{20,20}^{40,0}}{1536}-\frac{1}{768} \sqrt{\frac{35}{3}}
C_{20,20}^{40,1}-\frac{\sqrt{7} C_{22,00}^{42,0}}{4608}-\frac{\sqrt{7} C_{22,00}^{42,1}}{2304}+\frac{\sqrt{\frac{7}{3}} C_{22,20}^{42,0}}{1536}+\frac{1}{768}
\sqrt{\frac{7}{3}} C_{22,20}^{42,1}+\frac{\sqrt{35} C_{31,00}^{31,0}}{4608}+\frac{\sqrt{35} C_{31,00}^{31,1}}{2304}-\frac{\sqrt{\frac{35}{3}} C_{31,20}^{31,0}}{1536}-\frac{1}{768}
\sqrt{\frac{35}{3}} C_{31,20}^{31,1}-\frac{1}{512} \sqrt{\frac{3}{5}} C_{33,00}^{33,0}-\frac{1}{256} \sqrt{\frac{3}{5}} C_{33,00}^{33,1}+\frac{3
C_{33,20}^{33,0}}{512 \sqrt{5}}+\frac{3 C_{33,20}^{33,1}}{256 \sqrt{5}}\)

\noindent\(C_{22,2202,2}^{22,0000,1,\text{Loc}}= \frac{1}{256} \sqrt{\frac{35}{3}} C_{00,00}^{60,1}-\frac{1}{768} \sqrt{\frac{35}{3}}
C_{11,00}^{51,1}-\frac{1}{768} \sqrt{\frac{35}{3}} C_{20,00}^{40,1}+\frac{1}{768} \sqrt{\frac{7}{3}} C_{22,00}^{42,1}+\frac{1}{768} \sqrt{\frac{35}{3}}
C_{31,00}^{31,1}-\frac{3 C_{33,00}^{33,1}}{256 \sqrt{5}}\)

\noindent\(C_{22,2202,2}^{22,0000,1,\text{NonLoc}}= -\frac{1}{512} \sqrt{\frac{35}{3}} C_{00,00}^{60,0}+\frac{\sqrt{35} C_{00,20}^{60,0}}{512}-\frac{\sqrt{\frac{35}{3}}
C_{11,00}^{51,0}}{1536}+\frac{\sqrt{\frac{35}{3}} C_{20,00}^{40,0}}{1536}-\frac{\sqrt{35} C_{20,20}^{40,0}}{1536}-\frac{\sqrt{\frac{7}{3}} C_{22,00}^{42,0}}{1536}+\frac{\sqrt{7}
C_{22,20}^{42,0}}{1536}+\frac{\sqrt{\frac{35}{3}} C_{31,00}^{31,0}}{1536}-\frac{\sqrt{35} C_{31,20}^{31,0}}{1536}-\frac{3 C_{33,00}^{33,0}}{512 \sqrt{5}}+\frac{3}{512}
\sqrt{\frac{3}{5}} C_{33,20}^{33,0}\)

\noindent\(C_{22,2202,3}^{11,1112,0,\text{Loc}}= \frac{1}{48} \sqrt{\frac{7}{6}} C_{11,11}^{51,0}+\frac{1}{96} \sqrt{\frac{7}{6}} C_{11,11}^{51,1}-\frac{\sqrt{3}
C_{33,11}^{33,0}}{64}-\frac{\sqrt{3} C_{33,11}^{33,1}}{128}\)

\noindent\(C_{22,2202,3}^{11,1112,0,\text{NonLoc}}= \frac{1}{96} \sqrt{\frac{7}{6}} C_{11,11}^{51,0}+\frac{1}{48} \sqrt{\frac{7}{6}}
C_{11,11}^{51,1}-\frac{\sqrt{3} C_{33,11}^{33,0}}{128}-\frac{\sqrt{3} C_{33,11}^{33,1}}{64}\)

\noindent\(C_{22,2202,3}^{11,1112,1,\text{Loc}}= \frac{1}{96} \sqrt{\frac{7}{2}} C_{11,11}^{51,1}-\frac{3 C_{33,11}^{33,1}}{128}\)

\noindent\(C_{22,2202,3}^{11,1112,1,\text{NonLoc}}= \frac{1}{96} \sqrt{\frac{7}{2}} C_{11,11}^{51,0}-\frac{3 C_{33,11}^{33,0}}{128}\)

\noindent\(C_{22,2211,1}^{00,2011,0,\text{Loc}}= \frac{7 C_{00,20}^{60,0}}{192}+\frac{7 C_{00,20}^{60,1}}{384}+\frac{7 C_{00,22}^{62,0}}{128
\sqrt{5}}+\frac{7 C_{00,22}^{62,1}}{256 \sqrt{5}}+\frac{7 C_{11,20}^{31,0}}{576}+\frac{7 C_{11,20}^{31,1}}{1152}+\frac{7 C_{11,22}^{51,0}}{576 \sqrt{5}}+\frac{7
C_{11,22}^{51,1}}{1152 \sqrt{5}}+\frac{3 \sqrt{21} C_{11,22}^{53,0}}{1280}+\frac{3 \sqrt{21} C_{11,22}^{53,1}}{2560}-\frac{7 C_{20,20}^{40,0}}{576}-\frac{7
C_{20,20}^{40,1}}{1152}-\frac{7 C_{20,22}^{42,0}}{576 \sqrt{5}}-\frac{7 C_{20,22}^{42,1}}{1152 \sqrt{5}}+\frac{7 C_{22,20}^{42,0}}{576 \sqrt{5}}+\frac{7
C_{22,20}^{42,1}}{1152 \sqrt{5}}-\frac{7 \sqrt{5} C_{22,22}^{40,0}}{1152}-\frac{7 \sqrt{5} C_{22,22}^{40,1}}{2304}+\frac{7}{576} \sqrt{\frac{7}{5}}
C_{22,22}^{42,0}+\frac{7 \sqrt{\frac{7}{5}} C_{22,22}^{42,1}}{1152}-\frac{7 C_{31,20}^{31,0}}{576}-\frac{7 C_{31,20}^{31,1}}{1152}-\frac{7 C_{31,22}^{31,0}}{576
\sqrt{5}}-\frac{7 C_{31,22}^{31,1}}{1152 \sqrt{5}}-\frac{3 \sqrt{21} C_{31,22}^{33,0}}{1280}-\frac{3 \sqrt{21} C_{31,22}^{33,1}}{2560}+\frac{\sqrt{21}
C_{33,20}^{33,0}}{320}+\frac{\sqrt{21} C_{33,20}^{33,1}}{640}+\frac{3}{80} \sqrt{\frac{7}{10}} C_{33,22}^{33,0}+\frac{3}{160} \sqrt{\frac{7}{10}}
C_{33,22}^{33,1}\)

\noindent\(C_{22,2211,1}^{00,2011,0,\text{NonLoc}}= \frac{7 C_{00,00}^{60,0}}{256 \sqrt{3}}+\frac{7 C_{00,00}^{60,1}}{128 \sqrt{3}}+\frac{7
C_{00,20}^{60,0}}{768}+\frac{7 C_{00,20}^{60,1}}{384}-\frac{7 C_{00,22}^{62,0}}{256 \sqrt{5}}-\frac{7 C_{00,22}^{62,1}}{128 \sqrt{5}}-\frac{7 C_{11,00}^{51,0}}{768
\sqrt{3}}-\frac{7 C_{11,00}^{51,1}}{384 \sqrt{3}}-\frac{7 C_{11,20}^{31,0}}{2304}-\frac{7 C_{11,20}^{31,1}}{1152}+\frac{7 C_{11,22}^{51,0}}{1152
\sqrt{5}}+\frac{7 C_{11,22}^{51,1}}{576 \sqrt{5}}+\frac{3 \sqrt{21} C_{11,22}^{53,0}}{2560}+\frac{3 \sqrt{21} C_{11,22}^{53,1}}{1280}-\frac{7 C_{20,00}^{40,0}}{768
\sqrt{3}}-\frac{7 C_{20,00}^{40,1}}{384 \sqrt{3}}-\frac{7 C_{20,20}^{40,0}}{2304}-\frac{7 C_{20,20}^{40,1}}{1152}+\frac{7 C_{20,22}^{42,0}}{1152
\sqrt{5}}+\frac{7 C_{20,22}^{42,1}}{576 \sqrt{5}}+\frac{7 C_{22,00}^{42,0}}{768 \sqrt{15}}+\frac{7 C_{22,00}^{42,1}}{384 \sqrt{15}}+\frac{7 C_{22,20}^{42,0}}{2304
\sqrt{5}}+\frac{7 C_{22,20}^{42,1}}{1152 \sqrt{5}}+\frac{7 \sqrt{5} C_{22,22}^{40,0}}{2304}+\frac{7 \sqrt{5} C_{22,22}^{40,1}}{1152}-\frac{7 \sqrt{\frac{7}{5}}
C_{22,22}^{42,0}}{1152}-\frac{7}{576} \sqrt{\frac{7}{5}} C_{22,22}^{42,1}+\frac{7 C_{31,00}^{31,0}}{768 \sqrt{3}}+\frac{7 C_{31,00}^{31,1}}{384 \sqrt{3}}+\frac{7
C_{31,20}^{31,0}}{2304}+\frac{7 C_{31,20}^{31,1}}{1152}-\frac{7 C_{31,22}^{31,0}}{1152 \sqrt{5}}-\frac{7 C_{31,22}^{31,1}}{576 \sqrt{5}}-\frac{3
\sqrt{21} C_{31,22}^{33,0}}{2560}-\frac{3 \sqrt{21} C_{31,22}^{33,1}}{1280}-\frac{3 \sqrt{7} C_{33,00}^{33,0}}{1280}-\frac{3 \sqrt{7} C_{33,00}^{33,1}}{640}-\frac{\sqrt{21}
C_{33,20}^{33,0}}{1280}-\frac{\sqrt{21} C_{33,20}^{33,1}}{640}+\frac{3}{160} \sqrt{\frac{7}{10}} C_{33,22}^{33,0}+\frac{3}{80} \sqrt{\frac{7}{10}}
C_{33,22}^{33,1}\)

\noindent\(C_{22,2211,1}^{00,2011,1,\text{Loc}}= \frac{7 C_{00,20}^{60,1}}{128 \sqrt{3}}+\frac{7}{256} \sqrt{\frac{3}{5}} C_{00,22}^{62,1}+\frac{7
C_{11,20}^{31,1}}{384 \sqrt{3}}+\frac{7 C_{11,22}^{51,1}}{384 \sqrt{15}}+\frac{9 \sqrt{7} C_{11,22}^{53,1}}{2560}-\frac{7 C_{20,20}^{40,1}}{384 \sqrt{3}}-\frac{7
C_{20,22}^{42,1}}{384 \sqrt{15}}+\frac{7 C_{22,20}^{42,1}}{384 \sqrt{15}}-\frac{7}{768} \sqrt{\frac{5}{3}} C_{22,22}^{40,1}+\frac{7}{384} \sqrt{\frac{7}{15}}
C_{22,22}^{42,1}-\frac{7 C_{31,20}^{31,1}}{384 \sqrt{3}}-\frac{7 C_{31,22}^{31,1}}{384 \sqrt{15}}-\frac{9 \sqrt{7} C_{31,22}^{33,1}}{2560}+\frac{3
\sqrt{7} C_{33,20}^{33,1}}{640}+\frac{3}{160} \sqrt{\frac{21}{10}} C_{33,22}^{33,1}\)

\noindent\(C_{22,2211,1}^{00,2011,1,\text{NonLoc}}= \frac{7 C_{00,00}^{60,0}}{256}+\frac{7 C_{00,20}^{60,0}}{256 \sqrt{3}}-\frac{7}{256}
\sqrt{\frac{3}{5}} C_{00,22}^{62,0}-\frac{7 C_{11,00}^{51,0}}{768}-\frac{7 C_{11,20}^{31,0}}{768 \sqrt{3}}+\frac{7 C_{11,22}^{51,0}}{384 \sqrt{15}}+\frac{9
\sqrt{7} C_{11,22}^{53,0}}{2560}-\frac{7 C_{20,00}^{40,0}}{768}-\frac{7 C_{20,20}^{40,0}}{768 \sqrt{3}}+\frac{7 C_{20,22}^{42,0}}{384 \sqrt{15}}+\frac{7
C_{22,00}^{42,0}}{768 \sqrt{5}}+\frac{7 C_{22,20}^{42,0}}{768 \sqrt{15}}+\frac{7}{768} \sqrt{\frac{5}{3}} C_{22,22}^{40,0}-\frac{7}{384} \sqrt{\frac{7}{15}}
C_{22,22}^{42,0}+\frac{7 C_{31,00}^{31,0}}{768}+\frac{7 C_{31,20}^{31,0}}{768 \sqrt{3}}-\frac{7 C_{31,22}^{31,0}}{384 \sqrt{15}}-\frac{9 \sqrt{7}
C_{31,22}^{33,0}}{2560}-\frac{3 \sqrt{21} C_{33,00}^{33,0}}{1280}-\frac{3 \sqrt{7} C_{33,20}^{33,0}}{1280}+\frac{3}{160} \sqrt{\frac{21}{10}} C_{33,22}^{33,0}\)

\noindent\(C_{22,2211,1}^{00,2211,0,\text{Loc}}= \frac{7 C_{00,20}^{60,0}}{192 \sqrt{5}}+\frac{7 C_{00,20}^{60,1}}{384 \sqrt{5}}+\frac{21
C_{00,22}^{62,0}}{640}+\frac{21 C_{00,22}^{62,1}}{1280}+\frac{21 C_{11,22}^{51,0}}{640}+\frac{21 C_{11,22}^{51,1}}{1280}-\frac{3}{320} \sqrt{\frac{21}{5}}
C_{11,22}^{53,0}-\frac{3}{640} \sqrt{\frac{21}{5}} C_{11,22}^{53,1}-\frac{7 C_{20,20}^{40,0}}{576 \sqrt{5}}-\frac{7 C_{20,20}^{40,1}}{1152 \sqrt{5}}-\frac{21
C_{20,22}^{42,0}}{640}-\frac{21 C_{20,22}^{42,1}}{1280}+\frac{7 C_{22,20}^{42,0}}{2880}+\frac{7 C_{22,20}^{42,1}}{5760}+\frac{67 C_{22,22}^{40,0}}{1920}+\frac{67
C_{22,22}^{40,1}}{3840}-\frac{\sqrt{7} C_{22,22}^{42,0}}{960}-\frac{\sqrt{7} C_{22,22}^{42,1}}{1920}+\frac{3 C_{22,22}^{44,0}}{80 \sqrt{5}}+\frac{3
C_{22,22}^{44,1}}{160 \sqrt{5}}-\frac{7 C_{31,20}^{31,0}}{576 \sqrt{5}}-\frac{7 C_{31,20}^{31,1}}{1152 \sqrt{5}}-\frac{13 C_{31,22}^{31,0}}{384}-\frac{13
C_{31,22}^{31,1}}{768}+\frac{1}{32} \sqrt{\frac{7}{15}} C_{31,22}^{33,0}+\frac{1}{64} \sqrt{\frac{7}{15}} C_{31,22}^{33,1}+\frac{1}{320} \sqrt{\frac{21}{5}}
C_{33,20}^{33,0}+\frac{1}{640} \sqrt{\frac{21}{5}} C_{33,20}^{33,1}\)

\noindent\(C_{22,2211,1}^{00,2211,0,\text{NonLoc}}= \frac{7 C_{00,00}^{60,0}}{256 \sqrt{15}}+\frac{7 C_{00,00}^{60,1}}{128 \sqrt{15}}+\frac{7
C_{00,20}^{60,0}}{768 \sqrt{5}}+\frac{7 C_{00,20}^{60,1}}{384 \sqrt{5}}-\frac{21 C_{00,22}^{62,0}}{1280}-\frac{21 C_{00,22}^{62,1}}{640}-\frac{7
C_{11,00}^{51,0}}{768 \sqrt{15}}-\frac{7 C_{11,00}^{51,1}}{384 \sqrt{15}}+\frac{21 C_{11,22}^{51,0}}{1280}+\frac{21 C_{11,22}^{51,1}}{640}-\frac{3}{640}
\sqrt{\frac{21}{5}} C_{11,22}^{53,0}-\frac{3}{320} \sqrt{\frac{21}{5}} C_{11,22}^{53,1}-\frac{7 C_{20,00}^{40,0}}{768 \sqrt{15}}-\frac{7 C_{20,00}^{40,1}}{384
\sqrt{15}}-\frac{7 C_{20,20}^{40,0}}{2304 \sqrt{5}}-\frac{7 C_{20,20}^{40,1}}{1152 \sqrt{5}}+\frac{21 C_{20,22}^{42,0}}{1280}+\frac{21 C_{20,22}^{42,1}}{640}+\frac{7
C_{22,00}^{42,0}}{3840 \sqrt{3}}+\frac{7 C_{22,00}^{42,1}}{1920 \sqrt{3}}+\frac{7 C_{22,20}^{42,0}}{11520}+\frac{7 C_{22,20}^{42,1}}{5760}-\frac{67
C_{22,22}^{40,0}}{3840}-\frac{67 C_{22,22}^{40,1}}{1920}+\frac{\sqrt{7} C_{22,22}^{42,0}}{1920}+\frac{\sqrt{7} C_{22,22}^{42,1}}{960}-\frac{3 C_{22,22}^{44,0}}{160
\sqrt{5}}-\frac{3 C_{22,22}^{44,1}}{80 \sqrt{5}}+\frac{7 C_{31,00}^{31,0}}{768 \sqrt{15}}+\frac{7 C_{31,00}^{31,1}}{384 \sqrt{15}}+\frac{7 C_{31,20}^{31,0}}{2304
\sqrt{5}}+\frac{7 C_{31,20}^{31,1}}{1152 \sqrt{5}}-\frac{13 C_{31,22}^{31,0}}{768}-\frac{13 C_{31,22}^{31,1}}{384}+\frac{1}{64} \sqrt{\frac{7}{15}}
C_{31,22}^{33,0}+\frac{1}{32} \sqrt{\frac{7}{15}} C_{31,22}^{33,1}-\frac{3 \sqrt{\frac{7}{5}} C_{33,00}^{33,0}}{1280}-\frac{3}{640} \sqrt{\frac{7}{5}}
C_{33,00}^{33,1}-\frac{\sqrt{\frac{21}{5}} C_{33,20}^{33,0}}{1280}-\frac{1}{640} \sqrt{\frac{21}{5}} C_{33,20}^{33,1}\)

\noindent\(C_{22,2211,1}^{00,2211,1,\text{Loc}}= \frac{7 C_{00,20}^{60,1}}{128 \sqrt{15}}+\frac{21 \sqrt{3} C_{00,22}^{62,1}}{1280}+\frac{21
\sqrt{3} C_{11,22}^{51,1}}{1280}-\frac{9}{640} \sqrt{\frac{7}{5}} C_{11,22}^{53,1}-\frac{7 C_{20,20}^{40,1}}{384 \sqrt{15}}-\frac{21 \sqrt{3} C_{20,22}^{42,1}}{1280}+\frac{7
C_{22,20}^{42,1}}{1920 \sqrt{3}}+\frac{67 C_{22,22}^{40,1}}{1280 \sqrt{3}}-\frac{1}{640} \sqrt{\frac{7}{3}} C_{22,22}^{42,1}+\frac{3}{160} \sqrt{\frac{3}{5}}
C_{22,22}^{44,1}-\frac{7 C_{31,20}^{31,1}}{384 \sqrt{15}}-\frac{13 C_{31,22}^{31,1}}{256 \sqrt{3}}+\frac{1}{64} \sqrt{\frac{7}{5}} C_{31,22}^{33,1}+\frac{3}{640}
\sqrt{\frac{7}{5}} C_{33,20}^{33,1}\)

\noindent\(C_{22,2211,1}^{00,2211,1,\text{NonLoc}}= \frac{7 C_{00,00}^{60,0}}{256 \sqrt{5}}+\frac{7 C_{00,20}^{60,0}}{256 \sqrt{15}}-\frac{21
\sqrt{3} C_{00,22}^{62,0}}{1280}-\frac{7 C_{11,00}^{51,0}}{768 \sqrt{5}}+\frac{21 \sqrt{3} C_{11,22}^{51,0}}{1280}-\frac{9}{640} \sqrt{\frac{7}{5}}
C_{11,22}^{53,0}-\frac{7 C_{20,00}^{40,0}}{768 \sqrt{5}}-\frac{7 C_{20,20}^{40,0}}{768 \sqrt{15}}+\frac{21 \sqrt{3} C_{20,22}^{42,0}}{1280}+\frac{7
C_{22,00}^{42,0}}{3840}+\frac{7 C_{22,20}^{42,0}}{3840 \sqrt{3}}-\frac{67 C_{22,22}^{40,0}}{1280 \sqrt{3}}+\frac{1}{640} \sqrt{\frac{7}{3}} C_{22,22}^{42,0}-\frac{3}{160}
\sqrt{\frac{3}{5}} C_{22,22}^{44,0}+\frac{7 C_{31,00}^{31,0}}{768 \sqrt{5}}+\frac{7 C_{31,20}^{31,0}}{768 \sqrt{15}}-\frac{13 C_{31,22}^{31,0}}{256
\sqrt{3}}+\frac{1}{64} \sqrt{\frac{7}{5}} C_{31,22}^{33,0}-\frac{3 \sqrt{\frac{21}{5}} C_{33,00}^{33,0}}{1280}-\frac{3 \sqrt{\frac{7}{5}} C_{33,20}^{33,0}}{1280}\)

\noindent\(C_{22,2211,1}^{11,1101,0,\text{Loc}}= \frac{7 C_{11,11}^{51,0}}{1152}+\frac{7 C_{11,11}^{51,1}}{2304}+\frac{C_{31,11}^{31,0}}{384}+\frac{C_{31,11}^{31,1}}{768}-\frac{1}{160}
\sqrt{\frac{7}{2}} C_{33,11}^{33,0}-\frac{1}{320} \sqrt{\frac{7}{2}} C_{33,11}^{33,1}\)

\noindent\(C_{22,2211,1}^{11,1101,0,\text{NonLoc}}= \frac{7 C_{11,11}^{51,0}}{2304}+\frac{7 C_{11,11}^{51,1}}{1152}+\frac{C_{31,11}^{31,0}}{768}+\frac{C_{31,11}^{31,1}}{384}-\frac{1}{320}
\sqrt{\frac{7}{2}} C_{33,11}^{33,0}-\frac{1}{160} \sqrt{\frac{7}{2}} C_{33,11}^{33,1}\)

\noindent\(C_{22,2211,1}^{11,1101,1,\text{Loc}}= \frac{7 C_{11,11}^{51,1}}{768 \sqrt{3}}+\frac{C_{31,11}^{31,1}}{256 \sqrt{3}}-\frac{1}{320}
\sqrt{\frac{21}{2}} C_{33,11}^{33,1}\)

\noindent\(C_{22,2211,1}^{11,1101,1,\text{NonLoc}}= \frac{7 C_{11,11}^{51,0}}{768 \sqrt{3}}+\frac{C_{31,11}^{31,0}}{256 \sqrt{3}}-\frac{1}{320}
\sqrt{\frac{21}{2}} C_{33,11}^{33,0}\)

\noindent\(C_{22,2211,1}^{20,0011,0,\text{Loc}}= -\frac{7 C_{00,20}^{60,0}}{768}-\frac{7 C_{00,20}^{60,1}}{1536}-\frac{7 C_{00,22}^{62,0}}{512
\sqrt{5}}-\frac{7 C_{00,22}^{62,1}}{1024 \sqrt{5}}+\frac{7 C_{11,20}^{31,0}}{2304}+\frac{7 C_{11,20}^{31,1}}{4608}+\frac{7 C_{11,22}^{51,0}}{2304
\sqrt{5}}+\frac{7 C_{11,22}^{51,1}}{4608 \sqrt{5}}+\frac{3 \sqrt{21} C_{11,22}^{53,0}}{5120}+\frac{3 \sqrt{21} C_{11,22}^{53,1}}{10240}+\frac{7 C_{20,20}^{40,0}}{2304}+\frac{7
C_{20,20}^{40,1}}{4608}+\frac{7 C_{20,22}^{42,0}}{2304 \sqrt{5}}+\frac{7 C_{20,22}^{42,1}}{4608 \sqrt{5}}-\frac{7 C_{22,20}^{42,0}}{2304 \sqrt{5}}-\frac{7
C_{22,20}^{42,1}}{4608 \sqrt{5}}+\frac{7 \sqrt{5} C_{22,22}^{40,0}}{4608}+\frac{7 \sqrt{5} C_{22,22}^{40,1}}{9216}-\frac{7 \sqrt{\frac{7}{5}} C_{22,22}^{42,0}}{2304}-\frac{7
\sqrt{\frac{7}{5}} C_{22,22}^{42,1}}{4608}-\frac{7 C_{31,20}^{31,0}}{2304}-\frac{7 C_{31,20}^{31,1}}{4608}-\frac{7 C_{31,22}^{31,0}}{2304 \sqrt{5}}-\frac{7
C_{31,22}^{31,1}}{4608 \sqrt{5}}-\frac{3 \sqrt{21} C_{31,22}^{33,0}}{5120}-\frac{3 \sqrt{21} C_{31,22}^{33,1}}{10240}+\frac{\sqrt{21} C_{33,20}^{33,0}}{1280}+\frac{\sqrt{21}
C_{33,20}^{33,1}}{2560}+\frac{3}{320} \sqrt{\frac{7}{10}} C_{33,22}^{33,0}+\frac{3}{640} \sqrt{\frac{7}{10}} C_{33,22}^{33,1}\)

\noindent\(C_{22,2211,1}^{20,0011,0,\text{NonLoc}}= -\frac{7 C_{00,00}^{60,0}}{1024 \sqrt{3}}-\frac{7 C_{00,00}^{60,1}}{512 \sqrt{3}}-\frac{7
C_{00,20}^{60,0}}{3072}-\frac{7 C_{00,20}^{60,1}}{1536}+\frac{7 C_{00,22}^{62,0}}{1024 \sqrt{5}}+\frac{7 C_{00,22}^{62,1}}{512 \sqrt{5}}-\frac{7
C_{11,00}^{51,0}}{3072 \sqrt{3}}-\frac{7 C_{11,00}^{51,1}}{1536 \sqrt{3}}-\frac{7 C_{11,20}^{31,0}}{9216}-\frac{7 C_{11,20}^{31,1}}{4608}+\frac{7
C_{11,22}^{51,0}}{4608 \sqrt{5}}+\frac{7 C_{11,22}^{51,1}}{2304 \sqrt{5}}+\frac{3 \sqrt{21} C_{11,22}^{53,0}}{10240}+\frac{3 \sqrt{21} C_{11,22}^{53,1}}{5120}+\frac{7
C_{20,00}^{40,0}}{3072 \sqrt{3}}+\frac{7 C_{20,00}^{40,1}}{1536 \sqrt{3}}+\frac{7 C_{20,20}^{40,0}}{9216}+\frac{7 C_{20,20}^{40,1}}{4608}-\frac{7
C_{20,22}^{42,0}}{4608 \sqrt{5}}-\frac{7 C_{20,22}^{42,1}}{2304 \sqrt{5}}-\frac{7 C_{22,00}^{42,0}}{3072 \sqrt{15}}-\frac{7 C_{22,00}^{42,1}}{1536
\sqrt{15}}-\frac{7 C_{22,20}^{42,0}}{9216 \sqrt{5}}-\frac{7 C_{22,20}^{42,1}}{4608 \sqrt{5}}-\frac{7 \sqrt{5} C_{22,22}^{40,0}}{9216}-\frac{7 \sqrt{5}
C_{22,22}^{40,1}}{4608}+\frac{7 \sqrt{\frac{7}{5}} C_{22,22}^{42,0}}{4608}+\frac{7 \sqrt{\frac{7}{5}} C_{22,22}^{42,1}}{2304}+\frac{7 C_{31,00}^{31,0}}{3072
\sqrt{3}}+\frac{7 C_{31,00}^{31,1}}{1536 \sqrt{3}}+\frac{7 C_{31,20}^{31,0}}{9216}+\frac{7 C_{31,20}^{31,1}}{4608}-\frac{7 C_{31,22}^{31,0}}{4608
\sqrt{5}}-\frac{7 C_{31,22}^{31,1}}{2304 \sqrt{5}}-\frac{3 \sqrt{21} C_{31,22}^{33,0}}{10240}-\frac{3 \sqrt{21} C_{31,22}^{33,1}}{5120}-\frac{3 \sqrt{7}
C_{33,00}^{33,0}}{5120}-\frac{3 \sqrt{7} C_{33,00}^{33,1}}{2560}-\frac{\sqrt{21} C_{33,20}^{33,0}}{5120}-\frac{\sqrt{21} C_{33,20}^{33,1}}{2560}+\frac{3}{640}
\sqrt{\frac{7}{10}} C_{33,22}^{33,0}+\frac{3}{320} \sqrt{\frac{7}{10}} C_{33,22}^{33,1}\)

\noindent\(C_{22,2211,1}^{20,0011,1,\text{Loc}}= -\frac{7 C_{00,20}^{60,1}}{512 \sqrt{3}}-\frac{7 \sqrt{\frac{3}{5}} C_{00,22}^{62,1}}{1024}+\frac{7
C_{11,20}^{31,1}}{1536 \sqrt{3}}+\frac{7 C_{11,22}^{51,1}}{1536 \sqrt{15}}+\frac{9 \sqrt{7} C_{11,22}^{53,1}}{10240}+\frac{7 C_{20,20}^{40,1}}{1536
\sqrt{3}}+\frac{7 C_{20,22}^{42,1}}{1536 \sqrt{15}}-\frac{7 C_{22,20}^{42,1}}{1536 \sqrt{15}}+\frac{7 \sqrt{\frac{5}{3}} C_{22,22}^{40,1}}{3072}-\frac{7
\sqrt{\frac{7}{15}} C_{22,22}^{42,1}}{1536}-\frac{7 C_{31,20}^{31,1}}{1536 \sqrt{3}}-\frac{7 C_{31,22}^{31,1}}{1536 \sqrt{15}}-\frac{9 \sqrt{7} C_{31,22}^{33,1}}{10240}+\frac{3
\sqrt{7} C_{33,20}^{33,1}}{2560}+\frac{3}{640} \sqrt{\frac{21}{10}} C_{33,22}^{33,1}\)

\noindent\(C_{22,2211,1}^{20,0011,1,\text{NonLoc}}= -\frac{7 C_{00,00}^{60,0}}{1024}-\frac{7 C_{00,20}^{60,0}}{1024 \sqrt{3}}+\frac{7
\sqrt{\frac{3}{5}} C_{00,22}^{62,0}}{1024}-\frac{7 C_{11,00}^{51,0}}{3072}-\frac{7 C_{11,20}^{31,0}}{3072 \sqrt{3}}+\frac{7 C_{11,22}^{51,0}}{1536
\sqrt{15}}+\frac{9 \sqrt{7} C_{11,22}^{53,0}}{10240}+\frac{7 C_{20,00}^{40,0}}{3072}+\frac{7 C_{20,20}^{40,0}}{3072 \sqrt{3}}-\frac{7 C_{20,22}^{42,0}}{1536
\sqrt{15}}-\frac{7 C_{22,00}^{42,0}}{3072 \sqrt{5}}-\frac{7 C_{22,20}^{42,0}}{3072 \sqrt{15}}-\frac{7 \sqrt{\frac{5}{3}} C_{22,22}^{40,0}}{3072}+\frac{7
\sqrt{\frac{7}{15}} C_{22,22}^{42,0}}{1536}+\frac{7 C_{31,00}^{31,0}}{3072}+\frac{7 C_{31,20}^{31,0}}{3072 \sqrt{3}}-\frac{7 C_{31,22}^{31,0}}{1536
\sqrt{15}}-\frac{9 \sqrt{7} C_{31,22}^{33,0}}{10240}-\frac{3 \sqrt{21} C_{33,00}^{33,0}}{5120}-\frac{3 \sqrt{7} C_{33,20}^{33,0}}{5120}+\frac{3}{640}
\sqrt{\frac{21}{10}} C_{33,22}^{33,0}\)

\noindent\(C_{22,2211,1}^{22,0011,0,\text{Loc}}= -\frac{7 C_{00,20}^{60,0}}{768 \sqrt{5}}-\frac{7 C_{00,20}^{60,1}}{1536 \sqrt{5}}-\frac{21
C_{00,22}^{62,0}}{2560}-\frac{21 C_{00,22}^{62,1}}{5120}+\frac{3 \sqrt{\frac{21}{5}} C_{11,22}^{53,0}}{1024}+\frac{3 \sqrt{\frac{21}{5}} C_{11,22}^{53,1}}{2048}+\frac{7
C_{20,20}^{40,0}}{2304 \sqrt{5}}+\frac{7 C_{20,20}^{40,1}}{4608 \sqrt{5}}-\frac{7 C_{22,20}^{42,0}}{11520}-\frac{7 C_{22,20}^{42,1}}{23040}+\frac{23
C_{22,22}^{40,0}}{7680}+\frac{23 C_{22,22}^{40,1}}{15360}+\frac{\sqrt{7} C_{22,22}^{42,0}}{3840}+\frac{\sqrt{7} C_{22,22}^{42,1}}{7680}-\frac{3 C_{22,22}^{44,0}}{320
\sqrt{5}}-\frac{3 C_{22,22}^{44,1}}{640 \sqrt{5}}-\frac{7 C_{31,20}^{31,0}}{2304 \sqrt{5}}-\frac{7 C_{31,20}^{31,1}}{4608 \sqrt{5}}-\frac{C_{31,22}^{31,0}}{384}-\frac{C_{31,22}^{31,1}}{768}-\frac{\sqrt{\frac{7}{15}}
C_{31,22}^{33,0}}{1024}-\frac{\sqrt{\frac{7}{15}} C_{31,22}^{33,1}}{2048}+\frac{\sqrt{\frac{21}{5}} C_{33,20}^{33,0}}{1280}+\frac{\sqrt{\frac{21}{5}}
C_{33,20}^{33,1}}{2560}\)

\noindent\(C_{22,2211,1}^{22,0011,0,\text{NonLoc}}= -\frac{7 C_{00,00}^{60,0}}{1024 \sqrt{15}}-\frac{7 C_{00,00}^{60,1}}{512 \sqrt{15}}-\frac{7
C_{00,20}^{60,0}}{3072 \sqrt{5}}-\frac{7 C_{00,20}^{60,1}}{1536 \sqrt{5}}+\frac{21 C_{00,22}^{62,0}}{5120}+\frac{21 C_{00,22}^{62,1}}{2560}-\frac{7
C_{11,00}^{51,0}}{3072 \sqrt{15}}-\frac{7 C_{11,00}^{51,1}}{1536 \sqrt{15}}+\frac{3 \sqrt{\frac{21}{5}} C_{11,22}^{53,0}}{2048}+\frac{3 \sqrt{\frac{21}{5}}
C_{11,22}^{53,1}}{1024}+\frac{7 C_{20,00}^{40,0}}{3072 \sqrt{15}}+\frac{7 C_{20,00}^{40,1}}{1536 \sqrt{15}}+\frac{7 C_{20,20}^{40,0}}{9216 \sqrt{5}}+\frac{7
C_{20,20}^{40,1}}{4608 \sqrt{5}}-\frac{7 C_{22,00}^{42,0}}{15360 \sqrt{3}}-\frac{7 C_{22,00}^{42,1}}{7680 \sqrt{3}}-\frac{7 C_{22,20}^{42,0}}{46080}-\frac{7
C_{22,20}^{42,1}}{23040}-\frac{23 C_{22,22}^{40,0}}{15360}-\frac{23 C_{22,22}^{40,1}}{7680}-\frac{\sqrt{7} C_{22,22}^{42,0}}{7680}-\frac{\sqrt{7}
C_{22,22}^{42,1}}{3840}+\frac{3 C_{22,22}^{44,0}}{640 \sqrt{5}}+\frac{3 C_{22,22}^{44,1}}{320 \sqrt{5}}+\frac{7 C_{31,00}^{31,0}}{3072 \sqrt{15}}+\frac{7
C_{31,00}^{31,1}}{1536 \sqrt{15}}+\frac{7 C_{31,20}^{31,0}}{9216 \sqrt{5}}+\frac{7 C_{31,20}^{31,1}}{4608 \sqrt{5}}-\frac{C_{31,22}^{31,0}}{768}-\frac{C_{31,22}^{31,1}}{384}-\frac{\sqrt{\frac{7}{15}}
C_{31,22}^{33,0}}{2048}-\frac{\sqrt{\frac{7}{15}} C_{31,22}^{33,1}}{1024}-\frac{3 \sqrt{\frac{7}{5}} C_{33,00}^{33,0}}{5120}-\frac{3 \sqrt{\frac{7}{5}}
C_{33,00}^{33,1}}{2560}-\frac{\sqrt{\frac{21}{5}} C_{33,20}^{33,0}}{5120}-\frac{\sqrt{\frac{21}{5}} C_{33,20}^{33,1}}{2560}\)

\noindent\(C_{22,2211,1}^{22,0011,1,\text{Loc}}= -\frac{7 C_{00,20}^{60,1}}{512 \sqrt{15}}-\frac{21 \sqrt{3} C_{00,22}^{62,1}}{5120}+\frac{9
\sqrt{\frac{7}{5}} C_{11,22}^{53,1}}{2048}+\frac{7 C_{20,20}^{40,1}}{1536 \sqrt{15}}-\frac{7 C_{22,20}^{42,1}}{7680 \sqrt{3}}+\frac{23 C_{22,22}^{40,1}}{5120
\sqrt{3}}+\frac{\sqrt{\frac{7}{3}} C_{22,22}^{42,1}}{2560}-\frac{3}{640} \sqrt{\frac{3}{5}} C_{22,22}^{44,1}-\frac{7 C_{31,20}^{31,1}}{1536 \sqrt{15}}-\frac{C_{31,22}^{31,1}}{256
\sqrt{3}}-\frac{\sqrt{\frac{7}{5}} C_{31,22}^{33,1}}{2048}+\frac{3 \sqrt{\frac{7}{5}} C_{33,20}^{33,1}}{2560}\)

\noindent\(C_{22,2211,1}^{22,0011,1,\text{NonLoc}}= -\frac{7 C_{00,00}^{60,0}}{1024 \sqrt{5}}-\frac{7 C_{00,20}^{60,0}}{1024 \sqrt{15}}+\frac{21
\sqrt{3} C_{00,22}^{62,0}}{5120}-\frac{7 C_{11,00}^{51,0}}{3072 \sqrt{5}}+\frac{9 \sqrt{\frac{7}{5}} C_{11,22}^{53,0}}{2048}+\frac{7 C_{20,00}^{40,0}}{3072
\sqrt{5}}+\frac{7 C_{20,20}^{40,0}}{3072 \sqrt{15}}-\frac{7 C_{22,00}^{42,0}}{15360}-\frac{7 C_{22,20}^{42,0}}{15360 \sqrt{3}}-\frac{23 C_{22,22}^{40,0}}{5120
\sqrt{3}}-\frac{\sqrt{\frac{7}{3}} C_{22,22}^{42,0}}{2560}+\frac{3}{640} \sqrt{\frac{3}{5}} C_{22,22}^{44,0}+\frac{7 C_{31,00}^{31,0}}{3072 \sqrt{5}}+\frac{7
C_{31,20}^{31,0}}{3072 \sqrt{15}}-\frac{C_{31,22}^{31,0}}{256 \sqrt{3}}-\frac{\sqrt{\frac{7}{5}} C_{31,22}^{33,0}}{2048}-\frac{3 \sqrt{\frac{21}{5}}
C_{33,00}^{33,0}}{5120}-\frac{3 \sqrt{\frac{7}{5}} C_{33,20}^{33,0}}{5120}\)

\noindent\(C_{22,2211,2}^{00,2212,0,\text{Loc}}= -\frac{7 C_{00,20}^{60,0}}{192 \sqrt{3}}-\frac{7 C_{00,20}^{60,1}}{384 \sqrt{3}}+\frac{7
C_{00,22}^{62,0}}{128 \sqrt{15}}+\frac{7 C_{00,22}^{62,1}}{256 \sqrt{15}}+\frac{77 C_{11,22}^{51,0}}{1152 \sqrt{15}}+\frac{77 C_{11,22}^{51,1}}{2304
\sqrt{15}}-\frac{3 \sqrt{7} C_{11,22}^{53,0}}{640}-\frac{3 \sqrt{7} C_{11,22}^{53,1}}{1280}+\frac{7 C_{20,20}^{40,0}}{576 \sqrt{3}}+\frac{7 C_{20,20}^{40,1}}{1152
\sqrt{3}}-\frac{77 C_{20,22}^{42,0}}{1152 \sqrt{15}}-\frac{77 C_{20,22}^{42,1}}{2304 \sqrt{15}}-\frac{7 C_{22,20}^{42,0}}{576 \sqrt{15}}-\frac{7
C_{22,20}^{42,1}}{1152 \sqrt{15}}+\frac{11 \sqrt{\frac{5}{3}} C_{22,22}^{40,0}}{1152}+\frac{11 \sqrt{\frac{5}{3}} C_{22,22}^{40,1}}{2304}+\frac{7}{576}
\sqrt{\frac{7}{15}} C_{22,22}^{42,0}+\frac{7 \sqrt{\frac{7}{15}} C_{22,22}^{42,1}}{1152}+\frac{7 C_{31,20}^{31,0}}{576 \sqrt{3}}+\frac{7 C_{31,20}^{31,1}}{1152
\sqrt{3}}-\frac{59 C_{31,22}^{31,0}}{1152 \sqrt{15}}-\frac{59 C_{31,22}^{31,1}}{2304 \sqrt{15}}+\frac{\sqrt{7} C_{31,22}^{33,0}}{640}+\frac{\sqrt{7}
C_{31,22}^{33,1}}{1280}-\frac{\sqrt{7} C_{33,20}^{33,0}}{320}-\frac{\sqrt{7} C_{33,20}^{33,1}}{640}+\frac{1}{80} \sqrt{\frac{21}{10}} C_{33,22}^{33,0}+\frac{1}{160}
\sqrt{\frac{21}{10}} C_{33,22}^{33,1}\)

\noindent\(C_{22,2211,2}^{00,2212,0,\text{NonLoc}}= -\frac{7 C_{00,00}^{60,0}}{768}-\frac{7 C_{00,00}^{60,1}}{384}-\frac{7 C_{00,20}^{60,0}}{768
\sqrt{3}}-\frac{7 C_{00,20}^{60,1}}{384 \sqrt{3}}-\frac{7 C_{00,22}^{62,0}}{256 \sqrt{15}}-\frac{7 C_{00,22}^{62,1}}{128 \sqrt{15}}+\frac{7 C_{11,00}^{51,0}}{2304}+\frac{7
C_{11,00}^{51,1}}{1152}+\frac{77 C_{11,22}^{51,0}}{2304 \sqrt{15}}+\frac{77 C_{11,22}^{51,1}}{1152 \sqrt{15}}-\frac{3 \sqrt{7} C_{11,22}^{53,0}}{1280}-\frac{3
\sqrt{7} C_{11,22}^{53,1}}{640}+\frac{7 C_{20,00}^{40,0}}{2304}+\frac{7 C_{20,00}^{40,1}}{1152}+\frac{7 C_{20,20}^{40,0}}{2304 \sqrt{3}}+\frac{7
C_{20,20}^{40,1}}{1152 \sqrt{3}}+\frac{77 C_{20,22}^{42,0}}{2304 \sqrt{15}}+\frac{77 C_{20,22}^{42,1}}{1152 \sqrt{15}}-\frac{7 C_{22,00}^{42,0}}{2304
\sqrt{5}}-\frac{7 C_{22,00}^{42,1}}{1152 \sqrt{5}}-\frac{7 C_{22,20}^{42,0}}{2304 \sqrt{15}}-\frac{7 C_{22,20}^{42,1}}{1152 \sqrt{15}}-\frac{11 \sqrt{\frac{5}{3}}
C_{22,22}^{40,0}}{2304}-\frac{11 \sqrt{\frac{5}{3}} C_{22,22}^{40,1}}{1152}-\frac{7 \sqrt{\frac{7}{15}} C_{22,22}^{42,0}}{1152}-\frac{7}{576} \sqrt{\frac{7}{15}}
C_{22,22}^{42,1}-\frac{7 C_{31,00}^{31,0}}{2304}-\frac{7 C_{31,00}^{31,1}}{1152}-\frac{7 C_{31,20}^{31,0}}{2304 \sqrt{3}}-\frac{7 C_{31,20}^{31,1}}{1152
\sqrt{3}}-\frac{59 C_{31,22}^{31,0}}{2304 \sqrt{15}}-\frac{59 C_{31,22}^{31,1}}{1152 \sqrt{15}}+\frac{\sqrt{7} C_{31,22}^{33,0}}{1280}+\frac{\sqrt{7}
C_{31,22}^{33,1}}{640}+\frac{\sqrt{21} C_{33,00}^{33,0}}{1280}+\frac{\sqrt{21} C_{33,00}^{33,1}}{640}+\frac{\sqrt{7} C_{33,20}^{33,0}}{1280}+\frac{\sqrt{7}
C_{33,20}^{33,1}}{640}+\frac{1}{160} \sqrt{\frac{21}{10}} C_{33,22}^{33,0}+\frac{1}{80} \sqrt{\frac{21}{10}} C_{33,22}^{33,1}\)

\noindent\(C_{22,2211,2}^{00,2212,1,\text{Loc}}= -\frac{7 C_{00,20}^{60,1}}{384}+\frac{7 C_{00,22}^{62,1}}{256 \sqrt{5}}+\frac{77 C_{11,22}^{51,1}}{2304
\sqrt{5}}-\frac{3 \sqrt{21} C_{11,22}^{53,1}}{1280}+\frac{7 C_{20,20}^{40,1}}{1152}-\frac{77 C_{20,22}^{42,1}}{2304 \sqrt{5}}-\frac{7 C_{22,20}^{42,1}}{1152
\sqrt{5}}+\frac{11 \sqrt{5} C_{22,22}^{40,1}}{2304}+\frac{7 \sqrt{\frac{7}{5}} C_{22,22}^{42,1}}{1152}+\frac{7 C_{31,20}^{31,1}}{1152}-\frac{59 C_{31,22}^{31,1}}{2304
\sqrt{5}}+\frac{\sqrt{21} C_{31,22}^{33,1}}{1280}-\frac{\sqrt{21} C_{33,20}^{33,1}}{640}+\frac{3}{160} \sqrt{\frac{7}{10}} C_{33,22}^{33,1}\)

\noindent\(C_{22,2211,2}^{00,2212,1,\text{NonLoc}}= -\frac{7 C_{00,00}^{60,0}}{256 \sqrt{3}}-\frac{7 C_{00,20}^{60,0}}{768}-\frac{7
C_{00,22}^{62,0}}{256 \sqrt{5}}+\frac{7 C_{11,00}^{51,0}}{768 \sqrt{3}}+\frac{77 C_{11,22}^{51,0}}{2304 \sqrt{5}}-\frac{3 \sqrt{21} C_{11,22}^{53,0}}{1280}+\frac{7
C_{20,00}^{40,0}}{768 \sqrt{3}}+\frac{7 C_{20,20}^{40,0}}{2304}+\frac{77 C_{20,22}^{42,0}}{2304 \sqrt{5}}-\frac{7 C_{22,00}^{42,0}}{768 \sqrt{15}}-\frac{7
C_{22,20}^{42,0}}{2304 \sqrt{5}}-\frac{11 \sqrt{5} C_{22,22}^{40,0}}{2304}-\frac{7 \sqrt{\frac{7}{5}} C_{22,22}^{42,0}}{1152}-\frac{7 C_{31,00}^{31,0}}{768
\sqrt{3}}-\frac{7 C_{31,20}^{31,0}}{2304}-\frac{59 C_{31,22}^{31,0}}{2304 \sqrt{5}}+\frac{\sqrt{21} C_{31,22}^{33,0}}{1280}+\frac{3 \sqrt{7} C_{33,00}^{33,0}}{1280}+\frac{\sqrt{21}
C_{33,20}^{33,0}}{1280}+\frac{3}{160} \sqrt{\frac{7}{10}} C_{33,22}^{33,0}\)

\noindent\(C_{22,2211,2}^{11,1101,0,\text{Loc}}= \frac{7 C_{11,11}^{51,0}}{384 \sqrt{3}}+\frac{7 C_{11,11}^{51,1}}{768 \sqrt{3}}+\frac{C_{31,11}^{31,0}}{128
\sqrt{3}}+\frac{C_{31,11}^{31,1}}{256 \sqrt{3}}-\frac{1}{160} \sqrt{\frac{21}{2}} C_{33,11}^{33,0}-\frac{1}{320} \sqrt{\frac{21}{2}} C_{33,11}^{33,1}\)

\noindent\(C_{22,2211,2}^{11,1101,0,\text{NonLoc}}= \frac{7 C_{11,11}^{51,0}}{768 \sqrt{3}}+\frac{7 C_{11,11}^{51,1}}{384 \sqrt{3}}+\frac{C_{31,11}^{31,0}}{256
\sqrt{3}}+\frac{C_{31,11}^{31,1}}{128 \sqrt{3}}-\frac{1}{320} \sqrt{\frac{21}{2}} C_{33,11}^{33,0}-\frac{1}{160} \sqrt{\frac{21}{2}} C_{33,11}^{33,1}\)

\noindent\(C_{22,2211,2}^{11,1101,1,\text{Loc}}= \frac{7 C_{11,11}^{51,1}}{768}+\frac{C_{31,11}^{31,1}}{256}-\frac{3}{320} \sqrt{\frac{7}{2}}
C_{33,11}^{33,1}\)

\noindent\(C_{22,2211,2}^{11,1101,1,\text{NonLoc}}= \frac{7 C_{11,11}^{51,0}}{768}+\frac{C_{31,11}^{31,0}}{256}-\frac{3}{320} \sqrt{\frac{7}{2}}
C_{33,11}^{33,0}\)

\noindent\(C_{22,2211,2}^{22,0011,0,\text{Loc}}= \frac{7 C_{00,20}^{60,0}}{768 \sqrt{3}}+\frac{7 C_{00,20}^{60,1}}{1536 \sqrt{3}}-\frac{7
C_{00,22}^{62,0}}{512 \sqrt{15}}-\frac{7 C_{00,22}^{62,1}}{1024 \sqrt{15}}-\frac{7 C_{11,22}^{51,0}}{1152 \sqrt{15}}-\frac{7 C_{11,22}^{51,1}}{2304
\sqrt{15}}+\frac{9 \sqrt{7} C_{11,22}^{53,0}}{5120}+\frac{9 \sqrt{7} C_{11,22}^{53,1}}{10240}-\frac{7 C_{20,20}^{40,0}}{2304 \sqrt{3}}-\frac{7 C_{20,20}^{40,1}}{4608
\sqrt{3}}-\frac{7 C_{20,22}^{42,0}}{1152 \sqrt{15}}-\frac{7 C_{20,22}^{42,1}}{2304 \sqrt{15}}+\frac{7 C_{22,20}^{42,0}}{2304 \sqrt{15}}+\frac{7 C_{22,20}^{42,1}}{4608
\sqrt{15}}-\frac{C_{22,22}^{40,0}}{4608 \sqrt{15}}-\frac{C_{22,22}^{40,1}}{9216 \sqrt{15}}+\frac{17 \sqrt{\frac{7}{15}} C_{22,22}^{42,0}}{2304}+\frac{17
\sqrt{\frac{7}{15}} C_{22,22}^{42,1}}{4608}-\frac{\sqrt{3} C_{22,22}^{44,0}}{320}-\frac{\sqrt{3} C_{22,22}^{44,1}}{640}+\frac{7 C_{31,20}^{31,0}}{2304
\sqrt{3}}+\frac{7 C_{31,20}^{31,1}}{4608 \sqrt{3}}-\frac{C_{31,22}^{31,0}}{144 \sqrt{15}}-\frac{C_{31,22}^{31,1}}{288 \sqrt{15}}+\frac{13 \sqrt{7}
C_{31,22}^{33,0}}{15360}+\frac{13 \sqrt{7} C_{31,22}^{33,1}}{30720}-\frac{\sqrt{7} C_{33,20}^{33,0}}{1280}-\frac{\sqrt{7} C_{33,20}^{33,1}}{2560}-\frac{1}{160}
\sqrt{\frac{21}{10}} C_{33,22}^{33,0}-\frac{1}{320} \sqrt{\frac{21}{10}} C_{33,22}^{33,1}\)

\noindent\(C_{22,2211,2}^{22,0011,0,\text{NonLoc}}= \frac{7 C_{00,00}^{60,0}}{3072}+\frac{7 C_{00,00}^{60,1}}{1536}+\frac{7 C_{00,20}^{60,0}}{3072
\sqrt{3}}+\frac{7 C_{00,20}^{60,1}}{1536 \sqrt{3}}+\frac{7 C_{00,22}^{62,0}}{1024 \sqrt{15}}+\frac{7 C_{00,22}^{62,1}}{512 \sqrt{15}}+\frac{7 C_{11,00}^{51,0}}{9216}+\frac{7
C_{11,00}^{51,1}}{4608}-\frac{7 C_{11,22}^{51,0}}{2304 \sqrt{15}}-\frac{7 C_{11,22}^{51,1}}{1152 \sqrt{15}}+\frac{9 \sqrt{7} C_{11,22}^{53,0}}{10240}+\frac{9
\sqrt{7} C_{11,22}^{53,1}}{5120}-\frac{7 C_{20,00}^{40,0}}{9216}-\frac{7 C_{20,00}^{40,1}}{4608}-\frac{7 C_{20,20}^{40,0}}{9216 \sqrt{3}}-\frac{7
C_{20,20}^{40,1}}{4608 \sqrt{3}}+\frac{7 C_{20,22}^{42,0}}{2304 \sqrt{15}}+\frac{7 C_{20,22}^{42,1}}{1152 \sqrt{15}}+\frac{7 C_{22,00}^{42,0}}{9216
\sqrt{5}}+\frac{7 C_{22,00}^{42,1}}{4608 \sqrt{5}}+\frac{7 C_{22,20}^{42,0}}{9216 \sqrt{15}}+\frac{7 C_{22,20}^{42,1}}{4608 \sqrt{15}}+\frac{C_{22,22}^{40,0}}{9216
\sqrt{15}}+\frac{C_{22,22}^{40,1}}{4608 \sqrt{15}}-\frac{17 \sqrt{\frac{7}{15}} C_{22,22}^{42,0}}{4608}-\frac{17 \sqrt{\frac{7}{15}} C_{22,22}^{42,1}}{2304}+\frac{\sqrt{3}
C_{22,22}^{44,0}}{640}+\frac{\sqrt{3} C_{22,22}^{44,1}}{320}-\frac{7 C_{31,00}^{31,0}}{9216}-\frac{7 C_{31,00}^{31,1}}{4608}-\frac{7 C_{31,20}^{31,0}}{9216
\sqrt{3}}-\frac{7 C_{31,20}^{31,1}}{4608 \sqrt{3}}-\frac{C_{31,22}^{31,0}}{288 \sqrt{15}}-\frac{C_{31,22}^{31,1}}{144 \sqrt{15}}+\frac{13 \sqrt{7}
C_{31,22}^{33,0}}{30720}+\frac{13 \sqrt{7} C_{31,22}^{33,1}}{15360}+\frac{\sqrt{21} C_{33,00}^{33,0}}{5120}+\frac{\sqrt{21} C_{33,00}^{33,1}}{2560}+\frac{\sqrt{7}
C_{33,20}^{33,0}}{5120}+\frac{\sqrt{7} C_{33,20}^{33,1}}{2560}-\frac{1}{320} \sqrt{\frac{21}{10}} C_{33,22}^{33,0}-\frac{1}{160} \sqrt{\frac{21}{10}}
C_{33,22}^{33,1}\)

\noindent\(C_{22,2211,2}^{22,0011,1,\text{Loc}}= \frac{7 C_{00,20}^{60,1}}{1536}-\frac{7 C_{00,22}^{62,1}}{1024 \sqrt{5}}-\frac{7 C_{11,22}^{51,1}}{2304
\sqrt{5}}+\frac{9 \sqrt{21} C_{11,22}^{53,1}}{10240}-\frac{7 C_{20,20}^{40,1}}{4608}-\frac{7 C_{20,22}^{42,1}}{2304 \sqrt{5}}+\frac{7 C_{22,20}^{42,1}}{4608
\sqrt{5}}-\frac{C_{22,22}^{40,1}}{9216 \sqrt{5}}+\frac{17 \sqrt{\frac{7}{5}} C_{22,22}^{42,1}}{4608}-\frac{3 C_{22,22}^{44,1}}{640}+\frac{7 C_{31,20}^{31,1}}{4608}-\frac{C_{31,22}^{31,1}}{288
\sqrt{5}}+\frac{13 \sqrt{\frac{7}{3}} C_{31,22}^{33,1}}{10240}-\frac{\sqrt{21} C_{33,20}^{33,1}}{2560}-\frac{3}{320} \sqrt{\frac{7}{10}} C_{33,22}^{33,1}\)

\noindent\(C_{22,2211,2}^{22,0011,1,\text{NonLoc}}= \frac{7 C_{00,00}^{60,0}}{1024 \sqrt{3}}+\frac{7 C_{00,20}^{60,0}}{3072}+\frac{7
C_{00,22}^{62,0}}{1024 \sqrt{5}}+\frac{7 C_{11,00}^{51,0}}{3072 \sqrt{3}}-\frac{7 C_{11,22}^{51,0}}{2304 \sqrt{5}}+\frac{9 \sqrt{21} C_{11,22}^{53,0}}{10240}-\frac{7
C_{20,00}^{40,0}}{3072 \sqrt{3}}-\frac{7 C_{20,20}^{40,0}}{9216}+\frac{7 C_{20,22}^{42,0}}{2304 \sqrt{5}}+\frac{7 C_{22,00}^{42,0}}{3072 \sqrt{15}}+\frac{7
C_{22,20}^{42,0}}{9216 \sqrt{5}}+\frac{C_{22,22}^{40,0}}{9216 \sqrt{5}}-\frac{17 \sqrt{\frac{7}{5}} C_{22,22}^{42,0}}{4608}+\frac{3 C_{22,22}^{44,0}}{640}-\frac{7
C_{31,00}^{31,0}}{3072 \sqrt{3}}-\frac{7 C_{31,20}^{31,0}}{9216}-\frac{C_{31,22}^{31,0}}{288 \sqrt{5}}+\frac{13 \sqrt{\frac{7}{3}} C_{31,22}^{33,0}}{10240}+\frac{3
\sqrt{7} C_{33,00}^{33,0}}{5120}+\frac{\sqrt{21} C_{33,20}^{33,0}}{5120}-\frac{3}{320} \sqrt{\frac{7}{10}} C_{33,22}^{33,0}\)

\noindent\(C_{22,2211,3}^{00,2213,0,\text{Loc}}= \frac{1}{96} \sqrt{\frac{7}{15}} C_{00,20}^{60,0}+\frac{1}{192} \sqrt{\frac{7}{15}}
C_{00,20}^{60,1}+\frac{1}{40} \sqrt{\frac{7}{3}} C_{00,22}^{62,0}+\frac{1}{80} \sqrt{\frac{7}{3}} C_{00,22}^{62,1}+\frac{7 \sqrt{\frac{7}{3}} C_{11,22}^{51,0}}{2880}+\frac{7
\sqrt{\frac{7}{3}} C_{11,22}^{51,1}}{5760}+\frac{33 C_{11,22}^{53,0}}{640 \sqrt{5}}+\frac{33 C_{11,22}^{53,1}}{1280 \sqrt{5}}-\frac{1}{288} \sqrt{\frac{7}{15}}
C_{20,20}^{40,0}-\frac{1}{576} \sqrt{\frac{7}{15}} C_{20,20}^{40,1}-\frac{7 \sqrt{\frac{7}{3}} C_{20,22}^{42,0}}{2880}-\frac{7 \sqrt{\frac{7}{3}}
C_{20,22}^{42,1}}{5760}+\frac{\sqrt{\frac{7}{3}} C_{22,20}^{42,0}}{1440}+\frac{\sqrt{\frac{7}{3}} C_{22,20}^{42,1}}{2880}-\frac{11 \sqrt{\frac{7}{3}}
C_{22,22}^{40,0}}{1440}-\frac{11 \sqrt{\frac{7}{3}} C_{22,22}^{40,1}}{2880}+\frac{C_{22,22}^{42,0}}{720 \sqrt{3}}+\frac{C_{22,22}^{42,1}}{1440 \sqrt{3}}+\frac{9}{160}
\sqrt{\frac{3}{35}} C_{22,22}^{44,0}+\frac{9}{320} \sqrt{\frac{3}{35}} C_{22,22}^{44,1}-\frac{1}{288} \sqrt{\frac{7}{15}} C_{31,20}^{31,0}-\frac{1}{576}
\sqrt{\frac{7}{15}} C_{31,20}^{31,1}-\frac{5}{576} \sqrt{\frac{7}{3}} C_{31,22}^{31,0}-\frac{5 \sqrt{\frac{7}{3}} C_{31,22}^{31,1}}{1152}-\frac{C_{31,22}^{33,0}}{128
\sqrt{5}}-\frac{C_{31,22}^{33,1}}{256 \sqrt{5}}+\frac{C_{33,20}^{33,0}}{160 \sqrt{5}}+\frac{C_{33,20}^{33,1}}{320 \sqrt{5}}+\frac{1}{160} \sqrt{\frac{3}{2}}
C_{33,22}^{33,0}+\frac{1}{320} \sqrt{\frac{3}{2}} C_{33,22}^{33,1}\)

\noindent\(C_{22,2211,3}^{00,2213,0,\text{NonLoc}}= \frac{1}{384} \sqrt{\frac{7}{5}} C_{00,00}^{60,0}+\frac{1}{192} \sqrt{\frac{7}{5}}
C_{00,00}^{60,1}+\frac{1}{384} \sqrt{\frac{7}{15}} C_{00,20}^{60,0}+\frac{1}{192} \sqrt{\frac{7}{15}} C_{00,20}^{60,1}-\frac{1}{80} \sqrt{\frac{7}{3}}
C_{00,22}^{62,0}-\frac{1}{40} \sqrt{\frac{7}{3}} C_{00,22}^{62,1}-\frac{\sqrt{\frac{7}{5}} C_{11,00}^{51,0}}{1152}-\frac{1}{576} \sqrt{\frac{7}{5}}
C_{11,00}^{51,1}+\frac{7 \sqrt{\frac{7}{3}} C_{11,22}^{51,0}}{5760}+\frac{7 \sqrt{\frac{7}{3}} C_{11,22}^{51,1}}{2880}+\frac{33 C_{11,22}^{53,0}}{1280
\sqrt{5}}+\frac{33 C_{11,22}^{53,1}}{640 \sqrt{5}}-\frac{\sqrt{\frac{7}{5}} C_{20,00}^{40,0}}{1152}-\frac{1}{576} \sqrt{\frac{7}{5}} C_{20,00}^{40,1}-\frac{\sqrt{\frac{7}{15}}
C_{20,20}^{40,0}}{1152}-\frac{1}{576} \sqrt{\frac{7}{15}} C_{20,20}^{40,1}+\frac{7 \sqrt{\frac{7}{3}} C_{20,22}^{42,0}}{5760}+\frac{7 \sqrt{\frac{7}{3}}
C_{20,22}^{42,1}}{2880}+\frac{\sqrt{7} C_{22,00}^{42,0}}{5760}+\frac{\sqrt{7} C_{22,00}^{42,1}}{2880}+\frac{\sqrt{\frac{7}{3}} C_{22,20}^{42,0}}{5760}+\frac{\sqrt{\frac{7}{3}}
C_{22,20}^{42,1}}{2880}+\frac{11 \sqrt{\frac{7}{3}} C_{22,22}^{40,0}}{2880}+\frac{11 \sqrt{\frac{7}{3}} C_{22,22}^{40,1}}{1440}-\frac{C_{22,22}^{42,0}}{1440
\sqrt{3}}-\frac{C_{22,22}^{42,1}}{720 \sqrt{3}}-\frac{9}{320} \sqrt{\frac{3}{35}} C_{22,22}^{44,0}-\frac{9}{160} \sqrt{\frac{3}{35}} C_{22,22}^{44,1}+\frac{\sqrt{\frac{7}{5}}
C_{31,00}^{31,0}}{1152}+\frac{1}{576} \sqrt{\frac{7}{5}} C_{31,00}^{31,1}+\frac{\sqrt{\frac{7}{15}} C_{31,20}^{31,0}}{1152}+\frac{1}{576} \sqrt{\frac{7}{15}}
C_{31,20}^{31,1}-\frac{5 \sqrt{\frac{7}{3}} C_{31,22}^{31,0}}{1152}-\frac{5}{576} \sqrt{\frac{7}{3}} C_{31,22}^{31,1}-\frac{C_{31,22}^{33,0}}{256
\sqrt{5}}-\frac{C_{31,22}^{33,1}}{128 \sqrt{5}}-\frac{1}{640} \sqrt{\frac{3}{5}} C_{33,00}^{33,0}-\frac{1}{320} \sqrt{\frac{3}{5}} C_{33,00}^{33,1}-\frac{C_{33,20}^{33,0}}{640
\sqrt{5}}-\frac{C_{33,20}^{33,1}}{320 \sqrt{5}}+\frac{1}{320} \sqrt{\frac{3}{2}} C_{33,22}^{33,0}+\frac{1}{160} \sqrt{\frac{3}{2}} C_{33,22}^{33,1}\)

\noindent\(C_{22,2211,3}^{00,2213,1,\text{Loc}}= \frac{1}{192} \sqrt{\frac{7}{5}} C_{00,20}^{60,1}+\frac{\sqrt{7} C_{00,22}^{62,1}}{80}+\frac{7
\sqrt{7} C_{11,22}^{51,1}}{5760}+\frac{33 \sqrt{\frac{3}{5}} C_{11,22}^{53,1}}{1280}-\frac{1}{576} \sqrt{\frac{7}{5}} C_{20,20}^{40,1}-\frac{7 \sqrt{7}
C_{20,22}^{42,1}}{5760}+\frac{\sqrt{7} C_{22,20}^{42,1}}{2880}-\frac{11 \sqrt{7} C_{22,22}^{40,1}}{2880}+\frac{C_{22,22}^{42,1}}{1440}+\frac{27 C_{22,22}^{44,1}}{320
\sqrt{35}}-\frac{1}{576} \sqrt{\frac{7}{5}} C_{31,20}^{31,1}-\frac{5 \sqrt{7} C_{31,22}^{31,1}}{1152}-\frac{1}{256} \sqrt{\frac{3}{5}} C_{31,22}^{33,1}+\frac{1}{320}
\sqrt{\frac{3}{5}} C_{33,20}^{33,1}+\frac{3 C_{33,22}^{33,1}}{320 \sqrt{2}}\)

\noindent\(C_{22,2211,3}^{00,2213,1,\text{NonLoc}}= \frac{1}{128} \sqrt{\frac{7}{15}} C_{00,00}^{60,0}+\frac{1}{384} \sqrt{\frac{7}{5}}
C_{00,20}^{60,0}-\frac{\sqrt{7} C_{00,22}^{62,0}}{80}-\frac{1}{384} \sqrt{\frac{7}{15}} C_{11,00}^{51,0}+\frac{7 \sqrt{7} C_{11,22}^{51,0}}{5760}+\frac{33
\sqrt{\frac{3}{5}} C_{11,22}^{53,0}}{1280}-\frac{1}{384} \sqrt{\frac{7}{15}} C_{20,00}^{40,0}-\frac{\sqrt{\frac{7}{5}} C_{20,20}^{40,0}}{1152}+\frac{7
\sqrt{7} C_{20,22}^{42,0}}{5760}+\frac{\sqrt{\frac{7}{3}} C_{22,00}^{42,0}}{1920}+\frac{\sqrt{7} C_{22,20}^{42,0}}{5760}+\frac{11 \sqrt{7} C_{22,22}^{40,0}}{2880}-\frac{C_{22,22}^{42,0}}{1440}-\frac{27
C_{22,22}^{44,0}}{320 \sqrt{35}}+\frac{1}{384} \sqrt{\frac{7}{15}} C_{31,00}^{31,0}+\frac{\sqrt{\frac{7}{5}} C_{31,20}^{31,0}}{1152}-\frac{5 \sqrt{7}
C_{31,22}^{31,0}}{1152}-\frac{1}{256} \sqrt{\frac{3}{5}} C_{31,22}^{33,0}-\frac{3 C_{33,00}^{33,0}}{640 \sqrt{5}}-\frac{1}{640} \sqrt{\frac{3}{5}}
C_{33,20}^{33,0}+\frac{3 C_{33,22}^{33,0}}{320 \sqrt{2}}\)

\noindent\(C_{22,2211,3}^{22,0011,0,\text{Loc}}= -\frac{1}{384} \sqrt{\frac{7}{15}} C_{00,20}^{60,0}-\frac{1}{768} \sqrt{\frac{7}{15}}
C_{00,20}^{60,1}-\frac{1}{160} \sqrt{\frac{7}{3}} C_{00,22}^{62,0}-\frac{1}{320} \sqrt{\frac{7}{3}} C_{00,22}^{62,1}-\frac{\sqrt{\frac{7}{3}} C_{11,22}^{51,0}}{2304}-\frac{\sqrt{\frac{7}{3}}
C_{11,22}^{51,1}}{4608}+\frac{9 C_{11,22}^{53,0}}{512 \sqrt{5}}+\frac{9 C_{11,22}^{53,1}}{1024 \sqrt{5}}+\frac{\sqrt{\frac{7}{15}} C_{20,20}^{40,0}}{1152}+\frac{\sqrt{\frac{7}{15}}
C_{20,20}^{40,1}}{2304}-\frac{\sqrt{\frac{7}{3}} C_{20,22}^{42,0}}{2304}-\frac{\sqrt{\frac{7}{3}} C_{20,22}^{42,1}}{4608}-\frac{\sqrt{\frac{7}{3}}
C_{22,20}^{42,0}}{5760}-\frac{\sqrt{\frac{7}{3}} C_{22,20}^{42,1}}{11520}+\frac{11 \sqrt{\frac{7}{3}} C_{22,22}^{40,0}}{5760}+\frac{11 \sqrt{\frac{7}{3}}
C_{22,22}^{40,1}}{11520}+\frac{7 C_{22,22}^{42,0}}{1440 \sqrt{3}}+\frac{7 C_{22,22}^{42,1}}{2880 \sqrt{3}}-\frac{1}{320} \sqrt{\frac{21}{5}} C_{22,22}^{44,0}-\frac{1}{640}
\sqrt{\frac{21}{5}} C_{22,22}^{44,1}-\frac{\sqrt{\frac{7}{15}} C_{31,20}^{31,0}}{1152}-\frac{\sqrt{\frac{7}{15}} C_{31,20}^{31,1}}{2304}-\frac{5
\sqrt{\frac{7}{3}} C_{31,22}^{31,0}}{2304}-\frac{5 \sqrt{\frac{7}{3}} C_{31,22}^{31,1}}{4608}+\frac{C_{31,22}^{33,0}}{1536 \sqrt{5}}+\frac{C_{31,22}^{33,1}}{3072
\sqrt{5}}+\frac{C_{33,20}^{33,0}}{640 \sqrt{5}}+\frac{C_{33,20}^{33,1}}{1280 \sqrt{5}}-\frac{1}{320} \sqrt{\frac{3}{2}} C_{33,22}^{33,0}-\frac{1}{640}
\sqrt{\frac{3}{2}} C_{33,22}^{33,1}\)

\noindent\(C_{22,2211,3}^{22,0011,0,\text{NonLoc}}= -\frac{\sqrt{\frac{7}{5}} C_{00,00}^{60,0}}{1536}-\frac{1}{768} \sqrt{\frac{7}{5}}
C_{00,00}^{60,1}-\frac{\sqrt{\frac{7}{15}} C_{00,20}^{60,0}}{1536}-\frac{1}{768} \sqrt{\frac{7}{15}} C_{00,20}^{60,1}+\frac{1}{320} \sqrt{\frac{7}{3}}
C_{00,22}^{62,0}+\frac{1}{160} \sqrt{\frac{7}{3}} C_{00,22}^{62,1}-\frac{\sqrt{\frac{7}{5}} C_{11,00}^{51,0}}{4608}-\frac{\sqrt{\frac{7}{5}} C_{11,00}^{51,1}}{2304}-\frac{\sqrt{\frac{7}{3}}
C_{11,22}^{51,0}}{4608}-\frac{\sqrt{\frac{7}{3}} C_{11,22}^{51,1}}{2304}+\frac{9 C_{11,22}^{53,0}}{1024 \sqrt{5}}+\frac{9 C_{11,22}^{53,1}}{512 \sqrt{5}}+\frac{\sqrt{\frac{7}{5}}
C_{20,00}^{40,0}}{4608}+\frac{\sqrt{\frac{7}{5}} C_{20,00}^{40,1}}{2304}+\frac{\sqrt{\frac{7}{15}} C_{20,20}^{40,0}}{4608}+\frac{\sqrt{\frac{7}{15}}
C_{20,20}^{40,1}}{2304}+\frac{\sqrt{\frac{7}{3}} C_{20,22}^{42,0}}{4608}+\frac{\sqrt{\frac{7}{3}} C_{20,22}^{42,1}}{2304}-\frac{\sqrt{7} C_{22,00}^{42,0}}{23040}-\frac{\sqrt{7}
C_{22,00}^{42,1}}{11520}-\frac{\sqrt{\frac{7}{3}} C_{22,20}^{42,0}}{23040}-\frac{\sqrt{\frac{7}{3}} C_{22,20}^{42,1}}{11520}-\frac{11 \sqrt{\frac{7}{3}}
C_{22,22}^{40,0}}{11520}-\frac{11 \sqrt{\frac{7}{3}} C_{22,22}^{40,1}}{5760}-\frac{7 C_{22,22}^{42,0}}{2880 \sqrt{3}}-\frac{7 C_{22,22}^{42,1}}{1440
\sqrt{3}}+\frac{1}{640} \sqrt{\frac{21}{5}} C_{22,22}^{44,0}+\frac{1}{320} \sqrt{\frac{21}{5}} C_{22,22}^{44,1}+\frac{\sqrt{\frac{7}{5}} C_{31,00}^{31,0}}{4608}+\frac{\sqrt{\frac{7}{5}}
C_{31,00}^{31,1}}{2304}+\frac{\sqrt{\frac{7}{15}} C_{31,20}^{31,0}}{4608}+\frac{\sqrt{\frac{7}{15}} C_{31,20}^{31,1}}{2304}-\frac{5 \sqrt{\frac{7}{3}}
C_{31,22}^{31,0}}{4608}-\frac{5 \sqrt{\frac{7}{3}} C_{31,22}^{31,1}}{2304}+\frac{C_{31,22}^{33,0}}{3072 \sqrt{5}}+\frac{C_{31,22}^{33,1}}{1536 \sqrt{5}}-\frac{\sqrt{\frac{3}{5}}
C_{33,00}^{33,0}}{2560}-\frac{\sqrt{\frac{3}{5}} C_{33,00}^{33,1}}{1280}-\frac{C_{33,20}^{33,0}}{2560 \sqrt{5}}-\frac{C_{33,20}^{33,1}}{1280 \sqrt{5}}-\frac{1}{640}
\sqrt{\frac{3}{2}} C_{33,22}^{33,0}-\frac{1}{320} \sqrt{\frac{3}{2}} C_{33,22}^{33,1}\)

\noindent\(C_{22,2211,3}^{22,0011,1,\text{Loc}}= -\frac{1}{768} \sqrt{\frac{7}{5}} C_{00,20}^{60,1}-\frac{\sqrt{7} C_{00,22}^{62,1}}{320}-\frac{\sqrt{7}
C_{11,22}^{51,1}}{4608}+\frac{9 \sqrt{\frac{3}{5}} C_{11,22}^{53,1}}{1024}+\frac{\sqrt{\frac{7}{5}} C_{20,20}^{40,1}}{2304}-\frac{\sqrt{7} C_{20,22}^{42,1}}{4608}-\frac{\sqrt{7}
C_{22,20}^{42,1}}{11520}+\frac{11 \sqrt{7} C_{22,22}^{40,1}}{11520}+\frac{7 C_{22,22}^{42,1}}{2880}-\frac{3}{640} \sqrt{\frac{7}{5}} C_{22,22}^{44,1}-\frac{\sqrt{\frac{7}{5}}
C_{31,20}^{31,1}}{2304}-\frac{5 \sqrt{7} C_{31,22}^{31,1}}{4608}+\frac{C_{31,22}^{33,1}}{1024 \sqrt{15}}+\frac{\sqrt{\frac{3}{5}} C_{33,20}^{33,1}}{1280}-\frac{3
C_{33,22}^{33,1}}{640 \sqrt{2}}\)

\noindent\(C_{22,2211,3}^{22,0011,1,\text{NonLoc}}= -\frac{1}{512} \sqrt{\frac{7}{15}} C_{00,00}^{60,0}-\frac{\sqrt{\frac{7}{5}} C_{00,20}^{60,0}}{1536}+\frac{\sqrt{7}
C_{00,22}^{62,0}}{320}-\frac{\sqrt{\frac{7}{15}} C_{11,00}^{51,0}}{1536}-\frac{\sqrt{7} C_{11,22}^{51,0}}{4608}+\frac{9 \sqrt{\frac{3}{5}} C_{11,22}^{53,0}}{1024}+\frac{\sqrt{\frac{7}{15}}
C_{20,00}^{40,0}}{1536}+\frac{\sqrt{\frac{7}{5}} C_{20,20}^{40,0}}{4608}+\frac{\sqrt{7} C_{20,22}^{42,0}}{4608}-\frac{\sqrt{\frac{7}{3}} C_{22,00}^{42,0}}{7680}-\frac{\sqrt{7}
C_{22,20}^{42,0}}{23040}-\frac{11 \sqrt{7} C_{22,22}^{40,0}}{11520}-\frac{7 C_{22,22}^{42,0}}{2880}+\frac{3}{640} \sqrt{\frac{7}{5}} C_{22,22}^{44,0}+\frac{\sqrt{\frac{7}{15}}
C_{31,00}^{31,0}}{1536}+\frac{\sqrt{\frac{7}{5}} C_{31,20}^{31,0}}{4608}-\frac{5 \sqrt{7} C_{31,22}^{31,0}}{4608}+\frac{C_{31,22}^{33,0}}{1024 \sqrt{15}}-\frac{3
C_{33,00}^{33,0}}{2560 \sqrt{5}}-\frac{\sqrt{\frac{3}{5}} C_{33,20}^{33,0}}{2560}-\frac{3 C_{33,22}^{33,0}}{640 \sqrt{2}}\)

\noindent\(C_{22,2212,0}^{11,1101,0,\text{Loc}}= \frac{7}{192} \sqrt{\frac{5}{3}} C_{11,11}^{51,0}+\frac{7}{384} \sqrt{\frac{5}{3}}
C_{11,11}^{51,1}-\frac{5}{192} \sqrt{\frac{5}{3}} C_{31,11}^{31,0}-\frac{5}{384} \sqrt{\frac{5}{3}} C_{31,11}^{31,1}\)

\noindent\(C_{22,2212,0}^{11,1101,0,\text{NonLoc}}= \frac{7}{384} \sqrt{\frac{5}{3}} C_{11,11}^{51,0}+\frac{7}{192} \sqrt{\frac{5}{3}}
C_{11,11}^{51,1}-\frac{5}{384} \sqrt{\frac{5}{3}} C_{31,11}^{31,0}-\frac{5}{192} \sqrt{\frac{5}{3}} C_{31,11}^{31,1}\)

\noindent\(C_{22,2212,0}^{11,1101,1,\text{Loc}}= \frac{7 \sqrt{5} C_{11,11}^{51,1}}{384}-\frac{5 \sqrt{5} C_{31,11}^{31,1}}{384}\)

\noindent\(C_{22,2212,0}^{11,1101,1,\text{NonLoc}}= \frac{7 \sqrt{5} C_{11,11}^{51,0}}{384}-\frac{5 \sqrt{5} C_{31,11}^{31,0}}{384}\)

\noindent\(C_{22,2212,1}^{00,2011,0,\text{Loc}}= -\frac{7}{192} \sqrt{\frac{5}{3}} C_{00,20}^{60,0}-\frac{7}{384} \sqrt{\frac{5}{3}}
C_{00,20}^{60,1}+\frac{7 C_{00,22}^{62,0}}{128 \sqrt{3}}+\frac{7 C_{00,22}^{62,1}}{256 \sqrt{3}}-\frac{7}{576} \sqrt{\frac{5}{3}} C_{11,20}^{31,0}-\frac{7
\sqrt{\frac{5}{3}} C_{11,20}^{31,1}}{1152}+\frac{7 C_{11,22}^{51,0}}{576 \sqrt{3}}+\frac{7 C_{11,22}^{51,1}}{1152 \sqrt{3}}+\frac{3}{256} \sqrt{\frac{7}{5}}
C_{11,22}^{53,0}+\frac{3}{512} \sqrt{\frac{7}{5}} C_{11,22}^{53,1}+\frac{7}{576} \sqrt{\frac{5}{3}} C_{20,20}^{40,0}+\frac{7 \sqrt{\frac{5}{3}} C_{20,20}^{40,1}}{1152}-\frac{7
C_{20,22}^{42,0}}{576 \sqrt{3}}-\frac{7 C_{20,22}^{42,1}}{1152 \sqrt{3}}-\frac{7 C_{22,20}^{42,0}}{576 \sqrt{3}}-\frac{7 C_{22,20}^{42,1}}{1152 \sqrt{3}}-\frac{35
C_{22,22}^{40,0}}{1152 \sqrt{3}}-\frac{35 C_{22,22}^{40,1}}{2304 \sqrt{3}}+\frac{7}{576} \sqrt{\frac{7}{3}} C_{22,22}^{42,0}+\frac{7 \sqrt{\frac{7}{3}}
C_{22,22}^{42,1}}{1152}+\frac{7}{576} \sqrt{\frac{5}{3}} C_{31,20}^{31,0}+\frac{7 \sqrt{\frac{5}{3}} C_{31,20}^{31,1}}{1152}-\frac{7 C_{31,22}^{31,0}}{576
\sqrt{3}}-\frac{7 C_{31,22}^{31,1}}{1152 \sqrt{3}}-\frac{3}{256} \sqrt{\frac{7}{5}} C_{31,22}^{33,0}-\frac{3}{512} \sqrt{\frac{7}{5}} C_{31,22}^{33,1}-\frac{1}{64}
\sqrt{\frac{7}{5}} C_{33,20}^{33,0}-\frac{1}{128} \sqrt{\frac{7}{5}} C_{33,20}^{33,1}+\frac{1}{80} \sqrt{\frac{21}{2}} C_{33,22}^{33,0}+\frac{1}{160}
\sqrt{\frac{21}{2}} C_{33,22}^{33,1}\)

\noindent\(C_{22,2212,1}^{00,2011,0,\text{NonLoc}}= -\frac{7 \sqrt{5} C_{00,00}^{60,0}}{768}-\frac{7 \sqrt{5} C_{00,00}^{60,1}}{384}-\frac{7}{768}
\sqrt{\frac{5}{3}} C_{00,20}^{60,0}-\frac{7}{384} \sqrt{\frac{5}{3}} C_{00,20}^{60,1}-\frac{7 C_{00,22}^{62,0}}{256 \sqrt{3}}-\frac{7 C_{00,22}^{62,1}}{128
\sqrt{3}}+\frac{7 \sqrt{5} C_{11,00}^{51,0}}{2304}+\frac{7 \sqrt{5} C_{11,00}^{51,1}}{1152}+\frac{7 \sqrt{\frac{5}{3}} C_{11,20}^{31,0}}{2304}+\frac{7
\sqrt{\frac{5}{3}} C_{11,20}^{31,1}}{1152}+\frac{7 C_{11,22}^{51,0}}{1152 \sqrt{3}}+\frac{7 C_{11,22}^{51,1}}{576 \sqrt{3}}+\frac{3}{512} \sqrt{\frac{7}{5}}
C_{11,22}^{53,0}+\frac{3}{256} \sqrt{\frac{7}{5}} C_{11,22}^{53,1}+\frac{7 \sqrt{5} C_{20,00}^{40,0}}{2304}+\frac{7 \sqrt{5} C_{20,00}^{40,1}}{1152}+\frac{7
\sqrt{\frac{5}{3}} C_{20,20}^{40,0}}{2304}+\frac{7 \sqrt{\frac{5}{3}} C_{20,20}^{40,1}}{1152}+\frac{7 C_{20,22}^{42,0}}{1152 \sqrt{3}}+\frac{7 C_{20,22}^{42,1}}{576
\sqrt{3}}-\frac{7 C_{22,00}^{42,0}}{2304}-\frac{7 C_{22,00}^{42,1}}{1152}-\frac{7 C_{22,20}^{42,0}}{2304 \sqrt{3}}-\frac{7 C_{22,20}^{42,1}}{1152
\sqrt{3}}+\frac{35 C_{22,22}^{40,0}}{2304 \sqrt{3}}+\frac{35 C_{22,22}^{40,1}}{1152 \sqrt{3}}-\frac{7 \sqrt{\frac{7}{3}} C_{22,22}^{42,0}}{1152}-\frac{7}{576}
\sqrt{\frac{7}{3}} C_{22,22}^{42,1}-\frac{7 \sqrt{5} C_{31,00}^{31,0}}{2304}-\frac{7 \sqrt{5} C_{31,00}^{31,1}}{1152}-\frac{7 \sqrt{\frac{5}{3}}
C_{31,20}^{31,0}}{2304}-\frac{7 \sqrt{\frac{5}{3}} C_{31,20}^{31,1}}{1152}-\frac{7 C_{31,22}^{31,0}}{1152 \sqrt{3}}-\frac{7 C_{31,22}^{31,1}}{576
\sqrt{3}}-\frac{3}{512} \sqrt{\frac{7}{5}} C_{31,22}^{33,0}-\frac{3}{256} \sqrt{\frac{7}{5}} C_{31,22}^{33,1}+\frac{1}{256} \sqrt{\frac{21}{5}} C_{33,00}^{33,0}+\frac{1}{128}
\sqrt{\frac{21}{5}} C_{33,00}^{33,1}+\frac{1}{256} \sqrt{\frac{7}{5}} C_{33,20}^{33,0}+\frac{1}{128} \sqrt{\frac{7}{5}} C_{33,20}^{33,1}+\frac{1}{160}
\sqrt{\frac{21}{2}} C_{33,22}^{33,0}+\frac{1}{80} \sqrt{\frac{21}{2}} C_{33,22}^{33,1}\)

\noindent\(C_{22,2212,1}^{00,2011,1,\text{Loc}}= -\frac{7 \sqrt{5} C_{00,20}^{60,1}}{384}+\frac{7 C_{00,22}^{62,1}}{256}-\frac{7 \sqrt{5}
C_{11,20}^{31,1}}{1152}+\frac{7 C_{11,22}^{51,1}}{1152}+\frac{3}{512} \sqrt{\frac{21}{5}} C_{11,22}^{53,1}+\frac{7 \sqrt{5} C_{20,20}^{40,1}}{1152}-\frac{7
C_{20,22}^{42,1}}{1152}-\frac{7 C_{22,20}^{42,1}}{1152}-\frac{35 C_{22,22}^{40,1}}{2304}+\frac{7 \sqrt{7} C_{22,22}^{42,1}}{1152}+\frac{7 \sqrt{5}
C_{31,20}^{31,1}}{1152}-\frac{7 C_{31,22}^{31,1}}{1152}-\frac{3}{512} \sqrt{\frac{21}{5}} C_{31,22}^{33,1}-\frac{1}{128} \sqrt{\frac{21}{5}} C_{33,20}^{33,1}+\frac{3}{160}
\sqrt{\frac{7}{2}} C_{33,22}^{33,1}\)

\noindent\(C_{22,2212,1}^{00,2011,1,\text{NonLoc}}= -\frac{7}{256} \sqrt{\frac{5}{3}} C_{00,00}^{60,0}-\frac{7 \sqrt{5} C_{00,20}^{60,0}}{768}-\frac{7
C_{00,22}^{62,0}}{256}+\frac{7}{768} \sqrt{\frac{5}{3}} C_{11,00}^{51,0}+\frac{7 \sqrt{5} C_{11,20}^{31,0}}{2304}+\frac{7 C_{11,22}^{51,0}}{1152}+\frac{3}{512}
\sqrt{\frac{21}{5}} C_{11,22}^{53,0}+\frac{7}{768} \sqrt{\frac{5}{3}} C_{20,00}^{40,0}+\frac{7 \sqrt{5} C_{20,20}^{40,0}}{2304}+\frac{7 C_{20,22}^{42,0}}{1152}-\frac{7
C_{22,00}^{42,0}}{768 \sqrt{3}}-\frac{7 C_{22,20}^{42,0}}{2304}+\frac{35 C_{22,22}^{40,0}}{2304}-\frac{7 \sqrt{7} C_{22,22}^{42,0}}{1152}-\frac{7}{768}
\sqrt{\frac{5}{3}} C_{31,00}^{31,0}-\frac{7 \sqrt{5} C_{31,20}^{31,0}}{2304}-\frac{7 C_{31,22}^{31,0}}{1152}-\frac{3}{512} \sqrt{\frac{21}{5}} C_{31,22}^{33,0}+\frac{3}{256}
\sqrt{\frac{7}{5}} C_{33,00}^{33,0}+\frac{1}{256} \sqrt{\frac{21}{5}} C_{33,20}^{33,0}+\frac{3}{160} \sqrt{\frac{7}{2}} C_{33,22}^{33,0}\)

\noindent\(C_{22,2212,1}^{00,2211,0,\text{Loc}}= \frac{7 C_{00,20}^{60,0}}{192 \sqrt{3}}+\frac{7 C_{00,20}^{60,1}}{384 \sqrt{3}}-\frac{7
C_{00,22}^{62,0}}{128 \sqrt{15}}-\frac{7 C_{00,22}^{62,1}}{256 \sqrt{15}}-\frac{77 C_{11,22}^{51,0}}{1152 \sqrt{15}}-\frac{77 C_{11,22}^{51,1}}{2304
\sqrt{15}}+\frac{3 \sqrt{7} C_{11,22}^{53,0}}{640}+\frac{3 \sqrt{7} C_{11,22}^{53,1}}{1280}-\frac{7 C_{20,20}^{40,0}}{576 \sqrt{3}}-\frac{7 C_{20,20}^{40,1}}{1152
\sqrt{3}}+\frac{77 C_{20,22}^{42,0}}{1152 \sqrt{15}}+\frac{77 C_{20,22}^{42,1}}{2304 \sqrt{15}}+\frac{7 C_{22,20}^{42,0}}{576 \sqrt{15}}+\frac{7
C_{22,20}^{42,1}}{1152 \sqrt{15}}-\frac{11 \sqrt{\frac{5}{3}} C_{22,22}^{40,0}}{1152}-\frac{11 \sqrt{\frac{5}{3}} C_{22,22}^{40,1}}{2304}-\frac{7}{576}
\sqrt{\frac{7}{15}} C_{22,22}^{42,0}-\frac{7 \sqrt{\frac{7}{15}} C_{22,22}^{42,1}}{1152}-\frac{7 C_{31,20}^{31,0}}{576 \sqrt{3}}-\frac{7 C_{31,20}^{31,1}}{1152
\sqrt{3}}+\frac{59 C_{31,22}^{31,0}}{1152 \sqrt{15}}+\frac{59 C_{31,22}^{31,1}}{2304 \sqrt{15}}-\frac{\sqrt{7} C_{31,22}^{33,0}}{640}-\frac{\sqrt{7}
C_{31,22}^{33,1}}{1280}+\frac{\sqrt{7} C_{33,20}^{33,0}}{320}+\frac{\sqrt{7} C_{33,20}^{33,1}}{640}-\frac{1}{80} \sqrt{\frac{21}{10}} C_{33,22}^{33,0}-\frac{1}{160}
\sqrt{\frac{21}{10}} C_{33,22}^{33,1}\)

\noindent\(C_{22,2212,1}^{00,2211,0,\text{NonLoc}}= \frac{7 C_{00,00}^{60,0}}{768}+\frac{7 C_{00,00}^{60,1}}{384}+\frac{7 C_{00,20}^{60,0}}{768
\sqrt{3}}+\frac{7 C_{00,20}^{60,1}}{384 \sqrt{3}}+\frac{7 C_{00,22}^{62,0}}{256 \sqrt{15}}+\frac{7 C_{00,22}^{62,1}}{128 \sqrt{15}}-\frac{7 C_{11,00}^{51,0}}{2304}-\frac{7
C_{11,00}^{51,1}}{1152}-\frac{77 C_{11,22}^{51,0}}{2304 \sqrt{15}}-\frac{77 C_{11,22}^{51,1}}{1152 \sqrt{15}}+\frac{3 \sqrt{7} C_{11,22}^{53,0}}{1280}+\frac{3
\sqrt{7} C_{11,22}^{53,1}}{640}-\frac{7 C_{20,00}^{40,0}}{2304}-\frac{7 C_{20,00}^{40,1}}{1152}-\frac{7 C_{20,20}^{40,0}}{2304 \sqrt{3}}-\frac{7
C_{20,20}^{40,1}}{1152 \sqrt{3}}-\frac{77 C_{20,22}^{42,0}}{2304 \sqrt{15}}-\frac{77 C_{20,22}^{42,1}}{1152 \sqrt{15}}+\frac{7 C_{22,00}^{42,0}}{2304
\sqrt{5}}+\frac{7 C_{22,00}^{42,1}}{1152 \sqrt{5}}+\frac{7 C_{22,20}^{42,0}}{2304 \sqrt{15}}+\frac{7 C_{22,20}^{42,1}}{1152 \sqrt{15}}+\frac{11 \sqrt{\frac{5}{3}}
C_{22,22}^{40,0}}{2304}+\frac{11 \sqrt{\frac{5}{3}} C_{22,22}^{40,1}}{1152}+\frac{7 \sqrt{\frac{7}{15}} C_{22,22}^{42,0}}{1152}+\frac{7}{576} \sqrt{\frac{7}{15}}
C_{22,22}^{42,1}+\frac{7 C_{31,00}^{31,0}}{2304}+\frac{7 C_{31,00}^{31,1}}{1152}+\frac{7 C_{31,20}^{31,0}}{2304 \sqrt{3}}+\frac{7 C_{31,20}^{31,1}}{1152
\sqrt{3}}+\frac{59 C_{31,22}^{31,0}}{2304 \sqrt{15}}+\frac{59 C_{31,22}^{31,1}}{1152 \sqrt{15}}-\frac{\sqrt{7} C_{31,22}^{33,0}}{1280}-\frac{\sqrt{7}
C_{31,22}^{33,1}}{640}-\frac{\sqrt{21} C_{33,00}^{33,0}}{1280}-\frac{\sqrt{21} C_{33,00}^{33,1}}{640}-\frac{\sqrt{7} C_{33,20}^{33,0}}{1280}-\frac{\sqrt{7}
C_{33,20}^{33,1}}{640}-\frac{1}{160} \sqrt{\frac{21}{10}} C_{33,22}^{33,0}-\frac{1}{80} \sqrt{\frac{21}{10}} C_{33,22}^{33,1}\)

\noindent\(C_{22,2212,1}^{00,2211,1,\text{Loc}}= \frac{7 C_{00,20}^{60,1}}{384}-\frac{7 C_{00,22}^{62,1}}{256 \sqrt{5}}-\frac{77 C_{11,22}^{51,1}}{2304
\sqrt{5}}+\frac{3 \sqrt{21} C_{11,22}^{53,1}}{1280}-\frac{7 C_{20,20}^{40,1}}{1152}+\frac{77 C_{20,22}^{42,1}}{2304 \sqrt{5}}+\frac{7 C_{22,20}^{42,1}}{1152
\sqrt{5}}-\frac{11 \sqrt{5} C_{22,22}^{40,1}}{2304}-\frac{7 \sqrt{\frac{7}{5}} C_{22,22}^{42,1}}{1152}-\frac{7 C_{31,20}^{31,1}}{1152}+\frac{59 C_{31,22}^{31,1}}{2304
\sqrt{5}}-\frac{\sqrt{21} C_{31,22}^{33,1}}{1280}+\frac{\sqrt{21} C_{33,20}^{33,1}}{640}-\frac{3}{160} \sqrt{\frac{7}{10}} C_{33,22}^{33,1}\)

\noindent\(C_{22,2212,1}^{00,2211,1,\text{NonLoc}}= \frac{7 C_{00,00}^{60,0}}{256 \sqrt{3}}+\frac{7 C_{00,20}^{60,0}}{768}+\frac{7
C_{00,22}^{62,0}}{256 \sqrt{5}}-\frac{7 C_{11,00}^{51,0}}{768 \sqrt{3}}-\frac{77 C_{11,22}^{51,0}}{2304 \sqrt{5}}+\frac{3 \sqrt{21} C_{11,22}^{53,0}}{1280}-\frac{7
C_{20,00}^{40,0}}{768 \sqrt{3}}-\frac{7 C_{20,20}^{40,0}}{2304}-\frac{77 C_{20,22}^{42,0}}{2304 \sqrt{5}}+\frac{7 C_{22,00}^{42,0}}{768 \sqrt{15}}+\frac{7
C_{22,20}^{42,0}}{2304 \sqrt{5}}+\frac{11 \sqrt{5} C_{22,22}^{40,0}}{2304}+\frac{7 \sqrt{\frac{7}{5}} C_{22,22}^{42,0}}{1152}+\frac{7 C_{31,00}^{31,0}}{768
\sqrt{3}}+\frac{7 C_{31,20}^{31,0}}{2304}+\frac{59 C_{31,22}^{31,0}}{2304 \sqrt{5}}-\frac{\sqrt{21} C_{31,22}^{33,0}}{1280}-\frac{3 \sqrt{7} C_{33,00}^{33,0}}{1280}-\frac{\sqrt{21}
C_{33,20}^{33,0}}{1280}-\frac{3}{160} \sqrt{\frac{7}{10}} C_{33,22}^{33,0}\)

\noindent\(C_{22,2212,1}^{11,1101,0,\text{Loc}}= -\frac{7}{384} \sqrt{\frac{5}{3}} C_{11,11}^{51,0}-\frac{7}{768} \sqrt{\frac{5}{3}}
C_{11,11}^{51,1}+\frac{13}{384} \sqrt{\frac{5}{3}} C_{31,11}^{31,0}+\frac{13}{768} \sqrt{\frac{5}{3}} C_{31,11}^{31,1}-\frac{1}{32} \sqrt{\frac{21}{10}}
C_{33,11}^{33,0}-\frac{1}{64} \sqrt{\frac{21}{10}} C_{33,11}^{33,1}\)

\noindent\(C_{22,2212,1}^{11,1101,0,\text{NonLoc}}= -\frac{7}{768} \sqrt{\frac{5}{3}} C_{11,11}^{51,0}-\frac{7}{384} \sqrt{\frac{5}{3}}
C_{11,11}^{51,1}+\frac{13}{768} \sqrt{\frac{5}{3}} C_{31,11}^{31,0}+\frac{13}{384} \sqrt{\frac{5}{3}} C_{31,11}^{31,1}-\frac{1}{64} \sqrt{\frac{21}{10}}
C_{33,11}^{33,0}-\frac{1}{32} \sqrt{\frac{21}{10}} C_{33,11}^{33,1}\)

\noindent\(C_{22,2212,1}^{11,1101,1,\text{Loc}}= -\frac{7 \sqrt{5} C_{11,11}^{51,1}}{768}+\frac{13 \sqrt{5} C_{31,11}^{31,1}}{768}-\frac{3}{64}
\sqrt{\frac{7}{10}} C_{33,11}^{33,1}\)

\noindent\(C_{22,2212,1}^{11,1101,1,\text{NonLoc}}= -\frac{7 \sqrt{5} C_{11,11}^{51,0}}{768}+\frac{13 \sqrt{5} C_{31,11}^{31,0}}{768}-\frac{3}{64}
\sqrt{\frac{7}{10}} C_{33,11}^{33,0}\)

\noindent\(C_{22,2212,1}^{20,0011,0,\text{Loc}}= \frac{7}{768} \sqrt{\frac{5}{3}} C_{00,20}^{60,0}+\frac{7 \sqrt{\frac{5}{3}} C_{00,20}^{60,1}}{1536}-\frac{7
C_{00,22}^{62,0}}{512 \sqrt{3}}-\frac{7 C_{00,22}^{62,1}}{1024 \sqrt{3}}-\frac{7 \sqrt{\frac{5}{3}} C_{11,20}^{31,0}}{2304}-\frac{7 \sqrt{\frac{5}{3}}
C_{11,20}^{31,1}}{4608}+\frac{7 C_{11,22}^{51,0}}{2304 \sqrt{3}}+\frac{7 C_{11,22}^{51,1}}{4608 \sqrt{3}}+\frac{3 \sqrt{\frac{7}{5}} C_{11,22}^{53,0}}{1024}+\frac{3
\sqrt{\frac{7}{5}} C_{11,22}^{53,1}}{2048}-\frac{7 \sqrt{\frac{5}{3}} C_{20,20}^{40,0}}{2304}-\frac{7 \sqrt{\frac{5}{3}} C_{20,20}^{40,1}}{4608}+\frac{7
C_{20,22}^{42,0}}{2304 \sqrt{3}}+\frac{7 C_{20,22}^{42,1}}{4608 \sqrt{3}}+\frac{7 C_{22,20}^{42,0}}{2304 \sqrt{3}}+\frac{7 C_{22,20}^{42,1}}{4608
\sqrt{3}}+\frac{35 C_{22,22}^{40,0}}{4608 \sqrt{3}}+\frac{35 C_{22,22}^{40,1}}{9216 \sqrt{3}}-\frac{7 \sqrt{\frac{7}{3}} C_{22,22}^{42,0}}{2304}-\frac{7
\sqrt{\frac{7}{3}} C_{22,22}^{42,1}}{4608}+\frac{7 \sqrt{\frac{5}{3}} C_{31,20}^{31,0}}{2304}+\frac{7 \sqrt{\frac{5}{3}} C_{31,20}^{31,1}}{4608}-\frac{7
C_{31,22}^{31,0}}{2304 \sqrt{3}}-\frac{7 C_{31,22}^{31,1}}{4608 \sqrt{3}}-\frac{3 \sqrt{\frac{7}{5}} C_{31,22}^{33,0}}{1024}-\frac{3 \sqrt{\frac{7}{5}}
C_{31,22}^{33,1}}{2048}-\frac{1}{256} \sqrt{\frac{7}{5}} C_{33,20}^{33,0}-\frac{1}{512} \sqrt{\frac{7}{5}} C_{33,20}^{33,1}+\frac{1}{320} \sqrt{\frac{21}{2}}
C_{33,22}^{33,0}+\frac{1}{640} \sqrt{\frac{21}{2}} C_{33,22}^{33,1}\)

\noindent\(C_{22,2212,1}^{20,0011,0,\text{NonLoc}}= \frac{7 \sqrt{5} C_{00,00}^{60,0}}{3072}+\frac{7 \sqrt{5} C_{00,00}^{60,1}}{1536}+\frac{7
\sqrt{\frac{5}{3}} C_{00,20}^{60,0}}{3072}+\frac{7 \sqrt{\frac{5}{3}} C_{00,20}^{60,1}}{1536}+\frac{7 C_{00,22}^{62,0}}{1024 \sqrt{3}}+\frac{7 C_{00,22}^{62,1}}{512
\sqrt{3}}+\frac{7 \sqrt{5} C_{11,00}^{51,0}}{9216}+\frac{7 \sqrt{5} C_{11,00}^{51,1}}{4608}+\frac{7 \sqrt{\frac{5}{3}} C_{11,20}^{31,0}}{9216}+\frac{7
\sqrt{\frac{5}{3}} C_{11,20}^{31,1}}{4608}+\frac{7 C_{11,22}^{51,0}}{4608 \sqrt{3}}+\frac{7 C_{11,22}^{51,1}}{2304 \sqrt{3}}+\frac{3 \sqrt{\frac{7}{5}}
C_{11,22}^{53,0}}{2048}+\frac{3 \sqrt{\frac{7}{5}} C_{11,22}^{53,1}}{1024}-\frac{7 \sqrt{5} C_{20,00}^{40,0}}{9216}-\frac{7 \sqrt{5} C_{20,00}^{40,1}}{4608}-\frac{7
\sqrt{\frac{5}{3}} C_{20,20}^{40,0}}{9216}-\frac{7 \sqrt{\frac{5}{3}} C_{20,20}^{40,1}}{4608}-\frac{7 C_{20,22}^{42,0}}{4608 \sqrt{3}}-\frac{7 C_{20,22}^{42,1}}{2304
\sqrt{3}}+\frac{7 C_{22,00}^{42,0}}{9216}+\frac{7 C_{22,00}^{42,1}}{4608}+\frac{7 C_{22,20}^{42,0}}{9216 \sqrt{3}}+\frac{7 C_{22,20}^{42,1}}{4608
\sqrt{3}}-\frac{35 C_{22,22}^{40,0}}{9216 \sqrt{3}}-\frac{35 C_{22,22}^{40,1}}{4608 \sqrt{3}}+\frac{7 \sqrt{\frac{7}{3}} C_{22,22}^{42,0}}{4608}+\frac{7
\sqrt{\frac{7}{3}} C_{22,22}^{42,1}}{2304}-\frac{7 \sqrt{5} C_{31,00}^{31,0}}{9216}-\frac{7 \sqrt{5} C_{31,00}^{31,1}}{4608}-\frac{7 \sqrt{\frac{5}{3}}
C_{31,20}^{31,0}}{9216}-\frac{7 \sqrt{\frac{5}{3}} C_{31,20}^{31,1}}{4608}-\frac{7 C_{31,22}^{31,0}}{4608 \sqrt{3}}-\frac{7 C_{31,22}^{31,1}}{2304
\sqrt{3}}-\frac{3 \sqrt{\frac{7}{5}} C_{31,22}^{33,0}}{2048}-\frac{3 \sqrt{\frac{7}{5}} C_{31,22}^{33,1}}{1024}+\frac{\sqrt{\frac{21}{5}} C_{33,00}^{33,0}}{1024}+\frac{1}{512}
\sqrt{\frac{21}{5}} C_{33,00}^{33,1}+\frac{\sqrt{\frac{7}{5}} C_{33,20}^{33,0}}{1024}+\frac{1}{512} \sqrt{\frac{7}{5}} C_{33,20}^{33,1}+\frac{1}{640}
\sqrt{\frac{21}{2}} C_{33,22}^{33,0}+\frac{1}{320} \sqrt{\frac{21}{2}} C_{33,22}^{33,1}\)

\noindent\(C_{22,2212,1}^{20,0011,1,\text{Loc}}= \frac{7 \sqrt{5} C_{00,20}^{60,1}}{1536}-\frac{7 C_{00,22}^{62,1}}{1024}-\frac{7 \sqrt{5}
C_{11,20}^{31,1}}{4608}+\frac{7 C_{11,22}^{51,1}}{4608}+\frac{3 \sqrt{\frac{21}{5}} C_{11,22}^{53,1}}{2048}-\frac{7 \sqrt{5} C_{20,20}^{40,1}}{4608}+\frac{7
C_{20,22}^{42,1}}{4608}+\frac{7 C_{22,20}^{42,1}}{4608}+\frac{35 C_{22,22}^{40,1}}{9216}-\frac{7 \sqrt{7} C_{22,22}^{42,1}}{4608}+\frac{7 \sqrt{5}
C_{31,20}^{31,1}}{4608}-\frac{7 C_{31,22}^{31,1}}{4608}-\frac{3 \sqrt{\frac{21}{5}} C_{31,22}^{33,1}}{2048}-\frac{1}{512} \sqrt{\frac{21}{5}} C_{33,20}^{33,1}+\frac{3}{640}
\sqrt{\frac{7}{2}} C_{33,22}^{33,1}\)

\noindent\(C_{22,2212,1}^{20,0011,1,\text{NonLoc}}= \frac{7 \sqrt{\frac{5}{3}} C_{00,00}^{60,0}}{1024}+\frac{7 \sqrt{5} C_{00,20}^{60,0}}{3072}+\frac{7
C_{00,22}^{62,0}}{1024}+\frac{7 \sqrt{\frac{5}{3}} C_{11,00}^{51,0}}{3072}+\frac{7 \sqrt{5} C_{11,20}^{31,0}}{9216}+\frac{7 C_{11,22}^{51,0}}{4608}+\frac{3
\sqrt{\frac{21}{5}} C_{11,22}^{53,0}}{2048}-\frac{7 \sqrt{\frac{5}{3}} C_{20,00}^{40,0}}{3072}-\frac{7 \sqrt{5} C_{20,20}^{40,0}}{9216}-\frac{7 C_{20,22}^{42,0}}{4608}+\frac{7
C_{22,00}^{42,0}}{3072 \sqrt{3}}+\frac{7 C_{22,20}^{42,0}}{9216}-\frac{35 C_{22,22}^{40,0}}{9216}+\frac{7 \sqrt{7} C_{22,22}^{42,0}}{4608}-\frac{7
\sqrt{\frac{5}{3}} C_{31,00}^{31,0}}{3072}-\frac{7 \sqrt{5} C_{31,20}^{31,0}}{9216}-\frac{7 C_{31,22}^{31,0}}{4608}-\frac{3 \sqrt{\frac{21}{5}} C_{31,22}^{33,0}}{2048}+\frac{3
\sqrt{\frac{7}{5}} C_{33,00}^{33,0}}{1024}+\frac{\sqrt{\frac{21}{5}} C_{33,20}^{33,0}}{1024}+\frac{3}{640} \sqrt{\frac{7}{2}} C_{33,22}^{33,0}\)

\noindent\(C_{22,2212,1}^{22,0011,0,\text{Loc}}= -\frac{7 C_{00,20}^{60,0}}{768 \sqrt{3}}-\frac{7 C_{00,20}^{60,1}}{1536 \sqrt{3}}+\frac{7
C_{00,22}^{62,0}}{512 \sqrt{15}}+\frac{7 C_{00,22}^{62,1}}{1024 \sqrt{15}}-\frac{7 C_{11,22}^{51,0}}{2304 \sqrt{15}}-\frac{7 C_{11,22}^{51,1}}{4608
\sqrt{15}}-\frac{3 \sqrt{7} C_{11,22}^{53,0}}{5120}-\frac{3 \sqrt{7} C_{11,22}^{53,1}}{10240}+\frac{7 C_{20,20}^{40,0}}{2304 \sqrt{3}}+\frac{7 C_{20,20}^{40,1}}{4608
\sqrt{3}}-\frac{7 C_{20,22}^{42,0}}{2304 \sqrt{15}}-\frac{7 C_{20,22}^{42,1}}{4608 \sqrt{15}}-\frac{7 C_{22,20}^{42,0}}{2304 \sqrt{15}}-\frac{7 C_{22,20}^{42,1}}{4608
\sqrt{15}}-\frac{7 \sqrt{\frac{5}{3}} C_{22,22}^{40,0}}{4608}-\frac{7 \sqrt{\frac{5}{3}} C_{22,22}^{40,1}}{9216}+\frac{7 \sqrt{\frac{7}{15}} C_{22,22}^{42,0}}{2304}+\frac{7
\sqrt{\frac{7}{15}} C_{22,22}^{42,1}}{4608}-\frac{7 C_{31,20}^{31,0}}{2304 \sqrt{3}}-\frac{7 C_{31,20}^{31,1}}{4608 \sqrt{3}}+\frac{7 C_{31,22}^{31,0}}{2304
\sqrt{15}}+\frac{7 C_{31,22}^{31,1}}{4608 \sqrt{15}}+\frac{3 \sqrt{7} C_{31,22}^{33,0}}{5120}+\frac{3 \sqrt{7} C_{31,22}^{33,1}}{10240}+\frac{\sqrt{7}
C_{33,20}^{33,0}}{1280}+\frac{\sqrt{7} C_{33,20}^{33,1}}{2560}-\frac{1}{320} \sqrt{\frac{21}{10}} C_{33,22}^{33,0}-\frac{1}{640} \sqrt{\frac{21}{10}}
C_{33,22}^{33,1}\)

\noindent\(C_{22,2212,1}^{22,0011,0,\text{NonLoc}}= -\frac{7 C_{00,00}^{60,0}}{3072}-\frac{7 C_{00,00}^{60,1}}{1536}-\frac{7 C_{00,20}^{60,0}}{3072
\sqrt{3}}-\frac{7 C_{00,20}^{60,1}}{1536 \sqrt{3}}-\frac{7 C_{00,22}^{62,0}}{1024 \sqrt{15}}-\frac{7 C_{00,22}^{62,1}}{512 \sqrt{15}}-\frac{7 C_{11,00}^{51,0}}{9216}-\frac{7
C_{11,00}^{51,1}}{4608}-\frac{7 C_{11,22}^{51,0}}{4608 \sqrt{15}}-\frac{7 C_{11,22}^{51,1}}{2304 \sqrt{15}}-\frac{3 \sqrt{7} C_{11,22}^{53,0}}{10240}-\frac{3
\sqrt{7} C_{11,22}^{53,1}}{5120}+\frac{7 C_{20,00}^{40,0}}{9216}+\frac{7 C_{20,00}^{40,1}}{4608}+\frac{7 C_{20,20}^{40,0}}{9216 \sqrt{3}}+\frac{7
C_{20,20}^{40,1}}{4608 \sqrt{3}}+\frac{7 C_{20,22}^{42,0}}{4608 \sqrt{15}}+\frac{7 C_{20,22}^{42,1}}{2304 \sqrt{15}}-\frac{7 C_{22,00}^{42,0}}{9216
\sqrt{5}}-\frac{7 C_{22,00}^{42,1}}{4608 \sqrt{5}}-\frac{7 C_{22,20}^{42,0}}{9216 \sqrt{15}}-\frac{7 C_{22,20}^{42,1}}{4608 \sqrt{15}}+\frac{7 \sqrt{\frac{5}{3}}
C_{22,22}^{40,0}}{9216}+\frac{7 \sqrt{\frac{5}{3}} C_{22,22}^{40,1}}{4608}-\frac{7 \sqrt{\frac{7}{15}} C_{22,22}^{42,0}}{4608}-\frac{7 \sqrt{\frac{7}{15}}
C_{22,22}^{42,1}}{2304}+\frac{7 C_{31,00}^{31,0}}{9216}+\frac{7 C_{31,00}^{31,1}}{4608}+\frac{7 C_{31,20}^{31,0}}{9216 \sqrt{3}}+\frac{7 C_{31,20}^{31,1}}{4608
\sqrt{3}}+\frac{7 C_{31,22}^{31,0}}{4608 \sqrt{15}}+\frac{7 C_{31,22}^{31,1}}{2304 \sqrt{15}}+\frac{3 \sqrt{7} C_{31,22}^{33,0}}{10240}+\frac{3 \sqrt{7}
C_{31,22}^{33,1}}{5120}-\frac{\sqrt{21} C_{33,00}^{33,0}}{5120}-\frac{\sqrt{21} C_{33,00}^{33,1}}{2560}-\frac{\sqrt{7} C_{33,20}^{33,0}}{5120}-\frac{\sqrt{7}
C_{33,20}^{33,1}}{2560}-\frac{1}{640} \sqrt{\frac{21}{10}} C_{33,22}^{33,0}-\frac{1}{320} \sqrt{\frac{21}{10}} C_{33,22}^{33,1}\)

\noindent\(C_{22,2212,1}^{22,0011,1,\text{Loc}}= -\frac{7 C_{00,20}^{60,1}}{1536}+\frac{7 C_{00,22}^{62,1}}{1024 \sqrt{5}}-\frac{7
C_{11,22}^{51,1}}{4608 \sqrt{5}}-\frac{3 \sqrt{21} C_{11,22}^{53,1}}{10240}+\frac{7 C_{20,20}^{40,1}}{4608}-\frac{7 C_{20,22}^{42,1}}{4608 \sqrt{5}}-\frac{7
C_{22,20}^{42,1}}{4608 \sqrt{5}}-\frac{7 \sqrt{5} C_{22,22}^{40,1}}{9216}+\frac{7 \sqrt{\frac{7}{5}} C_{22,22}^{42,1}}{4608}-\frac{7 C_{31,20}^{31,1}}{4608}+\frac{7
C_{31,22}^{31,1}}{4608 \sqrt{5}}+\frac{3 \sqrt{21} C_{31,22}^{33,1}}{10240}+\frac{\sqrt{21} C_{33,20}^{33,1}}{2560}-\frac{3}{640} \sqrt{\frac{7}{10}}
C_{33,22}^{33,1}\)

\noindent\(C_{22,2212,1}^{22,0011,1,\text{NonLoc}}= -\frac{7 C_{00,00}^{60,0}}{1024 \sqrt{3}}-\frac{7 C_{00,20}^{60,0}}{3072}-\frac{7
C_{00,22}^{62,0}}{1024 \sqrt{5}}-\frac{7 C_{11,00}^{51,0}}{3072 \sqrt{3}}-\frac{7 C_{11,22}^{51,0}}{4608 \sqrt{5}}-\frac{3 \sqrt{21} C_{11,22}^{53,0}}{10240}+\frac{7
C_{20,00}^{40,0}}{3072 \sqrt{3}}+\frac{7 C_{20,20}^{40,0}}{9216}+\frac{7 C_{20,22}^{42,0}}{4608 \sqrt{5}}-\frac{7 C_{22,00}^{42,0}}{3072 \sqrt{15}}-\frac{7
C_{22,20}^{42,0}}{9216 \sqrt{5}}+\frac{7 \sqrt{5} C_{22,22}^{40,0}}{9216}-\frac{7 \sqrt{\frac{7}{5}} C_{22,22}^{42,0}}{4608}+\frac{7 C_{31,00}^{31,0}}{3072
\sqrt{3}}+\frac{7 C_{31,20}^{31,0}}{9216}+\frac{7 C_{31,22}^{31,0}}{4608 \sqrt{5}}+\frac{3 \sqrt{21} C_{31,22}^{33,0}}{10240}-\frac{3 \sqrt{7} C_{33,00}^{33,0}}{5120}-\frac{\sqrt{21}
C_{33,20}^{33,0}}{5120}-\frac{3}{640} \sqrt{\frac{7}{10}} C_{33,22}^{33,0}\)

\noindent\(C_{22,2212,2}^{00,2212,0,\text{Loc}}= \frac{1}{192} \sqrt{\frac{35}{3}} C_{00,20}^{60,0}+\frac{1}{384} \sqrt{\frac{35}{3}}
C_{00,20}^{60,1}-\frac{1}{128} \sqrt{\frac{7}{3}} C_{00,22}^{62,0}-\frac{1}{256} \sqrt{\frac{7}{3}} C_{00,22}^{62,1}-\frac{5 \sqrt{\frac{7}{3}} C_{11,22}^{51,0}}{1152}-\frac{5
\sqrt{\frac{7}{3}} C_{11,22}^{51,1}}{2304}-\frac{1}{576} \sqrt{\frac{35}{3}} C_{20,20}^{40,0}-\frac{\sqrt{\frac{35}{3}} C_{20,20}^{40,1}}{1152}+\frac{5
\sqrt{\frac{7}{3}} C_{20,22}^{42,0}}{1152}+\frac{5 \sqrt{\frac{7}{3}} C_{20,22}^{42,1}}{2304}+\frac{1}{576} \sqrt{\frac{7}{3}} C_{22,20}^{42,0}+\frac{\sqrt{\frac{7}{3}}
C_{22,20}^{42,1}}{1152}-\frac{13 \sqrt{\frac{7}{3}} C_{22,22}^{40,0}}{1152}-\frac{13 \sqrt{\frac{7}{3}} C_{22,22}^{40,1}}{2304}+\frac{17 C_{22,22}^{42,0}}{576
\sqrt{3}}+\frac{17 C_{22,22}^{42,1}}{1152 \sqrt{3}}-\frac{1}{16} \sqrt{\frac{3}{35}} C_{22,22}^{44,0}-\frac{1}{32} \sqrt{\frac{3}{35}} C_{22,22}^{44,1}-\frac{1}{576}
\sqrt{\frac{35}{3}} C_{31,20}^{31,0}-\frac{\sqrt{\frac{35}{3}} C_{31,20}^{31,1}}{1152}+\frac{11 \sqrt{\frac{7}{3}} C_{31,22}^{31,0}}{1152}+\frac{11
\sqrt{\frac{7}{3}} C_{31,22}^{31,1}}{2304}-\frac{7 C_{31,22}^{33,0}}{192 \sqrt{5}}-\frac{7 C_{31,22}^{33,1}}{384 \sqrt{5}}+\frac{C_{33,20}^{33,0}}{64
\sqrt{5}}+\frac{C_{33,20}^{33,1}}{128 \sqrt{5}}+\frac{1}{40} \sqrt{\frac{3}{2}} C_{33,22}^{33,0}+\frac{1}{80} \sqrt{\frac{3}{2}} C_{33,22}^{33,1}\)

\noindent\(C_{22,2212,2}^{00,2212,0,\text{NonLoc}}= \frac{\sqrt{35} C_{00,00}^{60,0}}{768}+\frac{\sqrt{35} C_{00,00}^{60,1}}{384}+\frac{1}{768}
\sqrt{\frac{35}{3}} C_{00,20}^{60,0}+\frac{1}{384} \sqrt{\frac{35}{3}} C_{00,20}^{60,1}+\frac{1}{256} \sqrt{\frac{7}{3}} C_{00,22}^{62,0}+\frac{1}{128}
\sqrt{\frac{7}{3}} C_{00,22}^{62,1}-\frac{\sqrt{35} C_{11,00}^{51,0}}{2304}-\frac{\sqrt{35} C_{11,00}^{51,1}}{1152}-\frac{5 \sqrt{\frac{7}{3}} C_{11,22}^{51,0}}{2304}-\frac{5
\sqrt{\frac{7}{3}} C_{11,22}^{51,1}}{1152}-\frac{\sqrt{35} C_{20,00}^{40,0}}{2304}-\frac{\sqrt{35} C_{20,00}^{40,1}}{1152}-\frac{\sqrt{\frac{35}{3}}
C_{20,20}^{40,0}}{2304}-\frac{\sqrt{\frac{35}{3}} C_{20,20}^{40,1}}{1152}-\frac{5 \sqrt{\frac{7}{3}} C_{20,22}^{42,0}}{2304}-\frac{5 \sqrt{\frac{7}{3}}
C_{20,22}^{42,1}}{1152}+\frac{\sqrt{7} C_{22,00}^{42,0}}{2304}+\frac{\sqrt{7} C_{22,00}^{42,1}}{1152}+\frac{\sqrt{\frac{7}{3}} C_{22,20}^{42,0}}{2304}+\frac{\sqrt{\frac{7}{3}}
C_{22,20}^{42,1}}{1152}+\frac{13 \sqrt{\frac{7}{3}} C_{22,22}^{40,0}}{2304}+\frac{13 \sqrt{\frac{7}{3}} C_{22,22}^{40,1}}{1152}-\frac{17 C_{22,22}^{42,0}}{1152
\sqrt{3}}-\frac{17 C_{22,22}^{42,1}}{576 \sqrt{3}}+\frac{1}{32} \sqrt{\frac{3}{35}} C_{22,22}^{44,0}+\frac{1}{16} \sqrt{\frac{3}{35}} C_{22,22}^{44,1}+\frac{\sqrt{35}
C_{31,00}^{31,0}}{2304}+\frac{\sqrt{35} C_{31,00}^{31,1}}{1152}+\frac{\sqrt{\frac{35}{3}} C_{31,20}^{31,0}}{2304}+\frac{\sqrt{\frac{35}{3}} C_{31,20}^{31,1}}{1152}+\frac{11
\sqrt{\frac{7}{3}} C_{31,22}^{31,0}}{2304}+\frac{11 \sqrt{\frac{7}{3}} C_{31,22}^{31,1}}{1152}-\frac{7 C_{31,22}^{33,0}}{384 \sqrt{5}}-\frac{7 C_{31,22}^{33,1}}{192
\sqrt{5}}-\frac{1}{256} \sqrt{\frac{3}{5}} C_{33,00}^{33,0}-\frac{1}{128} \sqrt{\frac{3}{5}} C_{33,00}^{33,1}-\frac{C_{33,20}^{33,0}}{256 \sqrt{5}}-\frac{C_{33,20}^{33,1}}{128
\sqrt{5}}+\frac{1}{80} \sqrt{\frac{3}{2}} C_{33,22}^{33,0}+\frac{1}{40} \sqrt{\frac{3}{2}} C_{33,22}^{33,1}\)

\noindent\(C_{22,2212,2}^{00,2212,1,\text{Loc}}= \frac{\sqrt{35} C_{00,20}^{60,1}}{384}-\frac{\sqrt{7} C_{00,22}^{62,1}}{256}-\frac{5
\sqrt{7} C_{11,22}^{51,1}}{2304}-\frac{\sqrt{35} C_{20,20}^{40,1}}{1152}+\frac{5 \sqrt{7} C_{20,22}^{42,1}}{2304}+\frac{\sqrt{7} C_{22,20}^{42,1}}{1152}-\frac{13
\sqrt{7} C_{22,22}^{40,1}}{2304}+\frac{17 C_{22,22}^{42,1}}{1152}-\frac{3 C_{22,22}^{44,1}}{32 \sqrt{35}}-\frac{\sqrt{35} C_{31,20}^{31,1}}{1152}+\frac{11
\sqrt{7} C_{31,22}^{31,1}}{2304}-\frac{7 C_{31,22}^{33,1}}{128 \sqrt{15}}+\frac{1}{128} \sqrt{\frac{3}{5}} C_{33,20}^{33,1}+\frac{3 C_{33,22}^{33,1}}{80
\sqrt{2}}\)

\noindent\(C_{22,2212,2}^{00,2212,1,\text{NonLoc}}= \frac{1}{256} \sqrt{\frac{35}{3}} C_{00,00}^{60,0}+\frac{\sqrt{35} C_{00,20}^{60,0}}{768}+\frac{\sqrt{7}
C_{00,22}^{62,0}}{256}-\frac{1}{768} \sqrt{\frac{35}{3}} C_{11,00}^{51,0}-\frac{5 \sqrt{7} C_{11,22}^{51,0}}{2304}-\frac{1}{768} \sqrt{\frac{35}{3}}
C_{20,00}^{40,0}-\frac{\sqrt{35} C_{20,20}^{40,0}}{2304}-\frac{5 \sqrt{7} C_{20,22}^{42,0}}{2304}+\frac{1}{768} \sqrt{\frac{7}{3}} C_{22,00}^{42,0}+\frac{\sqrt{7}
C_{22,20}^{42,0}}{2304}+\frac{13 \sqrt{7} C_{22,22}^{40,0}}{2304}-\frac{17 C_{22,22}^{42,0}}{1152}+\frac{3 C_{22,22}^{44,0}}{32 \sqrt{35}}+\frac{1}{768}
\sqrt{\frac{35}{3}} C_{31,00}^{31,0}+\frac{\sqrt{35} C_{31,20}^{31,0}}{2304}+\frac{11 \sqrt{7} C_{31,22}^{31,0}}{2304}-\frac{7 C_{31,22}^{33,0}}{128
\sqrt{15}}-\frac{3 C_{33,00}^{33,0}}{256 \sqrt{5}}-\frac{1}{256} \sqrt{\frac{3}{5}} C_{33,20}^{33,0}+\frac{3 C_{33,22}^{33,0}}{80 \sqrt{2}}\)

\noindent\(C_{22,2212,2}^{11,1101,0,\text{Loc}}= \frac{1}{128} \sqrt{\frac{35}{3}} C_{11,11}^{51,0}+\frac{1}{256} \sqrt{\frac{35}{3}}
C_{11,11}^{51,1}-\frac{1}{384} \sqrt{\frac{35}{3}} C_{31,11}^{31,0}-\frac{1}{768} \sqrt{\frac{35}{3}} C_{31,11}^{31,1}-\frac{1}{32} \sqrt{\frac{3}{10}}
C_{33,11}^{33,0}-\frac{1}{64} \sqrt{\frac{3}{10}} C_{33,11}^{33,1}\)

\noindent\(C_{22,2212,2}^{11,1101,0,\text{NonLoc}}= \frac{1}{256} \sqrt{\frac{35}{3}} C_{11,11}^{51,0}+\frac{1}{128} \sqrt{\frac{35}{3}}
C_{11,11}^{51,1}-\frac{1}{768} \sqrt{\frac{35}{3}} C_{31,11}^{31,0}-\frac{1}{384} \sqrt{\frac{35}{3}} C_{31,11}^{31,1}-\frac{1}{64} \sqrt{\frac{3}{10}}
C_{33,11}^{33,0}-\frac{1}{32} \sqrt{\frac{3}{10}} C_{33,11}^{33,1}\)

\noindent\(C_{22,2212,2}^{11,1101,1,\text{Loc}}= \frac{\sqrt{35} C_{11,11}^{51,1}}{256}-\frac{\sqrt{35} C_{31,11}^{31,1}}{768}-\frac{3
C_{33,11}^{33,1}}{64 \sqrt{10}}\)

\noindent\(C_{22,2212,2}^{11,1101,1,\text{NonLoc}}= \frac{\sqrt{35} C_{11,11}^{51,0}}{256}-\frac{\sqrt{35} C_{31,11}^{31,0}}{768}-\frac{3
C_{33,11}^{33,0}}{64 \sqrt{10}}\)

\noindent\(C_{22,2212,2}^{22,0011,0,\text{Loc}}= -\frac{1}{768} \sqrt{\frac{35}{3}} C_{00,20}^{60,0}-\frac{\sqrt{\frac{35}{3}} C_{00,20}^{60,1}}{1536}+\frac{1}{512}
\sqrt{\frac{7}{3}} C_{00,22}^{62,0}+\frac{\sqrt{\frac{7}{3}} C_{00,22}^{62,1}}{1024}-\frac{\sqrt{\frac{7}{3}} C_{11,22}^{51,0}}{2304}-\frac{\sqrt{\frac{7}{3}}
C_{11,22}^{51,1}}{4608}-\frac{3 C_{11,22}^{53,0}}{1024 \sqrt{5}}-\frac{3 C_{11,22}^{53,1}}{2048 \sqrt{5}}+\frac{\sqrt{\frac{35}{3}} C_{20,20}^{40,0}}{2304}+\frac{\sqrt{\frac{35}{3}}
C_{20,20}^{40,1}}{4608}-\frac{\sqrt{\frac{7}{3}} C_{20,22}^{42,0}}{2304}-\frac{\sqrt{\frac{7}{3}} C_{20,22}^{42,1}}{4608}-\frac{\sqrt{\frac{7}{3}}
C_{22,20}^{42,0}}{2304}-\frac{\sqrt{\frac{7}{3}} C_{22,20}^{42,1}}{4608}-\frac{5 \sqrt{\frac{7}{3}} C_{22,22}^{40,0}}{4608}-\frac{5 \sqrt{\frac{7}{3}}
C_{22,22}^{40,1}}{9216}+\frac{7 C_{22,22}^{42,0}}{2304 \sqrt{3}}+\frac{7 C_{22,22}^{42,1}}{4608 \sqrt{3}}-\frac{\sqrt{\frac{35}{3}} C_{31,20}^{31,0}}{2304}-\frac{\sqrt{\frac{35}{3}}
C_{31,20}^{31,1}}{4608}+\frac{\sqrt{\frac{7}{3}} C_{31,22}^{31,0}}{2304}+\frac{\sqrt{\frac{7}{3}} C_{31,22}^{31,1}}{4608}+\frac{3 C_{31,22}^{33,0}}{1024
\sqrt{5}}+\frac{3 C_{31,22}^{33,1}}{2048 \sqrt{5}}+\frac{C_{33,20}^{33,0}}{256 \sqrt{5}}+\frac{C_{33,20}^{33,1}}{512 \sqrt{5}}-\frac{1}{320} \sqrt{\frac{3}{2}}
C_{33,22}^{33,0}-\frac{1}{640} \sqrt{\frac{3}{2}} C_{33,22}^{33,1}\)

\noindent\(C_{22,2212,2}^{22,0011,0,\text{NonLoc}}= -\frac{\sqrt{35} C_{00,00}^{60,0}}{3072}-\frac{\sqrt{35} C_{00,00}^{60,1}}{1536}-\frac{\sqrt{\frac{35}{3}}
C_{00,20}^{60,0}}{3072}-\frac{\sqrt{\frac{35}{3}} C_{00,20}^{60,1}}{1536}-\frac{\sqrt{\frac{7}{3}} C_{00,22}^{62,0}}{1024}-\frac{1}{512} \sqrt{\frac{7}{3}}
C_{00,22}^{62,1}-\frac{\sqrt{35} C_{11,00}^{51,0}}{9216}-\frac{\sqrt{35} C_{11,00}^{51,1}}{4608}-\frac{\sqrt{\frac{7}{3}} C_{11,22}^{51,0}}{4608}-\frac{\sqrt{\frac{7}{3}}
C_{11,22}^{51,1}}{2304}-\frac{3 C_{11,22}^{53,0}}{2048 \sqrt{5}}-\frac{3 C_{11,22}^{53,1}}{1024 \sqrt{5}}+\frac{\sqrt{35} C_{20,00}^{40,0}}{9216}+\frac{\sqrt{35}
C_{20,00}^{40,1}}{4608}+\frac{\sqrt{\frac{35}{3}} C_{20,20}^{40,0}}{9216}+\frac{\sqrt{\frac{35}{3}} C_{20,20}^{40,1}}{4608}+\frac{\sqrt{\frac{7}{3}}
C_{20,22}^{42,0}}{4608}+\frac{\sqrt{\frac{7}{3}} C_{20,22}^{42,1}}{2304}-\frac{\sqrt{7} C_{22,00}^{42,0}}{9216}-\frac{\sqrt{7} C_{22,00}^{42,1}}{4608}-\frac{\sqrt{\frac{7}{3}}
C_{22,20}^{42,0}}{9216}-\frac{\sqrt{\frac{7}{3}} C_{22,20}^{42,1}}{4608}+\frac{5 \sqrt{\frac{7}{3}} C_{22,22}^{40,0}}{9216}+\frac{5 \sqrt{\frac{7}{3}}
C_{22,22}^{40,1}}{4608}-\frac{7 C_{22,22}^{42,0}}{4608 \sqrt{3}}-\frac{7 C_{22,22}^{42,1}}{2304 \sqrt{3}}+\frac{\sqrt{35} C_{31,00}^{31,0}}{9216}+\frac{\sqrt{35}
C_{31,00}^{31,1}}{4608}+\frac{\sqrt{\frac{35}{3}} C_{31,20}^{31,0}}{9216}+\frac{\sqrt{\frac{35}{3}} C_{31,20}^{31,1}}{4608}+\frac{\sqrt{\frac{7}{3}}
C_{31,22}^{31,0}}{4608}+\frac{\sqrt{\frac{7}{3}} C_{31,22}^{31,1}}{2304}+\frac{3 C_{31,22}^{33,0}}{2048 \sqrt{5}}+\frac{3 C_{31,22}^{33,1}}{1024
\sqrt{5}}-\frac{\sqrt{\frac{3}{5}} C_{33,00}^{33,0}}{1024}-\frac{1}{512} \sqrt{\frac{3}{5}} C_{33,00}^{33,1}-\frac{C_{33,20}^{33,0}}{1024 \sqrt{5}}-\frac{C_{33,20}^{33,1}}{512
\sqrt{5}}-\frac{1}{640} \sqrt{\frac{3}{2}} C_{33,22}^{33,0}-\frac{1}{320} \sqrt{\frac{3}{2}} C_{33,22}^{33,1}\)

\noindent\(C_{22,2212,2}^{22,0011,1,\text{Loc}}= -\frac{\sqrt{35} C_{00,20}^{60,1}}{1536}+\frac{\sqrt{7} C_{00,22}^{62,1}}{1024}-\frac{\sqrt{7}
C_{11,22}^{51,1}}{4608}-\frac{3 \sqrt{\frac{3}{5}} C_{11,22}^{53,1}}{2048}+\frac{\sqrt{35} C_{20,20}^{40,1}}{4608}-\frac{\sqrt{7} C_{20,22}^{42,1}}{4608}-\frac{\sqrt{7}
C_{22,20}^{42,1}}{4608}-\frac{5 \sqrt{7} C_{22,22}^{40,1}}{9216}+\frac{7 C_{22,22}^{42,1}}{4608}-\frac{\sqrt{35} C_{31,20}^{31,1}}{4608}+\frac{\sqrt{7}
C_{31,22}^{31,1}}{4608}+\frac{3 \sqrt{\frac{3}{5}} C_{31,22}^{33,1}}{2048}+\frac{1}{512} \sqrt{\frac{3}{5}} C_{33,20}^{33,1}-\frac{3 C_{33,22}^{33,1}}{640
\sqrt{2}}\)

\noindent\(C_{22,2212,2}^{22,0011,1,\text{NonLoc}}= -\frac{\sqrt{\frac{35}{3}} C_{00,00}^{60,0}}{1024}-\frac{\sqrt{35} C_{00,20}^{60,0}}{3072}-\frac{\sqrt{7}
C_{00,22}^{62,0}}{1024}-\frac{\sqrt{\frac{35}{3}} C_{11,00}^{51,0}}{3072}-\frac{\sqrt{7} C_{11,22}^{51,0}}{4608}-\frac{3 \sqrt{\frac{3}{5}} C_{11,22}^{53,0}}{2048}+\frac{\sqrt{\frac{35}{3}}
C_{20,00}^{40,0}}{3072}+\frac{\sqrt{35} C_{20,20}^{40,0}}{9216}+\frac{\sqrt{7} C_{20,22}^{42,0}}{4608}-\frac{\sqrt{\frac{7}{3}} C_{22,00}^{42,0}}{3072}-\frac{\sqrt{7}
C_{22,20}^{42,0}}{9216}+\frac{5 \sqrt{7} C_{22,22}^{40,0}}{9216}-\frac{7 C_{22,22}^{42,0}}{4608}+\frac{\sqrt{\frac{35}{3}} C_{31,00}^{31,0}}{3072}+\frac{\sqrt{35}
C_{31,20}^{31,0}}{9216}+\frac{\sqrt{7} C_{31,22}^{31,0}}{4608}+\frac{3 \sqrt{\frac{3}{5}} C_{31,22}^{33,0}}{2048}-\frac{3 C_{33,00}^{33,0}}{1024
\sqrt{5}}-\frac{\sqrt{\frac{3}{5}} C_{33,20}^{33,0}}{1024}-\frac{3 C_{33,22}^{33,0}}{640 \sqrt{2}}\)

\noindent\(C_{22,2212,3}^{00,2213,0,\text{Loc}}= -\frac{1}{48} \sqrt{\frac{7}{6}} C_{00,20}^{60,0}-\frac{1}{96} \sqrt{\frac{7}{6}}
C_{00,20}^{60,1}+\frac{1}{32} \sqrt{\frac{7}{30}} C_{00,22}^{62,0}+\frac{1}{64} \sqrt{\frac{7}{30}} C_{00,22}^{62,1}-\frac{1}{72} \sqrt{\frac{7}{30}}
C_{11,22}^{51,0}-\frac{1}{144} \sqrt{\frac{7}{30}} C_{11,22}^{51,1}+\frac{9 C_{11,22}^{53,0}}{320 \sqrt{2}}+\frac{9 C_{11,22}^{53,1}}{640 \sqrt{2}}+\frac{1}{144}
\sqrt{\frac{7}{6}} C_{20,20}^{40,0}+\frac{1}{288} \sqrt{\frac{7}{6}} C_{20,20}^{40,1}+\frac{1}{72} \sqrt{\frac{7}{30}} C_{20,22}^{42,0}+\frac{1}{144}
\sqrt{\frac{7}{30}} C_{20,22}^{42,1}-\frac{1}{144} \sqrt{\frac{7}{30}} C_{22,20}^{42,0}-\frac{1}{288} \sqrt{\frac{7}{30}} C_{22,20}^{42,1}-\frac{1}{288}
\sqrt{\frac{35}{6}} C_{22,22}^{40,0}-\frac{1}{576} \sqrt{\frac{35}{6}} C_{22,22}^{40,1}-\frac{C_{22,22}^{42,0}}{18 \sqrt{30}}-\frac{C_{22,22}^{42,1}}{36
\sqrt{30}}+\frac{1}{32} \sqrt{\frac{3}{14}} C_{22,22}^{44,0}+\frac{1}{64} \sqrt{\frac{3}{14}} C_{22,22}^{44,1}+\frac{1}{144} \sqrt{\frac{7}{6}} C_{31,20}^{31,0}+\frac{1}{288}
\sqrt{\frac{7}{6}} C_{31,20}^{31,1}-\frac{1}{144} \sqrt{\frac{7}{30}} C_{31,22}^{31,0}-\frac{1}{288} \sqrt{\frac{7}{30}} C_{31,22}^{31,1}+\frac{C_{31,22}^{33,0}}{960
\sqrt{2}}+\frac{C_{31,22}^{33,1}}{1920 \sqrt{2}}-\frac{C_{33,20}^{33,0}}{80 \sqrt{2}}-\frac{C_{33,20}^{33,1}}{160 \sqrt{2}}-\frac{7}{320} \sqrt{\frac{3}{5}}
C_{33,22}^{33,0}-\frac{7}{640} \sqrt{\frac{3}{5}} C_{33,22}^{33,1}\)

\noindent\(C_{22,2212,3}^{00,2213,0,\text{NonLoc}}= -\frac{1}{192} \sqrt{\frac{7}{2}} C_{00,00}^{60,0}-\frac{1}{96} \sqrt{\frac{7}{2}}
C_{00,00}^{60,1}-\frac{1}{192} \sqrt{\frac{7}{6}} C_{00,20}^{60,0}-\frac{1}{96} \sqrt{\frac{7}{6}} C_{00,20}^{60,1}-\frac{1}{64} \sqrt{\frac{7}{30}}
C_{00,22}^{62,0}-\frac{1}{32} \sqrt{\frac{7}{30}} C_{00,22}^{62,1}+\frac{1}{576} \sqrt{\frac{7}{2}} C_{11,00}^{51,0}+\frac{1}{288} \sqrt{\frac{7}{2}}
C_{11,00}^{51,1}-\frac{1}{144} \sqrt{\frac{7}{30}} C_{11,22}^{51,0}-\frac{1}{72} \sqrt{\frac{7}{30}} C_{11,22}^{51,1}+\frac{9 C_{11,22}^{53,0}}{640
\sqrt{2}}+\frac{9 C_{11,22}^{53,1}}{320 \sqrt{2}}+\frac{1}{576} \sqrt{\frac{7}{2}} C_{20,00}^{40,0}+\frac{1}{288} \sqrt{\frac{7}{2}} C_{20,00}^{40,1}+\frac{1}{576}
\sqrt{\frac{7}{6}} C_{20,20}^{40,0}+\frac{1}{288} \sqrt{\frac{7}{6}} C_{20,20}^{40,1}-\frac{1}{144} \sqrt{\frac{7}{30}} C_{20,22}^{42,0}-\frac{1}{72}
\sqrt{\frac{7}{30}} C_{20,22}^{42,1}-\frac{1}{576} \sqrt{\frac{7}{10}} C_{22,00}^{42,0}-\frac{1}{288} \sqrt{\frac{7}{10}} C_{22,00}^{42,1}-\frac{1}{576}
\sqrt{\frac{7}{30}} C_{22,20}^{42,0}-\frac{1}{288} \sqrt{\frac{7}{30}} C_{22,20}^{42,1}+\frac{1}{576} \sqrt{\frac{35}{6}} C_{22,22}^{40,0}+\frac{1}{288}
\sqrt{\frac{35}{6}} C_{22,22}^{40,1}+\frac{C_{22,22}^{42,0}}{36 \sqrt{30}}+\frac{C_{22,22}^{42,1}}{18 \sqrt{30}}-\frac{1}{64} \sqrt{\frac{3}{14}}
C_{22,22}^{44,0}-\frac{1}{32} \sqrt{\frac{3}{14}} C_{22,22}^{44,1}-\frac{1}{576} \sqrt{\frac{7}{2}} C_{31,00}^{31,0}-\frac{1}{288} \sqrt{\frac{7}{2}}
C_{31,00}^{31,1}-\frac{1}{576} \sqrt{\frac{7}{6}} C_{31,20}^{31,0}-\frac{1}{288} \sqrt{\frac{7}{6}} C_{31,20}^{31,1}-\frac{1}{288} \sqrt{\frac{7}{30}}
C_{31,22}^{31,0}-\frac{1}{144} \sqrt{\frac{7}{30}} C_{31,22}^{31,1}+\frac{C_{31,22}^{33,0}}{1920 \sqrt{2}}+\frac{C_{31,22}^{33,1}}{960 \sqrt{2}}+\frac{1}{320}
\sqrt{\frac{3}{2}} C_{33,00}^{33,0}+\frac{1}{160} \sqrt{\frac{3}{2}} C_{33,00}^{33,1}+\frac{C_{33,20}^{33,0}}{320 \sqrt{2}}+\frac{C_{33,20}^{33,1}}{160
\sqrt{2}}-\frac{7}{640} \sqrt{\frac{3}{5}} C_{33,22}^{33,0}-\frac{7}{320} \sqrt{\frac{3}{5}} C_{33,22}^{33,1}\)

\noindent\(C_{22,2212,3}^{00,2213,1,\text{Loc}}= -\frac{1}{96} \sqrt{\frac{7}{2}} C_{00,20}^{60,1}+\frac{1}{64} \sqrt{\frac{7}{10}}
C_{00,22}^{62,1}-\frac{1}{144} \sqrt{\frac{7}{10}} C_{11,22}^{51,1}+\frac{9}{640} \sqrt{\frac{3}{2}} C_{11,22}^{53,1}+\frac{1}{288} \sqrt{\frac{7}{2}}
C_{20,20}^{40,1}+\frac{1}{144} \sqrt{\frac{7}{10}} C_{20,22}^{42,1}-\frac{1}{288} \sqrt{\frac{7}{10}} C_{22,20}^{42,1}-\frac{1}{576} \sqrt{\frac{35}{2}}
C_{22,22}^{40,1}-\frac{C_{22,22}^{42,1}}{36 \sqrt{10}}+\frac{3 C_{22,22}^{44,1}}{64 \sqrt{14}}+\frac{1}{288} \sqrt{\frac{7}{2}} C_{31,20}^{31,1}-\frac{1}{288}
\sqrt{\frac{7}{10}} C_{31,22}^{31,1}+\frac{C_{31,22}^{33,1}}{640 \sqrt{6}}-\frac{1}{160} \sqrt{\frac{3}{2}} C_{33,20}^{33,1}-\frac{21 C_{33,22}^{33,1}}{640
\sqrt{5}}\)

\noindent\(C_{22,2212,3}^{00,2213,1,\text{NonLoc}}= -\frac{1}{64} \sqrt{\frac{7}{6}} C_{00,00}^{60,0}-\frac{1}{192} \sqrt{\frac{7}{2}}
C_{00,20}^{60,0}-\frac{1}{64} \sqrt{\frac{7}{10}} C_{00,22}^{62,0}+\frac{1}{192} \sqrt{\frac{7}{6}} C_{11,00}^{51,0}-\frac{1}{144} \sqrt{\frac{7}{10}}
C_{11,22}^{51,0}+\frac{9}{640} \sqrt{\frac{3}{2}} C_{11,22}^{53,0}+\frac{1}{192} \sqrt{\frac{7}{6}} C_{20,00}^{40,0}+\frac{1}{576} \sqrt{\frac{7}{2}}
C_{20,20}^{40,0}-\frac{1}{144} \sqrt{\frac{7}{10}} C_{20,22}^{42,0}-\frac{1}{192} \sqrt{\frac{7}{30}} C_{22,00}^{42,0}-\frac{1}{576} \sqrt{\frac{7}{10}}
C_{22,20}^{42,0}+\frac{1}{576} \sqrt{\frac{35}{2}} C_{22,22}^{40,0}+\frac{C_{22,22}^{42,0}}{36 \sqrt{10}}-\frac{3 C_{22,22}^{44,0}}{64 \sqrt{14}}-\frac{1}{192}
\sqrt{\frac{7}{6}} C_{31,00}^{31,0}-\frac{1}{576} \sqrt{\frac{7}{2}} C_{31,20}^{31,0}-\frac{1}{288} \sqrt{\frac{7}{10}} C_{31,22}^{31,0}+\frac{C_{31,22}^{33,0}}{640
\sqrt{6}}+\frac{3 C_{33,00}^{33,0}}{320 \sqrt{2}}+\frac{1}{320} \sqrt{\frac{3}{2}} C_{33,20}^{33,0}-\frac{21 C_{33,22}^{33,0}}{640 \sqrt{5}}\)

\noindent\(C_{22,2212,3}^{22,0011,0,\text{Loc}}= \frac{1}{192} \sqrt{\frac{7}{6}} C_{00,20}^{60,0}+\frac{1}{384} \sqrt{\frac{7}{6}}
C_{00,20}^{60,1}-\frac{1}{128} \sqrt{\frac{7}{30}} C_{00,22}^{62,0}-\frac{1}{256} \sqrt{\frac{7}{30}} C_{00,22}^{62,1}+\frac{1}{576} \sqrt{\frac{7}{30}}
C_{11,22}^{51,0}+\frac{\sqrt{\frac{7}{30}} C_{11,22}^{51,1}}{1152}+\frac{3 C_{11,22}^{53,0}}{1280 \sqrt{2}}+\frac{3 C_{11,22}^{53,1}}{2560 \sqrt{2}}-\frac{1}{576}
\sqrt{\frac{7}{6}} C_{20,20}^{40,0}-\frac{\sqrt{\frac{7}{6}} C_{20,20}^{40,1}}{1152}+\frac{1}{576} \sqrt{\frac{7}{30}} C_{20,22}^{42,0}+\frac{\sqrt{\frac{7}{30}}
C_{20,22}^{42,1}}{1152}+\frac{1}{576} \sqrt{\frac{7}{30}} C_{22,20}^{42,0}+\frac{\sqrt{\frac{7}{30}} C_{22,20}^{42,1}}{1152}+\frac{\sqrt{\frac{35}{6}}
C_{22,22}^{40,0}}{1152}+\frac{\sqrt{\frac{35}{6}} C_{22,22}^{40,1}}{2304}-\frac{7 C_{22,22}^{42,0}}{576 \sqrt{30}}-\frac{7 C_{22,22}^{42,1}}{1152
\sqrt{30}}+\frac{1}{576} \sqrt{\frac{7}{6}} C_{31,20}^{31,0}+\frac{\sqrt{\frac{7}{6}} C_{31,20}^{31,1}}{1152}-\frac{1}{576} \sqrt{\frac{7}{30}} C_{31,22}^{31,0}-\frac{\sqrt{\frac{7}{30}}
C_{31,22}^{31,1}}{1152}-\frac{3 C_{31,22}^{33,0}}{1280 \sqrt{2}}-\frac{3 C_{31,22}^{33,1}}{2560 \sqrt{2}}-\frac{C_{33,20}^{33,0}}{320 \sqrt{2}}-\frac{C_{33,20}^{33,1}}{640
\sqrt{2}}+\frac{1}{160} \sqrt{\frac{3}{5}} C_{33,22}^{33,0}+\frac{1}{320} \sqrt{\frac{3}{5}} C_{33,22}^{33,1}\)

\noindent\(C_{22,2212,3}^{22,0011,0,\text{NonLoc}}= \frac{1}{768} \sqrt{\frac{7}{2}} C_{00,00}^{60,0}+\frac{1}{384} \sqrt{\frac{7}{2}}
C_{00,00}^{60,1}+\frac{1}{768} \sqrt{\frac{7}{6}} C_{00,20}^{60,0}+\frac{1}{384} \sqrt{\frac{7}{6}} C_{00,20}^{60,1}+\frac{1}{256} \sqrt{\frac{7}{30}}
C_{00,22}^{62,0}+\frac{1}{128} \sqrt{\frac{7}{30}} C_{00,22}^{62,1}+\frac{\sqrt{\frac{7}{2}} C_{11,00}^{51,0}}{2304}+\frac{\sqrt{\frac{7}{2}} C_{11,00}^{51,1}}{1152}+\frac{\sqrt{\frac{7}{30}}
C_{11,22}^{51,0}}{1152}+\frac{1}{576} \sqrt{\frac{7}{30}} C_{11,22}^{51,1}+\frac{3 C_{11,22}^{53,0}}{2560 \sqrt{2}}+\frac{3 C_{11,22}^{53,1}}{1280
\sqrt{2}}-\frac{\sqrt{\frac{7}{2}} C_{20,00}^{40,0}}{2304}-\frac{\sqrt{\frac{7}{2}} C_{20,00}^{40,1}}{1152}-\frac{\sqrt{\frac{7}{6}} C_{20,20}^{40,0}}{2304}-\frac{\sqrt{\frac{7}{6}}
C_{20,20}^{40,1}}{1152}-\frac{\sqrt{\frac{7}{30}} C_{20,22}^{42,0}}{1152}-\frac{1}{576} \sqrt{\frac{7}{30}} C_{20,22}^{42,1}+\frac{\sqrt{\frac{7}{10}}
C_{22,00}^{42,0}}{2304}+\frac{\sqrt{\frac{7}{10}} C_{22,00}^{42,1}}{1152}+\frac{\sqrt{\frac{7}{30}} C_{22,20}^{42,0}}{2304}+\frac{\sqrt{\frac{7}{30}}
C_{22,20}^{42,1}}{1152}-\frac{\sqrt{\frac{35}{6}} C_{22,22}^{40,0}}{2304}-\frac{\sqrt{\frac{35}{6}} C_{22,22}^{40,1}}{1152}+\frac{7 C_{22,22}^{42,0}}{1152
\sqrt{30}}+\frac{7 C_{22,22}^{42,1}}{576 \sqrt{30}}-\frac{\sqrt{\frac{7}{2}} C_{31,00}^{31,0}}{2304}-\frac{\sqrt{\frac{7}{2}} C_{31,00}^{31,1}}{1152}-\frac{\sqrt{\frac{7}{6}}
C_{31,20}^{31,0}}{2304}-\frac{\sqrt{\frac{7}{6}} C_{31,20}^{31,1}}{1152}-\frac{\sqrt{\frac{7}{30}} C_{31,22}^{31,0}}{1152}-\frac{1}{576} \sqrt{\frac{7}{30}}
C_{31,22}^{31,1}-\frac{3 C_{31,22}^{33,0}}{2560 \sqrt{2}}-\frac{3 C_{31,22}^{33,1}}{1280 \sqrt{2}}+\frac{\sqrt{\frac{3}{2}} C_{33,00}^{33,0}}{1280}+\frac{1}{640}
\sqrt{\frac{3}{2}} C_{33,00}^{33,1}+\frac{C_{33,20}^{33,0}}{1280 \sqrt{2}}+\frac{C_{33,20}^{33,1}}{640 \sqrt{2}}+\frac{1}{320} \sqrt{\frac{3}{5}}
C_{33,22}^{33,0}+\frac{1}{160} \sqrt{\frac{3}{5}} C_{33,22}^{33,1}\)

\noindent\(C_{22,2212,3}^{22,0011,1,\text{Loc}}= \frac{1}{384} \sqrt{\frac{7}{2}} C_{00,20}^{60,1}-\frac{1}{256} \sqrt{\frac{7}{10}}
C_{00,22}^{62,1}+\frac{\sqrt{\frac{7}{10}} C_{11,22}^{51,1}}{1152}+\frac{3 \sqrt{\frac{3}{2}} C_{11,22}^{53,1}}{2560}-\frac{\sqrt{\frac{7}{2}} C_{20,20}^{40,1}}{1152}+\frac{\sqrt{\frac{7}{10}}
C_{20,22}^{42,1}}{1152}+\frac{\sqrt{\frac{7}{10}} C_{22,20}^{42,1}}{1152}+\frac{\sqrt{\frac{35}{2}} C_{22,22}^{40,1}}{2304}-\frac{7 C_{22,22}^{42,1}}{1152
\sqrt{10}}+\frac{\sqrt{\frac{7}{2}} C_{31,20}^{31,1}}{1152}-\frac{\sqrt{\frac{7}{10}} C_{31,22}^{31,1}}{1152}-\frac{3 \sqrt{\frac{3}{2}} C_{31,22}^{33,1}}{2560}-\frac{1}{640}
\sqrt{\frac{3}{2}} C_{33,20}^{33,1}+\frac{3 C_{33,22}^{33,1}}{320 \sqrt{5}}\)

\noindent\(C_{22,2212,3}^{22,0011,1,\text{NonLoc}}= \frac{1}{256} \sqrt{\frac{7}{6}} C_{00,00}^{60,0}+\frac{1}{768} \sqrt{\frac{7}{2}}
C_{00,20}^{60,0}+\frac{1}{256} \sqrt{\frac{7}{10}} C_{00,22}^{62,0}+\frac{1}{768} \sqrt{\frac{7}{6}} C_{11,00}^{51,0}+\frac{\sqrt{\frac{7}{10}} C_{11,22}^{51,0}}{1152}+\frac{3
\sqrt{\frac{3}{2}} C_{11,22}^{53,0}}{2560}-\frac{1}{768} \sqrt{\frac{7}{6}} C_{20,00}^{40,0}-\frac{\sqrt{\frac{7}{2}} C_{20,20}^{40,0}}{2304}-\frac{\sqrt{\frac{7}{10}}
C_{20,22}^{42,0}}{1152}+\frac{1}{768} \sqrt{\frac{7}{30}} C_{22,00}^{42,0}+\frac{\sqrt{\frac{7}{10}} C_{22,20}^{42,0}}{2304}-\frac{\sqrt{\frac{35}{2}}
C_{22,22}^{40,0}}{2304}+\frac{7 C_{22,22}^{42,0}}{1152 \sqrt{10}}-\frac{1}{768} \sqrt{\frac{7}{6}} C_{31,00}^{31,0}-\frac{\sqrt{\frac{7}{2}} C_{31,20}^{31,0}}{2304}-\frac{\sqrt{\frac{7}{10}}
C_{31,22}^{31,0}}{1152}-\frac{3 \sqrt{\frac{3}{2}} C_{31,22}^{33,0}}{2560}+\frac{3 C_{33,00}^{33,0}}{1280 \sqrt{2}}+\frac{\sqrt{\frac{3}{2}} C_{33,20}^{33,0}}{1280}+\frac{3
C_{33,22}^{33,0}}{320 \sqrt{5}}\)

\noindent\(C_{22,2213,1}^{00,2011,0,\text{Loc}}= \frac{7}{192} \sqrt{\frac{7}{3}} C_{00,20}^{60,0}+\frac{7}{384} \sqrt{\frac{7}{3}}
C_{00,20}^{60,1}+\frac{1}{64} \sqrt{\frac{7}{15}} C_{00,22}^{62,0}+\frac{1}{128} \sqrt{\frac{7}{15}} C_{00,22}^{62,1}+\frac{7}{576} \sqrt{\frac{7}{3}}
C_{11,20}^{31,0}+\frac{7 \sqrt{\frac{7}{3}} C_{11,20}^{31,1}}{1152}+\frac{1}{288} \sqrt{\frac{7}{15}} C_{11,22}^{51,0}+\frac{1}{576} \sqrt{\frac{7}{15}}
C_{11,22}^{51,1}+\frac{3 C_{11,22}^{53,0}}{640}+\frac{3 C_{11,22}^{53,1}}{1280}-\frac{7}{576} \sqrt{\frac{7}{3}} C_{20,20}^{40,0}-\frac{7 \sqrt{\frac{7}{3}}
C_{20,20}^{40,1}}{1152}-\frac{1}{288} \sqrt{\frac{7}{15}} C_{20,22}^{42,0}-\frac{1}{576} \sqrt{\frac{7}{15}} C_{20,22}^{42,1}+\frac{7}{576} \sqrt{\frac{7}{15}}
C_{22,20}^{42,0}+\frac{7 \sqrt{\frac{7}{15}} C_{22,20}^{42,1}}{1152}-\frac{1}{576} \sqrt{\frac{35}{3}} C_{22,22}^{40,0}-\frac{\sqrt{\frac{35}{3}}
C_{22,22}^{40,1}}{1152}+\frac{7 C_{22,22}^{42,0}}{288 \sqrt{15}}+\frac{7 C_{22,22}^{42,1}}{576 \sqrt{15}}-\frac{7}{576} \sqrt{\frac{7}{3}} C_{31,20}^{31,0}-\frac{7
\sqrt{\frac{7}{3}} C_{31,20}^{31,1}}{1152}-\frac{1}{288} \sqrt{\frac{7}{15}} C_{31,22}^{31,0}-\frac{1}{576} \sqrt{\frac{7}{15}} C_{31,22}^{31,1}-\frac{3
C_{31,22}^{33,0}}{640}-\frac{3 C_{31,22}^{33,1}}{1280}+\frac{7 C_{33,20}^{33,0}}{320}+\frac{7 C_{33,20}^{33,1}}{640}+\frac{1}{40} \sqrt{\frac{3}{10}}
C_{33,22}^{33,0}+\frac{1}{80} \sqrt{\frac{3}{10}} C_{33,22}^{33,1}\)

\noindent\(C_{22,2213,1}^{00,2011,0,\text{NonLoc}}= \frac{7 \sqrt{7} C_{00,00}^{60,0}}{768}+\frac{7 \sqrt{7} C_{00,00}^{60,1}}{384}+\frac{7}{768}
\sqrt{\frac{7}{3}} C_{00,20}^{60,0}+\frac{7}{384} \sqrt{\frac{7}{3}} C_{00,20}^{60,1}-\frac{1}{128} \sqrt{\frac{7}{15}} C_{00,22}^{62,0}-\frac{1}{64}
\sqrt{\frac{7}{15}} C_{00,22}^{62,1}-\frac{7 \sqrt{7} C_{11,00}^{51,0}}{2304}-\frac{7 \sqrt{7} C_{11,00}^{51,1}}{1152}-\frac{7 \sqrt{\frac{7}{3}}
C_{11,20}^{31,0}}{2304}-\frac{7 \sqrt{\frac{7}{3}} C_{11,20}^{31,1}}{1152}+\frac{1}{576} \sqrt{\frac{7}{15}} C_{11,22}^{51,0}+\frac{1}{288} \sqrt{\frac{7}{15}}
C_{11,22}^{51,1}+\frac{3 C_{11,22}^{53,0}}{1280}+\frac{3 C_{11,22}^{53,1}}{640}-\frac{7 \sqrt{7} C_{20,00}^{40,0}}{2304}-\frac{7 \sqrt{7} C_{20,00}^{40,1}}{1152}-\frac{7
\sqrt{\frac{7}{3}} C_{20,20}^{40,0}}{2304}-\frac{7 \sqrt{\frac{7}{3}} C_{20,20}^{40,1}}{1152}+\frac{1}{576} \sqrt{\frac{7}{15}} C_{20,22}^{42,0}+\frac{1}{288}
\sqrt{\frac{7}{15}} C_{20,22}^{42,1}+\frac{7 \sqrt{\frac{7}{5}} C_{22,00}^{42,0}}{2304}+\frac{7 \sqrt{\frac{7}{5}} C_{22,00}^{42,1}}{1152}+\frac{7
\sqrt{\frac{7}{15}} C_{22,20}^{42,0}}{2304}+\frac{7 \sqrt{\frac{7}{15}} C_{22,20}^{42,1}}{1152}+\frac{\sqrt{\frac{35}{3}} C_{22,22}^{40,0}}{1152}+\frac{1}{576}
\sqrt{\frac{35}{3}} C_{22,22}^{40,1}-\frac{7 C_{22,22}^{42,0}}{576 \sqrt{15}}-\frac{7 C_{22,22}^{42,1}}{288 \sqrt{15}}+\frac{7 \sqrt{7} C_{31,00}^{31,0}}{2304}+\frac{7
\sqrt{7} C_{31,00}^{31,1}}{1152}+\frac{7 \sqrt{\frac{7}{3}} C_{31,20}^{31,0}}{2304}+\frac{7 \sqrt{\frac{7}{3}} C_{31,20}^{31,1}}{1152}-\frac{1}{576}
\sqrt{\frac{7}{15}} C_{31,22}^{31,0}-\frac{1}{288} \sqrt{\frac{7}{15}} C_{31,22}^{31,1}-\frac{3 C_{31,22}^{33,0}}{1280}-\frac{3 C_{31,22}^{33,1}}{640}-\frac{7
\sqrt{3} C_{33,00}^{33,0}}{1280}-\frac{7 \sqrt{3} C_{33,00}^{33,1}}{640}-\frac{7 C_{33,20}^{33,0}}{1280}-\frac{7 C_{33,20}^{33,1}}{640}+\frac{1}{80}
\sqrt{\frac{3}{10}} C_{33,22}^{33,0}+\frac{1}{40} \sqrt{\frac{3}{10}} C_{33,22}^{33,1}\)

\noindent\(C_{22,2213,1}^{00,2011,1,\text{Loc}}= \frac{7 \sqrt{7} C_{00,20}^{60,1}}{384}+\frac{1}{128} \sqrt{\frac{7}{5}} C_{00,22}^{62,1}+\frac{7
\sqrt{7} C_{11,20}^{31,1}}{1152}+\frac{1}{576} \sqrt{\frac{7}{5}} C_{11,22}^{51,1}+\frac{3 \sqrt{3} C_{11,22}^{53,1}}{1280}-\frac{7 \sqrt{7} C_{20,20}^{40,1}}{1152}-\frac{1}{576}
\sqrt{\frac{7}{5}} C_{20,22}^{42,1}+\frac{7 \sqrt{\frac{7}{5}} C_{22,20}^{42,1}}{1152}-\frac{\sqrt{35} C_{22,22}^{40,1}}{1152}+\frac{7 C_{22,22}^{42,1}}{576
\sqrt{5}}-\frac{7 \sqrt{7} C_{31,20}^{31,1}}{1152}-\frac{1}{576} \sqrt{\frac{7}{5}} C_{31,22}^{31,1}-\frac{3 \sqrt{3} C_{31,22}^{33,1}}{1280}+\frac{7
\sqrt{3} C_{33,20}^{33,1}}{640}+\frac{3 C_{33,22}^{33,1}}{80 \sqrt{10}}\)

\noindent\(C_{22,2213,1}^{00,2011,1,\text{NonLoc}}= \frac{7}{256} \sqrt{\frac{7}{3}} C_{00,00}^{60,0}+\frac{7 \sqrt{7} C_{00,20}^{60,0}}{768}-\frac{1}{128}
\sqrt{\frac{7}{5}} C_{00,22}^{62,0}-\frac{7}{768} \sqrt{\frac{7}{3}} C_{11,00}^{51,0}-\frac{7 \sqrt{7} C_{11,20}^{31,0}}{2304}+\frac{1}{576} \sqrt{\frac{7}{5}}
C_{11,22}^{51,0}+\frac{3 \sqrt{3} C_{11,22}^{53,0}}{1280}-\frac{7}{768} \sqrt{\frac{7}{3}} C_{20,00}^{40,0}-\frac{7 \sqrt{7} C_{20,20}^{40,0}}{2304}+\frac{1}{576}
\sqrt{\frac{7}{5}} C_{20,22}^{42,0}+\frac{7}{768} \sqrt{\frac{7}{15}} C_{22,00}^{42,0}+\frac{7 \sqrt{\frac{7}{5}} C_{22,20}^{42,0}}{2304}+\frac{\sqrt{35}
C_{22,22}^{40,0}}{1152}-\frac{7 C_{22,22}^{42,0}}{576 \sqrt{5}}+\frac{7}{768} \sqrt{\frac{7}{3}} C_{31,00}^{31,0}+\frac{7 \sqrt{7} C_{31,20}^{31,0}}{2304}-\frac{1}{576}
\sqrt{\frac{7}{5}} C_{31,22}^{31,0}-\frac{3 \sqrt{3} C_{31,22}^{33,0}}{1280}-\frac{21 C_{33,00}^{33,0}}{1280}-\frac{7 \sqrt{3} C_{33,20}^{33,0}}{1280}+\frac{3
C_{33,22}^{33,0}}{80 \sqrt{10}}\)

\noindent\(C_{22,2213,1}^{00,2211,0,\text{Loc}}= \frac{1}{96} \sqrt{\frac{7}{15}} C_{00,20}^{60,0}+\frac{1}{192} \sqrt{\frac{7}{15}}
C_{00,20}^{60,1}+\frac{1}{40} \sqrt{\frac{7}{3}} C_{00,22}^{62,0}+\frac{1}{80} \sqrt{\frac{7}{3}} C_{00,22}^{62,1}+\frac{7 \sqrt{\frac{7}{3}} C_{11,22}^{51,0}}{2880}+\frac{7
\sqrt{\frac{7}{3}} C_{11,22}^{51,1}}{5760}+\frac{33 C_{11,22}^{53,0}}{640 \sqrt{5}}+\frac{33 C_{11,22}^{53,1}}{1280 \sqrt{5}}-\frac{1}{288} \sqrt{\frac{7}{15}}
C_{20,20}^{40,0}-\frac{1}{576} \sqrt{\frac{7}{15}} C_{20,20}^{40,1}-\frac{7 \sqrt{\frac{7}{3}} C_{20,22}^{42,0}}{2880}-\frac{7 \sqrt{\frac{7}{3}}
C_{20,22}^{42,1}}{5760}+\frac{\sqrt{\frac{7}{3}} C_{22,20}^{42,0}}{1440}+\frac{\sqrt{\frac{7}{3}} C_{22,20}^{42,1}}{2880}-\frac{11 \sqrt{\frac{7}{3}}
C_{22,22}^{40,0}}{1440}-\frac{11 \sqrt{\frac{7}{3}} C_{22,22}^{40,1}}{2880}+\frac{C_{22,22}^{42,0}}{720 \sqrt{3}}+\frac{C_{22,22}^{42,1}}{1440 \sqrt{3}}+\frac{9}{160}
\sqrt{\frac{3}{35}} C_{22,22}^{44,0}+\frac{9}{320} \sqrt{\frac{3}{35}} C_{22,22}^{44,1}-\frac{1}{288} \sqrt{\frac{7}{15}} C_{31,20}^{31,0}-\frac{1}{576}
\sqrt{\frac{7}{15}} C_{31,20}^{31,1}-\frac{5}{576} \sqrt{\frac{7}{3}} C_{31,22}^{31,0}-\frac{5 \sqrt{\frac{7}{3}} C_{31,22}^{31,1}}{1152}-\frac{C_{31,22}^{33,0}}{128
\sqrt{5}}-\frac{C_{31,22}^{33,1}}{256 \sqrt{5}}+\frac{C_{33,20}^{33,0}}{160 \sqrt{5}}+\frac{C_{33,20}^{33,1}}{320 \sqrt{5}}+\frac{1}{160} \sqrt{\frac{3}{2}}
C_{33,22}^{33,0}+\frac{1}{320} \sqrt{\frac{3}{2}} C_{33,22}^{33,1}\)

\noindent\(C_{22,2213,1}^{00,2211,0,\text{NonLoc}}= \frac{1}{384} \sqrt{\frac{7}{5}} C_{00,00}^{60,0}+\frac{1}{192} \sqrt{\frac{7}{5}}
C_{00,00}^{60,1}+\frac{1}{384} \sqrt{\frac{7}{15}} C_{00,20}^{60,0}+\frac{1}{192} \sqrt{\frac{7}{15}} C_{00,20}^{60,1}-\frac{1}{80} \sqrt{\frac{7}{3}}
C_{00,22}^{62,0}-\frac{1}{40} \sqrt{\frac{7}{3}} C_{00,22}^{62,1}-\frac{\sqrt{\frac{7}{5}} C_{11,00}^{51,0}}{1152}-\frac{1}{576} \sqrt{\frac{7}{5}}
C_{11,00}^{51,1}+\frac{7 \sqrt{\frac{7}{3}} C_{11,22}^{51,0}}{5760}+\frac{7 \sqrt{\frac{7}{3}} C_{11,22}^{51,1}}{2880}+\frac{33 C_{11,22}^{53,0}}{1280
\sqrt{5}}+\frac{33 C_{11,22}^{53,1}}{640 \sqrt{5}}-\frac{\sqrt{\frac{7}{5}} C_{20,00}^{40,0}}{1152}-\frac{1}{576} \sqrt{\frac{7}{5}} C_{20,00}^{40,1}-\frac{\sqrt{\frac{7}{15}}
C_{20,20}^{40,0}}{1152}-\frac{1}{576} \sqrt{\frac{7}{15}} C_{20,20}^{40,1}+\frac{7 \sqrt{\frac{7}{3}} C_{20,22}^{42,0}}{5760}+\frac{7 \sqrt{\frac{7}{3}}
C_{20,22}^{42,1}}{2880}+\frac{\sqrt{7} C_{22,00}^{42,0}}{5760}+\frac{\sqrt{7} C_{22,00}^{42,1}}{2880}+\frac{\sqrt{\frac{7}{3}} C_{22,20}^{42,0}}{5760}+\frac{\sqrt{\frac{7}{3}}
C_{22,20}^{42,1}}{2880}+\frac{11 \sqrt{\frac{7}{3}} C_{22,22}^{40,0}}{2880}+\frac{11 \sqrt{\frac{7}{3}} C_{22,22}^{40,1}}{1440}-\frac{C_{22,22}^{42,0}}{1440
\sqrt{3}}-\frac{C_{22,22}^{42,1}}{720 \sqrt{3}}-\frac{9}{320} \sqrt{\frac{3}{35}} C_{22,22}^{44,0}-\frac{9}{160} \sqrt{\frac{3}{35}} C_{22,22}^{44,1}+\frac{\sqrt{\frac{7}{5}}
C_{31,00}^{31,0}}{1152}+\frac{1}{576} \sqrt{\frac{7}{5}} C_{31,00}^{31,1}+\frac{\sqrt{\frac{7}{15}} C_{31,20}^{31,0}}{1152}+\frac{1}{576} \sqrt{\frac{7}{15}}
C_{31,20}^{31,1}-\frac{5 \sqrt{\frac{7}{3}} C_{31,22}^{31,0}}{1152}-\frac{5}{576} \sqrt{\frac{7}{3}} C_{31,22}^{31,1}-\frac{C_{31,22}^{33,0}}{256
\sqrt{5}}-\frac{C_{31,22}^{33,1}}{128 \sqrt{5}}-\frac{1}{640} \sqrt{\frac{3}{5}} C_{33,00}^{33,0}-\frac{1}{320} \sqrt{\frac{3}{5}} C_{33,00}^{33,1}-\frac{C_{33,20}^{33,0}}{640
\sqrt{5}}-\frac{C_{33,20}^{33,1}}{320 \sqrt{5}}+\frac{1}{320} \sqrt{\frac{3}{2}} C_{33,22}^{33,0}+\frac{1}{160} \sqrt{\frac{3}{2}} C_{33,22}^{33,1}\)

\noindent\(C_{22,2213,1}^{00,2211,1,\text{Loc}}= \frac{1}{192} \sqrt{\frac{7}{5}} C_{00,20}^{60,1}+\frac{\sqrt{7} C_{00,22}^{62,1}}{80}+\frac{7
\sqrt{7} C_{11,22}^{51,1}}{5760}+\frac{33 \sqrt{\frac{3}{5}} C_{11,22}^{53,1}}{1280}-\frac{1}{576} \sqrt{\frac{7}{5}} C_{20,20}^{40,1}-\frac{7 \sqrt{7}
C_{20,22}^{42,1}}{5760}+\frac{\sqrt{7} C_{22,20}^{42,1}}{2880}-\frac{11 \sqrt{7} C_{22,22}^{40,1}}{2880}+\frac{C_{22,22}^{42,1}}{1440}+\frac{27 C_{22,22}^{44,1}}{320
\sqrt{35}}-\frac{1}{576} \sqrt{\frac{7}{5}} C_{31,20}^{31,1}-\frac{5 \sqrt{7} C_{31,22}^{31,1}}{1152}-\frac{1}{256} \sqrt{\frac{3}{5}} C_{31,22}^{33,1}+\frac{1}{320}
\sqrt{\frac{3}{5}} C_{33,20}^{33,1}+\frac{3 C_{33,22}^{33,1}}{320 \sqrt{2}}\)

\noindent\(C_{22,2213,1}^{00,2211,1,\text{NonLoc}}= \frac{1}{128} \sqrt{\frac{7}{15}} C_{00,00}^{60,0}+\frac{1}{384} \sqrt{\frac{7}{5}}
C_{00,20}^{60,0}-\frac{\sqrt{7} C_{00,22}^{62,0}}{80}-\frac{1}{384} \sqrt{\frac{7}{15}} C_{11,00}^{51,0}+\frac{7 \sqrt{7} C_{11,22}^{51,0}}{5760}+\frac{33
\sqrt{\frac{3}{5}} C_{11,22}^{53,0}}{1280}-\frac{1}{384} \sqrt{\frac{7}{15}} C_{20,00}^{40,0}-\frac{\sqrt{\frac{7}{5}} C_{20,20}^{40,0}}{1152}+\frac{7
\sqrt{7} C_{20,22}^{42,0}}{5760}+\frac{\sqrt{\frac{7}{3}} C_{22,00}^{42,0}}{1920}+\frac{\sqrt{7} C_{22,20}^{42,0}}{5760}+\frac{11 \sqrt{7} C_{22,22}^{40,0}}{2880}-\frac{C_{22,22}^{42,0}}{1440}-\frac{27
C_{22,22}^{44,0}}{320 \sqrt{35}}+\frac{1}{384} \sqrt{\frac{7}{15}} C_{31,00}^{31,0}+\frac{\sqrt{\frac{7}{5}} C_{31,20}^{31,0}}{1152}-\frac{5 \sqrt{7}
C_{31,22}^{31,0}}{1152}-\frac{1}{256} \sqrt{\frac{3}{5}} C_{31,22}^{33,0}-\frac{3 C_{33,00}^{33,0}}{640 \sqrt{5}}-\frac{1}{640} \sqrt{\frac{3}{5}}
C_{33,20}^{33,0}+\frac{3 C_{33,22}^{33,0}}{320 \sqrt{2}}\)

\noindent\(C_{22,2213,1}^{11,1101,0,\text{Loc}}= \frac{1}{96} \sqrt{\frac{7}{3}} C_{11,11}^{51,0}+\frac{1}{192} \sqrt{\frac{7}{3}}
C_{11,11}^{51,1}-\frac{1}{96} \sqrt{\frac{7}{3}} C_{31,11}^{31,0}-\frac{1}{192} \sqrt{\frac{7}{3}} C_{31,11}^{31,1}+\frac{1}{160} \sqrt{\frac{3}{2}}
C_{33,11}^{33,0}+\frac{1}{320} \sqrt{\frac{3}{2}} C_{33,11}^{33,1}\)

\noindent\(C_{22,2213,1}^{11,1101,0,\text{NonLoc}}= \frac{1}{192} \sqrt{\frac{7}{3}} C_{11,11}^{51,0}+\frac{1}{96} \sqrt{\frac{7}{3}}
C_{11,11}^{51,1}-\frac{1}{192} \sqrt{\frac{7}{3}} C_{31,11}^{31,0}-\frac{1}{96} \sqrt{\frac{7}{3}} C_{31,11}^{31,1}+\frac{1}{320} \sqrt{\frac{3}{2}}
C_{33,11}^{33,0}+\frac{1}{160} \sqrt{\frac{3}{2}} C_{33,11}^{33,1}\)

\noindent\(C_{22,2213,1}^{11,1101,1,\text{Loc}}= \frac{\sqrt{7} C_{11,11}^{51,1}}{192}-\frac{\sqrt{7} C_{31,11}^{31,1}}{192}+\frac{3
C_{33,11}^{33,1}}{320 \sqrt{2}}\)

\noindent\(C_{22,2213,1}^{11,1101,1,\text{NonLoc}}= \frac{\sqrt{7} C_{11,11}^{51,0}}{192}-\frac{\sqrt{7} C_{31,11}^{31,0}}{192}+\frac{3
C_{33,11}^{33,0}}{320 \sqrt{2}}\)

\noindent\(C_{22,2213,1}^{20,0011,0,\text{Loc}}= -\frac{7}{768} \sqrt{\frac{7}{3}} C_{00,20}^{60,0}-\frac{7 \sqrt{\frac{7}{3}} C_{00,20}^{60,1}}{1536}-\frac{1}{256}
\sqrt{\frac{7}{15}} C_{00,22}^{62,0}-\frac{1}{512} \sqrt{\frac{7}{15}} C_{00,22}^{62,1}+\frac{7 \sqrt{\frac{7}{3}} C_{11,20}^{31,0}}{2304}+\frac{7
\sqrt{\frac{7}{3}} C_{11,20}^{31,1}}{4608}+\frac{\sqrt{\frac{7}{15}} C_{11,22}^{51,0}}{1152}+\frac{\sqrt{\frac{7}{15}} C_{11,22}^{51,1}}{2304}+\frac{3
C_{11,22}^{53,0}}{2560}+\frac{3 C_{11,22}^{53,1}}{5120}+\frac{7 \sqrt{\frac{7}{3}} C_{20,20}^{40,0}}{2304}+\frac{7 \sqrt{\frac{7}{3}} C_{20,20}^{40,1}}{4608}+\frac{\sqrt{\frac{7}{15}}
C_{20,22}^{42,0}}{1152}+\frac{\sqrt{\frac{7}{15}} C_{20,22}^{42,1}}{2304}-\frac{7 \sqrt{\frac{7}{15}} C_{22,20}^{42,0}}{2304}-\frac{7 \sqrt{\frac{7}{15}}
C_{22,20}^{42,1}}{4608}+\frac{\sqrt{\frac{35}{3}} C_{22,22}^{40,0}}{2304}+\frac{\sqrt{\frac{35}{3}} C_{22,22}^{40,1}}{4608}-\frac{7 C_{22,22}^{42,0}}{1152
\sqrt{15}}-\frac{7 C_{22,22}^{42,1}}{2304 \sqrt{15}}-\frac{7 \sqrt{\frac{7}{3}} C_{31,20}^{31,0}}{2304}-\frac{7 \sqrt{\frac{7}{3}} C_{31,20}^{31,1}}{4608}-\frac{\sqrt{\frac{7}{15}}
C_{31,22}^{31,0}}{1152}-\frac{\sqrt{\frac{7}{15}} C_{31,22}^{31,1}}{2304}-\frac{3 C_{31,22}^{33,0}}{2560}-\frac{3 C_{31,22}^{33,1}}{5120}+\frac{7
C_{33,20}^{33,0}}{1280}+\frac{7 C_{33,20}^{33,1}}{2560}+\frac{1}{160} \sqrt{\frac{3}{10}} C_{33,22}^{33,0}+\frac{1}{320} \sqrt{\frac{3}{10}} C_{33,22}^{33,1}\)

\noindent\(C_{22,2213,1}^{20,0011,0,\text{NonLoc}}= -\frac{7 \sqrt{7} C_{00,00}^{60,0}}{3072}-\frac{7 \sqrt{7} C_{00,00}^{60,1}}{1536}-\frac{7
\sqrt{\frac{7}{3}} C_{00,20}^{60,0}}{3072}-\frac{7 \sqrt{\frac{7}{3}} C_{00,20}^{60,1}}{1536}+\frac{1}{512} \sqrt{\frac{7}{15}} C_{00,22}^{62,0}+\frac{1}{256}
\sqrt{\frac{7}{15}} C_{00,22}^{62,1}-\frac{7 \sqrt{7} C_{11,00}^{51,0}}{9216}-\frac{7 \sqrt{7} C_{11,00}^{51,1}}{4608}-\frac{7 \sqrt{\frac{7}{3}}
C_{11,20}^{31,0}}{9216}-\frac{7 \sqrt{\frac{7}{3}} C_{11,20}^{31,1}}{4608}+\frac{\sqrt{\frac{7}{15}} C_{11,22}^{51,0}}{2304}+\frac{\sqrt{\frac{7}{15}}
C_{11,22}^{51,1}}{1152}+\frac{3 C_{11,22}^{53,0}}{5120}+\frac{3 C_{11,22}^{53,1}}{2560}+\frac{7 \sqrt{7} C_{20,00}^{40,0}}{9216}+\frac{7 \sqrt{7}
C_{20,00}^{40,1}}{4608}+\frac{7 \sqrt{\frac{7}{3}} C_{20,20}^{40,0}}{9216}+\frac{7 \sqrt{\frac{7}{3}} C_{20,20}^{40,1}}{4608}-\frac{\sqrt{\frac{7}{15}}
C_{20,22}^{42,0}}{2304}-\frac{\sqrt{\frac{7}{15}} C_{20,22}^{42,1}}{1152}-\frac{7 \sqrt{\frac{7}{5}} C_{22,00}^{42,0}}{9216}-\frac{7 \sqrt{\frac{7}{5}}
C_{22,00}^{42,1}}{4608}-\frac{7 \sqrt{\frac{7}{15}} C_{22,20}^{42,0}}{9216}-\frac{7 \sqrt{\frac{7}{15}} C_{22,20}^{42,1}}{4608}-\frac{\sqrt{\frac{35}{3}}
C_{22,22}^{40,0}}{4608}-\frac{\sqrt{\frac{35}{3}} C_{22,22}^{40,1}}{2304}+\frac{7 C_{22,22}^{42,0}}{2304 \sqrt{15}}+\frac{7 C_{22,22}^{42,1}}{1152
\sqrt{15}}+\frac{7 \sqrt{7} C_{31,00}^{31,0}}{9216}+\frac{7 \sqrt{7} C_{31,00}^{31,1}}{4608}+\frac{7 \sqrt{\frac{7}{3}} C_{31,20}^{31,0}}{9216}+\frac{7
\sqrt{\frac{7}{3}} C_{31,20}^{31,1}}{4608}-\frac{\sqrt{\frac{7}{15}} C_{31,22}^{31,0}}{2304}-\frac{\sqrt{\frac{7}{15}} C_{31,22}^{31,1}}{1152}-\frac{3
C_{31,22}^{33,0}}{5120}-\frac{3 C_{31,22}^{33,1}}{2560}-\frac{7 \sqrt{3} C_{33,00}^{33,0}}{5120}-\frac{7 \sqrt{3} C_{33,00}^{33,1}}{2560}-\frac{7
C_{33,20}^{33,0}}{5120}-\frac{7 C_{33,20}^{33,1}}{2560}+\frac{1}{320} \sqrt{\frac{3}{10}} C_{33,22}^{33,0}+\frac{1}{160} \sqrt{\frac{3}{10}} C_{33,22}^{33,1}\)

\noindent\(C_{22,2213,1}^{20,0011,1,\text{Loc}}= -\frac{7 \sqrt{7} C_{00,20}^{60,1}}{1536}-\frac{1}{512} \sqrt{\frac{7}{5}} C_{00,22}^{62,1}+\frac{7
\sqrt{7} C_{11,20}^{31,1}}{4608}+\frac{\sqrt{\frac{7}{5}} C_{11,22}^{51,1}}{2304}+\frac{3 \sqrt{3} C_{11,22}^{53,1}}{5120}+\frac{7 \sqrt{7} C_{20,20}^{40,1}}{4608}+\frac{\sqrt{\frac{7}{5}}
C_{20,22}^{42,1}}{2304}-\frac{7 \sqrt{\frac{7}{5}} C_{22,20}^{42,1}}{4608}+\frac{\sqrt{35} C_{22,22}^{40,1}}{4608}-\frac{7 C_{22,22}^{42,1}}{2304
\sqrt{5}}-\frac{7 \sqrt{7} C_{31,20}^{31,1}}{4608}-\frac{\sqrt{\frac{7}{5}} C_{31,22}^{31,1}}{2304}-\frac{3 \sqrt{3} C_{31,22}^{33,1}}{5120}+\frac{7
\sqrt{3} C_{33,20}^{33,1}}{2560}+\frac{3 C_{33,22}^{33,1}}{320 \sqrt{10}}\)

\noindent\(C_{22,2213,1}^{20,0011,1,\text{NonLoc}}= -\frac{7 \sqrt{\frac{7}{3}} C_{00,00}^{60,0}}{1024}-\frac{7 \sqrt{7} C_{00,20}^{60,0}}{3072}+\frac{1}{512}
\sqrt{\frac{7}{5}} C_{00,22}^{62,0}-\frac{7 \sqrt{\frac{7}{3}} C_{11,00}^{51,0}}{3072}-\frac{7 \sqrt{7} C_{11,20}^{31,0}}{9216}+\frac{\sqrt{\frac{7}{5}}
C_{11,22}^{51,0}}{2304}+\frac{3 \sqrt{3} C_{11,22}^{53,0}}{5120}+\frac{7 \sqrt{\frac{7}{3}} C_{20,00}^{40,0}}{3072}+\frac{7 \sqrt{7} C_{20,20}^{40,0}}{9216}-\frac{\sqrt{\frac{7}{5}}
C_{20,22}^{42,0}}{2304}-\frac{7 \sqrt{\frac{7}{15}} C_{22,00}^{42,0}}{3072}-\frac{7 \sqrt{\frac{7}{5}} C_{22,20}^{42,0}}{9216}-\frac{\sqrt{35} C_{22,22}^{40,0}}{4608}+\frac{7
C_{22,22}^{42,0}}{2304 \sqrt{5}}+\frac{7 \sqrt{\frac{7}{3}} C_{31,00}^{31,0}}{3072}+\frac{7 \sqrt{7} C_{31,20}^{31,0}}{9216}-\frac{\sqrt{\frac{7}{5}}
C_{31,22}^{31,0}}{2304}-\frac{3 \sqrt{3} C_{31,22}^{33,0}}{5120}-\frac{21 C_{33,00}^{33,0}}{5120}-\frac{7 \sqrt{3} C_{33,20}^{33,0}}{5120}+\frac{3
C_{33,22}^{33,0}}{320 \sqrt{10}}\)

\noindent\(C_{22,2213,1}^{22,0011,0,\text{Loc}}= -\frac{1}{384} \sqrt{\frac{7}{15}} C_{00,20}^{60,0}-\frac{1}{768} \sqrt{\frac{7}{15}}
C_{00,20}^{60,1}-\frac{1}{160} \sqrt{\frac{7}{3}} C_{00,22}^{62,0}-\frac{1}{320} \sqrt{\frac{7}{3}} C_{00,22}^{62,1}+\frac{7 \sqrt{\frac{7}{3}} C_{11,22}^{51,0}}{1152}+\frac{7
\sqrt{\frac{7}{3}} C_{11,22}^{51,1}}{2304}-\frac{3 C_{11,22}^{53,0}}{256 \sqrt{5}}-\frac{3 C_{11,22}^{53,1}}{512 \sqrt{5}}+\frac{\sqrt{\frac{7}{15}}
C_{20,20}^{40,0}}{1152}+\frac{\sqrt{\frac{7}{15}} C_{20,20}^{40,1}}{2304}+\frac{7 \sqrt{\frac{7}{3}} C_{20,22}^{42,0}}{1152}+\frac{7 \sqrt{\frac{7}{3}}
C_{20,22}^{42,1}}{2304}-\frac{\sqrt{\frac{7}{3}} C_{22,20}^{42,0}}{5760}-\frac{\sqrt{\frac{7}{3}} C_{22,20}^{42,1}}{11520}-\frac{17 \sqrt{\frac{7}{3}}
C_{22,22}^{40,0}}{2880}-\frac{17 \sqrt{\frac{7}{3}} C_{22,22}^{40,1}}{5760}-\frac{C_{22,22}^{42,0}}{2880 \sqrt{3}}-\frac{C_{22,22}^{42,1}}{5760 \sqrt{3}}-\frac{9}{640}
\sqrt{\frac{3}{35}} C_{22,22}^{44,0}-\frac{9 \sqrt{\frac{3}{35}} C_{22,22}^{44,1}}{1280}-\frac{\sqrt{\frac{7}{15}} C_{31,20}^{31,0}}{1152}-\frac{\sqrt{\frac{7}{15}}
C_{31,20}^{31,1}}{2304}-\frac{7 \sqrt{\frac{7}{3}} C_{31,22}^{31,0}}{1152}-\frac{7 \sqrt{\frac{7}{3}} C_{31,22}^{31,1}}{2304}+\frac{3 C_{31,22}^{33,0}}{256
\sqrt{5}}+\frac{3 C_{31,22}^{33,1}}{512 \sqrt{5}}+\frac{C_{33,20}^{33,0}}{640 \sqrt{5}}+\frac{C_{33,20}^{33,1}}{1280 \sqrt{5}}+\frac{1}{640} \sqrt{\frac{3}{2}}
C_{33,22}^{33,0}+\frac{\sqrt{\frac{3}{2}} C_{33,22}^{33,1}}{1280}\)

\noindent\(C_{22,2213,1}^{22,0011,0,\text{NonLoc}}= -\frac{\sqrt{\frac{7}{5}} C_{00,00}^{60,0}}{1536}-\frac{1}{768} \sqrt{\frac{7}{5}}
C_{00,00}^{60,1}-\frac{\sqrt{\frac{7}{15}} C_{00,20}^{60,0}}{1536}-\frac{1}{768} \sqrt{\frac{7}{15}} C_{00,20}^{60,1}+\frac{1}{320} \sqrt{\frac{7}{3}}
C_{00,22}^{62,0}+\frac{1}{160} \sqrt{\frac{7}{3}} C_{00,22}^{62,1}-\frac{\sqrt{\frac{7}{5}} C_{11,00}^{51,0}}{4608}-\frac{\sqrt{\frac{7}{5}} C_{11,00}^{51,1}}{2304}+\frac{7
\sqrt{\frac{7}{3}} C_{11,22}^{51,0}}{2304}+\frac{7 \sqrt{\frac{7}{3}} C_{11,22}^{51,1}}{1152}-\frac{3 C_{11,22}^{53,0}}{512 \sqrt{5}}-\frac{3 C_{11,22}^{53,1}}{256
\sqrt{5}}+\frac{\sqrt{\frac{7}{5}} C_{20,00}^{40,0}}{4608}+\frac{\sqrt{\frac{7}{5}} C_{20,00}^{40,1}}{2304}+\frac{\sqrt{\frac{7}{15}} C_{20,20}^{40,0}}{4608}+\frac{\sqrt{\frac{7}{15}}
C_{20,20}^{40,1}}{2304}-\frac{7 \sqrt{\frac{7}{3}} C_{20,22}^{42,0}}{2304}-\frac{7 \sqrt{\frac{7}{3}} C_{20,22}^{42,1}}{1152}-\frac{\sqrt{7} C_{22,00}^{42,0}}{23040}-\frac{\sqrt{7}
C_{22,00}^{42,1}}{11520}-\frac{\sqrt{\frac{7}{3}} C_{22,20}^{42,0}}{23040}-\frac{\sqrt{\frac{7}{3}} C_{22,20}^{42,1}}{11520}+\frac{17 \sqrt{\frac{7}{3}}
C_{22,22}^{40,0}}{5760}+\frac{17 \sqrt{\frac{7}{3}} C_{22,22}^{40,1}}{2880}+\frac{C_{22,22}^{42,0}}{5760 \sqrt{3}}+\frac{C_{22,22}^{42,1}}{2880 \sqrt{3}}+\frac{9
\sqrt{\frac{3}{35}} C_{22,22}^{44,0}}{1280}+\frac{9}{640} \sqrt{\frac{3}{35}} C_{22,22}^{44,1}+\frac{\sqrt{\frac{7}{5}} C_{31,00}^{31,0}}{4608}+\frac{\sqrt{\frac{7}{5}}
C_{31,00}^{31,1}}{2304}+\frac{\sqrt{\frac{7}{15}} C_{31,20}^{31,0}}{4608}+\frac{\sqrt{\frac{7}{15}} C_{31,20}^{31,1}}{2304}-\frac{7 \sqrt{\frac{7}{3}}
C_{31,22}^{31,0}}{2304}-\frac{7 \sqrt{\frac{7}{3}} C_{31,22}^{31,1}}{1152}+\frac{3 C_{31,22}^{33,0}}{512 \sqrt{5}}+\frac{3 C_{31,22}^{33,1}}{256
\sqrt{5}}-\frac{\sqrt{\frac{3}{5}} C_{33,00}^{33,0}}{2560}-\frac{\sqrt{\frac{3}{5}} C_{33,00}^{33,1}}{1280}-\frac{C_{33,20}^{33,0}}{2560 \sqrt{5}}-\frac{C_{33,20}^{33,1}}{1280
\sqrt{5}}+\frac{\sqrt{\frac{3}{2}} C_{33,22}^{33,0}}{1280}+\frac{1}{640} \sqrt{\frac{3}{2}} C_{33,22}^{33,1}\)

\noindent\(C_{22,2213,1}^{22,0011,1,\text{Loc}}= -\frac{1}{768} \sqrt{\frac{7}{5}} C_{00,20}^{60,1}-\frac{\sqrt{7} C_{00,22}^{62,1}}{320}+\frac{7
\sqrt{7} C_{11,22}^{51,1}}{2304}-\frac{3}{512} \sqrt{\frac{3}{5}} C_{11,22}^{53,1}+\frac{\sqrt{\frac{7}{5}} C_{20,20}^{40,1}}{2304}+\frac{7 \sqrt{7}
C_{20,22}^{42,1}}{2304}-\frac{\sqrt{7} C_{22,20}^{42,1}}{11520}-\frac{17 \sqrt{7} C_{22,22}^{40,1}}{5760}-\frac{C_{22,22}^{42,1}}{5760}-\frac{27
C_{22,22}^{44,1}}{1280 \sqrt{35}}-\frac{\sqrt{\frac{7}{5}} C_{31,20}^{31,1}}{2304}-\frac{7 \sqrt{7} C_{31,22}^{31,1}}{2304}+\frac{3}{512} \sqrt{\frac{3}{5}}
C_{31,22}^{33,1}+\frac{\sqrt{\frac{3}{5}} C_{33,20}^{33,1}}{1280}+\frac{3 C_{33,22}^{33,1}}{1280 \sqrt{2}}\)

\noindent\(C_{22,2213,1}^{22,0011,1,\text{NonLoc}}= -\frac{1}{512} \sqrt{\frac{7}{15}} C_{00,00}^{60,0}-\frac{\sqrt{\frac{7}{5}} C_{00,20}^{60,0}}{1536}+\frac{\sqrt{7}
C_{00,22}^{62,0}}{320}-\frac{\sqrt{\frac{7}{15}} C_{11,00}^{51,0}}{1536}+\frac{7 \sqrt{7} C_{11,22}^{51,0}}{2304}-\frac{3}{512} \sqrt{\frac{3}{5}}
C_{11,22}^{53,0}+\frac{\sqrt{\frac{7}{15}} C_{20,00}^{40,0}}{1536}+\frac{\sqrt{\frac{7}{5}} C_{20,20}^{40,0}}{4608}-\frac{7 \sqrt{7} C_{20,22}^{42,0}}{2304}-\frac{\sqrt{\frac{7}{3}}
C_{22,00}^{42,0}}{7680}-\frac{\sqrt{7} C_{22,20}^{42,0}}{23040}+\frac{17 \sqrt{7} C_{22,22}^{40,0}}{5760}+\frac{C_{22,22}^{42,0}}{5760}+\frac{27
C_{22,22}^{44,0}}{1280 \sqrt{35}}+\frac{\sqrt{\frac{7}{15}} C_{31,00}^{31,0}}{1536}+\frac{\sqrt{\frac{7}{5}} C_{31,20}^{31,0}}{4608}-\frac{7 \sqrt{7}
C_{31,22}^{31,0}}{2304}+\frac{3}{512} \sqrt{\frac{3}{5}} C_{31,22}^{33,0}-\frac{3 C_{33,00}^{33,0}}{2560 \sqrt{5}}-\frac{\sqrt{\frac{3}{5}} C_{33,20}^{33,0}}{2560}+\frac{3
C_{33,22}^{33,0}}{1280 \sqrt{2}}\)

\noindent\(C_{22,2213,2}^{00,2212,0,\text{Loc}}= \frac{1}{48} \sqrt{\frac{7}{6}} C_{00,20}^{60,0}+\frac{1}{96} \sqrt{\frac{7}{6}} C_{00,20}^{60,1}-\frac{1}{32}
\sqrt{\frac{7}{30}} C_{00,22}^{62,0}-\frac{1}{64} \sqrt{\frac{7}{30}} C_{00,22}^{62,1}+\frac{1}{72} \sqrt{\frac{7}{30}} C_{11,22}^{51,0}+\frac{1}{144}
\sqrt{\frac{7}{30}} C_{11,22}^{51,1}-\frac{9 C_{11,22}^{53,0}}{320 \sqrt{2}}-\frac{9 C_{11,22}^{53,1}}{640 \sqrt{2}}-\frac{1}{144} \sqrt{\frac{7}{6}}
C_{20,20}^{40,0}-\frac{1}{288} \sqrt{\frac{7}{6}} C_{20,20}^{40,1}-\frac{1}{72} \sqrt{\frac{7}{30}} C_{20,22}^{42,0}-\frac{1}{144} \sqrt{\frac{7}{30}}
C_{20,22}^{42,1}+\frac{1}{144} \sqrt{\frac{7}{30}} C_{22,20}^{42,0}+\frac{1}{288} \sqrt{\frac{7}{30}} C_{22,20}^{42,1}+\frac{1}{288} \sqrt{\frac{35}{6}}
C_{22,22}^{40,0}+\frac{1}{576} \sqrt{\frac{35}{6}} C_{22,22}^{40,1}+\frac{C_{22,22}^{42,0}}{18 \sqrt{30}}+\frac{C_{22,22}^{42,1}}{36 \sqrt{30}}-\frac{1}{32}
\sqrt{\frac{3}{14}} C_{22,22}^{44,0}-\frac{1}{64} \sqrt{\frac{3}{14}} C_{22,22}^{44,1}-\frac{1}{144} \sqrt{\frac{7}{6}} C_{31,20}^{31,0}-\frac{1}{288}
\sqrt{\frac{7}{6}} C_{31,20}^{31,1}+\frac{1}{144} \sqrt{\frac{7}{30}} C_{31,22}^{31,0}+\frac{1}{288} \sqrt{\frac{7}{30}} C_{31,22}^{31,1}-\frac{C_{31,22}^{33,0}}{960
\sqrt{2}}-\frac{C_{31,22}^{33,1}}{1920 \sqrt{2}}+\frac{C_{33,20}^{33,0}}{80 \sqrt{2}}+\frac{C_{33,20}^{33,1}}{160 \sqrt{2}}+\frac{7}{320} \sqrt{\frac{3}{5}}
C_{33,22}^{33,0}+\frac{7}{640} \sqrt{\frac{3}{5}} C_{33,22}^{33,1}\)

\noindent\(C_{22,2213,2}^{00,2212,0,\text{NonLoc}}= \frac{1}{192} \sqrt{\frac{7}{2}} C_{00,00}^{60,0}+\frac{1}{96} \sqrt{\frac{7}{2}}
C_{00,00}^{60,1}+\frac{1}{192} \sqrt{\frac{7}{6}} C_{00,20}^{60,0}+\frac{1}{96} \sqrt{\frac{7}{6}} C_{00,20}^{60,1}+\frac{1}{64} \sqrt{\frac{7}{30}}
C_{00,22}^{62,0}+\frac{1}{32} \sqrt{\frac{7}{30}} C_{00,22}^{62,1}-\frac{1}{576} \sqrt{\frac{7}{2}} C_{11,00}^{51,0}-\frac{1}{288} \sqrt{\frac{7}{2}}
C_{11,00}^{51,1}+\frac{1}{144} \sqrt{\frac{7}{30}} C_{11,22}^{51,0}+\frac{1}{72} \sqrt{\frac{7}{30}} C_{11,22}^{51,1}-\frac{9 C_{11,22}^{53,0}}{640
\sqrt{2}}-\frac{9 C_{11,22}^{53,1}}{320 \sqrt{2}}-\frac{1}{576} \sqrt{\frac{7}{2}} C_{20,00}^{40,0}-\frac{1}{288} \sqrt{\frac{7}{2}} C_{20,00}^{40,1}-\frac{1}{576}
\sqrt{\frac{7}{6}} C_{20,20}^{40,0}-\frac{1}{288} \sqrt{\frac{7}{6}} C_{20,20}^{40,1}+\frac{1}{144} \sqrt{\frac{7}{30}} C_{20,22}^{42,0}+\frac{1}{72}
\sqrt{\frac{7}{30}} C_{20,22}^{42,1}+\frac{1}{576} \sqrt{\frac{7}{10}} C_{22,00}^{42,0}+\frac{1}{288} \sqrt{\frac{7}{10}} C_{22,00}^{42,1}+\frac{1}{576}
\sqrt{\frac{7}{30}} C_{22,20}^{42,0}+\frac{1}{288} \sqrt{\frac{7}{30}} C_{22,20}^{42,1}-\frac{1}{576} \sqrt{\frac{35}{6}} C_{22,22}^{40,0}-\frac{1}{288}
\sqrt{\frac{35}{6}} C_{22,22}^{40,1}-\frac{C_{22,22}^{42,0}}{36 \sqrt{30}}-\frac{C_{22,22}^{42,1}}{18 \sqrt{30}}+\frac{1}{64} \sqrt{\frac{3}{14}}
C_{22,22}^{44,0}+\frac{1}{32} \sqrt{\frac{3}{14}} C_{22,22}^{44,1}+\frac{1}{576} \sqrt{\frac{7}{2}} C_{31,00}^{31,0}+\frac{1}{288} \sqrt{\frac{7}{2}}
C_{31,00}^{31,1}+\frac{1}{576} \sqrt{\frac{7}{6}} C_{31,20}^{31,0}+\frac{1}{288} \sqrt{\frac{7}{6}} C_{31,20}^{31,1}+\frac{1}{288} \sqrt{\frac{7}{30}}
C_{31,22}^{31,0}+\frac{1}{144} \sqrt{\frac{7}{30}} C_{31,22}^{31,1}-\frac{C_{31,22}^{33,0}}{1920 \sqrt{2}}-\frac{C_{31,22}^{33,1}}{960 \sqrt{2}}-\frac{1}{320}
\sqrt{\frac{3}{2}} C_{33,00}^{33,0}-\frac{1}{160} \sqrt{\frac{3}{2}} C_{33,00}^{33,1}-\frac{C_{33,20}^{33,0}}{320 \sqrt{2}}-\frac{C_{33,20}^{33,1}}{160
\sqrt{2}}+\frac{7}{640} \sqrt{\frac{3}{5}} C_{33,22}^{33,0}+\frac{7}{320} \sqrt{\frac{3}{5}} C_{33,22}^{33,1}\)

\noindent\(C_{22,2213,2}^{00,2212,1,\text{Loc}}= \frac{1}{96} \sqrt{\frac{7}{2}} C_{00,20}^{60,1}-\frac{1}{64} \sqrt{\frac{7}{10}}
C_{00,22}^{62,1}+\frac{1}{144} \sqrt{\frac{7}{10}} C_{11,22}^{51,1}-\frac{9}{640} \sqrt{\frac{3}{2}} C_{11,22}^{53,1}-\frac{1}{288} \sqrt{\frac{7}{2}}
C_{20,20}^{40,1}-\frac{1}{144} \sqrt{\frac{7}{10}} C_{20,22}^{42,1}+\frac{1}{288} \sqrt{\frac{7}{10}} C_{22,20}^{42,1}+\frac{1}{576} \sqrt{\frac{35}{2}}
C_{22,22}^{40,1}+\frac{C_{22,22}^{42,1}}{36 \sqrt{10}}-\frac{3 C_{22,22}^{44,1}}{64 \sqrt{14}}-\frac{1}{288} \sqrt{\frac{7}{2}} C_{31,20}^{31,1}+\frac{1}{288}
\sqrt{\frac{7}{10}} C_{31,22}^{31,1}-\frac{C_{31,22}^{33,1}}{640 \sqrt{6}}+\frac{1}{160} \sqrt{\frac{3}{2}} C_{33,20}^{33,1}+\frac{21 C_{33,22}^{33,1}}{640
\sqrt{5}}\)

\noindent\(C_{22,2213,2}^{00,2212,1,\text{NonLoc}}= \frac{1}{64} \sqrt{\frac{7}{6}} C_{00,00}^{60,0}+\frac{1}{192} \sqrt{\frac{7}{2}}
C_{00,20}^{60,0}+\frac{1}{64} \sqrt{\frac{7}{10}} C_{00,22}^{62,0}-\frac{1}{192} \sqrt{\frac{7}{6}} C_{11,00}^{51,0}+\frac{1}{144} \sqrt{\frac{7}{10}}
C_{11,22}^{51,0}-\frac{9}{640} \sqrt{\frac{3}{2}} C_{11,22}^{53,0}-\frac{1}{192} \sqrt{\frac{7}{6}} C_{20,00}^{40,0}-\frac{1}{576} \sqrt{\frac{7}{2}}
C_{20,20}^{40,0}+\frac{1}{144} \sqrt{\frac{7}{10}} C_{20,22}^{42,0}+\frac{1}{192} \sqrt{\frac{7}{30}} C_{22,00}^{42,0}+\frac{1}{576} \sqrt{\frac{7}{10}}
C_{22,20}^{42,0}-\frac{1}{576} \sqrt{\frac{35}{2}} C_{22,22}^{40,0}-\frac{C_{22,22}^{42,0}}{36 \sqrt{10}}+\frac{3 C_{22,22}^{44,0}}{64 \sqrt{14}}+\frac{1}{192}
\sqrt{\frac{7}{6}} C_{31,00}^{31,0}+\frac{1}{576} \sqrt{\frac{7}{2}} C_{31,20}^{31,0}+\frac{1}{288} \sqrt{\frac{7}{10}} C_{31,22}^{31,0}-\frac{C_{31,22}^{33,0}}{640
\sqrt{6}}-\frac{3 C_{33,00}^{33,0}}{320 \sqrt{2}}-\frac{1}{320} \sqrt{\frac{3}{2}} C_{33,20}^{33,0}+\frac{21 C_{33,22}^{33,0}}{640 \sqrt{5}}\)

\noindent\(C_{22,2213,2}^{11,1101,0,\text{Loc}}= -\frac{1}{96} \sqrt{\frac{7}{6}} C_{11,11}^{51,0}-\frac{1}{192} \sqrt{\frac{7}{6}}
C_{11,11}^{51,1}+\frac{1}{96} \sqrt{\frac{7}{6}} C_{31,11}^{31,0}+\frac{1}{192} \sqrt{\frac{7}{6}} C_{31,11}^{31,1}-\frac{\sqrt{3} C_{33,11}^{33,0}}{320}-\frac{\sqrt{3}
C_{33,11}^{33,1}}{640}\)

\noindent\(C_{22,2213,2}^{11,1101,0,\text{NonLoc}}= -\frac{1}{192} \sqrt{\frac{7}{6}} C_{11,11}^{51,0}-\frac{1}{96} \sqrt{\frac{7}{6}}
C_{11,11}^{51,1}+\frac{1}{192} \sqrt{\frac{7}{6}} C_{31,11}^{31,0}+\frac{1}{96} \sqrt{\frac{7}{6}} C_{31,11}^{31,1}-\frac{\sqrt{3} C_{33,11}^{33,0}}{640}-\frac{\sqrt{3}
C_{33,11}^{33,1}}{320}\)

\noindent\(C_{22,2213,2}^{11,1101,1,\text{Loc}}= -\frac{1}{192} \sqrt{\frac{7}{2}} C_{11,11}^{51,1}+\frac{1}{192} \sqrt{\frac{7}{2}}
C_{31,11}^{31,1}-\frac{3 C_{33,11}^{33,1}}{640}\)

\noindent\(C_{22,2213,2}^{11,1101,1,\text{NonLoc}}= -\frac{1}{192} \sqrt{\frac{7}{2}} C_{11,11}^{51,0}+\frac{1}{192} \sqrt{\frac{7}{2}}
C_{31,11}^{31,0}-\frac{3 C_{33,11}^{33,0}}{640}\)

\noindent\(C_{22,2213,2}^{22,0011,0,\text{Loc}}= -\frac{1}{192} \sqrt{\frac{7}{6}} C_{00,20}^{60,0}-\frac{1}{384} \sqrt{\frac{7}{6}}
C_{00,20}^{60,1}+\frac{1}{128} \sqrt{\frac{7}{30}} C_{00,22}^{62,0}+\frac{1}{256} \sqrt{\frac{7}{30}} C_{00,22}^{62,1}-\frac{11 \sqrt{\frac{7}{30}}
C_{11,22}^{51,0}}{1152}-\frac{11 \sqrt{\frac{7}{30}} C_{11,22}^{51,1}}{2304}+\frac{3 C_{11,22}^{53,0}}{640 \sqrt{2}}+\frac{3 C_{11,22}^{53,1}}{1280
\sqrt{2}}+\frac{1}{576} \sqrt{\frac{7}{6}} C_{20,20}^{40,0}+\frac{\sqrt{\frac{7}{6}} C_{20,20}^{40,1}}{1152}-\frac{11 \sqrt{\frac{7}{30}} C_{20,22}^{42,0}}{1152}-\frac{11
\sqrt{\frac{7}{30}} C_{20,22}^{42,1}}{2304}-\frac{1}{576} \sqrt{\frac{7}{30}} C_{22,20}^{42,0}-\frac{\sqrt{\frac{7}{30}} C_{22,20}^{42,1}}{1152}+\frac{13
\sqrt{\frac{7}{30}} C_{22,22}^{40,0}}{1152}+\frac{13 \sqrt{\frac{7}{30}} C_{22,22}^{40,1}}{2304}-\frac{C_{22,22}^{42,0}}{288 \sqrt{30}}-\frac{C_{22,22}^{42,1}}{576
\sqrt{30}}+\frac{3}{640} \sqrt{\frac{3}{14}} C_{22,22}^{44,0}+\frac{3 \sqrt{\frac{3}{14}} C_{22,22}^{44,1}}{1280}-\frac{1}{576} \sqrt{\frac{7}{6}}
C_{31,20}^{31,0}-\frac{\sqrt{\frac{7}{6}} C_{31,20}^{31,1}}{1152}+\frac{11 \sqrt{\frac{7}{30}} C_{31,22}^{31,0}}{1152}+\frac{11 \sqrt{\frac{7}{30}}
C_{31,22}^{31,1}}{2304}-\frac{3 C_{31,22}^{33,0}}{640 \sqrt{2}}-\frac{3 C_{31,22}^{33,1}}{1280 \sqrt{2}}+\frac{C_{33,20}^{33,0}}{320 \sqrt{2}}+\frac{C_{33,20}^{33,1}}{640
\sqrt{2}}+\frac{\sqrt{\frac{3}{5}} C_{33,22}^{33,0}}{1280}+\frac{\sqrt{\frac{3}{5}} C_{33,22}^{33,1}}{2560}\)

\noindent\(C_{22,2213,2}^{22,0011,0,\text{NonLoc}}= -\frac{1}{768} \sqrt{\frac{7}{2}} C_{00,00}^{60,0}-\frac{1}{384} \sqrt{\frac{7}{2}}
C_{00,00}^{60,1}-\frac{1}{768} \sqrt{\frac{7}{6}} C_{00,20}^{60,0}-\frac{1}{384} \sqrt{\frac{7}{6}} C_{00,20}^{60,1}-\frac{1}{256} \sqrt{\frac{7}{30}}
C_{00,22}^{62,0}-\frac{1}{128} \sqrt{\frac{7}{30}} C_{00,22}^{62,1}-\frac{\sqrt{\frac{7}{2}} C_{11,00}^{51,0}}{2304}-\frac{\sqrt{\frac{7}{2}} C_{11,00}^{51,1}}{1152}-\frac{11
\sqrt{\frac{7}{30}} C_{11,22}^{51,0}}{2304}-\frac{11 \sqrt{\frac{7}{30}} C_{11,22}^{51,1}}{1152}+\frac{3 C_{11,22}^{53,0}}{1280 \sqrt{2}}+\frac{3
C_{11,22}^{53,1}}{640 \sqrt{2}}+\frac{\sqrt{\frac{7}{2}} C_{20,00}^{40,0}}{2304}+\frac{\sqrt{\frac{7}{2}} C_{20,00}^{40,1}}{1152}+\frac{\sqrt{\frac{7}{6}}
C_{20,20}^{40,0}}{2304}+\frac{\sqrt{\frac{7}{6}} C_{20,20}^{40,1}}{1152}+\frac{11 \sqrt{\frac{7}{30}} C_{20,22}^{42,0}}{2304}+\frac{11 \sqrt{\frac{7}{30}}
C_{20,22}^{42,1}}{1152}-\frac{\sqrt{\frac{7}{10}} C_{22,00}^{42,0}}{2304}-\frac{\sqrt{\frac{7}{10}} C_{22,00}^{42,1}}{1152}-\frac{\sqrt{\frac{7}{30}}
C_{22,20}^{42,0}}{2304}-\frac{\sqrt{\frac{7}{30}} C_{22,20}^{42,1}}{1152}-\frac{13 \sqrt{\frac{7}{30}} C_{22,22}^{40,0}}{2304}-\frac{13 \sqrt{\frac{7}{30}}
C_{22,22}^{40,1}}{1152}+\frac{C_{22,22}^{42,0}}{576 \sqrt{30}}+\frac{C_{22,22}^{42,1}}{288 \sqrt{30}}-\frac{3 \sqrt{\frac{3}{14}} C_{22,22}^{44,0}}{1280}-\frac{3}{640}
\sqrt{\frac{3}{14}} C_{22,22}^{44,1}+\frac{\sqrt{\frac{7}{2}} C_{31,00}^{31,0}}{2304}+\frac{\sqrt{\frac{7}{2}} C_{31,00}^{31,1}}{1152}+\frac{\sqrt{\frac{7}{6}}
C_{31,20}^{31,0}}{2304}+\frac{\sqrt{\frac{7}{6}} C_{31,20}^{31,1}}{1152}+\frac{11 \sqrt{\frac{7}{30}} C_{31,22}^{31,0}}{2304}+\frac{11 \sqrt{\frac{7}{30}}
C_{31,22}^{31,1}}{1152}-\frac{3 C_{31,22}^{33,0}}{1280 \sqrt{2}}-\frac{3 C_{31,22}^{33,1}}{640 \sqrt{2}}-\frac{\sqrt{\frac{3}{2}} C_{33,00}^{33,0}}{1280}-\frac{1}{640}
\sqrt{\frac{3}{2}} C_{33,00}^{33,1}-\frac{C_{33,20}^{33,0}}{1280 \sqrt{2}}-\frac{C_{33,20}^{33,1}}{640 \sqrt{2}}+\frac{\sqrt{\frac{3}{5}} C_{33,22}^{33,0}}{2560}+\frac{\sqrt{\frac{3}{5}}
C_{33,22}^{33,1}}{1280}\)

\noindent\(C_{22,2213,2}^{22,0011,1,\text{Loc}}= -\frac{1}{384} \sqrt{\frac{7}{2}} C_{00,20}^{60,1}+\frac{1}{256} \sqrt{\frac{7}{10}}
C_{00,22}^{62,1}-\frac{11 \sqrt{\frac{7}{10}} C_{11,22}^{51,1}}{2304}+\frac{3 \sqrt{\frac{3}{2}} C_{11,22}^{53,1}}{1280}+\frac{\sqrt{\frac{7}{2}}
C_{20,20}^{40,1}}{1152}-\frac{11 \sqrt{\frac{7}{10}} C_{20,22}^{42,1}}{2304}-\frac{\sqrt{\frac{7}{10}} C_{22,20}^{42,1}}{1152}+\frac{13 \sqrt{\frac{7}{10}}
C_{22,22}^{40,1}}{2304}-\frac{C_{22,22}^{42,1}}{576 \sqrt{10}}+\frac{9 C_{22,22}^{44,1}}{1280 \sqrt{14}}-\frac{\sqrt{\frac{7}{2}} C_{31,20}^{31,1}}{1152}+\frac{11
\sqrt{\frac{7}{10}} C_{31,22}^{31,1}}{2304}-\frac{3 \sqrt{\frac{3}{2}} C_{31,22}^{33,1}}{1280}+\frac{1}{640} \sqrt{\frac{3}{2}} C_{33,20}^{33,1}+\frac{3
C_{33,22}^{33,1}}{2560 \sqrt{5}}\)

\noindent\(C_{22,2213,2}^{22,0011,1,\text{NonLoc}}= -\frac{1}{256} \sqrt{\frac{7}{6}} C_{00,00}^{60,0}-\frac{1}{768} \sqrt{\frac{7}{2}}
C_{00,20}^{60,0}-\frac{1}{256} \sqrt{\frac{7}{10}} C_{00,22}^{62,0}-\frac{1}{768} \sqrt{\frac{7}{6}} C_{11,00}^{51,0}-\frac{11 \sqrt{\frac{7}{10}}
C_{11,22}^{51,0}}{2304}+\frac{3 \sqrt{\frac{3}{2}} C_{11,22}^{53,0}}{1280}+\frac{1}{768} \sqrt{\frac{7}{6}} C_{20,00}^{40,0}+\frac{\sqrt{\frac{7}{2}}
C_{20,20}^{40,0}}{2304}+\frac{11 \sqrt{\frac{7}{10}} C_{20,22}^{42,0}}{2304}-\frac{1}{768} \sqrt{\frac{7}{30}} C_{22,00}^{42,0}-\frac{\sqrt{\frac{7}{10}}
C_{22,20}^{42,0}}{2304}-\frac{13 \sqrt{\frac{7}{10}} C_{22,22}^{40,0}}{2304}+\frac{C_{22,22}^{42,0}}{576 \sqrt{10}}-\frac{9 C_{22,22}^{44,0}}{1280
\sqrt{14}}+\frac{1}{768} \sqrt{\frac{7}{6}} C_{31,00}^{31,0}+\frac{\sqrt{\frac{7}{2}} C_{31,20}^{31,0}}{2304}+\frac{11 \sqrt{\frac{7}{10}} C_{31,22}^{31,0}}{2304}-\frac{3
\sqrt{\frac{3}{2}} C_{31,22}^{33,0}}{1280}-\frac{3 C_{33,00}^{33,0}}{1280 \sqrt{2}}-\frac{\sqrt{\frac{3}{2}} C_{33,20}^{33,0}}{1280}+\frac{3 C_{33,22}^{33,0}}{2560
\sqrt{5}}\)

\noindent\(C_{22,2213,3}^{00,2213,0,\text{Loc}}= \frac{1}{24} \sqrt{\frac{7}{10}} C_{00,20}^{60,0}+\frac{1}{48} \sqrt{\frac{7}{10}}
C_{00,20}^{60,1}+\frac{1}{160} \sqrt{\frac{7}{2}} C_{00,22}^{62,0}+\frac{1}{320} \sqrt{\frac{7}{2}} C_{00,22}^{62,1}-\frac{\sqrt{\frac{7}{2}} C_{11,22}^{51,0}}{1440}-\frac{\sqrt{\frac{7}{2}}
C_{11,22}^{51,1}}{2880}+\frac{3}{160} \sqrt{\frac{3}{10}} C_{11,22}^{53,0}+\frac{3}{320} \sqrt{\frac{3}{10}} C_{11,22}^{53,1}-\frac{1}{72} \sqrt{\frac{7}{10}}
C_{20,20}^{40,0}-\frac{1}{144} \sqrt{\frac{7}{10}} C_{20,20}^{40,1}+\frac{\sqrt{\frac{7}{2}} C_{20,22}^{42,0}}{1440}+\frac{\sqrt{\frac{7}{2}} C_{20,22}^{42,1}}{2880}+\frac{1}{360}
\sqrt{\frac{7}{2}} C_{22,20}^{42,0}+\frac{1}{720} \sqrt{\frac{7}{2}} C_{22,20}^{42,1}+\frac{\sqrt{\frac{7}{2}} C_{22,22}^{40,0}}{1440}+\frac{\sqrt{\frac{7}{2}}
C_{22,22}^{40,1}}{2880}-\frac{11 C_{22,22}^{42,0}}{720 \sqrt{2}}-\frac{11 C_{22,22}^{42,1}}{1440 \sqrt{2}}+\frac{9 C_{22,22}^{44,0}}{80 \sqrt{70}}+\frac{9
C_{22,22}^{44,1}}{160 \sqrt{70}}-\frac{1}{72} \sqrt{\frac{7}{10}} C_{31,20}^{31,0}-\frac{1}{144} \sqrt{\frac{7}{10}} C_{31,20}^{31,1}-\frac{1}{288}
\sqrt{\frac{7}{2}} C_{31,22}^{31,0}-\frac{1}{576} \sqrt{\frac{7}{2}} C_{31,22}^{31,1}+\frac{C_{31,22}^{33,0}}{32 \sqrt{30}}+\frac{C_{31,22}^{33,1}}{64
\sqrt{30}}+\frac{1}{40} \sqrt{\frac{3}{10}} C_{33,20}^{33,0}+\frac{1}{80} \sqrt{\frac{3}{10}} C_{33,20}^{33,1}-\frac{3 C_{33,22}^{33,0}}{160}-\frac{3
C_{33,22}^{33,1}}{320}\)

\noindent\(C_{22,2213,3}^{00,2213,0,\text{NonLoc}}= \frac{1}{32} \sqrt{\frac{7}{30}} C_{00,00}^{60,0}+\frac{1}{16} \sqrt{\frac{7}{30}}
C_{00,00}^{60,1}+\frac{1}{96} \sqrt{\frac{7}{10}} C_{00,20}^{60,0}+\frac{1}{48} \sqrt{\frac{7}{10}} C_{00,20}^{60,1}-\frac{1}{320} \sqrt{\frac{7}{2}}
C_{00,22}^{62,0}-\frac{1}{160} \sqrt{\frac{7}{2}} C_{00,22}^{62,1}-\frac{1}{96} \sqrt{\frac{7}{30}} C_{11,00}^{51,0}-\frac{1}{48} \sqrt{\frac{7}{30}}
C_{11,00}^{51,1}-\frac{\sqrt{\frac{7}{2}} C_{11,22}^{51,0}}{2880}-\frac{\sqrt{\frac{7}{2}} C_{11,22}^{51,1}}{1440}+\frac{3}{320} \sqrt{\frac{3}{10}}
C_{11,22}^{53,0}+\frac{3}{160} \sqrt{\frac{3}{10}} C_{11,22}^{53,1}-\frac{1}{96} \sqrt{\frac{7}{30}} C_{20,00}^{40,0}-\frac{1}{48} \sqrt{\frac{7}{30}}
C_{20,00}^{40,1}-\frac{1}{288} \sqrt{\frac{7}{10}} C_{20,20}^{40,0}-\frac{1}{144} \sqrt{\frac{7}{10}} C_{20,20}^{40,1}-\frac{\sqrt{\frac{7}{2}} C_{20,22}^{42,0}}{2880}-\frac{\sqrt{\frac{7}{2}}
C_{20,22}^{42,1}}{1440}+\frac{1}{480} \sqrt{\frac{7}{6}} C_{22,00}^{42,0}+\frac{1}{240} \sqrt{\frac{7}{6}} C_{22,00}^{42,1}+\frac{\sqrt{\frac{7}{2}}
C_{22,20}^{42,0}}{1440}+\frac{1}{720} \sqrt{\frac{7}{2}} C_{22,20}^{42,1}-\frac{\sqrt{\frac{7}{2}} C_{22,22}^{40,0}}{2880}-\frac{\sqrt{\frac{7}{2}}
C_{22,22}^{40,1}}{1440}+\frac{11 C_{22,22}^{42,0}}{1440 \sqrt{2}}+\frac{11 C_{22,22}^{42,1}}{720 \sqrt{2}}-\frac{9 C_{22,22}^{44,0}}{160 \sqrt{70}}-\frac{9
C_{22,22}^{44,1}}{80 \sqrt{70}}+\frac{1}{96} \sqrt{\frac{7}{30}} C_{31,00}^{31,0}+\frac{1}{48} \sqrt{\frac{7}{30}} C_{31,00}^{31,1}+\frac{1}{288}
\sqrt{\frac{7}{10}} C_{31,20}^{31,0}+\frac{1}{144} \sqrt{\frac{7}{10}} C_{31,20}^{31,1}-\frac{1}{576} \sqrt{\frac{7}{2}} C_{31,22}^{31,0}-\frac{1}{288}
\sqrt{\frac{7}{2}} C_{31,22}^{31,1}+\frac{C_{31,22}^{33,0}}{64 \sqrt{30}}+\frac{C_{31,22}^{33,1}}{32 \sqrt{30}}-\frac{3 C_{33,00}^{33,0}}{160 \sqrt{10}}-\frac{3
C_{33,00}^{33,1}}{80 \sqrt{10}}-\frac{1}{160} \sqrt{\frac{3}{10}} C_{33,20}^{33,0}-\frac{1}{80} \sqrt{\frac{3}{10}} C_{33,20}^{33,1}-\frac{3 C_{33,22}^{33,0}}{320}-\frac{3
C_{33,22}^{33,1}}{160}\)

\noindent\(C_{22,2213,3}^{00,2213,1,\text{Loc}}= \frac{1}{16} \sqrt{\frac{7}{30}} C_{00,20}^{60,1}+\frac{1}{320} \sqrt{\frac{21}{2}}
C_{00,22}^{62,1}-\frac{1}{960} \sqrt{\frac{7}{6}} C_{11,22}^{51,1}+\frac{9 C_{11,22}^{53,1}}{320 \sqrt{10}}-\frac{1}{48} \sqrt{\frac{7}{30}} C_{20,20}^{40,1}+\frac{1}{960}
\sqrt{\frac{7}{6}} C_{20,22}^{42,1}+\frac{1}{240} \sqrt{\frac{7}{6}} C_{22,20}^{42,1}+\frac{1}{960} \sqrt{\frac{7}{6}} C_{22,22}^{40,1}-\frac{11
C_{22,22}^{42,1}}{480 \sqrt{6}}+\frac{9}{160} \sqrt{\frac{3}{70}} C_{22,22}^{44,1}-\frac{1}{48} \sqrt{\frac{7}{30}} C_{31,20}^{31,1}-\frac{1}{192}
\sqrt{\frac{7}{6}} C_{31,22}^{31,1}+\frac{C_{31,22}^{33,1}}{64 \sqrt{10}}+\frac{3 C_{33,20}^{33,1}}{80 \sqrt{10}}-\frac{3 \sqrt{3} C_{33,22}^{33,1}}{320}\)

\noindent\(C_{22,2213,3}^{00,2213,1,\text{NonLoc}}= \frac{1}{32} \sqrt{\frac{7}{10}} C_{00,00}^{60,0}+\frac{1}{32} \sqrt{\frac{7}{30}}
C_{00,20}^{60,0}-\frac{1}{320} \sqrt{\frac{21}{2}} C_{00,22}^{62,0}-\frac{1}{96} \sqrt{\frac{7}{10}} C_{11,00}^{51,0}-\frac{1}{960} \sqrt{\frac{7}{6}}
C_{11,22}^{51,0}+\frac{9 C_{11,22}^{53,0}}{320 \sqrt{10}}-\frac{1}{96} \sqrt{\frac{7}{10}} C_{20,00}^{40,0}-\frac{1}{96} \sqrt{\frac{7}{30}} C_{20,20}^{40,0}-\frac{1}{960}
\sqrt{\frac{7}{6}} C_{20,22}^{42,0}+\frac{1}{480} \sqrt{\frac{7}{2}} C_{22,00}^{42,0}+\frac{1}{480} \sqrt{\frac{7}{6}} C_{22,20}^{42,0}-\frac{1}{960}
\sqrt{\frac{7}{6}} C_{22,22}^{40,0}+\frac{11 C_{22,22}^{42,0}}{480 \sqrt{6}}-\frac{9}{160} \sqrt{\frac{3}{70}} C_{22,22}^{44,0}+\frac{1}{96} \sqrt{\frac{7}{10}}
C_{31,00}^{31,0}+\frac{1}{96} \sqrt{\frac{7}{30}} C_{31,20}^{31,0}-\frac{1}{192} \sqrt{\frac{7}{6}} C_{31,22}^{31,0}+\frac{C_{31,22}^{33,0}}{64 \sqrt{10}}-\frac{3}{160}
\sqrt{\frac{3}{10}} C_{33,00}^{33,0}-\frac{3 C_{33,20}^{33,0}}{160 \sqrt{10}}-\frac{3 \sqrt{3} C_{33,22}^{33,0}}{320}\)

\noindent\(C_{22,2213,3}^{22,0011,0,\text{Loc}}= -\frac{1}{96} \sqrt{\frac{7}{10}} C_{00,20}^{60,0}-\frac{1}{192} \sqrt{\frac{7}{10}}
C_{00,20}^{60,1}-\frac{1}{640} \sqrt{\frac{7}{2}} C_{00,22}^{62,0}-\frac{\sqrt{\frac{7}{2}} C_{00,22}^{62,1}}{1280}+\frac{\sqrt{\frac{7}{2}} C_{11,22}^{51,0}}{1152}+\frac{\sqrt{\frac{7}{2}}
C_{11,22}^{51,1}}{2304}+\frac{1}{288} \sqrt{\frac{7}{10}} C_{20,20}^{40,0}+\frac{1}{576} \sqrt{\frac{7}{10}} C_{20,20}^{40,1}+\frac{\sqrt{\frac{7}{2}}
C_{20,22}^{42,0}}{1152}+\frac{\sqrt{\frac{7}{2}} C_{20,22}^{42,1}}{2304}-\frac{\sqrt{\frac{7}{2}} C_{22,20}^{42,0}}{1440}-\frac{\sqrt{\frac{7}{2}}
C_{22,20}^{42,1}}{2880}-\frac{\sqrt{\frac{7}{2}} C_{22,22}^{40,0}}{5760}-\frac{\sqrt{\frac{7}{2}} C_{22,22}^{40,1}}{11520}-\frac{C_{22,22}^{42,0}}{720
\sqrt{2}}-\frac{C_{22,22}^{42,1}}{1440 \sqrt{2}}-\frac{3 C_{22,22}^{44,0}}{640 \sqrt{70}}-\frac{3 C_{22,22}^{44,1}}{1280 \sqrt{70}}-\frac{1}{288}
\sqrt{\frac{7}{10}} C_{31,20}^{31,0}-\frac{1}{576} \sqrt{\frac{7}{10}} C_{31,20}^{31,1}-\frac{\sqrt{\frac{7}{2}} C_{31,22}^{31,0}}{1152}-\frac{\sqrt{\frac{7}{2}}
C_{31,22}^{31,1}}{2304}+\frac{1}{160} \sqrt{\frac{3}{10}} C_{33,20}^{33,0}+\frac{1}{320} \sqrt{\frac{3}{10}} C_{33,20}^{33,1}+\frac{3 C_{33,22}^{33,0}}{1280}+\frac{3
C_{33,22}^{33,1}}{2560}\)

\noindent\(C_{22,2213,3}^{22,0011,0,\text{NonLoc}}= -\frac{1}{128} \sqrt{\frac{7}{30}} C_{00,00}^{60,0}-\frac{1}{64} \sqrt{\frac{7}{30}}
C_{00,00}^{60,1}-\frac{1}{384} \sqrt{\frac{7}{10}} C_{00,20}^{60,0}-\frac{1}{192} \sqrt{\frac{7}{10}} C_{00,20}^{60,1}+\frac{\sqrt{\frac{7}{2}} C_{00,22}^{62,0}}{1280}+\frac{1}{640}
\sqrt{\frac{7}{2}} C_{00,22}^{62,1}-\frac{1}{384} \sqrt{\frac{7}{30}} C_{11,00}^{51,0}-\frac{1}{192} \sqrt{\frac{7}{30}} C_{11,00}^{51,1}+\frac{\sqrt{\frac{7}{2}}
C_{11,22}^{51,0}}{2304}+\frac{\sqrt{\frac{7}{2}} C_{11,22}^{51,1}}{1152}+\frac{1}{384} \sqrt{\frac{7}{30}} C_{20,00}^{40,0}+\frac{1}{192} \sqrt{\frac{7}{30}}
C_{20,00}^{40,1}+\frac{\sqrt{\frac{7}{10}} C_{20,20}^{40,0}}{1152}+\frac{1}{576} \sqrt{\frac{7}{10}} C_{20,20}^{40,1}-\frac{\sqrt{\frac{7}{2}} C_{20,22}^{42,0}}{2304}-\frac{\sqrt{\frac{7}{2}}
C_{20,22}^{42,1}}{1152}-\frac{\sqrt{\frac{7}{6}} C_{22,00}^{42,0}}{1920}-\frac{1}{960} \sqrt{\frac{7}{6}} C_{22,00}^{42,1}-\frac{\sqrt{\frac{7}{2}}
C_{22,20}^{42,0}}{5760}-\frac{\sqrt{\frac{7}{2}} C_{22,20}^{42,1}}{2880}+\frac{\sqrt{\frac{7}{2}} C_{22,22}^{40,0}}{11520}+\frac{\sqrt{\frac{7}{2}}
C_{22,22}^{40,1}}{5760}+\frac{C_{22,22}^{42,0}}{1440 \sqrt{2}}+\frac{C_{22,22}^{42,1}}{720 \sqrt{2}}+\frac{3 C_{22,22}^{44,0}}{1280 \sqrt{70}}+\frac{3
C_{22,22}^{44,1}}{640 \sqrt{70}}+\frac{1}{384} \sqrt{\frac{7}{30}} C_{31,00}^{31,0}+\frac{1}{192} \sqrt{\frac{7}{30}} C_{31,00}^{31,1}+\frac{\sqrt{\frac{7}{10}}
C_{31,20}^{31,0}}{1152}+\frac{1}{576} \sqrt{\frac{7}{10}} C_{31,20}^{31,1}-\frac{\sqrt{\frac{7}{2}} C_{31,22}^{31,0}}{2304}-\frac{\sqrt{\frac{7}{2}}
C_{31,22}^{31,1}}{1152}-\frac{3 C_{33,00}^{33,0}}{640 \sqrt{10}}-\frac{3 C_{33,00}^{33,1}}{320 \sqrt{10}}-\frac{1}{640} \sqrt{\frac{3}{10}} C_{33,20}^{33,0}-\frac{1}{320}
\sqrt{\frac{3}{10}} C_{33,20}^{33,1}+\frac{3 C_{33,22}^{33,0}}{2560}+\frac{3 C_{33,22}^{33,1}}{1280}\)

\noindent\(C_{22,2213,3}^{22,0011,1,\text{Loc}}= -\frac{1}{64} \sqrt{\frac{7}{30}} C_{00,20}^{60,1}-\frac{\sqrt{\frac{21}{2}} C_{00,22}^{62,1}}{1280}+\frac{1}{768}
\sqrt{\frac{7}{6}} C_{11,22}^{51,1}+\frac{1}{192} \sqrt{\frac{7}{30}} C_{20,20}^{40,1}+\frac{1}{768} \sqrt{\frac{7}{6}} C_{20,22}^{42,1}-\frac{1}{960}
\sqrt{\frac{7}{6}} C_{22,20}^{42,1}-\frac{\sqrt{\frac{7}{6}} C_{22,22}^{40,1}}{3840}-\frac{C_{22,22}^{42,1}}{480 \sqrt{6}}-\frac{3 \sqrt{\frac{3}{70}}
C_{22,22}^{44,1}}{1280}-\frac{1}{192} \sqrt{\frac{7}{30}} C_{31,20}^{31,1}-\frac{1}{768} \sqrt{\frac{7}{6}} C_{31,22}^{31,1}+\frac{3 C_{33,20}^{33,1}}{320
\sqrt{10}}+\frac{3 \sqrt{3} C_{33,22}^{33,1}}{2560}\)

\noindent\(C_{22,2213,3}^{22,0011,1,\text{NonLoc}}= -\frac{1}{128} \sqrt{\frac{7}{10}} C_{00,00}^{60,0}-\frac{1}{128} \sqrt{\frac{7}{30}}
C_{00,20}^{60,0}+\frac{\sqrt{\frac{21}{2}} C_{00,22}^{62,0}}{1280}-\frac{1}{384} \sqrt{\frac{7}{10}} C_{11,00}^{51,0}+\frac{1}{768} \sqrt{\frac{7}{6}}
C_{11,22}^{51,0}+\frac{1}{384} \sqrt{\frac{7}{10}} C_{20,00}^{40,0}+\frac{1}{384} \sqrt{\frac{7}{30}} C_{20,20}^{40,0}-\frac{1}{768} \sqrt{\frac{7}{6}}
C_{20,22}^{42,0}-\frac{\sqrt{\frac{7}{2}} C_{22,00}^{42,0}}{1920}-\frac{\sqrt{\frac{7}{6}} C_{22,20}^{42,0}}{1920}+\frac{\sqrt{\frac{7}{6}} C_{22,22}^{40,0}}{3840}+\frac{C_{22,22}^{42,0}}{480
\sqrt{6}}+\frac{3 \sqrt{\frac{3}{70}} C_{22,22}^{44,0}}{1280}+\frac{1}{384} \sqrt{\frac{7}{10}} C_{31,00}^{31,0}+\frac{1}{384} \sqrt{\frac{7}{30}}
C_{31,20}^{31,0}-\frac{1}{768} \sqrt{\frac{7}{6}} C_{31,22}^{31,0}-\frac{3}{640} \sqrt{\frac{3}{10}} C_{33,00}^{33,0}-\frac{3 C_{33,20}^{33,0}}{640
\sqrt{10}}+\frac{3 \sqrt{3} C_{33,22}^{33,0}}{2560}\)

\noindent\(C_{22,3101,1}^{00,1101,0,\text{Loc}}= \frac{7 C_{00,00}^{60,0}}{32 \sqrt{5}}+\frac{7 C_{00,00}^{60,1}}{64 \sqrt{5}}+\frac{7
C_{11,00}^{51,0}}{96 \sqrt{5}}+\frac{7 C_{11,00}^{51,1}}{192 \sqrt{5}}-\frac{7 C_{20,00}^{40,0}}{96 \sqrt{5}}-\frac{7 C_{20,00}^{40,1}}{192 \sqrt{5}}+\frac{7
C_{22,00}^{42,0}}{480}+\frac{7 C_{22,00}^{42,1}}{960}-\frac{7 C_{31,00}^{31,0}}{96 \sqrt{5}}-\frac{7 C_{31,00}^{31,1}}{192 \sqrt{5}}+\frac{3}{160}
\sqrt{\frac{21}{5}} C_{33,00}^{33,0}+\frac{3}{320} \sqrt{\frac{21}{5}} C_{33,00}^{33,1}\)

\noindent\(C_{22,3101,1}^{00,1101,0,\text{NonLoc}}= -\frac{7 C_{00,00}^{60,0}}{128 \sqrt{5}}-\frac{7 C_{00,00}^{60,1}}{64 \sqrt{5}}+\frac{7}{128}
\sqrt{\frac{3}{5}} C_{00,20}^{60,0}+\frac{7}{64} \sqrt{\frac{3}{5}} C_{00,20}^{60,1}+\frac{7 C_{11,00}^{51,0}}{384 \sqrt{5}}+\frac{7 C_{11,00}^{51,1}}{192
\sqrt{5}}-\frac{7 C_{11,20}^{31,0}}{128 \sqrt{15}}-\frac{7 C_{11,20}^{31,1}}{64 \sqrt{15}}+\frac{7 C_{20,00}^{40,0}}{384 \sqrt{5}}+\frac{7 C_{20,00}^{40,1}}{192
\sqrt{5}}-\frac{7 C_{20,20}^{40,0}}{128 \sqrt{15}}-\frac{7 C_{20,20}^{40,1}}{64 \sqrt{15}}-\frac{7 C_{22,00}^{42,0}}{1920}-\frac{7 C_{22,00}^{42,1}}{960}+\frac{7
C_{22,20}^{42,0}}{640 \sqrt{3}}+\frac{7 C_{22,20}^{42,1}}{320 \sqrt{3}}-\frac{7 C_{31,00}^{31,0}}{384 \sqrt{5}}-\frac{7 C_{31,00}^{31,1}}{192 \sqrt{5}}+\frac{7
C_{31,20}^{31,0}}{128 \sqrt{15}}+\frac{7 C_{31,20}^{31,1}}{64 \sqrt{15}}+\frac{3}{640} \sqrt{\frac{21}{5}} C_{33,00}^{33,0}+\frac{3}{320} \sqrt{\frac{21}{5}}
C_{33,00}^{33,1}-\frac{9}{640} \sqrt{\frac{7}{5}} C_{33,20}^{33,0}-\frac{9}{320} \sqrt{\frac{7}{5}} C_{33,20}^{33,1}\)

\noindent\(C_{22,3101,1}^{00,1101,1,\text{Loc}}= \frac{7}{64} \sqrt{\frac{3}{5}} C_{00,00}^{60,1}+\frac{7 C_{11,00}^{51,1}}{64 \sqrt{15}}-\frac{7
C_{20,00}^{40,1}}{64 \sqrt{15}}+\frac{7 C_{22,00}^{42,1}}{320 \sqrt{3}}-\frac{7 C_{31,00}^{31,1}}{64 \sqrt{15}}+\frac{9}{320} \sqrt{\frac{7}{5}}
C_{33,00}^{33,1}\)

\noindent\(C_{22,3101,1}^{00,1101,1,\text{NonLoc}}= -\frac{7}{128} \sqrt{\frac{3}{5}} C_{00,00}^{60,0}+\frac{21 C_{00,20}^{60,0}}{128
\sqrt{5}}+\frac{7 C_{11,00}^{51,0}}{128 \sqrt{15}}-\frac{7 C_{11,20}^{31,0}}{128 \sqrt{5}}+\frac{7 C_{20,00}^{40,0}}{128 \sqrt{15}}-\frac{7 C_{20,20}^{40,0}}{128
\sqrt{5}}-\frac{7 C_{22,00}^{42,0}}{640 \sqrt{3}}+\frac{7 C_{22,20}^{42,0}}{640}-\frac{7 C_{31,00}^{31,0}}{128 \sqrt{15}}+\frac{7 C_{31,20}^{31,0}}{128
\sqrt{5}}+\frac{9}{640} \sqrt{\frac{7}{5}} C_{33,00}^{33,0}-\frac{9}{640} \sqrt{\frac{21}{5}} C_{33,20}^{33,0}\)

\noindent\(C_{22,3101,1}^{11,0011,0,\text{Loc}}= -\frac{7 C_{11,11}^{51,0}}{384 \sqrt{5}}-\frac{7 C_{11,11}^{51,1}}{768 \sqrt{5}}-\frac{C_{31,11}^{31,0}}{128
\sqrt{5}}-\frac{C_{31,11}^{31,1}}{256 \sqrt{5}}+\frac{3}{160} \sqrt{\frac{7}{10}} C_{33,11}^{33,0}+\frac{3}{320} \sqrt{\frac{7}{10}} C_{33,11}^{33,1}\)

\noindent\(C_{22,3101,1}^{11,0011,0,\text{NonLoc}}= -\frac{7 C_{11,11}^{51,0}}{768 \sqrt{5}}-\frac{7 C_{11,11}^{51,1}}{384 \sqrt{5}}-\frac{C_{31,11}^{31,0}}{256
\sqrt{5}}-\frac{C_{31,11}^{31,1}}{128 \sqrt{5}}+\frac{3}{320} \sqrt{\frac{7}{10}} C_{33,11}^{33,0}+\frac{3}{160} \sqrt{\frac{7}{10}} C_{33,11}^{33,1}\)

\noindent\(C_{22,3101,1}^{11,0011,1,\text{Loc}}= -\frac{7 C_{11,11}^{51,1}}{256 \sqrt{15}}-\frac{1}{256} \sqrt{\frac{3}{5}} C_{31,11}^{31,1}+\frac{3}{320}
\sqrt{\frac{21}{10}} C_{33,11}^{33,1}\)

\noindent\(C_{22,3101,1}^{11,0011,1,\text{NonLoc}}= -\frac{7 C_{11,11}^{51,0}}{256 \sqrt{15}}-\frac{1}{256} \sqrt{\frac{3}{5}} C_{31,11}^{31,0}+\frac{3}{320}
\sqrt{\frac{21}{10}} C_{33,11}^{33,0}\)

\noindent\(C_{22,3101,2}^{11,0011,0,\text{Loc}}= -\frac{7 C_{11,11}^{51,0}}{128 \sqrt{15}}-\frac{7 C_{11,11}^{51,1}}{256 \sqrt{15}}-\frac{1}{128}
\sqrt{\frac{3}{5}} C_{31,11}^{31,0}-\frac{1}{256} \sqrt{\frac{3}{5}} C_{31,11}^{31,1}+\frac{3}{160} \sqrt{\frac{21}{10}} C_{33,11}^{33,0}+\frac{3}{320}
\sqrt{\frac{21}{10}} C_{33,11}^{33,1}\)

\noindent\(C_{22,3101,2}^{11,0011,0,\text{NonLoc}}= -\frac{7 C_{11,11}^{51,0}}{256 \sqrt{15}}-\frac{7 C_{11,11}^{51,1}}{128 \sqrt{15}}-\frac{1}{256}
\sqrt{\frac{3}{5}} C_{31,11}^{31,0}-\frac{1}{128} \sqrt{\frac{3}{5}} C_{31,11}^{31,1}+\frac{3}{320} \sqrt{\frac{21}{10}} C_{33,11}^{33,0}+\frac{3}{160}
\sqrt{\frac{21}{10}} C_{33,11}^{33,1}\)

\noindent\(C_{22,3101,2}^{11,0011,1,\text{Loc}}= -\frac{7 C_{11,11}^{51,1}}{256 \sqrt{5}}-\frac{3 C_{31,11}^{31,1}}{256 \sqrt{5}}+\frac{9}{320}
\sqrt{\frac{7}{10}} C_{33,11}^{33,1}\)

\noindent\(C_{22,3101,2}^{11,0011,1,\text{NonLoc}}= -\frac{7 C_{11,11}^{51,0}}{256 \sqrt{5}}-\frac{3 C_{31,11}^{31,0}}{256 \sqrt{5}}+\frac{9}{320}
\sqrt{\frac{7}{10}} C_{33,11}^{33,0}\)

\noindent\(C_{22,3110,2}^{00,1112,0,\text{Loc}}= -\frac{7 C_{00,20}^{60,0}}{96 \sqrt{5}}-\frac{7 C_{00,20}^{60,1}}{192 \sqrt{5}}-\frac{7
C_{00,22}^{62,0}}{160}-\frac{7 C_{00,22}^{62,1}}{320}-\frac{7 C_{11,20}^{31,0}}{288 \sqrt{5}}-\frac{7 C_{11,20}^{31,1}}{576 \sqrt{5}}-\frac{7 C_{11,22}^{51,0}}{2880}-\frac{7
C_{11,22}^{51,1}}{5760}-\frac{9}{640} \sqrt{\frac{21}{5}} C_{11,22}^{53,0}-\frac{9 \sqrt{\frac{21}{5}} C_{11,22}^{53,1}}{1280}+\frac{7 C_{20,20}^{40,0}}{288
\sqrt{5}}+\frac{7 C_{20,20}^{40,1}}{576 \sqrt{5}}+\frac{7 C_{20,22}^{42,0}}{2880}+\frac{7 C_{20,22}^{42,1}}{5760}-\frac{7 C_{22,20}^{42,0}}{1440}-\frac{7
C_{22,20}^{42,1}}{2880}+\frac{13 C_{22,22}^{40,0}}{720}+\frac{13 C_{22,22}^{40,1}}{1440}-\frac{\sqrt{7} C_{22,22}^{42,0}}{720}-\frac{\sqrt{7} C_{22,22}^{42,1}}{1440}-\frac{3
C_{22,22}^{44,0}}{80 \sqrt{5}}-\frac{3 C_{22,22}^{44,1}}{160 \sqrt{5}}+\frac{7 C_{31,20}^{31,0}}{288 \sqrt{5}}+\frac{7 C_{31,20}^{31,1}}{576 \sqrt{5}}+\frac{37
C_{31,22}^{31,0}}{2880}+\frac{37 C_{31,22}^{31,1}}{5760}+\frac{7}{640} \sqrt{\frac{7}{15}} C_{31,22}^{33,0}+\frac{7 \sqrt{\frac{7}{15}} C_{31,22}^{33,1}}{1280}-\frac{1}{160}
\sqrt{\frac{21}{5}} C_{33,20}^{33,0}-\frac{1}{320} \sqrt{\frac{21}{5}} C_{33,20}^{33,1}-\frac{3}{400} \sqrt{\frac{7}{2}} C_{33,22}^{33,0}-\frac{3}{800}
\sqrt{\frac{7}{2}} C_{33,22}^{33,1}\)

\noindent\(C_{22,3110,2}^{00,1112,0,\text{NonLoc}}= -\frac{7 C_{00,00}^{60,0}}{128 \sqrt{15}}-\frac{7 C_{00,00}^{60,1}}{64 \sqrt{15}}-\frac{7
C_{00,20}^{60,0}}{384 \sqrt{5}}-\frac{7 C_{00,20}^{60,1}}{192 \sqrt{5}}+\frac{7 C_{00,22}^{62,0}}{320}+\frac{7 C_{00,22}^{62,1}}{160}+\frac{7 C_{11,00}^{51,0}}{384
\sqrt{15}}+\frac{7 C_{11,00}^{51,1}}{192 \sqrt{15}}+\frac{7 C_{11,20}^{31,0}}{1152 \sqrt{5}}+\frac{7 C_{11,20}^{31,1}}{576 \sqrt{5}}-\frac{7 C_{11,22}^{51,0}}{5760}-\frac{7
C_{11,22}^{51,1}}{2880}-\frac{9 \sqrt{\frac{21}{5}} C_{11,22}^{53,0}}{1280}-\frac{9}{640} \sqrt{\frac{21}{5}} C_{11,22}^{53,1}+\frac{7 C_{20,00}^{40,0}}{384
\sqrt{15}}+\frac{7 C_{20,00}^{40,1}}{192 \sqrt{15}}+\frac{7 C_{20,20}^{40,0}}{1152 \sqrt{5}}+\frac{7 C_{20,20}^{40,1}}{576 \sqrt{5}}-\frac{7 C_{20,22}^{42,0}}{5760}-\frac{7
C_{20,22}^{42,1}}{2880}-\frac{7 C_{22,00}^{42,0}}{1920 \sqrt{3}}-\frac{7 C_{22,00}^{42,1}}{960 \sqrt{3}}-\frac{7 C_{22,20}^{42,0}}{5760}-\frac{7
C_{22,20}^{42,1}}{2880}-\frac{13 C_{22,22}^{40,0}}{1440}-\frac{13 C_{22,22}^{40,1}}{720}+\frac{\sqrt{7} C_{22,22}^{42,0}}{1440}+\frac{\sqrt{7} C_{22,22}^{42,1}}{720}+\frac{3
C_{22,22}^{44,0}}{160 \sqrt{5}}+\frac{3 C_{22,22}^{44,1}}{80 \sqrt{5}}-\frac{7 C_{31,00}^{31,0}}{384 \sqrt{15}}-\frac{7 C_{31,00}^{31,1}}{192 \sqrt{15}}-\frac{7
C_{31,20}^{31,0}}{1152 \sqrt{5}}-\frac{7 C_{31,20}^{31,1}}{576 \sqrt{5}}+\frac{37 C_{31,22}^{31,0}}{5760}+\frac{37 C_{31,22}^{31,1}}{2880}+\frac{7
\sqrt{\frac{7}{15}} C_{31,22}^{33,0}}{1280}+\frac{7}{640} \sqrt{\frac{7}{15}} C_{31,22}^{33,1}+\frac{3}{640} \sqrt{\frac{7}{5}} C_{33,00}^{33,0}+\frac{3}{320}
\sqrt{\frac{7}{5}} C_{33,00}^{33,1}+\frac{1}{640} \sqrt{\frac{21}{5}} C_{33,20}^{33,0}+\frac{1}{320} \sqrt{\frac{21}{5}} C_{33,20}^{33,1}-\frac{3}{800}
\sqrt{\frac{7}{2}} C_{33,22}^{33,0}-\frac{3}{400} \sqrt{\frac{7}{2}} C_{33,22}^{33,1}\)

\noindent\(C_{22,3110,2}^{00,1112,1,\text{Loc}}= -\frac{7 C_{00,20}^{60,1}}{64 \sqrt{15}}-\frac{7 \sqrt{3} C_{00,22}^{62,1}}{320}-\frac{7
C_{11,20}^{31,1}}{192 \sqrt{15}}-\frac{7 C_{11,22}^{51,1}}{1920 \sqrt{3}}-\frac{27 \sqrt{\frac{7}{5}} C_{11,22}^{53,1}}{1280}+\frac{7 C_{20,20}^{40,1}}{192
\sqrt{15}}+\frac{7 C_{20,22}^{42,1}}{1920 \sqrt{3}}-\frac{7 C_{22,20}^{42,1}}{960 \sqrt{3}}+\frac{13 C_{22,22}^{40,1}}{480 \sqrt{3}}-\frac{1}{480}
\sqrt{\frac{7}{3}} C_{22,22}^{42,1}-\frac{3}{160} \sqrt{\frac{3}{5}} C_{22,22}^{44,1}+\frac{7 C_{31,20}^{31,1}}{192 \sqrt{15}}+\frac{37 C_{31,22}^{31,1}}{1920
\sqrt{3}}+\frac{7 \sqrt{\frac{7}{5}} C_{31,22}^{33,1}}{1280}-\frac{3}{320} \sqrt{\frac{7}{5}} C_{33,20}^{33,1}-\frac{3}{800} \sqrt{\frac{21}{2}}
C_{33,22}^{33,1}\)

\noindent\(C_{22,3110,2}^{00,1112,1,\text{NonLoc}}= -\frac{7 C_{00,00}^{60,0}}{128 \sqrt{5}}-\frac{7 C_{00,20}^{60,0}}{128 \sqrt{15}}+\frac{7
\sqrt{3} C_{00,22}^{62,0}}{320}+\frac{7 C_{11,00}^{51,0}}{384 \sqrt{5}}+\frac{7 C_{11,20}^{31,0}}{384 \sqrt{15}}-\frac{7 C_{11,22}^{51,0}}{1920 \sqrt{3}}-\frac{27
\sqrt{\frac{7}{5}} C_{11,22}^{53,0}}{1280}+\frac{7 C_{20,00}^{40,0}}{384 \sqrt{5}}+\frac{7 C_{20,20}^{40,0}}{384 \sqrt{15}}-\frac{7 C_{20,22}^{42,0}}{1920
\sqrt{3}}-\frac{7 C_{22,00}^{42,0}}{1920}-\frac{7 C_{22,20}^{42,0}}{1920 \sqrt{3}}-\frac{13 C_{22,22}^{40,0}}{480 \sqrt{3}}+\frac{1}{480} \sqrt{\frac{7}{3}}
C_{22,22}^{42,0}+\frac{3}{160} \sqrt{\frac{3}{5}} C_{22,22}^{44,0}-\frac{7 C_{31,00}^{31,0}}{384 \sqrt{5}}-\frac{7 C_{31,20}^{31,0}}{384 \sqrt{15}}+\frac{37
C_{31,22}^{31,0}}{1920 \sqrt{3}}+\frac{7 \sqrt{\frac{7}{5}} C_{31,22}^{33,0}}{1280}+\frac{3}{640} \sqrt{\frac{21}{5}} C_{33,00}^{33,0}+\frac{3}{640}
\sqrt{\frac{7}{5}} C_{33,20}^{33,0}-\frac{3}{800} \sqrt{\frac{21}{2}} C_{33,22}^{33,0}\)

\noindent\(C_{22,3111,1}^{00,1111,0,\text{Loc}}= \frac{7 C_{00,20}^{60,0}}{64 \sqrt{15}}+\frac{7 C_{00,20}^{60,1}}{128 \sqrt{15}}-\frac{7
\sqrt{3} C_{00,22}^{62,0}}{640}-\frac{7 \sqrt{3} C_{00,22}^{62,1}}{1280}+\frac{7 C_{11,20}^{31,0}}{192 \sqrt{15}}+\frac{7 C_{11,20}^{31,1}}{384 \sqrt{15}}-\frac{7
C_{11,22}^{51,0}}{384 \sqrt{3}}-\frac{7 C_{11,22}^{51,1}}{768 \sqrt{3}}-\frac{7 C_{20,20}^{40,0}}{192 \sqrt{15}}-\frac{7 C_{20,20}^{40,1}}{384 \sqrt{15}}+\frac{7
C_{20,22}^{42,0}}{384 \sqrt{3}}+\frac{7 C_{20,22}^{42,1}}{768 \sqrt{3}}+\frac{7 C_{22,20}^{42,0}}{960 \sqrt{3}}+\frac{7 C_{22,20}^{42,1}}{1920 \sqrt{3}}+\frac{53
C_{22,22}^{40,0}}{1920 \sqrt{3}}+\frac{53 C_{22,22}^{40,1}}{3840 \sqrt{3}}-\frac{19}{960} \sqrt{\frac{7}{3}} C_{22,22}^{42,0}-\frac{19 \sqrt{\frac{7}{3}}
C_{22,22}^{42,1}}{1920}+\frac{3}{160} \sqrt{\frac{3}{5}} C_{22,22}^{44,0}+\frac{3}{320} \sqrt{\frac{3}{5}} C_{22,22}^{44,1}-\frac{7 C_{31,20}^{31,0}}{192
\sqrt{15}}-\frac{7 C_{31,20}^{31,1}}{384 \sqrt{15}}+\frac{C_{31,22}^{31,0}}{384 \sqrt{3}}+\frac{C_{31,22}^{31,1}}{768 \sqrt{3}}+\frac{1}{64} \sqrt{\frac{7}{5}}
C_{31,22}^{33,0}+\frac{1}{128} \sqrt{\frac{7}{5}} C_{31,22}^{33,1}+\frac{3}{320} \sqrt{\frac{7}{5}} C_{33,20}^{33,0}+\frac{3}{640} \sqrt{\frac{7}{5}}
C_{33,20}^{33,1}-\frac{3}{160} \sqrt{\frac{21}{2}} C_{33,22}^{33,0}-\frac{3}{320} \sqrt{\frac{21}{2}} C_{33,22}^{33,1}\)

\noindent\(C_{22,3111,1}^{00,1111,0,\text{NonLoc}}= \frac{7 C_{00,00}^{60,0}}{256 \sqrt{5}}+\frac{7 C_{00,00}^{60,1}}{128 \sqrt{5}}+\frac{7
C_{00,20}^{60,0}}{256 \sqrt{15}}+\frac{7 C_{00,20}^{60,1}}{128 \sqrt{15}}+\frac{7 \sqrt{3} C_{00,22}^{62,0}}{1280}+\frac{7 \sqrt{3} C_{00,22}^{62,1}}{640}-\frac{7
C_{11,00}^{51,0}}{768 \sqrt{5}}-\frac{7 C_{11,00}^{51,1}}{384 \sqrt{5}}-\frac{7 C_{11,20}^{31,0}}{768 \sqrt{15}}-\frac{7 C_{11,20}^{31,1}}{384 \sqrt{15}}-\frac{7
C_{11,22}^{51,0}}{768 \sqrt{3}}-\frac{7 C_{11,22}^{51,1}}{384 \sqrt{3}}-\frac{7 C_{20,00}^{40,0}}{768 \sqrt{5}}-\frac{7 C_{20,00}^{40,1}}{384 \sqrt{5}}-\frac{7
C_{20,20}^{40,0}}{768 \sqrt{15}}-\frac{7 C_{20,20}^{40,1}}{384 \sqrt{15}}-\frac{7 C_{20,22}^{42,0}}{768 \sqrt{3}}-\frac{7 C_{20,22}^{42,1}}{384 \sqrt{3}}+\frac{7
C_{22,00}^{42,0}}{3840}+\frac{7 C_{22,00}^{42,1}}{1920}+\frac{7 C_{22,20}^{42,0}}{3840 \sqrt{3}}+\frac{7 C_{22,20}^{42,1}}{1920 \sqrt{3}}-\frac{53
C_{22,22}^{40,0}}{3840 \sqrt{3}}-\frac{53 C_{22,22}^{40,1}}{1920 \sqrt{3}}+\frac{19 \sqrt{\frac{7}{3}} C_{22,22}^{42,0}}{1920}+\frac{19}{960} \sqrt{\frac{7}{3}}
C_{22,22}^{42,1}-\frac{3}{320} \sqrt{\frac{3}{5}} C_{22,22}^{44,0}-\frac{3}{160} \sqrt{\frac{3}{5}} C_{22,22}^{44,1}+\frac{7 C_{31,00}^{31,0}}{768
\sqrt{5}}+\frac{7 C_{31,00}^{31,1}}{384 \sqrt{5}}+\frac{7 C_{31,20}^{31,0}}{768 \sqrt{15}}+\frac{7 C_{31,20}^{31,1}}{384 \sqrt{15}}+\frac{C_{31,22}^{31,0}}{768
\sqrt{3}}+\frac{C_{31,22}^{31,1}}{384 \sqrt{3}}+\frac{1}{128} \sqrt{\frac{7}{5}} C_{31,22}^{33,0}+\frac{1}{64} \sqrt{\frac{7}{5}} C_{31,22}^{33,1}-\frac{3
\sqrt{\frac{21}{5}} C_{33,00}^{33,0}}{1280}-\frac{3}{640} \sqrt{\frac{21}{5}} C_{33,00}^{33,1}-\frac{3 \sqrt{\frac{7}{5}} C_{33,20}^{33,0}}{1280}-\frac{3}{640}
\sqrt{\frac{7}{5}} C_{33,20}^{33,1}-\frac{3}{320} \sqrt{\frac{21}{2}} C_{33,22}^{33,0}-\frac{3}{160} \sqrt{\frac{21}{2}} C_{33,22}^{33,1}\)

\noindent\(C_{22,3111,1}^{00,1111,1,\text{Loc}}= \frac{7 C_{00,20}^{60,1}}{128 \sqrt{5}}-\frac{21 C_{00,22}^{62,1}}{1280}+\frac{7 C_{11,20}^{31,1}}{384
\sqrt{5}}-\frac{7 C_{11,22}^{51,1}}{768}-\frac{7 C_{20,20}^{40,1}}{384 \sqrt{5}}+\frac{7 C_{20,22}^{42,1}}{768}+\frac{7 C_{22,20}^{42,1}}{1920}+\frac{53
C_{22,22}^{40,1}}{3840}-\frac{19 \sqrt{7} C_{22,22}^{42,1}}{1920}+\frac{9 C_{22,22}^{44,1}}{320 \sqrt{5}}-\frac{7 C_{31,20}^{31,1}}{384 \sqrt{5}}+\frac{C_{31,22}^{31,1}}{768}+\frac{1}{128}
\sqrt{\frac{21}{5}} C_{31,22}^{33,1}+\frac{3}{640} \sqrt{\frac{21}{5}} C_{33,20}^{33,1}-\frac{9}{320} \sqrt{\frac{7}{2}} C_{33,22}^{33,1}\)

\noindent\(C_{22,3111,1}^{00,1111,1,\text{NonLoc}}= \frac{7}{256} \sqrt{\frac{3}{5}} C_{00,00}^{60,0}+\frac{7 C_{00,20}^{60,0}}{256
\sqrt{5}}+\frac{21 C_{00,22}^{62,0}}{1280}-\frac{7 C_{11,00}^{51,0}}{256 \sqrt{15}}-\frac{7 C_{11,20}^{31,0}}{768 \sqrt{5}}-\frac{7 C_{11,22}^{51,0}}{768}-\frac{7
C_{20,00}^{40,0}}{256 \sqrt{15}}-\frac{7 C_{20,20}^{40,0}}{768 \sqrt{5}}-\frac{7 C_{20,22}^{42,0}}{768}+\frac{7 C_{22,00}^{42,0}}{1280 \sqrt{3}}+\frac{7
C_{22,20}^{42,0}}{3840}-\frac{53 C_{22,22}^{40,0}}{3840}+\frac{19 \sqrt{7} C_{22,22}^{42,0}}{1920}-\frac{9 C_{22,22}^{44,0}}{320 \sqrt{5}}+\frac{7
C_{31,00}^{31,0}}{256 \sqrt{15}}+\frac{7 C_{31,20}^{31,0}}{768 \sqrt{5}}+\frac{C_{31,22}^{31,0}}{768}+\frac{1}{128} \sqrt{\frac{21}{5}} C_{31,22}^{33,0}-\frac{9
\sqrt{\frac{7}{5}} C_{33,00}^{33,0}}{1280}-\frac{3 \sqrt{\frac{21}{5}} C_{33,20}^{33,0}}{1280}-\frac{9}{320} \sqrt{\frac{7}{2}} C_{33,22}^{33,0}\)

\noindent\(C_{22,3111,1}^{11,0000,0,\text{Loc}}= -\frac{7 C_{11,11}^{51,0}}{192 \sqrt{5}}-\frac{7 C_{11,11}^{51,1}}{384 \sqrt{5}}-\frac{C_{31,11}^{31,0}}{64
\sqrt{5}}-\frac{C_{31,11}^{31,1}}{128 \sqrt{5}}+\frac{3}{80} \sqrt{\frac{7}{10}} C_{33,11}^{33,0}+\frac{3}{160} \sqrt{\frac{7}{10}} C_{33,11}^{33,1}\)

\noindent\(C_{22,3111,1}^{11,0000,0,\text{NonLoc}}= -\frac{7 C_{11,11}^{51,0}}{384 \sqrt{5}}-\frac{7 C_{11,11}^{51,1}}{192 \sqrt{5}}-\frac{C_{31,11}^{31,0}}{128
\sqrt{5}}-\frac{C_{31,11}^{31,1}}{64 \sqrt{5}}+\frac{3}{160} \sqrt{\frac{7}{10}} C_{33,11}^{33,0}+\frac{3}{80} \sqrt{\frac{7}{10}} C_{33,11}^{33,1}\)

\noindent\(C_{22,3111,1}^{11,0000,1,\text{Loc}}= -\frac{7 C_{11,11}^{51,1}}{128 \sqrt{15}}-\frac{1}{128} \sqrt{\frac{3}{5}} C_{31,11}^{31,1}+\frac{3}{160}
\sqrt{\frac{21}{10}} C_{33,11}^{33,1}\)

\noindent\(C_{22,3111,1}^{11,0000,1,\text{NonLoc}}= -\frac{7 C_{11,11}^{51,0}}{128 \sqrt{15}}-\frac{1}{128} \sqrt{\frac{3}{5}} C_{31,11}^{31,0}+\frac{3}{160}
\sqrt{\frac{21}{10}} C_{33,11}^{33,0}\)

\noindent\(C_{22,3111,2}^{00,1112,0,\text{Loc}}= \frac{7 C_{00,20}^{60,0}}{64 \sqrt{5}}+\frac{7 C_{00,20}^{60,1}}{128 \sqrt{5}}-\frac{21
C_{00,22}^{62,0}}{640}-\frac{21 C_{00,22}^{62,1}}{1280}+\frac{7 C_{11,20}^{31,0}}{192 \sqrt{5}}+\frac{7 C_{11,20}^{31,1}}{384 \sqrt{5}}+\frac{7 C_{11,22}^{51,0}}{1920}+\frac{7
C_{11,22}^{51,1}}{3840}-\frac{9}{640} \sqrt{\frac{21}{5}} C_{11,22}^{53,0}-\frac{9 \sqrt{\frac{21}{5}} C_{11,22}^{53,1}}{1280}-\frac{7 C_{20,20}^{40,0}}{192
\sqrt{5}}-\frac{7 C_{20,20}^{40,1}}{384 \sqrt{5}}-\frac{7 C_{20,22}^{42,0}}{1920}-\frac{7 C_{20,22}^{42,1}}{3840}+\frac{7 C_{22,20}^{42,0}}{960}+\frac{7
C_{22,20}^{42,1}}{1920}+\frac{17 C_{22,22}^{40,0}}{1920}+\frac{17 C_{22,22}^{40,1}}{3840}+\frac{\sqrt{7} C_{22,22}^{42,0}}{192}+\frac{\sqrt{7} C_{22,22}^{42,1}}{384}-\frac{9
C_{22,22}^{44,0}}{160 \sqrt{5}}-\frac{9 C_{22,22}^{44,1}}{320 \sqrt{5}}-\frac{7 C_{31,20}^{31,0}}{192 \sqrt{5}}-\frac{7 C_{31,20}^{31,1}}{384 \sqrt{5}}+\frac{23
C_{31,22}^{31,0}}{1920}+\frac{23 C_{31,22}^{31,1}}{3840}-\frac{1}{640} \sqrt{\frac{21}{5}} C_{31,22}^{33,0}-\frac{\sqrt{\frac{21}{5}} C_{31,22}^{33,1}}{1280}+\frac{3}{320}
\sqrt{\frac{21}{5}} C_{33,20}^{33,0}+\frac{3}{640} \sqrt{\frac{21}{5}} C_{33,20}^{33,1}+\frac{9}{800} \sqrt{\frac{7}{2}} C_{33,22}^{33,0}+\frac{9
\sqrt{\frac{7}{2}} C_{33,22}^{33,1}}{1600}\)

\noindent\(C_{22,3111,2}^{00,1112,0,\text{NonLoc}}= \frac{7}{256} \sqrt{\frac{3}{5}} C_{00,00}^{60,0}+\frac{7}{128} \sqrt{\frac{3}{5}}
C_{00,00}^{60,1}+\frac{7 C_{00,20}^{60,0}}{256 \sqrt{5}}+\frac{7 C_{00,20}^{60,1}}{128 \sqrt{5}}+\frac{21 C_{00,22}^{62,0}}{1280}+\frac{21 C_{00,22}^{62,1}}{640}-\frac{7
C_{11,00}^{51,0}}{256 \sqrt{15}}-\frac{7 C_{11,00}^{51,1}}{128 \sqrt{15}}-\frac{7 C_{11,20}^{31,0}}{768 \sqrt{5}}-\frac{7 C_{11,20}^{31,1}}{384 \sqrt{5}}+\frac{7
C_{11,22}^{51,0}}{3840}+\frac{7 C_{11,22}^{51,1}}{1920}-\frac{9 \sqrt{\frac{21}{5}} C_{11,22}^{53,0}}{1280}-\frac{9}{640} \sqrt{\frac{21}{5}} C_{11,22}^{53,1}-\frac{7
C_{20,00}^{40,0}}{256 \sqrt{15}}-\frac{7 C_{20,00}^{40,1}}{128 \sqrt{15}}-\frac{7 C_{20,20}^{40,0}}{768 \sqrt{5}}-\frac{7 C_{20,20}^{40,1}}{384 \sqrt{5}}+\frac{7
C_{20,22}^{42,0}}{3840}+\frac{7 C_{20,22}^{42,1}}{1920}+\frac{7 C_{22,00}^{42,0}}{1280 \sqrt{3}}+\frac{7 C_{22,00}^{42,1}}{640 \sqrt{3}}+\frac{7
C_{22,20}^{42,0}}{3840}+\frac{7 C_{22,20}^{42,1}}{1920}-\frac{17 C_{22,22}^{40,0}}{3840}-\frac{17 C_{22,22}^{40,1}}{1920}-\frac{\sqrt{7} C_{22,22}^{42,0}}{384}-\frac{\sqrt{7}
C_{22,22}^{42,1}}{192}+\frac{9 C_{22,22}^{44,0}}{320 \sqrt{5}}+\frac{9 C_{22,22}^{44,1}}{160 \sqrt{5}}+\frac{7 C_{31,00}^{31,0}}{256 \sqrt{15}}+\frac{7
C_{31,00}^{31,1}}{128 \sqrt{15}}+\frac{7 C_{31,20}^{31,0}}{768 \sqrt{5}}+\frac{7 C_{31,20}^{31,1}}{384 \sqrt{5}}+\frac{23 C_{31,22}^{31,0}}{3840}+\frac{23
C_{31,22}^{31,1}}{1920}-\frac{\sqrt{\frac{21}{5}} C_{31,22}^{33,0}}{1280}-\frac{1}{640} \sqrt{\frac{21}{5}} C_{31,22}^{33,1}-\frac{9 \sqrt{\frac{7}{5}}
C_{33,00}^{33,0}}{1280}-\frac{9}{640} \sqrt{\frac{7}{5}} C_{33,00}^{33,1}-\frac{3 \sqrt{\frac{21}{5}} C_{33,20}^{33,0}}{1280}-\frac{3}{640} \sqrt{\frac{21}{5}}
C_{33,20}^{33,1}+\frac{9 \sqrt{\frac{7}{2}} C_{33,22}^{33,0}}{1600}+\frac{9}{800} \sqrt{\frac{7}{2}} C_{33,22}^{33,1}\)

\noindent\(C_{22,3111,2}^{00,1112,1,\text{Loc}}= \frac{7}{128} \sqrt{\frac{3}{5}} C_{00,20}^{60,1}-\frac{21 \sqrt{3} C_{00,22}^{62,1}}{1280}+\frac{7
C_{11,20}^{31,1}}{128 \sqrt{15}}+\frac{7 C_{11,22}^{51,1}}{1280 \sqrt{3}}-\frac{27 \sqrt{\frac{7}{5}} C_{11,22}^{53,1}}{1280}-\frac{7 C_{20,20}^{40,1}}{128
\sqrt{15}}-\frac{7 C_{20,22}^{42,1}}{1280 \sqrt{3}}+\frac{7 C_{22,20}^{42,1}}{640 \sqrt{3}}+\frac{17 C_{22,22}^{40,1}}{1280 \sqrt{3}}+\frac{1}{128}
\sqrt{\frac{7}{3}} C_{22,22}^{42,1}-\frac{9}{320} \sqrt{\frac{3}{5}} C_{22,22}^{44,1}-\frac{7 C_{31,20}^{31,1}}{128 \sqrt{15}}+\frac{23 C_{31,22}^{31,1}}{1280
\sqrt{3}}-\frac{3 \sqrt{\frac{7}{5}} C_{31,22}^{33,1}}{1280}+\frac{9}{640} \sqrt{\frac{7}{5}} C_{33,20}^{33,1}+\frac{9 \sqrt{\frac{21}{2}} C_{33,22}^{33,1}}{1600}\)

\noindent\(C_{22,3111,2}^{00,1112,1,\text{NonLoc}}= \frac{21 C_{00,00}^{60,0}}{256 \sqrt{5}}+\frac{7}{256} \sqrt{\frac{3}{5}} C_{00,20}^{60,0}+\frac{21
\sqrt{3} C_{00,22}^{62,0}}{1280}-\frac{7 C_{11,00}^{51,0}}{256 \sqrt{5}}-\frac{7 C_{11,20}^{31,0}}{256 \sqrt{15}}+\frac{7 C_{11,22}^{51,0}}{1280
\sqrt{3}}-\frac{27 \sqrt{\frac{7}{5}} C_{11,22}^{53,0}}{1280}-\frac{7 C_{20,00}^{40,0}}{256 \sqrt{5}}-\frac{7 C_{20,20}^{40,0}}{256 \sqrt{15}}+\frac{7
C_{20,22}^{42,0}}{1280 \sqrt{3}}+\frac{7 C_{22,00}^{42,0}}{1280}+\frac{7 C_{22,20}^{42,0}}{1280 \sqrt{3}}-\frac{17 C_{22,22}^{40,0}}{1280 \sqrt{3}}-\frac{1}{128}
\sqrt{\frac{7}{3}} C_{22,22}^{42,0}+\frac{9}{320} \sqrt{\frac{3}{5}} C_{22,22}^{44,0}+\frac{7 C_{31,00}^{31,0}}{256 \sqrt{5}}+\frac{7 C_{31,20}^{31,0}}{256
\sqrt{15}}+\frac{23 C_{31,22}^{31,0}}{1280 \sqrt{3}}-\frac{3 \sqrt{\frac{7}{5}} C_{31,22}^{33,0}}{1280}-\frac{9 \sqrt{\frac{21}{5}} C_{33,00}^{33,0}}{1280}-\frac{9
\sqrt{\frac{7}{5}} C_{33,20}^{33,0}}{1280}+\frac{9 \sqrt{\frac{21}{2}} C_{33,22}^{33,0}}{1600}\)

\noindent\(C_{22,3112,0}^{00,1110,0,\text{Loc}}= -\frac{7 C_{00,20}^{60,0}}{96 \sqrt{5}}-\frac{7 C_{00,20}^{60,1}}{192 \sqrt{5}}-\frac{7
C_{00,22}^{62,0}}{160}-\frac{7 C_{00,22}^{62,1}}{320}-\frac{7 C_{11,20}^{31,0}}{288 \sqrt{5}}-\frac{7 C_{11,20}^{31,1}}{576 \sqrt{5}}-\frac{7 C_{11,22}^{51,0}}{2880}-\frac{7
C_{11,22}^{51,1}}{5760}-\frac{9}{640} \sqrt{\frac{21}{5}} C_{11,22}^{53,0}-\frac{9 \sqrt{\frac{21}{5}} C_{11,22}^{53,1}}{1280}+\frac{7 C_{20,20}^{40,0}}{288
\sqrt{5}}+\frac{7 C_{20,20}^{40,1}}{576 \sqrt{5}}+\frac{7 C_{20,22}^{42,0}}{2880}+\frac{7 C_{20,22}^{42,1}}{5760}-\frac{7 C_{22,20}^{42,0}}{1440}-\frac{7
C_{22,20}^{42,1}}{2880}+\frac{13 C_{22,22}^{40,0}}{720}+\frac{13 C_{22,22}^{40,1}}{1440}-\frac{\sqrt{7} C_{22,22}^{42,0}}{720}-\frac{\sqrt{7} C_{22,22}^{42,1}}{1440}-\frac{3
C_{22,22}^{44,0}}{80 \sqrt{5}}-\frac{3 C_{22,22}^{44,1}}{160 \sqrt{5}}+\frac{7 C_{31,20}^{31,0}}{288 \sqrt{5}}+\frac{7 C_{31,20}^{31,1}}{576 \sqrt{5}}+\frac{37
C_{31,22}^{31,0}}{2880}+\frac{37 C_{31,22}^{31,1}}{5760}+\frac{7}{640} \sqrt{\frac{7}{15}} C_{31,22}^{33,0}+\frac{7 \sqrt{\frac{7}{15}} C_{31,22}^{33,1}}{1280}-\frac{1}{160}
\sqrt{\frac{21}{5}} C_{33,20}^{33,0}-\frac{1}{320} \sqrt{\frac{21}{5}} C_{33,20}^{33,1}-\frac{3}{400} \sqrt{\frac{7}{2}} C_{33,22}^{33,0}-\frac{3}{800}
\sqrt{\frac{7}{2}} C_{33,22}^{33,1}\)

\noindent\(C_{22,3112,0}^{00,1110,0,\text{NonLoc}}= -\frac{7 C_{00,00}^{60,0}}{128 \sqrt{15}}-\frac{7 C_{00,00}^{60,1}}{64 \sqrt{15}}-\frac{7
C_{00,20}^{60,0}}{384 \sqrt{5}}-\frac{7 C_{00,20}^{60,1}}{192 \sqrt{5}}+\frac{7 C_{00,22}^{62,0}}{320}+\frac{7 C_{00,22}^{62,1}}{160}+\frac{7 C_{11,00}^{51,0}}{384
\sqrt{15}}+\frac{7 C_{11,00}^{51,1}}{192 \sqrt{15}}+\frac{7 C_{11,20}^{31,0}}{1152 \sqrt{5}}+\frac{7 C_{11,20}^{31,1}}{576 \sqrt{5}}-\frac{7 C_{11,22}^{51,0}}{5760}-\frac{7
C_{11,22}^{51,1}}{2880}-\frac{9 \sqrt{\frac{21}{5}} C_{11,22}^{53,0}}{1280}-\frac{9}{640} \sqrt{\frac{21}{5}} C_{11,22}^{53,1}+\frac{7 C_{20,00}^{40,0}}{384
\sqrt{15}}+\frac{7 C_{20,00}^{40,1}}{192 \sqrt{15}}+\frac{7 C_{20,20}^{40,0}}{1152 \sqrt{5}}+\frac{7 C_{20,20}^{40,1}}{576 \sqrt{5}}-\frac{7 C_{20,22}^{42,0}}{5760}-\frac{7
C_{20,22}^{42,1}}{2880}-\frac{7 C_{22,00}^{42,0}}{1920 \sqrt{3}}-\frac{7 C_{22,00}^{42,1}}{960 \sqrt{3}}-\frac{7 C_{22,20}^{42,0}}{5760}-\frac{7
C_{22,20}^{42,1}}{2880}-\frac{13 C_{22,22}^{40,0}}{1440}-\frac{13 C_{22,22}^{40,1}}{720}+\frac{\sqrt{7} C_{22,22}^{42,0}}{1440}+\frac{\sqrt{7} C_{22,22}^{42,1}}{720}+\frac{3
C_{22,22}^{44,0}}{160 \sqrt{5}}+\frac{3 C_{22,22}^{44,1}}{80 \sqrt{5}}-\frac{7 C_{31,00}^{31,0}}{384 \sqrt{15}}-\frac{7 C_{31,00}^{31,1}}{192 \sqrt{15}}-\frac{7
C_{31,20}^{31,0}}{1152 \sqrt{5}}-\frac{7 C_{31,20}^{31,1}}{576 \sqrt{5}}+\frac{37 C_{31,22}^{31,0}}{5760}+\frac{37 C_{31,22}^{31,1}}{2880}+\frac{7
\sqrt{\frac{7}{15}} C_{31,22}^{33,0}}{1280}+\frac{7}{640} \sqrt{\frac{7}{15}} C_{31,22}^{33,1}+\frac{3}{640} \sqrt{\frac{7}{5}} C_{33,00}^{33,0}+\frac{3}{320}
\sqrt{\frac{7}{5}} C_{33,00}^{33,1}+\frac{1}{640} \sqrt{\frac{21}{5}} C_{33,20}^{33,0}+\frac{1}{320} \sqrt{\frac{21}{5}} C_{33,20}^{33,1}-\frac{3}{800}
\sqrt{\frac{7}{2}} C_{33,22}^{33,0}-\frac{3}{400} \sqrt{\frac{7}{2}} C_{33,22}^{33,1}\)

\noindent\(C_{22,3112,0}^{00,1110,1,\text{Loc}}= -\frac{7 C_{00,20}^{60,1}}{64 \sqrt{15}}-\frac{7 \sqrt{3} C_{00,22}^{62,1}}{320}-\frac{7
C_{11,20}^{31,1}}{192 \sqrt{15}}-\frac{7 C_{11,22}^{51,1}}{1920 \sqrt{3}}-\frac{27 \sqrt{\frac{7}{5}} C_{11,22}^{53,1}}{1280}+\frac{7 C_{20,20}^{40,1}}{192
\sqrt{15}}+\frac{7 C_{20,22}^{42,1}}{1920 \sqrt{3}}-\frac{7 C_{22,20}^{42,1}}{960 \sqrt{3}}+\frac{13 C_{22,22}^{40,1}}{480 \sqrt{3}}-\frac{1}{480}
\sqrt{\frac{7}{3}} C_{22,22}^{42,1}-\frac{3}{160} \sqrt{\frac{3}{5}} C_{22,22}^{44,1}+\frac{7 C_{31,20}^{31,1}}{192 \sqrt{15}}+\frac{37 C_{31,22}^{31,1}}{1920
\sqrt{3}}+\frac{7 \sqrt{\frac{7}{5}} C_{31,22}^{33,1}}{1280}-\frac{3}{320} \sqrt{\frac{7}{5}} C_{33,20}^{33,1}-\frac{3}{800} \sqrt{\frac{21}{2}}
C_{33,22}^{33,1}\)

\noindent\(C_{22,3112,0}^{00,1110,1,\text{NonLoc}}= -\frac{7 C_{00,00}^{60,0}}{128 \sqrt{5}}-\frac{7 C_{00,20}^{60,0}}{128 \sqrt{15}}+\frac{7
\sqrt{3} C_{00,22}^{62,0}}{320}+\frac{7 C_{11,00}^{51,0}}{384 \sqrt{5}}+\frac{7 C_{11,20}^{31,0}}{384 \sqrt{15}}-\frac{7 C_{11,22}^{51,0}}{1920 \sqrt{3}}-\frac{27
\sqrt{\frac{7}{5}} C_{11,22}^{53,0}}{1280}+\frac{7 C_{20,00}^{40,0}}{384 \sqrt{5}}+\frac{7 C_{20,20}^{40,0}}{384 \sqrt{15}}-\frac{7 C_{20,22}^{42,0}}{1920
\sqrt{3}}-\frac{7 C_{22,00}^{42,0}}{1920}-\frac{7 C_{22,20}^{42,0}}{1920 \sqrt{3}}-\frac{13 C_{22,22}^{40,0}}{480 \sqrt{3}}+\frac{1}{480} \sqrt{\frac{7}{3}}
C_{22,22}^{42,0}+\frac{3}{160} \sqrt{\frac{3}{5}} C_{22,22}^{44,0}-\frac{7 C_{31,00}^{31,0}}{384 \sqrt{5}}-\frac{7 C_{31,20}^{31,0}}{384 \sqrt{15}}+\frac{37
C_{31,22}^{31,0}}{1920 \sqrt{3}}+\frac{7 \sqrt{\frac{7}{5}} C_{31,22}^{33,0}}{1280}+\frac{3}{640} \sqrt{\frac{21}{5}} C_{33,00}^{33,0}+\frac{3}{640}
\sqrt{\frac{7}{5}} C_{33,20}^{33,0}-\frac{3}{800} \sqrt{\frac{21}{2}} C_{33,22}^{33,0}\)

\noindent\(C_{22,3112,1}^{00,1111,0,\text{Loc}}= -\frac{7 C_{00,20}^{60,0}}{64 \sqrt{5}}-\frac{7 C_{00,20}^{60,1}}{128 \sqrt{5}}+\frac{21
C_{00,22}^{62,0}}{640}+\frac{21 C_{00,22}^{62,1}}{1280}-\frac{7 C_{11,20}^{31,0}}{192 \sqrt{5}}-\frac{7 C_{11,20}^{31,1}}{384 \sqrt{5}}-\frac{7 C_{11,22}^{51,0}}{1920}-\frac{7
C_{11,22}^{51,1}}{3840}+\frac{9}{640} \sqrt{\frac{21}{5}} C_{11,22}^{53,0}+\frac{9 \sqrt{\frac{21}{5}} C_{11,22}^{53,1}}{1280}+\frac{7 C_{20,20}^{40,0}}{192
\sqrt{5}}+\frac{7 C_{20,20}^{40,1}}{384 \sqrt{5}}+\frac{7 C_{20,22}^{42,0}}{1920}+\frac{7 C_{20,22}^{42,1}}{3840}-\frac{7 C_{22,20}^{42,0}}{960}-\frac{7
C_{22,20}^{42,1}}{1920}-\frac{17 C_{22,22}^{40,0}}{1920}-\frac{17 C_{22,22}^{40,1}}{3840}-\frac{\sqrt{7} C_{22,22}^{42,0}}{192}-\frac{\sqrt{7} C_{22,22}^{42,1}}{384}+\frac{9
C_{22,22}^{44,0}}{160 \sqrt{5}}+\frac{9 C_{22,22}^{44,1}}{320 \sqrt{5}}+\frac{7 C_{31,20}^{31,0}}{192 \sqrt{5}}+\frac{7 C_{31,20}^{31,1}}{384 \sqrt{5}}-\frac{23
C_{31,22}^{31,0}}{1920}-\frac{23 C_{31,22}^{31,1}}{3840}+\frac{1}{640} \sqrt{\frac{21}{5}} C_{31,22}^{33,0}+\frac{\sqrt{\frac{21}{5}} C_{31,22}^{33,1}}{1280}-\frac{3}{320}
\sqrt{\frac{21}{5}} C_{33,20}^{33,0}-\frac{3}{640} \sqrt{\frac{21}{5}} C_{33,20}^{33,1}-\frac{9}{800} \sqrt{\frac{7}{2}} C_{33,22}^{33,0}-\frac{9
\sqrt{\frac{7}{2}} C_{33,22}^{33,1}}{1600}\)

\noindent\(C_{22,3112,1}^{00,1111,0,\text{NonLoc}}= -\frac{7}{256} \sqrt{\frac{3}{5}} C_{00,00}^{60,0}-\frac{7}{128} \sqrt{\frac{3}{5}}
C_{00,00}^{60,1}-\frac{7 C_{00,20}^{60,0}}{256 \sqrt{5}}-\frac{7 C_{00,20}^{60,1}}{128 \sqrt{5}}-\frac{21 C_{00,22}^{62,0}}{1280}-\frac{21 C_{00,22}^{62,1}}{640}+\frac{7
C_{11,00}^{51,0}}{256 \sqrt{15}}+\frac{7 C_{11,00}^{51,1}}{128 \sqrt{15}}+\frac{7 C_{11,20}^{31,0}}{768 \sqrt{5}}+\frac{7 C_{11,20}^{31,1}}{384 \sqrt{5}}-\frac{7
C_{11,22}^{51,0}}{3840}-\frac{7 C_{11,22}^{51,1}}{1920}+\frac{9 \sqrt{\frac{21}{5}} C_{11,22}^{53,0}}{1280}+\frac{9}{640} \sqrt{\frac{21}{5}} C_{11,22}^{53,1}+\frac{7
C_{20,00}^{40,0}}{256 \sqrt{15}}+\frac{7 C_{20,00}^{40,1}}{128 \sqrt{15}}+\frac{7 C_{20,20}^{40,0}}{768 \sqrt{5}}+\frac{7 C_{20,20}^{40,1}}{384 \sqrt{5}}-\frac{7
C_{20,22}^{42,0}}{3840}-\frac{7 C_{20,22}^{42,1}}{1920}-\frac{7 C_{22,00}^{42,0}}{1280 \sqrt{3}}-\frac{7 C_{22,00}^{42,1}}{640 \sqrt{3}}-\frac{7
C_{22,20}^{42,0}}{3840}-\frac{7 C_{22,20}^{42,1}}{1920}+\frac{17 C_{22,22}^{40,0}}{3840}+\frac{17 C_{22,22}^{40,1}}{1920}+\frac{\sqrt{7} C_{22,22}^{42,0}}{384}+\frac{\sqrt{7}
C_{22,22}^{42,1}}{192}-\frac{9 C_{22,22}^{44,0}}{320 \sqrt{5}}-\frac{9 C_{22,22}^{44,1}}{160 \sqrt{5}}-\frac{7 C_{31,00}^{31,0}}{256 \sqrt{15}}-\frac{7
C_{31,00}^{31,1}}{128 \sqrt{15}}-\frac{7 C_{31,20}^{31,0}}{768 \sqrt{5}}-\frac{7 C_{31,20}^{31,1}}{384 \sqrt{5}}-\frac{23 C_{31,22}^{31,0}}{3840}-\frac{23
C_{31,22}^{31,1}}{1920}+\frac{\sqrt{\frac{21}{5}} C_{31,22}^{33,0}}{1280}+\frac{1}{640} \sqrt{\frac{21}{5}} C_{31,22}^{33,1}+\frac{9 \sqrt{\frac{7}{5}}
C_{33,00}^{33,0}}{1280}+\frac{9}{640} \sqrt{\frac{7}{5}} C_{33,00}^{33,1}+\frac{3 \sqrt{\frac{21}{5}} C_{33,20}^{33,0}}{1280}+\frac{3}{640} \sqrt{\frac{21}{5}}
C_{33,20}^{33,1}-\frac{9 \sqrt{\frac{7}{2}} C_{33,22}^{33,0}}{1600}-\frac{9}{800} \sqrt{\frac{7}{2}} C_{33,22}^{33,1}\)

\noindent\(C_{22,3112,1}^{00,1111,1,\text{Loc}}= -\frac{7}{128} \sqrt{\frac{3}{5}} C_{00,20}^{60,1}+\frac{21 \sqrt{3} C_{00,22}^{62,1}}{1280}-\frac{7
C_{11,20}^{31,1}}{128 \sqrt{15}}-\frac{7 C_{11,22}^{51,1}}{1280 \sqrt{3}}+\frac{27 \sqrt{\frac{7}{5}} C_{11,22}^{53,1}}{1280}+\frac{7 C_{20,20}^{40,1}}{128
\sqrt{15}}+\frac{7 C_{20,22}^{42,1}}{1280 \sqrt{3}}-\frac{7 C_{22,20}^{42,1}}{640 \sqrt{3}}-\frac{17 C_{22,22}^{40,1}}{1280 \sqrt{3}}-\frac{1}{128}
\sqrt{\frac{7}{3}} C_{22,22}^{42,1}+\frac{9}{320} \sqrt{\frac{3}{5}} C_{22,22}^{44,1}+\frac{7 C_{31,20}^{31,1}}{128 \sqrt{15}}-\frac{23 C_{31,22}^{31,1}}{1280
\sqrt{3}}+\frac{3 \sqrt{\frac{7}{5}} C_{31,22}^{33,1}}{1280}-\frac{9}{640} \sqrt{\frac{7}{5}} C_{33,20}^{33,1}-\frac{9 \sqrt{\frac{21}{2}} C_{33,22}^{33,1}}{1600}\)

\noindent\(C_{22,3112,1}^{00,1111,1,\text{NonLoc}}= -\frac{21 C_{00,00}^{60,0}}{256 \sqrt{5}}-\frac{7}{256} \sqrt{\frac{3}{5}} C_{00,20}^{60,0}-\frac{21
\sqrt{3} C_{00,22}^{62,0}}{1280}+\frac{7 C_{11,00}^{51,0}}{256 \sqrt{5}}+\frac{7 C_{11,20}^{31,0}}{256 \sqrt{15}}-\frac{7 C_{11,22}^{51,0}}{1280
\sqrt{3}}+\frac{27 \sqrt{\frac{7}{5}} C_{11,22}^{53,0}}{1280}+\frac{7 C_{20,00}^{40,0}}{256 \sqrt{5}}+\frac{7 C_{20,20}^{40,0}}{256 \sqrt{15}}-\frac{7
C_{20,22}^{42,0}}{1280 \sqrt{3}}-\frac{7 C_{22,00}^{42,0}}{1280}-\frac{7 C_{22,20}^{42,0}}{1280 \sqrt{3}}+\frac{17 C_{22,22}^{40,0}}{1280 \sqrt{3}}+\frac{1}{128}
\sqrt{\frac{7}{3}} C_{22,22}^{42,0}-\frac{9}{320} \sqrt{\frac{3}{5}} C_{22,22}^{44,0}-\frac{7 C_{31,00}^{31,0}}{256 \sqrt{5}}-\frac{7 C_{31,20}^{31,0}}{256
\sqrt{15}}-\frac{23 C_{31,22}^{31,0}}{1280 \sqrt{3}}+\frac{3 \sqrt{\frac{7}{5}} C_{31,22}^{33,0}}{1280}+\frac{9 \sqrt{\frac{21}{5}} C_{33,00}^{33,0}}{1280}+\frac{9
\sqrt{\frac{7}{5}} C_{33,20}^{33,0}}{1280}-\frac{9 \sqrt{\frac{21}{2}} C_{33,22}^{33,0}}{1600}\)

\noindent\(C_{22,3112,2}^{00,1112,0,\text{Loc}}= -\frac{7}{192} \sqrt{\frac{7}{5}} C_{00,20}^{60,0}-\frac{7}{384} \sqrt{\frac{7}{5}}
C_{00,20}^{60,1}-\frac{\sqrt{7} C_{00,22}^{62,0}}{128}-\frac{\sqrt{7} C_{00,22}^{62,1}}{256}-\frac{7}{576} \sqrt{\frac{7}{5}} C_{11,20}^{31,0}-\frac{7
\sqrt{\frac{7}{5}} C_{11,20}^{31,1}}{1152}+\frac{11 \sqrt{7} C_{11,22}^{51,0}}{5760}+\frac{11 \sqrt{7} C_{11,22}^{51,1}}{11520}-\frac{9}{320} \sqrt{\frac{3}{5}}
C_{11,22}^{53,0}-\frac{9}{640} \sqrt{\frac{3}{5}} C_{11,22}^{53,1}+\frac{7}{576} \sqrt{\frac{7}{5}} C_{20,20}^{40,0}+\frac{7 \sqrt{\frac{7}{5}} C_{20,20}^{40,1}}{1152}-\frac{11
\sqrt{7} C_{20,22}^{42,0}}{5760}-\frac{11 \sqrt{7} C_{20,22}^{42,1}}{11520}-\frac{7 \sqrt{7} C_{22,20}^{42,0}}{2880}-\frac{7 \sqrt{7} C_{22,20}^{42,1}}{5760}+\frac{7
\sqrt{7} C_{22,22}^{40,0}}{5760}+\frac{7 \sqrt{7} C_{22,22}^{40,1}}{11520}+\frac{49 C_{22,22}^{42,0}}{2880}+\frac{49 C_{22,22}^{42,1}}{5760}-\frac{3}{160}
\sqrt{\frac{7}{5}} C_{22,22}^{44,0}-\frac{3}{320} \sqrt{\frac{7}{5}} C_{22,22}^{44,1}+\frac{7}{576} \sqrt{\frac{7}{5}} C_{31,20}^{31,0}+\frac{7 \sqrt{\frac{7}{5}}
C_{31,20}^{31,1}}{1152}+\frac{19 \sqrt{7} C_{31,22}^{31,0}}{5760}+\frac{19 \sqrt{7} C_{31,22}^{31,1}}{11520}-\frac{C_{31,22}^{33,0}}{40 \sqrt{15}}-\frac{C_{31,22}^{33,1}}{80
\sqrt{15}}-\frac{7}{320} \sqrt{\frac{3}{5}} C_{33,20}^{33,0}-\frac{7}{640} \sqrt{\frac{3}{5}} C_{33,20}^{33,1}+\frac{33 C_{33,22}^{33,0}}{800 \sqrt{2}}+\frac{33
C_{33,22}^{33,1}}{1600 \sqrt{2}}\)

\noindent\(C_{22,3112,2}^{00,1112,0,\text{NonLoc}}= -\frac{7}{256} \sqrt{\frac{7}{15}} C_{00,00}^{60,0}-\frac{7}{128} \sqrt{\frac{7}{15}}
C_{00,00}^{60,1}-\frac{7}{768} \sqrt{\frac{7}{5}} C_{00,20}^{60,0}-\frac{7}{384} \sqrt{\frac{7}{5}} C_{00,20}^{60,1}+\frac{\sqrt{7} C_{00,22}^{62,0}}{256}+\frac{\sqrt{7}
C_{00,22}^{62,1}}{128}+\frac{7}{768} \sqrt{\frac{7}{15}} C_{11,00}^{51,0}+\frac{7}{384} \sqrt{\frac{7}{15}} C_{11,00}^{51,1}+\frac{7 \sqrt{\frac{7}{5}}
C_{11,20}^{31,0}}{2304}+\frac{7 \sqrt{\frac{7}{5}} C_{11,20}^{31,1}}{1152}+\frac{11 \sqrt{7} C_{11,22}^{51,0}}{11520}+\frac{11 \sqrt{7} C_{11,22}^{51,1}}{5760}-\frac{9}{640}
\sqrt{\frac{3}{5}} C_{11,22}^{53,0}-\frac{9}{320} \sqrt{\frac{3}{5}} C_{11,22}^{53,1}+\frac{7}{768} \sqrt{\frac{7}{15}} C_{20,00}^{40,0}+\frac{7}{384}
\sqrt{\frac{7}{15}} C_{20,00}^{40,1}+\frac{7 \sqrt{\frac{7}{5}} C_{20,20}^{40,0}}{2304}+\frac{7 \sqrt{\frac{7}{5}} C_{20,20}^{40,1}}{1152}+\frac{11
\sqrt{7} C_{20,22}^{42,0}}{11520}+\frac{11 \sqrt{7} C_{20,22}^{42,1}}{5760}-\frac{7 \sqrt{\frac{7}{3}} C_{22,00}^{42,0}}{3840}-\frac{7 \sqrt{\frac{7}{3}}
C_{22,00}^{42,1}}{1920}-\frac{7 \sqrt{7} C_{22,20}^{42,0}}{11520}-\frac{7 \sqrt{7} C_{22,20}^{42,1}}{5760}-\frac{7 \sqrt{7} C_{22,22}^{40,0}}{11520}-\frac{7
\sqrt{7} C_{22,22}^{40,1}}{5760}-\frac{49 C_{22,22}^{42,0}}{5760}-\frac{49 C_{22,22}^{42,1}}{2880}+\frac{3}{320} \sqrt{\frac{7}{5}} C_{22,22}^{44,0}+\frac{3}{160}
\sqrt{\frac{7}{5}} C_{22,22}^{44,1}-\frac{7}{768} \sqrt{\frac{7}{15}} C_{31,00}^{31,0}-\frac{7}{384} \sqrt{\frac{7}{15}} C_{31,00}^{31,1}-\frac{7
\sqrt{\frac{7}{5}} C_{31,20}^{31,0}}{2304}-\frac{7 \sqrt{\frac{7}{5}} C_{31,20}^{31,1}}{1152}+\frac{19 \sqrt{7} C_{31,22}^{31,0}}{11520}+\frac{19
\sqrt{7} C_{31,22}^{31,1}}{5760}-\frac{C_{31,22}^{33,0}}{80 \sqrt{15}}-\frac{C_{31,22}^{33,1}}{40 \sqrt{15}}+\frac{21 C_{33,00}^{33,0}}{1280 \sqrt{5}}+\frac{21
C_{33,00}^{33,1}}{640 \sqrt{5}}+\frac{7 \sqrt{\frac{3}{5}} C_{33,20}^{33,0}}{1280}+\frac{7}{640} \sqrt{\frac{3}{5}} C_{33,20}^{33,1}+\frac{33 C_{33,22}^{33,0}}{1600
\sqrt{2}}+\frac{33 C_{33,22}^{33,1}}{800 \sqrt{2}}\)

\noindent\(C_{22,3112,2}^{00,1112,1,\text{Loc}}= -\frac{7}{128} \sqrt{\frac{7}{15}} C_{00,20}^{60,1}-\frac{\sqrt{21} C_{00,22}^{62,1}}{256}-\frac{7}{384}
\sqrt{\frac{7}{15}} C_{11,20}^{31,1}+\frac{11 \sqrt{\frac{7}{3}} C_{11,22}^{51,1}}{3840}-\frac{27 C_{11,22}^{53,1}}{640 \sqrt{5}}+\frac{7}{384} \sqrt{\frac{7}{15}}
C_{20,20}^{40,1}-\frac{11 \sqrt{\frac{7}{3}} C_{20,22}^{42,1}}{3840}-\frac{7 \sqrt{\frac{7}{3}} C_{22,20}^{42,1}}{1920}+\frac{7 \sqrt{\frac{7}{3}}
C_{22,22}^{40,1}}{3840}+\frac{49 C_{22,22}^{42,1}}{1920 \sqrt{3}}-\frac{3}{320} \sqrt{\frac{21}{5}} C_{22,22}^{44,1}+\frac{7}{384} \sqrt{\frac{7}{15}}
C_{31,20}^{31,1}+\frac{19 \sqrt{\frac{7}{3}} C_{31,22}^{31,1}}{3840}-\frac{C_{31,22}^{33,1}}{80 \sqrt{5}}-\frac{21 C_{33,20}^{33,1}}{640 \sqrt{5}}+\frac{33
\sqrt{\frac{3}{2}} C_{33,22}^{33,1}}{1600}\)

\noindent\(C_{22,3112,2}^{00,1112,1,\text{NonLoc}}= -\frac{7}{256} \sqrt{\frac{7}{5}} C_{00,00}^{60,0}-\frac{7}{256} \sqrt{\frac{7}{15}}
C_{00,20}^{60,0}+\frac{\sqrt{21} C_{00,22}^{62,0}}{256}+\frac{7}{768} \sqrt{\frac{7}{5}} C_{11,00}^{51,0}+\frac{7}{768} \sqrt{\frac{7}{15}} C_{11,20}^{31,0}+\frac{11
\sqrt{\frac{7}{3}} C_{11,22}^{51,0}}{3840}-\frac{27 C_{11,22}^{53,0}}{640 \sqrt{5}}+\frac{7}{768} \sqrt{\frac{7}{5}} C_{20,00}^{40,0}+\frac{7}{768}
\sqrt{\frac{7}{15}} C_{20,20}^{40,0}+\frac{11 \sqrt{\frac{7}{3}} C_{20,22}^{42,0}}{3840}-\frac{7 \sqrt{7} C_{22,00}^{42,0}}{3840}-\frac{7 \sqrt{\frac{7}{3}}
C_{22,20}^{42,0}}{3840}-\frac{7 \sqrt{\frac{7}{3}} C_{22,22}^{40,0}}{3840}-\frac{49 C_{22,22}^{42,0}}{1920 \sqrt{3}}+\frac{3}{320} \sqrt{\frac{21}{5}}
C_{22,22}^{44,0}-\frac{7}{768} \sqrt{\frac{7}{5}} C_{31,00}^{31,0}-\frac{7}{768} \sqrt{\frac{7}{15}} C_{31,20}^{31,0}+\frac{19 \sqrt{\frac{7}{3}}
C_{31,22}^{31,0}}{3840}-\frac{C_{31,22}^{33,0}}{80 \sqrt{5}}+\frac{21 \sqrt{\frac{3}{5}} C_{33,00}^{33,0}}{1280}+\frac{21 C_{33,20}^{33,0}}{1280
\sqrt{5}}+\frac{33 \sqrt{\frac{3}{2}} C_{33,22}^{33,0}}{1600}\)

\noindent\(C_{22,3303,1}^{00,1101,0,\text{Loc}}= \frac{1}{32} \sqrt{\frac{7}{3}} C_{00,00}^{60,0}+\frac{1}{64} \sqrt{\frac{7}{3}} C_{00,00}^{60,1}+\frac{1}{96}
\sqrt{\frac{7}{3}} C_{11,00}^{51,0}+\frac{1}{192} \sqrt{\frac{7}{3}} C_{11,00}^{51,1}-\frac{1}{96} \sqrt{\frac{7}{3}} C_{20,00}^{40,0}-\frac{1}{192}
\sqrt{\frac{7}{3}} C_{20,00}^{40,1}+\frac{1}{96} \sqrt{\frac{7}{15}} C_{22,00}^{42,0}+\frac{1}{192} \sqrt{\frac{7}{15}} C_{22,00}^{42,1}-\frac{1}{96}
\sqrt{\frac{7}{3}} C_{31,00}^{31,0}-\frac{1}{192} \sqrt{\frac{7}{3}} C_{31,00}^{31,1}+\frac{3 C_{33,00}^{33,0}}{160}+\frac{3 C_{33,00}^{33,1}}{320}\)

\noindent\(C_{22,3303,1}^{00,1101,0,\text{NonLoc}}= -\frac{1}{128} \sqrt{\frac{7}{3}} C_{00,00}^{60,0}-\frac{1}{64} \sqrt{\frac{7}{3}}
C_{00,00}^{60,1}+\frac{\sqrt{7} C_{00,20}^{60,0}}{128}+\frac{\sqrt{7} C_{00,20}^{60,1}}{64}+\frac{1}{384} \sqrt{\frac{7}{3}} C_{11,00}^{51,0}+\frac{1}{192}
\sqrt{\frac{7}{3}} C_{11,00}^{51,1}-\frac{\sqrt{7} C_{11,20}^{31,0}}{384}-\frac{\sqrt{7} C_{11,20}^{31,1}}{192}+\frac{1}{384} \sqrt{\frac{7}{3}}
C_{20,00}^{40,0}+\frac{1}{192} \sqrt{\frac{7}{3}} C_{20,00}^{40,1}-\frac{\sqrt{7} C_{20,20}^{40,0}}{384}-\frac{\sqrt{7} C_{20,20}^{40,1}}{192}-\frac{1}{384}
\sqrt{\frac{7}{15}} C_{22,00}^{42,0}-\frac{1}{192} \sqrt{\frac{7}{15}} C_{22,00}^{42,1}+\frac{1}{384} \sqrt{\frac{7}{5}} C_{22,20}^{42,0}+\frac{1}{192}
\sqrt{\frac{7}{5}} C_{22,20}^{42,1}-\frac{1}{384} \sqrt{\frac{7}{3}} C_{31,00}^{31,0}-\frac{1}{192} \sqrt{\frac{7}{3}} C_{31,00}^{31,1}+\frac{\sqrt{7}
C_{31,20}^{31,0}}{384}+\frac{\sqrt{7} C_{31,20}^{31,1}}{192}+\frac{3 C_{33,00}^{33,0}}{640}+\frac{3 C_{33,00}^{33,1}}{320}-\frac{3 \sqrt{3} C_{33,20}^{33,0}}{640}-\frac{3
\sqrt{3} C_{33,20}^{33,1}}{320}\)

\noindent\(C_{22,3303,1}^{00,1101,1,\text{Loc}}= \frac{\sqrt{7} C_{00,00}^{60,1}}{64}+\frac{\sqrt{7} C_{11,00}^{51,1}}{192}-\frac{\sqrt{7}
C_{20,00}^{40,1}}{192}+\frac{1}{192} \sqrt{\frac{7}{5}} C_{22,00}^{42,1}-\frac{\sqrt{7} C_{31,00}^{31,1}}{192}+\frac{3 \sqrt{3} C_{33,00}^{33,1}}{320}\)

\noindent\(C_{22,3303,1}^{00,1101,1,\text{NonLoc}}= -\frac{\sqrt{7} C_{00,00}^{60,0}}{128}+\frac{\sqrt{21} C_{00,20}^{60,0}}{128}+\frac{\sqrt{7}
C_{11,00}^{51,0}}{384}-\frac{1}{128} \sqrt{\frac{7}{3}} C_{11,20}^{31,0}+\frac{\sqrt{7} C_{20,00}^{40,0}}{384}-\frac{1}{128} \sqrt{\frac{7}{3}} C_{20,20}^{40,0}-\frac{1}{384}
\sqrt{\frac{7}{5}} C_{22,00}^{42,0}+\frac{1}{128} \sqrt{\frac{7}{15}} C_{22,20}^{42,0}-\frac{\sqrt{7} C_{31,00}^{31,0}}{384}+\frac{1}{128} \sqrt{\frac{7}{3}}
C_{31,20}^{31,0}+\frac{3 \sqrt{3} C_{33,00}^{33,0}}{640}-\frac{9 C_{33,20}^{33,0}}{640}\)

\noindent\(C_{22,3303,1}^{11,0011,0,\text{Loc}}= \frac{1}{96} \sqrt{\frac{7}{3}} C_{11,11}^{51,0}+\frac{1}{192} \sqrt{\frac{7}{3}}
C_{11,11}^{51,1}-\frac{1}{96} \sqrt{\frac{7}{3}} C_{31,11}^{31,0}-\frac{1}{192} \sqrt{\frac{7}{3}} C_{31,11}^{31,1}+\frac{1}{160} \sqrt{\frac{3}{2}}
C_{33,11}^{33,0}+\frac{1}{320} \sqrt{\frac{3}{2}} C_{33,11}^{33,1}\)

\noindent\(C_{22,3303,1}^{11,0011,0,\text{NonLoc}}= \frac{1}{192} \sqrt{\frac{7}{3}} C_{11,11}^{51,0}+\frac{1}{96} \sqrt{\frac{7}{3}}
C_{11,11}^{51,1}-\frac{1}{192} \sqrt{\frac{7}{3}} C_{31,11}^{31,0}-\frac{1}{96} \sqrt{\frac{7}{3}} C_{31,11}^{31,1}+\frac{1}{320} \sqrt{\frac{3}{2}}
C_{33,11}^{33,0}+\frac{1}{160} \sqrt{\frac{3}{2}} C_{33,11}^{33,1}\)

\noindent\(C_{22,3303,1}^{11,0011,1,\text{Loc}}= \frac{\sqrt{7} C_{11,11}^{51,1}}{192}-\frac{\sqrt{7} C_{31,11}^{31,1}}{192}+\frac{3
C_{33,11}^{33,1}}{320 \sqrt{2}}\)

\noindent\(C_{22,3303,1}^{11,0011,1,\text{NonLoc}}= \frac{\sqrt{7} C_{11,11}^{51,0}}{192}-\frac{\sqrt{7} C_{31,11}^{31,0}}{192}+\frac{3
C_{33,11}^{33,0}}{320 \sqrt{2}}\)

\noindent\(C_{22,3303,2}^{11,0011,0,\text{Loc}}= -\frac{1}{96} \sqrt{\frac{7}{6}} C_{11,11}^{51,0}-\frac{1}{192} \sqrt{\frac{7}{6}}
C_{11,11}^{51,1}+\frac{1}{96} \sqrt{\frac{7}{6}} C_{31,11}^{31,0}+\frac{1}{192} \sqrt{\frac{7}{6}} C_{31,11}^{31,1}-\frac{\sqrt{3} C_{33,11}^{33,0}}{320}-\frac{\sqrt{3}
C_{33,11}^{33,1}}{640}\)

\noindent\(C_{22,3303,2}^{11,0011,0,\text{NonLoc}}= -\frac{1}{192} \sqrt{\frac{7}{6}} C_{11,11}^{51,0}-\frac{1}{96} \sqrt{\frac{7}{6}}
C_{11,11}^{51,1}+\frac{1}{192} \sqrt{\frac{7}{6}} C_{31,11}^{31,0}+\frac{1}{96} \sqrt{\frac{7}{6}} C_{31,11}^{31,1}-\frac{\sqrt{3} C_{33,11}^{33,0}}{640}-\frac{\sqrt{3}
C_{33,11}^{33,1}}{320}\)

\noindent\(C_{22,3303,2}^{11,0011,1,\text{Loc}}= -\frac{1}{192} \sqrt{\frac{7}{2}} C_{11,11}^{51,1}+\frac{1}{192} \sqrt{\frac{7}{2}}
C_{31,11}^{31,1}-\frac{3 C_{33,11}^{33,1}}{640}\)

\noindent\(C_{22,3303,2}^{11,0011,1,\text{NonLoc}}= -\frac{1}{192} \sqrt{\frac{7}{2}} C_{11,11}^{51,0}+\frac{1}{192} \sqrt{\frac{7}{2}}
C_{31,11}^{31,0}-\frac{3 C_{33,11}^{33,0}}{640}\)

\noindent\(C_{22,3312,0}^{00,1110,0,\text{Loc}}= -\frac{1}{96} \sqrt{\frac{7}{3}} C_{00,20}^{60,0}-\frac{1}{192} \sqrt{\frac{7}{3}}
C_{00,20}^{60,1}-\frac{1}{32} \sqrt{\frac{7}{15}} C_{00,22}^{62,0}-\frac{1}{64} \sqrt{\frac{7}{15}} C_{00,22}^{62,1}-\frac{1}{288} \sqrt{\frac{7}{3}}
C_{11,20}^{31,0}-\frac{1}{576} \sqrt{\frac{7}{3}} C_{11,20}^{31,1}-\frac{1}{36} \sqrt{\frac{7}{15}} C_{11,22}^{51,0}-\frac{1}{72} \sqrt{\frac{7}{15}}
C_{11,22}^{51,1}+\frac{3 C_{11,22}^{53,0}}{320}+\frac{3 C_{11,22}^{53,1}}{640}+\frac{1}{288} \sqrt{\frac{7}{3}} C_{20,20}^{40,0}+\frac{1}{576} \sqrt{\frac{7}{3}}
C_{20,20}^{40,1}+\frac{1}{36} \sqrt{\frac{7}{15}} C_{20,22}^{42,0}+\frac{1}{72} \sqrt{\frac{7}{15}} C_{20,22}^{42,1}-\frac{1}{288} \sqrt{\frac{7}{15}}
C_{22,20}^{42,0}-\frac{1}{576} \sqrt{\frac{7}{15}} C_{22,20}^{42,1}-\frac{7}{288} \sqrt{\frac{7}{15}} C_{22,22}^{40,0}-\frac{7}{576} \sqrt{\frac{7}{15}}
C_{22,22}^{40,1}-\frac{C_{22,22}^{42,0}}{144 \sqrt{15}}-\frac{C_{22,22}^{42,1}}{288 \sqrt{15}}-\frac{1}{80} \sqrt{\frac{3}{7}} C_{22,22}^{44,0}-\frac{1}{160}
\sqrt{\frac{3}{7}} C_{22,22}^{44,1}+\frac{1}{288} \sqrt{\frac{7}{3}} C_{31,20}^{31,0}+\frac{1}{576} \sqrt{\frac{7}{3}} C_{31,20}^{31,1}+\frac{1}{36}
\sqrt{\frac{7}{15}} C_{31,22}^{31,0}+\frac{1}{72} \sqrt{\frac{7}{15}} C_{31,22}^{31,1}-\frac{3 C_{31,22}^{33,0}}{320}-\frac{3 C_{31,22}^{33,1}}{640}-\frac{C_{33,20}^{33,0}}{160}-\frac{C_{33,20}^{33,1}}{320}-\frac{1}{80}
\sqrt{\frac{3}{10}} C_{33,22}^{33,0}-\frac{1}{160} \sqrt{\frac{3}{10}} C_{33,22}^{33,1}\)

\noindent\(C_{22,3312,0}^{00,1110,0,\text{NonLoc}}= -\frac{\sqrt{7} C_{00,00}^{60,0}}{384}-\frac{\sqrt{7} C_{00,00}^{60,1}}{192}-\frac{1}{384}
\sqrt{\frac{7}{3}} C_{00,20}^{60,0}-\frac{1}{192} \sqrt{\frac{7}{3}} C_{00,20}^{60,1}+\frac{1}{64} \sqrt{\frac{7}{15}} C_{00,22}^{62,0}+\frac{1}{32}
\sqrt{\frac{7}{15}} C_{00,22}^{62,1}+\frac{\sqrt{7} C_{11,00}^{51,0}}{1152}+\frac{\sqrt{7} C_{11,00}^{51,1}}{576}+\frac{\sqrt{\frac{7}{3}} C_{11,20}^{31,0}}{1152}+\frac{1}{576}
\sqrt{\frac{7}{3}} C_{11,20}^{31,1}-\frac{1}{72} \sqrt{\frac{7}{15}} C_{11,22}^{51,0}-\frac{1}{36} \sqrt{\frac{7}{15}} C_{11,22}^{51,1}+\frac{3 C_{11,22}^{53,0}}{640}+\frac{3
C_{11,22}^{53,1}}{320}+\frac{\sqrt{7} C_{20,00}^{40,0}}{1152}+\frac{\sqrt{7} C_{20,00}^{40,1}}{576}+\frac{\sqrt{\frac{7}{3}} C_{20,20}^{40,0}}{1152}+\frac{1}{576}
\sqrt{\frac{7}{3}} C_{20,20}^{40,1}-\frac{1}{72} \sqrt{\frac{7}{15}} C_{20,22}^{42,0}-\frac{1}{36} \sqrt{\frac{7}{15}} C_{20,22}^{42,1}-\frac{\sqrt{\frac{7}{5}}
C_{22,00}^{42,0}}{1152}-\frac{1}{576} \sqrt{\frac{7}{5}} C_{22,00}^{42,1}-\frac{\sqrt{\frac{7}{15}} C_{22,20}^{42,0}}{1152}-\frac{1}{576} \sqrt{\frac{7}{15}}
C_{22,20}^{42,1}+\frac{7}{576} \sqrt{\frac{7}{15}} C_{22,22}^{40,0}+\frac{7}{288} \sqrt{\frac{7}{15}} C_{22,22}^{40,1}+\frac{C_{22,22}^{42,0}}{288
\sqrt{15}}+\frac{C_{22,22}^{42,1}}{144 \sqrt{15}}+\frac{1}{160} \sqrt{\frac{3}{7}} C_{22,22}^{44,0}+\frac{1}{80} \sqrt{\frac{3}{7}} C_{22,22}^{44,1}-\frac{\sqrt{7}
C_{31,00}^{31,0}}{1152}-\frac{\sqrt{7} C_{31,00}^{31,1}}{576}-\frac{\sqrt{\frac{7}{3}} C_{31,20}^{31,0}}{1152}-\frac{1}{576} \sqrt{\frac{7}{3}} C_{31,20}^{31,1}+\frac{1}{72}
\sqrt{\frac{7}{15}} C_{31,22}^{31,0}+\frac{1}{36} \sqrt{\frac{7}{15}} C_{31,22}^{31,1}-\frac{3 C_{31,22}^{33,0}}{640}-\frac{3 C_{31,22}^{33,1}}{320}+\frac{\sqrt{3}
C_{33,00}^{33,0}}{640}+\frac{\sqrt{3} C_{33,00}^{33,1}}{320}+\frac{C_{33,20}^{33,0}}{640}+\frac{C_{33,20}^{33,1}}{320}-\frac{1}{160} \sqrt{\frac{3}{10}}
C_{33,22}^{33,0}-\frac{1}{80} \sqrt{\frac{3}{10}} C_{33,22}^{33,1}\)

\noindent\(C_{22,3312,0}^{00,1110,1,\text{Loc}}= -\frac{\sqrt{7} C_{00,20}^{60,1}}{192}-\frac{1}{64} \sqrt{\frac{7}{5}} C_{00,22}^{62,1}-\frac{\sqrt{7}
C_{11,20}^{31,1}}{576}-\frac{1}{72} \sqrt{\frac{7}{5}} C_{11,22}^{51,1}+\frac{3 \sqrt{3} C_{11,22}^{53,1}}{640}+\frac{\sqrt{7} C_{20,20}^{40,1}}{576}+\frac{1}{72}
\sqrt{\frac{7}{5}} C_{20,22}^{42,1}-\frac{1}{576} \sqrt{\frac{7}{5}} C_{22,20}^{42,1}-\frac{7}{576} \sqrt{\frac{7}{5}} C_{22,22}^{40,1}-\frac{C_{22,22}^{42,1}}{288
\sqrt{5}}-\frac{3 C_{22,22}^{44,1}}{160 \sqrt{7}}+\frac{\sqrt{7} C_{31,20}^{31,1}}{576}+\frac{1}{72} \sqrt{\frac{7}{5}} C_{31,22}^{31,1}-\frac{3
\sqrt{3} C_{31,22}^{33,1}}{640}-\frac{\sqrt{3} C_{33,20}^{33,1}}{320}-\frac{3 C_{33,22}^{33,1}}{160 \sqrt{10}}\)

\noindent\(C_{22,3312,0}^{00,1110,1,\text{NonLoc}}= -\frac{1}{128} \sqrt{\frac{7}{3}} C_{00,00}^{60,0}-\frac{\sqrt{7} C_{00,20}^{60,0}}{384}+\frac{1}{64}
\sqrt{\frac{7}{5}} C_{00,22}^{62,0}+\frac{1}{384} \sqrt{\frac{7}{3}} C_{11,00}^{51,0}+\frac{\sqrt{7} C_{11,20}^{31,0}}{1152}-\frac{1}{72} \sqrt{\frac{7}{5}}
C_{11,22}^{51,0}+\frac{3 \sqrt{3} C_{11,22}^{53,0}}{640}+\frac{1}{384} \sqrt{\frac{7}{3}} C_{20,00}^{40,0}+\frac{\sqrt{7} C_{20,20}^{40,0}}{1152}-\frac{1}{72}
\sqrt{\frac{7}{5}} C_{20,22}^{42,0}-\frac{1}{384} \sqrt{\frac{7}{15}} C_{22,00}^{42,0}-\frac{\sqrt{\frac{7}{5}} C_{22,20}^{42,0}}{1152}+\frac{7}{576}
\sqrt{\frac{7}{5}} C_{22,22}^{40,0}+\frac{C_{22,22}^{42,0}}{288 \sqrt{5}}+\frac{3 C_{22,22}^{44,0}}{160 \sqrt{7}}-\frac{1}{384} \sqrt{\frac{7}{3}}
C_{31,00}^{31,0}-\frac{\sqrt{7} C_{31,20}^{31,0}}{1152}+\frac{1}{72} \sqrt{\frac{7}{5}} C_{31,22}^{31,0}-\frac{3 \sqrt{3} C_{31,22}^{33,0}}{640}+\frac{3
C_{33,00}^{33,0}}{640}+\frac{\sqrt{3} C_{33,20}^{33,0}}{640}-\frac{3 C_{33,22}^{33,0}}{160 \sqrt{10}}\)

\noindent\(C_{22,3312,1}^{00,1111,0,\text{Loc}}= \frac{1}{96} \sqrt{\frac{7}{3}} C_{00,20}^{60,0}+\frac{1}{192} \sqrt{\frac{7}{3}}
C_{00,20}^{60,1}-\frac{1}{64} \sqrt{\frac{7}{15}} C_{00,22}^{62,0}-\frac{1}{128} \sqrt{\frac{7}{15}} C_{00,22}^{62,1}+\frac{1}{288} \sqrt{\frac{7}{3}}
C_{11,20}^{31,0}+\frac{1}{576} \sqrt{\frac{7}{3}} C_{11,20}^{31,1}-\frac{7}{288} \sqrt{\frac{7}{15}} C_{11,22}^{51,0}-\frac{7}{576} \sqrt{\frac{7}{15}}
C_{11,22}^{51,1}+\frac{9 C_{11,22}^{53,0}}{640}+\frac{9 C_{11,22}^{53,1}}{1280}-\frac{1}{288} \sqrt{\frac{7}{3}} C_{20,20}^{40,0}-\frac{1}{576} \sqrt{\frac{7}{3}}
C_{20,20}^{40,1}+\frac{7}{288} \sqrt{\frac{7}{15}} C_{20,22}^{42,0}+\frac{7}{576} \sqrt{\frac{7}{15}} C_{20,22}^{42,1}+\frac{1}{288} \sqrt{\frac{7}{15}}
C_{22,20}^{42,0}+\frac{1}{576} \sqrt{\frac{7}{15}} C_{22,20}^{42,1}-\frac{19}{576} \sqrt{\frac{7}{15}} C_{22,22}^{40,0}-\frac{19 \sqrt{\frac{7}{15}}
C_{22,22}^{40,1}}{1152}+\frac{1}{288} \sqrt{\frac{5}{3}} C_{22,22}^{42,0}+\frac{1}{576} \sqrt{\frac{5}{3}} C_{22,22}^{42,1}-\frac{1}{80} \sqrt{\frac{3}{7}}
C_{22,22}^{44,0}-\frac{1}{160} \sqrt{\frac{3}{7}} C_{22,22}^{44,1}-\frac{1}{288} \sqrt{\frac{7}{3}} C_{31,20}^{31,0}-\frac{1}{576} \sqrt{\frac{7}{3}}
C_{31,20}^{31,1}+\frac{7}{288} \sqrt{\frac{7}{15}} C_{31,22}^{31,0}+\frac{7}{576} \sqrt{\frac{7}{15}} C_{31,22}^{31,1}-\frac{9 C_{31,22}^{33,0}}{640}-\frac{9
C_{31,22}^{33,1}}{1280}+\frac{C_{33,20}^{33,0}}{160}+\frac{C_{33,20}^{33,1}}{320}+\frac{1}{80} \sqrt{\frac{3}{10}} C_{33,22}^{33,0}+\frac{1}{160}
\sqrt{\frac{3}{10}} C_{33,22}^{33,1}\)

\noindent\(C_{22,3312,1}^{00,1111,0,\text{NonLoc}}= \frac{\sqrt{7} C_{00,00}^{60,0}}{384}+\frac{\sqrt{7} C_{00,00}^{60,1}}{192}+\frac{1}{384}
\sqrt{\frac{7}{3}} C_{00,20}^{60,0}+\frac{1}{192} \sqrt{\frac{7}{3}} C_{00,20}^{60,1}+\frac{1}{128} \sqrt{\frac{7}{15}} C_{00,22}^{62,0}+\frac{1}{64}
\sqrt{\frac{7}{15}} C_{00,22}^{62,1}-\frac{\sqrt{7} C_{11,00}^{51,0}}{1152}-\frac{\sqrt{7} C_{11,00}^{51,1}}{576}-\frac{\sqrt{\frac{7}{3}} C_{11,20}^{31,0}}{1152}-\frac{1}{576}
\sqrt{\frac{7}{3}} C_{11,20}^{31,1}-\frac{7}{576} \sqrt{\frac{7}{15}} C_{11,22}^{51,0}-\frac{7}{288} \sqrt{\frac{7}{15}} C_{11,22}^{51,1}+\frac{9
C_{11,22}^{53,0}}{1280}+\frac{9 C_{11,22}^{53,1}}{640}-\frac{\sqrt{7} C_{20,00}^{40,0}}{1152}-\frac{\sqrt{7} C_{20,00}^{40,1}}{576}-\frac{\sqrt{\frac{7}{3}}
C_{20,20}^{40,0}}{1152}-\frac{1}{576} \sqrt{\frac{7}{3}} C_{20,20}^{40,1}-\frac{7}{576} \sqrt{\frac{7}{15}} C_{20,22}^{42,0}-\frac{7}{288} \sqrt{\frac{7}{15}}
C_{20,22}^{42,1}+\frac{\sqrt{\frac{7}{5}} C_{22,00}^{42,0}}{1152}+\frac{1}{576} \sqrt{\frac{7}{5}} C_{22,00}^{42,1}+\frac{\sqrt{\frac{7}{15}} C_{22,20}^{42,0}}{1152}+\frac{1}{576}
\sqrt{\frac{7}{15}} C_{22,20}^{42,1}+\frac{19 \sqrt{\frac{7}{15}} C_{22,22}^{40,0}}{1152}+\frac{19}{576} \sqrt{\frac{7}{15}} C_{22,22}^{40,1}-\frac{1}{576}
\sqrt{\frac{5}{3}} C_{22,22}^{42,0}-\frac{1}{288} \sqrt{\frac{5}{3}} C_{22,22}^{42,1}+\frac{1}{160} \sqrt{\frac{3}{7}} C_{22,22}^{44,0}+\frac{1}{80}
\sqrt{\frac{3}{7}} C_{22,22}^{44,1}+\frac{\sqrt{7} C_{31,00}^{31,0}}{1152}+\frac{\sqrt{7} C_{31,00}^{31,1}}{576}+\frac{\sqrt{\frac{7}{3}} C_{31,20}^{31,0}}{1152}+\frac{1}{576}
\sqrt{\frac{7}{3}} C_{31,20}^{31,1}+\frac{7}{576} \sqrt{\frac{7}{15}} C_{31,22}^{31,0}+\frac{7}{288} \sqrt{\frac{7}{15}} C_{31,22}^{31,1}-\frac{9
C_{31,22}^{33,0}}{1280}-\frac{9 C_{31,22}^{33,1}}{640}-\frac{\sqrt{3} C_{33,00}^{33,0}}{640}-\frac{\sqrt{3} C_{33,00}^{33,1}}{320}-\frac{C_{33,20}^{33,0}}{640}-\frac{C_{33,20}^{33,1}}{320}+\frac{1}{160}
\sqrt{\frac{3}{10}} C_{33,22}^{33,0}+\frac{1}{80} \sqrt{\frac{3}{10}} C_{33,22}^{33,1}\)

\noindent\(C_{22,3312,1}^{00,1111,1,\text{Loc}}= \frac{\sqrt{7} C_{00,20}^{60,1}}{192}-\frac{1}{128} \sqrt{\frac{7}{5}} C_{00,22}^{62,1}+\frac{\sqrt{7}
C_{11,20}^{31,1}}{576}-\frac{7}{576} \sqrt{\frac{7}{5}} C_{11,22}^{51,1}+\frac{9 \sqrt{3} C_{11,22}^{53,1}}{1280}-\frac{\sqrt{7} C_{20,20}^{40,1}}{576}+\frac{7}{576}
\sqrt{\frac{7}{5}} C_{20,22}^{42,1}+\frac{1}{576} \sqrt{\frac{7}{5}} C_{22,20}^{42,1}-\frac{19 \sqrt{\frac{7}{5}} C_{22,22}^{40,1}}{1152}+\frac{\sqrt{5}
C_{22,22}^{42,1}}{576}-\frac{3 C_{22,22}^{44,1}}{160 \sqrt{7}}-\frac{\sqrt{7} C_{31,20}^{31,1}}{576}+\frac{7}{576} \sqrt{\frac{7}{5}} C_{31,22}^{31,1}-\frac{9
\sqrt{3} C_{31,22}^{33,1}}{1280}+\frac{\sqrt{3} C_{33,20}^{33,1}}{320}+\frac{3 C_{33,22}^{33,1}}{160 \sqrt{10}}\)

\noindent\(C_{22,3312,1}^{00,1111,1,\text{NonLoc}}= \frac{1}{128} \sqrt{\frac{7}{3}} C_{00,00}^{60,0}+\frac{\sqrt{7} C_{00,20}^{60,0}}{384}+\frac{1}{128}
\sqrt{\frac{7}{5}} C_{00,22}^{62,0}-\frac{1}{384} \sqrt{\frac{7}{3}} C_{11,00}^{51,0}-\frac{\sqrt{7} C_{11,20}^{31,0}}{1152}-\frac{7}{576} \sqrt{\frac{7}{5}}
C_{11,22}^{51,0}+\frac{9 \sqrt{3} C_{11,22}^{53,0}}{1280}-\frac{1}{384} \sqrt{\frac{7}{3}} C_{20,00}^{40,0}-\frac{\sqrt{7} C_{20,20}^{40,0}}{1152}-\frac{7}{576}
\sqrt{\frac{7}{5}} C_{20,22}^{42,0}+\frac{1}{384} \sqrt{\frac{7}{15}} C_{22,00}^{42,0}+\frac{\sqrt{\frac{7}{5}} C_{22,20}^{42,0}}{1152}+\frac{19
\sqrt{\frac{7}{5}} C_{22,22}^{40,0}}{1152}-\frac{\sqrt{5} C_{22,22}^{42,0}}{576}+\frac{3 C_{22,22}^{44,0}}{160 \sqrt{7}}+\frac{1}{384} \sqrt{\frac{7}{3}}
C_{31,00}^{31,0}+\frac{\sqrt{7} C_{31,20}^{31,0}}{1152}+\frac{7}{576} \sqrt{\frac{7}{5}} C_{31,22}^{31,0}-\frac{9 \sqrt{3} C_{31,22}^{33,0}}{1280}-\frac{3
C_{33,00}^{33,0}}{640}-\frac{\sqrt{3} C_{33,20}^{33,0}}{640}+\frac{3 C_{33,22}^{33,0}}{160 \sqrt{10}}\)

\noindent\(C_{22,3312,2}^{00,1112,0,\text{Loc}}= -\frac{C_{00,20}^{60,0}}{96 \sqrt{3}}-\frac{C_{00,20}^{60,1}}{192 \sqrt{3}}-\frac{1}{64}
\sqrt{\frac{5}{3}} C_{00,22}^{62,0}-\frac{1}{128} \sqrt{\frac{5}{3}} C_{00,22}^{62,1}-\frac{C_{11,20}^{31,0}}{288 \sqrt{3}}-\frac{C_{11,20}^{31,1}}{576
\sqrt{3}}-\frac{11 C_{11,22}^{51,0}}{288 \sqrt{15}}-\frac{11 C_{11,22}^{51,1}}{576 \sqrt{15}}-\frac{3 C_{11,22}^{53,0}}{640 \sqrt{7}}-\frac{3 C_{11,22}^{53,1}}{1280
\sqrt{7}}+\frac{C_{20,20}^{40,0}}{288 \sqrt{3}}+\frac{C_{20,20}^{40,1}}{576 \sqrt{3}}+\frac{11 C_{20,22}^{42,0}}{288 \sqrt{15}}+\frac{11 C_{20,22}^{42,1}}{576
\sqrt{15}}-\frac{C_{22,20}^{42,0}}{288 \sqrt{15}}-\frac{C_{22,20}^{42,1}}{576 \sqrt{15}}+\frac{C_{22,22}^{40,0}}{576 \sqrt{15}}+\frac{C_{22,22}^{40,1}}{1152
\sqrt{15}}-\frac{23 C_{22,22}^{42,0}}{288 \sqrt{105}}-\frac{23 C_{22,22}^{42,1}}{576 \sqrt{105}}-\frac{\sqrt{3} C_{22,22}^{44,0}}{560}-\frac{\sqrt{3}
C_{22,22}^{44,1}}{1120}+\frac{C_{31,20}^{31,0}}{288 \sqrt{3}}+\frac{C_{31,20}^{31,1}}{576 \sqrt{3}}+\frac{11 C_{31,22}^{31,0}}{288 \sqrt{15}}+\frac{11
C_{31,22}^{31,1}}{576 \sqrt{15}}+\frac{3 C_{31,22}^{33,0}}{640 \sqrt{7}}+\frac{3 C_{31,22}^{33,1}}{1280 \sqrt{7}}-\frac{C_{33,20}^{33,0}}{160 \sqrt{7}}-\frac{C_{33,20}^{33,1}}{320
\sqrt{7}}-\frac{1}{80} \sqrt{\frac{21}{10}} C_{33,22}^{33,0}-\frac{1}{160} \sqrt{\frac{21}{10}} C_{33,22}^{33,1}\)

\noindent\(C_{22,3312,2}^{00,1112,0,\text{NonLoc}}= -\frac{C_{00,00}^{60,0}}{384}-\frac{C_{00,00}^{60,1}}{192}-\frac{C_{00,20}^{60,0}}{384
\sqrt{3}}-\frac{C_{00,20}^{60,1}}{192 \sqrt{3}}+\frac{1}{128} \sqrt{\frac{5}{3}} C_{00,22}^{62,0}+\frac{1}{64} \sqrt{\frac{5}{3}} C_{00,22}^{62,1}+\frac{C_{11,00}^{51,0}}{1152}+\frac{C_{11,00}^{51,1}}{576}+\frac{C_{11,20}^{31,0}}{1152
\sqrt{3}}+\frac{C_{11,20}^{31,1}}{576 \sqrt{3}}-\frac{11 C_{11,22}^{51,0}}{576 \sqrt{15}}-\frac{11 C_{11,22}^{51,1}}{288 \sqrt{15}}-\frac{3 C_{11,22}^{53,0}}{1280
\sqrt{7}}-\frac{3 C_{11,22}^{53,1}}{640 \sqrt{7}}+\frac{C_{20,00}^{40,0}}{1152}+\frac{C_{20,00}^{40,1}}{576}+\frac{C_{20,20}^{40,0}}{1152 \sqrt{3}}+\frac{C_{20,20}^{40,1}}{576
\sqrt{3}}-\frac{11 C_{20,22}^{42,0}}{576 \sqrt{15}}-\frac{11 C_{20,22}^{42,1}}{288 \sqrt{15}}-\frac{C_{22,00}^{42,0}}{1152 \sqrt{5}}-\frac{C_{22,00}^{42,1}}{576
\sqrt{5}}-\frac{C_{22,20}^{42,0}}{1152 \sqrt{15}}-\frac{C_{22,20}^{42,1}}{576 \sqrt{15}}-\frac{C_{22,22}^{40,0}}{1152 \sqrt{15}}-\frac{C_{22,22}^{40,1}}{576
\sqrt{15}}+\frac{23 C_{22,22}^{42,0}}{576 \sqrt{105}}+\frac{23 C_{22,22}^{42,1}}{288 \sqrt{105}}+\frac{\sqrt{3} C_{22,22}^{44,0}}{1120}+\frac{\sqrt{3}
C_{22,22}^{44,1}}{560}-\frac{C_{31,00}^{31,0}}{1152}-\frac{C_{31,00}^{31,1}}{576}-\frac{C_{31,20}^{31,0}}{1152 \sqrt{3}}-\frac{C_{31,20}^{31,1}}{576
\sqrt{3}}+\frac{11 C_{31,22}^{31,0}}{576 \sqrt{15}}+\frac{11 C_{31,22}^{31,1}}{288 \sqrt{15}}+\frac{3 C_{31,22}^{33,0}}{1280 \sqrt{7}}+\frac{3 C_{31,22}^{33,1}}{640
\sqrt{7}}+\frac{1}{640} \sqrt{\frac{3}{7}} C_{33,00}^{33,0}+\frac{1}{320} \sqrt{\frac{3}{7}} C_{33,00}^{33,1}+\frac{C_{33,20}^{33,0}}{640 \sqrt{7}}+\frac{C_{33,20}^{33,1}}{320
\sqrt{7}}-\frac{1}{160} \sqrt{\frac{21}{10}} C_{33,22}^{33,0}-\frac{1}{80} \sqrt{\frac{21}{10}} C_{33,22}^{33,1}\)

\noindent\(C_{22,3312,2}^{00,1112,1,\text{Loc}}= -\frac{C_{00,20}^{60,1}}{192}-\frac{\sqrt{5} C_{00,22}^{62,1}}{128}-\frac{C_{11,20}^{31,1}}{576}-\frac{11
C_{11,22}^{51,1}}{576 \sqrt{5}}-\frac{3 \sqrt{\frac{3}{7}} C_{11,22}^{53,1}}{1280}+\frac{C_{20,20}^{40,1}}{576}+\frac{11 C_{20,22}^{42,1}}{576 \sqrt{5}}-\frac{C_{22,20}^{42,1}}{576
\sqrt{5}}+\frac{C_{22,22}^{40,1}}{1152 \sqrt{5}}-\frac{23 C_{22,22}^{42,1}}{576 \sqrt{35}}-\frac{3 C_{22,22}^{44,1}}{1120}+\frac{C_{31,20}^{31,1}}{576}+\frac{11
C_{31,22}^{31,1}}{576 \sqrt{5}}+\frac{3 \sqrt{\frac{3}{7}} C_{31,22}^{33,1}}{1280}-\frac{1}{320} \sqrt{\frac{3}{7}} C_{33,20}^{33,1}-\frac{3}{160}
\sqrt{\frac{7}{10}} C_{33,22}^{33,1}\)

\noindent\(C_{22,3312,2}^{00,1112,1,\text{NonLoc}}= -\frac{C_{00,00}^{60,0}}{128 \sqrt{3}}-\frac{C_{00,20}^{60,0}}{384}+\frac{\sqrt{5}
C_{00,22}^{62,0}}{128}+\frac{C_{11,00}^{51,0}}{384 \sqrt{3}}+\frac{C_{11,20}^{31,0}}{1152}-\frac{11 C_{11,22}^{51,0}}{576 \sqrt{5}}-\frac{3 \sqrt{\frac{3}{7}}
C_{11,22}^{53,0}}{1280}+\frac{C_{20,00}^{40,0}}{384 \sqrt{3}}+\frac{C_{20,20}^{40,0}}{1152}-\frac{11 C_{20,22}^{42,0}}{576 \sqrt{5}}-\frac{C_{22,00}^{42,0}}{384
\sqrt{15}}-\frac{C_{22,20}^{42,0}}{1152 \sqrt{5}}-\frac{C_{22,22}^{40,0}}{1152 \sqrt{5}}+\frac{23 C_{22,22}^{42,0}}{576 \sqrt{35}}+\frac{3 C_{22,22}^{44,0}}{1120}-\frac{C_{31,00}^{31,0}}{384
\sqrt{3}}-\frac{C_{31,20}^{31,0}}{1152}+\frac{11 C_{31,22}^{31,0}}{576 \sqrt{5}}+\frac{3 \sqrt{\frac{3}{7}} C_{31,22}^{33,0}}{1280}+\frac{3 C_{33,00}^{33,0}}{640
\sqrt{7}}+\frac{1}{640} \sqrt{\frac{3}{7}} C_{33,20}^{33,0}-\frac{3}{160} \sqrt{\frac{7}{10}} C_{33,22}^{33,0}\)

\noindent\(C_{22,3313,1}^{00,1111,0,\text{Loc}}= -\frac{1}{48} \sqrt{\frac{7}{6}} C_{00,20}^{60,0}-\frac{1}{96} \sqrt{\frac{7}{6}}
C_{00,20}^{60,1}+\frac{1}{32} \sqrt{\frac{7}{30}} C_{00,22}^{62,0}+\frac{1}{64} \sqrt{\frac{7}{30}} C_{00,22}^{62,1}-\frac{1}{144} \sqrt{\frac{7}{6}}
C_{11,20}^{31,0}-\frac{1}{288} \sqrt{\frac{7}{6}} C_{11,20}^{31,1}+\frac{1}{288} \sqrt{\frac{35}{6}} C_{11,22}^{51,0}+\frac{1}{576} \sqrt{\frac{35}{6}}
C_{11,22}^{51,1}+\frac{1}{144} \sqrt{\frac{7}{6}} C_{20,20}^{40,0}+\frac{1}{288} \sqrt{\frac{7}{6}} C_{20,20}^{40,1}-\frac{1}{288} \sqrt{\frac{35}{6}}
C_{20,22}^{42,0}-\frac{1}{576} \sqrt{\frac{35}{6}} C_{20,22}^{42,1}-\frac{1}{144} \sqrt{\frac{7}{30}} C_{22,20}^{42,0}-\frac{1}{288} \sqrt{\frac{7}{30}}
C_{22,20}^{42,1}+\frac{1}{288} \sqrt{\frac{7}{30}} C_{22,22}^{40,0}+\frac{1}{576} \sqrt{\frac{7}{30}} C_{22,22}^{40,1}+\frac{C_{22,22}^{42,0}}{36
\sqrt{30}}+\frac{C_{22,22}^{42,1}}{72 \sqrt{30}}+\frac{1}{160} \sqrt{\frac{3}{14}} C_{22,22}^{44,0}+\frac{1}{320} \sqrt{\frac{3}{14}} C_{22,22}^{44,1}+\frac{1}{144}
\sqrt{\frac{7}{6}} C_{31,20}^{31,0}+\frac{1}{288} \sqrt{\frac{7}{6}} C_{31,20}^{31,1}-\frac{1}{288} \sqrt{\frac{35}{6}} C_{31,22}^{31,0}-\frac{1}{576}
\sqrt{\frac{35}{6}} C_{31,22}^{31,1}-\frac{C_{33,20}^{33,0}}{80 \sqrt{2}}-\frac{C_{33,20}^{33,1}}{160 \sqrt{2}}+\frac{1}{64} \sqrt{\frac{3}{5}} C_{33,22}^{33,0}+\frac{1}{128}
\sqrt{\frac{3}{5}} C_{33,22}^{33,1}\)

\noindent\(C_{22,3313,1}^{00,1111,0,\text{NonLoc}}= -\frac{1}{192} \sqrt{\frac{7}{2}} C_{00,00}^{60,0}-\frac{1}{96} \sqrt{\frac{7}{2}}
C_{00,00}^{60,1}-\frac{1}{192} \sqrt{\frac{7}{6}} C_{00,20}^{60,0}-\frac{1}{96} \sqrt{\frac{7}{6}} C_{00,20}^{60,1}-\frac{1}{64} \sqrt{\frac{7}{30}}
C_{00,22}^{62,0}-\frac{1}{32} \sqrt{\frac{7}{30}} C_{00,22}^{62,1}+\frac{1}{576} \sqrt{\frac{7}{2}} C_{11,00}^{51,0}+\frac{1}{288} \sqrt{\frac{7}{2}}
C_{11,00}^{51,1}+\frac{1}{576} \sqrt{\frac{7}{6}} C_{11,20}^{31,0}+\frac{1}{288} \sqrt{\frac{7}{6}} C_{11,20}^{31,1}+\frac{1}{576} \sqrt{\frac{35}{6}}
C_{11,22}^{51,0}+\frac{1}{288} \sqrt{\frac{35}{6}} C_{11,22}^{51,1}+\frac{1}{576} \sqrt{\frac{7}{2}} C_{20,00}^{40,0}+\frac{1}{288} \sqrt{\frac{7}{2}}
C_{20,00}^{40,1}+\frac{1}{576} \sqrt{\frac{7}{6}} C_{20,20}^{40,0}+\frac{1}{288} \sqrt{\frac{7}{6}} C_{20,20}^{40,1}+\frac{1}{576} \sqrt{\frac{35}{6}}
C_{20,22}^{42,0}+\frac{1}{288} \sqrt{\frac{35}{6}} C_{20,22}^{42,1}-\frac{1}{576} \sqrt{\frac{7}{10}} C_{22,00}^{42,0}-\frac{1}{288} \sqrt{\frac{7}{10}}
C_{22,00}^{42,1}-\frac{1}{576} \sqrt{\frac{7}{30}} C_{22,20}^{42,0}-\frac{1}{288} \sqrt{\frac{7}{30}} C_{22,20}^{42,1}-\frac{1}{576} \sqrt{\frac{7}{30}}
C_{22,22}^{40,0}-\frac{1}{288} \sqrt{\frac{7}{30}} C_{22,22}^{40,1}-\frac{C_{22,22}^{42,0}}{72 \sqrt{30}}-\frac{C_{22,22}^{42,1}}{36 \sqrt{30}}-\frac{1}{320}
\sqrt{\frac{3}{14}} C_{22,22}^{44,0}-\frac{1}{160} \sqrt{\frac{3}{14}} C_{22,22}^{44,1}-\frac{1}{576} \sqrt{\frac{7}{2}} C_{31,00}^{31,0}-\frac{1}{288}
\sqrt{\frac{7}{2}} C_{31,00}^{31,1}-\frac{1}{576} \sqrt{\frac{7}{6}} C_{31,20}^{31,0}-\frac{1}{288} \sqrt{\frac{7}{6}} C_{31,20}^{31,1}-\frac{1}{576}
\sqrt{\frac{35}{6}} C_{31,22}^{31,0}-\frac{1}{288} \sqrt{\frac{35}{6}} C_{31,22}^{31,1}+\frac{1}{320} \sqrt{\frac{3}{2}} C_{33,00}^{33,0}+\frac{1}{160}
\sqrt{\frac{3}{2}} C_{33,00}^{33,1}+\frac{C_{33,20}^{33,0}}{320 \sqrt{2}}+\frac{C_{33,20}^{33,1}}{160 \sqrt{2}}+\frac{1}{128} \sqrt{\frac{3}{5}}
C_{33,22}^{33,0}+\frac{1}{64} \sqrt{\frac{3}{5}} C_{33,22}^{33,1}\)

\noindent\(C_{22,3313,1}^{00,1111,1,\text{Loc}}= -\frac{1}{96} \sqrt{\frac{7}{2}} C_{00,20}^{60,1}+\frac{1}{64} \sqrt{\frac{7}{10}}
C_{00,22}^{62,1}-\frac{1}{288} \sqrt{\frac{7}{2}} C_{11,20}^{31,1}+\frac{1}{576} \sqrt{\frac{35}{2}} C_{11,22}^{51,1}+\frac{1}{288} \sqrt{\frac{7}{2}}
C_{20,20}^{40,1}-\frac{1}{576} \sqrt{\frac{35}{2}} C_{20,22}^{42,1}-\frac{1}{288} \sqrt{\frac{7}{10}} C_{22,20}^{42,1}+\frac{1}{576} \sqrt{\frac{7}{10}}
C_{22,22}^{40,1}+\frac{C_{22,22}^{42,1}}{72 \sqrt{10}}+\frac{3 C_{22,22}^{44,1}}{320 \sqrt{14}}+\frac{1}{288} \sqrt{\frac{7}{2}} C_{31,20}^{31,1}-\frac{1}{576}
\sqrt{\frac{35}{2}} C_{31,22}^{31,1}-\frac{1}{160} \sqrt{\frac{3}{2}} C_{33,20}^{33,1}+\frac{3 C_{33,22}^{33,1}}{128 \sqrt{5}}\)

\noindent\(C_{22,3313,1}^{00,1111,1,\text{NonLoc}}= -\frac{1}{64} \sqrt{\frac{7}{6}} C_{00,00}^{60,0}-\frac{1}{192} \sqrt{\frac{7}{2}}
C_{00,20}^{60,0}-\frac{1}{64} \sqrt{\frac{7}{10}} C_{00,22}^{62,0}+\frac{1}{192} \sqrt{\frac{7}{6}} C_{11,00}^{51,0}+\frac{1}{576} \sqrt{\frac{7}{2}}
C_{11,20}^{31,0}+\frac{1}{576} \sqrt{\frac{35}{2}} C_{11,22}^{51,0}+\frac{1}{192} \sqrt{\frac{7}{6}} C_{20,00}^{40,0}+\frac{1}{576} \sqrt{\frac{7}{2}}
C_{20,20}^{40,0}+\frac{1}{576} \sqrt{\frac{35}{2}} C_{20,22}^{42,0}-\frac{1}{192} \sqrt{\frac{7}{30}} C_{22,00}^{42,0}-\frac{1}{576} \sqrt{\frac{7}{10}}
C_{22,20}^{42,0}-\frac{1}{576} \sqrt{\frac{7}{10}} C_{22,22}^{40,0}-\frac{C_{22,22}^{42,0}}{72 \sqrt{10}}-\frac{3 C_{22,22}^{44,0}}{320 \sqrt{14}}-\frac{1}{192}
\sqrt{\frac{7}{6}} C_{31,00}^{31,0}-\frac{1}{576} \sqrt{\frac{7}{2}} C_{31,20}^{31,0}-\frac{1}{576} \sqrt{\frac{35}{2}} C_{31,22}^{31,0}+\frac{3
C_{33,00}^{33,0}}{320 \sqrt{2}}+\frac{1}{320} \sqrt{\frac{3}{2}} C_{33,20}^{33,0}+\frac{3 C_{33,22}^{33,0}}{128 \sqrt{5}}\)

\noindent\(C_{22,3313,1}^{11,0000,0,\text{Loc}}= -\frac{1}{96} \sqrt{\frac{7}{2}} C_{11,11}^{51,0}-\frac{1}{192} \sqrt{\frac{7}{2}}
C_{11,11}^{51,1}+\frac{1}{96} \sqrt{\frac{7}{2}} C_{31,11}^{31,0}+\frac{1}{192} \sqrt{\frac{7}{2}} C_{31,11}^{31,1}-\frac{3 C_{33,11}^{33,0}}{320}-\frac{3
C_{33,11}^{33,1}}{640}\)

\noindent\(C_{22,3313,1}^{11,0000,0,\text{NonLoc}}= -\frac{1}{192} \sqrt{\frac{7}{2}} C_{11,11}^{51,0}-\frac{1}{96} \sqrt{\frac{7}{2}}
C_{11,11}^{51,1}+\frac{1}{192} \sqrt{\frac{7}{2}} C_{31,11}^{31,0}+\frac{1}{96} \sqrt{\frac{7}{2}} C_{31,11}^{31,1}-\frac{3 C_{33,11}^{33,0}}{640}-\frac{3
C_{33,11}^{33,1}}{320}\)

\noindent\(C_{22,3313,1}^{11,0000,1,\text{Loc}}= -\frac{1}{64} \sqrt{\frac{7}{6}} C_{11,11}^{51,1}+\frac{1}{64} \sqrt{\frac{7}{6}}
C_{31,11}^{31,1}-\frac{3 \sqrt{3} C_{33,11}^{33,1}}{640}\)

\noindent\(C_{22,3313,1}^{11,0000,1,\text{NonLoc}}= -\frac{1}{64} \sqrt{\frac{7}{6}} C_{11,11}^{51,0}+\frac{1}{64} \sqrt{\frac{7}{6}}
C_{31,11}^{31,0}-\frac{3 \sqrt{3} C_{33,11}^{33,0}}{640}\)

\noindent\(C_{22,3313,2}^{00,1112,0,\text{Loc}}= \frac{1}{96} \sqrt{\frac{7}{3}} C_{00,20}^{60,0}+\frac{1}{192} \sqrt{\frac{7}{3}}
C_{00,20}^{60,1}-\frac{1}{64} \sqrt{\frac{7}{15}} C_{00,22}^{62,0}-\frac{1}{128} \sqrt{\frac{7}{15}} C_{00,22}^{62,1}+\frac{1}{288} \sqrt{\frac{7}{3}}
C_{11,20}^{31,0}+\frac{1}{576} \sqrt{\frac{7}{3}} C_{11,20}^{31,1}+\frac{1}{576} \sqrt{\frac{7}{15}} C_{11,22}^{51,0}+\frac{\sqrt{\frac{7}{15}} C_{11,22}^{51,1}}{1152}-\frac{3
C_{11,22}^{53,0}}{320}-\frac{3 C_{11,22}^{53,1}}{640}-\frac{1}{288} \sqrt{\frac{7}{3}} C_{20,20}^{40,0}-\frac{1}{576} \sqrt{\frac{7}{3}} C_{20,20}^{40,1}-\frac{1}{576}
\sqrt{\frac{7}{15}} C_{20,22}^{42,0}-\frac{\sqrt{\frac{7}{15}} C_{20,22}^{42,1}}{1152}+\frac{1}{288} \sqrt{\frac{7}{15}} C_{22,20}^{42,0}+\frac{1}{576}
\sqrt{\frac{7}{15}} C_{22,20}^{42,1}+\frac{11}{576} \sqrt{\frac{7}{15}} C_{22,22}^{40,0}+\frac{11 \sqrt{\frac{7}{15}} C_{22,22}^{40,1}}{1152}-\frac{1}{144}
\sqrt{\frac{5}{3}} C_{22,22}^{42,0}-\frac{1}{288} \sqrt{\frac{5}{3}} C_{22,22}^{42,1}+\frac{1}{320} \sqrt{\frac{3}{7}} C_{22,22}^{44,0}+\frac{1}{640}
\sqrt{\frac{3}{7}} C_{22,22}^{44,1}-\frac{1}{288} \sqrt{\frac{7}{3}} C_{31,20}^{31,0}-\frac{1}{576} \sqrt{\frac{7}{3}} C_{31,20}^{31,1}-\frac{1}{576}
\sqrt{\frac{7}{15}} C_{31,22}^{31,0}-\frac{\sqrt{\frac{7}{15}} C_{31,22}^{31,1}}{1152}+\frac{3 C_{31,22}^{33,0}}{320}+\frac{3 C_{31,22}^{33,1}}{640}+\frac{C_{33,20}^{33,0}}{160}+\frac{C_{33,20}^{33,1}}{320}-\frac{11}{320}
\sqrt{\frac{3}{10}} C_{33,22}^{33,0}-\frac{11}{640} \sqrt{\frac{3}{10}} C_{33,22}^{33,1}\)

\noindent\(C_{22,3313,2}^{00,1112,0,\text{NonLoc}}= \frac{\sqrt{7} C_{00,00}^{60,0}}{384}+\frac{\sqrt{7} C_{00,00}^{60,1}}{192}+\frac{1}{384}
\sqrt{\frac{7}{3}} C_{00,20}^{60,0}+\frac{1}{192} \sqrt{\frac{7}{3}} C_{00,20}^{60,1}+\frac{1}{128} \sqrt{\frac{7}{15}} C_{00,22}^{62,0}+\frac{1}{64}
\sqrt{\frac{7}{15}} C_{00,22}^{62,1}-\frac{\sqrt{7} C_{11,00}^{51,0}}{1152}-\frac{\sqrt{7} C_{11,00}^{51,1}}{576}-\frac{\sqrt{\frac{7}{3}} C_{11,20}^{31,0}}{1152}-\frac{1}{576}
\sqrt{\frac{7}{3}} C_{11,20}^{31,1}+\frac{\sqrt{\frac{7}{15}} C_{11,22}^{51,0}}{1152}+\frac{1}{576} \sqrt{\frac{7}{15}} C_{11,22}^{51,1}-\frac{3
C_{11,22}^{53,0}}{640}-\frac{3 C_{11,22}^{53,1}}{320}-\frac{\sqrt{7} C_{20,00}^{40,0}}{1152}-\frac{\sqrt{7} C_{20,00}^{40,1}}{576}-\frac{\sqrt{\frac{7}{3}}
C_{20,20}^{40,0}}{1152}-\frac{1}{576} \sqrt{\frac{7}{3}} C_{20,20}^{40,1}+\frac{\sqrt{\frac{7}{15}} C_{20,22}^{42,0}}{1152}+\frac{1}{576} \sqrt{\frac{7}{15}}
C_{20,22}^{42,1}+\frac{\sqrt{\frac{7}{5}} C_{22,00}^{42,0}}{1152}+\frac{1}{576} \sqrt{\frac{7}{5}} C_{22,00}^{42,1}+\frac{\sqrt{\frac{7}{15}} C_{22,20}^{42,0}}{1152}+\frac{1}{576}
\sqrt{\frac{7}{15}} C_{22,20}^{42,1}-\frac{11 \sqrt{\frac{7}{15}} C_{22,22}^{40,0}}{1152}-\frac{11}{576} \sqrt{\frac{7}{15}} C_{22,22}^{40,1}+\frac{1}{288}
\sqrt{\frac{5}{3}} C_{22,22}^{42,0}+\frac{1}{144} \sqrt{\frac{5}{3}} C_{22,22}^{42,1}-\frac{1}{640} \sqrt{\frac{3}{7}} C_{22,22}^{44,0}-\frac{1}{320}
\sqrt{\frac{3}{7}} C_{22,22}^{44,1}+\frac{\sqrt{7} C_{31,00}^{31,0}}{1152}+\frac{\sqrt{7} C_{31,00}^{31,1}}{576}+\frac{\sqrt{\frac{7}{3}} C_{31,20}^{31,0}}{1152}+\frac{1}{576}
\sqrt{\frac{7}{3}} C_{31,20}^{31,1}-\frac{\sqrt{\frac{7}{15}} C_{31,22}^{31,0}}{1152}-\frac{1}{576} \sqrt{\frac{7}{15}} C_{31,22}^{31,1}+\frac{3
C_{31,22}^{33,0}}{640}+\frac{3 C_{31,22}^{33,1}}{320}-\frac{\sqrt{3} C_{33,00}^{33,0}}{640}-\frac{\sqrt{3} C_{33,00}^{33,1}}{320}-\frac{C_{33,20}^{33,0}}{640}-\frac{C_{33,20}^{33,1}}{320}-\frac{11}{640}
\sqrt{\frac{3}{10}} C_{33,22}^{33,0}-\frac{11}{320} \sqrt{\frac{3}{10}} C_{33,22}^{33,1}\)

\noindent\(C_{22,3313,2}^{00,1112,1,\text{Loc}}= \frac{\sqrt{7} C_{00,20}^{60,1}}{192}-\frac{1}{128} \sqrt{\frac{7}{5}} C_{00,22}^{62,1}+\frac{\sqrt{7}
C_{11,20}^{31,1}}{576}+\frac{\sqrt{\frac{7}{5}} C_{11,22}^{51,1}}{1152}-\frac{3 \sqrt{3} C_{11,22}^{53,1}}{640}-\frac{\sqrt{7} C_{20,20}^{40,1}}{576}-\frac{\sqrt{\frac{7}{5}}
C_{20,22}^{42,1}}{1152}+\frac{1}{576} \sqrt{\frac{7}{5}} C_{22,20}^{42,1}+\frac{11 \sqrt{\frac{7}{5}} C_{22,22}^{40,1}}{1152}-\frac{\sqrt{5} C_{22,22}^{42,1}}{288}+\frac{3
C_{22,22}^{44,1}}{640 \sqrt{7}}-\frac{\sqrt{7} C_{31,20}^{31,1}}{576}-\frac{\sqrt{\frac{7}{5}} C_{31,22}^{31,1}}{1152}+\frac{3 \sqrt{3} C_{31,22}^{33,1}}{640}+\frac{\sqrt{3}
C_{33,20}^{33,1}}{320}-\frac{33 C_{33,22}^{33,1}}{640 \sqrt{10}}\)

\noindent\(C_{22,3313,2}^{00,1112,1,\text{NonLoc}}= \frac{1}{128} \sqrt{\frac{7}{3}} C_{00,00}^{60,0}+\frac{\sqrt{7} C_{00,20}^{60,0}}{384}+\frac{1}{128}
\sqrt{\frac{7}{5}} C_{00,22}^{62,0}-\frac{1}{384} \sqrt{\frac{7}{3}} C_{11,00}^{51,0}-\frac{\sqrt{7} C_{11,20}^{31,0}}{1152}+\frac{\sqrt{\frac{7}{5}}
C_{11,22}^{51,0}}{1152}-\frac{3 \sqrt{3} C_{11,22}^{53,0}}{640}-\frac{1}{384} \sqrt{\frac{7}{3}} C_{20,00}^{40,0}-\frac{\sqrt{7} C_{20,20}^{40,0}}{1152}+\frac{\sqrt{\frac{7}{5}}
C_{20,22}^{42,0}}{1152}+\frac{1}{384} \sqrt{\frac{7}{15}} C_{22,00}^{42,0}+\frac{\sqrt{\frac{7}{5}} C_{22,20}^{42,0}}{1152}-\frac{11 \sqrt{\frac{7}{5}}
C_{22,22}^{40,0}}{1152}+\frac{\sqrt{5} C_{22,22}^{42,0}}{288}-\frac{3 C_{22,22}^{44,0}}{640 \sqrt{7}}+\frac{1}{384} \sqrt{\frac{7}{3}} C_{31,00}^{31,0}+\frac{\sqrt{7}
C_{31,20}^{31,0}}{1152}-\frac{\sqrt{\frac{7}{5}} C_{31,22}^{31,0}}{1152}+\frac{3 \sqrt{3} C_{31,22}^{33,0}}{640}-\frac{3 C_{33,00}^{33,0}}{640}-\frac{\sqrt{3}
C_{33,20}^{33,0}}{640}-\frac{33 C_{33,22}^{33,0}}{640 \sqrt{10}}\)

\noindent\(C_{22,3314,2}^{00,1112,0,\text{Loc}}= -\frac{C_{00,20}^{60,0}}{32}-\frac{C_{00,20}^{60,1}}{64}-\frac{C_{00,22}^{62,0}}{64
\sqrt{5}}-\frac{C_{00,22}^{62,1}}{128 \sqrt{5}}-\frac{C_{11,20}^{31,0}}{96}-\frac{C_{11,20}^{31,1}}{192}-\frac{C_{11,22}^{51,0}}{192 \sqrt{5}}-\frac{C_{11,22}^{51,1}}{384
\sqrt{5}}-\frac{1}{320} \sqrt{\frac{3}{7}} C_{11,22}^{53,0}-\frac{1}{640} \sqrt{\frac{3}{7}} C_{11,22}^{53,1}+\frac{C_{20,20}^{40,0}}{96}+\frac{C_{20,20}^{40,1}}{192}+\frac{C_{20,22}^{42,0}}{192
\sqrt{5}}+\frac{C_{20,22}^{42,1}}{384 \sqrt{5}}-\frac{C_{22,20}^{42,0}}{96 \sqrt{5}}-\frac{C_{22,20}^{42,1}}{192 \sqrt{5}}+\frac{C_{22,22}^{40,0}}{192
\sqrt{5}}+\frac{C_{22,22}^{40,1}}{384 \sqrt{5}}-\frac{C_{22,22}^{42,0}}{48 \sqrt{35}}-\frac{C_{22,22}^{42,1}}{96 \sqrt{35}}-\frac{C_{22,22}^{44,0}}{2240}-\frac{C_{22,22}^{44,1}}{4480}+\frac{C_{31,20}^{31,0}}{96}+\frac{C_{31,20}^{31,1}}{192}+\frac{C_{31,22}^{31,0}}{192
\sqrt{5}}+\frac{C_{31,22}^{31,1}}{384 \sqrt{5}}+\frac{1}{320} \sqrt{\frac{3}{7}} C_{31,22}^{33,0}+\frac{1}{640} \sqrt{\frac{3}{7}} C_{31,22}^{33,1}-\frac{3}{160}
\sqrt{\frac{3}{7}} C_{33,20}^{33,0}-\frac{3}{320} \sqrt{\frac{3}{7}} C_{33,20}^{33,1}-\frac{3}{320} \sqrt{\frac{7}{10}} C_{33,22}^{33,0}-\frac{3}{640}
\sqrt{\frac{7}{10}} C_{33,22}^{33,1}\)

\noindent\(C_{22,3314,2}^{00,1112,0,\text{NonLoc}}= -\frac{\sqrt{3} C_{00,00}^{60,0}}{128}-\frac{\sqrt{3} C_{00,00}^{60,1}}{64}-\frac{C_{00,20}^{60,0}}{128}-\frac{C_{00,20}^{60,1}}{64}+\frac{C_{00,22}^{62,0}}{128
\sqrt{5}}+\frac{C_{00,22}^{62,1}}{64 \sqrt{5}}+\frac{C_{11,00}^{51,0}}{128 \sqrt{3}}+\frac{C_{11,00}^{51,1}}{64 \sqrt{3}}+\frac{C_{11,20}^{31,0}}{384}+\frac{C_{11,20}^{31,1}}{192}-\frac{C_{11,22}^{51,0}}{384
\sqrt{5}}-\frac{C_{11,22}^{51,1}}{192 \sqrt{5}}-\frac{1}{640} \sqrt{\frac{3}{7}} C_{11,22}^{53,0}-\frac{1}{320} \sqrt{\frac{3}{7}} C_{11,22}^{53,1}+\frac{C_{20,00}^{40,0}}{128
\sqrt{3}}+\frac{C_{20,00}^{40,1}}{64 \sqrt{3}}+\frac{C_{20,20}^{40,0}}{384}+\frac{C_{20,20}^{40,1}}{192}-\frac{C_{20,22}^{42,0}}{384 \sqrt{5}}-\frac{C_{20,22}^{42,1}}{192
\sqrt{5}}-\frac{C_{22,00}^{42,0}}{128 \sqrt{15}}-\frac{C_{22,00}^{42,1}}{64 \sqrt{15}}-\frac{C_{22,20}^{42,0}}{384 \sqrt{5}}-\frac{C_{22,20}^{42,1}}{192
\sqrt{5}}-\frac{C_{22,22}^{40,0}}{384 \sqrt{5}}-\frac{C_{22,22}^{40,1}}{192 \sqrt{5}}+\frac{C_{22,22}^{42,0}}{96 \sqrt{35}}+\frac{C_{22,22}^{42,1}}{48
\sqrt{35}}+\frac{C_{22,22}^{44,0}}{4480}+\frac{C_{22,22}^{44,1}}{2240}-\frac{C_{31,00}^{31,0}}{128 \sqrt{3}}-\frac{C_{31,00}^{31,1}}{64 \sqrt{3}}-\frac{C_{31,20}^{31,0}}{384}-\frac{C_{31,20}^{31,1}}{192}+\frac{C_{31,22}^{31,0}}{384
\sqrt{5}}+\frac{C_{31,22}^{31,1}}{192 \sqrt{5}}+\frac{1}{640} \sqrt{\frac{3}{7}} C_{31,22}^{33,0}+\frac{1}{320} \sqrt{\frac{3}{7}} C_{31,22}^{33,1}+\frac{9
C_{33,00}^{33,0}}{640 \sqrt{7}}+\frac{9 C_{33,00}^{33,1}}{320 \sqrt{7}}+\frac{3}{640} \sqrt{\frac{3}{7}} C_{33,20}^{33,0}+\frac{3}{320} \sqrt{\frac{3}{7}}
C_{33,20}^{33,1}-\frac{3}{640} \sqrt{\frac{7}{10}} C_{33,22}^{33,0}-\frac{3}{320} \sqrt{\frac{7}{10}} C_{33,22}^{33,1}\)

\noindent\(C_{22,3314,2}^{00,1112,1,\text{Loc}}= -\frac{\sqrt{3} C_{00,20}^{60,1}}{64}-\frac{1}{128} \sqrt{\frac{3}{5}} C_{00,22}^{62,1}-\frac{C_{11,20}^{31,1}}{64
\sqrt{3}}-\frac{C_{11,22}^{51,1}}{128 \sqrt{15}}-\frac{3 C_{11,22}^{53,1}}{640 \sqrt{7}}+\frac{C_{20,20}^{40,1}}{64 \sqrt{3}}+\frac{C_{20,22}^{42,1}}{128
\sqrt{15}}-\frac{C_{22,20}^{42,1}}{64 \sqrt{15}}+\frac{C_{22,22}^{40,1}}{128 \sqrt{15}}-\frac{C_{22,22}^{42,1}}{32 \sqrt{105}}-\frac{\sqrt{3} C_{22,22}^{44,1}}{4480}+\frac{C_{31,20}^{31,1}}{64
\sqrt{3}}+\frac{C_{31,22}^{31,1}}{128 \sqrt{15}}+\frac{3 C_{31,22}^{33,1}}{640 \sqrt{7}}-\frac{9 C_{33,20}^{33,1}}{320 \sqrt{7}}-\frac{3}{640} \sqrt{\frac{21}{10}}
C_{33,22}^{33,1}\)

\noindent\(C_{22,3314,2}^{00,1112,1,\text{NonLoc}}= -\frac{3 C_{00,00}^{60,0}}{128}-\frac{\sqrt{3} C_{00,20}^{60,0}}{128}+\frac{1}{128}
\sqrt{\frac{3}{5}} C_{00,22}^{62,0}+\frac{C_{11,00}^{51,0}}{128}+\frac{C_{11,20}^{31,0}}{128 \sqrt{3}}-\frac{C_{11,22}^{51,0}}{128 \sqrt{15}}-\frac{3
C_{11,22}^{53,0}}{640 \sqrt{7}}+\frac{C_{20,00}^{40,0}}{128}+\frac{C_{20,20}^{40,0}}{128 \sqrt{3}}-\frac{C_{20,22}^{42,0}}{128 \sqrt{15}}-\frac{C_{22,00}^{42,0}}{128
\sqrt{5}}-\frac{C_{22,20}^{42,0}}{128 \sqrt{15}}-\frac{C_{22,22}^{40,0}}{128 \sqrt{15}}+\frac{C_{22,22}^{42,0}}{32 \sqrt{105}}+\frac{\sqrt{3} C_{22,22}^{44,0}}{4480}-\frac{C_{31,00}^{31,0}}{128}-\frac{C_{31,20}^{31,0}}{128
\sqrt{3}}+\frac{C_{31,22}^{31,0}}{128 \sqrt{15}}+\frac{3 C_{31,22}^{33,0}}{640 \sqrt{7}}+\frac{9}{640} \sqrt{\frac{3}{7}} C_{33,00}^{33,0}+\frac{9
C_{33,20}^{33,0}}{640 \sqrt{7}}-\frac{3}{640} \sqrt{\frac{21}{10}} C_{33,22}^{33,0}\)

\noindent\(C_{22,4011,1}^{00,0011,0,\text{Loc}}= \frac{21 C_{00,22}^{62,0}}{1280}+\frac{21 C_{00,22}^{62,1}}{2560}-\frac{7 C_{11,22}^{51,0}}{3840}-\frac{7
C_{11,22}^{51,1}}{7680}+\frac{9 \sqrt{\frac{21}{5}} C_{11,22}^{53,0}}{1280}+\frac{9 \sqrt{\frac{21}{5}} C_{11,22}^{53,1}}{2560}+\frac{7 C_{20,22}^{42,0}}{3840}+\frac{7
C_{20,22}^{42,1}}{7680}-\frac{17 C_{22,22}^{40,0}}{3840}-\frac{17 C_{22,22}^{40,1}}{7680}-\frac{\sqrt{7} C_{22,22}^{42,0}}{384}-\frac{\sqrt{7} C_{22,22}^{42,1}}{768}+\frac{9
C_{22,22}^{44,0}}{320 \sqrt{5}}+\frac{9 C_{22,22}^{44,1}}{640 \sqrt{5}}-\frac{23 C_{31,22}^{31,0}}{3840}-\frac{23 C_{31,22}^{31,1}}{7680}+\frac{\sqrt{\frac{21}{5}}
C_{31,22}^{33,0}}{1280}+\frac{\sqrt{\frac{21}{5}} C_{31,22}^{33,1}}{2560}-\frac{9 \sqrt{\frac{7}{2}} C_{33,22}^{33,0}}{1600}-\frac{9 \sqrt{\frac{7}{2}}
C_{33,22}^{33,1}}{3200}\)

\noindent\(C_{22,4011,1}^{00,0011,0,\text{NonLoc}}= -\frac{21 C_{00,22}^{62,0}}{2560}-\frac{21 C_{00,22}^{62,1}}{1280}-\frac{7 C_{11,22}^{51,0}}{7680}-\frac{7
C_{11,22}^{51,1}}{3840}+\frac{9 \sqrt{\frac{21}{5}} C_{11,22}^{53,0}}{2560}+\frac{9 \sqrt{\frac{21}{5}} C_{11,22}^{53,1}}{1280}-\frac{7 C_{20,22}^{42,0}}{7680}-\frac{7
C_{20,22}^{42,1}}{3840}+\frac{17 C_{22,22}^{40,0}}{7680}+\frac{17 C_{22,22}^{40,1}}{3840}+\frac{\sqrt{7} C_{22,22}^{42,0}}{768}+\frac{\sqrt{7} C_{22,22}^{42,1}}{384}-\frac{9
C_{22,22}^{44,0}}{640 \sqrt{5}}-\frac{9 C_{22,22}^{44,1}}{320 \sqrt{5}}-\frac{23 C_{31,22}^{31,0}}{7680}-\frac{23 C_{31,22}^{31,1}}{3840}+\frac{\sqrt{\frac{21}{5}}
C_{31,22}^{33,0}}{2560}+\frac{\sqrt{\frac{21}{5}} C_{31,22}^{33,1}}{1280}-\frac{9 \sqrt{\frac{7}{2}} C_{33,22}^{33,0}}{3200}-\frac{9 \sqrt{\frac{7}{2}}
C_{33,22}^{33,1}}{1600}\)

\noindent\(C_{22,4011,1}^{00,0011,1,\text{Loc}}= \frac{21 \sqrt{3} C_{00,22}^{62,1}}{2560}-\frac{7 C_{11,22}^{51,1}}{2560 \sqrt{3}}+\frac{27
\sqrt{\frac{7}{5}} C_{11,22}^{53,1}}{2560}+\frac{7 C_{20,22}^{42,1}}{2560 \sqrt{3}}-\frac{17 C_{22,22}^{40,1}}{2560 \sqrt{3}}-\frac{1}{256} \sqrt{\frac{7}{3}}
C_{22,22}^{42,1}+\frac{9}{640} \sqrt{\frac{3}{5}} C_{22,22}^{44,1}-\frac{23 C_{31,22}^{31,1}}{2560 \sqrt{3}}+\frac{3 \sqrt{\frac{7}{5}} C_{31,22}^{33,1}}{2560}-\frac{9
\sqrt{\frac{21}{2}} C_{33,22}^{33,1}}{3200}\)

\noindent\(C_{22,4011,1}^{00,0011,1,\text{NonLoc}}= -\frac{21 \sqrt{3} C_{00,22}^{62,0}}{2560}-\frac{7 C_{11,22}^{51,0}}{2560 \sqrt{3}}+\frac{27
\sqrt{\frac{7}{5}} C_{11,22}^{53,0}}{2560}-\frac{7 C_{20,22}^{42,0}}{2560 \sqrt{3}}+\frac{17 C_{22,22}^{40,0}}{2560 \sqrt{3}}+\frac{1}{256} \sqrt{\frac{7}{3}}
C_{22,22}^{42,0}-\frac{9}{640} \sqrt{\frac{3}{5}} C_{22,22}^{44,0}-\frac{23 C_{31,22}^{31,0}}{2560 \sqrt{3}}+\frac{3 \sqrt{\frac{7}{5}} C_{31,22}^{33,0}}{2560}-\frac{9
\sqrt{\frac{21}{2}} C_{33,22}^{33,0}}{3200}\)

\noindent\(C_{22,4202,0}^{00,0000,0,\text{Loc}}= -\frac{\sqrt{5} C_{00,00}^{60,0}}{64}-\frac{\sqrt{5} C_{00,00}^{60,1}}{128}-\frac{\sqrt{5}
C_{11,00}^{51,0}}{192}-\frac{\sqrt{5} C_{11,00}^{51,1}}{384}+\frac{\sqrt{5} C_{20,00}^{40,0}}{192}+\frac{\sqrt{5} C_{20,00}^{40,1}}{384}-\frac{C_{22,00}^{42,0}}{192}-\frac{C_{22,00}^{42,1}}{384}+\frac{\sqrt{5}
C_{31,00}^{31,0}}{192}+\frac{\sqrt{5} C_{31,00}^{31,1}}{384}-\frac{3}{64} \sqrt{\frac{3}{35}} C_{33,00}^{33,0}-\frac{3}{128} \sqrt{\frac{3}{35}}
C_{33,00}^{33,1}\)

\noindent\(C_{22,4202,0}^{00,0000,0,\text{NonLoc}}= \frac{\sqrt{5} C_{00,00}^{60,0}}{256}+\frac{\sqrt{5} C_{00,00}^{60,1}}{128}-\frac{\sqrt{15}
C_{00,20}^{60,0}}{256}-\frac{\sqrt{15} C_{00,20}^{60,1}}{128}-\frac{\sqrt{5} C_{11,00}^{51,0}}{768}-\frac{\sqrt{5} C_{11,00}^{51,1}}{384}+\frac{1}{256}
\sqrt{\frac{5}{3}} C_{11,20}^{31,0}+\frac{1}{128} \sqrt{\frac{5}{3}} C_{11,20}^{31,1}-\frac{\sqrt{5} C_{20,00}^{40,0}}{768}-\frac{\sqrt{5} C_{20,00}^{40,1}}{384}+\frac{1}{256}
\sqrt{\frac{5}{3}} C_{20,20}^{40,0}+\frac{1}{128} \sqrt{\frac{5}{3}} C_{20,20}^{40,1}+\frac{C_{22,00}^{42,0}}{768}+\frac{C_{22,00}^{42,1}}{384}-\frac{C_{22,20}^{42,0}}{256
\sqrt{3}}-\frac{C_{22,20}^{42,1}}{128 \sqrt{3}}+\frac{\sqrt{5} C_{31,00}^{31,0}}{768}+\frac{\sqrt{5} C_{31,00}^{31,1}}{384}-\frac{1}{256} \sqrt{\frac{5}{3}}
C_{31,20}^{31,0}-\frac{1}{128} \sqrt{\frac{5}{3}} C_{31,20}^{31,1}-\frac{3}{256} \sqrt{\frac{3}{35}} C_{33,00}^{33,0}-\frac{3}{128} \sqrt{\frac{3}{35}}
C_{33,00}^{33,1}+\frac{9 C_{33,20}^{33,0}}{256 \sqrt{35}}+\frac{9 C_{33,20}^{33,1}}{128 \sqrt{35}}\)

\noindent\(C_{22,4202,0}^{00,0000,1,\text{Loc}}= -\frac{\sqrt{15} C_{00,00}^{60,1}}{128}-\frac{1}{128} \sqrt{\frac{5}{3}} C_{11,00}^{51,1}+\frac{1}{128}
\sqrt{\frac{5}{3}} C_{20,00}^{40,1}-\frac{C_{22,00}^{42,1}}{128 \sqrt{3}}+\frac{1}{128} \sqrt{\frac{5}{3}} C_{31,00}^{31,1}-\frac{9 C_{33,00}^{33,1}}{128
\sqrt{35}}\)

\noindent\(C_{22,4202,0}^{00,0000,1,\text{NonLoc}}= \frac{\sqrt{15} C_{00,00}^{60,0}}{256}-\frac{3 \sqrt{5} C_{00,20}^{60,0}}{256}-\frac{1}{256}
\sqrt{\frac{5}{3}} C_{11,00}^{51,0}+\frac{\sqrt{5} C_{11,20}^{31,0}}{256}-\frac{1}{256} \sqrt{\frac{5}{3}} C_{20,00}^{40,0}+\frac{\sqrt{5} C_{20,20}^{40,0}}{256}+\frac{C_{22,00}^{42,0}}{256
\sqrt{3}}-\frac{C_{22,20}^{42,0}}{256}+\frac{1}{256} \sqrt{\frac{5}{3}} C_{31,00}^{31,0}-\frac{\sqrt{5} C_{31,20}^{31,0}}{256}-\frac{9 C_{33,00}^{33,0}}{256
\sqrt{35}}+\frac{9}{256} \sqrt{\frac{3}{35}} C_{33,20}^{33,0}\)

\noindent\(C_{22,4211,1}^{00,0011,0,\text{Loc}}= \frac{C_{00,20}^{60,0}}{64}+\frac{C_{00,20}^{60,1}}{128}+\frac{3 C_{00,22}^{62,0}}{128
\sqrt{5}}+\frac{3 C_{00,22}^{62,1}}{256 \sqrt{5}}+\frac{C_{11,20}^{31,0}}{192}+\frac{C_{11,20}^{31,1}}{384}+\frac{C_{11,22}^{51,0}}{192 \sqrt{5}}+\frac{C_{11,22}^{51,1}}{384
\sqrt{5}}+\frac{9 \sqrt{\frac{3}{7}} C_{11,22}^{53,0}}{1280}+\frac{9 \sqrt{\frac{3}{7}} C_{11,22}^{53,1}}{2560}-\frac{C_{20,20}^{40,0}}{192}-\frac{C_{20,20}^{40,1}}{384}-\frac{C_{20,22}^{42,0}}{192
\sqrt{5}}-\frac{C_{20,22}^{42,1}}{384 \sqrt{5}}+\frac{C_{22,20}^{42,0}}{192 \sqrt{5}}+\frac{C_{22,20}^{42,1}}{384 \sqrt{5}}-\frac{\sqrt{5} C_{22,22}^{40,0}}{384}-\frac{\sqrt{5}
C_{22,22}^{40,1}}{768}+\frac{1}{192} \sqrt{\frac{7}{5}} C_{22,22}^{42,0}+\frac{1}{384} \sqrt{\frac{7}{5}} C_{22,22}^{42,1}-\frac{C_{31,20}^{31,0}}{192}-\frac{C_{31,20}^{31,1}}{384}-\frac{C_{31,22}^{31,0}}{192
\sqrt{5}}-\frac{C_{31,22}^{31,1}}{384 \sqrt{5}}-\frac{9 \sqrt{\frac{3}{7}} C_{31,22}^{33,0}}{1280}-\frac{9 \sqrt{\frac{3}{7}} C_{31,22}^{33,1}}{2560}+\frac{3}{320}
\sqrt{\frac{3}{7}} C_{33,20}^{33,0}+\frac{3}{640} \sqrt{\frac{3}{7}} C_{33,20}^{33,1}+\frac{9 C_{33,22}^{33,0}}{80 \sqrt{70}}+\frac{9 C_{33,22}^{33,1}}{160
\sqrt{70}}\)

\noindent\(C_{22,4211,1}^{00,0011,0,\text{NonLoc}}= \frac{\sqrt{3} C_{00,00}^{60,0}}{256}+\frac{\sqrt{3} C_{00,00}^{60,1}}{128}+\frac{C_{00,20}^{60,0}}{256}+\frac{C_{00,20}^{60,1}}{128}-\frac{3
C_{00,22}^{62,0}}{256 \sqrt{5}}-\frac{3 C_{00,22}^{62,1}}{128 \sqrt{5}}-\frac{C_{11,00}^{51,0}}{256 \sqrt{3}}-\frac{C_{11,00}^{51,1}}{128 \sqrt{3}}-\frac{C_{11,20}^{31,0}}{768}-\frac{C_{11,20}^{31,1}}{384}+\frac{C_{11,22}^{51,0}}{384
\sqrt{5}}+\frac{C_{11,22}^{51,1}}{192 \sqrt{5}}+\frac{9 \sqrt{\frac{3}{7}} C_{11,22}^{53,0}}{2560}+\frac{9 \sqrt{\frac{3}{7}} C_{11,22}^{53,1}}{1280}-\frac{C_{20,00}^{40,0}}{256
\sqrt{3}}-\frac{C_{20,00}^{40,1}}{128 \sqrt{3}}-\frac{C_{20,20}^{40,0}}{768}-\frac{C_{20,20}^{40,1}}{384}+\frac{C_{20,22}^{42,0}}{384 \sqrt{5}}+\frac{C_{20,22}^{42,1}}{192
\sqrt{5}}+\frac{C_{22,00}^{42,0}}{256 \sqrt{15}}+\frac{C_{22,00}^{42,1}}{128 \sqrt{15}}+\frac{C_{22,20}^{42,0}}{768 \sqrt{5}}+\frac{C_{22,20}^{42,1}}{384
\sqrt{5}}+\frac{\sqrt{5} C_{22,22}^{40,0}}{768}+\frac{\sqrt{5} C_{22,22}^{40,1}}{384}-\frac{1}{384} \sqrt{\frac{7}{5}} C_{22,22}^{42,0}-\frac{1}{192}
\sqrt{\frac{7}{5}} C_{22,22}^{42,1}+\frac{C_{31,00}^{31,0}}{256 \sqrt{3}}+\frac{C_{31,00}^{31,1}}{128 \sqrt{3}}+\frac{C_{31,20}^{31,0}}{768}+\frac{C_{31,20}^{31,1}}{384}-\frac{C_{31,22}^{31,0}}{384
\sqrt{5}}-\frac{C_{31,22}^{31,1}}{192 \sqrt{5}}-\frac{9 \sqrt{\frac{3}{7}} C_{31,22}^{33,0}}{2560}-\frac{9 \sqrt{\frac{3}{7}} C_{31,22}^{33,1}}{1280}-\frac{9
C_{33,00}^{33,0}}{1280 \sqrt{7}}-\frac{9 C_{33,00}^{33,1}}{640 \sqrt{7}}-\frac{3 \sqrt{\frac{3}{7}} C_{33,20}^{33,0}}{1280}-\frac{3}{640} \sqrt{\frac{3}{7}}
C_{33,20}^{33,1}+\frac{9 C_{33,22}^{33,0}}{160 \sqrt{70}}+\frac{9 C_{33,22}^{33,1}}{80 \sqrt{70}}\)

\noindent\(C_{22,4211,1}^{00,0011,1,\text{Loc}}= \frac{\sqrt{3} C_{00,20}^{60,1}}{128}+\frac{3}{256} \sqrt{\frac{3}{5}} C_{00,22}^{62,1}+\frac{C_{11,20}^{31,1}}{128
\sqrt{3}}+\frac{C_{11,22}^{51,1}}{128 \sqrt{15}}+\frac{27 C_{11,22}^{53,1}}{2560 \sqrt{7}}-\frac{C_{20,20}^{40,1}}{128 \sqrt{3}}-\frac{C_{20,22}^{42,1}}{128
\sqrt{15}}+\frac{C_{22,20}^{42,1}}{128 \sqrt{15}}-\frac{1}{256} \sqrt{\frac{5}{3}} C_{22,22}^{40,1}+\frac{1}{128} \sqrt{\frac{7}{15}} C_{22,22}^{42,1}-\frac{C_{31,20}^{31,1}}{128
\sqrt{3}}-\frac{C_{31,22}^{31,1}}{128 \sqrt{15}}-\frac{27 C_{31,22}^{33,1}}{2560 \sqrt{7}}+\frac{9 C_{33,20}^{33,1}}{640 \sqrt{7}}+\frac{9}{160}
\sqrt{\frac{3}{70}} C_{33,22}^{33,1}\)

\noindent\(C_{22,4211,1}^{00,0011,1,\text{NonLoc}}= \frac{3 C_{00,00}^{60,0}}{256}+\frac{\sqrt{3} C_{00,20}^{60,0}}{256}-\frac{3}{256}
\sqrt{\frac{3}{5}} C_{00,22}^{62,0}-\frac{C_{11,00}^{51,0}}{256}-\frac{C_{11,20}^{31,0}}{256 \sqrt{3}}+\frac{C_{11,22}^{51,0}}{128 \sqrt{15}}+\frac{27
C_{11,22}^{53,0}}{2560 \sqrt{7}}-\frac{C_{20,00}^{40,0}}{256}-\frac{C_{20,20}^{40,0}}{256 \sqrt{3}}+\frac{C_{20,22}^{42,0}}{128 \sqrt{15}}+\frac{C_{22,00}^{42,0}}{256
\sqrt{5}}+\frac{C_{22,20}^{42,0}}{256 \sqrt{15}}+\frac{1}{256} \sqrt{\frac{5}{3}} C_{22,22}^{40,0}-\frac{1}{128} \sqrt{\frac{7}{15}} C_{22,22}^{42,0}+\frac{C_{31,00}^{31,0}}{256}+\frac{C_{31,20}^{31,0}}{256
\sqrt{3}}-\frac{C_{31,22}^{31,0}}{128 \sqrt{15}}-\frac{27 C_{31,22}^{33,0}}{2560 \sqrt{7}}-\frac{9 \sqrt{\frac{3}{7}} C_{33,00}^{33,0}}{1280}-\frac{9
C_{33,20}^{33,0}}{1280 \sqrt{7}}+\frac{9}{160} \sqrt{\frac{3}{70}} C_{33,22}^{33,0}\)

\noindent\(C_{22,4212,1}^{00,0011,0,\text{Loc}}= -\frac{1}{64} \sqrt{\frac{5}{3}} C_{00,20}^{60,0}-\frac{1}{128} \sqrt{\frac{5}{3}}
C_{00,20}^{60,1}+\frac{\sqrt{3} C_{00,22}^{62,0}}{128}+\frac{\sqrt{3} C_{00,22}^{62,1}}{256}-\frac{1}{192} \sqrt{\frac{5}{3}} C_{11,20}^{31,0}-\frac{1}{384}
\sqrt{\frac{5}{3}} C_{11,20}^{31,1}+\frac{C_{11,22}^{51,0}}{192 \sqrt{3}}+\frac{C_{11,22}^{51,1}}{384 \sqrt{3}}+\frac{9 C_{11,22}^{53,0}}{256 \sqrt{35}}+\frac{9
C_{11,22}^{53,1}}{512 \sqrt{35}}+\frac{1}{192} \sqrt{\frac{5}{3}} C_{20,20}^{40,0}+\frac{1}{384} \sqrt{\frac{5}{3}} C_{20,20}^{40,1}-\frac{C_{20,22}^{42,0}}{192
\sqrt{3}}-\frac{C_{20,22}^{42,1}}{384 \sqrt{3}}-\frac{C_{22,20}^{42,0}}{192 \sqrt{3}}-\frac{C_{22,20}^{42,1}}{384 \sqrt{3}}-\frac{5 C_{22,22}^{40,0}}{384
\sqrt{3}}-\frac{5 C_{22,22}^{40,1}}{768 \sqrt{3}}+\frac{1}{192} \sqrt{\frac{7}{3}} C_{22,22}^{42,0}+\frac{1}{384} \sqrt{\frac{7}{3}} C_{22,22}^{42,1}+\frac{1}{192}
\sqrt{\frac{5}{3}} C_{31,20}^{31,0}+\frac{1}{384} \sqrt{\frac{5}{3}} C_{31,20}^{31,1}-\frac{C_{31,22}^{31,0}}{192 \sqrt{3}}-\frac{C_{31,22}^{31,1}}{384
\sqrt{3}}-\frac{9 C_{31,22}^{33,0}}{256 \sqrt{35}}-\frac{9 C_{31,22}^{33,1}}{512 \sqrt{35}}-\frac{3 C_{33,20}^{33,0}}{64 \sqrt{35}}-\frac{3 C_{33,20}^{33,1}}{128
\sqrt{35}}+\frac{3}{80} \sqrt{\frac{3}{14}} C_{33,22}^{33,0}+\frac{3}{160} \sqrt{\frac{3}{14}} C_{33,22}^{33,1}\)

\noindent\(C_{22,4212,1}^{00,0011,0,\text{NonLoc}}= -\frac{\sqrt{5} C_{00,00}^{60,0}}{256}-\frac{\sqrt{5} C_{00,00}^{60,1}}{128}-\frac{1}{256}
\sqrt{\frac{5}{3}} C_{00,20}^{60,0}-\frac{1}{128} \sqrt{\frac{5}{3}} C_{00,20}^{60,1}-\frac{\sqrt{3} C_{00,22}^{62,0}}{256}-\frac{\sqrt{3} C_{00,22}^{62,1}}{128}+\frac{\sqrt{5}
C_{11,00}^{51,0}}{768}+\frac{\sqrt{5} C_{11,00}^{51,1}}{384}+\frac{1}{768} \sqrt{\frac{5}{3}} C_{11,20}^{31,0}+\frac{1}{384} \sqrt{\frac{5}{3}} C_{11,20}^{31,1}+\frac{C_{11,22}^{51,0}}{384
\sqrt{3}}+\frac{C_{11,22}^{51,1}}{192 \sqrt{3}}+\frac{9 C_{11,22}^{53,0}}{512 \sqrt{35}}+\frac{9 C_{11,22}^{53,1}}{256 \sqrt{35}}+\frac{\sqrt{5}
C_{20,00}^{40,0}}{768}+\frac{\sqrt{5} C_{20,00}^{40,1}}{384}+\frac{1}{768} \sqrt{\frac{5}{3}} C_{20,20}^{40,0}+\frac{1}{384} \sqrt{\frac{5}{3}} C_{20,20}^{40,1}+\frac{C_{20,22}^{42,0}}{384
\sqrt{3}}+\frac{C_{20,22}^{42,1}}{192 \sqrt{3}}-\frac{C_{22,00}^{42,0}}{768}-\frac{C_{22,00}^{42,1}}{384}-\frac{C_{22,20}^{42,0}}{768 \sqrt{3}}-\frac{C_{22,20}^{42,1}}{384
\sqrt{3}}+\frac{5 C_{22,22}^{40,0}}{768 \sqrt{3}}+\frac{5 C_{22,22}^{40,1}}{384 \sqrt{3}}-\frac{1}{384} \sqrt{\frac{7}{3}} C_{22,22}^{42,0}-\frac{1}{192}
\sqrt{\frac{7}{3}} C_{22,22}^{42,1}-\frac{\sqrt{5} C_{31,00}^{31,0}}{768}-\frac{\sqrt{5} C_{31,00}^{31,1}}{384}-\frac{1}{768} \sqrt{\frac{5}{3}}
C_{31,20}^{31,0}-\frac{1}{384} \sqrt{\frac{5}{3}} C_{31,20}^{31,1}-\frac{C_{31,22}^{31,0}}{384 \sqrt{3}}-\frac{C_{31,22}^{31,1}}{192 \sqrt{3}}-\frac{9
C_{31,22}^{33,0}}{512 \sqrt{35}}-\frac{9 C_{31,22}^{33,1}}{256 \sqrt{35}}+\frac{3}{256} \sqrt{\frac{3}{35}} C_{33,00}^{33,0}+\frac{3}{128} \sqrt{\frac{3}{35}}
C_{33,00}^{33,1}+\frac{3 C_{33,20}^{33,0}}{256 \sqrt{35}}+\frac{3 C_{33,20}^{33,1}}{128 \sqrt{35}}+\frac{3}{160} \sqrt{\frac{3}{14}} C_{33,22}^{33,0}+\frac{3}{80}
\sqrt{\frac{3}{14}} C_{33,22}^{33,1}\)

\noindent\(C_{22,4212,1}^{00,0011,1,\text{Loc}}= -\frac{\sqrt{5} C_{00,20}^{60,1}}{128}+\frac{3 C_{00,22}^{62,1}}{256}-\frac{\sqrt{5}
C_{11,20}^{31,1}}{384}+\frac{C_{11,22}^{51,1}}{384}+\frac{9}{512} \sqrt{\frac{3}{35}} C_{11,22}^{53,1}+\frac{\sqrt{5} C_{20,20}^{40,1}}{384}-\frac{C_{20,22}^{42,1}}{384}-\frac{C_{22,20}^{42,1}}{384}-\frac{5
C_{22,22}^{40,1}}{768}+\frac{\sqrt{7} C_{22,22}^{42,1}}{384}+\frac{\sqrt{5} C_{31,20}^{31,1}}{384}-\frac{C_{31,22}^{31,1}}{384}-\frac{9}{512} \sqrt{\frac{3}{35}}
C_{31,22}^{33,1}-\frac{3}{128} \sqrt{\frac{3}{35}} C_{33,20}^{33,1}+\frac{9 C_{33,22}^{33,1}}{160 \sqrt{14}}\)

\noindent\(C_{22,4212,1}^{00,0011,1,\text{NonLoc}}= -\frac{\sqrt{15} C_{00,00}^{60,0}}{256}-\frac{\sqrt{5} C_{00,20}^{60,0}}{256}-\frac{3
C_{00,22}^{62,0}}{256}+\frac{1}{256} \sqrt{\frac{5}{3}} C_{11,00}^{51,0}+\frac{\sqrt{5} C_{11,20}^{31,0}}{768}+\frac{C_{11,22}^{51,0}}{384}+\frac{9}{512}
\sqrt{\frac{3}{35}} C_{11,22}^{53,0}+\frac{1}{256} \sqrt{\frac{5}{3}} C_{20,00}^{40,0}+\frac{\sqrt{5} C_{20,20}^{40,0}}{768}+\frac{C_{20,22}^{42,0}}{384}-\frac{C_{22,00}^{42,0}}{256
\sqrt{3}}-\frac{C_{22,20}^{42,0}}{768}+\frac{5 C_{22,22}^{40,0}}{768}-\frac{\sqrt{7} C_{22,22}^{42,0}}{384}-\frac{1}{256} \sqrt{\frac{5}{3}} C_{31,00}^{31,0}-\frac{\sqrt{5}
C_{31,20}^{31,0}}{768}-\frac{C_{31,22}^{31,0}}{384}-\frac{9}{512} \sqrt{\frac{3}{35}} C_{31,22}^{33,0}+\frac{9 C_{33,00}^{33,0}}{256 \sqrt{35}}+\frac{3}{256}
\sqrt{\frac{3}{35}} C_{33,20}^{33,0}+\frac{9 C_{33,22}^{33,0}}{160 \sqrt{14}}\)

\noindent\(C_{22,4213,1}^{00,0011,0,\text{Loc}}= \frac{1}{64} \sqrt{\frac{7}{3}} C_{00,20}^{60,0}+\frac{1}{128} \sqrt{\frac{7}{3}}
C_{00,20}^{60,1}+\frac{1}{64} \sqrt{\frac{3}{35}} C_{00,22}^{62,0}+\frac{1}{128} \sqrt{\frac{3}{35}} C_{00,22}^{62,1}+\frac{1}{192} \sqrt{\frac{7}{3}}
C_{11,20}^{31,0}+\frac{1}{384} \sqrt{\frac{7}{3}} C_{11,20}^{31,1}+\frac{C_{11,22}^{51,0}}{96 \sqrt{105}}+\frac{C_{11,22}^{51,1}}{192 \sqrt{105}}+\frac{9
C_{11,22}^{53,0}}{4480}+\frac{9 C_{11,22}^{53,1}}{8960}-\frac{1}{192} \sqrt{\frac{7}{3}} C_{20,20}^{40,0}-\frac{1}{384} \sqrt{\frac{7}{3}} C_{20,20}^{40,1}-\frac{C_{20,22}^{42,0}}{96
\sqrt{105}}-\frac{C_{20,22}^{42,1}}{192 \sqrt{105}}+\frac{1}{192} \sqrt{\frac{7}{15}} C_{22,20}^{42,0}+\frac{1}{384} \sqrt{\frac{7}{15}} C_{22,20}^{42,1}-\frac{1}{192}
\sqrt{\frac{5}{21}} C_{22,22}^{40,0}-\frac{1}{384} \sqrt{\frac{5}{21}} C_{22,22}^{40,1}+\frac{C_{22,22}^{42,0}}{96 \sqrt{15}}+\frac{C_{22,22}^{42,1}}{192
\sqrt{15}}-\frac{1}{192} \sqrt{\frac{7}{3}} C_{31,20}^{31,0}-\frac{1}{384} \sqrt{\frac{7}{3}} C_{31,20}^{31,1}-\frac{C_{31,22}^{31,0}}{96 \sqrt{105}}-\frac{C_{31,22}^{31,1}}{192
\sqrt{105}}-\frac{9 C_{31,22}^{33,0}}{4480}-\frac{9 C_{31,22}^{33,1}}{8960}+\frac{3 C_{33,20}^{33,0}}{320}+\frac{3 C_{33,20}^{33,1}}{640}+\frac{3}{280}
\sqrt{\frac{3}{10}} C_{33,22}^{33,0}+\frac{3}{560} \sqrt{\frac{3}{10}} C_{33,22}^{33,1}\)

\noindent\(C_{22,4213,1}^{00,0011,0,\text{NonLoc}}= \frac{\sqrt{7} C_{00,00}^{60,0}}{256}+\frac{\sqrt{7} C_{00,00}^{60,1}}{128}+\frac{1}{256}
\sqrt{\frac{7}{3}} C_{00,20}^{60,0}+\frac{1}{128} \sqrt{\frac{7}{3}} C_{00,20}^{60,1}-\frac{1}{128} \sqrt{\frac{3}{35}} C_{00,22}^{62,0}-\frac{1}{64}
\sqrt{\frac{3}{35}} C_{00,22}^{62,1}-\frac{\sqrt{7} C_{11,00}^{51,0}}{768}-\frac{\sqrt{7} C_{11,00}^{51,1}}{384}-\frac{1}{768} \sqrt{\frac{7}{3}}
C_{11,20}^{31,0}-\frac{1}{384} \sqrt{\frac{7}{3}} C_{11,20}^{31,1}+\frac{C_{11,22}^{51,0}}{192 \sqrt{105}}+\frac{C_{11,22}^{51,1}}{96 \sqrt{105}}+\frac{9
C_{11,22}^{53,0}}{8960}+\frac{9 C_{11,22}^{53,1}}{4480}-\frac{\sqrt{7} C_{20,00}^{40,0}}{768}-\frac{\sqrt{7} C_{20,00}^{40,1}}{384}-\frac{1}{768}
\sqrt{\frac{7}{3}} C_{20,20}^{40,0}-\frac{1}{384} \sqrt{\frac{7}{3}} C_{20,20}^{40,1}+\frac{C_{20,22}^{42,0}}{192 \sqrt{105}}+\frac{C_{20,22}^{42,1}}{96
\sqrt{105}}+\frac{1}{768} \sqrt{\frac{7}{5}} C_{22,00}^{42,0}+\frac{1}{384} \sqrt{\frac{7}{5}} C_{22,00}^{42,1}+\frac{1}{768} \sqrt{\frac{7}{15}}
C_{22,20}^{42,0}+\frac{1}{384} \sqrt{\frac{7}{15}} C_{22,20}^{42,1}+\frac{1}{384} \sqrt{\frac{5}{21}} C_{22,22}^{40,0}+\frac{1}{192} \sqrt{\frac{5}{21}}
C_{22,22}^{40,1}-\frac{C_{22,22}^{42,0}}{192 \sqrt{15}}-\frac{C_{22,22}^{42,1}}{96 \sqrt{15}}+\frac{\sqrt{7} C_{31,00}^{31,0}}{768}+\frac{\sqrt{7}
C_{31,00}^{31,1}}{384}+\frac{1}{768} \sqrt{\frac{7}{3}} C_{31,20}^{31,0}+\frac{1}{384} \sqrt{\frac{7}{3}} C_{31,20}^{31,1}-\frac{C_{31,22}^{31,0}}{192
\sqrt{105}}-\frac{C_{31,22}^{31,1}}{96 \sqrt{105}}-\frac{9 C_{31,22}^{33,0}}{8960}-\frac{9 C_{31,22}^{33,1}}{4480}-\frac{3 \sqrt{3} C_{33,00}^{33,0}}{1280}-\frac{3
\sqrt{3} C_{33,00}^{33,1}}{640}-\frac{3 C_{33,20}^{33,0}}{1280}-\frac{3 C_{33,20}^{33,1}}{640}+\frac{3}{560} \sqrt{\frac{3}{10}} C_{33,22}^{33,0}+\frac{3}{280}
\sqrt{\frac{3}{10}} C_{33,22}^{33,1}\)

\noindent\(C_{22,4213,1}^{00,0011,1,\text{Loc}}= \frac{\sqrt{7} C_{00,20}^{60,1}}{128}+\frac{3 C_{00,22}^{62,1}}{128 \sqrt{35}}+\frac{\sqrt{7}
C_{11,20}^{31,1}}{384}+\frac{C_{11,22}^{51,1}}{192 \sqrt{35}}+\frac{9 \sqrt{3} C_{11,22}^{53,1}}{8960}-\frac{\sqrt{7} C_{20,20}^{40,1}}{384}-\frac{C_{20,22}^{42,1}}{192
\sqrt{35}}+\frac{1}{384} \sqrt{\frac{7}{5}} C_{22,20}^{42,1}-\frac{1}{384} \sqrt{\frac{5}{7}} C_{22,22}^{40,1}+\frac{C_{22,22}^{42,1}}{192 \sqrt{5}}-\frac{\sqrt{7}
C_{31,20}^{31,1}}{384}-\frac{C_{31,22}^{31,1}}{192 \sqrt{35}}-\frac{9 \sqrt{3} C_{31,22}^{33,1}}{8960}+\frac{3 \sqrt{3} C_{33,20}^{33,1}}{640}+\frac{9
C_{33,22}^{33,1}}{560 \sqrt{10}}\)

\noindent\(C_{22,4213,1}^{00,0011,1,\text{NonLoc}}= \frac{\sqrt{21} C_{00,00}^{60,0}}{256}+\frac{\sqrt{7} C_{00,20}^{60,0}}{256}-\frac{3
C_{00,22}^{62,0}}{128 \sqrt{35}}-\frac{1}{256} \sqrt{\frac{7}{3}} C_{11,00}^{51,0}-\frac{\sqrt{7} C_{11,20}^{31,0}}{768}+\frac{C_{11,22}^{51,0}}{192
\sqrt{35}}+\frac{9 \sqrt{3} C_{11,22}^{53,0}}{8960}-\frac{1}{256} \sqrt{\frac{7}{3}} C_{20,00}^{40,0}-\frac{\sqrt{7} C_{20,20}^{40,0}}{768}+\frac{C_{20,22}^{42,0}}{192
\sqrt{35}}+\frac{1}{256} \sqrt{\frac{7}{15}} C_{22,00}^{42,0}+\frac{1}{768} \sqrt{\frac{7}{5}} C_{22,20}^{42,0}+\frac{1}{384} \sqrt{\frac{5}{7}}
C_{22,22}^{40,0}-\frac{C_{22,22}^{42,0}}{192 \sqrt{5}}+\frac{1}{256} \sqrt{\frac{7}{3}} C_{31,00}^{31,0}+\frac{\sqrt{7} C_{31,20}^{31,0}}{768}-\frac{C_{31,22}^{31,0}}{192
\sqrt{35}}-\frac{9 \sqrt{3} C_{31,22}^{33,0}}{8960}-\frac{9 C_{33,00}^{33,0}}{1280}-\frac{3 \sqrt{3} C_{33,20}^{33,0}}{1280}+\frac{9 C_{33,22}^{33,0}}{560
\sqrt{10}}\)

\noindent\(C_{22,4413,1}^{00,0011,0,\text{Loc}}= \frac{C_{00,22}^{62,0}}{64 \sqrt{5}}+\frac{C_{00,22}^{62,1}}{128 \sqrt{5}}+\frac{C_{11,22}^{51,0}}{64
\sqrt{5}}+\frac{C_{11,22}^{51,1}}{128 \sqrt{5}}-\frac{1}{160} \sqrt{\frac{3}{7}} C_{11,22}^{53,0}-\frac{1}{320} \sqrt{\frac{3}{7}} C_{11,22}^{53,1}-\frac{C_{20,22}^{42,0}}{64
\sqrt{5}}-\frac{C_{20,22}^{42,1}}{128 \sqrt{5}}+\frac{C_{22,22}^{40,0}}{64 \sqrt{5}}+\frac{C_{22,22}^{40,1}}{128 \sqrt{5}}+\frac{C_{22,22}^{44,0}}{320}+\frac{C_{22,22}^{44,1}}{640}-\frac{C_{31,22}^{31,0}}{64
\sqrt{5}}-\frac{C_{31,22}^{31,1}}{128 \sqrt{5}}+\frac{1}{160} \sqrt{\frac{3}{7}} C_{31,22}^{33,0}+\frac{1}{320} \sqrt{\frac{3}{7}} C_{31,22}^{33,1}+\frac{3
C_{33,22}^{33,0}}{320 \sqrt{70}}+\frac{3 C_{33,22}^{33,1}}{640 \sqrt{70}}\)

\noindent\(C_{22,4413,1}^{00,0011,0,\text{NonLoc}}= -\frac{C_{00,22}^{62,0}}{128 \sqrt{5}}-\frac{C_{00,22}^{62,1}}{64 \sqrt{5}}+\frac{C_{11,22}^{51,0}}{128
\sqrt{5}}+\frac{C_{11,22}^{51,1}}{64 \sqrt{5}}-\frac{1}{320} \sqrt{\frac{3}{7}} C_{11,22}^{53,0}-\frac{1}{160} \sqrt{\frac{3}{7}} C_{11,22}^{53,1}+\frac{C_{20,22}^{42,0}}{128
\sqrt{5}}+\frac{C_{20,22}^{42,1}}{64 \sqrt{5}}-\frac{C_{22,22}^{40,0}}{128 \sqrt{5}}-\frac{C_{22,22}^{40,1}}{64 \sqrt{5}}-\frac{C_{22,22}^{44,0}}{640}-\frac{C_{22,22}^{44,1}}{320}-\frac{C_{31,22}^{31,0}}{128
\sqrt{5}}-\frac{C_{31,22}^{31,1}}{64 \sqrt{5}}+\frac{1}{320} \sqrt{\frac{3}{7}} C_{31,22}^{33,0}+\frac{1}{160} \sqrt{\frac{3}{7}} C_{31,22}^{33,1}+\frac{3
C_{33,22}^{33,0}}{640 \sqrt{70}}+\frac{3 C_{33,22}^{33,1}}{320 \sqrt{70}}\)

\noindent\(C_{22,4413,1}^{00,0011,1,\text{Loc}}= \frac{1}{128} \sqrt{\frac{3}{5}} C_{00,22}^{62,1}+\frac{1}{128} \sqrt{\frac{3}{5}}
C_{11,22}^{51,1}-\frac{3 C_{11,22}^{53,1}}{320 \sqrt{7}}-\frac{1}{128} \sqrt{\frac{3}{5}} C_{20,22}^{42,1}+\frac{1}{128} \sqrt{\frac{3}{5}} C_{22,22}^{40,1}+\frac{\sqrt{3}
C_{22,22}^{44,1}}{640}-\frac{1}{128} \sqrt{\frac{3}{5}} C_{31,22}^{31,1}+\frac{3 C_{31,22}^{33,1}}{320 \sqrt{7}}+\frac{3}{640} \sqrt{\frac{3}{70}}
C_{33,22}^{33,1}\)

\noindent\(C_{22,4413,1}^{00,0011,1,\text{NonLoc}}= -\frac{1}{128} \sqrt{\frac{3}{5}} C_{00,22}^{62,0}+\frac{1}{128} \sqrt{\frac{3}{5}}
C_{11,22}^{51,0}-\frac{3 C_{11,22}^{53,0}}{320 \sqrt{7}}+\frac{1}{128} \sqrt{\frac{3}{5}} C_{20,22}^{42,0}-\frac{1}{128} \sqrt{\frac{3}{5}} C_{22,22}^{40,0}-\frac{\sqrt{3}
C_{22,22}^{44,0}}{640}-\frac{1}{128} \sqrt{\frac{3}{5}} C_{31,22}^{31,0}+\frac{3 C_{31,22}^{33,0}}{320 \sqrt{7}}+\frac{3}{640} \sqrt{\frac{3}{70}}
C_{33,22}^{33,0}\)

\noindent\(C_{31,0000,1}^{00,3111,0,\text{Loc}}= -\frac{7 C_{11,11}^{51,0}}{640}-\frac{7 C_{11,11}^{51,1}}{1280}-\frac{3 C_{31,11}^{31,0}}{640}-\frac{3
C_{31,11}^{31,1}}{1280}+\frac{9}{800} \sqrt{\frac{7}{2}} C_{33,11}^{33,0}+\frac{9 \sqrt{\frac{7}{2}} C_{33,11}^{33,1}}{1600}\)

\noindent\(C_{31,0000,1}^{00,3111,0,\text{NonLoc}}= -\frac{7 C_{11,11}^{51,0}}{1280}-\frac{7 C_{11,11}^{51,1}}{640}-\frac{3 C_{31,11}^{31,0}}{1280}-\frac{3
C_{31,11}^{31,1}}{640}+\frac{9 \sqrt{\frac{7}{2}} C_{33,11}^{33,0}}{1600}+\frac{9}{800} \sqrt{\frac{7}{2}} C_{33,11}^{33,1}\)

\noindent\(C_{31,0000,1}^{00,3111,1,\text{Loc}}= -\frac{7 \sqrt{3} C_{11,11}^{51,1}}{1280}-\frac{3 \sqrt{3} C_{31,11}^{31,1}}{1280}+\frac{9
\sqrt{\frac{21}{2}} C_{33,11}^{33,1}}{1600}\)

\noindent\(C_{31,0000,1}^{00,3111,1,\text{NonLoc}}= -\frac{7 \sqrt{3} C_{11,11}^{51,0}}{1280}-\frac{3 \sqrt{3} C_{31,11}^{31,0}}{1280}+\frac{9
\sqrt{\frac{21}{2}} C_{33,11}^{33,0}}{1600}\)

\noindent\(C_{31,0000,1}^{11,2000,0,\text{Loc}}= -\frac{7 C_{00,00}^{60,0}}{256}-\frac{7 C_{00,00}^{60,1}}{512}+\frac{7 C_{11,00}^{51,0}}{768}+\frac{7
C_{11,00}^{51,1}}{1536}-\frac{5 C_{20,00}^{40,0}}{768}-\frac{5 C_{20,00}^{40,1}}{1536}+\frac{7 C_{22,00}^{42,0}}{384 \sqrt{5}}+\frac{7 C_{22,00}^{42,1}}{768
\sqrt{5}}-\frac{C_{31,00}^{31,0}}{768}-\frac{C_{31,00}^{31,1}}{1536}-\frac{3 \sqrt{21} C_{33,00}^{33,0}}{640}-\frac{3 \sqrt{21} C_{33,00}^{33,1}}{1280}\)

\noindent\(C_{31,0000,1}^{11,2000,0,\text{NonLoc}}= \frac{7 C_{00,00}^{60,0}}{1024}+\frac{7 C_{00,00}^{60,1}}{512}-\frac{7 \sqrt{3}
C_{00,20}^{60,0}}{1024}-\frac{7 \sqrt{3} C_{00,20}^{60,1}}{512}+\frac{7 C_{11,00}^{51,0}}{3072}+\frac{7 C_{11,00}^{51,1}}{1536}+\frac{5 C_{20,00}^{40,0}}{3072}+\frac{5
C_{20,00}^{40,1}}{1536}-\frac{5 C_{20,20}^{40,0}}{1024 \sqrt{3}}-\frac{5 C_{20,20}^{40,1}}{512 \sqrt{3}}-\frac{7 C_{22,00}^{42,0}}{1536 \sqrt{5}}-\frac{7
C_{22,00}^{42,1}}{768 \sqrt{5}}+\frac{7 C_{22,20}^{42,0}}{512 \sqrt{15}}+\frac{7 C_{22,20}^{42,1}}{256 \sqrt{15}}-\frac{C_{31,00}^{31,0}}{3072}-\frac{C_{31,00}^{31,1}}{1536}+\frac{C_{31,20}^{31,0}}{1024
\sqrt{3}}+\frac{C_{31,20}^{31,1}}{512 \sqrt{3}}-\frac{3 \sqrt{21} C_{33,00}^{33,0}}{2560}-\frac{3 \sqrt{21} C_{33,00}^{33,1}}{1280}+\frac{9 \sqrt{7}
C_{33,20}^{33,0}}{2560}+\frac{9 \sqrt{7} C_{33,20}^{33,1}}{1280}\)

\noindent\(C_{31,0000,1}^{11,2000,1,\text{Loc}}= -\frac{7 \sqrt{3} C_{00,00}^{60,1}}{512}+\frac{7 C_{11,00}^{51,1}}{512 \sqrt{3}}-\frac{5
C_{20,00}^{40,1}}{512 \sqrt{3}}+\frac{7 C_{22,00}^{42,1}}{256 \sqrt{15}}-\frac{C_{31,00}^{31,1}}{512 \sqrt{3}}-\frac{9 \sqrt{7} C_{33,00}^{33,1}}{1280}\)

\noindent\(C_{31,0000,1}^{11,2000,1,\text{NonLoc}}= \frac{7 \sqrt{3} C_{00,00}^{60,0}}{1024}-\frac{21 C_{00,20}^{60,0}}{1024}+\frac{7
C_{11,00}^{51,0}}{1024 \sqrt{3}}+\frac{5 C_{20,00}^{40,0}}{1024 \sqrt{3}}-\frac{5 C_{20,20}^{40,0}}{1024}-\frac{7 C_{22,00}^{42,0}}{512 \sqrt{15}}+\frac{7
C_{22,20}^{42,0}}{512 \sqrt{5}}-\frac{C_{31,00}^{31,0}}{1024 \sqrt{3}}+\frac{C_{31,20}^{31,0}}{1024}-\frac{9 \sqrt{7} C_{33,00}^{33,0}}{2560}+\frac{9
\sqrt{21} C_{33,20}^{33,0}}{2560}\)

\noindent\(C_{31,0000,1}^{11,2202,0,\text{Loc}}= -\frac{7 C_{00,00}^{60,0}}{128 \sqrt{5}}-\frac{7 C_{00,00}^{60,1}}{256 \sqrt{5}}+\frac{7
C_{11,00}^{51,0}}{384 \sqrt{5}}+\frac{7 C_{11,00}^{51,1}}{768 \sqrt{5}}+\frac{7 C_{20,00}^{40,0}}{384 \sqrt{5}}+\frac{7 C_{20,00}^{40,1}}{768 \sqrt{5}}-\frac{7
C_{22,00}^{42,0}}{1920}-\frac{7 C_{22,00}^{42,1}}{3840}-\frac{7 C_{31,00}^{31,0}}{384 \sqrt{5}}-\frac{7 C_{31,00}^{31,1}}{768 \sqrt{5}}+\frac{3}{640}
\sqrt{\frac{21}{5}} C_{33,00}^{33,0}+\frac{3 \sqrt{\frac{21}{5}} C_{33,00}^{33,1}}{1280}\)

\noindent\(C_{31,0000,1}^{11,2202,0,\text{NonLoc}}= \frac{7 C_{00,00}^{60,0}}{512 \sqrt{5}}+\frac{7 C_{00,00}^{60,1}}{256 \sqrt{5}}-\frac{7}{512}
\sqrt{\frac{3}{5}} C_{00,20}^{60,0}-\frac{7}{256} \sqrt{\frac{3}{5}} C_{00,20}^{60,1}+\frac{7 C_{11,00}^{51,0}}{1536 \sqrt{5}}+\frac{7 C_{11,00}^{51,1}}{768
\sqrt{5}}-\frac{7 C_{20,00}^{40,0}}{1536 \sqrt{5}}-\frac{7 C_{20,00}^{40,1}}{768 \sqrt{5}}+\frac{7 C_{20,20}^{40,0}}{512 \sqrt{15}}+\frac{7 C_{20,20}^{40,1}}{256
\sqrt{15}}+\frac{7 C_{22,00}^{42,0}}{7680}+\frac{7 C_{22,00}^{42,1}}{3840}-\frac{7 C_{22,20}^{42,0}}{2560 \sqrt{3}}-\frac{7 C_{22,20}^{42,1}}{1280
\sqrt{3}}-\frac{7 C_{31,00}^{31,0}}{1536 \sqrt{5}}-\frac{7 C_{31,00}^{31,1}}{768 \sqrt{5}}+\frac{7 C_{31,20}^{31,0}}{512 \sqrt{15}}+\frac{7 C_{31,20}^{31,1}}{256
\sqrt{15}}+\frac{3 \sqrt{\frac{21}{5}} C_{33,00}^{33,0}}{2560}+\frac{3 \sqrt{\frac{21}{5}} C_{33,00}^{33,1}}{1280}-\frac{9 \sqrt{\frac{7}{5}} C_{33,20}^{33,0}}{2560}-\frac{9
\sqrt{\frac{7}{5}} C_{33,20}^{33,1}}{1280}\)

\noindent\(C_{31,0000,1}^{11,2202,1,\text{Loc}}= -\frac{7}{256} \sqrt{\frac{3}{5}} C_{00,00}^{60,1}+\frac{7 C_{11,00}^{51,1}}{256 \sqrt{15}}+\frac{7
C_{20,00}^{40,1}}{256 \sqrt{15}}-\frac{7 C_{22,00}^{42,1}}{1280 \sqrt{3}}-\frac{7 C_{31,00}^{31,1}}{256 \sqrt{15}}+\frac{9 \sqrt{\frac{7}{5}} C_{33,00}^{33,1}}{1280}\)

\noindent\(C_{31,0000,1}^{11,2202,1,\text{NonLoc}}= \frac{7}{512} \sqrt{\frac{3}{5}} C_{00,00}^{60,0}-\frac{21 C_{00,20}^{60,0}}{512
\sqrt{5}}+\frac{7 C_{11,00}^{51,0}}{512 \sqrt{15}}-\frac{7 C_{20,00}^{40,0}}{512 \sqrt{15}}+\frac{7 C_{20,20}^{40,0}}{512 \sqrt{5}}+\frac{7 C_{22,00}^{42,0}}{2560
\sqrt{3}}-\frac{7 C_{22,20}^{42,0}}{2560}-\frac{7 C_{31,00}^{31,0}}{512 \sqrt{15}}+\frac{7 C_{31,20}^{31,0}}{512 \sqrt{5}}+\frac{9 \sqrt{\frac{7}{5}}
C_{33,00}^{33,0}}{2560}-\frac{9 \sqrt{\frac{21}{5}} C_{33,20}^{33,0}}{2560}\)

\noindent\(C_{31,0000,1}^{20,1111,0,\text{Loc}}= \frac{7 C_{11,11}^{51,0}}{1536}+\frac{7 C_{11,11}^{51,1}}{3072}-\frac{7 C_{22,11}^{42,0}}{256
\sqrt{15}}-\frac{7 C_{22,11}^{42,1}}{512 \sqrt{15}}+\frac{7 C_{31,11}^{31,0}}{1536}+\frac{7 C_{31,11}^{31,1}}{3072}+\frac{3}{640} \sqrt{\frac{7}{2}}
C_{33,11}^{33,0}+\frac{3 \sqrt{\frac{7}{2}} C_{33,11}^{33,1}}{1280}\)

\noindent\(C_{31,0000,1}^{20,1111,0,\text{NonLoc}}= \frac{7 C_{11,11}^{51,0}}{3072}+\frac{7 C_{11,11}^{51,1}}{1536}+\frac{7 C_{22,11}^{42,0}}{512
\sqrt{15}}+\frac{7 C_{22,11}^{42,1}}{256 \sqrt{15}}+\frac{7 C_{31,11}^{31,0}}{3072}+\frac{7 C_{31,11}^{31,1}}{1536}+\frac{3 \sqrt{\frac{7}{2}} C_{33,11}^{33,0}}{1280}+\frac{3}{640}
\sqrt{\frac{7}{2}} C_{33,11}^{33,1}\)

\noindent\(C_{31,0000,1}^{20,1111,1,\text{Loc}}= \frac{7 C_{11,11}^{51,1}}{1024 \sqrt{3}}-\frac{7 C_{22,11}^{42,1}}{512 \sqrt{5}}+\frac{7
C_{31,11}^{31,1}}{1024 \sqrt{3}}+\frac{3 \sqrt{\frac{21}{2}} C_{33,11}^{33,1}}{1280}\)

\noindent\(C_{31,0000,1}^{20,1111,1,\text{NonLoc}}= \frac{7 C_{11,11}^{51,0}}{1024 \sqrt{3}}+\frac{7 C_{22,11}^{42,0}}{512 \sqrt{5}}+\frac{7
C_{31,11}^{31,0}}{1024 \sqrt{3}}+\frac{3 \sqrt{\frac{21}{2}} C_{33,11}^{33,0}}{1280}\)

\noindent\(C_{31,0000,1}^{22,1111,0,\text{Loc}}= \frac{7 C_{11,11}^{51,0}}{768 \sqrt{5}}+\frac{7 C_{11,11}^{51,1}}{1536 \sqrt{5}}-\frac{7
C_{22,11}^{42,0}}{640 \sqrt{3}}-\frac{7 C_{22,11}^{42,1}}{1280 \sqrt{3}}+\frac{7 C_{31,11}^{31,0}}{768 \sqrt{5}}+\frac{7 C_{31,11}^{31,1}}{1536 \sqrt{5}}+\frac{3}{320}
\sqrt{\frac{7}{10}} C_{33,11}^{33,0}+\frac{3}{640} \sqrt{\frac{7}{10}} C_{33,11}^{33,1}\)

\noindent\(C_{31,0000,1}^{22,1111,0,\text{NonLoc}}= \frac{7 C_{11,11}^{51,0}}{1536 \sqrt{5}}+\frac{7 C_{11,11}^{51,1}}{768 \sqrt{5}}+\frac{7
C_{22,11}^{42,0}}{1280 \sqrt{3}}+\frac{7 C_{22,11}^{42,1}}{640 \sqrt{3}}+\frac{7 C_{31,11}^{31,0}}{1536 \sqrt{5}}+\frac{7 C_{31,11}^{31,1}}{768 \sqrt{5}}+\frac{3}{640}
\sqrt{\frac{7}{10}} C_{33,11}^{33,0}+\frac{3}{320} \sqrt{\frac{7}{10}} C_{33,11}^{33,1}\)

\noindent\(C_{31,0000,1}^{22,1111,1,\text{Loc}}= \frac{7 C_{11,11}^{51,1}}{512 \sqrt{15}}-\frac{7 C_{22,11}^{42,1}}{1280}+\frac{7 C_{31,11}^{31,1}}{512
\sqrt{15}}+\frac{3}{640} \sqrt{\frac{21}{10}} C_{33,11}^{33,1}\)

\noindent\(C_{31,0000,1}^{22,1111,1,\text{NonLoc}}= \frac{7 C_{11,11}^{51,0}}{512 \sqrt{15}}+\frac{7 C_{22,11}^{42,0}}{1280}+\frac{7
C_{31,11}^{31,0}}{512 \sqrt{15}}+\frac{3}{640} \sqrt{\frac{21}{10}} C_{33,11}^{33,0}\)

\noindent\(C_{31,0000,1}^{31,0000,0,\text{Loc}}= \frac{21 C_{00,00}^{60,0}}{10240}+\frac{21 C_{00,00}^{60,1}}{20480}-\frac{21 C_{11,00}^{51,0}}{10240}-\frac{21
C_{11,00}^{51,1}}{20480}+\frac{21 C_{20,00}^{40,0}}{10240}+\frac{21 C_{20,00}^{40,1}}{20480}+\frac{21 C_{22,00}^{42,0}}{5120 \sqrt{5}}+\frac{21 C_{22,00}^{42,1}}{10240
\sqrt{5}}-\frac{21 C_{31,00}^{31,0}}{10240}-\frac{21 C_{31,00}^{31,1}}{20480}-\frac{9 \sqrt{21} C_{33,00}^{33,0}}{25600}-\frac{9 \sqrt{21} C_{33,00}^{33,1}}{51200}\)

\noindent\(C_{31,0000,1}^{31,0000,0,\text{NonLoc}}= -\frac{21 C_{00,00}^{60,0}}{40960}-\frac{21 C_{00,00}^{60,1}}{20480}+\frac{21 \sqrt{3}
C_{00,20}^{60,0}}{40960}+\frac{21 \sqrt{3} C_{00,20}^{60,1}}{20480}-\frac{21 C_{11,00}^{51,0}}{40960}-\frac{21 C_{11,00}^{51,1}}{20480}-\frac{21
C_{20,00}^{40,0}}{40960}-\frac{21 C_{20,00}^{40,1}}{20480}+\frac{21 \sqrt{3} C_{20,20}^{40,0}}{40960}+\frac{21 \sqrt{3} C_{20,20}^{40,1}}{20480}-\frac{21
C_{22,00}^{42,0}}{20480 \sqrt{5}}-\frac{21 C_{22,00}^{42,1}}{10240 \sqrt{5}}+\frac{21 \sqrt{\frac{3}{5}} C_{22,20}^{42,0}}{20480}+\frac{21 \sqrt{\frac{3}{5}}
C_{22,20}^{42,1}}{10240}-\frac{21 C_{31,00}^{31,0}}{40960}-\frac{21 C_{31,00}^{31,1}}{20480}+\frac{21 \sqrt{3} C_{31,20}^{31,0}}{40960}+\frac{21
\sqrt{3} C_{31,20}^{31,1}}{20480}-\frac{9 \sqrt{21} C_{33,00}^{33,0}}{102400}-\frac{9 \sqrt{21} C_{33,00}^{33,1}}{51200}+\frac{27 \sqrt{7} C_{33,20}^{33,0}}{102400}+\frac{27
\sqrt{7} C_{33,20}^{33,1}}{51200}\)

\noindent\(C_{31,0000,1}^{31,0000,1,\text{Loc}}= \frac{21 \sqrt{3} C_{00,00}^{60,1}}{20480}-\frac{21 \sqrt{3} C_{11,00}^{51,1}}{20480}+\frac{21
\sqrt{3} C_{20,00}^{40,1}}{20480}+\frac{21 \sqrt{\frac{3}{5}} C_{22,00}^{42,1}}{10240}-\frac{21 \sqrt{3} C_{31,00}^{31,1}}{20480}-\frac{27 \sqrt{7}
C_{33,00}^{33,1}}{51200}\)

\noindent\(C_{31,0000,1}^{31,0000,1,\text{NonLoc}}= -\frac{21 \sqrt{3} C_{00,00}^{60,0}}{40960}+\frac{63 C_{00,20}^{60,0}}{40960}-\frac{21
\sqrt{3} C_{11,00}^{51,0}}{40960}-\frac{21 \sqrt{3} C_{20,00}^{40,0}}{40960}+\frac{63 C_{20,20}^{40,0}}{40960}-\frac{21 \sqrt{\frac{3}{5}} C_{22,00}^{42,0}}{20480}+\frac{63
C_{22,20}^{42,0}}{20480 \sqrt{5}}-\frac{21 \sqrt{3} C_{31,00}^{31,0}}{40960}+\frac{63 C_{31,20}^{31,0}}{40960}-\frac{27 \sqrt{7} C_{33,00}^{33,0}}{102400}+\frac{27
\sqrt{21} C_{33,20}^{33,0}}{102400}\)

\noindent\(C_{31,0011,0}^{11,2011,0,\text{Loc}}= \frac{7 C_{00,20}^{60,0}}{768}+\frac{7 C_{00,20}^{60,1}}{1536}+\frac{7 C_{00,22}^{62,0}}{256
\sqrt{5}}+\frac{7 C_{00,22}^{62,1}}{512 \sqrt{5}}-\frac{7 \sqrt{5} C_{11,22}^{51,0}}{2304}-\frac{7 \sqrt{5} C_{11,22}^{51,1}}{4608}+\frac{5 C_{20,20}^{40,0}}{2304}+\frac{5
C_{20,20}^{40,1}}{4608}+\frac{7 C_{20,22}^{42,0}}{2304 \sqrt{5}}+\frac{7 C_{20,22}^{42,1}}{4608 \sqrt{5}}-\frac{7 C_{22,20}^{42,0}}{1152 \sqrt{5}}-\frac{7
C_{22,20}^{42,1}}{2304 \sqrt{5}}+\frac{7 C_{22,22}^{40,0}}{2304 \sqrt{5}}+\frac{7 C_{22,22}^{40,1}}{4608 \sqrt{5}}-\frac{1}{576} \sqrt{\frac{7}{5}}
C_{22,22}^{42,0}-\frac{\sqrt{\frac{7}{5}} C_{22,22}^{42,1}}{1152}-\frac{3 C_{22,22}^{44,0}}{320}-\frac{3 C_{22,22}^{44,1}}{640}+\frac{C_{31,20}^{31,0}}{2304}+\frac{C_{31,20}^{31,1}}{4608}-\frac{7
C_{31,22}^{31,0}}{2304 \sqrt{5}}-\frac{7 C_{31,22}^{31,1}}{4608 \sqrt{5}}+\frac{1}{320} \sqrt{\frac{7}{3}} C_{31,22}^{33,0}+\frac{1}{640} \sqrt{\frac{7}{3}}
C_{31,22}^{33,1}+\frac{\sqrt{21} C_{33,20}^{33,0}}{640}+\frac{\sqrt{21} C_{33,20}^{33,1}}{1280}+\frac{3}{320} \sqrt{\frac{7}{10}} C_{33,22}^{33,0}+\frac{3}{640}
\sqrt{\frac{7}{10}} C_{33,22}^{33,1}\)

\noindent\(C_{31,0011,0}^{11,2011,0,\text{NonLoc}}= \frac{7 C_{00,00}^{60,0}}{1024 \sqrt{3}}+\frac{7 C_{00,00}^{60,1}}{512 \sqrt{3}}+\frac{7
C_{00,20}^{60,0}}{3072}+\frac{7 C_{00,20}^{60,1}}{1536}-\frac{7 C_{00,22}^{62,0}}{512 \sqrt{5}}-\frac{7 C_{00,22}^{62,1}}{256 \sqrt{5}}+\frac{7 C_{11,00}^{51,0}}{3072
\sqrt{3}}+\frac{7 C_{11,00}^{51,1}}{1536 \sqrt{3}}-\frac{7 \sqrt{5} C_{11,22}^{51,0}}{4608}-\frac{7 \sqrt{5} C_{11,22}^{51,1}}{2304}+\frac{5 C_{20,00}^{40,0}}{3072
\sqrt{3}}+\frac{5 C_{20,00}^{40,1}}{1536 \sqrt{3}}+\frac{5 C_{20,20}^{40,0}}{9216}+\frac{5 C_{20,20}^{40,1}}{4608}-\frac{7 C_{20,22}^{42,0}}{4608
\sqrt{5}}-\frac{7 C_{20,22}^{42,1}}{2304 \sqrt{5}}-\frac{7 C_{22,00}^{42,0}}{1536 \sqrt{15}}-\frac{7 C_{22,00}^{42,1}}{768 \sqrt{15}}-\frac{7 C_{22,20}^{42,0}}{4608
\sqrt{5}}-\frac{7 C_{22,20}^{42,1}}{2304 \sqrt{5}}-\frac{7 C_{22,22}^{40,0}}{4608 \sqrt{5}}-\frac{7 C_{22,22}^{40,1}}{2304 \sqrt{5}}+\frac{\sqrt{\frac{7}{5}}
C_{22,22}^{42,0}}{1152}+\frac{1}{576} \sqrt{\frac{7}{5}} C_{22,22}^{42,1}+\frac{3 C_{22,22}^{44,0}}{640}+\frac{3 C_{22,22}^{44,1}}{320}-\frac{C_{31,00}^{31,0}}{3072
\sqrt{3}}-\frac{C_{31,00}^{31,1}}{1536 \sqrt{3}}-\frac{C_{31,20}^{31,0}}{9216}-\frac{C_{31,20}^{31,1}}{4608}-\frac{7 C_{31,22}^{31,0}}{4608 \sqrt{5}}-\frac{7
C_{31,22}^{31,1}}{2304 \sqrt{5}}+\frac{1}{640} \sqrt{\frac{7}{3}} C_{31,22}^{33,0}+\frac{1}{320} \sqrt{\frac{7}{3}} C_{31,22}^{33,1}-\frac{3 \sqrt{7}
C_{33,00}^{33,0}}{2560}-\frac{3 \sqrt{7} C_{33,00}^{33,1}}{1280}-\frac{\sqrt{21} C_{33,20}^{33,0}}{2560}-\frac{\sqrt{21} C_{33,20}^{33,1}}{1280}+\frac{3}{640}
\sqrt{\frac{7}{10}} C_{33,22}^{33,0}+\frac{3}{320} \sqrt{\frac{7}{10}} C_{33,22}^{33,1}\)

\noindent\(C_{31,0011,0}^{11,2011,1,\text{Loc}}= \frac{7 C_{00,20}^{60,1}}{512 \sqrt{3}}+\frac{7}{512} \sqrt{\frac{3}{5}} C_{00,22}^{62,1}-\frac{7
\sqrt{\frac{5}{3}} C_{11,22}^{51,1}}{1536}+\frac{5 C_{20,20}^{40,1}}{1536 \sqrt{3}}+\frac{7 C_{20,22}^{42,1}}{1536 \sqrt{15}}-\frac{7 C_{22,20}^{42,1}}{768
\sqrt{15}}+\frac{7 C_{22,22}^{40,1}}{1536 \sqrt{15}}-\frac{1}{384} \sqrt{\frac{7}{15}} C_{22,22}^{42,1}-\frac{3 \sqrt{3} C_{22,22}^{44,1}}{640}+\frac{C_{31,20}^{31,1}}{1536
\sqrt{3}}-\frac{7 C_{31,22}^{31,1}}{1536 \sqrt{15}}+\frac{\sqrt{7} C_{31,22}^{33,1}}{640}+\frac{3 \sqrt{7} C_{33,20}^{33,1}}{1280}+\frac{3}{640}
\sqrt{\frac{21}{10}} C_{33,22}^{33,1}\)

\noindent\(C_{31,0011,0}^{11,2011,1,\text{NonLoc}}= \frac{7 C_{00,00}^{60,0}}{1024}+\frac{7 C_{00,20}^{60,0}}{1024 \sqrt{3}}-\frac{7}{512}
\sqrt{\frac{3}{5}} C_{00,22}^{62,0}+\frac{7 C_{11,00}^{51,0}}{3072}-\frac{7 \sqrt{\frac{5}{3}} C_{11,22}^{51,0}}{1536}+\frac{5 C_{20,00}^{40,0}}{3072}+\frac{5
C_{20,20}^{40,0}}{3072 \sqrt{3}}-\frac{7 C_{20,22}^{42,0}}{1536 \sqrt{15}}-\frac{7 C_{22,00}^{42,0}}{1536 \sqrt{5}}-\frac{7 C_{22,20}^{42,0}}{1536
\sqrt{15}}-\frac{7 C_{22,22}^{40,0}}{1536 \sqrt{15}}+\frac{1}{384} \sqrt{\frac{7}{15}} C_{22,22}^{42,0}+\frac{3 \sqrt{3} C_{22,22}^{44,0}}{640}-\frac{C_{31,00}^{31,0}}{3072}-\frac{C_{31,20}^{31,0}}{3072
\sqrt{3}}-\frac{7 C_{31,22}^{31,0}}{1536 \sqrt{15}}+\frac{\sqrt{7} C_{31,22}^{33,0}}{640}-\frac{3 \sqrt{21} C_{33,00}^{33,0}}{2560}-\frac{3 \sqrt{7}
C_{33,20}^{33,0}}{2560}+\frac{3}{640} \sqrt{\frac{21}{10}} C_{33,22}^{33,0}\)

\noindent\(C_{31,0011,0}^{11,2211,0,\text{Loc}}= \frac{7 C_{00,20}^{60,0}}{384 \sqrt{5}}+\frac{7 C_{00,20}^{60,1}}{768 \sqrt{5}}+\frac{7
C_{00,22}^{62,0}}{640}+\frac{7 C_{00,22}^{62,1}}{1280}-\frac{7 C_{11,22}^{51,0}}{11520}-\frac{7 C_{11,22}^{51,1}}{23040}-\frac{9 \sqrt{\frac{21}{5}}
C_{11,22}^{53,0}}{2560}-\frac{9 \sqrt{\frac{21}{5}} C_{11,22}^{53,1}}{5120}-\frac{7 C_{20,20}^{40,0}}{1152 \sqrt{5}}-\frac{7 C_{20,20}^{40,1}}{2304
\sqrt{5}}-\frac{7 C_{20,22}^{42,0}}{11520}-\frac{7 C_{20,22}^{42,1}}{23040}+\frac{7 C_{22,20}^{42,0}}{5760}+\frac{7 C_{22,20}^{42,1}}{11520}-\frac{13
C_{22,22}^{40,0}}{2880}-\frac{13 C_{22,22}^{40,1}}{5760}+\frac{\sqrt{7} C_{22,22}^{42,0}}{2880}+\frac{\sqrt{7} C_{22,22}^{42,1}}{5760}+\frac{3 C_{22,22}^{44,0}}{320
\sqrt{5}}+\frac{3 C_{22,22}^{44,1}}{640 \sqrt{5}}+\frac{7 C_{31,20}^{31,0}}{1152 \sqrt{5}}+\frac{7 C_{31,20}^{31,1}}{2304 \sqrt{5}}+\frac{37 C_{31,22}^{31,0}}{11520}+\frac{37
C_{31,22}^{31,1}}{23040}+\frac{7 \sqrt{\frac{7}{15}} C_{31,22}^{33,0}}{2560}+\frac{7 \sqrt{\frac{7}{15}} C_{31,22}^{33,1}}{5120}-\frac{1}{640} \sqrt{\frac{21}{5}}
C_{33,20}^{33,0}-\frac{\sqrt{\frac{21}{5}} C_{33,20}^{33,1}}{1280}-\frac{3 \sqrt{\frac{7}{2}} C_{33,22}^{33,0}}{1600}-\frac{3 \sqrt{\frac{7}{2}}
C_{33,22}^{33,1}}{3200}\)

\noindent\(C_{31,0011,0}^{11,2211,0,\text{NonLoc}}= \frac{7 C_{00,00}^{60,0}}{512 \sqrt{15}}+\frac{7 C_{00,00}^{60,1}}{256 \sqrt{15}}+\frac{7
C_{00,20}^{60,0}}{1536 \sqrt{5}}+\frac{7 C_{00,20}^{60,1}}{768 \sqrt{5}}-\frac{7 C_{00,22}^{62,0}}{1280}-\frac{7 C_{00,22}^{62,1}}{640}+\frac{7 C_{11,00}^{51,0}}{1536
\sqrt{15}}+\frac{7 C_{11,00}^{51,1}}{768 \sqrt{15}}-\frac{7 C_{11,22}^{51,0}}{23040}-\frac{7 C_{11,22}^{51,1}}{11520}-\frac{9 \sqrt{\frac{21}{5}}
C_{11,22}^{53,0}}{5120}-\frac{9 \sqrt{\frac{21}{5}} C_{11,22}^{53,1}}{2560}-\frac{7 C_{20,00}^{40,0}}{1536 \sqrt{15}}-\frac{7 C_{20,00}^{40,1}}{768
\sqrt{15}}-\frac{7 C_{20,20}^{40,0}}{4608 \sqrt{5}}-\frac{7 C_{20,20}^{40,1}}{2304 \sqrt{5}}+\frac{7 C_{20,22}^{42,0}}{23040}+\frac{7 C_{20,22}^{42,1}}{11520}+\frac{7
C_{22,00}^{42,0}}{7680 \sqrt{3}}+\frac{7 C_{22,00}^{42,1}}{3840 \sqrt{3}}+\frac{7 C_{22,20}^{42,0}}{23040}+\frac{7 C_{22,20}^{42,1}}{11520}+\frac{13
C_{22,22}^{40,0}}{5760}+\frac{13 C_{22,22}^{40,1}}{2880}-\frac{\sqrt{7} C_{22,22}^{42,0}}{5760}-\frac{\sqrt{7} C_{22,22}^{42,1}}{2880}-\frac{3 C_{22,22}^{44,0}}{640
\sqrt{5}}-\frac{3 C_{22,22}^{44,1}}{320 \sqrt{5}}-\frac{7 C_{31,00}^{31,0}}{1536 \sqrt{15}}-\frac{7 C_{31,00}^{31,1}}{768 \sqrt{15}}-\frac{7 C_{31,20}^{31,0}}{4608
\sqrt{5}}-\frac{7 C_{31,20}^{31,1}}{2304 \sqrt{5}}+\frac{37 C_{31,22}^{31,0}}{23040}+\frac{37 C_{31,22}^{31,1}}{11520}+\frac{7 \sqrt{\frac{7}{15}}
C_{31,22}^{33,0}}{5120}+\frac{7 \sqrt{\frac{7}{15}} C_{31,22}^{33,1}}{2560}+\frac{3 \sqrt{\frac{7}{5}} C_{33,00}^{33,0}}{2560}+\frac{3 \sqrt{\frac{7}{5}}
C_{33,00}^{33,1}}{1280}+\frac{\sqrt{\frac{21}{5}} C_{33,20}^{33,0}}{2560}+\frac{\sqrt{\frac{21}{5}} C_{33,20}^{33,1}}{1280}-\frac{3 \sqrt{\frac{7}{2}}
C_{33,22}^{33,0}}{3200}-\frac{3 \sqrt{\frac{7}{2}} C_{33,22}^{33,1}}{1600}\)

\noindent\(C_{31,0011,0}^{11,2211,1,\text{Loc}}= \frac{7 C_{00,20}^{60,1}}{256 \sqrt{15}}+\frac{7 \sqrt{3} C_{00,22}^{62,1}}{1280}-\frac{7
C_{11,22}^{51,1}}{7680 \sqrt{3}}-\frac{27 \sqrt{\frac{7}{5}} C_{11,22}^{53,1}}{5120}-\frac{7 C_{20,20}^{40,1}}{768 \sqrt{15}}-\frac{7 C_{20,22}^{42,1}}{7680
\sqrt{3}}+\frac{7 C_{22,20}^{42,1}}{3840 \sqrt{3}}-\frac{13 C_{22,22}^{40,1}}{1920 \sqrt{3}}+\frac{\sqrt{\frac{7}{3}} C_{22,22}^{42,1}}{1920}+\frac{3}{640}
\sqrt{\frac{3}{5}} C_{22,22}^{44,1}+\frac{7 C_{31,20}^{31,1}}{768 \sqrt{15}}+\frac{37 C_{31,22}^{31,1}}{7680 \sqrt{3}}+\frac{7 \sqrt{\frac{7}{5}}
C_{31,22}^{33,1}}{5120}-\frac{3 \sqrt{\frac{7}{5}} C_{33,20}^{33,1}}{1280}-\frac{3 \sqrt{\frac{21}{2}} C_{33,22}^{33,1}}{3200}\)

\noindent\(C_{31,0011,0}^{11,2211,1,\text{NonLoc}}= \frac{7 C_{00,00}^{60,0}}{512 \sqrt{5}}+\frac{7 C_{00,20}^{60,0}}{512 \sqrt{15}}-\frac{7
\sqrt{3} C_{00,22}^{62,0}}{1280}+\frac{7 C_{11,00}^{51,0}}{1536 \sqrt{5}}-\frac{7 C_{11,22}^{51,0}}{7680 \sqrt{3}}-\frac{27 \sqrt{\frac{7}{5}} C_{11,22}^{53,0}}{5120}-\frac{7
C_{20,00}^{40,0}}{1536 \sqrt{5}}-\frac{7 C_{20,20}^{40,0}}{1536 \sqrt{15}}+\frac{7 C_{20,22}^{42,0}}{7680 \sqrt{3}}+\frac{7 C_{22,00}^{42,0}}{7680}+\frac{7
C_{22,20}^{42,0}}{7680 \sqrt{3}}+\frac{13 C_{22,22}^{40,0}}{1920 \sqrt{3}}-\frac{\sqrt{\frac{7}{3}} C_{22,22}^{42,0}}{1920}-\frac{3}{640} \sqrt{\frac{3}{5}}
C_{22,22}^{44,0}-\frac{7 C_{31,00}^{31,0}}{1536 \sqrt{5}}-\frac{7 C_{31,20}^{31,0}}{1536 \sqrt{15}}+\frac{37 C_{31,22}^{31,0}}{7680 \sqrt{3}}+\frac{7
\sqrt{\frac{7}{5}} C_{31,22}^{33,0}}{5120}+\frac{3 \sqrt{\frac{21}{5}} C_{33,00}^{33,0}}{2560}+\frac{3 \sqrt{\frac{7}{5}} C_{33,20}^{33,0}}{2560}-\frac{3
\sqrt{\frac{21}{2}} C_{33,22}^{33,0}}{3200}\)

\noindent\(C_{31,0011,0}^{31,0011,0,\text{Loc}}= -\frac{7 C_{00,20}^{60,0}}{10240}-\frac{7 C_{00,20}^{60,1}}{20480}-\frac{21 C_{00,22}^{62,0}}{10240
\sqrt{5}}-\frac{21 C_{00,22}^{62,1}}{20480 \sqrt{5}}+\frac{21 C_{11,22}^{51,0}}{10240 \sqrt{5}}+\frac{21 C_{11,22}^{51,1}}{20480 \sqrt{5}}+\frac{9
\sqrt{21} C_{11,22}^{53,0}}{51200}+\frac{9 \sqrt{21} C_{11,22}^{53,1}}{102400}-\frac{7 C_{20,20}^{40,0}}{10240}-\frac{7 C_{20,20}^{40,1}}{20480}-\frac{21
C_{20,22}^{42,0}}{10240 \sqrt{5}}-\frac{21 C_{20,22}^{42,1}}{20480 \sqrt{5}}-\frac{7 C_{22,20}^{42,0}}{5120 \sqrt{5}}-\frac{7 C_{22,20}^{42,1}}{10240
\sqrt{5}}-\frac{21 C_{22,22}^{40,0}}{10240 \sqrt{5}}-\frac{21 C_{22,22}^{40,1}}{20480 \sqrt{5}}-\frac{3 \sqrt{\frac{7}{5}} C_{22,22}^{42,0}}{5120}-\frac{3
\sqrt{\frac{7}{5}} C_{22,22}^{42,1}}{10240}-\frac{9 C_{22,22}^{44,0}}{12800}-\frac{9 C_{22,22}^{44,1}}{25600}+\frac{7 C_{31,20}^{31,0}}{10240}+\frac{7
C_{31,20}^{31,1}}{20480}+\frac{21 C_{31,22}^{31,0}}{10240 \sqrt{5}}+\frac{21 C_{31,22}^{31,1}}{20480 \sqrt{5}}+\frac{9 \sqrt{21} C_{31,22}^{33,0}}{51200}+\frac{9
\sqrt{21} C_{31,22}^{33,1}}{102400}+\frac{3 \sqrt{21} C_{33,20}^{33,0}}{25600}+\frac{3 \sqrt{21} C_{33,20}^{33,1}}{51200}+\frac{9 \sqrt{\frac{7}{10}}
C_{33,22}^{33,0}}{12800}+\frac{9 \sqrt{\frac{7}{10}} C_{33,22}^{33,1}}{25600}\)

\noindent\(C_{31,0011,0}^{31,0011,0,\text{NonLoc}}= -\frac{7 \sqrt{3} C_{00,00}^{60,0}}{40960}-\frac{7 \sqrt{3} C_{00,00}^{60,1}}{20480}-\frac{7
C_{00,20}^{60,0}}{40960}-\frac{7 C_{00,20}^{60,1}}{20480}+\frac{21 C_{00,22}^{62,0}}{20480 \sqrt{5}}+\frac{21 C_{00,22}^{62,1}}{10240 \sqrt{5}}-\frac{7
\sqrt{3} C_{11,00}^{51,0}}{40960}-\frac{7 \sqrt{3} C_{11,00}^{51,1}}{20480}+\frac{21 C_{11,22}^{51,0}}{20480 \sqrt{5}}+\frac{21 C_{11,22}^{51,1}}{10240
\sqrt{5}}+\frac{9 \sqrt{21} C_{11,22}^{53,0}}{102400}+\frac{9 \sqrt{21} C_{11,22}^{53,1}}{51200}-\frac{7 \sqrt{3} C_{20,00}^{40,0}}{40960}-\frac{7
\sqrt{3} C_{20,00}^{40,1}}{20480}-\frac{7 C_{20,20}^{40,0}}{40960}-\frac{7 C_{20,20}^{40,1}}{20480}+\frac{21 C_{20,22}^{42,0}}{20480 \sqrt{5}}+\frac{21
C_{20,22}^{42,1}}{10240 \sqrt{5}}-\frac{7 \sqrt{\frac{3}{5}} C_{22,00}^{42,0}}{20480}-\frac{7 \sqrt{\frac{3}{5}} C_{22,00}^{42,1}}{10240}-\frac{7
C_{22,20}^{42,0}}{20480 \sqrt{5}}-\frac{7 C_{22,20}^{42,1}}{10240 \sqrt{5}}+\frac{21 C_{22,22}^{40,0}}{20480 \sqrt{5}}+\frac{21 C_{22,22}^{40,1}}{10240
\sqrt{5}}+\frac{3 \sqrt{\frac{7}{5}} C_{22,22}^{42,0}}{10240}+\frac{3 \sqrt{\frac{7}{5}} C_{22,22}^{42,1}}{5120}+\frac{9 C_{22,22}^{44,0}}{25600}+\frac{9
C_{22,22}^{44,1}}{12800}-\frac{7 \sqrt{3} C_{31,00}^{31,0}}{40960}-\frac{7 \sqrt{3} C_{31,00}^{31,1}}{20480}-\frac{7 C_{31,20}^{31,0}}{40960}-\frac{7
C_{31,20}^{31,1}}{20480}+\frac{21 C_{31,22}^{31,0}}{20480 \sqrt{5}}+\frac{21 C_{31,22}^{31,1}}{10240 \sqrt{5}}+\frac{9 \sqrt{21} C_{31,22}^{33,0}}{102400}+\frac{9
\sqrt{21} C_{31,22}^{33,1}}{51200}-\frac{9 \sqrt{7} C_{33,00}^{33,0}}{102400}-\frac{9 \sqrt{7} C_{33,00}^{33,1}}{51200}-\frac{3 \sqrt{21} C_{33,20}^{33,0}}{102400}-\frac{3
\sqrt{21} C_{33,20}^{33,1}}{51200}+\frac{9 \sqrt{\frac{7}{10}} C_{33,22}^{33,0}}{25600}+\frac{9 \sqrt{\frac{7}{10}} C_{33,22}^{33,1}}{12800}\)

\noindent\(C_{31,0011,0}^{31,0011,1,\text{Loc}}= -\frac{7 \sqrt{3} C_{00,20}^{60,1}}{20480}-\frac{21 \sqrt{\frac{3}{5}} C_{00,22}^{62,1}}{20480}+\frac{21
\sqrt{\frac{3}{5}} C_{11,22}^{51,1}}{20480}+\frac{27 \sqrt{7} C_{11,22}^{53,1}}{102400}-\frac{7 \sqrt{3} C_{20,20}^{40,1}}{20480}-\frac{21 \sqrt{\frac{3}{5}}
C_{20,22}^{42,1}}{20480}-\frac{7 \sqrt{\frac{3}{5}} C_{22,20}^{42,1}}{10240}-\frac{21 \sqrt{\frac{3}{5}} C_{22,22}^{40,1}}{20480}-\frac{3 \sqrt{\frac{21}{5}}
C_{22,22}^{42,1}}{10240}-\frac{9 \sqrt{3} C_{22,22}^{44,1}}{25600}+\frac{7 \sqrt{3} C_{31,20}^{31,1}}{20480}+\frac{21 \sqrt{\frac{3}{5}} C_{31,22}^{31,1}}{20480}+\frac{27
\sqrt{7} C_{31,22}^{33,1}}{102400}+\frac{9 \sqrt{7} C_{33,20}^{33,1}}{51200}+\frac{9 \sqrt{\frac{21}{10}} C_{33,22}^{33,1}}{25600}\)

\noindent\(C_{31,0011,0}^{31,0011,1,\text{NonLoc}}= -\frac{21 C_{00,00}^{60,0}}{40960}-\frac{7 \sqrt{3} C_{00,20}^{60,0}}{40960}+\frac{21
\sqrt{\frac{3}{5}} C_{00,22}^{62,0}}{20480}-\frac{21 C_{11,00}^{51,0}}{40960}+\frac{21 \sqrt{\frac{3}{5}} C_{11,22}^{51,0}}{20480}+\frac{27 \sqrt{7}
C_{11,22}^{53,0}}{102400}-\frac{21 C_{20,00}^{40,0}}{40960}-\frac{7 \sqrt{3} C_{20,20}^{40,0}}{40960}+\frac{21 \sqrt{\frac{3}{5}} C_{20,22}^{42,0}}{20480}-\frac{21
C_{22,00}^{42,0}}{20480 \sqrt{5}}-\frac{7 \sqrt{\frac{3}{5}} C_{22,20}^{42,0}}{20480}+\frac{21 \sqrt{\frac{3}{5}} C_{22,22}^{40,0}}{20480}+\frac{3
\sqrt{\frac{21}{5}} C_{22,22}^{42,0}}{10240}+\frac{9 \sqrt{3} C_{22,22}^{44,0}}{25600}-\frac{21 C_{31,00}^{31,0}}{40960}-\frac{7 \sqrt{3} C_{31,20}^{31,0}}{40960}+\frac{21
\sqrt{\frac{3}{5}} C_{31,22}^{31,0}}{20480}+\frac{27 \sqrt{7} C_{31,22}^{33,0}}{102400}-\frac{9 \sqrt{21} C_{33,00}^{33,0}}{102400}-\frac{9 \sqrt{7}
C_{33,20}^{33,0}}{102400}+\frac{9 \sqrt{\frac{21}{10}} C_{33,22}^{33,0}}{25600}\)

\noindent\(C_{31,0011,1}^{00,3101,0,\text{Loc}}= \frac{7 C_{11,11}^{51,0}}{640}+\frac{7 C_{11,11}^{51,1}}{1280}+\frac{3 C_{31,11}^{31,0}}{640}+\frac{3
C_{31,11}^{31,1}}{1280}-\frac{9}{800} \sqrt{\frac{7}{2}} C_{33,11}^{33,0}-\frac{9 \sqrt{\frac{7}{2}} C_{33,11}^{33,1}}{1600}\)

\noindent\(C_{31,0011,1}^{00,3101,0,\text{NonLoc}}= \frac{7 C_{11,11}^{51,0}}{1280}+\frac{7 C_{11,11}^{51,1}}{640}+\frac{3 C_{31,11}^{31,0}}{1280}+\frac{3
C_{31,11}^{31,1}}{640}-\frac{9 \sqrt{\frac{7}{2}} C_{33,11}^{33,0}}{1600}-\frac{9}{800} \sqrt{\frac{7}{2}} C_{33,11}^{33,1}\)

\noindent\(C_{31,0011,1}^{00,3101,1,\text{Loc}}= \frac{7 \sqrt{3} C_{11,11}^{51,1}}{1280}+\frac{3 \sqrt{3} C_{31,11}^{31,1}}{1280}-\frac{9
\sqrt{\frac{21}{2}} C_{33,11}^{33,1}}{1600}\)

\noindent\(C_{31,0011,1}^{00,3101,1,\text{NonLoc}}= \frac{7 \sqrt{3} C_{11,11}^{51,0}}{1280}+\frac{3 \sqrt{3} C_{31,11}^{31,0}}{1280}-\frac{9
\sqrt{\frac{21}{2}} C_{33,11}^{33,0}}{1600}\)

\noindent\(C_{31,0011,1}^{11,2011,0,\text{Loc}}= \frac{7 C_{00,20}^{60,0}}{256 \sqrt{3}}+\frac{7 C_{00,20}^{60,1}}{512 \sqrt{3}}-\frac{7}{512}
\sqrt{\frac{3}{5}} C_{00,22}^{62,0}-\frac{7 \sqrt{\frac{3}{5}} C_{00,22}^{62,1}}{1024}+\frac{7 \sqrt{\frac{5}{3}} C_{11,22}^{51,0}}{1536}+\frac{7
\sqrt{\frac{5}{3}} C_{11,22}^{51,1}}{3072}+\frac{5 C_{20,20}^{40,0}}{768 \sqrt{3}}+\frac{5 C_{20,20}^{40,1}}{1536 \sqrt{3}}-\frac{7 C_{20,22}^{42,0}}{1536
\sqrt{15}}-\frac{7 C_{20,22}^{42,1}}{3072 \sqrt{15}}-\frac{7 C_{22,20}^{42,0}}{384 \sqrt{15}}-\frac{7 C_{22,20}^{42,1}}{768 \sqrt{15}}-\frac{7 C_{22,22}^{40,0}}{1536
\sqrt{15}}-\frac{7 C_{22,22}^{40,1}}{3072 \sqrt{15}}+\frac{1}{384} \sqrt{\frac{7}{15}} C_{22,22}^{42,0}+\frac{1}{768} \sqrt{\frac{7}{15}} C_{22,22}^{42,1}+\frac{3
\sqrt{3} C_{22,22}^{44,0}}{640}+\frac{3 \sqrt{3} C_{22,22}^{44,1}}{1280}+\frac{C_{31,20}^{31,0}}{768 \sqrt{3}}+\frac{C_{31,20}^{31,1}}{1536 \sqrt{3}}+\frac{7
C_{31,22}^{31,0}}{1536 \sqrt{15}}+\frac{7 C_{31,22}^{31,1}}{3072 \sqrt{15}}-\frac{\sqrt{7} C_{31,22}^{33,0}}{640}-\frac{\sqrt{7} C_{31,22}^{33,1}}{1280}+\frac{3
\sqrt{7} C_{33,20}^{33,0}}{640}+\frac{3 \sqrt{7} C_{33,20}^{33,1}}{1280}-\frac{3}{640} \sqrt{\frac{21}{10}} C_{33,22}^{33,0}-\frac{3 \sqrt{\frac{21}{10}}
C_{33,22}^{33,1}}{1280}\)

\noindent\(C_{31,0011,1}^{11,2011,0,\text{NonLoc}}= \frac{7 C_{00,00}^{60,0}}{1024}+\frac{7 C_{00,00}^{60,1}}{512}+\frac{7 C_{00,20}^{60,0}}{1024
\sqrt{3}}+\frac{7 C_{00,20}^{60,1}}{512 \sqrt{3}}+\frac{7 \sqrt{\frac{3}{5}} C_{00,22}^{62,0}}{1024}+\frac{7}{512} \sqrt{\frac{3}{5}} C_{00,22}^{62,1}+\frac{7
C_{11,00}^{51,0}}{3072}+\frac{7 C_{11,00}^{51,1}}{1536}+\frac{7 \sqrt{\frac{5}{3}} C_{11,22}^{51,0}}{3072}+\frac{7 \sqrt{\frac{5}{3}} C_{11,22}^{51,1}}{1536}+\frac{5
C_{20,00}^{40,0}}{3072}+\frac{5 C_{20,00}^{40,1}}{1536}+\frac{5 C_{20,20}^{40,0}}{3072 \sqrt{3}}+\frac{5 C_{20,20}^{40,1}}{1536 \sqrt{3}}+\frac{7
C_{20,22}^{42,0}}{3072 \sqrt{15}}+\frac{7 C_{20,22}^{42,1}}{1536 \sqrt{15}}-\frac{7 C_{22,00}^{42,0}}{1536 \sqrt{5}}-\frac{7 C_{22,00}^{42,1}}{768
\sqrt{5}}-\frac{7 C_{22,20}^{42,0}}{1536 \sqrt{15}}-\frac{7 C_{22,20}^{42,1}}{768 \sqrt{15}}+\frac{7 C_{22,22}^{40,0}}{3072 \sqrt{15}}+\frac{7 C_{22,22}^{40,1}}{1536
\sqrt{15}}-\frac{1}{768} \sqrt{\frac{7}{15}} C_{22,22}^{42,0}-\frac{1}{384} \sqrt{\frac{7}{15}} C_{22,22}^{42,1}-\frac{3 \sqrt{3} C_{22,22}^{44,0}}{1280}-\frac{3
\sqrt{3} C_{22,22}^{44,1}}{640}-\frac{C_{31,00}^{31,0}}{3072}-\frac{C_{31,00}^{31,1}}{1536}-\frac{C_{31,20}^{31,0}}{3072 \sqrt{3}}-\frac{C_{31,20}^{31,1}}{1536
\sqrt{3}}+\frac{7 C_{31,22}^{31,0}}{3072 \sqrt{15}}+\frac{7 C_{31,22}^{31,1}}{1536 \sqrt{15}}-\frac{\sqrt{7} C_{31,22}^{33,0}}{1280}-\frac{\sqrt{7}
C_{31,22}^{33,1}}{640}-\frac{3 \sqrt{21} C_{33,00}^{33,0}}{2560}-\frac{3 \sqrt{21} C_{33,00}^{33,1}}{1280}-\frac{3 \sqrt{7} C_{33,20}^{33,0}}{2560}-\frac{3
\sqrt{7} C_{33,20}^{33,1}}{1280}-\frac{3 \sqrt{\frac{21}{10}} C_{33,22}^{33,0}}{1280}-\frac{3}{640} \sqrt{\frac{21}{10}} C_{33,22}^{33,1}\)

\noindent\(C_{31,0011,1}^{11,2011,1,\text{Loc}}= \frac{7 C_{00,20}^{60,1}}{512}-\frac{21 C_{00,22}^{62,1}}{1024 \sqrt{5}}+\frac{7 \sqrt{5}
C_{11,22}^{51,1}}{3072}+\frac{5 C_{20,20}^{40,1}}{1536}-\frac{7 C_{20,22}^{42,1}}{3072 \sqrt{5}}-\frac{7 C_{22,20}^{42,1}}{768 \sqrt{5}}-\frac{7
C_{22,22}^{40,1}}{3072 \sqrt{5}}+\frac{1}{768} \sqrt{\frac{7}{5}} C_{22,22}^{42,1}+\frac{9 C_{22,22}^{44,1}}{1280}+\frac{C_{31,20}^{31,1}}{1536}+\frac{7
C_{31,22}^{31,1}}{3072 \sqrt{5}}-\frac{\sqrt{21} C_{31,22}^{33,1}}{1280}+\frac{3 \sqrt{21} C_{33,20}^{33,1}}{1280}-\frac{9 \sqrt{\frac{7}{10}} C_{33,22}^{33,1}}{1280}\)

\noindent\(C_{31,0011,1}^{11,2011,1,\text{NonLoc}}= \frac{7 \sqrt{3} C_{00,00}^{60,0}}{1024}+\frac{7 C_{00,20}^{60,0}}{1024}+\frac{21
C_{00,22}^{62,0}}{1024 \sqrt{5}}+\frac{7 C_{11,00}^{51,0}}{1024 \sqrt{3}}+\frac{7 \sqrt{5} C_{11,22}^{51,0}}{3072}+\frac{5 C_{20,00}^{40,0}}{1024
\sqrt{3}}+\frac{5 C_{20,20}^{40,0}}{3072}+\frac{7 C_{20,22}^{42,0}}{3072 \sqrt{5}}-\frac{7 C_{22,00}^{42,0}}{512 \sqrt{15}}-\frac{7 C_{22,20}^{42,0}}{1536
\sqrt{5}}+\frac{7 C_{22,22}^{40,0}}{3072 \sqrt{5}}-\frac{1}{768} \sqrt{\frac{7}{5}} C_{22,22}^{42,0}-\frac{9 C_{22,22}^{44,0}}{1280}-\frac{C_{31,00}^{31,0}}{1024
\sqrt{3}}-\frac{C_{31,20}^{31,0}}{3072}+\frac{7 C_{31,22}^{31,0}}{3072 \sqrt{5}}-\frac{\sqrt{21} C_{31,22}^{33,0}}{1280}-\frac{9 \sqrt{7} C_{33,00}^{33,0}}{2560}-\frac{3
\sqrt{21} C_{33,20}^{33,0}}{2560}-\frac{9 \sqrt{\frac{7}{10}} C_{33,22}^{33,0}}{1280}\)

\noindent\(C_{31,0011,1}^{11,2211,0,\text{Loc}}= -\frac{7 C_{00,20}^{60,0}}{256 \sqrt{15}}-\frac{7 C_{00,20}^{60,1}}{512 \sqrt{15}}+\frac{7
\sqrt{3} C_{00,22}^{62,0}}{2560}+\frac{7 \sqrt{3} C_{00,22}^{62,1}}{5120}+\frac{7 C_{11,22}^{51,0}}{1920 \sqrt{3}}+\frac{7 C_{11,22}^{51,1}}{3840
\sqrt{3}}-\frac{27 \sqrt{\frac{7}{5}} C_{11,22}^{53,0}}{5120}-\frac{27 \sqrt{\frac{7}{5}} C_{11,22}^{53,1}}{10240}+\frac{7 C_{20,20}^{40,0}}{768
\sqrt{15}}+\frac{7 C_{20,20}^{40,1}}{1536 \sqrt{15}}+\frac{7 C_{20,22}^{42,0}}{1920 \sqrt{3}}+\frac{7 C_{20,22}^{42,1}}{3840 \sqrt{3}}-\frac{7 C_{22,20}^{42,0}}{3840
\sqrt{3}}-\frac{7 C_{22,20}^{42,1}}{7680 \sqrt{3}}+\frac{C_{22,22}^{40,0}}{7680 \sqrt{3}}+\frac{C_{22,22}^{40,1}}{15360 \sqrt{3}}-\frac{17 \sqrt{\frac{7}{3}}
C_{22,22}^{42,0}}{3840}-\frac{17 \sqrt{\frac{7}{3}} C_{22,22}^{42,1}}{7680}+\frac{3}{320} \sqrt{\frac{3}{5}} C_{22,22}^{44,0}+\frac{3}{640} \sqrt{\frac{3}{5}}
C_{22,22}^{44,1}-\frac{7 C_{31,20}^{31,0}}{768 \sqrt{15}}-\frac{7 C_{31,20}^{31,1}}{1536 \sqrt{15}}+\frac{C_{31,22}^{31,0}}{240 \sqrt{3}}+\frac{C_{31,22}^{31,1}}{480
\sqrt{3}}-\frac{13 \sqrt{\frac{7}{5}} C_{31,22}^{33,0}}{5120}-\frac{13 \sqrt{\frac{7}{5}} C_{31,22}^{33,1}}{10240}+\frac{3 \sqrt{\frac{7}{5}} C_{33,20}^{33,0}}{1280}+\frac{3
\sqrt{\frac{7}{5}} C_{33,20}^{33,1}}{2560}+\frac{3}{800} \sqrt{\frac{21}{2}} C_{33,22}^{33,0}+\frac{3 \sqrt{\frac{21}{2}} C_{33,22}^{33,1}}{1600}\)

\noindent\(C_{31,0011,1}^{11,2211,0,\text{NonLoc}}= -\frac{7 C_{00,00}^{60,0}}{1024 \sqrt{5}}-\frac{7 C_{00,00}^{60,1}}{512 \sqrt{5}}-\frac{7
C_{00,20}^{60,0}}{1024 \sqrt{15}}-\frac{7 C_{00,20}^{60,1}}{512 \sqrt{15}}-\frac{7 \sqrt{3} C_{00,22}^{62,0}}{5120}-\frac{7 \sqrt{3} C_{00,22}^{62,1}}{2560}-\frac{7
C_{11,00}^{51,0}}{3072 \sqrt{5}}-\frac{7 C_{11,00}^{51,1}}{1536 \sqrt{5}}+\frac{7 C_{11,22}^{51,0}}{3840 \sqrt{3}}+\frac{7 C_{11,22}^{51,1}}{1920
\sqrt{3}}-\frac{27 \sqrt{\frac{7}{5}} C_{11,22}^{53,0}}{10240}-\frac{27 \sqrt{\frac{7}{5}} C_{11,22}^{53,1}}{5120}+\frac{7 C_{20,00}^{40,0}}{3072
\sqrt{5}}+\frac{7 C_{20,00}^{40,1}}{1536 \sqrt{5}}+\frac{7 C_{20,20}^{40,0}}{3072 \sqrt{15}}+\frac{7 C_{20,20}^{40,1}}{1536 \sqrt{15}}-\frac{7 C_{20,22}^{42,0}}{3840
\sqrt{3}}-\frac{7 C_{20,22}^{42,1}}{1920 \sqrt{3}}-\frac{7 C_{22,00}^{42,0}}{15360}-\frac{7 C_{22,00}^{42,1}}{7680}-\frac{7 C_{22,20}^{42,0}}{15360
\sqrt{3}}-\frac{7 C_{22,20}^{42,1}}{7680 \sqrt{3}}-\frac{C_{22,22}^{40,0}}{15360 \sqrt{3}}-\frac{C_{22,22}^{40,1}}{7680 \sqrt{3}}+\frac{17 \sqrt{\frac{7}{3}}
C_{22,22}^{42,0}}{7680}+\frac{17 \sqrt{\frac{7}{3}} C_{22,22}^{42,1}}{3840}-\frac{3}{640} \sqrt{\frac{3}{5}} C_{22,22}^{44,0}-\frac{3}{320} \sqrt{\frac{3}{5}}
C_{22,22}^{44,1}+\frac{7 C_{31,00}^{31,0}}{3072 \sqrt{5}}+\frac{7 C_{31,00}^{31,1}}{1536 \sqrt{5}}+\frac{7 C_{31,20}^{31,0}}{3072 \sqrt{15}}+\frac{7
C_{31,20}^{31,1}}{1536 \sqrt{15}}+\frac{C_{31,22}^{31,0}}{480 \sqrt{3}}+\frac{C_{31,22}^{31,1}}{240 \sqrt{3}}-\frac{13 \sqrt{\frac{7}{5}} C_{31,22}^{33,0}}{10240}-\frac{13
\sqrt{\frac{7}{5}} C_{31,22}^{33,1}}{5120}-\frac{3 \sqrt{\frac{21}{5}} C_{33,00}^{33,0}}{5120}-\frac{3 \sqrt{\frac{21}{5}} C_{33,00}^{33,1}}{2560}-\frac{3
\sqrt{\frac{7}{5}} C_{33,20}^{33,0}}{5120}-\frac{3 \sqrt{\frac{7}{5}} C_{33,20}^{33,1}}{2560}+\frac{3 \sqrt{\frac{21}{2}} C_{33,22}^{33,0}}{1600}+\frac{3}{800}
\sqrt{\frac{21}{2}} C_{33,22}^{33,1}\)

\noindent\(C_{31,0011,1}^{11,2211,1,\text{Loc}}= -\frac{7 C_{00,20}^{60,1}}{512 \sqrt{5}}+\frac{21 C_{00,22}^{62,1}}{5120}+\frac{7
C_{11,22}^{51,1}}{3840}-\frac{27 \sqrt{\frac{21}{5}} C_{11,22}^{53,1}}{10240}+\frac{7 C_{20,20}^{40,1}}{1536 \sqrt{5}}+\frac{7 C_{20,22}^{42,1}}{3840}-\frac{7
C_{22,20}^{42,1}}{7680}+\frac{C_{22,22}^{40,1}}{15360}-\frac{17 \sqrt{7} C_{22,22}^{42,1}}{7680}+\frac{9 C_{22,22}^{44,1}}{640 \sqrt{5}}-\frac{7
C_{31,20}^{31,1}}{1536 \sqrt{5}}+\frac{C_{31,22}^{31,1}}{480}-\frac{13 \sqrt{\frac{21}{5}} C_{31,22}^{33,1}}{10240}+\frac{3 \sqrt{\frac{21}{5}} C_{33,20}^{33,1}}{2560}+\frac{9
\sqrt{\frac{7}{2}} C_{33,22}^{33,1}}{1600}\)

\noindent\(C_{31,0011,1}^{11,2211,1,\text{NonLoc}}= -\frac{7 \sqrt{\frac{3}{5}} C_{00,00}^{60,0}}{1024}-\frac{7 C_{00,20}^{60,0}}{1024
\sqrt{5}}-\frac{21 C_{00,22}^{62,0}}{5120}-\frac{7 C_{11,00}^{51,0}}{1024 \sqrt{15}}+\frac{7 C_{11,22}^{51,0}}{3840}-\frac{27 \sqrt{\frac{21}{5}}
C_{11,22}^{53,0}}{10240}+\frac{7 C_{20,00}^{40,0}}{1024 \sqrt{15}}+\frac{7 C_{20,20}^{40,0}}{3072 \sqrt{5}}-\frac{7 C_{20,22}^{42,0}}{3840}-\frac{7
C_{22,00}^{42,0}}{5120 \sqrt{3}}-\frac{7 C_{22,20}^{42,0}}{15360}-\frac{C_{22,22}^{40,0}}{15360}+\frac{17 \sqrt{7} C_{22,22}^{42,0}}{7680}-\frac{9
C_{22,22}^{44,0}}{640 \sqrt{5}}+\frac{7 C_{31,00}^{31,0}}{1024 \sqrt{15}}+\frac{7 C_{31,20}^{31,0}}{3072 \sqrt{5}}+\frac{C_{31,22}^{31,0}}{480}-\frac{13
\sqrt{\frac{21}{5}} C_{31,22}^{33,0}}{10240}-\frac{9 \sqrt{\frac{7}{5}} C_{33,00}^{33,0}}{5120}-\frac{3 \sqrt{\frac{21}{5}} C_{33,20}^{33,0}}{5120}+\frac{9
\sqrt{\frac{7}{2}} C_{33,22}^{33,0}}{1600}\)

\noindent\(C_{31,0011,1}^{11,2212,0,\text{Loc}}= \frac{7 C_{00,20}^{60,0}}{256 \sqrt{5}}+\frac{7 C_{00,20}^{60,1}}{512 \sqrt{5}}-\frac{21
C_{00,22}^{62,0}}{2560}-\frac{21 C_{00,22}^{62,1}}{5120}+\frac{7 C_{11,22}^{51,0}}{3840}+\frac{7 C_{11,22}^{51,1}}{7680}+\frac{9 \sqrt{\frac{21}{5}}
C_{11,22}^{53,0}}{5120}+\frac{9 \sqrt{\frac{21}{5}} C_{11,22}^{53,1}}{10240}-\frac{7 C_{20,20}^{40,0}}{768 \sqrt{5}}-\frac{7 C_{20,20}^{40,1}}{1536
\sqrt{5}}+\frac{7 C_{20,22}^{42,0}}{3840}+\frac{7 C_{20,22}^{42,1}}{7680}+\frac{7 C_{22,20}^{42,0}}{3840}+\frac{7 C_{22,20}^{42,1}}{7680}+\frac{7
C_{22,22}^{40,0}}{1536}+\frac{7 C_{22,22}^{40,1}}{3072}-\frac{7 \sqrt{7} C_{22,22}^{42,0}}{3840}-\frac{7 \sqrt{7} C_{22,22}^{42,1}}{7680}+\frac{7
C_{31,20}^{31,0}}{768 \sqrt{5}}+\frac{7 C_{31,20}^{31,1}}{1536 \sqrt{5}}-\frac{7 C_{31,22}^{31,0}}{3840}-\frac{7 C_{31,22}^{31,1}}{7680}-\frac{9
\sqrt{\frac{21}{5}} C_{31,22}^{33,0}}{5120}-\frac{9 \sqrt{\frac{21}{5}} C_{31,22}^{33,1}}{10240}-\frac{3 \sqrt{\frac{21}{5}} C_{33,20}^{33,0}}{1280}-\frac{3
\sqrt{\frac{21}{5}} C_{33,20}^{33,1}}{2560}+\frac{9 \sqrt{\frac{7}{2}} C_{33,22}^{33,0}}{1600}+\frac{9 \sqrt{\frac{7}{2}} C_{33,22}^{33,1}}{3200}\)

\noindent\(C_{31,0011,1}^{11,2212,0,\text{NonLoc}}= \frac{7 \sqrt{\frac{3}{5}} C_{00,00}^{60,0}}{1024}+\frac{7}{512} \sqrt{\frac{3}{5}}
C_{00,00}^{60,1}+\frac{7 C_{00,20}^{60,0}}{1024 \sqrt{5}}+\frac{7 C_{00,20}^{60,1}}{512 \sqrt{5}}+\frac{21 C_{00,22}^{62,0}}{5120}+\frac{21 C_{00,22}^{62,1}}{2560}+\frac{7
C_{11,00}^{51,0}}{1024 \sqrt{15}}+\frac{7 C_{11,00}^{51,1}}{512 \sqrt{15}}+\frac{7 C_{11,22}^{51,0}}{7680}+\frac{7 C_{11,22}^{51,1}}{3840}+\frac{9
\sqrt{\frac{21}{5}} C_{11,22}^{53,0}}{10240}+\frac{9 \sqrt{\frac{21}{5}} C_{11,22}^{53,1}}{5120}-\frac{7 C_{20,00}^{40,0}}{1024 \sqrt{15}}-\frac{7
C_{20,00}^{40,1}}{512 \sqrt{15}}-\frac{7 C_{20,20}^{40,0}}{3072 \sqrt{5}}-\frac{7 C_{20,20}^{40,1}}{1536 \sqrt{5}}-\frac{7 C_{20,22}^{42,0}}{7680}-\frac{7
C_{20,22}^{42,1}}{3840}+\frac{7 C_{22,00}^{42,0}}{5120 \sqrt{3}}+\frac{7 C_{22,00}^{42,1}}{2560 \sqrt{3}}+\frac{7 C_{22,20}^{42,0}}{15360}+\frac{7
C_{22,20}^{42,1}}{7680}-\frac{7 C_{22,22}^{40,0}}{3072}-\frac{7 C_{22,22}^{40,1}}{1536}+\frac{7 \sqrt{7} C_{22,22}^{42,0}}{7680}+\frac{7 \sqrt{7}
C_{22,22}^{42,1}}{3840}-\frac{7 C_{31,00}^{31,0}}{1024 \sqrt{15}}-\frac{7 C_{31,00}^{31,1}}{512 \sqrt{15}}-\frac{7 C_{31,20}^{31,0}}{3072 \sqrt{5}}-\frac{7
C_{31,20}^{31,1}}{1536 \sqrt{5}}-\frac{7 C_{31,22}^{31,0}}{7680}-\frac{7 C_{31,22}^{31,1}}{3840}-\frac{9 \sqrt{\frac{21}{5}} C_{31,22}^{33,0}}{10240}-\frac{9
\sqrt{\frac{21}{5}} C_{31,22}^{33,1}}{5120}+\frac{9 \sqrt{\frac{7}{5}} C_{33,00}^{33,0}}{5120}+\frac{9 \sqrt{\frac{7}{5}} C_{33,00}^{33,1}}{2560}+\frac{3
\sqrt{\frac{21}{5}} C_{33,20}^{33,0}}{5120}+\frac{3 \sqrt{\frac{21}{5}} C_{33,20}^{33,1}}{2560}+\frac{9 \sqrt{\frac{7}{2}} C_{33,22}^{33,0}}{3200}+\frac{9
\sqrt{\frac{7}{2}} C_{33,22}^{33,1}}{1600}\)

\noindent\(C_{31,0011,1}^{11,2212,1,\text{Loc}}= \frac{7}{512} \sqrt{\frac{3}{5}} C_{00,20}^{60,1}-\frac{21 \sqrt{3} C_{00,22}^{62,1}}{5120}+\frac{7
C_{11,22}^{51,1}}{2560 \sqrt{3}}+\frac{27 \sqrt{\frac{7}{5}} C_{11,22}^{53,1}}{10240}-\frac{7 C_{20,20}^{40,1}}{512 \sqrt{15}}+\frac{7 C_{20,22}^{42,1}}{2560
\sqrt{3}}+\frac{7 C_{22,20}^{42,1}}{2560 \sqrt{3}}+\frac{7 C_{22,22}^{40,1}}{1024 \sqrt{3}}-\frac{7 \sqrt{\frac{7}{3}} C_{22,22}^{42,1}}{2560}+\frac{7
C_{31,20}^{31,1}}{512 \sqrt{15}}-\frac{7 C_{31,22}^{31,1}}{2560 \sqrt{3}}-\frac{27 \sqrt{\frac{7}{5}} C_{31,22}^{33,1}}{10240}-\frac{9 \sqrt{\frac{7}{5}}
C_{33,20}^{33,1}}{2560}+\frac{9 \sqrt{\frac{21}{2}} C_{33,22}^{33,1}}{3200}\)

\noindent\(C_{31,0011,1}^{11,2212,1,\text{NonLoc}}= \frac{21 C_{00,00}^{60,0}}{1024 \sqrt{5}}+\frac{7 \sqrt{\frac{3}{5}} C_{00,20}^{60,0}}{1024}+\frac{21
\sqrt{3} C_{00,22}^{62,0}}{5120}+\frac{7 C_{11,00}^{51,0}}{1024 \sqrt{5}}+\frac{7 C_{11,22}^{51,0}}{2560 \sqrt{3}}+\frac{27 \sqrt{\frac{7}{5}} C_{11,22}^{53,0}}{10240}-\frac{7
C_{20,00}^{40,0}}{1024 \sqrt{5}}-\frac{7 C_{20,20}^{40,0}}{1024 \sqrt{15}}-\frac{7 C_{20,22}^{42,0}}{2560 \sqrt{3}}+\frac{7 C_{22,00}^{42,0}}{5120}+\frac{7
C_{22,20}^{42,0}}{5120 \sqrt{3}}-\frac{7 C_{22,22}^{40,0}}{1024 \sqrt{3}}+\frac{7 \sqrt{\frac{7}{3}} C_{22,22}^{42,0}}{2560}-\frac{7 C_{31,00}^{31,0}}{1024
\sqrt{5}}-\frac{7 C_{31,20}^{31,0}}{1024 \sqrt{15}}-\frac{7 C_{31,22}^{31,0}}{2560 \sqrt{3}}-\frac{27 \sqrt{\frac{7}{5}} C_{31,22}^{33,0}}{10240}+\frac{9
\sqrt{\frac{21}{5}} C_{33,00}^{33,0}}{5120}+\frac{9 \sqrt{\frac{7}{5}} C_{33,20}^{33,0}}{5120}+\frac{9 \sqrt{\frac{21}{2}} C_{33,22}^{33,0}}{3200}\)

\noindent\(C_{31,0011,1}^{20,1101,0,\text{Loc}}= -\frac{7 C_{11,11}^{51,0}}{1536}-\frac{7 C_{11,11}^{51,1}}{3072}+\frac{7 C_{22,11}^{42,0}}{256
\sqrt{15}}+\frac{7 C_{22,11}^{42,1}}{512 \sqrt{15}}-\frac{7 C_{31,11}^{31,0}}{1536}-\frac{7 C_{31,11}^{31,1}}{3072}-\frac{3}{640} \sqrt{\frac{7}{2}}
C_{33,11}^{33,0}-\frac{3 \sqrt{\frac{7}{2}} C_{33,11}^{33,1}}{1280}\)

\noindent\(C_{31,0011,1}^{20,1101,0,\text{NonLoc}}= -\frac{7 C_{11,11}^{51,0}}{3072}-\frac{7 C_{11,11}^{51,1}}{1536}-\frac{7 C_{22,11}^{42,0}}{512
\sqrt{15}}-\frac{7 C_{22,11}^{42,1}}{256 \sqrt{15}}-\frac{7 C_{31,11}^{31,0}}{3072}-\frac{7 C_{31,11}^{31,1}}{1536}-\frac{3 \sqrt{\frac{7}{2}} C_{33,11}^{33,0}}{1280}-\frac{3}{640}
\sqrt{\frac{7}{2}} C_{33,11}^{33,1}\)

\noindent\(C_{31,0011,1}^{20,1101,1,\text{Loc}}= -\frac{7 C_{11,11}^{51,1}}{1024 \sqrt{3}}+\frac{7 C_{22,11}^{42,1}}{512 \sqrt{5}}-\frac{7
C_{31,11}^{31,1}}{1024 \sqrt{3}}-\frac{3 \sqrt{\frac{21}{2}} C_{33,11}^{33,1}}{1280}\)

\noindent\(C_{31,0011,1}^{20,1101,1,\text{NonLoc}}= -\frac{7 C_{11,11}^{51,0}}{1024 \sqrt{3}}-\frac{7 C_{22,11}^{42,0}}{512 \sqrt{5}}-\frac{7
C_{31,11}^{31,0}}{1024 \sqrt{3}}-\frac{3 \sqrt{\frac{21}{2}} C_{33,11}^{33,0}}{1280}\)

\noindent\(C_{31,0011,1}^{22,1101,0,\text{Loc}}= \frac{7 C_{11,11}^{51,0}}{1536 \sqrt{5}}+\frac{7 C_{11,11}^{51,1}}{3072 \sqrt{5}}-\frac{7
C_{22,11}^{42,0}}{1280 \sqrt{3}}-\frac{7 C_{22,11}^{42,1}}{2560 \sqrt{3}}+\frac{7 C_{31,11}^{31,0}}{1536 \sqrt{5}}+\frac{7 C_{31,11}^{31,1}}{3072
\sqrt{5}}+\frac{3}{640} \sqrt{\frac{7}{10}} C_{33,11}^{33,0}+\frac{3 \sqrt{\frac{7}{10}} C_{33,11}^{33,1}}{1280}\)

\noindent\(C_{31,0011,1}^{22,1101,0,\text{NonLoc}}= \frac{7 C_{11,11}^{51,0}}{3072 \sqrt{5}}+\frac{7 C_{11,11}^{51,1}}{1536 \sqrt{5}}+\frac{7
C_{22,11}^{42,0}}{2560 \sqrt{3}}+\frac{7 C_{22,11}^{42,1}}{1280 \sqrt{3}}+\frac{7 C_{31,11}^{31,0}}{3072 \sqrt{5}}+\frac{7 C_{31,11}^{31,1}}{1536
\sqrt{5}}+\frac{3 \sqrt{\frac{7}{10}} C_{33,11}^{33,0}}{1280}+\frac{3}{640} \sqrt{\frac{7}{10}} C_{33,11}^{33,1}\)

\noindent\(C_{31,0011,1}^{22,1101,1,\text{Loc}}= \frac{7 C_{11,11}^{51,1}}{1024 \sqrt{15}}-\frac{7 C_{22,11}^{42,1}}{2560}+\frac{7
C_{31,11}^{31,1}}{1024 \sqrt{15}}+\frac{3 \sqrt{\frac{21}{10}} C_{33,11}^{33,1}}{1280}\)

\noindent\(C_{31,0011,1}^{22,1101,1,\text{NonLoc}}= \frac{7 C_{11,11}^{51,0}}{1024 \sqrt{15}}+\frac{7 C_{22,11}^{42,0}}{2560}+\frac{7
C_{31,11}^{31,0}}{1024 \sqrt{15}}+\frac{3 \sqrt{\frac{21}{10}} C_{33,11}^{33,0}}{1280}\)

\noindent\(C_{31,0011,1}^{31,0011,0,\text{Loc}}= -\frac{7 \sqrt{3} C_{00,20}^{60,0}}{10240}-\frac{7 \sqrt{3} C_{00,20}^{60,1}}{20480}+\frac{21
\sqrt{\frac{3}{5}} C_{00,22}^{62,0}}{20480}+\frac{21 \sqrt{\frac{3}{5}} C_{00,22}^{62,1}}{40960}-\frac{21 \sqrt{\frac{3}{5}} C_{11,22}^{51,0}}{20480}-\frac{21
\sqrt{\frac{3}{5}} C_{11,22}^{51,1}}{40960}-\frac{27 \sqrt{7} C_{11,22}^{53,0}}{102400}-\frac{27 \sqrt{7} C_{11,22}^{53,1}}{204800}-\frac{7 \sqrt{3}
C_{20,20}^{40,0}}{10240}-\frac{7 \sqrt{3} C_{20,20}^{40,1}}{20480}+\frac{21 \sqrt{\frac{3}{5}} C_{20,22}^{42,0}}{20480}+\frac{21 \sqrt{\frac{3}{5}}
C_{20,22}^{42,1}}{40960}-\frac{7 \sqrt{\frac{3}{5}} C_{22,20}^{42,0}}{5120}-\frac{7 \sqrt{\frac{3}{5}} C_{22,20}^{42,1}}{10240}+\frac{21 \sqrt{\frac{3}{5}}
C_{22,22}^{40,0}}{20480}+\frac{21 \sqrt{\frac{3}{5}} C_{22,22}^{40,1}}{40960}+\frac{3 \sqrt{\frac{21}{5}} C_{22,22}^{42,0}}{10240}+\frac{3 \sqrt{\frac{21}{5}}
C_{22,22}^{42,1}}{20480}+\frac{9 \sqrt{3} C_{22,22}^{44,0}}{25600}+\frac{9 \sqrt{3} C_{22,22}^{44,1}}{51200}+\frac{7 \sqrt{3} C_{31,20}^{31,0}}{10240}+\frac{7
\sqrt{3} C_{31,20}^{31,1}}{20480}-\frac{21 \sqrt{\frac{3}{5}} C_{31,22}^{31,0}}{20480}-\frac{21 \sqrt{\frac{3}{5}} C_{31,22}^{31,1}}{40960}-\frac{27
\sqrt{7} C_{31,22}^{33,0}}{102400}-\frac{27 \sqrt{7} C_{31,22}^{33,1}}{204800}+\frac{9 \sqrt{7} C_{33,20}^{33,0}}{25600}+\frac{9 \sqrt{7} C_{33,20}^{33,1}}{51200}-\frac{9
\sqrt{\frac{21}{10}} C_{33,22}^{33,0}}{25600}-\frac{9 \sqrt{\frac{21}{10}} C_{33,22}^{33,1}}{51200}\)

\noindent\(C_{31,0011,1}^{31,0011,0,\text{NonLoc}}= -\frac{21 C_{00,00}^{60,0}}{40960}-\frac{21 C_{00,00}^{60,1}}{20480}-\frac{7 \sqrt{3}
C_{00,20}^{60,0}}{40960}-\frac{7 \sqrt{3} C_{00,20}^{60,1}}{20480}-\frac{21 \sqrt{\frac{3}{5}} C_{00,22}^{62,0}}{40960}-\frac{21 \sqrt{\frac{3}{5}}
C_{00,22}^{62,1}}{20480}-\frac{21 C_{11,00}^{51,0}}{40960}-\frac{21 C_{11,00}^{51,1}}{20480}-\frac{21 \sqrt{\frac{3}{5}} C_{11,22}^{51,0}}{40960}-\frac{21
\sqrt{\frac{3}{5}} C_{11,22}^{51,1}}{20480}-\frac{27 \sqrt{7} C_{11,22}^{53,0}}{204800}-\frac{27 \sqrt{7} C_{11,22}^{53,1}}{102400}-\frac{21 C_{20,00}^{40,0}}{40960}-\frac{21
C_{20,00}^{40,1}}{20480}-\frac{7 \sqrt{3} C_{20,20}^{40,0}}{40960}-\frac{7 \sqrt{3} C_{20,20}^{40,1}}{20480}-\frac{21 \sqrt{\frac{3}{5}} C_{20,22}^{42,0}}{40960}-\frac{21
\sqrt{\frac{3}{5}} C_{20,22}^{42,1}}{20480}-\frac{21 C_{22,00}^{42,0}}{20480 \sqrt{5}}-\frac{21 C_{22,00}^{42,1}}{10240 \sqrt{5}}-\frac{7 \sqrt{\frac{3}{5}}
C_{22,20}^{42,0}}{20480}-\frac{7 \sqrt{\frac{3}{5}} C_{22,20}^{42,1}}{10240}-\frac{21 \sqrt{\frac{3}{5}} C_{22,22}^{40,0}}{40960}-\frac{21 \sqrt{\frac{3}{5}}
C_{22,22}^{40,1}}{20480}-\frac{3 \sqrt{\frac{21}{5}} C_{22,22}^{42,0}}{20480}-\frac{3 \sqrt{\frac{21}{5}} C_{22,22}^{42,1}}{10240}-\frac{9 \sqrt{3}
C_{22,22}^{44,0}}{51200}-\frac{9 \sqrt{3} C_{22,22}^{44,1}}{25600}-\frac{21 C_{31,00}^{31,0}}{40960}-\frac{21 C_{31,00}^{31,1}}{20480}-\frac{7 \sqrt{3}
C_{31,20}^{31,0}}{40960}-\frac{7 \sqrt{3} C_{31,20}^{31,1}}{20480}-\frac{21 \sqrt{\frac{3}{5}} C_{31,22}^{31,0}}{40960}-\frac{21 \sqrt{\frac{3}{5}}
C_{31,22}^{31,1}}{20480}-\frac{27 \sqrt{7} C_{31,22}^{33,0}}{204800}-\frac{27 \sqrt{7} C_{31,22}^{33,1}}{102400}-\frac{9 \sqrt{21} C_{33,00}^{33,0}}{102400}-\frac{9
\sqrt{21} C_{33,00}^{33,1}}{51200}-\frac{9 \sqrt{7} C_{33,20}^{33,0}}{102400}-\frac{9 \sqrt{7} C_{33,20}^{33,1}}{51200}-\frac{9 \sqrt{\frac{21}{10}}
C_{33,22}^{33,0}}{51200}-\frac{9 \sqrt{\frac{21}{10}} C_{33,22}^{33,1}}{25600}\)

\noindent\(C_{31,0011,1}^{31,0011,1,\text{Loc}}= -\frac{21 C_{00,20}^{60,1}}{20480}+\frac{63 C_{00,22}^{62,1}}{40960 \sqrt{5}}-\frac{63
C_{11,22}^{51,1}}{40960 \sqrt{5}}-\frac{27 \sqrt{21} C_{11,22}^{53,1}}{204800}-\frac{21 C_{20,20}^{40,1}}{20480}+\frac{63 C_{20,22}^{42,1}}{40960
\sqrt{5}}-\frac{21 C_{22,20}^{42,1}}{10240 \sqrt{5}}+\frac{63 C_{22,22}^{40,1}}{40960 \sqrt{5}}+\frac{9 \sqrt{\frac{7}{5}} C_{22,22}^{42,1}}{20480}+\frac{27
C_{22,22}^{44,1}}{51200}+\frac{21 C_{31,20}^{31,1}}{20480}-\frac{63 C_{31,22}^{31,1}}{40960 \sqrt{5}}-\frac{27 \sqrt{21} C_{31,22}^{33,1}}{204800}+\frac{9
\sqrt{21} C_{33,20}^{33,1}}{51200}-\frac{27 \sqrt{\frac{7}{10}} C_{33,22}^{33,1}}{51200}\)

\noindent\(C_{31,0011,1}^{31,0011,1,\text{NonLoc}}= -\frac{21 \sqrt{3} C_{00,00}^{60,0}}{40960}-\frac{21 C_{00,20}^{60,0}}{40960}-\frac{63
C_{00,22}^{62,0}}{40960 \sqrt{5}}-\frac{21 \sqrt{3} C_{11,00}^{51,0}}{40960}-\frac{63 C_{11,22}^{51,0}}{40960 \sqrt{5}}-\frac{27 \sqrt{21} C_{11,22}^{53,0}}{204800}-\frac{21
\sqrt{3} C_{20,00}^{40,0}}{40960}-\frac{21 C_{20,20}^{40,0}}{40960}-\frac{63 C_{20,22}^{42,0}}{40960 \sqrt{5}}-\frac{21 \sqrt{\frac{3}{5}} C_{22,00}^{42,0}}{20480}-\frac{21
C_{22,20}^{42,0}}{20480 \sqrt{5}}-\frac{63 C_{22,22}^{40,0}}{40960 \sqrt{5}}-\frac{9 \sqrt{\frac{7}{5}} C_{22,22}^{42,0}}{20480}-\frac{27 C_{22,22}^{44,0}}{51200}-\frac{21
\sqrt{3} C_{31,00}^{31,0}}{40960}-\frac{21 C_{31,20}^{31,0}}{40960}-\frac{63 C_{31,22}^{31,0}}{40960 \sqrt{5}}-\frac{27 \sqrt{21} C_{31,22}^{33,0}}{204800}-\frac{27
\sqrt{7} C_{33,00}^{33,0}}{102400}-\frac{9 \sqrt{21} C_{33,20}^{33,0}}{102400}-\frac{27 \sqrt{\frac{7}{10}} C_{33,22}^{33,0}}{51200}\)

\noindent\(C_{31,0011,2}^{11,2011,0,\text{Loc}}= \frac{7 \sqrt{5} C_{00,20}^{60,0}}{768}+\frac{7 \sqrt{5} C_{00,20}^{60,1}}{1536}+\frac{7
C_{00,22}^{62,0}}{2560}+\frac{7 C_{00,22}^{62,1}}{5120}-\frac{7 C_{11,22}^{51,0}}{4608}-\frac{7 C_{11,22}^{51,1}}{9216}+\frac{5 \sqrt{5} C_{20,20}^{40,0}}{2304}+\frac{5
\sqrt{5} C_{20,20}^{40,1}}{4608}+\frac{7 C_{20,22}^{42,0}}{23040}+\frac{7 C_{20,22}^{42,1}}{46080}-\frac{7 C_{22,20}^{42,0}}{1152}-\frac{7 C_{22,20}^{42,1}}{2304}+\frac{7
C_{22,22}^{40,0}}{23040}+\frac{7 C_{22,22}^{40,1}}{46080}-\frac{\sqrt{7} C_{22,22}^{42,0}}{5760}-\frac{\sqrt{7} C_{22,22}^{42,1}}{11520}-\frac{3
C_{22,22}^{44,0}}{640 \sqrt{5}}-\frac{3 C_{22,22}^{44,1}}{1280 \sqrt{5}}+\frac{\sqrt{5} C_{31,20}^{31,0}}{2304}+\frac{\sqrt{5} C_{31,20}^{31,1}}{4608}-\frac{7
C_{31,22}^{31,0}}{23040}-\frac{7 C_{31,22}^{31,1}}{46080}+\frac{1}{640} \sqrt{\frac{7}{15}} C_{31,22}^{33,0}+\frac{\sqrt{\frac{7}{15}} C_{31,22}^{33,1}}{1280}+\frac{1}{128}
\sqrt{\frac{21}{5}} C_{33,20}^{33,0}+\frac{1}{256} \sqrt{\frac{21}{5}} C_{33,20}^{33,1}+\frac{3 \sqrt{\frac{7}{2}} C_{33,22}^{33,0}}{3200}+\frac{3
\sqrt{\frac{7}{2}} C_{33,22}^{33,1}}{6400}\)

\noindent\(C_{31,0011,2}^{11,2011,0,\text{NonLoc}}= \frac{7 \sqrt{\frac{5}{3}} C_{00,00}^{60,0}}{1024}+\frac{7}{512} \sqrt{\frac{5}{3}}
C_{00,00}^{60,1}+\frac{7 \sqrt{5} C_{00,20}^{60,0}}{3072}+\frac{7 \sqrt{5} C_{00,20}^{60,1}}{1536}-\frac{7 C_{00,22}^{62,0}}{5120}-\frac{7 C_{00,22}^{62,1}}{2560}+\frac{7
\sqrt{\frac{5}{3}} C_{11,00}^{51,0}}{3072}+\frac{7 \sqrt{\frac{5}{3}} C_{11,00}^{51,1}}{1536}-\frac{7 C_{11,22}^{51,0}}{9216}-\frac{7 C_{11,22}^{51,1}}{4608}+\frac{5
\sqrt{\frac{5}{3}} C_{20,00}^{40,0}}{3072}+\frac{5 \sqrt{\frac{5}{3}} C_{20,00}^{40,1}}{1536}+\frac{5 \sqrt{5} C_{20,20}^{40,0}}{9216}+\frac{5 \sqrt{5}
C_{20,20}^{40,1}}{4608}-\frac{7 C_{20,22}^{42,0}}{46080}-\frac{7 C_{20,22}^{42,1}}{23040}-\frac{7 C_{22,00}^{42,0}}{1536 \sqrt{3}}-\frac{7 C_{22,00}^{42,1}}{768
\sqrt{3}}-\frac{7 C_{22,20}^{42,0}}{4608}-\frac{7 C_{22,20}^{42,1}}{2304}-\frac{7 C_{22,22}^{40,0}}{46080}-\frac{7 C_{22,22}^{40,1}}{23040}+\frac{\sqrt{7}
C_{22,22}^{42,0}}{11520}+\frac{\sqrt{7} C_{22,22}^{42,1}}{5760}+\frac{3 C_{22,22}^{44,0}}{1280 \sqrt{5}}+\frac{3 C_{22,22}^{44,1}}{640 \sqrt{5}}-\frac{\sqrt{\frac{5}{3}}
C_{31,00}^{31,0}}{3072}-\frac{\sqrt{\frac{5}{3}} C_{31,00}^{31,1}}{1536}-\frac{\sqrt{5} C_{31,20}^{31,0}}{9216}-\frac{\sqrt{5} C_{31,20}^{31,1}}{4608}-\frac{7
C_{31,22}^{31,0}}{46080}-\frac{7 C_{31,22}^{31,1}}{23040}+\frac{\sqrt{\frac{7}{15}} C_{31,22}^{33,0}}{1280}+\frac{1}{640} \sqrt{\frac{7}{15}} C_{31,22}^{33,1}-\frac{3}{512}
\sqrt{\frac{7}{5}} C_{33,00}^{33,0}-\frac{3}{256} \sqrt{\frac{7}{5}} C_{33,00}^{33,1}-\frac{1}{512} \sqrt{\frac{21}{5}} C_{33,20}^{33,0}-\frac{1}{256}
\sqrt{\frac{21}{5}} C_{33,20}^{33,1}+\frac{3 \sqrt{\frac{7}{2}} C_{33,22}^{33,0}}{6400}+\frac{3 \sqrt{\frac{7}{2}} C_{33,22}^{33,1}}{3200}\)

\noindent\(C_{31,0011,2}^{11,2011,1,\text{Loc}}= \frac{7}{512} \sqrt{\frac{5}{3}} C_{00,20}^{60,1}+\frac{7 \sqrt{3} C_{00,22}^{62,1}}{5120}-\frac{7
C_{11,22}^{51,1}}{3072 \sqrt{3}}+\frac{5 \sqrt{\frac{5}{3}} C_{20,20}^{40,1}}{1536}+\frac{7 C_{20,22}^{42,1}}{15360 \sqrt{3}}-\frac{7 C_{22,20}^{42,1}}{768
\sqrt{3}}+\frac{7 C_{22,22}^{40,1}}{15360 \sqrt{3}}-\frac{\sqrt{\frac{7}{3}} C_{22,22}^{42,1}}{3840}-\frac{3 \sqrt{\frac{3}{5}} C_{22,22}^{44,1}}{1280}+\frac{\sqrt{\frac{5}{3}}
C_{31,20}^{31,1}}{1536}-\frac{7 C_{31,22}^{31,1}}{15360 \sqrt{3}}+\frac{\sqrt{\frac{7}{5}} C_{31,22}^{33,1}}{1280}+\frac{3}{256} \sqrt{\frac{7}{5}}
C_{33,20}^{33,1}+\frac{3 \sqrt{\frac{21}{2}} C_{33,22}^{33,1}}{6400}\)

\noindent\(C_{31,0011,2}^{11,2011,1,\text{NonLoc}}= \frac{7 \sqrt{5} C_{00,00}^{60,0}}{1024}+\frac{7 \sqrt{\frac{5}{3}} C_{00,20}^{60,0}}{1024}-\frac{7
\sqrt{3} C_{00,22}^{62,0}}{5120}+\frac{7 \sqrt{5} C_{11,00}^{51,0}}{3072}-\frac{7 C_{11,22}^{51,0}}{3072 \sqrt{3}}+\frac{5 \sqrt{5} C_{20,00}^{40,0}}{3072}+\frac{5
\sqrt{\frac{5}{3}} C_{20,20}^{40,0}}{3072}-\frac{7 C_{20,22}^{42,0}}{15360 \sqrt{3}}-\frac{7 C_{22,00}^{42,0}}{1536}-\frac{7 C_{22,20}^{42,0}}{1536
\sqrt{3}}-\frac{7 C_{22,22}^{40,0}}{15360 \sqrt{3}}+\frac{\sqrt{\frac{7}{3}} C_{22,22}^{42,0}}{3840}+\frac{3 \sqrt{\frac{3}{5}} C_{22,22}^{44,0}}{1280}-\frac{\sqrt{5}
C_{31,00}^{31,0}}{3072}-\frac{\sqrt{\frac{5}{3}} C_{31,20}^{31,0}}{3072}-\frac{7 C_{31,22}^{31,0}}{15360 \sqrt{3}}+\frac{\sqrt{\frac{7}{5}} C_{31,22}^{33,0}}{1280}-\frac{3}{512}
\sqrt{\frac{21}{5}} C_{33,00}^{33,0}-\frac{3}{512} \sqrt{\frac{7}{5}} C_{33,20}^{33,0}+\frac{3 \sqrt{\frac{21}{2}} C_{33,22}^{33,0}}{6400}\)

\noindent\(C_{31,0011,2}^{11,2211,0,\text{Loc}}= \frac{7 C_{00,20}^{60,0}}{3840}+\frac{7 C_{00,20}^{60,1}}{7680}+\frac{77 C_{00,22}^{62,0}}{2560
\sqrt{5}}+\frac{77 C_{00,22}^{62,1}}{5120 \sqrt{5}}+\frac{7 C_{11,22}^{51,0}}{2880 \sqrt{5}}+\frac{7 C_{11,22}^{51,1}}{5760 \sqrt{5}}-\frac{63 \sqrt{21}
C_{11,22}^{53,0}}{25600}-\frac{63 \sqrt{21} C_{11,22}^{53,1}}{51200}-\frac{7 C_{20,20}^{40,0}}{11520}-\frac{7 C_{20,20}^{40,1}}{23040}+\frac{7 C_{20,22}^{42,0}}{2880
\sqrt{5}}+\frac{7 C_{20,22}^{42,1}}{5760 \sqrt{5}}+\frac{7 C_{22,20}^{42,0}}{11520 \sqrt{5}}+\frac{7 C_{22,20}^{42,1}}{23040 \sqrt{5}}-\frac{41 C_{22,22}^{40,0}}{4608
\sqrt{5}}-\frac{41 C_{22,22}^{40,1}}{9216 \sqrt{5}}-\frac{43 \sqrt{\frac{7}{5}} C_{22,22}^{42,0}}{11520}-\frac{43 \sqrt{\frac{7}{5}} C_{22,22}^{42,1}}{23040}+\frac{3
C_{22,22}^{44,0}}{320}+\frac{3 C_{22,22}^{44,1}}{640}+\frac{7 C_{31,20}^{31,0}}{11520}+\frac{7 C_{31,20}^{31,1}}{23040}+\frac{61 C_{31,22}^{31,0}}{5760
\sqrt{5}}+\frac{61 C_{31,22}^{31,1}}{11520 \sqrt{5}}-\frac{11 \sqrt{\frac{7}{3}} C_{31,22}^{33,0}}{25600}-\frac{11 \sqrt{\frac{7}{3}} C_{31,22}^{33,1}}{51200}-\frac{\sqrt{21}
C_{33,20}^{33,0}}{6400}-\frac{\sqrt{21} C_{33,20}^{33,1}}{12800}+\frac{3}{400} \sqrt{\frac{7}{10}} C_{33,22}^{33,0}+\frac{3}{800} \sqrt{\frac{7}{10}}
C_{33,22}^{33,1}\)

\noindent\(C_{31,0011,2}^{11,2211,0,\text{NonLoc}}= \frac{7 C_{00,00}^{60,0}}{5120 \sqrt{3}}+\frac{7 C_{00,00}^{60,1}}{2560 \sqrt{3}}+\frac{7
C_{00,20}^{60,0}}{15360}+\frac{7 C_{00,20}^{60,1}}{7680}-\frac{77 C_{00,22}^{62,0}}{5120 \sqrt{5}}-\frac{77 C_{00,22}^{62,1}}{2560 \sqrt{5}}+\frac{7
C_{11,00}^{51,0}}{15360 \sqrt{3}}+\frac{7 C_{11,00}^{51,1}}{7680 \sqrt{3}}+\frac{7 C_{11,22}^{51,0}}{5760 \sqrt{5}}+\frac{7 C_{11,22}^{51,1}}{2880
\sqrt{5}}-\frac{63 \sqrt{21} C_{11,22}^{53,0}}{51200}-\frac{63 \sqrt{21} C_{11,22}^{53,1}}{25600}-\frac{7 C_{20,00}^{40,0}}{15360 \sqrt{3}}-\frac{7
C_{20,00}^{40,1}}{7680 \sqrt{3}}-\frac{7 C_{20,20}^{40,0}}{46080}-\frac{7 C_{20,20}^{40,1}}{23040}-\frac{7 C_{20,22}^{42,0}}{5760 \sqrt{5}}-\frac{7
C_{20,22}^{42,1}}{2880 \sqrt{5}}+\frac{7 C_{22,00}^{42,0}}{15360 \sqrt{15}}+\frac{7 C_{22,00}^{42,1}}{7680 \sqrt{15}}+\frac{7 C_{22,20}^{42,0}}{46080
\sqrt{5}}+\frac{7 C_{22,20}^{42,1}}{23040 \sqrt{5}}+\frac{41 C_{22,22}^{40,0}}{9216 \sqrt{5}}+\frac{41 C_{22,22}^{40,1}}{4608 \sqrt{5}}+\frac{43
\sqrt{\frac{7}{5}} C_{22,22}^{42,0}}{23040}+\frac{43 \sqrt{\frac{7}{5}} C_{22,22}^{42,1}}{11520}-\frac{3 C_{22,22}^{44,0}}{640}-\frac{3 C_{22,22}^{44,1}}{320}-\frac{7
C_{31,00}^{31,0}}{15360 \sqrt{3}}-\frac{7 C_{31,00}^{31,1}}{7680 \sqrt{3}}-\frac{7 C_{31,20}^{31,0}}{46080}-\frac{7 C_{31,20}^{31,1}}{23040}+\frac{61
C_{31,22}^{31,0}}{11520 \sqrt{5}}+\frac{61 C_{31,22}^{31,1}}{5760 \sqrt{5}}-\frac{11 \sqrt{\frac{7}{3}} C_{31,22}^{33,0}}{51200}-\frac{11 \sqrt{\frac{7}{3}}
C_{31,22}^{33,1}}{25600}+\frac{3 \sqrt{7} C_{33,00}^{33,0}}{25600}+\frac{3 \sqrt{7} C_{33,00}^{33,1}}{12800}+\frac{\sqrt{21} C_{33,20}^{33,0}}{25600}+\frac{\sqrt{21}
C_{33,20}^{33,1}}{12800}+\frac{3}{800} \sqrt{\frac{7}{10}} C_{33,22}^{33,0}+\frac{3}{400} \sqrt{\frac{7}{10}} C_{33,22}^{33,1}\)

\noindent\(C_{31,0011,2}^{11,2211,1,\text{Loc}}= \frac{7 C_{00,20}^{60,1}}{2560 \sqrt{3}}+\frac{77 \sqrt{\frac{3}{5}} C_{00,22}^{62,1}}{5120}+\frac{7
C_{11,22}^{51,1}}{1920 \sqrt{15}}-\frac{189 \sqrt{7} C_{11,22}^{53,1}}{51200}-\frac{7 C_{20,20}^{40,1}}{7680 \sqrt{3}}+\frac{7 C_{20,22}^{42,1}}{1920
\sqrt{15}}+\frac{7 C_{22,20}^{42,1}}{7680 \sqrt{15}}-\frac{41 C_{22,22}^{40,1}}{3072 \sqrt{15}}-\frac{43 \sqrt{\frac{7}{15}} C_{22,22}^{42,1}}{7680}+\frac{3
\sqrt{3} C_{22,22}^{44,1}}{640}+\frac{7 C_{31,20}^{31,1}}{7680 \sqrt{3}}+\frac{61 C_{31,22}^{31,1}}{3840 \sqrt{15}}-\frac{11 \sqrt{7} C_{31,22}^{33,1}}{51200}-\frac{3
\sqrt{7} C_{33,20}^{33,1}}{12800}+\frac{3}{800} \sqrt{\frac{21}{10}} C_{33,22}^{33,1}\)

\noindent\(C_{31,0011,2}^{11,2211,1,\text{NonLoc}}= \frac{7 C_{00,00}^{60,0}}{5120}+\frac{7 C_{00,20}^{60,0}}{5120 \sqrt{3}}-\frac{77
\sqrt{\frac{3}{5}} C_{00,22}^{62,0}}{5120}+\frac{7 C_{11,00}^{51,0}}{15360}+\frac{7 C_{11,22}^{51,0}}{1920 \sqrt{15}}-\frac{189 \sqrt{7} C_{11,22}^{53,0}}{51200}-\frac{7
C_{20,00}^{40,0}}{15360}-\frac{7 C_{20,20}^{40,0}}{15360 \sqrt{3}}-\frac{7 C_{20,22}^{42,0}}{1920 \sqrt{15}}+\frac{7 C_{22,00}^{42,0}}{15360 \sqrt{5}}+\frac{7
C_{22,20}^{42,0}}{15360 \sqrt{15}}+\frac{41 C_{22,22}^{40,0}}{3072 \sqrt{15}}+\frac{43 \sqrt{\frac{7}{15}} C_{22,22}^{42,0}}{7680}-\frac{3 \sqrt{3}
C_{22,22}^{44,0}}{640}-\frac{7 C_{31,00}^{31,0}}{15360}-\frac{7 C_{31,20}^{31,0}}{15360 \sqrt{3}}+\frac{61 C_{31,22}^{31,0}}{3840 \sqrt{15}}-\frac{11
\sqrt{7} C_{31,22}^{33,0}}{51200}+\frac{3 \sqrt{21} C_{33,00}^{33,0}}{25600}+\frac{3 \sqrt{7} C_{33,20}^{33,0}}{25600}+\frac{3}{800} \sqrt{\frac{21}{10}}
C_{33,22}^{33,0}\)

\noindent\(C_{31,0011,2}^{11,2212,0,\text{Loc}}= -\frac{7 C_{00,20}^{60,0}}{256 \sqrt{15}}-\frac{7 C_{00,20}^{60,1}}{512 \sqrt{15}}+\frac{7
\sqrt{3} C_{00,22}^{62,0}}{2560}+\frac{7 \sqrt{3} C_{00,22}^{62,1}}{5120}-\frac{7 C_{11,22}^{51,0}}{3840 \sqrt{3}}-\frac{7 C_{11,22}^{51,1}}{7680
\sqrt{3}}-\frac{9 \sqrt{\frac{7}{5}} C_{11,22}^{53,0}}{5120}-\frac{9 \sqrt{\frac{7}{5}} C_{11,22}^{53,1}}{10240}+\frac{7 C_{20,20}^{40,0}}{768 \sqrt{15}}+\frac{7
C_{20,20}^{40,1}}{1536 \sqrt{15}}-\frac{7 C_{20,22}^{42,0}}{3840 \sqrt{3}}-\frac{7 C_{20,22}^{42,1}}{7680 \sqrt{3}}-\frac{7 C_{22,20}^{42,0}}{3840
\sqrt{3}}-\frac{7 C_{22,20}^{42,1}}{7680 \sqrt{3}}-\frac{7 C_{22,22}^{40,0}}{1536 \sqrt{3}}-\frac{7 C_{22,22}^{40,1}}{3072 \sqrt{3}}+\frac{7 \sqrt{\frac{7}{3}}
C_{22,22}^{42,0}}{3840}+\frac{7 \sqrt{\frac{7}{3}} C_{22,22}^{42,1}}{7680}-\frac{7 C_{31,20}^{31,0}}{768 \sqrt{15}}-\frac{7 C_{31,20}^{31,1}}{1536
\sqrt{15}}+\frac{7 C_{31,22}^{31,0}}{3840 \sqrt{3}}+\frac{7 C_{31,22}^{31,1}}{7680 \sqrt{3}}+\frac{9 \sqrt{\frac{7}{5}} C_{31,22}^{33,0}}{5120}+\frac{9
\sqrt{\frac{7}{5}} C_{31,22}^{33,1}}{10240}+\frac{3 \sqrt{\frac{7}{5}} C_{33,20}^{33,0}}{1280}+\frac{3 \sqrt{\frac{7}{5}} C_{33,20}^{33,1}}{2560}-\frac{3
\sqrt{\frac{21}{2}} C_{33,22}^{33,0}}{1600}-\frac{3 \sqrt{\frac{21}{2}} C_{33,22}^{33,1}}{3200}\)

\noindent\(C_{31,0011,2}^{11,2212,0,\text{NonLoc}}= -\frac{7 C_{00,00}^{60,0}}{1024 \sqrt{5}}-\frac{7 C_{00,00}^{60,1}}{512 \sqrt{5}}-\frac{7
C_{00,20}^{60,0}}{1024 \sqrt{15}}-\frac{7 C_{00,20}^{60,1}}{512 \sqrt{15}}-\frac{7 \sqrt{3} C_{00,22}^{62,0}}{5120}-\frac{7 \sqrt{3} C_{00,22}^{62,1}}{2560}-\frac{7
C_{11,00}^{51,0}}{3072 \sqrt{5}}-\frac{7 C_{11,00}^{51,1}}{1536 \sqrt{5}}-\frac{7 C_{11,22}^{51,0}}{7680 \sqrt{3}}-\frac{7 C_{11,22}^{51,1}}{3840
\sqrt{3}}-\frac{9 \sqrt{\frac{7}{5}} C_{11,22}^{53,0}}{10240}-\frac{9 \sqrt{\frac{7}{5}} C_{11,22}^{53,1}}{5120}+\frac{7 C_{20,00}^{40,0}}{3072 \sqrt{5}}+\frac{7
C_{20,00}^{40,1}}{1536 \sqrt{5}}+\frac{7 C_{20,20}^{40,0}}{3072 \sqrt{15}}+\frac{7 C_{20,20}^{40,1}}{1536 \sqrt{15}}+\frac{7 C_{20,22}^{42,0}}{7680
\sqrt{3}}+\frac{7 C_{20,22}^{42,1}}{3840 \sqrt{3}}-\frac{7 C_{22,00}^{42,0}}{15360}-\frac{7 C_{22,00}^{42,1}}{7680}-\frac{7 C_{22,20}^{42,0}}{15360
\sqrt{3}}-\frac{7 C_{22,20}^{42,1}}{7680 \sqrt{3}}+\frac{7 C_{22,22}^{40,0}}{3072 \sqrt{3}}+\frac{7 C_{22,22}^{40,1}}{1536 \sqrt{3}}-\frac{7 \sqrt{\frac{7}{3}}
C_{22,22}^{42,0}}{7680}-\frac{7 \sqrt{\frac{7}{3}} C_{22,22}^{42,1}}{3840}+\frac{7 C_{31,00}^{31,0}}{3072 \sqrt{5}}+\frac{7 C_{31,00}^{31,1}}{1536
\sqrt{5}}+\frac{7 C_{31,20}^{31,0}}{3072 \sqrt{15}}+\frac{7 C_{31,20}^{31,1}}{1536 \sqrt{15}}+\frac{7 C_{31,22}^{31,0}}{7680 \sqrt{3}}+\frac{7 C_{31,22}^{31,1}}{3840
\sqrt{3}}+\frac{9 \sqrt{\frac{7}{5}} C_{31,22}^{33,0}}{10240}+\frac{9 \sqrt{\frac{7}{5}} C_{31,22}^{33,1}}{5120}-\frac{3 \sqrt{\frac{21}{5}} C_{33,00}^{33,0}}{5120}-\frac{3
\sqrt{\frac{21}{5}} C_{33,00}^{33,1}}{2560}-\frac{3 \sqrt{\frac{7}{5}} C_{33,20}^{33,0}}{5120}-\frac{3 \sqrt{\frac{7}{5}} C_{33,20}^{33,1}}{2560}-\frac{3
\sqrt{\frac{21}{2}} C_{33,22}^{33,0}}{3200}-\frac{3 \sqrt{\frac{21}{2}} C_{33,22}^{33,1}}{1600}\)

\noindent\(C_{31,0011,2}^{11,2212,1,\text{Loc}}= -\frac{7 C_{00,20}^{60,1}}{512 \sqrt{5}}+\frac{21 C_{00,22}^{62,1}}{5120}-\frac{7
C_{11,22}^{51,1}}{7680}-\frac{9 \sqrt{\frac{21}{5}} C_{11,22}^{53,1}}{10240}+\frac{7 C_{20,20}^{40,1}}{1536 \sqrt{5}}-\frac{7 C_{20,22}^{42,1}}{7680}-\frac{7
C_{22,20}^{42,1}}{7680}-\frac{7 C_{22,22}^{40,1}}{3072}+\frac{7 \sqrt{7} C_{22,22}^{42,1}}{7680}-\frac{7 C_{31,20}^{31,1}}{1536 \sqrt{5}}+\frac{7
C_{31,22}^{31,1}}{7680}+\frac{9 \sqrt{\frac{21}{5}} C_{31,22}^{33,1}}{10240}+\frac{3 \sqrt{\frac{21}{5}} C_{33,20}^{33,1}}{2560}-\frac{9 \sqrt{\frac{7}{2}}
C_{33,22}^{33,1}}{3200}\)

\noindent\(C_{31,0011,2}^{11,2212,1,\text{NonLoc}}= -\frac{7 \sqrt{\frac{3}{5}} C_{00,00}^{60,0}}{1024}-\frac{7 C_{00,20}^{60,0}}{1024
\sqrt{5}}-\frac{21 C_{00,22}^{62,0}}{5120}-\frac{7 C_{11,00}^{51,0}}{1024 \sqrt{15}}-\frac{7 C_{11,22}^{51,0}}{7680}-\frac{9 \sqrt{\frac{21}{5}}
C_{11,22}^{53,0}}{10240}+\frac{7 C_{20,00}^{40,0}}{1024 \sqrt{15}}+\frac{7 C_{20,20}^{40,0}}{3072 \sqrt{5}}+\frac{7 C_{20,22}^{42,0}}{7680}-\frac{7
C_{22,00}^{42,0}}{5120 \sqrt{3}}-\frac{7 C_{22,20}^{42,0}}{15360}+\frac{7 C_{22,22}^{40,0}}{3072}-\frac{7 \sqrt{7} C_{22,22}^{42,0}}{7680}+\frac{7
C_{31,00}^{31,0}}{1024 \sqrt{15}}+\frac{7 C_{31,20}^{31,0}}{3072 \sqrt{5}}+\frac{7 C_{31,22}^{31,0}}{7680}+\frac{9 \sqrt{\frac{21}{5}} C_{31,22}^{33,0}}{10240}-\frac{9
\sqrt{\frac{7}{5}} C_{33,00}^{33,0}}{5120}-\frac{3 \sqrt{\frac{21}{5}} C_{33,20}^{33,0}}{5120}-\frac{9 \sqrt{\frac{7}{2}} C_{33,22}^{33,0}}{3200}\)

\noindent\(C_{31,0011,2}^{11,2213,0,\text{Loc}}= \frac{7}{640} \sqrt{\frac{7}{3}} C_{00,20}^{60,0}+\frac{7 \sqrt{\frac{7}{3}} C_{00,20}^{60,1}}{1280}+\frac{1}{640}
\sqrt{\frac{21}{5}} C_{00,22}^{62,0}+\frac{\sqrt{\frac{21}{5}} C_{00,22}^{62,1}}{1280}-\frac{1}{960} \sqrt{\frac{7}{15}} C_{11,22}^{51,0}-\frac{\sqrt{\frac{7}{15}}
C_{11,22}^{51,1}}{1920}-\frac{9 C_{11,22}^{53,0}}{6400}-\frac{9 C_{11,22}^{53,1}}{12800}-\frac{7 \sqrt{\frac{7}{3}} C_{20,20}^{40,0}}{1920}-\frac{7
\sqrt{\frac{7}{3}} C_{20,20}^{40,1}}{3840}-\frac{1}{960} \sqrt{\frac{7}{15}} C_{20,22}^{42,0}-\frac{\sqrt{\frac{7}{15}} C_{20,22}^{42,1}}{1920}+\frac{7
\sqrt{\frac{7}{15}} C_{22,20}^{42,0}}{1920}+\frac{7 \sqrt{\frac{7}{15}} C_{22,20}^{42,1}}{3840}-\frac{1}{384} \sqrt{\frac{7}{15}} C_{22,22}^{40,0}-\frac{1}{768}
\sqrt{\frac{7}{15}} C_{22,22}^{40,1}+\frac{7 C_{22,22}^{42,0}}{960 \sqrt{15}}+\frac{7 C_{22,22}^{42,1}}{1920 \sqrt{15}}+\frac{7 \sqrt{\frac{7}{3}}
C_{31,20}^{31,0}}{1920}+\frac{7 \sqrt{\frac{7}{3}} C_{31,20}^{31,1}}{3840}+\frac{1}{960} \sqrt{\frac{7}{15}} C_{31,22}^{31,0}+\frac{\sqrt{\frac{7}{15}}
C_{31,22}^{31,1}}{1920}+\frac{9 C_{31,22}^{33,0}}{6400}+\frac{9 C_{31,22}^{33,1}}{12800}-\frac{21 C_{33,20}^{33,0}}{3200}-\frac{21 C_{33,20}^{33,1}}{6400}-\frac{3}{400}
\sqrt{\frac{3}{10}} C_{33,22}^{33,0}-\frac{3}{800} \sqrt{\frac{3}{10}} C_{33,22}^{33,1}\)

\noindent\(C_{31,0011,2}^{11,2213,0,\text{NonLoc}}= \frac{7 \sqrt{7} C_{00,00}^{60,0}}{2560}+\frac{7 \sqrt{7} C_{00,00}^{60,1}}{1280}+\frac{7
\sqrt{\frac{7}{3}} C_{00,20}^{60,0}}{2560}+\frac{7 \sqrt{\frac{7}{3}} C_{00,20}^{60,1}}{1280}-\frac{\sqrt{\frac{21}{5}} C_{00,22}^{62,0}}{1280}-\frac{1}{640}
\sqrt{\frac{21}{5}} C_{00,22}^{62,1}+\frac{7 \sqrt{7} C_{11,00}^{51,0}}{7680}+\frac{7 \sqrt{7} C_{11,00}^{51,1}}{3840}-\frac{\sqrt{\frac{7}{15}}
C_{11,22}^{51,0}}{1920}-\frac{1}{960} \sqrt{\frac{7}{15}} C_{11,22}^{51,1}-\frac{9 C_{11,22}^{53,0}}{12800}-\frac{9 C_{11,22}^{53,1}}{6400}-\frac{7
\sqrt{7} C_{20,00}^{40,0}}{7680}-\frac{7 \sqrt{7} C_{20,00}^{40,1}}{3840}-\frac{7 \sqrt{\frac{7}{3}} C_{20,20}^{40,0}}{7680}-\frac{7 \sqrt{\frac{7}{3}}
C_{20,20}^{40,1}}{3840}+\frac{\sqrt{\frac{7}{15}} C_{20,22}^{42,0}}{1920}+\frac{1}{960} \sqrt{\frac{7}{15}} C_{20,22}^{42,1}+\frac{7 \sqrt{\frac{7}{5}}
C_{22,00}^{42,0}}{7680}+\frac{7 \sqrt{\frac{7}{5}} C_{22,00}^{42,1}}{3840}+\frac{7 \sqrt{\frac{7}{15}} C_{22,20}^{42,0}}{7680}+\frac{7 \sqrt{\frac{7}{15}}
C_{22,20}^{42,1}}{3840}+\frac{1}{768} \sqrt{\frac{7}{15}} C_{22,22}^{40,0}+\frac{1}{384} \sqrt{\frac{7}{15}} C_{22,22}^{40,1}-\frac{7 C_{22,22}^{42,0}}{1920
\sqrt{15}}-\frac{7 C_{22,22}^{42,1}}{960 \sqrt{15}}-\frac{7 \sqrt{7} C_{31,00}^{31,0}}{7680}-\frac{7 \sqrt{7} C_{31,00}^{31,1}}{3840}-\frac{7 \sqrt{\frac{7}{3}}
C_{31,20}^{31,0}}{7680}-\frac{7 \sqrt{\frac{7}{3}} C_{31,20}^{31,1}}{3840}+\frac{\sqrt{\frac{7}{15}} C_{31,22}^{31,0}}{1920}+\frac{1}{960} \sqrt{\frac{7}{15}}
C_{31,22}^{31,1}+\frac{9 C_{31,22}^{33,0}}{12800}+\frac{9 C_{31,22}^{33,1}}{6400}+\frac{21 \sqrt{3} C_{33,00}^{33,0}}{12800}+\frac{21 \sqrt{3} C_{33,00}^{33,1}}{6400}+\frac{21
C_{33,20}^{33,0}}{12800}+\frac{21 C_{33,20}^{33,1}}{6400}-\frac{3}{800} \sqrt{\frac{3}{10}} C_{33,22}^{33,0}-\frac{3}{400} \sqrt{\frac{3}{10}} C_{33,22}^{33,1}\)

\noindent\(C_{31,0011,2}^{11,2213,1,\text{Loc}}= \frac{7 \sqrt{7} C_{00,20}^{60,1}}{1280}+\frac{3 \sqrt{\frac{7}{5}} C_{00,22}^{62,1}}{1280}-\frac{\sqrt{\frac{7}{5}}
C_{11,22}^{51,1}}{1920}-\frac{9 \sqrt{3} C_{11,22}^{53,1}}{12800}-\frac{7 \sqrt{7} C_{20,20}^{40,1}}{3840}-\frac{\sqrt{\frac{7}{5}} C_{20,22}^{42,1}}{1920}+\frac{7
\sqrt{\frac{7}{5}} C_{22,20}^{42,1}}{3840}-\frac{1}{768} \sqrt{\frac{7}{5}} C_{22,22}^{40,1}+\frac{7 C_{22,22}^{42,1}}{1920 \sqrt{5}}+\frac{7 \sqrt{7}
C_{31,20}^{31,1}}{3840}+\frac{\sqrt{\frac{7}{5}} C_{31,22}^{31,1}}{1920}+\frac{9 \sqrt{3} C_{31,22}^{33,1}}{12800}-\frac{21 \sqrt{3} C_{33,20}^{33,1}}{6400}-\frac{9
C_{33,22}^{33,1}}{800 \sqrt{10}}\)

\noindent\(C_{31,0011,2}^{11,2213,1,\text{NonLoc}}= \frac{7 \sqrt{21} C_{00,00}^{60,0}}{2560}+\frac{7 \sqrt{7} C_{00,20}^{60,0}}{2560}-\frac{3
\sqrt{\frac{7}{5}} C_{00,22}^{62,0}}{1280}+\frac{7 \sqrt{\frac{7}{3}} C_{11,00}^{51,0}}{2560}-\frac{\sqrt{\frac{7}{5}} C_{11,22}^{51,0}}{1920}-\frac{9
\sqrt{3} C_{11,22}^{53,0}}{12800}-\frac{7 \sqrt{\frac{7}{3}} C_{20,00}^{40,0}}{2560}-\frac{7 \sqrt{7} C_{20,20}^{40,0}}{7680}+\frac{\sqrt{\frac{7}{5}}
C_{20,22}^{42,0}}{1920}+\frac{7 \sqrt{\frac{7}{15}} C_{22,00}^{42,0}}{2560}+\frac{7 \sqrt{\frac{7}{5}} C_{22,20}^{42,0}}{7680}+\frac{1}{768} \sqrt{\frac{7}{5}}
C_{22,22}^{40,0}-\frac{7 C_{22,22}^{42,0}}{1920 \sqrt{5}}-\frac{7 \sqrt{\frac{7}{3}} C_{31,00}^{31,0}}{2560}-\frac{7 \sqrt{7} C_{31,20}^{31,0}}{7680}+\frac{\sqrt{\frac{7}{5}}
C_{31,22}^{31,0}}{1920}+\frac{9 \sqrt{3} C_{31,22}^{33,0}}{12800}+\frac{63 C_{33,00}^{33,0}}{12800}+\frac{21 \sqrt{3} C_{33,20}^{33,0}}{12800}-\frac{9
C_{33,22}^{33,0}}{800 \sqrt{10}}\)

\noindent\(C_{31,0011,2}^{22,1101,0,\text{Loc}}= \frac{7 C_{11,11}^{51,0}}{512 \sqrt{15}}+\frac{7 C_{11,11}^{51,1}}{1024 \sqrt{15}}-\frac{7
C_{22,11}^{42,0}}{1280}-\frac{7 C_{22,11}^{42,1}}{2560}+\frac{7 C_{31,11}^{31,0}}{512 \sqrt{15}}+\frac{7 C_{31,11}^{31,1}}{1024 \sqrt{15}}+\frac{3}{640}
\sqrt{\frac{21}{10}} C_{33,11}^{33,0}+\frac{3 \sqrt{\frac{21}{10}} C_{33,11}^{33,1}}{1280}\)

\noindent\(C_{31,0011,2}^{22,1101,0,\text{NonLoc}}= \frac{7 C_{11,11}^{51,0}}{1024 \sqrt{15}}+\frac{7 C_{11,11}^{51,1}}{512 \sqrt{15}}+\frac{7
C_{22,11}^{42,0}}{2560}+\frac{7 C_{22,11}^{42,1}}{1280}+\frac{7 C_{31,11}^{31,0}}{1024 \sqrt{15}}+\frac{7 C_{31,11}^{31,1}}{512 \sqrt{15}}+\frac{3
\sqrt{\frac{21}{10}} C_{33,11}^{33,0}}{1280}+\frac{3}{640} \sqrt{\frac{21}{10}} C_{33,11}^{33,1}\)

\noindent\(C_{31,0011,2}^{22,1101,1,\text{Loc}}= \frac{7 C_{11,11}^{51,1}}{1024 \sqrt{5}}-\frac{7 \sqrt{3} C_{22,11}^{42,1}}{2560}+\frac{7
C_{31,11}^{31,1}}{1024 \sqrt{5}}+\frac{9 \sqrt{\frac{7}{10}} C_{33,11}^{33,1}}{1280}\)

\noindent\(C_{31,0011,2}^{22,1101,1,\text{NonLoc}}= \frac{7 C_{11,11}^{51,0}}{1024 \sqrt{5}}+\frac{7 \sqrt{3} C_{22,11}^{42,0}}{2560}+\frac{7
C_{31,11}^{31,0}}{1024 \sqrt{5}}+\frac{9 \sqrt{\frac{7}{10}} C_{33,11}^{33,0}}{1280}\)

\noindent\(C_{31,0011,2}^{31,0011,0,\text{Loc}}= -\frac{7 C_{00,20}^{60,0}}{2048 \sqrt{5}}-\frac{7 C_{00,20}^{60,1}}{4096 \sqrt{5}}-\frac{21
C_{00,22}^{62,0}}{102400}-\frac{21 C_{00,22}^{62,1}}{204800}+\frac{21 C_{11,22}^{51,0}}{102400}+\frac{21 C_{11,22}^{51,1}}{204800}+\frac{9 \sqrt{\frac{21}{5}}
C_{11,22}^{53,0}}{102400}+\frac{9 \sqrt{\frac{21}{5}} C_{11,22}^{53,1}}{204800}-\frac{7 C_{20,20}^{40,0}}{2048 \sqrt{5}}-\frac{7 C_{20,20}^{40,1}}{4096
\sqrt{5}}-\frac{21 C_{20,22}^{42,0}}{102400}-\frac{21 C_{20,22}^{42,1}}{204800}-\frac{7 C_{22,20}^{42,0}}{5120}-\frac{7 C_{22,20}^{42,1}}{10240}-\frac{21
C_{22,22}^{40,0}}{102400}-\frac{21 C_{22,22}^{40,1}}{204800}-\frac{3 \sqrt{7} C_{22,22}^{42,0}}{51200}-\frac{3 \sqrt{7} C_{22,22}^{42,1}}{102400}-\frac{9
C_{22,22}^{44,0}}{25600 \sqrt{5}}-\frac{9 C_{22,22}^{44,1}}{51200 \sqrt{5}}+\frac{7 C_{31,20}^{31,0}}{2048 \sqrt{5}}+\frac{7 C_{31,20}^{31,1}}{4096
\sqrt{5}}+\frac{21 C_{31,22}^{31,0}}{102400}+\frac{21 C_{31,22}^{31,1}}{204800}+\frac{9 \sqrt{\frac{21}{5}} C_{31,22}^{33,0}}{102400}+\frac{9 \sqrt{\frac{21}{5}}
C_{31,22}^{33,1}}{204800}+\frac{3 \sqrt{\frac{21}{5}} C_{33,20}^{33,0}}{5120}+\frac{3 \sqrt{\frac{21}{5}} C_{33,20}^{33,1}}{10240}+\frac{9 \sqrt{\frac{7}{2}}
C_{33,22}^{33,0}}{128000}+\frac{9 \sqrt{\frac{7}{2}} C_{33,22}^{33,1}}{256000}\)

\noindent\(C_{31,0011,2}^{31,0011,0,\text{NonLoc}}= -\frac{7 \sqrt{\frac{3}{5}} C_{00,00}^{60,0}}{8192}-\frac{7 \sqrt{\frac{3}{5}}
C_{00,00}^{60,1}}{4096}-\frac{7 C_{00,20}^{60,0}}{8192 \sqrt{5}}-\frac{7 C_{00,20}^{60,1}}{4096 \sqrt{5}}+\frac{21 C_{00,22}^{62,0}}{204800}+\frac{21
C_{00,22}^{62,1}}{102400}-\frac{7 \sqrt{\frac{3}{5}} C_{11,00}^{51,0}}{8192}-\frac{7 \sqrt{\frac{3}{5}} C_{11,00}^{51,1}}{4096}+\frac{21 C_{11,22}^{51,0}}{204800}+\frac{21
C_{11,22}^{51,1}}{102400}+\frac{9 \sqrt{\frac{21}{5}} C_{11,22}^{53,0}}{204800}+\frac{9 \sqrt{\frac{21}{5}} C_{11,22}^{53,1}}{102400}-\frac{7 \sqrt{\frac{3}{5}}
C_{20,00}^{40,0}}{8192}-\frac{7 \sqrt{\frac{3}{5}} C_{20,00}^{40,1}}{4096}-\frac{7 C_{20,20}^{40,0}}{8192 \sqrt{5}}-\frac{7 C_{20,20}^{40,1}}{4096
\sqrt{5}}+\frac{21 C_{20,22}^{42,0}}{204800}+\frac{21 C_{20,22}^{42,1}}{102400}-\frac{7 \sqrt{3} C_{22,00}^{42,0}}{20480}-\frac{7 \sqrt{3} C_{22,00}^{42,1}}{10240}-\frac{7
C_{22,20}^{42,0}}{20480}-\frac{7 C_{22,20}^{42,1}}{10240}+\frac{21 C_{22,22}^{40,0}}{204800}+\frac{21 C_{22,22}^{40,1}}{102400}+\frac{3 \sqrt{7}
C_{22,22}^{42,0}}{102400}+\frac{3 \sqrt{7} C_{22,22}^{42,1}}{51200}+\frac{9 C_{22,22}^{44,0}}{51200 \sqrt{5}}+\frac{9 C_{22,22}^{44,1}}{25600 \sqrt{5}}-\frac{7
\sqrt{\frac{3}{5}} C_{31,00}^{31,0}}{8192}-\frac{7 \sqrt{\frac{3}{5}} C_{31,00}^{31,1}}{4096}-\frac{7 C_{31,20}^{31,0}}{8192 \sqrt{5}}-\frac{7 C_{31,20}^{31,1}}{4096
\sqrt{5}}+\frac{21 C_{31,22}^{31,0}}{204800}+\frac{21 C_{31,22}^{31,1}}{102400}+\frac{9 \sqrt{\frac{21}{5}} C_{31,22}^{33,0}}{204800}+\frac{9 \sqrt{\frac{21}{5}}
C_{31,22}^{33,1}}{102400}-\frac{9 \sqrt{\frac{7}{5}} C_{33,00}^{33,0}}{20480}-\frac{9 \sqrt{\frac{7}{5}} C_{33,00}^{33,1}}{10240}-\frac{3 \sqrt{\frac{21}{5}}
C_{33,20}^{33,0}}{20480}-\frac{3 \sqrt{\frac{21}{5}} C_{33,20}^{33,1}}{10240}+\frac{9 \sqrt{\frac{7}{2}} C_{33,22}^{33,0}}{256000}+\frac{9 \sqrt{\frac{7}{2}}
C_{33,22}^{33,1}}{128000}\)

\noindent\(C_{31,0011,2}^{31,0011,1,\text{Loc}}= -\frac{7 \sqrt{\frac{3}{5}} C_{00,20}^{60,1}}{4096}-\frac{21 \sqrt{3} C_{00,22}^{62,1}}{204800}+\frac{21
\sqrt{3} C_{11,22}^{51,1}}{204800}+\frac{27 \sqrt{\frac{7}{5}} C_{11,22}^{53,1}}{204800}-\frac{7 \sqrt{\frac{3}{5}} C_{20,20}^{40,1}}{4096}-\frac{21
\sqrt{3} C_{20,22}^{42,1}}{204800}-\frac{7 \sqrt{3} C_{22,20}^{42,1}}{10240}-\frac{21 \sqrt{3} C_{22,22}^{40,1}}{204800}-\frac{3 \sqrt{21} C_{22,22}^{42,1}}{102400}-\frac{9
\sqrt{\frac{3}{5}} C_{22,22}^{44,1}}{51200}+\frac{7 \sqrt{\frac{3}{5}} C_{31,20}^{31,1}}{4096}+\frac{21 \sqrt{3} C_{31,22}^{31,1}}{204800}+\frac{27
\sqrt{\frac{7}{5}} C_{31,22}^{33,1}}{204800}+\frac{9 \sqrt{\frac{7}{5}} C_{33,20}^{33,1}}{10240}+\frac{9 \sqrt{\frac{21}{2}} C_{33,22}^{33,1}}{256000}\)

\noindent\(C_{31,0011,2}^{31,0011,1,\text{NonLoc}}= -\frac{21 C_{00,00}^{60,0}}{8192 \sqrt{5}}-\frac{7 \sqrt{\frac{3}{5}} C_{00,20}^{60,0}}{8192}+\frac{21
\sqrt{3} C_{00,22}^{62,0}}{204800}-\frac{21 C_{11,00}^{51,0}}{8192 \sqrt{5}}+\frac{21 \sqrt{3} C_{11,22}^{51,0}}{204800}+\frac{27 \sqrt{\frac{7}{5}}
C_{11,22}^{53,0}}{204800}-\frac{21 C_{20,00}^{40,0}}{8192 \sqrt{5}}-\frac{7 \sqrt{\frac{3}{5}} C_{20,20}^{40,0}}{8192}+\frac{21 \sqrt{3} C_{20,22}^{42,0}}{204800}-\frac{21
C_{22,00}^{42,0}}{20480}-\frac{7 \sqrt{3} C_{22,20}^{42,0}}{20480}+\frac{21 \sqrt{3} C_{22,22}^{40,0}}{204800}+\frac{3 \sqrt{21} C_{22,22}^{42,0}}{102400}+\frac{9
\sqrt{\frac{3}{5}} C_{22,22}^{44,0}}{51200}-\frac{21 C_{31,00}^{31,0}}{8192 \sqrt{5}}-\frac{7 \sqrt{\frac{3}{5}} C_{31,20}^{31,0}}{8192}+\frac{21
\sqrt{3} C_{31,22}^{31,0}}{204800}+\frac{27 \sqrt{\frac{7}{5}} C_{31,22}^{33,0}}{204800}-\frac{9 \sqrt{\frac{21}{5}} C_{33,00}^{33,0}}{20480}-\frac{9
\sqrt{\frac{7}{5}} C_{33,20}^{33,0}}{20480}+\frac{9 \sqrt{\frac{21}{2}} C_{33,22}^{33,0}}{256000}\)

\noindent\(C_{31,1101,0}^{11,1101,0,\text{Loc}}= \frac{7 C_{00,00}^{60,0}}{128}+\frac{7 C_{00,00}^{60,1}}{256}-\frac{7 C_{11,00}^{51,0}}{384}-\frac{7
C_{11,00}^{51,1}}{768}-\frac{C_{20,00}^{40,0}}{128}-\frac{C_{20,00}^{40,1}}{256}+\frac{5 C_{31,00}^{31,0}}{384}+\frac{5 C_{31,00}^{31,1}}{768}\)

\noindent\(C_{31,1101,0}^{11,1101,0,\text{NonLoc}}= -\frac{7 C_{00,00}^{60,0}}{512}-\frac{7 C_{00,00}^{60,1}}{256}+\frac{7 \sqrt{3}
C_{00,20}^{60,0}}{512}+\frac{7 \sqrt{3} C_{00,20}^{60,1}}{256}-\frac{7 C_{11,00}^{51,0}}{1536}-\frac{7 C_{11,00}^{51,1}}{768}+\frac{7 C_{11,20}^{31,0}}{512
\sqrt{3}}+\frac{7 C_{11,20}^{31,1}}{256 \sqrt{3}}+\frac{C_{20,00}^{40,0}}{512}+\frac{C_{20,00}^{40,1}}{256}-\frac{\sqrt{3} C_{20,20}^{40,0}}{512}-\frac{\sqrt{3}
C_{20,20}^{40,1}}{256}+\frac{5 C_{31,00}^{31,0}}{1536}+\frac{5 C_{31,00}^{31,1}}{768}-\frac{5 C_{31,20}^{31,0}}{512 \sqrt{3}}-\frac{5 C_{31,20}^{31,1}}{256
\sqrt{3}}\)

\noindent\(C_{31,1101,0}^{11,1101,1,\text{Loc}}= \frac{7 \sqrt{3} C_{00,00}^{60,1}}{256}-\frac{7 C_{11,00}^{51,1}}{256 \sqrt{3}}-\frac{\sqrt{3}
C_{20,00}^{40,1}}{256}+\frac{5 C_{31,00}^{31,1}}{256 \sqrt{3}}\)

\noindent\(C_{31,1101,0}^{11,1101,1,\text{NonLoc}}= -\frac{7 \sqrt{3} C_{00,00}^{60,0}}{512}+\frac{21 C_{00,20}^{60,0}}{512}-\frac{7
C_{11,00}^{51,0}}{512 \sqrt{3}}+\frac{7 C_{11,20}^{31,0}}{512}+\frac{\sqrt{3} C_{20,00}^{40,0}}{512}-\frac{3 C_{20,20}^{40,0}}{512}+\frac{5 C_{31,00}^{31,0}}{512
\sqrt{3}}-\frac{5 C_{31,20}^{31,0}}{512}\)

\noindent\(C_{31,1101,1}^{00,2011,0,\text{Loc}}= \frac{7 C_{11,11}^{51,0}}{384}+\frac{7 C_{11,11}^{51,1}}{768}+\frac{C_{31,11}^{31,0}}{128}+\frac{C_{31,11}^{31,1}}{256}-\frac{3}{160}
\sqrt{\frac{7}{2}} C_{33,11}^{33,0}-\frac{3}{320} \sqrt{\frac{7}{2}} C_{33,11}^{33,1}\)

\noindent\(C_{31,1101,1}^{00,2011,0,\text{NonLoc}}= \frac{7 C_{11,11}^{51,0}}{768}+\frac{7 C_{11,11}^{51,1}}{384}+\frac{C_{31,11}^{31,0}}{256}+\frac{C_{31,11}^{31,1}}{128}-\frac{3}{320}
\sqrt{\frac{7}{2}} C_{33,11}^{33,0}-\frac{3}{160} \sqrt{\frac{7}{2}} C_{33,11}^{33,1}\)

\noindent\(C_{31,1101,1}^{00,2011,1,\text{Loc}}= \frac{7 C_{11,11}^{51,1}}{256 \sqrt{3}}+\frac{\sqrt{3} C_{31,11}^{31,1}}{256}-\frac{3}{320}
\sqrt{\frac{21}{2}} C_{33,11}^{33,1}\)

\noindent\(C_{31,1101,1}^{00,2011,1,\text{NonLoc}}= \frac{7 C_{11,11}^{51,0}}{256 \sqrt{3}}+\frac{\sqrt{3} C_{31,11}^{31,0}}{256}-\frac{3}{320}
\sqrt{\frac{21}{2}} C_{33,11}^{33,0}\)

\noindent\(C_{31,1101,1}^{00,2211,0,\text{Loc}}= -\frac{7 C_{11,11}^{51,0}}{384 \sqrt{5}}-\frac{7 C_{11,11}^{51,1}}{768 \sqrt{5}}-\frac{C_{31,11}^{31,0}}{128
\sqrt{5}}-\frac{C_{31,11}^{31,1}}{256 \sqrt{5}}+\frac{3}{160} \sqrt{\frac{7}{10}} C_{33,11}^{33,0}+\frac{3}{320} \sqrt{\frac{7}{10}} C_{33,11}^{33,1}\)

\noindent\(C_{31,1101,1}^{00,2211,0,\text{NonLoc}}= -\frac{7 C_{11,11}^{51,0}}{768 \sqrt{5}}-\frac{7 C_{11,11}^{51,1}}{384 \sqrt{5}}-\frac{C_{31,11}^{31,0}}{256
\sqrt{5}}-\frac{C_{31,11}^{31,1}}{128 \sqrt{5}}+\frac{3}{320} \sqrt{\frac{7}{10}} C_{33,11}^{33,0}+\frac{3}{160} \sqrt{\frac{7}{10}} C_{33,11}^{33,1}\)

\noindent\(C_{31,1101,1}^{00,2211,1,\text{Loc}}= -\frac{7 C_{11,11}^{51,1}}{256 \sqrt{15}}-\frac{1}{256} \sqrt{\frac{3}{5}} C_{31,11}^{31,1}+\frac{3}{320}
\sqrt{\frac{21}{10}} C_{33,11}^{33,1}\)

\noindent\(C_{31,1101,1}^{00,2211,1,\text{NonLoc}}= -\frac{7 C_{11,11}^{51,0}}{256 \sqrt{15}}-\frac{1}{256} \sqrt{\frac{3}{5}} C_{31,11}^{31,0}+\frac{3}{320}
\sqrt{\frac{21}{10}} C_{33,11}^{33,0}\)

\noindent\(C_{31,1101,1}^{11,1101,0,\text{Loc}}= \frac{C_{20,00}^{40,0}}{32 \sqrt{3}}+\frac{C_{20,00}^{40,1}}{64 \sqrt{3}}-\frac{7
C_{22,00}^{42,0}}{128 \sqrt{15}}-\frac{7 C_{22,00}^{42,1}}{256 \sqrt{15}}-\frac{C_{31,00}^{31,0}}{64 \sqrt{3}}-\frac{C_{31,00}^{31,1}}{128 \sqrt{3}}+\frac{9
\sqrt{7} C_{33,00}^{33,0}}{640}+\frac{9 \sqrt{7} C_{33,00}^{33,1}}{1280}\)

\noindent\(C_{31,1101,1}^{11,1101,0,\text{NonLoc}}= -\frac{C_{20,00}^{40,0}}{128 \sqrt{3}}-\frac{C_{20,00}^{40,1}}{64 \sqrt{3}}+\frac{C_{20,20}^{40,0}}{128}+\frac{C_{20,20}^{40,1}}{64}+\frac{7
C_{22,00}^{42,0}}{512 \sqrt{15}}+\frac{7 C_{22,00}^{42,1}}{256 \sqrt{15}}-\frac{7 C_{22,20}^{42,0}}{512 \sqrt{5}}-\frac{7 C_{22,20}^{42,1}}{256 \sqrt{5}}-\frac{C_{31,00}^{31,0}}{256
\sqrt{3}}-\frac{C_{31,00}^{31,1}}{128 \sqrt{3}}+\frac{C_{31,20}^{31,0}}{256}+\frac{C_{31,20}^{31,1}}{128}+\frac{9 \sqrt{7} C_{33,00}^{33,0}}{2560}+\frac{9
\sqrt{7} C_{33,00}^{33,1}}{1280}-\frac{9 \sqrt{21} C_{33,20}^{33,0}}{2560}-\frac{9 \sqrt{21} C_{33,20}^{33,1}}{1280}\)

\noindent\(C_{31,1101,1}^{11,1101,1,\text{Loc}}= \frac{C_{20,00}^{40,1}}{64}-\frac{7 C_{22,00}^{42,1}}{256 \sqrt{5}}-\frac{C_{31,00}^{31,1}}{128}+\frac{9
\sqrt{21} C_{33,00}^{33,1}}{1280}\)

\noindent\(C_{31,1101,1}^{11,1101,1,\text{NonLoc}}= -\frac{C_{20,00}^{40,0}}{128}+\frac{\sqrt{3} C_{20,20}^{40,0}}{128}+\frac{7 C_{22,00}^{42,0}}{512
\sqrt{5}}-\frac{7}{512} \sqrt{\frac{3}{5}} C_{22,20}^{42,0}-\frac{C_{31,00}^{31,0}}{256}+\frac{\sqrt{3} C_{31,20}^{31,0}}{256}+\frac{9 \sqrt{21}
C_{33,00}^{33,0}}{2560}-\frac{27 \sqrt{7} C_{33,20}^{33,0}}{2560}\)

\noindent\(C_{31,1101,1}^{20,0011,0,\text{Loc}}= -\frac{7 C_{11,11}^{51,0}}{1536}-\frac{7 C_{11,11}^{51,1}}{3072}+\frac{7 C_{22,11}^{42,0}}{256
\sqrt{15}}+\frac{7 C_{22,11}^{42,1}}{512 \sqrt{15}}-\frac{7 C_{31,11}^{31,0}}{1536}-\frac{7 C_{31,11}^{31,1}}{3072}-\frac{3}{640} \sqrt{\frac{7}{2}}
C_{33,11}^{33,0}-\frac{3 \sqrt{\frac{7}{2}} C_{33,11}^{33,1}}{1280}\)

\noindent\(C_{31,1101,1}^{20,0011,0,\text{NonLoc}}= -\frac{7 C_{11,11}^{51,0}}{3072}-\frac{7 C_{11,11}^{51,1}}{1536}-\frac{7 C_{22,11}^{42,0}}{512
\sqrt{15}}-\frac{7 C_{22,11}^{42,1}}{256 \sqrt{15}}-\frac{7 C_{31,11}^{31,0}}{3072}-\frac{7 C_{31,11}^{31,1}}{1536}-\frac{3 \sqrt{\frac{7}{2}} C_{33,11}^{33,0}}{1280}-\frac{3}{640}
\sqrt{\frac{7}{2}} C_{33,11}^{33,1}\)

\noindent\(C_{31,1101,1}^{20,0011,1,\text{Loc}}= -\frac{7 C_{11,11}^{51,1}}{1024 \sqrt{3}}+\frac{7 C_{22,11}^{42,1}}{512 \sqrt{5}}-\frac{7
C_{31,11}^{31,1}}{1024 \sqrt{3}}-\frac{3 \sqrt{\frac{21}{2}} C_{33,11}^{33,1}}{1280}\)

\noindent\(C_{31,1101,1}^{20,0011,1,\text{NonLoc}}= -\frac{7 C_{11,11}^{51,0}}{1024 \sqrt{3}}-\frac{7 C_{22,11}^{42,0}}{512 \sqrt{5}}-\frac{7
C_{31,11}^{31,0}}{1024 \sqrt{3}}-\frac{3 \sqrt{\frac{21}{2}} C_{33,11}^{33,0}}{1280}\)

\noindent\(C_{31,1101,1}^{22,0011,0,\text{Loc}}= \frac{7 C_{11,11}^{51,0}}{1536 \sqrt{5}}+\frac{7 C_{11,11}^{51,1}}{3072 \sqrt{5}}-\frac{7
C_{22,11}^{42,0}}{1280 \sqrt{3}}-\frac{7 C_{22,11}^{42,1}}{2560 \sqrt{3}}+\frac{7 C_{31,11}^{31,0}}{1536 \sqrt{5}}+\frac{7 C_{31,11}^{31,1}}{3072
\sqrt{5}}+\frac{3}{640} \sqrt{\frac{7}{10}} C_{33,11}^{33,0}+\frac{3 \sqrt{\frac{7}{10}} C_{33,11}^{33,1}}{1280}\)

\noindent\(C_{31,1101,1}^{22,0011,0,\text{NonLoc}}= \frac{7 C_{11,11}^{51,0}}{3072 \sqrt{5}}+\frac{7 C_{11,11}^{51,1}}{1536 \sqrt{5}}+\frac{7
C_{22,11}^{42,0}}{2560 \sqrt{3}}+\frac{7 C_{22,11}^{42,1}}{1280 \sqrt{3}}+\frac{7 C_{31,11}^{31,0}}{3072 \sqrt{5}}+\frac{7 C_{31,11}^{31,1}}{1536
\sqrt{5}}+\frac{3 \sqrt{\frac{7}{10}} C_{33,11}^{33,0}}{1280}+\frac{3}{640} \sqrt{\frac{7}{10}} C_{33,11}^{33,1}\)

\noindent\(C_{31,1101,1}^{22,0011,1,\text{Loc}}= \frac{7 C_{11,11}^{51,1}}{1024 \sqrt{15}}-\frac{7 C_{22,11}^{42,1}}{2560}+\frac{7
C_{31,11}^{31,1}}{1024 \sqrt{15}}+\frac{3 \sqrt{\frac{21}{10}} C_{33,11}^{33,1}}{1280}\)

\noindent\(C_{31,1101,1}^{22,0011,1,\text{NonLoc}}= \frac{7 C_{11,11}^{51,0}}{1024 \sqrt{15}}+\frac{7 C_{22,11}^{42,0}}{2560}+\frac{7
C_{31,11}^{31,0}}{1024 \sqrt{15}}+\frac{3 \sqrt{\frac{21}{10}} C_{33,11}^{33,0}}{1280}\)

\noindent\(C_{31,1101,2}^{00,2212,0,\text{Loc}}= \frac{7 C_{11,11}^{51,0}}{128 \sqrt{15}}+\frac{7 C_{11,11}^{51,1}}{256 \sqrt{15}}+\frac{1}{128}
\sqrt{\frac{3}{5}} C_{31,11}^{31,0}+\frac{1}{256} \sqrt{\frac{3}{5}} C_{31,11}^{31,1}-\frac{3}{160} \sqrt{\frac{21}{10}} C_{33,11}^{33,0}-\frac{3}{320}
\sqrt{\frac{21}{10}} C_{33,11}^{33,1}\)

\noindent\(C_{31,1101,2}^{00,2212,0,\text{NonLoc}}= \frac{7 C_{11,11}^{51,0}}{256 \sqrt{15}}+\frac{7 C_{11,11}^{51,1}}{128 \sqrt{15}}+\frac{1}{256}
\sqrt{\frac{3}{5}} C_{31,11}^{31,0}+\frac{1}{128} \sqrt{\frac{3}{5}} C_{31,11}^{31,1}-\frac{3}{320} \sqrt{\frac{21}{10}} C_{33,11}^{33,0}-\frac{3}{160}
\sqrt{\frac{21}{10}} C_{33,11}^{33,1}\)

\noindent\(C_{31,1101,2}^{00,2212,1,\text{Loc}}= \frac{7 C_{11,11}^{51,1}}{256 \sqrt{5}}+\frac{3 C_{31,11}^{31,1}}{256 \sqrt{5}}-\frac{9}{320}
\sqrt{\frac{7}{10}} C_{33,11}^{33,1}\)

\noindent\(C_{31,1101,2}^{00,2212,1,\text{NonLoc}}= \frac{7 C_{11,11}^{51,0}}{256 \sqrt{5}}+\frac{3 C_{31,11}^{31,0}}{256 \sqrt{5}}-\frac{9}{320}
\sqrt{\frac{7}{10}} C_{33,11}^{33,0}\)

\noindent\(C_{31,1101,2}^{11,1101,0,\text{Loc}}= \frac{7 C_{00,00}^{60,0}}{64 \sqrt{5}}+\frac{7 C_{00,00}^{60,1}}{128 \sqrt{5}}-\frac{7
C_{11,00}^{51,0}}{192 \sqrt{5}}-\frac{7 C_{11,00}^{51,1}}{384 \sqrt{5}}+\frac{C_{20,00}^{40,0}}{64 \sqrt{5}}+\frac{C_{20,00}^{40,1}}{128 \sqrt{5}}-\frac{7
C_{22,00}^{42,0}}{640}-\frac{7 C_{22,00}^{42,1}}{1280}+\frac{C_{31,00}^{31,0}}{96 \sqrt{5}}+\frac{C_{31,00}^{31,1}}{192 \sqrt{5}}+\frac{9}{640} \sqrt{\frac{21}{5}}
C_{33,00}^{33,0}+\frac{9 \sqrt{\frac{21}{5}} C_{33,00}^{33,1}}{1280}\)

\noindent\(C_{31,1101,2}^{11,1101,0,\text{NonLoc}}= -\frac{7 C_{00,00}^{60,0}}{256 \sqrt{5}}-\frac{7 C_{00,00}^{60,1}}{128 \sqrt{5}}+\frac{7}{256}
\sqrt{\frac{3}{5}} C_{00,20}^{60,0}+\frac{7}{128} \sqrt{\frac{3}{5}} C_{00,20}^{60,1}-\frac{7 C_{11,00}^{51,0}}{768 \sqrt{5}}-\frac{7 C_{11,00}^{51,1}}{384
\sqrt{5}}+\frac{7 C_{11,20}^{31,0}}{256 \sqrt{15}}+\frac{7 C_{11,20}^{31,1}}{128 \sqrt{15}}-\frac{C_{20,00}^{40,0}}{256 \sqrt{5}}-\frac{C_{20,00}^{40,1}}{128
\sqrt{5}}+\frac{1}{256} \sqrt{\frac{3}{5}} C_{20,20}^{40,0}+\frac{1}{128} \sqrt{\frac{3}{5}} C_{20,20}^{40,1}+\frac{7 C_{22,00}^{42,0}}{2560}+\frac{7
C_{22,00}^{42,1}}{1280}-\frac{7 \sqrt{3} C_{22,20}^{42,0}}{2560}-\frac{7 \sqrt{3} C_{22,20}^{42,1}}{1280}+\frac{C_{31,00}^{31,0}}{384 \sqrt{5}}+\frac{C_{31,00}^{31,1}}{192
\sqrt{5}}-\frac{C_{31,20}^{31,0}}{128 \sqrt{15}}-\frac{C_{31,20}^{31,1}}{64 \sqrt{15}}+\frac{9 \sqrt{\frac{21}{5}} C_{33,00}^{33,0}}{2560}+\frac{9
\sqrt{\frac{21}{5}} C_{33,00}^{33,1}}{1280}-\frac{27 \sqrt{\frac{7}{5}} C_{33,20}^{33,0}}{2560}-\frac{27 \sqrt{\frac{7}{5}} C_{33,20}^{33,1}}{1280}\)

\noindent\(C_{31,1101,2}^{11,1101,1,\text{Loc}}= \frac{7}{128} \sqrt{\frac{3}{5}} C_{00,00}^{60,1}-\frac{7 C_{11,00}^{51,1}}{128 \sqrt{15}}+\frac{1}{128}
\sqrt{\frac{3}{5}} C_{20,00}^{40,1}-\frac{7 \sqrt{3} C_{22,00}^{42,1}}{1280}+\frac{C_{31,00}^{31,1}}{64 \sqrt{15}}+\frac{27 \sqrt{\frac{7}{5}} C_{33,00}^{33,1}}{1280}\)

\noindent\(C_{31,1101,2}^{11,1101,1,\text{NonLoc}}= -\frac{7}{256} \sqrt{\frac{3}{5}} C_{00,00}^{60,0}+\frac{21 C_{00,20}^{60,0}}{256
\sqrt{5}}-\frac{7 C_{11,00}^{51,0}}{256 \sqrt{15}}+\frac{7 C_{11,20}^{31,0}}{256 \sqrt{5}}-\frac{1}{256} \sqrt{\frac{3}{5}} C_{20,00}^{40,0}+\frac{3
C_{20,20}^{40,0}}{256 \sqrt{5}}+\frac{7 \sqrt{3} C_{22,00}^{42,0}}{2560}-\frac{21 C_{22,20}^{42,0}}{2560}+\frac{C_{31,00}^{31,0}}{128 \sqrt{15}}-\frac{C_{31,20}^{31,0}}{128
\sqrt{5}}+\frac{27 \sqrt{\frac{7}{5}} C_{33,00}^{33,0}}{2560}-\frac{27 \sqrt{\frac{21}{5}} C_{33,20}^{33,0}}{2560}\)

\noindent\(C_{31,1101,2}^{22,0011,0,\text{Loc}}= \frac{7 C_{11,11}^{51,0}}{512 \sqrt{15}}+\frac{7 C_{11,11}^{51,1}}{1024 \sqrt{15}}-\frac{7
C_{22,11}^{42,0}}{1280}-\frac{7 C_{22,11}^{42,1}}{2560}+\frac{7 C_{31,11}^{31,0}}{512 \sqrt{15}}+\frac{7 C_{31,11}^{31,1}}{1024 \sqrt{15}}+\frac{3}{640}
\sqrt{\frac{21}{10}} C_{33,11}^{33,0}+\frac{3 \sqrt{\frac{21}{10}} C_{33,11}^{33,1}}{1280}\)

\noindent\(C_{31,1101,2}^{22,0011,0,\text{NonLoc}}= \frac{7 C_{11,11}^{51,0}}{1024 \sqrt{15}}+\frac{7 C_{11,11}^{51,1}}{512 \sqrt{15}}+\frac{7
C_{22,11}^{42,0}}{2560}+\frac{7 C_{22,11}^{42,1}}{1280}+\frac{7 C_{31,11}^{31,0}}{1024 \sqrt{15}}+\frac{7 C_{31,11}^{31,1}}{512 \sqrt{15}}+\frac{3
\sqrt{\frac{21}{10}} C_{33,11}^{33,0}}{1280}+\frac{3}{640} \sqrt{\frac{21}{10}} C_{33,11}^{33,1}\)

\noindent\(C_{31,1101,2}^{22,0011,1,\text{Loc}}= \frac{7 C_{11,11}^{51,1}}{1024 \sqrt{5}}-\frac{7 \sqrt{3} C_{22,11}^{42,1}}{2560}+\frac{7
C_{31,11}^{31,1}}{1024 \sqrt{5}}+\frac{9 \sqrt{\frac{7}{10}} C_{33,11}^{33,1}}{1280}\)

\noindent\(C_{31,1101,2}^{22,0011,1,\text{NonLoc}}= \frac{7 C_{11,11}^{51,0}}{1024 \sqrt{5}}+\frac{7 \sqrt{3} C_{22,11}^{42,0}}{2560}+\frac{7
C_{31,11}^{31,0}}{1024 \sqrt{5}}+\frac{9 \sqrt{\frac{7}{10}} C_{33,11}^{33,0}}{1280}\)

\noindent\(C_{31,1110,1}^{11,1110,0,\text{Loc}}= -\frac{7 C_{00,20}^{60,0}}{384}-\frac{7 C_{00,20}^{60,1}}{768}-\frac{7 C_{00,22}^{62,0}}{128
\sqrt{5}}-\frac{7 C_{00,22}^{62,1}}{256 \sqrt{5}}+\frac{7 C_{11,20}^{31,0}}{1152}+\frac{7 C_{11,20}^{31,1}}{2304}-\frac{7 C_{11,22}^{51,0}}{1152
\sqrt{5}}-\frac{7 C_{11,22}^{51,1}}{2304 \sqrt{5}}+\frac{3 \sqrt{21} C_{11,22}^{53,0}}{640}+\frac{3 \sqrt{21} C_{11,22}^{53,1}}{1280}-\frac{5 C_{20,20}^{40,0}}{1152}-\frac{5
C_{20,20}^{40,1}}{2304}-\frac{7 C_{20,22}^{42,0}}{1152 \sqrt{5}}-\frac{7 C_{20,22}^{42,1}}{2304 \sqrt{5}}+\frac{7 C_{22,20}^{42,0}}{576 \sqrt{5}}+\frac{7
C_{22,20}^{42,1}}{1152 \sqrt{5}}+\frac{17 C_{22,22}^{40,0}}{1152 \sqrt{5}}+\frac{17 C_{22,22}^{40,1}}{2304 \sqrt{5}}+\frac{\sqrt{35} C_{22,22}^{42,0}}{576}+\frac{\sqrt{35}
C_{22,22}^{42,1}}{1152}-\frac{3 C_{22,22}^{44,0}}{160}-\frac{3 C_{22,22}^{44,1}}{320}-\frac{C_{31,20}^{31,0}}{1152}-\frac{C_{31,20}^{31,1}}{2304}-\frac{23
C_{31,22}^{31,0}}{1152 \sqrt{5}}-\frac{23 C_{31,22}^{31,1}}{2304 \sqrt{5}}+\frac{1}{640} \sqrt{\frac{7}{3}} C_{31,22}^{33,0}+\frac{\sqrt{\frac{7}{3}}
C_{31,22}^{33,1}}{1280}-\frac{\sqrt{21} C_{33,20}^{33,0}}{320}-\frac{\sqrt{21} C_{33,20}^{33,1}}{640}-\frac{3}{160} \sqrt{\frac{7}{10}} C_{33,22}^{33,0}-\frac{3}{320}
\sqrt{\frac{7}{10}} C_{33,22}^{33,1}\)

\noindent\(C_{31,1110,1}^{11,1110,0,\text{NonLoc}}= -\frac{7 C_{00,00}^{60,0}}{512 \sqrt{3}}-\frac{7 C_{00,00}^{60,1}}{256 \sqrt{3}}-\frac{7
C_{00,20}^{60,0}}{1536}-\frac{7 C_{00,20}^{60,1}}{768}+\frac{7 C_{00,22}^{62,0}}{256 \sqrt{5}}+\frac{7 C_{00,22}^{62,1}}{128 \sqrt{5}}-\frac{7 C_{11,00}^{51,0}}{1536
\sqrt{3}}-\frac{7 C_{11,00}^{51,1}}{768 \sqrt{3}}-\frac{7 C_{11,20}^{31,0}}{4608}-\frac{7 C_{11,20}^{31,1}}{2304}-\frac{7 C_{11,22}^{51,0}}{2304
\sqrt{5}}-\frac{7 C_{11,22}^{51,1}}{1152 \sqrt{5}}+\frac{3 \sqrt{21} C_{11,22}^{53,0}}{1280}+\frac{3 \sqrt{21} C_{11,22}^{53,1}}{640}-\frac{5 C_{20,00}^{40,0}}{1536
\sqrt{3}}-\frac{5 C_{20,00}^{40,1}}{768 \sqrt{3}}-\frac{5 C_{20,20}^{40,0}}{4608}-\frac{5 C_{20,20}^{40,1}}{2304}+\frac{7 C_{20,22}^{42,0}}{2304
\sqrt{5}}+\frac{7 C_{20,22}^{42,1}}{1152 \sqrt{5}}+\frac{7 C_{22,00}^{42,0}}{768 \sqrt{15}}+\frac{7 C_{22,00}^{42,1}}{384 \sqrt{15}}+\frac{7 C_{22,20}^{42,0}}{2304
\sqrt{5}}+\frac{7 C_{22,20}^{42,1}}{1152 \sqrt{5}}-\frac{17 C_{22,22}^{40,0}}{2304 \sqrt{5}}-\frac{17 C_{22,22}^{40,1}}{1152 \sqrt{5}}-\frac{\sqrt{35}
C_{22,22}^{42,0}}{1152}-\frac{\sqrt{35} C_{22,22}^{42,1}}{576}+\frac{3 C_{22,22}^{44,0}}{320}+\frac{3 C_{22,22}^{44,1}}{160}+\frac{C_{31,00}^{31,0}}{1536
\sqrt{3}}+\frac{C_{31,00}^{31,1}}{768 \sqrt{3}}+\frac{C_{31,20}^{31,0}}{4608}+\frac{C_{31,20}^{31,1}}{2304}-\frac{23 C_{31,22}^{31,0}}{2304 \sqrt{5}}-\frac{23
C_{31,22}^{31,1}}{1152 \sqrt{5}}+\frac{\sqrt{\frac{7}{3}} C_{31,22}^{33,0}}{1280}+\frac{1}{640} \sqrt{\frac{7}{3}} C_{31,22}^{33,1}+\frac{3 \sqrt{7}
C_{33,00}^{33,0}}{1280}+\frac{3 \sqrt{7} C_{33,00}^{33,1}}{640}+\frac{\sqrt{21} C_{33,20}^{33,0}}{1280}+\frac{\sqrt{21} C_{33,20}^{33,1}}{640}-\frac{3}{320}
\sqrt{\frac{7}{10}} C_{33,22}^{33,0}-\frac{3}{160} \sqrt{\frac{7}{10}} C_{33,22}^{33,1}\)

\noindent\(C_{31,1110,1}^{11,1110,1,\text{Loc}}= -\frac{7 C_{00,20}^{60,1}}{256 \sqrt{3}}-\frac{7}{256} \sqrt{\frac{3}{5}} C_{00,22}^{62,1}+\frac{7
C_{11,20}^{31,1}}{768 \sqrt{3}}-\frac{7 C_{11,22}^{51,1}}{768 \sqrt{15}}+\frac{9 \sqrt{7} C_{11,22}^{53,1}}{1280}-\frac{5 C_{20,20}^{40,1}}{768 \sqrt{3}}-\frac{7
C_{20,22}^{42,1}}{768 \sqrt{15}}+\frac{7 C_{22,20}^{42,1}}{384 \sqrt{15}}+\frac{17 C_{22,22}^{40,1}}{768 \sqrt{15}}+\frac{1}{384} \sqrt{\frac{35}{3}}
C_{22,22}^{42,1}-\frac{3 \sqrt{3} C_{22,22}^{44,1}}{320}-\frac{C_{31,20}^{31,1}}{768 \sqrt{3}}-\frac{23 C_{31,22}^{31,1}}{768 \sqrt{15}}+\frac{\sqrt{7}
C_{31,22}^{33,1}}{1280}-\frac{3 \sqrt{7} C_{33,20}^{33,1}}{640}-\frac{3}{320} \sqrt{\frac{21}{10}} C_{33,22}^{33,1}\)

\noindent\(C_{31,1110,1}^{11,1110,1,\text{NonLoc}}= -\frac{7 C_{00,00}^{60,0}}{512}-\frac{7 C_{00,20}^{60,0}}{512 \sqrt{3}}+\frac{7}{256}
\sqrt{\frac{3}{5}} C_{00,22}^{62,0}-\frac{7 C_{11,00}^{51,0}}{1536}-\frac{7 C_{11,20}^{31,0}}{1536 \sqrt{3}}-\frac{7 C_{11,22}^{51,0}}{768 \sqrt{15}}+\frac{9
\sqrt{7} C_{11,22}^{53,0}}{1280}-\frac{5 C_{20,00}^{40,0}}{1536}-\frac{5 C_{20,20}^{40,0}}{1536 \sqrt{3}}+\frac{7 C_{20,22}^{42,0}}{768 \sqrt{15}}+\frac{7
C_{22,00}^{42,0}}{768 \sqrt{5}}+\frac{7 C_{22,20}^{42,0}}{768 \sqrt{15}}-\frac{17 C_{22,22}^{40,0}}{768 \sqrt{15}}-\frac{1}{384} \sqrt{\frac{35}{3}}
C_{22,22}^{42,0}+\frac{3 \sqrt{3} C_{22,22}^{44,0}}{320}+\frac{C_{31,00}^{31,0}}{1536}+\frac{C_{31,20}^{31,0}}{1536 \sqrt{3}}-\frac{23 C_{31,22}^{31,0}}{768
\sqrt{15}}+\frac{\sqrt{7} C_{31,22}^{33,0}}{1280}+\frac{3 \sqrt{21} C_{33,00}^{33,0}}{1280}+\frac{3 \sqrt{7} C_{33,20}^{33,0}}{1280}-\frac{3}{320}
\sqrt{\frac{21}{10}} C_{33,22}^{33,0}\)

\noindent\(C_{31,1110,1}^{11,1112,0,\text{Loc}}= -\frac{7 C_{00,20}^{60,0}}{192 \sqrt{5}}-\frac{7 C_{00,20}^{60,1}}{384 \sqrt{5}}-\frac{7
C_{00,22}^{62,0}}{320}-\frac{7 C_{00,22}^{62,1}}{640}+\frac{7 C_{11,20}^{31,0}}{576 \sqrt{5}}+\frac{7 C_{11,20}^{31,1}}{1152 \sqrt{5}}+\frac{49 C_{11,22}^{51,0}}{5760}+\frac{49
C_{11,22}^{51,1}}{11520}+\frac{3 \sqrt{\frac{21}{5}} C_{11,22}^{53,0}}{1280}+\frac{3 \sqrt{\frac{21}{5}} C_{11,22}^{53,1}}{2560}+\frac{7 C_{20,20}^{40,0}}{576
\sqrt{5}}+\frac{7 C_{20,20}^{40,1}}{1152 \sqrt{5}}+\frac{7 C_{20,22}^{42,0}}{5760}+\frac{7 C_{20,22}^{42,1}}{11520}-\frac{7 C_{22,20}^{42,0}}{2880}-\frac{7
C_{22,20}^{42,1}}{5760}+\frac{7 C_{22,22}^{40,0}}{1440}+\frac{7 C_{22,22}^{40,1}}{2880}-\frac{\sqrt{7} C_{22,22}^{42,0}}{576}-\frac{\sqrt{7} C_{22,22}^{42,1}}{1152}+\frac{3
C_{22,22}^{44,0}}{160 \sqrt{5}}+\frac{3 C_{22,22}^{44,1}}{320 \sqrt{5}}-\frac{7 C_{31,20}^{31,0}}{576 \sqrt{5}}-\frac{7 C_{31,20}^{31,1}}{1152 \sqrt{5}}-\frac{7
C_{31,22}^{31,0}}{5760}-\frac{7 C_{31,22}^{31,1}}{11520}-\frac{17 \sqrt{\frac{7}{15}} C_{31,22}^{33,0}}{1280}-\frac{17 \sqrt{\frac{7}{15}} C_{31,22}^{33,1}}{2560}+\frac{1}{320}
\sqrt{\frac{21}{5}} C_{33,20}^{33,0}+\frac{1}{640} \sqrt{\frac{21}{5}} C_{33,20}^{33,1}+\frac{3}{800} \sqrt{\frac{7}{2}} C_{33,22}^{33,0}+\frac{3
\sqrt{\frac{7}{2}} C_{33,22}^{33,1}}{1600}\)

\noindent\(C_{31,1110,1}^{11,1112,0,\text{NonLoc}}= -\frac{7 C_{00,00}^{60,0}}{256 \sqrt{15}}-\frac{7 C_{00,00}^{60,1}}{128 \sqrt{15}}-\frac{7
C_{00,20}^{60,0}}{768 \sqrt{5}}-\frac{7 C_{00,20}^{60,1}}{384 \sqrt{5}}+\frac{7 C_{00,22}^{62,0}}{640}+\frac{7 C_{00,22}^{62,1}}{320}-\frac{7 C_{11,00}^{51,0}}{768
\sqrt{15}}-\frac{7 C_{11,00}^{51,1}}{384 \sqrt{15}}-\frac{7 C_{11,20}^{31,0}}{2304 \sqrt{5}}-\frac{7 C_{11,20}^{31,1}}{1152 \sqrt{5}}+\frac{49 C_{11,22}^{51,0}}{11520}+\frac{49
C_{11,22}^{51,1}}{5760}+\frac{3 \sqrt{\frac{21}{5}} C_{11,22}^{53,0}}{2560}+\frac{3 \sqrt{\frac{21}{5}} C_{11,22}^{53,1}}{1280}+\frac{7 C_{20,00}^{40,0}}{768
\sqrt{15}}+\frac{7 C_{20,00}^{40,1}}{384 \sqrt{15}}+\frac{7 C_{20,20}^{40,0}}{2304 \sqrt{5}}+\frac{7 C_{20,20}^{40,1}}{1152 \sqrt{5}}-\frac{7 C_{20,22}^{42,0}}{11520}-\frac{7
C_{20,22}^{42,1}}{5760}-\frac{7 C_{22,00}^{42,0}}{3840 \sqrt{3}}-\frac{7 C_{22,00}^{42,1}}{1920 \sqrt{3}}-\frac{7 C_{22,20}^{42,0}}{11520}-\frac{7
C_{22,20}^{42,1}}{5760}-\frac{7 C_{22,22}^{40,0}}{2880}-\frac{7 C_{22,22}^{40,1}}{1440}+\frac{\sqrt{7} C_{22,22}^{42,0}}{1152}+\frac{\sqrt{7} C_{22,22}^{42,1}}{576}-\frac{3
C_{22,22}^{44,0}}{320 \sqrt{5}}-\frac{3 C_{22,22}^{44,1}}{160 \sqrt{5}}+\frac{7 C_{31,00}^{31,0}}{768 \sqrt{15}}+\frac{7 C_{31,00}^{31,1}}{384 \sqrt{15}}+\frac{7
C_{31,20}^{31,0}}{2304 \sqrt{5}}+\frac{7 C_{31,20}^{31,1}}{1152 \sqrt{5}}-\frac{7 C_{31,22}^{31,0}}{11520}-\frac{7 C_{31,22}^{31,1}}{5760}-\frac{17
\sqrt{\frac{7}{15}} C_{31,22}^{33,0}}{2560}-\frac{17 \sqrt{\frac{7}{15}} C_{31,22}^{33,1}}{1280}-\frac{3 \sqrt{\frac{7}{5}} C_{33,00}^{33,0}}{1280}-\frac{3}{640}
\sqrt{\frac{7}{5}} C_{33,00}^{33,1}-\frac{\sqrt{\frac{21}{5}} C_{33,20}^{33,0}}{1280}-\frac{1}{640} \sqrt{\frac{21}{5}} C_{33,20}^{33,1}+\frac{3
\sqrt{\frac{7}{2}} C_{33,22}^{33,0}}{1600}+\frac{3}{800} \sqrt{\frac{7}{2}} C_{33,22}^{33,1}\)

\noindent\(C_{31,1110,1}^{11,1112,1,\text{Loc}}= -\frac{7 C_{00,20}^{60,1}}{128 \sqrt{15}}-\frac{7 \sqrt{3} C_{00,22}^{62,1}}{640}+\frac{7
C_{11,20}^{31,1}}{384 \sqrt{15}}+\frac{49 C_{11,22}^{51,1}}{3840 \sqrt{3}}+\frac{9 \sqrt{\frac{7}{5}} C_{11,22}^{53,1}}{2560}+\frac{7 C_{20,20}^{40,1}}{384
\sqrt{15}}+\frac{7 C_{20,22}^{42,1}}{3840 \sqrt{3}}-\frac{7 C_{22,20}^{42,1}}{1920 \sqrt{3}}+\frac{7 C_{22,22}^{40,1}}{960 \sqrt{3}}-\frac{1}{384}
\sqrt{\frac{7}{3}} C_{22,22}^{42,1}+\frac{3}{320} \sqrt{\frac{3}{5}} C_{22,22}^{44,1}-\frac{7 C_{31,20}^{31,1}}{384 \sqrt{15}}-\frac{7 C_{31,22}^{31,1}}{3840
\sqrt{3}}-\frac{17 \sqrt{\frac{7}{5}} C_{31,22}^{33,1}}{2560}+\frac{3}{640} \sqrt{\frac{7}{5}} C_{33,20}^{33,1}+\frac{3 \sqrt{\frac{21}{2}} C_{33,22}^{33,1}}{1600}\)

\noindent\(C_{31,1110,1}^{11,1112,1,\text{NonLoc}}= -\frac{7 C_{00,00}^{60,0}}{256 \sqrt{5}}-\frac{7 C_{00,20}^{60,0}}{256 \sqrt{15}}+\frac{7
\sqrt{3} C_{00,22}^{62,0}}{640}-\frac{7 C_{11,00}^{51,0}}{768 \sqrt{5}}-\frac{7 C_{11,20}^{31,0}}{768 \sqrt{15}}+\frac{49 C_{11,22}^{51,0}}{3840
\sqrt{3}}+\frac{9 \sqrt{\frac{7}{5}} C_{11,22}^{53,0}}{2560}+\frac{7 C_{20,00}^{40,0}}{768 \sqrt{5}}+\frac{7 C_{20,20}^{40,0}}{768 \sqrt{15}}-\frac{7
C_{20,22}^{42,0}}{3840 \sqrt{3}}-\frac{7 C_{22,00}^{42,0}}{3840}-\frac{7 C_{22,20}^{42,0}}{3840 \sqrt{3}}-\frac{7 C_{22,22}^{40,0}}{960 \sqrt{3}}+\frac{1}{384}
\sqrt{\frac{7}{3}} C_{22,22}^{42,0}-\frac{3}{320} \sqrt{\frac{3}{5}} C_{22,22}^{44,0}+\frac{7 C_{31,00}^{31,0}}{768 \sqrt{5}}+\frac{7 C_{31,20}^{31,0}}{768
\sqrt{15}}-\frac{7 C_{31,22}^{31,0}}{3840 \sqrt{3}}-\frac{17 \sqrt{\frac{7}{5}} C_{31,22}^{33,0}}{2560}-\frac{3 \sqrt{\frac{21}{5}} C_{33,00}^{33,0}}{1280}-\frac{3
\sqrt{\frac{7}{5}} C_{33,20}^{33,0}}{1280}+\frac{3 \sqrt{\frac{21}{2}} C_{33,22}^{33,0}}{1600}\)

\noindent\(C_{31,1111,0}^{00,2000,0,\text{Loc}}= \frac{7 C_{11,11}^{51,0}}{384}+\frac{7 C_{11,11}^{51,1}}{768}+\frac{C_{31,11}^{31,0}}{128}+\frac{C_{31,11}^{31,1}}{256}-\frac{3}{160}
\sqrt{\frac{7}{2}} C_{33,11}^{33,0}-\frac{3}{320} \sqrt{\frac{7}{2}} C_{33,11}^{33,1}\)

\noindent\(C_{31,1111,0}^{00,2000,0,\text{NonLoc}}= \frac{7 C_{11,11}^{51,0}}{768}+\frac{7 C_{11,11}^{51,1}}{384}+\frac{C_{31,11}^{31,0}}{256}+\frac{C_{31,11}^{31,1}}{128}-\frac{3}{320}
\sqrt{\frac{7}{2}} C_{33,11}^{33,0}-\frac{3}{160} \sqrt{\frac{7}{2}} C_{33,11}^{33,1}\)

\noindent\(C_{31,1111,0}^{00,2000,1,\text{Loc}}= \frac{7 C_{11,11}^{51,1}}{256 \sqrt{3}}+\frac{\sqrt{3} C_{31,11}^{31,1}}{256}-\frac{3}{320}
\sqrt{\frac{21}{2}} C_{33,11}^{33,1}\)

\noindent\(C_{31,1111,0}^{00,2000,1,\text{NonLoc}}= \frac{7 C_{11,11}^{51,0}}{256 \sqrt{3}}+\frac{\sqrt{3} C_{31,11}^{31,0}}{256}-\frac{3}{320}
\sqrt{\frac{21}{2}} C_{33,11}^{33,0}\)

\noindent\(C_{31,1111,0}^{11,1111,0,\text{Loc}}= -\frac{C_{20,20}^{40,0}}{96 \sqrt{3}}-\frac{C_{20,20}^{40,1}}{192 \sqrt{3}}+\frac{7
C_{20,22}^{42,0}}{384 \sqrt{15}}+\frac{7 C_{20,22}^{42,1}}{768 \sqrt{15}}+\frac{7 C_{22,20}^{42,0}}{384 \sqrt{15}}+\frac{7 C_{22,20}^{42,1}}{768
\sqrt{15}}+\frac{1}{192} \sqrt{\frac{5}{3}} C_{22,22}^{40,0}+\frac{1}{384} \sqrt{\frac{5}{3}} C_{22,22}^{40,1}-\frac{7}{384} \sqrt{\frac{7}{15}}
C_{22,22}^{42,0}-\frac{7}{768} \sqrt{\frac{7}{15}} C_{22,22}^{42,1}+\frac{C_{31,20}^{31,0}}{192 \sqrt{3}}+\frac{C_{31,20}^{31,1}}{384 \sqrt{3}}-\frac{C_{31,22}^{31,0}}{192
\sqrt{15}}-\frac{C_{31,22}^{31,1}}{384 \sqrt{15}}-\frac{\sqrt{7} C_{31,22}^{33,0}}{640}-\frac{\sqrt{7} C_{31,22}^{33,1}}{1280}-\frac{3 \sqrt{7} C_{33,20}^{33,0}}{640}-\frac{3
\sqrt{7} C_{33,20}^{33,1}}{1280}+\frac{3}{160} \sqrt{\frac{21}{10}} C_{33,22}^{33,0}+\frac{3}{320} \sqrt{\frac{21}{10}} C_{33,22}^{33,1}\)

\noindent\(C_{31,1111,0}^{11,1111,0,\text{NonLoc}}= -\frac{C_{20,00}^{40,0}}{384}-\frac{C_{20,00}^{40,1}}{192}-\frac{C_{20,20}^{40,0}}{384
\sqrt{3}}-\frac{C_{20,20}^{40,1}}{192 \sqrt{3}}-\frac{7 C_{20,22}^{42,0}}{768 \sqrt{15}}-\frac{7 C_{20,22}^{42,1}}{384 \sqrt{15}}+\frac{7 C_{22,00}^{42,0}}{1536
\sqrt{5}}+\frac{7 C_{22,00}^{42,1}}{768 \sqrt{5}}+\frac{7 C_{22,20}^{42,0}}{1536 \sqrt{15}}+\frac{7 C_{22,20}^{42,1}}{768 \sqrt{15}}-\frac{1}{384}
\sqrt{\frac{5}{3}} C_{22,22}^{40,0}-\frac{1}{192} \sqrt{\frac{5}{3}} C_{22,22}^{40,1}+\frac{7}{768} \sqrt{\frac{7}{15}} C_{22,22}^{42,0}+\frac{7}{384}
\sqrt{\frac{7}{15}} C_{22,22}^{42,1}-\frac{C_{31,00}^{31,0}}{768}-\frac{C_{31,00}^{31,1}}{384}-\frac{C_{31,20}^{31,0}}{768 \sqrt{3}}-\frac{C_{31,20}^{31,1}}{384
\sqrt{3}}-\frac{C_{31,22}^{31,0}}{384 \sqrt{15}}-\frac{C_{31,22}^{31,1}}{192 \sqrt{15}}-\frac{\sqrt{7} C_{31,22}^{33,0}}{1280}-\frac{\sqrt{7} C_{31,22}^{33,1}}{640}+\frac{3
\sqrt{21} C_{33,00}^{33,0}}{2560}+\frac{3 \sqrt{21} C_{33,00}^{33,1}}{1280}+\frac{3 \sqrt{7} C_{33,20}^{33,0}}{2560}+\frac{3 \sqrt{7} C_{33,20}^{33,1}}{1280}+\frac{3}{320}
\sqrt{\frac{21}{10}} C_{33,22}^{33,0}+\frac{3}{160} \sqrt{\frac{21}{10}} C_{33,22}^{33,1}\)

\noindent\(C_{31,1111,0}^{11,1111,1,\text{Loc}}= -\frac{C_{20,20}^{40,1}}{192}+\frac{7 C_{20,22}^{42,1}}{768 \sqrt{5}}+\frac{7 C_{22,20}^{42,1}}{768
\sqrt{5}}+\frac{\sqrt{5} C_{22,22}^{40,1}}{384}-\frac{7}{768} \sqrt{\frac{7}{5}} C_{22,22}^{42,1}+\frac{C_{31,20}^{31,1}}{384}-\frac{C_{31,22}^{31,1}}{384
\sqrt{5}}-\frac{\sqrt{21} C_{31,22}^{33,1}}{1280}-\frac{3 \sqrt{21} C_{33,20}^{33,1}}{1280}+\frac{9}{320} \sqrt{\frac{7}{10}} C_{33,22}^{33,1}\)

\noindent\(C_{31,1111,0}^{11,1111,1,\text{NonLoc}}= -\frac{C_{20,00}^{40,0}}{128 \sqrt{3}}-\frac{C_{20,20}^{40,0}}{384}-\frac{7 C_{20,22}^{42,0}}{768
\sqrt{5}}+\frac{7 C_{22,00}^{42,0}}{512 \sqrt{15}}+\frac{7 C_{22,20}^{42,0}}{1536 \sqrt{5}}-\frac{\sqrt{5} C_{22,22}^{40,0}}{384}+\frac{7}{768} \sqrt{\frac{7}{5}}
C_{22,22}^{42,0}-\frac{C_{31,00}^{31,0}}{256 \sqrt{3}}-\frac{C_{31,20}^{31,0}}{768}-\frac{C_{31,22}^{31,0}}{384 \sqrt{5}}-\frac{\sqrt{21} C_{31,22}^{33,0}}{1280}+\frac{9
\sqrt{7} C_{33,00}^{33,0}}{2560}+\frac{3 \sqrt{21} C_{33,20}^{33,0}}{2560}+\frac{9}{320} \sqrt{\frac{7}{10}} C_{33,22}^{33,0}\)

\noindent\(C_{31,1111,0}^{20,0000,0,\text{Loc}}= -\frac{7 C_{11,11}^{51,0}}{1536}-\frac{7 C_{11,11}^{51,1}}{3072}+\frac{7 C_{22,11}^{42,0}}{256
\sqrt{15}}+\frac{7 C_{22,11}^{42,1}}{512 \sqrt{15}}-\frac{7 C_{31,11}^{31,0}}{1536}-\frac{7 C_{31,11}^{31,1}}{3072}-\frac{3}{640} \sqrt{\frac{7}{2}}
C_{33,11}^{33,0}-\frac{3 \sqrt{\frac{7}{2}} C_{33,11}^{33,1}}{1280}\)

\noindent\(C_{31,1111,0}^{20,0000,0,\text{NonLoc}}= -\frac{7 C_{11,11}^{51,0}}{3072}-\frac{7 C_{11,11}^{51,1}}{1536}-\frac{7 C_{22,11}^{42,0}}{512
\sqrt{15}}-\frac{7 C_{22,11}^{42,1}}{256 \sqrt{15}}-\frac{7 C_{31,11}^{31,0}}{3072}-\frac{7 C_{31,11}^{31,1}}{1536}-\frac{3 \sqrt{\frac{7}{2}} C_{33,11}^{33,0}}{1280}-\frac{3}{640}
\sqrt{\frac{7}{2}} C_{33,11}^{33,1}\)

\noindent\(C_{31,1111,0}^{20,0000,1,\text{Loc}}= -\frac{7 C_{11,11}^{51,1}}{1024 \sqrt{3}}+\frac{7 C_{22,11}^{42,1}}{512 \sqrt{5}}-\frac{7
C_{31,11}^{31,1}}{1024 \sqrt{3}}-\frac{3 \sqrt{\frac{21}{2}} C_{33,11}^{33,1}}{1280}\)

\noindent\(C_{31,1111,0}^{20,0000,1,\text{NonLoc}}= -\frac{7 C_{11,11}^{51,0}}{1024 \sqrt{3}}-\frac{7 C_{22,11}^{42,0}}{512 \sqrt{5}}-\frac{7
C_{31,11}^{31,0}}{1024 \sqrt{3}}-\frac{3 \sqrt{\frac{21}{2}} C_{33,11}^{33,0}}{1280}\)

\noindent\(C_{31,1111,1}^{11,1111,0,\text{Loc}}= -\frac{7 C_{00,20}^{60,0}}{256}-\frac{7 C_{00,20}^{60,1}}{512}+\frac{21 C_{00,22}^{62,0}}{512
\sqrt{5}}+\frac{21 C_{00,22}^{62,1}}{1024 \sqrt{5}}+\frac{7 C_{11,20}^{31,0}}{768}+\frac{7 C_{11,20}^{31,1}}{1536}+\frac{7 C_{11,22}^{51,0}}{1536
\sqrt{5}}+\frac{7 C_{11,22}^{51,1}}{3072 \sqrt{5}}-\frac{9 \sqrt{21} C_{11,22}^{53,0}}{2560}-\frac{9 \sqrt{21} C_{11,22}^{53,1}}{5120}-\frac{C_{20,20}^{40,0}}{768}-\frac{C_{20,20}^{40,1}}{1536}-\frac{7
C_{20,22}^{42,0}}{1536 \sqrt{5}}-\frac{7 C_{20,22}^{42,1}}{3072 \sqrt{5}}+\frac{7 C_{22,20}^{42,0}}{768 \sqrt{5}}+\frac{7 C_{22,20}^{42,1}}{1536
\sqrt{5}}-\frac{37 C_{22,22}^{40,0}}{1536 \sqrt{5}}-\frac{37 C_{22,22}^{40,1}}{3072 \sqrt{5}}+\frac{1}{384} \sqrt{\frac{7}{5}} C_{22,22}^{42,0}+\frac{1}{768}
\sqrt{\frac{7}{5}} C_{22,22}^{42,1}+\frac{9 C_{22,22}^{44,0}}{640}+\frac{9 C_{22,22}^{44,1}}{1280}-\frac{C_{31,20}^{31,0}}{256}-\frac{C_{31,20}^{31,1}}{512}+\frac{9
C_{31,22}^{31,0}}{512 \sqrt{5}}+\frac{9 C_{31,22}^{31,1}}{1024 \sqrt{5}}+\frac{\sqrt{21} C_{31,22}^{33,0}}{2560}+\frac{\sqrt{21} C_{31,22}^{33,1}}{5120}-\frac{3
\sqrt{21} C_{33,20}^{33,0}}{1280}-\frac{3 \sqrt{21} C_{33,20}^{33,1}}{2560}-\frac{9}{640} \sqrt{\frac{7}{10}} C_{33,22}^{33,0}-\frac{9 \sqrt{\frac{7}{10}}
C_{33,22}^{33,1}}{1280}\)

\noindent\(C_{31,1111,1}^{11,1111,0,\text{NonLoc}}= -\frac{7 \sqrt{3} C_{00,00}^{60,0}}{1024}-\frac{7 \sqrt{3} C_{00,00}^{60,1}}{512}-\frac{7
C_{00,20}^{60,0}}{1024}-\frac{7 C_{00,20}^{60,1}}{512}-\frac{21 C_{00,22}^{62,0}}{1024 \sqrt{5}}-\frac{21 C_{00,22}^{62,1}}{512 \sqrt{5}}-\frac{7
C_{11,00}^{51,0}}{1024 \sqrt{3}}-\frac{7 C_{11,00}^{51,1}}{512 \sqrt{3}}-\frac{7 C_{11,20}^{31,0}}{3072}-\frac{7 C_{11,20}^{31,1}}{1536}+\frac{7
C_{11,22}^{51,0}}{3072 \sqrt{5}}+\frac{7 C_{11,22}^{51,1}}{1536 \sqrt{5}}-\frac{9 \sqrt{21} C_{11,22}^{53,0}}{5120}-\frac{9 \sqrt{21} C_{11,22}^{53,1}}{2560}-\frac{C_{20,00}^{40,0}}{1024
\sqrt{3}}-\frac{C_{20,00}^{40,1}}{512 \sqrt{3}}-\frac{C_{20,20}^{40,0}}{3072}-\frac{C_{20,20}^{40,1}}{1536}+\frac{7 C_{20,22}^{42,0}}{3072 \sqrt{5}}+\frac{7
C_{20,22}^{42,1}}{1536 \sqrt{5}}+\frac{7 C_{22,00}^{42,0}}{1024 \sqrt{15}}+\frac{7 C_{22,00}^{42,1}}{512 \sqrt{15}}+\frac{7 C_{22,20}^{42,0}}{3072
\sqrt{5}}+\frac{7 C_{22,20}^{42,1}}{1536 \sqrt{5}}+\frac{37 C_{22,22}^{40,0}}{3072 \sqrt{5}}+\frac{37 C_{22,22}^{40,1}}{1536 \sqrt{5}}-\frac{1}{768}
\sqrt{\frac{7}{5}} C_{22,22}^{42,0}-\frac{1}{384} \sqrt{\frac{7}{5}} C_{22,22}^{42,1}-\frac{9 C_{22,22}^{44,0}}{1280}-\frac{9 C_{22,22}^{44,1}}{640}+\frac{\sqrt{3}
C_{31,00}^{31,0}}{1024}+\frac{\sqrt{3} C_{31,00}^{31,1}}{512}+\frac{C_{31,20}^{31,0}}{1024}+\frac{C_{31,20}^{31,1}}{512}+\frac{9 C_{31,22}^{31,0}}{1024
\sqrt{5}}+\frac{9 C_{31,22}^{31,1}}{512 \sqrt{5}}+\frac{\sqrt{21} C_{31,22}^{33,0}}{5120}+\frac{\sqrt{21} C_{31,22}^{33,1}}{2560}+\frac{9 \sqrt{7}
C_{33,00}^{33,0}}{5120}+\frac{9 \sqrt{7} C_{33,00}^{33,1}}{2560}+\frac{3 \sqrt{21} C_{33,20}^{33,0}}{5120}+\frac{3 \sqrt{21} C_{33,20}^{33,1}}{2560}-\frac{9
\sqrt{\frac{7}{10}} C_{33,22}^{33,0}}{1280}-\frac{9}{640} \sqrt{\frac{7}{10}} C_{33,22}^{33,1}\)

\noindent\(C_{31,1111,1}^{11,1111,1,\text{Loc}}= -\frac{7 \sqrt{3} C_{00,20}^{60,1}}{512}+\frac{21 \sqrt{\frac{3}{5}} C_{00,22}^{62,1}}{1024}+\frac{7
C_{11,20}^{31,1}}{512 \sqrt{3}}+\frac{7 C_{11,22}^{51,1}}{1024 \sqrt{15}}-\frac{27 \sqrt{7} C_{11,22}^{53,1}}{5120}-\frac{C_{20,20}^{40,1}}{512 \sqrt{3}}-\frac{7
C_{20,22}^{42,1}}{1024 \sqrt{15}}+\frac{7 C_{22,20}^{42,1}}{512 \sqrt{15}}-\frac{37 C_{22,22}^{40,1}}{1024 \sqrt{15}}+\frac{1}{256} \sqrt{\frac{7}{15}}
C_{22,22}^{42,1}+\frac{9 \sqrt{3} C_{22,22}^{44,1}}{1280}-\frac{\sqrt{3} C_{31,20}^{31,1}}{512}+\frac{9 \sqrt{\frac{3}{5}} C_{31,22}^{31,1}}{1024}+\frac{3
\sqrt{7} C_{31,22}^{33,1}}{5120}-\frac{9 \sqrt{7} C_{33,20}^{33,1}}{2560}-\frac{9 \sqrt{\frac{21}{10}} C_{33,22}^{33,1}}{1280}\)

\noindent\(C_{31,1111,1}^{11,1111,1,\text{NonLoc}}= -\frac{21 C_{00,00}^{60,0}}{1024}-\frac{7 \sqrt{3} C_{00,20}^{60,0}}{1024}-\frac{21
\sqrt{\frac{3}{5}} C_{00,22}^{62,0}}{1024}-\frac{7 C_{11,00}^{51,0}}{1024}-\frac{7 C_{11,20}^{31,0}}{1024 \sqrt{3}}+\frac{7 C_{11,22}^{51,0}}{1024
\sqrt{15}}-\frac{27 \sqrt{7} C_{11,22}^{53,0}}{5120}-\frac{C_{20,00}^{40,0}}{1024}-\frac{C_{20,20}^{40,0}}{1024 \sqrt{3}}+\frac{7 C_{20,22}^{42,0}}{1024
\sqrt{15}}+\frac{7 C_{22,00}^{42,0}}{1024 \sqrt{5}}+\frac{7 C_{22,20}^{42,0}}{1024 \sqrt{15}}+\frac{37 C_{22,22}^{40,0}}{1024 \sqrt{15}}-\frac{1}{256}
\sqrt{\frac{7}{15}} C_{22,22}^{42,0}-\frac{9 \sqrt{3} C_{22,22}^{44,0}}{1280}+\frac{3 C_{31,00}^{31,0}}{1024}+\frac{\sqrt{3} C_{31,20}^{31,0}}{1024}+\frac{9
\sqrt{\frac{3}{5}} C_{31,22}^{31,0}}{1024}+\frac{3 \sqrt{7} C_{31,22}^{33,0}}{5120}+\frac{9 \sqrt{21} C_{33,00}^{33,0}}{5120}+\frac{9 \sqrt{7} C_{33,20}^{33,0}}{5120}-\frac{9
\sqrt{\frac{21}{10}} C_{33,22}^{33,0}}{1280}\)

\noindent\(C_{31,1111,1}^{11,1112,0,\text{Loc}}= \frac{7}{256} \sqrt{\frac{3}{5}} C_{00,20}^{60,0}+\frac{7}{512} \sqrt{\frac{3}{5}}
C_{00,20}^{60,1}-\frac{21 \sqrt{3} C_{00,22}^{62,0}}{2560}-\frac{21 \sqrt{3} C_{00,22}^{62,1}}{5120}-\frac{7 C_{11,20}^{31,0}}{256 \sqrt{15}}-\frac{7
C_{11,20}^{31,1}}{512 \sqrt{15}}+\frac{7 C_{11,22}^{51,0}}{512 \sqrt{3}}+\frac{7 C_{11,22}^{51,1}}{1024 \sqrt{3}}-\frac{7 C_{20,20}^{40,0}}{256 \sqrt{15}}-\frac{7
C_{20,20}^{40,1}}{512 \sqrt{15}}-\frac{7 C_{20,22}^{42,0}}{2560 \sqrt{3}}-\frac{7 C_{20,22}^{42,1}}{5120 \sqrt{3}}+\frac{7 C_{22,20}^{42,0}}{1280
\sqrt{3}}+\frac{7 C_{22,20}^{42,1}}{2560 \sqrt{3}}-\frac{7 C_{22,22}^{40,0}}{2560 \sqrt{3}}-\frac{7 C_{22,22}^{40,1}}{5120 \sqrt{3}}+\frac{1}{640}
\sqrt{\frac{7}{3}} C_{22,22}^{42,0}+\frac{\sqrt{\frac{7}{3}} C_{22,22}^{42,1}}{1280}+\frac{9}{640} \sqrt{\frac{3}{5}} C_{22,22}^{44,0}+\frac{9 \sqrt{\frac{3}{5}}
C_{22,22}^{44,1}}{1280}+\frac{7 C_{31,20}^{31,0}}{256 \sqrt{15}}+\frac{7 C_{31,20}^{31,1}}{512 \sqrt{15}}+\frac{7 C_{31,22}^{31,0}}{2560 \sqrt{3}}+\frac{7
C_{31,22}^{31,1}}{5120 \sqrt{3}}-\frac{3}{640} \sqrt{\frac{7}{5}} C_{31,22}^{33,0}-\frac{3 \sqrt{\frac{7}{5}} C_{31,22}^{33,1}}{1280}-\frac{9 \sqrt{\frac{7}{5}}
C_{33,20}^{33,0}}{1280}-\frac{9 \sqrt{\frac{7}{5}} C_{33,20}^{33,1}}{2560}-\frac{9 \sqrt{\frac{21}{2}} C_{33,22}^{33,0}}{3200}-\frac{9 \sqrt{\frac{21}{2}}
C_{33,22}^{33,1}}{6400}\)

\noindent\(C_{31,1111,1}^{11,1112,0,\text{NonLoc}}= \frac{21 C_{00,00}^{60,0}}{1024 \sqrt{5}}+\frac{21 C_{00,00}^{60,1}}{512 \sqrt{5}}+\frac{7
\sqrt{\frac{3}{5}} C_{00,20}^{60,0}}{1024}+\frac{7}{512} \sqrt{\frac{3}{5}} C_{00,20}^{60,1}+\frac{21 \sqrt{3} C_{00,22}^{62,0}}{5120}+\frac{21 \sqrt{3}
C_{00,22}^{62,1}}{2560}+\frac{7 C_{11,00}^{51,0}}{1024 \sqrt{5}}+\frac{7 C_{11,00}^{51,1}}{512 \sqrt{5}}+\frac{7 C_{11,20}^{31,0}}{1024 \sqrt{15}}+\frac{7
C_{11,20}^{31,1}}{512 \sqrt{15}}+\frac{7 C_{11,22}^{51,0}}{1024 \sqrt{3}}+\frac{7 C_{11,22}^{51,1}}{512 \sqrt{3}}-\frac{7 C_{20,00}^{40,0}}{1024
\sqrt{5}}-\frac{7 C_{20,00}^{40,1}}{512 \sqrt{5}}-\frac{7 C_{20,20}^{40,0}}{1024 \sqrt{15}}-\frac{7 C_{20,20}^{40,1}}{512 \sqrt{15}}+\frac{7 C_{20,22}^{42,0}}{5120
\sqrt{3}}+\frac{7 C_{20,22}^{42,1}}{2560 \sqrt{3}}+\frac{7 C_{22,00}^{42,0}}{5120}+\frac{7 C_{22,00}^{42,1}}{2560}+\frac{7 C_{22,20}^{42,0}}{5120
\sqrt{3}}+\frac{7 C_{22,20}^{42,1}}{2560 \sqrt{3}}+\frac{7 C_{22,22}^{40,0}}{5120 \sqrt{3}}+\frac{7 C_{22,22}^{40,1}}{2560 \sqrt{3}}-\frac{\sqrt{\frac{7}{3}}
C_{22,22}^{42,0}}{1280}-\frac{1}{640} \sqrt{\frac{7}{3}} C_{22,22}^{42,1}-\frac{9 \sqrt{\frac{3}{5}} C_{22,22}^{44,0}}{1280}-\frac{9}{640} \sqrt{\frac{3}{5}}
C_{22,22}^{44,1}-\frac{7 C_{31,00}^{31,0}}{1024 \sqrt{5}}-\frac{7 C_{31,00}^{31,1}}{512 \sqrt{5}}-\frac{7 C_{31,20}^{31,0}}{1024 \sqrt{15}}-\frac{7
C_{31,20}^{31,1}}{512 \sqrt{15}}+\frac{7 C_{31,22}^{31,0}}{5120 \sqrt{3}}+\frac{7 C_{31,22}^{31,1}}{2560 \sqrt{3}}-\frac{3 \sqrt{\frac{7}{5}} C_{31,22}^{33,0}}{1280}-\frac{3}{640}
\sqrt{\frac{7}{5}} C_{31,22}^{33,1}+\frac{9 \sqrt{\frac{21}{5}} C_{33,00}^{33,0}}{5120}+\frac{9 \sqrt{\frac{21}{5}} C_{33,00}^{33,1}}{2560}+\frac{9
\sqrt{\frac{7}{5}} C_{33,20}^{33,0}}{5120}+\frac{9 \sqrt{\frac{7}{5}} C_{33,20}^{33,1}}{2560}-\frac{9 \sqrt{\frac{21}{2}} C_{33,22}^{33,0}}{6400}-\frac{9
\sqrt{\frac{21}{2}} C_{33,22}^{33,1}}{3200}\)

\noindent\(C_{31,1111,1}^{11,1112,1,\text{Loc}}= \frac{21 C_{00,20}^{60,1}}{512 \sqrt{5}}-\frac{63 C_{00,22}^{62,1}}{5120}-\frac{7
C_{11,20}^{31,1}}{512 \sqrt{5}}+\frac{7 C_{11,22}^{51,1}}{1024}-\frac{7 C_{20,20}^{40,1}}{512 \sqrt{5}}-\frac{7 C_{20,22}^{42,1}}{5120}+\frac{7 C_{22,20}^{42,1}}{2560}-\frac{7
C_{22,22}^{40,1}}{5120}+\frac{\sqrt{7} C_{22,22}^{42,1}}{1280}+\frac{27 C_{22,22}^{44,1}}{1280 \sqrt{5}}+\frac{7 C_{31,20}^{31,1}}{512 \sqrt{5}}+\frac{7
C_{31,22}^{31,1}}{5120}-\frac{3 \sqrt{\frac{21}{5}} C_{31,22}^{33,1}}{1280}-\frac{9 \sqrt{\frac{21}{5}} C_{33,20}^{33,1}}{2560}-\frac{27 \sqrt{\frac{7}{2}}
C_{33,22}^{33,1}}{6400}\)

\noindent\(C_{31,1111,1}^{11,1112,1,\text{NonLoc}}= \frac{21 \sqrt{\frac{3}{5}} C_{00,00}^{60,0}}{1024}+\frac{21 C_{00,20}^{60,0}}{1024
\sqrt{5}}+\frac{63 C_{00,22}^{62,0}}{5120}+\frac{7 \sqrt{\frac{3}{5}} C_{11,00}^{51,0}}{1024}+\frac{7 C_{11,20}^{31,0}}{1024 \sqrt{5}}+\frac{7 C_{11,22}^{51,0}}{1024}-\frac{7
\sqrt{\frac{3}{5}} C_{20,00}^{40,0}}{1024}-\frac{7 C_{20,20}^{40,0}}{1024 \sqrt{5}}+\frac{7 C_{20,22}^{42,0}}{5120}+\frac{7 \sqrt{3} C_{22,00}^{42,0}}{5120}+\frac{7
C_{22,20}^{42,0}}{5120}+\frac{7 C_{22,22}^{40,0}}{5120}-\frac{\sqrt{7} C_{22,22}^{42,0}}{1280}-\frac{27 C_{22,22}^{44,0}}{1280 \sqrt{5}}-\frac{7
\sqrt{\frac{3}{5}} C_{31,00}^{31,0}}{1024}-\frac{7 C_{31,20}^{31,0}}{1024 \sqrt{5}}+\frac{7 C_{31,22}^{31,0}}{5120}-\frac{3 \sqrt{\frac{21}{5}} C_{31,22}^{33,0}}{1280}+\frac{27
\sqrt{\frac{7}{5}} C_{33,00}^{33,0}}{5120}+\frac{9 \sqrt{\frac{21}{5}} C_{33,20}^{33,0}}{5120}-\frac{27 \sqrt{\frac{7}{2}} C_{33,22}^{33,0}}{6400}\)

\noindent\(C_{31,1111,2}^{00,2202,0,\text{Loc}}= -\frac{7 C_{11,11}^{51,0}}{384 \sqrt{5}}-\frac{7 C_{11,11}^{51,1}}{768 \sqrt{5}}-\frac{C_{31,11}^{31,0}}{128
\sqrt{5}}-\frac{C_{31,11}^{31,1}}{256 \sqrt{5}}+\frac{3}{160} \sqrt{\frac{7}{10}} C_{33,11}^{33,0}+\frac{3}{320} \sqrt{\frac{7}{10}} C_{33,11}^{33,1}\)

\noindent\(C_{31,1111,2}^{00,2202,0,\text{NonLoc}}= -\frac{7 C_{11,11}^{51,0}}{768 \sqrt{5}}-\frac{7 C_{11,11}^{51,1}}{384 \sqrt{5}}-\frac{C_{31,11}^{31,0}}{256
\sqrt{5}}-\frac{C_{31,11}^{31,1}}{128 \sqrt{5}}+\frac{3}{320} \sqrt{\frac{7}{10}} C_{33,11}^{33,0}+\frac{3}{160} \sqrt{\frac{7}{10}} C_{33,11}^{33,1}\)

\noindent\(C_{31,1111,2}^{00,2202,1,\text{Loc}}= -\frac{7 C_{11,11}^{51,1}}{256 \sqrt{15}}-\frac{1}{256} \sqrt{\frac{3}{5}} C_{31,11}^{31,1}+\frac{3}{320}
\sqrt{\frac{21}{10}} C_{33,11}^{33,1}\)

\noindent\(C_{31,1111,2}^{00,2202,1,\text{NonLoc}}= -\frac{7 C_{11,11}^{51,0}}{256 \sqrt{15}}-\frac{1}{256} \sqrt{\frac{3}{5}} C_{31,11}^{31,0}+\frac{3}{320}
\sqrt{\frac{21}{10}} C_{33,11}^{33,0}\)

\noindent\(C_{31,1111,2}^{11,1111,0,\text{Loc}}= -\frac{7}{256} \sqrt{\frac{3}{5}} C_{00,20}^{60,0}-\frac{7}{512} \sqrt{\frac{3}{5}}
C_{00,20}^{60,1}+\frac{21 \sqrt{3} C_{00,22}^{62,0}}{2560}+\frac{21 \sqrt{3} C_{00,22}^{62,1}}{5120}+\frac{7 C_{11,20}^{31,0}}{256 \sqrt{15}}+\frac{7
C_{11,20}^{31,1}}{512 \sqrt{15}}+\frac{7 C_{11,22}^{51,0}}{2560 \sqrt{3}}+\frac{7 C_{11,22}^{51,1}}{5120 \sqrt{3}}-\frac{27 \sqrt{\frac{7}{5}} C_{11,22}^{53,0}}{2560}-\frac{27
\sqrt{\frac{7}{5}} C_{11,22}^{53,1}}{5120}-\frac{19 C_{20,20}^{40,0}}{768 \sqrt{15}}-\frac{19 C_{20,20}^{40,1}}{1536 \sqrt{15}}+\frac{7 C_{20,22}^{42,0}}{1536
\sqrt{3}}+\frac{7 C_{20,22}^{42,1}}{3072 \sqrt{3}}+\frac{49 C_{22,20}^{42,0}}{3840 \sqrt{3}}+\frac{49 C_{22,20}^{42,1}}{7680 \sqrt{3}}-\frac{31 C_{22,22}^{40,0}}{7680
\sqrt{3}}-\frac{31 C_{22,22}^{40,1}}{15360 \sqrt{3}}-\frac{11 \sqrt{\frac{7}{3}} C_{22,22}^{42,0}}{1920}-\frac{11 \sqrt{\frac{7}{3}} C_{22,22}^{42,1}}{3840}+\frac{9}{640}
\sqrt{\frac{3}{5}} C_{22,22}^{44,0}+\frac{9 \sqrt{\frac{3}{5}} C_{22,22}^{44,1}}{1280}-\frac{C_{31,20}^{31,0}}{768 \sqrt{15}}-\frac{C_{31,20}^{31,1}}{1536
\sqrt{15}}+\frac{13 C_{31,22}^{31,0}}{1536 \sqrt{3}}+\frac{13 C_{31,22}^{31,1}}{3072 \sqrt{3}}-\frac{1}{512} \sqrt{\frac{7}{5}} C_{31,22}^{33,0}-\frac{\sqrt{\frac{7}{5}}
C_{31,22}^{33,1}}{1024}-\frac{21 \sqrt{\frac{7}{5}} C_{33,20}^{33,0}}{1280}-\frac{21 \sqrt{\frac{7}{5}} C_{33,20}^{33,1}}{2560}+\frac{3}{640} \sqrt{\frac{21}{2}}
C_{33,22}^{33,0}+\frac{3 \sqrt{\frac{21}{2}} C_{33,22}^{33,1}}{1280}\)

\noindent\(C_{31,1111,2}^{11,1111,0,\text{NonLoc}}= -\frac{21 C_{00,00}^{60,0}}{1024 \sqrt{5}}-\frac{21 C_{00,00}^{60,1}}{512 \sqrt{5}}-\frac{7
\sqrt{\frac{3}{5}} C_{00,20}^{60,0}}{1024}-\frac{7}{512} \sqrt{\frac{3}{5}} C_{00,20}^{60,1}-\frac{21 \sqrt{3} C_{00,22}^{62,0}}{5120}-\frac{21 \sqrt{3}
C_{00,22}^{62,1}}{2560}-\frac{7 C_{11,00}^{51,0}}{1024 \sqrt{5}}-\frac{7 C_{11,00}^{51,1}}{512 \sqrt{5}}-\frac{7 C_{11,20}^{31,0}}{1024 \sqrt{15}}-\frac{7
C_{11,20}^{31,1}}{512 \sqrt{15}}+\frac{7 C_{11,22}^{51,0}}{5120 \sqrt{3}}+\frac{7 C_{11,22}^{51,1}}{2560 \sqrt{3}}-\frac{27 \sqrt{\frac{7}{5}} C_{11,22}^{53,0}}{5120}-\frac{27
\sqrt{\frac{7}{5}} C_{11,22}^{53,1}}{2560}-\frac{19 C_{20,00}^{40,0}}{3072 \sqrt{5}}-\frac{19 C_{20,00}^{40,1}}{1536 \sqrt{5}}-\frac{19 C_{20,20}^{40,0}}{3072
\sqrt{15}}-\frac{19 C_{20,20}^{40,1}}{1536 \sqrt{15}}-\frac{7 C_{20,22}^{42,0}}{3072 \sqrt{3}}-\frac{7 C_{20,22}^{42,1}}{1536 \sqrt{3}}+\frac{49
C_{22,00}^{42,0}}{15360}+\frac{49 C_{22,00}^{42,1}}{7680}+\frac{49 C_{22,20}^{42,0}}{15360 \sqrt{3}}+\frac{49 C_{22,20}^{42,1}}{7680 \sqrt{3}}+\frac{31
C_{22,22}^{40,0}}{15360 \sqrt{3}}+\frac{31 C_{22,22}^{40,1}}{7680 \sqrt{3}}+\frac{11 \sqrt{\frac{7}{3}} C_{22,22}^{42,0}}{3840}+\frac{11 \sqrt{\frac{7}{3}}
C_{22,22}^{42,1}}{1920}-\frac{9 \sqrt{\frac{3}{5}} C_{22,22}^{44,0}}{1280}-\frac{9}{640} \sqrt{\frac{3}{5}} C_{22,22}^{44,1}+\frac{C_{31,00}^{31,0}}{3072
\sqrt{5}}+\frac{C_{31,00}^{31,1}}{1536 \sqrt{5}}+\frac{C_{31,20}^{31,0}}{3072 \sqrt{15}}+\frac{C_{31,20}^{31,1}}{1536 \sqrt{15}}+\frac{13 C_{31,22}^{31,0}}{3072
\sqrt{3}}+\frac{13 C_{31,22}^{31,1}}{1536 \sqrt{3}}-\frac{\sqrt{\frac{7}{5}} C_{31,22}^{33,0}}{1024}-\frac{1}{512} \sqrt{\frac{7}{5}} C_{31,22}^{33,1}+\frac{21
\sqrt{\frac{21}{5}} C_{33,00}^{33,0}}{5120}+\frac{21 \sqrt{\frac{21}{5}} C_{33,00}^{33,1}}{2560}+\frac{21 \sqrt{\frac{7}{5}} C_{33,20}^{33,0}}{5120}+\frac{21
\sqrt{\frac{7}{5}} C_{33,20}^{33,1}}{2560}+\frac{3 \sqrt{\frac{21}{2}} C_{33,22}^{33,0}}{1280}+\frac{3}{640} \sqrt{\frac{21}{2}} C_{33,22}^{33,1}\)

\noindent\(C_{31,1111,2}^{11,1111,1,\text{Loc}}= -\frac{21 C_{00,20}^{60,1}}{512 \sqrt{5}}+\frac{63 C_{00,22}^{62,1}}{5120}+\frac{7
C_{11,20}^{31,1}}{512 \sqrt{5}}+\frac{7 C_{11,22}^{51,1}}{5120}-\frac{27 \sqrt{\frac{21}{5}} C_{11,22}^{53,1}}{5120}-\frac{19 C_{20,20}^{40,1}}{1536
\sqrt{5}}+\frac{7 C_{20,22}^{42,1}}{3072}+\frac{49 C_{22,20}^{42,1}}{7680}-\frac{31 C_{22,22}^{40,1}}{15360}-\frac{11 \sqrt{7} C_{22,22}^{42,1}}{3840}+\frac{27
C_{22,22}^{44,1}}{1280 \sqrt{5}}-\frac{C_{31,20}^{31,1}}{1536 \sqrt{5}}+\frac{13 C_{31,22}^{31,1}}{3072}-\frac{\sqrt{\frac{21}{5}} C_{31,22}^{33,1}}{1024}-\frac{21
\sqrt{\frac{21}{5}} C_{33,20}^{33,1}}{2560}+\frac{9 \sqrt{\frac{7}{2}} C_{33,22}^{33,1}}{1280}\)

\noindent\(C_{31,1111,2}^{11,1111,1,\text{NonLoc}}= -\frac{21 \sqrt{\frac{3}{5}} C_{00,00}^{60,0}}{1024}-\frac{21 C_{00,20}^{60,0}}{1024
\sqrt{5}}-\frac{63 C_{00,22}^{62,0}}{5120}-\frac{7 \sqrt{\frac{3}{5}} C_{11,00}^{51,0}}{1024}-\frac{7 C_{11,20}^{31,0}}{1024 \sqrt{5}}+\frac{7 C_{11,22}^{51,0}}{5120}-\frac{27
\sqrt{\frac{21}{5}} C_{11,22}^{53,0}}{5120}-\frac{19 C_{20,00}^{40,0}}{1024 \sqrt{15}}-\frac{19 C_{20,20}^{40,0}}{3072 \sqrt{5}}-\frac{7 C_{20,22}^{42,0}}{3072}+\frac{49
C_{22,00}^{42,0}}{5120 \sqrt{3}}+\frac{49 C_{22,20}^{42,0}}{15360}+\frac{31 C_{22,22}^{40,0}}{15360}+\frac{11 \sqrt{7} C_{22,22}^{42,0}}{3840}-\frac{27
C_{22,22}^{44,0}}{1280 \sqrt{5}}+\frac{C_{31,00}^{31,0}}{1024 \sqrt{15}}+\frac{C_{31,20}^{31,0}}{3072 \sqrt{5}}+\frac{13 C_{31,22}^{31,0}}{3072}-\frac{\sqrt{\frac{21}{5}}
C_{31,22}^{33,0}}{1024}+\frac{63 \sqrt{\frac{7}{5}} C_{33,00}^{33,0}}{5120}+\frac{21 \sqrt{\frac{21}{5}} C_{33,20}^{33,0}}{5120}+\frac{9 \sqrt{\frac{7}{2}}
C_{33,22}^{33,0}}{1280}\)

\noindent\(C_{31,1111,2}^{11,1112,0,\text{Loc}}= -\frac{7 C_{00,20}^{60,0}}{256 \sqrt{5}}-\frac{7 C_{00,20}^{60,1}}{512 \sqrt{5}}+\frac{21
C_{00,22}^{62,0}}{2560}+\frac{21 C_{00,22}^{62,1}}{5120}+\frac{7 C_{11,20}^{31,0}}{768 \sqrt{5}}+\frac{7 C_{11,20}^{31,1}}{1536 \sqrt{5}}-\frac{7
C_{11,22}^{51,0}}{1536}-\frac{7 C_{11,22}^{51,1}}{3072}+\frac{7 C_{20,20}^{40,0}}{768 \sqrt{5}}+\frac{7 C_{20,20}^{40,1}}{1536 \sqrt{5}}+\frac{7
C_{20,22}^{42,0}}{7680}+\frac{7 C_{20,22}^{42,1}}{15360}-\frac{7 C_{22,20}^{42,0}}{3840}-\frac{7 C_{22,20}^{42,1}}{7680}+\frac{7 C_{22,22}^{40,0}}{7680}+\frac{7
C_{22,22}^{40,1}}{15360}-\frac{\sqrt{7} C_{22,22}^{42,0}}{1920}-\frac{\sqrt{7} C_{22,22}^{42,1}}{3840}-\frac{9 C_{22,22}^{44,0}}{640 \sqrt{5}}-\frac{9
C_{22,22}^{44,1}}{1280 \sqrt{5}}-\frac{7 C_{31,20}^{31,0}}{768 \sqrt{5}}-\frac{7 C_{31,20}^{31,1}}{1536 \sqrt{5}}-\frac{7 C_{31,22}^{31,0}}{7680}-\frac{7
C_{31,22}^{31,1}}{15360}+\frac{1}{640} \sqrt{\frac{21}{5}} C_{31,22}^{33,0}+\frac{\sqrt{\frac{21}{5}} C_{31,22}^{33,1}}{1280}+\frac{3 \sqrt{\frac{21}{5}}
C_{33,20}^{33,0}}{1280}+\frac{3 \sqrt{\frac{21}{5}} C_{33,20}^{33,1}}{2560}+\frac{9 \sqrt{\frac{7}{2}} C_{33,22}^{33,0}}{3200}+\frac{9 \sqrt{\frac{7}{2}}
C_{33,22}^{33,1}}{6400}\)

\noindent\(C_{31,1111,2}^{11,1112,0,\text{NonLoc}}= -\frac{7 \sqrt{\frac{3}{5}} C_{00,00}^{60,0}}{1024}-\frac{7}{512} \sqrt{\frac{3}{5}}
C_{00,00}^{60,1}-\frac{7 C_{00,20}^{60,0}}{1024 \sqrt{5}}-\frac{7 C_{00,20}^{60,1}}{512 \sqrt{5}}-\frac{21 C_{00,22}^{62,0}}{5120}-\frac{21 C_{00,22}^{62,1}}{2560}-\frac{7
C_{11,00}^{51,0}}{1024 \sqrt{15}}-\frac{7 C_{11,00}^{51,1}}{512 \sqrt{15}}-\frac{7 C_{11,20}^{31,0}}{3072 \sqrt{5}}-\frac{7 C_{11,20}^{31,1}}{1536
\sqrt{5}}-\frac{7 C_{11,22}^{51,0}}{3072}-\frac{7 C_{11,22}^{51,1}}{1536}+\frac{7 C_{20,00}^{40,0}}{1024 \sqrt{15}}+\frac{7 C_{20,00}^{40,1}}{512
\sqrt{15}}+\frac{7 C_{20,20}^{40,0}}{3072 \sqrt{5}}+\frac{7 C_{20,20}^{40,1}}{1536 \sqrt{5}}-\frac{7 C_{20,22}^{42,0}}{15360}-\frac{7 C_{20,22}^{42,1}}{7680}-\frac{7
C_{22,00}^{42,0}}{5120 \sqrt{3}}-\frac{7 C_{22,00}^{42,1}}{2560 \sqrt{3}}-\frac{7 C_{22,20}^{42,0}}{15360}-\frac{7 C_{22,20}^{42,1}}{7680}-\frac{7
C_{22,22}^{40,0}}{15360}-\frac{7 C_{22,22}^{40,1}}{7680}+\frac{\sqrt{7} C_{22,22}^{42,0}}{3840}+\frac{\sqrt{7} C_{22,22}^{42,1}}{1920}+\frac{9 C_{22,22}^{44,0}}{1280
\sqrt{5}}+\frac{9 C_{22,22}^{44,1}}{640 \sqrt{5}}+\frac{7 C_{31,00}^{31,0}}{1024 \sqrt{15}}+\frac{7 C_{31,00}^{31,1}}{512 \sqrt{15}}+\frac{7 C_{31,20}^{31,0}}{3072
\sqrt{5}}+\frac{7 C_{31,20}^{31,1}}{1536 \sqrt{5}}-\frac{7 C_{31,22}^{31,0}}{15360}-\frac{7 C_{31,22}^{31,1}}{7680}+\frac{\sqrt{\frac{21}{5}} C_{31,22}^{33,0}}{1280}+\frac{1}{640}
\sqrt{\frac{21}{5}} C_{31,22}^{33,1}-\frac{9 \sqrt{\frac{7}{5}} C_{33,00}^{33,0}}{5120}-\frac{9 \sqrt{\frac{7}{5}} C_{33,00}^{33,1}}{2560}-\frac{3
\sqrt{\frac{21}{5}} C_{33,20}^{33,0}}{5120}-\frac{3 \sqrt{\frac{21}{5}} C_{33,20}^{33,1}}{2560}+\frac{9 \sqrt{\frac{7}{2}} C_{33,22}^{33,0}}{6400}+\frac{9
\sqrt{\frac{7}{2}} C_{33,22}^{33,1}}{3200}\)

\noindent\(C_{31,1111,2}^{11,1112,1,\text{Loc}}= -\frac{7}{512} \sqrt{\frac{3}{5}} C_{00,20}^{60,1}+\frac{21 \sqrt{3} C_{00,22}^{62,1}}{5120}+\frac{7
C_{11,20}^{31,1}}{512 \sqrt{15}}-\frac{7 C_{11,22}^{51,1}}{1024 \sqrt{3}}+\frac{7 C_{20,20}^{40,1}}{512 \sqrt{15}}+\frac{7 C_{20,22}^{42,1}}{5120
\sqrt{3}}-\frac{7 C_{22,20}^{42,1}}{2560 \sqrt{3}}+\frac{7 C_{22,22}^{40,1}}{5120 \sqrt{3}}-\frac{\sqrt{\frac{7}{3}} C_{22,22}^{42,1}}{1280}-\frac{9
\sqrt{\frac{3}{5}} C_{22,22}^{44,1}}{1280}-\frac{7 C_{31,20}^{31,1}}{512 \sqrt{15}}-\frac{7 C_{31,22}^{31,1}}{5120 \sqrt{3}}+\frac{3 \sqrt{\frac{7}{5}}
C_{31,22}^{33,1}}{1280}+\frac{9 \sqrt{\frac{7}{5}} C_{33,20}^{33,1}}{2560}+\frac{9 \sqrt{\frac{21}{2}} C_{33,22}^{33,1}}{6400}\)

\noindent\(C_{31,1111,2}^{11,1112,1,\text{NonLoc}}= -\frac{21 C_{00,00}^{60,0}}{1024 \sqrt{5}}-\frac{7 \sqrt{\frac{3}{5}} C_{00,20}^{60,0}}{1024}-\frac{21
\sqrt{3} C_{00,22}^{62,0}}{5120}-\frac{7 C_{11,00}^{51,0}}{1024 \sqrt{5}}-\frac{7 C_{11,20}^{31,0}}{1024 \sqrt{15}}-\frac{7 C_{11,22}^{51,0}}{1024
\sqrt{3}}+\frac{7 C_{20,00}^{40,0}}{1024 \sqrt{5}}+\frac{7 C_{20,20}^{40,0}}{1024 \sqrt{15}}-\frac{7 C_{20,22}^{42,0}}{5120 \sqrt{3}}-\frac{7 C_{22,00}^{42,0}}{5120}-\frac{7
C_{22,20}^{42,0}}{5120 \sqrt{3}}-\frac{7 C_{22,22}^{40,0}}{5120 \sqrt{3}}+\frac{\sqrt{\frac{7}{3}} C_{22,22}^{42,0}}{1280}+\frac{9 \sqrt{\frac{3}{5}}
C_{22,22}^{44,0}}{1280}+\frac{7 C_{31,00}^{31,0}}{1024 \sqrt{5}}+\frac{7 C_{31,20}^{31,0}}{1024 \sqrt{15}}-\frac{7 C_{31,22}^{31,0}}{5120 \sqrt{3}}+\frac{3
\sqrt{\frac{7}{5}} C_{31,22}^{33,0}}{1280}-\frac{9 \sqrt{\frac{21}{5}} C_{33,00}^{33,0}}{5120}-\frac{9 \sqrt{\frac{7}{5}} C_{33,20}^{33,0}}{5120}+\frac{9
\sqrt{\frac{21}{2}} C_{33,22}^{33,0}}{6400}\)

\noindent\(C_{31,1111,2}^{22,0000,0,\text{Loc}}= -\frac{7 C_{11,11}^{51,0}}{768 \sqrt{5}}-\frac{7 C_{11,11}^{51,1}}{1536 \sqrt{5}}+\frac{7
C_{22,11}^{42,0}}{640 \sqrt{3}}+\frac{7 C_{22,11}^{42,1}}{1280 \sqrt{3}}-\frac{7 C_{31,11}^{31,0}}{768 \sqrt{5}}-\frac{7 C_{31,11}^{31,1}}{1536 \sqrt{5}}-\frac{3}{320}
\sqrt{\frac{7}{10}} C_{33,11}^{33,0}-\frac{3}{640} \sqrt{\frac{7}{10}} C_{33,11}^{33,1}\)

\noindent\(C_{31,1111,2}^{22,0000,0,\text{NonLoc}}= -\frac{7 C_{11,11}^{51,0}}{1536 \sqrt{5}}-\frac{7 C_{11,11}^{51,1}}{768 \sqrt{5}}-\frac{7
C_{22,11}^{42,0}}{1280 \sqrt{3}}-\frac{7 C_{22,11}^{42,1}}{640 \sqrt{3}}-\frac{7 C_{31,11}^{31,0}}{1536 \sqrt{5}}-\frac{7 C_{31,11}^{31,1}}{768 \sqrt{5}}-\frac{3}{640}
\sqrt{\frac{7}{10}} C_{33,11}^{33,0}-\frac{3}{320} \sqrt{\frac{7}{10}} C_{33,11}^{33,1}\)

\noindent\(C_{31,1111,2}^{22,0000,1,\text{Loc}}= -\frac{7 C_{11,11}^{51,1}}{512 \sqrt{15}}+\frac{7 C_{22,11}^{42,1}}{1280}-\frac{7
C_{31,11}^{31,1}}{512 \sqrt{15}}-\frac{3}{640} \sqrt{\frac{21}{10}} C_{33,11}^{33,1}\)

\noindent\(C_{31,1111,2}^{22,0000,1,\text{NonLoc}}= -\frac{7 C_{11,11}^{51,0}}{512 \sqrt{15}}-\frac{7 C_{22,11}^{42,0}}{1280}-\frac{7
C_{31,11}^{31,0}}{512 \sqrt{15}}-\frac{3}{640} \sqrt{\frac{21}{10}} C_{33,11}^{33,0}\)

\noindent\(C_{31,1112,1}^{11,1110,0,\text{Loc}}= -\frac{7 C_{00,20}^{60,0}}{192 \sqrt{5}}-\frac{7 C_{00,20}^{60,1}}{384 \sqrt{5}}-\frac{7
C_{00,22}^{62,0}}{320}-\frac{7 C_{00,22}^{62,1}}{640}+\frac{7 C_{11,20}^{31,0}}{576 \sqrt{5}}+\frac{7 C_{11,20}^{31,1}}{1152 \sqrt{5}}+\frac{49 C_{11,22}^{51,0}}{5760}+\frac{49
C_{11,22}^{51,1}}{11520}+\frac{3 \sqrt{\frac{21}{5}} C_{11,22}^{53,0}}{1280}+\frac{3 \sqrt{\frac{21}{5}} C_{11,22}^{53,1}}{2560}+\frac{7 C_{20,20}^{40,0}}{576
\sqrt{5}}+\frac{7 C_{20,20}^{40,1}}{1152 \sqrt{5}}+\frac{7 C_{20,22}^{42,0}}{5760}+\frac{7 C_{20,22}^{42,1}}{11520}-\frac{7 C_{22,20}^{42,0}}{2880}-\frac{7
C_{22,20}^{42,1}}{5760}+\frac{7 C_{22,22}^{40,0}}{1440}+\frac{7 C_{22,22}^{40,1}}{2880}-\frac{\sqrt{7} C_{22,22}^{42,0}}{576}-\frac{\sqrt{7} C_{22,22}^{42,1}}{1152}+\frac{3
C_{22,22}^{44,0}}{160 \sqrt{5}}+\frac{3 C_{22,22}^{44,1}}{320 \sqrt{5}}-\frac{7 C_{31,20}^{31,0}}{576 \sqrt{5}}-\frac{7 C_{31,20}^{31,1}}{1152 \sqrt{5}}-\frac{7
C_{31,22}^{31,0}}{5760}-\frac{7 C_{31,22}^{31,1}}{11520}-\frac{17 \sqrt{\frac{7}{15}} C_{31,22}^{33,0}}{1280}-\frac{17 \sqrt{\frac{7}{15}} C_{31,22}^{33,1}}{2560}+\frac{1}{320}
\sqrt{\frac{21}{5}} C_{33,20}^{33,0}+\frac{1}{640} \sqrt{\frac{21}{5}} C_{33,20}^{33,1}+\frac{3}{800} \sqrt{\frac{7}{2}} C_{33,22}^{33,0}+\frac{3
\sqrt{\frac{7}{2}} C_{33,22}^{33,1}}{1600}\)

\noindent\(C_{31,1112,1}^{11,1110,0,\text{NonLoc}}= -\frac{7 C_{00,00}^{60,0}}{256 \sqrt{15}}-\frac{7 C_{00,00}^{60,1}}{128 \sqrt{15}}-\frac{7
C_{00,20}^{60,0}}{768 \sqrt{5}}-\frac{7 C_{00,20}^{60,1}}{384 \sqrt{5}}+\frac{7 C_{00,22}^{62,0}}{640}+\frac{7 C_{00,22}^{62,1}}{320}-\frac{7 C_{11,00}^{51,0}}{768
\sqrt{15}}-\frac{7 C_{11,00}^{51,1}}{384 \sqrt{15}}-\frac{7 C_{11,20}^{31,0}}{2304 \sqrt{5}}-\frac{7 C_{11,20}^{31,1}}{1152 \sqrt{5}}+\frac{49 C_{11,22}^{51,0}}{11520}+\frac{49
C_{11,22}^{51,1}}{5760}+\frac{3 \sqrt{\frac{21}{5}} C_{11,22}^{53,0}}{2560}+\frac{3 \sqrt{\frac{21}{5}} C_{11,22}^{53,1}}{1280}+\frac{7 C_{20,00}^{40,0}}{768
\sqrt{15}}+\frac{7 C_{20,00}^{40,1}}{384 \sqrt{15}}+\frac{7 C_{20,20}^{40,0}}{2304 \sqrt{5}}+\frac{7 C_{20,20}^{40,1}}{1152 \sqrt{5}}-\frac{7 C_{20,22}^{42,0}}{11520}-\frac{7
C_{20,22}^{42,1}}{5760}-\frac{7 C_{22,00}^{42,0}}{3840 \sqrt{3}}-\frac{7 C_{22,00}^{42,1}}{1920 \sqrt{3}}-\frac{7 C_{22,20}^{42,0}}{11520}-\frac{7
C_{22,20}^{42,1}}{5760}-\frac{7 C_{22,22}^{40,0}}{2880}-\frac{7 C_{22,22}^{40,1}}{1440}+\frac{\sqrt{7} C_{22,22}^{42,0}}{1152}+\frac{\sqrt{7} C_{22,22}^{42,1}}{576}-\frac{3
C_{22,22}^{44,0}}{320 \sqrt{5}}-\frac{3 C_{22,22}^{44,1}}{160 \sqrt{5}}+\frac{7 C_{31,00}^{31,0}}{768 \sqrt{15}}+\frac{7 C_{31,00}^{31,1}}{384 \sqrt{15}}+\frac{7
C_{31,20}^{31,0}}{2304 \sqrt{5}}+\frac{7 C_{31,20}^{31,1}}{1152 \sqrt{5}}-\frac{7 C_{31,22}^{31,0}}{11520}-\frac{7 C_{31,22}^{31,1}}{5760}-\frac{17
\sqrt{\frac{7}{15}} C_{31,22}^{33,0}}{2560}-\frac{17 \sqrt{\frac{7}{15}} C_{31,22}^{33,1}}{1280}-\frac{3 \sqrt{\frac{7}{5}} C_{33,00}^{33,0}}{1280}-\frac{3}{640}
\sqrt{\frac{7}{5}} C_{33,00}^{33,1}-\frac{\sqrt{\frac{21}{5}} C_{33,20}^{33,0}}{1280}-\frac{1}{640} \sqrt{\frac{21}{5}} C_{33,20}^{33,1}+\frac{3
\sqrt{\frac{7}{2}} C_{33,22}^{33,0}}{1600}+\frac{3}{800} \sqrt{\frac{7}{2}} C_{33,22}^{33,1}\)

\noindent\(C_{31,1112,1}^{11,1110,1,\text{Loc}}= -\frac{7 C_{00,20}^{60,1}}{128 \sqrt{15}}-\frac{7 \sqrt{3} C_{00,22}^{62,1}}{640}+\frac{7
C_{11,20}^{31,1}}{384 \sqrt{15}}+\frac{49 C_{11,22}^{51,1}}{3840 \sqrt{3}}+\frac{9 \sqrt{\frac{7}{5}} C_{11,22}^{53,1}}{2560}+\frac{7 C_{20,20}^{40,1}}{384
\sqrt{15}}+\frac{7 C_{20,22}^{42,1}}{3840 \sqrt{3}}-\frac{7 C_{22,20}^{42,1}}{1920 \sqrt{3}}+\frac{7 C_{22,22}^{40,1}}{960 \sqrt{3}}-\frac{1}{384}
\sqrt{\frac{7}{3}} C_{22,22}^{42,1}+\frac{3}{320} \sqrt{\frac{3}{5}} C_{22,22}^{44,1}-\frac{7 C_{31,20}^{31,1}}{384 \sqrt{15}}-\frac{7 C_{31,22}^{31,1}}{3840
\sqrt{3}}-\frac{17 \sqrt{\frac{7}{5}} C_{31,22}^{33,1}}{2560}+\frac{3}{640} \sqrt{\frac{7}{5}} C_{33,20}^{33,1}+\frac{3 \sqrt{\frac{21}{2}} C_{33,22}^{33,1}}{1600}\)

\noindent\(C_{31,1112,1}^{11,1110,1,\text{NonLoc}}= -\frac{7 C_{00,00}^{60,0}}{256 \sqrt{5}}-\frac{7 C_{00,20}^{60,0}}{256 \sqrt{15}}+\frac{7
\sqrt{3} C_{00,22}^{62,0}}{640}-\frac{7 C_{11,00}^{51,0}}{768 \sqrt{5}}-\frac{7 C_{11,20}^{31,0}}{768 \sqrt{15}}+\frac{49 C_{11,22}^{51,0}}{3840
\sqrt{3}}+\frac{9 \sqrt{\frac{7}{5}} C_{11,22}^{53,0}}{2560}+\frac{7 C_{20,00}^{40,0}}{768 \sqrt{5}}+\frac{7 C_{20,20}^{40,0}}{768 \sqrt{15}}-\frac{7
C_{20,22}^{42,0}}{3840 \sqrt{3}}-\frac{7 C_{22,00}^{42,0}}{3840}-\frac{7 C_{22,20}^{42,0}}{3840 \sqrt{3}}-\frac{7 C_{22,22}^{40,0}}{960 \sqrt{3}}+\frac{1}{384}
\sqrt{\frac{7}{3}} C_{22,22}^{42,0}-\frac{3}{320} \sqrt{\frac{3}{5}} C_{22,22}^{44,0}+\frac{7 C_{31,00}^{31,0}}{768 \sqrt{5}}+\frac{7 C_{31,20}^{31,0}}{768
\sqrt{15}}-\frac{7 C_{31,22}^{31,0}}{3840 \sqrt{3}}-\frac{17 \sqrt{\frac{7}{5}} C_{31,22}^{33,0}}{2560}-\frac{3 \sqrt{\frac{21}{5}} C_{33,00}^{33,0}}{1280}-\frac{3
\sqrt{\frac{7}{5}} C_{33,20}^{33,0}}{1280}+\frac{3 \sqrt{\frac{21}{2}} C_{33,22}^{33,0}}{1600}\)

\noindent\(C_{31,1112,1}^{11,1111,0,\text{Loc}}= \frac{7}{256} \sqrt{\frac{3}{5}} C_{00,20}^{60,0}+\frac{7}{512} \sqrt{\frac{3}{5}}
C_{00,20}^{60,1}-\frac{21 \sqrt{3} C_{00,22}^{62,0}}{2560}-\frac{21 \sqrt{3} C_{00,22}^{62,1}}{5120}-\frac{7 C_{11,20}^{31,0}}{256 \sqrt{15}}-\frac{7
C_{11,20}^{31,1}}{512 \sqrt{15}}+\frac{7 C_{11,22}^{51,0}}{512 \sqrt{3}}+\frac{7 C_{11,22}^{51,1}}{1024 \sqrt{3}}-\frac{7 C_{20,20}^{40,0}}{256 \sqrt{15}}-\frac{7
C_{20,20}^{40,1}}{512 \sqrt{15}}-\frac{7 C_{20,22}^{42,0}}{2560 \sqrt{3}}-\frac{7 C_{20,22}^{42,1}}{5120 \sqrt{3}}+\frac{7 C_{22,20}^{42,0}}{1280
\sqrt{3}}+\frac{7 C_{22,20}^{42,1}}{2560 \sqrt{3}}-\frac{7 C_{22,22}^{40,0}}{2560 \sqrt{3}}-\frac{7 C_{22,22}^{40,1}}{5120 \sqrt{3}}+\frac{1}{640}
\sqrt{\frac{7}{3}} C_{22,22}^{42,0}+\frac{\sqrt{\frac{7}{3}} C_{22,22}^{42,1}}{1280}+\frac{9}{640} \sqrt{\frac{3}{5}} C_{22,22}^{44,0}+\frac{9 \sqrt{\frac{3}{5}}
C_{22,22}^{44,1}}{1280}+\frac{7 C_{31,20}^{31,0}}{256 \sqrt{15}}+\frac{7 C_{31,20}^{31,1}}{512 \sqrt{15}}+\frac{7 C_{31,22}^{31,0}}{2560 \sqrt{3}}+\frac{7
C_{31,22}^{31,1}}{5120 \sqrt{3}}-\frac{3}{640} \sqrt{\frac{7}{5}} C_{31,22}^{33,0}-\frac{3 \sqrt{\frac{7}{5}} C_{31,22}^{33,1}}{1280}-\frac{9 \sqrt{\frac{7}{5}}
C_{33,20}^{33,0}}{1280}-\frac{9 \sqrt{\frac{7}{5}} C_{33,20}^{33,1}}{2560}-\frac{9 \sqrt{\frac{21}{2}} C_{33,22}^{33,0}}{3200}-\frac{9 \sqrt{\frac{21}{2}}
C_{33,22}^{33,1}}{6400}\)

\noindent\(C_{31,1112,1}^{11,1111,0,\text{NonLoc}}= \frac{21 C_{00,00}^{60,0}}{1024 \sqrt{5}}+\frac{21 C_{00,00}^{60,1}}{512 \sqrt{5}}+\frac{7
\sqrt{\frac{3}{5}} C_{00,20}^{60,0}}{1024}+\frac{7}{512} \sqrt{\frac{3}{5}} C_{00,20}^{60,1}+\frac{21 \sqrt{3} C_{00,22}^{62,0}}{5120}+\frac{21 \sqrt{3}
C_{00,22}^{62,1}}{2560}+\frac{7 C_{11,00}^{51,0}}{1024 \sqrt{5}}+\frac{7 C_{11,00}^{51,1}}{512 \sqrt{5}}+\frac{7 C_{11,20}^{31,0}}{1024 \sqrt{15}}+\frac{7
C_{11,20}^{31,1}}{512 \sqrt{15}}+\frac{7 C_{11,22}^{51,0}}{1024 \sqrt{3}}+\frac{7 C_{11,22}^{51,1}}{512 \sqrt{3}}-\frac{7 C_{20,00}^{40,0}}{1024
\sqrt{5}}-\frac{7 C_{20,00}^{40,1}}{512 \sqrt{5}}-\frac{7 C_{20,20}^{40,0}}{1024 \sqrt{15}}-\frac{7 C_{20,20}^{40,1}}{512 \sqrt{15}}+\frac{7 C_{20,22}^{42,0}}{5120
\sqrt{3}}+\frac{7 C_{20,22}^{42,1}}{2560 \sqrt{3}}+\frac{7 C_{22,00}^{42,0}}{5120}+\frac{7 C_{22,00}^{42,1}}{2560}+\frac{7 C_{22,20}^{42,0}}{5120
\sqrt{3}}+\frac{7 C_{22,20}^{42,1}}{2560 \sqrt{3}}+\frac{7 C_{22,22}^{40,0}}{5120 \sqrt{3}}+\frac{7 C_{22,22}^{40,1}}{2560 \sqrt{3}}-\frac{\sqrt{\frac{7}{3}}
C_{22,22}^{42,0}}{1280}-\frac{1}{640} \sqrt{\frac{7}{3}} C_{22,22}^{42,1}-\frac{9 \sqrt{\frac{3}{5}} C_{22,22}^{44,0}}{1280}-\frac{9}{640} \sqrt{\frac{3}{5}}
C_{22,22}^{44,1}-\frac{7 C_{31,00}^{31,0}}{1024 \sqrt{5}}-\frac{7 C_{31,00}^{31,1}}{512 \sqrt{5}}-\frac{7 C_{31,20}^{31,0}}{1024 \sqrt{15}}-\frac{7
C_{31,20}^{31,1}}{512 \sqrt{15}}+\frac{7 C_{31,22}^{31,0}}{5120 \sqrt{3}}+\frac{7 C_{31,22}^{31,1}}{2560 \sqrt{3}}-\frac{3 \sqrt{\frac{7}{5}} C_{31,22}^{33,0}}{1280}-\frac{3}{640}
\sqrt{\frac{7}{5}} C_{31,22}^{33,1}+\frac{9 \sqrt{\frac{21}{5}} C_{33,00}^{33,0}}{5120}+\frac{9 \sqrt{\frac{21}{5}} C_{33,00}^{33,1}}{2560}+\frac{9
\sqrt{\frac{7}{5}} C_{33,20}^{33,0}}{5120}+\frac{9 \sqrt{\frac{7}{5}} C_{33,20}^{33,1}}{2560}-\frac{9 \sqrt{\frac{21}{2}} C_{33,22}^{33,0}}{6400}-\frac{9
\sqrt{\frac{21}{2}} C_{33,22}^{33,1}}{3200}\)

\noindent\(C_{31,1112,1}^{11,1111,1,\text{Loc}}= \frac{21 C_{00,20}^{60,1}}{512 \sqrt{5}}-\frac{63 C_{00,22}^{62,1}}{5120}-\frac{7
C_{11,20}^{31,1}}{512 \sqrt{5}}+\frac{7 C_{11,22}^{51,1}}{1024}-\frac{7 C_{20,20}^{40,1}}{512 \sqrt{5}}-\frac{7 C_{20,22}^{42,1}}{5120}+\frac{7 C_{22,20}^{42,1}}{2560}-\frac{7
C_{22,22}^{40,1}}{5120}+\frac{\sqrt{7} C_{22,22}^{42,1}}{1280}+\frac{27 C_{22,22}^{44,1}}{1280 \sqrt{5}}+\frac{7 C_{31,20}^{31,1}}{512 \sqrt{5}}+\frac{7
C_{31,22}^{31,1}}{5120}-\frac{3 \sqrt{\frac{21}{5}} C_{31,22}^{33,1}}{1280}-\frac{9 \sqrt{\frac{21}{5}} C_{33,20}^{33,1}}{2560}-\frac{27 \sqrt{\frac{7}{2}}
C_{33,22}^{33,1}}{6400}\)

\noindent\(C_{31,1112,1}^{11,1111,1,\text{NonLoc}}= \frac{21 \sqrt{\frac{3}{5}} C_{00,00}^{60,0}}{1024}+\frac{21 C_{00,20}^{60,0}}{1024
\sqrt{5}}+\frac{63 C_{00,22}^{62,0}}{5120}+\frac{7 \sqrt{\frac{3}{5}} C_{11,00}^{51,0}}{1024}+\frac{7 C_{11,20}^{31,0}}{1024 \sqrt{5}}+\frac{7 C_{11,22}^{51,0}}{1024}-\frac{7
\sqrt{\frac{3}{5}} C_{20,00}^{40,0}}{1024}-\frac{7 C_{20,20}^{40,0}}{1024 \sqrt{5}}+\frac{7 C_{20,22}^{42,0}}{5120}+\frac{7 \sqrt{3} C_{22,00}^{42,0}}{5120}+\frac{7
C_{22,20}^{42,0}}{5120}+\frac{7 C_{22,22}^{40,0}}{5120}-\frac{\sqrt{7} C_{22,22}^{42,0}}{1280}-\frac{27 C_{22,22}^{44,0}}{1280 \sqrt{5}}-\frac{7
\sqrt{\frac{3}{5}} C_{31,00}^{31,0}}{1024}-\frac{7 C_{31,20}^{31,0}}{1024 \sqrt{5}}+\frac{7 C_{31,22}^{31,0}}{5120}-\frac{3 \sqrt{\frac{21}{5}} C_{31,22}^{33,0}}{1280}+\frac{27
\sqrt{\frac{7}{5}} C_{33,00}^{33,0}}{5120}+\frac{9 \sqrt{\frac{21}{5}} C_{33,20}^{33,0}}{5120}-\frac{27 \sqrt{\frac{7}{2}} C_{33,22}^{33,0}}{6400}\)

\noindent\(C_{31,1112,1}^{11,1112,0,\text{Loc}}= -\frac{119 C_{00,20}^{60,0}}{3840}-\frac{119 C_{00,20}^{60,1}}{7680}-\frac{49 C_{00,22}^{62,0}}{2560
\sqrt{5}}-\frac{49 C_{00,22}^{62,1}}{5120 \sqrt{5}}+\frac{119 C_{11,20}^{31,0}}{11520}+\frac{119 C_{11,20}^{31,1}}{23040}+\frac{203 C_{11,22}^{51,0}}{23040
\sqrt{5}}+\frac{203 C_{11,22}^{51,1}}{46080 \sqrt{5}}+\frac{3 \sqrt{21} C_{11,22}^{53,0}}{12800}+\frac{3 \sqrt{21} C_{11,22}^{53,1}}{25600}-\frac{C_{20,20}^{40,0}}{11520}-\frac{C_{20,20}^{40,1}}{23040}-\frac{91
C_{20,22}^{42,0}}{23040 \sqrt{5}}-\frac{91 C_{20,22}^{42,1}}{46080 \sqrt{5}}+\frac{91 C_{22,20}^{42,0}}{11520 \sqrt{5}}+\frac{91 C_{22,20}^{42,1}}{23040
\sqrt{5}}-\frac{17 C_{22,22}^{40,0}}{4608 \sqrt{5}}-\frac{17 C_{22,22}^{40,1}}{9216 \sqrt{5}}+\frac{19 \sqrt{\frac{7}{5}} C_{22,22}^{42,0}}{5760}+\frac{19
\sqrt{\frac{7}{5}} C_{22,22}^{42,1}}{11520}+\frac{3 C_{22,22}^{44,0}}{640}+\frac{3 C_{22,22}^{44,1}}{1280}-\frac{59 C_{31,20}^{31,0}}{11520}-\frac{59
C_{31,20}^{31,1}}{23040}+\frac{31 C_{31,22}^{31,0}}{23040 \sqrt{5}}+\frac{31 C_{31,22}^{31,1}}{46080 \sqrt{5}}-\frac{17 \sqrt{\frac{7}{3}} C_{31,22}^{33,0}}{12800}-\frac{17
\sqrt{\frac{7}{3}} C_{31,22}^{33,1}}{25600}-\frac{13 \sqrt{21} C_{33,20}^{33,0}}{6400}-\frac{13 \sqrt{21} C_{33,20}^{33,1}}{12800}-\frac{39 \sqrt{\frac{7}{10}}
C_{33,22}^{33,0}}{3200}-\frac{39 \sqrt{\frac{7}{10}} C_{33,22}^{33,1}}{6400}\)

\noindent\(C_{31,1112,1}^{11,1112,0,\text{NonLoc}}= -\frac{119 C_{00,00}^{60,0}}{5120 \sqrt{3}}-\frac{119 C_{00,00}^{60,1}}{2560 \sqrt{3}}-\frac{119
C_{00,20}^{60,0}}{15360}-\frac{119 C_{00,20}^{60,1}}{7680}+\frac{49 C_{00,22}^{62,0}}{5120 \sqrt{5}}+\frac{49 C_{00,22}^{62,1}}{2560 \sqrt{5}}-\frac{119
C_{11,00}^{51,0}}{15360 \sqrt{3}}-\frac{119 C_{11,00}^{51,1}}{7680 \sqrt{3}}-\frac{119 C_{11,20}^{31,0}}{46080}-\frac{119 C_{11,20}^{31,1}}{23040}+\frac{203
C_{11,22}^{51,0}}{46080 \sqrt{5}}+\frac{203 C_{11,22}^{51,1}}{23040 \sqrt{5}}+\frac{3 \sqrt{21} C_{11,22}^{53,0}}{25600}+\frac{3 \sqrt{21} C_{11,22}^{53,1}}{12800}-\frac{C_{20,00}^{40,0}}{15360
\sqrt{3}}-\frac{C_{20,00}^{40,1}}{7680 \sqrt{3}}-\frac{C_{20,20}^{40,0}}{46080}-\frac{C_{20,20}^{40,1}}{23040}+\frac{91 C_{20,22}^{42,0}}{46080 \sqrt{5}}+\frac{91
C_{20,22}^{42,1}}{23040 \sqrt{5}}+\frac{91 C_{22,00}^{42,0}}{15360 \sqrt{15}}+\frac{91 C_{22,00}^{42,1}}{7680 \sqrt{15}}+\frac{91 C_{22,20}^{42,0}}{46080
\sqrt{5}}+\frac{91 C_{22,20}^{42,1}}{23040 \sqrt{5}}+\frac{17 C_{22,22}^{40,0}}{9216 \sqrt{5}}+\frac{17 C_{22,22}^{40,1}}{4608 \sqrt{5}}-\frac{19
\sqrt{\frac{7}{5}} C_{22,22}^{42,0}}{11520}-\frac{19 \sqrt{\frac{7}{5}} C_{22,22}^{42,1}}{5760}-\frac{3 C_{22,22}^{44,0}}{1280}-\frac{3 C_{22,22}^{44,1}}{640}+\frac{59
C_{31,00}^{31,0}}{15360 \sqrt{3}}+\frac{59 C_{31,00}^{31,1}}{7680 \sqrt{3}}+\frac{59 C_{31,20}^{31,0}}{46080}+\frac{59 C_{31,20}^{31,1}}{23040}+\frac{31
C_{31,22}^{31,0}}{46080 \sqrt{5}}+\frac{31 C_{31,22}^{31,1}}{23040 \sqrt{5}}-\frac{17 \sqrt{\frac{7}{3}} C_{31,22}^{33,0}}{25600}-\frac{17 \sqrt{\frac{7}{3}}
C_{31,22}^{33,1}}{12800}+\frac{39 \sqrt{7} C_{33,00}^{33,0}}{25600}+\frac{39 \sqrt{7} C_{33,00}^{33,1}}{12800}+\frac{13 \sqrt{21} C_{33,20}^{33,0}}{25600}+\frac{13
\sqrt{21} C_{33,20}^{33,1}}{12800}-\frac{39 \sqrt{\frac{7}{10}} C_{33,22}^{33,0}}{6400}-\frac{39 \sqrt{\frac{7}{10}} C_{33,22}^{33,1}}{3200}\)

\noindent\(C_{31,1112,1}^{11,1112,1,\text{Loc}}= -\frac{119 C_{00,20}^{60,1}}{2560 \sqrt{3}}-\frac{49 \sqrt{\frac{3}{5}} C_{00,22}^{62,1}}{5120}+\frac{119
C_{11,20}^{31,1}}{7680 \sqrt{3}}+\frac{203 C_{11,22}^{51,1}}{15360 \sqrt{15}}+\frac{9 \sqrt{7} C_{11,22}^{53,1}}{25600}-\frac{C_{20,20}^{40,1}}{7680
\sqrt{3}}-\frac{91 C_{20,22}^{42,1}}{15360 \sqrt{15}}+\frac{91 C_{22,20}^{42,1}}{7680 \sqrt{15}}-\frac{17 C_{22,22}^{40,1}}{3072 \sqrt{15}}+\frac{19
\sqrt{\frac{7}{15}} C_{22,22}^{42,1}}{3840}+\frac{3 \sqrt{3} C_{22,22}^{44,1}}{1280}-\frac{59 C_{31,20}^{31,1}}{7680 \sqrt{3}}+\frac{31 C_{31,22}^{31,1}}{15360
\sqrt{15}}-\frac{17 \sqrt{7} C_{31,22}^{33,1}}{25600}-\frac{39 \sqrt{7} C_{33,20}^{33,1}}{12800}-\frac{39 \sqrt{\frac{21}{10}} C_{33,22}^{33,1}}{6400}\)

\noindent\(C_{31,1112,1}^{11,1112,1,\text{NonLoc}}= -\frac{119 C_{00,00}^{60,0}}{5120}-\frac{119 C_{00,20}^{60,0}}{5120 \sqrt{3}}+\frac{49
\sqrt{\frac{3}{5}} C_{00,22}^{62,0}}{5120}-\frac{119 C_{11,00}^{51,0}}{15360}-\frac{119 C_{11,20}^{31,0}}{15360 \sqrt{3}}+\frac{203 C_{11,22}^{51,0}}{15360
\sqrt{15}}+\frac{9 \sqrt{7} C_{11,22}^{53,0}}{25600}-\frac{C_{20,00}^{40,0}}{15360}-\frac{C_{20,20}^{40,0}}{15360 \sqrt{3}}+\frac{91 C_{20,22}^{42,0}}{15360
\sqrt{15}}+\frac{91 C_{22,00}^{42,0}}{15360 \sqrt{5}}+\frac{91 C_{22,20}^{42,0}}{15360 \sqrt{15}}+\frac{17 C_{22,22}^{40,0}}{3072 \sqrt{15}}-\frac{19
\sqrt{\frac{7}{15}} C_{22,22}^{42,0}}{3840}-\frac{3 \sqrt{3} C_{22,22}^{44,0}}{1280}+\frac{59 C_{31,00}^{31,0}}{15360}+\frac{59 C_{31,20}^{31,0}}{15360
\sqrt{3}}+\frac{31 C_{31,22}^{31,0}}{15360 \sqrt{15}}-\frac{17 \sqrt{7} C_{31,22}^{33,0}}{25600}+\frac{39 \sqrt{21} C_{33,00}^{33,0}}{25600}+\frac{39
\sqrt{7} C_{33,20}^{33,0}}{25600}-\frac{39 \sqrt{\frac{21}{10}} C_{33,22}^{33,0}}{6400}\)

\noindent\(C_{31,1112,2}^{00,2202,0,\text{Loc}}= -\frac{7 C_{11,11}^{51,0}}{128 \sqrt{15}}-\frac{7 C_{11,11}^{51,1}}{256 \sqrt{15}}-\frac{1}{128}
\sqrt{\frac{3}{5}} C_{31,11}^{31,0}-\frac{1}{256} \sqrt{\frac{3}{5}} C_{31,11}^{31,1}+\frac{3}{160} \sqrt{\frac{21}{10}} C_{33,11}^{33,0}+\frac{3}{320}
\sqrt{\frac{21}{10}} C_{33,11}^{33,1}\)

\noindent\(C_{31,1112,2}^{00,2202,0,\text{NonLoc}}= -\frac{7 C_{11,11}^{51,0}}{256 \sqrt{15}}-\frac{7 C_{11,11}^{51,1}}{128 \sqrt{15}}-\frac{1}{256}
\sqrt{\frac{3}{5}} C_{31,11}^{31,0}-\frac{1}{128} \sqrt{\frac{3}{5}} C_{31,11}^{31,1}+\frac{3}{320} \sqrt{\frac{21}{10}} C_{33,11}^{33,0}+\frac{3}{160}
\sqrt{\frac{21}{10}} C_{33,11}^{33,1}\)

\noindent\(C_{31,1112,2}^{00,2202,1,\text{Loc}}= -\frac{7 C_{11,11}^{51,1}}{256 \sqrt{5}}-\frac{3 C_{31,11}^{31,1}}{256 \sqrt{5}}+\frac{9}{320}
\sqrt{\frac{7}{10}} C_{33,11}^{33,1}\)

\noindent\(C_{31,1112,2}^{00,2202,1,\text{NonLoc}}= -\frac{7 C_{11,11}^{51,0}}{256 \sqrt{5}}-\frac{3 C_{31,11}^{31,0}}{256 \sqrt{5}}+\frac{9}{320}
\sqrt{\frac{7}{10}} C_{33,11}^{33,0}\)

\noindent\(C_{31,1112,2}^{11,1111,0,\text{Loc}}= -\frac{7 C_{00,20}^{60,0}}{256 \sqrt{5}}-\frac{7 C_{00,20}^{60,1}}{512 \sqrt{5}}+\frac{21
C_{00,22}^{62,0}}{2560}+\frac{21 C_{00,22}^{62,1}}{5120}+\frac{7 C_{11,20}^{31,0}}{768 \sqrt{5}}+\frac{7 C_{11,20}^{31,1}}{1536 \sqrt{5}}-\frac{7
C_{11,22}^{51,0}}{1536}-\frac{7 C_{11,22}^{51,1}}{3072}+\frac{7 C_{20,20}^{40,0}}{768 \sqrt{5}}+\frac{7 C_{20,20}^{40,1}}{1536 \sqrt{5}}+\frac{7
C_{20,22}^{42,0}}{7680}+\frac{7 C_{20,22}^{42,1}}{15360}-\frac{7 C_{22,20}^{42,0}}{3840}-\frac{7 C_{22,20}^{42,1}}{7680}+\frac{7 C_{22,22}^{40,0}}{7680}+\frac{7
C_{22,22}^{40,1}}{15360}-\frac{\sqrt{7} C_{22,22}^{42,0}}{1920}-\frac{\sqrt{7} C_{22,22}^{42,1}}{3840}-\frac{9 C_{22,22}^{44,0}}{640 \sqrt{5}}-\frac{9
C_{22,22}^{44,1}}{1280 \sqrt{5}}-\frac{7 C_{31,20}^{31,0}}{768 \sqrt{5}}-\frac{7 C_{31,20}^{31,1}}{1536 \sqrt{5}}-\frac{7 C_{31,22}^{31,0}}{7680}-\frac{7
C_{31,22}^{31,1}}{15360}+\frac{1}{640} \sqrt{\frac{21}{5}} C_{31,22}^{33,0}+\frac{\sqrt{\frac{21}{5}} C_{31,22}^{33,1}}{1280}+\frac{3 \sqrt{\frac{21}{5}}
C_{33,20}^{33,0}}{1280}+\frac{3 \sqrt{\frac{21}{5}} C_{33,20}^{33,1}}{2560}+\frac{9 \sqrt{\frac{7}{2}} C_{33,22}^{33,0}}{3200}+\frac{9 \sqrt{\frac{7}{2}}
C_{33,22}^{33,1}}{6400}\)

\noindent\(C_{31,1112,2}^{11,1111,0,\text{NonLoc}}= -\frac{7 \sqrt{\frac{3}{5}} C_{00,00}^{60,0}}{1024}-\frac{7}{512} \sqrt{\frac{3}{5}}
C_{00,00}^{60,1}-\frac{7 C_{00,20}^{60,0}}{1024 \sqrt{5}}-\frac{7 C_{00,20}^{60,1}}{512 \sqrt{5}}-\frac{21 C_{00,22}^{62,0}}{5120}-\frac{21 C_{00,22}^{62,1}}{2560}-\frac{7
C_{11,00}^{51,0}}{1024 \sqrt{15}}-\frac{7 C_{11,00}^{51,1}}{512 \sqrt{15}}-\frac{7 C_{11,20}^{31,0}}{3072 \sqrt{5}}-\frac{7 C_{11,20}^{31,1}}{1536
\sqrt{5}}-\frac{7 C_{11,22}^{51,0}}{3072}-\frac{7 C_{11,22}^{51,1}}{1536}+\frac{7 C_{20,00}^{40,0}}{1024 \sqrt{15}}+\frac{7 C_{20,00}^{40,1}}{512
\sqrt{15}}+\frac{7 C_{20,20}^{40,0}}{3072 \sqrt{5}}+\frac{7 C_{20,20}^{40,1}}{1536 \sqrt{5}}-\frac{7 C_{20,22}^{42,0}}{15360}-\frac{7 C_{20,22}^{42,1}}{7680}-\frac{7
C_{22,00}^{42,0}}{5120 \sqrt{3}}-\frac{7 C_{22,00}^{42,1}}{2560 \sqrt{3}}-\frac{7 C_{22,20}^{42,0}}{15360}-\frac{7 C_{22,20}^{42,1}}{7680}-\frac{7
C_{22,22}^{40,0}}{15360}-\frac{7 C_{22,22}^{40,1}}{7680}+\frac{\sqrt{7} C_{22,22}^{42,0}}{3840}+\frac{\sqrt{7} C_{22,22}^{42,1}}{1920}+\frac{9 C_{22,22}^{44,0}}{1280
\sqrt{5}}+\frac{9 C_{22,22}^{44,1}}{640 \sqrt{5}}+\frac{7 C_{31,00}^{31,0}}{1024 \sqrt{15}}+\frac{7 C_{31,00}^{31,1}}{512 \sqrt{15}}+\frac{7 C_{31,20}^{31,0}}{3072
\sqrt{5}}+\frac{7 C_{31,20}^{31,1}}{1536 \sqrt{5}}-\frac{7 C_{31,22}^{31,0}}{15360}-\frac{7 C_{31,22}^{31,1}}{7680}+\frac{\sqrt{\frac{21}{5}} C_{31,22}^{33,0}}{1280}+\frac{1}{640}
\sqrt{\frac{21}{5}} C_{31,22}^{33,1}-\frac{9 \sqrt{\frac{7}{5}} C_{33,00}^{33,0}}{5120}-\frac{9 \sqrt{\frac{7}{5}} C_{33,00}^{33,1}}{2560}-\frac{3
\sqrt{\frac{21}{5}} C_{33,20}^{33,0}}{5120}-\frac{3 \sqrt{\frac{21}{5}} C_{33,20}^{33,1}}{2560}+\frac{9 \sqrt{\frac{7}{2}} C_{33,22}^{33,0}}{6400}+\frac{9
\sqrt{\frac{7}{2}} C_{33,22}^{33,1}}{3200}\)

\noindent\(C_{31,1112,2}^{11,1111,1,\text{Loc}}= -\frac{7}{512} \sqrt{\frac{3}{5}} C_{00,20}^{60,1}+\frac{21 \sqrt{3} C_{00,22}^{62,1}}{5120}+\frac{7
C_{11,20}^{31,1}}{512 \sqrt{15}}-\frac{7 C_{11,22}^{51,1}}{1024 \sqrt{3}}+\frac{7 C_{20,20}^{40,1}}{512 \sqrt{15}}+\frac{7 C_{20,22}^{42,1}}{5120
\sqrt{3}}-\frac{7 C_{22,20}^{42,1}}{2560 \sqrt{3}}+\frac{7 C_{22,22}^{40,1}}{5120 \sqrt{3}}-\frac{\sqrt{\frac{7}{3}} C_{22,22}^{42,1}}{1280}-\frac{9
\sqrt{\frac{3}{5}} C_{22,22}^{44,1}}{1280}-\frac{7 C_{31,20}^{31,1}}{512 \sqrt{15}}-\frac{7 C_{31,22}^{31,1}}{5120 \sqrt{3}}+\frac{3 \sqrt{\frac{7}{5}}
C_{31,22}^{33,1}}{1280}+\frac{9 \sqrt{\frac{7}{5}} C_{33,20}^{33,1}}{2560}+\frac{9 \sqrt{\frac{21}{2}} C_{33,22}^{33,1}}{6400}\)

\noindent\(C_{31,1112,2}^{11,1111,1,\text{NonLoc}}= -\frac{21 C_{00,00}^{60,0}}{1024 \sqrt{5}}-\frac{7 \sqrt{\frac{3}{5}} C_{00,20}^{60,0}}{1024}-\frac{21
\sqrt{3} C_{00,22}^{62,0}}{5120}-\frac{7 C_{11,00}^{51,0}}{1024 \sqrt{5}}-\frac{7 C_{11,20}^{31,0}}{1024 \sqrt{15}}-\frac{7 C_{11,22}^{51,0}}{1024
\sqrt{3}}+\frac{7 C_{20,00}^{40,0}}{1024 \sqrt{5}}+\frac{7 C_{20,20}^{40,0}}{1024 \sqrt{15}}-\frac{7 C_{20,22}^{42,0}}{5120 \sqrt{3}}-\frac{7 C_{22,00}^{42,0}}{5120}-\frac{7
C_{22,20}^{42,0}}{5120 \sqrt{3}}-\frac{7 C_{22,22}^{40,0}}{5120 \sqrt{3}}+\frac{\sqrt{\frac{7}{3}} C_{22,22}^{42,0}}{1280}+\frac{9 \sqrt{\frac{3}{5}}
C_{22,22}^{44,0}}{1280}+\frac{7 C_{31,00}^{31,0}}{1024 \sqrt{5}}+\frac{7 C_{31,20}^{31,0}}{1024 \sqrt{15}}-\frac{7 C_{31,22}^{31,0}}{5120 \sqrt{3}}+\frac{3
\sqrt{\frac{7}{5}} C_{31,22}^{33,0}}{1280}-\frac{9 \sqrt{\frac{21}{5}} C_{33,00}^{33,0}}{5120}-\frac{9 \sqrt{\frac{7}{5}} C_{33,20}^{33,0}}{5120}+\frac{9
\sqrt{\frac{21}{2}} C_{33,22}^{33,0}}{6400}\)

\noindent\(C_{31,1112,2}^{11,1112,0,\text{Loc}}= -\frac{7 C_{00,20}^{60,0}}{256 \sqrt{15}}-\frac{7 C_{00,20}^{60,1}}{512 \sqrt{15}}+\frac{7
\sqrt{3} C_{00,22}^{62,0}}{2560}+\frac{7 \sqrt{3} C_{00,22}^{62,1}}{5120}+\frac{7 C_{11,20}^{31,0}}{768 \sqrt{15}}+\frac{7 C_{11,20}^{31,1}}{1536
\sqrt{15}}-\frac{77 C_{11,22}^{51,0}}{7680 \sqrt{3}}-\frac{77 C_{11,22}^{51,1}}{15360 \sqrt{3}}+\frac{9 \sqrt{\frac{7}{5}} C_{11,22}^{53,0}}{2560}+\frac{9
\sqrt{\frac{7}{5}} C_{11,22}^{53,1}}{5120}-\frac{11 C_{20,20}^{40,0}}{256 \sqrt{15}}-\frac{11 C_{20,20}^{40,1}}{512 \sqrt{15}}+\frac{7 C_{20,22}^{42,0}}{2560
\sqrt{3}}+\frac{7 C_{20,22}^{42,1}}{5120 \sqrt{3}}+\frac{7 \sqrt{3} C_{22,20}^{42,0}}{1280}+\frac{7 \sqrt{3} C_{22,20}^{42,1}}{2560}+\frac{17 C_{22,22}^{40,0}}{2560
\sqrt{3}}+\frac{17 C_{22,22}^{40,1}}{5120 \sqrt{3}}-\frac{1}{640} \sqrt{\frac{7}{3}} C_{22,22}^{42,0}-\frac{\sqrt{\frac{7}{3}} C_{22,22}^{42,1}}{1280}-\frac{9}{640}
\sqrt{\frac{3}{5}} C_{22,22}^{44,0}-\frac{9 \sqrt{\frac{3}{5}} C_{22,22}^{44,1}}{1280}+\frac{13 C_{31,20}^{31,0}}{768 \sqrt{15}}+\frac{13 C_{31,20}^{31,1}}{1536
\sqrt{15}}-\frac{41 C_{31,22}^{31,0}}{7680 \sqrt{3}}-\frac{41 C_{31,22}^{31,1}}{15360 \sqrt{3}}+\frac{7 \sqrt{\frac{7}{5}} C_{31,22}^{33,0}}{2560}+\frac{7
\sqrt{\frac{7}{5}} C_{31,22}^{33,1}}{5120}-\frac{27 \sqrt{\frac{7}{5}} C_{33,20}^{33,0}}{1280}-\frac{27 \sqrt{\frac{7}{5}} C_{33,20}^{33,1}}{2560}+\frac{9
\sqrt{\frac{21}{2}} C_{33,22}^{33,0}}{3200}+\frac{9 \sqrt{\frac{21}{2}} C_{33,22}^{33,1}}{6400}\)

\noindent\(C_{31,1112,2}^{11,1112,0,\text{NonLoc}}= -\frac{7 C_{00,00}^{60,0}}{1024 \sqrt{5}}-\frac{7 C_{00,00}^{60,1}}{512 \sqrt{5}}-\frac{7
C_{00,20}^{60,0}}{1024 \sqrt{15}}-\frac{7 C_{00,20}^{60,1}}{512 \sqrt{15}}-\frac{7 \sqrt{3} C_{00,22}^{62,0}}{5120}-\frac{7 \sqrt{3} C_{00,22}^{62,1}}{2560}-\frac{7
C_{11,00}^{51,0}}{3072 \sqrt{5}}-\frac{7 C_{11,00}^{51,1}}{1536 \sqrt{5}}-\frac{7 C_{11,20}^{31,0}}{3072 \sqrt{15}}-\frac{7 C_{11,20}^{31,1}}{1536
\sqrt{15}}-\frac{77 C_{11,22}^{51,0}}{15360 \sqrt{3}}-\frac{77 C_{11,22}^{51,1}}{7680 \sqrt{3}}+\frac{9 \sqrt{\frac{7}{5}} C_{11,22}^{53,0}}{5120}+\frac{9
\sqrt{\frac{7}{5}} C_{11,22}^{53,1}}{2560}-\frac{11 C_{20,00}^{40,0}}{1024 \sqrt{5}}-\frac{11 C_{20,00}^{40,1}}{512 \sqrt{5}}-\frac{11 C_{20,20}^{40,0}}{1024
\sqrt{15}}-\frac{11 C_{20,20}^{40,1}}{512 \sqrt{15}}-\frac{7 C_{20,22}^{42,0}}{5120 \sqrt{3}}-\frac{7 C_{20,22}^{42,1}}{2560 \sqrt{3}}+\frac{21 C_{22,00}^{42,0}}{5120}+\frac{21
C_{22,00}^{42,1}}{2560}+\frac{7 \sqrt{3} C_{22,20}^{42,0}}{5120}+\frac{7 \sqrt{3} C_{22,20}^{42,1}}{2560}-\frac{17 C_{22,22}^{40,0}}{5120 \sqrt{3}}-\frac{17
C_{22,22}^{40,1}}{2560 \sqrt{3}}+\frac{\sqrt{\frac{7}{3}} C_{22,22}^{42,0}}{1280}+\frac{1}{640} \sqrt{\frac{7}{3}} C_{22,22}^{42,1}+\frac{9 \sqrt{\frac{3}{5}}
C_{22,22}^{44,0}}{1280}+\frac{9}{640} \sqrt{\frac{3}{5}} C_{22,22}^{44,1}-\frac{13 C_{31,00}^{31,0}}{3072 \sqrt{5}}-\frac{13 C_{31,00}^{31,1}}{1536
\sqrt{5}}-\frac{13 C_{31,20}^{31,0}}{3072 \sqrt{15}}-\frac{13 C_{31,20}^{31,1}}{1536 \sqrt{15}}-\frac{41 C_{31,22}^{31,0}}{15360 \sqrt{3}}-\frac{41
C_{31,22}^{31,1}}{7680 \sqrt{3}}+\frac{7 \sqrt{\frac{7}{5}} C_{31,22}^{33,0}}{5120}+\frac{7 \sqrt{\frac{7}{5}} C_{31,22}^{33,1}}{2560}+\frac{27 \sqrt{\frac{21}{5}}
C_{33,00}^{33,0}}{5120}+\frac{27 \sqrt{\frac{21}{5}} C_{33,00}^{33,1}}{2560}+\frac{27 \sqrt{\frac{7}{5}} C_{33,20}^{33,0}}{5120}+\frac{27 \sqrt{\frac{7}{5}}
C_{33,20}^{33,1}}{2560}+\frac{9 \sqrt{\frac{21}{2}} C_{33,22}^{33,0}}{6400}+\frac{9 \sqrt{\frac{21}{2}} C_{33,22}^{33,1}}{3200}\)

\noindent\(C_{31,1112,2}^{11,1112,1,\text{Loc}}= -\frac{7 C_{00,20}^{60,1}}{512 \sqrt{5}}+\frac{21 C_{00,22}^{62,1}}{5120}+\frac{7
C_{11,20}^{31,1}}{1536 \sqrt{5}}-\frac{77 C_{11,22}^{51,1}}{15360}+\frac{9 \sqrt{\frac{21}{5}} C_{11,22}^{53,1}}{5120}-\frac{11 C_{20,20}^{40,1}}{512
\sqrt{5}}+\frac{7 C_{20,22}^{42,1}}{5120}+\frac{21 C_{22,20}^{42,1}}{2560}+\frac{17 C_{22,22}^{40,1}}{5120}-\frac{\sqrt{7} C_{22,22}^{42,1}}{1280}-\frac{27
C_{22,22}^{44,1}}{1280 \sqrt{5}}+\frac{13 C_{31,20}^{31,1}}{1536 \sqrt{5}}-\frac{41 C_{31,22}^{31,1}}{15360}+\frac{7 \sqrt{\frac{21}{5}} C_{31,22}^{33,1}}{5120}-\frac{27
\sqrt{\frac{21}{5}} C_{33,20}^{33,1}}{2560}+\frac{27 \sqrt{\frac{7}{2}} C_{33,22}^{33,1}}{6400}\)

\noindent\(C_{31,1112,2}^{11,1112,1,\text{NonLoc}}= -\frac{7 \sqrt{\frac{3}{5}} C_{00,00}^{60,0}}{1024}-\frac{7 C_{00,20}^{60,0}}{1024
\sqrt{5}}-\frac{21 C_{00,22}^{62,0}}{5120}-\frac{7 C_{11,00}^{51,0}}{1024 \sqrt{15}}-\frac{7 C_{11,20}^{31,0}}{3072 \sqrt{5}}-\frac{77 C_{11,22}^{51,0}}{15360}+\frac{9
\sqrt{\frac{21}{5}} C_{11,22}^{53,0}}{5120}-\frac{11 \sqrt{\frac{3}{5}} C_{20,00}^{40,0}}{1024}-\frac{11 C_{20,20}^{40,0}}{1024 \sqrt{5}}-\frac{7
C_{20,22}^{42,0}}{5120}+\frac{21 \sqrt{3} C_{22,00}^{42,0}}{5120}+\frac{21 C_{22,20}^{42,0}}{5120}-\frac{17 C_{22,22}^{40,0}}{5120}+\frac{\sqrt{7}
C_{22,22}^{42,0}}{1280}+\frac{27 C_{22,22}^{44,0}}{1280 \sqrt{5}}-\frac{13 C_{31,00}^{31,0}}{1024 \sqrt{15}}-\frac{13 C_{31,20}^{31,0}}{3072 \sqrt{5}}-\frac{41
C_{31,22}^{31,0}}{15360}+\frac{7 \sqrt{\frac{21}{5}} C_{31,22}^{33,0}}{5120}+\frac{81 \sqrt{\frac{7}{5}} C_{33,00}^{33,0}}{5120}+\frac{27 \sqrt{\frac{21}{5}}
C_{33,20}^{33,0}}{5120}+\frac{27 \sqrt{\frac{7}{2}} C_{33,22}^{33,0}}{6400}\)

\noindent\(C_{31,1112,3}^{11,1112,0,\text{Loc}}= -\frac{7}{320} \sqrt{\frac{7}{3}} C_{00,20}^{60,0}-\frac{7}{640} \sqrt{\frac{7}{3}}
C_{00,20}^{60,1}-\frac{1}{320} \sqrt{\frac{21}{5}} C_{00,22}^{62,0}-\frac{1}{640} \sqrt{\frac{21}{5}} C_{00,22}^{62,1}+\frac{7}{960} \sqrt{\frac{7}{3}}
C_{11,20}^{31,0}+\frac{7 \sqrt{\frac{7}{3}} C_{11,20}^{31,1}}{1920}+\frac{1}{480} \sqrt{\frac{7}{15}} C_{11,22}^{51,0}+\frac{1}{960} \sqrt{\frac{7}{15}}
C_{11,22}^{51,1}+\frac{9 C_{11,22}^{53,0}}{3200}+\frac{9 C_{11,22}^{53,1}}{6400}-\frac{1}{320} \sqrt{\frac{7}{3}} C_{20,20}^{40,0}-\frac{1}{640}
\sqrt{\frac{7}{3}} C_{20,20}^{40,1}-\frac{1}{640} \sqrt{\frac{7}{15}} C_{20,22}^{42,0}-\frac{\sqrt{\frac{7}{15}} C_{20,22}^{42,1}}{1280}+\frac{7}{640}
\sqrt{\frac{7}{15}} C_{22,20}^{42,0}+\frac{7 \sqrt{\frac{7}{15}} C_{22,20}^{42,1}}{1280}+\frac{7 C_{22,22}^{42,0}}{640 \sqrt{15}}+\frac{7 C_{22,22}^{42,1}}{1280
\sqrt{15}}-\frac{1}{480} \sqrt{\frac{7}{3}} C_{31,20}^{31,0}-\frac{1}{960} \sqrt{\frac{7}{3}} C_{31,20}^{31,1}-\frac{1}{960} \sqrt{\frac{7}{15}}
C_{31,22}^{31,0}-\frac{\sqrt{\frac{7}{15}} C_{31,22}^{31,1}}{1920}-\frac{C_{31,22}^{33,0}}{1600}-\frac{C_{31,22}^{33,1}}{3200}-\frac{63 C_{33,20}^{33,0}}{3200}-\frac{63
C_{33,20}^{33,1}}{6400}-\frac{9}{800} \sqrt{\frac{3}{10}} C_{33,22}^{33,0}-\frac{9 \sqrt{\frac{3}{10}} C_{33,22}^{33,1}}{1600}\)

\noindent\(C_{31,1112,3}^{11,1112,0,\text{NonLoc}}= -\frac{7 \sqrt{7} C_{00,00}^{60,0}}{1280}-\frac{7 \sqrt{7} C_{00,00}^{60,1}}{640}-\frac{7
\sqrt{\frac{7}{3}} C_{00,20}^{60,0}}{1280}-\frac{7}{640} \sqrt{\frac{7}{3}} C_{00,20}^{60,1}+\frac{1}{640} \sqrt{\frac{21}{5}} C_{00,22}^{62,0}+\frac{1}{320}
\sqrt{\frac{21}{5}} C_{00,22}^{62,1}-\frac{7 \sqrt{7} C_{11,00}^{51,0}}{3840}-\frac{7 \sqrt{7} C_{11,00}^{51,1}}{1920}-\frac{7 \sqrt{\frac{7}{3}}
C_{11,20}^{31,0}}{3840}-\frac{7 \sqrt{\frac{7}{3}} C_{11,20}^{31,1}}{1920}+\frac{1}{960} \sqrt{\frac{7}{15}} C_{11,22}^{51,0}+\frac{1}{480} \sqrt{\frac{7}{15}}
C_{11,22}^{51,1}+\frac{9 C_{11,22}^{53,0}}{6400}+\frac{9 C_{11,22}^{53,1}}{3200}-\frac{\sqrt{7} C_{20,00}^{40,0}}{1280}-\frac{\sqrt{7} C_{20,00}^{40,1}}{640}-\frac{\sqrt{\frac{7}{3}}
C_{20,20}^{40,0}}{1280}-\frac{1}{640} \sqrt{\frac{7}{3}} C_{20,20}^{40,1}+\frac{\sqrt{\frac{7}{15}} C_{20,22}^{42,0}}{1280}+\frac{1}{640} \sqrt{\frac{7}{15}}
C_{20,22}^{42,1}+\frac{7 \sqrt{\frac{7}{5}} C_{22,00}^{42,0}}{2560}+\frac{7 \sqrt{\frac{7}{5}} C_{22,00}^{42,1}}{1280}+\frac{7 \sqrt{\frac{7}{15}}
C_{22,20}^{42,0}}{2560}+\frac{7 \sqrt{\frac{7}{15}} C_{22,20}^{42,1}}{1280}-\frac{7 C_{22,22}^{42,0}}{1280 \sqrt{15}}-\frac{7 C_{22,22}^{42,1}}{640
\sqrt{15}}+\frac{\sqrt{7} C_{31,00}^{31,0}}{1920}+\frac{\sqrt{7} C_{31,00}^{31,1}}{960}+\frac{\sqrt{\frac{7}{3}} C_{31,20}^{31,0}}{1920}+\frac{1}{960}
\sqrt{\frac{7}{3}} C_{31,20}^{31,1}-\frac{\sqrt{\frac{7}{15}} C_{31,22}^{31,0}}{1920}-\frac{1}{960} \sqrt{\frac{7}{15}} C_{31,22}^{31,1}-\frac{C_{31,22}^{33,0}}{3200}-\frac{C_{31,22}^{33,1}}{1600}+\frac{63
\sqrt{3} C_{33,00}^{33,0}}{12800}+\frac{63 \sqrt{3} C_{33,00}^{33,1}}{6400}+\frac{63 C_{33,20}^{33,0}}{12800}+\frac{63 C_{33,20}^{33,1}}{6400}-\frac{9
\sqrt{\frac{3}{10}} C_{33,22}^{33,0}}{1600}-\frac{9}{800} \sqrt{\frac{3}{10}} C_{33,22}^{33,1}\)

\noindent\(C_{31,1112,3}^{11,1112,1,\text{Loc}}= -\frac{7 \sqrt{7} C_{00,20}^{60,1}}{640}-\frac{3}{640} \sqrt{\frac{7}{5}} C_{00,22}^{62,1}+\frac{7
\sqrt{7} C_{11,20}^{31,1}}{1920}+\frac{1}{960} \sqrt{\frac{7}{5}} C_{11,22}^{51,1}+\frac{9 \sqrt{3} C_{11,22}^{53,1}}{6400}-\frac{\sqrt{7} C_{20,20}^{40,1}}{640}-\frac{\sqrt{\frac{7}{5}}
C_{20,22}^{42,1}}{1280}+\frac{7 \sqrt{\frac{7}{5}} C_{22,20}^{42,1}}{1280}+\frac{7 C_{22,22}^{42,1}}{1280 \sqrt{5}}-\frac{\sqrt{7} C_{31,20}^{31,1}}{960}-\frac{\sqrt{\frac{7}{5}}
C_{31,22}^{31,1}}{1920}-\frac{\sqrt{3} C_{31,22}^{33,1}}{3200}-\frac{63 \sqrt{3} C_{33,20}^{33,1}}{6400}-\frac{27 C_{33,22}^{33,1}}{1600 \sqrt{10}}\)

\noindent\(C_{31,1112,3}^{11,1112,1,\text{NonLoc}}= -\frac{7 \sqrt{21} C_{00,00}^{60,0}}{1280}-\frac{7 \sqrt{7} C_{00,20}^{60,0}}{1280}+\frac{3}{640}
\sqrt{\frac{7}{5}} C_{00,22}^{62,0}-\frac{7 \sqrt{\frac{7}{3}} C_{11,00}^{51,0}}{1280}-\frac{7 \sqrt{7} C_{11,20}^{31,0}}{3840}+\frac{1}{960} \sqrt{\frac{7}{5}}
C_{11,22}^{51,0}+\frac{9 \sqrt{3} C_{11,22}^{53,0}}{6400}-\frac{\sqrt{21} C_{20,00}^{40,0}}{1280}-\frac{\sqrt{7} C_{20,20}^{40,0}}{1280}+\frac{\sqrt{\frac{7}{5}}
C_{20,22}^{42,0}}{1280}+\frac{7 \sqrt{\frac{21}{5}} C_{22,00}^{42,0}}{2560}+\frac{7 \sqrt{\frac{7}{5}} C_{22,20}^{42,0}}{2560}-\frac{7 C_{22,22}^{42,0}}{1280
\sqrt{5}}+\frac{1}{640} \sqrt{\frac{7}{3}} C_{31,00}^{31,0}+\frac{\sqrt{7} C_{31,20}^{31,0}}{1920}-\frac{\sqrt{\frac{7}{5}} C_{31,22}^{31,0}}{1920}-\frac{\sqrt{3}
C_{31,22}^{33,0}}{3200}+\frac{189 C_{33,00}^{33,0}}{12800}+\frac{63 \sqrt{3} C_{33,20}^{33,0}}{12800}-\frac{27 C_{33,22}^{33,0}}{1600 \sqrt{10}}\)

\noindent\(C_{31,2000,1}^{00,1111,0,\text{Loc}}= -\frac{7 C_{11,11}^{51,0}}{384}-\frac{7 C_{11,11}^{51,1}}{768}-\frac{C_{31,11}^{31,0}}{128}-\frac{C_{31,11}^{31,1}}{256}+\frac{3}{160}
\sqrt{\frac{7}{2}} C_{33,11}^{33,0}+\frac{3}{320} \sqrt{\frac{7}{2}} C_{33,11}^{33,1}\)

\noindent\(C_{31,2000,1}^{00,1111,0,\text{NonLoc}}= -\frac{7 C_{11,11}^{51,0}}{768}-\frac{7 C_{11,11}^{51,1}}{384}-\frac{C_{31,11}^{31,0}}{256}-\frac{C_{31,11}^{31,1}}{128}+\frac{3}{320}
\sqrt{\frac{7}{2}} C_{33,11}^{33,0}+\frac{3}{160} \sqrt{\frac{7}{2}} C_{33,11}^{33,1}\)

\noindent\(C_{31,2000,1}^{00,1111,1,\text{Loc}}= -\frac{7 C_{11,11}^{51,1}}{256 \sqrt{3}}-\frac{\sqrt{3} C_{31,11}^{31,1}}{256}+\frac{3}{320}
\sqrt{\frac{21}{2}} C_{33,11}^{33,1}\)

\noindent\(C_{31,2000,1}^{00,1111,1,\text{NonLoc}}= -\frac{7 C_{11,11}^{51,0}}{256 \sqrt{3}}-\frac{\sqrt{3} C_{31,11}^{31,0}}{256}+\frac{3}{320}
\sqrt{\frac{21}{2}} C_{33,11}^{33,0}\)

\noindent\(C_{31,2000,1}^{11,0000,0,\text{Loc}}= -\frac{7 C_{00,00}^{60,0}}{256}-\frac{7 C_{00,00}^{60,1}}{512}+\frac{7 C_{11,00}^{51,0}}{768}+\frac{7
C_{11,00}^{51,1}}{1536}-\frac{5 C_{20,00}^{40,0}}{768}-\frac{5 C_{20,00}^{40,1}}{1536}+\frac{7 C_{22,00}^{42,0}}{384 \sqrt{5}}+\frac{7 C_{22,00}^{42,1}}{768
\sqrt{5}}-\frac{C_{31,00}^{31,0}}{768}-\frac{C_{31,00}^{31,1}}{1536}-\frac{3 \sqrt{21} C_{33,00}^{33,0}}{640}-\frac{3 \sqrt{21} C_{33,00}^{33,1}}{1280}\)

\noindent\(C_{31,2000,1}^{11,0000,0,\text{NonLoc}}= \frac{7 C_{00,00}^{60,0}}{1024}+\frac{7 C_{00,00}^{60,1}}{512}-\frac{7 \sqrt{3}
C_{00,20}^{60,0}}{1024}-\frac{7 \sqrt{3} C_{00,20}^{60,1}}{512}+\frac{7 C_{11,00}^{51,0}}{3072}+\frac{7 C_{11,00}^{51,1}}{1536}-\frac{7 C_{11,20}^{31,0}}{1024
\sqrt{3}}-\frac{7 C_{11,20}^{31,1}}{512 \sqrt{3}}+\frac{5 C_{20,00}^{40,0}}{3072}+\frac{5 C_{20,00}^{40,1}}{1536}-\frac{5 C_{20,20}^{40,0}}{1024
\sqrt{3}}-\frac{5 C_{20,20}^{40,1}}{512 \sqrt{3}}-\frac{7 C_{22,00}^{42,0}}{1536 \sqrt{5}}-\frac{7 C_{22,00}^{42,1}}{768 \sqrt{5}}+\frac{7 C_{22,20}^{42,0}}{512
\sqrt{15}}+\frac{7 C_{22,20}^{42,1}}{256 \sqrt{15}}-\frac{C_{31,00}^{31,0}}{3072}-\frac{C_{31,00}^{31,1}}{1536}+\frac{C_{31,20}^{31,0}}{1024 \sqrt{3}}+\frac{C_{31,20}^{31,1}}{512
\sqrt{3}}-\frac{3 \sqrt{21} C_{33,00}^{33,0}}{2560}-\frac{3 \sqrt{21} C_{33,00}^{33,1}}{1280}+\frac{9 \sqrt{7} C_{33,20}^{33,0}}{2560}+\frac{9 \sqrt{7}
C_{33,20}^{33,1}}{1280}\)

\noindent\(C_{31,2000,1}^{11,0000,1,\text{Loc}}= -\frac{7 \sqrt{3} C_{00,00}^{60,1}}{512}+\frac{7 C_{11,00}^{51,1}}{512 \sqrt{3}}-\frac{5
C_{20,00}^{40,1}}{512 \sqrt{3}}+\frac{7 C_{22,00}^{42,1}}{256 \sqrt{15}}-\frac{C_{31,00}^{31,1}}{512 \sqrt{3}}-\frac{9 \sqrt{7} C_{33,00}^{33,1}}{1280}\)

\noindent\(C_{31,2000,1}^{11,0000,1,\text{NonLoc}}= \frac{7 \sqrt{3} C_{00,00}^{60,0}}{1024}-\frac{21 C_{00,20}^{60,0}}{1024}+\frac{7
C_{11,00}^{51,0}}{1024 \sqrt{3}}-\frac{7 C_{11,20}^{31,0}}{1024}+\frac{5 C_{20,00}^{40,0}}{1024 \sqrt{3}}-\frac{5 C_{20,20}^{40,0}}{1024}-\frac{7
C_{22,00}^{42,0}}{512 \sqrt{15}}+\frac{7 C_{22,20}^{42,0}}{512 \sqrt{5}}-\frac{C_{31,00}^{31,0}}{1024 \sqrt{3}}+\frac{C_{31,20}^{31,0}}{1024}-\frac{9
\sqrt{7} C_{33,00}^{33,0}}{2560}+\frac{9 \sqrt{21} C_{33,20}^{33,0}}{2560}\)

\noindent\(C_{31,2011,0}^{11,0011,0,\text{Loc}}= \frac{7 C_{00,20}^{60,0}}{768}+\frac{7 C_{00,20}^{60,1}}{1536}+\frac{7 C_{00,22}^{62,0}}{256
\sqrt{5}}+\frac{7 C_{00,22}^{62,1}}{512 \sqrt{5}}-\frac{7 C_{11,20}^{31,0}}{2304}-\frac{7 C_{11,20}^{31,1}}{4608}-\frac{7 \sqrt{5} C_{11,22}^{51,0}}{2304}-\frac{7
\sqrt{5} C_{11,22}^{51,1}}{4608}+\frac{5 C_{20,20}^{40,0}}{2304}+\frac{5 C_{20,20}^{40,1}}{4608}+\frac{7 C_{20,22}^{42,0}}{2304 \sqrt{5}}+\frac{7
C_{20,22}^{42,1}}{4608 \sqrt{5}}-\frac{7 C_{22,20}^{42,0}}{1152 \sqrt{5}}-\frac{7 C_{22,20}^{42,1}}{2304 \sqrt{5}}+\frac{7 C_{22,22}^{40,0}}{2304
\sqrt{5}}+\frac{7 C_{22,22}^{40,1}}{4608 \sqrt{5}}-\frac{1}{576} \sqrt{\frac{7}{5}} C_{22,22}^{42,0}-\frac{\sqrt{\frac{7}{5}} C_{22,22}^{42,1}}{1152}-\frac{3
C_{22,22}^{44,0}}{320}-\frac{3 C_{22,22}^{44,1}}{640}+\frac{C_{31,20}^{31,0}}{2304}+\frac{C_{31,20}^{31,1}}{4608}-\frac{7 C_{31,22}^{31,0}}{2304
\sqrt{5}}-\frac{7 C_{31,22}^{31,1}}{4608 \sqrt{5}}+\frac{1}{320} \sqrt{\frac{7}{3}} C_{31,22}^{33,0}+\frac{1}{640} \sqrt{\frac{7}{3}} C_{31,22}^{33,1}+\frac{\sqrt{21}
C_{33,20}^{33,0}}{640}+\frac{\sqrt{21} C_{33,20}^{33,1}}{1280}+\frac{3}{320} \sqrt{\frac{7}{10}} C_{33,22}^{33,0}+\frac{3}{640} \sqrt{\frac{7}{10}}
C_{33,22}^{33,1}\)

\noindent\(C_{31,2011,0}^{11,0011,0,\text{NonLoc}}= \frac{7 C_{00,00}^{60,0}}{1024 \sqrt{3}}+\frac{7 C_{00,00}^{60,1}}{512 \sqrt{3}}+\frac{7
C_{00,20}^{60,0}}{3072}+\frac{7 C_{00,20}^{60,1}}{1536}-\frac{7 C_{00,22}^{62,0}}{512 \sqrt{5}}-\frac{7 C_{00,22}^{62,1}}{256 \sqrt{5}}+\frac{7 C_{11,00}^{51,0}}{3072
\sqrt{3}}+\frac{7 C_{11,00}^{51,1}}{1536 \sqrt{3}}+\frac{7 C_{11,20}^{31,0}}{9216}+\frac{7 C_{11,20}^{31,1}}{4608}-\frac{7 \sqrt{5} C_{11,22}^{51,0}}{4608}-\frac{7
\sqrt{5} C_{11,22}^{51,1}}{2304}+\frac{5 C_{20,00}^{40,0}}{3072 \sqrt{3}}+\frac{5 C_{20,00}^{40,1}}{1536 \sqrt{3}}+\frac{5 C_{20,20}^{40,0}}{9216}+\frac{5
C_{20,20}^{40,1}}{4608}-\frac{7 C_{20,22}^{42,0}}{4608 \sqrt{5}}-\frac{7 C_{20,22}^{42,1}}{2304 \sqrt{5}}-\frac{7 C_{22,00}^{42,0}}{1536 \sqrt{15}}-\frac{7
C_{22,00}^{42,1}}{768 \sqrt{15}}-\frac{7 C_{22,20}^{42,0}}{4608 \sqrt{5}}-\frac{7 C_{22,20}^{42,1}}{2304 \sqrt{5}}-\frac{7 C_{22,22}^{40,0}}{4608
\sqrt{5}}-\frac{7 C_{22,22}^{40,1}}{2304 \sqrt{5}}+\frac{\sqrt{\frac{7}{5}} C_{22,22}^{42,0}}{1152}+\frac{1}{576} \sqrt{\frac{7}{5}} C_{22,22}^{42,1}+\frac{3
C_{22,22}^{44,0}}{640}+\frac{3 C_{22,22}^{44,1}}{320}-\frac{C_{31,00}^{31,0}}{3072 \sqrt{3}}-\frac{C_{31,00}^{31,1}}{1536 \sqrt{3}}-\frac{C_{31,20}^{31,0}}{9216}-\frac{C_{31,20}^{31,1}}{4608}-\frac{7
C_{31,22}^{31,0}}{4608 \sqrt{5}}-\frac{7 C_{31,22}^{31,1}}{2304 \sqrt{5}}+\frac{1}{640} \sqrt{\frac{7}{3}} C_{31,22}^{33,0}+\frac{1}{320} \sqrt{\frac{7}{3}}
C_{31,22}^{33,1}-\frac{3 \sqrt{7} C_{33,00}^{33,0}}{2560}-\frac{3 \sqrt{7} C_{33,00}^{33,1}}{1280}-\frac{\sqrt{21} C_{33,20}^{33,0}}{2560}-\frac{\sqrt{21}
C_{33,20}^{33,1}}{1280}+\frac{3}{640} \sqrt{\frac{7}{10}} C_{33,22}^{33,0}+\frac{3}{320} \sqrt{\frac{7}{10}} C_{33,22}^{33,1}\)

\noindent\(C_{31,2011,0}^{11,0011,1,\text{Loc}}= \frac{7 C_{00,20}^{60,1}}{512 \sqrt{3}}+\frac{7}{512} \sqrt{\frac{3}{5}} C_{00,22}^{62,1}-\frac{7
C_{11,20}^{31,1}}{1536 \sqrt{3}}-\frac{7 \sqrt{\frac{5}{3}} C_{11,22}^{51,1}}{1536}+\frac{5 C_{20,20}^{40,1}}{1536 \sqrt{3}}+\frac{7 C_{20,22}^{42,1}}{1536
\sqrt{15}}-\frac{7 C_{22,20}^{42,1}}{768 \sqrt{15}}+\frac{7 C_{22,22}^{40,1}}{1536 \sqrt{15}}-\frac{1}{384} \sqrt{\frac{7}{15}} C_{22,22}^{42,1}-\frac{3
\sqrt{3} C_{22,22}^{44,1}}{640}+\frac{C_{31,20}^{31,1}}{1536 \sqrt{3}}-\frac{7 C_{31,22}^{31,1}}{1536 \sqrt{15}}+\frac{\sqrt{7} C_{31,22}^{33,1}}{640}+\frac{3
\sqrt{7} C_{33,20}^{33,1}}{1280}+\frac{3}{640} \sqrt{\frac{21}{10}} C_{33,22}^{33,1}\)

\noindent\(C_{31,2011,0}^{11,0011,1,\text{NonLoc}}= \frac{7 C_{00,00}^{60,0}}{1024}+\frac{7 C_{00,20}^{60,0}}{1024 \sqrt{3}}-\frac{7}{512}
\sqrt{\frac{3}{5}} C_{00,22}^{62,0}+\frac{7 C_{11,00}^{51,0}}{3072}+\frac{7 C_{11,20}^{31,0}}{3072 \sqrt{3}}-\frac{7 \sqrt{\frac{5}{3}} C_{11,22}^{51,0}}{1536}+\frac{5
C_{20,00}^{40,0}}{3072}+\frac{5 C_{20,20}^{40,0}}{3072 \sqrt{3}}-\frac{7 C_{20,22}^{42,0}}{1536 \sqrt{15}}-\frac{7 C_{22,00}^{42,0}}{1536 \sqrt{5}}-\frac{7
C_{22,20}^{42,0}}{1536 \sqrt{15}}-\frac{7 C_{22,22}^{40,0}}{1536 \sqrt{15}}+\frac{1}{384} \sqrt{\frac{7}{15}} C_{22,22}^{42,0}+\frac{3 \sqrt{3} C_{22,22}^{44,0}}{640}-\frac{C_{31,00}^{31,0}}{3072}-\frac{C_{31,20}^{31,0}}{3072
\sqrt{3}}-\frac{7 C_{31,22}^{31,0}}{1536 \sqrt{15}}+\frac{\sqrt{7} C_{31,22}^{33,0}}{640}-\frac{3 \sqrt{21} C_{33,00}^{33,0}}{2560}-\frac{3 \sqrt{7}
C_{33,20}^{33,0}}{2560}+\frac{3}{640} \sqrt{\frac{21}{10}} C_{33,22}^{33,0}\)

\noindent\(C_{31,2011,1}^{00,1101,0,\text{Loc}}= \frac{7 C_{11,11}^{51,0}}{384}+\frac{7 C_{11,11}^{51,1}}{768}+\frac{C_{31,11}^{31,0}}{128}+\frac{C_{31,11}^{31,1}}{256}-\frac{3}{160}
\sqrt{\frac{7}{2}} C_{33,11}^{33,0}-\frac{3}{320} \sqrt{\frac{7}{2}} C_{33,11}^{33,1}\)

\noindent\(C_{31,2011,1}^{00,1101,0,\text{NonLoc}}= \frac{7 C_{11,11}^{51,0}}{768}+\frac{7 C_{11,11}^{51,1}}{384}+\frac{C_{31,11}^{31,0}}{256}+\frac{C_{31,11}^{31,1}}{128}-\frac{3}{320}
\sqrt{\frac{7}{2}} C_{33,11}^{33,0}-\frac{3}{160} \sqrt{\frac{7}{2}} C_{33,11}^{33,1}\)

\noindent\(C_{31,2011,1}^{00,1101,1,\text{Loc}}= \frac{7 C_{11,11}^{51,1}}{256 \sqrt{3}}+\frac{\sqrt{3} C_{31,11}^{31,1}}{256}-\frac{3}{320}
\sqrt{\frac{21}{2}} C_{33,11}^{33,1}\)

\noindent\(C_{31,2011,1}^{00,1101,1,\text{NonLoc}}= \frac{7 C_{11,11}^{51,0}}{256 \sqrt{3}}+\frac{\sqrt{3} C_{31,11}^{31,0}}{256}-\frac{3}{320}
\sqrt{\frac{21}{2}} C_{33,11}^{33,0}\)

\noindent\(C_{31,2011,1}^{11,0011,0,\text{Loc}}= \frac{7 C_{00,20}^{60,0}}{256 \sqrt{3}}+\frac{7 C_{00,20}^{60,1}}{512 \sqrt{3}}-\frac{7}{512}
\sqrt{\frac{3}{5}} C_{00,22}^{62,0}-\frac{7 \sqrt{\frac{3}{5}} C_{00,22}^{62,1}}{1024}-\frac{7 C_{11,20}^{31,0}}{768 \sqrt{3}}-\frac{7 C_{11,20}^{31,1}}{1536
\sqrt{3}}+\frac{7 \sqrt{\frac{5}{3}} C_{11,22}^{51,0}}{1536}+\frac{7 \sqrt{\frac{5}{3}} C_{11,22}^{51,1}}{3072}+\frac{5 C_{20,20}^{40,0}}{768 \sqrt{3}}+\frac{5
C_{20,20}^{40,1}}{1536 \sqrt{3}}-\frac{7 C_{20,22}^{42,0}}{1536 \sqrt{15}}-\frac{7 C_{20,22}^{42,1}}{3072 \sqrt{15}}-\frac{7 C_{22,20}^{42,0}}{384
\sqrt{15}}-\frac{7 C_{22,20}^{42,1}}{768 \sqrt{15}}-\frac{7 C_{22,22}^{40,0}}{1536 \sqrt{15}}-\frac{7 C_{22,22}^{40,1}}{3072 \sqrt{15}}+\frac{1}{384}
\sqrt{\frac{7}{15}} C_{22,22}^{42,0}+\frac{1}{768} \sqrt{\frac{7}{15}} C_{22,22}^{42,1}+\frac{3 \sqrt{3} C_{22,22}^{44,0}}{640}+\frac{3 \sqrt{3}
C_{22,22}^{44,1}}{1280}+\frac{C_{31,20}^{31,0}}{768 \sqrt{3}}+\frac{C_{31,20}^{31,1}}{1536 \sqrt{3}}+\frac{7 C_{31,22}^{31,0}}{1536 \sqrt{15}}+\frac{7
C_{31,22}^{31,1}}{3072 \sqrt{15}}-\frac{\sqrt{7} C_{31,22}^{33,0}}{640}-\frac{\sqrt{7} C_{31,22}^{33,1}}{1280}+\frac{3 \sqrt{7} C_{33,20}^{33,0}}{640}+\frac{3
\sqrt{7} C_{33,20}^{33,1}}{1280}-\frac{3}{640} \sqrt{\frac{21}{10}} C_{33,22}^{33,0}-\frac{3 \sqrt{\frac{21}{10}} C_{33,22}^{33,1}}{1280}\)

\noindent\(C_{31,2011,1}^{11,0011,0,\text{NonLoc}}= \frac{7 C_{00,00}^{60,0}}{1024}+\frac{7 C_{00,00}^{60,1}}{512}+\frac{7 C_{00,20}^{60,0}}{1024
\sqrt{3}}+\frac{7 C_{00,20}^{60,1}}{512 \sqrt{3}}+\frac{7 \sqrt{\frac{3}{5}} C_{00,22}^{62,0}}{1024}+\frac{7}{512} \sqrt{\frac{3}{5}} C_{00,22}^{62,1}+\frac{7
C_{11,00}^{51,0}}{3072}+\frac{7 C_{11,00}^{51,1}}{1536}+\frac{7 C_{11,20}^{31,0}}{3072 \sqrt{3}}+\frac{7 C_{11,20}^{31,1}}{1536 \sqrt{3}}+\frac{7
\sqrt{\frac{5}{3}} C_{11,22}^{51,0}}{3072}+\frac{7 \sqrt{\frac{5}{3}} C_{11,22}^{51,1}}{1536}+\frac{5 C_{20,00}^{40,0}}{3072}+\frac{5 C_{20,00}^{40,1}}{1536}+\frac{5
C_{20,20}^{40,0}}{3072 \sqrt{3}}+\frac{5 C_{20,20}^{40,1}}{1536 \sqrt{3}}+\frac{7 C_{20,22}^{42,0}}{3072 \sqrt{15}}+\frac{7 C_{20,22}^{42,1}}{1536
\sqrt{15}}-\frac{7 C_{22,00}^{42,0}}{1536 \sqrt{5}}-\frac{7 C_{22,00}^{42,1}}{768 \sqrt{5}}-\frac{7 C_{22,20}^{42,0}}{1536 \sqrt{15}}-\frac{7 C_{22,20}^{42,1}}{768
\sqrt{15}}+\frac{7 C_{22,22}^{40,0}}{3072 \sqrt{15}}+\frac{7 C_{22,22}^{40,1}}{1536 \sqrt{15}}-\frac{1}{768} \sqrt{\frac{7}{15}} C_{22,22}^{42,0}-\frac{1}{384}
\sqrt{\frac{7}{15}} C_{22,22}^{42,1}-\frac{3 \sqrt{3} C_{22,22}^{44,0}}{1280}-\frac{3 \sqrt{3} C_{22,22}^{44,1}}{640}-\frac{C_{31,00}^{31,0}}{3072}-\frac{C_{31,00}^{31,1}}{1536}-\frac{C_{31,20}^{31,0}}{3072
\sqrt{3}}-\frac{C_{31,20}^{31,1}}{1536 \sqrt{3}}+\frac{7 C_{31,22}^{31,0}}{3072 \sqrt{15}}+\frac{7 C_{31,22}^{31,1}}{1536 \sqrt{15}}-\frac{\sqrt{7}
C_{31,22}^{33,0}}{1280}-\frac{\sqrt{7} C_{31,22}^{33,1}}{640}-\frac{3 \sqrt{21} C_{33,00}^{33,0}}{2560}-\frac{3 \sqrt{21} C_{33,00}^{33,1}}{1280}-\frac{3
\sqrt{7} C_{33,20}^{33,0}}{2560}-\frac{3 \sqrt{7} C_{33,20}^{33,1}}{1280}-\frac{3 \sqrt{\frac{21}{10}} C_{33,22}^{33,0}}{1280}-\frac{3}{640} \sqrt{\frac{21}{10}}
C_{33,22}^{33,1}\)

\noindent\(C_{31,2011,1}^{11,0011,1,\text{Loc}}= \frac{7 C_{00,20}^{60,1}}{512}-\frac{21 C_{00,22}^{62,1}}{1024 \sqrt{5}}-\frac{7 C_{11,20}^{31,1}}{1536}+\frac{7
\sqrt{5} C_{11,22}^{51,1}}{3072}+\frac{5 C_{20,20}^{40,1}}{1536}-\frac{7 C_{20,22}^{42,1}}{3072 \sqrt{5}}-\frac{7 C_{22,20}^{42,1}}{768 \sqrt{5}}-\frac{7
C_{22,22}^{40,1}}{3072 \sqrt{5}}+\frac{1}{768} \sqrt{\frac{7}{5}} C_{22,22}^{42,1}+\frac{9 C_{22,22}^{44,1}}{1280}+\frac{C_{31,20}^{31,1}}{1536}+\frac{7
C_{31,22}^{31,1}}{3072 \sqrt{5}}-\frac{\sqrt{21} C_{31,22}^{33,1}}{1280}+\frac{3 \sqrt{21} C_{33,20}^{33,1}}{1280}-\frac{9 \sqrt{\frac{7}{10}} C_{33,22}^{33,1}}{1280}\)

\noindent\(C_{31,2011,1}^{11,0011,1,\text{NonLoc}}= \frac{7 \sqrt{3} C_{00,00}^{60,0}}{1024}+\frac{7 C_{00,20}^{60,0}}{1024}+\frac{21
C_{00,22}^{62,0}}{1024 \sqrt{5}}+\frac{7 C_{11,00}^{51,0}}{1024 \sqrt{3}}+\frac{7 C_{11,20}^{31,0}}{3072}+\frac{7 \sqrt{5} C_{11,22}^{51,0}}{3072}+\frac{5
C_{20,00}^{40,0}}{1024 \sqrt{3}}+\frac{5 C_{20,20}^{40,0}}{3072}+\frac{7 C_{20,22}^{42,0}}{3072 \sqrt{5}}-\frac{7 C_{22,00}^{42,0}}{512 \sqrt{15}}-\frac{7
C_{22,20}^{42,0}}{1536 \sqrt{5}}+\frac{7 C_{22,22}^{40,0}}{3072 \sqrt{5}}-\frac{1}{768} \sqrt{\frac{7}{5}} C_{22,22}^{42,0}-\frac{9 C_{22,22}^{44,0}}{1280}-\frac{C_{31,00}^{31,0}}{1024
\sqrt{3}}-\frac{C_{31,20}^{31,0}}{3072}+\frac{7 C_{31,22}^{31,0}}{3072 \sqrt{5}}-\frac{\sqrt{21} C_{31,22}^{33,0}}{1280}-\frac{9 \sqrt{7} C_{33,00}^{33,0}}{2560}-\frac{3
\sqrt{21} C_{33,20}^{33,0}}{2560}-\frac{9 \sqrt{\frac{7}{10}} C_{33,22}^{33,0}}{1280}\)

\noindent\(C_{31,2011,2}^{11,0011,0,\text{Loc}}= \frac{7 \sqrt{5} C_{00,20}^{60,0}}{768}+\frac{7 \sqrt{5} C_{00,20}^{60,1}}{1536}+\frac{7
C_{00,22}^{62,0}}{2560}+\frac{7 C_{00,22}^{62,1}}{5120}-\frac{7 \sqrt{5} C_{11,20}^{31,0}}{2304}-\frac{7 \sqrt{5} C_{11,20}^{31,1}}{4608}-\frac{7
C_{11,22}^{51,0}}{4608}-\frac{7 C_{11,22}^{51,1}}{9216}+\frac{5 \sqrt{5} C_{20,20}^{40,0}}{2304}+\frac{5 \sqrt{5} C_{20,20}^{40,1}}{4608}+\frac{7
C_{20,22}^{42,0}}{23040}+\frac{7 C_{20,22}^{42,1}}{46080}-\frac{7 C_{22,20}^{42,0}}{1152}-\frac{7 C_{22,20}^{42,1}}{2304}+\frac{7 C_{22,22}^{40,0}}{23040}+\frac{7
C_{22,22}^{40,1}}{46080}-\frac{\sqrt{7} C_{22,22}^{42,0}}{5760}-\frac{\sqrt{7} C_{22,22}^{42,1}}{11520}-\frac{3 C_{22,22}^{44,0}}{640 \sqrt{5}}-\frac{3
C_{22,22}^{44,1}}{1280 \sqrt{5}}+\frac{\sqrt{5} C_{31,20}^{31,0}}{2304}+\frac{\sqrt{5} C_{31,20}^{31,1}}{4608}-\frac{7 C_{31,22}^{31,0}}{23040}-\frac{7
C_{31,22}^{31,1}}{46080}+\frac{1}{640} \sqrt{\frac{7}{15}} C_{31,22}^{33,0}+\frac{\sqrt{\frac{7}{15}} C_{31,22}^{33,1}}{1280}+\frac{1}{128} \sqrt{\frac{21}{5}}
C_{33,20}^{33,0}+\frac{1}{256} \sqrt{\frac{21}{5}} C_{33,20}^{33,1}+\frac{3 \sqrt{\frac{7}{2}} C_{33,22}^{33,0}}{3200}+\frac{3 \sqrt{\frac{7}{2}}
C_{33,22}^{33,1}}{6400}\)

\noindent\(C_{31,2011,2}^{11,0011,0,\text{NonLoc}}= \frac{7 \sqrt{\frac{5}{3}} C_{00,00}^{60,0}}{1024}+\frac{7}{512} \sqrt{\frac{5}{3}}
C_{00,00}^{60,1}+\frac{7 \sqrt{5} C_{00,20}^{60,0}}{3072}+\frac{7 \sqrt{5} C_{00,20}^{60,1}}{1536}-\frac{7 C_{00,22}^{62,0}}{5120}-\frac{7 C_{00,22}^{62,1}}{2560}+\frac{7
\sqrt{\frac{5}{3}} C_{11,00}^{51,0}}{3072}+\frac{7 \sqrt{\frac{5}{3}} C_{11,00}^{51,1}}{1536}+\frac{7 \sqrt{5} C_{11,20}^{31,0}}{9216}+\frac{7 \sqrt{5}
C_{11,20}^{31,1}}{4608}-\frac{7 C_{11,22}^{51,0}}{9216}-\frac{7 C_{11,22}^{51,1}}{4608}+\frac{5 \sqrt{\frac{5}{3}} C_{20,00}^{40,0}}{3072}+\frac{5
\sqrt{\frac{5}{3}} C_{20,00}^{40,1}}{1536}+\frac{5 \sqrt{5} C_{20,20}^{40,0}}{9216}+\frac{5 \sqrt{5} C_{20,20}^{40,1}}{4608}-\frac{7 C_{20,22}^{42,0}}{46080}-\frac{7
C_{20,22}^{42,1}}{23040}-\frac{7 C_{22,00}^{42,0}}{1536 \sqrt{3}}-\frac{7 C_{22,00}^{42,1}}{768 \sqrt{3}}-\frac{7 C_{22,20}^{42,0}}{4608}-\frac{7
C_{22,20}^{42,1}}{2304}-\frac{7 C_{22,22}^{40,0}}{46080}-\frac{7 C_{22,22}^{40,1}}{23040}+\frac{\sqrt{7} C_{22,22}^{42,0}}{11520}+\frac{\sqrt{7}
C_{22,22}^{42,1}}{5760}+\frac{3 C_{22,22}^{44,0}}{1280 \sqrt{5}}+\frac{3 C_{22,22}^{44,1}}{640 \sqrt{5}}-\frac{\sqrt{\frac{5}{3}} C_{31,00}^{31,0}}{3072}-\frac{\sqrt{\frac{5}{3}}
C_{31,00}^{31,1}}{1536}-\frac{\sqrt{5} C_{31,20}^{31,0}}{9216}-\frac{\sqrt{5} C_{31,20}^{31,1}}{4608}-\frac{7 C_{31,22}^{31,0}}{46080}-\frac{7 C_{31,22}^{31,1}}{23040}+\frac{\sqrt{\frac{7}{15}}
C_{31,22}^{33,0}}{1280}+\frac{1}{640} \sqrt{\frac{7}{15}} C_{31,22}^{33,1}-\frac{3}{512} \sqrt{\frac{7}{5}} C_{33,00}^{33,0}-\frac{3}{256} \sqrt{\frac{7}{5}}
C_{33,00}^{33,1}-\frac{1}{512} \sqrt{\frac{21}{5}} C_{33,20}^{33,0}-\frac{1}{256} \sqrt{\frac{21}{5}} C_{33,20}^{33,1}+\frac{3 \sqrt{\frac{7}{2}}
C_{33,22}^{33,0}}{6400}+\frac{3 \sqrt{\frac{7}{2}} C_{33,22}^{33,1}}{3200}\)

\noindent\(C_{31,2011,2}^{11,0011,1,\text{Loc}}= \frac{7}{512} \sqrt{\frac{5}{3}} C_{00,20}^{60,1}+\frac{7 \sqrt{3} C_{00,22}^{62,1}}{5120}-\frac{7
\sqrt{\frac{5}{3}} C_{11,20}^{31,1}}{1536}-\frac{7 C_{11,22}^{51,1}}{3072 \sqrt{3}}+\frac{5 \sqrt{\frac{5}{3}} C_{20,20}^{40,1}}{1536}+\frac{7 C_{20,22}^{42,1}}{15360
\sqrt{3}}-\frac{7 C_{22,20}^{42,1}}{768 \sqrt{3}}+\frac{7 C_{22,22}^{40,1}}{15360 \sqrt{3}}-\frac{\sqrt{\frac{7}{3}} C_{22,22}^{42,1}}{3840}-\frac{3
\sqrt{\frac{3}{5}} C_{22,22}^{44,1}}{1280}+\frac{\sqrt{\frac{5}{3}} C_{31,20}^{31,1}}{1536}-\frac{7 C_{31,22}^{31,1}}{15360 \sqrt{3}}+\frac{\sqrt{\frac{7}{5}}
C_{31,22}^{33,1}}{1280}+\frac{3}{256} \sqrt{\frac{7}{5}} C_{33,20}^{33,1}+\frac{3 \sqrt{\frac{21}{2}} C_{33,22}^{33,1}}{6400}\)

\noindent\(C_{31,2011,2}^{11,0011,1,\text{NonLoc}}= \frac{7 \sqrt{5} C_{00,00}^{60,0}}{1024}+\frac{7 \sqrt{\frac{5}{3}} C_{00,20}^{60,0}}{1024}-\frac{7
\sqrt{3} C_{00,22}^{62,0}}{5120}+\frac{7 \sqrt{5} C_{11,00}^{51,0}}{3072}+\frac{7 \sqrt{\frac{5}{3}} C_{11,20}^{31,0}}{3072}-\frac{7 C_{11,22}^{51,0}}{3072
\sqrt{3}}+\frac{5 \sqrt{5} C_{20,00}^{40,0}}{3072}+\frac{5 \sqrt{\frac{5}{3}} C_{20,20}^{40,0}}{3072}-\frac{7 C_{20,22}^{42,0}}{15360 \sqrt{3}}-\frac{7
C_{22,00}^{42,0}}{1536}-\frac{7 C_{22,20}^{42,0}}{1536 \sqrt{3}}-\frac{7 C_{22,22}^{40,0}}{15360 \sqrt{3}}+\frac{\sqrt{\frac{7}{3}} C_{22,22}^{42,0}}{3840}+\frac{3
\sqrt{\frac{3}{5}} C_{22,22}^{44,0}}{1280}-\frac{\sqrt{5} C_{31,00}^{31,0}}{3072}-\frac{\sqrt{\frac{5}{3}} C_{31,20}^{31,0}}{3072}-\frac{7 C_{31,22}^{31,0}}{15360
\sqrt{3}}+\frac{\sqrt{\frac{7}{5}} C_{31,22}^{33,0}}{1280}-\frac{3}{512} \sqrt{\frac{21}{5}} C_{33,00}^{33,0}-\frac{3}{512} \sqrt{\frac{7}{5}} C_{33,20}^{33,0}+\frac{3
\sqrt{\frac{21}{2}} C_{33,22}^{33,0}}{6400}\)

\noindent\(C_{31,2202,1}^{00,1111,0,\text{Loc}}= \frac{7 C_{11,11}^{51,0}}{384 \sqrt{5}}+\frac{7 C_{11,11}^{51,1}}{768 \sqrt{5}}+\frac{C_{31,11}^{31,0}}{128
\sqrt{5}}+\frac{C_{31,11}^{31,1}}{256 \sqrt{5}}-\frac{3}{160} \sqrt{\frac{7}{10}} C_{33,11}^{33,0}-\frac{3}{320} \sqrt{\frac{7}{10}} C_{33,11}^{33,1}\)

\noindent\(C_{31,2202,1}^{00,1111,0,\text{NonLoc}}= \frac{7 C_{11,11}^{51,0}}{768 \sqrt{5}}+\frac{7 C_{11,11}^{51,1}}{384 \sqrt{5}}+\frac{C_{31,11}^{31,0}}{256
\sqrt{5}}+\frac{C_{31,11}^{31,1}}{128 \sqrt{5}}-\frac{3}{320} \sqrt{\frac{7}{10}} C_{33,11}^{33,0}-\frac{3}{160} \sqrt{\frac{7}{10}} C_{33,11}^{33,1}\)

\noindent\(C_{31,2202,1}^{00,1111,1,\text{Loc}}= \frac{7 C_{11,11}^{51,1}}{256 \sqrt{15}}+\frac{1}{256} \sqrt{\frac{3}{5}} C_{31,11}^{31,1}-\frac{3}{320}
\sqrt{\frac{21}{10}} C_{33,11}^{33,1}\)

\noindent\(C_{31,2202,1}^{00,1111,1,\text{NonLoc}}= \frac{7 C_{11,11}^{51,0}}{256 \sqrt{15}}+\frac{1}{256} \sqrt{\frac{3}{5}} C_{31,11}^{31,0}-\frac{3}{320}
\sqrt{\frac{21}{10}} C_{33,11}^{33,0}\)

\noindent\(C_{31,2202,1}^{11,0000,0,\text{Loc}}= -\frac{7 C_{00,00}^{60,0}}{128 \sqrt{5}}-\frac{7 C_{00,00}^{60,1}}{256 \sqrt{5}}+\frac{7
C_{11,00}^{51,0}}{384 \sqrt{5}}+\frac{7 C_{11,00}^{51,1}}{768 \sqrt{5}}+\frac{7 C_{20,00}^{40,0}}{384 \sqrt{5}}+\frac{7 C_{20,00}^{40,1}}{768 \sqrt{5}}-\frac{7
C_{22,00}^{42,0}}{1920}-\frac{7 C_{22,00}^{42,1}}{3840}-\frac{7 C_{31,00}^{31,0}}{384 \sqrt{5}}-\frac{7 C_{31,00}^{31,1}}{768 \sqrt{5}}+\frac{3}{640}
\sqrt{\frac{21}{5}} C_{33,00}^{33,0}+\frac{3 \sqrt{\frac{21}{5}} C_{33,00}^{33,1}}{1280}\)

\noindent\(C_{31,2202,1}^{11,0000,0,\text{NonLoc}}= \frac{7 C_{00,00}^{60,0}}{512 \sqrt{5}}+\frac{7 C_{00,00}^{60,1}}{256 \sqrt{5}}-\frac{7}{512}
\sqrt{\frac{3}{5}} C_{00,20}^{60,0}-\frac{7}{256} \sqrt{\frac{3}{5}} C_{00,20}^{60,1}+\frac{7 C_{11,00}^{51,0}}{1536 \sqrt{5}}+\frac{7 C_{11,00}^{51,1}}{768
\sqrt{5}}-\frac{7 C_{11,20}^{31,0}}{512 \sqrt{15}}-\frac{7 C_{11,20}^{31,1}}{256 \sqrt{15}}-\frac{7 C_{20,00}^{40,0}}{1536 \sqrt{5}}-\frac{7 C_{20,00}^{40,1}}{768
\sqrt{5}}+\frac{7 C_{20,20}^{40,0}}{512 \sqrt{15}}+\frac{7 C_{20,20}^{40,1}}{256 \sqrt{15}}+\frac{7 C_{22,00}^{42,0}}{7680}+\frac{7 C_{22,00}^{42,1}}{3840}-\frac{7
C_{22,20}^{42,0}}{2560 \sqrt{3}}-\frac{7 C_{22,20}^{42,1}}{1280 \sqrt{3}}-\frac{7 C_{31,00}^{31,0}}{1536 \sqrt{5}}-\frac{7 C_{31,00}^{31,1}}{768
\sqrt{5}}+\frac{7 C_{31,20}^{31,0}}{512 \sqrt{15}}+\frac{7 C_{31,20}^{31,1}}{256 \sqrt{15}}+\frac{3 \sqrt{\frac{21}{5}} C_{33,00}^{33,0}}{2560}+\frac{3
\sqrt{\frac{21}{5}} C_{33,00}^{33,1}}{1280}-\frac{9 \sqrt{\frac{7}{5}} C_{33,20}^{33,0}}{2560}-\frac{9 \sqrt{\frac{7}{5}} C_{33,20}^{33,1}}{1280}\)

\noindent\(C_{31,2202,1}^{11,0000,1,\text{Loc}}= -\frac{7}{256} \sqrt{\frac{3}{5}} C_{00,00}^{60,1}+\frac{7 C_{11,00}^{51,1}}{256 \sqrt{15}}+\frac{7
C_{20,00}^{40,1}}{256 \sqrt{15}}-\frac{7 C_{22,00}^{42,1}}{1280 \sqrt{3}}-\frac{7 C_{31,00}^{31,1}}{256 \sqrt{15}}+\frac{9 \sqrt{\frac{7}{5}} C_{33,00}^{33,1}}{1280}\)

\noindent\(C_{31,2202,1}^{11,0000,1,\text{NonLoc}}= \frac{7}{512} \sqrt{\frac{3}{5}} C_{00,00}^{60,0}-\frac{21 C_{00,20}^{60,0}}{512
\sqrt{5}}+\frac{7 C_{11,00}^{51,0}}{512 \sqrt{15}}-\frac{7 C_{11,20}^{31,0}}{512 \sqrt{5}}-\frac{7 C_{20,00}^{40,0}}{512 \sqrt{15}}+\frac{7 C_{20,20}^{40,0}}{512
\sqrt{5}}+\frac{7 C_{22,00}^{42,0}}{2560 \sqrt{3}}-\frac{7 C_{22,20}^{42,0}}{2560}-\frac{7 C_{31,00}^{31,0}}{512 \sqrt{15}}+\frac{7 C_{31,20}^{31,0}}{512
\sqrt{5}}+\frac{9 \sqrt{\frac{7}{5}} C_{33,00}^{33,0}}{2560}-\frac{9 \sqrt{\frac{21}{5}} C_{33,20}^{33,0}}{2560}\)

\noindent\(C_{31,2202,2}^{00,1112,0,\text{Loc}}= -\frac{7 C_{11,11}^{51,0}}{128 \sqrt{15}}-\frac{7 C_{11,11}^{51,1}}{256 \sqrt{15}}-\frac{1}{128}
\sqrt{\frac{3}{5}} C_{31,11}^{31,0}-\frac{1}{256} \sqrt{\frac{3}{5}} C_{31,11}^{31,1}+\frac{3}{160} \sqrt{\frac{21}{10}} C_{33,11}^{33,0}+\frac{3}{320}
\sqrt{\frac{21}{10}} C_{33,11}^{33,1}\)

\noindent\(C_{31,2202,2}^{00,1112,0,\text{NonLoc}}= -\frac{7 C_{11,11}^{51,0}}{256 \sqrt{15}}-\frac{7 C_{11,11}^{51,1}}{128 \sqrt{15}}-\frac{1}{256}
\sqrt{\frac{3}{5}} C_{31,11}^{31,0}-\frac{1}{128} \sqrt{\frac{3}{5}} C_{31,11}^{31,1}+\frac{3}{320} \sqrt{\frac{21}{10}} C_{33,11}^{33,0}+\frac{3}{160}
\sqrt{\frac{21}{10}} C_{33,11}^{33,1}\)

\noindent\(C_{31,2202,2}^{00,1112,1,\text{Loc}}= -\frac{7 C_{11,11}^{51,1}}{256 \sqrt{5}}-\frac{3 C_{31,11}^{31,1}}{256 \sqrt{5}}+\frac{9}{320}
\sqrt{\frac{7}{10}} C_{33,11}^{33,1}\)

\noindent\(C_{31,2202,2}^{00,1112,1,\text{NonLoc}}= -\frac{7 C_{11,11}^{51,0}}{256 \sqrt{5}}-\frac{3 C_{31,11}^{31,0}}{256 \sqrt{5}}+\frac{9}{320}
\sqrt{\frac{7}{10}} C_{33,11}^{33,0}\)

\noindent\(C_{31,2211,0}^{11,0011,0,\text{Loc}}= \frac{7 C_{00,20}^{60,0}}{384 \sqrt{5}}+\frac{7 C_{00,20}^{60,1}}{768 \sqrt{5}}+\frac{7
C_{00,22}^{62,0}}{640}+\frac{7 C_{00,22}^{62,1}}{1280}-\frac{7 C_{11,20}^{31,0}}{1152 \sqrt{5}}-\frac{7 C_{11,20}^{31,1}}{2304 \sqrt{5}}-\frac{7
C_{11,22}^{51,0}}{11520}-\frac{7 C_{11,22}^{51,1}}{23040}-\frac{9 \sqrt{\frac{21}{5}} C_{11,22}^{53,0}}{2560}-\frac{9 \sqrt{\frac{21}{5}} C_{11,22}^{53,1}}{5120}-\frac{7
C_{20,20}^{40,0}}{1152 \sqrt{5}}-\frac{7 C_{20,20}^{40,1}}{2304 \sqrt{5}}-\frac{7 C_{20,22}^{42,0}}{11520}-\frac{7 C_{20,22}^{42,1}}{23040}+\frac{7
C_{22,20}^{42,0}}{5760}+\frac{7 C_{22,20}^{42,1}}{11520}-\frac{13 C_{22,22}^{40,0}}{2880}-\frac{13 C_{22,22}^{40,1}}{5760}+\frac{\sqrt{7} C_{22,22}^{42,0}}{2880}+\frac{\sqrt{7}
C_{22,22}^{42,1}}{5760}+\frac{3 C_{22,22}^{44,0}}{320 \sqrt{5}}+\frac{3 C_{22,22}^{44,1}}{640 \sqrt{5}}+\frac{7 C_{31,20}^{31,0}}{1152 \sqrt{5}}+\frac{7
C_{31,20}^{31,1}}{2304 \sqrt{5}}+\frac{37 C_{31,22}^{31,0}}{11520}+\frac{37 C_{31,22}^{31,1}}{23040}+\frac{7 \sqrt{\frac{7}{15}} C_{31,22}^{33,0}}{2560}+\frac{7
\sqrt{\frac{7}{15}} C_{31,22}^{33,1}}{5120}-\frac{1}{640} \sqrt{\frac{21}{5}} C_{33,20}^{33,0}-\frac{\sqrt{\frac{21}{5}} C_{33,20}^{33,1}}{1280}-\frac{3
\sqrt{\frac{7}{2}} C_{33,22}^{33,0}}{1600}-\frac{3 \sqrt{\frac{7}{2}} C_{33,22}^{33,1}}{3200}\)

\noindent\(C_{31,2211,0}^{11,0011,0,\text{NonLoc}}= \frac{7 C_{00,00}^{60,0}}{512 \sqrt{15}}+\frac{7 C_{00,00}^{60,1}}{256 \sqrt{15}}+\frac{7
C_{00,20}^{60,0}}{1536 \sqrt{5}}+\frac{7 C_{00,20}^{60,1}}{768 \sqrt{5}}-\frac{7 C_{00,22}^{62,0}}{1280}-\frac{7 C_{00,22}^{62,1}}{640}+\frac{7 C_{11,00}^{51,0}}{1536
\sqrt{15}}+\frac{7 C_{11,00}^{51,1}}{768 \sqrt{15}}+\frac{7 C_{11,20}^{31,0}}{4608 \sqrt{5}}+\frac{7 C_{11,20}^{31,1}}{2304 \sqrt{5}}-\frac{7 C_{11,22}^{51,0}}{23040}-\frac{7
C_{11,22}^{51,1}}{11520}-\frac{9 \sqrt{\frac{21}{5}} C_{11,22}^{53,0}}{5120}-\frac{9 \sqrt{\frac{21}{5}} C_{11,22}^{53,1}}{2560}-\frac{7 C_{20,00}^{40,0}}{1536
\sqrt{15}}-\frac{7 C_{20,00}^{40,1}}{768 \sqrt{15}}-\frac{7 C_{20,20}^{40,0}}{4608 \sqrt{5}}-\frac{7 C_{20,20}^{40,1}}{2304 \sqrt{5}}+\frac{7 C_{20,22}^{42,0}}{23040}+\frac{7
C_{20,22}^{42,1}}{11520}+\frac{7 C_{22,00}^{42,0}}{7680 \sqrt{3}}+\frac{7 C_{22,00}^{42,1}}{3840 \sqrt{3}}+\frac{7 C_{22,20}^{42,0}}{23040}+\frac{7
C_{22,20}^{42,1}}{11520}+\frac{13 C_{22,22}^{40,0}}{5760}+\frac{13 C_{22,22}^{40,1}}{2880}-\frac{\sqrt{7} C_{22,22}^{42,0}}{5760}-\frac{\sqrt{7}
C_{22,22}^{42,1}}{2880}-\frac{3 C_{22,22}^{44,0}}{640 \sqrt{5}}-\frac{3 C_{22,22}^{44,1}}{320 \sqrt{5}}-\frac{7 C_{31,00}^{31,0}}{1536 \sqrt{15}}-\frac{7
C_{31,00}^{31,1}}{768 \sqrt{15}}-\frac{7 C_{31,20}^{31,0}}{4608 \sqrt{5}}-\frac{7 C_{31,20}^{31,1}}{2304 \sqrt{5}}+\frac{37 C_{31,22}^{31,0}}{23040}+\frac{37
C_{31,22}^{31,1}}{11520}+\frac{7 \sqrt{\frac{7}{15}} C_{31,22}^{33,0}}{5120}+\frac{7 \sqrt{\frac{7}{15}} C_{31,22}^{33,1}}{2560}+\frac{3 \sqrt{\frac{7}{5}}
C_{33,00}^{33,0}}{2560}+\frac{3 \sqrt{\frac{7}{5}} C_{33,00}^{33,1}}{1280}+\frac{\sqrt{\frac{21}{5}} C_{33,20}^{33,0}}{2560}+\frac{\sqrt{\frac{21}{5}}
C_{33,20}^{33,1}}{1280}-\frac{3 \sqrt{\frac{7}{2}} C_{33,22}^{33,0}}{3200}-\frac{3 \sqrt{\frac{7}{2}} C_{33,22}^{33,1}}{1600}\)

\noindent\(C_{31,2211,0}^{11,0011,1,\text{Loc}}= \frac{7 C_{00,20}^{60,1}}{256 \sqrt{15}}+\frac{7 \sqrt{3} C_{00,22}^{62,1}}{1280}-\frac{7
C_{11,20}^{31,1}}{768 \sqrt{15}}-\frac{7 C_{11,22}^{51,1}}{7680 \sqrt{3}}-\frac{27 \sqrt{\frac{7}{5}} C_{11,22}^{53,1}}{5120}-\frac{7 C_{20,20}^{40,1}}{768
\sqrt{15}}-\frac{7 C_{20,22}^{42,1}}{7680 \sqrt{3}}+\frac{7 C_{22,20}^{42,1}}{3840 \sqrt{3}}-\frac{13 C_{22,22}^{40,1}}{1920 \sqrt{3}}+\frac{\sqrt{\frac{7}{3}}
C_{22,22}^{42,1}}{1920}+\frac{3}{640} \sqrt{\frac{3}{5}} C_{22,22}^{44,1}+\frac{7 C_{31,20}^{31,1}}{768 \sqrt{15}}+\frac{37 C_{31,22}^{31,1}}{7680
\sqrt{3}}+\frac{7 \sqrt{\frac{7}{5}} C_{31,22}^{33,1}}{5120}-\frac{3 \sqrt{\frac{7}{5}} C_{33,20}^{33,1}}{1280}-\frac{3 \sqrt{\frac{21}{2}} C_{33,22}^{33,1}}{3200}\)

\noindent\(C_{31,2211,0}^{11,0011,1,\text{NonLoc}}= \frac{7 C_{00,00}^{60,0}}{512 \sqrt{5}}+\frac{7 C_{00,20}^{60,0}}{512 \sqrt{15}}-\frac{7
\sqrt{3} C_{00,22}^{62,0}}{1280}+\frac{7 C_{11,00}^{51,0}}{1536 \sqrt{5}}+\frac{7 C_{11,20}^{31,0}}{1536 \sqrt{15}}-\frac{7 C_{11,22}^{51,0}}{7680
\sqrt{3}}-\frac{27 \sqrt{\frac{7}{5}} C_{11,22}^{53,0}}{5120}-\frac{7 C_{20,00}^{40,0}}{1536 \sqrt{5}}-\frac{7 C_{20,20}^{40,0}}{1536 \sqrt{15}}+\frac{7
C_{20,22}^{42,0}}{7680 \sqrt{3}}+\frac{7 C_{22,00}^{42,0}}{7680}+\frac{7 C_{22,20}^{42,0}}{7680 \sqrt{3}}+\frac{13 C_{22,22}^{40,0}}{1920 \sqrt{3}}-\frac{\sqrt{\frac{7}{3}}
C_{22,22}^{42,0}}{1920}-\frac{3}{640} \sqrt{\frac{3}{5}} C_{22,22}^{44,0}-\frac{7 C_{31,00}^{31,0}}{1536 \sqrt{5}}-\frac{7 C_{31,20}^{31,0}}{1536
\sqrt{15}}+\frac{37 C_{31,22}^{31,0}}{7680 \sqrt{3}}+\frac{7 \sqrt{\frac{7}{5}} C_{31,22}^{33,0}}{5120}+\frac{3 \sqrt{\frac{21}{5}} C_{33,00}^{33,0}}{2560}+\frac{3
\sqrt{\frac{7}{5}} C_{33,20}^{33,0}}{2560}-\frac{3 \sqrt{\frac{21}{2}} C_{33,22}^{33,0}}{3200}\)

\noindent\(C_{31,2211,1}^{00,1101,0,\text{Loc}}= -\frac{7 C_{11,11}^{51,0}}{384 \sqrt{5}}-\frac{7 C_{11,11}^{51,1}}{768 \sqrt{5}}-\frac{C_{31,11}^{31,0}}{128
\sqrt{5}}-\frac{C_{31,11}^{31,1}}{256 \sqrt{5}}+\frac{3}{160} \sqrt{\frac{7}{10}} C_{33,11}^{33,0}+\frac{3}{320} \sqrt{\frac{7}{10}} C_{33,11}^{33,1}\)

\noindent\(C_{31,2211,1}^{00,1101,0,\text{NonLoc}}= -\frac{7 C_{11,11}^{51,0}}{768 \sqrt{5}}-\frac{7 C_{11,11}^{51,1}}{384 \sqrt{5}}-\frac{C_{31,11}^{31,0}}{256
\sqrt{5}}-\frac{C_{31,11}^{31,1}}{128 \sqrt{5}}+\frac{3}{320} \sqrt{\frac{7}{10}} C_{33,11}^{33,0}+\frac{3}{160} \sqrt{\frac{7}{10}} C_{33,11}^{33,1}\)

\noindent\(C_{31,2211,1}^{00,1101,1,\text{Loc}}= -\frac{7 C_{11,11}^{51,1}}{256 \sqrt{15}}-\frac{1}{256} \sqrt{\frac{3}{5}} C_{31,11}^{31,1}+\frac{3}{320}
\sqrt{\frac{21}{10}} C_{33,11}^{33,1}\)

\noindent\(C_{31,2211,1}^{00,1101,1,\text{NonLoc}}= -\frac{7 C_{11,11}^{51,0}}{256 \sqrt{15}}-\frac{1}{256} \sqrt{\frac{3}{5}} C_{31,11}^{31,0}+\frac{3}{320}
\sqrt{\frac{21}{10}} C_{33,11}^{33,0}\)

\noindent\(C_{31,2211,1}^{11,0011,0,\text{Loc}}= -\frac{7 C_{00,20}^{60,0}}{256 \sqrt{15}}-\frac{7 C_{00,20}^{60,1}}{512 \sqrt{15}}+\frac{7
\sqrt{3} C_{00,22}^{62,0}}{2560}+\frac{7 \sqrt{3} C_{00,22}^{62,1}}{5120}+\frac{7 C_{11,20}^{31,0}}{768 \sqrt{15}}+\frac{7 C_{11,20}^{31,1}}{1536
\sqrt{15}}+\frac{7 C_{11,22}^{51,0}}{1920 \sqrt{3}}+\frac{7 C_{11,22}^{51,1}}{3840 \sqrt{3}}-\frac{27 \sqrt{\frac{7}{5}} C_{11,22}^{53,0}}{5120}-\frac{27
\sqrt{\frac{7}{5}} C_{11,22}^{53,1}}{10240}+\frac{7 C_{20,20}^{40,0}}{768 \sqrt{15}}+\frac{7 C_{20,20}^{40,1}}{1536 \sqrt{15}}+\frac{7 C_{20,22}^{42,0}}{1920
\sqrt{3}}+\frac{7 C_{20,22}^{42,1}}{3840 \sqrt{3}}-\frac{7 C_{22,20}^{42,0}}{3840 \sqrt{3}}-\frac{7 C_{22,20}^{42,1}}{7680 \sqrt{3}}+\frac{C_{22,22}^{40,0}}{7680
\sqrt{3}}+\frac{C_{22,22}^{40,1}}{15360 \sqrt{3}}-\frac{17 \sqrt{\frac{7}{3}} C_{22,22}^{42,0}}{3840}-\frac{17 \sqrt{\frac{7}{3}} C_{22,22}^{42,1}}{7680}+\frac{3}{320}
\sqrt{\frac{3}{5}} C_{22,22}^{44,0}+\frac{3}{640} \sqrt{\frac{3}{5}} C_{22,22}^{44,1}-\frac{7 C_{31,20}^{31,0}}{768 \sqrt{15}}-\frac{7 C_{31,20}^{31,1}}{1536
\sqrt{15}}+\frac{C_{31,22}^{31,0}}{240 \sqrt{3}}+\frac{C_{31,22}^{31,1}}{480 \sqrt{3}}-\frac{13 \sqrt{\frac{7}{5}} C_{31,22}^{33,0}}{5120}-\frac{13
\sqrt{\frac{7}{5}} C_{31,22}^{33,1}}{10240}+\frac{3 \sqrt{\frac{7}{5}} C_{33,20}^{33,0}}{1280}+\frac{3 \sqrt{\frac{7}{5}} C_{33,20}^{33,1}}{2560}+\frac{3}{800}
\sqrt{\frac{21}{2}} C_{33,22}^{33,0}+\frac{3 \sqrt{\frac{21}{2}} C_{33,22}^{33,1}}{1600}\)

\noindent\(C_{31,2211,1}^{11,0011,0,\text{NonLoc}}= -\frac{7 C_{00,00}^{60,0}}{1024 \sqrt{5}}-\frac{7 C_{00,00}^{60,1}}{512 \sqrt{5}}-\frac{7
C_{00,20}^{60,0}}{1024 \sqrt{15}}-\frac{7 C_{00,20}^{60,1}}{512 \sqrt{15}}-\frac{7 \sqrt{3} C_{00,22}^{62,0}}{5120}-\frac{7 \sqrt{3} C_{00,22}^{62,1}}{2560}-\frac{7
C_{11,00}^{51,0}}{3072 \sqrt{5}}-\frac{7 C_{11,00}^{51,1}}{1536 \sqrt{5}}-\frac{7 C_{11,20}^{31,0}}{3072 \sqrt{15}}-\frac{7 C_{11,20}^{31,1}}{1536
\sqrt{15}}+\frac{7 C_{11,22}^{51,0}}{3840 \sqrt{3}}+\frac{7 C_{11,22}^{51,1}}{1920 \sqrt{3}}-\frac{27 \sqrt{\frac{7}{5}} C_{11,22}^{53,0}}{10240}-\frac{27
\sqrt{\frac{7}{5}} C_{11,22}^{53,1}}{5120}+\frac{7 C_{20,00}^{40,0}}{3072 \sqrt{5}}+\frac{7 C_{20,00}^{40,1}}{1536 \sqrt{5}}+\frac{7 C_{20,20}^{40,0}}{3072
\sqrt{15}}+\frac{7 C_{20,20}^{40,1}}{1536 \sqrt{15}}-\frac{7 C_{20,22}^{42,0}}{3840 \sqrt{3}}-\frac{7 C_{20,22}^{42,1}}{1920 \sqrt{3}}-\frac{7 C_{22,00}^{42,0}}{15360}-\frac{7
C_{22,00}^{42,1}}{7680}-\frac{7 C_{22,20}^{42,0}}{15360 \sqrt{3}}-\frac{7 C_{22,20}^{42,1}}{7680 \sqrt{3}}-\frac{C_{22,22}^{40,0}}{15360 \sqrt{3}}-\frac{C_{22,22}^{40,1}}{7680
\sqrt{3}}+\frac{17 \sqrt{\frac{7}{3}} C_{22,22}^{42,0}}{7680}+\frac{17 \sqrt{\frac{7}{3}} C_{22,22}^{42,1}}{3840}-\frac{3}{640} \sqrt{\frac{3}{5}}
C_{22,22}^{44,0}-\frac{3}{320} \sqrt{\frac{3}{5}} C_{22,22}^{44,1}+\frac{7 C_{31,00}^{31,0}}{3072 \sqrt{5}}+\frac{7 C_{31,00}^{31,1}}{1536 \sqrt{5}}+\frac{7
C_{31,20}^{31,0}}{3072 \sqrt{15}}+\frac{7 C_{31,20}^{31,1}}{1536 \sqrt{15}}+\frac{C_{31,22}^{31,0}}{480 \sqrt{3}}+\frac{C_{31,22}^{31,1}}{240 \sqrt{3}}-\frac{13
\sqrt{\frac{7}{5}} C_{31,22}^{33,0}}{10240}-\frac{13 \sqrt{\frac{7}{5}} C_{31,22}^{33,1}}{5120}-\frac{3 \sqrt{\frac{21}{5}} C_{33,00}^{33,0}}{5120}-\frac{3
\sqrt{\frac{21}{5}} C_{33,00}^{33,1}}{2560}-\frac{3 \sqrt{\frac{7}{5}} C_{33,20}^{33,0}}{5120}-\frac{3 \sqrt{\frac{7}{5}} C_{33,20}^{33,1}}{2560}+\frac{3
\sqrt{\frac{21}{2}} C_{33,22}^{33,0}}{1600}+\frac{3}{800} \sqrt{\frac{21}{2}} C_{33,22}^{33,1}\)

\noindent\(C_{31,2211,1}^{11,0011,1,\text{Loc}}= -\frac{7 C_{00,20}^{60,1}}{512 \sqrt{5}}+\frac{21 C_{00,22}^{62,1}}{5120}+\frac{7
C_{11,20}^{31,1}}{1536 \sqrt{5}}+\frac{7 C_{11,22}^{51,1}}{3840}-\frac{27 \sqrt{\frac{21}{5}} C_{11,22}^{53,1}}{10240}+\frac{7 C_{20,20}^{40,1}}{1536
\sqrt{5}}+\frac{7 C_{20,22}^{42,1}}{3840}-\frac{7 C_{22,20}^{42,1}}{7680}+\frac{C_{22,22}^{40,1}}{15360}-\frac{17 \sqrt{7} C_{22,22}^{42,1}}{7680}+\frac{9
C_{22,22}^{44,1}}{640 \sqrt{5}}-\frac{7 C_{31,20}^{31,1}}{1536 \sqrt{5}}+\frac{C_{31,22}^{31,1}}{480}-\frac{13 \sqrt{\frac{21}{5}} C_{31,22}^{33,1}}{10240}+\frac{3
\sqrt{\frac{21}{5}} C_{33,20}^{33,1}}{2560}+\frac{9 \sqrt{\frac{7}{2}} C_{33,22}^{33,1}}{1600}\)

\noindent\(C_{31,2211,1}^{11,0011,1,\text{NonLoc}}= -\frac{7 \sqrt{\frac{3}{5}} C_{00,00}^{60,0}}{1024}-\frac{7 C_{00,20}^{60,0}}{1024
\sqrt{5}}-\frac{21 C_{00,22}^{62,0}}{5120}-\frac{7 C_{11,00}^{51,0}}{1024 \sqrt{15}}-\frac{7 C_{11,20}^{31,0}}{3072 \sqrt{5}}+\frac{7 C_{11,22}^{51,0}}{3840}-\frac{27
\sqrt{\frac{21}{5}} C_{11,22}^{53,0}}{10240}+\frac{7 C_{20,00}^{40,0}}{1024 \sqrt{15}}+\frac{7 C_{20,20}^{40,0}}{3072 \sqrt{5}}-\frac{7 C_{20,22}^{42,0}}{3840}-\frac{7
C_{22,00}^{42,0}}{5120 \sqrt{3}}-\frac{7 C_{22,20}^{42,0}}{15360}-\frac{C_{22,22}^{40,0}}{15360}+\frac{17 \sqrt{7} C_{22,22}^{42,0}}{7680}-\frac{9
C_{22,22}^{44,0}}{640 \sqrt{5}}+\frac{7 C_{31,00}^{31,0}}{1024 \sqrt{15}}+\frac{7 C_{31,20}^{31,0}}{3072 \sqrt{5}}+\frac{C_{31,22}^{31,0}}{480}-\frac{13
\sqrt{\frac{21}{5}} C_{31,22}^{33,0}}{10240}-\frac{9 \sqrt{\frac{7}{5}} C_{33,00}^{33,0}}{5120}-\frac{3 \sqrt{\frac{21}{5}} C_{33,20}^{33,0}}{5120}+\frac{9
\sqrt{\frac{7}{2}} C_{33,22}^{33,0}}{1600}\)

\noindent\(C_{31,2211,2}^{11,0011,0,\text{Loc}}= \frac{7 C_{00,20}^{60,0}}{3840}+\frac{7 C_{00,20}^{60,1}}{7680}+\frac{77 C_{00,22}^{62,0}}{2560
\sqrt{5}}+\frac{77 C_{00,22}^{62,1}}{5120 \sqrt{5}}-\frac{7 C_{11,20}^{31,0}}{11520}-\frac{7 C_{11,20}^{31,1}}{23040}+\frac{7 C_{11,22}^{51,0}}{2880
\sqrt{5}}+\frac{7 C_{11,22}^{51,1}}{5760 \sqrt{5}}-\frac{63 \sqrt{21} C_{11,22}^{53,0}}{25600}-\frac{63 \sqrt{21} C_{11,22}^{53,1}}{51200}-\frac{7
C_{20,20}^{40,0}}{11520}-\frac{7 C_{20,20}^{40,1}}{23040}+\frac{7 C_{20,22}^{42,0}}{2880 \sqrt{5}}+\frac{7 C_{20,22}^{42,1}}{5760 \sqrt{5}}+\frac{7
C_{22,20}^{42,0}}{11520 \sqrt{5}}+\frac{7 C_{22,20}^{42,1}}{23040 \sqrt{5}}-\frac{41 C_{22,22}^{40,0}}{4608 \sqrt{5}}-\frac{41 C_{22,22}^{40,1}}{9216
\sqrt{5}}-\frac{43 \sqrt{\frac{7}{5}} C_{22,22}^{42,0}}{11520}-\frac{43 \sqrt{\frac{7}{5}} C_{22,22}^{42,1}}{23040}+\frac{3 C_{22,22}^{44,0}}{320}+\frac{3
C_{22,22}^{44,1}}{640}+\frac{7 C_{31,20}^{31,0}}{11520}+\frac{7 C_{31,20}^{31,1}}{23040}+\frac{61 C_{31,22}^{31,0}}{5760 \sqrt{5}}+\frac{61 C_{31,22}^{31,1}}{11520
\sqrt{5}}-\frac{11 \sqrt{\frac{7}{3}} C_{31,22}^{33,0}}{25600}-\frac{11 \sqrt{\frac{7}{3}} C_{31,22}^{33,1}}{51200}-\frac{\sqrt{21} C_{33,20}^{33,0}}{6400}-\frac{\sqrt{21}
C_{33,20}^{33,1}}{12800}+\frac{3}{400} \sqrt{\frac{7}{10}} C_{33,22}^{33,0}+\frac{3}{800} \sqrt{\frac{7}{10}} C_{33,22}^{33,1}\)

\noindent\(C_{31,2211,2}^{11,0011,0,\text{NonLoc}}= \frac{7 C_{00,00}^{60,0}}{5120 \sqrt{3}}+\frac{7 C_{00,00}^{60,1}}{2560 \sqrt{3}}+\frac{7
C_{00,20}^{60,0}}{15360}+\frac{7 C_{00,20}^{60,1}}{7680}-\frac{77 C_{00,22}^{62,0}}{5120 \sqrt{5}}-\frac{77 C_{00,22}^{62,1}}{2560 \sqrt{5}}+\frac{7
C_{11,00}^{51,0}}{15360 \sqrt{3}}+\frac{7 C_{11,00}^{51,1}}{7680 \sqrt{3}}+\frac{7 C_{11,20}^{31,0}}{46080}+\frac{7 C_{11,20}^{31,1}}{23040}+\frac{7
C_{11,22}^{51,0}}{5760 \sqrt{5}}+\frac{7 C_{11,22}^{51,1}}{2880 \sqrt{5}}-\frac{63 \sqrt{21} C_{11,22}^{53,0}}{51200}-\frac{63 \sqrt{21} C_{11,22}^{53,1}}{25600}-\frac{7
C_{20,00}^{40,0}}{15360 \sqrt{3}}-\frac{7 C_{20,00}^{40,1}}{7680 \sqrt{3}}-\frac{7 C_{20,20}^{40,0}}{46080}-\frac{7 C_{20,20}^{40,1}}{23040}-\frac{7
C_{20,22}^{42,0}}{5760 \sqrt{5}}-\frac{7 C_{20,22}^{42,1}}{2880 \sqrt{5}}+\frac{7 C_{22,00}^{42,0}}{15360 \sqrt{15}}+\frac{7 C_{22,00}^{42,1}}{7680
\sqrt{15}}+\frac{7 C_{22,20}^{42,0}}{46080 \sqrt{5}}+\frac{7 C_{22,20}^{42,1}}{23040 \sqrt{5}}+\frac{41 C_{22,22}^{40,0}}{9216 \sqrt{5}}+\frac{41
C_{22,22}^{40,1}}{4608 \sqrt{5}}+\frac{43 \sqrt{\frac{7}{5}} C_{22,22}^{42,0}}{23040}+\frac{43 \sqrt{\frac{7}{5}} C_{22,22}^{42,1}}{11520}-\frac{3
C_{22,22}^{44,0}}{640}-\frac{3 C_{22,22}^{44,1}}{320}-\frac{7 C_{31,00}^{31,0}}{15360 \sqrt{3}}-\frac{7 C_{31,00}^{31,1}}{7680 \sqrt{3}}-\frac{7
C_{31,20}^{31,0}}{46080}-\frac{7 C_{31,20}^{31,1}}{23040}+\frac{61 C_{31,22}^{31,0}}{11520 \sqrt{5}}+\frac{61 C_{31,22}^{31,1}}{5760 \sqrt{5}}-\frac{11
\sqrt{\frac{7}{3}} C_{31,22}^{33,0}}{51200}-\frac{11 \sqrt{\frac{7}{3}} C_{31,22}^{33,1}}{25600}+\frac{3 \sqrt{7} C_{33,00}^{33,0}}{25600}+\frac{3
\sqrt{7} C_{33,00}^{33,1}}{12800}+\frac{\sqrt{21} C_{33,20}^{33,0}}{25600}+\frac{\sqrt{21} C_{33,20}^{33,1}}{12800}+\frac{3}{800} \sqrt{\frac{7}{10}}
C_{33,22}^{33,0}+\frac{3}{400} \sqrt{\frac{7}{10}} C_{33,22}^{33,1}\)

\noindent\(C_{31,2211,2}^{11,0011,1,\text{Loc}}= \frac{7 C_{00,20}^{60,1}}{2560 \sqrt{3}}+\frac{77 \sqrt{\frac{3}{5}} C_{00,22}^{62,1}}{5120}-\frac{7
C_{11,20}^{31,1}}{7680 \sqrt{3}}+\frac{7 C_{11,22}^{51,1}}{1920 \sqrt{15}}-\frac{189 \sqrt{7} C_{11,22}^{53,1}}{51200}-\frac{7 C_{20,20}^{40,1}}{7680
\sqrt{3}}+\frac{7 C_{20,22}^{42,1}}{1920 \sqrt{15}}+\frac{7 C_{22,20}^{42,1}}{7680 \sqrt{15}}-\frac{41 C_{22,22}^{40,1}}{3072 \sqrt{15}}-\frac{43
\sqrt{\frac{7}{15}} C_{22,22}^{42,1}}{7680}+\frac{3 \sqrt{3} C_{22,22}^{44,1}}{640}+\frac{7 C_{31,20}^{31,1}}{7680 \sqrt{3}}+\frac{61 C_{31,22}^{31,1}}{3840
\sqrt{15}}-\frac{11 \sqrt{7} C_{31,22}^{33,1}}{51200}-\frac{3 \sqrt{7} C_{33,20}^{33,1}}{12800}+\frac{3}{800} \sqrt{\frac{21}{10}} C_{33,22}^{33,1}\)

\noindent\(C_{31,2211,2}^{11,0011,1,\text{NonLoc}}= \frac{7 C_{00,00}^{60,0}}{5120}+\frac{7 C_{00,20}^{60,0}}{5120 \sqrt{3}}-\frac{77
\sqrt{\frac{3}{5}} C_{00,22}^{62,0}}{5120}+\frac{7 C_{11,00}^{51,0}}{15360}+\frac{7 C_{11,20}^{31,0}}{15360 \sqrt{3}}+\frac{7 C_{11,22}^{51,0}}{1920
\sqrt{15}}-\frac{189 \sqrt{7} C_{11,22}^{53,0}}{51200}-\frac{7 C_{20,00}^{40,0}}{15360}-\frac{7 C_{20,20}^{40,0}}{15360 \sqrt{3}}-\frac{7 C_{20,22}^{42,0}}{1920
\sqrt{15}}+\frac{7 C_{22,00}^{42,0}}{15360 \sqrt{5}}+\frac{7 C_{22,20}^{42,0}}{15360 \sqrt{15}}+\frac{41 C_{22,22}^{40,0}}{3072 \sqrt{15}}+\frac{43
\sqrt{\frac{7}{15}} C_{22,22}^{42,0}}{7680}-\frac{3 \sqrt{3} C_{22,22}^{44,0}}{640}-\frac{7 C_{31,00}^{31,0}}{15360}-\frac{7 C_{31,20}^{31,0}}{15360
\sqrt{3}}+\frac{61 C_{31,22}^{31,0}}{3840 \sqrt{15}}-\frac{11 \sqrt{7} C_{31,22}^{33,0}}{51200}+\frac{3 \sqrt{21} C_{33,00}^{33,0}}{25600}+\frac{3
\sqrt{7} C_{33,20}^{33,0}}{25600}+\frac{3}{800} \sqrt{\frac{21}{10}} C_{33,22}^{33,0}\)

\noindent\(C_{31,2212,1}^{00,1101,0,\text{Loc}}= -\frac{7 C_{11,11}^{51,0}}{128 \sqrt{15}}-\frac{7 C_{11,11}^{51,1}}{256 \sqrt{15}}-\frac{1}{128}
\sqrt{\frac{3}{5}} C_{31,11}^{31,0}-\frac{1}{256} \sqrt{\frac{3}{5}} C_{31,11}^{31,1}+\frac{3}{160} \sqrt{\frac{21}{10}} C_{33,11}^{33,0}+\frac{3}{320}
\sqrt{\frac{21}{10}} C_{33,11}^{33,1}\)

\noindent\(C_{31,2212,1}^{00,1101,0,\text{NonLoc}}= -\frac{7 C_{11,11}^{51,0}}{256 \sqrt{15}}-\frac{7 C_{11,11}^{51,1}}{128 \sqrt{15}}-\frac{1}{256}
\sqrt{\frac{3}{5}} C_{31,11}^{31,0}-\frac{1}{128} \sqrt{\frac{3}{5}} C_{31,11}^{31,1}+\frac{3}{320} \sqrt{\frac{21}{10}} C_{33,11}^{33,0}+\frac{3}{160}
\sqrt{\frac{21}{10}} C_{33,11}^{33,1}\)

\noindent\(C_{31,2212,1}^{00,1101,1,\text{Loc}}= -\frac{7 C_{11,11}^{51,1}}{256 \sqrt{5}}-\frac{3 C_{31,11}^{31,1}}{256 \sqrt{5}}+\frac{9}{320}
\sqrt{\frac{7}{10}} C_{33,11}^{33,1}\)

\noindent\(C_{31,2212,1}^{00,1101,1,\text{NonLoc}}= -\frac{7 C_{11,11}^{51,0}}{256 \sqrt{5}}-\frac{3 C_{31,11}^{31,0}}{256 \sqrt{5}}+\frac{9}{320}
\sqrt{\frac{7}{10}} C_{33,11}^{33,0}\)

\noindent\(C_{31,2212,1}^{11,0011,0,\text{Loc}}= \frac{7 C_{00,20}^{60,0}}{256 \sqrt{5}}+\frac{7 C_{00,20}^{60,1}}{512 \sqrt{5}}-\frac{21
C_{00,22}^{62,0}}{2560}-\frac{21 C_{00,22}^{62,1}}{5120}-\frac{7 C_{11,20}^{31,0}}{768 \sqrt{5}}-\frac{7 C_{11,20}^{31,1}}{1536 \sqrt{5}}+\frac{7
C_{11,22}^{51,0}}{3840}+\frac{7 C_{11,22}^{51,1}}{7680}+\frac{9 \sqrt{\frac{21}{5}} C_{11,22}^{53,0}}{5120}+\frac{9 \sqrt{\frac{21}{5}} C_{11,22}^{53,1}}{10240}-\frac{7
C_{20,20}^{40,0}}{768 \sqrt{5}}-\frac{7 C_{20,20}^{40,1}}{1536 \sqrt{5}}+\frac{7 C_{20,22}^{42,0}}{3840}+\frac{7 C_{20,22}^{42,1}}{7680}+\frac{7
C_{22,20}^{42,0}}{3840}+\frac{7 C_{22,20}^{42,1}}{7680}+\frac{7 C_{22,22}^{40,0}}{1536}+\frac{7 C_{22,22}^{40,1}}{3072}-\frac{7 \sqrt{7} C_{22,22}^{42,0}}{3840}-\frac{7
\sqrt{7} C_{22,22}^{42,1}}{7680}+\frac{7 C_{31,20}^{31,0}}{768 \sqrt{5}}+\frac{7 C_{31,20}^{31,1}}{1536 \sqrt{5}}-\frac{7 C_{31,22}^{31,0}}{3840}-\frac{7
C_{31,22}^{31,1}}{7680}-\frac{9 \sqrt{\frac{21}{5}} C_{31,22}^{33,0}}{5120}-\frac{9 \sqrt{\frac{21}{5}} C_{31,22}^{33,1}}{10240}-\frac{3 \sqrt{\frac{21}{5}}
C_{33,20}^{33,0}}{1280}-\frac{3 \sqrt{\frac{21}{5}} C_{33,20}^{33,1}}{2560}+\frac{9 \sqrt{\frac{7}{2}} C_{33,22}^{33,0}}{1600}+\frac{9 \sqrt{\frac{7}{2}}
C_{33,22}^{33,1}}{3200}\)

\noindent\(C_{31,2212,1}^{11,0011,0,\text{NonLoc}}= \frac{7 \sqrt{\frac{3}{5}} C_{00,00}^{60,0}}{1024}+\frac{7}{512} \sqrt{\frac{3}{5}}
C_{00,00}^{60,1}+\frac{7 C_{00,20}^{60,0}}{1024 \sqrt{5}}+\frac{7 C_{00,20}^{60,1}}{512 \sqrt{5}}+\frac{21 C_{00,22}^{62,0}}{5120}+\frac{21 C_{00,22}^{62,1}}{2560}+\frac{7
C_{11,00}^{51,0}}{1024 \sqrt{15}}+\frac{7 C_{11,00}^{51,1}}{512 \sqrt{15}}+\frac{7 C_{11,20}^{31,0}}{3072 \sqrt{5}}+\frac{7 C_{11,20}^{31,1}}{1536
\sqrt{5}}+\frac{7 C_{11,22}^{51,0}}{7680}+\frac{7 C_{11,22}^{51,1}}{3840}+\frac{9 \sqrt{\frac{21}{5}} C_{11,22}^{53,0}}{10240}+\frac{9 \sqrt{\frac{21}{5}}
C_{11,22}^{53,1}}{5120}-\frac{7 C_{20,00}^{40,0}}{1024 \sqrt{15}}-\frac{7 C_{20,00}^{40,1}}{512 \sqrt{15}}-\frac{7 C_{20,20}^{40,0}}{3072 \sqrt{5}}-\frac{7
C_{20,20}^{40,1}}{1536 \sqrt{5}}-\frac{7 C_{20,22}^{42,0}}{7680}-\frac{7 C_{20,22}^{42,1}}{3840}+\frac{7 C_{22,00}^{42,0}}{5120 \sqrt{3}}+\frac{7
C_{22,00}^{42,1}}{2560 \sqrt{3}}+\frac{7 C_{22,20}^{42,0}}{15360}+\frac{7 C_{22,20}^{42,1}}{7680}-\frac{7 C_{22,22}^{40,0}}{3072}-\frac{7 C_{22,22}^{40,1}}{1536}+\frac{7
\sqrt{7} C_{22,22}^{42,0}}{7680}+\frac{7 \sqrt{7} C_{22,22}^{42,1}}{3840}-\frac{7 C_{31,00}^{31,0}}{1024 \sqrt{15}}-\frac{7 C_{31,00}^{31,1}}{512
\sqrt{15}}-\frac{7 C_{31,20}^{31,0}}{3072 \sqrt{5}}-\frac{7 C_{31,20}^{31,1}}{1536 \sqrt{5}}-\frac{7 C_{31,22}^{31,0}}{7680}-\frac{7 C_{31,22}^{31,1}}{3840}-\frac{9
\sqrt{\frac{21}{5}} C_{31,22}^{33,0}}{10240}-\frac{9 \sqrt{\frac{21}{5}} C_{31,22}^{33,1}}{5120}+\frac{9 \sqrt{\frac{7}{5}} C_{33,00}^{33,0}}{5120}+\frac{9
\sqrt{\frac{7}{5}} C_{33,00}^{33,1}}{2560}+\frac{3 \sqrt{\frac{21}{5}} C_{33,20}^{33,0}}{5120}+\frac{3 \sqrt{\frac{21}{5}} C_{33,20}^{33,1}}{2560}+\frac{9
\sqrt{\frac{7}{2}} C_{33,22}^{33,0}}{3200}+\frac{9 \sqrt{\frac{7}{2}} C_{33,22}^{33,1}}{1600}\)

\noindent\(C_{31,2212,1}^{11,0011,1,\text{Loc}}= \frac{7}{512} \sqrt{\frac{3}{5}} C_{00,20}^{60,1}-\frac{21 \sqrt{3} C_{00,22}^{62,1}}{5120}-\frac{7
C_{11,20}^{31,1}}{512 \sqrt{15}}+\frac{7 C_{11,22}^{51,1}}{2560 \sqrt{3}}+\frac{27 \sqrt{\frac{7}{5}} C_{11,22}^{53,1}}{10240}-\frac{7 C_{20,20}^{40,1}}{512
\sqrt{15}}+\frac{7 C_{20,22}^{42,1}}{2560 \sqrt{3}}+\frac{7 C_{22,20}^{42,1}}{2560 \sqrt{3}}+\frac{7 C_{22,22}^{40,1}}{1024 \sqrt{3}}-\frac{7 \sqrt{\frac{7}{3}}
C_{22,22}^{42,1}}{2560}+\frac{7 C_{31,20}^{31,1}}{512 \sqrt{15}}-\frac{7 C_{31,22}^{31,1}}{2560 \sqrt{3}}-\frac{27 \sqrt{\frac{7}{5}} C_{31,22}^{33,1}}{10240}-\frac{9
\sqrt{\frac{7}{5}} C_{33,20}^{33,1}}{2560}+\frac{9 \sqrt{\frac{21}{2}} C_{33,22}^{33,1}}{3200}\)

\noindent\(C_{31,2212,1}^{11,0011,1,\text{NonLoc}}= \frac{21 C_{00,00}^{60,0}}{1024 \sqrt{5}}+\frac{7 \sqrt{\frac{3}{5}} C_{00,20}^{60,0}}{1024}+\frac{21
\sqrt{3} C_{00,22}^{62,0}}{5120}+\frac{7 C_{11,00}^{51,0}}{1024 \sqrt{5}}+\frac{7 C_{11,20}^{31,0}}{1024 \sqrt{15}}+\frac{7 C_{11,22}^{51,0}}{2560
\sqrt{3}}+\frac{27 \sqrt{\frac{7}{5}} C_{11,22}^{53,0}}{10240}-\frac{7 C_{20,00}^{40,0}}{1024 \sqrt{5}}-\frac{7 C_{20,20}^{40,0}}{1024 \sqrt{15}}-\frac{7
C_{20,22}^{42,0}}{2560 \sqrt{3}}+\frac{7 C_{22,00}^{42,0}}{5120}+\frac{7 C_{22,20}^{42,0}}{5120 \sqrt{3}}-\frac{7 C_{22,22}^{40,0}}{1024 \sqrt{3}}+\frac{7
\sqrt{\frac{7}{3}} C_{22,22}^{42,0}}{2560}-\frac{7 C_{31,00}^{31,0}}{1024 \sqrt{5}}-\frac{7 C_{31,20}^{31,0}}{1024 \sqrt{15}}-\frac{7 C_{31,22}^{31,0}}{2560
\sqrt{3}}-\frac{27 \sqrt{\frac{7}{5}} C_{31,22}^{33,0}}{10240}+\frac{9 \sqrt{\frac{21}{5}} C_{33,00}^{33,0}}{5120}+\frac{9 \sqrt{\frac{7}{5}} C_{33,20}^{33,0}}{5120}+\frac{9
\sqrt{\frac{21}{2}} C_{33,22}^{33,0}}{3200}\)

\noindent\(C_{31,2212,2}^{11,0011,0,\text{Loc}}= -\frac{7 C_{00,20}^{60,0}}{256 \sqrt{15}}-\frac{7 C_{00,20}^{60,1}}{512 \sqrt{15}}+\frac{7
\sqrt{3} C_{00,22}^{62,0}}{2560}+\frac{7 \sqrt{3} C_{00,22}^{62,1}}{5120}+\frac{7 C_{11,20}^{31,0}}{768 \sqrt{15}}+\frac{7 C_{11,20}^{31,1}}{1536
\sqrt{15}}-\frac{7 C_{11,22}^{51,0}}{3840 \sqrt{3}}-\frac{7 C_{11,22}^{51,1}}{7680 \sqrt{3}}-\frac{9 \sqrt{\frac{7}{5}} C_{11,22}^{53,0}}{5120}-\frac{9
\sqrt{\frac{7}{5}} C_{11,22}^{53,1}}{10240}+\frac{7 C_{20,20}^{40,0}}{768 \sqrt{15}}+\frac{7 C_{20,20}^{40,1}}{1536 \sqrt{15}}-\frac{7 C_{20,22}^{42,0}}{3840
\sqrt{3}}-\frac{7 C_{20,22}^{42,1}}{7680 \sqrt{3}}-\frac{7 C_{22,20}^{42,0}}{3840 \sqrt{3}}-\frac{7 C_{22,20}^{42,1}}{7680 \sqrt{3}}-\frac{7 C_{22,22}^{40,0}}{1536
\sqrt{3}}-\frac{7 C_{22,22}^{40,1}}{3072 \sqrt{3}}+\frac{7 \sqrt{\frac{7}{3}} C_{22,22}^{42,0}}{3840}+\frac{7 \sqrt{\frac{7}{3}} C_{22,22}^{42,1}}{7680}-\frac{7
C_{31,20}^{31,0}}{768 \sqrt{15}}-\frac{7 C_{31,20}^{31,1}}{1536 \sqrt{15}}+\frac{7 C_{31,22}^{31,0}}{3840 \sqrt{3}}+\frac{7 C_{31,22}^{31,1}}{7680
\sqrt{3}}+\frac{9 \sqrt{\frac{7}{5}} C_{31,22}^{33,0}}{5120}+\frac{9 \sqrt{\frac{7}{5}} C_{31,22}^{33,1}}{10240}+\frac{3 \sqrt{\frac{7}{5}} C_{33,20}^{33,0}}{1280}+\frac{3
\sqrt{\frac{7}{5}} C_{33,20}^{33,1}}{2560}-\frac{3 \sqrt{\frac{21}{2}} C_{33,22}^{33,0}}{1600}-\frac{3 \sqrt{\frac{21}{2}} C_{33,22}^{33,1}}{3200}\)

\noindent\(C_{31,2212,2}^{11,0011,0,\text{NonLoc}}= -\frac{7 C_{00,00}^{60,0}}{1024 \sqrt{5}}-\frac{7 C_{00,00}^{60,1}}{512 \sqrt{5}}-\frac{7
C_{00,20}^{60,0}}{1024 \sqrt{15}}-\frac{7 C_{00,20}^{60,1}}{512 \sqrt{15}}-\frac{7 \sqrt{3} C_{00,22}^{62,0}}{5120}-\frac{7 \sqrt{3} C_{00,22}^{62,1}}{2560}-\frac{7
C_{11,00}^{51,0}}{3072 \sqrt{5}}-\frac{7 C_{11,00}^{51,1}}{1536 \sqrt{5}}-\frac{7 C_{11,20}^{31,0}}{3072 \sqrt{15}}-\frac{7 C_{11,20}^{31,1}}{1536
\sqrt{15}}-\frac{7 C_{11,22}^{51,0}}{7680 \sqrt{3}}-\frac{7 C_{11,22}^{51,1}}{3840 \sqrt{3}}-\frac{9 \sqrt{\frac{7}{5}} C_{11,22}^{53,0}}{10240}-\frac{9
\sqrt{\frac{7}{5}} C_{11,22}^{53,1}}{5120}+\frac{7 C_{20,00}^{40,0}}{3072 \sqrt{5}}+\frac{7 C_{20,00}^{40,1}}{1536 \sqrt{5}}+\frac{7 C_{20,20}^{40,0}}{3072
\sqrt{15}}+\frac{7 C_{20,20}^{40,1}}{1536 \sqrt{15}}+\frac{7 C_{20,22}^{42,0}}{7680 \sqrt{3}}+\frac{7 C_{20,22}^{42,1}}{3840 \sqrt{3}}-\frac{7 C_{22,00}^{42,0}}{15360}-\frac{7
C_{22,00}^{42,1}}{7680}-\frac{7 C_{22,20}^{42,0}}{15360 \sqrt{3}}-\frac{7 C_{22,20}^{42,1}}{7680 \sqrt{3}}+\frac{7 C_{22,22}^{40,0}}{3072 \sqrt{3}}+\frac{7
C_{22,22}^{40,1}}{1536 \sqrt{3}}-\frac{7 \sqrt{\frac{7}{3}} C_{22,22}^{42,0}}{7680}-\frac{7 \sqrt{\frac{7}{3}} C_{22,22}^{42,1}}{3840}+\frac{7 C_{31,00}^{31,0}}{3072
\sqrt{5}}+\frac{7 C_{31,00}^{31,1}}{1536 \sqrt{5}}+\frac{7 C_{31,20}^{31,0}}{3072 \sqrt{15}}+\frac{7 C_{31,20}^{31,1}}{1536 \sqrt{15}}+\frac{7 C_{31,22}^{31,0}}{7680
\sqrt{3}}+\frac{7 C_{31,22}^{31,1}}{3840 \sqrt{3}}+\frac{9 \sqrt{\frac{7}{5}} C_{31,22}^{33,0}}{10240}+\frac{9 \sqrt{\frac{7}{5}} C_{31,22}^{33,1}}{5120}-\frac{3
\sqrt{\frac{21}{5}} C_{33,00}^{33,0}}{5120}-\frac{3 \sqrt{\frac{21}{5}} C_{33,00}^{33,1}}{2560}-\frac{3 \sqrt{\frac{7}{5}} C_{33,20}^{33,0}}{5120}-\frac{3
\sqrt{\frac{7}{5}} C_{33,20}^{33,1}}{2560}-\frac{3 \sqrt{\frac{21}{2}} C_{33,22}^{33,0}}{3200}-\frac{3 \sqrt{\frac{21}{2}} C_{33,22}^{33,1}}{1600}\)

\noindent\(C_{31,2212,2}^{11,0011,1,\text{Loc}}= -\frac{7 C_{00,20}^{60,1}}{512 \sqrt{5}}+\frac{21 C_{00,22}^{62,1}}{5120}+\frac{7
C_{11,20}^{31,1}}{1536 \sqrt{5}}-\frac{7 C_{11,22}^{51,1}}{7680}-\frac{9 \sqrt{\frac{21}{5}} C_{11,22}^{53,1}}{10240}+\frac{7 C_{20,20}^{40,1}}{1536
\sqrt{5}}-\frac{7 C_{20,22}^{42,1}}{7680}-\frac{7 C_{22,20}^{42,1}}{7680}-\frac{7 C_{22,22}^{40,1}}{3072}+\frac{7 \sqrt{7} C_{22,22}^{42,1}}{7680}-\frac{7
C_{31,20}^{31,1}}{1536 \sqrt{5}}+\frac{7 C_{31,22}^{31,1}}{7680}+\frac{9 \sqrt{\frac{21}{5}} C_{31,22}^{33,1}}{10240}+\frac{3 \sqrt{\frac{21}{5}}
C_{33,20}^{33,1}}{2560}-\frac{9 \sqrt{\frac{7}{2}} C_{33,22}^{33,1}}{3200}\)

\noindent\(C_{31,2212,2}^{11,0011,1,\text{NonLoc}}= -\frac{7 \sqrt{\frac{3}{5}} C_{00,00}^{60,0}}{1024}-\frac{7 C_{00,20}^{60,0}}{1024
\sqrt{5}}-\frac{21 C_{00,22}^{62,0}}{5120}-\frac{7 C_{11,00}^{51,0}}{1024 \sqrt{15}}-\frac{7 C_{11,20}^{31,0}}{3072 \sqrt{5}}-\frac{7 C_{11,22}^{51,0}}{7680}-\frac{9
\sqrt{\frac{21}{5}} C_{11,22}^{53,0}}{10240}+\frac{7 C_{20,00}^{40,0}}{1024 \sqrt{15}}+\frac{7 C_{20,20}^{40,0}}{3072 \sqrt{5}}+\frac{7 C_{20,22}^{42,0}}{7680}-\frac{7
C_{22,00}^{42,0}}{5120 \sqrt{3}}-\frac{7 C_{22,20}^{42,0}}{15360}+\frac{7 C_{22,22}^{40,0}}{3072}-\frac{7 \sqrt{7} C_{22,22}^{42,0}}{7680}+\frac{7
C_{31,00}^{31,0}}{1024 \sqrt{15}}+\frac{7 C_{31,20}^{31,0}}{3072 \sqrt{5}}+\frac{7 C_{31,22}^{31,0}}{7680}+\frac{9 \sqrt{\frac{21}{5}} C_{31,22}^{33,0}}{10240}-\frac{9
\sqrt{\frac{7}{5}} C_{33,00}^{33,0}}{5120}-\frac{3 \sqrt{\frac{21}{5}} C_{33,20}^{33,0}}{5120}-\frac{9 \sqrt{\frac{7}{2}} C_{33,22}^{33,0}}{3200}\)

\noindent\(C_{31,2213,2}^{11,0011,0,\text{Loc}}= \frac{7}{640} \sqrt{\frac{7}{3}} C_{00,20}^{60,0}+\frac{7 \sqrt{\frac{7}{3}} C_{00,20}^{60,1}}{1280}+\frac{1}{640}
\sqrt{\frac{21}{5}} C_{00,22}^{62,0}+\frac{\sqrt{\frac{21}{5}} C_{00,22}^{62,1}}{1280}-\frac{7 \sqrt{\frac{7}{3}} C_{11,20}^{31,0}}{1920}-\frac{7
\sqrt{\frac{7}{3}} C_{11,20}^{31,1}}{3840}-\frac{1}{960} \sqrt{\frac{7}{15}} C_{11,22}^{51,0}-\frac{\sqrt{\frac{7}{15}} C_{11,22}^{51,1}}{1920}-\frac{9
C_{11,22}^{53,0}}{6400}-\frac{9 C_{11,22}^{53,1}}{12800}-\frac{7 \sqrt{\frac{7}{3}} C_{20,20}^{40,0}}{1920}-\frac{7 \sqrt{\frac{7}{3}} C_{20,20}^{40,1}}{3840}-\frac{1}{960}
\sqrt{\frac{7}{15}} C_{20,22}^{42,0}-\frac{\sqrt{\frac{7}{15}} C_{20,22}^{42,1}}{1920}+\frac{7 \sqrt{\frac{7}{15}} C_{22,20}^{42,0}}{1920}+\frac{7
\sqrt{\frac{7}{15}} C_{22,20}^{42,1}}{3840}-\frac{1}{384} \sqrt{\frac{7}{15}} C_{22,22}^{40,0}-\frac{1}{768} \sqrt{\frac{7}{15}} C_{22,22}^{40,1}+\frac{7
C_{22,22}^{42,0}}{960 \sqrt{15}}+\frac{7 C_{22,22}^{42,1}}{1920 \sqrt{15}}+\frac{7 \sqrt{\frac{7}{3}} C_{31,20}^{31,0}}{1920}+\frac{7 \sqrt{\frac{7}{3}}
C_{31,20}^{31,1}}{3840}+\frac{1}{960} \sqrt{\frac{7}{15}} C_{31,22}^{31,0}+\frac{\sqrt{\frac{7}{15}} C_{31,22}^{31,1}}{1920}+\frac{9 C_{31,22}^{33,0}}{6400}+\frac{9
C_{31,22}^{33,1}}{12800}-\frac{21 C_{33,20}^{33,0}}{3200}-\frac{21 C_{33,20}^{33,1}}{6400}-\frac{3}{400} \sqrt{\frac{3}{10}} C_{33,22}^{33,0}-\frac{3}{800}
\sqrt{\frac{3}{10}} C_{33,22}^{33,1}\)

\noindent\(C_{31,2213,2}^{11,0011,0,\text{NonLoc}}= \frac{7 \sqrt{7} C_{00,00}^{60,0}}{2560}+\frac{7 \sqrt{7} C_{00,00}^{60,1}}{1280}+\frac{7
\sqrt{\frac{7}{3}} C_{00,20}^{60,0}}{2560}+\frac{7 \sqrt{\frac{7}{3}} C_{00,20}^{60,1}}{1280}-\frac{\sqrt{\frac{21}{5}} C_{00,22}^{62,0}}{1280}-\frac{1}{640}
\sqrt{\frac{21}{5}} C_{00,22}^{62,1}+\frac{7 \sqrt{7} C_{11,00}^{51,0}}{7680}+\frac{7 \sqrt{7} C_{11,00}^{51,1}}{3840}+\frac{7 \sqrt{\frac{7}{3}}
C_{11,20}^{31,0}}{7680}+\frac{7 \sqrt{\frac{7}{3}} C_{11,20}^{31,1}}{3840}-\frac{\sqrt{\frac{7}{15}} C_{11,22}^{51,0}}{1920}-\frac{1}{960} \sqrt{\frac{7}{15}}
C_{11,22}^{51,1}-\frac{9 C_{11,22}^{53,0}}{12800}-\frac{9 C_{11,22}^{53,1}}{6400}-\frac{7 \sqrt{7} C_{20,00}^{40,0}}{7680}-\frac{7 \sqrt{7} C_{20,00}^{40,1}}{3840}-\frac{7
\sqrt{\frac{7}{3}} C_{20,20}^{40,0}}{7680}-\frac{7 \sqrt{\frac{7}{3}} C_{20,20}^{40,1}}{3840}+\frac{\sqrt{\frac{7}{15}} C_{20,22}^{42,0}}{1920}+\frac{1}{960}
\sqrt{\frac{7}{15}} C_{20,22}^{42,1}+\frac{7 \sqrt{\frac{7}{5}} C_{22,00}^{42,0}}{7680}+\frac{7 \sqrt{\frac{7}{5}} C_{22,00}^{42,1}}{3840}+\frac{7
\sqrt{\frac{7}{15}} C_{22,20}^{42,0}}{7680}+\frac{7 \sqrt{\frac{7}{15}} C_{22,20}^{42,1}}{3840}+\frac{1}{768} \sqrt{\frac{7}{15}} C_{22,22}^{40,0}+\frac{1}{384}
\sqrt{\frac{7}{15}} C_{22,22}^{40,1}-\frac{7 C_{22,22}^{42,0}}{1920 \sqrt{15}}-\frac{7 C_{22,22}^{42,1}}{960 \sqrt{15}}-\frac{7 \sqrt{7} C_{31,00}^{31,0}}{7680}-\frac{7
\sqrt{7} C_{31,00}^{31,1}}{3840}-\frac{7 \sqrt{\frac{7}{3}} C_{31,20}^{31,0}}{7680}-\frac{7 \sqrt{\frac{7}{3}} C_{31,20}^{31,1}}{3840}+\frac{\sqrt{\frac{7}{15}}
C_{31,22}^{31,0}}{1920}+\frac{1}{960} \sqrt{\frac{7}{15}} C_{31,22}^{31,1}+\frac{9 C_{31,22}^{33,0}}{12800}+\frac{9 C_{31,22}^{33,1}}{6400}+\frac{21
\sqrt{3} C_{33,00}^{33,0}}{12800}+\frac{21 \sqrt{3} C_{33,00}^{33,1}}{6400}+\frac{21 C_{33,20}^{33,0}}{12800}+\frac{21 C_{33,20}^{33,1}}{6400}-\frac{3}{800}
\sqrt{\frac{3}{10}} C_{33,22}^{33,0}-\frac{3}{400} \sqrt{\frac{3}{10}} C_{33,22}^{33,1}\)

\noindent\(C_{31,2213,2}^{11,0011,1,\text{Loc}}= \frac{7 \sqrt{7} C_{00,20}^{60,1}}{1280}+\frac{3 \sqrt{\frac{7}{5}} C_{00,22}^{62,1}}{1280}-\frac{7
\sqrt{7} C_{11,20}^{31,1}}{3840}-\frac{\sqrt{\frac{7}{5}} C_{11,22}^{51,1}}{1920}-\frac{9 \sqrt{3} C_{11,22}^{53,1}}{12800}-\frac{7 \sqrt{7} C_{20,20}^{40,1}}{3840}-\frac{\sqrt{\frac{7}{5}}
C_{20,22}^{42,1}}{1920}+\frac{7 \sqrt{\frac{7}{5}} C_{22,20}^{42,1}}{3840}-\frac{1}{768} \sqrt{\frac{7}{5}} C_{22,22}^{40,1}+\frac{7 C_{22,22}^{42,1}}{1920
\sqrt{5}}+\frac{7 \sqrt{7} C_{31,20}^{31,1}}{3840}+\frac{\sqrt{\frac{7}{5}} C_{31,22}^{31,1}}{1920}+\frac{9 \sqrt{3} C_{31,22}^{33,1}}{12800}-\frac{21
\sqrt{3} C_{33,20}^{33,1}}{6400}-\frac{9 C_{33,22}^{33,1}}{800 \sqrt{10}}\)

\noindent\(C_{31,2213,2}^{11,0011,1,\text{NonLoc}}= \frac{7 \sqrt{21} C_{00,00}^{60,0}}{2560}+\frac{7 \sqrt{7} C_{00,20}^{60,0}}{2560}-\frac{3
\sqrt{\frac{7}{5}} C_{00,22}^{62,0}}{1280}+\frac{7 \sqrt{\frac{7}{3}} C_{11,00}^{51,0}}{2560}+\frac{7 \sqrt{7} C_{11,20}^{31,0}}{7680}-\frac{\sqrt{\frac{7}{5}}
C_{11,22}^{51,0}}{1920}-\frac{9 \sqrt{3} C_{11,22}^{53,0}}{12800}-\frac{7 \sqrt{\frac{7}{3}} C_{20,00}^{40,0}}{2560}-\frac{7 \sqrt{7} C_{20,20}^{40,0}}{7680}+\frac{\sqrt{\frac{7}{5}}
C_{20,22}^{42,0}}{1920}+\frac{7 \sqrt{\frac{7}{15}} C_{22,00}^{42,0}}{2560}+\frac{7 \sqrt{\frac{7}{5}} C_{22,20}^{42,0}}{7680}+\frac{1}{768} \sqrt{\frac{7}{5}}
C_{22,22}^{40,0}-\frac{7 C_{22,22}^{42,0}}{1920 \sqrt{5}}-\frac{7 \sqrt{\frac{7}{3}} C_{31,00}^{31,0}}{2560}-\frac{7 \sqrt{7} C_{31,20}^{31,0}}{7680}+\frac{\sqrt{\frac{7}{5}}
C_{31,22}^{31,0}}{1920}+\frac{9 \sqrt{3} C_{31,22}^{33,0}}{12800}+\frac{63 C_{33,00}^{33,0}}{12800}+\frac{21 \sqrt{3} C_{33,20}^{33,0}}{12800}-\frac{9
C_{33,22}^{33,0}}{800 \sqrt{10}}\)

\noindent\(C_{31,3101,1}^{00,0011,0,\text{Loc}}= \frac{7 C_{11,11}^{51,0}}{640}+\frac{7 C_{11,11}^{51,1}}{1280}+\frac{3 C_{31,11}^{31,0}}{640}+\frac{3
C_{31,11}^{31,1}}{1280}-\frac{9}{800} \sqrt{\frac{7}{2}} C_{33,11}^{33,0}-\frac{9 \sqrt{\frac{7}{2}} C_{33,11}^{33,1}}{1600}\)

\noindent\(C_{31,3101,1}^{00,0011,0,\text{NonLoc}}= \frac{7 C_{11,11}^{51,0}}{1280}+\frac{7 C_{11,11}^{51,1}}{640}+\frac{3 C_{31,11}^{31,0}}{1280}+\frac{3
C_{31,11}^{31,1}}{640}-\frac{9 \sqrt{\frac{7}{2}} C_{33,11}^{33,0}}{1600}-\frac{9}{800} \sqrt{\frac{7}{2}} C_{33,11}^{33,1}\)

\noindent\(C_{31,3101,1}^{00,0011,1,\text{Loc}}= \frac{7 \sqrt{3} C_{11,11}^{51,1}}{1280}+\frac{3 \sqrt{3} C_{31,11}^{31,1}}{1280}-\frac{9
\sqrt{\frac{21}{2}} C_{33,11}^{33,1}}{1600}\)

\noindent\(C_{31,3101,1}^{00,0011,1,\text{NonLoc}}= \frac{7 \sqrt{3} C_{11,11}^{51,0}}{1280}+\frac{3 \sqrt{3} C_{31,11}^{31,0}}{1280}-\frac{9
\sqrt{\frac{21}{2}} C_{33,11}^{33,0}}{1600}\)

\noindent\(C_{31,3111,0}^{00,0000,0,\text{Loc}}= \frac{7 C_{11,11}^{51,0}}{640}+\frac{7 C_{11,11}^{51,1}}{1280}+\frac{3 C_{31,11}^{31,0}}{640}+\frac{3
C_{31,11}^{31,1}}{1280}-\frac{9}{800} \sqrt{\frac{7}{2}} C_{33,11}^{33,0}-\frac{9 \sqrt{\frac{7}{2}} C_{33,11}^{33,1}}{1600}\)

\noindent\(C_{31,3111,0}^{00,0000,0,\text{NonLoc}}= \frac{7 C_{11,11}^{51,0}}{1280}+\frac{7 C_{11,11}^{51,1}}{640}+\frac{3 C_{31,11}^{31,0}}{1280}+\frac{3
C_{31,11}^{31,1}}{640}-\frac{9 \sqrt{\frac{7}{2}} C_{33,11}^{33,0}}{1600}-\frac{9}{800} \sqrt{\frac{7}{2}} C_{33,11}^{33,1}\)

\noindent\(C_{31,3111,0}^{00,0000,1,\text{Loc}}= \frac{7 \sqrt{3} C_{11,11}^{51,1}}{1280}+\frac{3 \sqrt{3} C_{31,11}^{31,1}}{1280}-\frac{9
\sqrt{\frac{21}{2}} C_{33,11}^{33,1}}{1600}\)

\noindent\(C_{31,3111,0}^{00,0000,1,\text{NonLoc}}= \frac{7 \sqrt{3} C_{11,11}^{51,0}}{1280}+\frac{3 \sqrt{3} C_{31,11}^{31,0}}{1280}-\frac{9
\sqrt{\frac{21}{2}} C_{33,11}^{33,0}}{1600}\)

\noindent\(C_{33,0000,3}^{00,3313,0,\text{Loc}}= -\frac{1}{288} \sqrt{\frac{7}{2}} C_{11,11}^{51,0}-\frac{1}{576} \sqrt{\frac{7}{2}}
C_{11,11}^{51,1}+\frac{1}{288} \sqrt{\frac{7}{2}} C_{31,11}^{31,0}+\frac{1}{576} \sqrt{\frac{7}{2}} C_{31,11}^{31,1}-\frac{C_{33,11}^{33,0}}{320}-\frac{C_{33,11}^{33,1}}{640}\)

\noindent\(C_{33,0000,3}^{00,3313,0,\text{NonLoc}}= -\frac{1}{576} \sqrt{\frac{7}{2}} C_{11,11}^{51,0}-\frac{1}{288} \sqrt{\frac{7}{2}}
C_{11,11}^{51,1}+\frac{1}{576} \sqrt{\frac{7}{2}} C_{31,11}^{31,0}+\frac{1}{288} \sqrt{\frac{7}{2}} C_{31,11}^{31,1}-\frac{C_{33,11}^{33,0}}{640}-\frac{C_{33,11}^{33,1}}{320}\)

\noindent\(C_{33,0000,3}^{00,3313,1,\text{Loc}}= -\frac{1}{192} \sqrt{\frac{7}{6}} C_{11,11}^{51,1}+\frac{1}{192} \sqrt{\frac{7}{6}}
C_{31,11}^{31,1}-\frac{\sqrt{3} C_{33,11}^{33,1}}{640}\)

\noindent\(C_{33,0000,3}^{00,3313,1,\text{NonLoc}}= -\frac{1}{192} \sqrt{\frac{7}{6}} C_{11,11}^{51,0}+\frac{1}{192} \sqrt{\frac{7}{6}}
C_{31,11}^{31,0}-\frac{\sqrt{3} C_{33,11}^{33,0}}{640}\)

\noindent\(C_{33,0000,3}^{11,2202,0,\text{Loc}}= -\frac{1}{128} \sqrt{\frac{7}{3}} C_{00,00}^{60,0}-\frac{1}{256} \sqrt{\frac{7}{3}}
C_{00,00}^{60,1}+\frac{1}{384} \sqrt{\frac{7}{3}} C_{11,00}^{51,0}+\frac{1}{768} \sqrt{\frac{7}{3}} C_{11,00}^{51,1}+\frac{1}{384} \sqrt{\frac{7}{3}}
C_{20,00}^{40,0}+\frac{1}{768} \sqrt{\frac{7}{3}} C_{20,00}^{40,1}-\frac{1}{384} \sqrt{\frac{7}{15}} C_{22,00}^{42,0}-\frac{1}{768} \sqrt{\frac{7}{15}}
C_{22,00}^{42,1}-\frac{1}{384} \sqrt{\frac{7}{3}} C_{31,00}^{31,0}-\frac{1}{768} \sqrt{\frac{7}{3}} C_{31,00}^{31,1}+\frac{3 C_{33,00}^{33,0}}{640}+\frac{3
C_{33,00}^{33,1}}{1280}\)

\noindent\(C_{33,0000,3}^{11,2202,0,\text{NonLoc}}= \frac{1}{512} \sqrt{\frac{7}{3}} C_{00,00}^{60,0}+\frac{1}{256} \sqrt{\frac{7}{3}}
C_{00,00}^{60,1}-\frac{\sqrt{7} C_{00,20}^{60,0}}{512}-\frac{\sqrt{7} C_{00,20}^{60,1}}{256}+\frac{\sqrt{\frac{7}{3}} C_{11,00}^{51,0}}{1536}+\frac{1}{768}
\sqrt{\frac{7}{3}} C_{11,00}^{51,1}-\frac{\sqrt{\frac{7}{3}} C_{20,00}^{40,0}}{1536}-\frac{1}{768} \sqrt{\frac{7}{3}} C_{20,00}^{40,1}+\frac{\sqrt{7}
C_{20,20}^{40,0}}{1536}+\frac{\sqrt{7} C_{20,20}^{40,1}}{768}+\frac{\sqrt{\frac{7}{15}} C_{22,00}^{42,0}}{1536}+\frac{1}{768} \sqrt{\frac{7}{15}}
C_{22,00}^{42,1}-\frac{\sqrt{\frac{7}{5}} C_{22,20}^{42,0}}{1536}-\frac{1}{768} \sqrt{\frac{7}{5}} C_{22,20}^{42,1}-\frac{\sqrt{\frac{7}{3}} C_{31,00}^{31,0}}{1536}-\frac{1}{768}
\sqrt{\frac{7}{3}} C_{31,00}^{31,1}+\frac{\sqrt{7} C_{31,20}^{31,0}}{1536}+\frac{\sqrt{7} C_{31,20}^{31,1}}{768}+\frac{3 C_{33,00}^{33,0}}{2560}+\frac{3
C_{33,00}^{33,1}}{1280}-\frac{3 \sqrt{3} C_{33,20}^{33,0}}{2560}-\frac{3 \sqrt{3} C_{33,20}^{33,1}}{1280}\)

\noindent\(C_{33,0000,3}^{11,2202,1,\text{Loc}}= -\frac{\sqrt{7} C_{00,00}^{60,1}}{256}+\frac{\sqrt{7} C_{11,00}^{51,1}}{768}+\frac{\sqrt{7}
C_{20,00}^{40,1}}{768}-\frac{1}{768} \sqrt{\frac{7}{5}} C_{22,00}^{42,1}-\frac{\sqrt{7} C_{31,00}^{31,1}}{768}+\frac{3 \sqrt{3} C_{33,00}^{33,1}}{1280}\)

\noindent\(C_{33,0000,3}^{11,2202,1,\text{NonLoc}}= \frac{\sqrt{7} C_{00,00}^{60,0}}{512}-\frac{\sqrt{21} C_{00,20}^{60,0}}{512}+\frac{\sqrt{7}
C_{11,00}^{51,0}}{1536}-\frac{\sqrt{7} C_{20,00}^{40,0}}{1536}+\frac{1}{512} \sqrt{\frac{7}{3}} C_{20,20}^{40,0}+\frac{\sqrt{\frac{7}{5}} C_{22,00}^{42,0}}{1536}-\frac{1}{512}
\sqrt{\frac{7}{15}} C_{22,20}^{42,0}-\frac{\sqrt{7} C_{31,00}^{31,0}}{1536}+\frac{1}{512} \sqrt{\frac{7}{3}} C_{31,20}^{31,0}+\frac{3 \sqrt{3} C_{33,00}^{33,0}}{2560}-\frac{9
C_{33,20}^{33,0}}{2560}\)

\noindent\(C_{33,0000,3}^{22,1111,0,\text{Loc}}= \frac{1}{768} \sqrt{\frac{7}{3}} C_{11,11}^{51,0}+\frac{\sqrt{\frac{7}{3}} C_{11,11}^{51,1}}{1536}-\frac{1}{384}
\sqrt{\frac{7}{5}} C_{22,11}^{42,0}-\frac{1}{768} \sqrt{\frac{7}{5}} C_{22,11}^{42,1}+\frac{1}{768} \sqrt{\frac{7}{3}} C_{31,11}^{31,0}+\frac{\sqrt{\frac{7}{3}}
C_{31,11}^{31,1}}{1536}+\frac{1}{320} \sqrt{\frac{3}{2}} C_{33,11}^{33,0}+\frac{1}{640} \sqrt{\frac{3}{2}} C_{33,11}^{33,1}\)

\noindent\(C_{33,0000,3}^{22,1111,0,\text{NonLoc}}= \frac{\sqrt{\frac{7}{3}} C_{11,11}^{51,0}}{1536}+\frac{1}{768} \sqrt{\frac{7}{3}}
C_{11,11}^{51,1}+\frac{1}{768} \sqrt{\frac{7}{5}} C_{22,11}^{42,0}+\frac{1}{384} \sqrt{\frac{7}{5}} C_{22,11}^{42,1}+\frac{\sqrt{\frac{7}{3}} C_{31,11}^{31,0}}{1536}+\frac{1}{768}
\sqrt{\frac{7}{3}} C_{31,11}^{31,1}+\frac{1}{640} \sqrt{\frac{3}{2}} C_{33,11}^{33,0}+\frac{1}{320} \sqrt{\frac{3}{2}} C_{33,11}^{33,1}\)

\noindent\(C_{33,0000,3}^{22,1111,1,\text{Loc}}= \frac{\sqrt{7} C_{11,11}^{51,1}}{1536}-\frac{1}{256} \sqrt{\frac{7}{15}} C_{22,11}^{42,1}+\frac{\sqrt{7}
C_{31,11}^{31,1}}{1536}+\frac{3 C_{33,11}^{33,1}}{640 \sqrt{2}}\)

\noindent\(C_{33,0000,3}^{22,1111,1,\text{NonLoc}}= \frac{\sqrt{7} C_{11,11}^{51,0}}{1536}+\frac{1}{256} \sqrt{\frac{7}{15}} C_{22,11}^{42,0}+\frac{\sqrt{7}
C_{31,11}^{31,0}}{1536}+\frac{3 C_{33,11}^{33,0}}{640 \sqrt{2}}\)

\noindent\(C_{33,0000,3}^{33,0000,0,\text{Loc}}= \frac{\sqrt{\frac{7}{3}} C_{00,00}^{60,0}}{3072}+\frac{\sqrt{\frac{7}{3}} C_{00,00}^{60,1}}{6144}-\frac{\sqrt{\frac{7}{3}}
C_{11,00}^{51,0}}{3072}-\frac{\sqrt{\frac{7}{3}} C_{11,00}^{51,1}}{6144}+\frac{\sqrt{\frac{7}{3}} C_{20,00}^{40,0}}{3072}+\frac{\sqrt{\frac{7}{3}}
C_{20,00}^{40,1}}{6144}+\frac{\sqrt{\frac{7}{15}} C_{22,00}^{42,0}}{1536}+\frac{\sqrt{\frac{7}{15}} C_{22,00}^{42,1}}{3072}-\frac{\sqrt{\frac{7}{3}}
C_{31,00}^{31,0}}{3072}-\frac{\sqrt{\frac{7}{3}} C_{31,00}^{31,1}}{6144}-\frac{C_{33,00}^{33,0}}{2560}-\frac{C_{33,00}^{33,1}}{5120}\)

\noindent\(C_{33,0000,3}^{33,0000,0,\text{NonLoc}}= -\frac{\sqrt{\frac{7}{3}} C_{00,00}^{60,0}}{12288}-\frac{\sqrt{\frac{7}{3}} C_{00,00}^{60,1}}{6144}+\frac{\sqrt{7}
C_{00,20}^{60,0}}{12288}+\frac{\sqrt{7} C_{00,20}^{60,1}}{6144}-\frac{\sqrt{\frac{7}{3}} C_{11,00}^{51,0}}{12288}-\frac{\sqrt{\frac{7}{3}} C_{11,00}^{51,1}}{6144}-\frac{\sqrt{\frac{7}{3}}
C_{20,00}^{40,0}}{12288}-\frac{\sqrt{\frac{7}{3}} C_{20,00}^{40,1}}{6144}+\frac{\sqrt{7} C_{20,20}^{40,0}}{12288}+\frac{\sqrt{7} C_{20,20}^{40,1}}{6144}-\frac{\sqrt{\frac{7}{15}}
C_{22,00}^{42,0}}{6144}-\frac{\sqrt{\frac{7}{15}} C_{22,00}^{42,1}}{3072}+\frac{\sqrt{\frac{7}{5}} C_{22,20}^{42,0}}{6144}+\frac{\sqrt{\frac{7}{5}}
C_{22,20}^{42,1}}{3072}-\frac{\sqrt{\frac{7}{3}} C_{31,00}^{31,0}}{12288}-\frac{\sqrt{\frac{7}{3}} C_{31,00}^{31,1}}{6144}+\frac{\sqrt{7} C_{31,20}^{31,0}}{12288}+\frac{\sqrt{7}
C_{31,20}^{31,1}}{6144}-\frac{C_{33,00}^{33,0}}{10240}-\frac{C_{33,00}^{33,1}}{5120}+\frac{\sqrt{3} C_{33,20}^{33,0}}{10240}+\frac{\sqrt{3} C_{33,20}^{33,1}}{5120}\)

\noindent\(C_{33,0000,3}^{33,0000,1,\text{Loc}}= \frac{\sqrt{7} C_{00,00}^{60,1}}{6144}-\frac{\sqrt{7} C_{11,00}^{51,1}}{6144}+\frac{\sqrt{7}
C_{20,00}^{40,1}}{6144}+\frac{\sqrt{\frac{7}{5}} C_{22,00}^{42,1}}{3072}-\frac{\sqrt{7} C_{31,00}^{31,1}}{6144}-\frac{\sqrt{3} C_{33,00}^{33,1}}{5120}\)

\noindent\(C_{33,0000,3}^{33,0000,1,\text{NonLoc}}= -\frac{\sqrt{7} C_{00,00}^{60,0}}{12288}+\frac{\sqrt{\frac{7}{3}} C_{00,20}^{60,0}}{4096}-\frac{\sqrt{7}
C_{11,00}^{51,0}}{12288}-\frac{\sqrt{7} C_{20,00}^{40,0}}{12288}+\frac{\sqrt{\frac{7}{3}} C_{20,20}^{40,0}}{4096}-\frac{\sqrt{\frac{7}{5}} C_{22,00}^{42,0}}{6144}+\frac{\sqrt{\frac{7}{15}}
C_{22,20}^{42,0}}{2048}-\frac{\sqrt{7} C_{31,00}^{31,0}}{12288}+\frac{\sqrt{\frac{7}{3}} C_{31,20}^{31,0}}{4096}-\frac{\sqrt{3} C_{33,00}^{33,0}}{10240}+\frac{3
C_{33,20}^{33,0}}{10240}\)

\noindent\(C_{33,0011,2}^{11,2011,0,\text{Loc}}= \frac{1}{256} \sqrt{\frac{21}{5}} C_{00,22}^{62,0}+\frac{1}{512} \sqrt{\frac{21}{5}}
C_{00,22}^{62,1}-\frac{1}{768} \sqrt{\frac{35}{3}} C_{11,22}^{51,0}-\frac{\sqrt{\frac{35}{3}} C_{11,22}^{51,1}}{1536}+\frac{1}{768} \sqrt{\frac{7}{15}}
C_{20,22}^{42,0}+\frac{\sqrt{\frac{7}{15}} C_{20,22}^{42,1}}{1536}+\frac{1}{768} \sqrt{\frac{7}{15}} C_{22,22}^{40,0}+\frac{\sqrt{\frac{7}{15}} C_{22,22}^{40,1}}{1536}-\frac{C_{22,22}^{42,0}}{192
\sqrt{15}}-\frac{C_{22,22}^{42,1}}{384 \sqrt{15}}-\frac{3}{320} \sqrt{\frac{3}{7}} C_{22,22}^{44,0}-\frac{3}{640} \sqrt{\frac{3}{7}} C_{22,22}^{44,1}-\frac{1}{768}
\sqrt{\frac{7}{15}} C_{31,22}^{31,0}-\frac{\sqrt{\frac{7}{15}} C_{31,22}^{31,1}}{1536}+\frac{C_{31,22}^{33,0}}{320}+\frac{C_{31,22}^{33,1}}{640}+\frac{3}{320}
\sqrt{\frac{3}{10}} C_{33,22}^{33,0}+\frac{3}{640} \sqrt{\frac{3}{10}} C_{33,22}^{33,1}\)

\noindent\(C_{33,0011,2}^{11,2011,0,\text{NonLoc}}= -\frac{1}{512} \sqrt{\frac{21}{5}} C_{00,22}^{62,0}-\frac{1}{256} \sqrt{\frac{21}{5}}
C_{00,22}^{62,1}-\frac{\sqrt{\frac{35}{3}} C_{11,22}^{51,0}}{1536}-\frac{1}{768} \sqrt{\frac{35}{3}} C_{11,22}^{51,1}-\frac{\sqrt{\frac{7}{15}} C_{20,22}^{42,0}}{1536}-\frac{1}{768}
\sqrt{\frac{7}{15}} C_{20,22}^{42,1}-\frac{\sqrt{\frac{7}{15}} C_{22,22}^{40,0}}{1536}-\frac{1}{768} \sqrt{\frac{7}{15}} C_{22,22}^{40,1}+\frac{C_{22,22}^{42,0}}{384
\sqrt{15}}+\frac{C_{22,22}^{42,1}}{192 \sqrt{15}}+\frac{3}{640} \sqrt{\frac{3}{7}} C_{22,22}^{44,0}+\frac{3}{320} \sqrt{\frac{3}{7}} C_{22,22}^{44,1}-\frac{\sqrt{\frac{7}{15}}
C_{31,22}^{31,0}}{1536}-\frac{1}{768} \sqrt{\frac{7}{15}} C_{31,22}^{31,1}+\frac{C_{31,22}^{33,0}}{640}+\frac{C_{31,22}^{33,1}}{320}+\frac{3}{640}
\sqrt{\frac{3}{10}} C_{33,22}^{33,0}+\frac{3}{320} \sqrt{\frac{3}{10}} C_{33,22}^{33,1}\)

\noindent\(C_{33,0011,2}^{11,2011,1,\text{Loc}}= \frac{3}{512} \sqrt{\frac{7}{5}} C_{00,22}^{62,1}-\frac{\sqrt{35} C_{11,22}^{51,1}}{1536}+\frac{\sqrt{\frac{7}{5}}
C_{20,22}^{42,1}}{1536}+\frac{\sqrt{\frac{7}{5}} C_{22,22}^{40,1}}{1536}-\frac{C_{22,22}^{42,1}}{384 \sqrt{5}}-\frac{9 C_{22,22}^{44,1}}{640 \sqrt{7}}-\frac{\sqrt{\frac{7}{5}}
C_{31,22}^{31,1}}{1536}+\frac{\sqrt{3} C_{31,22}^{33,1}}{640}+\frac{9 C_{33,22}^{33,1}}{640 \sqrt{10}}\)

\noindent\(C_{33,0011,2}^{11,2011,1,\text{NonLoc}}= -\frac{3}{512} \sqrt{\frac{7}{5}} C_{00,22}^{62,0}-\frac{\sqrt{35} C_{11,22}^{51,0}}{1536}-\frac{\sqrt{\frac{7}{5}}
C_{20,22}^{42,0}}{1536}-\frac{\sqrt{\frac{7}{5}} C_{22,22}^{40,0}}{1536}+\frac{C_{22,22}^{42,0}}{384 \sqrt{5}}+\frac{9 C_{22,22}^{44,0}}{640 \sqrt{7}}-\frac{\sqrt{\frac{7}{5}}
C_{31,22}^{31,0}}{1536}+\frac{\sqrt{3} C_{31,22}^{33,0}}{640}+\frac{9 C_{33,22}^{33,0}}{640 \sqrt{10}}\)

\noindent\(C_{33,0011,2}^{11,2211,0,\text{Loc}}= \frac{1}{128} \sqrt{\frac{7}{15}} C_{00,20}^{60,0}+\frac{1}{256} \sqrt{\frac{7}{15}}
C_{00,20}^{60,1}+\frac{\sqrt{21} C_{00,22}^{62,0}}{1280}+\frac{\sqrt{21} C_{00,22}^{62,1}}{2560}-\frac{\sqrt{\frac{7}{3}} C_{11,22}^{51,0}}{1920}-\frac{\sqrt{\frac{7}{3}}
C_{11,22}^{51,1}}{3840}-\frac{9 C_{11,22}^{53,0}}{2560 \sqrt{5}}-\frac{9 C_{11,22}^{53,1}}{5120 \sqrt{5}}-\frac{1}{384} \sqrt{\frac{7}{15}} C_{20,20}^{40,0}-\frac{1}{768}
\sqrt{\frac{7}{15}} C_{20,20}^{40,1}-\frac{\sqrt{\frac{7}{3}} C_{20,22}^{42,0}}{1920}-\frac{\sqrt{\frac{7}{3}} C_{20,22}^{42,1}}{3840}+\frac{\sqrt{\frac{7}{3}}
C_{22,20}^{42,0}}{1920}+\frac{\sqrt{\frac{7}{3}} C_{22,20}^{42,1}}{3840}-\frac{1}{768} \sqrt{\frac{7}{3}} C_{22,22}^{40,0}-\frac{\sqrt{\frac{7}{3}}
C_{22,22}^{40,1}}{1536}+\frac{7 C_{22,22}^{42,0}}{1920 \sqrt{3}}+\frac{7 C_{22,22}^{42,1}}{3840 \sqrt{3}}+\frac{1}{384} \sqrt{\frac{7}{15}} C_{31,20}^{31,0}+\frac{1}{768}
\sqrt{\frac{7}{15}} C_{31,20}^{31,1}+\frac{\sqrt{\frac{7}{3}} C_{31,22}^{31,0}}{1920}+\frac{\sqrt{\frac{7}{3}} C_{31,22}^{31,1}}{3840}+\frac{9 C_{31,22}^{33,0}}{2560
\sqrt{5}}+\frac{9 C_{31,22}^{33,1}}{5120 \sqrt{5}}-\frac{3 C_{33,20}^{33,0}}{640 \sqrt{5}}-\frac{3 C_{33,20}^{33,1}}{1280 \sqrt{5}}-\frac{3}{800}
\sqrt{\frac{3}{2}} C_{33,22}^{33,0}-\frac{3 \sqrt{\frac{3}{2}} C_{33,22}^{33,1}}{1600}\)

\noindent\(C_{33,0011,2}^{11,2211,0,\text{NonLoc}}= \frac{1}{512} \sqrt{\frac{7}{5}} C_{00,00}^{60,0}+\frac{1}{256} \sqrt{\frac{7}{5}}
C_{00,00}^{60,1}+\frac{1}{512} \sqrt{\frac{7}{15}} C_{00,20}^{60,0}+\frac{1}{256} \sqrt{\frac{7}{15}} C_{00,20}^{60,1}-\frac{\sqrt{21} C_{00,22}^{62,0}}{2560}-\frac{\sqrt{21}
C_{00,22}^{62,1}}{1280}+\frac{\sqrt{\frac{7}{5}} C_{11,00}^{51,0}}{1536}+\frac{1}{768} \sqrt{\frac{7}{5}} C_{11,00}^{51,1}-\frac{\sqrt{\frac{7}{3}}
C_{11,22}^{51,0}}{3840}-\frac{\sqrt{\frac{7}{3}} C_{11,22}^{51,1}}{1920}-\frac{9 C_{11,22}^{53,0}}{5120 \sqrt{5}}-\frac{9 C_{11,22}^{53,1}}{2560
\sqrt{5}}-\frac{\sqrt{\frac{7}{5}} C_{20,00}^{40,0}}{1536}-\frac{1}{768} \sqrt{\frac{7}{5}} C_{20,00}^{40,1}-\frac{\sqrt{\frac{7}{15}} C_{20,20}^{40,0}}{1536}-\frac{1}{768}
\sqrt{\frac{7}{15}} C_{20,20}^{40,1}+\frac{\sqrt{\frac{7}{3}} C_{20,22}^{42,0}}{3840}+\frac{\sqrt{\frac{7}{3}} C_{20,22}^{42,1}}{1920}+\frac{\sqrt{7}
C_{22,00}^{42,0}}{7680}+\frac{\sqrt{7} C_{22,00}^{42,1}}{3840}+\frac{\sqrt{\frac{7}{3}} C_{22,20}^{42,0}}{7680}+\frac{\sqrt{\frac{7}{3}} C_{22,20}^{42,1}}{3840}+\frac{\sqrt{\frac{7}{3}}
C_{22,22}^{40,0}}{1536}+\frac{1}{768} \sqrt{\frac{7}{3}} C_{22,22}^{40,1}-\frac{7 C_{22,22}^{42,0}}{3840 \sqrt{3}}-\frac{7 C_{22,22}^{42,1}}{1920
\sqrt{3}}-\frac{\sqrt{\frac{7}{5}} C_{31,00}^{31,0}}{1536}-\frac{1}{768} \sqrt{\frac{7}{5}} C_{31,00}^{31,1}-\frac{\sqrt{\frac{7}{15}} C_{31,20}^{31,0}}{1536}-\frac{1}{768}
\sqrt{\frac{7}{15}} C_{31,20}^{31,1}+\frac{\sqrt{\frac{7}{3}} C_{31,22}^{31,0}}{3840}+\frac{\sqrt{\frac{7}{3}} C_{31,22}^{31,1}}{1920}+\frac{9 C_{31,22}^{33,0}}{5120
\sqrt{5}}+\frac{9 C_{31,22}^{33,1}}{2560 \sqrt{5}}+\frac{3 \sqrt{\frac{3}{5}} C_{33,00}^{33,0}}{2560}+\frac{3 \sqrt{\frac{3}{5}} C_{33,00}^{33,1}}{1280}+\frac{3
C_{33,20}^{33,0}}{2560 \sqrt{5}}+\frac{3 C_{33,20}^{33,1}}{1280 \sqrt{5}}-\frac{3 \sqrt{\frac{3}{2}} C_{33,22}^{33,0}}{1600}-\frac{3}{800} \sqrt{\frac{3}{2}}
C_{33,22}^{33,1}\)

\noindent\(C_{33,0011,2}^{11,2211,1,\text{Loc}}= \frac{1}{256} \sqrt{\frac{7}{5}} C_{00,20}^{60,1}+\frac{3 \sqrt{7} C_{00,22}^{62,1}}{2560}-\frac{\sqrt{7}
C_{11,22}^{51,1}}{3840}-\frac{9 \sqrt{\frac{3}{5}} C_{11,22}^{53,1}}{5120}-\frac{1}{768} \sqrt{\frac{7}{5}} C_{20,20}^{40,1}-\frac{\sqrt{7} C_{20,22}^{42,1}}{3840}+\frac{\sqrt{7}
C_{22,20}^{42,1}}{3840}-\frac{\sqrt{7} C_{22,22}^{40,1}}{1536}+\frac{7 C_{22,22}^{42,1}}{3840}+\frac{1}{768} \sqrt{\frac{7}{5}} C_{31,20}^{31,1}+\frac{\sqrt{7}
C_{31,22}^{31,1}}{3840}+\frac{9 \sqrt{\frac{3}{5}} C_{31,22}^{33,1}}{5120}-\frac{3 \sqrt{\frac{3}{5}} C_{33,20}^{33,1}}{1280}-\frac{9 C_{33,22}^{33,1}}{1600
\sqrt{2}}\)

\noindent\(C_{33,0011,2}^{11,2211,1,\text{NonLoc}}= \frac{1}{512} \sqrt{\frac{21}{5}} C_{00,00}^{60,0}+\frac{1}{512} \sqrt{\frac{7}{5}}
C_{00,20}^{60,0}-\frac{3 \sqrt{7} C_{00,22}^{62,0}}{2560}+\frac{1}{512} \sqrt{\frac{7}{15}} C_{11,00}^{51,0}-\frac{\sqrt{7} C_{11,22}^{51,0}}{3840}-\frac{9
\sqrt{\frac{3}{5}} C_{11,22}^{53,0}}{5120}-\frac{1}{512} \sqrt{\frac{7}{15}} C_{20,00}^{40,0}-\frac{\sqrt{\frac{7}{5}} C_{20,20}^{40,0}}{1536}+\frac{\sqrt{7}
C_{20,22}^{42,0}}{3840}+\frac{\sqrt{\frac{7}{3}} C_{22,00}^{42,0}}{2560}+\frac{\sqrt{7} C_{22,20}^{42,0}}{7680}+\frac{\sqrt{7} C_{22,22}^{40,0}}{1536}-\frac{7
C_{22,22}^{42,0}}{3840}-\frac{1}{512} \sqrt{\frac{7}{15}} C_{31,00}^{31,0}-\frac{\sqrt{\frac{7}{5}} C_{31,20}^{31,0}}{1536}+\frac{\sqrt{7} C_{31,22}^{31,0}}{3840}+\frac{9
\sqrt{\frac{3}{5}} C_{31,22}^{33,0}}{5120}+\frac{9 C_{33,00}^{33,0}}{2560 \sqrt{5}}+\frac{3 \sqrt{\frac{3}{5}} C_{33,20}^{33,0}}{2560}-\frac{9 C_{33,22}^{33,0}}{1600
\sqrt{2}}\)

\noindent\(C_{33,0011,2}^{11,2212,0,\text{Loc}}= \frac{\sqrt{7} C_{00,20}^{60,0}}{1152}+\frac{\sqrt{7} C_{00,20}^{60,1}}{2304}-\frac{1}{768}
\sqrt{\frac{7}{5}} C_{00,22}^{62,0}-\frac{\sqrt{\frac{7}{5}} C_{00,22}^{62,1}}{1536}+\frac{\sqrt{\frac{7}{5}} C_{11,22}^{51,0}}{3456}+\frac{\sqrt{\frac{7}{5}}
C_{11,22}^{51,1}}{6912}+\frac{\sqrt{3} C_{11,22}^{53,0}}{2560}+\frac{\sqrt{3} C_{11,22}^{53,1}}{5120}-\frac{\sqrt{7} C_{20,20}^{40,0}}{3456}-\frac{\sqrt{7}
C_{20,20}^{40,1}}{6912}+\frac{\sqrt{\frac{7}{5}} C_{20,22}^{42,0}}{3456}+\frac{\sqrt{\frac{7}{5}} C_{20,22}^{42,1}}{6912}+\frac{\sqrt{\frac{7}{5}}
C_{22,20}^{42,0}}{3456}+\frac{\sqrt{\frac{7}{5}} C_{22,20}^{42,1}}{6912}+\frac{\sqrt{35} C_{22,22}^{40,0}}{6912}+\frac{\sqrt{35} C_{22,22}^{40,1}}{13824}-\frac{7
C_{22,22}^{42,0}}{3456 \sqrt{5}}-\frac{7 C_{22,22}^{42,1}}{6912 \sqrt{5}}+\frac{\sqrt{7} C_{31,20}^{31,0}}{3456}+\frac{\sqrt{7} C_{31,20}^{31,1}}{6912}-\frac{\sqrt{\frac{7}{5}}
C_{31,22}^{31,0}}{3456}-\frac{\sqrt{\frac{7}{5}} C_{31,22}^{31,1}}{6912}-\frac{\sqrt{3} C_{31,22}^{33,0}}{2560}-\frac{\sqrt{3} C_{31,22}^{33,1}}{5120}-\frac{C_{33,20}^{33,0}}{640
\sqrt{3}}-\frac{C_{33,20}^{33,1}}{1280 \sqrt{3}}+\frac{C_{33,22}^{33,0}}{160 \sqrt{10}}+\frac{C_{33,22}^{33,1}}{320 \sqrt{10}}\)

\noindent\(C_{33,0011,2}^{11,2212,0,\text{NonLoc}}= \frac{\sqrt{\frac{7}{3}} C_{00,00}^{60,0}}{1536}+\frac{1}{768} \sqrt{\frac{7}{3}}
C_{00,00}^{60,1}+\frac{\sqrt{7} C_{00,20}^{60,0}}{4608}+\frac{\sqrt{7} C_{00,20}^{60,1}}{2304}+\frac{\sqrt{\frac{7}{5}} C_{00,22}^{62,0}}{1536}+\frac{1}{768}
\sqrt{\frac{7}{5}} C_{00,22}^{62,1}+\frac{\sqrt{\frac{7}{3}} C_{11,00}^{51,0}}{4608}+\frac{\sqrt{\frac{7}{3}} C_{11,00}^{51,1}}{2304}+\frac{\sqrt{\frac{7}{5}}
C_{11,22}^{51,0}}{6912}+\frac{\sqrt{\frac{7}{5}} C_{11,22}^{51,1}}{3456}+\frac{\sqrt{3} C_{11,22}^{53,0}}{5120}+\frac{\sqrt{3} C_{11,22}^{53,1}}{2560}-\frac{\sqrt{\frac{7}{3}}
C_{20,00}^{40,0}}{4608}-\frac{\sqrt{\frac{7}{3}} C_{20,00}^{40,1}}{2304}-\frac{\sqrt{7} C_{20,20}^{40,0}}{13824}-\frac{\sqrt{7} C_{20,20}^{40,1}}{6912}-\frac{\sqrt{\frac{7}{5}}
C_{20,22}^{42,0}}{6912}-\frac{\sqrt{\frac{7}{5}} C_{20,22}^{42,1}}{3456}+\frac{\sqrt{\frac{7}{15}} C_{22,00}^{42,0}}{4608}+\frac{\sqrt{\frac{7}{15}}
C_{22,00}^{42,1}}{2304}+\frac{\sqrt{\frac{7}{5}} C_{22,20}^{42,0}}{13824}+\frac{\sqrt{\frac{7}{5}} C_{22,20}^{42,1}}{6912}-\frac{\sqrt{35} C_{22,22}^{40,0}}{13824}-\frac{\sqrt{35}
C_{22,22}^{40,1}}{6912}+\frac{7 C_{22,22}^{42,0}}{6912 \sqrt{5}}+\frac{7 C_{22,22}^{42,1}}{3456 \sqrt{5}}-\frac{\sqrt{\frac{7}{3}} C_{31,00}^{31,0}}{4608}-\frac{\sqrt{\frac{7}{3}}
C_{31,00}^{31,1}}{2304}-\frac{\sqrt{7} C_{31,20}^{31,0}}{13824}-\frac{\sqrt{7} C_{31,20}^{31,1}}{6912}-\frac{\sqrt{\frac{7}{5}} C_{31,22}^{31,0}}{6912}-\frac{\sqrt{\frac{7}{5}}
C_{31,22}^{31,1}}{3456}-\frac{\sqrt{3} C_{31,22}^{33,0}}{5120}-\frac{\sqrt{3} C_{31,22}^{33,1}}{2560}+\frac{C_{33,00}^{33,0}}{2560}+\frac{C_{33,00}^{33,1}}{1280}+\frac{C_{33,20}^{33,0}}{2560
\sqrt{3}}+\frac{C_{33,20}^{33,1}}{1280 \sqrt{3}}+\frac{C_{33,22}^{33,0}}{320 \sqrt{10}}+\frac{C_{33,22}^{33,1}}{160 \sqrt{10}}\)

\noindent\(C_{33,0011,2}^{11,2212,1,\text{Loc}}= \frac{1}{768} \sqrt{\frac{7}{3}} C_{00,20}^{60,1}-\frac{1}{512} \sqrt{\frac{7}{15}}
C_{00,22}^{62,1}+\frac{\sqrt{\frac{7}{15}} C_{11,22}^{51,1}}{2304}+\frac{3 C_{11,22}^{53,1}}{5120}-\frac{\sqrt{\frac{7}{3}} C_{20,20}^{40,1}}{2304}+\frac{\sqrt{\frac{7}{15}}
C_{20,22}^{42,1}}{2304}+\frac{\sqrt{\frac{7}{15}} C_{22,20}^{42,1}}{2304}+\frac{\sqrt{\frac{35}{3}} C_{22,22}^{40,1}}{4608}-\frac{7 C_{22,22}^{42,1}}{2304
\sqrt{15}}+\frac{\sqrt{\frac{7}{3}} C_{31,20}^{31,1}}{2304}-\frac{\sqrt{\frac{7}{15}} C_{31,22}^{31,1}}{2304}-\frac{3 C_{31,22}^{33,1}}{5120}-\frac{C_{33,20}^{33,1}}{1280}+\frac{1}{320}
\sqrt{\frac{3}{10}} C_{33,22}^{33,1}\)

\noindent\(C_{33,0011,2}^{11,2212,1,\text{NonLoc}}= \frac{\sqrt{7} C_{00,00}^{60,0}}{1536}+\frac{\sqrt{\frac{7}{3}} C_{00,20}^{60,0}}{1536}+\frac{1}{512}
\sqrt{\frac{7}{15}} C_{00,22}^{62,0}+\frac{\sqrt{7} C_{11,00}^{51,0}}{4608}+\frac{\sqrt{\frac{7}{15}} C_{11,22}^{51,0}}{2304}+\frac{3 C_{11,22}^{53,0}}{5120}-\frac{\sqrt{7}
C_{20,00}^{40,0}}{4608}-\frac{\sqrt{\frac{7}{3}} C_{20,20}^{40,0}}{4608}-\frac{\sqrt{\frac{7}{15}} C_{20,22}^{42,0}}{2304}+\frac{\sqrt{\frac{7}{5}}
C_{22,00}^{42,0}}{4608}+\frac{\sqrt{\frac{7}{15}} C_{22,20}^{42,0}}{4608}-\frac{\sqrt{\frac{35}{3}} C_{22,22}^{40,0}}{4608}+\frac{7 C_{22,22}^{42,0}}{2304
\sqrt{15}}-\frac{\sqrt{7} C_{31,00}^{31,0}}{4608}-\frac{\sqrt{\frac{7}{3}} C_{31,20}^{31,0}}{4608}-\frac{\sqrt{\frac{7}{15}} C_{31,22}^{31,0}}{2304}-\frac{3
C_{31,22}^{33,0}}{5120}+\frac{\sqrt{3} C_{33,00}^{33,0}}{2560}+\frac{C_{33,20}^{33,0}}{2560}+\frac{1}{320} \sqrt{\frac{3}{10}} C_{33,22}^{33,0}\)

\noindent\(C_{33,0011,2}^{11,2213,0,\text{Loc}}= \frac{C_{00,20}^{60,0}}{1152 \sqrt{5}}+\frac{C_{00,20}^{60,1}}{2304 \sqrt{5}}+\frac{13
C_{00,22}^{62,0}}{1920}+\frac{13 C_{00,22}^{62,1}}{3840}-\frac{29 C_{11,22}^{51,0}}{4320}-\frac{29 C_{11,22}^{51,1}}{8640}+\frac{17 \sqrt{\frac{3}{35}}
C_{11,22}^{53,0}}{1280}+\frac{17 \sqrt{\frac{3}{35}} C_{11,22}^{53,1}}{2560}-\frac{C_{20,20}^{40,0}}{3456 \sqrt{5}}-\frac{C_{20,20}^{40,1}}{6912
\sqrt{5}}-\frac{29 C_{20,22}^{42,0}}{4320}-\frac{29 C_{20,22}^{42,1}}{8640}+\frac{C_{22,20}^{42,0}}{17280}+\frac{C_{22,20}^{42,1}}{34560}+\frac{23
C_{22,22}^{40,0}}{3456}+\frac{23 C_{22,22}^{40,1}}{6912}+\frac{C_{22,22}^{42,0}}{8640 \sqrt{7}}+\frac{C_{22,22}^{42,1}}{17280 \sqrt{7}}+\frac{3 C_{22,22}^{44,0}}{448
\sqrt{5}}+\frac{3 C_{22,22}^{44,1}}{896 \sqrt{5}}+\frac{C_{31,20}^{31,0}}{3456 \sqrt{5}}+\frac{C_{31,20}^{31,1}}{6912 \sqrt{5}}+\frac{29 C_{31,22}^{31,0}}{4320}+\frac{29
C_{31,22}^{31,1}}{8640}-\frac{17 \sqrt{\frac{3}{35}} C_{31,22}^{33,0}}{1280}-\frac{17 \sqrt{\frac{3}{35}} C_{31,22}^{33,1}}{2560}-\frac{C_{33,20}^{33,0}}{640
\sqrt{105}}-\frac{C_{33,20}^{33,1}}{1280 \sqrt{105}}-\frac{\sqrt{\frac{7}{2}} C_{33,22}^{33,0}}{1600}-\frac{\sqrt{\frac{7}{2}} C_{33,22}^{33,1}}{3200}\)

\noindent\(C_{33,0011,2}^{11,2213,0,\text{NonLoc}}= \frac{C_{00,00}^{60,0}}{1536 \sqrt{15}}+\frac{C_{00,00}^{60,1}}{768 \sqrt{15}}+\frac{C_{00,20}^{60,0}}{4608
\sqrt{5}}+\frac{C_{00,20}^{60,1}}{2304 \sqrt{5}}-\frac{13 C_{00,22}^{62,0}}{3840}-\frac{13 C_{00,22}^{62,1}}{1920}+\frac{C_{11,00}^{51,0}}{4608 \sqrt{15}}+\frac{C_{11,00}^{51,1}}{2304
\sqrt{15}}-\frac{29 C_{11,22}^{51,0}}{8640}-\frac{29 C_{11,22}^{51,1}}{4320}+\frac{17 \sqrt{\frac{3}{35}} C_{11,22}^{53,0}}{2560}+\frac{17 \sqrt{\frac{3}{35}}
C_{11,22}^{53,1}}{1280}-\frac{C_{20,00}^{40,0}}{4608 \sqrt{15}}-\frac{C_{20,00}^{40,1}}{2304 \sqrt{15}}-\frac{C_{20,20}^{40,0}}{13824 \sqrt{5}}-\frac{C_{20,20}^{40,1}}{6912
\sqrt{5}}+\frac{29 C_{20,22}^{42,0}}{8640}+\frac{29 C_{20,22}^{42,1}}{4320}+\frac{C_{22,00}^{42,0}}{23040 \sqrt{3}}+\frac{C_{22,00}^{42,1}}{11520
\sqrt{3}}+\frac{C_{22,20}^{42,0}}{69120}+\frac{C_{22,20}^{42,1}}{34560}-\frac{23 C_{22,22}^{40,0}}{6912}-\frac{23 C_{22,22}^{40,1}}{3456}-\frac{C_{22,22}^{42,0}}{17280
\sqrt{7}}-\frac{C_{22,22}^{42,1}}{8640 \sqrt{7}}-\frac{3 C_{22,22}^{44,0}}{896 \sqrt{5}}-\frac{3 C_{22,22}^{44,1}}{448 \sqrt{5}}-\frac{C_{31,00}^{31,0}}{4608
\sqrt{15}}-\frac{C_{31,00}^{31,1}}{2304 \sqrt{15}}-\frac{C_{31,20}^{31,0}}{13824 \sqrt{5}}-\frac{C_{31,20}^{31,1}}{6912 \sqrt{5}}+\frac{29 C_{31,22}^{31,0}}{8640}+\frac{29
C_{31,22}^{31,1}}{4320}-\frac{17 \sqrt{\frac{3}{35}} C_{31,22}^{33,0}}{2560}-\frac{17 \sqrt{\frac{3}{35}} C_{31,22}^{33,1}}{1280}+\frac{C_{33,00}^{33,0}}{2560
\sqrt{35}}+\frac{C_{33,00}^{33,1}}{1280 \sqrt{35}}+\frac{C_{33,20}^{33,0}}{2560 \sqrt{105}}+\frac{C_{33,20}^{33,1}}{1280 \sqrt{105}}-\frac{\sqrt{\frac{7}{2}}
C_{33,22}^{33,0}}{3200}-\frac{\sqrt{\frac{7}{2}} C_{33,22}^{33,1}}{1600}\)

\noindent\(C_{33,0011,2}^{11,2213,1,\text{Loc}}= \frac{C_{00,20}^{60,1}}{768 \sqrt{15}}+\frac{13 C_{00,22}^{62,1}}{1280 \sqrt{3}}-\frac{29
C_{11,22}^{51,1}}{2880 \sqrt{3}}+\frac{51 C_{11,22}^{53,1}}{2560 \sqrt{35}}-\frac{C_{20,20}^{40,1}}{2304 \sqrt{15}}-\frac{29 C_{20,22}^{42,1}}{2880
\sqrt{3}}+\frac{C_{22,20}^{42,1}}{11520 \sqrt{3}}+\frac{23 C_{22,22}^{40,1}}{2304 \sqrt{3}}+\frac{C_{22,22}^{42,1}}{5760 \sqrt{21}}+\frac{3}{896}
\sqrt{\frac{3}{5}} C_{22,22}^{44,1}+\frac{C_{31,20}^{31,1}}{2304 \sqrt{15}}+\frac{29 C_{31,22}^{31,1}}{2880 \sqrt{3}}-\frac{51 C_{31,22}^{33,1}}{2560
\sqrt{35}}-\frac{C_{33,20}^{33,1}}{1280 \sqrt{35}}-\frac{\sqrt{\frac{21}{2}} C_{33,22}^{33,1}}{3200}\)

\noindent\(C_{33,0011,2}^{11,2213,1,\text{NonLoc}}= \frac{C_{00,00}^{60,0}}{1536 \sqrt{5}}+\frac{C_{00,20}^{60,0}}{1536 \sqrt{15}}-\frac{13
C_{00,22}^{62,0}}{1280 \sqrt{3}}+\frac{C_{11,00}^{51,0}}{4608 \sqrt{5}}-\frac{29 C_{11,22}^{51,0}}{2880 \sqrt{3}}+\frac{51 C_{11,22}^{53,0}}{2560
\sqrt{35}}-\frac{C_{20,00}^{40,0}}{4608 \sqrt{5}}-\frac{C_{20,20}^{40,0}}{4608 \sqrt{15}}+\frac{29 C_{20,22}^{42,0}}{2880 \sqrt{3}}+\frac{C_{22,00}^{42,0}}{23040}+\frac{C_{22,20}^{42,0}}{23040
\sqrt{3}}-\frac{23 C_{22,22}^{40,0}}{2304 \sqrt{3}}-\frac{C_{22,22}^{42,0}}{5760 \sqrt{21}}-\frac{3}{896} \sqrt{\frac{3}{5}} C_{22,22}^{44,0}-\frac{C_{31,00}^{31,0}}{4608
\sqrt{5}}-\frac{C_{31,20}^{31,0}}{4608 \sqrt{15}}+\frac{29 C_{31,22}^{31,0}}{2880 \sqrt{3}}-\frac{51 C_{31,22}^{33,0}}{2560 \sqrt{35}}+\frac{\sqrt{\frac{3}{35}}
C_{33,00}^{33,0}}{2560}+\frac{C_{33,20}^{33,0}}{2560 \sqrt{35}}-\frac{\sqrt{\frac{21}{2}} C_{33,22}^{33,0}}{3200}\)

\noindent\(C_{33,0011,2}^{22,1101,0,\text{Loc}}= -\frac{\sqrt{7} C_{11,11}^{51,0}}{2304}-\frac{\sqrt{7} C_{11,11}^{51,1}}{4608}+\frac{1}{384}
\sqrt{\frac{7}{15}} C_{22,11}^{42,0}+\frac{1}{768} \sqrt{\frac{7}{15}} C_{22,11}^{42,1}-\frac{\sqrt{7} C_{31,11}^{31,0}}{2304}-\frac{\sqrt{7} C_{31,11}^{31,1}}{4608}-\frac{C_{33,11}^{33,0}}{320
\sqrt{2}}-\frac{C_{33,11}^{33,1}}{640 \sqrt{2}}\)

\noindent\(C_{33,0011,2}^{22,1101,0,\text{NonLoc}}= -\frac{\sqrt{7} C_{11,11}^{51,0}}{4608}-\frac{\sqrt{7} C_{11,11}^{51,1}}{2304}-\frac{1}{768}
\sqrt{\frac{7}{15}} C_{22,11}^{42,0}-\frac{1}{384} \sqrt{\frac{7}{15}} C_{22,11}^{42,1}-\frac{\sqrt{7} C_{31,11}^{31,0}}{4608}-\frac{\sqrt{7} C_{31,11}^{31,1}}{2304}-\frac{C_{33,11}^{33,0}}{640
\sqrt{2}}-\frac{C_{33,11}^{33,1}}{320 \sqrt{2}}\)

\noindent\(C_{33,0011,2}^{22,1101,1,\text{Loc}}= -\frac{\sqrt{\frac{7}{3}} C_{11,11}^{51,1}}{1536}+\frac{1}{768} \sqrt{\frac{7}{5}}
C_{22,11}^{42,1}-\frac{\sqrt{\frac{7}{3}} C_{31,11}^{31,1}}{1536}-\frac{1}{640} \sqrt{\frac{3}{2}} C_{33,11}^{33,1}\)

\noindent\(C_{33,0011,2}^{22,1101,1,\text{NonLoc}}= -\frac{\sqrt{\frac{7}{3}} C_{11,11}^{51,0}}{1536}-\frac{1}{768} \sqrt{\frac{7}{5}}
C_{22,11}^{42,0}-\frac{\sqrt{\frac{7}{3}} C_{31,11}^{31,0}}{1536}-\frac{1}{640} \sqrt{\frac{3}{2}} C_{33,11}^{33,0}\)

\noindent\(C_{33,0011,2}^{31,0011,0,\text{Loc}}= -\frac{3 \sqrt{\frac{21}{5}} C_{00,22}^{62,0}}{5120}-\frac{3 \sqrt{\frac{21}{5}} C_{00,22}^{62,1}}{10240}+\frac{3
\sqrt{\frac{21}{5}} C_{11,22}^{51,0}}{5120}+\frac{3 \sqrt{\frac{21}{5}} C_{11,22}^{51,1}}{10240}+\frac{27 C_{11,22}^{53,0}}{25600}+\frac{27 C_{11,22}^{53,1}}{51200}-\frac{3
\sqrt{\frac{21}{5}} C_{20,22}^{42,0}}{5120}-\frac{3 \sqrt{\frac{21}{5}} C_{20,22}^{42,1}}{10240}-\frac{3 \sqrt{\frac{21}{5}} C_{22,22}^{40,0}}{5120}-\frac{3
\sqrt{\frac{21}{5}} C_{22,22}^{40,1}}{10240}-\frac{3 \sqrt{\frac{3}{5}} C_{22,22}^{42,0}}{2560}-\frac{3 \sqrt{\frac{3}{5}} C_{22,22}^{42,1}}{5120}-\frac{9
\sqrt{\frac{3}{7}} C_{22,22}^{44,0}}{6400}-\frac{9 \sqrt{\frac{3}{7}} C_{22,22}^{44,1}}{12800}+\frac{3 \sqrt{\frac{21}{5}} C_{31,22}^{31,0}}{5120}+\frac{3
\sqrt{\frac{21}{5}} C_{31,22}^{31,1}}{10240}+\frac{27 C_{31,22}^{33,0}}{25600}+\frac{27 C_{31,22}^{33,1}}{51200}+\frac{9 \sqrt{\frac{3}{10}} C_{33,22}^{33,0}}{6400}+\frac{9
\sqrt{\frac{3}{10}} C_{33,22}^{33,1}}{12800}\)

\noindent\(C_{33,0011,2}^{31,0011,0,\text{NonLoc}}= \frac{3 \sqrt{\frac{21}{5}} C_{00,22}^{62,0}}{10240}+\frac{3 \sqrt{\frac{21}{5}}
C_{00,22}^{62,1}}{5120}+\frac{3 \sqrt{\frac{21}{5}} C_{11,22}^{51,0}}{10240}+\frac{3 \sqrt{\frac{21}{5}} C_{11,22}^{51,1}}{5120}+\frac{27 C_{11,22}^{53,0}}{51200}+\frac{27
C_{11,22}^{53,1}}{25600}+\frac{3 \sqrt{\frac{21}{5}} C_{20,22}^{42,0}}{10240}+\frac{3 \sqrt{\frac{21}{5}} C_{20,22}^{42,1}}{5120}+\frac{3 \sqrt{\frac{21}{5}}
C_{22,22}^{40,0}}{10240}+\frac{3 \sqrt{\frac{21}{5}} C_{22,22}^{40,1}}{5120}+\frac{3 \sqrt{\frac{3}{5}} C_{22,22}^{42,0}}{5120}+\frac{3 \sqrt{\frac{3}{5}}
C_{22,22}^{42,1}}{2560}+\frac{9 \sqrt{\frac{3}{7}} C_{22,22}^{44,0}}{12800}+\frac{9 \sqrt{\frac{3}{7}} C_{22,22}^{44,1}}{6400}+\frac{3 \sqrt{\frac{21}{5}}
C_{31,22}^{31,0}}{10240}+\frac{3 \sqrt{\frac{21}{5}} C_{31,22}^{31,1}}{5120}+\frac{27 C_{31,22}^{33,0}}{51200}+\frac{27 C_{31,22}^{33,1}}{25600}+\frac{9
\sqrt{\frac{3}{10}} C_{33,22}^{33,0}}{12800}+\frac{9 \sqrt{\frac{3}{10}} C_{33,22}^{33,1}}{6400}\)

\noindent\(C_{33,0011,2}^{31,0011,1,\text{Loc}}= -\frac{9 \sqrt{\frac{7}{5}} C_{00,22}^{62,1}}{10240}+\frac{9 \sqrt{\frac{7}{5}} C_{11,22}^{51,1}}{10240}+\frac{27
\sqrt{3} C_{11,22}^{53,1}}{51200}-\frac{9 \sqrt{\frac{7}{5}} C_{20,22}^{42,1}}{10240}-\frac{9 \sqrt{\frac{7}{5}} C_{22,22}^{40,1}}{10240}-\frac{9
C_{22,22}^{42,1}}{5120 \sqrt{5}}-\frac{27 C_{22,22}^{44,1}}{12800 \sqrt{7}}+\frac{9 \sqrt{\frac{7}{5}} C_{31,22}^{31,1}}{10240}+\frac{27 \sqrt{3}
C_{31,22}^{33,1}}{51200}+\frac{27 C_{33,22}^{33,1}}{12800 \sqrt{10}}\)

\noindent\(C_{33,0011,2}^{31,0011,1,\text{NonLoc}}= \frac{9 \sqrt{\frac{7}{5}} C_{00,22}^{62,0}}{10240}+\frac{9 \sqrt{\frac{7}{5}}
C_{11,22}^{51,0}}{10240}+\frac{27 \sqrt{3} C_{11,22}^{53,0}}{51200}+\frac{9 \sqrt{\frac{7}{5}} C_{20,22}^{42,0}}{10240}+\frac{9 \sqrt{\frac{7}{5}}
C_{22,22}^{40,0}}{10240}+\frac{9 C_{22,22}^{42,0}}{5120 \sqrt{5}}+\frac{27 C_{22,22}^{44,0}}{12800 \sqrt{7}}+\frac{9 \sqrt{\frac{7}{5}} C_{31,22}^{31,0}}{10240}+\frac{27
\sqrt{3} C_{31,22}^{33,0}}{51200}+\frac{27 C_{33,22}^{33,0}}{12800 \sqrt{10}}\)

\noindent\(C_{33,0011,2}^{33,0011,0,\text{Loc}}= -\frac{\sqrt{5} C_{00,20}^{60,0}}{9216}-\frac{\sqrt{5} C_{00,20}^{60,1}}{18432}-\frac{C_{00,22}^{62,0}}{7680}-\frac{C_{00,22}^{62,1}}{15360}+\frac{C_{11,22}^{51,0}}{7680}+\frac{C_{11,22}^{51,1}}{15360}+\frac{\sqrt{\frac{3}{35}}
C_{11,22}^{53,0}}{2560}+\frac{\sqrt{\frac{3}{35}} C_{11,22}^{53,1}}{5120}-\frac{\sqrt{5} C_{20,20}^{40,0}}{9216}-\frac{\sqrt{5} C_{20,20}^{40,1}}{18432}-\frac{C_{20,22}^{42,0}}{7680}-\frac{C_{20,22}^{42,1}}{15360}-\frac{C_{22,20}^{42,0}}{4608}-\frac{C_{22,20}^{42,1}}{9216}-\frac{C_{22,22}^{40,0}}{7680}-\frac{C_{22,22}^{40,1}}{15360}-\frac{C_{22,22}^{42,0}}{3840
\sqrt{7}}-\frac{C_{22,22}^{42,1}}{7680 \sqrt{7}}-\frac{C_{22,22}^{44,0}}{4480 \sqrt{5}}-\frac{C_{22,22}^{44,1}}{8960 \sqrt{5}}+\frac{\sqrt{5} C_{31,20}^{31,0}}{9216}+\frac{\sqrt{5}
C_{31,20}^{31,1}}{18432}+\frac{C_{31,22}^{31,0}}{7680}+\frac{C_{31,22}^{31,1}}{15360}+\frac{\sqrt{\frac{3}{35}} C_{31,22}^{33,0}}{2560}+\frac{\sqrt{\frac{3}{35}}
C_{31,22}^{33,1}}{5120}+\frac{C_{33,20}^{33,0}}{512 \sqrt{105}}+\frac{C_{33,20}^{33,1}}{1024 \sqrt{105}}+\frac{C_{33,22}^{33,0}}{3200 \sqrt{14}}+\frac{C_{33,22}^{33,1}}{6400
\sqrt{14}}\)

\noindent\(C_{33,0011,2}^{33,0011,0,\text{NonLoc}}= -\frac{\sqrt{\frac{5}{3}} C_{00,00}^{60,0}}{12288}-\frac{\sqrt{\frac{5}{3}} C_{00,00}^{60,1}}{6144}-\frac{\sqrt{5}
C_{00,20}^{60,0}}{36864}-\frac{\sqrt{5} C_{00,20}^{60,1}}{18432}+\frac{C_{00,22}^{62,0}}{15360}+\frac{C_{00,22}^{62,1}}{7680}-\frac{\sqrt{\frac{5}{3}}
C_{11,00}^{51,0}}{12288}-\frac{\sqrt{\frac{5}{3}} C_{11,00}^{51,1}}{6144}+\frac{C_{11,22}^{51,0}}{15360}+\frac{C_{11,22}^{51,1}}{7680}+\frac{\sqrt{\frac{3}{35}}
C_{11,22}^{53,0}}{5120}+\frac{\sqrt{\frac{3}{35}} C_{11,22}^{53,1}}{2560}-\frac{\sqrt{\frac{5}{3}} C_{20,00}^{40,0}}{12288}-\frac{\sqrt{\frac{5}{3}}
C_{20,00}^{40,1}}{6144}-\frac{\sqrt{5} C_{20,20}^{40,0}}{36864}-\frac{\sqrt{5} C_{20,20}^{40,1}}{18432}+\frac{C_{20,22}^{42,0}}{15360}+\frac{C_{20,22}^{42,1}}{7680}-\frac{C_{22,00}^{42,0}}{6144
\sqrt{3}}-\frac{C_{22,00}^{42,1}}{3072 \sqrt{3}}-\frac{C_{22,20}^{42,0}}{18432}-\frac{C_{22,20}^{42,1}}{9216}+\frac{C_{22,22}^{40,0}}{15360}+\frac{C_{22,22}^{40,1}}{7680}+\frac{C_{22,22}^{42,0}}{7680
\sqrt{7}}+\frac{C_{22,22}^{42,1}}{3840 \sqrt{7}}+\frac{C_{22,22}^{44,0}}{8960 \sqrt{5}}+\frac{C_{22,22}^{44,1}}{4480 \sqrt{5}}-\frac{\sqrt{\frac{5}{3}}
C_{31,00}^{31,0}}{12288}-\frac{\sqrt{\frac{5}{3}} C_{31,00}^{31,1}}{6144}-\frac{\sqrt{5} C_{31,20}^{31,0}}{36864}-\frac{\sqrt{5} C_{31,20}^{31,1}}{18432}+\frac{C_{31,22}^{31,0}}{15360}+\frac{C_{31,22}^{31,1}}{7680}+\frac{\sqrt{\frac{3}{35}}
C_{31,22}^{33,0}}{5120}+\frac{\sqrt{\frac{3}{35}} C_{31,22}^{33,1}}{2560}-\frac{C_{33,00}^{33,0}}{2048 \sqrt{35}}-\frac{C_{33,00}^{33,1}}{1024 \sqrt{35}}-\frac{C_{33,20}^{33,0}}{2048
\sqrt{105}}-\frac{C_{33,20}^{33,1}}{1024 \sqrt{105}}+\frac{C_{33,22}^{33,0}}{6400 \sqrt{14}}+\frac{C_{33,22}^{33,1}}{3200 \sqrt{14}}\)

\noindent\(C_{33,0011,2}^{33,0011,1,\text{Loc}}= -\frac{\sqrt{\frac{5}{3}} C_{00,20}^{60,1}}{6144}-\frac{C_{00,22}^{62,1}}{5120 \sqrt{3}}+\frac{C_{11,22}^{51,1}}{5120
\sqrt{3}}+\frac{3 C_{11,22}^{53,1}}{5120 \sqrt{35}}-\frac{\sqrt{\frac{5}{3}} C_{20,20}^{40,1}}{6144}-\frac{C_{20,22}^{42,1}}{5120 \sqrt{3}}-\frac{C_{22,20}^{42,1}}{3072
\sqrt{3}}-\frac{C_{22,22}^{40,1}}{5120 \sqrt{3}}-\frac{C_{22,22}^{42,1}}{2560 \sqrt{21}}-\frac{\sqrt{\frac{3}{5}} C_{22,22}^{44,1}}{8960}+\frac{\sqrt{\frac{5}{3}}
C_{31,20}^{31,1}}{6144}+\frac{C_{31,22}^{31,1}}{5120 \sqrt{3}}+\frac{3 C_{31,22}^{33,1}}{5120 \sqrt{35}}+\frac{C_{33,20}^{33,1}}{1024 \sqrt{35}}+\frac{\sqrt{\frac{3}{14}}
C_{33,22}^{33,1}}{6400}\)

\noindent\(C_{33,0011,2}^{33,0011,1,\text{NonLoc}}= -\frac{\sqrt{5} C_{00,00}^{60,0}}{12288}-\frac{\sqrt{\frac{5}{3}} C_{00,20}^{60,0}}{12288}+\frac{C_{00,22}^{62,0}}{5120
\sqrt{3}}-\frac{\sqrt{5} C_{11,00}^{51,0}}{12288}+\frac{C_{11,22}^{51,0}}{5120 \sqrt{3}}+\frac{3 C_{11,22}^{53,0}}{5120 \sqrt{35}}-\frac{\sqrt{5}
C_{20,00}^{40,0}}{12288}-\frac{\sqrt{\frac{5}{3}} C_{20,20}^{40,0}}{12288}+\frac{C_{20,22}^{42,0}}{5120 \sqrt{3}}-\frac{C_{22,00}^{42,0}}{6144}-\frac{C_{22,20}^{42,0}}{6144
\sqrt{3}}+\frac{C_{22,22}^{40,0}}{5120 \sqrt{3}}+\frac{C_{22,22}^{42,0}}{2560 \sqrt{21}}+\frac{\sqrt{\frac{3}{5}} C_{22,22}^{44,0}}{8960}-\frac{\sqrt{5}
C_{31,00}^{31,0}}{12288}-\frac{\sqrt{\frac{5}{3}} C_{31,20}^{31,0}}{12288}+\frac{C_{31,22}^{31,0}}{5120 \sqrt{3}}+\frac{3 C_{31,22}^{33,0}}{5120
\sqrt{35}}-\frac{\sqrt{\frac{3}{35}} C_{33,00}^{33,0}}{2048}-\frac{C_{33,20}^{33,0}}{2048 \sqrt{35}}+\frac{\sqrt{\frac{3}{14}} C_{33,22}^{33,0}}{6400}\)

\noindent\(C_{33,0011,3}^{00,3303,0,\text{Loc}}= \frac{1}{288} \sqrt{\frac{7}{2}} C_{11,11}^{51,0}+\frac{1}{576} \sqrt{\frac{7}{2}}
C_{11,11}^{51,1}-\frac{1}{288} \sqrt{\frac{7}{2}} C_{31,11}^{31,0}-\frac{1}{576} \sqrt{\frac{7}{2}} C_{31,11}^{31,1}+\frac{C_{33,11}^{33,0}}{320}+\frac{C_{33,11}^{33,1}}{640}\)

\noindent\(C_{33,0011,3}^{00,3303,0,\text{NonLoc}}= \frac{1}{576} \sqrt{\frac{7}{2}} C_{11,11}^{51,0}+\frac{1}{288} \sqrt{\frac{7}{2}}
C_{11,11}^{51,1}-\frac{1}{576} \sqrt{\frac{7}{2}} C_{31,11}^{31,0}-\frac{1}{288} \sqrt{\frac{7}{2}} C_{31,11}^{31,1}+\frac{C_{33,11}^{33,0}}{640}+\frac{C_{33,11}^{33,1}}{320}\)

\noindent\(C_{33,0011,3}^{00,3303,1,\text{Loc}}= \frac{1}{192} \sqrt{\frac{7}{6}} C_{11,11}^{51,1}-\frac{1}{192} \sqrt{\frac{7}{6}}
C_{31,11}^{31,1}+\frac{\sqrt{3} C_{33,11}^{33,1}}{640}\)

\noindent\(C_{33,0011,3}^{00,3303,1,\text{NonLoc}}= \frac{1}{192} \sqrt{\frac{7}{6}} C_{11,11}^{51,0}-\frac{1}{192} \sqrt{\frac{7}{6}}
C_{31,11}^{31,0}+\frac{\sqrt{3} C_{33,11}^{33,0}}{640}\)

\noindent\(C_{33,0011,3}^{11,2212,0,\text{Loc}}= \frac{1}{288} \sqrt{\frac{7}{2}} C_{00,20}^{60,0}+\frac{1}{576} \sqrt{\frac{7}{2}}
C_{00,20}^{60,1}-\frac{1}{192} \sqrt{\frac{7}{10}} C_{00,22}^{62,0}-\frac{1}{384} \sqrt{\frac{7}{10}} C_{00,22}^{62,1}+\frac{1}{864} \sqrt{\frac{7}{10}}
C_{11,22}^{51,0}+\frac{\sqrt{\frac{7}{10}} C_{11,22}^{51,1}}{1728}+\frac{1}{640} \sqrt{\frac{3}{2}} C_{11,22}^{53,0}+\frac{\sqrt{\frac{3}{2}} C_{11,22}^{53,1}}{1280}-\frac{1}{864}
\sqrt{\frac{7}{2}} C_{20,20}^{40,0}-\frac{\sqrt{\frac{7}{2}} C_{20,20}^{40,1}}{1728}+\frac{1}{864} \sqrt{\frac{7}{10}} C_{20,22}^{42,0}+\frac{\sqrt{\frac{7}{10}}
C_{20,22}^{42,1}}{1728}+\frac{1}{864} \sqrt{\frac{7}{10}} C_{22,20}^{42,0}+\frac{\sqrt{\frac{7}{10}} C_{22,20}^{42,1}}{1728}+\frac{\sqrt{\frac{35}{2}}
C_{22,22}^{40,0}}{1728}+\frac{\sqrt{\frac{35}{2}} C_{22,22}^{40,1}}{3456}-\frac{7 C_{22,22}^{42,0}}{864 \sqrt{10}}-\frac{7 C_{22,22}^{42,1}}{1728
\sqrt{10}}+\frac{1}{864} \sqrt{\frac{7}{2}} C_{31,20}^{31,0}+\frac{\sqrt{\frac{7}{2}} C_{31,20}^{31,1}}{1728}-\frac{1}{864} \sqrt{\frac{7}{10}} C_{31,22}^{31,0}-\frac{\sqrt{\frac{7}{10}}
C_{31,22}^{31,1}}{1728}-\frac{1}{640} \sqrt{\frac{3}{2}} C_{31,22}^{33,0}-\frac{\sqrt{\frac{3}{2}} C_{31,22}^{33,1}}{1280}-\frac{C_{33,20}^{33,0}}{160
\sqrt{6}}-\frac{C_{33,20}^{33,1}}{320 \sqrt{6}}+\frac{C_{33,22}^{33,0}}{80 \sqrt{5}}+\frac{C_{33,22}^{33,1}}{160 \sqrt{5}}\)

\noindent\(C_{33,0011,3}^{11,2212,0,\text{NonLoc}}= \frac{1}{384} \sqrt{\frac{7}{6}} C_{00,00}^{60,0}+\frac{1}{192} \sqrt{\frac{7}{6}}
C_{00,00}^{60,1}+\frac{\sqrt{\frac{7}{2}} C_{00,20}^{60,0}}{1152}+\frac{1}{576} \sqrt{\frac{7}{2}} C_{00,20}^{60,1}+\frac{1}{384} \sqrt{\frac{7}{10}}
C_{00,22}^{62,0}+\frac{1}{192} \sqrt{\frac{7}{10}} C_{00,22}^{62,1}+\frac{\sqrt{\frac{7}{6}} C_{11,00}^{51,0}}{1152}+\frac{1}{576} \sqrt{\frac{7}{6}}
C_{11,00}^{51,1}+\frac{\sqrt{\frac{7}{10}} C_{11,22}^{51,0}}{1728}+\frac{1}{864} \sqrt{\frac{7}{10}} C_{11,22}^{51,1}+\frac{\sqrt{\frac{3}{2}} C_{11,22}^{53,0}}{1280}+\frac{1}{640}
\sqrt{\frac{3}{2}} C_{11,22}^{53,1}-\frac{\sqrt{\frac{7}{6}} C_{20,00}^{40,0}}{1152}-\frac{1}{576} \sqrt{\frac{7}{6}} C_{20,00}^{40,1}-\frac{\sqrt{\frac{7}{2}}
C_{20,20}^{40,0}}{3456}-\frac{\sqrt{\frac{7}{2}} C_{20,20}^{40,1}}{1728}-\frac{\sqrt{\frac{7}{10}} C_{20,22}^{42,0}}{1728}-\frac{1}{864} \sqrt{\frac{7}{10}}
C_{20,22}^{42,1}+\frac{\sqrt{\frac{7}{30}} C_{22,00}^{42,0}}{1152}+\frac{1}{576} \sqrt{\frac{7}{30}} C_{22,00}^{42,1}+\frac{\sqrt{\frac{7}{10}} C_{22,20}^{42,0}}{3456}+\frac{\sqrt{\frac{7}{10}}
C_{22,20}^{42,1}}{1728}-\frac{\sqrt{\frac{35}{2}} C_{22,22}^{40,0}}{3456}-\frac{\sqrt{\frac{35}{2}} C_{22,22}^{40,1}}{1728}+\frac{7 C_{22,22}^{42,0}}{1728
\sqrt{10}}+\frac{7 C_{22,22}^{42,1}}{864 \sqrt{10}}-\frac{\sqrt{\frac{7}{6}} C_{31,00}^{31,0}}{1152}-\frac{1}{576} \sqrt{\frac{7}{6}} C_{31,00}^{31,1}-\frac{\sqrt{\frac{7}{2}}
C_{31,20}^{31,0}}{3456}-\frac{\sqrt{\frac{7}{2}} C_{31,20}^{31,1}}{1728}-\frac{\sqrt{\frac{7}{10}} C_{31,22}^{31,0}}{1728}-\frac{1}{864} \sqrt{\frac{7}{10}}
C_{31,22}^{31,1}-\frac{\sqrt{\frac{3}{2}} C_{31,22}^{33,0}}{1280}-\frac{1}{640} \sqrt{\frac{3}{2}} C_{31,22}^{33,1}+\frac{C_{33,00}^{33,0}}{640 \sqrt{2}}+\frac{C_{33,00}^{33,1}}{320
\sqrt{2}}+\frac{C_{33,20}^{33,0}}{640 \sqrt{6}}+\frac{C_{33,20}^{33,1}}{320 \sqrt{6}}+\frac{C_{33,22}^{33,0}}{160 \sqrt{5}}+\frac{C_{33,22}^{33,1}}{80
\sqrt{5}}\)

\noindent\(C_{33,0011,3}^{11,2212,1,\text{Loc}}= \frac{1}{192} \sqrt{\frac{7}{6}} C_{00,20}^{60,1}-\frac{1}{128} \sqrt{\frac{7}{30}}
C_{00,22}^{62,1}+\frac{1}{576} \sqrt{\frac{7}{30}} C_{11,22}^{51,1}+\frac{3 C_{11,22}^{53,1}}{1280 \sqrt{2}}-\frac{1}{576} \sqrt{\frac{7}{6}} C_{20,20}^{40,1}+\frac{1}{576}
\sqrt{\frac{7}{30}} C_{20,22}^{42,1}+\frac{1}{576} \sqrt{\frac{7}{30}} C_{22,20}^{42,1}+\frac{\sqrt{\frac{35}{6}} C_{22,22}^{40,1}}{1152}-\frac{7
C_{22,22}^{42,1}}{576 \sqrt{30}}+\frac{1}{576} \sqrt{\frac{7}{6}} C_{31,20}^{31,1}-\frac{1}{576} \sqrt{\frac{7}{30}} C_{31,22}^{31,1}-\frac{3 C_{31,22}^{33,1}}{1280
\sqrt{2}}-\frac{C_{33,20}^{33,1}}{320 \sqrt{2}}+\frac{1}{160} \sqrt{\frac{3}{5}} C_{33,22}^{33,1}\)

\noindent\(C_{33,0011,3}^{11,2212,1,\text{NonLoc}}= \frac{1}{384} \sqrt{\frac{7}{2}} C_{00,00}^{60,0}+\frac{1}{384} \sqrt{\frac{7}{6}}
C_{00,20}^{60,0}+\frac{1}{128} \sqrt{\frac{7}{30}} C_{00,22}^{62,0}+\frac{\sqrt{\frac{7}{2}} C_{11,00}^{51,0}}{1152}+\frac{1}{576} \sqrt{\frac{7}{30}}
C_{11,22}^{51,0}+\frac{3 C_{11,22}^{53,0}}{1280 \sqrt{2}}-\frac{\sqrt{\frac{7}{2}} C_{20,00}^{40,0}}{1152}-\frac{\sqrt{\frac{7}{6}} C_{20,20}^{40,0}}{1152}-\frac{1}{576}
\sqrt{\frac{7}{30}} C_{20,22}^{42,0}+\frac{\sqrt{\frac{7}{10}} C_{22,00}^{42,0}}{1152}+\frac{\sqrt{\frac{7}{30}} C_{22,20}^{42,0}}{1152}-\frac{\sqrt{\frac{35}{6}}
C_{22,22}^{40,0}}{1152}+\frac{7 C_{22,22}^{42,0}}{576 \sqrt{30}}-\frac{\sqrt{\frac{7}{2}} C_{31,00}^{31,0}}{1152}-\frac{\sqrt{\frac{7}{6}} C_{31,20}^{31,0}}{1152}-\frac{1}{576}
\sqrt{\frac{7}{30}} C_{31,22}^{31,0}-\frac{3 C_{31,22}^{33,0}}{1280 \sqrt{2}}+\frac{1}{640} \sqrt{\frac{3}{2}} C_{33,00}^{33,0}+\frac{C_{33,20}^{33,0}}{640
\sqrt{2}}+\frac{1}{160} \sqrt{\frac{3}{5}} C_{33,22}^{33,0}\)

\noindent\(C_{33,0011,3}^{11,2213,0,\text{Loc}}= \frac{\sqrt{7} C_{00,20}^{60,0}}{1152}+\frac{\sqrt{7} C_{00,20}^{60,1}}{2304}-\frac{1}{768}
\sqrt{\frac{7}{5}} C_{00,22}^{62,0}-\frac{\sqrt{\frac{7}{5}} C_{00,22}^{62,1}}{1536}+\frac{11 \sqrt{\frac{7}{5}} C_{11,22}^{51,0}}{6912}+\frac{11
\sqrt{\frac{7}{5}} C_{11,22}^{51,1}}{13824}-\frac{\sqrt{3} C_{11,22}^{53,0}}{1280}-\frac{\sqrt{3} C_{11,22}^{53,1}}{2560}-\frac{\sqrt{7} C_{20,20}^{40,0}}{3456}-\frac{\sqrt{7}
C_{20,20}^{40,1}}{6912}+\frac{11 \sqrt{\frac{7}{5}} C_{20,22}^{42,0}}{6912}+\frac{11 \sqrt{\frac{7}{5}} C_{20,22}^{42,1}}{13824}+\frac{\sqrt{\frac{7}{5}}
C_{22,20}^{42,0}}{3456}+\frac{\sqrt{\frac{7}{5}} C_{22,20}^{42,1}}{6912}-\frac{13 \sqrt{\frac{7}{5}} C_{22,22}^{40,0}}{6912}-\frac{13 \sqrt{\frac{7}{5}}
C_{22,22}^{40,1}}{13824}+\frac{C_{22,22}^{42,0}}{1728 \sqrt{5}}+\frac{C_{22,22}^{42,1}}{3456 \sqrt{5}}-\frac{3 C_{22,22}^{44,0}}{1280 \sqrt{7}}-\frac{3
C_{22,22}^{44,1}}{2560 \sqrt{7}}+\frac{\sqrt{7} C_{31,20}^{31,0}}{3456}+\frac{\sqrt{7} C_{31,20}^{31,1}}{6912}-\frac{11 \sqrt{\frac{7}{5}} C_{31,22}^{31,0}}{6912}-\frac{11
\sqrt{\frac{7}{5}} C_{31,22}^{31,1}}{13824}+\frac{\sqrt{3} C_{31,22}^{33,0}}{1280}+\frac{\sqrt{3} C_{31,22}^{33,1}}{2560}-\frac{C_{33,20}^{33,0}}{640
\sqrt{3}}-\frac{C_{33,20}^{33,1}}{1280 \sqrt{3}}-\frac{C_{33,22}^{33,0}}{1280 \sqrt{10}}-\frac{C_{33,22}^{33,1}}{2560 \sqrt{10}}\)

\noindent\(C_{33,0011,3}^{11,2213,0,\text{NonLoc}}= \frac{\sqrt{\frac{7}{3}} C_{00,00}^{60,0}}{1536}+\frac{1}{768} \sqrt{\frac{7}{3}}
C_{00,00}^{60,1}+\frac{\sqrt{7} C_{00,20}^{60,0}}{4608}+\frac{\sqrt{7} C_{00,20}^{60,1}}{2304}+\frac{\sqrt{\frac{7}{5}} C_{00,22}^{62,0}}{1536}+\frac{1}{768}
\sqrt{\frac{7}{5}} C_{00,22}^{62,1}+\frac{\sqrt{\frac{7}{3}} C_{11,00}^{51,0}}{4608}+\frac{\sqrt{\frac{7}{3}} C_{11,00}^{51,1}}{2304}+\frac{11 \sqrt{\frac{7}{5}}
C_{11,22}^{51,0}}{13824}+\frac{11 \sqrt{\frac{7}{5}} C_{11,22}^{51,1}}{6912}-\frac{\sqrt{3} C_{11,22}^{53,0}}{2560}-\frac{\sqrt{3} C_{11,22}^{53,1}}{1280}-\frac{\sqrt{\frac{7}{3}}
C_{20,00}^{40,0}}{4608}-\frac{\sqrt{\frac{7}{3}} C_{20,00}^{40,1}}{2304}-\frac{\sqrt{7} C_{20,20}^{40,0}}{13824}-\frac{\sqrt{7} C_{20,20}^{40,1}}{6912}-\frac{11
\sqrt{\frac{7}{5}} C_{20,22}^{42,0}}{13824}-\frac{11 \sqrt{\frac{7}{5}} C_{20,22}^{42,1}}{6912}+\frac{\sqrt{\frac{7}{15}} C_{22,00}^{42,0}}{4608}+\frac{\sqrt{\frac{7}{15}}
C_{22,00}^{42,1}}{2304}+\frac{\sqrt{\frac{7}{5}} C_{22,20}^{42,0}}{13824}+\frac{\sqrt{\frac{7}{5}} C_{22,20}^{42,1}}{6912}+\frac{13 \sqrt{\frac{7}{5}}
C_{22,22}^{40,0}}{13824}+\frac{13 \sqrt{\frac{7}{5}} C_{22,22}^{40,1}}{6912}-\frac{C_{22,22}^{42,0}}{3456 \sqrt{5}}-\frac{C_{22,22}^{42,1}}{1728
\sqrt{5}}+\frac{3 C_{22,22}^{44,0}}{2560 \sqrt{7}}+\frac{3 C_{22,22}^{44,1}}{1280 \sqrt{7}}-\frac{\sqrt{\frac{7}{3}} C_{31,00}^{31,0}}{4608}-\frac{\sqrt{\frac{7}{3}}
C_{31,00}^{31,1}}{2304}-\frac{\sqrt{7} C_{31,20}^{31,0}}{13824}-\frac{\sqrt{7} C_{31,20}^{31,1}}{6912}-\frac{11 \sqrt{\frac{7}{5}} C_{31,22}^{31,0}}{13824}-\frac{11
\sqrt{\frac{7}{5}} C_{31,22}^{31,1}}{6912}+\frac{\sqrt{3} C_{31,22}^{33,0}}{2560}+\frac{\sqrt{3} C_{31,22}^{33,1}}{1280}+\frac{C_{33,00}^{33,0}}{2560}+\frac{C_{33,00}^{33,1}}{1280}+\frac{C_{33,20}^{33,0}}{2560
\sqrt{3}}+\frac{C_{33,20}^{33,1}}{1280 \sqrt{3}}-\frac{C_{33,22}^{33,0}}{2560 \sqrt{10}}-\frac{C_{33,22}^{33,1}}{1280 \sqrt{10}}\)

\noindent\(C_{33,0011,3}^{11,2213,1,\text{Loc}}= \frac{1}{768} \sqrt{\frac{7}{3}} C_{00,20}^{60,1}-\frac{1}{512} \sqrt{\frac{7}{15}}
C_{00,22}^{62,1}+\frac{11 \sqrt{\frac{7}{15}} C_{11,22}^{51,1}}{4608}-\frac{3 C_{11,22}^{53,1}}{2560}-\frac{\sqrt{\frac{7}{3}} C_{20,20}^{40,1}}{2304}+\frac{11
\sqrt{\frac{7}{15}} C_{20,22}^{42,1}}{4608}+\frac{\sqrt{\frac{7}{15}} C_{22,20}^{42,1}}{2304}-\frac{13 \sqrt{\frac{7}{15}} C_{22,22}^{40,1}}{4608}+\frac{C_{22,22}^{42,1}}{1152
\sqrt{15}}-\frac{3 \sqrt{\frac{3}{7}} C_{22,22}^{44,1}}{2560}+\frac{\sqrt{\frac{7}{3}} C_{31,20}^{31,1}}{2304}-\frac{11 \sqrt{\frac{7}{15}} C_{31,22}^{31,1}}{4608}+\frac{3
C_{31,22}^{33,1}}{2560}-\frac{C_{33,20}^{33,1}}{1280}-\frac{\sqrt{\frac{3}{10}} C_{33,22}^{33,1}}{2560}\)

\noindent\(C_{33,0011,3}^{11,2213,1,\text{NonLoc}}= \frac{\sqrt{7} C_{00,00}^{60,0}}{1536}+\frac{\sqrt{\frac{7}{3}} C_{00,20}^{60,0}}{1536}+\frac{1}{512}
\sqrt{\frac{7}{15}} C_{00,22}^{62,0}+\frac{\sqrt{7} C_{11,00}^{51,0}}{4608}+\frac{11 \sqrt{\frac{7}{15}} C_{11,22}^{51,0}}{4608}-\frac{3 C_{11,22}^{53,0}}{2560}-\frac{\sqrt{7}
C_{20,00}^{40,0}}{4608}-\frac{\sqrt{\frac{7}{3}} C_{20,20}^{40,0}}{4608}-\frac{11 \sqrt{\frac{7}{15}} C_{20,22}^{42,0}}{4608}+\frac{\sqrt{\frac{7}{5}}
C_{22,00}^{42,0}}{4608}+\frac{\sqrt{\frac{7}{15}} C_{22,20}^{42,0}}{4608}+\frac{13 \sqrt{\frac{7}{15}} C_{22,22}^{40,0}}{4608}-\frac{C_{22,22}^{42,0}}{1152
\sqrt{15}}+\frac{3 \sqrt{\frac{3}{7}} C_{22,22}^{44,0}}{2560}-\frac{\sqrt{7} C_{31,00}^{31,0}}{4608}-\frac{\sqrt{\frac{7}{3}} C_{31,20}^{31,0}}{4608}-\frac{11
\sqrt{\frac{7}{15}} C_{31,22}^{31,0}}{4608}+\frac{3 C_{31,22}^{33,0}}{2560}+\frac{\sqrt{3} C_{33,00}^{33,0}}{2560}+\frac{C_{33,20}^{33,0}}{2560}-\frac{\sqrt{\frac{3}{10}}
C_{33,22}^{33,0}}{2560}\)

\noindent\(C_{33,0011,3}^{22,1101,0,\text{Loc}}= -\frac{\sqrt{\frac{7}{2}} C_{11,11}^{51,0}}{1152}-\frac{\sqrt{\frac{7}{2}} C_{11,11}^{51,1}}{2304}+\frac{1}{192}
\sqrt{\frac{7}{30}} C_{22,11}^{42,0}+\frac{1}{384} \sqrt{\frac{7}{30}} C_{22,11}^{42,1}-\frac{\sqrt{\frac{7}{2}} C_{31,11}^{31,0}}{1152}-\frac{\sqrt{\frac{7}{2}}
C_{31,11}^{31,1}}{2304}-\frac{C_{33,11}^{33,0}}{320}-\frac{C_{33,11}^{33,1}}{640}\)

\noindent\(C_{33,0011,3}^{22,1101,0,\text{NonLoc}}= -\frac{\sqrt{\frac{7}{2}} C_{11,11}^{51,0}}{2304}-\frac{\sqrt{\frac{7}{2}} C_{11,11}^{51,1}}{1152}-\frac{1}{384}
\sqrt{\frac{7}{30}} C_{22,11}^{42,0}-\frac{1}{192} \sqrt{\frac{7}{30}} C_{22,11}^{42,1}-\frac{\sqrt{\frac{7}{2}} C_{31,11}^{31,0}}{2304}-\frac{\sqrt{\frac{7}{2}}
C_{31,11}^{31,1}}{1152}-\frac{C_{33,11}^{33,0}}{640}-\frac{C_{33,11}^{33,1}}{320}\)

\noindent\(C_{33,0011,3}^{22,1101,1,\text{Loc}}= -\frac{1}{768} \sqrt{\frac{7}{6}} C_{11,11}^{51,1}+\frac{1}{384} \sqrt{\frac{7}{10}}
C_{22,11}^{42,1}-\frac{1}{768} \sqrt{\frac{7}{6}} C_{31,11}^{31,1}-\frac{\sqrt{3} C_{33,11}^{33,1}}{640}\)

\noindent\(C_{33,0011,3}^{22,1101,1,\text{NonLoc}}= -\frac{1}{768} \sqrt{\frac{7}{6}} C_{11,11}^{51,0}-\frac{1}{384} \sqrt{\frac{7}{10}}
C_{22,11}^{42,0}-\frac{1}{768} \sqrt{\frac{7}{6}} C_{31,11}^{31,0}-\frac{\sqrt{3} C_{33,11}^{33,0}}{640}\)

\noindent\(C_{33,0011,3}^{33,0011,0,\text{Loc}}= -\frac{\sqrt{7} C_{00,20}^{60,0}}{9216}-\frac{\sqrt{7} C_{00,20}^{60,1}}{18432}+\frac{\sqrt{\frac{7}{5}}
C_{00,22}^{62,0}}{6144}+\frac{\sqrt{\frac{7}{5}} C_{00,22}^{62,1}}{12288}-\frac{\sqrt{\frac{7}{5}} C_{11,22}^{51,0}}{6144}-\frac{\sqrt{\frac{7}{5}}
C_{11,22}^{51,1}}{12288}-\frac{\sqrt{3} C_{11,22}^{53,0}}{10240}-\frac{\sqrt{3} C_{11,22}^{53,1}}{20480}-\frac{\sqrt{7} C_{20,20}^{40,0}}{9216}-\frac{\sqrt{7}
C_{20,20}^{40,1}}{18432}+\frac{\sqrt{\frac{7}{5}} C_{20,22}^{42,0}}{6144}+\frac{\sqrt{\frac{7}{5}} C_{20,22}^{42,1}}{12288}-\frac{\sqrt{\frac{7}{5}}
C_{22,20}^{42,0}}{4608}-\frac{\sqrt{\frac{7}{5}} C_{22,20}^{42,1}}{9216}+\frac{\sqrt{\frac{7}{5}} C_{22,22}^{40,0}}{6144}+\frac{\sqrt{\frac{7}{5}}
C_{22,22}^{40,1}}{12288}+\frac{C_{22,22}^{42,0}}{3072 \sqrt{5}}+\frac{C_{22,22}^{42,1}}{6144 \sqrt{5}}+\frac{C_{22,22}^{44,0}}{2560 \sqrt{7}}+\frac{C_{22,22}^{44,1}}{5120
\sqrt{7}}+\frac{\sqrt{7} C_{31,20}^{31,0}}{9216}+\frac{\sqrt{7} C_{31,20}^{31,1}}{18432}-\frac{\sqrt{\frac{7}{5}} C_{31,22}^{31,0}}{6144}-\frac{\sqrt{\frac{7}{5}}
C_{31,22}^{31,1}}{12288}-\frac{\sqrt{3} C_{31,22}^{33,0}}{10240}-\frac{\sqrt{3} C_{31,22}^{33,1}}{20480}+\frac{C_{33,20}^{33,0}}{2560 \sqrt{3}}+\frac{C_{33,20}^{33,1}}{5120
\sqrt{3}}-\frac{C_{33,22}^{33,0}}{2560 \sqrt{10}}-\frac{C_{33,22}^{33,1}}{5120 \sqrt{10}}\)

\noindent\(C_{33,0011,3}^{33,0011,0,\text{NonLoc}}= -\frac{\sqrt{\frac{7}{3}} C_{00,00}^{60,0}}{12288}-\frac{\sqrt{\frac{7}{3}} C_{00,00}^{60,1}}{6144}-\frac{\sqrt{7}
C_{00,20}^{60,0}}{36864}-\frac{\sqrt{7} C_{00,20}^{60,1}}{18432}-\frac{\sqrt{\frac{7}{5}} C_{00,22}^{62,0}}{12288}-\frac{\sqrt{\frac{7}{5}} C_{00,22}^{62,1}}{6144}-\frac{\sqrt{\frac{7}{3}}
C_{11,00}^{51,0}}{12288}-\frac{\sqrt{\frac{7}{3}} C_{11,00}^{51,1}}{6144}-\frac{\sqrt{\frac{7}{5}} C_{11,22}^{51,0}}{12288}-\frac{\sqrt{\frac{7}{5}}
C_{11,22}^{51,1}}{6144}-\frac{\sqrt{3} C_{11,22}^{53,0}}{20480}-\frac{\sqrt{3} C_{11,22}^{53,1}}{10240}-\frac{\sqrt{\frac{7}{3}} C_{20,00}^{40,0}}{12288}-\frac{\sqrt{\frac{7}{3}}
C_{20,00}^{40,1}}{6144}-\frac{\sqrt{7} C_{20,20}^{40,0}}{36864}-\frac{\sqrt{7} C_{20,20}^{40,1}}{18432}-\frac{\sqrt{\frac{7}{5}} C_{20,22}^{42,0}}{12288}-\frac{\sqrt{\frac{7}{5}}
C_{20,22}^{42,1}}{6144}-\frac{\sqrt{\frac{7}{15}} C_{22,00}^{42,0}}{6144}-\frac{\sqrt{\frac{7}{15}} C_{22,00}^{42,1}}{3072}-\frac{\sqrt{\frac{7}{5}}
C_{22,20}^{42,0}}{18432}-\frac{\sqrt{\frac{7}{5}} C_{22,20}^{42,1}}{9216}-\frac{\sqrt{\frac{7}{5}} C_{22,22}^{40,0}}{12288}-\frac{\sqrt{\frac{7}{5}}
C_{22,22}^{40,1}}{6144}-\frac{C_{22,22}^{42,0}}{6144 \sqrt{5}}-\frac{C_{22,22}^{42,1}}{3072 \sqrt{5}}-\frac{C_{22,22}^{44,0}}{5120 \sqrt{7}}-\frac{C_{22,22}^{44,1}}{2560
\sqrt{7}}-\frac{\sqrt{\frac{7}{3}} C_{31,00}^{31,0}}{12288}-\frac{\sqrt{\frac{7}{3}} C_{31,00}^{31,1}}{6144}-\frac{\sqrt{7} C_{31,20}^{31,0}}{36864}-\frac{\sqrt{7}
C_{31,20}^{31,1}}{18432}-\frac{\sqrt{\frac{7}{5}} C_{31,22}^{31,0}}{12288}-\frac{\sqrt{\frac{7}{5}} C_{31,22}^{31,1}}{6144}-\frac{\sqrt{3} C_{31,22}^{33,0}}{20480}-\frac{\sqrt{3}
C_{31,22}^{33,1}}{10240}-\frac{C_{33,00}^{33,0}}{10240}-\frac{C_{33,00}^{33,1}}{5120}-\frac{C_{33,20}^{33,0}}{10240 \sqrt{3}}-\frac{C_{33,20}^{33,1}}{5120
\sqrt{3}}-\frac{C_{33,22}^{33,0}}{5120 \sqrt{10}}-\frac{C_{33,22}^{33,1}}{2560 \sqrt{10}}\)

\noindent\(C_{33,0011,3}^{33,0011,1,\text{Loc}}= -\frac{\sqrt{\frac{7}{3}} C_{00,20}^{60,1}}{6144}+\frac{\sqrt{\frac{7}{15}} C_{00,22}^{62,1}}{4096}-\frac{\sqrt{\frac{7}{15}}
C_{11,22}^{51,1}}{4096}-\frac{3 C_{11,22}^{53,1}}{20480}-\frac{\sqrt{\frac{7}{3}} C_{20,20}^{40,1}}{6144}+\frac{\sqrt{\frac{7}{15}} C_{20,22}^{42,1}}{4096}-\frac{\sqrt{\frac{7}{15}}
C_{22,20}^{42,1}}{3072}+\frac{\sqrt{\frac{7}{15}} C_{22,22}^{40,1}}{4096}+\frac{C_{22,22}^{42,1}}{2048 \sqrt{15}}+\frac{\sqrt{\frac{3}{7}} C_{22,22}^{44,1}}{5120}+\frac{\sqrt{\frac{7}{3}}
C_{31,20}^{31,1}}{6144}-\frac{\sqrt{\frac{7}{15}} C_{31,22}^{31,1}}{4096}-\frac{3 C_{31,22}^{33,1}}{20480}+\frac{C_{33,20}^{33,1}}{5120}-\frac{\sqrt{\frac{3}{10}}
C_{33,22}^{33,1}}{5120}\)

\noindent\(C_{33,0011,3}^{33,0011,1,\text{NonLoc}}= -\frac{\sqrt{7} C_{00,00}^{60,0}}{12288}-\frac{\sqrt{\frac{7}{3}} C_{00,20}^{60,0}}{12288}-\frac{\sqrt{\frac{7}{15}}
C_{00,22}^{62,0}}{4096}-\frac{\sqrt{7} C_{11,00}^{51,0}}{12288}-\frac{\sqrt{\frac{7}{15}} C_{11,22}^{51,0}}{4096}-\frac{3 C_{11,22}^{53,0}}{20480}-\frac{\sqrt{7}
C_{20,00}^{40,0}}{12288}-\frac{\sqrt{\frac{7}{3}} C_{20,20}^{40,0}}{12288}-\frac{\sqrt{\frac{7}{15}} C_{20,22}^{42,0}}{4096}-\frac{\sqrt{\frac{7}{5}}
C_{22,00}^{42,0}}{6144}-\frac{\sqrt{\frac{7}{15}} C_{22,20}^{42,0}}{6144}-\frac{\sqrt{\frac{7}{15}} C_{22,22}^{40,0}}{4096}-\frac{C_{22,22}^{42,0}}{2048
\sqrt{15}}-\frac{\sqrt{\frac{3}{7}} C_{22,22}^{44,0}}{5120}-\frac{\sqrt{7} C_{31,00}^{31,0}}{12288}-\frac{\sqrt{\frac{7}{3}} C_{31,20}^{31,0}}{12288}-\frac{\sqrt{\frac{7}{15}}
C_{31,22}^{31,0}}{4096}-\frac{3 C_{31,22}^{33,0}}{20480}-\frac{\sqrt{3} C_{33,00}^{33,0}}{10240}-\frac{C_{33,20}^{33,0}}{10240}-\frac{\sqrt{\frac{3}{10}}
C_{33,22}^{33,0}}{5120}\)

\noindent\(C_{33,0011,4}^{11,2213,0,\text{Loc}}= \frac{C_{00,20}^{60,0}}{128}+\frac{C_{00,20}^{60,1}}{256}+\frac{C_{00,22}^{62,0}}{256
\sqrt{5}}+\frac{C_{00,22}^{62,1}}{512 \sqrt{5}}-\frac{C_{11,22}^{51,0}}{768 \sqrt{5}}-\frac{C_{11,22}^{51,1}}{1536 \sqrt{5}}-\frac{\sqrt{\frac{3}{7}}
C_{11,22}^{53,0}}{1280}-\frac{\sqrt{\frac{3}{7}} C_{11,22}^{53,1}}{2560}-\frac{C_{20,20}^{40,0}}{384}-\frac{C_{20,20}^{40,1}}{768}-\frac{C_{20,22}^{42,0}}{768
\sqrt{5}}-\frac{C_{20,22}^{42,1}}{1536 \sqrt{5}}+\frac{C_{22,20}^{42,0}}{384 \sqrt{5}}+\frac{C_{22,20}^{42,1}}{768 \sqrt{5}}-\frac{C_{22,22}^{40,0}}{768
\sqrt{5}}-\frac{C_{22,22}^{40,1}}{1536 \sqrt{5}}+\frac{C_{22,22}^{42,0}}{192 \sqrt{35}}+\frac{C_{22,22}^{42,1}}{384 \sqrt{35}}+\frac{C_{22,22}^{44,0}}{8960}+\frac{C_{22,22}^{44,1}}{17920}+\frac{C_{31,20}^{31,0}}{384}+\frac{C_{31,20}^{31,1}}{768}+\frac{C_{31,22}^{31,0}}{768
\sqrt{5}}+\frac{C_{31,22}^{31,1}}{1536 \sqrt{5}}+\frac{\sqrt{\frac{3}{7}} C_{31,22}^{33,0}}{1280}+\frac{\sqrt{\frac{3}{7}} C_{31,22}^{33,1}}{2560}-\frac{3}{640}
\sqrt{\frac{3}{7}} C_{33,20}^{33,0}-\frac{3 \sqrt{\frac{3}{7}} C_{33,20}^{33,1}}{1280}-\frac{3 \sqrt{\frac{7}{10}} C_{33,22}^{33,0}}{1280}-\frac{3
\sqrt{\frac{7}{10}} C_{33,22}^{33,1}}{2560}\)

\noindent\(C_{33,0011,4}^{11,2213,0,\text{NonLoc}}= \frac{\sqrt{3} C_{00,00}^{60,0}}{512}+\frac{\sqrt{3} C_{00,00}^{60,1}}{256}+\frac{C_{00,20}^{60,0}}{512}+\frac{C_{00,20}^{60,1}}{256}-\frac{C_{00,22}^{62,0}}{512
\sqrt{5}}-\frac{C_{00,22}^{62,1}}{256 \sqrt{5}}+\frac{C_{11,00}^{51,0}}{512 \sqrt{3}}+\frac{C_{11,00}^{51,1}}{256 \sqrt{3}}-\frac{C_{11,22}^{51,0}}{1536
\sqrt{5}}-\frac{C_{11,22}^{51,1}}{768 \sqrt{5}}-\frac{\sqrt{\frac{3}{7}} C_{11,22}^{53,0}}{2560}-\frac{\sqrt{\frac{3}{7}} C_{11,22}^{53,1}}{1280}-\frac{C_{20,00}^{40,0}}{512
\sqrt{3}}-\frac{C_{20,00}^{40,1}}{256 \sqrt{3}}-\frac{C_{20,20}^{40,0}}{1536}-\frac{C_{20,20}^{40,1}}{768}+\frac{C_{20,22}^{42,0}}{1536 \sqrt{5}}+\frac{C_{20,22}^{42,1}}{768
\sqrt{5}}+\frac{C_{22,00}^{42,0}}{512 \sqrt{15}}+\frac{C_{22,00}^{42,1}}{256 \sqrt{15}}+\frac{C_{22,20}^{42,0}}{1536 \sqrt{5}}+\frac{C_{22,20}^{42,1}}{768
\sqrt{5}}+\frac{C_{22,22}^{40,0}}{1536 \sqrt{5}}+\frac{C_{22,22}^{40,1}}{768 \sqrt{5}}-\frac{C_{22,22}^{42,0}}{384 \sqrt{35}}-\frac{C_{22,22}^{42,1}}{192
\sqrt{35}}-\frac{C_{22,22}^{44,0}}{17920}-\frac{C_{22,22}^{44,1}}{8960}-\frac{C_{31,00}^{31,0}}{512 \sqrt{3}}-\frac{C_{31,00}^{31,1}}{256 \sqrt{3}}-\frac{C_{31,20}^{31,0}}{1536}-\frac{C_{31,20}^{31,1}}{768}+\frac{C_{31,22}^{31,0}}{1536
\sqrt{5}}+\frac{C_{31,22}^{31,1}}{768 \sqrt{5}}+\frac{\sqrt{\frac{3}{7}} C_{31,22}^{33,0}}{2560}+\frac{\sqrt{\frac{3}{7}} C_{31,22}^{33,1}}{1280}+\frac{9
C_{33,00}^{33,0}}{2560 \sqrt{7}}+\frac{9 C_{33,00}^{33,1}}{1280 \sqrt{7}}+\frac{3 \sqrt{\frac{3}{7}} C_{33,20}^{33,0}}{2560}+\frac{3 \sqrt{\frac{3}{7}}
C_{33,20}^{33,1}}{1280}-\frac{3 \sqrt{\frac{7}{10}} C_{33,22}^{33,0}}{2560}-\frac{3 \sqrt{\frac{7}{10}} C_{33,22}^{33,1}}{1280}\)

\noindent\(C_{33,0011,4}^{11,2213,1,\text{Loc}}= \frac{\sqrt{3} C_{00,20}^{60,1}}{256}+\frac{1}{512} \sqrt{\frac{3}{5}} C_{00,22}^{62,1}-\frac{C_{11,22}^{51,1}}{512
\sqrt{15}}-\frac{3 C_{11,22}^{53,1}}{2560 \sqrt{7}}-\frac{C_{20,20}^{40,1}}{256 \sqrt{3}}-\frac{C_{20,22}^{42,1}}{512 \sqrt{15}}+\frac{C_{22,20}^{42,1}}{256
\sqrt{15}}-\frac{C_{22,22}^{40,1}}{512 \sqrt{15}}+\frac{C_{22,22}^{42,1}}{128 \sqrt{105}}+\frac{\sqrt{3} C_{22,22}^{44,1}}{17920}+\frac{C_{31,20}^{31,1}}{256
\sqrt{3}}+\frac{C_{31,22}^{31,1}}{512 \sqrt{15}}+\frac{3 C_{31,22}^{33,1}}{2560 \sqrt{7}}-\frac{9 C_{33,20}^{33,1}}{1280 \sqrt{7}}-\frac{3 \sqrt{\frac{21}{10}}
C_{33,22}^{33,1}}{2560}\)

\noindent\(C_{33,0011,4}^{11,2213,1,\text{NonLoc}}= \frac{3 C_{00,00}^{60,0}}{512}+\frac{\sqrt{3} C_{00,20}^{60,0}}{512}-\frac{1}{512}
\sqrt{\frac{3}{5}} C_{00,22}^{62,0}+\frac{C_{11,00}^{51,0}}{512}-\frac{C_{11,22}^{51,0}}{512 \sqrt{15}}-\frac{3 C_{11,22}^{53,0}}{2560 \sqrt{7}}-\frac{C_{20,00}^{40,0}}{512}-\frac{C_{20,20}^{40,0}}{512
\sqrt{3}}+\frac{C_{20,22}^{42,0}}{512 \sqrt{15}}+\frac{C_{22,00}^{42,0}}{512 \sqrt{5}}+\frac{C_{22,20}^{42,0}}{512 \sqrt{15}}+\frac{C_{22,22}^{40,0}}{512
\sqrt{15}}-\frac{C_{22,22}^{42,0}}{128 \sqrt{105}}-\frac{\sqrt{3} C_{22,22}^{44,0}}{17920}-\frac{C_{31,00}^{31,0}}{512}-\frac{C_{31,20}^{31,0}}{512
\sqrt{3}}+\frac{C_{31,22}^{31,0}}{512 \sqrt{15}}+\frac{3 C_{31,22}^{33,0}}{2560 \sqrt{7}}+\frac{9 \sqrt{\frac{3}{7}} C_{33,00}^{33,0}}{2560}+\frac{9
C_{33,20}^{33,0}}{2560 \sqrt{7}}-\frac{3 \sqrt{\frac{21}{10}} C_{33,22}^{33,0}}{2560}\)

\noindent\(C_{33,0011,4}^{33,0011,0,\text{Loc}}= -\frac{C_{00,20}^{60,0}}{3072}-\frac{C_{00,20}^{60,1}}{6144}-\frac{C_{00,22}^{62,0}}{6144
\sqrt{5}}-\frac{C_{00,22}^{62,1}}{12288 \sqrt{5}}+\frac{C_{11,22}^{51,0}}{6144 \sqrt{5}}+\frac{C_{11,22}^{51,1}}{12288 \sqrt{5}}+\frac{\sqrt{\frac{3}{7}}
C_{11,22}^{53,0}}{10240}+\frac{\sqrt{\frac{3}{7}} C_{11,22}^{53,1}}{20480}-\frac{C_{20,20}^{40,0}}{3072}-\frac{C_{20,20}^{40,1}}{6144}-\frac{C_{20,22}^{42,0}}{6144
\sqrt{5}}-\frac{C_{20,22}^{42,1}}{12288 \sqrt{5}}-\frac{C_{22,20}^{42,0}}{1536 \sqrt{5}}-\frac{C_{22,20}^{42,1}}{3072 \sqrt{5}}-\frac{C_{22,22}^{40,0}}{6144
\sqrt{5}}-\frac{C_{22,22}^{40,1}}{12288 \sqrt{5}}-\frac{C_{22,22}^{42,0}}{3072 \sqrt{35}}-\frac{C_{22,22}^{42,1}}{6144 \sqrt{35}}-\frac{C_{22,22}^{44,0}}{17920}-\frac{C_{22,22}^{44,1}}{35840}+\frac{C_{31,20}^{31,0}}{3072}+\frac{C_{31,20}^{31,1}}{6144}+\frac{C_{31,22}^{31,0}}{6144
\sqrt{5}}+\frac{C_{31,22}^{31,1}}{12288 \sqrt{5}}+\frac{\sqrt{\frac{3}{7}} C_{31,22}^{33,0}}{10240}+\frac{\sqrt{\frac{3}{7}} C_{31,22}^{33,1}}{20480}+\frac{\sqrt{\frac{3}{7}}
C_{33,20}^{33,0}}{2560}+\frac{\sqrt{\frac{3}{7}} C_{33,20}^{33,1}}{5120}+\frac{C_{33,22}^{33,0}}{2560 \sqrt{70}}+\frac{C_{33,22}^{33,1}}{5120 \sqrt{70}}\)

\noindent\(C_{33,0011,4}^{33,0011,0,\text{NonLoc}}= -\frac{C_{00,00}^{60,0}}{4096 \sqrt{3}}-\frac{C_{00,00}^{60,1}}{2048 \sqrt{3}}-\frac{C_{00,20}^{60,0}}{12288}-\frac{C_{00,20}^{60,1}}{6144}+\frac{C_{00,22}^{62,0}}{12288
\sqrt{5}}+\frac{C_{00,22}^{62,1}}{6144 \sqrt{5}}-\frac{C_{11,00}^{51,0}}{4096 \sqrt{3}}-\frac{C_{11,00}^{51,1}}{2048 \sqrt{3}}+\frac{C_{11,22}^{51,0}}{12288
\sqrt{5}}+\frac{C_{11,22}^{51,1}}{6144 \sqrt{5}}+\frac{\sqrt{\frac{3}{7}} C_{11,22}^{53,0}}{20480}+\frac{\sqrt{\frac{3}{7}} C_{11,22}^{53,1}}{10240}-\frac{C_{20,00}^{40,0}}{4096
\sqrt{3}}-\frac{C_{20,00}^{40,1}}{2048 \sqrt{3}}-\frac{C_{20,20}^{40,0}}{12288}-\frac{C_{20,20}^{40,1}}{6144}+\frac{C_{20,22}^{42,0}}{12288 \sqrt{5}}+\frac{C_{20,22}^{42,1}}{6144
\sqrt{5}}-\frac{C_{22,00}^{42,0}}{2048 \sqrt{15}}-\frac{C_{22,00}^{42,1}}{1024 \sqrt{15}}-\frac{C_{22,20}^{42,0}}{6144 \sqrt{5}}-\frac{C_{22,20}^{42,1}}{3072
\sqrt{5}}+\frac{C_{22,22}^{40,0}}{12288 \sqrt{5}}+\frac{C_{22,22}^{40,1}}{6144 \sqrt{5}}+\frac{C_{22,22}^{42,0}}{6144 \sqrt{35}}+\frac{C_{22,22}^{42,1}}{3072
\sqrt{35}}+\frac{C_{22,22}^{44,0}}{35840}+\frac{C_{22,22}^{44,1}}{17920}-\frac{C_{31,00}^{31,0}}{4096 \sqrt{3}}-\frac{C_{31,00}^{31,1}}{2048 \sqrt{3}}-\frac{C_{31,20}^{31,0}}{12288}-\frac{C_{31,20}^{31,1}}{6144}+\frac{C_{31,22}^{31,0}}{12288
\sqrt{5}}+\frac{C_{31,22}^{31,1}}{6144 \sqrt{5}}+\frac{\sqrt{\frac{3}{7}} C_{31,22}^{33,0}}{20480}+\frac{\sqrt{\frac{3}{7}} C_{31,22}^{33,1}}{10240}-\frac{3
C_{33,00}^{33,0}}{10240 \sqrt{7}}-\frac{3 C_{33,00}^{33,1}}{5120 \sqrt{7}}-\frac{\sqrt{\frac{3}{7}} C_{33,20}^{33,0}}{10240}-\frac{\sqrt{\frac{3}{7}}
C_{33,20}^{33,1}}{5120}+\frac{C_{33,22}^{33,0}}{5120 \sqrt{70}}+\frac{C_{33,22}^{33,1}}{2560 \sqrt{70}}\)

\noindent\(C_{33,0011,4}^{33,0011,1,\text{Loc}}= -\frac{C_{00,20}^{60,1}}{2048 \sqrt{3}}-\frac{C_{00,22}^{62,1}}{4096 \sqrt{15}}+\frac{C_{11,22}^{51,1}}{4096
\sqrt{15}}+\frac{3 C_{11,22}^{53,1}}{20480 \sqrt{7}}-\frac{C_{20,20}^{40,1}}{2048 \sqrt{3}}-\frac{C_{20,22}^{42,1}}{4096 \sqrt{15}}-\frac{C_{22,20}^{42,1}}{1024
\sqrt{15}}-\frac{C_{22,22}^{40,1}}{4096 \sqrt{15}}-\frac{C_{22,22}^{42,1}}{2048 \sqrt{105}}-\frac{\sqrt{3} C_{22,22}^{44,1}}{35840}+\frac{C_{31,20}^{31,1}}{2048
\sqrt{3}}+\frac{C_{31,22}^{31,1}}{4096 \sqrt{15}}+\frac{3 C_{31,22}^{33,1}}{20480 \sqrt{7}}+\frac{3 C_{33,20}^{33,1}}{5120 \sqrt{7}}+\frac{\sqrt{\frac{3}{70}}
C_{33,22}^{33,1}}{5120}\)

\noindent\(C_{33,0011,4}^{33,0011,1,\text{NonLoc}}= -\frac{C_{00,00}^{60,0}}{4096}-\frac{C_{00,20}^{60,0}}{4096 \sqrt{3}}+\frac{C_{00,22}^{62,0}}{4096
\sqrt{15}}-\frac{C_{11,00}^{51,0}}{4096}+\frac{C_{11,22}^{51,0}}{4096 \sqrt{15}}+\frac{3 C_{11,22}^{53,0}}{20480 \sqrt{7}}-\frac{C_{20,00}^{40,0}}{4096}-\frac{C_{20,20}^{40,0}}{4096
\sqrt{3}}+\frac{C_{20,22}^{42,0}}{4096 \sqrt{15}}-\frac{C_{22,00}^{42,0}}{2048 \sqrt{5}}-\frac{C_{22,20}^{42,0}}{2048 \sqrt{15}}+\frac{C_{22,22}^{40,0}}{4096
\sqrt{15}}+\frac{C_{22,22}^{42,0}}{2048 \sqrt{105}}+\frac{\sqrt{3} C_{22,22}^{44,0}}{35840}-\frac{C_{31,00}^{31,0}}{4096}-\frac{C_{31,20}^{31,0}}{4096
\sqrt{3}}+\frac{C_{31,22}^{31,0}}{4096 \sqrt{15}}+\frac{3 C_{31,22}^{33,0}}{20480 \sqrt{7}}-\frac{3 \sqrt{\frac{3}{7}} C_{33,00}^{33,0}}{10240}-\frac{3
C_{33,20}^{33,0}}{10240 \sqrt{7}}+\frac{\sqrt{\frac{3}{70}} C_{33,22}^{33,0}}{5120}\)

\noindent\(C_{33,1101,2}^{00,2212,0,\text{Loc}}= \frac{\sqrt{7} C_{11,11}^{51,0}}{144}+\frac{\sqrt{7} C_{11,11}^{51,1}}{288}-\frac{\sqrt{7}
C_{31,11}^{31,0}}{144}-\frac{\sqrt{7} C_{31,11}^{31,1}}{288}+\frac{C_{33,11}^{33,0}}{80 \sqrt{2}}+\frac{C_{33,11}^{33,1}}{160 \sqrt{2}}\)

\noindent\(C_{33,1101,2}^{00,2212,0,\text{NonLoc}}= \frac{\sqrt{7} C_{11,11}^{51,0}}{288}+\frac{\sqrt{7} C_{11,11}^{51,1}}{144}-\frac{\sqrt{7}
C_{31,11}^{31,0}}{288}-\frac{\sqrt{7} C_{31,11}^{31,1}}{144}+\frac{C_{33,11}^{33,0}}{160 \sqrt{2}}+\frac{C_{33,11}^{33,1}}{80 \sqrt{2}}\)

\noindent\(C_{33,1101,2}^{00,2212,1,\text{Loc}}= \frac{1}{96} \sqrt{\frac{7}{3}} C_{11,11}^{51,1}-\frac{1}{96} \sqrt{\frac{7}{3}} C_{31,11}^{31,1}+\frac{1}{160}
\sqrt{\frac{3}{2}} C_{33,11}^{33,1}\)

\noindent\(C_{33,1101,2}^{00,2212,1,\text{NonLoc}}= \frac{1}{96} \sqrt{\frac{7}{3}} C_{11,11}^{51,0}-\frac{1}{96} \sqrt{\frac{7}{3}}
C_{31,11}^{31,0}+\frac{1}{160} \sqrt{\frac{3}{2}} C_{33,11}^{33,0}\)

\noindent\(C_{33,1101,2}^{11,1101,0,\text{Loc}}= \frac{1}{64} \sqrt{\frac{7}{3}} C_{00,00}^{60,0}+\frac{1}{128} \sqrt{\frac{7}{3}}
C_{00,00}^{60,1}-\frac{1}{192} \sqrt{\frac{7}{3}} C_{11,00}^{51,0}-\frac{1}{384} \sqrt{\frac{7}{3}} C_{11,00}^{51,1}-\frac{1}{192} \sqrt{\frac{7}{3}}
C_{20,00}^{40,0}-\frac{1}{384} \sqrt{\frac{7}{3}} C_{20,00}^{40,1}+\frac{1}{192} \sqrt{\frac{7}{15}} C_{22,00}^{42,0}+\frac{1}{384} \sqrt{\frac{7}{15}}
C_{22,00}^{42,1}+\frac{1}{192} \sqrt{\frac{7}{3}} C_{31,00}^{31,0}+\frac{1}{384} \sqrt{\frac{7}{3}} C_{31,00}^{31,1}-\frac{3 C_{33,00}^{33,0}}{320}-\frac{3
C_{33,00}^{33,1}}{640}\)

\noindent\(C_{33,1101,2}^{11,1101,0,\text{NonLoc}}= -\frac{1}{256} \sqrt{\frac{7}{3}} C_{00,00}^{60,0}-\frac{1}{128} \sqrt{\frac{7}{3}}
C_{00,00}^{60,1}+\frac{\sqrt{7} C_{00,20}^{60,0}}{256}+\frac{\sqrt{7} C_{00,20}^{60,1}}{128}-\frac{1}{768} \sqrt{\frac{7}{3}} C_{11,00}^{51,0}-\frac{1}{384}
\sqrt{\frac{7}{3}} C_{11,00}^{51,1}+\frac{1}{768} \sqrt{\frac{7}{3}} C_{20,00}^{40,0}+\frac{1}{384} \sqrt{\frac{7}{3}} C_{20,00}^{40,1}-\frac{\sqrt{7}
C_{20,20}^{40,0}}{768}-\frac{\sqrt{7} C_{20,20}^{40,1}}{384}-\frac{1}{768} \sqrt{\frac{7}{15}} C_{22,00}^{42,0}-\frac{1}{384} \sqrt{\frac{7}{15}}
C_{22,00}^{42,1}+\frac{1}{768} \sqrt{\frac{7}{5}} C_{22,20}^{42,0}+\frac{1}{384} \sqrt{\frac{7}{5}} C_{22,20}^{42,1}+\frac{1}{768} \sqrt{\frac{7}{3}}
C_{31,00}^{31,0}+\frac{1}{384} \sqrt{\frac{7}{3}} C_{31,00}^{31,1}-\frac{\sqrt{7} C_{31,20}^{31,0}}{768}-\frac{\sqrt{7} C_{31,20}^{31,1}}{384}-\frac{3
C_{33,00}^{33,0}}{1280}-\frac{3 C_{33,00}^{33,1}}{640}+\frac{3 \sqrt{3} C_{33,20}^{33,0}}{1280}+\frac{3 \sqrt{3} C_{33,20}^{33,1}}{640}\)

\noindent\(C_{33,1101,2}^{11,1101,1,\text{Loc}}= \frac{\sqrt{7} C_{00,00}^{60,1}}{128}-\frac{\sqrt{7} C_{11,00}^{51,1}}{384}-\frac{\sqrt{7}
C_{20,00}^{40,1}}{384}+\frac{1}{384} \sqrt{\frac{7}{5}} C_{22,00}^{42,1}+\frac{\sqrt{7} C_{31,00}^{31,1}}{384}-\frac{3 \sqrt{3} C_{33,00}^{33,1}}{640}\)

\noindent\(C_{33,1101,2}^{11,1101,1,\text{NonLoc}}= -\frac{\sqrt{7} C_{00,00}^{60,0}}{256}+\frac{\sqrt{21} C_{00,20}^{60,0}}{256}-\frac{\sqrt{7}
C_{11,00}^{51,0}}{768}+\frac{\sqrt{7} C_{20,00}^{40,0}}{768}-\frac{1}{256} \sqrt{\frac{7}{3}} C_{20,20}^{40,0}-\frac{1}{768} \sqrt{\frac{7}{5}} C_{22,00}^{42,0}+\frac{1}{256}
\sqrt{\frac{7}{15}} C_{22,20}^{42,0}+\frac{\sqrt{7} C_{31,00}^{31,0}}{768}-\frac{1}{256} \sqrt{\frac{7}{3}} C_{31,20}^{31,0}-\frac{3 \sqrt{3} C_{33,00}^{33,0}}{1280}+\frac{9
C_{33,20}^{33,0}}{1280}\)

\noindent\(C_{33,1101,2}^{22,0011,0,\text{Loc}}= -\frac{\sqrt{7} C_{11,11}^{51,0}}{2304}-\frac{\sqrt{7} C_{11,11}^{51,1}}{4608}+\frac{1}{384}
\sqrt{\frac{7}{15}} C_{22,11}^{42,0}+\frac{1}{768} \sqrt{\frac{7}{15}} C_{22,11}^{42,1}-\frac{\sqrt{7} C_{31,11}^{31,0}}{2304}-\frac{\sqrt{7} C_{31,11}^{31,1}}{4608}-\frac{C_{33,11}^{33,0}}{320
\sqrt{2}}-\frac{C_{33,11}^{33,1}}{640 \sqrt{2}}\)

\noindent\(C_{33,1101,2}^{22,0011,0,\text{NonLoc}}= -\frac{\sqrt{7} C_{11,11}^{51,0}}{4608}-\frac{\sqrt{7} C_{11,11}^{51,1}}{2304}-\frac{1}{768}
\sqrt{\frac{7}{15}} C_{22,11}^{42,0}-\frac{1}{384} \sqrt{\frac{7}{15}} C_{22,11}^{42,1}-\frac{\sqrt{7} C_{31,11}^{31,0}}{4608}-\frac{\sqrt{7} C_{31,11}^{31,1}}{2304}-\frac{C_{33,11}^{33,0}}{640
\sqrt{2}}-\frac{C_{33,11}^{33,1}}{320 \sqrt{2}}\)

\noindent\(C_{33,1101,2}^{22,0011,1,\text{Loc}}= -\frac{\sqrt{\frac{7}{3}} C_{11,11}^{51,1}}{1536}+\frac{1}{768} \sqrt{\frac{7}{5}}
C_{22,11}^{42,1}-\frac{\sqrt{\frac{7}{3}} C_{31,11}^{31,1}}{1536}-\frac{1}{640} \sqrt{\frac{3}{2}} C_{33,11}^{33,1}\)

\noindent\(C_{33,1101,2}^{22,0011,1,\text{NonLoc}}= -\frac{\sqrt{\frac{7}{3}} C_{11,11}^{51,0}}{1536}-\frac{1}{768} \sqrt{\frac{7}{5}}
C_{22,11}^{42,0}-\frac{\sqrt{\frac{7}{3}} C_{31,11}^{31,0}}{1536}-\frac{1}{640} \sqrt{\frac{3}{2}} C_{33,11}^{33,0}\)

\noindent\(C_{33,1101,3}^{00,2213,0,\text{Loc}}= \frac{1}{288} \sqrt{\frac{7}{2}} C_{11,11}^{51,0}+\frac{1}{576} \sqrt{\frac{7}{2}}
C_{11,11}^{51,1}-\frac{1}{288} \sqrt{\frac{7}{2}} C_{31,11}^{31,0}-\frac{1}{576} \sqrt{\frac{7}{2}} C_{31,11}^{31,1}+\frac{C_{33,11}^{33,0}}{320}+\frac{C_{33,11}^{33,1}}{640}\)

\noindent\(C_{33,1101,3}^{00,2213,0,\text{NonLoc}}= \frac{1}{576} \sqrt{\frac{7}{2}} C_{11,11}^{51,0}+\frac{1}{288} \sqrt{\frac{7}{2}}
C_{11,11}^{51,1}-\frac{1}{576} \sqrt{\frac{7}{2}} C_{31,11}^{31,0}-\frac{1}{288} \sqrt{\frac{7}{2}} C_{31,11}^{31,1}+\frac{C_{33,11}^{33,0}}{640}+\frac{C_{33,11}^{33,1}}{320}\)

\noindent\(C_{33,1101,3}^{00,2213,1,\text{Loc}}= \frac{1}{192} \sqrt{\frac{7}{6}} C_{11,11}^{51,1}-\frac{1}{192} \sqrt{\frac{7}{6}}
C_{31,11}^{31,1}+\frac{\sqrt{3} C_{33,11}^{33,1}}{640}\)

\noindent\(C_{33,1101,3}^{00,2213,1,\text{NonLoc}}= \frac{1}{192} \sqrt{\frac{7}{6}} C_{11,11}^{51,0}-\frac{1}{192} \sqrt{\frac{7}{6}}
C_{31,11}^{31,0}+\frac{\sqrt{3} C_{33,11}^{33,0}}{640}\)

\noindent\(C_{33,1101,3}^{22,0011,0,\text{Loc}}= -\frac{\sqrt{\frac{7}{2}} C_{11,11}^{51,0}}{1152}-\frac{\sqrt{\frac{7}{2}} C_{11,11}^{51,1}}{2304}+\frac{1}{192}
\sqrt{\frac{7}{30}} C_{22,11}^{42,0}+\frac{1}{384} \sqrt{\frac{7}{30}} C_{22,11}^{42,1}-\frac{\sqrt{\frac{7}{2}} C_{31,11}^{31,0}}{1152}-\frac{\sqrt{\frac{7}{2}}
C_{31,11}^{31,1}}{2304}-\frac{C_{33,11}^{33,0}}{320}-\frac{C_{33,11}^{33,1}}{640}\)

\noindent\(C_{33,1101,3}^{22,0011,0,\text{NonLoc}}= -\frac{\sqrt{\frac{7}{2}} C_{11,11}^{51,0}}{2304}-\frac{\sqrt{\frac{7}{2}} C_{11,11}^{51,1}}{1152}-\frac{1}{384}
\sqrt{\frac{7}{30}} C_{22,11}^{42,0}-\frac{1}{192} \sqrt{\frac{7}{30}} C_{22,11}^{42,1}-\frac{\sqrt{\frac{7}{2}} C_{31,11}^{31,0}}{2304}-\frac{\sqrt{\frac{7}{2}}
C_{31,11}^{31,1}}{1152}-\frac{C_{33,11}^{33,0}}{640}-\frac{C_{33,11}^{33,1}}{320}\)

\noindent\(C_{33,1101,3}^{22,0011,1,\text{Loc}}= -\frac{1}{768} \sqrt{\frac{7}{6}} C_{11,11}^{51,1}+\frac{1}{384} \sqrt{\frac{7}{10}}
C_{22,11}^{42,1}-\frac{1}{768} \sqrt{\frac{7}{6}} C_{31,11}^{31,1}-\frac{\sqrt{3} C_{33,11}^{33,1}}{640}\)

\noindent\(C_{33,1101,3}^{22,0011,1,\text{NonLoc}}= -\frac{1}{768} \sqrt{\frac{7}{6}} C_{11,11}^{51,0}-\frac{1}{384} \sqrt{\frac{7}{10}}
C_{22,11}^{42,0}-\frac{1}{768} \sqrt{\frac{7}{6}} C_{31,11}^{31,0}-\frac{\sqrt{3} C_{33,11}^{33,0}}{640}\)

\noindent\(C_{33,1110,3}^{11,1112,0,\text{Loc}}= -\frac{1}{192} \sqrt{\frac{7}{3}} C_{00,20}^{60,0}-\frac{1}{384} \sqrt{\frac{7}{3}}
C_{00,20}^{60,1}-\frac{1}{64} \sqrt{\frac{7}{15}} C_{00,22}^{62,0}-\frac{1}{128} \sqrt{\frac{7}{15}} C_{00,22}^{62,1}+\frac{7 \sqrt{\frac{7}{15}}
C_{11,22}^{51,0}}{1152}+\frac{7 \sqrt{\frac{7}{15}} C_{11,22}^{51,1}}{2304}+\frac{3 C_{11,22}^{53,0}}{1280}+\frac{3 C_{11,22}^{53,1}}{2560}+\frac{1}{576}
\sqrt{\frac{7}{3}} C_{20,20}^{40,0}+\frac{\sqrt{\frac{7}{3}} C_{20,20}^{40,1}}{1152}+\frac{\sqrt{\frac{7}{15}} C_{20,22}^{42,0}}{1152}+\frac{\sqrt{\frac{7}{15}}
C_{20,22}^{42,1}}{2304}-\frac{1}{576} \sqrt{\frac{7}{15}} C_{22,20}^{42,0}-\frac{\sqrt{\frac{7}{15}} C_{22,20}^{42,1}}{1152}+\frac{1}{288} \sqrt{\frac{7}{15}}
C_{22,22}^{40,0}+\frac{1}{576} \sqrt{\frac{7}{15}} C_{22,22}^{40,1}-\frac{1}{576} \sqrt{\frac{5}{3}} C_{22,22}^{42,0}-\frac{\sqrt{\frac{5}{3}} C_{22,22}^{42,1}}{1152}+\frac{1}{160}
\sqrt{\frac{3}{7}} C_{22,22}^{44,0}+\frac{1}{320} \sqrt{\frac{3}{7}} C_{22,22}^{44,1}-\frac{1}{576} \sqrt{\frac{7}{3}} C_{31,20}^{31,0}-\frac{\sqrt{\frac{7}{3}}
C_{31,20}^{31,1}}{1152}-\frac{\sqrt{\frac{7}{15}} C_{31,22}^{31,0}}{1152}-\frac{\sqrt{\frac{7}{15}} C_{31,22}^{31,1}}{2304}-\frac{17 C_{31,22}^{33,0}}{3840}-\frac{17
C_{31,22}^{33,1}}{7680}+\frac{C_{33,20}^{33,0}}{320}+\frac{C_{33,20}^{33,1}}{640}+\frac{1}{160} \sqrt{\frac{3}{10}} C_{33,22}^{33,0}+\frac{1}{320}
\sqrt{\frac{3}{10}} C_{33,22}^{33,1}\)

\noindent\(C_{33,1110,3}^{11,1112,0,\text{NonLoc}}= -\frac{\sqrt{7} C_{00,00}^{60,0}}{768}-\frac{\sqrt{7} C_{00,00}^{60,1}}{384}-\frac{1}{768}
\sqrt{\frac{7}{3}} C_{00,20}^{60,0}-\frac{1}{384} \sqrt{\frac{7}{3}} C_{00,20}^{60,1}+\frac{1}{128} \sqrt{\frac{7}{15}} C_{00,22}^{62,0}+\frac{1}{64}
\sqrt{\frac{7}{15}} C_{00,22}^{62,1}-\frac{\sqrt{7} C_{11,00}^{51,0}}{2304}-\frac{\sqrt{7} C_{11,00}^{51,1}}{1152}+\frac{7 \sqrt{\frac{7}{15}} C_{11,22}^{51,0}}{2304}+\frac{7
\sqrt{\frac{7}{15}} C_{11,22}^{51,1}}{1152}+\frac{3 C_{11,22}^{53,0}}{2560}+\frac{3 C_{11,22}^{53,1}}{1280}+\frac{\sqrt{7} C_{20,00}^{40,0}}{2304}+\frac{\sqrt{7}
C_{20,00}^{40,1}}{1152}+\frac{\sqrt{\frac{7}{3}} C_{20,20}^{40,0}}{2304}+\frac{\sqrt{\frac{7}{3}} C_{20,20}^{40,1}}{1152}-\frac{\sqrt{\frac{7}{15}}
C_{20,22}^{42,0}}{2304}-\frac{\sqrt{\frac{7}{15}} C_{20,22}^{42,1}}{1152}-\frac{\sqrt{\frac{7}{5}} C_{22,00}^{42,0}}{2304}-\frac{\sqrt{\frac{7}{5}}
C_{22,00}^{42,1}}{1152}-\frac{\sqrt{\frac{7}{15}} C_{22,20}^{42,0}}{2304}-\frac{\sqrt{\frac{7}{15}} C_{22,20}^{42,1}}{1152}-\frac{1}{576} \sqrt{\frac{7}{15}}
C_{22,22}^{40,0}-\frac{1}{288} \sqrt{\frac{7}{15}} C_{22,22}^{40,1}+\frac{\sqrt{\frac{5}{3}} C_{22,22}^{42,0}}{1152}+\frac{1}{576} \sqrt{\frac{5}{3}}
C_{22,22}^{42,1}-\frac{1}{320} \sqrt{\frac{3}{7}} C_{22,22}^{44,0}-\frac{1}{160} \sqrt{\frac{3}{7}} C_{22,22}^{44,1}+\frac{\sqrt{7} C_{31,00}^{31,0}}{2304}+\frac{\sqrt{7}
C_{31,00}^{31,1}}{1152}+\frac{\sqrt{\frac{7}{3}} C_{31,20}^{31,0}}{2304}+\frac{\sqrt{\frac{7}{3}} C_{31,20}^{31,1}}{1152}-\frac{\sqrt{\frac{7}{15}}
C_{31,22}^{31,0}}{2304}-\frac{\sqrt{\frac{7}{15}} C_{31,22}^{31,1}}{1152}-\frac{17 C_{31,22}^{33,0}}{7680}-\frac{17 C_{31,22}^{33,1}}{3840}-\frac{\sqrt{3}
C_{33,00}^{33,0}}{1280}-\frac{\sqrt{3} C_{33,00}^{33,1}}{640}-\frac{C_{33,20}^{33,0}}{1280}-\frac{C_{33,20}^{33,1}}{640}+\frac{1}{320} \sqrt{\frac{3}{10}}
C_{33,22}^{33,0}+\frac{1}{160} \sqrt{\frac{3}{10}} C_{33,22}^{33,1}\)

\noindent\(C_{33,1110,3}^{11,1112,1,\text{Loc}}= -\frac{\sqrt{7} C_{00,20}^{60,1}}{384}-\frac{1}{128} \sqrt{\frac{7}{5}} C_{00,22}^{62,1}+\frac{7
\sqrt{\frac{7}{5}} C_{11,22}^{51,1}}{2304}+\frac{3 \sqrt{3} C_{11,22}^{53,1}}{2560}+\frac{\sqrt{7} C_{20,20}^{40,1}}{1152}+\frac{\sqrt{\frac{7}{5}}
C_{20,22}^{42,1}}{2304}-\frac{\sqrt{\frac{7}{5}} C_{22,20}^{42,1}}{1152}+\frac{1}{576} \sqrt{\frac{7}{5}} C_{22,22}^{40,1}-\frac{\sqrt{5} C_{22,22}^{42,1}}{1152}+\frac{3
C_{22,22}^{44,1}}{320 \sqrt{7}}-\frac{\sqrt{7} C_{31,20}^{31,1}}{1152}-\frac{\sqrt{\frac{7}{5}} C_{31,22}^{31,1}}{2304}-\frac{17 C_{31,22}^{33,1}}{2560
\sqrt{3}}+\frac{\sqrt{3} C_{33,20}^{33,1}}{640}+\frac{3 C_{33,22}^{33,1}}{320 \sqrt{10}}\)

\noindent\(C_{33,1110,3}^{11,1112,1,\text{NonLoc}}= -\frac{1}{256} \sqrt{\frac{7}{3}} C_{00,00}^{60,0}-\frac{\sqrt{7} C_{00,20}^{60,0}}{768}+\frac{1}{128}
\sqrt{\frac{7}{5}} C_{00,22}^{62,0}-\frac{1}{768} \sqrt{\frac{7}{3}} C_{11,00}^{51,0}+\frac{7 \sqrt{\frac{7}{5}} C_{11,22}^{51,0}}{2304}+\frac{3
\sqrt{3} C_{11,22}^{53,0}}{2560}+\frac{1}{768} \sqrt{\frac{7}{3}} C_{20,00}^{40,0}+\frac{\sqrt{7} C_{20,20}^{40,0}}{2304}-\frac{\sqrt{\frac{7}{5}}
C_{20,22}^{42,0}}{2304}-\frac{1}{768} \sqrt{\frac{7}{15}} C_{22,00}^{42,0}-\frac{\sqrt{\frac{7}{5}} C_{22,20}^{42,0}}{2304}-\frac{1}{576} \sqrt{\frac{7}{5}}
C_{22,22}^{40,0}+\frac{\sqrt{5} C_{22,22}^{42,0}}{1152}-\frac{3 C_{22,22}^{44,0}}{320 \sqrt{7}}+\frac{1}{768} \sqrt{\frac{7}{3}} C_{31,00}^{31,0}+\frac{\sqrt{7}
C_{31,20}^{31,0}}{2304}-\frac{\sqrt{\frac{7}{5}} C_{31,22}^{31,0}}{2304}-\frac{17 C_{31,22}^{33,0}}{2560 \sqrt{3}}-\frac{3 C_{33,00}^{33,0}}{1280}-\frac{\sqrt{3}
C_{33,20}^{33,0}}{1280}+\frac{3 C_{33,22}^{33,0}}{320 \sqrt{10}}\)

\noindent\(C_{33,1111,2}^{00,2202,0,\text{Loc}}= \frac{1}{96} \sqrt{\frac{7}{3}} C_{11,11}^{51,0}+\frac{1}{192} \sqrt{\frac{7}{3}}
C_{11,11}^{51,1}-\frac{1}{96} \sqrt{\frac{7}{3}} C_{31,11}^{31,0}-\frac{1}{192} \sqrt{\frac{7}{3}} C_{31,11}^{31,1}+\frac{1}{160} \sqrt{\frac{3}{2}}
C_{33,11}^{33,0}+\frac{1}{320} \sqrt{\frac{3}{2}} C_{33,11}^{33,1}\)

\noindent\(C_{33,1111,2}^{00,2202,0,\text{NonLoc}}= \frac{1}{192} \sqrt{\frac{7}{3}} C_{11,11}^{51,0}+\frac{1}{96} \sqrt{\frac{7}{3}}
C_{11,11}^{51,1}-\frac{1}{192} \sqrt{\frac{7}{3}} C_{31,11}^{31,0}-\frac{1}{96} \sqrt{\frac{7}{3}} C_{31,11}^{31,1}+\frac{1}{320} \sqrt{\frac{3}{2}}
C_{33,11}^{33,0}+\frac{1}{160} \sqrt{\frac{3}{2}} C_{33,11}^{33,1}\)

\noindent\(C_{33,1111,2}^{00,2202,1,\text{Loc}}= \frac{\sqrt{7} C_{11,11}^{51,1}}{192}-\frac{\sqrt{7} C_{31,11}^{31,1}}{192}+\frac{3
C_{33,11}^{33,1}}{320 \sqrt{2}}\)

\noindent\(C_{33,1111,2}^{00,2202,1,\text{NonLoc}}= \frac{\sqrt{7} C_{11,11}^{51,0}}{192}-\frac{\sqrt{7} C_{31,11}^{31,0}}{192}+\frac{3
C_{33,11}^{33,0}}{320 \sqrt{2}}\)

\noindent\(C_{33,1111,2}^{11,1111,0,\text{Loc}}= \frac{\sqrt{7} C_{00,20}^{60,0}}{384}+\frac{\sqrt{7} C_{00,20}^{60,1}}{768}-\frac{1}{256}
\sqrt{\frac{7}{5}} C_{00,22}^{62,0}-\frac{1}{512} \sqrt{\frac{7}{5}} C_{00,22}^{62,1}-\frac{\sqrt{\frac{7}{5}} C_{11,22}^{51,0}}{2304}-\frac{\sqrt{\frac{7}{5}}
C_{11,22}^{51,1}}{4608}+\frac{3 \sqrt{3} C_{11,22}^{53,0}}{1280}+\frac{3 \sqrt{3} C_{11,22}^{53,1}}{2560}-\frac{\sqrt{7} C_{20,20}^{40,0}}{1152}-\frac{\sqrt{7}
C_{20,20}^{40,1}}{2304}+\frac{\sqrt{35} C_{20,22}^{42,0}}{2304}+\frac{\sqrt{35} C_{20,22}^{42,1}}{4608}+\frac{\sqrt{\frac{7}{5}} C_{22,20}^{42,0}}{1152}+\frac{\sqrt{\frac{7}{5}}
C_{22,20}^{42,1}}{2304}+\frac{11 \sqrt{\frac{7}{5}} C_{22,22}^{40,0}}{2304}+\frac{11 \sqrt{\frac{7}{5}} C_{22,22}^{40,1}}{4608}-\frac{C_{22,22}^{42,0}}{72
\sqrt{5}}-\frac{C_{22,22}^{42,1}}{144 \sqrt{5}}-\frac{3 C_{22,22}^{44,0}}{320 \sqrt{7}}-\frac{3 C_{22,22}^{44,1}}{640 \sqrt{7}}+\frac{\sqrt{7} C_{31,20}^{31,0}}{1152}+\frac{\sqrt{7}
C_{31,20}^{31,1}}{2304}-\frac{\sqrt{35} C_{31,22}^{31,0}}{2304}-\frac{\sqrt{35} C_{31,22}^{31,1}}{4608}-\frac{C_{31,22}^{33,0}}{256 \sqrt{3}}-\frac{C_{31,22}^{33,1}}{512
\sqrt{3}}-\frac{\sqrt{3} C_{33,20}^{33,0}}{640}-\frac{\sqrt{3} C_{33,20}^{33,1}}{1280}+\frac{3 C_{33,22}^{33,0}}{64 \sqrt{10}}+\frac{3 C_{33,22}^{33,1}}{128
\sqrt{10}}\)

\noindent\(C_{33,1111,2}^{11,1111,0,\text{NonLoc}}= \frac{1}{512} \sqrt{\frac{7}{3}} C_{00,00}^{60,0}+\frac{1}{256} \sqrt{\frac{7}{3}}
C_{00,00}^{60,1}+\frac{\sqrt{7} C_{00,20}^{60,0}}{1536}+\frac{\sqrt{7} C_{00,20}^{60,1}}{768}+\frac{1}{512} \sqrt{\frac{7}{5}} C_{00,22}^{62,0}+\frac{1}{256}
\sqrt{\frac{7}{5}} C_{00,22}^{62,1}+\frac{\sqrt{\frac{7}{3}} C_{11,00}^{51,0}}{1536}+\frac{1}{768} \sqrt{\frac{7}{3}} C_{11,00}^{51,1}-\frac{\sqrt{\frac{7}{5}}
C_{11,22}^{51,0}}{4608}-\frac{\sqrt{\frac{7}{5}} C_{11,22}^{51,1}}{2304}+\frac{3 \sqrt{3} C_{11,22}^{53,0}}{2560}+\frac{3 \sqrt{3} C_{11,22}^{53,1}}{1280}-\frac{\sqrt{\frac{7}{3}}
C_{20,00}^{40,0}}{1536}-\frac{1}{768} \sqrt{\frac{7}{3}} C_{20,00}^{40,1}-\frac{\sqrt{7} C_{20,20}^{40,0}}{4608}-\frac{\sqrt{7} C_{20,20}^{40,1}}{2304}-\frac{\sqrt{35}
C_{20,22}^{42,0}}{4608}-\frac{\sqrt{35} C_{20,22}^{42,1}}{2304}+\frac{\sqrt{\frac{7}{15}} C_{22,00}^{42,0}}{1536}+\frac{1}{768} \sqrt{\frac{7}{15}}
C_{22,00}^{42,1}+\frac{\sqrt{\frac{7}{5}} C_{22,20}^{42,0}}{4608}+\frac{\sqrt{\frac{7}{5}} C_{22,20}^{42,1}}{2304}-\frac{11 \sqrt{\frac{7}{5}} C_{22,22}^{40,0}}{4608}-\frac{11
\sqrt{\frac{7}{5}} C_{22,22}^{40,1}}{2304}+\frac{C_{22,22}^{42,0}}{144 \sqrt{5}}+\frac{C_{22,22}^{42,1}}{72 \sqrt{5}}+\frac{3 C_{22,22}^{44,0}}{640
\sqrt{7}}+\frac{3 C_{22,22}^{44,1}}{320 \sqrt{7}}-\frac{\sqrt{\frac{7}{3}} C_{31,00}^{31,0}}{1536}-\frac{1}{768} \sqrt{\frac{7}{3}} C_{31,00}^{31,1}-\frac{\sqrt{7}
C_{31,20}^{31,0}}{4608}-\frac{\sqrt{7} C_{31,20}^{31,1}}{2304}-\frac{\sqrt{35} C_{31,22}^{31,0}}{4608}-\frac{\sqrt{35} C_{31,22}^{31,1}}{2304}-\frac{C_{31,22}^{33,0}}{512
\sqrt{3}}-\frac{C_{31,22}^{33,1}}{256 \sqrt{3}}+\frac{3 C_{33,00}^{33,0}}{2560}+\frac{3 C_{33,00}^{33,1}}{1280}+\frac{\sqrt{3} C_{33,20}^{33,0}}{2560}+\frac{\sqrt{3}
C_{33,20}^{33,1}}{1280}+\frac{3 C_{33,22}^{33,0}}{128 \sqrt{10}}+\frac{3 C_{33,22}^{33,1}}{64 \sqrt{10}}\)

\noindent\(C_{33,1111,2}^{11,1111,1,\text{Loc}}= \frac{1}{256} \sqrt{\frac{7}{3}} C_{00,20}^{60,1}-\frac{1}{512} \sqrt{\frac{21}{5}}
C_{00,22}^{62,1}-\frac{\sqrt{\frac{7}{15}} C_{11,22}^{51,1}}{1536}+\frac{9 C_{11,22}^{53,1}}{2560}-\frac{1}{768} \sqrt{\frac{7}{3}} C_{20,20}^{40,1}+\frac{\sqrt{\frac{35}{3}}
C_{20,22}^{42,1}}{1536}+\frac{1}{768} \sqrt{\frac{7}{15}} C_{22,20}^{42,1}+\frac{11 \sqrt{\frac{7}{15}} C_{22,22}^{40,1}}{1536}-\frac{C_{22,22}^{42,1}}{48
\sqrt{15}}-\frac{3}{640} \sqrt{\frac{3}{7}} C_{22,22}^{44,1}+\frac{1}{768} \sqrt{\frac{7}{3}} C_{31,20}^{31,1}-\frac{\sqrt{\frac{35}{3}} C_{31,22}^{31,1}}{1536}-\frac{C_{31,22}^{33,1}}{512}-\frac{3
C_{33,20}^{33,1}}{1280}+\frac{3}{128} \sqrt{\frac{3}{10}} C_{33,22}^{33,1}\)

\noindent\(C_{33,1111,2}^{11,1111,1,\text{NonLoc}}= \frac{\sqrt{7} C_{00,00}^{60,0}}{512}+\frac{1}{512} \sqrt{\frac{7}{3}} C_{00,20}^{60,0}+\frac{1}{512}
\sqrt{\frac{21}{5}} C_{00,22}^{62,0}+\frac{\sqrt{7} C_{11,00}^{51,0}}{1536}-\frac{\sqrt{\frac{7}{15}} C_{11,22}^{51,0}}{1536}+\frac{9 C_{11,22}^{53,0}}{2560}-\frac{\sqrt{7}
C_{20,00}^{40,0}}{1536}-\frac{\sqrt{\frac{7}{3}} C_{20,20}^{40,0}}{1536}-\frac{\sqrt{\frac{35}{3}} C_{20,22}^{42,0}}{1536}+\frac{\sqrt{\frac{7}{5}}
C_{22,00}^{42,0}}{1536}+\frac{\sqrt{\frac{7}{15}} C_{22,20}^{42,0}}{1536}-\frac{11 \sqrt{\frac{7}{15}} C_{22,22}^{40,0}}{1536}+\frac{C_{22,22}^{42,0}}{48
\sqrt{15}}+\frac{3}{640} \sqrt{\frac{3}{7}} C_{22,22}^{44,0}-\frac{\sqrt{7} C_{31,00}^{31,0}}{1536}-\frac{\sqrt{\frac{7}{3}} C_{31,20}^{31,0}}{1536}-\frac{\sqrt{\frac{35}{3}}
C_{31,22}^{31,0}}{1536}-\frac{C_{31,22}^{33,0}}{512}+\frac{3 \sqrt{3} C_{33,00}^{33,0}}{2560}+\frac{3 C_{33,20}^{33,0}}{2560}+\frac{3}{128} \sqrt{\frac{3}{10}}
C_{33,22}^{33,0}\)

\noindent\(C_{33,1111,2}^{11,1112,0,\text{Loc}}= \frac{1}{384} \sqrt{\frac{7}{3}} C_{00,20}^{60,0}+\frac{1}{768} \sqrt{\frac{7}{3}}
C_{00,20}^{60,1}-\frac{1}{256} \sqrt{\frac{7}{15}} C_{00,22}^{62,0}-\frac{1}{512} \sqrt{\frac{7}{15}} C_{00,22}^{62,1}+\frac{\sqrt{\frac{35}{3}}
C_{11,22}^{51,0}}{2304}+\frac{\sqrt{\frac{35}{3}} C_{11,22}^{51,1}}{4608}-\frac{\sqrt{\frac{7}{3}} C_{20,20}^{40,0}}{1152}-\frac{\sqrt{\frac{7}{3}}
C_{20,20}^{40,1}}{2304}-\frac{\sqrt{\frac{7}{15}} C_{20,22}^{42,0}}{2304}-\frac{\sqrt{\frac{7}{15}} C_{20,22}^{42,1}}{4608}+\frac{\sqrt{\frac{7}{15}}
C_{22,20}^{42,0}}{1152}+\frac{\sqrt{\frac{7}{15}} C_{22,20}^{42,1}}{2304}-\frac{\sqrt{\frac{7}{15}} C_{22,22}^{40,0}}{2304}-\frac{\sqrt{\frac{7}{15}}
C_{22,22}^{40,1}}{4608}+\frac{C_{22,22}^{42,0}}{576 \sqrt{15}}+\frac{C_{22,22}^{42,1}}{1152 \sqrt{15}}+\frac{1}{320} \sqrt{\frac{3}{7}} C_{22,22}^{44,0}+\frac{1}{640}
\sqrt{\frac{3}{7}} C_{22,22}^{44,1}+\frac{\sqrt{\frac{7}{3}} C_{31,20}^{31,0}}{1152}+\frac{\sqrt{\frac{7}{3}} C_{31,20}^{31,1}}{2304}+\frac{\sqrt{\frac{7}{15}}
C_{31,22}^{31,0}}{2304}+\frac{\sqrt{\frac{7}{15}} C_{31,22}^{31,1}}{4608}-\frac{C_{31,22}^{33,0}}{960}-\frac{C_{31,22}^{33,1}}{1920}-\frac{C_{33,20}^{33,0}}{640}-\frac{C_{33,20}^{33,1}}{1280}-\frac{1}{320}
\sqrt{\frac{3}{10}} C_{33,22}^{33,0}-\frac{1}{640} \sqrt{\frac{3}{10}} C_{33,22}^{33,1}\)

\noindent\(C_{33,1111,2}^{11,1112,0,\text{NonLoc}}= \frac{\sqrt{7} C_{00,00}^{60,0}}{1536}+\frac{\sqrt{7} C_{00,00}^{60,1}}{768}+\frac{\sqrt{\frac{7}{3}}
C_{00,20}^{60,0}}{1536}+\frac{1}{768} \sqrt{\frac{7}{3}} C_{00,20}^{60,1}+\frac{1}{512} \sqrt{\frac{7}{15}} C_{00,22}^{62,0}+\frac{1}{256} \sqrt{\frac{7}{15}}
C_{00,22}^{62,1}+\frac{\sqrt{7} C_{11,00}^{51,0}}{4608}+\frac{\sqrt{7} C_{11,00}^{51,1}}{2304}+\frac{\sqrt{\frac{35}{3}} C_{11,22}^{51,0}}{4608}+\frac{\sqrt{\frac{35}{3}}
C_{11,22}^{51,1}}{2304}-\frac{\sqrt{7} C_{20,00}^{40,0}}{4608}-\frac{\sqrt{7} C_{20,00}^{40,1}}{2304}-\frac{\sqrt{\frac{7}{3}} C_{20,20}^{40,0}}{4608}-\frac{\sqrt{\frac{7}{3}}
C_{20,20}^{40,1}}{2304}+\frac{\sqrt{\frac{7}{15}} C_{20,22}^{42,0}}{4608}+\frac{\sqrt{\frac{7}{15}} C_{20,22}^{42,1}}{2304}+\frac{\sqrt{\frac{7}{5}}
C_{22,00}^{42,0}}{4608}+\frac{\sqrt{\frac{7}{5}} C_{22,00}^{42,1}}{2304}+\frac{\sqrt{\frac{7}{15}} C_{22,20}^{42,0}}{4608}+\frac{\sqrt{\frac{7}{15}}
C_{22,20}^{42,1}}{2304}+\frac{\sqrt{\frac{7}{15}} C_{22,22}^{40,0}}{4608}+\frac{\sqrt{\frac{7}{15}} C_{22,22}^{40,1}}{2304}-\frac{C_{22,22}^{42,0}}{1152
\sqrt{15}}-\frac{C_{22,22}^{42,1}}{576 \sqrt{15}}-\frac{1}{640} \sqrt{\frac{3}{7}} C_{22,22}^{44,0}-\frac{1}{320} \sqrt{\frac{3}{7}} C_{22,22}^{44,1}-\frac{\sqrt{7}
C_{31,00}^{31,0}}{4608}-\frac{\sqrt{7} C_{31,00}^{31,1}}{2304}-\frac{\sqrt{\frac{7}{3}} C_{31,20}^{31,0}}{4608}-\frac{\sqrt{\frac{7}{3}} C_{31,20}^{31,1}}{2304}+\frac{\sqrt{\frac{7}{15}}
C_{31,22}^{31,0}}{4608}+\frac{\sqrt{\frac{7}{15}} C_{31,22}^{31,1}}{2304}-\frac{C_{31,22}^{33,0}}{1920}-\frac{C_{31,22}^{33,1}}{960}+\frac{\sqrt{3}
C_{33,00}^{33,0}}{2560}+\frac{\sqrt{3} C_{33,00}^{33,1}}{1280}+\frac{C_{33,20}^{33,0}}{2560}+\frac{C_{33,20}^{33,1}}{1280}-\frac{1}{640} \sqrt{\frac{3}{10}}
C_{33,22}^{33,0}-\frac{1}{320} \sqrt{\frac{3}{10}} C_{33,22}^{33,1}\)

\noindent\(C_{33,1111,2}^{11,1112,1,\text{Loc}}= \frac{\sqrt{7} C_{00,20}^{60,1}}{768}-\frac{1}{512} \sqrt{\frac{7}{5}} C_{00,22}^{62,1}+\frac{\sqrt{35}
C_{11,22}^{51,1}}{4608}-\frac{\sqrt{7} C_{20,20}^{40,1}}{2304}-\frac{\sqrt{\frac{7}{5}} C_{20,22}^{42,1}}{4608}+\frac{\sqrt{\frac{7}{5}} C_{22,20}^{42,1}}{2304}-\frac{\sqrt{\frac{7}{5}}
C_{22,22}^{40,1}}{4608}+\frac{C_{22,22}^{42,1}}{1152 \sqrt{5}}+\frac{3 C_{22,22}^{44,1}}{640 \sqrt{7}}+\frac{\sqrt{7} C_{31,20}^{31,1}}{2304}+\frac{\sqrt{\frac{7}{5}}
C_{31,22}^{31,1}}{4608}-\frac{C_{31,22}^{33,1}}{640 \sqrt{3}}-\frac{\sqrt{3} C_{33,20}^{33,1}}{1280}-\frac{3 C_{33,22}^{33,1}}{640 \sqrt{10}}\)

\noindent\(C_{33,1111,2}^{11,1112,1,\text{NonLoc}}= \frac{1}{512} \sqrt{\frac{7}{3}} C_{00,00}^{60,0}+\frac{\sqrt{7} C_{00,20}^{60,0}}{1536}+\frac{1}{512}
\sqrt{\frac{7}{5}} C_{00,22}^{62,0}+\frac{\sqrt{\frac{7}{3}} C_{11,00}^{51,0}}{1536}+\frac{\sqrt{35} C_{11,22}^{51,0}}{4608}-\frac{\sqrt{\frac{7}{3}}
C_{20,00}^{40,0}}{1536}-\frac{\sqrt{7} C_{20,20}^{40,0}}{4608}+\frac{\sqrt{\frac{7}{5}} C_{20,22}^{42,0}}{4608}+\frac{\sqrt{\frac{7}{15}} C_{22,00}^{42,0}}{1536}+\frac{\sqrt{\frac{7}{5}}
C_{22,20}^{42,0}}{4608}+\frac{\sqrt{\frac{7}{5}} C_{22,22}^{40,0}}{4608}-\frac{C_{22,22}^{42,0}}{1152 \sqrt{5}}-\frac{3 C_{22,22}^{44,0}}{640 \sqrt{7}}-\frac{\sqrt{\frac{7}{3}}
C_{31,00}^{31,0}}{1536}-\frac{\sqrt{7} C_{31,20}^{31,0}}{4608}+\frac{\sqrt{\frac{7}{5}} C_{31,22}^{31,0}}{4608}-\frac{C_{31,22}^{33,0}}{640 \sqrt{3}}+\frac{3
C_{33,00}^{33,0}}{2560}+\frac{\sqrt{3} C_{33,20}^{33,0}}{2560}-\frac{3 C_{33,22}^{33,0}}{640 \sqrt{10}}\)

\noindent\(C_{33,1111,2}^{22,0000,0,\text{Loc}}= -\frac{1}{768} \sqrt{\frac{7}{3}} C_{11,11}^{51,0}-\frac{\sqrt{\frac{7}{3}} C_{11,11}^{51,1}}{1536}+\frac{1}{384}
\sqrt{\frac{7}{5}} C_{22,11}^{42,0}+\frac{1}{768} \sqrt{\frac{7}{5}} C_{22,11}^{42,1}-\frac{1}{768} \sqrt{\frac{7}{3}} C_{31,11}^{31,0}-\frac{\sqrt{\frac{7}{3}}
C_{31,11}^{31,1}}{1536}-\frac{1}{320} \sqrt{\frac{3}{2}} C_{33,11}^{33,0}-\frac{1}{640} \sqrt{\frac{3}{2}} C_{33,11}^{33,1}\)

\noindent\(C_{33,1111,2}^{22,0000,0,\text{NonLoc}}= -\frac{\sqrt{\frac{7}{3}} C_{11,11}^{51,0}}{1536}-\frac{1}{768} \sqrt{\frac{7}{3}}
C_{11,11}^{51,1}-\frac{1}{768} \sqrt{\frac{7}{5}} C_{22,11}^{42,0}-\frac{1}{384} \sqrt{\frac{7}{5}} C_{22,11}^{42,1}-\frac{\sqrt{\frac{7}{3}} C_{31,11}^{31,0}}{1536}-\frac{1}{768}
\sqrt{\frac{7}{3}} C_{31,11}^{31,1}-\frac{1}{640} \sqrt{\frac{3}{2}} C_{33,11}^{33,0}-\frac{1}{320} \sqrt{\frac{3}{2}} C_{33,11}^{33,1}\)

\noindent\(C_{33,1111,2}^{22,0000,1,\text{Loc}}= -\frac{\sqrt{7} C_{11,11}^{51,1}}{1536}+\frac{1}{256} \sqrt{\frac{7}{15}} C_{22,11}^{42,1}-\frac{\sqrt{7}
C_{31,11}^{31,1}}{1536}-\frac{3 C_{33,11}^{33,1}}{640 \sqrt{2}}\)

\noindent\(C_{33,1111,2}^{22,0000,1,\text{NonLoc}}= -\frac{\sqrt{7} C_{11,11}^{51,0}}{1536}-\frac{1}{256} \sqrt{\frac{7}{15}} C_{22,11}^{42,0}-\frac{\sqrt{7}
C_{31,11}^{31,0}}{1536}-\frac{3 C_{33,11}^{33,0}}{640 \sqrt{2}}\)

\noindent\(C_{33,1111,3}^{11,1112,0,\text{Loc}}= \frac{1}{96} \sqrt{\frac{7}{6}} C_{00,20}^{60,0}+\frac{1}{192} \sqrt{\frac{7}{6}}
C_{00,20}^{60,1}-\frac{1}{64} \sqrt{\frac{7}{30}} C_{00,22}^{62,0}-\frac{1}{128} \sqrt{\frac{7}{30}} C_{00,22}^{62,1}+\frac{1}{576} \sqrt{\frac{35}{6}}
C_{11,22}^{51,0}+\frac{\sqrt{\frac{35}{6}} C_{11,22}^{51,1}}{1152}-\frac{1}{288} \sqrt{\frac{7}{6}} C_{20,20}^{40,0}-\frac{1}{576} \sqrt{\frac{7}{6}}
C_{20,20}^{40,1}-\frac{1}{576} \sqrt{\frac{7}{30}} C_{20,22}^{42,0}-\frac{\sqrt{\frac{7}{30}} C_{20,22}^{42,1}}{1152}+\frac{1}{288} \sqrt{\frac{7}{30}}
C_{22,20}^{42,0}+\frac{1}{576} \sqrt{\frac{7}{30}} C_{22,20}^{42,1}-\frac{1}{576} \sqrt{\frac{7}{30}} C_{22,22}^{40,0}-\frac{\sqrt{\frac{7}{30}}
C_{22,22}^{40,1}}{1152}+\frac{C_{22,22}^{42,0}}{144 \sqrt{30}}+\frac{C_{22,22}^{42,1}}{288 \sqrt{30}}+\frac{1}{80} \sqrt{\frac{3}{14}} C_{22,22}^{44,0}+\frac{1}{160}
\sqrt{\frac{3}{14}} C_{22,22}^{44,1}+\frac{1}{288} \sqrt{\frac{7}{6}} C_{31,20}^{31,0}+\frac{1}{576} \sqrt{\frac{7}{6}} C_{31,20}^{31,1}+\frac{1}{576}
\sqrt{\frac{7}{30}} C_{31,22}^{31,0}+\frac{\sqrt{\frac{7}{30}} C_{31,22}^{31,1}}{1152}-\frac{C_{31,22}^{33,0}}{240 \sqrt{2}}-\frac{C_{31,22}^{33,1}}{480
\sqrt{2}}-\frac{C_{33,20}^{33,0}}{160 \sqrt{2}}-\frac{C_{33,20}^{33,1}}{320 \sqrt{2}}-\frac{1}{160} \sqrt{\frac{3}{5}} C_{33,22}^{33,0}-\frac{1}{320}
\sqrt{\frac{3}{5}} C_{33,22}^{33,1}\)

\noindent\(C_{33,1111,3}^{11,1112,0,\text{NonLoc}}= \frac{1}{384} \sqrt{\frac{7}{2}} C_{00,00}^{60,0}+\frac{1}{192} \sqrt{\frac{7}{2}}
C_{00,00}^{60,1}+\frac{1}{384} \sqrt{\frac{7}{6}} C_{00,20}^{60,0}+\frac{1}{192} \sqrt{\frac{7}{6}} C_{00,20}^{60,1}+\frac{1}{128} \sqrt{\frac{7}{30}}
C_{00,22}^{62,0}+\frac{1}{64} \sqrt{\frac{7}{30}} C_{00,22}^{62,1}+\frac{\sqrt{\frac{7}{2}} C_{11,00}^{51,0}}{1152}+\frac{1}{576} \sqrt{\frac{7}{2}}
C_{11,00}^{51,1}+\frac{\sqrt{\frac{35}{6}} C_{11,22}^{51,0}}{1152}+\frac{1}{576} \sqrt{\frac{35}{6}} C_{11,22}^{51,1}-\frac{\sqrt{\frac{7}{2}} C_{20,00}^{40,0}}{1152}-\frac{1}{576}
\sqrt{\frac{7}{2}} C_{20,00}^{40,1}-\frac{\sqrt{\frac{7}{6}} C_{20,20}^{40,0}}{1152}-\frac{1}{576} \sqrt{\frac{7}{6}} C_{20,20}^{40,1}+\frac{\sqrt{\frac{7}{30}}
C_{20,22}^{42,0}}{1152}+\frac{1}{576} \sqrt{\frac{7}{30}} C_{20,22}^{42,1}+\frac{\sqrt{\frac{7}{10}} C_{22,00}^{42,0}}{1152}+\frac{1}{576} \sqrt{\frac{7}{10}}
C_{22,00}^{42,1}+\frac{\sqrt{\frac{7}{30}} C_{22,20}^{42,0}}{1152}+\frac{1}{576} \sqrt{\frac{7}{30}} C_{22,20}^{42,1}+\frac{\sqrt{\frac{7}{30}} C_{22,22}^{40,0}}{1152}+\frac{1}{576}
\sqrt{\frac{7}{30}} C_{22,22}^{40,1}-\frac{C_{22,22}^{42,0}}{288 \sqrt{30}}-\frac{C_{22,22}^{42,1}}{144 \sqrt{30}}-\frac{1}{160} \sqrt{\frac{3}{14}}
C_{22,22}^{44,0}-\frac{1}{80} \sqrt{\frac{3}{14}} C_{22,22}^{44,1}-\frac{\sqrt{\frac{7}{2}} C_{31,00}^{31,0}}{1152}-\frac{1}{576} \sqrt{\frac{7}{2}}
C_{31,00}^{31,1}-\frac{\sqrt{\frac{7}{6}} C_{31,20}^{31,0}}{1152}-\frac{1}{576} \sqrt{\frac{7}{6}} C_{31,20}^{31,1}+\frac{\sqrt{\frac{7}{30}} C_{31,22}^{31,0}}{1152}+\frac{1}{576}
\sqrt{\frac{7}{30}} C_{31,22}^{31,1}-\frac{C_{31,22}^{33,0}}{480 \sqrt{2}}-\frac{C_{31,22}^{33,1}}{240 \sqrt{2}}+\frac{1}{640} \sqrt{\frac{3}{2}}
C_{33,00}^{33,0}+\frac{1}{320} \sqrt{\frac{3}{2}} C_{33,00}^{33,1}+\frac{C_{33,20}^{33,0}}{640 \sqrt{2}}+\frac{C_{33,20}^{33,1}}{320 \sqrt{2}}-\frac{1}{320}
\sqrt{\frac{3}{5}} C_{33,22}^{33,0}-\frac{1}{160} \sqrt{\frac{3}{5}} C_{33,22}^{33,1}\)

\noindent\(C_{33,1111,3}^{11,1112,1,\text{Loc}}= \frac{1}{192} \sqrt{\frac{7}{2}} C_{00,20}^{60,1}-\frac{1}{128} \sqrt{\frac{7}{10}}
C_{00,22}^{62,1}+\frac{\sqrt{\frac{35}{2}} C_{11,22}^{51,1}}{1152}-\frac{1}{576} \sqrt{\frac{7}{2}} C_{20,20}^{40,1}-\frac{\sqrt{\frac{7}{10}} C_{20,22}^{42,1}}{1152}+\frac{1}{576}
\sqrt{\frac{7}{10}} C_{22,20}^{42,1}-\frac{\sqrt{\frac{7}{10}} C_{22,22}^{40,1}}{1152}+\frac{C_{22,22}^{42,1}}{288 \sqrt{10}}+\frac{3 C_{22,22}^{44,1}}{160
\sqrt{14}}+\frac{1}{576} \sqrt{\frac{7}{2}} C_{31,20}^{31,1}+\frac{\sqrt{\frac{7}{10}} C_{31,22}^{31,1}}{1152}-\frac{C_{31,22}^{33,1}}{160 \sqrt{6}}-\frac{1}{320}
\sqrt{\frac{3}{2}} C_{33,20}^{33,1}-\frac{3 C_{33,22}^{33,1}}{320 \sqrt{5}}\)

\noindent\(C_{33,1111,3}^{11,1112,1,\text{NonLoc}}= \frac{1}{128} \sqrt{\frac{7}{6}} C_{00,00}^{60,0}+\frac{1}{384} \sqrt{\frac{7}{2}}
C_{00,20}^{60,0}+\frac{1}{128} \sqrt{\frac{7}{10}} C_{00,22}^{62,0}+\frac{1}{384} \sqrt{\frac{7}{6}} C_{11,00}^{51,0}+\frac{\sqrt{\frac{35}{2}} C_{11,22}^{51,0}}{1152}-\frac{1}{384}
\sqrt{\frac{7}{6}} C_{20,00}^{40,0}-\frac{\sqrt{\frac{7}{2}} C_{20,20}^{40,0}}{1152}+\frac{\sqrt{\frac{7}{10}} C_{20,22}^{42,0}}{1152}+\frac{1}{384}
\sqrt{\frac{7}{30}} C_{22,00}^{42,0}+\frac{\sqrt{\frac{7}{10}} C_{22,20}^{42,0}}{1152}+\frac{\sqrt{\frac{7}{10}} C_{22,22}^{40,0}}{1152}-\frac{C_{22,22}^{42,0}}{288
\sqrt{10}}-\frac{3 C_{22,22}^{44,0}}{160 \sqrt{14}}-\frac{1}{384} \sqrt{\frac{7}{6}} C_{31,00}^{31,0}-\frac{\sqrt{\frac{7}{2}} C_{31,20}^{31,0}}{1152}+\frac{\sqrt{\frac{7}{10}}
C_{31,22}^{31,0}}{1152}-\frac{C_{31,22}^{33,0}}{160 \sqrt{6}}+\frac{3 C_{33,00}^{33,0}}{640 \sqrt{2}}+\frac{1}{640} \sqrt{\frac{3}{2}} C_{33,20}^{33,0}-\frac{3
C_{33,22}^{33,0}}{320 \sqrt{5}}\)

\noindent\(C_{33,1112,1}^{11,1110,0,\text{Loc}}= -\frac{1}{192} \sqrt{\frac{7}{3}} C_{00,20}^{60,0}-\frac{1}{384} \sqrt{\frac{7}{3}}
C_{00,20}^{60,1}-\frac{1}{64} \sqrt{\frac{7}{15}} C_{00,22}^{62,0}-\frac{1}{128} \sqrt{\frac{7}{15}} C_{00,22}^{62,1}+\frac{7 \sqrt{\frac{7}{15}}
C_{11,22}^{51,0}}{1152}+\frac{7 \sqrt{\frac{7}{15}} C_{11,22}^{51,1}}{2304}+\frac{3 C_{11,22}^{53,0}}{1280}+\frac{3 C_{11,22}^{53,1}}{2560}+\frac{1}{576}
\sqrt{\frac{7}{3}} C_{20,20}^{40,0}+\frac{\sqrt{\frac{7}{3}} C_{20,20}^{40,1}}{1152}+\frac{\sqrt{\frac{7}{15}} C_{20,22}^{42,0}}{1152}+\frac{\sqrt{\frac{7}{15}}
C_{20,22}^{42,1}}{2304}-\frac{1}{576} \sqrt{\frac{7}{15}} C_{22,20}^{42,0}-\frac{\sqrt{\frac{7}{15}} C_{22,20}^{42,1}}{1152}+\frac{1}{288} \sqrt{\frac{7}{15}}
C_{22,22}^{40,0}+\frac{1}{576} \sqrt{\frac{7}{15}} C_{22,22}^{40,1}-\frac{1}{576} \sqrt{\frac{5}{3}} C_{22,22}^{42,0}-\frac{\sqrt{\frac{5}{3}} C_{22,22}^{42,1}}{1152}+\frac{1}{160}
\sqrt{\frac{3}{7}} C_{22,22}^{44,0}+\frac{1}{320} \sqrt{\frac{3}{7}} C_{22,22}^{44,1}-\frac{1}{576} \sqrt{\frac{7}{3}} C_{31,20}^{31,0}-\frac{\sqrt{\frac{7}{3}}
C_{31,20}^{31,1}}{1152}-\frac{\sqrt{\frac{7}{15}} C_{31,22}^{31,0}}{1152}-\frac{\sqrt{\frac{7}{15}} C_{31,22}^{31,1}}{2304}-\frac{17 C_{31,22}^{33,0}}{3840}-\frac{17
C_{31,22}^{33,1}}{7680}+\frac{C_{33,20}^{33,0}}{320}+\frac{C_{33,20}^{33,1}}{640}+\frac{1}{160} \sqrt{\frac{3}{10}} C_{33,22}^{33,0}+\frac{1}{320}
\sqrt{\frac{3}{10}} C_{33,22}^{33,1}\)

\noindent\(C_{33,1112,1}^{11,1110,0,\text{NonLoc}}= -\frac{\sqrt{7} C_{00,00}^{60,0}}{768}-\frac{\sqrt{7} C_{00,00}^{60,1}}{384}-\frac{1}{768}
\sqrt{\frac{7}{3}} C_{00,20}^{60,0}-\frac{1}{384} \sqrt{\frac{7}{3}} C_{00,20}^{60,1}+\frac{1}{128} \sqrt{\frac{7}{15}} C_{00,22}^{62,0}+\frac{1}{64}
\sqrt{\frac{7}{15}} C_{00,22}^{62,1}-\frac{\sqrt{7} C_{11,00}^{51,0}}{2304}-\frac{\sqrt{7} C_{11,00}^{51,1}}{1152}+\frac{7 \sqrt{\frac{7}{15}} C_{11,22}^{51,0}}{2304}+\frac{7
\sqrt{\frac{7}{15}} C_{11,22}^{51,1}}{1152}+\frac{3 C_{11,22}^{53,0}}{2560}+\frac{3 C_{11,22}^{53,1}}{1280}+\frac{\sqrt{7} C_{20,00}^{40,0}}{2304}+\frac{\sqrt{7}
C_{20,00}^{40,1}}{1152}+\frac{\sqrt{\frac{7}{3}} C_{20,20}^{40,0}}{2304}+\frac{\sqrt{\frac{7}{3}} C_{20,20}^{40,1}}{1152}-\frac{\sqrt{\frac{7}{15}}
C_{20,22}^{42,0}}{2304}-\frac{\sqrt{\frac{7}{15}} C_{20,22}^{42,1}}{1152}-\frac{\sqrt{\frac{7}{5}} C_{22,00}^{42,0}}{2304}-\frac{\sqrt{\frac{7}{5}}
C_{22,00}^{42,1}}{1152}-\frac{\sqrt{\frac{7}{15}} C_{22,20}^{42,0}}{2304}-\frac{\sqrt{\frac{7}{15}} C_{22,20}^{42,1}}{1152}-\frac{1}{576} \sqrt{\frac{7}{15}}
C_{22,22}^{40,0}-\frac{1}{288} \sqrt{\frac{7}{15}} C_{22,22}^{40,1}+\frac{\sqrt{\frac{5}{3}} C_{22,22}^{42,0}}{1152}+\frac{1}{576} \sqrt{\frac{5}{3}}
C_{22,22}^{42,1}-\frac{1}{320} \sqrt{\frac{3}{7}} C_{22,22}^{44,0}-\frac{1}{160} \sqrt{\frac{3}{7}} C_{22,22}^{44,1}+\frac{\sqrt{7} C_{31,00}^{31,0}}{2304}+\frac{\sqrt{7}
C_{31,00}^{31,1}}{1152}+\frac{\sqrt{\frac{7}{3}} C_{31,20}^{31,0}}{2304}+\frac{\sqrt{\frac{7}{3}} C_{31,20}^{31,1}}{1152}-\frac{\sqrt{\frac{7}{15}}
C_{31,22}^{31,0}}{2304}-\frac{\sqrt{\frac{7}{15}} C_{31,22}^{31,1}}{1152}-\frac{17 C_{31,22}^{33,0}}{7680}-\frac{17 C_{31,22}^{33,1}}{3840}-\frac{\sqrt{3}
C_{33,00}^{33,0}}{1280}-\frac{\sqrt{3} C_{33,00}^{33,1}}{640}-\frac{C_{33,20}^{33,0}}{1280}-\frac{C_{33,20}^{33,1}}{640}+\frac{1}{320} \sqrt{\frac{3}{10}}
C_{33,22}^{33,0}+\frac{1}{160} \sqrt{\frac{3}{10}} C_{33,22}^{33,1}\)

\noindent\(C_{33,1112,1}^{11,1110,1,\text{Loc}}= -\frac{\sqrt{7} C_{00,20}^{60,1}}{384}-\frac{1}{128} \sqrt{\frac{7}{5}} C_{00,22}^{62,1}+\frac{7
\sqrt{\frac{7}{5}} C_{11,22}^{51,1}}{2304}+\frac{3 \sqrt{3} C_{11,22}^{53,1}}{2560}+\frac{\sqrt{7} C_{20,20}^{40,1}}{1152}+\frac{\sqrt{\frac{7}{5}}
C_{20,22}^{42,1}}{2304}-\frac{\sqrt{\frac{7}{5}} C_{22,20}^{42,1}}{1152}+\frac{1}{576} \sqrt{\frac{7}{5}} C_{22,22}^{40,1}-\frac{\sqrt{5} C_{22,22}^{42,1}}{1152}+\frac{3
C_{22,22}^{44,1}}{320 \sqrt{7}}-\frac{\sqrt{7} C_{31,20}^{31,1}}{1152}-\frac{\sqrt{\frac{7}{5}} C_{31,22}^{31,1}}{2304}-\frac{17 C_{31,22}^{33,1}}{2560
\sqrt{3}}+\frac{\sqrt{3} C_{33,20}^{33,1}}{640}+\frac{3 C_{33,22}^{33,1}}{320 \sqrt{10}}\)

\noindent\(C_{33,1112,1}^{11,1110,1,\text{NonLoc}}= -\frac{1}{256} \sqrt{\frac{7}{3}} C_{00,00}^{60,0}-\frac{\sqrt{7} C_{00,20}^{60,0}}{768}+\frac{1}{128}
\sqrt{\frac{7}{5}} C_{00,22}^{62,0}-\frac{1}{768} \sqrt{\frac{7}{3}} C_{11,00}^{51,0}+\frac{7 \sqrt{\frac{7}{5}} C_{11,22}^{51,0}}{2304}+\frac{3
\sqrt{3} C_{11,22}^{53,0}}{2560}+\frac{1}{768} \sqrt{\frac{7}{3}} C_{20,00}^{40,0}+\frac{\sqrt{7} C_{20,20}^{40,0}}{2304}-\frac{\sqrt{\frac{7}{5}}
C_{20,22}^{42,0}}{2304}-\frac{1}{768} \sqrt{\frac{7}{15}} C_{22,00}^{42,0}-\frac{\sqrt{\frac{7}{5}} C_{22,20}^{42,0}}{2304}-\frac{1}{576} \sqrt{\frac{7}{5}}
C_{22,22}^{40,0}+\frac{\sqrt{5} C_{22,22}^{42,0}}{1152}-\frac{3 C_{22,22}^{44,0}}{320 \sqrt{7}}+\frac{1}{768} \sqrt{\frac{7}{3}} C_{31,00}^{31,0}+\frac{\sqrt{7}
C_{31,20}^{31,0}}{2304}-\frac{\sqrt{\frac{7}{5}} C_{31,22}^{31,0}}{2304}-\frac{17 C_{31,22}^{33,0}}{2560 \sqrt{3}}-\frac{3 C_{33,00}^{33,0}}{1280}-\frac{\sqrt{3}
C_{33,20}^{33,0}}{1280}+\frac{3 C_{33,22}^{33,0}}{320 \sqrt{10}}\)

\noindent\(C_{33,1112,1}^{11,1111,0,\text{Loc}}= -\frac{\sqrt{7} C_{00,20}^{60,0}}{384}-\frac{\sqrt{7} C_{00,20}^{60,1}}{768}+\frac{1}{256}
\sqrt{\frac{7}{5}} C_{00,22}^{62,0}+\frac{1}{512} \sqrt{\frac{7}{5}} C_{00,22}^{62,1}-\frac{\sqrt{35} C_{11,22}^{51,0}}{2304}-\frac{\sqrt{35} C_{11,22}^{51,1}}{4608}+\frac{\sqrt{7}
C_{20,20}^{40,0}}{1152}+\frac{\sqrt{7} C_{20,20}^{40,1}}{2304}+\frac{\sqrt{\frac{7}{5}} C_{20,22}^{42,0}}{2304}+\frac{\sqrt{\frac{7}{5}} C_{20,22}^{42,1}}{4608}-\frac{\sqrt{\frac{7}{5}}
C_{22,20}^{42,0}}{1152}-\frac{\sqrt{\frac{7}{5}} C_{22,20}^{42,1}}{2304}+\frac{\sqrt{\frac{7}{5}} C_{22,22}^{40,0}}{2304}+\frac{\sqrt{\frac{7}{5}}
C_{22,22}^{40,1}}{4608}-\frac{C_{22,22}^{42,0}}{576 \sqrt{5}}-\frac{C_{22,22}^{42,1}}{1152 \sqrt{5}}-\frac{3 C_{22,22}^{44,0}}{320 \sqrt{7}}-\frac{3
C_{22,22}^{44,1}}{640 \sqrt{7}}-\frac{\sqrt{7} C_{31,20}^{31,0}}{1152}-\frac{\sqrt{7} C_{31,20}^{31,1}}{2304}-\frac{\sqrt{\frac{7}{5}} C_{31,22}^{31,0}}{2304}-\frac{\sqrt{\frac{7}{5}}
C_{31,22}^{31,1}}{4608}+\frac{C_{31,22}^{33,0}}{320 \sqrt{3}}+\frac{C_{31,22}^{33,1}}{640 \sqrt{3}}+\frac{\sqrt{3} C_{33,20}^{33,0}}{640}+\frac{\sqrt{3}
C_{33,20}^{33,1}}{1280}+\frac{3 C_{33,22}^{33,0}}{320 \sqrt{10}}+\frac{3 C_{33,22}^{33,1}}{640 \sqrt{10}}\)

\noindent\(C_{33,1112,1}^{11,1111,0,\text{NonLoc}}= -\frac{1}{512} \sqrt{\frac{7}{3}} C_{00,00}^{60,0}-\frac{1}{256} \sqrt{\frac{7}{3}}
C_{00,00}^{60,1}-\frac{\sqrt{7} C_{00,20}^{60,0}}{1536}-\frac{\sqrt{7} C_{00,20}^{60,1}}{768}-\frac{1}{512} \sqrt{\frac{7}{5}} C_{00,22}^{62,0}-\frac{1}{256}
\sqrt{\frac{7}{5}} C_{00,22}^{62,1}-\frac{\sqrt{\frac{7}{3}} C_{11,00}^{51,0}}{1536}-\frac{1}{768} \sqrt{\frac{7}{3}} C_{11,00}^{51,1}-\frac{\sqrt{35}
C_{11,22}^{51,0}}{4608}-\frac{\sqrt{35} C_{11,22}^{51,1}}{2304}+\frac{\sqrt{\frac{7}{3}} C_{20,00}^{40,0}}{1536}+\frac{1}{768} \sqrt{\frac{7}{3}}
C_{20,00}^{40,1}+\frac{\sqrt{7} C_{20,20}^{40,0}}{4608}+\frac{\sqrt{7} C_{20,20}^{40,1}}{2304}-\frac{\sqrt{\frac{7}{5}} C_{20,22}^{42,0}}{4608}-\frac{\sqrt{\frac{7}{5}}
C_{20,22}^{42,1}}{2304}-\frac{\sqrt{\frac{7}{15}} C_{22,00}^{42,0}}{1536}-\frac{1}{768} \sqrt{\frac{7}{15}} C_{22,00}^{42,1}-\frac{\sqrt{\frac{7}{5}}
C_{22,20}^{42,0}}{4608}-\frac{\sqrt{\frac{7}{5}} C_{22,20}^{42,1}}{2304}-\frac{\sqrt{\frac{7}{5}} C_{22,22}^{40,0}}{4608}-\frac{\sqrt{\frac{7}{5}}
C_{22,22}^{40,1}}{2304}+\frac{C_{22,22}^{42,0}}{1152 \sqrt{5}}+\frac{C_{22,22}^{42,1}}{576 \sqrt{5}}+\frac{3 C_{22,22}^{44,0}}{640 \sqrt{7}}+\frac{3
C_{22,22}^{44,1}}{320 \sqrt{7}}+\frac{\sqrt{\frac{7}{3}} C_{31,00}^{31,0}}{1536}+\frac{1}{768} \sqrt{\frac{7}{3}} C_{31,00}^{31,1}+\frac{\sqrt{7}
C_{31,20}^{31,0}}{4608}+\frac{\sqrt{7} C_{31,20}^{31,1}}{2304}-\frac{\sqrt{\frac{7}{5}} C_{31,22}^{31,0}}{4608}-\frac{\sqrt{\frac{7}{5}} C_{31,22}^{31,1}}{2304}+\frac{C_{31,22}^{33,0}}{640
\sqrt{3}}+\frac{C_{31,22}^{33,1}}{320 \sqrt{3}}-\frac{3 C_{33,00}^{33,0}}{2560}-\frac{3 C_{33,00}^{33,1}}{1280}-\frac{\sqrt{3} C_{33,20}^{33,0}}{2560}-\frac{\sqrt{3}
C_{33,20}^{33,1}}{1280}+\frac{3 C_{33,22}^{33,0}}{640 \sqrt{10}}+\frac{3 C_{33,22}^{33,1}}{320 \sqrt{10}}\)

\noindent\(C_{33,1112,1}^{11,1111,1,\text{Loc}}= -\frac{1}{256} \sqrt{\frac{7}{3}} C_{00,20}^{60,1}+\frac{1}{512} \sqrt{\frac{21}{5}}
C_{00,22}^{62,1}-\frac{\sqrt{\frac{35}{3}} C_{11,22}^{51,1}}{1536}+\frac{1}{768} \sqrt{\frac{7}{3}} C_{20,20}^{40,1}+\frac{\sqrt{\frac{7}{15}} C_{20,22}^{42,1}}{1536}-\frac{1}{768}
\sqrt{\frac{7}{15}} C_{22,20}^{42,1}+\frac{\sqrt{\frac{7}{15}} C_{22,22}^{40,1}}{1536}-\frac{C_{22,22}^{42,1}}{384 \sqrt{15}}-\frac{3}{640} \sqrt{\frac{3}{7}}
C_{22,22}^{44,1}-\frac{1}{768} \sqrt{\frac{7}{3}} C_{31,20}^{31,1}-\frac{\sqrt{\frac{7}{15}} C_{31,22}^{31,1}}{1536}+\frac{C_{31,22}^{33,1}}{640}+\frac{3
C_{33,20}^{33,1}}{1280}+\frac{3}{640} \sqrt{\frac{3}{10}} C_{33,22}^{33,1}\)

\noindent\(C_{33,1112,1}^{11,1111,1,\text{NonLoc}}= -\frac{\sqrt{7} C_{00,00}^{60,0}}{512}-\frac{1}{512} \sqrt{\frac{7}{3}} C_{00,20}^{60,0}-\frac{1}{512}
\sqrt{\frac{21}{5}} C_{00,22}^{62,0}-\frac{\sqrt{7} C_{11,00}^{51,0}}{1536}-\frac{\sqrt{\frac{35}{3}} C_{11,22}^{51,0}}{1536}+\frac{\sqrt{7} C_{20,00}^{40,0}}{1536}+\frac{\sqrt{\frac{7}{3}}
C_{20,20}^{40,0}}{1536}-\frac{\sqrt{\frac{7}{15}} C_{20,22}^{42,0}}{1536}-\frac{\sqrt{\frac{7}{5}} C_{22,00}^{42,0}}{1536}-\frac{\sqrt{\frac{7}{15}}
C_{22,20}^{42,0}}{1536}-\frac{\sqrt{\frac{7}{15}} C_{22,22}^{40,0}}{1536}+\frac{C_{22,22}^{42,0}}{384 \sqrt{15}}+\frac{3}{640} \sqrt{\frac{3}{7}}
C_{22,22}^{44,0}+\frac{\sqrt{7} C_{31,00}^{31,0}}{1536}+\frac{\sqrt{\frac{7}{3}} C_{31,20}^{31,0}}{1536}-\frac{\sqrt{\frac{7}{15}} C_{31,22}^{31,0}}{1536}+\frac{C_{31,22}^{33,0}}{640}-\frac{3
\sqrt{3} C_{33,00}^{33,0}}{2560}-\frac{3 C_{33,20}^{33,0}}{2560}+\frac{3}{640} \sqrt{\frac{3}{10}} C_{33,22}^{33,0}\)

\noindent\(C_{33,1112,1}^{11,1112,0,\text{Loc}}= -\frac{1}{384} \sqrt{\frac{7}{15}} C_{00,20}^{60,0}-\frac{1}{768} \sqrt{\frac{7}{15}}
C_{00,20}^{60,1}-\frac{11 \sqrt{\frac{7}{3}} C_{00,22}^{62,0}}{1280}-\frac{11 \sqrt{\frac{7}{3}} C_{00,22}^{62,1}}{2560}+\frac{97 \sqrt{\frac{7}{3}}
C_{11,22}^{51,0}}{11520}+\frac{97 \sqrt{\frac{7}{3}} C_{11,22}^{51,1}}{23040}-\frac{21 C_{11,22}^{53,0}}{1280 \sqrt{5}}-\frac{21 C_{11,22}^{53,1}}{2560
\sqrt{5}}+\frac{\sqrt{\frac{7}{15}} C_{20,20}^{40,0}}{1152}+\frac{\sqrt{\frac{7}{15}} C_{20,20}^{40,1}}{2304}+\frac{91 \sqrt{\frac{7}{3}} C_{20,22}^{42,0}}{11520}+\frac{91
\sqrt{\frac{7}{3}} C_{20,22}^{42,1}}{23040}-\frac{\sqrt{\frac{7}{3}} C_{22,20}^{42,0}}{5760}-\frac{\sqrt{\frac{7}{3}} C_{22,20}^{42,1}}{11520}-\frac{19
\sqrt{\frac{7}{3}} C_{22,22}^{40,0}}{2304}-\frac{19 \sqrt{\frac{7}{3}} C_{22,22}^{40,1}}{4608}+\frac{C_{22,22}^{42,0}}{1440 \sqrt{3}}+\frac{C_{22,22}^{42,1}}{2880
\sqrt{3}}-\frac{1}{64} \sqrt{\frac{3}{35}} C_{22,22}^{44,0}-\frac{1}{128} \sqrt{\frac{3}{35}} C_{22,22}^{44,1}-\frac{\sqrt{\frac{7}{15}} C_{31,20}^{31,0}}{1152}-\frac{\sqrt{\frac{7}{15}}
C_{31,20}^{31,1}}{2304}-\frac{91 \sqrt{\frac{7}{3}} C_{31,22}^{31,0}}{11520}-\frac{91 \sqrt{\frac{7}{3}} C_{31,22}^{31,1}}{23040}+\frac{59 C_{31,22}^{33,0}}{3840
\sqrt{5}}+\frac{59 C_{31,22}^{33,1}}{7680 \sqrt{5}}+\frac{C_{33,20}^{33,0}}{640 \sqrt{5}}+\frac{C_{33,20}^{33,1}}{1280 \sqrt{5}}+\frac{\sqrt{\frac{3}{2}}
C_{33,22}^{33,0}}{1600}+\frac{\sqrt{\frac{3}{2}} C_{33,22}^{33,1}}{3200}\)

\noindent\(C_{33,1112,1}^{11,1112,0,\text{NonLoc}}= -\frac{\sqrt{\frac{7}{5}} C_{00,00}^{60,0}}{1536}-\frac{1}{768} \sqrt{\frac{7}{5}}
C_{00,00}^{60,1}-\frac{\sqrt{\frac{7}{15}} C_{00,20}^{60,0}}{1536}-\frac{1}{768} \sqrt{\frac{7}{15}} C_{00,20}^{60,1}+\frac{11 \sqrt{\frac{7}{3}}
C_{00,22}^{62,0}}{2560}+\frac{11 \sqrt{\frac{7}{3}} C_{00,22}^{62,1}}{1280}-\frac{\sqrt{\frac{7}{5}} C_{11,00}^{51,0}}{4608}-\frac{\sqrt{\frac{7}{5}}
C_{11,00}^{51,1}}{2304}+\frac{97 \sqrt{\frac{7}{3}} C_{11,22}^{51,0}}{23040}+\frac{97 \sqrt{\frac{7}{3}} C_{11,22}^{51,1}}{11520}-\frac{21 C_{11,22}^{53,0}}{2560
\sqrt{5}}-\frac{21 C_{11,22}^{53,1}}{1280 \sqrt{5}}+\frac{\sqrt{\frac{7}{5}} C_{20,00}^{40,0}}{4608}+\frac{\sqrt{\frac{7}{5}} C_{20,00}^{40,1}}{2304}+\frac{\sqrt{\frac{7}{15}}
C_{20,20}^{40,0}}{4608}+\frac{\sqrt{\frac{7}{15}} C_{20,20}^{40,1}}{2304}-\frac{91 \sqrt{\frac{7}{3}} C_{20,22}^{42,0}}{23040}-\frac{91 \sqrt{\frac{7}{3}}
C_{20,22}^{42,1}}{11520}-\frac{\sqrt{7} C_{22,00}^{42,0}}{23040}-\frac{\sqrt{7} C_{22,00}^{42,1}}{11520}-\frac{\sqrt{\frac{7}{3}} C_{22,20}^{42,0}}{23040}-\frac{\sqrt{\frac{7}{3}}
C_{22,20}^{42,1}}{11520}+\frac{19 \sqrt{\frac{7}{3}} C_{22,22}^{40,0}}{4608}+\frac{19 \sqrt{\frac{7}{3}} C_{22,22}^{40,1}}{2304}-\frac{C_{22,22}^{42,0}}{2880
\sqrt{3}}-\frac{C_{22,22}^{42,1}}{1440 \sqrt{3}}+\frac{1}{128} \sqrt{\frac{3}{35}} C_{22,22}^{44,0}+\frac{1}{64} \sqrt{\frac{3}{35}} C_{22,22}^{44,1}+\frac{\sqrt{\frac{7}{5}}
C_{31,00}^{31,0}}{4608}+\frac{\sqrt{\frac{7}{5}} C_{31,00}^{31,1}}{2304}+\frac{\sqrt{\frac{7}{15}} C_{31,20}^{31,0}}{4608}+\frac{\sqrt{\frac{7}{15}}
C_{31,20}^{31,1}}{2304}-\frac{91 \sqrt{\frac{7}{3}} C_{31,22}^{31,0}}{23040}-\frac{91 \sqrt{\frac{7}{3}} C_{31,22}^{31,1}}{11520}+\frac{59 C_{31,22}^{33,0}}{7680
\sqrt{5}}+\frac{59 C_{31,22}^{33,1}}{3840 \sqrt{5}}-\frac{\sqrt{\frac{3}{5}} C_{33,00}^{33,0}}{2560}-\frac{\sqrt{\frac{3}{5}} C_{33,00}^{33,1}}{1280}-\frac{C_{33,20}^{33,0}}{2560
\sqrt{5}}-\frac{C_{33,20}^{33,1}}{1280 \sqrt{5}}+\frac{\sqrt{\frac{3}{2}} C_{33,22}^{33,0}}{3200}+\frac{\sqrt{\frac{3}{2}} C_{33,22}^{33,1}}{1600}\)

\noindent\(C_{33,1112,1}^{11,1112,1,\text{Loc}}= -\frac{1}{768} \sqrt{\frac{7}{5}} C_{00,20}^{60,1}-\frac{11 \sqrt{7} C_{00,22}^{62,1}}{2560}+\frac{97
\sqrt{7} C_{11,22}^{51,1}}{23040}-\frac{21 \sqrt{\frac{3}{5}} C_{11,22}^{53,1}}{2560}+\frac{\sqrt{\frac{7}{5}} C_{20,20}^{40,1}}{2304}+\frac{91 \sqrt{7}
C_{20,22}^{42,1}}{23040}-\frac{\sqrt{7} C_{22,20}^{42,1}}{11520}-\frac{19 \sqrt{7} C_{22,22}^{40,1}}{4608}+\frac{C_{22,22}^{42,1}}{2880}-\frac{3
C_{22,22}^{44,1}}{128 \sqrt{35}}-\frac{\sqrt{\frac{7}{5}} C_{31,20}^{31,1}}{2304}-\frac{91 \sqrt{7} C_{31,22}^{31,1}}{23040}+\frac{59 C_{31,22}^{33,1}}{2560
\sqrt{15}}+\frac{\sqrt{\frac{3}{5}} C_{33,20}^{33,1}}{1280}+\frac{3 C_{33,22}^{33,1}}{3200 \sqrt{2}}\)

\noindent\(C_{33,1112,1}^{11,1112,1,\text{NonLoc}}= -\frac{1}{512} \sqrt{\frac{7}{15}} C_{00,00}^{60,0}-\frac{\sqrt{\frac{7}{5}} C_{00,20}^{60,0}}{1536}+\frac{11
\sqrt{7} C_{00,22}^{62,0}}{2560}-\frac{\sqrt{\frac{7}{15}} C_{11,00}^{51,0}}{1536}+\frac{97 \sqrt{7} C_{11,22}^{51,0}}{23040}-\frac{21 \sqrt{\frac{3}{5}}
C_{11,22}^{53,0}}{2560}+\frac{\sqrt{\frac{7}{15}} C_{20,00}^{40,0}}{1536}+\frac{\sqrt{\frac{7}{5}} C_{20,20}^{40,0}}{4608}-\frac{91 \sqrt{7} C_{20,22}^{42,0}}{23040}-\frac{\sqrt{\frac{7}{3}}
C_{22,00}^{42,0}}{7680}-\frac{\sqrt{7} C_{22,20}^{42,0}}{23040}+\frac{19 \sqrt{7} C_{22,22}^{40,0}}{4608}-\frac{C_{22,22}^{42,0}}{2880}+\frac{3 C_{22,22}^{44,0}}{128
\sqrt{35}}+\frac{\sqrt{\frac{7}{15}} C_{31,00}^{31,0}}{1536}+\frac{\sqrt{\frac{7}{5}} C_{31,20}^{31,0}}{4608}-\frac{91 \sqrt{7} C_{31,22}^{31,0}}{23040}+\frac{59
C_{31,22}^{33,0}}{2560 \sqrt{15}}-\frac{3 C_{33,00}^{33,0}}{2560 \sqrt{5}}-\frac{\sqrt{\frac{3}{5}} C_{33,20}^{33,0}}{2560}+\frac{3 C_{33,22}^{33,0}}{3200
\sqrt{2}}\)

\noindent\(C_{33,1112,2}^{00,2202,0,\text{Loc}}= -\frac{1}{96} \sqrt{\frac{7}{6}} C_{11,11}^{51,0}-\frac{1}{192} \sqrt{\frac{7}{6}}
C_{11,11}^{51,1}+\frac{1}{96} \sqrt{\frac{7}{6}} C_{31,11}^{31,0}+\frac{1}{192} \sqrt{\frac{7}{6}} C_{31,11}^{31,1}-\frac{\sqrt{3} C_{33,11}^{33,0}}{320}-\frac{\sqrt{3}
C_{33,11}^{33,1}}{640}\)

\noindent\(C_{33,1112,2}^{00,2202,0,\text{NonLoc}}= -\frac{1}{192} \sqrt{\frac{7}{6}} C_{11,11}^{51,0}-\frac{1}{96} \sqrt{\frac{7}{6}}
C_{11,11}^{51,1}+\frac{1}{192} \sqrt{\frac{7}{6}} C_{31,11}^{31,0}+\frac{1}{96} \sqrt{\frac{7}{6}} C_{31,11}^{31,1}-\frac{\sqrt{3} C_{33,11}^{33,0}}{640}-\frac{\sqrt{3}
C_{33,11}^{33,1}}{320}\)

\noindent\(C_{33,1112,2}^{00,2202,1,\text{Loc}}= -\frac{1}{192} \sqrt{\frac{7}{2}} C_{11,11}^{51,1}+\frac{1}{192} \sqrt{\frac{7}{2}}
C_{31,11}^{31,1}-\frac{3 C_{33,11}^{33,1}}{640}\)

\noindent\(C_{33,1112,2}^{00,2202,1,\text{NonLoc}}= -\frac{1}{192} \sqrt{\frac{7}{2}} C_{11,11}^{51,0}+\frac{1}{192} \sqrt{\frac{7}{2}}
C_{31,11}^{31,0}-\frac{3 C_{33,11}^{33,0}}{640}\)

\noindent\(C_{33,1112,2}^{11,1111,0,\text{Loc}}= -\frac{1}{192} \sqrt{\frac{7}{2}} C_{00,20}^{60,0}-\frac{1}{384} \sqrt{\frac{7}{2}}
C_{00,20}^{60,1}+\frac{1}{128} \sqrt{\frac{7}{10}} C_{00,22}^{62,0}+\frac{1}{256} \sqrt{\frac{7}{10}} C_{00,22}^{62,1}-\frac{\sqrt{\frac{35}{2}}
C_{11,22}^{51,0}}{1152}-\frac{\sqrt{\frac{35}{2}} C_{11,22}^{51,1}}{2304}+\frac{1}{576} \sqrt{\frac{7}{2}} C_{20,20}^{40,0}+\frac{\sqrt{\frac{7}{2}}
C_{20,20}^{40,1}}{1152}+\frac{\sqrt{\frac{7}{10}} C_{20,22}^{42,0}}{1152}+\frac{\sqrt{\frac{7}{10}} C_{20,22}^{42,1}}{2304}-\frac{1}{576} \sqrt{\frac{7}{10}}
C_{22,20}^{42,0}-\frac{\sqrt{\frac{7}{10}} C_{22,20}^{42,1}}{1152}+\frac{\sqrt{\frac{7}{10}} C_{22,22}^{40,0}}{1152}+\frac{\sqrt{\frac{7}{10}} C_{22,22}^{40,1}}{2304}-\frac{C_{22,22}^{42,0}}{288
\sqrt{10}}-\frac{C_{22,22}^{42,1}}{576 \sqrt{10}}-\frac{3 C_{22,22}^{44,0}}{160 \sqrt{14}}-\frac{3 C_{22,22}^{44,1}}{320 \sqrt{14}}-\frac{1}{576}
\sqrt{\frac{7}{2}} C_{31,20}^{31,0}-\frac{\sqrt{\frac{7}{2}} C_{31,20}^{31,1}}{1152}-\frac{\sqrt{\frac{7}{10}} C_{31,22}^{31,0}}{1152}-\frac{\sqrt{\frac{7}{10}}
C_{31,22}^{31,1}}{2304}+\frac{C_{31,22}^{33,0}}{160 \sqrt{6}}+\frac{C_{31,22}^{33,1}}{320 \sqrt{6}}+\frac{1}{320} \sqrt{\frac{3}{2}} C_{33,20}^{33,0}+\frac{1}{640}
\sqrt{\frac{3}{2}} C_{33,20}^{33,1}+\frac{3 C_{33,22}^{33,0}}{320 \sqrt{5}}+\frac{3 C_{33,22}^{33,1}}{640 \sqrt{5}}\)

\noindent\(C_{33,1112,2}^{11,1111,0,\text{NonLoc}}= -\frac{1}{256} \sqrt{\frac{7}{6}} C_{00,00}^{60,0}-\frac{1}{128} \sqrt{\frac{7}{6}}
C_{00,00}^{60,1}-\frac{1}{768} \sqrt{\frac{7}{2}} C_{00,20}^{60,0}-\frac{1}{384} \sqrt{\frac{7}{2}} C_{00,20}^{60,1}-\frac{1}{256} \sqrt{\frac{7}{10}}
C_{00,22}^{62,0}-\frac{1}{128} \sqrt{\frac{7}{10}} C_{00,22}^{62,1}-\frac{1}{768} \sqrt{\frac{7}{6}} C_{11,00}^{51,0}-\frac{1}{384} \sqrt{\frac{7}{6}}
C_{11,00}^{51,1}-\frac{\sqrt{\frac{35}{2}} C_{11,22}^{51,0}}{2304}-\frac{\sqrt{\frac{35}{2}} C_{11,22}^{51,1}}{1152}+\frac{1}{768} \sqrt{\frac{7}{6}}
C_{20,00}^{40,0}+\frac{1}{384} \sqrt{\frac{7}{6}} C_{20,00}^{40,1}+\frac{\sqrt{\frac{7}{2}} C_{20,20}^{40,0}}{2304}+\frac{\sqrt{\frac{7}{2}} C_{20,20}^{40,1}}{1152}-\frac{\sqrt{\frac{7}{10}}
C_{20,22}^{42,0}}{2304}-\frac{\sqrt{\frac{7}{10}} C_{20,22}^{42,1}}{1152}-\frac{1}{768} \sqrt{\frac{7}{30}} C_{22,00}^{42,0}-\frac{1}{384} \sqrt{\frac{7}{30}}
C_{22,00}^{42,1}-\frac{\sqrt{\frac{7}{10}} C_{22,20}^{42,0}}{2304}-\frac{\sqrt{\frac{7}{10}} C_{22,20}^{42,1}}{1152}-\frac{\sqrt{\frac{7}{10}} C_{22,22}^{40,0}}{2304}-\frac{\sqrt{\frac{7}{10}}
C_{22,22}^{40,1}}{1152}+\frac{C_{22,22}^{42,0}}{576 \sqrt{10}}+\frac{C_{22,22}^{42,1}}{288 \sqrt{10}}+\frac{3 C_{22,22}^{44,0}}{320 \sqrt{14}}+\frac{3
C_{22,22}^{44,1}}{160 \sqrt{14}}+\frac{1}{768} \sqrt{\frac{7}{6}} C_{31,00}^{31,0}+\frac{1}{384} \sqrt{\frac{7}{6}} C_{31,00}^{31,1}+\frac{\sqrt{\frac{7}{2}}
C_{31,20}^{31,0}}{2304}+\frac{\sqrt{\frac{7}{2}} C_{31,20}^{31,1}}{1152}-\frac{\sqrt{\frac{7}{10}} C_{31,22}^{31,0}}{2304}-\frac{\sqrt{\frac{7}{10}}
C_{31,22}^{31,1}}{1152}+\frac{C_{31,22}^{33,0}}{320 \sqrt{6}}+\frac{C_{31,22}^{33,1}}{160 \sqrt{6}}-\frac{3 C_{33,00}^{33,0}}{1280 \sqrt{2}}-\frac{3
C_{33,00}^{33,1}}{640 \sqrt{2}}-\frac{\sqrt{\frac{3}{2}} C_{33,20}^{33,0}}{1280}-\frac{1}{640} \sqrt{\frac{3}{2}} C_{33,20}^{33,1}+\frac{3 C_{33,22}^{33,0}}{640
\sqrt{5}}+\frac{3 C_{33,22}^{33,1}}{320 \sqrt{5}}\)

\noindent\(C_{33,1112,2}^{11,1111,1,\text{Loc}}= -\frac{1}{128} \sqrt{\frac{7}{6}} C_{00,20}^{60,1}+\frac{1}{256} \sqrt{\frac{21}{10}}
C_{00,22}^{62,1}-\frac{1}{768} \sqrt{\frac{35}{6}} C_{11,22}^{51,1}+\frac{1}{384} \sqrt{\frac{7}{6}} C_{20,20}^{40,1}+\frac{1}{768} \sqrt{\frac{7}{30}}
C_{20,22}^{42,1}-\frac{1}{384} \sqrt{\frac{7}{30}} C_{22,20}^{42,1}+\frac{1}{768} \sqrt{\frac{7}{30}} C_{22,22}^{40,1}-\frac{C_{22,22}^{42,1}}{192
\sqrt{30}}-\frac{3}{320} \sqrt{\frac{3}{14}} C_{22,22}^{44,1}-\frac{1}{384} \sqrt{\frac{7}{6}} C_{31,20}^{31,1}-\frac{1}{768} \sqrt{\frac{7}{30}}
C_{31,22}^{31,1}+\frac{C_{31,22}^{33,1}}{320 \sqrt{2}}+\frac{3 C_{33,20}^{33,1}}{640 \sqrt{2}}+\frac{3}{640} \sqrt{\frac{3}{5}} C_{33,22}^{33,1}\)

\noindent\(C_{33,1112,2}^{11,1111,1,\text{NonLoc}}= -\frac{1}{256} \sqrt{\frac{7}{2}} C_{00,00}^{60,0}-\frac{1}{256} \sqrt{\frac{7}{6}}
C_{00,20}^{60,0}-\frac{1}{256} \sqrt{\frac{21}{10}} C_{00,22}^{62,0}-\frac{1}{768} \sqrt{\frac{7}{2}} C_{11,00}^{51,0}-\frac{1}{768} \sqrt{\frac{35}{6}}
C_{11,22}^{51,0}+\frac{1}{768} \sqrt{\frac{7}{2}} C_{20,00}^{40,0}+\frac{1}{768} \sqrt{\frac{7}{6}} C_{20,20}^{40,0}-\frac{1}{768} \sqrt{\frac{7}{30}}
C_{20,22}^{42,0}-\frac{1}{768} \sqrt{\frac{7}{10}} C_{22,00}^{42,0}-\frac{1}{768} \sqrt{\frac{7}{30}} C_{22,20}^{42,0}-\frac{1}{768} \sqrt{\frac{7}{30}}
C_{22,22}^{40,0}+\frac{C_{22,22}^{42,0}}{192 \sqrt{30}}+\frac{3}{320} \sqrt{\frac{3}{14}} C_{22,22}^{44,0}+\frac{1}{768} \sqrt{\frac{7}{2}} C_{31,00}^{31,0}+\frac{1}{768}
\sqrt{\frac{7}{6}} C_{31,20}^{31,0}-\frac{1}{768} \sqrt{\frac{7}{30}} C_{31,22}^{31,0}+\frac{C_{31,22}^{33,0}}{320 \sqrt{2}}-\frac{3 \sqrt{\frac{3}{2}}
C_{33,00}^{33,0}}{1280}-\frac{3 C_{33,20}^{33,0}}{1280 \sqrt{2}}+\frac{3}{640} \sqrt{\frac{3}{5}} C_{33,22}^{33,0}\)

\noindent\(C_{33,1112,2}^{11,1112,0,\text{Loc}}= -\frac{1}{192} \sqrt{\frac{7}{6}} C_{00,20}^{60,0}-\frac{1}{384} \sqrt{\frac{7}{6}}
C_{00,20}^{60,1}+\frac{1}{128} \sqrt{\frac{7}{30}} C_{00,22}^{62,0}+\frac{1}{256} \sqrt{\frac{7}{30}} C_{00,22}^{62,1}-\frac{11 \sqrt{\frac{7}{30}}
C_{11,22}^{51,0}}{1152}-\frac{11 \sqrt{\frac{7}{30}} C_{11,22}^{51,1}}{2304}+\frac{3 C_{11,22}^{53,0}}{640 \sqrt{2}}+\frac{3 C_{11,22}^{53,1}}{1280
\sqrt{2}}+\frac{1}{576} \sqrt{\frac{7}{6}} C_{20,20}^{40,0}+\frac{\sqrt{\frac{7}{6}} C_{20,20}^{40,1}}{1152}-\frac{17 \sqrt{\frac{7}{30}} C_{20,22}^{42,0}}{1152}-\frac{17
\sqrt{\frac{7}{30}} C_{20,22}^{42,1}}{2304}-\frac{1}{576} \sqrt{\frac{7}{30}} C_{22,20}^{42,0}-\frac{\sqrt{\frac{7}{30}} C_{22,20}^{42,1}}{1152}+\frac{13
\sqrt{\frac{7}{30}} C_{22,22}^{40,0}}{1152}+\frac{13 \sqrt{\frac{7}{30}} C_{22,22}^{40,1}}{2304}+\frac{C_{22,22}^{42,0}}{144 \sqrt{30}}+\frac{C_{22,22}^{42,1}}{288
\sqrt{30}}+\frac{1}{80} \sqrt{\frac{3}{14}} C_{22,22}^{44,0}+\frac{1}{160} \sqrt{\frac{3}{14}} C_{22,22}^{44,1}-\frac{1}{576} \sqrt{\frac{7}{6}}
C_{31,20}^{31,0}-\frac{\sqrt{\frac{7}{6}} C_{31,20}^{31,1}}{1152}+\frac{17 \sqrt{\frac{7}{30}} C_{31,22}^{31,0}}{1152}+\frac{17 \sqrt{\frac{7}{30}}
C_{31,22}^{31,1}}{2304}-\frac{13 C_{31,22}^{33,0}}{1920 \sqrt{2}}-\frac{13 C_{31,22}^{33,1}}{3840 \sqrt{2}}+\frac{C_{33,20}^{33,0}}{320 \sqrt{2}}+\frac{C_{33,20}^{33,1}}{640
\sqrt{2}}-\frac{1}{160} \sqrt{\frac{3}{5}} C_{33,22}^{33,0}-\frac{1}{320} \sqrt{\frac{3}{5}} C_{33,22}^{33,1}\)

\noindent\(C_{33,1112,2}^{11,1112,0,\text{NonLoc}}= -\frac{1}{768} \sqrt{\frac{7}{2}} C_{00,00}^{60,0}-\frac{1}{384} \sqrt{\frac{7}{2}}
C_{00,00}^{60,1}-\frac{1}{768} \sqrt{\frac{7}{6}} C_{00,20}^{60,0}-\frac{1}{384} \sqrt{\frac{7}{6}} C_{00,20}^{60,1}-\frac{1}{256} \sqrt{\frac{7}{30}}
C_{00,22}^{62,0}-\frac{1}{128} \sqrt{\frac{7}{30}} C_{00,22}^{62,1}-\frac{\sqrt{\frac{7}{2}} C_{11,00}^{51,0}}{2304}-\frac{\sqrt{\frac{7}{2}} C_{11,00}^{51,1}}{1152}-\frac{11
\sqrt{\frac{7}{30}} C_{11,22}^{51,0}}{2304}-\frac{11 \sqrt{\frac{7}{30}} C_{11,22}^{51,1}}{1152}+\frac{3 C_{11,22}^{53,0}}{1280 \sqrt{2}}+\frac{3
C_{11,22}^{53,1}}{640 \sqrt{2}}+\frac{\sqrt{\frac{7}{2}} C_{20,00}^{40,0}}{2304}+\frac{\sqrt{\frac{7}{2}} C_{20,00}^{40,1}}{1152}+\frac{\sqrt{\frac{7}{6}}
C_{20,20}^{40,0}}{2304}+\frac{\sqrt{\frac{7}{6}} C_{20,20}^{40,1}}{1152}+\frac{17 \sqrt{\frac{7}{30}} C_{20,22}^{42,0}}{2304}+\frac{17 \sqrt{\frac{7}{30}}
C_{20,22}^{42,1}}{1152}-\frac{\sqrt{\frac{7}{10}} C_{22,00}^{42,0}}{2304}-\frac{\sqrt{\frac{7}{10}} C_{22,00}^{42,1}}{1152}-\frac{\sqrt{\frac{7}{30}}
C_{22,20}^{42,0}}{2304}-\frac{\sqrt{\frac{7}{30}} C_{22,20}^{42,1}}{1152}-\frac{13 \sqrt{\frac{7}{30}} C_{22,22}^{40,0}}{2304}-\frac{13 \sqrt{\frac{7}{30}}
C_{22,22}^{40,1}}{1152}-\frac{C_{22,22}^{42,0}}{288 \sqrt{30}}-\frac{C_{22,22}^{42,1}}{144 \sqrt{30}}-\frac{1}{160} \sqrt{\frac{3}{14}} C_{22,22}^{44,0}-\frac{1}{80}
\sqrt{\frac{3}{14}} C_{22,22}^{44,1}+\frac{\sqrt{\frac{7}{2}} C_{31,00}^{31,0}}{2304}+\frac{\sqrt{\frac{7}{2}} C_{31,00}^{31,1}}{1152}+\frac{\sqrt{\frac{7}{6}}
C_{31,20}^{31,0}}{2304}+\frac{\sqrt{\frac{7}{6}} C_{31,20}^{31,1}}{1152}+\frac{17 \sqrt{\frac{7}{30}} C_{31,22}^{31,0}}{2304}+\frac{17 \sqrt{\frac{7}{30}}
C_{31,22}^{31,1}}{1152}-\frac{13 C_{31,22}^{33,0}}{3840 \sqrt{2}}-\frac{13 C_{31,22}^{33,1}}{1920 \sqrt{2}}-\frac{\sqrt{\frac{3}{2}} C_{33,00}^{33,0}}{1280}-\frac{1}{640}
\sqrt{\frac{3}{2}} C_{33,00}^{33,1}-\frac{C_{33,20}^{33,0}}{1280 \sqrt{2}}-\frac{C_{33,20}^{33,1}}{640 \sqrt{2}}-\frac{1}{320} \sqrt{\frac{3}{5}}
C_{33,22}^{33,0}-\frac{1}{160} \sqrt{\frac{3}{5}} C_{33,22}^{33,1}\)

\noindent\(C_{33,1112,2}^{11,1112,1,\text{Loc}}= -\frac{1}{384} \sqrt{\frac{7}{2}} C_{00,20}^{60,1}+\frac{1}{256} \sqrt{\frac{7}{10}}
C_{00,22}^{62,1}-\frac{11 \sqrt{\frac{7}{10}} C_{11,22}^{51,1}}{2304}+\frac{3 \sqrt{\frac{3}{2}} C_{11,22}^{53,1}}{1280}+\frac{\sqrt{\frac{7}{2}}
C_{20,20}^{40,1}}{1152}-\frac{17 \sqrt{\frac{7}{10}} C_{20,22}^{42,1}}{2304}-\frac{\sqrt{\frac{7}{10}} C_{22,20}^{42,1}}{1152}+\frac{13 \sqrt{\frac{7}{10}}
C_{22,22}^{40,1}}{2304}+\frac{C_{22,22}^{42,1}}{288 \sqrt{10}}+\frac{3 C_{22,22}^{44,1}}{160 \sqrt{14}}-\frac{\sqrt{\frac{7}{2}} C_{31,20}^{31,1}}{1152}+\frac{17
\sqrt{\frac{7}{10}} C_{31,22}^{31,1}}{2304}-\frac{13 C_{31,22}^{33,1}}{1280 \sqrt{6}}+\frac{1}{640} \sqrt{\frac{3}{2}} C_{33,20}^{33,1}-\frac{3 C_{33,22}^{33,1}}{320
\sqrt{5}}\)

\noindent\(C_{33,1112,2}^{11,1112,1,\text{NonLoc}}= -\frac{1}{256} \sqrt{\frac{7}{6}} C_{00,00}^{60,0}-\frac{1}{768} \sqrt{\frac{7}{2}}
C_{00,20}^{60,0}-\frac{1}{256} \sqrt{\frac{7}{10}} C_{00,22}^{62,0}-\frac{1}{768} \sqrt{\frac{7}{6}} C_{11,00}^{51,0}-\frac{11 \sqrt{\frac{7}{10}}
C_{11,22}^{51,0}}{2304}+\frac{3 \sqrt{\frac{3}{2}} C_{11,22}^{53,0}}{1280}+\frac{1}{768} \sqrt{\frac{7}{6}} C_{20,00}^{40,0}+\frac{\sqrt{\frac{7}{2}}
C_{20,20}^{40,0}}{2304}+\frac{17 \sqrt{\frac{7}{10}} C_{20,22}^{42,0}}{2304}-\frac{1}{768} \sqrt{\frac{7}{30}} C_{22,00}^{42,0}-\frac{\sqrt{\frac{7}{10}}
C_{22,20}^{42,0}}{2304}-\frac{13 \sqrt{\frac{7}{10}} C_{22,22}^{40,0}}{2304}-\frac{C_{22,22}^{42,0}}{288 \sqrt{10}}-\frac{3 C_{22,22}^{44,0}}{160
\sqrt{14}}+\frac{1}{768} \sqrt{\frac{7}{6}} C_{31,00}^{31,0}+\frac{\sqrt{\frac{7}{2}} C_{31,20}^{31,0}}{2304}+\frac{17 \sqrt{\frac{7}{10}} C_{31,22}^{31,0}}{2304}-\frac{13
C_{31,22}^{33,0}}{1280 \sqrt{6}}-\frac{3 C_{33,00}^{33,0}}{1280 \sqrt{2}}-\frac{\sqrt{\frac{3}{2}} C_{33,20}^{33,0}}{1280}-\frac{3 C_{33,22}^{33,0}}{320
\sqrt{5}}\)

\noindent\(C_{33,1112,3}^{11,1112,0,\text{Loc}}= -\frac{1}{96} \sqrt{\frac{7}{10}} C_{00,20}^{60,0}-\frac{1}{192} \sqrt{\frac{7}{10}}
C_{00,20}^{60,1}-\frac{1}{320} \sqrt{\frac{7}{2}} C_{00,22}^{62,0}-\frac{1}{640} \sqrt{\frac{7}{2}} C_{00,22}^{62,1}+\frac{7 \sqrt{\frac{7}{2}} C_{11,22}^{51,0}}{2880}+\frac{7
\sqrt{\frac{7}{2}} C_{11,22}^{51,1}}{5760}-\frac{1}{320} \sqrt{\frac{3}{10}} C_{11,22}^{53,0}-\frac{1}{640} \sqrt{\frac{3}{10}} C_{11,22}^{53,1}+\frac{1}{288}
\sqrt{\frac{7}{10}} C_{20,20}^{40,0}+\frac{1}{576} \sqrt{\frac{7}{10}} C_{20,20}^{40,1}+\frac{\sqrt{\frac{7}{2}} C_{20,22}^{42,0}}{2880}+\frac{\sqrt{\frac{7}{2}}
C_{20,22}^{42,1}}{5760}-\frac{\sqrt{\frac{7}{2}} C_{22,20}^{42,0}}{1440}-\frac{\sqrt{\frac{7}{2}} C_{22,20}^{42,1}}{2880}-\frac{1}{576} \sqrt{\frac{7}{2}}
C_{22,22}^{40,0}-\frac{\sqrt{\frac{7}{2}} C_{22,22}^{40,1}}{1152}+\frac{C_{22,22}^{42,0}}{360 \sqrt{2}}+\frac{C_{22,22}^{42,1}}{720 \sqrt{2}}+\frac{C_{22,22}^{44,0}}{32
\sqrt{70}}+\frac{C_{22,22}^{44,1}}{64 \sqrt{70}}-\frac{1}{288} \sqrt{\frac{7}{10}} C_{31,20}^{31,0}-\frac{1}{576} \sqrt{\frac{7}{10}} C_{31,20}^{31,1}-\frac{\sqrt{\frac{7}{2}}
C_{31,22}^{31,0}}{2880}-\frac{\sqrt{\frac{7}{2}} C_{31,22}^{31,1}}{5760}-\frac{C_{31,22}^{33,0}}{320 \sqrt{30}}-\frac{C_{31,22}^{33,1}}{640 \sqrt{30}}+\frac{1}{160}
\sqrt{\frac{3}{10}} C_{33,20}^{33,0}+\frac{1}{320} \sqrt{\frac{3}{10}} C_{33,20}^{33,1}-\frac{9 C_{33,22}^{33,0}}{1600}-\frac{9 C_{33,22}^{33,1}}{3200}\)

\noindent\(C_{33,1112,3}^{11,1112,0,\text{NonLoc}}= -\frac{1}{128} \sqrt{\frac{7}{30}} C_{00,00}^{60,0}-\frac{1}{64} \sqrt{\frac{7}{30}}
C_{00,00}^{60,1}-\frac{1}{384} \sqrt{\frac{7}{10}} C_{00,20}^{60,0}-\frac{1}{192} \sqrt{\frac{7}{10}} C_{00,20}^{60,1}+\frac{1}{640} \sqrt{\frac{7}{2}}
C_{00,22}^{62,0}+\frac{1}{320} \sqrt{\frac{7}{2}} C_{00,22}^{62,1}-\frac{1}{384} \sqrt{\frac{7}{30}} C_{11,00}^{51,0}-\frac{1}{192} \sqrt{\frac{7}{30}}
C_{11,00}^{51,1}+\frac{7 \sqrt{\frac{7}{2}} C_{11,22}^{51,0}}{5760}+\frac{7 \sqrt{\frac{7}{2}} C_{11,22}^{51,1}}{2880}-\frac{1}{640} \sqrt{\frac{3}{10}}
C_{11,22}^{53,0}-\frac{1}{320} \sqrt{\frac{3}{10}} C_{11,22}^{53,1}+\frac{1}{384} \sqrt{\frac{7}{30}} C_{20,00}^{40,0}+\frac{1}{192} \sqrt{\frac{7}{30}}
C_{20,00}^{40,1}+\frac{\sqrt{\frac{7}{10}} C_{20,20}^{40,0}}{1152}+\frac{1}{576} \sqrt{\frac{7}{10}} C_{20,20}^{40,1}-\frac{\sqrt{\frac{7}{2}} C_{20,22}^{42,0}}{5760}-\frac{\sqrt{\frac{7}{2}}
C_{20,22}^{42,1}}{2880}-\frac{\sqrt{\frac{7}{6}} C_{22,00}^{42,0}}{1920}-\frac{1}{960} \sqrt{\frac{7}{6}} C_{22,00}^{42,1}-\frac{\sqrt{\frac{7}{2}}
C_{22,20}^{42,0}}{5760}-\frac{\sqrt{\frac{7}{2}} C_{22,20}^{42,1}}{2880}+\frac{\sqrt{\frac{7}{2}} C_{22,22}^{40,0}}{1152}+\frac{1}{576} \sqrt{\frac{7}{2}}
C_{22,22}^{40,1}-\frac{C_{22,22}^{42,0}}{720 \sqrt{2}}-\frac{C_{22,22}^{42,1}}{360 \sqrt{2}}-\frac{C_{22,22}^{44,0}}{64 \sqrt{70}}-\frac{C_{22,22}^{44,1}}{32
\sqrt{70}}+\frac{1}{384} \sqrt{\frac{7}{30}} C_{31,00}^{31,0}+\frac{1}{192} \sqrt{\frac{7}{30}} C_{31,00}^{31,1}+\frac{\sqrt{\frac{7}{10}} C_{31,20}^{31,0}}{1152}+\frac{1}{576}
\sqrt{\frac{7}{10}} C_{31,20}^{31,1}-\frac{\sqrt{\frac{7}{2}} C_{31,22}^{31,0}}{5760}-\frac{\sqrt{\frac{7}{2}} C_{31,22}^{31,1}}{2880}-\frac{C_{31,22}^{33,0}}{640
\sqrt{30}}-\frac{C_{31,22}^{33,1}}{320 \sqrt{30}}-\frac{3 C_{33,00}^{33,0}}{640 \sqrt{10}}-\frac{3 C_{33,00}^{33,1}}{320 \sqrt{10}}-\frac{1}{640}
\sqrt{\frac{3}{10}} C_{33,20}^{33,0}-\frac{1}{320} \sqrt{\frac{3}{10}} C_{33,20}^{33,1}-\frac{9 C_{33,22}^{33,0}}{3200}-\frac{9 C_{33,22}^{33,1}}{1600}\)

\noindent\(C_{33,1112,3}^{11,1112,1,\text{Loc}}= -\frac{1}{64} \sqrt{\frac{7}{30}} C_{00,20}^{60,1}-\frac{1}{640} \sqrt{\frac{21}{2}}
C_{00,22}^{62,1}+\frac{7 \sqrt{\frac{7}{6}} C_{11,22}^{51,1}}{1920}-\frac{3 C_{11,22}^{53,1}}{640 \sqrt{10}}+\frac{1}{192} \sqrt{\frac{7}{30}} C_{20,20}^{40,1}+\frac{\sqrt{\frac{7}{6}}
C_{20,22}^{42,1}}{1920}-\frac{1}{960} \sqrt{\frac{7}{6}} C_{22,20}^{42,1}-\frac{1}{384} \sqrt{\frac{7}{6}} C_{22,22}^{40,1}+\frac{C_{22,22}^{42,1}}{240
\sqrt{6}}+\frac{1}{64} \sqrt{\frac{3}{70}} C_{22,22}^{44,1}-\frac{1}{192} \sqrt{\frac{7}{30}} C_{31,20}^{31,1}-\frac{\sqrt{\frac{7}{6}} C_{31,22}^{31,1}}{1920}-\frac{C_{31,22}^{33,1}}{640
\sqrt{10}}+\frac{3 C_{33,20}^{33,1}}{320 \sqrt{10}}-\frac{9 \sqrt{3} C_{33,22}^{33,1}}{3200}\)

\noindent\(C_{33,1112,3}^{11,1112,1,\text{NonLoc}}= -\frac{1}{128} \sqrt{\frac{7}{10}} C_{00,00}^{60,0}-\frac{1}{128} \sqrt{\frac{7}{30}}
C_{00,20}^{60,0}+\frac{1}{640} \sqrt{\frac{21}{2}} C_{00,22}^{62,0}-\frac{1}{384} \sqrt{\frac{7}{10}} C_{11,00}^{51,0}+\frac{7 \sqrt{\frac{7}{6}}
C_{11,22}^{51,0}}{1920}-\frac{3 C_{11,22}^{53,0}}{640 \sqrt{10}}+\frac{1}{384} \sqrt{\frac{7}{10}} C_{20,00}^{40,0}+\frac{1}{384} \sqrt{\frac{7}{30}}
C_{20,20}^{40,0}-\frac{\sqrt{\frac{7}{6}} C_{20,22}^{42,0}}{1920}-\frac{\sqrt{\frac{7}{2}} C_{22,00}^{42,0}}{1920}-\frac{\sqrt{\frac{7}{6}} C_{22,20}^{42,0}}{1920}+\frac{1}{384}
\sqrt{\frac{7}{6}} C_{22,22}^{40,0}-\frac{C_{22,22}^{42,0}}{240 \sqrt{6}}-\frac{1}{64} \sqrt{\frac{3}{70}} C_{22,22}^{44,0}+\frac{1}{384} \sqrt{\frac{7}{10}}
C_{31,00}^{31,0}+\frac{1}{384} \sqrt{\frac{7}{30}} C_{31,20}^{31,0}-\frac{\sqrt{\frac{7}{6}} C_{31,22}^{31,0}}{1920}-\frac{C_{31,22}^{33,0}}{640
\sqrt{10}}-\frac{3}{640} \sqrt{\frac{3}{10}} C_{33,00}^{33,0}-\frac{3 C_{33,20}^{33,0}}{640 \sqrt{10}}-\frac{9 \sqrt{3} C_{33,22}^{33,0}}{3200}\)

\noindent\(C_{33,2011,2}^{11,0011,0,\text{Loc}}= \frac{1}{256} \sqrt{\frac{21}{5}} C_{00,22}^{62,0}+\frac{1}{512} \sqrt{\frac{21}{5}}
C_{00,22}^{62,1}-\frac{1}{768} \sqrt{\frac{35}{3}} C_{11,22}^{51,0}-\frac{\sqrt{\frac{35}{3}} C_{11,22}^{51,1}}{1536}+\frac{1}{768} \sqrt{\frac{7}{15}}
C_{20,22}^{42,0}+\frac{\sqrt{\frac{7}{15}} C_{20,22}^{42,1}}{1536}+\frac{1}{768} \sqrt{\frac{7}{15}} C_{22,22}^{40,0}+\frac{\sqrt{\frac{7}{15}} C_{22,22}^{40,1}}{1536}-\frac{C_{22,22}^{42,0}}{192
\sqrt{15}}-\frac{C_{22,22}^{42,1}}{384 \sqrt{15}}-\frac{3}{320} \sqrt{\frac{3}{7}} C_{22,22}^{44,0}-\frac{3}{640} \sqrt{\frac{3}{7}} C_{22,22}^{44,1}-\frac{1}{768}
\sqrt{\frac{7}{15}} C_{31,22}^{31,0}-\frac{\sqrt{\frac{7}{15}} C_{31,22}^{31,1}}{1536}+\frac{C_{31,22}^{33,0}}{320}+\frac{C_{31,22}^{33,1}}{640}+\frac{3}{320}
\sqrt{\frac{3}{10}} C_{33,22}^{33,0}+\frac{3}{640} \sqrt{\frac{3}{10}} C_{33,22}^{33,1}\)

\noindent\(C_{33,2011,2}^{11,0011,0,\text{NonLoc}}= -\frac{1}{512} \sqrt{\frac{21}{5}} C_{00,22}^{62,0}-\frac{1}{256} \sqrt{\frac{21}{5}}
C_{00,22}^{62,1}-\frac{\sqrt{\frac{35}{3}} C_{11,22}^{51,0}}{1536}-\frac{1}{768} \sqrt{\frac{35}{3}} C_{11,22}^{51,1}-\frac{\sqrt{\frac{7}{15}} C_{20,22}^{42,0}}{1536}-\frac{1}{768}
\sqrt{\frac{7}{15}} C_{20,22}^{42,1}-\frac{\sqrt{\frac{7}{15}} C_{22,22}^{40,0}}{1536}-\frac{1}{768} \sqrt{\frac{7}{15}} C_{22,22}^{40,1}+\frac{C_{22,22}^{42,0}}{384
\sqrt{15}}+\frac{C_{22,22}^{42,1}}{192 \sqrt{15}}+\frac{3}{640} \sqrt{\frac{3}{7}} C_{22,22}^{44,0}+\frac{3}{320} \sqrt{\frac{3}{7}} C_{22,22}^{44,1}-\frac{\sqrt{\frac{7}{15}}
C_{31,22}^{31,0}}{1536}-\frac{1}{768} \sqrt{\frac{7}{15}} C_{31,22}^{31,1}+\frac{C_{31,22}^{33,0}}{640}+\frac{C_{31,22}^{33,1}}{320}+\frac{3}{640}
\sqrt{\frac{3}{10}} C_{33,22}^{33,0}+\frac{3}{320} \sqrt{\frac{3}{10}} C_{33,22}^{33,1}\)

\noindent\(C_{33,2011,2}^{11,0011,1,\text{Loc}}= \frac{3}{512} \sqrt{\frac{7}{5}} C_{00,22}^{62,1}-\frac{\sqrt{35} C_{11,22}^{51,1}}{1536}+\frac{\sqrt{\frac{7}{5}}
C_{20,22}^{42,1}}{1536}+\frac{\sqrt{\frac{7}{5}} C_{22,22}^{40,1}}{1536}-\frac{C_{22,22}^{42,1}}{384 \sqrt{5}}-\frac{9 C_{22,22}^{44,1}}{640 \sqrt{7}}-\frac{\sqrt{\frac{7}{5}}
C_{31,22}^{31,1}}{1536}+\frac{\sqrt{3} C_{31,22}^{33,1}}{640}+\frac{9 C_{33,22}^{33,1}}{640 \sqrt{10}}\)

\noindent\(C_{33,2011,2}^{11,0011,1,\text{NonLoc}}= -\frac{3}{512} \sqrt{\frac{7}{5}} C_{00,22}^{62,0}-\frac{\sqrt{35} C_{11,22}^{51,0}}{1536}-\frac{\sqrt{\frac{7}{5}}
C_{20,22}^{42,0}}{1536}-\frac{\sqrt{\frac{7}{5}} C_{22,22}^{40,0}}{1536}+\frac{C_{22,22}^{42,0}}{384 \sqrt{5}}+\frac{9 C_{22,22}^{44,0}}{640 \sqrt{7}}-\frac{\sqrt{\frac{7}{5}}
C_{31,22}^{31,0}}{1536}+\frac{\sqrt{3} C_{31,22}^{33,0}}{640}+\frac{9 C_{33,22}^{33,0}}{640 \sqrt{10}}\)

\noindent\(C_{33,2202,1}^{00,1111,0,\text{Loc}}= -\frac{1}{96} \sqrt{\frac{7}{3}} C_{11,11}^{51,0}-\frac{1}{192} \sqrt{\frac{7}{3}}
C_{11,11}^{51,1}+\frac{1}{96} \sqrt{\frac{7}{3}} C_{31,11}^{31,0}+\frac{1}{192} \sqrt{\frac{7}{3}} C_{31,11}^{31,1}-\frac{1}{160} \sqrt{\frac{3}{2}}
C_{33,11}^{33,0}-\frac{1}{320} \sqrt{\frac{3}{2}} C_{33,11}^{33,1}\)

\noindent\(C_{33,2202,1}^{00,1111,0,\text{NonLoc}}= -\frac{1}{192} \sqrt{\frac{7}{3}} C_{11,11}^{51,0}-\frac{1}{96} \sqrt{\frac{7}{3}}
C_{11,11}^{51,1}+\frac{1}{192} \sqrt{\frac{7}{3}} C_{31,11}^{31,0}+\frac{1}{96} \sqrt{\frac{7}{3}} C_{31,11}^{31,1}-\frac{1}{320} \sqrt{\frac{3}{2}}
C_{33,11}^{33,0}-\frac{1}{160} \sqrt{\frac{3}{2}} C_{33,11}^{33,1}\)

\noindent\(C_{33,2202,1}^{00,1111,1,\text{Loc}}= -\frac{\sqrt{7} C_{11,11}^{51,1}}{192}+\frac{\sqrt{7} C_{31,11}^{31,1}}{192}-\frac{3
C_{33,11}^{33,1}}{320 \sqrt{2}}\)

\noindent\(C_{33,2202,1}^{00,1111,1,\text{NonLoc}}= -\frac{\sqrt{7} C_{11,11}^{51,0}}{192}+\frac{\sqrt{7} C_{31,11}^{31,0}}{192}-\frac{3
C_{33,11}^{33,0}}{320 \sqrt{2}}\)

\noindent\(C_{33,2202,1}^{11,0000,0,\text{Loc}}= -\frac{1}{128} \sqrt{\frac{7}{3}} C_{00,00}^{60,0}-\frac{1}{256} \sqrt{\frac{7}{3}}
C_{00,00}^{60,1}+\frac{1}{384} \sqrt{\frac{7}{3}} C_{11,00}^{51,0}+\frac{1}{768} \sqrt{\frac{7}{3}} C_{11,00}^{51,1}+\frac{1}{384} \sqrt{\frac{7}{3}}
C_{20,00}^{40,0}+\frac{1}{768} \sqrt{\frac{7}{3}} C_{20,00}^{40,1}-\frac{1}{384} \sqrt{\frac{7}{15}} C_{22,00}^{42,0}-\frac{1}{768} \sqrt{\frac{7}{15}}
C_{22,00}^{42,1}-\frac{1}{384} \sqrt{\frac{7}{3}} C_{31,00}^{31,0}-\frac{1}{768} \sqrt{\frac{7}{3}} C_{31,00}^{31,1}+\frac{3 C_{33,00}^{33,0}}{640}+\frac{3
C_{33,00}^{33,1}}{1280}\)

\noindent\(C_{33,2202,1}^{11,0000,0,\text{NonLoc}}= \frac{1}{512} \sqrt{\frac{7}{3}} C_{00,00}^{60,0}+\frac{1}{256} \sqrt{\frac{7}{3}}
C_{00,00}^{60,1}-\frac{\sqrt{7} C_{00,20}^{60,0}}{512}-\frac{\sqrt{7} C_{00,20}^{60,1}}{256}+\frac{\sqrt{\frac{7}{3}} C_{11,00}^{51,0}}{1536}+\frac{1}{768}
\sqrt{\frac{7}{3}} C_{11,00}^{51,1}-\frac{\sqrt{7} C_{11,20}^{31,0}}{1536}-\frac{\sqrt{7} C_{11,20}^{31,1}}{768}-\frac{\sqrt{\frac{7}{3}} C_{20,00}^{40,0}}{1536}-\frac{1}{768}
\sqrt{\frac{7}{3}} C_{20,00}^{40,1}+\frac{\sqrt{7} C_{20,20}^{40,0}}{1536}+\frac{\sqrt{7} C_{20,20}^{40,1}}{768}+\frac{\sqrt{\frac{7}{15}} C_{22,00}^{42,0}}{1536}+\frac{1}{768}
\sqrt{\frac{7}{15}} C_{22,00}^{42,1}-\frac{\sqrt{\frac{7}{5}} C_{22,20}^{42,0}}{1536}-\frac{1}{768} \sqrt{\frac{7}{5}} C_{22,20}^{42,1}-\frac{\sqrt{\frac{7}{3}}
C_{31,00}^{31,0}}{1536}-\frac{1}{768} \sqrt{\frac{7}{3}} C_{31,00}^{31,1}+\frac{\sqrt{7} C_{31,20}^{31,0}}{1536}+\frac{\sqrt{7} C_{31,20}^{31,1}}{768}+\frac{3
C_{33,00}^{33,0}}{2560}+\frac{3 C_{33,00}^{33,1}}{1280}-\frac{3 \sqrt{3} C_{33,20}^{33,0}}{2560}-\frac{3 \sqrt{3} C_{33,20}^{33,1}}{1280}\)

\noindent\(C_{33,2202,1}^{11,0000,1,\text{Loc}}= -\frac{\sqrt{7} C_{00,00}^{60,1}}{256}+\frac{\sqrt{7} C_{11,00}^{51,1}}{768}+\frac{\sqrt{7}
C_{20,00}^{40,1}}{768}-\frac{1}{768} \sqrt{\frac{7}{5}} C_{22,00}^{42,1}-\frac{\sqrt{7} C_{31,00}^{31,1}}{768}+\frac{3 \sqrt{3} C_{33,00}^{33,1}}{1280}\)

\noindent\(C_{33,2202,1}^{11,0000,1,\text{NonLoc}}= \frac{\sqrt{7} C_{00,00}^{60,0}}{512}-\frac{\sqrt{21} C_{00,20}^{60,0}}{512}+\frac{\sqrt{7}
C_{11,00}^{51,0}}{1536}-\frac{1}{512} \sqrt{\frac{7}{3}} C_{11,20}^{31,0}-\frac{\sqrt{7} C_{20,00}^{40,0}}{1536}+\frac{1}{512} \sqrt{\frac{7}{3}}
C_{20,20}^{40,0}+\frac{\sqrt{\frac{7}{5}} C_{22,00}^{42,0}}{1536}-\frac{1}{512} \sqrt{\frac{7}{15}} C_{22,20}^{42,0}-\frac{\sqrt{7} C_{31,00}^{31,0}}{1536}+\frac{1}{512}
\sqrt{\frac{7}{3}} C_{31,20}^{31,0}+\frac{3 \sqrt{3} C_{33,00}^{33,0}}{2560}-\frac{9 C_{33,20}^{33,0}}{2560}\)

\noindent\(C_{33,2202,2}^{00,1112,0,\text{Loc}}= -\frac{1}{96} \sqrt{\frac{7}{6}} C_{11,11}^{51,0}-\frac{1}{192} \sqrt{\frac{7}{6}}
C_{11,11}^{51,1}+\frac{1}{96} \sqrt{\frac{7}{6}} C_{31,11}^{31,0}+\frac{1}{192} \sqrt{\frac{7}{6}} C_{31,11}^{31,1}-\frac{\sqrt{3} C_{33,11}^{33,0}}{320}-\frac{\sqrt{3}
C_{33,11}^{33,1}}{640}\)

\noindent\(C_{33,2202,2}^{00,1112,0,\text{NonLoc}}= -\frac{1}{192} \sqrt{\frac{7}{6}} C_{11,11}^{51,0}-\frac{1}{96} \sqrt{\frac{7}{6}}
C_{11,11}^{51,1}+\frac{1}{192} \sqrt{\frac{7}{6}} C_{31,11}^{31,0}+\frac{1}{96} \sqrt{\frac{7}{6}} C_{31,11}^{31,1}-\frac{\sqrt{3} C_{33,11}^{33,0}}{640}-\frac{\sqrt{3}
C_{33,11}^{33,1}}{320}\)

\noindent\(C_{33,2202,2}^{00,1112,1,\text{Loc}}= -\frac{1}{192} \sqrt{\frac{7}{2}} C_{11,11}^{51,1}+\frac{1}{192} \sqrt{\frac{7}{2}}
C_{31,11}^{31,1}-\frac{3 C_{33,11}^{33,1}}{640}\)

\noindent\(C_{33,2202,2}^{00,1112,1,\text{NonLoc}}= -\frac{1}{192} \sqrt{\frac{7}{2}} C_{11,11}^{51,0}+\frac{1}{192} \sqrt{\frac{7}{2}}
C_{31,11}^{31,0}-\frac{3 C_{33,11}^{33,0}}{640}\)

\noindent\(C_{33,2211,2}^{11,0011,0,\text{Loc}}= \frac{1}{128} \sqrt{\frac{7}{15}} C_{00,20}^{60,0}+\frac{1}{256} \sqrt{\frac{7}{15}}
C_{00,20}^{60,1}+\frac{\sqrt{21} C_{00,22}^{62,0}}{1280}+\frac{\sqrt{21} C_{00,22}^{62,1}}{2560}-\frac{1}{384} \sqrt{\frac{7}{15}} C_{11,20}^{31,0}-\frac{1}{768}
\sqrt{\frac{7}{15}} C_{11,20}^{31,1}-\frac{\sqrt{\frac{7}{3}} C_{11,22}^{51,0}}{1920}-\frac{\sqrt{\frac{7}{3}} C_{11,22}^{51,1}}{3840}-\frac{9 C_{11,22}^{53,0}}{2560
\sqrt{5}}-\frac{9 C_{11,22}^{53,1}}{5120 \sqrt{5}}-\frac{1}{384} \sqrt{\frac{7}{15}} C_{20,20}^{40,0}-\frac{1}{768} \sqrt{\frac{7}{15}} C_{20,20}^{40,1}-\frac{\sqrt{\frac{7}{3}}
C_{20,22}^{42,0}}{1920}-\frac{\sqrt{\frac{7}{3}} C_{20,22}^{42,1}}{3840}+\frac{\sqrt{\frac{7}{3}} C_{22,20}^{42,0}}{1920}+\frac{\sqrt{\frac{7}{3}}
C_{22,20}^{42,1}}{3840}-\frac{1}{768} \sqrt{\frac{7}{3}} C_{22,22}^{40,0}-\frac{\sqrt{\frac{7}{3}} C_{22,22}^{40,1}}{1536}+\frac{7 C_{22,22}^{42,0}}{1920
\sqrt{3}}+\frac{7 C_{22,22}^{42,1}}{3840 \sqrt{3}}+\frac{1}{384} \sqrt{\frac{7}{15}} C_{31,20}^{31,0}+\frac{1}{768} \sqrt{\frac{7}{15}} C_{31,20}^{31,1}+\frac{\sqrt{\frac{7}{3}}
C_{31,22}^{31,0}}{1920}+\frac{\sqrt{\frac{7}{3}} C_{31,22}^{31,1}}{3840}+\frac{9 C_{31,22}^{33,0}}{2560 \sqrt{5}}+\frac{9 C_{31,22}^{33,1}}{5120
\sqrt{5}}-\frac{3 C_{33,20}^{33,0}}{640 \sqrt{5}}-\frac{3 C_{33,20}^{33,1}}{1280 \sqrt{5}}-\frac{3}{800} \sqrt{\frac{3}{2}} C_{33,22}^{33,0}-\frac{3
\sqrt{\frac{3}{2}} C_{33,22}^{33,1}}{1600}\)

\noindent\(C_{33,2211,2}^{11,0011,0,\text{NonLoc}}= \frac{1}{512} \sqrt{\frac{7}{5}} C_{00,00}^{60,0}+\frac{1}{256} \sqrt{\frac{7}{5}}
C_{00,00}^{60,1}+\frac{1}{512} \sqrt{\frac{7}{15}} C_{00,20}^{60,0}+\frac{1}{256} \sqrt{\frac{7}{15}} C_{00,20}^{60,1}-\frac{\sqrt{21} C_{00,22}^{62,0}}{2560}-\frac{\sqrt{21}
C_{00,22}^{62,1}}{1280}+\frac{\sqrt{\frac{7}{5}} C_{11,00}^{51,0}}{1536}+\frac{1}{768} \sqrt{\frac{7}{5}} C_{11,00}^{51,1}+\frac{\sqrt{\frac{7}{15}}
C_{11,20}^{31,0}}{1536}+\frac{1}{768} \sqrt{\frac{7}{15}} C_{11,20}^{31,1}-\frac{\sqrt{\frac{7}{3}} C_{11,22}^{51,0}}{3840}-\frac{\sqrt{\frac{7}{3}}
C_{11,22}^{51,1}}{1920}-\frac{9 C_{11,22}^{53,0}}{5120 \sqrt{5}}-\frac{9 C_{11,22}^{53,1}}{2560 \sqrt{5}}-\frac{\sqrt{\frac{7}{5}} C_{20,00}^{40,0}}{1536}-\frac{1}{768}
\sqrt{\frac{7}{5}} C_{20,00}^{40,1}-\frac{\sqrt{\frac{7}{15}} C_{20,20}^{40,0}}{1536}-\frac{1}{768} \sqrt{\frac{7}{15}} C_{20,20}^{40,1}+\frac{\sqrt{\frac{7}{3}}
C_{20,22}^{42,0}}{3840}+\frac{\sqrt{\frac{7}{3}} C_{20,22}^{42,1}}{1920}+\frac{\sqrt{7} C_{22,00}^{42,0}}{7680}+\frac{\sqrt{7} C_{22,00}^{42,1}}{3840}+\frac{\sqrt{\frac{7}{3}}
C_{22,20}^{42,0}}{7680}+\frac{\sqrt{\frac{7}{3}} C_{22,20}^{42,1}}{3840}+\frac{\sqrt{\frac{7}{3}} C_{22,22}^{40,0}}{1536}+\frac{1}{768} \sqrt{\frac{7}{3}}
C_{22,22}^{40,1}-\frac{7 C_{22,22}^{42,0}}{3840 \sqrt{3}}-\frac{7 C_{22,22}^{42,1}}{1920 \sqrt{3}}-\frac{\sqrt{\frac{7}{5}} C_{31,00}^{31,0}}{1536}-\frac{1}{768}
\sqrt{\frac{7}{5}} C_{31,00}^{31,1}-\frac{\sqrt{\frac{7}{15}} C_{31,20}^{31,0}}{1536}-\frac{1}{768} \sqrt{\frac{7}{15}} C_{31,20}^{31,1}+\frac{\sqrt{\frac{7}{3}}
C_{31,22}^{31,0}}{3840}+\frac{\sqrt{\frac{7}{3}} C_{31,22}^{31,1}}{1920}+\frac{9 C_{31,22}^{33,0}}{5120 \sqrt{5}}+\frac{9 C_{31,22}^{33,1}}{2560
\sqrt{5}}+\frac{3 \sqrt{\frac{3}{5}} C_{33,00}^{33,0}}{2560}+\frac{3 \sqrt{\frac{3}{5}} C_{33,00}^{33,1}}{1280}+\frac{3 C_{33,20}^{33,0}}{2560 \sqrt{5}}+\frac{3
C_{33,20}^{33,1}}{1280 \sqrt{5}}-\frac{3 \sqrt{\frac{3}{2}} C_{33,22}^{33,0}}{1600}-\frac{3}{800} \sqrt{\frac{3}{2}} C_{33,22}^{33,1}\)

\noindent\(C_{33,2211,2}^{11,0011,1,\text{Loc}}= \frac{1}{256} \sqrt{\frac{7}{5}} C_{00,20}^{60,1}+\frac{3 \sqrt{7} C_{00,22}^{62,1}}{2560}-\frac{1}{768}
\sqrt{\frac{7}{5}} C_{11,20}^{31,1}-\frac{\sqrt{7} C_{11,22}^{51,1}}{3840}-\frac{9 \sqrt{\frac{3}{5}} C_{11,22}^{53,1}}{5120}-\frac{1}{768} \sqrt{\frac{7}{5}}
C_{20,20}^{40,1}-\frac{\sqrt{7} C_{20,22}^{42,1}}{3840}+\frac{\sqrt{7} C_{22,20}^{42,1}}{3840}-\frac{\sqrt{7} C_{22,22}^{40,1}}{1536}+\frac{7 C_{22,22}^{42,1}}{3840}+\frac{1}{768}
\sqrt{\frac{7}{5}} C_{31,20}^{31,1}+\frac{\sqrt{7} C_{31,22}^{31,1}}{3840}+\frac{9 \sqrt{\frac{3}{5}} C_{31,22}^{33,1}}{5120}-\frac{3 \sqrt{\frac{3}{5}}
C_{33,20}^{33,1}}{1280}-\frac{9 C_{33,22}^{33,1}}{1600 \sqrt{2}}\)

\noindent\(C_{33,2211,2}^{11,0011,1,\text{NonLoc}}= \frac{1}{512} \sqrt{\frac{21}{5}} C_{00,00}^{60,0}+\frac{1}{512} \sqrt{\frac{7}{5}}
C_{00,20}^{60,0}-\frac{3 \sqrt{7} C_{00,22}^{62,0}}{2560}+\frac{1}{512} \sqrt{\frac{7}{15}} C_{11,00}^{51,0}+\frac{\sqrt{\frac{7}{5}} C_{11,20}^{31,0}}{1536}-\frac{\sqrt{7}
C_{11,22}^{51,0}}{3840}-\frac{9 \sqrt{\frac{3}{5}} C_{11,22}^{53,0}}{5120}-\frac{1}{512} \sqrt{\frac{7}{15}} C_{20,00}^{40,0}-\frac{\sqrt{\frac{7}{5}}
C_{20,20}^{40,0}}{1536}+\frac{\sqrt{7} C_{20,22}^{42,0}}{3840}+\frac{\sqrt{\frac{7}{3}} C_{22,00}^{42,0}}{2560}+\frac{\sqrt{7} C_{22,20}^{42,0}}{7680}+\frac{\sqrt{7}
C_{22,22}^{40,0}}{1536}-\frac{7 C_{22,22}^{42,0}}{3840}-\frac{1}{512} \sqrt{\frac{7}{15}} C_{31,00}^{31,0}-\frac{\sqrt{\frac{7}{5}} C_{31,20}^{31,0}}{1536}+\frac{\sqrt{7}
C_{31,22}^{31,0}}{3840}+\frac{9 \sqrt{\frac{3}{5}} C_{31,22}^{33,0}}{5120}+\frac{9 C_{33,00}^{33,0}}{2560 \sqrt{5}}+\frac{3 \sqrt{\frac{3}{5}} C_{33,20}^{33,0}}{2560}-\frac{9
C_{33,22}^{33,0}}{1600 \sqrt{2}}\)

\noindent\(C_{33,2212,1}^{00,1101,0,\text{Loc}}= -\frac{\sqrt{7} C_{11,11}^{51,0}}{144}-\frac{\sqrt{7} C_{11,11}^{51,1}}{288}+\frac{\sqrt{7}
C_{31,11}^{31,0}}{144}+\frac{\sqrt{7} C_{31,11}^{31,1}}{288}-\frac{C_{33,11}^{33,0}}{80 \sqrt{2}}-\frac{C_{33,11}^{33,1}}{160 \sqrt{2}}\)

\noindent\(C_{33,2212,1}^{00,1101,0,\text{NonLoc}}= -\frac{\sqrt{7} C_{11,11}^{51,0}}{288}-\frac{\sqrt{7} C_{11,11}^{51,1}}{144}+\frac{\sqrt{7}
C_{31,11}^{31,0}}{288}+\frac{\sqrt{7} C_{31,11}^{31,1}}{144}-\frac{C_{33,11}^{33,0}}{160 \sqrt{2}}-\frac{C_{33,11}^{33,1}}{80 \sqrt{2}}\)

\noindent\(C_{33,2212,1}^{00,1101,1,\text{Loc}}= -\frac{1}{96} \sqrt{\frac{7}{3}} C_{11,11}^{51,1}+\frac{1}{96} \sqrt{\frac{7}{3}}
C_{31,11}^{31,1}-\frac{1}{160} \sqrt{\frac{3}{2}} C_{33,11}^{33,1}\)

\noindent\(C_{33,2212,1}^{00,1101,1,\text{NonLoc}}= -\frac{1}{96} \sqrt{\frac{7}{3}} C_{11,11}^{51,0}+\frac{1}{96} \sqrt{\frac{7}{3}}
C_{31,11}^{31,0}-\frac{1}{160} \sqrt{\frac{3}{2}} C_{33,11}^{33,0}\)

\noindent\(C_{33,2212,1}^{11,0011,0,\text{Loc}}= -\frac{1}{384} \sqrt{\frac{7}{3}} C_{00,20}^{60,0}-\frac{1}{768} \sqrt{\frac{7}{3}}
C_{00,20}^{60,1}+\frac{1}{256} \sqrt{\frac{7}{15}} C_{00,22}^{62,0}+\frac{1}{512} \sqrt{\frac{7}{15}} C_{00,22}^{62,1}+\frac{\sqrt{\frac{7}{3}} C_{11,20}^{31,0}}{1152}+\frac{\sqrt{\frac{7}{3}}
C_{11,20}^{31,1}}{2304}-\frac{\sqrt{\frac{7}{15}} C_{11,22}^{51,0}}{1152}-\frac{\sqrt{\frac{7}{15}} C_{11,22}^{51,1}}{2304}-\frac{3 C_{11,22}^{53,0}}{2560}-\frac{3
C_{11,22}^{53,1}}{5120}+\frac{\sqrt{\frac{7}{3}} C_{20,20}^{40,0}}{1152}+\frac{\sqrt{\frac{7}{3}} C_{20,20}^{40,1}}{2304}-\frac{\sqrt{\frac{7}{15}}
C_{20,22}^{42,0}}{1152}-\frac{\sqrt{\frac{7}{15}} C_{20,22}^{42,1}}{2304}-\frac{\sqrt{\frac{7}{15}} C_{22,20}^{42,0}}{1152}-\frac{\sqrt{\frac{7}{15}}
C_{22,20}^{42,1}}{2304}-\frac{\sqrt{\frac{35}{3}} C_{22,22}^{40,0}}{2304}-\frac{\sqrt{\frac{35}{3}} C_{22,22}^{40,1}}{4608}+\frac{7 C_{22,22}^{42,0}}{1152
\sqrt{15}}+\frac{7 C_{22,22}^{42,1}}{2304 \sqrt{15}}-\frac{\sqrt{\frac{7}{3}} C_{31,20}^{31,0}}{1152}-\frac{\sqrt{\frac{7}{3}} C_{31,20}^{31,1}}{2304}+\frac{\sqrt{\frac{7}{15}}
C_{31,22}^{31,0}}{1152}+\frac{\sqrt{\frac{7}{15}} C_{31,22}^{31,1}}{2304}+\frac{3 C_{31,22}^{33,0}}{2560}+\frac{3 C_{31,22}^{33,1}}{5120}+\frac{C_{33,20}^{33,0}}{640}+\frac{C_{33,20}^{33,1}}{1280}-\frac{1}{160}
\sqrt{\frac{3}{10}} C_{33,22}^{33,0}-\frac{1}{320} \sqrt{\frac{3}{10}} C_{33,22}^{33,1}\)

\noindent\(C_{33,2212,1}^{11,0011,0,\text{NonLoc}}= -\frac{\sqrt{7} C_{00,00}^{60,0}}{1536}-\frac{\sqrt{7} C_{00,00}^{60,1}}{768}-\frac{\sqrt{\frac{7}{3}}
C_{00,20}^{60,0}}{1536}-\frac{1}{768} \sqrt{\frac{7}{3}} C_{00,20}^{60,1}-\frac{1}{512} \sqrt{\frac{7}{15}} C_{00,22}^{62,0}-\frac{1}{256} \sqrt{\frac{7}{15}}
C_{00,22}^{62,1}-\frac{\sqrt{7} C_{11,00}^{51,0}}{4608}-\frac{\sqrt{7} C_{11,00}^{51,1}}{2304}-\frac{\sqrt{\frac{7}{3}} C_{11,20}^{31,0}}{4608}-\frac{\sqrt{\frac{7}{3}}
C_{11,20}^{31,1}}{2304}-\frac{\sqrt{\frac{7}{15}} C_{11,22}^{51,0}}{2304}-\frac{\sqrt{\frac{7}{15}} C_{11,22}^{51,1}}{1152}-\frac{3 C_{11,22}^{53,0}}{5120}-\frac{3
C_{11,22}^{53,1}}{2560}+\frac{\sqrt{7} C_{20,00}^{40,0}}{4608}+\frac{\sqrt{7} C_{20,00}^{40,1}}{2304}+\frac{\sqrt{\frac{7}{3}} C_{20,20}^{40,0}}{4608}+\frac{\sqrt{\frac{7}{3}}
C_{20,20}^{40,1}}{2304}+\frac{\sqrt{\frac{7}{15}} C_{20,22}^{42,0}}{2304}+\frac{\sqrt{\frac{7}{15}} C_{20,22}^{42,1}}{1152}-\frac{\sqrt{\frac{7}{5}}
C_{22,00}^{42,0}}{4608}-\frac{\sqrt{\frac{7}{5}} C_{22,00}^{42,1}}{2304}-\frac{\sqrt{\frac{7}{15}} C_{22,20}^{42,0}}{4608}-\frac{\sqrt{\frac{7}{15}}
C_{22,20}^{42,1}}{2304}+\frac{\sqrt{\frac{35}{3}} C_{22,22}^{40,0}}{4608}+\frac{\sqrt{\frac{35}{3}} C_{22,22}^{40,1}}{2304}-\frac{7 C_{22,22}^{42,0}}{2304
\sqrt{15}}-\frac{7 C_{22,22}^{42,1}}{1152 \sqrt{15}}+\frac{\sqrt{7} C_{31,00}^{31,0}}{4608}+\frac{\sqrt{7} C_{31,00}^{31,1}}{2304}+\frac{\sqrt{\frac{7}{3}}
C_{31,20}^{31,0}}{4608}+\frac{\sqrt{\frac{7}{3}} C_{31,20}^{31,1}}{2304}+\frac{\sqrt{\frac{7}{15}} C_{31,22}^{31,0}}{2304}+\frac{\sqrt{\frac{7}{15}}
C_{31,22}^{31,1}}{1152}+\frac{3 C_{31,22}^{33,0}}{5120}+\frac{3 C_{31,22}^{33,1}}{2560}-\frac{\sqrt{3} C_{33,00}^{33,0}}{2560}-\frac{\sqrt{3} C_{33,00}^{33,1}}{1280}-\frac{C_{33,20}^{33,0}}{2560}-\frac{C_{33,20}^{33,1}}{1280}-\frac{1}{320}
\sqrt{\frac{3}{10}} C_{33,22}^{33,0}-\frac{1}{160} \sqrt{\frac{3}{10}} C_{33,22}^{33,1}\)

\noindent\(C_{33,2212,1}^{11,0011,1,\text{Loc}}= -\frac{\sqrt{7} C_{00,20}^{60,1}}{768}+\frac{1}{512} \sqrt{\frac{7}{5}} C_{00,22}^{62,1}+\frac{\sqrt{7}
C_{11,20}^{31,1}}{2304}-\frac{\sqrt{\frac{7}{5}} C_{11,22}^{51,1}}{2304}-\frac{3 \sqrt{3} C_{11,22}^{53,1}}{5120}+\frac{\sqrt{7} C_{20,20}^{40,1}}{2304}-\frac{\sqrt{\frac{7}{5}}
C_{20,22}^{42,1}}{2304}-\frac{\sqrt{\frac{7}{5}} C_{22,20}^{42,1}}{2304}-\frac{\sqrt{35} C_{22,22}^{40,1}}{4608}+\frac{7 C_{22,22}^{42,1}}{2304 \sqrt{5}}-\frac{\sqrt{7}
C_{31,20}^{31,1}}{2304}+\frac{\sqrt{\frac{7}{5}} C_{31,22}^{31,1}}{2304}+\frac{3 \sqrt{3} C_{31,22}^{33,1}}{5120}+\frac{\sqrt{3} C_{33,20}^{33,1}}{1280}-\frac{3
C_{33,22}^{33,1}}{320 \sqrt{10}}\)

\noindent\(C_{33,2212,1}^{11,0011,1,\text{NonLoc}}= -\frac{1}{512} \sqrt{\frac{7}{3}} C_{00,00}^{60,0}-\frac{\sqrt{7} C_{00,20}^{60,0}}{1536}-\frac{1}{512}
\sqrt{\frac{7}{5}} C_{00,22}^{62,0}-\frac{\sqrt{\frac{7}{3}} C_{11,00}^{51,0}}{1536}-\frac{\sqrt{7} C_{11,20}^{31,0}}{4608}-\frac{\sqrt{\frac{7}{5}}
C_{11,22}^{51,0}}{2304}-\frac{3 \sqrt{3} C_{11,22}^{53,0}}{5120}+\frac{\sqrt{\frac{7}{3}} C_{20,00}^{40,0}}{1536}+\frac{\sqrt{7} C_{20,20}^{40,0}}{4608}+\frac{\sqrt{\frac{7}{5}}
C_{20,22}^{42,0}}{2304}-\frac{\sqrt{\frac{7}{15}} C_{22,00}^{42,0}}{1536}-\frac{\sqrt{\frac{7}{5}} C_{22,20}^{42,0}}{4608}+\frac{\sqrt{35} C_{22,22}^{40,0}}{4608}-\frac{7
C_{22,22}^{42,0}}{2304 \sqrt{5}}+\frac{\sqrt{\frac{7}{3}} C_{31,00}^{31,0}}{1536}+\frac{\sqrt{7} C_{31,20}^{31,0}}{4608}+\frac{\sqrt{\frac{7}{5}}
C_{31,22}^{31,0}}{2304}+\frac{3 \sqrt{3} C_{31,22}^{33,0}}{5120}-\frac{3 C_{33,00}^{33,0}}{2560}-\frac{\sqrt{3} C_{33,20}^{33,0}}{2560}-\frac{3 C_{33,22}^{33,0}}{320
\sqrt{10}}\)

\noindent\(C_{33,2212,2}^{11,0011,0,\text{Loc}}= -\frac{1}{192} \sqrt{\frac{7}{6}} C_{00,20}^{60,0}-\frac{1}{384} \sqrt{\frac{7}{6}}
C_{00,20}^{60,1}+\frac{1}{128} \sqrt{\frac{7}{30}} C_{00,22}^{62,0}+\frac{1}{256} \sqrt{\frac{7}{30}} C_{00,22}^{62,1}+\frac{1}{576} \sqrt{\frac{7}{6}}
C_{11,20}^{31,0}+\frac{\sqrt{\frac{7}{6}} C_{11,20}^{31,1}}{1152}-\frac{1}{576} \sqrt{\frac{7}{30}} C_{11,22}^{51,0}-\frac{\sqrt{\frac{7}{30}} C_{11,22}^{51,1}}{1152}-\frac{3
C_{11,22}^{53,0}}{1280 \sqrt{2}}-\frac{3 C_{11,22}^{53,1}}{2560 \sqrt{2}}+\frac{1}{576} \sqrt{\frac{7}{6}} C_{20,20}^{40,0}+\frac{\sqrt{\frac{7}{6}}
C_{20,20}^{40,1}}{1152}-\frac{1}{576} \sqrt{\frac{7}{30}} C_{20,22}^{42,0}-\frac{\sqrt{\frac{7}{30}} C_{20,22}^{42,1}}{1152}-\frac{1}{576} \sqrt{\frac{7}{30}}
C_{22,20}^{42,0}-\frac{\sqrt{\frac{7}{30}} C_{22,20}^{42,1}}{1152}-\frac{\sqrt{\frac{35}{6}} C_{22,22}^{40,0}}{1152}-\frac{\sqrt{\frac{35}{6}} C_{22,22}^{40,1}}{2304}+\frac{7
C_{22,22}^{42,0}}{576 \sqrt{30}}+\frac{7 C_{22,22}^{42,1}}{1152 \sqrt{30}}-\frac{1}{576} \sqrt{\frac{7}{6}} C_{31,20}^{31,0}-\frac{\sqrt{\frac{7}{6}}
C_{31,20}^{31,1}}{1152}+\frac{1}{576} \sqrt{\frac{7}{30}} C_{31,22}^{31,0}+\frac{\sqrt{\frac{7}{30}} C_{31,22}^{31,1}}{1152}+\frac{3 C_{31,22}^{33,0}}{1280
\sqrt{2}}+\frac{3 C_{31,22}^{33,1}}{2560 \sqrt{2}}+\frac{C_{33,20}^{33,0}}{320 \sqrt{2}}+\frac{C_{33,20}^{33,1}}{640 \sqrt{2}}-\frac{1}{160} \sqrt{\frac{3}{5}}
C_{33,22}^{33,0}-\frac{1}{320} \sqrt{\frac{3}{5}} C_{33,22}^{33,1}\)

\noindent\(C_{33,2212,2}^{11,0011,0,\text{NonLoc}}= -\frac{1}{768} \sqrt{\frac{7}{2}} C_{00,00}^{60,0}-\frac{1}{384} \sqrt{\frac{7}{2}}
C_{00,00}^{60,1}-\frac{1}{768} \sqrt{\frac{7}{6}} C_{00,20}^{60,0}-\frac{1}{384} \sqrt{\frac{7}{6}} C_{00,20}^{60,1}-\frac{1}{256} \sqrt{\frac{7}{30}}
C_{00,22}^{62,0}-\frac{1}{128} \sqrt{\frac{7}{30}} C_{00,22}^{62,1}-\frac{\sqrt{\frac{7}{2}} C_{11,00}^{51,0}}{2304}-\frac{\sqrt{\frac{7}{2}} C_{11,00}^{51,1}}{1152}-\frac{\sqrt{\frac{7}{6}}
C_{11,20}^{31,0}}{2304}-\frac{\sqrt{\frac{7}{6}} C_{11,20}^{31,1}}{1152}-\frac{\sqrt{\frac{7}{30}} C_{11,22}^{51,0}}{1152}-\frac{1}{576} \sqrt{\frac{7}{30}}
C_{11,22}^{51,1}-\frac{3 C_{11,22}^{53,0}}{2560 \sqrt{2}}-\frac{3 C_{11,22}^{53,1}}{1280 \sqrt{2}}+\frac{\sqrt{\frac{7}{2}} C_{20,00}^{40,0}}{2304}+\frac{\sqrt{\frac{7}{2}}
C_{20,00}^{40,1}}{1152}+\frac{\sqrt{\frac{7}{6}} C_{20,20}^{40,0}}{2304}+\frac{\sqrt{\frac{7}{6}} C_{20,20}^{40,1}}{1152}+\frac{\sqrt{\frac{7}{30}}
C_{20,22}^{42,0}}{1152}+\frac{1}{576} \sqrt{\frac{7}{30}} C_{20,22}^{42,1}-\frac{\sqrt{\frac{7}{10}} C_{22,00}^{42,0}}{2304}-\frac{\sqrt{\frac{7}{10}}
C_{22,00}^{42,1}}{1152}-\frac{\sqrt{\frac{7}{30}} C_{22,20}^{42,0}}{2304}-\frac{\sqrt{\frac{7}{30}} C_{22,20}^{42,1}}{1152}+\frac{\sqrt{\frac{35}{6}}
C_{22,22}^{40,0}}{2304}+\frac{\sqrt{\frac{35}{6}} C_{22,22}^{40,1}}{1152}-\frac{7 C_{22,22}^{42,0}}{1152 \sqrt{30}}-\frac{7 C_{22,22}^{42,1}}{576
\sqrt{30}}+\frac{\sqrt{\frac{7}{2}} C_{31,00}^{31,0}}{2304}+\frac{\sqrt{\frac{7}{2}} C_{31,00}^{31,1}}{1152}+\frac{\sqrt{\frac{7}{6}} C_{31,20}^{31,0}}{2304}+\frac{\sqrt{\frac{7}{6}}
C_{31,20}^{31,1}}{1152}+\frac{\sqrt{\frac{7}{30}} C_{31,22}^{31,0}}{1152}+\frac{1}{576} \sqrt{\frac{7}{30}} C_{31,22}^{31,1}+\frac{3 C_{31,22}^{33,0}}{2560
\sqrt{2}}+\frac{3 C_{31,22}^{33,1}}{1280 \sqrt{2}}-\frac{\sqrt{\frac{3}{2}} C_{33,00}^{33,0}}{1280}-\frac{1}{640} \sqrt{\frac{3}{2}} C_{33,00}^{33,1}-\frac{C_{33,20}^{33,0}}{1280
\sqrt{2}}-\frac{C_{33,20}^{33,1}}{640 \sqrt{2}}-\frac{1}{320} \sqrt{\frac{3}{5}} C_{33,22}^{33,0}-\frac{1}{160} \sqrt{\frac{3}{5}} C_{33,22}^{33,1}\)

\noindent\(C_{33,2212,2}^{11,0011,1,\text{Loc}}= -\frac{1}{384} \sqrt{\frac{7}{2}} C_{00,20}^{60,1}+\frac{1}{256} \sqrt{\frac{7}{10}}
C_{00,22}^{62,1}+\frac{\sqrt{\frac{7}{2}} C_{11,20}^{31,1}}{1152}-\frac{\sqrt{\frac{7}{10}} C_{11,22}^{51,1}}{1152}-\frac{3 \sqrt{\frac{3}{2}} C_{11,22}^{53,1}}{2560}+\frac{\sqrt{\frac{7}{2}}
C_{20,20}^{40,1}}{1152}-\frac{\sqrt{\frac{7}{10}} C_{20,22}^{42,1}}{1152}-\frac{\sqrt{\frac{7}{10}} C_{22,20}^{42,1}}{1152}-\frac{\sqrt{\frac{35}{2}}
C_{22,22}^{40,1}}{2304}+\frac{7 C_{22,22}^{42,1}}{1152 \sqrt{10}}-\frac{\sqrt{\frac{7}{2}} C_{31,20}^{31,1}}{1152}+\frac{\sqrt{\frac{7}{10}} C_{31,22}^{31,1}}{1152}+\frac{3
\sqrt{\frac{3}{2}} C_{31,22}^{33,1}}{2560}+\frac{1}{640} \sqrt{\frac{3}{2}} C_{33,20}^{33,1}-\frac{3 C_{33,22}^{33,1}}{320 \sqrt{5}}\)

\noindent\(C_{33,2212,2}^{11,0011,1,\text{NonLoc}}= -\frac{1}{256} \sqrt{\frac{7}{6}} C_{00,00}^{60,0}-\frac{1}{768} \sqrt{\frac{7}{2}}
C_{00,20}^{60,0}-\frac{1}{256} \sqrt{\frac{7}{10}} C_{00,22}^{62,0}-\frac{1}{768} \sqrt{\frac{7}{6}} C_{11,00}^{51,0}-\frac{\sqrt{\frac{7}{2}} C_{11,20}^{31,0}}{2304}-\frac{\sqrt{\frac{7}{10}}
C_{11,22}^{51,0}}{1152}-\frac{3 \sqrt{\frac{3}{2}} C_{11,22}^{53,0}}{2560}+\frac{1}{768} \sqrt{\frac{7}{6}} C_{20,00}^{40,0}+\frac{\sqrt{\frac{7}{2}}
C_{20,20}^{40,0}}{2304}+\frac{\sqrt{\frac{7}{10}} C_{20,22}^{42,0}}{1152}-\frac{1}{768} \sqrt{\frac{7}{30}} C_{22,00}^{42,0}-\frac{\sqrt{\frac{7}{10}}
C_{22,20}^{42,0}}{2304}+\frac{\sqrt{\frac{35}{2}} C_{22,22}^{40,0}}{2304}-\frac{7 C_{22,22}^{42,0}}{1152 \sqrt{10}}+\frac{1}{768} \sqrt{\frac{7}{6}}
C_{31,00}^{31,0}+\frac{\sqrt{\frac{7}{2}} C_{31,20}^{31,0}}{2304}+\frac{\sqrt{\frac{7}{10}} C_{31,22}^{31,0}}{1152}+\frac{3 \sqrt{\frac{3}{2}} C_{31,22}^{33,0}}{2560}-\frac{3
C_{33,00}^{33,0}}{1280 \sqrt{2}}-\frac{\sqrt{\frac{3}{2}} C_{33,20}^{33,0}}{1280}-\frac{3 C_{33,22}^{33,0}}{320 \sqrt{5}}\)

\noindent\(C_{33,2213,0}^{11,0011,0,\text{Loc}}= \frac{1}{384} \sqrt{\frac{7}{3}} C_{00,20}^{60,0}+\frac{1}{768} \sqrt{\frac{7}{3}}
C_{00,20}^{60,1}+\frac{1}{128} \sqrt{\frac{7}{15}} C_{00,22}^{62,0}+\frac{1}{256} \sqrt{\frac{7}{15}} C_{00,22}^{62,1}-\frac{\sqrt{\frac{7}{3}} C_{11,20}^{31,0}}{1152}-\frac{\sqrt{\frac{7}{3}}
C_{11,20}^{31,1}}{2304}-\frac{1}{144} \sqrt{\frac{7}{15}} C_{11,22}^{51,0}-\frac{1}{288} \sqrt{\frac{7}{15}} C_{11,22}^{51,1}+\frac{3 C_{11,22}^{53,0}}{1280}+\frac{3
C_{11,22}^{53,1}}{2560}-\frac{\sqrt{\frac{7}{3}} C_{20,20}^{40,0}}{1152}-\frac{\sqrt{\frac{7}{3}} C_{20,20}^{40,1}}{2304}-\frac{1}{144} \sqrt{\frac{7}{15}}
C_{20,22}^{42,0}-\frac{1}{288} \sqrt{\frac{7}{15}} C_{20,22}^{42,1}+\frac{\sqrt{\frac{7}{15}} C_{22,20}^{42,0}}{1152}+\frac{\sqrt{\frac{7}{15}} C_{22,20}^{42,1}}{2304}+\frac{7
\sqrt{\frac{7}{15}} C_{22,22}^{40,0}}{1152}+\frac{7 \sqrt{\frac{7}{15}} C_{22,22}^{40,1}}{2304}+\frac{C_{22,22}^{42,0}}{576 \sqrt{15}}+\frac{C_{22,22}^{42,1}}{1152
\sqrt{15}}+\frac{1}{320} \sqrt{\frac{3}{7}} C_{22,22}^{44,0}+\frac{1}{640} \sqrt{\frac{3}{7}} C_{22,22}^{44,1}+\frac{\sqrt{\frac{7}{3}} C_{31,20}^{31,0}}{1152}+\frac{\sqrt{\frac{7}{3}}
C_{31,20}^{31,1}}{2304}+\frac{1}{144} \sqrt{\frac{7}{15}} C_{31,22}^{31,0}+\frac{1}{288} \sqrt{\frac{7}{15}} C_{31,22}^{31,1}-\frac{3 C_{31,22}^{33,0}}{1280}-\frac{3
C_{31,22}^{33,1}}{2560}-\frac{C_{33,20}^{33,0}}{640}-\frac{C_{33,20}^{33,1}}{1280}-\frac{1}{320} \sqrt{\frac{3}{10}} C_{33,22}^{33,0}-\frac{1}{640}
\sqrt{\frac{3}{10}} C_{33,22}^{33,1}\)

\noindent\(C_{33,2213,0}^{11,0011,0,\text{NonLoc}}= \frac{\sqrt{7} C_{00,00}^{60,0}}{1536}+\frac{\sqrt{7} C_{00,00}^{60,1}}{768}+\frac{\sqrt{\frac{7}{3}}
C_{00,20}^{60,0}}{1536}+\frac{1}{768} \sqrt{\frac{7}{3}} C_{00,20}^{60,1}-\frac{1}{256} \sqrt{\frac{7}{15}} C_{00,22}^{62,0}-\frac{1}{128} \sqrt{\frac{7}{15}}
C_{00,22}^{62,1}+\frac{\sqrt{7} C_{11,00}^{51,0}}{4608}+\frac{\sqrt{7} C_{11,00}^{51,1}}{2304}+\frac{\sqrt{\frac{7}{3}} C_{11,20}^{31,0}}{4608}+\frac{\sqrt{\frac{7}{3}}
C_{11,20}^{31,1}}{2304}-\frac{1}{288} \sqrt{\frac{7}{15}} C_{11,22}^{51,0}-\frac{1}{144} \sqrt{\frac{7}{15}} C_{11,22}^{51,1}+\frac{3 C_{11,22}^{53,0}}{2560}+\frac{3
C_{11,22}^{53,1}}{1280}-\frac{\sqrt{7} C_{20,00}^{40,0}}{4608}-\frac{\sqrt{7} C_{20,00}^{40,1}}{2304}-\frac{\sqrt{\frac{7}{3}} C_{20,20}^{40,0}}{4608}-\frac{\sqrt{\frac{7}{3}}
C_{20,20}^{40,1}}{2304}+\frac{1}{288} \sqrt{\frac{7}{15}} C_{20,22}^{42,0}+\frac{1}{144} \sqrt{\frac{7}{15}} C_{20,22}^{42,1}+\frac{\sqrt{\frac{7}{5}}
C_{22,00}^{42,0}}{4608}+\frac{\sqrt{\frac{7}{5}} C_{22,00}^{42,1}}{2304}+\frac{\sqrt{\frac{7}{15}} C_{22,20}^{42,0}}{4608}+\frac{\sqrt{\frac{7}{15}}
C_{22,20}^{42,1}}{2304}-\frac{7 \sqrt{\frac{7}{15}} C_{22,22}^{40,0}}{2304}-\frac{7 \sqrt{\frac{7}{15}} C_{22,22}^{40,1}}{1152}-\frac{C_{22,22}^{42,0}}{1152
\sqrt{15}}-\frac{C_{22,22}^{42,1}}{576 \sqrt{15}}-\frac{1}{640} \sqrt{\frac{3}{7}} C_{22,22}^{44,0}-\frac{1}{320} \sqrt{\frac{3}{7}} C_{22,22}^{44,1}-\frac{\sqrt{7}
C_{31,00}^{31,0}}{4608}-\frac{\sqrt{7} C_{31,00}^{31,1}}{2304}-\frac{\sqrt{\frac{7}{3}} C_{31,20}^{31,0}}{4608}-\frac{\sqrt{\frac{7}{3}} C_{31,20}^{31,1}}{2304}+\frac{1}{288}
\sqrt{\frac{7}{15}} C_{31,22}^{31,0}+\frac{1}{144} \sqrt{\frac{7}{15}} C_{31,22}^{31,1}-\frac{3 C_{31,22}^{33,0}}{2560}-\frac{3 C_{31,22}^{33,1}}{1280}+\frac{\sqrt{3}
C_{33,00}^{33,0}}{2560}+\frac{\sqrt{3} C_{33,00}^{33,1}}{1280}+\frac{C_{33,20}^{33,0}}{2560}+\frac{C_{33,20}^{33,1}}{1280}-\frac{1}{640} \sqrt{\frac{3}{10}}
C_{33,22}^{33,0}-\frac{1}{320} \sqrt{\frac{3}{10}} C_{33,22}^{33,1}\)

\noindent\(C_{33,2213,0}^{11,0011,1,\text{Loc}}= \frac{\sqrt{7} C_{00,20}^{60,1}}{768}+\frac{1}{256} \sqrt{\frac{7}{5}} C_{00,22}^{62,1}-\frac{\sqrt{7}
C_{11,20}^{31,1}}{2304}-\frac{1}{288} \sqrt{\frac{7}{5}} C_{11,22}^{51,1}+\frac{3 \sqrt{3} C_{11,22}^{53,1}}{2560}-\frac{\sqrt{7} C_{20,20}^{40,1}}{2304}-\frac{1}{288}
\sqrt{\frac{7}{5}} C_{20,22}^{42,1}+\frac{\sqrt{\frac{7}{5}} C_{22,20}^{42,1}}{2304}+\frac{7 \sqrt{\frac{7}{5}} C_{22,22}^{40,1}}{2304}+\frac{C_{22,22}^{42,1}}{1152
\sqrt{5}}+\frac{3 C_{22,22}^{44,1}}{640 \sqrt{7}}+\frac{\sqrt{7} C_{31,20}^{31,1}}{2304}+\frac{1}{288} \sqrt{\frac{7}{5}} C_{31,22}^{31,1}-\frac{3
\sqrt{3} C_{31,22}^{33,1}}{2560}-\frac{\sqrt{3} C_{33,20}^{33,1}}{1280}-\frac{3 C_{33,22}^{33,1}}{640 \sqrt{10}}\)

\noindent\(C_{33,2213,0}^{11,0011,1,\text{NonLoc}}= \frac{1}{512} \sqrt{\frac{7}{3}} C_{00,00}^{60,0}+\frac{\sqrt{7} C_{00,20}^{60,0}}{1536}-\frac{1}{256}
\sqrt{\frac{7}{5}} C_{00,22}^{62,0}+\frac{\sqrt{\frac{7}{3}} C_{11,00}^{51,0}}{1536}+\frac{\sqrt{7} C_{11,20}^{31,0}}{4608}-\frac{1}{288} \sqrt{\frac{7}{5}}
C_{11,22}^{51,0}+\frac{3 \sqrt{3} C_{11,22}^{53,0}}{2560}-\frac{\sqrt{\frac{7}{3}} C_{20,00}^{40,0}}{1536}-\frac{\sqrt{7} C_{20,20}^{40,0}}{4608}+\frac{1}{288}
\sqrt{\frac{7}{5}} C_{20,22}^{42,0}+\frac{\sqrt{\frac{7}{15}} C_{22,00}^{42,0}}{1536}+\frac{\sqrt{\frac{7}{5}} C_{22,20}^{42,0}}{4608}-\frac{7 \sqrt{\frac{7}{5}}
C_{22,22}^{40,0}}{2304}-\frac{C_{22,22}^{42,0}}{1152 \sqrt{5}}-\frac{3 C_{22,22}^{44,0}}{640 \sqrt{7}}-\frac{\sqrt{\frac{7}{3}} C_{31,00}^{31,0}}{1536}-\frac{\sqrt{7}
C_{31,20}^{31,0}}{4608}+\frac{1}{288} \sqrt{\frac{7}{5}} C_{31,22}^{31,0}-\frac{3 \sqrt{3} C_{31,22}^{33,0}}{2560}+\frac{3 C_{33,00}^{33,0}}{2560}+\frac{\sqrt{3}
C_{33,20}^{33,0}}{2560}-\frac{3 C_{33,22}^{33,0}}{640 \sqrt{10}}\)

\noindent\(C_{33,2213,1}^{00,1101,0,\text{Loc}}= \frac{1}{288} \sqrt{\frac{7}{2}} C_{11,11}^{51,0}+\frac{1}{576} \sqrt{\frac{7}{2}}
C_{11,11}^{51,1}-\frac{1}{288} \sqrt{\frac{7}{2}} C_{31,11}^{31,0}-\frac{1}{576} \sqrt{\frac{7}{2}} C_{31,11}^{31,1}+\frac{C_{33,11}^{33,0}}{320}+\frac{C_{33,11}^{33,1}}{640}\)

\noindent\(C_{33,2213,1}^{00,1101,0,\text{NonLoc}}= \frac{1}{576} \sqrt{\frac{7}{2}} C_{11,11}^{51,0}+\frac{1}{288} \sqrt{\frac{7}{2}}
C_{11,11}^{51,1}-\frac{1}{576} \sqrt{\frac{7}{2}} C_{31,11}^{31,0}-\frac{1}{288} \sqrt{\frac{7}{2}} C_{31,11}^{31,1}+\frac{C_{33,11}^{33,0}}{640}+\frac{C_{33,11}^{33,1}}{320}\)

\noindent\(C_{33,2213,1}^{00,1101,1,\text{Loc}}= \frac{1}{192} \sqrt{\frac{7}{6}} C_{11,11}^{51,1}-\frac{1}{192} \sqrt{\frac{7}{6}}
C_{31,11}^{31,1}+\frac{\sqrt{3} C_{33,11}^{33,1}}{640}\)

\noindent\(C_{33,2213,1}^{00,1101,1,\text{NonLoc}}= \frac{1}{192} \sqrt{\frac{7}{6}} C_{11,11}^{51,0}-\frac{1}{192} \sqrt{\frac{7}{6}}
C_{31,11}^{31,0}+\frac{\sqrt{3} C_{33,11}^{33,0}}{640}\)

\noindent\(C_{33,2213,1}^{11,0011,0,\text{Loc}}= \frac{1}{192} \sqrt{\frac{7}{6}} C_{00,20}^{60,0}+\frac{1}{384} \sqrt{\frac{7}{6}}
C_{00,20}^{60,1}-\frac{1}{128} \sqrt{\frac{7}{30}} C_{00,22}^{62,0}-\frac{1}{256} \sqrt{\frac{7}{30}} C_{00,22}^{62,1}-\frac{1}{576} \sqrt{\frac{7}{6}}
C_{11,20}^{31,0}-\frac{\sqrt{\frac{7}{6}} C_{11,20}^{31,1}}{1152}+\frac{11 \sqrt{\frac{7}{30}} C_{11,22}^{51,0}}{1152}+\frac{11 \sqrt{\frac{7}{30}}
C_{11,22}^{51,1}}{2304}-\frac{3 C_{11,22}^{53,0}}{640 \sqrt{2}}-\frac{3 C_{11,22}^{53,1}}{1280 \sqrt{2}}-\frac{1}{576} \sqrt{\frac{7}{6}} C_{20,20}^{40,0}-\frac{\sqrt{\frac{7}{6}}
C_{20,20}^{40,1}}{1152}+\frac{11 \sqrt{\frac{7}{30}} C_{20,22}^{42,0}}{1152}+\frac{11 \sqrt{\frac{7}{30}} C_{20,22}^{42,1}}{2304}+\frac{1}{576} \sqrt{\frac{7}{30}}
C_{22,20}^{42,0}+\frac{\sqrt{\frac{7}{30}} C_{22,20}^{42,1}}{1152}-\frac{13 \sqrt{\frac{7}{30}} C_{22,22}^{40,0}}{1152}-\frac{13 \sqrt{\frac{7}{30}}
C_{22,22}^{40,1}}{2304}+\frac{C_{22,22}^{42,0}}{288 \sqrt{30}}+\frac{C_{22,22}^{42,1}}{576 \sqrt{30}}-\frac{3}{640} \sqrt{\frac{3}{14}} C_{22,22}^{44,0}-\frac{3
\sqrt{\frac{3}{14}} C_{22,22}^{44,1}}{1280}+\frac{1}{576} \sqrt{\frac{7}{6}} C_{31,20}^{31,0}+\frac{\sqrt{\frac{7}{6}} C_{31,20}^{31,1}}{1152}-\frac{11
\sqrt{\frac{7}{30}} C_{31,22}^{31,0}}{1152}-\frac{11 \sqrt{\frac{7}{30}} C_{31,22}^{31,1}}{2304}+\frac{3 C_{31,22}^{33,0}}{640 \sqrt{2}}+\frac{3
C_{31,22}^{33,1}}{1280 \sqrt{2}}-\frac{C_{33,20}^{33,0}}{320 \sqrt{2}}-\frac{C_{33,20}^{33,1}}{640 \sqrt{2}}-\frac{\sqrt{\frac{3}{5}} C_{33,22}^{33,0}}{1280}-\frac{\sqrt{\frac{3}{5}}
C_{33,22}^{33,1}}{2560}\)

\noindent\(C_{33,2213,1}^{11,0011,0,\text{NonLoc}}= \frac{1}{768} \sqrt{\frac{7}{2}} C_{00,00}^{60,0}+\frac{1}{384} \sqrt{\frac{7}{2}}
C_{00,00}^{60,1}+\frac{1}{768} \sqrt{\frac{7}{6}} C_{00,20}^{60,0}+\frac{1}{384} \sqrt{\frac{7}{6}} C_{00,20}^{60,1}+\frac{1}{256} \sqrt{\frac{7}{30}}
C_{00,22}^{62,0}+\frac{1}{128} \sqrt{\frac{7}{30}} C_{00,22}^{62,1}+\frac{\sqrt{\frac{7}{2}} C_{11,00}^{51,0}}{2304}+\frac{\sqrt{\frac{7}{2}} C_{11,00}^{51,1}}{1152}+\frac{\sqrt{\frac{7}{6}}
C_{11,20}^{31,0}}{2304}+\frac{\sqrt{\frac{7}{6}} C_{11,20}^{31,1}}{1152}+\frac{11 \sqrt{\frac{7}{30}} C_{11,22}^{51,0}}{2304}+\frac{11 \sqrt{\frac{7}{30}}
C_{11,22}^{51,1}}{1152}-\frac{3 C_{11,22}^{53,0}}{1280 \sqrt{2}}-\frac{3 C_{11,22}^{53,1}}{640 \sqrt{2}}-\frac{\sqrt{\frac{7}{2}} C_{20,00}^{40,0}}{2304}-\frac{\sqrt{\frac{7}{2}}
C_{20,00}^{40,1}}{1152}-\frac{\sqrt{\frac{7}{6}} C_{20,20}^{40,0}}{2304}-\frac{\sqrt{\frac{7}{6}} C_{20,20}^{40,1}}{1152}-\frac{11 \sqrt{\frac{7}{30}}
C_{20,22}^{42,0}}{2304}-\frac{11 \sqrt{\frac{7}{30}} C_{20,22}^{42,1}}{1152}+\frac{\sqrt{\frac{7}{10}} C_{22,00}^{42,0}}{2304}+\frac{\sqrt{\frac{7}{10}}
C_{22,00}^{42,1}}{1152}+\frac{\sqrt{\frac{7}{30}} C_{22,20}^{42,0}}{2304}+\frac{\sqrt{\frac{7}{30}} C_{22,20}^{42,1}}{1152}+\frac{13 \sqrt{\frac{7}{30}}
C_{22,22}^{40,0}}{2304}+\frac{13 \sqrt{\frac{7}{30}} C_{22,22}^{40,1}}{1152}-\frac{C_{22,22}^{42,0}}{576 \sqrt{30}}-\frac{C_{22,22}^{42,1}}{288 \sqrt{30}}+\frac{3
\sqrt{\frac{3}{14}} C_{22,22}^{44,0}}{1280}+\frac{3}{640} \sqrt{\frac{3}{14}} C_{22,22}^{44,1}-\frac{\sqrt{\frac{7}{2}} C_{31,00}^{31,0}}{2304}-\frac{\sqrt{\frac{7}{2}}
C_{31,00}^{31,1}}{1152}-\frac{\sqrt{\frac{7}{6}} C_{31,20}^{31,0}}{2304}-\frac{\sqrt{\frac{7}{6}} C_{31,20}^{31,1}}{1152}-\frac{11 \sqrt{\frac{7}{30}}
C_{31,22}^{31,0}}{2304}-\frac{11 \sqrt{\frac{7}{30}} C_{31,22}^{31,1}}{1152}+\frac{3 C_{31,22}^{33,0}}{1280 \sqrt{2}}+\frac{3 C_{31,22}^{33,1}}{640
\sqrt{2}}+\frac{\sqrt{\frac{3}{2}} C_{33,00}^{33,0}}{1280}+\frac{1}{640} \sqrt{\frac{3}{2}} C_{33,00}^{33,1}+\frac{C_{33,20}^{33,0}}{1280 \sqrt{2}}+\frac{C_{33,20}^{33,1}}{640
\sqrt{2}}-\frac{\sqrt{\frac{3}{5}} C_{33,22}^{33,0}}{2560}-\frac{\sqrt{\frac{3}{5}} C_{33,22}^{33,1}}{1280}\)

\noindent\(C_{33,2213,1}^{11,0011,1,\text{Loc}}= \frac{1}{384} \sqrt{\frac{7}{2}} C_{00,20}^{60,1}-\frac{1}{256} \sqrt{\frac{7}{10}}
C_{00,22}^{62,1}-\frac{\sqrt{\frac{7}{2}} C_{11,20}^{31,1}}{1152}+\frac{11 \sqrt{\frac{7}{10}} C_{11,22}^{51,1}}{2304}-\frac{3 \sqrt{\frac{3}{2}}
C_{11,22}^{53,1}}{1280}-\frac{\sqrt{\frac{7}{2}} C_{20,20}^{40,1}}{1152}+\frac{11 \sqrt{\frac{7}{10}} C_{20,22}^{42,1}}{2304}+\frac{\sqrt{\frac{7}{10}}
C_{22,20}^{42,1}}{1152}-\frac{13 \sqrt{\frac{7}{10}} C_{22,22}^{40,1}}{2304}+\frac{C_{22,22}^{42,1}}{576 \sqrt{10}}-\frac{9 C_{22,22}^{44,1}}{1280
\sqrt{14}}+\frac{\sqrt{\frac{7}{2}} C_{31,20}^{31,1}}{1152}-\frac{11 \sqrt{\frac{7}{10}} C_{31,22}^{31,1}}{2304}+\frac{3 \sqrt{\frac{3}{2}} C_{31,22}^{33,1}}{1280}-\frac{1}{640}
\sqrt{\frac{3}{2}} C_{33,20}^{33,1}-\frac{3 C_{33,22}^{33,1}}{2560 \sqrt{5}}\)

\noindent\(C_{33,2213,1}^{11,0011,1,\text{NonLoc}}= \frac{1}{256} \sqrt{\frac{7}{6}} C_{00,00}^{60,0}+\frac{1}{768} \sqrt{\frac{7}{2}}
C_{00,20}^{60,0}+\frac{1}{256} \sqrt{\frac{7}{10}} C_{00,22}^{62,0}+\frac{1}{768} \sqrt{\frac{7}{6}} C_{11,00}^{51,0}+\frac{\sqrt{\frac{7}{2}} C_{11,20}^{31,0}}{2304}+\frac{11
\sqrt{\frac{7}{10}} C_{11,22}^{51,0}}{2304}-\frac{3 \sqrt{\frac{3}{2}} C_{11,22}^{53,0}}{1280}-\frac{1}{768} \sqrt{\frac{7}{6}} C_{20,00}^{40,0}-\frac{\sqrt{\frac{7}{2}}
C_{20,20}^{40,0}}{2304}-\frac{11 \sqrt{\frac{7}{10}} C_{20,22}^{42,0}}{2304}+\frac{1}{768} \sqrt{\frac{7}{30}} C_{22,00}^{42,0}+\frac{\sqrt{\frac{7}{10}}
C_{22,20}^{42,0}}{2304}+\frac{13 \sqrt{\frac{7}{10}} C_{22,22}^{40,0}}{2304}-\frac{C_{22,22}^{42,0}}{576 \sqrt{10}}+\frac{9 C_{22,22}^{44,0}}{1280
\sqrt{14}}-\frac{1}{768} \sqrt{\frac{7}{6}} C_{31,00}^{31,0}-\frac{\sqrt{\frac{7}{2}} C_{31,20}^{31,0}}{2304}-\frac{11 \sqrt{\frac{7}{10}} C_{31,22}^{31,0}}{2304}+\frac{3
\sqrt{\frac{3}{2}} C_{31,22}^{33,0}}{1280}+\frac{3 C_{33,00}^{33,0}}{1280 \sqrt{2}}+\frac{\sqrt{\frac{3}{2}} C_{33,20}^{33,0}}{1280}-\frac{3 C_{33,22}^{33,0}}{2560
\sqrt{5}}\)

\noindent\(C_{33,2213,2}^{11,0011,0,\text{Loc}}= \frac{1}{192} \sqrt{\frac{7}{10}} C_{00,20}^{60,0}+\frac{1}{384} \sqrt{\frac{7}{10}}
C_{00,20}^{60,1}+\frac{1}{640} \sqrt{\frac{7}{2}} C_{00,22}^{62,0}+\frac{\sqrt{\frac{7}{2}} C_{00,22}^{62,1}}{1280}-\frac{1}{576} \sqrt{\frac{7}{10}}
C_{11,20}^{31,0}-\frac{\sqrt{\frac{7}{10}} C_{11,20}^{31,1}}{1152}-\frac{7 \sqrt{\frac{7}{2}} C_{11,22}^{51,0}}{5760}-\frac{7 \sqrt{\frac{7}{2}}
C_{11,22}^{51,1}}{11520}+\frac{1}{640} \sqrt{\frac{3}{10}} C_{11,22}^{53,0}+\frac{\sqrt{\frac{3}{10}} C_{11,22}^{53,1}}{1280}-\frac{1}{576} \sqrt{\frac{7}{10}}
C_{20,20}^{40,0}-\frac{\sqrt{\frac{7}{10}} C_{20,20}^{40,1}}{1152}-\frac{7 \sqrt{\frac{7}{2}} C_{20,22}^{42,0}}{5760}-\frac{7 \sqrt{\frac{7}{2}}
C_{20,22}^{42,1}}{11520}+\frac{\sqrt{\frac{7}{2}} C_{22,20}^{42,0}}{2880}+\frac{\sqrt{\frac{7}{2}} C_{22,20}^{42,1}}{5760}+\frac{\sqrt{\frac{7}{2}}
C_{22,22}^{40,0}}{1152}+\frac{\sqrt{\frac{7}{2}} C_{22,22}^{40,1}}{2304}+\frac{C_{22,22}^{42,0}}{1440 \sqrt{2}}+\frac{C_{22,22}^{42,1}}{2880 \sqrt{2}}+\frac{C_{22,22}^{44,0}}{128
\sqrt{70}}+\frac{C_{22,22}^{44,1}}{256 \sqrt{70}}+\frac{1}{576} \sqrt{\frac{7}{10}} C_{31,20}^{31,0}+\frac{\sqrt{\frac{7}{10}} C_{31,20}^{31,1}}{1152}+\frac{7
\sqrt{\frac{7}{2}} C_{31,22}^{31,0}}{5760}+\frac{7 \sqrt{\frac{7}{2}} C_{31,22}^{31,1}}{11520}-\frac{1}{640} \sqrt{\frac{3}{10}} C_{31,22}^{33,0}-\frac{\sqrt{\frac{3}{10}}
C_{31,22}^{33,1}}{1280}-\frac{1}{320} \sqrt{\frac{3}{10}} C_{33,20}^{33,0}-\frac{1}{640} \sqrt{\frac{3}{10}} C_{33,20}^{33,1}-\frac{9 C_{33,22}^{33,0}}{6400}-\frac{9
C_{33,22}^{33,1}}{12800}\)

\noindent\(C_{33,2213,2}^{11,0011,0,\text{NonLoc}}= \frac{1}{256} \sqrt{\frac{7}{30}} C_{00,00}^{60,0}+\frac{1}{128} \sqrt{\frac{7}{30}}
C_{00,00}^{60,1}+\frac{1}{768} \sqrt{\frac{7}{10}} C_{00,20}^{60,0}+\frac{1}{384} \sqrt{\frac{7}{10}} C_{00,20}^{60,1}-\frac{\sqrt{\frac{7}{2}} C_{00,22}^{62,0}}{1280}-\frac{1}{640}
\sqrt{\frac{7}{2}} C_{00,22}^{62,1}+\frac{1}{768} \sqrt{\frac{7}{30}} C_{11,00}^{51,0}+\frac{1}{384} \sqrt{\frac{7}{30}} C_{11,00}^{51,1}+\frac{\sqrt{\frac{7}{10}}
C_{11,20}^{31,0}}{2304}+\frac{\sqrt{\frac{7}{10}} C_{11,20}^{31,1}}{1152}-\frac{7 \sqrt{\frac{7}{2}} C_{11,22}^{51,0}}{11520}-\frac{7 \sqrt{\frac{7}{2}}
C_{11,22}^{51,1}}{5760}+\frac{\sqrt{\frac{3}{10}} C_{11,22}^{53,0}}{1280}+\frac{1}{640} \sqrt{\frac{3}{10}} C_{11,22}^{53,1}-\frac{1}{768} \sqrt{\frac{7}{30}}
C_{20,00}^{40,0}-\frac{1}{384} \sqrt{\frac{7}{30}} C_{20,00}^{40,1}-\frac{\sqrt{\frac{7}{10}} C_{20,20}^{40,0}}{2304}-\frac{\sqrt{\frac{7}{10}} C_{20,20}^{40,1}}{1152}+\frac{7
\sqrt{\frac{7}{2}} C_{20,22}^{42,0}}{11520}+\frac{7 \sqrt{\frac{7}{2}} C_{20,22}^{42,1}}{5760}+\frac{\sqrt{\frac{7}{6}} C_{22,00}^{42,0}}{3840}+\frac{\sqrt{\frac{7}{6}}
C_{22,00}^{42,1}}{1920}+\frac{\sqrt{\frac{7}{2}} C_{22,20}^{42,0}}{11520}+\frac{\sqrt{\frac{7}{2}} C_{22,20}^{42,1}}{5760}-\frac{\sqrt{\frac{7}{2}}
C_{22,22}^{40,0}}{2304}-\frac{\sqrt{\frac{7}{2}} C_{22,22}^{40,1}}{1152}-\frac{C_{22,22}^{42,0}}{2880 \sqrt{2}}-\frac{C_{22,22}^{42,1}}{1440 \sqrt{2}}-\frac{C_{22,22}^{44,0}}{256
\sqrt{70}}-\frac{C_{22,22}^{44,1}}{128 \sqrt{70}}-\frac{1}{768} \sqrt{\frac{7}{30}} C_{31,00}^{31,0}-\frac{1}{384} \sqrt{\frac{7}{30}} C_{31,00}^{31,1}-\frac{\sqrt{\frac{7}{10}}
C_{31,20}^{31,0}}{2304}-\frac{\sqrt{\frac{7}{10}} C_{31,20}^{31,1}}{1152}+\frac{7 \sqrt{\frac{7}{2}} C_{31,22}^{31,0}}{11520}+\frac{7 \sqrt{\frac{7}{2}}
C_{31,22}^{31,1}}{5760}-\frac{\sqrt{\frac{3}{10}} C_{31,22}^{33,0}}{1280}-\frac{1}{640} \sqrt{\frac{3}{10}} C_{31,22}^{33,1}+\frac{3 C_{33,00}^{33,0}}{1280
\sqrt{10}}+\frac{3 C_{33,00}^{33,1}}{640 \sqrt{10}}+\frac{\sqrt{\frac{3}{10}} C_{33,20}^{33,0}}{1280}+\frac{1}{640} \sqrt{\frac{3}{10}} C_{33,20}^{33,1}-\frac{9
C_{33,22}^{33,0}}{12800}-\frac{9 C_{33,22}^{33,1}}{6400}\)

\noindent\(C_{33,2213,2}^{11,0011,1,\text{Loc}}= \frac{1}{128} \sqrt{\frac{7}{30}} C_{00,20}^{60,1}+\frac{\sqrt{\frac{21}{2}} C_{00,22}^{62,1}}{1280}-\frac{1}{384}
\sqrt{\frac{7}{30}} C_{11,20}^{31,1}-\frac{7 \sqrt{\frac{7}{6}} C_{11,22}^{51,1}}{3840}+\frac{3 C_{11,22}^{53,1}}{1280 \sqrt{10}}-\frac{1}{384} \sqrt{\frac{7}{30}}
C_{20,20}^{40,1}-\frac{7 \sqrt{\frac{7}{6}} C_{20,22}^{42,1}}{3840}+\frac{\sqrt{\frac{7}{6}} C_{22,20}^{42,1}}{1920}+\frac{1}{768} \sqrt{\frac{7}{6}}
C_{22,22}^{40,1}+\frac{C_{22,22}^{42,1}}{960 \sqrt{6}}+\frac{1}{256} \sqrt{\frac{3}{70}} C_{22,22}^{44,1}+\frac{1}{384} \sqrt{\frac{7}{30}} C_{31,20}^{31,1}+\frac{7
\sqrt{\frac{7}{6}} C_{31,22}^{31,1}}{3840}-\frac{3 C_{31,22}^{33,1}}{1280 \sqrt{10}}-\frac{3 C_{33,20}^{33,1}}{640 \sqrt{10}}-\frac{9 \sqrt{3} C_{33,22}^{33,1}}{12800}\)

\noindent\(C_{33,2213,2}^{11,0011,1,\text{NonLoc}}= \frac{1}{256} \sqrt{\frac{7}{10}} C_{00,00}^{60,0}+\frac{1}{256} \sqrt{\frac{7}{30}}
C_{00,20}^{60,0}-\frac{\sqrt{\frac{21}{2}} C_{00,22}^{62,0}}{1280}+\frac{1}{768} \sqrt{\frac{7}{10}} C_{11,00}^{51,0}+\frac{1}{768} \sqrt{\frac{7}{30}}
C_{11,20}^{31,0}-\frac{7 \sqrt{\frac{7}{6}} C_{11,22}^{51,0}}{3840}+\frac{3 C_{11,22}^{53,0}}{1280 \sqrt{10}}-\frac{1}{768} \sqrt{\frac{7}{10}} C_{20,00}^{40,0}-\frac{1}{768}
\sqrt{\frac{7}{30}} C_{20,20}^{40,0}+\frac{7 \sqrt{\frac{7}{6}} C_{20,22}^{42,0}}{3840}+\frac{\sqrt{\frac{7}{2}} C_{22,00}^{42,0}}{3840}+\frac{\sqrt{\frac{7}{6}}
C_{22,20}^{42,0}}{3840}-\frac{1}{768} \sqrt{\frac{7}{6}} C_{22,22}^{40,0}-\frac{C_{22,22}^{42,0}}{960 \sqrt{6}}-\frac{1}{256} \sqrt{\frac{3}{70}}
C_{22,22}^{44,0}-\frac{1}{768} \sqrt{\frac{7}{10}} C_{31,00}^{31,0}-\frac{1}{768} \sqrt{\frac{7}{30}} C_{31,20}^{31,0}+\frac{7 \sqrt{\frac{7}{6}}
C_{31,22}^{31,0}}{3840}-\frac{3 C_{31,22}^{33,0}}{1280 \sqrt{10}}+\frac{3 \sqrt{\frac{3}{10}} C_{33,00}^{33,0}}{1280}+\frac{3 C_{33,20}^{33,0}}{1280
\sqrt{10}}-\frac{9 \sqrt{3} C_{33,22}^{33,0}}{12800}\)

\noindent\(C_{33,3303,1}^{00,0011,0,\text{Loc}}= \frac{1}{288} \sqrt{\frac{7}{2}} C_{11,11}^{51,0}+\frac{1}{576} \sqrt{\frac{7}{2}}
C_{11,11}^{51,1}-\frac{1}{288} \sqrt{\frac{7}{2}} C_{31,11}^{31,0}-\frac{1}{576} \sqrt{\frac{7}{2}} C_{31,11}^{31,1}+\frac{C_{33,11}^{33,0}}{320}+\frac{C_{33,11}^{33,1}}{640}\)

\noindent\(C_{33,3303,1}^{00,0011,0,\text{NonLoc}}= \frac{1}{576} \sqrt{\frac{7}{2}} C_{11,11}^{51,0}+\frac{1}{288} \sqrt{\frac{7}{2}}
C_{11,11}^{51,1}-\frac{1}{576} \sqrt{\frac{7}{2}} C_{31,11}^{31,0}-\frac{1}{288} \sqrt{\frac{7}{2}} C_{31,11}^{31,1}+\frac{C_{33,11}^{33,0}}{640}+\frac{C_{33,11}^{33,1}}{320}\)

\noindent\(C_{33,3303,1}^{00,0011,1,\text{Loc}}= \frac{1}{192} \sqrt{\frac{7}{6}} C_{11,11}^{51,1}-\frac{1}{192} \sqrt{\frac{7}{6}}
C_{31,11}^{31,1}+\frac{\sqrt{3} C_{33,11}^{33,1}}{640}\)

\noindent\(C_{33,3303,1}^{00,0011,1,\text{NonLoc}}= \frac{1}{192} \sqrt{\frac{7}{6}} C_{11,11}^{51,0}-\frac{1}{192} \sqrt{\frac{7}{6}}
C_{31,11}^{31,0}+\frac{\sqrt{3} C_{33,11}^{33,0}}{640}\)

\noindent\(C_{33,3313,0}^{00,0000,0,\text{Loc}}= \frac{1}{288} \sqrt{\frac{7}{2}} C_{11,11}^{51,0}+\frac{1}{576} \sqrt{\frac{7}{2}}
C_{11,11}^{51,1}-\frac{1}{288} \sqrt{\frac{7}{2}} C_{31,11}^{31,0}-\frac{1}{576} \sqrt{\frac{7}{2}} C_{31,11}^{31,1}+\frac{C_{33,11}^{33,0}}{320}+\frac{C_{33,11}^{33,1}}{640}\)

\noindent\(C_{33,3313,0}^{00,0000,0,\text{NonLoc}}= \frac{1}{576} \sqrt{\frac{7}{2}} C_{11,11}^{51,0}+\frac{1}{288} \sqrt{\frac{7}{2}}
C_{11,11}^{51,1}-\frac{1}{576} \sqrt{\frac{7}{2}} C_{31,11}^{31,0}-\frac{1}{288} \sqrt{\frac{7}{2}} C_{31,11}^{31,1}+\frac{C_{33,11}^{33,0}}{640}+\frac{C_{33,11}^{33,1}}{320}\)

\noindent\(C_{33,3313,0}^{00,0000,1,\text{Loc}}= \frac{1}{192} \sqrt{\frac{7}{6}} C_{11,11}^{51,1}-\frac{1}{192} \sqrt{\frac{7}{6}}
C_{31,11}^{31,1}+\frac{\sqrt{3} C_{33,11}^{33,1}}{640}\)

\noindent\(C_{33,3313,0}^{00,0000,1,\text{NonLoc}}= \frac{1}{192} \sqrt{\frac{7}{6}} C_{11,11}^{51,0}-\frac{1}{192} \sqrt{\frac{7}{6}}
C_{31,11}^{31,0}+\frac{\sqrt{3} C_{33,11}^{33,0}}{640}\)

\noindent\(C_{40,0000,0}^{00,2000,0,\text{Loc}}= \frac{7 C_{00,00}^{60,0}}{1024}+\frac{7 C_{00,00}^{60,1}}{2048}-\frac{7 C_{11,00}^{51,0}}{3072}-\frac{7
C_{11,00}^{51,1}}{6144}+\frac{5 C_{20,00}^{40,0}}{3072}+\frac{5 C_{20,00}^{40,1}}{6144}-\frac{7 C_{22,00}^{42,0}}{1536 \sqrt{5}}-\frac{7 C_{22,00}^{42,1}}{3072
\sqrt{5}}+\frac{C_{31,00}^{31,0}}{3072}+\frac{C_{31,00}^{31,1}}{6144}+\frac{3 \sqrt{21} C_{33,00}^{33,0}}{2560}+\frac{3 \sqrt{21} C_{33,00}^{33,1}}{5120}\)

\noindent\(C_{40,0000,0}^{00,2000,0,\text{NonLoc}}= -\frac{7 C_{00,00}^{60,0}}{4096}-\frac{7 C_{00,00}^{60,1}}{2048}+\frac{7 \sqrt{3}
C_{00,20}^{60,0}}{4096}+\frac{7 \sqrt{3} C_{00,20}^{60,1}}{2048}-\frac{7 C_{11,00}^{51,0}}{12288}-\frac{7 C_{11,00}^{51,1}}{6144}+\frac{7 C_{11,20}^{31,0}}{8192
\sqrt{3}}+\frac{7 C_{11,20}^{31,1}}{4096 \sqrt{3}}-\frac{5 C_{20,00}^{40,0}}{12288}-\frac{5 C_{20,00}^{40,1}}{6144}+\frac{5 C_{20,20}^{40,0}}{4096
\sqrt{3}}+\frac{5 C_{20,20}^{40,1}}{2048 \sqrt{3}}+\frac{7 C_{22,00}^{42,0}}{6144 \sqrt{5}}+\frac{7 C_{22,00}^{42,1}}{3072 \sqrt{5}}-\frac{7 C_{22,20}^{42,0}}{2048
\sqrt{15}}-\frac{7 C_{22,20}^{42,1}}{1024 \sqrt{15}}+\frac{C_{31,00}^{31,0}}{12288}+\frac{C_{31,00}^{31,1}}{6144}-\frac{C_{31,20}^{31,0}}{4096 \sqrt{3}}-\frac{C_{31,20}^{31,1}}{2048
\sqrt{3}}+\frac{3 \sqrt{21} C_{33,00}^{33,0}}{10240}+\frac{3 \sqrt{21} C_{33,00}^{33,1}}{5120}-\frac{9 \sqrt{7} C_{33,20}^{33,0}}{10240}-\frac{9
\sqrt{7} C_{33,20}^{33,1}}{5120}\)

\noindent\(C_{40,0000,0}^{00,2000,1,\text{Loc}}= \frac{7 \sqrt{3} C_{00,00}^{60,1}}{2048}-\frac{7 C_{11,00}^{51,1}}{2048 \sqrt{3}}+\frac{5
C_{20,00}^{40,1}}{2048 \sqrt{3}}-\frac{7 C_{22,00}^{42,1}}{1024 \sqrt{15}}+\frac{C_{31,00}^{31,1}}{2048 \sqrt{3}}+\frac{9 \sqrt{7} C_{33,00}^{33,1}}{5120}\)

\noindent\(C_{40,0000,0}^{00,2000,1,\text{NonLoc}}= -\frac{7 \sqrt{3} C_{00,00}^{60,0}}{4096}+\frac{21 C_{00,20}^{60,0}}{4096}-\frac{7
C_{11,00}^{51,0}}{4096 \sqrt{3}}+\frac{7 C_{11,20}^{31,0}}{8192}-\frac{5 C_{20,00}^{40,0}}{4096 \sqrt{3}}+\frac{5 C_{20,20}^{40,0}}{4096}+\frac{7
C_{22,00}^{42,0}}{2048 \sqrt{15}}-\frac{7 C_{22,20}^{42,0}}{2048 \sqrt{5}}+\frac{C_{31,00}^{31,0}}{4096 \sqrt{3}}-\frac{C_{31,20}^{31,0}}{4096}+\frac{9
\sqrt{7} C_{33,00}^{33,0}}{10240}-\frac{9 \sqrt{21} C_{33,20}^{33,0}}{10240}\)

\noindent\(C_{40,0000,0}^{11,1111,0,\text{Loc}}= -\frac{7 C_{11,11}^{51,0}}{3072}-\frac{7 C_{11,11}^{51,1}}{6144}+\frac{7 C_{22,11}^{42,0}}{512
\sqrt{15}}+\frac{7 C_{22,11}^{42,1}}{1024 \sqrt{15}}-\frac{7 C_{31,11}^{31,0}}{3072}-\frac{7 C_{31,11}^{31,1}}{6144}-\frac{3 \sqrt{\frac{7}{2}} C_{33,11}^{33,0}}{1280}-\frac{3
\sqrt{\frac{7}{2}} C_{33,11}^{33,1}}{2560}\)

\noindent\(C_{40,0000,0}^{11,1111,0,\text{NonLoc}}= -\frac{7 C_{11,11}^{51,0}}{6144}-\frac{7 C_{11,11}^{51,1}}{3072}-\frac{7 C_{22,11}^{42,0}}{1024
\sqrt{15}}-\frac{7 C_{22,11}^{42,1}}{512 \sqrt{15}}-\frac{7 C_{31,11}^{31,0}}{6144}-\frac{7 C_{31,11}^{31,1}}{3072}-\frac{3 \sqrt{\frac{7}{2}} C_{33,11}^{33,0}}{2560}-\frac{3
\sqrt{\frac{7}{2}} C_{33,11}^{33,1}}{1280}\)

\noindent\(C_{40,0000,0}^{11,1111,1,\text{Loc}}= -\frac{7 C_{11,11}^{51,1}}{2048 \sqrt{3}}+\frac{7 C_{22,11}^{42,1}}{1024 \sqrt{5}}-\frac{7
C_{31,11}^{31,1}}{2048 \sqrt{3}}-\frac{3 \sqrt{\frac{21}{2}} C_{33,11}^{33,1}}{2560}\)

\noindent\(C_{40,0000,0}^{11,1111,1,\text{NonLoc}}= -\frac{7 C_{11,11}^{51,0}}{2048 \sqrt{3}}-\frac{7 C_{22,11}^{42,0}}{1024 \sqrt{5}}-\frac{7
C_{31,11}^{31,0}}{2048 \sqrt{3}}-\frac{3 \sqrt{\frac{21}{2}} C_{33,11}^{33,0}}{2560}\)

\noindent\(C_{40,0000,0}^{20,0000,0,\text{Loc}}= -\frac{7 C_{00,00}^{60,0}}{4096}-\frac{7 C_{00,00}^{60,1}}{8192}+\frac{7 C_{11,00}^{51,0}}{4096}+\frac{7
C_{11,00}^{51,1}}{8192}-\frac{7 C_{20,00}^{40,0}}{4096}-\frac{7 C_{20,00}^{40,1}}{8192}-\frac{7 C_{22,00}^{42,0}}{2048 \sqrt{5}}-\frac{7 C_{22,00}^{42,1}}{4096
\sqrt{5}}+\frac{7 C_{31,00}^{31,0}}{4096}+\frac{7 C_{31,00}^{31,1}}{8192}+\frac{3 \sqrt{21} C_{33,00}^{33,0}}{10240}+\frac{3 \sqrt{21} C_{33,00}^{33,1}}{20480}\)

\noindent\(C_{40,0000,0}^{20,0000,0,\text{NonLoc}}= \frac{7 C_{00,00}^{60,0}}{16384}+\frac{7 C_{00,00}^{60,1}}{8192}-\frac{7 \sqrt{3}
C_{00,20}^{60,0}}{16384}-\frac{7 \sqrt{3} C_{00,20}^{60,1}}{8192}+\frac{7 C_{11,00}^{51,0}}{16384}+\frac{7 C_{11,00}^{51,1}}{8192}+\frac{7 C_{20,00}^{40,0}}{16384}+\frac{7
C_{20,00}^{40,1}}{8192}-\frac{7 \sqrt{3} C_{20,20}^{40,0}}{16384}-\frac{7 \sqrt{3} C_{20,20}^{40,1}}{8192}+\frac{7 C_{22,00}^{42,0}}{8192 \sqrt{5}}+\frac{7
C_{22,00}^{42,1}}{4096 \sqrt{5}}-\frac{7 \sqrt{\frac{3}{5}} C_{22,20}^{42,0}}{8192}-\frac{7 \sqrt{\frac{3}{5}} C_{22,20}^{42,1}}{4096}+\frac{7 C_{31,00}^{31,0}}{16384}+\frac{7
C_{31,00}^{31,1}}{8192}-\frac{7 \sqrt{3} C_{31,20}^{31,0}}{16384}-\frac{7 \sqrt{3} C_{31,20}^{31,1}}{8192}+\frac{3 \sqrt{21} C_{33,00}^{33,0}}{40960}+\frac{3
\sqrt{21} C_{33,00}^{33,1}}{20480}-\frac{9 \sqrt{7} C_{33,20}^{33,0}}{40960}-\frac{9 \sqrt{7} C_{33,20}^{33,1}}{20480}\)

\noindent\(C_{40,0000,0}^{20,0000,1,\text{Loc}}= -\frac{7 \sqrt{3} C_{00,00}^{60,1}}{8192}+\frac{7 \sqrt{3} C_{11,00}^{51,1}}{8192}-\frac{7
\sqrt{3} C_{20,00}^{40,1}}{8192}-\frac{7 \sqrt{\frac{3}{5}} C_{22,00}^{42,1}}{4096}+\frac{7 \sqrt{3} C_{31,00}^{31,1}}{8192}+\frac{9 \sqrt{7} C_{33,00}^{33,1}}{20480}\)

\noindent\(C_{40,0000,0}^{20,0000,1,\text{NonLoc}}= \frac{7 \sqrt{3} C_{00,00}^{60,0}}{16384}-\frac{21 C_{00,20}^{60,0}}{16384}+\frac{7
\sqrt{3} C_{11,00}^{51,0}}{16384}+\frac{7 \sqrt{3} C_{20,00}^{40,0}}{16384}-\frac{21 C_{20,20}^{40,0}}{16384}+\frac{7 \sqrt{\frac{3}{5}} C_{22,00}^{42,0}}{8192}-\frac{21
C_{22,20}^{42,0}}{8192 \sqrt{5}}+\frac{7 \sqrt{3} C_{31,00}^{31,0}}{16384}-\frac{21 C_{31,20}^{31,0}}{16384}+\frac{9 \sqrt{7} C_{33,00}^{33,0}}{40960}-\frac{9
\sqrt{21} C_{33,20}^{33,0}}{40960}\)

\noindent\(C_{40,0011,1}^{00,2011,0,\text{Loc}}= -\frac{7 C_{00,20}^{60,0}}{1024}-\frac{7 C_{00,20}^{60,1}}{2048}+\frac{7 C_{11,20}^{31,0}}{6144}+\frac{7
C_{11,20}^{31,1}}{12288}-\frac{5 C_{20,20}^{40,0}}{3072}-\frac{5 C_{20,20}^{40,1}}{6144}+\frac{7 C_{22,20}^{42,0}}{1536 \sqrt{5}}+\frac{7 C_{22,20}^{42,1}}{3072
\sqrt{5}}-\frac{C_{31,20}^{31,0}}{3072}-\frac{C_{31,20}^{31,1}}{6144}-\frac{3 \sqrt{21} C_{33,20}^{33,0}}{2560}-\frac{3 \sqrt{21} C_{33,20}^{33,1}}{5120}\)

\noindent\(C_{40,0011,1}^{00,2011,0,\text{NonLoc}}= -\frac{7 \sqrt{3} C_{00,00}^{60,0}}{4096}-\frac{7 \sqrt{3} C_{00,00}^{60,1}}{2048}-\frac{7
C_{00,20}^{60,0}}{4096}-\frac{7 C_{00,20}^{60,1}}{2048}-\frac{7 C_{11,00}^{51,0}}{4096 \sqrt{3}}-\frac{7 C_{11,00}^{51,1}}{2048 \sqrt{3}}-\frac{7
C_{11,20}^{31,0}}{24576}-\frac{7 C_{11,20}^{31,1}}{12288}-\frac{5 C_{20,00}^{40,0}}{4096 \sqrt{3}}-\frac{5 C_{20,00}^{40,1}}{2048 \sqrt{3}}-\frac{5
C_{20,20}^{40,0}}{12288}-\frac{5 C_{20,20}^{40,1}}{6144}+\frac{7 C_{22,00}^{42,0}}{2048 \sqrt{15}}+\frac{7 C_{22,00}^{42,1}}{1024 \sqrt{15}}+\frac{7
C_{22,20}^{42,0}}{6144 \sqrt{5}}+\frac{7 C_{22,20}^{42,1}}{3072 \sqrt{5}}+\frac{C_{31,00}^{31,0}}{4096 \sqrt{3}}+\frac{C_{31,00}^{31,1}}{2048 \sqrt{3}}+\frac{C_{31,20}^{31,0}}{12288}+\frac{C_{31,20}^{31,1}}{6144}+\frac{9
\sqrt{7} C_{33,00}^{33,0}}{10240}+\frac{9 \sqrt{7} C_{33,00}^{33,1}}{5120}+\frac{3 \sqrt{21} C_{33,20}^{33,0}}{10240}+\frac{3 \sqrt{21} C_{33,20}^{33,1}}{5120}\)

\noindent\(C_{40,0011,1}^{00,2011,1,\text{Loc}}= -\frac{7 \sqrt{3} C_{00,20}^{60,1}}{2048}+\frac{7 C_{11,20}^{31,1}}{4096 \sqrt{3}}-\frac{5
C_{20,20}^{40,1}}{2048 \sqrt{3}}+\frac{7 C_{22,20}^{42,1}}{1024 \sqrt{15}}-\frac{C_{31,20}^{31,1}}{2048 \sqrt{3}}-\frac{9 \sqrt{7} C_{33,20}^{33,1}}{5120}\)

\noindent\(C_{40,0011,1}^{00,2011,1,\text{NonLoc}}= -\frac{21 C_{00,00}^{60,0}}{4096}-\frac{7 \sqrt{3} C_{00,20}^{60,0}}{4096}-\frac{7
C_{11,00}^{51,0}}{4096}-\frac{7 C_{11,20}^{31,0}}{8192 \sqrt{3}}-\frac{5 C_{20,00}^{40,0}}{4096}-\frac{5 C_{20,20}^{40,0}}{4096 \sqrt{3}}+\frac{7
C_{22,00}^{42,0}}{2048 \sqrt{5}}+\frac{7 C_{22,20}^{42,0}}{2048 \sqrt{15}}+\frac{C_{31,00}^{31,0}}{4096}+\frac{C_{31,20}^{31,0}}{4096 \sqrt{3}}+\frac{9
\sqrt{21} C_{33,00}^{33,0}}{10240}+\frac{9 \sqrt{7} C_{33,20}^{33,0}}{10240}\)

\noindent\(C_{40,0011,1}^{00,2211,0,\text{Loc}}= -\frac{21 C_{00,22}^{62,0}}{5120}-\frac{21 C_{00,22}^{62,1}}{10240}-\frac{7 C_{11,22}^{51,0}}{15360}-\frac{7
C_{11,22}^{51,1}}{30720}+\frac{9 \sqrt{\frac{21}{5}} C_{11,22}^{53,0}}{5120}+\frac{9 \sqrt{\frac{21}{5}} C_{11,22}^{53,1}}{10240}-\frac{7 C_{20,22}^{42,0}}{15360}-\frac{7
C_{20,22}^{42,1}}{30720}+\frac{17 C_{22,22}^{40,0}}{15360}+\frac{17 C_{22,22}^{40,1}}{30720}+\frac{\sqrt{7} C_{22,22}^{42,0}}{1536}+\frac{\sqrt{7}
C_{22,22}^{42,1}}{3072}-\frac{9 C_{22,22}^{44,0}}{1280 \sqrt{5}}-\frac{9 C_{22,22}^{44,1}}{2560 \sqrt{5}}-\frac{23 C_{31,22}^{31,0}}{15360}-\frac{23
C_{31,22}^{31,1}}{30720}+\frac{\sqrt{\frac{21}{5}} C_{31,22}^{33,0}}{5120}+\frac{\sqrt{\frac{21}{5}} C_{31,22}^{33,1}}{10240}-\frac{9 \sqrt{\frac{7}{2}}
C_{33,22}^{33,0}}{6400}-\frac{9 \sqrt{\frac{7}{2}} C_{33,22}^{33,1}}{12800}\)

\noindent\(C_{40,0011,1}^{00,2211,0,\text{NonLoc}}= \frac{21 C_{00,22}^{62,0}}{10240}+\frac{21 C_{00,22}^{62,1}}{5120}-\frac{7 C_{11,22}^{51,0}}{30720}-\frac{7
C_{11,22}^{51,1}}{15360}+\frac{9 \sqrt{\frac{21}{5}} C_{11,22}^{53,0}}{10240}+\frac{9 \sqrt{\frac{21}{5}} C_{11,22}^{53,1}}{5120}+\frac{7 C_{20,22}^{42,0}}{30720}+\frac{7
C_{20,22}^{42,1}}{15360}-\frac{17 C_{22,22}^{40,0}}{30720}-\frac{17 C_{22,22}^{40,1}}{15360}-\frac{\sqrt{7} C_{22,22}^{42,0}}{3072}-\frac{\sqrt{7}
C_{22,22}^{42,1}}{1536}+\frac{9 C_{22,22}^{44,0}}{2560 \sqrt{5}}+\frac{9 C_{22,22}^{44,1}}{1280 \sqrt{5}}-\frac{23 C_{31,22}^{31,0}}{30720}-\frac{23
C_{31,22}^{31,1}}{15360}+\frac{\sqrt{\frac{21}{5}} C_{31,22}^{33,0}}{10240}+\frac{\sqrt{\frac{21}{5}} C_{31,22}^{33,1}}{5120}-\frac{9 \sqrt{\frac{7}{2}}
C_{33,22}^{33,0}}{12800}-\frac{9 \sqrt{\frac{7}{2}} C_{33,22}^{33,1}}{6400}\)

\noindent\(C_{40,0011,1}^{00,2211,1,\text{Loc}}= -\frac{21 \sqrt{3} C_{00,22}^{62,1}}{10240}-\frac{7 C_{11,22}^{51,1}}{10240 \sqrt{3}}+\frac{27
\sqrt{\frac{7}{5}} C_{11,22}^{53,1}}{10240}-\frac{7 C_{20,22}^{42,1}}{10240 \sqrt{3}}+\frac{17 C_{22,22}^{40,1}}{10240 \sqrt{3}}+\frac{\sqrt{\frac{7}{3}}
C_{22,22}^{42,1}}{1024}-\frac{9 \sqrt{\frac{3}{5}} C_{22,22}^{44,1}}{2560}-\frac{23 C_{31,22}^{31,1}}{10240 \sqrt{3}}+\frac{3 \sqrt{\frac{7}{5}}
C_{31,22}^{33,1}}{10240}-\frac{9 \sqrt{\frac{21}{2}} C_{33,22}^{33,1}}{12800}\)

\noindent\(C_{40,0011,1}^{00,2211,1,\text{NonLoc}}= \frac{21 \sqrt{3} C_{00,22}^{62,0}}{10240}-\frac{7 C_{11,22}^{51,0}}{10240 \sqrt{3}}+\frac{27
\sqrt{\frac{7}{5}} C_{11,22}^{53,0}}{10240}+\frac{7 C_{20,22}^{42,0}}{10240 \sqrt{3}}-\frac{17 C_{22,22}^{40,0}}{10240 \sqrt{3}}-\frac{\sqrt{\frac{7}{3}}
C_{22,22}^{42,0}}{1024}+\frac{9 \sqrt{\frac{3}{5}} C_{22,22}^{44,0}}{2560}-\frac{23 C_{31,22}^{31,0}}{10240 \sqrt{3}}+\frac{3 \sqrt{\frac{7}{5}}
C_{31,22}^{33,0}}{10240}-\frac{9 \sqrt{\frac{21}{2}} C_{33,22}^{33,0}}{12800}\)

\noindent\(C_{40,0011,1}^{11,1101,0,\text{Loc}}= -\frac{7 C_{11,11}^{51,0}}{3072}-\frac{7 C_{11,11}^{51,1}}{6144}+\frac{7 C_{22,11}^{42,0}}{512
\sqrt{15}}+\frac{7 C_{22,11}^{42,1}}{1024 \sqrt{15}}-\frac{7 C_{31,11}^{31,0}}{3072}-\frac{7 C_{31,11}^{31,1}}{6144}-\frac{3 \sqrt{\frac{7}{2}} C_{33,11}^{33,0}}{1280}-\frac{3
\sqrt{\frac{7}{2}} C_{33,11}^{33,1}}{2560}\)

\noindent\(C_{40,0011,1}^{11,1101,0,\text{NonLoc}}= -\frac{7 C_{11,11}^{51,0}}{6144}-\frac{7 C_{11,11}^{51,1}}{3072}-\frac{7 C_{22,11}^{42,0}}{1024
\sqrt{15}}-\frac{7 C_{22,11}^{42,1}}{512 \sqrt{15}}-\frac{7 C_{31,11}^{31,0}}{6144}-\frac{7 C_{31,11}^{31,1}}{3072}-\frac{3 \sqrt{\frac{7}{2}} C_{33,11}^{33,0}}{2560}-\frac{3
\sqrt{\frac{7}{2}} C_{33,11}^{33,1}}{1280}\)

\noindent\(C_{40,0011,1}^{11,1101,1,\text{Loc}}= -\frac{7 C_{11,11}^{51,1}}{2048 \sqrt{3}}+\frac{7 C_{22,11}^{42,1}}{1024 \sqrt{5}}-\frac{7
C_{31,11}^{31,1}}{2048 \sqrt{3}}-\frac{3 \sqrt{\frac{21}{2}} C_{33,11}^{33,1}}{2560}\)

\noindent\(C_{40,0011,1}^{11,1101,1,\text{NonLoc}}= -\frac{7 C_{11,11}^{51,0}}{2048 \sqrt{3}}-\frac{7 C_{22,11}^{42,0}}{1024 \sqrt{5}}-\frac{7
C_{31,11}^{31,0}}{2048 \sqrt{3}}-\frac{3 \sqrt{\frac{21}{2}} C_{33,11}^{33,0}}{2560}\)

\noindent\(C_{40,0011,1}^{20,0011,0,\text{Loc}}= \frac{7 C_{00,20}^{60,0}}{4096}+\frac{7 C_{00,20}^{60,1}}{8192}+\frac{7 C_{20,20}^{40,0}}{4096}+\frac{7
C_{20,20}^{40,1}}{8192}+\frac{7 C_{22,20}^{42,0}}{2048 \sqrt{5}}+\frac{7 C_{22,20}^{42,1}}{4096 \sqrt{5}}-\frac{7 C_{31,20}^{31,0}}{4096}-\frac{7
C_{31,20}^{31,1}}{8192}-\frac{3 \sqrt{21} C_{33,20}^{33,0}}{10240}-\frac{3 \sqrt{21} C_{33,20}^{33,1}}{20480}\)

\noindent\(C_{40,0011,1}^{20,0011,0,\text{NonLoc}}= \frac{7 \sqrt{3} C_{00,00}^{60,0}}{16384}+\frac{7 \sqrt{3} C_{00,00}^{60,1}}{8192}+\frac{7
C_{00,20}^{60,0}}{16384}+\frac{7 C_{00,20}^{60,1}}{8192}+\frac{7 \sqrt{3} C_{11,00}^{51,0}}{16384}+\frac{7 \sqrt{3} C_{11,00}^{51,1}}{8192}+\frac{7
\sqrt{3} C_{20,00}^{40,0}}{16384}+\frac{7 \sqrt{3} C_{20,00}^{40,1}}{8192}+\frac{7 C_{20,20}^{40,0}}{16384}+\frac{7 C_{20,20}^{40,1}}{8192}+\frac{7
\sqrt{\frac{3}{5}} C_{22,00}^{42,0}}{8192}+\frac{7 \sqrt{\frac{3}{5}} C_{22,00}^{42,1}}{4096}+\frac{7 C_{22,20}^{42,0}}{8192 \sqrt{5}}+\frac{7 C_{22,20}^{42,1}}{4096
\sqrt{5}}+\frac{7 \sqrt{3} C_{31,00}^{31,0}}{16384}+\frac{7 \sqrt{3} C_{31,00}^{31,1}}{8192}+\frac{7 C_{31,20}^{31,0}}{16384}+\frac{7 C_{31,20}^{31,1}}{8192}+\frac{9
\sqrt{7} C_{33,00}^{33,0}}{40960}+\frac{9 \sqrt{7} C_{33,00}^{33,1}}{20480}+\frac{3 \sqrt{21} C_{33,20}^{33,0}}{40960}+\frac{3 \sqrt{21} C_{33,20}^{33,1}}{20480}\)

\noindent\(C_{40,0011,1}^{20,0011,1,\text{Loc}}= \frac{7 \sqrt{3} C_{00,20}^{60,1}}{8192}+\frac{7 \sqrt{3} C_{20,20}^{40,1}}{8192}+\frac{7
\sqrt{\frac{3}{5}} C_{22,20}^{42,1}}{4096}-\frac{7 \sqrt{3} C_{31,20}^{31,1}}{8192}-\frac{9 \sqrt{7} C_{33,20}^{33,1}}{20480}\)

\noindent\(C_{40,0011,1}^{20,0011,1,\text{NonLoc}}= \frac{21 C_{00,00}^{60,0}}{16384}+\frac{7 \sqrt{3} C_{00,20}^{60,0}}{16384}+\frac{21
C_{11,00}^{51,0}}{16384}+\frac{21 C_{20,00}^{40,0}}{16384}+\frac{7 \sqrt{3} C_{20,20}^{40,0}}{16384}+\frac{21 C_{22,00}^{42,0}}{8192 \sqrt{5}}+\frac{7
\sqrt{\frac{3}{5}} C_{22,20}^{42,0}}{8192}+\frac{21 C_{31,00}^{31,0}}{16384}+\frac{7 \sqrt{3} C_{31,20}^{31,0}}{16384}+\frac{9 \sqrt{21} C_{33,00}^{33,0}}{40960}+\frac{9
\sqrt{7} C_{33,20}^{33,0}}{40960}\)

\noindent\(C_{40,0011,1}^{22,0011,0,\text{Loc}}= \frac{21 C_{00,22}^{62,0}}{20480}+\frac{21 C_{00,22}^{62,1}}{40960}-\frac{21 C_{11,22}^{51,0}}{20480}-\frac{21
C_{11,22}^{51,1}}{40960}-\frac{9 \sqrt{\frac{21}{5}} C_{11,22}^{53,0}}{20480}-\frac{9 \sqrt{\frac{21}{5}} C_{11,22}^{53,1}}{40960}+\frac{21 C_{20,22}^{42,0}}{20480}+\frac{21
C_{20,22}^{42,1}}{40960}+\frac{21 C_{22,22}^{40,0}}{20480}+\frac{21 C_{22,22}^{40,1}}{40960}+\frac{3 \sqrt{7} C_{22,22}^{42,0}}{10240}+\frac{3 \sqrt{7}
C_{22,22}^{42,1}}{20480}+\frac{9 C_{22,22}^{44,0}}{5120 \sqrt{5}}+\frac{9 C_{22,22}^{44,1}}{10240 \sqrt{5}}-\frac{21 C_{31,22}^{31,0}}{20480}-\frac{21
C_{31,22}^{31,1}}{40960}-\frac{9 \sqrt{\frac{21}{5}} C_{31,22}^{33,0}}{20480}-\frac{9 \sqrt{\frac{21}{5}} C_{31,22}^{33,1}}{40960}-\frac{9 \sqrt{\frac{7}{2}}
C_{33,22}^{33,0}}{25600}-\frac{9 \sqrt{\frac{7}{2}} C_{33,22}^{33,1}}{51200}\)

\noindent\(C_{40,0011,1}^{22,0011,0,\text{NonLoc}}= -\frac{21 C_{00,22}^{62,0}}{40960}-\frac{21 C_{00,22}^{62,1}}{20480}-\frac{21 C_{11,22}^{51,0}}{40960}-\frac{21
C_{11,22}^{51,1}}{20480}-\frac{9 \sqrt{\frac{21}{5}} C_{11,22}^{53,0}}{40960}-\frac{9 \sqrt{\frac{21}{5}} C_{11,22}^{53,1}}{20480}-\frac{21 C_{20,22}^{42,0}}{40960}-\frac{21
C_{20,22}^{42,1}}{20480}-\frac{21 C_{22,22}^{40,0}}{40960}-\frac{21 C_{22,22}^{40,1}}{20480}-\frac{3 \sqrt{7} C_{22,22}^{42,0}}{20480}-\frac{3 \sqrt{7}
C_{22,22}^{42,1}}{10240}-\frac{9 C_{22,22}^{44,0}}{10240 \sqrt{5}}-\frac{9 C_{22,22}^{44,1}}{5120 \sqrt{5}}-\frac{21 C_{31,22}^{31,0}}{40960}-\frac{21
C_{31,22}^{31,1}}{20480}-\frac{9 \sqrt{\frac{21}{5}} C_{31,22}^{33,0}}{40960}-\frac{9 \sqrt{\frac{21}{5}} C_{31,22}^{33,1}}{20480}-\frac{9 \sqrt{\frac{7}{2}}
C_{33,22}^{33,0}}{51200}-\frac{9 \sqrt{\frac{7}{2}} C_{33,22}^{33,1}}{25600}\)

\noindent\(C_{40,0011,1}^{22,0011,1,\text{Loc}}= \frac{21 \sqrt{3} C_{00,22}^{62,1}}{40960}-\frac{21 \sqrt{3} C_{11,22}^{51,1}}{40960}-\frac{27
\sqrt{\frac{7}{5}} C_{11,22}^{53,1}}{40960}+\frac{21 \sqrt{3} C_{20,22}^{42,1}}{40960}+\frac{21 \sqrt{3} C_{22,22}^{40,1}}{40960}+\frac{3 \sqrt{21}
C_{22,22}^{42,1}}{20480}+\frac{9 \sqrt{\frac{3}{5}} C_{22,22}^{44,1}}{10240}-\frac{21 \sqrt{3} C_{31,22}^{31,1}}{40960}-\frac{27 \sqrt{\frac{7}{5}}
C_{31,22}^{33,1}}{40960}-\frac{9 \sqrt{\frac{21}{2}} C_{33,22}^{33,1}}{51200}\)

\noindent\(C_{40,0011,1}^{22,0011,1,\text{NonLoc}}= -\frac{21 \sqrt{3} C_{00,22}^{62,0}}{40960}-\frac{21 \sqrt{3} C_{11,22}^{51,0}}{40960}-\frac{27
\sqrt{\frac{7}{5}} C_{11,22}^{53,0}}{40960}-\frac{21 \sqrt{3} C_{20,22}^{42,0}}{40960}-\frac{21 \sqrt{3} C_{22,22}^{40,0}}{40960}-\frac{3 \sqrt{21}
C_{22,22}^{42,0}}{20480}-\frac{9 \sqrt{\frac{3}{5}} C_{22,22}^{44,0}}{10240}-\frac{21 \sqrt{3} C_{31,22}^{31,0}}{40960}-\frac{27 \sqrt{\frac{7}{5}}
C_{31,22}^{33,0}}{40960}-\frac{9 \sqrt{\frac{21}{2}} C_{33,22}^{33,0}}{51200}\)

\noindent\(C_{40,1101,1}^{00,1101,0,\text{Loc}}= -\frac{7 C_{00,00}^{60,0}}{512}-\frac{7 C_{00,00}^{60,1}}{1024}+\frac{7 C_{11,00}^{51,0}}{1536}+\frac{7
C_{11,00}^{51,1}}{3072}-\frac{5 C_{20,00}^{40,0}}{1536}-\frac{5 C_{20,00}^{40,1}}{3072}+\frac{7 C_{22,00}^{42,0}}{768 \sqrt{5}}+\frac{7 C_{22,00}^{42,1}}{1536
\sqrt{5}}-\frac{C_{31,00}^{31,0}}{1536}-\frac{C_{31,00}^{31,1}}{3072}-\frac{3 \sqrt{21} C_{33,00}^{33,0}}{1280}-\frac{3 \sqrt{21} C_{33,00}^{33,1}}{2560}\)

\noindent\(C_{40,1101,1}^{00,1101,0,\text{NonLoc}}= \frac{7 C_{00,00}^{60,0}}{2048}+\frac{7 C_{00,00}^{60,1}}{1024}-\frac{7 \sqrt{3}
C_{00,20}^{60,0}}{2048}-\frac{7 \sqrt{3} C_{00,20}^{60,1}}{1024}+\frac{7 C_{11,00}^{51,0}}{6144}+\frac{7 C_{11,00}^{51,1}}{3072}-\frac{7 C_{11,20}^{31,0}}{2048
\sqrt{3}}-\frac{7 C_{11,20}^{31,1}}{1024 \sqrt{3}}+\frac{5 C_{20,00}^{40,0}}{6144}+\frac{5 C_{20,00}^{40,1}}{3072}-\frac{5 C_{20,20}^{40,0}}{2048
\sqrt{3}}-\frac{5 C_{20,20}^{40,1}}{1024 \sqrt{3}}-\frac{7 C_{22,00}^{42,0}}{3072 \sqrt{5}}-\frac{7 C_{22,00}^{42,1}}{1536 \sqrt{5}}+\frac{7 C_{22,20}^{42,0}}{1024
\sqrt{15}}+\frac{7 C_{22,20}^{42,1}}{512 \sqrt{15}}-\frac{C_{31,00}^{31,0}}{6144}-\frac{C_{31,00}^{31,1}}{3072}+\frac{C_{31,20}^{31,0}}{2048 \sqrt{3}}+\frac{C_{31,20}^{31,1}}{1024
\sqrt{3}}-\frac{3 \sqrt{21} C_{33,00}^{33,0}}{5120}-\frac{3 \sqrt{21} C_{33,00}^{33,1}}{2560}+\frac{9 \sqrt{7} C_{33,20}^{33,0}}{5120}+\frac{9 \sqrt{7}
C_{33,20}^{33,1}}{2560}\)

\noindent\(C_{40,1101,1}^{00,1101,1,\text{Loc}}= -\frac{7 \sqrt{3} C_{00,00}^{60,1}}{1024}+\frac{7 C_{11,00}^{51,1}}{1024 \sqrt{3}}-\frac{5
C_{20,00}^{40,1}}{1024 \sqrt{3}}+\frac{7 C_{22,00}^{42,1}}{512 \sqrt{15}}-\frac{C_{31,00}^{31,1}}{1024 \sqrt{3}}-\frac{9 \sqrt{7} C_{33,00}^{33,1}}{2560}\)

\noindent\(C_{40,1101,1}^{00,1101,1,\text{NonLoc}}= \frac{7 \sqrt{3} C_{00,00}^{60,0}}{2048}-\frac{21 C_{00,20}^{60,0}}{2048}+\frac{7
C_{11,00}^{51,0}}{2048 \sqrt{3}}-\frac{7 C_{11,20}^{31,0}}{2048}+\frac{5 C_{20,00}^{40,0}}{2048 \sqrt{3}}-\frac{5 C_{20,20}^{40,0}}{2048}-\frac{7
C_{22,00}^{42,0}}{1024 \sqrt{15}}+\frac{7 C_{22,20}^{42,0}}{1024 \sqrt{5}}-\frac{C_{31,00}^{31,0}}{2048 \sqrt{3}}+\frac{C_{31,20}^{31,0}}{2048}-\frac{9
\sqrt{7} C_{33,00}^{33,0}}{5120}+\frac{9 \sqrt{21} C_{33,20}^{33,0}}{5120}\)

\noindent\(C_{40,1101,1}^{11,0011,0,\text{Loc}}= -\frac{7 C_{11,11}^{51,0}}{3072}-\frac{7 C_{11,11}^{51,1}}{6144}+\frac{7 C_{22,11}^{42,0}}{512
\sqrt{15}}+\frac{7 C_{22,11}^{42,1}}{1024 \sqrt{15}}-\frac{7 C_{31,11}^{31,0}}{3072}-\frac{7 C_{31,11}^{31,1}}{6144}-\frac{3 \sqrt{\frac{7}{2}} C_{33,11}^{33,0}}{1280}-\frac{3
\sqrt{\frac{7}{2}} C_{33,11}^{33,1}}{2560}\)

\noindent\(C_{40,1101,1}^{11,0011,0,\text{NonLoc}}= -\frac{7 C_{11,11}^{51,0}}{6144}-\frac{7 C_{11,11}^{51,1}}{3072}-\frac{7 C_{22,11}^{42,0}}{1024
\sqrt{15}}-\frac{7 C_{22,11}^{42,1}}{512 \sqrt{15}}-\frac{7 C_{31,11}^{31,0}}{6144}-\frac{7 C_{31,11}^{31,1}}{3072}-\frac{3 \sqrt{\frac{7}{2}} C_{33,11}^{33,0}}{2560}-\frac{3
\sqrt{\frac{7}{2}} C_{33,11}^{33,1}}{1280}\)

\noindent\(C_{40,1101,1}^{11,0011,1,\text{Loc}}= -\frac{7 C_{11,11}^{51,1}}{2048 \sqrt{3}}+\frac{7 C_{22,11}^{42,1}}{1024 \sqrt{5}}-\frac{7
C_{31,11}^{31,1}}{2048 \sqrt{3}}-\frac{3 \sqrt{\frac{21}{2}} C_{33,11}^{33,1}}{2560}\)

\noindent\(C_{40,1101,1}^{11,0011,1,\text{NonLoc}}= -\frac{7 C_{11,11}^{51,0}}{2048 \sqrt{3}}-\frac{7 C_{22,11}^{42,0}}{1024 \sqrt{5}}-\frac{7
C_{31,11}^{31,0}}{2048 \sqrt{3}}-\frac{3 \sqrt{\frac{21}{2}} C_{33,11}^{33,0}}{2560}\)

\noindent\(C_{40,1110,0}^{00,1110,0,\text{Loc}}= \frac{7 C_{00,20}^{60,0}}{1536}+\frac{7 C_{00,20}^{60,1}}{3072}+\frac{7 C_{00,22}^{62,0}}{512
\sqrt{5}}+\frac{7 C_{00,22}^{62,1}}{1024 \sqrt{5}}-\frac{7 C_{11,20}^{31,0}}{4608}-\frac{7 C_{11,20}^{31,1}}{9216}+\frac{7 C_{11,22}^{51,0}}{4608
\sqrt{5}}+\frac{7 C_{11,22}^{51,1}}{9216 \sqrt{5}}-\frac{3 \sqrt{21} C_{11,22}^{53,0}}{2560}-\frac{3 \sqrt{21} C_{11,22}^{53,1}}{5120}+\frac{5 C_{20,20}^{40,0}}{4608}+\frac{5
C_{20,20}^{40,1}}{9216}+\frac{7 C_{20,22}^{42,0}}{4608 \sqrt{5}}+\frac{7 C_{20,22}^{42,1}}{9216 \sqrt{5}}-\frac{7 C_{22,20}^{42,0}}{2304 \sqrt{5}}-\frac{7
C_{22,20}^{42,1}}{4608 \sqrt{5}}-\frac{17 C_{22,22}^{40,0}}{4608 \sqrt{5}}-\frac{17 C_{22,22}^{40,1}}{9216 \sqrt{5}}-\frac{\sqrt{35} C_{22,22}^{42,0}}{2304}-\frac{\sqrt{35}
C_{22,22}^{42,1}}{4608}+\frac{3 C_{22,22}^{44,0}}{640}+\frac{3 C_{22,22}^{44,1}}{1280}+\frac{C_{31,20}^{31,0}}{4608}+\frac{C_{31,20}^{31,1}}{9216}+\frac{23
C_{31,22}^{31,0}}{4608 \sqrt{5}}+\frac{23 C_{31,22}^{31,1}}{9216 \sqrt{5}}-\frac{\sqrt{\frac{7}{3}} C_{31,22}^{33,0}}{2560}-\frac{\sqrt{\frac{7}{3}}
C_{31,22}^{33,1}}{5120}+\frac{\sqrt{21} C_{33,20}^{33,0}}{1280}+\frac{\sqrt{21} C_{33,20}^{33,1}}{2560}+\frac{3}{640} \sqrt{\frac{7}{10}} C_{33,22}^{33,0}+\frac{3
\sqrt{\frac{7}{10}} C_{33,22}^{33,1}}{1280}\)

\noindent\(C_{40,1110,0}^{00,1110,0,\text{NonLoc}}= \frac{7 C_{00,00}^{60,0}}{2048 \sqrt{3}}+\frac{7 C_{00,00}^{60,1}}{1024 \sqrt{3}}+\frac{7
C_{00,20}^{60,0}}{6144}+\frac{7 C_{00,20}^{60,1}}{3072}-\frac{7 C_{00,22}^{62,0}}{1024 \sqrt{5}}-\frac{7 C_{00,22}^{62,1}}{512 \sqrt{5}}+\frac{7
C_{11,00}^{51,0}}{6144 \sqrt{3}}+\frac{7 C_{11,00}^{51,1}}{3072 \sqrt{3}}+\frac{7 C_{11,20}^{31,0}}{18432}+\frac{7 C_{11,20}^{31,1}}{9216}+\frac{7
C_{11,22}^{51,0}}{9216 \sqrt{5}}+\frac{7 C_{11,22}^{51,1}}{4608 \sqrt{5}}-\frac{3 \sqrt{21} C_{11,22}^{53,0}}{5120}-\frac{3 \sqrt{21} C_{11,22}^{53,1}}{2560}+\frac{5
C_{20,00}^{40,0}}{6144 \sqrt{3}}+\frac{5 C_{20,00}^{40,1}}{3072 \sqrt{3}}+\frac{5 C_{20,20}^{40,0}}{18432}+\frac{5 C_{20,20}^{40,1}}{9216}-\frac{7
C_{20,22}^{42,0}}{9216 \sqrt{5}}-\frac{7 C_{20,22}^{42,1}}{4608 \sqrt{5}}-\frac{7 C_{22,00}^{42,0}}{3072 \sqrt{15}}-\frac{7 C_{22,00}^{42,1}}{1536
\sqrt{15}}-\frac{7 C_{22,20}^{42,0}}{9216 \sqrt{5}}-\frac{7 C_{22,20}^{42,1}}{4608 \sqrt{5}}+\frac{17 C_{22,22}^{40,0}}{9216 \sqrt{5}}+\frac{17 C_{22,22}^{40,1}}{4608
\sqrt{5}}+\frac{\sqrt{35} C_{22,22}^{42,0}}{4608}+\frac{\sqrt{35} C_{22,22}^{42,1}}{2304}-\frac{3 C_{22,22}^{44,0}}{1280}-\frac{3 C_{22,22}^{44,1}}{640}-\frac{C_{31,00}^{31,0}}{6144
\sqrt{3}}-\frac{C_{31,00}^{31,1}}{3072 \sqrt{3}}-\frac{C_{31,20}^{31,0}}{18432}-\frac{C_{31,20}^{31,1}}{9216}+\frac{23 C_{31,22}^{31,0}}{9216 \sqrt{5}}+\frac{23
C_{31,22}^{31,1}}{4608 \sqrt{5}}-\frac{\sqrt{\frac{7}{3}} C_{31,22}^{33,0}}{5120}-\frac{\sqrt{\frac{7}{3}} C_{31,22}^{33,1}}{2560}-\frac{3 \sqrt{7}
C_{33,00}^{33,0}}{5120}-\frac{3 \sqrt{7} C_{33,00}^{33,1}}{2560}-\frac{\sqrt{21} C_{33,20}^{33,0}}{5120}-\frac{\sqrt{21} C_{33,20}^{33,1}}{2560}+\frac{3
\sqrt{\frac{7}{10}} C_{33,22}^{33,0}}{1280}+\frac{3}{640} \sqrt{\frac{7}{10}} C_{33,22}^{33,1}\)

\noindent\(C_{40,1110,0}^{00,1110,1,\text{Loc}}= \frac{7 C_{00,20}^{60,1}}{1024 \sqrt{3}}+\frac{7 \sqrt{\frac{3}{5}} C_{00,22}^{62,1}}{1024}-\frac{7
C_{11,20}^{31,1}}{3072 \sqrt{3}}+\frac{7 C_{11,22}^{51,1}}{3072 \sqrt{15}}-\frac{9 \sqrt{7} C_{11,22}^{53,1}}{5120}+\frac{5 C_{20,20}^{40,1}}{3072
\sqrt{3}}+\frac{7 C_{20,22}^{42,1}}{3072 \sqrt{15}}-\frac{7 C_{22,20}^{42,1}}{1536 \sqrt{15}}-\frac{17 C_{22,22}^{40,1}}{3072 \sqrt{15}}-\frac{\sqrt{\frac{35}{3}}
C_{22,22}^{42,1}}{1536}+\frac{3 \sqrt{3} C_{22,22}^{44,1}}{1280}+\frac{C_{31,20}^{31,1}}{3072 \sqrt{3}}+\frac{23 C_{31,22}^{31,1}}{3072 \sqrt{15}}-\frac{\sqrt{7}
C_{31,22}^{33,1}}{5120}+\frac{3 \sqrt{7} C_{33,20}^{33,1}}{2560}+\frac{3 \sqrt{\frac{21}{10}} C_{33,22}^{33,1}}{1280}\)

\noindent\(C_{40,1110,0}^{00,1110,1,\text{NonLoc}}= \frac{7 C_{00,00}^{60,0}}{2048}+\frac{7 C_{00,20}^{60,0}}{2048 \sqrt{3}}-\frac{7
\sqrt{\frac{3}{5}} C_{00,22}^{62,0}}{1024}+\frac{7 C_{11,00}^{51,0}}{6144}+\frac{7 C_{11,20}^{31,0}}{6144 \sqrt{3}}+\frac{7 C_{11,22}^{51,0}}{3072
\sqrt{15}}-\frac{9 \sqrt{7} C_{11,22}^{53,0}}{5120}+\frac{5 C_{20,00}^{40,0}}{6144}+\frac{5 C_{20,20}^{40,0}}{6144 \sqrt{3}}-\frac{7 C_{20,22}^{42,0}}{3072
\sqrt{15}}-\frac{7 C_{22,00}^{42,0}}{3072 \sqrt{5}}-\frac{7 C_{22,20}^{42,0}}{3072 \sqrt{15}}+\frac{17 C_{22,22}^{40,0}}{3072 \sqrt{15}}+\frac{\sqrt{\frac{35}{3}}
C_{22,22}^{42,0}}{1536}-\frac{3 \sqrt{3} C_{22,22}^{44,0}}{1280}-\frac{C_{31,00}^{31,0}}{6144}-\frac{C_{31,20}^{31,0}}{6144 \sqrt{3}}+\frac{23 C_{31,22}^{31,0}}{3072
\sqrt{15}}-\frac{\sqrt{7} C_{31,22}^{33,0}}{5120}-\frac{3 \sqrt{21} C_{33,00}^{33,0}}{5120}-\frac{3 \sqrt{7} C_{33,20}^{33,0}}{5120}+\frac{3 \sqrt{\frac{21}{10}}
C_{33,22}^{33,0}}{1280}\)

\noindent\(C_{40,1111,1}^{00,1111,0,\text{Loc}}= \frac{7 C_{00,20}^{60,0}}{512 \sqrt{3}}+\frac{7 C_{00,20}^{60,1}}{1024 \sqrt{3}}-\frac{7
\sqrt{\frac{3}{5}} C_{00,22}^{62,0}}{1024}-\frac{7 \sqrt{\frac{3}{5}} C_{00,22}^{62,1}}{2048}-\frac{7 C_{11,20}^{31,0}}{1536 \sqrt{3}}-\frac{7 C_{11,20}^{31,1}}{3072
\sqrt{3}}-\frac{7 C_{11,22}^{51,0}}{3072 \sqrt{15}}-\frac{7 C_{11,22}^{51,1}}{6144 \sqrt{15}}+\frac{9 \sqrt{7} C_{11,22}^{53,0}}{5120}+\frac{9 \sqrt{7}
C_{11,22}^{53,1}}{10240}+\frac{5 C_{20,20}^{40,0}}{1536 \sqrt{3}}+\frac{5 C_{20,20}^{40,1}}{3072 \sqrt{3}}-\frac{7 C_{20,22}^{42,0}}{3072 \sqrt{15}}-\frac{7
C_{20,22}^{42,1}}{6144 \sqrt{15}}-\frac{7 C_{22,20}^{42,0}}{768 \sqrt{15}}-\frac{7 C_{22,20}^{42,1}}{1536 \sqrt{15}}+\frac{17 C_{22,22}^{40,0}}{3072
\sqrt{15}}+\frac{17 C_{22,22}^{40,1}}{6144 \sqrt{15}}+\frac{\sqrt{\frac{35}{3}} C_{22,22}^{42,0}}{1536}+\frac{\sqrt{\frac{35}{3}} C_{22,22}^{42,1}}{3072}-\frac{3
\sqrt{3} C_{22,22}^{44,0}}{1280}-\frac{3 \sqrt{3} C_{22,22}^{44,1}}{2560}+\frac{C_{31,20}^{31,0}}{1536 \sqrt{3}}+\frac{C_{31,20}^{31,1}}{3072 \sqrt{3}}-\frac{23
C_{31,22}^{31,0}}{3072 \sqrt{15}}-\frac{23 C_{31,22}^{31,1}}{6144 \sqrt{15}}+\frac{\sqrt{7} C_{31,22}^{33,0}}{5120}+\frac{\sqrt{7} C_{31,22}^{33,1}}{10240}+\frac{3
\sqrt{7} C_{33,20}^{33,0}}{1280}+\frac{3 \sqrt{7} C_{33,20}^{33,1}}{2560}-\frac{3 \sqrt{\frac{21}{10}} C_{33,22}^{33,0}}{1280}-\frac{3 \sqrt{\frac{21}{10}}
C_{33,22}^{33,1}}{2560}\)

\noindent\(C_{40,1111,1}^{00,1111,0,\text{NonLoc}}= \frac{7 C_{00,00}^{60,0}}{2048}+\frac{7 C_{00,00}^{60,1}}{1024}+\frac{7 C_{00,20}^{60,0}}{2048
\sqrt{3}}+\frac{7 C_{00,20}^{60,1}}{1024 \sqrt{3}}+\frac{7 \sqrt{\frac{3}{5}} C_{00,22}^{62,0}}{2048}+\frac{7 \sqrt{\frac{3}{5}} C_{00,22}^{62,1}}{1024}+\frac{7
C_{11,00}^{51,0}}{6144}+\frac{7 C_{11,00}^{51,1}}{3072}+\frac{7 C_{11,20}^{31,0}}{6144 \sqrt{3}}+\frac{7 C_{11,20}^{31,1}}{3072 \sqrt{3}}-\frac{7
C_{11,22}^{51,0}}{6144 \sqrt{15}}-\frac{7 C_{11,22}^{51,1}}{3072 \sqrt{15}}+\frac{9 \sqrt{7} C_{11,22}^{53,0}}{10240}+\frac{9 \sqrt{7} C_{11,22}^{53,1}}{5120}+\frac{5
C_{20,00}^{40,0}}{6144}+\frac{5 C_{20,00}^{40,1}}{3072}+\frac{5 C_{20,20}^{40,0}}{6144 \sqrt{3}}+\frac{5 C_{20,20}^{40,1}}{3072 \sqrt{3}}+\frac{7
C_{20,22}^{42,0}}{6144 \sqrt{15}}+\frac{7 C_{20,22}^{42,1}}{3072 \sqrt{15}}-\frac{7 C_{22,00}^{42,0}}{3072 \sqrt{5}}-\frac{7 C_{22,00}^{42,1}}{1536
\sqrt{5}}-\frac{7 C_{22,20}^{42,0}}{3072 \sqrt{15}}-\frac{7 C_{22,20}^{42,1}}{1536 \sqrt{15}}-\frac{17 C_{22,22}^{40,0}}{6144 \sqrt{15}}-\frac{17
C_{22,22}^{40,1}}{3072 \sqrt{15}}-\frac{\sqrt{\frac{35}{3}} C_{22,22}^{42,0}}{3072}-\frac{\sqrt{\frac{35}{3}} C_{22,22}^{42,1}}{1536}+\frac{3 \sqrt{3}
C_{22,22}^{44,0}}{2560}+\frac{3 \sqrt{3} C_{22,22}^{44,1}}{1280}-\frac{C_{31,00}^{31,0}}{6144}-\frac{C_{31,00}^{31,1}}{3072}-\frac{C_{31,20}^{31,0}}{6144
\sqrt{3}}-\frac{C_{31,20}^{31,1}}{3072 \sqrt{3}}-\frac{23 C_{31,22}^{31,0}}{6144 \sqrt{15}}-\frac{23 C_{31,22}^{31,1}}{3072 \sqrt{15}}+\frac{\sqrt{7}
C_{31,22}^{33,0}}{10240}+\frac{\sqrt{7} C_{31,22}^{33,1}}{5120}-\frac{3 \sqrt{21} C_{33,00}^{33,0}}{5120}-\frac{3 \sqrt{21} C_{33,00}^{33,1}}{2560}-\frac{3
\sqrt{7} C_{33,20}^{33,0}}{5120}-\frac{3 \sqrt{7} C_{33,20}^{33,1}}{2560}-\frac{3 \sqrt{\frac{21}{10}} C_{33,22}^{33,0}}{2560}-\frac{3 \sqrt{\frac{21}{10}}
C_{33,22}^{33,1}}{1280}\)

\noindent\(C_{40,1111,1}^{00,1111,1,\text{Loc}}= \frac{7 C_{00,20}^{60,1}}{1024}-\frac{21 C_{00,22}^{62,1}}{2048 \sqrt{5}}-\frac{7
C_{11,20}^{31,1}}{3072}-\frac{7 C_{11,22}^{51,1}}{6144 \sqrt{5}}+\frac{9 \sqrt{21} C_{11,22}^{53,1}}{10240}+\frac{5 C_{20,20}^{40,1}}{3072}-\frac{7
C_{20,22}^{42,1}}{6144 \sqrt{5}}-\frac{7 C_{22,20}^{42,1}}{1536 \sqrt{5}}+\frac{17 C_{22,22}^{40,1}}{6144 \sqrt{5}}+\frac{\sqrt{35} C_{22,22}^{42,1}}{3072}-\frac{9
C_{22,22}^{44,1}}{2560}+\frac{C_{31,20}^{31,1}}{3072}-\frac{23 C_{31,22}^{31,1}}{6144 \sqrt{5}}+\frac{\sqrt{21} C_{31,22}^{33,1}}{10240}+\frac{3
\sqrt{21} C_{33,20}^{33,1}}{2560}-\frac{9 \sqrt{\frac{7}{10}} C_{33,22}^{33,1}}{2560}\)

\noindent\(C_{40,1111,1}^{00,1111,1,\text{NonLoc}}= \frac{7 \sqrt{3} C_{00,00}^{60,0}}{2048}+\frac{7 C_{00,20}^{60,0}}{2048}+\frac{21
C_{00,22}^{62,0}}{2048 \sqrt{5}}+\frac{7 C_{11,00}^{51,0}}{2048 \sqrt{3}}+\frac{7 C_{11,20}^{31,0}}{6144}-\frac{7 C_{11,22}^{51,0}}{6144 \sqrt{5}}+\frac{9
\sqrt{21} C_{11,22}^{53,0}}{10240}+\frac{5 C_{20,00}^{40,0}}{2048 \sqrt{3}}+\frac{5 C_{20,20}^{40,0}}{6144}+\frac{7 C_{20,22}^{42,0}}{6144 \sqrt{5}}-\frac{7
C_{22,00}^{42,0}}{1024 \sqrt{15}}-\frac{7 C_{22,20}^{42,0}}{3072 \sqrt{5}}-\frac{17 C_{22,22}^{40,0}}{6144 \sqrt{5}}-\frac{\sqrt{35} C_{22,22}^{42,0}}{3072}+\frac{9
C_{22,22}^{44,0}}{2560}-\frac{C_{31,00}^{31,0}}{2048 \sqrt{3}}-\frac{C_{31,20}^{31,0}}{6144}-\frac{23 C_{31,22}^{31,0}}{6144 \sqrt{5}}+\frac{\sqrt{21}
C_{31,22}^{33,0}}{10240}-\frac{9 \sqrt{7} C_{33,00}^{33,0}}{5120}-\frac{3 \sqrt{21} C_{33,20}^{33,0}}{5120}-\frac{9 \sqrt{\frac{7}{10}} C_{33,22}^{33,0}}{2560}\)

\noindent\(C_{40,1111,1}^{11,0000,0,\text{Loc}}= \frac{7 C_{11,11}^{51,0}}{3072}+\frac{7 C_{11,11}^{51,1}}{6144}-\frac{7 C_{22,11}^{42,0}}{512
\sqrt{15}}-\frac{7 C_{22,11}^{42,1}}{1024 \sqrt{15}}+\frac{7 C_{31,11}^{31,0}}{3072}+\frac{7 C_{31,11}^{31,1}}{6144}+\frac{3 \sqrt{\frac{7}{2}} C_{33,11}^{33,0}}{1280}+\frac{3
\sqrt{\frac{7}{2}} C_{33,11}^{33,1}}{2560}\)

\noindent\(C_{40,1111,1}^{11,0000,0,\text{NonLoc}}= \frac{7 C_{11,11}^{51,0}}{6144}+\frac{7 C_{11,11}^{51,1}}{3072}+\frac{7 C_{22,11}^{42,0}}{1024
\sqrt{15}}+\frac{7 C_{22,11}^{42,1}}{512 \sqrt{15}}+\frac{7 C_{31,11}^{31,0}}{6144}+\frac{7 C_{31,11}^{31,1}}{3072}+\frac{3 \sqrt{\frac{7}{2}} C_{33,11}^{33,0}}{2560}+\frac{3
\sqrt{\frac{7}{2}} C_{33,11}^{33,1}}{1280}\)

\noindent\(C_{40,1111,1}^{11,0000,1,\text{Loc}}= \frac{7 C_{11,11}^{51,1}}{2048 \sqrt{3}}-\frac{7 C_{22,11}^{42,1}}{1024 \sqrt{5}}+\frac{7
C_{31,11}^{31,1}}{2048 \sqrt{3}}+\frac{3 \sqrt{\frac{21}{2}} C_{33,11}^{33,1}}{2560}\)

\noindent\(C_{40,1111,1}^{11,0000,1,\text{NonLoc}}= \frac{7 C_{11,11}^{51,0}}{2048 \sqrt{3}}+\frac{7 C_{22,11}^{42,0}}{1024 \sqrt{5}}+\frac{7
C_{31,11}^{31,0}}{2048 \sqrt{3}}+\frac{3 \sqrt{\frac{21}{2}} C_{33,11}^{33,0}}{2560}\)

\noindent\(C_{40,1112,2}^{00,1112,0,\text{Loc}}= \frac{7 \sqrt{5} C_{00,20}^{60,0}}{1536}+\frac{7 \sqrt{5} C_{00,20}^{60,1}}{3072}+\frac{7
C_{00,22}^{62,0}}{5120}+\frac{7 C_{00,22}^{62,1}}{10240}-\frac{7 \sqrt{5} C_{11,20}^{31,0}}{4608}-\frac{7 \sqrt{5} C_{11,20}^{31,1}}{9216}+\frac{7
C_{11,22}^{51,0}}{46080}+\frac{7 C_{11,22}^{51,1}}{92160}-\frac{3 \sqrt{\frac{21}{5}} C_{11,22}^{53,0}}{5120}-\frac{3 \sqrt{\frac{21}{5}} C_{11,22}^{53,1}}{10240}+\frac{5
\sqrt{5} C_{20,20}^{40,0}}{4608}+\frac{5 \sqrt{5} C_{20,20}^{40,1}}{9216}+\frac{7 C_{20,22}^{42,0}}{46080}+\frac{7 C_{20,22}^{42,1}}{92160}-\frac{7
C_{22,20}^{42,0}}{2304}-\frac{7 C_{22,20}^{42,1}}{4608}-\frac{17 C_{22,22}^{40,0}}{46080}-\frac{17 C_{22,22}^{40,1}}{92160}-\frac{\sqrt{7} C_{22,22}^{42,0}}{4608}-\frac{\sqrt{7}
C_{22,22}^{42,1}}{9216}+\frac{3 C_{22,22}^{44,0}}{1280 \sqrt{5}}+\frac{3 C_{22,22}^{44,1}}{2560 \sqrt{5}}+\frac{\sqrt{5} C_{31,20}^{31,0}}{4608}+\frac{\sqrt{5}
C_{31,20}^{31,1}}{9216}+\frac{23 C_{31,22}^{31,0}}{46080}+\frac{23 C_{31,22}^{31,1}}{92160}-\frac{\sqrt{\frac{7}{15}} C_{31,22}^{33,0}}{5120}-\frac{\sqrt{\frac{7}{15}}
C_{31,22}^{33,1}}{10240}+\frac{1}{256} \sqrt{\frac{21}{5}} C_{33,20}^{33,0}+\frac{1}{512} \sqrt{\frac{21}{5}} C_{33,20}^{33,1}+\frac{3 \sqrt{\frac{7}{2}}
C_{33,22}^{33,0}}{6400}+\frac{3 \sqrt{\frac{7}{2}} C_{33,22}^{33,1}}{12800}\)

\noindent\(C_{40,1112,2}^{00,1112,0,\text{NonLoc}}= \frac{7 \sqrt{\frac{5}{3}} C_{00,00}^{60,0}}{2048}+\frac{7 \sqrt{\frac{5}{3}} C_{00,00}^{60,1}}{1024}+\frac{7
\sqrt{5} C_{00,20}^{60,0}}{6144}+\frac{7 \sqrt{5} C_{00,20}^{60,1}}{3072}-\frac{7 C_{00,22}^{62,0}}{10240}-\frac{7 C_{00,22}^{62,1}}{5120}+\frac{7
\sqrt{\frac{5}{3}} C_{11,00}^{51,0}}{6144}+\frac{7 \sqrt{\frac{5}{3}} C_{11,00}^{51,1}}{3072}+\frac{7 \sqrt{5} C_{11,20}^{31,0}}{18432}+\frac{7 \sqrt{5}
C_{11,20}^{31,1}}{9216}+\frac{7 C_{11,22}^{51,0}}{92160}+\frac{7 C_{11,22}^{51,1}}{46080}-\frac{3 \sqrt{\frac{21}{5}} C_{11,22}^{53,0}}{10240}-\frac{3
\sqrt{\frac{21}{5}} C_{11,22}^{53,1}}{5120}+\frac{5 \sqrt{\frac{5}{3}} C_{20,00}^{40,0}}{6144}+\frac{5 \sqrt{\frac{5}{3}} C_{20,00}^{40,1}}{3072}+\frac{5
\sqrt{5} C_{20,20}^{40,0}}{18432}+\frac{5 \sqrt{5} C_{20,20}^{40,1}}{9216}-\frac{7 C_{20,22}^{42,0}}{92160}-\frac{7 C_{20,22}^{42,1}}{46080}-\frac{7
C_{22,00}^{42,0}}{3072 \sqrt{3}}-\frac{7 C_{22,00}^{42,1}}{1536 \sqrt{3}}-\frac{7 C_{22,20}^{42,0}}{9216}-\frac{7 C_{22,20}^{42,1}}{4608}+\frac{17
C_{22,22}^{40,0}}{92160}+\frac{17 C_{22,22}^{40,1}}{46080}+\frac{\sqrt{7} C_{22,22}^{42,0}}{9216}+\frac{\sqrt{7} C_{22,22}^{42,1}}{4608}-\frac{3
C_{22,22}^{44,0}}{2560 \sqrt{5}}-\frac{3 C_{22,22}^{44,1}}{1280 \sqrt{5}}-\frac{\sqrt{\frac{5}{3}} C_{31,00}^{31,0}}{6144}-\frac{\sqrt{\frac{5}{3}}
C_{31,00}^{31,1}}{3072}-\frac{\sqrt{5} C_{31,20}^{31,0}}{18432}-\frac{\sqrt{5} C_{31,20}^{31,1}}{9216}+\frac{23 C_{31,22}^{31,0}}{92160}+\frac{23
C_{31,22}^{31,1}}{46080}-\frac{\sqrt{\frac{7}{15}} C_{31,22}^{33,0}}{10240}-\frac{\sqrt{\frac{7}{15}} C_{31,22}^{33,1}}{5120}-\frac{3 \sqrt{\frac{7}{5}}
C_{33,00}^{33,0}}{1024}-\frac{3}{512} \sqrt{\frac{7}{5}} C_{33,00}^{33,1}-\frac{\sqrt{\frac{21}{5}} C_{33,20}^{33,0}}{1024}-\frac{1}{512} \sqrt{\frac{21}{5}}
C_{33,20}^{33,1}+\frac{3 \sqrt{\frac{7}{2}} C_{33,22}^{33,0}}{12800}+\frac{3 \sqrt{\frac{7}{2}} C_{33,22}^{33,1}}{6400}\)

\noindent\(C_{40,1112,2}^{00,1112,1,\text{Loc}}= \frac{7 \sqrt{\frac{5}{3}} C_{00,20}^{60,1}}{1024}+\frac{7 \sqrt{3} C_{00,22}^{62,1}}{10240}-\frac{7
\sqrt{\frac{5}{3}} C_{11,20}^{31,1}}{3072}+\frac{7 C_{11,22}^{51,1}}{30720 \sqrt{3}}-\frac{9 \sqrt{\frac{7}{5}} C_{11,22}^{53,1}}{10240}+\frac{5
\sqrt{\frac{5}{3}} C_{20,20}^{40,1}}{3072}+\frac{7 C_{20,22}^{42,1}}{30720 \sqrt{3}}-\frac{7 C_{22,20}^{42,1}}{1536 \sqrt{3}}-\frac{17 C_{22,22}^{40,1}}{30720
\sqrt{3}}-\frac{\sqrt{\frac{7}{3}} C_{22,22}^{42,1}}{3072}+\frac{3 \sqrt{\frac{3}{5}} C_{22,22}^{44,1}}{2560}+\frac{\sqrt{\frac{5}{3}} C_{31,20}^{31,1}}{3072}+\frac{23
C_{31,22}^{31,1}}{30720 \sqrt{3}}-\frac{\sqrt{\frac{7}{5}} C_{31,22}^{33,1}}{10240}+\frac{3}{512} \sqrt{\frac{7}{5}} C_{33,20}^{33,1}+\frac{3 \sqrt{\frac{21}{2}}
C_{33,22}^{33,1}}{12800}\)

\noindent\(C_{40,1112,2}^{00,1112,1,\text{NonLoc}}= \frac{7 \sqrt{5} C_{00,00}^{60,0}}{2048}+\frac{7 \sqrt{\frac{5}{3}} C_{00,20}^{60,0}}{2048}-\frac{7
\sqrt{3} C_{00,22}^{62,0}}{10240}+\frac{7 \sqrt{5} C_{11,00}^{51,0}}{6144}+\frac{7 \sqrt{\frac{5}{3}} C_{11,20}^{31,0}}{6144}+\frac{7 C_{11,22}^{51,0}}{30720
\sqrt{3}}-\frac{9 \sqrt{\frac{7}{5}} C_{11,22}^{53,0}}{10240}+\frac{5 \sqrt{5} C_{20,00}^{40,0}}{6144}+\frac{5 \sqrt{\frac{5}{3}} C_{20,20}^{40,0}}{6144}-\frac{7
C_{20,22}^{42,0}}{30720 \sqrt{3}}-\frac{7 C_{22,00}^{42,0}}{3072}-\frac{7 C_{22,20}^{42,0}}{3072 \sqrt{3}}+\frac{17 C_{22,22}^{40,0}}{30720 \sqrt{3}}+\frac{\sqrt{\frac{7}{3}}
C_{22,22}^{42,0}}{3072}-\frac{3 \sqrt{\frac{3}{5}} C_{22,22}^{44,0}}{2560}-\frac{\sqrt{5} C_{31,00}^{31,0}}{6144}-\frac{\sqrt{\frac{5}{3}} C_{31,20}^{31,0}}{6144}+\frac{23
C_{31,22}^{31,0}}{30720 \sqrt{3}}-\frac{\sqrt{\frac{7}{5}} C_{31,22}^{33,0}}{10240}-\frac{3 \sqrt{\frac{21}{5}} C_{33,00}^{33,0}}{1024}-\frac{3 \sqrt{\frac{7}{5}}
C_{33,20}^{33,0}}{1024}+\frac{3 \sqrt{\frac{21}{2}} C_{33,22}^{33,0}}{12800}\)

\noindent\(C_{40,2000,0}^{00,0000,0,\text{Loc}}= \frac{7 C_{00,00}^{60,0}}{1024}+\frac{7 C_{00,00}^{60,1}}{2048}-\frac{7 C_{11,00}^{51,0}}{3072}-\frac{7
C_{11,00}^{51,1}}{6144}+\frac{5 C_{20,00}^{40,0}}{3072}+\frac{5 C_{20,00}^{40,1}}{6144}-\frac{7 C_{22,00}^{42,0}}{1536 \sqrt{5}}-\frac{7 C_{22,00}^{42,1}}{3072
\sqrt{5}}+\frac{C_{31,00}^{31,0}}{3072}+\frac{C_{31,00}^{31,1}}{6144}+\frac{3 \sqrt{21} C_{33,00}^{33,0}}{2560}+\frac{3 \sqrt{21} C_{33,00}^{33,1}}{5120}\)

\noindent\(C_{40,2000,0}^{00,0000,0,\text{NonLoc}}= -\frac{7 C_{00,00}^{60,0}}{4096}-\frac{7 C_{00,00}^{60,1}}{2048}+\frac{7 \sqrt{3}
C_{00,20}^{60,0}}{4096}+\frac{7 \sqrt{3} C_{00,20}^{60,1}}{2048}-\frac{7 C_{11,00}^{51,0}}{12288}-\frac{7 C_{11,00}^{51,1}}{6144}+\frac{7 C_{11,20}^{31,0}}{4096
\sqrt{3}}+\frac{7 C_{11,20}^{31,1}}{2048 \sqrt{3}}-\frac{5 C_{20,00}^{40,0}}{12288}-\frac{5 C_{20,00}^{40,1}}{6144}+\frac{5 C_{20,20}^{40,0}}{4096
\sqrt{3}}+\frac{5 C_{20,20}^{40,1}}{2048 \sqrt{3}}+\frac{7 C_{22,00}^{42,0}}{6144 \sqrt{5}}+\frac{7 C_{22,00}^{42,1}}{3072 \sqrt{5}}-\frac{7 C_{22,20}^{42,0}}{2048
\sqrt{15}}-\frac{7 C_{22,20}^{42,1}}{1024 \sqrt{15}}+\frac{C_{31,00}^{31,0}}{12288}+\frac{C_{31,00}^{31,1}}{6144}-\frac{C_{31,20}^{31,0}}{4096 \sqrt{3}}-\frac{C_{31,20}^{31,1}}{2048
\sqrt{3}}+\frac{3 \sqrt{21} C_{33,00}^{33,0}}{10240}+\frac{3 \sqrt{21} C_{33,00}^{33,1}}{5120}-\frac{9 \sqrt{7} C_{33,20}^{33,0}}{10240}-\frac{9
\sqrt{7} C_{33,20}^{33,1}}{5120}\)

\noindent\(C_{40,2000,0}^{00,0000,1,\text{Loc}}= \frac{7 \sqrt{3} C_{00,00}^{60,1}}{2048}-\frac{7 C_{11,00}^{51,1}}{2048 \sqrt{3}}+\frac{5
C_{20,00}^{40,1}}{2048 \sqrt{3}}-\frac{7 C_{22,00}^{42,1}}{1024 \sqrt{15}}+\frac{C_{31,00}^{31,1}}{2048 \sqrt{3}}+\frac{9 \sqrt{7} C_{33,00}^{33,1}}{5120}\)

\noindent\(C_{40,2000,0}^{00,0000,1,\text{NonLoc}}= -\frac{7 \sqrt{3} C_{00,00}^{60,0}}{4096}+\frac{21 C_{00,20}^{60,0}}{4096}-\frac{7
C_{11,00}^{51,0}}{4096 \sqrt{3}}+\frac{7 C_{11,20}^{31,0}}{4096}-\frac{5 C_{20,00}^{40,0}}{4096 \sqrt{3}}+\frac{5 C_{20,20}^{40,0}}{4096}+\frac{7
C_{22,00}^{42,0}}{2048 \sqrt{15}}-\frac{7 C_{22,20}^{42,0}}{2048 \sqrt{5}}+\frac{C_{31,00}^{31,0}}{4096 \sqrt{3}}-\frac{C_{31,20}^{31,0}}{4096}+\frac{9
\sqrt{7} C_{33,00}^{33,0}}{10240}-\frac{9 \sqrt{21} C_{33,20}^{33,0}}{10240}\)

\noindent\(C_{40,2011,1}^{00,0011,0,\text{Loc}}= -\frac{7 C_{00,20}^{60,0}}{1024}-\frac{7 C_{00,20}^{60,1}}{2048}+\frac{7 C_{11,20}^{31,0}}{3072}+\frac{7
C_{11,20}^{31,1}}{6144}-\frac{5 C_{20,20}^{40,0}}{3072}-\frac{5 C_{20,20}^{40,1}}{6144}+\frac{7 C_{22,20}^{42,0}}{1536 \sqrt{5}}+\frac{7 C_{22,20}^{42,1}}{3072
\sqrt{5}}-\frac{C_{31,20}^{31,0}}{3072}-\frac{C_{31,20}^{31,1}}{6144}-\frac{3 \sqrt{21} C_{33,20}^{33,0}}{2560}-\frac{3 \sqrt{21} C_{33,20}^{33,1}}{5120}\)

\noindent\(C_{40,2011,1}^{00,0011,0,\text{NonLoc}}= -\frac{7 \sqrt{3} C_{00,00}^{60,0}}{4096}-\frac{7 \sqrt{3} C_{00,00}^{60,1}}{2048}-\frac{7
C_{00,20}^{60,0}}{4096}-\frac{7 C_{00,20}^{60,1}}{2048}-\frac{7 C_{11,00}^{51,0}}{4096 \sqrt{3}}-\frac{7 C_{11,00}^{51,1}}{2048 \sqrt{3}}-\frac{7
C_{11,20}^{31,0}}{12288}-\frac{7 C_{11,20}^{31,1}}{6144}-\frac{5 C_{20,00}^{40,0}}{4096 \sqrt{3}}-\frac{5 C_{20,00}^{40,1}}{2048 \sqrt{3}}-\frac{5
C_{20,20}^{40,0}}{12288}-\frac{5 C_{20,20}^{40,1}}{6144}+\frac{7 C_{22,00}^{42,0}}{2048 \sqrt{15}}+\frac{7 C_{22,00}^{42,1}}{1024 \sqrt{15}}+\frac{7
C_{22,20}^{42,0}}{6144 \sqrt{5}}+\frac{7 C_{22,20}^{42,1}}{3072 \sqrt{5}}+\frac{C_{31,00}^{31,0}}{4096 \sqrt{3}}+\frac{C_{31,00}^{31,1}}{2048 \sqrt{3}}+\frac{C_{31,20}^{31,0}}{12288}+\frac{C_{31,20}^{31,1}}{6144}+\frac{9
\sqrt{7} C_{33,00}^{33,0}}{10240}+\frac{9 \sqrt{7} C_{33,00}^{33,1}}{5120}+\frac{3 \sqrt{21} C_{33,20}^{33,0}}{10240}+\frac{3 \sqrt{21} C_{33,20}^{33,1}}{5120}\)

\noindent\(C_{40,2011,1}^{00,0011,1,\text{Loc}}= -\frac{7 \sqrt{3} C_{00,20}^{60,1}}{2048}+\frac{7 C_{11,20}^{31,1}}{2048 \sqrt{3}}-\frac{5
C_{20,20}^{40,1}}{2048 \sqrt{3}}+\frac{7 C_{22,20}^{42,1}}{1024 \sqrt{15}}-\frac{C_{31,20}^{31,1}}{2048 \sqrt{3}}-\frac{9 \sqrt{7} C_{33,20}^{33,1}}{5120}\)

\noindent\(C_{40,2011,1}^{00,0011,1,\text{NonLoc}}= -\frac{21 C_{00,00}^{60,0}}{4096}-\frac{7 \sqrt{3} C_{00,20}^{60,0}}{4096}-\frac{7
C_{11,00}^{51,0}}{4096}-\frac{7 C_{11,20}^{31,0}}{4096 \sqrt{3}}-\frac{5 C_{20,00}^{40,0}}{4096}-\frac{5 C_{20,20}^{40,0}}{4096 \sqrt{3}}+\frac{7
C_{22,00}^{42,0}}{2048 \sqrt{5}}+\frac{7 C_{22,20}^{42,0}}{2048 \sqrt{15}}+\frac{C_{31,00}^{31,0}}{4096}+\frac{C_{31,20}^{31,0}}{4096 \sqrt{3}}+\frac{9
\sqrt{21} C_{33,00}^{33,0}}{10240}+\frac{9 \sqrt{7} C_{33,20}^{33,0}}{10240}\)

\noindent\(C_{40,2211,1}^{00,0011,0,\text{Loc}}= -\frac{21 C_{00,22}^{62,0}}{5120}-\frac{21 C_{00,22}^{62,1}}{10240}-\frac{7 C_{11,22}^{51,0}}{15360}-\frac{7
C_{11,22}^{51,1}}{30720}+\frac{9 \sqrt{\frac{21}{5}} C_{11,22}^{53,0}}{5120}+\frac{9 \sqrt{\frac{21}{5}} C_{11,22}^{53,1}}{10240}-\frac{7 C_{20,22}^{42,0}}{15360}-\frac{7
C_{20,22}^{42,1}}{30720}+\frac{17 C_{22,22}^{40,0}}{15360}+\frac{17 C_{22,22}^{40,1}}{30720}+\frac{\sqrt{7} C_{22,22}^{42,0}}{1536}+\frac{\sqrt{7}
C_{22,22}^{42,1}}{3072}-\frac{9 C_{22,22}^{44,0}}{1280 \sqrt{5}}-\frac{9 C_{22,22}^{44,1}}{2560 \sqrt{5}}-\frac{23 C_{31,22}^{31,0}}{15360}-\frac{23
C_{31,22}^{31,1}}{30720}+\frac{\sqrt{\frac{21}{5}} C_{31,22}^{33,0}}{5120}+\frac{\sqrt{\frac{21}{5}} C_{31,22}^{33,1}}{10240}-\frac{9 \sqrt{\frac{7}{2}}
C_{33,22}^{33,0}}{6400}-\frac{9 \sqrt{\frac{7}{2}} C_{33,22}^{33,1}}{12800}\)

\noindent\(C_{40,2211,1}^{00,0011,0,\text{NonLoc}}= \frac{21 C_{00,22}^{62,0}}{10240}+\frac{21 C_{00,22}^{62,1}}{5120}-\frac{7 C_{11,22}^{51,0}}{30720}-\frac{7
C_{11,22}^{51,1}}{15360}+\frac{9 \sqrt{\frac{21}{5}} C_{11,22}^{53,0}}{10240}+\frac{9 \sqrt{\frac{21}{5}} C_{11,22}^{53,1}}{5120}+\frac{7 C_{20,22}^{42,0}}{30720}+\frac{7
C_{20,22}^{42,1}}{15360}-\frac{17 C_{22,22}^{40,0}}{30720}-\frac{17 C_{22,22}^{40,1}}{15360}-\frac{\sqrt{7} C_{22,22}^{42,0}}{3072}-\frac{\sqrt{7}
C_{22,22}^{42,1}}{1536}+\frac{9 C_{22,22}^{44,0}}{2560 \sqrt{5}}+\frac{9 C_{22,22}^{44,1}}{1280 \sqrt{5}}-\frac{23 C_{31,22}^{31,0}}{30720}-\frac{23
C_{31,22}^{31,1}}{15360}+\frac{\sqrt{\frac{21}{5}} C_{31,22}^{33,0}}{10240}+\frac{\sqrt{\frac{21}{5}} C_{31,22}^{33,1}}{5120}-\frac{9 \sqrt{\frac{7}{2}}
C_{33,22}^{33,0}}{12800}-\frac{9 \sqrt{\frac{7}{2}} C_{33,22}^{33,1}}{6400}\)

\noindent\(C_{40,2211,1}^{00,0011,1,\text{Loc}}= -\frac{21 \sqrt{3} C_{00,22}^{62,1}}{10240}-\frac{7 C_{11,22}^{51,1}}{10240 \sqrt{3}}+\frac{27
\sqrt{\frac{7}{5}} C_{11,22}^{53,1}}{10240}-\frac{7 C_{20,22}^{42,1}}{10240 \sqrt{3}}+\frac{17 C_{22,22}^{40,1}}{10240 \sqrt{3}}+\frac{\sqrt{\frac{7}{3}}
C_{22,22}^{42,1}}{1024}-\frac{9 \sqrt{\frac{3}{5}} C_{22,22}^{44,1}}{2560}-\frac{23 C_{31,22}^{31,1}}{10240 \sqrt{3}}+\frac{3 \sqrt{\frac{7}{5}}
C_{31,22}^{33,1}}{10240}-\frac{9 \sqrt{\frac{21}{2}} C_{33,22}^{33,1}}{12800}\)

\noindent\(C_{40,2211,1}^{00,0011,1,\text{NonLoc}}= \frac{21 \sqrt{3} C_{00,22}^{62,0}}{10240}-\frac{7 C_{11,22}^{51,0}}{10240 \sqrt{3}}+\frac{27
\sqrt{\frac{7}{5}} C_{11,22}^{53,0}}{10240}+\frac{7 C_{20,22}^{42,0}}{10240 \sqrt{3}}-\frac{17 C_{22,22}^{40,0}}{10240 \sqrt{3}}-\frac{\sqrt{\frac{7}{3}}
C_{22,22}^{42,0}}{1024}+\frac{9 \sqrt{\frac{3}{5}} C_{22,22}^{44,0}}{2560}-\frac{23 C_{31,22}^{31,0}}{10240 \sqrt{3}}+\frac{3 \sqrt{\frac{7}{5}}
C_{31,22}^{33,0}}{10240}-\frac{9 \sqrt{\frac{21}{2}} C_{33,22}^{33,0}}{12800}\)

\noindent\(C_{42,0000,2}^{00,2202,0,\text{Loc}}= \frac{\sqrt{5} C_{00,00}^{60,0}}{256}+\frac{\sqrt{5} C_{00,00}^{60,1}}{512}-\frac{\sqrt{5}
C_{11,00}^{51,0}}{768}-\frac{\sqrt{5} C_{11,00}^{51,1}}{1536}-\frac{\sqrt{5} C_{20,00}^{40,0}}{768}-\frac{\sqrt{5} C_{20,00}^{40,1}}{1536}+\frac{C_{22,00}^{42,0}}{768}+\frac{C_{22,00}^{42,1}}{1536}+\frac{\sqrt{5}
C_{31,00}^{31,0}}{768}+\frac{\sqrt{5} C_{31,00}^{31,1}}{1536}-\frac{3}{256} \sqrt{\frac{3}{35}} C_{33,00}^{33,0}-\frac{3}{512} \sqrt{\frac{3}{35}}
C_{33,00}^{33,1}\)

\noindent\(C_{42,0000,2}^{00,2202,0,\text{NonLoc}}= -\frac{\sqrt{5} C_{00,00}^{60,0}}{1024}-\frac{\sqrt{5} C_{00,00}^{60,1}}{512}+\frac{\sqrt{15}
C_{00,20}^{60,0}}{1024}+\frac{\sqrt{15} C_{00,20}^{60,1}}{512}-\frac{\sqrt{5} C_{11,00}^{51,0}}{3072}-\frac{\sqrt{5} C_{11,00}^{51,1}}{1536}+\frac{\sqrt{5}
C_{20,00}^{40,0}}{3072}+\frac{\sqrt{5} C_{20,00}^{40,1}}{1536}-\frac{\sqrt{\frac{5}{3}} C_{20,20}^{40,0}}{1024}-\frac{1}{512} \sqrt{\frac{5}{3}}
C_{20,20}^{40,1}-\frac{C_{22,00}^{42,0}}{3072}-\frac{C_{22,00}^{42,1}}{1536}+\frac{C_{22,20}^{42,0}}{1024 \sqrt{3}}+\frac{C_{22,20}^{42,1}}{512 \sqrt{3}}+\frac{\sqrt{5}
C_{31,00}^{31,0}}{3072}+\frac{\sqrt{5} C_{31,00}^{31,1}}{1536}-\frac{\sqrt{\frac{5}{3}} C_{31,20}^{31,0}}{1024}-\frac{1}{512} \sqrt{\frac{5}{3}}
C_{31,20}^{31,1}-\frac{3 \sqrt{\frac{3}{35}} C_{33,00}^{33,0}}{1024}-\frac{3}{512} \sqrt{\frac{3}{35}} C_{33,00}^{33,1}+\frac{9 C_{33,20}^{33,0}}{1024
\sqrt{35}}+\frac{9 C_{33,20}^{33,1}}{512 \sqrt{35}}\)

\noindent\(C_{42,0000,2}^{00,2202,1,\text{Loc}}= \frac{\sqrt{15} C_{00,00}^{60,1}}{512}-\frac{1}{512} \sqrt{\frac{5}{3}} C_{11,00}^{51,1}-\frac{1}{512}
\sqrt{\frac{5}{3}} C_{20,00}^{40,1}+\frac{C_{22,00}^{42,1}}{512 \sqrt{3}}+\frac{1}{512} \sqrt{\frac{5}{3}} C_{31,00}^{31,1}-\frac{9 C_{33,00}^{33,1}}{512
\sqrt{35}}\)

\noindent\(C_{42,0000,2}^{00,2202,1,\text{NonLoc}}= -\frac{\sqrt{15} C_{00,00}^{60,0}}{1024}+\frac{3 \sqrt{5} C_{00,20}^{60,0}}{1024}-\frac{\sqrt{\frac{5}{3}}
C_{11,00}^{51,0}}{1024}+\frac{\sqrt{\frac{5}{3}} C_{20,00}^{40,0}}{1024}-\frac{\sqrt{5} C_{20,20}^{40,0}}{1024}-\frac{C_{22,00}^{42,0}}{1024 \sqrt{3}}+\frac{C_{22,20}^{42,0}}{1024}+\frac{\sqrt{\frac{5}{3}}
C_{31,00}^{31,0}}{1024}-\frac{\sqrt{5} C_{31,20}^{31,0}}{1024}-\frac{9 C_{33,00}^{33,0}}{1024 \sqrt{35}}+\frac{9 \sqrt{\frac{3}{35}} C_{33,20}^{33,0}}{1024}\)

\noindent\(C_{42,0000,2}^{11,1111,0,\text{Loc}}= -\frac{\sqrt{5} C_{11,11}^{51,0}}{768}-\frac{\sqrt{5} C_{11,11}^{51,1}}{1536}+\frac{C_{22,11}^{42,0}}{128
\sqrt{3}}+\frac{C_{22,11}^{42,1}}{256 \sqrt{3}}-\frac{\sqrt{5} C_{31,11}^{31,0}}{768}-\frac{\sqrt{5} C_{31,11}^{31,1}}{1536}-\frac{3 C_{33,11}^{33,0}}{64
\sqrt{70}}-\frac{3 C_{33,11}^{33,1}}{128 \sqrt{70}}\)

\noindent\(C_{42,0000,2}^{11,1111,0,\text{NonLoc}}= -\frac{\sqrt{5} C_{11,11}^{51,0}}{1536}-\frac{\sqrt{5} C_{11,11}^{51,1}}{768}-\frac{C_{22,11}^{42,0}}{256
\sqrt{3}}-\frac{C_{22,11}^{42,1}}{128 \sqrt{3}}-\frac{\sqrt{5} C_{31,11}^{31,0}}{1536}-\frac{\sqrt{5} C_{31,11}^{31,1}}{768}-\frac{3 C_{33,11}^{33,0}}{128
\sqrt{70}}-\frac{3 C_{33,11}^{33,1}}{64 \sqrt{70}}\)

\noindent\(C_{42,0000,2}^{11,1111,1,\text{Loc}}= -\frac{1}{512} \sqrt{\frac{5}{3}} C_{11,11}^{51,1}+\frac{C_{22,11}^{42,1}}{256}-\frac{1}{512}
\sqrt{\frac{5}{3}} C_{31,11}^{31,1}-\frac{3}{128} \sqrt{\frac{3}{70}} C_{33,11}^{33,1}\)

\noindent\(C_{42,0000,2}^{11,1111,1,\text{NonLoc}}= -\frac{1}{512} \sqrt{\frac{5}{3}} C_{11,11}^{51,0}-\frac{C_{22,11}^{42,0}}{256}-\frac{1}{512}
\sqrt{\frac{5}{3}} C_{31,11}^{31,0}-\frac{3}{128} \sqrt{\frac{3}{70}} C_{33,11}^{33,0}\)

\noindent\(C_{42,0000,2}^{22,0000,0,\text{Loc}}= -\frac{\sqrt{5} C_{00,00}^{60,0}}{1024}-\frac{\sqrt{5} C_{00,00}^{60,1}}{2048}+\frac{\sqrt{5}
C_{11,00}^{51,0}}{1024}+\frac{\sqrt{5} C_{11,00}^{51,1}}{2048}-\frac{\sqrt{5} C_{20,00}^{40,0}}{1024}-\frac{\sqrt{5} C_{20,00}^{40,1}}{2048}-\frac{C_{22,00}^{42,0}}{512}-\frac{C_{22,00}^{42,1}}{1024}+\frac{\sqrt{5}
C_{31,00}^{31,0}}{1024}+\frac{\sqrt{5} C_{31,00}^{31,1}}{2048}+\frac{3}{512} \sqrt{\frac{3}{35}} C_{33,00}^{33,0}+\frac{3 \sqrt{\frac{3}{35}} C_{33,00}^{33,1}}{1024}\)

\noindent\(C_{42,0000,2}^{22,0000,0,\text{NonLoc}}= \frac{\sqrt{5} C_{00,00}^{60,0}}{4096}+\frac{\sqrt{5} C_{00,00}^{60,1}}{2048}-\frac{\sqrt{15}
C_{00,20}^{60,0}}{4096}-\frac{\sqrt{15} C_{00,20}^{60,1}}{2048}+\frac{\sqrt{5} C_{11,00}^{51,0}}{4096}+\frac{\sqrt{5} C_{11,00}^{51,1}}{2048}+\frac{\sqrt{5}
C_{20,00}^{40,0}}{4096}+\frac{\sqrt{5} C_{20,00}^{40,1}}{2048}-\frac{\sqrt{15} C_{20,20}^{40,0}}{4096}-\frac{\sqrt{15} C_{20,20}^{40,1}}{2048}+\frac{C_{22,00}^{42,0}}{2048}+\frac{C_{22,00}^{42,1}}{1024}-\frac{\sqrt{3}
C_{22,20}^{42,0}}{2048}-\frac{\sqrt{3} C_{22,20}^{42,1}}{1024}+\frac{\sqrt{5} C_{31,00}^{31,0}}{4096}+\frac{\sqrt{5} C_{31,00}^{31,1}}{2048}-\frac{\sqrt{15}
C_{31,20}^{31,0}}{4096}-\frac{\sqrt{15} C_{31,20}^{31,1}}{2048}+\frac{3 \sqrt{\frac{3}{35}} C_{33,00}^{33,0}}{2048}+\frac{3 \sqrt{\frac{3}{35}} C_{33,00}^{33,1}}{1024}-\frac{9
C_{33,20}^{33,0}}{2048 \sqrt{35}}-\frac{9 C_{33,20}^{33,1}}{1024 \sqrt{35}}\)

\noindent\(C_{42,0000,2}^{22,0000,1,\text{Loc}}= -\frac{\sqrt{15} C_{00,00}^{60,1}}{2048}+\frac{\sqrt{15} C_{11,00}^{51,1}}{2048}-\frac{\sqrt{15}
C_{20,00}^{40,1}}{2048}-\frac{\sqrt{3} C_{22,00}^{42,1}}{1024}+\frac{\sqrt{15} C_{31,00}^{31,1}}{2048}+\frac{9 C_{33,00}^{33,1}}{1024 \sqrt{35}}\)

\noindent\(C_{42,0000,2}^{22,0000,1,\text{NonLoc}}= \frac{\sqrt{15} C_{00,00}^{60,0}}{4096}-\frac{3 \sqrt{5} C_{00,20}^{60,0}}{4096}+\frac{\sqrt{15}
C_{11,00}^{51,0}}{4096}+\frac{\sqrt{15} C_{20,00}^{40,0}}{4096}-\frac{3 \sqrt{5} C_{20,20}^{40,0}}{4096}+\frac{\sqrt{3} C_{22,00}^{42,0}}{2048}-\frac{3
C_{22,20}^{42,0}}{2048}+\frac{\sqrt{15} C_{31,00}^{31,0}}{4096}-\frac{3 \sqrt{5} C_{31,20}^{31,0}}{4096}+\frac{9 C_{33,00}^{33,0}}{2048 \sqrt{35}}-\frac{9
\sqrt{\frac{3}{35}} C_{33,20}^{33,0}}{2048}\)

\noindent\(C_{42,0011,1}^{00,2011,0,\text{Loc}}= -\frac{3 C_{00,22}^{62,0}}{512}-\frac{3 C_{00,22}^{62,1}}{1024}+\frac{5 C_{11,22}^{51,0}}{1536}+\frac{5
C_{11,22}^{51,1}}{3072}-\frac{C_{20,22}^{42,0}}{1536}-\frac{C_{20,22}^{42,1}}{3072}-\frac{C_{22,22}^{40,0}}{1536}-\frac{C_{22,22}^{40,1}}{3072}+\frac{C_{22,22}^{42,0}}{384
\sqrt{7}}+\frac{C_{22,22}^{42,1}}{768 \sqrt{7}}+\frac{9 C_{22,22}^{44,0}}{896 \sqrt{5}}+\frac{9 C_{22,22}^{44,1}}{1792 \sqrt{5}}+\frac{C_{31,22}^{31,0}}{1536}+\frac{C_{31,22}^{31,1}}{3072}-\frac{1}{128}
\sqrt{\frac{3}{35}} C_{31,22}^{33,0}-\frac{1}{256} \sqrt{\frac{3}{35}} C_{31,22}^{33,1}-\frac{9 C_{33,22}^{33,0}}{640 \sqrt{14}}-\frac{9 C_{33,22}^{33,1}}{1280
\sqrt{14}}\)

\noindent\(C_{42,0011,1}^{00,2011,0,\text{NonLoc}}= \frac{3 C_{00,22}^{62,0}}{1024}+\frac{3 C_{00,22}^{62,1}}{512}+\frac{5 C_{11,22}^{51,0}}{3072}+\frac{5
C_{11,22}^{51,1}}{1536}+\frac{C_{20,22}^{42,0}}{3072}+\frac{C_{20,22}^{42,1}}{1536}+\frac{C_{22,22}^{40,0}}{3072}+\frac{C_{22,22}^{40,1}}{1536}-\frac{C_{22,22}^{42,0}}{768
\sqrt{7}}-\frac{C_{22,22}^{42,1}}{384 \sqrt{7}}-\frac{9 C_{22,22}^{44,0}}{1792 \sqrt{5}}-\frac{9 C_{22,22}^{44,1}}{896 \sqrt{5}}+\frac{C_{31,22}^{31,0}}{3072}+\frac{C_{31,22}^{31,1}}{1536}-\frac{1}{256}
\sqrt{\frac{3}{35}} C_{31,22}^{33,0}-\frac{1}{128} \sqrt{\frac{3}{35}} C_{31,22}^{33,1}-\frac{9 C_{33,22}^{33,0}}{1280 \sqrt{14}}-\frac{9 C_{33,22}^{33,1}}{640
\sqrt{14}}\)

\noindent\(C_{42,0011,1}^{00,2011,1,\text{Loc}}= -\frac{3 \sqrt{3} C_{00,22}^{62,1}}{1024}+\frac{5 C_{11,22}^{51,1}}{1024 \sqrt{3}}-\frac{C_{20,22}^{42,1}}{1024
\sqrt{3}}-\frac{C_{22,22}^{40,1}}{1024 \sqrt{3}}+\frac{C_{22,22}^{42,1}}{256 \sqrt{21}}+\frac{9 \sqrt{\frac{3}{5}} C_{22,22}^{44,1}}{1792}+\frac{C_{31,22}^{31,1}}{1024
\sqrt{3}}-\frac{3 C_{31,22}^{33,1}}{256 \sqrt{35}}-\frac{9 \sqrt{\frac{3}{14}} C_{33,22}^{33,1}}{1280}\)

\noindent\(C_{42,0011,1}^{00,2011,1,\text{NonLoc}}= \frac{3 \sqrt{3} C_{00,22}^{62,0}}{1024}+\frac{5 C_{11,22}^{51,0}}{1024 \sqrt{3}}+\frac{C_{20,22}^{42,0}}{1024
\sqrt{3}}+\frac{C_{22,22}^{40,0}}{1024 \sqrt{3}}-\frac{C_{22,22}^{42,0}}{256 \sqrt{21}}-\frac{9 \sqrt{\frac{3}{5}} C_{22,22}^{44,0}}{1792}+\frac{C_{31,22}^{31,0}}{1024
\sqrt{3}}-\frac{3 C_{31,22}^{33,0}}{256 \sqrt{35}}-\frac{9 \sqrt{\frac{3}{14}} C_{33,22}^{33,0}}{1280}\)

\noindent\(C_{42,0011,1}^{00,2211,0,\text{Loc}}= -\frac{C_{00,20}^{60,0}}{256}-\frac{C_{00,20}^{60,1}}{512}-\frac{3 C_{00,22}^{62,0}}{512
\sqrt{5}}-\frac{3 C_{00,22}^{62,1}}{1024 \sqrt{5}}+\frac{C_{11,22}^{51,0}}{768 \sqrt{5}}+\frac{C_{11,22}^{51,1}}{1536 \sqrt{5}}+\frac{9 \sqrt{\frac{3}{7}}
C_{11,22}^{53,0}}{5120}+\frac{9 \sqrt{\frac{3}{7}} C_{11,22}^{53,1}}{10240}+\frac{C_{20,20}^{40,0}}{768}+\frac{C_{20,20}^{40,1}}{1536}+\frac{C_{20,22}^{42,0}}{768
\sqrt{5}}+\frac{C_{20,22}^{42,1}}{1536 \sqrt{5}}-\frac{C_{22,20}^{42,0}}{768 \sqrt{5}}-\frac{C_{22,20}^{42,1}}{1536 \sqrt{5}}+\frac{\sqrt{5} C_{22,22}^{40,0}}{1536}+\frac{\sqrt{5}
C_{22,22}^{40,1}}{3072}-\frac{1}{768} \sqrt{\frac{7}{5}} C_{22,22}^{42,0}-\frac{\sqrt{\frac{7}{5}} C_{22,22}^{42,1}}{1536}-\frac{C_{31,20}^{31,0}}{768}-\frac{C_{31,20}^{31,1}}{1536}-\frac{C_{31,22}^{31,0}}{768
\sqrt{5}}-\frac{C_{31,22}^{31,1}}{1536 \sqrt{5}}-\frac{9 \sqrt{\frac{3}{7}} C_{31,22}^{33,0}}{5120}-\frac{9 \sqrt{\frac{3}{7}} C_{31,22}^{33,1}}{10240}+\frac{3
\sqrt{\frac{3}{7}} C_{33,20}^{33,0}}{1280}+\frac{3 \sqrt{\frac{3}{7}} C_{33,20}^{33,1}}{2560}+\frac{9 C_{33,22}^{33,0}}{320 \sqrt{70}}+\frac{9 C_{33,22}^{33,1}}{640
\sqrt{70}}\)

\noindent\(C_{42,0011,1}^{00,2211,0,\text{NonLoc}}= -\frac{\sqrt{3} C_{00,00}^{60,0}}{1024}-\frac{\sqrt{3} C_{00,00}^{60,1}}{512}-\frac{C_{00,20}^{60,0}}{1024}-\frac{C_{00,20}^{60,1}}{512}+\frac{3
C_{00,22}^{62,0}}{1024 \sqrt{5}}+\frac{3 C_{00,22}^{62,1}}{512 \sqrt{5}}-\frac{C_{11,00}^{51,0}}{1024 \sqrt{3}}-\frac{C_{11,00}^{51,1}}{512 \sqrt{3}}+\frac{C_{11,22}^{51,0}}{1536
\sqrt{5}}+\frac{C_{11,22}^{51,1}}{768 \sqrt{5}}+\frac{9 \sqrt{\frac{3}{7}} C_{11,22}^{53,0}}{10240}+\frac{9 \sqrt{\frac{3}{7}} C_{11,22}^{53,1}}{5120}+\frac{C_{20,00}^{40,0}}{1024
\sqrt{3}}+\frac{C_{20,00}^{40,1}}{512 \sqrt{3}}+\frac{C_{20,20}^{40,0}}{3072}+\frac{C_{20,20}^{40,1}}{1536}-\frac{C_{20,22}^{42,0}}{1536 \sqrt{5}}-\frac{C_{20,22}^{42,1}}{768
\sqrt{5}}-\frac{C_{22,00}^{42,0}}{1024 \sqrt{15}}-\frac{C_{22,00}^{42,1}}{512 \sqrt{15}}-\frac{C_{22,20}^{42,0}}{3072 \sqrt{5}}-\frac{C_{22,20}^{42,1}}{1536
\sqrt{5}}-\frac{\sqrt{5} C_{22,22}^{40,0}}{3072}-\frac{\sqrt{5} C_{22,22}^{40,1}}{1536}+\frac{\sqrt{\frac{7}{5}} C_{22,22}^{42,0}}{1536}+\frac{1}{768}
\sqrt{\frac{7}{5}} C_{22,22}^{42,1}+\frac{C_{31,00}^{31,0}}{1024 \sqrt{3}}+\frac{C_{31,00}^{31,1}}{512 \sqrt{3}}+\frac{C_{31,20}^{31,0}}{3072}+\frac{C_{31,20}^{31,1}}{1536}-\frac{C_{31,22}^{31,0}}{1536
\sqrt{5}}-\frac{C_{31,22}^{31,1}}{768 \sqrt{5}}-\frac{9 \sqrt{\frac{3}{7}} C_{31,22}^{33,0}}{10240}-\frac{9 \sqrt{\frac{3}{7}} C_{31,22}^{33,1}}{5120}-\frac{9
C_{33,00}^{33,0}}{5120 \sqrt{7}}-\frac{9 C_{33,00}^{33,1}}{2560 \sqrt{7}}-\frac{3 \sqrt{\frac{3}{7}} C_{33,20}^{33,0}}{5120}-\frac{3 \sqrt{\frac{3}{7}}
C_{33,20}^{33,1}}{2560}+\frac{9 C_{33,22}^{33,0}}{640 \sqrt{70}}+\frac{9 C_{33,22}^{33,1}}{320 \sqrt{70}}\)

\noindent\(C_{42,0011,1}^{00,2211,1,\text{Loc}}= -\frac{\sqrt{3} C_{00,20}^{60,1}}{512}-\frac{3 \sqrt{\frac{3}{5}} C_{00,22}^{62,1}}{1024}+\frac{C_{11,22}^{51,1}}{512
\sqrt{15}}+\frac{27 C_{11,22}^{53,1}}{10240 \sqrt{7}}+\frac{C_{20,20}^{40,1}}{512 \sqrt{3}}+\frac{C_{20,22}^{42,1}}{512 \sqrt{15}}-\frac{C_{22,20}^{42,1}}{512
\sqrt{15}}+\frac{\sqrt{\frac{5}{3}} C_{22,22}^{40,1}}{1024}-\frac{1}{512} \sqrt{\frac{7}{15}} C_{22,22}^{42,1}-\frac{C_{31,20}^{31,1}}{512 \sqrt{3}}-\frac{C_{31,22}^{31,1}}{512
\sqrt{15}}-\frac{27 C_{31,22}^{33,1}}{10240 \sqrt{7}}+\frac{9 C_{33,20}^{33,1}}{2560 \sqrt{7}}+\frac{9}{640} \sqrt{\frac{3}{70}} C_{33,22}^{33,1}\)

\noindent\(C_{42,0011,1}^{00,2211,1,\text{NonLoc}}= -\frac{3 C_{00,00}^{60,0}}{1024}-\frac{\sqrt{3} C_{00,20}^{60,0}}{1024}+\frac{3
\sqrt{\frac{3}{5}} C_{00,22}^{62,0}}{1024}-\frac{C_{11,00}^{51,0}}{1024}+\frac{C_{11,22}^{51,0}}{512 \sqrt{15}}+\frac{27 C_{11,22}^{53,0}}{10240
\sqrt{7}}+\frac{C_{20,00}^{40,0}}{1024}+\frac{C_{20,20}^{40,0}}{1024 \sqrt{3}}-\frac{C_{20,22}^{42,0}}{512 \sqrt{15}}-\frac{C_{22,00}^{42,0}}{1024
\sqrt{5}}-\frac{C_{22,20}^{42,0}}{1024 \sqrt{15}}-\frac{\sqrt{\frac{5}{3}} C_{22,22}^{40,0}}{1024}+\frac{1}{512} \sqrt{\frac{7}{15}} C_{22,22}^{42,0}+\frac{C_{31,00}^{31,0}}{1024}+\frac{C_{31,20}^{31,0}}{1024
\sqrt{3}}-\frac{C_{31,22}^{31,0}}{512 \sqrt{15}}-\frac{27 C_{31,22}^{33,0}}{10240 \sqrt{7}}-\frac{9 \sqrt{\frac{3}{7}} C_{33,00}^{33,0}}{5120}-\frac{9
C_{33,20}^{33,0}}{5120 \sqrt{7}}+\frac{9}{640} \sqrt{\frac{3}{70}} C_{33,22}^{33,0}\)

\noindent\(C_{42,0011,1}^{11,1101,0,\text{Loc}}= \frac{\sqrt{5} C_{11,11}^{51,0}}{1536}+\frac{\sqrt{5} C_{11,11}^{51,1}}{3072}-\frac{C_{22,11}^{42,0}}{256
\sqrt{3}}-\frac{C_{22,11}^{42,1}}{512 \sqrt{3}}+\frac{\sqrt{5} C_{31,11}^{31,0}}{1536}+\frac{\sqrt{5} C_{31,11}^{31,1}}{3072}+\frac{3 C_{33,11}^{33,0}}{128
\sqrt{70}}+\frac{3 C_{33,11}^{33,1}}{256 \sqrt{70}}\)

\noindent\(C_{42,0011,1}^{11,1101,0,\text{NonLoc}}= \frac{\sqrt{5} C_{11,11}^{51,0}}{3072}+\frac{\sqrt{5} C_{11,11}^{51,1}}{1536}+\frac{C_{22,11}^{42,0}}{512
\sqrt{3}}+\frac{C_{22,11}^{42,1}}{256 \sqrt{3}}+\frac{\sqrt{5} C_{31,11}^{31,0}}{3072}+\frac{\sqrt{5} C_{31,11}^{31,1}}{1536}+\frac{3 C_{33,11}^{33,0}}{256
\sqrt{70}}+\frac{3 C_{33,11}^{33,1}}{128 \sqrt{70}}\)

\noindent\(C_{42,0011,1}^{11,1101,1,\text{Loc}}= \frac{\sqrt{\frac{5}{3}} C_{11,11}^{51,1}}{1024}-\frac{C_{22,11}^{42,1}}{512}+\frac{\sqrt{\frac{5}{3}}
C_{31,11}^{31,1}}{1024}+\frac{3}{256} \sqrt{\frac{3}{70}} C_{33,11}^{33,1}\)

\noindent\(C_{42,0011,1}^{11,1101,1,\text{NonLoc}}= \frac{\sqrt{\frac{5}{3}} C_{11,11}^{51,0}}{1024}+\frac{C_{22,11}^{42,0}}{512}+\frac{\sqrt{\frac{5}{3}}
C_{31,11}^{31,0}}{1024}+\frac{3}{256} \sqrt{\frac{3}{70}} C_{33,11}^{33,0}\)

\noindent\(C_{42,0011,1}^{20,0011,0,\text{Loc}}= \frac{3 C_{00,22}^{62,0}}{2048}+\frac{3 C_{00,22}^{62,1}}{4096}-\frac{3 C_{11,22}^{51,0}}{2048}-\frac{3
C_{11,22}^{51,1}}{4096}-\frac{9 \sqrt{\frac{3}{35}} C_{11,22}^{53,0}}{2048}-\frac{9 \sqrt{\frac{3}{35}} C_{11,22}^{53,1}}{4096}+\frac{3 C_{20,22}^{42,0}}{2048}+\frac{3
C_{20,22}^{42,1}}{4096}+\frac{3 C_{22,22}^{40,0}}{2048}+\frac{3 C_{22,22}^{40,1}}{4096}+\frac{3 C_{22,22}^{42,0}}{1024 \sqrt{7}}+\frac{3 C_{22,22}^{42,1}}{2048
\sqrt{7}}+\frac{9 C_{22,22}^{44,0}}{3584 \sqrt{5}}+\frac{9 C_{22,22}^{44,1}}{7168 \sqrt{5}}-\frac{3 C_{31,22}^{31,0}}{2048}-\frac{3 C_{31,22}^{31,1}}{4096}-\frac{9
\sqrt{\frac{3}{35}} C_{31,22}^{33,0}}{2048}-\frac{9 \sqrt{\frac{3}{35}} C_{31,22}^{33,1}}{4096}-\frac{9 C_{33,22}^{33,0}}{2560 \sqrt{14}}-\frac{9
C_{33,22}^{33,1}}{5120 \sqrt{14}}\)

\noindent\(C_{42,0011,1}^{20,0011,0,\text{NonLoc}}= -\frac{3 C_{00,22}^{62,0}}{4096}-\frac{3 C_{00,22}^{62,1}}{2048}-\frac{3 C_{11,22}^{51,0}}{4096}-\frac{3
C_{11,22}^{51,1}}{2048}-\frac{9 \sqrt{\frac{3}{35}} C_{11,22}^{53,0}}{4096}-\frac{9 \sqrt{\frac{3}{35}} C_{11,22}^{53,1}}{2048}-\frac{3 C_{20,22}^{42,0}}{4096}-\frac{3
C_{20,22}^{42,1}}{2048}-\frac{3 C_{22,22}^{40,0}}{4096}-\frac{3 C_{22,22}^{40,1}}{2048}-\frac{3 C_{22,22}^{42,0}}{2048 \sqrt{7}}-\frac{3 C_{22,22}^{42,1}}{1024
\sqrt{7}}-\frac{9 C_{22,22}^{44,0}}{7168 \sqrt{5}}-\frac{9 C_{22,22}^{44,1}}{3584 \sqrt{5}}-\frac{3 C_{31,22}^{31,0}}{4096}-\frac{3 C_{31,22}^{31,1}}{2048}-\frac{9
\sqrt{\frac{3}{35}} C_{31,22}^{33,0}}{4096}-\frac{9 \sqrt{\frac{3}{35}} C_{31,22}^{33,1}}{2048}-\frac{9 C_{33,22}^{33,0}}{5120 \sqrt{14}}-\frac{9
C_{33,22}^{33,1}}{2560 \sqrt{14}}\)

\noindent\(C_{42,0011,1}^{20,0011,1,\text{Loc}}= \frac{3 \sqrt{3} C_{00,22}^{62,1}}{4096}-\frac{3 \sqrt{3} C_{11,22}^{51,1}}{4096}-\frac{27
C_{11,22}^{53,1}}{4096 \sqrt{35}}+\frac{3 \sqrt{3} C_{20,22}^{42,1}}{4096}+\frac{3 \sqrt{3} C_{22,22}^{40,1}}{4096}+\frac{3 \sqrt{\frac{3}{7}} C_{22,22}^{42,1}}{2048}+\frac{9
\sqrt{\frac{3}{5}} C_{22,22}^{44,1}}{7168}-\frac{3 \sqrt{3} C_{31,22}^{31,1}}{4096}-\frac{27 C_{31,22}^{33,1}}{4096 \sqrt{35}}-\frac{9 \sqrt{\frac{3}{14}}
C_{33,22}^{33,1}}{5120}\)

\noindent\(C_{42,0011,1}^{20,0011,1,\text{NonLoc}}= -\frac{3 \sqrt{3} C_{00,22}^{62,0}}{4096}-\frac{3 \sqrt{3} C_{11,22}^{51,0}}{4096}-\frac{27
C_{11,22}^{53,0}}{4096 \sqrt{35}}-\frac{3 \sqrt{3} C_{20,22}^{42,0}}{4096}-\frac{3 \sqrt{3} C_{22,22}^{40,0}}{4096}-\frac{3 \sqrt{\frac{3}{7}} C_{22,22}^{42,0}}{2048}-\frac{9
\sqrt{\frac{3}{5}} C_{22,22}^{44,0}}{7168}-\frac{3 \sqrt{3} C_{31,22}^{31,0}}{4096}-\frac{27 C_{31,22}^{33,0}}{4096 \sqrt{35}}-\frac{9 \sqrt{\frac{3}{14}}
C_{33,22}^{33,0}}{5120}\)

\noindent\(C_{42,0011,1}^{22,0011,0,\text{Loc}}= \frac{C_{00,20}^{60,0}}{1024}+\frac{C_{00,20}^{60,1}}{2048}+\frac{3 C_{00,22}^{62,0}}{2048
\sqrt{5}}+\frac{3 C_{00,22}^{62,1}}{4096 \sqrt{5}}-\frac{3 C_{11,22}^{51,0}}{2048 \sqrt{5}}-\frac{3 C_{11,22}^{51,1}}{4096 \sqrt{5}}-\frac{9 \sqrt{\frac{3}{7}}
C_{11,22}^{53,0}}{10240}-\frac{9 \sqrt{\frac{3}{7}} C_{11,22}^{53,1}}{20480}+\frac{C_{20,20}^{40,0}}{1024}+\frac{C_{20,20}^{40,1}}{2048}+\frac{3
C_{20,22}^{42,0}}{2048 \sqrt{5}}+\frac{3 C_{20,22}^{42,1}}{4096 \sqrt{5}}+\frac{C_{22,20}^{42,0}}{512 \sqrt{5}}+\frac{C_{22,20}^{42,1}}{1024 \sqrt{5}}+\frac{3
C_{22,22}^{40,0}}{2048 \sqrt{5}}+\frac{3 C_{22,22}^{40,1}}{4096 \sqrt{5}}+\frac{3 C_{22,22}^{42,0}}{1024 \sqrt{35}}+\frac{3 C_{22,22}^{42,1}}{2048
\sqrt{35}}+\frac{9 C_{22,22}^{44,0}}{17920}+\frac{9 C_{22,22}^{44,1}}{35840}-\frac{C_{31,20}^{31,0}}{1024}-\frac{C_{31,20}^{31,1}}{2048}-\frac{3
C_{31,22}^{31,0}}{2048 \sqrt{5}}-\frac{3 C_{31,22}^{31,1}}{4096 \sqrt{5}}-\frac{9 \sqrt{\frac{3}{7}} C_{31,22}^{33,0}}{10240}-\frac{9 \sqrt{\frac{3}{7}}
C_{31,22}^{33,1}}{20480}-\frac{3 \sqrt{\frac{3}{7}} C_{33,20}^{33,0}}{2560}-\frac{3 \sqrt{\frac{3}{7}} C_{33,20}^{33,1}}{5120}-\frac{9 C_{33,22}^{33,0}}{2560
\sqrt{70}}-\frac{9 C_{33,22}^{33,1}}{5120 \sqrt{70}}\)

\noindent\(C_{42,0011,1}^{22,0011,0,\text{NonLoc}}= \frac{\sqrt{3} C_{00,00}^{60,0}}{4096}+\frac{\sqrt{3} C_{00,00}^{60,1}}{2048}+\frac{C_{00,20}^{60,0}}{4096}+\frac{C_{00,20}^{60,1}}{2048}-\frac{3
C_{00,22}^{62,0}}{4096 \sqrt{5}}-\frac{3 C_{00,22}^{62,1}}{2048 \sqrt{5}}+\frac{\sqrt{3} C_{11,00}^{51,0}}{4096}+\frac{\sqrt{3} C_{11,00}^{51,1}}{2048}-\frac{3
C_{11,22}^{51,0}}{4096 \sqrt{5}}-\frac{3 C_{11,22}^{51,1}}{2048 \sqrt{5}}-\frac{9 \sqrt{\frac{3}{7}} C_{11,22}^{53,0}}{20480}-\frac{9 \sqrt{\frac{3}{7}}
C_{11,22}^{53,1}}{10240}+\frac{\sqrt{3} C_{20,00}^{40,0}}{4096}+\frac{\sqrt{3} C_{20,00}^{40,1}}{2048}+\frac{C_{20,20}^{40,0}}{4096}+\frac{C_{20,20}^{40,1}}{2048}-\frac{3
C_{20,22}^{42,0}}{4096 \sqrt{5}}-\frac{3 C_{20,22}^{42,1}}{2048 \sqrt{5}}+\frac{\sqrt{\frac{3}{5}} C_{22,00}^{42,0}}{2048}+\frac{\sqrt{\frac{3}{5}}
C_{22,00}^{42,1}}{1024}+\frac{C_{22,20}^{42,0}}{2048 \sqrt{5}}+\frac{C_{22,20}^{42,1}}{1024 \sqrt{5}}-\frac{3 C_{22,22}^{40,0}}{4096 \sqrt{5}}-\frac{3
C_{22,22}^{40,1}}{2048 \sqrt{5}}-\frac{3 C_{22,22}^{42,0}}{2048 \sqrt{35}}-\frac{3 C_{22,22}^{42,1}}{1024 \sqrt{35}}-\frac{9 C_{22,22}^{44,0}}{35840}-\frac{9
C_{22,22}^{44,1}}{17920}+\frac{\sqrt{3} C_{31,00}^{31,0}}{4096}+\frac{\sqrt{3} C_{31,00}^{31,1}}{2048}+\frac{C_{31,20}^{31,0}}{4096}+\frac{C_{31,20}^{31,1}}{2048}-\frac{3
C_{31,22}^{31,0}}{4096 \sqrt{5}}-\frac{3 C_{31,22}^{31,1}}{2048 \sqrt{5}}-\frac{9 \sqrt{\frac{3}{7}} C_{31,22}^{33,0}}{20480}-\frac{9 \sqrt{\frac{3}{7}}
C_{31,22}^{33,1}}{10240}+\frac{9 C_{33,00}^{33,0}}{10240 \sqrt{7}}+\frac{9 C_{33,00}^{33,1}}{5120 \sqrt{7}}+\frac{3 \sqrt{\frac{3}{7}} C_{33,20}^{33,0}}{10240}+\frac{3
\sqrt{\frac{3}{7}} C_{33,20}^{33,1}}{5120}-\frac{9 C_{33,22}^{33,0}}{5120 \sqrt{70}}-\frac{9 C_{33,22}^{33,1}}{2560 \sqrt{70}}\)

\noindent\(C_{42,0011,1}^{22,0011,1,\text{Loc}}= \frac{\sqrt{3} C_{00,20}^{60,1}}{2048}+\frac{3 \sqrt{\frac{3}{5}} C_{00,22}^{62,1}}{4096}-\frac{3
\sqrt{\frac{3}{5}} C_{11,22}^{51,1}}{4096}-\frac{27 C_{11,22}^{53,1}}{20480 \sqrt{7}}+\frac{\sqrt{3} C_{20,20}^{40,1}}{2048}+\frac{3 \sqrt{\frac{3}{5}}
C_{20,22}^{42,1}}{4096}+\frac{\sqrt{\frac{3}{5}} C_{22,20}^{42,1}}{1024}+\frac{3 \sqrt{\frac{3}{5}} C_{22,22}^{40,1}}{4096}+\frac{3 \sqrt{\frac{3}{35}}
C_{22,22}^{42,1}}{2048}+\frac{9 \sqrt{3} C_{22,22}^{44,1}}{35840}-\frac{\sqrt{3} C_{31,20}^{31,1}}{2048}-\frac{3 \sqrt{\frac{3}{5}} C_{31,22}^{31,1}}{4096}-\frac{27
C_{31,22}^{33,1}}{20480 \sqrt{7}}-\frac{9 C_{33,20}^{33,1}}{5120 \sqrt{7}}-\frac{9 \sqrt{\frac{3}{70}} C_{33,22}^{33,1}}{5120}\)

\noindent\(C_{42,0011,1}^{22,0011,1,\text{NonLoc}}= \frac{3 C_{00,00}^{60,0}}{4096}+\frac{\sqrt{3} C_{00,20}^{60,0}}{4096}-\frac{3
\sqrt{\frac{3}{5}} C_{00,22}^{62,0}}{4096}+\frac{3 C_{11,00}^{51,0}}{4096}-\frac{3 \sqrt{\frac{3}{5}} C_{11,22}^{51,0}}{4096}-\frac{27 C_{11,22}^{53,0}}{20480
\sqrt{7}}+\frac{3 C_{20,00}^{40,0}}{4096}+\frac{\sqrt{3} C_{20,20}^{40,0}}{4096}-\frac{3 \sqrt{\frac{3}{5}} C_{20,22}^{42,0}}{4096}+\frac{3 C_{22,00}^{42,0}}{2048
\sqrt{5}}+\frac{\sqrt{\frac{3}{5}} C_{22,20}^{42,0}}{2048}-\frac{3 \sqrt{\frac{3}{5}} C_{22,22}^{40,0}}{4096}-\frac{3 \sqrt{\frac{3}{35}} C_{22,22}^{42,0}}{2048}-\frac{9
\sqrt{3} C_{22,22}^{44,0}}{35840}+\frac{3 C_{31,00}^{31,0}}{4096}+\frac{\sqrt{3} C_{31,20}^{31,0}}{4096}-\frac{3 \sqrt{\frac{3}{5}} C_{31,22}^{31,0}}{4096}-\frac{27
C_{31,22}^{33,0}}{20480 \sqrt{7}}+\frac{9 \sqrt{\frac{3}{7}} C_{33,00}^{33,0}}{10240}+\frac{9 C_{33,20}^{33,0}}{10240 \sqrt{7}}-\frac{9 \sqrt{\frac{3}{70}}
C_{33,22}^{33,0}}{5120}\)

\noindent\(C_{42,0011,2}^{00,2212,0,\text{Loc}}= -\frac{1}{256} \sqrt{\frac{5}{3}} C_{00,20}^{60,0}-\frac{1}{512} \sqrt{\frac{5}{3}}
C_{00,20}^{60,1}+\frac{\sqrt{3} C_{00,22}^{62,0}}{512}+\frac{\sqrt{3} C_{00,22}^{62,1}}{1024}-\frac{C_{11,22}^{51,0}}{768 \sqrt{3}}-\frac{C_{11,22}^{51,1}}{1536
\sqrt{3}}-\frac{9 C_{11,22}^{53,0}}{1024 \sqrt{35}}-\frac{9 C_{11,22}^{53,1}}{2048 \sqrt{35}}+\frac{1}{768} \sqrt{\frac{5}{3}} C_{20,20}^{40,0}+\frac{\sqrt{\frac{5}{3}}
C_{20,20}^{40,1}}{1536}-\frac{C_{20,22}^{42,0}}{768 \sqrt{3}}-\frac{C_{20,22}^{42,1}}{1536 \sqrt{3}}-\frac{C_{22,20}^{42,0}}{768 \sqrt{3}}-\frac{C_{22,20}^{42,1}}{1536
\sqrt{3}}-\frac{5 C_{22,22}^{40,0}}{1536 \sqrt{3}}-\frac{5 C_{22,22}^{40,1}}{3072 \sqrt{3}}+\frac{1}{768} \sqrt{\frac{7}{3}} C_{22,22}^{42,0}+\frac{\sqrt{\frac{7}{3}}
C_{22,22}^{42,1}}{1536}-\frac{1}{768} \sqrt{\frac{5}{3}} C_{31,20}^{31,0}-\frac{\sqrt{\frac{5}{3}} C_{31,20}^{31,1}}{1536}+\frac{C_{31,22}^{31,0}}{768
\sqrt{3}}+\frac{C_{31,22}^{31,1}}{1536 \sqrt{3}}+\frac{9 C_{31,22}^{33,0}}{1024 \sqrt{35}}+\frac{9 C_{31,22}^{33,1}}{2048 \sqrt{35}}+\frac{3 C_{33,20}^{33,0}}{256
\sqrt{35}}+\frac{3 C_{33,20}^{33,1}}{512 \sqrt{35}}-\frac{3}{320} \sqrt{\frac{3}{14}} C_{33,22}^{33,0}-\frac{3}{640} \sqrt{\frac{3}{14}} C_{33,22}^{33,1}\)

\noindent\(C_{42,0011,2}^{00,2212,0,\text{NonLoc}}= -\frac{\sqrt{5} C_{00,00}^{60,0}}{1024}-\frac{\sqrt{5} C_{00,00}^{60,1}}{512}-\frac{\sqrt{\frac{5}{3}}
C_{00,20}^{60,0}}{1024}-\frac{1}{512} \sqrt{\frac{5}{3}} C_{00,20}^{60,1}-\frac{\sqrt{3} C_{00,22}^{62,0}}{1024}-\frac{\sqrt{3} C_{00,22}^{62,1}}{512}-\frac{\sqrt{5}
C_{11,00}^{51,0}}{3072}-\frac{\sqrt{5} C_{11,00}^{51,1}}{1536}-\frac{C_{11,22}^{51,0}}{1536 \sqrt{3}}-\frac{C_{11,22}^{51,1}}{768 \sqrt{3}}-\frac{9
C_{11,22}^{53,0}}{2048 \sqrt{35}}-\frac{9 C_{11,22}^{53,1}}{1024 \sqrt{35}}+\frac{\sqrt{5} C_{20,00}^{40,0}}{3072}+\frac{\sqrt{5} C_{20,00}^{40,1}}{1536}+\frac{\sqrt{\frac{5}{3}}
C_{20,20}^{40,0}}{3072}+\frac{\sqrt{\frac{5}{3}} C_{20,20}^{40,1}}{1536}+\frac{C_{20,22}^{42,0}}{1536 \sqrt{3}}+\frac{C_{20,22}^{42,1}}{768 \sqrt{3}}-\frac{C_{22,00}^{42,0}}{3072}-\frac{C_{22,00}^{42,1}}{1536}-\frac{C_{22,20}^{42,0}}{3072
\sqrt{3}}-\frac{C_{22,20}^{42,1}}{1536 \sqrt{3}}+\frac{5 C_{22,22}^{40,0}}{3072 \sqrt{3}}+\frac{5 C_{22,22}^{40,1}}{1536 \sqrt{3}}-\frac{\sqrt{\frac{7}{3}}
C_{22,22}^{42,0}}{1536}-\frac{1}{768} \sqrt{\frac{7}{3}} C_{22,22}^{42,1}+\frac{\sqrt{5} C_{31,00}^{31,0}}{3072}+\frac{\sqrt{5} C_{31,00}^{31,1}}{1536}+\frac{\sqrt{\frac{5}{3}}
C_{31,20}^{31,0}}{3072}+\frac{\sqrt{\frac{5}{3}} C_{31,20}^{31,1}}{1536}+\frac{C_{31,22}^{31,0}}{1536 \sqrt{3}}+\frac{C_{31,22}^{31,1}}{768 \sqrt{3}}+\frac{9
C_{31,22}^{33,0}}{2048 \sqrt{35}}+\frac{9 C_{31,22}^{33,1}}{1024 \sqrt{35}}-\frac{3 \sqrt{\frac{3}{35}} C_{33,00}^{33,0}}{1024}-\frac{3}{512} \sqrt{\frac{3}{35}}
C_{33,00}^{33,1}-\frac{3 C_{33,20}^{33,0}}{1024 \sqrt{35}}-\frac{3 C_{33,20}^{33,1}}{512 \sqrt{35}}-\frac{3}{640} \sqrt{\frac{3}{14}} C_{33,22}^{33,0}-\frac{3}{320}
\sqrt{\frac{3}{14}} C_{33,22}^{33,1}\)

\noindent\(C_{42,0011,2}^{00,2212,1,\text{Loc}}= -\frac{\sqrt{5} C_{00,20}^{60,1}}{512}+\frac{3 C_{00,22}^{62,1}}{1024}-\frac{C_{11,22}^{51,1}}{1536}-\frac{9
\sqrt{\frac{3}{35}} C_{11,22}^{53,1}}{2048}+\frac{\sqrt{5} C_{20,20}^{40,1}}{1536}-\frac{C_{20,22}^{42,1}}{1536}-\frac{C_{22,20}^{42,1}}{1536}-\frac{5
C_{22,22}^{40,1}}{3072}+\frac{\sqrt{7} C_{22,22}^{42,1}}{1536}-\frac{\sqrt{5} C_{31,20}^{31,1}}{1536}+\frac{C_{31,22}^{31,1}}{1536}+\frac{9 \sqrt{\frac{3}{35}}
C_{31,22}^{33,1}}{2048}+\frac{3}{512} \sqrt{\frac{3}{35}} C_{33,20}^{33,1}-\frac{9 C_{33,22}^{33,1}}{640 \sqrt{14}}\)

\noindent\(C_{42,0011,2}^{00,2212,1,\text{NonLoc}}= -\frac{\sqrt{15} C_{00,00}^{60,0}}{1024}-\frac{\sqrt{5} C_{00,20}^{60,0}}{1024}-\frac{3
C_{00,22}^{62,0}}{1024}-\frac{\sqrt{\frac{5}{3}} C_{11,00}^{51,0}}{1024}-\frac{C_{11,22}^{51,0}}{1536}-\frac{9 \sqrt{\frac{3}{35}} C_{11,22}^{53,0}}{2048}+\frac{\sqrt{\frac{5}{3}}
C_{20,00}^{40,0}}{1024}+\frac{\sqrt{5} C_{20,20}^{40,0}}{3072}+\frac{C_{20,22}^{42,0}}{1536}-\frac{C_{22,00}^{42,0}}{1024 \sqrt{3}}-\frac{C_{22,20}^{42,0}}{3072}+\frac{5
C_{22,22}^{40,0}}{3072}-\frac{\sqrt{7} C_{22,22}^{42,0}}{1536}+\frac{\sqrt{\frac{5}{3}} C_{31,00}^{31,0}}{1024}+\frac{\sqrt{5} C_{31,20}^{31,0}}{3072}+\frac{C_{31,22}^{31,0}}{1536}+\frac{9
\sqrt{\frac{3}{35}} C_{31,22}^{33,0}}{2048}-\frac{9 C_{33,00}^{33,0}}{1024 \sqrt{35}}-\frac{3 \sqrt{\frac{3}{35}} C_{33,20}^{33,0}}{1024}-\frac{9
C_{33,22}^{33,0}}{640 \sqrt{14}}\)

\noindent\(C_{42,0011,2}^{11,1101,0,\text{Loc}}= \frac{1}{512} \sqrt{\frac{5}{3}} C_{11,11}^{51,0}+\frac{\sqrt{\frac{5}{3}} C_{11,11}^{51,1}}{1024}-\frac{C_{22,11}^{42,0}}{256}-\frac{C_{22,11}^{42,1}}{512}+\frac{1}{512}
\sqrt{\frac{5}{3}} C_{31,11}^{31,0}+\frac{\sqrt{\frac{5}{3}} C_{31,11}^{31,1}}{1024}+\frac{3}{128} \sqrt{\frac{3}{70}} C_{33,11}^{33,0}+\frac{3}{256}
\sqrt{\frac{3}{70}} C_{33,11}^{33,1}\)

\noindent\(C_{42,0011,2}^{11,1101,0,\text{NonLoc}}= \frac{\sqrt{\frac{5}{3}} C_{11,11}^{51,0}}{1024}+\frac{1}{512} \sqrt{\frac{5}{3}}
C_{11,11}^{51,1}+\frac{C_{22,11}^{42,0}}{512}+\frac{C_{22,11}^{42,1}}{256}+\frac{\sqrt{\frac{5}{3}} C_{31,11}^{31,0}}{1024}+\frac{1}{512} \sqrt{\frac{5}{3}}
C_{31,11}^{31,1}+\frac{3}{256} \sqrt{\frac{3}{70}} C_{33,11}^{33,0}+\frac{3}{128} \sqrt{\frac{3}{70}} C_{33,11}^{33,1}\)

\noindent\(C_{42,0011,2}^{11,1101,1,\text{Loc}}= \frac{\sqrt{5} C_{11,11}^{51,1}}{1024}-\frac{\sqrt{3} C_{22,11}^{42,1}}{512}+\frac{\sqrt{5}
C_{31,11}^{31,1}}{1024}+\frac{9 C_{33,11}^{33,1}}{256 \sqrt{70}}\)

\noindent\(C_{42,0011,2}^{11,1101,1,\text{NonLoc}}= \frac{\sqrt{5} C_{11,11}^{51,0}}{1024}+\frac{\sqrt{3} C_{22,11}^{42,0}}{512}+\frac{\sqrt{5}
C_{31,11}^{31,0}}{1024}+\frac{9 C_{33,11}^{33,0}}{256 \sqrt{70}}\)

\noindent\(C_{42,0011,2}^{22,0011,0,\text{Loc}}= \frac{\sqrt{\frac{5}{3}} C_{00,20}^{60,0}}{1024}+\frac{\sqrt{\frac{5}{3}} C_{00,20}^{60,1}}{2048}-\frac{\sqrt{3}
C_{00,22}^{62,0}}{2048}-\frac{\sqrt{3} C_{00,22}^{62,1}}{4096}+\frac{\sqrt{3} C_{11,22}^{51,0}}{2048}+\frac{\sqrt{3} C_{11,22}^{51,1}}{4096}+\frac{9
C_{11,22}^{53,0}}{2048 \sqrt{35}}+\frac{9 C_{11,22}^{53,1}}{4096 \sqrt{35}}+\frac{\sqrt{\frac{5}{3}} C_{20,20}^{40,0}}{1024}+\frac{\sqrt{\frac{5}{3}}
C_{20,20}^{40,1}}{2048}-\frac{\sqrt{3} C_{20,22}^{42,0}}{2048}-\frac{\sqrt{3} C_{20,22}^{42,1}}{4096}+\frac{C_{22,20}^{42,0}}{512 \sqrt{3}}+\frac{C_{22,20}^{42,1}}{1024
\sqrt{3}}-\frac{\sqrt{3} C_{22,22}^{40,0}}{2048}-\frac{\sqrt{3} C_{22,22}^{40,1}}{4096}-\frac{\sqrt{\frac{3}{7}} C_{22,22}^{42,0}}{1024}-\frac{\sqrt{\frac{3}{7}}
C_{22,22}^{42,1}}{2048}-\frac{3 \sqrt{\frac{3}{5}} C_{22,22}^{44,0}}{3584}-\frac{3 \sqrt{\frac{3}{5}} C_{22,22}^{44,1}}{7168}-\frac{\sqrt{\frac{5}{3}}
C_{31,20}^{31,0}}{1024}-\frac{\sqrt{\frac{5}{3}} C_{31,20}^{31,1}}{2048}+\frac{\sqrt{3} C_{31,22}^{31,0}}{2048}+\frac{\sqrt{3} C_{31,22}^{31,1}}{4096}+\frac{9
C_{31,22}^{33,0}}{2048 \sqrt{35}}+\frac{9 C_{31,22}^{33,1}}{4096 \sqrt{35}}-\frac{3 C_{33,20}^{33,0}}{512 \sqrt{35}}-\frac{3 C_{33,20}^{33,1}}{1024
\sqrt{35}}+\frac{3 \sqrt{\frac{3}{14}} C_{33,22}^{33,0}}{2560}+\frac{3 \sqrt{\frac{3}{14}} C_{33,22}^{33,1}}{5120}\)

\noindent\(C_{42,0011,2}^{22,0011,0,\text{NonLoc}}= \frac{\sqrt{5} C_{00,00}^{60,0}}{4096}+\frac{\sqrt{5} C_{00,00}^{60,1}}{2048}+\frac{\sqrt{\frac{5}{3}}
C_{00,20}^{60,0}}{4096}+\frac{\sqrt{\frac{5}{3}} C_{00,20}^{60,1}}{2048}+\frac{\sqrt{3} C_{00,22}^{62,0}}{4096}+\frac{\sqrt{3} C_{00,22}^{62,1}}{2048}+\frac{\sqrt{5}
C_{11,00}^{51,0}}{4096}+\frac{\sqrt{5} C_{11,00}^{51,1}}{2048}+\frac{\sqrt{3} C_{11,22}^{51,0}}{4096}+\frac{\sqrt{3} C_{11,22}^{51,1}}{2048}+\frac{9
C_{11,22}^{53,0}}{4096 \sqrt{35}}+\frac{9 C_{11,22}^{53,1}}{2048 \sqrt{35}}+\frac{\sqrt{5} C_{20,00}^{40,0}}{4096}+\frac{\sqrt{5} C_{20,00}^{40,1}}{2048}+\frac{\sqrt{\frac{5}{3}}
C_{20,20}^{40,0}}{4096}+\frac{\sqrt{\frac{5}{3}} C_{20,20}^{40,1}}{2048}+\frac{\sqrt{3} C_{20,22}^{42,0}}{4096}+\frac{\sqrt{3} C_{20,22}^{42,1}}{2048}+\frac{C_{22,00}^{42,0}}{2048}+\frac{C_{22,00}^{42,1}}{1024}+\frac{C_{22,20}^{42,0}}{2048
\sqrt{3}}+\frac{C_{22,20}^{42,1}}{1024 \sqrt{3}}+\frac{\sqrt{3} C_{22,22}^{40,0}}{4096}+\frac{\sqrt{3} C_{22,22}^{40,1}}{2048}+\frac{\sqrt{\frac{3}{7}}
C_{22,22}^{42,0}}{2048}+\frac{\sqrt{\frac{3}{7}} C_{22,22}^{42,1}}{1024}+\frac{3 \sqrt{\frac{3}{5}} C_{22,22}^{44,0}}{7168}+\frac{3 \sqrt{\frac{3}{5}}
C_{22,22}^{44,1}}{3584}+\frac{\sqrt{5} C_{31,00}^{31,0}}{4096}+\frac{\sqrt{5} C_{31,00}^{31,1}}{2048}+\frac{\sqrt{\frac{5}{3}} C_{31,20}^{31,0}}{4096}+\frac{\sqrt{\frac{5}{3}}
C_{31,20}^{31,1}}{2048}+\frac{\sqrt{3} C_{31,22}^{31,0}}{4096}+\frac{\sqrt{3} C_{31,22}^{31,1}}{2048}+\frac{9 C_{31,22}^{33,0}}{4096 \sqrt{35}}+\frac{9
C_{31,22}^{33,1}}{2048 \sqrt{35}}+\frac{3 \sqrt{\frac{3}{35}} C_{33,00}^{33,0}}{2048}+\frac{3 \sqrt{\frac{3}{35}} C_{33,00}^{33,1}}{1024}+\frac{3
C_{33,20}^{33,0}}{2048 \sqrt{35}}+\frac{3 C_{33,20}^{33,1}}{1024 \sqrt{35}}+\frac{3 \sqrt{\frac{3}{14}} C_{33,22}^{33,0}}{5120}+\frac{3 \sqrt{\frac{3}{14}}
C_{33,22}^{33,1}}{2560}\)

\noindent\(C_{42,0011,2}^{22,0011,1,\text{Loc}}= \frac{\sqrt{5} C_{00,20}^{60,1}}{2048}-\frac{3 C_{00,22}^{62,1}}{4096}+\frac{3 C_{11,22}^{51,1}}{4096}+\frac{9
\sqrt{\frac{3}{35}} C_{11,22}^{53,1}}{4096}+\frac{\sqrt{5} C_{20,20}^{40,1}}{2048}-\frac{3 C_{20,22}^{42,1}}{4096}+\frac{C_{22,20}^{42,1}}{1024}-\frac{3
C_{22,22}^{40,1}}{4096}-\frac{3 C_{22,22}^{42,1}}{2048 \sqrt{7}}-\frac{9 C_{22,22}^{44,1}}{7168 \sqrt{5}}-\frac{\sqrt{5} C_{31,20}^{31,1}}{2048}+\frac{3
C_{31,22}^{31,1}}{4096}+\frac{9 \sqrt{\frac{3}{35}} C_{31,22}^{33,1}}{4096}-\frac{3 \sqrt{\frac{3}{35}} C_{33,20}^{33,1}}{1024}+\frac{9 C_{33,22}^{33,1}}{5120
\sqrt{14}}\)

\noindent\(C_{42,0011,2}^{22,0011,1,\text{NonLoc}}= \frac{\sqrt{15} C_{00,00}^{60,0}}{4096}+\frac{\sqrt{5} C_{00,20}^{60,0}}{4096}+\frac{3
C_{00,22}^{62,0}}{4096}+\frac{\sqrt{15} C_{11,00}^{51,0}}{4096}+\frac{3 C_{11,22}^{51,0}}{4096}+\frac{9 \sqrt{\frac{3}{35}} C_{11,22}^{53,0}}{4096}+\frac{\sqrt{15}
C_{20,00}^{40,0}}{4096}+\frac{\sqrt{5} C_{20,20}^{40,0}}{4096}+\frac{3 C_{20,22}^{42,0}}{4096}+\frac{\sqrt{3} C_{22,00}^{42,0}}{2048}+\frac{C_{22,20}^{42,0}}{2048}+\frac{3
C_{22,22}^{40,0}}{4096}+\frac{3 C_{22,22}^{42,0}}{2048 \sqrt{7}}+\frac{9 C_{22,22}^{44,0}}{7168 \sqrt{5}}+\frac{\sqrt{15} C_{31,00}^{31,0}}{4096}+\frac{\sqrt{5}
C_{31,20}^{31,0}}{4096}+\frac{3 C_{31,22}^{31,0}}{4096}+\frac{9 \sqrt{\frac{3}{35}} C_{31,22}^{33,0}}{4096}+\frac{9 C_{33,00}^{33,0}}{2048 \sqrt{35}}+\frac{3
\sqrt{\frac{3}{35}} C_{33,20}^{33,0}}{2048}+\frac{9 C_{33,22}^{33,0}}{5120 \sqrt{14}}\)

\noindent\(C_{42,0011,3}^{00,2213,0,\text{Loc}}= -\frac{1}{256} \sqrt{\frac{7}{3}} C_{00,20}^{60,0}-\frac{1}{512} \sqrt{\frac{7}{3}}
C_{00,20}^{60,1}-\frac{1}{256} \sqrt{\frac{3}{35}} C_{00,22}^{62,0}-\frac{1}{512} \sqrt{\frac{3}{35}} C_{00,22}^{62,1}+\frac{C_{11,22}^{51,0}}{384
\sqrt{105}}+\frac{C_{11,22}^{51,1}}{768 \sqrt{105}}+\frac{9 C_{11,22}^{53,0}}{17920}+\frac{9 C_{11,22}^{53,1}}{35840}+\frac{1}{768} \sqrt{\frac{7}{3}}
C_{20,20}^{40,0}+\frac{\sqrt{\frac{7}{3}} C_{20,20}^{40,1}}{1536}+\frac{C_{20,22}^{42,0}}{384 \sqrt{105}}+\frac{C_{20,22}^{42,1}}{768 \sqrt{105}}-\frac{1}{768}
\sqrt{\frac{7}{15}} C_{22,20}^{42,0}-\frac{\sqrt{\frac{7}{15}} C_{22,20}^{42,1}}{1536}+\frac{1}{768} \sqrt{\frac{5}{21}} C_{22,22}^{40,0}+\frac{\sqrt{\frac{5}{21}}
C_{22,22}^{40,1}}{1536}-\frac{C_{22,22}^{42,0}}{384 \sqrt{15}}-\frac{C_{22,22}^{42,1}}{768 \sqrt{15}}-\frac{1}{768} \sqrt{\frac{7}{3}} C_{31,20}^{31,0}-\frac{\sqrt{\frac{7}{3}}
C_{31,20}^{31,1}}{1536}-\frac{C_{31,22}^{31,0}}{384 \sqrt{105}}-\frac{C_{31,22}^{31,1}}{768 \sqrt{105}}-\frac{9 C_{31,22}^{33,0}}{17920}-\frac{9
C_{31,22}^{33,1}}{35840}+\frac{3 C_{33,20}^{33,0}}{1280}+\frac{3 C_{33,20}^{33,1}}{2560}+\frac{3 \sqrt{\frac{3}{10}} C_{33,22}^{33,0}}{1120}+\frac{3
\sqrt{\frac{3}{10}} C_{33,22}^{33,1}}{2240}\)

\noindent\(C_{42,0011,3}^{00,2213,0,\text{NonLoc}}= -\frac{\sqrt{7} C_{00,00}^{60,0}}{1024}-\frac{\sqrt{7} C_{00,00}^{60,1}}{512}-\frac{\sqrt{\frac{7}{3}}
C_{00,20}^{60,0}}{1024}-\frac{1}{512} \sqrt{\frac{7}{3}} C_{00,20}^{60,1}+\frac{1}{512} \sqrt{\frac{3}{35}} C_{00,22}^{62,0}+\frac{1}{256} \sqrt{\frac{3}{35}}
C_{00,22}^{62,1}-\frac{\sqrt{7} C_{11,00}^{51,0}}{3072}-\frac{\sqrt{7} C_{11,00}^{51,1}}{1536}+\frac{C_{11,22}^{51,0}}{768 \sqrt{105}}+\frac{C_{11,22}^{51,1}}{384
\sqrt{105}}+\frac{9 C_{11,22}^{53,0}}{35840}+\frac{9 C_{11,22}^{53,1}}{17920}+\frac{\sqrt{7} C_{20,00}^{40,0}}{3072}+\frac{\sqrt{7} C_{20,00}^{40,1}}{1536}+\frac{\sqrt{\frac{7}{3}}
C_{20,20}^{40,0}}{3072}+\frac{\sqrt{\frac{7}{3}} C_{20,20}^{40,1}}{1536}-\frac{C_{20,22}^{42,0}}{768 \sqrt{105}}-\frac{C_{20,22}^{42,1}}{384 \sqrt{105}}-\frac{\sqrt{\frac{7}{5}}
C_{22,00}^{42,0}}{3072}-\frac{\sqrt{\frac{7}{5}} C_{22,00}^{42,1}}{1536}-\frac{\sqrt{\frac{7}{15}} C_{22,20}^{42,0}}{3072}-\frac{\sqrt{\frac{7}{15}}
C_{22,20}^{42,1}}{1536}-\frac{\sqrt{\frac{5}{21}} C_{22,22}^{40,0}}{1536}-\frac{1}{768} \sqrt{\frac{5}{21}} C_{22,22}^{40,1}+\frac{C_{22,22}^{42,0}}{768
\sqrt{15}}+\frac{C_{22,22}^{42,1}}{384 \sqrt{15}}+\frac{\sqrt{7} C_{31,00}^{31,0}}{3072}+\frac{\sqrt{7} C_{31,00}^{31,1}}{1536}+\frac{\sqrt{\frac{7}{3}}
C_{31,20}^{31,0}}{3072}+\frac{\sqrt{\frac{7}{3}} C_{31,20}^{31,1}}{1536}-\frac{C_{31,22}^{31,0}}{768 \sqrt{105}}-\frac{C_{31,22}^{31,1}}{384 \sqrt{105}}-\frac{9
C_{31,22}^{33,0}}{35840}-\frac{9 C_{31,22}^{33,1}}{17920}-\frac{3 \sqrt{3} C_{33,00}^{33,0}}{5120}-\frac{3 \sqrt{3} C_{33,00}^{33,1}}{2560}-\frac{3
C_{33,20}^{33,0}}{5120}-\frac{3 C_{33,20}^{33,1}}{2560}+\frac{3 \sqrt{\frac{3}{10}} C_{33,22}^{33,0}}{2240}+\frac{3 \sqrt{\frac{3}{10}} C_{33,22}^{33,1}}{1120}\)

\noindent\(C_{42,0011,3}^{00,2213,1,\text{Loc}}= -\frac{\sqrt{7} C_{00,20}^{60,1}}{512}-\frac{3 C_{00,22}^{62,1}}{512 \sqrt{35}}+\frac{C_{11,22}^{51,1}}{768
\sqrt{35}}+\frac{9 \sqrt{3} C_{11,22}^{53,1}}{35840}+\frac{\sqrt{7} C_{20,20}^{40,1}}{1536}+\frac{C_{20,22}^{42,1}}{768 \sqrt{35}}-\frac{\sqrt{\frac{7}{5}}
C_{22,20}^{42,1}}{1536}+\frac{\sqrt{\frac{5}{7}} C_{22,22}^{40,1}}{1536}-\frac{C_{22,22}^{42,1}}{768 \sqrt{5}}-\frac{\sqrt{7} C_{31,20}^{31,1}}{1536}-\frac{C_{31,22}^{31,1}}{768
\sqrt{35}}-\frac{9 \sqrt{3} C_{31,22}^{33,1}}{35840}+\frac{3 \sqrt{3} C_{33,20}^{33,1}}{2560}+\frac{9 C_{33,22}^{33,1}}{2240 \sqrt{10}}\)

\noindent\(C_{42,0011,3}^{00,2213,1,\text{NonLoc}}= -\frac{\sqrt{21} C_{00,00}^{60,0}}{1024}-\frac{\sqrt{7} C_{00,20}^{60,0}}{1024}+\frac{3
C_{00,22}^{62,0}}{512 \sqrt{35}}-\frac{\sqrt{\frac{7}{3}} C_{11,00}^{51,0}}{1024}+\frac{C_{11,22}^{51,0}}{768 \sqrt{35}}+\frac{9 \sqrt{3} C_{11,22}^{53,0}}{35840}+\frac{\sqrt{\frac{7}{3}}
C_{20,00}^{40,0}}{1024}+\frac{\sqrt{7} C_{20,20}^{40,0}}{3072}-\frac{C_{20,22}^{42,0}}{768 \sqrt{35}}-\frac{\sqrt{\frac{7}{15}} C_{22,00}^{42,0}}{1024}-\frac{\sqrt{\frac{7}{5}}
C_{22,20}^{42,0}}{3072}-\frac{\sqrt{\frac{5}{7}} C_{22,22}^{40,0}}{1536}+\frac{C_{22,22}^{42,0}}{768 \sqrt{5}}+\frac{\sqrt{\frac{7}{3}} C_{31,00}^{31,0}}{1024}+\frac{\sqrt{7}
C_{31,20}^{31,0}}{3072}-\frac{C_{31,22}^{31,0}}{768 \sqrt{35}}-\frac{9 \sqrt{3} C_{31,22}^{33,0}}{35840}-\frac{9 C_{33,00}^{33,0}}{5120}-\frac{3
\sqrt{3} C_{33,20}^{33,0}}{5120}+\frac{9 C_{33,22}^{33,0}}{2240 \sqrt{10}}\)

\noindent\(C_{42,0011,3}^{22,0011,0,\text{Loc}}= \frac{\sqrt{\frac{7}{3}} C_{00,20}^{60,0}}{1024}+\frac{\sqrt{\frac{7}{3}} C_{00,20}^{60,1}}{2048}+\frac{\sqrt{\frac{3}{35}}
C_{00,22}^{62,0}}{1024}+\frac{\sqrt{\frac{3}{35}} C_{00,22}^{62,1}}{2048}-\frac{\sqrt{\frac{3}{35}} C_{11,22}^{51,0}}{1024}-\frac{\sqrt{\frac{3}{35}}
C_{11,22}^{51,1}}{2048}-\frac{9 C_{11,22}^{53,0}}{35840}-\frac{9 C_{11,22}^{53,1}}{71680}+\frac{\sqrt{\frac{7}{3}} C_{20,20}^{40,0}}{1024}+\frac{\sqrt{\frac{7}{3}}
C_{20,20}^{40,1}}{2048}+\frac{\sqrt{\frac{3}{35}} C_{20,22}^{42,0}}{1024}+\frac{\sqrt{\frac{3}{35}} C_{20,22}^{42,1}}{2048}+\frac{1}{512} \sqrt{\frac{7}{15}}
C_{22,20}^{42,0}+\frac{\sqrt{\frac{7}{15}} C_{22,20}^{42,1}}{1024}+\frac{\sqrt{\frac{3}{35}} C_{22,22}^{40,0}}{1024}+\frac{\sqrt{\frac{3}{35}} C_{22,22}^{40,1}}{2048}+\frac{\sqrt{\frac{3}{5}}
C_{22,22}^{42,0}}{3584}+\frac{\sqrt{\frac{3}{5}} C_{22,22}^{42,1}}{7168}+\frac{3 \sqrt{\frac{3}{7}} C_{22,22}^{44,0}}{8960}+\frac{3 \sqrt{\frac{3}{7}}
C_{22,22}^{44,1}}{17920}-\frac{\sqrt{\frac{7}{3}} C_{31,20}^{31,0}}{1024}-\frac{\sqrt{\frac{7}{3}} C_{31,20}^{31,1}}{2048}-\frac{\sqrt{\frac{3}{35}}
C_{31,22}^{31,0}}{1024}-\frac{\sqrt{\frac{3}{35}} C_{31,22}^{31,1}}{2048}-\frac{9 C_{31,22}^{33,0}}{35840}-\frac{9 C_{31,22}^{33,1}}{71680}-\frac{3
C_{33,20}^{33,0}}{2560}-\frac{3 C_{33,20}^{33,1}}{5120}-\frac{3 \sqrt{\frac{3}{10}} C_{33,22}^{33,0}}{8960}-\frac{3 \sqrt{\frac{3}{10}} C_{33,22}^{33,1}}{17920}\)

\noindent\(C_{42,0011,3}^{22,0011,0,\text{NonLoc}}= \frac{\sqrt{7} C_{00,00}^{60,0}}{4096}+\frac{\sqrt{7} C_{00,00}^{60,1}}{2048}+\frac{\sqrt{\frac{7}{3}}
C_{00,20}^{60,0}}{4096}+\frac{\sqrt{\frac{7}{3}} C_{00,20}^{60,1}}{2048}-\frac{\sqrt{\frac{3}{35}} C_{00,22}^{62,0}}{2048}-\frac{\sqrt{\frac{3}{35}}
C_{00,22}^{62,1}}{1024}+\frac{\sqrt{7} C_{11,00}^{51,0}}{4096}+\frac{\sqrt{7} C_{11,00}^{51,1}}{2048}-\frac{\sqrt{\frac{3}{35}} C_{11,22}^{51,0}}{2048}-\frac{\sqrt{\frac{3}{35}}
C_{11,22}^{51,1}}{1024}-\frac{9 C_{11,22}^{53,0}}{71680}-\frac{9 C_{11,22}^{53,1}}{35840}+\frac{\sqrt{7} C_{20,00}^{40,0}}{4096}+\frac{\sqrt{7} C_{20,00}^{40,1}}{2048}+\frac{\sqrt{\frac{7}{3}}
C_{20,20}^{40,0}}{4096}+\frac{\sqrt{\frac{7}{3}} C_{20,20}^{40,1}}{2048}-\frac{\sqrt{\frac{3}{35}} C_{20,22}^{42,0}}{2048}-\frac{\sqrt{\frac{3}{35}}
C_{20,22}^{42,1}}{1024}+\frac{\sqrt{\frac{7}{5}} C_{22,00}^{42,0}}{2048}+\frac{\sqrt{\frac{7}{5}} C_{22,00}^{42,1}}{1024}+\frac{\sqrt{\frac{7}{15}}
C_{22,20}^{42,0}}{2048}+\frac{\sqrt{\frac{7}{15}} C_{22,20}^{42,1}}{1024}-\frac{\sqrt{\frac{3}{35}} C_{22,22}^{40,0}}{2048}-\frac{\sqrt{\frac{3}{35}}
C_{22,22}^{40,1}}{1024}-\frac{\sqrt{\frac{3}{5}} C_{22,22}^{42,0}}{7168}-\frac{\sqrt{\frac{3}{5}} C_{22,22}^{42,1}}{3584}-\frac{3 \sqrt{\frac{3}{7}}
C_{22,22}^{44,0}}{17920}-\frac{3 \sqrt{\frac{3}{7}} C_{22,22}^{44,1}}{8960}+\frac{\sqrt{7} C_{31,00}^{31,0}}{4096}+\frac{\sqrt{7} C_{31,00}^{31,1}}{2048}+\frac{\sqrt{\frac{7}{3}}
C_{31,20}^{31,0}}{4096}+\frac{\sqrt{\frac{7}{3}} C_{31,20}^{31,1}}{2048}-\frac{\sqrt{\frac{3}{35}} C_{31,22}^{31,0}}{2048}-\frac{\sqrt{\frac{3}{35}}
C_{31,22}^{31,1}}{1024}-\frac{9 C_{31,22}^{33,0}}{71680}-\frac{9 C_{31,22}^{33,1}}{35840}+\frac{3 \sqrt{3} C_{33,00}^{33,0}}{10240}+\frac{3 \sqrt{3}
C_{33,00}^{33,1}}{5120}+\frac{3 C_{33,20}^{33,0}}{10240}+\frac{3 C_{33,20}^{33,1}}{5120}-\frac{3 \sqrt{\frac{3}{10}} C_{33,22}^{33,0}}{17920}-\frac{3
\sqrt{\frac{3}{10}} C_{33,22}^{33,1}}{8960}\)

\noindent\(C_{42,0011,3}^{22,0011,1,\text{Loc}}= \frac{\sqrt{7} C_{00,20}^{60,1}}{2048}+\frac{3 C_{00,22}^{62,1}}{2048 \sqrt{35}}-\frac{3
C_{11,22}^{51,1}}{2048 \sqrt{35}}-\frac{9 \sqrt{3} C_{11,22}^{53,1}}{71680}+\frac{\sqrt{7} C_{20,20}^{40,1}}{2048}+\frac{3 C_{20,22}^{42,1}}{2048
\sqrt{35}}+\frac{\sqrt{\frac{7}{5}} C_{22,20}^{42,1}}{1024}+\frac{3 C_{22,22}^{40,1}}{2048 \sqrt{35}}+\frac{3 C_{22,22}^{42,1}}{7168 \sqrt{5}}+\frac{9
C_{22,22}^{44,1}}{17920 \sqrt{7}}-\frac{\sqrt{7} C_{31,20}^{31,1}}{2048}-\frac{3 C_{31,22}^{31,1}}{2048 \sqrt{35}}-\frac{9 \sqrt{3} C_{31,22}^{33,1}}{71680}-\frac{3
\sqrt{3} C_{33,20}^{33,1}}{5120}-\frac{9 C_{33,22}^{33,1}}{17920 \sqrt{10}}\)

\noindent\(C_{42,0011,3}^{22,0011,1,\text{NonLoc}}= \frac{\sqrt{21} C_{00,00}^{60,0}}{4096}+\frac{\sqrt{7} C_{00,20}^{60,0}}{4096}-\frac{3
C_{00,22}^{62,0}}{2048 \sqrt{35}}+\frac{\sqrt{21} C_{11,00}^{51,0}}{4096}-\frac{3 C_{11,22}^{51,0}}{2048 \sqrt{35}}-\frac{9 \sqrt{3} C_{11,22}^{53,0}}{71680}+\frac{\sqrt{21}
C_{20,00}^{40,0}}{4096}+\frac{\sqrt{7} C_{20,20}^{40,0}}{4096}-\frac{3 C_{20,22}^{42,0}}{2048 \sqrt{35}}+\frac{\sqrt{\frac{21}{5}} C_{22,00}^{42,0}}{2048}+\frac{\sqrt{\frac{7}{5}}
C_{22,20}^{42,0}}{2048}-\frac{3 C_{22,22}^{40,0}}{2048 \sqrt{35}}-\frac{3 C_{22,22}^{42,0}}{7168 \sqrt{5}}-\frac{9 C_{22,22}^{44,0}}{17920 \sqrt{7}}+\frac{\sqrt{21}
C_{31,00}^{31,0}}{4096}+\frac{\sqrt{7} C_{31,20}^{31,0}}{4096}-\frac{3 C_{31,22}^{31,0}}{2048 \sqrt{35}}-\frac{9 \sqrt{3} C_{31,22}^{33,0}}{71680}+\frac{9
C_{33,00}^{33,0}}{10240}+\frac{3 \sqrt{3} C_{33,20}^{33,0}}{10240}-\frac{9 C_{33,22}^{33,0}}{17920 \sqrt{10}}\)

\noindent\(C_{42,1101,1}^{00,1101,0,\text{Loc}}= -\frac{\sqrt{5} C_{00,00}^{60,0}}{128}-\frac{\sqrt{5} C_{00,00}^{60,1}}{256}+\frac{\sqrt{5}
C_{11,00}^{51,0}}{384}+\frac{\sqrt{5} C_{11,00}^{51,1}}{768}+\frac{\sqrt{5} C_{20,00}^{40,0}}{384}+\frac{\sqrt{5} C_{20,00}^{40,1}}{768}-\frac{C_{22,00}^{42,0}}{384}-\frac{C_{22,00}^{42,1}}{768}-\frac{\sqrt{5}
C_{31,00}^{31,0}}{384}-\frac{\sqrt{5} C_{31,00}^{31,1}}{768}+\frac{3}{128} \sqrt{\frac{3}{35}} C_{33,00}^{33,0}+\frac{3}{256} \sqrt{\frac{3}{35}}
C_{33,00}^{33,1}\)

\noindent\(C_{42,1101,1}^{00,1101,0,\text{NonLoc}}= \frac{\sqrt{5} C_{00,00}^{60,0}}{512}+\frac{\sqrt{5} C_{00,00}^{60,1}}{256}-\frac{\sqrt{15}
C_{00,20}^{60,0}}{512}-\frac{\sqrt{15} C_{00,20}^{60,1}}{256}+\frac{\sqrt{5} C_{11,00}^{51,0}}{1536}+\frac{\sqrt{5} C_{11,00}^{51,1}}{768}-\frac{1}{512}
\sqrt{\frac{5}{3}} C_{11,20}^{31,0}-\frac{1}{256} \sqrt{\frac{5}{3}} C_{11,20}^{31,1}-\frac{\sqrt{5} C_{20,00}^{40,0}}{1536}-\frac{\sqrt{5} C_{20,00}^{40,1}}{768}+\frac{1}{512}
\sqrt{\frac{5}{3}} C_{20,20}^{40,0}+\frac{1}{256} \sqrt{\frac{5}{3}} C_{20,20}^{40,1}+\frac{C_{22,00}^{42,0}}{1536}+\frac{C_{22,00}^{42,1}}{768}-\frac{C_{22,20}^{42,0}}{512
\sqrt{3}}-\frac{C_{22,20}^{42,1}}{256 \sqrt{3}}-\frac{\sqrt{5} C_{31,00}^{31,0}}{1536}-\frac{\sqrt{5} C_{31,00}^{31,1}}{768}+\frac{1}{512} \sqrt{\frac{5}{3}}
C_{31,20}^{31,0}+\frac{1}{256} \sqrt{\frac{5}{3}} C_{31,20}^{31,1}+\frac{3}{512} \sqrt{\frac{3}{35}} C_{33,00}^{33,0}+\frac{3}{256} \sqrt{\frac{3}{35}}
C_{33,00}^{33,1}-\frac{9 C_{33,20}^{33,0}}{512 \sqrt{35}}-\frac{9 C_{33,20}^{33,1}}{256 \sqrt{35}}\)

\noindent\(C_{42,1101,1}^{00,1101,1,\text{Loc}}= -\frac{\sqrt{15} C_{00,00}^{60,1}}{256}+\frac{1}{256} \sqrt{\frac{5}{3}} C_{11,00}^{51,1}+\frac{1}{256}
\sqrt{\frac{5}{3}} C_{20,00}^{40,1}-\frac{C_{22,00}^{42,1}}{256 \sqrt{3}}-\frac{1}{256} \sqrt{\frac{5}{3}} C_{31,00}^{31,1}+\frac{9 C_{33,00}^{33,1}}{256
\sqrt{35}}\)

\noindent\(C_{42,1101,1}^{00,1101,1,\text{NonLoc}}= \frac{\sqrt{15} C_{00,00}^{60,0}}{512}-\frac{3 \sqrt{5} C_{00,20}^{60,0}}{512}+\frac{1}{512}
\sqrt{\frac{5}{3}} C_{11,00}^{51,0}-\frac{\sqrt{5} C_{11,20}^{31,0}}{512}-\frac{1}{512} \sqrt{\frac{5}{3}} C_{20,00}^{40,0}+\frac{\sqrt{5} C_{20,20}^{40,0}}{512}+\frac{C_{22,00}^{42,0}}{512
\sqrt{3}}-\frac{C_{22,20}^{42,0}}{512}-\frac{1}{512} \sqrt{\frac{5}{3}} C_{31,00}^{31,0}+\frac{\sqrt{5} C_{31,20}^{31,0}}{512}+\frac{9 C_{33,00}^{33,0}}{512
\sqrt{35}}-\frac{9}{512} \sqrt{\frac{3}{35}} C_{33,20}^{33,0}\)

\noindent\(C_{42,1101,1}^{11,0011,0,\text{Loc}}= \frac{\sqrt{5} C_{11,11}^{51,0}}{1536}+\frac{\sqrt{5} C_{11,11}^{51,1}}{3072}-\frac{C_{22,11}^{42,0}}{256
\sqrt{3}}-\frac{C_{22,11}^{42,1}}{512 \sqrt{3}}+\frac{\sqrt{5} C_{31,11}^{31,0}}{1536}+\frac{\sqrt{5} C_{31,11}^{31,1}}{3072}+\frac{3 C_{33,11}^{33,0}}{128
\sqrt{70}}+\frac{3 C_{33,11}^{33,1}}{256 \sqrt{70}}\)

\noindent\(C_{42,1101,1}^{11,0011,0,\text{NonLoc}}= \frac{\sqrt{5} C_{11,11}^{51,0}}{3072}+\frac{\sqrt{5} C_{11,11}^{51,1}}{1536}+\frac{C_{22,11}^{42,0}}{512
\sqrt{3}}+\frac{C_{22,11}^{42,1}}{256 \sqrt{3}}+\frac{\sqrt{5} C_{31,11}^{31,0}}{3072}+\frac{\sqrt{5} C_{31,11}^{31,1}}{1536}+\frac{3 C_{33,11}^{33,0}}{256
\sqrt{70}}+\frac{3 C_{33,11}^{33,1}}{128 \sqrt{70}}\)

\noindent\(C_{42,1101,1}^{11,0011,1,\text{Loc}}= \frac{\sqrt{\frac{5}{3}} C_{11,11}^{51,1}}{1024}-\frac{C_{22,11}^{42,1}}{512}+\frac{\sqrt{\frac{5}{3}}
C_{31,11}^{31,1}}{1024}+\frac{3}{256} \sqrt{\frac{3}{70}} C_{33,11}^{33,1}\)

\noindent\(C_{42,1101,1}^{11,0011,1,\text{NonLoc}}= \frac{\sqrt{\frac{5}{3}} C_{11,11}^{51,0}}{1024}+\frac{C_{22,11}^{42,0}}{512}+\frac{\sqrt{\frac{5}{3}}
C_{31,11}^{31,0}}{1024}+\frac{3}{256} \sqrt{\frac{3}{70}} C_{33,11}^{33,0}\)

\noindent\(C_{42,1101,2}^{11,0011,0,\text{Loc}}= \frac{1}{512} \sqrt{\frac{5}{3}} C_{11,11}^{51,0}+\frac{\sqrt{\frac{5}{3}} C_{11,11}^{51,1}}{1024}-\frac{C_{22,11}^{42,0}}{256}-\frac{C_{22,11}^{42,1}}{512}+\frac{1}{512}
\sqrt{\frac{5}{3}} C_{31,11}^{31,0}+\frac{\sqrt{\frac{5}{3}} C_{31,11}^{31,1}}{1024}+\frac{3}{128} \sqrt{\frac{3}{70}} C_{33,11}^{33,0}+\frac{3}{256}
\sqrt{\frac{3}{70}} C_{33,11}^{33,1}\)

\noindent\(C_{42,1101,2}^{11,0011,0,\text{NonLoc}}= \frac{\sqrt{\frac{5}{3}} C_{11,11}^{51,0}}{1024}+\frac{1}{512} \sqrt{\frac{5}{3}}
C_{11,11}^{51,1}+\frac{C_{22,11}^{42,0}}{512}+\frac{C_{22,11}^{42,1}}{256}+\frac{\sqrt{\frac{5}{3}} C_{31,11}^{31,0}}{1024}+\frac{1}{512} \sqrt{\frac{5}{3}}
C_{31,11}^{31,1}+\frac{3}{256} \sqrt{\frac{3}{70}} C_{33,11}^{33,0}+\frac{3}{128} \sqrt{\frac{3}{70}} C_{33,11}^{33,1}\)

\noindent\(C_{42,1101,2}^{11,0011,1,\text{Loc}}= \frac{\sqrt{5} C_{11,11}^{51,1}}{1024}-\frac{\sqrt{3} C_{22,11}^{42,1}}{512}+\frac{\sqrt{5}
C_{31,11}^{31,1}}{1024}+\frac{9 C_{33,11}^{33,1}}{256 \sqrt{70}}\)

\noindent\(C_{42,1101,2}^{11,0011,1,\text{NonLoc}}= \frac{\sqrt{5} C_{11,11}^{51,0}}{1024}+\frac{\sqrt{3} C_{22,11}^{42,0}}{512}+\frac{\sqrt{5}
C_{31,11}^{31,0}}{1024}+\frac{9 C_{33,11}^{33,0}}{256 \sqrt{70}}\)

\noindent\(C_{42,1110,2}^{00,1112,0,\text{Loc}}= \frac{\sqrt{5} C_{00,20}^{60,0}}{384}+\frac{\sqrt{5} C_{00,20}^{60,1}}{768}+\frac{C_{00,22}^{62,0}}{128}+\frac{C_{00,22}^{62,1}}{256}-\frac{\sqrt{5}
C_{11,20}^{31,0}}{1152}-\frac{\sqrt{5} C_{11,20}^{31,1}}{2304}-\frac{7 C_{11,22}^{51,0}}{2304}-\frac{7 C_{11,22}^{51,1}}{4608}-\frac{3}{512} \sqrt{\frac{3}{35}}
C_{11,22}^{53,0}-\frac{3 \sqrt{\frac{3}{35}} C_{11,22}^{53,1}}{1024}-\frac{\sqrt{5} C_{20,20}^{40,0}}{1152}-\frac{\sqrt{5} C_{20,20}^{40,1}}{2304}-\frac{C_{20,22}^{42,0}}{2304}-\frac{C_{20,22}^{42,1}}{4608}+\frac{C_{22,20}^{42,0}}{1152}+\frac{C_{22,20}^{42,1}}{2304}-\frac{C_{22,22}^{40,0}}{576}-\frac{C_{22,22}^{40,1}}{1152}+\frac{5
C_{22,22}^{42,0}}{1152 \sqrt{7}}+\frac{5 C_{22,22}^{42,1}}{2304 \sqrt{7}}-\frac{3 C_{22,22}^{44,0}}{448 \sqrt{5}}-\frac{3 C_{22,22}^{44,1}}{896 \sqrt{5}}+\frac{\sqrt{5}
C_{31,20}^{31,0}}{1152}+\frac{\sqrt{5} C_{31,20}^{31,1}}{2304}+\frac{C_{31,22}^{31,0}}{2304}+\frac{C_{31,22}^{31,1}}{4608}+\frac{17 C_{31,22}^{33,0}}{512
\sqrt{105}}+\frac{17 C_{31,22}^{33,1}}{1024 \sqrt{105}}-\frac{1}{128} \sqrt{\frac{3}{35}} C_{33,20}^{33,0}-\frac{1}{256} \sqrt{\frac{3}{35}} C_{33,20}^{33,1}-\frac{3
C_{33,22}^{33,0}}{320 \sqrt{14}}-\frac{3 C_{33,22}^{33,1}}{640 \sqrt{14}}\)

\noindent\(C_{42,1110,2}^{00,1112,0,\text{NonLoc}}= \frac{1}{512} \sqrt{\frac{5}{3}} C_{00,00}^{60,0}+\frac{1}{256} \sqrt{\frac{5}{3}}
C_{00,00}^{60,1}+\frac{\sqrt{5} C_{00,20}^{60,0}}{1536}+\frac{\sqrt{5} C_{00,20}^{60,1}}{768}-\frac{C_{00,22}^{62,0}}{256}-\frac{C_{00,22}^{62,1}}{128}+\frac{\sqrt{\frac{5}{3}}
C_{11,00}^{51,0}}{1536}+\frac{1}{768} \sqrt{\frac{5}{3}} C_{11,00}^{51,1}+\frac{\sqrt{5} C_{11,20}^{31,0}}{4608}+\frac{\sqrt{5} C_{11,20}^{31,1}}{2304}-\frac{7
C_{11,22}^{51,0}}{4608}-\frac{7 C_{11,22}^{51,1}}{2304}-\frac{3 \sqrt{\frac{3}{35}} C_{11,22}^{53,0}}{1024}-\frac{3}{512} \sqrt{\frac{3}{35}} C_{11,22}^{53,1}-\frac{\sqrt{\frac{5}{3}}
C_{20,00}^{40,0}}{1536}-\frac{1}{768} \sqrt{\frac{5}{3}} C_{20,00}^{40,1}-\frac{\sqrt{5} C_{20,20}^{40,0}}{4608}-\frac{\sqrt{5} C_{20,20}^{40,1}}{2304}+\frac{C_{20,22}^{42,0}}{4608}+\frac{C_{20,22}^{42,1}}{2304}+\frac{C_{22,00}^{42,0}}{1536
\sqrt{3}}+\frac{C_{22,00}^{42,1}}{768 \sqrt{3}}+\frac{C_{22,20}^{42,0}}{4608}+\frac{C_{22,20}^{42,1}}{2304}+\frac{C_{22,22}^{40,0}}{1152}+\frac{C_{22,22}^{40,1}}{576}-\frac{5
C_{22,22}^{42,0}}{2304 \sqrt{7}}-\frac{5 C_{22,22}^{42,1}}{1152 \sqrt{7}}+\frac{3 C_{22,22}^{44,0}}{896 \sqrt{5}}+\frac{3 C_{22,22}^{44,1}}{448 \sqrt{5}}-\frac{\sqrt{\frac{5}{3}}
C_{31,00}^{31,0}}{1536}-\frac{1}{768} \sqrt{\frac{5}{3}} C_{31,00}^{31,1}-\frac{\sqrt{5} C_{31,20}^{31,0}}{4608}-\frac{\sqrt{5} C_{31,20}^{31,1}}{2304}+\frac{C_{31,22}^{31,0}}{4608}+\frac{C_{31,22}^{31,1}}{2304}+\frac{17
C_{31,22}^{33,0}}{1024 \sqrt{105}}+\frac{17 C_{31,22}^{33,1}}{512 \sqrt{105}}+\frac{3 C_{33,00}^{33,0}}{512 \sqrt{35}}+\frac{3 C_{33,00}^{33,1}}{256
\sqrt{35}}+\frac{1}{512} \sqrt{\frac{3}{35}} C_{33,20}^{33,0}+\frac{1}{256} \sqrt{\frac{3}{35}} C_{33,20}^{33,1}-\frac{3 C_{33,22}^{33,0}}{640 \sqrt{14}}-\frac{3
C_{33,22}^{33,1}}{320 \sqrt{14}}\)

\noindent\(C_{42,1110,2}^{00,1112,1,\text{Loc}}= \frac{1}{256} \sqrt{\frac{5}{3}} C_{00,20}^{60,1}+\frac{\sqrt{3} C_{00,22}^{62,1}}{256}-\frac{1}{768}
\sqrt{\frac{5}{3}} C_{11,20}^{31,1}-\frac{7 C_{11,22}^{51,1}}{1536 \sqrt{3}}-\frac{9 C_{11,22}^{53,1}}{1024 \sqrt{35}}-\frac{1}{768} \sqrt{\frac{5}{3}}
C_{20,20}^{40,1}-\frac{C_{20,22}^{42,1}}{1536 \sqrt{3}}+\frac{C_{22,20}^{42,1}}{768 \sqrt{3}}-\frac{C_{22,22}^{40,1}}{384 \sqrt{3}}+\frac{5 C_{22,22}^{42,1}}{768
\sqrt{21}}-\frac{3}{896} \sqrt{\frac{3}{5}} C_{22,22}^{44,1}+\frac{1}{768} \sqrt{\frac{5}{3}} C_{31,20}^{31,1}+\frac{C_{31,22}^{31,1}}{1536 \sqrt{3}}+\frac{17
C_{31,22}^{33,1}}{1024 \sqrt{35}}-\frac{3 C_{33,20}^{33,1}}{256 \sqrt{35}}-\frac{3}{640} \sqrt{\frac{3}{14}} C_{33,22}^{33,1}\)

\noindent\(C_{42,1110,2}^{00,1112,1,\text{NonLoc}}= \frac{\sqrt{5} C_{00,00}^{60,0}}{512}+\frac{1}{512} \sqrt{\frac{5}{3}} C_{00,20}^{60,0}-\frac{\sqrt{3}
C_{00,22}^{62,0}}{256}+\frac{\sqrt{5} C_{11,00}^{51,0}}{1536}+\frac{\sqrt{\frac{5}{3}} C_{11,20}^{31,0}}{1536}-\frac{7 C_{11,22}^{51,0}}{1536 \sqrt{3}}-\frac{9
C_{11,22}^{53,0}}{1024 \sqrt{35}}-\frac{\sqrt{5} C_{20,00}^{40,0}}{1536}-\frac{\sqrt{\frac{5}{3}} C_{20,20}^{40,0}}{1536}+\frac{C_{20,22}^{42,0}}{1536
\sqrt{3}}+\frac{C_{22,00}^{42,0}}{1536}+\frac{C_{22,20}^{42,0}}{1536 \sqrt{3}}+\frac{C_{22,22}^{40,0}}{384 \sqrt{3}}-\frac{5 C_{22,22}^{42,0}}{768
\sqrt{21}}+\frac{3}{896} \sqrt{\frac{3}{5}} C_{22,22}^{44,0}-\frac{\sqrt{5} C_{31,00}^{31,0}}{1536}-\frac{\sqrt{\frac{5}{3}} C_{31,20}^{31,0}}{1536}+\frac{C_{31,22}^{31,0}}{1536
\sqrt{3}}+\frac{17 C_{31,22}^{33,0}}{1024 \sqrt{35}}+\frac{3}{512} \sqrt{\frac{3}{35}} C_{33,00}^{33,0}+\frac{3 C_{33,20}^{33,0}}{512 \sqrt{35}}-\frac{3}{640}
\sqrt{\frac{3}{14}} C_{33,22}^{33,0}\)

\noindent\(C_{42,1111,1}^{00,1111,0,\text{Loc}}= -\frac{1}{256} \sqrt{\frac{5}{3}} C_{00,20}^{60,0}-\frac{1}{512} \sqrt{\frac{5}{3}}
C_{00,20}^{60,1}+\frac{\sqrt{3} C_{00,22}^{62,0}}{512}+\frac{\sqrt{3} C_{00,22}^{62,1}}{1024}+\frac{1}{768} \sqrt{\frac{5}{3}} C_{11,20}^{31,0}+\frac{\sqrt{\frac{5}{3}}
C_{11,20}^{31,1}}{1536}+\frac{C_{11,22}^{51,0}}{1536 \sqrt{3}}+\frac{C_{11,22}^{51,1}}{3072 \sqrt{3}}-\frac{9 C_{11,22}^{53,0}}{512 \sqrt{35}}-\frac{9
C_{11,22}^{53,1}}{1024 \sqrt{35}}+\frac{1}{768} \sqrt{\frac{5}{3}} C_{20,20}^{40,0}+\frac{\sqrt{\frac{5}{3}} C_{20,20}^{40,1}}{1536}-\frac{5 C_{20,22}^{42,0}}{1536
\sqrt{3}}-\frac{5 C_{20,22}^{42,1}}{3072 \sqrt{3}}-\frac{C_{22,20}^{42,0}}{768 \sqrt{3}}-\frac{C_{22,20}^{42,1}}{1536 \sqrt{3}}-\frac{11 C_{22,22}^{40,0}}{1536
\sqrt{3}}-\frac{11 C_{22,22}^{40,1}}{3072 \sqrt{3}}+\frac{C_{22,22}^{42,0}}{48 \sqrt{21}}+\frac{C_{22,22}^{42,1}}{96 \sqrt{21}}+\frac{3}{896} \sqrt{\frac{3}{5}}
C_{22,22}^{44,0}+\frac{3 \sqrt{\frac{3}{5}} C_{22,22}^{44,1}}{1792}-\frac{1}{768} \sqrt{\frac{5}{3}} C_{31,20}^{31,0}-\frac{\sqrt{\frac{5}{3}} C_{31,20}^{31,1}}{1536}+\frac{5
C_{31,22}^{31,0}}{1536 \sqrt{3}}+\frac{5 C_{31,22}^{31,1}}{3072 \sqrt{3}}+\frac{1}{512} \sqrt{\frac{5}{7}} C_{31,22}^{33,0}+\frac{\sqrt{\frac{5}{7}}
C_{31,22}^{33,1}}{1024}+\frac{3 C_{33,20}^{33,0}}{256 \sqrt{35}}+\frac{3 C_{33,20}^{33,1}}{512 \sqrt{35}}-\frac{3}{128} \sqrt{\frac{3}{14}} C_{33,22}^{33,0}-\frac{3}{256}
\sqrt{\frac{3}{14}} C_{33,22}^{33,1}\)

\noindent\(C_{42,1111,1}^{00,1111,0,\text{NonLoc}}= -\frac{\sqrt{5} C_{00,00}^{60,0}}{1024}-\frac{\sqrt{5} C_{00,00}^{60,1}}{512}-\frac{\sqrt{\frac{5}{3}}
C_{00,20}^{60,0}}{1024}-\frac{1}{512} \sqrt{\frac{5}{3}} C_{00,20}^{60,1}-\frac{\sqrt{3} C_{00,22}^{62,0}}{1024}-\frac{\sqrt{3} C_{00,22}^{62,1}}{512}-\frac{\sqrt{5}
C_{11,00}^{51,0}}{3072}-\frac{\sqrt{5} C_{11,00}^{51,1}}{1536}-\frac{\sqrt{\frac{5}{3}} C_{11,20}^{31,0}}{3072}-\frac{\sqrt{\frac{5}{3}} C_{11,20}^{31,1}}{1536}+\frac{C_{11,22}^{51,0}}{3072
\sqrt{3}}+\frac{C_{11,22}^{51,1}}{1536 \sqrt{3}}-\frac{9 C_{11,22}^{53,0}}{1024 \sqrt{35}}-\frac{9 C_{11,22}^{53,1}}{512 \sqrt{35}}+\frac{\sqrt{5}
C_{20,00}^{40,0}}{3072}+\frac{\sqrt{5} C_{20,00}^{40,1}}{1536}+\frac{\sqrt{\frac{5}{3}} C_{20,20}^{40,0}}{3072}+\frac{\sqrt{\frac{5}{3}} C_{20,20}^{40,1}}{1536}+\frac{5
C_{20,22}^{42,0}}{3072 \sqrt{3}}+\frac{5 C_{20,22}^{42,1}}{1536 \sqrt{3}}-\frac{C_{22,00}^{42,0}}{3072}-\frac{C_{22,00}^{42,1}}{1536}-\frac{C_{22,20}^{42,0}}{3072
\sqrt{3}}-\frac{C_{22,20}^{42,1}}{1536 \sqrt{3}}+\frac{11 C_{22,22}^{40,0}}{3072 \sqrt{3}}+\frac{11 C_{22,22}^{40,1}}{1536 \sqrt{3}}-\frac{C_{22,22}^{42,0}}{96
\sqrt{21}}-\frac{C_{22,22}^{42,1}}{48 \sqrt{21}}-\frac{3 \sqrt{\frac{3}{5}} C_{22,22}^{44,0}}{1792}-\frac{3}{896} \sqrt{\frac{3}{5}} C_{22,22}^{44,1}+\frac{\sqrt{5}
C_{31,00}^{31,0}}{3072}+\frac{\sqrt{5} C_{31,00}^{31,1}}{1536}+\frac{\sqrt{\frac{5}{3}} C_{31,20}^{31,0}}{3072}+\frac{\sqrt{\frac{5}{3}} C_{31,20}^{31,1}}{1536}+\frac{5
C_{31,22}^{31,0}}{3072 \sqrt{3}}+\frac{5 C_{31,22}^{31,1}}{1536 \sqrt{3}}+\frac{\sqrt{\frac{5}{7}} C_{31,22}^{33,0}}{1024}+\frac{1}{512} \sqrt{\frac{5}{7}}
C_{31,22}^{33,1}-\frac{3 \sqrt{\frac{3}{35}} C_{33,00}^{33,0}}{1024}-\frac{3}{512} \sqrt{\frac{3}{35}} C_{33,00}^{33,1}-\frac{3 C_{33,20}^{33,0}}{1024
\sqrt{35}}-\frac{3 C_{33,20}^{33,1}}{512 \sqrt{35}}-\frac{3}{256} \sqrt{\frac{3}{14}} C_{33,22}^{33,0}-\frac{3}{128} \sqrt{\frac{3}{14}} C_{33,22}^{33,1}\)

\noindent\(C_{42,1111,1}^{00,1111,1,\text{Loc}}= -\frac{\sqrt{5} C_{00,20}^{60,1}}{512}+\frac{3 C_{00,22}^{62,1}}{1024}+\frac{\sqrt{5}
C_{11,20}^{31,1}}{1536}+\frac{C_{11,22}^{51,1}}{3072}-\frac{9 \sqrt{\frac{3}{35}} C_{11,22}^{53,1}}{1024}+\frac{\sqrt{5} C_{20,20}^{40,1}}{1536}-\frac{5
C_{20,22}^{42,1}}{3072}-\frac{C_{22,20}^{42,1}}{1536}-\frac{11 C_{22,22}^{40,1}}{3072}+\frac{C_{22,22}^{42,1}}{96 \sqrt{7}}+\frac{9 C_{22,22}^{44,1}}{1792
\sqrt{5}}-\frac{\sqrt{5} C_{31,20}^{31,1}}{1536}+\frac{5 C_{31,22}^{31,1}}{3072}+\frac{\sqrt{\frac{15}{7}} C_{31,22}^{33,1}}{1024}+\frac{3}{512}
\sqrt{\frac{3}{35}} C_{33,20}^{33,1}-\frac{9 C_{33,22}^{33,1}}{256 \sqrt{14}}\)

\noindent\(C_{42,1111,1}^{00,1111,1,\text{NonLoc}}= -\frac{\sqrt{15} C_{00,00}^{60,0}}{1024}-\frac{\sqrt{5} C_{00,20}^{60,0}}{1024}-\frac{3
C_{00,22}^{62,0}}{1024}-\frac{\sqrt{\frac{5}{3}} C_{11,00}^{51,0}}{1024}-\frac{\sqrt{5} C_{11,20}^{31,0}}{3072}+\frac{C_{11,22}^{51,0}}{3072}-\frac{9
\sqrt{\frac{3}{35}} C_{11,22}^{53,0}}{1024}+\frac{\sqrt{\frac{5}{3}} C_{20,00}^{40,0}}{1024}+\frac{\sqrt{5} C_{20,20}^{40,0}}{3072}+\frac{5 C_{20,22}^{42,0}}{3072}-\frac{C_{22,00}^{42,0}}{1024
\sqrt{3}}-\frac{C_{22,20}^{42,0}}{3072}+\frac{11 C_{22,22}^{40,0}}{3072}-\frac{C_{22,22}^{42,0}}{96 \sqrt{7}}-\frac{9 C_{22,22}^{44,0}}{1792 \sqrt{5}}+\frac{\sqrt{\frac{5}{3}}
C_{31,00}^{31,0}}{1024}+\frac{\sqrt{5} C_{31,20}^{31,0}}{3072}+\frac{5 C_{31,22}^{31,0}}{3072}+\frac{\sqrt{\frac{15}{7}} C_{31,22}^{33,0}}{1024}-\frac{9
C_{33,00}^{33,0}}{1024 \sqrt{35}}-\frac{3 \sqrt{\frac{3}{35}} C_{33,20}^{33,0}}{1024}-\frac{9 C_{33,22}^{33,0}}{256 \sqrt{14}}\)

\noindent\(C_{42,1111,1}^{11,0000,0,\text{Loc}}= \frac{\sqrt{5} C_{11,11}^{51,0}}{768}+\frac{\sqrt{5} C_{11,11}^{51,1}}{1536}-\frac{C_{22,11}^{42,0}}{128
\sqrt{3}}-\frac{C_{22,11}^{42,1}}{256 \sqrt{3}}+\frac{\sqrt{5} C_{31,11}^{31,0}}{768}+\frac{\sqrt{5} C_{31,11}^{31,1}}{1536}+\frac{3 C_{33,11}^{33,0}}{64
\sqrt{70}}+\frac{3 C_{33,11}^{33,1}}{128 \sqrt{70}}\)

\noindent\(C_{42,1111,1}^{11,0000,0,\text{NonLoc}}= \frac{\sqrt{5} C_{11,11}^{51,0}}{1536}+\frac{\sqrt{5} C_{11,11}^{51,1}}{768}+\frac{C_{22,11}^{42,0}}{256
\sqrt{3}}+\frac{C_{22,11}^{42,1}}{128 \sqrt{3}}+\frac{\sqrt{5} C_{31,11}^{31,0}}{1536}+\frac{\sqrt{5} C_{31,11}^{31,1}}{768}+\frac{3 C_{33,11}^{33,0}}{128
\sqrt{70}}+\frac{3 C_{33,11}^{33,1}}{64 \sqrt{70}}\)

\noindent\(C_{42,1111,1}^{11,0000,1,\text{Loc}}= \frac{1}{512} \sqrt{\frac{5}{3}} C_{11,11}^{51,1}-\frac{C_{22,11}^{42,1}}{256}+\frac{1}{512}
\sqrt{\frac{5}{3}} C_{31,11}^{31,1}+\frac{3}{128} \sqrt{\frac{3}{70}} C_{33,11}^{33,1}\)

\noindent\(C_{42,1111,1}^{11,0000,1,\text{NonLoc}}= \frac{1}{512} \sqrt{\frac{5}{3}} C_{11,11}^{51,0}+\frac{C_{22,11}^{42,0}}{256}+\frac{1}{512}
\sqrt{\frac{5}{3}} C_{31,11}^{31,0}+\frac{3}{128} \sqrt{\frac{3}{70}} C_{33,11}^{33,0}\)

\noindent\(C_{42,1111,2}^{00,1112,0,\text{Loc}}= -\frac{\sqrt{5} C_{00,20}^{60,0}}{256}-\frac{\sqrt{5} C_{00,20}^{60,1}}{512}+\frac{3
C_{00,22}^{62,0}}{512}+\frac{3 C_{00,22}^{62,1}}{1024}+\frac{\sqrt{5} C_{11,20}^{31,0}}{768}+\frac{\sqrt{5} C_{11,20}^{31,1}}{1536}-\frac{5 C_{11,22}^{51,0}}{1536}-\frac{5
C_{11,22}^{51,1}}{3072}+\frac{\sqrt{5} C_{20,20}^{40,0}}{768}+\frac{\sqrt{5} C_{20,20}^{40,1}}{1536}+\frac{C_{20,22}^{42,0}}{1536}+\frac{C_{20,22}^{42,1}}{3072}-\frac{C_{22,20}^{42,0}}{768}-\frac{C_{22,20}^{42,1}}{1536}+\frac{C_{22,22}^{40,0}}{1536}+\frac{C_{22,22}^{40,1}}{3072}-\frac{C_{22,22}^{42,0}}{384
\sqrt{7}}-\frac{C_{22,22}^{42,1}}{768 \sqrt{7}}-\frac{9 C_{22,22}^{44,0}}{896 \sqrt{5}}-\frac{9 C_{22,22}^{44,1}}{1792 \sqrt{5}}-\frac{\sqrt{5} C_{31,20}^{31,0}}{768}-\frac{\sqrt{5}
C_{31,20}^{31,1}}{1536}-\frac{C_{31,22}^{31,0}}{1536}-\frac{C_{31,22}^{31,1}}{3072}+\frac{1}{128} \sqrt{\frac{3}{35}} C_{31,22}^{33,0}+\frac{1}{256}
\sqrt{\frac{3}{35}} C_{31,22}^{33,1}+\frac{3}{256} \sqrt{\frac{3}{35}} C_{33,20}^{33,0}+\frac{3}{512} \sqrt{\frac{3}{35}} C_{33,20}^{33,1}+\frac{9
C_{33,22}^{33,0}}{640 \sqrt{14}}+\frac{9 C_{33,22}^{33,1}}{1280 \sqrt{14}}\)

\noindent\(C_{42,1111,2}^{00,1112,0,\text{NonLoc}}= -\frac{\sqrt{15} C_{00,00}^{60,0}}{1024}-\frac{\sqrt{15} C_{00,00}^{60,1}}{512}-\frac{\sqrt{5}
C_{00,20}^{60,0}}{1024}-\frac{\sqrt{5} C_{00,20}^{60,1}}{512}-\frac{3 C_{00,22}^{62,0}}{1024}-\frac{3 C_{00,22}^{62,1}}{512}-\frac{\sqrt{\frac{5}{3}}
C_{11,00}^{51,0}}{1024}-\frac{1}{512} \sqrt{\frac{5}{3}} C_{11,00}^{51,1}-\frac{\sqrt{5} C_{11,20}^{31,0}}{3072}-\frac{\sqrt{5} C_{11,20}^{31,1}}{1536}-\frac{5
C_{11,22}^{51,0}}{3072}-\frac{5 C_{11,22}^{51,1}}{1536}+\frac{\sqrt{\frac{5}{3}} C_{20,00}^{40,0}}{1024}+\frac{1}{512} \sqrt{\frac{5}{3}} C_{20,00}^{40,1}+\frac{\sqrt{5}
C_{20,20}^{40,0}}{3072}+\frac{\sqrt{5} C_{20,20}^{40,1}}{1536}-\frac{C_{20,22}^{42,0}}{3072}-\frac{C_{20,22}^{42,1}}{1536}-\frac{C_{22,00}^{42,0}}{1024
\sqrt{3}}-\frac{C_{22,00}^{42,1}}{512 \sqrt{3}}-\frac{C_{22,20}^{42,0}}{3072}-\frac{C_{22,20}^{42,1}}{1536}-\frac{C_{22,22}^{40,0}}{3072}-\frac{C_{22,22}^{40,1}}{1536}+\frac{C_{22,22}^{42,0}}{768
\sqrt{7}}+\frac{C_{22,22}^{42,1}}{384 \sqrt{7}}+\frac{9 C_{22,22}^{44,0}}{1792 \sqrt{5}}+\frac{9 C_{22,22}^{44,1}}{896 \sqrt{5}}+\frac{\sqrt{\frac{5}{3}}
C_{31,00}^{31,0}}{1024}+\frac{1}{512} \sqrt{\frac{5}{3}} C_{31,00}^{31,1}+\frac{\sqrt{5} C_{31,20}^{31,0}}{3072}+\frac{\sqrt{5} C_{31,20}^{31,1}}{1536}-\frac{C_{31,22}^{31,0}}{3072}-\frac{C_{31,22}^{31,1}}{1536}+\frac{1}{256}
\sqrt{\frac{3}{35}} C_{31,22}^{33,0}+\frac{1}{128} \sqrt{\frac{3}{35}} C_{31,22}^{33,1}-\frac{9 C_{33,00}^{33,0}}{1024 \sqrt{35}}-\frac{9 C_{33,00}^{33,1}}{512
\sqrt{35}}-\frac{3 \sqrt{\frac{3}{35}} C_{33,20}^{33,0}}{1024}-\frac{3}{512} \sqrt{\frac{3}{35}} C_{33,20}^{33,1}+\frac{9 C_{33,22}^{33,0}}{1280
\sqrt{14}}+\frac{9 C_{33,22}^{33,1}}{640 \sqrt{14}}\)

\noindent\(C_{42,1111,2}^{00,1112,1,\text{Loc}}= -\frac{\sqrt{15} C_{00,20}^{60,1}}{512}+\frac{3 \sqrt{3} C_{00,22}^{62,1}}{1024}+\frac{1}{512}
\sqrt{\frac{5}{3}} C_{11,20}^{31,1}-\frac{5 C_{11,22}^{51,1}}{1024 \sqrt{3}}+\frac{1}{512} \sqrt{\frac{5}{3}} C_{20,20}^{40,1}+\frac{C_{20,22}^{42,1}}{1024
\sqrt{3}}-\frac{C_{22,20}^{42,1}}{512 \sqrt{3}}+\frac{C_{22,22}^{40,1}}{1024 \sqrt{3}}-\frac{C_{22,22}^{42,1}}{256 \sqrt{21}}-\frac{9 \sqrt{\frac{3}{5}}
C_{22,22}^{44,1}}{1792}-\frac{1}{512} \sqrt{\frac{5}{3}} C_{31,20}^{31,1}-\frac{C_{31,22}^{31,1}}{1024 \sqrt{3}}+\frac{3 C_{31,22}^{33,1}}{256 \sqrt{35}}+\frac{9
C_{33,20}^{33,1}}{512 \sqrt{35}}+\frac{9 \sqrt{\frac{3}{14}} C_{33,22}^{33,1}}{1280}\)

\noindent\(C_{42,1111,2}^{00,1112,1,\text{NonLoc}}= -\frac{3 \sqrt{5} C_{00,00}^{60,0}}{1024}-\frac{\sqrt{15} C_{00,20}^{60,0}}{1024}-\frac{3
\sqrt{3} C_{00,22}^{62,0}}{1024}-\frac{\sqrt{5} C_{11,00}^{51,0}}{1024}-\frac{\sqrt{\frac{5}{3}} C_{11,20}^{31,0}}{1024}-\frac{5 C_{11,22}^{51,0}}{1024
\sqrt{3}}+\frac{\sqrt{5} C_{20,00}^{40,0}}{1024}+\frac{\sqrt{\frac{5}{3}} C_{20,20}^{40,0}}{1024}-\frac{C_{20,22}^{42,0}}{1024 \sqrt{3}}-\frac{C_{22,00}^{42,0}}{1024}-\frac{C_{22,20}^{42,0}}{1024
\sqrt{3}}-\frac{C_{22,22}^{40,0}}{1024 \sqrt{3}}+\frac{C_{22,22}^{42,0}}{256 \sqrt{21}}+\frac{9 \sqrt{\frac{3}{5}} C_{22,22}^{44,0}}{1792}+\frac{\sqrt{5}
C_{31,00}^{31,0}}{1024}+\frac{\sqrt{\frac{5}{3}} C_{31,20}^{31,0}}{1024}-\frac{C_{31,22}^{31,0}}{1024 \sqrt{3}}+\frac{3 C_{31,22}^{33,0}}{256 \sqrt{35}}-\frac{9
\sqrt{\frac{3}{35}} C_{33,00}^{33,0}}{1024}-\frac{9 C_{33,20}^{33,0}}{1024 \sqrt{35}}+\frac{9 \sqrt{\frac{3}{14}} C_{33,22}^{33,0}}{1280}\)

\noindent\(C_{42,1112,0}^{00,1110,0,\text{Loc}}= \frac{\sqrt{5} C_{00,20}^{60,0}}{384}+\frac{\sqrt{5} C_{00,20}^{60,1}}{768}+\frac{C_{00,22}^{62,0}}{128}+\frac{C_{00,22}^{62,1}}{256}-\frac{\sqrt{5}
C_{11,20}^{31,0}}{1152}-\frac{\sqrt{5} C_{11,20}^{31,1}}{2304}-\frac{7 C_{11,22}^{51,0}}{2304}-\frac{7 C_{11,22}^{51,1}}{4608}-\frac{3}{512} \sqrt{\frac{3}{35}}
C_{11,22}^{53,0}-\frac{3 \sqrt{\frac{3}{35}} C_{11,22}^{53,1}}{1024}-\frac{\sqrt{5} C_{20,20}^{40,0}}{1152}-\frac{\sqrt{5} C_{20,20}^{40,1}}{2304}-\frac{C_{20,22}^{42,0}}{2304}-\frac{C_{20,22}^{42,1}}{4608}+\frac{C_{22,20}^{42,0}}{1152}+\frac{C_{22,20}^{42,1}}{2304}-\frac{C_{22,22}^{40,0}}{576}-\frac{C_{22,22}^{40,1}}{1152}+\frac{5
C_{22,22}^{42,0}}{1152 \sqrt{7}}+\frac{5 C_{22,22}^{42,1}}{2304 \sqrt{7}}-\frac{3 C_{22,22}^{44,0}}{448 \sqrt{5}}-\frac{3 C_{22,22}^{44,1}}{896 \sqrt{5}}+\frac{\sqrt{5}
C_{31,20}^{31,0}}{1152}+\frac{\sqrt{5} C_{31,20}^{31,1}}{2304}+\frac{C_{31,22}^{31,0}}{2304}+\frac{C_{31,22}^{31,1}}{4608}+\frac{17 C_{31,22}^{33,0}}{512
\sqrt{105}}+\frac{17 C_{31,22}^{33,1}}{1024 \sqrt{105}}-\frac{1}{128} \sqrt{\frac{3}{35}} C_{33,20}^{33,0}-\frac{1}{256} \sqrt{\frac{3}{35}} C_{33,20}^{33,1}-\frac{3
C_{33,22}^{33,0}}{320 \sqrt{14}}-\frac{3 C_{33,22}^{33,1}}{640 \sqrt{14}}\)

\noindent\(C_{42,1112,0}^{00,1110,0,\text{NonLoc}}= \frac{1}{512} \sqrt{\frac{5}{3}} C_{00,00}^{60,0}+\frac{1}{256} \sqrt{\frac{5}{3}}
C_{00,00}^{60,1}+\frac{\sqrt{5} C_{00,20}^{60,0}}{1536}+\frac{\sqrt{5} C_{00,20}^{60,1}}{768}-\frac{C_{00,22}^{62,0}}{256}-\frac{C_{00,22}^{62,1}}{128}+\frac{\sqrt{\frac{5}{3}}
C_{11,00}^{51,0}}{1536}+\frac{1}{768} \sqrt{\frac{5}{3}} C_{11,00}^{51,1}+\frac{\sqrt{5} C_{11,20}^{31,0}}{4608}+\frac{\sqrt{5} C_{11,20}^{31,1}}{2304}-\frac{7
C_{11,22}^{51,0}}{4608}-\frac{7 C_{11,22}^{51,1}}{2304}-\frac{3 \sqrt{\frac{3}{35}} C_{11,22}^{53,0}}{1024}-\frac{3}{512} \sqrt{\frac{3}{35}} C_{11,22}^{53,1}-\frac{\sqrt{\frac{5}{3}}
C_{20,00}^{40,0}}{1536}-\frac{1}{768} \sqrt{\frac{5}{3}} C_{20,00}^{40,1}-\frac{\sqrt{5} C_{20,20}^{40,0}}{4608}-\frac{\sqrt{5} C_{20,20}^{40,1}}{2304}+\frac{C_{20,22}^{42,0}}{4608}+\frac{C_{20,22}^{42,1}}{2304}+\frac{C_{22,00}^{42,0}}{1536
\sqrt{3}}+\frac{C_{22,00}^{42,1}}{768 \sqrt{3}}+\frac{C_{22,20}^{42,0}}{4608}+\frac{C_{22,20}^{42,1}}{2304}+\frac{C_{22,22}^{40,0}}{1152}+\frac{C_{22,22}^{40,1}}{576}-\frac{5
C_{22,22}^{42,0}}{2304 \sqrt{7}}-\frac{5 C_{22,22}^{42,1}}{1152 \sqrt{7}}+\frac{3 C_{22,22}^{44,0}}{896 \sqrt{5}}+\frac{3 C_{22,22}^{44,1}}{448 \sqrt{5}}-\frac{\sqrt{\frac{5}{3}}
C_{31,00}^{31,0}}{1536}-\frac{1}{768} \sqrt{\frac{5}{3}} C_{31,00}^{31,1}-\frac{\sqrt{5} C_{31,20}^{31,0}}{4608}-\frac{\sqrt{5} C_{31,20}^{31,1}}{2304}+\frac{C_{31,22}^{31,0}}{4608}+\frac{C_{31,22}^{31,1}}{2304}+\frac{17
C_{31,22}^{33,0}}{1024 \sqrt{105}}+\frac{17 C_{31,22}^{33,1}}{512 \sqrt{105}}+\frac{3 C_{33,00}^{33,0}}{512 \sqrt{35}}+\frac{3 C_{33,00}^{33,1}}{256
\sqrt{35}}+\frac{1}{512} \sqrt{\frac{3}{35}} C_{33,20}^{33,0}+\frac{1}{256} \sqrt{\frac{3}{35}} C_{33,20}^{33,1}-\frac{3 C_{33,22}^{33,0}}{640 \sqrt{14}}-\frac{3
C_{33,22}^{33,1}}{320 \sqrt{14}}\)

\noindent\(C_{42,1112,0}^{00,1110,1,\text{Loc}}= \frac{1}{256} \sqrt{\frac{5}{3}} C_{00,20}^{60,1}+\frac{\sqrt{3} C_{00,22}^{62,1}}{256}-\frac{1}{768}
\sqrt{\frac{5}{3}} C_{11,20}^{31,1}-\frac{7 C_{11,22}^{51,1}}{1536 \sqrt{3}}-\frac{9 C_{11,22}^{53,1}}{1024 \sqrt{35}}-\frac{1}{768} \sqrt{\frac{5}{3}}
C_{20,20}^{40,1}-\frac{C_{20,22}^{42,1}}{1536 \sqrt{3}}+\frac{C_{22,20}^{42,1}}{768 \sqrt{3}}-\frac{C_{22,22}^{40,1}}{384 \sqrt{3}}+\frac{5 C_{22,22}^{42,1}}{768
\sqrt{21}}-\frac{3}{896} \sqrt{\frac{3}{5}} C_{22,22}^{44,1}+\frac{1}{768} \sqrt{\frac{5}{3}} C_{31,20}^{31,1}+\frac{C_{31,22}^{31,1}}{1536 \sqrt{3}}+\frac{17
C_{31,22}^{33,1}}{1024 \sqrt{35}}-\frac{3 C_{33,20}^{33,1}}{256 \sqrt{35}}-\frac{3}{640} \sqrt{\frac{3}{14}} C_{33,22}^{33,1}\)

\noindent\(C_{42,1112,0}^{00,1110,1,\text{NonLoc}}= \frac{\sqrt{5} C_{00,00}^{60,0}}{512}+\frac{1}{512} \sqrt{\frac{5}{3}} C_{00,20}^{60,0}-\frac{\sqrt{3}
C_{00,22}^{62,0}}{256}+\frac{\sqrt{5} C_{11,00}^{51,0}}{1536}+\frac{\sqrt{\frac{5}{3}} C_{11,20}^{31,0}}{1536}-\frac{7 C_{11,22}^{51,0}}{1536 \sqrt{3}}-\frac{9
C_{11,22}^{53,0}}{1024 \sqrt{35}}-\frac{\sqrt{5} C_{20,00}^{40,0}}{1536}-\frac{\sqrt{\frac{5}{3}} C_{20,20}^{40,0}}{1536}+\frac{C_{20,22}^{42,0}}{1536
\sqrt{3}}+\frac{C_{22,00}^{42,0}}{1536}+\frac{C_{22,20}^{42,0}}{1536 \sqrt{3}}+\frac{C_{22,22}^{40,0}}{384 \sqrt{3}}-\frac{5 C_{22,22}^{42,0}}{768
\sqrt{21}}+\frac{3}{896} \sqrt{\frac{3}{5}} C_{22,22}^{44,0}-\frac{\sqrt{5} C_{31,00}^{31,0}}{1536}-\frac{\sqrt{\frac{5}{3}} C_{31,20}^{31,0}}{1536}+\frac{C_{31,22}^{31,0}}{1536
\sqrt{3}}+\frac{17 C_{31,22}^{33,0}}{1024 \sqrt{35}}+\frac{3}{512} \sqrt{\frac{3}{35}} C_{33,00}^{33,0}+\frac{3 C_{33,20}^{33,0}}{512 \sqrt{35}}-\frac{3}{640}
\sqrt{\frac{3}{14}} C_{33,22}^{33,0}\)

\noindent\(C_{42,1112,1}^{00,1111,0,\text{Loc}}= \frac{\sqrt{5} C_{00,20}^{60,0}}{256}+\frac{\sqrt{5} C_{00,20}^{60,1}}{512}-\frac{3
C_{00,22}^{62,0}}{512}-\frac{3 C_{00,22}^{62,1}}{1024}-\frac{\sqrt{5} C_{11,20}^{31,0}}{768}-\frac{\sqrt{5} C_{11,20}^{31,1}}{1536}+\frac{5 C_{11,22}^{51,0}}{1536}+\frac{5
C_{11,22}^{51,1}}{3072}-\frac{\sqrt{5} C_{20,20}^{40,0}}{768}-\frac{\sqrt{5} C_{20,20}^{40,1}}{1536}-\frac{C_{20,22}^{42,0}}{1536}-\frac{C_{20,22}^{42,1}}{3072}+\frac{C_{22,20}^{42,0}}{768}+\frac{C_{22,20}^{42,1}}{1536}-\frac{C_{22,22}^{40,0}}{1536}-\frac{C_{22,22}^{40,1}}{3072}+\frac{C_{22,22}^{42,0}}{384
\sqrt{7}}+\frac{C_{22,22}^{42,1}}{768 \sqrt{7}}+\frac{9 C_{22,22}^{44,0}}{896 \sqrt{5}}+\frac{9 C_{22,22}^{44,1}}{1792 \sqrt{5}}+\frac{\sqrt{5} C_{31,20}^{31,0}}{768}+\frac{\sqrt{5}
C_{31,20}^{31,1}}{1536}+\frac{C_{31,22}^{31,0}}{1536}+\frac{C_{31,22}^{31,1}}{3072}-\frac{1}{128} \sqrt{\frac{3}{35}} C_{31,22}^{33,0}-\frac{1}{256}
\sqrt{\frac{3}{35}} C_{31,22}^{33,1}-\frac{3}{256} \sqrt{\frac{3}{35}} C_{33,20}^{33,0}-\frac{3}{512} \sqrt{\frac{3}{35}} C_{33,20}^{33,1}-\frac{9
C_{33,22}^{33,0}}{640 \sqrt{14}}-\frac{9 C_{33,22}^{33,1}}{1280 \sqrt{14}}\)

\noindent\(C_{42,1112,1}^{00,1111,0,\text{NonLoc}}= \frac{\sqrt{15} C_{00,00}^{60,0}}{1024}+\frac{\sqrt{15} C_{00,00}^{60,1}}{512}+\frac{\sqrt{5}
C_{00,20}^{60,0}}{1024}+\frac{\sqrt{5} C_{00,20}^{60,1}}{512}+\frac{3 C_{00,22}^{62,0}}{1024}+\frac{3 C_{00,22}^{62,1}}{512}+\frac{\sqrt{\frac{5}{3}}
C_{11,00}^{51,0}}{1024}+\frac{1}{512} \sqrt{\frac{5}{3}} C_{11,00}^{51,1}+\frac{\sqrt{5} C_{11,20}^{31,0}}{3072}+\frac{\sqrt{5} C_{11,20}^{31,1}}{1536}+\frac{5
C_{11,22}^{51,0}}{3072}+\frac{5 C_{11,22}^{51,1}}{1536}-\frac{\sqrt{\frac{5}{3}} C_{20,00}^{40,0}}{1024}-\frac{1}{512} \sqrt{\frac{5}{3}} C_{20,00}^{40,1}-\frac{\sqrt{5}
C_{20,20}^{40,0}}{3072}-\frac{\sqrt{5} C_{20,20}^{40,1}}{1536}+\frac{C_{20,22}^{42,0}}{3072}+\frac{C_{20,22}^{42,1}}{1536}+\frac{C_{22,00}^{42,0}}{1024
\sqrt{3}}+\frac{C_{22,00}^{42,1}}{512 \sqrt{3}}+\frac{C_{22,20}^{42,0}}{3072}+\frac{C_{22,20}^{42,1}}{1536}+\frac{C_{22,22}^{40,0}}{3072}+\frac{C_{22,22}^{40,1}}{1536}-\frac{C_{22,22}^{42,0}}{768
\sqrt{7}}-\frac{C_{22,22}^{42,1}}{384 \sqrt{7}}-\frac{9 C_{22,22}^{44,0}}{1792 \sqrt{5}}-\frac{9 C_{22,22}^{44,1}}{896 \sqrt{5}}-\frac{\sqrt{\frac{5}{3}}
C_{31,00}^{31,0}}{1024}-\frac{1}{512} \sqrt{\frac{5}{3}} C_{31,00}^{31,1}-\frac{\sqrt{5} C_{31,20}^{31,0}}{3072}-\frac{\sqrt{5} C_{31,20}^{31,1}}{1536}+\frac{C_{31,22}^{31,0}}{3072}+\frac{C_{31,22}^{31,1}}{1536}-\frac{1}{256}
\sqrt{\frac{3}{35}} C_{31,22}^{33,0}-\frac{1}{128} \sqrt{\frac{3}{35}} C_{31,22}^{33,1}+\frac{9 C_{33,00}^{33,0}}{1024 \sqrt{35}}+\frac{9 C_{33,00}^{33,1}}{512
\sqrt{35}}+\frac{3 \sqrt{\frac{3}{35}} C_{33,20}^{33,0}}{1024}+\frac{3}{512} \sqrt{\frac{3}{35}} C_{33,20}^{33,1}-\frac{9 C_{33,22}^{33,0}}{1280
\sqrt{14}}-\frac{9 C_{33,22}^{33,1}}{640 \sqrt{14}}\)

\noindent\(C_{42,1112,1}^{00,1111,1,\text{Loc}}= \frac{\sqrt{15} C_{00,20}^{60,1}}{512}-\frac{3 \sqrt{3} C_{00,22}^{62,1}}{1024}-\frac{1}{512}
\sqrt{\frac{5}{3}} C_{11,20}^{31,1}+\frac{5 C_{11,22}^{51,1}}{1024 \sqrt{3}}-\frac{1}{512} \sqrt{\frac{5}{3}} C_{20,20}^{40,1}-\frac{C_{20,22}^{42,1}}{1024
\sqrt{3}}+\frac{C_{22,20}^{42,1}}{512 \sqrt{3}}-\frac{C_{22,22}^{40,1}}{1024 \sqrt{3}}+\frac{C_{22,22}^{42,1}}{256 \sqrt{21}}+\frac{9 \sqrt{\frac{3}{5}}
C_{22,22}^{44,1}}{1792}+\frac{1}{512} \sqrt{\frac{5}{3}} C_{31,20}^{31,1}+\frac{C_{31,22}^{31,1}}{1024 \sqrt{3}}-\frac{3 C_{31,22}^{33,1}}{256 \sqrt{35}}-\frac{9
C_{33,20}^{33,1}}{512 \sqrt{35}}-\frac{9 \sqrt{\frac{3}{14}} C_{33,22}^{33,1}}{1280}\)

\noindent\(C_{42,1112,1}^{00,1111,1,\text{NonLoc}}= \frac{3 \sqrt{5} C_{00,00}^{60,0}}{1024}+\frac{\sqrt{15} C_{00,20}^{60,0}}{1024}+\frac{3
\sqrt{3} C_{00,22}^{62,0}}{1024}+\frac{\sqrt{5} C_{11,00}^{51,0}}{1024}+\frac{\sqrt{\frac{5}{3}} C_{11,20}^{31,0}}{1024}+\frac{5 C_{11,22}^{51,0}}{1024
\sqrt{3}}-\frac{\sqrt{5} C_{20,00}^{40,0}}{1024}-\frac{\sqrt{\frac{5}{3}} C_{20,20}^{40,0}}{1024}+\frac{C_{20,22}^{42,0}}{1024 \sqrt{3}}+\frac{C_{22,00}^{42,0}}{1024}+\frac{C_{22,20}^{42,0}}{1024
\sqrt{3}}+\frac{C_{22,22}^{40,0}}{1024 \sqrt{3}}-\frac{C_{22,22}^{42,0}}{256 \sqrt{21}}-\frac{9 \sqrt{\frac{3}{5}} C_{22,22}^{44,0}}{1792}-\frac{\sqrt{5}
C_{31,00}^{31,0}}{1024}-\frac{\sqrt{\frac{5}{3}} C_{31,20}^{31,0}}{1024}+\frac{C_{31,22}^{31,0}}{1024 \sqrt{3}}-\frac{3 C_{31,22}^{33,0}}{256 \sqrt{35}}+\frac{9
\sqrt{\frac{3}{35}} C_{33,00}^{33,0}}{1024}+\frac{9 C_{33,20}^{33,0}}{1024 \sqrt{35}}-\frac{9 \sqrt{\frac{3}{14}} C_{33,22}^{33,0}}{1280}\)

\noindent\(C_{42,1112,2}^{00,1112,0,\text{Loc}}= \frac{\sqrt{35} C_{00,20}^{60,0}}{768}+\frac{\sqrt{35} C_{00,20}^{60,1}}{1536}+\frac{5
C_{00,22}^{62,0}}{512 \sqrt{7}}+\frac{5 C_{00,22}^{62,1}}{1024 \sqrt{7}}-\frac{\sqrt{35} C_{11,20}^{31,0}}{2304}-\frac{\sqrt{35} C_{11,20}^{31,1}}{4608}-\frac{31
C_{11,22}^{51,0}}{4608 \sqrt{7}}-\frac{31 C_{11,22}^{51,1}}{9216 \sqrt{7}}+\frac{3 \sqrt{\frac{3}{5}} C_{11,22}^{53,0}}{3584}+\frac{3 \sqrt{\frac{3}{5}}
C_{11,22}^{53,1}}{7168}-\frac{\sqrt{35} C_{20,20}^{40,0}}{2304}-\frac{\sqrt{35} C_{20,20}^{40,1}}{4608}+\frac{11 C_{20,22}^{42,0}}{4608 \sqrt{7}}+\frac{11
C_{20,22}^{42,1}}{9216 \sqrt{7}}+\frac{\sqrt{7} C_{22,20}^{42,0}}{2304}+\frac{\sqrt{7} C_{22,20}^{42,1}}{4608}+\frac{17 C_{22,22}^{40,0}}{4608 \sqrt{7}}+\frac{17
C_{22,22}^{40,1}}{9216 \sqrt{7}}-\frac{C_{22,22}^{42,0}}{576}-\frac{C_{22,22}^{42,1}}{1152}-\frac{3 C_{22,22}^{44,0}}{128 \sqrt{35}}-\frac{3 C_{22,22}^{44,1}}{256
\sqrt{35}}+\frac{\sqrt{35} C_{31,20}^{31,0}}{2304}+\frac{\sqrt{35} C_{31,20}^{31,1}}{4608}-\frac{11 C_{31,22}^{31,0}}{4608 \sqrt{7}}-\frac{11 C_{31,22}^{31,1}}{9216
\sqrt{7}}+\frac{19 C_{31,22}^{33,0}}{3584 \sqrt{15}}+\frac{19 C_{31,22}^{33,1}}{7168 \sqrt{15}}-\frac{1}{256} \sqrt{\frac{3}{5}} C_{33,20}^{33,0}-\frac{1}{512}
\sqrt{\frac{3}{5}} C_{33,20}^{33,1}+\frac{33 C_{33,22}^{33,0}}{4480 \sqrt{2}}+\frac{33 C_{33,22}^{33,1}}{8960 \sqrt{2}}\)

\noindent\(C_{42,1112,2}^{00,1112,0,\text{NonLoc}}= \frac{\sqrt{\frac{35}{3}} C_{00,00}^{60,0}}{1024}+\frac{1}{512} \sqrt{\frac{35}{3}}
C_{00,00}^{60,1}+\frac{\sqrt{35} C_{00,20}^{60,0}}{3072}+\frac{\sqrt{35} C_{00,20}^{60,1}}{1536}-\frac{5 C_{00,22}^{62,0}}{1024 \sqrt{7}}-\frac{5
C_{00,22}^{62,1}}{512 \sqrt{7}}+\frac{\sqrt{\frac{35}{3}} C_{11,00}^{51,0}}{3072}+\frac{\sqrt{\frac{35}{3}} C_{11,00}^{51,1}}{1536}+\frac{\sqrt{35}
C_{11,20}^{31,0}}{9216}+\frac{\sqrt{35} C_{11,20}^{31,1}}{4608}-\frac{31 C_{11,22}^{51,0}}{9216 \sqrt{7}}-\frac{31 C_{11,22}^{51,1}}{4608 \sqrt{7}}+\frac{3
\sqrt{\frac{3}{5}} C_{11,22}^{53,0}}{7168}+\frac{3 \sqrt{\frac{3}{5}} C_{11,22}^{53,1}}{3584}-\frac{\sqrt{\frac{35}{3}} C_{20,00}^{40,0}}{3072}-\frac{\sqrt{\frac{35}{3}}
C_{20,00}^{40,1}}{1536}-\frac{\sqrt{35} C_{20,20}^{40,0}}{9216}-\frac{\sqrt{35} C_{20,20}^{40,1}}{4608}-\frac{11 C_{20,22}^{42,0}}{9216 \sqrt{7}}-\frac{11
C_{20,22}^{42,1}}{4608 \sqrt{7}}+\frac{\sqrt{\frac{7}{3}} C_{22,00}^{42,0}}{3072}+\frac{\sqrt{\frac{7}{3}} C_{22,00}^{42,1}}{1536}+\frac{\sqrt{7}
C_{22,20}^{42,0}}{9216}+\frac{\sqrt{7} C_{22,20}^{42,1}}{4608}-\frac{17 C_{22,22}^{40,0}}{9216 \sqrt{7}}-\frac{17 C_{22,22}^{40,1}}{4608 \sqrt{7}}+\frac{C_{22,22}^{42,0}}{1152}+\frac{C_{22,22}^{42,1}}{576}+\frac{3
C_{22,22}^{44,0}}{256 \sqrt{35}}+\frac{3 C_{22,22}^{44,1}}{128 \sqrt{35}}-\frac{\sqrt{\frac{35}{3}} C_{31,00}^{31,0}}{3072}-\frac{\sqrt{\frac{35}{3}}
C_{31,00}^{31,1}}{1536}-\frac{\sqrt{35} C_{31,20}^{31,0}}{9216}-\frac{\sqrt{35} C_{31,20}^{31,1}}{4608}-\frac{11 C_{31,22}^{31,0}}{9216 \sqrt{7}}-\frac{11
C_{31,22}^{31,1}}{4608 \sqrt{7}}+\frac{19 C_{31,22}^{33,0}}{7168 \sqrt{15}}+\frac{19 C_{31,22}^{33,1}}{3584 \sqrt{15}}+\frac{3 C_{33,00}^{33,0}}{1024
\sqrt{5}}+\frac{3 C_{33,00}^{33,1}}{512 \sqrt{5}}+\frac{\sqrt{\frac{3}{5}} C_{33,20}^{33,0}}{1024}+\frac{1}{512} \sqrt{\frac{3}{5}} C_{33,20}^{33,1}+\frac{33
C_{33,22}^{33,0}}{8960 \sqrt{2}}+\frac{33 C_{33,22}^{33,1}}{4480 \sqrt{2}}\)

\noindent\(C_{42,1112,2}^{00,1112,1,\text{Loc}}= \frac{1}{512} \sqrt{\frac{35}{3}} C_{00,20}^{60,1}+\frac{5 \sqrt{\frac{3}{7}} C_{00,22}^{62,1}}{1024}-\frac{\sqrt{\frac{35}{3}}
C_{11,20}^{31,1}}{1536}-\frac{31 C_{11,22}^{51,1}}{3072 \sqrt{21}}+\frac{9 C_{11,22}^{53,1}}{7168 \sqrt{5}}-\frac{\sqrt{\frac{35}{3}} C_{20,20}^{40,1}}{1536}+\frac{11
C_{20,22}^{42,1}}{3072 \sqrt{21}}+\frac{\sqrt{\frac{7}{3}} C_{22,20}^{42,1}}{1536}+\frac{17 C_{22,22}^{40,1}}{3072 \sqrt{21}}-\frac{C_{22,22}^{42,1}}{384
\sqrt{3}}-\frac{3}{256} \sqrt{\frac{3}{35}} C_{22,22}^{44,1}+\frac{\sqrt{\frac{35}{3}} C_{31,20}^{31,1}}{1536}-\frac{11 C_{31,22}^{31,1}}{3072 \sqrt{21}}+\frac{19
C_{31,22}^{33,1}}{7168 \sqrt{5}}-\frac{3 C_{33,20}^{33,1}}{512 \sqrt{5}}+\frac{33 \sqrt{\frac{3}{2}} C_{33,22}^{33,1}}{8960}\)

\noindent\(C_{42,1112,2}^{00,1112,1,\text{NonLoc}}= \frac{\sqrt{35} C_{00,00}^{60,0}}{1024}+\frac{\sqrt{\frac{35}{3}} C_{00,20}^{60,0}}{1024}-\frac{5
\sqrt{\frac{3}{7}} C_{00,22}^{62,0}}{1024}+\frac{\sqrt{35} C_{11,00}^{51,0}}{3072}+\frac{\sqrt{\frac{35}{3}} C_{11,20}^{31,0}}{3072}-\frac{31 C_{11,22}^{51,0}}{3072
\sqrt{21}}+\frac{9 C_{11,22}^{53,0}}{7168 \sqrt{5}}-\frac{\sqrt{35} C_{20,00}^{40,0}}{3072}-\frac{\sqrt{\frac{35}{3}} C_{20,20}^{40,0}}{3072}-\frac{11
C_{20,22}^{42,0}}{3072 \sqrt{21}}+\frac{\sqrt{7} C_{22,00}^{42,0}}{3072}+\frac{\sqrt{\frac{7}{3}} C_{22,20}^{42,0}}{3072}-\frac{17 C_{22,22}^{40,0}}{3072
\sqrt{21}}+\frac{C_{22,22}^{42,0}}{384 \sqrt{3}}+\frac{3}{256} \sqrt{\frac{3}{35}} C_{22,22}^{44,0}-\frac{\sqrt{35} C_{31,00}^{31,0}}{3072}-\frac{\sqrt{\frac{35}{3}}
C_{31,20}^{31,0}}{3072}-\frac{11 C_{31,22}^{31,0}}{3072 \sqrt{21}}+\frac{19 C_{31,22}^{33,0}}{7168 \sqrt{5}}+\frac{3 \sqrt{\frac{3}{5}} C_{33,00}^{33,0}}{1024}+\frac{3
C_{33,20}^{33,0}}{1024 \sqrt{5}}+\frac{33 \sqrt{\frac{3}{2}} C_{33,22}^{33,0}}{8960}\)

\noindent\(C_{42,2011,1}^{00,0011,0,\text{Loc}}= -\frac{3 C_{00,22}^{62,0}}{512}-\frac{3 C_{00,22}^{62,1}}{1024}+\frac{5 C_{11,22}^{51,0}}{1536}+\frac{5
C_{11,22}^{51,1}}{3072}-\frac{C_{20,22}^{42,0}}{1536}-\frac{C_{20,22}^{42,1}}{3072}-\frac{C_{22,22}^{40,0}}{1536}-\frac{C_{22,22}^{40,1}}{3072}+\frac{C_{22,22}^{42,0}}{384
\sqrt{7}}+\frac{C_{22,22}^{42,1}}{768 \sqrt{7}}+\frac{9 C_{22,22}^{44,0}}{896 \sqrt{5}}+\frac{9 C_{22,22}^{44,1}}{1792 \sqrt{5}}+\frac{C_{31,22}^{31,0}}{1536}+\frac{C_{31,22}^{31,1}}{3072}-\frac{1}{128}
\sqrt{\frac{3}{35}} C_{31,22}^{33,0}-\frac{1}{256} \sqrt{\frac{3}{35}} C_{31,22}^{33,1}-\frac{9 C_{33,22}^{33,0}}{640 \sqrt{14}}-\frac{9 C_{33,22}^{33,1}}{1280
\sqrt{14}}\)

\noindent\(C_{42,2011,1}^{00,0011,0,\text{NonLoc}}= \frac{3 C_{00,22}^{62,0}}{1024}+\frac{3 C_{00,22}^{62,1}}{512}+\frac{5 C_{11,22}^{51,0}}{3072}+\frac{5
C_{11,22}^{51,1}}{1536}+\frac{C_{20,22}^{42,0}}{3072}+\frac{C_{20,22}^{42,1}}{1536}+\frac{C_{22,22}^{40,0}}{3072}+\frac{C_{22,22}^{40,1}}{1536}-\frac{C_{22,22}^{42,0}}{768
\sqrt{7}}-\frac{C_{22,22}^{42,1}}{384 \sqrt{7}}-\frac{9 C_{22,22}^{44,0}}{1792 \sqrt{5}}-\frac{9 C_{22,22}^{44,1}}{896 \sqrt{5}}+\frac{C_{31,22}^{31,0}}{3072}+\frac{C_{31,22}^{31,1}}{1536}-\frac{1}{256}
\sqrt{\frac{3}{35}} C_{31,22}^{33,0}-\frac{1}{128} \sqrt{\frac{3}{35}} C_{31,22}^{33,1}-\frac{9 C_{33,22}^{33,0}}{1280 \sqrt{14}}-\frac{9 C_{33,22}^{33,1}}{640
\sqrt{14}}\)

\noindent\(C_{42,2011,1}^{00,0011,1,\text{Loc}}= -\frac{3 \sqrt{3} C_{00,22}^{62,1}}{1024}+\frac{5 C_{11,22}^{51,1}}{1024 \sqrt{3}}-\frac{C_{20,22}^{42,1}}{1024
\sqrt{3}}-\frac{C_{22,22}^{40,1}}{1024 \sqrt{3}}+\frac{C_{22,22}^{42,1}}{256 \sqrt{21}}+\frac{9 \sqrt{\frac{3}{5}} C_{22,22}^{44,1}}{1792}+\frac{C_{31,22}^{31,1}}{1024
\sqrt{3}}-\frac{3 C_{31,22}^{33,1}}{256 \sqrt{35}}-\frac{9 \sqrt{\frac{3}{14}} C_{33,22}^{33,1}}{1280}\)

\noindent\(C_{42,2011,1}^{00,0011,1,\text{NonLoc}}= \frac{3 \sqrt{3} C_{00,22}^{62,0}}{1024}+\frac{5 C_{11,22}^{51,0}}{1024 \sqrt{3}}+\frac{C_{20,22}^{42,0}}{1024
\sqrt{3}}+\frac{C_{22,22}^{40,0}}{1024 \sqrt{3}}-\frac{C_{22,22}^{42,0}}{256 \sqrt{21}}-\frac{9 \sqrt{\frac{3}{5}} C_{22,22}^{44,0}}{1792}+\frac{C_{31,22}^{31,0}}{1024
\sqrt{3}}-\frac{3 C_{31,22}^{33,0}}{256 \sqrt{35}}-\frac{9 \sqrt{\frac{3}{14}} C_{33,22}^{33,0}}{1280}\)

\noindent\(C_{42,2202,0}^{00,0000,0,\text{Loc}}= \frac{\sqrt{5} C_{00,00}^{60,0}}{256}+\frac{\sqrt{5} C_{00,00}^{60,1}}{512}-\frac{\sqrt{5}
C_{11,00}^{51,0}}{768}-\frac{\sqrt{5} C_{11,00}^{51,1}}{1536}-\frac{\sqrt{5} C_{20,00}^{40,0}}{768}-\frac{\sqrt{5} C_{20,00}^{40,1}}{1536}+\frac{C_{22,00}^{42,0}}{768}+\frac{C_{22,00}^{42,1}}{1536}+\frac{\sqrt{5}
C_{31,00}^{31,0}}{768}+\frac{\sqrt{5} C_{31,00}^{31,1}}{1536}-\frac{3}{256} \sqrt{\frac{3}{35}} C_{33,00}^{33,0}-\frac{3}{512} \sqrt{\frac{3}{35}}
C_{33,00}^{33,1}\)

\noindent\(C_{42,2202,0}^{00,0000,0,\text{NonLoc}}= -\frac{\sqrt{5} C_{00,00}^{60,0}}{1024}-\frac{\sqrt{5} C_{00,00}^{60,1}}{512}+\frac{\sqrt{15}
C_{00,20}^{60,0}}{1024}+\frac{\sqrt{15} C_{00,20}^{60,1}}{512}-\frac{\sqrt{5} C_{11,00}^{51,0}}{3072}-\frac{\sqrt{5} C_{11,00}^{51,1}}{1536}+\frac{\sqrt{\frac{5}{3}}
C_{11,20}^{31,0}}{1024}+\frac{1}{512} \sqrt{\frac{5}{3}} C_{11,20}^{31,1}+\frac{\sqrt{5} C_{20,00}^{40,0}}{3072}+\frac{\sqrt{5} C_{20,00}^{40,1}}{1536}-\frac{\sqrt{\frac{5}{3}}
C_{20,20}^{40,0}}{1024}-\frac{1}{512} \sqrt{\frac{5}{3}} C_{20,20}^{40,1}-\frac{C_{22,00}^{42,0}}{3072}-\frac{C_{22,00}^{42,1}}{1536}+\frac{C_{22,20}^{42,0}}{1024
\sqrt{3}}+\frac{C_{22,20}^{42,1}}{512 \sqrt{3}}+\frac{\sqrt{5} C_{31,00}^{31,0}}{3072}+\frac{\sqrt{5} C_{31,00}^{31,1}}{1536}-\frac{\sqrt{\frac{5}{3}}
C_{31,20}^{31,0}}{1024}-\frac{1}{512} \sqrt{\frac{5}{3}} C_{31,20}^{31,1}-\frac{3 \sqrt{\frac{3}{35}} C_{33,00}^{33,0}}{1024}-\frac{3}{512} \sqrt{\frac{3}{35}}
C_{33,00}^{33,1}+\frac{9 C_{33,20}^{33,0}}{1024 \sqrt{35}}+\frac{9 C_{33,20}^{33,1}}{512 \sqrt{35}}\)

\noindent\(C_{42,2202,0}^{00,0000,1,\text{Loc}}= \frac{\sqrt{15} C_{00,00}^{60,1}}{512}-\frac{1}{512} \sqrt{\frac{5}{3}} C_{11,00}^{51,1}-\frac{1}{512}
\sqrt{\frac{5}{3}} C_{20,00}^{40,1}+\frac{C_{22,00}^{42,1}}{512 \sqrt{3}}+\frac{1}{512} \sqrt{\frac{5}{3}} C_{31,00}^{31,1}-\frac{9 C_{33,00}^{33,1}}{512
\sqrt{35}}\)

\noindent\(C_{42,2202,0}^{00,0000,1,\text{NonLoc}}= -\frac{\sqrt{15} C_{00,00}^{60,0}}{1024}+\frac{3 \sqrt{5} C_{00,20}^{60,0}}{1024}-\frac{\sqrt{\frac{5}{3}}
C_{11,00}^{51,0}}{1024}+\frac{\sqrt{5} C_{11,20}^{31,0}}{1024}+\frac{\sqrt{\frac{5}{3}} C_{20,00}^{40,0}}{1024}-\frac{\sqrt{5} C_{20,20}^{40,0}}{1024}-\frac{C_{22,00}^{42,0}}{1024
\sqrt{3}}+\frac{C_{22,20}^{42,0}}{1024}+\frac{\sqrt{\frac{5}{3}} C_{31,00}^{31,0}}{1024}-\frac{\sqrt{5} C_{31,20}^{31,0}}{1024}-\frac{9 C_{33,00}^{33,0}}{1024
\sqrt{35}}+\frac{9 \sqrt{\frac{3}{35}} C_{33,20}^{33,0}}{1024}\)

\noindent\(C_{42,2211,1}^{00,0011,0,\text{Loc}}= -\frac{C_{00,20}^{60,0}}{256}-\frac{C_{00,20}^{60,1}}{512}-\frac{3 C_{00,22}^{62,0}}{512
\sqrt{5}}-\frac{3 C_{00,22}^{62,1}}{1024 \sqrt{5}}+\frac{C_{11,20}^{31,0}}{768}+\frac{C_{11,20}^{31,1}}{1536}+\frac{C_{11,22}^{51,0}}{768 \sqrt{5}}+\frac{C_{11,22}^{51,1}}{1536
\sqrt{5}}+\frac{9 \sqrt{\frac{3}{7}} C_{11,22}^{53,0}}{5120}+\frac{9 \sqrt{\frac{3}{7}} C_{11,22}^{53,1}}{10240}+\frac{C_{20,20}^{40,0}}{768}+\frac{C_{20,20}^{40,1}}{1536}+\frac{C_{20,22}^{42,0}}{768
\sqrt{5}}+\frac{C_{20,22}^{42,1}}{1536 \sqrt{5}}-\frac{C_{22,20}^{42,0}}{768 \sqrt{5}}-\frac{C_{22,20}^{42,1}}{1536 \sqrt{5}}+\frac{\sqrt{5} C_{22,22}^{40,0}}{1536}+\frac{\sqrt{5}
C_{22,22}^{40,1}}{3072}-\frac{1}{768} \sqrt{\frac{7}{5}} C_{22,22}^{42,0}-\frac{\sqrt{\frac{7}{5}} C_{22,22}^{42,1}}{1536}-\frac{C_{31,20}^{31,0}}{768}-\frac{C_{31,20}^{31,1}}{1536}-\frac{C_{31,22}^{31,0}}{768
\sqrt{5}}-\frac{C_{31,22}^{31,1}}{1536 \sqrt{5}}-\frac{9 \sqrt{\frac{3}{7}} C_{31,22}^{33,0}}{5120}-\frac{9 \sqrt{\frac{3}{7}} C_{31,22}^{33,1}}{10240}+\frac{3
\sqrt{\frac{3}{7}} C_{33,20}^{33,0}}{1280}+\frac{3 \sqrt{\frac{3}{7}} C_{33,20}^{33,1}}{2560}+\frac{9 C_{33,22}^{33,0}}{320 \sqrt{70}}+\frac{9 C_{33,22}^{33,1}}{640
\sqrt{70}}\)

\noindent\(C_{42,2211,1}^{00,0011,0,\text{NonLoc}}= -\frac{\sqrt{3} C_{00,00}^{60,0}}{1024}-\frac{\sqrt{3} C_{00,00}^{60,1}}{512}-\frac{C_{00,20}^{60,0}}{1024}-\frac{C_{00,20}^{60,1}}{512}+\frac{3
C_{00,22}^{62,0}}{1024 \sqrt{5}}+\frac{3 C_{00,22}^{62,1}}{512 \sqrt{5}}-\frac{C_{11,00}^{51,0}}{1024 \sqrt{3}}-\frac{C_{11,00}^{51,1}}{512 \sqrt{3}}-\frac{C_{11,20}^{31,0}}{3072}-\frac{C_{11,20}^{31,1}}{1536}+\frac{C_{11,22}^{51,0}}{1536
\sqrt{5}}+\frac{C_{11,22}^{51,1}}{768 \sqrt{5}}+\frac{9 \sqrt{\frac{3}{7}} C_{11,22}^{53,0}}{10240}+\frac{9 \sqrt{\frac{3}{7}} C_{11,22}^{53,1}}{5120}+\frac{C_{20,00}^{40,0}}{1024
\sqrt{3}}+\frac{C_{20,00}^{40,1}}{512 \sqrt{3}}+\frac{C_{20,20}^{40,0}}{3072}+\frac{C_{20,20}^{40,1}}{1536}-\frac{C_{20,22}^{42,0}}{1536 \sqrt{5}}-\frac{C_{20,22}^{42,1}}{768
\sqrt{5}}-\frac{C_{22,00}^{42,0}}{1024 \sqrt{15}}-\frac{C_{22,00}^{42,1}}{512 \sqrt{15}}-\frac{C_{22,20}^{42,0}}{3072 \sqrt{5}}-\frac{C_{22,20}^{42,1}}{1536
\sqrt{5}}-\frac{\sqrt{5} C_{22,22}^{40,0}}{3072}-\frac{\sqrt{5} C_{22,22}^{40,1}}{1536}+\frac{\sqrt{\frac{7}{5}} C_{22,22}^{42,0}}{1536}+\frac{1}{768}
\sqrt{\frac{7}{5}} C_{22,22}^{42,1}+\frac{C_{31,00}^{31,0}}{1024 \sqrt{3}}+\frac{C_{31,00}^{31,1}}{512 \sqrt{3}}+\frac{C_{31,20}^{31,0}}{3072}+\frac{C_{31,20}^{31,1}}{1536}-\frac{C_{31,22}^{31,0}}{1536
\sqrt{5}}-\frac{C_{31,22}^{31,1}}{768 \sqrt{5}}-\frac{9 \sqrt{\frac{3}{7}} C_{31,22}^{33,0}}{10240}-\frac{9 \sqrt{\frac{3}{7}} C_{31,22}^{33,1}}{5120}-\frac{9
C_{33,00}^{33,0}}{5120 \sqrt{7}}-\frac{9 C_{33,00}^{33,1}}{2560 \sqrt{7}}-\frac{3 \sqrt{\frac{3}{7}} C_{33,20}^{33,0}}{5120}-\frac{3 \sqrt{\frac{3}{7}}
C_{33,20}^{33,1}}{2560}+\frac{9 C_{33,22}^{33,0}}{640 \sqrt{70}}+\frac{9 C_{33,22}^{33,1}}{320 \sqrt{70}}\)

\noindent\(C_{42,2211,1}^{00,0011,1,\text{Loc}}= -\frac{\sqrt{3} C_{00,20}^{60,1}}{512}-\frac{3 \sqrt{\frac{3}{5}} C_{00,22}^{62,1}}{1024}+\frac{C_{11,20}^{31,1}}{512
\sqrt{3}}+\frac{C_{11,22}^{51,1}}{512 \sqrt{15}}+\frac{27 C_{11,22}^{53,1}}{10240 \sqrt{7}}+\frac{C_{20,20}^{40,1}}{512 \sqrt{3}}+\frac{C_{20,22}^{42,1}}{512
\sqrt{15}}-\frac{C_{22,20}^{42,1}}{512 \sqrt{15}}+\frac{\sqrt{\frac{5}{3}} C_{22,22}^{40,1}}{1024}-\frac{1}{512} \sqrt{\frac{7}{15}} C_{22,22}^{42,1}-\frac{C_{31,20}^{31,1}}{512
\sqrt{3}}-\frac{C_{31,22}^{31,1}}{512 \sqrt{15}}-\frac{27 C_{31,22}^{33,1}}{10240 \sqrt{7}}+\frac{9 C_{33,20}^{33,1}}{2560 \sqrt{7}}+\frac{9}{640}
\sqrt{\frac{3}{70}} C_{33,22}^{33,1}\)

\noindent\(C_{42,2211,1}^{00,0011,1,\text{NonLoc}}= -\frac{3 C_{00,00}^{60,0}}{1024}-\frac{\sqrt{3} C_{00,20}^{60,0}}{1024}+\frac{3
\sqrt{\frac{3}{5}} C_{00,22}^{62,0}}{1024}-\frac{C_{11,00}^{51,0}}{1024}-\frac{C_{11,20}^{31,0}}{1024 \sqrt{3}}+\frac{C_{11,22}^{51,0}}{512 \sqrt{15}}+\frac{27
C_{11,22}^{53,0}}{10240 \sqrt{7}}+\frac{C_{20,00}^{40,0}}{1024}+\frac{C_{20,20}^{40,0}}{1024 \sqrt{3}}-\frac{C_{20,22}^{42,0}}{512 \sqrt{15}}-\frac{C_{22,00}^{42,0}}{1024
\sqrt{5}}-\frac{C_{22,20}^{42,0}}{1024 \sqrt{15}}-\frac{\sqrt{\frac{5}{3}} C_{22,22}^{40,0}}{1024}+\frac{1}{512} \sqrt{\frac{7}{15}} C_{22,22}^{42,0}+\frac{C_{31,00}^{31,0}}{1024}+\frac{C_{31,20}^{31,0}}{1024
\sqrt{3}}-\frac{C_{31,22}^{31,0}}{512 \sqrt{15}}-\frac{27 C_{31,22}^{33,0}}{10240 \sqrt{7}}-\frac{9 \sqrt{\frac{3}{7}} C_{33,00}^{33,0}}{5120}-\frac{9
C_{33,20}^{33,0}}{5120 \sqrt{7}}+\frac{9}{640} \sqrt{\frac{3}{70}} C_{33,22}^{33,0}\)

\noindent\(C_{42,2212,1}^{00,0011,0,\text{Loc}}= \frac{1}{256} \sqrt{\frac{5}{3}} C_{00,20}^{60,0}+\frac{1}{512} \sqrt{\frac{5}{3}}
C_{00,20}^{60,1}-\frac{\sqrt{3} C_{00,22}^{62,0}}{512}-\frac{\sqrt{3} C_{00,22}^{62,1}}{1024}-\frac{1}{768} \sqrt{\frac{5}{3}} C_{11,20}^{31,0}-\frac{\sqrt{\frac{5}{3}}
C_{11,20}^{31,1}}{1536}+\frac{C_{11,22}^{51,0}}{768 \sqrt{3}}+\frac{C_{11,22}^{51,1}}{1536 \sqrt{3}}+\frac{9 C_{11,22}^{53,0}}{1024 \sqrt{35}}+\frac{9
C_{11,22}^{53,1}}{2048 \sqrt{35}}-\frac{1}{768} \sqrt{\frac{5}{3}} C_{20,20}^{40,0}-\frac{\sqrt{\frac{5}{3}} C_{20,20}^{40,1}}{1536}+\frac{C_{20,22}^{42,0}}{768
\sqrt{3}}+\frac{C_{20,22}^{42,1}}{1536 \sqrt{3}}+\frac{C_{22,20}^{42,0}}{768 \sqrt{3}}+\frac{C_{22,20}^{42,1}}{1536 \sqrt{3}}+\frac{5 C_{22,22}^{40,0}}{1536
\sqrt{3}}+\frac{5 C_{22,22}^{40,1}}{3072 \sqrt{3}}-\frac{1}{768} \sqrt{\frac{7}{3}} C_{22,22}^{42,0}-\frac{\sqrt{\frac{7}{3}} C_{22,22}^{42,1}}{1536}+\frac{1}{768}
\sqrt{\frac{5}{3}} C_{31,20}^{31,0}+\frac{\sqrt{\frac{5}{3}} C_{31,20}^{31,1}}{1536}-\frac{C_{31,22}^{31,0}}{768 \sqrt{3}}-\frac{C_{31,22}^{31,1}}{1536
\sqrt{3}}-\frac{9 C_{31,22}^{33,0}}{1024 \sqrt{35}}-\frac{9 C_{31,22}^{33,1}}{2048 \sqrt{35}}-\frac{3 C_{33,20}^{33,0}}{256 \sqrt{35}}-\frac{3 C_{33,20}^{33,1}}{512
\sqrt{35}}+\frac{3}{320} \sqrt{\frac{3}{14}} C_{33,22}^{33,0}+\frac{3}{640} \sqrt{\frac{3}{14}} C_{33,22}^{33,1}\)

\noindent\(C_{42,2212,1}^{00,0011,0,\text{NonLoc}}= \frac{\sqrt{5} C_{00,00}^{60,0}}{1024}+\frac{\sqrt{5} C_{00,00}^{60,1}}{512}+\frac{\sqrt{\frac{5}{3}}
C_{00,20}^{60,0}}{1024}+\frac{1}{512} \sqrt{\frac{5}{3}} C_{00,20}^{60,1}+\frac{\sqrt{3} C_{00,22}^{62,0}}{1024}+\frac{\sqrt{3} C_{00,22}^{62,1}}{512}+\frac{\sqrt{5}
C_{11,00}^{51,0}}{3072}+\frac{\sqrt{5} C_{11,00}^{51,1}}{1536}+\frac{\sqrt{\frac{5}{3}} C_{11,20}^{31,0}}{3072}+\frac{\sqrt{\frac{5}{3}} C_{11,20}^{31,1}}{1536}+\frac{C_{11,22}^{51,0}}{1536
\sqrt{3}}+\frac{C_{11,22}^{51,1}}{768 \sqrt{3}}+\frac{9 C_{11,22}^{53,0}}{2048 \sqrt{35}}+\frac{9 C_{11,22}^{53,1}}{1024 \sqrt{35}}-\frac{\sqrt{5}
C_{20,00}^{40,0}}{3072}-\frac{\sqrt{5} C_{20,00}^{40,1}}{1536}-\frac{\sqrt{\frac{5}{3}} C_{20,20}^{40,0}}{3072}-\frac{\sqrt{\frac{5}{3}} C_{20,20}^{40,1}}{1536}-\frac{C_{20,22}^{42,0}}{1536
\sqrt{3}}-\frac{C_{20,22}^{42,1}}{768 \sqrt{3}}+\frac{C_{22,00}^{42,0}}{3072}+\frac{C_{22,00}^{42,1}}{1536}+\frac{C_{22,20}^{42,0}}{3072 \sqrt{3}}+\frac{C_{22,20}^{42,1}}{1536
\sqrt{3}}-\frac{5 C_{22,22}^{40,0}}{3072 \sqrt{3}}-\frac{5 C_{22,22}^{40,1}}{1536 \sqrt{3}}+\frac{\sqrt{\frac{7}{3}} C_{22,22}^{42,0}}{1536}+\frac{1}{768}
\sqrt{\frac{7}{3}} C_{22,22}^{42,1}-\frac{\sqrt{5} C_{31,00}^{31,0}}{3072}-\frac{\sqrt{5} C_{31,00}^{31,1}}{1536}-\frac{\sqrt{\frac{5}{3}} C_{31,20}^{31,0}}{3072}-\frac{\sqrt{\frac{5}{3}}
C_{31,20}^{31,1}}{1536}-\frac{C_{31,22}^{31,0}}{1536 \sqrt{3}}-\frac{C_{31,22}^{31,1}}{768 \sqrt{3}}-\frac{9 C_{31,22}^{33,0}}{2048 \sqrt{35}}-\frac{9
C_{31,22}^{33,1}}{1024 \sqrt{35}}+\frac{3 \sqrt{\frac{3}{35}} C_{33,00}^{33,0}}{1024}+\frac{3}{512} \sqrt{\frac{3}{35}} C_{33,00}^{33,1}+\frac{3
C_{33,20}^{33,0}}{1024 \sqrt{35}}+\frac{3 C_{33,20}^{33,1}}{512 \sqrt{35}}+\frac{3}{640} \sqrt{\frac{3}{14}} C_{33,22}^{33,0}+\frac{3}{320} \sqrt{\frac{3}{14}}
C_{33,22}^{33,1}\)

\noindent\(C_{42,2212,1}^{00,0011,1,\text{Loc}}= \frac{\sqrt{5} C_{00,20}^{60,1}}{512}-\frac{3 C_{00,22}^{62,1}}{1024}-\frac{\sqrt{5}
C_{11,20}^{31,1}}{1536}+\frac{C_{11,22}^{51,1}}{1536}+\frac{9 \sqrt{\frac{3}{35}} C_{11,22}^{53,1}}{2048}-\frac{\sqrt{5} C_{20,20}^{40,1}}{1536}+\frac{C_{20,22}^{42,1}}{1536}+\frac{C_{22,20}^{42,1}}{1536}+\frac{5
C_{22,22}^{40,1}}{3072}-\frac{\sqrt{7} C_{22,22}^{42,1}}{1536}+\frac{\sqrt{5} C_{31,20}^{31,1}}{1536}-\frac{C_{31,22}^{31,1}}{1536}-\frac{9 \sqrt{\frac{3}{35}}
C_{31,22}^{33,1}}{2048}-\frac{3}{512} \sqrt{\frac{3}{35}} C_{33,20}^{33,1}+\frac{9 C_{33,22}^{33,1}}{640 \sqrt{14}}\)

\noindent\(C_{42,2212,1}^{00,0011,1,\text{NonLoc}}= \frac{\sqrt{15} C_{00,00}^{60,0}}{1024}+\frac{\sqrt{5} C_{00,20}^{60,0}}{1024}+\frac{3
C_{00,22}^{62,0}}{1024}+\frac{\sqrt{\frac{5}{3}} C_{11,00}^{51,0}}{1024}+\frac{\sqrt{5} C_{11,20}^{31,0}}{3072}+\frac{C_{11,22}^{51,0}}{1536}+\frac{9
\sqrt{\frac{3}{35}} C_{11,22}^{53,0}}{2048}-\frac{\sqrt{\frac{5}{3}} C_{20,00}^{40,0}}{1024}-\frac{\sqrt{5} C_{20,20}^{40,0}}{3072}-\frac{C_{20,22}^{42,0}}{1536}+\frac{C_{22,00}^{42,0}}{1024
\sqrt{3}}+\frac{C_{22,20}^{42,0}}{3072}-\frac{5 C_{22,22}^{40,0}}{3072}+\frac{\sqrt{7} C_{22,22}^{42,0}}{1536}-\frac{\sqrt{\frac{5}{3}} C_{31,00}^{31,0}}{1024}-\frac{\sqrt{5}
C_{31,20}^{31,0}}{3072}-\frac{C_{31,22}^{31,0}}{1536}-\frac{9 \sqrt{\frac{3}{35}} C_{31,22}^{33,0}}{2048}+\frac{9 C_{33,00}^{33,0}}{1024 \sqrt{35}}+\frac{3
\sqrt{\frac{3}{35}} C_{33,20}^{33,0}}{1024}+\frac{9 C_{33,22}^{33,0}}{640 \sqrt{14}}\)

\noindent\(C_{42,2213,1}^{00,0011,0,\text{Loc}}= -\frac{1}{256} \sqrt{\frac{7}{3}} C_{00,20}^{60,0}-\frac{1}{512} \sqrt{\frac{7}{3}}
C_{00,20}^{60,1}-\frac{1}{256} \sqrt{\frac{3}{35}} C_{00,22}^{62,0}-\frac{1}{512} \sqrt{\frac{3}{35}} C_{00,22}^{62,1}+\frac{1}{768} \sqrt{\frac{7}{3}}
C_{11,20}^{31,0}+\frac{\sqrt{\frac{7}{3}} C_{11,20}^{31,1}}{1536}+\frac{C_{11,22}^{51,0}}{384 \sqrt{105}}+\frac{C_{11,22}^{51,1}}{768 \sqrt{105}}+\frac{9
C_{11,22}^{53,0}}{17920}+\frac{9 C_{11,22}^{53,1}}{35840}+\frac{1}{768} \sqrt{\frac{7}{3}} C_{20,20}^{40,0}+\frac{\sqrt{\frac{7}{3}} C_{20,20}^{40,1}}{1536}+\frac{C_{20,22}^{42,0}}{384
\sqrt{105}}+\frac{C_{20,22}^{42,1}}{768 \sqrt{105}}-\frac{1}{768} \sqrt{\frac{7}{15}} C_{22,20}^{42,0}-\frac{\sqrt{\frac{7}{15}} C_{22,20}^{42,1}}{1536}+\frac{1}{768}
\sqrt{\frac{5}{21}} C_{22,22}^{40,0}+\frac{\sqrt{\frac{5}{21}} C_{22,22}^{40,1}}{1536}-\frac{C_{22,22}^{42,0}}{384 \sqrt{15}}-\frac{C_{22,22}^{42,1}}{768
\sqrt{15}}-\frac{1}{768} \sqrt{\frac{7}{3}} C_{31,20}^{31,0}-\frac{\sqrt{\frac{7}{3}} C_{31,20}^{31,1}}{1536}-\frac{C_{31,22}^{31,0}}{384 \sqrt{105}}-\frac{C_{31,22}^{31,1}}{768
\sqrt{105}}-\frac{9 C_{31,22}^{33,0}}{17920}-\frac{9 C_{31,22}^{33,1}}{35840}+\frac{3 C_{33,20}^{33,0}}{1280}+\frac{3 C_{33,20}^{33,1}}{2560}+\frac{3
\sqrt{\frac{3}{10}} C_{33,22}^{33,0}}{1120}+\frac{3 \sqrt{\frac{3}{10}} C_{33,22}^{33,1}}{2240}\)

\noindent\(C_{42,2213,1}^{00,0011,0,\text{NonLoc}}= -\frac{\sqrt{7} C_{00,00}^{60,0}}{1024}-\frac{\sqrt{7} C_{00,00}^{60,1}}{512}-\frac{\sqrt{\frac{7}{3}}
C_{00,20}^{60,0}}{1024}-\frac{1}{512} \sqrt{\frac{7}{3}} C_{00,20}^{60,1}+\frac{1}{512} \sqrt{\frac{3}{35}} C_{00,22}^{62,0}+\frac{1}{256} \sqrt{\frac{3}{35}}
C_{00,22}^{62,1}-\frac{\sqrt{7} C_{11,00}^{51,0}}{3072}-\frac{\sqrt{7} C_{11,00}^{51,1}}{1536}-\frac{\sqrt{\frac{7}{3}} C_{11,20}^{31,0}}{3072}-\frac{\sqrt{\frac{7}{3}}
C_{11,20}^{31,1}}{1536}+\frac{C_{11,22}^{51,0}}{768 \sqrt{105}}+\frac{C_{11,22}^{51,1}}{384 \sqrt{105}}+\frac{9 C_{11,22}^{53,0}}{35840}+\frac{9
C_{11,22}^{53,1}}{17920}+\frac{\sqrt{7} C_{20,00}^{40,0}}{3072}+\frac{\sqrt{7} C_{20,00}^{40,1}}{1536}+\frac{\sqrt{\frac{7}{3}} C_{20,20}^{40,0}}{3072}+\frac{\sqrt{\frac{7}{3}}
C_{20,20}^{40,1}}{1536}-\frac{C_{20,22}^{42,0}}{768 \sqrt{105}}-\frac{C_{20,22}^{42,1}}{384 \sqrt{105}}-\frac{\sqrt{\frac{7}{5}} C_{22,00}^{42,0}}{3072}-\frac{\sqrt{\frac{7}{5}}
C_{22,00}^{42,1}}{1536}-\frac{\sqrt{\frac{7}{15}} C_{22,20}^{42,0}}{3072}-\frac{\sqrt{\frac{7}{15}} C_{22,20}^{42,1}}{1536}-\frac{\sqrt{\frac{5}{21}}
C_{22,22}^{40,0}}{1536}-\frac{1}{768} \sqrt{\frac{5}{21}} C_{22,22}^{40,1}+\frac{C_{22,22}^{42,0}}{768 \sqrt{15}}+\frac{C_{22,22}^{42,1}}{384 \sqrt{15}}+\frac{\sqrt{7}
C_{31,00}^{31,0}}{3072}+\frac{\sqrt{7} C_{31,00}^{31,1}}{1536}+\frac{\sqrt{\frac{7}{3}} C_{31,20}^{31,0}}{3072}+\frac{\sqrt{\frac{7}{3}} C_{31,20}^{31,1}}{1536}-\frac{C_{31,22}^{31,0}}{768
\sqrt{105}}-\frac{C_{31,22}^{31,1}}{384 \sqrt{105}}-\frac{9 C_{31,22}^{33,0}}{35840}-\frac{9 C_{31,22}^{33,1}}{17920}-\frac{3 \sqrt{3} C_{33,00}^{33,0}}{5120}-\frac{3
\sqrt{3} C_{33,00}^{33,1}}{2560}-\frac{3 C_{33,20}^{33,0}}{5120}-\frac{3 C_{33,20}^{33,1}}{2560}+\frac{3 \sqrt{\frac{3}{10}} C_{33,22}^{33,0}}{2240}+\frac{3
\sqrt{\frac{3}{10}} C_{33,22}^{33,1}}{1120}\)

\noindent\(C_{42,2213,1}^{00,0011,1,\text{Loc}}= -\frac{\sqrt{7} C_{00,20}^{60,1}}{512}-\frac{3 C_{00,22}^{62,1}}{512 \sqrt{35}}+\frac{\sqrt{7}
C_{11,20}^{31,1}}{1536}+\frac{C_{11,22}^{51,1}}{768 \sqrt{35}}+\frac{9 \sqrt{3} C_{11,22}^{53,1}}{35840}+\frac{\sqrt{7} C_{20,20}^{40,1}}{1536}+\frac{C_{20,22}^{42,1}}{768
\sqrt{35}}-\frac{\sqrt{\frac{7}{5}} C_{22,20}^{42,1}}{1536}+\frac{\sqrt{\frac{5}{7}} C_{22,22}^{40,1}}{1536}-\frac{C_{22,22}^{42,1}}{768 \sqrt{5}}-\frac{\sqrt{7}
C_{31,20}^{31,1}}{1536}-\frac{C_{31,22}^{31,1}}{768 \sqrt{35}}-\frac{9 \sqrt{3} C_{31,22}^{33,1}}{35840}+\frac{3 \sqrt{3} C_{33,20}^{33,1}}{2560}+\frac{9
C_{33,22}^{33,1}}{2240 \sqrt{10}}\)

\noindent\(C_{42,2213,1}^{00,0011,1,\text{NonLoc}}= -\frac{\sqrt{21} C_{00,00}^{60,0}}{1024}-\frac{\sqrt{7} C_{00,20}^{60,0}}{1024}+\frac{3
C_{00,22}^{62,0}}{512 \sqrt{35}}-\frac{\sqrt{\frac{7}{3}} C_{11,00}^{51,0}}{1024}-\frac{\sqrt{7} C_{11,20}^{31,0}}{3072}+\frac{C_{11,22}^{51,0}}{768
\sqrt{35}}+\frac{9 \sqrt{3} C_{11,22}^{53,0}}{35840}+\frac{\sqrt{\frac{7}{3}} C_{20,00}^{40,0}}{1024}+\frac{\sqrt{7} C_{20,20}^{40,0}}{3072}-\frac{C_{20,22}^{42,0}}{768
\sqrt{35}}-\frac{\sqrt{\frac{7}{15}} C_{22,00}^{42,0}}{1024}-\frac{\sqrt{\frac{7}{5}} C_{22,20}^{42,0}}{3072}-\frac{\sqrt{\frac{5}{7}} C_{22,22}^{40,0}}{1536}+\frac{C_{22,22}^{42,0}}{768
\sqrt{5}}+\frac{\sqrt{\frac{7}{3}} C_{31,00}^{31,0}}{1024}+\frac{\sqrt{7} C_{31,20}^{31,0}}{3072}-\frac{C_{31,22}^{31,0}}{768 \sqrt{35}}-\frac{9
\sqrt{3} C_{31,22}^{33,0}}{35840}-\frac{9 C_{33,00}^{33,0}}{5120}-\frac{3 \sqrt{3} C_{33,20}^{33,0}}{5120}+\frac{9 C_{33,22}^{33,0}}{2240 \sqrt{10}}\)

\noindent\(C_{44,0011,3}^{00,2213,0,\text{Loc}}= -\frac{C_{00,22}^{62,0}}{256 \sqrt{5}}-\frac{C_{00,22}^{62,1}}{512 \sqrt{5}}+\frac{C_{11,22}^{51,0}}{256
\sqrt{5}}+\frac{C_{11,22}^{51,1}}{512 \sqrt{5}}-\frac{1}{640} \sqrt{\frac{3}{7}} C_{11,22}^{53,0}-\frac{\sqrt{\frac{3}{7}} C_{11,22}^{53,1}}{1280}+\frac{C_{20,22}^{42,0}}{256
\sqrt{5}}+\frac{C_{20,22}^{42,1}}{512 \sqrt{5}}-\frac{C_{22,22}^{40,0}}{256 \sqrt{5}}-\frac{C_{22,22}^{40,1}}{512 \sqrt{5}}-\frac{C_{22,22}^{44,0}}{1280}-\frac{C_{22,22}^{44,1}}{2560}-\frac{C_{31,22}^{31,0}}{256
\sqrt{5}}-\frac{C_{31,22}^{31,1}}{512 \sqrt{5}}+\frac{1}{640} \sqrt{\frac{3}{7}} C_{31,22}^{33,0}+\frac{\sqrt{\frac{3}{7}} C_{31,22}^{33,1}}{1280}+\frac{3
C_{33,22}^{33,0}}{1280 \sqrt{70}}+\frac{3 C_{33,22}^{33,1}}{2560 \sqrt{70}}\)

\noindent\(C_{44,0011,3}^{00,2213,0,\text{NonLoc}}= \frac{C_{00,22}^{62,0}}{512 \sqrt{5}}+\frac{C_{00,22}^{62,1}}{256 \sqrt{5}}+\frac{C_{11,22}^{51,0}}{512
\sqrt{5}}+\frac{C_{11,22}^{51,1}}{256 \sqrt{5}}-\frac{\sqrt{\frac{3}{7}} C_{11,22}^{53,0}}{1280}-\frac{1}{640} \sqrt{\frac{3}{7}} C_{11,22}^{53,1}-\frac{C_{20,22}^{42,0}}{512
\sqrt{5}}-\frac{C_{20,22}^{42,1}}{256 \sqrt{5}}+\frac{C_{22,22}^{40,0}}{512 \sqrt{5}}+\frac{C_{22,22}^{40,1}}{256 \sqrt{5}}+\frac{C_{22,22}^{44,0}}{2560}+\frac{C_{22,22}^{44,1}}{1280}-\frac{C_{31,22}^{31,0}}{512
\sqrt{5}}-\frac{C_{31,22}^{31,1}}{256 \sqrt{5}}+\frac{\sqrt{\frac{3}{7}} C_{31,22}^{33,0}}{1280}+\frac{1}{640} \sqrt{\frac{3}{7}} C_{31,22}^{33,1}+\frac{3
C_{33,22}^{33,0}}{2560 \sqrt{70}}+\frac{3 C_{33,22}^{33,1}}{1280 \sqrt{70}}\)

\noindent\(C_{44,0011,3}^{00,2213,1,\text{Loc}}= -\frac{1}{512} \sqrt{\frac{3}{5}} C_{00,22}^{62,1}+\frac{1}{512} \sqrt{\frac{3}{5}}
C_{11,22}^{51,1}-\frac{3 C_{11,22}^{53,1}}{1280 \sqrt{7}}+\frac{1}{512} \sqrt{\frac{3}{5}} C_{20,22}^{42,1}-\frac{1}{512} \sqrt{\frac{3}{5}} C_{22,22}^{40,1}-\frac{\sqrt{3}
C_{22,22}^{44,1}}{2560}-\frac{1}{512} \sqrt{\frac{3}{5}} C_{31,22}^{31,1}+\frac{3 C_{31,22}^{33,1}}{1280 \sqrt{7}}+\frac{3 \sqrt{\frac{3}{70}} C_{33,22}^{33,1}}{2560}\)

\noindent\(C_{44,0011,3}^{00,2213,1,\text{NonLoc}}= \frac{1}{512} \sqrt{\frac{3}{5}} C_{00,22}^{62,0}+\frac{1}{512} \sqrt{\frac{3}{5}}
C_{11,22}^{51,0}-\frac{3 C_{11,22}^{53,0}}{1280 \sqrt{7}}-\frac{1}{512} \sqrt{\frac{3}{5}} C_{20,22}^{42,0}+\frac{1}{512} \sqrt{\frac{3}{5}} C_{22,22}^{40,0}+\frac{\sqrt{3}
C_{22,22}^{44,0}}{2560}-\frac{1}{512} \sqrt{\frac{3}{5}} C_{31,22}^{31,0}+\frac{3 C_{31,22}^{33,0}}{1280 \sqrt{7}}+\frac{3 \sqrt{\frac{3}{70}} C_{33,22}^{33,0}}{2560}\)

\noindent\(C_{44,0011,3}^{22,0011,0,\text{Loc}}= \frac{C_{00,22}^{62,0}}{1024 \sqrt{5}}+\frac{C_{00,22}^{62,1}}{2048 \sqrt{5}}-\frac{C_{11,22}^{51,0}}{1024
\sqrt{5}}-\frac{C_{11,22}^{51,1}}{2048 \sqrt{5}}-\frac{3 \sqrt{\frac{3}{7}} C_{11,22}^{53,0}}{5120}-\frac{3 \sqrt{\frac{3}{7}} C_{11,22}^{53,1}}{10240}+\frac{C_{20,22}^{42,0}}{1024
\sqrt{5}}+\frac{C_{20,22}^{42,1}}{2048 \sqrt{5}}+\frac{C_{22,22}^{40,0}}{1024 \sqrt{5}}+\frac{C_{22,22}^{40,1}}{2048 \sqrt{5}}+\frac{C_{22,22}^{42,0}}{512
\sqrt{35}}+\frac{C_{22,22}^{42,1}}{1024 \sqrt{35}}+\frac{3 C_{22,22}^{44,0}}{8960}+\frac{3 C_{22,22}^{44,1}}{17920}-\frac{C_{31,22}^{31,0}}{1024
\sqrt{5}}-\frac{C_{31,22}^{31,1}}{2048 \sqrt{5}}-\frac{3 \sqrt{\frac{3}{7}} C_{31,22}^{33,0}}{5120}-\frac{3 \sqrt{\frac{3}{7}} C_{31,22}^{33,1}}{10240}-\frac{3
C_{33,22}^{33,0}}{1280 \sqrt{70}}-\frac{3 C_{33,22}^{33,1}}{2560 \sqrt{70}}\)

\noindent\(C_{44,0011,3}^{22,0011,0,\text{NonLoc}}= -\frac{C_{00,22}^{62,0}}{2048 \sqrt{5}}-\frac{C_{00,22}^{62,1}}{1024 \sqrt{5}}-\frac{C_{11,22}^{51,0}}{2048
\sqrt{5}}-\frac{C_{11,22}^{51,1}}{1024 \sqrt{5}}-\frac{3 \sqrt{\frac{3}{7}} C_{11,22}^{53,0}}{10240}-\frac{3 \sqrt{\frac{3}{7}} C_{11,22}^{53,1}}{5120}-\frac{C_{20,22}^{42,0}}{2048
\sqrt{5}}-\frac{C_{20,22}^{42,1}}{1024 \sqrt{5}}-\frac{C_{22,22}^{40,0}}{2048 \sqrt{5}}-\frac{C_{22,22}^{40,1}}{1024 \sqrt{5}}-\frac{C_{22,22}^{42,0}}{1024
\sqrt{35}}-\frac{C_{22,22}^{42,1}}{512 \sqrt{35}}-\frac{3 C_{22,22}^{44,0}}{17920}-\frac{3 C_{22,22}^{44,1}}{8960}-\frac{C_{31,22}^{31,0}}{2048 \sqrt{5}}-\frac{C_{31,22}^{31,1}}{1024
\sqrt{5}}-\frac{3 \sqrt{\frac{3}{7}} C_{31,22}^{33,0}}{10240}-\frac{3 \sqrt{\frac{3}{7}} C_{31,22}^{33,1}}{5120}-\frac{3 C_{33,22}^{33,0}}{2560 \sqrt{70}}-\frac{3
C_{33,22}^{33,1}}{1280 \sqrt{70}}\)

\noindent\(C_{44,0011,3}^{22,0011,1,\text{Loc}}= \frac{\sqrt{\frac{3}{5}} C_{00,22}^{62,1}}{2048}-\frac{\sqrt{\frac{3}{5}} C_{11,22}^{51,1}}{2048}-\frac{9
C_{11,22}^{53,1}}{10240 \sqrt{7}}+\frac{\sqrt{\frac{3}{5}} C_{20,22}^{42,1}}{2048}+\frac{\sqrt{\frac{3}{5}} C_{22,22}^{40,1}}{2048}+\frac{\sqrt{\frac{3}{35}}
C_{22,22}^{42,1}}{1024}+\frac{3 \sqrt{3} C_{22,22}^{44,1}}{17920}-\frac{\sqrt{\frac{3}{5}} C_{31,22}^{31,1}}{2048}-\frac{9 C_{31,22}^{33,1}}{10240
\sqrt{7}}-\frac{3 \sqrt{\frac{3}{70}} C_{33,22}^{33,1}}{2560}\)

\noindent\(C_{44,0011,3}^{22,0011,1,\text{NonLoc}}= -\frac{\sqrt{\frac{3}{5}} C_{00,22}^{62,0}}{2048}-\frac{\sqrt{\frac{3}{5}} C_{11,22}^{51,0}}{2048}-\frac{9
C_{11,22}^{53,0}}{10240 \sqrt{7}}-\frac{\sqrt{\frac{3}{5}} C_{20,22}^{42,0}}{2048}-\frac{\sqrt{\frac{3}{5}} C_{22,22}^{40,0}}{2048}-\frac{\sqrt{\frac{3}{35}}
C_{22,22}^{42,0}}{1024}-\frac{3 \sqrt{3} C_{22,22}^{44,0}}{17920}-\frac{\sqrt{\frac{3}{5}} C_{31,22}^{31,0}}{2048}-\frac{9 C_{31,22}^{33,0}}{10240
\sqrt{7}}-\frac{3 \sqrt{\frac{3}{70}} C_{33,22}^{33,0}}{2560}\)

\noindent\(C_{44,1112,2}^{00,1112,0,\text{Loc}}= \frac{C_{00,22}^{62,0}}{128 \sqrt{5}}+\frac{C_{00,22}^{62,1}}{256 \sqrt{5}}-\frac{C_{11,22}^{51,0}}{128
\sqrt{5}}-\frac{C_{11,22}^{51,1}}{256 \sqrt{5}}+\frac{1}{320} \sqrt{\frac{3}{7}} C_{11,22}^{53,0}+\frac{1}{640} \sqrt{\frac{3}{7}} C_{11,22}^{53,1}-\frac{C_{20,22}^{42,0}}{128
\sqrt{5}}-\frac{C_{20,22}^{42,1}}{256 \sqrt{5}}+\frac{C_{22,22}^{40,0}}{128 \sqrt{5}}+\frac{C_{22,22}^{40,1}}{256 \sqrt{5}}+\frac{C_{22,22}^{44,0}}{640}+\frac{C_{22,22}^{44,1}}{1280}+\frac{C_{31,22}^{31,0}}{128
\sqrt{5}}+\frac{C_{31,22}^{31,1}}{256 \sqrt{5}}-\frac{1}{320} \sqrt{\frac{3}{7}} C_{31,22}^{33,0}-\frac{1}{640} \sqrt{\frac{3}{7}} C_{31,22}^{33,1}-\frac{3
C_{33,22}^{33,0}}{640 \sqrt{70}}-\frac{3 C_{33,22}^{33,1}}{1280 \sqrt{70}}\)

\noindent\(C_{44,1112,2}^{00,1112,0,\text{NonLoc}}= -\frac{C_{00,22}^{62,0}}{256 \sqrt{5}}-\frac{C_{00,22}^{62,1}}{128 \sqrt{5}}-\frac{C_{11,22}^{51,0}}{256
\sqrt{5}}-\frac{C_{11,22}^{51,1}}{128 \sqrt{5}}+\frac{1}{640} \sqrt{\frac{3}{7}} C_{11,22}^{53,0}+\frac{1}{320} \sqrt{\frac{3}{7}} C_{11,22}^{53,1}+\frac{C_{20,22}^{42,0}}{256
\sqrt{5}}+\frac{C_{20,22}^{42,1}}{128 \sqrt{5}}-\frac{C_{22,22}^{40,0}}{256 \sqrt{5}}-\frac{C_{22,22}^{40,1}}{128 \sqrt{5}}-\frac{C_{22,22}^{44,0}}{1280}-\frac{C_{22,22}^{44,1}}{640}+\frac{C_{31,22}^{31,0}}{256
\sqrt{5}}+\frac{C_{31,22}^{31,1}}{128 \sqrt{5}}-\frac{1}{640} \sqrt{\frac{3}{7}} C_{31,22}^{33,0}-\frac{1}{320} \sqrt{\frac{3}{7}} C_{31,22}^{33,1}-\frac{3
C_{33,22}^{33,0}}{1280 \sqrt{70}}-\frac{3 C_{33,22}^{33,1}}{640 \sqrt{70}}\)

\noindent\(C_{44,1112,2}^{00,1112,1,\text{Loc}}= \frac{1}{256} \sqrt{\frac{3}{5}} C_{00,22}^{62,1}-\frac{1}{256} \sqrt{\frac{3}{5}}
C_{11,22}^{51,1}+\frac{3 C_{11,22}^{53,1}}{640 \sqrt{7}}-\frac{1}{256} \sqrt{\frac{3}{5}} C_{20,22}^{42,1}+\frac{1}{256} \sqrt{\frac{3}{5}} C_{22,22}^{40,1}+\frac{\sqrt{3}
C_{22,22}^{44,1}}{1280}+\frac{1}{256} \sqrt{\frac{3}{5}} C_{31,22}^{31,1}-\frac{3 C_{31,22}^{33,1}}{640 \sqrt{7}}-\frac{3 \sqrt{\frac{3}{70}} C_{33,22}^{33,1}}{1280}\)

\noindent\(C_{44,1112,2}^{00,1112,1,\text{NonLoc}}= -\frac{1}{256} \sqrt{\frac{3}{5}} C_{00,22}^{62,0}-\frac{1}{256} \sqrt{\frac{3}{5}}
C_{11,22}^{51,0}+\frac{3 C_{11,22}^{53,0}}{640 \sqrt{7}}+\frac{1}{256} \sqrt{\frac{3}{5}} C_{20,22}^{42,0}-\frac{1}{256} \sqrt{\frac{3}{5}} C_{22,22}^{40,0}-\frac{\sqrt{3}
C_{22,22}^{44,0}}{1280}+\frac{1}{256} \sqrt{\frac{3}{5}} C_{31,22}^{31,0}-\frac{3 C_{31,22}^{33,0}}{640 \sqrt{7}}-\frac{3 \sqrt{\frac{3}{70}} C_{33,22}^{33,0}}{1280}\)

\noindent\(C_{44,2213,1}^{00,0011,0,\text{Loc}}= -\frac{C_{00,22}^{62,0}}{256 \sqrt{5}}-\frac{C_{00,22}^{62,1}}{512 \sqrt{5}}+\frac{C_{11,22}^{51,0}}{256
\sqrt{5}}+\frac{C_{11,22}^{51,1}}{512 \sqrt{5}}-\frac{1}{640} \sqrt{\frac{3}{7}} C_{11,22}^{53,0}-\frac{\sqrt{\frac{3}{7}} C_{11,22}^{53,1}}{1280}+\frac{C_{20,22}^{42,0}}{256
\sqrt{5}}+\frac{C_{20,22}^{42,1}}{512 \sqrt{5}}-\frac{C_{22,22}^{40,0}}{256 \sqrt{5}}-\frac{C_{22,22}^{40,1}}{512 \sqrt{5}}-\frac{C_{22,22}^{44,0}}{1280}-\frac{C_{22,22}^{44,1}}{2560}-\frac{C_{31,22}^{31,0}}{256
\sqrt{5}}-\frac{C_{31,22}^{31,1}}{512 \sqrt{5}}+\frac{1}{640} \sqrt{\frac{3}{7}} C_{31,22}^{33,0}+\frac{\sqrt{\frac{3}{7}} C_{31,22}^{33,1}}{1280}+\frac{3
C_{33,22}^{33,0}}{1280 \sqrt{70}}+\frac{3 C_{33,22}^{33,1}}{2560 \sqrt{70}}\)

\noindent\(C_{44,2213,1}^{00,0011,0,\text{NonLoc}}= \frac{C_{00,22}^{62,0}}{512 \sqrt{5}}+\frac{C_{00,22}^{62,1}}{256 \sqrt{5}}+\frac{C_{11,22}^{51,0}}{512
\sqrt{5}}+\frac{C_{11,22}^{51,1}}{256 \sqrt{5}}-\frac{\sqrt{\frac{3}{7}} C_{11,22}^{53,0}}{1280}-\frac{1}{640} \sqrt{\frac{3}{7}} C_{11,22}^{53,1}-\frac{C_{20,22}^{42,0}}{512
\sqrt{5}}-\frac{C_{20,22}^{42,1}}{256 \sqrt{5}}+\frac{C_{22,22}^{40,0}}{512 \sqrt{5}}+\frac{C_{22,22}^{40,1}}{256 \sqrt{5}}+\frac{C_{22,22}^{44,0}}{2560}+\frac{C_{22,22}^{44,1}}{1280}-\frac{C_{31,22}^{31,0}}{512
\sqrt{5}}-\frac{C_{31,22}^{31,1}}{256 \sqrt{5}}+\frac{\sqrt{\frac{3}{7}} C_{31,22}^{33,0}}{1280}+\frac{1}{640} \sqrt{\frac{3}{7}} C_{31,22}^{33,1}+\frac{3
C_{33,22}^{33,0}}{2560 \sqrt{70}}+\frac{3 C_{33,22}^{33,1}}{1280 \sqrt{70}}\)

\noindent\(C_{44,2213,1}^{00,0011,1,\text{Loc}}= -\frac{1}{512} \sqrt{\frac{3}{5}} C_{00,22}^{62,1}+\frac{1}{512} \sqrt{\frac{3}{5}}
C_{11,22}^{51,1}-\frac{3 C_{11,22}^{53,1}}{1280 \sqrt{7}}+\frac{1}{512} \sqrt{\frac{3}{5}} C_{20,22}^{42,1}-\frac{1}{512} \sqrt{\frac{3}{5}} C_{22,22}^{40,1}-\frac{\sqrt{3}
C_{22,22}^{44,1}}{2560}-\frac{1}{512} \sqrt{\frac{3}{5}} C_{31,22}^{31,1}+\frac{3 C_{31,22}^{33,1}}{1280 \sqrt{7}}+\frac{3 \sqrt{\frac{3}{70}} C_{33,22}^{33,1}}{2560}\)

\noindent\(C_{44,2213,1}^{00,0011,1,\text{NonLoc}}= \frac{1}{512} \sqrt{\frac{3}{5}} C_{00,22}^{62,0}+\frac{1}{512} \sqrt{\frac{3}{5}}
C_{11,22}^{51,0}-\frac{3 C_{11,22}^{53,0}}{1280 \sqrt{7}}-\frac{1}{512} \sqrt{\frac{3}{5}} C_{20,22}^{42,0}+\frac{1}{512} \sqrt{\frac{3}{5}} C_{22,22}^{40,0}+\frac{\sqrt{3}
C_{22,22}^{44,0}}{2560}-\frac{1}{512} \sqrt{\frac{3}{5}} C_{31,22}^{31,0}+\frac{3 C_{31,22}^{33,0}}{1280 \sqrt{7}}+\frac{3 \sqrt{\frac{3}{70}} C_{33,22}^{33,0}}{2560}\)

\noindent\(C_{51,0000,1}^{00,1111,0,\text{Loc}}= \frac{C_{11,11}^{51,0}}{1024}+\frac{C_{11,11}^{51,1}}{2048}-\frac{1}{512} \sqrt{\frac{3}{5}}
C_{22,11}^{42,0}-\frac{\sqrt{\frac{3}{5}} C_{22,11}^{42,1}}{1024}+\frac{C_{31,11}^{31,0}}{1024}+\frac{C_{31,11}^{31,1}}{2048}+\frac{9 C_{33,11}^{33,0}}{1280
\sqrt{14}}+\frac{9 C_{33,11}^{33,1}}{2560 \sqrt{14}}\)

\noindent\(C_{51,0000,1}^{00,1111,0,\text{NonLoc}}= \frac{C_{11,11}^{51,0}}{2048}+\frac{C_{11,11}^{51,1}}{1024}+\frac{\sqrt{\frac{3}{5}}
C_{22,11}^{42,0}}{1024}+\frac{1}{512} \sqrt{\frac{3}{5}} C_{22,11}^{42,1}+\frac{C_{31,11}^{31,0}}{2048}+\frac{C_{31,11}^{31,1}}{1024}+\frac{9 C_{33,11}^{33,0}}{2560
\sqrt{14}}+\frac{9 C_{33,11}^{33,1}}{1280 \sqrt{14}}\)

\noindent\(C_{51,0000,1}^{00,1111,1,\text{Loc}}= \frac{\sqrt{3} C_{11,11}^{51,1}}{2048}-\frac{3 C_{22,11}^{42,1}}{1024 \sqrt{5}}+\frac{\sqrt{3}
C_{31,11}^{31,1}}{2048}+\frac{9 \sqrt{\frac{3}{14}} C_{33,11}^{33,1}}{2560}\)

\noindent\(C_{51,0000,1}^{00,1111,1,\text{NonLoc}}= \frac{\sqrt{3} C_{11,11}^{51,0}}{2048}+\frac{3 C_{22,11}^{42,0}}{1024 \sqrt{5}}+\frac{\sqrt{3}
C_{31,11}^{31,0}}{2048}+\frac{9 \sqrt{\frac{3}{14}} C_{33,11}^{33,0}}{2560}\)

\noindent\(C_{51,0000,1}^{11,0000,0,\text{Loc}}= \frac{3 C_{00,00}^{60,0}}{2048}+\frac{3 C_{00,00}^{60,1}}{4096}-\frac{3 C_{11,00}^{51,0}}{2048}-\frac{3
C_{11,00}^{51,1}}{4096}+\frac{3 C_{20,00}^{40,0}}{2048}+\frac{3 C_{20,00}^{40,1}}{4096}+\frac{3 C_{22,00}^{42,0}}{1024 \sqrt{5}}+\frac{3 C_{22,00}^{42,1}}{2048
\sqrt{5}}-\frac{3 C_{31,00}^{31,0}}{2048}-\frac{3 C_{31,00}^{31,1}}{4096}-\frac{9 \sqrt{\frac{3}{7}} C_{33,00}^{33,0}}{5120}-\frac{9 \sqrt{\frac{3}{7}}
C_{33,00}^{33,1}}{10240}\)

\noindent\(C_{51,0000,1}^{11,0000,0,\text{NonLoc}}= -\frac{3 C_{00,00}^{60,0}}{8192}-\frac{3 C_{00,00}^{60,1}}{4096}+\frac{3 \sqrt{3}
C_{00,20}^{60,0}}{8192}+\frac{3 \sqrt{3} C_{00,20}^{60,1}}{4096}-\frac{3 C_{11,00}^{51,0}}{8192}-\frac{3 C_{11,00}^{51,1}}{4096}+\frac{3 \sqrt{3}
C_{11,20}^{31,0}}{8192}+\frac{3 \sqrt{3} C_{11,20}^{31,1}}{4096}-\frac{3 C_{20,00}^{40,0}}{8192}-\frac{3 C_{20,00}^{40,1}}{4096}+\frac{3 \sqrt{3}
C_{20,20}^{40,0}}{8192}+\frac{3 \sqrt{3} C_{20,20}^{40,1}}{4096}-\frac{3 C_{22,00}^{42,0}}{4096 \sqrt{5}}-\frac{3 C_{22,00}^{42,1}}{2048 \sqrt{5}}+\frac{3
\sqrt{\frac{3}{5}} C_{22,20}^{42,0}}{4096}+\frac{3 \sqrt{\frac{3}{5}} C_{22,20}^{42,1}}{2048}-\frac{3 C_{31,00}^{31,0}}{8192}-\frac{3 C_{31,00}^{31,1}}{4096}+\frac{3
\sqrt{3} C_{31,20}^{31,0}}{8192}+\frac{3 \sqrt{3} C_{31,20}^{31,1}}{4096}-\frac{9 \sqrt{\frac{3}{7}} C_{33,00}^{33,0}}{20480}-\frac{9 \sqrt{\frac{3}{7}}
C_{33,00}^{33,1}}{10240}+\frac{27 C_{33,20}^{33,0}}{20480 \sqrt{7}}+\frac{27 C_{33,20}^{33,1}}{10240 \sqrt{7}}\)

\noindent\(C_{51,0000,1}^{11,0000,1,\text{Loc}}= \frac{3 \sqrt{3} C_{00,00}^{60,1}}{4096}-\frac{3 \sqrt{3} C_{11,00}^{51,1}}{4096}+\frac{3
\sqrt{3} C_{20,00}^{40,1}}{4096}+\frac{3 \sqrt{\frac{3}{5}} C_{22,00}^{42,1}}{2048}-\frac{3 \sqrt{3} C_{31,00}^{31,1}}{4096}-\frac{27 C_{33,00}^{33,1}}{10240
\sqrt{7}}\)

\noindent\(C_{51,0000,1}^{11,0000,1,\text{NonLoc}}= -\frac{3 \sqrt{3} C_{00,00}^{60,0}}{8192}+\frac{9 C_{00,20}^{60,0}}{8192}-\frac{3
\sqrt{3} C_{11,00}^{51,0}}{8192}+\frac{9 C_{11,20}^{31,0}}{8192}-\frac{3 \sqrt{3} C_{20,00}^{40,0}}{8192}+\frac{9 C_{20,20}^{40,0}}{8192}-\frac{3
\sqrt{\frac{3}{5}} C_{22,00}^{42,0}}{4096}+\frac{9 C_{22,20}^{42,0}}{4096 \sqrt{5}}-\frac{3 \sqrt{3} C_{31,00}^{31,0}}{8192}+\frac{9 C_{31,20}^{31,0}}{8192}-\frac{27
C_{33,00}^{33,0}}{20480 \sqrt{7}}+\frac{27 \sqrt{\frac{3}{7}} C_{33,20}^{33,0}}{20480}\)

\noindent\(C_{51,0011,0}^{11,0011,0,\text{Loc}}= -\frac{C_{00,20}^{60,0}}{2048}-\frac{C_{00,20}^{60,1}}{4096}-\frac{3 C_{00,22}^{62,0}}{2048
\sqrt{5}}-\frac{3 C_{00,22}^{62,1}}{4096 \sqrt{5}}+\frac{C_{11,20}^{31,0}}{2048}+\frac{C_{11,20}^{31,1}}{4096}+\frac{3 C_{11,22}^{51,0}}{2048 \sqrt{5}}+\frac{3
C_{11,22}^{51,1}}{4096 \sqrt{5}}+\frac{9 \sqrt{\frac{3}{7}} C_{11,22}^{53,0}}{10240}+\frac{9 \sqrt{\frac{3}{7}} C_{11,22}^{53,1}}{20480}-\frac{C_{20,20}^{40,0}}{2048}-\frac{C_{20,20}^{40,1}}{4096}-\frac{3
C_{20,22}^{42,0}}{2048 \sqrt{5}}-\frac{3 C_{20,22}^{42,1}}{4096 \sqrt{5}}-\frac{C_{22,20}^{42,0}}{1024 \sqrt{5}}-\frac{C_{22,20}^{42,1}}{2048 \sqrt{5}}-\frac{3
C_{22,22}^{40,0}}{2048 \sqrt{5}}-\frac{3 C_{22,22}^{40,1}}{4096 \sqrt{5}}-\frac{3 C_{22,22}^{42,0}}{1024 \sqrt{35}}-\frac{3 C_{22,22}^{42,1}}{2048
\sqrt{35}}-\frac{9 C_{22,22}^{44,0}}{17920}-\frac{9 C_{22,22}^{44,1}}{35840}+\frac{C_{31,20}^{31,0}}{2048}+\frac{C_{31,20}^{31,1}}{4096}+\frac{3
C_{31,22}^{31,0}}{2048 \sqrt{5}}+\frac{3 C_{31,22}^{31,1}}{4096 \sqrt{5}}+\frac{9 \sqrt{\frac{3}{7}} C_{31,22}^{33,0}}{10240}+\frac{9 \sqrt{\frac{3}{7}}
C_{31,22}^{33,1}}{20480}+\frac{3 \sqrt{\frac{3}{7}} C_{33,20}^{33,0}}{5120}+\frac{3 \sqrt{\frac{3}{7}} C_{33,20}^{33,1}}{10240}+\frac{9 C_{33,22}^{33,0}}{2560
\sqrt{70}}+\frac{9 C_{33,22}^{33,1}}{5120 \sqrt{70}}\)

\noindent\(C_{51,0011,0}^{11,0011,0,\text{NonLoc}}= -\frac{\sqrt{3} C_{00,00}^{60,0}}{8192}-\frac{\sqrt{3} C_{00,00}^{60,1}}{4096}-\frac{C_{00,20}^{60,0}}{8192}-\frac{C_{00,20}^{60,1}}{4096}+\frac{3
C_{00,22}^{62,0}}{4096 \sqrt{5}}+\frac{3 C_{00,22}^{62,1}}{2048 \sqrt{5}}-\frac{\sqrt{3} C_{11,00}^{51,0}}{8192}-\frac{\sqrt{3} C_{11,00}^{51,1}}{4096}-\frac{C_{11,20}^{31,0}}{8192}-\frac{C_{11,20}^{31,1}}{4096}+\frac{3
C_{11,22}^{51,0}}{4096 \sqrt{5}}+\frac{3 C_{11,22}^{51,1}}{2048 \sqrt{5}}+\frac{9 \sqrt{\frac{3}{7}} C_{11,22}^{53,0}}{20480}+\frac{9 \sqrt{\frac{3}{7}}
C_{11,22}^{53,1}}{10240}-\frac{\sqrt{3} C_{20,00}^{40,0}}{8192}-\frac{\sqrt{3} C_{20,00}^{40,1}}{4096}-\frac{C_{20,20}^{40,0}}{8192}-\frac{C_{20,20}^{40,1}}{4096}+\frac{3
C_{20,22}^{42,0}}{4096 \sqrt{5}}+\frac{3 C_{20,22}^{42,1}}{2048 \sqrt{5}}-\frac{\sqrt{\frac{3}{5}} C_{22,00}^{42,0}}{4096}-\frac{\sqrt{\frac{3}{5}}
C_{22,00}^{42,1}}{2048}-\frac{C_{22,20}^{42,0}}{4096 \sqrt{5}}-\frac{C_{22,20}^{42,1}}{2048 \sqrt{5}}+\frac{3 C_{22,22}^{40,0}}{4096 \sqrt{5}}+\frac{3
C_{22,22}^{40,1}}{2048 \sqrt{5}}+\frac{3 C_{22,22}^{42,0}}{2048 \sqrt{35}}+\frac{3 C_{22,22}^{42,1}}{1024 \sqrt{35}}+\frac{9 C_{22,22}^{44,0}}{35840}+\frac{9
C_{22,22}^{44,1}}{17920}-\frac{\sqrt{3} C_{31,00}^{31,0}}{8192}-\frac{\sqrt{3} C_{31,00}^{31,1}}{4096}-\frac{C_{31,20}^{31,0}}{8192}-\frac{C_{31,20}^{31,1}}{4096}+\frac{3
C_{31,22}^{31,0}}{4096 \sqrt{5}}+\frac{3 C_{31,22}^{31,1}}{2048 \sqrt{5}}+\frac{9 \sqrt{\frac{3}{7}} C_{31,22}^{33,0}}{20480}+\frac{9 \sqrt{\frac{3}{7}}
C_{31,22}^{33,1}}{10240}-\frac{9 C_{33,00}^{33,0}}{20480 \sqrt{7}}-\frac{9 C_{33,00}^{33,1}}{10240 \sqrt{7}}-\frac{3 \sqrt{\frac{3}{7}} C_{33,20}^{33,0}}{20480}-\frac{3
\sqrt{\frac{3}{7}} C_{33,20}^{33,1}}{10240}+\frac{9 C_{33,22}^{33,0}}{5120 \sqrt{70}}+\frac{9 C_{33,22}^{33,1}}{2560 \sqrt{70}}\)

\noindent\(C_{51,0011,0}^{11,0011,1,\text{Loc}}= -\frac{\sqrt{3} C_{00,20}^{60,1}}{4096}-\frac{3 \sqrt{\frac{3}{5}} C_{00,22}^{62,1}}{4096}+\frac{\sqrt{3}
C_{11,20}^{31,1}}{4096}+\frac{3 \sqrt{\frac{3}{5}} C_{11,22}^{51,1}}{4096}+\frac{27 C_{11,22}^{53,1}}{20480 \sqrt{7}}-\frac{\sqrt{3} C_{20,20}^{40,1}}{4096}-\frac{3
\sqrt{\frac{3}{5}} C_{20,22}^{42,1}}{4096}-\frac{\sqrt{\frac{3}{5}} C_{22,20}^{42,1}}{2048}-\frac{3 \sqrt{\frac{3}{5}} C_{22,22}^{40,1}}{4096}-\frac{3
\sqrt{\frac{3}{35}} C_{22,22}^{42,1}}{2048}-\frac{9 \sqrt{3} C_{22,22}^{44,1}}{35840}+\frac{\sqrt{3} C_{31,20}^{31,1}}{4096}+\frac{3 \sqrt{\frac{3}{5}}
C_{31,22}^{31,1}}{4096}+\frac{27 C_{31,22}^{33,1}}{20480 \sqrt{7}}+\frac{9 C_{33,20}^{33,1}}{10240 \sqrt{7}}+\frac{9 \sqrt{\frac{3}{70}} C_{33,22}^{33,1}}{5120}\)

\noindent\(C_{51,0011,0}^{11,0011,1,\text{NonLoc}}= -\frac{3 C_{00,00}^{60,0}}{8192}-\frac{\sqrt{3} C_{00,20}^{60,0}}{8192}+\frac{3
\sqrt{\frac{3}{5}} C_{00,22}^{62,0}}{4096}-\frac{3 C_{11,00}^{51,0}}{8192}-\frac{\sqrt{3} C_{11,20}^{31,0}}{8192}+\frac{3 \sqrt{\frac{3}{5}} C_{11,22}^{51,0}}{4096}+\frac{27
C_{11,22}^{53,0}}{20480 \sqrt{7}}-\frac{3 C_{20,00}^{40,0}}{8192}-\frac{\sqrt{3} C_{20,20}^{40,0}}{8192}+\frac{3 \sqrt{\frac{3}{5}} C_{20,22}^{42,0}}{4096}-\frac{3
C_{22,00}^{42,0}}{4096 \sqrt{5}}-\frac{\sqrt{\frac{3}{5}} C_{22,20}^{42,0}}{4096}+\frac{3 \sqrt{\frac{3}{5}} C_{22,22}^{40,0}}{4096}+\frac{3 \sqrt{\frac{3}{35}}
C_{22,22}^{42,0}}{2048}+\frac{9 \sqrt{3} C_{22,22}^{44,0}}{35840}-\frac{3 C_{31,00}^{31,0}}{8192}-\frac{\sqrt{3} C_{31,20}^{31,0}}{8192}+\frac{3
\sqrt{\frac{3}{5}} C_{31,22}^{31,0}}{4096}+\frac{27 C_{31,22}^{33,0}}{20480 \sqrt{7}}-\frac{9 \sqrt{\frac{3}{7}} C_{33,00}^{33,0}}{20480}-\frac{9
C_{33,20}^{33,0}}{20480 \sqrt{7}}+\frac{9 \sqrt{\frac{3}{70}} C_{33,22}^{33,0}}{5120}\)

\noindent\(C_{51,0011,1}^{00,1101,0,\text{Loc}}= -\frac{C_{11,11}^{51,0}}{1024}-\frac{C_{11,11}^{51,1}}{2048}+\frac{1}{512} \sqrt{\frac{3}{5}}
C_{22,11}^{42,0}+\frac{\sqrt{\frac{3}{5}} C_{22,11}^{42,1}}{1024}-\frac{C_{31,11}^{31,0}}{1024}-\frac{C_{31,11}^{31,1}}{2048}-\frac{9 C_{33,11}^{33,0}}{1280
\sqrt{14}}-\frac{9 C_{33,11}^{33,1}}{2560 \sqrt{14}}\)

\noindent\(C_{51,0011,1}^{00,1101,0,\text{NonLoc}}= -\frac{C_{11,11}^{51,0}}{2048}-\frac{C_{11,11}^{51,1}}{1024}-\frac{\sqrt{\frac{3}{5}}
C_{22,11}^{42,0}}{1024}-\frac{1}{512} \sqrt{\frac{3}{5}} C_{22,11}^{42,1}-\frac{C_{31,11}^{31,0}}{2048}-\frac{C_{31,11}^{31,1}}{1024}-\frac{9 C_{33,11}^{33,0}}{2560
\sqrt{14}}-\frac{9 C_{33,11}^{33,1}}{1280 \sqrt{14}}\)

\noindent\(C_{51,0011,1}^{00,1101,1,\text{Loc}}= -\frac{\sqrt{3} C_{11,11}^{51,1}}{2048}+\frac{3 C_{22,11}^{42,1}}{1024 \sqrt{5}}-\frac{\sqrt{3}
C_{31,11}^{31,1}}{2048}-\frac{9 \sqrt{\frac{3}{14}} C_{33,11}^{33,1}}{2560}\)

\noindent\(C_{51,0011,1}^{00,1101,1,\text{NonLoc}}= -\frac{\sqrt{3} C_{11,11}^{51,0}}{2048}-\frac{3 C_{22,11}^{42,0}}{1024 \sqrt{5}}-\frac{\sqrt{3}
C_{31,11}^{31,0}}{2048}-\frac{9 \sqrt{\frac{3}{14}} C_{33,11}^{33,0}}{2560}\)

\noindent\(C_{51,0011,1}^{11,0011,0,\text{Loc}}= -\frac{\sqrt{3} C_{00,20}^{60,0}}{2048}-\frac{\sqrt{3} C_{00,20}^{60,1}}{4096}+\frac{3
\sqrt{\frac{3}{5}} C_{00,22}^{62,0}}{4096}+\frac{3 \sqrt{\frac{3}{5}} C_{00,22}^{62,1}}{8192}+\frac{\sqrt{3} C_{11,20}^{31,0}}{2048}+\frac{\sqrt{3}
C_{11,20}^{31,1}}{4096}-\frac{3 \sqrt{\frac{3}{5}} C_{11,22}^{51,0}}{4096}-\frac{3 \sqrt{\frac{3}{5}} C_{11,22}^{51,1}}{8192}-\frac{27 C_{11,22}^{53,0}}{20480
\sqrt{7}}-\frac{27 C_{11,22}^{53,1}}{40960 \sqrt{7}}-\frac{\sqrt{3} C_{20,20}^{40,0}}{2048}-\frac{\sqrt{3} C_{20,20}^{40,1}}{4096}+\frac{3 \sqrt{\frac{3}{5}}
C_{20,22}^{42,0}}{4096}+\frac{3 \sqrt{\frac{3}{5}} C_{20,22}^{42,1}}{8192}-\frac{\sqrt{\frac{3}{5}} C_{22,20}^{42,0}}{1024}-\frac{\sqrt{\frac{3}{5}}
C_{22,20}^{42,1}}{2048}+\frac{3 \sqrt{\frac{3}{5}} C_{22,22}^{40,0}}{4096}+\frac{3 \sqrt{\frac{3}{5}} C_{22,22}^{40,1}}{8192}+\frac{3 \sqrt{\frac{3}{35}}
C_{22,22}^{42,0}}{2048}+\frac{3 \sqrt{\frac{3}{35}} C_{22,22}^{42,1}}{4096}+\frac{9 \sqrt{3} C_{22,22}^{44,0}}{35840}+\frac{9 \sqrt{3} C_{22,22}^{44,1}}{71680}+\frac{\sqrt{3}
C_{31,20}^{31,0}}{2048}+\frac{\sqrt{3} C_{31,20}^{31,1}}{4096}-\frac{3 \sqrt{\frac{3}{5}} C_{31,22}^{31,0}}{4096}-\frac{3 \sqrt{\frac{3}{5}} C_{31,22}^{31,1}}{8192}-\frac{27
C_{31,22}^{33,0}}{20480 \sqrt{7}}-\frac{27 C_{31,22}^{33,1}}{40960 \sqrt{7}}+\frac{9 C_{33,20}^{33,0}}{5120 \sqrt{7}}+\frac{9 C_{33,20}^{33,1}}{10240
\sqrt{7}}-\frac{9 \sqrt{\frac{3}{70}} C_{33,22}^{33,0}}{5120}-\frac{9 \sqrt{\frac{3}{70}} C_{33,22}^{33,1}}{10240}\)

\noindent\(C_{51,0011,1}^{11,0011,0,\text{NonLoc}}= -\frac{3 C_{00,00}^{60,0}}{8192}-\frac{3 C_{00,00}^{60,1}}{4096}-\frac{\sqrt{3}
C_{00,20}^{60,0}}{8192}-\frac{\sqrt{3} C_{00,20}^{60,1}}{4096}-\frac{3 \sqrt{\frac{3}{5}} C_{00,22}^{62,0}}{8192}-\frac{3 \sqrt{\frac{3}{5}} C_{00,22}^{62,1}}{4096}-\frac{3
C_{11,00}^{51,0}}{8192}-\frac{3 C_{11,00}^{51,1}}{4096}-\frac{\sqrt{3} C_{11,20}^{31,0}}{8192}-\frac{\sqrt{3} C_{11,20}^{31,1}}{4096}-\frac{3 \sqrt{\frac{3}{5}}
C_{11,22}^{51,0}}{8192}-\frac{3 \sqrt{\frac{3}{5}} C_{11,22}^{51,1}}{4096}-\frac{27 C_{11,22}^{53,0}}{40960 \sqrt{7}}-\frac{27 C_{11,22}^{53,1}}{20480
\sqrt{7}}-\frac{3 C_{20,00}^{40,0}}{8192}-\frac{3 C_{20,00}^{40,1}}{4096}-\frac{\sqrt{3} C_{20,20}^{40,0}}{8192}-\frac{\sqrt{3} C_{20,20}^{40,1}}{4096}-\frac{3
\sqrt{\frac{3}{5}} C_{20,22}^{42,0}}{8192}-\frac{3 \sqrt{\frac{3}{5}} C_{20,22}^{42,1}}{4096}-\frac{3 C_{22,00}^{42,0}}{4096 \sqrt{5}}-\frac{3 C_{22,00}^{42,1}}{2048
\sqrt{5}}-\frac{\sqrt{\frac{3}{5}} C_{22,20}^{42,0}}{4096}-\frac{\sqrt{\frac{3}{5}} C_{22,20}^{42,1}}{2048}-\frac{3 \sqrt{\frac{3}{5}} C_{22,22}^{40,0}}{8192}-\frac{3
\sqrt{\frac{3}{5}} C_{22,22}^{40,1}}{4096}-\frac{3 \sqrt{\frac{3}{35}} C_{22,22}^{42,0}}{4096}-\frac{3 \sqrt{\frac{3}{35}} C_{22,22}^{42,1}}{2048}-\frac{9
\sqrt{3} C_{22,22}^{44,0}}{71680}-\frac{9 \sqrt{3} C_{22,22}^{44,1}}{35840}-\frac{3 C_{31,00}^{31,0}}{8192}-\frac{3 C_{31,00}^{31,1}}{4096}-\frac{\sqrt{3}
C_{31,20}^{31,0}}{8192}-\frac{\sqrt{3} C_{31,20}^{31,1}}{4096}-\frac{3 \sqrt{\frac{3}{5}} C_{31,22}^{31,0}}{8192}-\frac{3 \sqrt{\frac{3}{5}} C_{31,22}^{31,1}}{4096}-\frac{27
C_{31,22}^{33,0}}{40960 \sqrt{7}}-\frac{27 C_{31,22}^{33,1}}{20480 \sqrt{7}}-\frac{9 \sqrt{\frac{3}{7}} C_{33,00}^{33,0}}{20480}-\frac{9 \sqrt{\frac{3}{7}}
C_{33,00}^{33,1}}{10240}-\frac{9 C_{33,20}^{33,0}}{20480 \sqrt{7}}-\frac{9 C_{33,20}^{33,1}}{10240 \sqrt{7}}-\frac{9 \sqrt{\frac{3}{70}} C_{33,22}^{33,0}}{10240}-\frac{9
\sqrt{\frac{3}{70}} C_{33,22}^{33,1}}{5120}\)

\noindent\(C_{51,0011,1}^{11,0011,1,\text{Loc}}= -\frac{3 C_{00,20}^{60,1}}{4096}+\frac{9 C_{00,22}^{62,1}}{8192 \sqrt{5}}+\frac{3
C_{11,20}^{31,1}}{4096}-\frac{9 C_{11,22}^{51,1}}{8192 \sqrt{5}}-\frac{27 \sqrt{\frac{3}{7}} C_{11,22}^{53,1}}{40960}-\frac{3 C_{20,20}^{40,1}}{4096}+\frac{9
C_{20,22}^{42,1}}{8192 \sqrt{5}}-\frac{3 C_{22,20}^{42,1}}{2048 \sqrt{5}}+\frac{9 C_{22,22}^{40,1}}{8192 \sqrt{5}}+\frac{9 C_{22,22}^{42,1}}{4096
\sqrt{35}}+\frac{27 C_{22,22}^{44,1}}{71680}+\frac{3 C_{31,20}^{31,1}}{4096}-\frac{9 C_{31,22}^{31,1}}{8192 \sqrt{5}}-\frac{27 \sqrt{\frac{3}{7}}
C_{31,22}^{33,1}}{40960}+\frac{9 \sqrt{\frac{3}{7}} C_{33,20}^{33,1}}{10240}-\frac{27 C_{33,22}^{33,1}}{10240 \sqrt{70}}\)

\noindent\(C_{51,0011,1}^{11,0011,1,\text{NonLoc}}= -\frac{3 \sqrt{3} C_{00,00}^{60,0}}{8192}-\frac{3 C_{00,20}^{60,0}}{8192}-\frac{9
C_{00,22}^{62,0}}{8192 \sqrt{5}}-\frac{3 \sqrt{3} C_{11,00}^{51,0}}{8192}-\frac{3 C_{11,20}^{31,0}}{8192}-\frac{9 C_{11,22}^{51,0}}{8192 \sqrt{5}}-\frac{27
\sqrt{\frac{3}{7}} C_{11,22}^{53,0}}{40960}-\frac{3 \sqrt{3} C_{20,00}^{40,0}}{8192}-\frac{3 C_{20,20}^{40,0}}{8192}-\frac{9 C_{20,22}^{42,0}}{8192
\sqrt{5}}-\frac{3 \sqrt{\frac{3}{5}} C_{22,00}^{42,0}}{4096}-\frac{3 C_{22,20}^{42,0}}{4096 \sqrt{5}}-\frac{9 C_{22,22}^{40,0}}{8192 \sqrt{5}}-\frac{9
C_{22,22}^{42,0}}{4096 \sqrt{35}}-\frac{27 C_{22,22}^{44,0}}{71680}-\frac{3 \sqrt{3} C_{31,00}^{31,0}}{8192}-\frac{3 C_{31,20}^{31,0}}{8192}-\frac{9
C_{31,22}^{31,0}}{8192 \sqrt{5}}-\frac{27 \sqrt{\frac{3}{7}} C_{31,22}^{33,0}}{40960}-\frac{27 C_{33,00}^{33,0}}{20480 \sqrt{7}}-\frac{9 \sqrt{\frac{3}{7}}
C_{33,20}^{33,0}}{20480}-\frac{27 C_{33,22}^{33,0}}{10240 \sqrt{70}}\)

\noindent\(C_{51,0011,2}^{11,0011,0,\text{Loc}}= -\frac{\sqrt{5} C_{00,20}^{60,0}}{2048}-\frac{\sqrt{5} C_{00,20}^{60,1}}{4096}-\frac{3
C_{00,22}^{62,0}}{20480}-\frac{3 C_{00,22}^{62,1}}{40960}+\frac{\sqrt{5} C_{11,20}^{31,0}}{2048}+\frac{\sqrt{5} C_{11,20}^{31,1}}{4096}+\frac{3 C_{11,22}^{51,0}}{20480}+\frac{3
C_{11,22}^{51,1}}{40960}+\frac{9 \sqrt{\frac{3}{35}} C_{11,22}^{53,0}}{20480}+\frac{9 \sqrt{\frac{3}{35}} C_{11,22}^{53,1}}{40960}-\frac{\sqrt{5}
C_{20,20}^{40,0}}{2048}-\frac{\sqrt{5} C_{20,20}^{40,1}}{4096}-\frac{3 C_{20,22}^{42,0}}{20480}-\frac{3 C_{20,22}^{42,1}}{40960}-\frac{C_{22,20}^{42,0}}{1024}-\frac{C_{22,20}^{42,1}}{2048}-\frac{3
C_{22,22}^{40,0}}{20480}-\frac{3 C_{22,22}^{40,1}}{40960}-\frac{3 C_{22,22}^{42,0}}{10240 \sqrt{7}}-\frac{3 C_{22,22}^{42,1}}{20480 \sqrt{7}}-\frac{9
C_{22,22}^{44,0}}{35840 \sqrt{5}}-\frac{9 C_{22,22}^{44,1}}{71680 \sqrt{5}}+\frac{\sqrt{5} C_{31,20}^{31,0}}{2048}+\frac{\sqrt{5} C_{31,20}^{31,1}}{4096}+\frac{3
C_{31,22}^{31,0}}{20480}+\frac{3 C_{31,22}^{31,1}}{40960}+\frac{9 \sqrt{\frac{3}{35}} C_{31,22}^{33,0}}{20480}+\frac{9 \sqrt{\frac{3}{35}} C_{31,22}^{33,1}}{40960}+\frac{3
\sqrt{\frac{3}{35}} C_{33,20}^{33,0}}{1024}+\frac{3 \sqrt{\frac{3}{35}} C_{33,20}^{33,1}}{2048}+\frac{9 C_{33,22}^{33,0}}{25600 \sqrt{14}}+\frac{9
C_{33,22}^{33,1}}{51200 \sqrt{14}}\)

\noindent\(C_{51,0011,2}^{11,0011,0,\text{NonLoc}}= -\frac{\sqrt{15} C_{00,00}^{60,0}}{8192}-\frac{\sqrt{15} C_{00,00}^{60,1}}{4096}-\frac{\sqrt{5}
C_{00,20}^{60,0}}{8192}-\frac{\sqrt{5} C_{00,20}^{60,1}}{4096}+\frac{3 C_{00,22}^{62,0}}{40960}+\frac{3 C_{00,22}^{62,1}}{20480}-\frac{\sqrt{15}
C_{11,00}^{51,0}}{8192}-\frac{\sqrt{15} C_{11,00}^{51,1}}{4096}-\frac{\sqrt{5} C_{11,20}^{31,0}}{8192}-\frac{\sqrt{5} C_{11,20}^{31,1}}{4096}+\frac{3
C_{11,22}^{51,0}}{40960}+\frac{3 C_{11,22}^{51,1}}{20480}+\frac{9 \sqrt{\frac{3}{35}} C_{11,22}^{53,0}}{40960}+\frac{9 \sqrt{\frac{3}{35}} C_{11,22}^{53,1}}{20480}-\frac{\sqrt{15}
C_{20,00}^{40,0}}{8192}-\frac{\sqrt{15} C_{20,00}^{40,1}}{4096}-\frac{\sqrt{5} C_{20,20}^{40,0}}{8192}-\frac{\sqrt{5} C_{20,20}^{40,1}}{4096}+\frac{3
C_{20,22}^{42,0}}{40960}+\frac{3 C_{20,22}^{42,1}}{20480}-\frac{\sqrt{3} C_{22,00}^{42,0}}{4096}-\frac{\sqrt{3} C_{22,00}^{42,1}}{2048}-\frac{C_{22,20}^{42,0}}{4096}-\frac{C_{22,20}^{42,1}}{2048}+\frac{3
C_{22,22}^{40,0}}{40960}+\frac{3 C_{22,22}^{40,1}}{20480}+\frac{3 C_{22,22}^{42,0}}{20480 \sqrt{7}}+\frac{3 C_{22,22}^{42,1}}{10240 \sqrt{7}}+\frac{9
C_{22,22}^{44,0}}{71680 \sqrt{5}}+\frac{9 C_{22,22}^{44,1}}{35840 \sqrt{5}}-\frac{\sqrt{15} C_{31,00}^{31,0}}{8192}-\frac{\sqrt{15} C_{31,00}^{31,1}}{4096}-\frac{\sqrt{5}
C_{31,20}^{31,0}}{8192}-\frac{\sqrt{5} C_{31,20}^{31,1}}{4096}+\frac{3 C_{31,22}^{31,0}}{40960}+\frac{3 C_{31,22}^{31,1}}{20480}+\frac{9 \sqrt{\frac{3}{35}}
C_{31,22}^{33,0}}{40960}+\frac{9 \sqrt{\frac{3}{35}} C_{31,22}^{33,1}}{20480}-\frac{9 C_{33,00}^{33,0}}{4096 \sqrt{35}}-\frac{9 C_{33,00}^{33,1}}{2048
\sqrt{35}}-\frac{3 \sqrt{\frac{3}{35}} C_{33,20}^{33,0}}{4096}-\frac{3 \sqrt{\frac{3}{35}} C_{33,20}^{33,1}}{2048}+\frac{9 C_{33,22}^{33,0}}{51200
\sqrt{14}}+\frac{9 C_{33,22}^{33,1}}{25600 \sqrt{14}}\)

\noindent\(C_{51,0011,2}^{11,0011,1,\text{Loc}}= -\frac{\sqrt{15} C_{00,20}^{60,1}}{4096}-\frac{3 \sqrt{3} C_{00,22}^{62,1}}{40960}+\frac{\sqrt{15}
C_{11,20}^{31,1}}{4096}+\frac{3 \sqrt{3} C_{11,22}^{51,1}}{40960}+\frac{27 C_{11,22}^{53,1}}{40960 \sqrt{35}}-\frac{\sqrt{15} C_{20,20}^{40,1}}{4096}-\frac{3
\sqrt{3} C_{20,22}^{42,1}}{40960}-\frac{\sqrt{3} C_{22,20}^{42,1}}{2048}-\frac{3 \sqrt{3} C_{22,22}^{40,1}}{40960}-\frac{3 \sqrt{\frac{3}{7}} C_{22,22}^{42,1}}{20480}-\frac{9
\sqrt{\frac{3}{5}} C_{22,22}^{44,1}}{71680}+\frac{\sqrt{15} C_{31,20}^{31,1}}{4096}+\frac{3 \sqrt{3} C_{31,22}^{31,1}}{40960}+\frac{27 C_{31,22}^{33,1}}{40960
\sqrt{35}}+\frac{9 C_{33,20}^{33,1}}{2048 \sqrt{35}}+\frac{9 \sqrt{\frac{3}{14}} C_{33,22}^{33,1}}{51200}\)

\noindent\(C_{51,0011,2}^{11,0011,1,\text{NonLoc}}= -\frac{3 \sqrt{5} C_{00,00}^{60,0}}{8192}-\frac{\sqrt{15} C_{00,20}^{60,0}}{8192}+\frac{3
\sqrt{3} C_{00,22}^{62,0}}{40960}-\frac{3 \sqrt{5} C_{11,00}^{51,0}}{8192}-\frac{\sqrt{15} C_{11,20}^{31,0}}{8192}+\frac{3 \sqrt{3} C_{11,22}^{51,0}}{40960}+\frac{27
C_{11,22}^{53,0}}{40960 \sqrt{35}}-\frac{3 \sqrt{5} C_{20,00}^{40,0}}{8192}-\frac{\sqrt{15} C_{20,20}^{40,0}}{8192}+\frac{3 \sqrt{3} C_{20,22}^{42,0}}{40960}-\frac{3
C_{22,00}^{42,0}}{4096}-\frac{\sqrt{3} C_{22,20}^{42,0}}{4096}+\frac{3 \sqrt{3} C_{22,22}^{40,0}}{40960}+\frac{3 \sqrt{\frac{3}{7}} C_{22,22}^{42,0}}{20480}+\frac{9
\sqrt{\frac{3}{5}} C_{22,22}^{44,0}}{71680}-\frac{3 \sqrt{5} C_{31,00}^{31,0}}{8192}-\frac{\sqrt{15} C_{31,20}^{31,0}}{8192}+\frac{3 \sqrt{3} C_{31,22}^{31,0}}{40960}+\frac{27
C_{31,22}^{33,0}}{40960 \sqrt{35}}-\frac{9 \sqrt{\frac{3}{35}} C_{33,00}^{33,0}}{4096}-\frac{9 C_{33,20}^{33,0}}{4096 \sqrt{35}}+\frac{9 \sqrt{\frac{3}{14}}
C_{33,22}^{33,0}}{51200}\)

\noindent\(C_{51,1101,1}^{00,0011,0,\text{Loc}}= -\frac{C_{11,11}^{51,0}}{1024}-\frac{C_{11,11}^{51,1}}{2048}+\frac{1}{512} \sqrt{\frac{3}{5}}
C_{22,11}^{42,0}+\frac{\sqrt{\frac{3}{5}} C_{22,11}^{42,1}}{1024}-\frac{C_{31,11}^{31,0}}{1024}-\frac{C_{31,11}^{31,1}}{2048}-\frac{9 C_{33,11}^{33,0}}{1280
\sqrt{14}}-\frac{9 C_{33,11}^{33,1}}{2560 \sqrt{14}}\)

\noindent\(C_{51,1101,1}^{00,0011,0,\text{NonLoc}}= -\frac{C_{11,11}^{51,0}}{2048}-\frac{C_{11,11}^{51,1}}{1024}-\frac{\sqrt{\frac{3}{5}}
C_{22,11}^{42,0}}{1024}-\frac{1}{512} \sqrt{\frac{3}{5}} C_{22,11}^{42,1}-\frac{C_{31,11}^{31,0}}{2048}-\frac{C_{31,11}^{31,1}}{1024}-\frac{9 C_{33,11}^{33,0}}{2560
\sqrt{14}}-\frac{9 C_{33,11}^{33,1}}{1280 \sqrt{14}}\)

\noindent\(C_{51,1101,1}^{00,0011,1,\text{Loc}}= -\frac{\sqrt{3} C_{11,11}^{51,1}}{2048}+\frac{3 C_{22,11}^{42,1}}{1024 \sqrt{5}}-\frac{\sqrt{3}
C_{31,11}^{31,1}}{2048}-\frac{9 \sqrt{\frac{3}{14}} C_{33,11}^{33,1}}{2560}\)

\noindent\(C_{51,1101,1}^{00,0011,1,\text{NonLoc}}= -\frac{\sqrt{3} C_{11,11}^{51,0}}{2048}-\frac{3 C_{22,11}^{42,0}}{1024 \sqrt{5}}-\frac{\sqrt{3}
C_{31,11}^{31,0}}{2048}-\frac{9 \sqrt{\frac{3}{14}} C_{33,11}^{33,0}}{2560}\)

\noindent\(C_{51,1111,0}^{00,0000,0,\text{Loc}}= -\frac{C_{11,11}^{51,0}}{1024}-\frac{C_{11,11}^{51,1}}{2048}+\frac{1}{512} \sqrt{\frac{3}{5}}
C_{22,11}^{42,0}+\frac{\sqrt{\frac{3}{5}} C_{22,11}^{42,1}}{1024}-\frac{C_{31,11}^{31,0}}{1024}-\frac{C_{31,11}^{31,1}}{2048}-\frac{9 C_{33,11}^{33,0}}{1280
\sqrt{14}}-\frac{9 C_{33,11}^{33,1}}{2560 \sqrt{14}}\)

\noindent\(C_{51,1111,0}^{00,0000,0,\text{NonLoc}}= -\frac{C_{11,11}^{51,0}}{2048}-\frac{C_{11,11}^{51,1}}{1024}-\frac{\sqrt{\frac{3}{5}}
C_{22,11}^{42,0}}{1024}-\frac{1}{512} \sqrt{\frac{3}{5}} C_{22,11}^{42,1}-\frac{C_{31,11}^{31,0}}{2048}-\frac{C_{31,11}^{31,1}}{1024}-\frac{9 C_{33,11}^{33,0}}{2560
\sqrt{14}}-\frac{9 C_{33,11}^{33,1}}{1280 \sqrt{14}}\)

\noindent\(C_{51,1111,0}^{00,0000,1,\text{Loc}}= -\frac{\sqrt{3} C_{11,11}^{51,1}}{2048}+\frac{3 C_{22,11}^{42,1}}{1024 \sqrt{5}}-\frac{\sqrt{3}
C_{31,11}^{31,1}}{2048}-\frac{9 \sqrt{\frac{3}{14}} C_{33,11}^{33,1}}{2560}\)

\noindent\(C_{51,1111,0}^{00,0000,1,\text{NonLoc}}= -\frac{\sqrt{3} C_{11,11}^{51,0}}{2048}-\frac{3 C_{22,11}^{42,0}}{1024 \sqrt{5}}-\frac{\sqrt{3}
C_{31,11}^{31,0}}{2048}-\frac{9 \sqrt{\frac{3}{14}} C_{33,11}^{33,0}}{2560}\)

\noindent\(C_{53,0011,2}^{11,0011,0,\text{Loc}}= -\frac{\sqrt{\frac{7}{15}} C_{00,22}^{62,0}}{1024}-\frac{\sqrt{\frac{7}{15}} C_{00,22}^{62,1}}{2048}+\frac{\sqrt{\frac{7}{15}}
C_{11,22}^{51,0}}{1024}+\frac{\sqrt{\frac{7}{15}} C_{11,22}^{51,1}}{2048}+\frac{3 C_{11,22}^{53,0}}{5120}+\frac{3 C_{11,22}^{53,1}}{10240}-\frac{\sqrt{\frac{7}{15}}
C_{20,22}^{42,0}}{1024}-\frac{\sqrt{\frac{7}{15}} C_{20,22}^{42,1}}{2048}-\frac{\sqrt{\frac{7}{15}} C_{22,22}^{40,0}}{1024}-\frac{\sqrt{\frac{7}{15}}
C_{22,22}^{40,1}}{2048}-\frac{C_{22,22}^{42,0}}{512 \sqrt{15}}-\frac{C_{22,22}^{42,1}}{1024 \sqrt{15}}-\frac{\sqrt{\frac{3}{7}} C_{22,22}^{44,0}}{1280}-\frac{\sqrt{\frac{3}{7}}
C_{22,22}^{44,1}}{2560}+\frac{\sqrt{\frac{7}{15}} C_{31,22}^{31,0}}{1024}+\frac{\sqrt{\frac{7}{15}} C_{31,22}^{31,1}}{2048}+\frac{3 C_{31,22}^{33,0}}{5120}+\frac{3
C_{31,22}^{33,1}}{10240}+\frac{\sqrt{\frac{3}{10}} C_{33,22}^{33,0}}{1280}+\frac{\sqrt{\frac{3}{10}} C_{33,22}^{33,1}}{2560}\)

\noindent\(C_{53,0011,2}^{11,0011,0,\text{NonLoc}}= \frac{\sqrt{\frac{7}{15}} C_{00,22}^{62,0}}{2048}+\frac{\sqrt{\frac{7}{15}} C_{00,22}^{62,1}}{1024}+\frac{\sqrt{\frac{7}{15}}
C_{11,22}^{51,0}}{2048}+\frac{\sqrt{\frac{7}{15}} C_{11,22}^{51,1}}{1024}+\frac{3 C_{11,22}^{53,0}}{10240}+\frac{3 C_{11,22}^{53,1}}{5120}+\frac{\sqrt{\frac{7}{15}}
C_{20,22}^{42,0}}{2048}+\frac{\sqrt{\frac{7}{15}} C_{20,22}^{42,1}}{1024}+\frac{\sqrt{\frac{7}{15}} C_{22,22}^{40,0}}{2048}+\frac{\sqrt{\frac{7}{15}}
C_{22,22}^{40,1}}{1024}+\frac{C_{22,22}^{42,0}}{1024 \sqrt{15}}+\frac{C_{22,22}^{42,1}}{512 \sqrt{15}}+\frac{\sqrt{\frac{3}{7}} C_{22,22}^{44,0}}{2560}+\frac{\sqrt{\frac{3}{7}}
C_{22,22}^{44,1}}{1280}+\frac{\sqrt{\frac{7}{15}} C_{31,22}^{31,0}}{2048}+\frac{\sqrt{\frac{7}{15}} C_{31,22}^{31,1}}{1024}+\frac{3 C_{31,22}^{33,0}}{10240}+\frac{3
C_{31,22}^{33,1}}{5120}+\frac{\sqrt{\frac{3}{10}} C_{33,22}^{33,0}}{2560}+\frac{\sqrt{\frac{3}{10}} C_{33,22}^{33,1}}{1280}\)

\noindent\(C_{53,0011,2}^{11,0011,1,\text{Loc}}= -\frac{\sqrt{\frac{7}{5}} C_{00,22}^{62,1}}{2048}+\frac{\sqrt{\frac{7}{5}} C_{11,22}^{51,1}}{2048}+\frac{3
\sqrt{3} C_{11,22}^{53,1}}{10240}-\frac{\sqrt{\frac{7}{5}} C_{20,22}^{42,1}}{2048}-\frac{\sqrt{\frac{7}{5}} C_{22,22}^{40,1}}{2048}-\frac{C_{22,22}^{42,1}}{1024
\sqrt{5}}-\frac{3 C_{22,22}^{44,1}}{2560 \sqrt{7}}+\frac{\sqrt{\frac{7}{5}} C_{31,22}^{31,1}}{2048}+\frac{3 \sqrt{3} C_{31,22}^{33,1}}{10240}+\frac{3
C_{33,22}^{33,1}}{2560 \sqrt{10}}\)

\noindent\(C_{53,0011,2}^{11,0011,1,\text{NonLoc}}= \frac{\sqrt{\frac{7}{5}} C_{00,22}^{62,0}}{2048}+\frac{\sqrt{\frac{7}{5}} C_{11,22}^{51,0}}{2048}+\frac{3
\sqrt{3} C_{11,22}^{53,0}}{10240}+\frac{\sqrt{\frac{7}{5}} C_{20,22}^{42,0}}{2048}+\frac{\sqrt{\frac{7}{5}} C_{22,22}^{40,0}}{2048}+\frac{C_{22,22}^{42,0}}{1024
\sqrt{5}}+\frac{3 C_{22,22}^{44,0}}{2560 \sqrt{7}}+\frac{\sqrt{\frac{7}{5}} C_{31,22}^{31,0}}{2048}+\frac{3 \sqrt{3} C_{31,22}^{33,0}}{10240}+\frac{3
C_{33,22}^{33,0}}{2560 \sqrt{10}}\)

\noindent\(C_{60,0000,0}^{00,0000,0,\text{Loc}}= -\frac{C_{00,00}^{60,0}}{4096}-\frac{C_{00,00}^{60,1}}{8192}+\frac{C_{11,00}^{51,0}}{4096}+\frac{C_{11,00}^{51,1}}{8192}-\frac{C_{20,00}^{40,0}}{4096}-\frac{C_{20,00}^{40,1}}{8192}-\frac{C_{22,00}^{42,0}}{2048
\sqrt{5}}-\frac{C_{22,00}^{42,1}}{4096 \sqrt{5}}+\frac{C_{31,00}^{31,0}}{4096}+\frac{C_{31,00}^{31,1}}{8192}+\frac{3 \sqrt{\frac{3}{7}} C_{33,00}^{33,0}}{10240}+\frac{3
\sqrt{\frac{3}{7}} C_{33,00}^{33,1}}{20480}\)

\noindent\(C_{60,0000,0}^{00,0000,0,\text{NonLoc}}= \frac{C_{00,00}^{60,0}}{16384}+\frac{C_{00,00}^{60,1}}{8192}-\frac{\sqrt{3} C_{00,20}^{60,0}}{16384}-\frac{\sqrt{3}
C_{00,20}^{60,1}}{8192}+\frac{C_{11,00}^{51,0}}{16384}+\frac{C_{11,00}^{51,1}}{8192}-\frac{\sqrt{3} C_{11,20}^{31,0}}{16384}-\frac{\sqrt{3} C_{11,20}^{31,1}}{8192}+\frac{C_{20,00}^{40,0}}{16384}+\frac{C_{20,00}^{40,1}}{8192}-\frac{\sqrt{3}
C_{20,20}^{40,0}}{16384}-\frac{\sqrt{3} C_{20,20}^{40,1}}{8192}+\frac{C_{22,00}^{42,0}}{8192 \sqrt{5}}+\frac{C_{22,00}^{42,1}}{4096 \sqrt{5}}-\frac{\sqrt{\frac{3}{5}}
C_{22,20}^{42,0}}{8192}-\frac{\sqrt{\frac{3}{5}} C_{22,20}^{42,1}}{4096}+\frac{C_{31,00}^{31,0}}{16384}+\frac{C_{31,00}^{31,1}}{8192}-\frac{\sqrt{3}
C_{31,20}^{31,0}}{16384}-\frac{\sqrt{3} C_{31,20}^{31,1}}{8192}+\frac{3 \sqrt{\frac{3}{7}} C_{33,00}^{33,0}}{40960}+\frac{3 \sqrt{\frac{3}{7}} C_{33,00}^{33,1}}{20480}-\frac{9
C_{33,20}^{33,0}}{40960 \sqrt{7}}-\frac{9 C_{33,20}^{33,1}}{20480 \sqrt{7}}\)

\noindent\(C_{60,0000,0}^{00,0000,1,\text{Loc}}= -\frac{\sqrt{3} C_{00,00}^{60,1}}{8192}+\frac{\sqrt{3} C_{11,00}^{51,1}}{8192}-\frac{\sqrt{3}
C_{20,00}^{40,1}}{8192}-\frac{\sqrt{\frac{3}{5}} C_{22,00}^{42,1}}{4096}+\frac{\sqrt{3} C_{31,00}^{31,1}}{8192}+\frac{9 C_{33,00}^{33,1}}{20480 \sqrt{7}}\)

\noindent\(C_{60,0000,0}^{00,0000,1,\text{NonLoc}}= \frac{\sqrt{3} C_{00,00}^{60,0}}{16384}-\frac{3 C_{00,20}^{60,0}}{16384}+\frac{\sqrt{3}
C_{11,00}^{51,0}}{16384}-\frac{3 C_{11,20}^{31,0}}{16384}+\frac{\sqrt{3} C_{20,00}^{40,0}}{16384}-\frac{3 C_{20,20}^{40,0}}{16384}+\frac{\sqrt{\frac{3}{5}}
C_{22,00}^{42,0}}{8192}-\frac{3 C_{22,20}^{42,0}}{8192 \sqrt{5}}+\frac{\sqrt{3} C_{31,00}^{31,0}}{16384}-\frac{3 C_{31,20}^{31,0}}{16384}+\frac{9
C_{33,00}^{33,0}}{40960 \sqrt{7}}-\frac{9 \sqrt{\frac{3}{7}} C_{33,20}^{33,0}}{40960}\)

\noindent\(C_{60,0011,1}^{00,0011,0,\text{Loc}}= \frac{C_{00,20}^{60,0}}{4096}+\frac{C_{00,20}^{60,1}}{8192}-\frac{C_{11,20}^{31,0}}{4096}-\frac{C_{11,20}^{31,1}}{8192}+\frac{C_{20,20}^{40,0}}{4096}+\frac{C_{20,20}^{40,1}}{8192}+\frac{C_{22,20}^{42,0}}{2048
\sqrt{5}}+\frac{C_{22,20}^{42,1}}{4096 \sqrt{5}}-\frac{C_{31,20}^{31,0}}{4096}-\frac{C_{31,20}^{31,1}}{8192}-\frac{3 \sqrt{\frac{3}{7}} C_{33,20}^{33,0}}{10240}-\frac{3
\sqrt{\frac{3}{7}} C_{33,20}^{33,1}}{20480}\)

\noindent\(C_{60,0011,1}^{00,0011,0,\text{NonLoc}}= \frac{\sqrt{3} C_{00,00}^{60,0}}{16384}+\frac{\sqrt{3} C_{00,00}^{60,1}}{8192}+\frac{C_{00,20}^{60,0}}{16384}+\frac{C_{00,20}^{60,1}}{8192}+\frac{\sqrt{3}
C_{11,00}^{51,0}}{16384}+\frac{\sqrt{3} C_{11,00}^{51,1}}{8192}+\frac{C_{11,20}^{31,0}}{16384}+\frac{C_{11,20}^{31,1}}{8192}+\frac{\sqrt{3} C_{20,00}^{40,0}}{16384}+\frac{\sqrt{3}
C_{20,00}^{40,1}}{8192}+\frac{C_{20,20}^{40,0}}{16384}+\frac{C_{20,20}^{40,1}}{8192}+\frac{\sqrt{\frac{3}{5}} C_{22,00}^{42,0}}{8192}+\frac{\sqrt{\frac{3}{5}}
C_{22,00}^{42,1}}{4096}+\frac{C_{22,20}^{42,0}}{8192 \sqrt{5}}+\frac{C_{22,20}^{42,1}}{4096 \sqrt{5}}+\frac{\sqrt{3} C_{31,00}^{31,0}}{16384}+\frac{\sqrt{3}
C_{31,00}^{31,1}}{8192}+\frac{C_{31,20}^{31,0}}{16384}+\frac{C_{31,20}^{31,1}}{8192}+\frac{9 C_{33,00}^{33,0}}{40960 \sqrt{7}}+\frac{9 C_{33,00}^{33,1}}{20480
\sqrt{7}}+\frac{3 \sqrt{\frac{3}{7}} C_{33,20}^{33,0}}{40960}+\frac{3 \sqrt{\frac{3}{7}} C_{33,20}^{33,1}}{20480}\)

\noindent\(C_{60,0011,1}^{00,0011,1,\text{Loc}}= \frac{\sqrt{3} C_{00,20}^{60,1}}{8192}-\frac{\sqrt{3} C_{11,20}^{31,1}}{8192}+\frac{\sqrt{3}
C_{20,20}^{40,1}}{8192}+\frac{\sqrt{\frac{3}{5}} C_{22,20}^{42,1}}{4096}-\frac{\sqrt{3} C_{31,20}^{31,1}}{8192}-\frac{9 C_{33,20}^{33,1}}{20480 \sqrt{7}}\)

\noindent\(C_{60,0011,1}^{00,0011,1,\text{NonLoc}}= \frac{3 C_{00,00}^{60,0}}{16384}+\frac{\sqrt{3} C_{00,20}^{60,0}}{16384}+\frac{3
C_{11,00}^{51,0}}{16384}+\frac{\sqrt{3} C_{11,20}^{31,0}}{16384}+\frac{3 C_{20,00}^{40,0}}{16384}+\frac{\sqrt{3} C_{20,20}^{40,0}}{16384}+\frac{3
C_{22,00}^{42,0}}{8192 \sqrt{5}}+\frac{\sqrt{\frac{3}{5}} C_{22,20}^{42,0}}{8192}+\frac{3 C_{31,00}^{31,0}}{16384}+\frac{\sqrt{3} C_{31,20}^{31,0}}{16384}+\frac{9
\sqrt{\frac{3}{7}} C_{33,00}^{33,0}}{40960}+\frac{9 C_{33,20}^{33,0}}{40960 \sqrt{7}}\)

\noindent\(C_{62,0011,1}^{00,0011,0,\text{Loc}}= \frac{C_{00,22}^{62,0}}{4096}+\frac{C_{00,22}^{62,1}}{8192}-\frac{C_{11,22}^{51,0}}{4096}-\frac{C_{11,22}^{51,1}}{8192}-\frac{3
\sqrt{\frac{3}{35}} C_{11,22}^{53,0}}{4096}-\frac{3 \sqrt{\frac{3}{35}} C_{11,22}^{53,1}}{8192}+\frac{C_{20,22}^{42,0}}{4096}+\frac{C_{20,22}^{42,1}}{8192}+\frac{C_{22,22}^{40,0}}{4096}+\frac{C_{22,22}^{40,1}}{8192}+\frac{C_{22,22}^{42,0}}{2048
\sqrt{7}}+\frac{C_{22,22}^{42,1}}{4096 \sqrt{7}}+\frac{3 C_{22,22}^{44,0}}{7168 \sqrt{5}}+\frac{3 C_{22,22}^{44,1}}{14336 \sqrt{5}}-\frac{C_{31,22}^{31,0}}{4096}-\frac{C_{31,22}^{31,1}}{8192}-\frac{3
\sqrt{\frac{3}{35}} C_{31,22}^{33,0}}{4096}-\frac{3 \sqrt{\frac{3}{35}} C_{31,22}^{33,1}}{8192}-\frac{3 C_{33,22}^{33,0}}{5120 \sqrt{14}}-\frac{3
C_{33,22}^{33,1}}{10240 \sqrt{14}}\)

\noindent\(C_{62,0011,1}^{00,0011,0,\text{NonLoc}}= -\frac{C_{00,22}^{62,0}}{8192}-\frac{C_{00,22}^{62,1}}{4096}-\frac{C_{11,22}^{51,0}}{8192}-\frac{C_{11,22}^{51,1}}{4096}-\frac{3
\sqrt{\frac{3}{35}} C_{11,22}^{53,0}}{8192}-\frac{3 \sqrt{\frac{3}{35}} C_{11,22}^{53,1}}{4096}-\frac{C_{20,22}^{42,0}}{8192}-\frac{C_{20,22}^{42,1}}{4096}-\frac{C_{22,22}^{40,0}}{8192}-\frac{C_{22,22}^{40,1}}{4096}-\frac{C_{22,22}^{42,0}}{4096
\sqrt{7}}-\frac{C_{22,22}^{42,1}}{2048 \sqrt{7}}-\frac{3 C_{22,22}^{44,0}}{14336 \sqrt{5}}-\frac{3 C_{22,22}^{44,1}}{7168 \sqrt{5}}-\frac{C_{31,22}^{31,0}}{8192}-\frac{C_{31,22}^{31,1}}{4096}-\frac{3
\sqrt{\frac{3}{35}} C_{31,22}^{33,0}}{8192}-\frac{3 \sqrt{\frac{3}{35}} C_{31,22}^{33,1}}{4096}-\frac{3 C_{33,22}^{33,0}}{10240 \sqrt{14}}-\frac{3
C_{33,22}^{33,1}}{5120 \sqrt{14}}\)

\noindent\(C_{62,0011,1}^{00,0011,1,\text{Loc}}= \frac{\sqrt{3} C_{00,22}^{62,1}}{8192}-\frac{\sqrt{3} C_{11,22}^{51,1}}{8192}-\frac{9
C_{11,22}^{53,1}}{8192 \sqrt{35}}+\frac{\sqrt{3} C_{20,22}^{42,1}}{8192}+\frac{\sqrt{3} C_{22,22}^{40,1}}{8192}+\frac{\sqrt{\frac{3}{7}} C_{22,22}^{42,1}}{4096}+\frac{3
\sqrt{\frac{3}{5}} C_{22,22}^{44,1}}{14336}-\frac{\sqrt{3} C_{31,22}^{31,1}}{8192}-\frac{9 C_{31,22}^{33,1}}{8192 \sqrt{35}}-\frac{3 \sqrt{\frac{3}{14}}
C_{33,22}^{33,1}}{10240}\)

\noindent\(C_{62,0011,1}^{00,0011,1,\text{NonLoc}}= -\frac{\sqrt{3} C_{00,22}^{62,0}}{8192}-\frac{\sqrt{3} C_{11,22}^{51,0}}{8192}-\frac{9
C_{11,22}^{53,0}}{8192 \sqrt{35}}-\frac{\sqrt{3} C_{20,22}^{42,0}}{8192}-\frac{\sqrt{3} C_{22,22}^{40,0}}{8192}-\frac{\sqrt{\frac{3}{7}} C_{22,22}^{42,0}}{4096}-\frac{3
\sqrt{\frac{3}{5}} C_{22,22}^{44,0}}{14336}-\frac{\sqrt{3} C_{31,22}^{31,0}}{8192}-\frac{9 C_{31,22}^{33,0}}{8192 \sqrt{35}}-\frac{3 \sqrt{\frac{3}{14}}
C_{33,22}^{33,0}}{10240}\)

\vspace{1.0 cm}
\noindent\ \textbf{76 vanishing sixth-order coupling constants of the nonlocal EDF.}
\vspace{1.0 cm}

C_{11,3110,1}^{11,1111,0,\text{Loc}}

C_{11,3110,1}^{11,1111,0,\text{NonLoc}}

C_{11,3110,1}^{11,1111,1,\text{Loc}}

C_{11,3110,1}^{11,1111,1,\text{NonLoc}}

C_{11,3111,1}^{11,1110,0,\text{Loc}}

C_{11,3111,1}^{11,1110,0,\text{NonLoc}}

C_{11,3111,1}^{11,1110,1,\text{Loc}}

C_{11,3111,1}^{11,1110,1,\text{NonLoc}}

C_{22,0000,2}^{11,3112,0,\text{Loc}}

C_{22,0000,2}^{11,3112,0,\text{NonLoc}}

C_{22,0000,2}^{11,3112,1,\text{Loc}}

C_{22,0000,2}^{11,3112,1,\text{NonLoc}}

C_{22,0000,2}^{11,3312,0,\text{Loc}}

C_{22,0000,2}^{11,3312,0,\text{NonLoc}}

C_{22,0000,2}^{11,3312,1,\text{Loc}}

C_{22,0000,2}^{11,3312,1,\text{NonLoc}}

C_{22,1110,2}^{11,2202,0,\text{Loc}}

C_{22,1110,2}^{11,2202,0,\text{NonLoc}}

C_{22,1110,2}^{11,2202,1,\text{Loc}}

C_{22,1110,2}^{11,2202,1,\text{NonLoc}}

C_{22,1111,2}^{22,1110,0,\text{Loc}}

C_{22,1111,2}^{22,1110,0,\text{NonLoc}}

C_{22,1111,2}^{22,1110,1,\text{Loc}}

C_{22,1111,2}^{22,1110,1,\text{NonLoc}}

C_{22,1112,1}^{11,2000,0,\text{Loc}}

C_{22,1112,1}^{11,2000,0,\text{NonLoc}}

C_{22,1112,1}^{11,2000,1,\text{Loc}}

C_{22,1112,1}^{11,2000,1,\text{NonLoc}}

C_{22,2000,2}^{11,1112,0,\text{Loc}}

C_{22,2000,2}^{11,1112,0,\text{NonLoc}}

C_{22,2000,2}^{11,1112,1,\text{Loc}}

C_{22,2000,2}^{11,1112,1,\text{NonLoc}}

C_{22,2202,1}^{11,1110,0,\text{Loc}}

C_{22,2202,1}^{11,1110,0,\text{NonLoc}}

C_{22,2202,1}^{11,1110,1,\text{Loc}}

C_{22,2202,1}^{11,1110,1,\text{NonLoc}}

C_{22,3112,1}^{11,0000,0,\text{Loc}}

C_{22,3112,1}^{11,0000,0,\text{NonLoc}}

C_{22,3112,1}^{11,0000,1,\text{Loc}}

C_{22,3112,1}^{11,0000,1,\text{NonLoc}}

C_{22,3312,1}^{11,0000,0,\text{Loc}}

C_{22,3312,1}^{11,0000,0,\text{NonLoc}}

C_{22,3312,1}^{11,0000,1,\text{Loc}}

C_{22,3312,1}^{11,0000,1,\text{NonLoc}}

C_{31,0000,1}^{22,1112,0,\text{Loc}}

C_{31,0000,1}^{22,1112,0,\text{NonLoc}}

C_{31,0000,1}^{22,1112,1,\text{Loc}}

C_{31,0000,1}^{22,1112,1,\text{NonLoc}}

C_{31,1110,1}^{11,1111,0,\text{Loc}}

C_{31,1110,1}^{11,1111,0,\text{NonLoc}}

C_{31,1110,1}^{11,1111,1,\text{Loc}}

C_{31,1110,1}^{11,1111,1,\text{NonLoc}}

C_{31,1111,1}^{11,1110,0,\text{Loc}}

C_{31,1111,1}^{11,1110,0,\text{NonLoc}}

C_{31,1111,1}^{11,1110,1,\text{Loc}}

C_{31,1111,1}^{11,1110,1,\text{NonLoc}}

C_{31,1112,2}^{22,0000,0,\text{Loc}}

C_{31,1112,2}^{22,0000,0,\text{NonLoc}}

C_{31,1112,2}^{22,0000,1,\text{Loc}}

C_{31,1112,2}^{22,0000,1,\text{NonLoc}}

C_{33,0000,3}^{22,1112,0,\text{Loc}}

C_{33,0000,3}^{22,1112,0,\text{NonLoc}}

C_{33,0000,3}^{22,1112,1,\text{Loc}}

C_{33,0000,3}^{22,1112,1,\text{NonLoc}}

C_{33,1112,2}^{22,0000,0,\text{Loc}}

C_{33,1112,2}^{22,0000,0,\text{NonLoc}}

C_{33,1112,2}^{22,0000,1,\text{Loc}}

C_{33,1112,2}^{22,0000,1,\text{NonLoc}}

C_{42,0000,2}^{11,1112,0,\text{Loc}}

C_{42,0000,2}^{11,1112,0,\text{NonLoc}}

C_{42,0000,2}^{11,1112,1,\text{Loc}}

C_{42,0000,2}^{11,1112,1,\text{NonLoc}}

C_{42,1112,1}^{11,0000,0,\text{Loc}}

C_{42,1112,1}^{11,0000,0,\text{NonLoc}}

C_{42,1112,1}^{11,0000,1,\text{Loc}}

C_{42,1112,1}^{11,0000,1,\text{NonLoc}}


\noindent\ \textbf{72 second-order isoscalar and isovector coupling constants of the nonlocal EDF expressed by 14 second-order finite-range pseudopotential parameters.}

\hspace{1.0cm}

\noindent\(C_{00,1101,1}^{00,1101,0,\text{Loc}}= -\frac{C_{00,00}^{20,0}}{4}-\frac{C_{00,00}^{20,1}}{8}-\frac{C_{11,00}^{11,0}}{4}-\frac{C_{11,00}^{11,1}}{8}\)

\noindent\(C_{00,1101,1}^{00,1101,0,\text{NonLoc}}= \frac{C_{00,00}^{20,0}}{16}+\frac{C_{00,00}^{20,1}}{8}-\frac{\sqrt{3} C_{00,20}^{20,0}}{16}-\frac{\sqrt{3}
C_{00,20}^{20,1}}{8}-\frac{C_{11,00}^{11,0}}{16}-\frac{C_{11,00}^{11,1}}{8}+\frac{\sqrt{3} C_{11,20}^{11,0}}{16}+\frac{\sqrt{3} C_{11,20}^{11,1}}{8}\)

\noindent\(C_{00,1101,1}^{00,1101,1,\text{Loc}}= -\frac{\sqrt{3} C_{00,00}^{20,1}}{8}-\frac{\sqrt{3} C_{11,00}^{11,1}}{8}\)

\noindent\(C_{00,1101,1}^{00,1101,1,\text{NonLoc}}= \frac{\sqrt{3} C_{00,00}^{20,0}}{16}-\frac{3 C_{00,20}^{20,0}}{16}-\frac{\sqrt{3}
C_{11,00}^{11,0}}{16}+\frac{3 C_{11,20}^{11,0}}{16}\)

\noindent\(C_{00,1110,0}^{00,1110,0,\text{Loc}}= \frac{C_{00,20}^{20,0}}{12}+\frac{C_{00,20}^{20,1}}{24}+\frac{\sqrt{5} C_{00,22}^{22,0}}{12}+\frac{\sqrt{5}
C_{00,22}^{22,1}}{24}+\frac{C_{11,20}^{11,0}}{12}+\frac{C_{11,20}^{11,1}}{24}+\frac{\sqrt{5} C_{11,22}^{11,0}}{12}+\frac{\sqrt{5} C_{11,22}^{11,1}}{24}\)

\noindent\(C_{00,1110,0}^{00,1110,0,\text{NonLoc}}= \frac{C_{00,00}^{20,0}}{16 \sqrt{3}}+\frac{C_{00,00}^{20,1}}{8 \sqrt{3}}+\frac{C_{00,20}^{20,0}}{48}+\frac{C_{00,20}^{20,1}}{24}-\frac{\sqrt{5}
C_{00,22}^{22,0}}{24}-\frac{\sqrt{5} C_{00,22}^{22,1}}{12}-\frac{C_{11,00}^{11,0}}{16 \sqrt{3}}-\frac{C_{11,00}^{11,1}}{8 \sqrt{3}}-\frac{C_{11,20}^{11,0}}{48}-\frac{C_{11,20}^{11,1}}{24}+\frac{\sqrt{5}
C_{11,22}^{11,0}}{24}+\frac{\sqrt{5} C_{11,22}^{11,1}}{12}\)

\noindent\(C_{00,1110,0}^{00,1110,1,\text{Loc}}= \frac{C_{00,20}^{20,1}}{8 \sqrt{3}}+\frac{1}{8} \sqrt{\frac{5}{3}} C_{00,22}^{22,1}+\frac{C_{11,20}^{11,1}}{8
\sqrt{3}}+\frac{1}{8} \sqrt{\frac{5}{3}} C_{11,22}^{11,1}\)

\noindent\(C_{00,1110,0}^{00,1110,1,\text{NonLoc}}= \frac{C_{00,00}^{20,0}}{16}+\frac{C_{00,20}^{20,0}}{16 \sqrt{3}}-\frac{1}{8} \sqrt{\frac{5}{3}}
C_{00,22}^{22,0}-\frac{C_{11,00}^{11,0}}{16}-\frac{C_{11,20}^{11,0}}{16 \sqrt{3}}+\frac{1}{8} \sqrt{\frac{5}{3}} C_{11,22}^{11,0}\)

\noindent\(C_{00,1111,1}^{00,1111,0,\text{Loc}}= \frac{C_{00,20}^{20,0}}{4 \sqrt{3}}+\frac{C_{00,20}^{20,1}}{8 \sqrt{3}}-\frac{1}{8}
\sqrt{\frac{5}{3}} C_{00,22}^{22,0}-\frac{1}{16} \sqrt{\frac{5}{3}} C_{00,22}^{22,1}+\frac{C_{11,20}^{11,0}}{4 \sqrt{3}}+\frac{C_{11,20}^{11,1}}{8
\sqrt{3}}-\frac{1}{8} \sqrt{\frac{5}{3}} C_{11,22}^{11,0}-\frac{1}{16} \sqrt{\frac{5}{3}} C_{11,22}^{11,1}\)

\noindent\(C_{00,1111,1}^{00,1111,0,\text{NonLoc}}= \frac{C_{00,00}^{20,0}}{16}+\frac{C_{00,00}^{20,1}}{8}+\frac{C_{00,20}^{20,0}}{16
\sqrt{3}}+\frac{C_{00,20}^{20,1}}{8 \sqrt{3}}+\frac{1}{16} \sqrt{\frac{5}{3}} C_{00,22}^{22,0}+\frac{1}{8} \sqrt{\frac{5}{3}} C_{00,22}^{22,1}-\frac{C_{11,00}^{11,0}}{16}-\frac{C_{11,00}^{11,1}}{8}-\frac{C_{11,20}^{11,0}}{16
\sqrt{3}}-\frac{C_{11,20}^{11,1}}{8 \sqrt{3}}-\frac{1}{16} \sqrt{\frac{5}{3}} C_{11,22}^{11,0}-\frac{1}{8} \sqrt{\frac{5}{3}} C_{11,22}^{11,1}\)

\noindent\(C_{00,1111,1}^{00,1111,1,\text{Loc}}= \frac{C_{00,20}^{20,1}}{8}-\frac{\sqrt{5} C_{00,22}^{22,1}}{16}+\frac{C_{11,20}^{11,1}}{8}-\frac{\sqrt{5}
C_{11,22}^{11,1}}{16}\)

\noindent\(C_{00,1111,1}^{00,1111,1,\text{NonLoc}}= \frac{\sqrt{3} C_{00,00}^{20,0}}{16}+\frac{C_{00,20}^{20,0}}{16}+\frac{\sqrt{5}
C_{00,22}^{22,0}}{16}-\frac{\sqrt{3} C_{11,00}^{11,0}}{16}-\frac{C_{11,20}^{11,0}}{16}-\frac{\sqrt{5} C_{11,22}^{11,0}}{16}\)

\noindent\(C_{00,1112,2}^{00,1112,0,\text{Loc}}= \frac{\sqrt{5} C_{00,20}^{20,0}}{12}+\frac{\sqrt{5} C_{00,20}^{20,1}}{24}+\frac{C_{00,22}^{22,0}}{24}+\frac{C_{00,22}^{22,1}}{48}+\frac{\sqrt{5}
C_{11,20}^{11,0}}{12}+\frac{\sqrt{5} C_{11,20}^{11,1}}{24}+\frac{C_{11,22}^{11,0}}{24}+\frac{C_{11,22}^{11,1}}{48}\)

\noindent\(C_{00,1112,2}^{00,1112,0,\text{NonLoc}}= \frac{1}{16} \sqrt{\frac{5}{3}} C_{00,00}^{20,0}+\frac{1}{8} \sqrt{\frac{5}{3}}
C_{00,00}^{20,1}+\frac{\sqrt{5} C_{00,20}^{20,0}}{48}+\frac{\sqrt{5} C_{00,20}^{20,1}}{24}-\frac{C_{00,22}^{22,0}}{48}-\frac{C_{00,22}^{22,1}}{24}-\frac{1}{16}
\sqrt{\frac{5}{3}} C_{11,00}^{11,0}-\frac{1}{8} \sqrt{\frac{5}{3}} C_{11,00}^{11,1}-\frac{\sqrt{5} C_{11,20}^{11,0}}{48}-\frac{\sqrt{5} C_{11,20}^{11,1}}{24}+\frac{C_{11,22}^{11,0}}{48}+\frac{C_{11,22}^{11,1}}{24}\)

\noindent\(C_{00,1112,2}^{00,1112,1,\text{Loc}}= \frac{1}{8} \sqrt{\frac{5}{3}} C_{00,20}^{20,1}+\frac{C_{00,22}^{22,1}}{16 \sqrt{3}}+\frac{1}{8}
\sqrt{\frac{5}{3}} C_{11,20}^{11,1}+\frac{C_{11,22}^{11,1}}{16 \sqrt{3}}\)

\noindent\(C_{00,1112,2}^{00,1112,1,\text{NonLoc}}= \frac{\sqrt{5} C_{00,00}^{20,0}}{16}+\frac{1}{16} \sqrt{\frac{5}{3}} C_{00,20}^{20,0}-\frac{C_{00,22}^{22,0}}{16
\sqrt{3}}-\frac{\sqrt{5} C_{11,00}^{11,0}}{16}-\frac{1}{16} \sqrt{\frac{5}{3}} C_{11,20}^{11,0}+\frac{C_{11,22}^{11,0}}{16 \sqrt{3}}\)

\noindent\(C_{00,2000,0}^{00,0000,0,\text{Loc}}= \frac{C_{00,00}^{20,0}}{4}+\frac{C_{00,00}^{20,1}}{8}+\frac{C_{11,00}^{11,0}}{4}+\frac{C_{11,00}^{11,1}}{8}\)

\noindent\(C_{00,2000,0}^{00,0000,0,\text{NonLoc}}= -\frac{C_{00,00}^{20,0}}{16}-\frac{C_{00,00}^{20,1}}{8}+\frac{\sqrt{3} C_{00,20}^{20,0}}{16}+\frac{\sqrt{3}
C_{00,20}^{20,1}}{8}+\frac{C_{11,00}^{11,0}}{16}+\frac{C_{11,00}^{11,1}}{8}-\frac{\sqrt{3} C_{11,20}^{11,0}}{16}-\frac{\sqrt{3} C_{11,20}^{11,1}}{8}\)

\noindent\(C_{00,2000,0}^{00,0000,1,\text{Loc}}= \frac{\sqrt{3} C_{00,00}^{20,1}}{8}+\frac{\sqrt{3} C_{11,00}^{11,1}}{8}\)

\noindent\(C_{00,2000,0}^{00,0000,1,\text{NonLoc}}= -\frac{\sqrt{3} C_{00,00}^{20,0}}{16}+\frac{3 C_{00,20}^{20,0}}{16}+\frac{\sqrt{3}
C_{11,00}^{11,0}}{16}-\frac{3 C_{11,20}^{11,0}}{16}\)

\noindent\(C_{00,2011,1}^{00,0011,0,\text{Loc}}= -\frac{C_{00,20}^{20,0}}{4}-\frac{C_{00,20}^{20,1}}{8}-\frac{C_{11,20}^{11,0}}{4}-\frac{C_{11,20}^{11,1}}{8}\)

\noindent\(C_{00,2011,1}^{00,0011,0,\text{NonLoc}}= -\frac{\sqrt{3} C_{00,00}^{20,0}}{16}-\frac{\sqrt{3} C_{00,00}^{20,1}}{8}-\frac{C_{00,20}^{20,0}}{16}-\frac{C_{00,20}^{20,1}}{8}+\frac{\sqrt{3}
C_{11,00}^{11,0}}{16}+\frac{\sqrt{3} C_{11,00}^{11,1}}{8}+\frac{C_{11,20}^{11,0}}{16}+\frac{C_{11,20}^{11,1}}{8}\)

\noindent\(C_{00,2011,1}^{00,0011,1,\text{Loc}}= -\frac{\sqrt{3} C_{00,20}^{20,1}}{8}-\frac{\sqrt{3} C_{11,20}^{11,1}}{8}\)

\noindent\(C_{00,2011,1}^{00,0011,1,\text{NonLoc}}= -\frac{3 C_{00,00}^{20,0}}{16}-\frac{\sqrt{3} C_{00,20}^{20,0}}{16}+\frac{3 C_{11,00}^{11,0}}{16}+\frac{\sqrt{3}
C_{11,20}^{11,0}}{16}\)

\noindent\(C_{00,2211,1}^{00,0011,0,\text{Loc}}= -\frac{C_{00,22}^{22,0}}{4}-\frac{C_{00,22}^{22,1}}{8}-\frac{C_{11,22}^{11,0}}{4}-\frac{C_{11,22}^{11,1}}{8}\)

\noindent\(C_{00,2211,1}^{00,0011,0,\text{NonLoc}}= \frac{C_{00,22}^{22,0}}{8}+\frac{C_{00,22}^{22,1}}{4}-\frac{C_{11,22}^{11,0}}{8}-\frac{C_{11,22}^{11,1}}{4}\)

\noindent\(C_{00,2211,1}^{00,0011,1,\text{Loc}}= -\frac{\sqrt{3} C_{00,22}^{22,1}}{8}-\frac{\sqrt{3} C_{11,22}^{11,1}}{8}\)

\noindent\(C_{00,2211,1}^{00,0011,1,\text{NonLoc}}= \frac{\sqrt{3} C_{00,22}^{22,0}}{8}-\frac{\sqrt{3} C_{11,22}^{11,0}}{8}\)

\noindent\(C_{11,0000,1}^{00,1111,0,\text{Loc}}= \frac{C_{11,11}^{11,0}}{4}+\frac{C_{11,11}^{11,1}}{8}\)

\noindent\(C_{11,0000,1}^{00,1111,0,\text{NonLoc}}= \frac{C_{11,11}^{11,0}}{8}+\frac{C_{11,11}^{11,1}}{4}\)

\noindent\(C_{11,0000,1}^{00,1111,1,\text{Loc}}= \frac{\sqrt{3} C_{11,11}^{11,1}}{8}\)

\noindent\(C_{11,0000,1}^{00,1111,1,\text{NonLoc}}= \frac{\sqrt{3} C_{11,11}^{11,0}}{8}\)

\noindent\(C_{11,0000,1}^{11,0000,0,\text{Loc}}= \frac{C_{00,00}^{20,0}}{16}+\frac{C_{00,00}^{20,1}}{32}-\frac{C_{11,00}^{11,0}}{16}-\frac{C_{11,00}^{11,1}}{32}\)

\noindent\(C_{11,0000,1}^{11,0000,0,\text{NonLoc}}= -\frac{C_{00,00}^{20,0}}{64}-\frac{C_{00,00}^{20,1}}{32}+\frac{\sqrt{3} C_{00,20}^{20,0}}{64}+\frac{\sqrt{3}
C_{00,20}^{20,1}}{32}-\frac{C_{11,00}^{11,0}}{64}-\frac{C_{11,00}^{11,1}}{32}+\frac{\sqrt{3} C_{11,20}^{11,0}}{64}+\frac{\sqrt{3} C_{11,20}^{11,1}}{32}\)

\noindent\(C_{11,0000,1}^{11,0000,1,\text{Loc}}= \frac{\sqrt{3} C_{00,00}^{20,1}}{32}-\frac{\sqrt{3} C_{11,00}^{11,1}}{32}\)

\noindent\(C_{11,0000,1}^{11,0000,1,\text{NonLoc}}= -\frac{\sqrt{3} C_{00,00}^{20,0}}{64}+\frac{3 C_{00,20}^{20,0}}{64}-\frac{\sqrt{3}
C_{11,00}^{11,0}}{64}+\frac{3 C_{11,20}^{11,0}}{64}\)

\noindent\(C_{11,0011,0}^{11,0011,0,\text{Loc}}= -\frac{C_{00,20}^{20,0}}{48}-\frac{C_{00,20}^{20,1}}{96}-\frac{\sqrt{5} C_{00,22}^{22,0}}{48}-\frac{\sqrt{5}
C_{00,22}^{22,1}}{96}+\frac{C_{11,20}^{11,0}}{48}+\frac{C_{11,20}^{11,1}}{96}+\frac{\sqrt{5} C_{11,22}^{11,0}}{48}+\frac{\sqrt{5} C_{11,22}^{11,1}}{96}\)

\noindent\(C_{11,0011,0}^{11,0011,0,\text{NonLoc}}= -\frac{C_{00,00}^{20,0}}{64 \sqrt{3}}-\frac{C_{00,00}^{20,1}}{32 \sqrt{3}}-\frac{C_{00,20}^{20,0}}{192}-\frac{C_{00,20}^{20,1}}{96}+\frac{\sqrt{5}
C_{00,22}^{22,0}}{96}+\frac{\sqrt{5} C_{00,22}^{22,1}}{48}-\frac{C_{11,00}^{11,0}}{64 \sqrt{3}}-\frac{C_{11,00}^{11,1}}{32 \sqrt{3}}-\frac{C_{11,20}^{11,0}}{192}-\frac{C_{11,20}^{11,1}}{96}+\frac{\sqrt{5}
C_{11,22}^{11,0}}{96}+\frac{\sqrt{5} C_{11,22}^{11,1}}{48}\)

\noindent\(C_{11,0011,0}^{11,0011,1,\text{Loc}}= -\frac{C_{00,20}^{20,1}}{32 \sqrt{3}}-\frac{1}{32} \sqrt{\frac{5}{3}} C_{00,22}^{22,1}+\frac{C_{11,20}^{11,1}}{32
\sqrt{3}}+\frac{1}{32} \sqrt{\frac{5}{3}} C_{11,22}^{11,1}\)

\noindent\(C_{11,0011,0}^{11,0011,1,\text{NonLoc}}= -\frac{C_{00,00}^{20,0}}{64}-\frac{C_{00,20}^{20,0}}{64 \sqrt{3}}+\frac{1}{32}
\sqrt{\frac{5}{3}} C_{00,22}^{22,0}-\frac{C_{11,00}^{11,0}}{64}-\frac{C_{11,20}^{11,0}}{64 \sqrt{3}}+\frac{1}{32} \sqrt{\frac{5}{3}} C_{11,22}^{11,0}\)

\noindent\(C_{11,0011,1}^{00,1101,0,\text{Loc}}= -\frac{C_{11,11}^{11,0}}{4}-\frac{C_{11,11}^{11,1}}{8}\)

\noindent\(C_{11,0011,1}^{00,1101,0,\text{NonLoc}}= -\frac{C_{11,11}^{11,0}}{8}-\frac{C_{11,11}^{11,1}}{4}\)

\noindent\(C_{11,0011,1}^{00,1101,1,\text{Loc}}= -\frac{\sqrt{3} C_{11,11}^{11,1}}{8}\)

\noindent\(C_{11,0011,1}^{00,1101,1,\text{NonLoc}}= -\frac{\sqrt{3} C_{11,11}^{11,0}}{8}\)

\noindent\(C_{11,0011,1}^{11,0011,0,\text{Loc}}= -\frac{C_{00,20}^{20,0}}{16 \sqrt{3}}-\frac{C_{00,20}^{20,1}}{32 \sqrt{3}}+\frac{1}{32}
\sqrt{\frac{5}{3}} C_{00,22}^{22,0}+\frac{1}{64} \sqrt{\frac{5}{3}} C_{00,22}^{22,1}+\frac{C_{11,20}^{11,0}}{16 \sqrt{3}}+\frac{C_{11,20}^{11,1}}{32
\sqrt{3}}-\frac{1}{32} \sqrt{\frac{5}{3}} C_{11,22}^{11,0}-\frac{1}{64} \sqrt{\frac{5}{3}} C_{11,22}^{11,1}\)

\noindent\(C_{11,0011,1}^{11,0011,0,\text{NonLoc}}= -\frac{C_{00,00}^{20,0}}{64}-\frac{C_{00,00}^{20,1}}{32}-\frac{C_{00,20}^{20,0}}{64
\sqrt{3}}-\frac{C_{00,20}^{20,1}}{32 \sqrt{3}}-\frac{1}{64} \sqrt{\frac{5}{3}} C_{00,22}^{22,0}-\frac{1}{32} \sqrt{\frac{5}{3}} C_{00,22}^{22,1}-\frac{C_{11,00}^{11,0}}{64}-\frac{C_{11,00}^{11,1}}{32}-\frac{C_{11,20}^{11,0}}{64
\sqrt{3}}-\frac{C_{11,20}^{11,1}}{32 \sqrt{3}}-\frac{1}{64} \sqrt{\frac{5}{3}} C_{11,22}^{11,0}-\frac{1}{32} \sqrt{\frac{5}{3}} C_{11,22}^{11,1}\)

\noindent\(C_{11,0011,1}^{11,0011,1,\text{Loc}}= -\frac{C_{00,20}^{20,1}}{32}+\frac{\sqrt{5} C_{00,22}^{22,1}}{64}+\frac{C_{11,20}^{11,1}}{32}-\frac{\sqrt{5}
C_{11,22}^{11,1}}{64}\)

\noindent\(C_{11,0011,1}^{11,0011,1,\text{NonLoc}}= -\frac{\sqrt{3} C_{00,00}^{20,0}}{64}-\frac{C_{00,20}^{20,0}}{64}-\frac{\sqrt{5}
C_{00,22}^{22,0}}{64}-\frac{\sqrt{3} C_{11,00}^{11,0}}{64}-\frac{C_{11,20}^{11,0}}{64}-\frac{\sqrt{5} C_{11,22}^{11,0}}{64}\)

\noindent\(C_{11,0011,2}^{11,0011,0,\text{Loc}}= -\frac{\sqrt{5} C_{00,20}^{20,0}}{48}-\frac{\sqrt{5} C_{00,20}^{20,1}}{96}-\frac{C_{00,22}^{22,0}}{96}-\frac{C_{00,22}^{22,1}}{192}+\frac{\sqrt{5}
C_{11,20}^{11,0}}{48}+\frac{\sqrt{5} C_{11,20}^{11,1}}{96}+\frac{C_{11,22}^{11,0}}{96}+\frac{C_{11,22}^{11,1}}{192}\)

\noindent\(C_{11,0011,2}^{11,0011,0,\text{NonLoc}}= -\frac{1}{64} \sqrt{\frac{5}{3}} C_{00,00}^{20,0}-\frac{1}{32} \sqrt{\frac{5}{3}}
C_{00,00}^{20,1}-\frac{\sqrt{5} C_{00,20}^{20,0}}{192}-\frac{\sqrt{5} C_{00,20}^{20,1}}{96}+\frac{C_{00,22}^{22,0}}{192}+\frac{C_{00,22}^{22,1}}{96}-\frac{1}{64}
\sqrt{\frac{5}{3}} C_{11,00}^{11,0}-\frac{1}{32} \sqrt{\frac{5}{3}} C_{11,00}^{11,1}-\frac{\sqrt{5} C_{11,20}^{11,0}}{192}-\frac{\sqrt{5} C_{11,20}^{11,1}}{96}+\frac{C_{11,22}^{11,0}}{192}+\frac{C_{11,22}^{11,1}}{96}\)

\noindent\(C_{11,0011,2}^{11,0011,1,\text{Loc}}= -\frac{1}{32} \sqrt{\frac{5}{3}} C_{00,20}^{20,1}-\frac{C_{00,22}^{22,1}}{64 \sqrt{3}}+\frac{1}{32}
\sqrt{\frac{5}{3}} C_{11,20}^{11,1}+\frac{C_{11,22}^{11,1}}{64 \sqrt{3}}\)

\noindent\(C_{11,0011,2}^{11,0011,1,\text{NonLoc}}= -\frac{\sqrt{5} C_{00,00}^{20,0}}{64}-\frac{1}{64} \sqrt{\frac{5}{3}} C_{00,20}^{20,0}+\frac{C_{00,22}^{22,0}}{64
\sqrt{3}}-\frac{\sqrt{5} C_{11,00}^{11,0}}{64}-\frac{1}{64} \sqrt{\frac{5}{3}} C_{11,20}^{11,0}+\frac{C_{11,22}^{11,0}}{64 \sqrt{3}}\)

\noindent\(C_{11,1101,1}^{00,0011,0,\text{Loc}}= -\frac{C_{11,11}^{11,0}}{4}-\frac{C_{11,11}^{11,1}}{8}\)

\noindent\(C_{11,1101,1}^{00,0011,0,\text{NonLoc}}= -\frac{C_{11,11}^{11,0}}{8}-\frac{C_{11,11}^{11,1}}{4}\)

\noindent\(C_{11,1101,1}^{00,0011,1,\text{Loc}}= -\frac{\sqrt{3} C_{11,11}^{11,1}}{8}\)

\noindent\(C_{11,1101,1}^{00,0011,1,\text{NonLoc}}= -\frac{\sqrt{3} C_{11,11}^{11,0}}{8}\)

\noindent\(C_{11,1111,0}^{00,0000,0,\text{Loc}}= -\frac{C_{11,11}^{11,0}}{4}-\frac{C_{11,11}^{11,1}}{8}\)

\noindent\(C_{11,1111,0}^{00,0000,0,\text{NonLoc}}= -\frac{C_{11,11}^{11,0}}{8}-\frac{C_{11,11}^{11,1}}{4}\)

\noindent\(C_{11,1111,0}^{00,0000,1,\text{Loc}}= -\frac{\sqrt{3} C_{11,11}^{11,1}}{8}\)

\noindent\(C_{11,1111,0}^{00,0000,1,\text{NonLoc}}= -\frac{\sqrt{3} C_{11,11}^{11,0}}{8}\)

\noindent\(C_{20,0000,0}^{00,0000,0,\text{Loc}}= -\frac{C_{00,00}^{20,0}}{16}-\frac{C_{00,00}^{20,1}}{32}+\frac{C_{11,00}^{11,0}}{16}+\frac{C_{11,00}^{11,1}}{32}\)

\noindent\(C_{20,0000,0}^{00,0000,0,\text{NonLoc}}= \frac{C_{00,00}^{20,0}}{64}+\frac{C_{00,00}^{20,1}}{32}-\frac{\sqrt{3} C_{00,20}^{20,0}}{64}-\frac{\sqrt{3}
C_{00,20}^{20,1}}{32}+\frac{C_{11,00}^{11,0}}{64}+\frac{C_{11,00}^{11,1}}{32}-\frac{\sqrt{3} C_{11,20}^{11,0}}{64}-\frac{\sqrt{3} C_{11,20}^{11,1}}{32}\)

\noindent\(C_{20,0000,0}^{00,0000,1,\text{Loc}}= -\frac{\sqrt{3} C_{00,00}^{20,1}}{32}+\frac{\sqrt{3} C_{11,00}^{11,1}}{32}\)

\noindent\(C_{20,0000,0}^{00,0000,1,\text{NonLoc}}= \frac{\sqrt{3} C_{00,00}^{20,0}}{64}-\frac{3 C_{00,20}^{20,0}}{64}+\frac{\sqrt{3}
C_{11,00}^{11,0}}{64}-\frac{3 C_{11,20}^{11,0}}{64}\)

\noindent\(C_{20,0011,1}^{00,0011,0,\text{Loc}}= \frac{C_{00,20}^{20,0}}{16}+\frac{C_{00,20}^{20,1}}{32}-\frac{C_{11,20}^{11,0}}{16}-\frac{C_{11,20}^{11,1}}{32}\)

\noindent\(C_{20,0011,1}^{00,0011,0,\text{NonLoc}}= \frac{\sqrt{3} C_{00,00}^{20,0}}{64}+\frac{\sqrt{3} C_{00,00}^{20,1}}{32}+\frac{C_{00,20}^{20,0}}{64}+\frac{C_{00,20}^{20,1}}{32}+\frac{\sqrt{3}
C_{11,00}^{11,0}}{64}+\frac{\sqrt{3} C_{11,00}^{11,1}}{32}+\frac{C_{11,20}^{11,0}}{64}+\frac{C_{11,20}^{11,1}}{32}\)

\noindent\(C_{20,0011,1}^{00,0011,1,\text{Loc}}= \frac{\sqrt{3} C_{00,20}^{20,1}}{32}-\frac{\sqrt{3} C_{11,20}^{11,1}}{32}\)

\noindent\(C_{20,0011,1}^{00,0011,1,\text{NonLoc}}= \frac{3 C_{00,00}^{20,0}}{64}+\frac{\sqrt{3} C_{00,20}^{20,0}}{64}+\frac{3 C_{11,00}^{11,0}}{64}+\frac{\sqrt{3}
C_{11,20}^{11,0}}{64}\)

\noindent\(C_{22,0011,1}^{00,0011,0,\text{Loc}}= \frac{C_{00,22}^{22,0}}{16}+\frac{C_{00,22}^{22,1}}{32}-\frac{C_{11,22}^{11,0}}{16}-\frac{C_{11,22}^{11,1}}{32}\)

\noindent\(C_{22,0011,1}^{00,0011,0,\text{NonLoc}}= -\frac{C_{00,22}^{22,0}}{32}-\frac{C_{00,22}^{22,1}}{16}-\frac{C_{11,22}^{11,0}}{32}-\frac{C_{11,22}^{11,1}}{16}\)

\noindent\(C_{22,0011,1}^{00,0011,1,\text{Loc}}= \frac{\sqrt{3} C_{00,22}^{22,1}}{32}-\frac{\sqrt{3} C_{11,22}^{11,1}}{32}\)

\noindent\(C_{22,0011,1}^{00,0011,1,\text{NonLoc}}= -\frac{\sqrt{3} C_{00,22}^{22,0}}{32}-\frac{\sqrt{3} C_{11,22}^{11,0}}{32}\)